\documentclass[a4paper,11pt]{article}
\usepackage{jcappub}
\usepackage{amsmath}
\usepackage{graphicx}

\usepackage{xspace}

\usepackage{float}

\usepackage{longtable}

\usepackage{slashed}

\makeatletter
\gdef\@fpheader{}
\makeatother

\newcommand{\ef}[1]{\tilde{#1}}
\newcommand{\jf}[1]{\bar{#1}}

\newcommand{\ie}{i.e.\xspace}
\newcommand{\eg}{e.g.\xspace}

\newlength{\wsingfig}
\newlength{\wdblefig}
\newlength{\wappfig}
\newcommand{\Refc}[1]{Ref.~\cite{#1}}
\newcommand{\Refcs}[1]{Refs.~\cite{#1}}
\newcommand{\Eq}[1]{Eq.~\eqref{#1}}
\newcommand{\Eqs}[1]{Eqs.~\eqref{#1}}
\newcommand{\Fig}[1]{Fig.~\ref{#1}}
\newcommand{\Figs}[1]{Figs.~\ref{#1}}
\newcommand{\sectionc}[1]{section~\ref{#1}}
\newcommand{\sectioncs}[1]{sections~\ref{#1}}

\newcommand{\order}[1]{\mathcal{O}\!\left(#1\right)}
\newcommand{\Vol}[1]{\mathrm{Vol}\!\left(#1\right)}
\newcommand{\Lambert}[1]{\mathrm{W}_{\negthinspace #1}}
\newcommand{\Fgauss}{{}_2F_1}
\newcommand{\Li}[1]{\mathrm{Li}_{#1}}
\DeclareMathOperator{\Ei}{Ei}
\DeclareMathOperator{\sech}{sech}
\DeclareMathOperator{\csch}{csch}
\DeclareMathOperator{\arcosh}{arccosh}
\DeclareMathOperator{\arsinh}{arcsinh}
\DeclareMathOperator{\artanh}{arctanh}
\DeclareMathOperator{\arsech}{arcsech}
\DeclareMathOperator{\arcoth}{arccoth} 
\newcommand{\arccosh}{\arcosh}
\newcommand{\arcsinh}{\arsinh}
\newcommand{\arctanh}{\artanh}
\newcommand{\arccoth}{\arcoth}

\newcommand{\Tr}{\mathrm{Tr}}
\newcommand{\diag}{\mathrm{diag}}

\newcommand{\dd}{\mathrm{d}}
\newcommand{\ee}{e}
\newcommand{\efold}{$\ee$-fold\xspace}
\newcommand{\efolds}{$\ee$-folds\xspace}
\newcommand{\efolding}{$\ee$-folding\xspace}

\newcommand{\sss}[1]{{\scriptscriptstyle{#1}}}
\newcommand{\boldmathsymbol}[1]{{\ensuremath{\boldsymbol{#1}}}}

\newcommand{\Nmodel}{118}

\newcommand{\uPl}{\mathrm{Pl}}
\newcommand{\usr}{\mathrm{sr}}
\newcommand{\uin}{\mathrm{in}}
\newcommand{\utip}{\mathrm{tip}}
\newcommand{\uend}{\mathrm{end}}
\newcommand{\uini}{\mathrm{ini}}
\newcommand{\ueff}{\mathrm{eff}}

\newcommand{\ureh}{\mathrm{reh}}
\newcommand{\urad}{\mathrm{rad}}
\newcommand{\umat}{\mathrm{mat}}
\newcommand{\utot}{\mathrm{tot}}
\newcommand{\uthroat}{\mathrm{throat}}

\newcommand{\ustg}{\mathrm{stg}}
\newcommand{\utop}{\mathrm{top}}
\newcommand{\ucri}{\mathrm{cri}}
\newcommand{\unuc}{\mathrm{nuc}}

\newcommand{\uloc}{\mathrm{loc}}
\newcommand{\uortho}{\mathrm{ortho}}
\newcommand{\ueq}{\mathrm{eq}}
\newcommand{\ue}{\mathrm{e}}
\newcommand{\uf}{\mathrm{f}}
\newcommand{\uc}{\mathrm{c}}

\newcommand{\ub}{\mathrm{b}}
\newcommand{\ut}{\mathrm{t}}
\newcommand{\us}{\mathrm{s}}
\newcommand{\uS}{\mathrm{S}}

\newcommand{\uR}{\mathrm{R}}
\newcommand{\uL}{\mathrm{L}}
\newcommand{\uUV}{\mathrm{UV}}
\newcommand{\uY}{\mathrm{Y}}
\newcommand{\um}{\mathrm{mat}}
\newcommand{\uPSP}{(\uS)}
\newcommand{\uGUM}{\mathrm{GUM}}
\newcommand{\uOI}{\mathrm{OI}}

\newcommand{\uT}{\mathrm{T}}
\newcommand{\uPTP}{(\uT)}
\newcommand{\uE}{\mathrm{E}}
\newcommand{\uH}{\mathrm{H}}
\newcommand{\uBD}{\mathrm{BD}}

\newcommand{\usssS}{\sss{\uS}}
\newcommand{\usssPSP}{\sss{\uPSP}}
\newcommand{\usssT}{\sss{\uT}}
\newcommand{\usssPTP}{\sss{\uPTP}}
\newcommand{\usssUV}{\sss{\uUV}}

\newcommand{\uGEP}{\mathrm{GEP}}
\newcommand{\usssGEP}{\sss{\uGEP}}
\newcommand{\uRIPI}{\mathrm{RIPI}}
\newcommand{\usssRIPI}{\sss{\uRIPI}}
\newcommand{\usssOI}{\sss{\uOI}}
\newcommand{\usssGUM}{\sss{\uGUM}}
\newcommand{\usssBD}{\sss{\uBD}}

\newcommand{\usssPl}{\sss{\uPl}}

\newcommand{\usssE}{\sss{\uE}}
\newcommand{\nS}{n_\usssS}
\newcommand{\nT}{n_\usssT}
\newcommand{\alphaS}{\alpha_\usssS}
\newcommand{\alphaT}{\alpha_\usssT}
\newcommand{\zero}{{_0}}

\newcommand{\udm}{\mathrm{dm}}
\newcommand{\ud}{\mathrm{d}}
\newcommand{\ug}{\mathrm{g}}
\newcommand{\uu}{\mathrm{u}}
\newcommand{\muS}{\mu_\usssS}
\newcommand{\muT}{\mu_\usssT}

\newcommand{\uNL}{\mathrm{NL}}
\newcommand{\umon}{\mathrm{mon}}

\newcommand{\bmk}{\boldmathsymbol{k}}
\newcommand{\bmx}{\boldmathsymbol{x}}

\newcommand{\calH}{\mathcal{H}}
\newcommand{\calP}{\mathcal{P}}

\newcommand{\calO}{\mathcal{O}}
\newcommand{\calI}{\mathcal{I}}
\newcommand{\calN}{\mathcal{N}}
\newcommand{\calL}{\mathcal{L}}
\newcommand{\calV}{\mathcal{V}}
\newcommand{\calM}{\mathcal{M}}
\newcommand{\calK}{\mathcal{K}}
\newcommand{\calF}{\mathcal{F}}
\newcommand{\calA}{\mathcal{A}}
\newcommand{\calS}{\mathcal{S}}
\newcommand{\calW}{\mathcal{W}}
\newcommand{\calD}{\mathcal{D}}

\newcommand{\eV}{\mathrm{eV}}

\newcommand{\MeV}{\mathrm{MeV}}
\newcommand{\GeV}{\mathrm{GeV}}
\newcommand{\TeV}{\mathrm{TeV}}
\newcommand{\cm}{\mathrm{cm}}

\newcommand{\Mpc}{\mathrm{Mpc}}
\newcommand{\K}{\mathrm{K}}

\newcommand{\OmegaL}{\Omega_\Lambda}
\newcommand{\OmegaCDM}{\Omega_\udm}

\newcommand{\OmegaR}{\Omega_\gamma}
\newcommand{\Omegaend}{\Omega_\uend}
\newcommand{\Qrms}{Q_{\mathrm{rms-PS}}}
\newcommand{\Qstar}{Q_*}

\newcommand{\Ms}{M_\us}
\newcommand{\ells}{\ell_\us}
\newcommand{\gstrings}{g_\us}

\newcommand{\rlss}{r_{\mathrm{\ell ss}}}
\newcommand{\omegaBD}{\omega_{\usssBD}}

\newcommand{\Clt}{C_{\ell}^{\mathrm{th}}}
\newcommand{\Clm}{C_{\ell}^{\mathrm{meas}}}
\newcommand{\tetas}{\theta_{\mathrm{stand}}}
\newcommand{\tetai}{\theta_{\mathrm{inf}}}

\newcommand{\mpl}{m_\usssPl}
\newcommand{\Mp}{M_\usssPl}
\newcommand{\Mg}{M_{\ug}}
\newcommand{\vev}{\textit{vev}\xspace}
\newcommand{\vevs}{\textit{vevs}\xspace}
\newcommand{\mGUM}{m_\usssGUM}

\newcommand{\urlaspic}{\url{https://github.com/cosmicinflation/aspic}}

\newcommand{\ASPIC}{\texttt{ASPIC}\xspace}
\newcommand{\EI}{\textit{Encyclop{\ae}dia Inflationaris}\xspace}

\newcommand{\BINGO}{\texttt{BINGO}\xspace}
\newcommand{\FieldInf}{\texttt{FieldInf}\xspace}

\newcommand{\COBE}{COBE\xspace}
\newcommand{\DetailedData}{Planck 2018 + Bicep-Keck\xspace}
\newcommand{\data}{\DetailedData}

\newcommand{\kstar}{k_*}
\newcommand{\etastar}{\eta_*}
\newcommand{\Pstar}{P_*}
\newcommand{\Hstar}{H_*}

\newcommand{\fnl}{f_{_\uNL}}
\newcommand{\fnlloc}{f_\uNL^\uloc}
\newcommand{\fnleq}{f_\uNL^\ueq}
\newcommand{\fnlortho}{f_\uNL^\uortho}
\newcommand{\Sm}{S_\um}
\newcommand{\SBD}{S_\usssBD}
\newcommand{\wreh}{w_\ureh}
\newcommand{\wrehbar}{\overline{w}_{\ureh}}
\newcommand{\rdof}{\mathcal{Q}}
\newcommand{\rdofreh}{\rdof_\ureh}
\newcommand{\gzero}{g_\sss{0}}
\newcommand{\greh}{g_\ureh}
\newcommand{\gs}{q}
\newcommand{\gszero}{\gs_\sss{0}}
\newcommand{\gsreh}{\gs_\ureh}
\newcommand{\gyuk}{g_\uf}
\newcommand{\vbar}{\bar{v}}

\newcommand{\epsilonV}{\epsilon^{V}}
\newcommand{\epsilonH}{\epsilon}
\newcommand{\etaH}{\eta}
\newcommand{\AI}{A_{_{\mathrm{I}}}}
\newcommand{\hI}{h_{_{\mathrm{I}}}}
\newcommand{\xbks}{x_{_{\mathrm{BKS}}}}
\newcommand{\cs}{c_{_\uS}}

\newcommand{\Rrad}{R_\urad}
\newcommand{\Rradbar}{\bar{R}_\urad}

\newcommand{\Treh}{T_\ureh}
\newcommand{\Vend}{V_\uend}

\newcommand{\Vmin}{V_{\min}}
\newcommand{\Vstar}{V_*}
\newcommand{\Vgep}{V_\usssGEP}
\newcommand{\rhoreh}{\rho_\ureh}
\newcommand{\rhoend}{\rho_\uend}
\newcommand{\rhorehbar}{\bar{\rho}_\ureh}
\newcommand{\rhoendbar}{\bar{\rho}_\uend}

\newcommand{\rhozero}{\rho_\zero}
\newcommand{\rhogamma}{\rho_\gamma}
\newcommand{\rhotildegamma}{\tilde{\rho}_\gamma}
\newcommand{\rhotildegammabar}{\bar{\tilde{\rho}}_\gamma}
\newcommand{\rhonuc}{\rho_\unuc}
\newcommand{\rhonucbar}{\bar{\rho}_\unuc}
\newcommand{\Areh}{A_\ureh}
\newcommand{\Aend}{A_\uend}
\newcommand{\Vthroat}{V^{\uthroat}}
\newcommand{\Vtot}{V^{\utot}}

\newcommand{\iUV}{_\usssUV}
\newcommand{\iVtop}{_\utop}
\newcommand{\iVmax}{_{\sss{V}^{\max}}}
\newcommand{\iVmin}{_{\sss{V}^{\min}}}
\newcommand{\iVzero}{_{V=0}}
\newcommand{\ifplus}{_{f^{+}}}
\newcommand{\ifminus}{_{f^{-}}}
\newcommand{\iVinfty}{_{V\rightarrow \infty}}
\newcommand{\iVzeroPlus}{\iVzero^+}
\newcommand{\iVzeroMinus}{\iVzero^-}

\newcommand{\iVzeroPM}{\iVzero^\pm}

\newcommand{\idVzero}{_{V'=0}}
\newcommand{\idVzeroPlus}{\idVzero^+}
\newcommand{\idVzeroMinus}{\idVzero^-}
\newcommand{\idVzeroPM}{\idVzero^\pm}
\newcommand{\iddVzero}{_{V''=0}}
\newcommand{\iddVzeroPlus}{\iddVzero^+}
\newcommand{\iddVzeroMinus}{\iddVzero^-}
\newcommand{\iddVzeroPM}{\iddVzero^\pm}

\newcommand{\ieps}{_{\epsilon}}

\newcommand{\iepsoneOne}{_{\epsilon_1=1}}

\newcommand{\iepsoneOneZero}{\iepsoneOne^\zero}
\newcommand{\iepsoneOnePlus}{\iepsoneOne^+}
\newcommand{\iepsoneOneMinus}{\iepsoneOne^-}
\newcommand{\iepsoneOnePM}{\iepsoneOne^\pm}
\newcommand{\iepsoneOneMP}{\iepsoneOne^\mp}
\newcommand{\iepstwoMinusOne}{_{\epsilon_2=-1}}
\newcommand{\iepstwoZero}{_{\epsilon_2=0}}
\newcommand{\iepstwoMax}{_{\epsilon_2^{\max}}}
\newcommand{\iepstwoMin}{_{\epsilon_2^{\min}}}
\newcommand{\iepstwoZeroPlus}{\iepstwoZero^+}
\newcommand{\iepstwoZeroMinus}{\iepstwoZero^-}
\newcommand{\iepstwoZeroPM}{\iepstwoZero^\pm}

\newcommand{\iepstwoMinusOneMinus}{\iepstwoMinusOne^-}
\newcommand{\iepstwoMinusOnePM}{\iepstwoMinusOne^\pm}
\newcommand{\iepstwoOne}{_{\epsilon_2=1}}
\newcommand{\iepsthreeOne}{_{\epsilon_3=1}}
\newcommand{\iepsthreeMinusOne}{_{\epsilon_3=-1}}
\newcommand{\iepsthreeMinusOnePM}{\iepsthreeMinusOne^\pm}

\newcommand{\iepsoneMax}{_{\epsilon_1^{\max}}}
\newcommand{\iepsoneMin}{_{\epsilon_1^{\min}}}

\newcommand{\epsthreeEnd}{\epsilon_3^\uend}
\newcommand{\epstwoEnd}{\epsilon_2^{\uend}}
\newcommand{\epsonestar}{\epsilon_{1*}}
\newcommand{\epstwostar}{\epsilon_{2*}}
\newcommand{\epsthreestar}{\epsilon_{3*}}
\newcommand{\epsoneend}{\epsilon_{1\uend}}
\newcommand{\epsonemin}{\epsilon_{1}^{\min}}
\newcommand{\epsonemax}{\epsilon_{1}^{\max}}
\newcommand{\epstwomin}{\epsilon_{2}^{\min}}
\newcommand{\epstwotop}{\epsilon_{2}^{\utop}}

\newcommand{\epstwoMin}{\epstwomin}
\newcommand{\epsthreeMin}{\epsilon_{3}^{\min}}

\newcommand{\epstwotopMax}{\epsilon_{2,\utop}^{\max}}
\newcommand{\epsoneminPlus}{\epsilon_{1+}^{\min}}
\newcommand{\epsoneminMinus}{\epsilon_{1-}^{\min}}

\newcommand{\epsonestarcase}[1]{\epsonestar^{(#1)}}
\newcommand{\epstwostarcase}[1]{\epstwostar^{(#1)}}
\newcommand{\epsthreestarcase}[1]{\epsthreestar^{(#1)}}

\newcommand{\mueps}{\mu\ieps}
\newcommand{\FT}{F_\usssT}
\newcommand{\phiend}{\phi_\uend}
\newcommand{\phiini}{\phi_\uini}
\newcommand{\phistar}{\phi_*}

\newcommand{\phizero}{\phi_\zero}
\newcommand{\phimax}{\phi_{\max}}
\newcommand{\phistarnuc}{\phi_*^\unuc}
\newcommand{\phistarend}{\phi_*^\uend}
\newcommand{\phiUV}{\phi\iUV}
\newcommand{\phistg}{\phi_\ustg}

\newcommand{\phiVtop}{\phi\iVtop}
\newcommand{\phiVzero}{\phi\iVzero}
\newcommand{\phiVmax}{\phi\iVmax}
\newcommand{\phiddVzero}{\phi\iddVzero}
\newcommand{\phiddVzeroMinus}{\phi\iddVzeroMinus}
\newcommand{\phiVmin}{\phi\iVmin}

\newcommand{\phiepstwoOne}{\phi\iepstwoOne}

\newcommand{\Tzero}{T_\zero}
\newcommand{\Tbar}{\bar{T}}
\newcommand{\betasr}{\beta_{\usr}}
\newcommand{\betam}{\beta_{\um}}
\newcommand{\mloop}{m_{\mathrm{loop}}}

\newcommand{\PsiR}{\Psi_\uR}
\newcommand{\PsiL}{\Psi_\uL}
\newcommand{\PsiLR}{\Psi_{\uL\uR}}

\newcommand{\rzero}{r_\zero}
\newcommand{\rtip}{r_{\utip}}
\newcommand{\rUV}{r\iUV}
\newcommand{\rstg}{r_\ustg}

\newcommand{\barh}{\bar{h}}
\newcommand{\barv}{\bar{v}}
\newcommand{\barhend}{\barh_\uend}
\newcommand{\barhstar}{\barh_*}
\newcommand{\barhVmax}{\barh\iVmax}

\newcommand{\chiVmax}{\chi\iVmax}
\newcommand{\chiend}{\chi_\uend}

\newcommand{\xistar}{\xi_*}
\newcommand{\xizero}{\xi_{\zero}}

\newcommand{\xend}{x_\uend}
\newcommand{\xendhat}{\hat{x}_\uend}
\newcommand{\xendcase}[1]{\xend^{(#1)}}
\newcommand{\xini}{x_\uini}
\newcommand{\xzero}{x_\zero}
\newcommand{\xstar}{x_*}
\newcommand{\xUV}{x\iUV}
\newcommand{\xstg}{x_\ustg}
\newcommand{\xstarstg}{x_*^\ustg}
\newcommand{\xVtop}{x\iVtop}
\newcommand{\xVmax}{x\iVmax}
\newcommand{\xVmin}{x\iVmin}
\newcommand{\xfplus}{x\ifplus}
\newcommand{\xfminus}{x\ifminus}
\newcommand{\xinimax}{\xini^{\max}}

\newcommand{\xendmax}{\xend^{\max}}
\newcommand{\xendmin}{\xend^{\min}}
\newcommand{\xVinfty}{x\iVinfty}
\newcommand{\xVzero}{x\iVzero}
\newcommand{\xVzeroPlus}{x\iVzeroPlus}
\newcommand{\xVzeroMinus}{x\iVzeroMinus}
\newcommand{\xVzeroPM}{x\iVzeroPM}
\newcommand{\xdVzero}{x\idVzero}
\newcommand{\xdVzeroPlus}{x\idVzeroPlus}
\newcommand{\xdVzeroMinus}{x\idVzeroMinus}
\newcommand{\xdVzeroPM}{x\idVzeroPM}
\newcommand{\xddVzero}{x\iddVzero}
\newcommand{\xddVzeroPlus}{x\iddVzeroPlus}
\newcommand{\xddVzeroMinus}{x\iddVzeroMinus}
\newcommand{\xddVzeroPM}{x\iddVzeroPM}
\newcommand{\xepsoneMax}{x\iepsoneMax}
\newcommand{\xepsoneMin}{x\iepsoneMin}

\newcommand{\xepsoneOne}{x\iepsoneOne}

\newcommand{\xepsoneOnePlus}{x\iepsoneOnePlus}
\newcommand{\xepsoneOneMinus}{x\iepsoneOneMinus}
\newcommand{\xepsoneOnePM}{x\iepsoneOnePM}
\newcommand{\xepsoneOnePMcase}[1]{\xepsoneOne^{\pm (#1)}}

\newcommand{\xepsoneOneA}{x\iepsoneOneZero}
\newcommand{\xepsoneOneBC}{x\iepsoneOneMP}
\newcommand{\xepsoneOneC}{x\iepsoneOnePlus}
\newcommand{\xepstwoOne}{x\iepstwoOne}
\newcommand{\xepstwoMax}{x\iepstwoMax}
\newcommand{\xepstwoMin}{x\iepstwoMin}
\newcommand{\xepstwoZero}{x\iepstwoZero}
\newcommand{\xepstwoZeroPlus}{x\iepstwoZeroPlus}
\newcommand{\xepstwoZeroMinus}{x\iepstwoZeroMinus}
\newcommand{\xepstwoZeroPM}{x\iepstwoZeroPM}

\newcommand{\xepstwoMinusOneMinus}{x\iepstwoMinusOneMinus}
\newcommand{\xepstwoMinusOnePM}{x\iepstwoMinusOnePM}
\newcommand{\xepsthreeOne}{x\iepsthreeOne}
\newcommand{\xepsthreeMinusOnePM}{x\iepsthreeMinusOnePM}

\newcommand{\xiniepstwo}{\xini^{\epsilon_2}}
\newcommand{\xinistg}{\xini^{\ustg}}

\newcommand{\yend}{y_\uend}

\newcommand{\ystar}{y_*}
\newcommand{\ymax}{y_{\max}}
\newcommand{\yepsoneone}{y\iepsoneOne}
\newcommand{\yUV}{y\iUV}

\newcommand{\mend}{m_\uend}

\newcommand{\mmon}{m_\umon}
\newcommand{\mstar}{m_*}
\newcommand{\Kstar}{K_*}
\newcommand{\Estar}{E_*}
\newcommand{\Kpstar}{K_*'}

\newcommand{\Kend}{K_\uend}
\newcommand{\Eend}{E_\uend}
\newcommand{\Kpend}{K_{\uend}'}
\newcommand{\Epend}{E_\uend'}
\newcommand{\nuend}{\nu_\uend}
\newcommand{\dnuend}{\dot{\nu}_\uend}

\newcommand{\azero}{a_\zero}
\newcommand{\aend}{a_\uend}

\newcommand{\areh}{a_\ureh}
\newcommand{\zend}{z_\uend}

\newcommand{\Nini}{N_\uini}
\newcommand{\Ntot}{N_\utot}
\newcommand{\Nend}{N_\uend}
\newcommand{\Nreh}{N_\ureh}
\newcommand{\Nstar}{N_*}
\newcommand{\Nzero}{N_\zero}
\newcommand{\Nmax}{N_{\max}}
\newcommand{\Nmin}{N_{\min}}
\newcommand{\Ntop}{N_{\utop}}
\newcommand{\Nstarnuc}{N_*^\unuc}
\newcommand{\Nstarend}{N_*^\uend}
\newcommand{\Nc}{N_\uc}
\newcommand{\Nf}{N_\uf}
\newcommand{\tzero}{t_\zero}

\newcommand{\alphamin}{\alpha_{\min}}
\newcommand{\alphamax}{\alpha_{\max}}
\newcommand{\alphaMinus}{\alpha_1}
\newcommand{\alphaPlus}{\alpha_2}
\newcommand{\alphaRIPI}{\alpha_\usssRIPI}
\newcommand{\alphaOI}{\alpha_\usssOI}
\newcommand{\alphahat}{\hat{\alpha}}
\newcommand{\alphazero}{\alpha_\zero}
\newcommand{\mumax}{\mu_{\max}}
\newcommand{\mumin}{\mu_{\min}}
\newcommand{\mhiggs}{m_\uH}

\subheader{Opiparous Edition}
\title{\it Encyclop{\ae}dia Inflationaris}

\author[a]{J\'er\^ome Martin,}
\author[b,a]{Christophe Ringeval,}
\author[c,a]{Vincent Vennin}

\affiliation[a]{Institut d'Astrophysique de Paris, UMR
7095-CNRS, Universit\'e Pierre et Marie Curie, 98bis boulevard Arago,
75014 Paris (France)}

\affiliation[b]{Cosmology, Universe and Relativity at Louvain,
  Institute of Mathematics and Physics, Louvain University, 2 Chemin
  du Cyclotron, 1348 Louvain-la-Neuve (Belgium)}
  
\affiliation[c]{Laboratoire de Physique de
  l'\'Ecole Normale Sup\'erieure, ENS, CNRS, Universit\'e PSL,
  Sorbonne Universit\'e, Universit\'e Paris Cit\'e, 75005 Paris (France)}

\emailAdd{jmartin@iap.fr}
\emailAdd{christophe.ringeval@uclouvain.be}
\emailAdd{vincent.vennin@ens.fr}

\date{today}

\begin{document}

\abstract{The current flow of high-accuracy astrophysical data, among
  which are the Cosmic Microwave Background (CMB) measurements by the
  Planck satellite, offers an unprecedented opportunity to constrain
  the inflationary theory. This is however a challenging project given
  the size of the inflationary landscape which contains hundreds of
  different scenarios. Given that there is currently no observational
  evidence for primordial non-Gaussianities, isocurvature
  perturbations or any other non-minimal extension of the inflationary
  paradigm, a reasonable approach is to consider the simplest models
  first, namely the slow-roll single-field models with minimal kinetic
  terms. This still leaves us with a very populated landscape, the
  exploration of which requires new and efficient strategies. It has
  been customary to tackle this problem by means of approximate
  model-independent methods while a more ambitious alternative is to
  study the inflationary scenarios one by one. We have developed the
  publicly available runtime library \ASPIC\footnote{\urlaspic} to
  implement this last approach. The \ASPIC code provides all routines
  needed to quickly derive reheating-consistent observable predictions
  within this class of scenarios. \ASPIC has been designed as an
  evolutive code which presently supports $\Nmodel$ different
  models. In this paper, for each of the \ASPIC models, we present and
  collect new results in a systematic manner, thereby constituting the
  first \emph{Encyclop{\ae}dia Inflationaris}.}
    

\arxivnumber{1303.3787}


\maketitle

\newpage

\vspace*{\fill}
\begin{flushleft}
Dedicated to \textbf{Jean Le Rond d'Alembert} (1717--1783) and
  \textbf{Denis Diderot} (1713--1784).
  \begin{quote}
   \emph{ Le but d'une encyclop\'edie est de rassembler les
   connaissances \'eparses sur la surface de la terre; d'en exposer le
   syst\`eme g\'en\'eral aux hommes avec qui nous vivons, et de les
   transmettre aux hommes qui viendront apr\`es nous; afin que les
   travaux des siècles passés n'aient pas été des travaux inutiles
   pour les siècles qui succéderont; que nos neveux, devenant plus
   instruits, deviennent en même temps plus vertueux et plus heureux,
   et que nous ne mourions pas sans avoir bien mérité du genre
   humain.}
    \end{quote}
\end{flushleft}
\vspace*{\fill}

\newpage

\section*{Preface to the new edition}

The first edition of \emph{Encyclop{\ae}dia
Inflationaris}~\cite{EIOriginal} compiles the single-field slow-roll
models of inflation that were proposed prior to 2013. It contains
accurate reheating-consistent slow-roll calculations of the dynamics
of the background universe in these models, as well as of cosmological
scalar and tensor perturbations. Although it is often seen as a
review, it goes beyond the mere gathering of known results, since a
fair fraction of these models had been studied only under rough
approximations, which do not match the need for high accuracy from
current and forthcoming cosmological data. The ambition of the
encyclopaedic project is instead to provide an (almost) exact
treatment of these models, using only the slow-roll approximation, and
incorporating the effects of reheating on the scale correspondence in
a consistent way.  All these results are incorporated in a public
library, {\ASPIC}~\cite{aspic}, standing for ``Accurate Slow-roll
Predictions for Inflationary Cosmology'', which has been used in a
number of subsequent works, as for instance in the \emph{Particle Data
Group}~\cite{ParticleDataGroup:2022pth}. Bayesian model comparison
using the Cosmic Microwave Background (CMB) data has been presented in
\Refcs{Martin:2013nzq, Martin:2014lra} while the constraints derived
on the reheating epoch have been separately presented in
\Refcs{Martin:2014nya, Martin:2016oyk, Martin:2024qnn}.

Ten years later, inflation remains the most favored scenario of the
early universe~\cite{Chowdhury:2019otk} but several developments call
for the release of a new edition. First, at the theoretical level, new
models have been proposed. Some of them boil down to one of the
functional forms of the inflationary potentials already encoded in
{\ASPIC}, in which case they have been added in the relevant
sections\footnote{These models may nonetheless come with different
values for the parameters describing the potential, \emph{i.e.}
different priors in the framework of a Bayesian analysis.}. Some other
models give rise to new inflationary potentials, and therefore
constitute new sections of {\EI}, as well as new entries in the
{\ASPIC} library. There are 24 such new potential functions in the new
edition (here ordered alphabetically): Axion Hilltop Inflation
(\hyperref[sec:ahi]{AHI}), Cubicly Corrected Starobinsky Inflation
(\hyperref[sec:ccsi]{CCSI}), Double Exponential Inflation
(\hyperref[sec:dei]{DEI}), Dual Inflation (\hyperref[sec:di]{DI}),
Fibre Inflation (\hyperref[sec:fi]{FI}), Generalized Double Well
Inflation (\hyperref[sec:gdwi]{GDWI}), Hyperbolic Inflation
(\hyperref[sec:hbi]{HBI}), Hybrid Natural Inflation
(\hyperref[sec:hni]{HNI}), Non-Renormalizable Corrected Loop
Inflation (\hyperref[sec:ncli]{NCLI}), N-Formalism Inflation
(\hyperref[sec:nfi]{NFI}), Non-Minimal Large Field Inflation
(\hyperref[sec:nmlfi]{NMLFI}), Pure Arctan Inflation
(\hyperref[sec:pai]{PAI}), Radiatively Corrected Inflection Point
Inflation (\hyperref[sec:rcipi]{RCIPI}), Radiatively Corrected Large
Field Inflation (\hyperref[sec:rclfi]{RCLFI}), String Axion Inflation
I (\hyperref[sec:saii]{SAII}), String Axion Inflation II
(\hyperref[sec:saiii]{SAIII}), Super-conformal Alpha Attractor A
Inflation (\hyperref[sec:saai]{SAAI}), T-Model Inflation
(\hyperref[sec:tmi]{TMI}), Super-conformal Alpha Attractor B
Inflation (\hyperref[sec:sabi]{SABI}), Super-conformal Alpha
Attractor T Inflation (\hyperref[sec:sati]{SATI}), Symmetry Breaking
K\"ahler Inflation (\hyperref[sec:sbki]{SBKI}), S-Dual Inflation
(\hyperref[sec:sdi]{SDI}), Smeared Higgs Inflation
(\hyperref[sec:shi]{SHI}), Mukhanov Inflation
(\hyperref[sec:vfmi]{VFMI}).  The inclusion of these models allows
the new edition to provide an up-to-date landscape of all single-field
slow-roll inflationary models, bringing the number of models included
in the {\ASPIC} library to {\Nmodel}.

Second, at the observational level, the first edition compared the
predictions of single-field models with the early release of the
Planck 2013 data. Since then, additional data has been collected, and
the second edition features second-order slow-roll constraints from
the full {\DetailedData} data combination. The Bayesian evidence of
all models has been re-assessed with the latest data sets, and the
results are presented in a separate publication~\cite{Martin:2024qnn}.

Third, as the accuracy of the recent data releases has kept improving,
several projects are on their way that should deliver even more
accurate cosmological data in the years to come. In particular, let us
mention ground-based experiments that are currently operating such as
BICEP3, Keck array~\cite{BicepKeck:2021ybl, BICEP:2021xfz} and
SPT~\cite{DES:2022urg, SPT-3G:2022hvq} in Antarctica,
QUIJOTE~\cite{QUIJOTE:2015npn} in the Canary islands, and
CLASS~\cite{CLASS:2018mcw, CLASS:2020zyk}, ACT~\cite{ACTPol:2016kmo,
  ACT:2023kun} and POLARBEAR/SIMONS~\cite{POLARBEAR:2015ixw,
  Polarbear:2020lii,POLARBEAR:2022dxa, SimonsObservatory:2018koc} in
the Atacama desert. They will soon be joined by
QUBIC~\cite{QUBIC:2022sua}. In space, the EUCLID
satellite~\cite{Euclid:2021icp, Tutusaus:2022aef} is taking data and
will provide unprecedented measurements on the matter power spectrum,
down to very small scales. The LiteBIRD
satellite~\cite{LiteBIRD:2022cnt} is planned to be launched in 2028
and should allow us to further constrain the $B$-mode signal in the
polarization of the CMB. These prospects of ever increasing precision
confirm the relevance of the original {\EI} and {\ASPIC} projects,
namely the need for accurate predictions on a model-to-model
basis. For this reason, we have continued to pay special attention to
solve the inflationary dynamics exactly, without any other
approximations than those contained in the slow-roll framework. This
one has indeed been shown to be sufficiently accurate for the Planck
CMB data~\cite{Ringeval:2013lea, Martin:2016iqo} and can be extended
to arbitrary precision if needed~\cite{Auclair:2022yxs}. On the
contrary, other commonly-employed approximations are now too imprecise
to allow for a fair comparison with the data. As discussed in
Ref.~\cite{Martin:2016iqo}, this also applies to a number of
``model-independent'' approaches.

The role of reheating is also carefully taken care of, for two main
reasons. First, since the reheating expansion history determines the
part of the inflationary potential being probed by cosmological
measurements, it can now substantially affect the preference shown by
the data for a given model (in technical terms, the Bayesian
evidence). This is one of the reasons why the Starobinsky model (SI)
and Higgs Inflation (HI) are now treated as distinct models, since
they come with different reheating histories. Moreover, even though
they are often treated as sharing the same potential, this is only
correct at leading order in an expansion with respect to the inverse
of the non-minimal coupling of the field. Differences arise at
next-to-leading order that we now account for in an exact
manner. Second, taking the optimistic view that the data will narrow
the number of viable models down to a few (which could for instance
happen as a result of a detection of the gravitational-wave background
associated with inflation), the next question will clearly be to
further constrain the parameters of the happy chosen models, among
which are the reheating parameter. Coming back to the example of
Starobinsky and Higgs inflation, once the energy scale of the
potential has been set to match the amplitude of the CMB power
spectrum, the only parameter of the model that is left describes the
kinematics of reheating. Therefore, improved measurements of the
primordial power spectra will allow us to directly constrain the
reheating parameter, giving access to sectors of the theory describing
the coupling between the inflaton and other degrees of freedom that we
could not probe before. This again confirms the strategy adopted since
the early days of {\EI} and {\ASPIC} to derive reheating-consistent
predictions.

Let us note that in its new edition, the format of {\EI} has been
purposely kept similar to its original version. In particular, the
introduction (section~2) has been essentially left untouched in order
to keep track of our original motivations, and of the main
considerations that were discussed in the field at that time. We have
nevertheless removed a section that was listing the new analytical results
derived in the first 46 potentials, given that we think it has
already become clear that {\EI} is more than a review indeed.

Finally, about the benefit of a second edition, let us quote Jean Le Rond
d'Alembert (in a letter to Voltaire, June 23, 1766):
\begin{flushleft}
  \begin{quote}
   \emph{Quant \`a l'ouvrage, il est maigre, mais il est ais\'e de lui donner de
l'embonpoint dans une seconde \'edition.}
    \end{quote}
\end{flushleft}

We would like to thank all colleagues who wrote to us since
2013 to point out new models, ask for clarifications, and helped us to
improve the content of both {\EI} and {\ASPIC}. This clearly
contributed to making these tools fully updated and best suited for
analyzing current and forthcoming data.

\begin{flushright}
Paris \& Louvain-la-Neuve, May 2024.\\
J.~Martin, C.~Ringeval, V.~Vennin.
\end{flushright} 

\newpage

\section{List of potentials}
\label{aspiclist}

The following table shows the acronym of the models contained in the
current release of the {\ASPIC} library. For each model, an hyperlink
points to the adequate section in {\EI}, we give the number of free
parameters characterizing the potential, the number of sub-models and
the functional shape of the potential. The total number of models is
$\Nmodel$.

\begin{center}
\begin{longtable}{||c||c||c||c||}
\hline
  Name & Parameters & Sub-models  & $V(\phi)$ \\
    \hline \hline
  \hyperref[sec:si]{SI} & 0 & 1 & $M^4\left(1-\ee^{-\sqrt{2/3}\phi/\Mp}\right)^2$\\
  \hline \hline
  \hyperref[sec:hi]{HI} & 0 & 1 & $M^4\left(\frac{\barh^2 -
    \barv^2}{1+\barh^2}\right)^2$ \\
  &  &  & $\begin{aligned}
      \dfrac{\phi}{\Mp} & =  \sqrt{6 + 1/\xi} \ln \left[ \sqrt{1+(1+6\xi) \barh^2}
  + \sqrt{(1+6\xi) \barh^2} \right] \\ & + \sqrt{6} \ln
\left[\frac{\sqrt{1+ \barh^2}}{\sqrt{1+(1+6\xi)\barh^2} + \sqrt{6\xi
      \barh^2}} \right]
    \end{aligned}$ \\
  \hline \hline
  \hyperref[sec:rchi]{RCHI} & 1 & 1 & $M^4\left(1-2\ee^{-\sqrt{2/3}\phi/\Mp}+\frac{\AI}{16\pi^2}
 \frac{\phi}{\sqrt{6}\Mp}\right)$\\
  \hline \hline
  \hyperref[sec:lfi]{LFI} & 1 & 1 & $M^4\left(\frac{\phi}{\Mp}\right)^p$\\
  \hline \hline
  \hyperref[sec:mlfi]{MLFI} & 1 & 1 & $M^4\frac{\phi^2}{\Mp^2} \left(1 + \alpha \frac{\phi^2}{\Mp^2}\right)$\\
  \hline \hline
  \hyperref[sec:rcmi]{RCMI} & 1 & 1 & $M^4\left(\frac{\phi}{\Mp}\right)^2
\left[1-2\alpha\frac{\phi^2}{\Mp^2}\ln \left(\frac{\phi}{\Mp}\right)\right]$\\
  \hline \hline
  \hyperref[sec:rcqi]{RCQI} & 1 & 1 & $M^4\left(\frac{\phi}{\Mp}\right)^4
\left[1-\alpha \ln\left(\frac{\phi}{\Mp}\right)\right]$\\
  \hline \hline
  \hyperref[sec:ni]{NI} & 1 & 1 & $M^4\left[1+\cos\left(\frac{\phi}{f}\right)\right]$\\
  \hline \hline
  \hyperref[sec:esi]{ESI} & 1 & 1 & $M^4\left(1-\ee^{-q\phi/\Mp}\right)$\\
  \hline \hline
  \hyperref[sec:pli]{PLI} & 1 & 1 & $M^4\ee^{-\alpha \phi/\Mp}$\\
  \hline \hline
  \hyperref[sec:kmii]{KMII} & 1 & 2 & $M^4\left(1-\alpha\frac{\phi}{\Mp}\ee^{-\phi/\Mp}\right)$\\
  \hline \hline
  \hyperref[sec:hf1i]{HF1I} & 1 & 1 & $M^4 \left(1+A_1 \dfrac{\phi}{\Mp}\right)^2\left[1-\frac{2}{3}
\left(\frac{A_1}{1+A_1\phi/\Mp}\right)^2\right]$\\
  \hline \hline
  \hyperref[sec:cwi]{CWI} & 1 & 1 & $M^4\left[1 +
\alpha\left(\frac{\phi}{Q}\right)^4 \ln
\left(\frac{\phi}{Q}\right)\right]$\\
  \hline \hline
  \hyperref[sec:li]{LI} & 1 & 2 & $M^4\left[1
+\alpha\ln \left(\frac{\phi}{\Mp}\right)\right]$\\
  \hline \hline
  \hyperref[sec:rpi]{RpI} & 1 & 3 & $M^4 \ee^{-2 \sqrt{2/3}\phi/\Mp} \left|\ee^{\sqrt{2/3}\phi/\Mp} 
 - 1 \right|^{2p/(2p-1)}$ \\
  \hline \hline
  \hyperref[sec:dwi]{DWI} & 1 & 1 & $M^4\left[\left(\frac{\phi}{\phizero}\right)^2-1\right]^2$\\
  \hline \hline
  \hyperref[sec:mhi]{MHI} & 1 & 1 & $M^4 \left[1-{\sech} \left(\frac{\phi}{\mu} \right) \right]$\\
  \hline \hline
  \hyperref[sec:rgi]{RGI} & 1 & 1 & $M^4\frac{\left(\phi/\Mp\right)^2}{\alpha+\left(\phi/\Mp\right)^2}$\\
  \hline \hline
  \hyperref[sec:mssmi]{MSSMI} & 1 & 1 & $M^4\left[\left(\frac{\phi}{\phizero}\right)^2-\frac{2}{3}
  \left(\frac{\phi}{\phizero}\right)^6+\frac{1}{5}\left(
  \frac{\phi}{\phizero}\right)^{10}\right]$\\
  \hline \hline
  \hyperref[sec:ripi]{RIPI} & 1 & 1 & $M^4 \left[ \left(\frac{\phi}{\phizero}\right)^2 -
   \frac{4}{3} \left( \frac{\phi}{\phizero} \right)^3 + \frac{1}{2}
   \left( \frac{\phi}{\phizero} \right)^4 \right]$\\
  \hline \hline
  \hyperref[sec:ai]{AI} & 1 & 1 & $M^4\left[1-\frac{2}{\pi}
    \arctan\left(\frac{\phi}{\mu}\right)\right]$\\
  \hline \hline
  \hyperref[sec:cnai]{CNAI} & 1 & 1 & $M^4\left[3-\left(3+\alpha^2 \right) \tanh^2
    \left( \frac{\alpha}{\sqrt{2}} \frac{\phi}{\Mp} \right) \right]$\\
  \hline \hline
  \hyperref[sec:cnbi]{CNBI} & 1 & 1 & $M^4\left[\left(3-\alpha^2\right) \tan^2
    \left(\frac{\alpha}{\sqrt{2}}\frac{\phi}{\Mp} \right)-3\right]$\\
  \hline \hline
  \hyperref[sec:osti]{OSTI} & 1 & 1 & $-M^4\left(\frac{\phi}{\phizero}\right)^2\ln\left[\left(\frac{\phi}{\phizero}\right)^2\right]$\\
    \hline \hline
  \hyperref[sec:wri]{WRI} & 1 & 1 & $M^4\ln\left(\frac{\phi}{\phizero}\right)^2$\\
\hline \hline
   \hyperref[sec:di]{DI} & 1 & 1 & $M^4 \left[ 1 + V_0(f) -
  2\frac{K-E}{m K} - \frac{\pi \nu^2 \Theta(\nu)}{m K K'} \right]$ \\
   &  &  & with $\nu = 1 - \frac{8 \sqrt{2}}{\pi^2 f}
  \frac{K}{m^{1/2}(E'-K')^2}$ \\
  & & & and $\frac{\ud \phi}{\ud m} = -\frac{2 \sqrt{2}}{\pi}
  \frac{\sqrt{K K'}}{m^{3/2}}$ \\
  \hline \hline
   \hyperref[sec:ccsi]{CCSI} & 1 & 3 & $M^4 \left(1 - e^{-\sqrt{2/3} \phi/\Mp}
\right)^2$\\ &  &  & $\times \left[1 + \sqrt{1 +
      3\alpha\left(e^{\sqrt{2/3} \phi/\Mp} -
      1\right)}\right]^{-3}$ \\  &  &  & $\times \bigg[1 + \sqrt{1 +
    3\alpha\left(e^{\sqrt{2/3} \phi/\Mp}-1\right)}$ \\
  & & & $+
  2\alpha \left(e^{\sqrt{2/3} \phi/\Mp}-1
  \right)\bigg]$ \\
  \hline \hline
  \hyperref[sec:sbki]{SBKI} & 1 & 1 & $M^4\left(\frac{\phi}{\Mp}\right)^2
\exp\left[\alpha\left(\frac{\phi}{\Mp}\right)^2
+\frac{\alpha^2}{6}\left(\frac{\phi}{\Mp}\right)^4\right]$ \\
\hline\hline
\hyperref[sec:ahi]{AHI} & 1 & 1 & $M^4\left[\nu_0 - 2 \cos\left(\frac{\phi}{f}\right) +
  \left(\pi-\frac{\phi}{f}\right)
  \sin\left(\frac{\phi}{f}\right)\right]$ \\
\hline \hline
\hyperref[sec:pai]{PAI} & 1 & 1 & $M^4 \arctan\left(\frac{\phi}{\mu}\right)$ \\
\hline \hline
\hyperref[sec:saai]{SAAI} & 1 & 1 & $M^4 \left(1 - e^{-\sqrt{\frac{2}{3\alpha}} \frac{\phi}{\Mp}} \right)^2$ \\
\hline \hline
\hyperref[sec:tmi]{TMI} & 1 & 1 & $M^4 \tanh^{2n}\negthinspace\left(\frac{\phi}{\Mp\sqrt{6}}\right)$ \\
\hline \hline
\hyperref[sec:sfi]{SFI} & 2 & 1 & $M^4 \left[1 -\left(\frac{\phi}{\mu}\right)^{p}\right]$\\
\hline \hline
\hyperref[sec:ii]{II} & 2 & 1 & $M^4\left(\frac{\phi-\phizero}{\Mp}\right)^{-\beta}
-M^4\frac{\beta ^2}{6}\left(\frac{\phi-\phizero}
{\Mp}\right)^{-\beta-2}$ \\
\hline \hline
\hyperref[sec:kmiii]{KMIII} & 2 & 1 & $M^4\left(1-\alpha\frac{\phi}{\Mp} e^{-\beta\frac{\phi}{\Mp}}\right)$\\
\hline \hline
\hyperref[sec:lmi]{LMI} & 2 & 2 & $M^4\left(\frac{\phi}{\Mp}\right)^{4(1-\gamma)}\exp\left[-\beta (\phi/\Mp)^{\gamma}\right]$\\
\hline \hline
\hyperref[sec:twi]{TWI} & 2 & 1 & $M^4\left[1-A\left(\frac{\phi}{\phizero} \right)^2\ee^{-\phi/\phizero} \right]$\\
\hline \hline
\hyperref[sec:gmssmi]{GMSSMI} & 2 & 2 & $M^4\left[\left(\frac{\phi}{\phizero}\right)^2-\frac{2}
{3}\alpha\left(\frac{\phi}{\phizero}\right)^6+\frac{\alpha}
{5}\left(\frac{\phi}{\phizero}\right)^{10}\right]$\\
  \hline \hline
\hyperref[sec:gripi]{GRIPI} & 2 & 2 & $M^4\left[\left(\frac{\phi}{\phizero}\right)^2-\frac{4}
{3}\alpha\left(\frac{\phi}{\phizero}\right)^3+\frac{\alpha}
{2}\left(\frac{\phi}{\phizero}\right)^{4}\right]$\\
  \hline \hline
\hyperref[sec:bsusybi]{BSUSYBI} & 2 & 1 & $M^4\left(\ee^{\sqrt{6}\frac{\phi}{\Mp}} + \ee^{\sqrt{6} \gamma
  \frac{\phi}{\Mp}} \right)$\\
  \hline \hline
\hyperref[sec:ti]{TI} & 2 & 3 & $M^4\left(1+\cos\frac{\phi}{\mu}+\alpha\sin^2\frac{\phi}{\mu}\right)$\\
  \hline \hline
\hyperref[sec:bei]{BEI} & 2 & 1 & $M^4\exp_{1-\beta}\left(-\lambda \frac{\phi}{\Mp}\right)$\\
  \hline \hline
\hyperref[sec:psni]{PSNI} & 2 & 1 & $M^4\left[1+\alpha \ln \left(\cos\frac{\phi}{f}\right)\right]$\\
  \hline \hline
\hyperref[sec:ncki]{NCKI} & 2 & 2 & $M^4\left[1+\alpha \ln \left(\frac{\phi}{\Mp}\right) + \beta
    \left(\frac{\phi}{\Mp}\right)^2\right]$\\
  \hline \hline
\hyperref[sec:csi]{CSI} & 2 & 1 & $\frac{M^4}{\left( 1-\alpha\frac{\phi}{\Mp} \right)^2}$\\
  \hline \hline
\hyperref[sec:oi]{OI} & 2 & 1 & $M^4 \left(\frac{\phi}{\phizero} \right)^{4}\left[
 \left(\ln \frac{\phi}{\phizero} \right)^2- \alpha \right]$\\
  \hline \hline
\hyperref[sec:cnci]{CNCI} & 2 & 1 & $M^4\left[ \left( 3+\alpha^2 \right) \coth^2
    \left(\frac{\alpha}{\sqrt{2}}\frac{\phi}{\Mp}\right)- 3 \right]$\\
  \hline \hline
\hyperref[sec:sbi]{SBI} & 2 & 2 & $M^4\left\lbrace 1 + \left[ -\alpha + \beta\ln
  \left( \frac{\phi}{\Mp} \right) \right] \left( \frac{\phi}{\Mp}
\right)^4 \right \rbrace$\\
  \hline \hline
\hyperref[sec:ssbi]{SSBI} & 2 & 6 & $M^4\left[1 + \alpha\left(\frac{\phi}{\Mp}\right)^2
  + \beta\left( \frac{\phi}{\Mp} \right)^4 \right]$\\
  \hline \hline
\hyperref[sec:imi]{IMI} & 2 & 1 & $M^4\left(\frac{\phi}{\Mp}\right)^{-p}$\\
  \hline \hline
\hyperref[sec:bi]{BI} & 2 & 2 & $M^4 \left[1 -\left(\frac{\phi}{\mu}\right)^{-p}\right]$ \\
\hline \hline
\hyperref[sec:saii]{SAII} & 2 & 2 & $M^4 \left[1 - \cos\left(\frac{\phi}{\mu}\right) +
  \alpha \frac{\phi}{\mu} \sin\left(\frac{\phi}{\mu}\right)\right]$ \\
\hline \hline
\hyperref[sec:vfmi]{VFMI}  & 2 & 1 & $M^4\left\{1 - \beta \left[2
  \left(1+\frac{2-\alpha}{2\sqrt{3\beta}}\frac{\phi}{\Mp}\right)^{\frac{2\alpha}{2-\alpha}} \right]^{-1}
  \right\}$ \\ &  &  &  $\times \exp\left\{ \frac{3 \beta}{1-\alpha}
\left[\left(1+\frac{2-\alpha}{2\sqrt{3\beta}}\frac{\phi}{\Mp}\right)^{\frac{2(1-\alpha)}{2-\alpha}}
  - 1 \right] \right\}$ \\
\hline \hline
\hyperref[sec:fi]{FI} & 2 & 1 & $M^4\left[\left(1+\frac{2}{3}\delta\right)
\ee^{-\frac{4}{\sqrt{3}}\frac{\phi}{\Mp}} \right. $
 \\ &  &  &  $\left. -4\left(1+\frac{\delta}{6}\right)
\ee^{-\frac{1}{\sqrt{3}}\frac{\phi}{\Mp}}
+\frac{\delta}{1+n}
\ee^{\frac{2(1+n)}{\sqrt{3}}\frac{\phi}{\Mp}}+3-\frac{\delta}{1+n}\right]$\\
\hline \hline
\hyperref[sec:hbi]{HBI} & 2 & 1 & $M^4\sinh^n\left(\frac{\phi}{\mu}\right)$\\
\hline \hline
\hyperref[sec:shi]{SHI} & 2 & 1 & $M^4\left\lbrace \left[1-\left(\frac{\phi}{\phizero}\right)^2\right]^2 
+ \alpha\left(\frac{\phi}{\phizero}\right)^4
\left[\ln\left(\frac{\phi}{\phizero}\right)-\frac{1}{4}\right]
+\frac{\alpha}{4}\right\rbrace$ \\
\hline \hline
\hyperref[sec:dei]{DEI} & 2 & 2 & $M^4\left(\ee^{\beta\frac{\phi}{\phizero}}-\beta^2\ee^{\frac{1}{\beta}\frac{\phi}{\phizero}}\right)$\\
\hline \hline
\hyperref[sec:sdi]{SDI} & 2 & 1 & $\frac{M^4}{\cosh\left(\frac{\phi}{\mu}\right)}$ \\
\hline \hline
\hyperref[sec:gdwi]{GDWI} & 2 & 1 & $M^4 \left[\left(\frac{\phi}{\phizero}\right)^{2p} -
  1\right]^2$ \\
\hline
\newpage
\hline
\hyperref[sec:nmlfi]{NMLFI} & 2 & 4 & $M^4 \frac{\barh^p}{1+ \barh^2}$
\\
  &  &  & $\begin{aligned}
      \dfrac{\phi}{\Mp} & =  \sqrt{6 + 1/\xi} \ln \left[ \sqrt{1+(1+6\xi) \barh^2}
  + \sqrt{(1+6\xi) \barh^2} \right] \\ & + \sqrt{6} \ln
\left[\frac{\sqrt{1+ \barh^2}}{\sqrt{1+(1+6\xi)\barh^2} + \sqrt{6\xi
      \barh^2}} \right]
    \end{aligned}$ \\
\hline \hline
\hyperref[sec:sabi]{SABI} & 2 & 1 & $M^4 \left(1 - e^{-\sqrt{\frac{2}{3\alpha}} x}\right)^{2n}$ \\
\hline \hline
\hyperref[sec:sati]{SATI} & 2 & 1 & $M^4 \tanh^{2n}\negthinspace\left(\frac{\phi}{\Mp\sqrt{6\alpha}}\right)$ \\
\hline \hline
\hyperref[sec:rmi]{RMI} & 3 & 4 & $M^4\left[1-\frac{c}{2}\left(-\frac{1}{2} +\ln
\frac{\phi }{\phizero}\right)\frac{\phi ^2}{\Mp^2}\right]$\\
  \hline \hline
\hyperref[sec:vhi]{VHI} & 3 & 1 & $M^4\left[1 +\left(\frac{\phi}{\mu} \right)^{p} \right]$\\
  \hline \hline
\hyperref[sec:dsi]{DSI} & 3 & 1 & $M^4\left[ 1+\left(\frac{\phi}{\mu} \right)^{-p} \right]$\\
  \hline \hline
\hyperref[sec:gmlfi]{GMLFI} & 3 & 1 & $M^4\left(\frac{\phi}{\Mp}\right)^p \left[1 + \alpha\left(
  \frac{\phi}{\Mp} \right)^q \right]$\\
  \hline \hline
\hyperref[sec:lpi]{LPI} & 3 & 3 & $M^4\left(\frac{\phi}{\phizero}\right)^{p}
  \left(\ln \frac{\phi}{\phizero}\right)^q$\\
  \hline \hline
\hyperref[sec:cndi]{CNDI} & 3 & 3 & $\frac{M^4}{ \left\lbrace 1 + \beta\cos\left[
    \alpha \left( \frac{\phi-\phizero}{\Mp} \right) \right] \right \rbrace^2}$
\\  \hline \hline
\hyperref[sec:saiii]{SAIII} & 3 & 6 & $M^4 \left\{1 - \cos\left(\frac{\phi}{\mu}\right) +
  \alpha \left[ \frac{\phi}{\mu} \sin\left(\frac{\phi}{\mu}\right) +
    \frac{1}{2} \beta \left(\frac{\phi}{\mu} \right)^2 \right]\right\}$ \\
\hline \hline
\hyperref[sec:rclfi]{RCLFI} & 3  & 4  & $M^4 \left[ \left(\frac{\phi}{\mu}\right)^p + \alpha
  \left(\frac{\phi}{\mu}\right)^4 \ln \left(\frac{\phi}{\mu}\right) \right]$ \\
\hline \hline
\hyperref[sec:ncli]{NCLI} & 3 & 1 & $M^4\left[1+\alpha\ln\left(\frac{\phi}{\Mp}\right)
+\left(\frac{\phi}{\phizero}\right)^{4+2n}\right]$\\    
\hline \hline
\hyperref[sec:hni]{HNI} & 3 & 2 & $M^4\left[1+\alpha
  \cos\left(\frac{\phi}{\Mp}\right) \right]$\\
\hline \hline
\hyperref[sec:nfi]{NFI} & 3 & 4 & $M^4 \exp\left[-a \left(\frac{\phi}{\Mp}\right)^b \right]$\\    
\hline \hline
\hyperref[sec:rcipi]{RCIPI} & 3 & 2 & $M^4 \left(\frac{\phi}{\Mp}\right)^p
\left[1 + \alpha \ln\left(\frac{\phi}{\Mp}\right) + \beta
  \ln^2\left(\frac{\phi}{\Mp}\right) \right]$ \\
\hline
\end{longtable}
\end{center}

\newpage

\section{Introduction}
\label{sec:introduction}

\begin{figure}
\begin{center}
\includegraphics[width=\wsingfig]{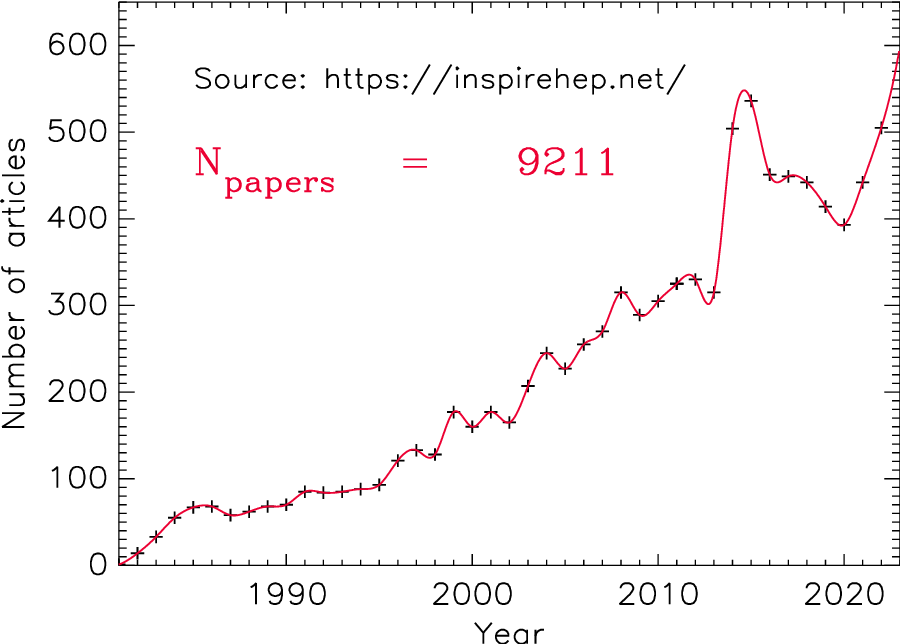}
\caption{Number of articles containing the word ``inflation'' and its
  variations (\ie ``inflating'', ``inflationary'', etc \dots ) in its
  title published each year since the advent of inflation. The total
  number exceeds $9200$ papers.}
\label{fig:statinf}
\end{center}
\end{figure}

The theory of inflation~\cite{Guth:1980zm, Linde:1981mu,
  Albrecht:1982wi, Linde:1983gd} represents a cornerstone of the
standard model of modern cosmology (the ``hot Big-Bang model'' of
Lema\^{\i}tre and Friedmann )~\cite{Linde:2007fr, Martin:2003bt,
  Martin:2004um, Martin:2007bw}. By definition, it is a phase of
accelerated expansion which is supposed to take place in the very
early universe, at very high energy, between Big-Bang Nucleosynthesis
(BBN) and $10^{15}~\GeV$. Inflation allows us to understand several
puzzles that plagued the pre-inflationary standard model (before
$1981$) and that could not be understood otherwise. Without inflation,
the standard model of cosmology would remain incomplete and highly
unsatisfactory. The most spectacular achievement of inflation is that,
combined with Quantum Mechanics, it provides a convincing mechanism
for the origin of the cosmological fluctuations (the seeds of the
galaxies and of the Cosmic Microwave Background - CMB - anisotropies)
and predicts that their spectrum should be almost scale invariant (\ie
equal power on all spatial scales)~\cite{Starobinsky:1979ty,
  Mukhanov:1981xt, Hawking:1982cz, Starobinsky:1982ee, Guth:1982ec,
  Bardeen:1983qw,Stewart:1993bc, Mukhanov:1990me, Liddle:1994dx} which
is fully consistent with the observations. Let us notice in passing
that this part of the scenario is particularly remarkable since it
combines General Relativity and Quantum
Mechanics~\cite{Grishchuk:1990bj, Polarski:1995jg, Kiefer:1998qe,
  Martin:2004um, Martin:2007bw, Kiefer:2008ku, Sudarsky:2009za,
  Martin:2012pea, Martin:2012ua}. Given all these spectacular
successes and given the fact that, despite many efforts, inflation has
not been superseded by its various challengers~\cite{Alexander:2000xv,
  Steinhardt:2001st, Khoury:2001bz, Khoury:2001wf, Martin:2001ue,
  Steinhardt:2002ih, Finelli:2001sr, Brandenberger:2001kj,
  Kallosh:2001ai, Martin:2002ar, Peter:2002cn, Tsujikawa:2002qc,
  Kofman:2002cj, Khoury:2003rt, Martin:2003bp, Martin:2003sf,
  Martin:2004pm, Nayeri:2005ck, Peter:2006hx, Finelli:2007tr,
  Abramo:2007mp, Falciano:2008gt, Linde:2009mc, Abramo:2009qk,
  Brandenberger:2009yt, Brandenberger:2011et, Brandenberger:2012zb,
  Cai:2012va, Cai:2013vm}, this scenario has gradually become a
crucial part of modern cosmology. As can be seen in \Fig{fig:statinf},
the number of papers devoted to this topic and published each year is
inflating since the advent of inflation.

In order to produce a phase of inflation within General Relativity,
the matter content of the universe has to be dominated by a fluid with
negative pressure. At very high energy, matter is described by field
theory, the prototypical example being a scalar field since it is
compatible with the symmetries implied by the cosmological
principle. Quite remarkably, if the potential of this scalar field is
sufficiently flat (in fact, more precisely, its logarithm) so that the
field moves slowly, then the corresponding pressure is negative. This
is why it is believed that inflation is driven by one (or several)
scalar field(s). For obvious reasons, this scalar field was given the
name ``inflaton''. However, the physical nature of the inflaton and
its relation with the standard model of particle physics and its
extensions remain elusive. Moreover the shape of its potential is not
known except that it must be sufficiently flat. This is not so
surprising since, as mentioned above, the inflationary mechanism is
supposed to take place at very high energies in a regime where
particle physics is not known and has not been tested in accelerators.

Another crucial aspect of the inflationary scenario is how it ends and
how it is connected to the subsequent hot Big-Bang phase. It is
believed that, after the slow-roll period, the field oscillates at the
bottom of its potential, or undergoes tachyonic preheating, but
finally decays into radiation. In this way, inflation is smoothly
connected to the radiation-dominated epoch~\cite{Turner:1983he,
  Kofman:1997yn, Bassett:2005xm, Mazumdar:2010sa, Finelli:1998bu,
  Bassett:1998wg, Finelli:2000ya,Jedamzik:2010dq, Jedamzik:2010hq,
  Easther:2010mr}. Unfortunately, very little is observationally known
on this so-called reheating period. Let us stress that adiabatic
initial conditions, as favored from the current CMB measurements,
naturally stem from such a setup within single field models. Another
hard bound is that the reheating energy, $\Treh$, must be higher
than the nucleosynthesis scale (\ie a few $\MeV$). The very first
constraints on $\Treh$ using CMB data (from the WMAP satellite) were
derived in \Refc{Martin:2010kz}, but these were concerning a few
models only.

We see that, despite the fact that it has become a cornerstone, the
inflationary era is not as observationally known as the other parts of
the standard model of Cosmology. However, there is now a flow of
increasingly accurate astrophysical data which gives us a unique
opportunity to learn more about inflation. In particular, the Planck
satellite data~\cite{2010AA520A9L,Ade:2013ktc,Planck:2018nkj} play a
crucial role in this process. The mission complements and improves upon
observations made by the NASA WMAP satellite~\cite{Bennett:2012fp,
  Hinshaw:2012fq} and is a major source of information relevant to
several cosmological issues including inflation~\cite{Ade:2013uln,
  Ade:2013ydc, Planck:2018jri, Planck:2018vyg}. As shown in
\Refcs{Martin:2014nya, Martin:2016oyk, Martin:2024qnn}, these data
allow us to gain more than one bit of information on the reheating era
when averaged over all the models presented in this paper. But the
flow of new data does not only concern the CMB. The Supernovae
projects~\cite{Tonry:2003zg, Riess:2004nr, Riess:2006fw, Riess:2011yx,
  SDSS:2014iwm, Pan-STARRS1:2017jku} continue to measure the distances
to the nearby exploding SN1A stars while the large scale galaxy
surveys such as the Sloan Digital Sky Survey
(SDSS)~\cite{AdelmanMcCarthy:2007aa, Abazajian:2008wr} are providing
an unprecedented picture of the structure of the universe. Galaxy
surveys allow for measuring the so-called Baryonic Acoustic
Oscillations (BAO)~\cite{Beutler:2011hx, Ross:2014qpa}. They are the
red-shifted version of the acoustic oscillations observed in the CMB
anisotropies which have been transferred to the galaxy power
spectrum~\cite{BOSS:2016wmc, Bautista:2017zgn, Carter:2018vce,
  deSainteAgathe:2019voe}. The ``lever arm'' in length scales between
CMB and galaxy power spectra increases the sensitivity to the small
deviations from scale invariance, and thus should be extremely
powerful to constrain inflationary models. For this reason, the data
from the Euclid satellite will be another step forward in our
understanding of inflation~\cite{Amiaux:2012bt, Euclid:2019clj,
  Euclid:2021qvm}. Let us also mention the possibility of direct
detection of the primordial gravitational waves for high energy
inflationary models on large scales~\cite{Turner:1993vb,
  Maggiore:1999vm, Kudoh:2005as, Kuroyanagi:2009br, Kawamura:2011zz,
  AmaroSeoane:2012km, Kuroyanagi:2013ns} and also on small
scales~\cite{Jedamzik:2010hq, Gorbunov:2012ns}.

The CMB small angular scales
of Planck are already complemented by ground-based microwave
telescopes such as the Atacama Cosmology Telescope
(ACT)~\cite{Dunkley:2013vu, Sievers:2013wk, ACTPol:2016kmo} or the South Pole
Telescope (SPT)~\cite{Hou:2012xq, Story:2012wx, SPT:2014tmz,
  SPT:2017jdf, Hou:2017smn} while ultra-sensitive
polarization dedicated experiments are on their
way~\cite{Baumann:2008aq, Crill:2008rd}. In a foreseeable future, the
last bit of yet unexplored length scales are expected to be unveiled
by the $21 \cm$ cosmological telescopes. These ones will be sensitive
to the red-shifted $21 \cm$ line absorbed by hydrogen clouds before
the formation of galaxies~\cite{Zaldarriaga:2003du, Lewis:2007kz,
  Tegmark:2008au, Barger:2008ii, Mao:2008ug, Adshead:2010mc,
  Clesse:2012th}. With such data, we will have a complete tomography
of the universe history from the time of CMB emission at the surface
of last scattering to the distribution of galaxies today.

The main goal of this article is to develop methods that will allow us
to constrain the inflationary scenario at a level matching the
accuracy of these new data. Since we have now entered the era of
massive multi-data analysis, the project aims at a change of scale
compared to previous approaches. In particular, one way to deal with
this question is to perform systematic and ``industrial'' studies of
this issue. Our ability to see through the inflationary window turns
the early universe into a laboratory for ultra-high energy physics, at
scales entirely inaccessible to conventional experimentation. In other
words, this window offers a unique opportunity to learn about the very
early universe and about physics in a regime that cannot be tested
otherwise, even in accelerators such as the Large Hadron Collider
(LHC) and its possible successors.

\subsection{Methodology}
\label{sec:method}
Let us now discuss how, in practice, the above described goals can be
reached. One issue often raised is that, since there are (literally) a
few hundreds different scenarios, it is difficult to falsify
inflation. This is, however, not a very convincing argument since
different models belong to different classes and usually do differ in
their observable predictions. They can thus be observationally
distinguished. A natural way to proceed is therefore to test
inflationary models step by step, starting with the simplest
scenarios. This is consistent with the Occam's razor point of view and
the way inference is achieved within Bayesian statistics (see
below). With this in mind, we can classify models in three different
broad categories: single-field inflation (category I), multiple-field
inflation (category II) and models where matter is not described by a
scalar field as, for instance, vector
inflation~\cite{Golovnev:2008cf}, chromo-natural
inflation~\cite{Adshead:2012kp} and/or
gauge-flation~\cite{Maleknejad:2011jw,Maleknejad:2011sq,Maleknejad:2012fw}
(category III). Within each category, one could further identify
various sub-categories. For example, within category I, the scalar
field can possess a minimal kinetic term and a smooth potential
(category IA), a minimal kinetic term and a potential with features
(category IB), a non-minimal kinetic term with a smooth potential
(category IC) or a non-minimal kinetic term and a potential with
features (category ID, see for instance \Refc{Avila:2013ela}) (a fifth
category could be models of warm inflation~\cite{Berera:1995ie,
  Yokoyama:1998ju, BasteroGil:2010pb, Bartrum:2013oka}). The same four
sub-categories can also be defined within category II [for instance,
  multiple Dirac Born Infeld (DBI) field
  inflation~\cite{Alishahiha:2004eh, Langlois:2008qf, Langlois:2009ej}
  belongs to category IIC] and so on. As already mentioned, each
category leads to different predictions. For instance, all models of
category IA predict a negligible level of non-Gaussianities,
$\fnlloc=5(1-\nS)/12\simeq 0.017$~\cite{Gangui:1993tt, Gangui:1994yr,
  Gangui:1999vg, Wang:1999vf, Maldacena:2002vr, Acquaviva:2002ud,
  Creminelli:2004yq, Cheung:2007sv,Ganc:2010ff, DeFelice:2013ar}
while, on the contrary, models of categories IB-ID yield
non-negligible non-Gaussianities~\cite{Seery:2005wm, Chen:2005fe,
  Chen:2006nt, Chen:2010xka, Chen:2006xjb, Chen:2008wn,
  Hannestad:2009yx, Flauger:2010ja, Adshead:2011jq, Martin:2011sn,
  Chen:2010bka, Gangui:2002qc, Holman:2007na, Xue:2008mk,
  Meerburg:2009ys, Ashoorioon:2010xg}; models belonging to IB and to
IC, or II, may not predict exactly the same type of
non-Gaussianities~\cite{Lehners:2009ja, RenauxPetel:2011uk}, etc\dots
In this context, as already mentioned, a crucial step was the recent
release of the Planck data~\cite{Ade:2013ktc, Planck:2013kta,
  Ade:2013ydc, Ade:2013xla, Planck:2019nip, Planck:2019evm}. Thanks to
the polarization, lensing and BAO, they are compatible with a
negligible (and slow-roll compatible) running
$\dd \nS /\dd \ln k=0.0011\pm 0.0099$ and a negligible running of the
running $\dd ^2 \nS/\dd \ln^2 k=0.009\pm 0.012$, with a pivot scale
chosen at $\kstar=0.05\,\Mpc^{-1}$. These data are also compatible
with adiabaticity at $95\%$ CL such that there is no evidence for
isocurvature modes, although the analysis is done with one
isocurvature mode at a time only. The Planck data do not find evidence
for primordial non-Gaussianity, namely \Refc{Planck:2019kim}
reports $\fnlloc=-0.9\pm 5.1$, $\fnleq=-26\pm 47$ and $\fnlortho=-38\pm
24$. Therefore, at this stage, everything seems to be well described
by simplest scenarios of inflation and, as consequence, a reasonable
method is to start with the IA-models. Following category IA, if the
present observational situation evolves in the future, one should then
treat categories IB-ID, then category II and so on. In this way, one
can falsify inflation step by step, in a Bayesian motivated fashion.

Bayesian inference for inflation requires some cosmological data that
are sensitive to it, such as the ones enumerated above. For the
purpose of illustration, let us consider the CMB angular power
spectrum. Cosmological measurements give us a set of numbers, $\Clm$,
that we are able to calculate theoretically within an inflationary
model. This means that we know the functions $\Clt\equiv
\Clt(\tetas,\tetai)$, where $\tetas$ represents a set of parameters
describing post-inflationary physics, \ie $\tetas=(h,\OmegaL,
\OmegaCDM, \cdots)$ and $\tetai$ a set of parameters describing
inflationary physics. We are interested in constraining the values of
those parameters, especially the $\tetai$'s. Within a given
experiment, one is given a likelihood, or an effective chi-squared
$\chi^2\left(\tetas,\tetai\right)$, encoding all the underlying
uncertainties. In a frequentist approach, the searched values of
$\tetas$ and $\tetai$ would be chosen at the best fit, \ie those
verifying $\partial \chi^2/\partial \theta=0$. In a Bayesian
approach~\cite{Trotta:2008qt}, we are interested in determining the
posterior distributions of the parameters, using Bayes's theorem
\begin{equation}
\label{eq:postdistri}
P\left(\tetas,\tetai\vert \Clm\right)
=\frac{1}{\calN}\calL\left(\Clm\vert \tetas,\tetai \right)
\pi\left(\tetas,\tetai\right),
\end{equation}
where $\calL\left(\Clm\vert \tetas,\tetai
\right)=\ee^{-\chi^2\left(\tetas,\tetai\right)/2}$ is the likelihood
function, $\pi\left(\tetas,\tetai\right)$ the prior distribution,
describing our prejudices about the values of the parameters before
our information is updated, and $\calN$ a normalization factor, also
called Bayesian evidence. Because we are interested in the
inflationary parameters, one has to integrate over the
post-inflationary parameters in order to obtain the marginalized
probability distribution $P\left(\tetai\vert \Clm\right)=\int
P\left(\tetas,\tetai\vert \Clm\right)\dd \tetas$. CMB physics also
tells us that the multipole moment $\Clt$ can be written as
\begin{equation}
\Clt\left(\tetas,\tetai\right)=\int_0^{+\infty}
\frac{\dd k}{k}j_{\ell}(k\rlss)T(k;\tetas)\calP_\zeta(k;\tetai),
\end{equation}
where $j_{\ell}$ is a spherical Bessel function, $T(k;\tetas)$ is the
transfer function which describes the evolution of cosmological
perturbations during the standard Friedmann-Lema\^{\i}tre eras and
$\calP_\zeta$ is the inflationary power spectrum. As a result, the
process of constraining inflation from the $\Clm$ reduces to the
calculation of $\calP_\zeta$. The same lines of reasoning could be
generalized to any other cosmological observables sourced during
inflation, such as higher order correlation functions. 

At this stage, there are, a priori, two possibilities (it is also
worth noticing that yet another approach is the reconstruction
program~\cite{Lidsey:1995np,deOliveira:2005mf}). Either one uses a
model-independent, necessarily approximate, shape for $\calP_\zeta$
or, on the contrary, one scans the inflationary landscape, model by
model, and for each of them, calculates $\calP_\zeta$ exactly.

The advantage of working with a model-independent technique is
obvious. However, it often requires an approximation scheme that may
not be available for all models. In practice, an approximate method,
the slow-roll approach, is known for the category IA and for the
category IC, see the recent papers~\cite{Martin:2013uma,
  Jimenez:2013xwa, Boyanovsky:2005pw, Destri:2007pv, Burigana:2010hg,
  Boyanovsky:2009xh,Auclair:2022yxs}. In this case, the set of
inflationary parameters $\tetai$ becomes the Hubble flow functions:
$\tetai=\{\epsilon_n\}$ where the $\epsilon_n$ are defined in
\Eq{eq:defhf} and the corresponding expression of
$\calP_\zeta(k;\epsilon_n)$ is provided in \Eqs{spectrumsr},
(\ref{eqn:as0}), (\ref{eqn:as1}) and~(\ref{eqn:as2}). Assuming some
priors $\pi(\epsilon_n)$ on the Hubble flow functions, this method
yields the posterior distributions $P\left(\epsilon_n|\Clm\right)$ for
the Hubble flow functions evaluated at the pivot scale. This approach
has already been successfully implemented for the WMAP data in
\Refcs{Leach:2003us, Martin:2006rs, Lorenz:2008je, Finelli:2009bs,
  Martin:2010kz}.

The second approach is more ambitious. It consists in treating exactly
all the inflationary models that have been proposed so far and in a
systematic manner. For each model, the power spectrum is determined
exactly by means of a mode by mode numerical integration, for instance
using the exact integration routines provided by the {\FieldInf}
code~\cite{fieldinf}. Such an approach can also be used with the
higher correlation functions with, for instance, the recent release of
the \BINGO code calculating the inflationary
bispectrum~\cite{Hazra:2012yn}.

In this case, the set of parameters $\tetai$ differs according to the
model considered. For instance, Large Field Inflation (LFI) for which
$V(\phi)=M^4\left(\phi/\Mp\right)^p$, has $\tetai=(M,p)$ while Small
Field Inflation (SFI) with
$V(\phi)=M^4\left[1-\left(\phi/\mu\right)^p\right]$ has
$\tetai=(M,p,\mu)$. From \FieldInf one can then compute
$\calP_\zeta(k;M,p)$ for LFI and $\calP_\zeta(k;M,p,\mu)$ for SFI
without any other assumptions than linear perturbation theory and
General Relativity. Starting from some priors on the model parameters,
\eg in the case of LFI, $\pi(M,p)$, this method allows us to determine
the posterior distributions $P(M\vert \Clm)$ and $P(p\vert \Clm)$,
thereby providing parameter inference about the corresponding
inflationary model. This approach, which was successfully implemented
for the first time in \Refcs{Ringeval:2005yn, Martin:2006rs,
  Ringeval:2007am, Lorenz:2007ze}, and subsequently used in
\Refc{Mortonson:2010er}, has several advantages that we now discuss.

\begin{figure}
\begin{center}
\includegraphics[width=\wsingfig]{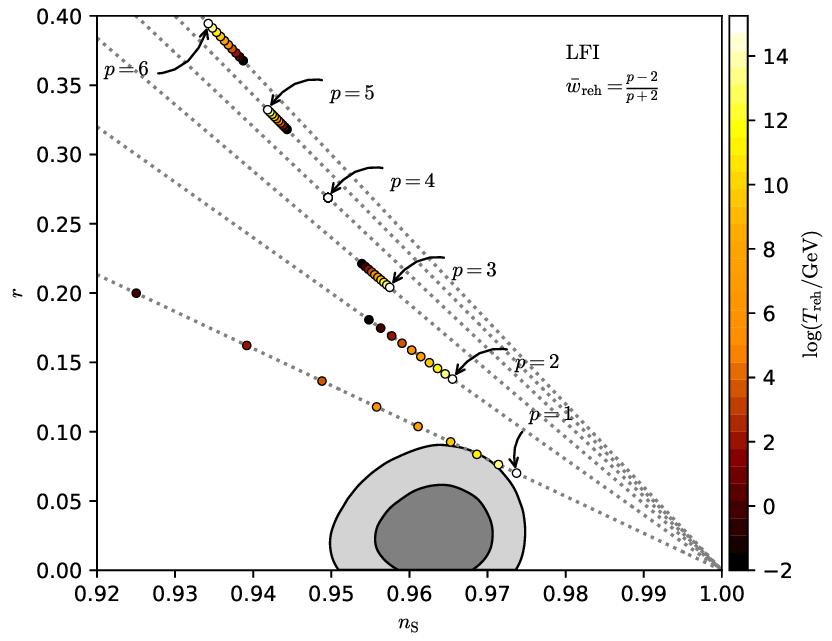}
\caption{Observational predictions for the LFI models, $V(\phi)
  \propto \phi ^p$, in the plane $(\nS,r)$ (\ie scalar spectral index
  and gravity wave contribution) compared to the {\DetailedData}
  data~\cite{Ade:2013ktc, Planck:2013kta, Ade:2013ydc,
    Ade:2013xla,Ade:2013uln,Ade:2013zuv}. Each continuous line and
  each color represent a different value of $p$. Along each line, each
  point (\ie each small ``circle'') denotes a different reheating
  temperature compatible with the constraint
  $\rhoend>\rhoreh>\rhonuc$. We see that the details of the reheating
  stage now matter: along a given line, some reheating temperatures
  are compatible with the observational constraints while others are
  not. This means that the CMB observations can now put constraints on
  $\Treh$. The mean equation of state parameter is defined in
  \Eq{eq:wrehbardef}.}
\label{fig:nsRLFIintro}
\end{center}
\end{figure}

Firstly, the most obvious advantage is that the result is exact. The
slow-roll method is an approximation and, for this reason, remains
somehow limited. As mentioned before, there are plethora of models,
such as single field models with features or multiple field scenarios,
for which a numerical integration is mandatory.

A second reason is that a full numerical approach permits a new
treatment of reheating. In the previous approaches, the influence of
the reheating is only marginally taken into account. Any observable
predictions depend on the number of \efolds associated with a
reheating era. From the fact that the reheating must proceed after the
end of inflation and before the Big-Bang Nucleosynthesis scale, one
can put an order of magnitude bound on this number of
\efolds~\cite{Liddle:2003as}. This causes small uncertainties in the
inflationary predictions that were not crucial in the past. However,
with the accuracy of the present and future data this question now
matters. This is illustrated in \Fig{fig:nsRLFIintro} which represents
the slow-roll predictions of LFI for which $V(\phi)\propto
\phi^p$. Each colored segment represents the range of observable
predictions for a given value of $p$, each point within a segment
corresponding to a given number of \efolds for the reheating or,
equivalently, to a given reheating energy $\Treh$. We see that, for
relatively small values of $p$, it is necessary to know the number of
\efolds the Universe reheated to decide whether the model is
compatible with the data or not. Conversely, the data are becoming so
accurate that one can start constraining the reheating
epoch~\cite{Martin:2016oyk}. Therefore, instead of viewing the
reheating parameters as external source of uncertainties, it is more
accurate to include them in the numerical approach and consider they
are part of the inflationary model. In its simplest description, the
reheating epoch can be modeled as a cosmological fluid with a mean
equation of state $\wrehbar>-1/3$. Notice that $\wreh$, the
instantaneous equation of state parameter, does not need to be
constant (see section~\ref{subsec:reheating}). For a simple quadratic
potential, and a parametric reheating, one would have for instance
$\wrehbar=0$. In this way, both $\wrehbar$ and $\Treh$ are added to
the inflationary parameters, \eg we now have
$\tetai=(M,p,\Treh,\wrehbar)$ for LFI, and \FieldInf computes
$\calP_\zeta(k;M,p,\Treh,\wrehbar)$. Starting from some priors
$\pi(\Treh,\wrehbar)$ one can then obtain the corresponding posterior
distributions $P\left(\Treh | \Clm\right)$ and $P\left(\wrehbar |
\Clm\right)$. The feasibility of this method has already been
demonstrated in \Refcs{Martin:2006rs, Martin:2010kz, Martin:2010hh}
where constraints on the reheating temperature for LFI and SFI have
been derived for the first time (see also \Refc{Easther:2011yq} and
later works of \Refcs{Tsujikawa:2013ila, Martin:2014nya,
  Martin:2016oyk}). In view of the expected accuracy of the future
data, the preheating/reheating era should become a compulsory element
of inflationary model testing. This issue plays an important role in
the proposal put forward in this article. In addition, let us also
emphasize that a proper treatment of the reheating and preheating
stages is mandatory in multiple field inflation because they can
affect the evolution of $\calP_\zeta$ on large
scales~\cite{Martin:2021frd}. Only a numerical approach can presently
deal with this problem.

A third advantage of the numerical approach is to address the question
of the priors choice in a particularly well-defined way. A crucial
aspect of Bayesian statistics is that the result depends on the choice
of the priors. Therefore, these ones must be chosen and discussed
carefully. In the slow-roll (approximated) approach described before,
the priors are chosen on the slow-roll parameters themselves. For
instance, a Jeffreys' prior is typically chosen on $\epsilon_1$ (\ie
uniform prior on $\log \epsilon_1$), as appropriate when the order of
magnitude of a parameter is not known. However, from a physical point
of view, it is better to choose the priors directly on the parameters
of the model, \eg the parameters entering the potential. For instance,
several potentials that we will treat are the results of a one-loop
calculation, namely a perturbative calculation with the coupling
constant playing the role of the small parameter. It is clear that the
prior must encode the fact that this parameter is small. With the
numerical approach, this is very conveniently done since we directly
compute the power spectrum from the potential itself. As another
example, let us consider the case of LFI where $\epsilon_1\simeq
p/\left(4\Delta \Nstar+p/4\right)$ ($\Delta \Nstar$ is the number of
\efolds between Hubble exit and the end of inflation, see
below). Owing to the non-trivial relation between the first slow-roll
parameter and $p$, a Jeffreys' prior $\pi(\epsilon_1)$ on $\epsilon_1$
implies a complicated prior $\pi(p)$ on $p$ while a natural choice
would be a flat prior. Again, implementing the priors directly on the
parameters of the model is a more theoretically justified
choice. Conversely, who could dispute that, beside the posterior
$P(\epsilon_1 | \Clm)$, it is theoretically interesting to know the
posterior distribution of $p$, \ie $P(p|\Clm)$. The exact numerical
integration is a reliable technique to obtain such distributions.

The numerical approach, however, has also some disadvantages. Firstly,
one needs to specify the inflationary scenarios explicitly and,
therefore, the constraints obtained are not
model-independent. Although this shortcoming can in fact never be
avoided (we always need to make some assumptions even in the slow-roll
approach) it may be partially overcome by scanning the complete
inflationary landscape. Secondly, and more importantly, it is time
consuming since the exact integration of the cosmological
perturbations and of the corresponding correlation functions is heavy
and can take up to a few minutes for complicated models. Finally, one
should expect multiple degeneracies for models having a high number of
inflationary parameters since the data have a limited sensitivity to
the shape of the primordial observables.

Based on the previous considerations, we conclude that it would be
very interesting to have an intermediate method that would allow us to
get most of the results that can be derived using the exact numerical
approach while being less time consuming and immune to the problems
induced by high parameter degeneracies. This is what we suggest in the
following. Our strategy is to use the slow-roll approximation in order
to skip the numerical calculation of the power spectrum, while being
combined with a systematic scan of the whole inflationary landscape
and reheating properties. As argued before, the Planck data drive us
towards testing inflation with the simplest models first and such a
method would therefore need to be implemented for the class of
scenarios IA only. More precisely, inferring the posterior
distributions of the Hubble flow parameters $\epsilon_n$ becomes a
first step, from which we take advantage of the fact that the
$\epsilon_n$ can be computed in terms of the more fundamental
parameters describing the reheating and $V(\phi)$. The degeneracies
between the inflationary parameters being confined to the computation
of the $\epsilon_n$, the data analysis used to determine the
$\epsilon_n$'s posteriors remains immune to this issue. Then, in a
second step, we use the knowledge of the functionals
$\epsilon_n(\tetai,\Treh,\wrehbar)$ to fastly derive the posteriors for the
inflationary and reheating parameters from the ones of the
$\epsilon_n$. High degeneracies may still be present, but in the
second step only. As shown in \Refc{Ringeval:2013lea}, for each model,
the method permits a quick, efficient and accurate extraction of the
posterior distributions of the inflationary and reheating parameters.

In our opinion, this third technique should not be viewed as a
competitor of the two others mentioned earlier but rather as
complementary and the corresponding results should be compared. Let us
also notice that, if, in order to scan all the inflationary scenarios,
the full exact numerical approach needs to be carried out at some
point, this would by no means render the results derived in the
present article useless. Indeed, the slow-roll approach is often a
very useful guide of which kind of physics one should expect for a
given model (initial conditions, range of the parameters, etc \dots
). In particular it allows us to understand any eventual parameter
degeneracies within the primordial observables. In other words, the
slow-roll method is an ideal tool to prepare a full numerical study.

\begin{figure}
\begin{center}
\includegraphics[width=\wsingfig]{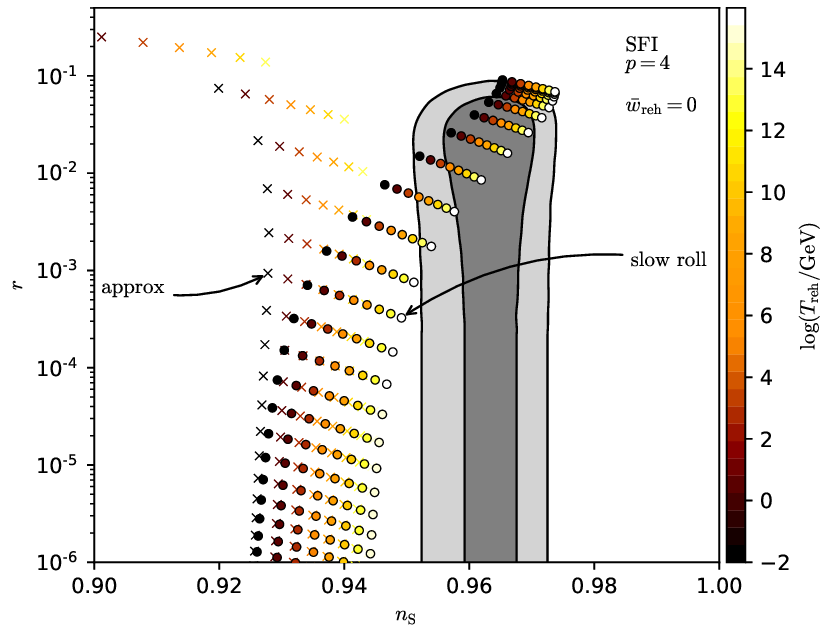}
\caption{Exact slow-roll predictions for SFI models, $V(\phi) \propto
  1-\left(\phi/\mu\right)^4$, compared to the \data
  data~\cite{Ade:2013ktc, Planck:2013kta, Ade:2013ydc,
    Ade:2013xla,Ade:2013uln,Ade:2013zuv}. Each colored segment
  represents a different value of $\mu$ and within each segment the
  color traces the reheating temperature. The segments made with
  ``crosses'', systematically on the left, represent some extra
  approximations usually made in the literature on top of slow roll,
  valid for $\mu/\Mp \ll 1$, see \Eqs{eq:sfiminivev}. We see that
  both coincide at very small values of $r$ but differ already for $r
  \gtrsim 10^{-3}$ where the extra approximations become
  inaccurate. Moreover, these approximations would indicate that this
  class of models is disfavored while the correct slow-roll
  predictions show that, on the contrary, they remain compatible with
  the data.}
\label{fig:sfiapproxintro}
\end{center}
\end{figure}

\begin{figure}
\begin{center}
\includegraphics[width=\wsingfig]{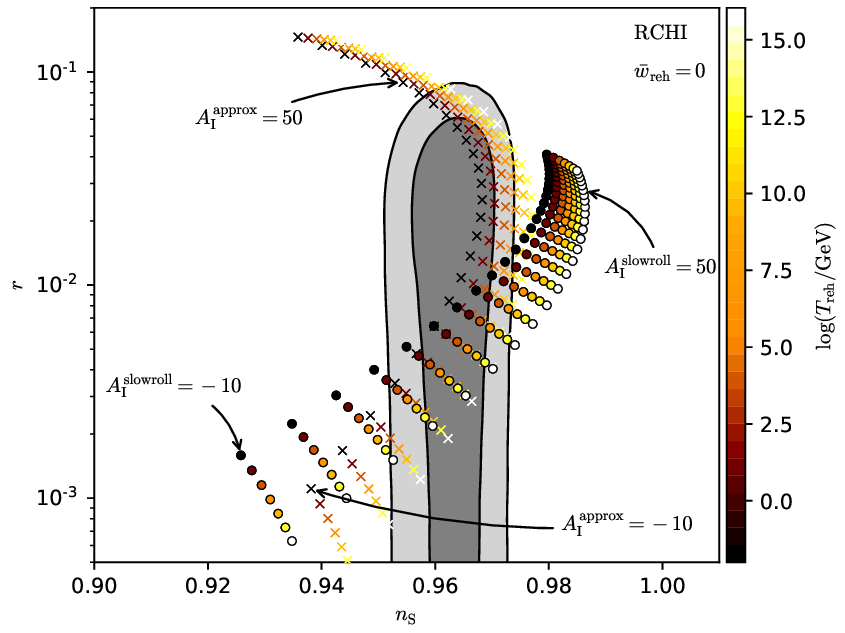}
\caption{Predictions of the RCHI model in the plane $(\nS,r)$ together
  with \data data~\cite{Ade:2013ktc, Planck:2013kta, Ade:2013ydc,
    Ade:2013xla,Ade:2013uln,Ade:2013zuv}. These predictions depend on
  one free parameter, $\AI$, for details see \sectionc{sec:rchi}. The
  segments made of small circles represent the slow-roll predictions
  (same conventions as in \Fig{fig:sfiapproxintro}), obtained when the
  coefficients $a_i=a_i\left[\epsilon_n\left(\tetai,\Treh,\wrehbar\right)\right]$ are
  numerically evaluated. On the contrary, the segments made with
  crosses are obtained with some approximated predictions. We see that
  there is a significant difference at both negatively and positively
  large values of $\AI$.}
\label{fig:rchiapproxintro}
\end{center}
\end{figure}

At this point, it is worth making the following remark. The method put
forward in this article uses an approximate shape for the power
spectrum, namely ($\kstar$ is the pivot scale)
\begin{equation}
  \calP_\zeta(k) \propto a_0\left(\epsilon_n\right) 
+ a_1\left(\epsilon_n\right) 
\ln \left(\dfrac{k}{\kstar}\right) 
  + \frac{1}{2}a_2\left(\epsilon_n\right) \ln^2\left(\dfrac{k}{\kstar}\right)
  + \dots \, ,
\end{equation}
in order to shortcut a numerical integration of $\calP_\zeta$ but is
otherwise completely self-consistent. In other words, once the
slow-roll approximation is accepted, no additional approximation
should be made. This may still require some numerical calculations,
however, in order to determine the coefficients $a_i$, or more
precisely the explicit expression, at Hubble crossing, of
$a_i=a_i\left[\epsilon_n\left(\tetai,\Treh,\wrehbar\right)\right]$. This is an
important issue given the accuracy of the current data as it is
illustrated in \Fig{fig:sfiapproxintro} (see also
\Refc{Martin:2006rs}). In this figure, we have represented the
slow-roll predictions of a SFI model, $V(\phi)\propto
1-(\phi/\mu)^4$. Each colored segment represents the \emph{exact}
slow-roll predictions of a model given the parameter $\mu$ and for
different numbers of \efolds during the reheating. These predictions
have been computed by solving numerically the slow-roll
equations. But, in the same plot, there are also other segments, on
the left, and represented in yellow only. They are predictions for
different values of $\mu$ but based on widespread \emph{approximate}
slow-roll formulas used in the literature. We see that, given the
accuracy of the data, the approximated formulas are no longer accurate
enough: the approximate results would predict that models with
$\mu/\Mp>1$ are strongly disfavored while the correct slow-roll
results show that they are still compatible with the data. Another
textbook example is provided by Higgs inflation with radiative
corrections (RCHI) and is presented in \Fig{fig:rchiapproxintro}. This
scenario is studied in detail in \sectionc{sec:rchi} and depends on
one free parameter, $\AI$. The segments made of colored circles
represent the exact predictions for different values of $\AI$ (see the
color bar on the side of the plot). The colored ``crosses'' indicate
predictions based on a commonly used approximate equation for the
coefficients $a_i=a_i\left(\epsilon_n\right)$. We see that this is no
longer sufficient as soon as $|\AI|$ becomes large. From these two
examples, we conclude that it is safer to use the slow-roll
approximation (which is usually extremely good) and nothing else, in
particular no extra approximation on top of the slow-roll
approximation. The fact that we may still need to use numerical
calculations to establish the observational predictions of a model
does not make our approach useless. Indeed, the numerics needed to
estimate $a_i=a_i\left[\epsilon_n\left(\tetai\right)\right]$ are, by
far, much easier than those needed to exactly compute
$\calP_\zeta$. Therefore, the gain in computational time mentioned
above is huge and allows for a fast and reliable method to constrain
the inflationary landscape.

\subsection{The \ASPIC library}

In order to implement the project described before, we have coded a
public runtime library, named \ASPIC~\cite{aspic} for ``Accurate Slow-roll
Predictions for Inflationary Cosmology'', which is meant to contain
all the inflationary models that can be treated with the method
described in \sectionc{sec:method}. \ASPIC already has $\Nmodel$
different inflationary scenarios, a number that should be compared to
the three or four models that are usually considered. The \ASPIC
library is an open-source evolutive project and, although it already
contains all the most popular inflationary scenarios, it aims at
including more models. In this way, it will converge towards a
situation where all the category IA models published since the advent
of inflation are implemented thereby allowing us to exhaustively scan
this part of the inflationary landscape. The list of the $\Nmodel$
\ASPIC models, as well as their acronym, is presented in
\sectionc{aspiclist}. If future cosmological data force us to move to
more complicated scenarios, the \ASPIC library will be upgraded
accordingly. It can, moreover, already be interfaced with the exact
field integration routines provided by \FieldInf thereby allowing for
a full numerical approach, if needed. This would be especially
relevant for all the single field models with modified kinetic terms
(category IB) such as DBI models, models with features (category IC)
such as the Starobinsky model~\cite{Starobinsky:1992ts} or the
multiple-field inflationary scenarios of category II, see for instance
\Refcs{Silk:1986vc, Peter:1994dx, Polarski:1995zn,
  Parkinson:2004yx,Tsujikawa:2002qx, Linde:1993cn, Lyth:1996kt,
  Liddle:1998jc, Ashoorioon:2009wa, Ashoorioon:2009sr,
  Ashoorioon:2011ki}. However, if the data continue to favor simple
models, such as those producing negligible non-Gaussianities and
isocurvature perturbations, the \ASPIC library in its present form
already contains the most relevant inflationary scenarios.

The \ASPIC library contains all the necessary routines to compare the
predictions of any of the $\Nmodel$ different models to high-accuracy
data. It is programmed in modern fortran and contains an exhaustive
documentation of its interfaces. The source files amount to more than
$100000$ lines of code. They are publicly available and distributed
under the GNU General Public License~\cite{gpl} at
\begin{center}
  {\urlaspic}
\end{center}

This paper presents the general architecture of the \ASPIC project and
all the calculations needed to understand and write these codes. For
each model, we give the theoretical framework, the calculation of the
three first slow-roll potential parameters, a discussion on how
inflation ends, the calculation of the associated field value, a
discussion of the priors, a calculation of the relevant range of
variation of the reheating temperature and an exact integration of the
slow-roll trajectory. Then, using our coding within the \ASPIC
library, we work out some basic reheating-consistent theoretical
predictions in the planes $(\epsilon_1,\epsilon_2)$ and $(\nS,r)$. Let
us stress again that, beside slow-roll, no other approximation is used
in the numerical codes of \ASPIC.

\subsection{New results}
\label{sec:NewResults}

Most of the \ASPIC models have already been partially studied in the
literature but let us emphasize that, for each of them, this paper
contains new results, and, sometimes, corrects predictions that were
not accurate enough for a proper use of the current data. In other
words, it does not aim at being a mere review and, therefore, the
presentation of already derived results has been kept to the minimal.

For all the models studied, this is the first time that their
observational predictions are worked out when the constraints on the
reheating phase are accurately taken into account. As explained in
\Refc{Martin:2010kz}, and briefly reviewed in~\sectionc{sec:basics},
it has become too inaccurate to derive the predictions of a model by
simply assuming a fixed range for $\Delta \Nstar$. For instance, this
could lead to a reheating energy density larger than the energy
density at the end of inflation, which is physically impossible (a
concrete example for LFI being discussed in
Ref.~\cite{Martin:2010kz}). Therefore, the predictions have been
re-worked in a consistent fashion (for the LFI and SFI models,
reheating-consistent predictions were already derived in
\Refc{Martin:2010kz}). This already constitutes a significant result
which goes beyond the current state-of-the-art. In the appendix, we
present a series of plots which give some basic predictions of the
various \ASPIC models in the planes $(\nS,r)$ and
$(\epsilon_1,\epsilon_2)$ for different values of the free parameters
characterizing each potential and for different reheating
energies. Most often, this is the first time that these predictions
are worked out for such a wide range of parameters and, moreover, this
is also the first time that these predictions are presented in this
fashion. In some sense, our paper can be viewed as the first
\textit{Encyclop{\ae}dia Inflationaris}.

Because no other approximations than slow roll have been used, the
paper also contains new expressions of the slow-roll trajectories for
many models, such as Higgs Inflation (HI), Starobinsky Inflation (SI),
Radiatively Corrected Higgs Inflation (RCHI), Mixed Large Field
Inflation (MLFI), Natural Inflation (NI) and others. As a consequence,
the predictions made using our non-approximated formulas often correct
previous results, as illustrated for RCHI in
\Fig{fig:rchiapproxintro}.  For many models, we are providing, for the
first time, exact expressions of the potential slow-roll parameters
$\epsilon_2$ and $\epsilon_3$, both entering into the determination of
running of the spectral index. The field value at which inflation ends
is also worked out without any other approximation than slow roll and
this, sometimes, has forced us to consider additional mechanisms to
stop inflation, such as a tachyonic instability, see Cubicly Corrected
Starobinsky Inflation 2 (CCSI2), for instance.  Some other models have
required a complete new treatment to deal with slow roll while
enforcing the internal consistency of their theoretical
framework. See, for instance, Double Well Inflation (DWI), K\"ahler
Moduli Inflation I (KMII) and K\"ahler Moduli Inflation II (KMIII),
Coleman-Weinberg Inflation (CWI), Dual Inflation (DI) or Radiatively
Corrected Large Field Inflation (RCLFI). Some new regimes of inflation
were also unveiled and this has motivated us to propose new classes of
inflation models, as, for instance, Generalized Double Well Inflation
(GDWI) or Non-Minimal Large Field Inflation (NMLFI).  Among other
consequences are the analysis of new subclasses of models, for a given
potential, their number being reported in Table~\ref{aspiclist}. Many
other new results are given in this article but we cannot summarize
all of them. They are explicitly presented in each section of this
manuscript.

\subsection{Scope}

Before concluding this introduction, let us stress that this article
and the \ASPIC library are the first step to carry out the final goal
which consists in assessing how good a model is and in comparing the
various inflationary models.

This very problem can be dealt with within Bayesian inference, using model
comparison. It allows one to determine, in a statistically
well-defined way, what ``the best models of inflation'' are given some
data sets. For this purpose, one has to calculate, for each model, the
global likelihood which is obtained by integrating the usual
likelihood over all of the model parameter values, weighted by their
respective prior probability distribution. The resulting quantity is a
number associated with each model which gives the ``evidence'' that
the model explains the data [this is the number $\calN$ in
  \Eq{eq:postdistri}]. Their respective ratios give the odds that one
model explains all data compared to the others. Bayesian methods have
the advantage to automatically incorporate the ``Occam's razor'':
complicated inflationary models will be assigned large probability
only if the complexity is required by the data. On the practical side,
these two steps can be implemented by the use of
Markov--Chains--Monte--Carlo (MCMC) methods, which are especially well
suited with the approach advocated here.

The complete Bayesian data analysis and model comparison of all the
models presented here have been achieved, using the {\ASPIC} library,
in various other works, and for different data sets, see
\Refcs{Martin:2013nzq, Martin:2014lra, Martin:2014nya, Martin:2016oyk,
  Martin:2024qnn, Sorensen:2024ezb}. For this reason, we will not
discuss here what the favored models are and we refer, instead, the
interested readers to these references. Nonetheless, because it is
interesting to qualitatively assess if a model can have predictions
compatible with the current cosmological data, we have reported, in
all the figures of appendix~\ref{sec:predictions}, the one- and
two-sigma confidence contours in the planes $(\epsilon_1,\epsilon_2)$
and $(\nS,r)$. These confidence regions have been derived from a
robust minimal set of cosmological data, {\data}, which concern CMB
anisotropies (temperature, $E$ and $B$-mode polarization), the
derivation of which being detailed in \Refc{Martin:2024qnn}.

This article is organized as follows. In the next section,
\sectionc{sec:basics}, we briefly summarize slow-roll inflation and
give the equations needed for the rest of this article. We also
discuss the reheating stage and explains how it can be
implemented. Then, in \sectionc{sec:zerop}, we study inflationary
models which, up to the potential normalization, do not contain any
free parameter (concretely, at this stage, Starobinsky and Higgs
inflation). In \sectioncs{sec:onep}, \ref{sec:twop}
and~\ref{sec:threep}, we analyze scenarios characterized by one, two
and three free parameters, respectively. Finally, in
\sectionc{sec:conclusion}, we present our conclusions and discuss
future works. In appendix~\ref{sec:predictions}, we give, in the
planes $(\nS,r)$ and $(\epsilon_1,\epsilon_2)$, the
reheating-consistent slow-roll predictions of all the $\Nmodel$ \ASPIC
models.

\section{Basic Equations}
\label{sec:basics}

In this section, we very briefly recall the theoretical foundations of
inflation and we present the main tools and equations that will be
used in the rest of this paper. We start by reviewing the slow-roll
phase, where the cosmological fluctuations are generated and, then, we
describe how the end of inflation and the transition to the standard
hot Big Bang phase can be modeled.

\subsection{The slow-roll phase}
\label{subsec:slowroll}

Let us consider a single-field inflationary model with a minimal
kinetic term and a potential $V(\phi)$. The behavior of the system is
controlled by the Friedmann-Lema\^{\i}tre and Klein-Gordon equations,
namely
\begin{align}
\label{eq:friedman}
H^2 &=\frac{1}{3\Mp^2}\left[\frac{\dot{\phi}^2}{2}+V(\phi)\right], \\
\label{eq:kg}
\ddot{\phi} & +3H\dot{\phi}+V_{\phi} = 0,
\end{align}
where $H\equiv \dot{a}/a$ denotes the Hubble parameter, $a(t)$ being
the Friedmann-Lema\^{\i}tre-Robertson Walker (FLRW) scale factor and
$\dot{a}$ its derivative with respect to cosmic time $t$. $\Mp = 8 \pi
G$ denotes the reduced Planck mass. A subscript $\phi$ means a
derivative with respect to the inflaton field. In order to describe
the evolution of the background, it is convenient to introduce the
Hubble flow functions $\epsilon_n$ defined by~\cite{Hoffman:2000ue,
  Schwarz:2001vv}
\begin{equation}
\label{eq:defhf}
\epsilon_{n+1} \equiv \frac{\dd \ln \left \vert \epsilon_n \right \vert}{\dd N}, 
\quad n\ge 0,
\end{equation}
where $\epsilon_0\equiv H_\uini/H$ and $N\equiv \ln(a/a_\uini)$ is the
number of \efolds. By definition, inflation is a phase of accelerated
expansion, $\ddot{a}/a>0$, or, equivalently, $\epsilon_1<1$. As a
consequence, the end of inflation is defined by the condition
$\epsilon_1=1$.  On the other hand, the slow-roll conditions (or
slow-roll approximation) refer to a situation where all the
$\epsilon_n$'s satisfy $\epsilon_n\ll 1$. If this is the case, then
the parameters $\epsilon_n$ can also be expressed in terms of the
successive derivatives of the potential, namely~\cite{Liddle:1994dx}
\begin{align} 
\label{eq:eps1}
\epsilon_1 & \simeq
\frac{\Mp^2}{2}\left(\frac{V_\phi}{V}\right)^2, \\ 
\label{eq:eps2}
\epsilon_2 & \simeq
2\Mp^2\left[\left(\frac{V_\phi}{V}\right)^2-\frac{V_{\phi \phi}}{V}\right], \\
\label{eq:eps3}
\epsilon_2\epsilon_3 & \simeq 2\Mp^4\left[
\frac{V_{\phi \phi \phi}V_\phi}{V^2}-3\frac{V_{\phi \phi}}{V}
\left(\frac{V_\phi}{V}\right)^2
+2\left(\frac{V_\phi}{V}\right)^4\right].
\end{align} 
Therefore, a measurement of the $\epsilon_n$'s also provides
information with regards to the shape of the inflationary potential.

In terms of the number of \efolds, one can decouple \Eqs{eq:friedman}
and \eqref{eq:kg} to get the field evolution
\begin{equation}
\dfrac{1}{3- \epsilon_1} \frac{\ud^2 \phi}{\ud N^2} + \dfrac{\ud
  \phi}{\ud N} = - \Mp^2 \dfrac{\ud \ln V}{\ud \phi}\,,
\label{eq:KGefolds}
\end{equation}
showing that the potential driving the field in FLRW spacetime is
$\ln[V(\phi)]$. This equation can be further simplified by using the
definition of $\epsilon_1$ and $\epsilon_2$ to get ride of the second
order derivatives. From
\begin{equation}
\epsilon_1 = \dfrac{1}{2 \Mp^2} \left(\dfrac{\ud \phi}{\ud N} \right)^2,
\label{eq:defeps1}
\end{equation}
one gets
\begin{equation}
\left(1 + \dfrac{\epsilon_2}{6 - 2\epsilon_1}\right)
\dfrac{\ud \phi}{\ud N} = - \Mp^2 \dfrac{\ud \ln V}{\ud \phi}\,.
\end{equation}
As a result, in the slow-roll approximation, one has
\begin{equation}
\dfrac{\ud \phi}{\ud N} \simeq - \Mp^2 \dfrac{\ud \ln V}{\ud \phi}\,.
\label{eq:phidot2}
\end{equation}
This equation can be integrated to give an explicit expression of the
classical trajectory. One arrives at
\begin{equation}
\label{eq:srtrajectory}
N-\Nini=-\frac{1}{\Mp^2}\int_{\phiini}^{\phi}\frac{V(\chi)}
{V_\chi(\chi)}\, \ud \chi \, .
\end{equation} 
In this article, for each model, we provide the expressions of the
first three Hubble flow parameters, a determination of $\phiend$, the
value of the field at which inflation comes to an end (and the
corresponding discussion) and an explicit expression of the slow-roll
trajectory \Eq{eq:srtrajectory}.

Let us now consider the behavior of inflationary cosmological
perturbations. The evolution of scalar (density) perturbations can be
reduced to the study of a single variable, the so-called
Mukhanov--Sasaki variable $v_\bmk$. In Fourier space, its equation of
motion can be expressed
as~\cite{Mukhanov:1990me,Martin:2003bt,Martin:2004um,Martin:2007bw}
\begin{equation}
\label{eq:eqmotv}
v_\bmk''+\left[k^2-\frac{\left(a\sqrt{\epsilon_1}\right)''}
{a\sqrt{\epsilon_1}}\right]v_{\bmk}=0.
\end{equation} 
Here, a prime denotes a derivative with respect to conformal time and
the quantity $k$ is the comoving wave number modulus of the Fourier mode under
consideration. This equation is the equation of a parametric
oscillator, \ie an oscillator with a time-dependent frequency. The
time-dependence of the effective frequency is controlled by the
dynamics of the background, more precisely by the scale factor and its
derivatives (up to fourth order). The quantity $v_\bmk$ is related to
the curvature perturbation $\zeta_\bmk$ through the following
expression:
\begin{equation}
\label{eq:zetavsv}
\zeta_\bmk=\frac{1}{\Mp}\frac{v_\bmk}{a\sqrt{2\epsilon _1}}\,.
\end{equation} 
The importance of $\zeta_\bmk $ lies in the fact that it can be viewed
as a ``tracer'' of the fluctuations on super-Hubble scales, \ie for
all $k \eta \ll 1$, where $\eta$ denotes the conformal time. Indeed,
in the case of single-field inflation, this quantity becomes constant
in this limit. Therefore, it can be used to ``propagate'' the
perturbations from inflation to the subsequent cosmological eras. The
statistical properties of the fluctuations can be characterized by the
$n$-point correlation functions of $\zeta_\bmk$. In particular, the
two-point correlation function can be written as an integral over wave
numbers (in a logarithmic interval) of the power spectrum
$\calP_{\zeta}(k)$, which can be expressed as
\begin{equation} 
\calP_{\zeta}(k)\equiv \frac{k^3}{2\pi ^2} \left\vert
\zeta_\bmk\right\vert ^2 =\frac{k^3}{4\pi^2 \Mp^2}\left\vert
\frac{v_\bmk}{a\sqrt{\epsilon _1}}\right\vert ^2\, .
\label{Pzeta}
\end{equation} 
In order to calculate $\calP_{\zeta}(k)$, one needs to integrate
\Eq{eq:eqmotv}, which requires the knowledge of the initial
conditions for the mode function $v_\bmk$. Since, at the beginning of
inflation, all the modes of cosmological interest today were much
smaller than the Hubble radius, the initial conditions are chosen to
be the Bunch-Davis vacuum which amounts to, up to a phase,
\begin{equation} \lim_{k \eta  \rightarrow -\infty}
v_\bmk=\frac{1}{\sqrt{2k}} \ee^{-ik\eta }\, ,
\label{eq:initial}
\end{equation}
where $\calH=aH$ is the conformal Hubble parameter.

The evolution of tensor perturbations (or primordial gravity waves)
can also be reduced to the study of a parametric oscillator. The
amplitude of each transverse Fourier mode of the gravity wave,
$\mu_\bmk(\eta)$, obeys the following equation
\begin{equation}
\label{eq:eomtensor}
\mu_\bmk''+\left(k^2-\frac{a''}{a}\right)\mu_\bmk=0.
\end{equation}
We notice that the time-dependence of the effective frequency differs
from that of the scalar case and now involves the derivative of the
scale factor up to second order only. It is then straightforward to
determine the resulting power spectrum. From a calculation of the
two-point correlation function, one obtains
\begin{equation}
\label{eq:specgw}
\calP_h(k) = \frac{2k^3}{\pi ^2}\left \vert \frac{\mu_\bmk}{a}
\right \vert ^2 .
\end{equation}
In order to calculate this quantity, the equation of motion
\Eq{eq:eomtensor} needs to be solved. As it is the case for density
perturbations, the initial state is chosen to be the Bunch-Davies
vacuum.

The power spectra can be computed exactly by means of a mode by mode
integration of \Eqs{eq:eqmotv} and (\ref{eq:eomtensor}), which also
requires an exact integration of the background, \ie of
\Eqs{eq:friedman} and (\ref{eq:kg}). As discussed in the introduction,
this can be done with the help of publicly available codes such as
\FieldInf. We have seen above that the slow-roll approximation can be
used to calculate the classical background trajectory. Quite
remarkably, the same approximation also permits the derivation of the
scalar and tensor power spectra. This involves a double expansion. The
power spectra are expanded around a chosen pivot scale $\kstar$ such that
\begin{equation} 
\label{spectrumsr}
\frac{\calP(k)}{\calP_\zero} = a_0 + a_1 \ln \left(\dfrac{k}{\kstar}\right) 
 + \frac{a_2}{2} \ln^2\left(\dfrac{k}{\kstar}\right)
 + \dots \, ,
\end{equation}
where
\begin{equation}
\calP_{\zeta \zero} =\frac{H^2}{8 \pi^2 \epsilon_1 \Mp^2}\,,
\qquad \calP_{h \zero} =\frac{2 H^2}{\pi^2 \Mp^2}\,,
\end{equation}
and, then, the coefficients $a_i$ are determined in terms of the
Hubble flow functions. For scalar
perturbations, one gets~\cite{Schwarz:2001vv,Martin:2002vn,
  Casadio:2004ru, Casadio:2005xv, Casadio:2005em, Gong:2001he,
  Choe:2004zg, Gong:2001he, Leach:2002ar, Martin:2013uma,
  Jimenez:2013xwa}
\begin{eqnarray}
\label{eqn:as0}
a_{0}^{\usssPSP} &=& 1 - 2\left(C + 1\right)\epsilon_1 - C \epsilon_2
+ \left(2C^2 + 2C + \frac{\pi^2}{2} - f\right) \epsilon_1^2 \nonumber
\\ & + & \left(C^2 - C + \frac{7\pi^2}{12} - g\right)
\epsilon_1\epsilon_2 + \left(\frac12 C^2 + \frac{\pi^2}{8} -
1\right)\epsilon_2^2 \nonumber \\ & + & \left(-\frac12 C^2  +
\frac{\pi^2}{24}\right)  \epsilon_2\epsilon_3 \, , \\
\label{eqn:as1}
a_{1}^{\usssPSP} & = & - 2\epsilon_1 - \epsilon_2 + 2(2C+1)\epsilon_1^2
+ (2C - 1)\epsilon_1\epsilon_2 + C\epsilon_2^2 - C\epsilon_2\epsilon_3
\, ,\\ a_{2}^{\usssPSP} &=& 4\epsilon_1^2 + 2\epsilon_1\epsilon_2 +
\epsilon_2^2 - \epsilon_2\epsilon_3 \, ,
\label{eqn:as2}
\end{eqnarray}
where $C \equiv \gamma_{\usssE} + \ln 2 - 2 \approx -0.7296$, $\gamma
_\usssE$ being the Euler constant, $f=5$ and $g=7$. For the
gravitational waves, the coefficients $a_i$ read
\begin{eqnarray}
a_{0}^{\usssPTP} &=&
 1 - 2\left(C + 1\right)\epsilon_1 
 + \left(2C^2 + 2C + \frac{\pi^2}{2} - f\right) 
 \epsilon_1^2 \nonumber \\ & + & \left(-C^2 - 2C + \frac{\pi^2}{12} - 2\right) 
 \epsilon_1\epsilon_2 \, ,\\
a_{1}^{\usssPTP} &=& 
 - 2\epsilon_1 + 2(2C + 1)\epsilon_1^2 
 - 2(C + 1)\epsilon_1\epsilon_2 \, ,
\label{eqn:at1} \\
a_{2}^{\usssPTP} &=& 4\epsilon_1^2 - 2\epsilon_1\epsilon_2 \, .
\label{eqn:at2}
\end{eqnarray}
These expressions are actually known at one more order, namely third
order in the Hubble flow functions, and can be found in
\Refc{Auclair:2022yxs}. The Hubble flow functions are time-dependent
quantities such that in the above expression, it is understood that
they should be evaluated at the time at which the pivot scale crosses
the Hubble radius during inflation, \ie at a time $\etastar$ such that
$\kstar = \calH(\etastar)$. Let us notice that setting the pivot at
another time affects the previous expression. For instance, setting
$\etastar$ such that $\kstar \etastar=-1$ would set $f=3$ and
$g=6$. We will see below that this introduces a dependence in the
parameters describing the reheating stage.

The properties of the power spectra can also be characterized by the
spectral indices and their ``running''. They are defined by the
coefficients of the Taylor expansions of the power spectra logarithm
with respect to $\ln k$, evaluated at the pivot scale $\kstar$. This
gives
\begin{equation}
\label{n}
\nS -1 \equiv \left. \frac{\dd \ln \calP_\zeta}{\dd \ln k}
\right \vert_{\kstar},
\qquad
\nT \equiv \left.\frac{\dd \ln \calP_h}{\dd \ln k}
\right \vert_{\kstar}.
\end{equation}
For the runnings, one similarly has the two following expressions
\begin{equation}
\label{alpha}
\alphaS \equiv \left .\frac{\dd^2 \ln \calP_\zeta}{\dd (\ln k)^2}
\right\vert_{\kstar}, \qquad \alphaT \equiv \left. \frac{\dd^2 \ln
  \calP_h}{\dd ( \ln k)^2} \right \vert_{\kstar},
\end{equation}
and, in principle, we could also define the running of the running and
so on. The slow-roll approximation allows us to calculate the
quantities defined above. For instance, we have at first order in the
Hubble flow parameters
\begin{equation}
\nS=1-2\epsilon_1-\epsilon_2, \quad \nT=-2\epsilon_1.
\end{equation}
Let us also notice that the tensor-to-scalar ratio at leading order
can be expressed as
\begin{equation}
r \equiv \frac{\calP_h}{\calP_\zeta}=16\epsilon_1.
\end{equation}
In the rest of this article, we give the observational predictions of
each inflationary model of the \ASPIC library in the planes
$(\epsilon_1,\epsilon_2)$ but also $(\nS,r)$.

Each inflationary model must also be CMB normalized, that is to say
the amplitude of the power spectra, say at $k=\kstar$, is completely
fixed by the amplitude of the CMB anisotropies measured today. On the
largest length scales, this is given to a good approximation by the
CMB quadrupole $\Qrms/T\equiv \sqrt{5C_2/(4\pi)}\simeq 6\times
10^{-6}$, where $T\simeq 2.725\, \K$ is the CMB blackbody
temperature. This is achieved if $\calP_{\zeta \zero}\simeq
60\, \Qrms^2/T^2$. Using the slow-roll approximation of the
Friedmann-Lema\^{\i}tre equation and writing the potential as
$V(\phi)=M^4v(\phi)$, such that the mass scale $M$ is singled out, one
arrives at
\begin{equation}
\left(\frac{M}{\Mp}\right)^4=1440\pi^2\frac{\epsilon_{1*}}{v(\phistar)}
\frac{\Qrms^2}{T^2}\,.
\label{eq:cobenorm}
\end{equation}
This is a model-depend expression (it depends on $v$) in which we have
rendered explicit the dependence in the pivot time. On a more robust
basis, CMB data are strongly constraining the value of $\Pstar \equiv
\calP_\zeta(\kstar)$, from the {\DetailedData} data one gets the
one-sigma confidence interval
\begin{equation}
\ln\left(10^{10} \Pstar \right) = 3.05 \pm 0.016\,,
\end{equation}
at $\kstar = 0.05\,\Mpc^{-1}$. This constraint and the one- and
two-sigma contours in the planes $(\epsilon_1,\epsilon_2)$ and
$(\nS,r)$ represented in all the figures have been obtained from a
slow-roll analysis of the \data data. Since the analysis is in all
point identical to the one of the WMAP seven years data performed in
\Refc{Martin:2010kz}, we do not repeat it here. The interested reader
can find all the details in the appendix B of
\Refc{Martin:2010kz}. Moreover, in order to get a robust inference, we
have used the second order expression for the power
spectra. Therefore, all the results presented below are marginalized
over the second order slow-roll parameters.

Since at leading order in the slow-roll expansion we have $\Pstar
\simeq \Hstar^2/(8 \pi^2 \epsonestar \Mp^2$), the
Friedmann--Lema\^{\i}tre equation allows us to derive the relation
\begin{equation}
\left(\dfrac{M}{\Mp} \right)^4 = 24 \pi^2
\dfrac{\epsonestar}{v(\phistar)} \, \Pstar\,,
\label{eq:pstarnorm}
\end{equation}
which is, as expected, formally identical to \Eq{eq:cobenorm} with
\begin{equation}
  \dfrac{\Qrms^2}{T^2} = \dfrac{\Pstar}{60}\,.
\end{equation}
It has however the advantage of using $\Pstar$ which is a well
inferred quantity because it is fitted against all the $C_\ell$. In
the following we will make no-distinction between the so-called \COBE
normalization and the CMB normalization, both being identical provided
the above equation is used. For each inflationary model, these
expressions will completely fix the allowed values for $M$.

We have shown how to calculate the two point correlation functions in
the slow-roll approximation. The next logical step would be to
determine the higher correlation functions. However, for the type of
models considered here (\ie category IA models), it is well-known that
the corresponding signal is so small that it will stay out of reach
for a while~\cite{Gangui:1993tt, Gangui:1994yr, Gangui:1999vg,
  Wang:1999vf, Maldacena:2002vr}. Therefore, we now consider the
question of how to calculate the values of the $\epsilon_n$ when the
pivot scale exits the Hubble radius and how this result depends on the
details of the reheating period.

\subsection{The reheating phase}
\label{subsec:reheating}

In the last subsection, we have seen that the power
spectrum in \Eq{spectrumsr} can be calculated with the help of the
slow-roll approximation and expressed in terms of the Hubble flow
parameters evaluated at Hubble radius crossing. Here, we briefly
explain how these Hubble flow parameters can be determined. It is easy
to calculate $\epsilon_1$, $\epsilon_2$ and $\epsilon_3$ as a function
of $\phi$ from \Eqs{eq:eps1}, (\ref{eq:eps2})
and~(\ref{eq:eps3}). Then, from the
trajectory in \Eq{eq:srtrajectory}, one can calculate $\Nend$, the
total number of \efolds during inflation and $\Nstar$, the number of
\efolds at the point when the pivot scale crosses the Hubble
radius. If we denote by $\calI$ the following primitive
\begin{equation}
\calI(\phi) = \int^\phi \dfrac{V(\psi)}{V_\psi(\psi)} \ud \psi,
\label{eq:trajdef}
\end{equation}
which is also the slow-roll trajectory in \Eq{eq:srtrajectory}, then
we have
\begin{equation}
\Nend=-\frac{1}{\Mp^2}\left[\calI(\phiend)-\calI(\phiini)\right],
\quad
\Nstar=-\frac{1}{\Mp^2}\left[\calI(\phistar)-\calI(\phiini)\right],
\label{eq:NofPhi}
\end{equation}
where $\phistar$ is the vacuum expectation value of the field, again
evaluated when the pivot scale crosses the Hubble radius. From these
two expressions, it follows that
\begin{equation}
\phistar=\calI^{-1}\left[\calI(\phiend)+\Mp^2\Delta\Nstar\right],
\end{equation}
where
\begin{equation}
  \Delta \Nstar\equiv \Nend-\Nstar.
\end{equation}
Inserting this formula into the expressions of the Hubble flow
parameters allows us to find $\epsilon_{n*}$ and, therefore, $r$ and
$\nS$.

However, in order to make the above-described calculation concrete, we
need to say something about the quantity $\Delta \Nstar$. As was
explained in details in \Refc{Martin:2010kz}, this requires to
take into account the reheating stage. Let $\rho $ and $P$ be the
energy density and pressure of the effective fluid dominating the
Universe during reheating. Conservation of energy implies that
\begin{equation}
\label{eq:rhorehend} \rho\left(N\right)=\rhoend
\exp\left\{-3\int_{\Nend}^{N}\left[1+ \wreh \left(n\right)\right]{\ud}
n\right\},
\end{equation} 
where $\wreh\equiv P/\rho$ is the ``instantaneous'' equation of state
during reheating. One can also define the mean equation of state
parameter, $\wrehbar$, by\footnote{In the figures, $\wrehbar$ has been
denoted by $w$ for simplicity.}
\begin{equation} 
\wrehbar \equiv \frac{1}{\Delta N}\int_{\Nend}^{\Nreh} \wreh(n)\dd n,
\label{eq:wrehbardef}
\end{equation}
where
\begin{equation} 
\Delta N \equiv \Nreh - \Nend,
\end{equation} 
is the total number of \efolds during reheating, $\Nreh$ being the
number of \efolds at which reheating is completed and the radiation
dominated era begins. Then, one introduces a new parameter
\begin{equation}
\Rrad \equiv \dfrac{\aend}{\areh} \left(\dfrac{\rhoend}{\rhoreh}\right)^{1/4},
\label{eq:Rrad} 
\end{equation}
where $\rhoreh$ has to be understood as the energy density at the end
of the reheating era, \ie $\rho(\Nreh)$. This definition shows that
$\Rrad$ encodes any deviations the reheating may have compared to a
pure radiation era. In fact, $\Rrad$ completely characterizes the
reheating stage and can be expressed in terms of
\begin{equation}
\label{eq:RradN} 
\ln \Rrad \equiv \frac{\Delta N}{4}\left(-1+3\, \wrehbar\right),
\end{equation} 
which renders explicit that if $\wrehbar=1/3$, \ie the effective fluid
during reheating is equivalent to radiation, then reheating cannot be
distinguished from the subsequent radiation dominated era. In this
case, one simply has $\Rrad=1$. Let us notice that it is also possible
to express (or define) $\ln \Rrad$ as
\begin{equation}
\label{eq:Rradw} 
\ln \Rrad = \frac{1-3\, \wrehbar}{12(1+\wrehbar)}\ln
\left( \frac{\rhoreh}{\rhoend}\right).
\end{equation} 
Using entropy conservation till the beginning of the radiation era,
the redshift at which inflation ended can be expressed in terms of
$\Rrad$ as
\begin{equation}
\label{eq:zend}
  1+\zend = \dfrac{1}{\Rrad} \left(\dfrac{\rhoend}{\rhotildegamma}
  \right)^{1/4}, \qquad \rhotildegamma \equiv \rdofreh \rhogamma.
\end{equation}
The quantity $\rhogamma=3 H_0^2 \Mp^2 \OmegaR$ is the total energy
density of radiation today ($\OmegaR \simeq 2.471\times
10^{-5}h^{-2}$) while $\rdofreh \equiv \gszero^{4/3}
\greh/(\gsreh^{4/3} \gzero)$ is the measure of the change of
relativistic degrees of freedom between the reheating epoch and
today. In this expression $\gs$ and $g$ respectively denotes the
number of entropy and energetic relativistic degrees of freedom. In
view of the current CMB data, the precise value for $\rdofreh$ is
unimportant as this factor has only a minimal effect. At most it can
shift the values of $\ln \Rrad$ by a $\order{1}$ number.

Then, straightforward considerations~\cite{Martin:2010kz,
  Ringeval:2013hfa} show that the quantities $\Delta \Nstar$ and
$\Rrad$ are related by
\begin{equation}
\Delta\Nstar = \ln \Rrad - \Nzero - \dfrac{1}{4}
\ln\left[\dfrac{9}{\epsonestar(3-\epsoneend)} \dfrac{\Vend}{\Vstar}\right] +
\dfrac{1}{4} \ln (8 \pi^2 \Pstar),
\label{eq:dnstarlnrad}
\end{equation}
where we have defined\footnote{One may also wonder about the influence
  of the cosmological constant on this result. In fact, one can show
  that it leads to a negligible correction. Indeed, it simply amounts
  to redefining $\Nzero$ by
\begin{equation}
\Nzero \rightarrow \Nzero+\frac{1}{3}\ln\left[1-\frac{\OmegaL\OmegaR^3}{\OmegaCDM^4}
\left(\frac{g_\mathrm{eq}}{g_0}\right)^3\left(\frac{q_0}{q_\mathrm{eq}}\right)^4\right].
\end{equation}
which is clearly a very tiny modification (the subscript ``eq''
denotes quantities at the equivalence time between radiation and
matter).}
\begin{equation}
\Nzero \equiv \ln
\left(\dfrac{\kstar/\azero}{\rhotildegamma^{1/4}}\right),
\end{equation}
which roughly measures the number of \efolds of deceleration of the
Friedmann-Lema\^{\i}tre model. From \Eq{eq:Rradw}, we see that the
quantity $\ln \Rrad$ is not arbitrary since $-1/3<\wrehbar<1$ and
$\rhonuc<\rhoreh<\rhoend$. Notice that the range allowed for
$\wrehbar$ might be extended to smaller values if one allows a phase
of acceleration to take place at lower energy than $\rhoend$, such as
in thermal or multistage inflation~\cite{Lyth:1995ka,
  Biswas:2005vz}. The quantity $\Delta \Nstar$ is
also constrained to vary in a given range, \ie $\Delta \Nstar \in
[\Delta \Nstarnuc,\Delta \Nstarend]$. Moreover, this range is
model-dependent since $\rhoend$ or $\Vend/\Vstar$ differ for different
inflationary scenarios. In fact, for each allowed value of $\ln
\Rrad$, \Eq{eq:dnstarlnrad} must be viewed as an algebraic equation
allowing us to determine the corresponding $\phistar$. Explicitly,
using \Eq{eq:NofPhi}, this equation reads
\begin{equation}
  \frac{1}{\Mp^2}\left[\calI(\phistar)-\calI(\phiend)\right] = \ln
  \Rrad - \Nzero - \dfrac{1}{4}
  \ln\left\{\dfrac{9}{\epsilon_1(\phistar)[3-\epsilon_1(\phiend)]}
    \dfrac{V(\phiend)}{V(\phistar)}\right\} + \dfrac{1}{4} \ln (8 \pi^2
  \Pstar)\,.
\label{eq:phistarlnrrad}
\end{equation}
In general, this equation cannot be solved explicitly (except for LFI
models, see \Refc{Martin:2010kz}) and we have to rely on numerical
calculations. Solving for each allowed value of $\ln \Rrad$, one can
determine the range of variation of $\phistar
\in[\phistarnuc,\phistarend]$ and, therefore, find the corresponding
dispersion in $r$ and $\nS$. In this paper, this task is carried out
for all the models of the \ASPIC library. Let us notice that it is
compulsory to do so otherwise, assuming blindly say $\Delta \Nstar \in
[40,60]$, would lead to inconsistent reheating energy densities,
either larger than $\rhoend$ or smaller than $\rhonuc$. Clearly, this
method also allows us to put model-dependent constraints on the
reheating temperature. Indeed, for some values of $\rhoreh$, the
corresponding $\epsilon_{n}(\phistar)$ will turn out to be outside the
$1\sigma$ or $2\sigma$ contours (depending on the criterion one wishes
to adopt) thus signaling some tension with the data, see the
discussion in the Introduction and \Fig{fig:nsRLFIintro}.

Let us emphasize that the parametrization presented in this section is
independent on the microphysics of reheating and we do not need to
specify explicitly the couplings of the inflaton field with the rest
of the world. In particular, preheating effects on the background
evolution are already taken into account with the present
framework. Furthermore, at the perturbed level, they cannot influence
the shape of the large scale power spectrum for the class of models
considered here~\cite{Finelli:1998bu}.

Before closing this section, let us remind that, for each inflationary
model, \ASPIC gives the expression of the first three Hubble flow
parameters, a discussion of the mechanism that ends inflation and the
value of $\phiend$, the classical trajectory $\calI(\phi)$, the CMB
normalization $M/\Mp$ and a determination of the exact range
$[\phistarnuc,\phistarend]$. Then all these information are compared
to CMB data in the planes $(\epsilon_1,\epsilon_2)$ and
$(\nS,r)$. This provides a powerful tool to systematically derive the
predictions for the \ASPIC models and, therefore, to scan the
inflationary landscape. In the next section, we start the systematic
exploration of the category IA models that have been studied in the
literature since the advent of inflation.

\section{Zero Parameter Models}
\label{sec:zerop}

\subsection{Starobinsky Inflation (SI)}
\label{sec:si}

\subsubsection{Original Theoretical Justifications}
\label{sec:theorysi}

One of the very first models of inflation was proposed by Alexei
Starobinsky in 1980 in \Refc{Starobinsky:1980te}. The idea is to
generate inflation through a purely quantum-gravitational effect, by
considering the case of a Friedmann-Lema\^{\i}tre-Robertson-Walker
universe filled with massless conformally-covariant quantum
fields. Because of conformal invariance, these massless fields do not
undergo particle creation, so the stress-energy tensor is only made of
terms that arise in the regularization process, {\ie} from the
interaction of quantum free matter fields with a classical
gravitational field. Those terms are quadratic in the space-time
curvature~\cite{Davies:1977ze, Bunch:1977sq}, and give rise to a
non-vanishing expectation value for the stress-energy tensor, $\langle
T_{\mu\nu}\rangle$, which, in the context of semi-classical gravity,
sources the Einstein equations. It was then realized in
\Refc{Vilenkin:1985md} that the same stress-energy tensor can be
obtained by varying the action
\begin{equation}
S=\frac{\Mg^2}{2}\int\dd^4 \bmx \sqrt{- g}\,  f(R)\,
,\quad\quad\text{where}\quad\quad f(R)=R+\frac{R^2}{\mu^2}\, ,
  \label{eq:actionf(R)}
\end{equation}
where $\mu$ is a mass parameter that depends on the (conformal and
massless) field content. The mass scale of gravity is denoted $\Mg$
here. From the point of view of effective theories, \Eq{eq:actionf(R)}
may be merely seen as the leading correction to General Relativity in
the class of $f(R)$ theories. Indeed, at low energy, {\ie} when $R$ is
small, the leading term in a generic Taylor expansion of the $f(R)$
function dominates and one recovers an action looking like General
Relativity and matching Newtonian gravity provided the numerical value
of $\Mg^2 \simeq (4/3)\Mp^2$ (see \sectionc{subsubsec:theoryhi} for
more details on the scalar-tensor theories). Here one considers the
first correction, that may play an important role at the high energies
at which inflation proceeds. It is also worth mentioning that when
that first correction is not quadratic but of a different order, one
obtains $R+R^{2p}/\mu^{4p-2}$ Inflation (RpI), presented in
\sectionc{sec:rpi}, while, when the next-to-next-to-leading correction
is included, {\ie} when a term $R^3$ is also considered in the $f(R)$
function [namely, $f(R)=R+R^2/\mu^2+\alpha R^3/\mu^4$], one obtains
the Cubicly Corrected Starobinsky Inflation model (CCSI), discussed in
\sectionc{sec:ccsi}.

Let us first establish some general equations in the case where the
action describing gravity is given by \Eq{eq:actionf(R)} to which we
add a contribution representing matter fields, namely
\begin{equation}
\begin{aligned}
S\left(\psi,g_{\mu \nu}\right)=\frac{\Mg^2}{2}\int\dd^4 \bmx \sqrt{-g}\, f(R)
  + \int\dd^4 \bmx \, \calL_\umat
  \left(\psi,g_{\mu \nu}\right),
\end{aligned}
\label{eq:actionf(R)mat}
\end{equation}
where $\calL_\umat\left(\psi,g_{\mu \nu}\right)$ is the Lagrangian of
matter. The field $\psi$ being, ``symbolically'', a matter field and
we are implicitly assuming that $\calL_\umat$ contains the covariant
volume factor $\sqrt{-g}$. Including the matter action in our
considerations will be important when we deal with reheating. If
viewed as exact, the above theory can be seen as a generalization of
Einstein gravity. A maybe more realistic point of view, as already
sketched above, is to interpret this framework as an effective theory
of gravity taking into higher order operators into account, \ie
$f(R)=R+R^2/\mu^2 +\cdots $. In this last point of view, however, one
could also ask why other terms, such as $R_{\mu \nu}R^{\mu \nu}$, are
not included (one could also add the contraction of the Riemann tensor
but this can always be re-expressed in terms of the scalar curvature,
the contraction of the Ricci tensor and the Gauss-Bonnet term, which
is topological in four dimensions). Varying the above
action, \Eq{eq:actionf(R)mat}, with respect to the metric tensor lead
to the following equations of motion
\begin{equation}
\begin{aligned}
  \Sigma_{\mu \nu}=F(R)R_{\mu \nu}-\frac12 f(R)g_{\mu \nu}
  -\nabla_\mu\nabla_\nu F(R)+g_{\mu \nu}\Box F(R)=\frac{1}{\Mg^2}
  T_{\mu \nu},
\end{aligned}
\label{eq:eomjordan}
\end{equation}
where $F(R)=\partial f/\partial R$ and $T_{\mu \nu}$ is the
stress-energy tensor of matter, namely
\begin{equation}
  T_{\mu \nu}=-\frac{2}{\sqrt{-g}}\frac{\delta \calL_\umat}
  {\delta g^{\mu \nu}}.
\end{equation}
The tensor $T_{\mu \nu}$ is conserved and one can check that this is
also the case for $\Sigma_{\mu \nu}$, $\nabla_{\mu}\Sigma^{\mu \nu}=0$
which is evidently required for consistency of the equations of
motion, see \Eq{eq:eomjordan}.

So far, we have worked in the so-called Jordan frame. However, as is
well-known, see for instance \Refcs{Maeda:1988ab, DeFelice:2010aj},
the above $f(R)$ theory can also be cast in different equivalent
formulations. For instance, it is equivalent to the Brans-Dicke theory
the action of which is given by
\begin{equation}
\begin{aligned}
  \SBD\left(\phi,\psi,g_{\mu \nu}\right)&=\Mg^2\int\dd^4 \bmx \sqrt{- g}
  \left[\frac12 \phi R-\frac{\omegaBD}{\phi}
    \frac12 g^{\mu \nu}\partial_\mu \phi\partial _\nu \phi -V(\phi)\right]
  +
  \int\dd^4 \bmx \, \calL_\umat
  \left(\psi,g_{\mu \nu}\right),
\end{aligned}
\end{equation}
where $\phi$ is a (dimensionless) scalar field,
$\omegaBD$ the (dimensionless) Brans-Dicke parameter and
$V(\phi)$ a (dimension $2$) potential. In order to prove the
equivalence between the $f(R)$ theory and the Brans-Dicke theory, let
us consider the following action
\begin{equation}
\begin{aligned}
  S\left(\chi,\psi,g_{\mu \nu}\right) & =\frac{\Mg^2}{2}\int\dd^4 \bmx \sqrt{- g}
  \left[f(\chi)+ \left(R-\chi\right) \frac{\partial f}{\partial \chi}\right]+
  \int\dd^4 \bmx \, \calL_\umat
  \left(\psi,g_{\mu \nu}\right),
\end{aligned}
\label{eq:actionLmultiplier}
\end{equation}
where $\chi$ is a new field of dimension $M^2$. The function $f(\chi)$
is of same dimension since one can expand it as $f(\chi)=\chi
+\cdots$. Varying this action with respect to $\chi$, one obtains
\begin{equation}
  (R-\chi) \dfrac{\partial^2 f}{\partial \chi^2}=0,
\label{eq:chionshell}
\end{equation}
which, provided $f''(\chi)\ne0$ implies $\chi=R$ and
\Eq{eq:actionLmultiplier} reduces to \Eq{eq:actionf(R)}.  Notice that
for $f''(\chi)=0$, one would simply recover General Relativity with a
cosmological constant. The next step consists in introducing the
dimensionless field $\phi$ defined by $\phi=\partial f/\partial \chi$
so that $\chi=\chi(\phi)$. Using this definition in
\Eq{eq:actionLmultiplier}, one arrives at
\begin{equation}
\begin{aligned}
  \SBD\left(\phi,\psi,g_{\mu \nu}\right)&=\Mg^2\int\dd^4 \bmx \sqrt{- g}
  \left(\frac{1}{2}\phi R-\frac{1}{2}
  \left\{\chi(\phi)\phi-f[\chi(\phi)]\right\}\right) + 
  \int\dd^4 \bmx \,\calL_\umat
  \left(\psi,g_{\mu \nu}\right),
\end{aligned}
\end{equation}
which is exactly of the Brans-Dicke form with
$\omegaBD=0$ and
\begin{equation}
  V(\phi)=\dfrac{1}{2} \left\{\chi(\phi)\phi-f[\chi(\phi)]\right\} = \dfrac{\mu^2}{8}\left(\phi-1\right)^2,
\end{equation}
the last equality holding for the Starobinsky model only.

One can also obtain another description of the same theory by using
conformal transformations. For this purpose, let us rewrite the
action~\eqref{eq:actionf(R)} with the Lagrange multiplier introduced
in \Eq{eq:actionLmultiplier}. Defining
\begin{equation}
F(\chi) \equiv \dfrac{\partial f(\chi)}{\partial \chi},
\end{equation}
one has
\begin{equation}
\begin{aligned}
 S\left(\psi,g_{\mu \nu}\right)&= \int\dd^4 \bmx \sqrt{- g}
 \left\{\frac{\Mg^2}{2}F(\chi)R-\frac{\Mg^2}{2}\left[F(\chi) \chi - f(\chi)\right]\right\}
 + \int\dd^4 \bmx \, \calL_\umat \left(\psi,g_{\mu \nu}\right).
\end{aligned}
\label{eq:preparationactionEin}
\end{equation}
Let us now perform a conformal transformation induced by a
dimensionless scalar field, say $\sigma$, and rewrite this action in
the so-called Einstein frame of metric
\begin{equation}
\ef{g}_{\mu \nu}=e^{-2\sigma} g_{\mu \nu}.
\end{equation}
In this section, quantities with a ``tilde'' will refer to the
Einstein frame while quantities without are written in the Jordan
frame. Under this conformal transformation, the scalar curvature
changes according to
\begin{equation}
\begin{aligned}
  \label{eq:transR}
  R=e^{-2\sigma}\left(\ef{R}-6\ef{g}^{\mu \nu} \ef{\nabla}_\mu
  \partial_\nu \sigma
-6 \ef{g}^{\mu \nu}\partial_\mu \sigma \partial_\nu \sigma\right).
\end{aligned}
\end{equation}
As a consequence, if we now express the action given by
\Eq{eq:preparationactionEin} in terms of quantities written in the
Einstein frame, using the above transformation~\eqref{eq:transR} for
the scalar curvature, then one is led to the following expression
\begin{equation}
\begin{aligned}
  S\left(\sigma,\psi,\ef{g}_{\mu \nu}\right)&=
  \int\dd^4 \bmx  \, \sqrt{- \ef{g}}
\, \biggl\{ e^{4\sigma}\frac{\Mg^2}{2} F(\chi) e^{-2\sigma}\left(\ef{R}
-6\ef{g}^{\mu \nu}\ef{\nabla}_\mu
  \partial_\nu \sigma
  -6 \ef{g}^{\mu \nu}\partial_\mu \sigma \partial_\nu \sigma\right)
  \\ & - e^{4\sigma}\frac{\Mg^2}{2}\left[F(\chi)\chi-f(\chi)\right]\biggr\}
+
  \int\dd^4 \bmx \, \calL_\umat
  \left(\psi,e^{2\sigma}\ef{g}_{\mu \nu}\right).
\end{aligned}
\end{equation}
Since, by definition of the Einstein frame, we want a theory the
action of which is linear in $\ef{R}$, we see that one must choose
$\sigma(\chi)$ such that $e^{-2\sigma}=F(\chi)$. Then, the term containing
the second derivative of $\sigma$ is a total derivative and can be
discarded. Indeed, for any metric tensor $\ef{g}_{\alpha \beta}$, one has
\begin{equation}
    \ef{g}^{\alpha \beta}\nabla_\alpha \nabla_\beta \sigma
    =\frac{1}{\sqrt{-\ef{g}}}\partial_\alpha\left(
    \sqrt{-\ef{g}}\ef{g}^{\alpha \beta}\partial_\beta \sigma\right).
\end{equation}
Moreover, if one defines the new scalar degree of freedom $\phi$ by
\begin{equation}
\frac{\phi}{\Mg} \equiv \sqrt{\frac{3}{2}} \ln\left[F(\chi)
\right]=-\sqrt{6}\,\sigma(\chi) ,
\label{eq:si:scalarfielddef}
\end{equation}
then, one arrives at
\begin{align}
S\left(\phi,\psi,\ef{g}_{\mu \nu}\right)&=  \int\dd^4 \bmx \, \sqrt{- \ef{g}}
\, \biggl[ \frac{\Mg^2}{2}\ef{R}
-\frac{1}{2}\ef{g}^{\mu \nu}\partial_\mu\phi
  \partial_\nu \phi-V(\phi)\biggr]
+ \int\dd^4 \bmx \, \calL_\umat
  \left(\psi,e^{2\sigma}\ef{g}_{\mu \nu}\right),
  \end{align}
with
\begin{align}
 V\left(\phi\right)= \dfrac{\Mg^2}{2} \frac{\chi F(\chi) - f(\chi)}{F^2(\chi)}\,.
\end{align}
This potential is sometimes written in terms of $R$ instead of
$\chi(\phi)$. Indeed, on shell, the Lagrange multiplier $\chi$ being
the solution of \Eq{eq:chionshell}, one has $\chi=R$. One therefore
obtains, in the Einstein frame, Einstein gravity plus a canonically
normalized scalar field $\phi$. This is an additional scalar mode
propagating in the theory, and, an important ingredient for inflation,
the coupling to matter is no longer universal due to the presence of
the combination $e^{2\sigma}\ef{g}_{\mu \nu}$ in the matter action. As
already mentioned, this will play an important role for reheating. Let
us mention that this additional scalar degree of freedom may modify
the measured gravitational constant and, in general, one cannot
identify $\Mg$ and $\Mp$~\cite{Damour:1992we,
  EspositoFarese:2000ij}. However, if one assumes that, after
inflation, this scalar degree of freedom relaxes to very small values
(which is the case here since $\chi=R$), then, for all
post-inflationary physics, $\Mp \simeq \Mg$.

Now, let us apply the previous considerations to the Starobinsky
model. In that case, one has $F(\chi) \equiv \partial f/\partial \chi
= 1+2\chi/\mu^2$, and the field $\phi$ evolves in the potential given
by
\begin{equation}
  V\left(\phi\right)=
  \frac{\Mg^2}{2\mu^2} \dfrac{\chi^2}{\left(1+2\dfrac{\chi}{\mu^2}\right)^2}\,.
\label{eq:potsiR}
\end{equation}
Using the relationship~\eqref{eq:si:scalarfielddef} between the
Lagrange multiplier $\chi$ and the field $\phi$, one gets
\begin{equation}
  \chi=\dfrac{\mu^2}{2}\left(e^{\sqrt{2/3}\phi/\Mg}-1\right),
\end{equation}
and the potential is explicitly given by
\begin{equation}
  V\left(\phi\right)= M^4 \left(
  1-\ee^{-\sqrt{\frac{2}{3}}\frac{\phi}{\Mg}}\right)^2,
\label{eq:potsi}
\end{equation}
with $M^4 \equiv \Mg^2 \mu^2/8$. 

\subsubsection{Other Theoretical Justifications}
\label{sec:othertheosi}

Many authors have tried to realize Starobinsky inflation in the
framework of supersymmetry and
supergravity~\cite{Wess:1978bu,Ferrara:1977mv,Wess:1977fn}. One of the
earliest attempt was based on models containing physical multiplets
that are not chiral but vector or linear~\cite{Cecotti:1987qe,
  Ferrara:2010in, Ferrara:2013rsa, Farakos:2013cqa,
  Aldabergenov:2020pry}. A great advantage of this type of approaches
(compared to formulations using chiral multiplets) is that there is no
need to stabilize additional scalar fields during inflation simply
because there is none; indeed there is only one scalar field which is
interpreted as the inflaton field. The extra fields are typically
vector fields and they do not acquire a vacuum expectation value
during inflation. The bosonic action obtained from the action of a
massive vector field $V$ was derived in \Refc{Cecotti:1987qe}. It
reads
\begin{align}
  \label{eq:lagrangiansupervector}
  \calL=-\frac{R}{2}-\frac14 F_{\mu \nu}F^{\mu \nu}
  +\frac{g^2}{2}J_{CC}B_{\mu}B^{\mu}
  +\frac12 J_{CC}\partial _\mu C\partial^{\mu }C-\frac{g^2}{2}
  J_C^2,
\end{align}
where $C$ is the scalar field present in the vector multiplet,
$B_{\mu}$ is the vector in the vector multiplet and the subscript
``$C$'' denotes a derivative with respect to the field $C$. The
arbitrary function $J$ is written $J=3/2 \ln \Phi$ where $\Phi$ is a
function of $C$. Finally, the quantity $g$ is the gauge coupling.

As mentioned above, $B_\mu$ does not acquire a vacuum expectation
value during inflation and, therefore, in
\Eq{eq:lagrangiansupervector}, we are left with the action of gravity
plus a non-canonically normalized scalar field $C$. If one chooses the
function $\Phi$ such that
\begin{align}
  \label{eq:defc}
  \Phi(C)=-C e^C,
\end{align}
and canonically normalize $C$ with $C=-e^{\sqrt{2/3}\phi/\Mg}$, then
the potential, which corresponds to the last term in
\Eq{eq:lagrangiansupervector}, reads
\begin{align}
  V(\phi)=\frac{9g^2}{4}\left(1-e^{-\sqrt{2/3}\phi/\Mg}\right)^2.
\end{align}
One recognizes the Starobinsky potential already given in
\Eq{eq:potsi}.

More recently, various other theoretical constructions have been
proposed that also give a potential matching \Eq{eq:potsi}.

In \Refc{Ellis:2013xoa}, a supergravity realization of this model was
presented that we now briefly review. The model is based on no-scale
supergravity and has two fields, a modulus $T$ and the inflaton
$\phi$. The K\"ahler and super-potential are given by
\begin{equation}
  \begin{aligned}
  K &= -3\Mg^2\ln
  \left(\dfrac{T}{\Mg}+\dfrac{T^{\dagger}}{\Mg} - \dfrac{\vert \phi \vert^2}{3\Mg^2}\right),\\
  W & = \hat{\mu}\phi^2-\dfrac{\lambda}{3}\phi^3,
  \end{aligned}
\end{equation}
where $\hat{\mu}$ is of dimension 1 and $\lambda $ dimensionless
(recall that the K\"ahler potential is of dimension 2 while the
super-potential is of dimension 3), respectively. The quantities
$\hat{\mu}$ and $\lambda$ are constants characterizing the model. It
follows that the K\"ahler matrix and its inverse\footnote{The inverse
of the K\"ahler matrix is $K^{\bar{j} k}$ so $K^{k \bar{j}}$ is the
transpose of the inverse} can be written as
\begin{align}
K_{i\bar{\jmath}} &=\frac{3}{\left[T/\Mg+T^{\dagger}/\Mg-\vert \phi\vert^2/(3\Mg^2)\right]^2}
\begin{bmatrix}
 \left(T+T^{\dagger}\right)/(3\Mg) & -\phi^{\dagger}/(3\Mg) \\
-\phi/(3\Mg) & 1
\end{bmatrix}, \\
K^{k\bar{\jmath}} &= \left(\frac{T}{\Mg}+\frac{T^{\dagger}}{\Mg}
-\frac{\vert \phi\vert^2}{3\Mg^2}\right)
\begin{bmatrix}
1 & \phi/(3\Mg) \\
\phi^{\dagger}/(3\Mg) & (T+T^{\dagger})/(3\Mg)
\end{bmatrix}.
\end{align}
Then, assuming that the modulus is stabilized such $\langle
T+T^{\dagger}\rangle =c\Mg$ and $\langle T-T^{\dagger}\rangle =0$, one
obtains the effective Lagrangian
\begin{equation}
  \calL_\ueff = -\dfrac{c}{\Delta^2} \left \vert \partial_\mu\phi
  \right \vert^2 - \dfrac{1}{\Delta^2} \left\vert\dfrac{\partial
W}{\partial\phi}\right\vert^2,
\end{equation}
where $\Delta\equiv c-\vert \phi\vert^2/(3\Mg^2)$. The next step
consists in introducing the fields $x$ and $y$ defined by
\begin{equation}
  \dfrac{\phi}{\Mg} \equiv \sqrt{3c}\tanh\left(\dfrac{x+iy }{\Mg\sqrt{3}}\right).
\end{equation}
Expressed in terms of these two fields, the previous Lagrangian takes
the following form
\begin{equation}
  \begin{aligned}
\calL_\ueff &= -\frac{1}{2\cos^2\left[\sqrt{2/3}(y/\Mg)\right]}
\left[\left(\partial_\mu x\right)^2+\left(\partial_\mu
  y\right)^2\right] \\
&- \frac{\mu^2}{2}\frac{1}{2\cos^2\left[\sqrt{2/3}(y/\Mg)\right]}
\ee^{-\sqrt{2/3}x}\left[\cosh\left(\sqrt{\frac{2}{3}}\frac{x}{\Mg}\right)
-\cos\left(\sqrt{\frac{2}{3}}\frac{y}{\Mg}\right)\right],
  \end{aligned}
\end{equation}
where $\mu\equiv \hat{\mu}\sqrt{3/c}$. In order to obtain this formula, 
we have crucially assumed that
\begin{equation}
\lambda=\frac{\mu}{3\Mg}\,.
\end{equation}
The form of the effective Lagrangian has also been studied in
\Refc{Ellis:2013xoa} in the case where this relation is no longer
valid. The last step consists in remarking that $y=0$ during
inflation. If we expand the above Lagrangian about $y=0$, then the
field $x$ is canonically normalized and the potential becomes
precisely the one of \Eq{eq:potsi}. As such, it constitutes another
scenario where this potential arises.

Let us also notice that other approaches based on superconformal
D-term inflation also lead to the same
potential~\cite{Buchmuller:2013zfa}. Various multifield extensions
have also been studied in which the inflationary phase can still be
described by the one-field Higgs potential~\cite{Lerner:2009xg,
  EliasMiro:2012ay, Arina:2012fb}.

More recently, the Starobinsky model has also been derived from
theories that are conformally invariant (with spontaneous symmetry
breaking). Moreover, the supersymmetric version of these theories,
superconformal theories (with spontaneous breaking of the
superconformal symmetry) have also been shown to lead to the
Starobinsky model, thus providing another supergravity description of
this model. In the following, we present these considerations which
are based on \Refc{Kallosh:2013hoa}. In order to understand the
context in which these models have been developed, it is useful to
first consider the action given by the following expression
\begin{equation}
\label{eq:actioninvaconf}
S\left(g_{\mu \nu},\chi\right)=\frac{\Mg^2}{2}\int \dd ^4 \bmx \, \sqrt{-g}\, 
\left(\frac{\chi^2}{6}R +
g^{\mu \nu}\partial _\mu \chi \partial _\nu \chi
  -\frac{\lambda}{2}\chi^4\right),
\end{equation}
where $\lambda$ is a coupling constant of dimension 2 (here, the field
$\chi$ is dimensionless). It should be noticed that the sign of the
kinetic term for the dimensionless field $\chi$ is the ``wrong''
one. Then, the fundamental remark is that the above action is
invariant under the conformal transformation $\ef{g}_{\mu
  \nu}=e^{-2\sigma}g_{\mu \nu}$ and $\ef{\chi}=e^{\sigma}\chi$, where
$\sigma$ is a dimensionless field. Indeed, if one inserts the previous
transformation into the action~\eqref{eq:actioninvaconf}, one obtains
\begin{equation}
\begin{aligned}
  S\left(g_{\mu \nu},\chi\right)&=\frac{\Mg^2}{2}\int \dd ^4 \bmx \, e^{4\sigma}
  \sqrt{-\ef{g}} \, \biggl\{e^{-2\sigma}\frac{\ef{\chi}^2}{6}
  e^{-2\sigma}\biggl[\ef{R}-6\ef{g}^{\mu \nu}
    \ef{\nabla}_\mu \partial_\nu \sigma
    -6 \ef{g}^{\mu \nu}\partial_\mu \sigma \partial_\nu \sigma \biggr] \\
  &+ e^{-2\sigma}\ef{g}^{\mu \nu}\partial_\mu \left(e^{-\sigma}\ef{\chi}
  \right)\partial_\nu \left(e^{-\sigma}\ef{\chi}\right)
  -\frac{\lambda}{2}e^{-4\sigma}\ef{\chi}^4\biggr\}
  \\
  &=\frac{\Mg^2}{2}\int \dd ^4 \bmx \, \sqrt{-\ef{g}}\,
\left[\frac{\ef{\chi}^2}{6}\ef{R} +
\ef{g}^{\mu \nu}\partial _\mu \ef{\chi} \partial _\nu \ef{\chi}
-\frac{\lambda}{2}\ef{\chi}^4\right]
-\frac{\Mg^2}{2}
\int \dd ^4 \bmx \, \sqrt{-\ef{g}}\biggl[2\ef{g}^{\mu \nu}
  \ef{\chi}\partial_\mu \sigma \partial \ef{\chi} \\
    &+ \ef{\chi}^2\ef{g}^{\mu \nu}\ef{\nabla}_\mu \partial_\nu \sigma
  \biggr],
\end{aligned}
\end{equation}
where we have used that the transformation of the scalar curvature is
given by \Eq{eq:transR}. Using $\ef{\chi}^2\ef{\nabla}_\mu
\partial_\nu \sigma=\ef{\nabla}_\mu
\left(\ef{\chi}^2\partial_\nu\sigma\right)
-\ef{\nabla}_\mu\left(\ef{\chi}^2\right)\partial_\nu \sigma$, the
second term in the above expression reduces to a total derivative,
thus showing that, indeed, the action is invariant, see
\Eq{eq:actioninvaconf}.

The fact that the field $\chi$ has a kinetic term with a ``wrong''
sign is not problematic because, as explained in
\Refc{Kallosh:2013hoa}, it can be removed by fixing its value. If one
takes $\chi=\sqrt{6}$, \Eq{eq:actioninvaconf} reduces to
\begin{equation}
  S\left(g_{\mu \nu}\right)=\int \dd ^4 \bmx \, \sqrt{-g}\,
  \left(\frac{\Mg^2}{2}R-9\lambda \Mg^2\right).
  \label{eq:actionconformon}
\end{equation}
The field $\chi$ is called the ``conformon'' because it is used to
break the conformal symmetry. This last equation is nothing but the
action of GR with a cosmological constant. If this one is positive,
the homogeneous and isotropic solution is de Sitter, the prototype of
a Universe undergoing a phase of inflation.

The second step in~\Refc{Kallosh:2013hoa} consists in introducing a
two-field model, which is a generalization of the
\Eq{eq:actioninvaconf} and is described by the following expression
\begin{equation}
\begin{aligned}
S\left(g_{\mu \nu},\chi,\phi\right) &= \frac{\Mg^2}{2}\int \dd ^4 \bmx \, \sqrt{-g}\, 
\left[\frac{\chi^2}{6}R +
g^{\mu \nu}\partial _\mu \chi \partial _\nu \chi
-\frac{\phi^2}{6}R -
g^{\mu \nu}\partial _\mu \phi \partial _\nu \phi
-\frac{\lambda}{2}\left(\phi^2-\chi^2\right)^2\right].
\label{eq:actioninvaconf2field}
\end{aligned}
\end{equation}
Obviously, this action ressembles \Eq{eq:actioninvaconf}. The field
$\chi$ is still a conformon since its kinetic term has the ``wrong''
sign but we notice that this is not the case for the field $\phi$. It
is also clear that \Eq{eq:actioninvaconf2field} is
invariant under the conformal transformation $\ef{g}_{\mu \nu}=e^{-2
  \sigma}g_{\mu \nu}$, $\ef{\phi}=e^{\sigma}\phi$ and
$\ef{\chi}=e^{\sigma}\chi$. This action possesses an additional
symmetry: it is invariant under global $\mathrm{SO}(1,1)$
transformations in the fields $\phi$ and $\chi$. Let us recall that
this group can be represented by the two-by-two matrices $M$ of the
form
\begin{equation}
  M=
  \begin{pmatrix}
    a & b\\
    b & a
  \end{pmatrix},
\end{equation}
where $a$ and $b$ are real numbers such that $a^2-b^2=1$. If
\begin{equation}
\begin{pmatrix} \ef{\phi}
  \\ \ef{\chi}\end{pmatrix}=M\begin{pmatrix} \phi
\\ \chi \end{pmatrix},
\end{equation}
then $\phi^2-\chi^2$ is a $\mathrm{SO}(1,1)$-invariant and this makes
the invariance of the action~\eqref{eq:actioninvaconf2field} under
$\mathrm{SO}(1,1)$ explicit.

As before, the next step consists in fixing the conformal gauge. A
first example is the so-called ``rapidity'' gauge defined by
$\chi^2-\phi^2=6$, which is $\mathrm{SO}(1,1)$ invariant. Such a gauge
condition does not completely fix the value of the conformon but only
constrains its relationship with the field $\phi$. This constraint can
also be enforced by introducing an additional field $\varphi$ and
demanding that
\begin{equation}
  \begin{aligned}
    \chi & =\sqrt{6}\cosh \left(
      \dfrac{\varphi}{\sqrt{6}\Mg} \right), \qquad 
    \phi &=\sqrt{6}\sinh
    \left(\dfrac{\varphi}{\sqrt{6}\Mg}\right).
  \end{aligned}
\label{eq:rapidityfields}
\end{equation}
Then, the \Eq{eq:actioninvaconf2field} becomes
\begin{equation}
  S\left(g_{\mu \nu},\varphi\right)=\int \dd ^4 \bmx \, \sqrt{-g}\,
  \left(\frac{\Mg^2}{2}R -\frac12 g^{\mu \nu}\partial_\mu \varphi \,
  \partial_\nu \varphi -9\lambda \Mg^2\right).
  \label{eq:actionconformon2fieldgaugefixed1}
\end{equation}
One recovers the action~\eqref{eq:actionconformon} but, this time,
with one additional degree of freedom described by the field $\varphi$
which has a constant potential.

The idea to obtain a less trivial theory is to break the
$\mathrm{SO}(1,1)$ symmetry and to consider the potential
$\lambda\phi^2\left(\phi-\chi\right)^2/4$ instead of
$\lambda(\phi^2-\chi^2)^2/2$ in \Eq{eq:actioninvaconf2field}. In the
rapidity gauge, the potential of the field $\varphi$ now reads
\begin{align}
  V(\varphi)=\frac{9\lambda\Mg^2}{4}\left(1-e^{-\sqrt{2/3}\varphi/\Mg}\right)^2,
\end{align}
which is exactly the Starobinsky model.

The above considerations were generalized to a supersymmetric
framework (superconformal theory and supergravity) in
\Refc{Kallosh:2013lkr}. As is well-known, standard supergravity
depends on two functions, the K\"ahler and super potentials. The
K\"ahler potential leads to the kinetic terms of the fields while the
superpotential allows us to calculate the scalar potential of the
theory. Standard supergravity implies that all the scalars in the
theory are minimaly coupled to gravity. However, supergravity can be
reformulated, leading to conformal supergravity, in such a way that
scalars can non-minimally couple to gravity and, in the following, we
will be interested in this class of models. As it is the case for
standard supergravity, conformal supergravity also depends on two
functions, $\calN$, the embedding K\"ahler potential, and a
superpotential $\calW$. The quantity $\calN$ which appears in front of
the scalar curvature is also used to calculate the kinetic terms of
the fields in the model, namely $G_{I\bar{J}}=\partial^2
\calN/(\partial X^I \partial \bar{X}^{\bar{J}})$. In this context, the
Lagrangian of the theory can be written as
\begin{align}
  \label{eq:superconflagrangian}
  \calL=\sqrt{-g}\left(-\frac{\calN}{6}R-G_{I\bar{J}}\, \partial^\mu X^I
  \partial_{\mu} \bar{X}^{\bar{J}}-V\right).
  \end{align}
In order to implement the Starobinsky model, we use a version where
there are three fields: the so-called compensator field $X^0$, the
inflaton field $X^1=\Phi$ and the so-called Goldstino superfield
$X^2=S$. The compensator field $X^0$ is also called the conformon
field because the superconformal theory becomes supergravity after the
conformal symmetry has been broken which can be achieved when the
conformal field acquires a constant value,
$X^0=\bar{X}^{\bar{0}}=\sqrt{3}\Mg$ (this particular value is chosen
in order to correctly normalize gravity). Then, the corresponding
$N=1$ supergravity theory is described by the following K\"ahler and
super potential
\begin{equation}
  \begin{aligned}
  \calN \left(X^I,\bar{X}^{\bar{I}}\right)\biggl
  \vert_{X^0=\bar{X}^{\bar{0}} = \sqrt{3}\Mg} &= -3\, \Mg^2 \,
  e^{-\frac{K(\Phi,\bar{\Phi},S\bar{S})}{3\Mg^2}},
  \\ \calW\left(X^I\right)\biggl \vert_{X^0=\sqrt{3}\Mg}&= W(\Phi,S).
\end{aligned}
  \label{eq:embedpot}
\end{equation}
The Starobinsky model can be obtained by assuming the following form
for the potential of the embedding manifold and the superpotential
\begin{equation}
  \begin{aligned}
\label{eq:superconformembedpot}
    \calN \left(X^I,\bar{X}^{\bar{I}}\right)&=
  -\vert X^0\vert^2 \exp\left[
    -\frac{\vert S\vert^2}{\vert X^0\vert^2}+\frac12
    \left(\frac{\Phi}{X^0}-\frac{\bar{\Phi}}{\bar{X}^{\bar{0}}}\right)^2
    +\zeta \frac{\vert S\vert^4}{\vert X^0\vert^4}\right],  \\
    \calW \left(X^I\right )&= \frac{M}{2\sqrt{3}\Mg}S(X^0)^2
    \left(1-e^{-2\Phi/X^0}\right),
  \end{aligned}
\end{equation}
where $M$ is a mass scale and $\zeta $ a dimensionless
parameter. Using \Eq{eq:embedpot}, after breaking the conformal
symmetry, one finds that the corresponding K\"ahler and super
potentials are given by
\begin{equation}
\begin{aligned}
  K&=\vert S\vert^2-\frac12 (\Phi-\bar{\Phi})^2-\frac{\zeta}{3}
  \frac{\vert S\vert^4}{\Mg^2}, \\
 W&=\frac{M \Mg \sqrt{3}}{2}S\left[1-e^{-2\Phi/(\sqrt{3}\Mg)}\right].
\end{aligned}
\end{equation}
It is important to notice that the superpotential has the form $W=S
f(\Phi)$. This particular form will play an important role in the
following, in particular in rendering the calculations much
simpler.

From the above expression of the K\"ahler potential, one can calculate
the K\"ahler matrix which reads
\begin{equation}
  G_{A\bar{B}}=\frac{1}{\Mg^2}
  \begin{pmatrix}
    \displaystyle
    1-\frac43 \zeta \frac{S \bar{S}}{\Mg^2} & 0 \\ \\
    0 & 1
  \end{pmatrix}.
\end{equation}
Then, the $F$-term scalar potential can be inferred. It contains two
terms corresponding to the two non-vanishing component of the K\"ahler
matrix and can be expressed as
\begin{equation}
\begin{aligned}
  V &=\frac{e^{K/\Mg^2}}{1-\frac43 \zeta \frac{S\bar{S}}{\Mg^2}}
  \Biggl(\frac{1}{\Mg^4} WW^\dagger \frac{\partial K}{\partial \bar{S}}
 \frac{\partial K}{\partial S} 
+\frac{1}{\Mg^2}W^{\dagger}\frac{\partial K}{\partial \bar{S}}
\frac{\partial W}{\partial S}
+\frac{1}{\Mg^2} W \frac{\partial K}{\partial S}
\frac{\partial W^{\dagger}}{\partial \bar{S}} +\frac{\partial W}{\partial S}
\frac{\partial W^{\dagger}}{\partial \bar{S}}\Biggr) \\ 
&+ e^{K/\Mg^2}
  \Biggl(\frac{1}{\Mg^4} WW^\dagger \frac{\partial K}{\partial \bar{\Phi}}
 \frac{\partial K}{\partial \Phi} 
+\frac{1}{\Mg^2} W^{\dagger}\frac{\partial K}{\partial \bar{\Phi}}
\frac{\partial W}{\partial \Phi}
+\frac{1}{\Mg^2} W \frac{\partial K}{\partial \Phi}
\frac{\partial W^{\dagger}}{\partial \bar{\Phi}} +\frac{\partial W}{\partial \Phi}
\frac{\partial W^{\dagger}}{\partial \bar{\Phi}}\Biggr).
\end{aligned}
\end{equation}
Using the explicit form of the K\"ahler and super potentials, one
finally gets
\begin{equation}
\begin{aligned}
V &= \frac{e^{K/\Mg^2}}{1-\frac43 \zeta \frac{S\bar{S}}{\Mg^2}}
\Biggl[\frac{1}{\Mg^4} S\bar{S} f(\Phi)f(\bar{\Phi})
  \left(S-\frac{2}{3}\zeta \frac{S^2\bar{S}}{\Mg^2}\right)
\left(\bar{S}-\frac{2}{3}\zeta \frac{S\bar{S}^2}{\Mg^2}\right)
 \\ &
+\frac{1}{\Mg^2}\bar{S}f(\bar{\Phi})\left(S-\frac23
\zeta\frac{S^2\bar{S}}{\Mg^2}
\right) f(\Phi)
+\frac{1}{\Mg^2}Sf(\Phi)\left(\bar{S}-\frac23
\zeta\frac{S\bar{S}^2}{\Mg^2}\right) f(\Phi)
+f(\Phi)f(\bar{\Phi})
\Biggr]
\nonumber \\ &
+e^{K/\Mg^2}\Biggl[-\frac{1}{\Mg^4} S\bar{S}
  f(\Phi)f(\bar{\Phi})(\Phi-\bar{\Phi})^2
  +\frac{1}{\Mg^2}\bar{S}f(\bar{\Phi})(\Phi-\bar{\Phi})
  S\frac{\partial f}{\partial \Phi}
\nonumber \\ &
-\frac{1}{\Mg^4} S f(\Phi)(\Phi-\bar{\Phi})\bar{S}
\frac{\partial f}{\partial \bar{\Phi}}
+S \bar{S}\frac{\partial f}{\partial \Phi}\frac{\partial f}
{\partial \bar{\Phi}}\Biggr].
\end{aligned}
\end{equation}
Now if one considers the trajectory $S=0$ and $\alpha=0$, where
$\Phi=(\varphi +i\alpha)/\sqrt{2}$, the potential reduces to
\begin{align}
  V(\varphi)=\frac34 M^2\Mg^2\left(1-e^{-\sqrt{2/3}\varphi/\Mg}\right)^2,
\end{align}
which is exactly the Starobinsky model. We notice that, in the full
scalar potential, all the terms vanish thanks to $S=0$, except
$f(\Phi)f(\bar{\Phi})$, which gives rise the Starobinsky potential. As
already mentioned, this is because the superpotential is of the form
$W=Sf(\Phi)$.

Finally, it is important to notice that the inflationary trajectory
considered above $S=0$ is stable. In fact this is the whole purpose of
introducing the term proportional to the parameter $\zeta $ in
\Eq{eq:superconformembedpot}: it gives a positive mass to the field
$S$ and renders the whole scenario consistent. More details on this
class of models can be found in \Refc{Kallosh:2013lkr}.

\subsubsection{Slow-Roll Analysis}
\label{sec:sisr}

\begin{figure}
\begin{center}
\includegraphics[width=\wdblefig]{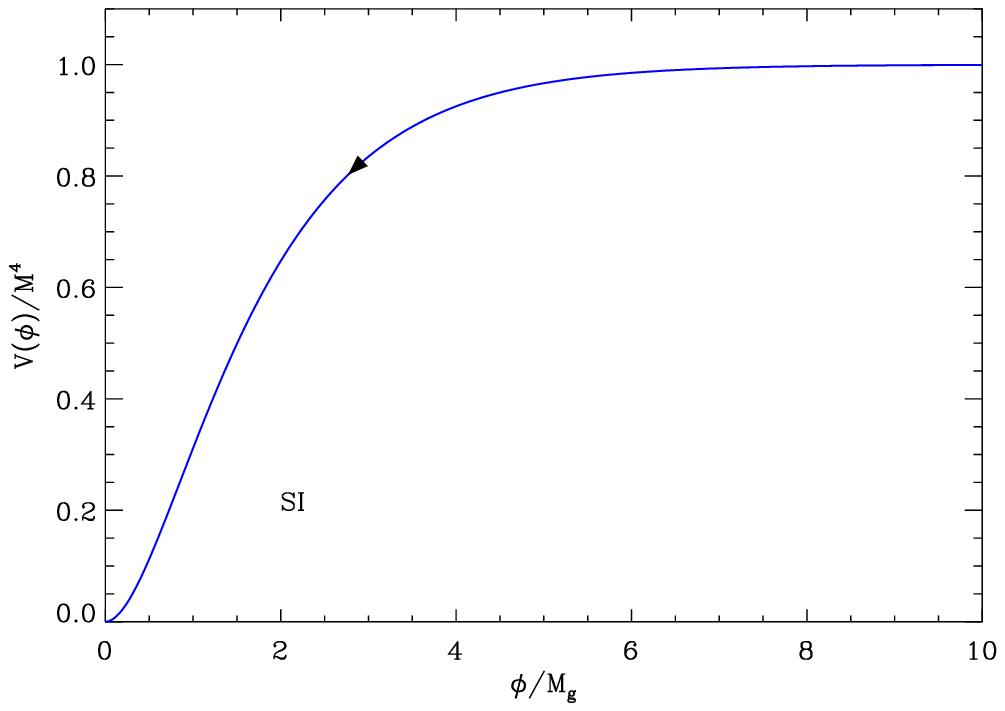}
\includegraphics[width=\wdblefig]{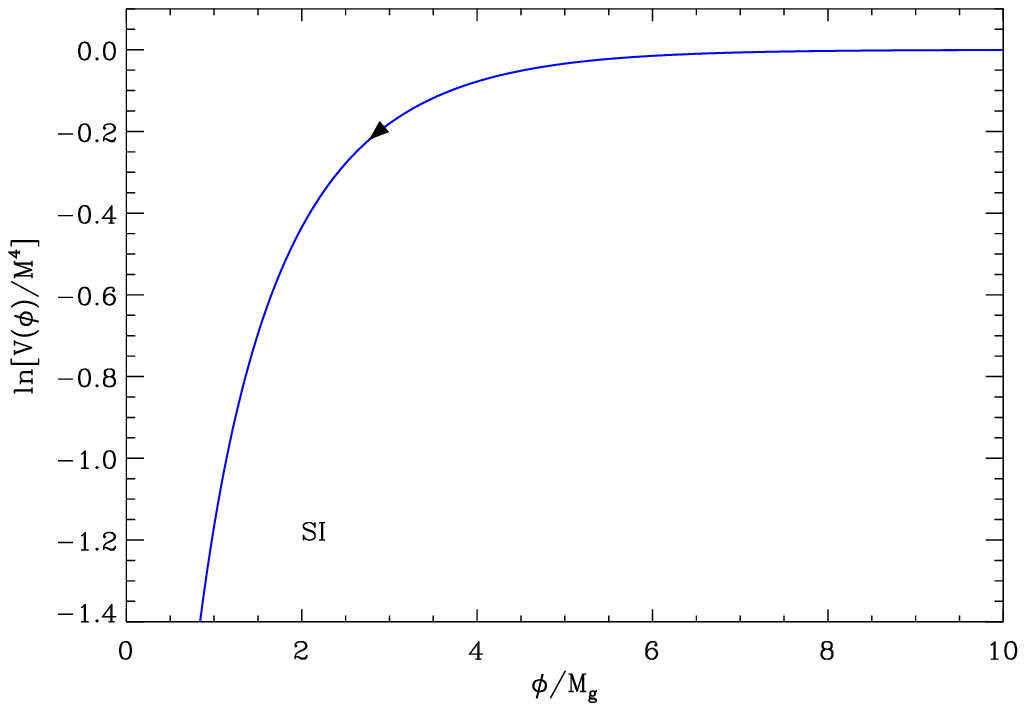}
\includegraphics[width=\wdblefig]{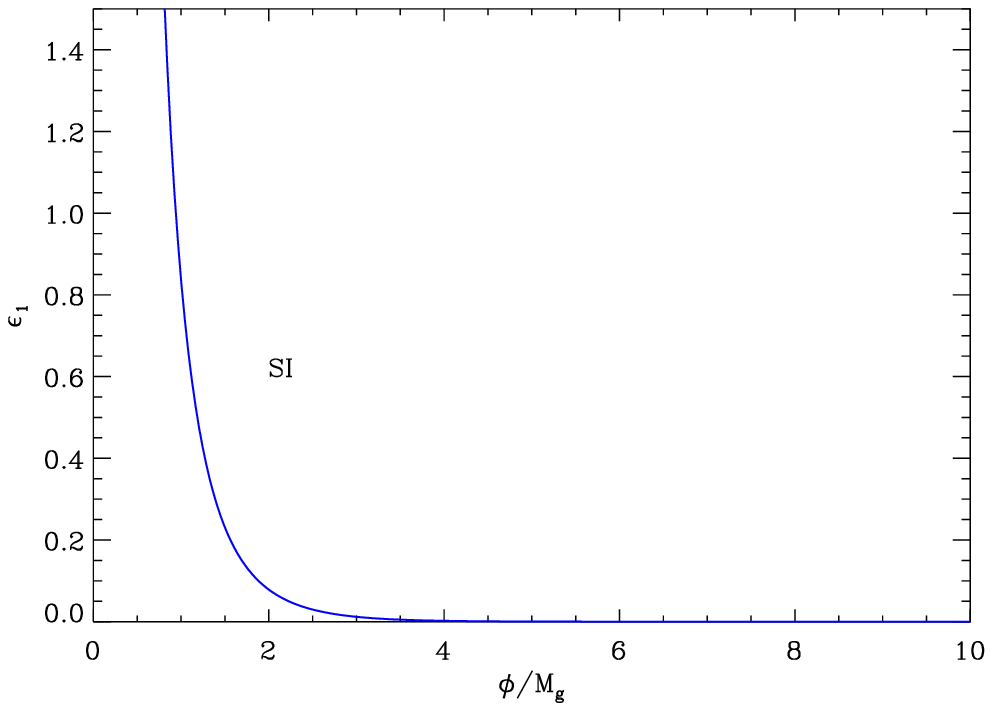}
\includegraphics[width=\wdblefig]{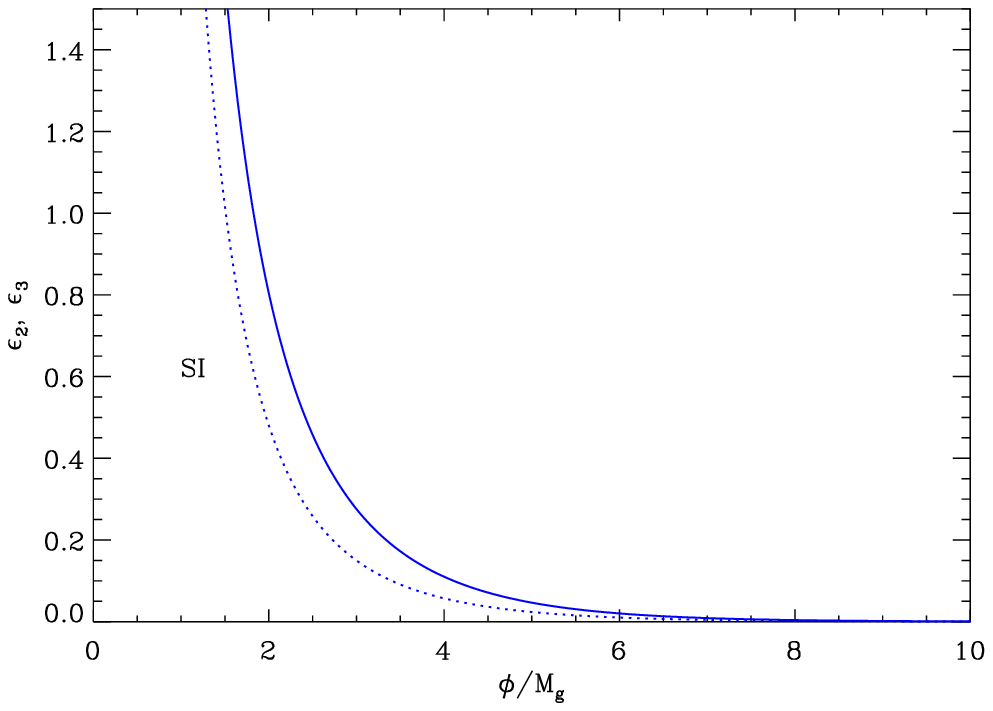}
\caption{Starobinsky Inflation (SI). Top left panel: Starobinsky potential
  corresponding to \Eq{eq:potsi}.  Top right panel:
  logarithm of the potential. It is clear from these two plots
  that inflation proceeds from the right to the left. Bottom left
  panel: slow-roll parameter $\epsilon _1$ as a function of the field
  $\phi$. Inflation ends when $\epsilon_1$ becomes larger than unity.
Bottom right panel: slow-roll parameters
  $\epsilon _2$ (solid line) and $\epsilon _3$ (dotted line) for the
  same potential.}
\label{potsi}
\end{center}
\end{figure}

Let us move back to the original notation and denote by $\phi$ the
inflaton field for the Starobinsky model in the Einstein frame. The
potential is given by \Eq{eq:potsi} and, defining $x\equiv\phi/\Mg$,
the first three slow-roll parameters are given by
\begin{equation}
\begin{aligned}
  \epsilon_1 &= \frac{4}{3}\left(1-\ee^{\sqrt{2/3}x}\right)^{-2}, \qquad
  \epsilon_2 =
  \frac{2}{3}\left[\sinh\left(\frac{x}{\sqrt{6}}\right)\right]^{-2}, \\
  \epsilon_3 & = \frac{2}{3} \left[\coth \left(\frac{x}{\sqrt{6}}\right)
    -1\right]\coth\left(\frac{x}{\sqrt{6}}\right).
\end{aligned}
\label{eq:srparametershiggs}
\end{equation}
Notice that \Eqs{eq:eps1} to \eqref{eq:eps3} are still applicable here
with the formal replacement $\Mp \to \Mg$. These quantities are represented
in \Fig{potsi} (left and right bottom panels) together with the
potential and its logarithm. The minimum of the potential being at
$x=0$, after inflation the numerical value of $\Mg \simeq \Mp$.

In this model, as can be noticed on these plots, inflation stops by
violation of the slow-roll conditions. The condition $\epsilon_1=1$
occurs for $x = \xend$ where $\xend$ can be expressed as
\begin{equation}
\label{eq:endhiggs}
\xend=\sqrt{\frac{3}{2}}\ln\left(1+\frac{2}{\sqrt{3}}\right)\simeq
0.94\, .
\end{equation}
In fact, before the end of inflation, the slow-roll approximation
breaks down when $\epsilon_2$ becomes greater than one. This happens
for $x=\xepstwoOne$ where
\begin{equation}
 \xepstwoOne = \sqrt{6}\,\arsinh
  \left(\sqrt{\frac{2}{3}}\right) \simeq 1.83\, .
\end{equation}
The third slow-roll parameter $\epsilon_3$ also becomes greater than
one before the end of inflation (but after the second slow-roll
parameter has become unity). The corresponding vacuum expectation
value can be written as
\begin{equation}
  \xepsthreeOne = \sqrt{6}\,
  \artanh \left(\frac{2}{1+\sqrt{7}}\right) \simeq 1.51\, .
\end{equation}
\begin{figure}
\begin{center}
\includegraphics[width=\wsingfig]{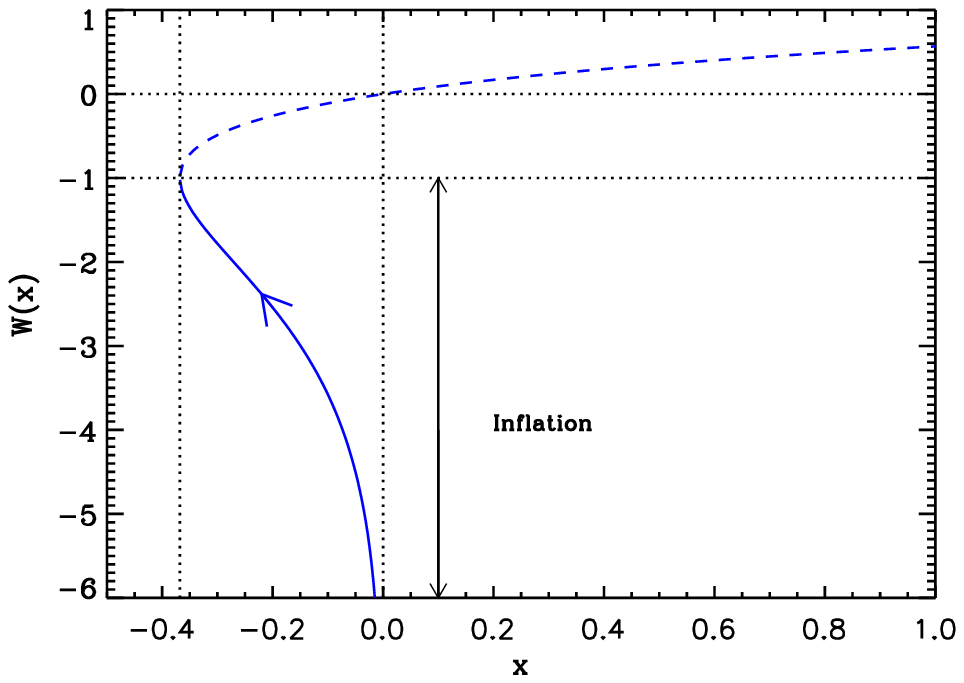}
\caption{Lambert functions $\Lambert{0}(x)$ (dashed line) and
  $\Lambert{-1}(x)$ (solid line). During Starobinsky inflation, inflation
  proceeds along the ``$-1$'' branch in the direction specified by the
  arrow in the figure.}
\label{fig:hi:lambert}
\end{center}
\end{figure}

We can calculate the slow-roll trajectory exactly. Using
\Eq{eq:potsi}, it can be integrated and yields
\begin{equation}
\label{eq:hi:traj}
\Nend-N=\frac12\sqrt{\frac32}\left(\xend-x\right) +
\frac34\left(\ee^{\sqrt{\frac23} x}-\ee^{\sqrt{\frac23}\xend}\right).
\end{equation}
In the regime where $x\gg 1$, the last term is dominant. The
trajectory can be inverted and expressed in term of the
``$-1$-branch'' of the Lambert function $\Lambert{-1}$, leading to
\begin{equation}
\begin{aligned}
  x &= \sqrt\frac32\left\lbrace -\frac43
    \Delta N +\sqrt\frac23 \xend-\ee^{\sqrt{\frac23}\xend} - \Lambert{-1}
    \left[-\exp\left(-\frac43 \Delta N +\sqrt\frac23
        \xend-\ee^{\sqrt{\frac23}\xend} \right)\right]\right\rbrace,
    \label{eq:hi:trajinverted} 
\end{aligned}
\end{equation}
where $\Delta N = \Nend - N$. The fact that inflation proceeds on the
$-1$ branch of the Lambert function $\Lambert{-1}$, as can be seen in
\Fig{fig:hi:lambert}, can be justified by the following
considerations. When $\Delta N=0$, the value taken by the Lambert
function is $-\exp(\sqrt{2/3}\xend)$, which is smaller than $-1$. On
the other hand, if $x=0$, the value given for $\Delta N$ by \Eq{eq:hi:traj}
can be inserted in \Eq{eq:hi:trajinverted} and one finds that the
argument of the Lambert function is $-1$, \ie the connection point
between the $-1$ branch and the $0$ branch.  Therefore inflation takes
place between these two points.

Finally, the value of the inflaton field, $\xstar$, at which the pivot
mode crossed the Hubble radius is related to the number {\efolds}
before the end of inflation by
\begin{equation}
\begin{aligned}
\label{eq:xstarsi}
  \xstar & =\sqrt\frac32\left(-\frac43\Delta
    \Nstar + \ln\left( 1+\frac{2}{\sqrt{3}} \right) - \left(
      1+\frac{2}{\sqrt{3}} \right)
  \right. \\ &\left.
    -\Lambert{-1}\left\lbrace - \exp\left[ -\frac43\Delta
        \Nstar+\ln\left(1+\frac{2}{\sqrt{3}} \right) - \left(
          1+\frac{2}{\sqrt{3}} \right)
      \right] \right \rbrace \right)\, .
\end{aligned}
\end{equation}

Assuming that $\xstar$ is known, the energy scale of the potential is
fixed by the CMB normalization and one obtains
\begin{equation}
\label{eq:si:COBE}
\frac{M^4}{\Mg^4}=1920\pi^2\left(1-\ee^{\sqrt{\frac23}\xstar}\right)^{-4}
\ee^{2\sqrt{\frac23}\xstar} \frac{\Qrms^2}{T^2}\, .
\end{equation}
Upon using the trajectory given by \Eq{eq:xstarsi}, for the fiducial
value $\Delta \Nstar=55$, one gets $M\simeq 3.3\times 10^{-3}\Mg$,
\ie, roughly speaking, inflation takes place at the GUT scale. This
also implies that the mass scale $\mu$ is of the order $\mu\simeq
10^{-5}\Mg\simeq 10^{13}\GeV$.

The actual values of $\Delta\Nstar$ and $\xstar$ are obtained by
solving the reheating equation. However, in the Einstein frame, this
one is no longer given by \Eq{eq:dnstarlnrad}. Indeed, in this frame,
matter is not universally coupled to the metric tensor and, therefore,
it is compulsory to re-consider the parametrization presented in
\sectionc{subsec:reheating}. This is the subject of
\sectionc{sec:EFreheating}.

\subsubsection{Reheating in the Einstein frame}
\label{sec:EFreheating}

In an Einstein frame of metric $g_{\mu\nu}$, the matter action is
given by $\Sm [\psi, A^2(\phi)g_{\mu \nu}]$, where $\psi $ denotes
some generic matter field and $g_{\mu \nu}\equiv F(\chi)
\jf{g}_{\mu\nu}$ with~\cite{EspositoFarese:2000ij}
\begin{equation}
  A \equiv \dfrac{1}{\Omega} = \dfrac{1}{\sqrt{F}}\,.
\end{equation}
Here, $\jf{g}_{\mu\nu}$ denotes the metric in the Jordan frame.  As
most of the inflationary predictions are derived in the Einstein
frame, in this section, we will be using the convenient convention
that quantities in the Jordan frame have a ``bar'' whereas quantities
in the Einstein frame are left untagged. Notice that this differs from
the convention we have used for the theoretical motivations presented
in \sectionc{sec:si}.

In the Jordan frame, the energy density of a (conserved) fluid with a
constant equation of state $w=\jf{p}/\jf{\rho}$ scales as
$\jf{\rho}\propto \jf{a}^{-3(1+w)}$ while, in the Einstein frame,
$\rho \propto A^4 \jf{\rho}\propto A^{1-3w}a^{-3(1+w)}$ since the
scale factors in the two frames are related by $\jf{a}=A a$. As
explained in \Refc{Martin:2010kz} and briefly reviewed in
\sectionc{subsec:reheating}, the dependence of the observational
predictions on reheating originates from the gradient term $k/\calH$
present in the Mukhanov-Sasaki variable equation of motion. In order
to evaluate concretely this term, one must relate the comoving
wave-number $k$ during inflation with physical scales measured
now. Clearly, this depends on the whole history of the universe and,
therefore, explains why the final result depends on the reheating
duration. In the Einstein frame, one can show that the gradient term
takes the standard form, namely
\begin{equation}
\label{eq:gradeinsteinframe}
\frac{\kstar}{\calH}=\frac{\ee^{\Nend-\Nstar}}{H}\frac{\kstar}{\azero}
\left(\frac{\rhoend}{\rhotildegamma}\right)^{1/4}
\frac{1}{\Rrad}, 
\end{equation}
with 
\begin{equation}
\label{eq:Rradeinstein}
\ln \Rrad = \frac{1-3 \wreh}{12(1+\wreh)}\ln \left(\frac{\rhoreh}{\rhoend}\right)
-\frac{1-3\wreh}{3(1+\wreh)}\ln \left(\frac{\Areh}{\Aend}\right),
\end{equation}
where $\wreh$ is the equation of state of the effective dominant fluid
during reheating. In the above expressions, it is important to
emphasize that all the quantities are defined in the Einstein frame
and that the non-standard scaling of the various energy densities
(pressure-less matter and radiation) has been systematically taken
into account. All the extra terms cancel out except in the expression
of the parameter $\Rrad$ where there is an additional term
that involves the function $A$. Remarkably, this additional term is
exactly such that the parameter $\Rrad$ can be re-expressed in
terms of the energy densities in the Jordan frame only, namely
\begin{equation}
\label{eq:lnRradbar}
\ln\Rrad = \ln \Rradbar \equiv \frac{1-3 \wreh}{12(1+\wreh)}\ln 
\left(\frac{\rhorehbar}{\rhoendbar}\right).
\end{equation}
In other words, this is exactly the parameter $\Rradbar$ that one
would have defined by looking only at energy densities in the Jordan
frame. Let us stress again that the above equation has an unusual
form: it is a quantity used in the Einstein frame but expressed in
terms of quantities defined in the Jordan frame.

It is also important to notice an additional limitation compared to
the standard case: in presence of non-minimal coupling to gravity, our
parametrization of the reheating stage works only for a constant
equation of state $\wreh$ while in \Refc{Martin:2010kz} it was valid for
any $\wreh$. We now explain the origin of this limitation. In the Einstein
frame, the general expression of the parameter $\Rrad $ is given by
\begin{equation}
\frac{1}{\Rrad}=\left(\frac{\rhoreh}{\rhoend}\right)^{1/4}
\frac{\areh}{\aend}.
\end{equation}
In order to obtain \Eq{eq:Rradeinstein} from that formula, one should
express the Einstein frame scale factor in term of the energy density
$\rho$. If the equation of state $\wreh$ is a constant, then $a\propto
A^{(1-3\wreh)/(3+3\wreh)}a^{-1/(3+3\wreh)}$. This is what has been
used above and this led to \Eqs{eq:Rradeinstein}
and~(\ref{eq:lnRradbar}). But let us now assume that $\wreh$ is
not a constant (notice that one always has $w=\jf{w}$ since, in the Einstein frame, the
energy density and the pressure scale with the same power of the
function $A$). Then, $\rho $ and $a$ are related
by
\begin{equation}
\frac{\dd \rho}{\rho}=\left(1-3\wreh\right)\frac{\dd A}{A}
-3\left(1+\wreh\right)\frac{\dd a}{a}.
\end{equation}
If $A$ is a constant, one can always write~\cite{Martin:2010kz}
\begin{equation}
\frac{\areh}{\aend}=
\left(\frac{\rhoreh}{\rhoend}\right)^{-1/(3+3\wrehbar)},
\end{equation}
where $\wrehbar$ is the mean equation of state during reheating, namely
\begin{equation}
\wrehbar\equiv \frac{1}{\Nreh-\Nend}\int _{\Nend}^{\Nreh}\wreh(n)\dd n.
\end{equation}
If $A$ and $\wreh$, however, are not constant, it is no longer
possible to express the final formula in terms of $\wrehbar$. In
particular, we do not obtain a term $A^{1-3\wrehbar}$ as
desired. Therefore, in what follows, we restrict our considerations to
the case where the effective fluid dominating the matter content of
the Universe has a constant equation of state.

Then, from \Eqs{eq:gradeinsteinframe} and \eqref{eq:lnRradbar}, at a
given $\Rradbar$, the remaining terms can be re-expressed in
terms of quantities defined at Hubble radius crossing by using the
Friedmann-Lema\^{\i}tre equations. In particular the energy density at
the end of inflation in the Einstein frame reads
\begin{equation}
\dfrac{\rhoend}{\Mp^4} = \dfrac{3\Hstar^2}{\Mp^2}
\dfrac{\Vend}{\Vstar} \dfrac{3-\epsonestar}{3-\epsoneend}\,,
\label{eq:rhoendFL}
\end{equation}
from which one obtains
\begin{equation}
\label{eq:dnstareinstein}
\Delta \Nstar=\ln \Rradbar -\ln \left(\frac{\kstar/\azero}{\rhotildegamma^{1/4}}\right)
+\frac14 \ln \left(\frac{\Hstar^2}{\Mp^2\epsonestar}\right)
-\frac{1}{4}\ln \left(\frac{3}{\epsonestar}\frac{\Vend}{\Vstar}
\frac{3-\epsonestar}{3-\epsoneend}\right).
\end{equation}
In this last equation we have voluntarily made explicit the term in
$\Hstar^2/(\Mp^2 \epsonestar) = 8\pi^2 \Pstar$, the amplitude of the
primordial power spectrum at the pivot scale, a well measured
quantity. Of course, this equation resembles a lot \Eq{eq:dnstarlnrad}
but one has to realize that it involves quantities defined both in the
Einstein frame and in the Jordan frame. The term
\begin{equation}
  \dfrac{\kstar/\azero}{\rhotildegamma^{1/4}} =
  \dfrac{\kstar/\jf{a}_\zero}{\rhotildegammabar^{1/4}} = e^{\Nzero},
\end{equation}
and, therefore, its numerical value, is the standard one. The other
quantities appearing in this equation are obtained using our standard
procedures since they refer to the inflaton sector only.

However, the fact that $\ln\Rradbar$, defined with energies in the
Jordan frame, appears in \Eq{eq:dnstareinstein}, has various
implications. For instance, the range of variation of $\Delta \Nstar$
in \Eq{eq:dnstareinstein} is determined by putting limits on $\ln
\Rradbar$ coming from the fact that reheating must proceed between
the end of inflation and BBN. This means that the physical value
of the energy density at the end of reheating, that is to say
$\rhorehbar$, must be such that $\rhonucbar\equiv \left(10
\MeV\right)^4\le \rhorehbar \le \rhoendbar$. We emphasize that
physical limits must refer to quantities defined in the Jordan
frame. The possible range for $\Delta \Nstar$ is $\left[\Delta
  \Nstarnuc,\Delta \Nstarend\right]$. The upper bound is obtained from
the saturating value $\rhorehbar=\rhoendbar$, which implies that
$\ln\Rradbar = 0$, and then
\begin{equation}
\Delta \Nstarend=-\Nzero + \frac{1}{4} \ln \left(8\pi^2 \Pstar\right)
-\frac{1}{4}\ln \left(\frac{3}{\epsonestar}\frac{\Vend}{\Vstar}
\frac{3-\epsonestar}{3-\epsoneend}\right).
\end{equation}
All the quantities in the above equation are calculated in the
Einstein frame and are therefore unchanged compared to their standard
value. The other limit is $\rhorehbar=\rhonucbar$ and gives
\begin{equation}
\begin{aligned}
  \Delta \Nstarnuc & =-\Nzero + \frac{1}{4} \ln \left(8\pi^2 \Pstar\right)
-\frac{1}{4}\ln \left(\frac{3}{\epsonestar}\frac{\Vend}{\Vstar}
\frac{3-\epsonestar}{3-\epsoneend}\right)
+\frac{1-3\wreh}{12(1+\wreh)}\ln \left(\frac{\rhonucbar}{\Mp^4}\right)
\\ & - \frac{1-3\wreh}{12(1+\wreh)}\ln \left(\frac{\rhoendbar}{\Mp^4}\right).
\end{aligned}
\label{eq:DeltaNnucEF}
\end{equation}
The quantity $\rhonucbar$ is defined in the Jordan frame but its
value is explicitly known, see above. On the other hand, we need to
evaluate $\rhoendbar$ since the Friedmann-Lema\^{\i}tre equations only
determine $\rhoend$ by \Eq{eq:rhoendFL}. By
definition, we have
\begin{equation}
\rhoendbar=\dfrac{\rhoend}{A_\uend^4} = \Omegaend^4 \rhoend.
\end{equation}
Plugging this expression into \Eq{eq:DeltaNnucEF} and making use of
\Eq{eq:rhoendFL} one gets
\begin{equation}
\begin{aligned}
  \Delta \Nstarnuc & =-\Nzero + \dfrac{3 \wreh+1}{6(1+\wreh)} \ln \left(8\pi^2 \Pstar\right)
-\frac{1}{3(1+\wreh)}\ln \left[\frac{3}{\epsonestar^{(1+3\wreh)/2}} \frac{\Vend}{\Vstar}
\frac{3-\epsonestar}{3-\epsoneend}\right]
\\ & +\frac{1-3\wreh}{12(1+\wreh)}\ln
\left(\frac{\rhonucbar}{\Mp^4}\right) - \frac{1-3\wreh}{3(1+\wreh)}
\ln | \Omegaend |.
\end{aligned}
\label{eq:DeltaNnucEFfull}
\end{equation}
All terms but the last one are standard. The scalar-tensor effects appear
in the term containing $\ln|\Omegaend|$. In most cases of interest, it
is a very small correction which, for SI, amounts to
\begin{equation}
\ln|\Omegaend| = \dfrac{1}{2} \ln|F(\xend)| = \dfrac{\xend}{\sqrt{6}} =
\dfrac{1}{2} \ln\left(1+\dfrac{2}{\sqrt{3}}\right) \simeq 0.38.
\end{equation}
Even though it is a small effect, the scalar-tensor corrections on
reheating are all included in the {\ASPIC} library when the
inflationary models are solved in the Einsten frame.

The reheating-consistent observational predictions of Starobinsky
Inflation are represented in \Fig{fig:CMBHI} where we have displayed
their dependence on the reheating temperature.

\subsection{Higgs Inflation (HI)}
\label{sec:hi}

\subsubsection{Non-minimal gravity}
\label{sec:nonmingrav}

We start this section with some general considerations about
non-minimal gravity or scalar-tensor theories. Non-minimal gravity
plays an important role throughout this article for various
reasons. Among them is the fact that the extra terms (compared to
Einstein gravity) that characterize non-minimal gravity seem to be
generated ``automatically'' by quantum corrections. From an effective
field theory point of view, these models are therefore very
well-motivated. Regarding inflation, as it will be discussed in
details later on, non-minimal gravity can be used to ``save'' a model
of inflation, that is to say a model can be ruled out when considered
in the framework of Einstein gravity but compatible with the data when
studied in a non-minimal setup. Several examples illustrating this
claim will be studied in the following. Finally, in the inflationary
context and contrary to what the name suggests, non-minimal gravity
can be viewed as a framework which is as simple as Einstein
gravity. Indeed, as we have already seen with Starobinsky inflation,
non-minimal gravity alone has the same field content as Einstein
gravity plus an additional scalar field. A simple inflationary model
can thus be built only from the scalar-tensor action (of course,
matter is needed when reheating is investigated but, again, the field
content can be the same in both approaches). For all these reasons,
inflationary scenarios based on scalar-tensor theories play an
important role in the current efforts to understand the model building
problem of inflation.

We have already discussed the $f(R)$ theory and how it is in fact
equivalent either to the Brans-Dicke theory or to Einstein gravity
plus a scalar field. Here, we discuss the same question by starting
straightaway from a scalar-tensor theory, which can also be reduced to
Einstein gravity plus a scalar field.

Let us consider the general action defining a scalar-tensor theory in
a Jordan frame of metric $\jf{g}_{\mu\nu}$. Here, in order to avoid
any confusion, we will explicitly follow the conventions of
\sectioncs{sec:theorysi} and \ref{sec:EFreheating} and denote
quantities in the Jordan frame with a ``bar'' and quantities in the
Einstein frame with a ``tilde''. Such an action reads
\begin{equation}
\begin{aligned}
  S\left(\jf{g}^{\mu \nu},\jf{\phi},\psi \right)&= \frac{\Mg^2}{2}\int \dd^4 \bmx
  \sqrt{-\jf{g}} \left[F(\jf{\phi})\jf{R}-Z(\jf{\phi})\jf{g}^{\mu \nu}\partial_\mu \jf{\phi}
    \partial_\nu \jf{\phi} -2U(\jf{\phi})\right] + \int\dd^4 \bmx \, \calL_\umat
  \left(\psi,\jf{g}_{\mu \nu}\right).
\end{aligned}
  \label{eq:actionst}
\end{equation}
The gravity sector is characterized by three functions,
$F(\jf{\phi})$, $Z(\jf{\phi})$ and $U(\jf{\phi})$ and the mass scale
$\Mg$. Different representations can be used, for instance the
Brans-Dicke representation where $F(\jf{\phi})=\jf{\phi}$ and
$Z(\jf{\phi})=\omega(\jf{\phi})/\jf{\phi}$ or the simple
representation where, after having canonically renormalized the field
$\jf{\phi}$, one has $F(\jf{\phi})$ arbitrary and
$Z(\jf{\phi})=1$. However, sometimes, this representation can be
pathological and, in the most general situation, one has to keep the
three functions.

Let us now consider the following conformal transformation,
$\ef{g}_{\mu \nu}=F(\jf{\phi})\jf{g}_{\mu \nu}$, where $\ef{g}_{\mu
  \nu}$ is the metric tensor in the Einstein frame. Using
\Eq{eq:transR}, the action becomes
\begin{equation}
\begin{aligned}
  S\left(\ef{g}^{\mu \nu},\jf{\phi},\psi \right)&= \int \dd^4 \bmx
  \sqrt{-\ef{g}}\biggl[\frac{\Mg^2}{2}\ef{R} +\frac{\Mg^2}{6} \ef{g}^{\mu
      \nu}\ef{\nabla}_\mu \partial _\nu \left(\ln F\right)
    -\frac{3\Mg^2}{4F^2}\ef{g}^{\mu \nu}\partial _\mu F\partial_\nu F
    \\ & -\frac{\Mg^2Z(\jf{\phi})}{2F(\jf{\phi})}\ef{g}^{\mu
      \nu}\partial_\mu \jf{\phi} \partial_\nu \jf{\phi}
    -\Mg^2\frac{U(\jf{\phi})}{F^2(\jf{\phi})}\biggr] + \int\dd^4 \bmx \,
  \calL_\umat \left[\psi,F^{-1}(\jf{\phi})\ef{g}_{\mu \nu}\right].
\end{aligned}
\end{equation}
The term which contains $\ln F$ is a total derivative and,
therefore, can be discarded. If one introduces the field
$\ef{\phi}$ defined by the relation
\begin{align}
  \label{eq:dphisquare}
  \left(\frac{\dd \ef{\phi}}{\dd \jf{\phi}}\right)^2=2
  \left[\frac{3\Mg^2}{4F^2(\jf{\phi})}\left(\frac{\dd F}{\dd \jf{\phi}}\right)^2
    +\frac{\Mg^2Z(\jf{\phi})}{2F(\jf{\phi})}\right],
  \end{align}
and if one defines a new potential $V=\Mg^2U/F^2$, then
\Eq{eq:actionst} takes the following form
\begin{align}
  S\left(\ef{g}^{\mu \nu},\ef{\phi},\psi \right)&=  \int \dd^4 \bmx
  \sqrt{-\ef{g}}\biggl[\frac{\Mg^2}{2}\ef{R}
-\frac{1}{2}\ef{g}^{\mu \nu}\partial _\mu \ef{\phi}\partial_\nu \ef{\phi}
-V(\ef{\phi})\biggr]+
  \int\dd^4 \bmx \, \calL_\umat
  \left[\psi,F^{-1}(\ef{\phi})\ef{g}_{\mu \nu}\right].
\end{align}
As announced before, one recognizes the action of Einstein gravity
plus a scalar field with a minimal kinetic term. As it was the case
for the $f(R)$ theory, one also notices that matter is no longer
universally coupled to the metric tensor, as revealed by the term
$F^{-1}(\ef{\phi})\ef{g}_{\mu \nu}$ in the matter action. As already
mentioned, this has implications for reheating and these have been
discussed in \sectionc{sec:EFreheating}.

\subsubsection{Theoretical Justifications}
\label{subsubsec:theoryhi}

Having briefly discussed non-minimal gravity, let us now apply it to
Higgs inflation (the previous calculations will be useful in many
other contexts throughout this article). This model has been proposed
in \Refcs{Bezrukov:2007ep, Bezrukov:2008ej, Bezrukov:2009db,
  GarciaBellido:2011de} and postulates that the inflaton field is the
Higgs field $\Sigma$ (discovered in 2012 at the Large Hadron
Collider~\cite{:2012gk,:2012gu}) non-minimally coupled to
gravity. Indeed, one can argue that, in curved spacetime, the simplest
model compatible with our knowledge of particle physics is described
by a Lagrangian which is the standard model Lagrangian plus an extra
term of the form $\xi \Sigma^\dagger \Sigma R$. As already argued,
this last term is natural since, in curved spacetime, it should be
automatically generated by quantum
corrections~\cite{Birrell:1982ix}. In the Jordan frame, the action of
the model can be written as
\begin{equation}
\label{eq:actionjordan}
S=\frac{\Mg^2}{2}\int \dd ^4 \bmx \sqrt{-\jf{g}}
\left[F\left(h\right)\jf{R} -Z\left(h\right)
\jf{g}^{\mu \nu}\partial _\mu h \partial _\nu h
  -2U\left(h\right)\right],
\end{equation}
where we have defined, in the unitary gauge,
\begin{equation}
  \Sigma = \dfrac{\Mg}{\sqrt{2}} \left(
  \begin{array}{c}
    0\\h
  \end{array}
  \right),
\end{equation}
and the quantity $\Mg$ is a mass scale that, for the moment, is not
identified with the Planck scale. As before, the tensor $\jf{g}_{\mu \nu}$
denotes the metric in the Jordan frame. The three functions $F(h)$,
$Z(h)$ and $U(h)$ completely characterize the model and are chosen to
be
\begin{equation}
\label{eq:Higgs:Jordan:pot:TreeLevel}
  F(h)=1+\xi h^2, \quad Z(h)=1, \quad U(h)=\Mg^2\frac{\lambda}{4}
  \left(h^2-\frac{v^2}{\Mg^2}\right)^2,
\end{equation}
where $\xi $ is a new dimensionless parameter. The quantity $U(h)$ is
the standard Higgs boson potential with $v \simeq 246\,\GeV$, the
Higgs vacuum expectation value, and $\lambda \simeq 0.13$ the
self-interacting coupling constant. Here, the field $h$ is
dimensionless (as the functions $F$ and $Z$) while the potential $U$
is of dimension two. The effective gravitational constant (measured in
Cavendish-type experiments) is affected by the new scalar degree of
freedom and given by~\cite{EspositoFarese:2000ij}
\begin{equation}
\frac{1}{\Mp^2}=\frac{1}{\Mg^2}\frac{1+\xi h^2+8 \xi^2h^2}
{(1+\xi h^2)\left(1+\xi h^2+6\xi^2h^2\right)}\,.
\label{eq:cavendish}
\end{equation}
Since, today, one has $h\simeq v/\Mg\ll 1$, it follows that $\Mg\simeq
\Mp$ at an accuracy far greater than the uncertainties associated with
the measurements of the gravity coupling constant. As a result, from
now on, we will take $\Mp$ as the numerical value of $\Mg$ in HI (see
\sectionc{sec:nmlfi} for a model in which the equality is not
always satisfied).

The above-described model can also be written in the Einstein frame
where the corresponding slow-roll analysis is easier. For clarity, let
us drop the ``tilde'' above Einstein frame quantities and denote in
the following the metric tensor in this frame by $g_{\mu \nu}$. The
action now takes the form
\begin{equation}
S= \Mg^2 \int \dd ^4 \bmx \sqrt{-g}\left[\frac{R}{2}
-\frac{1}{2} g^{\mu \nu}\partial _\mu \chi \partial _\nu \chi
- W\left(\chi\right)\right],
\end{equation}
where the fields $h$ and $\chi$ are related by
\begin{equation}
\label{eq:relejframes}
\frac{\dd \chi}{\dd h}=\frac{\sqrt{1+\xi(1+6\xi)h^2}}{1+\xi h^2}\,,
\end{equation}
and the potential is given $V\equiv \Mg^2 W =\Mg^2 U/F^2$. Notice
also that the canonically normalized field in the Einstein frame is
simply given by $\phi \equiv \Mg \chi$. It is also important to
recall that, in the Einstein frame, matter is now explicitly coupled
to the scalar field $\phi$. The consequences for reheating are
discussed in \sectionc{sec:EFreheating} and
\Refcs{GarciaBellido:2008ab,Bertolami:2010ke,Motohashi:2012tt}. The
differential obtained in \Eq{eq:relejframes} can be integrated exactly
and the result reads
\begin{equation}
\label{eq:chihexplicit}
\chi=\sqrt{\frac{1+6 \xi}{\xi}}\arsinh 
\left[h\sqrt{\xi(1+6\xi)}\right]
-\sqrt{6}\artanh\left[\frac{\xi \sqrt{6} h}
{\sqrt{1+\xi(1+6\xi)h^2}}\right].
\end{equation}
The inverse hyperbolic tangent is always well-defined since its
argument is always smaller than one. This exact formula between the
Einstein and Jordan frame fields was also derived in
\Refc{GarciaBellido:2008ab}. Using the identities
\begin{equation}
  \arsinh(x)=\ln\left(x+\sqrt{1+x^2}\right), \qquad
  \artanh(x)=\dfrac{1}{2} \ln\left(\dfrac{1+x}{1-x}\right),
\end{equation}
and defining
\begin{equation}
\barh \equiv \sqrt{\xi}\,h,
\end{equation}
\Eq{eq:chihexplicit} can be further simplified as
\begin{equation}
  \chi = \sqrt{6 + \dfrac{1}{\xi}} \ln \left[ \sqrt{1+(1+6\xi) \barh^2}
  + \sqrt{(1+6\xi) \barh^2} \right] + \sqrt{6} \ln
\left[\dfrac{\sqrt{1+ \barh^2}}{\sqrt{1+(1+6\xi)\barh^2} + \sqrt{6\xi
      \barh^2}} \right].
\label{eq:chifrombarh}
\end{equation}
Higgs Inflation is usually considered in the large coupling limit $\xi
\gg 1$ from which one gets
\begin{equation}
\chi \simeq \sqrt{6} \ln(2 \barh \sqrt{6\xi}) + \sqrt{6}
\ln\left( \dfrac{\sqrt{1+\barh^2}}{2 \barh \sqrt{6\xi}} \right) = 
\sqrt{6} \ln\left(\sqrt{1+\barh^2}\right) = \sqrt{\dfrac{3}{2}}
\ln\left(1+ \xi h^2\right).
\label{eq:chihapprox}
\end{equation}

The same expression can also be directly derived from
\Eq{eq:relejframes} which, for $\xi \gg 1$, can be
approximated as
\begin{equation}
\label{eq:frameapprox}
\frac{\dd \chi}{\dd h}\simeq \frac{\sqrt{6}\xi h}{1+\xi h^2}\,.
\end{equation}
The solution to this equation is exactly \Eq{eq:chihapprox}. The last
step consists in inserting the approximate expression of $h$ in terms of $\chi$
(and, therefore, in terms of $\phi$) into the definition of the
potential $V$ in the Einstein frame. This leads to the following
expression
\begin{equation}
\label{eq:pothiggs}
  V(\phi) \simeq \frac{\Mg^4\lambda}{4\xi^2}
\left(1-\ee^{-\sqrt{2/3}\phi/\Mg}\right)^2 ,
\end{equation}
{\ie} one obtains the same potential as in Starobinsky Inflation, see
\Eq{eq:potsi}. Interestingly enough, the parameters $\xi$ and $\lambda
$ enter the approximate potential only through its overall
amplitude. In the following, we define
\begin{equation}
  M^4 \equiv \dfrac{\Mg^4 \lambda}{4\xi^2} \simeq \dfrac{\Mp^4 \lambda}{4\xi^2}\,.
\label{eq:hiM4}
\end{equation}
In this sense, Higgs inflation is, as for Starobinsky inflation, a
``zero-parameter model'' since the scale $M$, and the parameter $\xi$,
are entirely determined by the amplitude of the CMB anisotropies.

Let us stress, however, that the above potential is only approximate
for Higgs inflation whereas \Eq{eq:potsi} is \emph{exact} for
Starobinsky inflation. The two models match at leading order in
$1/\xi$ only but are not strictly identical. In the slow-roll analysis below,
we will be providing both a next-to-leading order and parametric exact
treatment of Higgs inflation to quantity by how much the observable
quantities between the two scenarios can differ.

Finally, let us also notice that other approaches based on
superconformal D-term inflation also lead to the same
potential~\cite{Buchmuller:2013zfa}. Various multifield extensions
have also been studied in which the inflationary phase can still be
described by the one-field Higgs potential~\cite{Lerner:2009xg,
  EliasMiro:2012ay, Arina:2012fb}.

\subsubsection{Next-to-Leading Order Slow-Roll Analysis}
\label{subsubsec:leadhisr}

As explained above, the leading-order potential of Higgs Inflation,
\Eq{eq:pothiggs}, is the same as the one in Starobinsky Inflation, see
\Eq{eq:potsi}, for which the slow-roll analysis was already performed
in \sectionc{sec:sisr}. Therefore, the results derived in
\sectionc{sec:sisr} also apply to Higgs inflation at leading order.

\begin{figure}
\begin{center}
\includegraphics[width=\wdblefig]{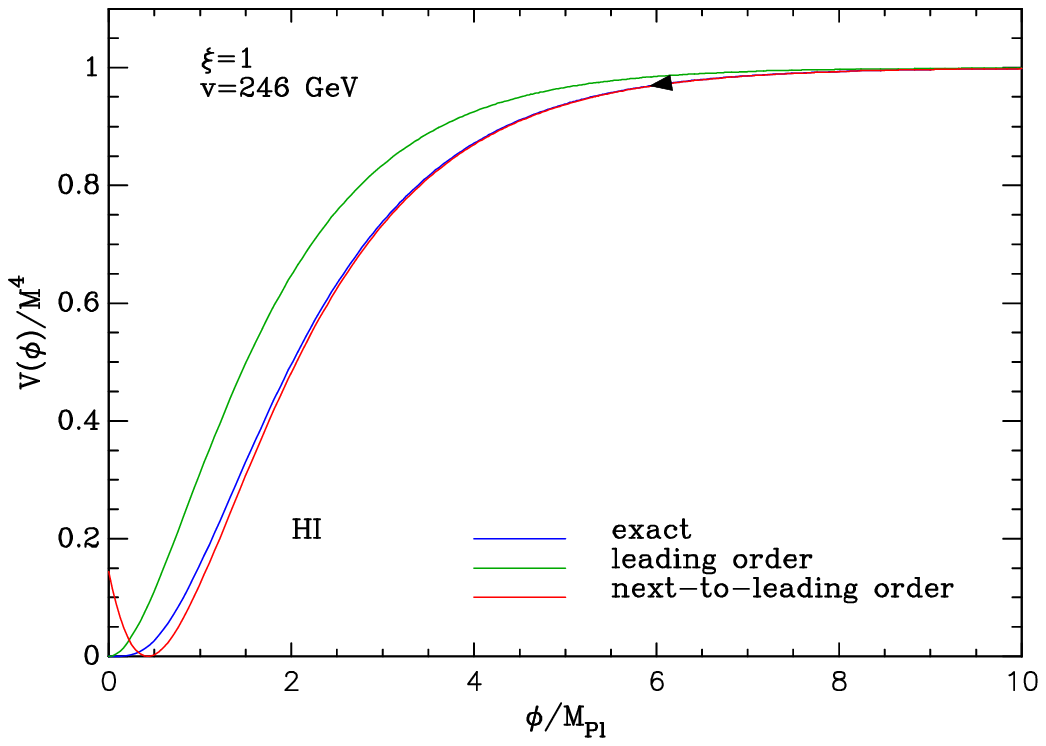}
\includegraphics[width=\wdblefig]{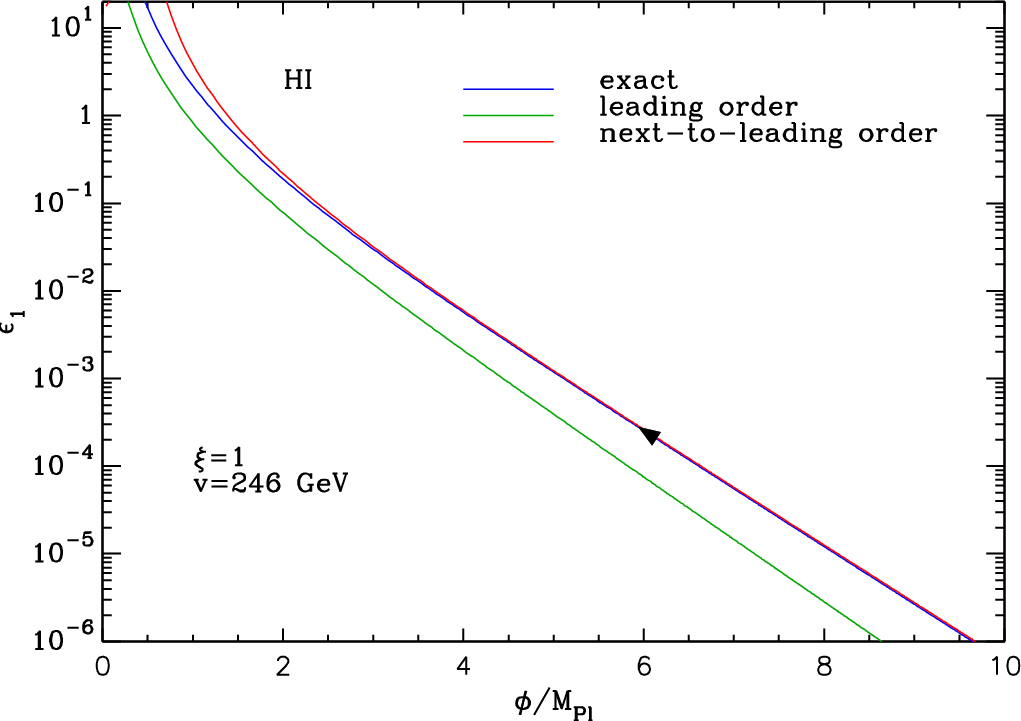}
\caption{Higgs Inflation (HI). Left panel: full Higgs potential for
  $\xi=1$ and $v=246\,\GeV$ (blue line) compared with a leading-order
  expansion in $\xi$ [green line, corresponding to \Eq{eq:pothiggs}]
  and to a next-to-leading expansion in $\xi$ [red line, corresponding
    to \Eq{eq:pothiggs:nlo}]. The value $\xi=1$ is unrealistically
  small but it has be chosen to make the difference between the three
  curves visible. Right panel: first Hubble-flow parameter in the same
  situation as in the left panel.}
\label{fig:pothi}
\end{center}
\end{figure}
There are however two differences between these models that we now
further discuss. In Higgs inflation, \Eq{eq:pothiggs} is only an
approximation to the full potential, since the limit $\xi\gg 1$ was
taken when inverting \Eq{eq:chihexplicit}. In general, Higgs Inflation
thus depends on two additional parameters, namely $\xi$ (which
otherwise only appears in the overall normalization of the potential)
and $v$ (the current vacuum expectation value of the Higgs field). It
is displayed in \Fig{fig:pothi} for $\xi=1$ and $v=246\,\GeV$, and
compared with the approximation derived in \Eq{eq:pothiggs}. As we
will see below, the value $\xi=1$ is unrealistically small, but it
allows one to distinguish between the two curves, otherwise the
difference would not be visible by eye. One can check that, at
large-field values, \Eq{eq:pothiggs} indeed provides a good
approximation to the full potential.

In order to better assess the reliability of this approximation, let
us carry it out at next-to-leading order. Expanding
\Eq{eq:chihexplicit} at order $\xi^{-1}$, thus at next-to-leading
order compared to \Eq{eq:pothiggs}, one obtains for the inflationary
potential
\begin{align}
\label{eq:pothiggs:nlo}
V=\frac{\lambda \Mg^4}{4\xi^2} \left(1-\ee^{- \sqrt{\frac{2}{3}}x}\right)^2,
\qquad\text{with}\qquad
x=  \frac{\phi/\Mg}{1+1/(12\xi)}- \sqrt{\frac{3}{2}}\frac{1+\ln(24 \xi)}{1+12\xi}\, .
\end{align}
When $\xi\gg 1$, one has $x\simeq \phi/\Mg$ so \Eq{eq:pothiggs} is
recovered. Let us note that we have neglected terms of order
$v/\Mg \simeq 10^{-16}$ when deriving this expression. It is displayed
with the red solid line in \Fig{fig:pothi} and one can see that already
with $\xi=1$, it provides an excellent approximation to the full
potential in the inflating region. In the right panel of \Fig{fig:pothi},
the first Hubble-flow parameter is also displayed. At next-to-leading
order, from \Eq{eq:pothiggs:nlo}, it is given by
\begin{align}
\epsilon_1 = \frac{4}{3}\left[\frac{1-\ee^{\sqrt{2/3}x}}{1+1/(12 \xi)}\right]^2 ,
\end{align}
where we recall that $x$ is defined in \Eq{eq:pothiggs:nlo}. One can
see in \Fig{fig:pothi} that, already with $\xi=1$, this again provides an
excellent approximation to the full first Hubble-flow parameter. This
can be used to better estimate the error made when using
\Eq{eq:srparametershiggs} to compute the first Hubble-flow
parameters. In the large-field regime where $\phi\gg \Mg$, one indeed
has
\begin{align}
\label{eq:hiigs:eps1:nlo}
\frac{\epsilon_1^{\mathrm{nlo}}}{\epsilon_1^{\mathrm{lo}}} \simeq
\frac{1}{\left(1+12\xi\right)^2}\exp
\left\lbrace\frac{2}{1+12\xi}\left[1+\ln(24
  \xi)+\sqrt{\frac{2}{3}}\frac{\phi}{\Mg}\right]\right\rbrace,
\end{align}
where $\epsilon_1^{\mathrm{lo}}$ stands for \Eq{eq:srparametershiggs},
and $\epsilon_1^{\mathrm{nlo}}$ for \Eq{eq:hiigs:eps1:nlo}. For this
ratio to remain close to unity, hence for the relative error on the
slow-roll parameters to remain small, one not only needs to impose
$\xi\gg 1$ but also
\begin{align}
\label{eq:hi:cond:xi}
\xi \gg \frac{\phi}{\Mg}\, .
\end{align}
In particular, there is always a region far away in the plateau where
the relative precision of the leading-order expressions breaks
down. When $\xi$ is sufficiently large, this region is however removed
out of the observable phase of the inflationary dynamics, and the
leading-order expressions derived in \sectionc{sec:sisr} can be safely
employed.

The reheating-consistent observational predictions of Higgs inflation
are represented in \Fig{fig:CMBHI} (there are almost the same as for
Starobinsky Inflation) where we have displayed their dependence on the
reheating energy defined in the Jordan frame by $\Treh =
\rhoreh^{1/4}$.  Notice that, a priori, the reheating temperature can
be calculated exactly in Higgs inflation since all the couplings
between the Higgs and the other fields in the standard model are
known~\cite{GarciaBellido:2008ab}. This gives a spectral index which
is in good agreement with the data and a small contribution of gravitational
waves. At this stage, in the Higgs case, the constraints on the
parameter $\xi$ come from the amplitude of the CMB anisotropies, \ie
from \Eq{eq:si:COBE}. As explained below \Eq{eq:si:COBE}, for the
fiducial value $\Delta \Nstar=55$, one gets $M\simeq 3.3\times
10^{-3}\Mg$, \ie, inflation takes place at the GUT scale. Then, using
the expression of $M$, one obtains the following condition for
the parameter $\xi$,
\begin{equation}
\xistar \simeq 46000 \sqrt{\lambda}\,.
\end{equation}
The value of $\xistar$ matching the amplitude of the CMB anisotropies
thus depends on the self-interacting coupling constant $\lambda$ and,
for $\lambda \simeq 0.13$, it satisfies \Eq{eq:hi:cond:xi} across the
field range that is of observational relevance. These considerations
are in agreement with the conclusions obtained in
\Refcs{Bezrukov:2007ep,Bezrukov:2008ej,Bezrukov:2009db}. Notice that
such a large value for the coupling constant $\xi$ is sometimes
considered as problematic~\cite{Barbon:2009ya}.

If we now consider the supergravity realization of the model described
in \sectionc{sec:othertheosi}, one obtains a
constraint on the parameter $\hat{\mu}$, hence, if one takes $c=1$,
one obtains a constraint on $\mu$ and $\lambda$, see
\Refc{Ellis:2013xoa}.

\subsubsection{Exact Slow-Roll and Reheating Analysis}

\label{sec:hiexact}

We give in this section the exact slow-roll analysis of Higgs
inflation as it is coded in the {\ASPIC} library. Such a treatment is
required for other non-minimal gravity models, such as Non-Minimal
Large Field Inflation discussed in \sectionc{sec:nmlfi} and is
analogous to models in which a non-canonical K\"ahler metric prevents
the kinetic term of the inflaton to be explicitly normalized, as in
Dual Inflation presented in \sectionc{sec:di}.

Because \Eq{eq:chifrombarh} cannot be inverted
to obtain an explicit potential $V(\phi)$, the analysis
is parametric and uses as a proxy the dimensionless field $\barh =
\sqrt{\xi} h$. From \Eq{eq:Higgs:Jordan:pot:TreeLevel}, the parametric
potential of Higgs inflation reads
\begin{equation}
V(\barh) = M^4 \left(\dfrac{\barh^2 - \barv^2 }{1+\barh^2}\right)^2\,,
\label{eq:pothiparam}
\end{equation}
where $M$ is still defined by \Eq{eq:hiM4} and we have introduced the
rescaled and dimensionless vacuum expectation value
\begin{equation}
\barv \equiv \sqrt{\xi} \dfrac{v}{\Mg}\,.
\label{eq:hibarv}
\end{equation}
The canonically normalized field $\phi(\barh) = \Mg \chi(\barh)$ is
explicit and given by \Eq{eq:chifrombarh}.

The first Hubble flow function in the slow-roll approximation is given
by
\begin{equation}
\epsilon_1 = \dfrac{1}{2} \left(\dfrac{\ud \ln V}{\ud \chi} \right)^2
= \dfrac{\left(\dfrac{\ud
\ln V}{\ud \barh} \right)^2}{2 \left(\dfrac{\ud \chi}{\ud \barh} \right)^2}\,.
\label{eq:eps1defparam}
\end{equation}
Using \Eqs{eq:relejframes} and \eqref{eq:pothiparam} gives the explicit parametric
expression
\begin{equation}
\epsilon_1(\barh) = \dfrac{8 \xi \barh^2 \left(1+\barv^2\right)^2}{\left(\barh^2 - \barv^2\right)^2
  \left( 1+\barh^2 + 6 \xi \barh^2\right)}\,.
\label{eq:eps1hiparam}
\end{equation}
One can also obtain explicit expressions for the other Hubble-flow
functions. The second Hubble-flow function reads
\begin{equation}
\epsilon_2(\barh) = \dfrac{8 \xi \left(1+\barh^2 \right)\left(1 +
  \barv^2 \right) \left[\barh^2 + \barv^2 + 2(1+6\xi) \barh^4
    \right]}{\left(\barh^2 - \barv^2 \right)^2 \left[1 + (1+6\xi)
  \barh^2 \right]^2}\, ,
\label{eq:eps2hiparam}
\end{equation}
while the third one is given by
\begin{equation}
  \begin{aligned}
    \epsilon_3(\barh) & = \dfrac{8 \xi \barh^2 \left(1+\barv^2
      \right)}{\left(\barh^2 - \barv^2 \right)^2 \left[1 + (1+6\xi)
        \barh^2 \right]^2 \left[\barh^2 + \barv^2 + 2(1+6\xi) \barh^4
        \right]} \\ & \times \bigg\{ 3 \barv^2 - (1+12 \xi) \barv^4 + \left[1 +
        2(5+21\xi) \barv^2 - (1+6\xi)\barv^4 \right]\barh^2 \\ &+ 3 (1+6\xi) \left(1+3\barv^2
      \right) \barh^4 + 2(1+6\xi)^2
      \left(2+\barv^2\right) \barh^6 + 2(1+6\xi)^2 \barh^8 \bigg\}.
  \end{aligned}
\label{eq:eps3hiparam}
\end{equation}
These expressions make explicit that the observable quantities of
Higgs inflation, such as the spectral index and tensor-to-scalar
ratio, depend on $\xi$. On the contrary, in Starobinsky Inflation,
the same quantities do not depend on the new energy scale $\mu$. In
other words, at same potential normalization and same reheating
history, Higgs Inflation and Starobinsky Inflation predict a slightly
different spectral index and tensor-to-scalar ratio. As detailed in
the previous section, the differences in the observable range do not
exceed $\order{1/\xi} \simeq 10^{-4}$.

The parametric field value $\barhend$ at which Higgs Inflation ends
can also be determined analytically by solving
$\epsilon_1(\barh)=1$. This is a cubic polynomial equation in
$\barh^2$ admitting a real root in the relevant domain ($\barh >
\barv$)
\begin{equation}
\begin{aligned}
 \barhend^2 &= \dfrac{1}{12(1+6\xi)} \Big\{-4 + 8 \barv^2(1+6\xi) -
 2i(i+\sqrt{3}) \left[P(\xi,\barv)\right]^{1/3} - 2i
 (i-\sqrt{3})\left[P(\xi,\barv)\right]^{-1/3} \\ & \times
 \left[(1+12\xi)^2+2\barv^2(1+6\xi)(1+24\xi)
   \barv^4(1+6\xi)(1+30\xi)\right] \Big\},
\label{eq:hi:barhend}
\end{aligned}
\end{equation}
where
\begin{equation}
\begin{aligned}
  P(\xi,\barv) & \equiv 1 + 3\barv^2 + 3 \barv^4 + \barv^6 + 36\xi +18
  \xi \barv^2 - 72 \xi \barv^4 - 54 \xi \barv^6 + 216 \xi^2 - 432
  \xi^2 \barv^2 - 1404 \barv^4 \xi^2 \\ & - 756 \xi^2 \barv^6 - 2592
  \xi^3 \barv^2 - 5184 \xi^3 \barv^4 - 2376 \xi^3 \barv^6 + 6
  \sqrt{6\xi} (1+\barv^2)(1+6\xi) \\ & \times \left[-\barv^2
    \left(1+\barv^2\right)^3 - 2 \xi \left(1+\barv^2\right)^2
    \left(1+20\barv^2 + \barv^4 \right) + 4 \xi^2
    \left(1+\barv^2\right) \right. \\ & \left. \times
    \left(-16-108\barv^2 - 60 \barv^4+5\barv^6\right) -24 \xi^3
    \left(4+8\barv^2+\barv^4\right)^2 \right]^{1/2}.
\end{aligned}
\end{equation}

The parametric slow-roll trajectory can be integrated
analytically. Expressing \Eq{eq:phidot2} in terms of $\chi$ one gets
\begin{equation}
\dfrac{\ud \chi}{\ud N} \simeq - \dfrac{\ud \ln V}{ \ud \chi} = -
\dfrac{\ud \barh}{\ud \chi} \dfrac{\ud \ln V}{\ud \barh}\,,
\label{eq:hitrajparamsplit}
\end{equation}
which can be used to have an explicit expression for
\begin{equation}
\dfrac{\ud \barh}{\ud N} = \dfrac{\ud \barh}{\ud \chi} \dfrac{\ud
  \chi}{\ud N} \simeq - \dfrac{1}{\left(\dfrac{\ud \chi}{\ud
    \barh}\right)^2} \dfrac{\ud \ln V}{\ud \barh} = - \dfrac{4 \xi \left(1+ \barv^2
  \right) \barh \left(1+ \barh^2\right) }{\left(\barh^2 - \barv^2 \right) \left[1 + (1+6\xi) \barh^2 \right]}\,, 
\end{equation}
where ``$\simeq$'' refers to the use of the slow-roll approximation
when replacing $\dd\chi/\dd N\simeq -\dd\ln V/\dd \chi$.  This
equation can be analytically integrated to get the slow-roll
trajectory as
\begin{equation}
\begin{aligned}
  \Nend - N & = \dfrac{1}{8\xi\left(1+\barv^2\right)}
\left[(1+6\xi)\left(\barh^2 - \barhend^2\right) - \barv^2
  \ln\left(\dfrac{\barh^2}{\barhend^2}\right) - 6\xi
  \left(1+\barv^2\right) \ln
  \left(\dfrac{1+\barh^2}{1+\barhend^2}\right) \right],
\end{aligned}
\label{eq:hitrajparam}
\end{equation}
where $\barhend$ is given in \Eq{eq:hi:barhend}.

In order to determine the parametric field value $\barhstar$ at which
the pivot scale crosses the Hubble radius during inflation, one has to
solve the reheating equation, in the Einstein frame, as we are in
presence of a scalar-tensor inflaton. However, without making the
approximations discussed in the previous sections, the slow-roll
parameters, and the potential, depend on $\xi$. Its actual
value, say $\xistar$, being obtained from the amplitude of the CMB
anisotropies, one ends up having a system of two coupled non-linear
algebraic equations, the solution of which giving both $\barhstar$ and
$\xistar$. The equation fixing the normalization of the potential is
given by \Eq{eq:pstarnorm} and reads
\begin{equation}
\dfrac{\lambda}{4 \xistar^2} = 24 \pi^2 \Pstar
\dfrac{\epsilon_1(\barhstar,\xistar)}{V(\barhstar,\xistar)/M^4}\,,
\label{eq:cobehiparam}
\end{equation}
where $V(\barhstar,\xistar)$ is given in \Eq{eq:pothiparam} and
$\epsilon_1(\barhstar,\xistar)$ in \Eq{eq:eps1hiparam}. The reheating
equation in the Einstein frame has been derived in \Eq{eq:dnstareinstein} and
reads
\begin{equation}
\Delta \Nstar=\ln \Rradbar - \Nzero
+\frac{1}{4} \ln \left(8 \pi^2 \Pstar \right)
-\frac{1}{4}\ln \left(\frac{3}{\epsonestar}\frac{\Vend}{\Vstar}
\frac{3-\epsonestar}{3-\epsoneend}\right),
\label{eq:reheathiparam}
\end{equation}
where it is understood that $\Vend = V(\barhend,\xistar)$,
$\Vstar=V(\barhstar,\xistar)$, $\epsoneend=1$ and
\begin{equation}
\Delta\Nstar(\barhstar,\xistar) = \Nend - \Nstar,
\end{equation}
which is given in \Eq{eq:hitrajparam}. Let us notice that from
\Eq{eq:hi:barhend} one has $\barhend(\xistar)$ and that $\barv$ is
also an explicit function of $\xistar$ given by \Eq{eq:hibarv}. For $v
\ne 0$, there is no analytical solution to the algebraic system made
of \Eqs{eq:cobehiparam} and \eqref{eq:reheathiparam} and it is solved
numerically in the {\ASPIC} code. We do not plot the numerical
solutions as they would be indistinguishable from the next-to-leading
order analysis presented in \sectionc{subsubsec:leadhisr}.

The reheating-consistent observational predictions of Higgs
Inflation are represented in \Fig{fig:CMBHI} where we have displayed
their dependence on the reheating temperature.

\section{One Parameter Models}
\label{sec:onep}

\subsection{Radiatively Corrected Higgs Inflation (RCHI)}
\label{sec:rchi}

\subsubsection{Theoretical Justifications}
\label{subsubsec:theoryrchi}

\begin{figure}
\begin{center}
\includegraphics[width=\wdblefig]{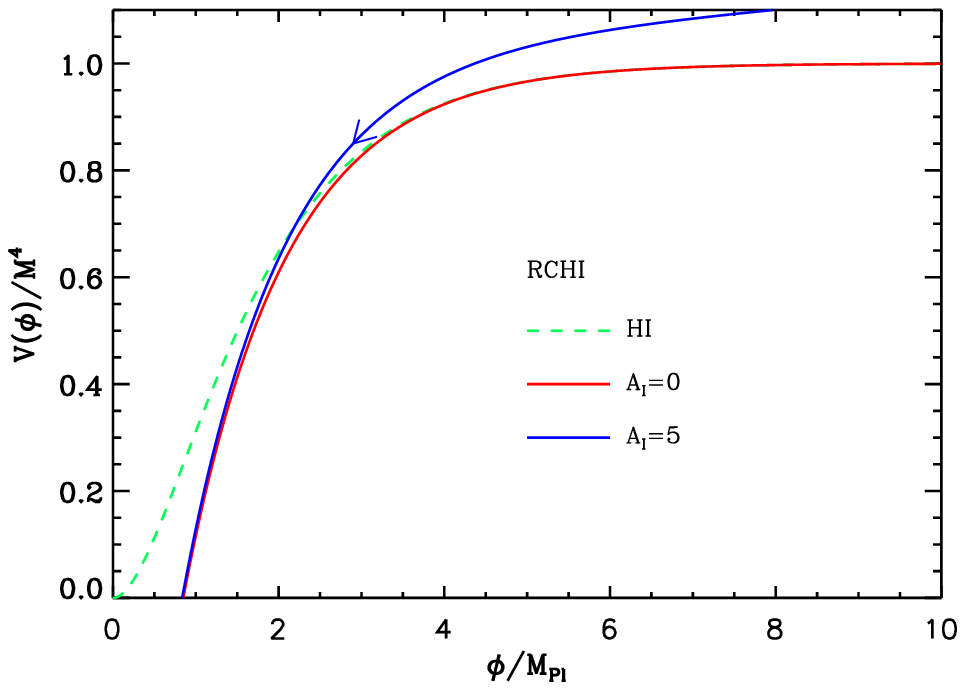}
\includegraphics[width=\wdblefig]{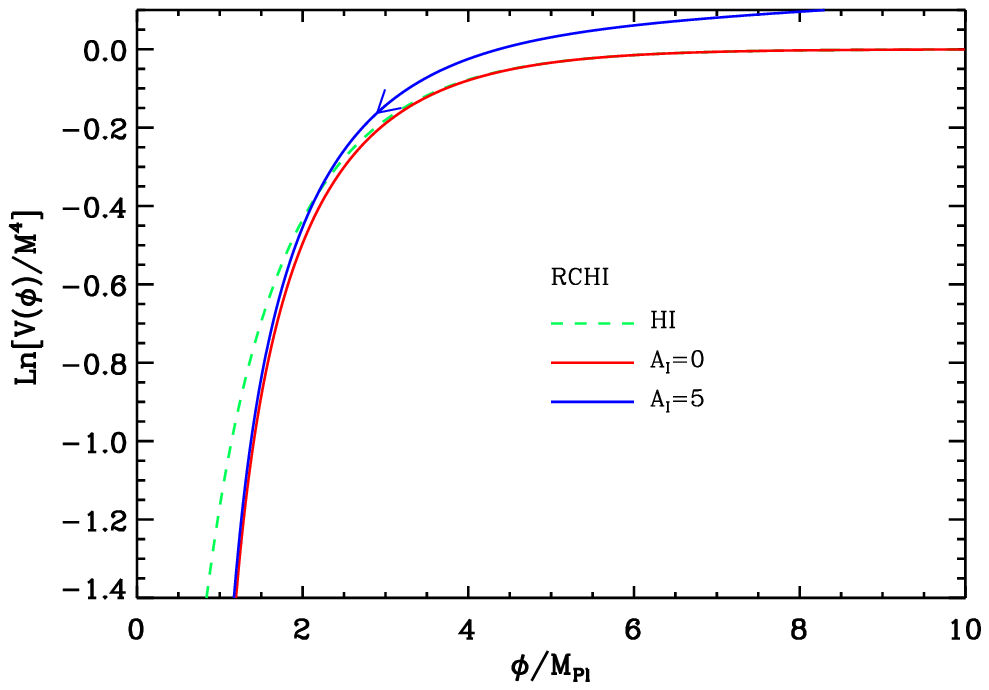}
\includegraphics[width=\wdblefig]{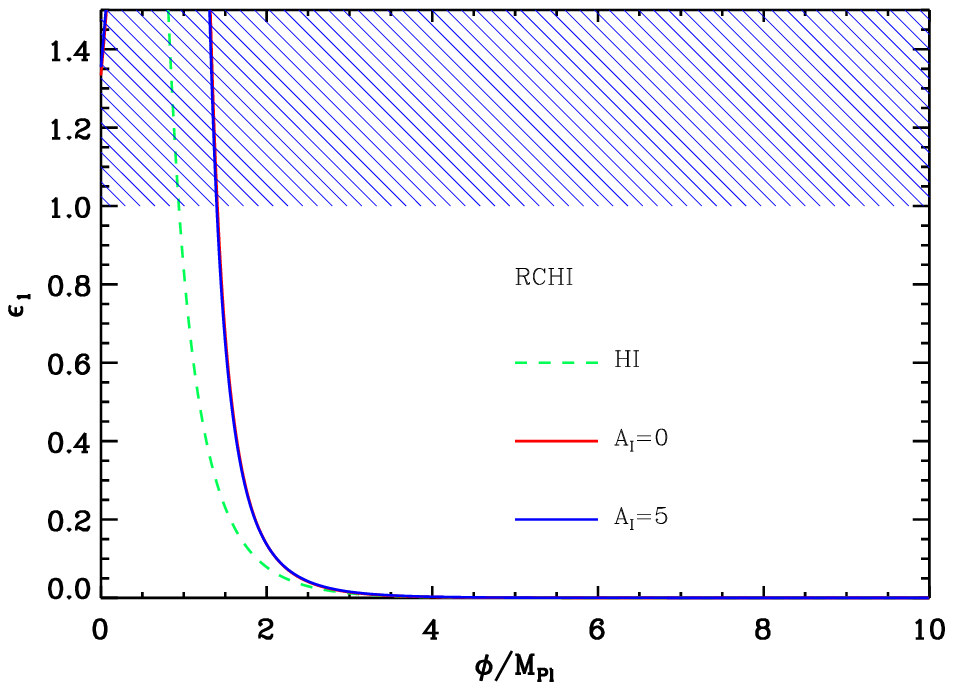}
\includegraphics[width=\wdblefig]{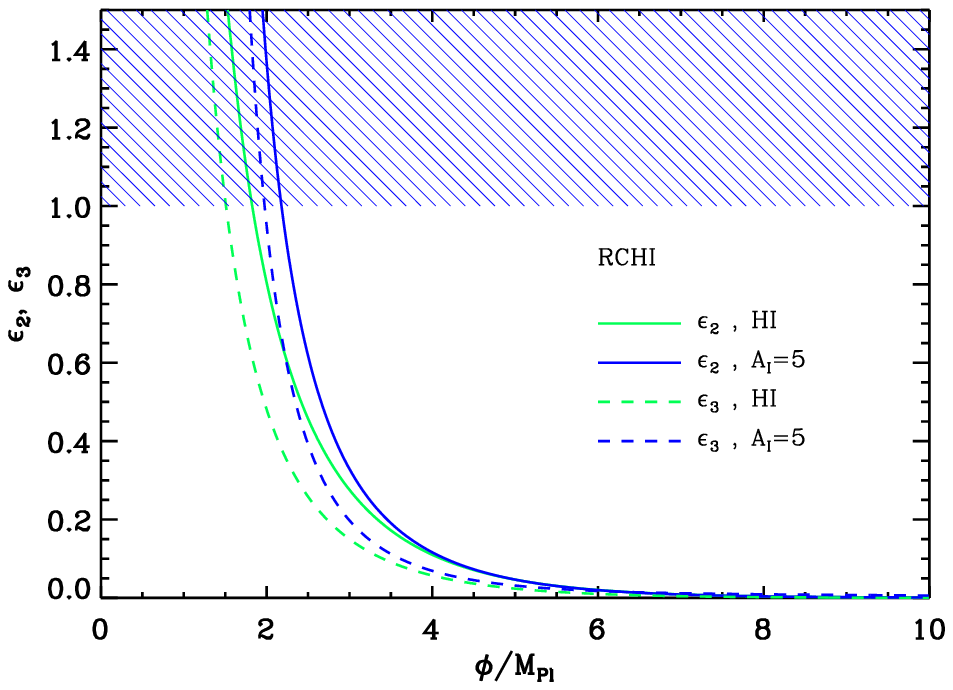}
\caption{Top left panel: the solid blue line represents the
  radiatively corrected Higgs potential, see \Eq{eq:higgsAIpot}, with
  $\AI=5$. It is compared to the tree level potential given by
  \Eq{eq:pothiggs} (dashed green line) and to \Eq{eq:higgsAIpot} with
  $\AI=0$ (solid red line) which is supposed to be a good
  approximation of the tree level potential. It is obvious that this
  is indeed the case in the regime of interest, where the \vev of the
  Higgs field is not too small. Top right panel: logarithm of
  potential, the three lines and the color code having the same
  meaning as in the top left panel. Bottom left panel: slow-roll
  parameter $\epsilon _1$ as a function of the field $\phi$, still
  with the same convention. As can be seen in this plot, even in
  presence of radiative corrections, the end of inflation occurs by
  violation of the slow-roll condition. Bottom right panel: slow-roll
  parameters $\epsilon _2$ (solid blue line) and $\epsilon _3$ (dashed
  blue line) for $\AI=5$ compared to their tree level counterparts
  (solid and dashed green lines, respectively).}
\label{potrchi}
\end{center}
\end{figure}

Let us consider again the model given by \Eq{eq:actionjordan}. The
three functions describing this action are modified when quantum
corrections are taken into account. As a consequence, the potential
which supports inflation is also modified and this leads to a new
inflationary scenario that we call Radiatively Corrected Higgs
Inflation (RCHI). This scenario has been studied in
\Refcs{Barvinsky:2008ia, DeSimone:2008ei, Barvinsky:2009ii,
  Bezrukov:2010jz, Steinwachs:2013tr,Bezrukov:2012hx}. At first order,
$\Mg \simeq \Mp$, the corrections to the function $Z(h)$ can be
neglected while the corrections to $F(h)$ and to $U(h)$ read
\begin{align}
F(h)&=1+\xi h^2 +\frac{C}{16\pi^2}h^2\ln \left(\frac{\Mp^2h^2}{\mu^2}\right), 
\\
U(h)&= \Mp^2\frac{\lambda}{4}\left(h^2-\frac{v^2}{\Mp^2}\right)^2
+\frac{\lambda A}{128 \pi^2}\Mp^2h^4\ln \left(\frac{\Mp^2h^2}{\mu^2}\right),
\end{align}
where $\mu$ is the renormalization scale and $A$ and $C$ are two new
constants given by
\begin{align}
A&=\frac{3}{8\lambda}\left[2g^4+\left(g^2+g'^2\right)-16y_\ut^4\right]
+6\lambda +\calO\left(\xi^{-2}\right), \\
C&=3\xi \lambda +\calO\left(\xi^0\right),
\end{align}
$y_\ut$ being the Yukawa coupling of the top quark and $g$ and $g'$
the coupling constants of the $\mathrm{SU}(2)_{\mathrm{L}}$ and
$\mathrm{U}(1)_{\mathrm{Y}}$ groups.  The presence of quantum
corrections modifies the relation between the Jordan and the Einstein
frames and changes the shape of the potential in the Einstein
frame. Assuming the smallness of $A/(32\pi^2)\ll 1$ and $C/(8\pi ^2
\xi)\ll 1$, which is necessary for the consistence of the one-loop
calculation (the second condition is in fact equivalent to
$C\lambda/(8\pi^2)\ll 1$ because $C$ is proportional to $\xi$), one
obtains the following expression
\begin{equation}
V\simeq \frac{\Mp^4\lambda}{4 \xi^2}
\frac{\xi^2 h^4}{\left(1+\xi h^2\right)^2}
\left[1-\frac{\xi h^2}{1+\xi h^2}\frac{C}{8\pi^2 \xi}
\ln \left(\frac{\Mp^2h^2}{\mu^2}\right)
+\frac{A}{32\pi^2}\ln \left(\frac{\Mp^2h^2}{\mu^2}\right)\right].
\end{equation}
Of course, if $A=C=0$, one checks that this potential reduces to the
potential of the previous section. Notice that, at this stage, we have
not assumed that $\xi h^2\gg 1$. If we further postulate that $\xi
h^2\gg 1$ and approximate $\xi^2h^4/\left(1+\xi h^2\right)^2\simeq
1-2/(\xi h^2)$, then the above formula reduces to
\begin{equation}
\label{eq:approxpot}
V\simeq \frac{\Mp^4\lambda}{4 \xi^2}
\left[1-\frac{2}{\xi h^2}
+\frac{\AI}{16\pi^2}\ln \left(\frac{\Mp h }{\mu}\right)\right], 
\end{equation}
where $\AI\equiv A-12\lambda$ is the inflationary anomalous
scaling. This formula coincides with Eq.~(6) of
\Refc{Barvinsky:2009ii} and Eq.~(9) of
\Refc{Steinwachs:2013tr}. Although the above formulas give $V$ in the
Einstein frame, it is still expressed in term of $h$. The expression
for the field in the Einstein frame, $\chi$, remains to be
established. Assuming the smallness of the loop corrections, we obtain
\begin{equation}
\frac{\dd \chi}{\dd h}\simeq \frac{\sqrt{3}h\xi}{(1+\xi h^2)}
\left[1+\frac{C}{16\pi^2 \xi}
+\frac{C}{8\pi^2\xi}\frac{1}{1+\xi h^2}
\ln \left(\frac{\Mp h}{\mu}\right)\right].
\end{equation}
Notice that, in order to obtain this equation, we have neglected
a term proportional to $1/(\xi h)^2\ll 1$. Contrary to the assumption
$\xi h^2\gg 1$, the condition $\left(\xi h\right)^2 \gg 1 $ was also
used in \sectionc{sec:hi}. Then, the integration of this differential
equation leads to
\begin{equation}
\chi \simeq \frac{\sqrt{3}}{2}\ln \left(1+\xi h^2\right)
+\frac{\sqrt{3}C}{16 \pi^2 \xi}
\left[\ln h -\frac{1}{1+\xi h^2}
\ln \left(\frac{\Mp h}{\mu}\right)\right].
\end{equation}
Using only now the limit $\xi h^2\gg 1$, this expression reduces to
\begin{equation}
\chi \simeq \frac{\sqrt{3}}{2}\ln \left(\xi h^2\right)
+\frac{\sqrt{3}C}{16\pi^2\xi}\ln h .
\end{equation}
As expected the relation between the Jordan frame field $h$ and the
Einstein frame field $\chi $ is modified by the quantum
corrections. Inverting the above formula gives
\begin{equation}
\label{eq:joreinfield}
\xi^{1/2}h\simeq \ee ^{\chi/\sqrt{3}}-\frac{C}{16 \pi^2\xi}\ee ^{\chi /\sqrt{3}}
\left(\frac{\chi}{\sqrt{3}}-\frac12 \ln \xi \right).
\end{equation}
This equation allows us to find the expression of the potential in the
Einstein frame. Inserting \Eq{eq:joreinfield} into \Eq{eq:approxpot}
and introducing the canonically normalized field $\phi\equiv
\sqrt{2}\Mp \chi$, one obtains
\begin{align}
V(\phi) & \simeq  \frac{\Mp^4 \lambda}{4 \xi^2}
\biggl[1-2\ee^{-2\phi/(\sqrt{6}\Mp)}
-\frac{C}{4\pi^2\xi}
\ee^{-2\phi/(\sqrt{6}\Mp)}
\left(\frac{\phi}{\sqrt{6}\Mp}-\frac12 \ln \xi \right)
\nonumber \\
& +\frac{\AI}{16\pi^2}
\ln \left(\frac{\Mp}{\mu \sqrt{\xi}}\right)
+\frac{\AI}{16 \pi^2}
\frac{\phi}{\sqrt{6}\Mp}\biggr]
\nonumber \\ 
\label{eq:higgsAIpot}
& \simeq 
\frac{\Mp^4 \lambda}{4 \xi^2}
\left[1-2\ee^{-2\phi/(\sqrt{6}\Mp)}
+\frac{\AI}{16 \pi^2}
\frac{\phi}{\sqrt{6}\Mp}\right].
\end{align}
We see that we now deal with a ``one parameter model'', $\AI$, since
the mass scale $M^4\equiv \Mp^4 \lambda /(4 \xi^2)$ is determined by
the COBE normalization. In the case $\AI=0$, it is also interesting to
compare the above potential with the one given by \Eq{eq:pothiggs}. We
see that this corresponds to assuming that the exponential $\ee
^{-2\phi/(\sqrt{6}\Mp)}\ll 1$ (or, equivalently, $\phi/\Mp\gg 1$) and
to expand the corresponding expression at first order in this small
parameter. This leads to the following formula: $V\simeq
M^4\left[1-2\ee ^{-2\phi/(\sqrt{6}\Mp)}\right]$, \ie exactly
\Eq{eq:higgsAIpot} for $\AI=0$. It is worth remarking that this
approximation is not very good towards the end of inflation. Indeed,
it is easy to show that (see below), for the potential~\eqref{eq:higgsAIpot} with $\AI=0$, $\phiend/\Mp=\sqrt{3/2}\ln
\left(2+2/\sqrt{3}\right)\simeq 1.4$ which should be compared with
\Eq{eq:endhiggs} for the potential~\eqref{eq:pothiggs} according to
which $\phiend/\Mp\simeq 0.94$.

\subsubsection{Slow-Roll Analysis}
\label{subsubsec:srrchi}

Given \Eq{eq:higgsAIpot}, namely
\begin{equation}
\label{eq:potrchisr}
V(\phi)=M^4
\left[1-2\ee^{-2\phi/(\sqrt{6}\Mp)}
+\frac{\AI}{16 \pi^2}
\frac{\phi}{\sqrt{6}\Mp}\right],
\end{equation}
we can now proceed to the slow-roll analysis.  The
potential~(\ref{eq:potrchisr}) is represented and compared with its
tree level counterpart in \Fig{potrchi}. Defining $x\equiv \phi/\Mp $,
the three first slow-roll parameters can be written as
\begin{align}
\label{eq:eps1rchi}
\epsilon_1 &=\frac{1}{12}\left[\frac{4\ee^{-\sqrt{2/3}x}+\AI/(16\pi^2)}
{1-2\ee^{-\sqrt{2/3}x}+\AI/(32\pi^2)\sqrt{2/3}x}\right]^2,
\end{align}
\begin{align}
\label{eq:eps2rchi}
\epsilon_2 &= \frac{1}{3}
\frac{8\ee ^{-\sqrt{2/3}x}\left[1+\AI/(16\pi^2)+\AI/(32\pi^2)\sqrt{2/3}x\right]
+\AI^2/(256\pi^4)}{\left[1-2\ee^{-\sqrt{2/3}x}+\AI/(32\pi^2)\sqrt{2/3}x\right]^2},
\end{align}
and 
\begin{align}
\epsilon_3 &= 12\left(4+\frac{\AI}{16 \pi^2}\ee^{\sqrt{2/3}x}\right)
\Biggl\{48+8\frac{\AI}{16\pi^2}\left(9+\sqrt{6}x\right)
+3\frac{\AI^3}{4096\pi^6}\ee^{2\sqrt{2/3}x}
\nonumber \\ &
+2\ee^{\sqrt{2/3}x}
\left[12+18\frac{\AI}{16\pi^2}\left(1+\frac{\AI}{16\pi^2}\right)
+\sqrt{6}\frac{\AI}{16\pi^2}\left(4+3\frac{\AI}{16\pi^2}\right)x
+2\frac{\AI^2}{256\pi^4}x^2\right]\Biggr\}
\nonumber \\ & \times
\left[24+\frac{\AI}{16\pi^2}\left(24+4\sqrt{6}x+3\frac{\AI}{16\pi^2}
\ee^{\sqrt{2/3}x}\right)\right]^{-1}
\left[-12+\ee^{\sqrt{2/3}x}\left(6+\sqrt{6}\frac{\AI}{16\pi^2}x\right)\right]^{-2}.
\end{align}
These three slow-roll parameters are represented in \Fig{potrchi}
(bottom panels). It is interesting to compare these formulas with the
expressions derived in \Refc{Barvinsky:2008ia} [see Eqs.~(22) and~(23)
of that paper]. An approximate equation for the first slow-roll
parameter is obtained by neglecting the second and third terms in the
denominator of \Eq{eq:eps1rchi}, which, as a matter of fact, consists
in writing $V(\phi)\simeq M^4$. Then, it follows that
\begin{equation}
\label{eq:approxeps1rchi}
\epsilon_1\simeq \frac{4}{3}\ee^{-2\sqrt{2/3}x}\left(1+\frac{\AI}{64 \pi^2}
\ee^{\sqrt{2/3}x}\right)^2\simeq \frac{4}{3}\frac{1}{\xi^2h^4}\left(1+
\frac{h^2}{\hI^2}\right)^2,
\end{equation}
where we have defined $\hI^2\equiv 64\pi^2/(\xi \AI)$ in agreement
with \Refc{Barvinsky:2008ia}. The same approximation is made for the
second slow-roll parameter (except that \Refc{Barvinsky:2008ia}
calculates $\hat{\eta}\equiv \Mp^2V_{\phi \phi}/V$ rather than
$\epsilon_2$). The second field derivative of the potential can be
written as $V_{\phi \phi}=-4M^4\ee^{-\sqrt{2/3}x}/(3\Mp^2)$ and,
therefore, if one considers that $V(\phi)\simeq M^4$, then
$\hat{\eta}\simeq -4/(3\xi h^2)$. We conclude that our expressions of
$\epsilon_1$ and $\epsilon_2$ reproduce Eqs.~(22) and~(23) of
\Refc{Barvinsky:2008ia} in the limit where $V(\phi)\simeq M^4$.

Let us now study how inflation ends in this model. From \Fig{potrchi},
it is clear that this occurs by violation of the slow-roll
conditions. Working out the condition $\epsilon_1=1$, it follows that
\begin{align}
\xend &=\frac{1}{\sqrt{2}}
-\sqrt{\frac{3}{2}}\frac{32\pi^2}{\AI}
+\sqrt{\frac{3}{2}}
\Lambert{{}^{\ 0}_{-1}}
\left[\frac{64\pi^2}{\AI}\left(1+\frac{1}{\sqrt{3}}\right)
\ee^{32\pi^2/\AI-1/\sqrt{3}}\right],
\end{align}
where, if $\AI>0$, $\Lambert{{}^{\ 0}_{-1}}=\Lambert{0}$ while, if
$\AI<0$, $\Lambert{{}^{\ 0}_{-1}}=\Lambert{-1}$.

We now turn to the slow-roll trajectory. It can be integrated exactly
and straightforward manipulations lead to the following expression
\begin{align}
N-\Nini &=\sqrt{\frac{3}{2}}\, x
-\frac{48\pi^2}{\AI}\left[1+\frac{\AI}{32\pi^2}
\left(1+\sqrt{\frac{2}{3}}\, x \right)\right]
\ln \left(1+\frac{\AI}{64 \pi^2}\ee^{\sqrt{2/3}\,x}
\right)
\nonumber \\ &
-\frac32\Li{2}\left(-\frac{\AI}{64 \pi^2}
\ee^{\sqrt{2/3}\, x}\right)
-\sqrt{\frac{3}{2}} \xini
+\frac{48\pi^2}{\AI}\left[1+\frac{\AI}{32\pi^2}
\left(1+\sqrt{\frac{2}{3}}\, \xini \right)\right]
\nonumber \\ & \times
\ln \left(1+\frac{\AI}{64 \pi^2}\ee^{\sqrt{2/3}\, \xini}
\right)
+\frac32\Li{2}\left(-\frac{\AI}{64 \pi^2}
\ee^{\sqrt{2/3} \,\xini}\right),
\end{align}
where $\Li{2}$ denotes the dilogarithm
function~\cite{Abramovitz:1970aa, Gradshteyn:1965aa}. Let us also
notice that if we use the approximation $V(\phi)\simeq M^4$ already
discussed before, then one can obtain a much simpler formula, namely
\begin{align}
\label{eq:approxtrajecrchi}
N-\Nini &=
-\frac{48\pi^2}{\AI}
\ln \left(1+\frac{\AI}{64 \pi^2}\ee^{\sqrt{2/3} \, x}
\right)
+\frac{48\pi^2}{\AI}
\ln \left(1+\frac{\AI}{64 \pi^2}\ee^{\sqrt{2/3} \,\xini}
\right).
\end{align}
This expression is in agreement with Eq.~(24) of
\Refc{Barvinsky:2008ia}. In this case, the trajectory can even be
inverted and the corresponding expression for the field $\phi$ reads
\begin{equation}
x=\sqrt{\frac{3}{2}}
\ln \left[\left(\frac{64 \pi^2}{\AI}
+\ee^{\sqrt{2/3}\, \xini}\right)\ee^{\AI(N-\Nini)/(48 \pi^2)}
-\frac{64 \pi^2}{\AI}\right].
\end{equation}

\begin{figure}
\begin{center}
\includegraphics[width=\wsingfig]{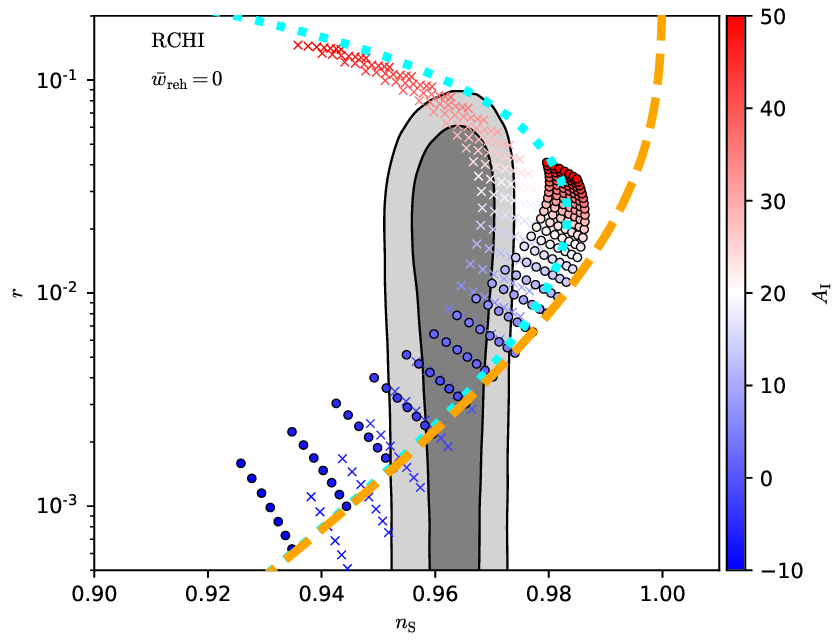}
\caption{Predictions of the RCHI model in the plane $(\nS,r)$. The
  exact slow-roll predictions (colored circles starting in dark blue
  at the bottom/left part of the plot and ending in red slightly on
  the right of the allowed contours) are compared to various
  approximations represented by the second collection of segments made
  of ``crosses'', by the orange thick dashed line and by the
  aquamarine dotted line, see the text for a detailed explanation. In
  the regime of large $|\AI|$, the exact predictions may significantly
  differ from the approximate ones.}
\label{fig:approxrchi}
\end{center}
\end{figure}

We are now in a position where the predictions of the models can be
calculated. They are presented in \Fig{fig:CMBHI}. We see that very
negative values of $\AI$ are incompatible with the CMB while large
values of $\AI$ remain close to the allowed contours. Of
course $\vert \AI \vert $ cannot be too large since we have required
$\AI/(64\pi^2)\ll 1$. We have chosen the upper bound in
\Fig{fig:CMBHI} to be $\AI=100$ for which $\AI/(64 \pi^2)\simeq
0.16$, \ie still a reasonable number. It is interesting to compare
these findings with the existing literature. Using the approximate
trajectory in \Eq{eq:approxtrajecrchi} and neglecting the contribution
originating from the end of inflation, one obtains
\begin{equation}
\xstar=\sqrt{\frac{3}{2}}\ln \left[\frac{64 \pi^2}{\AI}
\left(\ee ^{\xbks}-1\right)\right],
\end{equation}
where $\xbks\equiv \AI \Delta \Nstar/(48\pi^2)$ ($\xbks$ is denoted
$x$ in \Refc{Barvinsky:2008ia}).
Then, from \Eq{eq:approxeps1rchi} and the fact that
$\epsilon_2=4\epsilon_1-2\hat{\eta}$, it follows that
\begin{align}
\epsilon_1 &= \frac{4}{3}\left(\frac{\AI}{64\pi^2}\right)^2
\left(\frac{\ee^{\xbks}}{\ee ^{\xbks}-1}
\right)^2=\frac{3}{4\Delta \Nstar^2}\left(\frac{\xbks \ee^{\xbks}}{\ee ^{\xbks}-1}
\right)^2, \\
\epsilon_2&=4\epsilon_1+\frac{8}{3}\frac{\AI}{64\pi^2}
\frac{1}{\ee ^{\xbks}-1}=4\epsilon_1+\frac{2}{\Delta \Nstar}
\frac{\xbks}{\ee ^{\xbks}-1}\,.
\end{align}
{}From these two expressions, one deduces that
\begin{align}
\label{eq:srsimple}
\nS &=1-\frac{2}{\Delta \Nstar}
\frac{\xbks}{\ee ^{\xbks}-1}, \quad 
r= \frac{12}{\Delta \Nstar^2}\left(\frac{\xbks \ee^{\xbks}}{\ee ^{\xbks}-1}
\right)^2.
\end{align}
Notice that, in the formula giving the spectral index, the
contribution originating from $\epsilon_1$ has been neglected since it
scales $\propto 1/\Delta \Nstar^2$. These approximate expressions
match Eqs.~(32) and~(34) of \Refc{Barvinsky:2008ia}. For $\Delta
\Nstar=60$, they can be represented as a line $r=r(\nS)$ in the plane
$(\nS,r)$, the parameter along the curve being $\AI$. This line has
been plotted in \Fig{fig:approxrchi} for $-30<\AI<100$ (thick orange
dashed line). Requiring $\nS$ to be within the $2\sigma$ {\data}
suggests that $-8\lesssim \AI\lesssim 4$ (or $-12\lesssim \AI\lesssim
14$ with WMAP, again in agreement with \Refc{Barvinsky:2008ia}). These
predictions are compared to the exact slow-roll predictions of
\Fig{fig:CMBHI}. As before, the slow-roll predictions are represented
by a collection of segments made of circles, each segment
corresponding to different values of $\AI$ and each point of a given
segment being in one-to-one correspondence with a given reheating
temperature. The exact slow-roll predictions are such that, for
$\AI<0$, the segments go to the bottom left side of the figure while
for $\AI\rightarrow 100$, the segments remain close to the allowed
contours (see also \Fig{fig:CMBHI}). In the limit of ``large''
positive values of $\AI$, the exact slow-roll predictions and the
predictions based on \Eqs{eq:srsimple} significantly differ.

Let us try to identify the origin of this discrepancy more
precisely. In order to investigate this issue, we have also
represented in \Fig{fig:approxrchi}, the predictions obtained when the
approximate trajectory of \Eq{eq:approxtrajecrchi}, the approximate
expression of the first slow-roll parameter in \Eq{eq:approxeps1rchi}
and the relation $\epsilon_2=4\epsilon_1-2\hat{\eta}$ but, now,
without neglecting $\epsilon_1$, are used together with an exact
expression for $\phiend$. They are represented by the second
collections of segments made of crosses in \Fig{fig:approxrchi}. We
see that for $\AI\gtrsim 0$, they differ from the orange dashed thick
line and bend towards the upper left part of the plot which is also the
direction taken by the exact predictions. This suggests that
neglecting the term $4\epsilon_1$ in the expression of $\epsilon_2$
causes a non-negligible error. This is confirmed if, instead of using
\Eq{eq:srsimple} for $\nS$, we now take
\begin{align}
\label{eq:nssimple}
\nS &=1-\frac{9}{2\Delta \Nstar^2}
\left(\frac{\xbks \ee^{\xbks}}{\ee ^{\xbks}-1}
\right)^2-\frac{2}{\Delta \Nstar}
\frac{\xbks}{\ee ^{\xbks}-1}\,,
\end{align}
and plot again the line $r=r(\nS)$. This gives the orange dotted curve
which follows the second collection of segments. If, however, we
compare the red segments, namely those with $\AI$ ``large'',
corresponding the exact predictions to the approximate red ones, we
see that including the term $4\epsilon_1$ is not sufficient. We
conclude that RCHI represents a textbook case for \ASPIC. It
illustrates that, sometimes, ``approximating the slow-roll
approximation'' can lead to too drastic conclusions, especially given
the current accuracy of the data. It is an additional motivation to
use the slow-roll method without any other scheme of approximations
and this is the essence of the \ASPIC project presented in this
article.

A last word is in order concerning the constraints on the parameter
$\AI$. Particle physics implies that $-48\lesssim \AI\lesssim -20$ and
the previously discussed inaccuracies were concerning only a weaker
upper limit on $\AI$. Therefore, when particle physics and
cosmological data are simultaneously taken into account, the
conclusions of \Refc{Barvinsky:2008ia} are unchanged and RCHI remains
disfavored.

Finally, the scale $M$ can be determined from the CMB normalization
and this leads to the following expression
\begin{equation}
  \frac{M^4}{\Mp^4}=120\pi^2\frac{\Qrms^2}{T^2}
  \frac{\left[4\, \ee^{-\sqrt{2/3}\xstar}+\AI/(16 \pi^2)\right]^2}
  {\left[1-2\, \ee^{-\sqrt{2/3}\xstar}+\AI/(32\pi^2)\sqrt{2/3}\xstar\right]^3}\,.
\end{equation}
The knowledge of $\phistar$ allows us to find the posterior
distribution of $M$, that is to say of $\lambda/\xi^2$ or $\xi$, since
the Higgs self coupling, $\lambda=\mhiggs/v$, is now known.

\subsection{Large Field Inflation (LFI)}
\label{sec:lfi}

\subsubsection{Theoretical Justifications}
\label{subsubsec:theorylfi}

\begin{figure}
\begin{center}
\includegraphics[width=\wdblefig]{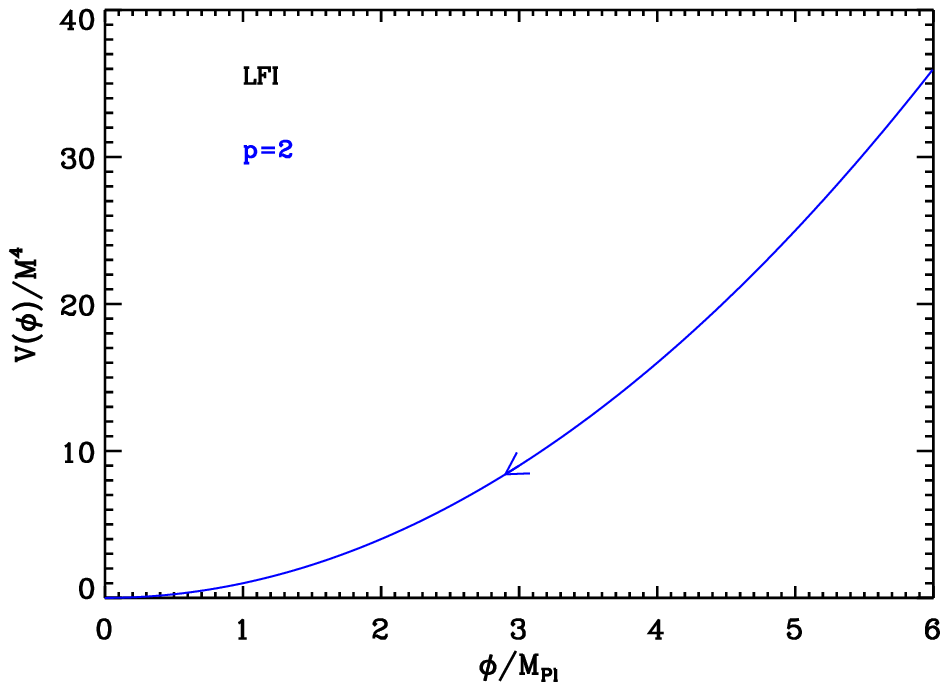}
\includegraphics[width=\wdblefig]{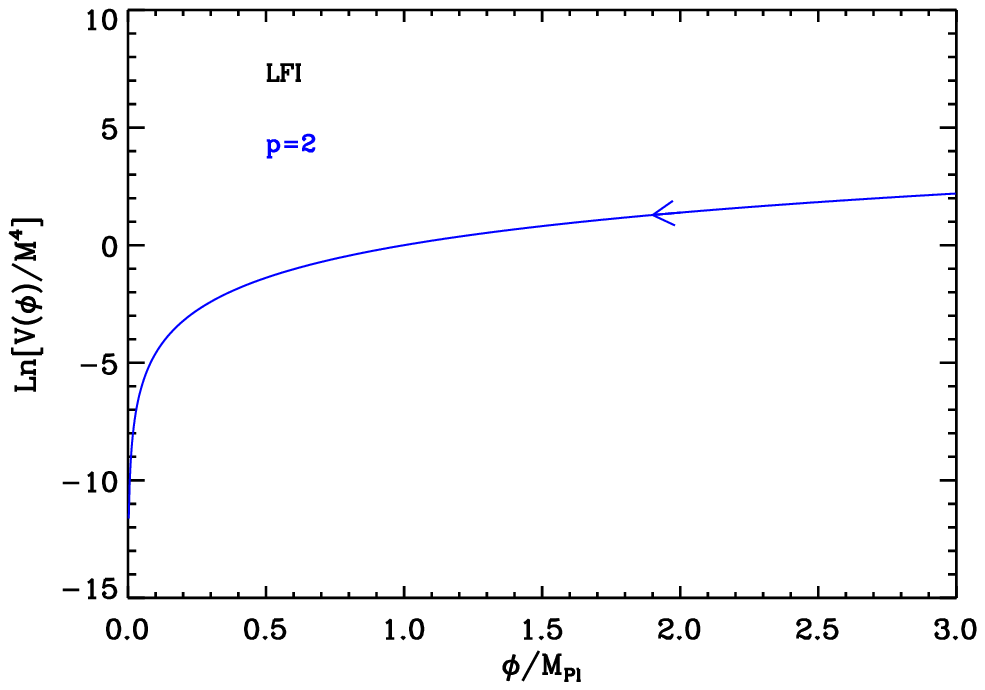}
\includegraphics[width=\wdblefig]{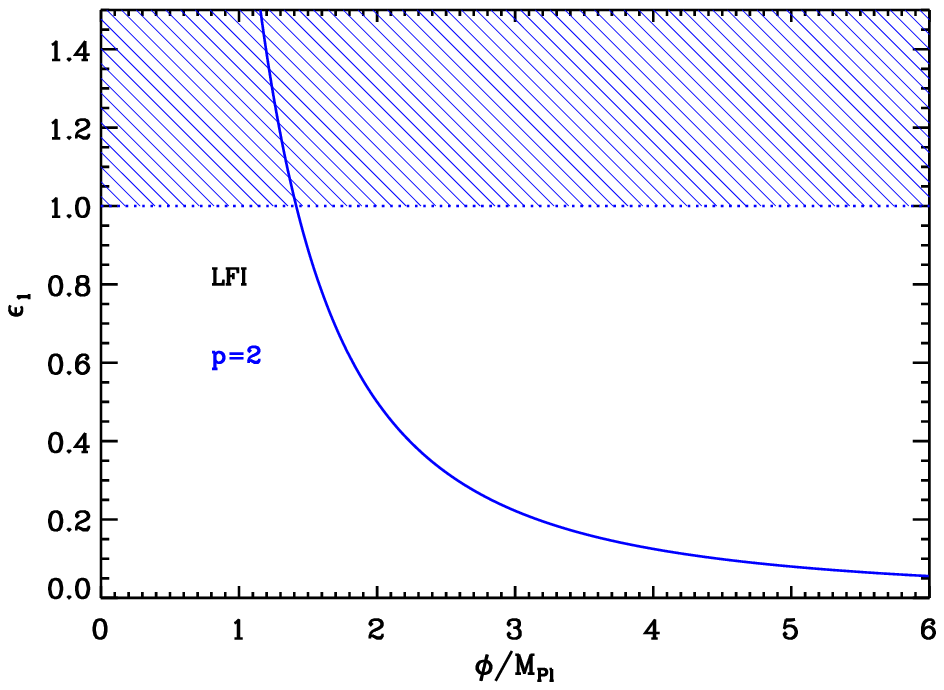}
\includegraphics[width=\wdblefig]{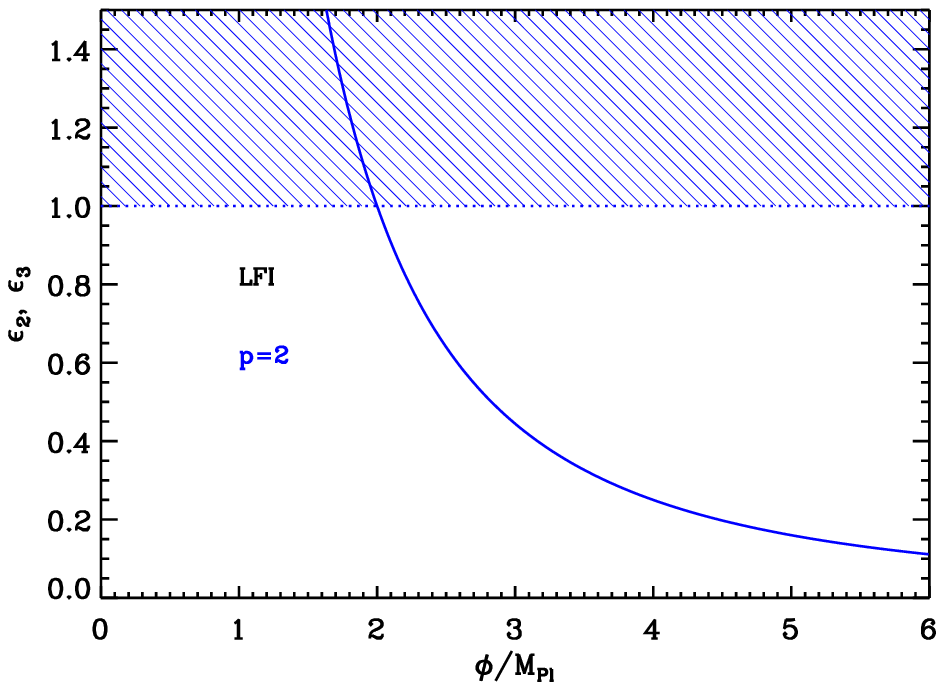}
\caption{Large Field Inflation (LFI). Top left panel: large field
  potential for $p=2$. Top right panel: logarithm of the potential for
  the same value of $p$. The required flatness of the potential
  becomes obvious on this plot. Bottom left panel: slow-roll parameter
  $\epsilon _1$ for a large field potential with $p=2$. The shaded
  area indicates where acceleration stops. Bottom right panel:
  slow-roll parameters $\epsilon _2$ and $\epsilon _3$ for a large
  field potential with $p=2$. Only one curve appears because $\epsilon
  _2=\epsilon_3$. On this plot, the shaded region signals the
  breakdown of the slow-roll approximation, which is not necessarily
  the end of the accelerated phase.}
\label{potLFI}
\end{center}
\end{figure}

Large fields models, also referred to as chaotic
inflation~\cite{Vilenkin:2004vx}, are characterized by the monomial
potential~\cite{Linde:1984st,Madsen:1988xe,
  Lazarides:1993fi,Kofman:1994rk,Lazarides:1995he} $V(\phi) \propto
M^4 \phi ^p$. The number $p$ is the only model parameter, in addition
to the normalization $M$ of the potential. The index $p$ is usually a
positive integer (and it was recently realized in
\Refc{Kawasaki:2000yn} that this type of scenario can emerge in the
context of supergravity) but various models have been proposed in
which it can also be a rational number~\cite{Baumann:2007ah,
  Silverstein:2008sg, Brandenberger:2008kn,
  Nakayama:2010sk,Takahashi:2010ky,Nakayama:2010kt}. It is interesting
to briefly discuss concrete models where this is actually the
case. Here, we follow \Refcs{Takahashi:2010ky,Nakayama:2010kt}. These
models are supergravity models where one assumes that the K\"ahler
potential is invariant under a generalization of the shift symmetry
(usually needed in order to avoid the so called $\eta$-problem). In
the present case, the transformation is taken to be $\chi^n
\rightarrow \chi^n+\alpha$ where $\alpha$ is a real number and $\chi$
a chiral superfield. This means that the K\"ahler potential should be
a function of $\chi^n-\chi^{\dagger}{}^n$ only. In addition, we allow
the presence of a small breaking term in the K\"ahler potential of the
form $b\chi\chi^{\dagger}$ where $b\ll 1$. We also assume that the
superpotential breaks the generalized shift symmetry. Summarizing, we
assume that
\begin{eqnarray}
K &=& b\chi\chi^{\dagger}+c_1\kappa^{(n-1)/2}\left(\chi^n-\chi^{\dagger}{}^n\right)
-\frac{\kappa^{n-1}}{2}\left(\chi^n-\chi^{\dagger}{}^n\right)^2+XX^{\dagger}, \\
W &=& \lambda X\chi^m,
\end{eqnarray}
where $X$ is another superfield and $\lambda $ and $c_1$ (notice that
it is pure imaginary) are constant. The model is parametrized by the
quantities $n$ and $m$ and $\kappa\equiv 1/\Mp^2$. If, during
inflation, $X$ acquires a large mass compared to the Hubble parameter
and is stabilized at the origin, $\langle X\rangle =0$, then it is not
difficult to show that this supergravity model can be described by the
following effective Lagrangian
\begin{eqnarray}
\calL &=& -\left[b+n^2\kappa^{n-1}\left(\chi \chi^{\dagger}\right)^{n-1}\right]
\partial_\mu \chi\partial^{\mu}\chi^{\dagger}
\nonumber \\ & &
-\exp\left[b\kappa\vert \chi\vert^2
+c_1\kappa^{n/2}\left(\chi^n-\chi^{\dagger}{}^n\right)
-\frac{\kappa^n}{2}\left(\chi^n-\chi^{\dagger}{}^n\right)^2\right]
\lambda^2\left(\chi\chi^{\dagger}\right)^m.
\end{eqnarray}
Then, one can write the field $\chi$ in polar form, $\chi\equiv
\alpha\ee^{i\beta}$ ($\alpha $ is of dimension one and $\beta$
dimensionless) and the above potential takes the form
\begin{equation}
V=\lambda^2\alpha^{2m}\exp\left[b\kappa \alpha^2
+2ic_1\kappa^{n/2}\alpha^n\sin\left(n\beta\right)
+2\kappa^n\alpha^{2n}\sin^2\left(n\beta\right)\right].
\end{equation}
Writing $\partial V/\partial \beta =0$, one obtains the condition
$2i\kappa^{n/2}\alpha^n\sin(n\beta)=-ic_1$ or
$\kappa^{n/2}\left(\chi^n-\chi^{\dagger}{}^n\right)=c_1$. It is thus
natural to assume that the inflaton field rolls along that
direction. As a consequence, the effective Lagrangian takes the form
\begin{eqnarray}
\calL &=& -\left[b+n^2\kappa^{n-1}\left(\chi \chi^{\dagger}\right)^{n-1}\right]
\partial_\mu \chi\partial^{\mu}\chi^{\dagger}
-\ee^{b\kappa\vert \chi\vert^2
+c_1^2/2}
\lambda^2\left(\chi\chi^{\dagger}\right)^m.
\end{eqnarray}
Now, in the regime $b\kappa\vert \chi\vert^2\ll 1$, the exponential
becomes essentially independent of the field $\chi$ and the
coefficient $b$ in the kinetic term becomes negligible. It is
therefore natural to define a new quantity $\theta\equiv
\kappa^{(n-1)/2}\chi^n$ for which one obtains the Lagrangian of a
canonically normalized field, namely
\begin{eqnarray}
\calL &=& -
\partial_\mu \theta\partial^{\mu}\theta^{\dagger}
-\ee^{c_1^2/2}
\lambda^2\left(\theta\theta^{\dagger}\right)^{m/n}.
\end{eqnarray}
Finally, we take the imaginary part of $\theta$ to be stabilized to
$c_1$ in order to satisfy the condition discussed above and we define
the real field $\phi$ by $\theta=\phi/\sqrt{2}+c_1/2$. As a
consequence, it follows
\begin{eqnarray}
\calL &\simeq & -
\frac{1}{2}\partial_\mu \phi\partial^{\mu}\phi^{\dagger}
-\ee^{c_1^2/2}
\lambda^2\phi^{2m/n}.
\end{eqnarray}
Therefore, we have obtained a LFI model with $p=2m/n$ (neglecting a
term $\vert c_1\vert^2$ in $V$). In \Refc{Takahashi:2010ky}, the case
$n=2$ and $m=1$ was considered and we see that this leads to a linear
potential. In \Refc{Nakayama:2010kt}, the generalized case considered
before was introduced and studied. It is worth mentioning that, when
the condition $b\kappa\vert \chi\vert^2\ll 1$ is not satisfied, the
potential remains of the LFI form but with a different $p$, see
\Refc{Nakayama:2010kt}. For instance, as shown in
\Refc{Takahashi:2010ky}, if $n=2$ and $m=1$, the potential is in fact
quadratic at the origin. This means that the standard relation between
$p$ (in the inflationary regime) and the mean equation of state during
reheating namely, $\wrehbar=(p-2)/(p+2)$~\cite{Turner:1983he}, is no
longer valid in that case.

\subsubsection{Slow-Roll Analysis}
\label{subsubsec:srlfi}

Having studied how the LFI model can be implemented in high energy
physics, we now turn to the inflationary analysis. In the following,
we write $V(\phi)$ as
\begin{equation} 
 V(\phi) = M^4\left(\frac{\phi}{\Mp}\right)^p.
\end{equation} 
This potential is represented in \Fig{potLFI} for $p=2$. The three
Hubble flow functions are straightforwardly obtained from
\Eqs{eq:eps1}, \eqref{eq:eps2} and \eqref{eq:eps3}. Defining $x\equiv
\phi/\Mp$, one gets
\begin{equation}
\label{eq:epslfi}
  \epsilon _1 = \frac{p^2}{2 x^2}\,
  ,\qquad \epsilon_2= \frac{2p }{x^2}\,, \qquad \epsilon
  _3=\epsilon_2\, .
\end{equation}
These functions are represented in the two bottom panels of
\Fig{potLFI}. They are monotonic decreasing functions of $\phi$. One
can immediately deduce that, for a given $p$, the model in the plane
$(\epsilon _1,\epsilon _2)$ is contained in the line $\epsilon
_1=(p/4)\epsilon _2$.

The slow-roll trajectory is completely explicit and obtained by
quadrature from \Eq{eq:srtrajectory}
\begin{equation}
  N-\Nend = -\frac{1}{\Mp^2}\int _{\phiend}^{\phi}
  \frac{V(\chi )}{V'(\chi )} \dd \chi = -\frac{1}{p}
  \int_{\phiend/\Mp}^{\phi/\Mp} x \dd x =\dfrac{1}{2p}
  \left(\xend^2 -
    x^2 \right).
\label{eq:trajlf}
\end{equation}
This expression can be inverted and reads
\begin{equation}
x = \sqrt{\xend^2 -2p\left( N - \Nend\right)}\, .
\label{eq:lfi:InvertedTraj}
\end{equation}

For the large field models, inflation ends naturally when
$\epsilon_1=1$ (see \sectionc{sec:introduction}). Along the $\phi>0$
branch of the potential, this leads to
\begin{equation}
\xend=\frac{p}{\sqrt{2}}\, .
\label{eq:phiendlf} 
\end{equation} 
This expression also allows us to obtain the total number of
\efolds. Plugging \Eq{eq:phiendlf} into \Eq{eq:trajlf}, one
arrives at
\begin{equation}
  \Nend-\Nini = \frac{1}{2p} \xini^2 
  - \frac{p}{4}\, ,
\end{equation}
which can be very large if the initial field value is
super-Planckian. Notice that this does not imply that the energy
density is close to the Planck scale as this one is typically given by
the potential and proportional to $M^4$. In fact, the model remains
under control only if the initial energy density is smaller than
$\Mp^4$ and this imposes a constraint on both $\phiini$ and $M$ which
reads
\begin{equation}
  \xini=\dfrac{\phiini}{\Mp} \lesssim
  \left(\frac{\Mp}{M}\right)^{4/p}.
\end{equation}
Let us notice that, when the inflaton energy density approaches the
Planck energy density, quantum effects become important. In this case,
the stochastic inflation formalism must be used~\cite{Vilenkin:1983xp,
  Vilenkin:1983xq, Goncharov:1987ir, Linde:1993xx, Starobinsky:1986fx,
  Martin:2005ir, Martin:2005hb}.

We now turn to the explicit determination of the slow-roll
parameters. We have seen that the model is represented by the
trajectory $\epsilon _1=(p/4)\epsilon _2$ but observable models only
lie in a limited portion of this straight line. Indeed, the
Hubble flow parameters should be evaluated when the scales of
astrophysical interest today left the Hubble radius during
inflation. Following the discussion of \sectionc{subsec:reheating}, we
assume the pivot mode crossed the Hubble radius for $\phi=\phistar$ at
the \efold number $\Nstar$. From the trajectory,
we have
\begin{equation}
  \xstar^2 = 2 p \left(\Delta
    \Nstar+\frac{p}{4}\right),
\end{equation}
and the slow-roll parameters read
\begin{equation}
\label{eq:lfi:epsstar}
  \epsonestar =\frac{p}{4\left(\Delta
      \Nstar+p/4\right)}\, ,\qquad \epstwostar=\frac{1}{\Delta 
      \Nstar+p/4}\,,
  \qquad \epsthreestar=\epstwostar\, .
\end{equation}
Solving \Eq{eq:phistarlnrrad} for $\phistar$ yields the slow-roll 
predictions represented in
\Fig{fig:CMBLFI}. As expected, the whole family lies in the region
$\epsilon _2>0$ and verifies $\epsilon _1=p/4\epsilon _2$.  From
\Fig{fig:CMBLFI}, we see that all the models with $p \gtrsim 3$ lie
outside the $2\sigma $ contour. The quadratic (or massive) model
is under great pressure since it predicts quite a high
contribution of gravitational waves, up to $r \simeq 15\%$ level.

Finally, the parameter $M$ can be determined from the amplitude of the 
CMB anisotropies, and one gets
\begin{equation}
  \frac{\Qrms^2}{T^2}
  =\frac{1}{480\pi^2\epsilon_{1*}}\frac{H_*^2}{\Mp^2} =\frac{1}{1440
    \pi^2\epsilon_{1*}}\frac{\Vstar}{\Mp^4}\,.
\label{multi} 
\end{equation}
In the case of large fields model, this implies
\begin{equation}
  \left(\frac{M}{\Mp}\right)^4=\frac{720 \pi^2 p^2} {\left( \xstar^2
    \right)^{p/2+1}}  \frac{\Qrms^2}{T^2}\,,
\label{eq:scaleMlf} 
\end{equation}
and given the constraints on $p$ and $\Delta \Nstar$, this leads to
$M/\Mp \simeq 3\times 10^{-3}$. We recover the conclusion that, for
large field models, inflation takes place close to the Grand Unified
Theory (GUT) scale.

\subsection{Mixed Large Field Inflation (MLFI)}
\label{sec:mlfi}

\begin{figure}
\begin{center}
\includegraphics[width=\wdblefig]{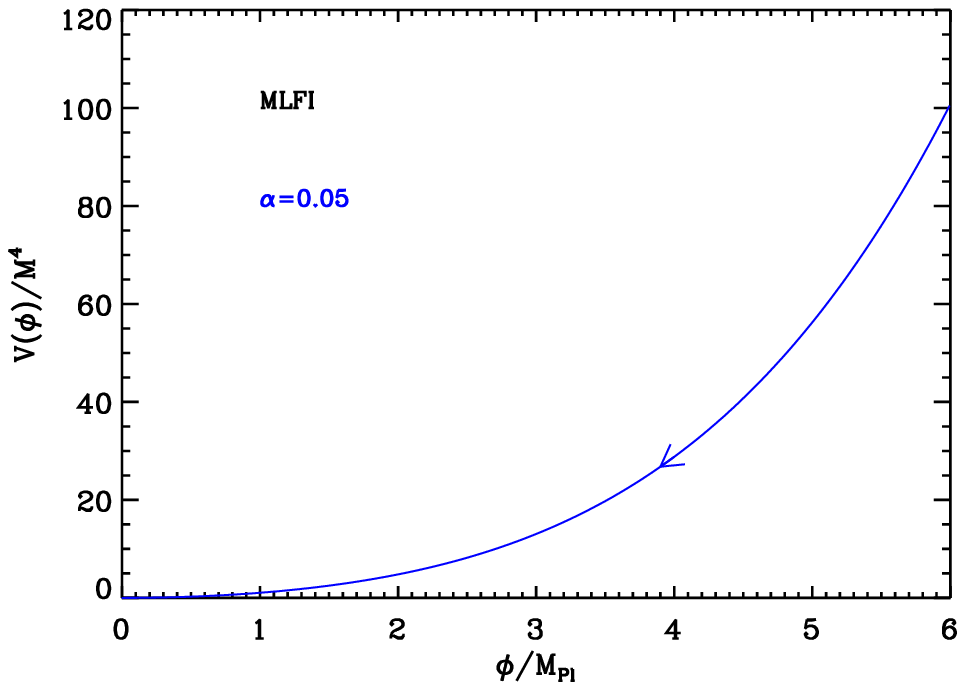}
\includegraphics[width=\wdblefig]{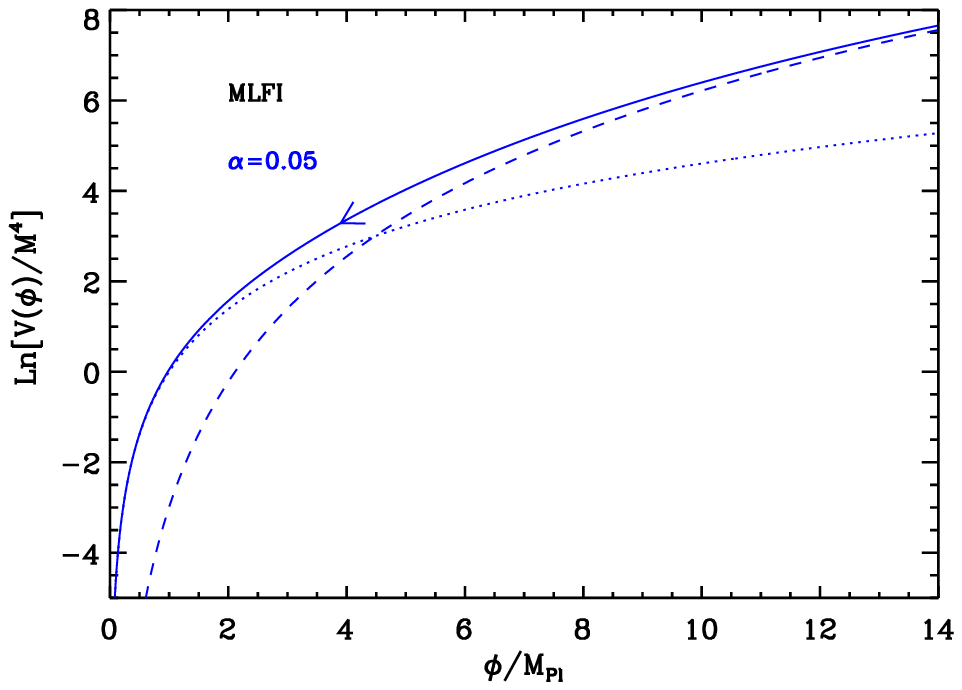}
\includegraphics[width=\wdblefig]{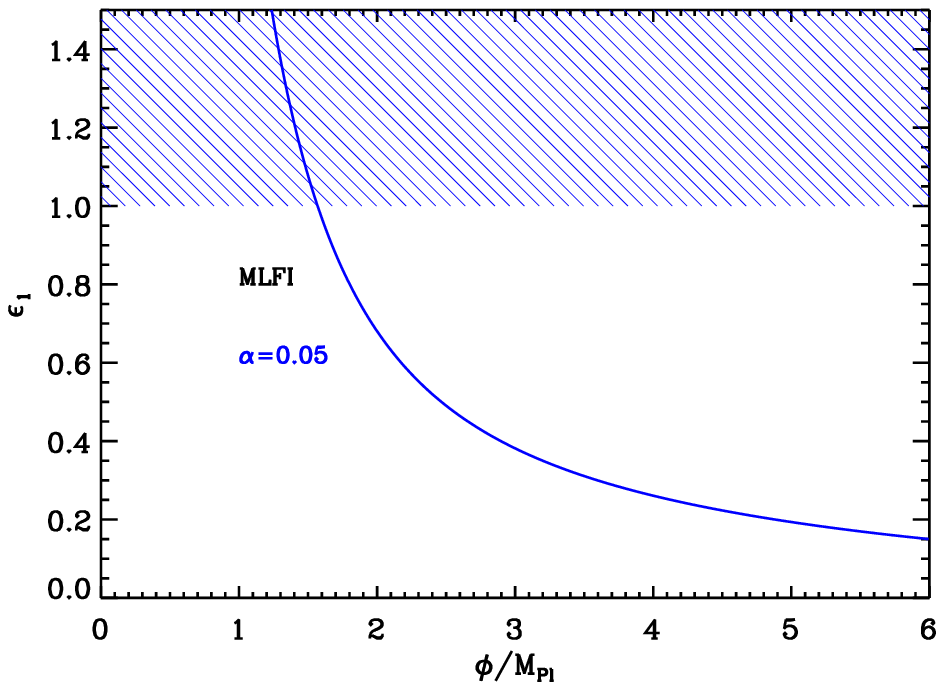}
\includegraphics[width=\wdblefig]{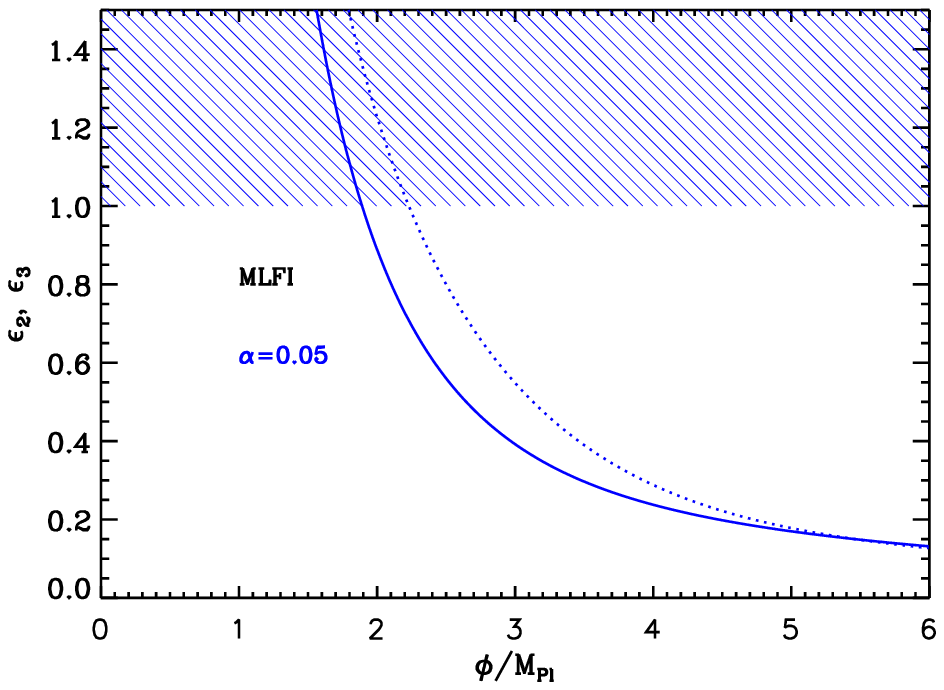}
\caption{Top left panel: mixed large field (MLFI) potential, see
  \Eq{eq:mlfipot}, for $\alpha=0.05$.  Top right panel: logarithm of
  the potential for the same value of $\alpha$. The dotted line
  indicates the potential $V\left(\phi\right)\simeq M^4 \phi^2/\Mp^2$
  which is the limit of the MLFI potential in the regime $\phi/\Mp\ll
  1/\sqrt{\alpha}$ while the dashed line represents the expression
  $V\left(\phi\right)\simeq M^4\alpha \phi^4/\Mp^4$, the limit of
  $V(\phi)$ when $\phi/\Mp\gg 1/\sqrt{\alpha}$.  For $\alpha=0.05$ the
  two lines meet at the following value, $1/\sqrt{\alpha}\simeq 4.5$,
  as can be directly checked in the figure. The arrow in the top left
  and right panels indicate in which direction inflation
  proceeds. Bottom left panel: slow-roll parameter $\epsilon _1$ for a
  mixed large field potential with $\alpha=0.05$. Bottom right panel:
  slow-roll parameters $\epsilon _2$ (solid line) and $\epsilon _3$
  (dotted line) still for $\alpha=0.05$.}
\label{potMLFI}
\end{center}
\end{figure}

This model is a generalization of the LFI model
$V(\phi)\propto\phi^p$, see \sectionc{sec:lfi}, where two monomials
$\propto\phi^2$ and $\propto\phi^4$ are added. The MLFI potential
reads
\begin{equation}
\label{eq:mlfipot}
  V(\phi) = M^4\frac{\phi^2}{\Mp^2} \left(1 
+ \alpha \frac{\phi^2}{\Mp^2}\right),
\end{equation}
where $\alpha $ is a positive dimensionless parameter. If $\phi/\Mp\ll
1/\sqrt{\alpha}$, then the potential is of the LFI type with $p=2$,
\ie $V\left(\phi\right)\simeq M^4 \phi^2/\Mp^2$, whereas if
$\phi/\Mp\gg 1/\sqrt{\alpha}$, the potential is of the LFI type with
$p=4$, \ie $V\left(\phi\right)\simeq M^4\alpha \phi^4/\Mp^4$. Clearly,
the interesting regime is when $\phi/\Mp\simeq 1/\sqrt{\alpha}$, where
the two terms are of equal importance. The potential and its logarithm
are displayed in \Fig{potMLFI}. We notice that $V(\phi)$ is an
increasing function of the field \vev and, as a consequence, that
inflation proceeds from the right to the left.

This model has been investigated in different contexts. Of course, the
shape of the potential appears to be natural and well-motivated since
it just represents a free theory (with particles of mass $2M^4/\Mp^2$)
corrected by the usual self-interacting quartic term. Therefore, it
does not come as a surprise that this potential has been used in many
different works. In \Refc{Mohapatra:2000cm}, this model is studied in
the case where a bulk scalar field is driving inflation in large extra
dimensions. In \Refc{Cao:2002qe}, it is considered in a situation
where inflation is driven by highly excited quantum states. In
\Refcs{Bellini:2001ka, Bellini:2002zr, Bellini:2002vm}, the MLFI
potential is utilized in the context of ``fresh inflation''. The same
potential was again considered in \Refc{Chen:2010uc} where the role of
inflaton is played by the Higgs triplet in a model where the type II
seesaw mechanism is used to generate the small masses of left-handed
neutrinos. Finally, it is also studied in \Refc{Bouaouda:2010zz} where
supersymmetric hybrid inflation (in the framework of the
Randall-Sundrum type II Braneworld model) is considered. The only
constraint on the parameters of the model that is (sometimes) required
is that the self-interacting term should be sub-dominant. This leads
to the condition $\alpha M^4/\Mp^4\ll 1$. Given the typical values
imposed by CMB normalization, \ie $M/\Mp\simeq 10^{-3}$ [see
  \Eq{eq:scaleMlf}], this is not very stringent and $\alpha $ can in
fact vary in a quite large range of values.

Defining $x\equiv\phi/\Mp$, the three first slow-roll parameters can
be expressed as
\begin{equation}
  \epsilon_1 
  = \frac{2}{x^2} \left( \frac{1+2\alpha
      x^2 }{1 +
      \alpha x^2} \right)^2,\quad 
  \epsilon_2  =
  \frac{4}{x^2} \frac{1+\alpha x^2+2\alpha^2
    x^4}{\left(1+\alpha x^2 \right)^2}\,,
\end{equation}
and
\begin{equation}
  \epsilon_3 = \frac{\Mp^2}{x^2} \frac{1 + 2\alpha x^2}
  {\left(1 + \alpha x^2 \right)^2}
  \frac{4 + 12\alpha x^2 + 8\alpha^3 x^6}
  {1+\alpha x^2 + 2\alpha^2 x^4} \, .
\end{equation}
They are displayed in \Fig{potMLFI}. We see that the three slow-roll
parameters are decreasing functions of the field \vev, which means
that they are all increasing functions during inflation. As a
consequence, inflation can stop by violation of the slow-roll
conditions at $\xend$ given by $\epsilon_1=1$ (see below). We also
notice that $\epsilon_2$ and $\epsilon_3$ are larger than one at
$\xend$. This means that the slow-roll approximation breaks down
slightly before the end of inflation and that the last few \efolds
of inflation may be not properly described by the slow-roll
approximation.

\begin{figure}
\begin{center}
\includegraphics[width=\wsingfig]{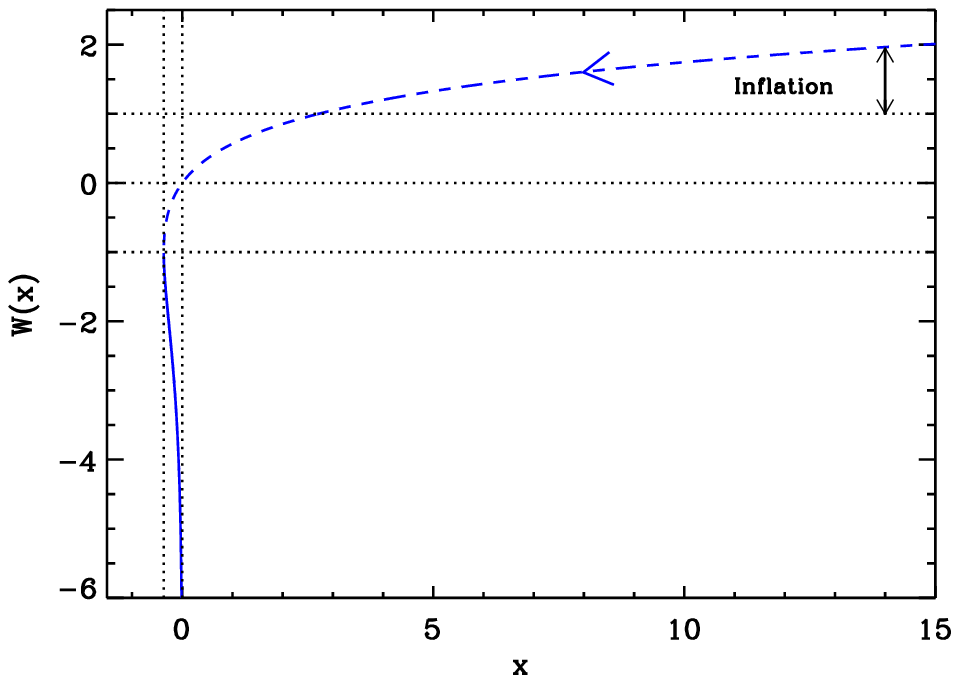}
\caption{Lambert functions $\Lambert{0}(x)$ (dashed line) and
  $\Lambert{-1}(x)$ (solid line). During Mixed Large Field inflation,
  inflation proceeds along the ``$0$'' branch above the line $W=1$ in
  the direction specified by the arrow.}
\label{fig:potlambertMLFI}
\end{center}
\end{figure}

Let us now study the slow-roll trajectory. It is given by
\begin{equation}
   \Nend -N=-\frac{1}{8}\left[\xend^2+\frac{1}{2\alpha }\ln
    \left( 1+2\alpha \xend^2\right)-x^2
    -\frac{1}{2\alpha }\ln \left( 1+2\alpha x^2\right)\right],
\end{equation}
where $\Nend$ is the number of \efolds at the end of
inflation. One can check that this expression is asymptotically
correct. Indeed, when $\alpha\ll 1$, the slow-roll trajectory reduces
to
\begin{equation}
\xend^2=x^2-4\left(\Nend-N\right),
\end{equation}
which is the trajectory in the massive case, \ie LFI with $p=2$, see
\Eq{eq:trajlf}.  On the other hand, in the limit $\alpha \rightarrow
\infty$, one obtains
\begin{equation}
\xend^2=x^2-8\left(\Nend -N\right),
\end{equation}
which is, as expected, the slow-roll trajectory in the quartic case,
\ie LFI with $p=4$. In general, the trajectory can be inverted and
expressed in terms of the Lambert function. Straightforward
manipulations lead to
\begin{equation} 
\label{eq:mlfi:InvertedTraj}
x=\frac{1}{\sqrt{2\alpha }}
\sqrt{-1+\Lambert{0}\left[\ee^{1+2\alpha \xend^2}
\left(1+2\alpha \xend^2\right)\ee^{-16\alpha
\left(N-\Nend\right)}\right]}\, .
\end{equation}
The corresponding Lambert function is displayed in
\Fig{fig:potlambertMLFI}, together with the region where inflation
proceeds.

We have seen that, in MLFI, inflation stops by violation of the
slow-roll condition. Let us therefore determine the corresponding \vev
of the field. The condition $\epsilon_1=1$ leads to
\begin{equation} \alpha\xend^3
-2\sqrt{2}\alpha\xend^2
+\xend-\sqrt{2}=0.
\end{equation}
This is a cubic algebraic equation that can be solved exactly. In the
limit $\alpha \gg 1$, the solution reads $\xend\simeq 2\sqrt{2}$ which
is indeed the solution for the quartic case, see \Eq{eq:phiendlf}. On
the other hand, if $\alpha \ll 1$, then $\xend\simeq\sqrt{2}$ which is
also the correct result for the quadratic case. The general solution
is
\begin{equation}
\begin{aligned}
\label{eq:mlfi:xend}
  \xend&= \frac{2\sqrt{2}}{3} +\frac{1}{3\alpha
  }\Biggl\{\frac{1}{4\sqrt{2}}\Biggl[ 4\alpha^2\left(32\alpha+9\right)
  +2\alpha \sqrt{4\alpha ^2\left(32\alpha+9\right)^2
    -8\alpha\left(8\alpha -3\right)^3}\Biggr]\Biggr\}^{1/3} \\ &+
  \frac{1}{3}\left(8\alpha-3\right)\Biggl\{\frac{1}{4\sqrt{2}}\Biggl[
  4\alpha^2\left(32\alpha+9\right) +2\alpha \sqrt{4\alpha
    ^2\left(32\alpha+9\right)^2 -8\alpha\left(8\alpha
      -3\right)^3}\Biggr]\Biggr\}^{-1/3}\,,
\end{aligned}
\end{equation}
which is the one used in the \ASPIC library.

Finally, the parameter $M$ can be determined from the amplitude of the
CMB anisotropies, and one gets
\begin{equation}
  \left(\frac{M}{\Mp}\right)^4
  =\frac{2880\pi^2}{x^4}\frac{\left(1+2\alpha \xstar^2
    \right)^{2}}{\left(1+\alpha \xstar^2
    \right)^{3}}\frac{\Qrms^2}{T^2}\, .
\end{equation} 
Similarly to LFI (see \sectionc{sec:lfi}), this gives rise to
$M/\Mp\simeq 10^{-3}$. The reheating consistent slow-roll predictions
for the MLFI models are displayed in \Fig{fig:CMBMLFI}. The reheating
equation of state parameter $\wrehbar$ has been taken to $0$ which is
consistent with the fact that the potential is quadratic close to its
minimum.  As expected, when $\alpha\ll 1$ the predictions of the model
match those of LFI with $p=2$ and are aligned along the
$\epsilon_1=\epsilon_2/2$ line.  On the other hand, if $\alpha\gg 1$,
then the predictions are consistent with those of LFI with $p=4$ and
are aligned along the $\epsilon_1=\epsilon_2$ line. In the
intermediate regime, it is interesting to notice that the MLFI
predictions continuously interpolate between these two asymptotic
solutions but do not remain inside the domain delimited by those two
lines. Indeed, when $\alpha$ is larger than some value, one has
$\epsilon_1>\epsilon_2$. This means that, if one starts from a pure
quartic potential (LFI with $p=4$) and adds a small quadratic term,
this extra term has the effect of increasing the ``effective value''
of $p$, which is quite counter intuitive. On the other hand, since the
quadratic model fits better the data than the quartic one, small
values for the parameter $\alpha$ are favored (all the models with
$\alpha>10^{-3}$ lie outside the $2\sigma $ contour of the \data
data). High reheating temperatures are also preferred.

\subsection{Radiatively Corrected Massive Inflation (RCMI)}
\label{sec:rcmi}

\begin{figure}
\begin{center}
\includegraphics[width=\wdblefig]{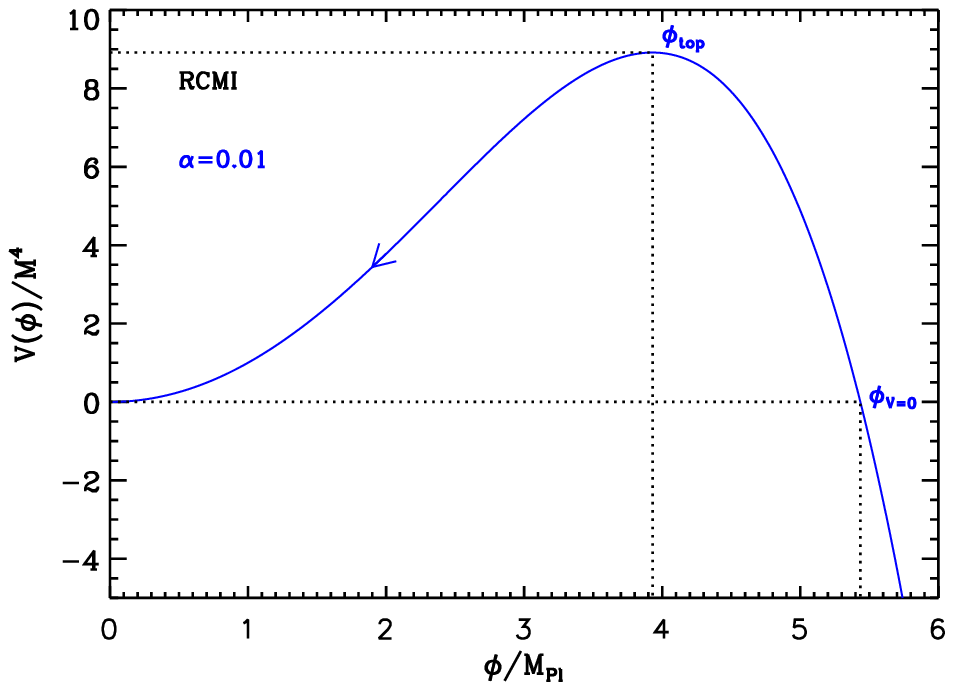}
\includegraphics[width=\wdblefig]{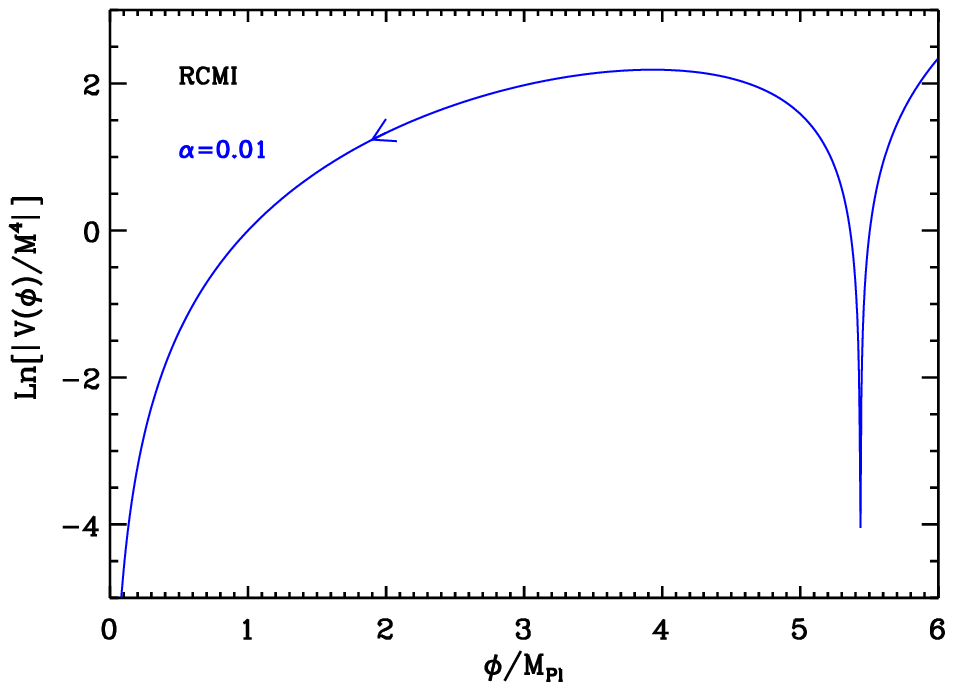}
\includegraphics[width=\wdblefig]{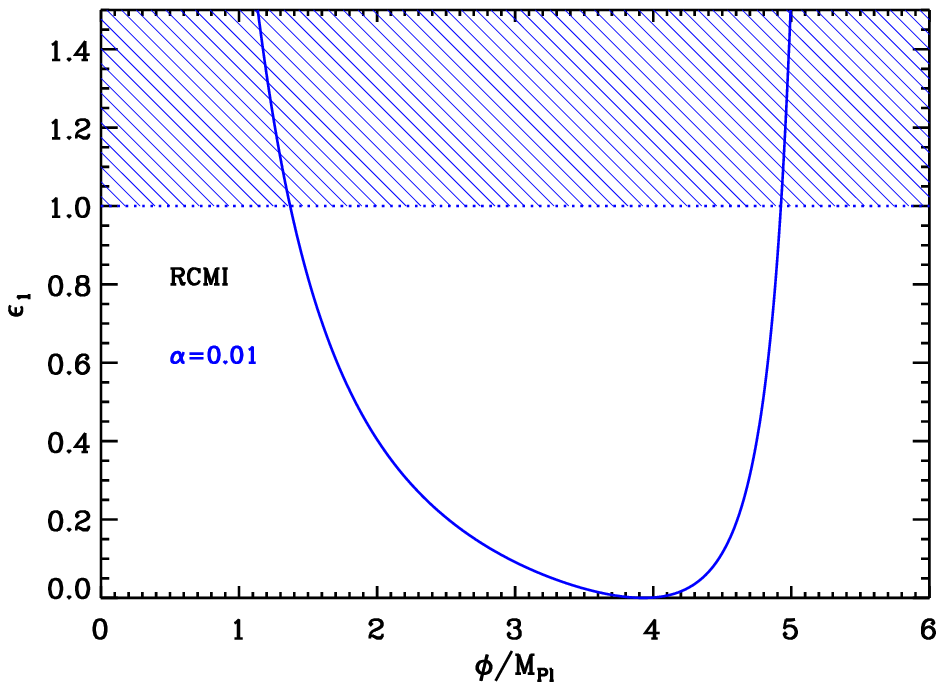}
\includegraphics[width=\wdblefig]{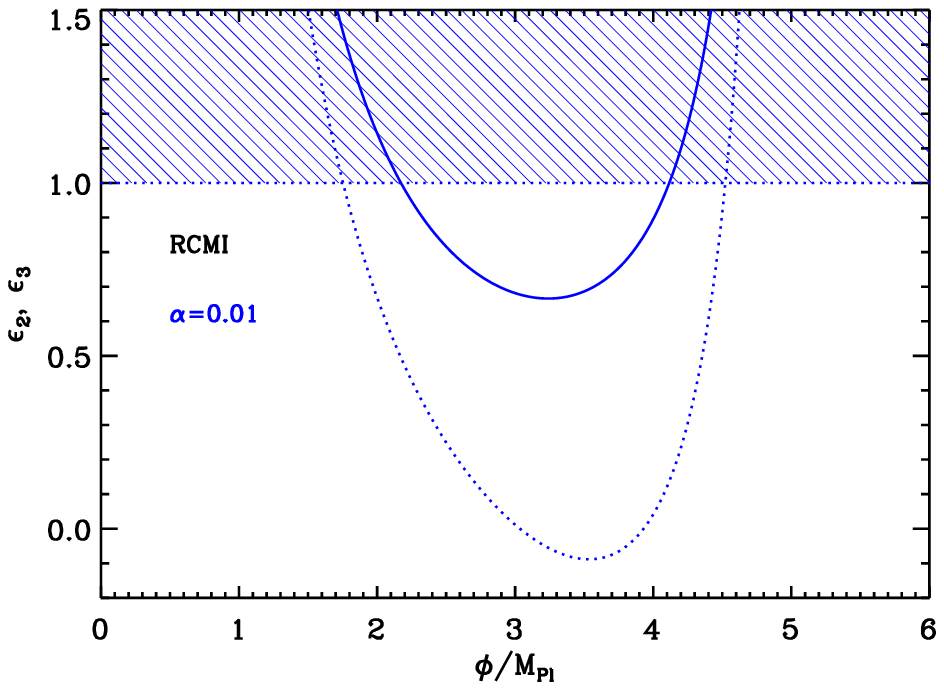}
\caption{Radiatively Corrected Massive Inflation (RCMI) for
  $\alpha=0.01$. Top panels: potential (left) and logarithm of the
  potential (right). Bottom left panel: slow-roll parameter
  $\epsilon_1$ with respect to field values. The shaded area indicates
  where inflation stops. Bottom right panel: slow-roll parameters
  $\epsilon_2$ (solid line) and $\epsilon_3$ (dotted line).}
\label{potRCMI}
\end{center}
\end{figure}

This model is based on \Refc{NeferSenoguz:2008nn} and implements
radiative corrections due to fermion couplings over the massive
($p=2$) large field model (see \sectionc{sec:lfi}). With an
appropriate choice of the renormalization scale $\mu=g \Mp$, $g$
denoting the Yukawa coupling, the potential is given by
\begin{equation}
\begin{aligned}
V(\phi)&=\frac{1}{2}m^2\phi ^2-\frac{g^4}{16\pi ^2}\phi ^4
\ln \left(\frac{\phi}{\Mp}\right) = M^4\left(\frac{\phi}{\Mp}\right)^2
\left[1-2\alpha\frac{\phi^2}{\Mp^2}\ln \left(\frac{\phi}{\Mp}\right)\right],
\end{aligned}
\label{eq:potrcmi}
\end{equation}
where 
\begin{equation}
  M^4 \equiv \dfrac{1}{2} m^2 \Mp^2, 
\qquad \alpha \equiv \frac{g^4\Mp^2}{16\pi^2m^2}\,.
\end{equation}
This expression is obtained in the large field regime $\phi \gg m/g$
(this condition coming from the requirement that the fermion loop
contribution dominates over the self-interaction loop contribution),
i.e. assuming that the inflationary regime takes place under the
condition
\begin{equation}
\dfrac{\phi^4}{\Mp^4} \gg \dfrac{1}{8 \pi^2 \alpha} \dfrac{M^4}{\Mp^4}\,.
\label{eq:alphaminrcmi}
\end{equation}
Defining $x \equiv \phi/\Mp$, the Hubble flow functions are given by
\begin{equation}
\begin{aligned}
  \epsilon_1 = \dfrac{2}{x^2} \left(\frac{1-\alpha
      x^2 - 4\alpha x^2 \ln x}{1-2\alpha x^2
      \ln x} \right)^2\,,
\end{aligned}
\end{equation}
\begin{equation}
\begin{aligned}
& \epsilon_2  = \dfrac{4}{x^2} \dfrac{\left(1 + \alpha  x^2\right)
    \left(1 + 2 \alpha x^2\right) - 2 \alpha  x^2 \ln x \left(1 - \alpha
      x^2 - 4 \alpha  x^2 \ln x\right)}{\left(1 - 2 \alpha 
 x^2 \ln x \right)^2}\,,
 \end{aligned}
\end{equation}
and
\begin{equation}
\begin{aligned}
& \epsilon_3 =  \dfrac{4}{x^2} \dfrac{1 - \alpha  x^2 - 4 \alpha  x^2 \ln
  x }{\left(1 - 2 \alpha  x^2 \ln x \right)^2 }  \\ & \times
\dfrac{1 -
   \alpha  x^2 \left[\alpha  x^2 \left(4 \alpha 
   x^2+9\right)+1\right] -  \alpha  x^2 \ln x \left[4
   \alpha ^2 x^4 \ln x (4 \ln x+1)+\left(\alpha 
   x^2+3\right) \left(6 \alpha 
   x^2 + 2\right)\right]} {\left(1 + \alpha  x^2 \right) \left(1 + 2
   \alpha  x^2 \right) - 2 \alpha  x^2 \ln x \left(1 - \alpha
    x^2 - 4 \alpha  x^2 \ln x\right)}\,.
\end{aligned}
\end{equation}
If $\alpha=0$, one recovers the slow-roll parameters of the massive
case (namely LFI with $p=2$, see \sectionc{sec:lfi}) as expected. The
potential and Hubble-flow functions have been represented in
\Fig{potRCMI}.

Let us now discuss the field domains in which inflation can take
place. It is clear that the above potential is not positive definite
for all field values. It becomes negative at the point
\begin{equation}
\xVzero = \dfrac{\phiVzero}{\Mp} = \sqrt{\frac{1}{\alpha \Lambert{0}
    \left(1/\alpha\right)}}\,,
\end{equation}
where $\Lambert{0}$ is the $0$-branch of the Lambert function. The
model is defined only in the regime $\phi<\phiVzero$. On the other
hand, the top of the potential, where $V^\prime=0$ (or equivalently
$\epsilon_1=0$), is given by
\begin{equation}
\xVtop = \dfrac{\phiVtop}{\Mp} = \sqrt{\frac{1}{2\alpha \Lambert{0}
\left(\dfrac{\sqrt{e}}{2\alpha}\right)}}\,.
\label{eq:xtoprcmi}
\end{equation}
As the model makes sense only if the logarithmic terms do not dominate
the potential, the acceptable regime is $\phi<\phiVtop<\phiVzero$, and
a large field region only exists for $\phiVtop/\Mp\gg 1$. From the
above expression, this means that we must be in the regime $\alpha \ll
1$. For $\phi<\phiVtop$ one can check from \Eqs{eq:potrcmi} and
\eqref{eq:xtoprcmi} that the loop corrections never exceed $\alpha/e$.

Let us now turn to the slow-roll trajectory. It is given by
\begin{equation}
N-\Nend=-\frac{1}{2}\int _{\phiend/\Mp}^{\phi/\Mp}
\frac{x-2\alpha x^3\ln x}{1-\alpha x^2-4\alpha x^2\ln x}
\dd x,
\end{equation}
an integral that cannot be performed analytically and which is
numerically evaluated in \ASPIC. For the purpose of this section, we
can nevertheless make an expansion in $\alpha$ to obtain an
approximate expression
\begin{equation}
\begin{aligned}
N - \Nend & = - \dfrac{x^2}{4} \left[1 + \alpha
  \dfrac{x^2}{4} \left(1 + 4 \ln x \right)
\right]  + \dfrac{\xend^2}{4} \left[1 + \alpha
  \dfrac{\xend^2}{4} \left(1 + 4 \ln \xend \right)
\right] + \order{\alpha^2}\, .
\end{aligned}
\label{eq:trajapproxrcmi}
\end{equation}
Inflation stops close to the minimum of the potential when
$\epsilon_1=1$. This last equation cannot be solved analytically but
we can also perform an expansion at first order in $\alpha$ and one gets
\begin{equation}
  \xend = \frac{\phiend}{\Mp} \simeq\dfrac{1}{\sqrt{2 \alpha
      \Lambert{0} \left[\dfrac{e^{1+1/(4\alpha)}}{2 \alpha}\right]}}
      \simeq\sqrt{2}-2\sqrt{2}\alpha\,.
\label{eq:xendapproxrcmi}
\end{equation}
In the limit $\alpha \rightarrow 0$, we recover the large field result
for $p=2$, i.e. $\xend \rightarrow \sqrt{2}$. The maximum total number
of \efolds one can realize between $\phi=\phiVtop$ and $\phi=\phiend$
can be calculated from the previous expressions. It reads
\begin{equation}
\begin{aligned}
  \Delta \Nmax = \Nend-\Ntop &  =  \dfrac{5}{32 \alpha
    \Lambert{0}\left(\dfrac{\sqrt{e}}{2 \alpha} \right)} 
  +  \dfrac{1+2\alpha - 20 \alpha \Lambert{0} \left[
      \dfrac{\ee^{1+1/(4\alpha)}}{2 \alpha} \right]} {128 \alpha^2 
    \Lambert{0}^2\left[
      \dfrac{\ee^{1+1/(4\alpha)}}{2 \alpha} \right]}\\
   &\simeq-\frac{5}{32\alpha\ln\left(\alpha\right)}   
      \,.
\end{aligned}
\end{equation}
This is a decreasing function of $\alpha$, so that $\alpha$ has to be
small enough if one wants a sufficiently high number of \efolds to
take place.  Indeed, if one wants at least $\Delta\Nmin$ \efolds to 
occur, one needs to work with
\begin{equation}
\alpha<\frac{5}{32\Delta\Nmin}\frac{1}{\ln\left(\frac{32\Delta\Nmin}{10}\right)}\, .
\end{equation}
For example, $\Delta \Nmin=50$ imposes $\alpha<6
\times 10^{-4}$. The fact that $\alpha$ is bounded from above can be 
directly checked in \Fig{fig:CMBRCMI}. The field $\phistar$ value at which 
the pivot mode crossed the Hubble radius during inflation is obtained from
\Eq{eq:phistarlnrrad} whereas the corresponding \efold number can be
obtained from the trajectory.

Finally, the parameter $M$ can be determined from the amplitude of the 
CMB anisotropies, and one gets
\begin{equation}
\begin{aligned}
  \left(\frac{M}{\Mp}\right)^4 &= \dfrac{2880\pi^2}{\xstar^4}
  \dfrac{\left(1 - 2\alpha \xstar^2 \ln \xstar \right)^3}{\left( 1 -\alpha
      \xstar^2 -4 \alpha \xstar^2 \ln \xstar \right)^{2}}
  \frac{\Qrms^2}{T^2}\, .
\end{aligned}
\end{equation}
The reheating consistent slow-roll predictions for the RCMI models are 
represented in \Fig{fig:CMBRCMI}. As expected, the LFI quadratic model 
case is properly recovered for $\alpha\rightarrow0$.  From this figure, we 
see that all models having $\alpha\gtrsim 10^{-3.7}$ lie outside the $2\sigma$
contour. Let us emphasize that the value of $\alpha$ cannot be
infinitely small due to \Eq{eq:alphaminrcmi}. At zero order, one has
$\phi > \phiend \simeq \sqrt{2} \Mp$ such that \Eq{eq:alphaminrcmi}
can be recast into
\begin{equation}
\alpha > \dfrac{M^4}{8 \pi^2 \Mp^4}=\frac{m^2}{16\pi^2\Mp^2}\,.
\end{equation}
{}From the \COBE normalization, and in the limit of small $\alpha$, one
gets $M/\Mp \gtrsim 10^{-3}$ and the lower bound reads $\alpha>
10^{-15}$.

\subsection{Radiatively Corrected Quartic Inflation (RCQI)}
\label{sec:rcqi}

\begin{figure}
\begin{center}
\includegraphics[width=\wdblefig]{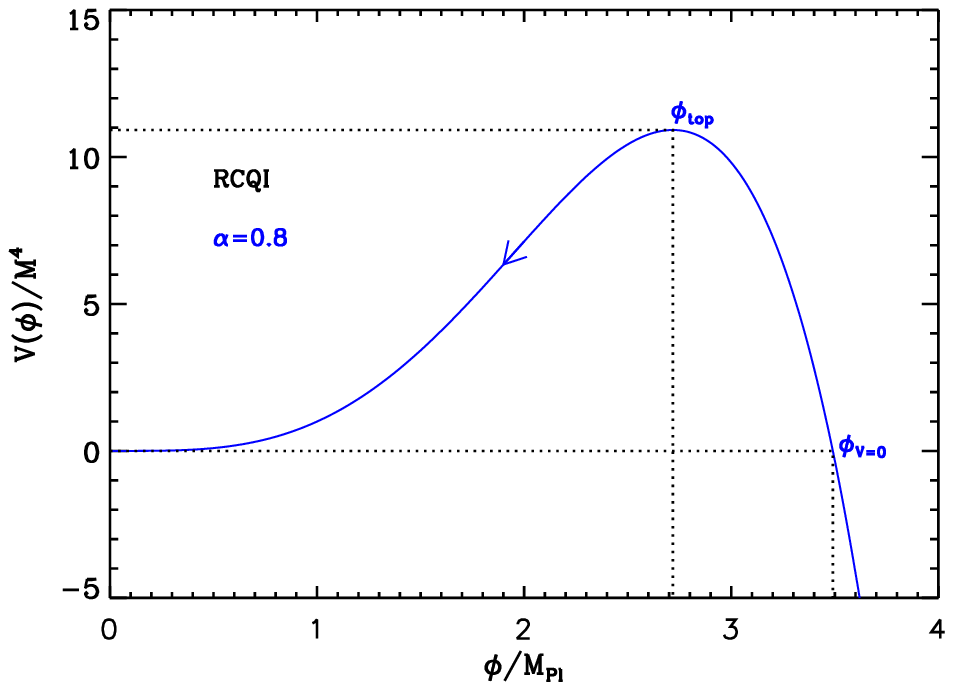}
\includegraphics[width=\wdblefig]{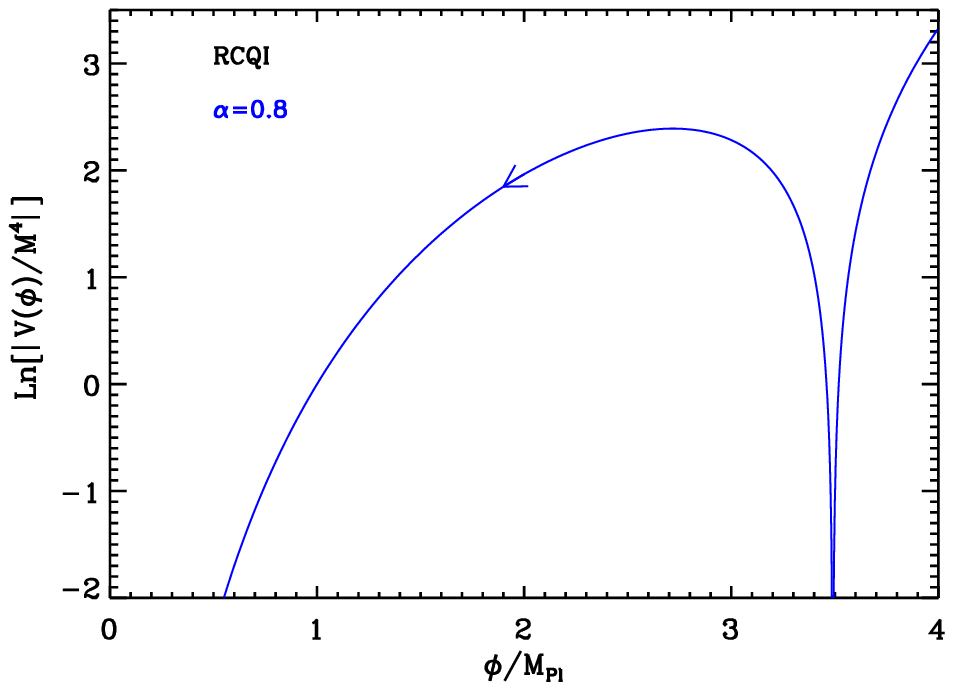}
\includegraphics[width=\wdblefig]{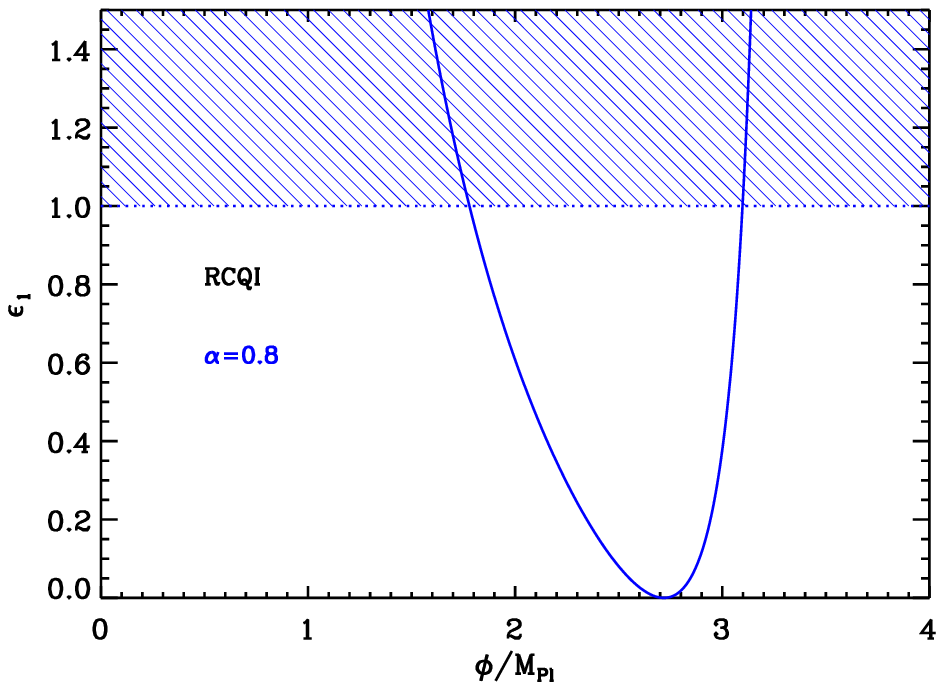}
\includegraphics[width=\wdblefig]{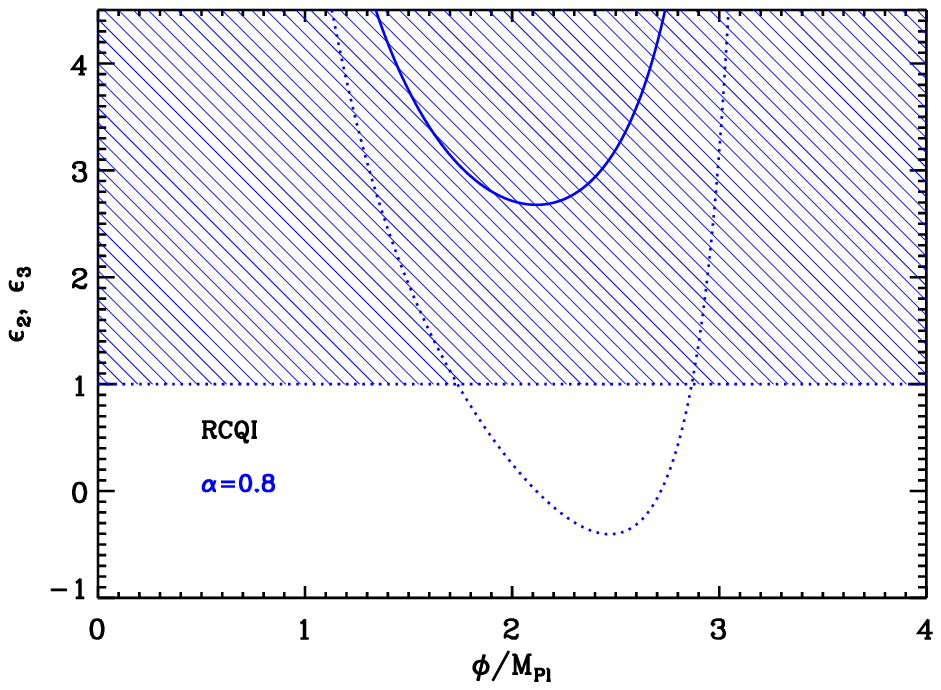}
\caption{Radiatively Corrected Quartic Inflation (RCQI) for
  $\alpha=0.8$.  Top panels: the potential and its logarithm as a
  function of the field values. Bottom left panel: slow-roll parameter
  $\epsilon _1$. The shaded area indicates where inflation
  stops. Bottom right panel: slow-roll parameters $\epsilon _2$ (solid
  line) and $\epsilon _3$ (dotted line). The shaded region for
  $\epsilon_2$ and $\epsilon_3$ shows where the slow-roll
  approximation is violated for that value of $\alpha$.}
\label{potRCQI}
\end{center}
\end{figure}

This model is similar to RCMI discussed in \sectionc{sec:rmi} but
implements radiative corrections due to fermion couplings over a
quartic ($p=4$) large field model~\cite{NeferSenoguz:2008nn}
(see \sectionc{sec:lfi}). The potential is given by
\begin{eqnarray}
  V=\lambda \phi^4-\frac{g^4}{16\pi^2}
\phi^4\ln\left(\frac{\phi}{\Mp}\right)
=M^4\left(\frac{\phi}{\Mp}\right)^4
\left[1-\alpha \ln\left(\frac{\phi}{\Mp}\right)\right],
\label{eq:potrcqi}
\end{eqnarray}
where
\begin{equation}
M^4 = \lambda \Mp^4,\qquad \alpha \equiv \frac{g^4}{16 \pi^2\lambda}\,.
\end{equation}
Defining $x = \phi/\Mp$, the Hubble flow functions in the slow-roll
approximation read
\begin{equation}
\begin{aligned}
  \epsilon_1 = \dfrac{8}{x^2} \left(\dfrac{1- \dfrac{\alpha}{4} -\alpha \ln x}
    {1-\alpha \ln x}\right)^2, \qquad \epsilon_2 = \frac{8}{x^2}
  \dfrac{1 + \dfrac{\alpha}{4} (\alpha-1) + \alpha
    \left(\dfrac{\alpha}{4}  -2 \right) \ln x + \alpha ^2 \ln^2
    x}{\left(1 - \alpha \ln x \right)^2}\,,
\end{aligned}
\end{equation}
and
\begin{equation}
\begin{aligned}
  \epsilon_3 &= \dfrac{8}{x^2} \frac{(1 -\dfrac{\alpha}{2} - \alpha
    \ln x) (1 - \dfrac{\alpha}{4} - \alpha \ln x) \left[1 +
      \dfrac{\alpha^2}{2} + \dfrac{\alpha}{4} - \alpha \left(2 +
        \dfrac{\alpha}{4} - \alpha \ln x \right)\ln x \right] }{(1 -
    \alpha \ln x)^2 \left[1 + \dfrac{\alpha}{4} (\alpha -1) - \alpha
      \left(2 - \dfrac{\alpha}{4} - \alpha \ln x \right) \ln x
    \right]}\,.
\end{aligned}
\end{equation}
The shape of the potential and the Hubble flow functions are very
similar to the ones of the RCMI model and have been represented in
\Fig{potRCQI}.
In particular, the potential is vanishing and maximal at the field values
\begin{equation}
\xVzero = \dfrac{\phiVzero}{\Mp} = \ee^{1/\alpha }, \quad 
\xVtop = \dfrac{\phiVtop}{\Mp} = \ee^{1/\alpha-1/4},
\label{eq:xVrcqi}
\end{equation}
respectively. As the model makes sense only if the corrections are
small compared to the quartic term, one should consider $\alpha \ll 1$
and not too large super-Planckian field values.

The slow-roll trajectory can integrated analytically from
\Eqs{eq:srtrajectory} and \eqref{eq:potrcqi} and one gets
\begin{equation}
\begin{aligned}
  N-\Nend  = -\frac{1}{16}\Biggl[& 2 x^2 -\ee^{-1/2+2/\alpha} \Ei
  \left(\frac{1}{2}-\frac{2}{\alpha}
    +2\ln x \right) \\
  & -2 \xend^2 +\ee^{-1/2+2/\alpha} \Ei
  \left(\frac{1}{2}-\frac{2}{\alpha} +2\ln \xend \right)\Biggr],
\end{aligned}
\end{equation}
where the exponential integral function is defined by
\begin{equation}
\Ei(x)\equiv -\int _{-x}^{+\infty}\frac{\ee^{-t}}{t}\dd t.
\end{equation}
The quartic limit $\alpha \rightarrow 0$ is recovered by noticing that
\begin{equation}
  \Ei(-2/\alpha) \underset{\alpha\rightarrow 0}{\sim} -\dfrac{\alpha}{2} e^{-2/\alpha}.
\end{equation}
Contrary to the RCMI model, the top of the potential is flat enough to
support inflation. Indeed, one sees from \Eq{eq:xVrcqi} that the
argument of the exponential integral function vanishes at
$x=\xVtop$. Since for $y\rightarrow 0$, one has $\Ei(y) \sim \gamma +
\ln y$, whatever the value of $\xend$ the total number of \efolds is
divergent. This means that it is always possible to realize the
required $\Delta \Nstar$ number of \efolds provided inflation starts
close enough to the top of the potential.

As for RCMI, inflation stops at $\epsilon_1=1$ but this equation can
only be solved numerically. For illustrative purpose, one can
nevertheless solve it at first order in $\alpha$ to get
\begin{equation}
\xend = \frac{\phiend}{\Mp} \simeq 2 \sqrt{2}-\frac{\sqrt{2}}{2}\alpha.
\end{equation}
The link between $\phistar$ and $\Delta \Nstar$ is given by the
slow-roll trajectory with $\phistar$ given by \Eq{eq:phistarlnrrad}.

Finally, the parameter $M$ can be determined from the amplitude of the 
CMB anisotropies, and one gets
\begin{equation}
  \lambda=\frac{M^4}{\Mp^4}= \dfrac{11520 \pi^2}{\xstar^6}
  \dfrac{\left(1-\frac{\alpha}{4}-\alpha\ln \xstar \right)^{2}
}{\left(1-\alpha\ln \xstar\right)^3}
\frac{\Qrms^2}{T^2}\, .
\end{equation}
The slow-roll predictions for RCQI are represented in
\Fig{fig:CMBRCQI} and \ref{fig:CMBRCQIb}. As expected, the quartic
model case is properly recovered in the limit $\alpha \rightarrow
0$. From \Fig{fig:CMBRCQI}, we see that all the models seem to lie
outside the $2\sigma$ contour for $\wrehbar=0$. As the reheating phase
takes place at the bottom of a quartic-like potential, we have also
represented the prediction for $\wrehbar=1/3$ in
\Fig{fig:CMBRCQIb}. For a radiation-dominated reheating, $\Delta
\Nstar$ is fixed and for each value of $\alpha$ one has only a single
point. In that situation, all these models are still disfavored at the
two-sigma level.

\subsection{Natural Inflation (NI)}
\label{sec:ni}

\subsubsection{Theoretical Justifications}
\label{subsubsec:theoryni}

Natural inflation was first proposed as an attempt to solve the
so-called ``fine-tuning'' problem of inflation. In particular, in
order to obtain sufficient inflation and the correct normalization for
the microwave background anisotropies, the potential $V(\phi)$ of the
inflaton must be sufficiently flat. It is usually argued that, on
general grounds, such a flatness is not robust under radiative
corrections, unless it is protected by some symmetry. This is the
reason that has motivated \Refcs{Freese:1990rb,Adams:1992bn} to put
forward Natural Inflation, in which the inflaton potential is flat due
to shift symmetries. The model makes use of Nambu-Goldstone
bosons~\cite{Peccei:1977ur,Peccei:1977hh} which arise whenever a
global symmetry is spontaneously broken. The main idea can be very
simply illustrated with the following action
\begin{eqnarray}
\label{eq:actionni}
S &=& -\int \dd \bmx \sqrt{-g}
\biggl[g^{\mu \nu}\partial _\mu\Phi^{\dagger}\partial_\nu \Phi
+i\bar{\Psi} \gamma^{\mu}\partial _\mu \Psi+
\lambda \left(\Phi^{\dagger}\Phi-\frac{f^2}{2}\right)^2
\nonumber \\ & & 
+\gyuk \bar{\Psi}_\uL \Phi \PsiR
+\gyuk \bar{\Psi}_\uR \Phi^{\dagger} \PsiL
\biggr],
\end{eqnarray}
where $\Phi$ is a complex scalar field, $\Psi $ a Dirac spinor and
$\PsiLR=\left(1\pm \gamma_5\right)/2\Psi$. The quantity $f$ is the
energy scale at which the symmetry is spontaneously broken, $\lambda$
is a dimensionless coupling constant and $\gyuk$ a dimensionless
Yukawa coupling. This action is invariant under the $\mathrm{U}(1)$
transformation: $\Phi\rightarrow \ee^{i\alpha}\Phi$, $\PsiL\rightarrow
\ee^{i\alpha/2}\PsiL$ and $\PsiR\rightarrow \ee^{-i\alpha/2}\PsiR$,
where $\alpha$ is an arbitrary constant. Due to the ``Mexican hat''
potential for the scalar field, this symmetry is spontaneously broken
below the scale $f$ and the scalar field acquires the \vev $\langle
\Phi \rangle =f/\sqrt{2}\ee^{i \phi/f}$. The field $\phi$ corresponds
to an ``angular variable'' and is a Goldstone boson. Below the scale
of broken symmetry, the effective Lagrangian can be expressed as
\begin{equation}
\calL=\frac{1}{2}\partial_\mu\phi\partial^{\mu}\phi
+i\bar{\Psi}\gamma^{\mu}\partial_\mu\Psi+ \gyuk \frac{f}{\sqrt{2}}
\left(\bar{\Psi}_\uL \PsiR\ee^{i\phi/f}
+\bar{\Psi}_\uR \PsiL \ee^{-i\phi/f}\right).
\end{equation}
It is now invariant under $\phi\rightarrow \phi+2\pi f$,
$\PsiL\rightarrow \ee^{i\alpha/2}\PsiL$ and $\PsiR\rightarrow
\ee^{-i\alpha/2}\PsiR$. Then, we assume that an explicit symmetry
breaking takes place, for instance through the appearance of a fermion
condensate for which $\langle \bar{\Psi}\Psi\rangle \simeq \Ms^3$
where $\Ms<f$ is the scale at which this symmetry breaking occurs. As
a consequence, the effective Lagrangian takes the form
\begin{equation}
\calL=
\frac{1}{2}\partial_\mu\phi\partial^{\mu}\phi
+2\gyuk \Ms^3\frac{f}{\sqrt{2}}
\cos \left(\frac{\phi}{f}\right).
\end{equation}
We see that the Nambu-Goldstone boson has acquired a cosine potential
and the overall scale of the potential is given by $M^4\simeq
\gyuk \Ms^3f$. Therefore, if one takes $f\simeq \Mp$, $\Ms$ slightly below
the GUT scale and a Yukawa coupling of order one, one can
``naturally'' generate a small ratio $M/f$. A last remark is in order
on this model. Suppose that quantum gravity effects generate
non-renormalizable higher order terms in the
action~\eqref{eq:actionni} like
\begin{equation}
\Delta V=a_{mn}\frac{\vert \Phi\vert ^{2m}}{\Mp^{2m+n-4}}
\left(\Phi^n+\Phi^{\dagger}{}^n\right),
\end{equation}
where $a_{mn}$ are a priori unknown coefficients. After symmetry
breaking, one would therefore obtain a correction of the form
\begin{equation}
\Delta V =a_{mn}\Mp^4 \left(\frac{f}{\Mp}\right)^{2m+n}
\cos\left(n\frac{\phi}{f}\right).
\end{equation}
If $f\gtrsim \Mp$, as favored by current cosmological data (see
below) these terms should dominate unless the coefficients $a_{mn}$
are fine-tuned to very small values. Notice that the overall scale of
the potential is now given by $a_{mn}\Mp^4$, which also demands that
$a_{mn}\lesssim 10^{-15}$ in order to have the correct CMB
normalization. These terms are therefore dangerous for the consistency
and the natural character of the model. This model has been studied in
more details in \Refcs{Lyth:1991ub, Knox:1993zn, GarciaBellido:1996qt,
  Lyth:1998xn, Tsujikawa:1999ni, Wang:2002hf, Freese:2004un,
  Savage:2006tr, Panotopoulos:2007pt, Grimm:2007hs, Freese:2008if,
  Mohanty:2008ab, Ashoorioon:2008pj, Olsson:2007he, Maity:2012dx}.

Many other types of candidates have subsequently been explored in
order to produce scenarios similar to that of Natural Inflation.  For
example, in \Refc{Freese:1994fp}, it was suggested to use a
pseudo-Nambu Goldstone boson as the rolling field in double field
inflation. Then, NI potentials generated by radiative corrections in
models with explicitly broken Abelian \cite{Kinney:1995xv} and
non-Abelian \cite{Kinney:1995cc,Ross:2009hg} symmetries were
considered, showing that NI models with $f \simeq \Mp$ and $f \ll \Mp$
can both be generated. In \Refcs{German:2001sm}, the field $\phi$ is
considered to be a Polonyi field~\cite{Bailin:1994qt} and the model
predicts that
$f=\Mp$. Refs.~\cite{ArkaniHamed:2003wu,ArkaniHamed:2003mz} have
examined natural inflation in the context of extra dimensions and
\Refc{Kaplan:2003aj} has used pseudo-Nambu Goldstone bosons from
little Higgs models to drive hybrid inflation. Also,
\Refcs{Firouzjahi:2003zy, Hsu:2004hi} have used the natural inflation
idea of pseudo-Nambu Goldstone bosons in the context of braneworld
scenarios to drive inflation, \Refc{GonzalezFelipe:2007uy} has studied
the model in 5-$D$ warped backgrounds. The same potential has also
been obtained and studied in \Refc{Ovrut:1991iw} when studying
instantons in non-linear sigma models, and in \Refc{Kim:1998kx} as
providing quintessential inflation. In some of these references the
potential is sometimes found with the minus sign in front of the
cosine term, which is, up to a shift in the field \vev
$\phi/f\rightarrow\phi/f+\pi$, the same potential as already studied
before. This last model has also been derived and studied in
\Refcs{ArkaniHamed:2003wu, ArkaniHamed:2003mz, Park:2007sp} in the
context of orbifold GUT inflation, where the potential is given by
\begin{equation}
V(\phi) = M^4\left[F\left(\frac{\phi}{\phizero}\right)
+F\left(2\frac{\phi}{\phizero}\right)
+\frac{F(0)}{2}\right],
\end{equation}
with
\begin{equation}
F(x) = -\sum_{n=1}^{\infty}\frac{\cos\left(n\pi x\right)}{n^5}\, .
\end{equation}
This potential must be studied in its increasing branch, and in the
small field limit. At leading order, one recovers the cosine
potential. 

Finally, an important question is whether a situation where $f>\Mp$
makes sense from the high energy physics and effective field theory
point of view. In fact, it was shown in \Refcs{Preskill:1982cy,
  Abbott:1982af, Dine:1982ah} that $f\lessapprox 10^{12}\GeV$ in order
for the corresponding energy density not to exceed the critical energy
density. But this constraint applies to the post inflationary Universe
and, during inflation, \Refc{Linde:1987bx} has argued that it is not
relevant. However, it remains the question of whether $f>\Mp$ makes
sense or not. To address this issue, an interesting mechanism has been
proposed in \Refc{Kim:2004rp} (see also \Refc{Dimopoulos:2005ac})
which shows that two axion fields at sub-Planckian scales can have an
effective dynamics similar to the one field Natural Inflation model
with $f > \Mp$.

Let us consider a model with two axions, $\theta $ and $\rho$ the
effective Lagrangian of which is given by
\begin{equation}
\calL=\frac{1}{2}\partial_\mu\theta\partial^{\mu}\theta
+\frac{1}{2}\partial_\mu \rho\partial^{\mu}\rho
+M_1^4\left[1-\cos\left(\frac{\theta}{f}+\frac{\rho}{g_1}\right)\right]
+M_2^4\left[1-\cos\left(\frac{\theta}{f}+\frac{\rho}{g_2}\right)\right],
\end{equation}
where $M_1$ and $M_2$, $f$, $g_1$ and $g_2$ are constant, a priori,
arbitrary scales. The same model can be re-written in terms of the 
fields $\psi$ and $\xi$ defined by
\begin{equation}
  \psi=\frac{fg_1}{\sqrt{f^2+g_1^2}}\left(\frac{\theta}{f}+\frac{\rho}{g_1}\right),
  \qquad 
  \xi=\frac{fg_1}{\sqrt{f^2+g_1^2}}\left(-\frac{\theta}{g_1}+\frac{\rho}{f}\right).
\end{equation}
It is easy to show that this leads to 
\begin{eqnarray}
\calL&=&\frac{1}{2}\partial_\mu\psi\partial^{\mu}\psi
+\frac{1}{2}\partial_\mu \xi\partial^{\mu}\xi
+M_1^4\left[1-\cos\left(\frac{\sqrt{f^2+g_1^2}}{fg_1}\psi\right)\right]
\nonumber \\ & &
+M_2^4\left[1-\cos\left(\frac{f^2+g_1g_2}{fg_2\sqrt{f^2+g_1^2}}\psi
+\frac{g_1-g_2}{g_2\sqrt{f^2+g_1^2}}\xi
\right)\right].
\end{eqnarray}
Moreover, the mass of the two fields $\psi$ and $\xi $ can be expressed as
\begin{equation}
m_{\psi}^2=\left(\frac{1}{f^2}+\frac{1}{g_1^2}\right)M_1^4, 
\qquad
m_{\xi}^2=\frac{(g_1-g_2)^2}{g_2^2\left(f^2+g_1^2\right)}M_2^4.
\end{equation}
If $g_1$ is very close to $g_2$, then the field $\xi$ will be light
and, therefore, will have a non-trivial dynamics. In addition, if the
field $\psi$ is sufficiently heavy (compared to the Hubble parameter),
then its \vev will be frozen at $\psi=0$. In this case, we see that
the original two fields model effectively reduces to a one field NI
model with a scale $f_\xi$ given by
\begin{equation}
f_\xi=\frac{g_2\sqrt{f^2+g_1^2}}{g_1-g_2}.
\end{equation}
But, since, $g_1$ is close to $g_2$, the scale $f_\xi$ will be large
even if the fundamental scales $f$, $g_1$ and/or $g_2$ are
sub-Planckian. In this way, one can generate super-Planckian values
for the scale $f$ and, at the same time, have a theory which can be
consistent from the effective field theory point of view.

\begin{figure}
\begin{center}
\includegraphics[width=\wdblefig]{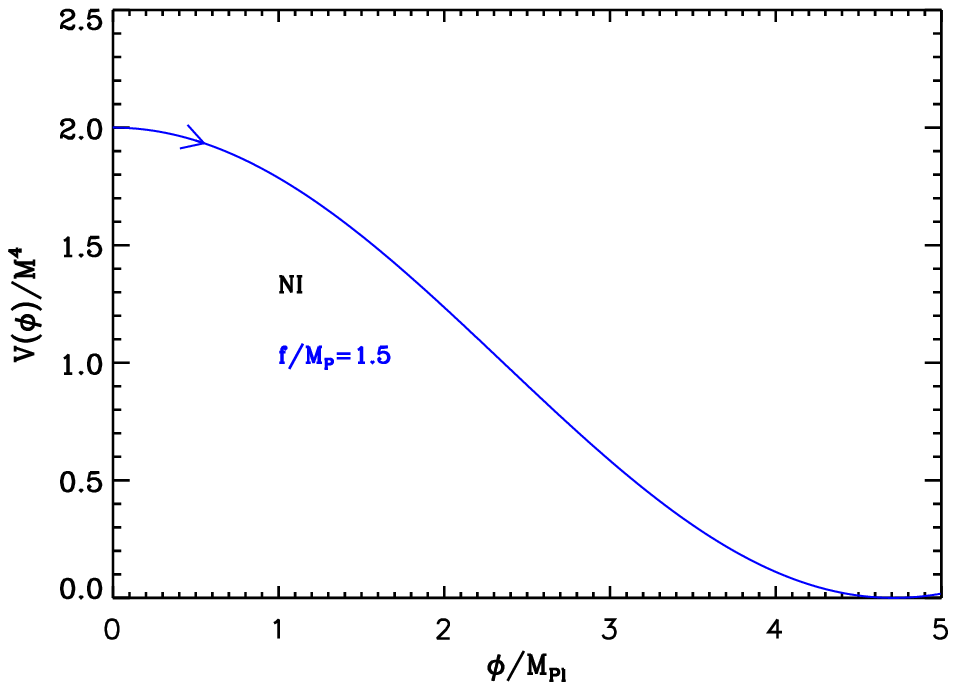}
\includegraphics[width=\wdblefig]{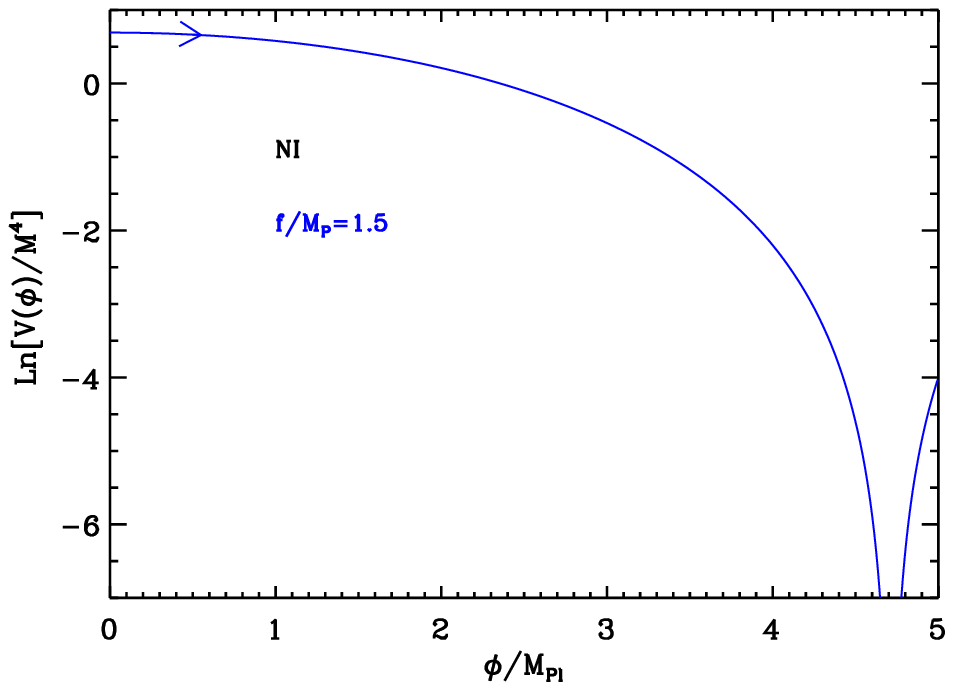}
\includegraphics[width=\wdblefig]{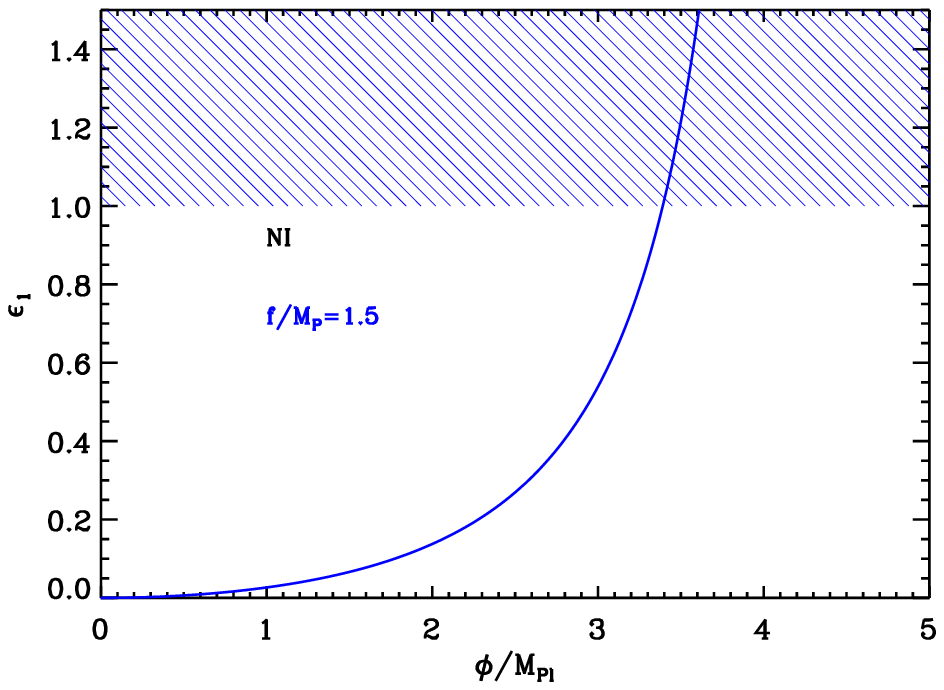}
\includegraphics[width=\wdblefig]{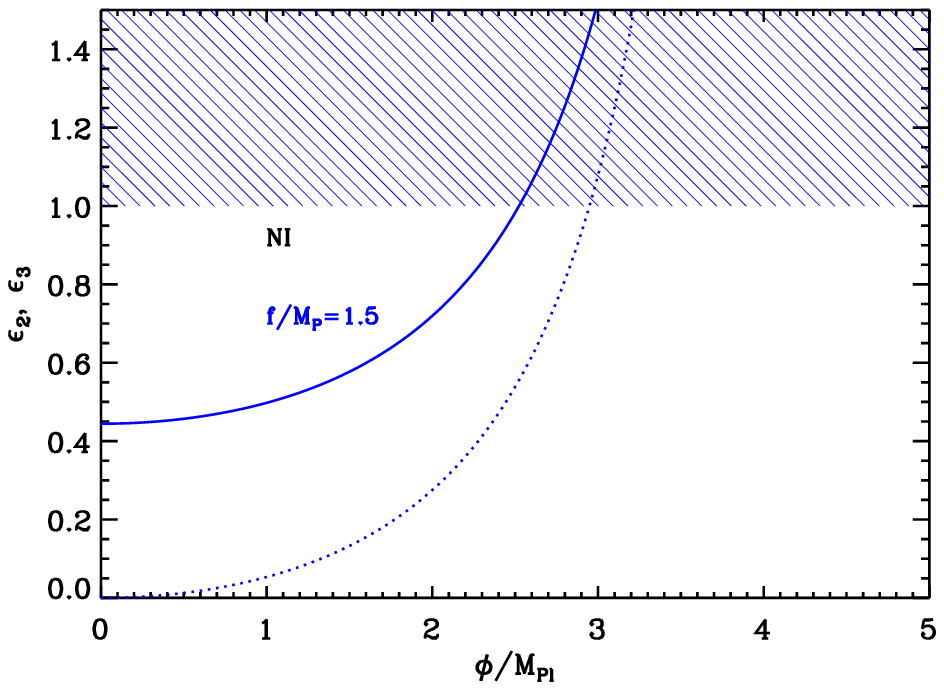}
\caption{Natural Inflation (NI).
  Top left panel: potential for $f/\Mp=1.5$. Top right panel:
  logarithm of the potential for the same value of $f$. Bottom left
  panel: slow-roll parameter $\epsilon _1$ for a potential with
  $f/\Mp=1.5$. The shaded area indicates the breakdown of the
  slow-roll inflation (strictly speaking when the acceleration
  stops). Bottom right panel: slow-roll parameters $\epsilon _2$
  (solid line) and $\epsilon _3$ (dotted line) for a potential with
  $f/\Mp=1.5$.}
\label{potNI}
\end{center}
\end{figure}

\subsubsection{Slow-Roll Analysis}
\label{subsubsec:srni}

Summarizing the above discussion, the model that we consider in this
section makes use of a potential that can be written as
\begin{equation}
\label{eq:ni:pot}
V(\phi)=M^4\left[1+\cos\left(\frac{\phi}{f}\right)\right].
\end{equation}
The scale $M$ is determined by the CMB normalization and the potential
depends on one parameter: the a priori unknown scale $f$. The
potential of \Eq{eq:ni:pot} is displayed with its logarithm in
\Fig{potNI}.  Since it is a periodic and even function of the field
\vev $\phi$, it is enough to study it in the range $\phi \in [0,\pi
f]$ where inflation proceeds from the left to the right. If one lets
$x\equiv\phi/f$, the slow-roll parameters can be expressed as
\begin{equation}
  \epsilon_1 = \frac{\Mp^2}{2f^2}
  \frac{\sin ^2 x }{\left(1+
      \cos x \right)^2}\, , \qquad
  \epsilon_2  = \frac{2\Mp ^2}{f^2}
  \frac{1}{1+
    \cos x }\,,\qquad
  \epsilon_3 = 2\epsilon_1\, .
\end{equation}
They are displayed in \Fig{potNI}, where one can see that they are all
increasing functions of the field \vev, which means that they all
increase during inflation. Inflation stops at the position $\xend$
given by $\epsilon_1=1$ (see below), and one can see that $\epsilon_2$
and $\epsilon_3$ are already greater than one at this point. This
means that the slow-roll approximation stops being valid slightly
before the end of inflation, and the few last \efolds may not be
properly described in this frame of approximations. Another remark to
be made is the fact that one generically has
\begin{equation}
\epsilon_2>\frac{\Mp^2}{f^2}\,.
\end{equation}
This means that in order for the slow-roll approximation to be valid,
one must require $f/\Mp\gg 1$ which is not necessarily problematic
from a high energy physics point of view (see the above discussion).

The end of inflation occurs when $\epsilon _1=1$, \ie at a position given by
\begin{equation}
  \xend = \arccos\left(\frac{1-2f^2/\Mp^2}
  {1+2f^2/\Mp^2}\right).
\label{eq:ni:xend}
\end{equation}
{}From this expression, one can calculate the value of the other slow
roll parameters at the end of inflation, namely
$\epstwoEnd=2+\Mp^2/f^2$ and $\epsthreeEnd=2\epstwoEnd$, which
confirms that the last few \efolds may not be described properly
in the slow-roll approximation.

Let us now calculate the slow-roll trajectory. It is given by
\begin{equation}
  \Nend -N =\frac{f^2}{\Mp^2}\ln \left(
    \frac{1- \cos \xend}
    {1- \cos x } \right),
\end{equation}
where $\Nend$ is the number of \efolds at the end of inflation, and
$N$ is the number of \efolds at some point when the scaled field \vev
is $x$.  This trajectory can be inverted and one obtains
\begin{equation}
x = \arccos \left\lbrace 1-\left(1-\cos \xend \right)
\exp\left[-\frac{\Mp ^2}{f^2}\left(\Nend-N\right)\right]\right\rbrace\, .
\end{equation}
Replacing $\xend$ by its value [see \Eq{eq:ni:xend}] gives
\begin{equation}
\label{eq:ni:traj}
x = \arccos \left\lbrace 1-\frac{4f^2}{\Mp^2+2f^2}
\exp\left[-\frac{\Mp ^2}{f^2}\left(\Nend-N\right)\right]\right\rbrace\, .
\end{equation}

Finally, the amplitude of the CMB anisotropies fixes the parameter $M$
to
\begin{equation}
\left(\frac{M}{\Mp}\right)^4=720 \pi^2\frac{\Qrms^2}{T^2}
\frac{\Mp^2}{f^2}\frac{\sin ^2 \xstar}
{\left(1+\cos \xstar \right)^3}\, .
\end{equation}
If $f/\Mp = \order{1}$, this expression simplifies to
\begin{equation}
\left(\frac{M}{\Mp}\right)^4\simeq 720 \pi^2\frac{\Qrms^2}{T^2}
\frac{\ee^{-2\Mp^ 2/f^ 2\Delta\Nstar}}{1+2f^2/\Mp^2}\, ,
\end{equation}
which gives rise to $M/\Mp\simeq 10^{-13}$. On the contrary, if
$f/\Mp\gg 1$ one has
\begin{equation}
\left(\frac{M}{\Mp}\right)^4\simeq 360 \pi^2\frac{\Qrms^2}{T^2}
\left(\frac{f}{\Mp}\right)^2\frac{1}{\Delta\Nstar^2}\, ,
\end{equation}
and the potential energy scale goes up. For instance, if $f/\Mp=10^2$
one has $M/\Mp\simeq 10^{-2}$.

The reheating consistent slow-roll predictions for the natural
inflation models are displayed in \Fig{fig:CMBNI}.  The reheating
equation of state parameter $\wrehbar$ has been taken to $0$ since the
potential is quadratic close to its minimum.  In the limit
$f/\Mp\rightarrow\infty$, the quadratic model predictions (LFI with
$p=2$, see \sectionc{sec:lfi}) seem to be recovered.  Indeed, from the
above formula, one can check that in this limit both $\xend$ and
$\xstar$ approach $\pi$ and the potential is, at leading order, a
parabola. More precisely, one can check from \Eq{eq:ni:traj} that in
the limit $f/\Mp\rightarrow\infty$, one has $\cos\xstar
\simeq-1+\left(1+2\Delta N_*\right)\Mp^2/f^2$, from which one deduces
that $\epsonestar\simeq 1/\left(1+2\Delta \Nstar\right)$ and
$\epstwostar \simeq 2/ \left( 1 + 2 \Delta \Nstar \right) \simeq 2
\epsonestar$. These relations are characteristic of the LFI quadratic
models, see \Eq{eq:lfi:epsstar}. However, one has
$\epsthreestar=2\epstwostar$ which differs from the LFI quadratic
relationship $\epsthreestar=\epstwostar$, and therefore quantities
sensitive to $\epsilon_3$, such as the running $\alphaS$, would break
the degeneracy between NI and the LFI quadratic model. As expected,
large values of $f/\Mp$ seem to be favored by the data (as well as
high reheating temperatures), and in practice, $f/\Mp<4$ appears to be
disfavored at the $2\sigma $ level by the \data data.

\subsection{Exponential SUSY Inflation (ESI)}
\label{sec:esi}

\subsubsection{Theoretical Justifications}
\label{subsubsec:theoryesi}

This model has been discussed in \Refc{Obukhov:1993fd} in the context
of spin-driven inflation and derived in \Refc{Stewart:1994ts} in the
context of supergravity and superstrings. The potential is given by
$V(\phi)\propto \left(1-\ee^{-q\phi/\Mp}\right)$. The same potential
also appears in \Refc{Dvali:1998pa} in the context of brane inflation,
in \Refc{Cicoli:2008gp} in the context of type IIB string
compactification as fibre inflation and more recently in
\Refc{Giudice:2010ka} as unitarized Higgs inflation models. This type
of models can be obtained under very general considerations. Suppose
that one has a supergravity model with a K\"ahler potential depending
on one field $\psi$ given by $K=-\beta/\kappa \ln\left(1-\alpha \kappa
  \psi\psi^{\dagger}\right)$, where $\alpha $ and $\beta $ are two
free parameters. This model leads to a scalar potential but for a
field which is not canonically normalized. The canonically normalized
field $\theta $ is given by
\begin{equation}
\kappa^{1/2}\theta \simeq \frac{1}{\sqrt{\alpha}}\left(1
-2\ee^{-\sqrt{2/\beta}\kappa^{1/2}\psi}\right),
\end{equation}
where we have assumed that inflation takes place at relatively large
$\psi$ \vev's. Then, suppose that the superpotential leads to a given
function $V=f(\theta)$. One can always expand $f$ such that
\begin{equation}
\label{eq:potesitheory}
V(\phi)\simeq V_0\left(1-\ee^{-\sqrt{2/\beta}\kappa^{1/2}\phi}\right)+\cdots,
\end{equation}
where $\kappa ^{1/2}\phi\equiv \kappa^{1/2}\theta+\sqrt{\beta/2}\ln
\left[2f_\theta/(\sqrt{\alpha}f)\right]$ and $V_0$ is just the
function $f$ evaluated at $1/\sqrt{\alpha}$. We see that one obtains
exactly the ESI potential with $q=\sqrt{2/\beta}$. Preferred choices
for $\beta $ are $\beta =1$ or $\beta =3$ leading to $q=\sqrt{2}$ or
$q=\sqrt{2/3}$. In absence of any more further guidance, it seems
reasonable to assume that $\beta $, and hence $q$, is just a number of
order one.

\begin{figure}
\begin{center}
\includegraphics[width=\wdblefig]{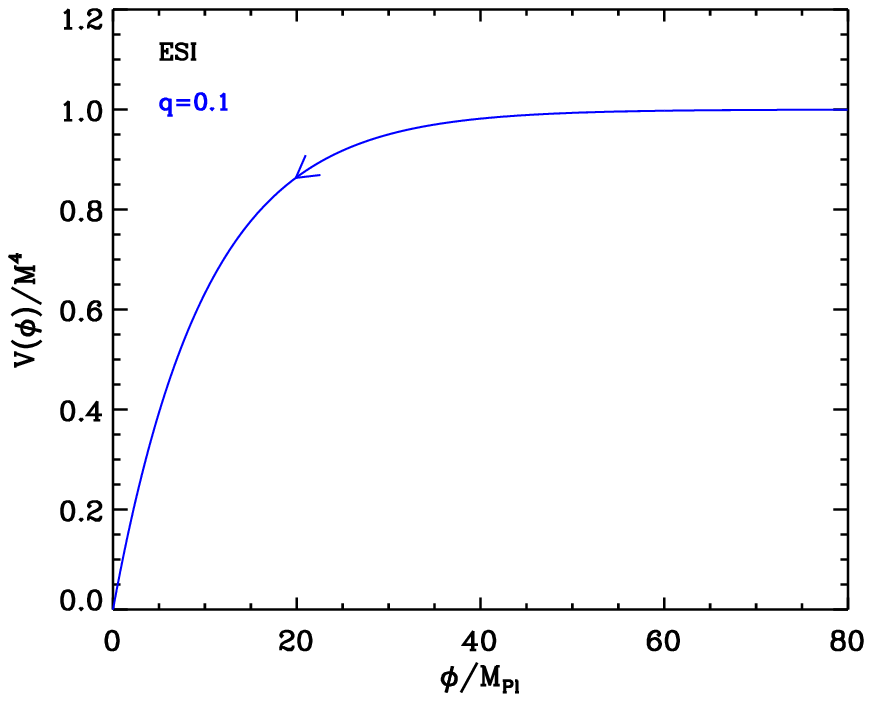}
\includegraphics[width=\wdblefig]{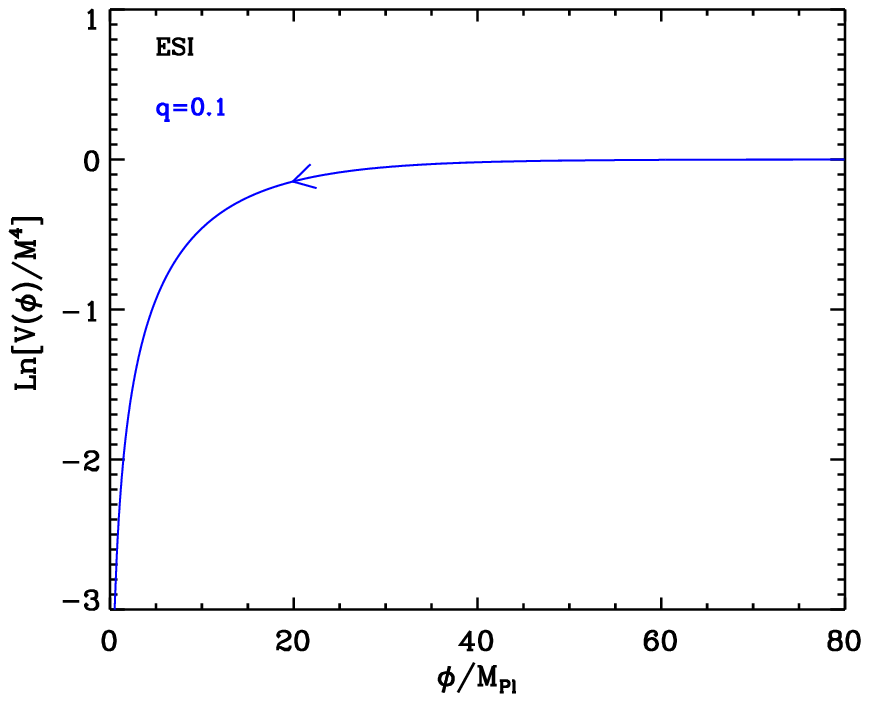}
\includegraphics[width=\wdblefig]{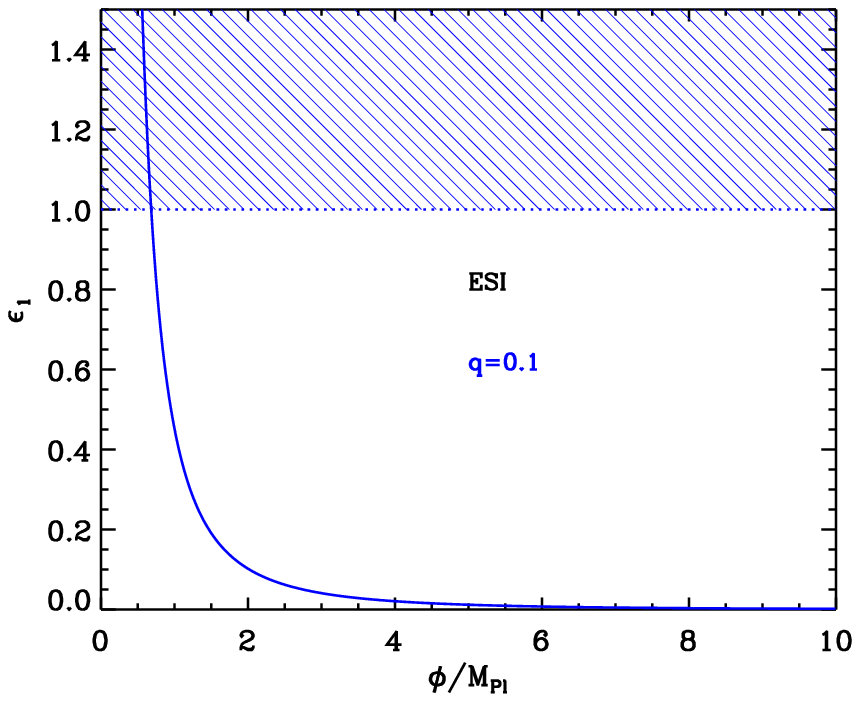}
\includegraphics[width=\wdblefig]{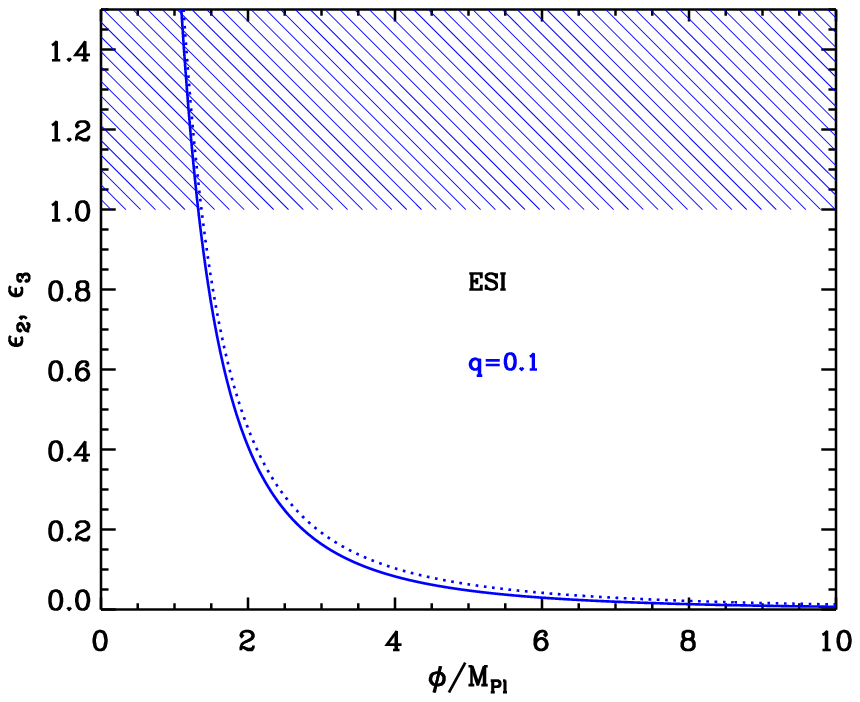}
\caption{Exponential SUSY Inflation (ESI) for $q=0.1$. Top
  panels: the potential and its logarithm. Bottom left panel:
  slow-roll parameter $\epsilon _1$. The shaded area indicates where
  acceleration stops. Bottom right panel: slow-roll parameters
  $\epsilon _2$ (solid line) and $\epsilon _3$ (dotted line). For
  those, the shaded region signals the breakdown of the slow-roll
  approximation but not necessarily the end of the accelerated
  expansion.}
\label{potlambertESI}
\end{center}
\end{figure}

\subsubsection{Slow-roll Analysis}
\label{subsubsec:sresi}

Based on the previous considerations, we now study the following
potential
\begin{equation}
\label{eq:potesi}
V(\phi)=M^4\left(1-\ee^{-q\phi/\Mp}\right),
\end{equation}
where $q$ is a positive dimensionless parameter and inflation proceeds
at decreasing field values in the region where $\phi/\Mp>0$. Defining
$x \equiv \phi/\Mp$, the Hubble flow functions in the slow-roll
approximation read
\begin{equation}
\epsilon _1 = \frac{q^2}{2}\frac{\ee^{-2q x}}
{\left(1-\ee^{-q x}\right)^2}\, ,
\qquad
\epsilon _2 = 2q^2\frac{\ee^{-q x}}
{\left(1-\ee^{-q x}\right)^2}\, ,
\qquad
\epsilon_3 = q^2\frac{\ee^{-q x}\left(1+\ee^{-q x}\right)}
{\left(1-\ee^{-q x}\right)^2}\,  .
\end{equation}
The potential and the Hubble flow functions with respect to the field
values are represented in \Fig{potlambertESI}.

\begin{figure}
\begin{center}
\includegraphics[width=\wsingfig]{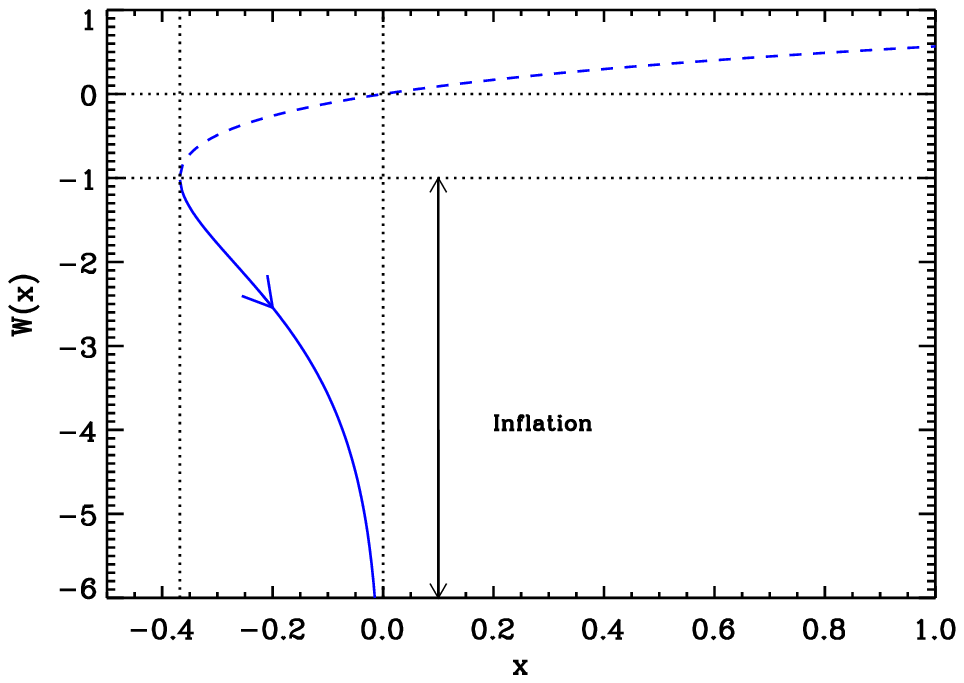}
\caption{Lambert functions $\Lambert{0}(x)$ (dashed line) and
  $\Lambert{-1}(x)$ (solid line). During Exponential SUSY inflation,
  inflation proceeds along the ``$-1$'' branch in the direction
  specified by the arrow on the figure.}
\label{fig:potlambertESI}
\end{center}
\end{figure}
The slow-roll trajectory can be integrated analytically from
\Eq{eq:srtrajectory} and one finds
\begin{equation}
  N - \Nend = -\dfrac{\ee^{q x} -q x}{q^2} + \dfrac{\ee^{q \xend} -q
    \xend}{q^2}\,.
\end{equation}
This equation can also be inverted in terms of the Lambert function to get
the field value in terms of the number of \efolds:
\begin{equation}
\label{eq:trajecexpsusy}
x = q (N-\Nend)  -\dfrac{\ee^{q\xend} -q \xend}{q} -
\frac{1}{q} \Lambert{-1}\left\{-\exp\left[ q^2 (N - \Nend) - 
(\ee^{q\xend} -q \xend)\right]\right\} .
\end{equation}
The fact that one should choose the branch $\Lambert{-1}$ is justified
below. The argument of the Lambert function is always negative as the
exponential is always positive. Moreover, since $\xend>0$ and $N<
\Nend$, the maximal value of exponential argument is saturated for
$\xend \rightarrow 0$, i.e. for a Lambert function argument equals
to $-1/e$. As the result the Lambert function argument varies, at
most, in $[-1/e,0]$. Finally, since $x>0$, we see directly from
\Eq{eq:trajecexpsusy} that the Lambert function values have to be
negative thereby ensuring that inflation proceeds only along the
``$-1$''-branch (see \Fig{fig:potlambertESI}).

With such a potential, inflation ends naturally at $\epsilon_1=1$,
i.e. at the field value
\begin{equation}
\xend = \frac{1}{q}\ln \left(1
+\frac{q}{\sqrt{2}}\right).
\end{equation}
{}From this equation and the trajectory, we have an explicit relation
between the field value $\phistar$ at which the pivot mode crossed the
Hubble radius during inflation and the corresponding \efold number
$\Delta \Nstar$.

Finally, the parameter $M$ can be determined from the amplitude of the 
CMB anisotropies, and one gets
\begin{equation}
\left(\frac{M}{\Mp}\right)^4=720 q^2 \pi^2
\dfrac{\ee^{-2q \xstar}}{\left(1-\ee^{-q \xstar}\right)^{3}} \frac{\Qrms^2}{T^2}\, ,
\end{equation}
where the value of $\phistar$ (or $\Delta \Nstar$) is obtained from
\Eq{eq:phistarlnrrad}. The reheating consistent slow-roll prediction
for the exponential Susy models are represented in \Figs{fig:CMBESI}
and \ref{fig:CMBESIb}. In the limit $q\rightarrow 0$, we
recover the same prediction as a linear large field model. From
\Fig{fig:CMBESI}, we see that all the models remains compatible with
the current data. These figures correspond to $\wrehbar=0$, but one could
argue that $\wrehbar \gtrsim -1/3$ make more sense if a parametric
reheating would feel the linear shape of the potential. This quite
extreme situation is represented in \Fig{fig:CMBESIb}. In that case,
the low reheating temperatures are clearly disfavored.

\subsection{Power Law Inflation (PLI)}
\label{sec:pli}

\begin{figure}
\begin{center}
\includegraphics[width=\wdblefig]{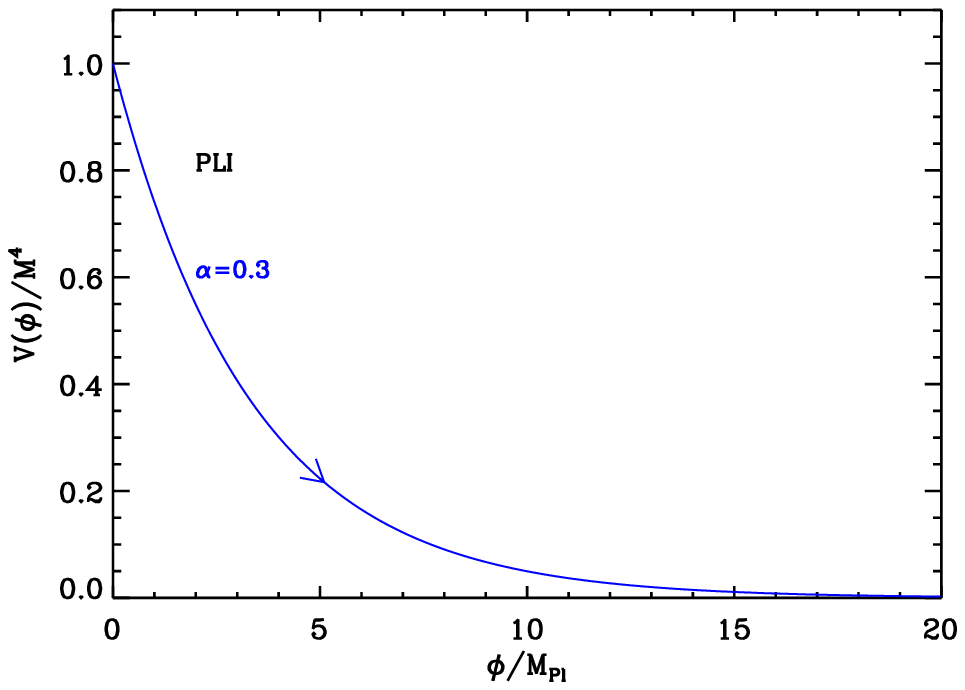}
\includegraphics[width=\wdblefig]{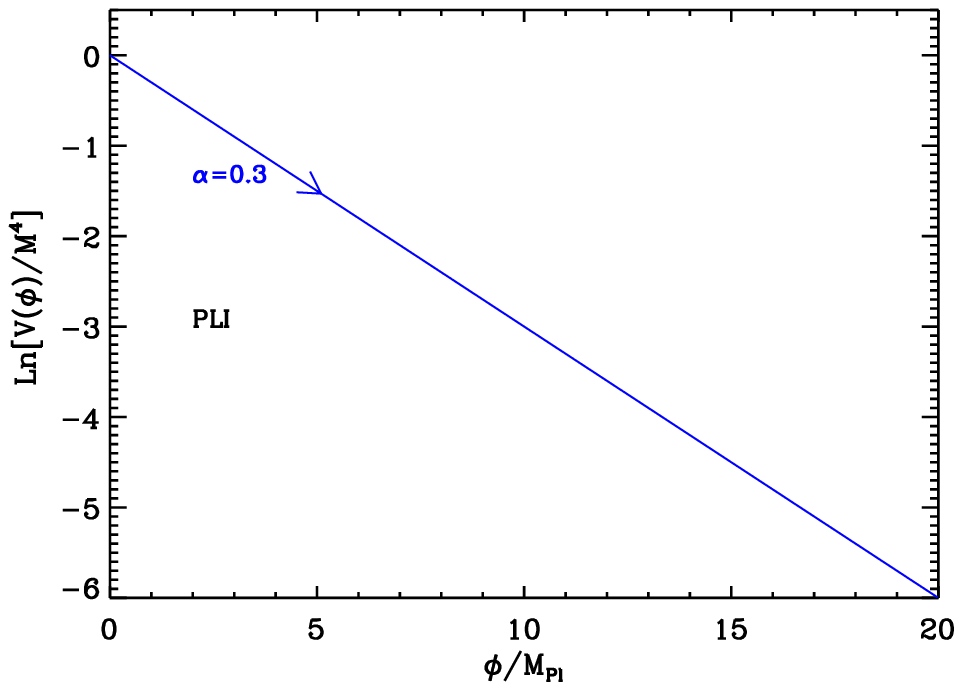}
\includegraphics[width=\wdblefig]{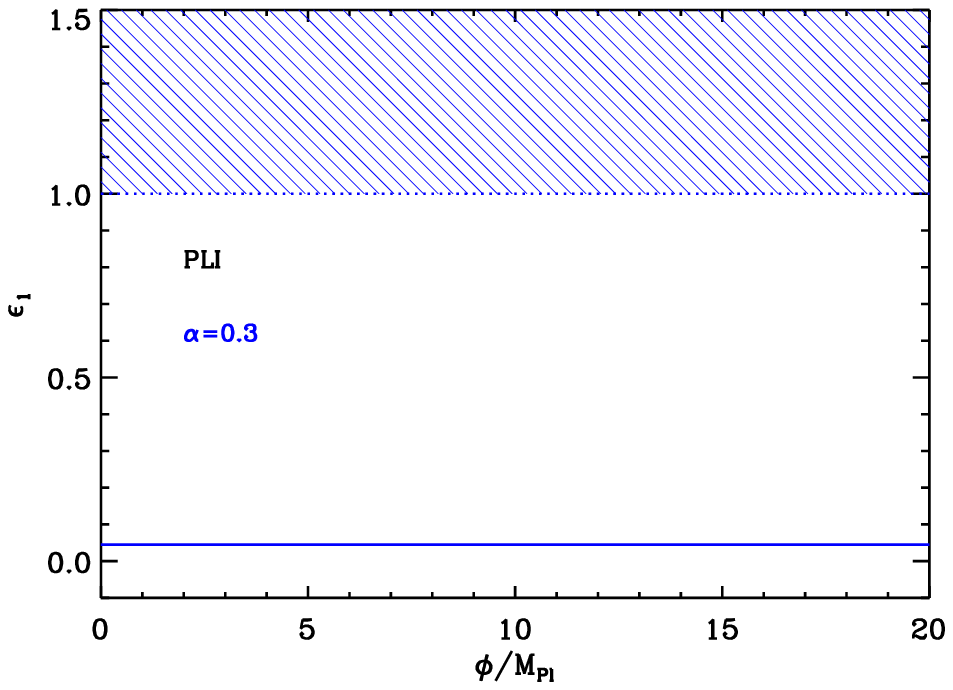}
\includegraphics[width=\wdblefig]{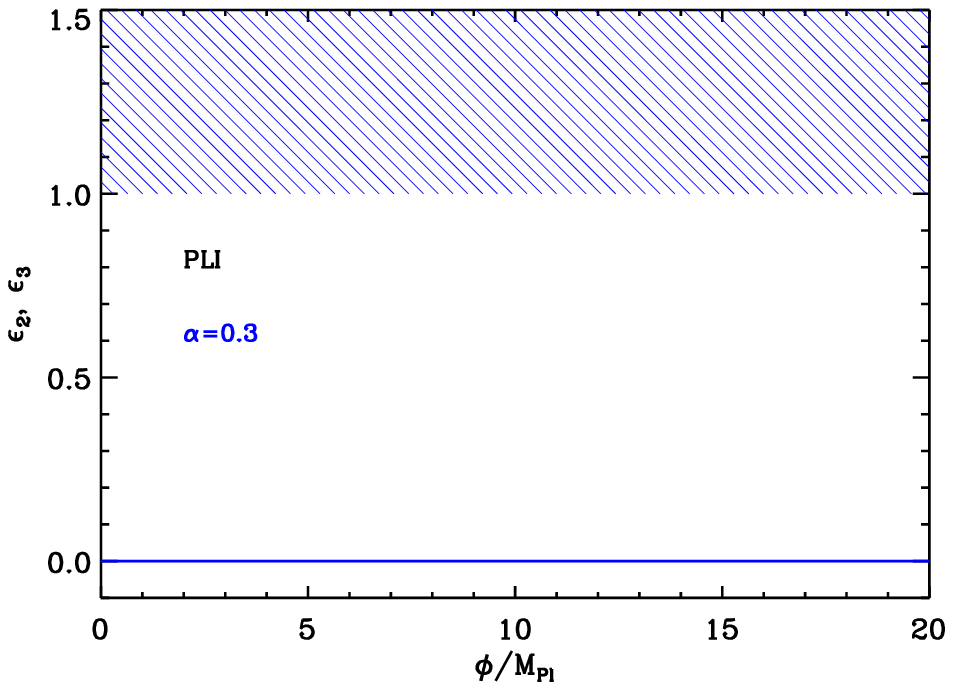}
\caption{Power Law Inflation (PLI) for $\alpha=0.3$.  Top panels:
  power law potential (left) and its logarithm (right). Bottom left
  panel: slow-roll parameter $\epsilon_1$.  Bottom right panel:
  slow-roll parameters $\epsilon_2=\epsilon_3=0$. On these plots, the
  shaded area indicates the region where slow-roll is violated.}
\label{potPLI}
\end{center}
\end{figure}

These models refer to inflationary potentials of the form
\begin{equation}
\label{eq:pli:pot}
V(\phi)=M^4\ee^{-\alpha \phi/\Mp}\, ,
\end{equation}
where $\alpha$ is a dimensionless parameter. They have been
intensively studied since they lead to an exact inflationary dynamics,
of the power law form, hence their name. Moreover, the power spectrum
can also be determined exactly in this case. The background solution
reads $a\propto (t/t_0)^{2/\alpha^2}$ and $\phi=\phizero +2\Mp/\alpha
\ln\left(t/t_0\right)$ with
$t_0^2=2\Mp^2/(\alpha^2M^4)(6/\alpha^2-1)\ee^{\alpha\phizero/\Mp}$. We
see that we have inflation provided $\alpha\in
\left[0,\sqrt{2}\right]$.

This scenario was introduced in \Refc{Lucchin:1984yf, Abbott:1984fp}
where the two point correlation function of the cosmological
fluctuations was calculated for the first time (see also
\Refcs{Sahni:1988zb,Sahni:1990tx}). The predictions of this model were
recently compared to the Planck data in
\Refc{Unnikrishnan:2013vga}. Soon after \Refc{Abbott:1984fp}, it was
also considered in \Refcs{Ratra:1987rm, Ferreira:1997hj} but in the
context of quintessence, \ie for models of dark energy in which the
energy density of the scalar field redshifts as a power law of the
scale factor $\rho \propto a^{-q}$. In that case, one has $\alpha
=\sqrt{q/2}$. The same potential also arises in the case where large
field inflation is considered (LFI, see \sectionc{sec:lfi}) but with a
non-minimal coupling of the inflaton to the gravity sector, see
\Refcs{La:1989za, Kolb:1990ey} (the exponential potential appears
after the transformation to the Einstein frame). In
\Refc{Kitada:1991ih}, a cosmic no-hair theorem for Bianchi models was
proven assuming that the potential of the inflaton is of the same type
as in \Eq{eq:pli:pot}.  It was shown that one must have
$0<\alpha<\sqrt{2/3}$ so that the isotropic power law solution is the
unique attractor for any initially expanding Bianchi type model
(except type IX). In \Refc{Mendes:1990eb}, the potential
\eqref{eq:pli:pot} has been studied in the Kantowski-Sachs metric, and
it was found that the production of particles by the scalar field acts
as viscous forces which enlarges the range of initial conditions
leading to successful inflation. In \Refc{Banerjee:1998qn}, the nature
of the potential $V(\phi)$ relevant to having inflation in presence of
a minimally coupled scalar field together with a causal viscous fluid
was investigated. It was shown that this leads to an exponential
potential. In \Refcs{Fairbairn:2002yp, Sami:2002fs, Cardenas:2006py},
the exponential potential was used to describe the dynamics of a
tachyonic matter field (\ie with a non-minimal kinetic term). In
\Refc{Aguirregabiria:2003uh}, the general transformations that leave
unchanged the form of the field equations for Bianchi V cosmologies
were investigated, and it was found that they admit asymptotic stable
points that lead to power law solutions of the type
\eqref{eq:pli:pot}. In \Refc{Becker:2005sg}, inflation was studied in
the context of M-theory on $S^1/\mathbb{Z}_2$ via the non-perturbative
dynamics of M5-branes. The open membrane instanton interactions
between the branes give rise to potentials of the type
\eqref{eq:pli:pot}. Within the same framework,
\Refc{Ashoorioon:2006wc} has discussed a realization of cascade
inflation as assisted inflation built upon a succession of power law
inflationary eras. \Refc{Bennai:2006th} has used the exponential
potential \eqref{eq:pli:pot} in the context of Randall-Sandrum type II
Braneworld model. Finally, the general dynamics of power law inflation
was studied in detail in \Refcs{Lucchin:1984yf, Yokoyama:1987an,
  Liddle:1988tb,Ratra:1989uz, Ratra:1989uv, Schmidt:1990gb,
  Maartens:1995uz, Copeland:1997et, Hirai:2005tn, Heinzle:2005fz},
where various aspects of its phenomenology were highlighted.

The potential and its logarithm are displayed in \Fig{potPLI}. They
are decreasing functions of the field, hence inflation proceeds from
the left to the right. The slow-roll parameters take a simple form
given by
\begin{equation}
\epsilon_1=\frac{\alpha ^2}{2}, \quad 
\epsilon _{i> 1}=0.
\end{equation} 
Since the first slow-roll parameter is constant, inflation cannot stop
by slow-roll violation and one has to assume that, at some \vev
$\phiend$, a tachyonic instability is triggered. A priori, this means
that the model has in fact an additional new free parameter. However,
because the slow-roll parameters do not depend on $\phi$, as well as
all the other properties of the inflationary dynamics (even when the
slow-roll approximation is not satisfied, see below), the
observational predictions of the model cannot depend on $\phiend$ and
this parameter turns out to be irrelevant.

The Hubble flow hierarchy being almost trivial, the exact dynamics of
the model can be worked out even if the slow-roll approximation is
violated. Indeed, let us first notice that the slow-roll trajectory
can be explicitly integrated, and gives
\begin{equation}
\label{eq:pli:traj}
\frac{\phi}{\Mp}=\frac{\phiend}{\Mp}+\alpha \left(N - \Nend \right).
\end{equation}
Then, one can remark that this trajectory is also a solution of the
exact Klein-Gordon equation of motion, which reads in terms of the
number of \efolds $N$,
\begin{equation}
\label{eq:pli:kg}
H^2\frac{\partial^2\phi}{\partial N^2}
+\left(3H^2+H\partial\dfrac{\partial H}{\partial N}\right)
\dfrac{\partial \phi}{\partial N}+\dfrac{\dd V}{\dd \phi}=0.
\end{equation}
Indeed, the first term vanishes, and the second term requires
\begin{equation}
H^2=\frac{V+\dot{\phi}^2/2}{3\Mp^2}=
\dfrac{V+\dfrac{H^2}{2}\left(\dfrac{\partial\phi}{\partial N}\right)^2}{3\Mp^2}=
\dfrac{V+\dfrac{H^2}{2}\alpha^2\Mp^2}{3\Mp^2}\, ,
\end{equation}
from which one gets
\begin{equation}
\label{eq:pli:H(phi)}
H^2=\dfrac{V}{3\Mp^2}\dfrac{1}{1-\alpha^2/6}\, .
\end{equation}
{}From there, one can evaluate all terms in the Klein-Gordon equation,
and verify that \Eq{eq:pli:traj} is indeed a solution of
\Eq{eq:pli:kg}. Since it is a second order differential equation,
other solutions exist, but it can be shown~\cite{Ratra:1987rm,
  Ferreira:1997hj} that the exact solution is an attractor.  Let us 
also notice that combining \Eq{eq:pli:H(phi)} with \Eq{eq:pli:traj} gives rise to 
\begin{equation}
H = H_\uend \left(\dfrac{\aend}{a} \right)^{\alpha^2/2},
\end{equation}
which can be integrated and gives
\begin{equation}
a(t)= \aend \left(\frac{t}{t_\uend}\right)^{2/\alpha ^2}.
\end{equation}
One recovers the solution mentioned at the beginning of this
section. Finally, the equation of state $w=P/\rho$ can also be worked
out exactly and one gets
\begin{equation}
w=-1+\frac{\alpha ^2}{3}\,.
\end{equation}
Again, all the previous expressions are valid even if the slow-roll
approximation is not satisfied. One can see that pure de Sitter
corresponds to $\alpha =0$. In this case the potential is constant,
the equation of state is $-1$ and the scale factor expands
exponentially.

Another nice feature of power-law inflation is that the spectrum of
the perturbations can be computed exactly without relying on any
approximation. Defining the parameter $\beta\leq -2$ from
$\alpha^2/2=(\beta+2)/(\beta+1)$, the primordial scalar power spectrum
is given by
\begin{equation}
{\calP}_{\zeta}=\frac{\Hstar^2}{\pi \epsilon_1(8\pi \Mp^2)}f(\beta)
\left(\frac{k}{\kstar}\right)^{2\beta +4},
\end{equation}
where 
\begin{equation}
f(\beta)\equiv \frac{1}{\pi}\left[\frac{(1+\beta)^{1+\beta}}{2^{1+\beta }}
\Gamma\left(\frac12+\beta\right)\right]^2.
\end{equation}
In particular, $f(\beta =-2)=1$.  The power spectrum of gravitational
waves can also be obtained remarking that we have $\muS=\muT$ for
power law inflation. From
\begin{equation}
{\calP}_\zeta=\frac{k^3}{8\pi^2}\left \vert \frac{\muS}{a\sqrt{\epsilon_1}}
\right \vert ^2, \qquad {\calP}_h=\frac{2k^3}{\pi^2}
\left \vert \frac{\muT}{a}
\right \vert ^2,
\end{equation}
one gets
\begin{equation}
r\equiv \frac{{\calP}_h}{{\calP}_\zeta}=16\epsilon_1=\frac{16\nT}{\nT-2}\,,
\end{equation}
since $\nT=\nS-1=2\beta +4$.

Finally, the overall amplitude of the CMB anisotropies leads to a
determination of the scale $M$, namely
\begin{equation}
\left(\frac{M}{\Mp}\right)^4=720\pi^2\alpha ^2
\ee^{\alpha\phistar/\Mp}\frac{\Qrms^2}{T^2}\, .
\end{equation}
Obviously, this normalization depends on the value of $\phiend$, and it 
is more relevant to express it in terms of the potential energy, say, 
at the end of inflation:
\begin{equation}
\frac{V_\uend}{\Mp^4}=720\pi^2\alpha^2
\ee^{-\alpha^2\Delta\Nstar}\frac{\Qrms^2}{T^2}\, ,
\end{equation}
from which one typically gets $V_\uend^{1/4}/\Mp\simeq 10^{-4}$.

The reheating consistent slow-roll predictions for the 
power law inflation models are displayed in \Fig{fig:CMBPLI}.
Because the slow-roll parameters are constant during inflation,
one can check that the predictions of the models do not depend on the
energy scale at which the power law reheating ends.
One has $\nS=1-\alpha^2$ and $r=8\alpha^2$, and from the \data
constraints, all the models are disfavored at more than
two-sigma confidence level.

\subsection{K\"ahler Moduli Inflation I (KMII)}
\label{sec:kmii}

\begin{figure}
\begin{center}
\includegraphics[width=\wdblefig]{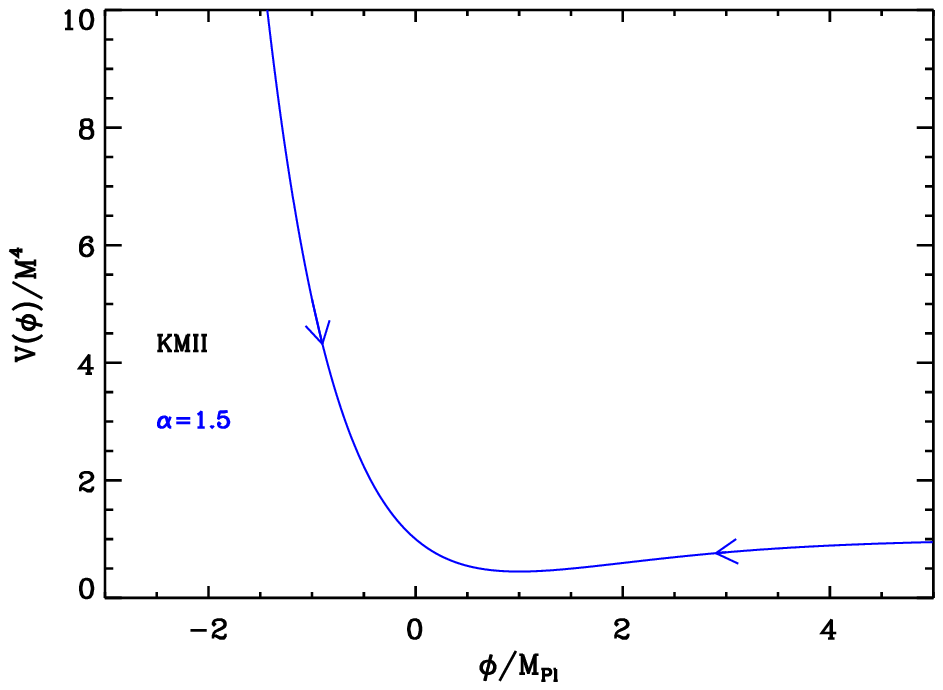}
\includegraphics[width=\wdblefig]{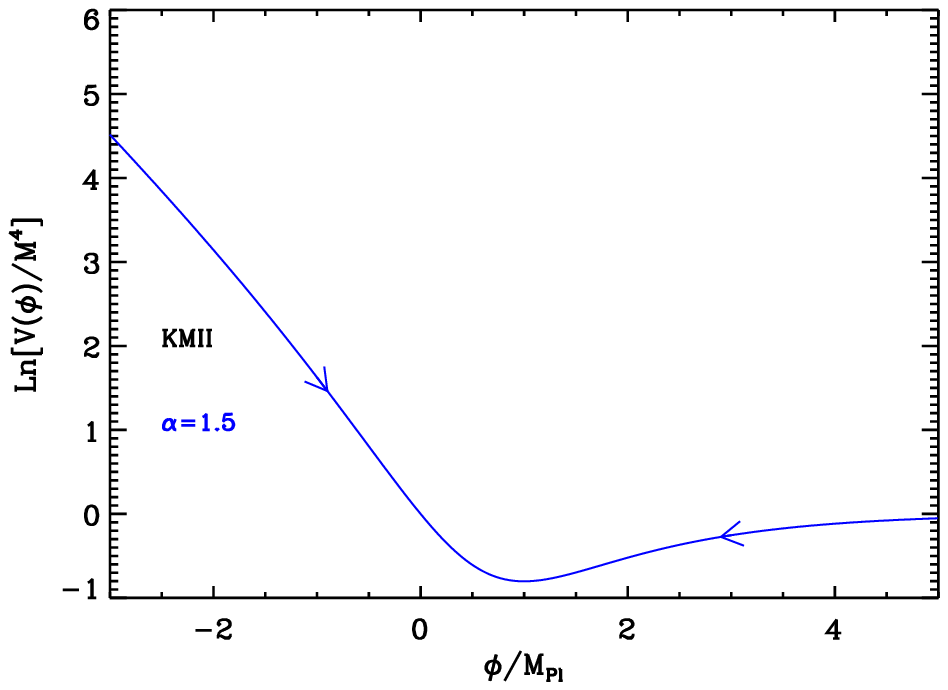}
\includegraphics[width=\wdblefig]{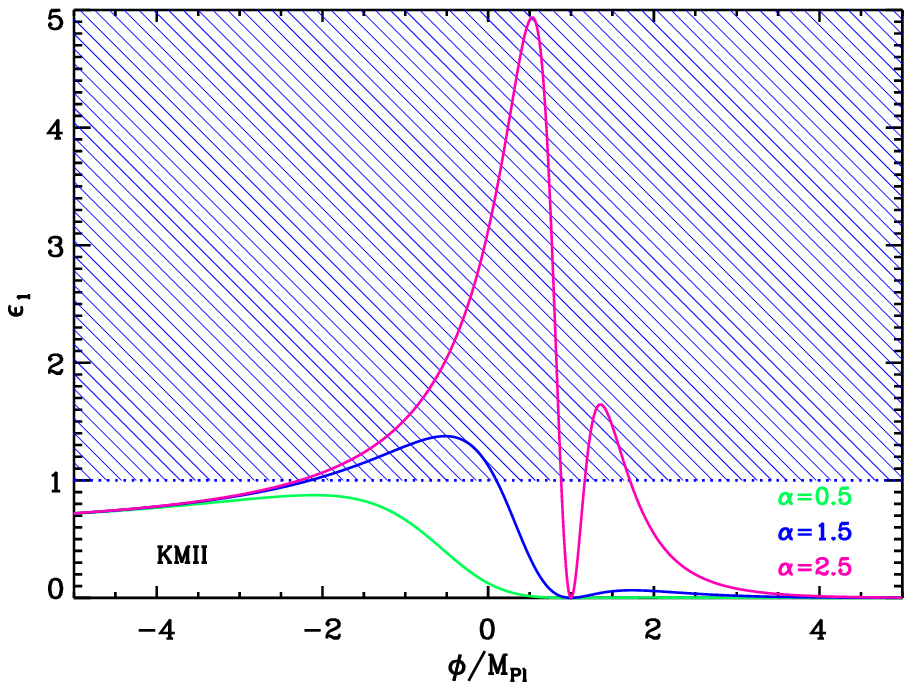}
\includegraphics[width=\wdblefig]{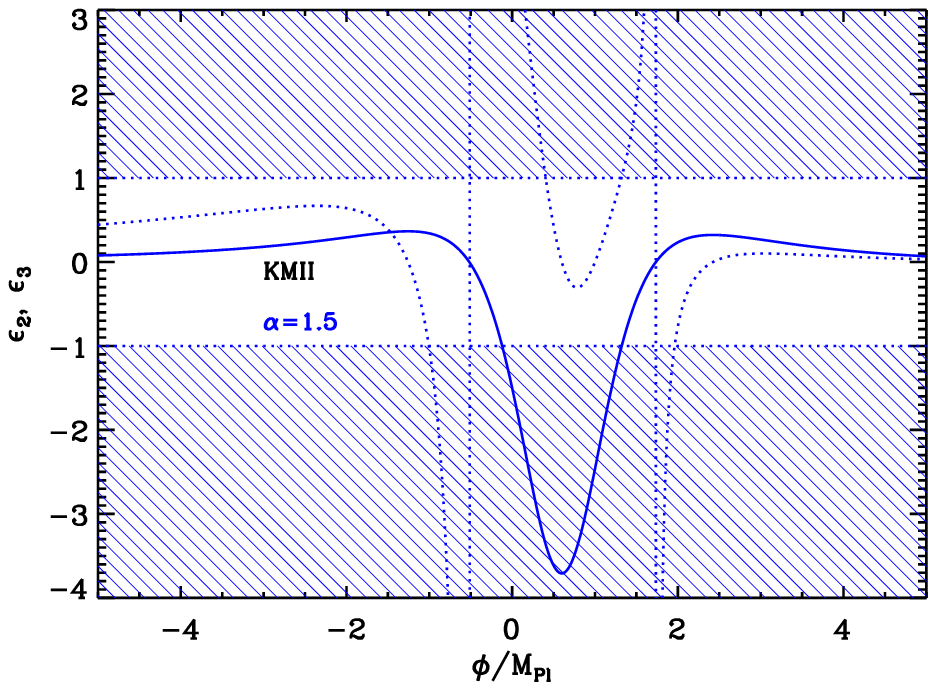}
\caption{Top left panel: K\"ahler moduli inflation (KMII) potential
  for $\alpha=1.5$. The two arrows indicate the two regions of the
  potential where inflation can take place. Top right panel: logarithm
  of the potential for the same value of $\alpha$. Bottom left panel:
  slow-roll parameter $\epsilon _1$ for $\alpha=0.5$ (solid green
  line), $\alpha=1.5$ (solid blue line) and $\alpha=2.5$ (solid pink
  line). Obviously, the number of solutions of the equation
  $\epsilon_1=1$ depends on the value of $\alpha$. Bottom right panel:
  slow-roll parameters $\epsilon _2$ (solid line) and $\epsilon _3$
  (dotted line) for $\alpha=1.5$.}
\label{potkmii}
\end{center}
\end{figure}
 
These models are stringy models and arise when type IIB string
theories via Calabi-Yau flux compactification are used. KMII scenarios
have been derived and studied in \Refcs{Conlon:2005jm,Bond:2006nc,
  Yang:2008ns,Krippendorf:2009zza,BlancoPillado:2009nw,
  Kawasaki:2010ux,Lee:2010tk}. More specifically, when internal spaces
are weighted projective spaces, one of the K\"ahler moduli can play
the role of an inflaton field and its potential, in the large field
limit, reads
\begin{equation}
\label{eq:kmii:pot}
V(\phi)=M^4\left(1-\alpha\frac{\phi}{\Mp}\ee^{-\phi/\Mp}\right) ,
\end{equation}
$\alpha$ being a positive dimensionless parameter. Actually, since we
deal with a modulus, $\phi$ usually possesses a non-minimal kinetic
term. Then, once the inflaton field has been canonically normalized,
$\phi$ has to be replaced with $\propto\phi^{4/3}$. The corresponding
corrected potential is studied as ``K\"ahler Moduli Inflation II"
(KMIII) in \sectionc{sec:kmiii}. However, sometimes, the
potential~(\ref{eq:kmii:pot}) (with $\phi$ already canonically
normalized) is also studied as a toy model (notably in
\Refc{Lee:2010tk}), the hope being that it can give a simpler
description of the physics that naturally appears in the context of
moduli inflation. Therefore, in this section, we also consider this
scenario.

The potential in \Eq{eq:kmii:pot} depends on one free parameter,
$\alpha$. A priori, there does not exist any bound on its
value. However, as explained below, in order for slow-roll inflation
to occur, one must restrict the range of possible values for
$\alpha$. Within this range, we will show that the predictions of the
model turn out to be almost independent of $\alpha$ (in fact, they
logarithmically depend on $\alpha$). The potential~(\ref{eq:kmii:pot})
and its logarithm are displayed in \Fig{potkmii}. It decreases from
$\phi=0$ (where it blows up), reaches a minimum at $\phi=\Mp$, and
then increases to the asymptotic value $V=M^4$ when $\phi\rightarrow
+\infty$.  Therefore, two regimes of inflation may a priori exist:
either inflation proceeds from the left to the right in the decreasing
$\phi<\Mp$ branch of the potential (in this branch the \vev $\phi $
increases during inflation) or it proceeds from the right to the left
in the increasing $\phi>\Mp$ branch of the potential (and the \vev
decreases during inflation). However, one should keep in mind that the
potential is derived under the large field assumption and,
consequently, only the second regime is in fact meaningful. As a toy
model, one might nevertheless want to study both regimes but it turns
out that, in the first one, inflation could not stop by violation of
the slow-roll conditions. This is why we will mainly focus on the
second regime in the rest of this section. Let us also notice that the
minimum value of the potential is located at $\phi=\Mp$ and is
$V_{\mathrm{min}}=M^4\left(1-\alpha/e\right)$. Therefore, if one
requires the potential to be positive definite everywhere, then one
must have $0<\alpha<e\simeq 2.72$. However, this condition may also be
ignored if one considers that the potential~(\ref{eq:kmii:pot}) is in
any case not valid at $\phi/\Mp\lesssim 1$.

Defining $x\equiv\phi/\Mp$, the three first slow-roll parameters can
be expressed as
\begin{equation}
\begin{aligned}
\epsilon_1 = \frac{\alpha ^2}{2}\ee^{-2 x}
\frac{\left(1-x\right)^2}{\left(1-\alpha \ee^{-x}x
\right)^2}\, , \quad\quad
\epsilon_2  =  \frac{2\alpha \ee^{-x}}
{\left(1-\alpha \ee^{-x}x
\right)^2}\left(\alpha \ee^{-x}+
x-2\right),
\end{aligned}
\label{srsf}
\end{equation}
and
\begin{equation}
\begin{aligned}
  \epsilon_3 = \frac{\alpha \ee^{-x} \left(x-1\right)}
  {\left(1-\alpha \ee^{-x}x \right)^2\left(\alpha
      \ee^{-x}+x-2\right)} 
  \Biggl[x-3+\alpha\ee^{-x}
  \left(x^2-3x+6\right)
  -2\alpha^2\ee^{-2x}\Biggr]\, .
\end{aligned}
\label{srsf3}
\end{equation}

Let us now study in more detail how inflation stops in this model. As
can be seen in \Fig{potkmii}, the number of solutions of
$\epsilon_1=1$ depends on the value of $\alpha$. We now define the
numbers $\alphaMinus$ and $\alphaPlus$ by
\begin{equation}
  \alphaMinus \equiv
  \dfrac{\sqrt{2}}{\sqrt{2}-1}\ee^{\frac{2-\sqrt{2}}{1-\sqrt{2}}} \simeq
  0.83, \qquad \alphaPlus \equiv \dfrac{\sqrt{2}}{\sqrt{2}+1}
\ee^{\frac{2+\sqrt{2}}{1+\sqrt{2}}} \simeq 2.41.
\end{equation}
If $0 < \alpha < \alphaMinus$, then there is no solution (this
corresponds to the green line in the bottom left panel in
\Fig{potkmii}). The inflaton field eventually oscillates around the
minimum of its potential but remains in a region where inflation
continues forever. In this case, in order to stop inflation, one must
add an auxiliary field to the model such that a tachyonic instability
is triggered at some value $\xend$. This of course increases the
number of parameters of this model. If $\alphaMinus <\alpha <
\alphaPlus $ (which corresponds to the blue line in \Fig{potkmii}),
then two solutions appear:
\begin{align}
\label{eq:sol2kmii}
\xepsoneOneMinus\vert_{x<1}&=\xend\vert_{x<1}=\frac{1}{1-\sqrt{2}}-\Lambert{0}
\left(\frac{\sqrt{2}}{1-\sqrt{2}}\frac{\ee^{\frac{1}{1-\sqrt{2}}}}{\alpha}
\right)\simeq-2.4-\Lambert{0}\left(-\frac{0.3}{\alpha}\right),\\
\xepsoneOnePlus\vert_{x<1}&=\frac{1}{1-\sqrt{2}}-\Lambert{-1}
\left(\frac{\sqrt{2}}{1-\sqrt{2}}\frac{\ee^{\frac{1}{1-\sqrt{2}}}}{\alpha}
\right)\simeq-2.4-\Lambert{-1}\left(-\frac{0.3}{\alpha}\right),
\label{eq:sol2kmii2}
\end{align}
where $\Lambert{0}$ and $\Lambert{-1}$ denotes the ``$0$-branch'' and
the ``$-1$-branch'' of the Lambert function respectively. These two
solutions are both smaller than one so that they both lie in the
decreasing branch of the potential. Correspondingly, two regimes of
inflation exist. The first one proceeds from the left to the right and
stops at $\xend\vert_{x<1}$. However, using the expression for the
slow-roll parameters~(\ref{srsf}), it is easy to see that $\epsilon_1$
is always larger than $1/2$ in this domain. Therefore, the slow-roll
approximation breaks down in this case. The second regime takes place
in the $\phi/\Mp>1$ branch of the potential but inflation cannot stop
by slow-roll violation. Finally, if $\alphaPlus < \alpha $ (this
situation corresponds to the pink line in the bottom left panel in
\Fig{potkmii}), then four solutions exist: two were already given in
\Eqs{eq:sol2kmii}, (\ref{eq:sol2kmii2}) and the two new ones read
\begin{align}
\xepsoneOneMinus\vert_{x>1}&=\frac{1}{1+\sqrt{2}}-\Lambert{0}
\left(-\frac{\sqrt{2}}{1+\sqrt{2}}\frac{\ee^{\frac{1}{1+\sqrt{2}}}}{\alpha}
\right)\simeq 0.4-\Lambert{0}\left(\frac{-0.9}{\alpha}\right),\\
\xepsoneOnePlus\vert_{x>1}&=\xend\vert_{x>1}=\frac{1}{1+\sqrt{2}}-\Lambert{-1}
\left(-\frac{\sqrt{2}}{1+\sqrt{2}}\frac{\ee^{\frac{1}{1+\sqrt{2}}}}{\alpha}
\right)\simeq 0.4-\Lambert{-1}\left(\frac{-0.9}{\alpha}\right).
\end{align}
The two new solutions are greater than one and therefore lie in the
increasing branch of the potential. Thus two regimes exist in this
situation. The first one is the same as before, proceeds again from
the left to right, stops at $\xend\vert_{x<1}$ and suffers from the
fact that $\epsilon_1$ is always larger than $1/2$. The second one
proceeds from the right to the left and ends at $\xend\vert_{x>1}$. We
conclude that this regime is the regime of interest for the KMII model
and that we must therefore require $\alpha > \alphaPlus$.

\begin{figure}
\begin{center}
\includegraphics[width=\wsingfig]{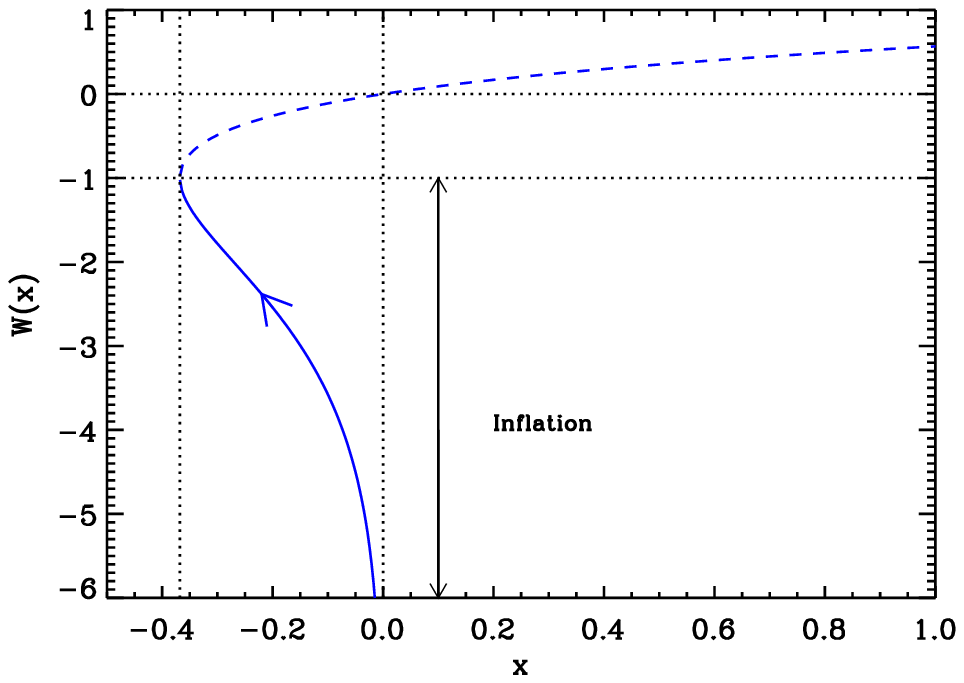}
\caption{Lambert functions $\Lambert{0}(x)$ (dashed line) and
  $\Lambert{-1}(x)$ (solid line). During K\"ahler moduli inflation,
  inflation proceeds along the ``$-1$'' branch in the direction
  specified by the arrow.}
\label{potlambertKMII}
\end{center}
\end{figure}

Let us now study the slow-roll trajectory. It can be integrated
exactly and its expression can be written as
\begin{equation}
\begin{aligned}
\Nend -N &= \xend-\frac{e}{\alpha}\Ei\left(\xend -1\right)
+\ln \left(\xend-1\right)\\&
-x+\frac{e}{\alpha}\Ei\left(x
-1\right) -\ln \left(x-1\right),
\end{aligned}
\end{equation}
where $\Ei$ is the exponential integral
function~\cite{Abramovitz:1970aa,Gradshteyn:1965aa}.  At this point, a
few remarks are in order. Firstly, let us notice that $N$ goes to
$\infty$ when $x$ tends to $1$. This means that, in the slow-roll
approximation, the field can never cross the minimum of its
potential. In particular, if $\alpha < \alphaPlus$, that is to say if
one starts from the $\phi/\Mp<1$ branch and rolls down from the left
to the right, then one can never reach the physical $\phi/\Mp>1$
branch of the potential and inflation can never come to an
end. Secondly, when $x\gg 1$, the trajectory can be approximated by
\begin{equation}
\Nend- N \simeq \frac{e}{\alpha}\left(\frac{\ee^{x}}{x}
-\frac{\ee^{\xend}}{\xend}\right) .
\end{equation}
Moreover, in this approximation, it can be inverted exactly and one
obtains
\begin{equation}
  x\simeq -\Lambert{-1}\left[-\frac{1}{\alpha\left(\Nend-N\right)/\ee
      +\ee^{\xend}/\xend}\right],
\end{equation}
in agreement with what was obtained in \Refc{Lee:2010tk}. In the above
expression, $\Lambert{-1}$ is the $-1$ branch of the Lambert
function. Let us also notice that, in \Refc{Lee:2010tk}, the branch of
the Lambert function was in fact incorrectly chosen. The fact that the
$-1$ branch of the Lambert function has to be considered comes from
the following argument. When $\Nend-N\rightarrow\infty$, the argument
of the Lambert function goes to $0^-$ and, therefore, since $x$ must
tend towards $+\infty$ in this limit, the $-1$ branch must be
chosen. In addition, if $\Nend-N\rightarrow 0$, then one must have
$x\rightarrow\xend>1$ which is also the case if the $-1$ branch is
retained. This is represented in \Fig{potlambertKMII} where the arrow
indicates the direction along which inflation proceeds. In the third
place, since, when $x\rightarrow\infty$, one has
$\Nend-N\rightarrow\infty$, a sufficient number of \efolds can
always be realized in this model. Finally, it is inaccurate to assume
that $\xend\gg 1$ and, therefore, the above approximated trajectory is
not so useful. However, if one only assumes that $x\gg 1$ (which can
be checked to be a good approximation, especially at $x=\xstar$) but not
$\xend \gg 1$, then one can write
\begin{equation}
\Nend- N \simeq \frac{e}{\alpha}\frac{\ee^{x}}{x}
+\xend-\frac{e}{\alpha}\Ei\left(\xend-1\right),
\end{equation}
which, moreover, can be inverted into
\begin{equation}
\label{eq:kmii:xstar}
  x\simeq -\Lambert{-1}\left[-\frac{1}{\alpha\left(\Nend-N\right)e
      +\Ei\left(\xend-1\right)-\alpha\xend/e}\right],
\end{equation}
and which is valid whenever $x\gg 1$. However, one should keep in mind
that, now, and contrary to the former approximated trajectory, taking
the limit $N\rightarrow\Nend$ in the above expression is meaningless.

The energy scale $M$ is, as before, given by the CMB normalization and
one obtains the following expression
\begin{equation}
\label{eq:cmbkmii}
  \left(\frac{M}{\Mp}\right)^4=720\pi^2\alpha^2
  \frac{\left(1-\xstar\right)^2}{\left(1
      -\alpha \xstar\ee^{-\xstar}\right)^3}
  \ee^{-2\xstar}
  \frac{\Qrms^2}{T^2}\, .
\end{equation}
If one uses the $\xstar\gg 1$ approximation, then \Eq{eq:kmii:xstar}
tells us that $\xstar\simeq\ln\left(\alpha\Delta\Nstar\right)$ and
\Eq{eq:cmbkmii} can be re-written as
\begin{equation}
  \left(\frac{M}{\Mp}\right)^4=\mathcal{O}(1)\,720\frac{\pi^2}
  {\Delta\Nstar^2}\frac{\Qrms^2}{T^2}\, .
\end{equation}
It is remarkable that this equation does not depend on $\alpha$. Using
a fiducial value for $\Delta \Nstar$, one typically gets $M/\Mp\sim
10^{-3}$.

The predictions of KMII models are displayed in \Fig{fig:CMBKMII}, for
$\alpha>\alphaPlus$. The reheating equation of state parameter $\wrehbar$ has
been taken to $0$ since the potential is quadratic close to its
minimum [but, it should be reminded that, in principle, the potential
\Eq{eq:kmii:pot} cannot be trusted close to its minimum]. One can see
that, as announced at the beginning of this section, the predictions
depend on $\alpha$ in a very mild way, a conclusion which is in
agreement with \Refcs{Lee:2010tk,Conlon:2005jm}. This can be
understood as follows. If one assumes that $\xstar\gg 1$, then we have
already noticed that \Eq{eq:kmii:xstar} implies that
$\xstar\simeq\ln\left(\alpha\Delta\Nstar\right)$. From this result,
one obtains that
\begin{equation}
\begin{aligned}
\epsonestar \simeq \frac{1}{2\Delta \Nstar^2}
\ln^2\left(\alpha\Delta\Nstar\right)\, ,\quad
\epstwostar \simeq \frac{2}{\Delta \Nstar}
\ln\left(\alpha\Delta\Nstar\right)\, ,\quad
\epsthreestar \simeq \frac{1}{\Delta \Nstar}
\ln\left(\alpha\Delta\Nstar\right)\, .
\end{aligned}
\end{equation}
In these expressions, we notice that the slow-roll parameters (at
Hubble crossing) logarithmically depend on $\alpha$. This explains the
weak $\alpha$ dependence observed in \Fig{fig:CMBKMII}. Of course,
one can also calculate the corresponding expressions of the spectral
index, tensor to scalar ratio and running. One arrives at
\begin{equation}
\begin{aligned}
  \nS \simeq 1-2\frac{\ln\left(\alpha\Delta\Nstar\right)}
  {\Delta\Nstar}\, ,\quad r \simeq 8\frac{\ln^
    2\left(\alpha\Delta\Nstar\right)} {\Delta\Nstar^2}\, ,\quad
  \alphaS \simeq -2\frac{\ln^2\left(\alpha\Delta\Nstar\right)}
  {\Delta\Nstar^2}\, .
\end{aligned}
\end{equation}
These expressions are in accordance with the estimates derived in
\Refcs{Lee:2010tk,Conlon:2005jm}. However, contrary to what is claimed
in \Refcs{Lee:2010tk}, the predicted value of the running is not
excluded by the CMB observations since, according to
the Planck results~\cite{Ade:2013zuv}, one has
$\alphaS = -0.013 \pm 0.009$.

\subsection{Horizon Flow Inflation at first order (HF1I)}
\label{sec:hf1i}

The horizon flow models have been introduced in \Refc{Liddle:2003py}
and consist into designing field potentials to exactly produce a
truncated Taylor expansion of the Hubble parameter with respect to the
field. As such they constitute a whole class of phenomenological
inflationary models. Here, we are considering a potential designed
such that $H(\phi)=H_0(1+A_1\phi/\Mp)$, where $A_1$ is a free
dimensionless parameter. The shape of the potential
reads~\cite{Liddle:2003py}
\begin{equation}
  V(\phi)=M^4 \left(1+A_1 \dfrac{\phi}{\Mp}\right)^2\left[1-\frac{2}{3}
\left(\frac{A_1}{1+A_1\dfrac{\phi}{\Mp}}\right)^2\right].
\label{eq:pothf1i}
\end{equation}
Denoting $x \equiv \phi/\Mp$, the potential admits a global minimum at
$\xVmin=-1/A_1$, which is negative
\begin{equation}
\Vmin = V\left(\phiVmin\right) = -\dfrac{2}{3} M^4 A_1^2<0.
\end{equation}
As a result, there are two disconnected field domains in which the
potential remains definite positive, either $x>\xVzeroPlus$ or
$x<\xVzeroMinus$ where $\xVzeroPM$ are the two roots of $V(\xVzeroPM)
= 0$, i.e.
\begin{equation}
\xVzeroPlus = \sqrt{\dfrac{2}{3}} - \dfrac{1}{A_1}\,, \qquad
\xVzeroMinus = -\sqrt{\dfrac{2}{3}} - \dfrac{1}{A_1}\,.
\end{equation}

\begin{figure}
\begin{center}
\includegraphics[width=\wdblefig]{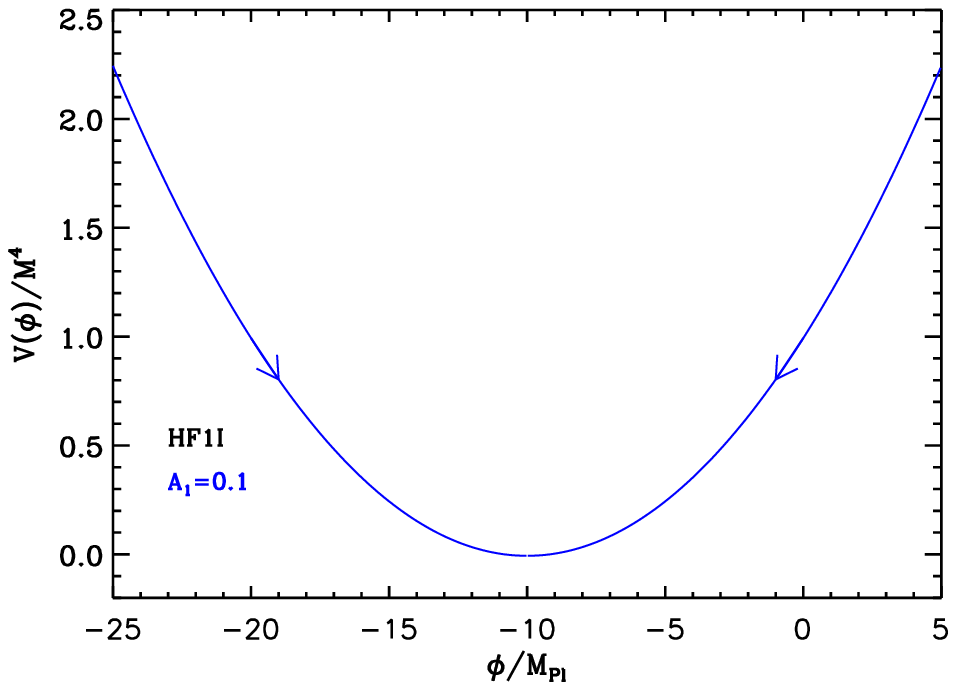}
\includegraphics[width=\wdblefig]{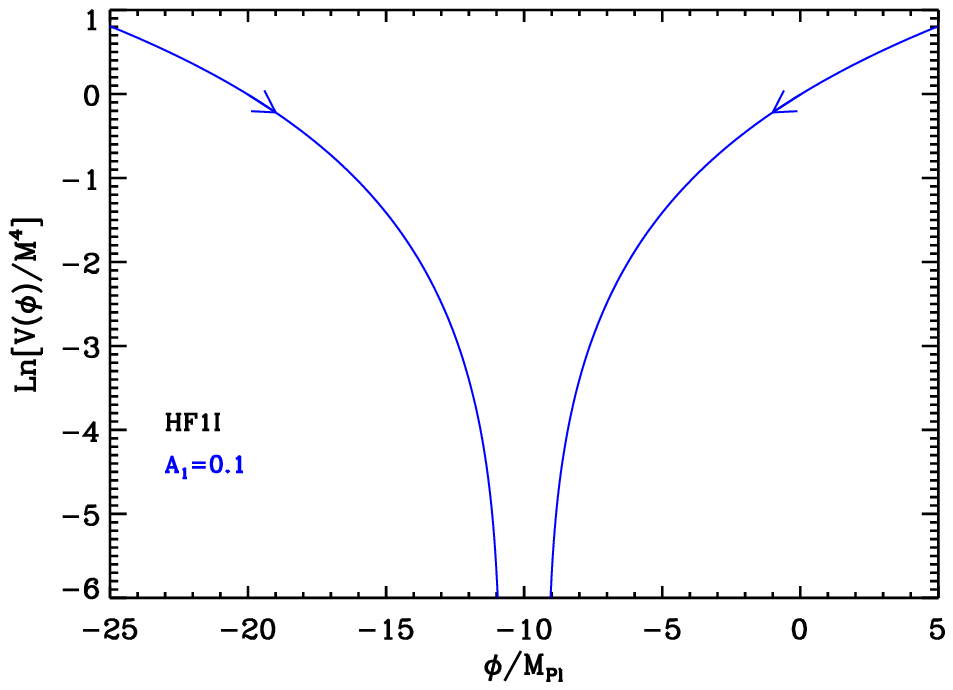}
\includegraphics[width=\wdblefig]{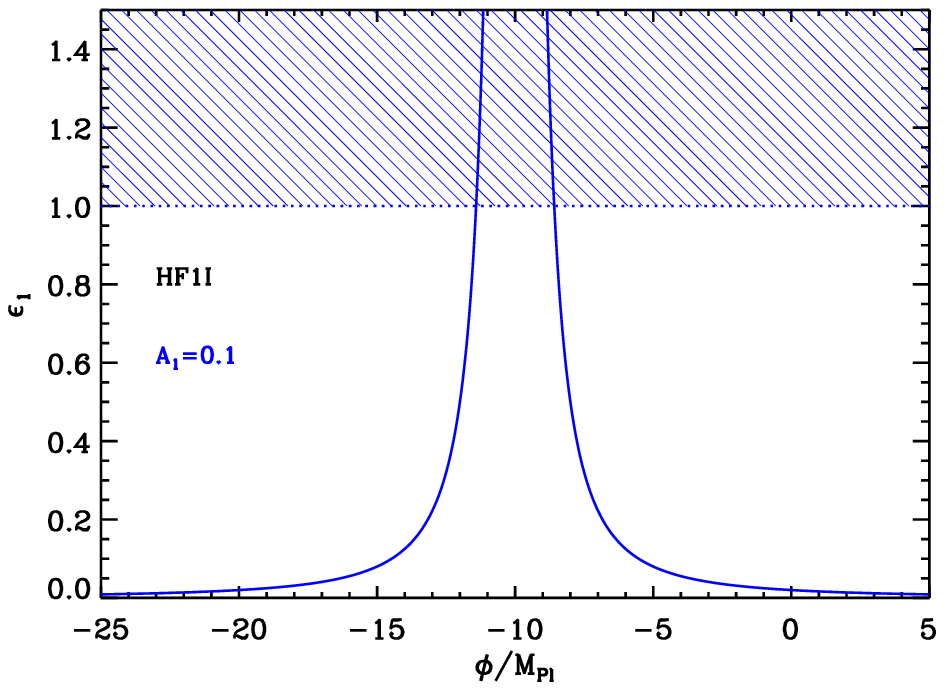}
\includegraphics[width=\wdblefig]{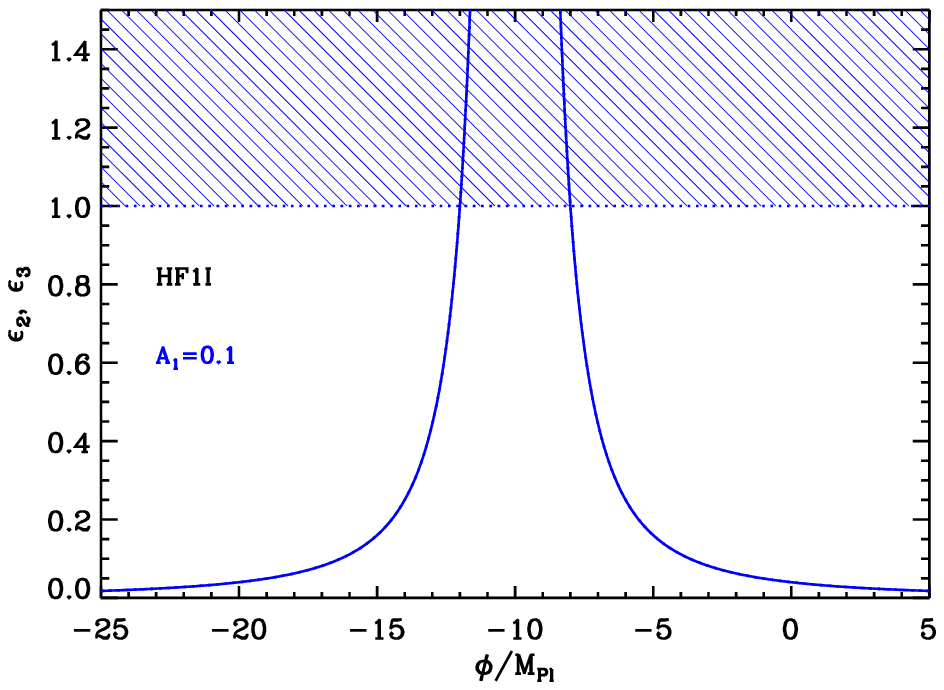}
\caption{Top left panel: Horizon Flow Inflation at first order
  potential for $A_1=0.1$. Top panels: the potential and its logarithm
  with respect to the field values. Bottom left panel: the first
  Hubble flow function $\epsilonH_1$ (exact) and the corresponding
  shaded area where inflation stops. Bottom right panel: Hubble flow
  functions $\epsilonH_2$ (solid line) and $\epsilonH_3$ (dotted line)
  for the same potential. These two functions are equal to
  $2\epsilonH_1$.}
\label{pothf1i}
\end{center}
\end{figure}

An interesting consequence of the horizon flow approach is that the
Hubble flow functions can be calculated exactly, \ie without the
slow-roll approximation because $H(\phi)$ is exactly known. As
discussed in \Refcs{Liddle:1994dx, Copeland:1998fz}, one could compare
them with the other hierarchy of parameters, $\epsilonV_i$, that are
defined by the successive logarithmic derivatives of the potential. In
the slow-roll approximation, one precisely uses the potential
derivatives to approximate the Hubble flow functions. {}From $H
\propto 1 +A_1 x$, one gets the exact Hubble flow functions
\begin{equation}
\epsilonH_1 = 2 \left(\dfrac{A_1}{1+A_1 x}\right)^2, \qquad \epsilonH_2
= \epsilonH_3 = 2 \epsilonH_1,
\label{eq:epsHhf1i}
\end{equation}
whereas the slow-roll functions associated with
the potential are
\begin{equation}
\begin{aligned}
\epsilonV_1 & = \frac{18 A_1^2 (A_1 x+1)^2}{\left[3 + 6 A_1 x + A^2 \left(3 x^2 -
      2\right)  \right]^2}\,, \qquad
\epsilonV_2 = \frac{12 A_1^2 \left[3 + 6 A_1
   x + A_1^2 \left(3 x^2 + 2\right) \right]}{\left[3 + 6 A_1 x +  A_1^2 \left(3 x^2 - 2 \right) \right]^2}\,,
\end{aligned}
\end{equation}
and
\begin{equation}
\begin{aligned}
  \epsilonV_3 & = \frac{108 A_1^2 (A_1 x+1)^2 \left[1 + 2 A_1 x +
      A_1^2 \left( x^2 + 2\right)\right]}{\left[3 + 6 A_1 x + A_1^2
      \left(3 x^2-2\right)\right]^2 \left[3 + 6 A_1 x + A_1^2 \left(3
        x^2+2 \right) \right]}\,.
\end{aligned}
\end{equation}
As shown in \Refc{Liddle:1994dx}, the link between the two hierarchies
can be made explicit and one has
\begin{equation}
\epsilonV_1 = \epsilonH_1 \left(\dfrac{1 - \etaH/3}{1 - \epsilonH_1/3}
\right)^2.
\label{eq:epsVH}
\end{equation}
The $\etaH$ parameter is defined as
\begin{equation}
\etaH \equiv \dfrac{2}{H} \dfrac{\ud^2 H}{\ud x^2}\,,
\label{eq:}
\end{equation}
and vanishes in our case. As a result, provided $\epsilonH_1 \ll 1$,
i.e. we are in the slow-roll approximation, both hierarchies give the
same results at first order. In order to establish \Eq{eq:epsVH}, one has
to show first that
\begin{equation}
  \etaH = \epsilonH_1 + \dfrac{1}{\sqrt{2 \epsilonH_1}} \dfrac{\ud \epsilonH_1}{\ud x}\,,
\end{equation}
and then that\footnote{A sign in these two equations differs from the
  ones typeset in \Refc{Liddle:1994dx}, most probably due to a
  misprint.}
\begin{equation}
\dfrac{\ud \epsilonH_1}{\ud x} = \left(\epsilonH_1 -3 \right)
\left(\dfrac{\ud \ln V}{\ud x} - \sqrt{2 \epsilonH_1} \right).
\end{equation}
The potential and the exact Hubble flow functions have been
represented in \Fig{pothf1i}.

Inflation can take place inside the two positive definite domains of
the potential, i.e. at negative or positive field values. However, the
Hubble parameter has to be positive such that $H_0$ has to be chosen
negative if $1+ A_1 x < 0$ along the field trajectory. Since the
potential is completely symmetric with respect to its minimum
$\xVmin$, we can study in full generality only the $x>\xVzeroPlus$
branch. In particular, as the Hubble flow functions are exact, we can
also derive the exact field trajectory
\begin{equation}
\label{eq:trajecHF1}
N-\Nend = -\frac{1}{2A_1}\left(x + \frac{1}{2} A_1 x^2 - \xend -
  \frac{1}{2} A_1 \xend^2 \right).
\end{equation}
Let us notice that, in the slow-roll approximation, one would have
derived the trajectory from $\epsilonV_1$. Doing so, one would have
obtained
\begin{equation}
  N - \Nend = -\frac{1}{2A_1}\left(x +\frac{1}{2} A_1 x^2  - \xend - \frac{1}{2}
    A_1 \xend^2 -\frac{2}{3} A_1 \ln \left\vert\frac{1+A_1 x}{1 + A_1 \xend} \right\vert\right).
\label{eq:trajecHF1b} 
\end{equation}
It is amusing to remark that here, the simplest formula is not given
by the slow-roll derived one, but rather by the exact one. From this
remark one should keep in mind that, in order to simplify trajectories
integration, one can always add factors of order
$\order{\epsilon_1}$. The exact trajectory \eqref{eq:trajecHF1} can
be inverted and one finds
\begin{equation}
x = - \dfrac{1}{A_1} + \dfrac{1}{A_1} \sqrt{1 + 2 A_1 \xend + A_1^2
  \left[\xend^2 -4 (N-\Nend) \right]}\,.
\label{eq:invtrajhf1i}
\end{equation}

Along both the positive and negative branch of the potential,
inflation ends naturally at $\epsilonH_1=1$, that is at
\begin{equation}
\xepsoneOnePM = \dfrac{-1 \pm \sqrt{2} A_1}{A_1}\,.
\end{equation}
Along the positive branch we are interested in, we therefore have
\begin{equation}
\xend = \xepsoneOnePlus = \dfrac{-1 + \sqrt{2} A_1}{A_1}\,.
\end{equation}
Plugging this expression into \Eq{eq:invtrajhf1i} gives  the
field value $\xstar$ at which the pivot mode crossed the Hubble radius
during inflation in terms of the \efold number $\Delta \Nstar =
\Nend-\Nstar$. Let us remember that solving for $\xstar$ (or $\Delta
\Nstar$) is made through \Eq{eq:phistarlnrrad}. From \Eq{eq:epsHhf1i},
one gets
\begin{equation}
\epsilonH_{1*} = \dfrac{1}{1 + 2 \Delta \Nstar}\,
\end{equation}
which, together with $\epsilonH_2 = 2 \epsilonH_1$, yields
\begin{equation}
\nS-1=2\nT, \qquad r=4(1-\nS).
\end{equation}
Notice that this relation is different from the power law case and
consistent with \Refc{Ramirez:2005cy}. In that reference, the authors
mention that the horizon flow models predicts $r \simeq 4.8(1-\nS)$ as a
result of Monte-Carlo simulations.

Finally, the potential parameter $M$ can be determined from the CMB
normalization
\begin{equation}
  \left(\frac{M}{\Mp}\right)^4 = 960
  \pi^2 \dfrac{A_1^2}{(1 + A_1 \xstar)^4} \dfrac{\Qrms^2}{T^2}\, .
\end{equation}

It is interesting to notice that the typical energy scale of inflation
in these models does not depend on $A_1$. The previous equation indeed
leads to
\begin{equation}
\frac{V\left(\xstar\right)}{\Mp^4}=\frac{480\pi^2}{1+2\Delta\Nstar}
\dfrac{\Qrms^2}{T^2}\left(1-\frac{1}{3+6\Delta\Nstar}\right)
\simeq 10^{-9}\, .
\end{equation}

The reheating consistent (exact) predictions for the horizon flow
inflation I models are represented in \Fig{fig:CMBHF1I}. As expected,
the relation $\epsilonH_2 = 2\epsilonH_1$, which is the same as for
the LFI quadratic case, is properly recovered. The predictions do not
depend much on the potential parameter $A_1$.

\subsection{Colemann-Weinberg Inflation (CWI)}
\label{sec:cwi}

\subsubsection{Theoretical Justifications}

The potential of this model was first introduced by Coleman and
Weinberg in \Refc{Coleman:1973jx}, in the context of spontaneous
symmetry breaking generated by radiative corrections. The starting
point of this work is to calculate the effective potential for a
massless charged meson minimally coupled to the electrodynamic
field.

In that reference, the effective action is explicitly constructed from
a Legendre transform of the partition function, and expanded into
one-particle-irreducible Feynman diagrams with $n$ external lines
(and summing up over $n$). The exact knowledge of the effective
potential requires an infinite summation of all these Feynman
diagrams, which is in practice intractable. It is thus made use of the
one loop expansion method where all diagrams with no closed loops are
first summed, then all diagrams with one closed loop are added, and
all higher loops diagrams neglected.  Starting with a quartic
interacting scalar field, and requiring that the renormalized mass
vanishes, one obtains a potential of the form
\begin{equation} 
V(\phi) \propto 1 +
\alpha\left(\frac{\phi}{Q}\right)^4 \ln
\left(\frac{\phi}{Q}\right).
\end{equation}
Let us emphasize that another useful frame of approximation is the
Gaussian effective potential method.  The Gaussian effective potential
is a non-perturbative approach to quantum field theory
\cite{Stevenson:1984rt, Stevenson:1985zy, Stevenson:1985kr,
  Stevenson:1986na, Stevenson:1986bq, Stevenson:1986nb,
  Stevenson:1986sb, Hajj:1987gk, IbanezMeier:1992fp}, originally
developed in the context of quantum mechanics, and generalized to
field theory afterwards. In quantum mechanics, when studying systems
governed by Hamiltonians of the form $H=p^2/2+V(\phi)$, the idea is to
calculate en effective potential $\Vgep$ defined as
\begin{equation}
\Vgep\left(\phizero\right)=\min_\Omega\left[\left\langle\psi\right\vert 
H \left\vert\psi\right\rangle\, ,\, 
\psi\left(\phi\right)=\left(\frac{\Omega}{\hbar\pi}\right)^{1/4}
\ee^{-\Omega\left(\phi-\phizero\right)^2/(2\hbar)}\right],
\end{equation}
\ie the minimum possible quantum mean energy of a Gaussian
wavefunction centered over $\phizero$. Such an object turns out to be a
powerful tool to addressing the effects of quantum fluctuations on the
physical behavior of a system in a non-perturbative way. It can be
easily generalized to quantum field theories, expanding the field
operator $\Phi$ only over $\Omega$-massive excitations around the
classical value $\phizero$ in $d$ dimensions,
\begin{equation}
\Phi\left(t,\bmx\right)=\phizero+\left(2\pi\right)^{(1-d)/2}
\int\frac{\dd^{d-1}\bmk}{\sqrt{2\sqrt{k^2+\Omega^2}}}
\left(a_\bmk\ee^{-i\sqrt{k^2+\Omega^2}t+i\bmk\cdot\bmx}
+a_\bmk^\dagger\ee^{i\sqrt{k^2+\Omega^2}t-i\bmk\cdot\bmx}
\right),
\end{equation}
where $a_\bmk^\dagger$ and $a_\bmk$ are the usual creation and
annihilation operators, and minimizing the quantum mean value of the
Hamiltonian density over $\Omega$. In \Refc{Stevenson:1985zy}, the
quartic interacting scalar field has been worked out with this method,
\ie starting from $V(\phi)=m^2\phi^2/2+\lambda\phi^4$. The Gaussian
effective potential $\Vgep$ obtained in this way can expanded
in power of $\hbar$ to show that the first order terms match with the
potential of Coleman and Weinberg. This is not surprising as this is
equivalent of performing a one loop expansion over the effective
action. However, it should be stressed that the Gaussian effective
potential method provides a much more general expression for the
potential, that is valid beyond this perturbative limit and that can
address regimes where quantum diffusion dominates the dynamics of the
scalar field. 

The model is defined such that inflation ends by violation of the
slow-roll conditions, and is followed by a preheating stage in which
the inflaton field oscillates at the bottom of its potential.
Therefore this potential minimum must be set to zero, which implies
\begin{equation}
\alpha=4e\, .
\end{equation}
One is thus left with one mass parameter, $Q$, which sets the typical
\vev at which inflation takes place. On the other hand, the value
taken for $Q$ also depends on the underlying high energy model from
which the CW potential emerges.

The CWI potential appears in various other contexts and, in fact,
historically, it was the first model of inflation ever
proposed~\cite{Guth:1980zm} (also known as ``old inflation''). The
idea was that inflation occurs while the field is trapped in a false
vacuum state $\langle \phi \rangle=0$. Then, inflation comes to an end
when the field tunnels from this state to the symmetry breaking true
minimum. Unfortunately, this models was quickly realized to be ruled
out since the above mentioned process is accompanied by bubble
formation and these bubbles, while colliding, produce too large
inhomogeneities. Then, this problem was solved by a modification of
the old inflation scenario called ``new
inflation''~\cite{Linde:1981mu,Albrecht:1982wi}. The main idea is that
inflation does not occur while the field is trapped but when the field
is rolling down from the origin to its true minimum. Bubbles are also
formed but there are so big that our entire universe is contained in
one of them. As a consequence, we do not observe bubble collisions and
our universe is extremely homogeneous as indicated by the
observations. This new inflationary scenario was explicitly
implemented in \Refc{Linde:1981mu} where the
$\mathrm{SU}(5)\rightarrow \mathrm{SU}(3)\times \mathrm{SU}(2)\times
\mathrm{U}(1)$ phase transition in GUTs is investigated. The model
makes use of a CWI potential that can be described by
\begin{equation}
\label{eq:cwi:potLinde}
V(\phi) = \frac{5625}{512\pi^2}g^4 \left[ \phi^4 \ln\left(
  \frac{\phi}{\phizero} \right)- \frac{\phi^4}{4} + \frac{\phizero^4}{4}
  \right],
\end{equation}
where $\phizero\simeq 10^{14}-10^{15}\, \GeV$, representing the GUT
symmetry breaking scale, and $g^2\simeq 1/3$ is the $\mathrm{SU}(5)$
gauge coupling constant. However, as noticed afterwards in
\Refcs{Abbott:1981rg, Ellis:1982ws, Albrecht:1983ad, Shafi:1983bd,
  Albrecht:1984qt}, this model has also a fatal flaw. Indeed, one sees
in \Eq{eq:cwi:potLinde} that the overall normalization of the
potential reads $M^4=5625g^4\phizero^4/(2048 \pi^2)$ and that,
therefore, the amplitude of the fluctuations is in fact already
fixed. Using the value of the $\mathrm{SU}(5)$ coupling constant and
$Q/\Mp=\ee^{1/4}\phizero/\Mp\simeq 5\times10^{-5}-5\times 10^{-4}$,
one arrives at $M^4 \simeq \left(10^{-13}-10^{-17}\right)\Mp^4$. This
turns out to be incompatible with the CMB normalization [see
  \Eq{eq:cwi:COBE} below]. However, the same model was re-considered
in \Refcs{Shafi:1983bd,Rehman:2008qs} (see also
\Refc{Langbein:1993ym}), but with additional fields and couplings. It
was then shown that the scale $M$ acquires a different form and can
scale as the inverse of the coupling constants. Since these ones are
small, it becomes possible to obtain a higher value for $M$ and to
correctly CMB normalize the model. In what follows, we will therefore
consider the scale $M$ as a free parameter fixed by the overall
amplitude of the cosmological fluctuations.

We also notice that, in \Refc{GonzalezDiaz:1986bu}, the CWI potential
is obtained in the context of Kaluza-Klein inflation, \ie in higher
dimensions and with higher derivative terms and logarithmic dependence
on the curvature scalar. Again, the typical value for $Q\simeq
10^{15}\, \GeV$. The CWI potential appears also in
\Refc{Yokoyama:1998rw}, but the value used for $Q$ is rather
different, $Q=0.223\Mp$, and is fine-tuned in order to have two phases
of inflation, a ``chaotic inflationary'' phase followed by a ``new
inflationary'' phase. Finally, in \Refc{Gong:1998nf}, the
Coleman-Weinberg potential is studied in the framework of
Einstein-Brans-Dicke gravity, with the same typical value for $Q\simeq
10^{15}\, \GeV$ and the same typical value for $M^4/\Mp^4\simeq
10^{-15}$ as in the original paper.

\subsubsection{Slow-Roll Analysis}

\begin{figure}
\begin{center}
\includegraphics[width=\wdblefig]{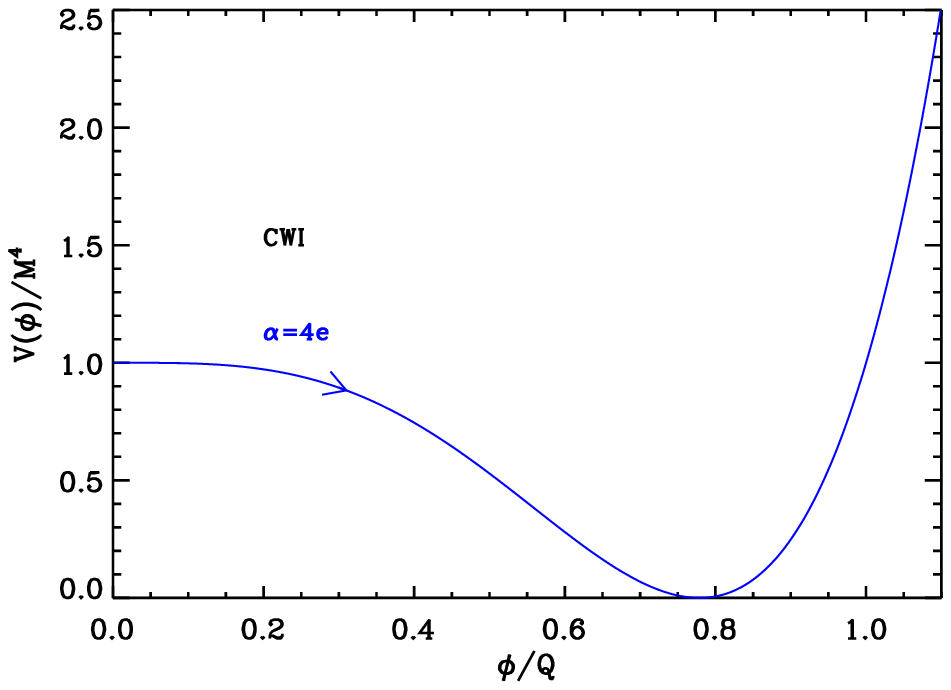}
\includegraphics[width=\wdblefig]{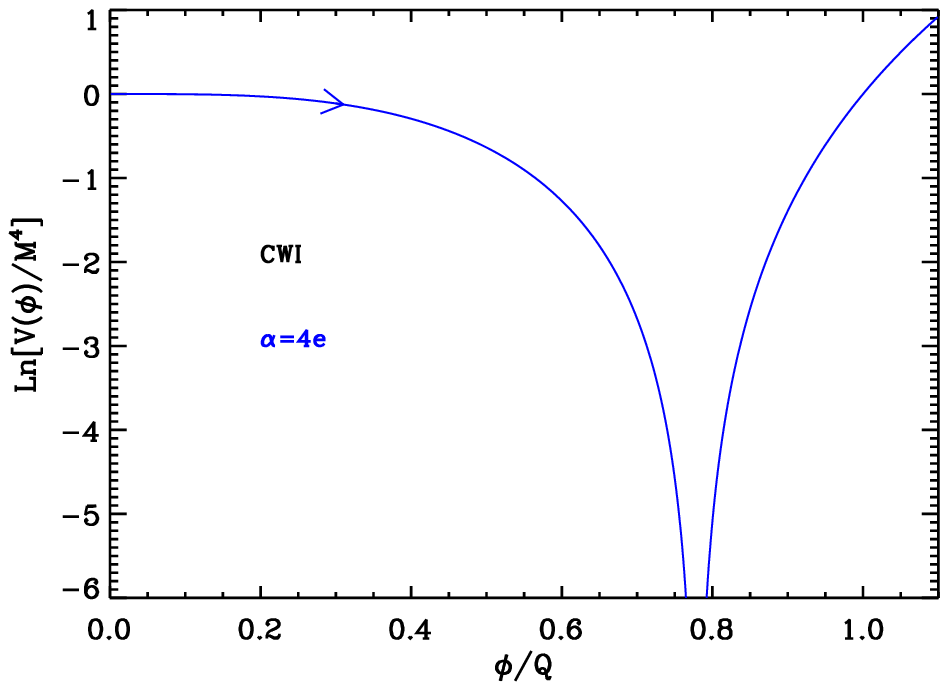}
\includegraphics[width=\wdblefig]{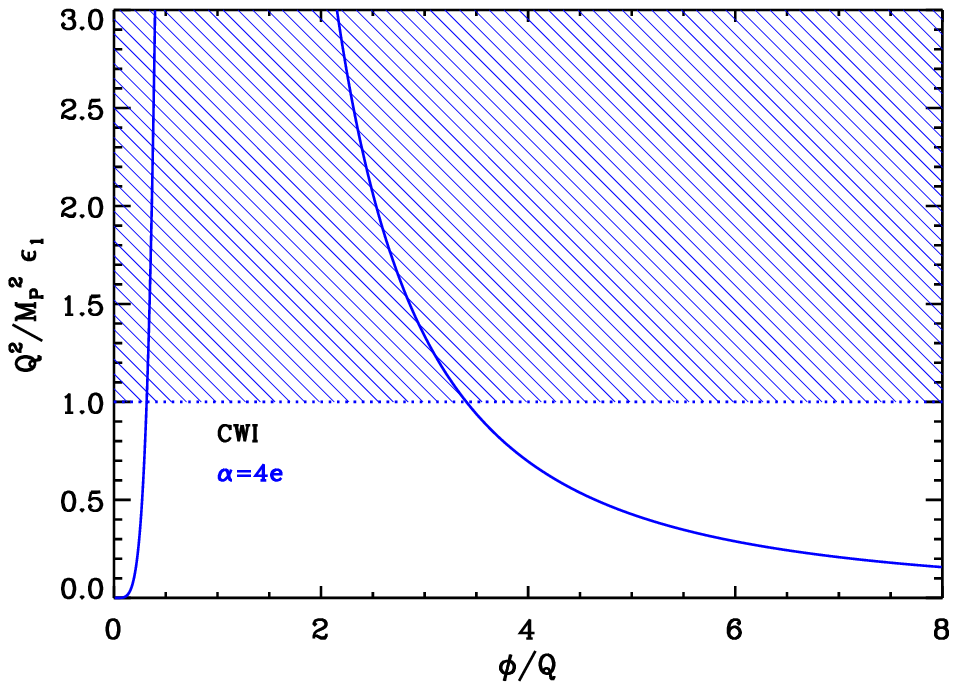}
\includegraphics[width=\wdblefig]{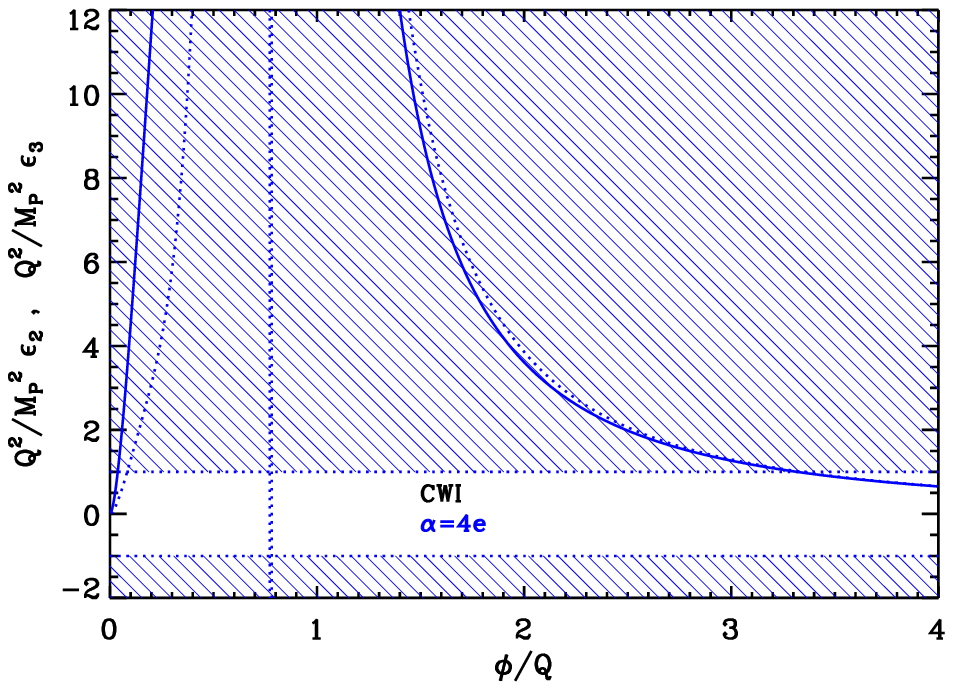}
\caption{Colemann-Weinberg Inflation (CWI) for $\alpha=4e$.  Top left
  panel: Colemann-Weinberg Inflation potential as a function of
  $\phi/Q$.  Top right panel: logarithm of the potential for the same
  value of $\alpha$.  Bottom left panel: normalized first slow-roll
  parameter $Q^2/\Mp^2\,\epsilon_1$. The shaded area indicates the
  where inflation stops if $Q=\Mp$. Bottom right panel: normalized
  second and third slow-roll parameters $Q^2/\Mp^2\,\epsilon _2$
  (solid line) and $Q^2/\Mp^2\,\epsilon _3$ (dotted line) for the same
  potential.}
\label{potcwi}
\end{center}
\end{figure}

Considering the previous considerations, we take the potential to be
\begin{equation} 
V(\phi)=M^4\left[1 +
\alpha\left(\frac{\phi}{Q}\right)^4 \ln
\left(\frac{\phi}{Q}\right)\right],
\end{equation}
with a parameter $Q/\Mp$ in the range $\left[10^{-5},10^{-3}\right]$
and $\alpha=4 \ee$. As already mentioned, the mass parameter $M$ will
be viewed as free and fixed by the normalization to the amplitude of
the CMB anisotropies. The potential is displayed \Fig{potcwi}. It
starts decreasing with the inflaton \vev at $\phi=0$, reaches a
minimum at $\phi/Q=\ee^{-1/4}$ where it vanishes, and then increases
and diverges as $\phi$ goes to $\infty$. As mentioned above, inflation
proceeds along the decreasing branch of the potential, in the
direction specified by the arrow in the figure.

\begin{figure}
\begin{center}
\includegraphics[width=\wsingfig]{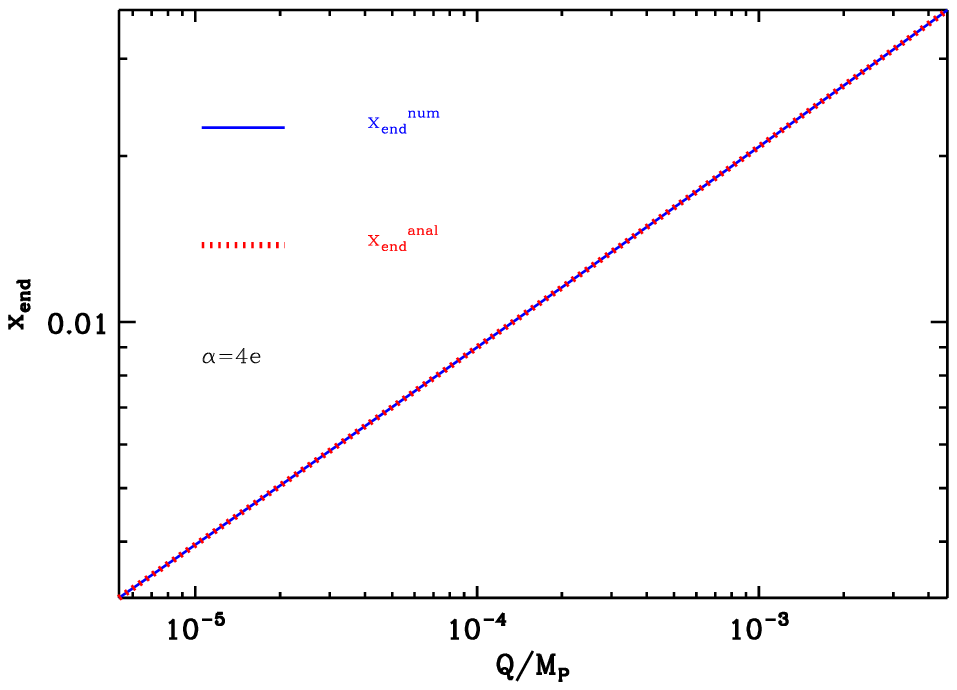}
\caption{End of inflation in Coleman-Weinberg inflation. The
  approximated formula of \Eq{eq:cwi:xend} for $\xend$ (red dashed
  line) is compared with the exact numerical solution of
  $\epsilon_1=1$ (blue solid line), for $\alpha=4e$, in the physically
  relevant range of values for $Q/\Mp$. The agreement is obviously
  excellent.}
\label{fig:cwi:xend}
\end{center}
\end{figure}

Let us now derive the first slow-roll parameters. Defining
$x\equiv\phi/Q$, they are given by
\begin{equation}
  \epsilon_1 = \frac{\Mp^2}{Q^2}\frac{\alpha^2}{2}x^6
  \left(\frac{1+4\ln x } {1+\alpha
      x^4\ln x }\right)^2,
\end{equation}
while
\begin{equation}
\begin{aligned}
  \epsilon_2 = 2\frac{\Mp^2}{Q^2}\alpha x^2  
\dfrac{ -7-12\ln x  +\alpha x^4 +\alpha
   x^4\ln x  +4\alpha x^4
   \ln^2 x }{\left(1+\alpha x^4
     \ln x \right)^{2}}\,,
\end{aligned}
\end{equation}
and finally
\begin{equation}
\begin{aligned} \epsilon_3 &= \frac{\Mp^2}{Q^2}\left(
-26\alpha x^2 +21\alpha^2 x^6-2\alpha^3 x^{10} -128\alpha x^2\ln x
\right. \\ & \left. +
152\alpha^2 x^6\ln x  -11\alpha^3 x^{10}\ln x
-96\alpha x^2\ln ^2 x
\right. \\ & \left. + 
368 \alpha^2 x^6\ln ^2 x -14\alpha^3 x^{10}\ln^2 x
+384\alpha^2 x^6\ln^3 x
\right. \\ & \left. - 
16 \alpha^3 x^{10}\ln^3 x  -32\alpha^3 x^{10}\ln^4 x \right)
\left(1+\alpha x^4\ln x  \right)^{-2} 
\\ & \times
\left(7-\alpha x^{4} +12\ln x -\alpha x^{4}\ln x
-4\alpha x^{4}\ln^2 x \right)^{-1}.
\end{aligned}
\end{equation}
The three of them have the same general behavior. They vanish at
$x=0$, increase with $x$ in the decreasing branch of the potential and
diverge at the minimum of the potential. Then they decrease from
infinity in the increasing branch of the potential, and reach
asymptotically vanishing values when the field \vev goes to
infinity. Inflation stops by slow-roll violation when
$\epsilon_1=1$. The value of $x$ at which this happens needs to be
determined numerically, but in the limit $Q/\Mp\ll 1$ (remember that
$Q/\Mp\simeq 10^{-4}$) where one expects $\xend\ll 1$, one can derive
an analytic approximated formula, namely
\begin{equation}
\label{eq:cwi:xend}
\xend\simeq\ee^{-1/4}\exp\left[\Lambert{-1}
\left(-\frac{3\sqrt{2}}{4\alpha}\frac{Q}{\Mp}\ee^{3/4}\right)\right],
\end{equation}
where $\Lambert{-1}$ is the $-1$ branch of the Lambert function. A
comparison between this approximated formula and the numerical
solution for $\xend$ is displayed in \Fig{fig:cwi:xend}. The agreement
is excellent.

Let us now calculate the slow-roll trajectory from
\Eq{eq:srtrajectory}. It is given by
\begin{equation}
\begin{aligned}
  \Nend-N & =\frac{Q^2}{\Mp^2} \frac{\sqrt{\ee}}{4\alpha}
  \left[ \Ei\left(-\frac{1}{2} -
      2\ln x  \right) 
    -\Ei\left(-\frac{1}{2} -
      2\ln \xend  \right) \right ]
  \\
  & +\frac{Q^2}{\Mp^2} \frac{1}{16\sqrt{\ee}}
  \left[\Ei\left(\frac{1}{2} + 2\ln \xend  \right) 
    -\Ei \left(\frac{1}{2} + 2\ln x  \right)
  \right]
  \\
  &+\frac{1}{8}\frac{Q^2}{\Mp^2} \left( x^2-\xend^2 \right),\\
\end{aligned}
\end{equation}
where $\Ei$ is the exponential integral function, $\Nend$ is the
number of \efolds at the end of inflation and $N$ is the number of
\efolds corresponding to the scaled field \vev $x$.  In the
$Q/\Mp\ll 1$ limit where $x\ll 1$, the first term of this expression
dominates. Since $\alpha=4e$, the previous expression can be slightly
simplified:
\begin{equation}
\begin{aligned}
  \Nend-N  &=\frac{Q^2}{\Mp^2}\frac{1}{16\sqrt{e}}
  \left[\Ei\left(-\frac{1}{2} -
      2\ln x \right) 
    -\Ei\left(-\frac{1}{2} -
      2\ln \xend \right)\right. \\
  & + \left.\Ei\left(\frac{1}{2}+2\ln \xend  \right) 
    -\Ei\left(\frac{1}{2} + 2\ln x 
    \right) \right]
  +\frac{1}{8}\frac{Q^2}{\Mp^2} \left( \xend^2-x^2\right).\\
\end{aligned}
\end{equation}

After having solved the above equation for $\xstar$, the field value
at which the pivot scale crossed the Hubble radius during inflation,
$M$ is fixed by the amplitude of the CMB anisotropies to
\begin{equation}
\label{eq:cwi:COBE}
  \left(\frac{M}{\Mp}\right)^4=720\pi^2\alpha^2\frac{\Mp^2}{Q^2}
  \xstar^6
  \left(1+4\ln \xstar\right)^2
  \left(1 +\alpha\xstar^4
    \ln \xstar  \right)^{-3}
    \frac{\Qrms^2}{T^2}\, .
\end{equation}

The reheating consistent slow-roll predictions of the Coleman-Weinberg
models are displayed \Fig{fig:CMBCWI1} in the physical range
$Q/\Mp\in\left[10^{-5},10^{-3}\right]$.  The reheating equation of
state parameter $\wrehbar$ has been taken to $0$ since the potential is
quadratic close to its minimum $V\left(x\right)\simeq 2 \alpha
M^4\ee^{-1/2}\left(x-\ee^{-1/4}\right)^2$.  The typical predicted
amount of gravitational waves is extremely small, and a non-negligible
deviation from $\nS=1$ is noticed.  Also, one could choose to relax
the constraint on the parameter $Q$ and study the Coleman-Weinberg
potential in general. This was done for instance in
\Refc{Rehman:2008qs} where the Coleman-Weinberg potential predictions
are compared with the WMAP observations on general grounds. It is
found that the potential normalization should be of the order
$M\simeq10^{16}\, \GeV$, and that $Q\simeq 10\,\Mp$ in order to match
$\nS\simeq 0.96$. For this reason the reheating consistent slow-roll
predictions are displayed in \Fig{fig:CMBCWI2} in the extended range
$Q/\Mp\in\left[1,100\right]$. In the limit $Q/\Mp\gg 1$, the model is
well approximated by a quadratic potential around its minimum, and one
asymptotically approaches the LFI predictions with $p=2$ (see
\sectionc{sec:lfi}).

\subsection{Loop Inflation (LI)}
\label{sec:li}

\begin{figure}
\begin{center}
\includegraphics[width=\wdblefig]{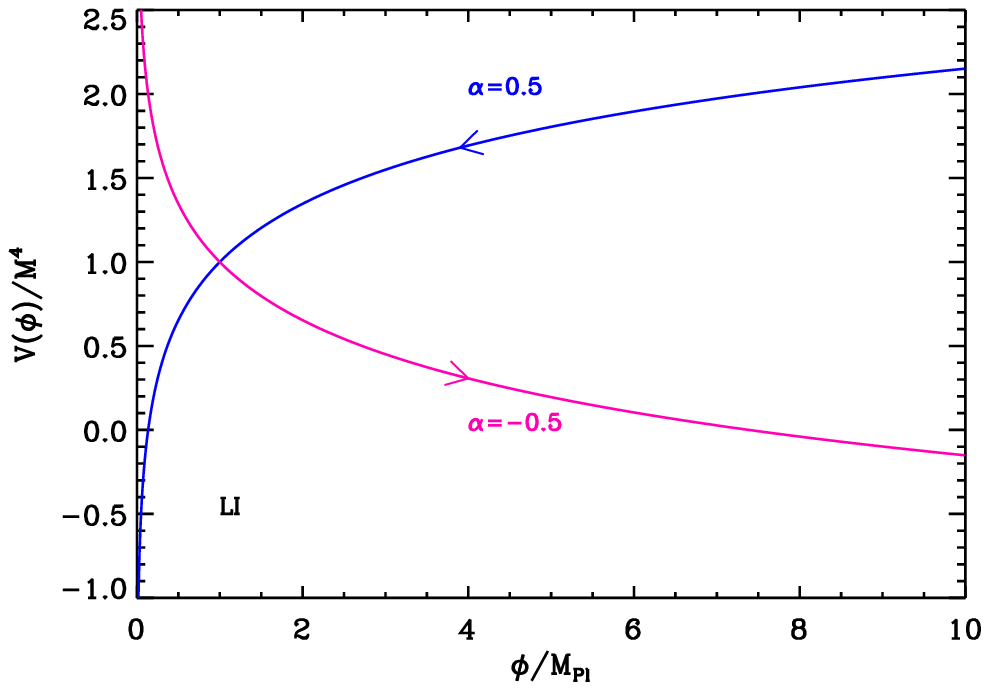}
\includegraphics[width=\wdblefig]{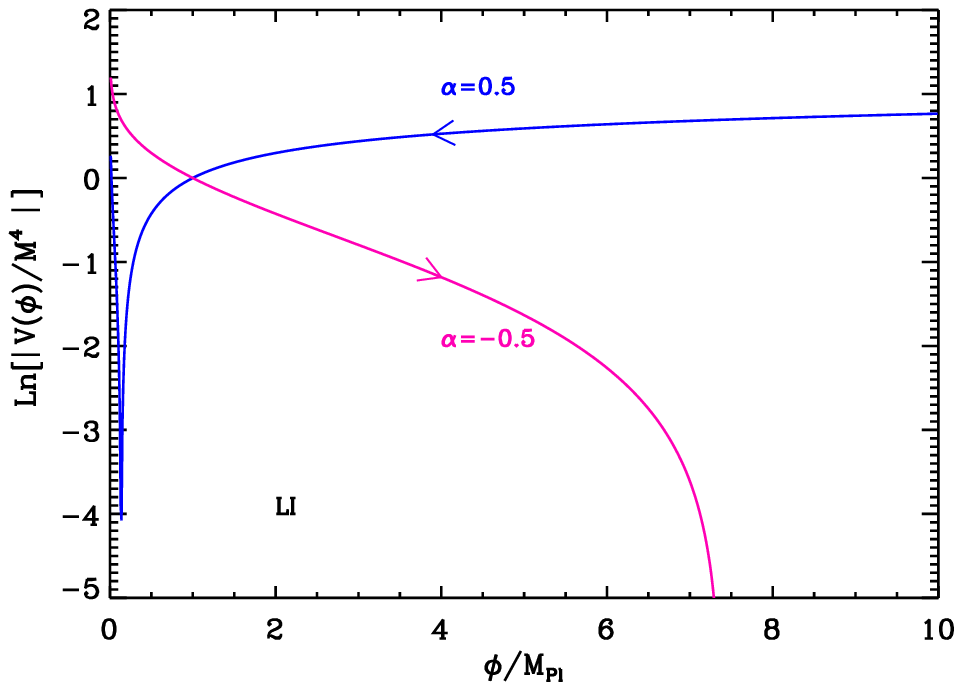}
\includegraphics[width=\wdblefig]{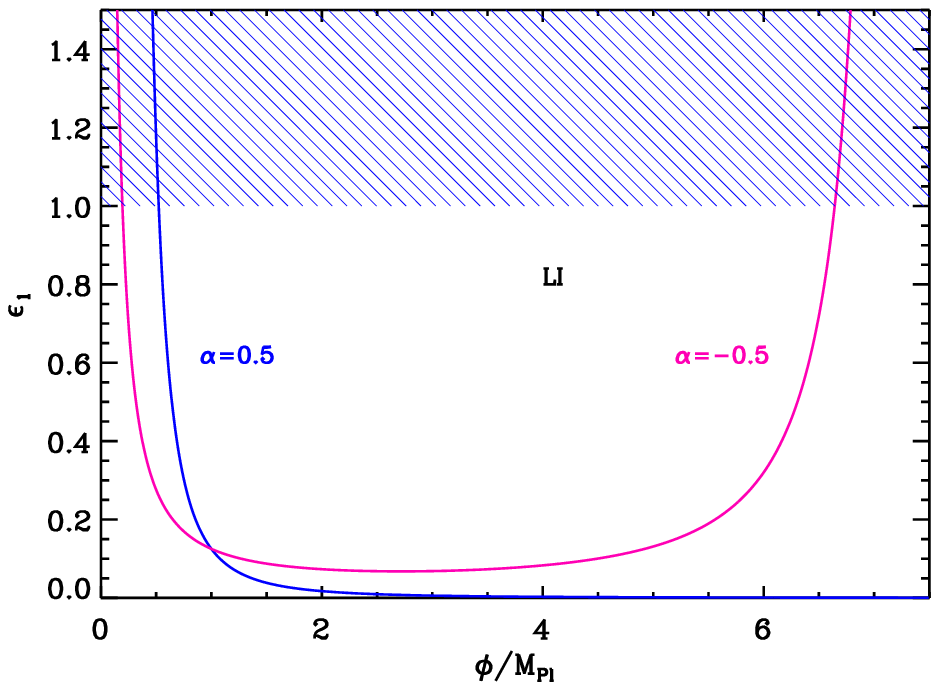}
\includegraphics[width=\wdblefig]{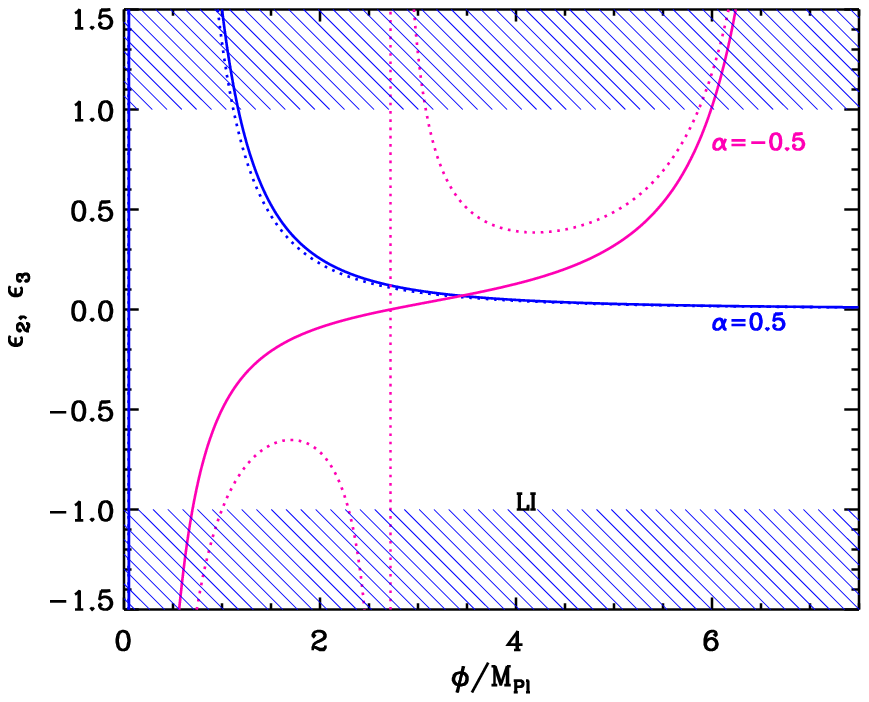}
\caption{Loop Inflation (LI). Top left panel: Loop Inflation potential
  for $\alpha=\pm 0.5$, the case $\alpha=0.5$ being displayed in blue
  and the case $\alpha=-0.5$ being displayed in pink.  Top right
  panel: logarithm of the potential for the same values of
  $\alpha$. Bottom left panel: slow-roll parameter $\epsilon _1$ with
  the same values of $\alpha$. The shaded area indicates where
  inflation stops. Bottom right panel: slow-roll parameters $\epsilon
  _2$ (solid line) and $\epsilon _3$ (dotted line) for the same values
  of $\alpha$.}
\label{potLI}
\end{center}
\end{figure}

\subsubsection{Theoretical Justifications}
\label{subsubsec:theoryli}

The flatness of an inflationary potential is in general altered by
radiative corrections. One loop order corrections generically take the
form of a logarithmic function, $\ln(\phi/\mu)$, where $\mu$ is a
renormalization scale. Starting from a perfectly flat potential, one
obtains a potential of the form $V(\phi)=M^4\left[1 +\alpha\ln
  \left(\phi /\Mp\right)\right]$ where $\alpha$ is a dimensionless
parameter that tunes the strength of the radiative effects.  Studying
such potentials is therefore a simple way to discuss in which cases
the quantum correction ``spoil'' the flatness of a potential, and how
this happens.

In fact, this type of scenarios were invented in the context of $F$
and $D$-term inflation in
\Refcs{Binetruy:1996xj,Halyo:1996pp,Dvali:1996ub,Dvali:2001fw}. The
original motivation was to build an inflationary model in supersymmetry
but without the $\eta$-problem that appears in the $F$-term
approach. Indeed, if one considers a simple superpotential
$W=f/2X\phi^2-\mu^2X$ where $\phi $ and $X$ are two superfields, then
it is easy to obtain the supersymmetric potential assuming a minimal
K\"ahler potential: $V=\vert f\phi^2/2-\mu^2\vert^2 +f^2\vert
X\vert^2\vert \phi\vert^2$. There is a flat direction for $\phi=0$
along the $X$ direction with $V=\mu^4$. Lifting this direction with a
one loop correction leads to the LI potential which is suitable for
inflation. However, considering non-minimal term in the K\"ahler
potential destroys the flatness of $V$. The $D$-term approach was
shown to be a viable alternative. The idea is to consider a theory
with a $\mathrm{U}(1)$ symmetry and three chiral superfields, $X$,
$\phi_+$ and $\phi_-$ with charges $0$, $+1$ and $-1$ respectively. It
then follows that the superpotential has the form $W=\lambda
X\phi_+\phi_-$. If we compute the corresponding potential in global
supersymmetry, one arrives at
\begin{equation}
V=\lambda^2 \vert X\vert^2\left(\vert \phi_-\vert^2+\vert\phi_+\vert^2\right)
+\lambda^2\vert \phi_+\phi_-\vert^2+\frac{g^2}{2}\left(\vert \phi_+\vert^2
-\vert\phi_-\vert^2+\xi\right)^2,
\end{equation}
where the part proportional to $g$ ($g$ being the gauge coupling)
represents the $D$-part of $V$. In this expression $\xi$ is a
Fayet-Iliopoulos term. There is a unique supersymmetric vacuum at
$X=\phi_+=0$ and $\vert\phi_-\vert=\sqrt{\xi}$ and a flat direction
along the $X$ direction with $\phi_+=\phi_-=0$ where the potential
$V=g^2\xi^2/2$ can drive inflation. Since supersymmetry is broken
along the flat direction, this produces one loop corrections and we
obtain
\begin{equation}
\label{eq:dtermpot}
V=\frac{g^2}{2}\xi^2\left[1+\frac{g^2}{16\pi^2}
\ln\left(\frac{\lambda^2\vert X\vert^2}{\mu ^2}\right)\right],
\end{equation}
where $\mu$ is a renormalization scale. We see that this potential has
exactly the form of an LI potential where the scale $M$ is
related to the Fayet-Iliopoulos term $\xi$ and where $\alpha$ is in
fact the square of the gauge coupling. In particular, this implies
that $\alpha >0$ in this context. One can also reproduce the above
calculation in supergravity (with minimal K\"ahler potentials) and
show that the $D$-part of the theory leads to the same potential which
is free of the $\eta$ problem.

After these initial works on $D$-term inflation, many other papers
addressing different issues were published. Observational constraints
on this type of scenarios were discussed in
\Refcs{Covi:2000gx,Safsafi:2012zz}. \Refc{Matsuda:1997ys} has discussed
how to produce $D$-term inflation and to stabilize the moduli at the
same time. Then, in \Refcs{Espinosa:1998ks,
  Kolda:1998yt,Halyo:1999bq}, it was shown that the stringy
implementation of $D$-term inflation is problematic. We have seen that
the scale $M$ is essentially controlled by the value of the
Fayet-Iliopoulos term $\xi$. Therefore, the CMB normalization allows
us to calculate the value of $\xi$. Anticipating the calculation at
the end of this section, if one uses the equation after
\Eq{eq:cmbli} with $M^4=g^2\xi^2/2$ and $\alpha=g^2/(8\pi^2)$
[from \Eq{eq:dtermpot}], then one arrives at
\begin{equation}
\label{eq:xilicmb}
\xi \simeq \left[\left(\frac{90}{\Delta\Nstar}\right)^{1/4}
\left(\frac{\Qrms}{T}\right)^{1/2}\Mp\right]^2\simeq 
\left(6.9\times 10^{15}\GeV\right)^2,
\end{equation}
where we have taken the fiducial value $\Delta \Nstar\simeq 50$. As
noticed in \Refcs{Espinosa:1998ks, Kolda:1998yt,Halyo:1999bq}, in
string theory, one typically obtains $\xi=(\Tr Q)\Ms^2/(192\pi^2)$
where $\Ms$ is the string scale and $\Tr Q\simeq 100$ sums the
$\mathrm{U}(1)$ charges of all massless states. This leads to
$\xi\simeq (\mathrm{few}\times 10^{17}\GeV)^2$ and, therefore, does
not match the CMB normalization~(\ref{eq:xilicmb}). Then,
\Refcs{Suematsu:2002hj,Davis:1999tk} studied more complicated models
in the supersymmetric context in order to fix the problem we have just
discussed. Other scenarios were also investigated in
\Refcs{Urrestilla:2004eh,Lin:2006xta,Lin:2007va,Kawasaki:2003dd}. $D$-term
inflation in the context of string theory and brane inflation was also
discussed in \Refc{GomezReino:2002fs, Halyo:2004kd, Bouaouda:2010zz,
  Hebecker:2011hk,Jones:2002cv,Halyo:2003wd, Dasgupta:2004dw}. The
same topic was also addressed in
\Refcs{McDonald:2002bd,Panotopoulos:2005hc} but in the context where
the Friedmann equations receives quadratic corrections. Finally,
\Refc{Halyo:2010mq} studied LI potentials in the case of Wess--Zumino
models. Let us emphasize again that, in all these models, the constant
$\alpha$ is positive and given in terms of the square of a gauge
coupling.

The LI potential was also derived in a different framework in
\Refc{Vayonakis:1982uc}. This article uses the O'Raifeartaigh-Witten
model that will be studied in more detail in
\sectionc{sec:wri}. Therefore, we do not give the details here and
only quote results that will be reviewed in that section. In
particular, we will see in \Eq{eq:potsu5wri} that the only
difference is that the parameter $\alpha$ is now given in terms of
three coupling constants and has a rather involved form which allows
for negative $\alpha$ values. For this reason we will not fix the sign
of $\alpha$ in the following.

\subsubsection{Slow-Roll Analysis}
\label{subsubsec:srli}

Let us now turn to the slow-roll study of loop inflation. We recall 
that the potential takes the following form
\begin{equation}
\label{eq:li:pot}
V(\phi)=M^4\left[1
+\alpha\ln \left(\frac{\phi}{\Mp}\right)\right],
\end{equation}
where $\alpha$ is a dimensionless parameter, that can a priori be
either positive or negative (see the above discussion). Let us define
the quantity $x\equiv\phi/\Mp$. The potential \Eq{eq:li:pot},
as well as its logarithm, is displayed in \Fig{potLI}. If $\alpha>0$,
it is an increasing function of the field \vev, and vanishes at
\begin{equation}
\label{eq:li:xVzero}
\xVzero=\ee^{-1/\alpha}\, .
\end{equation}
Hence inflation proceeds from the right to the left at $x>\xVzero$ in
that case. If $\alpha<0$ however, the potential is a decreasing
function of the field, which vanishes at $\xVzero$, still given by
\Eq{eq:li:xVzero}, hence inflation proceeds from the left to the right
at $x<\xVzero$.

The three first Hubble flow functions in the slow-roll
approximation are given by
\begin{equation}
  \epsilon _1 = \frac{\alpha^2}{2}\frac{1}{x^2}
  \left(1
    +\alpha\ln x \right)^{-2}\, , \qquad
  \epsilon _2 = 2 \alpha \frac{1}{x^2} \dfrac{1+\alpha
    +\alpha\ln x}{\left(1
      +\alpha\ln x \right)^{2}}\, ,
\end{equation}
and
\begin{equation}
\begin{aligned}
  \epsilon_3 = &2\alpha\frac{1}{x^2} \left(1 +\alpha\ln
    x \right)^{-2}
    \left(1+\alpha
  +\alpha\ln x\right)^{-1}\times\\ &
  \left[1+\frac{3\alpha}{2}+\alpha^2
  +\left(2\alpha+\frac{3}{2}\alpha^2\right)
  \ln x +\alpha^2\ln^2 x \right].
\end{aligned}
\end{equation}
If $\alpha>0$, the first slow-roll parameter is a decreasing function
of the field \vev, which diverges at $\xVzero$ and vanishes when
$x\rightarrow\infty$. Therefore inflation stops by slow-roll
violation, at the point $\xend$ satisfying $\epsilon_1=1$ and given by
\begin{equation}
\label{eq:li:xend1}
\xend=\frac{1}{\sqrt{2}}\left[\Lambert{0}
\left(\frac{\ee^{1/\alpha}}{\sqrt{2}}\right)\right]^{-1},
\end{equation}
where $\Lambert{0}$ is the $0$-branch of the Lambert function. One can
check that since $\Lambert{0}(y)<y$ for any $y$, one always has $\xend
>\xVzero$, as required. When $\alpha\ll 1$, one has
$\xend\simeq\alpha/\sqrt{2}.$ If $\alpha<0$ on the other hand, the
first slow-roll parameter diverges at $x=0$, decreases with $x$,
reaches a minimum at $\xepstwoZero=\exp\left(-1-1/\alpha\right)$, then
increases with $x$ and diverges at $\xVzero$. The minimum value of
$\epsilon_1$ equals
$\epsilon_1\left(\xepstwoZero\right)=\exp(2+2/\alpha)/2$ which is
smaller than unity only if $\alpha>2/(\ln 2-2)\simeq-1.53$. Otherwise
$\epsilon_1(x)>1$ all over the domain and inflation cannot take
place. If $\alpha>2/(\ln 2-2)$, the inflationary domain lies between
$\xepsoneOneMinus$ and $\xend=\xepsoneOnePlus$, with
\begin{equation}
\xepsoneOneMinus=-\frac{1}{\sqrt{2}}\left[\Lambert{-1}
\left(\frac{-\ee^{1/\alpha}}{\sqrt{2}}\right)\right]^{-1},
\qquad 
\xend=\xepsoneOnePlus=-\frac{1}{\sqrt{2}}\left[\Lambert{0}
\left(\frac{-\ee^{1/\alpha}}{\sqrt{2}}\right)\right]^{-1} ,
\label{eq:li:xend2}
\end{equation}
and where $\Lambert{-1}$ is the $-1$-branch of the Lambert
function. When $\vert\alpha\vert\ll 1$, one has
$\xend\simeq\ee^{-1/\alpha}-1/\sqrt{2}\gg 1$. Let us notice that the
end of inflation occurs in the region $\phi\gg\Mp$, where
\Eq{eq:li:pot} may not be well defined. Therefore, depending on the
underlying theoretical setting, the end of inflation by slow-roll
violation may not be meaningful.

\begin{figure}
\begin{center}
\includegraphics[width=\wdblefig]{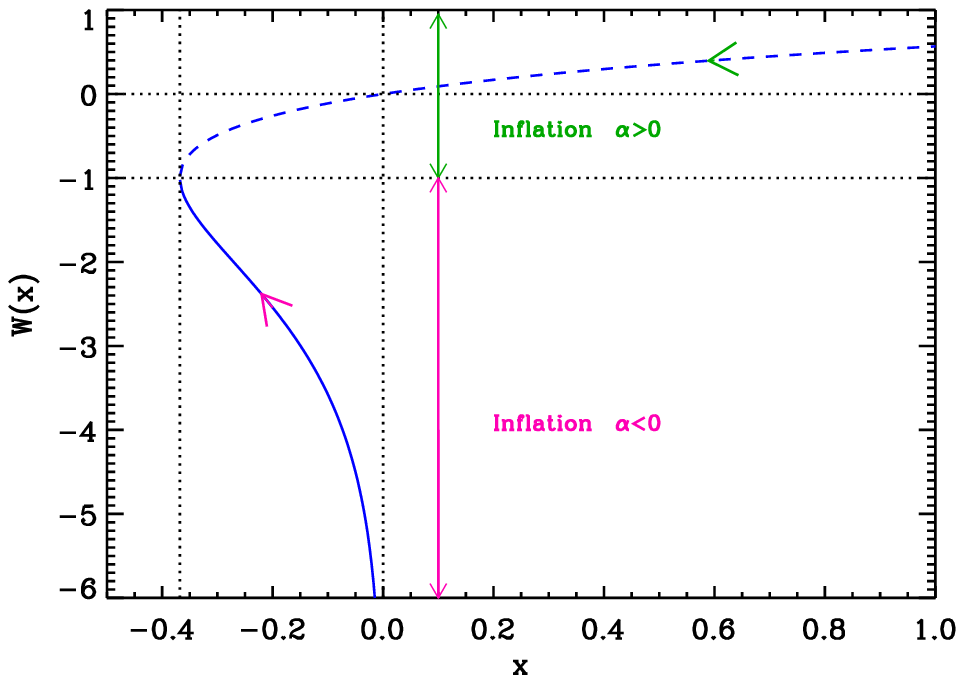}
\includegraphics[width=\wdblefig]{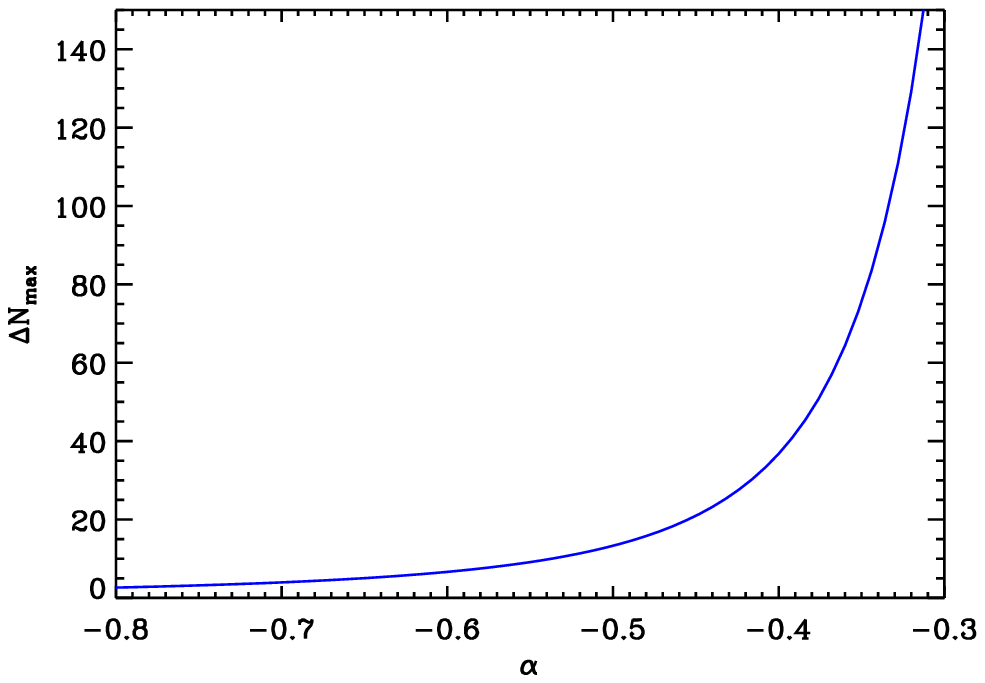}
\caption{Left panel: Lambert functions $\Lambert0(x)$ (dashed line) and
  $\Lambert{-1}(x)$ (solid line). During loop inflation, inflation
  proceeds along the ``$0$'' branch in the direction specified by the
  green arrow on the figure if $\alpha>0$, and along the ``$-1$'' branch in 
  the direction specified by the pink arrow on the figure if $\alpha<0$.
  Right panel: Maximal number of $e$-folds $\Delta\Nmax$ one can realize when $\alpha<0$,
  between $\xepsoneOneMinus$ and $\xepsoneOnePlus$, as a function of $\alpha$. }
\label{fig:li:LambertAndAlphamin}
\end{center}
\end{figure}

Let us now turn to the slow-roll trajectory. It can be integrated, giving 
rise to
\begin{equation}
\label{eq:li:traj}
\Nend-N=\frac{x^2}{2}\left(\ln x +\frac{1}{\alpha}-\frac{1}
{2}\right)-\frac{\xend^2}{2}\left(\ln \xend +\frac{1}
{\alpha}-\frac{1}{2}\right).
\end{equation}
When $\vert\alpha\vert\ll 1$, it approximately takes the form 
$2\alpha\left(\Nend-N\right)=x^2-\xend^2$. The trajectory \Eq{eq:li:traj} 
can be inverted making use of the Lambert function, and one obtains
\begin{equation}
\begin{aligned}
\label{eq:trajecsusyloop}
x^2&= \dfrac{
  4\left(\Nend-N\right)-\xend^2\left[1-\dfrac{2}{\alpha}
    -\ln\left(\xend^2\right)\right]}
{\Lambert{{}^{\ 0}_{-1}}\Biggl\{4\left(\Nend-
N\right)\ee^{-\left(1-2/\alpha\right)}
  -\left[1-\dfrac{2}{\alpha}
    -\ln\left(\xend^2\right)\right]
  \exp\left[-1+\dfrac{2}{\alpha}+\ln\left(\xend^2
    \right)\right]\Biggr\}}\, ,
\end{aligned}
\end{equation}
where the $0$ branch of the Lambert function must be chosen if
$\alpha>0$, while the $-1$ branch must be chosen if $\alpha<0$. The
Lambert function is displayed in the left panel of
\Fig{fig:li:LambertAndAlphamin}, together with the regions in which
inflation proceeds. Let us now comment and check that this expression
is valid. Firstly, if $N=\Nend$, the Lambert function is of the form
$\Lambert{}(-\zend\ee ^{-\zend})=-\zend$, where $z\equiv
(1-2/\alpha)-\ln(x^2)$, and this automatically cancels the numerator
such that one has indeed $x=\xend$. Secondly, if $\alpha>0$, the
condition $\xend>\xVzero$ implies that $\zend<1$, and the Lambert
function at $\Nend$ is equal to $-\zend>-1$. Therefore, at the end of
inflation, one should use the zero branch of the Lambert
function. Finally, as inflation is under way, the argument of the
Lambert function is decreasing which implies that the whole
inflationary stage takes place on the zero branch. On the other hand,
if $\alpha<0$ using similar arguments, the whole inflationary stage
can be shown to take place on the $-1$ branch.

In this later case ($\alpha<0$), it is also interesting to notice that
the total number of $e$-folds is bounded, since inflation can only
proceed between $\xepsoneOneMinus$ and $\xepsoneOnePlus$. The
corresponding maximal number of $e$-folds $\Delta \Nmax$ is displayed,
as a function of $\alpha$, in the right panel of
\Fig{fig:li:LambertAndAlphamin}. One can see that when
$\alpha\lesssim-0.35$, not a sufficient number of $e$-folds can be
realized. For such values of $\alpha$, one already has
$\xend>10$. Since inflation is supposed to take place at sub-Planckian
\vevs, it means that this regime of inflation is a priori
forbidden. If one allows slightly super-Planckian field \vevs, up to
$x\simeq 100$ or $x\simeq 1000$, this implies that
$\alpha<-0.1$. Therefore even in this case, $\alpha$ must lie in the
rather narrow range $-0.3<\alpha<-0.1$.

Making use of the approximated trajectories and expressions for
$\xend$, some analytic predictions can be derived in the case
$\alpha>0$. The observable field value $\xstar$, and its associated
number of \efolds $\Delta \Nstar = \Nend - \Nstar$ at which the
pivot mode crossed the Hubble radius during inflation are obtained
from the above equations together with \Eq{eq:phistarlnrrad}. In the
limit $\alpha\ll 1$, one obtains the approximate expressions
\begin{equation}
\label{eq:li:predic}
\epsilon_{1*}\simeq\frac{\alpha}{4\Delta\Nstar}\, ,
\qquad
\epsilon_{2*}\simeq\epsilon_{3*}\simeq\frac{1}{\Delta\Nstar}\, ,
\end{equation}
hence
\begin{equation}
r\simeq\frac{\alpha}{64\Delta\Nstar}\, ,
\qquad
\nS-1\simeq-\frac{1}{\Delta\Nstar}\, ,
\qquad
\alphaS\simeq\frac{1}{\Delta\Nstar^2}\, .
\end{equation}
Finally, the parameter $M$ can be determined from the amplitude of the 
CMB anisotropies, and one gets
\begin{equation}
\label{eq:cmbli}
\left(\frac{M}{\Mp}\right)^4=720\pi^2\frac{\alpha^2}{\xstar^2}
\frac{\Qrms^2}{T^2}\left(1+\alpha\ln \xstar \right)^{-3}\, .
\end{equation}
In the small $\vert\alpha\vert$ limit, one obtains $M^4/\Mp^4\simeq
360\pi^2\alpha/\Delta\Nstar\Qrms^2/T^2$ for $\alpha>0$, and
$M^4/\Mp^4\simeq 720\pi^2\alpha^2\ee^{2/\alpha}\Qrms^2/T^2$ for
negative values of $\alpha$.

The reheating consistent slow-roll predictions of the loop inflation
models are displayed in \Fig{fig:CMBLIalphaPositive} for $\alpha>0$,
and in \Fig{fig:CMBLIalphaNegative} for $\alpha<0$. For $\alpha>0$ and
$\alpha\ll 1$, the approximations in \Eqs{eq:li:predic} give a good
description of what is numerically obtained, namely a deviation from
scale invariance which almost does not depend on $\alpha$, and an
amount of gravitational waves which grows linearly with $\alpha$. For
$\alpha<0$, the predictions blow out of the observational one- and
two-sigma contours when $\alpha$ approaches the upper bound derived
above, as expected.  Correspondingly, the parameter $\alpha$ does not
seem to be much constrained when it is positive, whereas close-to-zero
values are favored when it is negative.

\subsection{\texorpdfstring{$(R+R^{2p})$}{RpI} Inflation (RpI)}
\label{sec:rpi}

This model is the Einstein frame description of a scalar-tensor theory
equivalent to $f(R)=R+\epsilon R^{2p}/\mu^{4p-2}$, where $\mu$ is a
mass scale, $\epsilon=\pm 1$, and $p>1/2$ (otherwise the expansion is
meaningless). It generalizes the original Starobinsky
model~\cite{Starobinsky:1980te} obtained for $p=1$. Such theories are
quite generic and appear as limiting cases of more general modified
gravity theories~\cite{Stelle:1977ry, Teyssandier:1983zz,
  Maeda:1988ab, Wands:1993uu, DeFelice:2011jm} (see
\Refc{DeFelice:2010aj} for a review).

Following \Refcs{Maeda:1988ab, DeFelice:2010aj}, one can introduce the
scalar degree of freedom $\phi$ defined by
\begin{equation}
\label{eq:rpi:scalarfielddef}
\frac{\phi}{\Mg}=\sqrt{\frac{3}{2}} \ln\left(\left| F(R) \right| \right) \,,
\end{equation}
where $F(R) \equiv \partial f/\partial R$. For the sake of clarify, we
identify the Lagrange multiplier field $\chi$ with its on shell value
$\chi=R$ and drop the ``tilde'' over Einstein frame quantities, see
\sectionc{sec:theorysi} for a detailed discussion of this class of
models.

The quantity $F\equiv\Omega^2$ is also the square of the conformal
factor inducing the transformation from the Jordan frame to the
Einstein frame. In the Einstein frame, the field $\phi$ evolves in a
potential given by
\begin{equation}
  V\left(\phi\right)= \dfrac{\Mg^2}{2} \dfrac{|F|}{F} \dfrac{R F - f}{F^2}\,.
\label{eq:potrpiR}
\end{equation}
In the present case, one has
\begin{equation}
F(R) = 1 + 2 \epsilon p  \left(\dfrac{R}{\mu^{2}}\right)^{2p -1}\,,
\label{eq:Frpi}
\end{equation}
which, for small departures with respect to the Einstein-Hilbert
action $R\ll \mu^2$, implies that $F(R)>0$ as needed. Let us notice
that in the opposite situation, accelerated (and super-accelerated)
solutions have been shown to exist~\cite{DeFelice:2010aj}. Defining
the quantity $y$ by
\begin{equation}
  y \equiv \sqrt{\dfrac{2}{3}} \dfrac{\phi}{\Mg}\,,
\label{eq:rpidefy}
\end{equation}
and inserting \Eq{eq:Frpi} into \Eq{eq:potrpiR} one obtains the
Einstein frame potential
\begin{equation}
V = M^4 \ee^{-2 y} \left|\ee^y - 1 \right|^{2p/(2p-1)}\,.
\end{equation}
The normalization constant $M^4$ is related to the modified gravity
scale $\mu$ through the following expression
\begin{equation}
M^4 = \dfrac{2p -1}{4p} \dfrac{\Mg^2 \mu^2}{(2 p)^{1/(2p-1)}}\,.
\end{equation}
For $F(R)>0$, \Eq{eq:rpi:scalarfielddef} implies that for
$\epsilon=1$, the model is defined in the domain $y>0$, whereas for
$\epsilon=-1$ one should consider the domain $y<0$ only. Such a
potential has also been studied in \Refc{Kofman:1985aw} for $p=1$, in
\Refcs{Maeda:1988ab, Kaneda:2010qv} for $p=4$ and in
\Refc{Ketov:2010qz} for $p=2$. Let us notice that the case $p=1$
corresponds to the Higgs inflation potential studied
in \sectionc{sec:hi}. The case $p=1/2$ is singular since one recovers
$f(R) \propto R$. Taking the limit $p\rightarrow \infty$, the
potential asymptotes $V \rightarrow M^4 \ee^{-2y}
\left|\ee^{y}-1\right|$ and varying $p$ allows us to explore different
potential shapes.

Let us first consider the case $y>0$ ($\epsilon=1$). If $p>1$, the potential admits
a maximum at
\begin{equation}
 \ymax = \ln\left(\frac{2p-1}{p-1} \right),
\end{equation}
such that inflation can proceed either for $0<y<\ymax$ or
$y>\ymax$. We respectively call these regimes RpI1 and RpI2.  If
$p<1$, the potential is an increasing function of $y$, hence inflation
proceeds from the right to the left. We call this regime RpI3.  The
case $p=1$ is singular and again, it corresponds to the Higgs
inflation potential studied in
\sectionc{sec:hi}.

The Hubble flow functions in the slow-roll approximation
read
\begin{equation}
  \epsilon_1 =\frac{4}{3}
  \frac{\left[1+\left(p-1\right)\ee^{y}-2p\right]^2}
  {\left(2p-1\right)^2\left(\ee^y-1\right)^2} \, , \qquad
  \epsilon_2 = \frac{8}{3}\dfrac{p \, \ee^y}{(2p-1)\left(\ee^y-1\right)^2}\,,
\label{eq:rpieps12}
\end{equation}
and
\begin{equation}
  \epsilon_3 =-\frac{4}{3}\frac{\left(\ee^y+1\right)
    \left[1+\left(p-1\right)\ee^y-2p\right]}
  {\left(2p-1\right)\left(\ee^y-1\right)^2} \,.
\label{eq:rpieps3}
\end{equation}

\begin{figure}
\begin{center}
\includegraphics[width=\wdblefig]{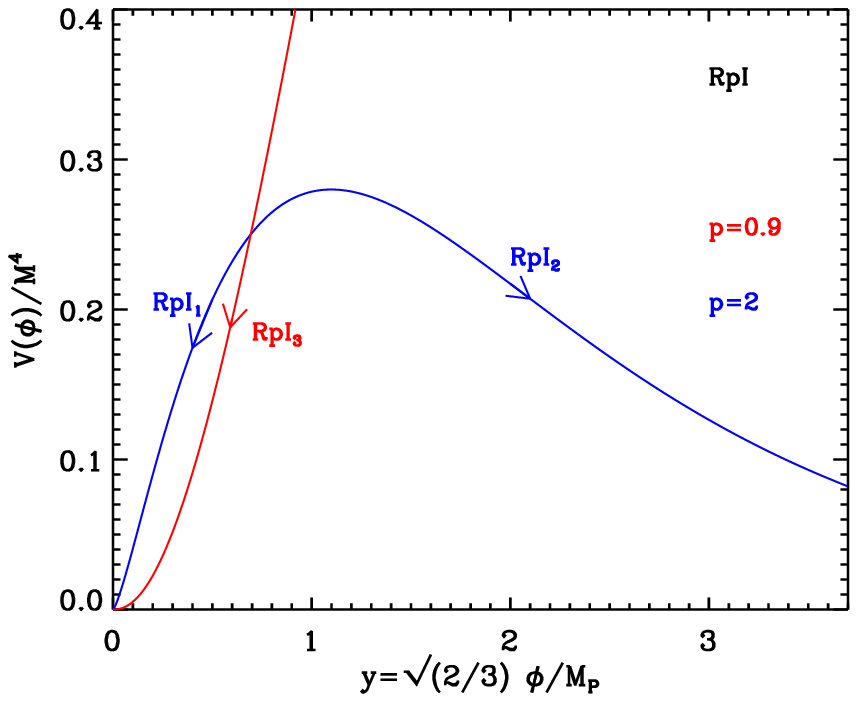}
\includegraphics[width=\wdblefig]{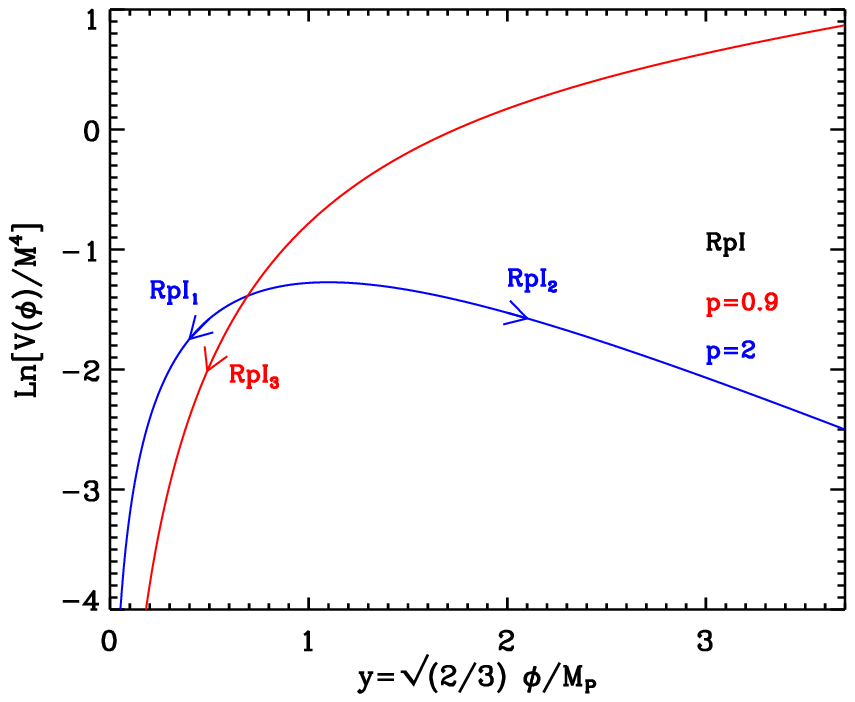}
\includegraphics[width=\wdblefig]{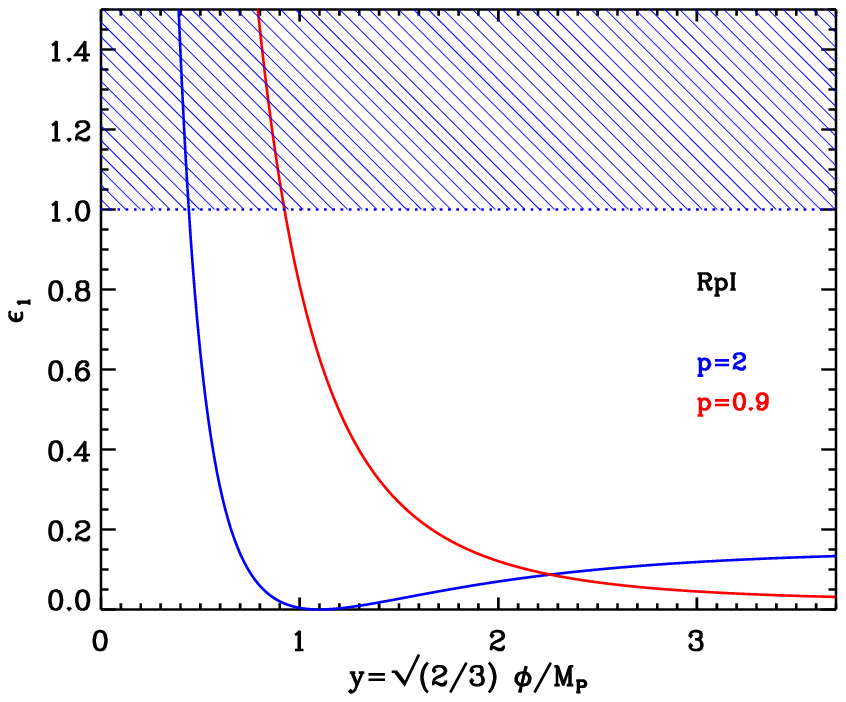}
\includegraphics[width=\wdblefig]{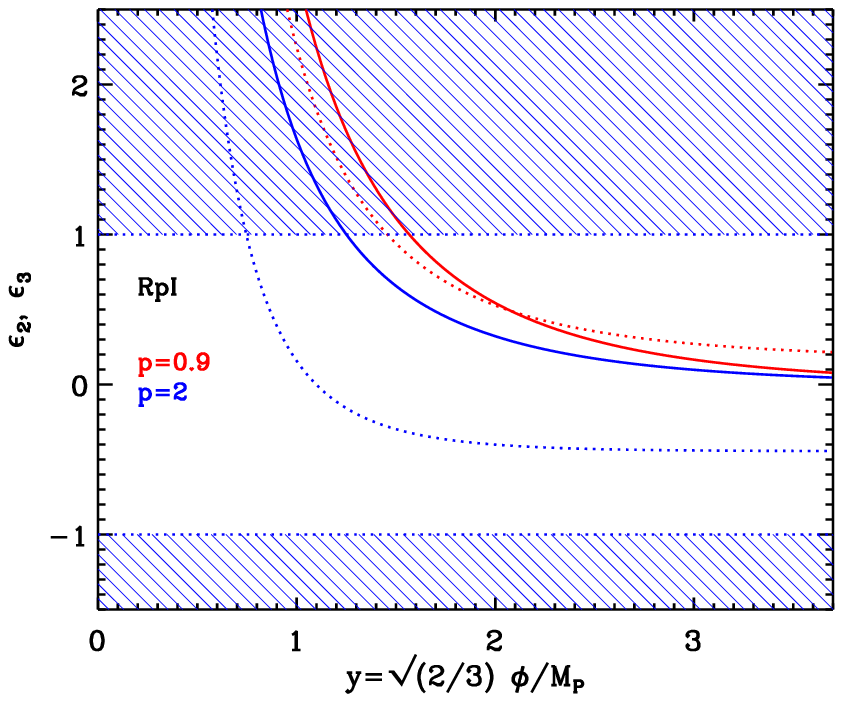}
\caption{$(R+R^{2p})$ Inflation (RpI) in the Einstein frame for $p=2$
  (RpI1 and RpI2), and $p=0.9$ (RpI3) (assuming $\Mg \simeq \Mp$). Top
  panels: the potential and its logarithm.  Bottom left panel:
  slow-roll parameter $\epsilon_1$ with the region in which inflation
  stops (shaded area). In the RpI2 regime, inflation never stops and
  one has to consider an extra-mechanism to end inflation. For this
  special case, one does not longer have $\Mg \simeq \Mp$ and $y$ is
  defined by \Eq{eq:rpidefy}. Bottom right panel: slow-roll
  parameters $\epsilon_2$ (solid line) and $\epsilon_3$ (dotted
  line).}
\label{potrpi}
\end{center}
\end{figure}

The potential and the Hubble flow functions for $y>0$ have been
represented in \Fig{potrpi}. As one can check on these figures,
inflation never stops in the RpI2 regime and one needs to complement
the model with a mechanism that can end inflation, as for instance
with an extra-field and a tachyonic instability.  This adds one
additional parameter $\yend$ to the model.  When this parameter is
large, all the three Hubble flow functions admit asymptotically
constant values:
\begin{equation}
\label{ref:eq:rpi:epsAsympt}
\lim_{y \rightarrow \infty} \epsilon_1 = \dfrac{4}{3}
\left(\dfrac{p-1}{2p-1}\right)^2, \qquad \lim_{y \rightarrow \infty}
\epsilon_2 = 0, \qquad \lim_{y \rightarrow \infty}
\epsilon_3 = -\frac{4}{3}\frac{p-1}{2p-1}.
\end{equation}
If $p$ is an integer, except for the special case $p=1$
(see \sectionc{sec:hi}), these values are always smaller that unity,
but not particularly small. As such, all these models predict large
deviation from scale invariance. Indeed, the spectral index at first
order is given by
\begin{equation}
\nS - 1 \simeq -\dfrac{8}{3} \left(\dfrac{p-1}{2p-1} \right)^2,
\end{equation}
which, for $p \ge 2$, remains always smaller than $-8/27\simeq -0.3$. This is
strongly disfavored by current CMB measurements. Therefore, only the
models such that $p$ is close enough to $1$ are to be considered (\ie
non integer values of $p$.)

If inflation proceeds in the RpI1 regime, then inflation stops
naturally when $\epsilon_1=1$, \ie at the field value
\begin{equation}
  \yend = \ln\left[ (2p-1) \dfrac{1 + 2 p(\sqrt{3}+1)}{8 p^2 - 4p -1} \right].
\label{eq:rpiyend}
\end{equation}
However, the second Hubble flow function can only take relatively
large value. From \Eq{eq:rpieps12}, since $y<\ymax$, one gets
\begin{equation}
  \epsilon_2> \epsilon_2(\ymax) = \frac{8}{3}\frac{p-1}{p}\,.
\end{equation}
For $p \ge 2$, we are in a situation where $\epsilon_2>4/3$ and again,
the models are ruled out by a simple slow roll analysis. Therefore, as
already noticed before, $p$ must take (non integer) close enough to
$1$ values for the models to be viable.

Finally, in the RpI3 regime, inflation stops naturally when
$\epsilon_1=1$, with $\yend$ still given by \Eq{eq:rpiyend}.  This
expression is defined only if $p>(1+\sqrt{3})/2\simeq 0.68$ but the
first slow roll parameter continuously decreases with $y$, and its
asymptotic value is again given by
\Eq{ref:eq:rpi:epsAsympt}. Therefore, this regime is viable only when
$p$ is close enough to unity.

Let us now turn to the slow-roll trajectory. It is given by
\begin{equation}
 N-\Nend=\frac{3}{4}\left\{\frac{p}{p-1} \ln\left[
\dfrac{(p-1) \ee^y + 1 - 2 p}{(p-1)\ee^{\yend} +1 - 2 p} \right]
+y-\yend\right\}.
\label{eq:rpi:traj}
\end{equation}
This expression is not properly defined for $p=1$ but this case has
already been considered in the section on the Higgs inflation
model. When $p>1$, if $y=\ymax$, the argument of the logarithm
vanishes and the total number of \efolds diverges. As a result,
provided inflation starts close enough to the top of the potential, it
is always possible to find a long enough inflationary period. For
$p<1$, the number of $\ee$-folds diverges when $y\rightarrow
\infty$. The slow-roll trajectory cannot be analytically inverted, but
using the same reheating model as in
\sectionc{sec:hi}, one can solve for the field value $\ystar$ at which
the pivot mode crossed out the Hubble radius. The corresponding number
of \efold $\Delta \Nstar= \Nend - \Nstar$ being given by
\Eq{eq:rpi:traj}.

Concerning the case $\epsilon=-1$, \ie the domain $y<0$, all of the
previous formula still apply but the potential is now a monotonic
decreasing function of the field \vev which is too steep to support
inflation. In particular, over the whole negative domain,
\Eq{eq:rpieps12} implies that $\epsilon_1(y<0) >
\epsilon_1(y\rightarrow -\infty)=4/3$, independently on whether $p>1$
or $p<1$.

Finally, the constant $M$ can be determined from the amplitude of the
CMB anisotropies. It follows that
\begin{equation}
  \frac{M^4}{\Mg^4}=1920\pi^2\frac{\left[1+\left(p - 1\right)
      \ee^{\ystar} - 2p\right]^2
    \ee^{2\ystar}}{\left(2p-1\right)^2\left(\ee^{\ystar} - 1
    \right)^{\frac{6p-2}{2p-1}}}\frac{\Qrms^2}{T^2}\, .
\end{equation}

The reheating consistent slow-roll predictions of the RpI models are
displayed in \Fig{fig:CMBRpI1} for the RpI1 regime, in
\Fig{fig:CMBRpI2} for the RpI2 regime, and in \Fig{fig:CMBRpI3} for
the RpI3 regime. In the RpI1 regime, the Higgs inflation model
predictions (see \Fig{fig:CMBHI}) are recovered when $p\rightarrow 1$,
and one can see that $p<1.02$ is a necessary condition for the
spectral index not to be too red. For RpI2 the limit $p \rightarrow 1$
is such that one does not reproduce the Higgs inflation results and
for $\yend\rightarrow\infty$ the predictions lie on the line
$\epsilon_{2*} = 0$. Moreover, one can see that when $p>1.1$, the
models predict too much gravity waves to be compatible with the CMB
data. Let us stress that since $\yend$ is not necessarily small for
RpI2, we are in a situation where the numerical value of $\Mg$ may be
significantly different than $\Mp$ (see
\sectionc{subsubsec:theoryhi}). Finally for the RpI3 regimes, the Higgs
inflation model predictions (see \Fig{fig:CMBHI}) are recovered when
$p\rightarrow 1$, and they remain compatible with the data within the
two-sigma contours provided $p>0.99$.

\subsection{Double-Well Inflation (DWI)}
\label{sec:dwi}

\begin{figure}
\begin{center}
\includegraphics[width=\wdblefig]{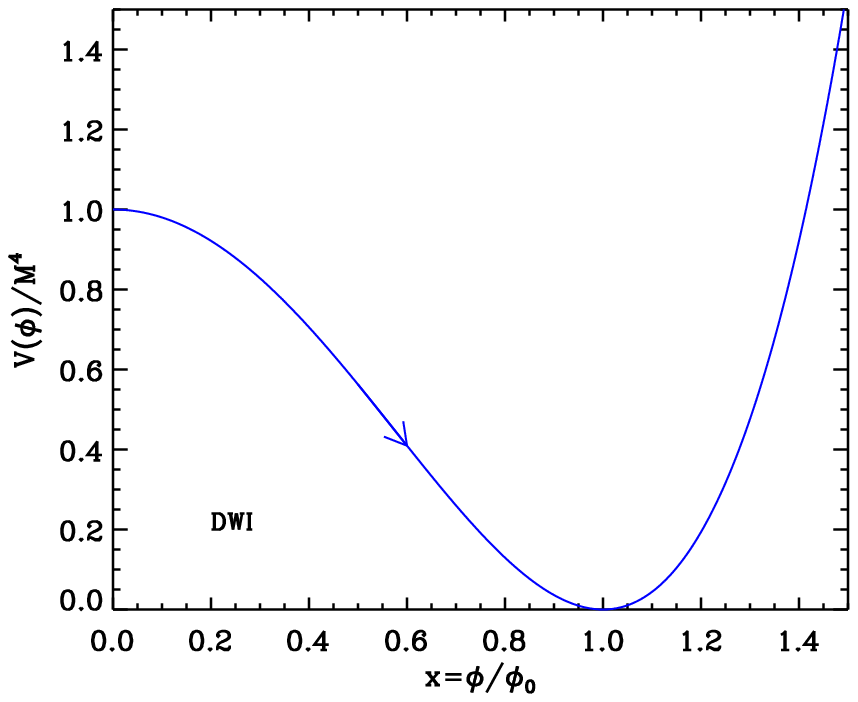}
\includegraphics[width=\wdblefig]{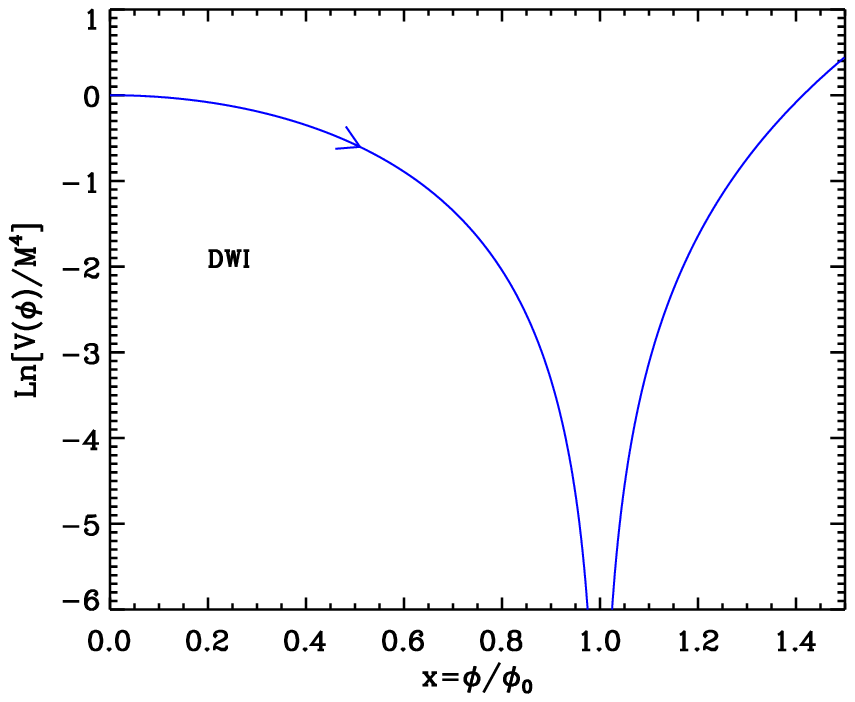}
\includegraphics[width=\wdblefig]{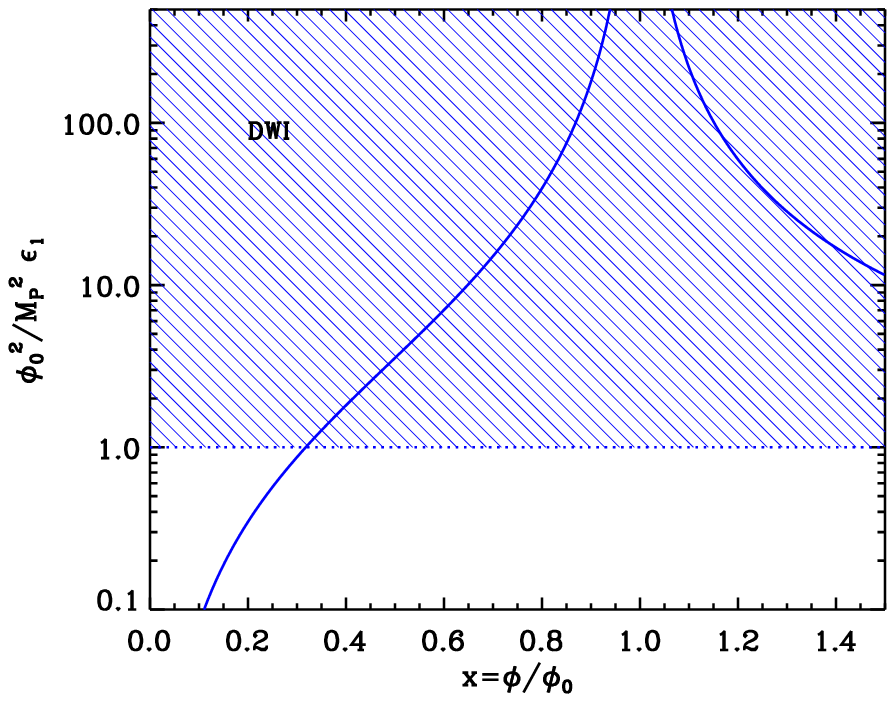}
\includegraphics[width=\wdblefig]{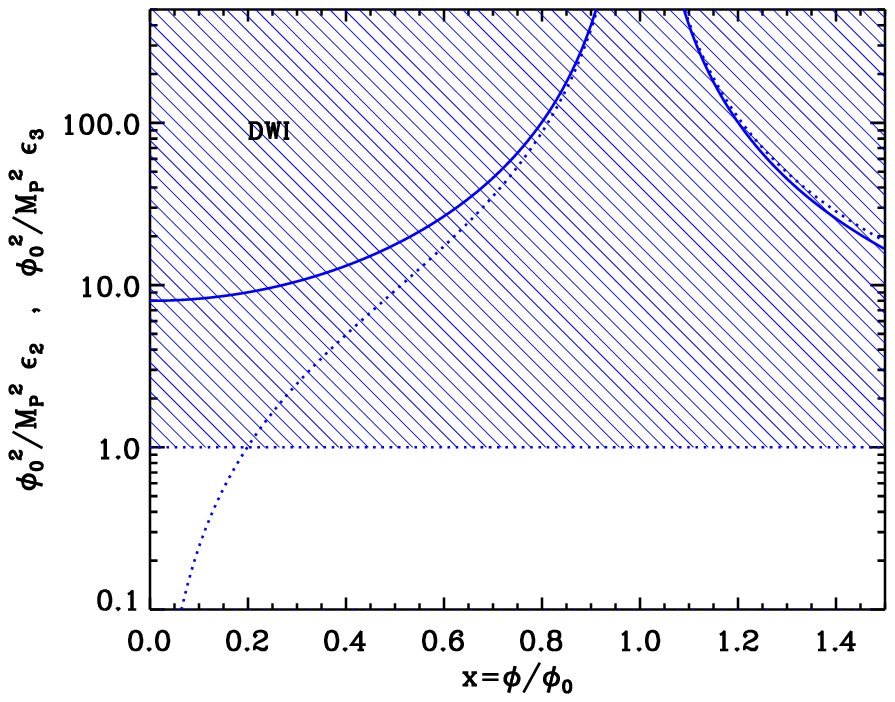}
\caption{Top left panel: Double Well Inflation (DWI) potential as a
  function of $\phi/\phizero$. Only the $\phi>0$ region is displayed
  since the potential is symmetric under $\phi\rightarrow -\phi$.  Top
  right panel: logarithm of the potential. The arrow indicates in
  which direction inflation can proceed. Bottom left panel: slow-roll
  parameter $\epsilon _1$, rescaled by the quantity
  $\Mp^2/\phizero^2$, such that the corresponding expression becomes
  universal, \ie independent of $\phizero$. Bottom right panel:
  slow-roll parameters $\epsilon _2$ (solid line) and $\epsilon _3$
  (dotted line), rescaled by $\Mp^2/\phizero^2$ for the same reason as
  mentioned before.}
\label{potdwi}
\end{center}
\end{figure}

In this section, we study the famous ``Mexican hat'' potential given
by
\begin{equation}
\label{eq:potdwi}
V(\phi)=M^4\left[\left(\frac{\phi}{\phizero}\right)^2-1\right]^2 .
\end{equation}
Except for the mass $M$ determined by the CMB normalization, it
depends on one parameter, the \vev $\phizero$. Historically, this
potential was first introduced by Goldstone in \Refc{Goldstone:1961eq}
as a toy model for dynamical symmetry breaking. In cosmology, it is of
course utilized to investigate the formation and the microscopic
structure of topological defects~\cite{Witten:1984eb, Peter:1993tm,
  Carter:1994hn, Peter:1995ks, Peter:2000sw, Ringeval:2000kz,
  Ringeval:2001xd}. In the context of inflation, it was first used to
construct scenarios of topological
inflation~\cite{Linde:1994wt,Vilenkin:1994pv}. In this case, it is
made use of the fact that the discrete $\mathbb{Z}_2$ symmetry,
$\phi\rightarrow -\phi$, makes the state $\phi=0$ unstable. Therefore,
the Universe will split into two different regions separated by a
domain wall. One can then show that inflation takes place within this
topological defect. More precisely, the potential is usually written
as $V=\lambda/4\left(\phi^2-\eta^2\right)^2$ where $\eta$ represents
the position of the minima of the potential. Then,
\Refcs{Linde:1994wt,Vilenkin:1994pv} show that topological inflation
occurs if $\eta>\Mp$. On the other hand, if one writes
\Eq{eq:potdwi} as
$V=M^4/\phizero^4\left(\phi^2-\phizero^2\right)^2$, one sees that one
can identify $\eta$ with $\phizero$. And we will precisely show that
agreement with the CMB observations requires $\phizero>\Mp$. The
potential~(\ref{eq:potdwi}) was also used in
\Refcs{Green:1996xe,GarciaBellido:1996gd} in the context of open
inflation. In a rather different theoretical framework, \Eq{eq:potdwi}
was studied in \Refcs{Linde:1983fq,Linde:1984cd} where it was derived
in $N=1$ supergravity coupled to matter. It is also interesting to
notice that it was obtained using various stringy constructions as
early as the $80$'s, see \Refcs{Casas:1988du,Casas:1988pa}. More
recently, this potential was found to be relevant in a large number of
different physical situations~\cite{CervantesCota:1995tz,
  Alexander:2001ks, Easther:2004qs, Gong:2006hf, Kallosh:2007wm,
  Lazarides:2007ii, Rehman:2008qs, Rehman:2010es, Bauer:2010jg,
  Barvinsky:2010yb, Barenboim:2008ds, Kallosh:2010ug}. The same potential was also obtained in the context of M-flation, see
\Refcs{Ashoorioon:2009wa, Ashoorioon:2009sr, Ashoorioon:2011ki}. Let us finally
mention that this model is sometimes viewed as a realistic version of
Small Field Inflation (SFI) with $p=2$ (see \sectionc{sec:sfi}), the
extra quartic term preventing the potential from becoming negative.
However, as will be shown in the following, these two classes of
models should actually be described separately since their predictions
differ in the relevant range of parameters.

The parameter $\phizero$ sets the typical \vev at which inflation
proceeds and depends on the symmetry breaking scale one considers. In
principle, it could vary over a wide range of values, from $\phizero\sim
10^ {15}\, \GeV$ for GUT symmetry breaking schemes to super-Planckian
\vev in a stringy or supergravity context. As will be shown in the
following, it is in fact constrained to be large (super-Planckian) in
order for the predictions of the model to be compatible with the CMB
data.  The DWI potential is displayed in \Fig{potdwi} together with
its logarithm. One has represented the region $\phi>0$ only because
the potential is symmetric under $\phi\rightarrow -\phi$. We see that
it decreases for $\phi<\phizero$, vanishes at $\phizero$ and then
increases for $\phi>\phizero$. As was already mentioned before, this
potential is used to describe dynamical symmetry breaking and, as a
consequence, inflation should proceed from the left to the right at
$\phi<\phizero$, in the direction specified by the arrow in
\Fig{potdwi}.

Let us now calculate the slow-roll parameters. If one defines $x
\equiv \phi/\phizero$ they are given by
\begin{equation}
  \epsilon_1 =\left( \frac{\Mp}{\phizero} \right)^2 \frac{8 x^2}{
    \left(x^2-1\right)^2 }\,
  , \qquad 
  \epsilon_2 =\left( \frac{\Mp}{\phizero} \right)^2 \frac{8(1+x^2)}{
    \left(x^2-1\right)^2 }\, ,\qquad
  \epsilon_3 = \left( \frac{\Mp}{\phizero} \right)^2 \frac{8(x^4+3x^2)}{
    \left(x^2-1\right)^2\left(x^2+1\right) }\, .
\end{equation}
The behavior of these parameters is represented in \Fig{potdwi}. The
first slow-roll parameter $\epsilon_1$ is an increasing function of
$\phi$ in the range $x\in [0,1]$. It vanishes at $x=0$ and blows up at
$x=1$. Then, for $x>1$, it becomes a decreasing function going to zero
when $x$ goes to infinity. We see in \Fig{potdwi} that inflation stops
by violation of the slow-roll conditions. The slow roll parameters
$\epsilon_2$ and $\epsilon_3$ have similar behaviors, except that
$\epsilon_2$ does not vanish when $x=0$ but is equal to
$\epsilon_2(x=0)=8\left(\Mp/\phizero\right)^2$. Therefore, in order
for slow-roll to be valid, this last value should be less than one,
which amounts to
\begin{equation}
\label{eq:lowerlimitdwi}
\frac{\phizero}{\Mp}>2\sqrt{2}\, .
\end{equation}
This constraint on the parameter $\phizero$ shows that the symmetry
breaking scale needs to be super-Planckian. If this last condition is
verified, then $\epsilon_2$ becomes greater than one during inflation
at $\phiepstwoOne$ defined by
\begin{equation}
\xepstwoOne=\sqrt{1+4\left(\frac{\Mp}{\phizero}\right)^2
\left[1-\sqrt{1+\left(\frac{\phizero}{\Mp}\right)^2}\right]}\, .
\end{equation}
This happens before the end of inflation ($\epsilon_1=1$) which occurs
at the following value of the field
\begin{equation}
\xend=\sqrt{2+\left(\frac{\phizero}{\Mp}\right)^2}-\sqrt{2}\, .
\end{equation}

\begin{figure}
\begin{center}
\includegraphics[width=\wsingfig]{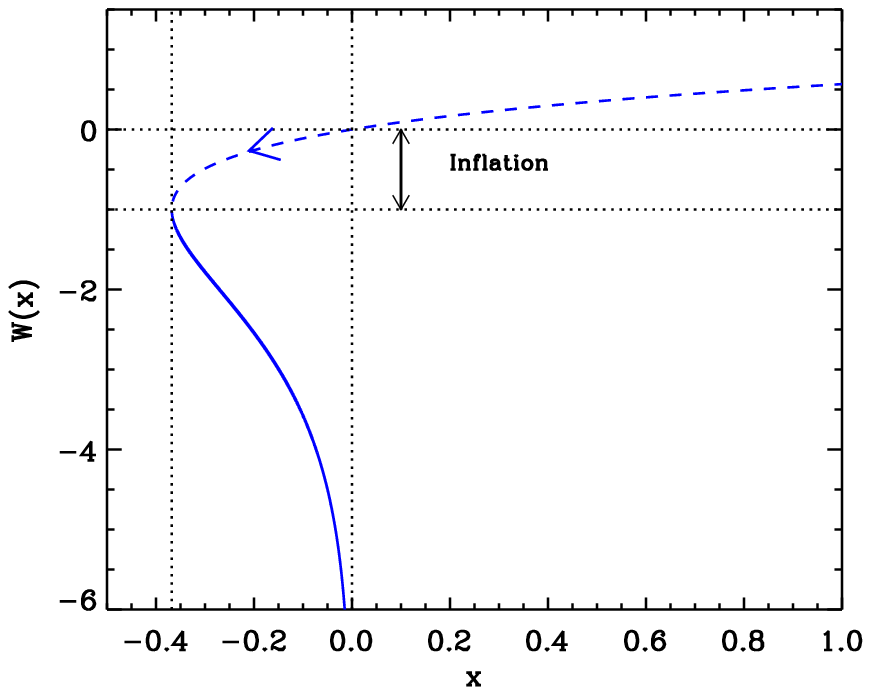}
\caption{Lambert functions $\Lambert0(x)$ (dashed line) and
  $\Lambert{-1}(x)$ (solid line). In DWI, inflation proceeds along
  the negative part of the ``$0$'' branch in the direction specified
  by the arrow.}
\label{plotlambertDWI}
\end{center}
\end{figure}

Let us now turn to the slow-roll trajectory. It can be integrated
exactly and yields the following formula
\begin{equation}
\Nend -N=\frac{1}{4}\left(\frac{\phizero}{\Mp}\right)^2\left[
\ln\left(\frac{\xend}{x}\right)-\frac{1}{2}\left(\xend^2-x^2\right)\right],
\end{equation}
where $\Nend$ is the number of \efolds at the end of
inflation. Using the $0$-branch of the Lambert function $\Lambert{0}$,
this trajectory can be inverted. One obtains
\begin{equation}
  x=\sqrt{-\Lambert{0} \left[-\xend^2\ee^{-\xend^2}
      \ee^{8\left(\frac{\Mp}{\phizero}\right)^2\left(N-\Nend\right)} \right] }\, .
      \label{eq:dwi:traj}
\end{equation}
The fact that the $0$-branch of the Lambert function should be chosen
comes from the requirement that $x<1$. The corresponding
``trajectory'' along the Lambert curve is displayed in
\Fig{plotlambertDWI}, the arrow indicating in which direction
inflation proceeds. This trajectory is remarkably similar to the one
of SFI with $p=2$, see \sectionc{sec:sfi} and \Eq{eq:sfip2traj}, the
only difference being that the factor $8$ in front of $N-\Nend$ is
just $4$ in the case of SFI. Therefore not only these two potentials
coincide at small fields, but they also give rise to the same kind of
slow-roll trajectory. This is why these two models are sometimes
identified, DWI being considered as a realistic realization of
SFI. However, as shown below, the observations favors super-Planckian
values of $\phizero$ and, in this limit, the two models are not
equivalent (of course, this also has something to do with the debate
about whether having super-Planckian \vev is meaningful or not). In
fact, in the regime $\phizero/\Mp\gg 1$, one can write
\begin{equation}
  \xstar\simeq 1-\sqrt{2}\frac{\Mp}{\phizero} \sqrt{1+2\Delta\Nstar}
  +\frac{1}{3}\left(\frac{\Mp}{\phizero}\right)^2 \left(1+2\Delta\Nstar
    + \frac{2}{\sqrt{1+2\Delta\Nstar}}\right)+\ldots\, .
\end{equation}
{}From this expression it is clear that, for super-Planckian values of 
$\phizero$, $\phistar$ 
is close to the minimum of the potential where the quartic term 
plays an important role and, consequently, where the SFI potential 
is not a good approximation. A calculation of the Hubble flow parameters 
at Hubble crossing confirms this conclusion. They are given by
\begin{equation}
\epsilon_{1*}\simeq\frac{1}{1+2\Delta\Nstar}
\,,\quad\quad
\epsilon_{2*}\simeq\frac{2}{1+2\Delta\Nstar}
\,,\quad\quad
\epsilon_{3*}\simeq\frac{2}{1+2\Delta\Nstar}
\,.
\end{equation}
This allows us to establish the corresponding expressions of the
tensor to scalar ratio, spectral index and running. One obtains
\begin{equation}
r\simeq\frac{16}{1+2\Delta\Nstar}
\,,\quad\quad
\nS-1\simeq-\frac{4}{1+2\Delta\Nstar}
\,,\quad\quad
\alphaS\simeq-\frac{8}{1+2\Delta\Nstar}\,.
\end{equation}
These expressions should be compared with \Eqs{eq:sfimaxivev}. We see
that the first Hubble flow parameter for SFI and DWI differ by a
factor close to $4$ and that the $\epsilon_2$ roughly differ by a factor of
$2$. As a consequence, as can be checked in \Fig{fig:CMBDWI},
the DWI predictions are such that $\epsilon_{2*}=2\epsilon_{1*}$ [or
equivalently, $r=4(1-\nS)$], whereas, as can be checked in
\Fig{fig:CMBSFI2}, we have $\epsilon_{2*}=4\epsilon_{1*}$ for SFI [or
equivalently, $r=8/3(1-\nS)$]. This explains why the two models can in
fact lead to quite different predictions and why DWI cannot be simply
viewed as a mere realistic continuation of SFI.

Finally, it is also interesting to constrain the energy scale $M$. For
this purpose, we use the CMB normalization which gives
\begin{equation}
  \frac{M^4}{\Mp^4} = 11520\pi^2\left(\frac{\Mp}{\phizero}\right)^2
  \frac{\xstar^2} {\left(\xstar^2-1\right)^4}\frac{\Qrms^2}{T^2}\, .
\end{equation}
Then, using the approximated trajectory $\xstar\simeq
1-\sqrt{2+4\Delta\Nstar}\Mp/\phizero$ in the above formula, one obtains
the following expression
\begin{equation}
  \frac{M^4}{\Mp^4} \simeq 1440\pi^2
\left(\frac{\phizero}{\Mp}\right)^2\frac{1}{\left(1+2\Delta\Nstar\right)^2}
\frac{\Qrms^2}{T^2}\, .
\end{equation}
Then, requiring that $M<\Mp$ leads to the following upper bound on the
value of $\phizero$, $\phizero/\Mp\lesssim 1.5\times 10^5$. Combined
with the lower limit of \Eq{eq:lowerlimitdwi}, we see that the
possible range of variation of $\phizero$ is quite large.

The reheating consistent slow-roll predictions for the DWI models are
displayed in \Fig{fig:CMBDWI}. The reheating equation of state
parameter $\wrehbar$ has been chosen to be $0$ since the potential is
quadratic close to its minimum $V\left(\phi\right)\simeq 4
M^4/\phizero^2\left(\phi-\phizero\right)^2$.  As claimed before, one
can check that only super-Planckian values of the symmetry breaking
scale $\phizero$ are compatible with the data. Actually, this is also
true for the SFI models, see \sectionc{sec:sfi} and
\Fig{fig:CMBSFI2}. As already mentioned before, in this regime, the
two models differ while, as expected, they are very similar for
sub-Planckian values of the field \vev.

\subsection{Mutated Hilltop Inflation (MHI)}
\label{sec:mhi}

This model belongs to the class of hilltop
models~\cite{Boubekeur:2005zm, Tzirakis:2007bf}. In this type of
scenarios, inflation is supposed to occur at the top of the
potential. In particular, it was shown in \Refcs{Boubekeur:2005zm,
  Tzirakis:2007bf} that, by adding the contributions coming from
higher order operators, $F$ or $D$ term inflation can be turned into
hilltop models. Here, we consider mutated hilltop inflation which was
first introduced and discussed in \Refcs{Pal:2009sd, Pal:2010eb}. The
potential is phenomenological only and given by
\begin{equation}
V = M^4 \left[1-{\sech} \left(\frac{\phi}{\mu} \right) \right],
\end{equation}
with $\sech x =1/\cosh x$. As argued in \Refcs{Pal:2009sd,
  Pal:2010eb}, it can be viewed as small field inflation (hilltop
inflation) completed by an infinite number of higher order operators,
these operators giving rise to a power series responsible for the
appearance of the $\sech$ function. From an effective field theory
point of view, reasonable values of the parameter $\mu$ seem to be
such that $\mu<\Mp$ but in other contexts such a restriction may not
be necessary. This is why although the model is studied for any value
of $\mu$, approximated formula will also be derived in the $\mu\ll\Mp$
approximation.

\begin{figure}
\begin{center}
\includegraphics[width=\wdblefig]{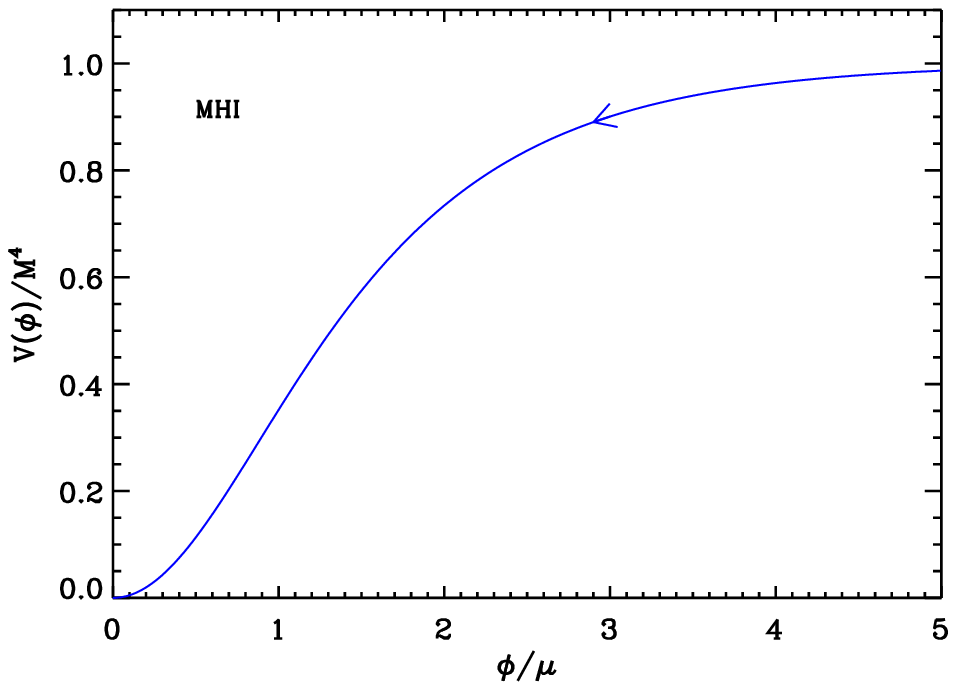}
\includegraphics[width=\wdblefig]{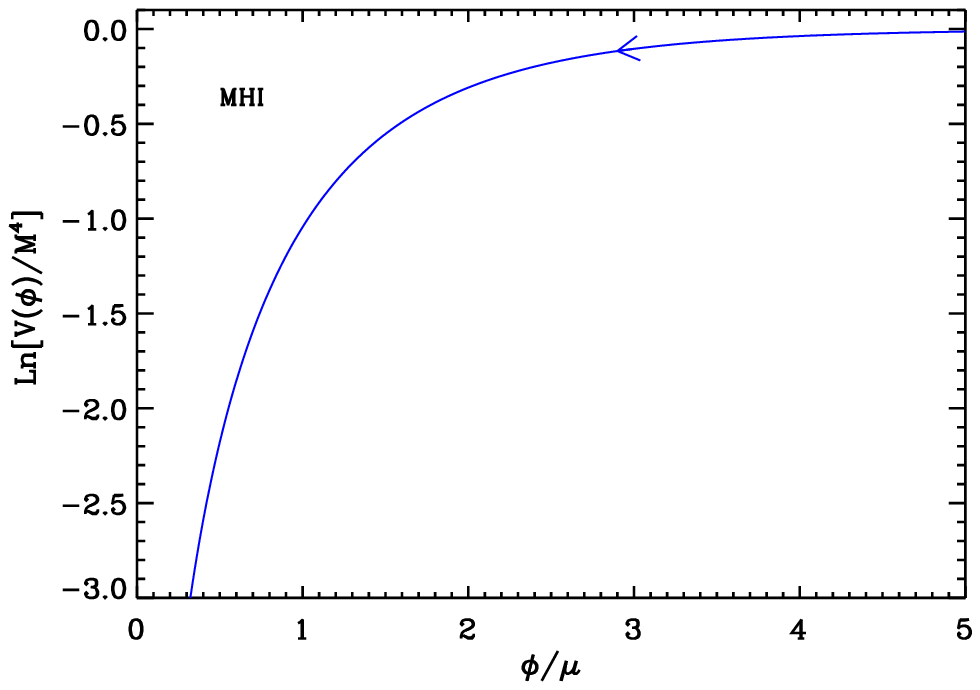}
\includegraphics[width=\wdblefig]{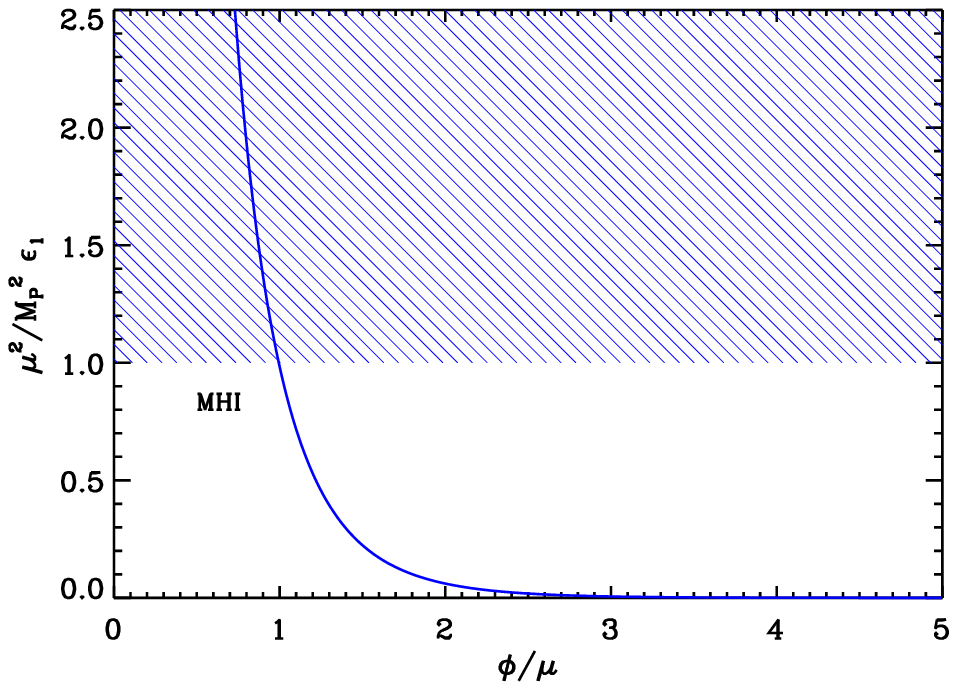}
\includegraphics[width=\wdblefig]{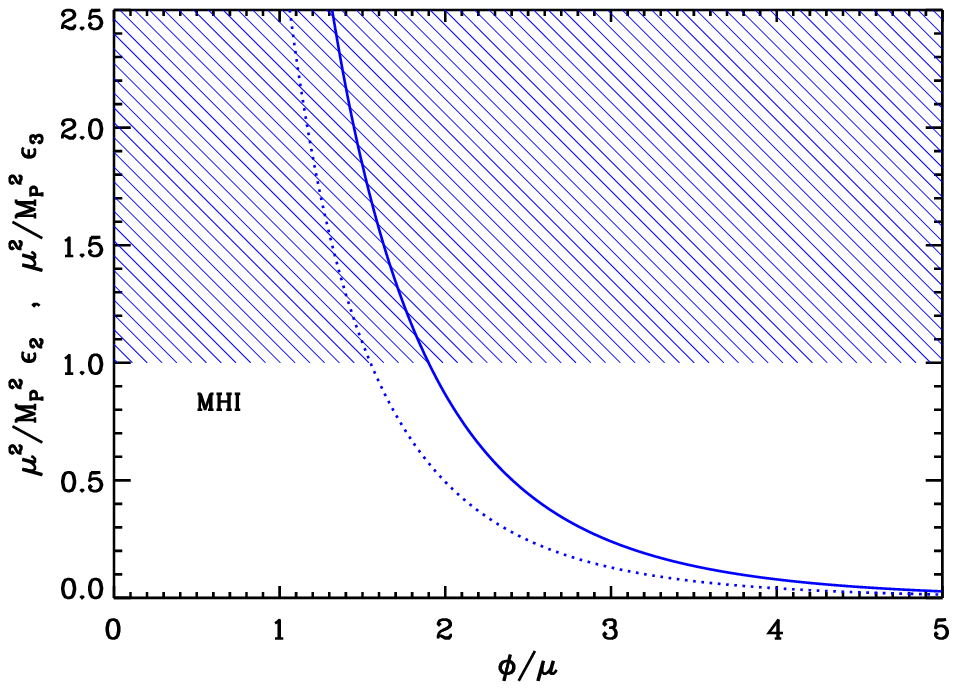}
\caption{Mutated Hilltop Inflation (MHI).  The top panels show the
  potential and its logarithm as a function of $x=\phi/\mu$. Bottom
  left panel: Rescaled slow-roll parameter $\epsilon_1$ (divided by
  $\Mp^2/\mu^2$). The shaded area represents the region in which
  inflation stops if $\mu=\Mp$. It should be accordingly rescaled for
  other values of $\mu$. Bottom right panel: slow-roll parameters
  $\epsilon_2$ (solid line) and $\epsilon_3$ (dotted line), again
  rescaled by $\Mp^2/\mu^2$ together with the region of slow-roll
  violation for $\mu=\Mp$.}
\label{potmhi}
\end{center}
\end{figure}

Defining $x\equiv\phi/\mu$, the three first Hubble flow functions in the
slow-roll approximation are given by
\begin{equation}
  \epsilon_1= \frac{\Mp^2}{2\mu^2}\coth^2 \left(\frac{x}{2}\right)
  \sech^2 x, \qquad
  \epsilon_2 =
  \frac{\Mp^2}{\mu^2}\left[\csch^2\left(\frac{x}{2}\right)+2\,\sech^2x \right],
\label{eq:mhieps12}
\end{equation}
\begin{equation}
  \epsilon_3 = \frac{\Mp^2}{\mu^2} \dfrac{\cosh x
    \coth^2\left(\dfrac{x}{2}\right) +2 \tanh^2 x}{\cosh x + \sinh^2 x}\,.
\end{equation}
where $\csch x =1/\sinh x $. These three quantities are monotonic
decreasing functions of the field values and inflation proceeds from
large field values towards small field values. Together with the
potential, they are represented as a function of $x$ in \Fig{potmhi}.

\begin{figure}
\begin{center}
\includegraphics[width=\wsingfig]{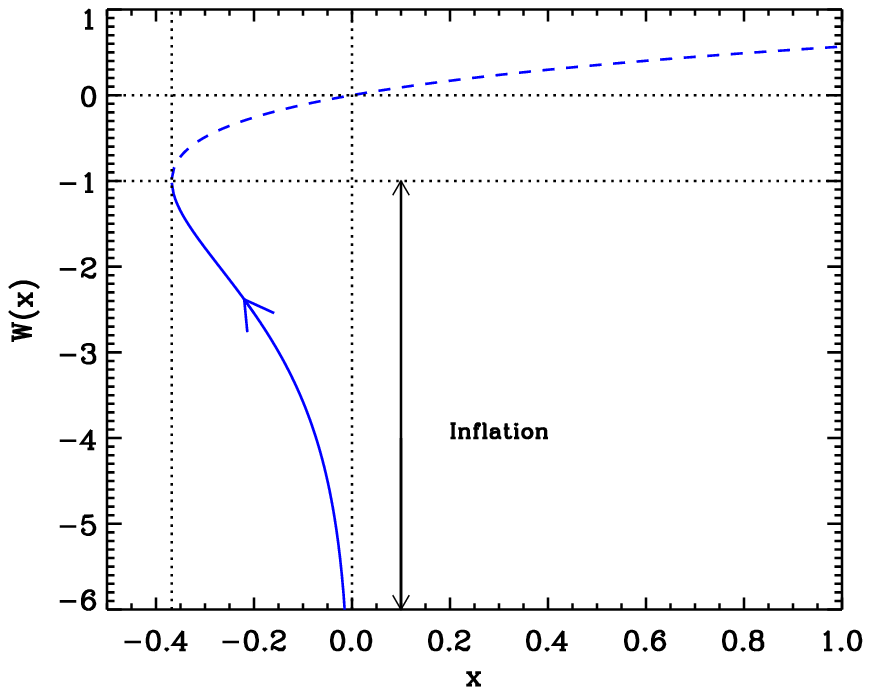}
\caption{Lambert functions $\Lambert0(x)$ (dashed line) and
  $\Lambert{-1}(x)$ (solid line). During Mutated Hilltop inflation,
  inflation proceeds along the ``$-1$'' branch in the direction
  specified by the arrow on the figure.}
\label{plotlambertMHI}
\end{center}
\end{figure}

The slow-roll trajectory can be integrated exactly from
\Eq{eq:srtrajectory} and reads
\begin{equation}
  N-\Nend = \frac{\mu^2}{\Mp^2}\left\lbrace 2\ln \left[
      \frac{\cosh \left(x/2\right)}{\cosh\left(\xend/2\right)} \right]
    -\cosh x +\cosh \xend  \right \rbrace.
\end{equation}
It can also be inverted analytically to give the field values in terms
of the number of \efolds using the Lambert function
$\Lambert{-1}$. One obtains
\begin{equation}
  x=\arcosh \left(-1-\Lambert{-1} \left\lbrace - \left(1 + \cosh \xend
      \right)
      \exp\left[\dfrac{\Mp^2}{\mu^2}(N-\Nend)
        -1-\cosh\xend \right]
             \right\rbrace\right).
\end{equation}
Since $N-\Nend<0$ and the function $y \ee^{-y}$ has a global maximum
equals to $1/e$, inflation proceeds along the $-1$ branch of the
Lambert function as represented in \Fig{plotlambertMHI}. Note that in the 
$\mu\ll\Mp$ limit, this trajectory simply becomes 
$N-\Nend\simeq\mu^2/(2\Mp^2)\left(\ee^{\xend}-\ee^{x}\right)$.

For MHI, inflation naturally stops when $\epsilon_1=1$, which has an
unique solution given by
\begin{equation}
\begin{aligned}
\label{eq:mhi:xend}
  \xend  = \arsech & \left[-\frac{1}{3} + \dfrac{1}{3} \left(1 -
      6\dfrac{\mu^2}{\Mp^2}\right) \left(-1+ 36\frac{\mu^2}{\Mp^2}
        +3\sqrt{6}\frac{\mu}{\Mp} \sqrt{4\frac{\mu^4}{\Mp^4} + 22
          \frac{\mu^2}{\Mp^2}-1}
      \right)^{-1/3} \right. \\
    &\ \ + \left. \frac{1}{3} \left( -1+36\frac{\mu^2}{\Mp^2} +
        3\sqrt{6}\frac{\mu}{\Mp} \sqrt{4 \frac{\mu^4}{\Mp^4}
          +22\frac{\mu^2}{\Mp^2}-1} \right)^{1/3}\right],
\end{aligned}
\end{equation}
and with $\arsech x=\arcosh(1/x)$. One should note that the previous
equation is always well defined, regardless of the sign of the square
root argument by analytic continuation. Let us notice that from
\Eq{eq:mhieps12} one has
\begin{equation}
\epsilon_2 - \epsilon_1 = \dfrac{1}{2} \csch^2\left(\dfrac{x}{2}
  \right) + \sech x + \dfrac{5}{2} \sech^2 x  > 0.
\end{equation}
Consequently, the slow-roll approximation may become inaccurate before
the end of inflation because $\epsilon_2>1$ occurs just before
$\epsilon_1=1$. However, one can check that this happens during a
negligible number of \efolds and the observable predictions for MHI
remain mostly unaffected. Also, in the limit $\mu\ll\Mp$, \Eq{eq:mhi:xend} 
gives $\xend\simeq\ln\left(\sqrt{2}\Mp/\mu\right)$. 

The value $\xstar=\phistar/\mu$ at which the pivot mode crossed the
Hubble radius during inflation is obtained by solving
\Eq{eq:phistarlnrrad} for a given reheating energy. In terms of 
$\Delta\Nstar$, and in the limit $\mu\ll\Mp$, one has 
$\xstar\simeq\ln\left(2\Delta\Nstar\Mp^2/\mu^2\right)$. This enables to give 
estimates for the slow-roll parameters at Hubble crossing, namely
\begin{equation}
\begin{aligned}
\epsilon_{1*}\simeq\frac{1}{2\Delta\Nstar^2}\left(\frac{\mu}{\Mp}\right)^2
,\qquad
\epsilon_{2*}\simeq\frac{2}{\Delta\Nstar}
\, ,\qquad
\epsilon_{3*}\simeq\frac{1}{\Delta\Nstar}\, ,
\end{aligned}
\end{equation}
hence, at first order in slow-roll
\begin{equation}
\begin{aligned}
r\simeq\frac{8}{\Delta\Nstar^2}\left(\frac{\mu}{\Mp}\right)^2
\, ,\qquad
\nS-1\simeq-\frac{2}{\Delta\Nstar}
\, ,\qquad
\alphaS\simeq-\frac{2}{\Delta\Nstar^2}\, .
\end{aligned}
\end{equation}
One can see that for $\mu/\Mp\ll 1$, the typical predicted amount of 
gravitational waves is very small, and the deviation from scale invariance
almost does not depend on $\mu$.

Finally, the constant $M$ can be determined from the amplitude of the
CMB anisotropies
\begin{equation}
  \frac{M^4}{\Mp^4} = 90 \pi^2 \dfrac{\Mp^2}{\mu^2} \csch^6 \left(
    \frac{\xstar}{2} \right)
  \sinh \xstar  \tanh \xstar 
  \frac{\Qrms^2}{T^2}\, .
\end{equation}
In the $\mu/\Mp\ll 1$ limit, one obtains
\begin{equation}
  \frac{M^4}{\Mp^4} \simeq \frac{720\pi^2}{\Delta\Nstar^2}
  \dfrac{\mu^2}{\Mp^2}\frac{\Qrms^2}{T^2}\, .
\end{equation}
Typically, for $\mu/\Mp\simeq 10^{-2}$, one has $M/\Mp\simeq 10^{-4}$.

The reheating consistent slow-roll predictions for MHI have been
represented in \Fig{fig:CMBMHI}. As expected, for small values of
$\mu/\Mp$, the predicted amount of gravitational waves is extremely
small and the deviation from scale invariance almost does not depend on
$\mu$.

\subsection{Radion Gauge Inflation (RGI)}
\label{sec:rgi}

\begin{figure}
\begin{center}
\includegraphics[width=\wdblefig]{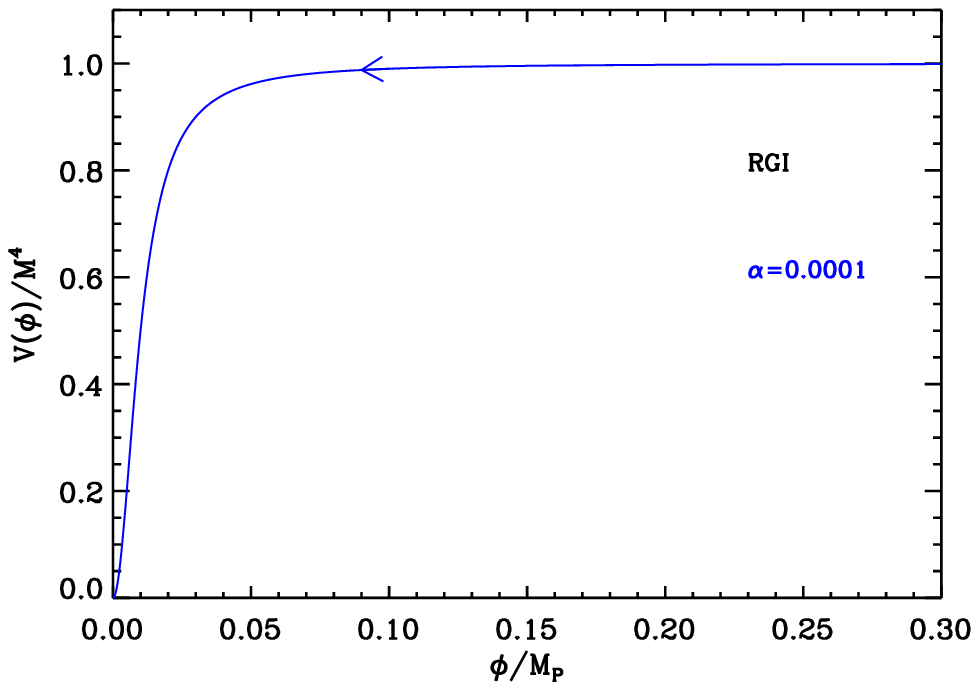}
\includegraphics[width=\wdblefig]{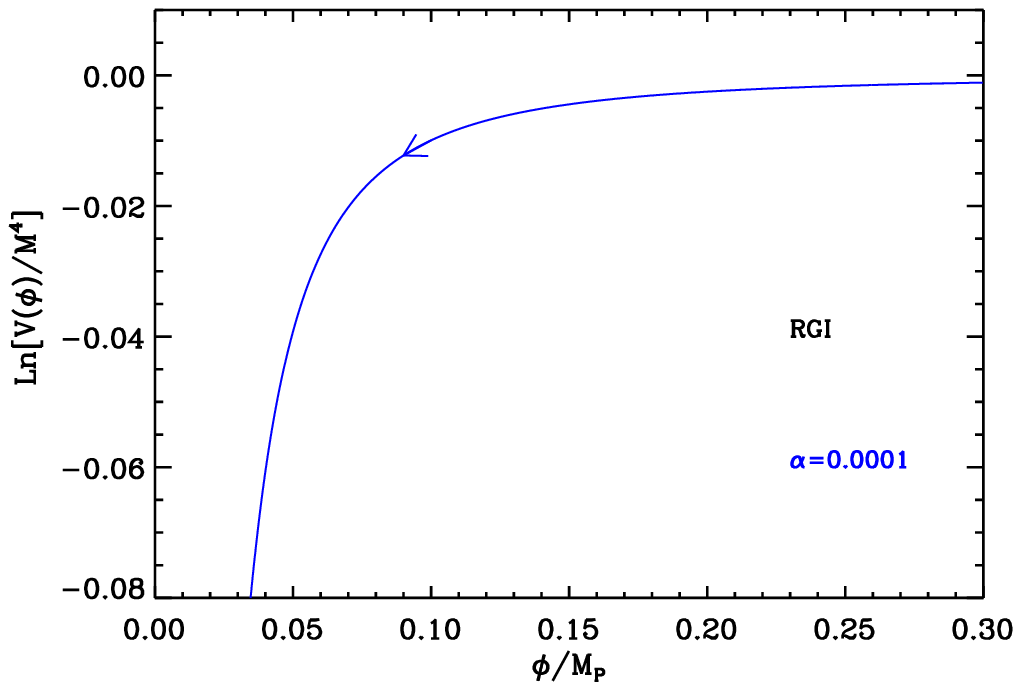}
\includegraphics[width=\wdblefig]{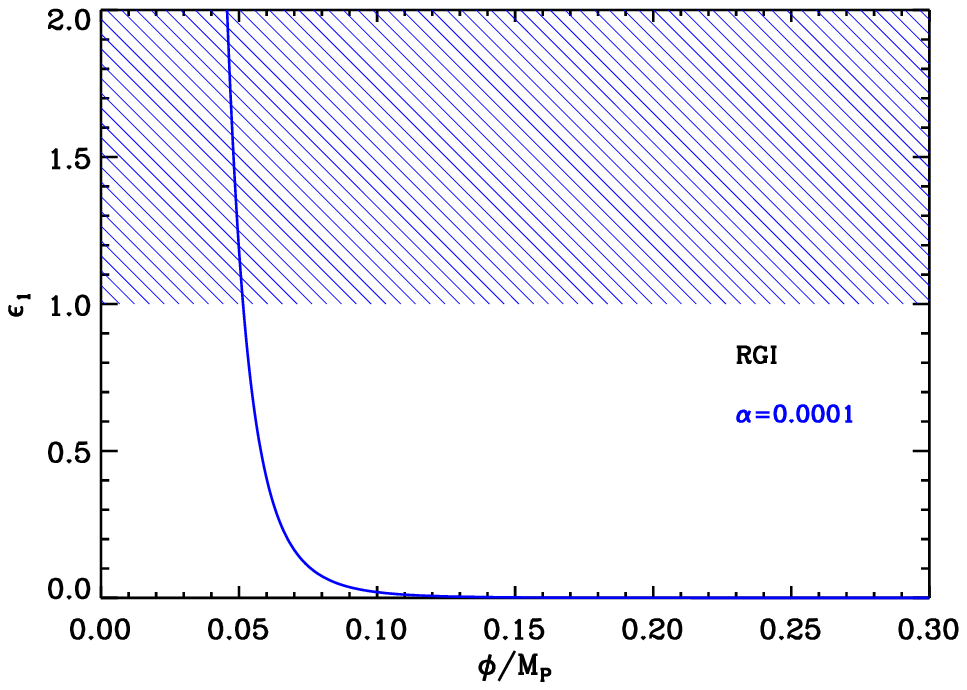}
\includegraphics[width=\wdblefig]{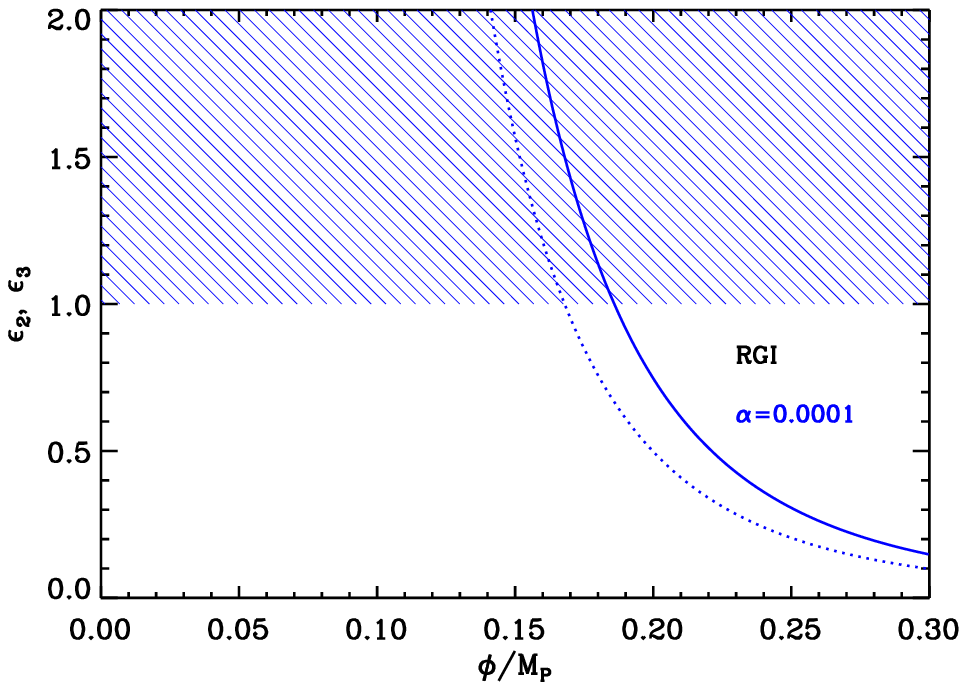}
\caption{Radion Inflation (RGI) for $\alpha=10^{-4}$. Top frames: the
  potential and its logarithm. Bottom left panel: slow-roll parameter
  $\epsilon_1$ and the shaded area in which inflation stops
  ($\epsilon_1>1$). Bottom right panel: slow-roll parameters
  $\epsilon_2$ (solid line) and $\epsilon _3$ (dotted line).}
\label{potrgi}
\end{center}
\end{figure}

This model was studied in \Refc{Fairbairn:2003yx}. It is an extension
of the gauge inflation scenario in which the radius modulus field
around which the Wilson loop is wrapped assists inflation as it
shrinks~\cite{Freese:1990rb}. Assuming that the radion field value is
such that the potential energy is minimal, for each value of the
inflaton field $\phi$, one can derive an effective potential
\begin{equation}
V(\phi)=M^4\frac{\left(\phi/\Mp\right)^2}{\alpha+\left(\phi/\Mp\right)^2}\, ,
\end{equation}
where $\alpha$ is a dimensionless positive parameter. In the context
of \Refc{Fairbairn:2003yx}, the model is natural for $\alpha<1$ but
larger than unity values are not forbidden. The same potential has
been obtained in \Refc{delaMacorra:1995qh} in the context of S-dual
superstring models. In that case, $\alpha$ represents a typical {\vev}
for the inflaton, in Planck units. Defining $x=\phi/\Mp$, the first
three slow-roll parameters read
\begin{equation}
  \epsilon_1 = \frac{2\alpha^2}{x^2 \left( \alpha+x^2 \right)^2}\,, \qquad
  \epsilon_2  =
  4 \alpha \frac{\alpha+3x^2}{x^2 \left( \alpha+x^2 \right)^2}\,, \qquad 
  \epsilon_3 =  4 \alpha \frac{\alpha^2+3\alpha
    x^2+6x^4}{ x^2 \left( \alpha + x^2 \right)^2 \left(\alpha +
      3x^2\right)} \,.
\label{eq:rgieps123}
\end{equation}
The potential, its logarithm, and the Hubble flow functions are
represented in \Fig{potrgi}.

The slow-roll trajectory can be integrated analytically from
\Eq{eq:srtrajectory} to obtain
\begin{equation}
  N - \Nend=\frac{\xend^2}{4} + \frac{\xend^4}{8\alpha} -
  \frac{x^2}{4} - \frac{x^4}{8\alpha}\, .
\end{equation}
Moreover, it can be inverted explicitly to give the field values in
terms of the number of \efolds as
\begin{equation}
  x =\sqrt{ -\alpha + \sqrt{-8 \alpha (N-\Nend) + (\alpha + \xend^2)^2}}\,.
\end{equation}

The end of inflation naturally occurs for $\epsilon_1=1$, i.e., from
\Eq{eq:rgieps123}, at the field value $\xend$ given by
\begin{equation}
\label{eq:rgi:xend}
\xend = \dfrac{- \sqrt[3]{6}\, \alpha + \left[9 \alpha + 
    \sqrt{3 \alpha ^2 (2 \alpha +27)} \right]^{2/3} }
{162^{1/6} \left[9 \alpha + \sqrt{3 \alpha ^2 (2 \alpha
      +27)}\right]^{1/3}}\,.
\end{equation} 
As for the MHI models, one should pay attention that
\begin{equation}
\epsilon_2 - \epsilon_1 = 2 \alpha \dfrac{\alpha + 6
  x^2}{x^2(\alpha+x^2)^2} > 0,
\end{equation}
for any positive values of $\alpha$. As a result, slow-roll violation,
i.e. $\epsilon_2>1$, occurs in RGI before inflation ends. However,
since the first Hubble flow function is monotonic, this is not very
problematic as it happens only during a negligible number of \efolds
and only around $\Nend$. The slow-roll observable predictions
therefore remain accurate.

As before, the observable field value $\xstar$ is obtained by solving
\Eq{eq:phistarlnrrad} for a given reheating model and allows the
determination of the parameter $M$ from the amplitude of the CMB
anisotropies. One gets
\begin{equation}
  \frac{M^4}{\Mp^4} = \frac{2880\pi^2\alpha^2}{\xstar^4 \left(\alpha +
      \xstar^2\right)} \frac{\Qrms^2}{T^2}\, .
\end{equation}
The reheating consistent slow-roll predictions for these models are
displayed in \Fig{fig:CMBRGI}. Large values of $\alpha$ give back the
same predictions as the large field models with $p=2$
(see \sectionc{sec:lfi}) having $\epsilon_{2*} = 2\epsilon_{1*}$.

\subsection{MSSM Inflation (MSSMI)}
\label{sec:mssmi}

\subsubsection{Theoretical Justifications}
\label{subsubsec:theorymssm}

The Minimal Supersymmetric Standard Model (MSSM) is an extension of
the Standard Model of particle physics. Its Lagrangian is
characterized by the following super potential
\begin{equation}
\label{eq:mssmsuperpot}
  W_{_\mathrm{MSSM}}=\lambda_u^{ij}Q_i\cdot H_uU^{\uc}_j
  +\lambda_d^{ij}Q_i\cdot H_dD^{\uc}_j+\lambda_e^{ij}L_i\cdot H_dE^{\uc}_j
  +\mu H_u\cdot H_d.
\end{equation}
The quantity $Q_i$ denotes a doublet of left handed quarks superfields
where $i$ is a family index. In practice this means that
\begin{equation}
Q_1=\left(\begin{matrix}
U \\ D\end{matrix}\right), \quad 
Q_2=\left(\begin{matrix}
C \\ S \end{matrix}\right), \quad 
Q_3=\left(\begin{matrix}
T \\ B \end{matrix}\right),
\end{equation}
where the components of the doublets are superfields. For instance,
the scalar part of $U$ is the $\tilde{u}$ squark and its fermionic
part is the ordinary $u$ quark. Of course, there is also a color index
$a=1,2,3$ and, in fact, one should write the corresponding doublet as
$Q_{ia}$. Moreover, one can also introduce a third
$\mathrm{SU}(2)_{\uL}$ index $\alpha=1,2$ and write $Q_{ia\alpha}$
with, for instance, $Q_{1a1}=U$ and $Q_{1a2}=D$. On the other hand,
the quantities $U^{\uc}_j$ and $D^{\uc}_j$ denotes the right handed
superfields where $j$ is the family index (and the color index has
been ignored in order to simplify the notation): for instance,
$U^{\uc}_2$ means the right handed charm quark superfield which is a
singlet under $\mathrm{SU}(2)_\uL$.

In the same fashion, $L_i$ denotes a doublet of left handed lepton
superfields
\begin{equation}
L_1=\left(\begin{matrix}
N_e \\ E_e\end{matrix}\right), \quad 
Q_2=\left(\begin{matrix}
N_\mu \\ E_\mu \end{matrix}\right), \quad 
Q_3=\left(\begin{matrix}
N_\tau \\ E_\tau \end{matrix}\right),
\end{equation}
where, for instance, $N_e$ denotes the electronic neutrino superfield
(the scalar part being the neutralino and the fermionic part the
electronic neutrino itself) while $E_e$ denotes the electron
superfield. On the other hand, the quantities $E^{\uc}_j$ denote the
right handed superfields that are singlet under $\mathrm{SU}(2)_{\uL}$
(for instance, $E^{\uc}_2$ is the right handed muonic superfield). In
the superpotential~\eqref{eq:mssmsuperpot}, there are two terms
involving the quarks and only one involving the leptons because, as
well-known, there is no right handed neutrinos in the standard model.

The last term in \Eq{eq:mssmsuperpot} describes the Higgs sector with
two Higgs doublet $H_u$ and $H_d$. The quantity $\mu$ is a new
dimensionful (of dimension one) parameter of the model. The dot
indicates an $\mathrm{SU}(2)$ invariant product. Finally, $\lambda_u$,
$\lambda_d$, $\lambda_e$ are the $3\times3$ Yukawa matrices.

From the superpotential~\eqref{eq:mssmsuperpot}, one can determine the
scalar potential of the theory by means of the usual supersymmetric
machinery. As is well-known, the scalar potential is made of two
pieces, the $F$-term part and the $D$-term part. Clearly, given the
number of fields in the theory, the scalar potential is a complicated
object. For inflation, we are especially interested in the flat
directions of this potential.  A flat direction is a direction such
that the $F$ and $D$-terms vanish, that is to say such that $V_F=0$,
$V_D=0$ and, therefore, $V\equiv V_F+V_D=0$. It was shown that the
MSSM scalar potential contains nearly $300$ gauge invariant flat
directions~\cite{Gherghetta:1995dv,Enqvist:2003gh,Mazumdar:2010sa}. Finding
these directions is a non-trivial task and we now very briefly explain
how this can be done. Usually, it consists in putting all the fields
to zero except a few ones, these few ones being carefully chosen such
that cancellations occur in such a way that the potential exactly
vanishes. We now illustrate this method on a particular case. Let us
first recall that the general formula giving the $D$-term potential is
\begin{equation}
V_D=\frac{1}{2}\sum_ag_a^2D^aD^a,
\end{equation}
where $D^a=\phi^{\dagger}T^a\phi$, $T^a$ being the generator of the
group and $\phi$ denoting a generic field (of course, the index $a$
should not be confused with the color index discussed above). For the
standard model, we have the group $\mathrm{SU}(2)_{\uL}\times
\mathrm{U}(1)_{\uY}$ and, therefore, the explicit expression of the
$D$-term reads
\begin{equation}
V_D=\frac{g^2}{2}\left(D_1^2+D_2^2+D_3^2\right)+\frac{g_{_Y}}{2}D_Y^2,
\end{equation}
$g$ and $g_{_Y}$ being the coupling constants of the two groups. For
the $\mathrm{SU}(2)$ group, the generators $T^a$ are nothing but the
Pauli matrices and, therefore, $T^a=\sigma^a/2$. Following
\Refcs{Gherghetta:1995dv,Dine:1995kz}, let us consider a situation
where all the fields in the MSSM are assumed to have a vanishing \vev
except $L_i$ and $E^{\uc}_j$ where we remind that $i$ and $j$ are
family indices. If we write $L_i^{\uparrow}$ and $L_i^{\downarrow}$ as
respectively the upper and lower component of the doublet $L_i$, then
one has (\ie we put $\phi=L_i$ in the general formula expressing
$D^a$)
\begin{eqnarray}
D_1&=&\frac{1}{2}\sum_{i=1}^3\left(L_i^{\uparrow}{}^* L_i^{\downarrow}
+L_i^{\downarrow}{}^*L_i^{\uparrow}\right), \quad 
D_2=\frac{i}{2}\sum_{i=1}^3\left(L_i^{\uparrow}{}^*L_i^{\downarrow}
-L_i^{\downarrow}{}^*L_i^{\uparrow}\right), \\ 
D_3&=&\frac{1}{2}\sum_{i=1}^3\left(\left\vert L_i^{\uparrow}\right\vert^2
-\left\vert L_i^{\downarrow}\right\vert^2\right).
\end{eqnarray}
The quantity $E^{\uc}$ being a $\mathrm{SU}(2)$ singlet does not
participate to the above expression. On the other hand, the
contribution from the $\mathrm{U}(1)$ group reads
\begin{equation}
D_Y=\frac{1}{2}\sum_{i=1}^3\left(2\vert e_i\vert^2-
\left\vert L_i^{\uparrow}\right\vert^2
-\left\vert L_i^{\downarrow}\right\vert^2\right),
\end{equation}
where $e_i$ denotes the scalar field of the $E_i^{\uc}$
supersymmetric multiplet. We see that, if we take
\begin{equation}
\label{eq:flatDmssm}
L_i=\left(\begin{matrix}
\phi \\ 0\end{matrix}\right), \quad 
L_j=\left(\begin{matrix}
0 \\ \phi\end{matrix}\right), \quad 
e_k=\phi,
\end{equation}
then we have $V_D=0$. 

The next step consists in calculating the $F$-term for the
choice made in \Eq{eq:flatDmssm}. It is easy to check that
$V_F=0$. Therefore, we have identified a flat direction. It is denoted
$L_iL_je_k$ or \textbf{LLe} to recall that all family combination are
possible. This direction is represented by a ``composite operator
$X_m$'' formed by the product of the superfields making up the flat
direction. In our case $X_3=L_iL_je_k=\phi^3$ and $m=3$ since we have
three operators participating to the definition of $X_3$. This
direction has been proposed in \Refc{Allahverdi:2006iq} as a possible
candidate for the inflaton field. Let us also remark that another
choice put forward in that reference was \textbf{udd}.

We have just seen how to identify flat directions in the MSSM
potential. However, this flatness is usually spoiled by the presence
of higher order non-renormalizable operators appearing in the MSSM
(viewed here as a low energy effective field) and by supersymmetry
breaking~\cite{Gherghetta:1995dv,Enqvist:2003gh,Mazumdar:2010sa}. Higher
order operators are described by the following superpotential
\begin{equation}
  \label{eq:liftmssm}
W=\frac{\lambda_n}{n}\frac{X_m^k}{\Mp^{mk-3}},
\end{equation}
where $\lambda_n$ is a coupling constant, $n\equiv mk$ and $k=1$ or
$k=2$ depending on whether the flat direction is even or odd under
R-parity. Recall that $Q$, $L$, $U^{\uc}$, $D^{\uc}$ and $E^{\uc}$
have R-parity $-1$ and $H_u$, $H_d$ have R-parity $+1$.  It follows
that \textbf{LLe} (for instance) has odd R-parity and, therefore, that
$k=2$. For the directions \textbf{LLe} (this is also true for
\textbf{uud}), this means that
\begin{equation}
n\equiv mk=6.
\end{equation}
The above superpotential~\eqref{eq:liftmssm} will produce a term
$\vert \partial W/\partial \phi \vert^2\propto \phi^{2(km-1)}$ in the
scalar potential. Then, we have the contributions originating from
supersymmetry breaking. They can be easily calculated if, for
instance, we assume that we have an independent hidden sector where
supersymmetry is broken and that this breaking is mediated by gravity
only. This gives two types of soft terms, one proportional to $\phi^2$
and another, the so-called ``$A$-term'', proportional to
$\left(\phi\partial W/\partial \phi +\mathrm{cc}\right)$ that is to say,
given \Eq{eq:liftmssm}, proportional to $\phi^{mk}$.

More generally, if one starts from a flat direction with a given $n$,
then the superpotential has the form $W=\lambda_n/n\Phi^n\Mp^{3-n}$,
where $\Phi=\phi\ee^{i\theta}$ is the superfield which contains the
flat direction. Then, the scalar potential takes the form
\begin{equation}
\label{eq:mssmi:pot}
V(\phi)=\frac{1}{2}m_\phi^ 2\phi^2+A\cos(n\theta+\theta_0)
\frac{\lambda_n}{n}\frac{\phi^n}{\Mp^{n-3}}+\lambda_n^2
\frac{\phi^{2(n-1)}}{\Mp^{2(n-3)}}\, ,
\end{equation}
where the second term involves the angular part of the superfield
via a term $\cos(n\theta+\theta_0)$, which in practice is fixed at
$-1$ to maximize its contribution. As explained below, the fact that
the second term appears with a negative coefficient plays a crucial
role in making this scenario a credible inflationary one.

Together with the global minimum at $\phi=0$, under the condition
$A^2\geq 8(n-1)m_\phi^2$, the potential has a secondary minimum at
$\phizero \simeq \left(m_\phi\Mp^{n-3}\right)^{1/(n-2)}$. If $A^2\gg
8(n-1)m_\phi^2$, this secondary minimum becomes the deepest one and
thus the true one. The curvature of the potential at this minimum is
of the order $m_\phi^2$. If inflation occurs there, one gets $H \simeq
m_\phi (m_\phi/\Mp)^{1/(n-2)}$, which is much smaller than the
potential curvature for $m_\phi\ll\Mp$. This implies that the
potential is too steep for quantum effects during inflaton to kick
$\phi$ out of the false minimum. Such a situation is similar to the
old inflationary scenario. However, this barrier disappears if one
saturates the previous inequality and takes
\begin{equation}
A^2 = 8(n-1)m_\phi^2.
\end{equation}
In that case, the potential has a flat inflection point at $\phizero$
and inflation can proceed between this plateau and $\phi=0$. This is
the case we study in this section. This model (and its
generalizations) has also been studied in \Refcs{GarciaBellido:2006fd,
  Allahverdi:2006rt, Lyth:2006ec, Allahverdi:2006wt,
  Allahverdi:2007vy, Enqvist:2007tf, Allahverdi:2008bt, Kamada:2009hy,
  Allahverdi:2010zp, Enqvist:2010vd, Kohri:2010sj,Choudhury:2013jya,
  Weymann-Despres:2023wly}. Its generalizations will be investigated
in more details in \sectionc{sec:gmssmi} and \sectionc{sec:gripi}. Let
also us notice that when $n=3$, the same potential appears in
\Refcs{Linde:1995rv, Linde:1998iw} as ``Generalized Chaotic
Inflation'', and later in \Refcs{Jain:2008dw, Jain:2009pm,
  Lowe:2010np} as ``Punctuated Inflation''. In these references, it is
shown that slow-roll inflation is briefly interrupted when the
inflaton crosses the flat inflection point and this can produce
step-like features in the primordial power spectra. These effects are
outside the scope of the following slow-roll analysis as we will be
dealing with the last slow-roll inflationary stage within this
scenario.

\begin{figure}
\begin{center}
\includegraphics[width=\wdblefig]{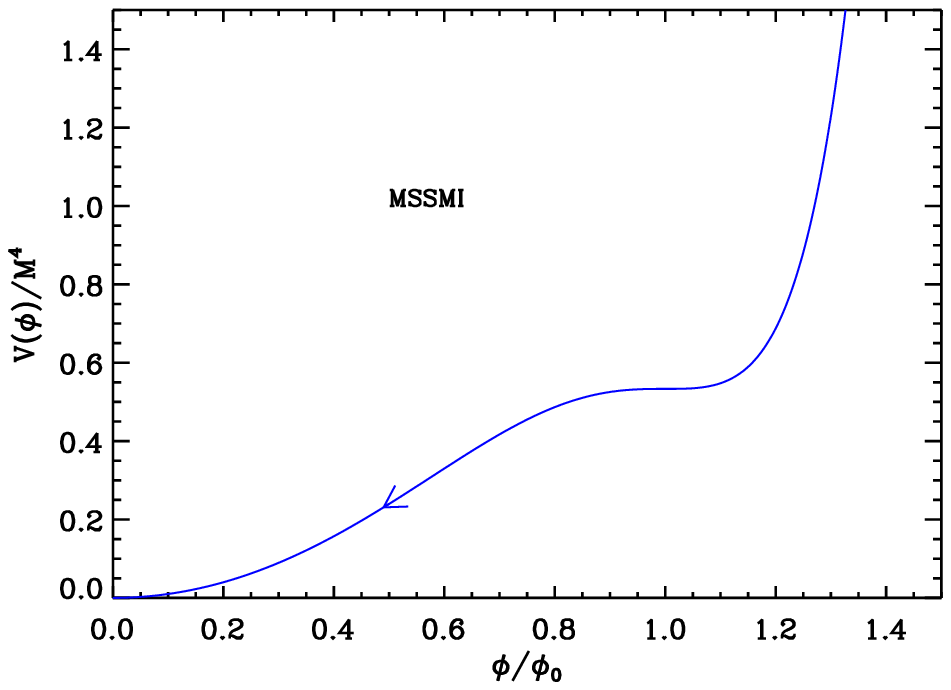}
\includegraphics[width=\wdblefig]{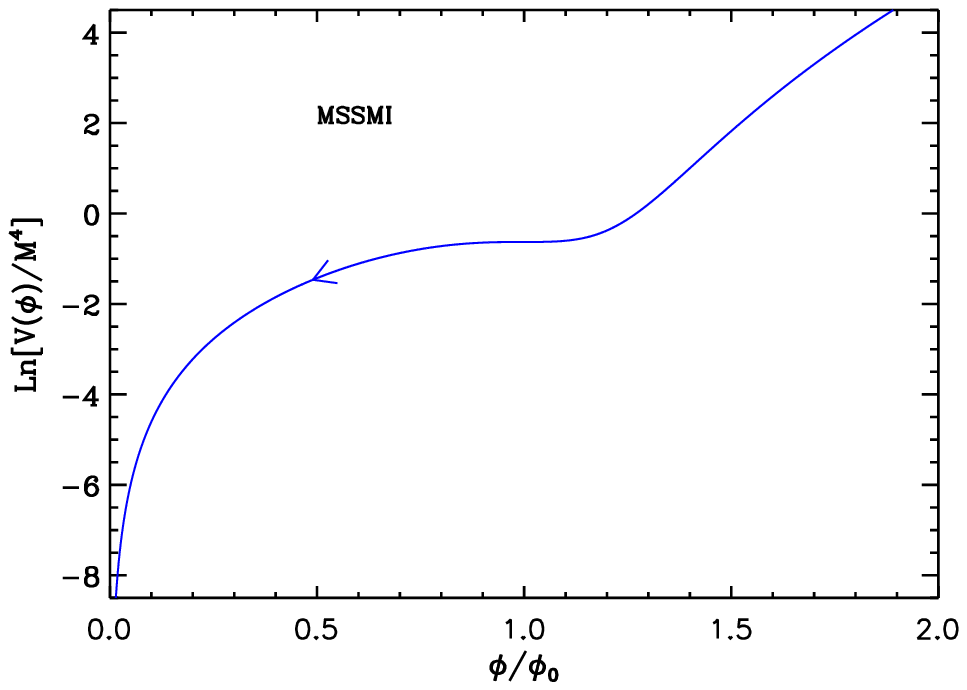}
\includegraphics[width=\wdblefig]{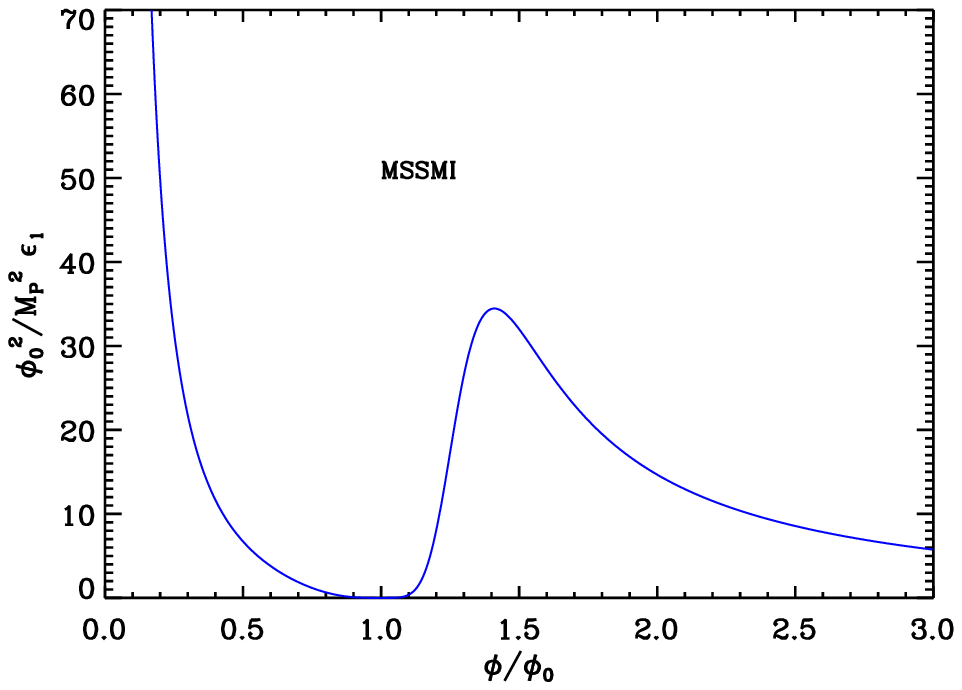}
\includegraphics[width=\wdblefig]{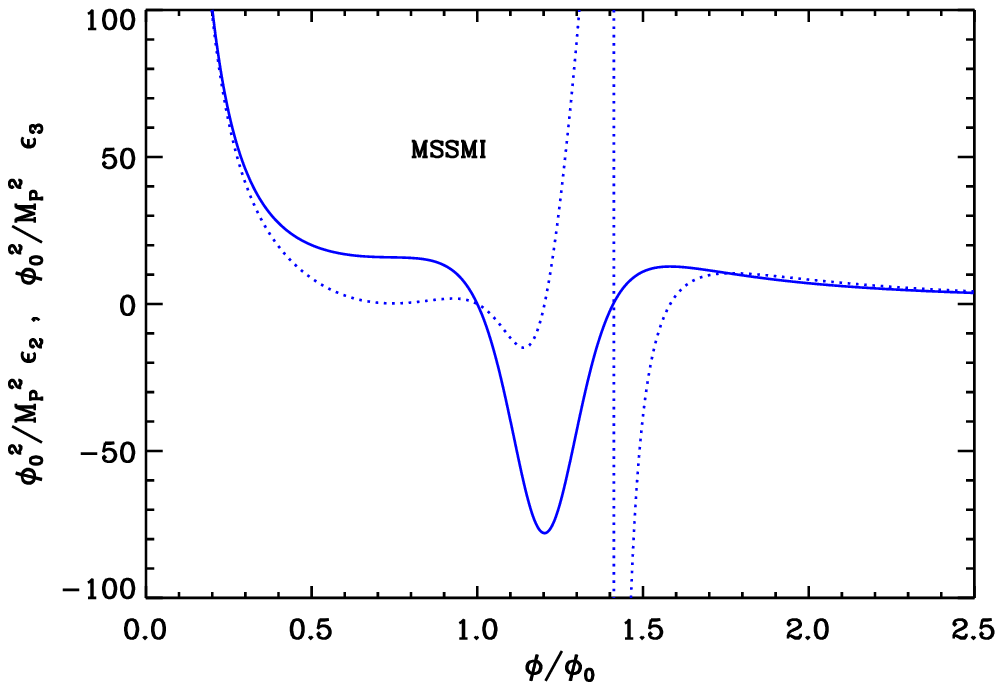}
\caption{MSSM Inflation (MSSMI).
  Top left panel: MSSM Inflation potential \Eq{eq:mssmi:pot2} 
  as a function of $\phi/\phizero$. Top right panel:
  logarithm of the potential. Bottom left panel: slow-roll 
  parameter $\epsilon _1$
  scaled by $\phizero^2/\Mp^2$.  
  Bottom right panel: slow-roll parameters $\epsilon _2$ (solid line) and 
  $\epsilon_3$ (dotted line) scaled by $\phizero^2/\Mp^2$.}
\label{potMSSMI}
\end{center}
\end{figure}

\subsubsection{Slow-Roll Analysis}
\label{subsubsec:srmssmi}
 
We now turn to the slow-roll analysis of MSSM inflation. As discussed
before, we assume that the inflaton is the flat direction \textbf{LLe}
or \textbf{uud}. This implies that $n=6$ in \Eq{eq:mssmi:pot}. Then,
rewriting the potential~(\ref{eq:mssmi:pot}) in a more convenient
fashion, one arrives at
\begin{equation}
\label{eq:mssmi:pot2}
  V(\phi)=M^4\left[\left(\frac{\phi}{\phizero}\right)^2-\frac{2}{3}
  \left(\frac{\phi}{\phizero}\right)^6+\frac{1}{5}\left(
  \frac{\phi}{\phizero}\right)^{10}\right],
\end{equation} 
where we have defined new parameters according to
\begin{equation}
\label{eq:parammssmi}
M^8 = \dfrac{\Mp^3 m_\phi^5}{4 \sqrt{10}\lambda_6}\,, \qquad \phizero^8 =
\dfrac{\Mp^6 m_\phi^2}{10\lambda_6^2}\,.
\end{equation}
These definitions and the value of the coefficients ensure that
$\phizero$ is the location of a flat inflection point. Since
$m_\phi^2\phi^2$ is a soft SUSY breaking term, we typically expect
that $m_\phi\simeq 1\,\TeV$ and this is the reason why, in what
follows, typical values of the field are taken to be
\begin{equation}
\phizero \simeq 10^ {14}\, \GeV, 
\end{equation}
in agreement with the second of \Eqs{eq:parammssmi} (the
coupling constant $\lambda_6$ is taken to be of order one). An
interesting feature of this model is that it provides inflation at
sub-Planckian \vev and at low scale $V\simeq (10^{9}\,\GeV)^ 4$.  As
noticed in \Refc{Allahverdi:2006iq}, higher values than $n=6$ would
produce too small amplitude for the scalar perturbations. This is why
the model is commonly studied with $n=6$ (with $n=3$, this is RIPI,
see \sectionc{sec:ripi}).

The potential in \Eq{eq:mssmi:pot2} is displayed in \Fig{potMSSMI},
together with its logarithm. It is an increasing function of the
field, the derivative of which vanishes at $\phi=0$ and at its second
inflection point $\phi=\phizero$, the position of the first inflection
point being given by $\phiddVzeroMinus=\phizero/\sqrt{3}$.  Inflation
proceeds in the region $\phi\in [0,\phizero]$, in the direction
specified by the arrow in \Fig{potMSSMI}.

Defining the dimensionless quantity $x$ by
\begin{equation}
x\equiv \dfrac{\phi}{\phizero}\,,
\end{equation}
the first three Hubble flow functions in the slow-roll approximation
are given by
\begin{equation}
  \epsilon_1=450\frac{\Mp^2}{\phizero^2}\frac{\left(x^4-1\right)^4}
          {x^2\left(3x^8-10x^4+15\right)^2}\, , \qquad \epsilon_2 =
          60\frac{\Mp^2}{\phizero^2}\frac{3x^{16}-58x^8+40x^4+15}
          {x^2\left(3x^8-10x^4+15\right)^2}\,,
\end{equation}
and
\begin{equation}
\begin{aligned}
\epsilon_3 & = \frac{\Mp^2}{\phizero^2} \frac{60}{x^2} \left(
-225 + 1575 x^4 - 3165 x^8 + 395 x^{12} + 2605 x^{16} - 1275 x^{20} + 
 81 x^{24} + 9 x^{28}
\right) \\
& \times \left(3x^8-10x^4+15\right)^{-2}
\times\left(-15 - 55 x^4 + 3 x^8 + 3 x^{12}\right)^{-1}\, .
\end{aligned}
\end{equation}
These two slow-roll parameters diverge when the field \vev goes to
$0$, and vanish when the field \vev goes to infinity.  The first slow
roll parameter $\epsilon_1$ first decreases, vanishes at the flat
inflection point where $\epsilon_2$ vanishes too, then increases to
reach a local maximum where $\epsilon_2$ vanishes again, and
eventually decreases again, to vanish at infinity where $\epsilon_2$
also goes to zero. Denoting by $\xepstwoZeroPlus$ the position of the
second extremum, one has
\begin{equation}
\label{eq:mssmi:xepsilon2=0}
\xepstwoZeroPlus =\left(\frac{1}{3}\right)^{1/4}
\left[2^{4/3}\left(i \sqrt{685}-1\right)^{1/3} +14\times 2^{2/3}
  \left(i\sqrt{685} -1\right)^{-1/3} -1\right]^{1/4}\simeq 1.41022.
\end{equation}
In between the two local extrema of $\epsilon_1$, the second slow-roll
parameter $\epsilon_2$ is negative whereas it is positive
elsewhere. The value of $\epsilon_1$ at its local maximum is given by
\begin{equation}
\epsonemax = \epsilon_1\left(\xepstwoZeroPlus\right)\simeq 34.459
\dfrac{\Mp^2}{\phizero^2}\,.
\end{equation}
With the typical above-mentioned value for $\phizero\simeq
10^{14}\GeV$, one has $\Mp^2/\phizero^2\simeq 10^{8}$ and
$\epsonemax>1$. This means that if inflation proceeds for \vev's
larger than that of the flat inflection point, it can naturally stop
by slow-roll violation.  However, if this happens, inflation proceeds
at $x \gg 1$ and the potential is effectively very close to a large
field model one (LFI, see \sectionc{sec:lfi}) with $p=10$.

\begin{figure}
\begin{center}
\includegraphics[width=\wsingfig]{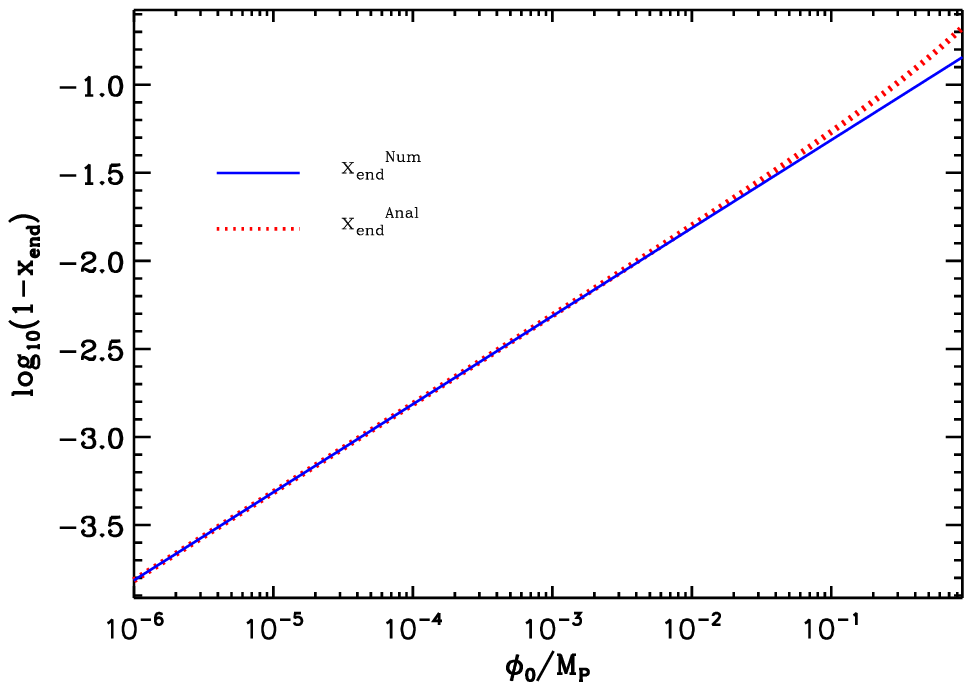}
\caption{Location of the slow-roll violation induced end of inflation
  $\xend=\phiend/\phi$ for the MSSM inflation models, as a function of
  $\phizero/\Mp$. The blue solid curve represents a numerical solution
  of $\epsilon_1=1$, while the red dotted curve corresponds to the
  approximated analytic solution \Eq{eq:mssmi:xendappr}.  For
  physical values $\phizero\simeq 10^{-4}\Mp$, the agreement is
  obviously excellent.}
\label{fig:mssmi:xend}
\end{center}
\end{figure}

For this reason, we will be focused to the case in which inflation
occurs for \vev's smaller than that of the flat inflection point. In
this case, the value of $\xend$ at which inflation stops by slow-roll
violation must be determined numerically. In the limit
$\phizero/\Mp\ll 1$ however, one has $\xend\simeq 1$ and an
approximate analytic formula can be derived
\begin{equation}
\label{eq:mssmi:xendappr}
\xend\simeq 1-\frac{1}{2^{3/4}\sqrt{15}}\sqrt{\frac{\phizero}{\Mp}}\, .
\end{equation}
A comparison between this expression and the numerical solution of
$\epsilon_1=1$ is displayed in \Fig{fig:mssmi:xend}. For physical
values $\phizero \simeq 10^{-4}\Mp$, the agreement is excellent.

Let us now turn to the slow-roll trajectory. It can be integrated from
\Eq{eq:srtrajectory} and leads to
\begin{equation}
\begin{aligned}
  \Nend-N & =  \left(\frac{\phizero}{\Mp}\right)^2\left\lbrace
   \frac{x^2-\xend^2}{20}
  +\frac{1}{15}\left(\frac{\xend^2}{\xend^4-1} -
  \frac{x^2}{x^4-1}\right) 
  \right. \nonumber\\ &- \left.
  \frac{2}{15}\left[\artanh\left(\xend^2\right) - 
  \artanh\left(x^2\right)\right]\right\rbrace ,
  \label{eq:mssmi:traj}
\end{aligned}
\end{equation}
where $\Nend$ is the number of \efolds at the end of inflation and $N$
is the number of \efolds at some point when the scaled field \vev is
$x$. A few remarks are in order.  Firstly, when $x\simeq 1$, the second
term of the previous expression dominates, and one has $\Nend-N\simeq
1/15\, (\phizero/\Mp)^2[1/(\xend^4-1)-1/(x^4-1)]$, which can be
inverted and gives
\begin{equation}
x\simeq 1-\frac{1}{4}\left[2^{-5/4}\sqrt{15}\sqrt{\frac{\Mp}{\phizero}}
+15\frac{\Mp^2}{\phizero^2}\left(\Nend-N\right)\right]^{-1}\, .
\end{equation}
Secondly, one could wonder if a sufficient number of \efolds can be 
realized in the regime studied here. When $x\rightarrow 1$, the 
corresponding number of \efolds diverges, but in practice, the 
inflationary dynamics close to the flat inflection point is governed by 
the quantum diffusion and the classical equation of motion cannot be 
trusted in this domain. 

If one introduces the ratio $\eta$ between the quantum kicks amplitude
$H/(2\pi)$ and the classical drift $\Mp^2V_\phi/V$, when $x\simeq 1$,
one has 
\begin{equation}
\eta \simeq \dfrac{1}{90\sqrt{30}\pi}
M^2\phizero\Mp^{-3}\left(x-1\right)^{-2} \simeq \dfrac{4\sqrt{10}
}{\pi \sqrt{3} } M^2 \Mp \phizero^{-3} \left(\Nend-N\right)^2,
\end{equation}
where the last equality comes from the approximate trajectory. In
order to estimate the value of $\eta$,  one needs the value of $M$
which is fixed by the amplitude of the CMB anisotropies. With $\xstar$
the observable field value associated with $\Delta\Nstar = \Nend -
\Nstar$, one gets
\begin{equation}
  \left(\frac{M}{\Mp}\right)^4 = 2880\pi^2\dfrac{\Mp^2}{\phizero^2}
  \dfrac{\left(1-\xstar^4\right)^4}
  {\xstar^4\left(1 -\dfrac{2}{3} \xstar^4+\dfrac{1}{5} \xstar^8 \right)^3}
  \frac{\Qrms^2}{T^2}\, .
\end{equation}
In the $\xstar\simeq 1$ approximation, this gives
\begin{equation}
\dfrac{M^4}{\Mp^4} \simeq \dfrac{3}{8} \pi^2 \dfrac{\Qrms^2}{T^2}
\dfrac{\phizero^6}{\Mp^6\left(\Nend-\Nstar\right)^{4}}\,,
\end{equation}
and thus
\begin{equation}
\eta\simeq \sqrt{20\frac{\Qrms^2}{T^2}}\left(\frac{\Nend-N}
           {\Delta\Nstar}\right)^2\, .
\end{equation}
It is quite remarkable that this formula does not depend on $\phizero$
anymore but only on the ratio $(\Nend-N)/\Delta\Nstar$. From
$\Qrms/T\simeq 6\times 10^{-6}$, one has $\Nend- \Nmin \simeq 10^{4}$
in the classical regime~\cite{Allahverdi:2006iq}. For $\phizero\simeq
10^{14}\,\GeV$, one obtains $M\simeq 10^{8}\GeV$, in agreement with what
was announced earlier.

Finally, it can be interesting to write down the approximated
slow-roll parameters at Hubble crossing and in the limit
$\phizero/\Mp\ll 1$. One obtains
\begin{equation}
\label{eq:mssmi:predic}
\epsonestar\simeq\left(\frac{\phizero}{\Mp}\right)^6\frac{1}
{7200\Delta\Nstar^4}\, ,\qquad
\epstwostar\simeq\frac{4}{\Delta\Nstar}\, ,\qquad
\epsthreestar\simeq\frac{1}{\Delta\Nstar}\, ,
\end{equation}
hence
\begin{equation}
r\simeq\left(\frac{\phizero}{\Mp}\right)^6\frac{1}
{450\Delta\Nstar^4}\, ,
\qquad
\nS\simeq 1-\frac{4}{\Delta\Nstar}\, ,\qquad
\alphaS\simeq-\frac{4}{\Delta\Nstar^2}\, .
\end{equation}
They are similar with the typical predictions of the RIPI models [see
  \Eq{eq:ripi:predic}].

  The reheating consistent slow-roll predictions of the MSSMI models
  are displayed in \Fig{fig:CMBMSSMI}. The reheating equation of state
  parameter $\wrehbar$ has been taken to $0$ since the potential is
  quadratic in the vicinity of its minimum. One can check that, in the
  limit $\phizero/\Mp\ll 1$, the first slow-roll parameter is indeed
  extremely small, while the second slow-roll parameter does not
  depend much on $\phizero$.  Remembering that $\phizero/\Mp\simeq
  10^{-4}$, one can see that these models seem to be disfavored by the
  data since they predict a too large deviation from scale
  invariance. In order to better reproduce the constraints on the
  spectral index, these models should be such that $\phizero/\Mp\gg
  1$, for which they become similar to large field models
  (LFI, see \sectionc{sec:lfi}). This can be seen from the
  previous formulas in the limit $x\gg 1$. Unfortunately, such values
  for $\phizero$ are not compatible with the MSSM. Finally, comparing
  \Fig{fig:CMBMSSMI} with \Fig{fig:CMBRIPI}, one can see that the
  general features of MSSMI are very similar to the RIPI ones, and
  that the conclusions drawn here are rather robust against a change
  in $n$ appearing in \Eq{eq:mssmi:pot}.

\subsection{Renormalizable Inflection Point Inflation (RIPI)}
\label{sec:ripi}

\subsubsection{Theoretical Justifications}
\label{subsubsec:theoryripi}

In \sectionc{sec:mssmi} inflation is implemented within the Minimal
Supersymmetric Standard Model (MSSM) around a flat inflection
point. Here, we consider a similar model but with $n=3$ instead of
$n=6$. Such a scenario can emerge in the following situation, see
\Refcs{Allahverdi:2007wt,Allahverdi:2006cx}. Let us consider the MSSM
with three additional superfields $N_i$ representing three
right-handed neutrinos. These fields are singlet under the standard
model gauge group but this one can be extended to
$\mathrm{SU}(3)_{_{\uc}}\times\mathrm{SU}(2)_{_\uL}\times
\mathrm{U}(1)_{_{\uY}}\times \mathrm{U}(1)_{_{\mathrm{B-L}}}$ and
the $N_i$ are assumed to be charged under the extra
$\mathrm{U}(1)_{_\mathrm{B-L}}$. Then, we postulate the following
superpotential
\begin{equation}
W=W_{_\mathrm{MSSM}}+hNH_uL,
\end{equation}
where $h\lesssim 10^{-12}$ in order to explain the neutrino mass,
$m_\nu\simeq \order{0.1}\, \eV$. It follows that $NH_uL$ is a
$D$-flat direction of the potential and we parametrize this direction
by $\phi$. As a consequence, if one now calculates the corresponding
potential, one finds that
\begin{equation}
\label{eq:potripihep}
  V=\frac{1}{2}m_\phi^2\phi^2-\frac{Ah}{6\sqrt{3}}\phi^3+\frac{h^2}{12}\phi^4,
\end{equation}
where, as usual, we have included the soft supersymmetry breaking
terms (since $W\propto \phi^3$, the $A$-term, proportional to
$\phi \partial W/\partial \phi $ is, this time, cubic) and have
minimized $V$ along the angular direction. If $A$ is chosen such that
$A=4m_\phi$, then we have a flat inflection point at
$\phizero=\sqrt{3}m_\phi/h$. A discussion on the fine-tuning required
to get a flat inflection point can be found in \sectionc{sec:gripi},

\subsubsection{Slow-Roll Analysis}
\label{subsubsec:srripi}

We now turn to the slow-roll analysis of the potential given in
\Eq{eq:potripihep}. For this purpose, it is more convenient to
re-write it as
\begin{equation}
\label{eq:ripi:pot}
  V \left(\phi\right) = M^4 \left[ \left(\frac{\phi}{\phizero}\right)^2 -
   \frac{4}{3} \left( \frac{\phi}{\phizero} \right)^3 + \frac{1}{2}
    \left( \frac{\phi}{\phizero} \right)^4 \right],
\end{equation}
where we have defined the quantities $M$ and $\phizero$ by
\begin{equation}
M^4=\frac{1}{2}m_\phi^2\phizero^2, \qquad \phizero=\sqrt{3}\frac{m_\phi}{h}.
\end{equation}
Relevant values of $m_\phi$ range from $100\,\GeV$ to $10\,\TeV$ and
$h\simeq 10^{-12}$. This means
that~\cite{Allahverdi:2007wt,Allahverdi:2006cx}
\begin{equation}
\phizero \simeq 10^{14}\,\GeV,
\end{equation}
a value that turns out to be similar to the one considered in the
MSSMI case (see \sectionc{sec:mssmi}).

\begin{figure}
\begin{center}
\includegraphics[width=\wdblefig]{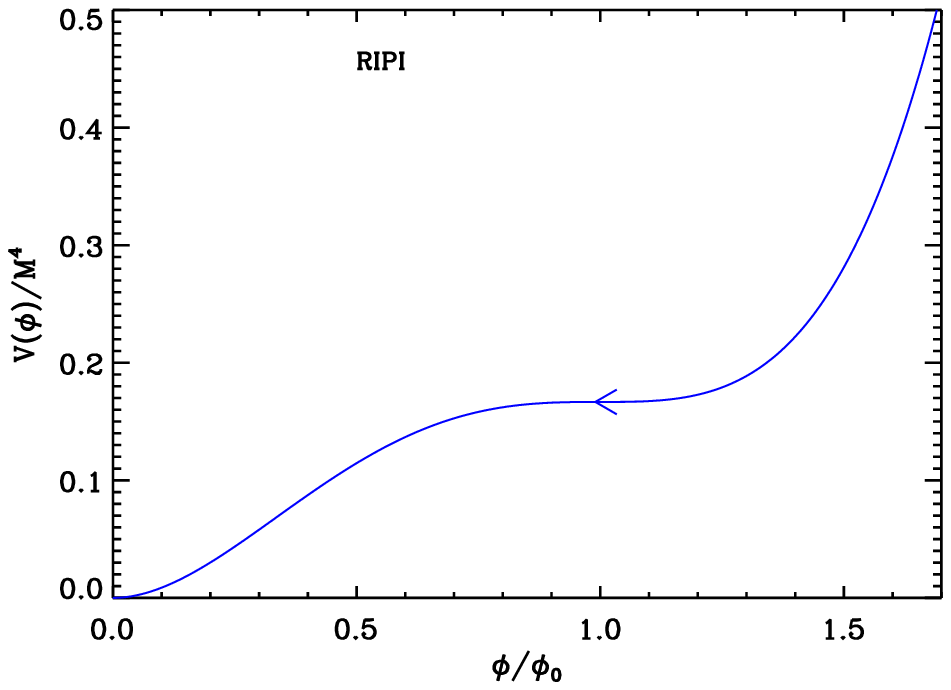}
\includegraphics[width=\wdblefig]{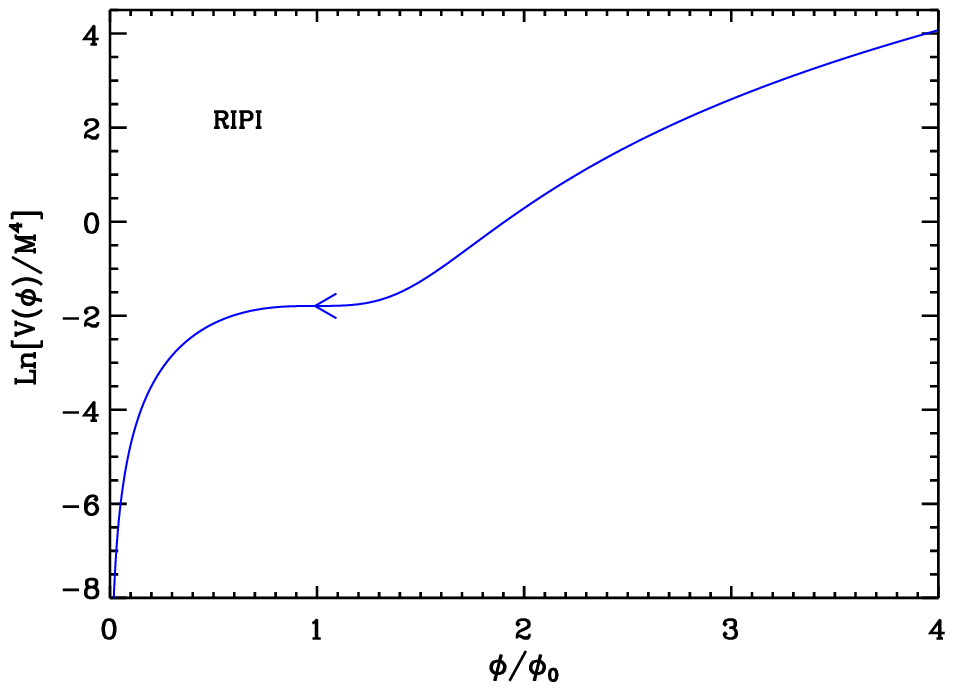}
\includegraphics[width=\wdblefig]{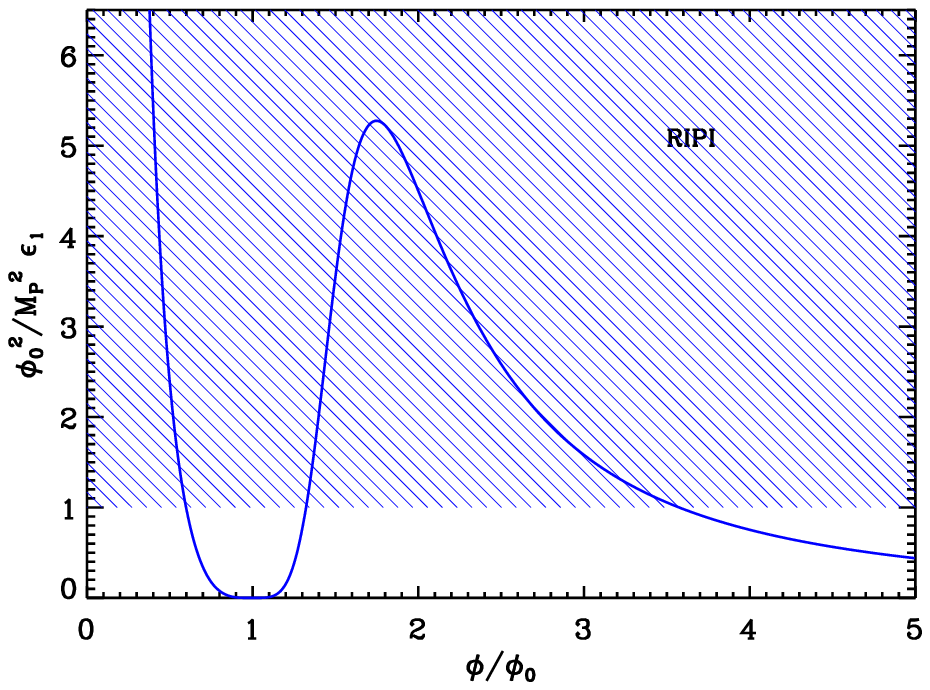}
\includegraphics[width=\wdblefig]{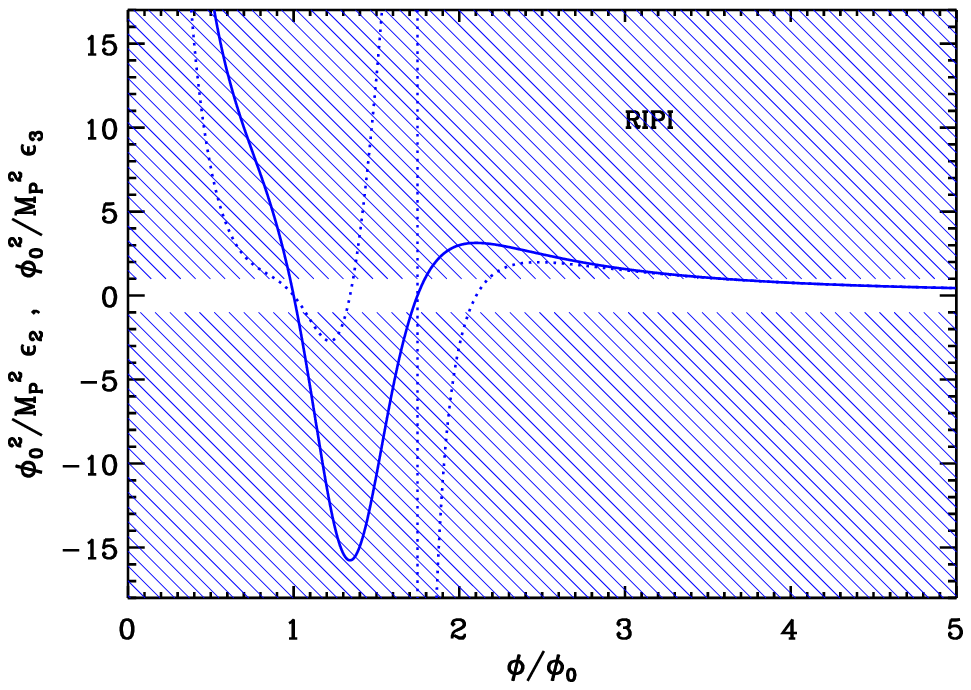}
\caption{Renormalizable Inflection Point Inflation (RIPI).  Top left
  panel: renormalizable inflection point inflation potential as a
  function of $\phi/\phizero$. Top right panel: logarithm of the
  potential, the required flatness of the potential close to its
  inflection point becomes obvious on this plot. Bottom left panel:
  slow-roll parameter $\epsilon_1$ normalized by
  $\Mp^2/\phizero^2$. The shaded area indicates the region in which
  $\epsilon_1>1$ and thus where inflation stops (this has to be
  rescaled for $\phizero \ne \Mp$). Bottom right panel: slow-roll
  parameters $\epsilon _2$ (solid line) and $\epsilon_3$ (dotted
  line), normalized by $\Mp^2/\phizero^2$.}
\label{potRIPI}
\end{center}
\end{figure}

Let us now define the quantity $x$ by the following expression
\begin{equation}
x \equiv \dfrac{\phi}{\phizero}\,.
\end{equation}
The potential is an increasing function of the field \vev, hence
inflation proceeds from the right to the left. It has two inflection
points $\xddVzeroPM$, given by
\begin{equation}
\xddVzeroMinus=\frac{1}{3}\,\quad \textrm{and}\quad
\xddVzeroPlus=1,
\end{equation}
the second one being a flat inflection point [\ie
  $V^\prime\left(\xddVzeroPlus\right)=0$], close to which inflation
takes place. This potential is displayed in \Fig{potRIPI}, together
with its logarithm.

Let us now turn to the slow-roll parameters. The first three Hubble
flow functions in the slow-roll approximation are given by
\begin{equation}
  \epsilon_1 =72\frac{\Mp^2}{\phizero^2}\frac{\left(x-1\right)^4}
  {\left(3x^3-8 x^2 + 6 x \right)^2}\,
  ,\qquad \epsilon_2=24\frac{\Mp^2}{\phizero^2}\left(x-1\right)
  \frac{3x^3-9x^2+10x-6}
  {\left(3x^3-8x^2+6x\right)^2}\, ,\\
\end{equation}
and
\begin{equation}
\begin{aligned}
  \epsilon_3 =& 24\frac{\Mp^2}{\phizero^2} \left(x-1\right) 
  \left(36-144x+246x^2-236x^3+144x^4-54x^5+9x^6\right)
  \nonumber\\& \times
  \left(6x-8x^2+3x^3\right)^{-2} 
  \left(10x-9x^2+3x^3-6\right)^{-1}\,.
\end{aligned}
\end{equation}
Both $\epsilon_1(x)$ and $\epsilon_2(x)$ diverge when the field \vev
goes to $0$, and vanish when the field \vev goes to infinity.  The
first slow-roll parameter $\epsilon_1$ first decreases, vanishes at
$\xddVzeroPlus$ where $\epsilon_2$ vanishes too,
$\xepstwoZeroMinus=\xddVzeroPlus$, then increases to reach a local
maximum at $\xepstwoZeroPlus$ where $\epsilon_2$ vanishes again, and
eventually decreases again. The value of $\xepstwoZeroPlus$ is given
by
\begin{equation}
  \label{eq:rip:xepsilon2=0}
  \xepstwoZeroPlus=1-\frac{1}{3\left(9+\sqrt{82}\right)^{1/3}}
+\frac{1}{3}\left(9+\sqrt{82}\right)^{1/3}\simeq 1.75\, .
\end{equation}
In between these two local extrema of $\epsilon_1$, the second slow
roll parameter $\epsilon_2$ is negative, and it is positive
elsewhere. The value of $\epsilon_1$ at its local maximum,
$\epsonemax$, is given by
\begin{equation}
\begin{aligned}
  \epsilon_1^{\max} 
  \simeq 5.2753\frac{\Mp^2}{\phizero^2}\,. 
 \label{eq:ripi:criticalalpha}
 \end{aligned}
\end{equation}
Therefore, if $\phizero/\Mp\lesssim  2.3$,
inflation can stop by slow-roll violation in the region corresponding
to \vev's larger than that of the second inflection point
$\xepstwoZeroPlus$. Remembering that typically $\phizero\simeq
10^{14}\, \GeV \simeq 4\times10^{-5}\Mp$, this condition is easily
satisfied. In that case, an expression for the \vev at which inflation
ends, $\xepsoneOnePlus$, can be obtained but is does not add much to
the discussion since for reasonable values of $\phizero$, it is
extremely far from the flat inflection point (\eg for $\phizero/\Mp=
10^{-4}$, one has $\xepsoneOnePlus\simeq 28285$). Since the potential
is introduced in order to study inflation in the vicinity of the flat
inflection point, it should be studied in the other regime, as it is
the case for MSSM inflation (see
\sectionc{sec:mssmi}), \ie when inflation takes place between $x=0$
and the second inflection point $\xepstwoZeroMinus$. In that
situation, it ends at
\begin{align}
 \xend = \xepsoneOneMinus & = \frac{1}{9}\frac{\Mp}{\phizero}\Bigg[6\sqrt{2} 
  + 8 \frac{\phizero}{\Mp}+2\left(-36 + 6 \sqrt{2}\frac{\phizero}
  {\Mp} -5 \frac{\phizero^2}
  {\Mp^2}\right) \nonumber  \\ & \times
  \left(216\frac{\phizero}{\Mp} - 99 \sqrt{2} 
  \frac{\phizero^2}{\Mp^2} + 136 \frac{\phizero^3}
  {\Mp^3}-432\sqrt{2}
    \right.\nonumber  \\& \left.
    +27\sqrt{2}\sqrt{-72 \sqrt{2}\frac{\phizero^3}{\Mp^3} 
  +33\frac{\phizero^4}{\Mp^4}-16 \sqrt{2}\frac{\phizero^5}{\Mp^5} + 12 
  \frac{\phizero^6}{\Mp^6}}\right)^{-1/3}
  \nonumber   \\& 
    -\left(216\frac{\phizero}{\Mp} - 99 \sqrt{2} 
  \frac{\phizero^2}{\Mp^2} + 136 \frac{\phizero^3}{\Mp^3}-432\sqrt{2}
      \right.\nonumber  \\& \left.
      +27\sqrt{2}\sqrt{-72 \sqrt{2}\frac{\phizero^3}{\Mp^3} + 
  33\frac{\phizero^4}{\Mp^4} - 16 \sqrt{2}\frac{\phizero^5}{\Mp^5} + 12 
  \frac{\phizero^6}{\Mp^6}}\right)^{1/3}\Bigg]\, .
\end{align}
For $\phizero/\Mp\ll 1$, one can numerically check that this expression is 
very close to the flat inflection point location $\xepstwoZeroMinus$, 
namely
\begin{equation}
\label{eq:ripi:xendappr}
\xend \simeq 1-\sqrt{6\sqrt{2}\frac{\phizero}{\Mp}}\, .
\end{equation}
The whole inflationary stage therefore proceeds in the vicinity of
this point.

The slow-roll trajectory is obtained from \Eq{eq:srtrajectory} and reads
\begin{equation}
\begin{aligned}
\label{eq:ripi:traj}
\Nend-N & =\frac{\phizero^2}{\Mp^2}  \left[
  -\frac{x}{6}+\frac{x^2}{8}
  +\frac{1}{12\left(1-x\right)}
  -\frac{\ln\left(1-x\right)}{12}
 \right. \\ & \left.
 +\frac{\xend}{6}-\frac{\xend^2}{8}
 - \frac{1}{12\left(1-\xend\right)}
  +\frac{\ln\left(1-\xend\right)}{12}\right].
\end{aligned}
\end{equation}
Several remarks are in order. Firstly, from this expression, one can
see that the number of \efolds diverges when the field approaches
the inflection point of the potential. This means that this point is
never crossed and that, if inflation proceeds for \vev's larger than
that of this inflection point, then the field approaches it
asymptotically but never actually reaches it. However, an exact
numerical integration of the equations of motion reveals that, if the
field approaches the inflection point in such a way that the slow-roll
conditions are not satisfied, then it can cross it. This is typically
the case if its speed is large enough. On the other hand, the field
dynamics at the exact location of the inflection point is dominated by
quantum diffusion, and a more careful study must be carried out to
describe what exactly happens there. Following the considerations of
\sectionc{sec:mssmi}, we focus on the inflationary regime only in the
region where the \vev of $\phi$ is smaller than that of the flat
inflection and where deviations from slow-roll and quantum diffusion
plays a negligible role. Since for $\phizero/\Mp\ll 1$ inflation takes
place relatively close to the inflection point, the two last terms of
\Eq{eq:ripi:traj} dominate over the two first ones. In this limit, the
trajectory can be inverted to get
\begin{equation}
\begin{aligned}
\xstar\simeq 1-W^{-1}_0\left\lbrace 
\exp\left[12\left(\frac{\Mp}
{\phizero}\right)^2\Delta\Nstar+\frac{1}
{1-\xend}-\ln\left(1-\xend\right)\right]\right\rbrace .
\end{aligned}
\end{equation}
Making use of \Eq{eq:ripi:xendappr}, and keeping only the dominant terms in 
$\phizero/\Mp$, one obtains
\begin{equation}
\label{eq:ripi:trajappr}
\xstar\simeq 1-\frac{1}{12}\left(\frac{\phizero}
{\Mp}\right)^2\frac{1}{\Delta\Nstar}\, .
\end{equation}
This expression can be useful to determine typical values for the
slow-roll parameters evaluated at Hubble crossing. One obtains
\begin{equation}
\label{eq:ripi:predic}
\epsilon_{1*}\simeq\frac{1}{288}\frac{1}{\Delta\Nstar^4}\frac{\phizero^6}
{\Mp^6}\, ,\qquad \epsilon_{2*}\simeq\frac{4}{\Delta\Nstar}\,\qquad
\epsilon_{3*}\simeq\frac{1}{\Delta\Nstar}\, ,
\end{equation}
hence
\begin{equation}
r\simeq\frac{1}{18}\frac{1}{\Delta\Nstar^4}\frac{\phizero^6}
{\Mp^6}\, ,\qquad \nS-1\simeq-\frac{4}{\Delta\Nstar}\,, \qquad
\alphaS\simeq-\frac{4}{\Delta\Nstar^2}\, .
\end{equation}
One can see that these models typically predict a tiny amount of 
gravitational waves, but a substantial deviation from scale invariance 
$\nS-1\simeq-4/\Delta\Nstar\simeq 0.1$. The similarity with 
\Eqs{eq:mssmi:predic} is obvious.

Finally, the parameter $M$ can be determined from the amplitude of the
CMB anisotropies and the observable field value $\xstar=x(\Nstar)$ by
\begin{equation}
  \left(\frac{M}{\Mp}\right)^4=622080\frac{\Mp^2}{\phizero^2} \pi^2 
  \frac{\left(\xstar-1\right)^4}{\xstar^4\left(3 \xstar^2-8
    \xstar+3\right)^3} \frac{\Qrms^2}{T^2}\, .
\end{equation}
For $\phizero/\Mp\ll 1$, one can make use of \Eq{eq:ripi:trajappr} to
get the approximate expression
\begin{equation}
  \left(\frac{M}{\Mp}\right)^4\simeq 30 \frac{\pi^2}
  {\Delta\Nstar^4}\left(\frac{\phizero}{\Mp}\right)^6\frac{\Qrms^2}{T^2}\, .
\end{equation}
Using the typical value $\phizero\simeq 10^{14}\,\GeV$, one gets
$M/\Mp \simeq 5\times 10^{-11}$.

The reheating consistent slow-roll predictions of the renormalizable
inflection point models are displayed in \Fig{fig:CMBRIPI}. The
reheating equation of state parameter $\wrehbar$ has been taken to $0$
since the potential is quadratic close to its minimum. One can check
that in the limit $\phizero/\Mp\ll 1$, the first slow-roll parameter
is indeed extremely small, while the second slow-roll parameter does
not depend much on $\phizero$. Remembering that $\phizero/\Mp \simeq
10^{-4}$, one can see that these models are disfavored by the CMB data
since they predict a too large deviation from scale invariance. In
order to remain inside the two-sigma confidence intervals, these
models should be such that $\phizero/\Mp\gg 1$, for which they are
close to the large field models (LFI, see
\sectionc{sec:lfi}). However, such values for $\phizero$ are, a
priori, outside the range of validity of the RIPI scenario. Finally,
comparing \Fig{fig:CMBMSSMI} with \Fig{fig:CMBRIPI}, one can see that
the general features of RIPI are very close to the MSSMI ones, and
that the conclusions drawn before are therefore robust against the
precise value of the power index $n$ in \Eq{eq:mssmi:pot}.

\subsection{Arctan Inflation (AI)}
\label{sec:ai}

\begin{figure}
\begin{center}
\includegraphics[width=\wdblefig]{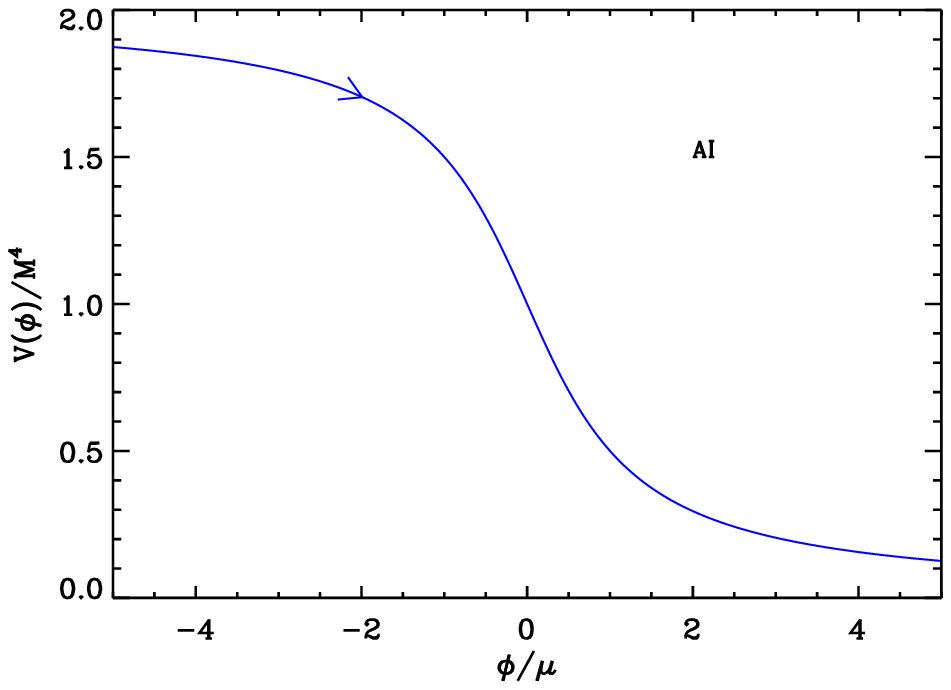}
\includegraphics[width=\wdblefig]{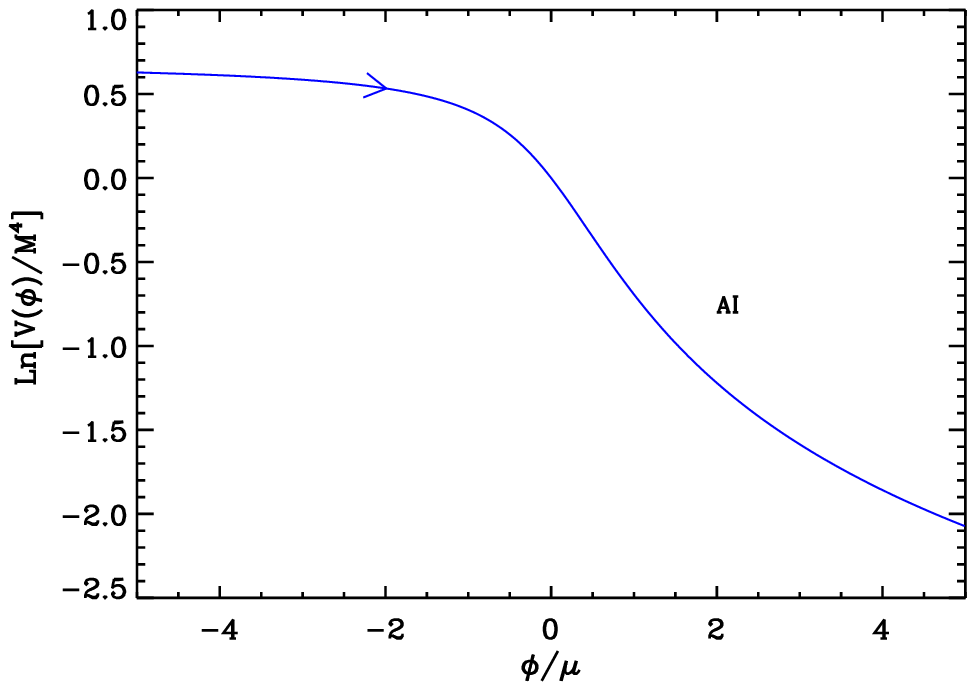}
\includegraphics[width=\wdblefig]{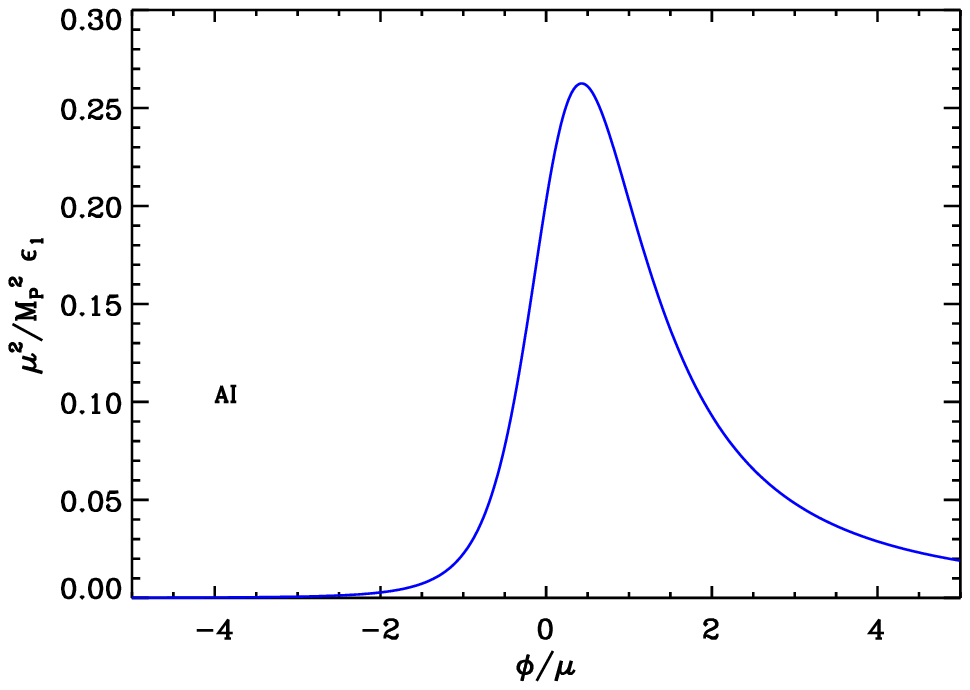}
\includegraphics[width=\wdblefig]{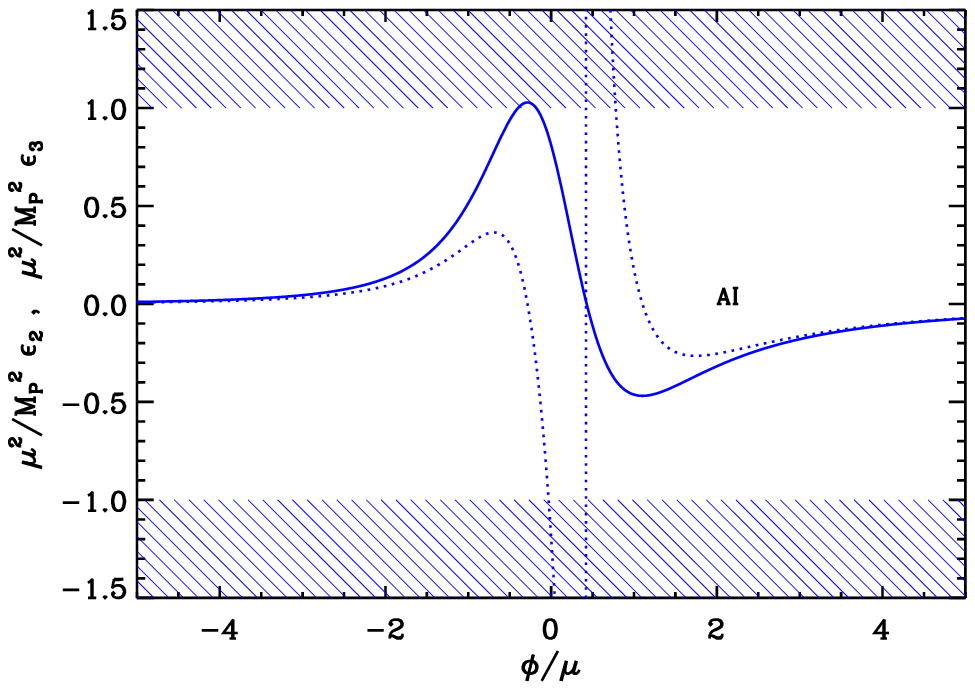}
\caption{Top left panel: Arctan Inflation (AI) potential as a function
  of $\phi/\mu$.  Top right panel: logarithm of the potential.  Bottom
  left panel: slow-roll parameter $\epsilon _1$ rescaled by
  $\Mp^2/\mu^2$ which renders the corresponding expression
  ``universal'', \ie independent of the free parameter $\mu$.  Bottom
  right panel: slow-roll parameters $\epsilon _2$ (solid line) and
  $\epsilon _3$ (dotted line) rescaled by $\Mp^2/\mu^2$ (for the same
  reason as mentioned before).}
\label{potAI}
\end{center}
\end{figure}

This scenario was originally introduced in \Refc{Wang:1997cw} as a toy
model where the equation of state changes rapidly around $\phi=0$. The
potential reads
\begin{equation}
\label{eq:ai:pot}
  V\left(\phi\right) = M^4\left[1-\frac{2}{\pi}
    \arctan\left(\frac{\phi}{\mu}\right)
  \right] ,
\end{equation}
and depends on one free parameter, $\mu$. This model was considered in
order to test the reliability of different computational methods and
schemes of approximation used in the calculations of the inflationary
cosmological perturbations power spectrum, see
\Refc{Wang:1997cw}. More precisely, in \Refc{Leach:2002ar}, it was
also used to study with which accuracy the first and second slow-roll
order power spectra can approximate the actual power spectrum of the
fluctuations in the case where the underlying model has both quite
large tilt and running. This potential was considered again in
\Refcs{Drees:2011yz,Drees:2012sz} in order to study whether it can
lead to the formation of long-lived primordial black holes. In the
following slow-roll analysis, $\mu$ will be viewed as a free parameter
with no restricted range of variation. Let us notice, however, that
since it characterizes the typical \vev at which inflation takes
place, it could also be limited to the sub-Planckian regime if one
wants inflation to proceed in a small field regime. As a matter of
fact, it will be shown below that this needs to be the case to end
inflation by slow-roll violation.

The potential~(\ref{eq:ai:pot}), as well as its logarithm, are
displayed in \Fig{potAI}. They are decreasing functions of the field
and, hence, inflation proceed from the left to the right, in the
direction specified by the arrow in \Fig{potAI}. 

Let us now compute the three first slow-roll parameters. If one
defines $x\equiv\phi/\mu$, their expressions are given by
\begin{equation}
\begin{aligned}
  \epsilon_1&= \frac{\Mp^2}{\mu^2}\frac{2}{\left(1+x^2\right)^2 \left(
      \pi-2\arctan x \right)^2}\,, \qquad
\epsilon_2 &= 8\frac{\Mp^2}{\mu^2}\frac{1-\pi x+2 x
  \arctan x }
        {\left(1+x^2\right)^2\left(\pi-2\arctan x \right)^2}\,,
\end{aligned}
\end{equation}
and
\begin{equation}
\begin{aligned}
\epsilon_3 &= 2\frac{\Mp^2}{\mu^2}\left[ -4+6\pi x+\pi^2\left(1-3
  x^2\right) +4\left(3\pi x^2-3 x -\pi\right)\arctan x 
  \right. \\ & + \left. 4 \left(1-3x^2\right) \arctan^2 x 
\right] \left[
\left(1+x^2\right)^2\left(\pi-2\arctan x \right)^2
 \left(-1+\pi x-2 x \arctan x \right) \right]^{-1}.
\end{aligned}
\end{equation}
They are displayed in \Fig{potAI}.  The first slow-roll parameter
$\epsilon_1$ increases during inflation, reaches a maximum at
$\xepsoneMax$ and then decreases. Whether inflation can stop by
violation of slow-roll or not depends on the value of $\epsilon_1$ at
its maximum: $\epsonemax$. This value is a solution of the following
equation
\begin{equation}
2 \xepsoneMax \arctan\left(\xepsoneMax\right)+1=\pi \xepsoneMax,
\end{equation}
which can only be solved numerically. One gets $\xepsoneMax\simeq
0.428978$, from which one deduces that
\begin{equation}
\epsonemax\simeq 0.262531 \frac{\Mp^2}{\mu^2}\,.
\end{equation}
Therefore, in order for inflation to end by slow-roll violation, one
needs to work under the assumption that $\mu/\Mp<0.512378$. In that
case, inflation proceeds along the plateau located at values of $x$
such that $x<\xepsoneMax$, in the direction specified by the arrow in
\Fig{potAI} (\ie from the left to the right). Otherwise, if one wants
inflation to occur in other parts of the potential and/or for values
of $\mu$ such that $\mu/\Mp>0.512378$, another mechanism needs to be
consider in order to stop it (typically, we imagine a tachyonic
instability in another direction in field space). This means that we
also need to introduce an extra parameter $\xend$ which gives the
location of the \vev at which the tachyonic instability is
triggered. Let us remark that we could also consider a model where the
inflaton starts at $x<\xepsoneMax$, then crosses the region where
$\epsilon_1$ has its maximum and then causes the end of inflation by
tachyonic instability. This case would give a bump in the power
spectrum and, clearly, cannot be properly described in the slow-roll
framework. In this article, we restrict ourselves to the first version
of the scenario mentioned above. In this situation $\xend$ is given by
the smallest solution of the equation $\epsilon_1=1$ and needs to be
computed numerically. Before inflation stops, one can see in
\Fig{potAI} that the second slow-roll parameter $\epsilon_2$ reaches a
maximum, the location of which can be numerically computed to be
$\xepstwoMax\simeq -0.28539 <\xepsoneMax$. At this point, one has
$\epsilon_2^{\mathrm{max}}\simeq 1.02827\Mp^2/\mu^2>\epsonemax$. As a
consequence, the slow-roll approximation breaks down before the end of
inflation. This conclusion is reinforced by the fact that $\epsilon_3$
diverges at $\xepsoneMax$. This means that the last \efolds of
inflation cannot be properly described in the slow-roll framework.

Let us now turn to the slow-roll trajectory. It can be integrated
exactly and yields the following expression
\begin{equation}
\begin{aligned}
  \Nend-N = \frac{\mu^2}{\Mp^2}&\left[\frac{\pi \xend}{2} + 
  \frac{\xend^2}{6}
    + \frac{\pi \xend^3}{6} -\left(1+\frac{\xend^2}{3}\right) \xend
    \arctan \xend  + \frac{1}{3} \ln 
    \left(1+\xend^2\right)\right. \\ & \ -
    \left. \frac{\pi x}{2}-\frac{x^2}{6} -\frac{\pi x^3}{6} +\left( 1 + 
    \frac{x^2}{3}\right)
    x \arctan x  + \frac{1}{3}\ln\left( 1+x^2
    \right) \right],
\end{aligned}
\end{equation}
where $\Nend$ is the number of \efolds at the end of inflation. In
the vacuum dominated approximation where the potential is just given
by $V(\phi)\simeq M^4$, this trajectory can be approximated by
$\Nend-N = \mu^2/ \Mp^2 (\pi\xend + \xend^2/6 + \pi x^3/3 - \pi
x-x^2/6-\pi x^3/3)$, which can be inverted exactly if needed.  This
formula is valid if $\mu/\Mp\ll 1$, since in that case,
$\xend\simeq-\sqrt{\Mp/\left(\mu\pi\sqrt{2}\right)}\ll -1$. Under this
assumption, one has
$\xstar^3\simeq-3\Mp^2/\left(\pi\mu^2\right)\Delta\Nstar$, from which one
can approximate the values of the three first Hubble flow parameters
at Hubble radius crossing
\begin{equation}
\begin{aligned}
\epsilon_{1*}=\frac{\left(\mu/\Mp\right)^{2/3}}
{2\left(\pi\Delta\Nstar^2\right)^{2/3}}\, ,\qquad
\epsilon_{2*}=\frac{4}{3\Delta\Nstar}\, ,\qquad
\epsilon_{3*}=\frac{1}{\Delta\Nstar}\, ,
\end{aligned}
\end{equation}
Then, one can calculate the tensor-to-scalar ratio, the spectral index
and the running. One obtains the following expressions
\begin{equation}
\begin{aligned}
r=\frac{8\left(\mu/\Mp\right)^{2/3}}
{\left(\pi\Delta\Nstar^2\right)^{2/3}}\, ,\qquad
\nS-1=-\frac{4}{3\Delta\Nstar}\simeq-0.03\, ,\qquad
\alphaS=-\frac{4}{3\Delta\Nstar^2}\simeq-5\times 10^{-4}\, .
\label{eq:ai:roughpredictions}
\end{aligned}
\end{equation}
These formulas are in agreement with the consistency relation
$\alphaS=-3/4\left(\nS-1\right)^2$ obtained in \Refc{Drees:2011yz}.

Finally, it is interesting to estimate the energy scale $M$ from the
CMB normalization. This leads to
\begin{equation}
  \left( \frac{M}{\Mp}\right)^4=\frac{2880 \pi^3 \Mp^2/\mu^2}
       {\left(1+\xstar^2\right)^2\left(\pi-2\arctan \xstar \right)^3}
       \frac{\Qrms^2}{T^2}\, .
\end{equation}
Under the vacuum dominated approximation ($\mu/\Mp\ll 1$), the above
equation can be re-expressed as
\begin{equation}
  \left( \frac{M}{\Mp}\right)^4\simeq\frac{40\times 3^{2/3}\pi^{4/3}}
  {\Delta\Nstar}\left(\frac{\mu}{\Mp}\right)^{2/3}
       \frac{\Qrms^2}{T^2}\, .
\end{equation}
The requirement $M<\Mp$ is always satisfied form sub-Planckian values
of $\mu$. The typical value $M/\Mp\simeq 10^{-3}$ corresponds to
$\mu/\Mp\simeq 10^{-2}$.

The slow-roll predictions of the AI models are displayed in
\Fig{fig:CMBAI}, in the range $\mu/\Mp<0.512378$ (so that inflation
can end by slow-roll violation). The reheating equation of state
parameter $\wrehbar$ has been taken to be $0$ but since there is no potential
minimum around which the inflaton field can oscillate at the end of
inflation, this parameter is a priori unspecified. One can see that
this model typically predicts a small amount of gravitational waves,
and a deviation from scale invariance which is in accordance with the
observations. The predictions in the planes $(\nS,r)$ are
qualitatively well described by the vacuum dominated
analysis presented before, see \Eq{eq:ai:roughpredictions}.

\subsection{Constant \texorpdfstring{$\nS$}{nS} A Inflation (CNAI)}
\label{sec:cnai}

\begin{figure}
\begin{center}
\includegraphics[width=\wdblefig]{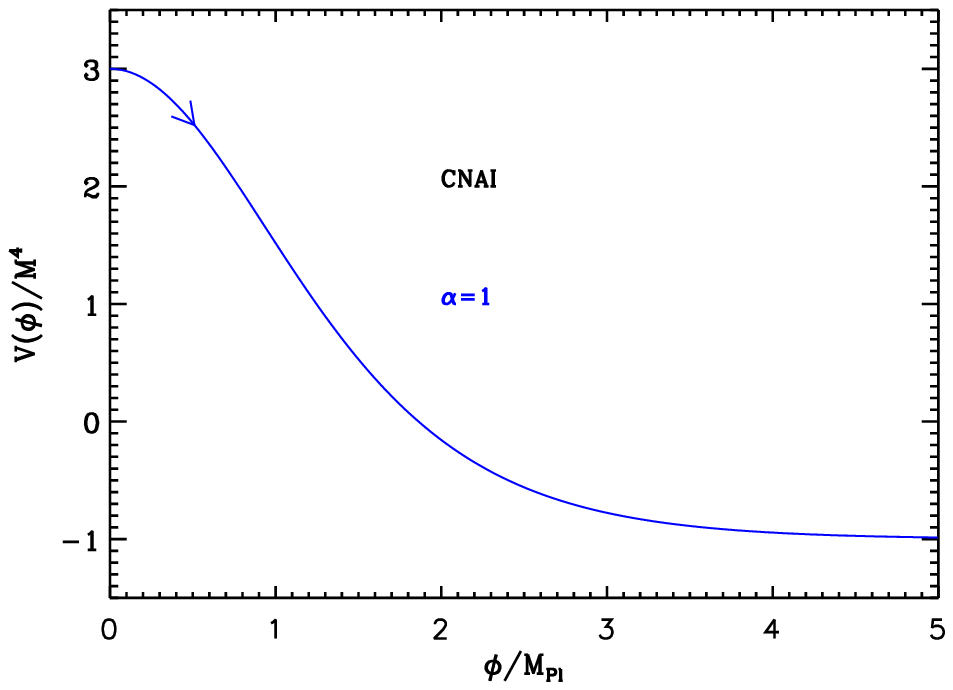}
\includegraphics[width=\wdblefig]{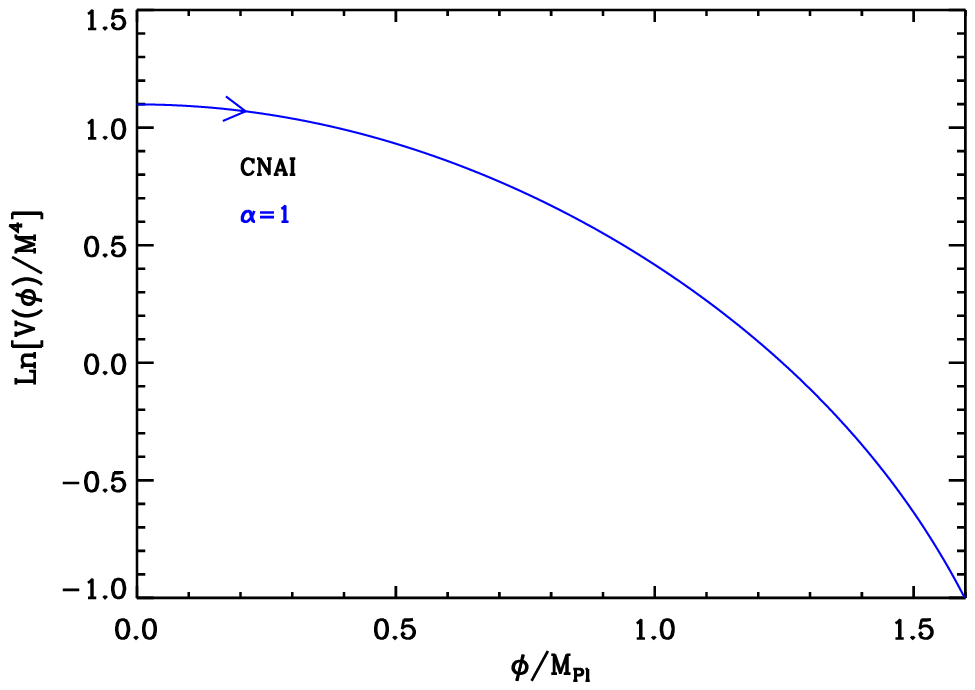}
\includegraphics[width=\wdblefig]{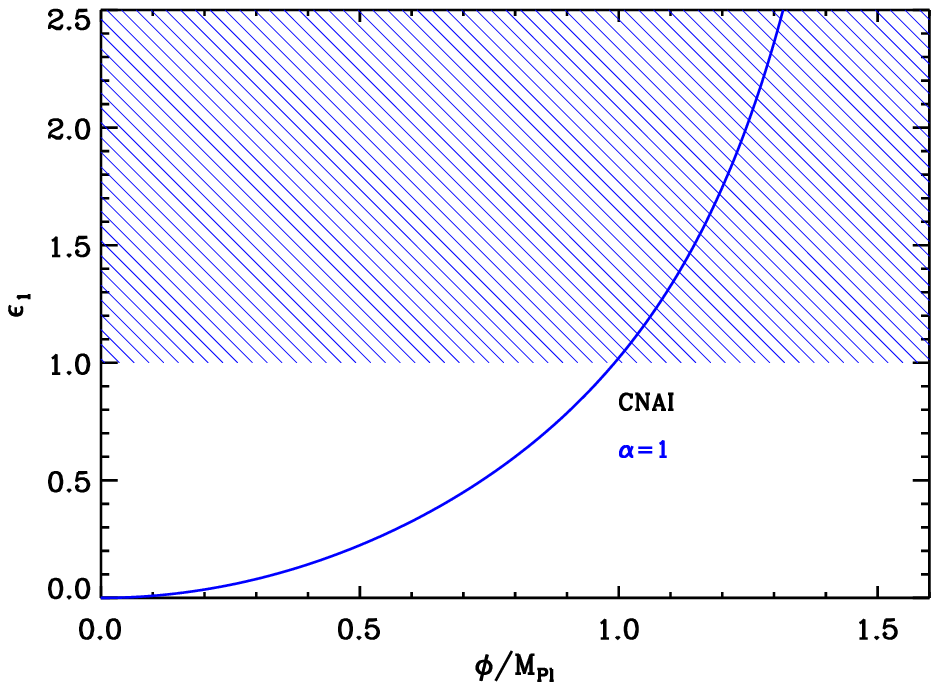}
\includegraphics[width=\wdblefig]{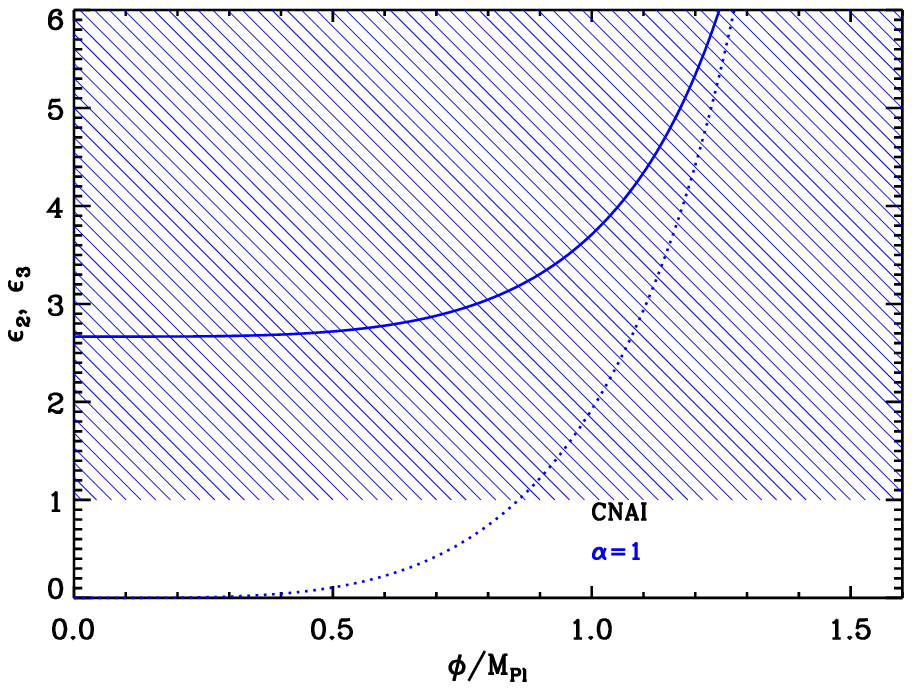}
\caption{Constant $\nS$ A Inflation (CNAI) potential and slow-roll
  parameters versus the vacuum expectation value of the inflaton
  field.  Top left panel: Constant $\nS$ A Inflation potential for
  $\alpha=1$. Top right panel: logarithm of the potential for the same
  value of $\alpha$. Bottom left panel: slow-roll parameter $\epsilon
  _1$ (same value of $\alpha$): inflation stops when $\epsilon_1=1$ in
  this model. Bottom right panel: slow-roll parameters $\epsilon _2$
  and $\epsilon_3$ ($\alpha=1$).}
\label{potcnai}
\end{center}
\end{figure}

This class of models is designed in order to produce power spectra
with constant spectral index. It was studied for the first time in
\Refc{Vallinotto:2003vf}. The rational behind this approach is that,
so far, no evidence for a significant running has been found in the
cosmological data. Since, from a Bayesian point of view, one should
avoid introducing parameters that are unnecessary in order to
reproduce the observations, it makes sense to consider models which
lead to exact power-law power spectra. This is of course the case for
power-law inflation as discussed in \sectionc{sec:pli} and we will see
other examples in \sectioncs{sec:cnbi}, \ref{sec:cnci} and
\ref{sec:cndi}. In fact, in \Refc{Vallinotto:2003vf}, a systematic
analysis of potentials that yield constant spectral index was carried
out. It was found that the following potential belongs to this
category of models
\begin{equation}
  V\left(\phi\right) = M^4\left[3-\left(3+\alpha^2 \right) \tanh^2
    \left( \frac{\alpha}{\sqrt{2}} \frac{\phi}{\Mp} \right) \right],
\end{equation}
where $\alpha$ is a positive massless parameter (denoted $n_0^2$ in
\Refc{Vallinotto:2003vf}) and, in this section, we study this
case. This potential is represented in \Fig{potcnai} and, since it is
symmetrical under the transformation $\phi\rightarrow -\phi$, only the
$\phi>0$ part is displayed. The potential is a decreasing function of
the field \vev and, therefore, inflation proceeds from the left to the
right. It is positive provided $\phi<\phizero$, where
\begin{equation}
\frac{\phizero}{\Mp}=\frac{\sqrt{2}}{\alpha} \artanh
\left(\sqrt{\frac{3}{3+\alpha^2}} \right).
\end{equation}
There is no value of $\alpha$ for which the potential is always
positive. Defining $x=\phi/\Mp$, the slow-roll parameters are given by
\begin{equation}
\label{eq:eps1cnai}
  \epsilon_1 =
  \frac{4\alpha^2\left(3+\alpha^2\right)^2\tanh^2\left(\frac{\alpha
      x}{\sqrt{2}}\right)} {\left[ 6 +
      \alpha^2-\alpha^2\cosh\left(\sqrt{2}\alpha x \right)
      \right]^2}\, ,
\end{equation}
\begin{equation}
  \epsilon_2 = \frac{2\alpha^2 \left(3 + \alpha^2\right)
    \left[12+\alpha^2 -2\alpha^2\cosh\left(\sqrt{2}\alpha
      x\right)+\alpha^2\cosh\left(2\sqrt{2}\alpha x \right) \right]}
          {\left[6+\alpha^2-\alpha^2\cosh\left(\sqrt{2}\alpha
              x\right)\right]^2\cosh^2\left(\frac{\alpha x}{\sqrt{2}}
            \right)}\,,
\end{equation}
\begin{equation}
\begin{aligned}
  \epsilon_3 & = 2 \alpha^2 \left(3+\alpha^2\right)
  \tanh^2\left(\frac{\alpha}{\sqrt{2}}x\right) \left[ 6\left(
    -24+2\alpha^2 - \alpha^4\right) +
    \left(120\alpha^2+7\alpha^4\right)\cosh\left(\sqrt{2}\alpha
    x\right) \right. \\ & \left. - 2 \alpha^2 \left(\alpha^2-6\right)
    \cosh\left(2\sqrt{2} \alpha x\right) +
    \alpha^4\cosh\left(3\sqrt{2}\alpha x\right)\right] \\ & \times
  \left[6+\alpha^2-\alpha^2\cosh\left(\sqrt{2}\alpha
    x\right)\right]^{-2} \left[ 12 + \alpha^2 - 2\alpha^2
    \cosh\left(\sqrt{2} \alpha x\right) +
    \alpha^2\cosh\left(2\sqrt{2}\alpha x \right) \right]^{-1}.
\end{aligned}
\end{equation}
These slow-roll parameters are displayed in \Fig{potcnai}. They all
increase as inflation proceeds and diverge when the field approaches
$\phizero$. Hence inflation ends by slow-roll violation. Notice that
the equation $\epsilon_1=1$ can be solved analytically. If we define
$y\equiv \sinh^2(\alpha x /\sqrt{2})$, then one has to solve the
following cubic equation $\alpha^4y^3+(\alpha^4-6\alpha^2)y^2
+[9-6\alpha^2-\alpha^2(3+\alpha^2)]y+9=0$. The relevant solution reads
\begin{equation}
\label{eq:phiendcnai}
\yend=\frac{6-\alpha^2}{3\alpha^2}-\frac{1-i\sqrt{3}}{3 
\times 2^{1/3}}(3+\alpha^2)^2(1+3\alpha^2)
P^{-1/3}-\frac{1+i\sqrt{3}}{6 \times 2^{1/3}\alpha^4}P^{1/3},
\end{equation}
where we have defined $P$ by
\begin{align}
P\equiv &-\alpha^6\left(3+\alpha^2\right)^2\left(6-52 \alpha^2+9\alpha^4\right)
\nonumber \\ &
+\sqrt{-27\alpha^{14}\left(3+\alpha^2\right)^4
\left(36-60\alpha^2+96\alpha^4+25\alpha^6+4\alpha^8\right)}\,.
\end{align}
The slow-roll parameters $\epsilon_1$ and $\epsilon_3$ both vanish
when the field \vev goes to $0$, whereas $\epsilon_2$ has a
non-vanishing minimum value, given by $\epsilon_2\rightarrow
2\alpha^2\left(3+\alpha^2\right)/3$ when $x=0$. Therefore, if $\alpha$
is larger than some maximum value
\begin{equation}
\alphamax=\sqrt{\frac{1}{2}\left(\sqrt{15}-3\right)}\simeq 0.66,
\end{equation}
then $\epsilon_2$ is larger than $1$ in the whole inflationary regime
and the slow-roll approximation does not hold. It is therefore
necessary to work under the assumption $\alpha<\alphamax$ which we
assume in the following.

Let now us check that the spectral index
$\nS-1=-2\epsilon_1-\epsilon_{2}$ (at first order in slow-roll), can
be made constant, as announced previously. Expanding the slow-roll
parameters $\epsilon_1$ and $\epsilon_2$ in small values of $\alpha$,
and crucially assuming that $\alpha \xstar$ remains small, one obtains
$\epsilon_1= \order{\alpha^4}$ and $\epsilon_2 = 2\alpha^2 +
\order{\alpha^4}$, so that $\nS-1=-2\alpha^2+
\order{\alpha^4}$. Therefore, the corresponding expression is indeed a
constant (\ie does no depend on $\phistar$). Since we have $\vert
\nS-1\vert \ll 1$, this implies that $\alpha$ should be small which is
consistent with the condition $\alpha<\alphamax$ derived above.

\begin{figure}
\begin{center}
\includegraphics[width=\wsingfig]{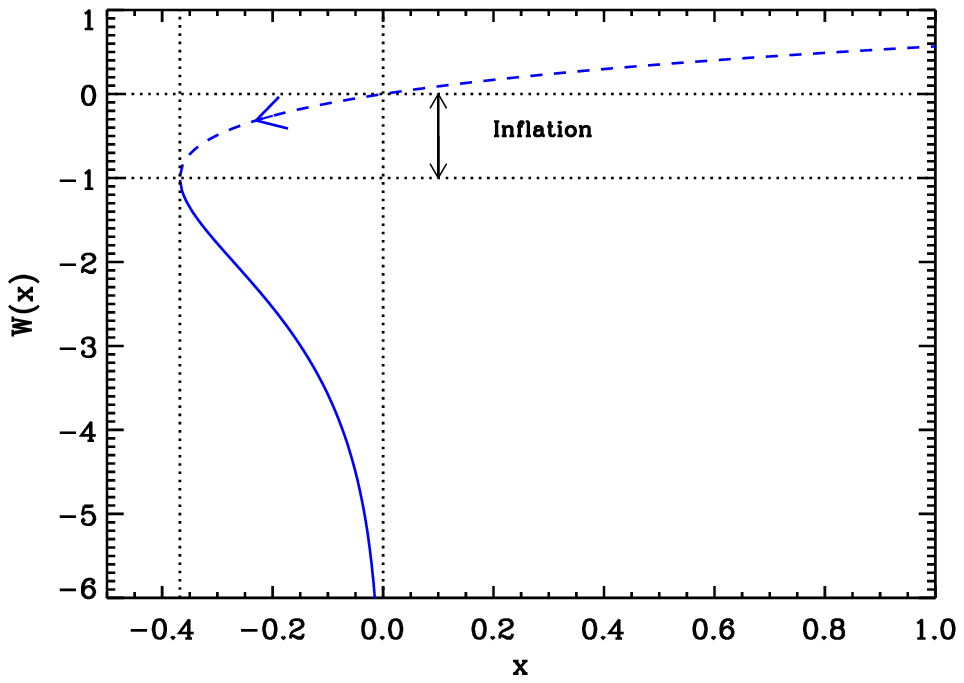}
\caption{Lambert functions $\Lambert0(x)$ (dashed line) and
  $\Lambert{-1}(x)$ (solid line). During CNAI inflation, inflation
  proceeds along the ``$0$'' branch in the direction specified by the
  arrow on the figure.}
\label{fig:cnai:lambert}
\end{center}
\end{figure}

Let us now study the slow-roll trajectory of the system. This one can
be integrated exactly leading to the following formula
\begin{eqnarray}
  N-\Nend & = & \frac{1}{\alpha^2 \left(3+\alpha^2\right)} \left \lbrace
    3\ln \left[\sinh\left(\frac{\alpha}{\sqrt{2}}x\right)\right] -
    \frac{\alpha^2}{2}
    \sinh^2\left(\frac{\alpha}{\sqrt{2}} x \right)
  \right.\nonumber\\& &\left.
    -3 \ln\left[\sinh \left(\frac{\alpha}{\sqrt{2}}\xend\right)
    \right] + \frac{\alpha^2}{2} \sinh^2 \left(\frac{\alpha}{\sqrt{2}}
      \xend\right)\right\rbrace .
\end{eqnarray}
Moreover, this trajectory can be inverted which allows us to
explicitly express the \vev of the inflaton field in terms of the
\efolds number. One obtains
\begin{equation}
\label{eq:trajeccnai}
\begin{aligned}
  x & = \frac{\sqrt{2}}{\alpha} \arsinh \left[ -\frac{3}{\alpha^2}
    \Lambert{0}\left(-\frac{\alpha^2}{3} \exp
      \left\lbrace\frac{2}{3}\alpha^2\left(3+\alpha^2\right)
        \left(N-\Nend\right)
      \right. \right. \right. \\
  & + \left. \left. \left. 2 \ln\left[ \sinh\left(
            \frac{\alpha}{\sqrt{2}} \xend\right)\right] -
        \frac{\alpha^2}{3} \sinh^2 \left(\frac{\alpha}{\sqrt{2}}
          \xend\right) \right \rbrace \right) \right]^{1/2},
\end{aligned}
\end{equation}
where $\Lambert{0}$ is the $0$ branch of the Lambert function as
required since $x\left(N\right)$ is an increasing function of $N$. It
is displayed in \Fig{fig:cnai:lambert} where the CNAI trajectory takes
place between $\phi/\Mp=0$ at the origin of the plot, and
$x=\phizero/\Mp$ at the junction between the $-1$ branch and the $0$
branch.

The slow-roll predictions of the CNAI models are displayed in
\Fig{fig:CMBCNAI}.  When $\alpha$ is small (but not too small), the
value of $\nS$ is indeed constant (and compatible with the
considerations presented above) but, unfortunately, too far from scale
invariance to be compatible with CMB data. When $\alpha\ll 10^{-1}$,
the predictions become roughly compatible with the data but, clearly,
$\nS$ is no longer constant and no longer given by $-2\alpha^2$. At
first sight, this is surprising since we expect the spectral index to
tend towards $-2\alpha^2 $ when $\alpha$ goes to zero (see above). In
order to understand this point, let us remark that, in the limit where
$\alpha$ vanishes, one can expand \Eq{eq:phiendcnai} to find $\yend
\simeq 3/\alpha^2-3/\alpha+\calO\left(\alpha\right)$ (the term at
order $\alpha^0$ is absent and this plays an important role in what
follows). This leads to $\xend \simeq (\sqrt{2}/\alpha)\ln\left(2
\sqrt{3}/\alpha\right)-1/\sqrt{2}+\calO\left(\alpha\right)$. Notice
that this last equation is compatible with the behavior of the first
Hubble-flow parameter~(\ref{eq:eps1cnai}) in the vicinity of
$\phizero$: $\epsilon_1\simeq
\Mp^2/[2(\phi-\phizero)^2]$. Therefore, the expression of $\xend $
found before corresponds in fact to writing $\epsilon_1=1$ with this
approximated $\epsilon_1$. Then, using the slow-roll
trajectory~(\ref{eq:trajeccnai}), one gets
\begin{equation}
\label{eq:sinhxstarcnai}
\sinh^2\left(\frac{\alpha \xstar}{\sqrt{2}}\right)
=-\frac{3}{\alpha^2}\Lambert{0}\left(-\frac{\alpha^2}{3}\ee^{-2A/3}\right),
\end{equation} 
where $A$ is given by the following expression
\begin{equation}
A\equiv \alpha^2\left(3+\alpha^2\right)\Delta \Nstar
-3\ln \left[ \sinh\left(\frac{\alpha \xend}{\sqrt{2}}\right)\right]
+\frac{\alpha^2}{2}
\sinh^2\left(\frac{\alpha \xend}{\sqrt{2}}\right).
\end{equation}
This quantity can be expanded in $\alpha$ using the equation for $\yend$
derived above and, at leading order, one obtains
\begin{equation}
\label{eq:2A/3}
-\frac{2}{3}A\simeq -\frac{2}{3}\alpha^2\Delta \Nstar
+\ln \left(\frac{3}{\alpha^2}\right)-1-\frac{\alpha^2}{2}.
\end{equation}
For simplicity, the last term in the previous expression can be
ignored since $2\Delta \Nstar\gg 1/2$. It follows that, introducing
the formula for $-2A/3$ into \Eq{eq:sinhxstarcnai}, one arrives at
\begin{equation}
\sinh^2\left(\frac{\alpha \xstar}{\sqrt{2}}\right)
=-\frac{3}{\alpha^2}\Lambert{0}\left(-\frac{1}{\ee}
\ee^{-2\alpha^2\Delta \Nstar}\right).
\end{equation} 
If we ignore the exponential in the argument of the Lambert function
(since $\alpha \ll 1$) and use the identity
$\arsinh(x)=\ln(x+\sqrt{x^2+1})$, one finally arrives at
\begin{equation}
\alpha \xstar \underset{\alpha \rightarrow 0}{\sim} \sqrt{2}\ln \left( \dfrac{2\sqrt{3}}{\alpha}
\right).
\end{equation}
We now understand why, in the limit $\alpha \rightarrow 0$, the
spectral index is no longer constant. The naive expression $\nS\simeq
-2\alpha^2$ is obtained by expanding the expressions of $\epsilon_1$
and $\epsilon_2$ in $\alpha$, including the hyperbolic function of
argument $\alpha \xstar$. But we have just shown that, when $\alpha
\ll 1$, $\alpha \xstar$ is not small and, therefore, the Taylor
expansion of those terms is no longer justified. This is why, in
\Fig{fig:CMBCNAI}, we see a deviation from $\nS$ constant at very
small values of $\alpha$. In fact, this questions the interest of this
model since the condition of constant spectral index is obtained only
for values of $\nS$ that are already ruled out by the CMB data. On the
other hand, when $\alpha\ll 1$, the model seems compatible with the
data and, therefore, represents a legitimate inflationary scenario
even if the spectral index is not constant in this case.

Finally, it is also interesting to study the energy scale at which
inflation takes place in this model. The CMB normalization gives
\begin{equation}
\label{eq:cobecnai}
  \left(\frac{M}{\Mp}\right)^4 = \frac{11520\pi^2\alpha^2 \left(\alpha^2
      + 3\right)^2
    \sinh^2\left(\frac{\alpha}{\sqrt{2}}\xstar\right)}
  {\left[\alpha^2+6-\alpha^2\cosh\left(\sqrt{2}\alpha \xstar\right)\right]^3}
  \frac{\Qrms^2}{T^2}\, .
\end{equation}
Since we have established the expression of $\xstar$ above, it is
sufficient to use it in the above formula. We have, however, to be
careful about the calculation of the denominator. Indeed, if we
neglect again the exponential in the argument of the Lambert function,
\Eq{eq:sinhxstarcnai}, then $\sinh^2(\alpha\xstar/\sqrt{2})\simeq
3/\alpha^2$ and the denominator in \Eq{eq:cobecnai}
vanishes. Therefore, one needs to evaluate the Lambert function more
precisely and to keep the corrections proportional to $\Delta
\Nstar$. This can be done with the help of Eq.~(33) of
\Refc{DBLP:journals/corr/abs-1209-0735} which implies that $
\sinh^2(\alpha\xstar/\sqrt{2})\simeq 3/\alpha^2-6\sqrt{\Delta
  \Nstar}/\alpha$. Using this expression, one arrives at
\begin{equation}
\frac{M}{\Mp}\simeq 0.016\, \alpha^{-3/4}\left(\Delta \Nstar\right )^{-3/8}.
\end{equation}
For an order of magnitude estimate, one can use the fiducial value
$\Delta \Nstar\simeq 55$. This leads to $M/\Mp\simeq 0.0035 \,
\alpha^{-3/4}$. Requiring $M<\Mp$ puts a lower bound on the parameter
$\alpha$, namely $\alpha\gtrsim 5\times 10^{-4}$. This roughly
corresponds to the range studied in \Fig{fig:CMBCNAI}. 

\subsection{Constant \texorpdfstring{$\nS$}{nS} B Inflation (CNBI)}
\label{sec:cnbi}

\begin{figure}
\begin{center}
\includegraphics[width=\wdblefig]{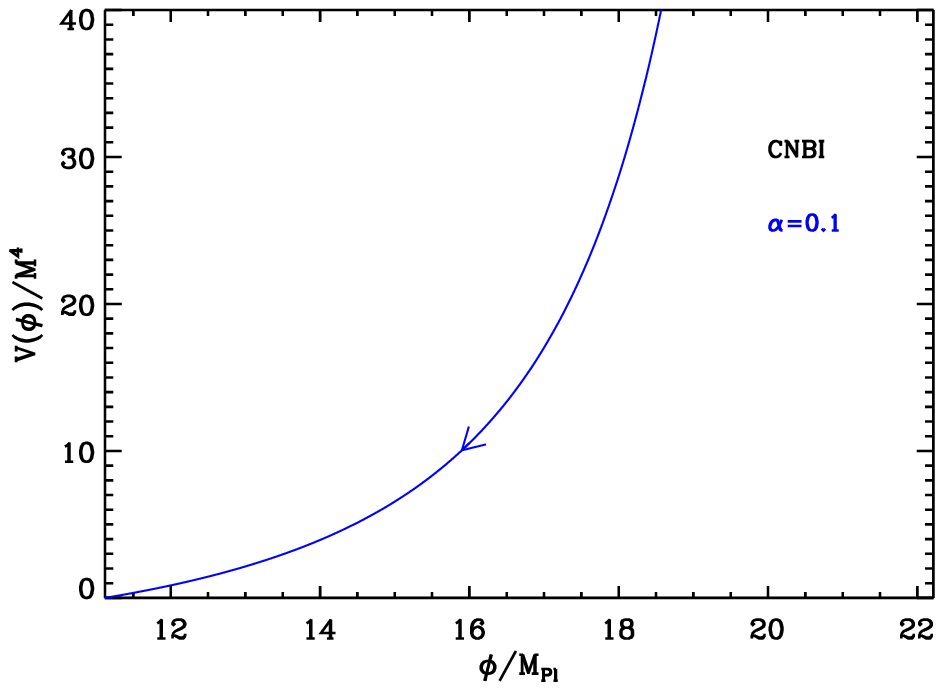}
\includegraphics[width=\wdblefig]{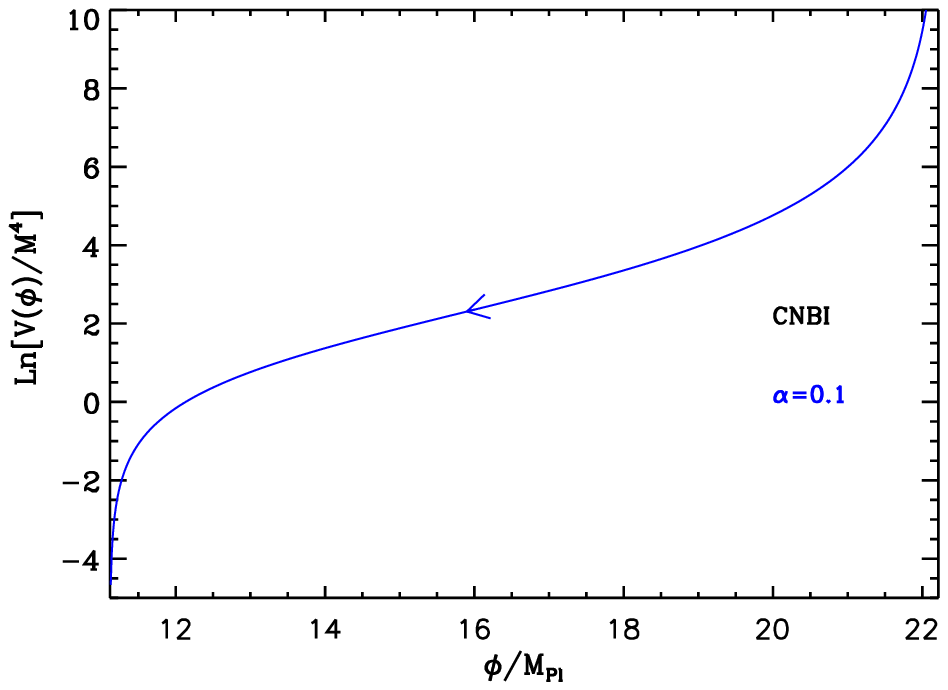}
\includegraphics[width=\wdblefig]{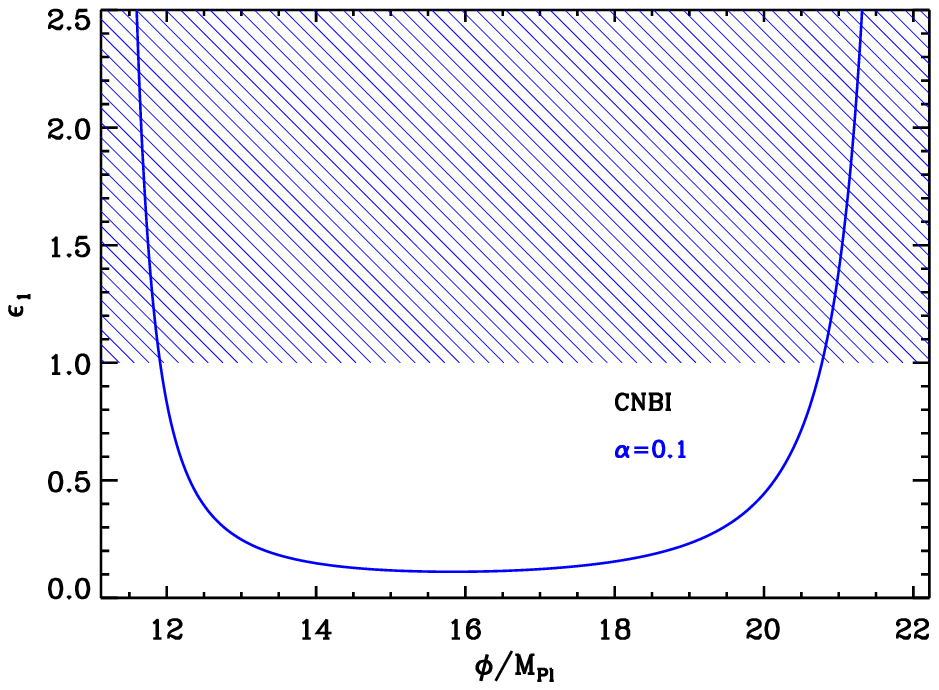}
\includegraphics[width=\wdblefig]{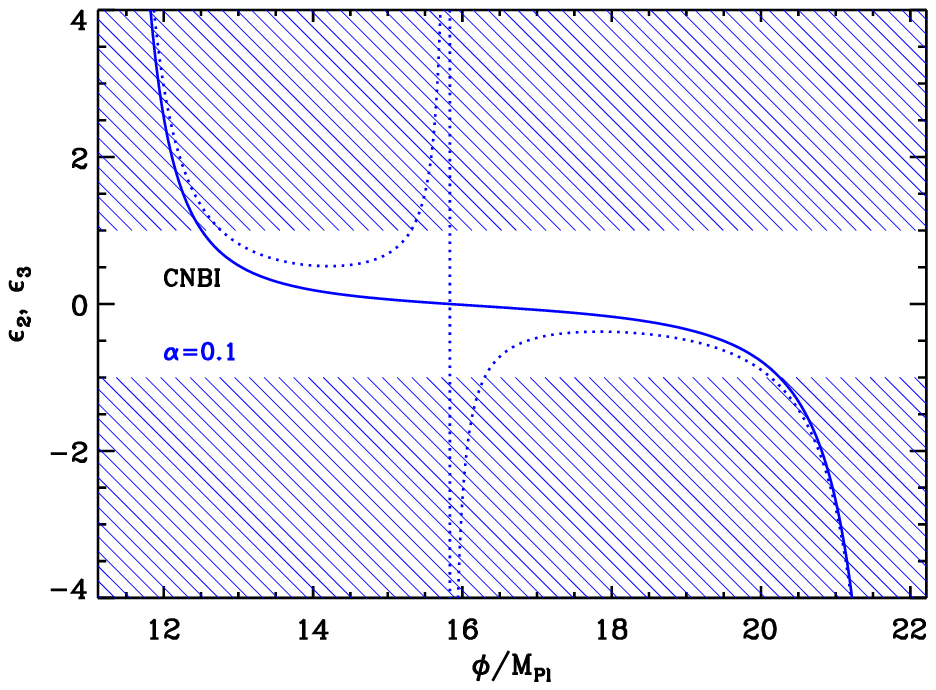}
\caption{Top left panel: constant $\nS$ T Inflation (CNBI) potential
  for $\alpha=0.1$, see \Eq{eq:potcnbi}. Top right panel: logarithm of
  this potential (for the same value of $\alpha$). Bottom left panel:
  slow-roll parameter $\epsilon _1$ still for $\alpha=0.1$.  Bottom
  right panel: slow-roll parameters $\epsilon _2$ and $\epsilon_3$
  again for $\alpha=0.1$.}
\label{potcnbi}
\end{center}
\end{figure}

This model is another representative of the class of scenarios studied
in \Refc{Vallinotto:2003vf}. As was already discussed in
\sectionc{sec:cnai}, it is designed such that the corresponding power
spectrum has a constant spectral index. The potential is given by
\begin{equation}
\label{eq:potcnbi}
  V\left(\phi\right) = M^4\left[\left(3-\alpha^2\right) \tan^2
    \left(\frac{\alpha}{\sqrt{2}}\frac{\phi}{\Mp} \right)-3\right],
\end{equation}
where $\alpha$ is a positive dimensionless
parameter~\cite{Vallinotto:2003vf}. Since the potential is periodic
with period $\pi\sqrt{2}/\alpha$ and, moreover, invariant under
$\phi\rightarrow -\phi$, one can restrict ourselves to the range
$0<\phi/\Mp<\pi/\left(\sqrt{2}\alpha\right)$ without loss of
generality. The potential is an increasing function of the field and,
as a consequence, inflation proceeds from the right to the
left. Finally, $V(\phi)$ is positive provided $\phi>\phizero$, where
\begin{equation}
  \frac{\phizero}{\Mp} = \frac{\sqrt{2}}{\alpha}
  \arctan\left(\sqrt{\frac{3}{3 - \alpha^2}}\right).
\end{equation}
Obviously, in order for the potential not to be negative everywhere,
one needs to impose that $\alpha<\sqrt{3}$ and, as a result, the
previous expression is well defined. The potential (and its logarithm)
is displayed in \Fig{potcnbi}, in the relevant range
$\phizero/\Mp<\phi/\Mp<\pi/\left(\sqrt{2}\alpha\right)$. 

Then, defining $x=\phi/\Mp$, the slow-roll parameters are given by
\begin{equation}
  \epsilon_1 = \frac{4\alpha^2\left(\alpha^2 - 3\right)^2 
    \tan^2 \left(\dfrac{\alpha}{\sqrt{2}}x\right)}
  {\left[\alpha^2 + \left(6-\alpha^2\right) \cos\left(\sqrt{2}\alpha
        x\right) \right]^2}\,,
\end{equation}
\begin{equation}
  \epsilon_2 = \dfrac{\alpha^2\left(3-\alpha^2\right) \left[ 6 +
      \alpha^2 + 2\left(6-\alpha^2\right) \cos\left(\sqrt{2} \alpha
        x\right)+\left(\alpha^2-6\right)\cos\left(2\sqrt{2}\alpha
        x\right)\right]}
  {2\cos^6 \left(\dfrac{\alpha}{\sqrt{2}}x\right) \left[3 +
      \left(\alpha^2-3\right) \tan^2\left(\dfrac{\alpha
          x}{\sqrt{2}}\right) \right]^2}\, ,
\end{equation}
and
\begin{equation}
\begin{aligned}
  \epsilon_3 &= 2\alpha^2 \left(\alpha^2-3\right)
  \tan^2\left(\frac{\alpha}{\sqrt{2}}x\right) \left[6
    \left(-72+14\alpha^2 -\alpha^4\right) + \left(\alpha^2-6\right)
    \left(7\alpha^2 + 78\right) \cos\left(\sqrt{2}\alpha x\right)
  \right. \\ &\left.
    -2\left(\alpha^4-18\alpha^2+72\right)\cos\left(2\sqrt{2}\alpha
      x\right)+\left(\alpha^2-6\right)^2\cos\left(3\sqrt{2}\alpha
      x\right)\right] \\
  & \times
  \left[\alpha^2+\left(6-\alpha^2\right)\cos\left(\sqrt{2}\alpha
      x\right)\right]^{-2} \left[
    6+\alpha^2+2\left(6-\alpha^2\right)\cos\left(\sqrt{2}\alpha
      x\right) \right. \\
  & + \left. \left(\alpha^2-6\right)\cos\left(2\sqrt{2}\alpha x
    \right) \right]^{-1} .
\end{aligned}
\end{equation}

These slow-roll parameters are displayed in \Fig{potcnbi} (bottom
panels). The first slow-roll parameter $\epsilon_1$ first decreases as
the field \vev increases and reaches a minimum value at $\xepstwoZero$
where $\epsilon_2$ vanishes and then increases. The value of
$\xepstwoZero$ is given by
\begin{equation}
  \xepstwoZero = \frac{1}{\alpha\sqrt{2}}
  \arccos\left[\frac{\alpha^2 - 6 + \sqrt{\alpha^4 - 36\alpha^2+180}}{
      2\left(\alpha^2-6\right)}\right]\, .
\end{equation}
The second slow-roll parameter, $\epsilon_2$, always decreases as
inflation proceeds, crossing $\epsilon_2=0$ at $\xepstwoZero$. The
third slow-roll parameter, $\epsilon_3$, is positive for
$x<\xepstwoZero$. In this domain, it decreases to reach a minimum and
then increases and diverges when $x$ approaches $\xepstwoZero$. On the
contrary, for $x>\xepstwoZero$, $\epsilon_3$ becomes negative. It
first increases and reaches a local maximum, then decreases and goes
to $-\infty$ at $x=\pi/\left(\sqrt{2}\alpha\right)$. The three slow
roll parameters diverge when $\phi$ goes to $\phizero$ and to
$\Mp\pi/\left(\sqrt{2}\alpha\right)$.

The minimum value of $\epsilon_1$ at $\xepstwoZero$ turns out to be
smaller than $1$ only if $\alpha<\alphamax \simeq 0.2975$. A (rather
long) analytic expression for $\alphamax$ can be derived, but it
does not provide much information to the present
discussion. Therefore, one must require $\alpha<0.2975$ in order to
realize slow-roll inflation in this model. Then, assuming this is the
case, it is clear from \Fig{potcnbi} and from the previous
considerations that inflation ends by slow-roll violation. If we
define $y\equiv \sin^2(\alpha x /\sqrt{2})$, then the condition
$\epsilon_1=1$ is equivalent to
$4(6-\alpha^2)^2y^3-4(12-\alpha^2)(6-\alpha^2)y^2
+4(45+3\alpha^2-6\alpha^4+\alpha^6)y-36=0$. The relevant solution is
given by
\begin{align}
\yend &= \frac{12-\alpha^2}{3(6-\alpha^2)}
+\frac{4}{3}2^{-2/3}\left(1-i\sqrt{3}\right)
\frac{\left(3\alpha^2-1\right)\left(18-9\alpha^2+\alpha^4\right)^2}
{\left(6-\alpha^2\right)^2}P^{-1/3}
\nonumber \\ &
-\left(1+i\sqrt{3}\right)\frac{2^{-1/3}}{24\left(6-\alpha^2\right)^2}P^{1/3},
\end{align}
where we have defined the quantity $P$ by
\begin{align}
P & \equiv 64\left(-6+\alpha^2\right)^3\left(-3+\alpha^2\right)^2
\biggl(-6+110\alpha^2-9\alpha^4+3\alpha \sqrt{3}
\nonumber \\ & \times
\sqrt{-36+408\alpha^2-12\alpha^4-25\alpha^6+4\alpha^8}\biggr).
\end{align}
If $\alpha \ll 1$, then $\yend \simeq 1/2$ and $\xend\simeq
\sqrt{2}/\alpha \arcsin(1/\sqrt{2})=\pi/(2\sqrt{2}\alpha)$.

As for the CNAI model, the spectral index
$\nS-1=-2\epsilon_1-\epsilon_{2}$, at first order in slow-roll, can be
made constant in some limit. Expanding the slow-roll parameters in
$\alpha$, while assuming $\alpha x$ to be small, gives $\epsilon_1=
x^2\alpha^4/2 + \order{\alpha^6}$ and $\epsilon_2=2\alpha^2 +
\order{\alpha^4}$, so that $\nS-1=-2\alpha^2 +
\order{\alpha^4}$. Therefore, approximate scale-invariance, $\vert
\nS-1\vert \ll 1$, implies $\alpha$ small.

\begin{figure}
\begin{center}
\includegraphics[width=\wsingfig]{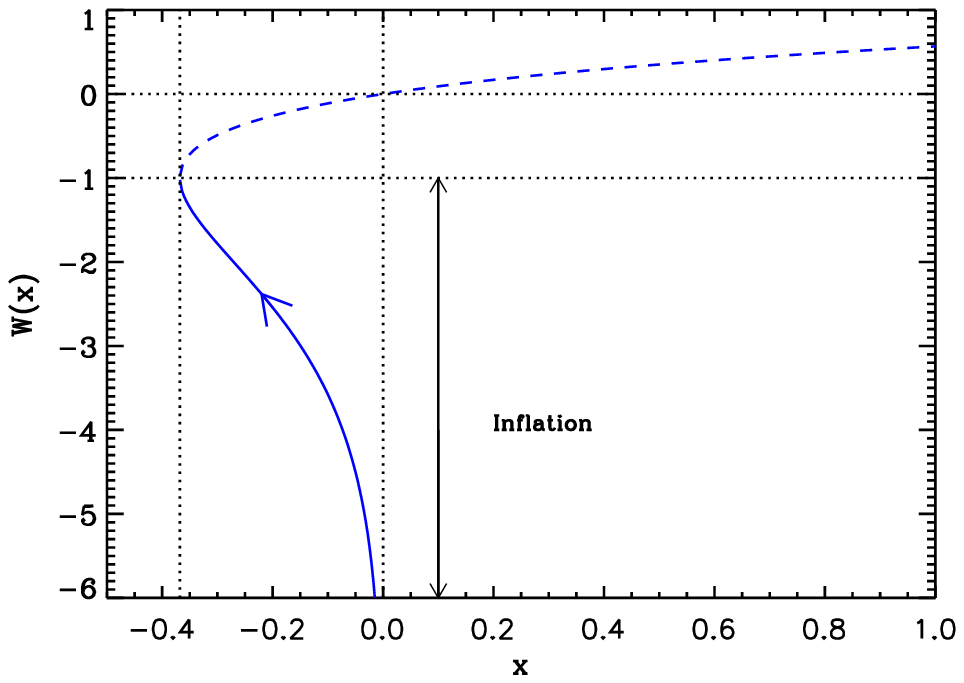}
\caption{Lambert functions $\Lambert{0}(x)$ (dashed line) and
  $\Lambert{-1}(x)$ (solid line). During Constant $\nS$ T Inflation,
  inflation proceeds along the ``$-1$'' branch in the direction
  specified by the arrow.}
\label{fig:cnbi:lambert}
\end{center}
\end{figure}

Let us now turn to the slow-roll trajectory. This one can be
integrated exactly, leading to the following formula
\begin{eqnarray}
\label{eq:trajeccnbi}
  N-\Nend &=& \frac{1}{\alpha^2 \left(3-\alpha^2 \right)}
  \left\lbrace3\ln\left[\sin\left(\frac{\alpha}{\sqrt{2}}x \right)
    \right] - \frac{6-\alpha^2}{2}
  \sin^2\left(\frac{\alpha}{\sqrt{2}}x \right) \right.\nonumber\\&
  &\left.  -3\ln \left[ \sin \left(\frac{\alpha}{\sqrt{2}}\xend
    \right) \right] + \frac{6-\alpha^2}{2}\sin^2
  \left(\frac{\alpha}{\sqrt{2}}\xend \right) \right \rbrace .
\end{eqnarray}
This formula can be inverted and $x$ can be expressed explicitly in
terms of the \efolds number. One obtains
\begin{eqnarray}
  x&=& \frac{\sqrt{2}}{\alpha}
  \arcsin\left[-\frac{3}{6-\alpha^2} \Lambert{-1}\left(-\frac{6-\alpha^2}{3}\exp
      \left\lbrace\frac{2}{3} \alpha^2 \left(3-\alpha^2\right) \left(N-\Nend\right)
      \right.\right.\right. \nonumber\\& & \left.\left.\left.
        +2 \ln\left[ \sin\left(\frac{\alpha}{\sqrt{2}}\xend\right) \right]
        - \frac{6-\alpha^2}{3} \sin^2 \left(\frac{\alpha}{\sqrt{2}}
          \xend\right) \right\rbrace \right) \right]^{1/2} ,
\end{eqnarray}
where $\Lambert{-1}$ is the $-1$ branch of the Lambert function. It is
displayed in \Fig{fig:cnbi:lambert}.  When
$x=\pi/\left(\sqrt{2}\alpha\right)$, the argument of the Lambert
function is $\left(\alpha^ 2-6\right)\exp\left(\alpha^ 2/3-2\right)/3$
which is always larger than $-1/\ee$ for any value of $\alpha$ (this
expression decreases with $\alpha$ when $\alpha<\sqrt{3}$), whereas
when $x=\phizero/\Mp$, the argument of the Lambert function is just
given by $-1/\ee$. For $x>\phizero/\Mp$, the value taken by the
Lambert function must be less than $-1$ which indicates that the $-1$
branch is the relevant one. Therefore, inflation proceeds in the
domain displayed in \Fig{fig:cnbi:lambert} in which one easily checks
that the above trajectory is always well defined.

The slow-roll predictions of the CNBI models are displayed in
\Fig{fig:CMBCNBI} for the range $10^{-5}\lesssim \alpha \lesssim
10^{-1.3}$. For very small values of $\alpha$, the predictions are in
agreement with the data with a value of $\nS$ centered around the
constant value $\nS\simeq 0.97$ and an amount of gravitational waves
such that $r \gtrsim 0.07$. But one also notices that the spectral
index is not really constant. In fact, it does not come as a surprise
that the same phenomenon highlighted in \sectionc{sec:cnai} is at work
here. Indeed, using the slow-roll trajectory~(\ref{eq:trajeccnbi}),
one has
\begin{equation}
\label{eq:sinxstarcnbi}
\sin^2\left(\frac{\alpha \xstar}{\sqrt{2}}\right)
=-\frac{3}{6-\alpha^2}\Lambert{-1}\left(-\frac{6-\alpha^2}{3}\ee^{-2A/3}\right),
\end{equation} 
where $A$ is given by the following expression
\begin{equation}
A\equiv \alpha^2\left(3-\alpha^2\right)\Delta \Nstar
-3\ln \left[\sin \left(\frac{\alpha \xend}{\sqrt{2}}\right)\right]
+\frac{6-\alpha^2}{2}
\sin^2\left(\frac{\alpha \xend}{\sqrt{2}}\right).
\end{equation}
Using the formula for $\xend$ derived above, one obtains, in the limit
$\alpha \ll 1$ and at this order of approximation that $\xstar\simeq
\xend$. Therefore, as in \sectionc{sec:cnai}, $\alpha \xstar$ is not a
small quantity and one cannot always Taylor expand the trigonometric
functions that appear in the expressions of the slow-roll
parameters. This explains why, in the limit $\alpha \ll 1$, the
spectral index is in fact not constant (see section~\ref{sec:cnai}).

Finally, the CMB normalization gives
\begin{equation}
  \left(\frac{M}{\Mp}\right)^4=\frac{11520\pi^2\alpha^2\left(3-\alpha^2\right)^2
    \sin^2\left(\frac{\alpha}{\sqrt{2}}\xstar\right)}
       {\left[\left(\alpha^2-6\right)\cos\left(\sqrt{2}\alpha
           \xstar\right)-\alpha^2\right]^3} \frac{\Qrms^2}{T^2}\, .
\end{equation}
In the limit $\alpha \ll 1$ we are interested in (since we have seen
that, if $\alpha$ is not small, then the model is ruled out), the
above expression takes the form $M/\Mp\simeq 0.02 \, \alpha^{-1/4}\,
\left(\Delta \Nstar\right)^{-3/8}$. We obtain almost exactly the same
result as for CNAI, see \Eq{eq:cobecnai}, except that the power of
$\alpha $ is different. Taking the value $\Delta \Nstar=55$, it
follows that $M/\Mp\simeq 0.0044 \, \alpha^{-1/4}$ and requiring
$M<\Mp$, one obtains the following lower bound, $\alpha \gtrsim
3.8\times 10^{-10}$.

\subsection{Open String Tachyonic Inflation (OSTI)}
\label{sec:osti}

\subsubsection{Theoretical Justifications}
\label{subsubsec:theoryosti}

In this section, we consider tachyon inflation. It was shown in
\Refcs{Witten:1992qy, Witten:1992cr, Gerasimov:2000zp, Kutasov:2000qp}
that, in bosonic string theory, the four-dimensional action for a
tachyon field $T$ on a D$3$-brane can be approximated as
\cite{Gerasimov:2000zp,Kutasov:2000qp}
\begin{equation}
\label{eq:lagrangianosti}
S_T=T_3\int \dd ^4\bmx \sqrt{-g}\left[\alpha'\ee^{-T/\Tzero}
\partial _\mu\left(\frac{T}{\Tzero}\right)\partial ^{\mu}
\left(\frac{T}{\Tzero}\right)+\left(1+\frac{T}
{\Tzero}\right)\ee^{-T/\Tzero}\right],
\end{equation}
where higher derivative terms have been ignored. In this stringy
setting, $\Tzero$ is of the order of the string scale $\Tzero\simeq
\Ms=\ells^{-1}=1/\sqrt{\alpha'}$, where $\ells$ is the string
length. The constant $T_3$ is the brane tension which can be expressed
as $T_3\propto \Ms^4/\gstrings$, $\gstrings$ being the string
coupling. The tachyon is assumed to be minimally coupled to Einstein
gravity and the Planck mass in four dimensions can be written as
$\Mp^2=\Ms^2v/\gstrings^2$, where $v=(\Ms r)^d/\pi$, $r$ being a
radius of compactification and $d$ the number of compactified
dimensions. This four dimensional approximation is valid provided
$r\gg \ells$ or $v\gg 1$. The action~(\ref{eq:lagrangianosti}) can be
viewed as a truncated version of the action
\begin{equation}
\label{eq:tachyonXinf}
S_{\Tbar}=\int \dd^4 \bmx \sqrt{-g} \, V(\Tbar)\sqrt{1+\alpha'
\partial _\mu\left(\frac{\Tbar}{\Tzero}\right)\partial ^{\mu}
\left(\frac{\Tbar}{\Tzero}\right)}\,.
\end{equation}
Indeed, following
\Refcs{Kofman:2002rh,Choudhury:2002xu,Fairbairn:2002yp}, redefining
the field $\Tbar$ by $\Tbar/\Tzero\equiv \sqrt{8(1+T/\Tzero)}$
with $V\left[\Tbar(T)\right] \equiv T_3 (1+T/\Tzero)
\exp{(-T/\Tzero)}$, it is straightforward to show that the leading terms
of \Eq{eq:tachyonXinf} give back \Eq{eq:lagrangianosti}. Conversely,
the full action of tachyonic inflation, under the assumptions
discussed previously, can thus be described in terms of $\Tbar$
by \Eq{eq:tachyonXinf} with~\cite{Kofman:2002rh}
\begin{equation}
V(\Tbar)=\frac{T_3\ee}{8}\frac{\Tbar^2}{\Tzero^2}
\ee^{-\Tbar^2/(8\Tzero^2)}.
\end{equation}
Because the action~(\ref{eq:tachyonXinf}) is a particular case of
k-inflation for which $S=\int \dd ^4 \bmx \sqrt{-g}P(T,X)$ with
$X\equiv -g^{\mu \nu}\partial_\mu T\partial_\nu T/2$ and, here,
$P(T,X)=\sqrt{1-2X}$, tachyonic inflation could produce observable
non-Gaussianities. Therefore, one may wonder how accurate is the
truncated action to describe the observable features of the model. On
the theoretical point of view, knowing whether the truncated action is
a faithful representation of the actual action is a complicated
question since even an exact derivation of the complete action is
still an open problem. On a more phenomenological point of view,
non-Gaussianities are not observed by Planck~\cite{Ade:2013ydc}. More
precisely, the parameter $\fnl$ (equilateral configuration)
characterizing the amplitude of the bispectrum in Fourier space can be
written as~\cite{Chen:2006nt,Lidsey:2006ia}
\begin{equation}
\fnl=\frac{35}{108}\left(\frac{1}{\cs^2}-1\right)-
\frac{5}{81}\left(\frac{1}{\cs^2}-1-2\Lambda\right),
\end{equation}
where, in our case, $\cs^2=1-2X$ and $1/\cs^2-1=2\Lambda$ so that the
last term in the above equation cancels out~\cite{Lidsey:2006ia}. This
leads to $\fnl=35X/[54(1-2X)]$. In the range of interest
$X\in[0,1/2]$, the Planck constraint~\cite{Ade:2013ydc}, $\fnl=-42\pm
75$, yields $X\lesssim 0.495$. As a result, departures from the
leading order action of \Eq{eq:lagrangianosti} are, a priori, still allowed
by the CMB data. We will see at the end of this section that tachyonic
inflation has however other problems. For the moment, given that
\Eq{eq:lagrangianosti} can always be seen as a phenomenological
model, we can continue to work with this action in order to see if, at
least, this can lead to an inflationary scenario compatible with the
CMB data.

\subsubsection{Slow-Roll Analysis}
\label{subsubsec:srosti}

The inflationary dynamics can be studied directly from
\Eq{eq:lagrangianosti} but since it is linear in $X$, the field
can be canonically normalized. Performing the change of variable
$\ee^{- T/\Tzero}\equiv\left(\phi/\Tzero\right)^2/8$, the Lagrangian can be
re-written with an ordinary kinetic term, as a function of the field
$\phi$ and with a potential given by
\begin{equation}
\label{eq:potosti}
V\left(\phi\right)=-M^4\left(\frac{\phi}{\phizero} 
\right)^2\ln\left[\left(\frac{\phi}{\phizero}\right)^2\right],
\end{equation}
where $M^4\equiv\ee T_3$ and $\phizero^2\equiv 8\ee \Tzero^2$. We
notice that it corresponds to a particular case of LPI discussed in
\sectionc{sec:lpi}, with $q=1$ and $p=2$. Such a potential was also
introduced in \Refc{Minahan:2000ff} as a toy model of tachyon
condensation. Let us also comment on the parameter $\phizero$. In the
original model $\phizero\simeq \Ms$ and, as such, it is a
zero-parameter scenario. Here, given the issues discussed before (see
also the end of this section) we consider $\phizero$ as a free
parameter. If necessary, one can always recover the situation where
$\phizero$ is fixed to the string scale by assuming the corresponding
prior $\phizero=\Ms$.

\begin{figure}
\begin{center}
\includegraphics[width=\wdblefig]{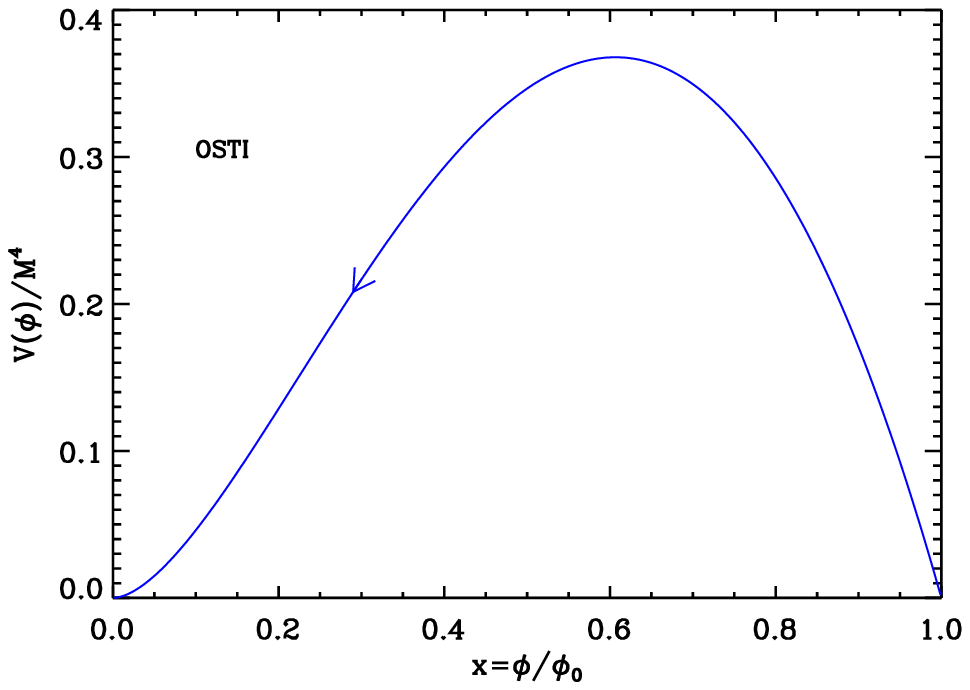}
\includegraphics[width=\wdblefig]{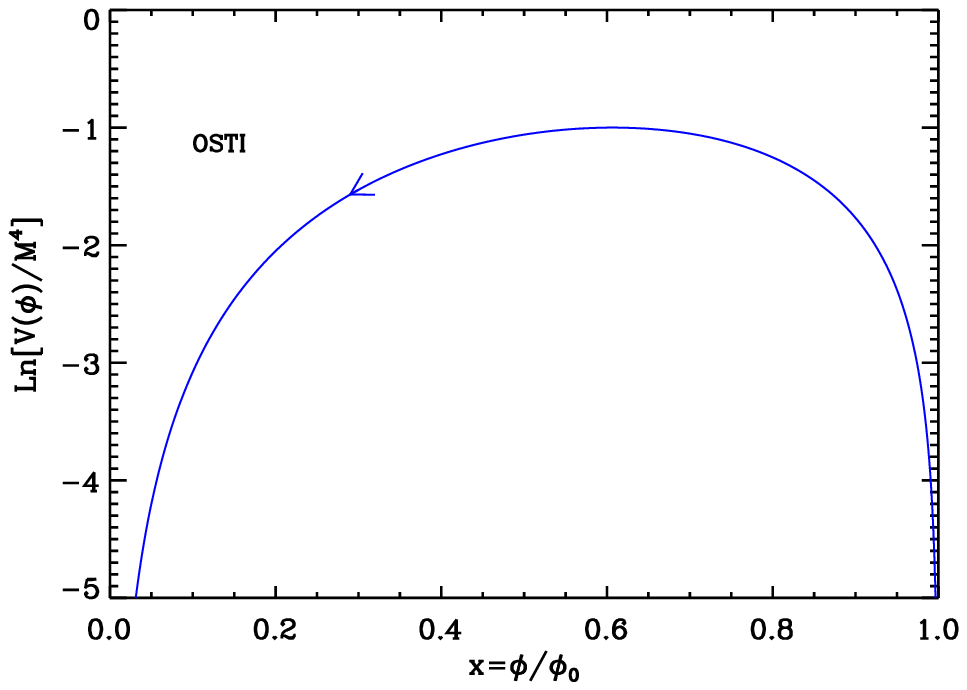}
\includegraphics[width=\wdblefig]{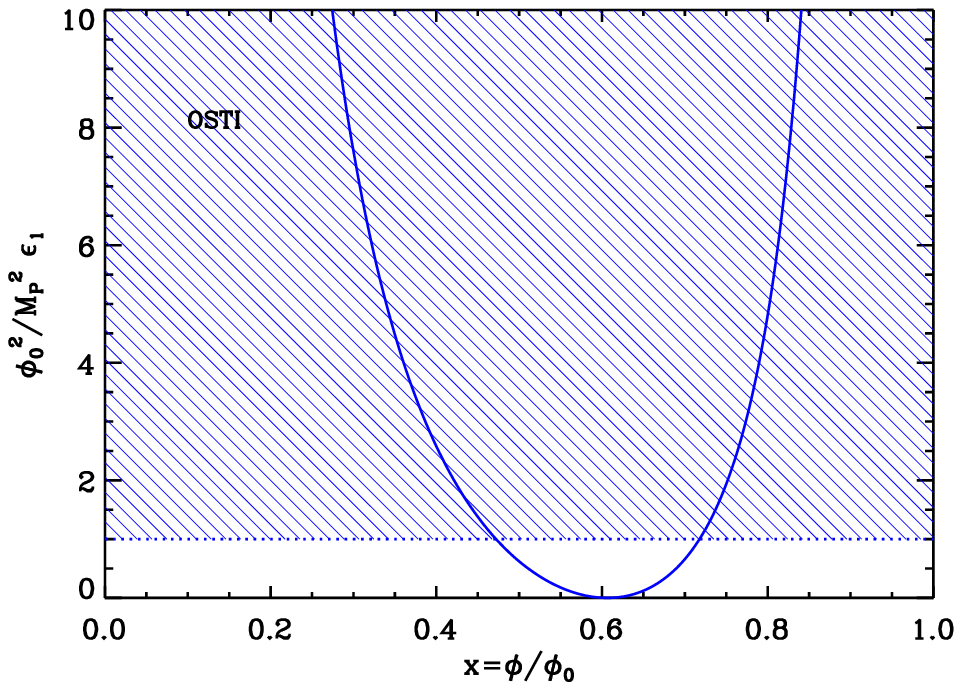}
\includegraphics[width=\wdblefig]{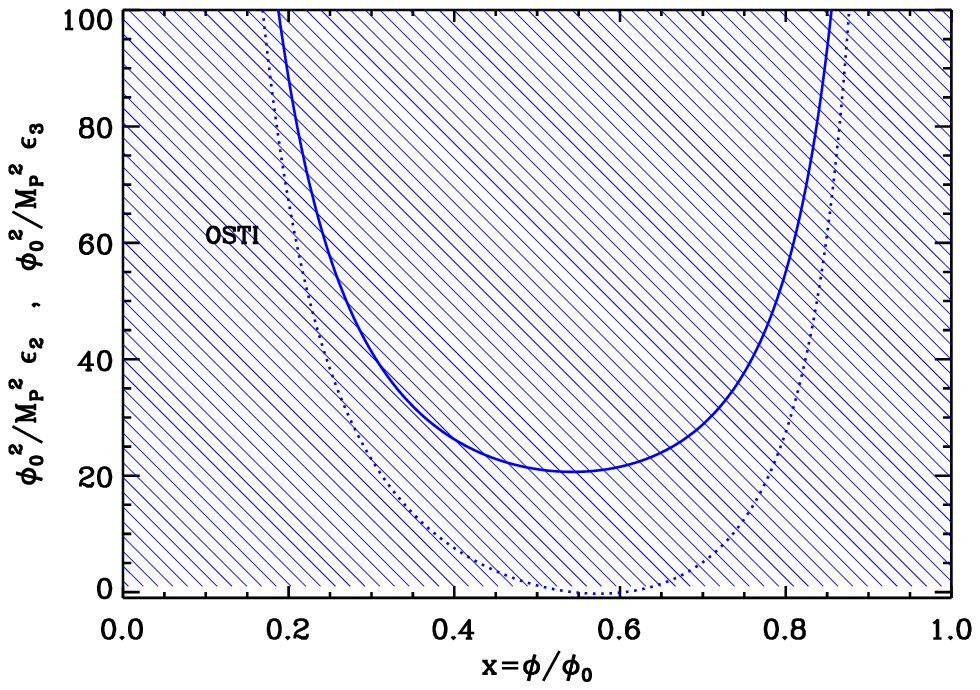}
\caption{Top left panel: Open String Tachyonic Inflation (OSTI) 
potential as a function of $\phi/\phizero$.  Top
  right panel: logarithm of the potential. The arrow indicates in
  which direction inflation proceeds. Bottom left panel: slow-roll
  parameter $\epsilon _1$, rescaled by the quantity
  $\Mp^2/\phizero^2$, such that the corresponding expression becomes
  universal, \ie independent of $\phizero$. Bottom right panel:
  slow-roll parameters $\epsilon _2$ (solid line) and $\epsilon _3$
  (dotted line), rescaled by $\Mp^2/\phizero^2$ for the same reason as
  mentioned before.}
\label{potosti}
\end{center}
\end{figure}

The potential~(\ref{eq:potosti}) in represented in \Fig{potosti},
together with its logarithm (top panels), as a function of
$x\equiv\phi/\phizero$. Since it is invariant under $x\rightarrow -x$,
and since it is positive definite only if $x^2<1$, it is only
displayed in the range $0<x<1$. The potential vanishes at $x=0$,
increases with $x$, reaches a maximum at $\xdVzero=\ee^{-1/2}$, then
decreases with $x$ and vanishes at $\xVzero=1$. Inflation is supposed
to take place between $\xdVzero$, where the effective mass of the
inflaton is negative $m_\phi^2=-4\phizero^2$, and $x=0$, where the
effective mass is positive and infinite $m_\phi^2\rightarrow
+\infty$. Hence it proceeds from the right to the left, at decreasing
field values (see \Fig{potosti}).

Let us now calculate the three first slow-roll parameters. They are 
given by
\begin{equation}
  \epsilon_1 = 2\left(\frac{\Mp}{\phizero}\right)^2
  \left[\frac{1+\ln\left(x^2\right)}
  {x\ln\left(x^2\right)}\right]^2,
\end{equation}
\begin{equation}
  \epsilon_2 = 4\left(\frac{\Mp}{\phizero}\right)^2
   \frac{2+\ln\left(x^2\right)+\ln^2\left(x^2\right)}
   {x^2\ln^2\left(x^2\right)}\, ,
\end{equation}
and
\begin{equation}
\begin{aligned}
  \epsilon_3 &= 4\left(\frac{\Mp}{\phizero}\right)^2
  \frac{1+\ln\left(x^2\right)}
  {x^2\ln^2\left(x^2\right)}
  \frac{4+3\ln\left(x^2\right)+\ln^2\left(x^2\right)
  +\ln^3\left(x^2\right)}{2+\ln\left(x^2\right)
  +\ln^2\left(x^2\right)}\, .
  \end{aligned}
\end{equation}
They are displayed in the bottom panels of \Fig{potosti}. The first
slow-roll parameter $\epsilon_1$ diverges when $x\rightarrow 0$,
decreases with $x$, vanishes at $\xdVzero$ and then increases with $x$
and diverges when $x\rightarrow\xVzero$. As a consequence, inflation
stops by slow-roll violation at a point $\xend$ where $\epsilon_1=1$
that needs to be determined numerically. The second slow-roll
parameter $\epsilon_2$ has the same kind of behavior, except that it
has a non-vanishing minimum located at a point $\xepstwoMin$, which is
such that $0<\xepstwoMin<\xVzero$. An analytic expression for
$\xepstwoMin$ can be derived but it does not add much to the
discussion. It yields $\epstwoMin\simeq 20.65\,\Mp^2/\phizero^2$. This
means that in order for a slow-roll inflationary regime to take place,
$\epstwoMin\ll 1$ requires that the parameter $\phizero$ be sufficiently
super-Planckian. Finally, the third slow-roll parameter has the same
behavior as the two previous ones, except that it has a negative
minimum $\epsthreeMin\simeq-0.2733\,\Mp^2/\phizero^2$, located between
$\xepstwoMin$ and $\xdVzero$ where it vanishes.

Let us now turn to the slow-roll trajectory. It can be integrated, 
and gives rise to
\begin{equation}
\label{eq:osti:traj}
\Nend-N=\frac{1}{4}\left(\frac{\phizero}{\Mp}\right)^2
\left[x^2-\frac{1}{\ee}\Ei\left(1+\ln x^2\right)-\xend^2+\frac{1}
{\ee}\Ei\left(1+\ln \xend^2\right)\right],
\end{equation}
where $\Ei$ is the exponential integral
function~\cite{Abramovitz:1970aa,Gradshteyn:1965aa} and $\Nend$ is the
number of \efolds at the end of inflation. This trajectory can
only be inverted numerically to obtain $\phi(N)$.

Finally, it is interesting to constrain the value of the scale $M$
with the CMB normalization. It follows that
\begin{equation}
\left(\frac{M}{\Mp}\right)^4=2880\pi^2\left(\frac{\Mp}{\phizero}\right)^2
\frac{\left[1+\ln \left(\xstar^2\right)\right]^2}{\xstar^4\left\vert
  \ln\left(\xstar^2\right)\right\vert^3} \frac{\Qrms^2}{T^2}\, .
\end{equation}
The reheating consistent slow-roll predictions of the open string
tachyonic inflation models are displayed in \Fig{fig:CMBOSTI}. It is
interesting to notice that, as expected, these models are compatible
with the CMB data only for super-Planckian values of $\phizero$,
$\phizero/\Mp\gg 1$. In this limit, one has
$\xend\simeq\sqrt{2}\Mp/\phizero$, the quadratic terms in the slow
roll trajectory \Eq{eq:osti:traj} dominate over the exponential
integral ones, such that one has $\xstar\simeq 2\Mp/\phizero
\sqrt{\Delta\Nstar+\frac{1}{2}}$. It follows that
\begin{equation}
\epsilon_{1*}\simeq\frac{1}{2\Delta\Nstar+1}
\, ,\quad\quad
\epsilon_{2*}\simeq\epsilon_{3*}\simeq 2\epsilon_{1*}\, ,
\end{equation}
hence
\begin{equation}
r\simeq\frac{16}{2\Delta\Nstar+1}
\, ,\quad\quad
1-\nS\simeq\frac{4}{2\Delta\Nstar+1}\, ,
\quad\mathrm{and}\quad
\alphaS\simeq-\frac{8}{\left(2\Delta\Nstar+1\right)^2}\, .
\end{equation}
One can check that indeed, in the $\phizero/\Mp\gg 1$ limit, the
prediction points lie in the line $\epsilon_2=2\epsilon_1$, or
equivalently, $1-\nS=r/4$.

Finally, let us close this section by some additional considerations
on the difficulties that tachyonic inflation
faces~\cite{Kofman:2002rh}. Using the above equations, it is easy to
show that
\begin{equation}
  \left(\frac{M}{\Mp}\right)^4\simeq \frac{2880\pi^2}{16\Delta \Nstar}
  \frac{\Qrms^2}{T^2}\frac{\phizero^2}{\Mp^2}
  \frac{\left[5-2\ln (\phizero/\Mp)\right]^2}
  {\left[4-2\ln \left(\phizero/\Mp\right)\right]^3}\ll 1.
\end{equation}
Given that $T_3 \simeq M^4$, this implies that $\gstrings^3\ll v^2$. On the
other hand, we have seen that the model is compatible with the CMB
data only if $\phizero/\Mp=(g/v)^{1/2}\gg 1$. This last inequality is
consistent with $\gstrings^3\ll v^2$ only if $v\ll 1$. But $v\ll 1$ is
in contradiction with the assumption that $r\gg \ells$, which implies
that $v\gg 1$. Therefore, it seems that the constraints obtained from
the CMB data invalidates the use of an effective four-dimensional
approach to describe tachyonic inflation~\cite{Kofman:2002rh}. On the
other hand, this can also justify our approach which just considers
this scenario as a phenomenological model.

\subsection{Witten-O'Raifeartaigh Inflation (WRI)}
\label{sec:wri}

\subsubsection{Theoretical Justifications}
\label{subsubsec:theorywri}

This model arises in different contexts and we now briefly review one
of its theoretical motivation. The first situation originates from
supersymmetric theories aimed at explaining the gauge hierarchy
problem (that is to say why the GUT scale differs so much from the
weak scale). In the supersymmetric scenario of \Refc{Witten:1981kv},
three chiral superfields $A$, $X$ and $Y$ are considered in a
superpotential of the O'Raifeartaigh
type~\cite{O'Raifeartaigh:1975pr},
\begin{equation}
\label{eq:superpotwri}
W=\lambda X(A^2-m^2)+gYA,
\end{equation}
where $m$ and $g$ are constant of mass dimension. The corresponding
(global) supersymmetric potential can be expressed as
\begin{equation}
V=\lambda^2\vert
A^2-m^2\vert^2+g^2\vert A\vert^2+\vert 2\lambda XA+gY\vert^2.
\end{equation}
The minimum of this potential is given by $\langle Y\rangle=-2\lambda
\langle X\rangle \langle A\rangle /g$ and $\langle A\rangle =0$ [there
is also another minimum at $\langle A\rangle
=\sqrt{m^2-g^2/(2\lambda^2)}$]. Clearly, the potential is minimized
regardless of $\langle X\rangle $, that is to say we have a flat
direction along $X$. Along that direction, $V=\lambda^2m^4$ and
supersymmetry is broken since $F_X\equiv \partial W/\partial X \neq
0$. As a consequence, the mass of the real part and imaginary parts of
$A$ are split and are given by $4\lambda^2\vert X\vert^2+g^2\pm
2m^2\lambda^2$. The mass of the fermion field $\psi_A$ is
$4\lambda^2\vert X\vert^2+g^2$. The fact that supersymmetry is broken
implies that the potential will receive corrections: as is well-known,
if supersymmetry is preserved, the corrections originating from bosons
and fermions exactly cancel out. Here, this is not the case and the
amplitude of the corrections will be determined by the split between
the bosonic and fermionic masses that we have just evaluated before. A
simple calculation leads to
\begin{equation}
V=\lambda^2m^4\left[1+\frac{\lambda^2}{8\pi^2}\ln\left(\frac{\vert
    X\vert^2}{\mu^2}\right)\right],
\end{equation}
where $\mu$ is the renormalization scale. Therefore, one obtains an
increasing function of the field \vev and this implies that $X$ cannot
become large because it cannot climb its potential. As a consequence,
one cannot generate a large hierarchy in this scenario. In fact, as
explained in Ref.~\cite{Witten:1981kv}, this is due to the fact that
the one loop correction is positive, as appropriate in a theory with
scalars and fermions. This can also be understood from the
renormalization group perspective where the appearance of the
logarithm in the above expression of $V(X)$ can be viewed as the
renormalization of the coupling constant such that $\lambda^2
\rightarrow \lambda^2\left[1+\lambda^2/(8\pi^2)\ln\left(\vert
    X\vert^2/\mu^2\right)\right]$. The conclusion of
Ref.~\cite{Witten:1981kv} is that if $m$ is the small scale (the weak
scale) and $\langle X\rangle $ the large one (the GUT scale), a large
hierarchy cannot be achieved in this approach.

However, it is well-known that asymptotic freedom is possible in
non-Abelian gauge theories. This means that the renormalization group
equations have to produce \emph{negative} one loop corrections. In
such a situation, the field could run to infinity, in the
non-perturbative regime. For this reason, it is interesting to
re-consider the previous model in the framework of a non-Abelian gauge
group such as in Grand Unified $\mathrm{SU}(5)$
theories. \Refcs{Witten:1981nf,Dimopoulos:1982gm} consider two matter
fields $A_a^b$ and $Z_a^b$ in the adjoint representation of
$\mathrm{SU}(5)$ and one singlet $X$ in a superpotential given by
\begin{equation}
W=\lambda_1\Tr(ZA^2)+\lambda_2
X\left[\Tr\left(A^2\right)-m^2\right],
\end{equation}
which is the non-Abelian generalization of \Eq{eq:superpotwri}. One
can show that supersymmetry is again necessarily broken\footnote{For
this purpose, it is convenient to write that
$A_d^c=\left(\phi_A\right)_a^b\left(T_b^a\right)_d^c$ and
$Z_d^c=\left(\phi_Z\right)_a^b\left(T_b^a\right)_d^c$, where $T_a^b$,
$a,b=1,\cdots, 5$ is a basis of $\mathrm{SU}(5)$
generators. Concretely, one has
$\left(T_b^a\right)_d^c=\delta_b^c\delta_d^a-\delta_b^a\delta_d^c/5$. As
a consequence, the three F-term can be expressed as
$F_X=\lambda_2\left[\Tr\left(\phi_A^2\right)-m^2\right]$,
$F_Z=\lambda_1\left[\phi_A^2-\Tr\left(\phi_A^2\right)\mathbb{1}/5\right]$
and
$F_A=\lambda_1\left[\phi_Z\phi_A+\phi_A\phi_Z-2\Tr\left(\phi_Z\phi_A\right)
  \mathbb{1}/5\right]+2\lambda_2 \phi_X\phi_A$. These expressions are
obtained by explicitly writing the superpotential in terms of the
components $\left(\phi_A\right)^a_b$ and $\left(\phi_Z\right)^a_b$ and
differentiating $W$ with respect to them. From $F_X=0$ it follows that
$\Tr\left(\phi_A^2\right)=m^2$ and, therefore, $F_Z=0$ implies that
$\phi_A^2=m^2\mathbb{1}/5$. This last relation is compatible with
$\Tr\left(\phi_A^2\right)=m^2$ but not with $\Tr\left(\phi_A\right)=0$
in five dimensions. The conditions $F_X=0$ and $F_Z=0$ are thus
incompatible and supersymmetry is spontaneously broken in this model.}
and that the potential exhibits a flat direction with the value
$V=\lambda_1^2\lambda_2^2m^4/(30\lambda_2^2+\lambda_1^2)$. As it was
the case in the first simple example presented above, and since
supersymmetry is broken, quantum corrections modify the potential. At
the one loop order, one obtains the following
expression~\cite{Witten:1981nf}
\begin{equation}
\label{eq:potsu5wri}
V(X)=\frac{\lambda_1^2\lambda_2^2m^4}{30\lambda_2^2+\lambda_1^2}
\left(1+\frac{\lambda_2^2}{\lambda_2^2+\lambda_1^2/30}
\frac{29\lambda_1^2-50g^2}{80\pi^2}\ln \vert X\vert^2\right),
\end{equation}
where $g$ is the $\mathrm{SU}(5)$ gauge coupling constant. If
$29\lambda_1^2<50g^2$, the correction is \emph{negative} contrary
to the case studied before. Again, this is precisely because we deal
with non-Abelian gauge interaction. The field $X$ will grow and can
reach a point where the perturbative approach is no longer
valid. However, asymptotic freedom tells us that the potential could
develop a minimum in this regime in which $X$ could be stabilized, hence
the original motivation for this scenario: the scale $m$ can be taken
to be relatively small while $\langle X\rangle $ can now be very large
thereby addressing the gauge hierarchy problem.

This class of model was considered in \Refc{Albrecht:1983ib} in order
to build a new inflationary scenario. The idea is to start from a
potential of the form derived above, namely
$V(\phi)=M^4\left(1+\tilde{b}\ln \phi \right)$ with a negative
coefficient $\tilde{b}$. Therefore, the field is driven towards a
regime where higher corrections must become important. Typically, one
expects $\tilde{b}$ to acquire a logarithmic dependence in $\phi$ and
the potential to develop a minimum at, say $\phi=\mGUM$. Therefore,
this leads to $V(\phi)=M^4\left[1+b\ln^2(\phi/\mGUM)\right]$ where $b$
is a constant. Moreover, if one requires the potential to vanish at
the minimum, we are led to $V(\phi) \propto \ln^2(\phi/\mGUM)$ and
this is the potential studied in this section. In
\Refc{Albrecht:1983ib}, it is argued that $\mGUM\simeq \Mp$ and that,
initially, $\phi\simeq \mu\simeq
\left(m_{\mathrm{weak}}\mGUM\right)^{1/2}\simeq 10^{12}\GeV$. We will
come back to these conditions in what follows.

Another way to obtain the same potential is based on
\Refc{Papantonopoulos:1986gc,Pollock:1987qc} in which one consider the
following action
\begin{equation}
S=-\int \dd^4 \bmx \sqrt{-g}\left[\hat{g}_{A\bar{B}}
\left(z^C,\bar{z}^{\bar{C}}\right) g^{\mu \nu}
\partial_{\mu}z^A\partial_{\nu}\bar{z}^{\bar{B}}
-V\left(z^C,\bar{z}^{\bar{C}}\right)\right].
\end{equation}
The $z^A$'s are complex scalar fields and $\hat{g}_{A\bar{B}}$ is the
K\"ahler metric. The corresponding equations of motion can be
expressed as
\begin{equation}
g^{\mu\nu}\nabla_{\mu}\nabla_{\nu}\bar{z}^{\bar{D}}
+\Gamma_{\bar{A}\bar{B}}^{\bar{D}}g^{\mu \nu}\partial_\mu
\bar{z}^{\bar{A}}\partial_{\nu}\bar{z}^{\bar{B}}
-\hat{g}^{C\bar{D}}\frac{\partial V}{\partial z^C}=0,
\end{equation}
where $\Gamma_{\bar{A}\bar{B}}^{\bar{D}}\equiv \hat{g}^{C\bar{D}}
\partial_{\bar{A}}\hat{g}_{C\bar{B}}$. If we restrict ourselves to
cosmological spacetimes, the above equation becomes
$\ddot{\bar{z}}^{\bar{D}}
+3H\dot{z}^{\bar{D}}+\Gamma^{\bar{D}}_{\bar{A}\bar{B}}
\dot{\bar{z}}^{\bar{A}}\dot{\bar{z}}^{\bar{B}}+\hat{g}^{C\bar{D}}\partial
V/\partial z^C=0$, where $H$ is the Hubble parameter. Then, for
simplicity, we assume that there is only one field $Z$ and we denote
its real part as $u$ and its imaginary part as $v$. We also assume
that the potential is flat in the $v$-direction and take
$V=V(z+\bar{z})$, $\hat{g}_{Z\bar{Z}}\equiv \hat{g}(Z+\bar{Z})$. It
follows that
\begin{eqnarray}
\ddot{u}+3H\dot{u}+\Gamma(u)\left(\dot{u}^2-\dot{v}^2\right)
+\partial_u V/(2\hat{g}) &=& 0,
\\
\ddot{v}+3H\dot{v}+2\Gamma(u)\dot{u}\dot{v} &=& 0,
\end{eqnarray}
with $\Gamma=\partial_u\hat{g}/(2\hat{g})$. The second differential
equation can be integrated and one obtains $\dot{v}=Qa^{-3}/\hat{g}$,
where $Q$ is a constant. The next step consists in defining the field
$\phi$ by $\dot{\phi}\equiv \sqrt{\hat{g}}\dot{u}$. As a consequence,
the first differential equation can be re-written as
$\ddot{\phi}+3H\dot{\phi}+
\partial_\phi\left[V+Q^2/(\hat{g}a^6)\right]=0$, that is to say $\phi$
is now canonically normalized and its evolution is controlled by the
effective potential $V(\phi)+Q^2/(\hat{g}a^6)$. One can show that the
presence of the additional term proportional to $Q^2$ is not
crucial~\cite{Papantonopoulos:1986gc,Pollock:1987qc}. Initially, it
dominates because $a$ is small but, quickly, since it is proportional
to $a^{-6}$, it goes to zero as the universe expands. As a
consequence, one is left with $V(\phi)$ only.  A specific version of
this scenario has been studied in details in
\Refc{Papantonopoulos:1986gc}. In that article, it is assumed that
$\hat{g}=\ee^{-2u}/2$ and $V=0$. This corresponds to the bosonic
action of a model which is superconformal
invariant~\cite{Kobayashi:1985jj}. Then, this invariance is softly
broken by adding a term $m^2u^2/2$ and, through the redefinition of
the field, one can check that this leads to a potential proportional
to $m^2\left(\ln \phi\right)^2$, that is to say of the type studied in
this section. Moreover, one can also verifies that, in the regime
discussed above where the term $Q^2/(\hat{g}a^6)$ dominates, an exact
solution can be found and reads: $a=a_0t^{1/3}$ and
$\phi^2(t)=E^2\left(\ln t+C\right)^2+4Q^2/(a_0^6E^2)$, where $E$ and
$C$ are two integration constants. As a consequence, when the universe
expands, $Q^2/(\hat{g}a^6)$ goes to zero and one is left with the
logarithmic potential only.

\subsubsection{Slow-Roll Analysis}
\label{subsubsec:srwri}

\begin{figure}
\begin{center}
\includegraphics[width=\wdblefig]{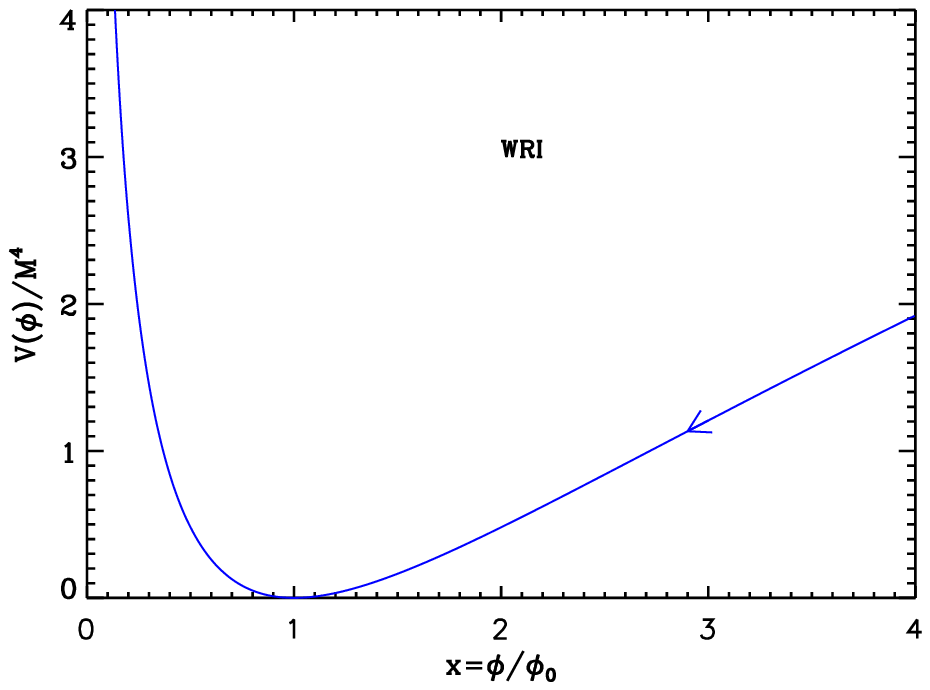}
\includegraphics[width=\wdblefig]{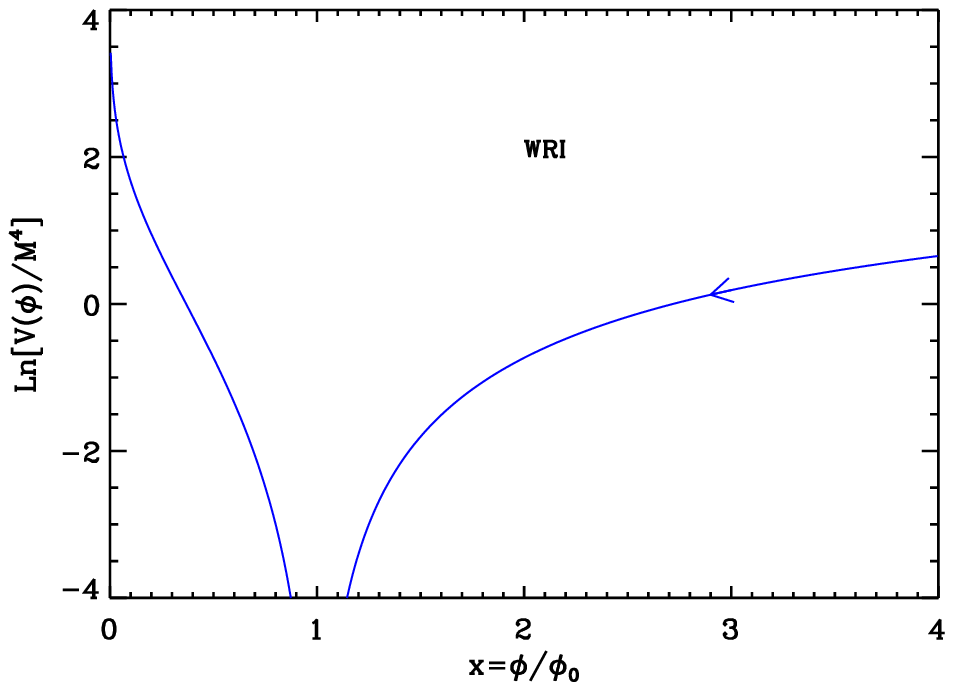}
\includegraphics[width=\wdblefig]{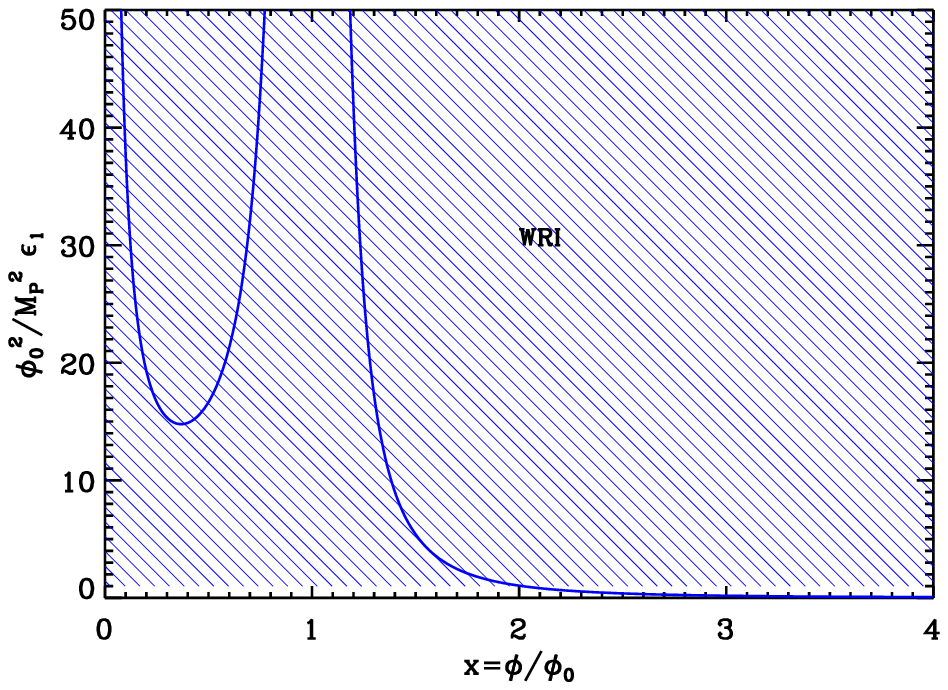}
\includegraphics[width=\wdblefig]{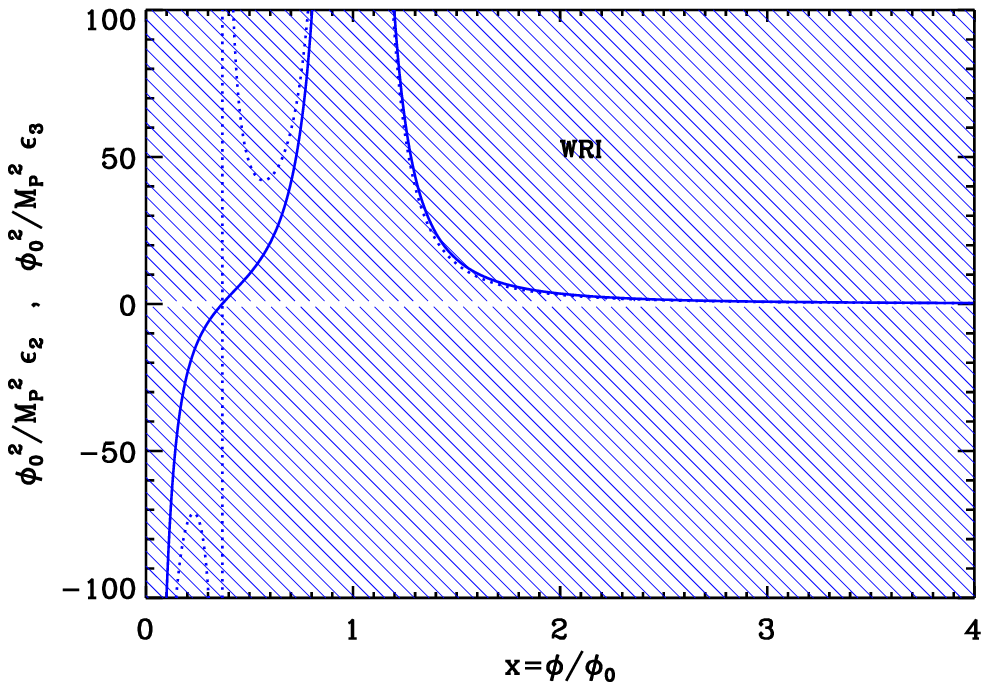}
\caption{Witten-O'Raifeartaigh Inflation (WRI) 
potential as a function of $\phi/\phizero$.  Top
  right panel: logarithm of the potential. The arrow indicates in
  which direction inflation proceeds. Bottom left panel: slow-roll
  parameter $\epsilon_1$, rescaled by the quantity
  $\Mp^2/\phizero^2$, such that the corresponding expression becomes
  universal, \ie independent of $\phizero$. Bottom right panel:
  slow-roll parameters $\epsilon _2$ (solid line) and $\epsilon _3$
  (dotted line), rescaled by $\Mp^2/\phizero^2$ for the same reason 
  as mentioned before.}
\label{potwri}
\end{center}
\end{figure}

Based on the previous considerations, we study the WRI potential
\begin{equation}
\label{eq:wri:pot}
V(\phi) = M^4\ln^2\left(\frac{\phi}{\phizero}\right),
\end{equation}
where $\phizero$ is viewed as a free parameter but we also keep in
mind that a natural prior is $\phizero=\Mp$. The potential
\Eq{eq:wri:pot} is displayed in \Fig{potwri}, together with its
logarithm (top panels). The arrow indicates that inflation proceeds
from the right to the left. Let us now calculate the Hubble flow
parameters. If one defines $x\equiv\phi/\phizero$, they are given by
\begin{equation}
\label{eq:wri:eps1}
\epsilon_1=2\frac{\Mp^2}{\phizero^2}\frac{1}{x^2\ln^2 x }\, ,
\end{equation}
\begin{equation}
\label{eq:wri:eps2}
\epsilon_2=4\frac{\Mp^2}{\phizero^2}\frac{1+\ln x }{x^2\ln^2 x }\,
,
\end{equation}
and
\begin{equation}
\epsilon_3=2\frac{\Mp^2}{\phizero^2}\frac{2+3\ln x+2\ln^2 x }
        {x^2\ln^2 x \left(1+\ln x \right)}\, .
\end{equation}
They are displayed in the bottom panels of \Fig{potwri}. One can see
that they all vanish when $x\rightarrow\infty$, that they increase as
inflation proceed, diverging when $x\rightarrow 1$. At this stage, a
remark is in order about \Refc{Albrecht:1983ib}. As already mentioned
above, a natural prior is $\phizero=\Mp$. This means that if, initially,
one has $\phi\simeq \mu$, one is in fact in the decreasing branch of
the potential and, as a matter of fact, one cannot have inflation
since $\epsilon_1>1$ always. Clearly, the only way to have inflation
in this branch is to assume that $\phizero\gg \Mp$, a case which appears
to be difficult to justify in this context. Here, we do not consider
this case. In the increasing branch of the potential, inflation stops
by slow-roll violation when $\epsilon_1=1$, at a \vev $\xend$ given by
\begin{equation}
\label{eq:wri:xend}
\xend=\exp\left[\Lambert{0}\left(\sqrt{2}\frac{\Mp}
{\phizero}\right)\right]\, ,
\end{equation}
where $\Lambert{0}$ is the $0$-branch of the Lambert function, which
must be chosen in order to have $x>1$.

\begin{figure}
\begin{center}
\includegraphics[width=\wsingfig]{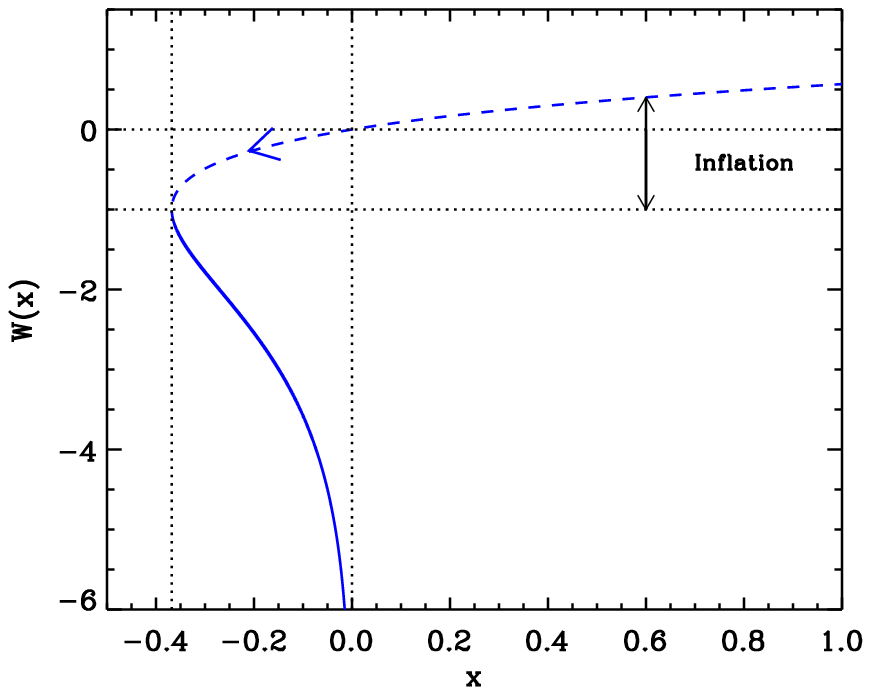}
\caption{Lambert functions $\Lambert{0}(x)$ (dashed line) and
  $\Lambert{-1}(x)$ (solid line). During Witten-O'Raifeartaigh
  inflation, inflation proceeds along the ``$0$'' branch in the
   direction specified by the arrow.}
\label{fig:potlambertWRI}
\end{center}
\end{figure}

Let us now turn to the slow-roll trajectory. It can be integrated
exactly and this leads to the following expression
\begin{equation}
\Nend-N=\frac{1}{4}\frac{\phizero^2}
{\Mp^2}\left(x^2\ln x -\frac{x^2}
{2}-\xend^2\ln \xend +\frac{\xend^2}{2}\right),
\end{equation}
where $\Nend$ is the number of \efolds at the end of
inflation. Interestingly enough, this trajectory can be inverted, and
one obtains
\begin{equation}
\label{eq:wri:InvertedTraj}
x=\exp\left\lbrace\frac{1}{2}\Lambert{0}\left[\frac{8}
{\ee}\frac{\Mp^2}{\phizero^2}\left(\Nend-N\right)+\frac{2}
{\ee}\xend^2\ln \xend -\frac{\xend^2}{\ee}\right]	
+\frac{1}{2}\right\rbrace ,
\end{equation}
where $\Lambert{0}$ is still the $0$-branch of the Lambert
function. It is displayed in \Fig{fig:potlambertWRI}, together with
the region where inflation proceeds.

Finally, it is interesting to constrain the value of the scale $M$ 
with the CMB normalization. It follows that
\begin{equation}
\left(\frac{M}{\Mp}\right)^4=2880\pi^2\left(\frac{\Mp}
     {\phizero}\right)^2\frac{1}
     {\xstar^2\ln^4\xstar}\frac{\Qrms^2}{T^2}\, .
\end{equation}

The reheating consistent slow-roll predictions of the Witten-
O'Raifeartaigh inflation models are displayed in
\Fig{fig:CMBWRI}. One should remember that in principle,
$\phizero\simeq\Mp$, even if a wider range of values for $\phizero$ is
displayed in order to understand how the predictions depend on this
parameter. In particular, when $\phizero\gg\Mp$, the predictions lie
along the line $\epsilon_2=2\epsilon_1$. Indeed, in this limit,
\Eq{eq:wri:xend} shows that $\xend\rightarrow 1$ while
\Eq{eq:wri:InvertedTraj} indicates that $\xstar\rightarrow 1$. As a
consequence, one obtains $\epsilon_{2*}\simeq\epsilon_{1*}$ from
\Eqs{eq:wri:eps1}~and~(\ref{eq:wri:eps2}).

\subsection{Dual Inflation (DI)}
\label{sec:di}
\subsubsection{Theoretical Justifications}
This model finds its roots in the $N=2$ supersymmetry $SU(2)$
Yang-Mills theories \`a la Seiberg-Witten~\cite{Seiberg:1994rs,
  Seiberg:1994aj}. If $\phi^i$ ($i \in \{1,2,3\}$) denote the scalars
belonging to the $N=2$ vector multiplets $\calA^i$ in the adjoint
representation, the classical potential of the theory is given
by~\cite{Seiberg:1994rs}
\begin{equation}
V =\frac{1}{g^2} \Tr\left[\phi,\phi^\dagger\right]^2.
\label{eq:Vsw}
\end{equation}
It exhibits a set of inequivalent vacua defined by the vanishing
commutator $\left[\phi,\phi^\dagger\right]=0$. Up to a gauge transformation, the
minima of \Eq{eq:Vsw} can be chosen along $\phi \equiv \frac{1}{2} a
\sigma_3$, where $\sigma_3=\diag(1,-1)$ and $a$ is a complex scalar. A
gauge invariant representation of this moduli space can be restored by
using the complex variable $u=\frac{1}{2}a^2 = \Tr \phi^2$, such that
for $u \ne 0$, $SU(2)$ is spontaneously broken to a residual
$U(1)$. Integrating out the multiplets which acquire a mass for
$u\ne0$ gives a low-energy effective theory describing a single $N=2$
vector multiplet $\calA$ whose scalar component is the order parameter
$a$. As shown in \Refcs{Seiberg:1994rs, Seiberg:1994aj}, expressed in
terms of $N=1$ supersymmetry, $\calA$ contains a $N=1$ chiral
multiplet $A$ and a vector multiplet $W_\alpha$ whose dynamics in
superspace is described by the Lagrangian
\begin{equation}
\calL = \dfrac{1}{4\pi} \Im \negthickspace \left[ \int \ud^4 \theta \dfrac{\partial
    \calF(A)}{\partial A} \overline{A} + \int \ud^2 \theta \dfrac{1}{2}
  \dfrac{\partial^2 \calF(A)}{\partial A^2} W_\alpha W^{\alpha} \right].
\label{eq:Lsw}
\end{equation}
From an exhaustive analysis of the monodromies that must exist within
the moduli space of the quantum vacua, \Refcs{Seiberg:1994rs,
  Seiberg:1994aj} were able to give an explicit expression for the
prepotential $\calF(A)$. Remarkably, it includes perturbative and
non-perturbative quantum corrections as a function of the only free
parameter of the model: the dynamical generated mass scale $\Lambda$
(which naturally appears at one loop). The spectrum of the theory
contains various magnetic and electric charges, dyons and
monopoles. As shown in \Refc{Seiberg:1994rs}, adding a $N=1$
preserving-mass term to the Lagrangian softly breaks $N=2$ to $N=1$
supersymmetry and triggers monopole condensation. Because the strongly
coupled regime of the theory is dual to weakly coupled
monopoles~\cite{Seiberg:1994pq}, this condensation has been shown to
explicitly describe confinement of the electric
charges~\cite{Seiberg:1994rs, Seiberg:1994aj}.

In this context, and motivated by the actual QCD vacuum structure,
\'Alvarez-Gaum\'e et al. have shown in \Refc{AlvarezGaume:1996gd} that
the only possible way to soft-break $N=2$ supersymmetry directly to
$N=0$ while preserving the analyticity properties of the
Seiberg-Witten model is to promote the dynamic scale $\Lambda$ to a
function of a new $N=2$ vector multiplet $\calS$. Once the scalar and
auxiliary components of this superfield are frozen, $N=2$ is softly
broken to $N=0$ while the non-perturbative scalar potential of the low
energy effective action still remains uniquely determined by the
analyticity properties of the moduli space of quantum vacua. Denoting
$a_0=s$ and $a_1=a$ the order parameters of the two multiplets, up to
small terms, the low-energy effective potential for the vacuum with
$S$ frozen reads~\cite{AlvarezGaume:1996gd}
\begin{equation}
V(a) = -\dfrac{2}{b_{11}} \rho^4 - \dfrac{\det(b)}{b_{11}} f_0^2, \qquad
\rho^2 = \sup\left\{-b_{11} |a|^2 + \dfrac{|b_{01}|
  f_0}{\sqrt{2}},0\right\} .
\label{eq:Vag}
\end{equation}
In this expression, $f_0 < \Lambda$ stands for the {\vev} of $S$'s
auxiliary field once frozen. The $\rho^4$ term is non vanishing only
when the monopoles acquire a non-vanishing {\vev}, which then lowers
the vacuum energy ($b_{11}>0$, see below) and shows that confinement
is favored~\cite{AlvarezGaume:1996gd}. The $b_{ij}$ are the imaginary
parts of the coupling matrix $\tau_{ij}$
\begin{equation}
b_{ij} \equiv \dfrac{1}{4 \pi} \Im(\tau_{ij}) = \dfrac{1}{4 \pi}
\Im\negthickspace \left( \dfrac{\partial^2 \calF}{\partial a_i
  \partial a_j} \right),
\label{eq:dicouplings}
\end{equation}
and everything can be expressed in terms of the complex variable
$u$. One gets~\cite{Seiberg:1994rs, Seiberg:1994aj,
  AlvarezGaume:1996gd}
\begin{equation}
a=\dfrac{4i \Lambda}{\pi k}(E'-K'), \quad \tau_{11} = i\dfrac{K}{K'}\,,
\quad \tau_{01}=i \dfrac{2 \Lambda}{k K'}\, \quad \tau_{00}= i \dfrac{8
  \Lambda^2}{\pi}\left(\dfrac{E'}{k^2 K'} - \dfrac{1}{2} \right),
\label{eq:dimonodromy}
\end{equation}
where $E(k)$, $K(k)$, $E'(k)$ and $K'(k)$ are the first and second
kind complete elliptic integrals~\cite{Gradshteyn:1965aa}. The
elliptic modulus $k$ is related to $u$ by
\begin{equation}
k^2 \equiv \dfrac{2}{1+\dfrac{u}{\Lambda^2}}\,.
\label{eq:dik2u}
\end{equation}
The potential of \Eq{eq:Vag} is thus defined over a two-dimensional
K\"ahler manifold, which in terms of $u$, has a non-canonical K\"ahler metric
\begin{equation}
\ud s^2 = \Im \negthickspace \left(\dfrac{\partial^2 \calF}{\partial
  a^2}\right) \ud a \ud \bar{a} = 4\pi b_{11} \left| \dfrac{\ud a}{\ud
  u} \right|^2 \ud u \ud \bar{u}.
\label{eq:dikahler}
\end{equation}
As remarked by Garcia-Bellido in \Refc{GarciaBellido:1997mq}, far from
the monopole condensation region, the potential $V(u)$ exhibits a flat
valley along the $\Re(u)$-direction and can support an inflationary
period. Because inflation ends within the monopole condensation
region, reheating can naturally occur by exciting the monopoles which
are the confined states associated with the electric charges of
theory. In this picture, inflation appears as a consequence of a
Yang-Mills phase transition from asymptotic freedom to confinement and
the inflaton is the order parameter. As discussed in
\Refc{GarciaBellido:1997mq}, the potential \eqref{eq:Vag} is not yet
completely satisfactory for cosmological purposes as it admits a
negative minimum and has to be uplifted. Assuming $f_0 \ll \Lambda$,
the minimum occurs at $k^2 \simeq 1$ and the uplifting constant to add
to the potential is
\begin{equation}
V_0 \simeq 1.
\end{equation}
Under these assumptions, dual inflation has been shown in
\Refcs{GarciaBellido:1997mq,GarciaBellido:1997gx} to generically yield
a spectral index in the range $\nS=0.9 \pm 0.1$ while being compatible
with the measured amplitude of the CMB anisotropies. In the next
section, we extend these papers to second order in slow roll, without
any other approximation, and then calculate how the model predictions
are affected by all the possible reheating histories. As detailed in
section~\ref{sec:direheat}, Dual Inflation shares with
Starobinsky/Higgs Inflation the remarkable feature to predict the
overall amplitude of the CMB anisotropies such that some care should
be taken when solving for the reheating equations.

\subsubsection{Slow-Roll Analysis}
\label{sec:disr}

Focussing on the inflationary valley defined to be on the real axis of
the complex moduli plane $u$ with $|u|>|\Lambda^2|$, and assuming
without loss of generality that $\Lambda = |\Lambda|$ is also real,
some simplifications can be made. As explicit in \Eq{eq:dik2u}, the
elliptic parameter $m \equiv k^2$, is bounded to $0<m<1$ and all the
complete elliptic integrals are real. Plugging \Eqs{eq:dimonodromy}
and \eqref{eq:dicouplings} into \Eq{eq:Vag} yields, after some algebra
\begin{equation}
\begin{aligned}
V(m) & = \dfrac{f_0^2 \Lambda^2}{\pi^2} \left\{1 + V_0 - 2 \dfrac{K-E}{m K}
- \dfrac{\pi}{m K K'} \nu^2(m) \Theta[\nu(m)] \right\},\\
\nu(m) & \equiv 1 - \dfrac{8 \sqrt{2}}{\pi^2} \dfrac{\Lambda}{f_0}
\dfrac{K (E'-K')^2}{m^{1/2}}\,,
\end{aligned}
\label{eq:dipot}
\end{equation}
where $\Theta(x)$ stands for the Heaviside function, $V_0$ is the
uplifting constant, and the terms in $\nu^2(m)$ correspond to the
monopole condensation previously discussed. They are non-vanishing for
large values of $m>\mmon$ where $\nu(\mmon) = 0$. All elliptic
integrals are implicitly assumed to be evaluated at the modulus
$k=\sqrt{m}$. This potential alone does not encode the inflationary
dynamics because $m$ is not the canonical scalar degree of
freedom. From the K\"ahler metric in \Eq{eq:dikahler}, along the $u$
(and $m$) real direction, the canonical scalar field $\phi$ can be
defined as
\begin{equation}
\left(\dfrac{\ud \phi}{\ud m}\right)^2 = \dfrac{8
  \Lambda^2}{\pi^2} \dfrac{K K'}{m^3} \impliedby
\phi(m) = \Lambda \dfrac{2\sqrt{2}}{\pi} \int_m^1
\dfrac{\sqrt{K(p^{1/2}) K'(p^{1/2})}}{p^{3/2}} \, \ud p,
\label{eq:diphi}
\end{equation}
in which $\phi(1)=0$ and $\phi(m) \to \infty$ for $m \to 0$. The above
expression cannot be explicitly integrated, neither inverted, such
that the potential of dual inflation is only known parametrically as
$\phi(m)$ and $V(m)$. Nonetheless, \Eqs{eq:dipot} and \eqref{eq:diphi}
show that dual inflation depends on two parameters $f_0$ and
$\Lambda$. They appear in the definition of $\nu(m)$ as the ratio
$f_0/\Lambda$ such that the shape of the potential actually depends
only on $f_0/\Lambda$. The combination $f_0^2\Lambda^2$ is an overall
multiplicative constant rescaling the potential as a whole whereas
$\Lambda$ alone rescales the field values. For these reasons, it is
more convenient to introduce the equivalent set of parameters, and
dimensionless field values, defined by
\begin{equation}
f \equiv \dfrac{f_0}{\Lambda}<1, \qquad M^4 \equiv 
\dfrac{f^2 \Lambda^4}{\pi^2}\,, \qquad x \equiv \dfrac{\phi}{\Lambda}\,.
\label{eq:fM4xdi}
\end{equation}
Because the parameter $M^4$ can be determined by the amplitude of the
CMB anisotropies, see \Eq{eq:pstarnorm}, dual inflation is a
one-parameter model parametrized by $f$. Once $f$ is chosen, the
value of $\Lambda$ can be obtained from $M^4$ using the above
equation. Let us notice that the uplifting term $V_0$ must be such
such that the minimum of the potential is exactly vanishing. As such,
although $V_0$ is of order unity for small values of $f$, it is a
non-trivial function of the parameter $f$ given by
\begin{equation}
V_0(f) = -1 + \left. \left(2  \dfrac{K-E}{m K}  + 
\dfrac{\pi \nu^2}{m K K'} \right)\right|_{m=m_{\min}},
\end{equation}
where $m_{\min}$ is the elliptic parameter at which $V(m)$ is
minimal. The logarithmic derivative of \Eq{eq:dipot} reads
\begin{equation}
\dfrac{\ud \ln V}{\ud m} = \dfrac{\pi \nu \Theta(\nu) \left\{ -4 m
  (m-1) \dot{\nu} K K'- \nu \left[K' (E+K) -E' K
    \right]\right \} + 2 {K'}^2 \left[E^2 + (m-1) K^2 \right]}{2
  m(m-1) K K' \left\{K' [2 E+K (m V_0 + m - 2)] - \pi \nu^2
  \Theta(\nu) \right\}}\,,
\label{eq:didlnVdm}
\end{equation}
and $m_{\min}$ is obtained by (numerically) solving
\begin{equation}
\pi \nu \left\{-4 m(m-1) \dot{\nu} K K' - \nu \left[K'(E+K)-E'K
  \right]\right\} + 2 {K'}^2 \left[E^2 + (m-1) K^2 \right] = 0,
\end{equation}
where $\dot{\nu}$ stands for the derivative of $\nu(m)$ with respect to $m$:
\begin{equation}
\dot{\nu}(m) = \dfrac{4 \sqrt{2}}{\pi^2 f} \dfrac{(E'-K') \{ E
  E'-K'[E + 2 (m-1)K ] \} }{(m-1) m^{3/2}}\,.
\label{eq:dinudot}
\end{equation}

\begin{figure}
\begin{center}
\includegraphics[width=\wdblefig]{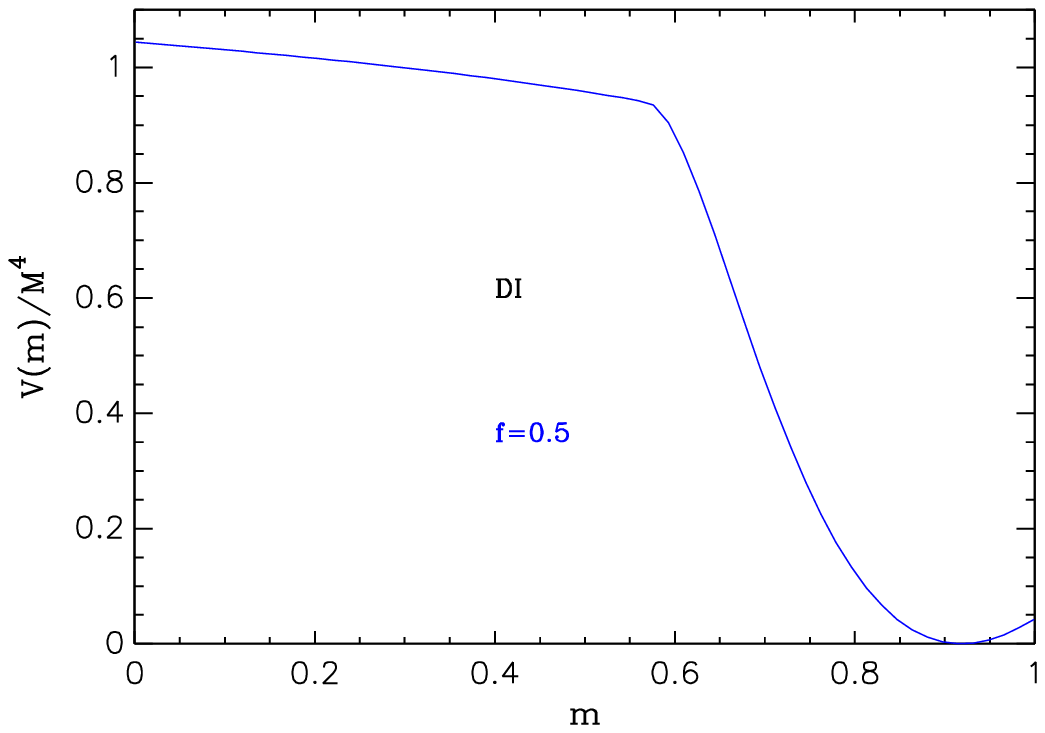}
\includegraphics[width=\wdblefig]{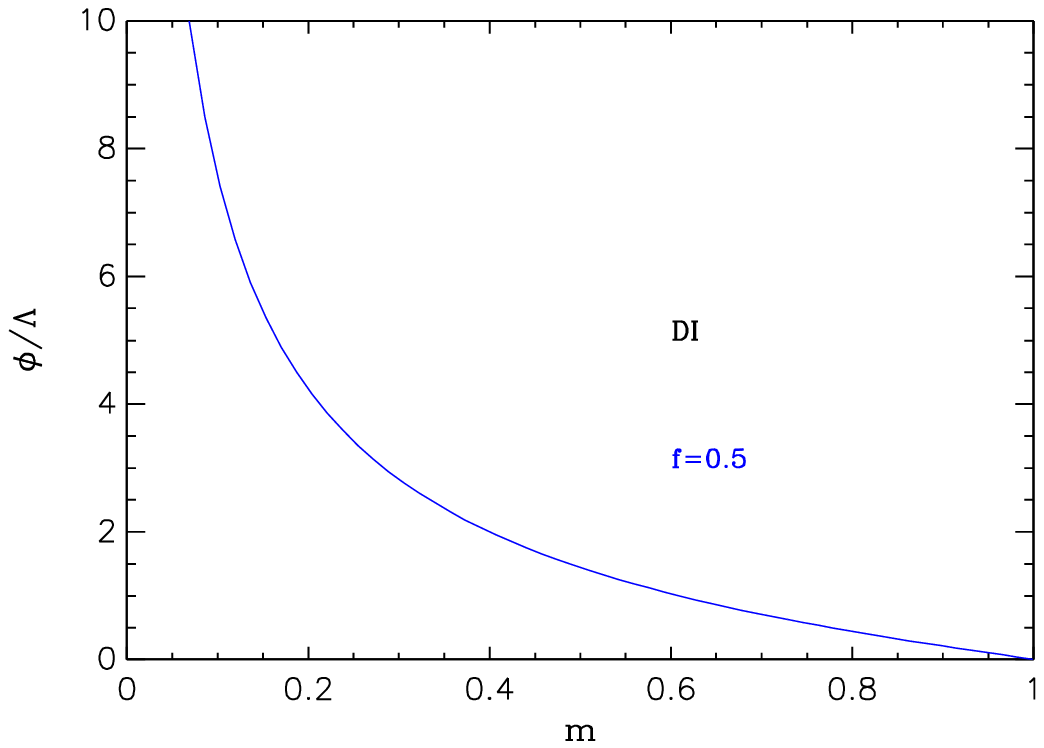}
\includegraphics[width=\wdblefig]{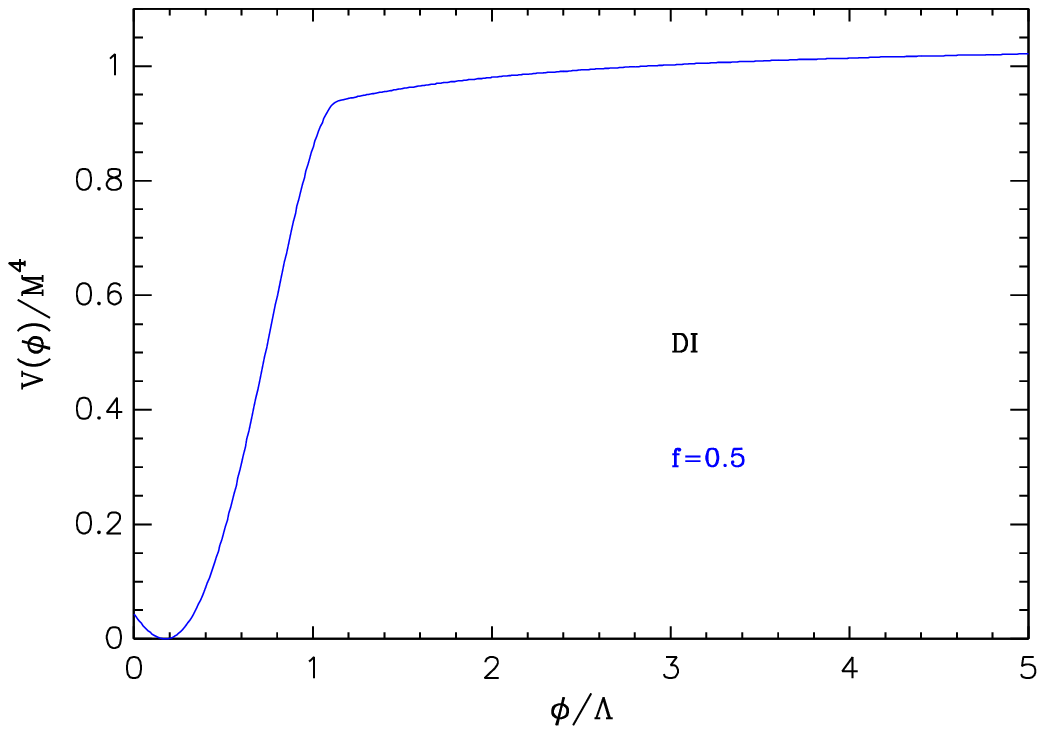}
\includegraphics[width=\wdblefig]{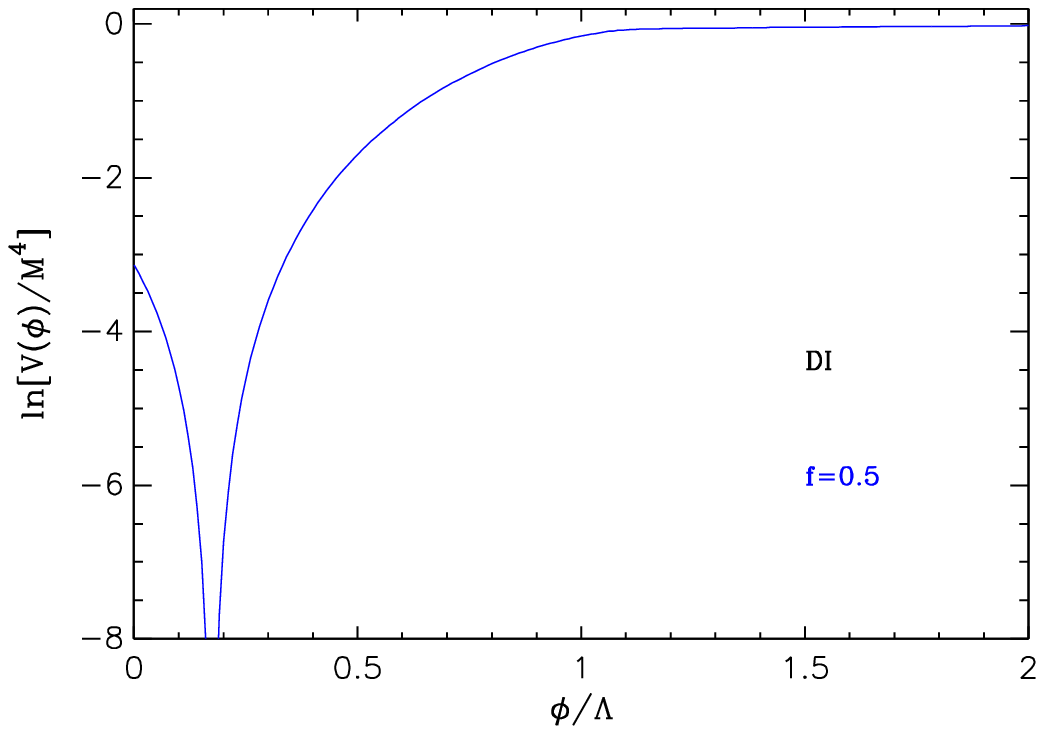}
\caption{Dual Inflation (DI) for $f=0.5$. Upper panels: the parametric
  potential $V(m)$ and field values $\phi(m)/\Lambda$ as a function of
  the elliptic parameter $m \equiv k^2$. Bottom panels: the potential
  $V(\phi)$ and its logarithm as a function of the canonically
  normalized field values $\phi/\Lambda$.}
\label{fig:potdi}
\end{center}
\end{figure}

The potentials $V(m)$ and $V(\phi)$, as well as the field values
$x(m)$ have been represented in figure~\ref{fig:potdi} for
$f=0.5$. Inflation proceeds from large field values $x$ to small field
values, or, equivalently, from small elliptic parameter values $m$ to
values close to unity.

In order to calculate the slow-roll parameters, one can use the chain
rule to introduce the parametric trajectory as a proxy. From the
definition of the Hubble flow functions in \Eqs{eq:defhf} and
\eqref{eq:defeps1}, one has
\begin{equation}
\begin{aligned}
\epsilon_1 & = \dfrac{1}{2 \Mp^2} \left(\dfrac{\ud \phi}{\ud m}
\right)^2 \left(\dfrac{\ud m}{\ud N} \right)^2 = \dfrac{4
  \Lambda^2}{\pi^2 \Mp^2} \dfrac{K K'}{m^3} \left(\dfrac{\ud m}{\ud N}
\right)^2, \\
\epsilon_{n+1} & = \dfrac{\ud \ln |\epsilon_n|}{\ud N} = \dfrac{\ud \ln
  |\epsilon_n|}{\ud m} \dfrac{\ud m}{\ud N}\,,
\end{aligned}
\end{equation}
such that the derivation of all $\epsilon_n$ requires only the
determination of the parametric trajectory $m(N)$. Assuming slow-roll
for the inflaton $\phi$, one has, at leading order, from
\Eq{eq:phidot2}
\begin{equation}
\dfrac{\ud \phi}{\ud N} = - \Mp^2 \dfrac{\ud \ln V}{\ud \phi}
\implies \dfrac{\ud m}{\ud N} = - \dfrac{\Mp^2 \pi^2}{8
  \Lambda^2} \dfrac{m^3}{K K'} \dfrac{\ud \ln V}{\ud m}\,,
\label{eq:diparamsr}
\end{equation}
such that the parametric slow-roll expressions for $\epsilon_n(m)$ simplify to
\begin{equation}
\begin{aligned}
\epsilon_1 & = \dfrac{\Mp^2 \pi^2}{16 \Lambda^2} \dfrac{m^3}{K
  K'} \left(\dfrac{\ud \ln V}{\ud m} \right)^2, \\
\epsilon_n & = -\dfrac{\Mp^2 \pi^2}{8 \Lambda^2} \dfrac{m^3}{K K'}
\left(\dfrac{\ud \ln V}{\ud m}\right) \left( \dfrac{\ud
  \ln|\epsilon_n|}{\ud m} \right).
\end{aligned}
\label{eq:diphf}
\end{equation}
Let us notice that all the slow-roll parameters have the same
dependency in the parameter $\Lambda$, namely $\epsilon_n \propto
1/\Lambda^2$. Because $\Lambda$ also enters into the definition of
$M^4$, it means that the field value at which inflation ends depends
on how the potential normalization matches the amplitude of the CMB
anisotropies, which itself depends on the reheating, and the reheating
quantities depend on the field value at which inflation ends. The
consistent way to simultaneously solve these conditions is discussed
in section~\ref{sec:direheat}. For the time being, plugging
\Eqs{eq:dipot} into \Eqs{eq:diphf} gives an explicit expression for
the parametric slow-roll functions. For the first one, one gets
\begin{equation}
\begin{aligned}
 & \epsilon_1(m) = \dfrac{\Mp^2 \pi^2}{16 \Lambda^2} \dfrac{m^3}{K
  K'} \\ & \times \left(\dfrac{\pi \nu \Theta(\nu) \left\{ -4 m
  (m-1) \dot{\nu} K K'- \nu \left[K' (E+K) -E' K
    \right]\right \} + 2 {K'}^2 \left[E^2 + (m-1) K^2 \right]}{2
  m(m-1) K K' \left\{K' [2 E+K (m V_0 + m - 2)] - \pi \nu^2
  \Theta(\nu) \right\}} \right)^2,
\end{aligned}
\label{eq:dieps1}
\end{equation}
in which $\nu(m)$ and $\dot{\nu}(m)$ are given in \Eqs{eq:dipot} and
\eqref{eq:dinudot}, respectively. The second Hubble flow function is a
bit more involved and reads
\begin{equation}
\begin{aligned}
& \epsilon_2(m) = \dfrac{\Mp^2 \pi^2}{8 \Lambda^2} \dfrac{m^4}{4
    (m-1)^2 K^4 {K'}^4 \left\{K' [2 E+K (m V_0+m-2)]-\pi \Theta(\nu) \nu
    ^2\right\}^2} \\  & \times \left(4 K K' \left\{K' [2 E+K (m
    V_0+m-2)]-\pi \Theta(\nu) \nu^2\right\} \right.  \\ & \times \left. 
  \left(\pi \Theta(\nu) \left\{K' \left[K (m-1) \left\{\nu \left[4
      \ddot{\nu} (m-1) m^2-\nu \right]+4 \dot{\nu}^2 (m-1) m^2+4
    \dot{\nu} m^2 \nu \right\}-E \nu ^2\right]
  \right. \right. \right. \\  & + \left.\left.\left.   E \nu ^2 E'\right\}
   + 2 K' \left\{K' [E+K
    (m-1)]^2-E' \left[E^2+K^2(m-1)\right]\right\} \right)
   \right. \\ & + \left.  \left\{K
  \left[(4 m-5) K'-3 E'\right]+3 E K'\right\} \right. \\ & \times
  \left.  \left\{ \pi \Theta(\nu) \nu \left(K \left\{K' [4 \dot{\nu}
    (m-1) m+\nu]-\nu E'\right\} + E \nu K'\right)- 2 {K'}^2
  \left[E^2+K^2(m-1)\right] \right\} \right. \\ & \times \left. 
  \left\{K' [2 E+K (m V_0+m-2)]-\pi \Theta(\nu) \nu ^2\right\} +2 K K'
  \right.  \\ & \times \left.  \left\{\pi \Theta(\nu) \nu \left[K \left\{K'
    [4 \dot{\nu} (m-1) m+\nu]-\nu E'\right\} + E\nu K'\right] - 2
        {K'}^2 \left[E^2+K^2(m-1)\right]\right\} \right. \\ & \times
        \left.  \left\{4 \pi \dot{\nu} \Theta(\nu) (m-1) m \nu -E' [2
          E+K (m V_0+m-2)]+m K' [E (V_0+1)+K (V_0-1)] \right\} \right),
\end{aligned}
\label{eq:dieps2}
\end{equation}
in which $\ddot{\nu}$ stands for the double derivative of $\nu(m)$. It
reads
\begin{equation}
\begin{aligned}
\ddot{\nu}(m) & = \dfrac{2 \sqrt{2}}{\pi^2 f}  \dfrac{1}{(m-1)^2
  m^{5/2}} \left\{2 E' K' [E (4 m-2)+K m (4 m-5)+K] \vphantom{{E'}^2}\right.\\
 &+ \left. {E'}^2 [-2 E m-3 K
   (m-1)] + {K'}^2 [E (4-6 m)-K (m-1) (10 m-7)] \right\}.
\end{aligned}
\end{equation}
Let us notice the appearance of Heaviside functions $\Theta(\nu)$ alone
in the expression of $\epsilon_2(m)$ showing that it is discontinuous,
but finite, when the monopole terms become non-vanishing. Finally,
after some algebra, one straightforwardly gets the third Hubble flow
function: 
{
\allowdisplaybreaks
\begin{align}
 \epsilon_3(m) & = -\dfrac{\Mp^2 \pi^2}{8 \Lambda^2} \dfrac{m^3}{K K'}
 \nonumber \\ & \times \left(
 \vphantom{\left[\left(\left\{\nu^2\right\}^{-2}\right)^{-1}\right]} 4
 K K' \left\{[2 E+K (V_0 m+m-2)] K'-\nu ^2 \pi \Theta(\nu) \right\}
 \left\{2 K' \left\{[E+K (m-1)]^2 K' \right. \right. \right.
 \nonumber \\ &- \left.\left.\left. \left[E^2+K^2 (m-1)\right]
 E'\right\}+\pi \Theta(\nu) \left[E E' \nu ^2 + \left(K (m-1) \left\{4
   \dot{\nu}^2 (m-1) m^2+4 \dot{\nu} \nu m^2
   \right. \right. \right.\right. \right.  \nonumber \\ & +
   \left.\left.\left. \left.\left. \left[4 \ddot{\nu} (m-1) m^2-\nu
     \right] \nu \right\} - E \nu ^2\right) K'\right] \right\} + 2 K
 K' \left\{4 \dot{\nu} (m-1) m \nu \pi \Theta(\nu) \right. \right.
 \nonumber \\ & - \left.\left.  [2 E+K (V_0 m+m-2)] E'+m [K (V_0-1)+E
   (V_0+1)] K'\right\} \right.  \nonumber \\ & \times
 \left. \left\{\nu \pi \Theta(\nu) \left(E \nu K'+K \left\{ [4
   \dot{\nu} (m-1) m+\nu ] K'-\nu E'\right\} \right) - 2 \left[E^2+K^2
   (m-1)\right] K'^2\right\} \right.  \nonumber \\ & + \left.\left[2
   E+K (V_0 m+m-2)] K'-\nu ^2 \pi \Theta(\nu) \right\} \left\{3 E K'+K
 \left[(4 m-5) K'-3 E'\right]\right\} \right.  \nonumber \\ & \times
 \left. \left\{\nu \pi \Theta(\nu) \left(E \nu K'+K \left\{[4
   \dot{\nu} (m-1) m+\nu ] K'-\nu E'\right\} \right) - 2 \left[E^2+K^2
   (m-1)\right] K'^2\right\}
 \vphantom{\left[\left(\left\{\nu^2\right\}^{-2}\right)^{-1}\right]}
 \right) \nonumber \\ & \times \left(
 \vphantom{\left[\left(\left\{\nu^2\right\}^{-2}\right)^{-1}\right]} 2
 K (m-1) m K' \left\{ \left[2 E+K (V_0 m+m-2)\right] K'-\nu ^2 \pi
 \Theta(\nu) \right\} \right.  \nonumber \\ & \times \left.  \left\{ 2
 \left[E^2+K^2 (m-1)\right] K'^2- \nu \pi \Theta(\nu) \left(E \nu K'+K
 \left\{[4 \dot{\nu} (m-1) m+\nu ] K'-\nu E'\right\} \right) \right\}
 \right.  \nonumber \\ & \times \left.  \left\{
 \vphantom{\left(\left\{\nu^2\right\}^{-2}\right)^{-1}} \left[
   \left\{[2 E+K (V_0 m+m-2)] K'-\nu ^2 \pi \Theta(\nu) \right\}
   \left\{3 E K'+K \left[(4 m-5) K'-3 E'\right]\right\}
   \right. \right. \right. \nonumber \\ & \times \left. \left. \left.
   \left\{-2 \left( 6 K' E^4 + K \left\{[m (5 V_0+13)-24] K'-2
   E'\right\} E^3-K^2 \left\{(V_0 m+m-2) E'
   \right.\right.\right.\right.\right.\right.  \nonumber \\ & +
   \left.\left. \left. \left. \left.\left.\left[-4 (V_0+1) m^2+(7
     V_0+29) m-28\right] K'\right\} E^2-K^3 (m-1)
   \right. \right.\right.\right.\right.  \nonumber \\ & \times
   \left.\left.\left.\left.\left.  \left\{2 E'+[3 m (V_0+1)-8]
   K'\right\} E-K^4 (m-1) \left\{(V_0 m+m-2) E'
   \right. \right.\right.\right.\right.\right.  \nonumber \\ &-
   \left.\left.\left.\left.\left.\left. [m (V_0-3)+2]
   K'\right\}\right) K'^3+\pi \Theta(\nu) \left\{12 \nu ^2 K'^2 E^3+K
   \nu K' \left\{[8 \dot{\nu} (m-1) m
     \right. \right.\right.\right.\right.\right.  \nonumber \\ &+
     \left.\left.\left.\left.\left.\left. \nu (5 V_0 m+21 m-36)] K'-6
   \nu E'\right\} E^2+2 K^2 \left[5 \nu ^2 E'^2+\nu \{\nu [4-m (3
       V_0+7)] \right.\right.\right.\right.\right.\right.  \nonumber
     \\ &- \left.\left.\left.\left.\left.\left. 20 \dot{\nu} (m-1) m\}
     K' E' + \left(16 \dot{\nu}^2 (m-1)^2 m^2+2 \dot{\nu} (m-1) \nu [m
       (5 V_0+21)-16] m
     \right.\right.\right.\right.\right.\right.\right.  \nonumber \\ &
     +\left.\left.\left.\left.\left.\left.\left. \nu \left\{16
     \ddot{\nu} (m-1)^2 m^2+\nu \left[2 (V_0+1) m^2-(V_0+12)
       m+10\right]\right\} \right) K'^2\right] E
   \right.\right.\right.\right.\right.  \nonumber \\ &+
   \left.\left.\left.\left.\left. K^3 \left[5 \nu ^2 (V_0 m+m-2)
     E'^2-2 \nu \left\{10 \dot{\nu} (m-1) m (V_0 m+m-2)
     \right.\right.\right.\right.\right.\right.\right.  \nonumber \\ &
     + \left.\left.\left.\left.\left.\left.\left. \nu \left[2 (V_0+1)
       m^2+(V_0-6) m+1\right]\right\} K' E'+\left(16 \dot{\nu}^2
     (m-1)^2 (V_0 m+m-2) m^2
     \right.\right.\right.\right.\right.\right.\right.  \nonumber
     \\ &+ \left.\left.\left.\left.\left.\left.\left. 4 \dot{\nu}
     (m-1) \nu \left[8 (V_0+1) m^2-3 (V_0+9) m+14\right] m
     \right.\right.\right.\right.\right.\right.\right.  \nonumber \\ &
     +\left.\left.\left.\left.\left.\left.\left. \nu \left\{16
     \ddot{\nu} (m-1)^2 (V_0 m+m-2) m^2+\nu [m
       (V_0-5)+4]\right\}\right) K'^2\right]\right\} K'+\nu ^2 \pi ^2
   \Theta(\nu) ^2 \right.\right.\right.\right.  \nonumber \\ & \times
   \left.\left.\left.\left.  \left[\left(-3 \nu ^2 E'^2+2 \nu [2
       \dot{\nu} (m-1) m+(2 m-1) \nu ] K' E'+\left\{16 \dot{\nu}^2
     (m-1)^2 m^2 \right.\right.\right.\right.\right.\right.\right.
     \nonumber \\ & - \left.\left.\left.\left.\left.\left.\left. 4
     \dot{\nu} \left(8 m^2-15 m+7\right) \nu m+\nu \left[\nu -16
       \ddot{\nu} (m-1)^2 m^2\right] \right\} K'^2\right) K^2+2 E \nu
     K' \right.\right.\right.\right.\right.  \nonumber \\ & \times
     \left. \left.\left.\left. \left.  \left\{\nu E'+[(3-2 m) \nu -2
       \dot{\nu} (m-1) m] K'\right\} K-3 E^2 \nu ^2 K'^2\right]
   \right\} \right.\right.\right.  \nonumber \\ & -
   \left.\left. \left.  2 K K' \left\{-4 \dot{\nu} (m-1) m \nu \pi
   \Theta(\nu) +[2 E+K (V_0 m+m-2)] E' \right.\right.\right.\right.
   \nonumber \\ & - \left.\left.\left. \left. m [K (V_0-1)+E (V_0+1)]
   K'\right\} \left(4 K K' \left\{[2 E+K (V_0 m+m-2)] K'-\nu ^2 \pi
   \Theta(\nu) \right\} \right.\right.\right.\right.  \nonumber \\ &
   \times \left.\left.\left. \left.  \left\{2 K' \left\{[E+K (m-1)]^2
   K'-\left[E^2+K^2 (m-1)\right] E'\right\}
   \right.\right.\right.\right.\right.  \nonumber \\ & +
   \left.\left.\left. \left.\left. \pi \Theta(\nu) \left[E E' \nu
     ^2+\left(K (m-1) \left\{4 \dot{\nu}^2 (m-1) m^2+4 \dot{\nu} \nu
     m^2+\left[4 \ddot{\nu} (m-1) m^2-\nu \right] \nu \right\}
     \right.\right.\right.\right.\right.\right.\right.  \nonumber \\ &
     - \left.\left.\left. \left.\left.\left. \left. E \nu ^2\right)
     K'\right] \right\} +2 K K' \left\{4 \dot{\nu} (m-1) m \nu \pi
   \Theta(\nu) -[2 E+K (V_0 m+m-2)] E'
   \right.\right.\right.\right.\right.  \nonumber \\ & +
   \left.\left.\left. \left.\left. m [K (V_0-1)+E (V_0+1)] K'\right\}
   \left(\nu \pi \Theta(\nu) \left\{E \nu K'+K \left[(4 \dot{\nu}
     (m-1) m+\nu ) K'-\nu E'\right]\right\}
   \right.\right.\right.\right.\right.  \nonumber \\ & -
   \left.\left.\left. \left.\left. 2 \left[E^2+K^2 (m-1)\right]
   K'^2\right) + \left\{[2 E+K (V_0 m+m-2)] K'-\nu ^2 \pi \Theta(\nu)
   \right\} \right.\right.\right.\right.  \nonumber \\ & \times
   \left.\left.\left. \left.  \left(3 E K'+K \left[(4 m-5) K'-3
     E'\right]\right) \left\{\nu \pi \Theta(\nu) \left(E \nu K'+K
   \left\{[4 \dot{\nu} (m-1) m+\nu ] K'
   \right.\right.\right.\right.\right.\right.\right.  \nonumber \\ & -
   \left.\left.\left. \left.\left.\left. \left. \nu E'\right\}\right)
   - 2 \left[E^2+K^2 (m-1)\right] K'^2\right\}\right) \right]
 \right.\right.  \nonumber \\ & \times \left.\left.  \left(8 K^4
 (m-1)^3 K'^4 \left\{[2 E+K (V_0 m+m-2)] K'-\nu ^2 \pi \Theta(\nu)
 \right\}^3\right)^{-1} \right.\right.  \nonumber \\ & + \left.\left.
 \left\{m \left[ 4 K K' \left\{[2 E+K (V_0 m+m-2)] K'-\nu ^2 \pi
   \Theta(\nu) \right\} \left(4 K (m-1)^2 m^2 \pi \delta (\nu ) K'
   \dot{\nu}^3 \right.\right. \right.\right.\right.  \nonumber \\ & +
   \left. \left.\left. \left.\left.  \left[m(m-1)\right]^{-1}
   \left\{-\left[E^2+K^2 (m-1)\right] E'^2+2 \left[-(m-2) E^2+4 K
     (m-1) E \right.\right. \right.\right.\right.\right.\right.
     \nonumber \\ & +
     \left. \left.\left. \left. \left.\left.\left. K^2 \left(m^2-3
     m+2\right)\right] K' E'- \left[E^2+2 K (m-1) E-K^2 (m-1)\right] m
   K'^2 \right.\right. \right.\right.\right.\right.  \nonumber \\ & +
   \left.\left. \left.\left.\left.\left. (m-1) \pi \Theta(\nu) \left[
     \left(E \nu (2 \dot{\nu} m+\nu )+K \left\{2 \dot{\nu}^2 (m-1)
     m^2+2 \dot{\nu} \nu m^2
     \right.\right. \right.\right.\right.\right.\right.\right.\right.
     \nonumber \\ & +
     \left.\left. \left.\left.\left.\left. \left.\left.\left.  \left[2
       \ddot{\nu} (m-1) m^2-\nu \right] \nu \right\}\right) E'+2 m
     \left(K \left\{m \left(8 m^2-11 m+3\right) \dot{\nu}^2 +
     \left[\nu (1-2 m)^2
       \right. \right. \right. \right. \right. \right.\right. \right. \right. \right. \nonumber
       \\ & +
       \left. \left. \left.\left.\left.\left. \left.\left.\left.\left.
       6 \ddot{\nu} (m-1)^2 m^2\right] \dot{\nu}+(m-1) m [2
       \dddot{\nu} (m-1) m+\ddot{\nu} (8 m-3)] \nu \right\}
     \right. \right.\right.\right.\right.\right.\right.\right.
     \nonumber \\ & -
     \left.\left. \left.\left.\left.\left. \left.\left. E \left[(m-1)
       m \dot{\nu}^2+(m+1) \nu \dot{\nu}+\ddot{\nu} (m-1) m \nu
       \right] \right) K'\right] \right\} \right)
   +[2(m-1)]^{-1}\left(m^{-1} \right.\right.\right.\right.\right.
   \nonumber \\ & \times \left.  \left.\left.\left.\left.  \left[4 K
     K' \left\{-4 \dot{\nu} (m-1) m \nu \pi \Theta(\nu) +[2 E+K (V_0
       m+m-2)] E'\right.\right.\right.\right.\right.\right.\right.
     \nonumber \\ & - \left.\left.\left.\left.\left.\left.\left. m [K
       (V_0-1)+E (V_0+1)] K'\right\} \left\{2 K' \left\{[E+K(m-1)]^2
     K'- \left[E^2+K^2 (m-1)\right] E'\right\}
     \right.\right.\right.\right.\right.\right.\right.  \nonumber \\ &
     + \left.\left.\left.\left.\left.\left.\left. \pi \Theta(\nu)
     \left[E E' \nu ^2+\left(K (m-1) \left\{4 \dot{\nu}^2 (m-1) m^2+4
       \dot{\nu} \nu m^2 + \left[4 \ddot{\nu} (m-1) m^2-\nu \right]
       \nu \right\}
       \right.\right.\right.\right.\right.\right. \right.\right.\right.
       \nonumber \\ &-
       \left. \left.\left. \left.\left.\left.\left.\left.\left. E \nu
       ^2\right) K'\right]\right\} + \left(4 K K' \left\{4 \dot{\nu}
     (m-1) m \nu \pi \Theta(\nu) -[2 E+K (V_0 m+m-2)] E'
     \right.\right.\right.\right.\right. \right.\right.\right.
     \nonumber \\ & +
     \left.\left. \left.\left.\left.\left.\left.\left. m [K (V_0-1)+E
       (V_0+1)] K'\right\} \left\{2 K' \left\{[E+K (m-1)]^2
     K'-\left[E^2+K^2 (m-1)\right] E'\right\}
     \right.\right.\right.\right.\right. \right.\right.\right.
     \nonumber \\ & +
     \left.\left. \left.\left.\left.\left.\left.\left. \pi \Theta(\nu)
     \left[E E' \nu ^2+\left(K (m-1) \left\{4 \dot{\nu}^2 (m-1) m^2+4
       \dot{\nu} \nu m^2+\left[4 \ddot{\nu} (m-1) m^2-\nu \right] \nu
       \right\}
       \right.\right.\right.\right.\right. \right.\right.\right.\right.\right.
       \nonumber \\ & -
       \left.\left. \left.\left.\left.\left.\left.\left.\left.\left. E
       \nu ^2\right) K'\right]\right\}\right) +\left(2 \left\{[2 E+K
       (V_0 m+m-2)] K'-\nu ^2 \pi \Theta(\nu) \right\} \left\{3 E K' +
     K \right.\right.\right.\right.\right. \right.\right.\right.
     \nonumber \\ & \times
     \left.\left. \left.\left.\left.\left.\left.\left. \left[(4 m-5)
       K'-3 E'\right]\right\} \left\{2 K' \left\{[E+K (m-1)]^2
     K'-\left[E^2+K^2 (m-1)\right] E'\right\} + \pi \Theta(\nu)
     \right.\right.\right.\right.\right. \right.\right.\right.
     \nonumber \\ & \times
     \left.\left. \left.\left.\left.\left.\left.\left. \left[E E' \nu
       ^2+\left(K (m-1) \left\{4 \dot{\nu}^2 (m-1) m^2+4 \dot{\nu} \nu
       m^2+\left[4 \ddot{\nu} (m-1) m^2-\nu \right] \nu \right\}
       \right.\right.\right.\right.\right. \right.\right.\right.\right.\right.
       \nonumber \\ & -
       \left.\left. \left.\left.\left.\left.\left.\left.\left.\left. E
       \nu ^2\right) K'\right]\right\}\right)\right] + 4 K K' \left\{4
   (m-1) \left[(m-1) m \dot{\nu}^2+(2 m-1) \nu \dot{\nu}+\ddot{\nu}
     (m-1) m \nu \right] \pi \Theta(\nu)
   \right.\right.\right.\right.\right. \right.  \nonumber \\ & +
   \left.\left. \left.\left.\left.\left. [E (V_0-1)-K (m-1) (V_0+1)]
   E'+E (V_0 m+m-2 V_0) K'\right\} \left\{\nu \pi \Theta(\nu) \left( E
   \nu K' \right.\right.\right.\right.\right. \right.\right.
   \nonumber \\ & + \left.\left. \left.\left.\left.\left.\left. K
   \left\{[4 \dot{\nu} (m-1) m+\nu] K'-\nu E'\right\} \right) - 2
   \left[E^2+K^2 (m-1)\right] K'^2 \right\} + m^{-1}
   \right.\right.\right.\right.\right.  \nonumber \\ & \times
   \left.\left. \left.\left.\left. \left\{ \left\{-4 \dot{\nu} (m-1) m
   \nu \pi \Theta(\nu) + [2 E+K (V_0 m+m-2)] E'-m [K (V_0-1)
     \right.\right.\right.\right.\right.\right.\right.  \nonumber \\ &
     + \left.\left. \left.\left.\left.\left.\left. E (V_0+1)]
   K'\right\} \left\{3 E K'+K \left[(4 m-5) K'-3 E'\right] \right\}
   \left[\nu \pi \Theta(\nu) \left(E \nu K'
     \right.\right.\right.\right.\right.\right.\right. \right.
     \nonumber \\ & +
     \left.\left. \left.\left.\left.\left.\left. \left.  K \left\{[4
       \dot{\nu} (m-1) m+\nu ] K'-\nu E'\right\} \right) - 2
     \left[E^2+K^2 (m-1)\right] K'^2\right]
   \right.\right.\right.\right.\right.\right.  \nonumber \\ & +
   \left.\left.\left.\left. \left.\left. \left[2 \left(-2 [E+K (m-1)]
     K' \left\{ [2 E+K (V_0 m+m-2)] K'-\nu ^2 \pi \Theta(\nu) \right\}
     \right.\right.\right.\right.\right. \right. \right.\right.
     \nonumber \\ & \times
     \left.\left.\left.\left. \left. \left. \left. \left.  \left\{2 K'
     \left\{ [E+K (m-1)]^2 K'-\left[E^2+K^2 (m-1)\right] E'\right\} +
     \pi \Theta(\nu)
     \right.\right.\right.\right. \right. \right.\right.\right.\right.
     \nonumber \\ & \times
     \left. \left. \left.\left.\left.\left. \left. \left. \left.
     \left[E E' \nu ^2+\left(K (m-1) \left\{4 \dot{\nu}^2 (m-1) m^2+4
       \dot{\nu} \nu m^2+\left[4 \ddot{\nu} (m-1) m^2-\nu \right] \nu
       \right\}
       \right.\right.\right.\right. \right. \right.\right. \right.\right.\right.\right.
       \nonumber \\ & -
       \left. \left. \left.\left.\left.\left. \left. \left. \left.\left. \left.
       E \nu ^2\right) K'\right] \right\} + 2 K \left(E'-m K'\right)
     \left\{ [2 E+K (V_0 m+m-2)] K'-\nu ^2 \pi \Theta(\nu) \right\}
     \right.\right.\right.\right. \right. \right.\right. \right.
     \nonumber \\ & \times
     \left. \left. \left.\left.\left.\left. \left. \left.  \left\{2 K'
     \left\{[E+K (m-1)]^2 K'-\left[E^2+K^2 (m-1)\right] E'\right\} +
     \pi \Theta(\nu) \left[E E' \nu ^2 + \left(K (m-1)
       \right.\right.\right.\right. \right. \right.\right. \right.\right.\right.\right.
       \nonumber \\ & \times
       \left. \left. \left.\left.\left.\left. \left. \left. \left.\left. \left.
       \left\{4 \dot{\nu}^2 (m-1) m^2+4 \dot{\nu} \nu m^2+\left[4
         \ddot{\nu} (m-1) m^2-\nu \right] \nu \right\} - E
       \nu^2\right) K'\right] \right\} + K \left(E'-m K'\right)
     \right.\right.\right.\right. \right. \right.\right. \right.
     \nonumber \\ & \times
     \left. \left. \left. \left. \left. \left. \left. \left. \left\{-4
     \dot{\nu} (m-1) m \nu \pi \Theta(\nu) +[2 E+K (V_0 m+m-2)] E'-m
     \left[K (V_0-1)+E (V_0+1)\right]
     \right.\right.\right.\right. \right. \right.\right. \right. \right.
     \nonumber \\ & \times
     \left. \left. \left. \left. \left. \left. \left. \left. \left.
     K'\right\} \left[2 \left[E^2+K^2 (m-1)\right] K'^2-\nu \pi
       \Theta(\nu) \left(E \nu K'+ K \left\{ [4 \dot{\nu} (m-1) m+\nu
       ] K'
       \right.\right.\right.\right. \right. \right.\right. \right. \right. \right.\right.
       \nonumber \\ & -
       \left. \left. \left. \left. \left. \left. \left. \left. \left.\left. \left.\nu
       E'\right\} \right) \right] + [E+K (m-1)] K' \left\{4 \dot{\nu}
     (m-1) m \nu \pi \Theta(\nu) -[2 E+K (V_0 m+m-2)] E'
     \right.\right.\right.\right. \right. \right.\right. \right. \right.
     \nonumber \\ & +
     \left. \left. \left. \left. \left. \left. \left. \left. \left. m
     \left[K (V_0-1)+E (V_0+1)\right] K'\right\} \left[ 2
       \left[E^2+K^2 (m-1)\right] K'^2 - \nu \pi \Theta(\nu)
       \right.\right.\right.\right. \right. \right.\right. \right.\right.
       \nonumber \\ & \times
       \left. \left. \left. \left. \left. \left. \left. \left. \left. \left(E
       \nu K'+K \left\{ [4 \dot{\nu} (m-1) m+\nu ] K'-\nu E'\right\}
       \right) \right] + \left\{[3 E+2 K (m-2)] E'
     \right.\right.\right.\right. \right. \right.\right. \right.\right.
     \nonumber \\ & +
     \left. \left. \left. \left. \left. \left. \left. \left. \left.
     \left[E (1-2 m)+K (3 m-1)\right] K'\right\} \left\{[2 E+K (V_0
       m+m-2)] K'-\nu ^2 \pi \Theta(\nu) \right\}
     \right.\right.\right.\right. \right. \right.\right.\right.
     \nonumber \\ & \times
     \left. \left. \left. \left. \left. \left. \left. \left.
     \left\{\nu \pi \Theta(\nu) \left(E \nu K'+K \left\{ [4 \dot{\nu}
       (m-1) m+\nu ] K'-\nu E'\right\} \right)
     \right.\right.\right.\right. \right. \right.\right.\right.\right.
     \nonumber \\ & -
     \left.\left. \left. \left. \left. \left. \left. \left. \left. 2
     \left[E^2+K^2 (m-1)\right] K'^2\right\} \right) \right] \right\}
   \right) \right] \right\} \left(4 K^3 (m-1)^2 K'^3
 \vphantom{\left\{\nu^2\right\}^{2}} \right.\right.\right.  \nonumber
 \\ & \times \left. \left. \left. \left\{[2 E+K (V_0 m+m-2)] K'-\nu ^2
 \pi \Theta(\nu) \right\}^2\right)^{-1} \right\} \right).
\label{eq:dieps3}
\end{align}
}
In this expression, $\dddot{\nu}$ is the third derivative of $\nu(m)$
with respect to $m$ and reads
\begin{equation}
\begin{aligned}
\dddot{\nu} & = \dfrac{\sqrt{2}}{\pi ^2 f} \dfrac{1}{(m-1)^3 m^{7/2}}
\left( -2 E' K' \left\{E [m (26 m-23)+5]+K (m-1) [m (24 m-13)-3]
\right\} \right. \\ &+ \left. E'^2  \left\{E [m (8 m+7)-7]+4 K (m-1) (7
m-5) \right\} \right. \\ & + \left. K'^2 \left[E \left(50 m^2-65
   m+23\right )+ 2 K (m-1) \left(36 m^2-51 m+19\right) \right] \right).
\end{aligned}
\label{eq:dinudddot}
\end{equation}
\begin{figure}
\begin{center}
\includegraphics[width=\wdblefig]{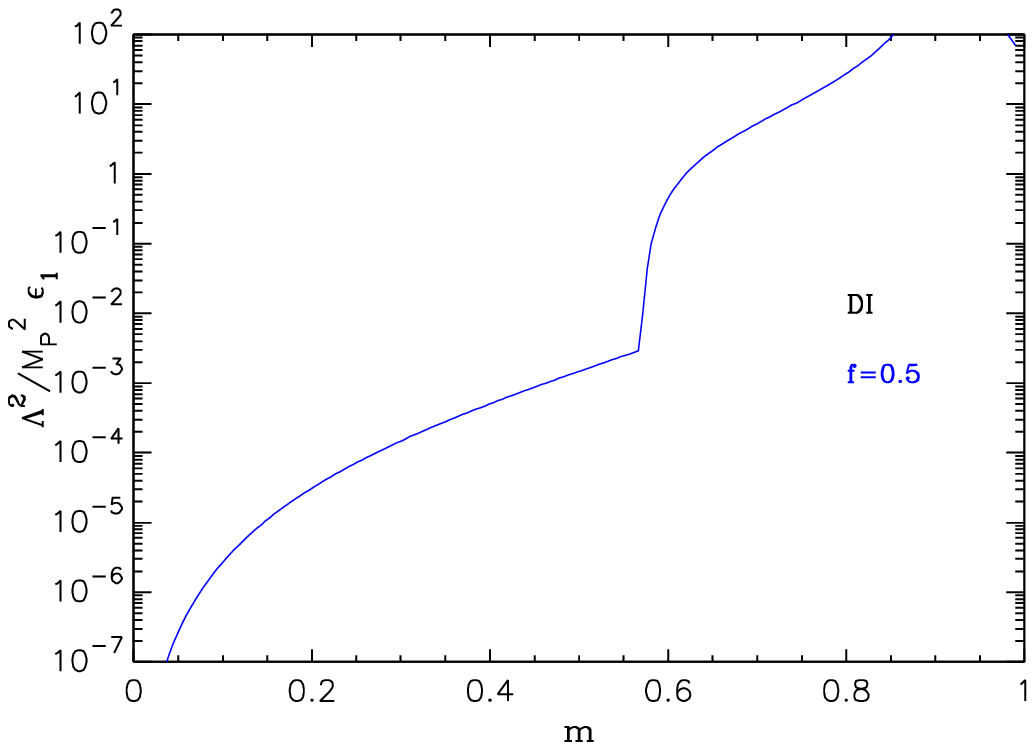}
\includegraphics[width=\wdblefig]{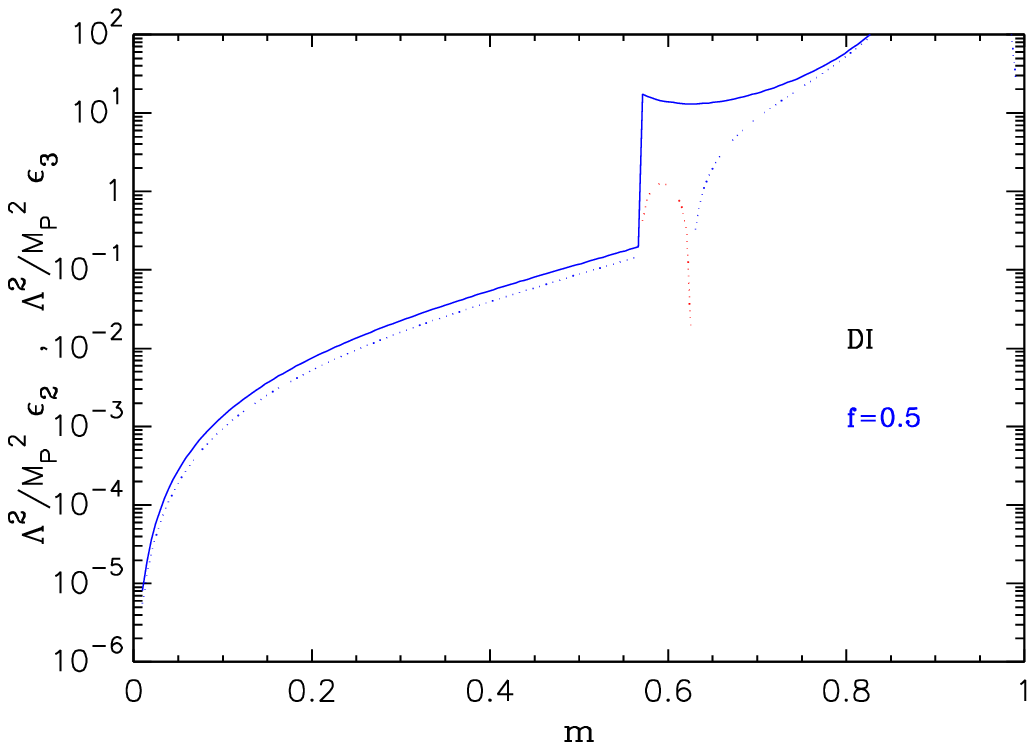}
\caption{Left panel: rescaled parametric slow-roll function
  $\epsilon_1(m)$ (multiplied by $\Lambda^2/\Mp^2$) with respect to
  the elliptic parameter $m$ for dual inflation (the potential is
  represented in \Fig{fig:potdi}). Right panel: rescaled slow-roll
  functions $\epsilon_2(m)$ (solid line) and $\epsilon_3(m)$ (dotted
  lines), both multiplied by $\Lambda^2/\Mp^2$. Although
  $\epsilon_2(m)$ is discontinuous at $m=\mmon$, it remains finite. The
  parameter $\epsilon_3(m)$ is however singular at $m=\mmon$ and
  changes sign (negative values have been represented in red on the
  logarithmic scale). Because $\Lambda<\Mp$, inflation always occurs
  in the domain $m < \mmon$ for which $\epsilon_1(m) < 1$.}
\label{fig:psr123di}
\end{center}
\end{figure}
As expected, the expression of $\epsilon_3(m)$ contains a Dirac
distribution $\delta(\nu)$ and is therefore singular at the elliptic
parameter value $\mmon$ for which $\nu(\mmon)=0$. As explicit in
\Eq{eq:dipot}, the potential contains a feature for $m=\mmon$ and the
slow-roll approximation breaks down, in a way similar to the
Starobinsky model~\cite{Starobinsky:1992ts}. There are however two
differences. The first one is that this feature generically occurs after,
or at the end of, inflation for $\Lambda<\Mp$ (see
\Fig{fig:psr123di}), namely during the reheating stage. Therefore, it
is not observable and cannot affect the slow-roll predictions which
are confined withing the inflationary valley at $m \ll \mmon$. The
second difference with respect to \Refc{Starobinsky:1992ts} is that
the singular behavior induced by the feature only appears at the
perturbative level. Indeed, since both $\epsilon_1(m)$ and
$\epsilon_2(m)$ remain finite at $m=\mmon$, slow roll is not
necessarily violated for the background field trajectory [see
  \Eq{eq:phidot2}]. However, the singularity in $\epsilon_3(m)$
necessarily implies that the perturbations generated at $m=\mmon$
cannot be of the slow-roll kind. Again, this is not problematic for
our purpose as this concerns very small length scales but such a
feature might have some interesting consequences concerning reheating
microphysics and black-hole formation~\cite{Clesse:2015wea},
especially if the model is viewed as a toy realization of the QCD-like
phase transition~\cite{Carr:2019hud}. The parametric Hubble flow
functions have been represented in \Fig{fig:psr123di} as a function of
the elliptic parameter $m$.

\begin{figure}
\begin{center}
\includegraphics[width=\wsingfig]{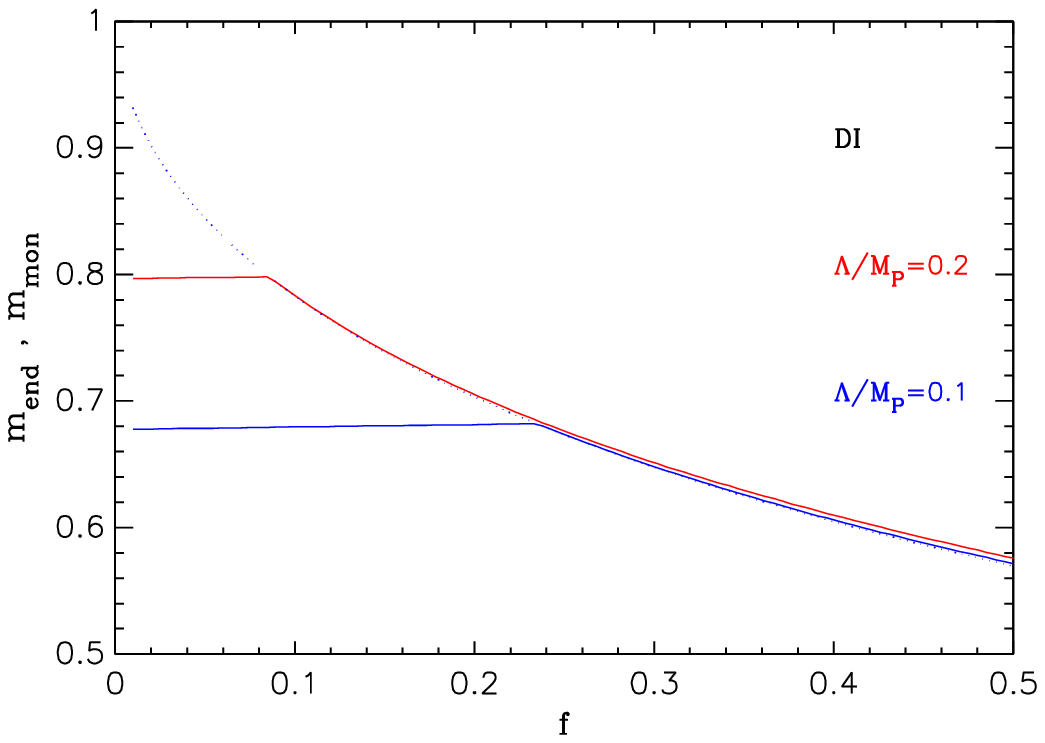}
\caption{Elliptic parameter $\mend$ at which inflation ends (solid
  lines), and $\mmon$ at which monopole condensation occurs (dotted
  line), as a function of $f$. The upper red curve shows $\mend$ for
  $\Lambda/\Mp=0.2$ while the lower blue one for $\Lambda/\Mp=0.1$
  ($\mmon$ does not depend on $\Lambda$). For small values of $f$,
  inflation always ends before monopole condensation $\mend <
  \mmon$. However, for Planckian-like values of $\Lambda$, inflation
  may end in the monopole condensation if $f$ is sufficiently
  large. As these curves show, in that situation, inflation does no
  last much longer, the potential with monopole terms switched on
  being steep, one has $\mend \gtrsim \mmon$ independently of $f$ (see
  also $\epsilon_1$ in \Fig{fig:psr123di}).}
\label{fig:modsDI}
\end{center}
\end{figure}

Owing to the parametric expression of the slow-roll functions given in
\Eqs{eq:dieps1}, \eqref{eq:dieps2} and \eqref{eq:dieps3}, observable
predictions can be directly extracted from the parametric trajectory
$m(N)$. From \Eq{eq:diparamsr}, one gets
\begin{equation}
\Nend - N = \dfrac{16 \Lambda^2}{\Mp^2 \pi^2} \left[\calI(m,f) -
  \calI(\mend,f)\right],
\label{eq:trajdi}
\end{equation}
where
\begin{equation}
\begin{aligned}
 & \calI(m,f) \\ &= \int_{}^m \dfrac{ (p-1) \left(K K'\right)^2
    \left\{K' [2 E+K (p V_0 + p - 2)] - \pi \nu^2 \Theta(\nu)
    \right\}}{\pi \nu \Theta(\nu) p^2 \left\{ -4 p (p-1) \dot{\nu} K
    K'- \nu \left[K' (E+K) -E' K \right]\right \} + 2 p^2 K'^2 \left[E^2
      + (p-1) K^2 \right]}\, \ud p.
\end{aligned}
\label{eq:diprim}
\end{equation}
In practice, provided one remains focused on the inflationary dynamics
for $m < \mmon$, this expression can be simplified into
\begin{equation}
\calI(m < \mmon,f) = \int_{\mmon}^m \dfrac{ (p-1) \left(K K'\right)^2
    \left\{K' [2 E+K (p V_0 + p - 2)]\right\}}{2 p^2 K'^2 \left[E^2
      + (p-1) K^2 \right]}\, \ud p.
\end{equation}
There is no explicit analytic solutions for these two integrals and
$\calI(m,f)$ has to be evaluated numerically. One should notice the
overall $\Lambda^2$ factor in \Eq{eq:trajdi} showing that the number
of {\efolds}, at given elliptic parameter $m$, indirectly depends on
the amplitude of the CMB anisotropies (see
section~\ref{sec:direheat}). This expression has to be completed by
the value of the elliptic parameter $\mend$ at which inflation
ends. As can be seen in \Fig{fig:psr123di}, dual inflation gracefully
exits as soon as $\epsilon_1(\mend)=1$, i.e. for
\begin{equation}
\begin{aligned}
 \dfrac{4 \Lambda}{\pi \Mp}  & = \dfrac{\mend^{1/2}\left(\Kend
      \Kpend\right)^{-3/2}}{2 (\mend-1) 
    \left\{\Kpend [2 \Eend + \Kend (\mend V_0 + \mend - 2)] - \pi
    \nuend^2 \Theta(\nuend) \right\}} \\ & \times  \left( \pi \nuend
 \Theta(\nuend) \left\{ -4 \mend
    (\mend-1) \dnuend \Kend \Kpend \right. \right. \\ & -
 \left. \left. \nuend \left[\Kpend (\Eend+\Kend) -
      \Epend \Kend \right]\right \} + 2 \Kpend^2 \left[\Eend^2 +
      (\mend-1) \Kend^2 \right] \right) \,.
\end{aligned}
\label{eq:dimend}
\end{equation}
There is neither analytical solution to this equation and it has to be
solved for $\mend(\Lambda,f)$ numerically. In \Fig{fig:modsDI}, we
have represented the values of $\mend(\Lambda,f)$ and $\mmon(f)$ for
$\Lambda=0.1\Mp$ and $\Lambda=0.2\Mp$. At small $f$, inflation always
ends before monopole condensation ($\mend < \mmon$). Only for
Planckian-like values of $\Lambda$ there exists a domain, corresponding
to values of $f$ or order unity, in which $\mend \gtrsim \mmon$. As
the solid curves of \Fig{fig:modsDI} show, in this situation, $\mend$
still remains very close to $\mmon$ because the potential for $m>\mmon$
is very steep and inflation ends \emph{almost} instantaneously. From
an observational point of view, the value of $f$ at which the two
curves intersect correspond to a transition between two typical
paradigms. For small values of $f$, $\epsilon_1$ smoothly increases
during inflation to reach unity before the reheating starts. Such a
situation is reminiscent with the parametric reheating paradigm. On
the contrary, for large values of $f$, corresponding to $\mend \gtrsim
\mmon$, the monopole condensation stops inflation almost
instantaneously, i.e., the first Hubble flow function $\epsilon_1(m)$
can still be small just before the end of inflation. The latter
situation actually mimics the effect of a tachyonic reheating.

In the next section, we discuss how to determine $\mstar$, the
elliptic parameter value at which the pivot mode crosses the Hubble
radius during inflation, in a consistent way, by taking into account
the CMB normalization of the parameter $M^4$.

\subsubsection{Reheating consistent observable predictions}
\label{sec:direheat}

From \Eqs{eq:pstarnorm} and \eqref{eq:fM4xdi}, at a given parameter
$f$, the value of $\Lambda$ matching the amplitude of the CMB
anisotropies verifies
\begin{equation}
\dfrac{\Lambda^4}{\Mp^4} = \dfrac{24 \pi^4 \Pstar}{f^2}
\dfrac{\epsilon_1(\mstar)}{v(\mstar)}\,,
\end{equation}
where, because $\mstar < \mmon$, one has
\begin{equation}
v(\mstar) \equiv \dfrac{V(\mstar)}{M^4} = 1 + V_0(f) - 2
\dfrac{K(\mstar^{1/2}) - E(\mstar^{1/2})}{\mstar K(\mstar^{1/2})}\,.
\label{eq:divstar}
\end{equation}
From the expression of $\epsilon_1$ given in \Eq{eq:dieps1}, using again the
fact that $\mstar < \mmon$, this formula can be simplified and one gets
\begin{equation}
\begin{aligned}
\dfrac{\Lambda^6}{\Mp^6} & = \dfrac{3 \pi^6 \Pstar}{2 f^2}
\dfrac{\mstar \left[\Estar^2 + (\mstar-1) \Kstar^2
    \right]^2}{(\mstar-1)^2 \Kstar^3 \Kpstar [2 \Estar +\Kstar
    (\mstar V_0 + \mstar - 2)]^2 \left[1+V_0(f)
    -2\dfrac{\Kstar - \Estar}{\mstar \Kstar}\right]}\,,
\end{aligned}
\label{eq:dilambdastar}
\end{equation}
in which all the elliptic integrals have a ``$*$'' because they are
evaluated at $m=\mstar$. Difficulties arise because $\mstar$ must
solve the reheating equation~\eqref{eq:dnstarlnrad}. Plugging
\Eq{eq:trajdi} into \Eq{eq:dnstarlnrad} gives a new ``parametric''
reheating equation to solve,
\begin{equation}
\begin{aligned}
\dfrac{16 \Lambda^2}{\Mp^2 \pi^2} \left[\calI(\mstar,f) -
\calI(\mend,f) \right] = \ln\Rrad - \Nzero - \dfrac{1}{4}
\ln\left[\dfrac{9}{2 \epsilon_1(\mstar)}
  \dfrac{v(\mend)}{v(\mstar)}\right]
+ \dfrac{1}{4} \ln\left(8 \pi^2 \Pstar\right),
\end{aligned}
\label{eq:direheat}
\end{equation}
where $\calI(m,f)$ is given in \Eq{eq:diprim}, $\epsilon_1(m)$ in
\Eq{eq:dieps1} and $\mend(\Lambda,f)$ is the solution of
\Eq{eq:dimend}. In this expression, although $v(\mstar)$ is given by
\Eq{eq:divstar}, $v(\mend)$ requires the full
expression of \Eq{eq:dipot}
\begin{equation}
v(\mend) = 1 + V_0(f) - 2 \dfrac{\Kend-\Eend}{\mend \Kend} -
\dfrac{\pi}{\mend \Kend \Kpend} \nuend^2 \Theta\left(\nuend\right).
\end{equation}
The parametric reheating equation~\eqref{eq:direheat} may be compared
to the ordinary one in \Eq{eq:phistarlnrrad}. We see that the price to
pay for having worked with the non-canonically normalized variable $m$
is an explicit dependence on $\Lambda/\Mp$. Let us emphasize that this
factor is not a mere rescaling of $\Mp^2$ but an explicit function of
$\mstar$, thereby rendering \Eq{eq:direheat} fundamentally different
than the usual reheating equation.

\begin{figure}
\begin{center}
\includegraphics[width=\wdblefig]{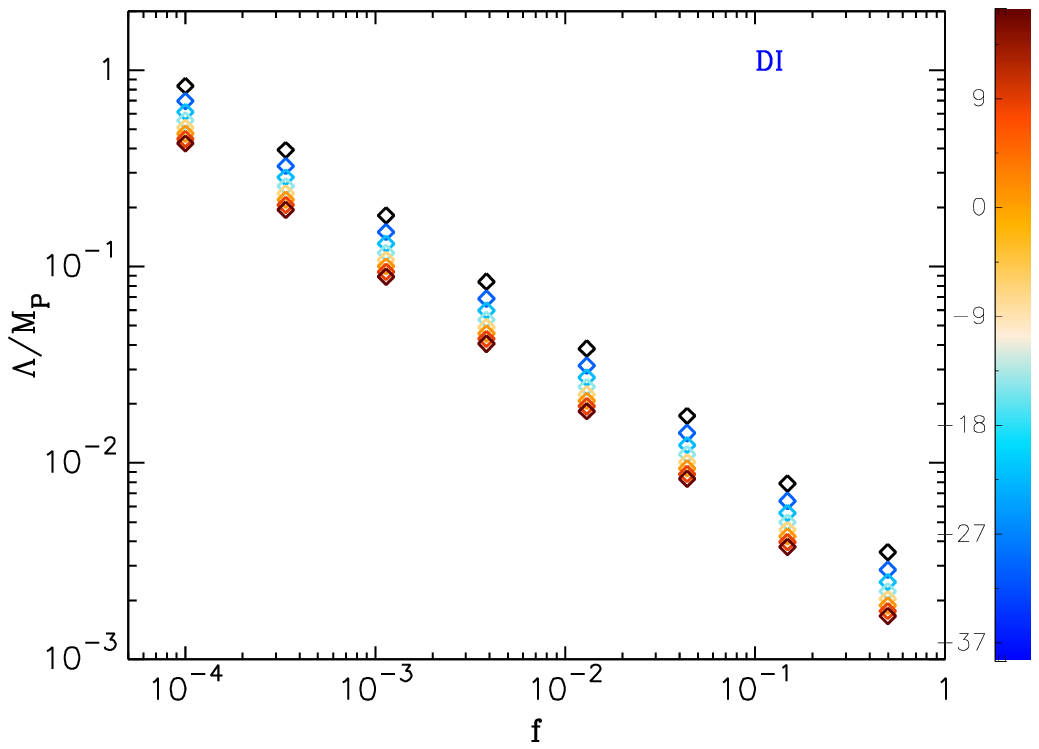}
\includegraphics[width=\wdblefig]{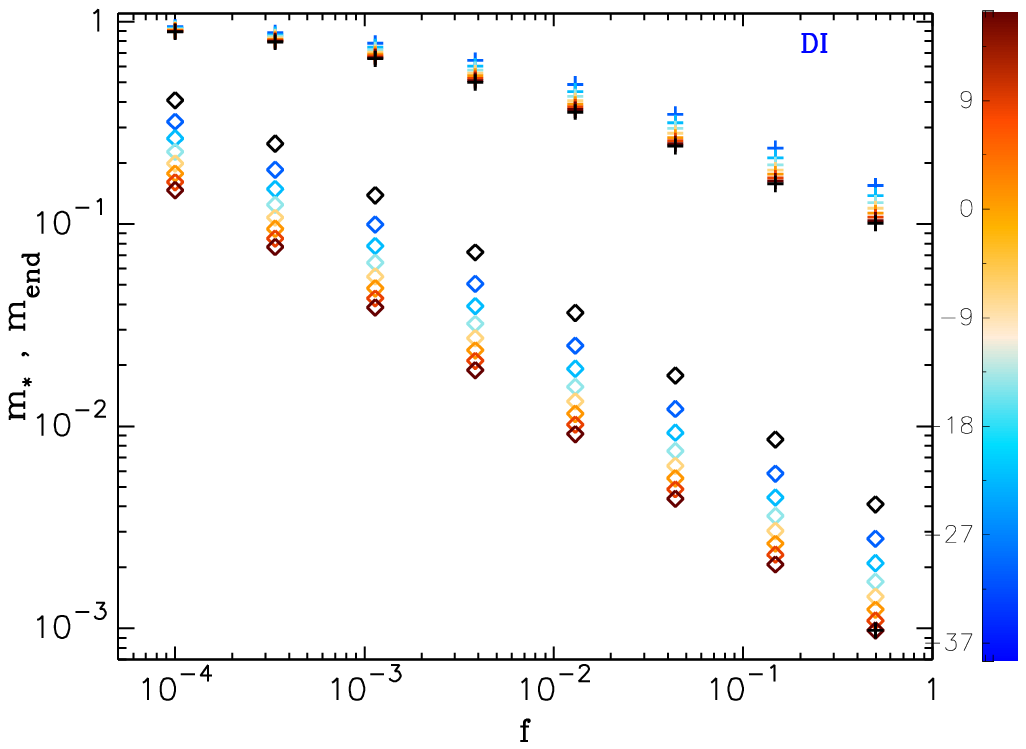}
\caption{Left panel: CMB normalized values of $\Lambda/\Mp$ with
  respect to the parameter $f$ for all possible reheating
  histories. The color bar encodes the values of $\ln \Rrad$. Right panel:
  CMB normalized values of $\mstar$ (diamonds) and $\mend$ (plus
  signs) as a function of $f$ and for various values of $\ln \Rrad$
  (color bar). For sub-Planckian values of $\Lambda$, the parameter $f$
  cannot take arbitrarily small values, \ie $f> \order{10^{-5}}$.}
\label{fig:diallstar}
\end{center}
\end{figure}

Plugging the expression of $\Lambda/\Mp$ given in \Eqs{eq:dilambdastar}
into \Eqs{eq:dimend} and \eqref{eq:direheat} yields a closed
algebraic system of two equations whose unknowns are $\mend$ and
$\mstar$. For an input value of $f$ and of the reheating parameter $\ln
\Rrad$, this system can be numerically solved to get the values of $\mend$
and $\mstar$. From $\mstar$, one can finally use \Eq{eq:dilambdastar}
to get the actual value of $\Lambda/\Mp$. In \Fig{fig:diallstar}, we
have represented $\mend$, $\mstar$ and $\Lambda/\Mp$ as a function of
$f$ and for all possible values of the reheating parameters $\ln
\Rrad$. The left panel of this figure shows that, in order to match
the amplitude of the CMB anisotropies, the dynamic scale $\Lambda/\Mp$
has to take larger values for smaller values of $f$. This is expected as
the overall potential normalization scales as $M^4 = f^2
\Lambda^4/\pi^2$. Requiring that $\Lambda/\Mp < 1$ imposes a lower bound on
$f$. As can be seen in this plot, the reheating modulates this bound
by, typically, a factor of five showing that $f > \order{10^{-5}}$.

Let us finally remark that, from $\mstar$, the observable values of
the slow-roll parameters, spectral index, tensor-to-scalar ratio and
running can be immediately read-off from the expression of the
$\epsilon_n(\mstar)$ given earlier. The reheating consistent slow-roll
predictions for dual inflation are represented in \Fig{fig:CMBDI}.

\subsection{Cubicly Corrected Starobinsky Inflation (CCSI)}
\label{sec:ccsi}

\subsubsection{Theoretical Justifications}

We have already encounter one class of corrections to the Starobinsky
model~\cite{Starobinsky:1980te} with RpI in \sectionc{sec:rpi}. Another
possibility which has been discussed in
\Refcs{Artymowski:2015mva,Artymowski:2015ida,Artymowski:2015pna,
  Chowdhury:2019otk} is to consider Starobinsky Inflation as the
leading correction of a Taylor-expanded $f(R)$ modified gravity
theory~\cite{Maeda:1988ab, Wands:1993uu, DeFelice:2010aj,
  DeFelice:2011jm}. As such, the next-to-leading order correction
would include a cubic dependency in $R$ such that
\begin{equation}
f(R) = R + \dfrac{R^2}{\mu^2} + \alpha \dfrac{R^3}{\mu^4}\,,
\end{equation}
where $\mu$ is the mass scale introduced in the Starobinsky model and
$\alpha$ a new dimensionless parameter encoding the strength of the
next-to-leading order corrections.

Following the same notation and methodology as for RpI, one can introduce the
scalar degree of freedom $\phi$ defined by
\begin{equation}
\frac{\phi}{\Mg}=\sqrt{\frac{3}{2}} \ln\left(\left| F(R) \right| \right),
\label{eq:ccsi:scalarfielddef}
\end{equation}
where $F(R) \equiv \partial f/\partial R$. This is also the square of
the conformal factor allowing to recast all the equations in the
Einstein frame and one finds the associated potential for the field
$\phi$ to be
\begin{equation}
  V\left(\phi\right)= \dfrac{\Mg^2}{2} \dfrac{|F|}{F} \dfrac{R F - f}{F^2}\,.
\label{eq:potccsiR}
\end{equation}
This is the same expression as for RpI, see \Eq{eq:potrpiR}, but the
function $F$ now reads
\begin{equation}
F = 1 + 2 \dfrac{R}{\mu^2} + 3 \alpha \dfrac{R^2}{\mu^4}\,.
\label{eq:Fccsi}
\end{equation}
Defining the quantity $y$ by
\begin{equation}
  y \equiv \sqrt{\dfrac{2}{3}} \dfrac{\phi}{\Mg}\,,
\end{equation}
one gets from \Eq{eq:ccsi:scalarfielddef} that $F = e^{y}$. Solving
\Eq{eq:Fccsi} for $R$ gives a quadratic equation with two
solutions, but only one allows us to recover the Starobinsky model in
the limit $\alpha \to 0$. It reads
\begin{equation}
\dfrac{R}{\mu^2} = \dfrac{\sqrt{1 + 3\alpha\left(e^y-1\right)}-1}{3\alpha}\,.
\label{eq:ccsi:phiofy}
\end{equation}
Plugging this expression into \Eq{eq:potccsiR} yields a closed
expression for the potential of CCSI in the Einstein frame
\begin{equation}
V(\phi) = \dfrac{\Mg^2 \mu^2}{2} \left(1 - e^{-y} \right)^2
\dfrac{1+\sqrt{1+3\alpha\left(e^y-1\right)} + 2\alpha \left(e^y-1
  \right)}{\left[1 + \sqrt{1 + 3\alpha\left(e^y - 1\right)}\right]^3}\,.
\end{equation}
The multiplicative constant terms can be absorbed into the potential
normalization and we define
\begin{equation}
  M^4 \equiv \dfrac{1}{2} \Mg^2 \mu^2.
\label{eq:ccsi:potnorm}
\end{equation}
The limiting case $\alpha \to 0$ gives
\begin{equation}
\lim_{\alpha \to 0}V(\phi) = \dfrac{M^4}{4} \left(1-e^{-y}\right)^2,
\end{equation}
which is, up to a different definition of $M^4$, the potential of the
Starobinsky model and also matches the one of Higgs inflation, see
\Eq{eq:potsi} and \eqref{eq:pothiggs}.

\subsubsection{Slow-Roll Analysis}

\begin{figure}
\begin{center}
\includegraphics[width=\wdblefig]{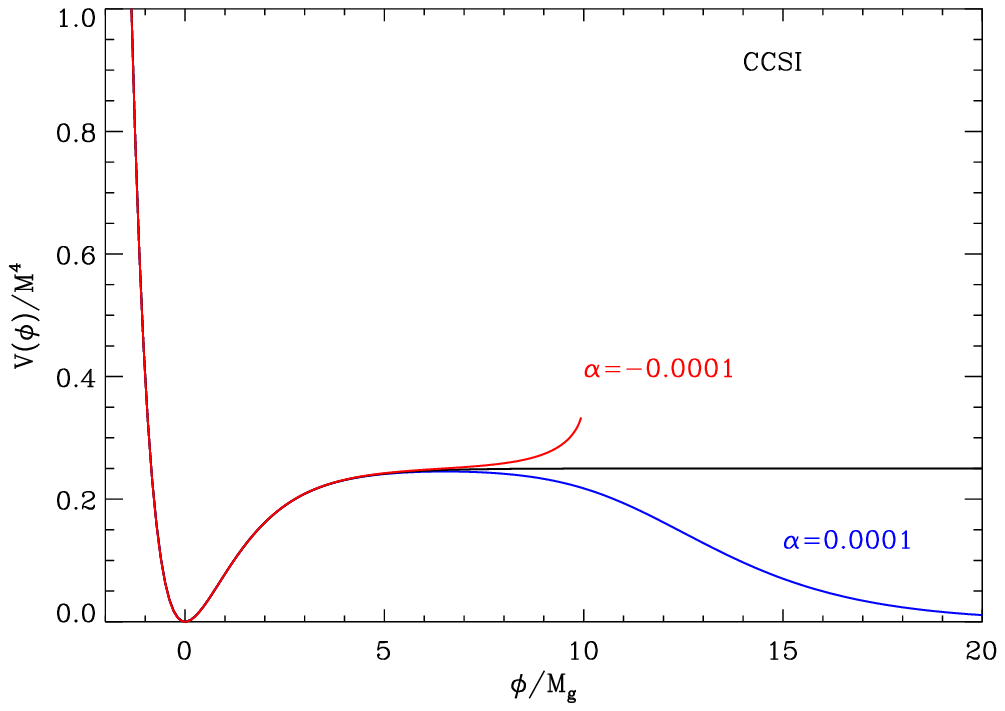}
\includegraphics[width=\wdblefig]{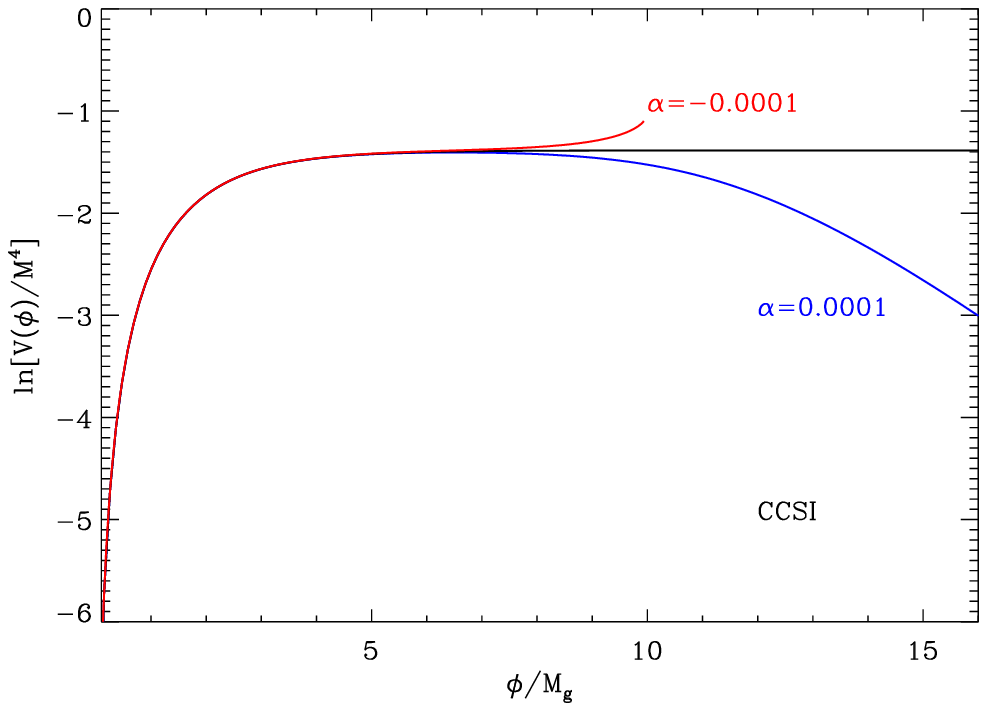}
\includegraphics[width=\wdblefig]{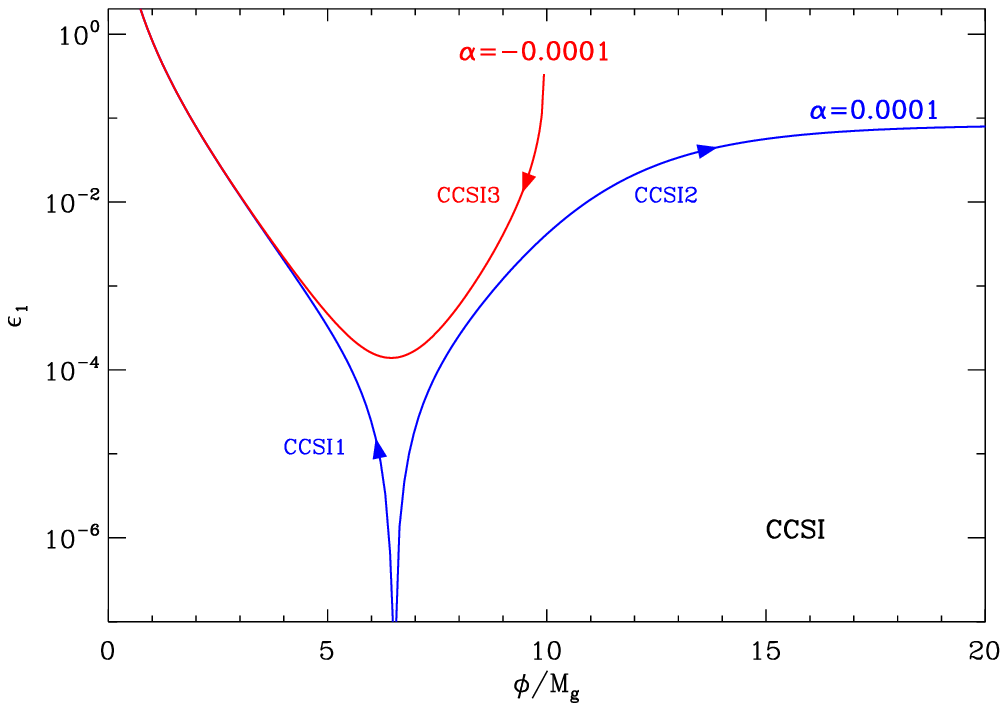}
\includegraphics[width=\wdblefig]{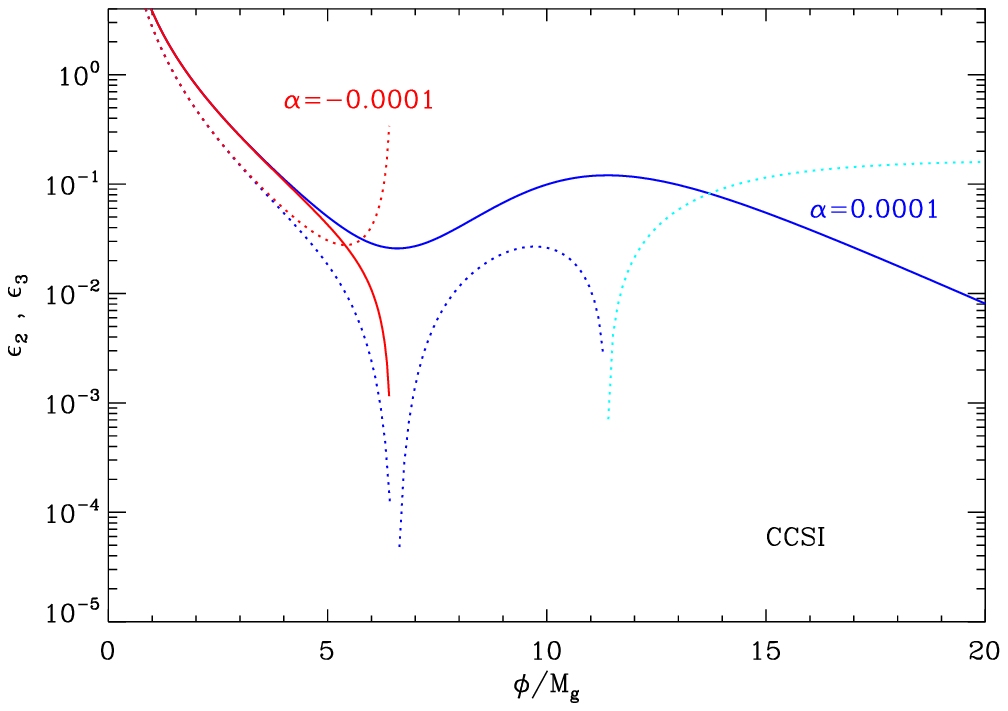}
\caption{Cubicly Corrected Starobinsky Inflation (CCSI) in the
  Einstein frame for $\alpha=\pm 10^{-4}$. Top panels: the potential
  and its logarithm.  Bottom left panel: slow-roll parameter
  $\epsilon_1$ with the different inflationary regimes of CCSI
  annotated with an arrow indicating the direction to which the field
  evolves. Notice that, for the CCSI2 regime, inflation never stops
  and one has to consider an extra-mechanism to end inflation. Bottom
  right panel: slow-roll parameters $\epsilon_2$ (solid line) and
  $\epsilon_3$ (dotted line). For CCSI2, $\epsilon_3$ becomes negative
  at large field values and this is represented as a cyan dotted line
  while the blue dotted line corresponds to positive values.}
\label{fig:potccsi}
\end{center}
\end{figure}

In terms of the canonically normalized field $\phi$, the potential of
CCSI depends on only one parameter $\alpha$ and reads
\begin{equation}
V(\phi) = M^4 \left(1 - e^{-\sqrt{\frac{2}{3}}\frac{\phi}{\Mg}}
\right)^2 \dfrac{1 + \sqrt{1 +
    3\alpha\left(e^{\sqrt{\frac{2}{3}}\frac{\phi}{\Mg}}-1\right)} +
  2\alpha \left(e^{\sqrt{\frac{2}{3}}\frac{\phi}{\Mg}}-1
  \right)}{\left[1 + \sqrt{1 +
      3\alpha\left(e^{\sqrt{\frac{2}{3}}\frac{\phi}{\Mg}} -
      1\right)}\right]^3}\,.
\label{eq:potccsi}
\end{equation}
Because the sign of $\alpha$ changes the shape and the domain of
definition of $V(\phi)$, this leads us to consider three different
regimes. For $\alpha > 0$, the potential is defined over all field
values and exhibits a maximum at
\begin{equation}
\phiVmax = \sqrt{\dfrac{3}{2}} \Mg \ln\left(\dfrac{2 +
  4\sqrt{\alpha}}{\sqrt{\alpha}} \right).
\label{eq:ccsimax}
\end{equation}
For $0<\phi < \phiVmax$, inflation can proceed at decreasing field
values and this regime will be referred to as CCSI1. For $\phi >
\phiVmax$, it proceeds at increasing field values and we call this
regime CCSI2. Finally, if $\alpha < 0$, the potential is defined only
in the domain $\phi <\phiUV$ where
\begin{equation}
\phiUV = \sqrt{\dfrac{3}{2}} \Mg \ln \left(1 - \dfrac{1}{3\alpha} \right).
\label{eq:ccsiuv}
\end{equation}
Inflation here proceeds at decreasing field values and will be
referred to as CCSI3. The potential and its logarithm have been
represented in the top panels of \Fig{fig:potccsi}.

In terms of the dimensionless field $y$, the Hubble flow functions in
the slow-roll approximation read
\begin{equation}
\begin{aligned}
  \epsilon_1 &= \dfrac{1}{3} \left\{\dfrac{2 - 8\alpha - e^y\left[1+
      2\alpha \left(e^y-5\right) - \sqrt{1+3\alpha\left(e^y-1\right)}
      \right]}{ \left(e^y-1\right) \left[1 + 4 \alpha
      \left(e^y-1\right) \right]} \right\}^2,\\
  \epsilon_2 & = \dfrac{2 e^y}{3 \left(e^y-1\right)^2 \left[1+4 \alpha
      \left(e^y-1\right)\right]^2\sqrt{3 \alpha \left(e^y-1\right)+1}}
  \\ & \times \bigg(12 \alpha ^2 \left(e^y-1\right)^2  \left[4 \sqrt{3
      \alpha \left(e^y-1\right)+1}+e^y+2\right] +2
  \left[\sqrt{3 \alpha \left(e^y-1\right)+1}+1\right] \\ & - \alpha
  \left(e^y-1\right) \left\{e^y \left[4 \sqrt{3 \alpha
      \left(e^y-1\right)+1}-5\right] -2 \left[10
    \sqrt{3 \alpha \left(e^y-1\right)+1}+7\right]\right\}\bigg),
  \end{aligned}
\end{equation}
and
{
\allowdisplaybreaks
\begin{align}
\epsilon_3 & = -\Bigg[3 \left(e^y-1\right)^2
   \left[3 \alpha 
   \left(e^y-1\right)+1\right] \left[4
   \alpha 
   \left(e^y-1\right)+1\right]^2
   \bigg(2 \left[\sqrt{3 \alpha 
     \left(e^y-1\right)+1}+1\right] \nonumber \\ & + 12 \alpha ^2
   \left(e^y-1\right)^2 \left[4
   \sqrt{3 \alpha 
   \left(e^y-1\right)+1}+e^y+2\right]-
   \alpha  \left(e^y-1\right)
   \left\{e^y \left[4 \sqrt{3 \alpha 
   \left(e^y-1\right)+1}-5\right]\right. \nonumber \\ & - \left. 2
   \left[10 \sqrt{3 \alpha 
   \left(e^y-1\right)+1}+7\right]\right\}\bigg) \Bigg]^{-1} \nonumber \\ & \times
\left\{8 \alpha +e^y \left[2 \alpha  \left(e^y-5\right)-\sqrt{3
    \alpha \left(e^y-1\right)+1}+1\right]-2\right\} \nonumber \\ & \times
\Bigg\{144 \alpha ^4 e^{6 y}+24
   \alpha ^3 e^{5 y} \left\{12 \alpha 
   \left[4 \sqrt{3 \alpha 
   \left(e^y-1\right)+1}+3\right]-4
   \sqrt{3 \alpha \left(e^y-1\right)+1}+3\right\} \nonumber \\
   & - \alpha ^2 e^{4 y} \left(24 \alpha 
   \left\{36 \alpha  \left[4 \sqrt{3\alpha 
   \left(e^y-1\right)+1}+5\right]-48
   \sqrt{3 \alpha \left(e^y-1\right)+1}-55\right\} \right.\nonumber  \\ &
   \left. + 8 \sqrt{3 \alpha \left(e^y-1\right)+1}+23\right)
   +3 (4 \alpha -1) \alpha  e^{2 y}
   \left(\alpha  \left\{12 \alpha 
   \left[16 \sqrt{3 \alpha 
   \left(e^y-1\right)+1}-15\right] \right.\right. \nonumber \\ & \left.\left. - 16
   \sqrt{3 \alpha 
   \left(e^y-1\right)+1}+165\right\}-4
   \left[3 \sqrt{3 \alpha 
     \left(e^y-1\right)+1}+8\right] \right) \nonumber \\
   &+ 2 \alpha  e^{3 y} \left[2 \alpha
    \left(6 \alpha  \left\{48 \alpha 
   \left[2 \sqrt{3 \alpha 
   \left(e^y-1\right)+1}+5\right]-72
   \sqrt{3 \alpha \left(e^y-1\right)+1}-155\right\} \right. \right.
  \nonumber  \\ & + \left. \left. 68 \sqrt{3 \alpha \left(e^y-1\right)+1}+155\right)
   +4 \sqrt{3 \alpha 
     \left(e^y-1\right)+1}-5\right] \nonumber \\
   & + 4 (1-4 \alpha )^2 (3 \alpha -1)
   \left\{\alpha  \left[6 \sqrt{3
   \alpha \left(e^y-1\right)+1}+3\right]-\sqrt{3 \alpha  \left(e^y-1\right)+1}-1\right\}
   \nonumber \\ &- 2 (4 \alpha -1) e^y \left[ \alpha  \left(3 \alpha  \left\{36 \alpha
    \left[4 \sqrt{3 \alpha 
   \left(e^y-1\right)+1}+1\right]-72
   \sqrt{3 \alpha 
   \left(e^y-1\right)+1}-11\right\} \right. \right. \nonumber \\ & \left. \left. + 20
   \sqrt{3 \alpha \left(e^y-1\right)+1}-7\right) + 2 \sqrt{3 \alpha
     \left(e^y-1\right)+1}+2\right]
   \Bigg\}.
\end{align}
}
They have been represented in the bottom panel of
\Fig{fig:potccsi}. Notice that $\epsilon_3(y)$ changes sign, which is
represented by the cyan blue dotted line in the figure (logarithmic
scale). One also remarks that in the regime CCSI2, obtained for
$\alpha > 0$ and $\phi > \phiVmax$, inflation does not end and one has
to introduce another mechanism to end the accelerated expansion, as
for instance a tachyonic instability triggered by another
field. Therefore, CCSI2 has an additional parameter, say $\yend$ (or
$\phiend$), the field value at which inflation ends. For the two other
models, CCSI1 and CCSI3, inflation naturally stops for $\yend$
solution of $\epsilon_1(\yend)=1$, namely
\begin{equation}
\yend = \ln \left(\dfrac{-15 - 14 \sqrt{3} + 176 \alpha  + 132 \sqrt{3}
  \alpha +\sqrt{813 + 420 \sqrt{3} + 4444 \alpha +2728\sqrt{3}
    \alpha}}{242 \alpha} \right).
\label{eq:ccsi:yend}
\end{equation}
This formula is also valid for CCSI3 ($\alpha < 0$), although the
equation $\epsilon_1=1$ admits a second root in that case,
$\yepsoneone$, which bounds the inflationary domain to $\yend < y<
\yepsoneone$ ($\yepsoneone < \yUV$) with
\begin{equation}
\yepsoneone = \ln \left(\dfrac{-15 - 14 \sqrt{3} + 176 \alpha  + 132 \sqrt{3}
  \alpha - \sqrt{813 + 420 \sqrt{3} + 4444 \alpha +2728\sqrt{3}
    \alpha}}{242 \alpha} \right).
\label{eq:ccsi:yepsoneone}
\end{equation}

\begin{figure}
\begin{center}
\includegraphics[width=\wsingfig]{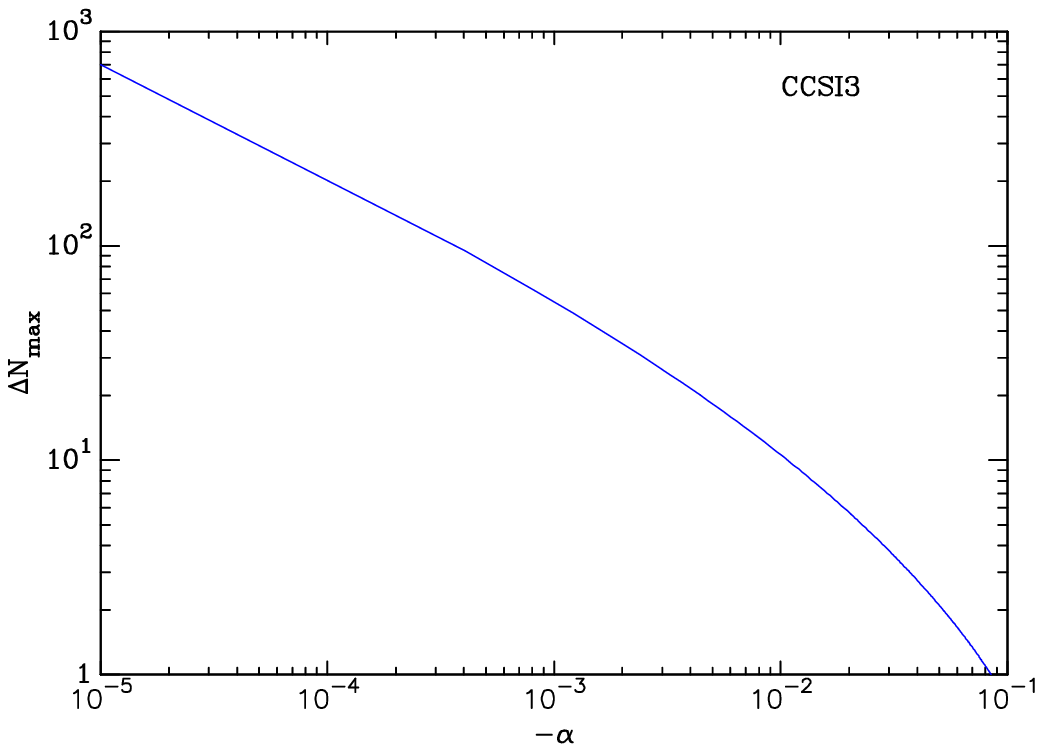}
\caption{Maximal number of {\efolds} of inflation within the CCSI3
  regime as a function of $\alpha<0$. In order to have at least $60$
  {\efolds} of inflation, $|\alpha|$ should be smaller than $|\alpha|
  < 8 \times 10^{-4}$.}
\label{fig:ccsi:ntot}
\end{center}
\end{figure}

Let us now turn to the slow-roll trajectory, it can be integrated
analytically and one gets
\begin{equation}
\begin{aligned}
\Nend - N  & = -\dfrac{3}{4}\left(y - \yend \right) - \dfrac{9}{8}
 \log\left[\dfrac{4-\alpha \left(e^y-4\right)^2}{4-\alpha
     \left(e^{\yend}-4\right)^2}\right] \\ & + \dfrac{3 + 9
   \sqrt{\alpha}}{4\sqrt{\alpha}} \left\{\arccoth\left[\frac{3 \sqrt{\alpha
     }+1}{\sqrt{3 \alpha \left(e^y-1\right)+1}}\right] - \arccoth\left[\frac{3 \sqrt{\alpha
     }+1}{\sqrt{3 \alpha \left(e^{\yend}-1\right)+1}}\right] \right\} \\ & +
 \dfrac{3}{4\sqrt{\alpha}} \left\{ \arctanh\left[\frac{1}{2} \sqrt{\alpha }
   \left(e^y-4\right)\right] - \arctanh\left[\frac{1}{2} \sqrt{\alpha }
   \left(e^{\yend}-4\right)\right]\right\} \\ & -
 \dfrac{3}{4\sqrt{\alpha}} \left\{
 \arccoth\left[\frac{1-3 \sqrt{\alpha }}{\sqrt{3 \alpha
       \left(e^y-1\right)+1}}\right] - \arccoth\left[\frac{1-3 \sqrt{\alpha }}{\sqrt{3 \alpha
       \left(e^{\yend}-1\right)+1}}\right] \right\} \\ & -
 \dfrac{9}{4} \left\{
 \arccoth\left[\frac{3 \sqrt{\alpha }-1}{\sqrt{3 \alpha
       \left(e^y-1\right)+1}}\right] - \arccoth\left[\frac{3 \sqrt{\alpha }-1}{\sqrt{3 \alpha
       \left(e^y-1\right)+1}}\right]\right\}.
\label{eq:ccsi:traj}
\end{aligned}
\end{equation}
This expression implicitly assumes complex numbers, as for instance in
the CCSI3 regime where $\alpha < 0$, but the result is a real
number. Moreover, because the inflationary domain for CCSI3 is bounded
within $\yend <y<\yepsoneone$, the \emph{total} number of e-folds of
inflation $\Delta\Nmax$ is also bounded and a function of $\alpha$
only. Its value can be obtained by the formal replacement $y \to
\yepsoneone$ in \Eq{eq:ccsi:traj} and using \Eqs{eq:ccsi:yend} and
\eqref{eq:ccsi:yepsoneone}. The expression being not particularly
illuminating, we have plotted $\Delta\Nmax$ as a function of $\alpha$
in \Fig{fig:ccsi:ntot}. As this plot shows, $|\alpha|$ should be small
for inflation to be long enough, typically $|\alpha| <
\order{10^{-3}}$ to get more than $60$ {\efolds} of accelerated
expansion.

The slow-roll trajectory giving $y$ as a function of $\Delta N =
\Nend-N$ cannot be analytically inverted from \Eq{eq:ccsi:traj}, but
as for RpI, $\ystar$, the dimensionless field value at which the pivot
mode crossed the Hubble radius, is uniquely determined from the
reheating model described in \sectionc{sec:hi}. The corresponding
number of {\efold} $\Delta \Nstar= \Nend - \Nstar$ is given by
\Eq{eq:ccsi:traj}.

Finally, the potential normalization $M$ is determined from the
amplitude of the CMB anistropies and satisfies
\begin{equation}
\begin{aligned}
\left(\dfrac{M}{\Mg}\right)^4 & = 480 \pi^2 \dfrac{\left\{2 - 8\alpha - e^{\ystar}\left[1+
      2\alpha \left(e^{\ystar}-5\right) - \sqrt{1+3\alpha\left(e^{\ystar}-1\right)}
      \right]\right\}^2}{e^{-2 {\ystar}}\left(e^{\ystar}-1\right)^4 \left[1 + 4
      \alpha\left(e^{\ystar} - 1\right)\right]^2} \\
  & \times  \dfrac{\left[1 + \sqrt{1+3 \alpha \left(e^{\ystar}-1\right)}
      \right]^3}{1 + \sqrt{1 + 3 \alpha \left(e^{\ystar}-1 \right)} + 2\alpha
    \left(e^{\ystar}-1 \right)} \dfrac{\Qrms^2}{T^2}\,.
\end{aligned}
\end{equation}

The reheating consistent slow-roll predictions for the CCSI model in
its different regimes are represented in \Fig{fig:CMBCCSI1} to
\Fig{fig:CMBCCSI3}. For CCSI1 and CCSI3, the limit $\alpha \to 0$
gives back the model predictions of Starobinsky Inflation (and Higgs
Inflation), but not for the CCSI2 regime for which the model
predictions strongly depend on the value of the new parameter
$\yend$. Such a situation is reminiscent with the RpI2 case in
\sectionc{sec:rpi} and, here as well, for CCSI2 one does not longer
have numerical equality between $\Mg$ and $\Mp$.

\subsection{Symmetry Breaking K\"ahler Inflation (SBKI)}
\label{sec:sbki}

\subsubsection{Theoretical Justifications}

This model was proposed in \Refc{Harigaya:2014qza}, in the context of
the supergravity constructions of Large Field Inflation (see
\sectionc{subsubsec:theorylfi}). In supergravity, the scalar potential
$V$ of a chiral multiplet $\phi^i$ is determined by the K\"ahler
potential $K(\phi^i, \phi^{*\bar{\imath}})$ and the superpotential
$W(\phi^i)$ according to
\begin{equation}
\label{eq:sbki:potSUGRA}
V=\ee^K\left[ K^{\bar{\imath} i}\left(W_i+K_i
  W\right)\left(W_{\bar{\imath}}^*+K_{\bar{\imath}} W^*\right)-3\left\vert
  W\right\vert^2\right],
\end{equation}
where $\phi^{*\bar{\imath}}$ is the conjugate multiplet and the $D$-term
contributions are omitted. As explained in \sectionc{sec:lfi},
Large-Field Inflation with $p=2$ (LFI2) can be achieved by
introducing two chiral multiplets $\Phi$ and $X$ and by considering
that the K\"ahler potential and the superpotential are respectively
given by
\begin{align}
K&=X X^*+\frac{1}{2}\left(\Phi+\Phi^*\right)^2 ,\\
W&=m\Phi X.
\end{align}
The K\"ahler potential enjoys a shift symmetry under the
transformation $\Phi \to \Phi + i c$, so if the inflaton $\phi$ is
identified with the imaginary part of $\Phi$, its potential does not
receive any exponential contribution [that would otherwise be present,
  due to the prefactor $\ee^K$ in \Eq{eq:sbki:potSUGRA}]. It ends up
being of the quadratic form $V(\phi)=m^2\phi^2/2$.

In \Refc{Harigaya:2014qza}, it is pointed out that the shift symmetry,
which is here broken by the superpotential, could also be broken at
the level of the K\"ahler potential by means of a non-holomorphic
spurious field $\mathcal{E}$. The K\"ahler potential is then expanded
around the origin $\mathcal{E}=0$ according to
\begin{equation}
K=X X^*+\frac{1}{2}\left(\Phi+\Phi^*\right)^2-\frac{ \mathcal{E}}{2}
\left(\Phi-\Phi^*\right)^2 +\frac{\mathcal{E}^2}{4!
}\left(\Phi-\Phi^*\right)^4+\cdots\, ,
\end{equation}
where the dots denote higher-order terms in the $\mathcal{E}$
expansion. These extra-terms modify the potential of LFI2 according to
\begin{equation}
V(\phi) =
\exp\left(\mathcal{E}\phi^2+\frac{\mathcal{E}^2}{6}\phi^4+\cdots\right)
\frac{m^2}{2}\phi^2, 
\end{equation}
and this expression will define the potential of SBKI.

\subsubsection{Slow-Roll Analysis}

\begin{figure}
\begin{center}
\includegraphics[width=\wdblefig]{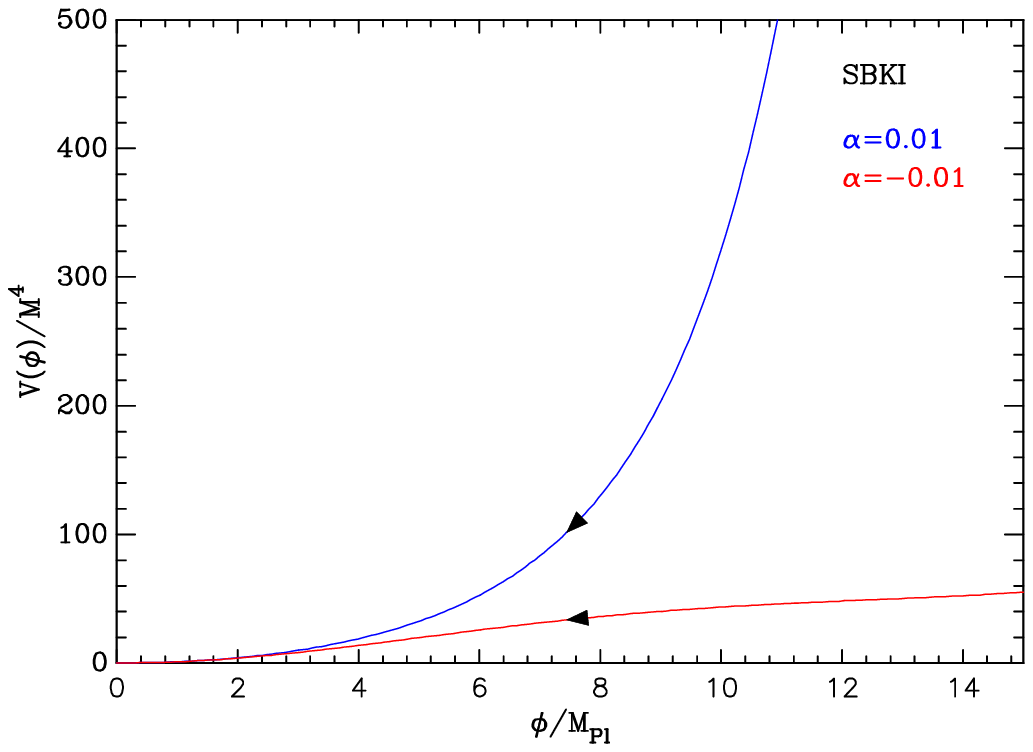}
\includegraphics[width=\wdblefig]{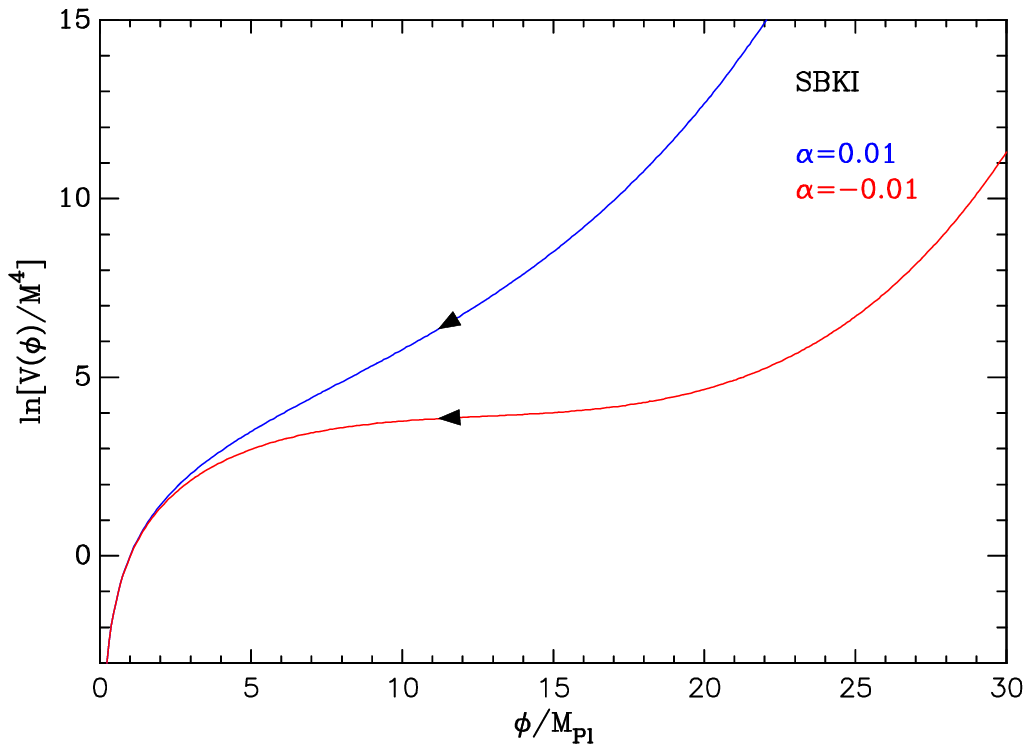}
\includegraphics[width=\wdblefig]{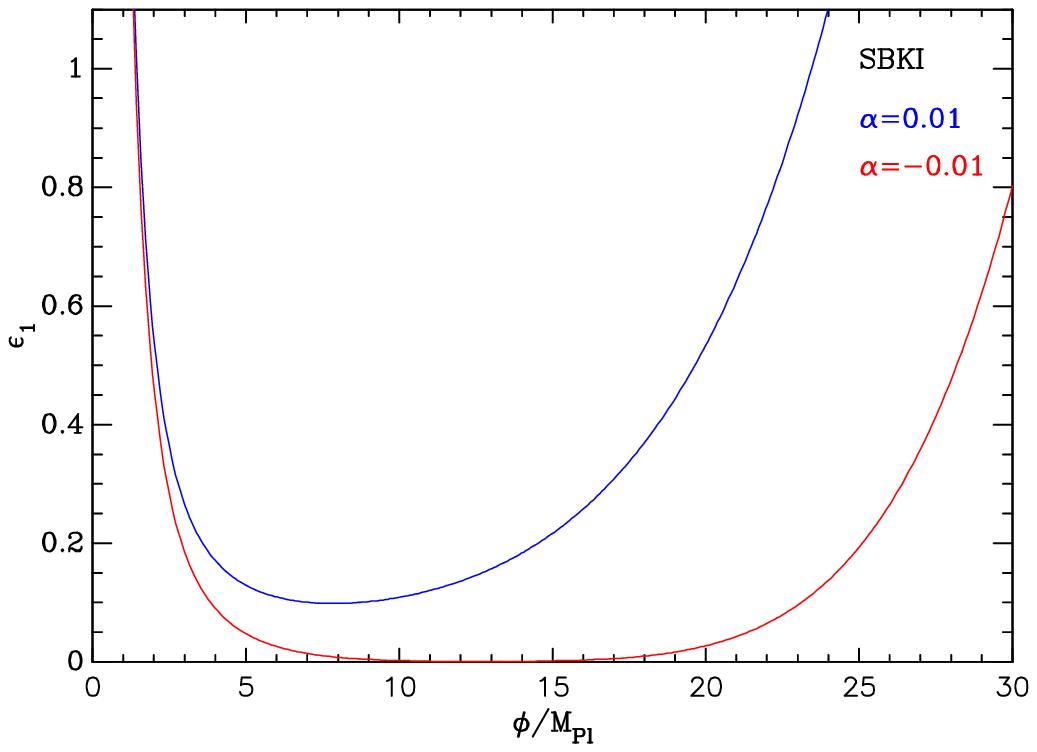}
\includegraphics[width=\wdblefig]{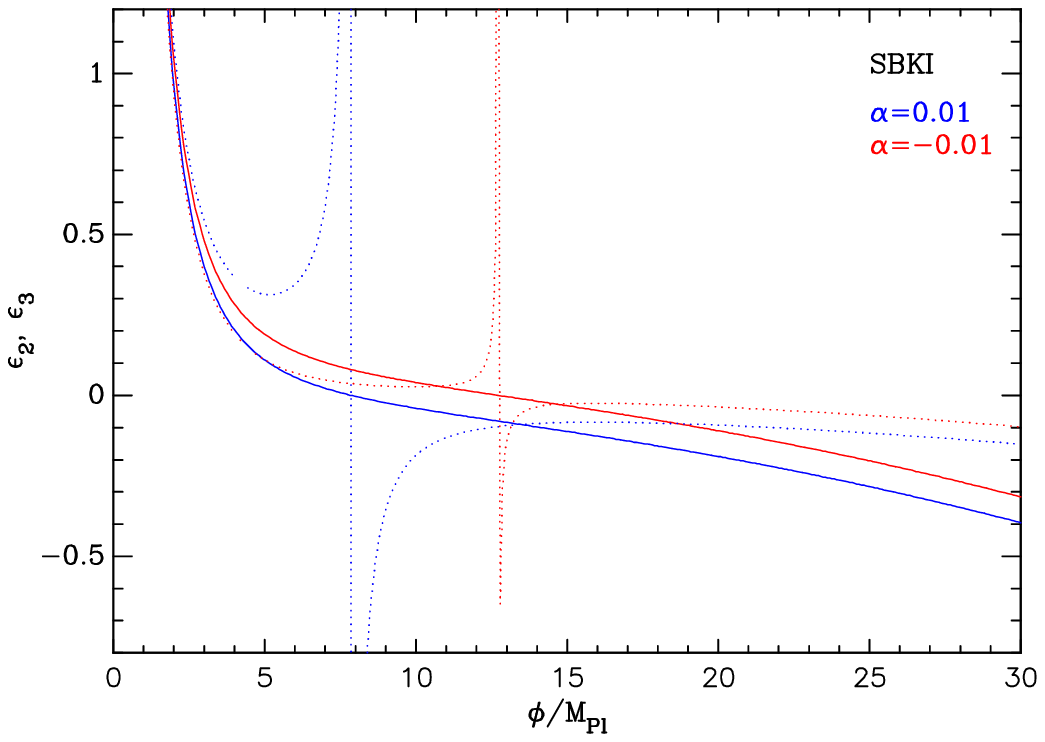}
\caption{Symmetry Breaking K\"ahler Inflation (SBKI) for
  $\alpha=0.01$ (blue curves) and $\alpha=-0.01$ (red curves). Upper
  panels: the potential and its logarithm as a function of
  $\phi/\Mp$. Bottom left panel: slow-roll parameter $\epsilon
  _1$. Bottom right panel: slow-roll parameters $\epsilon_2$ (solid
  lines) and $\epsilon_3$ (dotted lines).}
\label{potsbki}
\end{center}
\end{figure}
It is more convenient to rewrite the potential as
\begin{equation}
V=M^4\left(\frac{\phi}{\Mp}\right)^2
\exp\left[\alpha\left(\frac{\phi}{\Mp}\right)^2
+\frac{\alpha^2}{6}\left(\frac{\phi}{\Mp}\right)^4\right],
\end{equation}
where $M^4=m^2\Mp^2/2$ and $\alpha=\mathcal{E}/\Mp$, and where we have
restored the Planck masses. The parameter $\alpha$ can be positive or
negative. Since it is an even function of the field value, it is
enough to restrict the analysis to the region $\phi>0$. The derivative
of the potential is proportional to
$3+3\alpha\phi^2+\alpha^2\phi^4$. Seen as a polynomial in $\phi$, the
discriminant of that expression is equal to $-3\alpha^2$ and it is
always negative, so the polynomial [hence $V'(\phi)$] is always
positive. This shows that the potential is an increasing function of
the field value, regardless of the sign of $\alpha$, and that
inflation proceeds at decreasing field value. The potential and its
logarithm are displayed in \Fig{potsbki}

Let us define the dimensionless field value
\begin{equation}
x \equiv \dfrac{\phi}{\Mp}\,.
\end{equation}
The first and second Hubble-flow functions, in the slow-roll
approximation, are given by
\begin{equation}
\epsilon_1 = 2\left[\dfrac{1}{x}+\alpha x
+\dfrac{\alpha^2}{3} x^{3}\right]^2, \qquad
\epsilon_2 = \dfrac{4}{x^2} -4 \alpha -4 \alpha^2 x^2,
\label{eq:eps1skbi}
\end{equation}
while the third reads
\begin{equation}
  \epsilon_3 = -\frac{4 \left(1 + \alpha ^2 x^4\right) \left[\alpha 
   x^2 \left(\alpha  x^2+3\right)+3\right]}{3 x^2
   \left(\alpha ^2 x^4+\alpha  x^2-1\right)}\,.
\end{equation}
They are also displayed in the lower panels of \Fig{potsbki}. The
first Hubble-flow parameter diverges when $x \to 0$ and also for $x
\to \infty$. In between, it reaches a minimum at a field value where
the second Hubble-flow parameter vanishes and that we denote
$\xepstwoZero$. Its expression depends on the sign of $\alpha$,
and one finds
\begin{equation}
  \xepstwoZeroPlus = \sqrt{\dfrac{-1+\sqrt{5}}{2\alpha}}  \quad
  \text{for $\alpha>0$}, \qquad
  \xepstwoZeroMinus =  \sqrt{\dfrac{1+\sqrt{5}}{2\vert \alpha\vert }}
  \quad \text{for $\alpha<0$}.\\
\end{equation}
The corresponding values of $\epsilon_1$ can be evaluated at its minimum, and one obtains
\begin{equation}
\epsoneminPlus = \dfrac{4}{9}\left(5\sqrt{5} + 11\right) \alpha, \qquad
\epsoneminMinus = \dfrac{4}{9}\left(5\sqrt{5} - 11 \right) \left|\alpha\right|.
\end{equation}
For inflation to proceed, one must impose that $\epsonemin < 1$, which
requires that $\alpha$ lies in the range $[\alphamin,\alphamax]$ with
\begin{equation}
\label{eq:sbki:cond:alpha}
\alphamin \equiv -\frac{9}{16}\left(5\sqrt{5} + 11\right), \qquad \alphamax
\equiv \frac{9}{16}\left(5\sqrt{5} - 11\right) \, .
\end{equation}
In order to identify where inflation may proceed, and ends, one has to
find the roots of the polynomial equation $\epsilon_1=1$, \ie
\begin{align}
\frac{\alpha^2}{3} x^4
+\alpha x^2 -\frac{x}{\sqrt{2}} + 1=0.
\end{align}
This equation can be solved exactly but the explicit form of the
solution (which only depends on $\alpha$) is not especially
illuminating. Here, we just remark that the above equation has four
solutions, two of which being positive. We denote these two solutions
as $\xepsoneOnePlus(\alpha)$ for the largest one, and as
$\xepsoneOneMinus(\alpha)$ for the smallest positive. Having
$\xepsoneOnePlus > \xepsoneOneMinus$, this implies that inflation
starts for $x <\xepsoneOnePlus$ and ends at
$\xend=\xepsoneOneMinus$. In the small coupling limit, $\vert \alpha
\vert \ll 1$, one has the limiting forms
\begin{align}
\label{eq:sbki:phiend:appr}
\xend &=
\sqrt{2}+2\sqrt{2}\alpha+\frac{28}{3}\sqrt{2}\alpha^2+\order{\alpha^3},\\
\xepsoneOnePlus &= \frac{3^{1/3}}{2^{1/6} \left\vert \alpha
  \right\vert^{2/3}}-\mathrm{sign}(\alpha)\frac{2^{1/6}}{3^{1/3}\left\vert
  \alpha \right\vert^{1/3}} -\frac{\sqrt{2}}{3}+\order{\alpha
  ^{1/3}},
\end{align}
while the exact expressions can be found in the {\ASPIC} library.

The slow-roll trajectory can be integrated explicitly and reads
\begin{align}
\label{eq:sbki:Ntraj}
\Nend-N =\frac{\sqrt{3}}{2\alpha} \left[ \arctan\left(\frac{3+2\alpha
    x^2}{\sqrt{3}}\right) - \arctan\left(\frac{3+2\alpha
    \xend^2}{\sqrt{3}}\right) \right],
\end{align}
and it can be inverted to get the field value as a function of the
number of {\efolds} as
\begin{equation}
\label{eq:sbki:xtraj}
x =\sqrt{- \dfrac{3}{2\alpha} +\dfrac{\sqrt{3}}{2\alpha}\tan\left[
    \frac{2\alpha}{\sqrt{3}}\left(\Nend-N \right) +
    \arctan\left(\frac{3+2\alpha \xend^2}{\sqrt{3}}\right) \right] }\,.
\end{equation}

\begin{figure}
\begin{center}
\includegraphics[width=\wsingfig]{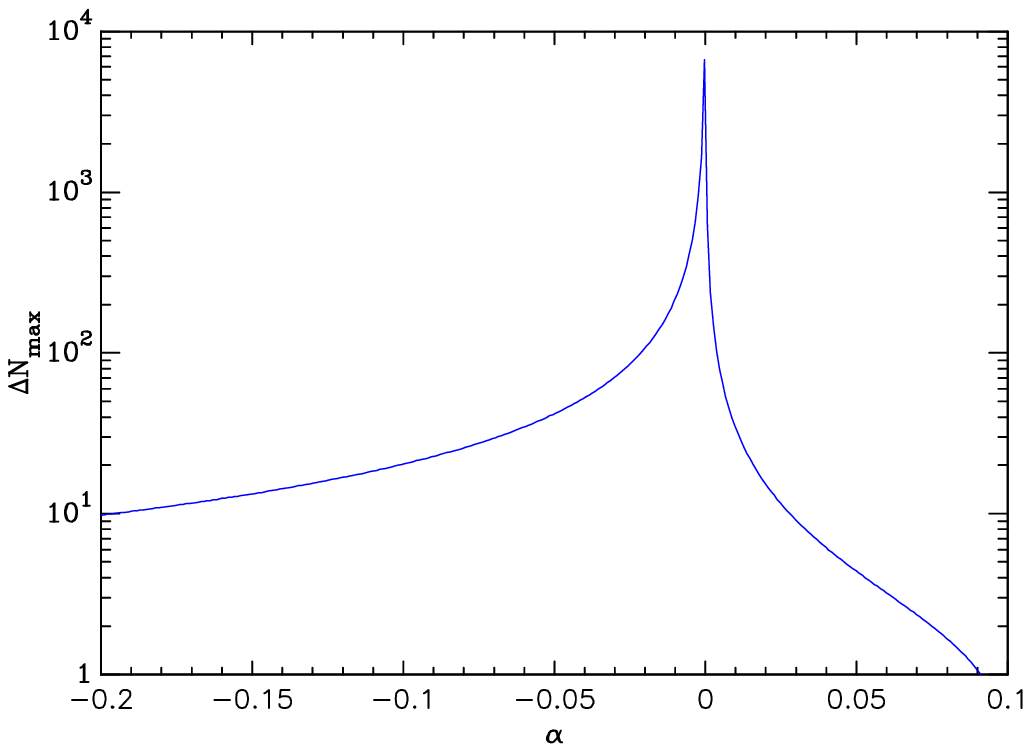}
\caption{Maximal total number of {\efolds} $\Delta\Nmax$ achievable
  within SBKI, from $\xepsoneOnePlus$ to $\xepsoneOneMinus$, as a
  function of the parameter $\alpha$. This one has to be small for
  inflation to last long enough.}
\label{fig:sbkiefoldmax}
\end{center}
\end{figure}

Since the first Hubble-flow parameter is below unity only over a
finite field range of field values, the total number of inflationary
{\efolds} that can be realized in SBKI, which we denote $\Delta\Nmax$,
is bounded. It can be estimated by evaluating \Eq{eq:sbki:Ntraj} with
$\xend=\xepsoneOneMinus$ and $\xini=\xepsoneOnePlus$. In practice, one
must check that this number is large enough to solve the FLRW
problems. Although it is possible to give an exact expression, for
clarity, we here only report the result expanded at first order in
$\alpha$. The condition $\Delta\Nmax > \Delta \Nmin$, where $\Delta \Nmin$ is the
minimum number of {\efolds} required, translates into
\begin{equation}
\label{eq:sbki:cond:alpha:2}
-\dfrac{5\pi \sqrt{3}}{12 \Delta \Nmin} \lesssim \alpha\lesssim \dfrac{\pi
  \sqrt{3}}{12 \Delta \Nmin}\, .
\end{equation}
With $\Delta \Nmin = 60$, this implies that $-0.038<\alpha<0.0075$,
which is a more stringent condition than \Eq{eq:sbki:cond:alpha} and
justifies, a posterior, the Taylor expansion in $\alpha$. The exact
expression has been plotted in \Fig{fig:sbkiefoldmax} as a function of
$\alpha$ (assumed to be in the interval $]\alphamin,\alphamax[$). As
expected from the above estimate, $\alpha$ should be small for
$\Delta\Nmax$ to be large.

Solving for the reheating equation~\eqref{eq:phistarlnrrad}, together
with the field value $\xend$ at which inflation ends,
\Eq{eq:sbki:xtraj} uniquely determines $\xstar$, the field value at
which the pivot mode crossed the Hubble radius during inflation. The
CMB normalization fixes the overall scale of the potential to
\begin{equation}
\left(\dfrac{M}{\Mp}\right)^4 = 2880 \pi^2 \dfrac{\left[
    \dfrac{1}{\xstar^2} + \alpha + \dfrac{\alpha^2}{3} \xstar^2
    \right]^2}{\exp{\left(\alpha \xstar^2 + \dfrac{\alpha^2}{6}
    \xstar^4 \right)}} \dfrac{\Qrms^2}{T^2}\,.
\end{equation}

The reheating-consistent slow-roll predictions of the model are shown
in \Fig{fig:CMBSBKI}. For $\alpha=0$, one recovers the same
predictions as LFI2, as already stressed. When $\alpha$ increases away
from $0$, the first Hubble-flow function, hence the tensor-to-scalar
ratio $r$, increases, which makes the model disfavored by the
data. However, when $\alpha$ decreases away from $0$, the
tensor-to-scalar ratio decreases, making the model in better agreement
with the data, before the spectral index becomes too blue.

The behavior displayed in these figures can also be analytically
recovered by expanding both \Eqs{eq:sbki:phiend:appr}
and~\eqref{eq:sbki:xtraj} in $\alpha$. One obtains the approximate
expression
\begin{equation}
\xstar =2\sqrt{\Delta \Nstar} + 2 \Delta\Nstar^{3/2}\alpha + \order{\alpha^2},
\end{equation}
where we have also used that $\Delta\Nstar\gg 1$. Expanding
\Eqs{eq:eps1skbi} in the same manner, one obtains the approximate
expressions of the Hubble flow functions
\begin{equation}
\epsilon_{1*}\simeq \frac{1}{2\Delta \Nstar}+3\alpha , \qquad 
\epsilon_{2*} \simeq \frac{1}{\Delta \Nstar}-6\alpha .
\end{equation} 
As expected, the leading order in $\alpha$ gives back the predictions
of LFI2, see \Eq{eq:lfi:epsstar}.

\subsection{Axion Hilltop Inflation (AHI)}
\label{sec:ahi}

\subsubsection{Theoretical justifications}

This model is a generalization of Natural Inflation (NI), see
\sectionc{sec:ni}, where two oscillatory functions in the potential
are considered~\cite{Czerny:2014wza}. A supergravity realization of
this idea was proposed in \Refc{Czerny:2014xja}. It introduces an
axion chiral superfield $\Phi$ having K\"ahler potential and superpotential
respectively given by
\begin{equation}
K = \dfrac{\lambda^2}{2}\left(\Phi+\Phi^*\right)^2, \qquad W = W_0 + A
e^{-a \Phi} + B e^{-b \Phi}.
\end{equation}
The scalar component of $\Phi$ is $\sigma + i \varphi$ where the axion
$\varphi$ will play the role of the inflaton. The choice of the
K\"ahler potential ensures a shift symmetry in the axion direction
($K$ is invariant under $\varphi \rightarrow\varphi + C$) and gives a
mass to the saxion $\sigma$. This one is then stabilized and will
remain a spectator field during inflation. Indeed, in Planck units,
provided $|A| \ll |W_0|$, $|B| \ll |W_0|$, the shift symmetry is only
weakly broken by the superpotential and one can show that the axion
mass remains much smaller than the saxion mass. Therefore, taking the
saxion at vanishing expectation value, the potential for the
canonically normalized axion $\phi = \sqrt{2} \lambda \varphi$ reads,
up to an overall constant~\cite{Czerny:2014xja},
\begin{equation}
\begin{aligned}
  V(\phi) & \simeq 6 |A| |W_0|
  \left[1-\cos\left(\dfrac{\phi}{\lambda_1}\right) \right] +
  6 |B| |W_0| \left[1 -
    \cos\left(\dfrac{\phi}{\lambda_2}+\theta\right) \right] \\ & - 2 |A|
  |B|\left(\dfrac{2}{\lambda_1 \lambda_2} -3 \right)\left[1 -
    \cos\left(\dfrac{\lambda_2 -\lambda_1}{\lambda_1 \lambda_2}\phi -
    \theta\right) \right],
\end{aligned}
\label{eq:fullpotahi}
\end{equation}
where $B=|B|e^{i\theta}$, $\lambda_1=\sqrt{2}\lambda/a$ and
$\lambda_2=\sqrt{2}\lambda/b$. Having a saxion more massive than the
Hubble scale during inflation and forging a plateau-like shape for
this potential requires some adjustment between the parameters. As
discussed in \Refc{Czerny:2014xja}, this amounts to fixing the axion
value at which the potential is maximal at $\phimax=\pi\lambda_1$,
having $A/\lambda_1^2 = B/\lambda_2^2$ and fixing $\theta=-\pi
\lambda_1/\lambda_2$. A first possible choice is to set $\lambda_1 = 2
\lambda_2$, the leading terms in \Eq{eq:fullpotahi} reduce to $V
\simeq V_0 + C \phi^4$, which is a small-field inflation model with a
power law index $p=4$ (SFI4, see \sectionc{sec:sfi}).  The other
choice considered in \Refc{Czerny:2014xja} is to consider the limit
$\lambda_1 \to \lambda_2$, for which the leading terms of the
potential simplify to
\begin{equation}
V = 6 |A| |W_0| \dfrac{\lambda_1-\lambda_2}{\lambda_2} \left[\nu_0 -
  2 \cos\left(\dfrac{\phi}{\lambda_1} \right) + \left(\pi -
  \dfrac{\phi}{\lambda_1} \right)
  \sin\left(\dfrac{\phi}{\lambda_1}\right) \right] + \cdots,
\end{equation}
where $\nu_0 \simeq 4.82$ is an constant ensuring that the potential
vanishes at its minimum. The remaining terms are, at most, of order
$(\lambda_1-\lambda_2)^2/\lambda_1^2$ and will be neglected in the
following.

\subsubsection{Slow-roll Analysis}

\begin{figure}
\begin{center}
\includegraphics[width=\wdblefig]{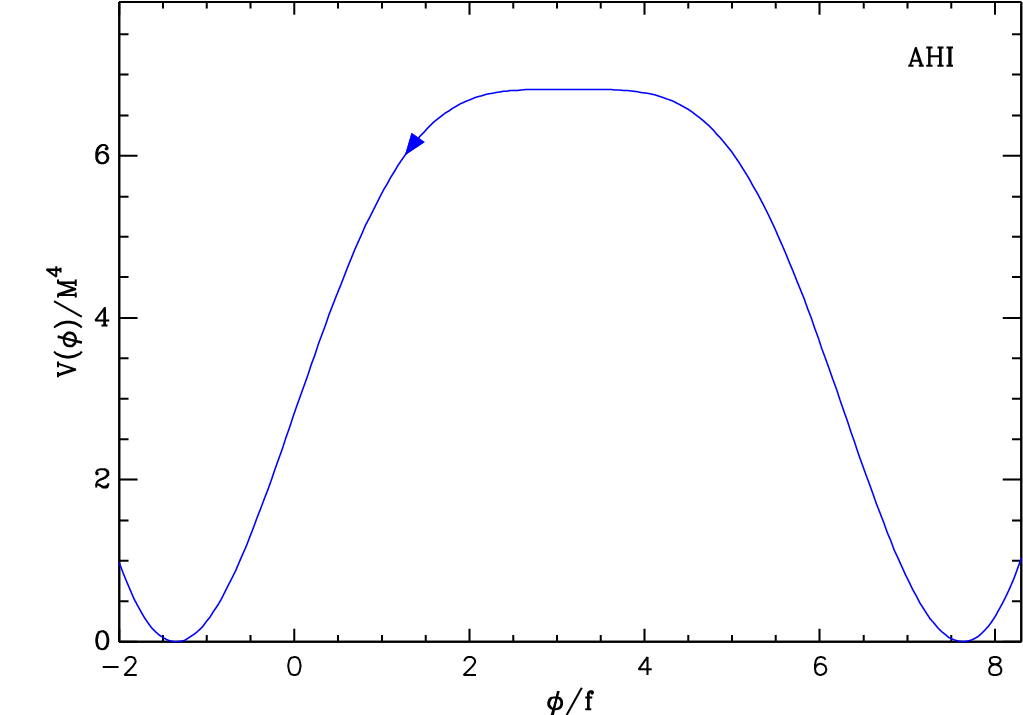}
\includegraphics[width=\wdblefig]{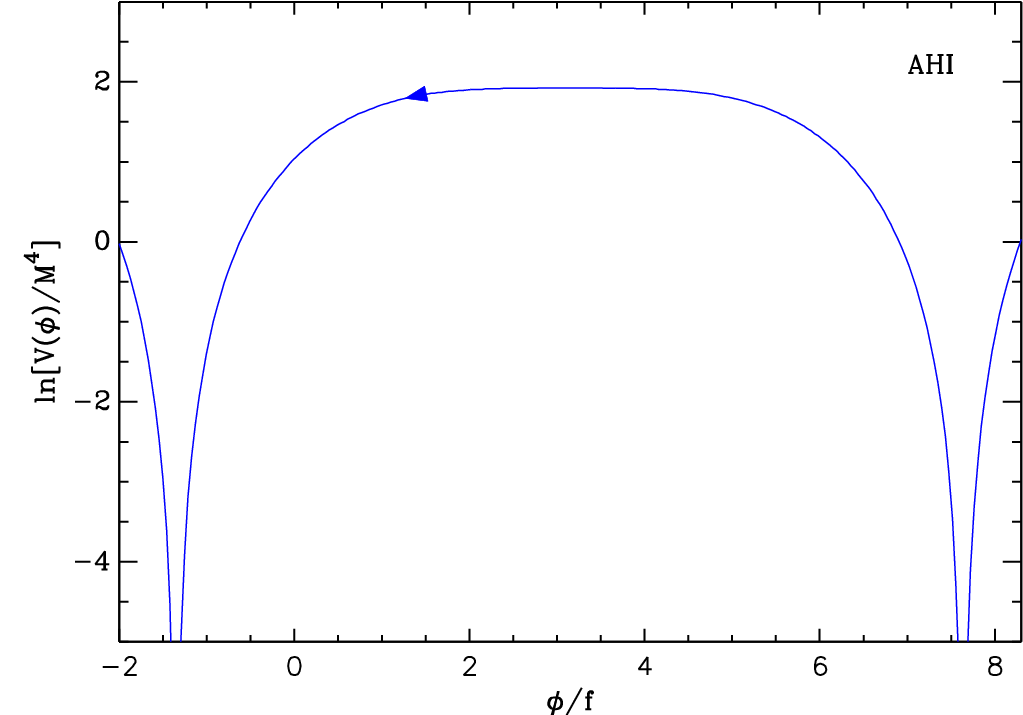}
\includegraphics[width=\wdblefig]{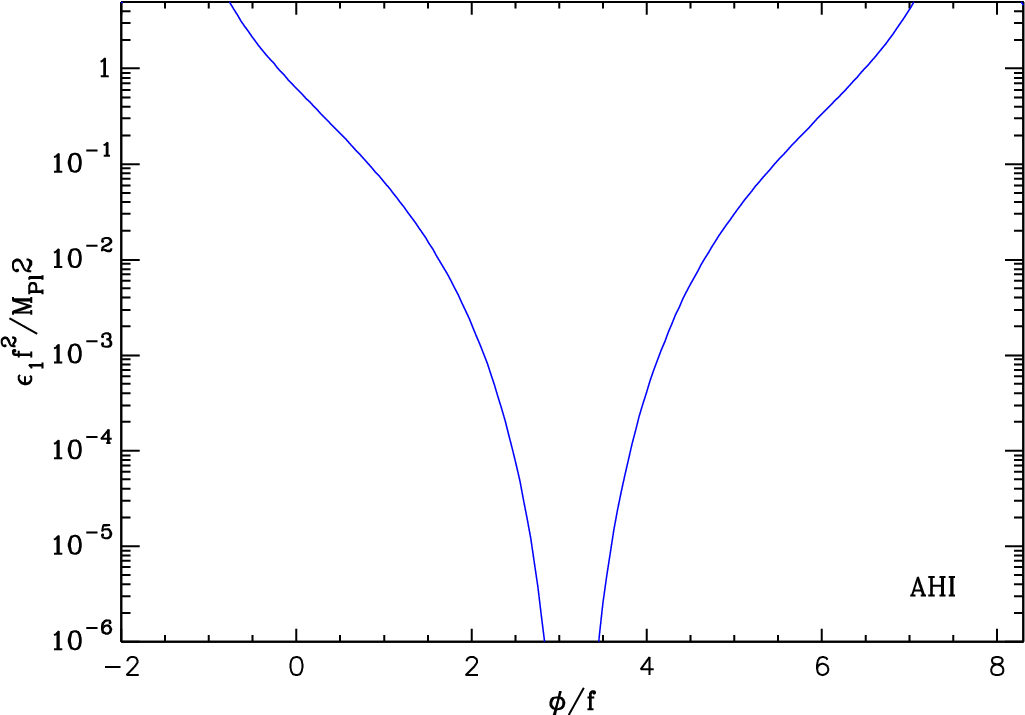}
\includegraphics[width=\wdblefig]{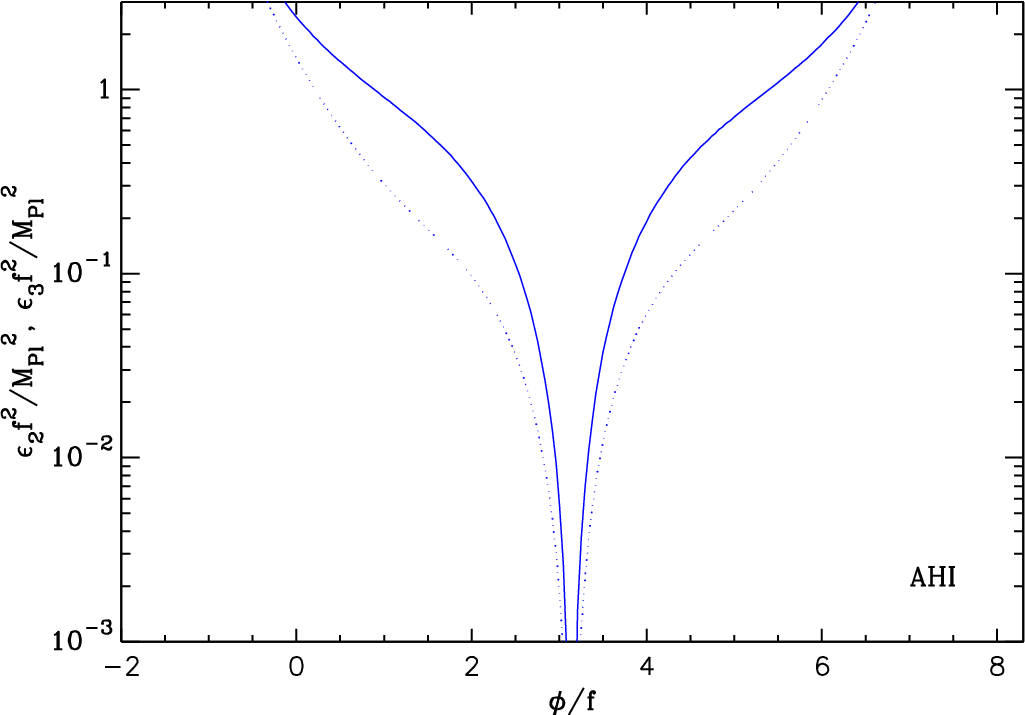}
\caption{Axion Hilltop Inflation. Top panels: the potential and its
  logarithm as a function of $\phi/f$. Bottom left panel: slow-roll
  parameter $\epsilon_1 f^2/\Mp^2$. There are two inflationary domains
  corresponding to the field running away from the top of the
  potential. Because they are symmetrical, we only study the regime
  annotated with an arrow in which inflation proceeds at decreasing
  field value. Bottom right panel: slow-roll parameters $\epsilon_2
  f^2/\Mp^2$ (solid line) and $\epsilon_3 f^2/\Mp^2$ (dotted line).}
\label{fig:potahi}
\end{center}
\end{figure}

\begin{figure}
\begin{center}
\includegraphics[width=\wsingfig]{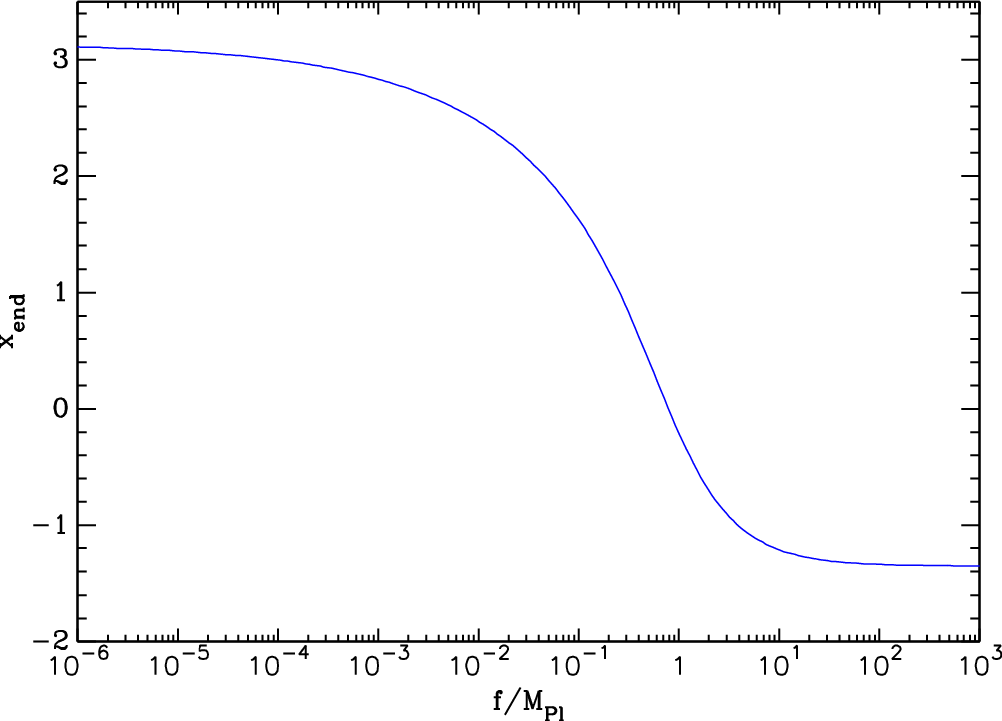}
\caption{The dimensionless field value $\xend=\phiend/f$ at which
  inflation ends as a function of $f/\Mp$. For $f \ll \Mp$, inflation
  is confined around the top of the potential ($\xVmax=\pi$).}
\label{fig:xendahi}
\end{center}
\end{figure}

From the previous discussion, we can rewrite the potential of Axion
Hilltop Inflation as
\begin{equation}
V(\phi) = M^4 \left[\nu_0 - 2 \cos\left(\dfrac{\phi}{f} \right) +
\left(\pi - \dfrac{\phi}{f} \right) \sin\left(\dfrac{\phi}{f}\right) \right],
\label{eq:potadhi}
\end{equation}
where we have introduced the parameter $f \equiv \lambda_1$ and $M^4
\equiv 6 |A| |W_0| (\lambda_1 - \lambda_2)/\lambda_2$. This is a
periodic function and, in the following, the analysis is restricted to
the region nearby the origin where the potential exhibits a
plateau-like maximum.

As already mentioned, the constant $\nu_0$ is chosen such that the
potential vanishes at its minimum. Therefore, the zeros of the
potential are also solutions of $V'(x)=0$, i.e.,
\begin{equation}
\tan(x)  = x - \pi.
\end{equation}
where we have defined
\begin{equation}
  x \equiv \dfrac{\phi}{f}\,.
\end{equation}
There is one obvious solution to this equation at $x=\pi$, but the others
have to be determined numerically. The geometrical interpretation is
however clear. The solutions are the intersections between the line
$y=x-\pi$ and the function $y=\tan(x)$ in the Euclidean plane
$(x,y)$. Among the three solutions closest to the origin, two of them
corresponds to the minima of the potential, $\xVzeroPM$, the other,
$\xVmax$, is the field value at which the potential is maximal. Their
numerical values read
\begin{equation}
\xVzeroMinus \simeq -1.35, \qquad \xVzeroPlus \simeq 7.63, \qquad
\xVmax = \pi.
\end{equation}
Let us notice that the actual values of $\xVzeroPM$ uniquely
determine the value of $\nu_0$. Enforcing $V(\xVzeroPM)=0$, one gets
\begin{equation}
\nu_0 = 2 \cos\left(\xVzeroPM\right) + (\xVzeroPM-\pi) \sin\left(\xVzeroPM\right),
\end{equation}
which is numerically $\nu_0 \simeq 4.82$, matching the value given above. Because $\xVzeroMinus
<\xVmax < \xVzeroPlus$, there are two inflationary domains. Either
$\xVmax< x < \xVzeroPlus$ and inflation proceeds at increasing field
value, or $\xVzeroMinus <x<\xVmax$ and inflation proceeds at
decreasing field values. However the potential is symmetric with
respect to $\xVmax$ and both of these regimes lead to identical
observable predictions. For this reason, in the following, we focus
the analysis on the first domain.

The first Hubble-flow function, in the slow-roll approximation, reads
\begin{equation}
\epsilon_1 = \dfrac{\Mp^2}{2 f^2} \left[\dfrac{\sin(x) +
    (\pi-x)\cos(x)}{\nu_0 - 2\cos(x) + (\pi-x)\sin(x)} \right]^2,
\label{eq:ahisr1}
\end{equation}
while the second Hubble-flow function is given by
\begin{equation}
\epsilon_2 = \dfrac{\Mp^2}{f^2} \dfrac{1+2(\pi-x)^2 - \cos(2 x) +
  2\nu_0(\pi-x)\sin(x)}{\left[\nu_0 - 2\cos(x)+ (\pi-x) \sin(x)
    \right]^2}\,,
\label{eq:ahisr2}
\end{equation}
and the third one by
\begin{equation}
\begin{aligned}
  \epsilon_3 & = \dfrac{\Mp^2}{f^2} \dfrac{(\pi-x)\cos(x)
    +\sin(x)}{\left[\nu_0 - 2\cos(x)+(\pi-x)\sin(x)\right]^2\left[1 +
      2(\pi-x)^2-\cos(2x)+2\nu_0(\pi-x)\sin(x)\right]} \\ & \times
  \Big\{ 9 \nu_0(\pi-x) - \left[8+2\nu_0^2-4(\pi-x)^2
      \right](\pi-x)\cos(x) - \nu_0(\pi-x)\cos(2x) \\ & +
    \left[5+2\nu_0^2+8(\pi-x)^2 \right]\sin(x) - \nu_0\left[4 -
      (\pi-x)^2\right]\sin(2x) + \sin(3x) \Big\}.
\end{aligned}
\label{eq:ahisr3}
\end{equation}
They have been plotted, along with the potential and its logarithm, in
\Fig{fig:potahi}. We have also represented the inflationary regime for
$x<\xVmax$ with an arrow.

Because the potential vanishes at its minimum, this guarantees that
inflation gracefully ends at a field value for which $\epsilon(x)=1$ in the domain
$]\xVzeroMinus,\xVmax[$, i.e., for which
\begin{equation}
\left[ \sin(x) + (\pi-x)\cos(x) \right]^2 - \dfrac{2f^2}{\Mp^2}
\left[\nu_0 -2 \cos(x) +(\pi-x)\sin(x)\right]^2 = 0.
\label{eq:ahixendsolve}
\end{equation}
This equation has to be solved numerically and we denote the solution $\xend$. Its
dependence with respect to $f/\Mp$ is been represented in \Fig{fig:xendahi}.

The slow-roll trajectory cannot be integrated analytically and one has
to numerically solve the following equation
\begin{equation}
\Nend - N = \dfrac{f^2}{\Mp^2}\int_{\xend}^x \dfrac{\nu_0 - 2\cos(y)+
  (\pi-y)\sin(y)}{ \sin(y) + (\pi-y) \cos(y) }\,\ud y.
\label{eq:ahitraj}
\end{equation}
The denominator of this equation vanishes at the top of the potential,
for $x\to \xVmax=\pi$. This ensures that a sufficient
number of {\efolds} can always be made in that region.
As can be seen in \Fig{fig:xendahi}, for $f\ll\Mp$
on has $\xend \to \xVmax$ such that the inflationary domain is
actually confined around the top of the potential. For
$x\to\xVmax$, one can derive an approximate expression for the
trajectory by expanding both the numerator and denominator of \Eq{eq:ahitraj}. One gets
\begin{equation}
\Nend - N \simeq \dfrac{6+3 \nu_0}{2} \left[\dfrac{1}{(x-\pi)^2} -
  \dfrac{1}{(\xend-\pi)^2} \right],
\end{equation}
which can be inverted into
\begin{equation}
x \simeq \pi - \dfrac{1}{\sqrt{\dfrac{1}{(\xend-\pi)^2} + \dfrac{2}{6+3
        \nu_0} \dfrac{\Mp^2}{f^2}\left(\Nend - N \right)}}\,.
\label{eq:ahixtrajapprox}
\end{equation}
The trajectory of \Eq{eq:ahitraj} combined with the reheating
equation~\eqref{eq:phistarlnrrad}, numerically determine $\xstar$, the
field value at which the pivot mode crossed the Hubble radius during
inflation. The mass scale $M$ of the potential is then determined from
the CMB normalization and one finds
\begin{equation}
\left(\dfrac{M}{\Mp}\right)^4 = 720 \pi^2 \dfrac{\Mp^2}{f^2}
\dfrac{\left[\sin(\xstar) + (\pi-\xstar)\cos(\xstar)\right]^2}{\left[\nu_0 -
    2\cos(\xstar) + (\pi-\xstar) \sin(\xstar) \right]^3} \dfrac{\Qrms^2}{T^2}\,.
\end{equation}
The reheating consistent observable predictions have been represented
in \Fig{fig:CMBAHI_0}. For $f/\Mp \ll 1$, $\epsilon_1$ becomes very
small while $\epsilon_2$ reaches a constant value. It is possible to
understand this limit by using the approximate trajectory of
\Eq{eq:ahixtrajapprox}. This equation needs as an input $\xend$. In
the limit $f\ll\Mp$, we have shown that $\xend \to \xVmax=\pi$ such
that an approximate solution can be found by expanding
\Eq{eq:ahixendsolve} around $\pi$. One finds
\begin{equation}
\xend \simeq \pi - \left[\sqrt{2} (3 \nu_0+6)\right]^{1/3}
\left(\dfrac{f}{\Mp} \right)^{1/3},
\end{equation}
and $(\xend-\pi)^{-2} \propto (\Mp/f)^{2/3}$ remains negligible in
\Eq{eq:ahixtrajapprox} for small-enough $f$. Therefore, one has
\begin{equation}
  \xstar \simeq \pi - \sqrt{ \dfrac{6+3\nu_0}{2 \Delta \Nstar} } \dfrac{f}{\Mp}\,,
\end{equation}
which gives, from \Eqs{eq:ahisr1} and \eqref{eq:ahisr2},
\begin{equation}
\epsilon_1 \simeq \dfrac{6+3\nu_0}{16} \dfrac{f^4}{\Mp^4}
\dfrac{1}{\Delta\Nstar^3}\,, \qquad \epsilon_2 \simeq
\dfrac{3}{\Delta\Nstar} - \dfrac{6+3 \nu_0}{4}  \dfrac{f^2}{\Mp^2}
\dfrac{1}{\Delta \Nstar^4}\,.
\end{equation}
At fixed reheating history, the model therefore asymptotes to a
constant spectral index with vanishing tensor-to-scalar ratio.

\subsection{Pure Arctan Inflation (PAI)}
\label{sec:pai}

\begin{figure}
\begin{center}
\includegraphics[width=\wdblefig]{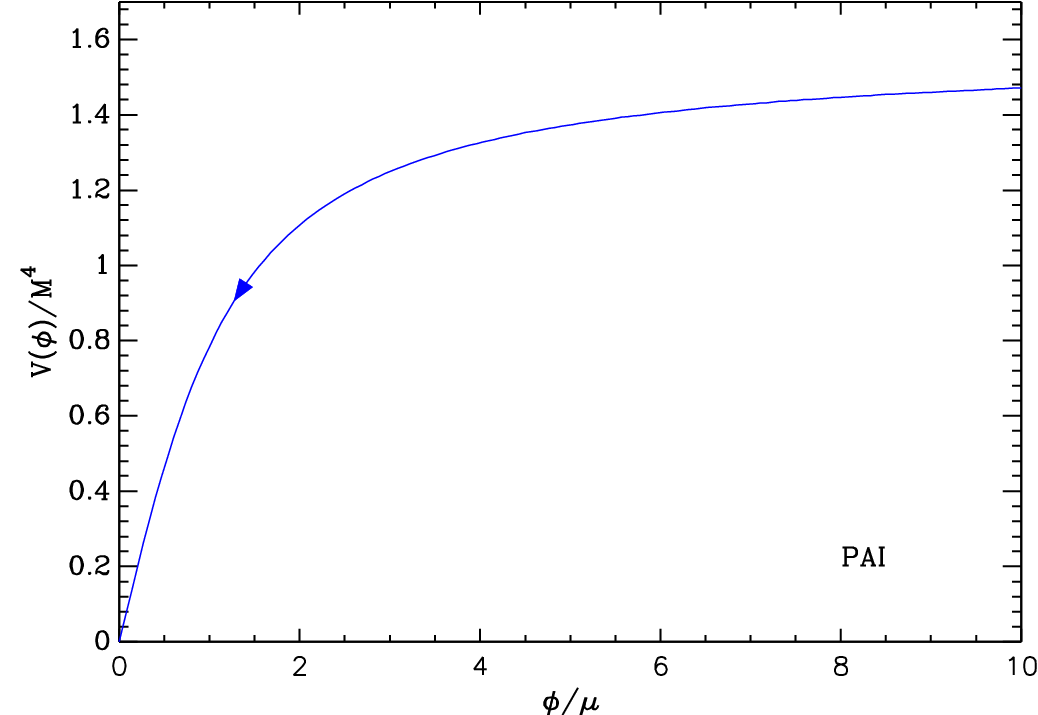}
\includegraphics[width=\wdblefig]{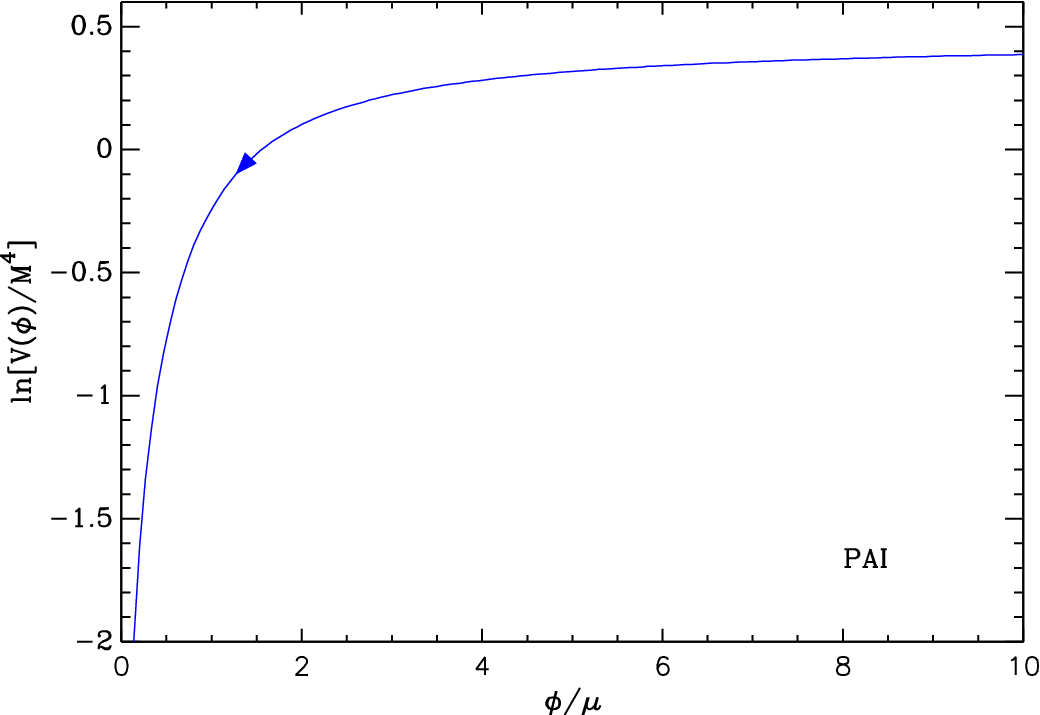}
\includegraphics[width=\wdblefig]{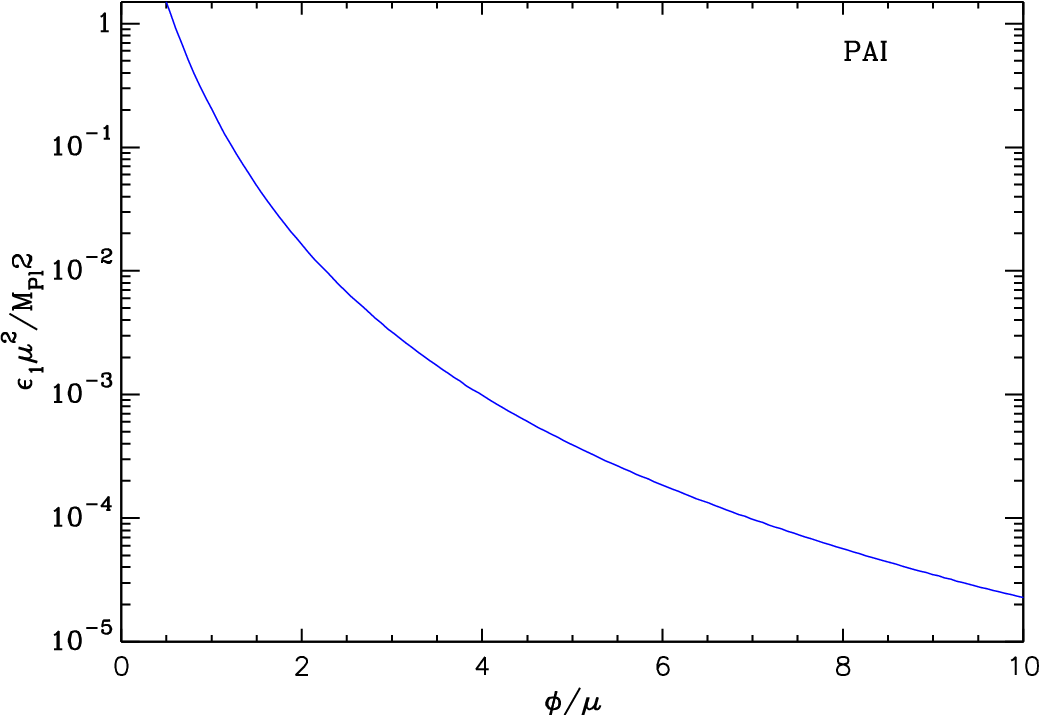}
\includegraphics[width=\wdblefig]{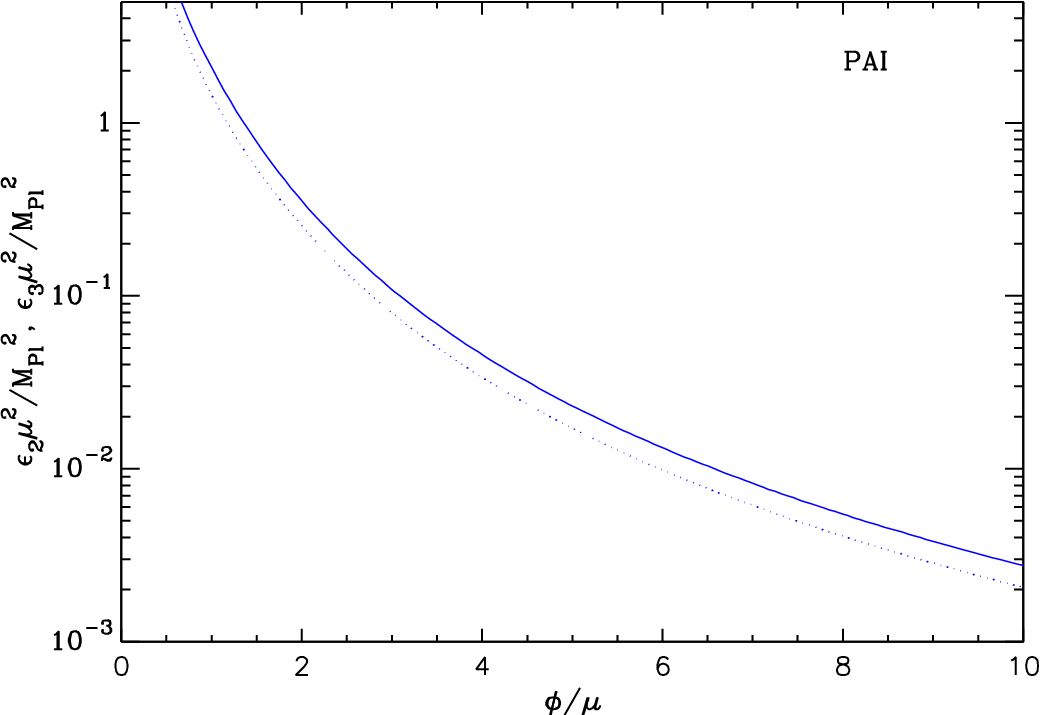}
\caption{Pure Arctan Inflation. Top panels: the potential and its
  logarithm as a function of $\phi/\mu$. Bottom left panel: slow-roll
  parameter $\epsilon_1 \mu^2/\Mp^2$. Bottom right panel: slow-roll
  parameters $\epsilon_2 \mu^2/\Mp^2$ (solid line) and $\epsilon_3
  \mu^2/\Mp^2$ (dotted line).}
\label{fig:potpai}
\end{center}
\end{figure}

This model has been proposed and discussed in \Refcs{Neves:2020lru,
  Neves:2020anh} in the context of brane inflation within a
five-dimensional bulk (see also \sectionc{sec:bi}). In this reference,
it is argued that the interaction of bulk particles with a
four-dimensional domain wall, assumed to be our universe, can trigger
an accelerated expansion. From a four-dimensional point of view,
inflation is driven by the so-called radion field, associated with the
position of the wall in the fifth dimension, and the effective
potential reads
\begin{equation}
V(\phi) = M^4 \arctan\left(\dfrac{\phi}{\mu}\right).
\label{eq:potpai}
\end{equation}
The functional shape of this potential can be obtained from the one of
Arctan Inflation (AI), see \sectionc{sec:ai}, by the transformation
$\phi \rightarrow 1/\phi$. However, such a transformation on the field
cannot be reabsorbed into a redefinition of some constants and Pure
Arctan Inflation is a different model than Arctan Inflation. The
potential is negative for $\phi<0$ where it does not describe a physical
situation in the context of brane inflation. We therefore restrict our
analysis to the positive domain only.

The Hubble-flow functions associated
with the potential~\eqref{eq:potpai}, in the slow-roll approximation, are
\begin{equation}
\epsilon_1 = \dfrac{\Mp^2}{2 \mu^2} \dfrac{1}{\arctan^2(x) \left(1+x^2
  \right)^2}\,, \qquad \epsilon_2 = \dfrac{2 \Mp^2}{\mu^2} \dfrac{1 + 2
  x \arctan(x)}{\arctan^2(x) \left(1 + x^2\right)^2}\,,
\label{eq:paisr12}
\end{equation}
and
\begin{equation}
\epsilon_3 = \dfrac{2\Mp^2}{\mu^2} \dfrac{1+ 3x \arctan(x) -
  \left(1-3x^2\right) \arctan^2(x)}{\arctan^2(x) \left(1+x^2\right)^2
  \left[1 + 2 x \arctan(x) \right]}\,,
\label{eq:paisr3}
\end{equation}
where we have defined the dimensionless field
\begin{equation}
x \equiv \dfrac{\phi}{\mu}\,.
\end{equation}
They have been represented, together with the potential, as a function
of $x$ in \Fig{fig:potpai}. The potential vanishes at $x=0$ and this
triggers a divergence of $\epsilon_1$ ensuring that inflation
gracefully ends within the positive domain. Notice that, in the region
close to the origin, one also has $\epsilon_2$ and $\epsilon_3$ larger
than unity showing that slow-roll violations may also occur just
before the end of inflation. The value of the inflaton field at which inflation ends, that
we denote by $\xend$, is the positive root of $\epsilon_1(x)=1$ and
can be obtained by solving the following equation:
\begin{equation}
\arctan(x) \left(1+ x^2 \right) = \dfrac{\Mp}{\sqrt{2} \mu}\,.
\label{eq:paixend}
\end{equation}
There is not analytical solution to this equation, which has to be solved
numerically. However, in the limits $\mu \ll \Mp$ and $\mu \gg \Mp$, the solution can be approximated as
\begin{equation}
\xend \simeq  
\begin{cases}
\sqrt{\dfrac{\sqrt{2} \Mp}{\pi\mu}}\quad\text{if}\quad\mu \ll \Mp\\ \\
\dfrac{\Mp}{\sqrt{2} \mu}\quad\text{if}\quad\mu \gg \Mp
\end{cases}
\, .
\label{eq:paixendapprox}
\end{equation}

The slow-roll trajectory can be integrated analytically and one
obtains
\begin{equation}
\begin{aligned}
  \Nend - N & = \dfrac{\mu^2}{6 \Mp^2} \Bigg[2 x \left(x^2+3 \right)
  \arctan(x) - 2 \xend \left(\xend^2 + 3 \right) \arctan(\xend)
\\ &  + (\xend^2
  - x^2) + 2 \ln\left(\dfrac{1+\xend^2}{1+x^2} \right) \Bigg].
\end{aligned}
\label{eq:paitraj}
\end{equation}
This trajectory, together with the reheating
equation~\eqref{eq:phistarlnrrad} and $\xend$ from \Eq{eq:paixend},
allow us to determine the field value $\xstar$ at which the pivot mode
crossed the Hubble radius during inflation. This also fixes the
energy scale of the potential by the CMB normalization and one gets
\begin{equation}
\left(\dfrac{M}{\Mp} \right)^4 = 720 \pi^2 \dfrac{\Mp^2}{\mu^2}
\dfrac{1}{\arctan^3(\xstar) \left(1+\xstar^2 \right)^2}
\dfrac{\Qrms^2}{T^2}\,.
\end{equation}
The reheating consistent slow-roll prediction for Pure Arctan
inflation have been represented in \Fig{fig:CMBPAI_0}. The two regimes
$\mu\ll \Mp$ and $\mu\gg \Mp$ can be understood as follows. In these
two limits, the trajectory~\eqref{eq:paitraj} can be inverted as
\begin{equation}
\xstar \simeq
\begin{cases}
\left( \xend^3+ \dfrac{6 \Delta\Nstar}{\pi}\dfrac{\Mp^2}{\mu^2}  \right)^{1/3}\quad\mu \ll \Mp\\
\sqrt{\xend^2+2\Delta\Nstar\frac{\Mp^2}{\mu^2}}  \quad\mu \gg \Mp
 \end{cases}\, .
\end{equation}
Using the approximate expressions for $\xend$ given in
\Eq{eq:paixendapprox}, and plugging the resulting expressions 
of $\xstar$ into
\Eqs{eq:paisr12} and~\eqref{eq:paisr3}, properly expanded in the relevant limit for $\mu$, one finally gets
\begin{align}
 \epsonestar &\simeq  2\left(\dfrac{\mu}{\pi\Mp}\right)^{2/3}
\dfrac{1}{\left(6 \Delta \Nstar\right)^{4/3}}
\,, \quad  \epstwostar \simeq  \dfrac{4}{3 \Delta \Nstar}
\,, \quad \epsthreestar \simeq \frac{1}{\Delta\Nstar}
\quad\text{if}\quad\mu\ll\Mp\, ,\\
\epsonestar &\simeq \frac{8\mu^2}{\pi^2\left(1+4\Delta\Nstar\right)^2\Mp^2}
\,, \quad  \epstwostar \simeq \frac{16\sqrt{2}\mu}{\left(1+4\Delta\Nstar\right)^{3/2}\pi\Mp} 
\,, \quad \epsthreestar  \simeq \frac{3\epsilon_{2*}}{4}
\quad\text{if}\quad\mu\gg\Mp\, .
\end{align}
One can see that, in the $\mu\ll\Mp$ limit, $\epsilon_1$ becomes very
small while $\epsilon_2$ remains constant at fixed $\Delta \Nstar$,
while in the limit $\mu\gg\Mp$, the first three slow-roll parameters
become large, and the model is excluded.

\subsection{Superconformal \texorpdfstring{$\alpha$}{alpha}-Attractor A Inflation (SAAI)}
\label{sec:saai}

\subsubsection{Theoretical Justifications}
\label{subsubsec:theorysaai}

The model is based on the vector multiplet Lagrangian introduced in
\eqref{eq:lagrangiansupervector} which depends on one arbitrary
function, $J(C)=3/2 \ln \Phi$. In \sectionc{sec:si}, it was shown that
the choice given by \eqref{eq:defc}, namely $\Phi=-Ce^C$, leads to the
Starobinsky model. In Refs.~\cite{Ferrara:2013rsa}
and~\cite{Kallosh:2013yoa}, an extension based on the choice
\begin{align}
  \Phi(C)=\left(-C\right)^{\alpha}e^{\beta C},
\end{align}
where $\alpha $ and $\beta $ are two new free parameters, was
considered. It leads to the so-called ``$\alpha-\beta $ model''. Using
\eqref{eq:lagrangiansupervector}, it is easy to show that
\begin{align}
  V(C)=\frac{9}{8} g^2\left(\beta +\frac{\alpha }{C}\right)^2,
\end{align}
where the field $C$ is not canonically normalized. In terms of the
canonically normalized field $\phi$,
$C=-\exp[\sqrt{2/(3\alpha)}\phi/\Mg]$, the potential acquires the
following form
\begin{align}
  V(\phi)=\frac98 g^2\left(\beta -\alpha \, e^{\sqrt{\frac{2}{3\alpha}}
    \frac{\phi}{\Mg}}\right)^2.
  \end{align}
Writing $M^4=9g^2\beta^2/8$ and shifting the field
$\phi-\phizero\rightarrow \phi$, where
$\exp\left(-\sqrt{\frac{2}{3\alpha}}\frac{\phizero}{\Mg}\right)=\alpha/\beta$,
one arrives at the potential~(\ref{eq:potsaai}). It is also
interesting to notice that the potential~(\ref{eq:potsaai})
interpolates between the Starobinsky model ($\alpha=1$) and the
quadratic LFI model. Indeed, when $\alpha \rightarrow +\infty$, one
has $V\sim 2M^4/(3\alpha)(\phi/\Mg)^2$.

Let us finally notice that the above potential can also be obtained in
the context of the models discussed in \sectionc{sec:sabi}.

\subsubsection{Slow-Roll Analysis}
\label{subsubsec:srsaai}

\begin{figure}
\begin{center}
\includegraphics[width=\wdblefig]{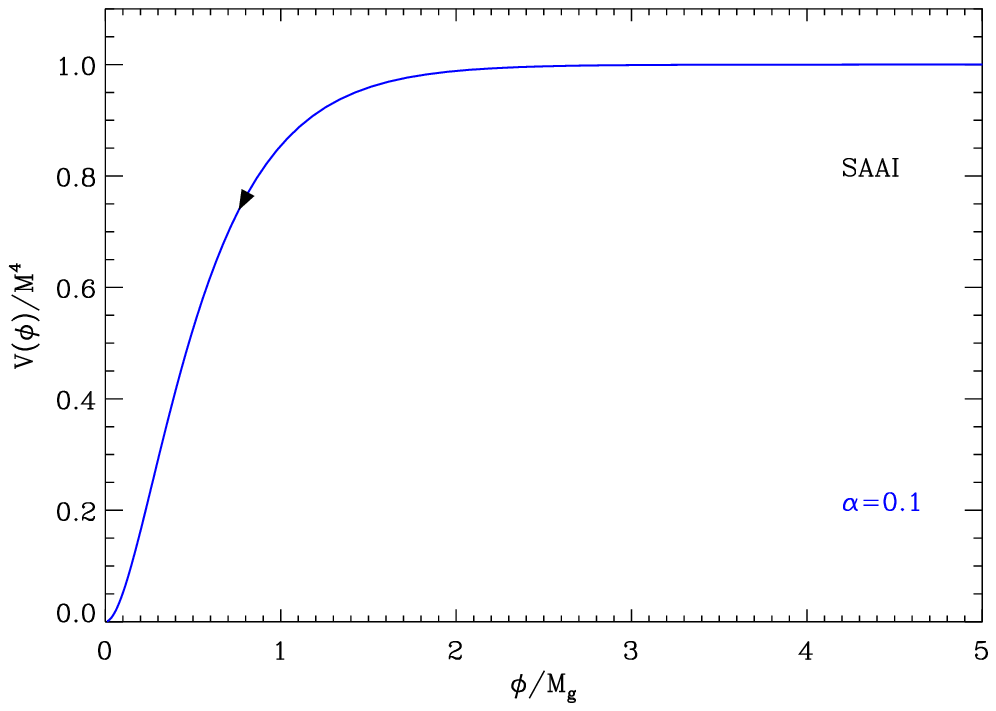}
\includegraphics[width=\wdblefig]{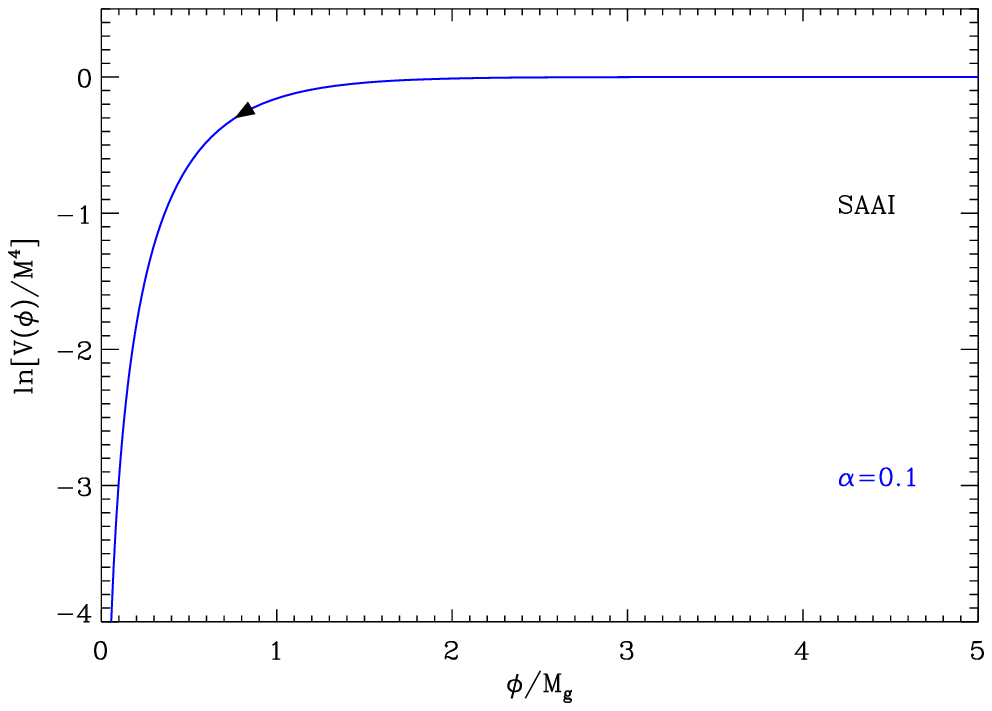}
\includegraphics[width=\wdblefig]{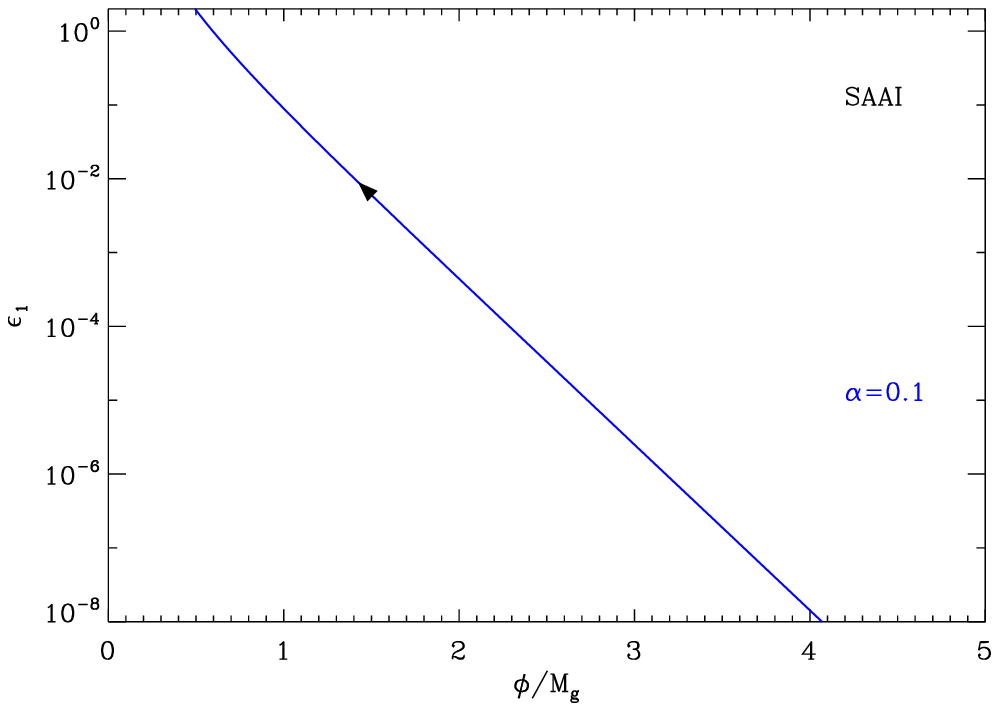}
\includegraphics[width=\wdblefig]{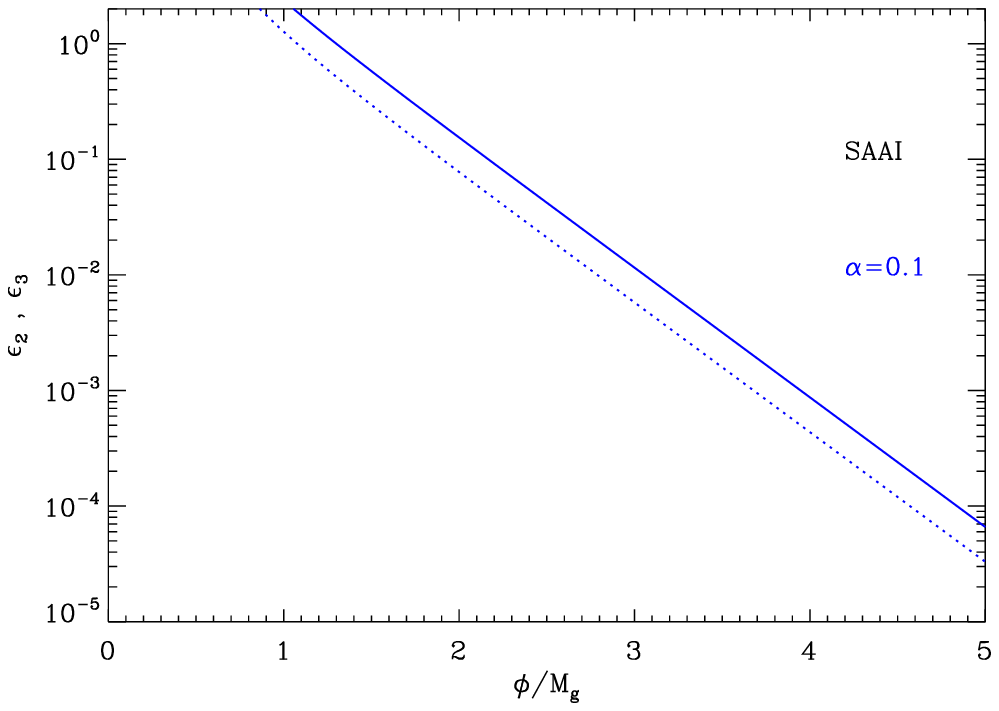}
\caption{Superconformal $\alpha$-attractor A Inflation (SAAI). Top
  left panel: the potential as a function of $\phi/\Mg$.  Top right
  panel: logarithm of the potential. Bottom left panel: the first
  slow-roll parameter $\epsilon_1$. Bottom right panel: slow-roll
  parameters $\epsilon_2$ (solid line) and $\epsilon_3$ (dotted
  line).}
\label{fig:potsaai}
\end{center}
\end{figure}

The potential of $\alpha$-attractor models can be written as
\begin{equation}
  \label{eq:potsaai}
 V(\phi) = M^4\left(1-e^{-\sqrt{\frac{2}{3\alpha}}\frac{\phi}{\Mg}}\right)^2.
\end{equation}
It clearly bears resemblance with the Starobinsky/Higgs models. In
fact, if $\alpha=1$, it exactly reduces to these models. It can thus
be seen as a generalization of these scenarios. The
potential~(\ref{eq:potsaai}) is represented in \Fig{fig:potsaai} for
different values of $\alpha$.

The three Hubble flow functions can be easily calculated and, defining
$x\equiv \phi/\Mg$, one obtains
\begin{align}
\label{eq:epssaai}
\epsilon _1 &= \frac{4}{3\alpha}
\left(1-e^{\sqrt{\frac{2}{3\alpha}}x}\right)^{-2}
\, 
,\quad
\epsilon_2= \frac{2}{3\alpha}
\left[\sinh\left(\frac{x}{\sqrt{6\alpha}}\right)\right]^{-2}
\,, \\
\epsilon  _3&=
\frac{2}{3\alpha} \left[\coth \left(\frac{x}{\sqrt{6\alpha}}\right)
    -1\right]\coth\left(\frac{x}{\sqrt{6\alpha}}\right)\, .
\end{align}
Evidently, when $\alpha=1$, these expressions reduce to
\Eqs{eq:srparametershiggs}.

In this scenario, inflation ends by violation of the slow-roll
conditions. Inflation stops when $x=\xend$ with
\begin{align}
  \xend=\sqrt{\frac{3\alpha}{2}}\ln \left(1+
  \frac{2}{\sqrt{3\alpha}}\right).
  \end{align}
However, as it was the case for Higgs inflation, see
\sectionc{sec:hi}, violation of the slow-roll conditions can occur
before. Indeed, one has $\epsilon_2=1$ for
\begin{equation}
 \xepstwoOne = \sqrt{6\alpha}\,\arsinh
  \left(\sqrt{\frac{2}{3\alpha}}\right)\, ,
\end{equation}
and $\epsilon_3=1$ if
\begin{equation}
  \xepsthreeOne = \sqrt{6\alpha}\,
  \artanh \left(\frac{2}{1+\sqrt{1+6\alpha}}\right)\, .
\end{equation}
In fact as inflation proceeds, the field reaches first the value
$\xepstwoOne$, then $\xepsthreeOne$ and, finally, $\xend$. It is
interesting to notice that this hierarchy does not depend on $\alpha$.

The next step consists in calculating the slow-roll
trajectory. Straightforward manipulations lead to the following
expression
\begin{equation}
\label{eq:saai:traj}
\Nend-N=\frac12\sqrt{\frac{3\alpha}{2}}\left(\xend-x\right) +
\frac{3\alpha}{4}\left(\ee^{\sqrt{\frac{2}{3\alpha}} x}
-\ee^{\sqrt{\frac{2}{3\alpha}}\xend}\right).
\end{equation}
Of course, one can check that, for $\alpha=1$, the above formula
reduces to the trajectory found in the Higgs scenario. For large
values of $x$, $x\gg 1$, the last term is dominant. This trajectory
can be inverted and expressed in term of the ``$-1$-branch'' of the
Lambert function $\Lambert{-1}$. One obtains
\begin{align}
  x =& \sqrt{\frac{3\alpha}{2}}\Biggl\lbrace -\frac{4}{3\alpha}
  \Delta N +\sqrt{\frac{2}{3\alpha}}\xend
  -\ee^{\sqrt{\frac{2}{3\alpha}}\xend} 
\nonumber \\ &-
  \Lambert{-1}
    \left[-\exp\left(-\frac{4}{3\alpha} \Delta N +\sqrt\frac{2}{3\alpha}
        \xend-\ee^{\sqrt{\frac{2}{3\alpha}}\xend} \right)\right]\Biggr\rbrace,
    \label{eq:saai:trajinverted} 
\end{align}
where, as usual, $\Delta N = \Nend - N$. The reason that inflation
proceeds along the $-1$ branch of the Lambert function can be
understood by means of arguments similar to those presented in
\sectionc{sec:hi}. The Lambert function in \Eq{eq:saai:trajinverted}
can be written as $\Lambert{-1}\{\exp[-4\Delta N/(3\alpha)]ye^y\}$
with $y=-\exp[\sqrt{2/(3\alpha)}\xend]$. At the end of inflation, by
definition, one has $\Delta N=0$. Using the property that
$\Lambert{-1}\left(y\ee^y\right)=-e^y$ (if $y<-1$), one has therefore
that the Lambert function equals $-\exp(\sqrt{2/(3\alpha)}\xend$,
which is smaller than $-1$; and, as can be seen in
\Fig{fig:hi:lambert}, a value of the Lambert function smaller than
$-1$ necessarily corresponds to the $-1$ branch. The previous argument
is valid at the end of inflation only. On more general grounds, for
$N$'s before the end of inflation, $\Delta N>0$ becomes large and,
therefore, the argument of the Lambert function becomes small. In
order, for $x$ to say positive in \Eq{eq:saai:trajinverted}, the
Lambert function must be large and negative in this limit. This
immediately implies that the branch $-1$ is the relevant one.

Finally, the value of $\xstar$, at which the pivot
mode crossed out the Hubble radius during inflation can be expressed as
\begin{equation}
\begin{aligned}
\label{eq:xstarsaai}
  \xstar & =\sqrt\frac{3\alpha}{2}\left(-\frac{4}{3\alpha}\Delta
    \Nstar + \ln\left( 1+\frac{2}{\sqrt{3\alpha}} \right) - \left(
      1+\frac{2}{\sqrt{3\alpha}} \right)
  \right. \\ &\left.
    -\Lambert{-1}\left\lbrace - \exp\left[ -\frac{4}{3\alpha}\Delta
        \Nstar+\ln\left(1+\frac{2}{\sqrt{3\alpha}} \right) - \left(
          1+\frac{2}{\sqrt{3\alpha}} \right)
      \right] \right \rbrace \right)\, ,
\end{aligned}
\end{equation}
where, in this expression, we have used the value of $\xend$ derived
above. From the knowledge of $\xstar$, the energy scale $M$ of the
potential can be inferred and one obtains
\begin{equation}
\label{eq:saai:COBE}
\frac{M^4}{\Mg^4}=1920\frac{\pi^2}{\alpha}
\left(1-\ee^{\sqrt{\frac{2}{3\alpha}}\xstar}\right)^{-4}
\ee^{2\sqrt{\frac{2}{3\alpha}}\xstar} \frac{\Qrms^2}{T^2}\, .
\end{equation}

The reheating consistent slow-roll prediction for Superconformal
$\alpha$-attractor A Inflation have been represented in \Fig{fig:CMBSAAI_0}.

\subsection{T-Model Inflation (TMI)}
\label{sec:tmi}

\subsubsection{Theoretical Justifications}
\label{sec:theorytmi}

The theoretical motivations underlying these scenarios, which were
named ``T-models'' in \Refc{Kallosh:2013hoa}, find their roots in the
superconformal context already presented for Starobinsky Inflation
(SI) in \sectionc{sec:othertheosi}. As discussed in that section, the
corresponding scenarios have some attractive features. In particular,
they are conformally and $\mathrm{SO}(1,1)$ invariant. If the
conformal invariance is broken in the rapidity gauge,
$\chi^2-\phi^2=6$, it was shown that the model reduces to the standard
action of a scalar field, minimally coupled to gravity, with a
constant potential, see \Eq{eq:actionconformon2fieldgaugefixed1}. The
corresponding solution is de Sitter spacetime, thus showing a
connection between the action of \Eq{eq:actioninvaconf2field} and the
theory of inflation. Let us add that the way conformal invariance is
broken is a choice and other instances are possible, but leading to the
same conclusions. For instance, the conformal gauge $\chi=\sqrt{6}$ is
a good illustration of the above claim since it is particularly
simple. Indeed, in that case, the
action~\eqref{eq:actioninvaconf2field} takes the form
\begin{equation}
S\left(g_{\mu \nu},\phi\right)=\frac{\Mg^2}{2}\int \dd ^4 \bmx \,
\sqrt{-g}\, \left[\left(1-\frac{\phi^2}{6}\right)R - g^{\mu
    \nu}\partial _\mu \phi \, \partial _\nu \phi
  -\frac{\lambda}{2}(\phi^2-6)^2\right].
\label{eq:actioninvaconf2fieldgaugefixed2}
\end{equation}
Using the notations presented in \sectionc{sec:nonmingrav}, while
dropping the ``bar'' over Jordan frame quantities here, we see that
this is the action of a scalar tensor theory with
$F(\phi)=1-\phi^2/6$, $Z(\phi)=1$ and $U=\lambda(\phi^2-6)^2/4$, see
\Eq{eq:actionst}. Expressed in the Einstein frame, the action
exactly reduces to \Eq{eq:actionconformon2fieldgaugefixed1} since the
potential reads $V=\Mg^2U/F^2=9\lambda \Mg^2$. Theferore, this
confirms that the final result is the same regardless of the gauge is
used to break conformal invariance.

The above preliminary considerations help us to understand the context
in which the TMI model is designed. Indeed, let us now consider the
following action which is a generalization of the
\Eq{eq:actioninvaconf2field},
\begin{align}
\label{eq:actioninvaconf2fieldtli}
S\left(g_{\mu \nu},\chi,\phi\right)&=\frac{\Mg^2}{2}\int \dd ^4 \bmx
\, \sqrt{-g}\, 
\biggl[\frac{\chi^2}{6}R +
g^{\mu \nu}\partial _\mu \chi \partial _\nu \chi
-\frac{\phi^2}{6}R -
g^{\mu \nu}\partial _\mu \phi \partial _\nu \phi
\nonumber \\ &
-\frac{1}{36}\FT\left(\frac{\phi}{\chi}\right)
\left(\phi^2-\chi^2\right)^2\biggr],
\end{align}
where $\FT(.)$, the new ingredient of the model, is a priori an
arbitrary function. Given that $\FT=1$ leads to de Sitter, deviations
from this case should lead to non-trivial models of inflation. It is
also important to notice that the
action~\eqref{eq:actioninvaconf2fieldtli} is still conformally
invariant, thanks to the dependence of $\FT(.)$ in
$\phi/\chi$. However, the symmetry $\mathrm{SO}(1,1)$ is broken,
unless $\FT$ is a constant which was of course precisely the case in
\Eq{eq:actioninvaconf2field}. Notice also that, in \sectionc{sec:si},
it was shown that the above model with a term $\sim
\phi^2(\phi-\chi)^2$, or, equivalently, $\FT \sim
(\phi/\chi)^2/(1+\phi/\chi)^2$ leads to the Starobinsky
model. This important model is therefore included in the class of
models studied in this section.

Then, in order to proceed, one needs to choose a gauge. Taking
$\chi=\sqrt{6}$, \Eq{eq:actioninvaconf2fieldtli} takes the form
\begin{equation}
  S\left(g_{\mu \nu},\chi,\phi\right) =\frac{\Mg^2}{2}\int \dd ^4
  \bmx \, \sqrt{-g}\, 
\biggl[\left(1-\frac{\phi^2}{6}\right)R
-g^{\mu \nu}\partial _\mu \phi \partial _\nu \phi
-\frac{1}{36}\FT\left(\frac{\phi}{\sqrt{6}}\right)
\left(\phi^2-6\right)^2\biggr].
\label{eq:actioninvaconf2fieldtli2}
\end{equation}
Again, one recognizes a scalar tensor theory with
$F(\phi)=1-\phi^2/6$, $Z(\phi)=1$ and
$2U=\FT(\phi/\sqrt{6})(\phi^2/6-1)^2$. The potential,
$V=\Mg^2 U/F^2$, is then given by
$V=\Mg^2\FT(\phi/\sqrt{6})/2$. In addition, the
relationship between the fields $\phi$ and $\ef{\phi}$ is given by
\Eq{eq:dphisquare} and leads to
\begin{equation}
  \dfrac{\ef{\phi}}{\Mg} =-\sqrt{\dfrac{3}{2}} \ln
  \left(\dfrac{\sqrt{6}-\phi}{\sqrt{6}+\phi}\right).
\end{equation}
This relation can be inverted and one arrives at
\begin{align}
  \phi=\sqrt{6}\tanh \left(\frac{\ef{\phi}}{\Mg \sqrt{6}}\right).
\end{align}
As a consequence, in the Einstein frame, the action takes the form
\begin{equation}
  S\left(\ef{g}_{\mu \nu},\ef{\phi}\right)=\int \dd ^4 \bmx \, \sqrt{-\ef{g}}\,
  \left\{\frac{\Mg^2}{2}\ef{R}
  -\frac12 \ef{g}^{\mu \nu}\partial_\mu\ef{\phi} \partial_\nu \ef{\phi}
  -\frac{\Mg^2}{2}
  \FT\left[\tanh \left(\frac{\ef{\phi}}
    {\sqrt{6}\Mg}\right)\right]\right\}.
  \label{eq:actionconformon2field}
\end{equation}
One can also use the rapidity condition $\chi^2-\phi^2=6$. Then, using
the parametrization defined in \Eq{eq:rapidityfields} for this gauge,
one can check that the action~\eqref{eq:actionconformon2field} is
directly recovered.

The next question is of course which function
$\FT(.)$ should be chosen? In absence of any deep reasons,
\Refc{Kallosh:2013hoa} argues that a reasonable and interesting choice
is simply to take a power-law, namely
\begin{equation}
  \label{eq:ftpowerlaw}
  \FT\left(\frac{\phi}{\chi}\right)=\lambda
  \left(\frac{\phi}{\chi}\right)^{2n}.
  \end{equation}
This leads to the potential studied in the next section. One remarks
that for large values of the field,
$\FT\{\tanh[\bar{\phi}/(\sqrt{6}\Mg)]\}$ tends to a constant and, in this
regime, the $\mathrm{SO}(1,1)$ symmetry is restored.

In~\Refc{Kallosh:2013hoa}, the model is also discussed in the
framework of conformal supergravity, the action of which was given by
\Eq{eq:superconflagrangian}. In that case, the embedding K\"ahler
potential can be chosen as
\begin{equation}
  \label{eq:embkalone}
  \calN \left(X^I,\bar{X}^{\bar{I}}\right)
=- \left\vert X^0 \right\vert^2+ \left\vert
X^1\right\vert^2+\left\vert S \right\vert^2
-3\zeta \frac{\left(S\bar{S}\right)^2}{ \left\vert X^0 \right\vert ^2-\left\vert X^1\right\vert ^2}\,.
\end{equation}
Here, $X^0$ is a conformon, $X^1=\Phi$ is the inflaton and $X^3=S$ is
a Goldstino and $\zeta$ is a dimensionless parameter. The above
K\"ahler potential has a SU(1,1) symmetry between the conformon and
the inflaton. As it was the case in \sectionc{sec:si}, the last term,
proportional to the parameter $\zeta$, is introduced to stabilize the
inflationary trajectory. The superpotential is taken to be
\begin{equation}
  \label{eq:embwalone}
  \calW=S\left[\left(X^0\right)^2-\left(X^1\right)^2\right]f\left(\frac{X^1}{X^0}\right),
\end{equation}
where, at this stage, the function $f$ is arbitrary. It is interesting
to emphasize the difference with the superconformal model of
\sectionc{sec:si}. In that section, the embedding and superpotential
were both given in terms of an exponential of the fields, leading to a
power-law K\"ahler potential while, here, the embedding potential is
expressed directly in terms of powers of the fields and, as a
consequence, the corresponding K\"ahler potential will be given by a
logarithm, a structure reminiscent of no scale supergravity.

Then, as usual, we fix the conformon field and choose the gauge where
$X^0=\bar{X}^{\bar{0}}=\sqrt{3}\Mg$. As a consequence, looking at
Eqs.~(\ref{eq:embedpot}), the above choices imply that a (ordinary)
supergravity description of the model can be expressed, as announced
above, in terms of the following logarithmic K\"ahler potential
\begin{equation}
  K=-3\Mg^2 \ln \left(1+k\right),
  \label{eq:Kahlertmodel}
\end{equation}
where $k$ is a function given by
\begin{equation}
  \label{eq:defkkahlerembed}
  k=-\frac{\Phi \bar{\Phi}}{3\Mg^2}-\frac{S\bar{S}}{3\Mg^2}
  +\frac{\zeta}{\Mg^2}\frac{S^2\bar{S}^2}{3\Mg^2-\Phi \bar{\Phi}}\,.
\end{equation}
The exact scalar potential corresponding to the theory we have just
described is quite complicated. Indeed, as already mentioned several
times, in general, it can be written as
\begin{equation}
  V=\Mg^4e^G\left(G^{\bar{B}C}G_{\bar{B}}G_C-3\right),
\end{equation}
with
\begin{equation}
  G= \dfrac{K}{\Mg^2} + \ln\left(\dfrac{WW^\dagger}{\Mg^6}\right).
\end{equation}
Expanding this formula, one obtains the following complicated
expression
\begin{equation}
  \begin{aligned}
  V &=\frac{e^{K/\Mg^2}}{\Mg^2}G^{\bar{B}C}
  \biggl(\frac{WW^\dagger}{\Mg^4}\frac{\partial K}{\partial \bar{X}^{\bar{B}}}
\frac{\partial K}{\partial X^{C}}
+\frac{W^{\dagger}}{\Mg^2}
\frac{\partial K}{\partial \bar{X}^{\bar{B}}}
\frac{\partial W}{\partial X^{C}}
+\frac{W}{\Mg^2}
\frac{\partial W^\dagger}{\partial \bar{X}^{\bar{B}}}
\frac{\partial K}{\partial X^{C}}
 \\ & +
\frac{\partial W^\dagger}{\partial \bar{X}^{\bar{B}}}
\frac{\partial W}{\partial X^{C}}
-3WW^\dagger\biggr).
\end{aligned}
\end{equation}

However, a crucial aspect of this scenario is that the superpotential,
after symmetry breaking, can be expressed as
\begin{equation}
  W=S(3\Mg^2-\Phi^2)\, f\left(\dfrac{\Phi}{\sqrt{3}\Mg}\right),
\end{equation}
and we notice that, as it was the case in \sectionc{sec:si}, this
superpotential is again of the form $W=Sf(\Phi)$. Moreover, a
fundamental remark made in~\Refc{Kallosh:2013hoa} is that inflation
takes place at $S=0$ and in the direction where $\Phi$ is real. This
greatly simplifies the calculations. Indeed, the simplification comes
from the fact that, if $S=0$, then $W=0$ and the previous expression
for the scalar potential can be reduced to a compact formula, namely
\begin{equation}
  \label{eq:potSvanish}
V =\frac{e^{K/\Mg^2}}{\Mg^2} G^{\bar{B}C}
\frac{\partial W^\dagger}{\partial \bar{X}^{\bar{B}}}
\frac{\partial W}{\partial X^{C}}
=\frac{e^{K/\Mg^2}}{\Mg^2} G^{\bar{S}S}
\frac{\partial W^\dagger}{\partial \bar{S}}
\frac{\partial W}{\partial S}
\,.
\end{equation}
To go further, we must now calculate the K\"ahler matrix. Using
\Eq{eq:Kahlertmodel}, one obtains that
\begin{equation}
\label{eq:Kmatrixalone}
  G_{A\bar{B}}=-\frac{3}{1+k} \frac{\partial ^2 k}{\partial X^A
    \partial \bar{X}^{\bar{B}}}+\frac{3}{(1+k)^2}
  \frac{\partial k}{\partial X^A}
  \frac{\partial k}{\partial \bar{X}^{\bar{B}}}\,.
  \end{equation}
It follows that, along the inflationary direction, the K\"ahler matrix
can be simplified as
\begin{equation}
  \label{eq:Kmatrixalone2}
  G_{A\bar{B}}=\frac{1}{\Mg^2(1+k)^2}
  \begin{pmatrix}
    1 & 0 \\
    0 & 1+k
    \end{pmatrix}.
\end{equation}
Given that, along the inflationary direction, $k=-\Phi^2/(3\Mg^2)$,
the above form of the K\"ahler matrix implies that the canonically
normalized field $\varphi$ is related to $\Phi$ by the following
formula
\begin{equation}
  \label{eq:fieldnormalalone}
  \Phi=\sqrt{6}\Mg \tanh\left(\dfrac{\varphi}{\sqrt{6}\Mg}\right).
\end{equation}
Finally, by noticing that $\partial W/\partial S=3\Mg^2(1+k)f$, it
follows that the potential of the canonically normalized field
$\varphi$ reads
\begin{align}
  \label{eq:Tpot}
  V(\varphi)=9\Mg^4\left\vert
  f\left[\tanh\left(\frac{\varphi}{\sqrt{6}\Mg}\right)
    \right]\right\vert^2.
\end{align}
As a consequence, the T-model can be reproduced by choosing the
function $f$ appropriately, namely proportional to $\FT$
in \Eq{eq:actionconformon2field}.

\subsubsection{Slow-Roll Analysis}
\label{subsubsec:srtmi}

\begin{figure}
\begin{center}
\includegraphics[width=\wdblefig]{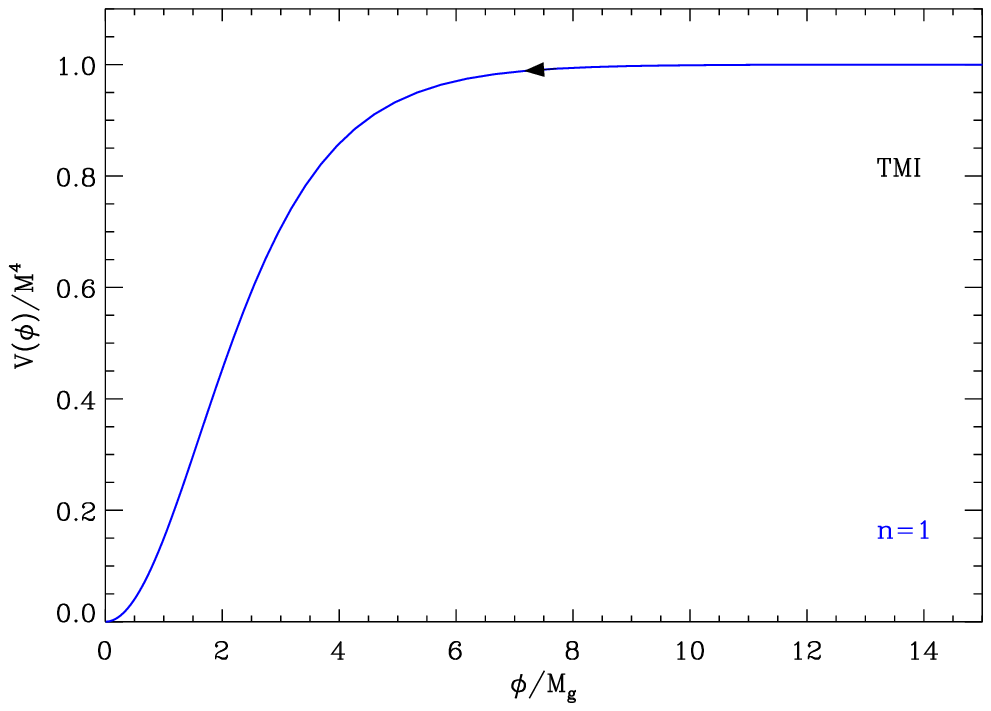}
\includegraphics[width=\wdblefig]{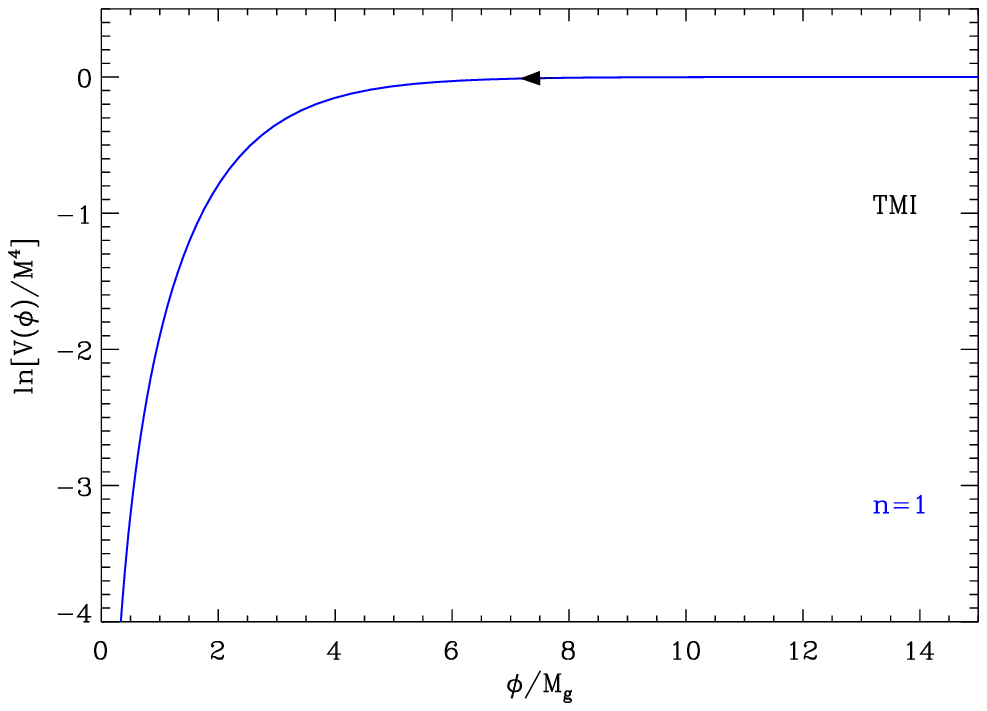}
\includegraphics[width=\wdblefig]{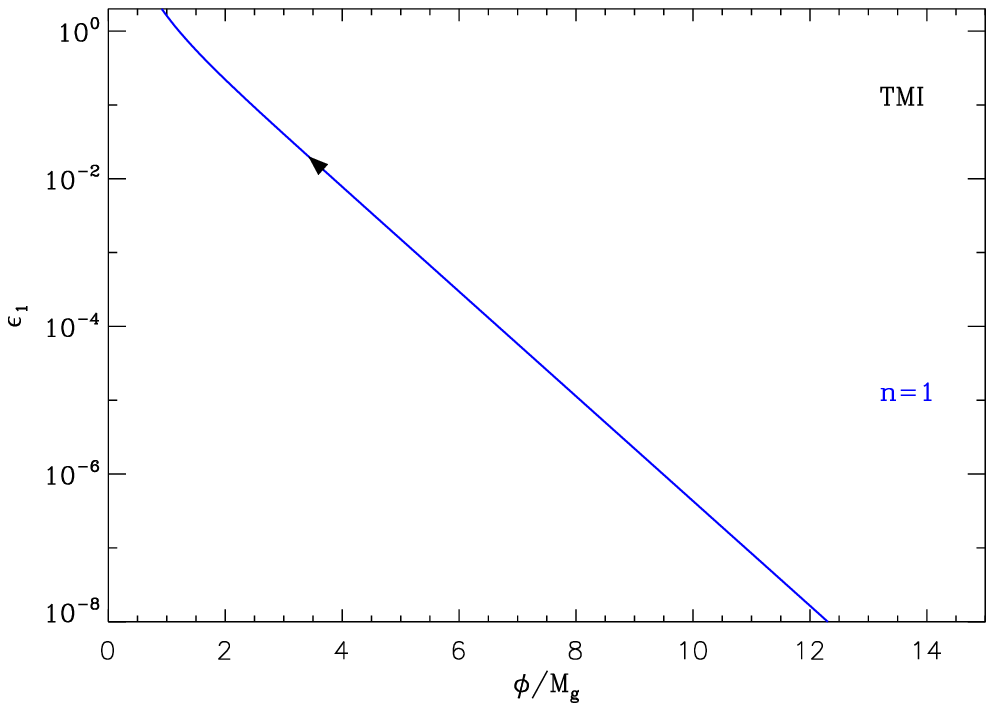}
\includegraphics[width=\wdblefig]{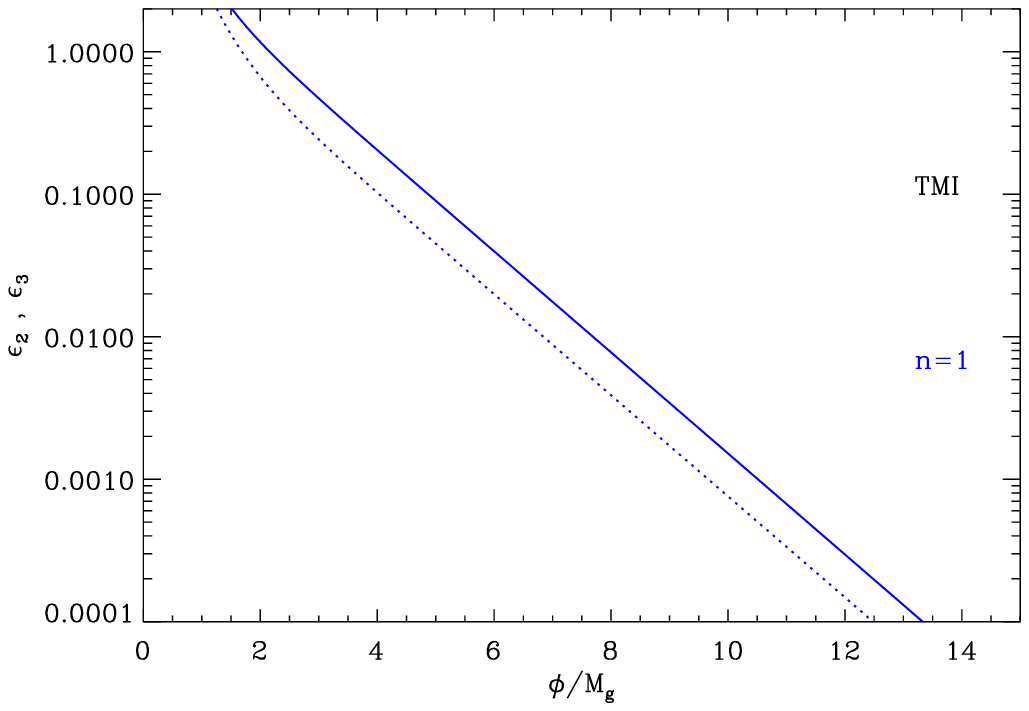}
\caption{T-Model Inflation (TMI). Top left panel: the potential as a
  function of $\phi/\Mg$.  Top right panel: logarithm of the
  potential. Bottom left panel: the first slow-roll parameter
  $\epsilon_1$. Bottom right panel: slow-roll parameters $\epsilon_2$
  (solid line) and $\epsilon_3$ (dotted line).}
\label{fig:pottmi}
\end{center}
\end{figure}

Going back to our usual notation for the inflaton, we now denote by
$\phi$ the canonically normalized field (noted $\varphi$ in the
previous section, and not be confused with the Jordan frame field of
\sectionc{sec:theorytmi}).

The potential of the TMI can therefore be written as~\cite{Kallosh:2013yoa}
\begin{equation}
  V(\phi)=M^4\left[\tanh \left(\frac{\phi}{\sqrt{6}\Mg}\right)\right]^{2n},
  \label{eq:pottmi}
\end{equation}
and describes a one parameter model, the parameter being $n$.

From the above potential, we can obtain the Hubble flow
functions. Defining $x\equiv \phi/\Mg$, one gets
\begin{equation}
\begin{aligned}
\epsilon _1 &= \frac{4n^2}{3}\sinh^{-2}\left(\frac{2x}{\sqrt{6}}\right)
,\quad
\epsilon_2= \frac{8n}{3}
\frac{\cosh\left(\dfrac{2x}{\sqrt{6}}\right)}
     {\sinh^2\left(\dfrac{2x}{\sqrt{6}}\right)}
\,, \quad
\epsilon_3=\frac{2n}{3}\frac{3+\cosh\left(
  \dfrac{4 x}{\sqrt{6}}\right)}{\sinh^2\left(\dfrac{2 x}{\sqrt{6}}\right)
  \cosh\left(\dfrac{2 x}{\sqrt{6}}\right)}\,.
\end{aligned}
\label{eq:epstmi}
\end{equation}
The potential and the Hubble-flow functions have been plotted in \Fig{fig:pottmi}.

In this scenario, inflation stops by violation of the slow-roll
conditions. This happens when $\epsilon_1=1$ corresponding to
following vacuum expectation value of the field
\begin{equation}
  \xend=\frac{\sqrt{6}}{2}\arcsinh\left(\frac{2n}{\sqrt{3}}\right).
\end{equation}

The slow-roll trajectory can be integrated and one gets
\begin{equation}
  \Nend-N=\frac{3}{4n}\left[\cosh\left(\frac{2x}{\sqrt{6}}\right)
    -\cosh\left(\frac{2\xend}{\sqrt{6}}\right)\right].
\end{equation}
This trajectory can be inverted leading to an explicit formula for
$\phi/\Mg$ during slow-roll inflation
\begin{equation}
  x=\frac{\sqrt{6}}{2}\arccosh\left(\sqrt{1+\frac{4n^2}{3}}+\frac{4n}{3}
    \Delta N\right),
\label{eq:tmitraj}
\end{equation}
where $\Delta N = \Nend - N$. The value of $\xstar$, namely the value
of the field when the pivot scale crosses out the Hubble radius during
inflation is given just given by the above expression with $\Delta
N=\Delta N_*$.

Finally, the mass scale $M$ that normalizes the potential can be expressed as
\begin{equation}
\label{eq:tmi:COBE}
\frac{M^4}{\Mg^4}=\frac{1920\pi^2n^2}{
\sinh^2\left(\displaystyle\frac{2\xstar}{\sqrt{6}}\right)
     \left[\tanh\left(\displaystyle\frac{\xstar}{\sqrt{6}}\right)\right]^{2n}}
\frac{\Qrms^2}{T^2}\, .
\end{equation}

The reheating consistent observable predictions for TMI have been
represented in \Fig{fig:CMBTMI_0} for various values of $n$. One
notices that the dependence on $n$ of the spectral index and of the
tensor-to-scalar ratio is very small. Indeed, provided the quantity $n
\Delta N$ dominates in \Eq{eq:tmitraj}, one has
\begin{equation}
  \xstar \simeq \dfrac{\sqrt{6}}{2} \arccosh\left(\dfrac{4n}{3}
  \Delta\Nstar\right).
\end{equation}
Plugging this approximation into \Eq{eq:epstmi} gives
\begin{equation}
\epsonestar \simeq \dfrac{3}{4 \Delta\Nstar^2}\,, \quad \epstwostar
\simeq \dfrac{2}{\Delta\Nstar}\,, \quad \epsthreestar \simeq
\dfrac{1}{\Delta\Nstar}\,,
\end{equation}
and the Hubble-flow functions are independent of $n$ in the large
$\Delta\Nstar$ limit.

\section{Two Parameters Models}
\label{sec:twop}

\subsection{Small Field Inflation (SFI)}
\label{sec:sfi}

\begin{figure}
\begin{center}
\includegraphics[width=\wdblefig]{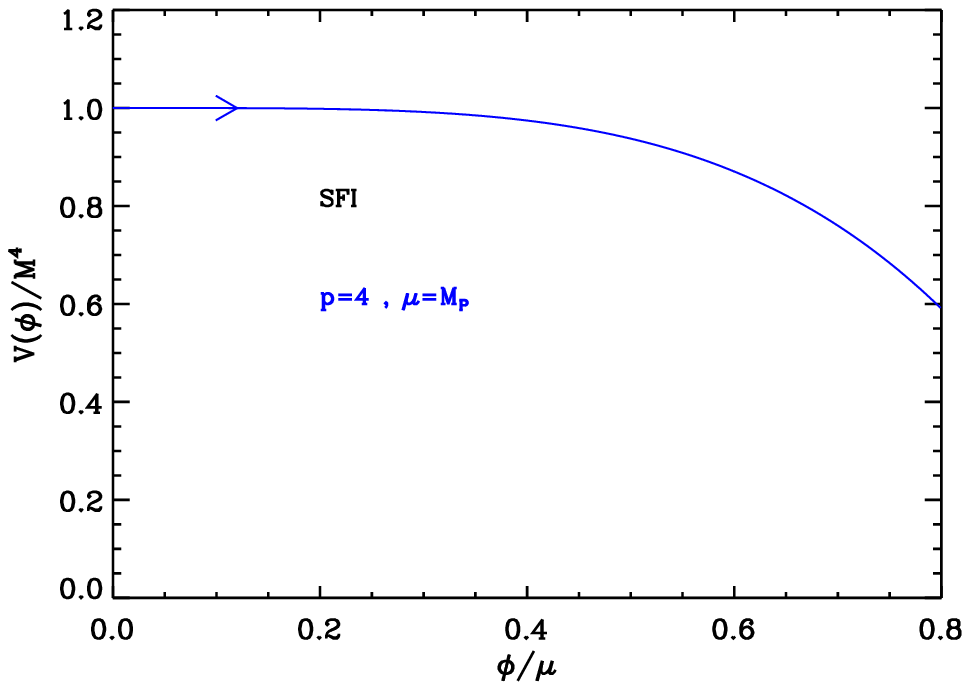}
\includegraphics[width=\wdblefig]{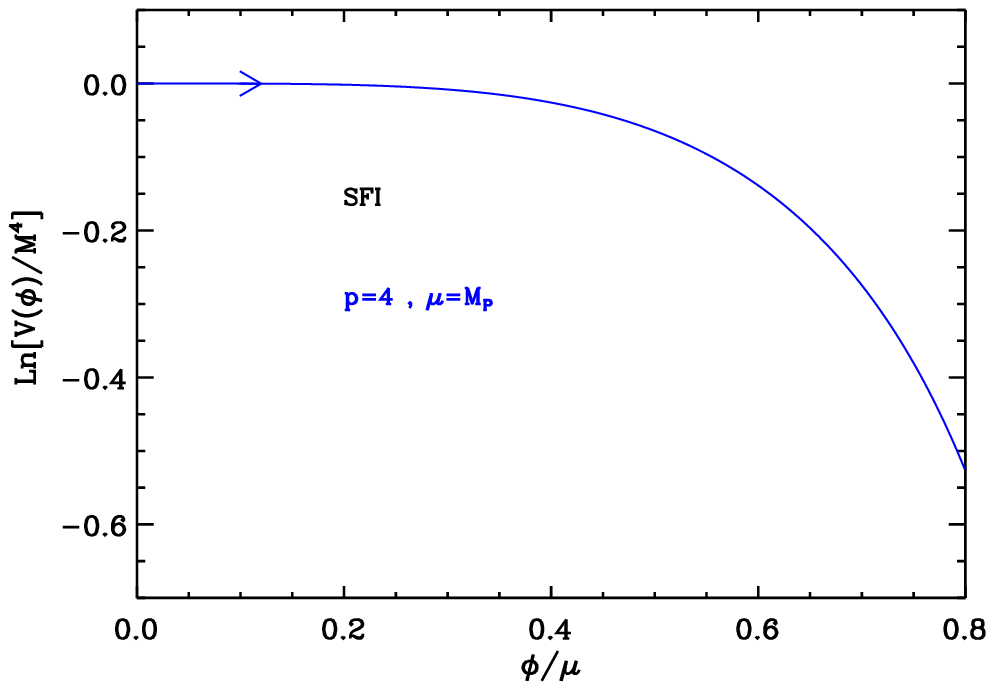}
\includegraphics[width=\wdblefig]{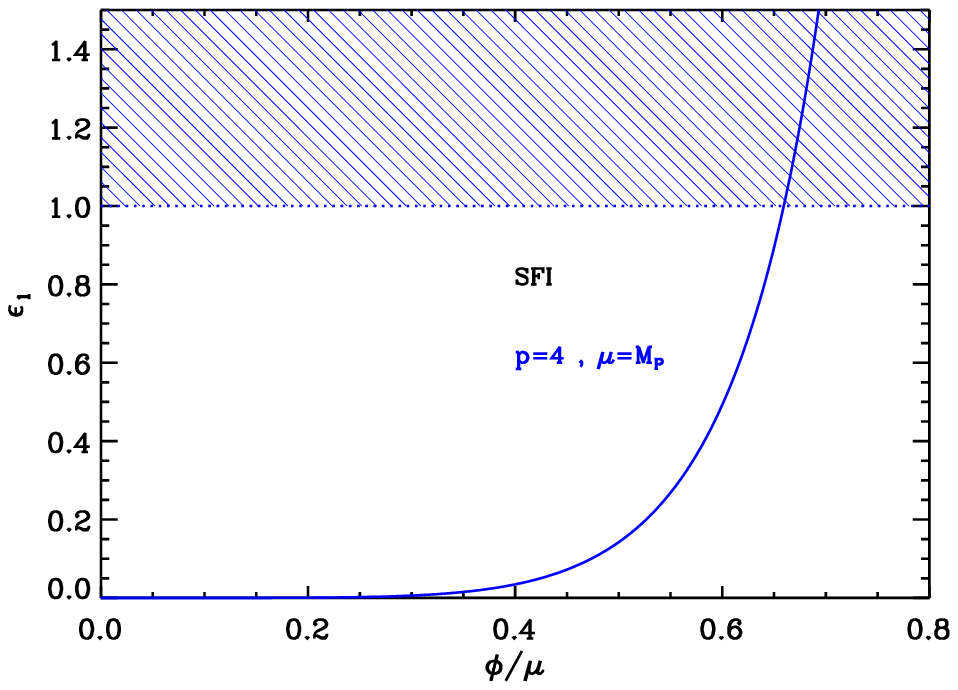}
\includegraphics[width=\wdblefig]{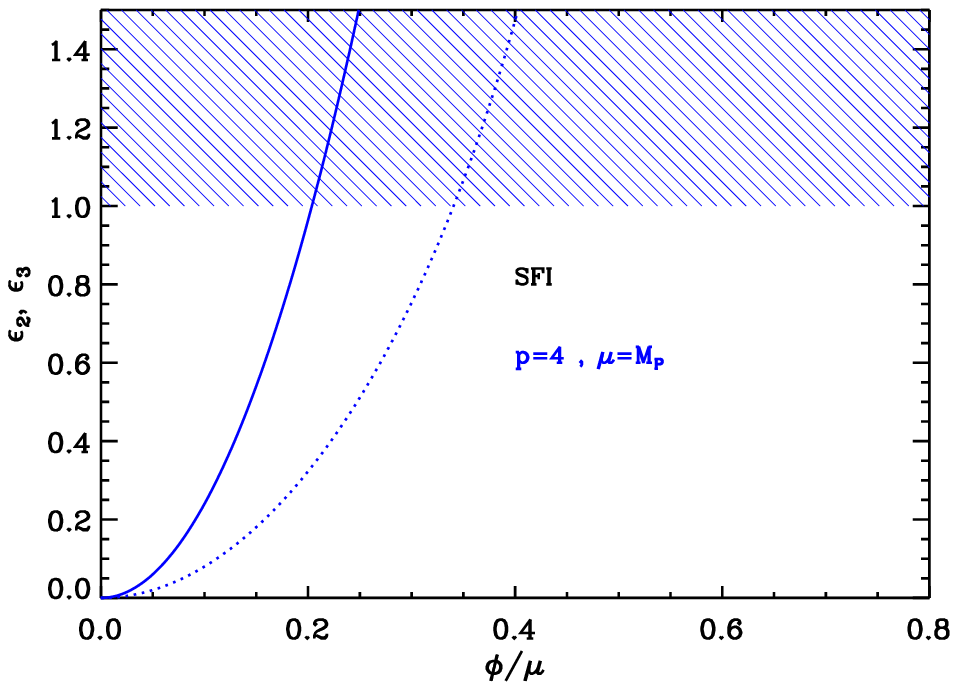}
\caption{Small Field Inflation (SFI) for $p=4$ and $\mu=\Mp$. Upper
  panels: the potential and its logarithm as a function of
  $\phi/\mu$. Bottom left panel: slow-roll parameter $\epsilon _1$,
  the shaded area indicates where inflation stops. Bottom right panel:
  slow-roll parameters $\epsilon_2$ (solid line) and $\epsilon_3$
  (dotted line).}
\label{potsfi}
\end{center}
\end{figure}

This model is proto-typical of inflation occurring at the top of a
flat-enough potential. As such it appears in very different
contexts. It has been introduced in \Refc{Linde:1981mu, Linde:1984cd}
and derived in \Refc{Albrecht:1982wi} in the context of radiatively
induced symmetry breaking. It appears within superstring
models~\cite{Binetruy:1986ss}, low scale symmetry
breaking~\cite{Kinney:1995cc, Kinney:1995ki},
supersymmetry~\cite{Covi:2000gx, Kawasaki:2001as} and
supergravity~\cite{Freese:1990rb, Adams:1992bn, Knox:1993zn,
  Kumekawa:1994gx, Kinney:1995xv, Adams:1996yd, Izawa:1996dv,
  Izawa:1998rh, Buchmuller:2004tm}. It is also obtained in non-linear
sigma models~\cite{Ovrut:1991iw} or using moduli as
inflatons~\cite{Banks:1995dp}. It has been discussed in braneworld
cosmology in \Refcs{Himemoto:2000nd, Sago:2001gi, Chen:2005ad} and is
more recently referred to as ``hilltop inflation'' from
\Refc{Boubekeur:2005zm, Tzirakis:2007bf}. The potential is given by
\begin{equation}
\label{eq:potsfi}
V(\phi) = M^4 \left[1 -\left(\frac{\phi}{\mu}\right)^{p}\right],
\end{equation}
and has two parameters in addition to the overall normalization $M$: a
typical {\vev} $\mu$ and the power index $p$. As this potential can be
associated with very different physical frameworks, $\mu$ can take any
values while $p>0$ for being at the top of a potential (in the small
field limit, namely $\phi\ll\mu$). In particular, we will allow
super-Planckian values for $\mu$ even though, in the supergravity
context, one would require $\mu<\Mp$. Let us stress that
\Eq{eq:potsfi} is defined only in the domain $\phi<\mu$ as one assumes
that the small field potential describes only the field dynamics
during inflation. The equation of state during reheating is thus not
specified by \Eq{eq:potsfi}. Defining
\begin{equation}
x\equiv \dfrac{\phi}{\mu}\,,
\end{equation}
the first three Hubble flow functions read
\begin{equation}
\label{SFI:sr}
\epsilon_1 = \frac{p^2}{2}\left(\frac{\Mp}{\mu }\right)^2 \dfrac{
  x^{2p-2}}{ \left(1- x^p \right)^2}\,, \qquad \epsilon_2 =
2p \left(\frac{\Mp}{\mu }\right)^2 x^{p-2} \frac{p-1 + x^{p}}{\left(
    1 - x^p \right)^2} \,,
\end{equation}
and
\begin{equation}
\begin{aligned}
  \epsilon_3 & = p \left( \frac{\Mp}{\mu }\right)^2 \frac{ x^{p-2}
    \left[2 x^{2p} + (p-1)(p+4)x^p +(p-1)(p-2)\right]}{ \left( 1-
      x^p\right)^2 \left( p-1 + x^p\right)}\,.
\label{SFI:sr3}
\end{aligned}
\end{equation}
They are monotonic functions of the field value but also decreasing
functions of the {\vev} $\mu$. The potential, its logarithm and the
Hubble flow functions are represented in \Fig{potsfi}.

The slow-roll trajectory is obtained by integrating
\Eq{eq:srtrajectory} to get
\begin{equation}
  N-\Nend = \dfrac{1}{2p} \dfrac{\mu^2}{\Mp^2} \left[-x^2 + \xend^2 +
    \dfrac{2}{2-p} \left(x^{2-p} - \xend^{2-p} \right) \right].
\label{eq:sfitraj}
\end{equation}
This equation seems to be well-defined only for $p\ne2$. However, the
particular case $p=2$ can be directly obtained from
\Eqs{eq:srtrajectory} and \eqref{eq:potsfi} to get
\begin{equation}
N-\Nend = \dfrac{1}{4} \dfrac{\mu^2}{\Mp^2} \left[-x^2 + \xend^2 +
    2 \ln\left(\dfrac{x}{\xend}\right) \right].
\label{eq:sfip2traj}
\end{equation}
This expression can also be viewed as the limit of \Eq{eq:sfitraj} for
$p \rightarrow 2$. In general, the trajectory cannot be analytically
inverted to give the field value $x(N)$ but one can find some
analytic form for almost all integer values of $p$ (\eg for $p=1$,
$p=2$, $p=3$, $p=4$, $p=6$) that we do not write down for the sake of
clarity.

{}From the potential \Eq{eq:potsfi}, inflation can stop naturally at
$\epsilon_1(\xend)=1$ with $\xend<1$. This condition gives the
algebraic equation
\begin{equation}
\label{eq:xendsfi}
\xend^p + \dfrac{p }{\sqrt{2}} \dfrac{\Mp}{\mu} \xend^{p-1} =1,
\end{equation}
which cannot be solved analytically in full generality. As for the
trajectory, there are however explicit solutions for almost all
integer values of $p$, the first two being
\begin{equation}
\xendcase{p=1} = 1 - \dfrac{\Mp}{\sqrt{2} \mu}, \qquad
\xendcase{p=2} = \dfrac{\Mp}{\sqrt{2} \mu} \left(-1 + \sqrt{1 + 2
  \dfrac{\mu^2}{\Mp^2}}\right).
\end{equation}

Together with \Eq{eq:phistarlnrrad}, these equations are enough to
allow the determination of the field value $\xstar$ at which the
observable modes crossed the Hubble radius during inflation. This
fixes the value of the parameter $M$ to match the observed amplitude
of the CMB anisotropies at
\begin{equation}
  \dfrac{M^4}{\Mp^4} = 720 \pi^2 p^2 \dfrac{\Mp^2}{\mu^2}
  \dfrac{ \xstar^{2p-2} }{ \left(1-\xstar^p\right)^3 } \dfrac{\Qrms^2}{T^2}\,.
\end{equation}
The reheating consistent slow-roll predictions for the small field
models are represented in \Figs{fig:CMBSFI1} to \ref{fig:CMBSFI4}
for $p=1$, $p=2$ and $p=4$. The $p=1$ case is trivial since one then
has $\epsilon_{2*} = 4 \epsilon_{1*}$. For $p=2$ or $p=4$, one sees that
the reheating temperature is limited from below to fit in the
observable range. For instance, with $p=2$, values of $\mu$ such that
$\mu/\Mp<10$ are clearly disfavored. Let us notice that the relation
$\epsilon_{2*} = 4 \epsilon_{1*}$ is recovered in the limit $\mu/\Mp\gg
1$ whereas one clearly observes a systematic shift in $\nS$ (or
$\epsilon_2$) when $\mu \ll \Mp$. These behaviors can in fact be
understood analytically.

Small field models in the supergravity context are commonly studied in
the limit $\mu \ll \Mp$. In this situation it is possible to find some
approximate solution to both the trajectory and $\xend$. Keeping only
the dominant term in \Eq{eq:xendsfi}, one gets
\begin{equation}
  \xendcase{p\ne1} \simeq \left(\dfrac{\sqrt{2}}{p} \dfrac{\mu}{\Mp}
  \right)^{1/(p-1)},
\end{equation}
the case $p \le 1$ being incompatible with the limit $\mu \ll \Mp$
and the consistency requirement that $\xend < 1$. The small \vev limit
can also be used to invert \Eq{eq:sfitraj}. Assuming $\mu \ll \Mp$ and
$\xend \ll 1$, neglecting the quadratic terms for $p>1$, the
approximate trajectory reads
\begin{equation}
N - \Nend \simeq \dfrac{\mu^2}{\Mp^2}\dfrac{x^{2-p} -
  \xend^{2-p}}{p(2-p)}\,,
\end{equation}
which can be inverted to
\begin{equation}
  x \simeq \left[\xend^{2-p} - \dfrac{\Mp^2}{\mu^2}p(2-p) \left(\Nend
      - N \right)  \right]^{1/(2-p)}.
\label{eq:sfixtrajapprox}
\end{equation}
Notice that far from the end of inflation, \ie $N \ll \Nend$, the
first term can be neglected (for $p > 2$) since $\xend<1$ and $\Mp/\mu
\gg 1$. Defining $\Delta\Nstar = \Nend -\Nstar$, one can now plug this
expression for $\xstar$ into the Hubble flow functions of \Eqs{SFI:sr}
and \eqref{SFI:sr3} to get their observable values:
\begin{equation}
\epsilon_{1*}  \simeq  \frac{p^2}{2}\left(\frac{\Mp }{\mu }\right)^2
\left[\Delta \Nstar p(p-2)\left(\frac{\Mp }{\mu
    }\right)^2\right]^{-\frac{2(p-1)}{p-2}}, \qquad
\epsilon _{2*}  \simeq 
\frac{2}{\Delta \Nstar}\frac{p-1}{p-2}\,, \qquad
\epsilon _{3*}  \simeq  \frac{1}{\Delta \Nstar}.
\label{eq:sfiminivev}
\end{equation}
It is crucial to keep in mind that the above formulas are valid only
in the limit $\mu \ll \Mp$ and $p > 2$. As before, the limiting case
$p\rightarrow 2$ has to be taken with care and, starting with
\Eq{eq:sfip2traj}, one obtains
\begin{equation}
  \epsonestarcase{p=2} = \exp\left(-4 \dfrac{\Mp^2}{\mu^2}
    \Delta\Nstar  \right), \qquad \epstwostarcase{p=2} = 4
  \dfrac{\Mp^2}{\mu^2}\,, \qquad \epsthreestarcase{p=2} =
  6 \epsonestarcase{p=2}.
\label{eq:sfip2minivev}
\end{equation}
Both \Eqs{eq:sfiminivev} and \eqref{eq:sfip2minivev} describes the
observed behavior in \Figs{fig:CMBSFI1} to \ref{fig:CMBSFI4} when
$\mu/\Mp \rightarrow 0$ but they do fail in the intermediate region as
we have discussed in the introduction (see \Fig{fig:sfiapproxintro}).

If the theoretical motivations underlying the
potential~\ref{eq:potsfi} do not require the \vev to be small, one can
similarly derive approximate expressions for the observables in the
limit $\mu/\Mp \gg 1$ (but still with $x < 1$). Defining $\varepsilon
\equiv \Mp/\mu$, one has $\xend(\varepsilon)$ and we can search for
a Taylor expanded solution of \Eq{eq:xendsfi} to get
\begin{equation}
  \xend = 1 -\dfrac{\varepsilon}{\sqrt{2}} + \dfrac{p-1}{4}
  \varepsilon^2 
  + \order{\varepsilon^3}.
\end{equation}
Similarly one can search for a Taylor expanded solution for the
trajectory~\Eq{eq:sfitraj}, plugging in the previous expression for
$\xend$. Doing so yields
\begin{equation}
  \xstar = 1 - \varepsilon \sqrt{\dfrac{1}{2} + 2 \Delta\Nstar} 
  + \order{\varepsilon^2}.
\end{equation}
{}From this, one gets the corresponding Hubble flow functions
\begin{equation}
  \epsilon_{1*} \simeq \frac{1}{4\Delta \Nstar+1}\, \qquad
  \epsilon_{2*} \simeq  4 \epsilon_{1*}, 
  \qquad \epsilon_{3*} \simeq \epsilon_1\, .
\label{eq:sfimaxivev}
\end{equation}
This result is quite remarkable since the observable slow-roll
parameters become $\mu$ and $p$ independent. Performing the same
calculation in the singular case $p \rightarrow 2$ yields exactly the
same result. The spectral index, tensor-to-scalar ratio
and running are immediately obtained from \Eq{eq:sfimaxivev} with $r=16
\epsilon_{1*}$, $\nS -1 \simeq -3r/8$ and $\alpha \simeq -r$. Again,
these expressions match with \Figs{fig:CMBSFI1} to \ref{fig:CMBSFI4} when
$\mu/\Mp \rightarrow \infty$.

\subsection{Intermediate Inflation (II)}
\label{sec:ii}

This model was introduced in \Refcs{Barrow:1990vx, Barrow:1990td,
  Barrow:1993zq, Barrow:2006dh} as an implementation of an equation of
state of the form
\begin{equation}
\label{eq:ii:w}
\rho+p=\gamma\rho^\lambda\, ,
\end{equation}
where $\rho$ stands for the energy density and $p$ the pressure. Both
$\gamma>0$ and $\lambda>1$ are dimensionless constants. As will be
made explicit, this equation of state leads to a scale factor which is
given by $a(t)\propto \exp\left(At^f\right)$ where $0<f<1$. In some
sense the expansion is thus faster than power law but slower than de
Sitter, hence the name of the model. The pure de Sitter case
corresponds to $f=1$. Inserting the Friedmann-Lema\^{\i}tre equation,
$3\Mp^2H^2=\rho$ as well as the equation of state \Eq{eq:ii:w} into
the equation of conservation $\dot{\rho}+3H\left(\rho+p\right)=0$, one
obtains a closed equation for $\rho$ which is solved by
\begin{equation}
\rho= \rhozero \left[3\gamma\left(\lambda-1\right)\ln\left(\frac{a}
  {\azero}\right)\right]^{1/(1-\lambda)},
\end{equation}
where $\rhozero$ and $\azero$ are positive constants. Making use of
the Friedmann-Lema\^{\i}tre equation again, one deduces the behavior
for $a$,
\begin{equation}
\ln\left(\frac{a}
{\azero}\right)=3^{\lambda/(1-2\lambda)}\gamma^{1/(1-2\lambda)}
\frac{\left(\lambda-\frac{1}{2}\right)^{(1-\lambda)/(1-2\lambda)}}
{\lambda-1}
\left(\dfrac{t}{\tzero}\right)^{(1-\lambda)/(1-2\lambda)}\, ,
\end{equation}
\ie the announced form  $a(t)\propto \exp\left(At^f\right)$, with 
$f=2(1-\lambda)/(1-2\lambda)$. Since $\lambda>1$, this means that 
$0<f<1$. Then, one can notice that it is possible to reinterpret the 
matter source as that of a scalar field with the potential $V(\phi)$ 
given by
\begin{equation}
\begin{aligned}
V(\phi) & = 3A^2f^2\Mp^4\left[\frac{\phi-\phizero}
{\Mp\sqrt{8A\left(f^{-1}-1\right)}}\right]^{4\left(1-1/f\right)}
\\&
-\Mp^4Af\left(1-f\right)\left[\frac{\phi-\phizero}
{\Mp\sqrt{8A\left(f^{-1}-1\right)}}\right]^{2-4/f}.
\end{aligned}
\end{equation}
Indeed, starting from this potential, the Klein-Gordon equation with
$H=Aft^{f-1}$, has an exact non-trivial solution given by
\begin{equation}
\label{eq:ii:traj1}
\phi=\phizero+\Mp\sqrt{8A\left(f^{-1}-1\right)}\,
\left(\dfrac{t}{\tzero}\right) ^{f/2}\, .
\end{equation}
It is then straightforward to calculate $\rho=\dot{\phi}^2/2+V$ and  
$p=\dot{\phi}^2/2-V$, and to show that they satisfy the equation of 
state \Eq{eq:ii:w}. The potential can be recast in the form
\begin{equation}
\label{eq:ii:pot}
V(\phi)=M^4\left(\frac{\phi-\phizero}{\Mp}\right)^{-\beta}
-M^4\frac{\beta ^2}{6}\left(\frac{\phi-\phizero}
{\Mp}\right)^{-\beta-2}\, ,
\end{equation}
with $\beta =4(1/f-1)$. The constraint $0<f<1$ means that $\beta > 0$. 
Defining
\begin{equation}
x\equiv \dfrac{\phi-\phizero}{\Mp}\,,
\end{equation}
it is shown below that the model predictions do not depend on
$\phizero$. Therefore Intermediate Inflation is a priori a one
parameter family of models, but as explained below, one needs an extra
parameter $\xend$ specifying the field value at which an unspecified
mechanism is triggered to end of inflation. It is thus a two
parameters model.

This potential appears in the earlier work of \Refc{Barrow:1988yc} as
a solution for a cosmological model containing a string creation term.
It is also discussed in the context of tachyon fields in
\Refcs{delCampo:2009ma, Farajollahi:2011fu}. Warm intermediate
inflation was considered in \Refcs{delCampo:2009xi, delCampo:2008fc},
intermediate inflation within a Gauss-Bonnet braneworld was studied in
\Refc{Herrera:2010vv}, and with Jordan-Brans-Dicke theory in
\Refcs{Cid:1900zz, Cid:2012kw}.

\begin{figure}
\begin{center}
\includegraphics[width=\wdblefig]{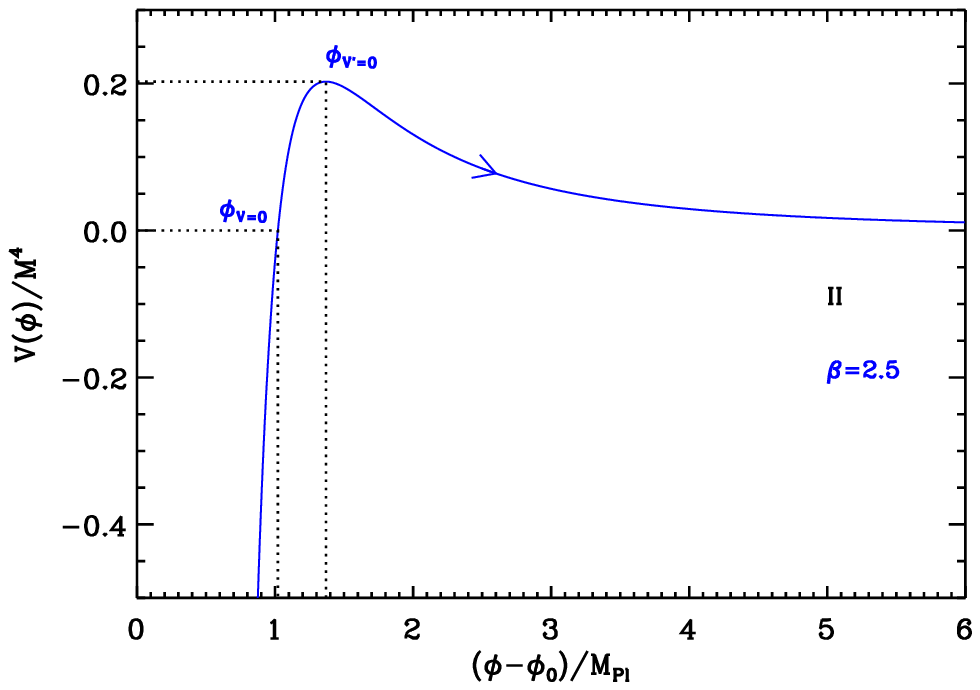}
\includegraphics[width=\wdblefig]{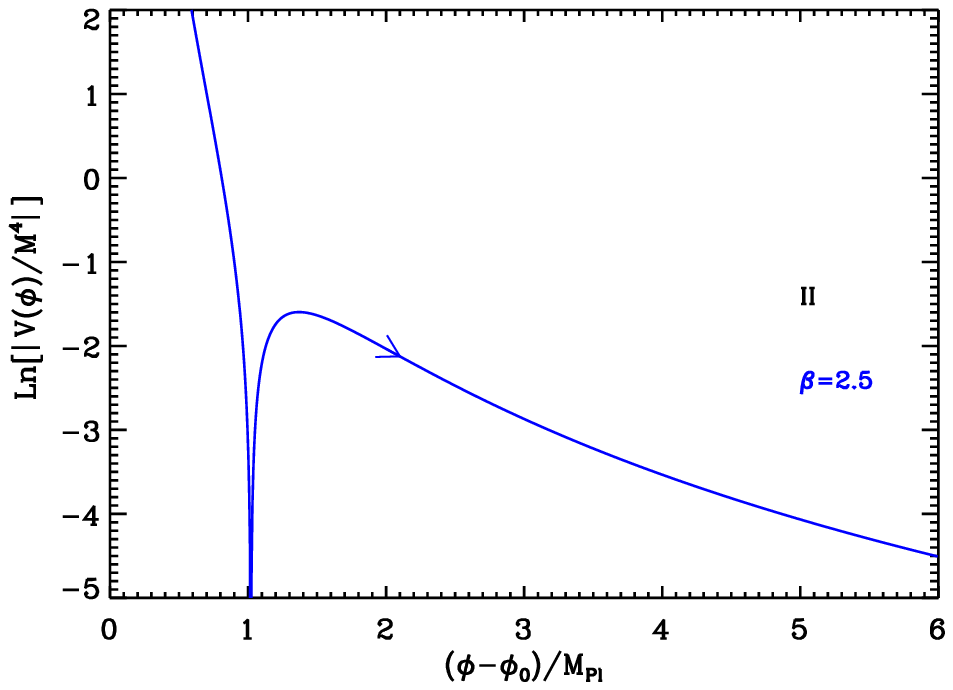}
\includegraphics[width=\wdblefig]{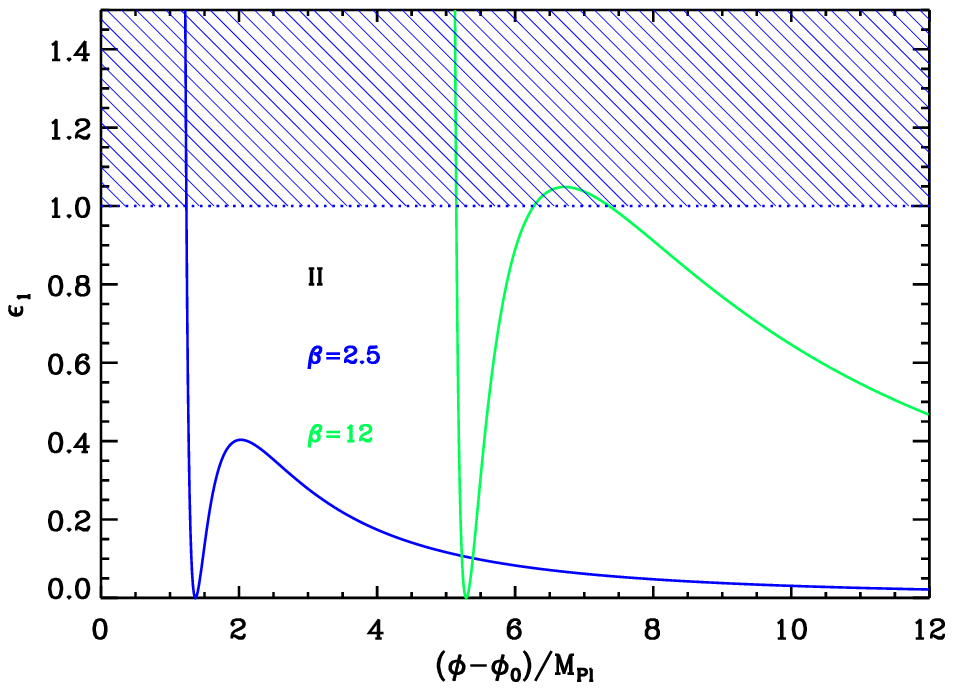}
\includegraphics[width=\wdblefig]{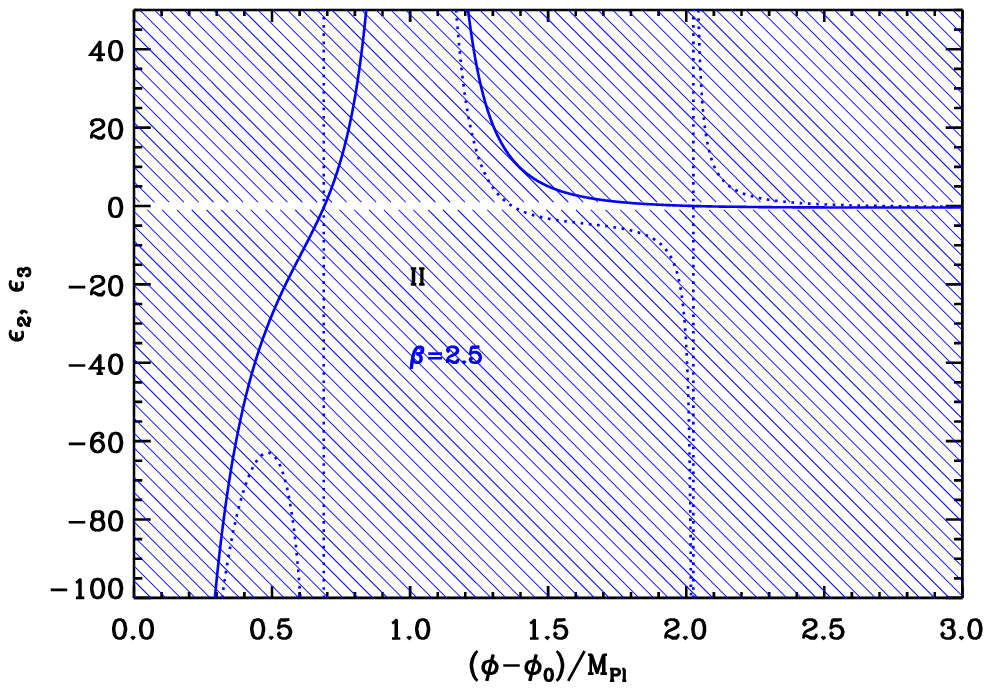}
\caption{Intermediate Inflation (II).  Upper panels: the potential and
  its logarithm for $\beta=2.5$. Bottom left panel: slow-roll
  parameter $\epsilon _1$ for a potential with $\beta=2.5$ and
  $\beta=12$. The position of the maximum of $\epsilon_1$ with respect
  to one depends on $\beta$. The shaded area indicates where inflation
  stops.. Bottom right panel: slow-roll parameters $\epsilon_2$ (solid
  line) and $\epsilon_3$ (dotted line) for a potential with
  $\beta=2.5$.}
\label{potII}
\end{center}
\end{figure}

The potential \eqref{eq:ii:pot}, as well as its logarithm, are
displayed in \Fig{potII}. It is positive definite for $x>\xVzero
\equiv\beta /\sqrt{6}$. Therefore, one must restrict the inflaton \vev
to lie beyond this value. The potential increases with $x$, reaches a
maximum at $x_{V'=0} \equiv\sqrt{\beta (\beta +2)/6}$, then decreases
with $x$ to asymptotically vanish when $x$ goes to
infinity. Therefore, a priori, two regimes of inflation exist. Either
inflation proceeds at $x<\xdVzero$ from the right to the left, either
it proceeds at $x>\xdVzero$ from the left to the right. However, in
\Eq{eq:ii:traj1}, one can see that the inflaton \vev has to increase
with time. Therefore only the branch $x>\xdVzero$ can produce an
equation of state of the form of \Eq{eq:ii:w}, which is where the
model will be studied in the following.

Let us now turn to the slow-roll parameters. The first three Hubble
flow functions in the slow-roll approximation are given by
\begin{equation}
\epsilon_1 =\dfrac{1}{2}\left[\dfrac{\beta ^2(\beta+2)-6\beta x^2}
  {-\beta ^2x+6x^3}\right]^2, \qquad
\epsilon_2 = \dfrac{-2\beta
  x^4+\dfrac{\beta^2}{3}\left(2\beta+6\right)x^2
  -\dfrac{\beta^4}{18}\left(\beta+2\right)}{\left(x^3 - \dfrac{\beta^2
    x}{6} \right)^2} \, ,
\end{equation}
and
\begin{equation}
\begin{aligned}
\epsilon_3 & = \dfrac{ \beta
  \left[6x^2-\beta\left(2+\beta\right)\right]
  \left[\dfrac{\beta^5}{18}\left(2+\beta\right) - \beta^3 \left(
    2+\beta\right) x^2 +6 \beta\left(4+\beta \right) x^4- 12x^6
    \right] }{ \left(x^3-\dfrac{\beta^2}{6}x\right)^2
  \left[\beta^3\left(\beta+2\right)-12\beta\left(\beta+3\right)x^2
    +36x^4\right]}\, .
\end{aligned}
\end{equation}
They are displayed in \Fig{potII}. The first slow-roll parameter 
diverges where the potential vanishes at $x_{V=0}$, decreases from here 
and vanishes at the maximum of the potential $\xdVzero$. Then it 
increases again, reaches a local maximum at $\xepsoneMax$, and 
decreases to asymptotically vanish when $x$ goes to infinity. The 
location $\xepsoneMax$ is given by
\begin{equation}
\label{eq:ii:epsonemax}
\xepsoneMax = \sqrt{\frac{\beta }{2}
\left(1+\frac{\beta }{3}+\sqrt{1+\frac{4\beta }{9}}\right)}\,.
\end{equation}
At this point, the maximum value of $\epsilon_1$ is
\begin{equation}
  \epsilon_1^{\max}
  =\frac{\beta }{9}
  \frac{\left(1+3\sqrt{1+4\beta /9}\right)^2}
  {\left(1+\sqrt{1+4\beta/9}\right)^2\left(1+\beta/3+\sqrt{1+4\beta 
  /9}\right)}\,.
\end{equation}
If $\beta<9/2\left(1+\sqrt{2}\right)\simeq 10.86$, this maximum value
is smaller than one. In this case inflation cannot stop by slow-roll
violation in the decreasing branch of the potential and an extra
parameter $\xend$ must be added to the model to specify the location
where another mechanism such as \eg tachyonic instability could
trigger the end of inflation. If
$\beta>9/2\left(1+\sqrt{2}\right)\simeq 10.86$, the local maximum
value of $\epsilon_1$ is higher than one and in the decreasing branch
of the potential, either inflation takes place between $\xdVzero$ and
the first solution of $\epsilon_1=1$, either it takes place between
the second solution of $\epsilon_1=1$ and $x=\infty$. As will be shown
below, only the latter case is consistent with the exact trajectory
\Eq{eq:ii:traj1} which allows for an equation of state of the form of
\Eq{eq:ii:w}.

The slow-roll trajectory of the model can be obtained from
\Eq{eq:srtrajectory}. However, as already mentioned, a non-trivial
and exact field evolution is given by \Eq{eq:ii:traj1}.  Written in
terms of the number of \efolds $N-\Nzero = \ln(a/\azero)=A(t^f-
\tzero^f)$, one obtains
\begin{equation}
\label{eq:ii:traj}
x=\sqrt{\xend^2+
2\beta\left(N-\Nend\right)}\, .
\end{equation}
This expression is exact and does not involve any approximations. It
can be compared to slow-roll trajectory which reads
\begin{equation}
\begin{aligned}
\label{eq:ii:trajSR}
\Nend-N & = \frac{1}{2\beta}\left(\xend^2-x^2\right)
 +\frac{1}{6}\ln \left[\xend^2
-\frac{\beta\left(\beta+2\right)}{6}\right]
 -\ln \left[x^2-\frac{\beta\left(\beta+2\right)}{6}\right] ,
\end{aligned}
\end{equation}
where $\Nend$ is the number of \efolds at the end of inflation and $N$
is the number of \efolds at some point when the scaled field \vev is
$x$. As mentioned above, the slow-roll trajectory should match the
exact one in the decreasing branch of the potential. For $x
\rightarrow \infty$, one can neglect the logarithmic terms in
\Eq{eq:ii:trajSR} and one indeed recovers \Eq{eq:ii:traj}. This is
expected since in this limit, the slow-roll parameters all go to zero
and the slow-roll approximation becomes increasingly accurate. As a
result, the domain of validity lies at $x\gg \xdVzero$, \ie between
the second solution of $\epsilon_1=1$ and $x=\infty$ and inflation
cannot stop by slow-roll violation. This justifies the need of the
extra-parameter $\xend$. This parameter is thus constrained to $\xend
> \xdVzero$ and should be large enough to allow for a sufficient
number of \efolding. In order to get $\Nend-\Nini$ \efolds, making use
of \Eq{eq:ii:traj}, one gets
\begin{equation}
\xend =\sqrt{\xini^2+2 \beta (\Nend-\Nini)}\, .
\end{equation}
If $\beta>9/2\left(1+\sqrt{2}\right)\simeq 10.86$, $\xini$ is bounded
from below by the highest solution of the equation
$\epsilon_1=1$. This equation admits three solutions which, from the
smallest to the biggest, are given by
\begin{align}
\xepsoneOneA & = -\frac{\beta}{3\sqrt{2}}
+\frac{\sqrt{2}}{3}\frac{\beta^{4/3}}
{\sqrt[3]{9+2\beta+i\sqrt{-81-36\beta+4\beta^2}}} \nonumber \\ &
+\frac{\beta^{2/3}}{3\sqrt{2}} \sqrt[3]{9+2\beta +
  i\sqrt{-81-36\beta+4\beta^2}}\, ,\\
\xepsoneOneBC &=\frac{\beta}{3\sqrt{2}} + \frac{1\mp
  i\sqrt{3}}{3\sqrt{2}} \frac{\beta^{4/3}} {\sqrt[3]{9+2 \beta + i
    \sqrt{-81-36 \beta + 4\beta^2}}} \nonumber \\ & + \left(1\pm
i\sqrt{3} \right) \frac{\beta^{2/3}}{6\sqrt{2}}
\sqrt[3]{9+2\beta+i\sqrt{-81-36\beta+4\beta^2}}\, .
\end{align}
The first solution is located below the maximum of the potential 
$\xepsoneOneA<\xdVzero$, while the two others are located beyond it 
$\xepsoneOneBC>\xdVzero$. Using the larger solution as a lower bound 
for $\xini$, one gets
\begin{equation}
\label{eq:II:prior1}
\xend >\sqrt{\left(\xepsoneOneC \right)^2+2\beta (\Nend-\Nini)}\, .
\end{equation}

\begin{figure}
\begin{center}
\includegraphics[width=\wsingfig]{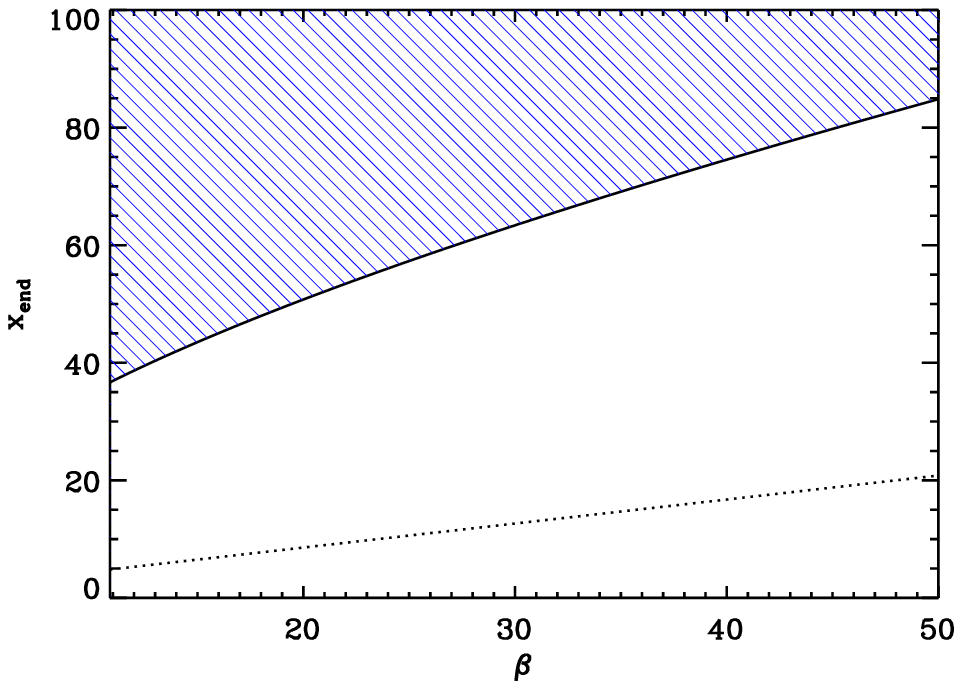}
\caption{Prior space on $\xend$ derived from \Eq{eq:II:prior1} with
  $\Nend-\Nini=60$, as a function of $\beta
  >9/2\left(1+\sqrt{2}\right)$ (black solid line). The black dotted 
  line corresponds to $x_{V'=0}$. For $\beta <
  9/2\left(1+\sqrt{2}\right)$, provided some fine-tuning on the initial
  conditions, $\xend$ can take any values. The dashed area corresponds
  to parameters for the model which produce at least the required
  number of \efolds.}
\label{II:prior}
\end{center}
\end{figure}

If $\beta<9/2\left(1+\sqrt{2}\right)$, only one solution to
$\epsilon_1=1$ exists,
\begin{equation}
\xepsoneOne = -\frac{\beta}{3\sqrt{2}}
+\frac{\sqrt{2}}{3}\frac{\beta^{4/3}}
{\sqrt[3]{9+2\beta+\sqrt{81+36\beta-4\beta^2}}}
+\frac{\beta^{2/3}}{3\sqrt{2}}
\sqrt[3]{9+2\beta+\sqrt{81+36\beta-4\beta^2}}\, ,
\end{equation}
which is located below the maximum of the potential
$\xepsoneOneA<\xdVzero$. In principle $\xini$ is now only bounded from
below by $\xdVzero$ and one can check from \Eq{eq:ii:trajSR} that the
total number of \efolds diverges close to $\xdVzero$. As a result,
provided $\xini$ is fine-tuned to the top of the potential, there is
no bound on $\xend$. The prior space described by these relations is
displayed in \Fig{II:prior}.

According to the previous discussion, the observable field value, at
which the pivot mode crossed the Hubble radius during inflation, is
such that $\xstar \gg 1$. In this limit, it is possible to approximate
the slow-roll parameters at Hubble crossing with
\begin{equation}
\label{eq:ii:epsstar}
\epsilon_1^*\simeq\frac{\beta^2}{2\xstar^2}\, ,\qquad
\epsilon_2^*\simeq\epsilon_3^*\simeq-\frac{2\beta}{2\xstar^2}\, ,
\end{equation}
hence
\begin{equation}
\label{eq:ii:rnsas}
r\simeq\frac{8\beta^2}{\xstar^2}\, ,\qquad
\nS-1\simeq\frac{\beta\left(2-\beta\right)}{\xstar^2}\, ,\qquad
\alphaS=\frac{2\beta^2\left(\beta-2\right)}{\xstar^4}.
\end{equation}
These estimates match with those of \Refc{Barrow:2006dh}. Finally, the
parameter $M$ is obtained from the amplitude of the CMB anisotropies
\begin{equation}
\left(\frac{M}{\Mp}\right)^4 = 720\pi^2
\left[\frac{\beta^2\left(\beta+2\right)}{6}-\beta \xstar^2\right]^2
\left(\xstar^3-\frac{\beta^2\xstar}{6}\right)^{-2}
\left(\xstar^{-\beta}-\frac{\beta^2}{6}\xstar^{-\beta-2}\right)
\frac{\Qrms^2}{T^2}\, .
\end{equation}
In the $\xstar\gg 1$ limit, this gives
\begin{equation}
\dfrac{M^4}{\Mp^4} \simeq 720\pi^2\beta^2\xstar^{-2-\beta}
\dfrac{\Qrms^2}{T^2}\,,
\end{equation}
which yields $M/\Mp\lesssim 10^{-2}$.

The reheating consistent slow-roll predictions for the intermediate
inflation models are displayed in \Fig{fig:CMBII}, for different
values of $\beta>0$, and for $\xend$ describing the prior space
displayed in \Fig{II:prior}. The reheating equation of state parameter
$\wrehbar$ has been taken to $0$ but since there is no potential
minimum around which the inflaton field can oscillate at the end of
inflation, this parameter is a priori unspecified and can take
different values. In any case the reheating temperature is fully
degenerate with the parameter $\xend$, and therefore these two
parameters cannot be constrained independently.  However one can see
that $\xend$ is clearly limited from below as expected. The black
solid lines represent the locus of the points such that
$\epsilon_1^*=-\beta/4\epsilon_2^*$, or equivalently, $\nS-1 =
\left(1/\beta-1/2\right)r/4$, these consistency relations arising from
\Eqs{eq:ii:epsstar}. One can check that they provide a good
qualitative description of the model predictions. In particular, they
explain why, for $\beta<2$, one has a blue tilt $\nS>1$.

\subsection{K\"ahler Moduli Inflation II (KMIII)}
\label{sec:kmiii}

\subsubsection{Theoretical Justifications}
\label{subsubsec:theorykmiii}

These models are string motivated scenarios. They arise in the context
of type IIB string theory via Calabi-Yau flux compactification. They
have been derived and studied in \Refcs{Conlon:2005jm,Bond:2006nc,
  Yang:2008ns,Krippendorf:2009zza,BlancoPillado:2009nw,
  Kawasaki:2010ux,Lee:2010tk}, and a two-field generalization of this
model has been investigated in \Refcs{Bond:2006nc,
  Yang:2008ns,Krippendorf:2009zza,BlancoPillado:2009nw,
  Kawasaki:2010ux}. They can be understood in the context of
supergravity, viewed as an effective theory. In this framework, one
starts with the following superpotential for the moduli $T_i$
\begin{equation}
W=W_0+\sum_{i=2}^n A_i\ee ^{-a_iT_i},
\end{equation}
where $a_i=2\pi/(\gstrings N)$, $N$ being a positive integer (not to
be confused with the \efold number), $\gstrings$ the string coupling,
and $W_0$ and $A_i$ are model dependent constants. The K\"ahler
potential can be written as
\begin{equation}
K=-2\Mp^2\ln \left(\frac{\calV}{2\ells^6}+\frac{\xi}{2}\right),
\end{equation}
where the constant $\xi$ is given by
$\xi=-\zeta(3)\chi(M)/[2(2\pi)^2]$, $\chi(M)$ being the Euler
characteristic of the compactification manifold. The quantity $\calV$
represents the overall volume of the Calabi-Yau manifold and can be
taken to be
\begin{equation}
\calV=\frac{\gamma\ells^6}{2\sqrt{2}}
\left[\left(T_1+T_1^{\dagger}\right)^{3/2}
-\sum_{i=2}^n\lambda_i\left(T_i+T_i^{\dagger}\right)^{3/2}\right],
\end{equation}
where $\gamma$ and $\lambda_i$ are positive constants and depend on
the details of the model. From the expression of the K\"ahler and
superpotentials, it is then straightforward to calculate the
corresponding F-term potential which is a relatively complex
expression that can be found in \Refc{BlancoPillado:2009nw}. If,
however, one consider the limit $\calV\gg 1$ (and $T_1\gg T_i$), then
the F-term simplifies a lot and gives rise to the following equation
\begin{equation}
\label{eq:interpotkmiii}
V(\tau_i)\simeq \frac{3\xi W_0^2}{4\Mp^2\calV_\us^3}+\sum_{i=2}^n
\left[\frac{4W_0a_iA_i}{\Mp^2\calV_\us^2}\tau_i\ee^{-a_i\tau_i}
\cos \left(a_i\theta_i\right)
+\frac{8\left(a_iA_i\right)^2}{3\Mp^2\gamma\lambda_i\calV_\us
}\sqrt{\tau_i}\ee^{-2a_i\tau_i}
\right],
\end{equation}
where we have written $T_i=\tau_i+i\theta_i$ and $\calV_\us\equiv
\calV/\ells^6$. We see that all the constants introduced before,
namely $a_i$, $A_i$, $W_0$, $\xi$, $\gamma$ and $\lambda_i$
participate to the expression of the potential. From
\Eq{eq:interpotkmiii}, solving $\partial V/\partial \tau_i=0$, one can
estimate the value of each $\tau_i$ at the global minimum of the
potential. In the following, we denote this quantity by
$\tau_i^{_\mathrm{min}}$. Then, one can also calculate the value of
the potential at this minimum. One finds [where, as usual, we have
taken $\cos\left(a_i\theta_i\right)=-1$]
\begin{equation}
V_{\min}\simeq \frac{3\xi W_0^2}{4\Mp^2\calV_\us^3}
-\frac{3W_0^2\gamma}{2\Mp^2\calV_\us^3}\sum_{i=2}^n\frac{\lambda_i}{a_i^{3/2}}
\left(a_i\tau_i^{_\mathrm{min}}\right)^{3/2}.
\end{equation}
As a consequence, if for one of the fields, say $\tau_n$, one has
$\left(\lambda_n/a_n^{3/2}\right)/\left[\sum_{i=2}^{n-1}(\lambda_i/a_i^{3/2})\right]
\ll 1$, then the value of $V_{\min}$ is not modified even if one
displaces $\tau_n$ from $\tau_n^{_\mathrm{min}}$. In other words, we
have an inflationary valley along the $\tau_n$ direction and one can
use it to produce inflation. In that case, the potential can be
re-written as
\begin{equation}
  V(\tau_n)\simeq \frac{BW_0^2}{\Mp^2\calV_\us^3}
  -\frac{4W_0a_nA_n}{\Mp^2\calV_\us^2}\tau_n\ee^{-a_n\tau_n},
\end{equation}
where the second exponential in \Eq{eq:interpotkmiii} has been
neglected, thanks to the condition $a_n\tau_n\gg 1$ and $B$ is a
constant that includes the constant term in \Eq{eq:interpotkmiii} as
well as the contributions of the other fields at their minimum, \ie
$B=3\xi/4+\cdots $. It is important to notice that the assumption of
large volume translates into a condition on the \vev of $\tau_n$. The
above potential is of the form of the toy model studied as ``K\"ahler
Moduli Inflation I (KMII)'' in
\sectionc{sec:kmii}. The field is however not canonically normalized
since it is a modulus. It is therefore necessary to first canonically
normalize it and, then, re-derive the corresponding potential. Using
the form of the K\"ahler potential given above, denoting by $\phi$ the
canonical field, one arrives at
\begin{equation}
  \tau_n=\left(\frac{3\calV_\us}{4\gamma \lambda_n}\right)^{2/3}
\left(\frac{\phi}{\Mp}\right)^{4/3}.
\end{equation}
As a consequence, the final form of the inflaton's potential is given by
\begin{equation}
  V(\phi)=\frac{BW_0^2}{\Mp^2\calV_\us^3}
  -\frac{4W_0a_nA_n}{\Mp^2\calV_\us^2}
  \left(\frac{3\calV_\us}{4\gamma \lambda_n}\right)^{2/3}
  \left(\frac{\phi}{\Mp}\right)^{4/3}
  \exp\left[-a_n\left(\frac{3\calV_\us}{4\gamma
\lambda_n}\right)^{2/3}
    \left(\frac{\phi}{\Mp}\right)^{4/3}\right].
\end{equation}
Let us now see what are the typical values that the parameters
appearing in the above potential can take. As already mentioned, the
quantity $\calV_\us$ represents the Calabi-Yau volume and is supposed
to be such that $\calV_\us\gg 1$ or $\calV\gg \ells^6$. In
\Refc{Lee:2010tk} the typical value $\calV_\us\simeq 3\times 10^{6}$
was chosen. The parameter $A_n$ depends on the complex structure
moduli and is typically of order
$\mathcal{O}\left(\ells^3\right)$. This is also the case for
$W_0$. One has $a_n=2\pi/N$, where $N$ is a positive integer (for
$D3$-brane instantons, one has $N=1$).  The dimensionless parameter
$\lambda_n$ is model dependent but is considered to be of order
$\mathcal{O}(1)$. The quantity $\xi=\zeta(3)\chi
/\left[2(2\pi)^3\right]$, where $\chi$ is the Euler number of the
internal Calabi-Yau space, is also of order $\mathcal{O}(1)$ as well
as the coefficient $\gamma$. This means that $B$ is of order
$\mathcal{O}(1)$.

\subsubsection{Slow-Roll Analysis}
\label{subsubsec:srkmiii}

\begin{figure}
\begin{center}
\includegraphics[width=\wdblefig]{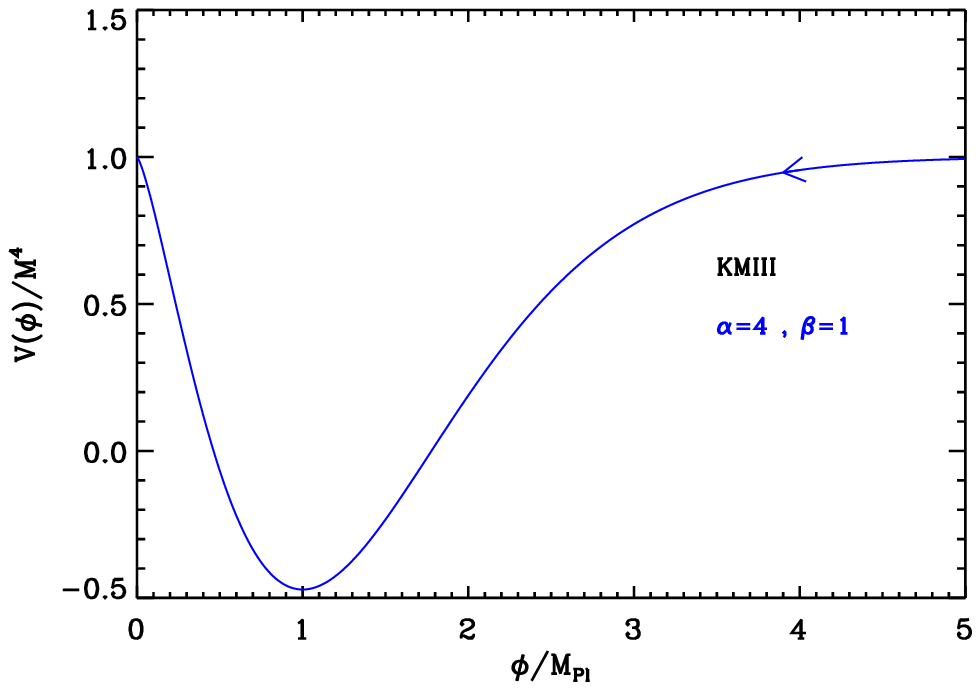}
\includegraphics[width=\wdblefig]{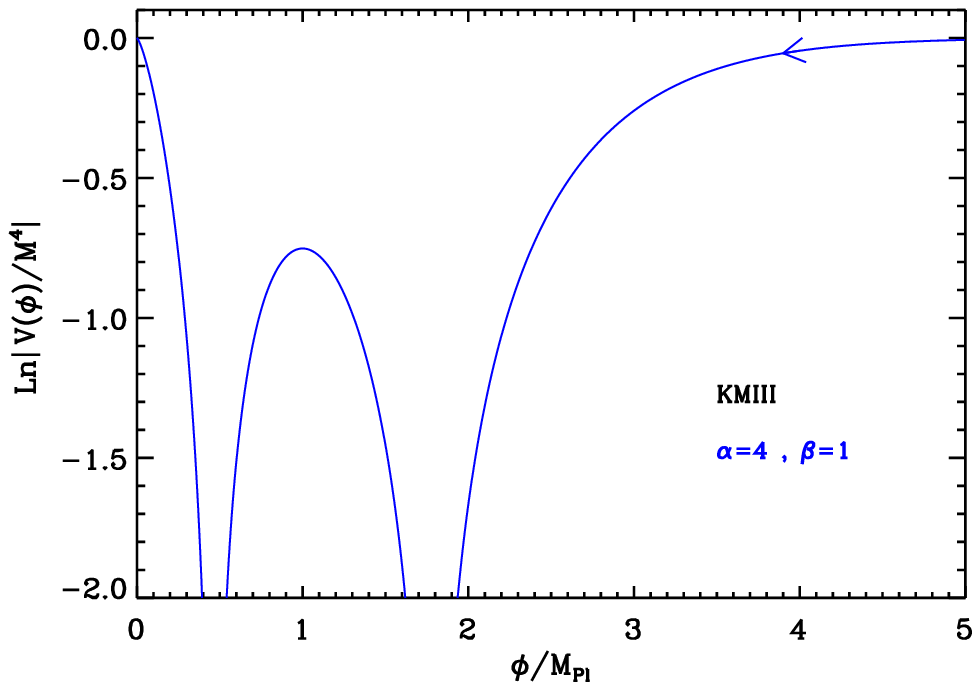}
\includegraphics[width=\wdblefig]{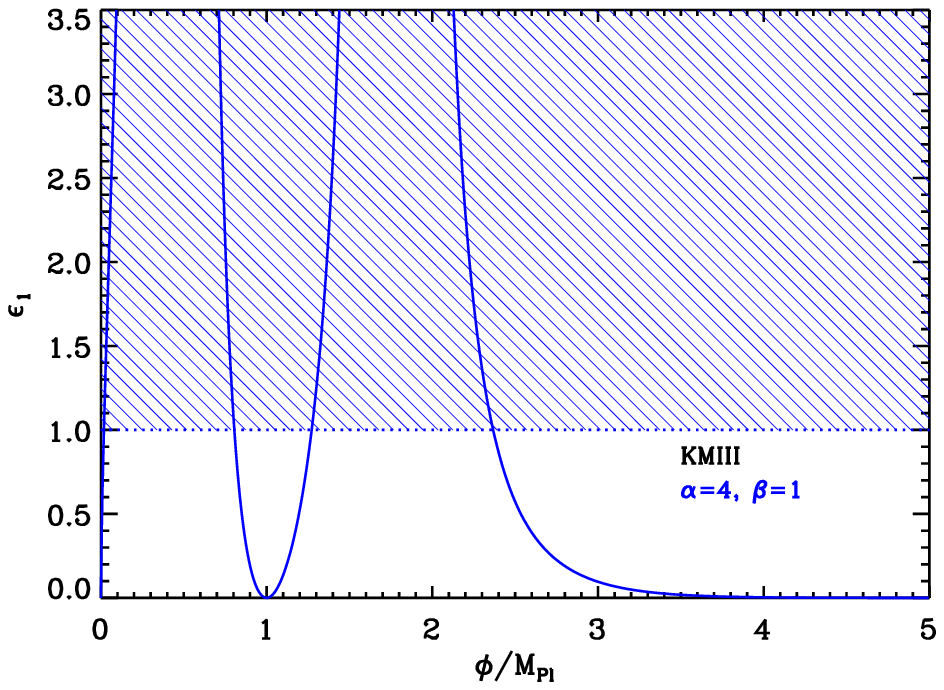}
\includegraphics[width=\wdblefig]{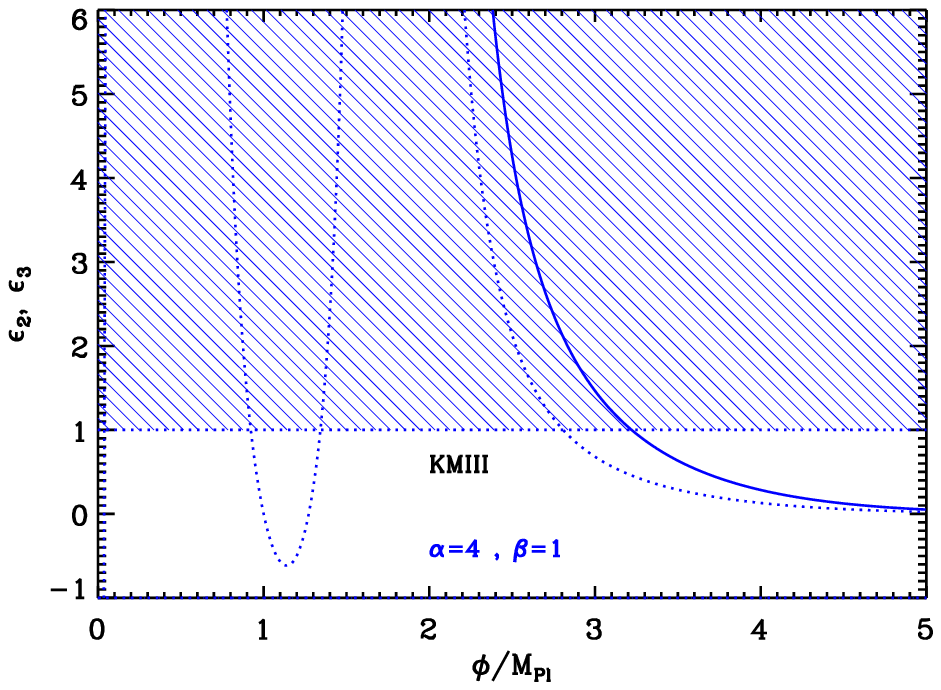}
\caption{Top left panel: K\"ahler moduli inflation II (KMIII)
  potential for $\alpha=4$ and $\beta=1$. These parameters are
  not physical but they are used for display convenience.
  Top right panel: logarithm
  of the potential for the same value of $\alpha$ and $\beta$. Bottom
  left panel: slow-roll parameter $\epsilon _1$ for a potential with
  $\alpha=4$ and $\beta=1$. The shaded area indicates the breakdown of the slow-roll
  inflation (strictly speaking when the acceleration stops). Bottom
  right panel: slow-roll parameters $\epsilon _2$ (solid line) and
  $\epsilon _3$ (dotted line) for $\alpha=4$ and $\beta=1$.}
\label{potkmiii}
\end{center}
\end{figure}

We now study the inflationary scenario based on the potential
derived above. Re-writing $V(\phi)$ in a more convenient way, we have
\begin{equation}
\label{eq:kmiii:pot}
V(\phi)=M^4\left[1-\alpha \left(\frac{\phi}{\Mp}\right)^{4/3}
\ee^{-\beta(\phi/\Mp)^{4/3}}\right]\, .
\end{equation}
where we have defined the parameters $M$, $\alpha$ and $\beta$ by
\begin{equation}
\begin{aligned}
\label{eq:kmiii:paramdef}
M^4=\frac{BW_0^2}{\Mp^2\calV_\us^3}\, , \quad
\alpha=\frac{16\calV_\us a_n}{3}\frac{A_n}{W_0}
\left(\frac{3\calV_\us}{4\gamma\lambda_n}\right)^{2/3},
\quad
\beta=a_n\left(\frac{3\calV_\us}{4\gamma\lambda_n}\right)^{2/3}.
\end{aligned}
\end{equation}
Making use of the typical orders of magnitude for the various
quantities entering these expression, one sees that
\begin{equation}
\begin{aligned}
\label{eq:kmiii:priors}
\alpha &= \mathcal{O}\left(\mathcal{V}_s^{5/3}\right),\qquad
\beta &= \mathcal{O}\left(\mathcal{V}_s^{2/3}\right),
\end{aligned}
\end{equation}
with $\calV_\us\gg 1$.

The potential~(\ref{eq:kmiii:pot}) and its logarithm are displayed in
\Fig{potkmiii}. $V(\phi)$ decreases from $V/M^4=1$ at $\phi=0$,
reaches a minimum at $\phi/\Mp=\beta^{-3/4}$, and then increases to
the asymptotic value $V/M^4=1$ when $\phi/\Mp\rightarrow +\infty$.
However, since the potential is derived under the large field
assumption, only the increasing branch of the potential is relevant.
Inflation proceeds from the right to the left along this branch.  The
minimum value of the potential at $\phi=\Mp\beta^{-3/4}$ is given by
$V_{\mathrm{min}}=M^4\left[1-\alpha/\left(\beta e\right)\right]$.
Therefore, if one wants the potential to be definite positive
everywhere, one must have $\alpha/\beta<e$. However, from
\Eq{eq:kmiii:priors}, we see that this condition cannot be satisfied
since $\alpha/\beta=\mathcal{O}(\calV_\us)\gg 1$. This means
that the potential necessarily vanishes at some point. In the
increasing branch of the potential, this occurs for a \vev given by
\begin{equation}
\xVzero \equiv \frac{\phiVzero}{\Mp}=\left[-\frac{1}{\beta}
\Lambert{-1}
\left(-\frac{\beta}{\alpha}\right)\right]^ {3/4}.
\end{equation}
Anyway, since the potential~(\ref{eq:kmiii:pot}) is only valid in the
large field region, this criterion does not play an important role in
what follows.

Let us now calculate the three first Hubble flow parameters. Defining
$x\equiv\phi/\Mp$, they are given by
\begin{equation}
\label{eq:kmiii:eps1}
\epsilon_1 = \frac{8\alpha ^2}{9} x^{2/3} \ee^{-2\beta x^{4/3}} \left(
\dfrac{1-\beta x^{4/3}}{1-\alpha x^{4/3} \ee^{-\beta x^{4/3}}}
\right)^2,
\end{equation}
\begin{equation}
\label{eq:kmiii:eps2}
\epsilon_2 = \frac{8\alpha}{9} x^{-2/3} \ee^{-2\beta x^{4/3}}
\dfrac{3\alpha x^{4/3} +\alpha\beta x^{8/3} +\ee^{\beta x^{4/3}}
  \left(1-9\beta x^{4/3} +4\beta^2 x^{8/3}\right)}{\left(1-\alpha
  x^{4/3} \ee^{-\beta x^{4/3}}\right)^{2}} ,
\end{equation}
and
\begin{equation}
\begin{aligned}
\label{eq:kmiii:eps3}
\epsilon_3 &=
\Biggl\{8\alpha\left(1-\beta x^{4/3}\right)
\biggr[\alpha^2 {x}^{8/3}
\left(9+\beta {x}^{4/3}\right)-
2\alpha\ee^{\beta x^{4/3}}{x}^{4/3}
\biggl(-4+19\beta {x}^{4/3}-9\beta^2{x}^{8/3}
 \\
&  +4\beta^3{x}^4\biggr)-\ee^{2\beta {x}^{4/3}}
\left(1+11\beta{x}^{4/3} -30\beta^2 {x}^{8/3}
  +8\beta^3{x}^4\right)\biggr]\Biggr\}
\Biggl\{9{x}^{2/3}
\left(\ee^{\beta x^{4/3}}-\alpha{x}^{4/3}\right)^2
\\ &  \times
\left[\alpha{x}^{4/3}\left(3+\beta{x}^{4/3}\right)
  +\ee^{\beta x^{4/3}}
  \left(1-9\beta {x}^{4/3}+4\beta^2{x}^{8/3}\right)\right]\Biggr\}^{-1} .
\end{aligned}
\end{equation}

\begin{figure}
\begin{center}
  \includegraphics[width=\wsingfig]{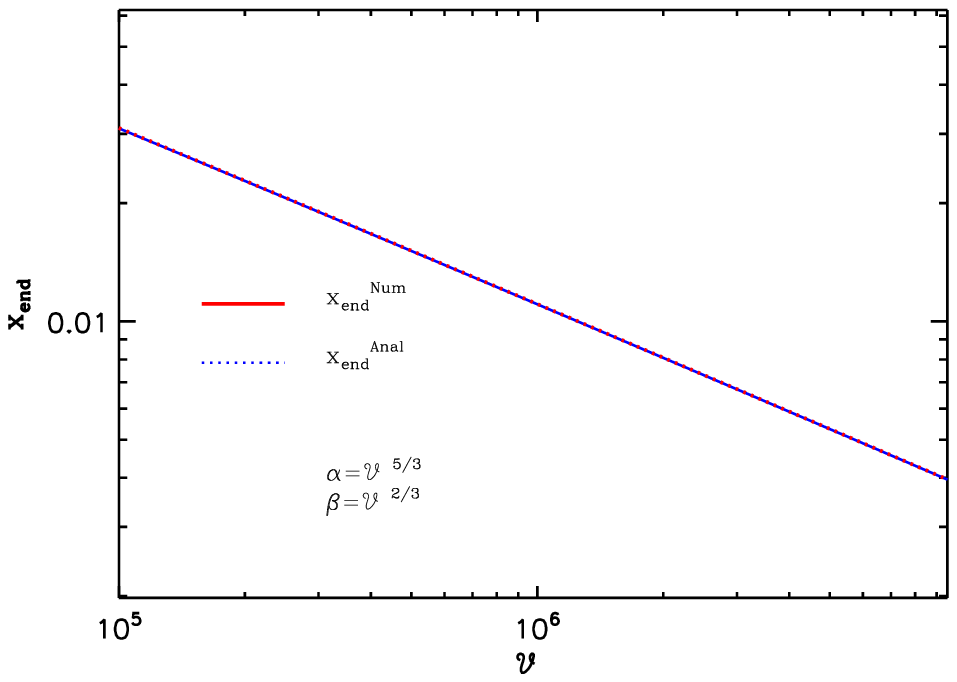}
  \caption{Comparison between the
    exact numerical value of $\xend(\alpha,\beta)$ (blue solid line), and
    the approximated formula given by \Eq{eq:kmiii:xend} (red dotted
    line) for $\alpha=\calV^{5/3}$ and $\beta=\calV^{2/3}$. The agreement is excellent
    but a numerical calculation is used in \ASPIC anyway.}
\label{fig:priorKMIII}
\end{center}
\end{figure}

Inflation stops when $\epsilon_1(\xend)=1$. As can be seen in \Fig{potkmiii},
for $\alpha/\beta\gg 1$, the first slow-roll parameter $\epsilon_1$
starts increasing from $\epsilon_1=0$ at $x=0$, diverges at a \vev
that we do not need to compute here, and then decreases to vanish at
$x=\beta^{-3/4}$. Then, it increases again, blows up at $\xVzero$ and,
finally, asymptotically vanishes when $x\rightarrow\infty$. Since
inflation proceeds at $x>\xVzero$ it always stops by violation of the
slow-roll conditions. Unfortunately is not possible to find an
analytic expression for $\xend$ but one can provide the following
approximated formula,
\begin{equation}
\label{eq:kmiii:xend}
\xend\simeq \left[-\frac{5}{4\beta}\Lambert{-1}
\left(-\frac{4 \times 9^{2/5}}{5 \times 8^{2/5}}\alpha^{-4/5}
\beta^{1/5}\right) \right]^ {3/4},
\end{equation}
where $\Lambert{-1}$ is the Lambert function. It is compared to the
numerical solution for $\xend$ implemented in the \ASPIC code in
\Fig{fig:priorKMIII}. The agreement is excellent.

\begin{figure}
\begin{center}
\includegraphics[width=\wsingfig]{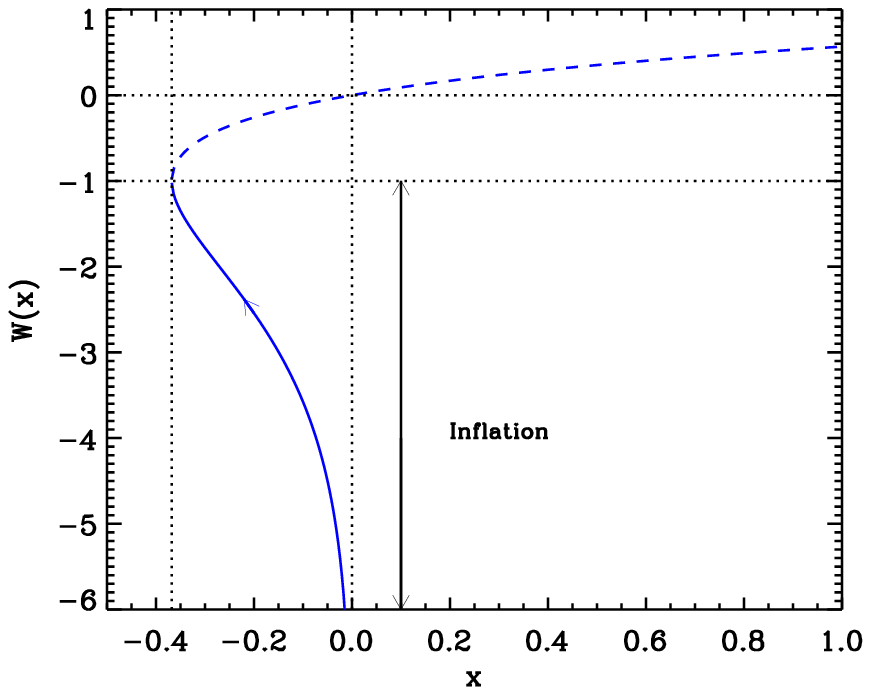}
\caption{Lambert functions $\Lambert{0}(x)$ (dashed line) and
  $\Lambert{-1}(x)$ (solid line). During K\"ahler moduli inflation II,
  inflation proceeds along the ``$-1$'' branch in the direction
  specified by the arrow.}
\label{potlambertKMIII}
\end{center}
\end{figure}

Let us now turn to the slow-roll trajectory. Unfortunately, KMIII is
one of the rare cases for which it cannot be integrated by
quadrature. As such, in the \ASPIC library, the slow-roll trajectory
is numerically integrated. However, in the large field limit
$x\gg\beta^{-3/4}$, one can obtain an approximate analytic formula
given by
\begin{equation}
\Nend-N \simeq  \frac{9}{16
  \alpha\beta^2}\left(\frac{\ee^{\beta x^{4/3}}}
  {x^2}-\frac{\ee^{\beta\xend^{4/3}}}
  {\xend^2} \right),
\end{equation}
from which one deduces that
\begin{equation}
x \simeq \left(-\frac{3}{2\beta}\Lambert{-1}
\left\lbrace-\frac{2}{3}\beta\left[\frac{\ee^{\beta\xend^{4/3}}}
{\xend^2}+\frac{16\alpha\beta^2}{9}\left(\Nend-N\right)
\right]^{-2/3}\right\rbrace\right)^{3/4}.
\end{equation}
This approximation is in agreement with what was obtained in
\Refc{Lee:2010tk}, up to an incorrect choice of the Lambert function
branch. The Lambert function is displayed in \Fig{potlambertKMIII} and
the part of the curve where inflation proceeds is indicated by the
arrow. The fact that the $-1$ branch of the Lambert function has to be
chosen comes from the fact that, when $\Nend-N\rightarrow\infty$, one
must have $x\rightarrow \infty$. On the other hand, when
$\Nend-N\rightarrow 0$, $x\rightarrow\xend>\beta^{-3/4}$ and this is
again consistent with the choice of the $-1$ branch.

Finally, one can use the CMB normalization to calculate the mass scale
$M$. Without any approximation on top of slow-roll, this leads to the
following expression
\begin{equation}
\left(\frac{M}{\Mp}\right)^4 =
1280\pi^2\alpha^2\xstar^{2/3}\ee^{-2\beta \xstar^{4/3}} \left(1-\beta
\xstar^{4/3}\right)^2 \left(1-\alpha x^{4/3}\ee^{-\beta
  \xstar^{4/3}}\right)^{-2} \frac{\Qrms^2}{T^2}\, .
\end{equation}
Making use of the approximated trajectory and of the expression for
the scale $M$, one roughly obtains
\begin{equation}
\calV_\us \simeq \frac{\Delta \Nstar}{\pi \sqrt{720}}
\frac{1}{(\Mp \ells)^3}\left[
\frac{4Ba_n\left(W_0\ells^3\right)^2}{3\gamma \lambda_n}
\right]
\ln^{-5/4}\left(\frac{16\alpha\beta^2}{9}\Delta\Nstar\right)
\frac{T}{\Qrms}\,.
\end{equation}
Given that $a_n$, $B$, $\gamma$, $\lambda_n$, $W_0\ells^3$ are a
priori coefficients of order one, we see that the above expression
roughly implies that $\calV$ is of the order $10^6\ells$.

The reheating consistent slow-roll predictions for the K\"ahler moduli
inflation II models are displayed in \Fig{fig:CMBKMIII}, for
$\calV\in[10^{5},10^{7}]$, and taking $\alpha=\calV^{5/3}$ and
$\beta=\calV^{2/3}$. One can check that even if one adds $\order{1}$
factors in these relations, the slow-roll predictions do not depend
significantly on them. Also, we notice that $\epsilon_1$ is typically
extremely small and that $\epsilon_2$ is almost independent of
$\calV$. These effects can be analytically understood. Working out
\Eq{eq:kmiii:xend} and \Eqs{eq:kmiii:eps1}, (\ref{eq:kmiii:eps2}), and
(\ref{eq:kmiii:eps3}) in the large field limit, one obtains
\begin{equation}
\begin{aligned}
\epsonestar \simeq \frac{1}{324 \beta^{3/2}(\Delta\Nstar)^2}
\ln^{5/2}\left(16\sqrt{\frac{9}{8}}\alpha\beta^{1/2}\Delta\Nstar\right) ,\qquad
\epstwostar \simeq \frac{2}{\Delta\Nstar}\, ,\qquad
\epsthreestar \simeq \frac{1}{\Delta\Nstar}\, ,
\end{aligned}
\end{equation}
from which one deduces that
\begin{equation}
\begin{aligned}
  \nS \simeq 1-\frac{2}{\Delta\Nstar}\, ,\qquad
  r \simeq \frac{4}{81\beta^{3/2}(\Delta\Nstar)^2}
\ln^{5/2}\left(16\sqrt{\frac{9}{8}}\alpha\beta^{1/2}\Delta\Nstar\right),\qquad
  \alphaS \simeq -\frac{2}{\Delta\Nstar^2}\, .
\end{aligned}
\end{equation}
Firstly, we see that the slow-roll parameters at Hubble crossing
depend on $\alpha$ logarithmically only. This explains the weak
dependence in the $\order{1}$ factors mentioned above.
Secondly, we also notice
that $\epstwostar$ and $\epsthreestar$ do not depend on $\beta$. In a
third place, $\epsilon_1$ is a very small number since it is
proportional to the inverse of $\beta^{3/2}$. This also means that,
when $\calV$ increases, $\epsilon_1$ decreases. All these
considerations can be checked in \Fig{fig:CMBKMIII} and the amount of
gravitational waves predicted by this model is very small. This is in
agreement with the rough estimates given in \Refcs{Conlon:2005jm,
  Krippendorf:2009zza, BlancoPillado:2009nw, Lee:2010tk}.  However,
contrary to what is claimed in \Refc{Lee:2010tk}, the predicted value
for the running of the spectral index is not excluded by observations
since, according to the Planck results~\cite{Planck:2019nip, Planck:2019evm},
$\alphaS=0.0011\pm 0.0099$ while, for the fiducial value $\Delta \Nstar
\simeq 55$, one obtains $\alphaS\simeq -0.0006$.

\subsection{Logamediate Inflation (LMI)}
\label{sec:lmi}

Logamediate inflation has been discussed in \Refcs{Barrow:2007zr,
  Parsons:1995ew} and refers to inflationary scenarios in which the scale
factor evolves according to
\begin{equation}
  a\left( t\right)=\azero\exp\left[A\left(\ln
    \frac{t}{t_0}\right)^\lambda\right],
\label{eq:lmi:scalefactor}
\end{equation}
where $A$ and $\lambda$ are two dimensionless parameters and where
$t_0$ has the dimension of a cosmic time. This evolution form for the
scale factor is required to occur ``at late times'', \ie when $t\gg
t_0$. If $\lambda=1$, one recovers the power law model (see
\sectionc{sec:pli}), and in that case, $t_0$ can be absorbed in a
rescaling of the scale factor. Otherwise, these three parameters are
relevant and one therefore expects LMI to be a two parameters models
according to our classification. Following \Refc{Barrow:2007zr}, from
\Eq{eq:lmi:scalefactor}, one has
\begin{equation}
H\equiv\frac{\dot{a}}{a} = \frac{A\lambda}{t} \left(\ln \frac{t}{t_0}
\right)^{\lambda-1} ,
\label{eq:lmi:H}
\end{equation}
from which one deduces that $A\lambda>0$ in order to have expansion
($H>0$).  From \Eq{eq:lmi:scalefactor}, one can also establish
that
\begin{equation}
  \frac{\ddot{a}}{a} = \frac{A\lambda}{t^2}\left(\ln \frac{t}{t_0}
  \right)^{\lambda-1} \left[\left(\lambda-1\right) \left(\ln
    \frac{t}{t_0} \right)^{-1}-1 + A\lambda\left(\ln \frac{t}{t_0}
    \right)^{\lambda-1}\right] ,
\end{equation}
from which one deduces that in order to have inflation at late times
(when $t\gg t_0$), one must have $\lambda>1$, or if $\lambda=1$,
$A>1$. If this inflationary scenario is implemented within a single
minimally coupled scalar field $\phi$, one can derive the
corresponding potential. From the Friedmann-Lema\^{\i}tre and
Klein-Gordon equations one can show that~\cite{Barrow:2007zr}
\begin{equation}
  \frac{\dot{\phi}\left(t\right)}{\Mp} = \frac{\sqrt{2A\lambda}}{t}
  \left(\ln \frac{t}{t_0}\right)^{\frac{\lambda-1}{2}}\, .
\end{equation}
This equation can easily be integrated into
\begin{equation}
\label{eq:lmi:phiVSt}
\frac{\phi\left(t\right)}{\Mp} =
\frac{\phizero}{\Mp}+2\frac{\sqrt{2A\lambda}}{\lambda+1}
\left(\ln \frac{t}{t_0}\right)^{\frac{\lambda+1}{2}} .
\end{equation}
Combining the Friedmann-Lema\^{\i}tre equation
$3\Mp^2H^2=V(\phi)+\dot{\phi}^2/2$ and the relation
$2\Mp^2\dot{H}=-\dot{\phi}^2$, one obtains
$V(\phi)=3\Mp^2H^2+\Mp^2\dot{H}$, namely
\begin{equation}
V(\phi) =
\frac{3\Mp^2A^2\lambda^2}{t^2}\left(\ln\frac{t}{t_0}\right)^{2(\lambda-1)}
+\frac{\Mp^2A\lambda}{t^2} \left(\lambda-1\right) \left(\ln
\frac{t}{t_0} \right)^{\lambda-2} - \frac{\Mp^2A\lambda}{t^2}\left(\ln
\frac{t}{t_0}\right)^{\lambda-1}.
\label{eq:lmipotint}
\end{equation}
Together with \Eq{eq:lmi:phiVSt}, this gives a parametric
representation of the field potential in terms of $t$. It can be
further simplified since the Logamediate regime occurs in the limit $t
\gg t_0$. If $\lambda>1$, the first term of this expression dominates
at late times and one has $V(\phi)=3\Mp^2A^2\lambda^2\left(\ln
t/t_0\right)^{2(\lambda-1)}/t^2$. Defining $x\equiv
\left(\phi-\phizero\right)/\Mp$, one makes use of \Eq{eq:lmi:phiVSt}
to obtain
\begin{equation}
\label{eq:lmi:pot}
V(\phi)=M^4x^{\alpha}\exp\left(-\beta x^{\gamma}\right),
\end{equation}
where the new parameters are defined by
\begin{equation}
\alpha = 4\frac{\lambda-1}{\lambda+1}\, , \qquad
\beta =
2\left(\frac{\lambda+1}{2\sqrt{2A\lambda}}\right)^{2/\left(\lambda+1\right)}
,\qquad
\gamma = \frac{2}{\lambda+1}\, ,
\label{eq:lmi:abg}
\end{equation}
and
\begin{equation}
\frac{M^4}{\Mp^4} = \frac{3A^2\lambda^2}{\Mp^2t_0^2}\left(
\frac{\lambda+1}{2\sqrt{2A\lambda}}
\right)^{4\frac{\lambda-1}{\lambda+1}}.
\end{equation}
The same potential has been studied for $\alpha=2$, $\beta=1/8$ and
$\gamma=2$ within tachyon inflation models in \Refc{Kofman:2002rh}.
The case $\lambda=1$ is particular. At late times, the first term and
the last term must be kept in \Eq{eq:lmipotint}, such that
$V(\phi)=(3A-1)A\Mp^2/t^2$. In that situation, one has $x=\sqrt{2A}\ln
t/t_0$, and the derived potential shares the same expressions for
$\alpha$, $\beta$ and $\gamma$ as in \Eq{eq:lmi:abg} but evaluated at
$\lambda=1$. There is a difference however because $M^4$ now reads
$M^4=(3A-1)A\Mp^2/t_0^2$. We recover explicitly that $\lambda=1$
corresponds to power law inflation and has already been treated in
\sectionc{sec:pli}.

In the following, we will work only with the derived parameters
$\beta$, $\gamma$ and $M^4$, noticing that
\begin{equation}
  \alpha=4\left(1-\gamma\right).
\end{equation}
The restrictions $A\lambda>0$ and $\lambda\geq 1$ translates into the
conditions $0<\gamma\leq1 $ and $\beta>0$. Following
\Refc{Parsons:1995ew}, since there is no fundamental reasons preventing
it, we will generalize this model to any possible values of these
parameters supporting inflation.

\begin{figure}
\begin{center}
\includegraphics[width=\wdblefig]{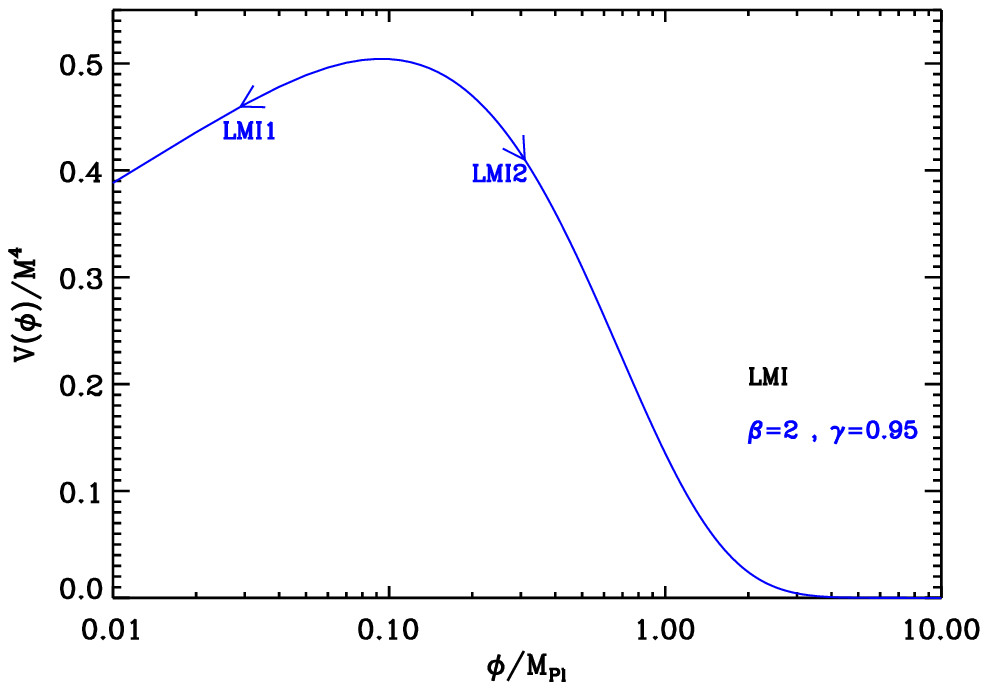}
\includegraphics[width=\wdblefig]{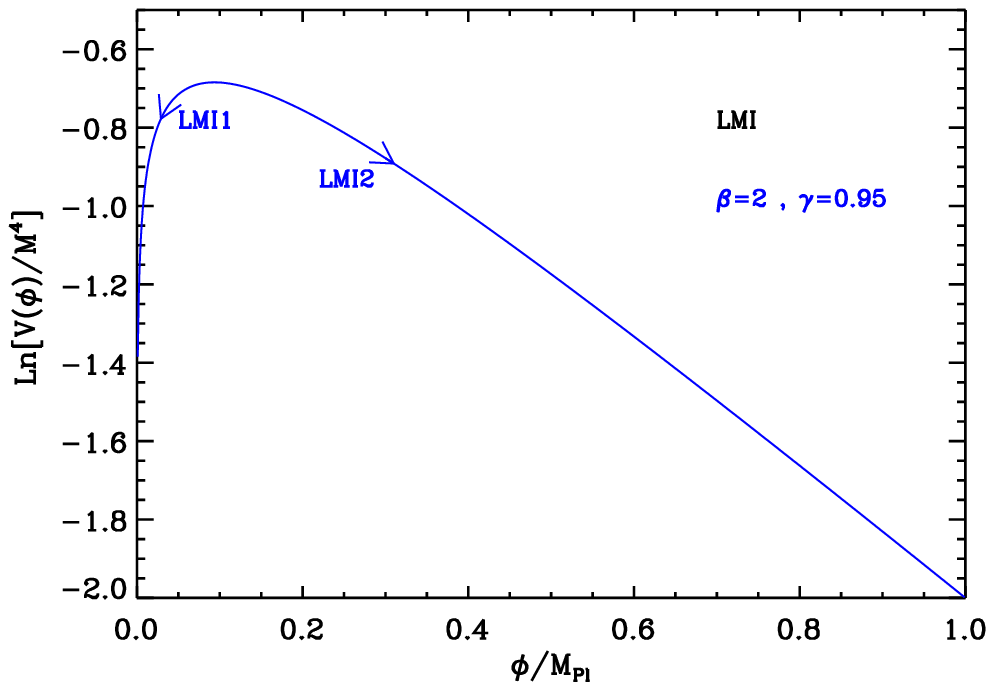}
\includegraphics[width=\wdblefig]{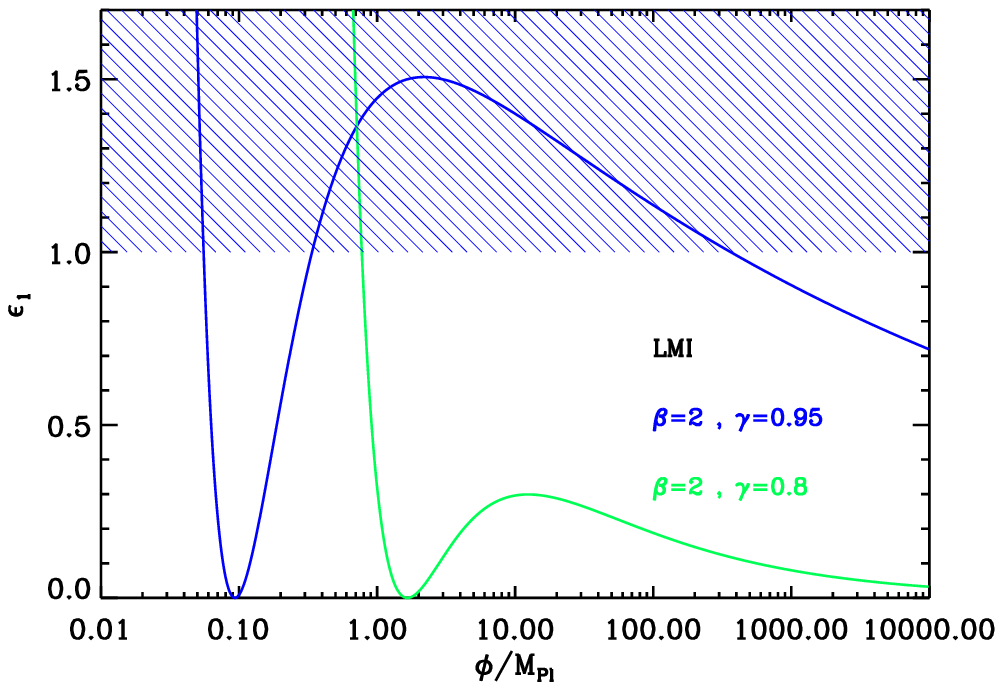}
\includegraphics[width=\wdblefig]{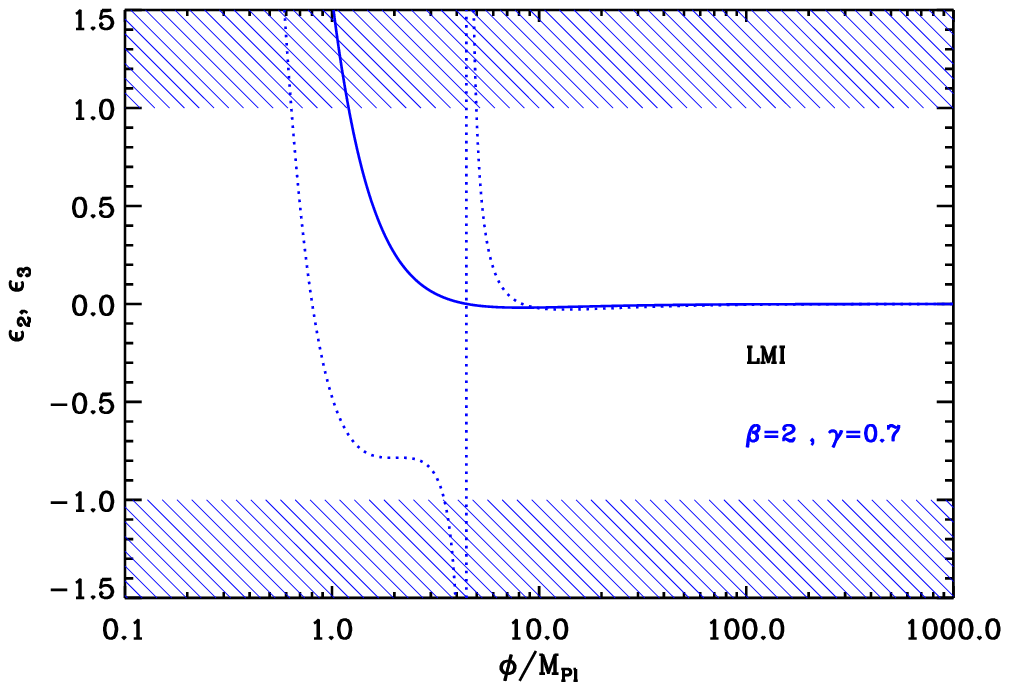}
\caption{Logamediate Inflation (LMI).  Upper panels: the potential and
  its logarithm for $\beta=2,\gamma=0.95$. Bottom left panel: Hubble
  flow function $\epsilon_1$ for a potential with
  $\beta=2,\gamma=0.95$ (blue curve) and $\beta=2,\gamma=0.8$ (green
  curve). The position of the maximum of $\epsilon_1$ with respect to
  one depends on $\gamma$. The shaded region indicates where inflation
  stops. Bottom right panel: slow-roll parameters $\epsilon_2$ (solid
  line) and $\epsilon _3$ (dotted line) for a potential with
  $\beta=2,\gamma=0.7$.}
\label{fig:potLMI}
\end{center}
\end{figure}

The three first Hubble flow functions in the slow-roll approximation
read
\begin{equation}
\epsilon_1 = \frac{\left(\alpha-\beta\gamma x^{\gamma}\right)^2}{2x^2}\, ,\qquad
\epsilon_2 = \frac{2}{x^2}\left[\alpha + \beta\left(\gamma-1
  \right)\gamma x^{\gamma} \right],
\end{equation}
\begin{equation}
\epsilon_3 =\frac{\alpha-\beta\gamma x^{\gamma}}{x^2}
\dfrac{2\alpha-\beta\left(\gamma-2 \right) \left(\gamma-1
  \right)\gamma x^{\gamma}}{\alpha+\beta\left(\gamma-1 \right)\gamma
  x^{\gamma}}\, .
\end{equation}
The potential and the Hubble flow functions in the slow-roll
approximation have been represented in \Fig{fig:potLMI}.

Inflation can proceed in two regimes: either at decreasing field
values, left to the top of the potential (LMI1), or at increasing
field values, right to the top of the potential (LMI2). Notice that
from \Eq{eq:lmi:phiVSt}, $\phi$ has to increase with time to reproduce
the scale factor expansion \Eq{eq:lmi:scalefactor} and this happens
only in the regime LMI2 for large values of $x$. As can be seen in
\Fig{fig:potLMI}, the slow-roll parameter $\epsilon_1$ diverges when
$x$ approaches zero, it vanishes at the top of the potential for
$x=\xVmax$ and it is maximal at $x=\xepsoneMax$ with
\begin{equation}
\xVmax \equiv \left(
\dfrac{\alpha}{\beta\gamma}\right)^{1/\gamma}, \qquad \xepsoneMax
=\left[\frac{\alpha}{\beta\gamma\left(1-\gamma\right)}\right]^{1/\gamma}.
\end{equation}
Finally it asymptotes to zero for large values of the field. The value
of the local maximum of $\epsilon_1$ reads
\begin{equation}
\epsilon_1^\mathrm{max}=\frac{\alpha^2}{2}
\left[\frac{\beta\gamma\left(1 - \gamma\right)}{\alpha}
  \right]^{\frac{2}{\gamma}} \left(\frac{\gamma}{1-\gamma}\right)^2 .
\label{eq:lmi:eps1max}
\end{equation}
Thus in the regime LMI1, inflation always stops naturally as
$\epsilon_1$ becomes larger than unity whereas in the regime LMI2,
this may occur only if $\epsonemax>1$ and if inflation has started
from $\xini<\xepsoneMax$. Otherwise, if inflation starts with $\xini >
\xepsoneMax$, or if $\epsonemax<1$, one needs to add an
extra-parameter $\xend$ encoding an unspecified mechanism to end
inflation. In that situation, the model becomes a three parameters
one. If one makes use of $\alpha=4\left(1-\gamma\right)$, one obtains
$\epsonemax=8\gamma^2\left(\beta\gamma/4\right)^{2/\gamma}$. Solving
$\epsonemax \ge 1$ for $\beta$ gives
\begin{equation}
\beta \ge \dfrac{4}{\gamma \left(8 \gamma^2 \right)^{\gamma/2}}\,.
\end{equation}
This condition is therefore required for the model LMI2, if one wants
inflation to end naturally. As we will see below, LMI2 inflating in
the domain $\xVmax < x < \xepsoneMax$ is a very fine-tuned situation
which is strongly disfavored by the observations. Notice that if one
assumes $0<\gamma \le 1$, this conditions translates into
$\beta>\sqrt{2}$.

Finally, let us notice that for the value of $\epsilon_2$ at the top of
the potential to be smaller than some maximal value $\epstwotopMax$,
one needs to impose the condition
\begin{equation}
\beta<\beta^{\max}\left(\gamma,\epstwotopMax\right)
=2^{2-3\gamma/2}\left(\epstwotopMax\right)^{\gamma/2}
\frac{\left(1-\gamma\right)^{1-\gamma/2}}{\gamma^{1+\gamma/2}}\, .
\end{equation}
In the LMI1 model, a slow roll regime of inflation can proceed only if
such a condition is verified (with typically $\epstwotopMax\simeq
10^{-1}$).

The slow-roll trajectory can be integrated thanks to the
hypergeometric function \cite{Abramovitz:1970aa,Gradshteyn:1965aa}
$\Fgauss$, leading to
\begin{equation}
  N-\Nend=\frac{\xend^2}{2\alpha}\, \Fgauss
  \left[1,\frac{2}{\gamma},\frac{2}{\gamma}+1,\left(\frac{\xend}{\xVmax}
    \right)^\gamma \right] -\frac{x^2}{2\alpha}\,
  \Fgauss\left[1,\frac{2}{\gamma},\frac{2}{\gamma}+1,\left(\frac{x}{\xVmax}
    \right)^\gamma \right] .
\label{eq:lmi:traj0}
\end{equation}
One can notice that inserting $\alpha=4(1-\gamma)$, as a function of
$x/\xVmax$, this trajectory only involves $\gamma$. Plugging
$x=\xVmax$ into \Eq{eq:lmi:traj0} one gets an infinite number of
\efolds. This means that the required number of \efolds to solve the
problems of the standard Big-Bang scenario can always be realized,
both in the decreasing branch of the potential and the increasing one,
provided that inflation starts close enough to $\xVmax$. However, it
can numerically be checked that in the case of LMI2 with
$\epsonemax>1$ and inside the $\xVmax<x<\xepsoneMax$ region, one has
to fine-tune $\xini$ and $\xstar$ extremely close to $\xVmax$. In
that situation $\nS=1$, with vanishing $r$ and vanishing running of
the spectral index, can be considered as generic predictions of the
model. For this reason, it is more natural to consider LMI2 in the
large field regime, namely $x>\max(\xVmax,\xepsoneMax)$, together with
the extra-parameter $\xend$.

The trajectory in \Eq{eq:lmi:traj0} cannot be inverted
analytically. However, one can perform some consistency checks in the
limit $x/\xVmax\gg 1$ in which
\begin{equation}
  N-\Nend \simeq \frac{1}{\beta\gamma\left(2-\gamma\right)}\left(x^{2-\gamma}
    - \xend^{2-\gamma}\right),
\end{equation}
and
\begin{equation}
  x \simeq \left[\xend^{2-\gamma} + \beta\gamma\left(2 -
      \gamma\right)(N-\Nend)
  \right]^{\frac{1}{2-\gamma}}.
\label{eq:lmi:traj1}
\end{equation}
These expressions can be compared to \Eq{eq:lmi:phiVSt}
\begin{equation}
  x=2\frac{\sqrt{2 A\lambda}}{\lambda+1} \left(\ln \frac{t}{t_0} \right)^{\frac{\lambda+1}{2}}\,,
\end{equation}
where $t$ in terms of the number of \efolds $N$ can be obtained
from \Eq{eq:lmi:scalefactor}. With $N-\Nzero=A\left(\ln
t/t_0\right)^\lambda$, one gets
\begin{equation}
  x = 2\frac{\sqrt{2A\lambda}}{\lambda +1} \left(\frac{N-\Nzero}{A}
  \right)^{\frac{\lambda +1}{2\lambda}}.
\label{eq:lmi:traj2}
\end{equation}
The previous calculations are consistent since, making use of
\Eq{eq:lmi:abg}, \Eq{eq:lmi:traj1} and \Eq{eq:lmi:traj2} are the same
when setting the constants $\Nzero=\Nini$ and $x_0=\xini=0$. This
means that in the late times limit $x/\xVmax\gg 1$, the slow-roll
trajectory coincides with the exact one, as expected.

The amplitude of the CMB anisotropies fixes the value of the parameter
$M$ according to
\begin{equation}
  \frac{M^4}{\Mp^4} = 720\pi^2\left(\alpha-\beta\gamma
  \xstar^\gamma\right)^2 \ee^{\beta \xstar^\gamma} \xstar^{-\alpha-2}
  \frac{\Qrms^2}{T^2}\,,
  \label{eq:lmi:COBE}
\end{equation}
where $\xstar$ is the observable field value obtained by solving
\Eq{eq:phistarlnrrad} given some assumptions on the reheating. The
reheating consistent slow-roll predictions for the models LMI1 and
LMI2 (at $x>\xepsoneMax$) are displayed in
\Figs{fig:CMBLMI1beta10PowerMinus3}, \ref{fig:CMBLMI1beta1}, and
\ref{fig:CMBLMI1beta50} for LMI1, and in
\Figs{fig:CMBLMI2beta10PowerMinus1}, \ref{fig:CMBLMI2beta1}, and
\ref{fig:CMBLMI2beta10} for LMI2. In the case of LMI2, the turning
points in the plots precisely correspond to the case where inflation
occurs in the fine-tuned domain $\xVmax < \xstar < \xepsoneMax$ and in
which the model behaves like a pure de Sitter era.

\subsection{Twisted Inflation (TWI)}
\label{sec:twi}

\subsubsection{Theoretical Justifications}
\label{subsubsec:theorytwi}

\begin{figure}
\begin{center}
\includegraphics[width=\wdblefig]{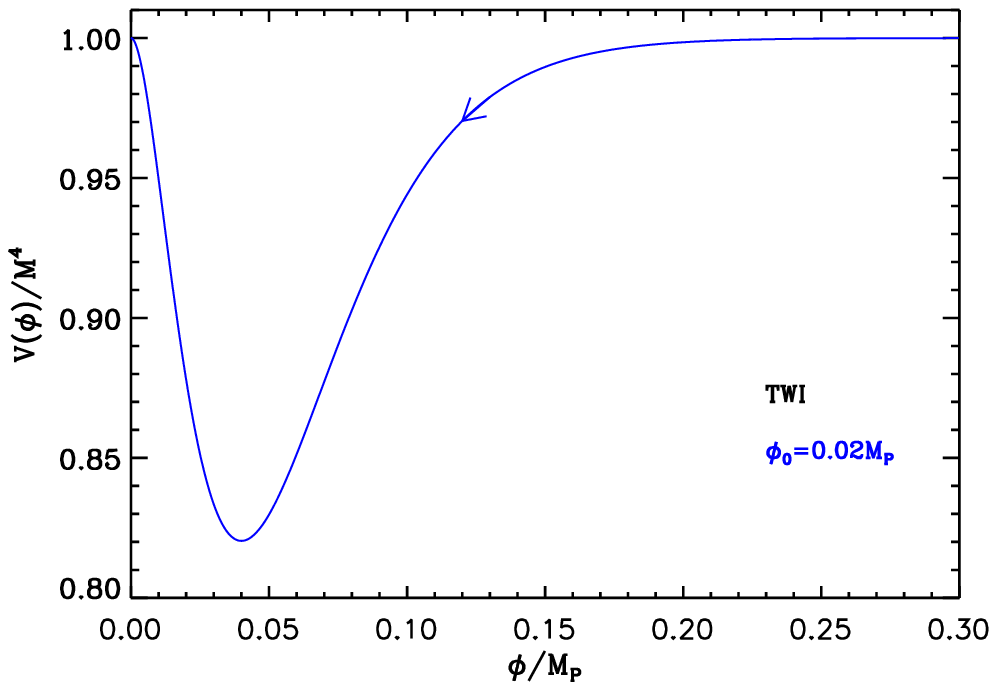}
\includegraphics[width=\wdblefig]{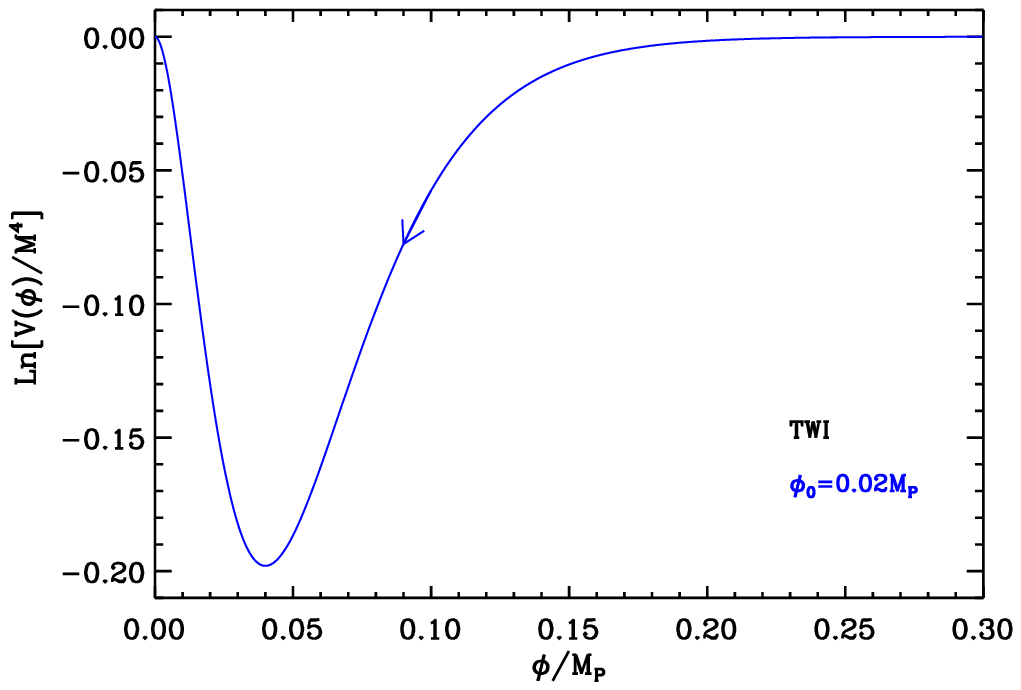}
\includegraphics[width=\wdblefig]{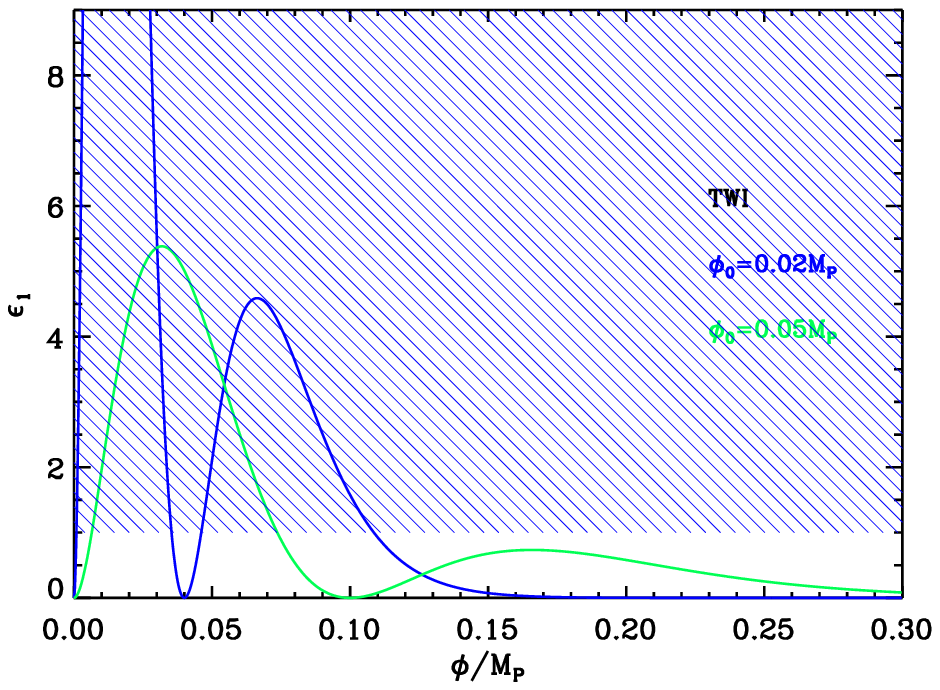}
\includegraphics[width=\wdblefig]{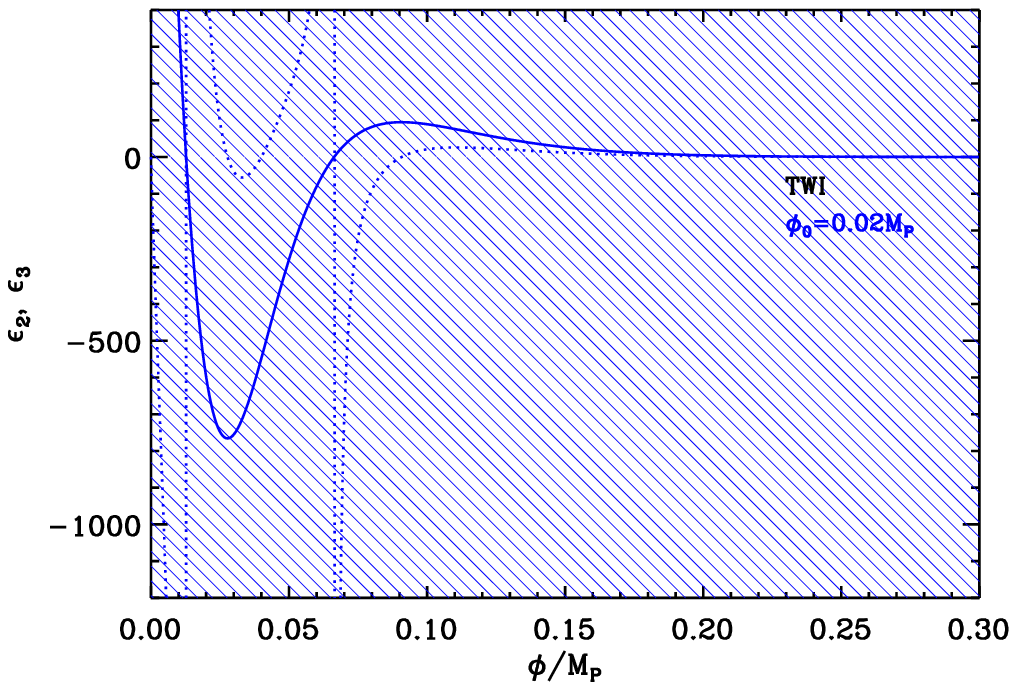}
\caption{Top left panel: Twisted Potential Inflation (TWI) for
  $\phizero=0.02\Mp$. Top right panel: logarithm of the potential for
  the same value of $\phizero$. Bottom left panel: slow-roll parameter
  $\epsilon _1$ with $\phizero=0.02\Mp$ (solid blue line) and
  $\phizero=0.05\Mp$ (solid green line). The shaded area indicates the
  non-inflationary region. Bottom right panel: slow-roll parameters
  $\epsilon _2$ (solid line) and $\epsilon _3$ (dotted line) with
  $\phizero=0.02\Mp$.}
\label{pottwi}
\end{center}
\end{figure}

This model was introduced in \Refc{Davis:2010it} and is based on
higher dimensional supersymmetric gauge theories. The idea is to
assume that, in higher dimensions, we have a flat direction $\phi$ in
the potential. Since the theory is supersymmetric, this flat direction
will not receive corrections because the bosonic and fermionic
contributions exactly cancel out. Then, we compactify the theory down
to $3+1$ dimensions but with boundary conditions that break
supersymmetry. The typical example given in \Refc{Davis:2010it} is
``twisted'' circle compactification, hence the name of the
model. Since supersymmetry is broken, the ``Kaluza-Klein'' masses of
bosons and fermions will differ. Typically, they can be written as
\begin{equation}
m_\ub=\sqrt{\phi^2+\frac{n^2}{R^2}}, \qquad 
m_\uf=\sqrt{\phi^2+\frac{(n+1/2)^2}{R^2}}\,,
\end{equation}
where $R$ is the radius of compactification and $n$ an integer. Since
$m_\ub\neq m_\uf$, this time, the potential will receive
one loop corrections which lift the potential. However,
it is clear that, when $\phi R\gg n$, one has approximately
$m_\ub\simeq m_\uf$. Therefore, in this regime, we
expect the corrections to vanish and the flat direction to remain
flat. This is thus particularly well-suited for inflation. In practice,
the higher dimensional model considered to implement the above
discussed mechanism is a maximally supersymmetric $4+1$
$\mathrm{U}(\calN)$ Yang-Mills theory compactified on a circle of radius
$R$. A priori, we have therefore two parameters: $\calN$ and the
compactification scale $R$.

\subsubsection{Slow-Roll Analysis}
\label{subsubsec:srtwi}

As shown in \Refc{Davis:2010it}, the above considerations leads to the
following expression for the inflaton potential
\begin{equation}
  V(\phi)=M^4\left[1-A\left(\frac{\phi}{\phizero}
    \right)^2\ee^{-\phi/\phizero} \right],
\label{eq:twi:pot}
\end{equation}
where $A$ is a constant parameter given by
 \begin{equation}
A\equiv\frac{32}{93\zeta\left(5\right)}\simeq 0.33\, ,
\end{equation}
and where $\phizero$ is related to the compactification scale $R$
through the following equation
\begin{equation}
\frac{\phizero}{\Mp}=\frac{1}{2\pi R\Mp}.
\end{equation}
Since the radius $R$ must be larger than the Planck length, \ie
$R\Mp\gg 1$, this implies that $\phizero/\Mp\ll 1$. On the other hand,
the overall normalization can be expressed as
\begin{equation}
M^4=\frac{8\calN}{A\pi^2(2\pi R)^4}\,.
\end{equation}
We see that the scale $M$ depends on the compactification radius $R$
but also on the number $\calN$. In addition, one must have
$\phi<\sqrt{3/\calN}\Mp$ or $\phi\ll \Mp$ which guarantees that the
higher order Planck suppressed operators do not alter the
potential. The potential~(\ref{eq:twi:pot}) is the small coupling
limit of the model, while the strong coupling limit corresponds to a
BI model with $p=3$, see \sectionc{sec:bi}.

The potential \Eq{eq:twi:pot}, as well as its logarithm, is displayed
in \Fig{pottwi}. Inflation is supposed to take place for \vev 's
larger than the scale $\phizero$, \ie for $\phi>\phizero$, in the
increasing branch of the potential. This means that it proceeds from
the right to the left in the direction indicated by the arrow. The
minimum of the potential is located at $\phi/\phizero=2$.

Let us now turn to the calculation of the Hubble flow parameters. If
one defines $x$ by $x\equiv\phi/\phizero$, then they are given by
\begin{equation}
\label{eq:twi:eps1}
\epsilon_1 = \frac{A^2}{2}\left(\frac{\Mp}{\phizero}\right)^2\ee^{-2x}
\left[\frac{x\left(x-2\right)}{1-Ax^2\ee^{-x}}\right]^2\, ,\qquad
\epsilon_2 = 2A\left(\frac{\Mp}{\phizero}\right)^2\ee^{-2x}
\frac{2Ax^2+\ee^{x}\left(x^2-4x+2\right)}
{\left(1-Ax^2\ee^{-x}\right)^2}\, ,
\end{equation}
and
\begin{equation}
\begin{aligned}
\epsilon_3 & = A \left( \frac{\Mp}{\phizero}\right)^2x\left(2-x
\right)\ee^{-2x} \dfrac{ 4A^2x^3-\ee^{2x} \left(x^2-6x+6 \right) -A x
  \ee^{x}\left(x^3-6x^2+18x-12 \right)}{\left(1-Ax^2\ee^{-x}\right)^2
  \left[2Ax^2+\ee^{x}\left(x^2-4x+2\right) \right]}\, .
\end{aligned}
\end{equation}
They are displayed in \Fig{pottwi}. The first slow-roll parameter
$\epsilon_1$ vanishes at the minimum of the potential when $x=2$, then
increases with $x$ and reaches a maximum at $\xepsoneMax$, and finally
decreases to zero when $x$ goes to infinity. The value of $\epsilon_1$
at this local maximum is larger than one if $\phizero$ is smaller than
some value that can only be determined numerically. We find
\begin{equation}
\label{condtwiphizero}
\phizero< 0.04228 \Mp\, .
\end{equation}
Therefore, a priori, inflation could stop by slow-roll violation.
\begin{figure}
\begin{center}
\includegraphics[width=\wdblefig]{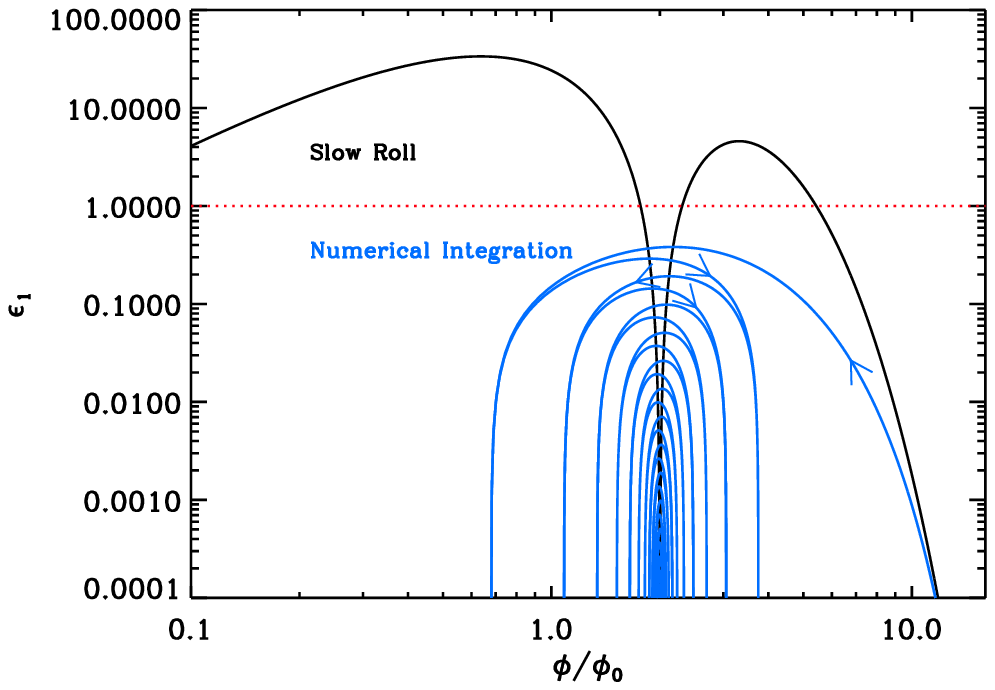}
\includegraphics[width=\wdblefig]{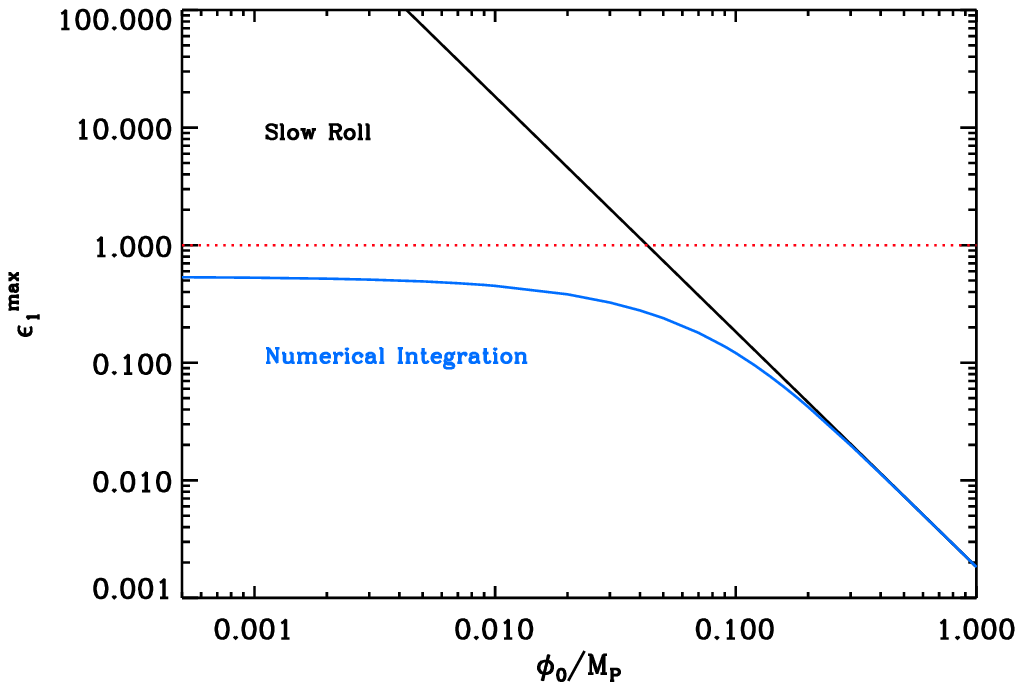}
\caption{Left panel: slow-roll parameter $\epsilon_1$ as a function of
  the field \vev $\phi/\phizero$, for $\phizero/\Mp=0.02<0.04228$, see
  \Eq{condtwiphizero}. The solid black line corresponds to the
  approximated slow-roll formula of \Eq{eq:twi:eps1}, \ie
  $\epsilon_1^V=\Mp^2/2\,V_\phi^2/V^2$, while the solid blue line
  represents the exact $\epsilon_1^H=-\dot{H}/H^2$ obtained from a
  numerical integration starting at $\phiini/\Mp=0.33$ and vanishing
  initial velocity. We see that the exact $\epsilon_1^H$ remains in
  fact always smaller than one and that inflation never stops. The
  inflaton eventually oscillates around the minimum of its potential
  located at $\phi=2\phizero$ (the arrows indicate the direction of
  the first oscillations). Right panel: Maximum value taken by
  $\epsilon_1^V$ (solid black line) and $\epsilon_1^H$ (solid blue
  line) for different values of $\phizero$. One can see that
  $\epsilon_1^H$ remains smaller than one for any value of
  $\phizero$. When $\phizero$ increases, the slow-roll parameters,
  which scale proportional to $\Mp^2/\phizero^2$, decrease so that the
  slow-roll approximation becomes more and more efficient and
  eventually starts matching the numerical exact predictions.}
\label{eps1twi}
\end{center}
\end{figure}
However, by numerically integrating the exact trajectory (\ie if one
does not make use of the slow-roll approximation), one realizes that,
in fact, the first Hubble flow function, which is defined by
$\epsilon_1^H=-\dot{H}/H^2$, remains smaller than one for all field
values, see \Fig{eps1twi}. This is due to the fact that while the
inflaton rolls down its potential and approaches its minimum, the
slow-roll parameters continuously increase and the slow-roll
approximation is broken before $\epsilon_1$ becomes
$\order{1}$. Usually, this leads only to small corrections at the end
of inflation. However, in the case of twisted inflation, this leads to
a radically different picture because the potential does not vanish at
its minimum and, therefore, acts as a cosmological constant. In
practice, the numerical calculations indicate that the field
oscillates around its minimum but always such that $\epsilon_1^H<1$
and independently on the value of $\phizero$, see \Fig{eps1twi}. In
principle, inflation can never stops in this model since the final
stage of the evolution corresponds to an inflaton field sitting for
ever at the bottom of the potential and, as already mentioned, it acts
as a cosmological constant. However, as explained in
\Refc{Davis:2010it}, the interactions of the inflaton field with the
other degrees of freedom of the standard model starts to play a role
in this regime. As a consequence, the energy contained in the inflaton
field should quickly be transferred to other fields and a phase of
reheating starts. The details of this process are complicated and are
discussed in \Refc{Davis:2010it}. In order to model the end of
inflation, we therefore introduce the extra parameter $\xend$ giving
the \vev at which inflation stops. As a consequence, TWI is in fact a
two parameter model, $\phizero$ and $\phiend$.

Let us now turn to the slow-roll trajectory. It can be integrated
exactly and leads to the following expression
\begin{equation}
\begin{aligned}
  \Nend-N  =\left(\frac{\phizero}{\Mp}\right)^2 &\left \lbrace
    \frac{1}{2A} \left[ \Ei\left(\xend\right)-\Ei\left(x\right) 
    \right]
    -\frac{\ee^2}{2A} \left[\Ei\left(\xend-2\right)-\Ei\left(x-2
      \right) \right]
  \right. \\ & \quad
    + \left. \xend-x+2\ln\left(\frac{\xend-2}{x-2} \right)
  \right\rbrace ,
\end{aligned}
\end{equation}
where $\Nend$ is the number of \efolds at the end of inflation and
$\Ei$ is the exponential integral
function~\cite{Abramovitz:1970aa,Gradshteyn:1965aa}. This expression
is the one used in the \ASPIC library. However, if one makes the
vacuum dominated approximation, $x\gg 1$, then a simpler formula can
be derived for the trajectory, namely
\begin{equation}
\Nend-N\simeq \dfrac{1}{A} \left(\dfrac{\phizero}{\Mp}\right)^2
\left(\dfrac{\ee^x}{x^2} - \dfrac{\ee^{\xend}}{\xend^2} \right).
\end{equation}
This allows us to obtain an approximated expression for the \vev of
the field at Hubble radius crossing which reads
\begin{equation}
\xstar \simeq \ln\left[4A\Delta\Nstar\left(\frac{\Mp}
{\phizero}\right)^2\right].
\label{eq:twi:xstar}
\end{equation}
It is valid provided $\phizero/\Mp\ll 1$, \ie precisely in the regime
for which the TWI potential was derived. Using this formula, one can
estimate the value of the three first Hubble flow parameters at Hubble
crossing. One arrives at
\begin{equation}
\begin{aligned}
\epsilon_{1*}&\simeq\frac{A^2}{2}\left(\frac{\Mp}
{\phizero}\right)^2\ee^{-2\xstar}\xstar^4\simeq\frac{1}
{32\Delta\Nstar^2}\left(\frac{\phizero}{\Mp}\right)^2,\\
\epsilon_{2*}&\simeq\frac{\epsilon_{3*}}{2}\simeq 2A\left(\frac{\Mp}
{\phizero}\right)^2\ee^{-\xstar}\xstar^2\simeq\frac{1}{2\Delta\Nstar}\, .
\end{aligned}
\end{equation}
Finally, we can derive an expression for the tensor-to-scalar ratio,
the spectral index
\begin{equation}
\label{eq:twi:roughpred}
r \simeq 8A^2 \left(\frac{\Mp}
{\phizero}\right)^2\ee^{-2\xstar}\xstar^4\simeq\frac{1}
{2\Delta\Nstar^2}\left(\frac{\phizero}{\Mp}\right)^2,
\qquad
\nS-1\simeq-2A\left(\frac{\Mp}
{\phizero}\right)^2\xstar^2\ee^{-\xstar}\simeq\frac{1}{2\Delta\Nstar}
\, ,
\end{equation}
and the running
\begin{equation}
\alphaS \simeq-2A^2\left(\frac{\Mp}{\phizero} \right)^4 \xstar^4
\ee^{-2\xstar} \simeq -\frac{1}{8\Delta\Nstar^2} \, .
\end{equation}
These estimates are in agreement with the ones of \Refc{Davis:2010it},
up to a missing factor $4$ in \Eq{eq:twi:xstar}. However, we have
checked that this does not affect the predictions in a significant
way.

It is also interesting to discuss the value of the scale $M$ since
this is important from the model building point of view. The CMB
normalization gives
\begin{eqnarray}
\frac{M^4}{\Mp^4}&=&720\pi^2 A^2\left(\frac{\Mp}{\phizero}\right)^2
\frac{\left[\ee^{-\xstar}\xstar\left(\xstar-2\right)\right]^2}
{\left(1-A\xstar^2\ee^{-\xstar}\right)^3}
\frac{\Qrms^2}{T^2}\, .
\end{eqnarray}
In the vacuum dominated approximation, the above expression simplifies
and gives $M^4/\Mp^4 \simeq
45\pi^2/\Delta\Nstar^2\phizero^2/\Mp^2\Qrms^2/T^2$. This leads to
\begin{equation}
\Mp R=\sqrt{\frac{2\calN}{45 A}}\frac{\Delta \Nstar}{\pi^3}
\frac{T}{\Qrms}\simeq 1.2\times 10^5 \sqrt{\calN}\,,
\end{equation}
where we have taken $\Delta \Nstar\simeq 60$. This also implies that
\begin{equation}
\frac{\phizero}{\Mp}\simeq \frac{1.35}{\sqrt{\calN}}\times 10^{-5}.
\end{equation}
Therefore, we have a rough determination of the compactification
radius. The model seems consistent since we obtain that $\Mp R\gg 1$,
in agreement with the assumptions made at the beginning of this
section.

The predictions for TWI are presented in \Fig{fig:CMBTWI}. The
reheating equation of state parameter $\wrehbar$ has been taken to be
$0$ since the potential is quadratic close to its minimum.  However,
since the details of reheating depend on the details of the
interactions between the inflaton field and the others degrees of
freedom in the theory, this parameter is a priori unspecified and
could very well take different values. In the \ASPIC code, $\wrehbar$
can be freely chosen. Anyway, since the reheating temperature is in
fact fully degenerate with the parameter $\xend$, these two parameters
cannot be constrained independently. One can check that the rough
description provided by \Eqs{eq:twi:roughpred} is correct: the model
typically predicts a small amount of gravitational waves which
increases with $\phizero$, and a deviation from scale invariance which
does not significantly depends on $\phizero$. When $\phizero/\Mp =
\order{1}$, however, one notices a turning point (at fixed values of
$\phizero$). This corresponds to the separation between two regimes,
one where $\xstar<\xepsoneMax$ and $\epsilon_1$ is an increasing
function of $x$ (hence $\epsilon_{1*}$ increases with $\xend$) and
another where $\xstar>\xepsoneMax$ and $\epsilon_1$ is a decreasing
function of $x$ (hence $\epsilon_{1*}$ decreases with $\xend$). If a
sufficient number of \efolds can be realized in the
$2<x<\xepsoneMax$ part of the potential, then $\epsilon_{2*}$ can
become negative. However, this mostly happens for fine-tuned values of
$\xend \simeq 2$.

\subsection{Generalized MSSM Inflation (GMSSMI)}
\label{sec:gmssmi}

As for the MSSMI models, see \sectionc{sec:mssmi}, GMSSMI scenarios
are based on the Minimal Supersymmetric Model (MSSM) in which a flat
direction is lifted by soft supersymmetry breaking terms and
by superpotential corrections. The potential is of the form
\begin{equation}
\label{eq:gmssmi:pot}
V(\phi)=\frac{1}{2}m_\phi^ 2\phi^2-A\frac{\lambda_n}{n}\frac{\phi^n}
{\Mp^{n-3}}+\lambda_n^2\frac{\phi^{2(n-1)}}{\Mp^{2(n-3)}}\,.
\end{equation}
The MSSMI model corresponds to $n=6$ and $A^2= 8(n-1)m_\phi^2$. This
last relation is of crucial importance since it implies an exactly flat
inflection point. Following \Refcs{Lyth:2006ec,Lyth:2006nx,
  BuenoSanchez:2006xk, Allahverdi:2006we,
  Allahverdi:2006wt,Allahverdi:2008bt,Chatterjee:2011qr}, one may
wonder whether the model is robust when this relation is not exactly
satisfied. In order to investigate this question, we therefore relax
the condition $A^2=8(n-1)m_\phi^2$. In this more general case, the
potential can be reparametrized in the form
\begin{equation}
\label{eq:gmssmi:pot2}
V(\phi)=M^4\left[\left(\frac{\phi}{\phizero}\right)^2-\frac{2}
{3}\alpha\left(\frac{\phi}{\phizero}\right)^6+\frac{\alpha}
{5}\left(\frac{\phi}{\phizero}\right)^{10}\right],
\end{equation}
where $\phizero\simeq 10^{14}\, \GeV$, this value being the same as
the one found in \sectionc{sec:mssmi}. The positive dimensionless
parameter $\alpha$ encodes any deviations from the MSSM case for which
it equals unity, $\alpha_{\mathrm{MSSM}}=1$.

\begin{figure}
\begin{center}
\includegraphics[width=\wdblefig]{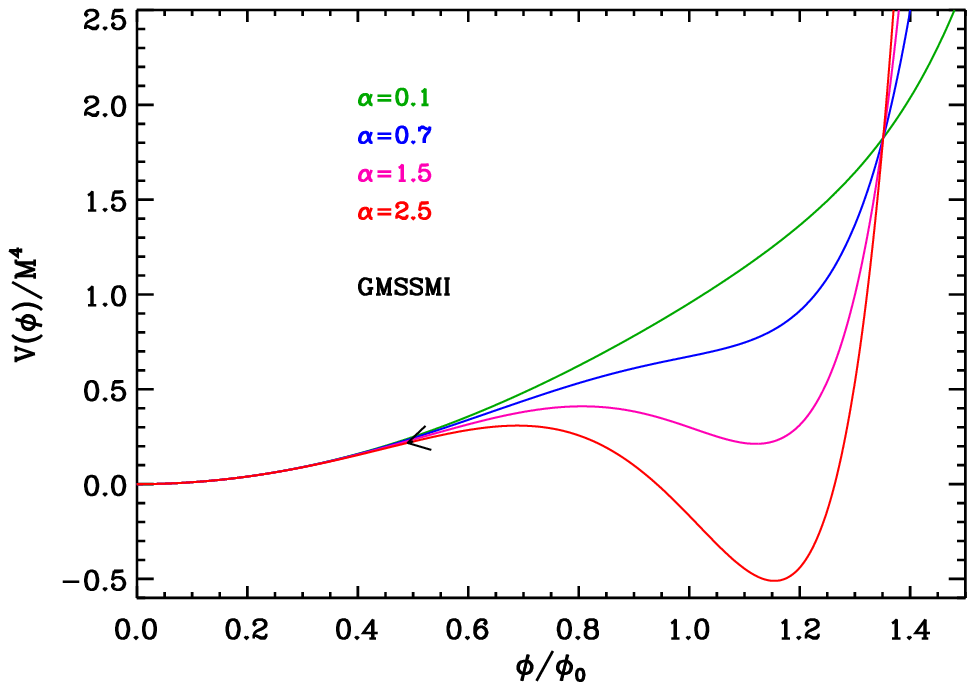}
\includegraphics[width=\wdblefig]{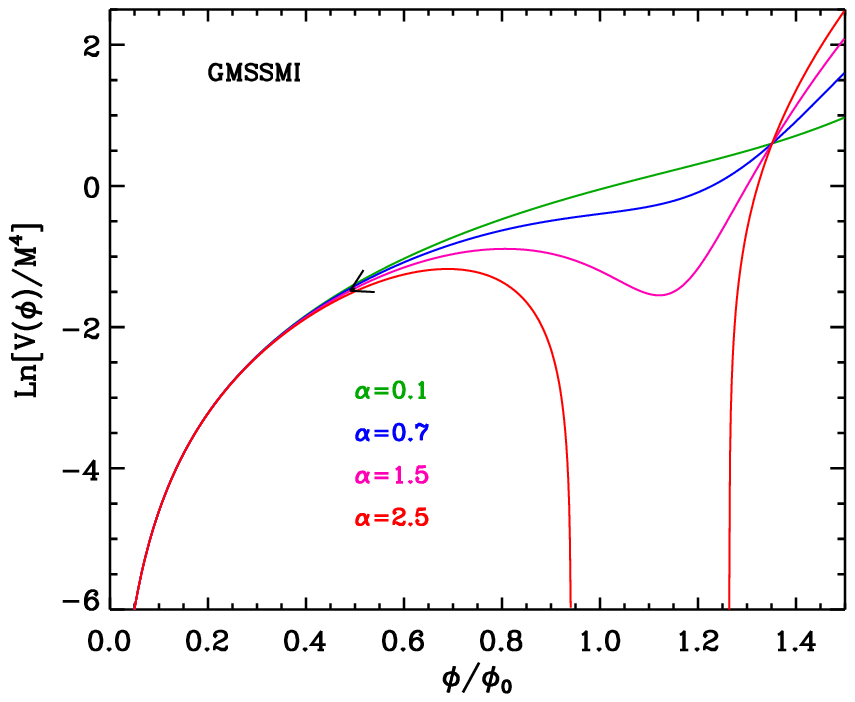}
\includegraphics[width=\wdblefig]{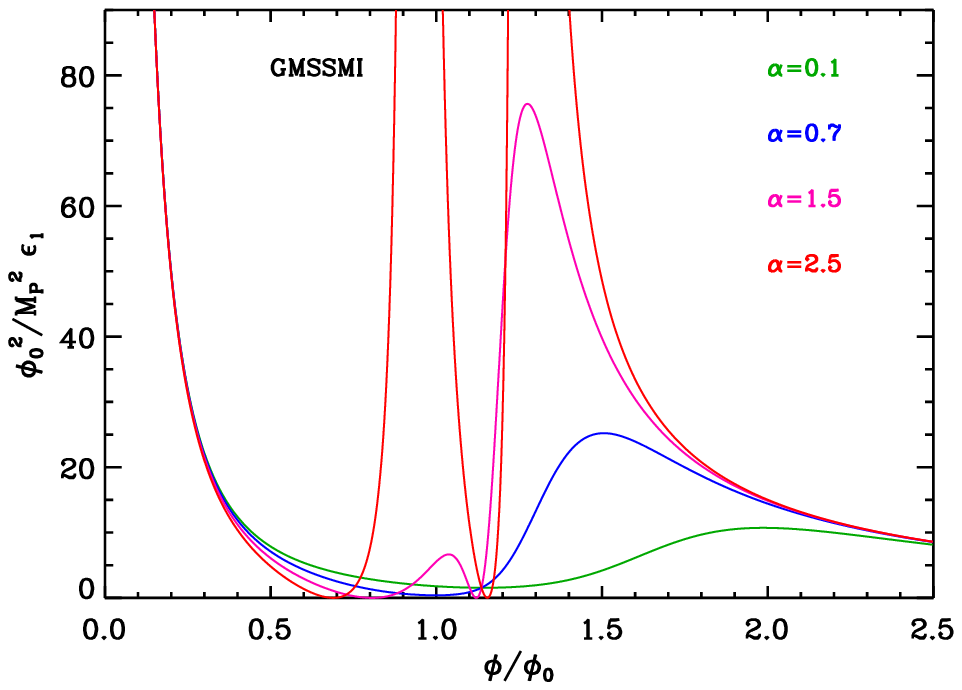}
\includegraphics[width=\wdblefig]{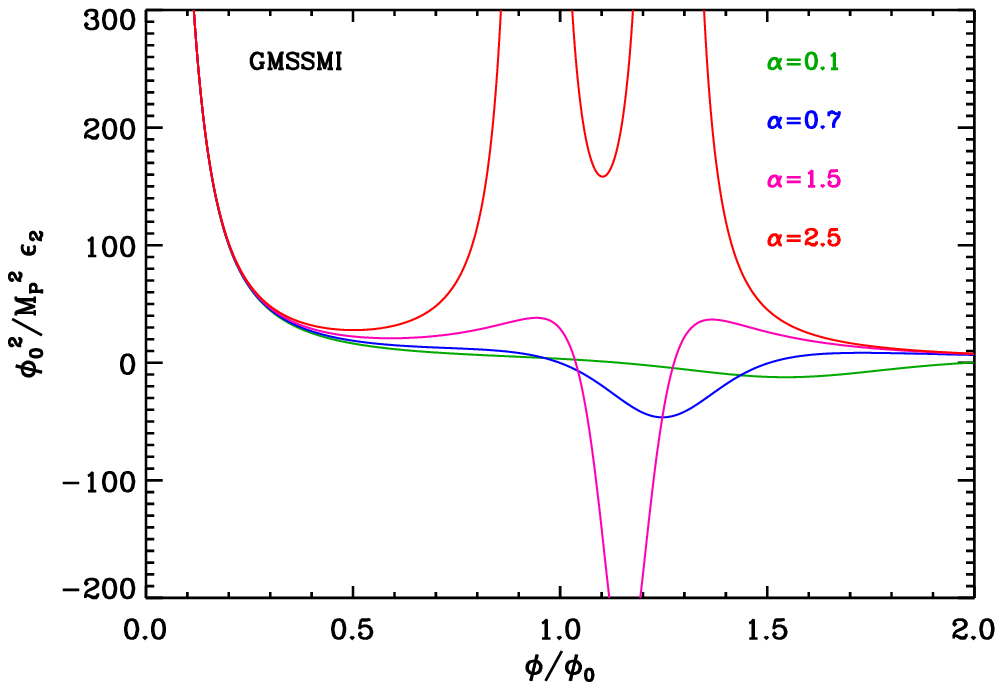}
\caption{GMSSM Inflation (GMSSMI). Top left panel: GMSSM Inflation
  potential \Eq{eq:gmssmi:pot2} for $\alpha=0.1,0.7,1.5,2.5$, 
  as a function of $\phi/\phizero$. Top right panel: logarithm of the  
  potentials for the same value of $\alpha$. Bottom left panel:
  slow-roll parameter $\epsilon _1$ for a potential with the same
  values of $\alpha$.  Bottom right panel: slow-roll
  parameter $\epsilon _2$ for a potential with the same values of
  $\alpha$. See discussion in the text body.}
\label{potGMSSMI}
\end{center}
\end{figure}

The potential is displayed in \Fig{potGMSSMI}, where four cases can be
distinguished. In the following, we define the quantity $x$ by the
expression
\begin{equation}
x \equiv \dfrac{\phi}{\phizero}\,.
\end{equation}
If $\alpha<9/25$, the second derivative of the potential does not
vanish and the potential is convex everywhere. This corresponds to the
case $\alpha=0.1$ case in \Fig{potGMSSMI}. If $9/25<\alpha<1$, the
potential has two inflection points $\xddVzeroPM$ and is concave in
between. It remains an increasing function of the field since its
first derivative never vanishes. This is illustrated with the case
$\alpha=0.7$ in \Fig{potGMSSMI}. If $\alpha=1$, this is the MSSM
inflation models (see \sectionc{sec:mssmi}) where the potential has a
flat inflection point. If $1<\alpha<9/5$, the potential decreases in
between $\xdVzeroPM$ but remains positive everywhere. This is
exemplified by the case $\alpha=1.5$ in \Fig{potGMSSMI}. Finally, if
$\alpha>9/5$, the potential becomes negative (hence is not properly
defined) between the two points $\xVzeroPM$ (see $\alpha=2.5$ in
\Fig{potGMSSMI}). The values of the field \vev's appearing in this
discussion are given by the following formulas:
\begin{equation}
  \xddVzeroPM = \left[\frac{5}{9} \left( 1\pm\sqrt{1 -
      \frac{9}{25\alpha}} \right) \right]^{1/4} ,\qquad
  \xdVzeroPM = \left(1\pm\sqrt{1
      -\frac{1}{\alpha}} \right)^{1/4} ,
      \label{eq:gmssmi:xdVzeroPM}
\end{equation}
and
\begin{equation}
  \xVzeroPM = \left[\frac{5}{3} \left(1\pm\sqrt{1
      -\frac{9}{5\alpha}} \right) \right]^{1/4}.
\end{equation}

Let us now calculate the first three Hubble flow functions in the
slow-roll approximation. They are given by
\begin{equation}
\begin{aligned}
  \epsilon_1 &= 450\left(\frac{\Mp}
  {\phizero}\right)^2\frac{\left(1-2\alpha x^4+\alpha x^8 \right)^2}{
    x^2\left(15-10\alpha x^4 +3\alpha x^8 \right)^2}\, ,\\
  \epsilon_2 &= 60\left(\frac{\Mp}{\phizero}\right)^2\frac{15+40\alpha 
  x^4+\alpha\left(20\alpha-78\right) x^8+3\alpha^2 x^{16}}{
   x^2\left(15-10\alpha x^4 +3\alpha x^8 \right)^2}\, ,
   \end{aligned}
\end{equation}
and
\begin{equation}
\begin{aligned}
  \epsilon_3 & = 60\left(\frac{\Mp}{\phizero}\right)^2
  \left[225 - 1800 \alpha x^4 
  + 60\alpha\left(69  + 10 \alpha\right) x^8
- 40 \left(189 -100 \alpha\right) \alpha^2 x^{12}
\right.\\& \left.
+10\alpha^2\left(243-504\alpha+402\alpha^2\right)x^{16} 
 +40\alpha^3\left(117-20\alpha\right) x^{20} 
 + 12\alpha^3\left(10\alpha-123\right) x^{24}
 \right.\\& \left.
  + 72 \alpha^4 x^{28} + 9 \alpha^4 x^32 \right] 
  \times \left[3375 x^2 + 4500 \alpha x^6 
  - 600\alpha\left(27+10 \alpha\right) x^{10}
   \right.\\& \left.
+100 \alpha^2 \left(261-20\alpha\right) x^{14}
 +10 \alpha^2 \left(200\alpha^2-840\alpha-621\right)x^{18} 
 + 60\alpha^3\left(69-20\alpha\right) x^{22}
  \right.\\& \left.
  +48\alpha^3 \left(10\alpha-9\right) x^{26} - 
 180 \alpha^4 x^{30} + 27 \alpha^4 x^{34}\right]^{-1}.
\end{aligned}
\end{equation}
The first two slow-roll parameters diverge when $x \rightarrow 0$ and
vanish asymptotically. In between, their shape depends on $\alpha$ as
it is represented in \Fig{potGMSSMI}. If $\alpha<1$, $\epsilon_1$
first decreases, reaches a local non-zero minimum where $\epsilon_2$
vanishes, then increases to reach a local maximum where $\epsilon_2$
vanishes again, and eventually decreases again. Let $\xepstwoZeroPM$
be the position of these two local extrema. From Ferrari's solutions
for depressed quartic equations one gets
\begin{equation}
\label{eq:gmssmi:xepsilon2=0}
\xepstwoZeroPM=\left[\frac{1}{2\alpha}\sqrt{\frac{5}{3}}\left(
    \sqrt{\Sigma}\pm2\sqrt{\frac{39}{5}\alpha-2\alpha^2-\frac{\Sigma}
    {4}- \frac{12}{\sqrt{15\Sigma}}\alpha^2}\right)\right]^{1/4},
\end{equation}
where
\begin{equation}
\begin{aligned}
  \delta &= \frac{736 \alpha^2}{25} - \frac{208 \alpha^3}{15} + 
  \frac{16 \alpha^4}{9}\, ,\\ 
  \Delta &= -\frac{430336 \alpha^4}{625}+\frac{612352 \alpha^5}{1125} - 
  \frac{20992 \alpha^6}{225} + \frac{256 \alpha^8}{243}\, ,\\ 
  \sigma &= -\frac{12896 \alpha^3}{125} + \frac{2944 \alpha^4}{25} - 
  \frac{416 \alpha^5}{15} 
  + \frac{64 \alpha^6}{27} + \frac{6}{5} \sqrt{15\Delta} \alpha\, ,\\
  \Sigma &= \frac{52 \alpha}{5} - \frac{8 \alpha^2}{3}+\frac{\delta}
  {\sigma^{1/3}}  +\sigma^{1/3},
\end{aligned}
\end{equation}
are intermediate quantities introduced solely to reduce the size of
\Eq{eq:gmssmi:xepsilon2=0}.  If $\alpha>1$, $\epsilon_1$ has two local
minimums located at $\xdVzero^{\pm}$ where it vanishes. In between it
reaches a local maximum or may even diverges for $\alpha>9/5$ (see
\Fig{potGMSSMI}). The slow-roll parameter $\epsilon_2$ vanishes when
$\epsilon_1$ reaches these local maxima, or diverge when $\epsilon_1$
does (for $\alpha>9/5$). As explained in \sectionc{sec:mssmi},
inflation is meant to proceed at $\phi\lesssim\phizero$. Let us assume
that inflation can end for $\epsilon_1>1$ between $x=0$ and the
position of the first minimum $\xepsoneMin$. Following the previous
considerations, this latter location is defined as
\begin{equation}
\label{eq:mssmi:xeps1min}
\xepsoneMin=\left\lbrace
\begin{aligned}
& \xepstwoZeroMinus \quad \textrm{if} \quad  \alpha<1 \\
& \xdVzeroMinus  \quad \textrm{if} \quad  \alpha>1 \\
\end{aligned} 
\right.,
\end{equation}
and provides an upper bound to $\xend$ the solution of
$\epsilon_1(\xend)=1$. This one can only be determined numerically.
\begin{figure}
\begin{center}
\includegraphics[width=\wdblefig]{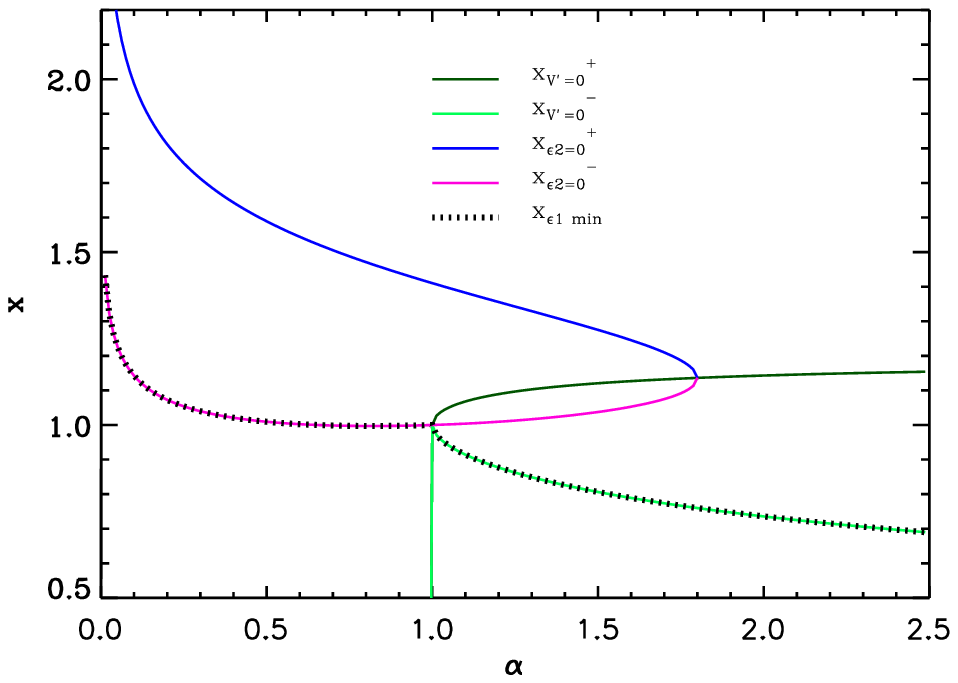}
\includegraphics[width=\wdblefig]{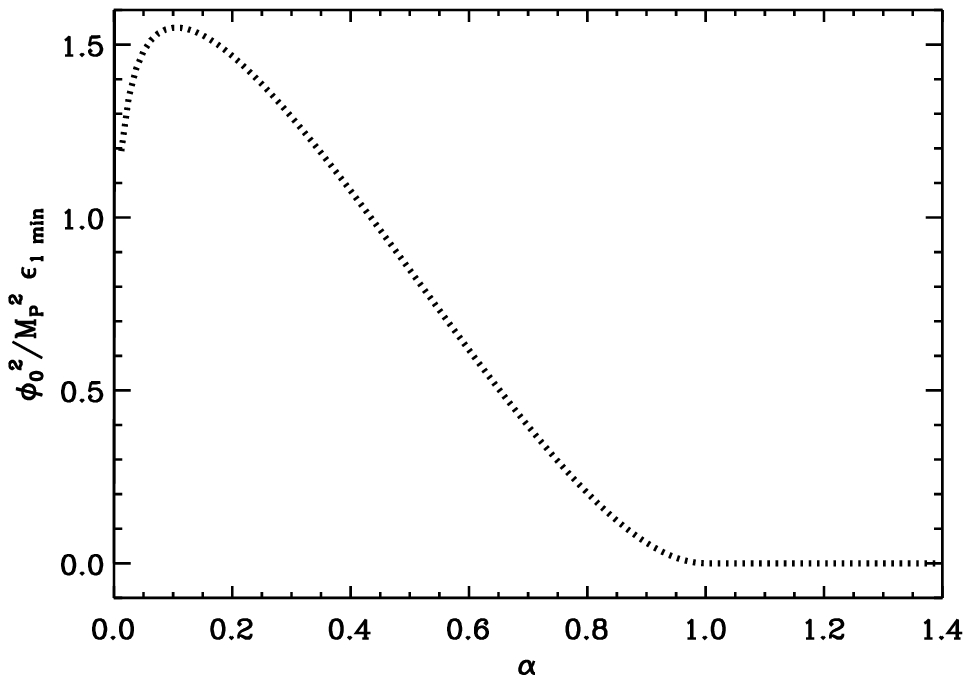}
\caption{GMSSM Inflation (GMSSMI).  Left panel: $\xepstwoZeroPM$
  defined in \Eq{eq:gmssmi:xepsilon2=0} and $\xdVzeroPM$ defined in
  \Eq{eq:gmssmi:xdVzeroPM} together with $\xepsoneMin$ [see
  \Eq{eq:mssmi:xeps1min}] as a function of $\alpha$.  Right panel:
  minimal value of the slow-roll parameter $\epsilon_1$ (rescaled by
  $\phizero^2/\Mp^2$) as a function of $\alpha$. When it is greater than
  unity, inflation cannot occur.}
\label{fig:GMSSMI:eps1min}
\end{center}
\end{figure}
The values of $\xepstwoZeroPM$ and $\xdVzeroPM$ in terms of $\alpha$
are displayed in the left panel in \Fig{fig:GMSSMI:eps1min} together
with $\xepsoneMin$. The right panel of \Fig{fig:GMSSMI:eps1min}
represents the value of the first slow-roll parameter at this minimum,
$\epsonemin = \epsilon_1(\xepsoneMin)$. For $\alpha<1$, one can see
that $\epsonemin<1$ only if the parameter $\alpha \lesssim 1$. This
defines a minimum value for $\alpha$, which depends on $\phizero$,
such that inflation can take place within this domain. When $\alpha
\simeq 1$, one can derive an approximated version of
\Eq{eq:gmssmi:xepsilon2=0}, namely,
$\xepstwoZero^-\simeq1-(1-\alpha)/32$. Plugging it into the expression
for $\epsilon_1$ one obtains
\begin{equation}
\epsonemin \simeq \dfrac{225}{32} (\alpha-1)^2 \dfrac{\Mp^2}{\phizero^2}\,,
\end{equation}
from which one gets
\begin{equation}
\label{eq:gmssmi:alphamin}
\alpha> 1-\frac{4\sqrt{2}}{15}\frac{\phizero}{\Mp}\, .
\end{equation}
For the value suggested in \Refc{Allahverdi:2006iq},
$\phizero/\Mp\simeq 10^{-4}$, one obtains $\alpha> 1-10^{-5}$, which
is in agreement with \Refc{Lyth:2006nx}, and shows that the model
needs to be sufficiently fine-tuned (\ie sufficiently close to regular
MSSM inflation) in order to be a viable inflationary model.

On top of that, as shall be seen now, the constraints on $\alpha$ are 
even tighter if one wants a sufficient number of \efolds to be 
produced. Let us thus turn to the slow-roll trajectory. It can be 
integrated, and leads to
\begin{equation}
\begin{aligned}
\Nend - N & = \frac{\phizero^2}{\Mp^2}\left\lbrace
-\frac{\xend^2-x^2}{20} - \frac{b_+}{10\sqrt{a_+}}
\left[\arctan \left(\sqrt{a_+}\xend^2 \right) -
  \arctan\left(\sqrt{a_+}x^2 \right) \right]
  \right. \\ - & \left.
\frac{b_-}{10\sqrt{a_-}}\left[\arctan \left(\sqrt{a_-} \xend^2 \right)-
  \arctan \left(\sqrt{a_-}x^2 \right)\right]\right\rbrace ,
\end{aligned}
\label{eq:gmssmi:traj}
\end{equation}
where
\begin{equation}
a_\pm = -\alpha\pm \sqrt{\alpha^2-\alpha}\,
,\qquad b_\pm = 2\frac{a_\pm+\alpha/3}{a_\pm-a_\mp}\, ,
\end{equation}
A few remarks are in order. Firstly, even if the terms appearing in
the previous expression are complex, their imaginary contributions
cancel out and the resulting expression is truly a real quantity.
Then, one can check that formally, when $\alpha\rightarrow 0$, one has
$a_\pm\rightarrow0$ and $b_{\pm}\rightarrow 1$, hence $N\simeq
-\left(x^2-\xini^2\right)/4$, which is precisely the LFI slow-roll
trajectory for $p=2$, see \sectionc{sec:lfi}. This is just a formal
check since $\alpha$ is meant to be tuned close to $1$ in the GMSSMI
scenario. Finally, let us notice that, in the case $\alpha<1$, and
contrary to the MSSM models ($\alpha=1$), the number of \efolds
never diverges at a given point $x$. Therefore, the total number of
\efolds is bounded from above for the field \vev's considered
here. Working out the limit of \Eq{eq:gmssmi:traj} when
$\alpha\rightarrow 1$, one has
\begin{equation}
\Nend - \Nini \leq
\left(\frac{\phizero}{\Mp}\right)^2\frac{\pi}{30}\frac{1}
     {\sqrt{1-\alpha}}\, .
\end{equation}
If one require at least $\Delta N=\Nend-\Nini$ \efolds during
inflation, then $\alpha$ has to be fine-tuned to
\begin{equation}
\alpha>1-\left(\frac{\phizero}{\Mp}\right)^4\frac{\pi^2}{900 \Delta 
N^2}\, .
\end{equation}
Remembering that the small parameter here is $\phizero/\Mp$, one can
see that it is a much tighter constraint than the one of
\Eq{eq:gmssmi:alphamin}. Taking $\phizero/\Mp \simeq 10^{-4}$ and
$\Delta N \simeq 50$, one obtains $\alpha> 1-10^{-22}$. This is
clearly an extreme fine-tuning which can even make the numerical
investigation of the model challenging\footnote{This exceeds the usual
  $64$ bits precision on floating point numbers (FP64).}. As explained
below, the same condition $\left\vert\alpha-1\right\vert<\phizero^
4/\Mp^4/\Delta N^2$ also applies to the case $\alpha>1$ in order to
maintain an acceptable deviation from scale invariance. This makes
GMSSM inflation a severely fine-tuned scenario. Let us also notice
that our parameter $\alpha$ is related to the parameter $\delta$ of
\Refc{BuenoSanchez:2006xk} by
$\delta=\sqrt{\alpha^{-2}-1}$. \Refc{BuenoSanchez:2006xk} finds that,
in order for the model to be compatible with the data, $\delta\simeq
10^{-20}$. Therefore, although our method slightly differs from that
of \Refc{BuenoSanchez:2006xk}, our results are in broad agreement.

Finally, the amplitude of the CMB anisotropies fixes the parameter $M$
to
\begin{equation}
  \left(\frac{M}{\Mp}\right)^4=2880\pi^2\frac{\Mp^2}{\phizero^2} 
  \frac{\left(1-2\alpha
    x^4_*+\alpha \xstar^8\right)^2}{x^4_*\left(1-\frac{2}{3}\alpha 
    x^4_*+\frac{\alpha}{5}
    x^8_*\right)^3} \frac{\Qrms^2}{T^2}\, .
\end{equation}
As explained in \sectionc{sec:mssmi}, this leads to $M/\Mp\simeq
10^8\,\GeV$ for $\phizero/\Mp\simeq 10^{-4}$.

The reheating consistent slow-roll predictions of the GMSSMI models
are displayed in \Figs{fig:CMBGMSSMIalpha>1},
\ref{fig:CMBGMSSMIalpha<1}, for $\alpha>1$ and $\alpha<1$,
respectively. The reheating equation of state parameter $\wrehbar$ has
been taken to $0$ since the potential is quadratic close to its
minimum. In both cases, one can see that in the limit
$\alpha\rightarrow 1$, the standard MSSM predictions are recovered,
see \Fig{fig:CMBMSSMI}. The amount of gravitational waves $r$ seems to
be quite independent on $\alpha$ and, therefore, is similar to its
regular MSSM counterpart. On the other hand, the spectral index $\nS$
strongly depends on $\alpha$. In the case $\alpha>1$, larger values of
$\alpha -1$ worsens the spectral index problem, already present in
standard MSSMI. These models are therefore strongly disfavored by the
data. In the case $\alpha<1$ however, there is a very narrow range of
acceptable values for $\alpha$. They are well inside the
$\left\vert\alpha-1\right\vert<\phizero^ 4/\Mp^4/\Delta N^2$ condition
and the spectral index is inside the two-sigma confidence
intervals. But, as can be seen in \Fig{fig:CMBGMSSMIalpha<1}, the
spectral index varies so quickly with $\alpha$ that one has to
fine-tune the power of the fine-tuning to remain inside the two-sigma
contours. In \Refcs{Lyth:2006nx, BuenoSanchez:2006xk,
  Chatterjee:2011qr, Allahverdi:2006we, Allahverdi:2006wt}, it is
argued that, since the flat saddle point condition is robust against
radiative corrections, such a fine-tuning may not be a
problem. However, as explained here and in \sectionc{sec:mssmi}, if
the flat saddle point condition is exactly satisfied, the model is
disfavored by the observations because the spectral index is too
red. The only way out is therefore to detune the condition $\alpha=1$
at an extremely fine-tuned level.

\subsection{Generalized Renormalizable Point Inflation (GRIPI)}
\label{sec:gripi}

As for the MSSMI models (see \sectionc{sec:mssmi}) and for the RIPI
models (see \sectionc{sec:ripi}), the GRIPI models have a potential of
the form
\begin{equation}
\label{eq:gripi:pot}
V(\phi)=\frac{1}{2}m_\phi^ 2\phi^2-A\frac{\lambda_n}{n}\frac{\phi^n}
{\Mp^{n-3}}+\lambda_n^2\frac{\phi^{2(n-1)}}{\Mp^{2(n-3)}}\,.
\end{equation}
In \sectionc{sec:ripi}, the particular example $n=3$ is discussed in
the case where the potential has a flat inflection point, \ie when
$A^2= 16m_\phi^2$. Then, as studied in \sectionc{sec:gmssmi} for
MSSMI, comes the question of what happens when we relax this
condition. To address this issue, it is convenient to reparametrize
the potential as
\begin{equation}
\label{eq:gripi:pot2}
V(\phi)=M^4\left[\left(\frac{\phi}{\phizero}\right)^2-\frac{4}
{3}\alpha\left(\frac{\phi}{\phizero}\right)^3+\frac{\alpha}
{2}\left(\frac{\phi}{\phizero}\right)^{4}\right],
\end{equation}
where the positive dimensionless parameter $\alpha$ encodes the
deviation from the RIPI case (that is to say $\alphaRIPI=1$). This
model was studied in \Refc{Hotchkiss:2011am} and in
\Refcs{Aulakh:2012st, Charanjit:2012aa}. In the first reference, the
mass $m_\phi$ is fixed by the soft supersymmetry breaking terms and,
in \sectionc{sec:ripi}, it was shown that this leads to
$\phizero\simeq 10^{14}\, \GeV$. However, in \Refcs{Aulakh:2012st,
  Charanjit:2012aa}, the scale $m_\phi$ is no longer controlled by the
soft supersymmetry breaking terms but by the right-handed neutrino
mass in Type I supersymmetric seesaw and this leads to a different
value for $\phizero$, namely $\phizero\simeq 10^{17}\,
\GeV$. Therefore, in what follows, we will use both values.

\begin{figure}
\begin{center}
\includegraphics[width=\wdblefig]{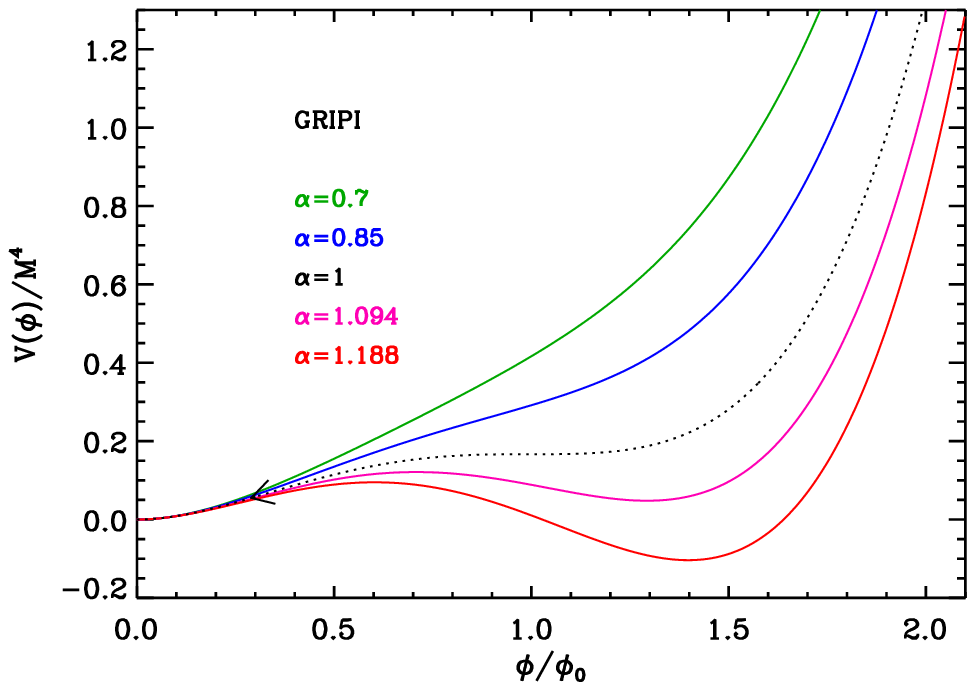}
\includegraphics[width=\wdblefig]{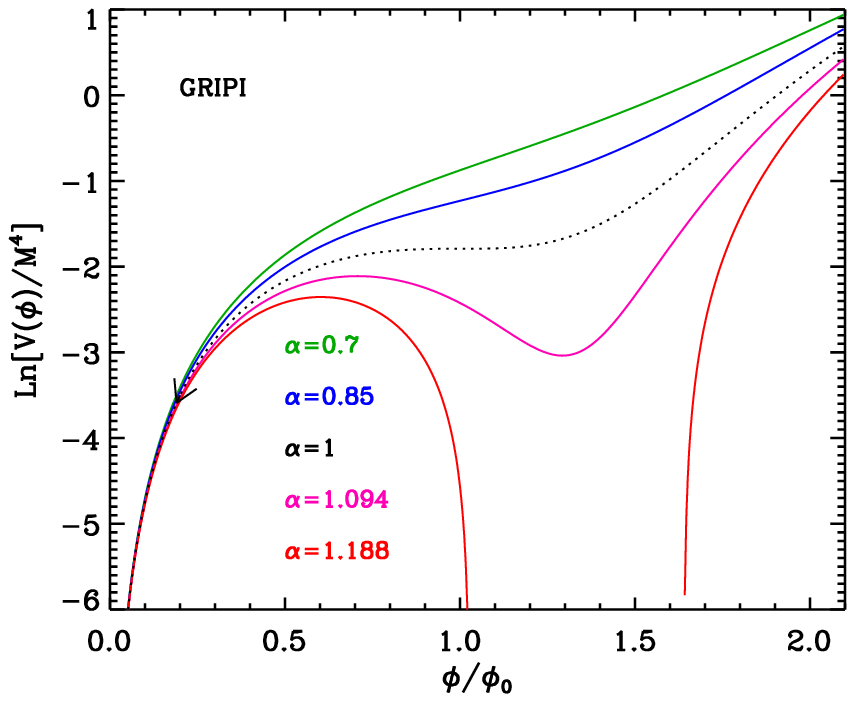}
\includegraphics[width=\wdblefig]{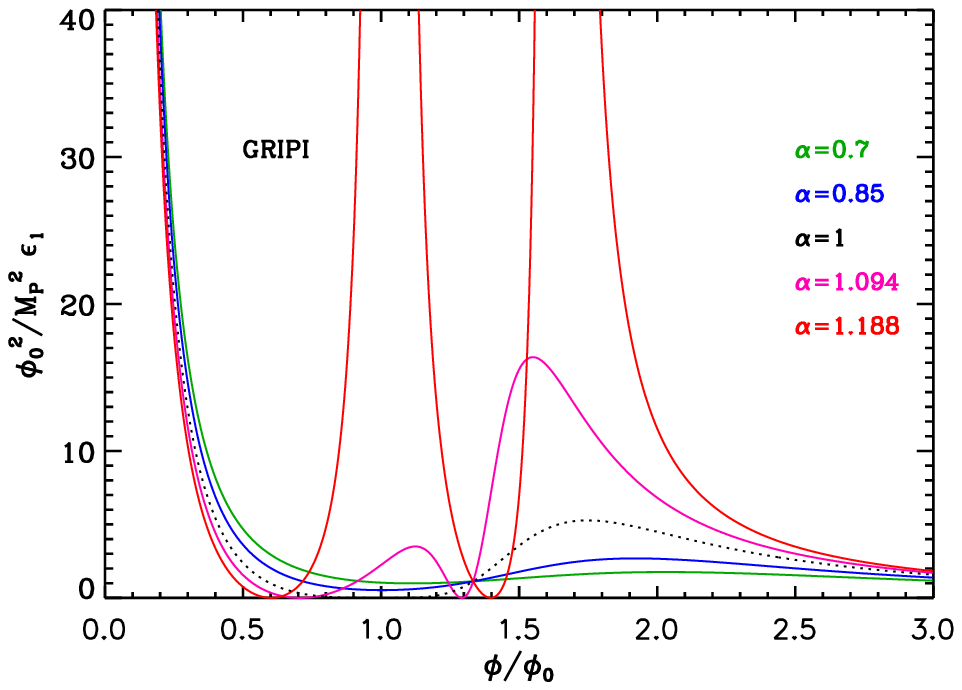}
\includegraphics[width=\wdblefig]{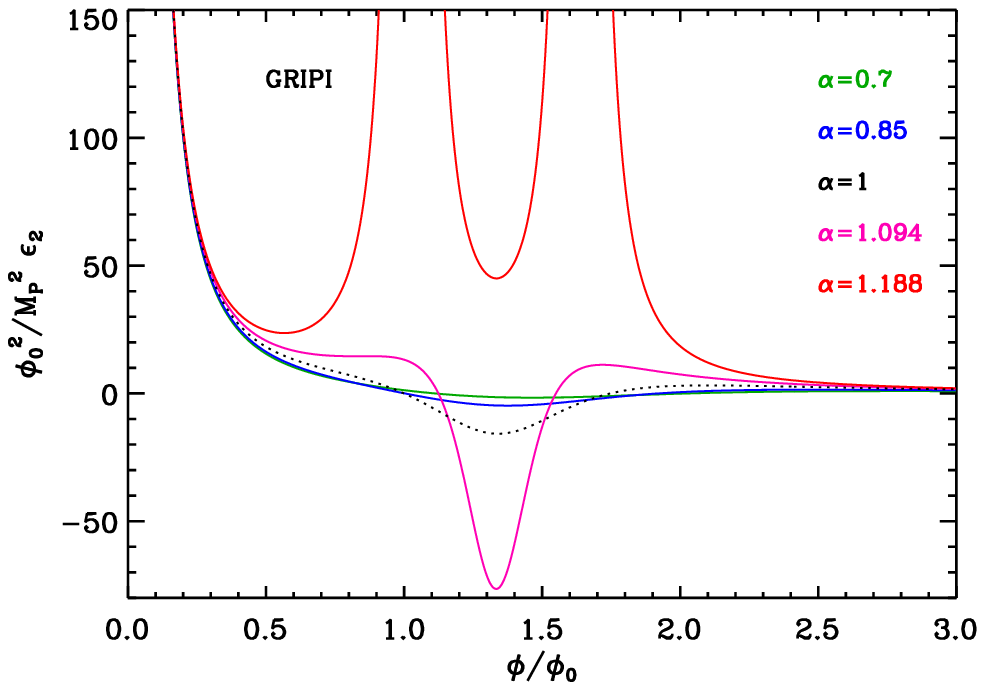}
\caption{Top left panel: Generalized Renormalizable Point Inflation
  (GRIPI) potential given by \Eq{eq:gripi:pot2} for
  $\alpha=0.7,0.85,1,1.094,1.188$, as a function of
  $\phi/\phizero$. Top right panel: logarithm of the potentials for
  the same values of $\alpha$. Bottom left panel: slow-roll parameter
  $\epsilon _1$ rescaled by $\Mp^2/\phizero^2$, for GRIPI models with
  the same values of $\alpha$.  Bottom right panel: slow-roll
  parameter $\epsilon _2$, rescaled by $\Mp^2/\phizero^2$. A
  description of these various quantities can be found in the text.}
\label{potGRIPI}
\end{center}
\end{figure}

The potential is displayed in \Fig{potGRIPI}, where four cases can be
distinguished. In the following, for convenience, we use the quantity
$x$ defined by
\begin{equation}
x \equiv \dfrac{\phi}{\phizero}\,.
\end{equation}
If $\alpha<3/4$, the second derivative of the potential does not
vanish and the potential is convex everywhere. This corresponds to the
case $\alpha=0.7$ case in \Fig{potGRIPI}. If $3/4<\alpha<1$, the
potential has two inflection points $\xddVzeroPM$ and is concave in
between. It remains an increasing function of the field since its
first derivative never vanishes. This is illustrated by the case
$\alpha=0.85$ in \Fig{potGRIPI}. If $\alpha=1$, then this is the RIPI
model (see \sectionc{sec:ripi}) where the potential has a flat
inflection point. If $1<\alpha<9/8$, then the potential decreases
between the two values of $x$, $\xdVzeroPM$, for which the derivative
is zero, but remains positive everywhere. Typically, this corresponds
to the case $\alpha=1.094$ in \Fig{potGRIPI}. Finally, if
$\alpha>9/8$, then the potential becomes negative (hence is not
properly defined everywhere) between $\xVzeroPM$ (see the case
$\alpha=1.188$ in \Fig{potGRIPI}). The values of the field \vev in
this discussion are given by the following formulas:
\begin{equation}
  \xddVzeroPM = \frac{2}{3}\left(1\pm\sqrt{1-\frac{3}{4\alpha}}
  \right), \qquad
  \xdVzeroPM = 1\pm\sqrt{\frac{\alpha-1}{\alpha}}\, ,
      \label{eq:gripi:xdVzeroPM}
\end{equation}
and
\begin{equation}
  \xVzeroPM = \frac{4}{3}\left(1\pm\sqrt{1-\frac{9}{8\alpha}}\right).
\end{equation}

Let us now calculate the first Hubble flow functions in the slow-roll
approximation. They are given by
\begin{equation}
\begin{aligned}
  \epsilon_1 &= 72\left(\frac{\Mp}
  {\phizero}\right)^2\frac{\left(1-2\alpha x+\alpha x^2 \right)^2}{
    x^2\left( 6-8\alpha x +3\alpha x^2\right)^2}\, ,\\
  \epsilon_2 &= 24\left(\frac{\Mp}{\phizero}\right)^2\frac{6-16\alpha 
  x+\left(3+16\alpha\right)\alpha x^2-12\alpha^2 x^{3}+3\alpha^2 x^4}{
    x^2\left( 6-8\alpha x +3\alpha x^2\right)^2}\, ,
   \end{aligned}
\end{equation}
and
\begin{equation}
\begin{aligned}
  \epsilon_3 = & 24\left(\frac{\Mp}{\phizero}\right)^2
  \left[36 - 216 \alpha x+ 30\alpha\left(3  + 16 \alpha\right) x^2
- 8\left(45+64 \alpha\right) \alpha^2 x^{3}
\right.\\ &\left.
+2\left(27+276\alpha+128\alpha^2\right)\alpha^2x^{4} 
 -2\left( 208\alpha+81\right)\alpha^3 x^{5} 
 +9\left(1+28\alpha\right)\alpha^3 x^{6}
 \right.\\ &\left.
  - 72 \alpha^4 x^{7} + 9 \alpha^4 x^8 \right] 
  \times \left[x^2\left(6-8\alpha x+3\alpha x^2\right)^2
  \left(6-16\alpha x + 3\alpha x^2+16\alpha^2 x^2
 \right. \right. \\& \left.\left. 
  -12\alpha^2 x^3 + 3 \alpha^2 x^4\right)\right]^{-1} .
\end{aligned}
\end{equation}
The first two slow-roll parameters diverge when $x \rightarrow 0$ and
asymptotically goes to zero when $x \rightarrow \infty$. In between,
their behavior depends on $\alpha$ as can be seen in
\Fig{potGRIPI}. If $\alpha<\alphazero$, where
\begin{equation}
\alphazero=\frac{3}{16}\left[ 5-3^{2/3}
\left(6-2\sqrt{3}\right)^{-1/3}-2^{-2/3} 
\left(9-3\sqrt{3}\right)^{1/3}\right]\simeq 0.4671 ,
\end{equation}
$\epsilon_1$ monotonically decreases with $x$. If $\alphazero<\alpha<1$,
$\epsilon_1$ first decreases, reaches a local non-vanishing minimum at
a value of $x$ for which $\epsilon_2$ vanishes, then increases to
reach a local maximum where $\epsilon_2$ vanishes again, and
eventually decreases for $x\rightarrow \infty$, as already
mentioned. Let $\xepstwoZeroPM$ be the position of these two local
extrema. Similarly to \Eq{eq:gmssmi:xepsilon2=0} for the generalized
MSSM inflation models, analytic expressions can be obtained for
these two quantities using Ferrari's solutions for depressed quartic
equations. They are implemented in \ASPIC but are not displayed here
since this does not add much to the discussion. If $\alpha>1$,
$\epsilon_1$ has two local minima located at $\xdVzero^{\pm}$ where it
vanishes. In between it reaches a local maximum or may even diverge
for $\alpha>9/8$ (see \Fig{potGRIPI}). The slow-roll parameter
$\epsilon_2$ vanishes when $\epsilon_1$ reaches these local maxima, or
diverge when $\epsilon_1$ itself diverges (for $\alpha>9/8$). 

As explained in \sectionc{sec:ripi}, inflation is supposed to proceed
at $\phi\lesssim\phizero$. Let us assume that inflation ends by
violation of slow-roll between $x=0$ and the position of the first
minimum $\xepsoneMin$. Following the previous considerations, this
latter value of $x$ is defined by
\begin{equation}
\label{eq:gripi:xeps1min}
\xepsoneMin=\left\lbrace
\begin{aligned}
& \xepstwoZeroMinus \quad \textrm{if} \quad  \alphazero<\alpha<1 \\
& \xdVzeroMinus  \quad \textrm{if} \quad  \alpha>1 \\
\end{aligned} 
\right.,
\end{equation}
and, moreover, provides an upper bound to determine $\xend$ [\ie the
solution of the equation $\epsilon_1(\xend)=1$]. Let us emphasize that
this one can only be determined numerically.
\begin{figure}
\begin{center}
\includegraphics[width=\wdblefig]{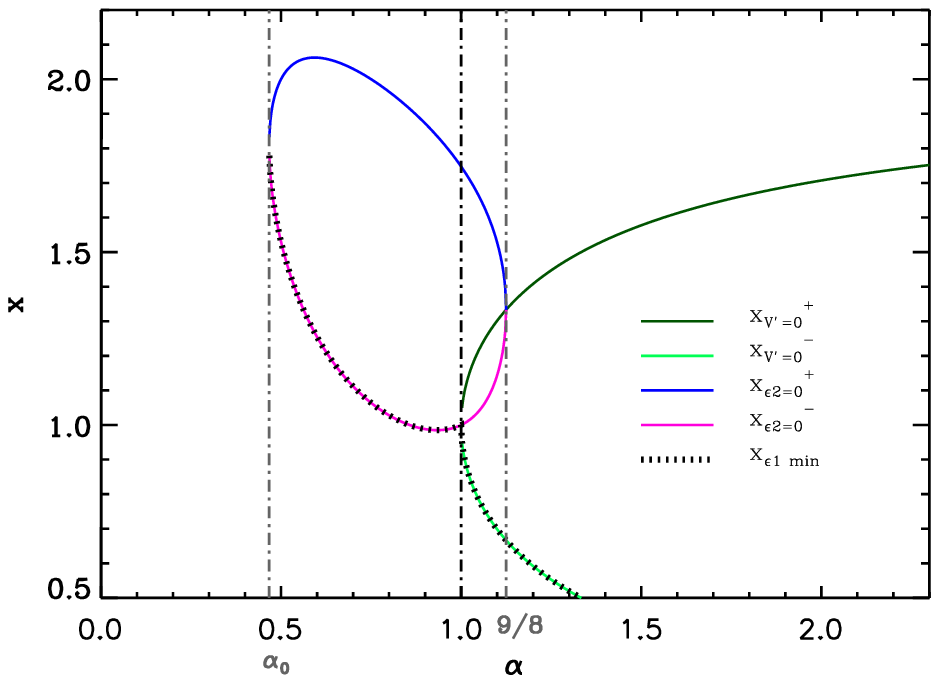}
\includegraphics[width=\wdblefig]{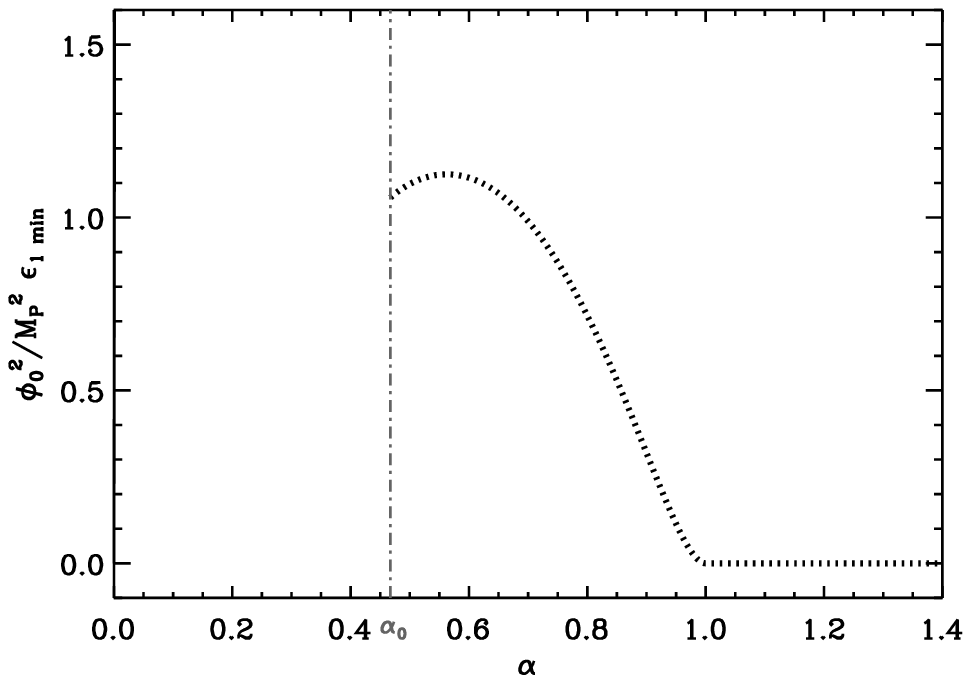}
\caption{Left panel: $\xepstwoZeroPM$ and $\xdVzeroPM$ [defined in
    \Eq{eq:gripi:xdVzeroPM}] together with $\xepsoneMin$ [see
    \Eq{eq:gripi:xeps1min}] as a function of $\alpha$.  Right panel:
  minimal value of the slow-roll parameter $\epsilon_1$, \ie
  $\epsilon_1(\xepsoneMin)$, rescaled by $\phizero^2/\Mp^2$, as a
  function of $\alpha$. When it is greater than unity, inflation
  cannot occur.}
\label{fig:GRIPI:eps1min}
\end{center}
\end{figure}
The values of $\xepstwoZeroPM$ and $\xdVzeroPM$ in terms of $\alpha$
are displayed in the left panel of \Fig{fig:GRIPI:eps1min} together
with $\xepsoneMin$. The right panel of \Fig{fig:GRIPI:eps1min}
represents the value of the first slow-roll parameter at this minimum,
$\epsonemin = \epsilon_1(\xepsoneMin)$. For $\alpha<\alphazero$, one has
$\epsilon_1(x=1)>1.5\Mp^2/\phizero^2$ and, recalling that typically
$\phizero\simeq 10^{14}\,\GeV$ or $\phizero\simeq 10^{17}\,\GeV$, one sees
that inflation cannot proceed in this case. For $\alphazero<\alpha<1$,
one has $\epsonemin<1$ only if the parameter $\alpha \lesssim 1$. This
defines a minimum value for $\alpha$, which depends on $\phizero$,
allowing for inflation to take place. When $\alpha \simeq 1$, one can
derive an approximated formula for $\xepstwoZero^-$, namely,
$\xepstwoZero^-\simeq1-(1-\alpha)/2$. Plugging it into the expression
for $\epsilon_1$ one obtains
\begin{equation}
\epsonemin \simeq 72 (\alpha-1)^2 \dfrac{\Mp^2}{\phizero^2}\,,
\end{equation}
from which it follows that
\begin{equation}
\label{eq:gripi:alphamin}
\alpha> 1-\frac{\sqrt{2}}{12}\frac{\phizero}{\Mp}\, .
\end{equation}
With $\phizero/\Mp\simeq 10^{-1}$, one obtains $\alpha> 0.99$, which
shows that the model needs to be sufficiently fine-tuned such that it
becomes very similar to the regular RIPI scenario. If, on the other
hand, $\phizero/\Mp\simeq 10^{-4}$, the constraint is much
tighter. As discussed in \Refcs{Aulakh:2012st, Charanjit:2012aa}, one
of the main advantage of the model studied in those references is that
a value $\phizero \simeq 10^{17}\GeV$ leads to a less severe fine
tuning problem than $\phizero \simeq 10^{14}\GeV$.

However, the constraints on $\alpha$ are tighter to get a
sufficient number of \efolds. Let us therefore now
turn to the determination of the slow-roll trajectory. It can be
integrated exactly to give
\begin{equation}
\begin{aligned}
\Nend - N & = \frac{\phizero^2}{\Mp^2}\left\lbrace
\frac{5-4\alpha}{12\sqrt{\alpha\left(1-\alpha\right)}}
\arctan\left(\frac{x-1}{\sqrt{1/\alpha-1}}\right)
+\frac{x}{2}\left(\frac{x}{4}-\frac{1}{3}\right)
\right. \\&\left.
+\left(\frac{1}{8\alpha}-\frac{1}{6}\right)
\ln\left[1+\alpha x\left(x-2\right)\right]
-\frac{5-4\alpha}{12\sqrt{\alpha\left(1-\alpha\right)}}
\arctan\left(\frac{\xend-1}{\sqrt{1/\alpha-1}}\right)
\right. \\&\left.
-\frac{\xend}{2}\left(\frac{\xend}{4}-\frac{1}{3}\right)
-\left(\frac{1}{8\alpha}-\frac{1}{6}\right)
\ln\left[1+\alpha \xend\left(\xend-2\right)\right]
\right\rbrace .
\end{aligned}
\label{eq:gripi:traj}
\end{equation}
Exactly the same remarks we have made for the GMSSMI model also applies
here (see section~\ref{sec:gmssmi}). In particular, for $\alpha<1$,
and contrary to the RIPI models ($\alpha=1$), the number of
\efolds never diverges at a given point $x$. Therefore, the total
number of \efolds is bounded by some maximal finite value. From
\Eq{eq:gripi:traj} when $\alpha\rightarrow 1$, one has
\begin{equation}
\Nend - \Nini \leq
\left(\frac{\phizero}{\Mp}\right)^2\frac{\pi}{24}\frac{1}
     {\sqrt{1-\alpha}}\, .
\end{equation}
Therefore, if one require at least $\Delta N=\Nend-\Nini$ \efolds, one
has to fine-tune $\alpha$ to
\begin{equation}
\alpha>1-\left(\frac{\phizero}{\Mp}\right)^4\frac{\pi^2}{576 \Delta 
N^2}\, .
\end{equation}
Remembering that the small parameter here is $\phizero/\Mp$, one can
see that it is a much tighter constraint than the one of
\Eq{eq:gripi:alphamin}. Taking $\phizero/\Mp \simeq 10^{-1}$ and
$\Delta N \simeq 50$, one obtains $\alpha> 1-10^{-10}$. This makes the
fine-tuning quite important and, as explained below, the same
condition $\left\vert\alpha-1\right\vert<\phizero^ 4/\Mp^4/\Delta N^2$
also applies to the case $\alpha>1$ to maintain an acceptable
deviation from scale invariance, making the whole class of models
fine-tuned. However, as already mentioned above, the value
$\phizero\simeq 10^{17}\GeV$ makes the fine-tuning issue easier to
accept than the value $\phizero\simeq 10^{14}\GeV$.

Finally, the amplitude of the CMB anisotropies fixes the parameter $M$
to
\begin{equation}
  \left(\frac{M}{\Mp}\right)^4=622080\pi^2\frac{\Mp^2}{\phizero^2} 
  \frac{\left(1-2\alpha
    \xstar+\alpha \xstar^2\right)^2}{x^4_*\left(6-8\alpha 
    \xstar+3\alpha
    \xstar^2 \right)^3} \frac{\Qrms^2}{T^2}\, .
\end{equation}
As explained in \sectionc{sec:mssmi}, this leads to $M/\Mp\simeq
10^{13}\,\GeV$ for $\phizero/\Mp\simeq 10^{-4}$.

The reheating consistent slow-roll predictions of the GRIPI models are
displayed in \Figs{fig:CMBGRIPIalpha>1}, \ref{fig:CMBGRIPIalpha<1},
for $\alpha>1$ and $\alpha<1$ respectively, and for values of $\phizero$
such that $\phizero\simeq 10^{17}\,\GeV$:
$\phizero/\Mp=10^{-2},\,10^{-1.5},\,10^{-1},\,10^{-0.5}, \, 1$. The
reheating equation of state parameter $\wrehbar$ has been taken to $0$
since the potential is quadratic close to its minimum. In both cases,
one can see that in the limit $\alpha\rightarrow 1$, the standard RIPI
predictions are recovered, see \Fig{fig:CMBRIPI}. The amount of
gravitational waves $r$ seems to be quite independent on $\alpha$
while the spectral index $\nS$ strongly depends on it. In the
case $\alpha>1$, the fine-tuning is as important as in the case
$\alpha<1$ as mentioned above. Considering values of $\alpha$ very
different from $1$ worsens the spectral index problem, already present
in standard RIPI. These models are therefore strongly disfavored by
the data. In the case $\alpha<1$ however, there is a very narrow range
of acceptable values for $\alpha$. They are well inside the
$\left\vert\alpha-1\right\vert<\phizero^ 4/\Mp^4/\Delta N^2$ condition
and the spectral index is inside the two-sigma confidence
intervals. But as can be seen in \Fig{fig:CMBGRIPIalpha<1}, the
spectral index varies so quickly with $\alpha$ that, even if the
fine-tuning is less problematic than in the GMSSMI case (due to the
different value of $\phizero$), it is still very important.

\subsection{Brane SUSY Breaking Inflation (BSUSYBI)}
\label{sec:bsusybi}

This model has been studied in \Refc{Dudas:2012vv} in the context of
superstrings models\footnote{see Eq.~(1.1) and Eq.~(2.9) in that
  reference.}. The potential is a sum of two exponential terms
\begin{equation}
V(\phi)=M^4\left(\ee^{\sqrt{6}\frac{\phi}{\Mp}} + \ee^{\sqrt{6} \gamma
  \frac{\phi}{\Mp}} \right),
\end{equation}
one is a ``hard'' exponential brought about by a SUSY breaking
mechanism and the other is a ``slow-roll term'' having
$0<\gamma<1/\sqrt{3}$ and that dominates the eventual inflationary
dynamics. It was shown in \Refc{Dudas:2012vv} that the inflationary
dynamics can also generate superimposed oscillations in the primordial
power spectrum but we will not focus on this case since, obviously,
slow-roll is not satisfied in this situation~\cite{Martin:2003sg,
  Martin:2004iv, Martin:2004yi}. Let us also notice that if the term
in $\sqrt{6}$ in the first exponential function is relaxed to be a
free parameter, the potential becomes as in \Refc{Trudeau:2011ew}, \ie
a general exponential brane potential. Defining
\begin{equation}
x \equiv \dfrac{\phi}{\Mp}\,,
\end{equation}
the first three Hubble flow functions in the slow-roll approximation
read
\begin{equation}
\epsilon_1 = 3 \left(\dfrac{\ee^{\sqrt{6}x}+ \gamma\ee^{\sqrt{6}\gamma
    x}} {\ee^{\sqrt{6}x}+\ee^{\sqrt{6}\gamma x} }\right)^2 ,\qquad
\epsilon_2 = -12\left(\gamma-1\right)^2
\frac{\ee^{\sqrt{6}\left(\gamma +1 \right)x}}
     {\left(\ee^{\sqrt{6}x}+\ee^{\sqrt{6}\gamma x} \right)^2}\,,
\end{equation}
and
\begin{equation}
\epsilon_3 = 6\left(1-\gamma \right) \frac{
       \left(\ee^{\sqrt{6}x} - \ee^{\sqrt{6}\gamma x}\right)
       \left(\ee^{\sqrt{6}x}+\gamma\ee^{\sqrt{6}\gamma x}\right)}
     {\left(\ee^{\sqrt{6}x}+\ee^{\sqrt{6}\gamma x}\right)^2}\, .
\end{equation}

\begin{figure}
\begin{center}
\includegraphics[width=\wdblefig]{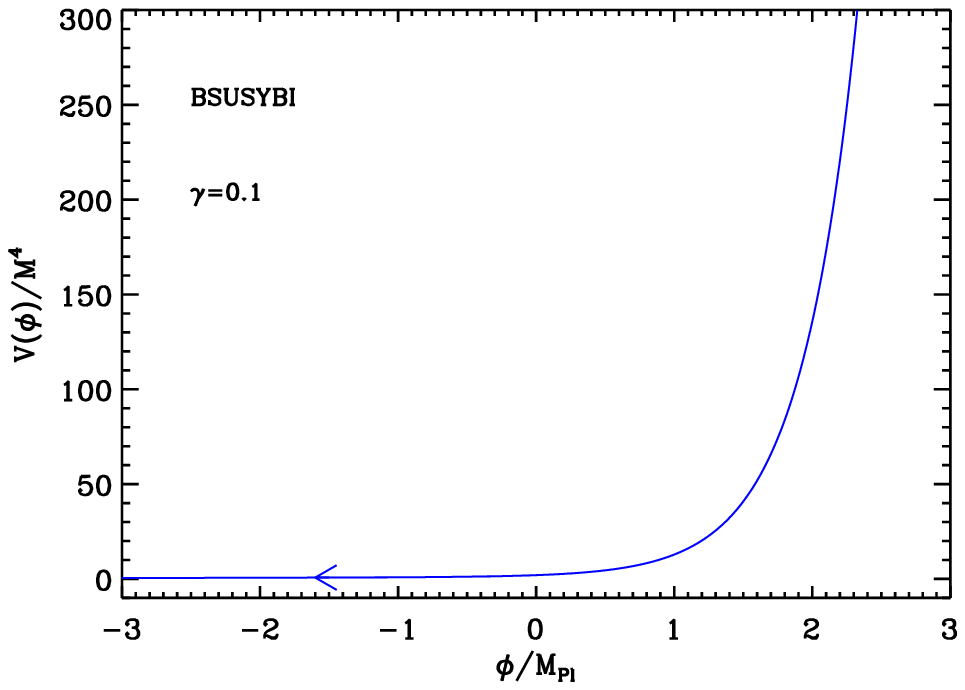}
\includegraphics[width=\wdblefig]{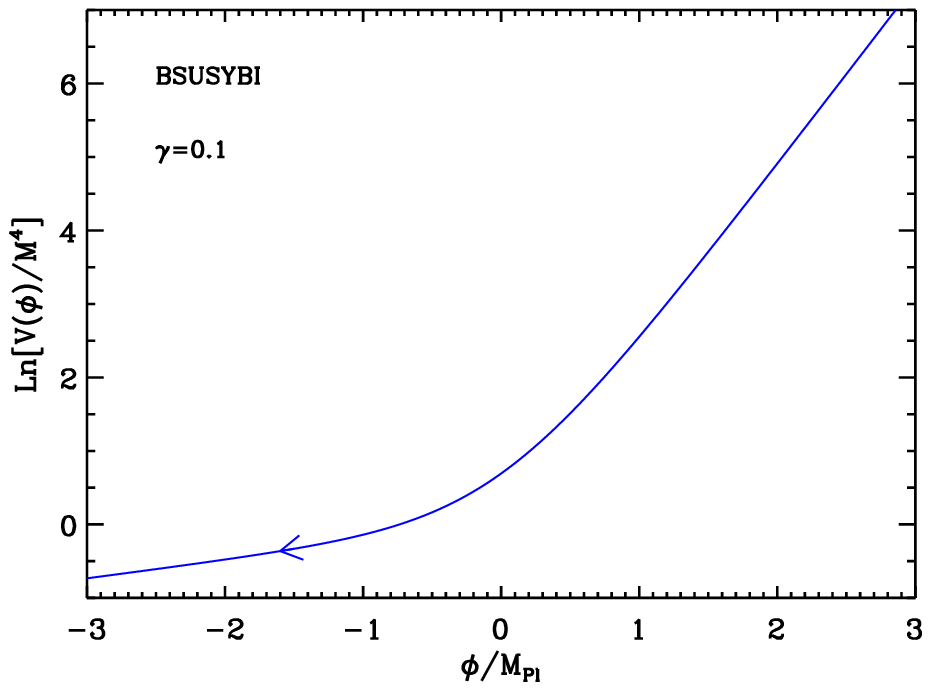}
\includegraphics[width=\wdblefig]{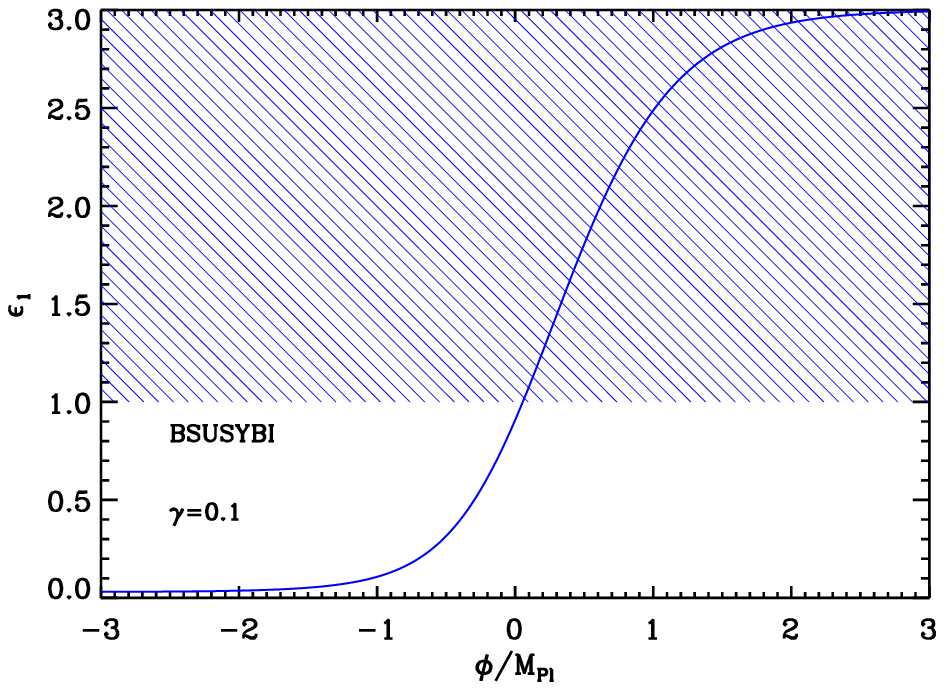}
\includegraphics[width=\wdblefig]{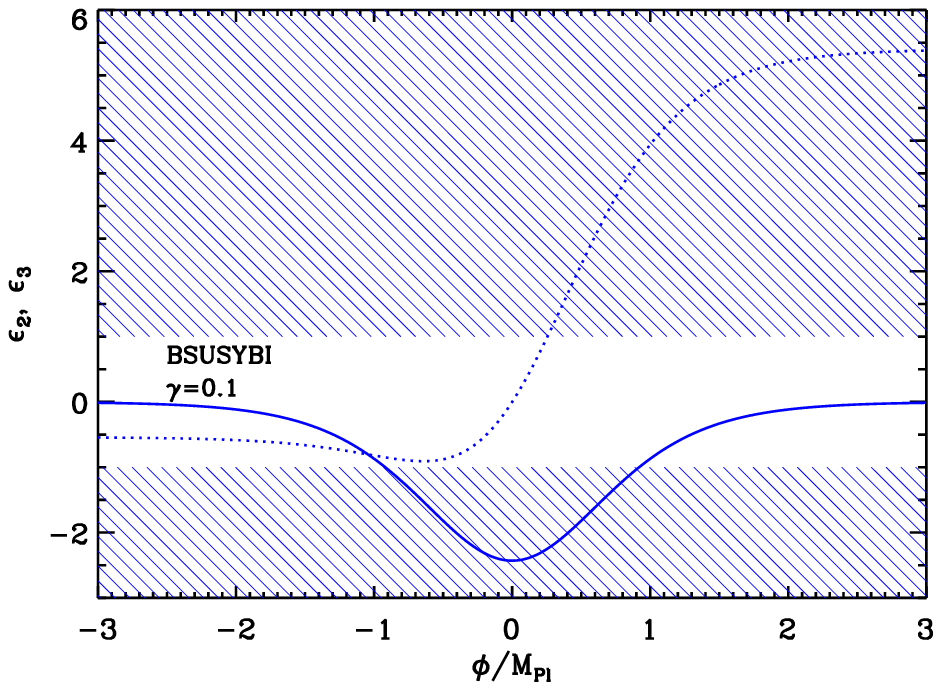}
\caption{Brane SUSY Breaking Inflation (BSUSYBI) for
  $\gamma=0.1$. Upper panels: the potential and its logarithm. Bottom
  left panel: the first slow-roll parameter $\epsilon_1$ as a function
  of the field value, the shaded area indicates where inflation
  stops. Bottom right panel: slow-roll parameter $\epsilon_2$ and
  $\epsilon_3$.}
\label{potBSUSYBI}
\end{center}
\end{figure}
These functions together with the potential are displayed in
\Fig{potBSUSYBI}. The two exponential components are clearly visible
on the plot of the logarithm of the potential. The required flatness
of the potential is realized only along the $\gamma$ branch and for
negative values of $x$. The first Hubble flow function $\epsilon_1$ is
an increasing function of $x$ which varies between its asymptotic
values:
\begin{equation}
\lim_{x\rightarrow -\infty} \epsilon_1 = 3\gamma^2 ,\qquad \lim_{x
  \rightarrow +\infty} \epsilon_1 = 3.
\end{equation}
For $\gamma$ small enough ($\gamma<1/\sqrt{3}$), there is a regime
where it is less than unity. This regime is given by the condition
$x<\xepsoneOne$ with
\begin{equation}
\label{eq:bsusybi:xeps1}
\xepsoneOne=\frac{1}{\sqrt{6}\left(\gamma-1\right)}
\ln\left(\frac{\sqrt{3}-1}{1-\gamma\sqrt{3}}\right).
\end{equation}
As a result, inflation can only proceed in the domain $x<\xepsoneOne$
and it never stops. Hence the need for an extra-parameter
$\xend$ encoding the field value at which some unspecified mechanism
(such as a tachyonic instability) is triggered and stops
inflation. Let us notice that the slow-roll parameter $\epsilon_2$ is
always negative and goes to zero at large $|x|$ with a local minimum
in $x=0$ equals to $\epstwoMin=-3\left(\gamma-1\right)^2$.  Finally, the
slow-roll parameter $\epsilon_3$ vanishes when $x=0$ and shares the
same sign as $x$. Its asymptotic values are
\begin{equation}
  \lim_{x\rightarrow -\infty} \epsilon_3 = 6\gamma
  \left(\gamma-1\right), \qquad \lim_{x\rightarrow +\infty} \epsilon_3
  = 6\left(1-\gamma\right).
\end{equation}

\begin{figure}
\begin{center}
\includegraphics[width=\wsingfig]{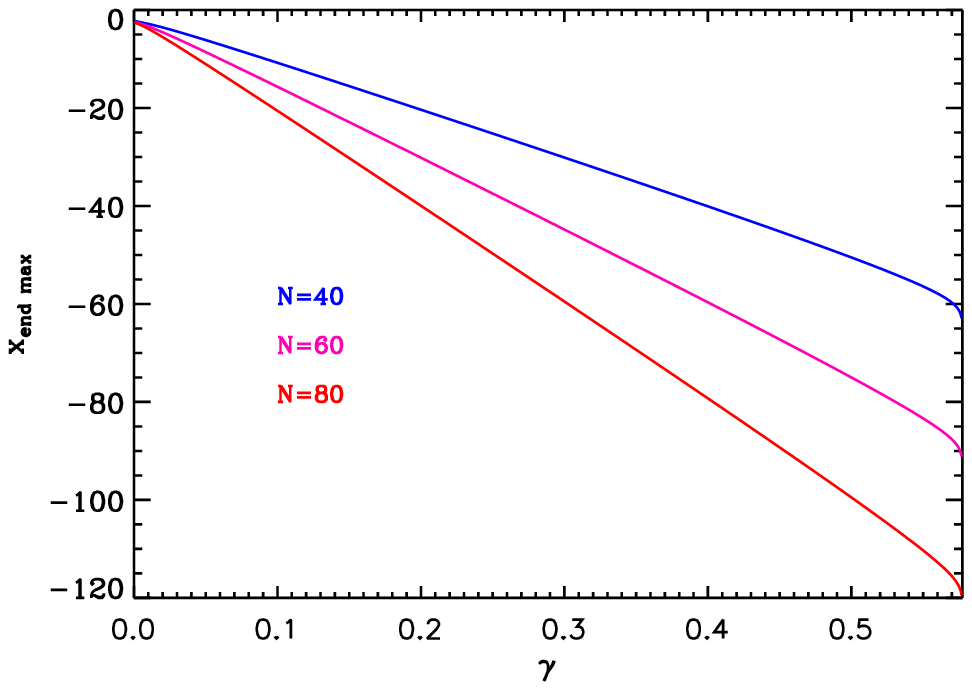}
\caption{Maximum value of $\xend$ in order to realize $N$ \efolds of
  inflation between $\xepsoneOne$ and $\xend$ as a function of
  $0<\gamma<1/\sqrt{3}$. This condition defines a prior for the model
  parameter $\xend$, which is the region lying under the curves on the
  figure.}
\label{fig:bsusybi:prior}
\end{center}
\end{figure}

The slow-roll trajectory can be integrated and gives
\begin{equation}
N-\Nend=-\frac{1}{\sqrt{6}}\left(x-\xend\right)
+\frac{1}{6\gamma}\ln\left[\frac{1+\gamma\ee^{\sqrt{6}\left(\gamma-1\right)x}}
{1+\gamma\ee^{\sqrt{6}\left(\gamma-1\right)\xend}}\right].
\end{equation}
This equation cannot be analytically inverted but since inflation
requires $x<\xepsoneOne$, it shows that $\xend$ should not be too
close to $\xepsoneOne$ in order to realize enough \efolds of
inflation. This puts some upper bound on $\xend$, that can be computed
numerically and that is displayed in \Fig{fig:bsusybi:prior}.  This
value $\xendmax$ defines a prior for the model parameter $\xend$,
which is the region lying under the curves on the figure.

Integrating \Eq{eq:phistarlnrrad} finally gives the field value
$\xstar$ at which the pivot mode crossed the Hubble radius during
inflation. The parameter $M$ being fixed by the amplitude of the CMB
anisotropies
\begin{eqnarray}
\left(\frac{M}{\Mp}\right)^4&=&4320\pi^2
\frac{\left(\ee^{\sqrt{6}\xstar}+\gamma\ee^{\sqrt{6}\gamma \xstar}\right)^2}
{\left(\ee^{\sqrt{6}\xstar}+\ee^{\sqrt{6}\gamma \xstar}\right)^3}
\frac{\Qrms^2}{T^2}\, .
\end{eqnarray}
The reheating consistent slow-roll predictions of the BSUSYBI models
have been plotted in \Fig{fig:CMBBSUSYBI}. The parameter $\xend$
varies between $2 \xendmax < \xend < \xendmax$ with $\xendmax<0$,
under which the predictions of the model coincide with those of PLI
(see \sectionc{sec:pli}). Large values for the parameter $\gamma$ are
disfavored and it has to be smaller than $\lesssim 5\times 10^{-2}$ to
generate a reasonable amount of gravitational waves.

\subsection{Tip Inflation (TI)}
\label{sec:ti}

\subsubsection{Theoretical Justifications}
\label{subsubsec:theoryti}

This model is a scenario based on string theory in which the motion of
branes in extra-dimensions causes the four-dimensional spacetime to
inflate, see for instance \Refcs{Burgess:2006cb, Lorenz:2007ze,
  Krause:2007jk, Baumann:2007ah, Baumann:2007np, DeWolfe:2007hd,
  Pajer:2008uy, Chen:2008au}. Let us assume string theory with flux
compactification. In this situation, the six-dimensional Calabi-Yau
space has generically the shape of a bulk with warped throat(s)
attached to it. The metric in the bulk is usually not known but, along
the throat, explicit examples are available. A representative case is
the Klebanov-Strassler throat~\cite{Klebanov:2000hb} for which one can
write the metric as
\begin{equation}
\dd s^2=h^{-1/2}(r)\eta_{\mu \nu}\dd x^{\mu}\dd x^{\nu}
+h^{1/2}(r)\left(\dd r^2+r^2\dd s_5^2\right).
\label{eq:ksthroat}
\end{equation}
The function $h(r)$ describes the warping along the radial coordinate
$r$ of the throat. We see that the throat is in fact a cone with
five-dimensional sections given by the metric $\dd s_5^2$. For a
conifold, these sections are two spheres $S_2\times S_3$ which shrink
to zero at the tip of the cone~\cite{Candelas:1989js}. Let us recall
that a conifold can also be defined by the equation
$\sum_{A=1}^4\left(Z_A\right)^2=0$, \ie a six-dimensional (or three
complex dimension) surface in $\mathbb{C}^4$. However, if one has a
deformed conifold, then, at the tip the $S_2$ sphere shrinks to zero
but the $S_3$ remains finite~\cite{Candelas:1989js}. A deformed
conifold can similarly be defined by the equation
$\sum_{A=1}^4\left(Z_A\right)^2=\varepsilon^2$ and, at the tip, one
has $\sum_{A=1}^4\left\vert Z_A\right\vert ^2=\varepsilon^2$. Usually
brane inflation takes place when a brane is moving along the radial
direction of the throat, see \sectionc{sec:bi}. Here, following
\Refc{Pajer:2008uy}, we will consider a different situation, namely
the case of a brane moving at the tip of the deformed conifold. In
addition, we will not only consider radial motion only but also
angular motion.

Technically, the above model can be described in the framework of
supergravity (viewed, in this context, as a low energy effective field
theory). Let us assume that there is a $D$3-brane moving at the tip
and that complex structure moduli and the dilaton are stabilized,
thanks to the presence of fluxes. Furthermore, following
\Refc{Pajer:2008uy}, we suppose that there is only one volume modulus,
$\rho$, plus three fields $z_i$, $i=1,\cdots,3$ describing the $D3$-brane
position. It follows that the corresponding K\"ahler potential is
given by
\begin{equation}
K\left(\rho,z_i,z_i^{\dagger}\right)=-3\Mp^2\ln \left[\rho+\rho^{\dagger}
-\gamma k\left(z_i,z^{\dagger}_i\right)\right],
\end{equation}
where $k$ is a function of the brane coordinates and $\gamma$ is a
constant (of mass dimension $-2$) related to the brane tension $T_3$,
an approximate expression of which will be given below. In the
vicinity of the deformed conifold tip, the function $k$ takes the form
\begin{equation}
k\left(z_i,z^{\dagger}_i\right)=k_0+c\varepsilon^{-2/3}
\left(\sum_{A=1}^4\left \vert Z_A\right \vert^2-\varepsilon^2\right).
\end{equation}
Here $c$ is a numerical constant $c=2^{1/6}/3^{1/3}\simeq 0.77$ and
$k_0$ stands for the value of the function $k$ at the tip. The
quantity $\varepsilon^{2/3}=\rtip$ can be viewed as the radius of the
tip as illustrated in Figs.~1 and 2 of \Refc{Pajer:2008uy}.

The last ingredient of the model is a stack of $n$ $D7$-branes placed
far from the tip. Then, the superpotential (Kuperstein
embedding~\cite{Kuperstein:2004hy}) can be written as
\begin{equation}
W=W_0+A(z_1)\ee^{-a \rho}=W_0+A_0\left(1-\frac{z_1}{\mu}\right)^{1/n}\ee^{-a \rho}.
\end{equation}
In this expression, $\mu^{2/3}$ represents the distance between the
stack of $D7$-branes and the tip (see Fig.~2 of \Refc{Pajer:2008uy}
for an illustration). We always assume that this distance is much
larger than the size of the tip, \ie $\epsilon/\mu\ll 1$. The
quantities $W_0$, $A_0$ and $a$ are constants. It is interesting to
remark that the above superpotential only depends on $z_1$ and
therefore breaks the symmetry of the tip.

We are now in a position where the potential and the kinetic term can
be calculated for the fields $z_i$ and $\rho$. The $F$-term potential
reads
\begin{eqnarray}
V(\sigma,x_1) &=& \frac{2a\ee^{-a\sigma}}{\Mp^2U^2}
\left(\frac{aU}{6}\vert A\vert^2\ee^{-a\sigma}
+\vert A\vert^2\ee^{-a\sigma}-\vert W_0A\vert\right)
\nonumber \\
& & +\frac{\ee^{-2a\sigma}}{3\Mp^2\gamma U^2}
\frac{\vert A\vert^2}{n^2\mu^2}\frac{\varepsilon^{2/3}}{c}
\left(1-\frac{x_1^2}{\varepsilon^2}\right)
\left(1-\frac{x_1}{\mu}\right)^{-2}+\frac{D}{U^b},
\end{eqnarray}
where we have taken, from the definition $z_i=x_i+iy_i$, $z_1=x_1$ at
the tip. Because of our choice of the superpotential, $V$ no longer
depends on $x_2$, $x_3$. In the above expression, we have defined
$\rho=\sigma+i\tau$ and $\tau$ is chosen such that $V$ is minimal. The
quantity $U$ is defined by $U=\rho+\rho^{\dagger}-k=2\sigma-k_0$ at
the tip. Finally, the last term $D/U^b$, with $D$ and $b$ constant, is
an uplifting term which is added in order to avoid having an anti-de
Sitter minimum. In practice, uplifting potentials generically have
$b=3$~\cite{Kachru:2003aw}.

The calculation of the kinetic term is difficult since the K\"ahler
matrix mixes all the fields $z_i$. For this reason, it is easier to
use another parametrization such where $z_1=\varepsilon \cos
\varphi$, $z_2=\varepsilon \sin \varphi \cos \theta$, $z_3=\varepsilon
\sin \varphi \sin \theta \cos \psi$ and $z_4=\varepsilon \sin \varphi
\sin \theta \sin \psi$, as appropriate since the tip of the deformed
conifold is $S_3$. In this case, the K\"ahler matrix becomes diagonal
and expanding everything in the small parameter $\epsilon/\mu\ll 1$,
one obtains
\begin{equation}
\label{eq:pottiinter1}
V(\sigma,\varphi)=\Lambda(\sigma)+B(\sigma)\cos \varphi+C(\sigma)\sin^2\varphi 
+\cdots ,
\end{equation}
where 
\begin{eqnarray}
\Lambda (\sigma)&=&  \frac{2a\vert A_0\vert\ee^{-a\sigma}}{\Mp^2U^2}
\left(\frac{aU}{6}\vert A_0\vert \ee^{-a\sigma}
+\vert A_0\vert \ee^{-a\sigma}-\vert W_0\vert\right)+\frac{D}{U^b}\,, \\
B(\sigma) &=& \frac{2a\vert A_0\vert\ee^{-a\sigma}}{\Mp^2U^2n}
\frac{\varepsilon}{\mu}\left(-\frac{aU\vert A_0\vert}{3}\ee^{-a\sigma}
-2\vert A_0\vert \ee^{-a\sigma}+\vert W_0\vert\right),\\
C(\sigma) &=& \frac{\vert A_0\vert^2\ee^{-2a\sigma}}{3\Mp^2U^2\gamma\mu^2n^2}
\frac{\varepsilon^{2/3}}{c}.
\end{eqnarray}

Let us now discuss this result. If one ignores, for the moment, all
terms depending on the brane position, it remains only the term
$\Lambda(\sigma)$ which is nothing but the
Kachru-Kallosh-Linde-Trivedi (KKLT) potential for the volume
modulus~\cite{Kachru:2003aw}. We see that in absence of the uplifting
term $D/U^b$, its minimum given by $\partial \Lambda/\partial \sigma
=0$ would be located at $\sigma =\sigma_0$, solution of the implicit
equation
\begin{equation}
W_0=-A_0\left[1+\frac{a}{3}\left(2\sigma_0-k_0\right)\right]\ee^{-a\sigma_0}.
\end{equation}
The corresponding value of the potential would actually be negative
(anti-de Sitter) and given by
\begin{equation}
\Lambda(\sigma_0)=-\frac{a^2\vert A_0\vert^2}{3\Mp^2U}\ee^{-2a\sigma_0}<0.
\end{equation}
Hence the required uplifting term from which one can find a new
minimum at which $V$ is positive. This is precisely how KKLT managed
to find a de Sitter minimum instead of an anti-de Sitter one for the
first time in string theory~\cite{Kachru:2003aw}.

If the position of the minimum were not changed by adding the uplifting
term, one would obtain a vanishing value of $V$ for
\begin{equation}
D_0 = \frac{a^2\vert A_0\vert^2U^{b-1}(\sigma_0)}{3\Mp^2}\ee^{-2a\sigma_0}.
\end{equation}
This suggests to introduce a new parameter $\beta$, defined by
\begin{equation}
\beta \equiv D \dfrac{3 \Mp^2}{a^2\vert A_0\vert^2U^{b-1}(\sigma_0)}
\ee^{2 a \sigma_0},
\end{equation}
such that one can trade $D$ for $\beta$ in all the uplifting
terms. Therefore, $\beta=1$ represents a situation in which the
potential is uplifted while the position of its minimum is
unchanged. In general, as expected in presence of the brane, the KKLT
minimum $\sigma_0$ of $\Lambda(\sigma)$ will be shifted. The
correction due to the uplifting terms can be evaluated perturbatively
and one obtains the following expression
\begin{equation}
\sigma_{\min}=\sigma_0 + \frac{b \beta}{2a^2\sigma_0}+\cdots,
\end{equation}
valid provided $b\beta/(2a^2\sigma_0)\ll 1$. For $\beta=0$, one
recovers that $\sigma_{\min}=\sigma_0$ as expected without uplifting
terms (and with a negative minimum for $V$). There are other
corrections to the position of the minimum due to the presence of the
brane but one can show that they do not play an important role (they
are calculated in \Refc{Pajer:2008uy}). The final argument consists in
considering that the modulus is stabilized at this minimum. Then, one
obtains a single field model
$V(\varphi)=V(\sigma_\mathrm{min},\varphi)$ where the coefficients in
\Eq{eq:pottiinter1} are now given by
\begin{eqnarray}
\Lambda
\left(\sigma_\mathrm{min}\right)\equiv \Lambda 
&\simeq &\frac{a^2\vert A_0\vert^2\ee^{-2a \sigma_0}}
{6\Mp^2\sigma_0}\left[\left(\beta-1\right)+\cdots \right], \\
\label{eq:coefB}
B\left(\sigma_\mathrm{min}\right) \equiv B &\simeq& 
\frac{a\vert A_0\vert^2\varepsilon\ee^{-2a\sigma_0}}
{6\Mp^2n\mu \sigma_0^2}
\left[\left(b\beta-3\right)+\frac{b\beta}{4a\sigma_0}\left(14-3b\beta\right)
+\cdots\right], \\
C\left(\sigma_\mathrm{min}\right) \equiv C &\simeq & 
\frac{\vert A_0\vert^2\varepsilon^{2/3}\ee^{-2a\sigma_0}}
{12\Mp^2n^2\mu^2 \sigma_0^2\gamma c}+\cdots .
\end{eqnarray}
The above relations express the parameters of the potential in terms
of the stringy parameters. We see that, if $\beta>1$, we have that the
KKLT potential is positive at the minimum that could account for a
cosmological constant today for
$\beta-1=\order{\sigma_0^{-2}}$~\cite{Pajer:2008uy}.

Finally, the kinetic term for $\varphi $ remains to be
calculated. Using the explicit form of the K\"ahler metric, one
obtains
\begin{equation}
K_{I\bar{J}}\partial_\mu z^I\partial^{\mu}z^{\bar{J}}\simeq \frac{3\Mp^2}{U}\gamma 
c\varepsilon^{4/3}
\partial_{\mu}\varphi\partial^{\mu}\varphi+\cdots ,
\end{equation}
where, at the minimum, one has
\begin{equation}
\gamma \simeq \dfrac{\sigma_0 T_3}{3\Mp^2}\,,
\end{equation}
$T_3$ being the brane tension. Therefore, in the large volume limit,
the canonical field $\phi$ is $\phi=
\sqrt{T_3c}\varepsilon^{2/3}\varphi$. As a consequence, the final form
of the potential reads
\begin{equation}
V(\phi)=\Lambda+B\cos\left(\frac{\phi}{\sqrt{T_3c}\varepsilon^{2/3}}\right)
+C\sin^2\left(\frac{\phi}{\sqrt{T_3c}\varepsilon^{2/3}}\right).
\end{equation}
To end this section, it is interesting to discuss the orders of
magnitude of the parameters appearing in the above potential. For this
purpose, it is useful to recall that $\sigma_0$, being a volume
modulus, is related to the size (or volume) of the extra-dimensions,
$V_6\simeq \sigma_0^{3/2}\alpha'^3$. The brane tension can be written
as $T_3=(2\pi)^{-3}\gstrings^{-1}\alpha'^{-2}$ while the Planck mass
takes the form $\Mp^2=2(2\pi)^{-7}V_6\gstrings^{-2}\alpha'^{-4}$
($\gstrings$ is the string coupling). As already mentioned, the
distance $\mu^{2/3}$ can be viewed as the distance between the stack
of $D7$-branes and the tip. It is therefore of the order of the size
of the throat which allows us to write that $\mu\simeq (27\pi
\gstrings \calN\alpha'^2/4)^{3/8}$ where the positive integer $\calN$
is the total background Ramond-Ramond charge.

In order to have a successful slow-roll scenario, we must assume that
the potential vanishes at its minimum. This amounts to take
$\Lambda=B$ which can always be achieved by choosing
$\beta=\betasr$ such that (with $b=3$, see before)
\begin{equation}
\betasr=1+\frac{45\varepsilon}{4n \mu a^2\sigma_0^2}+\cdots ,
\end{equation}
where we have performed a large volume expansion. Then, at the top of
the potential, one has $\partial ^2V/\partial\phi^2\simeq 2C-\Lambda$
and if one wants a flat potential $2C-\Lambda=2C-B$ must be a very
small quantity, i.e. $C/B \simeq 1/2$. Using the equations established
above, one can write
\begin{equation}
\frac{C}{B}=\Upsilon \frac{\sigma_0^{3/2}}{\gstrings(\gstrings\pi \calN)^{3/8}}
\left(\frac{\rtip}{\ells}\right)^{-1/2},
\end{equation}
where the numerical factor $\Upsilon=(12/15)\times
(4/27)^{3/8}/[(2\pi)^4nc]\simeq 5\times 10^{-5}$ and $\rtip \equiv
\varepsilon^{2/3}$. The string length is given by
$\ells=\sqrt{\alpha'}$. Let us also recall that we have taken
$b=3$. We see in the above expressions, especially
\Eq{eq:coefB}, that this case is special because
$\betasr \simeq 1$ and we have an additional suppression.
It is also interesting to discuss the mass scale which appears in the
arguments of the trigonometric functions. Straightforward calculations
lead to
\begin{equation}
\frac{\sqrt{T_3c}\varepsilon^{2/3}}{\Mp}=(2\pi)^2\sqrt{\frac{c}{2}}
\gstrings^{1/2}\sigma_0^{-3/4}
\left(\frac{\rtip}{\ells}\right).
\end{equation}
For fixed $\gstrings$ and $\calN$, the two inflationary parameters
$C/B$ and $\sqrt{T_3c}\varepsilon^{2/3}/\Mp$ are in fact controlled by
the radius of the tip and the volume of the extra-dimensions. 

Finally, if one requires $C/B=1/2$, as appropriate in a slow-roll 
analysis, then the above equations imply that 
\begin{equation}
\frac{\sqrt{T_3c}\varepsilon^{2/3}}{\Mp}
\simeq 2\times 10^{8}\sigma_0^{9/4}.
\end{equation}
This equation is relevant for the question of the priors that should
be put on the model parameters.

\subsubsection{Slow-roll Analysis}
\label{subsubsec:srti}

We now turn to the slow-roll analysis of the model. For the
canonically normalized inflaton field, we have just seen that the
potential is given by
\begin{equation}
\label{eq:pot:ti}
V=M^4\left(1+\cos\frac{\phi}{\mu}+\alpha\sin^2\frac{\phi}{\mu}\right),
\end{equation}
where inflation proceeds in the region $0<\phi/\mu<\pi$. Here, we have
written $\Lambda=M^4$, $C/B=\alpha$ and
$\mu=\sqrt{T_3c}\varepsilon^{2/3}$ (not to be confused with the scale
$\mu$ introduced above and related to the distance between the stack
of branes and the tip). When $\alpha\ll 1$, the potential reduces to
the natural inflation (NI) one. Yet, it was shown in \sectionc{sec:ni}
that only super-Planckian decay constants $\mu/\Mp>\order{1}$ could
make the natural inflation models compatible with observations (see
\eg \Fig{fig:CMBNI}). As noticed in \Refc{Pajer:2008uy}, this means
that tip inflation models with $\alpha\ll 1$ are not viable. On the
other hand, as was discussed in detail in the previous sub-section, if
$\alpha$ is fine-tuned to $\alpha\simeq 1/2$, then the potential of
\Eq{eq:pot:ti} becomes very flat at the top and a phenomenologically
successful slow-roll inflationary stage could occur. This is why, in
the following, these models are studied with $\alpha\simeq 1/2$.

\begin{figure}
\begin{center}
\includegraphics[width=\wdblefig]{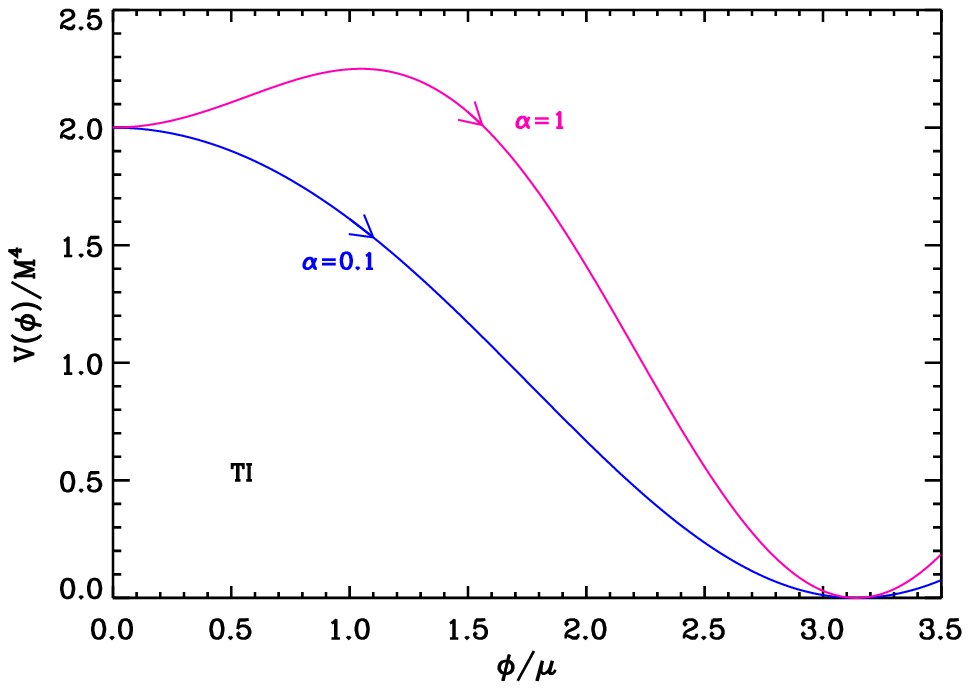}
\includegraphics[width=\wdblefig]{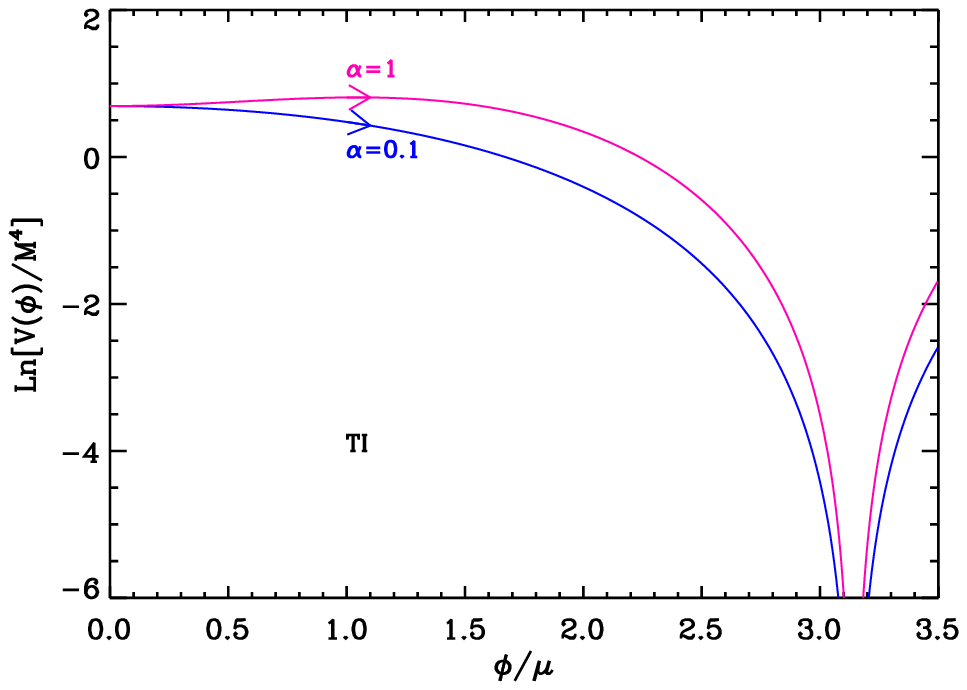}
\includegraphics[width=\wdblefig]{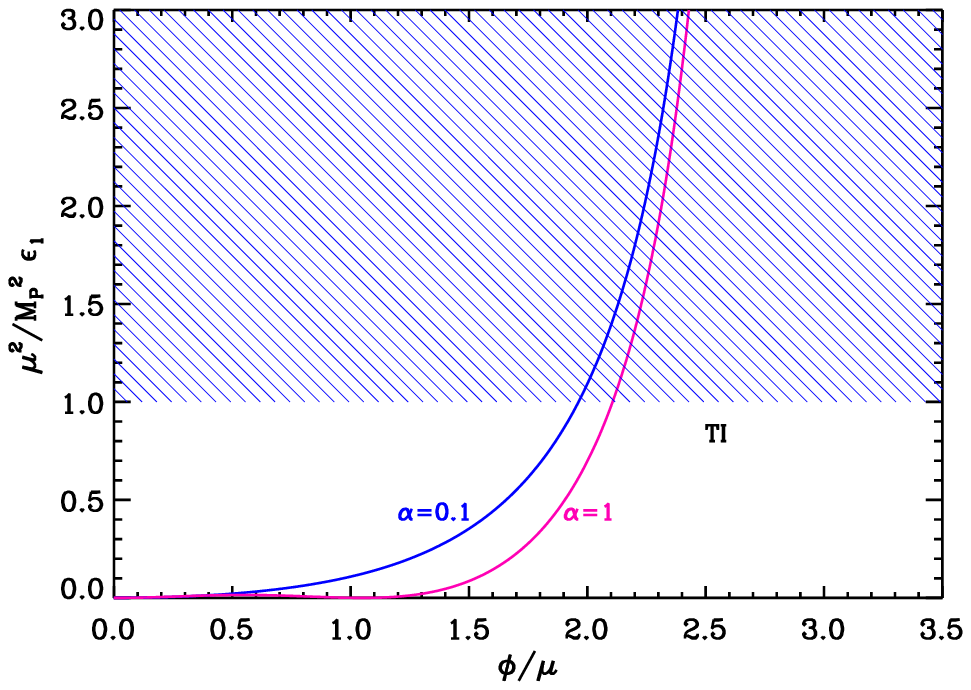}
\includegraphics[width=\wdblefig]{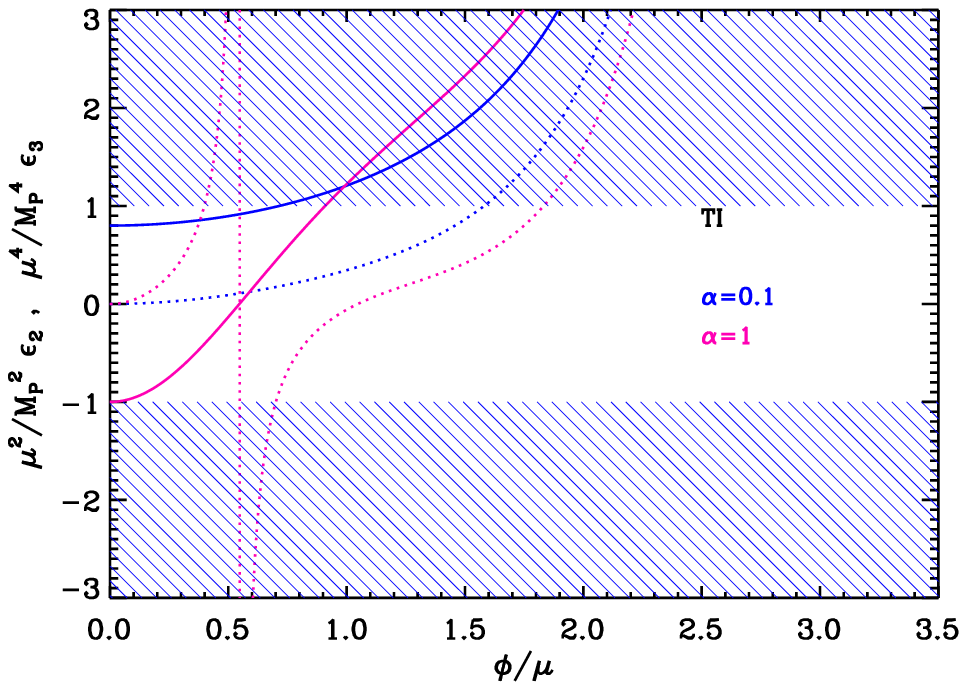}
\caption{Tip Inflation (TI).  Upper panels: Tip Inflation potential
  and its logarithm for $\alpha=0.1$ (blue line) and $\alpha=1$ (pink
  line), as a function of $\phi/\mu$. Bottom left panel: slow-roll
  parameter $\epsilon_1$ normalized by $\Mp^2/\mu^2$. The shaded area
  indicates the breakdown of the slow-roll inflation if $\mu=\Mp$
  (strictly speaking when the acceleration stops).  Bottom right
  panel: slow-roll parameter $\epsilon_2$ (solid line) and
  $\epsilon_3$ (dotted line), again rescaled by $\Mp^2/\mu^2$.}
\label{potTI}
\end{center}
\end{figure}

Defining
\begin{equation}
x \equiv \dfrac{\phi}{\mu}\,,
\end{equation}
the potential of \Eq{eq:pot:ti} and its logarithm with respect to $x$
are displayed in \Fig{potTI}. Its general shape depends on the value
of $\alpha$. If $\alpha<1/2$, it is a decreasing function of the field
\vev, hence inflation proceeds from the left to the right, and it has
a vanishing minimum at $x=\pi$. Its first derivative vanishes at the
top of the potential for $x=0$ while its second derivative
$V^{\prime\prime}(x=0) \propto2\alpha-1$. It vanishes there when
$\alpha=1/2$ and the potential becomes flat enough to support
inflation. If $\alpha>1/2$, the potential maximum is not located at
$x=0$ anymore but at $x=\arccos\left[ 1/(2\alpha)\right]$. Let us thus
define
\begin{equation}
\xdVzero=\left\lbrace
\begin{aligned}
& 0\quad &\textrm{if}\quad\alpha<1/2,\\
& \arccos\left(\frac{1}{2\alpha}\right)\quad &\textrm{if}\quad\alpha>1/2.
\end{aligned}
\right.
\end{equation}
If $\alpha>1/2$, the potential decreases with the field \vev in the
range $\xdVzero<x<\pi$, where inflation proceeds from the left to the
right. Again, the first derivative of the potential vanishes at the
top of the potential while its second derivative $V^
{\prime\prime}(x=\xdVzero)\propto 1/(2\alpha) - 2\alpha$ again
vanishes when $\alpha=1/2$. This is why $\alpha$ must be close enough
to $1/2$ in order for a viable slow-roll inflationary regime to take
place.

Let us calculate the Hubble flow functions within the slow-roll
approximation. They read
\begin{equation}
\epsilon_1 = \frac{\Mp^2}{\mu^2} \frac{\left(1 - 2\alpha\cos
  x\right)^2 \sin^2x}{2\left(1+\cos x + \alpha\sin^2x\right)^2}\,,
\end{equation}
\begin{equation}
\epsilon_2 = \frac{\Mp^2}{\mu^2}\frac{2\cos^2\frac{x}{2}}{\left(1+\cos
    x+\alpha\sin^2x\right)^2}
\left[2+\alpha\left(3+4\alpha\right)-2\alpha\left(3+2\alpha\right)\cos
  x-\alpha\cos \left(2x\right)\right],
\end{equation}
and
\begin{equation}
\begin{aligned}
\epsilon_3  =\frac{\Mp^2}{\mu^2} & \left\lbrace
-2-\frac{2+4\alpha}{\left(1+\alpha-\alpha\cos x\right)^2}
+\frac{5+3\alpha}{1+\alpha-\alpha\cos x}+\frac{1}{\cos^2\left(\frac{x}{2}\right)}
\right. \\ &\ \ \left.
+\frac{4\left(1+\alpha+3\alpha^2\right)-2\alpha\left(7+4\alpha\right)\cos 
x}{\alpha\left[\cos \left(2x\right)+\left(6+4\alpha\right)\cos x-3-4\alpha\right]-2}
\right\rbrace .
\end{aligned}
\end{equation}
They are displayed in \Fig{potTI} and are increasing functions of the
field \vev in the inflationary domain $\xdVzero<x<\pi$. Notice that
they diverge when $x\rightarrow\pi$.  The first and third slow-roll
parameters $\epsilon_1$ and $\epsilon_3$ vanish at the potential
maximum. However, the second slow-roll parameter $\epsilon_2$ takes a
non-vanishing positive value given by
\begin{equation}
\epsilon_2 \left(x=\xdVzero\right)=\left\lbrace
\begin{aligned}
 \frac{\Mp^2}{\mu^2} \left(1-2\alpha\right)
\quad\textrm{if}\quad\alpha<1/2,\\
 4\frac{\Mp^2}{\mu^2} \frac{2\alpha-1}{2\alpha+1}
 \quad\textrm{if}\quad\alpha>1/2.
\end{aligned}
\right.
\end{equation}
Requiring $\left\vert\epsilon_2\right\vert<1$ implies again to adjust
$\alpha$ close to $1/2$ such that
$\vert\alpha-1/2\vert\ll\mu^2/\Mp^2\ll 1$.

Inflation stops when $\epsilon_1=1$ at the position $\xend$ given by
\begin{equation}
\label{ti:eq:xend}
\xend = \arccos\left[\Sigma+\frac{\left(1+i\sqrt{3}\right)\sigma}
  {3\times 2^{2/3}\left(\delta+\sqrt{\Delta}\right)^{1/3}} 
-\frac{\left(1-i\sqrt{3}\right)\sigma^\prime}{6\times 2^{1/3}}
\left(\delta+\sqrt{\Delta}\right)^{1/3}\right].
\end{equation}
In this formula, we have defined
\begin{equation}
\begin{aligned}
\Delta & =-864 \alpha^6\left(2\alpha+1\right)^3\frac{\mu^2}{\Mp^2}
\left(\frac{\mu^2}{\Mp^2}+2\right)^2 \\ 
& \times \Bigg\lbrace
\left(2\alpha-1\right)^3 +2\left(2\alpha+1\right)
\left[\left(\alpha-10\right)\alpha-2\right]\frac{\mu^2}{\Mp^2}
-4\left(2\alpha+1\right)^2\frac{\mu^4}{\Mp^4}\Bigg\rbrace ,
\end{aligned}
\end{equation}
and
\begin{equation}
\begin{aligned}
\delta & = 8\alpha^3 \left[2 \left(2\alpha-1\right)^3 - 3\left(1+2\alpha\right)
\left(5 + 2\alpha\right) \left(1+4\alpha\right) \frac{\mu^2}{\Mp^2}
\right. \\
& \
- 15\left(1+\alpha\right) \left(1+2\alpha\right)^2 \frac{\mu^4}{\Mp^4}
- \left. 2 \left(1+2\alpha\right)^3 \frac{\mu^6}{\Mp^6}\right],
\end{aligned}
\end{equation}
together with
\begin{equation}
\sigma = 3 + 4\alpha\left(1-\alpha\right) -2\frac{\mu^2}{\Mp^2} \left(1+2\alpha\right)^2
-\frac{8}{2+\dfrac{\mu^2}{\Mp^2}}\,, \qquad
\sigma' = \dfrac{1}{2\alpha^2\left(2 + \dfrac{\mu^2}{\Mp^2}\right)}\,.
\end{equation}

Let us now turn to the slow-roll trajectory. It can be integrated
explicitly, leading to
\begin{equation}
  \Nend-N=\frac{\mu^2}{\Mp^2}\frac{1}{2\alpha-1} \ln \left(\frac{1-\cos x}
    {1-\cos \xend}\right)-\frac{\mu^2}{2\Mp^2} \frac{2\alpha+1}{2\alpha-1}
  \ln\left(\frac{1-2\alpha\cos x}{1-2\alpha\cos\xend}\right).
\end{equation}
For $\alpha=1/2$, this expression is singular, and one has
\begin{equation}
  \Nend-N = 
  \frac{\mu^2}{\Mp^2} \left[\frac{1}{1-\cos x}-\frac{1}
  {1-\cos \xend}-\frac{1}{2}\ln\left(\frac{1-\cos x}
  {1-\cos \xend }\right)\right].
\end{equation}

Finally, the parameter $M$ can be determined from the amplitude of the
CMB anisotropies and the observable field value $\xstar$ [see
  \Eq{eq:phistarlnrrad}], and one gets
\begin{equation}
\left(\frac{M}{\Mp}\right)^4 = 720\pi^2
\frac{\Mp^2}{\mu^2}
\frac{\left(1-2\alpha\cos \xstar\right)^2\sin^2\xstar}
{\left(1+\cos \xstar+\alpha\sin^2\xstar\right)^3}
\frac{\Qrms^2}{T^2}\, .
\end{equation}

The reheating consistent slow-roll predictions of the TI models are
displayed in \Fig{fig:CMBTIalphaLTonehalf} for $\alpha<1/2$ and in
\Fig{fig:CMBTIalphaGTonehalf} for $\alpha>1/2$, with
$\mu/\Mp=10^{-6}$, $10^{-4}$ and $10^{-2}$. In both cases, one can see
that $\alpha$ needs to be sufficiently adjusted to $1/2$, namely
$\vert 2\alpha-1\vert\ll\mu^2/\Mp^2$, otherwise the deviation from
scale invariance is too important. The typical amount of gravitational
waves is very small. To see how $\mu/\Mp$ is constrained, the
slow-roll predictions are displayed for $\alpha=1/2$ in
\Fig{fig:CMBTIalphaEQonehalf}, and with $\mu$ varying. One can see
that even if one allows values of $\mu$ larger than the typical ones
($\mu/\Mp\simeq 10^{-4}$) these models are disfavored by the
observations since they deviate too much from scale invariance.

\subsection{\texorpdfstring{$\beta$}{Beta} Exponential Inflation (BEI)}
\label{sec:bei}

\begin{figure}
\begin{center}
\includegraphics[width=\wdblefig]{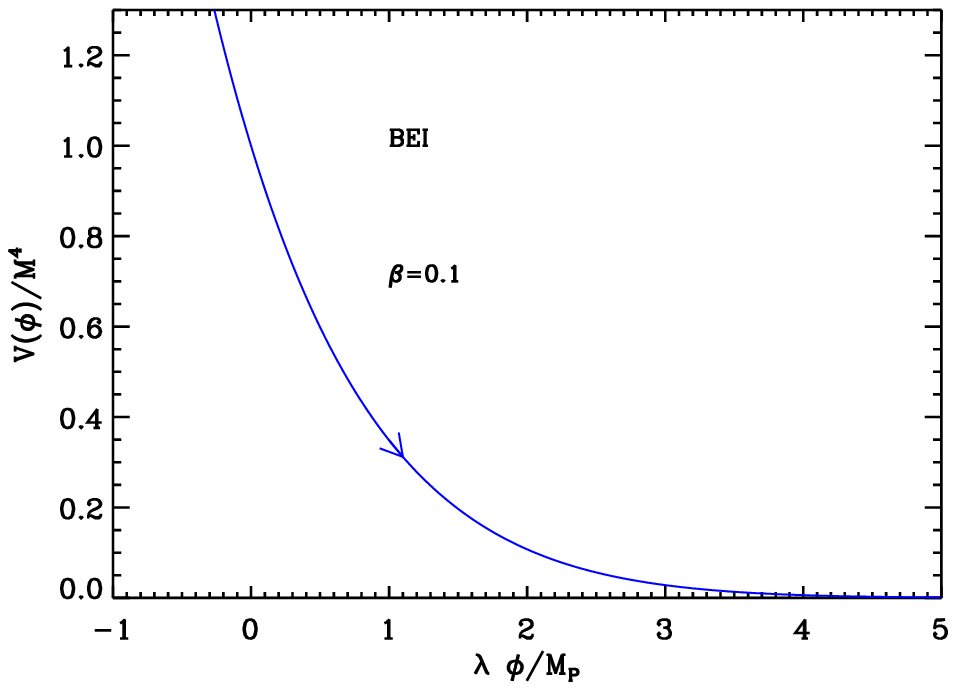}
\includegraphics[width=\wdblefig]{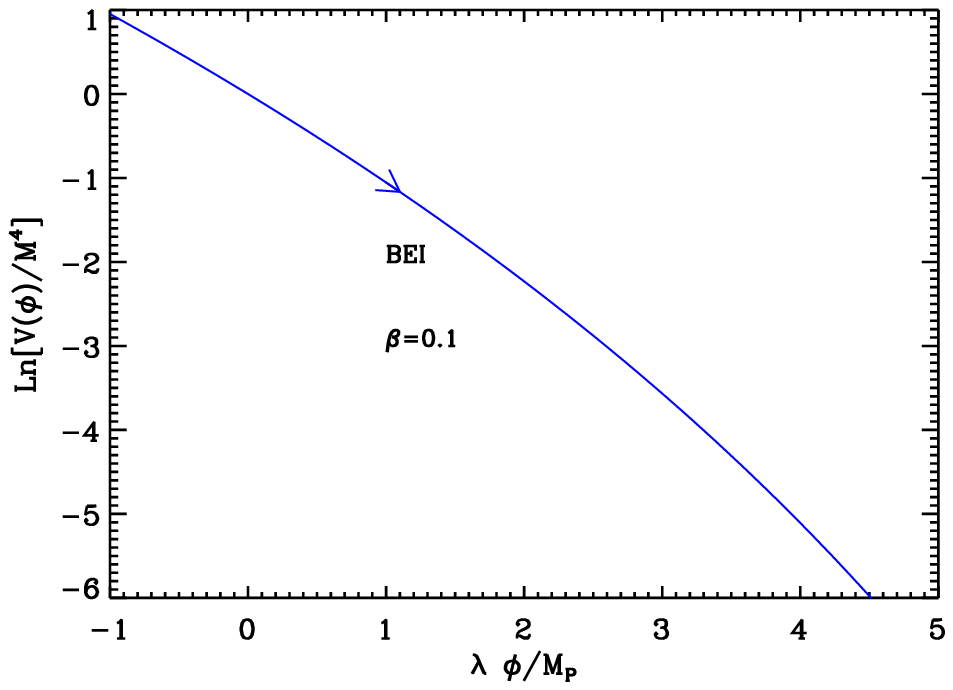}
\includegraphics[width=\wdblefig]{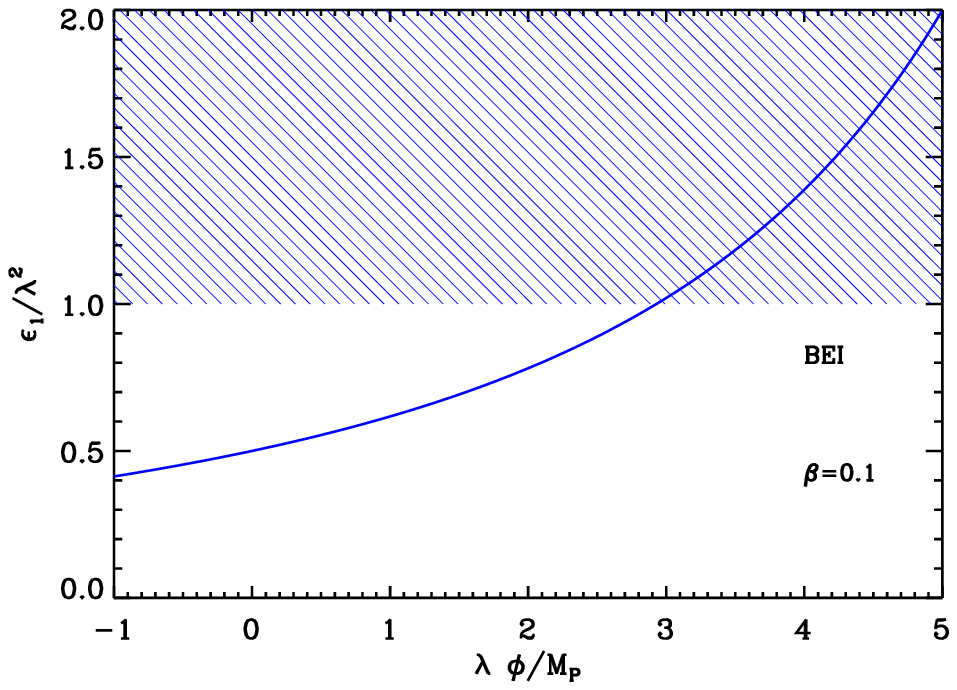}
\includegraphics[width=\wdblefig]{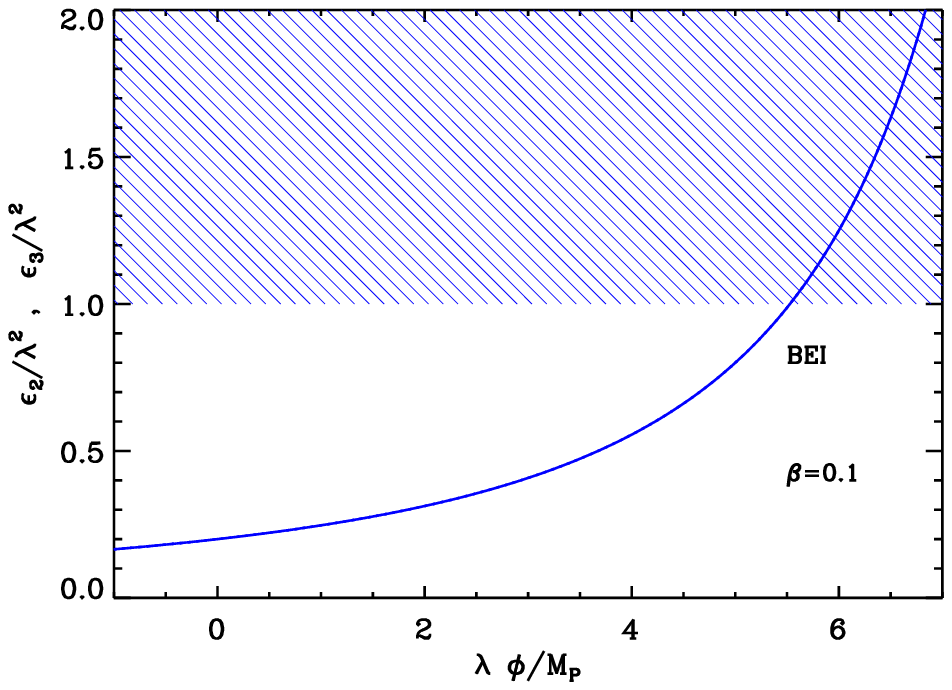}
\caption{$\beta $ Exponential Inflation (BEI) for $\beta=0.1$. Upper
  panels: the potential and its logarithm. Bottom left panel:
  slow-roll parameter $\epsilon_1$ with respect to the field
  values. The shaded area indicates where inflation stops if
  $\lambda=1$. Bottom right panel: slow-roll parameters
  $\epsilon_2=\epsilon_3$.}
\label{potBEI}
\end{center}
\end{figure}

This model was introduced and studied in \Refc{Alcaniz:2006nu} as a
phenomenological generalization of the PLI exponential potential (see
\sectionc{sec:pli}). The potential is given by
\begin{equation}
V(\phi)=M^4\exp_{1-\beta}\left(-\lambda \frac{\phi}{\Mp}\right),
\end{equation}
 where the generalized exponential function $\exp_{1-\beta}$ is defined by
\begin{eqnarray}
\exp_{1-\beta}\left(f\right)=\left\lbrace
\begin{array}{cc}
\left(1+\beta f\right)^{1/\beta} & \quad \textrm{for}\ \ 1+\beta f >0\, ,\\
0 & \quad \textrm{otherwise}\, .
\end{array}
\right.
\end{eqnarray}
As discussed in \Refc{Alcaniz:2006nu}, for $f>0$ and $g>0$, this
function satisfies the following identities:
\begin{equation}
\exp_{1-\beta} \left[\ln_{1-\beta}\left(f\right)\right] = f, \qquad
\ln_{1-\beta}\left(f\right)+\ln_{1-\beta}\left(g\right)=
\ln_{1-\beta}\left(fg\right)-\beta\left[\ln_{1-\beta}\left(f\right)
\ln_{1-\beta}\left(g\right)\right],
\end{equation}
where $\ln_{1-\beta}\left(f\right)=\left(f^\beta-1\right)/\beta$ is
the generalized logarithmic function. In the limit $\beta\rightarrow
0$, all the above expressions reproduce the usual exponential and
logarithm properties. Therefore, the limit $\beta\rightarrow 0$
reproduces the PLI potential (see \sectionc{sec:pli}). However, as
discussed below, this is not the case for the observable predictions
which remain different. Defining the quantity $x$ by
\begin{equation}
x\equiv \dfrac{\phi}{\Mp}\,,
\end{equation}
the range of field \vev for which inflation occurs depends on the sign
of $\beta$. For $\beta>0$, the field values are such that
$x<1/(\beta\lambda)$, whereas if $\beta<0$, the potential is defined
for $x>1/(\beta\lambda)$. In both cases, inflation proceeds from the
left to the right. The first three Hubble flow functions in the
slow-roll approximation are given by
\begin{equation}
\epsilon_1 = \frac{\lambda^2}{2\left(1-\beta\lambda
  x\right)^2}\,,\qquad \epsilon_2 =
\frac{2\beta\lambda^2}{\left(1-\beta\lambda x\right)^2} =
4\beta\epsilon_1 , \qquad \epsilon_3 = \epsilon_2 .
\end{equation}
Together with the potential, they are represented in \Fig{potBEI}.

One immediately sees that $\epsilon_1$ is an increasing function of
$x$ only for the case where $\beta>0$. Therefore inflation can
naturally stop at $\xend$ such that $\epsilon_1(\xend)=1$. In the
opposite situation, namely $\beta<0$, inflation has to be ended by
some additional mechanism and $\xend$ would become an
extra-parameter. Since this model is purely phenomenological, in the
following, we restrict ourselves to the case $\beta>0$ for which
\begin{equation}
\label{eq:bei:xend}
\xend = \frac{1}{\beta} \left(\frac{1}{\lambda} -
\frac{1}{\sqrt{2}}\right).
\end{equation}

The next step consists in determining the slow-roll trajectory. It can
be integrated explicitly and the result reads
\begin{equation}
N-\Nend = \frac{1}{\lambda} \left(x-\xend\right) - \frac{\beta}{2}
\left(x^2 - \xend^2\right).
\end{equation}
It can also be inverted and one obtains the following expression for
$x$ as a function of the \efolds number
\begin{equation}
  x = \frac{1}{\lambda\beta} - \sqrt{\left(
    \xend-\frac{1}{\lambda\beta} \right)^2 - \frac{2}{\beta}(N-\Nend)}\, .
\end{equation}
Using these expressions, the observable field value $\xstar$ can be
related to the number of \efolds $\Delta \Nstar = \Nend - \Nstar$
at which the pivot scale crossed out the Hubble radius during
inflation. Making use of \Eq{eq:bei:xend}, one gets
\begin{equation}
\xstar = \frac{1}{\lambda\beta} - \sqrt{\frac{1}{2\beta^2} +
  \frac{2}{\beta} \Delta \Nstar} \,.
\label{eq:xstarbei}
\end{equation}
Inserting this expression into the slow-roll parameters formulas yields
\begin{equation}
\epsilon_{1*} = \frac{1}{1+4\beta\Delta \Nstar}\, ,\qquad
\epsilon_{2*} = \epsilon_{3*} = 4\beta\epsilon_{1*}\, .
\end{equation}
Therefore, the slow-roll predictions of these models do not depend on
the parameter $\lambda$.  Moreover, the limit $\beta\rightarrow 0$
does not give the same observable predictions as for the PLI models
due to the singular behavior of $\xend$. These models can therefore be
viewed as a completely different class.

Finally, the amplitude of the CMB anisotropies fixes the parameter $M$
with
\begin{eqnarray}
\left(\frac{M}{\Mp}\right)^4&=&720\pi^2 \lambda^2\left(1-\beta\lambda
\xstar\right)^{-2-\frac{1}{\beta}} \frac{\Qrms^2}{T^2}\, .
\end{eqnarray}
Notice that, from \Eq{eq:xstarbei}, the above expression can be written
in terms of $\Delta \Nstar$ and that it does not depend on $\lambda$
anymore. The reheating consistent slow-roll predictions for the BEI
models are displayed in \Fig{fig:CMBBEI}. The parameter $\beta$ must
be such that $\beta \gtrsim 0.6$ in order for the predictions of the
model to remain inside the two-sigma confidence intervals, while the
parameter $\lambda$ remains totally unconstrained.

\subsection{Pseudo Natural Inflation (PSNI)}
\label{sec:psni}

\subsubsection{Theoretical Justifications}
\label{subsubsec:theorypsni}

Pseudo Natural Inflation (PSNI) was introduced and studied in
\Refc{ArkaniHamed:2003mz}. This model has common points with NI, see
\sectionc{sec:ni}. Indeed, in PSNI, the inflaton field is also a
pseudo-Nambu Goldstone boson which appears after symmetry
breaking. The corresponding potential is nearly flat which is
well-suited for inflation. The main ideas behind this construction are
reviewed in \sectionc{sec:ni}. The main difference with respect to
natural inflation, for which the broken symmetry is a shift symmetry,
is that in pseudo natural inflation the broken symmetry is now a
$U(1)$ one. A concrete implementation of this idea has been proposed
in \Refc{ArkaniHamed:2003mz} and starts with the following
supersymmetric hybrid superpotential
\begin{equation}
W\left(S,X,\varphi,\psi_1,\psi_2\right)=\lambda_0 S
\left(\psi_1^2+\psi_2^2-f^2\right)+\frac{\lambda_1}
{2}\psi_1\varphi^2+\lambda_2X \left(\varphi^2-v^2\right),
\end{equation}
with $\lambda_1^2f^2>2\lambda_2^2v^2$, where $S$, $X$, $\psi_1$,
$\psi_2$ and $\varphi$ are scalar fields and $\lambda_0$, $\lambda_1$
and $\lambda_2$ are coupling constants. We see that the $U(1)$
symmetry is explicitly broken by the term proportional to
$\lambda_1$. The corresponding potential can be written as
\begin{equation}
V=\lambda_0^2\left\vert \psi_1^2+\psi_2^2-f^2\right\vert^2
+\left\vert 2\lambda_0S\psi_1+\frac{\lambda_1}{2}\varphi^2\right\vert^2
+4\lambda_0^2\left\vert S\psi_2\right\vert^2
+\vert \varphi\vert^2\left\vert\lambda_1\psi_1+2\lambda_2X\right\vert^2
+\lambda_2^2\left\vert\varphi^2-v^2\right\vert^2.
\end{equation}
The flat directions of this superpotential can be reparametrized as
\begin{equation}
\psi_1 + i\psi_2 \equiv \left(f+\sigma\right) \ee^{i\phi/f}
\, ,\quad
\psi_1 - i \psi_2 \equiv \left(f-\sigma\right)\ee^{-i\phi/f} ,
\end{equation}
where $\phi$ is the Nambu-Goldstone boson associated to the broken
$U(1)$ symmetry and $\sigma$ is a modulus. One can assume that
$\sigma$ is stabilized and sits at $\sigma=0$, the minimum of a
potential originating from supersymmetry breaking. The field $\phi$
plays the role of the inflaton.  Using the above expressions and the
condition $\sigma =0$, one obtains that
$\psi_1=f\cos\left(\phi/f\right)$ and
$\psi_2=f\sin\left(\phi/f\right)$. In that case, a flat direction for
$\phi$ is obtained for $\varphi=0$ and $S=0$ since then we have
\begin{equation}
V=\lambda_2^2v^4.
\end{equation}
Notice that SUSY is broken because $F_X\equiv \langle \partial
W/\partial X\rangle= \lambda_2 v^2\neq 0 $. As a consequence, the
corresponding vacuum energy density is indeed given by $V_0\simeq\vert
F_X\vert^ 2=\lambda_2^2v^4$.

This tree level potential is corrected by two kind of
contributions. First, supergravity induces a soft SUSY breaking mass
of order $H$ for every scalar, but since $\phi$ is a pseudo
Nambu-Goldstone boson, it only receives a potential due to the
explicit breaking term proportional to $\lambda_1$. The corresponding
contribution is loop suppressed, $m_\phi^2\simeq
3\lambda_1^2H^2/(16\pi^2)$, as soon as $\lambda_1\lesssim 1$ which
will be assumed.  Second, the potential receives a direct Yukawa
mediated contribution through a $\varphi$ loop and
\Refc{ArkaniHamed:2003mz} has shown that it takes the form
\begin{equation}
\label{eq:potpsnihe}
V(\phi)\simeq V_0\left(1+\frac{\lambda_2^2}
{4\pi^2}\ln\frac{\lambda_1\psi_1}{\mu}\right)
=V_0\left[1+\frac{\lambda_2^2}{4\pi^2}\ln\frac{\cos\left(\phi/f\right)}
{\mu/f}\right].
\end{equation}
where $\mu$ is some renormalization scale. The above formula gives
rise to a new type of potential that we study in the next sub-section.

\subsubsection{Slow-Roll Analysis}
\label{subsubsec:srpsni}

\begin{figure}
\begin{center}
\includegraphics[width=\wdblefig]{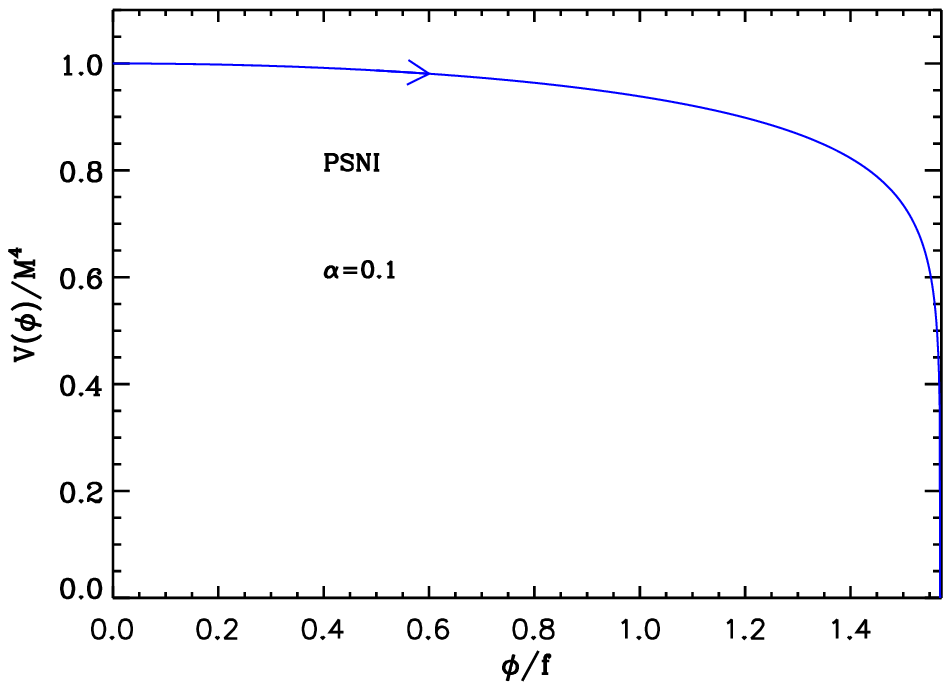}
\includegraphics[width=\wdblefig]{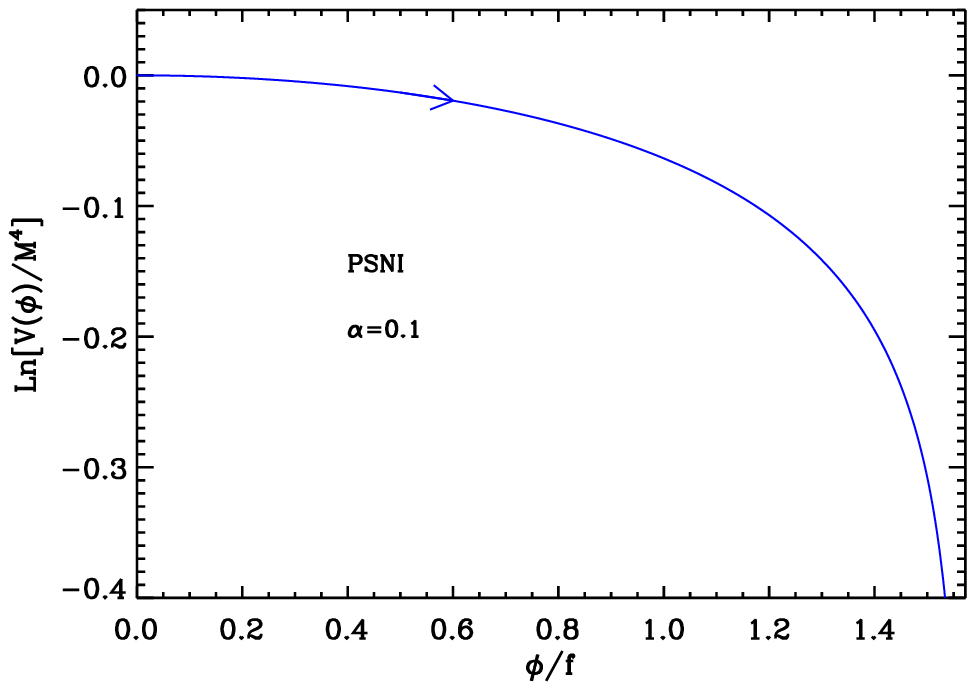}
\includegraphics[width=\wdblefig]{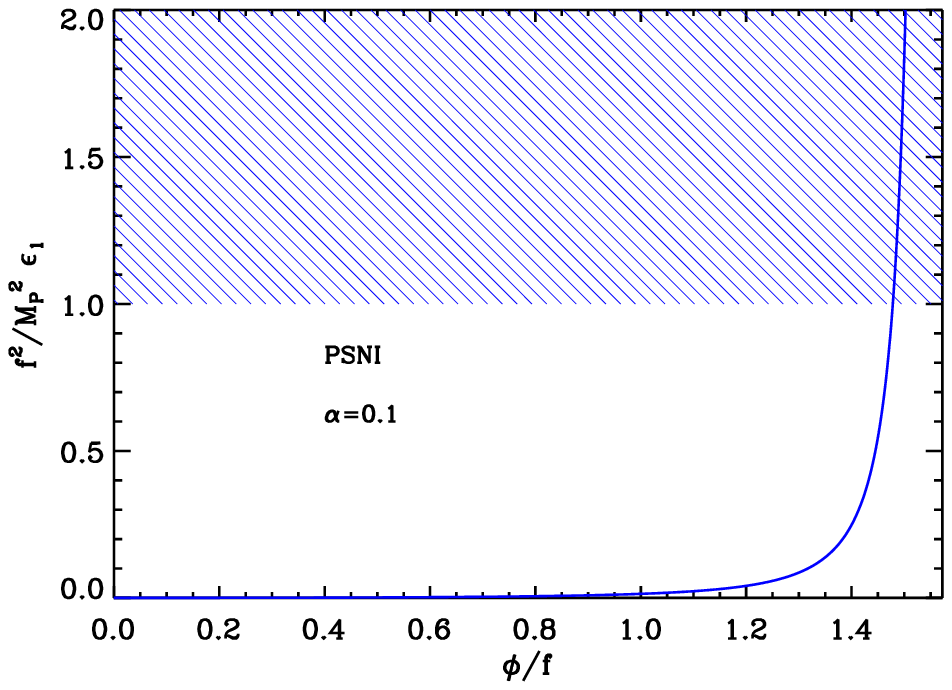}
\includegraphics[width=\wdblefig]{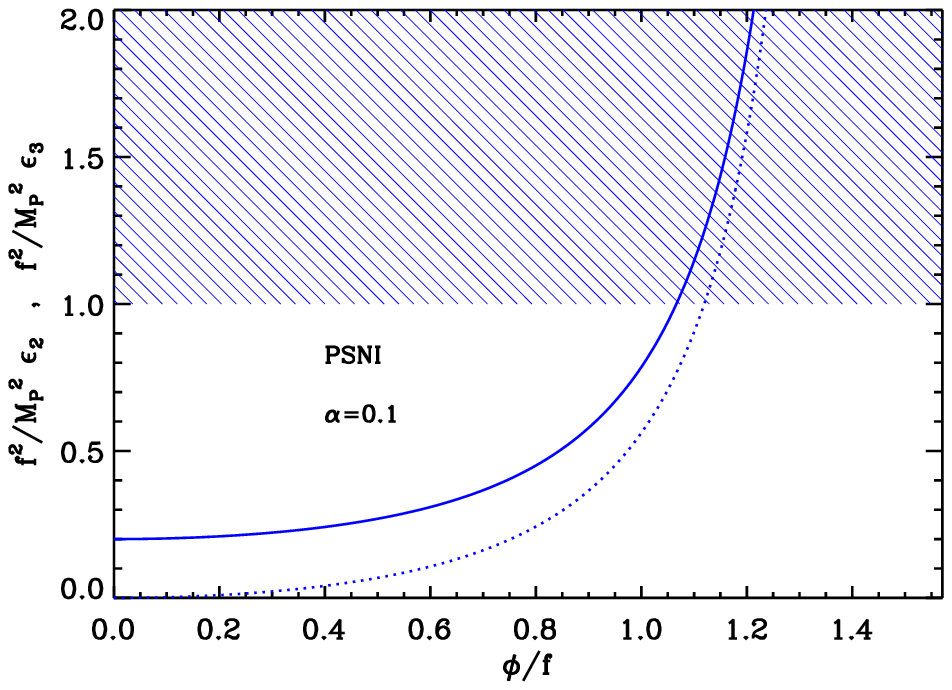}
\caption{Top left panel: Pseudo Natural Inflation (PSNI) potential,
  for $\alpha=0.1$, as a function of $\phi/f$.  Top right panel:
  logarithm of the potential for the same value of $\alpha$. Bottom
  left panel: slow-roll parameter $\epsilon _1$, rescaled by the
  quantity $\Mp^2/f^2$ such that it acquires a universal form, for the
  same value of $\alpha$. Bottom right panel: slow-roll parameter
  $\epsilon _2$ (solid line) and $\epsilon_3$ (dotted line), rescaled
  by the quantity $\Mp^2/f^2$, still for the same value of $\alpha$.}
\label{potPSNI}
\end{center}
\end{figure}

We now turn to the slow-roll analysis of the PSNI model. Using more
friendly notations, the potential~(\ref{eq:potpsnihe}) can be
re-expressed as
\begin{equation}
\label{eq:psni:pot}
V=M^4\left[1+\alpha \ln \left(\cos\frac{\phi}{f}\right)\right],
\end{equation}
with the following definitions
\begin{equation}
\begin{aligned}
M^4=\lambda_2^2v^4\left[1+\frac{\lambda_2^2}
{4\pi^2}\ln\left(\frac{\lambda_1 f}{\mu}\right)\right], \qquad
\alpha=\frac{\lambda_2^2/\left(4\pi^2\right)}{1+\lambda_2^2
/\left(4\pi^2\right)\ln\left(\frac{\lambda_1 f}{\mu}\right)}\, .
\end{aligned}
\end{equation}
Therefore, one typically has $\alpha\ll 1$, and the scale $f$ should a
priori be such that $f\lesssim\Mp$ in order to avoid the usual
problems of natural inflation.

The potential~(\ref{eq:psni:pot}) as well as its logarithm are
displayed in \Fig{potPSNI}. Since $\phi$ is assumed to be such that
$\phi\simeq 0$ initially, the potential must be studied in the range
$\phi/f\in\left[0,\pi/2\right]$. It is positive definite in the range
$\phi/f\in\left[0,\arccos\left(\ee^{-1/\alpha}\right)\right]$. We see
that it is a decreasing function of the inflaton \vev, which means
that inflation proceeds from the left to the right in the direction
specified by the arrow in \Fig{potPSNI}.

Let us now turn to the slow-roll parameters. If one defines
$x\equiv\phi/f$, then the three first Hubble flow parameters are given
by
\begin{equation}
\epsilon_1 = \frac{\Mp^2}{2f^2}\frac{\alpha^2\tan^2 x}
{\left(1+\alpha\ln\cos x\right)^2}\, ,\qquad
\epsilon_2 = 2\alpha\frac{\Mp^2}{f^2}\frac{1+\alpha+\alpha\ln\cos 
x-\alpha\cos^2 x}{\cos^2 x
\left(1+\alpha\ln\cos x\right)^2},
\end{equation}
\begin{equation}
\begin{aligned}
\epsilon_3 & = \alpha\frac{\Mp^2}{f^2} (\tan x)^2 \dfrac{2+3\alpha +
  \alpha^ 2 - \alpha^ 2\cos \left(2x\right) + \left(4+3\alpha\right) \alpha \ln\cos
  x + 2\alpha^2\ln^2\cos x}{\left(1+\alpha\ln\cos x \right)^2 \left(
  1+\alpha\ln\cos x+\alpha\sin^2 x \right)}\, .
\end{aligned}
\end{equation}
They are displayed in \Fig{potPSNI}. We see on this plot that the
slow-roll parameters $\epsilon_1$ and $\epsilon_3$ vanish when $x$
goes to $0$ and diverge when $x$ goes to $\pi/2$. On the other hand,
the slow-roll parameter $\epsilon_2$ has a non-zero limit when $x$
goes to $0$, namely
\begin{equation}
\lim_{ x \rightarrow 0} \epsilon_2 = 2 \frac{\Mp^2}{f^2} \alpha .
\end{equation}
This quantity should be small in order for slow-roll to be valid. This
means that, at a fixed scale $f$, the parameter $\alpha$ needs to be
smaller than $f^2/\Mp^2$. From the monotonic behavior of $\epsilon_1$,
one also notices that inflation naturally stops at
$\epsilon_1=1$. Unfortunately, this equation cannot be solved exactly
and the solution needs to be determined numerically. However, since we
are in a regime where $f/\Mp\ll 1$ and $\alpha\Mp^2/f^2\ll 1$, $\xend$
must be close to $\pi/2$. One can derive a better approximation by
solving the equation $\epsilon_1=1$ using an expansion in the small
quantities of the problem. One arrives at
\begin{equation}
\label{eq:psni:xend}
\xend\simeq\frac{\pi}{2}-\frac{\alpha}{\sqrt{2}}\frac{\Mp}{f}\, ,
\end{equation}
that is to say the first correction to $\pi/2$ is linear in $\alpha
\Mp/f$ and, as expected, negative. As usual, the \ASPIC code makes use
of the complete slow-roll solution.

Let us now turn to the slow-roll trajectory. It can be integrated
exactly in terms of the dilogarithm function $\Li{2}$ (also referred
to as Spence's function, or Joncqui\`ere function). This function was
already used in this paper, for instance in \sectionc{sec:rchi}. The
explicit expression of the trajectory reads
\begin{eqnarray}
\Nend-N &=& \frac{f^2}{\alpha\Mp^2}\left[\left(1+\alpha\ln\cos
  \xend\right)\ln\sin \xend +\frac{\alpha}{4} \Li{2} \left(\cos^2 \xend 
  \right)
  \right] \nonumber\\& & -\frac{f^2}{\alpha\Mp^2} \left[ \left(
  1+\alpha\ln\cos x \right)\ln \sin x +\frac{\alpha}{4} \Li{2}
  \left( \cos^2 x \right) \right],
\end{eqnarray}
where $\Nend$ is the number of \efolds at the end of
inflation. Unfortunately, this trajectory cannot be inverted
analytically. However, if one uses the two conditions $f/\Mp\ll 1$ and
$\alpha \Mp^2/f^2\ll 1$, one can simplify a lot its expression. In
particular, at Hubble crossing, one can write
\begin{equation}
\Delta\Nstar\simeq\frac{f^2}{2\alpha\Mp^2}\left[\left(\xstar-\frac{\pi}
{2}\right)^2-\left(\xend-\frac{\pi}{2}\right)^2\right],
\end{equation}
from which one can obtain an explicit formula for $\xstar$
\begin{equation}
\xstar\simeq\frac{\pi}{2}-\sqrt{2\alpha\Delta\Nstar}\frac{\Mp}{f}\, .
\end{equation}
Then, this also allows us to derive useful approximated equations for
the first three Hubble flow parameters, namely
\begin{equation}
\label{eq:psni:predic}
\epsilon_{1*}\simeq\frac{\alpha}{4\Delta\Nstar}\, ,\qquad
\epsilon_{2*}\simeq \epsilon_{3*}\simeq  \frac{1}{\Delta\Nstar}\, .
\end{equation}
The expressions of the tensor-to-scalar ratio, spectral index and
running are
\begin{equation}
r\simeq \frac{4\alpha}{\Delta\Nstar}\, ,\qquad
\nS-1\simeq\alphaS\simeq-\frac{1}{\Delta\Nstar}\, ,
\end{equation}
These formulas are in agreement with the estimates given in
\Refc{ArkaniHamed:2003mz}. Interestingly enough, we see that these
predictions are independent of the scale $f$ and that the spectral
index (and the running) is even independent of $\alpha$.

The last step consists in using the CMB normalization in order to
extract the mass scale $M$. Straightforward manipulations lead to
\begin{equation}
\left(\frac{M}{\Mp}\right)^4=720\pi^2
\alpha^2\frac{\Mp^2}{f^2}\frac{\tan^2 \xstar} {\left(1+\alpha\ln\cos
  \xstar\right)^3} \frac{\Qrms^2}{T^2}\, .
\end{equation}
Under the two conditions $f/\Mp\ll 1$ and $\alpha\Mp^2/f^2\ll 1$ and
using the same method as before, this leads to
\begin{equation}
\left(\frac{M}{\Mp}\right)^4\simeq \frac{360\pi^2\alpha}{\Delta\Nstar} 
\frac{\Qrms^2}{T^2}\, .
\end{equation}
Requiring $M<\Mp$ is easily achieved since, for the fiducial value
$\Delta \Nstar\simeq 55$, this is equivalent to $\alpha \lesssim 2580$
whereas we have $\alpha \ll 1$. Taking the more realistic value
$\alpha\simeq 10^{-6}$ and $\Delta\Nstar\simeq 55$, one typically obtains
that $M/\Mp\simeq 10^{-3}$.

The predictions of the PSNI models are displayed in \Fig{fig:CMBPSNI}
for $f/\Mp=10^{-3},10^{-1},10$ respectively (although this last value
is considered just for the purpose of illustration since
super-Planckian values of $f$ are not very physical). The reheating
equation of state parameter $\wrehbar$ has been taken to $0$ but since there
is no potential minimum around which the inflaton field can oscillate
at the end of inflation, this parameter is a priori unspecified and
can take different values (in the \ASPIC code, this parameter can be
freely chosen). One can see that the rough description provided by
\Eqs{eq:psni:predic} is correct: when $\alpha\Mp^2/f^2\ll 1$, the
deviation from scale invariance does not depend on the model
parameters and is of the order of $\nS\simeq 1-1/\Delta\Nstar\simeq
0.975$, while $r\simeq 4\alpha/\Delta\Nstar$ is typically very small.

\subsection{Non Canonical K\"ahler Inflation (NCKI)}
\label{sec:ncki}

\subsubsection{Theoretical Justifications}
\label{subsubsec:theoryncki}

This model was introduced and studied in \Refc{Boubekeur:2005zm} as a
way to model hilltop inflation. The idea is to consider $F$ or $D$
term inflation in which we have a flat direction lifted by one loop
corrections. This gives rise to loop inflation as discussed in
\sectionc{sec:li}. The LI potential has been obtained, however, under
the assumption of a minimal K\"ahler potential. Now, corrections
originating from higher order operators, always present in the
K\"ahler potential, should typically produce a mass term and,
therefore, the scalar potential gets modified and takes the form
\begin{equation}
\label{eq:potnckihep}
  V\left(\phi\right)\simeq V_0+\alpha \ln\left(\frac{\phi}{Q}\right)
+b \phi^2,
\end{equation}
where $Q$ is a renormalization scale. This is the model we study in
this section. Let us notice that the coefficient $b$ can be positive
or negative. The case $b>0$ has been investigated in
\Refcs{Panagiotakopoulos:1997ej,Panagiotakopoulos:1997qd} as ``hybrid
inflation with quasi-canonical supergravity'' and the case $b<0$ was
studied in \Refc{Boubekeur:2005zm}. For $b>0$, the
potential~(\ref{eq:potnckihep}) can be viewed as a valley hybrid
potential [VHI, see \sectionc{sec:vhi} and \Eq{eq:vhi:pot}] plus
logarithmic radiative corrections. Therefore, a consistency check of
our calculations will be that, when $\alpha \rightarrow 0$, all the
formulas derived below must reproduce those derived in
\sectionc{sec:vhi}. Finally, let us mention that the
potential~(\ref{eq:potnckihep}) has also been studied in
\Refc{Hall:2007qw} for $b<0$ under the name ``SUSY breaking
potential'' and in \Refc{Kyae:2009jm} in the context of supersymmetric
hybrid inflation.

\subsubsection{Slow-Roll Analysis}
\label{subsubsec:srncki}

In this sub-section, we now turn to the slow-roll analysis of the NCKI
scenario. For this purpose, it is convenient to re-write the
potential~(\ref{eq:potnckihep}) under the following form
\begin{equation}
\label{eq:ncki:pot}
  V=M^4\left[1+\alpha \ln \left(\frac{\phi}{\Mp}\right) + \beta
    \left(\frac{\phi}{\Mp}\right)^2\right],
\end{equation}
where $\alpha$ is a small positive dimensionless parameter and $\beta$
a dimensionless parameter of order $\order{1}$ which can be either
positive or negative. Notice that the coefficient $\alpha $ has be
redefined and that $\beta$ is directly related to $b$.

\begin{figure}
\begin{center}
\includegraphics[width=\wdblefig]{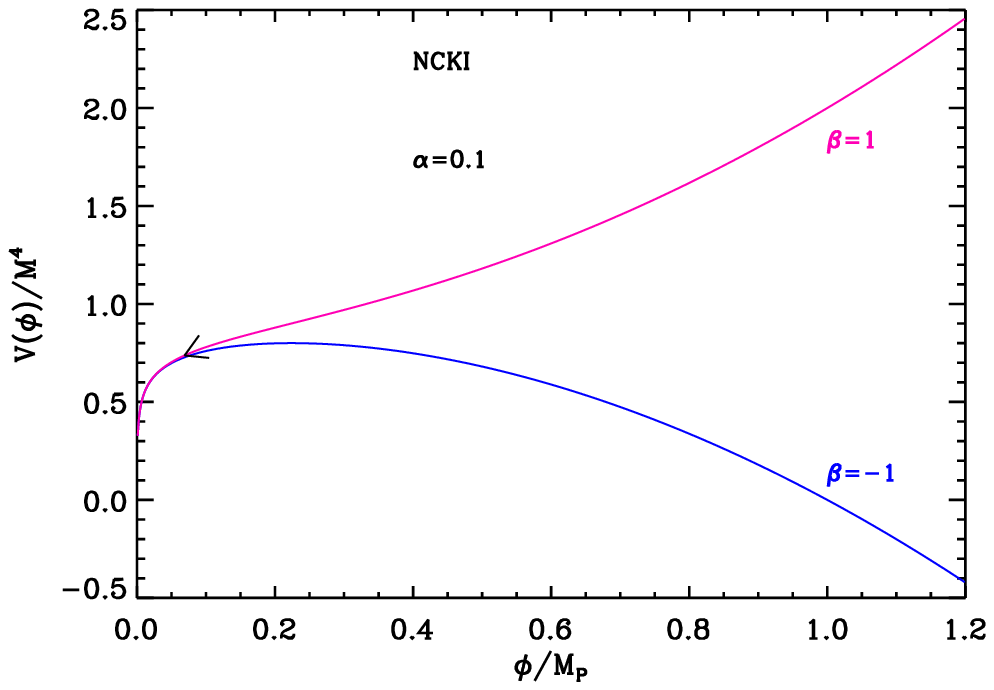}
\includegraphics[width=\wdblefig]{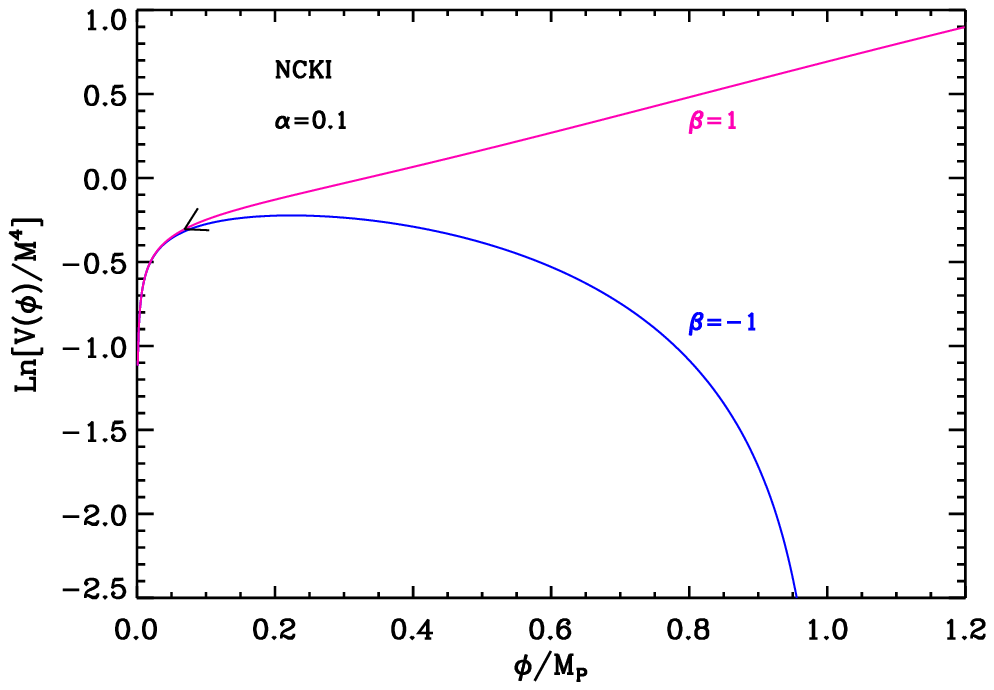}
\includegraphics[width=\wdblefig]{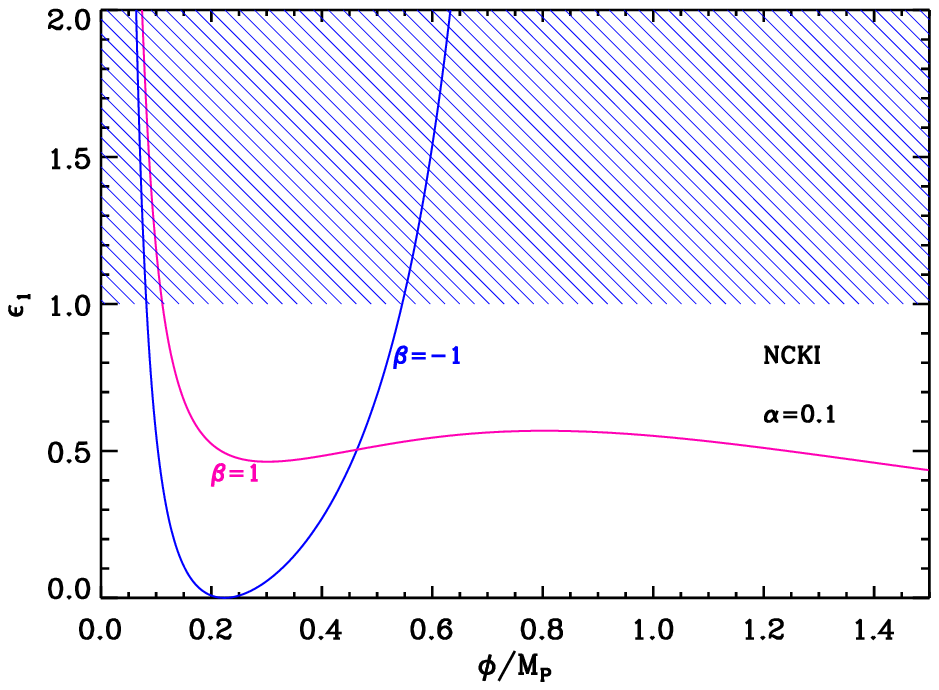}
\includegraphics[width=\wdblefig]{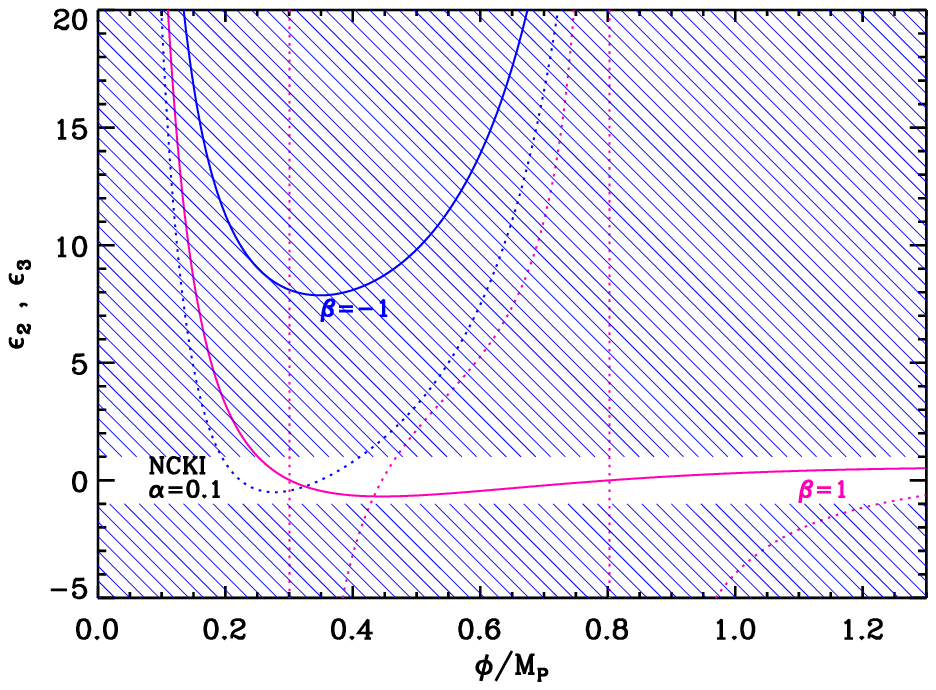}
\caption{Top left panel: Non Canonical K\"ahler Inflation (NCKI)
  potential for $\alpha=0.1$ and $\beta=\pm 1$. The solid blue line
  represents the case $\beta=-1$ while the solid pink line represents
  the case $\beta=1$. Top right panel: logarithm of the potential for
  the same values of $\alpha$ and $\beta$. Bottom left panel:
  slow-roll parameter $\epsilon_1$, for a potential with the same
  values of $\alpha$ and $\beta$ and the same color code.  The shaded
  area indicates the region where inflation is not possible.  Bottom
  right panel: slow-roll parameters $\epsilon _2$ (solid blue and pink
  lines) and $\epsilon_3$ (dotted blue and pink lines), for a
  potential with the values of $\alpha$ and $\beta$ already considered
  in the other panels.}
\label{potNCKI}
\end{center}
\end{figure}

The potential~(\ref{eq:ncki:pot}), as well as its logarithm, are
displayed in \Fig{potNCKI}. We now describe its shape. For this
purpose, let us first define the quantity $x\equiv\phi/\Mp$. If
$\beta>0$, the potential is definite positive provided
$x>\xVzeroMinus$, where
\begin{equation}
\label{eq:ncki:xV0Minus}
\xVzeroMinus=\left[\frac{\alpha}{2\beta} \Lambert{0} \left( \frac{2\beta}
{\alpha}  \ee^{-2/\alpha} \right) \right]^{1/2},
\end{equation}
and where $\Lambert{0}$ is the ``$0$''-branch of the Lambert
function. In this case, the potential is an increasing function of the
field \vev and, therefore, inflation proceeds from the right to the
left in the direction indicated by the arrow in \Fig{potNCKI}. Let us
also notice that, in this case, the potential has an inflection point
located at $\xddVzero = \sqrt{\alpha/\left(2\beta\right)}$. If
$\beta<0$, we must have $2\beta/\alpha\exp\left(1-2/\alpha\right)>-1$
in order to avoid the situation where the potential is everywhere
negative. This implies that either $\beta>-1$ or $\beta<-1$ and, in
this last case,
$\alpha<-2/\Lambert{-1}\left[1/\left(\ee\beta\right)\right]$ or
$\alpha>-2/\Lambert{0}\left[1/\left(\ee\beta\right)\right]$.  If one
of these conditions is satisfied (which is generically the case when
$\alpha\ll 1$), the potential is positive provided $\xVzeroMinus < x
<\xVzeroPlus$, where $\xVzeroMinus$ is defined in
\Eq{eq:ncki:xV0Minus} and where
\begin{equation}
\xVzeroPlus = \left[
  \frac{\alpha}{2\beta} \Lambert{-1}\left( \frac{2\beta}{\alpha}
  \ee^{-2/\alpha} \right) \right]^{1/2},
\end{equation}
$\Lambert{-1}$ being the $-1$ branch of the Lambert function. In this
case, the potential is a concave function of the field \vev, with a
maximum located at $\xdVzero =\sqrt{-\alpha/\left(2\beta\right)}$.
Typically, inflation proceeds from the right to the left at small
values of the field \vev compared to the Planck mass.

The Hubble flow functions in the slow-roll approximation are given by
\begin{equation}
\epsilon_1 = \frac{\left(\alpha+2\beta
  x^2\right)^2}{2x^2\left(1+\alpha\ln x+\beta x^2\right)^2},
\end{equation}
\begin{equation}
\epsilon_2 = 2
\dfrac{\alpha\left(\alpha+1\right)+\left(5\alpha-2\right)\beta
    x^2 +2\beta^2 x^4+\alpha\left(\alpha-2\beta x^2\right)\ln
    x}{x^2\left(1+\alpha\ln x+\beta x^2\right)^2} ,
\end{equation}
and
\begin{equation}
\begin{aligned}
\epsilon_3 &= \frac{1}{x^2}\left[
\frac{2\left(\alpha+2\beta x^2\right)^2}{\left(1+\alpha\ln x+\beta 
x^2\right)^2}
+\frac{\alpha-2\beta x^2}{1+\alpha\ln x+\beta x^2}
\right. \\& \left.
+\frac{\alpha^2+8\alpha\beta x^2-4\beta^2 x^4}
{\alpha\left(\alpha+1\right)+\left(5\alpha-2\right)\beta x^2+2\beta^2 x^4
+\alpha\left(\alpha-2\beta x^2\right)\ln x}\right].
\end{aligned}
\end{equation}
The are displayed in the bottom panels in \Fig{potNCKI}. If $\beta>0$,
the first slow-roll parameter $\epsilon_1$ diverges when $x\rightarrow
\xVzeroMinus$. For $x> \xVzeroMinus$, it first decreases, then reaches
a minimum, then increases and reaches a local maximum. Finally, from
this maximum, it decreases again and vanishes at infinity. Therefore,
inflation stops at a \vev $\xend$ solution of $\epsilon_1(\xend)=1$,
which cannot be solved analytically. It can be noticed that the value of
$\epsilon_1$ as its local maximum increases when $\alpha$ decreases. In the 
limit $\alpha\ll 1$, one has
\begin{equation}
  \epsilon_{1}^\mathrm{max}\simeq \frac{\beta}{2}\, ,
\end{equation}
which is reached at $x_{\epsilon_{1}^\mathrm{max}}\simeq
1/\sqrt{\beta}$ (still in the limit of very small $\beta$). This sets
an upper bound on $\beta$ in order for this local maximum to satisfy
$\epsilon_1\ll 1$. If not, inflation would proceed in the part of the
potential beyond its inflection point, corresponding to ``large
values'' of the field \vev and the model would formally be equivalent
to a quadratic model (LFI${}_2$, see \sectionc{sec:lfi}).
 
If $\beta<0$, the first slow-roll parameter diverges when
$x\rightarrow \xVzeroMinus$. For $x> \xVzeroMinus$, $\epsilon_1$
decreases, vanishes at the potential local maximum $\xdVzero$, and
then increases to blow up when $x\rightarrow x_{V=0}^+$. At the same
time, the second slow-roll parameter $\epsilon_2$ decreases in the
inflationary range $\xVzeroMinus < x <\xdVzero$. Let us also notice
that, since $\epsilon_2(\xdVzero)\propto
2\alpha-\alpha^2+\alpha^2\ln\left[-\alpha/(2\beta)\right]$, one has
$\epsilon_2>0$, thanks to the condition
$2\beta/\alpha\exp\left(1-2/\alpha\right)>-1$. Therefore the minimum
value of $\epsilon_2$ in the increasing branch of the potential is
reached at the potential maximum and is given by
\begin{equation}
\epstwoMin=\frac{-16\beta}
{2-\alpha\left[1+\ln\left(-2\frac{\beta}{\alpha}\right)\right]}\, .
\end{equation}
For $\alpha<-2\beta/e$ (which is generically the case since $\alpha\ll
1$), this number is such that $\epstwoMin>-8\beta$,
which puts a lower bound on $\beta$ in order for $\epsilon_2$ to
remain small and slow-roll to be satisfied. As it was the case for
$\beta>0$, inflation also ends when $\epsilon_1=1$. Notice that the
exact calculations are implemented in the \ASPIC routines.

Let us now turn to the slow-roll trajectory. It can be analytically
integrated using the dilogarithm function $\Li{2}$ and the
corresponding expression reads
\begin{equation}
\begin{aligned}
\Nend-N  = \left( 1-\frac{\alpha}{2}+\alpha\ln x \right) \frac{\ln
  \left(\alpha+2\beta x^2\right)}{4\beta} +\frac{x^2}{4} -
\frac{\alpha}{4\beta} \ln \alpha \ln x
+ \frac{\alpha}{8\beta} \Li{2} \left(-2\frac{\beta}{\alpha}x^2\right)
\\  - \left(1-\frac{\alpha}{2}+\alpha\ln \xend\right) \frac{\ln \left(
  \alpha + 2\beta \xend^2 \right)}{4\beta}
-\frac{\xend^2}{4} +\frac{\alpha}{4\beta}\ln \alpha \ln \xend
-\frac{\alpha}{8\beta} \Li{2} \left(-2\frac{\beta}{\alpha}\xend^2\right)
,
\end{aligned}
\end{equation}
where $\Nend$ is the number of \efolds at the end of inflation. An
approximate and simpler expression can be derived in the limit $\alpha
\ll 1$. In that limit, one obtains
$\Nend-N=x^2/4+\ln(x)/(2\beta)-\xend^ 2/4-\ln(\xend)/(2\beta)$, which
is precisely the slow-roll trajectory for the VHI models with
$\mu=\Mp/\sqrt{\beta}$ and $p=2$, see \Eq{eq:vhi:trajp2}. For
$\alpha\neq 0$, the exact trajectory cannot be inverted analytically.

Finally, the parameter $M$ can be determined from the CMB
normalization. One obtains the following expression
\begin{eqnarray}
  \left(\frac{M}{\Mp}\right)^4&=&720\pi^2
  \frac{\left(\alpha+2\beta \xstar^2\right)^2}
  {\xstar^2\left(1+\alpha\ln \xstar+\beta \xstar^2\right)^3}
  \frac{\Qrms^2}{T^2}\, .
\end{eqnarray}

The slow-roll predictions of the NCKI models are displayed in
\Fig{fig:CMBNCKIbetaGT0} and \Fig{fig:CMBNCKIbetaLT0} for $\beta>0$
and $\beta<0$, respectively. The reheating equation of state parameter
$\wrehbar$ has been taken to be $0$ but, since there is no potential
minimum around which the inflaton field can oscillate at the end of
inflation, this parameter is in fact unspecified. Some remarks are in
order at this point. Firstly, when $\beta>0$, we notice that
$\epsilon_2$ at Hubble crossing is either positive or negative while,
when $\beta<0$, it is always positive. This is in agreement with what
we have discussed before. Secondly, when $\beta>0$ and $\alpha\ll 1$,
one can check that the predictions of the models are similar to the
VHI ones with $p=2$ (compare with \Fig{fig:CMBVHIpEQ2}). Again, this
is consistent with the previous considerations. Thirdly, when
$\vert\beta\vert\gtrsim \order{1}$, the predictions of the models do
not depend much on $\beta$ . Finally, as expected, when
$\beta\rightarrow 0$, one recovers the predictions of the LI models,
see \sectionc{sec:li} and \Fig{fig:CMBLIalphaPositive}. Now, in the
regime $\vert\beta\vert= \order{1}$ and $\alpha\ll 1$,
\Fig{fig:CMBNCKIbetaGT0} and \Fig{fig:CMBNCKIbetaLT0} indicate that
the case $\beta>0$ is disfavored by the observations. The situation is
even worst for $\beta<0$, the deviation from scale invariance being
clearly too important to satisfy the observational constraints.

\subsection{Constant Spectrum Inflation (CSI)}
\label{sec:csi}

\begin{figure}
\begin{center}
\includegraphics[width=\wdblefig]{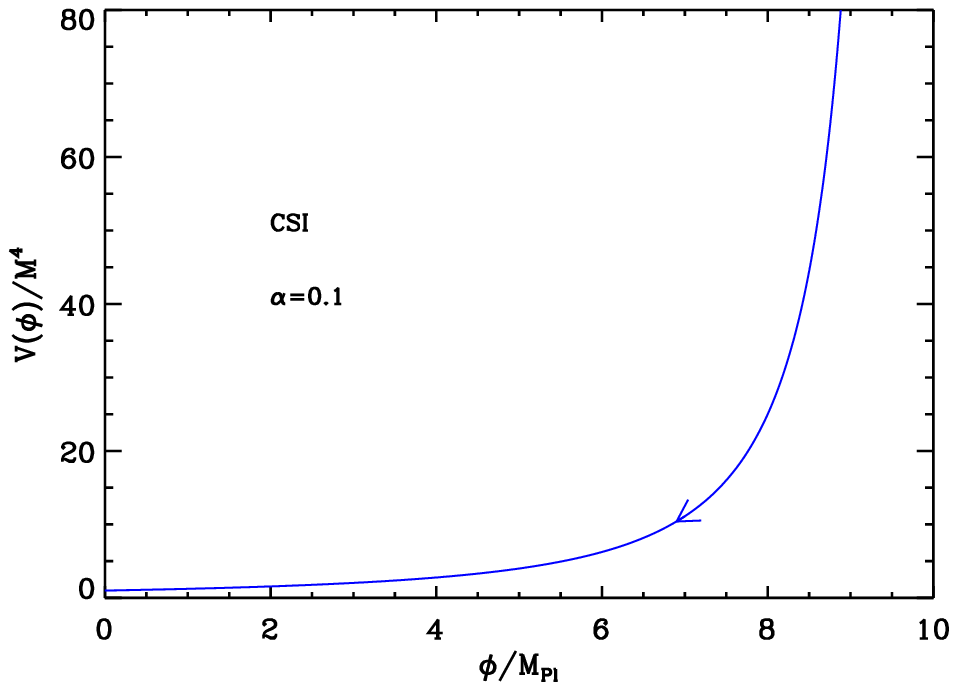}
\includegraphics[width=\wdblefig]{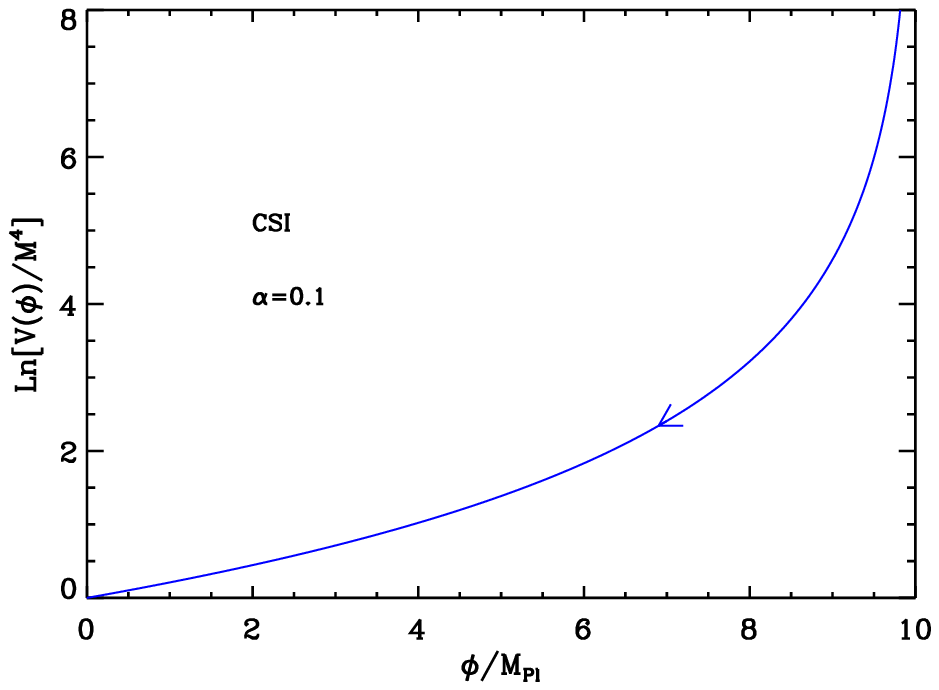}
\includegraphics[width=\wdblefig]{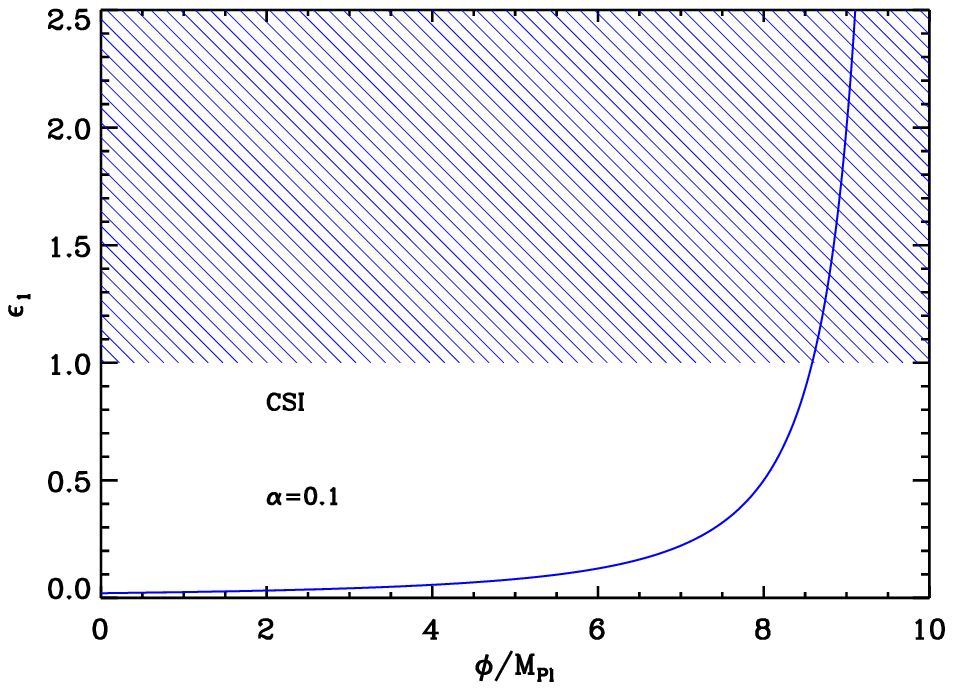}
\includegraphics[width=\wdblefig]{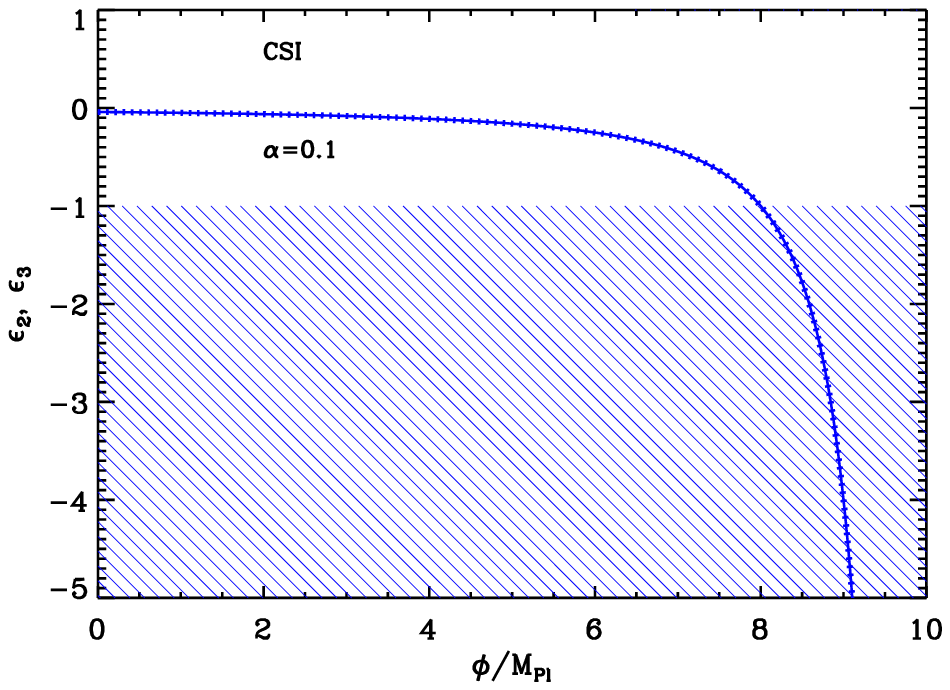}
\caption{Constant Spectrum Inflation (CSI) for $\alpha=0.1$.  Upper
  panels: the potential and its logarithm along the branch
  $x<1/\alpha$. Bottom left panel: slow-roll parameter $\epsilon_1$
  together with the region in which it is larger than unity and in
  which inflation cannot occur (shaded). Bottom right panel: slow-roll
  parameter $\epsilon_2=\epsilon _3$ along the same branch
  $x<1/\alpha$.}
\label{fig:potCSI}
\end{center}
\end{figure}

This potential belongs to the class of models discussed in
\Refc{Hodges:1990bf} and is constructed in order to produce a power
spectrum $P\left(k\right)\propto k^0$ for the primordial density
fluctuations, \ie a power spectrum with constant spectral index such
that $\nS=1$ (exact scale invariance). It reads
\begin{equation}
V\left(\phi\right) = \frac{M^4}{\left( 1-\alpha \dfrac{\phi}{\Mp} \right)^2}\, .
\end{equation}
There is a symmetry for $\phi/\Mp\rightarrow 2/\alpha-\phi/\Mp$ and
inflation can proceed indifferently in the branch $\phi/\Mp<1/\alpha$
or in the branch $\phi/\Mp>1/\alpha$, leading to the same physical
predictions. For this reason, in the following, we will be interested
in the branch $\phi/\Mp<1/\alpha$. Defining the quantity $x$ by
\begin{equation}
x \equiv \dfrac{\phi}{\Mp}\,,
\end{equation}
the first three Hubble flow functions in the slow-roll approximation
are given by
\begin{equation}
\epsilon_1 = \frac{2\alpha^2}{\left(\alpha x-1\right)^2}\, ,\qquad
\epsilon_2 = \epsilon_3=-2 \epsilon_1.
\end{equation}
The previous relation $\epsilon_2=-2\epsilon_1$ means that, at first
order in slow-roll, the spectral index is indeed equals to unity,
$\nS-1=0$. Recall that the potential of this model is precisely
constructed in order for this relation to be true. Let us notice,
however, that, at second order in slow-roll,
$\epsilon_2=\epsilon_3=-2\epsilon_1$ yields
$\nS-1=4\epsilon_1^2>0$. One should note that another way to realize
$\nS-1=0$ at first order in slow-roll is to take the large field
inflation potential LFI (see \sectionc{sec:lfi}) with a negative power
index $p=-2$. In that case one also has
$\epsilon_2=\epsilon_3=-2\epsilon_1$ and, at second order,
$\nS-1=4\epsilon_1^2$ is also verified. However, since the explicit
expressions of $\epsilon_1$ for CSI and LFI ($p=-2$) are different,
the actual value of the spectral index at second order is also
different. The potential and the Hubble flow functions have been
represented in \Fig{fig:potCSI}.

As can be checked in this figure, $\epsilon_1$ is a monotonic
function of $x$ in both branches of the potential. It diverges at
$x=1/\alpha$ and vanishes for $x\rightarrow \pm\infty$. Inflation can
therefore take place in the region $x<\xepsoneOneMinus$ for the branch
$x<1/\alpha$ (or $x>\xepsoneOnePlus$ for the branch $x>1/\alpha$),
where $\xepsoneOnePM$ are the field values at which $\epsilon_1=1$:
\begin{equation}
\xepsoneOnePM = \frac{1\pm\sqrt{2}\alpha}{\alpha}\, .
\end{equation}
Since the field evolution proceeds from the right to the left from
$\xepsoneOnePM$, inflation does not stop by slow-roll violation and an
extra mechanism parametrized by $\xend$ should be considered in order
to end it. For this reason, CSI is in fact a two parameters model. Let
us also notice that the slow-roll parameters $\epsilon_2=\epsilon_3$
are negative monotonic functions of $x$ in both branches of the
potential and cross the line $\epsilon_2=\epsilon_3=-1$ at
\begin{equation}
\xepstwoMinusOnePM = \xepsthreeMinusOnePM = \frac{1\pm 2\alpha}{\alpha}\, .
\end{equation}
As a result, there is a small domain $\xepstwoMinusOneMinus
<x<\xepsoneOneMinus$ where we have inflation but where the slow-roll
approximation is violated (this is also true for the other
branch). This is not problematic since the system is driven away from
this regime towards a situation in which all the Hubble flow functions
become small (see \Fig{fig:potCSI}).

The slow-roll trajectory can be integrated explicitly and reads
\begin{equation}
\Nend -N = -\frac{x^2}{4} + \frac{x}{2\alpha} + \frac{\xend^2}{4} -
\frac{\xend}{2\alpha}\,.
\end{equation}
It can also be  inverted analytically and it follows that
\begin{equation}
x=\frac{ 1\pm\sqrt{1-2\alpha \xend + \alpha^2 \xend^2 +
    4\alpha^2\left(N-\Nend\right)}}{\alpha}\,.
\end{equation}
The sign $\mp$ depends on whether one works in the $x<1/\alpha$ branch
or in the $x>1/\alpha$ branch, respectively. A consequence of this
formula is the fact that, if one requires $\Nend-\Nini$ \efolds
during inflation, then $\xend$ should be smaller than some value
$\xendmax$ given by
\begin{equation}
\xendmax=\dfrac{1}{\alpha} - \sqrt{2+4\left(\Nend-\Nini\right)}\,,
\end{equation}
in the $x<1/\alpha$ branch. Equivalently, taking the minus sign in
this expression would lead to $\xendmin$ for the branch $x>1/\alpha$.

Finally, the observable field value $\xstar$ is obtained by solving
\Eq{eq:phistarlnrrad} while the amplitude of the CMB anisotropies
fixes the parameter $M$ to
\begin{equation}
\left(\frac{M}{\Mp}\right)^4=2880\pi^2\alpha^2 \frac{\Qrms^2}{T^2}\, .
\end{equation}
Interestingly enough, it only depends on $\alpha$, and not on $\xstar$
(\ie it has no explicit dependence on the reheating). The reheating
consistent slow-roll predictions for the CSI models are represented in
\Figs{fig:CMBCSIalphaEQ10PowerMinus3} and \ref{fig:CMBCSIalphaEQ1} for
$\alpha=10^{-3}$ and $\alpha=1$, respectively.

\subsection{Orientifold Inflation (OI)}
\label{sec:oi}

\subsubsection{Theoretical Justifications}

The model is based on the following considerations. Let us start with
a $N=1$ supersymmetric Yang-Mills gauge theory the Lagrangian of which
can be written as
\begin{equation}
\calL=-\frac{1}{4}F_{\mu \nu}^aF^{a\mu \nu}
+\frac{i}{2}\bar{\lambda}^a\slashed D _{ab}\lambda^b,
\end{equation}
with $a=1, \cdots, N_\uc^2$, $N_\uc$ being the number characterizing
the group $\mathrm{SU}(N_\uc)$. $F_{\mu \nu}^a$ is the field strength,
$\lambda^a$ a spinor field and $\slashed{D}$ a covariant derivative. A
is a composite scalar field, i.e. a bound state denoted by $\varphi
\simeq \lambda \bar{\lambda}$, can actually appear in the theory if a
strongly interacting regime takes place. The effective Lagrangian
aimed at describing its dynamics has been derived in
\Refc{Veneziano:1982ah} and reads
\begin{equation}
\calL_{\mathrm{YV}} =-\frac{N_\uc^2}{\alphaOI}\left(\varphi\varphi^{\dagger}\right)^{-2/3}
\partial_{\mu}\varphi\partial^{\mu}\varphi^{\dagger}
-\frac{4\alphaOI N_\uc^2}{9}\left(\varphi\varphi^{\dagger}\right)^{2/3}
\ln\left(\frac{\varphi}{\Lambda^3}\right)
\ln\left(\frac{\varphi^{\dagger}}{\Lambda^3}\right),
\end{equation}
where $\alphaOI$ is a constant and $\Lambda$ a mass scale. This class of
theories are discussed in more detail in \sectionc{sec:lpi}. However,
in \Refc{Channuie:2012bv}, it was argued that in ``orientifold
theories'', the above Lagrangian can be slightly deformed and now takes
the form
\begin{equation}
\label{eq:lagrangeoi}
\calL_{\uOI} =-\frac{N_\uc^2}{\alphaOI}\left(\varphi\varphi^{\dagger}\right)^{-2/3}
\partial_{\mu}\varphi\partial^{\mu}\varphi^{\dagger}
-\frac{4\alphaOI N_\uc^2}{9}\left(\varphi\varphi^{\dagger}\right)^{2/3}
\left[\ln\left(\frac{\varphi}{\Lambda^3}\right)
\ln\left(\frac{\varphi^{\dagger}}{\Lambda^3}\right)
-\beta\right],
\end{equation}
where $\beta=\order{1/N_\uc}$. \Refc{Channuie:2012bv} raised the
possibility that $\varphi$ (or, rather, its canonically conjugated
version) could be the inflaton. In fact, in order to study
this question, one must also specify the gravitational coupling. In
\Refc{Channuie:2012bv}, the scalar field $\varphi $ is non-minimally
coupled to gravity such that, in the Jordan frame,
\begin{equation}
S=\int \dd^4\bmx \sqrt{-g}\left[-\frac{M^2+N_\uc^2\xi
\left(\varphi\varphi^{\dagger}\right)^{1/3}}{2}R+\calL_{\uOI}\right],
\end{equation}
where $M$ is a mass scale. There is a new parameter in the problem,
$\xi$, which describes the strength of the non-minimal coupling to
gravity (as it was the case for Higgs inflation,
see \sectionc{sec:hi}). Then, in the Einstein frame, one can write the
above model as~\Refc{Channuie:2012bv}
\begin{eqnarray}
S &=&\int \dd^4 \bmx\sqrt{-g}
\biggl\{-\frac{1}{2}\Mp^2R-\frac{N_\uc^2}{\alphaOI}
\Omega^{-2}\left[1+\frac{\alphaOI N_\uc^2\xi^2}{3\Mp^2}\Omega^{-2}
\left(\varphi\varphi^{\dagger}\right)^{1/3}\right]
\left(\varphi\varphi^{\dagger}\right)^{-2/3}
\partial_{\mu}\varphi\partial^{\mu}\varphi^{\dagger}
\nonumber \\ & &
-\Omega^{-4}V_{\uOI}\biggr\}.
\end{eqnarray}
In this expression, $V_{\uOI}$ refers to the second term in
\Eq{eq:lagrangeoi} and
\begin{equation}
\Omega^2\equiv \dfrac{M^2+N_\uc^2\xi
  \left(\varphi\varphi^{\dagger}\right)^{1/3}}{\Mp^2}\,.
\end{equation}
In the following, we consider two situations: the case where $\xi\neq
0$ such that $\Omega ^2\simeq N_\uc^2\xi\varphi^{2/3}/\Mp^2$, \ie the
second term in the definition of $\Omega^2$ dominates (the large field
limit) and the case $\xi=0$. In the first case, taking
$\varphi=\varphi^{\dagger}$ and canonically normalizing the field one
finds
\begin{equation}
V(\varphi)=\frac{4\alphaOI \Mp^4}{9N_{\uc}^2\xi^2}
\left[\left(\ln \frac{\varphi}{\Lambda^3}\right)^2-\beta\right].
\end{equation}
The canonically normalized field is $\phi/\Mp\propto \ln
\varphi$. Since $\beta$ is a small number, it can be neglected and
this model is in fact a LFI model with $V(\phi)\propto \phi^2$ which
was already studied in \sectionc{sec:lfi}. For the second case, it is
sufficient to restart from \Eq{eq:lagrangeoi}. Then, the
canonically normalized field reads
\begin{equation}
\frac{\varphi}{\Lambda^3}=\left(\frac{\phi}{\phizero}\right)^3,
\end{equation}
with
\begin{equation}
\label{eq:defphizero}
\phizero=3N_\uc\left(\frac{2}{\alphaOI}\right)^{1/3}\Lambda.
\end{equation}
It follows that the potential can be written as
\begin{equation}
\label{eq:potoi}
  V=\alphaOI N_{\uc}^2\Lambda^4 \left(\frac{\phi}{\phizero}\right)^4
  \left[\ln^2\left(\frac{\phi}{\phizero}\right)-\frac{\beta}{9}\right].
\end{equation}
This model is studied in detail in the next subsection. The case
$\beta=0$ will also be investigated in \sectionc{sec:lpi}.

\subsubsection{Slow-Roll Analysis}
\label{subsubsec:sroi}

\begin{figure}
\begin{center}
\includegraphics[width=\wdblefig]{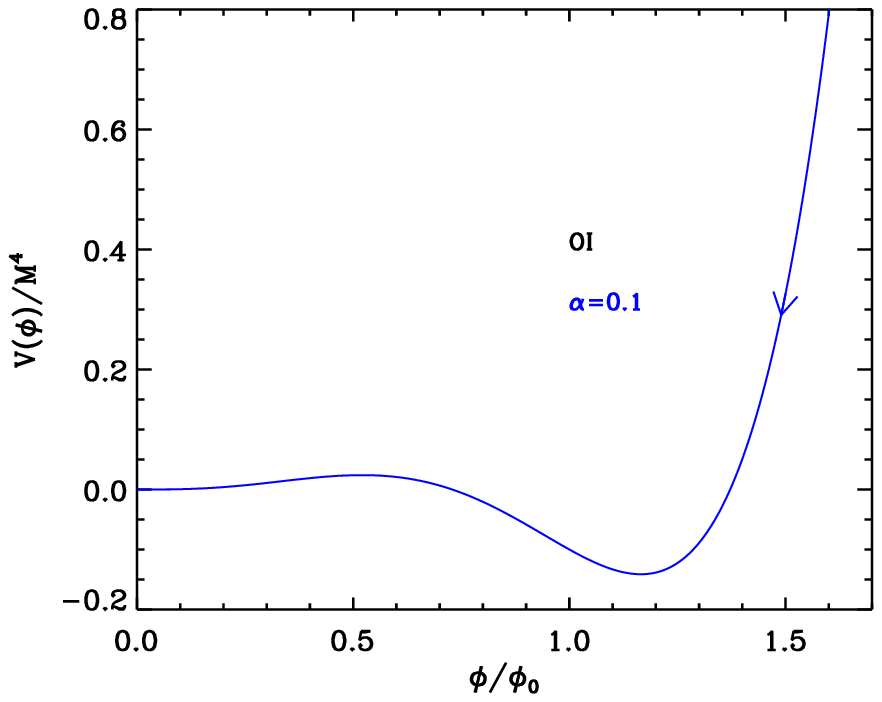}
\includegraphics[width=\wdblefig]{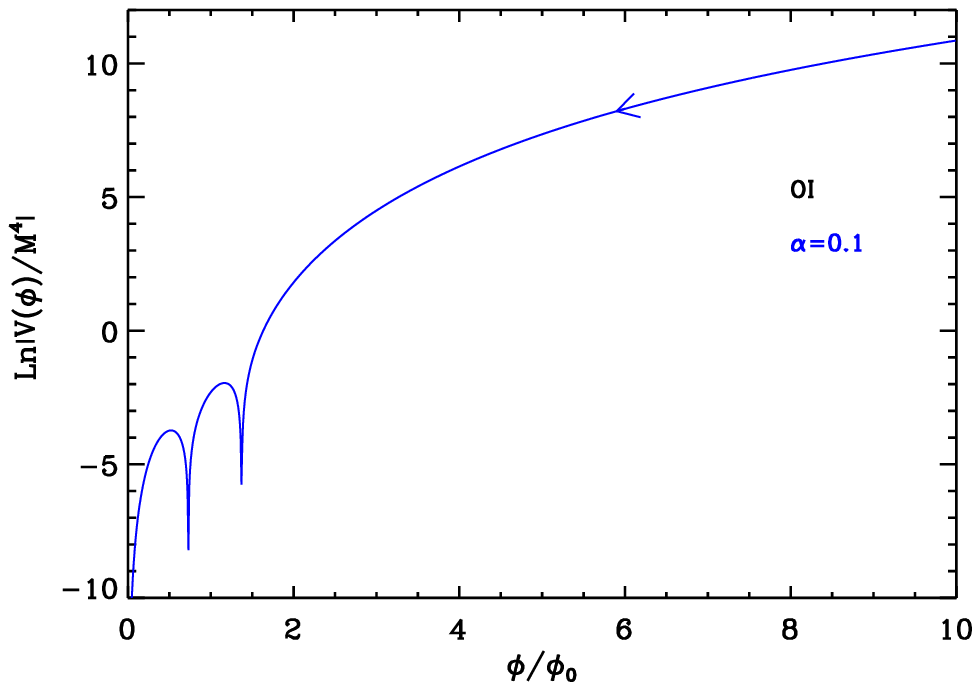}
\includegraphics[width=\wdblefig]{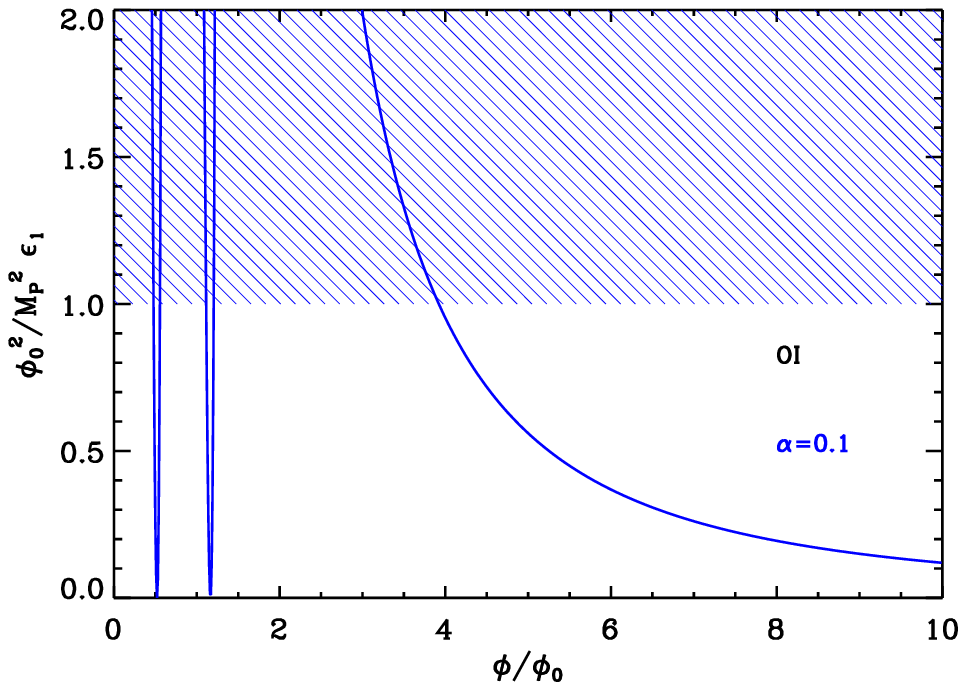}
\includegraphics[width=\wdblefig]{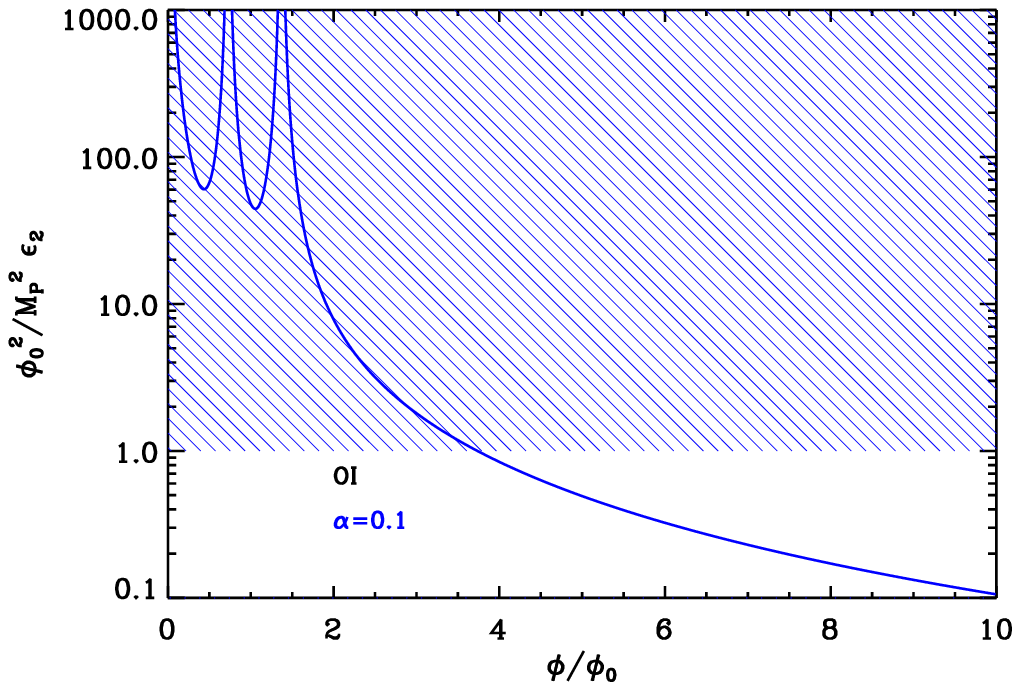}
\caption{Orientifold Inflation (OI) for $\alpha=0.1$. Upper panels:
  the potential and its logarithm. Bottom left panel: slow-roll
  parameter $\epsilon_1$, rescaled by the factor
  $\phizero^2/\Mp^2$. The shaded area indicates where inflation cannot
  occur (for $\phizero=\Mp$).  Bottom right panel: rescaled slow-roll
  parameter $\epsilon_2$.}
\label{fig:potOI}
\end{center}
\end{figure}

We now turn to the slow-roll study of the potential derived previously
in \Eq{eq:potoi}. This one can be re-written as
\begin{equation}
V\left(\phi\right) = M^4 \left(\frac{\phi}{\phizero} \right)^{4}\left[
 \left(\ln \frac{\phi}{\phizero} \right)^2- \alpha \right],
\end{equation}
where we have defined 
\begin{equation}
M^4=\alphaOI N_{\uc}^2\Lambda^4, \qquad \alpha\equiv \frac{\beta}{9}\,.
\end{equation}
One should be careful that $\alphaOI$ appearing in the first of the
two above equations stems from the Lagrangian used in the previous
subsection while the observable constant $\alpha$ only refers to the
quantity $\beta/9 = \order{1/\Nc} \ll 1$. The scale
$\phizero$ is defined in \Eq{eq:defphizero} and will be chosen
such that $\phizero\simeq 10^{16}\,\GeV$. The potential as well as its
logarithm are displayed in \Fig{fig:potOI}.

Defining the quantity $x$ by the following expression
\begin{equation}
x\equiv \dfrac{\phi}{\phizero}\,,
\end{equation}
the potential remains positive provided $x<\xVzeroMinus$ or
$x>\xVzeroPlus$, where
\begin{equation}
\xVzeroPM = \ee^{\pm{\sqrt{\alpha}}} .
\end{equation}
It vanishes at $x=0$, then increases to reach a local maximum at
$\xdVzeroMinus$, decreases again to become negative at $\xVzeroMinus$,
reaches a local minimum at $\xdVzeroPlus$, then increases again to
become positive at $\xVzeroPlus$ and diverges asymptotically. The
values of $\xdVzeroMinus$ and $\xdVzeroPlus$ are given by
\begin{equation}
x_{V^\prime=0}^\pm=\ee^{-\frac{1}{4}\pm\sqrt{\frac{1}{16}+\alpha}} .
\label{eq:oi:xvprimeNul}
\end{equation}
A priori three regimes of inflation may exist: $x<\xdVzeroMinus$ and
inflation proceeds from the right to the left,
$\xdVzeroMinus<x<\xVzeroMinus$ and inflation proceeds from the left to
the right, $\xVzeroPlus<x$ and inflation proceeds from the right to
the left in the direction specified by the arrow in
\Fig{fig:potOI}. As explained below, only the third possibility allows
us to have a slow-roll inflationary regime.

Let us now calculate the quantities $\epsilon_n$. The first three
Hubble flow functions in the slow-roll approximation are given by
\begin{equation}
\epsilon_1 = 2\frac{\Mp^2}{\phizero^2}
\left(\frac{2\ln^2 x +\ln x -2\alpha}
{x \ln^2 x -\alpha x } \right)^2 ,
\label{eq:sr1oi}
\end{equation}
\begin{equation}
\epsilon_2 = 4\frac{\Mp^2}{\phizero^2}
\frac{2\ln^4 x +\ln^3 x
+\left(1-4\alpha\right)\ln^2 x - \alpha\ln x +\alpha+2\alpha^2}
{\left(x \ln^2 x -\alpha x\right)^2} \,,
\label{eq:sr2oi}
\end{equation}
and
\begin{equation}
\begin{aligned}
\epsilon_3 & = 2\frac{\Mp^2}{\phizero^2} \left[8\alpha^4+6\alpha^3-\alpha^2
  \left(8\alpha + 15\right) \ln x  + 2 \alpha \left(3
  -16\alpha^2 - 2 \alpha \right) \ln^2 x  \right. \\ &
  + 8 \alpha \left(3\alpha + 1\right) \ln^3 x   + \left. 2
  \left(24\alpha^2 - 5 \alpha + 1\right) \ln^4 x  +
  \left(7 -24\alpha \right) \ln^5 x  + 8
  \left(1 -4\alpha \right)\ln^6 x  \right. \\ & +
  \left. 8 \ln^7 x + 8
  \ln^8 x \right] \left(x\ln^2 x - \alpha x
  \right)^{-2} \\ & \times \left[2\alpha^2 + \alpha - \alpha 
  \ln x  +
  \left(1-4\alpha\right) \ln^2 x  + 
  \ln^3 x  + 2 \ln^4 x \right]^{-1}.
\end{aligned}
\label{eq:sr3oi}
\end{equation}
They have been represented in \Fig{fig:potOI}. One can see that the
slow-roll regime can only take place in the $x>\xVzeroPlus$ region,
where $\epsilon_1$ continuously increase as inflation proceeds from
the right to the left, and diverges at $\xVzeroPlus$. In the other
domains, $\epsilon_2$ remains too large to support slow-roll
inflation. Within the $x>\xVzeroPlus$ domain, inflation naturally ends
by slow-roll violation, but the field value $\xend$ at which this
occurs has to be determined numerically. However, since $\phizero
\simeq 10^{16}\,\GeV$, one can derive an approximated formula for
$\xend$ in the $\phizero\ll\Mp$ limit, namely
\begin{equation}
\xend\simeq2\sqrt{2}\frac{\Mp}{\phizero}\, .
\end{equation}

The next step is to derive the slow-roll trajectory. It can be
obtained from \Eq{eq:srtrajectory} and reads
\begin{equation}
\begin{aligned}
& \Nend - N = -\frac{\phizero^2}{\Mp^2}\Bigg\lbrace\frac{\xend^2-x^2}{8}
+\frac{\ln^2\left(\xdVzeroPlus\right)-\alpha}{2\sqrt{1+16\alpha}} \left(
\xdVzeroPlus\right)^2 \left[ \Ei\left(2\ln \frac{\xend}{\xdVzeroPlus}
  \right) \right. \\ & - \left. \Ei\left(2\ln
  \frac{x}{\xdVzeroPlus}\right)\right] - \frac{
  \ln^2\left(\xdVzeroMinus\right)-\alpha}{2\sqrt{1+16\alpha}} \left(
\xdVzeroMinus \right)^2 \left[\Ei\left( 2\ln \frac{\xend}{\xdVzeroMinus}
  \right) -\Ei\left(2\ln \frac{x}{\xdVzeroMinus} \right) \right]
\Bigg\rbrace\, ,
\end{aligned}
\end{equation}
where $\Ei$ is the exponential integral function, and where
$\xdVzeroPM$ have been defined in \Eq{eq:oi:xvprimeNul}. In the
$\phizero\ll\Mp$ limit, this trajectory reduces to
$\Delta\Nstar\simeq\phizero^2/(8\Mp^2)(\xstar^2-\xend^2)$, where we have
introduced the observable field value $\xstar$ at which the pivot
scale crossed the Hubble radius during inflation. It can be inverted
to give $\xstar$ in terms of $\Delta \Nstar=\Nend - \Nstar$ and one
gets
\begin{equation}
\xstar\simeq2\sqrt{2}\frac{\Mp}{\phizero}\sqrt{\Delta\Nstar+1}\, .
\end{equation}
Plugging this into \Eqs{eq:sr1oi}, \eqref{eq:sr2oi} and
\eqref{eq:sr3oi} gives the approximated expressions
\begin{equation}
\label{eq:oi:predic}
\epsilon_{1*} \simeq \epsilon_{2*} \simeq \epsilon_{3*}
\simeq\frac{1}{\Delta\Nstar+1}\, ,
\end{equation}
hence
\begin{equation}
r\simeq\frac{16}{\Delta\Nstar+1}\, ,\quad\quad\quad\quad
\nS-1\simeq-\frac{3}{\Delta\Nstar+1}\, ,\quad\quad\quad\quad
\alphaS\simeq-\frac{3}{\left(\Delta\Nstar+1\right)^2}\, .
\end{equation}

{}From $\xstar$, the parameter $M$ is fixed by the amplitude of the
CMB anisotropies and one obtains
\begin{equation}
\left(\frac{M}{\Mp}\right)^4=\frac{2880\pi^2
  \left(2\ln^2 \xstar +\ln \xstar -2\alpha\right)^2}
     {\xstar^6\left(\ln^2 \xstar -\alpha\right)^3}
     \dfrac{\Mp^2}{\phizero^2} \dfrac{\Qrms^2}{T^2}\, .
\end{equation}
In the $\phizero\ll\Mp$ limit, the previous expression reduces to the
following formula
\begin{equation}
\left(\frac{M}{\Mp}\right)^4\simeq \dfrac{45\pi^2}{2 \left(
  \Delta\Nstar + 1\right)^3} \left(\dfrac{\phizero}{\Mp}
\right)^4\frac{1}{ \ln^ 2 \left(2\sqrt{2}\dfrac{\Mp}{\phizero}
  \sqrt{\Delta\Nstar+1}\right)} \dfrac{\Qrms^2}{T^2}\, .
\end{equation}
With $\phizero\simeq 10^{16}\, \GeV$, this typically gives $M/\Mp
\simeq 5\times 10^{-4}$.

The reheating consistent slow-roll predictions for the orientifold
inflation models are displayed in \Fig{fig:CMBOI}, for
$\phizero/\Mp=10^{-4}$,$10^{-2}$, and $1$. Let us recall that natural
values are around $\phizero \simeq 10^{16}\,\GeV$ and $\alpha \in
\left[10^{-3},1 \right]$.  The reheating equation of state parameter
has been fixed to $\wrehbar=0$ since the potential is quadratic in the
vicinity of its minimum. According to the rough picture provided by
\Eq{eq:oi:predic}, the predictions of these models almost do not
depend on its parameters $\phizero$ and $\alpha$, which is why all the
points in \Fig{fig:CMBOI} are superimposed. In particular, one can see
that these models generically predict an important amount of
gravitational waves which is disfavored by the observations.

\subsection{Constant \texorpdfstring{$\nS$}{nS} C Inflation (CNCI)}
\label{sec:cnci}

\begin{figure}
\begin{center}
\includegraphics[width=\wdblefig]{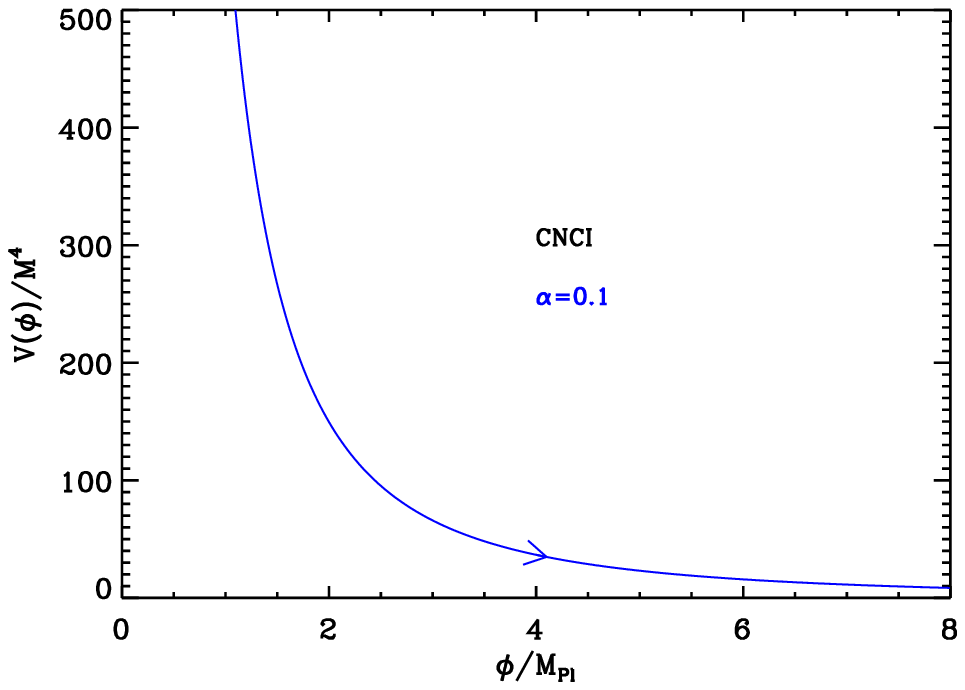}
\includegraphics[width=\wdblefig]{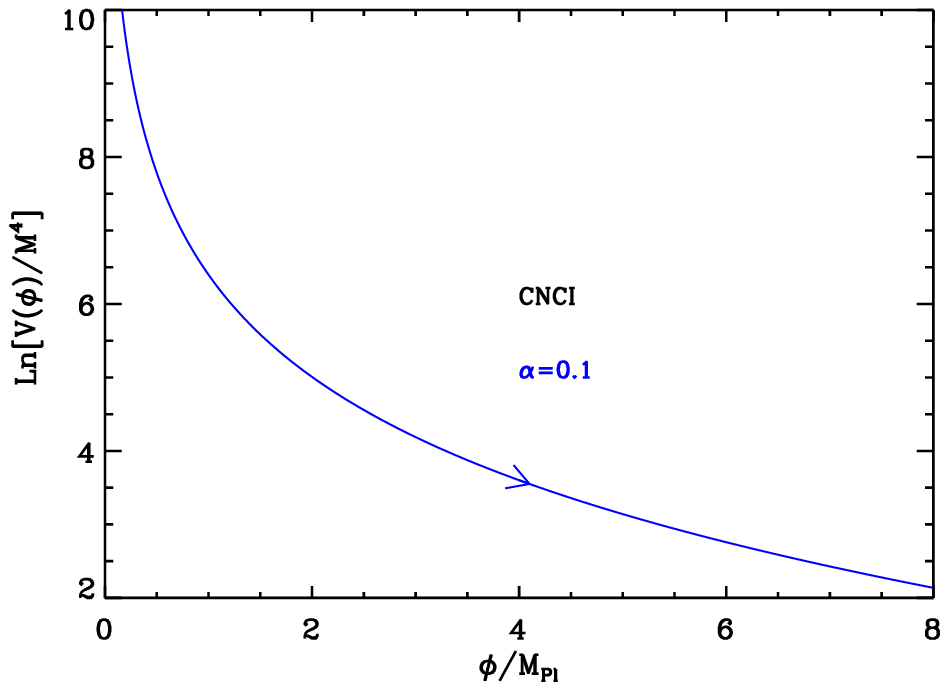}
\includegraphics[width=\wdblefig]{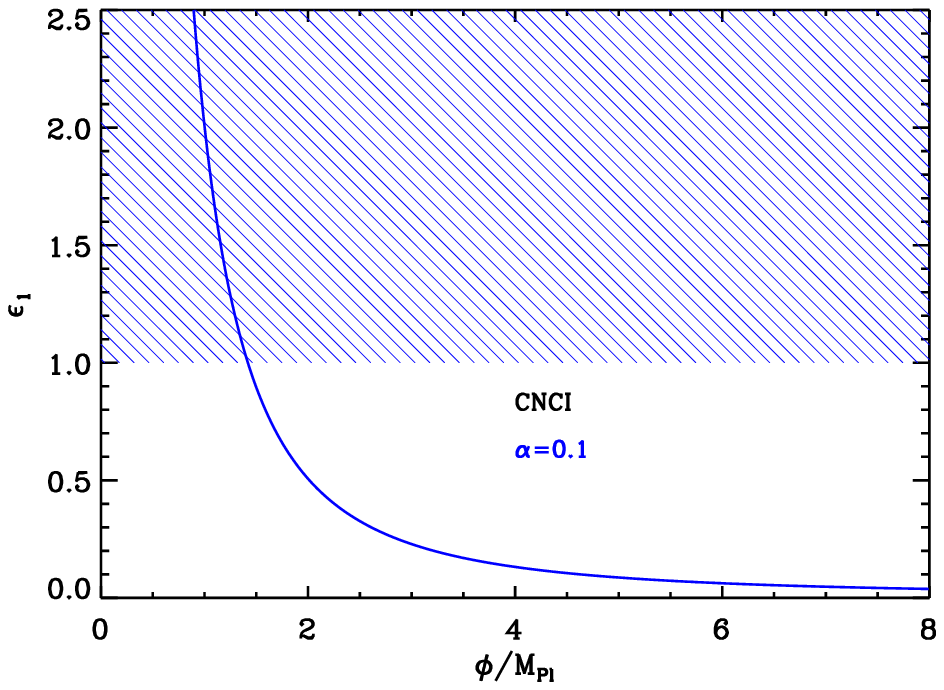}
\includegraphics[width=\wdblefig]{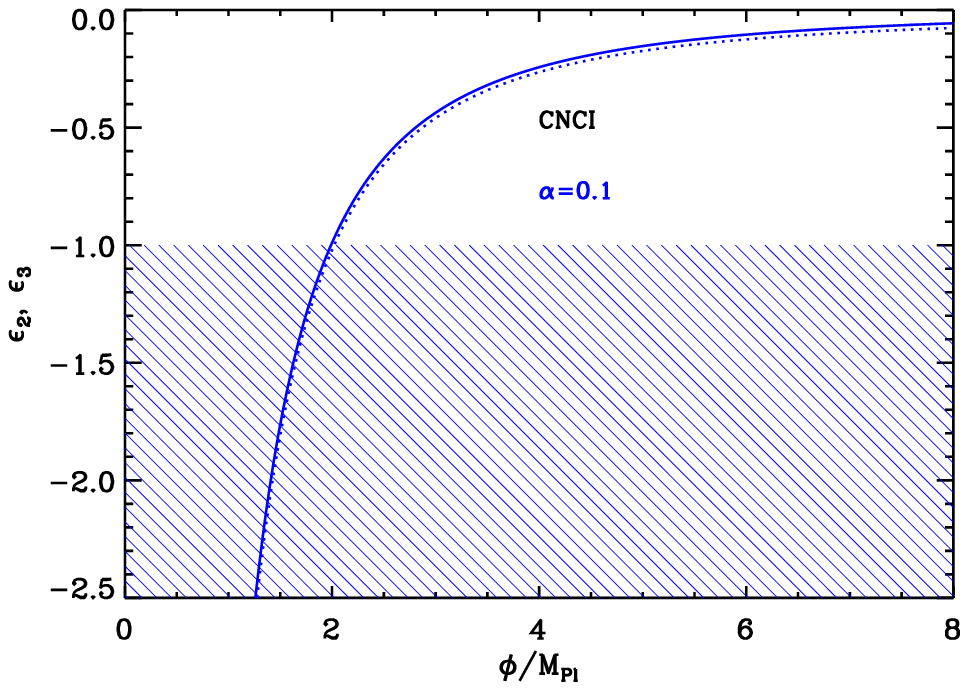}
\caption{Top left panel: Constant $\nS$ C inflaton potential for
  $\alpha=0.1$.  Inflation proceeds from the left to the right as
  indicated by the arrow. Top right panel: logarithm of the potential
  for the same value of $\alpha$. Bottom left panel: the first
  slow-roll parameter $\epsilon _1$ for $\alpha=0.1$. Bottom right
  panel: slow-roll parameters $\epsilon _2$ and $\epsilon _3$, still
  for $\alpha=0.1$.}
\label{fig:potCNCI}
\end{center}
\end{figure}

This model has been obtained in \Refc{Vallinotto:2003vf} and is the
third example of a class of scenarios already studied in
\sectioncs{sec:cnai} and~\ref{sec:cnbi}. As explained in those
sections, the corresponding potential is designed in order to produce
a power spectrum with constant spectral index. The potential studied
in this section reads
\begin{equation}
\label{eq:potCNCI}
  V\left(\phi\right) = M^4\left[ \left( 3+\alpha^2 \right) \coth^2
    \left(\frac{\alpha}{\sqrt{2}}\frac{\phi}{\Mp}\right)- 3 \right],
\end{equation}
where $\alpha$ is a positive dimensionless parameter (denoted $n_0$ in
\Refc{Vallinotto:2003vf}). The potential being symmetrical in
$\phi\rightarrow -\phi$, only the $\phi>0$ part is displayed in
\Fig{fig:potCNCI}. It is a decreasing function of the field \vev, and
its asymptotic value when $\phi/\Mp$ goes to infinity is given by
$\alpha^2M^4$, hence the potential is always positive.

Defining $x=\phi/\Mp$, the three first slow-roll parameters are given
by
\begin{equation}
  \epsilon_1 = \frac{ 4\alpha^2 \left( 3+\alpha^2 \right)^2 \coth^2
    \left( \dfrac{\alpha x}{\sqrt{2}} \right)}{ \left[
      6+\alpha^2+\alpha^2\cosh\left( \sqrt{2}\alpha x \right)
      \right]^2}\,,
\end{equation}
\begin{equation}
\epsilon_2 = -\frac{2\alpha^2\left(3+\alpha^2\right) \left[
    12+\alpha^2+2\alpha^2\cosh\left( \sqrt{2}\alpha x \right) +
    \alpha^2\cosh\left( 2\sqrt{2}\alpha x \right) \right]} {\left[
    6+\alpha^2+\alpha^2\cosh\left( \sqrt{2}\alpha x \right) \right]^2
  \sinh^2\left( \dfrac{\alpha x}{\sqrt{2}} \right)} \, ,
\end{equation}
and
\begin{equation}
\begin{aligned}
  \epsilon_3 &= -2\alpha^2\left( 3+\alpha^2 \right)\left[
    6\left(24-2\alpha^2+\alpha^4 \right) + \left(
    120\alpha^2+7\alpha^4 \right) \cosh\left( \sqrt{2}\alpha x \right)
    \right. \\ & + \left. 2\alpha^2\left( \alpha^2-6
    \right)\cosh\left( 2\sqrt{2}\alpha x \right)+\alpha^4\cosh\left(
    3\sqrt{2}\alpha x \right) \right]
  \coth^2\left(\frac{\alpha}{\sqrt{2}}x \right) \\ & \times  \left[
    6+\alpha^2+\alpha^2\cosh \left( \sqrt{2}\alpha x \right)
    \right]^{-2} \left[ 12+\alpha^2+2\alpha^2\cosh \left(
    \sqrt{2}\alpha x \right) + \alpha^2\cosh\left( 2\sqrt{2}\alpha x
    \right) \right]^{-1}.
\end{aligned}
\end{equation}
These slow-roll parameters are displayed in \Fig{fig:potCNCI} (bottom
panels). We see that the first slow-roll parameters monotonically
decreases during inflation. It blows up as the field \vev approaches
zero and tends to zero when the field \vev goes to infinity. On the
contrary, the second and third slow-roll parameters monotonically
increase from $-\infty$ to zero as inflation proceeds.

Given the above described behavior of $\epsilon_1$, it is clear that
inflation cannot stop by slow-roll violation. Therefore, it should be
stopped by instability which means that an extra parameter $\xend$
should be added to the model.

As for CNAI and CNBI, the spectral index
$\nS-1=-2\epsilon_1-\epsilon_{2}$ at first order in slow-roll, can be
made constant in some limit. Expanding the slow-roll parameters
$\epsilon_1$ and $\epsilon_2$ in $\alpha$, assuming that $x \alpha$
remains small, one obtains $\epsilon_1 = 2/x^2 + 2\alpha^2 /3 +
\order{\alpha^4}$ and $\epsilon_2= -4/x^2 + 2\alpha^2 /3 +
\order{\alpha^4}$, so that $\nS-1 = -2\alpha^2 + \order{\alpha^4}$. As
for the similar calculations performed in \sectioncs{sec:cnai}
and~\ref{sec:cnbi}, one should remark that, if $\xend$ is such that
$\alpha \xstar \gtrsim 1$, the previous expansion can be inaccurate
and some deviations from constant $\nS$ may appear.

Let us now consider the slow-roll trajectory. It can be integrated
analytically and is given by the following formula
\begin{equation}
\begin{aligned}
\label{eq:trajecCNCI}
  N-\Nend & = \frac{1}{\alpha^2\left(3+\alpha^2 \right)} \left\lbrace
  3 \ln\left[ \cosh\left( \frac{\alpha}{\sqrt{2}}x \right) \right] +
  \frac{\alpha^2}{2} \cosh^2\left( \frac{\alpha}{\sqrt{2}}x \right)
  \right. \\ & - \left. 3
  \ln\left[ \cosh\left(\frac{\alpha}{\sqrt{2}} \xend \right) \right]
   - \frac{\alpha^2}{2}\cosh^2\left( \frac{\alpha}{\sqrt{2}}\xend
  \right) \right \rbrace  .
\end{aligned}
\end{equation}
Moreover, this expression can be explicitly inverted. As a
consequence, the function $x(N)$ can be written as
\begin{equation}
\begin{aligned}
x & = \frac{\sqrt{2}}{\alpha} \arcosh \left[\frac{3}{\alpha^2}
  \Lambert{0} \left(\frac{\alpha^2}{3} \exp \left \lbrace\frac{2}{3}
  \alpha^2\left( 3+\alpha^2 \right) \left(N-\Nend\right)
  \right. \right. \right. \\ & + \left. \left. \left.  2 \ln\left[
    \cosh\left(\frac{\alpha}{\sqrt{2}} \xend \right) \right] +
  \frac{\alpha^2}{3} \cosh^2\left( \frac{\alpha}{\sqrt{2}}\xend
  \right) \right \rbrace\right)\right]^{1/2},
\end{aligned}
\end{equation}
where $\Lambert{0}$ is the Lambert function. The fact that we deal
with the $0$-branch is obvious since the argument of this function is
positive definite.

The predictions of the CNCI models are displayed in \Fig{fig:CMBCNCI},
for $\alpha=10^{-3}$, $0.1$ and $0.2$. The thin black solid lines are
the lines such that $\nS-1=-2\alpha^2$. We see that, for very small
values of $\alpha$, the predictions are indeed such that the spectral
index is constant. For $\alpha$ not too small, however, we also notice
deviations from this law and the larger $\alpha$ the stronger these
deviations. This is reminiscent with the phenomenon observed in
\sectioncs{sec:cnai} and~\ref{sec:cnbi} but now $\xend$ is a free
parameter and, for a given value of $\alpha$, the deviations from
$\nS-1=-2\alpha^2$ become larger when $\xend$ increase (\ie when the
line becomes redder in \Fig{fig:CMBCNCI}). In this case, the Taylor
expansion of the trigonometric functions which appear in the
expressions of the slow-roll parameters is no longer valid because a larger
$\xend$ implies a larger $\xstar$. This has for consequence that CNCI
inflation is only marginally consistent with the data. Indeed, it is
precisely in the region where $\nS-1=-2\alpha^2$ would be compatible
with the observations that the deviations play an important role and
push the predictions away from the allowed contours. In fact, these
properties can be better illustrated by deriving explicitly
$\xstar$. Using \Eq{eq:trajecCNCI}, one gets
\begin{equation}
\label{eq:cosh2cnci}
\cosh^2\left(\frac{\alpha \xstar}{\sqrt{2}}\right)
=\frac{3}{\alpha^2}\Lambert{0}\left(\frac{\alpha^2}{3}\ee^{2A/3}\right),
\end{equation}
where we have defined the quantity $A$ by 
\begin{equation}
A\equiv -\alpha^2\left(3+\alpha^2\right)\Delta \Nstar
+3\ln \left[\cosh\left(\frac{\alpha \xend}{\sqrt{2}}\right)\right]
+\frac{\alpha^2}{2}
\cosh^2\left(\frac{\alpha \xend}{\sqrt{2}}\right).
\end{equation}
In the regime where both $\alpha\ll 1$ and $\alpha\xend\ll 1$, the
previous expression reduces to $\xstar^2\simeq \xend^2-4\Delta
\Nstar$. This last formula is identical to the slow-roll trajectory
for LFI provided $p=-2$, see \Eq{eq:trajlf}. At the beginning of this
section, we have show that, at leading order $\epsilon_1\simeq 2/x^2$
and $\epsilon_2\simeq -4/x^2$ and, comparing with \Eqs{eq:epslfi}, we
notice that these are also the slow-roll parameters for LFI with
$p=-2$. In fact, expanding \Eq{eq:potCNCI}, one sees that
$V(\phi)\propto \phi^{-2}$ which confirms the previous
considerations. In the regime where $\alpha\ll 1$ and $\alpha\xend\ll
1$, the model is very close to LFI with $p=-2$. On the contrary, if
$\alpha \xend$ is not small, then the above relation does not hold
anymore and one does not recover a constant spectral index.

Finally, we conclude this section by discussing how the mass scale $M$
can be chosen. The CMB normalization gives
\begin{equation}
\label{eq:cobecnci}
\left(\frac{M}{\Mp}\right)^4 = \frac{11520 \pi^2 \alpha^2
  \left(3+\alpha^2 \right)^2 \cosh^2 \left(
  \frac{\alpha}{\sqrt{2}}\xstar \right)} {\left[6+\alpha^2 + \alpha^2
    \cosh\left(\sqrt{2} \alpha \xstar\right)\right]^3}
\frac{\Qrms^2}{T^2}\, .
\end{equation}
From \Eq{eq:cosh2cnci}, one deduces that $\cosh^2(\alpha
\xstar/\sqrt{2})\simeq 1-2\alpha^2\Delta
\Nstar+\alpha^2\xend^2/2\simeq 1$. Inserting this formula into
\Eq{eq:cobecnci}, and taking the leading order in $\alpha$, one
obtains $M/\Mp\simeq 0.02\sqrt{\alpha}$. This implies that $M<\Mp$ if
$\alpha \lesssim 2420$, which is largely the case for the predictions
displayed in \Fig{fig:CMBCNCI}.

\subsection{Supergravity Brane Inflation (SBI)}
\label{sec:sbi}

\subsubsection{Theoretical Justifications}
\label{subsubsec:theorysbi}

This model can emerge in different contexts. Following
\Refc{NeferSenoguz:2008nn}, let us consider a model with a scalar
field and a massive fermion interacting through a Yukawa type term
(with a coupling constant $g$). The corresponding Lagrangian can be
written as
\begin{equation}
-\calL=\frac{1}{2}\partial_{\mu}\phi \partial^{\mu}\phi
+\frac{i}{2}\bar{\psi}\gamma^{\mu}\partial_{\mu}\psi
+\frac12m^2\phi^2+\frac{\lambda}{4!}\phi^4+m_\uf\bar{\psi}\psi
+\frac{1}{2}g\phi \bar{\psi}\psi,
\end{equation}
where we have assumed the most general renormalizable scalar
potential. At one loop level, the potential takes the form
\begin{equation}
\begin{aligned}
V(\phi)&= V_0+\frac12m^2\phi^2+\frac{\lambda}{4!}\phi^4
+\frac{1}{64\pi^2}\left(m^2+\frac{\lambda}{2}\phi^2\right)^2
\ln \left(\frac{m^2+\lambda\phi^2/2}{\mu^2}\right)
 \\ 
 & - \frac{2}{64\pi^2}\left(g\phi+m_\uf\right)^4
\ln \left[\frac{\left(g\phi+m_\uf\right)^2}{\mu^2}\right],
\end{aligned}
\end{equation}
where $\mu$ is a renormalization scale. Then, assuming that, for some
reason, the bosonic and fermionic massive terms are negligible, the
potential can be expressed as
\begin{equation}
\label{eq:potsbihep}
V(\phi)\simeq V_0+\left[\frac{\lambda}{4!}+\frac{\lambda^2}
{256\pi^2}\ln\left(\frac{\lambda}{2}\right) -\frac{g^4}{16\pi^2}\ln g\right]\phi^4
+\frac{1}{64\pi^2}\left(\frac{\lambda^2}{2}-\frac{g^4}{4}\right)
\phi^4\ln\left(\frac{\phi}{\mu}\right).
\end{equation}
This is the type of potential that we study in this section. Notice
that a change in the renormalization scale $\mu $ is in fact
equivalent to a change in the coefficient of the terms $\propto \phi^4
$ and $\propto \phi\ln(\phi/\mu)$. This potential was also studied in
\Refc{Shafi:2006cs} but the coefficient of the $\phi^4$ term was chosen
such that, at its minimum, the potential exactly vanishes. This
particular case will also be treated in what follows. Finally, it is
interesting to remark that this model was also proposed in
\Refcs{Choudhury:2011sq, Choudhury:2012ib} in the context of brane
cosmology within a supergravity bulk spacetime.

\subsubsection{Slow-Roll Analysis}
\label{subsubsec:srsbi}

Let us now turn to the slow-roll analysis of the potential given by
\Eq{eq:potsbihep}. It is more convenient to write it under the
following form
\begin{equation}
V\left(\phi\right) = M^4\left\lbrace 1 + \left[ -\alpha + \beta\ln
  \left( \frac{\phi}{\Mp} \right) \right] \left( \frac{\phi}{\Mp}
\right)^4 \right \rbrace ,
\label{eq:potSBI}
\end{equation}
where $\alpha$ and $\beta$ are dimensionless quantities that must be
considered as small quantities since they are typically proportional
to coupling constants, see \Eq{eq:potsbihep}. It is worth noticing
that setting $\alpha=0$ in the above expression allows us to recover
the Coleman-Weinberg CWI models already studied
in \sectionc{sec:cwi}. Defining the quantity $x$ by the following
expression
\begin{equation}
 x \equiv \dfrac{\phi}{\Mp} \,,
\end{equation}
one sees that the potential decreases from $x=0$ to reach a minimum
located at $x=\xdVzero$, then increases and diverges when $x$ goes to
infinity. The value of $\xdVzero$ is given by
\begin{equation}
\xdVzero = \exp\left(\frac{\alpha}{\beta}-\frac{1}{4}\right).
\end{equation}
Since the logarithm terms in \Eq{eq:potSBI} are one loop corrections,
they should not dominate the leading order terms. As a result,
inflation can take place only in the domain $x<\xdVzero$ if one wants
the model to be such that additional corrections to $V(\phi)$ are
negligible. The value of the potential at the minimum reads
\begin{equation}
\Vmin = V(\xdVzero) = M^4 \left( 1 - \dfrac{\beta}{4} \ee^{4
  \alpha/\beta -1} \right),
\end{equation}
which is negative or vanishing if the following condition is satisfied
\begin{equation}
\alpha \ge \alphamin\left(\beta\right)= \frac{\beta}{4} \left[1 -
\ln\left(\dfrac{\beta}{4}\right) \right] .
\label{eq:sbialphamin}
\end{equation}
Inflation proceeds from the left to the right in the range $0<x<
\xVzero < \xdVzero$ where $\xVzero$ is the value at which the
potential vanishes. It is given by
\begin{equation}
\xVzero = \left[\frac{-4/\beta}{\Lambert{-1}
    \left(-4/\beta \ee^{-4\alpha/\beta}\right)} \right]^{1/4},
\end{equation}
where $\Lambert{-1}$ is the $-1$ branch of the Lambert function. In
this situation, inflation stops by slow-roll violation at
$x=\xVzero$. As noticed above, the case $\alpha=\alphamin(\beta)$ is
also interesting. It corresponds to tuning the parameters $\alpha$ and
$\beta $ such that the minimum of the potential exactly vanishes. When
this condition is satisfied the previous formula reduces to
$\xVzero=\xdVzero=(\beta/4)^{-1/4}$. Then, the first slow roll
parameter $\epsilon_1$ diverges at this point (see below) and, as a
consequence, inflation also ends by slow roll violation.

\begin{figure}
\begin{center}
\includegraphics[width=\wdblefig]{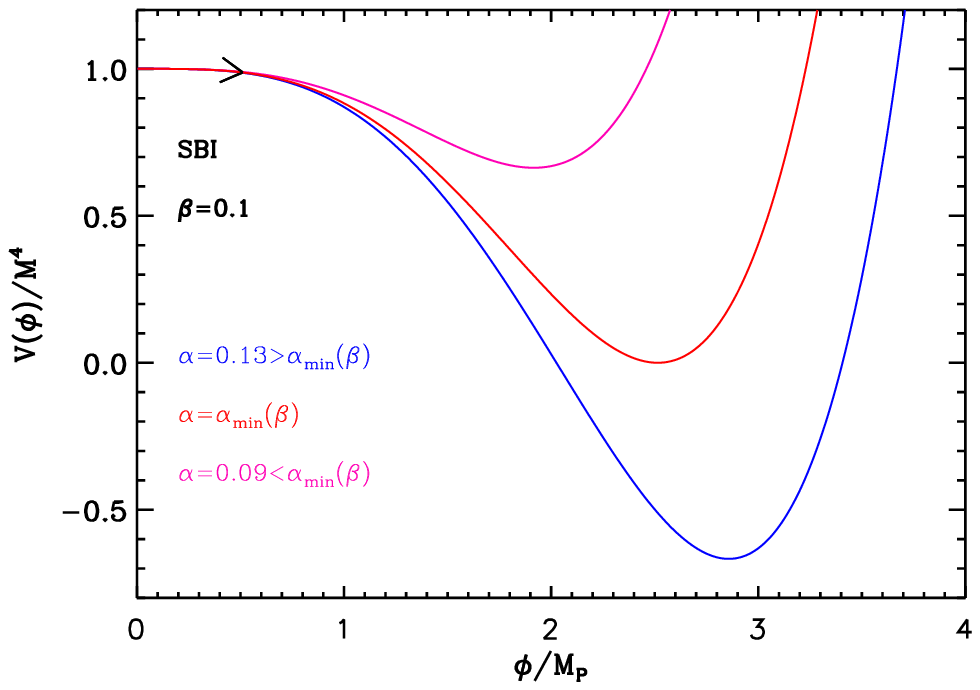}
\includegraphics[width=\wdblefig]{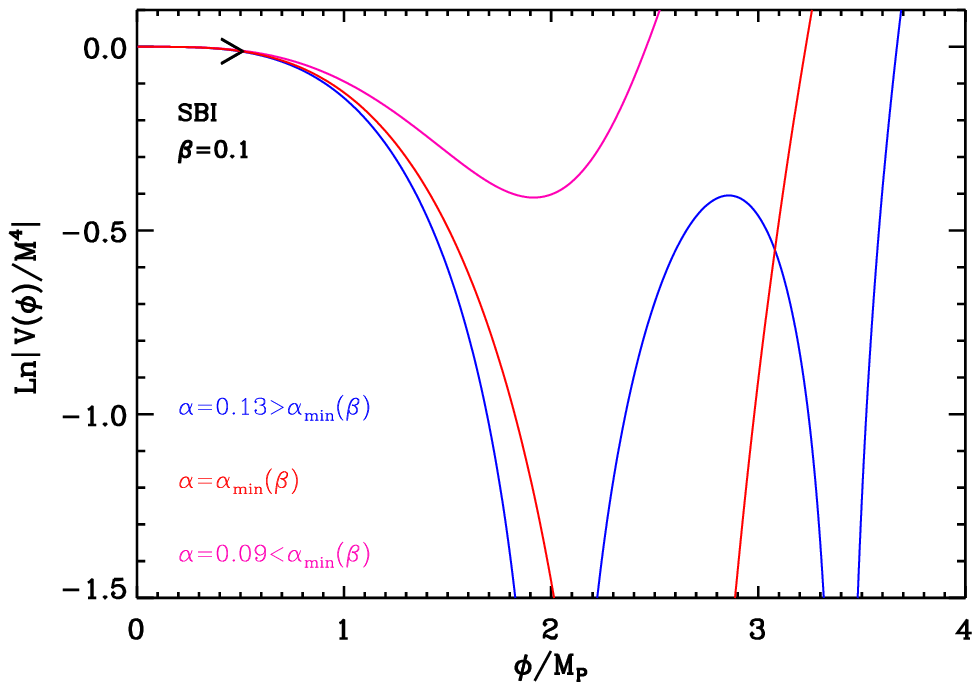}
\includegraphics[width=\wdblefig]{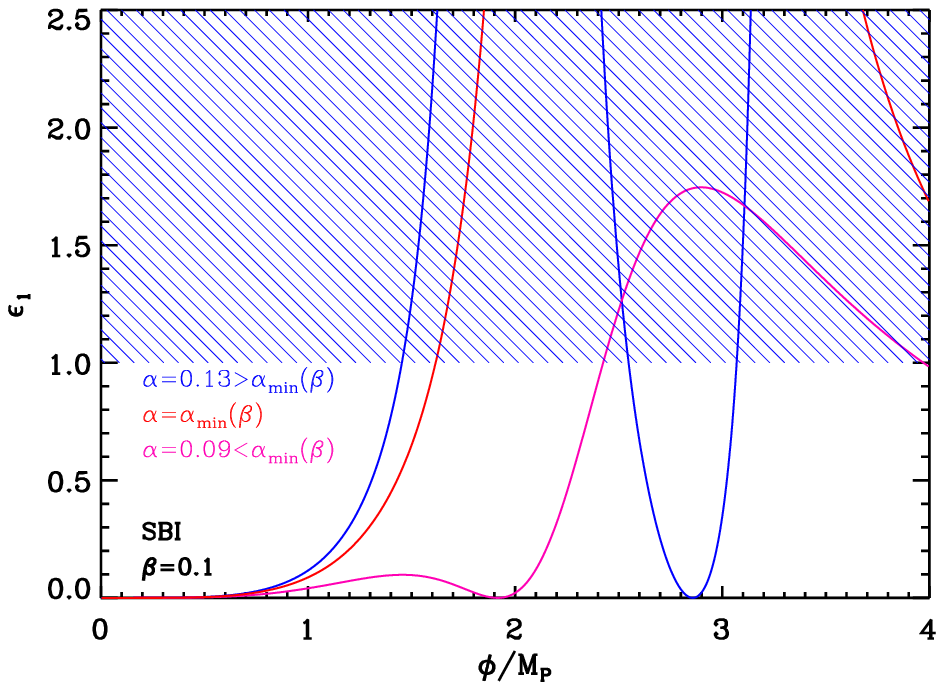}
\includegraphics[width=\wdblefig]{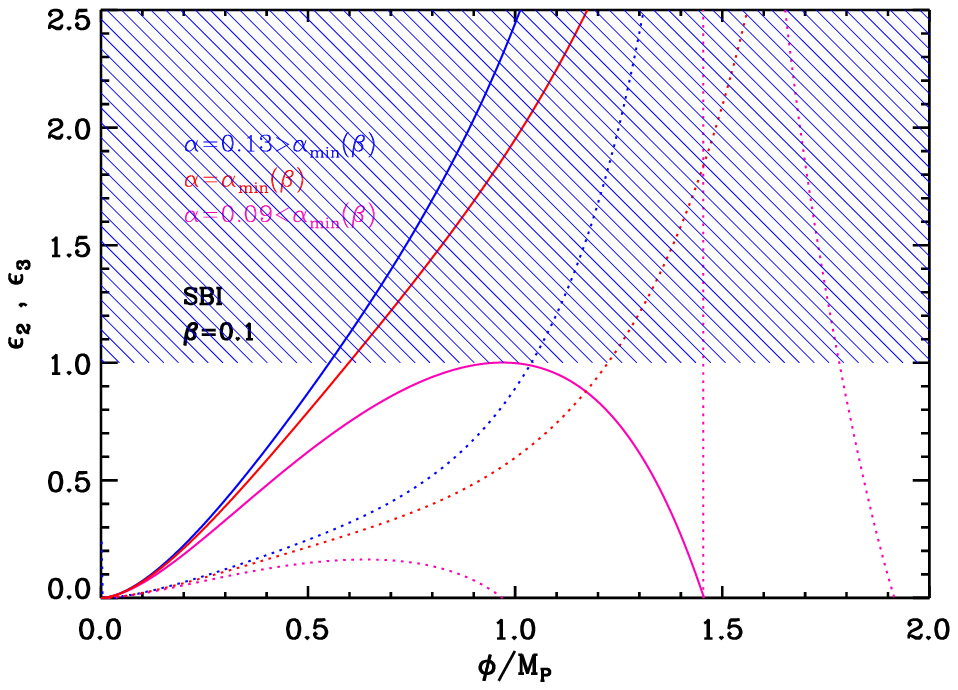}
\caption{Supergravity Brane Inflation (SBI) for $\beta=0.7$ and
  $\alpha=0.13>\alphamin(\beta)$, $\alpha=\alphamin(\beta)$, and 
  $\alpha=0.09<\alphamin(\beta)$ (where $\alphamin$ is defined
  in \Eq{eq:sbialphamin}).  Upper panels: the potential and its logarithm.
  Inflation proceeds in the place and direction labeled by the arrow.
  Bottom left panel: slow-roll parameter
  $\epsilon_1$. The shaded area indicates where inflation stops.
  Bottom right panel: slow-roll parameters $\epsilon_2$ (solid line)
  and $\epsilon_3$ (dotted line), only displayed in the branch of
  the potential where inflation proceeds.}
\label{fig:potSBI}
\end{center}
\end{figure}

The first three Hubble flow functions in the slow-roll approximation
are given by
\begin{equation}
  \epsilon_1 = \frac{x^6\left(-4\alpha+\beta+4\beta\ln x
    \right)^2}{2\left(1-\alpha x^4+\beta x^4\ln x \right)^2}\, ,
\label{eq:sr1SBI}
\end{equation}
\begin{equation}
  \epsilon_2 = 2 \dfrac{\left(12\alpha-7\beta-12\beta\ln x\right)x^2+
    \left(4\alpha^2-\alpha\beta+\beta^ 2+\beta^ 2\ln x-8\alpha\beta\ln
    x+4\beta^2\ln^2 x\right)x^6}{ \left[1+x^4\left(-\alpha+\beta\ln x
      \right)\right]^2}\, ,
\end{equation}
\begin{equation}
\begin{aligned}
  \epsilon_3 & = \frac{8}{x^2}+2\frac{\left(-4+\beta
    x^4\right)^2}{x^2\left(1-\alpha x^4+\beta x^4\ln
    x\right)^2}+\frac{1}{x^2}\frac{-52+9\beta x^4}{1-\alpha x^4+\beta
    x^4\ln x} \\ & + \dfrac{144\alpha -84\beta + \left( 28\alpha -11\beta
    \right) \beta x^4-4\beta\left(36+7\beta x^4\right)\ln x}{
  \left(12\alpha-7\beta-12\beta\ln x\right)x^2 + \left(
    4\alpha^2 - \alpha\beta + \beta^2 - 8\alpha\beta\ln x + \beta^2\ln
    x + 4\beta^2\ln^2 x \right)x^6 }\, .
\end{aligned}
\end{equation}
Together with the potential, they are represented in \Fig{fig:potSBI}
for the physical branch $0<x<\xVzero$.

As already mentioned, inflation stops by violation of the slow-roll
conditions. This happens when $x=\xend$ where $\xend$ is the solution
of $\epsilon_1(\xend) = 1$. We see in \Eq{eq:sr1SBI} that there is no
simple analytic solution for $\xend$ and this equation must in fact be
solved numerically. We have, however, already stressed that, when
$\alpha\leq\alphamin(\beta)$, $\epsilon_1$ diverges for $x\rightarrow
\xVzero$, and therefore one already knows that $\xend < \xVzero$.

Let us now consider the slow-roll trajectory. It can be integrated
analytically and one obtains the following expression
\begin{equation}
\begin{aligned}
N-\Nend & = \frac{ \ee^{2\frac{\alpha}{\beta} - \frac12}}{16} \left[
  \Ei\left(\frac{1}{2} - 2\frac{\alpha}{\beta} + 2\ln x\right) -
  \Ei\left(\frac{1}{2}-2\frac{\alpha}{\beta} + 2\ln
  \xend\right)\right] \\ &-
\frac{\ee^{\frac12-2\frac{\alpha}{\beta}}}{4\beta} \left[ \Ei\left(
  -\frac{1}{2} + 2\frac{\alpha}{\beta} - 2\ln x\right) - \Ei\left(
  -\frac{1}{2}+2\frac{\alpha}{\beta}-2\ln \xend \right) \right] -
\frac{x^2-\xend^2}{8}\, .
\end{aligned}
\end{equation}
The field value $\xstar$ at which the pivot scale crossed the Hubble
radius during inflation is obtained by solving
\Eq{eq:phistarlnrrad}. Clearly, it must also been done numerically and
those calculations are implemented in the corresponding \ASPIC
routines.

Finally, the parameter $M$ is fixed by the amplitude of the CMB
anisotropies and one obtains
\begin{equation}
\left(\frac{M}{\Mp}\right)^4=\frac{720\pi^2\left(4\alpha-\beta-4\beta\ln
  \xstar\right)^2} {\left(1-\alpha \xstar^4+\beta \xstar^4\ln
  \xstar\right)^3} \frac{\Qrms^2}{T^2}\, .
\end{equation}
The reheating consistent slow-roll predictions for the SBI models are
displayed in \Figs{fig:CMBSBIbetaEQ5x10PowerMinus5} and
\ref{fig:CMBSBIbetaEQ10PowerMinus3}, for $\beta=5\times 10^{-5}$ and
$\beta=10^{-3}$, respectively, and with $\alpha\leq\alphamin(\beta)$.
These plots show that the larger values of $\beta$, the more
negligible the amount of gravitational waves. The predictions for the
special case $\alpha=\alphamin(\beta)$ are also displayed in
\Fig{fig:CMBSBIalphamin}, where it is clear that smaller values of
$\beta$ are preferred.

\subsection{Spontaneous Symmetry Breaking Inflation (SSBI)}
\label{sec:ssbi}

\subsubsection{Theoretical Justifications}
\label{subsubsec:theoryssbi}

The potential that we study in this section is given by the following
expression
\begin{equation}
\label{eq:potssbihep}
V\left(\phi\right)=V_0+a\phi^2+b\phi^4,
\end{equation}
where $a$ and $b$ are constant coefficients the sign of which is not a
priori determined. Before turning to the slow-roll analysis, it is
interesting to study in which context such a potential can arise.

First of all, it is clear that this potential is very general since it
is just made of the three first terms of a general Taylor
expansion. Therefore, it can just be considered as a phenomenological
description of a generic inflaton potential. This view was for
instance adopted in \Refc{Albrecht:1984qt}, where this potential was
used as a toy model to implement ``new inflation''. In the same
fashion, it was also considered in \Refc{Moss:1985wn} (with the
assumptions $a<0$ and $b>0$) in the framework of models with
spontaneous symmetry breaking where $\phi$ represents one of the
components of a Higgs field. In \Refc{Hu:1986exa}, it was also studied
in the context of ``mixmaster inflation''.

However, there are also models where this specific shape explicitly
arises and, here, when necessary, we also briefly review them.

The first example is given by \Refcs{Dine:1997kf, Riotto:1997iv}. In
these articles, inflation was investigated in the context of gauge
mediated SUSY breaking scenarios. One of the basic idea of this
approach is that the inflaton field should not be an extra field added
to the theory on purpose but rather a field which is already present
in known high energy theories. In the MSSM, see also
\sectionc{sec:mssmi}, we know that the Higgs sector superpotential
contains the term $\mu H_\uu\cdot H_\ud$ where $\mu$ should be of the
order of the electroweak scale, that is to say far from the Planck
scale. This is the so-called $\mu$-problem. One possible solution is
to consider that this term dynamically arises due to the presence of
another superfield (usually a singlet), $S$, in the
theory. \Refcs{Dine:1997kf, Riotto:1997iv} take advantage of this fact
and build a model where $S$ can also play the role of the
inflaton. Since the model is also formulated in the framework of
gauge-mediated supersymmetry breaking scenarios, there is an
additional superfield $X$ such that its scalar component (also denoted
$X$) and auxiliary component $F_X$ acquire non-vanishing \vev. Let us
now consider the following super-potential
\begin{equation}
  W=-\beta \frac{XS^4}{\Mp^2}+\frac{S^5}{\Mp^2}
  +\lambda \frac{S^2}{\Mp}H_\uu \cdot H_\ud + \bar{W},
\end{equation}
where the function $\bar{W}$ describes all the other extra terms in
$W$ and, crucially, is assumed to be independent of $S$. The
quantities $\lambda $ and $\beta$ are constant coefficients. As argued
in \Refcs{Dine:1997kf, Riotto:1997iv}, this form of $W$ can be
enforced by discrete symmetries. In particular, we notice the absence
of a term $SH_\uu\cdot H_\ud$. Another important ingredient of the
model is the assumption that the \vev $F_X$ comes from the extra-terms
in the above superpotential, \ie $F_X\simeq \partial \bar{W}/\partial
X$. Then, the scalar potential reads
\begin{equation}
V=\left(F_X-\beta \frac{S^4}{\Mp^2}\right)^2+\left(5\frac{S^4}{\Mp^2}
-4\beta\frac{X}{\Mp^2}S^3\right)^2.
\end{equation}
Taking into account supergravity corrections, which are typically of
the form $(\partial W/\partial X)/\Mp^2$, \ie $m^2=a F_X^2/\Mp^2$,
where $a$ is a coefficient of order one we are led to
\begin{eqnarray}
V\simeq F_X^2-a\frac{F_X^2}{\Mp^2}S^2-2\beta F_X\frac{S^4}{\Mp^2}
+16\beta^2\frac{X^2}{\Mp^4}S^6
-40\beta \frac{X}{\Mp^4}S^7
+(25+\beta^2)\frac{S^8}{\Mp^4}.
\end{eqnarray}
In addition, making the reasonable assumption that the field $X$ is
stabilized at a \vev such that $X/\Mp\ll 1$, one can neglect higher
order terms in this expression. Then, we see that $S$ can play the
role of the inflaton with a potential of the form given by
\Eq{eq:potssbihep}, namely
\begin{equation}
V\simeq F_X^2\left(1-a\frac{S^2}{\Mp^2}
-\frac{2\beta\Mp^2}{F_X}\frac{S^4}{\Mp^4}\right).
\end{equation}
At the minimum of the potential, $S^4\simeq \Mp^2F_X$ and this implies
a $\mu$ term for the MSSM of the form $\mu\simeq \lambda
\sqrt{F_X}$. As explained before, this model dynamically produces the
$\mu$ term while obtaining a candidate for the inflaton
field. Finally, let us remark that the CMB normalization will
determine the scale $F_X$ and that the spectrum of the superparticles
depends on the ratio $F_X/X$. Therefore, given a value of $F_X$, one
can always choose $X$ in order to obtain reasonable values for the
superparticle masses.

The SSBI potential was also used, as a toy model, in
\Refcs{Cormier:1998nt, Cormier:1999ia} to study a model of ``Spinodal
Inflation''. After the $90$'s, it was considered again several times:
in the context of the Randall-Sundrum model in
\Refc{Bhattacharya:2001wq} (but within the framework of Brans-Dicke
theories), in the context of the little Higgs model in
\Refc{ArkaniHamed:2003mz} and in the context of induced gravity
inflation in \Refc{Wang:2004ka}. In this last reference, a potential
of the form~(\ref{eq:potssbihep}) was considered but in the Jordan
frame. Since the potential is different in the Einstein frame, in
fact, this model does not belong to the class of scenarios studied
here. Finally, it was also considered in the context of electroweak
inflation in \Refc{Fukuyama:2005dp}.

In \Refc{Antusch:2006gh}, an inflationary scenario was studied in
which the superpartner of the right-handed neutrino plays the role of
the inflaton field. Let us denote by $N$ the singlet neutrino
superfield, $\phi$ the super waterfall field (that can be put to zero
during inflation) and $S$ another singlet superfield (which can also
be put to zero during inflation). Then, on very general grounds, the
K\"ahler potential can be written as
\begin{eqnarray}
K &=& \vert S\vert^2+\vert\phi\vert^2+\vert N\vert^2
+\kappa_S\frac{\vert S\vert^4}{4\Mp^2}
+\kappa_N\frac{\vert N\vert^4}{4\Mp^2}
+\kappa_\phi\frac{\vert \phi\vert^4}{4\Mp^2}
+\kappa_{S\phi}\frac{\vert S\vert^2\vert \phi\vert^2}{\Mp^2}
+\kappa_{SN}\frac{\vert S\vert^2\vert N\vert^2}{\Mp^2}
\nonumber \\ & &
+\kappa_{N\phi}\frac{\vert N\vert^2\vert \phi\vert^2}{\Mp^2}+\cdots,
\end{eqnarray}
where the dimensionless coefficients $\kappa$ are a priori of order
one. The superpotential can be expressed as
\begin{equation}
W=\kappa S\left(\frac{\phi^4}{M'^2}-M^2\right)
+\frac{\lambda}{M_*}N^2\phi^2 +\cdots,
\end{equation}
where $M$, $M'$ and $M_*$ are three mass scales and $\kappa$ and
$\lambda$ are coupling constants. Since the three fields introduced
before are singlets the potential does not contain $D$-term
contributions. As a consequence, for $S\simeq0$ and $\phi\simeq 0$, we
are left with the $F$-term potential only and this one can be written
as
\begin{equation}
V(N)\simeq \kappa^2M^4\left[1+\left(1-\kappa_{SN}\right)\frac{N^2}{\Mp^2}
+\left(\frac12+\frac{\kappa_N}{4}-\kappa_{SN}+\kappa_{SN}^2\right)
\frac{N^4}{\Mp^4}+\cdots \right].
\end{equation}
We see that it has the form of
\Eq{eq:potssbihep}. \Refc{Antusch:2006gh} also discusses how to stop
inflation by tachyonic instability. Since the field $\phi$ is viewed
as the waterfall field, one has to calculate his mass to see when the
instability is triggered. This can be done by evaluating the quadratic
correction in $\phi$ to the potential calculated before. This leads to
\begin{equation}
m_\phi^2=\left(1+\kappa_{N\phi}\frac{N^2}{\Mp^2}
-\kappa_{S\phi}\right)\frac{\kappa^2M^4}{\Mp^2}
+4\frac{\lambda^2}{M_*^2}N^4.
\end{equation}
Neglecting the term $N^2/\Mp^2\ll 1$ in this expression, the effective
mass vanishes for
\begin{equation}
N_\ucri \simeq \frac{\kappa M^2M_*}{2\lambda \Mp}
\sqrt{-(1-\kappa_{S\phi})}\,.
\end{equation}
We see that this requires $1-\kappa_{S\phi}<0$. On the other hand,
this model also provides an expression for the coefficients $a$ and
$b$ in terms of the fundamental coefficients of the K\"ahler
potential. Except from the above mentioned condition, there is no
other constraint on the coefficients $\kappa$ and, as a consequence,
the sign of $a$ and $b$ is, a priori, not fixed in this scenario.

Another context in which \Eq{eq:potssbihep} arises is ``racetrack
inflation''~\cite{BlancoPillado:2004ns, Brax:2007fz}. Racetrack
inflation is a string inspired inflationary scenario where the
inflaton is a volume modulus. Therefore, this model belongs to the
same class as KMIII, see \sectionc{sec:kmiii}. The K\"ahler and super
potentials are given by standard formulas, namely
\begin{eqnarray}
K &=& -\frac{3}{\kappa}\ln \left(T+T^{\dagger}\right), \qquad
W =W_0+A\ee^{-aT}+B\ee^{-bT}.
\end{eqnarray}
Writing $T=X+iY$, it follows that the scalar $F$-term potential
reduces to
\begin{equation}
\begin{aligned}
V(X,Y) &= \frac{\kappa}{6X^2}\biggl\{aA^2\left(3+aX\right)\ee^{-2aX}
+bB^2\left(3+bX\right)\ee^{-2bX}
+3aAW_0\ee^{-aX}\cos\left(aY\right)
\\ & +3bBW_0\ee^{-bX}\cos\left(bY\right)
+AB\left[2abX+3\left(a+b\right)\right]
\ee^{-(a+b)X}\cos\left[\left(a-b\right)Y\right]\biggr\}
 +\frac{E}{X^{\alpha}} \,,
\end{aligned}
\end{equation}
where an uplifting term $\propto X^{-\alpha}$ has been added. Let us
mention that $X$ and $Y$ are not canonically normalized and their
kinetic term reads $3[(\partial_\mu X)^2+(\partial_\mu Y)^2]/(4\kappa
X^2)$. The above potential has a very rich structure and for $W_0=0$
and $a=b$, we have a flat direction in $Y$. Moreover, for $Y=0$, one
can find a minimum in the $X$ direction. If we then combine the two
above remarks, then it is clear that there exists a choice of
parameters such that one has a saddle point around $Y=0$ (a specific
example was exhibited in \Refc{BlancoPillado:2004ns}). This point
seems suitable for inflation. Around such a point, it is argued in
\Refc{Brax:2007fz} that one can write
\begin{equation}
V(Y)=V_0\left(1+\frac{\eta_0}{2}y^2+\frac{C}{4}y^4+\cdots \right),
\end{equation}
where $y$ is now the canonically normalized field when $X$ is
stabilized. This is again a potential of the type given by
\Eq{eq:potssbihep}. In order to phenomenologically reproduce racetrack
inflation, one should have $\eta_0$ small and negative and $C$
large and positive.

The potential of \Eq{eq:potssbihep} was also used, as a toy model, in
the context of minimal left-right symmetric models with spontaneous
$D$-parity breaking in \Refc{Gong:2007yv} and in the context of
hilltop supernatural inflation in \Refcs{Lee:2009mj, Lin:2009yt,
  Lin:2010zzk}. A justification based on high energy physics was
offered and the idea is to assume that the full potential has a SUSY
flat direction. The approach is therefore similar to what was already
investigated in \sectionc{sec:mssmi}. In that situation, one can write
$V(\phi)$ as
\begin{equation}
V=V_0+\frac{1}{2}m^2\phi^2-A\frac{\lambda_p\phi^p}{p\Mp^{p-3}}
+\lambda_p^2\frac{\phi^{2p-2}}{\Mp^{2p-6}} \, ,
\end{equation}
where the term $V_0$ is added by hand. If one chooses $p=4$ and
neglects the last term (for instance if $\phi\ll \Mp$), then one
arrives at
\begin{equation}
V(\phi)\simeq V_0+\frac{1}{2}m^2\phi^2-\frac{\lambda_4A}{4\Mp}\phi^4,
\end{equation}
which is of the form of \Eq{eq:potssbihep}. In this framework, $m$ and
$A$ are SUSY soft terms and, therefore, should be taken of
$\order{\TeV}$. The term $V_0=\Ms^4$ where $\Ms$ is the SUSY breaking
scale, $\Ms\simeq 10^{11}\GeV$.

Finally, let us mention that SSBI was also considered in the context
of a supersymmetric $B$-$L$ extension of the standard model in
\Refcs{Khalil:2011kd,Khalil:2012zz} and in the context of
K\"ahler-driven ``tribrid inflation'' in \Refc{Antusch:2012jc}. In
this last case, one obtains a situation very similar to the one
discussed above for sneutrino inflation. In particular, the
coefficients $a$ and $b$ can be expressed in terms of the coefficients
appearing in the K\"ahler potential. To end this part, let us notice
that the potential~(\ref{eq:potssbihep}) also arises in the context of
Higgs inflation in a false vacuum, as shown in \Refcs{Masina:2011un,
  Masina:2011aa, Masina:2012yd}.

As already mentioned above, these works differ on the signs of
$\alpha$ and $\beta$. Summarizing, \Refcs{Hu:1986exa,Antusch:2006gh}
require $\alpha>0$, $\beta>0$ while \Refcs{Albrecht:1984qt,
  Moss:1985wn, Cormier:1998nt, Cormier:1999ia, ArkaniHamed:2003mz,
  Wang:2004ka, Fukuyama:2005dp, Gong:2007yv, Brax:2007fz} assume
$\alpha<0$, $\beta>0$. On the other hand,
\Refcs{Lee:2009mj,Lin:2009yt, Lin:2010zzk} consider that $\alpha>0$
and $\beta<0$ and \Refcs{Dine:1997kf, Riotto:1997iv, Masina:2011un,
  Masina:2011aa, Masina:2012yd} have $\alpha<0$, $\beta<0$. We see
that the four possible combinations have all been studied. Also, in
\Refcs{Khalil:2011kd,Khalil:2012zz}, one has $\alpha,\beta \lesssim
\order{1}$ and inflation only takes place in the increasing branches
of the potential (see below). Finally, in \Refcs{Bhattacharya:2001wq,
  Antusch:2012jc}, $\beta $ is taken to be positive and the sign of
$\alpha$ is left unspecified.

\subsubsection{Slow-Roll Analysis}
\label{subsubsec:srssbi}

\begin{figure}
\begin{center}
\includegraphics[width=\wdblefig]{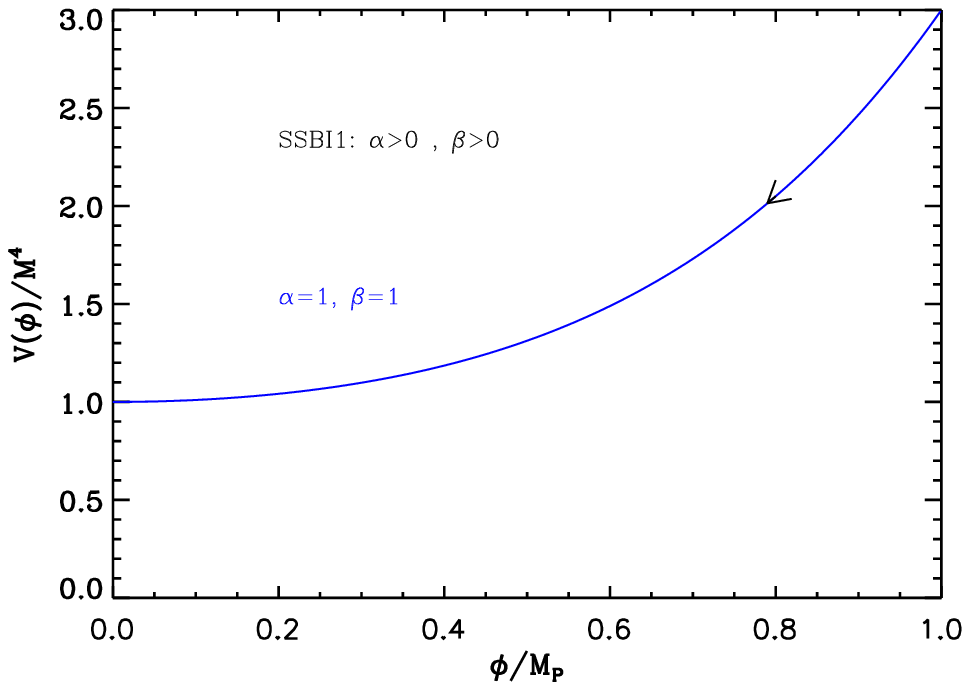}
\includegraphics[width=\wdblefig]{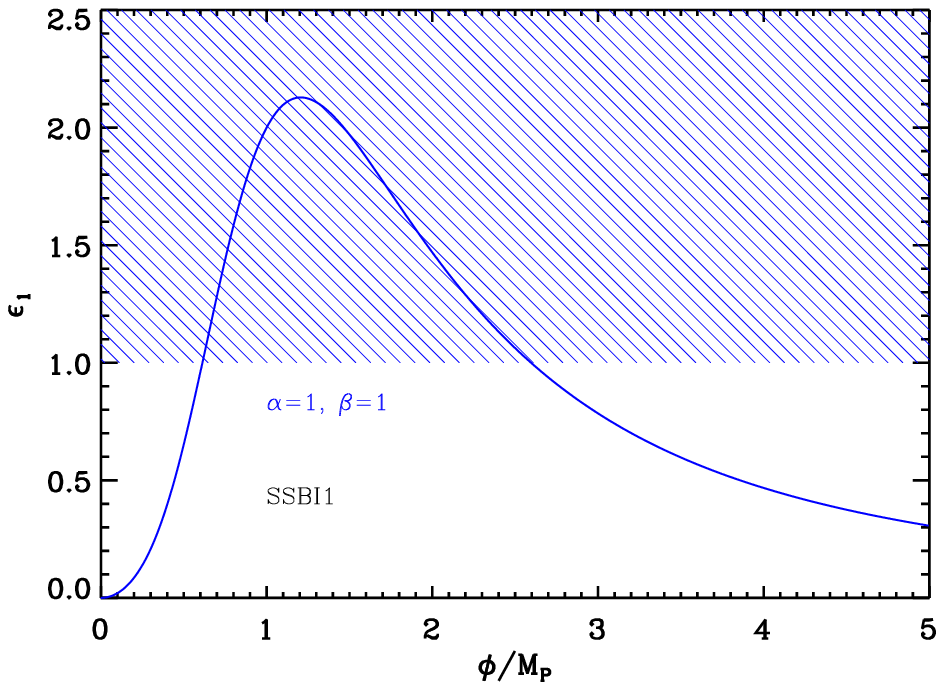}
\includegraphics[width=\wdblefig]{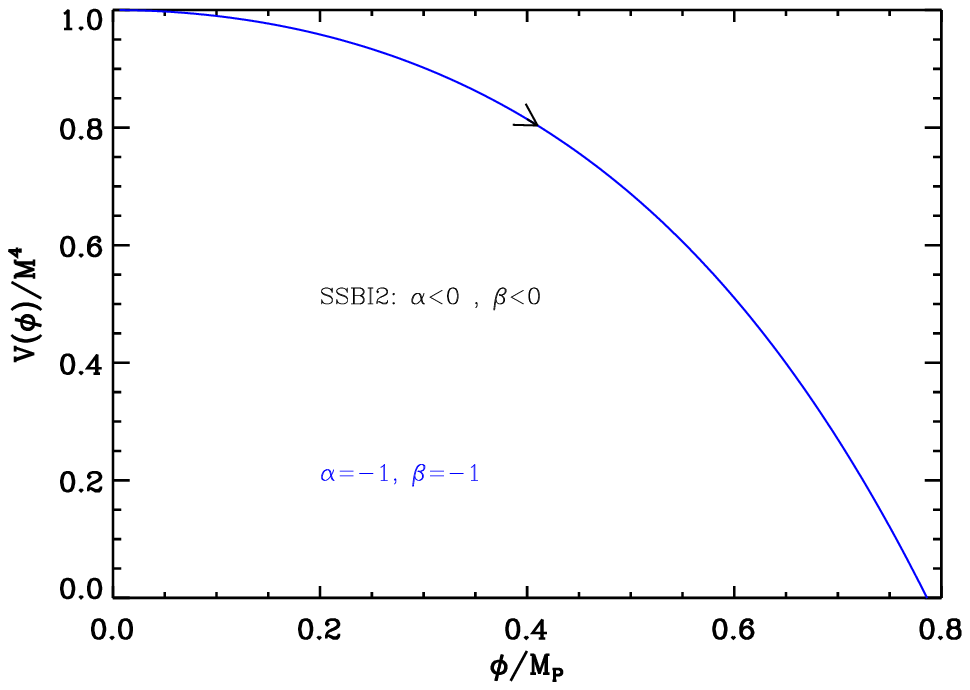}
\includegraphics[width=\wdblefig]{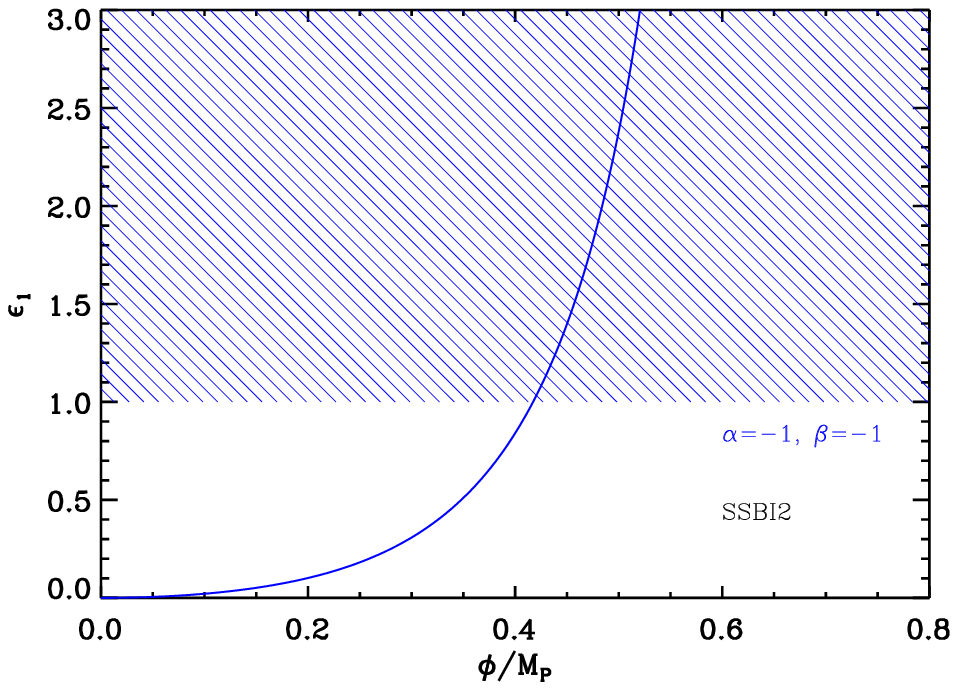}
\caption{Spontaneous Symmetry Breaking Inflation (SSBI) potential and
  the corresponding Hubble flow parameter $\epsilon_1$ for the two
  cases $\alpha>0$, $\beta>0$ (SSBI1), and $\alpha<0$, $\beta<0$
  (SSBI2). The values of the parameters are chosen to be $\alpha,
  \beta=\pm 1$. The four other possibilities, namely SSBI3, SSBI4,
  SSBI5, SSBI6 are displayed in \Fig{fig:potssbib}.}
\label{fig:potssbia}
\end{center}
\end{figure}

\begin{figure}
\begin{center}
\includegraphics[width=\wdblefig]{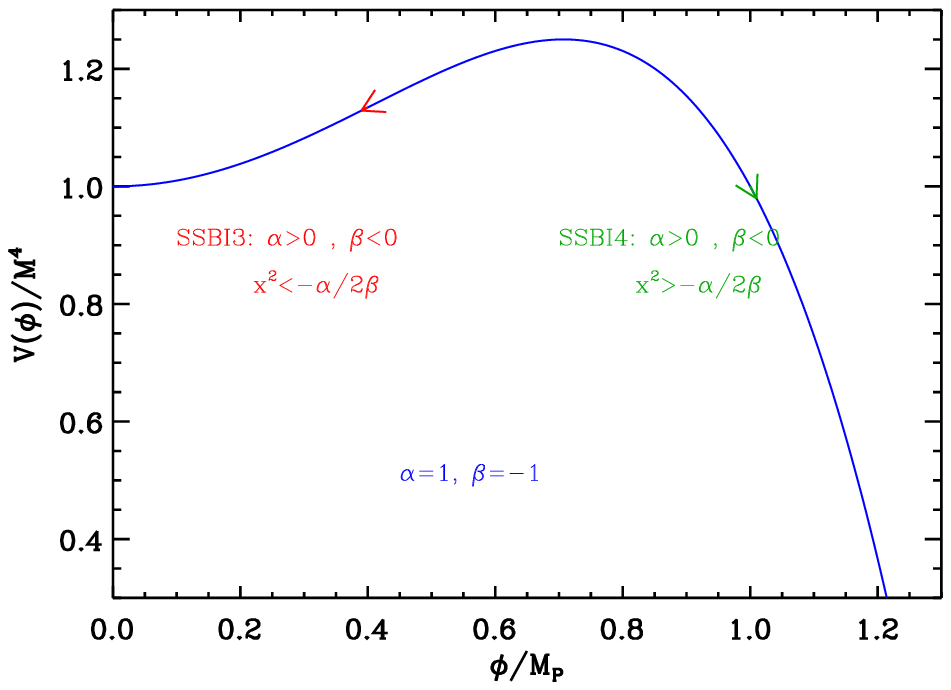}
\includegraphics[width=\wdblefig]{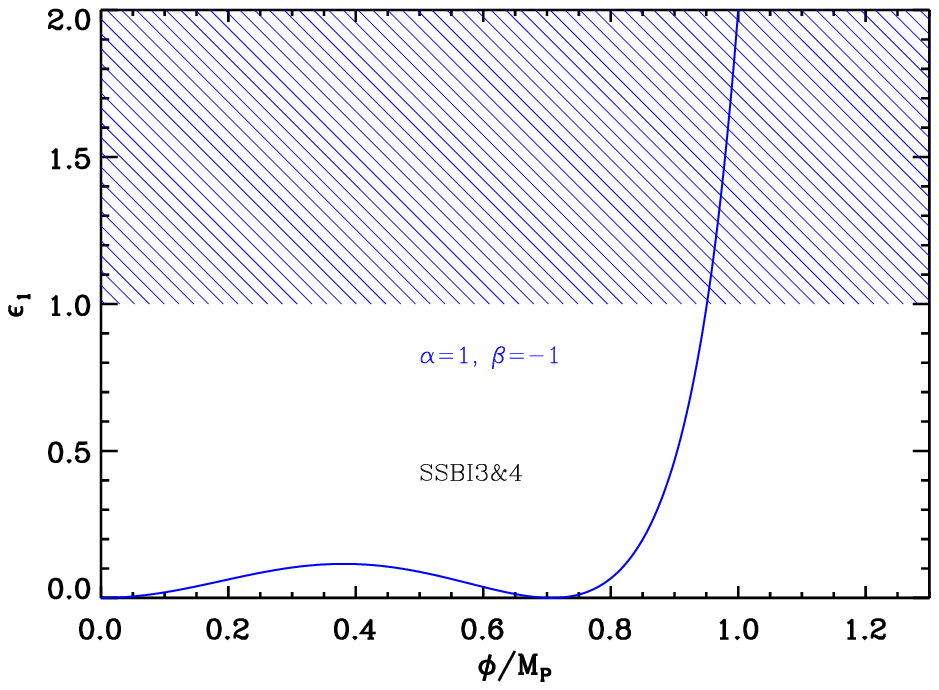}
\includegraphics[width=\wdblefig]{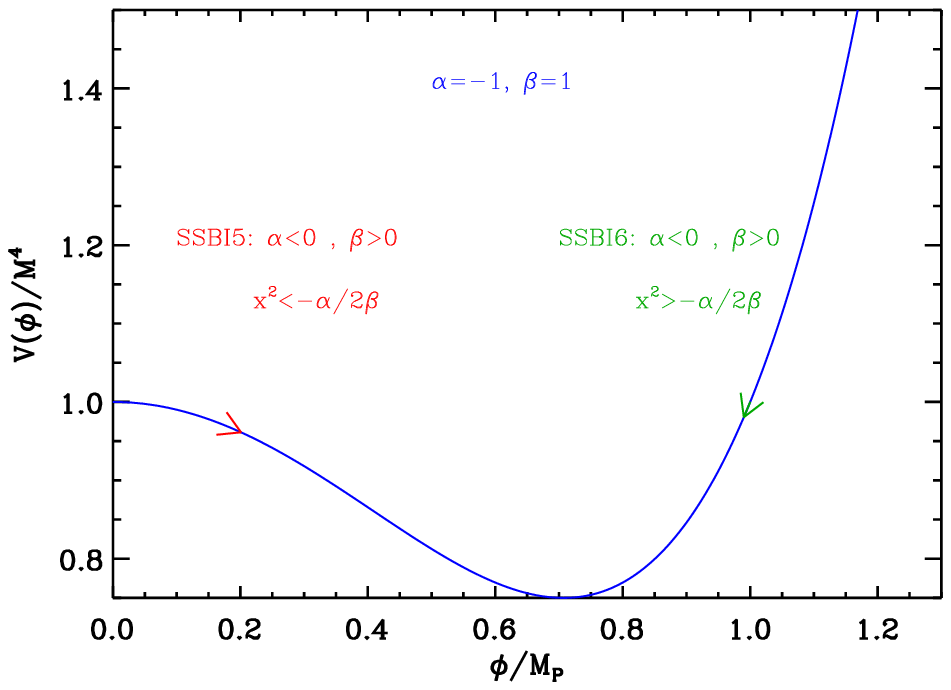}
\includegraphics[width=\wdblefig]{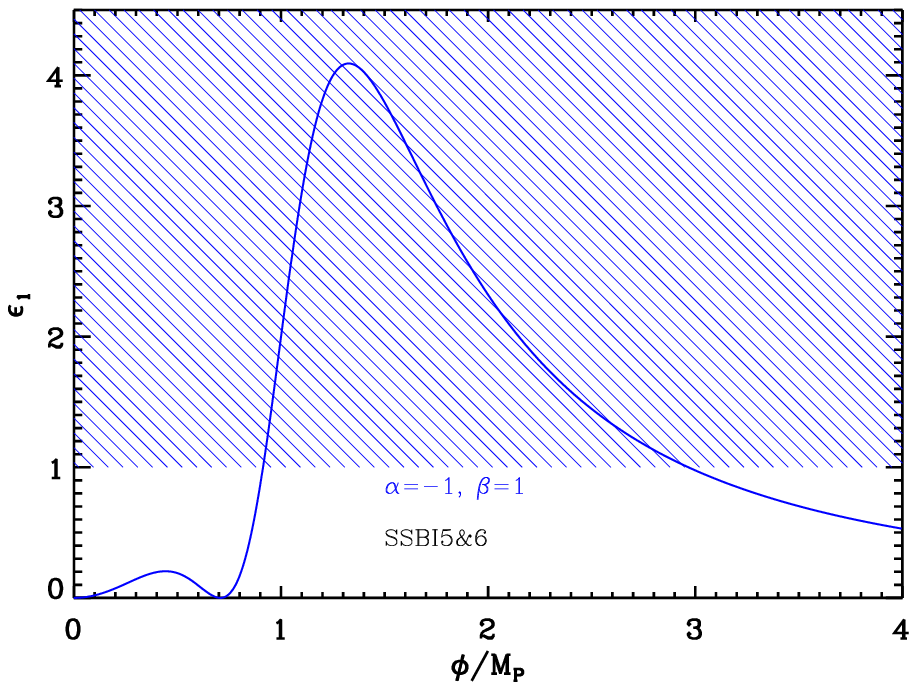}
\caption{Spontaneous Symmetry Breaking Inflation (SSBI) potential and
  the corresponding Hubble flow parameter $\epsilon_1$ for the two
  cases $\alpha>0$, $\beta<0$ (corresponding to SSBI3 to SSBI4) and
  $\alpha<0$, $\beta>0$ (corresponding to SSBI5 and to SSBI6). In each
  of these cases, the direction in which inflation proceeds is
  indicated by the arrow.}
\label{fig:potssbib}
\end{center}
\end{figure}

Let us now turn to the slow-roll analysis of SSBI. For this purpose,
it is more convenient to rewrite the potential~(\ref{eq:potssbihep}) as
\begin{equation}
V\left(\phi\right)=M^4\left[1 + \alpha\left(\frac{\phi}{\Mp}\right)^2
  + \beta\left( \frac{\phi}{\Mp} \right)^4 \right],
\end{equation}
where $\alpha$ and $\beta$ are two dimensionless parameters. Based on
the previous brief review of the literature, we conclude that it is
necessary to study the model in full generality and, therefore, in
what follows, we investigate all possible situations. As mentioned
above, four cases should be distinguished: $\alpha>0,\, \beta>0$;
$\alpha<0,\, \beta<0$; $\alpha>0,\, \beta<0$ and $\alpha<0,\,
\beta>0$, with two possible domains of inflation in the two latter
cases. Therefore we have six regimes of inflation that we label SSBI1,
SSBI2, SSBI3, SSBI4, SSBI5 and SSBI6. The different potentials and
inflationary regimes are displayed and defined in \Fig{fig:potssbia}
and \Fig{fig:potssbib}. Since the potential is symmetric under
$\phi/\Mp\rightarrow -\phi/\Mp$, it is only displayed and studied for
$\phi>0$.

\begin{figure}
\begin{center}
\includegraphics[width=\wdblefig]{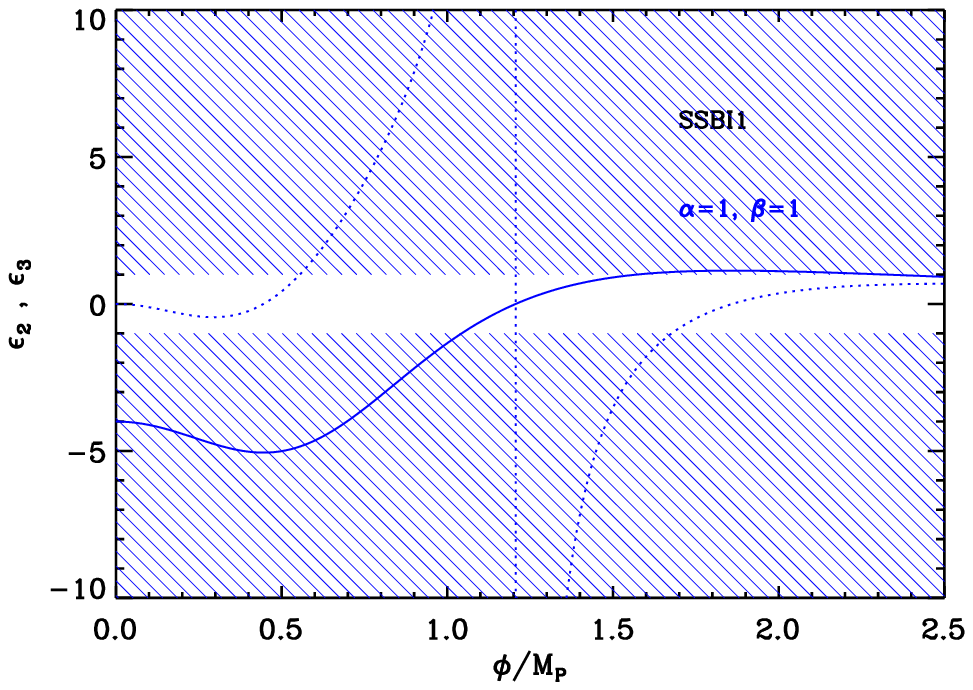}
\includegraphics[width=\wdblefig]{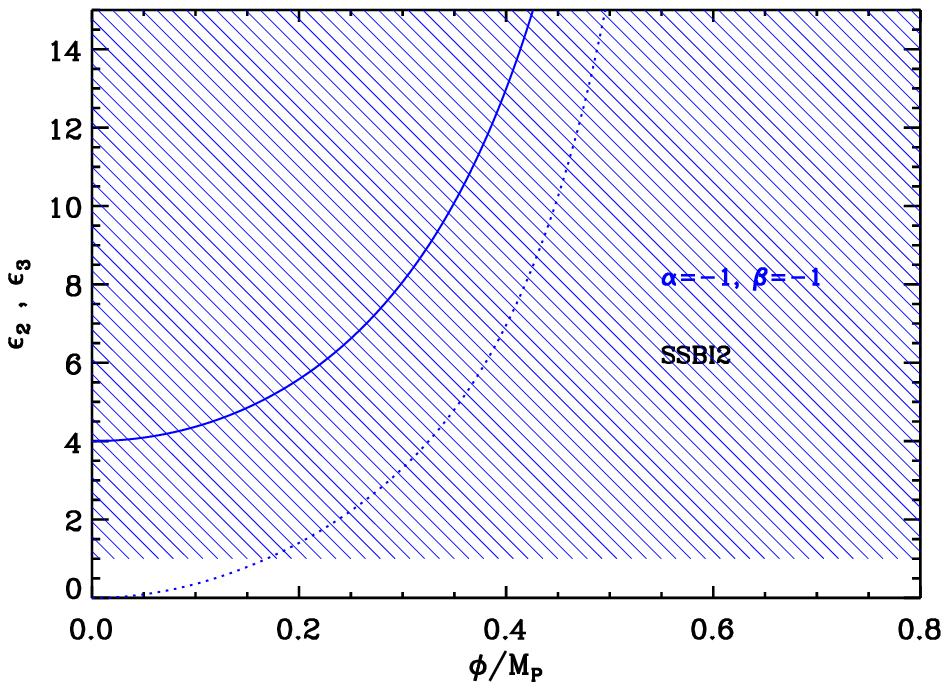}
\includegraphics[width=\wdblefig]{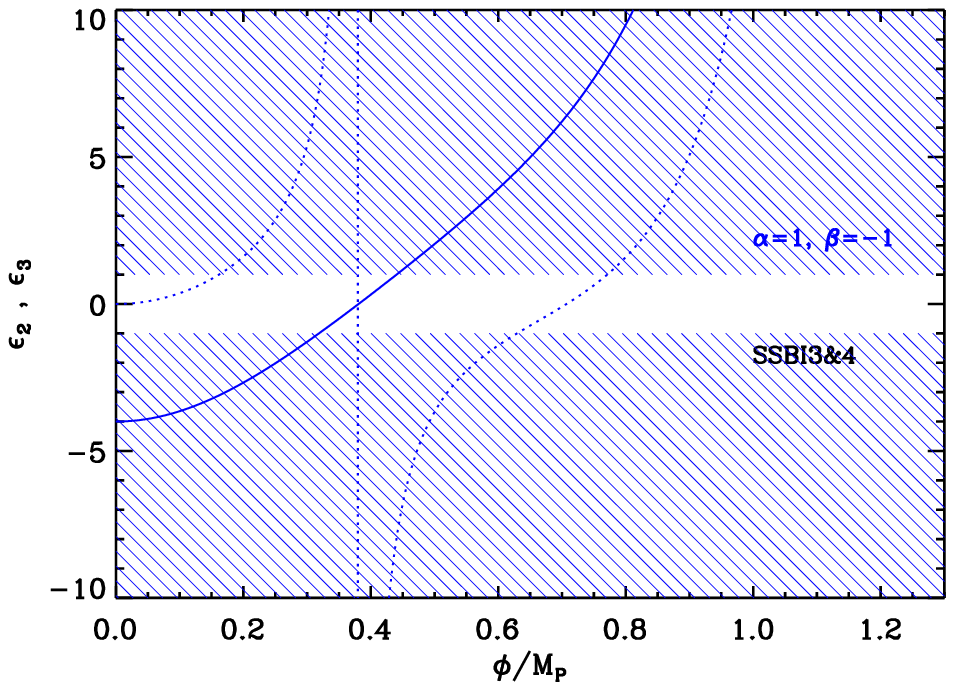}
\includegraphics[width=\wdblefig]{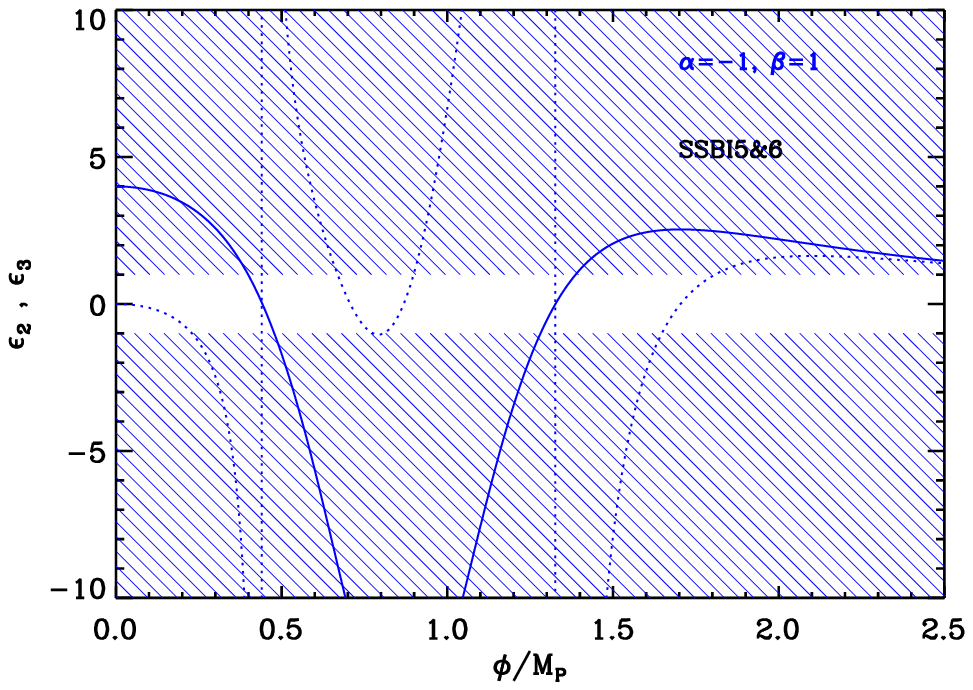}
\caption{Second slow-roll parameter $\epsilon_2$ (solid line) and
  third slow-roll parameter $\epsilon_3$ (dotted line), for the six
  SSBI models studied in this section. The free parameters of the
  models are chosen to be $\alpha,\beta=\pm 1$.}
\label{fig:sr23ssbi}
\end{center}
\end{figure}

\begin{figure}
\begin{center}
\includegraphics[width=\wdblefig]{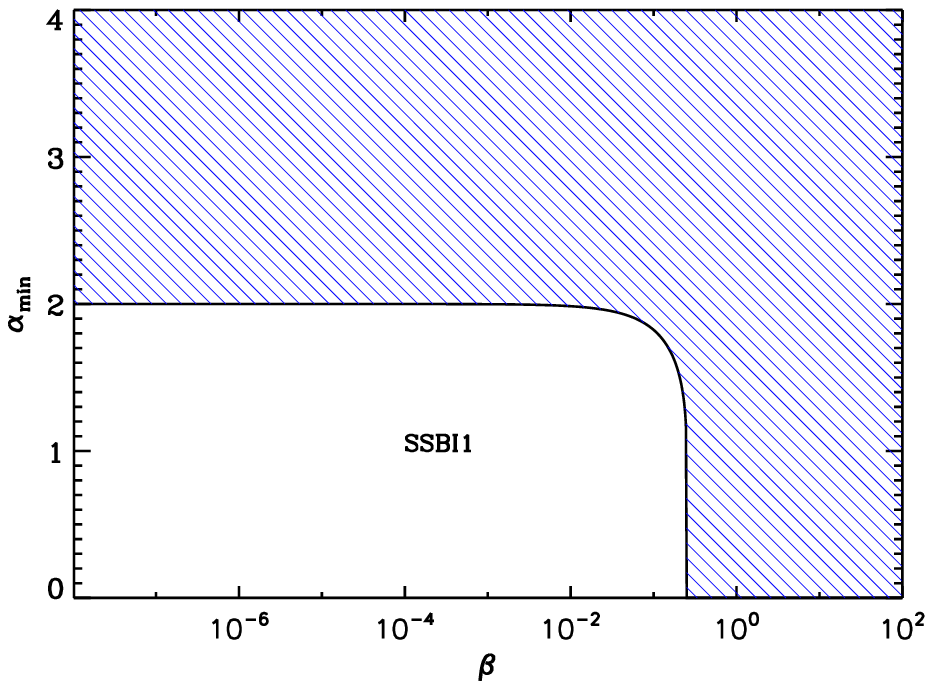}
\includegraphics[width=\wdblefig]{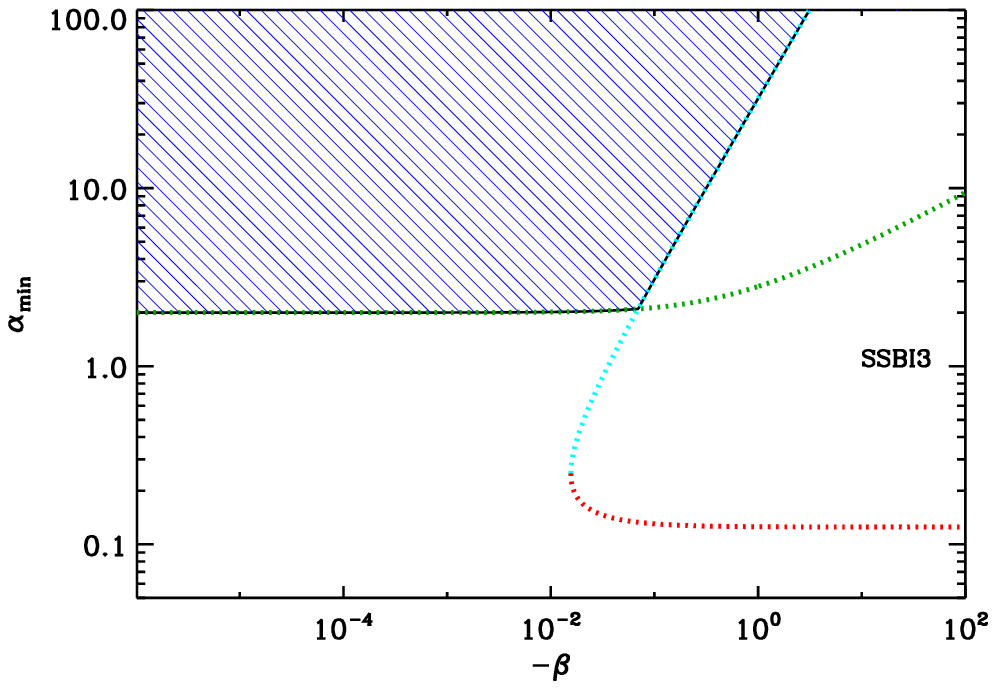}
\includegraphics[width=\wdblefig]{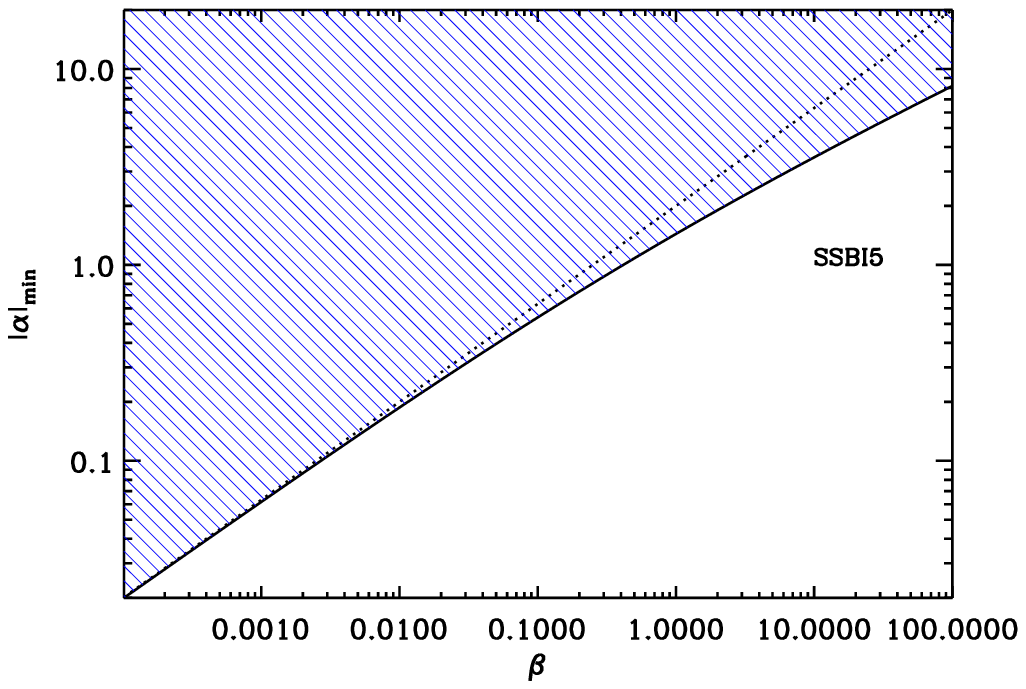}
\includegraphics[width=\wdblefig]{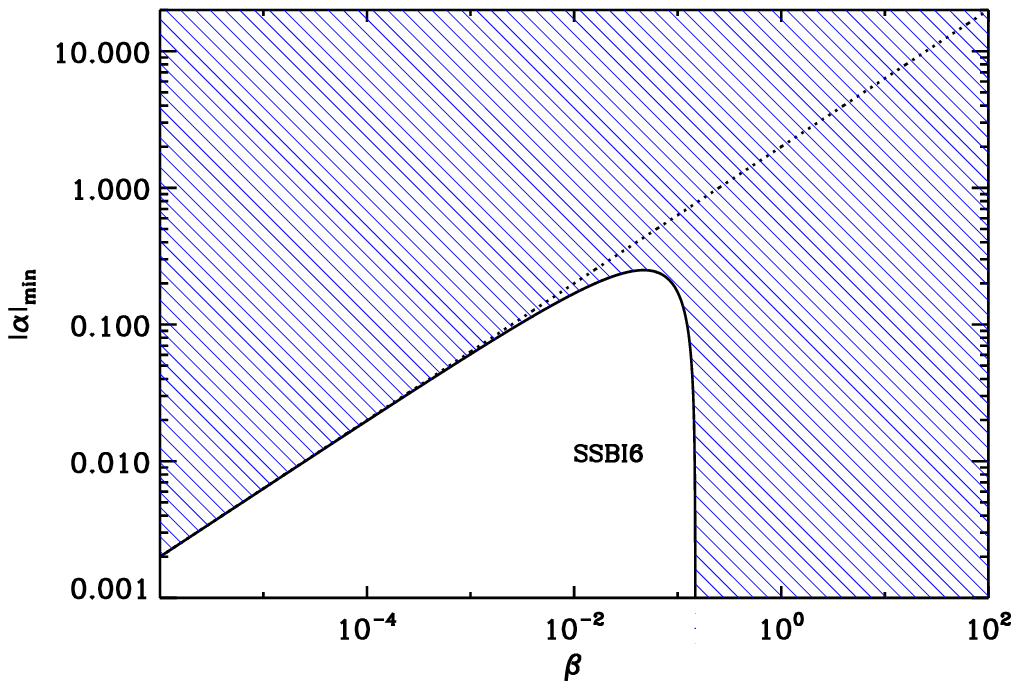}
\caption{The black solid line gives the minimum value of
  $\vert\alpha\vert$, denoted here by $\alphamin$, as a function of
  $\beta$ in order for inflation to stop by slow-roll violation for
  SSBI1 (top left panel), SSBI5 (bottom left panel) and SSBI6 (bottom
  right panel). For SSBI3 (top right panel), the green dotted line
  denotes the minimum value of $\alpha$ for inflation to stop by
  slow-roll violation, and the cyan and red dotted line restrict the
  values of $\alpha$ for which $\epsilon_2^{\mathrm{top}}>1$ (defined
  only for $\beta<-1/64$).  In the bottom panels, the dotted lines
  correspond to $\alpha^2=4\beta$, see the discussion in the text.  In
  all the panels, the region above the black solid curve (shaded
  region) represents the allowed region (\ie the one where a slow roll
  regime of inflation stops because $\epsilon_1$ reaches one). For
  SSBI1, when $\beta\gtrapprox0.25$, this is always the case.  For
  SSBI1 and SSBI3, $\alphamin$ approaches the asymptotic value
  $\alphamin= 2$ when $\vert\beta\vert\ll 1$. For SSBI5 and SSBI6,
  inflation stops by slow-roll violation when
  $\alpha<-\vert\alphamin\vert$.}
\label{fig:ssbi:alphaminssbi}
\end{center}
\end{figure}

Let us now calculate the slow-roll parameters. If one defines $x$ by
$x\equiv\phi/\Mp$, then the three first Hubble parameters are given by
the following expressions
\begin{equation}
\epsilon_1 = \frac{2\left(\alpha x + 2\beta
  x^3\right)^2}{\left(1+\alpha x^2+\beta x^4\right)^2}\, , \qquad
\epsilon_2 = \frac{4 \left[ -\alpha+\left(\alpha^2 - 6 \beta \right)
    x^2 + \alpha\beta x^4 + 2\beta^2 x^6 \right] }{ \left(1+\alpha x^2
  + \beta x^4 \right)^2} \, ,
\end{equation}
and
\begin{equation}
\epsilon_3 = \dfrac{4x^2\left(\alpha+2\beta 
x^2\right)\left[-3\alpha^2+6\beta
  + \alpha \left(\alpha^2 - 12 \beta\right) x^2 +3 \left(\alpha^2 - 8
  \beta\right) \beta x^4 + 2 \beta^3 x^8\right]}{\left(1
+ \alpha x^2 + \beta x^4\right)^2 \left[-\alpha + \left(\alpha^2 - 6
  \beta\right) x^2 + \alpha \beta x^4 + 2 \beta^2 x^6\right] }\,.
\end{equation}
The first slow-roll parameter $\epsilon_1$ is displayed in the right
panels of \Figs{fig:potssbia} and~\ref{fig:potssbib} while the second and
third slow-roll parameters $\epsilon_2$ and $\epsilon_3$ are displayed
in \Fig{fig:sr23ssbi}. Let us describe the behavior of these slow-roll
parameters, for the six models under consideration.
For SSBI1, $\epsilon_1$ vanishes at $x=0$, reaches a maximum at
$\xepstwoZero^{\mathrm{SSBI1}}$ (where $\epsilon_2$ vanishes and
$\epsilon_3$ diverges) and then decreases to asymptotically vanish
when $x$ goes to infinity. The value of
$\xepstwoZero^{\mathrm{SSBI1}}$ is given by
\begin{equation}
\begin{aligned}
\xepstwoZero^{\mathrm{SSBI1\&3\&6}} & = \left\lbrace
-\frac{\alpha}{6\beta} + \frac{1}{6 \beta}
\left[8\alpha^3+\sqrt{64\alpha^6 + \left(5\alpha^2-36\beta\right)^3}
  \right]^{1/3} \right. \\ & + \left. \frac{36\beta-5\alpha^2}{6\beta}
\left[8\alpha^3+\sqrt{64\alpha^6+\left(5\alpha^2-36\beta\right)^3}
  \right]^{-1/3} \right\rbrace^{1/2} .
\end{aligned}
\end{equation}
Whether the maximum of $\epsilon_1$ at this point is larger or smaller
than $1$ depends on $\alpha$ and $\beta$. In the following, we
restrict ourselves to the physical regime where $\alpha,\beta \lesssim
\order{1}$. For each value of $\beta$, there is a
minimum value of $\alpha$, denoted $\alphamin$, above which the
maximum is larger than $1$. The line $\alphamin(\beta)$ is
displayed in \Fig{fig:ssbi:alphaminssbi} and the shaded area in this
plot represents the region in the parameter space where inflation
stops by slow-roll violation. When $\beta \ll 1$,
$\alphamin(\beta)$ approaches $2$ as can be noticed in the
figure. In addition, for $\beta\gtrsim 0.25$, the maximum value for
$\epsilon_1$ becomes larger than $1$ for any value of $\alpha$.

For SSBI2, the three first slow-roll parameters are monotonic
increasing functions of the field \vev and diverge when the potential
vanishes at
\begin{equation}
\xVzero^{\mathrm{SSBI2\&4\&5}}=\sqrt{-\frac{\alpha+\sqrt{\alpha^2-4\beta}}
{2\beta}}\, .
\end{equation}
Hence inflation ends by slow-roll violation at $\xend$. Unfortunately,
the corresponding \vev cannot be found exactly and one has to rely on
numerical calculations. Let us also notice that, while the first and
third slow-roll parameters $\epsilon_1$ and $\epsilon_3$ vanish at
$x=0$, $\epsilon_2$ is equal to $\epstwomin=-4\alpha$
at this point. Therefore, in order for the slow-roll approximation to
be valid, one needs to work with $\vert\alpha\vert\ll 1$.

For SSBI3, the first slow-roll parameter $\epsilon_1$ vanishes at
$x=0$ and at $x=\sqrt{-\alpha/\left(2\beta\right)}$. In between, it
reaches a maximum located at
\begin{equation}
\xepstwoZero^{\mathrm{SSBI3}}=\xepstwoZero^{\mathrm{SSBI1}}\, ,
\end{equation}
a point where $\epsilon_2$ vanishes and $\epsilon_3$ diverges.
Whether the maximum of $\epsilon_1$ at this point is larger or smaller
than $1$ depends again on $\alpha$ and $\beta$. For each value of
$\beta$, there is a minimum value for $\alpha$ above which inflation
stops by slow-roll violation, similarly to the SSBI1 case. This
corresponds to the green dotted line in \Fig{fig:ssbi:alphaminssbi}
(top right panel). One way to estimate whether a slow roll regime of
inflation can occur in the decreasing branch of $\epsilon_1$ is to
look at the value of $\epsilon_2$ at the top of the potential.  It is
given by
\begin{equation}
\epstwotop=\frac{-32\alpha\beta}{\alpha^2-4\beta}\,.
\end{equation}
This number is smaller than one when $\beta<-1/64$, or when $\alpha$
lies outside the range with limits given by
$-16\beta\pm\sqrt{\beta(1+64\beta)}$, displayed in
\Fig{fig:ssbi:alphaminssbi} with the red and cyan dotted lines.
Therefore, requiring that $\epstwotop<1$ and that
inflation stops by slow roll violation leads to the allowed space
$\alpha>\alphamin$, represented by the shaded region in
\Fig{fig:ssbi:alphaminssbi}.

For SSBI4, the three first slow-roll parameters are monotonic
increasing functions of the field \vev and diverge when the potential
vanishes at $\xVzero^{\mathrm{SSBI2\&4}}$. The first and third
slow-roll parameters $\epsilon_1$ and $\epsilon_3$ vanish when
$x=\sqrt{-\alpha/\left(2\beta\right)}$ while $\epsilon_2$ has a
non-zero value $\epstwomin=8\alpha\beta/ (\beta^2-\alpha^2/4)$ at this
point. From the above discussion, it is clear that, in this version of
the scenario, inflation also stops by violation of the slow-roll
condition. As for SSBI2, however, the corresponding \vev cannot be
determined exactly and a numerical calculation is needed.

For SSBI5, the behavior of the slow-roll parameters depend on
$\alpha^2/\beta$. If $\alpha^2/\beta\geq 4$, the minimum of the
potential at $x=\sqrt{-\alpha/\left(2\beta\right)}$ is negative. The
potential vanishes at $\xVzero^{\mathrm{SSBI2\&4\&5}}$ and the three
first slow-roll parameters continuously increase between $x=0$ where
they vanish (except $\epsilon_2$ for which
$\epstwomin=-4\alpha$) and
$\xVzero^{\mathrm{SSBI2\&4\&5}}$ where they diverge. Inflation ends by
slow-roll violation at some point $\xend$ that needs to be determined
numerically. On the other hand, if $\alpha^2/\beta\leq 4$,
$\epsilon_1$ vanishes at $x=0$, reaches a maximum at
$\xepstwoZero^{\mathrm{SSBI5}}$ (where $\epsilon_2$ vanishes and
$\epsilon_3$ diverges), then decreases and finally vanishes at
$x=\sqrt{-\alpha/\left(2\beta\right)}$. The value of
$\xepstwoZero^{\mathrm{SSBI5}}$ is given by
\begin{equation}
\begin{aligned}
\xepstwoZero^{\mathrm{SSBI5}} & = \left\lbrace - \frac{\alpha}{6\beta}
- \frac{1+i\sqrt{3}}{12 \beta} \left[ 8\alpha^3+\sqrt{64\alpha^6 +
    \left(5\alpha^2-36\beta\right)^3} \right]^{1/3} \right. \\ & +
\left. \frac{5\alpha^2-36\beta}{12\beta} \left(1-i\sqrt{3} \right)
\left[8\alpha^3+\sqrt{64\alpha^6+\left(5\alpha^2-36\beta\right)^3}\right]^{-1/3}
\right \rbrace^{1/2}.
\end{aligned}
\end{equation}
Whether the maximum of $\epsilon_1$ at this point is larger or smaller
than $1$ depends on $\alpha$ and $\beta$ and is again similar to what
has already been discussed before. The region in the parameter space
where inflation ends by slow-roll violation is displayed in
\Fig{fig:ssbi:alphaminssbi} and corresponds to the points such that
$\alpha<-\vert\alphamin\vert$. In this plot, the dotted line
represents the curve $\alpha^2=4\beta$, above which one is sure that
inflation ends by slow-roll violation since the minimum of the
potential is negative in this case. For values of $\beta\ll 1$, one
can see that $\vert\alphamin\vert\simeq 2\sqrt{\beta}$ and the allowed
region becomes negligible.

Finally the case SSBI6 remains to be treated. The behavior of the slow
roll parameters depend on $\alpha^2/\beta$ in the same way as before.
If $\alpha^2/\beta\geq 4$, the minimum of the potential at
$x=\sqrt{-\alpha/\left(2\beta\right)}$ is negative. The potential
vanishes at $\xVzero^{\mathrm{SSBI6}}$ and the slow-roll parameters
continuously decrease from this value (where they blow up) and go to
zero at infinity. The value of $\xVzero^{\mathrm{SSBI6}}$ can be
expressed as
\begin{equation}
  \xVzero^{\mathrm{SSBI6}} = \sqrt{
    \frac{-\alpha+\sqrt{\alpha^2-4\beta}}{2\beta} }\, .
\end{equation}
On the other hand, if $\alpha^2/\beta\leq 4$, $\epsilon_1$ vanishes at
$x=\sqrt{-\alpha/\left(2\beta\right)}$, reaches a maximum at
$\xepstwoZero^{\mathrm{SSBI6}}$ and then decreases. At infinity, it
goes to zero. The value of $\xepstwoZero^{\mathrm{SSBI6}}$ is given by
\begin{equation}
\xepstwoZero^{\mathrm{SSBI6}}= \xepstwoZero^{\mathrm{SSBI3}}= 
\xepstwoZero^{\mathrm{SSBI1}}.
\end{equation}
Whether the maximum of $\epsilon_1$ at this point is larger or smaller
than $1$ depends on $\alpha$ and $\beta$. The corresponding region in
the parameter space is displayed in \Fig{fig:ssbi:alphaminssbi} and
corresponds to the inequality $\alpha<-\vert\alphamin\vert$. The
dotted line represents the law $\alpha^2=4\beta$. Above this line, one
is sure that inflation can stop by slow-roll violation since, in this
case, the potential becomes negative at some point. It is also
interesting to notice that, when $\beta\gtrsim 1.48$, the maximum
value of $\epsilon_1$ is larger than $1$ for any value of $\alpha$. On
the other hand, if $\beta \ll 1$, the allowed region shrinks to zero.

Let us now turn to the slow-roll trajectory. This one can be
integrated analytically to get
\begin{equation}
  \Nend-N = -\frac{1}{2\alpha} \ln\left( \frac{\xend}{x} \right) -
  \frac{\xend^2-x^2}{8} - \frac{\alpha^2-4\beta}{16\alpha\beta} \ln
  \left( \dfrac{1 + \dfrac{2\beta}{\alpha}\xend^2 }{1 +
    \dfrac{2\beta}{\alpha}x^2} \right),
\end{equation}
where $\Nend$ is the number of \efolds at the end of inflation. It
is important to notice that the argument of the logarithm is always
positive. This trajectory cannot be inverted analytically. But,
numerically, it is easy to use this expression in order to determine
$\xstar$, the value of $x$ at Hubble radius crossing.

Finally, it is interesting to constrain the value of the scale $M$ with
the CMB normalization. It follows that
\begin{equation}
\left(\frac{M}{\Mp}\right)^4=\frac{2880\left(\alpha \xstar+2\beta
  \xstar^3\right)^2\pi^2}{\left(1+\alpha \xstar^2+\beta
  \xstar^4\right)^3} \frac{\Qrms^2}{T^2}\, .
\end{equation}

We are now in a position where we can discuss the predictions of the
six versions of this model. The reheating consistent slow-roll
predictions for the SSBI1 models are displayed in
\Figs{fig:CMBSSBI1betaEQ10PowerMinus3},
\ref{fig:CMBSSBI1betaEQ10PowerMinus1} and~\ref{fig:CMBSSBI1betaEQ10}
for $\beta=10^{-3}$, $\beta=10^{-1}$ and $\beta=10$,
respectively. SSBI1 seems to be disfavored by the observations. The
predictions of SSBI2 models are displayed in \Fig{fig:CMBSSBI2} for
different values of $\beta $ and $\alpha $. We notice that they depend
on the parameter $\alpha$ quite strongly. The spectral index is
clearly red and, for values of $\beta$ of order one, the contribution
of gravity waves becomes very small. For SSBI3, the predictions are
presented in \Figs{fig:CMBSSBI3betaEQMinus10PowerMinus3},
\ref{fig:CMBSSBI3betaEQMinus5x10PowerMinus3} and
\ref{fig:CMBSSBI3betaEQMinus10PowerMinus2} for $\beta=-10^{-3}$,
$\beta=-5\times 10^{-3}$ and $\beta=-10^{-2}$, respectively. As we
increase $\beta $, the points start spreading in the plane
$(\nS,r)$. For this class of models, the spectrum is red and the level
of gravity waves quite important. The predictions for the SSBI4 models
are displayed in \Figs{fig:CMBSSBI4betaEQMinus10PowerMinus5},
\ref{fig:CMBSSBI4betaEQMinus10PowerMinus4}, and
\ref{fig:CMBSSBI4betaEQMinus10PowerMinus3} for $\beta=-10^{-5}$,
$\beta=-10^{-4}$, $\beta=-10^{-3}$, respectively. One can notice that
the typical predicted values for $\epsilon_1$ decrease with the
absolute value of $\beta$. As before the spread of the points
increases with $\beta$. The tilt is still red and the contribution of
gravity waves is small for small values of $\alpha$. The predictions
for the SSBI5 models are displayed in
\Figs{fig:CMBSSBI5betaEQ10PowerMinus6},
\ref{fig:CMBSSBI5betaEQ10PowerMinus5} and
\ref{fig:CMBSSBI5betaEQ10PowerMinus4} for $\beta=10^{-6}$,
$\beta=10^{-5}$ and $\beta=10^{-4}$, respectively. Once again, for
$\order{1}$ values of $\beta$, one can see that the model predict a
small amount of gravitational waves but has a deviation from scale
invariance strongly disfavored by the observational
constraints. Finally, the reheating consistent slow-roll predictions
for the SSBI6 models are displayed in
\Figs{fig:CMBSSBI6betaEQ10PowerMinus5},
\ref{fig:CMBSSBI6betaEQ10PowerMinus1} and \ref{fig:CMBSSBI6betaEQ1}
for $\beta=10^{-6}$, $\beta=10^{-1}$ and $\beta=1$, respectively. When
$\beta\ll 1$ the predictions of the model do not depend on
$\beta$. Moreover, for values of $\beta $ of order one, the
predictions become almost independent of the two parameters of the
model.

\subsection{Inverse Monomial Inflation (IMI)}
\label{sec:imi}

These models are characterized by the inverse monomial potential given
by
\begin{equation}
\label{eq:potimi}
V\left(\phi\right)=M^4\left(\frac{\phi}{\Mp}\right)^{-p},
\end{equation}
where $p$ is a positive number. This scenario has been studied in many
different situations: in
\Refcs{Peebles:1987ek,Ratra:1987rm,Huey:2001ae} it was considered in
the context of quintessential inflation, in
\Refcs{Feinstein:2002aj,Sami:2002zy,Wang:2002sz,Abramo:2003cp} in the
context of tachyon inflation, in \Refcs{Barrow:1990td,Barrow:2006dh}
in the context of intermediate inflation and in \Refc{Bennai:2006th}
in the context of Randall-Sundrum braneworld models. In all these
articles, the potential was just postulated. An attempt to derive this
potential from high energy considerations was made in
\Refcs{Binetruy:1998rz,Brax:1999yv} in the context of supersymmetric
QCD. Let us, however, notice that this was done in order to build a
model of quintessence and not of inflation. The model uses the group
$\mathrm{SU}(\Nc)$ and has $\Nf$ flavors. The quarks
$Q^i$, $i=1,\cdots, \Nf$ are placed in the fundamental
representation of $\mathrm{SU}(\Nc)$ and the anti-quarks
$Q_i^{\dagger}$ in the conjugate
representation~\cite{Binetruy:1998rz}. At scales below the gauge
breaking scale $\Lambda$, the relevant degrees of freedom are the
pions $\pi^i_j=Q^iQ_j^{\dagger}$ and one can show that the
corresponding superpotential is given
by~\cite{Taylor:1982bp,Affleck:1984xz}
\begin{equation}
W=\left(\Nc-\Nf\right)
\frac{\Lambda^{3(\Nc-\Nf)/(\Nc-\Nf)}}
{\left(\det \pi\right)^{1/(\Nc-\Nf)}}\,.
\end{equation}
The potential~(\ref{eq:potimi}) then follows from the F-term
associated to the above superpotential.

\begin{figure}
\begin{center}
\includegraphics[width=\wdblefig]{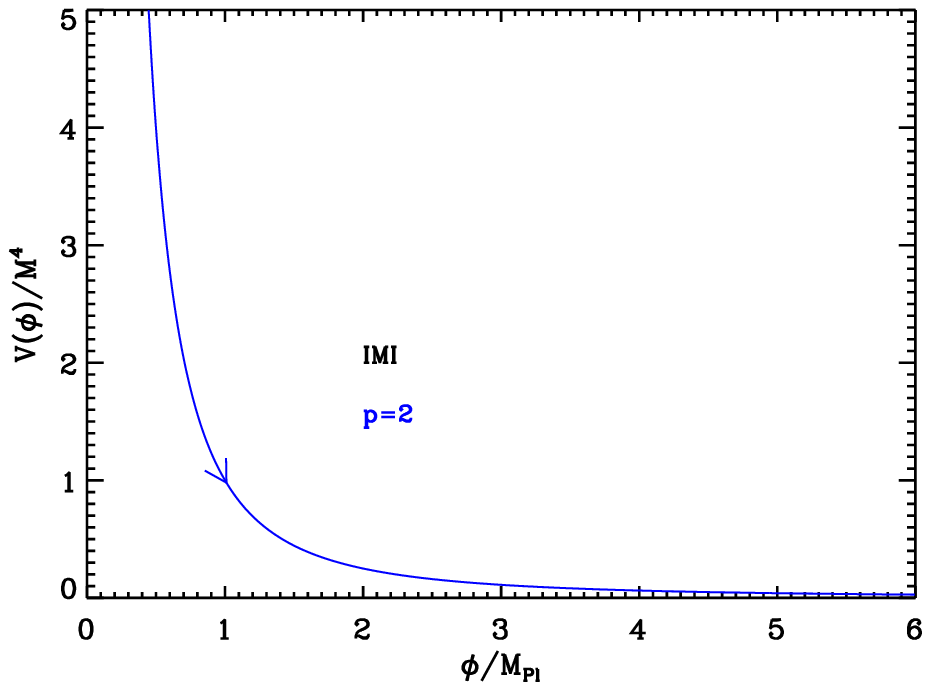}
\includegraphics[width=\wdblefig]{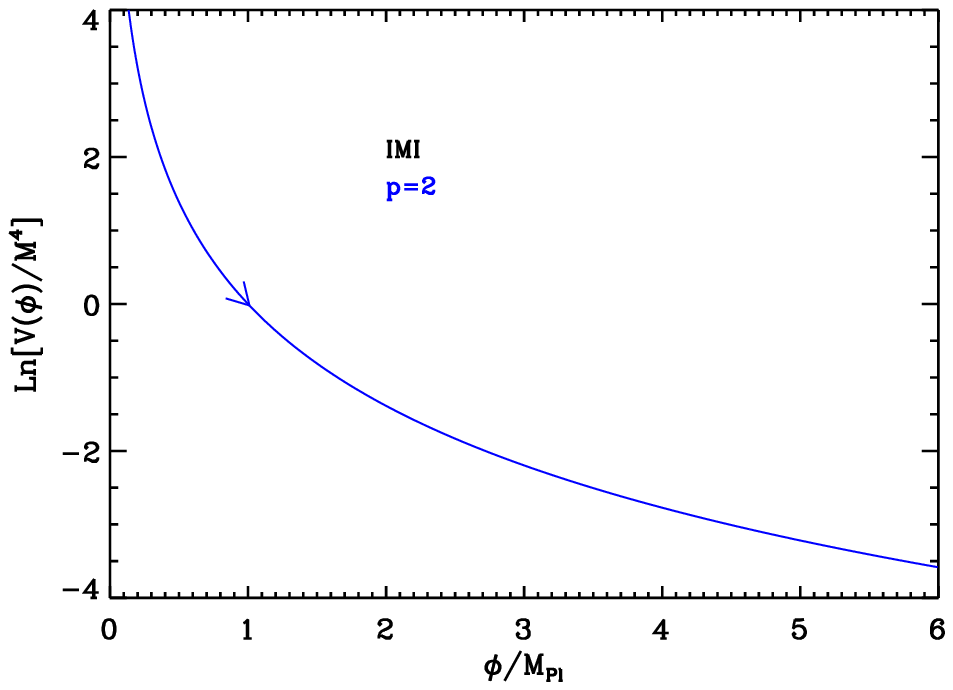}
\includegraphics[width=\wdblefig]{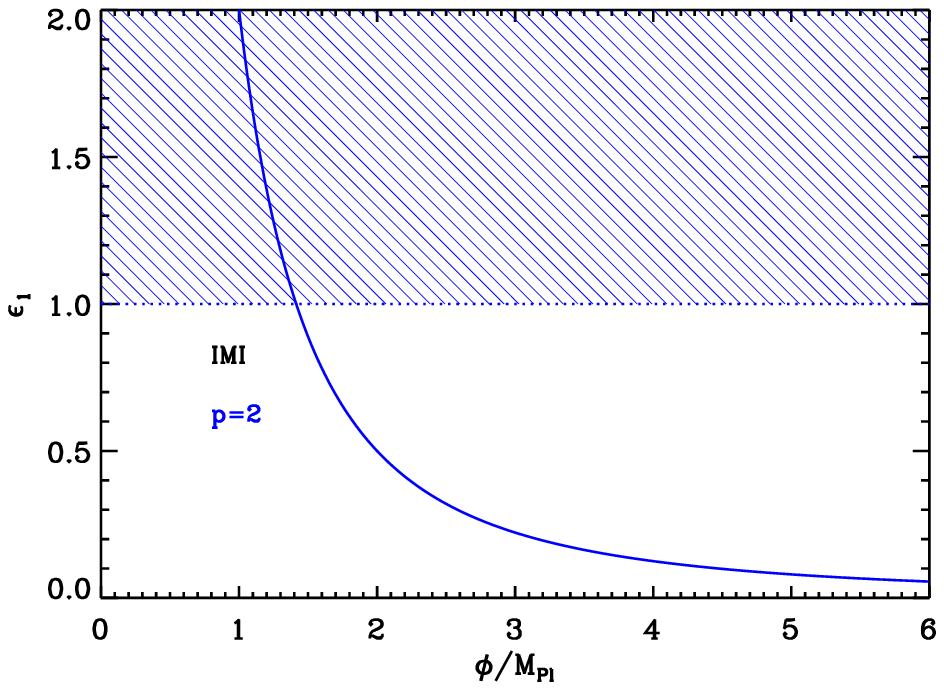}
\includegraphics[width=\wdblefig]{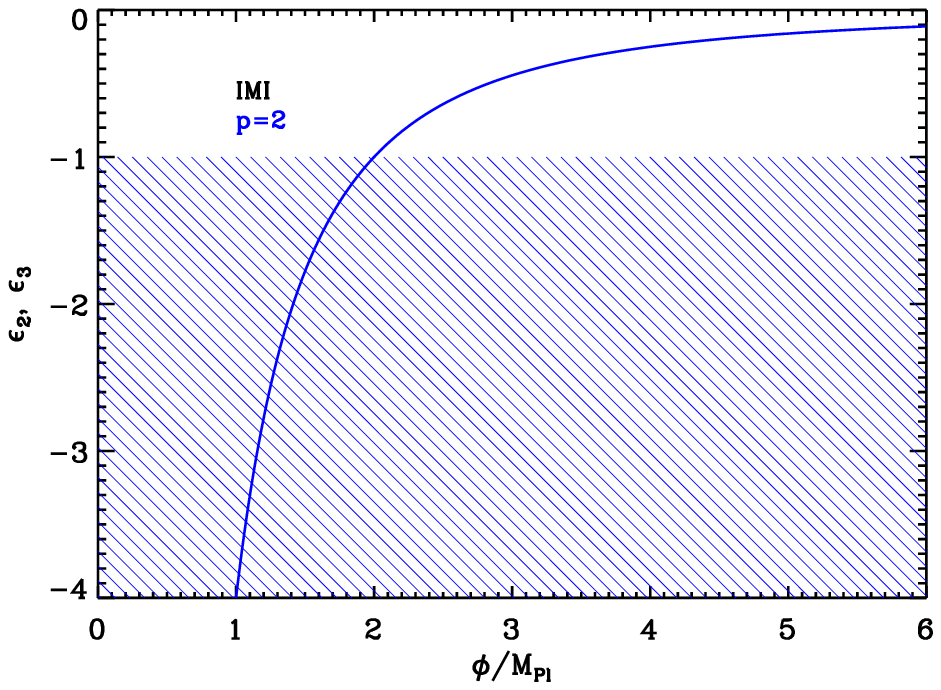}
\caption{Top left panel: Inverse Monomial Inflation (IMI) potential
  for $p=2$. Top right panel: logarithm of the potential for the same
  value of $p$. Bottom left panel: slow-roll parameter $\epsilon_1$
  for $p=2$.  Bottom right panel: slow-roll parameters $\epsilon_2$
  and $\epsilon_3$ for $p=2$. Only one line appears because
  $\epsilon_2=\epsilon_3$. On these plots, the shaded region
  represents the region where the slow-roll approximation breaks
  down.}
\label{potIMI}
\end{center}
\end{figure}

The potential is represented in \Fig{potIMI} for $p=2$. It is a
decreasing function of the field \vev and, hence, inflation proceeds
from the left to the right, in the direction specified by the arrow in
the figure.

The three Hubble flow functions are straightforwardly obtained from
\Eqs{eq:eps1}, \eqref{eq:eps2} and \eqref{eq:eps3}. Defining $x\equiv
\phi/\Mp$, one gets
\begin{equation}
\label{eq:imi:eps}
  \epsilon _1 = \frac{p^2}{2 x^2}\,
  ,\qquad \epsilon_2= -\frac{2p }{x^2}\,, \qquad \epsilon
  _3=\epsilon_2.
\end{equation}
These functions are represented in the two bottom panels in
\Fig{potIMI}. The first slow-roll parameter is a monotonic decreasing
function of $\phi$ while $\epsilon_2$ and $\epsilon_3$ are negative
increasing functions. From these expressions, one can also immediately
deduce that, for a given $p$, the model in the plane $(\epsilon
_1,\epsilon _2)$ is represented by the line $\epsilon_1=-(p/4)\epsilon
_2$. Since inflation proceeds from the left to the right, it cannot
stop by slow-roll violation. As a consequence, an extra-mechanism, such
as \eg tachyonic instability, must be implemented to end
inflation. Let us denote $\xend$ the position at which such a process
occurs. The model has therefore two free parameters: $p$ and $\xend$.

The slow-roll trajectory can be obtained by quadrature from
\Eq{eq:srtrajectory}, and one obtains
\begin{equation}
  N-\Nend =\dfrac{1}{2p}
  \left(x^2 -
    \xend^2 \right).
\label{eq:imi:traj}
\end{equation}
This expression can be inverted and reads
\begin{equation}
\label{eq:imi:InvertedTraj}
x = \sqrt{\xend^2 +2p\left( N - \Nend\right)}\, .
\end{equation}
Let us now derive some prior condition on $\xend$. One can notice that
when $x<\xepsoneOne=p/\sqrt{2}$, one has $\epsilon_1>1$ and inflation
cannot take place. This means that inflation can only proceed between
$\xepsoneOne$ and $\xend$, where the maximum number of \efolds is,
using \Eq{eq:imi:traj}, $\Delta \Nmax\left(\xend\right) =
\left(\xend^2 -\xepsoneOne^2 \right)/(2p)$. Put it differently, if one
wants to realize at least $\Delta N$ \efolds, then one has to work
with $\xend>\xendmin$ where
\begin{equation}
\xendmin\left(\Delta N\right)=\sqrt{p^2/2 +2p\Delta N}\, .
\end{equation}
This defines a prior condition on $\xend$.

Finally, the parameter $M$ can be determined from the amplitude of the 
CMB anisotropies, and it follows that
\begin{equation}
  \left(\frac{M}{\Mp}\right)^4=720 \pi^2 p^2 \xstar^{p-2}
  \frac{\Qrms^2}{T^2}\,.
\label{eq:imi:cobe} 
\end{equation}
The reheating consistent slow-roll predictions for the IMI models are
displayed in \Fig{fig:CMBIMI}. For a given value of $p$, they lie
along the line $\left(1-2/p\right)r=8\left(1-\nS\right)$, \ie
$\epsilon_1=-(p/4)\epsilon_2$. As expected, large values of $\xend$,
or small values of the reheating temperature (these two parameters
being degenerate), are preferred.

\subsection{Brane Inflation (BI)}
\label{sec:bi}

\subsubsection{Theoretical Justifications}
\label{subsubsec:theorybi}

This section is devoted to brane inflation, a class of models widely
discussed in the literature~\cite{Alexander:2001ks, Burgess:2001fx,
  Dvali:2001fw, Shiu:2001sy, Jones:2002cv, GarciaBellido:2003wu,
  Pogosian:2003mz, Matsuda:2003ct, Matsuda:2004qi, Yang:2005ed,
  Pogosian:2003mz, Huang:2006ra, Bean:2007hc, Lorenz:2007ze,
  Battye:2007si, HenryTye:2006uv, Brandenberger:2007ca, Lorenz:2008pc,
  Ma:2008rf}. The idea is that inflation is caused by branes moving in
the extra dimensions as it was already the case in TI, see
\sectionc{sec:ti}. For this reason, the setup is very similar to the
one considered in that section. One starts from type IIB superstring
theory where six dimensions are compactified. The effective, low
energy, description of the model contains various fields among which
are the dilaton, the axion and the (tensorial) gravitational
field. One also has anti-symmetric fields with their corresponding
field strength. The compact dimensions form a Calabi-Yau space and,
generically, this Calabi-Yau space is made of a bulk plus throats
attached to it. Along a given throat, a solution for the
ten-dimensional metric is given by the conifold already discussed in
\sectionc{sec:ti} whose metric is given in \Eq{eq:ksthroat}. In this
equation, the metric $\dd s_5^2$ lives on the five-dimensional section
$\Sigma_5$ and $r$ is the ``radial'' coordinate. In the following, we
will denote by $\rUV$ the radial coordinate at which the cone is glued
to the bulk and $\rzero$ the coordinate at the tip of the cone. The
volume of the cone section is denoted by $\Vol{\Sigma_5}$ and
will be measured in terms of the volume of the five-dimensional
sphere, namely
\begin{equation}
v\equiv \frac{\Vol{\Sigma_5}}{\Vol{S_5}}\,.
\end{equation}
The geometry of the section $\Sigma_5$ depends on the background
fluxes, denoted by $\calM$ and $\calK$, that are quantities related to
the values of the anti-symmetric fields. If these fluxes vanish then
the five-dimensional sections are simply given by $S_2\times S_3$. In
that case, the conifold can be written as $\sum_{i=1}^4w_i^2=0$ where
$w_i$ are four complex coordinates, see
also \sectionc{sec:ti}. Moreover, an exact expression for the warp
function $h(r)$ can be found and reads
\begin{equation}
\label{eq:warpbi}
h(r)=C_2+\frac{C_1}{r^4}\,,
\end{equation}
$C_1$ and $C_2$ being constants. On the other hand, if the fluxes are
turned on, then the background geometry responses accordingly and, as
a consequence, the geometry of the cone is modified. It is now given
by a ``deformed conifold'', $\sum_{i=1}^4w_i^2=z$, where $z$ is a
number which depends on $\calM $ and $\calK$. The warp function
acquires a more complicated form and, obviously, becomes
$z$-dependent, \ie $h(r,z)$. The explicit form of this warp function
is not needed here but it is interesting to notice that, far from the
tip, one has $h(r,z)\simeq h(r)$. In other words, the modification of
the extra-dimensional geometry due to the fluxes is significant only
in the vicinity of the tip. Notice that, provided the depth of the
throat is comparable to its width, the radial coordinate
$\rUV$ can be expressed in terms of the quantity
$\calN\equiv \calM \calK$. One obtains~\cite{Baumann:2006cd}
\begin{equation}
\rUV^4=4\pi \gstrings \alpha'^2\frac{\calN}{v},
\end{equation}
where $\gstrings$ is the string coupling and $\alpha'\equiv \ells^2$,
$\ells$ being the string length.

Finally, an anti-$D3$ brane is placed at the tip of the conifold, \ie
at the bottom of the throat. This brane is heavy and is supposed to
slightly disturb the geometry of the throat in a way that has been
calculated for instance in \Refcs{Lorenz:2007ze, Lorenz:2008pc,
  Lorenz:2010nx}. Then, in this geometry, one studies the motion of a
light $D3$ brane with tension
\begin{equation}
T_3=\frac{1}{(2\pi)^3\gstrings \alpha'^2} \,. 
\end{equation}
This brane is attracted by the anti-$D3$ brane and as a consequence
moves radially along the throat. In principle it possesses a DBI
kinetic term but one can show that, in the regime considered here,
it always reduces to an ordinary, minimal, kinetic term, see
\Refc{Lorenz:2007ze}. If $r$ represents the distance between the two
branes, then the effective Lagrangian of the system can be expressed
as
\begin{equation}
\label{eq:lagrangianbi}
\calL=-\frac12 \left(\frac{\partial \phi}{\partial t}\right)^2
-\frac{2T_3\rzero^4}{\rUV^4}\left(1-
\frac{\rzero^4T_3^2}{\calN}\frac{1}{\phi^4}
\right),
\end{equation}
where $\phi \equiv \sqrt{T_3}r$. The shape of the potential is now
completely fixed and the behavior $\propto \phi^{-4}$ is of course due
to the particular scaling $\propto r^{-4}$ of the warp function given
by \Eq{eq:warpbi}.

In order to be valid, the effective model described above must satisfy
some conditions that we now discuss in more detail. Defining
$\phizero \equiv \sqrt{T_3}\rzero$ and $\phiUV\equiv
\sqrt{T_3}\rUV$, it is clear that the presence of the
brane in the throat implies that
$\phizero<\phi<\phiUV$. In addition, as discussed for
instance in \Refc{Lorenz:2007ze}, from the trivial fact that the
volume of the throat, $\Vthroat_6=2\pi^4\gstrings\calN
\alpha'^2\rUV^2$, cannot be bigger than the volume of the
total Calabi-Yau manifold $\Vtot_6$, one can derive the
bound
\begin{equation}
\label{eq:conditionuv}
\phiUV<\frac{\mpl}{\sqrt{2\pi \calN}}\,,
\end{equation}
where the Planck mass can be expressed as $\mpl^2=8\pi
\Vtot_6/\kappa_{10}$ and
$\kappa_{10}=(2\pi)^7\gstrings^2\alpha'^4/2$. Another constraint comes
from the fact that the effective model is valid only if the proper
distance between the two branes is larger than the Planck length. One
can show, see \Refc{Lorenz:2007ze}, that this means
$r>\rstg$ where
\begin{equation}
\rstg\equiv \rzero\ee^{\sqrt{\alpha'}/\rUV}.
\end{equation}
In particular, as will be seen in the following, the value of
$\rstg$ plays an important role regarding the mechanism
ending inflation. In the next section, we carry out the slow-roll
analysis of this model.

Let us also mention that the same potential arises in the context of
tachyon inflation~\cite{Panigrahi:2007sq, Kwon:2011wc}, in the context
of SQCD inflation~\cite{Brax:2009yd} and in the context of the strong
coupling limit of twisted models of SQCD inflation, (see TWI,
\sectionc{sec:twi} and \Refc{Davis:2010it}). It is also worth noticing
that the same kind of inverse power law potential is sometimes used in
quintessence models~\cite{Peebles:1987ek, Ratra:1987rm, Huey:2001ae}.
The brane inflation potential can also receive power law
corrections~\cite{Bean:2007eh} with either positive (UV models) or
negative sign (IR models). The UV case is similar to RIPI models while
the IR corresponds to SFI models.

\subsubsection{Slow-Roll Analysis}
\label{subsubsec:srbi}

We now turn to the slow-roll analysis of BI. For this purpose, it is
more convenient to re-write the potential appearing in
\Eq{eq:lagrangianbi} in the following way
\begin{equation}
\label{eq:bi:pot}
V(\phi) = M^4\left[ 1-\left(\frac{\phi}{\mu} \right)^{-p} \right],
\end{equation}
where $\mu$ and $p$ are free parameters. Compared to
\Eq{eq:lagrangianbi}, we have generalized by hand the expression of
$V(\phi)$ by considering an arbitrary $p$. In such a way, this
potential can be viewed as a generalization of the small field models
to negative values of $p$ (see \sectionc{sec:sfi}). In the following,
we will also consider the non-approximated KKLT potential 
\begin{equation}
\label{eq:kklti:pot}
V(\phi) = \frac{M^4}{1+\left(\dfrac{\phi}{\mu}\right)^{-p}}\,,
\end{equation}
from which \eqref{eq:bi:pot} is the $\mu\ll\Mp$ limit.

In the context of the brane inflationary scenario, the value $p=4$ is
special in the sense that, as explained above, it corresponds to the
motion of a test $D3$ brane in a warped throat and is, therefore, a
case of physical interest. Let us notice that the parameters of the
potential are related to their stringy counterparts by
\begin{equation}
M^4=\frac{2T_3\rzero^4}{\rUV^4}=\frac{4\pi^2v}{\calN}\phizero^4, 
\qquad \mu^4=\frac{T_3^2\rzero^4}{\calN} = \dfrac{M^4}{4 \pi^2 v}\,.
\label{eq:bi:stgparams}
\end{equation}
Moreover, brane inflation proceeds under the condition $\mu/\Mp\ll
1$. Indeed, using the formulas established in the previous subsection,
it is easy to show that
\begin{equation}
\frac{\mu^4}{\Mp^4}=\frac{1}{\calN}\left(\frac{\phizero}{\Mp}\right)^4
<\frac{1}{\calN}\left(\frac{\phiUV}{\Mp}\right)^4<\frac{16}{\calN ^3}
\ll 1,
\end{equation}
where we have used the condition $\phizero<\phiUV$ and
\Eq{eq:conditionuv}. Finally, let us stress that the brane motion in
the throat ends by a tachyonic instabilities at $\phi = \phistg$. As
we discuss below, the observable predictions of the model crucially
depends on whether the universe is still inflating at $\phi \gtrsim
\phistg$, or not. Therefore, in the context of string theory, we
necessarily have $\mu/\Mp\ll 1$, $p=4$ and an additional model
parameter $\phistg$.

In the following, we will first consider arbitrary values for $\mu$
and $p$ viewing \Eq{eq:bi:pot} as a phenomenological potential in
which $\phistg$ has no meaning, and then, the discussion will be
focused on the stringy scenario. BI is another proto-typical case
exemplifying how two models having exactly the same potential can lead
to different observable predictions. Here this will be due to the
mechanism ending inflation.

\begin{figure}
\begin{center}
\includegraphics[width=\wdblefig]{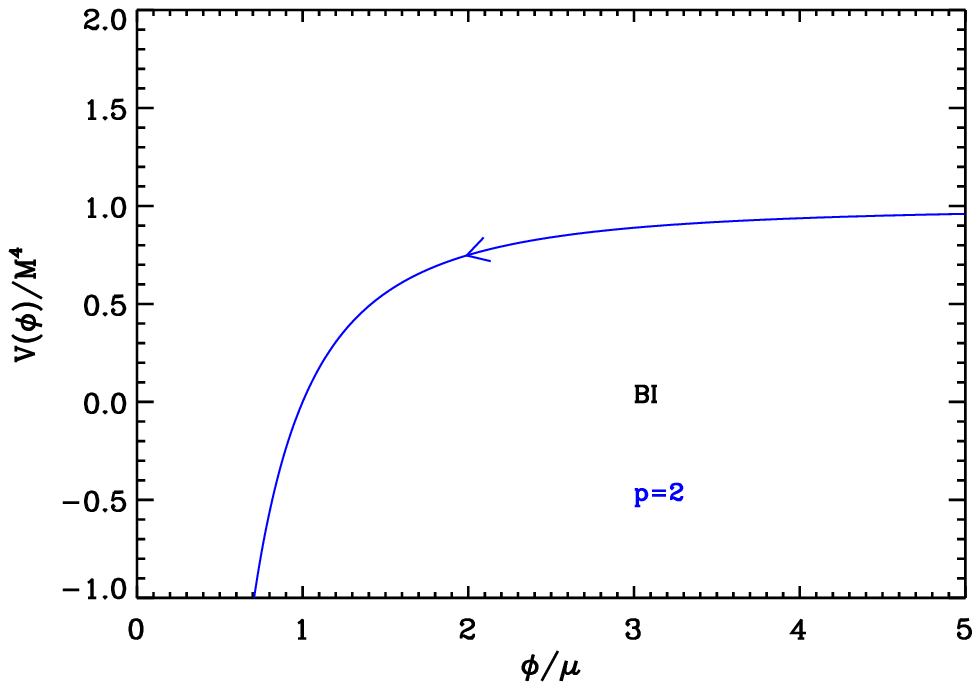}
\includegraphics[width=\wdblefig]{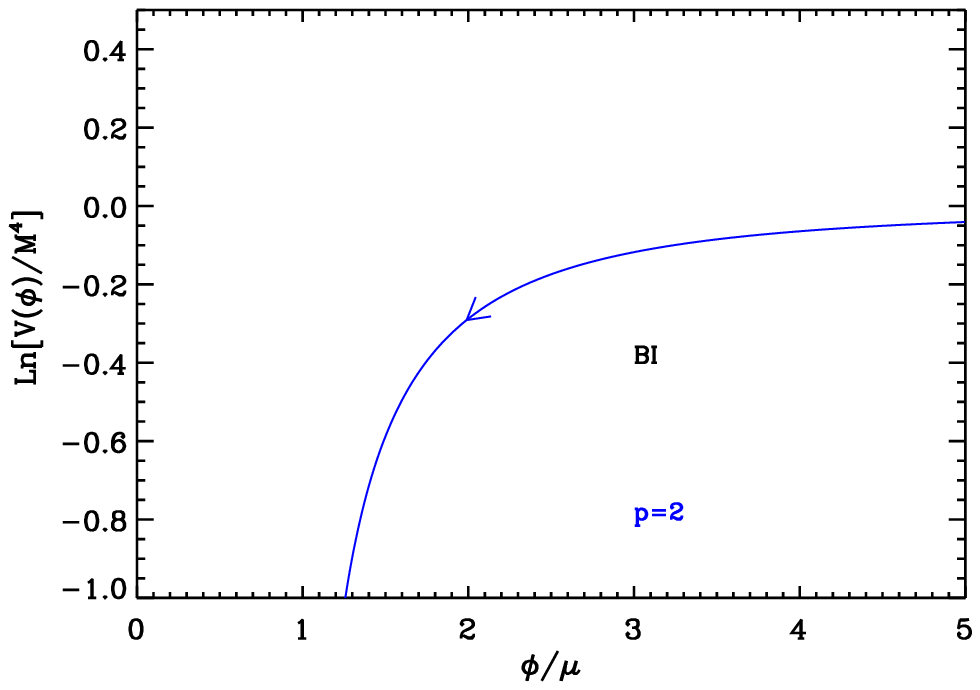}
\includegraphics[width=\wdblefig]{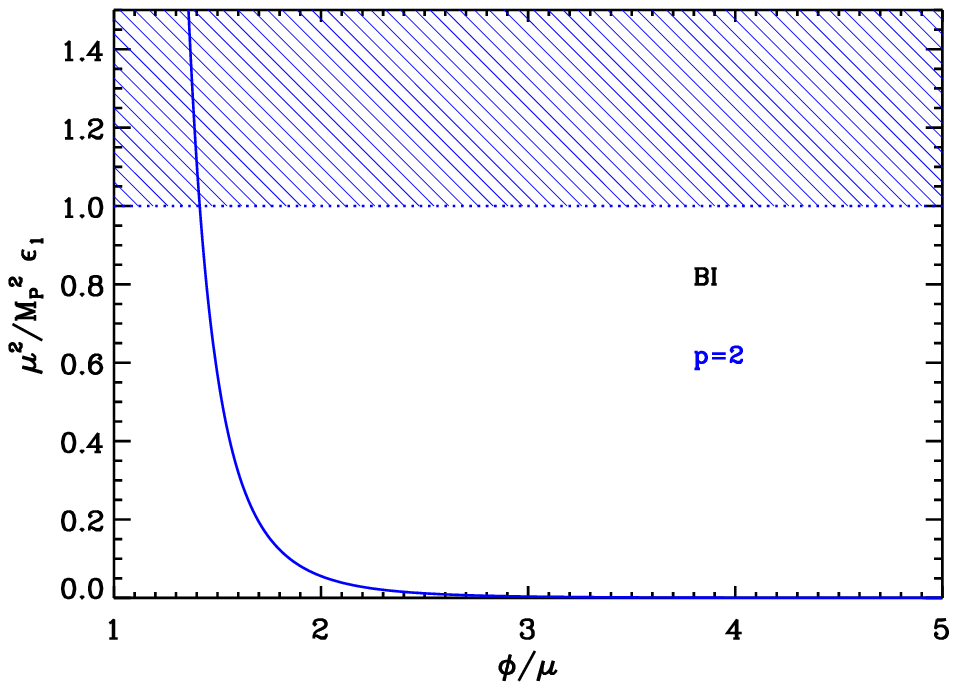}
\includegraphics[width=\wdblefig]{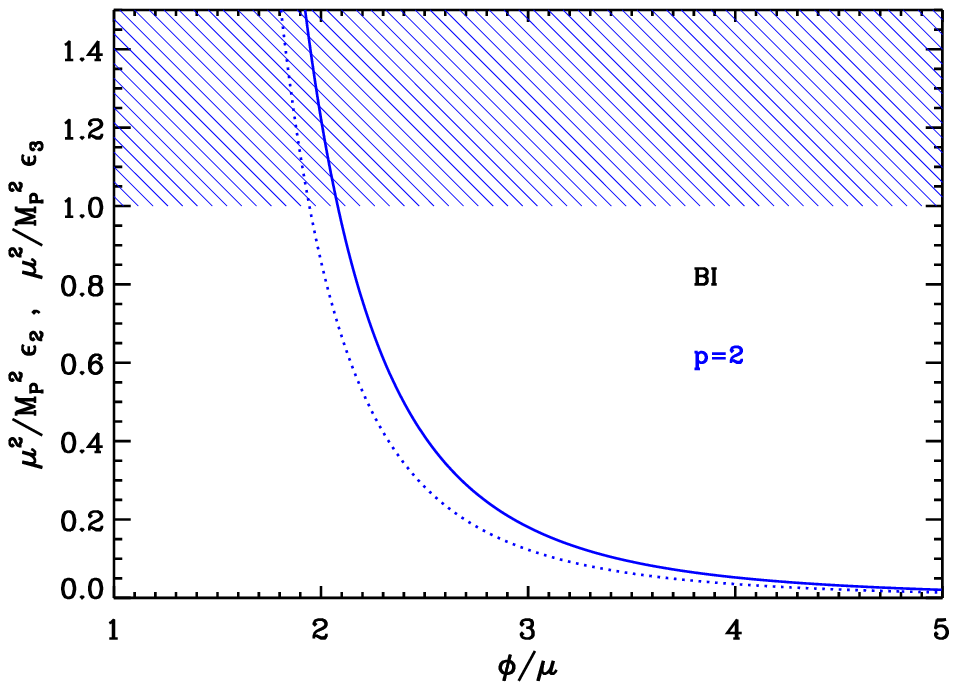}
\caption{Brane Inflation (BI) for $p=2$. Upper
  panels: the potential and its logarithm as a function of
  $\phi/\mu$. Bottom left panel: slow-roll parameter $\epsilon_1$
  divided by $\Mp^2/\mu^2$. The shaded area indicates the region in
  which inflation cannot occur for $\mu = \Mp$. Bottom right panel:
  slow-roll parameters $\epsilon_2$ (solid line) and $\epsilon_3$
  (dotted line), divided by $\Mp^2/\mu^2$.}
\label{potbi}
\end{center}
\end{figure}

The potential~(\ref{eq:bi:pot}), as well as its logarithm, are
displayed in \Fig{potbi}. It is an increasing function of the field,
hence inflation proceeds from the right to the left. It vanishes for
$\phi/\mu=1$ and, hence, it should be studied in the $\phi/\mu>1$ region only. Let
us calculate the slow-roll parameters. Defining the quantity $x$ by
the following expression
\begin{equation}
x \equiv \dfrac{\phi}{\mu}\,,
\end{equation}
one can express the first three Hubble flow functions in the slow-roll
approximation as
\begin{equation}
\label{eq:epsbi}
\epsilon_1 = \left(\dfrac{\Mp}{\mu }\right)^2\dfrac{p^2}
{2x^2\left(1-x^p\right)^2} \,, \qquad
\epsilon_2 = 2p\left(\dfrac{\Mp}{\mu }\right)^2 
\frac{\left(1+p\right)x^p-1}{x^2\left(1-x^p\right)^2}\, ,
\end{equation}
and
\begin{equation}
\epsilon_3 = p\left(\dfrac{\Mp}{\mu }\right)^2
\frac{2+\left(p-4\right)\left(p+1\right)x^p+
\left(1+p\right)\left(2+p\right)x^{2p}}
{x^2\left(1-x^p\right)^2\left[\left(1+p\right)x^p-1\right]}\,.
\end{equation}
These functions are displayed in \Fig{potbi}. They become very small
at large fields $x\gg 1$, and diverge when the potential vanishes at
$x\rightarrow 1$. Therefore inflation can naturally end with slow-roll
violation at a field value $\xend$, solution of $\epsilon_1(\xend)=1$,
\ie, verifying
\begin{equation}
\xend^{p+1} -  \xend = \dfrac{p}{\sqrt{2}} \dfrac{\Mp}{\mu}\,.
\end{equation}
Unless $p$ takes integer values, this equation has to be solved
numerically (see also \sectionc{sec:sfi}).

However, in the limits
$\mu/\Mp\ll 1$ and $\mu/\Mp\gg 1$ we can find an approximate
expression for $\xend$. Solving perturbatively the equation
$\epsilon_1=1$, one obtains
\begin{equation}
\xend\underset{\mu\ll \Mp}{\simeq}
\left(\frac{p\Mp}{\sqrt{2}\mu}\right)^{\frac{1}{p+1}}
+\frac{1}{p+1}\left(\frac{p\Mp}{\sqrt{2}\mu}\right)^{\frac{1-p}{1+p}}, \qquad
\xend\underset{\mu \gg \Mp}{\simeq} 1
+\frac{1}{\sqrt{2}}\frac{\Mp}{\mu}-\frac{p+1}{4}\frac{\Mp^2}{\mu^2}\,.
\label{eq:bi:xend}
\end{equation}
It is also interesting to find the solution of $\epsilon_2=1$. As
before, this cannot be done exactly but, perturbatively, one obtains
\begin{equation}
\xepstwoOne \underset{\mu\ll \Mp}{\simeq}
\left[2p(1+p)\left(\frac{\Mp}{\mu}\right)^2\right]^{\frac{1}{p+2}}, \qquad
\xepstwoOne \underset{\mu \gg \Mp}{\simeq}1
+\sqrt{2}\frac{\Mp}{\mu} \,.
\label{eq:bi:xeps2one}
\end{equation}
From the above expressions, we deduce that slow-roll violation always
occurs before the end of inflation, that is to say $\epsilon_2$
becomes unity before $\epsilon_1$. This has not effect on the
observable predictions since only a few \efolds of inflation are spent
in this regime (see \Fig{potbi}).

The slow-roll trajectory can be integrated explicitly from
\Eq{eq:srtrajectory} and one obtains
\begin{equation}
\begin{aligned}
\label{eq:bi:traj}
\Nend-N & = \frac{\mu^2}{2p\Mp^2}\left(\xend^2
  -\frac{2}{p+2} \xend^{p+2} - x^2 +\frac{2}{p+2} x^{p+2} \right),
\end{aligned}
\end{equation}
an expression which cannot be inverted in general. However, in the
$\mu \ll \Mp$ and $\mu \gg \Mp$ limits, one has $x\gg1$ and $x \simeq
1$ respectively and the previous equation can be approximately
inverted leading to the following expressions
\begin{equation}
\xstar \underset{\mu \ll \Mp}{\simeq} \left[p(p+2) \frac{\Mp^2}{\mu^2}
  \Delta\Nstar+ \xend^{p+2} \right]^{\frac{1}{p+2}}, \qquad \xstar
\underset{\mu \gg \Mp}{\simeq} 1 + \dfrac{\Mp}{\mu} \sqrt{\dfrac{1}{2} 
+ 2 \Delta\Nstar}\,,
\label{eq:bi:xstar}
\end{equation}
where use has been made of \Eq{eq:bi:xend}. Also, making use of the full
KKLT potential \eqref{eq:kklti:pot}, the slow roll trajectory reads
\begin{equation}
\begin{aligned}
\Nend-N & = \frac{\mu^2}{2p\Mp^2}\left(-\xend^2
  -\frac{2}{p+2} \xend^{p+2} + x^2 +\frac{2}{p+2} x^{p+2} \right),
\end{aligned}
\end{equation}
which coincides with \eqref{eq:bi:traj} in the limit $\mu\ll\Mp$.

The mass scale $M$ is given by the CMB normalization and verifies
\begin{equation}
  \left(\frac{M}{\Mp}\right)^4 = 720\pi^2p^2
  \left(\frac{\Mp}{\mu}\right)^2 \dfrac{\xstar^{p-2}}{\left(\xstar^{p}
    - 1\right)^{3}} \frac{\Qrms^2}{T^2}\,.
\label{eq:bi:cmb}
\end{equation}
which can be further simplified in the appropriate limits using
\Eqs{eq:bi:xend} and \eqref{eq:bi:xstar}.

The reheating consistent slow-roll predictions for the
phenomenological models are displayed in \Figs{fig:CMBBI2},
\ref{fig:CMBBI3}, \ref{fig:CMBBI4} for $p=2$, $p=3$ and $p=4$,
respectively, and with $\mu/\Mp\in\left[10^{-3},10^3\right]$.  The
reheating equation of state parameter $\wrehbar=0$ but since the shape
of the potential is unknown at $x<1$, this parameter is a priori
unspecified and could take different values. For small values of
$\mu$, we see that $\nS\simeq 0.96$ and $r \ll 1$. In the opposite
case, $\mu \gg \Mp$, the model predictions lie around
$\epsilon_2\simeq 4\epsilon_1$ with $\nS\simeq 0.97$ and $r\simeq
0.08$. These behaviors can be recovered by plugging the approximated
expressions given in \Eqs{eq:bi:xend} and \eqref{eq:bi:xstar} into the
Hubble flow functions. For $\mu \ll \Mp$, one obtains
\begin{equation}
\begin{aligned}
\epsilon_{1*} \simeq\frac{p^2}{2}\left[p\left(p+2\right)
\Delta\Nstar\right]^{-\frac{2p+2}{p+2}}\left(\frac{\mu}{\Mp}\right)^\frac{2p}{p+2}\, ,\qquad
\epsilon_{2*} \simeq\frac{2}{\Delta \Nstar}\frac{p+1}{p+2}, \qquad
\epsilon_{3*} \simeq\frac{1}{\Delta \Nstar}\,,
\end{aligned}
\label{eq:bi:epssmall}
\end{equation}
and the spectral index is of the order $\nS \simeq
1-2/\Delta\Nstar(p+1)/(p+2)\sim 0.96$ with $r\ll 1$. Similarly, for
$\mu \gg \Mp$ limit, the Hubble flow parameters at Hubble crossing
behave as
\begin{equation}
\epsilon_{1*}\simeq\frac{1}{4\Delta \Nstar}, \qquad
\epsilon_{2*}\simeq\frac{1}{\Delta \Nstar}, \qquad
\epsilon_{3*}\simeq\frac{1}{\Delta \Nstar}.
\end{equation}
Therefore, the predicted level of gravity waves is now of the order
$r\simeq 4/\Delta\Nstar\simeq 0.08$ and the spectral index is
$\nS\simeq 1-3/(2\Delta\Nstar)\simeq 0.97$, which is again in
agreement with the numerical results.

Finally, the predictions for the KKLTI models, \ie using the full
potential \eqref{eq:kklti:pot}, are displayed in \Figs{fig:CMBKKLTI2},
\ref{fig:CMBKKLTI3}, \ref{fig:CMBKKLTI4} for the same parameters. One
can see that they deviate from the ones of brane inflation only when
$\mu\gg\Mp$.

\subsubsection{Slow-Roll Analysis of the Stringy Scenario}

\begin{figure}
\begin{center}
\includegraphics[width=\wsingfig]{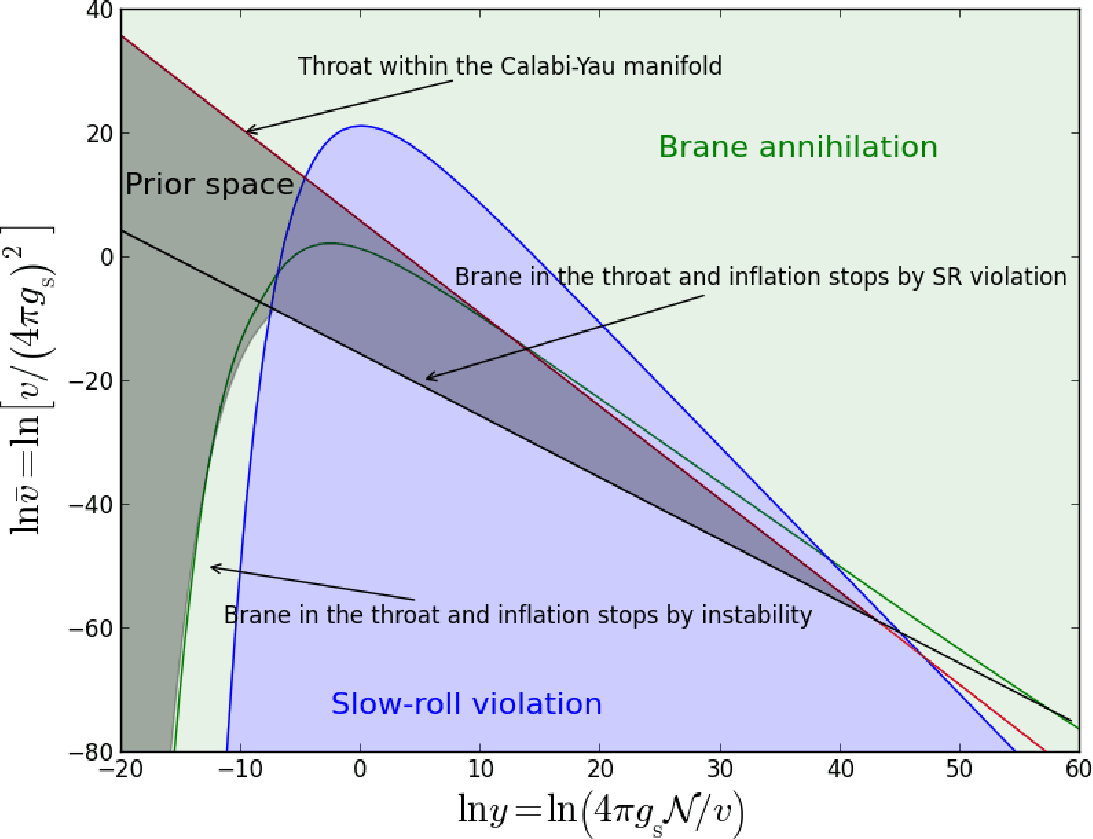}
\caption{Theoretical prior space for the stringy scenario of brane
  inflation~\cite{Lorenz:2007ze} in the plane of the ``universal''
  coordinates $(y,\vbar)$. The solid blue line is the frontier above
  which inflation ends by tachyonic pre-heating triggered by brane
  annihilation (light green region). Only in the region enclosed by
  this curve (light blue region), inflation ends by slow-roll
  violation. The upper thick red line is the volume bound of
  \Eq{eq:bi:volume}. The lower black straight line is the ``UV'' limit
  given by \eqref{eq:bi:uveps} and is relevant only if inflation stops
  by slow-roll violation. The solid green curve is given by
  \eqref{eq:bi:uvstg} and also represents the ``UV'' limit but, this
  time, in the regime where inflation stops when the two branes
  collide. As a consequence, the admissible region is the one shaded
  in light black. We see that, even in this allowed region, inflation
  can either end by tachyonic instability or slow-roll violation
  depending on the string parameter values. In principle, the blue,
  black and green lines should cross at a single point. Due to the
  approximations used here, we see that this is true only
  approximately. In order to give a more faithful description of the
  allowed region, the light black area has been slightly deformed
  around the crossing point (see Ref.~\cite{Lorenz:2007ze} for an
  exact determination of these frontiers).}
\label{fig:bi:stgbounds}
\end{center}
\end{figure}

In the case where the model is interpreted as a stringy scenario, with
$p=4$, we have seen before that the low energy description is valid
provided $r>\rstg$, or $x>\xstg$ with
\begin{equation}
\xstg \equiv \dfrac{\sqrt{T_3} \, \rstg}{\mu} =  \calN^{1/4}\exp
\left[\left(4\pi \gstrings \frac{\calN}{v}\right)^{-1/4}\right].
\label{eq:bi:xstg}
\end{equation}
If slow-roll violation occurs before the system reaches $\xstg$, then
the effective string description is always valid and the observable
predictions will be exactly the same as those derived in the previous
paragraph (for $p=4$ and $\mu \ll \Mp$). However, if, on the contrary,
slow-roll violation occurs after the field crosses the value $\xstg$,
then inflation stops by instability at $\xstg$ instead of the naively
expected $\xend$. Indeed, in this case, a tachyon appears and triggers
the process of branes annihilation. Therefore, the mechanism ending
inflation in this model depends on whether slow-roll violation occurs
in a regime where the distance between the branes is larger or smaller
than the string length. And this question depends on the value of the
parameters characterizing BI. One can determine the two regimes by
evaluating the ratio
\begin{equation}
\label{eq:ratiox}
\frac{\xepstwoOne}{\xstg} =
40^{1/6}\left(\frac{M}{\Mp}\right)^{-1/3} \calN
^{-1/4}(4 \pi^2 v)^{1/12} \exp \left[-\left(4\pi \gstrings
  \frac{\calN}{v}\right)^{-1/4}\right],
\end{equation}
in which we have used \Eqs{eq:bi:stgparams}, \eqref{eq:bi:xeps2one}
and \eqref{eq:bi:xstg} (with $p=4$ and $\mu \ll \Mp$). If this ratio
is larger than one, inflation stops by slow-roll violation and if it
is smaller than one by instability. The complicated part of the
analysis lies in the fact that the above equation depends on the mass
scale $M$. In order to have an explicit expression of $M$ in terms of
the parameters of the model, one must first CMB normalize the model
which, in turn, requires the knowledge of the mechanism ending
inflation. However, we are interested in calculating the frontier
where $\xepstwoOne=\xstg$ and, therefore, the two possible mechanisms
for stopping inflation coincide in that case. Replacing $\xend$ by
$\xstg=\xepstwoOne$ in \Eq{eq:bi:xstar} yields
\begin{equation}
\xstar^\uf\simeq \left[24 \frac{\Mp^2}{\mu^2}
\left(\Delta\Nstar+\frac{5}{3}\right)\right]^{1/6},
\end{equation}
from which one can obtain an explicit formula for the first
slow-roll coefficient~(\ref{eq:epsbi}) at Hubble radius crossing
\begin{equation}
\begin{aligned}
\epsilon_{1*}^\uf &\simeq 8 \left[24 \left(\Delta\Nstar+\frac{5}{3}\right)
\right]^{-5/3}\left(\frac{\mu}{\Mp}\right)^{4/3}.
\end{aligned}
\end{equation}
Comparing this expression to \Eq{eq:bi:epssmall}, we see that there is
a very small shift by $5/3$ in $\Delta\Nstar$. It accounts for the
difference of \efolds between the time at which slow-roll violations
occur, \ie for $x=\xepstwoOne$, and the end of inflation at
$\xend$. As argued before, we see that these effects are too small to
be observable and completely degenerated with the reheating
duration. Plugging this expression into the CMB normalization, and
using the relation $M^4=4\pi^2v\mu^4$, one arrives at the following
expression for $M$
\begin{equation}
\frac{M}{\Mp}=C(4\pi^2v)^{-1/8}\left(\Delta\Nstar+\frac{5}{3}\right)^{-5/8},
\label{eq:bi:Meps}
\end{equation}
where we have defined
\begin{equation}
C \equiv 3^{-5/8}(8\pi^2 \Qstar)^{3/8}, \qquad \Qstar \equiv 45
\dfrac{\Qrms^2}{T^2} = 2700 \Pstar.
\end{equation}
We can now insert this expression of $M$ in \Eq{eq:ratiox} to get the
equation defining the frontier in the string parameter space, namely
\begin{equation} 
\label{eq:ratioxfinal}
\left. \frac{\xepstwoOne}{\xstg}\right|_\uf = 1 =
\left(\dfrac{40}{C^2}\right)^{1/6} \left(\Delta\Nstar+\dfrac{5}{3}\right)^{5/24}
(4\pi^2v)^{1/8}\calN^{-1/4} \exp \left[-\left(4\pi \gstrings
  \frac{\calN}{v}\right)^{-1/4}\right].
\end{equation}
Following \Refc{Lorenz:2007ze}, if one defines the two following
rescaled stringy parameters
\begin{equation}
y\equiv 4\pi\gstrings \frac{\calN}{v}, \qquad 
\vbar \equiv \frac{v}{(4\pi \gstrings)^2}, 
\label{eq:bi:univparams}
\end{equation}
then the frontier~(\ref{eq:ratioxfinal}) is defined by the following
``universal'' form
\begin{equation}
y^{1/4}\ee^{y^{-1/4}}\vbar^{1/8}  - \left(\dfrac{40}{C^2}\right)^{1/6}
\left(\Delta\Nstar+\dfrac{5}{3}\right)^{5/24}\left(4\pi^2\right)^{1/8} = 0,
\label{eq:bi:frontiere}
\end{equation}
which is independent of the string coupling $\gstrings$. As
represented in \Fig{fig:bi:stgbounds}, in the plane $(y,\vbar)$, this
relation is a curve that separates the region where inflation stops by
slow-roll violation (below the curve) and the region where inflation
stops by instability due to brane annihilation (above the curve).

The requirement of having the throat contained within the Calabi-Yau
manifold can equally be written in terms of the universal
variables. From \Eqs{eq:conditionuv} and \eqref{eq:bi:univparams}, one
gets
\begin{equation}
y^{3/2} \vbar  < 8 \pi^2 \Mp^2 \ells^2,
\label{eq:bi:volume}
\end{equation}
which therefore depends on the string length $\ells=\sqrt{\alpha'}$
but not on the string coupling $\gstrings$.

Finally, the last theoretical prior comes from requiring that the
brane motion remains located inside the throat, \ie $x < \xUV$ with
\begin{equation}
\xUV \equiv \dfrac{\sqrt{T_3} \rUV}{\mu} = \dfrac{\Mp}{M}
\left( \dfrac{\calN}{4 \pi^3 \alpha'^2 \gstrings} \right)^{1/4}.
\end{equation}
Since during inflation $x$ decreases, this condition gives an upper
limit on the admissible initial field values. However, the initial
field values depends on the \emph{total} number of \efolds of
inflation, say $\Delta\Ntot$, and on the field value at which
inflation ends, \ie either $\xstg$ or $\xepstwoOne$ depending on if
brane annihilation occurs before slow-roll violations.

Let us first assume that brane annihilation occurs well after the end
of inflation, \ie we are in lower part of the string parameter space
$(y,\vbar)$ separated by \Eq{eq:bi:frontiere}. For the relevant limit,
$\mu \ll \Mp$, the initial field value is given by
\begin{equation}
\xiniepstwo \simeq \left[24 \dfrac{\Mp^2}{\mu^2}\left(\Delta\Ntot +
  \dfrac{5}{3}\right) \right]^{1/6}.
\label{eq:bi:xinieps2}
\end{equation}
This expression involves $\mu$ and therefore $M$ through
\Eq{eq:bi:stgparams}. Again, one has to determine $M$ using the CMB
normalization and we are assuming that inflation ends at $\xepstwoOne$, \ie
exactly \Eq{eq:bi:Meps}. Plugging everything together and making use
of the universal variables, one gets
\begin{equation}
y \vbar \underset{\xstg < \xepstwoOne}{>} C^{8/3} \pi^2 \Mp^2 \ells^4 \left[24
  \left(\Delta\Ntot+\frac{5}{3}\right) \right]^{2/3}
\left(\Delta\Nstar + \dfrac{5}{3} \right)^{-5/3}.
\label{eq:bi:uveps}
\end{equation}

If inflation ends by brane annihilation at $x=\xstg$, \ie the string
parameters $(y,\vbar)$ lie above the curve given by \Eq{eq:bi:Meps},
then $\xini$ and $\xstar$ are accordingly modified. For $\mu \ll \Mp$,
their new expressions are however still given by \Eq{eq:bi:xstar}, up
to the replacement $\xend \rightarrow \xstg$, \ie
\begin{equation}
\xinistg \simeq \left(24 \dfrac{\Mp^2}{\mu^2} \Delta\Ntot +
  \xstg^6\right)^{1/6}, \qquad \xstarstg \simeq \left(24
  \dfrac{\Mp^2}{\mu^2} \Delta\Nstar + \xstg^6\right)^{1/6}.
\label{eq:bi:xinistarstg}
\end{equation}
As before, $\xinistg$ and $\xstarstg$ depend on $\mu$ and therefore on
$M$, which is determined by the CMB normalization. However, since
inflation now ends by tachyonic instability this one has to be
re-determined by plugging $\xstarstg$ into \Eq{eq:bi:cmb}. Doing so
gives an implicit expression for $M$
\begin{equation}
\dfrac{M}{\Mp} \simeq C(4 \pi^2 v)^{-1/8} \left(\Delta\Nstar +
\dfrac{\mu^2}{\Mp^2} \dfrac{\xstg^6}{24} \right)^{-5/8} =
C(4 \pi^2 v)^{-1/8} \left[\Delta\Nstar + \dfrac{5}{3}
  \left(\dfrac{\xstg}{\xepstwoOne}\right)^6 \right]^{-5/8},
\end{equation}
where use has been made of \Eq{eq:bi:xeps2one}, for $\mu \ll
\Mp$. This equation cannot be analytically solved for $M$ because
$\mu$, and $\xepstwoOne$, depends on $M$. However, if brane
annihilation occurs well before slow-roll violation, one has $\xstg
\gg \xepstwoOne$ such that the term in $\Delta\Nstar$ can be
neglected. In that situation, from $\mu^4=M^4/(4 \pi^2 v)$, one gets
the approximate expression
\begin{equation}
\dfrac{M}{\Mp} \underset{\xstg \gg \xepstwoOne}{\simeq} 24^{5/18}
C^{4/9} (4 \pi^2 v)^{1/12} \xstg^{-5/3}.
\label{eq:bi:cmbstg}
\end{equation}
Requiring $\xinistg < \xUV$ finally yields
\begin{equation}
y^{19/6} \vbar^{7/3} \exp{\left(\frac{20}{3} y^{-1/4}\right)}
\underset{\xstg \gg \xepstwoOne}{>} \left(8 \pi^2 \ells^2 \right)^3
\Qstar \left[y^{2/3} \vbar^{1/3} \exp{\left(\frac{8}{3}
    y^{-1/4}\right)} + \dfrac{6 \Delta\Ntot}{\Qstar^{1/3}} \right],
\label{eq:bi:uvstg}
\end{equation}
which completes the bounds coming from $\xUV$.

Brane inflation within the string scenario has therefore a rather
involved set of priors. In addition to have $p=4$ and $\mu \ll \Mp$,
the model parameters should simultaneously verify \Eq{eq:bi:volume}
and either \Eq{eq:bi:uveps}, or \Eq{eq:bi:uvstg}, according to the
sign of the left hand side of \Eq{eq:bi:frontiere}. All these
equations involve the amplitude of the CMB anisotropies, which is well
measured, the total number of \efolds $\Delta\Ntot$, which is an
unknown quantity, and the number of \efolds $\Delta\Nstar$ before the
end of inflation at which the pivot mode crossed the Hubble radius. As
discussed in \sectionc{subsec:reheating}, $\Delta\Nstar$ can only be
obtained by solving \Eq{eq:dnstarlnrad}, \ie after having specified
the reheating parameter. As the result, the reheating slow-roll
predictions for the string scenario can only be sorted out
numerically, paying attention that for a given reheating history, all
of the previous theoretical constraints are satisfied. As an
illustration, we have plotted in \Fig{fig:bi:stgbounds} the bounds for
the typical values $\Delta\Nstar=50$ and $\Delta\Ntot=60$ with
$\alpha' \Mp^2 \simeq 1/4$~\cite{Kachru:2003sx, Lorenz:2007ze}.

The reheating consistent slow-roll predictions for the
string models are displayed in \Figs{fig:CMBBIstg} for a
set of realistic fundamental parameters. Also, making
use of the full potential \eqref{eq:kklti:pot}, the predictions
of the corresponding KKLT inflation models are displayed in
\Figs{fig:CMBKKLTIstg}. One can check that they match perfectly.

\subsection{String Axion Inflation I (SAII)}
\label{sec:saii}

The model emerges from geometrical compactifications on Calibi-Yau
manifold, in presence of fluxes, and in the framework of type IIB
superstring theory. It has been proposed in
\Refc{Kobayashi:2015aaa} and shares some similarities with the
KKLT construction of \sectionc{sec:bi}. However, here, the inflaton
is identified with an axion of the complex structure moduli while its
potential comes from worldsheet instanton effects~\cite{Gukov:1999ya}
that have been derived in \Refc{Kobayashi:2015aaa}. The potential
reads
\begin{equation}
V(\phi) = M^4 \left[ 1 - \cos\left(\dfrac{\phi}{\mu}\right) + \alpha
  \dfrac{\phi}{\mu} \sin\left(\dfrac{\phi}{\mu}\right) \right],
\label{eq:pot:saii}
\end{equation}
where $\mu$ is a {\vev} and $\alpha$ a dimensionless constant that is
not required to be small. The first two terms in \Eq{eq:pot:saii}
match, up to a field redefinition, the potential of Natural Inflation
(NI) in \sectionc{sec:ni}. Therefore, the potential of SAII can be
viewed as a modulated addition to NI, a situation also discussed in
\Refc{Czerny:2014wua}. Let us stress, however, that, depending on the
values of $\alpha$, the predictions of SAII can be quite different
from the ones of NI. \Refc{Kobayashi:2015aaa} also considers
higher-order terms in the instanton effects and, under some
assumptions, these ones can generate an additional mass term in the
potential. This case corresponds to the model String Axion Inflation
II (SAIII), which is analyzed in \sectionc{sec:saiii}.

\begin{figure}
\begin{center}
\includegraphics[width=\wdblefig]{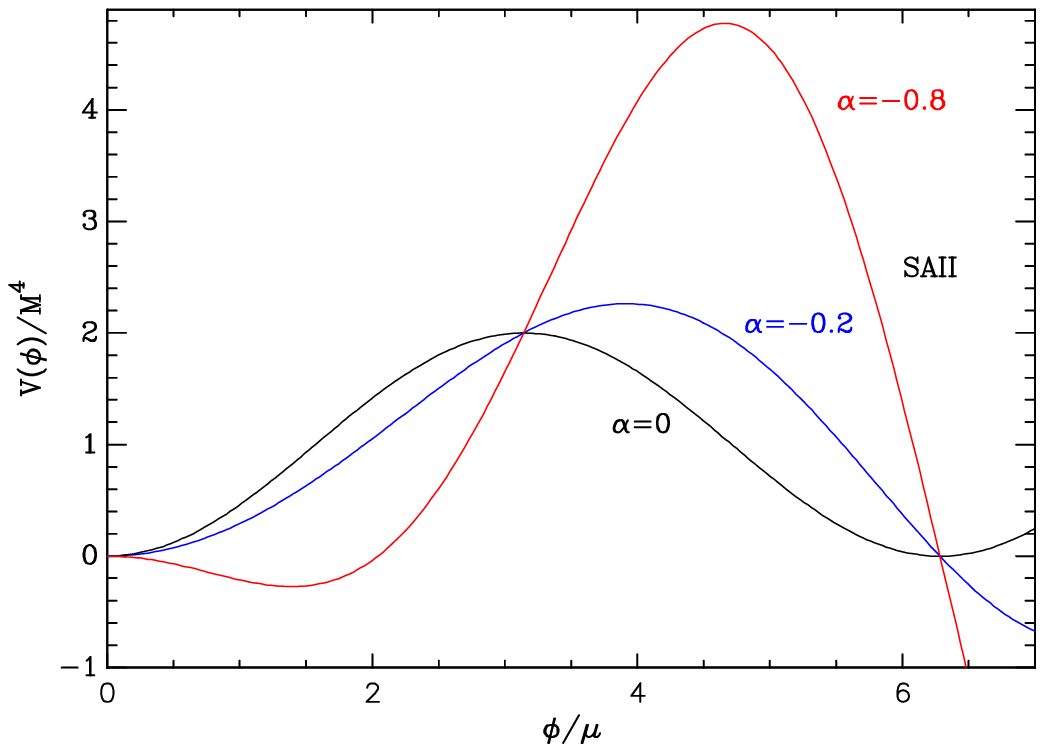}
\includegraphics[width=\wdblefig]{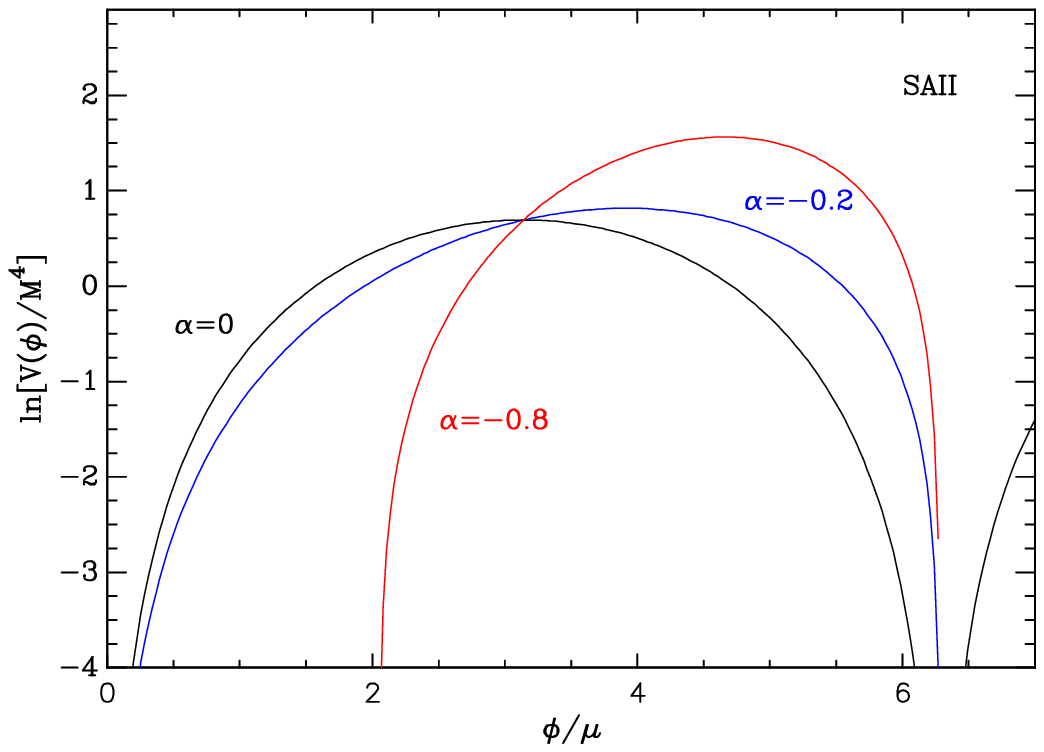}
\includegraphics[width=\wdblefig]{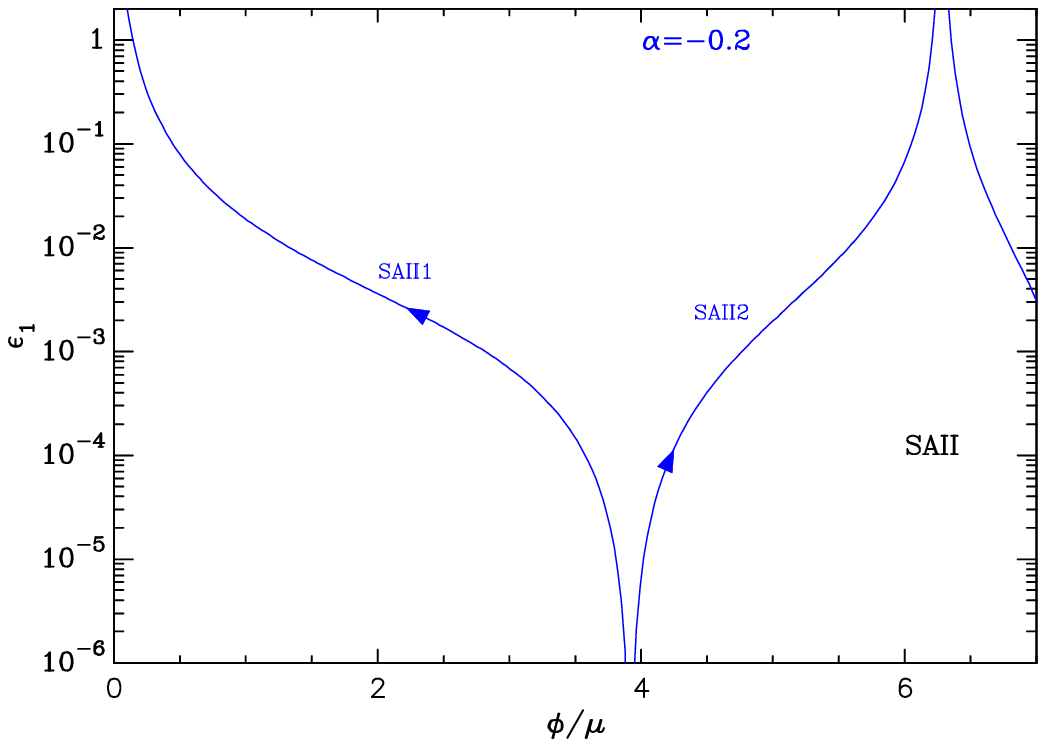}
\includegraphics[width=\wdblefig]{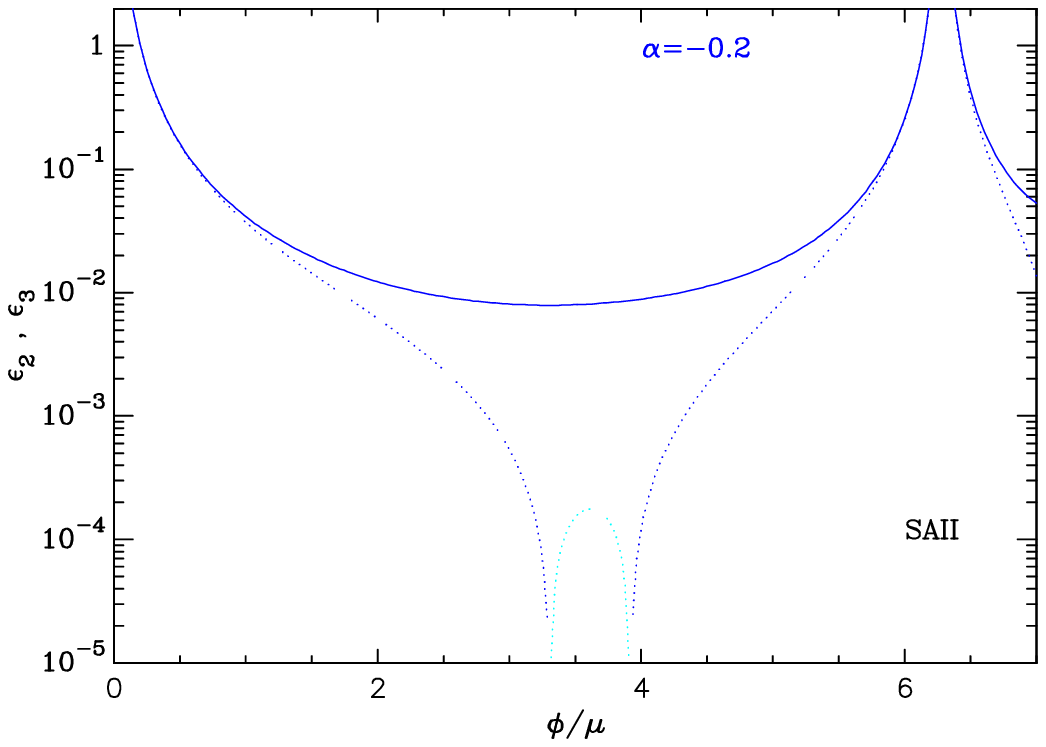}
\caption{String Axion Inflation I (SAII) for $\alpha=-0.2$ (blue
  curve) and $\alpha=-0.8$ (red curve). Top panels: the potential and
  its logarithm, for comparison the potential of Natural Inflation
  ($\alpha=0$, black curve) is represented. Bottom left panel: slow-roll parameter
  $\epsilon_1$ for $\alpha=-0.2$ and $\mu = 10\,\Mp$, with the
  different inflationary regimes of SAII annotated with an arrow
  indicating the direction to which the field evolves. Bottom right
  panel: slow-roll parameters $\epsilon_2$ (solid line) and
  $\epsilon_3$ (dotted line) for the same parameters value. When
  $\epsilon_3$ becomes negative, the plot shows $|\epsilon_3|$ as a
  cyan dotted line, the blue dotted line corresponds to positive
  values.}
\label{fig:potsaii}
\end{center}
\end{figure}

The potential in \Eq{eq:pot:saii} depends on two parameters,
$\mu$ and $\alpha$, that can take any value. It is symmetric with
respect to $\phi=0$, and we can therefore restrict the analysis to the $\phi\geq 0$ region. 
As soon as $\alpha$ is non-vanishing, the potential
becomes negative in some regions, and for $\alpha <
-1/2$, this occurs around the origin (see below). As a result,
slow-roll inflation can take place only within some limited field
range, that depends on $\alpha$, and around the maxima of the
potential. The potential and its logarithm are displayed in the top
panels of \Fig{fig:potsaii}.

Defining $x \equiv \phi/\mu$, the smallest local maximum of the
potential, denoted $x=\xVmax$, is a solution of
\begin{equation}
(1+ \alpha) \sin(x)  + \alpha x \cos (x) = 0.
\label{eq:saii:xvmax}
\end{equation}
This is a transcendental equation, which has to be solved numerically
for each value of $\alpha$. Here, we are interested in the smallest
positive solution of this equation for which $V(\xVmax) > 0$.
Expanding \Eq{eq:pot:saii} around the origin, one gets
\begin{equation}
\dfrac{V(x)}{M^4} = \left(\alpha+\dfrac{1}{2}\right) x^2 + \order{x^4},
\label{eq:saii:vzero}
\end{equation}
which implies that, for $\alpha > -1/2$, the potential is a positive
increasing function of $x$ in a neirboorhood of $x=0$, up to its first local
maximum at $x=\xVmax$. Therefore, inflation can take place within the
domain $x\in[0,\xVmax]$, at decreasing field values, and this regime
will be referred to as SAII1. The potential is minimal at the origin,
with $V(x=0)=0$, such that reheating after inflation ends up in a
Minkowski vacuum.
For $\alpha < -1/2$, \Eq{eq:saii:vzero} shows that the
potential is decreasing towards a negative minimum around the origin,
and then becomes positive to reach its first local maximum at $x=\xVmax$. In that
situation, the SAII1 inflationary domain lies within
$x\in[\xVzeroMinus,\xVmax]$, where $\xVzeroMinus$ is the smallest
strictly positive solution of $V(x)=0$, \ie
\begin{equation}
1 - \cos(x) + \alpha x \sin(x) = 0.
\label{eq:saii:xvzero}
\end{equation}
This equation is again transcendantal and has to be solved
numerically. However, there are some trivial solutions, namely $x=2\pi
n$ where $n$ is an integer number. Unfortunately, for $\alpha < -1/2$,
$\xVzeroMinus < 2\pi$ does not belong to this subset of solutions. Let
us also notice that, in this situation, the reheating would proceed
after inflation around an anti-de Sitter minimum, which should thus be
lifted somehow. If one is ready to accept to rely on such a mechanism,
then one should also consider the inflating regime at $x > \xVmax$,
where inflation proceeds at increasing field values and for which
reheating also occurs around an anti-de Sitter minimum. This regime
will be referred to as SAII2, see \Fig{fig:potsaii}. Strictly
speaking, there are an infinite numbers of negative minima for the
potential at larger values of $x$, but the value of $V$ at the minimum
becomes negatively larger for each of them. We will be therefore
ignore these in the following.

\begin{figure}
\begin{center}
\includegraphics[width=\wsingfig]{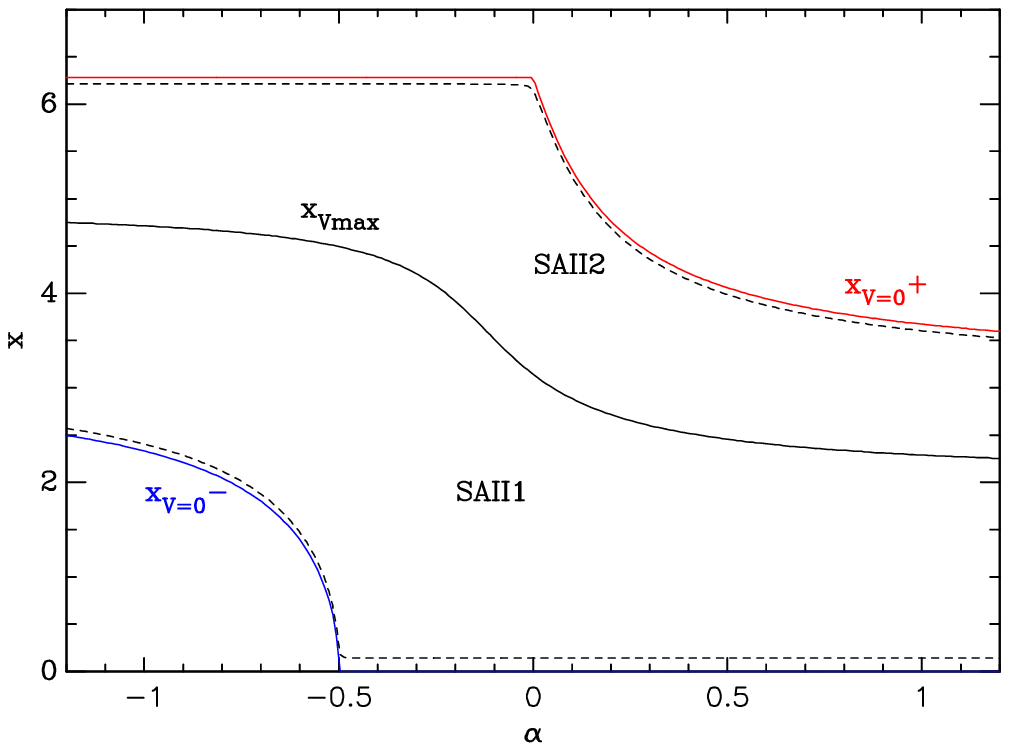}
\caption{Field domains in which SAII1 and SAII2 are defined. The black
  central curve shows the field value $\xVmax$ (in unit of $\mu$) at
  which the potential is maximal. SAII1 takes place at decreasing
  field value from $\xVmax$ while SAII2 inflates from $\xVmax$ at
  increasing field values. The curves $\xVzeroMinus(\alpha)>0$ and
  $\xVzeroPlus(\alpha)$ are the separatrix under which (and,
  respectively, above which) the potential becomes negative. In both
  regimes, inflation gracefully ends before the separatrix is reached
  at $x=\xend$, where $\xend$ is the relevant solution of
  $\epsilon_1(x)=1$. The dashed lines show the value of
  $\xend(\alpha$) for SAII1 (lower dashed curve) and SAII2 (top dashed
  curve) when $\mu=10\,\Mp$.}
\label{fig:xvsaii}
\end{center}
\end{figure}

For all values of $\alpha$, SAII2 occurs in the range
$x\in[\xVmax,\xVzeroPlus]$ where $\xVzeroPlus$ is the smallest
solution of \Eq{eq:saii:xvzero} satisfying $\xVzeroPlus >
\xVmax$. This time, if $\alpha \le 0$, one has $\xVzeroPlus = 2\pi$,
but for $\alpha > 0$, \Eq{eq:saii:xvzero} has to be solved
numerically in order to determine $\xVzeroPlus$. The situation is summarized in
\Fig{fig:xvsaii} where we have plotted $\xVmax(\alpha)$,
$\xVzeroMinus(\alpha)$ and $\xVzeroPlus(\alpha)$.

The first two slow-roll parameters read
\begin{equation}
  \begin{aligned}
    \epsilon_1 & = \dfrac{1}{2 \mu^2} \left[\dfrac{(1+\alpha) \sin(x) +
        \alpha x \cos (x)}{1-\cos (x) + \alpha x \sin (x)}\right]^2,
    \\ \epsilon_2 & = \dfrac{1}{\mu^2} \dfrac{2 + 2 \alpha x \sin(x) -
      2(1+2\alpha)\cos(x) - \alpha^2 \cos(2x) + \alpha(4+\alpha+2\alpha
      x^2)}{\left[1-\cos (x) + \alpha x \sin (x)\right]^2}\,,
    \label{eq:saiieps12}
  \end{aligned}
\end{equation}
while the third one is given by
\begin{equation}
  \begin{aligned}
    \epsilon_3 & = - \dfrac{1}{\mu^2 [1 - \cos(x) + \alpha x \sin(x)]^2}
    \\ & \times \dfrac{(\alpha +1)\sin(x) + \alpha x \cos(x)}{\alpha ^2+4
      \alpha +2 \alpha ^2 x^2-\alpha ^2 \cos (2 x)+2 \alpha x \sin(x)-2 (2
      \alpha +1) \cos(x) + 2 } \\ & \times \Big[-4 \alpha ^2 x^2
      \sin(x)-\alpha ^2 x^2 \sin(2 x) - 2 \alpha x \left(6 \alpha +2
      \alpha ^2 x^2+1\right) \cos(x) - 3 \alpha^3 \sin(x) \\
      & + \alpha ^3
      \sin(3 x) + 9 \alpha^2 x - 12 \alpha ^2 \sin(x) + 6 \alpha ^2 \sin(2
      x) - 6 \alpha \sin(x) + 3 \alpha \sin(2 x) \\ & + (3 \alpha + 2) \alpha x
      \cos(2 x) - 2 \sin(x) + \sin(2 x)\Big].
    \label{eq:saiieps3}
  \end{aligned}
\end{equation}
The denominator of $\epsilon_1(x)$ in \Eq{eq:saiieps12}
diverges for $x \to 0$, $x \to \xVzeroMinus$ and $x
\to \xVzeroPlus$, which ensures that inflation gracefully ends
before the potential becomes negative. Moreover, all three
slow-roll parameters feature an overall scaling $\propto \mu^{-2}$. This implies that, for any value of
$\alpha$, at fixed $x$, the slow-roll parameters can be made arbitrarily small by
choosing large values of $\mu$. Conversely, for small values of
$\mu$, one expects inflation to be of the hilltop kind and confined
around $\xVmax$. The three slow-roll parameters have been plotted for
$\alpha=-0.2$ in the bottom panels of \Fig{fig:potsaii}.

Both SAII1 and SAII2 inflationary regimes end for
$\epsilon_1(\xend)=1$ at a field value $\xend$ that is solution of
\begin{equation}
(1+\alpha) \sin(x) + \alpha x \cos(x) = \pm \mu \sqrt{2} \left[1 -
    \cos(x) + \alpha x \sin(x) \right]\, ,
\label{eq:saii:xend}
\end{equation}
which needs to be solved in the range $\xend \in [\xVzeroMinus,\xVmax]$ for SAII1
and $\xend \in [\xVmax,\xVzeroPlus]$ for SAII2. Again,
\Eq{eq:saii:xend} being transcendental, it has to be solved
numerically and the solutions have been represented as dashed curves
in \Fig{fig:xvsaii}.

The slow-roll trajectory also requires a numerical integration and
reads
\begin{equation}
\Nend - N =\mu^2 \int_{\xend}^x \dfrac{1 - \cos(y) + \alpha y
  \sin(y)}{(1+\alpha) \sin(y) + \alpha y \cos(y)}\,\ud y,
\end{equation}
with $\xend$ the relevant solution of \Eq{eq:saii:xend},
depending on the regime under consideration. This expression diverges
for $x\to\xVmax$ and the total number of \efolds is therefore unbounded. 

Finally, the normalization of the potential $M^4$ can be obtained from
the amplitude of the CMB anisotropies once one has determined the
field value $\xstar$ at which the pivot mode crossed the Hubble
radius. It reads
\begin{equation}
\left(\dfrac{M}{\Mp}\right)^4 = \dfrac{720 \pi^2}{\mu^2}
  \dfrac{\left[(1+\alpha) \sin(\xstar) + \alpha \xstar \cos(\xstar)
      \right]^2}{\left[1 - \cos(\xstar) + \alpha \xstar
      \sin(\xstar)\right]^3}\,.
\end{equation}
The reheating consistent slow-roll predictions for SAII1 and SAII2 are
represented in \Figs{fig:CMBSAII1} to \ref{fig:CMBSAII2_2} for
various values of $\alpha$ and $\mu$.

\subsection{Mukhanov Inflation (VFMI)}
\label{sec:vfmi}

This model has been proposed by Mukhanov in
\Refc{Mukhanov:2013tua} and relies on a hydrodynamical
representation of the inflationary background evolution. Instead of
specifying a potential, one postulates the form of the equation-of-state parameter
$w$ as a function of the number of {\efolds} elapsed before the end of inflation
$\Nend-N$. VFMI is defined by
\begin{equation}
W(N) \equiv 1+w(N) \equiv \dfrac{\beta}{\left(C + \Nend - N \right)^\alpha}\,,
\label{eq:vfmiw}
\end{equation}
where $\alpha>0$ and $\beta>0$ are the model parameters and $C$ is a constant that has been
set to unity in \Refc{Mukhanov:2013tua}. Because inflation
ends at $N=\Nend$ and, by definition, when the universe stops
accelerating, i.e., for $w(0)=-1/3$, one has in fact
\begin{equation}
\label{eq:vfmi:C:beta}
C = \left(\dfrac{3 \beta}{2} \right)^{1/\alpha}.
\end{equation}

At the perturbative level, cosmological fluctuations are still assumed
to be of quantum-mechanical origin, adiabatic, and conserved on super-Hubble
scales, \ie coupled with a single scalar field.

\subsubsection{Equivalence with a scalar field theory}
\label{eq:vfmi}
On general grounds, giving the functional form of $W(
N)=1+w( N)$ is strictly equivalent to specifying a parametric
potential for a canonically normalized scalar field $\phi$. This can
be seen from the hydrodynamical Friedmann-Lema\^itre equations
\begin{equation}
H^2=\dfrac{\rho}{3\Mp^2}, \qquad \dfrac{\ddot{a}}{a} = -\dfrac{\rho}{6
  \Mp^2} \left(1+ 3 \dfrac{P}{\rho} \right).
\end{equation}
From $P=w \rho$, and in terms of the number of {\efolds} $N$, the second
equation reads
\begin{equation}
H \dfrac{\ud H}{\ud N} + H^2 =- \dfrac{\rho}{6 \Mp^2} (1 + 3 w).
\label{eq:eosiFL}
\end{equation}
Plugging the first Friedmann-Lema\^{\i}tre equation $\rho= 3 \Mp^2 H^2$,
and dividing \Eq{eq:eosiFL} by $H^2$, one gets
\begin{equation}
-\dfrac{1}{H}\dfrac{\ud H}{\ud N} = \dfrac{3}{2}(1+w).
\end{equation}
By definition of the Hubble-flow functions in \Eq{eq:defhf}, this
equation reduces to
\begin{equation}
\epsilon_1(N) = \dfrac{3}{2}\left[1 + w(N) \right] = \dfrac{3
  \beta}{2\left(C + \Nend -N\right)^{\alpha}}\,.
\label{eq:eosieps1}
\end{equation}
As a result, postulating the equation of state during inflation is
exactly equivalent to postulating the first Hubble-flow function
$\epsilon_1(N)$. Therefore, the complete hierarchy of Hubble-flow
functions is exactly obtained by taking the successive logarithmic
derivatives of \Eq{eq:eosieps1}. For instance,
\begin{equation}
\epsilon_2 = \dfrac{1}{W} \dfrac{\ud W}{\ud N} = \dfrac{\alpha}{C +
  \Nend -N}\,, \qquad \epsilon_3 =
\dfrac{1}{(\ud W/\ud N)}\dfrac{\ud^2 W}{\ud N^2}  - \dfrac{1}{W}
\dfrac{\ud W}{\ud N} =\dfrac{1}{C + \Nend - N}\,.
\label{eq:eosieps23}
\end{equation}

Comparing these expressions with the ones associated with a homogeneous
scalar field, see \Eqs{eq:KGefolds} and \eqref{eq:defeps1}, one
obtains
\begin{equation}
\left\{
\begin{aligned}
\left(\dfrac{\ud \phi}{\ud N} \right)^2 & = 3 \Mp^2  W,\\
\dfrac{\ud \ln V}{\ud N} & = -3 W + \dfrac{\ud \ln(2-W)}{\ud N}\,.
\end{aligned}
\right.
\label{eq:eosi}
\end{equation}
They can be formally integrated into the exact field trajectory and
parametric potential
\begin{equation}
\begin{aligned}
\phi(N) & = \pm \sqrt{3 } \Mp \int^N W^{1/2}(x)\, \ud x + \phizero, \\
V(N) & = M^4 \left[1 - \frac{W(N)}{2}\right] \exp\left[{\displaystyle -3 \int^N
    W(x) \ud x}\right],
\end{aligned}
\label{eq:eosisols}
\end{equation}
where $\phizero$ and $M^4$ are two expected integration
constants. Because specifying the equation of state is equivalent to
postulate $\epsilon_1$, which is also the field velocity in {\efolds},
there is a hard-coded shift symmetry in the problem and the field
values are defined up to a constant. The ambiguity of sign in the
trajectory is reminiscent with the problems associated with horizon-flow inflation~\cite{Vennin:2014xta}: one of the two exact solutions
would actually climb up the potential and is strongly unstable. The
integration constant $M^4$ fixes the energy scale of inflation, which
remains obviously unspecified by postulating only the equation of
state parameter.

\subsubsection{Exact field trajectory and potential}

The field trajectory and potential of \Eq{eq:eosisols} can be exactly
integrated for the VFMI equation of state given in
\Eq{eq:vfmiw}. Defining
\begin{equation}
x \equiv \dfrac{\phi}{\Mp}\,,
\end{equation}
one gets
\begin{equation}
x  = \dfrac{\sqrt{3 \beta}}{1 - \alpha/2} \left[\left(C
  +\Nend - N\right)^{1-\alpha/2}-1\right],
\label{eq:vfmitraj}
\end{equation}
in which the integration constant $\phizero$ has been chosen such that the
limiting case $\alpha=2$ takes the simple form
\begin{equation}
x_{(\alpha=2)} = \sqrt{3 \beta} \ln\left(C + \Nend-N \right).
\end{equation}
The field value at the end of inflation can be immediately
read off for $N=\Nend$
\begin{equation}
\xend = \dfrac{\sqrt{3 \beta}}{1 - \alpha/2} \left[
\left(\dfrac{3\beta}{2}\right)^{\frac{2-\alpha}{2\alpha}}-1\right],
\label{eq:vfmixend}
\end{equation}
where \Eq{eq:vfmi:C:beta} has also been used.
Similarly, the parametric potential $V(N)$ reads
\begin{equation}
V(N) = M^4 \left[1 - \dfrac{\beta}{2(C+\Nend-N)^\alpha}\right]
\exp\left\{\dfrac{3 \beta}{1 - \alpha} \left[ \left(C + \Nend
  -N\right)^{1-\alpha} - 1\right]\right\},
\label{eq:vfmiparametricpot}
\end{equation}
which, reduces to 
\begin{equation}
V_{(\alpha=1)}(N)  = M^4 \left[1 - \dfrac{\beta}{2\left(C+\Nend-N\right)}\right]
\left(C+\Nend-N\right)^{3\beta}
\end{equation}
in the limiting case $\alpha=1$.
Inverting the field trajectory in \Eq{eq:vfmitraj} gives
\begin{equation}
C + \Nend - N = \left(1 + \dfrac{2-\alpha}{2\sqrt{3\beta}} x\right)^{\frac{2}{2-\alpha}},
\label{eq:vfmiefolds}
\end{equation}
which can be plugged into \Eq{eq:vfmiparametricpot} to obtain the
exact field potential for VFMI
\begin{equation}
V(\phi) = M^4\left[1 - \dfrac{\beta}{2
    \left(1+\dfrac{2-\alpha}{2\sqrt{3\beta}}\dfrac{\phi}{\Mp}\right)^{\frac{2\alpha}{2-\alpha}}}
  \right] \exp\left\{ \dfrac{3 \beta}{1-\alpha}
\left[\left(1+\dfrac{2-\alpha}{2\sqrt{3\beta}}\dfrac{\phi}{\Mp}\right)^{\frac{2(1-\alpha)}{2-\alpha}}
  - 1 \right] \right\}.
\label{eq:vfmipot}
\end{equation}
According to the values of $\alpha$, the potential smoothly
interpolates between various typical inflationary
regimes~\cite{Mukhanov:2013tua}. For $\alpha \le 1$, the potential is
unbounded for $x \to \infty$, which is reminiscent with Large Field
Inflation (LFI). For $1 <\alpha \le 2$, the potential is of the
plateau-type and takes a finite value at large $x$. For $\alpha > 2$
inflation takes place at the top of the potential, around $x=\xVmax$
with
\begin{equation}
\xVmax = \dfrac{2\sqrt{3\beta}}{\alpha - 2}\,, \qquad (\alpha > 2),
\end{equation}
which is similar to Small Field Inflation (SFI). Let us notice that
for $\alpha>2$, the left hand side of \Eq{eq:vfmiefolds} becomes
infinite for $x\to\xVmax$ and the maximal number of {\efolds} in VFMI
for $\alpha>2$ is unbounded. Within our choice of sign for the field
trajectory \eqref{eq:vfmitraj}, inflation always proceeds from large
field values towards small field values and stops at $x=\xend$.

\begin{figure}
\begin{center}
\includegraphics[width=\wdblefig]{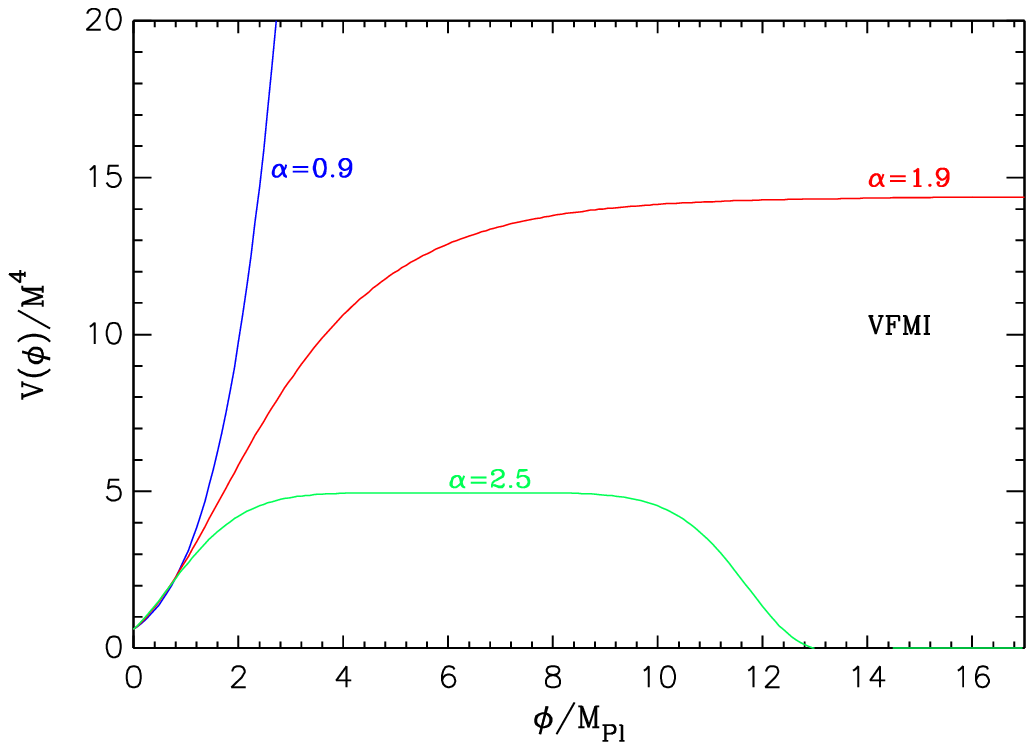}
\includegraphics[width=\wdblefig]{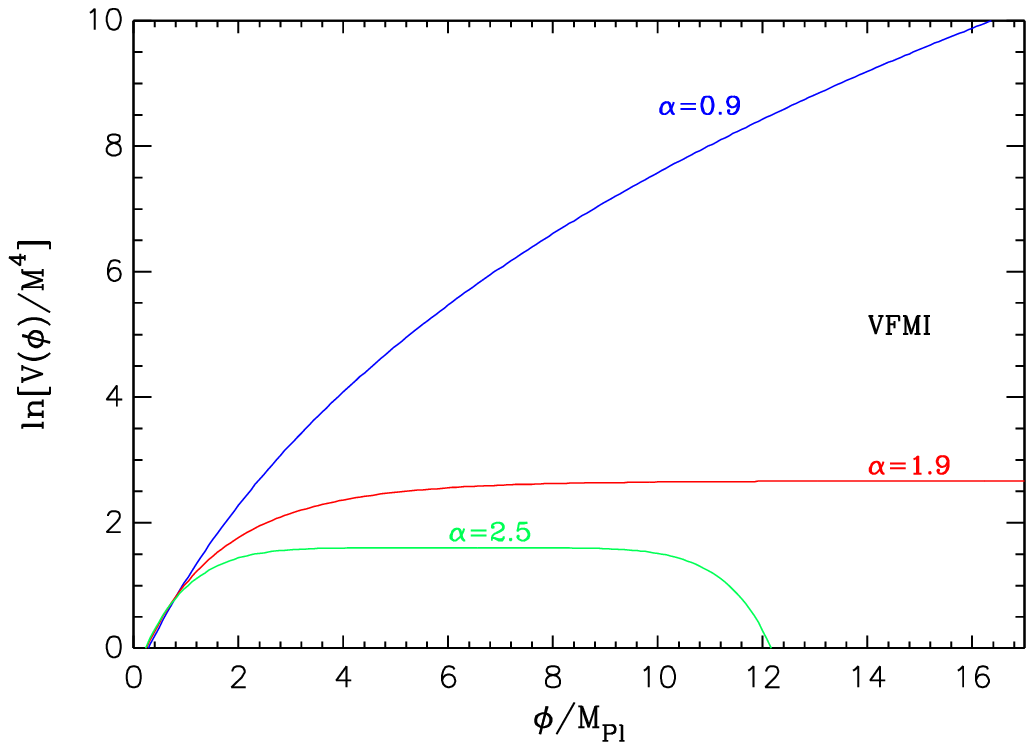}
\includegraphics[width=\wdblefig]{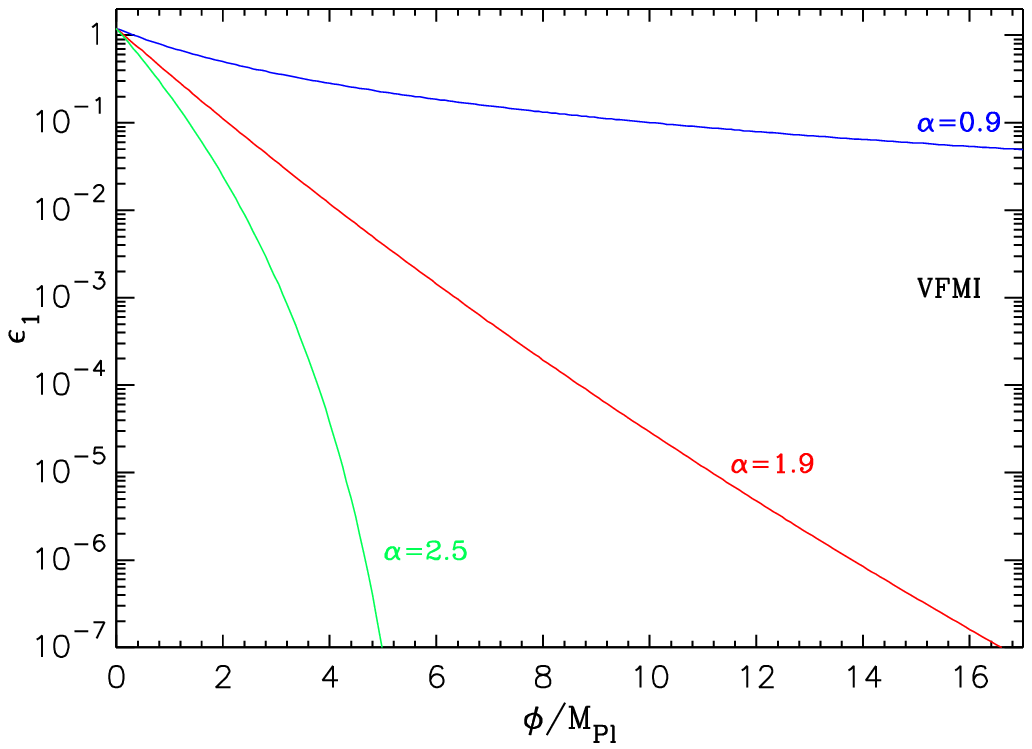}
\includegraphics[width=\wdblefig]{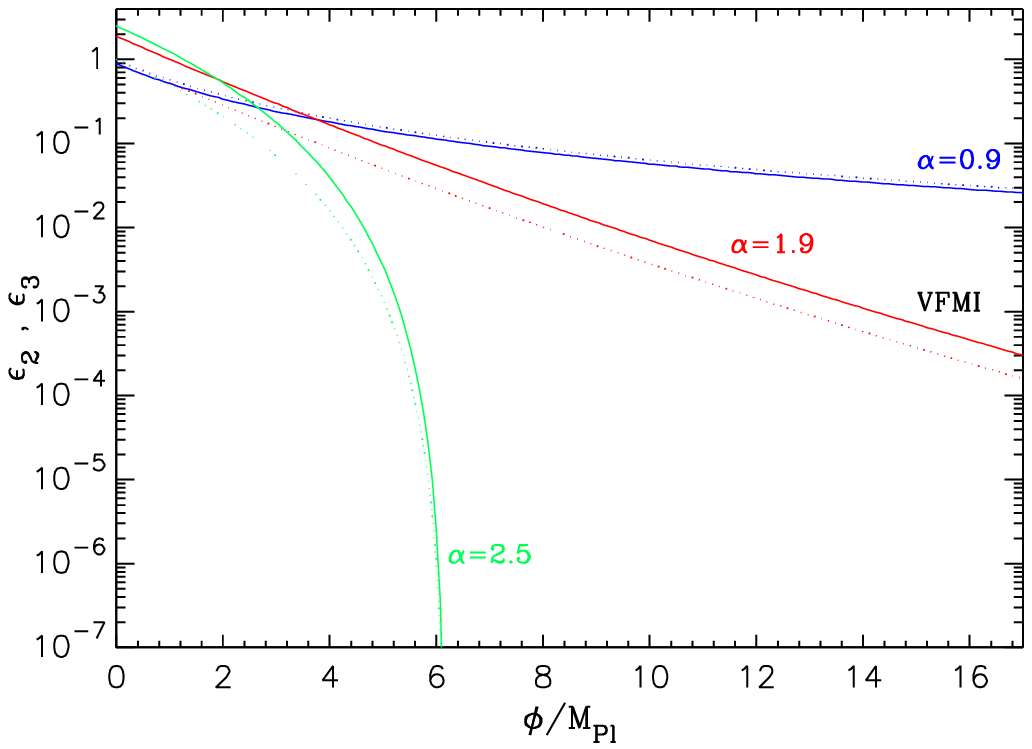}
\caption{Mukhanov Inflation (VFMI) for three
  typical values of $\alpha\le1$, $1 < \alpha \le 2$ and $\alpha>2$
  and $\beta=0.8$. Upper panels: the potential $V(\phi)$ and its
  logarithm as a function of the normalized field values
  $\phi/\Mp$. Bottom left panel: first Hubble-flow function
  $\epsilon_1$, for the same three values of $\alpha$. Bottom right
  panels: second Hubble-flow function $\epsilon_2$ (solid lines) and
  third Hubble flow functions $\epsilon_3$ (dotted lines). In all
  cases inflation proceeds towards small field values. For the cases
  in which $\alpha > 2$, only the small-field inflationary regime
  $x<\xVmax$ is plotted, $x>\xVmax$ being a symmetric case.}
\label{fig:potvfmi}
\end{center}
\end{figure}

The exact Hubble flow functions have been derived in \Eqs{eq:eosieps1}
and \eqref{eq:eosieps23} in terms of the number of {\efolds}. From
\Eq{eq:vfmiefolds}, they can be expressed in terms of field values and
read
\begin{equation}
\epsilon_1 = \dfrac{3\beta}{2 \left(1 +
  \dfrac{2-\alpha}{2\sqrt{3\beta}} \,
  x\right)^{\frac{2\alpha}{2-\alpha}}}\,, \qquad \epsilon_2 =
\dfrac{\alpha}{\left(1 + \dfrac{2-\alpha}{2 \sqrt{3
      \beta}}\,x\right)^{\frac{2}{2-\alpha}}}\,,\qquad \epsilon_{n\ge
  3} = \dfrac{\epsilon_2}{\alpha}\,.
\end{equation}
Together with the potential, they have been reprensented in
\Fig{fig:potvfmi} for three typical values of $\alpha$.

The field value $\xstar$, or the {\efolds} number $\Delta\Nstar$, at
which the observable pivot scale crossed the Hubble radius during
inflation are obtained by solving the reheating equations
\Eq{eq:phistarlnrrad} and \Eq{eq:dnstarlnrad}, respectively. The
observed values of the Hubble-flow functions are immediately given by
\Eqs{eq:eosieps1} and \eqref{eq:eosieps23}, i.e.,
\begin{equation}
\epsilon_{1*} = \dfrac{3 \beta}{2(C + \Delta\Nstar)^\alpha}, \qquad
\epsilon_{2*} = \dfrac{\alpha}{C + \Delta\Nstar}\,, \qquad \epsilon_{n*\ge3}
= \dfrac{\epsilon_{2*}}{\alpha}\,.
\end{equation}
Finally, the integration constant $M^4$ fixing the energy scale of
inflation is inferred from the amplitude of the CMB anisotropies. One
gets
\begin{equation}
\left(\dfrac{M}{\Mp}\right)^4 = \dfrac{4320 \pi^2
  \beta}{2(C+\Delta\Nstar)^\alpha - \beta} \exp\left\{\dfrac{3
    \beta}{1-\alpha}\left[(C+\Delta\Nstar)^{1-\alpha} - 1 \right]
\right\}
\dfrac{\Qrms^2}{T^2}\,.
\end{equation}
Let us remark that within any equation of state parametrization of the
inflationary background, $M^4$ being an integration constant, its
value \emph{cannot} be a theoretical output of the model. This has to
be contrasted with the more usual specification of a field potential
in which the value of $M^4$ \emph{may} very well be predicted, as it
is the case in Higgs/Starobinsky inflation (HI), the original
Colemann-Weinberg model (CWI) and Dual Inflation (DI). The reheating
consistent predictions for VMFI have been represented in
\Fig{fig:CMBVFMI} for various values of $\alpha$ and $\beta$.

\subsection{Fibre Inflation (FI)}
\label{sec:fi}

This model was proposed in \Refc{Cicoli:2013oba} in the context of
string theory, where inflation is driven by a closed string modulus
that parameterizes the size of the extra dimensions. This imposes that
$\phi>0$, and the potential is given by
\begin{equation}
\begin{aligned}
  V(\phi)=M^4\left[\left(1+\frac{2}{3}\delta\right)
\ee^{-\frac{4}{\sqrt{3}}\frac{\phi}{\Mp}}
-4\left(1+\frac{\delta}{6}\right)
\ee^{-\frac{1}{\sqrt{3}}\frac{\phi}{\Mp}}
+\frac{\delta}{1+n}
\ee^{\frac{2(1+n)}{\sqrt{3}}\frac{\phi}{\Mp}}+3-\frac{\delta}{1+n}\right], 
\end{aligned}
\end{equation}
where $n$ is a non-negative integer number (when $n=0$, the model was
studied in \Refc{Cicoli:2008gp}), and $\delta$ is a parametrically
small positive number that is related to the string coupling
$\gs$ via $\delta\propto \gs^{4(1+n/3)}$.

\begin{figure}
\begin{center}
\includegraphics[width=\wdblefig]{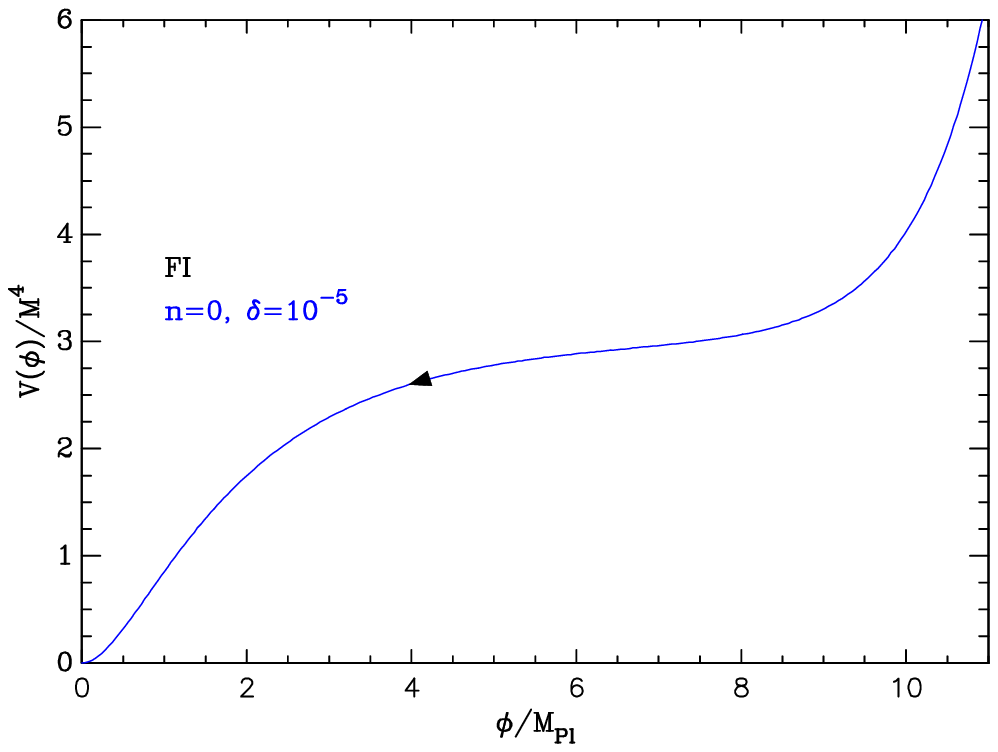}
\includegraphics[width=\wdblefig]{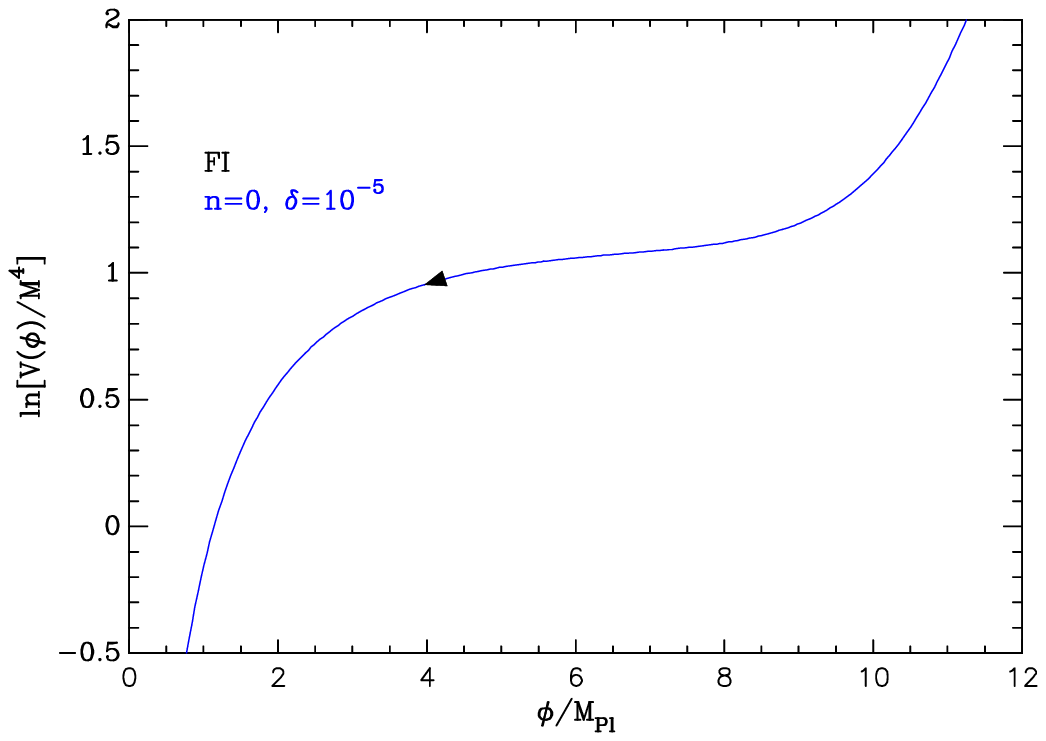}
\includegraphics[width=\wdblefig]{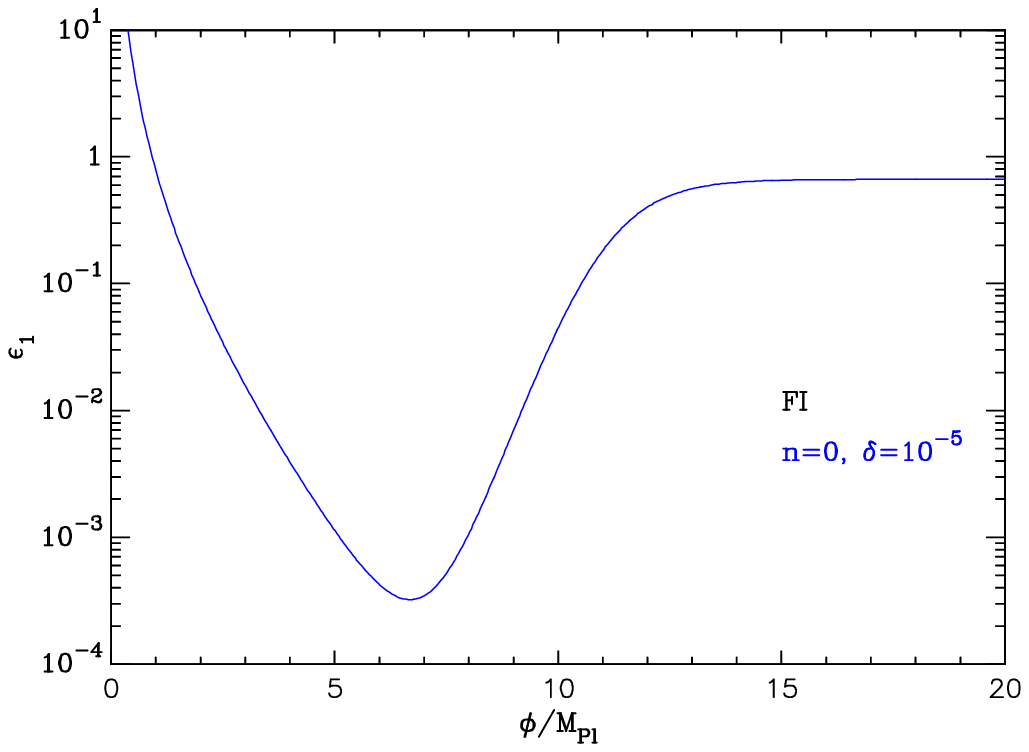}
\includegraphics[width=\wdblefig]{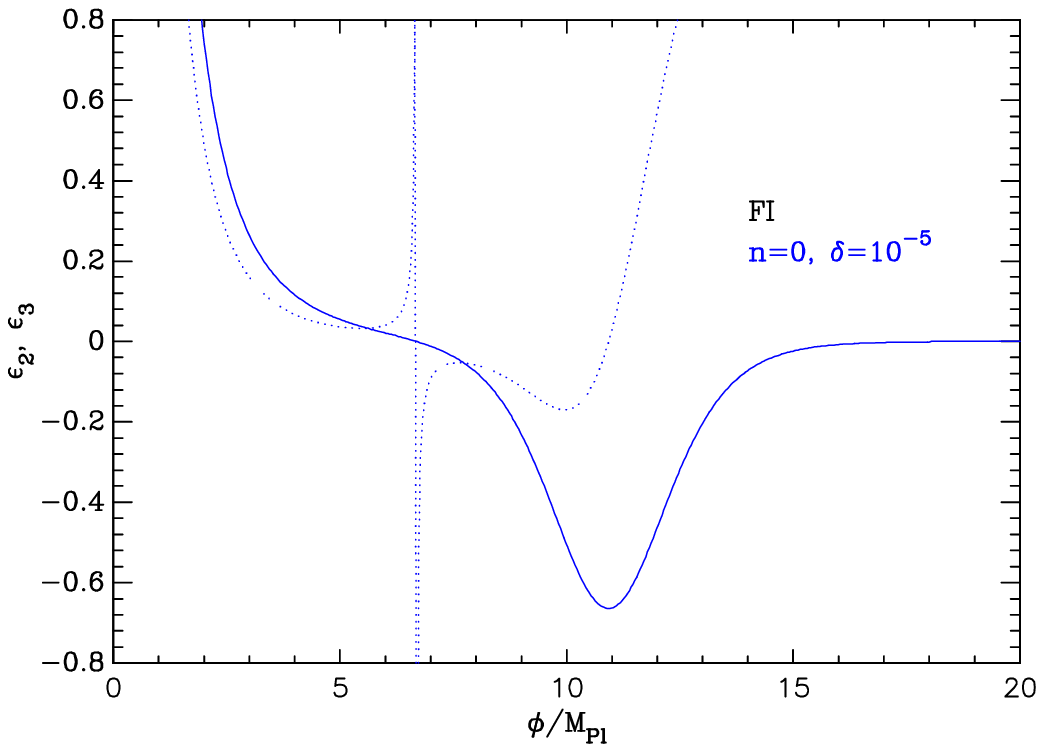}
\caption{Top left panel: Fibre Inflation potential for $n=0$ and $\delta=10^{-5}$. Top right
  panel: logarithm of the potentials for the same values of $n$ and
  $\delta$. Bottom left panel: slow-roll parameter $\epsilon _1$. Bottom right
  panel: slow-roll parameters $\epsilon _2$ (solid line) and $\epsilon
  _3$ (dotted line) for the same values of $n$ and
  $\delta$.}
\label{potfi}
\end{center}
\end{figure}
The potential is displayed in \Fig{potfi}. It vanishes at the origin
$\phi=0$, where the derivative of the potential vanishes too, and it
is a monotonic increasing function of the field. Indeed, the equation
$V'(\phi)=0$ reduces to
\begin{align}
\label{eq:fi:Vprime_eq_0}
\left(1+\frac{\delta }{6}\right)\ee^{\sqrt{3}\frac{\phi}{\Mp }}
+\frac{\delta }{2}\ee^{\frac{2(3+n)}{\sqrt{3}}\frac{\phi}{\Mp}}=1+\frac{2}{3}\delta,
\end{align}
which can be satisfied only if $\phi=0$ (since the left-hand side of
\Eq{eq:fi:Vprime_eq_0} is a manifestly increasing function of $\phi$).
When $\delta=0$, the potential asymptotes a constant, so it has a
plateau shape. When $\delta>0$, this plateau is broken at some field
value (which can be estimating by equating the constant term with the
exponential growing term in the potential function) above which the
potential grows exponentially.

Defining
\begin{equation}
  x \equiv \dfrac{\phi}{\Mp}\,,
\end{equation}
the first Hubble flow function in the slow-roll approximation reads
\begin{equation}
\begin{aligned}
\epsilon_1 &= \frac{16}{6}
\left[\dfrac{-1-\frac{2}{3}\delta+\left(1+\frac{\delta}{6}\right)\ee^{\sqrt{3}x}
+\frac{\delta}{2}\ee^{\frac{6+2n}{\sqrt{3}} x}}{
1+\frac{2}{3}\delta-4\left(1+\frac{\delta }{6}\right)\ee^{\sqrt{3}x}
+\frac{\delta}{n+1} \ee^{\frac{6+2n}{\sqrt{3}}x}
+\left(3-\frac{\delta}{n+1}\right)\ee^{\frac{4}{\sqrt{3}}x}} \right]^{2},
\end{aligned}
\end{equation}
while the second is given by
\begin{equation}
\begin{aligned}
\epsilon_2 &= 
\frac{4}{9(1+n)}e^{-\frac{8}{\sqrt{3}}x}
\Biggl[
3(1+n)(6+\delta)(3+2\delta)
e^{\sqrt{3}x}
-2\delta (3+2\delta)(3+n)^2e^{\frac{2(3+n)}{\sqrt{3}}x}
 \\ & +\delta(6+\delta)(3+2n)^2e^{\frac{9+2n}{\sqrt{3}}x}
+8(3+2\delta)\left(\delta-3-3n\right)
e^{\frac{4}{\sqrt{3}}x}
-(6+\delta)\left(\delta-3-3n\right)
e^{\frac{7}{\sqrt{3}}x}
 \\ &
+6(n+1)\delta \left(\delta-3-3n\right)
e^{\frac{2(5+n)}{\sqrt{3}}x}\Biggr]
\\ & \times
\left[3+\left(1+\frac{2\delta}{3}\right)
e^{-\frac{4}{\sqrt{3}}x}
-\frac23(6+\delta)e^{-\frac{1}{\sqrt{3}}x}
-\frac{\delta}{1+n}+\frac{\delta}{1+n}
e^{\frac{2(1+n)}{\sqrt{3}}x}\right]^{-2}\, ,
\end{aligned}
\end{equation}
and, finally, the third one reads
\begin{equation}
\begin{aligned}
\epsilon_3 &=\Biggl[2(1+n)e^{-\frac{8}{\sqrt{3}}x} 
\left\{6\left(e^{\sqrt{3}x}-1\right)
+\delta \left[-4+e^{\sqrt{3}x}
+3e^{\frac{2(3+n)}{\sqrt{3}}x}\right]\right\}
 \\ & \times
\Biggl(8\Biggl\{
6\left(e^{\sqrt{3}x}-1\right)
+\delta \left[-4+e^{\sqrt{3}x}
+3e^{\frac{2(3+n)}{\sqrt{3}}x}\right]
\Biggr\}^3
-\frac{6}{1+n}
\Biggl\{
6\left(e^{\sqrt{3}x}-1\right)
 \\ & 
+\delta \left[-4+e^{\sqrt{3}x}
+3e^{\frac{2(3+n)}{\sqrt{3}}x}\right]
\Biggr\}
\Biggl\{3(1+n)\left(-1-3e^{\frac{4}{\sqrt{3}}x}
+4e^{\sqrt{3}x}\right)
 \\ & 
+\delta \left[3e^{\frac{4}{\sqrt{3}}x}
-3e^{\frac{2(3+n)}{\sqrt{3}}x}
-2(1+n)
+2(1+n)e^{\sqrt{3}x}\right]\Biggr\}
\Biggl\{6\left(-4+e^{\sqrt{3}x}\right)
 \\ &
+\delta \left[-16+e^{\sqrt{3}x}
-6(1+n)e^{\frac{2(3+n)}{\sqrt{3}}x}\right]\Biggr\}
+9e^{\frac{8}{\sqrt{3}}x}\biggl[3+\left(1+\frac{2\delta}{3}\right)
e^{-\frac{4}{\sqrt{3}}x}
 \\ & 
-\frac23(6+\delta)e^{-\frac{1}{\sqrt{3}}x}
-\frac{\delta}{1+n}+\frac{\delta}{1+n}e^{\frac{2(1+n)}{\sqrt{3}}x}
\biggr]^2
\Biggl\{6\left(-16+e^{\sqrt{3}x}\right)
+\delta \biggl[-64+e^{\sqrt{3}x}
 \\ &
+12(1+n)^2
e^{\frac{2(3+n)}{\sqrt{3}}x}\biggr]\Biggr\}\Biggr)\Biggr]
\Biggl(81\biggl[3+\left(1+\frac{2\delta}{3}\right)
e^{-\frac{4}{\sqrt{3}}x}
-\frac23(6+\delta)e^{-\frac{1}{\sqrt{3}}x}
-\frac{\delta}{1+n}
 \\ &
+\frac{\delta}{1+n}e^{\frac{2(1+n)}{\sqrt{3}}x}
\biggr]^2
\Biggl\{3(6+\delta)(3+2\delta)(1+n)e^{\sqrt{3}x}
-2\delta(3+2\delta)(3+n)^2e^{\frac{2(3+n)}{\sqrt{3}}x}
 \\ & 
+\delta(6+\delta)(3+2n)^2
e^{\frac{9+2n}{\sqrt{3}}x}
+8(3+2\delta)\left[\delta-3(1+n)\right]
e^{\frac{4}{\sqrt{3}}x}
 \\ & 
-(6+\delta)\left[\delta-3(1+n)\right]e^{\frac{7}{\sqrt{3}}x}
+6\delta(1+n)\left[\delta-3(1+n)\right]
e^{\frac{2(5+n)}{\sqrt{3}} x}\Biggr\}\Biggr)^{-1}.
\end{aligned}
\end{equation}
They are displayed in the lower panels of \Fig{potfi}. One can see
that the first slow-roll parameter diverges at $x=0$, decreases until
a field value that we denote $\xepstwoZero$ and that, in practice,
needs to be computed numerically, and then increases again to reach
the asymptotic value
\begin{equation}
\epsilon_{1,\infty} = \frac{2}{3}\left(1+n\right)^2.
\end{equation}
This indicates that inflation stops by violation of the slow-roll
conditions, at a field value $\xend$ that needs to be determined
numerically by solving the equation $\epsilon_1=1$.

The slow-roll trajectory
\begin{equation}
\begin{aligned}
\label{eq:fi:SlowRollTraj}
\Nend-N &= \frac{\sqrt{3}}{4}
\int _{\xend}^x
\biggl[\left(1+\frac{2\delta}{3}\right)\ee^{-\frac{4}{\sqrt{3}} y}
-4\left(1+\frac{\delta}{6}\right)\ee^{-\frac{1}{\sqrt{3}} y}
+\frac{\delta}{1+n}\ee^{\frac{2(1+n)}{\sqrt{3}} y}
+3
 \\ &
-\frac{\delta}{1+n}\biggr]
\biggl[-\left(1+\frac{2\delta}{3}\right)\ee^{-\frac{4}{3} y }
+\left(1+\frac{\delta}{6}\right)\ee^{-\frac{1}{\sqrt{3}} y}
+\frac{\delta}{2} \ee^{\frac{2(1+n)}{\sqrt{3}} y}\biggr]^{-1}
\dd y,
\end{aligned}
\end{equation}
also needs to be integrated numerically.

\begin{figure}
\begin{center}
\includegraphics[width=\wsingfig]{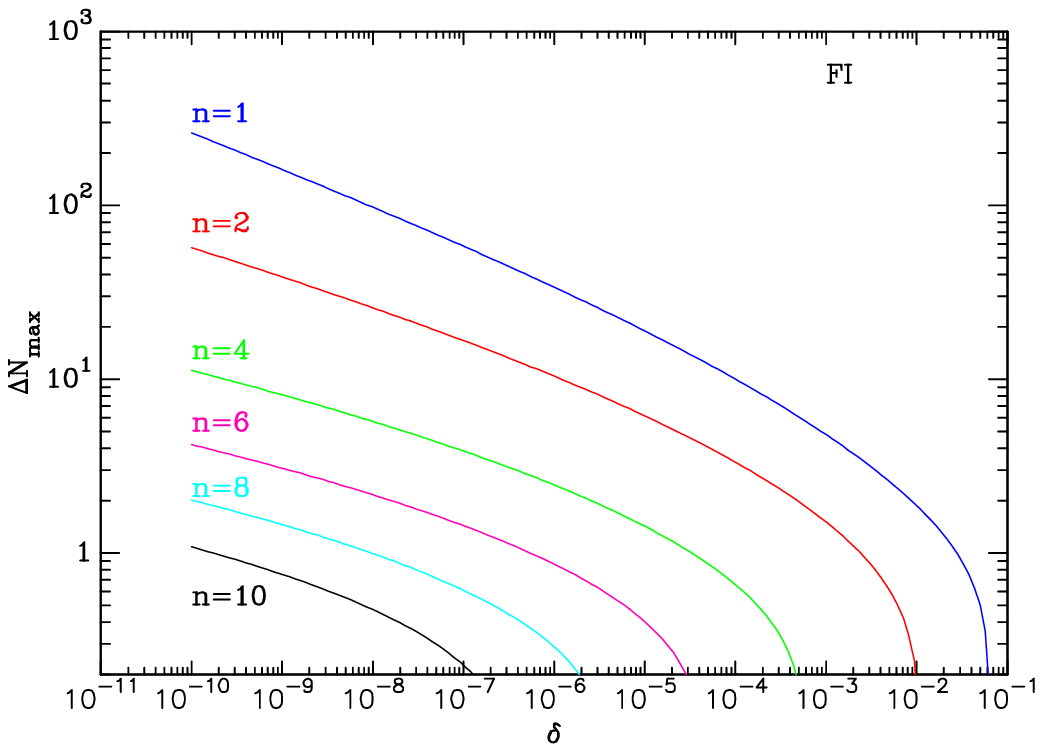}
\caption{Maximum number of inflationary {\efolds} realized in Fibre
  Inflation, as a function of $\delta$ and for a few values of
  $n$. When $n\geq 1$, the first Hubble-flow parameter is smaller than
  one across a finite field range, and $\Delta\Nmax$ corresponds to the
  number of {\efolds}, \Eq{eq:fi:SlowRollTraj}, realized between the
  two boundaries of that interval, the location of which are obtained
  by numerically solving the equation $\epsilon_1=1$.}
\label{fig:FI:Nmax}
\end{center}
\end{figure}

Let us notice that, when $n\geq 1$, $\epsilon_{1,\infty} >1$ so there is
a finite range of field value where the first slow-roll parameter is
below unity. As a consequence, only a finite number of \efolds~can be
realized in such cases, between the two field values where
$\epsilon_1=1$. This number is displayed in \Fig{fig:FI:Nmax}, where
one can check that, for a sufficiently long inflationary phase to take
place, $\delta$ needs to be small enough, and the upper bound on
$\delta$ is smaller for larger $n$.  These considerations are based on
a slow-roll analysis and one may wonder whether, in the large-field
region, inflation can take place outside the slow-roll regime. This is
not the case since in the large-field limit, the potential function
becomes approximately exponential, and is of the same form as in Power
Law Inflation (PLI, see \sectionc{sec:pli}). In this model, the
dynamics can be solved exactly, \ie~without resorting to the slow-roll
approximation, and in \sectionc{sec:pli} it is shown that inflation
requires the coefficient in the exponential to be smaller than
$\sqrt{2}$ (when the field is expressed in reduced Planck mass
units). This is not the case for $n\geq 1$ in the present model, which
confirms the validity of the above discussion beyond the slow-roll
approximation.

Finally, the normalization of the potential $M^4$ can be obtained from
the amplitude of the CMB anisotropies once one has determined the
field value $\xstar$ at which the pivot mode crossed the Hubble
radius. It reads
\begin{equation}
\left(\dfrac{M}{\Mp}\right)^4 = 
3840\pi^2\dfrac{\ee^{\frac{4}{\sqrt{3}}\xstar}
\left[-1-\frac{2}{3}\delta+\left(1+\frac{\delta}{6}\right)\ee^{\sqrt{3}x}
+\frac{\delta}{2}\ee^{\frac{6+2n}{\sqrt{3}} x}\right]^2}{\left[
1+\frac{2}{3}\delta-4\left(1+\frac{\delta }{6}\right)\ee^{\sqrt{3}x}
+\frac{\delta}{n+1} \ee^{\frac{6+2n}{\sqrt{3}}x}
+\left(3-\frac{\delta}{n+1}\right)\ee^{\frac{4}{\sqrt{3}}x}\right]^{3}} 
\dfrac{\Qrms^2}{T^2}
\,.
\end{equation}

The reheating consistent predictions for FI have been represented in
\Figs{fig:CMBFI_0} and~\ref{fig:CMBFI_1} for $n=0$ and $n=1$
respectively. One can see that, in the small-$\delta$ limit, the model
is in good agreement with the data since the potential has a plateau
shape. One should also note that the potential (hence the predictions
of the model) is independent of $n$ in that limit.

\subsection{Hyperbolic Inflation (HBI)}
\label{sec:hbi}

\subsubsection{Theoretical Justifications}
\label{subsubsec:theoryhbi}

In \Refc{Rubano:2001xi}, the cosmological evolution driven by a scalar
field $\phi$ in the presence of a perfect fluid (denoted ``f''
hereafter) is discussed, in the case where both the scalar field and
the perfect fluid follow scaling solutions, \ie $\rho_\phi=C_\phi
a^{-n_\phi}$ and $\rho_{\uf} = C_\uf a^{-n_{\uf}}$. In these
expressions, $C_\phi$, $C_{\uf}$, $n_\phi$ and $n_{\uf}$ are constant,
and $n_\phi>n_{\uf}$ for the scalar field to dominate at early
time\footnote{This is the choice made in \Refc{Basilakos:2015sza},
  but the opposite assumption can also be made, if the scalar field
  accounts for late-time acceleration rather than inflation, see
  \Refc{Rubano:2001xi}.}. In that case, the inflaton field $\phi$ still
follows the Klein-Gordon equation~\eqref{eq:kg}, but the Friedmann-Lema\^{\i}re
equation~\eqref{eq:friedman} becomes
\begin{equation}
\label{eq:hbi:Friedman:mod}
H^2=\frac{\rho_\phi+\rho_{\uf}}{3\Mp^2}\, ,
\end{equation}
where $\rho_\phi=V(\phi)+\dot{\phi}^2/2$. The scalar-field potential
$V(\phi)$ leading to such a solution is derived in
\Refc{Rubano:2001xi}, and the inflationary model associated to that
potential is compared with the Planck 2015 data in
\Refc{Basilakos:2015sza}.

Let us see how the potential function $V(\phi)$ can be obtained. By
making use of the Klein-Gordon equation~\eqref{eq:kg}, the time
derivative of the energy density associated to the inflaton field
reads $\dot{\rho}_\phi = -3 H\dot{\phi}^2$. Since $\rho_\phi=C_\phi
a^{-n_\phi}$, one also has $\dot{\rho}_\phi = -H n_\phi C_\phi a^{-n_\phi}
$, hence the above implies that $\dot{\phi}^2=n_\phi C_\phi a^{-n_\phi}/3$,
so the kinetic energy scales in the same way as the total energy
density associated to $\phi$. This is simply because, for the inflaton
to follow a scaling solution, its equation-of-state parameter must be
constant, hence the ratio between its potential energy and its kinetic
energy is constant too, and both the potential and the kinetic energy
follow the same scaling solution.

Since $\dd \phi/\dd a = \dot{\phi}/(aH)$, using the modified Friedmann-Lema\^{\i}tre
equation~\eqref{eq:hbi:Friedman:mod}, one has
\begin{equation}
\label{eq:hbni:dphi:da}
\dfrac{\dd\phi}{\dd a} = \pm
\frac{\Mp}{a}\sqrt{\dfrac{n_\phi}{1+\dfrac{C_{\uf}}{C_\phi}a^{n_\phi-n_{\uf}}}}\,
.
\end{equation}
In the absence of perfect fluid, \ie~if $C_{\uf}=0$, this can be
readily integrated as $\phi/\Mp= \pm \sqrt{n_\phi} (N- \Nend) +
\phiend$, which is the dynamics of Power Law Inflation (PLI), see
\Eq{eq:pli:traj} in \sectionc{sec:pli}. Indeed, it is shown that PLI
(where no perfect fluid is considered) yields scaling solutions. Even
if $C_{\uf}\neq 0$, \Eq{eq:hbni:dphi:da} can still be integrated, and
one obtains
\begin{equation}
\phi(a)=\phiend \pm
\frac{2\sqrt{n_\phi}}{n_{\phi}-n_\uf}\Mp\left[\arcsinh
  \left(\sqrt{\frac{C_\phi}{C_{\uf} \aend^{n_{\phi}-n_\uf}}
  }\right)-\arcsinh\left(\sqrt{\frac{C_\phi}{C_{\uf} a
    ^{n_{\phi}-n_\uf}} }\right)\right].
\label{eq:hbni:phia}
\end{equation}
This can be inverted to yield the function $a(\phi)$. Since
$V=\rho_\phi-\dot{\phi^2}/2$, one has $V(a)=(1-n/6)C_\phi a^{-n_\phi}$
and an explicit form of the potential can be obtained as
\begin{equation}
\label{eq:hbi:pot:th}
V(\phi) = \left(1-\frac{n_\phi}{6}\right) C_\phi
\left(\frac{C_\uf}{C_{\phi}}\right)^{\frac{n_\phi}{n_{\phi}-n_\uf}}
\sinh^{\frac{2n_\phi}{n_\phi-n_{\uf}}}
\left[\frac{n_{\phi}-n_\uf}{2\sqrt{n_\phi}}
  \frac{\phi-\phiend}{\Mp}+\arcsinh\left(\sqrt{\frac{C_\phi}{C_{\uf} 
    \aend^{n_{\phi}-n_\uf}}}\right)\right].
\end{equation}
For $n_\phi > n_\uf$, we have chosen a growing potential with respect
to $\phi$, implying a rolling down trajectory, and thus a negative
sign in \Eq{eq:hbni:dphi:da}.  One can check that when $C_{\uf}$ goes
to zero (\ie in the absence of perfect fluid), the argument of the
$\arcsinh(\,)$ function in \Eq{eq:hbni:phia} becomes very large, hence the
scale factor $a(\phi)$ approaches an exponential function, and one
recovers the potential of Power Law Inflation.

Using the shift symmetry of the problem, one can absorb the constant
term in the argument of the hyperbolic sine function in
\Eq{eq:hbi:pot:th} and rewrite the potential as $V \propto
\sinh^n(\phi/\mu)$ where
\begin{equation}
n = \dfrac{2 n_\phi}{n_\phi-n_\uf}\,, \qquad \mu = 2\Mp \dfrac{\sqrt{n_\phi}}{n_\phi-n_{\uf}}\,.
\label{eq:hbinandmu}
\end{equation}
Since $n_\phi$ and $n_{\uf}$ are of order one, with $n_\phi >
n_{\uf}$, the consistency of the model demands $n>2$. This also
implies that the typical {\vev} $\mu$ of the field is of the order of
the Planck mass. In \Refc{Rubano:2001xi}, this potential is studied in
the context of a single scalar field theory, \ie by neglecting
completely the presence of the perfect fluid and making use of the
Friedmann-Lema\^{\i}tre equation~\eqref{eq:friedman} rather than
\Eq{eq:hbi:Friedman:mod}. At early times, with $n_\phi>n_{\uf}$, the
contribution of the perfect fluid may indeed be negligible with
respect to the one of the field. However, as inflation proceeds, the
fluid energy density becomes more and more important and such an
assumption will eventually break down. As such, without an explicit
description of the fluid, one cannot guarantee that the usage of the
non-modified FL equations is justified till the end of inflation. Let
us also notice that the problem could be alleviated by assuming $C_\uf
\to 0$, but, in that case, the model predictions would be
undistinguishable from PLI.

In the next section, these issues will be ignored and we will take a
phenomenological approach to explore the observable predictions of the
HBI potential.

\subsubsection{Slow-Roll Analysis}

\begin{figure}
\begin{center}
\includegraphics[width=\wdblefig]{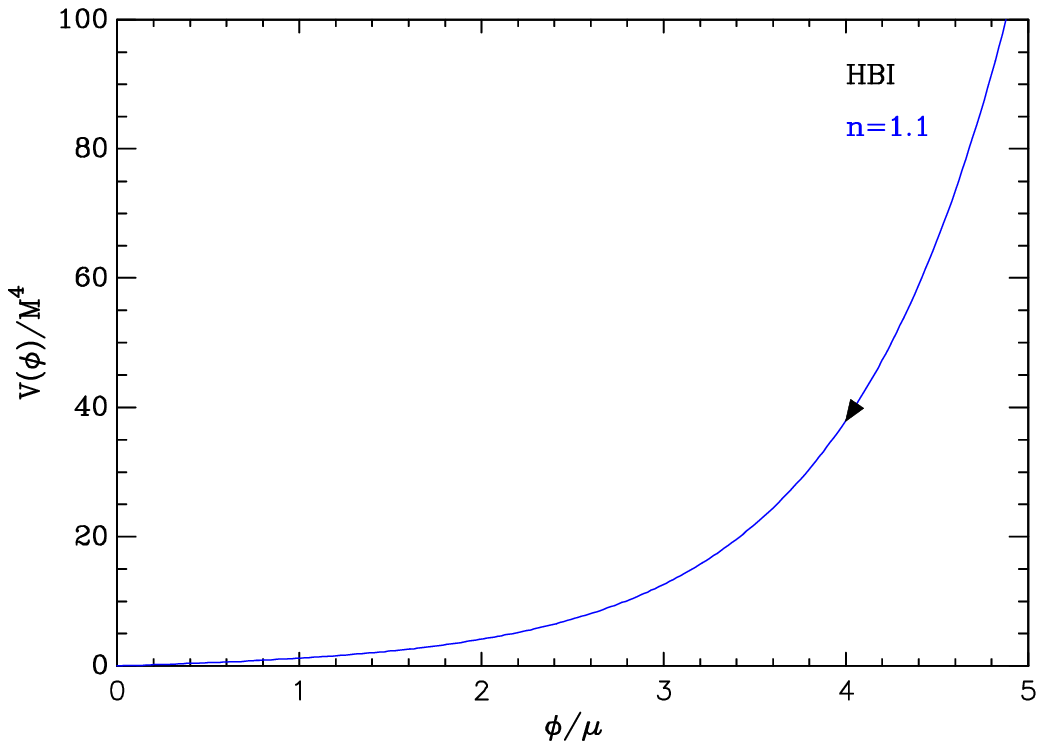}
\includegraphics[width=\wdblefig]{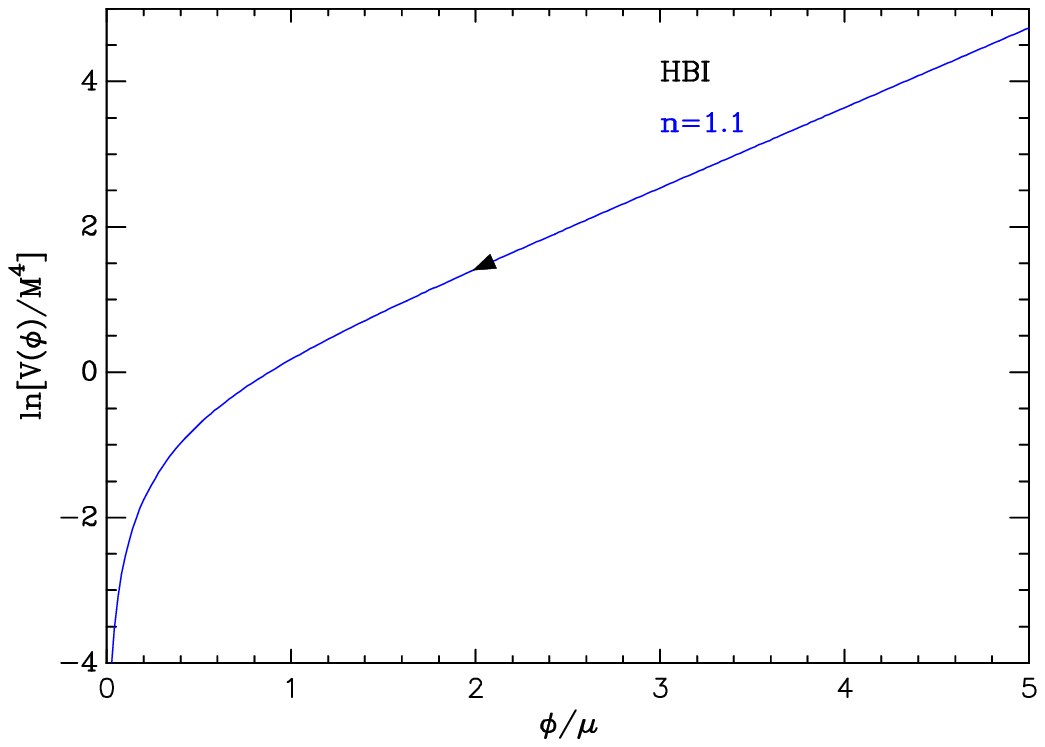}
\includegraphics[width=\wdblefig]{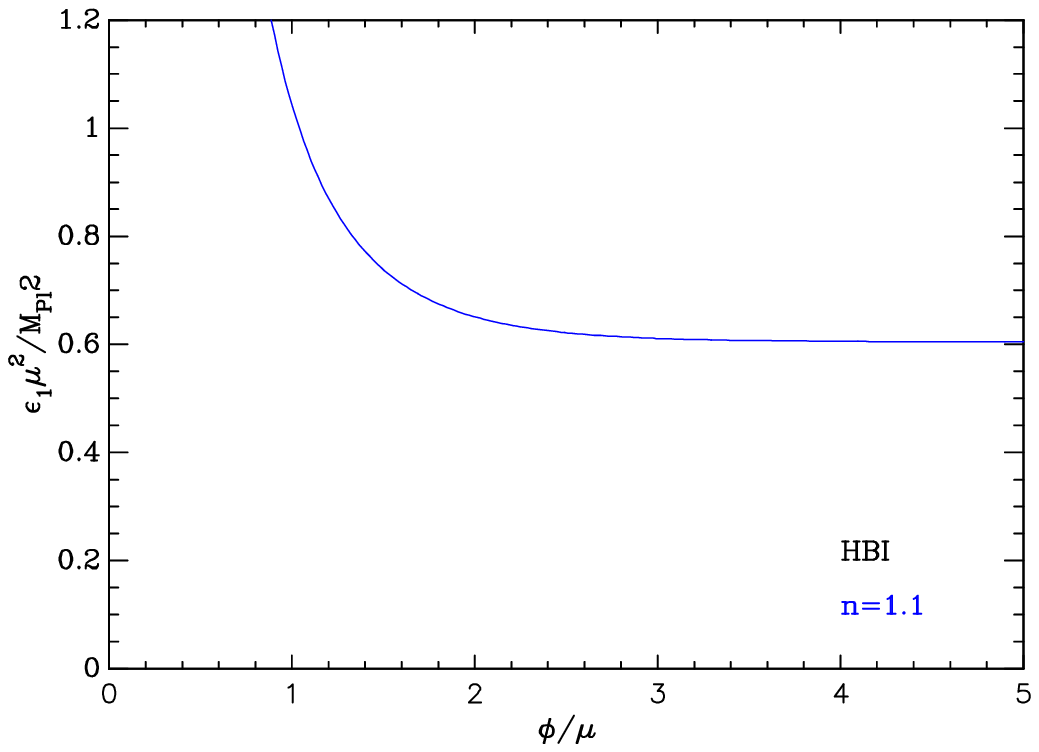}
\includegraphics[width=\wdblefig]{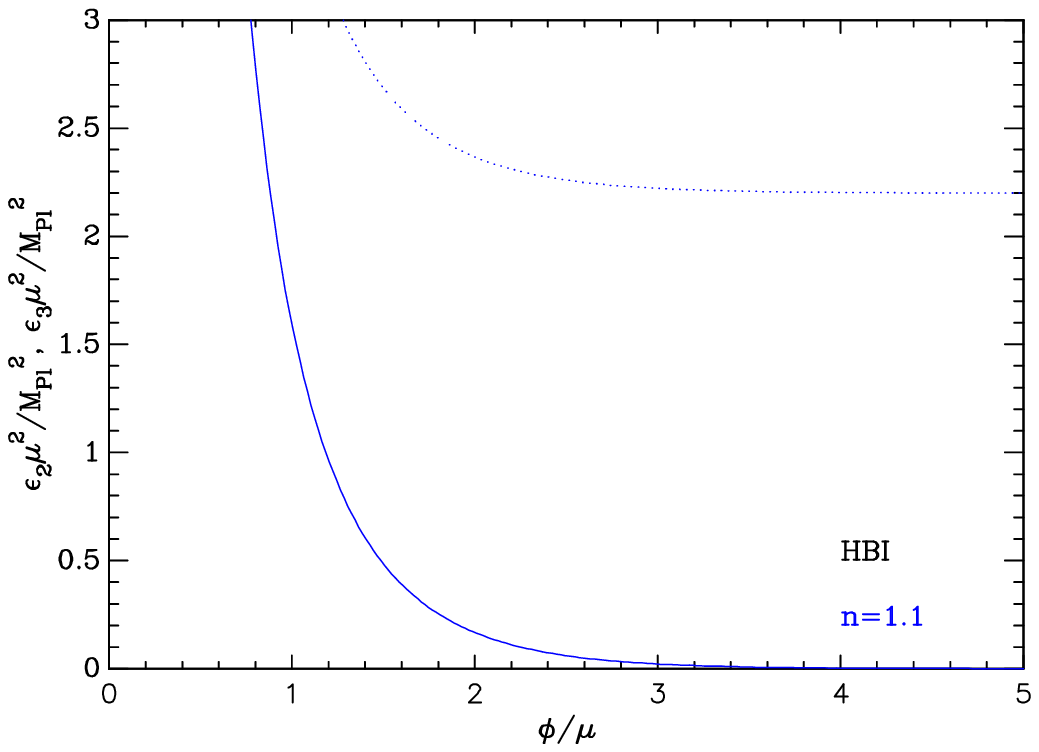}
\caption{Hyperbolic Inflation potential (HBI) for $n=1.1$. Top left
  panel: the potential as a function of $\phi/\mu$.  Top right panel:
  logarithm of the potential. Bottom left panel: rescaled slow-roll
  parameter $\epsilon _1 \mu^2/\Mp^2$. Bottom right panel: rescaled
  slow-roll parameters $\epsilon_2 \mu^2/\Mp^2$ (solid line) and
  $\epsilon_3 \mu^2/\Mp^2$ (dotted line). HBI can only inflate for
  Planckian and super-Planckian values of $\mu$.}
\label{pothbi}
\end{center}
\end{figure}

From the previous discussion, we take the potential of Hyperbolic
Inflation as
\begin{align}
\label{eq:HBI:pot}
V=M^4\sinh^n\left(\frac{\phi}{\mu}\right),
\end{align}
where $M$ is a mass scale. Because $n$ is not necessarily an even
integer, we will consider inflation to occur only in the branch
$\phi>0$. The potential is a monotonic increasing function of the
field values, so inflation proceeds at decreasing field value. It is
represented in \Fig{pothbi}.

Defining
\begin{equation}
x \equiv \dfrac{\phi}{\mu}\,,
\end{equation}
the Hubble-flow functions in the slow-roll approximation are given by
\begin{equation}
\epsilon_1 =\frac{n^2\Mp^2}{2\mu^2 \tanh^2(x)}\,, \qquad \epsilon_2
=\frac{2n \Mp^2}{\mu^2 \sinh^2(x)}\,, \qquad \epsilon_3 =
\frac{2n\Mp^2}{\mu^2 \tanh^2(x)}\,,
\label{eq:hbi:eps}
\end{equation}
and are displayed in the lower panels of \Fig{pothbi}. They all
decrease with the field value $x$, hence they increase as inflation
proceeds, and diverge in the limit $x\to 0$. In the opposite limit,
when $x \to +\infty$, $\epsilon_1$ approaches a constant value
\begin{equation}
\epsonemin = \dfrac{n^2 \Mp^2}{2\mu^2}\,.
\end{equation}
In order for slow-roll inflation to occur, one must have $\epsonemin <
1$ and this gives a lower bound on $\mu$ given by\footnote{From
  \Eq{eq:hbinandmu}, the bound $\mu > \mumin$ implies that
  $n_\phi<2$. Since $n_\uf<n_\phi$, this also implies that
  $n_{\uf}<2$ and, as noted in \Refc{Rubano:2001xi}, this
  excludes the possibility that the perfect fluid is pressureless
  matter ($n_{\uf}=3$) or radiation ($n_{\uf}=4$).}
\begin{equation}
\mumin = \frac{n}{\sqrt{2}} \Mp.
\label{eq:hbimumin}
\end{equation}

For $\mu > \mumin$, Hyperbolic Inflation gracefully ends when
$\epsilon_1=1$, at a field value given by
\begin{align}
\label{eq:hbiend}
\xend= \arctanh\left(\frac{n\Mp}{\sqrt{2}\mu}\right).
\end{align}
The slow-roll trajectory can be integrated, and one obtains
\begin{align}
\Nend - N = \frac{\mu^2}{n \Mp^2} \ln \left[\dfrac{\cosh(x)}{\cosh{(\xend)}}\right].
\label{eq:hbitrajN}
\end{align}
This trajectory can be explicitly inverted to get the field values $x$
as a function of the number of {\efolds} as
\begin{equation}
x =  \arccosh \negthickspace \left[\cosh(\xend) e^{-n
    \frac{\Mp^2}{\mu^2}(\Nend-N)}\right]=\arccosh \negthickspace \left(\dfrac{e^{-n
      \frac{\Mp^2}{\mu^2}\Delta N}}{\sqrt{1 - \dfrac{n^2 \Mp^2}{2
        \mu^2}}} \right),
\label{eq:hbitrajx}
\end{equation}
with $\Delta N=\Nend - N$ and where \Eq{eq:hbiend} has been used to
express $\xend$. The trajectory~\eqref{eq:hbitrajN}, combined with the
reheating equation~\eqref{eq:phistarlnrrad}, allows us to determine
$\xstar$, the field value at which the pivot mode crossed the Hubble
radius during inflation. In turn, this determines the mass scale $M$
of the potential from the CMB normalization and one finds
\begin{equation}
\left(\dfrac{M}{\Mp}\right)^4 = 720 \pi^2 \dfrac{n^2 \Mp^2}{\mu^2
  \tanh^2(\xstar) \sinh^n(\xstar)} \dfrac{\Qrms^2}{T^2}\,.
\end{equation}

The reheating-consistent slow-roll predictions of HBI are displayed in
\Figs{fig:CMBHBI_0} to \ref{fig:CMBHBI_2}, for $n=0.5$, $n=1$ and
$n=1.5$, and various values of $\mu$. As these plots suggest, HBI
produces a significant amount of tensor modes putting it under
pressure.

The model predictions can be analytically understood by plugging
\Eq{eq:hbitrajx} into the Hubble flow functions in
\Eqs{eq:hbi:eps}. One obtains
\begin{equation}
\epsilon_{1*} =
\dfrac{\dfrac{n^2}{2}\dfrac{\Mp^2}{\mu^2}}{1-\left(1-\dfrac{n^2}{2}\dfrac{\Mp^2}{\mu^2}\right)\ee^{-2
    n \frac{\Mp^2}{\mu^2}\Delta \Nstar}}\, ,\qquad \epsilon_{2*} =
2n\dfrac{\Mp^2}{\mu^2}
\frac{1-\dfrac{n^2}{2}\dfrac{\Mp^2}{\mu^2}}{\ee^{-2 n
    \frac{\Mp^2}{\mu^2}\Delta \Nstar}-1+\dfrac{n^2}{2}\dfrac{\Mp^2}{\mu^2}}\,,
\end{equation}
and $\epsilon_{3*}=(4/n) \epsilon_{1*}$. As stressed above, for
slow-roll inflation to take place, $\mu$ must be super-Planckian,
which is why it is interesting to evaluate these formulas in the limit
$\mu/\Mp\gg \sqrt{\Delta \Nstar}$, where one has
\begin{equation}
\epsilon_{1*}\simeq \dfrac{n}{4\left(\Delta \Nstar + \dfrac{n}{4}\right)}\,, \qquad 
\epsilon_{2*} \simeq \epsilon_{3*}\simeq \frac{1}{\Delta \Nstar+ \dfrac{n}{4}}\,.
\end{equation}
One notices that these expressions are the same as the ones obtained
for Large Field Inflation (LFI), see \Eq{eq:lfi:epsstar}, where the
parameter denoted $p$ in LFI is identified with $n$ in HBI. This is
because, in the regime $\mu\gg \Mp$, the last \efolds~of inflation
proceed at $\phi\ll\mu$, where the potential
function~\eqref{eq:HBI:pot} can be Taylor expanded, yielding $V\simeq
M^4 (\phi/\mu)^p$. Note that the condition $n>2$ therefore implies
that the original fluid model is disfavored.

\subsection{Smeared Higgs Inflation (SHI)}
\label{sec:shi}

\subsubsection{Theoretical Justifications}
\label{subsubsec:theoryshi}

In \Refc{Senoguz:2015lba}, an extension of the Colemann-Weinberg model
is considered, see \sectionc{sec:cwi}, in the context of SU(5)
GUT. The two fields in presence are a Higgs gauge singlet $\phi$ and a
field $\chi$ that breaks the SU(5) group. After taking radiative
corrections into account, the potential is given by
\begin{equation}
V= - \frac{m^2}{2}\phi^2 + \frac{\lambda}{4}\phi^4 -
\frac{\beta^2}{2}\phi^2\chi^2+\frac{a}{4}\chi^4+A\phi^4\left[\ln\left(\frac{\phi}{\phizero}\right)+C\right]+V_0\,
.
\end{equation}
The situation where $m^2=0$ was considered in \Refc{Albrecht:1984qt}
and, as we will see below, it leads to Colemann-Weinberg inflation
(CWI). In that case, the fields $\phi$ and $\chi$ have vanishing
masses and quartic dimensionless self-coupling $\lambda$ and $a$
respectively. The parameter $\beta$ is also dimensionless and couples
the two fields, while $A\sim \beta^4/(16\pi^2)$, $C$ and $\phizero$ are
renormalization parameters. The potential energy at vanishing field
configurations is denoted $V_0$. The additional contribution
considered in \Refc{Senoguz:2015lba} is a negative squared mass for
$\phi$, and we now explain how it modifies the Colemann-Weinberg
potential.

The first step is to set the field $\chi$ at the minimum of its
effective potential, so $\chi=\beta\phi/\sqrt{a}$. Then, the parameter
$C$ can be set such that the resulting effective potential for $\phi$
is minimum at $\phi=\phizero$. This leads to $4A C =
m^2/\phizero^2+\beta^4/a-\lambda-A$. Finally, $V_0$ is set such that the
potential vanishes at this minimum, which gives rise to $4 V_0=A
\phizero^4+m^2 \phizero^2$. Only three parameters remain, namely $m$,
$\phizero$ and $A$, in terms of which the potential reads
\begin{equation}
\label{eq:pot:th:shi}
V=\frac{m^2
  \phizero^2}{4}\left[1-\left(\frac{\phi}{\phizero}\right)^2\right]^2 + A
\phi^4
\left[\ln\left(\frac{\phi}{\phizero}\right)-\frac{1}{4}\right]+\frac{A
  \phizero^4}{4}\, .
\end{equation}
As announced above, when $m^2=0$, one recovers the Colemann-Weinberg
potential, see \Eq{eq:cwi:potLinde}. In the opposite limit where $m^2$
is very large, one obtains the Higgs tree-level potential, see
\Eq{eq:Higgs:Jordan:pot:TreeLevel}, which corresponds to the
double-well inflation model (DWI) studied in \sectionc{sec:dwi}. One
therefore expects SHI to interpolate between these two limits, CWI and
DWI. It can either be viewed as a generalization of CWI, or as a
radiatively-corrected version of DWI.

\subsubsection{Slow-Roll Analysis}
\label{subsubsec:srshi}
\begin{figure}
\begin{center}
\includegraphics[width=\wdblefig]{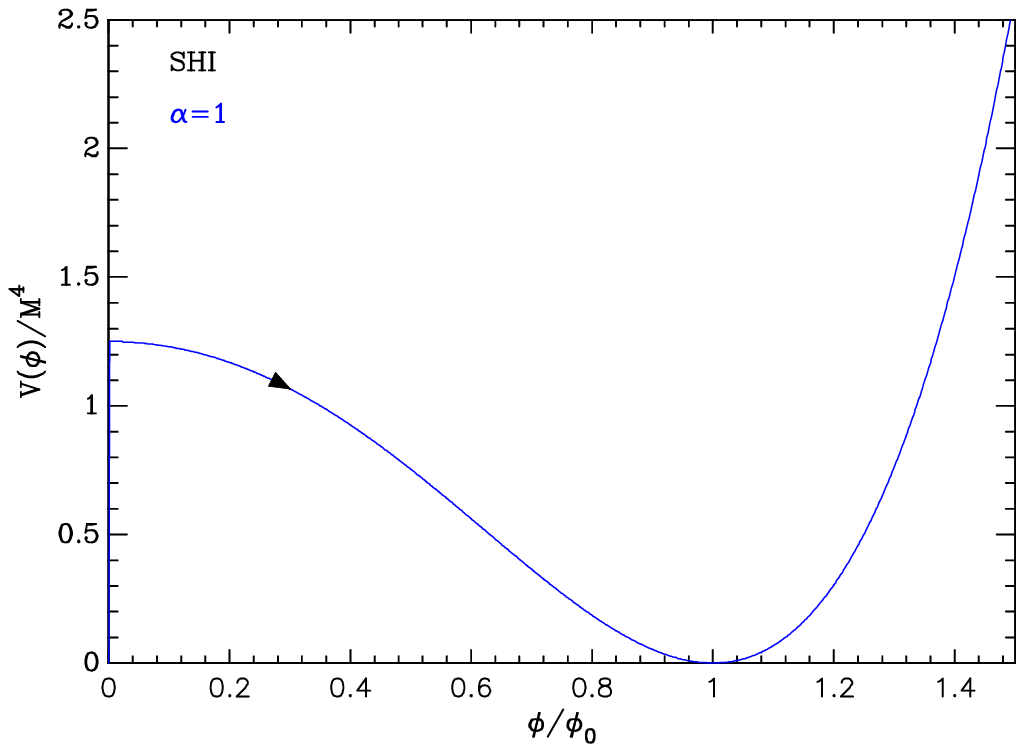}
\includegraphics[width=\wdblefig]{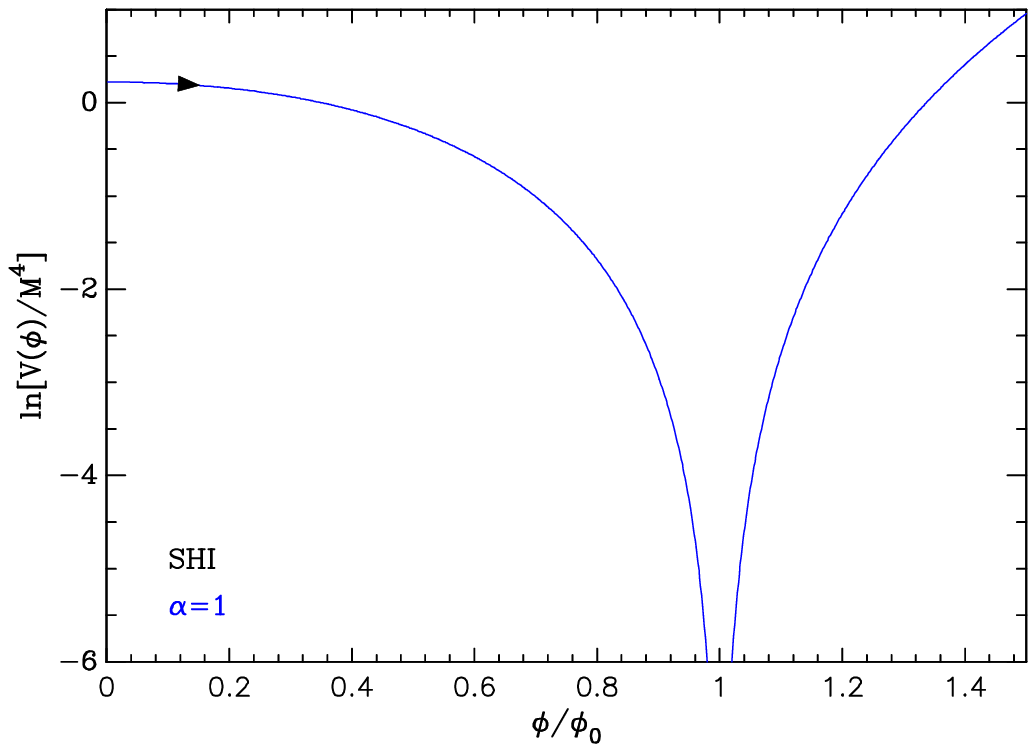}
\includegraphics[width=\wdblefig]{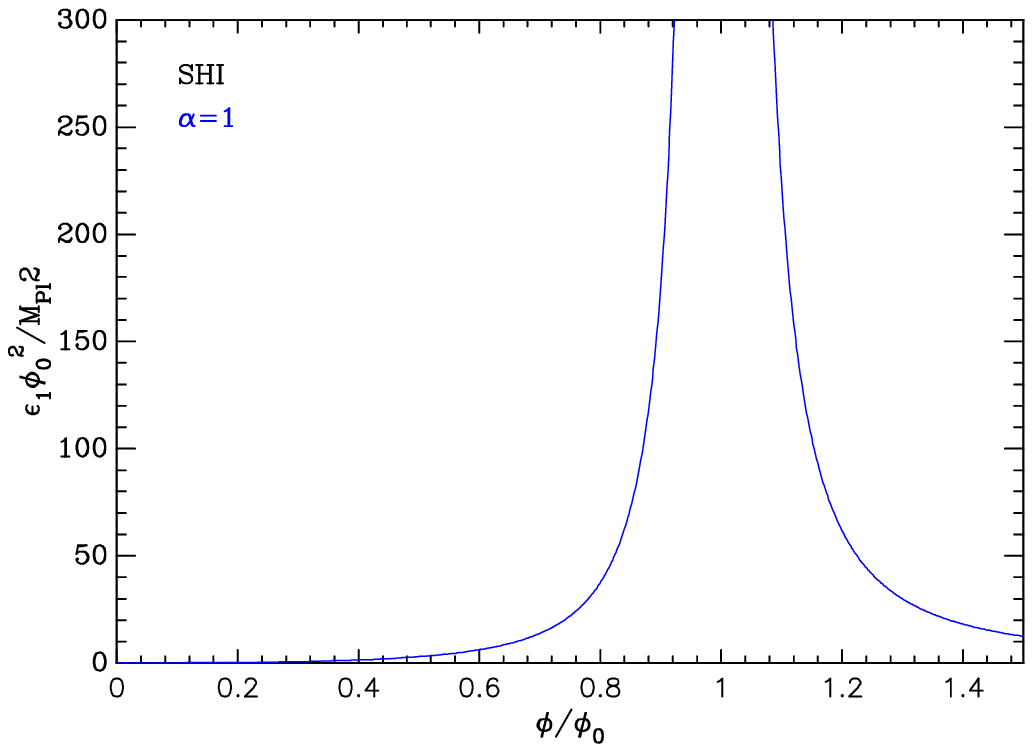}
\includegraphics[width=\wdblefig]{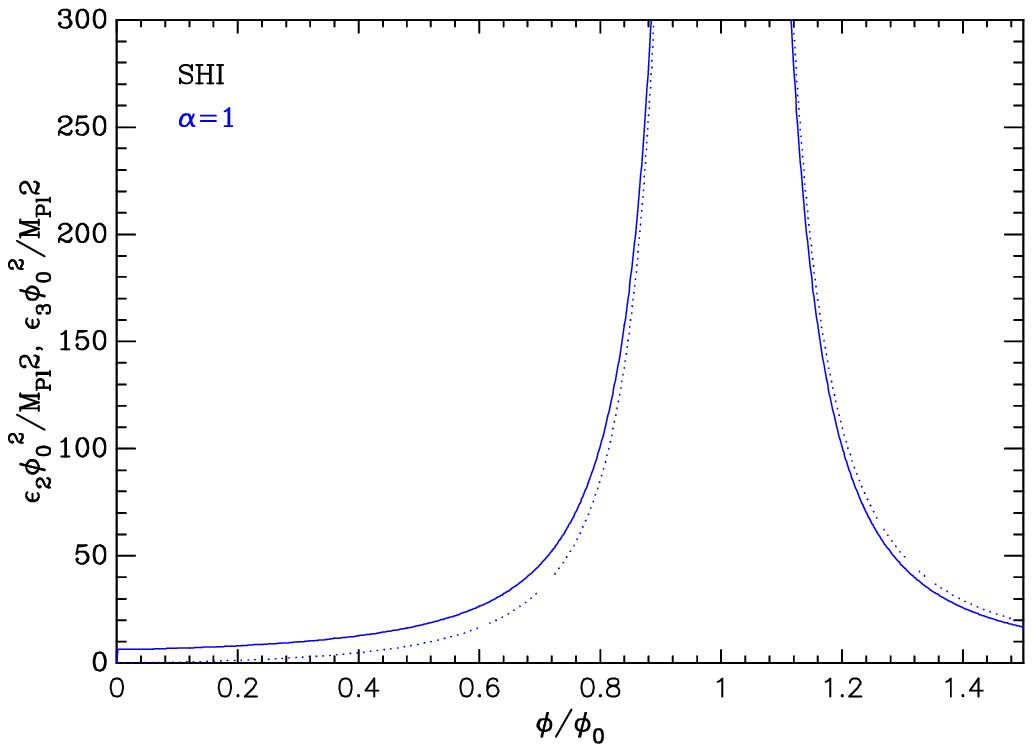}
\caption{Smeared Higgs Inflation potential (SHI) for $\alpha=1$. Top left
  panel: the potential as a function of $\phi/\phizero$.  Top right panel:
  logarithm of the potential. Bottom left panel: rescaled slow-roll
  parameter $\epsilon _1 \phizero^2/\Mp^2$. Bottom right panel: rescaled
  slow-roll parameters $\epsilon_2 \phizero^2/\Mp^2$ (solid line) and
  $\epsilon_3 \phizero^2/\Mp^2$ (dotted line).}
\label{potshi}
\end{center}
\end{figure}

The potential~\eqref{eq:pot:th:shi} can be more conveniently rewritten as 
\begin{equation}
V=M^4\left\lbrace \left[1-\left(\frac{\phi}{\phizero}\right)^2\right]^2 
+ \alpha\left(\frac{\phi}{\phizero}\right)^4
\left[\ln\left(\frac{\phi}{\phizero}\right)-\frac{1}{4}\right]
+\frac{\alpha}{4}\right\rbrace,
\end{equation}
where we have introduced $M\equiv m^2 \phizero^2/4$ and $\alpha\equiv
A\phizero^4/M^4$. It is a two-parameter potential, defined for
$\phi>0$, where the DWI-limit now corresponds to $\alpha\to 0$ and the
CWI-limit to $\alpha\to \infty$. It is represented in
\Fig{potshi}. Starting from $\phi=0$ where $V'=0$, it decreases until
$\phi=\phizero$ where it vanishes, and then it increases with $\phi$
at $\phi>\phizero$. As a consequence, there are a priori two relevant
regimes for inflation: a hilltop regime at $\phi<\phizero$, and a
large-field regime at $\phi>\phizero$. However, since the model was
introduced in \Refc{Senoguz:2015lba} as a hilltop model, we will focus
on the first regime. Moreover, in the large-field regime, the
potential is approximately quartic, which is strongly disfavored by
CMB observations (see the discussion regarding LFI${}_4$ in
\sectionc{sec:lfi} and \Fig{fig:CMBLFI}).

Defining 
\begin{equation}
x\equiv \frac{\phi}{\phizero}\, ,
\end{equation}
the Hubble-flow functions in the slow-roll approximation are given by
\begin{equation}
\label{eq:shi:eps1}
\epsilon_1 = \frac{8\Mp^2}{\phizero^2}
x^2\frac{
\left[-1+x^2+\alpha x^2\ln (x)\right]^2}
{\{\left(1-x^2\right)^2+\alpha x^4\left[\ln(x)-\frac{1}{4}\right]
+\frac{\alpha}{4}\}^2} ,
\end{equation}
\begin{align}
\epsilon_2 =&
\frac{32\Mp^2}{\phizero^2}
\Biggl\{\left(1-x^2\right)
\left[4+\alpha -\alpha \left(6+\alpha\right) x^2
-\left(4-\alpha+\alpha^2\right)x^4\right]
\nonumber \\ &  
-\alpha x^2
\left[\left(\alpha-8\right)x^4
+4x^2+3(4+\alpha)\right]
\ln \left(x\right)
+4\alpha^2 x^6
\ln ^2\left(x\right)
\Biggr\}
\nonumber \\ &   \times
\Biggl[4+\alpha-(\alpha-4)x^4
-8x^2
+4\alpha x^4
\ln \left(x\right)\Biggr]^{-2} ,
\end{align}
\begin{align}
\epsilon_3 =& 16 \frac{\Mp^2}{\phizero^2}
x^2
\left[-1+x^2+\alpha x^2
\ln \left(x\right)\right]
\Biggl\{
(4+\alpha)(8-6\alpha+5\alpha^2)x^8
+32\alpha(\alpha+5)x^6
\nonumber \\  &
-2(96+96\alpha+30\alpha^2+5\alpha^3)x^4
-32(\alpha^2+\alpha-8)x^2
+(4+\alpha)(6+\alpha)(5\alpha -4)
\nonumber \\   &
+2\alpha \ln \left(x\right)
\Biggl[(48-16\alpha+7\alpha^2)x^8
+80\alpha x^6
-2(144+68\alpha+5\alpha^2)x^4
+48(4+\alpha)x^2
+3(4+\alpha)^2\Biggr]
\nonumber \\   &
+16\alpha^2 x^4 \ln ^2\left(x\right)
\Biggl[-6(\alpha+4)+(6-\alpha)x^4
+2\alpha x^4
\ln \left(x\right)\Biggr]\Biggr\}
\Biggl\{\Biggl[(\alpha-4)x^4
+8x^2
\nonumber \\  &
-(\alpha+4)
-4\alpha x^4
\ln \left(x\right)\Biggr]^2
\Biggl[(4-\alpha+\alpha^2)x^6
+(7\alpha-4)
x^4
-(4+7\alpha+\alpha^2)x^2
\nonumber \\  &
+4+\alpha +(8-\alpha)\alpha x^6
\ln \left(x\right)
-4\alpha x^4\ln \left(x\right)
-3\alpha(4+\alpha)x^2
\ln \left(x\right)
+4\alpha^2x^6\ln ^2\left(x\right)
\Biggr]\Biggr\}^{-1} ,
\end{align}
and are displayed in the lower panels of \Fig{potshi}. When $\phi<\phizero$, \ie $x<1$, they all increase with the field value $x$, hence they increase as inflation proceeds, and diverge in the limit $x\to 1$. In the opposite limit, when $x\to 0$, $\epsilon_1$ and $\epsilon_3$ vanish, while $\epsilon_2$ approaches a constant value
\begin{equation}
\label{eq:shi:eps2min}
\epstwomin =  \frac{32 \Mp^2}{(\alpha +4)\phizero^2}\, .
\end{equation}
As a consequence, slow-roll inflation requires $\phizero^2/\Mp^2 \gg 1/(\alpha+4)$. Inflation ends when $\epsilon_1=1$, at a field value $\phiend$ that needs to be determined numerically. The slow-roll trajectory,
\begin{equation}
\label{eq:shi:traj}
\Nend - N = \frac{\phizero^2}{\Mp^2}
\int _{\phiend/\phizero}^{\phi/\phizero}
{\dd} x \frac{(1-x^2)^2+\alpha x^4(\ln x-1/4)+\frac{\alpha}{4}}
{4x(-1+x^2+\alpha x^2\ln x)}\, ,
\end{equation}
also needs to be integrated and inverted numerically. Combined with the
reheating equation~\eqref{eq:phistarlnrrad}, this allows us to determine
$\xstar$, the field value at which the pivot mode crosses out the Hubble
radius during inflation. In turn, this determines the mass scale $M$
of the potential from the CMB normalization and one finds
\begin{equation}
\left(\dfrac{M}{\Mp}\right)^4 = 11520 \pi^2 \dfrac{\Mp^2}{\phizero^2}
\xstar^2\frac{ \left[-1+\xstar^2+\alpha \xstar^2\ln
    \left(\xstar\right)\right]^2} {\{\left(1-\xstar^2\right)^2+\alpha
  \xstar^4\left[\ln\left(\xstar\right)-\frac{1}{4}\right]
  +\frac{\alpha}{4}\}^3} \dfrac{\Qrms^2}{T^2}\,.
\end{equation}
Let us note that inflation necessarily explores the regime where $x$
is of order one, such that one cannot use Taylor expansions in $x$ in
order to approximate the slow-roll trajectory. Indeed, if one assume
that $\xend\ll 1$, \Eq{eq:shi:eps1} gives rise to
$\xend\simeq(4+\alpha)\phizero/(8\sqrt{2}\Mp)$, which is much smaller
than one provided $\phizero^2/\Mp^2\ll 1/(4+\alpha)^2$. This leads to
$\epstwomin \gg 1$, hence it discards this possibility.

The reheating-consistent slow-roll predictions of SHI are displayed in
\Figs{fig:CMBSHI_0} to \ref{fig:CMBSHI_2}, for $\phizero/\Mp=10,\,
15,\,20$ and $25$ respectively, and various values of $\alpha$. One
notices that when $\phizero/\Mp\gg 1$, the model's predictions approach
the ones of LFI${}_2$ (see \sectionc{sec:lfi} and \Fig{fig:CMBLFI}).
This is because, in that regime, the last \efolds~of inflation are
realized close to the quadratic minimum of the potential at
$x=1$. Indeed, from \Eq{eq:shi:eps1}, one can check that inflation
ends at $\phiend = \phizero-\sqrt{2}\Mp$ in this limit (so
$\xend \simeq 1-\sqrt{2}\Mp/\phizero$ is close to one), which coincides
with \Eq{eq:phiendlf} with $p=2$ and when the field value is displayed
by $\phizero$. The slow-roll trajectory~\eqref{eq:shi:traj} can then be
integrated as
\begin{equation}
\xstar\simeq 1-2 \frac{\Mp}{\phizero} \sqrt{\frac{1}{2}+\Delta\Nstar}\, ,
\end{equation}
which is again close to one when $\phizero \gg \Mp$.
The slow-roll parameters at Hubble crossing of the pivot scale are given by
\begin{equation}
  \epsonestar \simeq\frac{1}{2\left(\Delta
      \Nstar+1/2\right)}\, ,\qquad \epstwostar\simeq\frac{1}{\Delta 
      \Nstar+1/2}\,,
  \qquad \epsthreestar\simeq \epstwostar\, ,
\end{equation}
which coincides with \Eq{eq:lfi:epsstar} when $p=2$.

\subsection{Double Exponential Inflation (DEI)}
\label{sec:dei}

\subsubsection{Theoretical Justifications}
\label{subsubsec:theorydei}

The model was proposed in \Refc{Cadoni:2015iua} as a phenomenological
realization of hilltop inflation by means of a single-field potential
containing two exponential terms,
\begin{equation}
V(\phi)= \Lambda^4\left(\alpha_1 \ee^{\beta_1\frac{\phi}{\Mp}} + \alpha_2 \ee^{\beta_2\frac{\phi}{\Mp}}\right).
\end{equation}
In this expression, $\alpha_1$, $\beta_1$, $\alpha_2$ and $\beta_2$
are dimensionless parameters, and $\Lambda$ sets the overall scale of
the potential. Without loss of generality, one can set parameters such
that the top of the ``hill'' (\ie the local maximum of the potential)
corresponds to $\phi=0$, which amounts to imposing $V(0)>0$, $V'(0)=0$
and $V''(0)<0$. The condition $V''(0)<0$ implies that
$\alpha_1\beta_1^2 + \alpha_2\beta_2^2<0$, so $\alpha_1$ and
$\alpha_2$ have different signs. Since the ordering of the two
exponential terms is arbitrary, one can take $\alpha_1>0$ and
$\alpha_2<0$. The condition $V'(0)=0$ then leads to
$\alpha_1\beta_1+\alpha_2\beta_2=0$, so $\beta_1/\beta_2 =
-\alpha_2/\alpha_1\equiv \beta^2$, which defines the parameter $\beta$
and where we have used that $\alpha_1$ and $\alpha_2$ have different
signs. The condition $V(0)>0$, \ie $\alpha_1+\alpha_2>0$, implies that
$\beta^2<1$. Upon introducing $M^4\equiv \Lambda^4\alpha_1$ and
$\phizero\equiv \Mp\beta/\beta_1$, the potential reads
\begin{equation}
\label{eq:pot:dei}
V= M^4\left(\ee^{\beta\frac{\phi}{\phizero}} -\beta^2
\ee^{\frac{1}{\beta}\frac{\phi}{\phizero}}\right).
\end{equation}

\subsubsection{Slow-Roll Analysis}
\label{subsubsec:srdei}
\begin{figure}
\begin{center}
\includegraphics[width=\wdblefig]{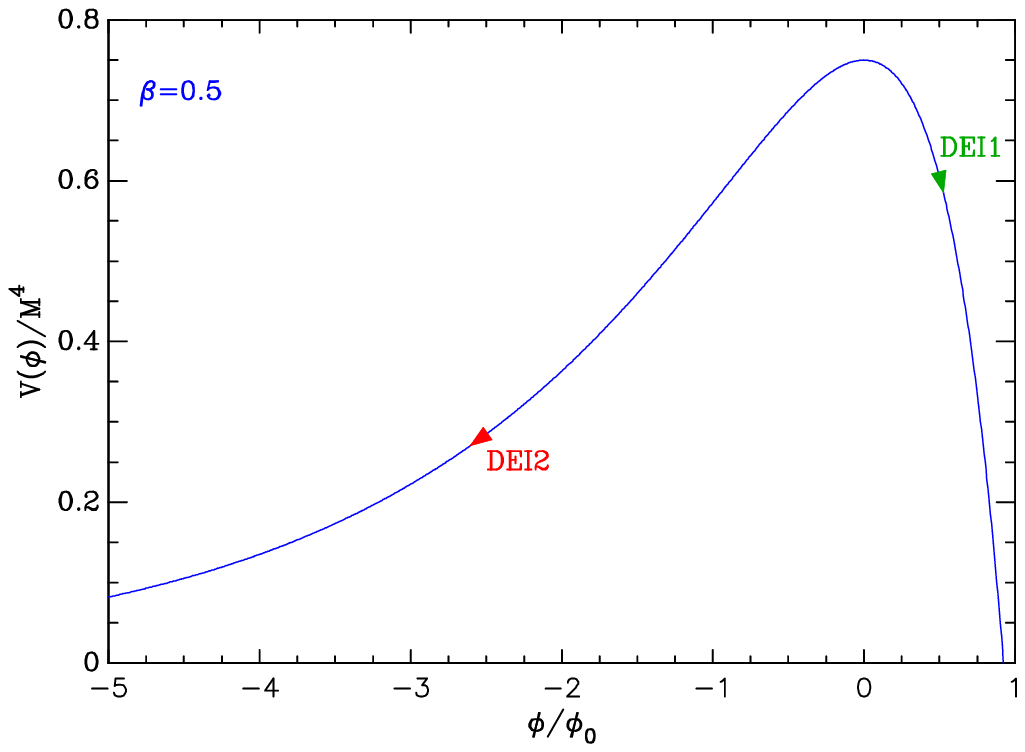}
\includegraphics[width=\wdblefig]{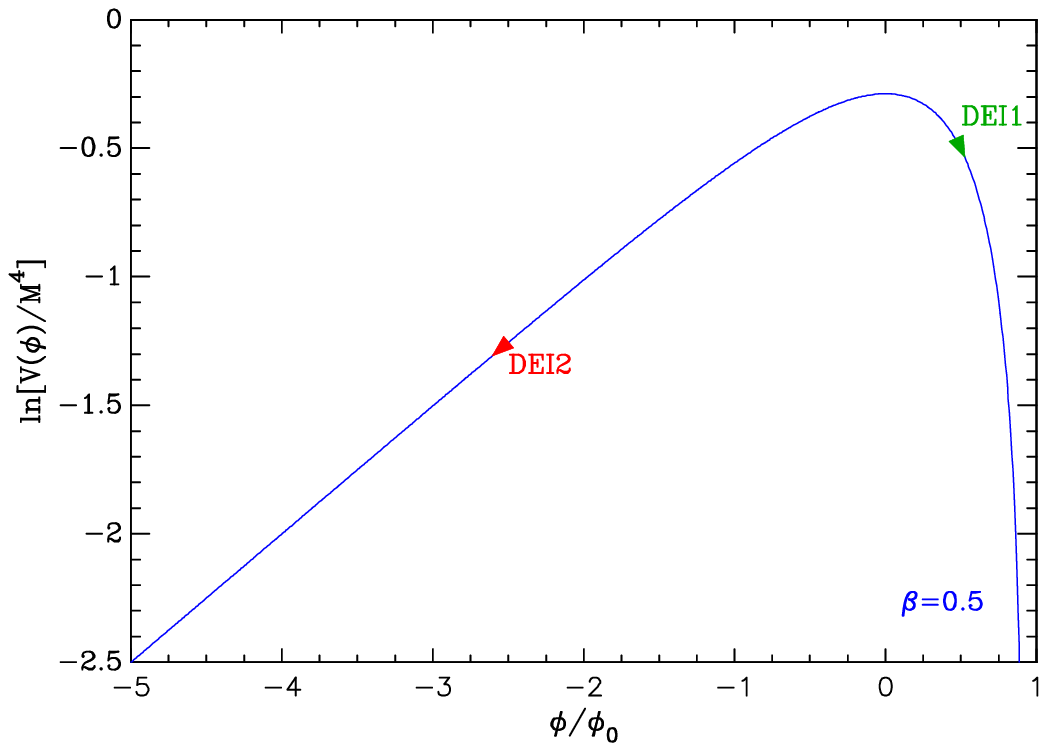}
\includegraphics[width=\wdblefig]{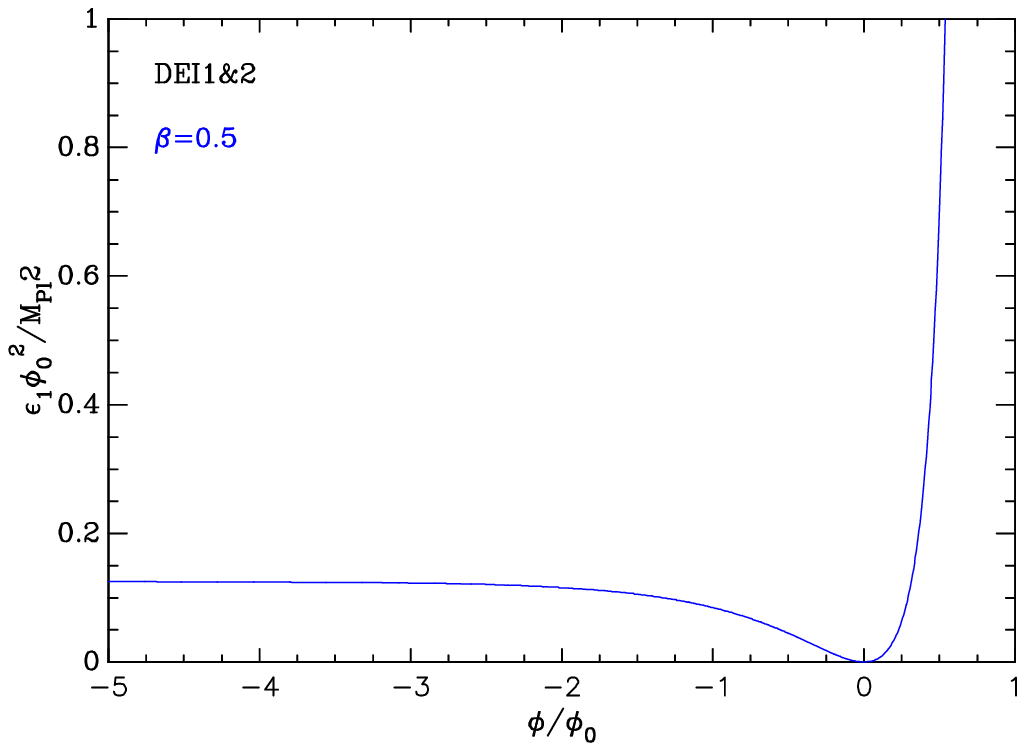}
\includegraphics[width=\wdblefig]{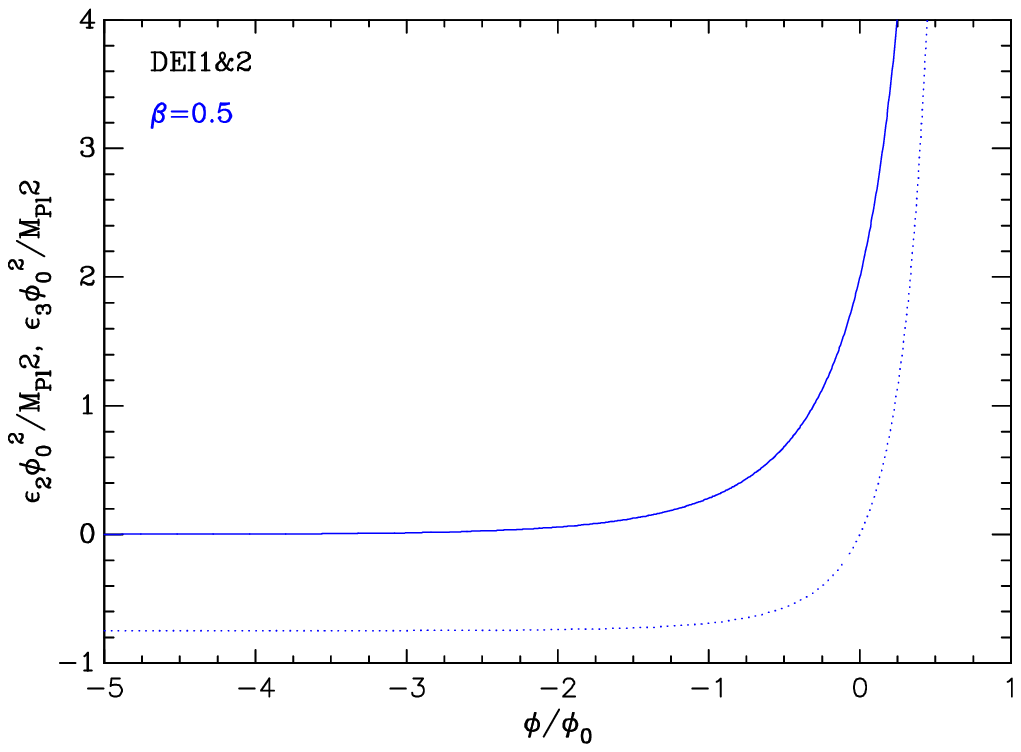}
\caption{Double Exponential Inflation potential (DEI) for $\beta=0.5$. Top left
  panel: the potential as a function of $\phi/\phizero$.  Top right panel:
  logarithm of the potential. Bottom left panel: rescaled slow-roll
  parameter $\epsilon _1 \phizero^2/\Mp^2$. Bottom right panel: rescaled
  slow-roll parameters $\epsilon_2 \phizero^2/\Mp^2$ (solid line) and
  $\epsilon_3 \phizero^2/\Mp^2$ (dotted line).}
\label{potdei}
\end{center}
\end{figure}

Let us now perform the slow-roll analysis of the
potential~\eqref{eq:pot:dei}. We recall that $\beta^2<1$, so
$-1<\beta<1$. However, since the potential is invariant under the
transformation $\beta\to - \beta$ and $\phi\to -\phi$, our
considerations can be restricted to the interval $0<\beta<1$. The
potential is displayed in \Fig{potdei}. It is maximal at $\phizero$,
and possesses two regimes of inflation. When $\phi>0$, the potential
decreases with $\phi$ until it vanishes at
\begin{equation}
\phiVzero=2\phizero\beta \frac{\ln \beta}{\beta ^2-1}\, ,
\end{equation}
above which it is negative. One can check that, for $0<\beta<1$, one
has $\phiVzero>0$. This is why there is a first regime of inflation,
at $0<\phi<\phiVzero$, that we denote DEI1. When $\phi<0$, the
potential decreases as $\phi$ decreases and asymptotes $0$ at $\phi\to
-\infty$. This second regime of inflation will be denoted DEI2.

Defining 
\begin{equation}
x\equiv \frac{\phi}{\phizero}\, ,
\end{equation}
the Hubble-flow functions in the slow-roll approximation are given by
\begin{equation}
\label{eq:dei:sr1}
\epsilon_1 = \frac{\beta ^2}{2}\frac{\Mp^2}{\phizero^2}
\frac{\left(e^{\beta x}-
    e^{x/\beta}\right)^2}
{\left(e^{\beta x}-\beta ^2
    e^{x/\beta}\right)^2}\,,
\end{equation}
\begin{equation}
\label{eq:dei:sr2}
\epsilon_2 = 2\frac{\Mp^2}{\phizero^2}
\left(\beta^2-1\right)^2  
\frac{e^{(1+\beta ^2)x/\beta }}
{\left(e^{\beta x}-\beta ^2
    e^{x/\beta}\right)^2}\,,
\end{equation}
\begin{equation}
\label{eq:dei:sr3}
\epsilon_3 = \frac{\Mp^2}{\phizero^2}
\left(\beta ^2-1\right)
\frac{\left(e^{\beta x}-e^{x/\beta}\right)
\left(e^{\beta x}+\beta^2e^{x/\beta}\right)}
{\left(e^{\beta x}-\beta ^2e^{x/\beta}\right)^2}\,,
\end{equation}
and are displayed in the lower panels of \Fig{potdei}. Let us describe
their behavior in the two regimes of interest.

If $x>0$, the three Hubble-flow parameters increase with $x$,
and diverge as $x$ approaches $\xVzero$. Inflation terminates by
slow-roll violation when $\epsilon_1(x)=1$, at a positive field value given by
\begin{equation}
\xend = \xepsoneOnePlus = \frac{\beta }{\beta ^2-1}
\ln \left(\frac{\beta \sqrt{2}\frac{\phizero}{\Mp}+1}
{ \frac{\sqrt{2}}{\beta}\frac{\phizero}{\Mp}+1}\right) .
\end{equation}
The first and third Hubble-flow parameters vanish at $x=0$, while
the second Hubble-flow parameter is
\begin{equation}
  \epstwomin(x>0) = 2\frac{\Mp^2}{\phizero^2}\, .
\end{equation}
For this reason, slow-roll inflation in DEI1 requires that $\phizero\gg 1$. 

If $x<0$, $\epsilon_1$ increases away from $0$ as $x$ decreases, and reaches the asymptotic value
\begin{equation}
\epsonemax(x<0)= \frac{\beta^2}{2}\frac{\Mp^2}{\phizero^2}\,,
\label{eq:dei:epsonemaxneg}
\end{equation}
when $x\to -\infty$. Whether or not inflation ends by slow-roll
violation in DEI2 thus depends on the values of $\beta$ and
$\phizero$. More precisely, if $\beta>\sqrt{2}\phizero/\Mp$, then
\Eq{eq:dei:epsonemaxneg} becomes larger than unity and inflation ends
at the field value solution of $\epsilon_1(x)=1$, in the negative
field domain, given by
\begin{equation}
\label{eq:dei2:xend}
\xepsoneOneMinus = \frac{\beta }{\beta ^2-1}
\ln \left(\frac{\beta \sqrt{2}\frac{\phizero}{\Mp}-1}
{ \frac{\sqrt{2}}{\beta}\frac{\phizero}{\Mp}-1}\right) .
\end{equation}
Otherwise, if $\beta<\sqrt{2}\phizero/\Mp$, inflation does not stop by
violation of the slow-roll conditions and one needs to invoke other
mechanisms, which results in the introduction of another free
parameter $\xend$. This possibility is not discussed in
\Refc{Cadoni:2015iua}, and would otherwise corresponds to a PLI
regime. For these reasons, we do not consider it either, and impose
the condition $\beta>\sqrt{2}\phizero/\Mp$. Note that since $\beta<1$,
for this regime to exist, one needs to assume
$\phizero/\Mp<1/\sqrt{2}$.

Since $\epsilon_2$ decreases as inflation proceeds in DEI2, its
minimum value is obtained by evaluating \Eq{eq:dei:sr2} at
$\xepsoneOneMinus$ given by \Eq{eq:dei2:xend}. Given that
$\phizero/\Mp<1/\sqrt{2}$, the resulting expression can be evaluated
in the limit $\phizero\ll\Mp$, and one obtains
\begin{equation}
\epstwomin(x<0) \simeq 2\frac{\Mp^2}{\phizero^2}\, ,
\end{equation}
which coincides with $\epstwomin(x>0)$, \ie with the value of
$\epsilon_2$ at the maximum of the potential. One has therefore
$\epstwomin \gg 1$ in this regime, which excludes the possibility
to realize slow-roll inflation. The only remaining solution would be
to fine tune $\beta$ close to $\sqrt{2}\phizero/\Mp$. In that case,
however, inflation would end at very large negative values of $x$,
where the potential is dominated by its first exponential branch. The
predictions of the model become again close to the ones of Power-Law
Inflation (PLI, see \sectionc{sec:pli}) in this regime. This is why we
will not further consider the regime DEI2, and will focus on DEI1
hereafter.

The slow-roll trajectory can be integrated, and one obtains
\begin{equation}
\label{eq:deitrajec}
\Nend - N = \frac{1+\beta ^2}{\beta}\frac{\phizero^2}{\Mp^2}
\left(x-\xend\right)
+\frac{\phizero^2}{\Mp^2}
\ln \left(\frac{e^{\xend/\beta}-e^{\beta \xend}}
{e^{x/\beta}-e^{\beta x}}
\right).
\end{equation}
When $x \to 0$, the number of {\efolds} diverges, which indicates that
one can always realize a sufficient number of {\efolds} by starting
close enough to the maximum of the potential. Unfortunately, this
trajectory needs to be inverted numerically. Combined with the
reheating equation~\eqref{eq:phistarlnrrad}, this allows us to
determine $\xstar$, the field value at which the pivot mode crosses
out the Hubble radius during inflation. In turn, this determines the
mass scale $M$ of the potential from the CMB normalization and one
finds
\begin{equation}
\left(\dfrac{M}{\Mp}\right)^4 =  720\pi^2 \beta^2\dfrac{\Qrms^2}{T^2}\frac{\Mp^2}{\phizero^2}
\frac{\left(e^{\beta \xstar}-
    e^{\xstar/\beta}\right)^2}
{\left(e^{\beta \xstar}-\beta ^2
    e^{\xstar/\beta}\right)^3}\,.
\end{equation}

The reheating-consistent slow-roll predictions of DEI1 are displayed
in \Figs{fig:CMBDEI_0} to \ref{fig:CMBDEI_3} for $\phizero/\Mp=10,\,
20,\, 50$ and $100$ respectively.  In DEI1, as argued above, slow-roll
inflation requires $\phizero\gg \Mp$. This is why, in order to gain
some analytical insight, it is useful to expand the above expressions
in this limit. In this regime (more precisely, under the condition
$\phizero/\Mp \gg 1/\beta$), one has $\xend \simeq
2\beta\ln(\beta)/(\beta^2-1)$, and the slow-roll trajectory can be
approximated as $\xstar \simeq \xend -\sqrt{2}\Delta
\Nstar\Mp/\phizero $, which gives rise to
\begin{equation}
  \epsonestar \simeq\frac{1}{\left(1+2\Delta
      \Nstar\right)^2}\, ,\qquad \epstwostar\simeq\epsthreestar\simeq 4\epsonestar \, .
\end{equation}

\subsection{S-Dual Inflation (SDI)}

\label{sec:sdi}

\begin{figure}
\begin{center}
\includegraphics[width=\wdblefig]{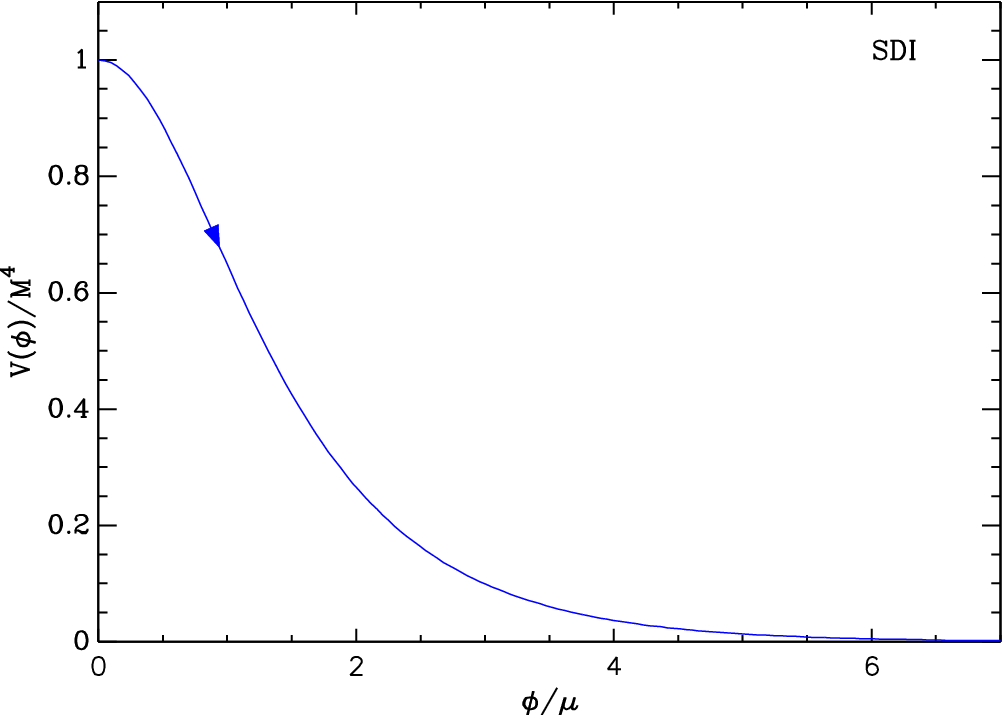}
\includegraphics[width=\wdblefig]{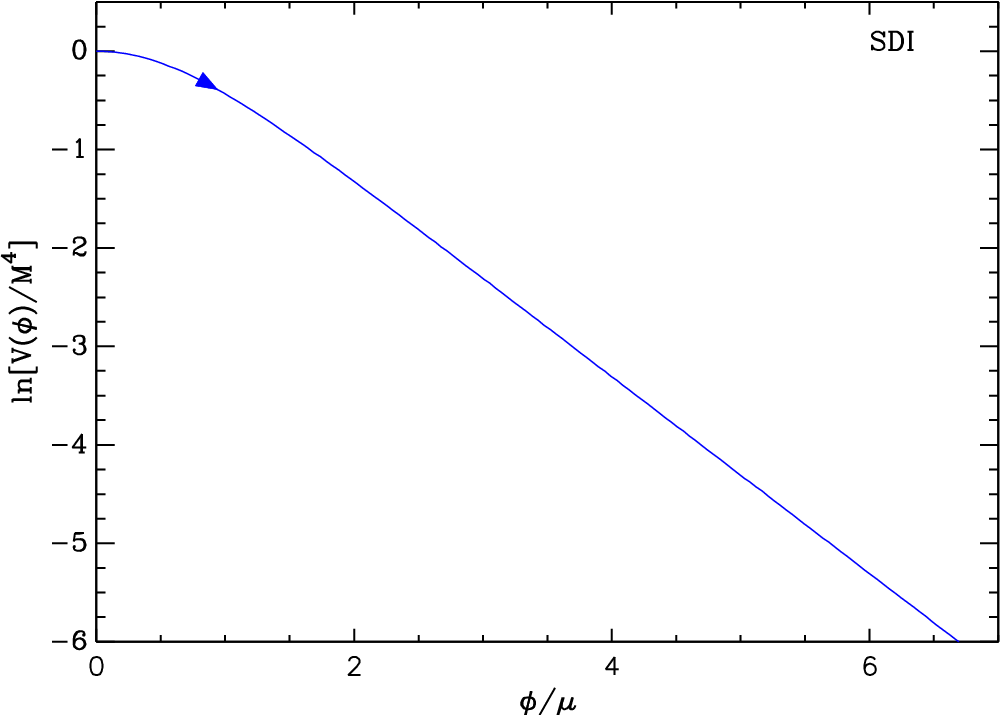}
\includegraphics[width=\wdblefig]{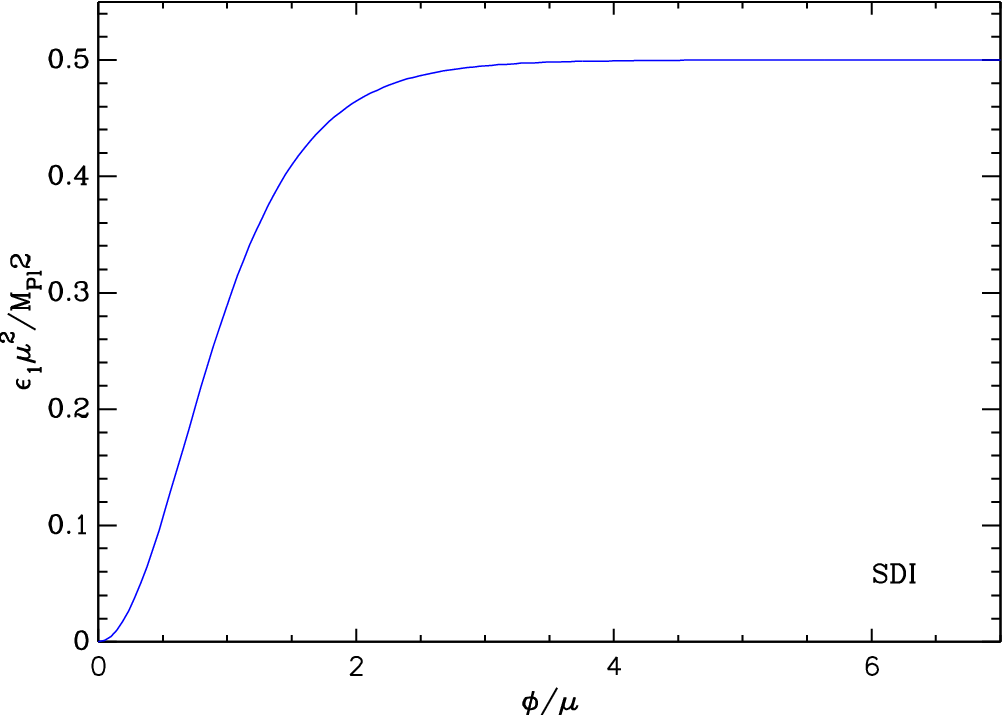}
\includegraphics[width=\wdblefig]{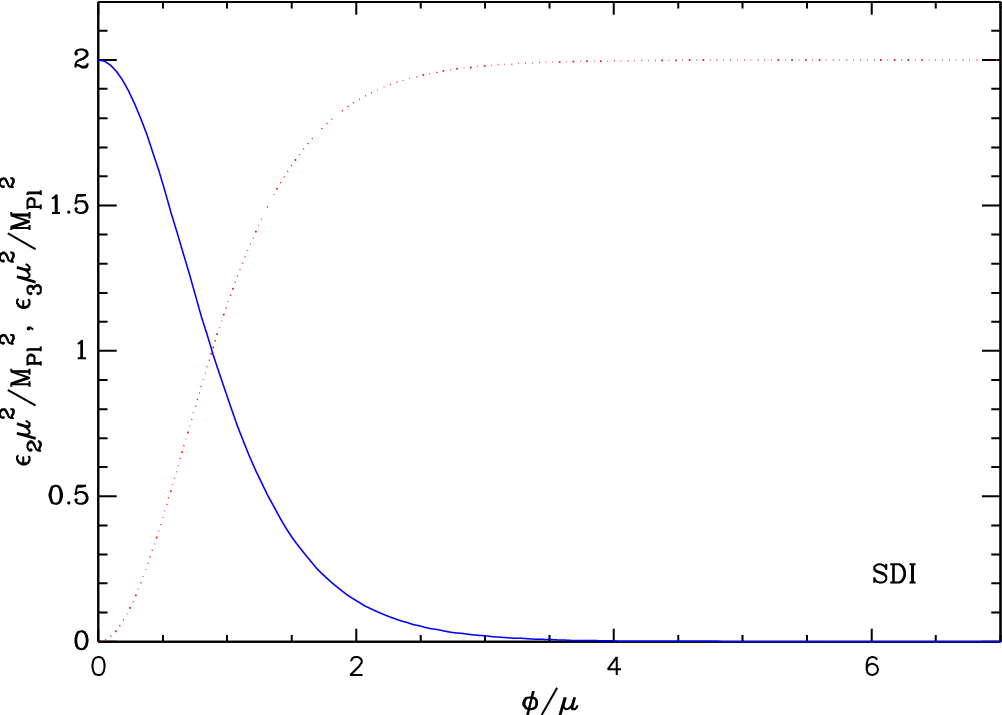}
\caption{S-Dual Inflation (SHI). Top left panel: the potential as a
  function of $\phi/\mu$.  Top right panel: logarithm of the
  potential. Bottom left panel: rescaled slow-roll parameter
  $\epsilon_1 \mu^2/\Mp^2$. Notice that $\epsilon_1$ is always smaller
  than unity for super-Planckian $\mu$, in which case inflation does not
  gracefully ends. Bottom right panel: rescaled slow-roll parameters
  $\epsilon_2 \mu^2/\Mp^2$ (solid line) and $-\epsilon_3 \mu^2/\Mp^2$
  (red dotted line).}
\label{fig:potsdi}
\end{center}
\end{figure}

This scenario has been proposed in \Refc{Anchordoqui:2014uua} and
motivated by the wish to have inflation producing a significant amount
of tensor modes while having a concave potential. It is loosely
motivated by the S-duality in String Theory as the inflaton is
considered to be a dilaton field. Because the string coupling constant
is given by $g \propto e^{\phi/\mu}$, symmetry under the S-duality
transformation $g \rightarrow 1/g$ requires the potential to be
symmetric under parity $\phi \rightarrow -\phi$. Moreover, since a
low-energy effective action should be an expansion in the string
coupling constant $g$, the potential should be made of exponential
terms. From these motivations, \Refc{Anchordoqui:2014uua} considers a
potential of the form
\begin{equation}
V(\phi)  = \dfrac{M^4}{\cosh\left(\dfrac{\phi}{\mu} \right)}\,,
\label{eq:potsdi}
\end{equation}
where $\mu$ is a typical vacuum expectation value for the dilaton field.

The potential is even, by construction, so we can restrict our
analysis to positive field values only. It is a monotonic decreasing
function of the field and inflation proceeds at increasing field
values. Defining
\begin{equation}
  x \equiv \dfrac{\phi}{\mu}\,,
\end{equation}
the Hubble flow functions in the slow-roll approximation read
\begin{equation}
\epsilon_1 = \dfrac{\Mp^2}{2 \mu^2} \tanh^2(x), \qquad \epsilon_2 = 
\dfrac{2\Mp^2}{\mu^2} \dfrac{1}{\cosh^2(x)}\,, \qquad \epsilon_3 =
-\dfrac{2\Mp^2}{\mu^2} \tanh^2(x) = -4 \epsilon_1\,.
\label{eq:sdisr123}
\end{equation}
The potential and the Hubble-flow functions have been presented in
\Fig{fig:potsdi}. As this figure emphasizes, the first Hubble-flow
function asymptotes to
\begin{equation}
\epsonemax = \dfrac{2 \Mp^2}{\mu^2}\,,
\end{equation}
at large field values. As a result, inflation ends naturally only for
$\mu < \sqrt{2} \Mp$ and at a field value given by
\begin{equation}
  \xepsoneOne =   \arctanh\left(\sqrt{2} \dfrac{\mu}{\Mp} \right).
\end{equation}
In this regime, inflation proceeds at increasing field value within the domain
$0<x<\xepsoneOne$. However, as can be seen in the bottom-right panel of
\Fig{fig:potsdi}, $\epsilon_2$ may be larger than unity in this
region. More precisely, one has $\epsilon_2(x)=1$ at the field value
\begin{equation}
  \xepstwoOne = \arccosh\left( \dfrac{\sqrt{2} \Mp}{\mu} \right).
\end{equation}
For all $\mu < \sqrt{2/5}\Mp$, one has $\xepstwoOne > \xepsoneOne$,
and since $\epsilon_2$ is a decreasing function of the field value,
this implies that slow roll is violated $\epsilon_2>1$ over the whole
inflating domain. One may want to restrict $\mu$ to the range
$\sqrt{2/5}< \mu/\Mp < 1/\sqrt{2}$, for which $\epsilon_2$ is smaller
than one when inflation ends, but one can check that the values of
$\epsilon_2$ in the relevant part of the inflationary dynamics are
still too large to produce a viable inflationary scenario. For these
reasons, we now consider only the super-Planckian values of $\mu >
\Mp/\sqrt{2}$, for which an additional mechanism has to be invoked to
end inflation. This could be, for instance, a tachyonic instability
triggered by an additional field. We denote the field value at which
inflation ends by $\xend = \phiend/\mu$ making SDI a two-parameter
model.

The slow-roll trajectory can be integrated analytically and reads
\begin{equation}
\Nend - N = \dfrac{\mu^2}{\Mp^2} \ln\left[\dfrac{\sinh(\xend)}{\sinh(x)}\right],
\label{eq:sditraj}
\end{equation}
which can be inverted as
\begin{equation}
x = \arcsinh\left[ e^{-\frac{\Mp^2(\Nend-N)}{\mu^2}} \sinh(\xend)\right].
\label{eq:sditrajx}
\end{equation}
Combined with the reheating equation~\eqref{eq:phistarlnrrad}, one can
determine uniquely $\xstar$, the field value at which the pivot mode
crossed the Hubble radius during inflation. The mass scale of the
potential is then given by the CMB normalization and one finds
\begin{equation}
\left(\dfrac{M}{\Mp}\right)^4 = 720 \pi^2 \dfrac{\Mp^2}{\mu^2}
\sinh(\xstar) \tanh(\xstar) \dfrac{\Qrms^2}{T^2} .
\end{equation}
The reheating consistent slow-roll predictions for SDI have been
plotted in \Fig{fig:CMBSDI_0}. At small values of $\xend$, the model
predictions asymptote a $\mu$-dependent constant spectral index with a
very small amount of gravitational waves. This can be immediately
understood from \Eq{eq:sdisr123}. The inflationary domain being at
$x<\xend$, in the limit of small $x$ one has $\epsilon_1 \to 0$ and
$\epsilon_2 \to 2\Mp^2/\mu^2$, which is typical of a small-field model
with non-vanishing mass (see SFI2 in \sectionc{sec:sfi}). At large
values of $\xend$, one can check that, for mildly super-Planckian
values of $\mu$, a substantial amount of gravitational waves can be
produced (as mentioned above this was one of the original motivations
for this model, although it occurs in the convexe region of the
potential), since $\epsilon_1$ asymptotes a constant at large-field
values and the tensor-to-scalar ratio is controlled by $\epsilon_1$ at
leading order in slow roll.

\subsection{Generalized Double Well Inflation (GDWI)}
\label{sec:gdwi}

These models are a generalization of Double Well Inflation (DWI)
discussed in \sectionc{sec:dwi} and are of the ``Mexican-hat'' type. The
potential is given by
\begin{equation}
V(\phi) = M^4 \left[\left(\dfrac{\phi}{\phizero}\right)^{2p} - 1\right]^2,
\label{eq:potgdwi}
\end{equation}
where $\phizero$ is a {\vev} and $p > 1$ is the power index. The case
$p=1$ corresponds to DWI, which is presented in \sectionc{sec:dwi}. There,
it is shown that DWI has different observable predictions than the
quadratic Small Field Inflation (SFI) model of \sectionc{sec:sfi} and,
as such, cannot be simply viewed as a large-field regularization of the
SFI potential. This is due to the fact that both DWI and quadractic
SFI support slow-roll inflation only for $\phizero > \Mp$ for which
they significantly differ. Indeed, at the top of the potential, the
second Hubble-flow function for these two models is given by
$\epsilon_2 = 8 \Mp^2/\phizero^2$ and it would exceed unity for
$\phizero < \Mp$. Such a feature comes from the fact that a quadratic
term in the potential implies a non-vanishing effective mass at the
top of the potential.

For $p>1$ the effective mass term vanishes and GDWI can
support slow-roll inflation for both $\phizero < \Mp$ and $\phizero
\ge \Mp$. As such, for sub-Planckian $\phizero$, GDWI with $p>1$ can be
considered as a UV completion of the SFI models with the same power
index $p$ (see \sectionc{sec:sfi}). In particular, the case $p=2$ has
been numerically studied in \Refc{Chowdhury:2019otk} and shown to
smoothly regularize the quartic SFI model while reducing some of
its fine-tuning issues at very small {\vev} $\phizero \ll \Mp$.

\begin{figure}
\begin{center}
\includegraphics[width=\wdblefig]{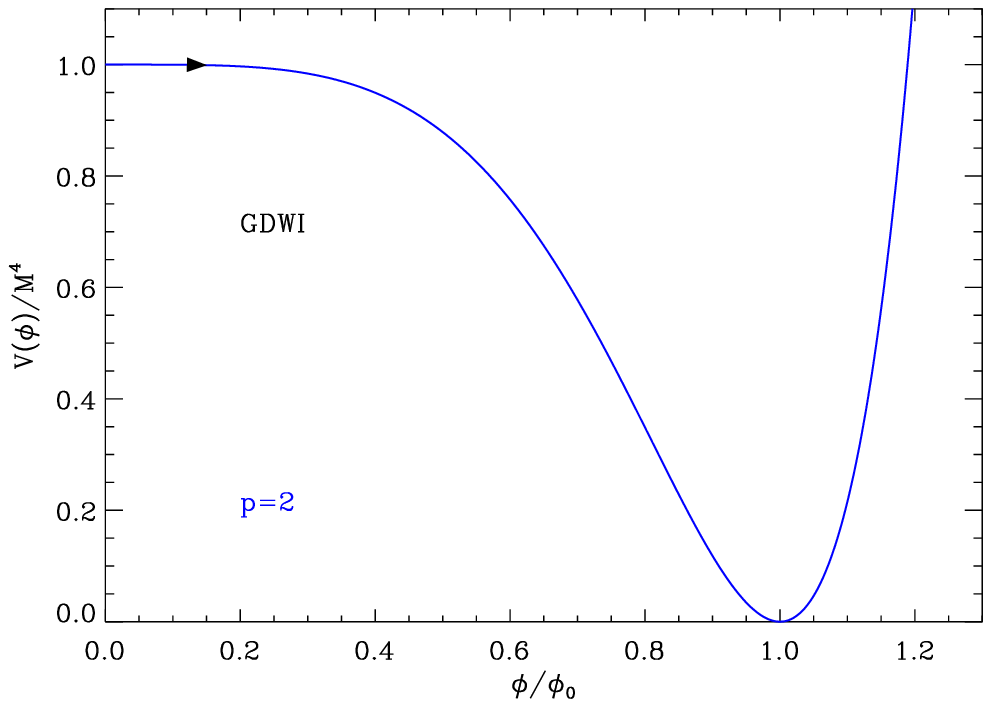}
\includegraphics[width=\wdblefig]{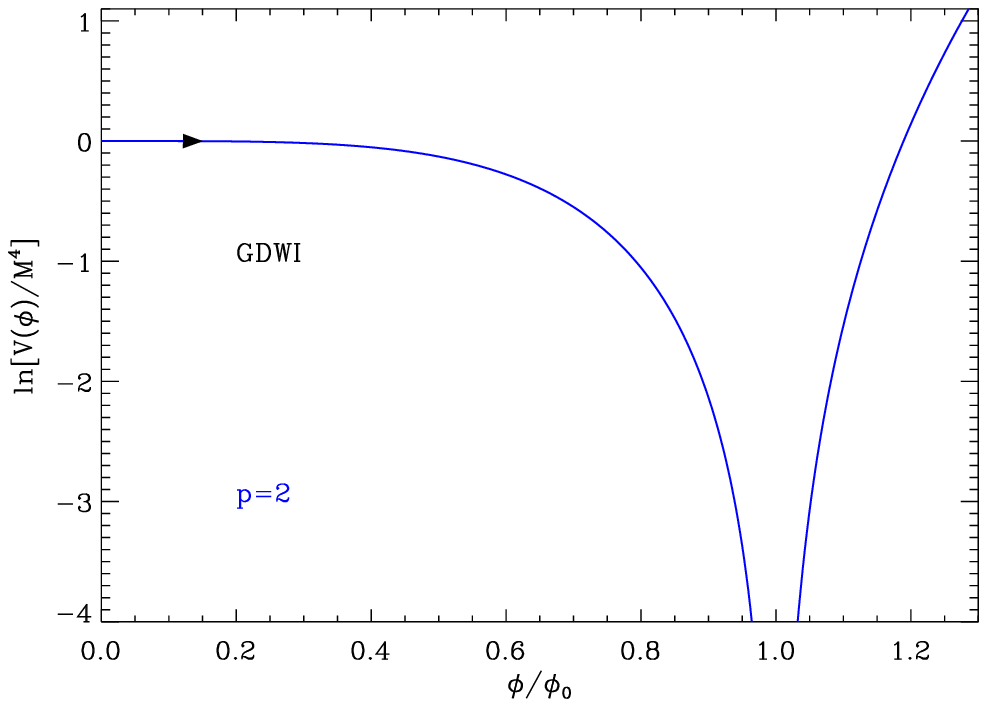}
\includegraphics[width=\wdblefig]{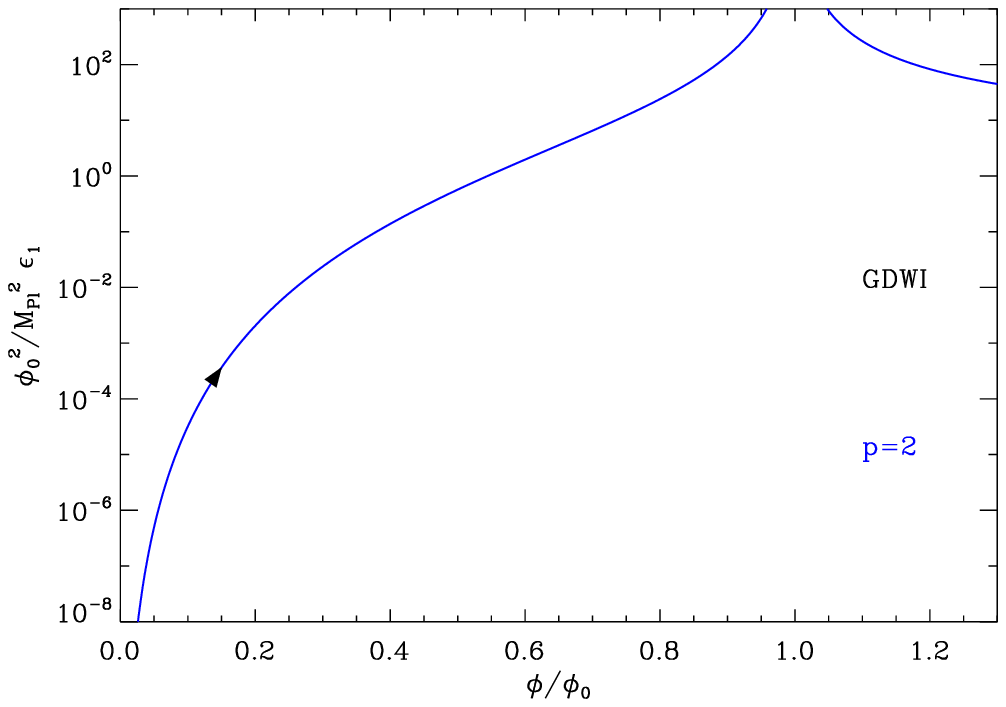}
\includegraphics[width=\wdblefig]{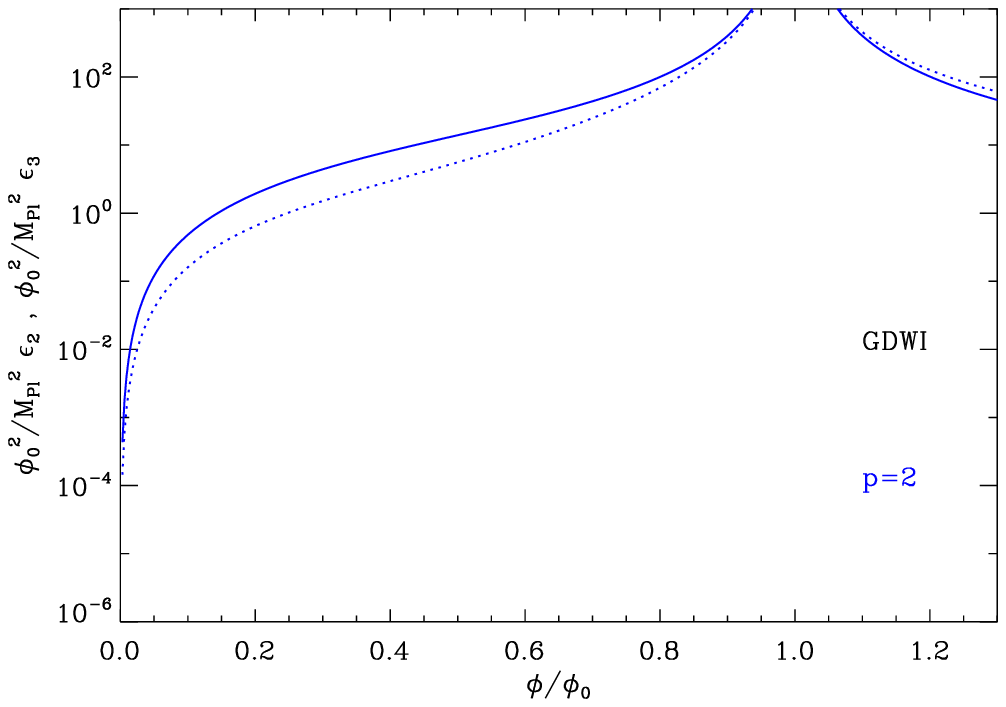}
\caption{Generalized Double Well Inflation for $p=2$. Upper panels:
  the potential and its logarithm as a function of
  $\phi/\phizero$. Only the positive domains are represented as the
  potential is symmetric under $\phi \to - \phi$. Bottom left panel:
  first Hubble-flow function $\epsilon_1$, divided by
  $\Mp^2/\phizero^2$, as a function of $\phi/\phizero$. The bottom
  right panel shows $\epsilon_2$ (solid line) and $\epsilon_3$ (dotted
  line), both divided by $\Mp^2/\phizero^2$, as a function of
  $\phi/\phizero$.}
\label{potgdwi}
\end{center}
\end{figure}

The potential of GDWI, and its logarithm, have been represented in
\Fig{potgdwi} as a function of $\phi/\phizero$. The negative domain is
not represented as the potential is symmetric under the mapping
$\phi\to-\phi$. Moreover, we consider only the inflationary domain
$\phi < \phizero$. The potential can also inflate at large field
values but, in this region, it behaves exactly as Large Field
Inflation (LFI) with a power index of $4p$ (see \sectionc{sec:lfi}).

Defining the dimensionless field value
\begin{equation}
x \equiv \dfrac{\phi}{\phizero}\,,
\end{equation}
the Hubble-flow functions, in the slow-roll approximation, reduce to
\begin{equation}
\epsilon_1 =  8 p^2 \left(\dfrac{\Mp}{\phizero}\right)^2 \dfrac{
  x^{2(2p-1)}}{\left(x^{2p}-1\right)^2}\,, \qquad \epsilon_2 =
8 p \left(\dfrac{\Mp}{\phizero}\right)^2 \dfrac{x^{2p} \left(x^{2p} +
  2p -1 \right)}{x^2\left(x^{2p} - 1\right)^2}\,,
\label{gdwisr12}
\end{equation}
and
\begin{equation}
\epsilon_3 = 8 p \left(\dfrac{\Mp}{\phizero}\right)^2 \dfrac{x^{2p}
  \left[x^{4p}+\left(2p-1\right)\left(p+2\right)x^{2p} + 2p^2 -3p
    +1\right]}{x^2\left(x^{2p}-1\right)^2\left(x^{2p}+2p-1\right)}\,.
\label{gdwisr3}
\end{equation}
They have been represented in the lower panels of \Fig{potgdwi}, and,
as can be seen in these plots, inflation proceeds at the top of the
hill towards increasing field values. It gracefully ends when
$\epsilon_1(\xend)=1$ and, from \Eq{gdwisr12}, $\xend$ is the solution of
\begin{equation}
\xend^{2p} - 2 \sqrt{2} p \frac{\Mp}{\phizero} \xend^{2p-1} = 1,
\label{eq:gdwixend}
\end{equation}
in the small field domain $0<x<1$. This equation cannot be solved
analytically for arbitrary values of $p$ and, in general, $\xend$ has
to be determined numerically.

The slow-roll trajectory can, however, be determined by quadrature
from \Eq{eq:srtrajectory} and reads
\begin{equation}
\Nend - N = \dfrac{\phizero^2}{8 p(p-1)\Mp^2} \left[ x^2 \left(p - 1 +
  x^{-2p}\right) -  \xend^2 \left(p - 1 +
  \xend^{-2p}\right)\right].
\end{equation}
Combined with the reheating equation~\eqref{eq:phistarlnrrad}, and the
numerical solution of \Eq{eq:gdwixend}, one can numerically determine
$\xstar$, the field value at which the pivot mode crossed the Hubble
radius during inflation.

The mass scale $M$ of the potential is then given by the CMB normalization
and one finds
\begin{equation}
\left(\dfrac{M}{\Mp}\right)^4 = 11520 \pi^2 p^2 \dfrac{\Mp^2}{\phizero^2}
\dfrac{\xstar^{2(2p-1)}}{\left(\xstar^{2p} - 1\right)^4} \dfrac{\Qrms^2}{T^2}\,.
\end{equation}
The reheating consistent slow-roll predictions for GDWI are
represented in \Figs{fig:CMBGDWI_0} to~\ref{fig:CMBGDWI_2}.

\subsection{Non-Minimal Large Field Inflation (NMLFI)}
\label{sec:nmlfi}

\subsubsection{Theoretical justification}

We consider again the conformal model described by
\Eq{eq:actioninvaconf2field}, except that the potential is now given
by a quartic power of the field $\phi$, namely
\begin{equation}
  S\left(g_{\mu \nu},\chi,\phi\right)=\frac{\Mg^2}{2}\int \dd ^4 \bmx
  \, \sqrt{-g}\, 
\left[\frac{\chi^2}{6}R +
g^{\mu \nu}\partial _\mu \chi \partial _\nu \chi
-\frac{\phi^2}{6}R -
g^{\mu \nu}\partial _\mu \phi \partial _\nu \phi
-\frac{\lambda}{2}\phi^4\right].
\label{eq:actioninvaconf2fieldnmlfi}
\end{equation}
Here again, for the sake of clarify, we drop the ``bar'' over Jordan
frame quantities. This model was considered in
\Refc{Kallosh:2013pby}. The above action resembles the action of the
T-model, see \sectionc{sec:tmi}. However, instead of a potential term
$\propto \FT(\phi/\chi)(\phi^2-\chi^2)^2$, we now have a potential
which no longer depends on $\chi$ and is simply given by $\phi^4$.

As noticed after \Eq{eq:actioninvaconf2field}, this action is
invariant under the transformation,
$\ef{g}_{\mu\nu}=e^{-2\sigma}g_{\mu \nu}$, $\ef\phi=e^{\sigma}\phi$
and $\ef{\chi}=e^{\sigma}\chi$. Notice that the conformal invariance
requires the potential to be proportional to $\phi^4$ (if it does not
depend on $\chi$). As before, the sign of the kinetic term of $\chi$
(the ``conformon'') is the ``wrong'' one. However, as before, this is
not a problem because we can always fix the conformal gauge, for
instance by choosing $\chi=\sqrt{6}$. In that case, the action can be
re-written as
\begin{equation}
S\left(g_{\mu \nu},\phi\right)=\frac{\Mg^2}{2}\int \dd ^4 \bmx \, \sqrt{-g}\, 
\left[
\left(1-\frac{\phi^2}{6}\right)R -
g^{\mu \nu}\partial _\mu \phi \partial _\nu \phi
-\frac{\lambda}{2}\phi^4\right].
  \label{eq:actionstnmlf}
\end{equation}
The above action corresponds to a scalar-tensor theory. Using
\Eqs{eq:actionst}, \eqref{eq:dphisquare} and the equation
for the potential in the text below \Eq{eq:dphisquare}, the model can be
rewritten in the Einstein frame, with a potential
\begin{equation}
  V(\ef{\phi})=\frac{\Mg^2\lambda}{4}\frac{\phi^4 }
  {\left(1-\dfrac{\phi^2}{6}\right)^2}\,,
\end{equation}
where $\phi(\ef{\phi})$ is given in terms of the canonically normalized field
$\ef{\phi}$ in the Einstein frame by
\begin{equation}
  \frac{\phi}{\sqrt{6}}=\frac{1-e^{-\sqrt{\frac23}\frac{\ef{\phi}}{\Mg}}}
       {1+e^{-\sqrt{\frac23}\frac{\ef{\phi}}{\Mg}}}\,.
       \label{eq:normalizednmlf}
\end{equation}
As noticed in Ref.~\cite{Kallosh:2013pby}, the coefficient in front of
the term $\phi^2 R$ in the action is fixed by the requirement of
maintaining the conformal invariance of the model. However, if the
model is embedded in conformal supergravity, this restriction can be
avoided. We now discuss this case.

As discussed at the end of \sectionc{sec:othertheosi}, conformal
supergravity depends on two functions, $\calN$ and the potential
$\calW$. The Lagrangian density of conformal supergravity was already
given in \Eq{eq:superconflagrangian} and reads
\begin{equation}
  \calL =\sqrt{-g}\left[-\frac{1}{6}\calN \left(X,\bar{X}\right)R - G_{I\bar{J}}
    \calD^{\mu } X^I\calD_\mu \bar{X}^{\bar{J}} - V\left(X,\bar{X}\right)
    \right],
  \label{eq:defLconfsugra}
\end{equation}
with $\calD_\mu X^I=\partial_\mu X^I-iA_\mu X^I$. The fact that the
superfields $X^I$ are now charged is a difference compared with
\Eq{eq:superconflagrangian}. Here, we consider a model where
the function $\calN(X,\bar{X})$ is defined by
\begin{equation}
  \calN\left(X^0,X^1\right)=-\left \vert X^0\right \vert ^2
  +\left \vert X^1\right\vert ^2-3\Delta \left\vert X^0\right\vert ^2
  \left[\left(\frac{X^1}{X^0}\right)^2
    +\left(\frac{\bar{X}^{\bar{1}}}{\bar{X}^{\bar{0}}}\right)^2\right],
  \end{equation}
where $\Delta $ is a dimensionless parameter, $X^0$ the conformon and
$X^1$ the inflaton. Notice that, when $\Delta=0$, the embedding
potential has an enhanced $\mathrm{SU}(1,1)$ symmetry. Compared to the
superconformal model of \sectioncs{sec:si} and~\ref{sec:tmi}, we see
that only two fields are present, the conformon $X^0$ and the inflaton
$X^1$. The Goldstino $S$ is now absent.
Straightforward calculations lead to the kinetic terms of those
fields and one obtains
\begin{equation}
\begin{aligned}
  G_{0 \bar{0}}&=-1+3\Delta \left[\left(\frac{X^1}{X^0}\right)^2
    +\left(\frac{\bar{X}^{\bar{1}}}{\bar{X}^{\bar{0}}}\right)^2\right],
  \quad 
  G_{0\bar{1}}&=-6\Delta \frac{\bar{X}^{\bar{1}}}{\bar{X}^{\bar{0}}}, \quad 
  G_{1\bar{0}}=-6\Delta \frac{X^1}{X^0}, \quad 
  G_{1 \bar{1}}=1.
\end{aligned}
\label{eq:fieldmetricsugra}
\end{equation}
We see that the parameter $\Delta $ is proportional to the curvature
of the K\"ahler internal manifold since $\Delta=0$ implies that
$G_{I\bar{J}}=\delta _{I\bar{J}}$.

An important property of the above action is that it is invariant
under the following transformation
\begin{equation}
  \ef{g}_{\mu \nu}=e^{-2 \sigma} g_{\mu \nu}, \quad
  \ef{X}^I=e^{\sigma+i\Lambda}X^I, \quad
  \ef{\bar{X}}^{\bar{J}}=e^{\sigma-i\Lambda}\bar{X}^{\bar{J}},
  \quad \tilde{A}_\mu=A_\mu+\partial_\mu \Lambda,
  \label{eq:transgfield}
\end{equation}
as we are going to show explicitly. Let us first notice that the
transformations~(\ref{eq:transgfield}) imply that
\begin{equation}
  \calD_\mu X^I=e^{-\sigma-i\Lambda }\tilde\calD_\mu
  \tilde{X}^I-e^{-\sigma-i\Lambda}\tilde{X}^I\partial_\mu\sigma.
\end{equation}
Let us split the Lagrangian in two parts $\calL=\calL_1+\calL_2$ with
\begin{equation}
\begin{aligned}
  \calL_1 &= -\frac16 \sqrt{-g}\left[\left(-\left \vert
    X^0\right\vert^2 +\left \vert X^1\right \vert^2\right)-\left \vert
    X^0\right\vert^2 \left(G_{0\bar{0}}+1\right)\right]R, \\ \calL_2
  &=-\sqrt{-g}\left(G_{I\bar{J}}g^{\mu \nu} \calD_\mu X^I{\cal
    D}_\nu \bar{X}^{\bar{J}}+V\right).
\end{aligned}
\end{equation}
By the transformation~\eqref{eq:transgfield}, $\calL_1$ becomes
\begin{equation}
  \begin{aligned}
  \ef{\calL}_1 &= -\frac{1}{6} e^{4\sigma}\sqrt{-\ef{g}}\left[e^{-2\sigma}
  \left(-\left \vert \ef X^0\right\vert ^2+\left \vert \ef{X}^1\right\vert^2
  \right)
-e^{-2\sigma}\left\vert \ef{X}^0\right\vert ^2
\left(\ef{G}_{0\bar{0}}+1\right)\right] \\ & \times
  e^{-2\sigma}\left(\ef{R}-6\ef{g}^{\mu \nu}\ef{\nabla}_\mu \partial_\nu \sigma
  -6\ef{g}^{\mu \nu}\partial_\mu \sigma \partial_\nu \sigma\right),
  \end{aligned}
\end{equation}
where we have used the fact that the components of $G_{I\bar{J}}$ are
invariant under \Eq{eq:transgfield}. We notice that the exponential
terms exactly cancel out. Then, the transformation of the term
$\calL_2$ can be expressed as
\begin{equation}
\begin{aligned}
  \ef\calL_2&=-e^{4\sigma}\sqrt{-\ef{g}}\left\{\ef{G}_{I\bar{J}}e^{-2\sigma}
      \ef{g}^{\mu \nu}
\left[e^{-2\sigma}\ef\calD_\mu \ef{X}^I\ef\calD_\nu\ef{\bar{X}}^{\bar{J}}
  -e^{-2\sigma}\left(\ef\calD_\mu\ef{X}^I\right)\ef{\bar{X}}^{\bar{J}}\partial_\nu
  \sigma \right. \right.
   \\ & \left. \left.
-e^{-2\sigma}\ef{X}^I\left(\ef\calD_\nu\ef{\bar{X}}^{\bar{J}}\right)\partial_\mu \sigma
+e^{-2\sigma}\ef{X}^I\ef{\bar{X}}^{\bar{J}}\partial_\mu \sigma \partial_{\nu }\sigma \right]
+V\right\}
\\
&=
-\sqrt{-\ef{g}}\, \ef{G}_{I\bar{J}}\,
      \ef{g}^{\mu \nu}\left[
\ef\calD_\mu \ef{X}^I\ef\calD_\nu\ef{\bar{X}}^{\bar{J}}
-\left(\partial_\mu\ef{X}^I-i\ef{A}_\mu\ef{X}^I\right)\ef{\bar{X}}^{\bar{J}}
\partial_\nu \sigma
-\ef{X}^I\left(\partial_\nu\ef{\bar{X}}^{\bar{J}}
+i\ef{A}_\mu \ef{\bar{X}}^{\bar{J}}\right)\partial_\mu \sigma
\right. \\ & \left.
 +\ef{X}^I\ef{\bar{X}}^{\bar{J}}\partial_\mu \sigma \partial_{\nu }\sigma \right]
-\sqrt{-\ef{g}} e^{4\sigma}V.
\end{aligned}
\end{equation}
The two terms proportional to the gauge fields $\ef{A}_\mu$ cancel out
while the second and third terms can be rewritten as a total
derivative\footnote{For any vector $V^{\mu}$, $\sqrt{-g}\,
\nabla_\mu V^\mu=\partial_\mu (\sqrt{-g}V^\mu)$.}, namely
\begin{equation}
\begin{aligned}
  \partial_\mu\left[\sqrt{-\ef{g}}\, \ef{g}^{\mu
      \nu}\ef{G}_{I\bar{J}}\ef{X}^I
    \ef{\bar{X}}^{\bar{J}}\left(\partial_\nu \sigma \right)\right]
  &=\sqrt{-\ef{g}} \left[ \ef{\nabla}_\mu\left(\ef{g}^{\mu
      \nu}\partial_\nu \sigma\right)
    \ef{G}_{I\bar{J}}\ef{X}^I\ef{\bar{X}}^{\bar{J}} + \ef{g}^{\mu
      \nu}\left(\partial_\nu \sigma\right) \left(\ef{\nabla}_\mu
    \ef{G}_{I\bar{J}}\right)\ef{X}^I\ef{\bar{X}}^{\bar{J}}
    \right. \\ & \left.  + \ef{g}^{\mu \nu}\left(\partial_\nu
    \sigma\right)
    \ef{G}_{I\bar{J}}\ef{\nabla}_\mu\left(\ef{X}^I\ef{\bar{X}}^{\bar{J}}\right)
    \right].
\end{aligned}
\end{equation}
Collecting the various terms in $\calL = \calL_1+\calL_2$, one obtains
\begin{equation}
  \begin{aligned}
  \ef\calL&=
  \sqrt{-\ef{g}}\left[-\frac{1}{6}
    \calN\left(\ef{X},\ef{\bar{X}}\right) \ef{R} - \ef{G}_{I\bar{J}}\, \ef g^{\mu \nu}\,
    \ef{\calD}_\mu \ef{X}^I \ef{\calD}_\nu \ef{\bar{X}}^{\bar{J}}-
    e^{4\sigma}V\right]
 \\ &
  +\sqrt{-\ef{g}}\left[\left(-\left \vert \ef{X}^0\right\vert ^2
    +\left \vert \ef X^1\right\vert^2\right)-\left \vert \ef X^0\right\vert^2
    \left(\ef{G}_{0\bar{0}}+1\right)-\ef{G}_{I\bar{J}}\ef{X}^I\ef{\bar{X}}^{\bar{J}}\right]
\left[\ef{g}^{\mu \nu}\partial_\mu \sigma \partial_\nu \sigma +\ef{g}^{\mu \nu}
  \ef{\nabla}_\mu \left(\partial _\nu \sigma\right)\right]
 \\ &
+ \partial_\mu\left[\sqrt{-\ef g}\, \ef{g}^{\mu \nu}\ef{G}_{I\bar{J}}\ef{X}^I
  \ef{\bar{X}}^{\bar{J}}\left(\partial_\nu \sigma \right)\right]
-\sqrt{-\ef{g}}\, \ef{g}^{\mu \nu}\left(\partial_\nu \sigma\right)
\left(\partial_\mu \ef{G}_{I\bar{J}}\right)\ef{X}^I\ef{\bar{X}}^{\bar{J}}.
  \end{aligned}
  \end{equation}
Using the internal metric of \Eq{eq:fieldmetricsugra}, one finds that
the first term of the second line, the one within square brackets,
vanishes. It remains the last term. Using the definition of the
internal metric, one has
\begin{equation}
  \partial_\mu
\tilde{G}_{I\bar{J}}=\partial_\mu \calN_{I\bar{J}}=\calN_{KI\bar{J}}
\partial_\mu X^{K}+{\cal N}_{\bar{K}I\bar{J}}\partial_\mu
\bar{X}^{\bar{K}},
\end{equation}
from which
\begin{equation}
  \left(\partial_\mu G_{I\bar{J}} \right)
  X^I\bar{X}^{\bar{J}}=\calN_{KI\bar{J}}X^I\bar{X}^{\bar{J}}\partial_\mu
  X^K+{\cal N}_{\bar{K}I\bar{J}}X^I\bar{X}^{\bar{J}}\partial_\mu
  \bar{X}^{\bar{K}}=0.
\end{equation}
If $V(X,\bar{X})$ is homogeneous and of second degree in $X$ and
$\bar{X}$ then, as announced, the Lagrangian~(\ref{eq:defLconfsugra})
is indeed invariant under the
transformation~(\ref{eq:transgfield}). An explicit example is given in
\Refc{Kallosh:2013pby} with $V=\lambda(X^1\bar{X}^{\bar{1}})^2$.

Let us now return to \Eq{eq:defLconfsugra} and fix the conformon by
assuming $X^0=\bar{X}^{\bar{0}}=\sqrt{3}\Mg$. Then, the Lagrangian
becomes
\begin{equation}
  \begin{aligned}
  \left(\sqrt{-g}\right)^{-1}\calL&= \left\{\frac{\Mg^2}{2}
  -\frac{1}{6}\left\vert X^1\right\vert^2+\frac{\Delta}{2}
  \left[\left(X^1\right)^2+\left(\bar{X}^{\bar{1}}\right)^2\right]
  \right\} R
  -\partial_\mu X^1 \partial^\mu \bar{X}^{\bar{1}}-\lambda \left(X^1 \bar{X}^{\bar{1}}\right)^2.
  \end{aligned}
\end{equation}
Decomposing $X^1=\left(\varphi_1+i\varphi_2\right)/\sqrt{2}$, the
Lagrangian reads
\begin{equation}
\begin{aligned}
  \left(\sqrt{-g}\right)^{-1}\calL & = \left[\frac{\Mg^2}{2}+\frac12\left(\Delta -\frac16\right)\varphi_1^2
    -\frac{1}{2} \left(\Delta +\frac{1}{6}\right)\varphi_2^2\right] R
  -\frac{1}{2} \partial_\mu \varphi_1\partial^{\mu}\varphi_1
   \\ &
  -\frac{1}{2} \partial_\mu \varphi_2\partial^{\mu} \varphi_2
  -\frac{\lambda}{4}\left(\varphi_1^2+\varphi_2^2\right)^2.
\end{aligned}
\end{equation}
In particular, notice that the model is invariant by changing the
sign of $\Delta$ and swapping the fields $\varphi_1$ and
$\varphi_2$. If one focuses on the choice $\Delta>0$, then the ground
state of the system is obtained for $\varphi_2=0$ and, renaming
$\varphi_1\equiv \phi$, one has
\begin{equation}
  \left(\sqrt{-g}\right)^{-1}\calL
  =\frac{\Mg^2}{2}R-\frac12 \partial_\mu \phi \partial^\mu \phi
    -\frac12 \left(\frac16-\Delta \right)\phi^2 R-\frac{\lambda}{4}\phi^4.
\end{equation}
As announced, one therefore obtains a non-minimally coupled large
field quartic model but, contrary to \Eq{eq:actionstnmlf}, there is
now an arbitrary coefficient in front of the $\phi^2 R$ term. This
Lagrangian describes a scalar-tensor theory in the Jordan frame, from
which one obtains the Einstein frame's potential (see
\sectionc{subsubsec:theoryhi}).
\begin{equation}
  V(\ef{\phi})=\frac{\lambda}{8} \dfrac{\phi^4}{\left(1+\xi \phi^2/\Mg^2\right)^2}\,,
\end{equation}
where we have defined $\xi\equiv \Delta-1/6$ and where the canonically
normalized field $\ef{\phi}$ can be expressed as
\begin{equation}
  \frac{\ef{\phi}}{\Mg}=\sqrt{\frac{1+6\xi}{\xi}}
  \arcsinh\left[\sqrt{\xi(1+6\xi)}\frac{\phi}{\Mg}\right]
  -\sqrt{6}\arctanh\left[\frac{\xi \sqrt{6}\phi/\Mg}
    {\sqrt{1+\xi(1+6\xi)\phi^2/\Mg^2}}\right].
\end{equation}
As one may expect, this relation is exactly the same as
\Eq{eq:chihexplicit} for Higgs Inflation (with the identification
$h=\phi/\Mg$ and $\chi = \ef{\phi}/\Mg$). As it is the case for
HI, it is not possible to analytically invert this relation to obtain
an explicit expression for the potential $V(\ef{\phi})$.
Let us notice that in the limit where $\Delta \rightarrow 0$, one
finds
\begin{align}
  \frac{\ef{\phi}}{\Mg}=\sqrt{6}\arctanh\left(\frac{\phi} {\Mg
    \sqrt{6}}\right) +\order{\Delta},
\end{align}
which gives back \Eq{eq:normalizednmlf}, as expected.

\subsubsection{Slow-roll Analysis}

\begin{figure}
\begin{center}
\includegraphics[width=\wdblefig]{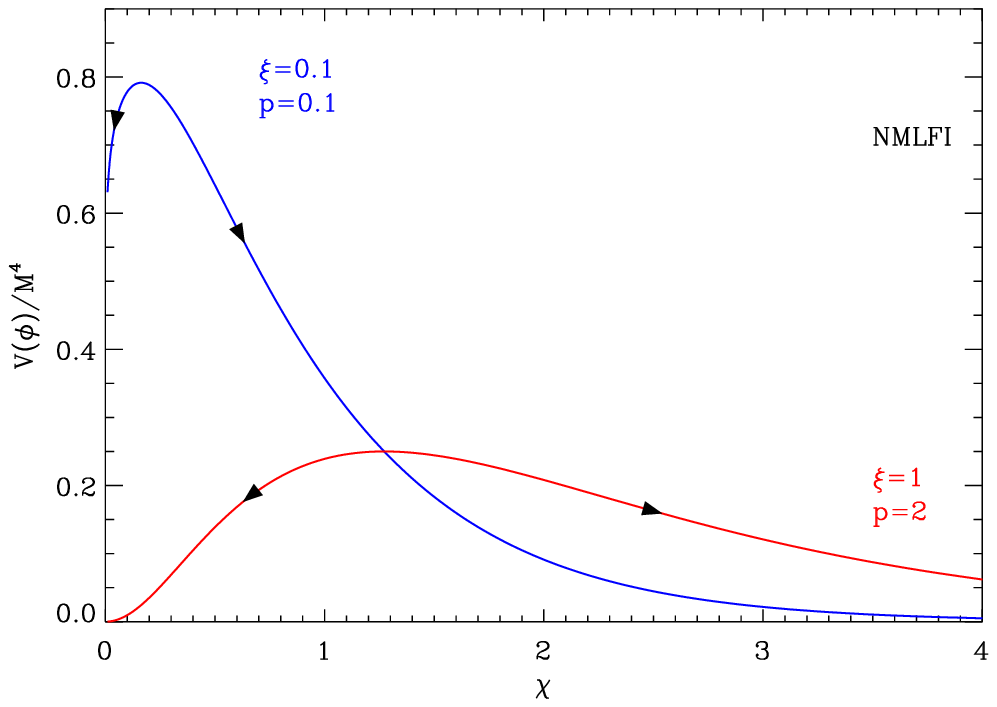}
\includegraphics[width=\wdblefig]{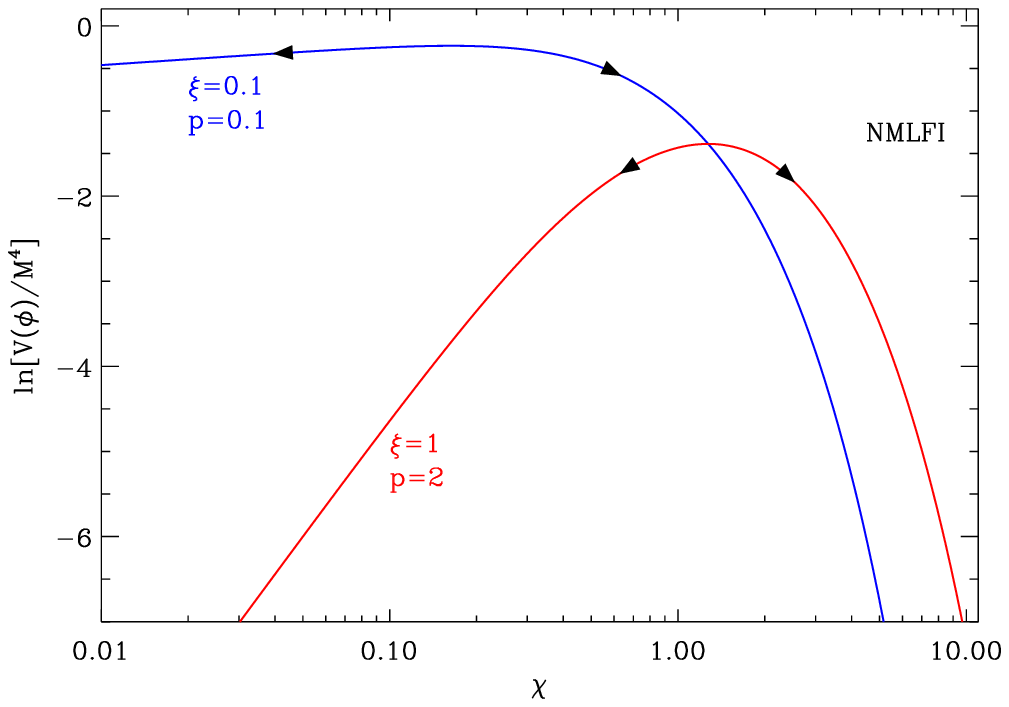}
\includegraphics[width=\wdblefig]{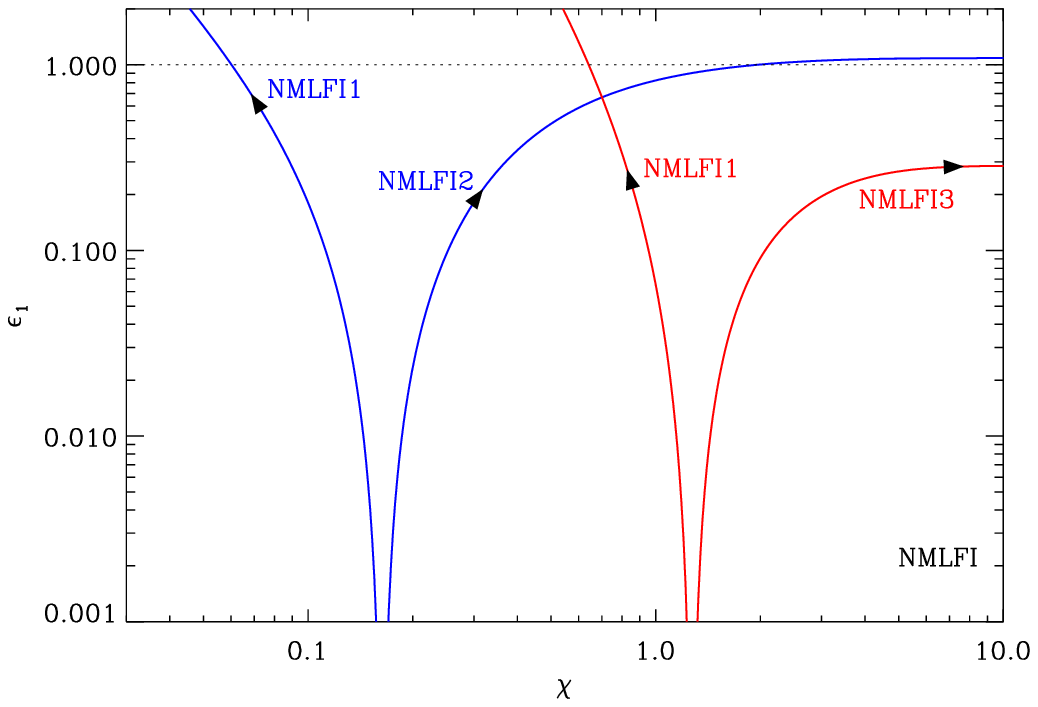}
\includegraphics[width=\wdblefig]{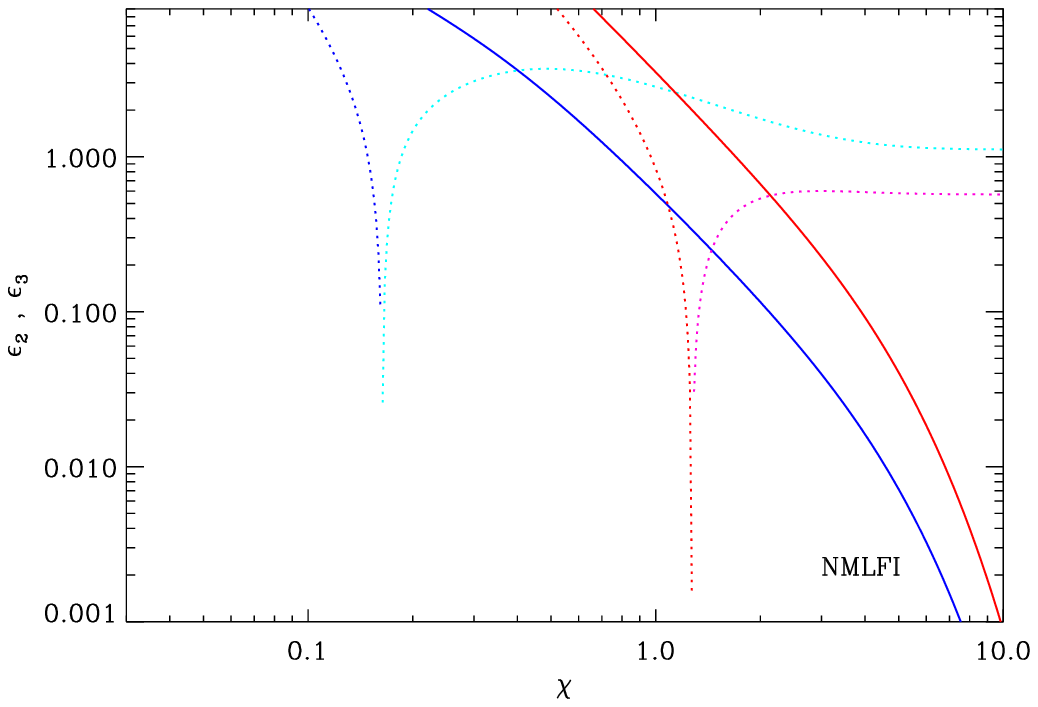}
\caption{Non-Minimal Large Field Inflation (NMLFI) for $p<4$, the
  potential develops a maximum at intermediate field values. The top
  left panel shows the potential as a function of $\chi=\ef{\phi}/\Mp$
  for the two sets of parameter values reported in the figure. Top
  right panel: logarithm of the potential. For $\xi=0.1$ and $p=0.1$
  (blue curves), both the NMLFI1 (left of the maximum) and NMLFI2
  (right of the maximum) inflationary regimes ends when the first
  Hubble flow function (lower left panel) exceeds unity. The slow-roll
  parameters $\epsilon_2$ (solid curve) and $|\epsilon_3|$ (dotted
  curve) are represented in the bottom right panel. For $\xi=1$ and
  $p=2$ (red curves), NMLFI1 still gracefuly ends whereas inflation on
  the right of the potential maximum, NMLFI3, needs an additional
  mechanism to end inflation. The potential for $p>4$ is represented
  in \Fig{fig:potnmlfinotop}.}
\label{fig:potnmlfitop}
\end{center}
\end{figure}

\begin{figure}
\begin{center}
\includegraphics[width=\wdblefig]{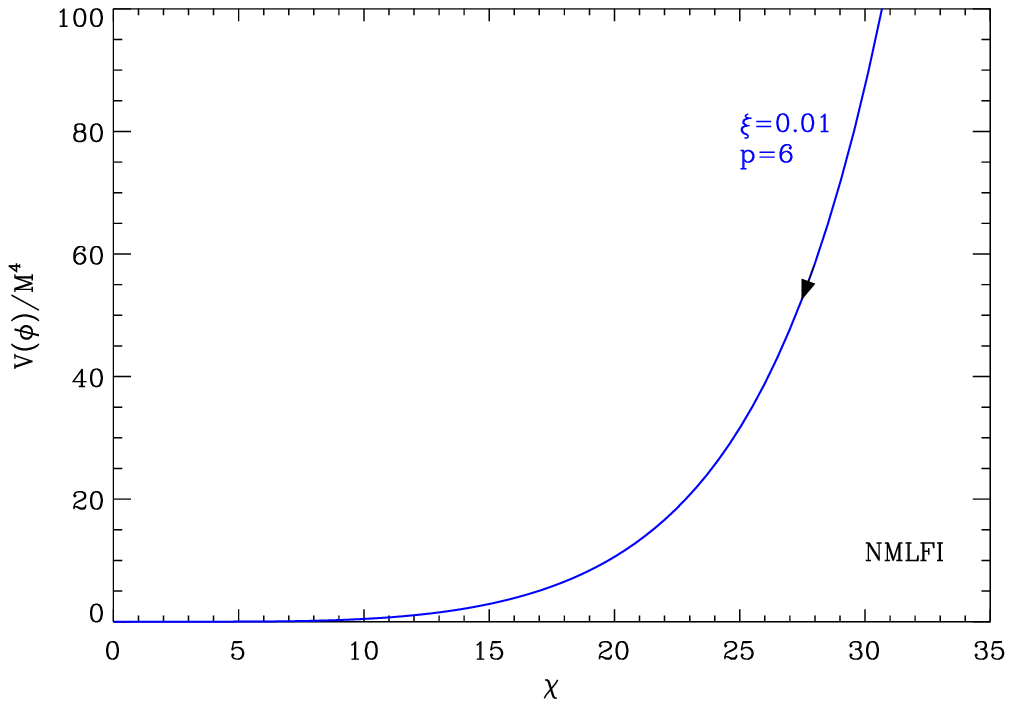}
\includegraphics[width=\wdblefig]{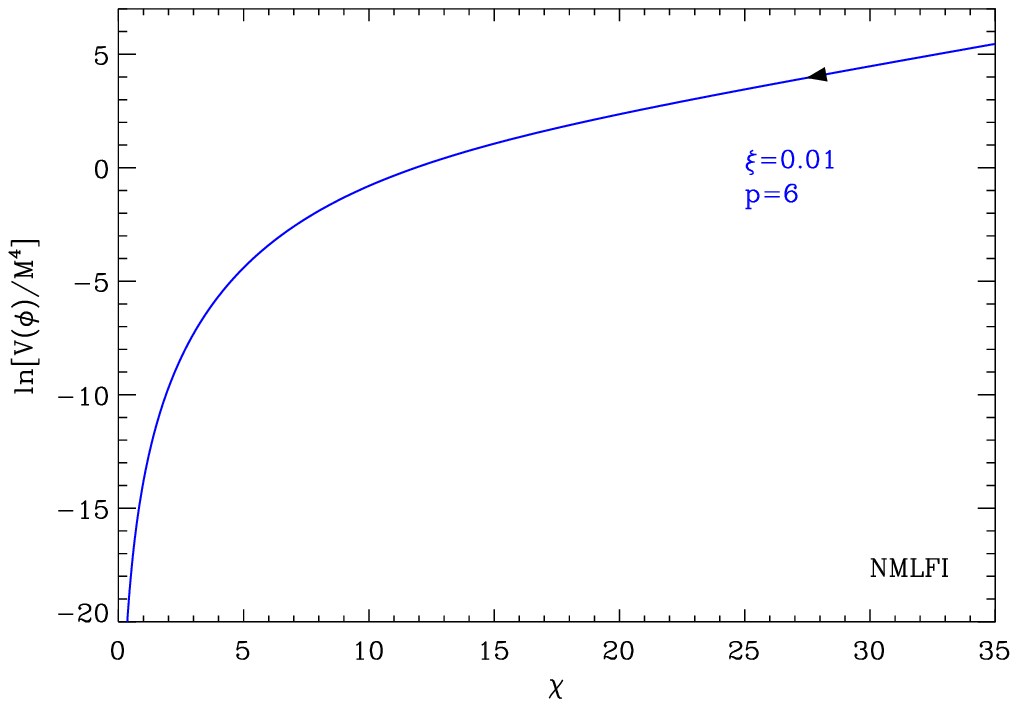}
\includegraphics[width=\wdblefig]{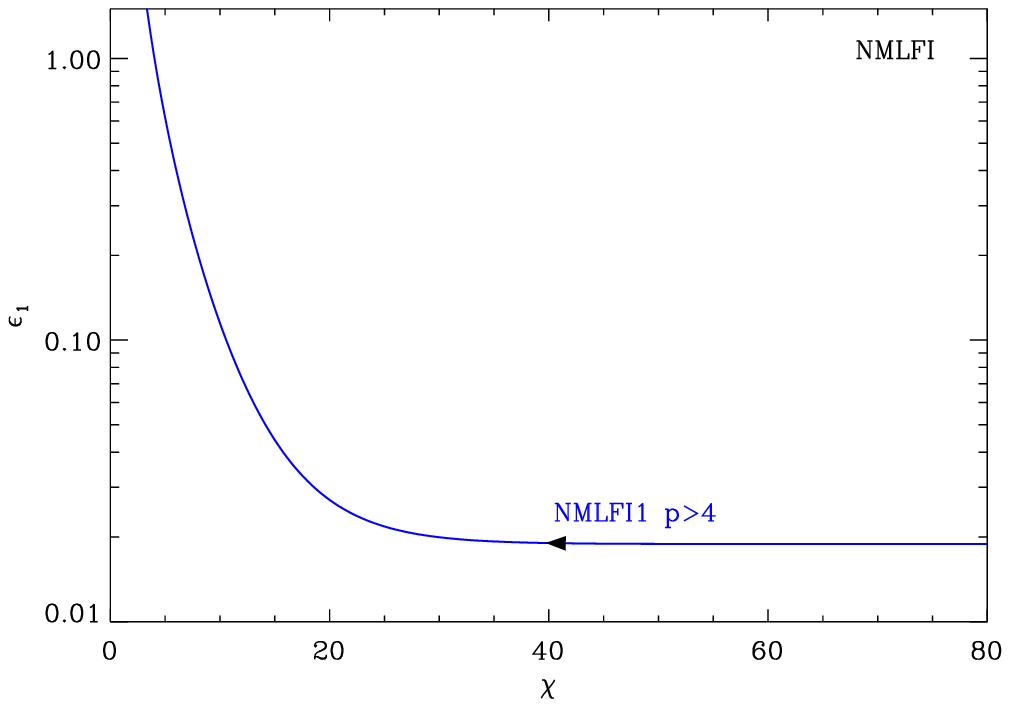}
\includegraphics[width=\wdblefig]{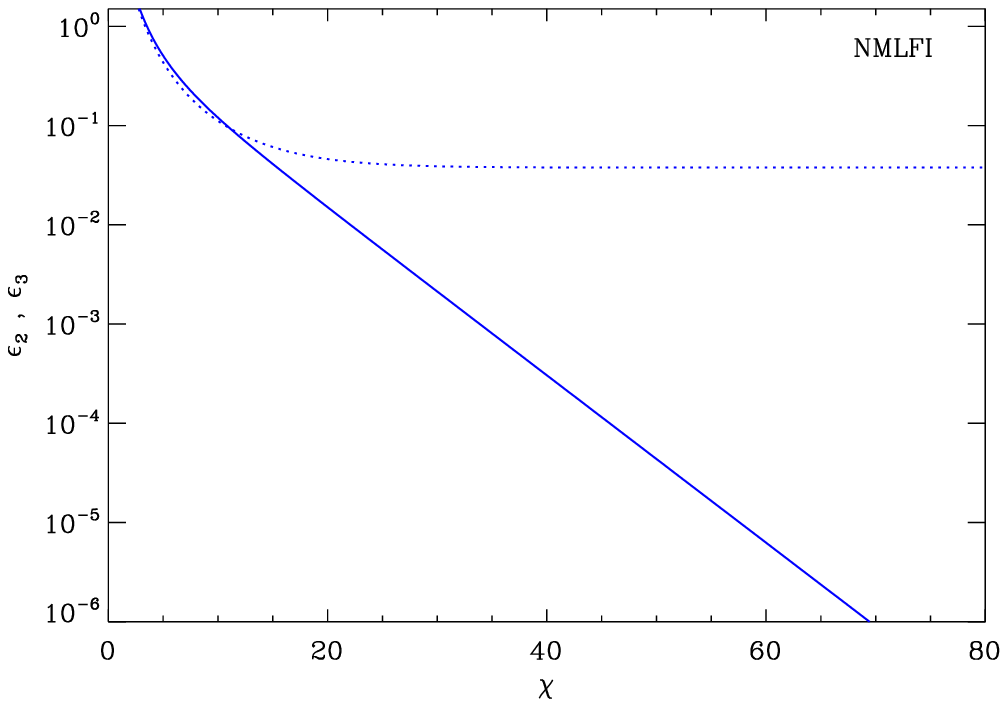}
\caption{Non-Minimal Large Field Inflation (NMLFI) for $p>4$, the
  potential has no maximum and only the NMLF1 regime exists. For
  $p=4$, the potential has a plateau and one recovers the potential of
  Higgs Inflation for any value of $\xi$, see \Fig{fig:pothi}. The top
  left panel shows the potential as a function of
  $\chi=\ef{\phi}/\Mp$. Top right panel: logarithm of the
  potential. Bottom left panel: the first slow-roll parameter
  $\epsilon_1$. At large field value $\chi$, $\epsilon_1$ is
  constant. Bottom right panel: slow-roll parameters $\epsilon_2$
  (solid curve) and $\epsilon_3$ (dotted curve). The NMLFI potential
  for $p<4$ is represented in \Fig{fig:potnmlfitop}.}
\label{fig:potnmlfinotop}
\end{center}
\end{figure}

From the previous theoretical motivations, we consider the class of
Non-Minimal Large Field Inflation (NMLFI) models defined as having a
potential in the Jordan frame identical to the LFI models of
\sectionc{sec:lfi}, i.e., $U(\jf{\phi}) = \jf{\lambda}
\Mg^2(\jf{\phi}/\Mg)^p$~\cite{Okada:2010jf, Bezrukov:2013fca}, where
we now denotes by $\jf{\phi}$ the Jordan frame real scalar field. Here
$\Mg$ is the gravitational mass scale in the Jordan frame and
$\jf{\lambda}$ a dimensionless coupling constant. As explained in
\sectionc{subsubsec:theoryhi}, only if the vacuum state of the theory
after inflation is at $\jf{\phi}/\Mg \simeq 0$, one can identify the
numerical value of $\Mg \simeq \Mp$, see \Eq{eq:cavendish}. The
non-minimal coupling functions appearing in the scalar-tensor action
of \Eq{eq:actionst} are of the form
\begin{equation}
  F(h) = 1 + \xi \left(\dfrac{\jf{\phi}}{\Mg}\right)^2, \qquad Z(h) = 1,
\end{equation}
with $\xi \ge 0$. As for Higgs Inflation in \sectionc{sec:hi}, we
introduce the two dimensionless fields
\begin{equation}
\barh \equiv \sqrt{\xi} \dfrac{\jf{\phi}}{\Mg}, \qquad \chi \equiv \dfrac{\ef{\phi}}{\Mg}\,,
\end{equation}
where $\ef{\phi}$ is the canonically normalized scalar field in the
Einstein frame. The potential of NMLFI in the Einstein frame can only be given in the
parametric way and reads
\begin{equation}
  \begin{aligned}
    V(\ef{\phi}) & = M^4 \dfrac{\barh^p}{\left(1+ \barh^2\right)^2}\,,
  \end{aligned}
  \label{eq:potnmlfi}
\end{equation}
where $M$ is the inflationary mass scale in the Einstein frame and
verifies $M^4 = \Mg^4 \jf{\lambda}/\xi^{p/2}$, see \Eq{eq:hiM4}. The
function $\barh(\chi)$ is the solution of \Eq{eq:chifrombarh}, namely
\begin{equation}
  \begin{aligned}
\chi & = \sqrt{6 + \dfrac{1}{\xi}} \ln \left[ \sqrt{1+(1+6\xi) \barh^2}
  + \sqrt{(1+6\xi) \barh^2} \right] + \sqrt{6} \ln
\left[\dfrac{\sqrt{1+ \barh^2}}{\sqrt{1+(1+6\xi)\barh^2} + \sqrt{6\xi
      \barh^2}} \right],
  \end{aligned}
\label{eq:nmlfichiofbarh}
\end{equation}
which cannot be inverted explicitly.  This is not an issue as the
Hubble flow functions can nevertheless be determined in a parametric
form. Following what has been done for HI in \sectionc{sec:hiexact},
from \Eq{eq:eps1defparam}, one gets
\begin{equation}
\begin{aligned}
  \epsilon_1(\barh) & = \dfrac{\xi}{2 \barh^2} \, \dfrac{\left[p+
   (p-4) \barh^2 \right]^2}{1 +(1+6\xi) \barh^2} \,, \qquad
\epsilon_2(\barh) = 2 \xi \, \dfrac{1+\barh^2}{\barh^2} \, \dfrac{p +
  \left(p+4+12 p \xi\right) \barh^2}{\left[1+(1+6\xi) \barh^2
    \right]^2}\,,
\end{aligned}
\label{eq:nmlfisr12}
\end{equation}
and
\begin{equation}
  \begin{aligned}
\epsilon_3(\barh) & = 2 \xi \, \dfrac{p + (p-4) \barh^2}{\barh^2} \\ & \times
\dfrac{p + 3p(1+6\xi) \barh^2 + \left[4 + 3p + 48 \xi + 36 p \xi
    (1+4\xi)\right]\barh^4 + (1+6\xi)(4 + p + 12
  p\xi)\barh^6}{\left[1+(1+6\xi) \barh^2 \right]^2 \left[p + (4 + p +
    12 p\xi) \barh^2 \right]}\,.
  \end{aligned}
  \label{eq:nmlfisr3}
\end{equation}
The potential and the Hubble flow functions have been plotted in
\Fig{fig:potnmlfitop} and \Fig{fig:potnmlfinotop}, for various values
of $\xi$ and $p$, in terms of the dimensionless canonically normalized
field $\chi$.

As can be seen on this figure, the potential admits a local maximum
for all values of $p < 4$. Solving for $\epsilon_1=0$, the maximum
occurs at the parameter value $\barhVmax$ given by
\begin{equation}
\barhVmax = \sqrt{\dfrac{p}{4-p}}\,.
\label{eq:barhVmax}
\end{equation}
The corresponding value of the canonically normalized field $\chiVmax$
can be obtained by plugging \Eq{eq:barhVmax} into
\Eq{eq:nmlfichiofbarh}. For $p<4$, inflation can then occur in two
regions. Either at decreasing parametric field values, for $\barh <
\barhVmax$, in a regime that will be referred to as NMLFI1, or at
increasing parametric field values for $\barh > \barhVmax$. Since
$\epsilon_1(\barh)$ increases when $\barh$ decreases, NMLFI1 always
gracefully ends at a parametric field value close to zero. In the
other domain, at large $\barh$ values, \Eq{eq:nmlfisr12} implies
\begin{equation}
\lim_{\barh \to \infty}\epsilon_1 = \dfrac{(p-4)^2}{2} \dfrac{\xi}{1+6\xi}\,.
\label{eq:nmlfieps1infty}
\end{equation}
We immediately see that if $\xi$ is too small, the asymptotic limit of
$\epsilon_1<1$ and inflation never ends. More specifically, let us define
\begin{equation}
\xizero(p) \equiv \dfrac{2}{p(p-8)+4} = \dfrac{2}{(p-p_-)(p-p_+)}\,,
\label{eq:nmlfixizero}
\end{equation}
where $p_\pm$ are the two roots of the quadratic denominator:
\begin{equation}
p_- \equiv 2 \left(2-\sqrt{3}\right) \simeq 0.54, \qquad p_+ \equiv 2 \left(2+\sqrt{3}\right) \simeq 7.46.
\label{eq:nmlfippm}
\end{equation}
From \Eq{eq:nmlfieps1infty}, one has
$\lim_{\barh\to\infty}\epsilon_1>1$ provided two conditions are
satisfied: $p < p_-$ and $\xi>\xizero(p)$. Under these conditions, in the
domain $\barh > \barhVmax$ inflation stops at the field
value where $\epsilon_1$ reaches unity. This regime will be refereed to
as NMLFI2.

Still in the domain $\barh > \barhVmax$, for $p<p_-$ and
$\xi<\xizero(p)$, but also for $p_-<p<4$, the asymptotic limit of
$\epsilon_1 < 1$ and inflation never ends. One therefore needs an
additional mechanism to stop inflation, as for instance a tachyonic
instability triggered by an extra field. This inflationary regime is
then a model with three parameters, $p$, $\xi$ and $\chiend$ (or
$\barhend$), that we refer to as NMLFI3.

For $p>4$ the potential has no maximum, inflation can only proceed at
decreasing field values and this regime will be also referred to as
NMLFI1. The limiting case $p=4$ is exactly Higgs Inflation (HI) with
$v=0$ , a vanishing {\vev}, and unconstrained values for $\xi$, see
\sectionc{sec:hiexact}.

Let us notice that since both NMLFI2 and NMLFI3 can explore some part
of the large field region, one should pay attention to the actual
value of $\barhend$, the parametric field value after inflation, in
order to determine how much the numerical value of $\Mg$ differs from
$\Mp$. From \Eq{eq:cavendish}, one indeed has, at the end of inflation
\begin{equation}
\Mg^2 =\dfrac{\Mp^2}{1 +
  \barhend^2} \dfrac{1 + \barhend^2 + 8\xi \barhend^2}{1 + \barhend^2 + 6 \xi \barhend^2}\,.
\end{equation}
For large enough $\barhend$, and provided $\barh$ remains frozen after
inflation, one has $\Mg < \Mp$ and these models can potentially
address the Planck mass hierarchy problem~\cite{Ringeval:2019bob}.

For both NMLFI1 and NMLFI2, the parametric field value at which
inflation stops is solution of $\epsilon_1=1$. From \Eq{eq:nmlfisr12},
this equation admits, a priori, two solutions
\begin{equation}
\barh_{\pm}^2 = \dfrac{p (p-4)\xi -1 \pm \sqrt{(1+2p\xi)(1+6p\xi)}}{2
  - \xi \left[p(p-8)+4\right]}\,.
\label{eq:nmlfibarhend}
\end{equation}

For $p_-<p<p_+$ the denominator is always positive. Therefore, one has
$\barh_+^2>0$ whereas $\barh_-^2 < 0$ and the end of inflation for NMLFI1
occurs at the parametric field value $\barhend= \barh_+$.

For $p>p_+$, one always have $\barh_-^2 < 0$ whereas $\barh_+^2 > 0$ under the
additional condition that $\xi < \xizero(p)$. Let us mention that
$\xizero(p)$ is also a root of the numerator in \Eq{eq:nmlfibarhend}
such that determining its sign requires some attention. For these
cases, NMLFI1 ends again at $\barhend = \barh_+$. If $\xi>\xizero(p)$
(still for $p>p_+$), one has $\epsilon_1 > 1$ for all the values of
$\barh$ and the potential ends up being too steep to support inflation
at all.

At last, for $p<p_-$ one always have $\barh_+^2>0$ whereas $\barh_-^2 >0$ only
under the additional condition of having $\xi > \xizero(p)$. As a
result, for $p<p_-$, NMLFI1, which proceeds at $\barh < \barhVmax$,
ends once more at $\barhend = \barh_+$ whereas NMLFI2, which proceeds
at $\barh > \barhVmax$, ends at $\barhend = \barh_-$ but only if $\xi
>\xizero(p)$. For $\xi < \xizero(p)$, as already discussed, inflation
does not end by itself and only NMLFI3 exists in the domain $\barh >\barhVmax$.

The parametric slow-roll trajectory can be determined as done for
Higgs Inflation in \Eq{eq:hitrajparamsplit}, and, can be analytically
integrated. The case $p=4$ requires special attention and one gets
\begin{equation}
\begin{aligned}
 \Delta N & = \dfrac{2 + 3 \xi p}{4 \xi (p-4)}
 \ln\left[\dfrac{p+(p-4)\barh^2}{p+(p-4)\barhend^2}\right] -
 \dfrac{3}{4} \ln\left(\dfrac{1 + \barh^2}{1+\barhend^2}\right), &
 \textrm{for $p \ne 4$},\\
 \Delta N & = \dfrac{1+6\xi}{8\xi} \left(
 \barh^2 - \barhend^2\right) - \dfrac{3}{4} \ln
 \left(\dfrac{1+\barh^2}{1+\barhend^2}\right), & \textrm{for
   $p=4$},
\end{aligned}
\label{eq:nmlfitraj}
\end{equation}
where $\Delta N = \Nend - N$. For $p \ne 4$, the trajectory cannot be
inverted analytically and one has to solve \Eq{eq:nmlfitraj}
numerically to determine $\barh(\Delta N)$, and hence $\chi(\Delta N)$
from \Eq{eq:nmlfichiofbarh}. The special case $p=4$ can nevertheless
be inverted in terms of a Lambert function as
\begin{equation}
\barh^2(\Delta N) = -1 - \dfrac{6\xi}{1+6\xi}
\Lambert{-1}\left\{-\dfrac{(1+6\xi)\left(1+\barhend^2\right)}{6\xi}
  e^{-\frac{1}{6\xi}\left[(1+6\xi)(1+\barhend^2) +8 \xi \Delta
      N\right]} \right\}, \quad \textrm{for $p=4$}.
\end{equation}

From \Eq{eq:nmlfitraj}, one can check that, for $p<4$, $\Delta N \to
\infty$ for $\barh \to \barhVmax$ and an infinite number of e-folds
can be realized at the top of the potential. However, the divergence
is only logarithmic and this implies that NMLFI2, which inflates in
the domain $]\barhVmax,\barhend[$ is a very fine-tuned, if not
ruled-out, model. Indeed, it exists only for large enough values
of $\xi > \xizero(p)$ and this implies that $\Delta N$ can be made
larger than, say, $60$ {\efolds} only if the initial field value
is exponentially fine-tuned to the top of the potential. Then,
\Eq{eq:nmlfisr12} implies that
\begin{equation}
  \epsilon_2(\barhVmax) = \dfrac{8 \xi(4-p)}{2+3p\xi} >  \dfrac{8}{1-p}\,,
\end{equation}
where the last inequality comes from $\xi > \xizero(p)$. Because
NMLFI2 requires $p<p_-$, one gets $\epsilon_2(\barhVmax) > 8$ and
since all of the {\efolds} of inflation are done at the top of the
potential, NMLFI2 violates slow-roll and is hardly compatible with
cosmological observations. For these reasons, it is no longer
considered in the following.

For NMLFI1 and NMLFI3, the previous equations allow us to determine
the parameter $\barhstar$ at which the pivot scale crosses the Hubble
radius during inflation by solving the Einstein frame's reheating
equation of \sectionc{sec:EFreheating}. Notice that, contrary to HI,
the value of the coupling $\jf{\lambda}$ is not set by the underlying
model and the non-minimal coupling $\xi$ is a model parameter not
fixed by the amplitude of the CMB anisotropies. As such, one can
forget about $\jf{\lambda}$ and trade it for the Einstein frame mass
scale $M$. Once $\barhstar$ is determined, $M$ is simply given by the
normalization of the potential
\begin{equation}
\dfrac{M^4}{\Mg^4} = 720 \pi^2 \xi \dfrac{\left(1+\barhstar^2\right)^2
  \left[p+(p-4)\barhstar^2\right]^2}{\barhstar^{p+2}\left[1 + (1+6\xi)
    \barhstar^2 \right]} \dfrac{\Qrms^2}{T^2}\,.
\end{equation}
Let us notice that in a situation where $\jf{\lambda}$ would be fixed
by the underlying theory, then, one should solve instead the coupled
reheating equations as explained in \sectionc{sec:hiexact}.

The reheating-consistent slow-roll prediction for NMLFI1 have been
represented in \Figs{fig:CMBNMLFI1_0} to \ref{fig:CMBNMLFI1_5} for
various values of $p$ and $\xi$, in the two regimes $p<4$ where it is
a small field model and for $p>4$ where it becomes large field-like.
Predictions for NMLFI3 can be found in \Figs{fig:CMBNMLFI3_0} to
\ref{fig:CMBNMLFI3_17} for various values of the three parameters $p$,
$\xi$ and $\chiend$. Here as well, we have split the parameter
domains into a ``small'' region, for $p<p_-$ with $ \xi<\xizero(p)$
and a ``large'' region for $p_-<p<4$. Let us notice that, for NMLFI3,
the values of $\chiend$ reported on these plots imply that the
numerical value of $\Mp$ could be up to three orders of magnitude
larger than the numerical value of $\Mg$, after inflation. Another
remark is that the case $p=4$ (HI) is unique in the sense that only
for a quartic power index $p$ the potential in the Einstein frame is
of the plateau-type. For any other values of $p$, one ends up with
potentials radically different than the plateau-type.

\subsection{Superconformal \texorpdfstring{$\alpha$}{alpha}-Attactor B Inflation (SABI)}
\label{sec:sabi}

\subsubsection{Theoretical Justifications}

In this section, we consider a generalization of the ``T-models''
(TMI, see \sectionc{sec:tmi}), which leads to another family of
$\alpha$-attractor models. The idea is to introduce a new embedding
K\"ahler potential and a new superconformal superpotential
given by
\begin{align}
  \calN \left(X^I,\bar{X}^{\bar{I}}\right)&=-\left \vert X^0
  \right\vert^2 \left[1-\frac{\left\vert X^1\right\vert^2+ \vert
      S\vert^2}{\left\vert X^0 \right\vert^2} +3\zeta \frac{\vert
      S\vert^4} {\left\vert X^0\right\vert^2\left(\left\vert X^0
      \right\vert^2- \left\vert X^1\right\vert^2\right)}
    \right]^{\alpha}, \\ \calW &=S\left(X^0\right)^2
  f\left(\frac{X^1}{X^0}\right)
  \left[1-\frac{\left(X^1\right)^2}{\left(X^0\right)^2}\right]^{\frac{3\alpha-1}{2}}.
    \end{align}
The field content of the model is the same as for the TMI model,
namely a conformon $X^0$, the inflaton $X^1=\Phi$ and a Goldstino
$X^3=S$. As before, the term proportional to the parameter $\zeta$ is
introduced to make sure that the inflationary trajectory is stable.

The main new aspect of the model is the presence of the parameter
$\alpha$. For $\alpha=1$, one recovers the embedding K\"ahler
potential~\eqref{eq:embkalone} and the superconformal
superpotential~\eqref{eq:embwalone}. Then, the conformal symmetry is
fixed by choosing $X^0=\bar{X}^{\bar{0}}=\sqrt{3}\,\Mg$ and, using
\Eq{eq:embedpot}, one obtains the K\"ahler and superpotential, namely
\begin{align}
  K &=-3\alpha \Mg^2\ln\left(1+k\right),
  \\
  W &=3\Mg^2 S f\left(\frac{\Phi}{\sqrt{3}\Mg}\right)
  \left(1-\frac{\Phi ^2}{3\Mg^2}\right)^{\frac{3\alpha-1}{2}},
\end{align}
where the function $k$ has already been defined in
\Eq{eq:defkkahlerembed}. Compared to the K\"ahler potential of TMI,
\sectionc{sec:tmi}, we see that it is still given by a logarithm of
the function $1+k$ but, now, multiplied by the parameter $\alpha$.

From the above expressions of $K$ and $W$, one can then calculate the
kinetic term and the potential of the inflaton field. As done in
\sectionc{sec:tmi}, this calculation is carried out along the
inflationary trajectory $S=0$. One obtains
\begin{equation}
  G_{A\bar{B}}=-\frac{3\alpha }{1+k} \frac{\partial ^2 k}{\partial X^A
    \partial \bar{X}^{\bar{B}}}+\frac{3\alpha}{(1+k)^2}
  \frac{\partial k}{\partial X^A}
  \frac{\partial k}{\partial \bar{X}^{\bar{B}}}\,,
  \end{equation}
that is to say 
\begin{equation}
  G_{A\bar{B}}=\frac{\alpha}{\Mg^2(1+k)^2}
  \begin{pmatrix}
    1 & 0 \\
    0 & 1+k
    \end{pmatrix}.
\end{equation}
These two expressions are the generalization of \Eqs{eq:Kmatrixalone}
and~\eqref{eq:Kmatrixalone2}. One immediately deduces that the
canonically normalized inflaton field $\varphi$ can be expressed in
terms of $\Phi$ in the following way
\begin{equation}
  \label{eq:sabi:fieldnormalalone}
  \Phi=\sqrt{6}\Mg \tanh\left(\dfrac{\varphi}{\sqrt{6\alpha}\Mg}\right),
\end{equation}
and this expression should be compared to its TMI's
counterpart in \Eq{eq:fieldnormalalone}. Finally, by using
\Eq{eq:potSvanish}, the potential of the canonically normalized field
$\varphi$ can been derived and one obtains
\begin{equation}
  \label{eq:pottanhal}
  V(\varphi)=9\Mg^4\left\vert
  f\left[\tanh\left(\frac{\varphi}{\sqrt{6\alpha}\Mg}\right)
    \right]\right\vert^2.
\end{equation}
As a consequence, one obtains a potential which ressembles a lot the
potential of the T-models, the difference being that the argument of
the function $f$ now involves $\varphi/(\sqrt{6\alpha}\Mg)$ instead of
$\varphi/(\sqrt{6}\Mg)$. We have therefore obtained a mixed model,
combining aspects of the T-model with aspects of the
$\alpha$-attractor.

In order to have an explicit potential, one must choose the function
$f$, which is equivalent to the choice of the function $\FT(.)$ in
\sectionc{sec:tmi}. As noticed in \sectioncs{sec:othertheosi} and
\ref{sec:tmi}, the choice
$\FT(\phi/\chi)=(\phi/\chi)^2/(1+\phi/\chi)^2$ leads to the
Starobinsky model for $\alpha=1$. If $\alpha \neq 1$, this leads to an
alternative derivation of the $\alpha$-attractor model, as mentioned
at the end of \sectionc{sec:saai}. One can also choose the more
general form
\begin{equation}
  \FT\left(\frac{\phi}{\chi}\right)
  =\left(\frac{\phi/\chi}{1+\phi/\chi}\right)^{2n},
\end{equation}
where $n$ is a new free index. This leads to the potential
\begin{equation}
V(\phi) = 9M^4 \left[ \dfrac{\tanh\left(\dfrac{\phi}{\Mg
      \sqrt{6\alpha}} \right)}{ 1+ \tanh\left(\dfrac{\phi}{\Mg
      \sqrt{6\alpha}} \right) } \right]^{2n},
\end{equation}
Clearly, this potential is a direct generalization of the
$\alpha$-attractor potential of \sectionc{sec:saai}.

\subsubsection{Slow-roll Analysis}

\begin{figure}
\begin{center}
\includegraphics[width=\wdblefig]{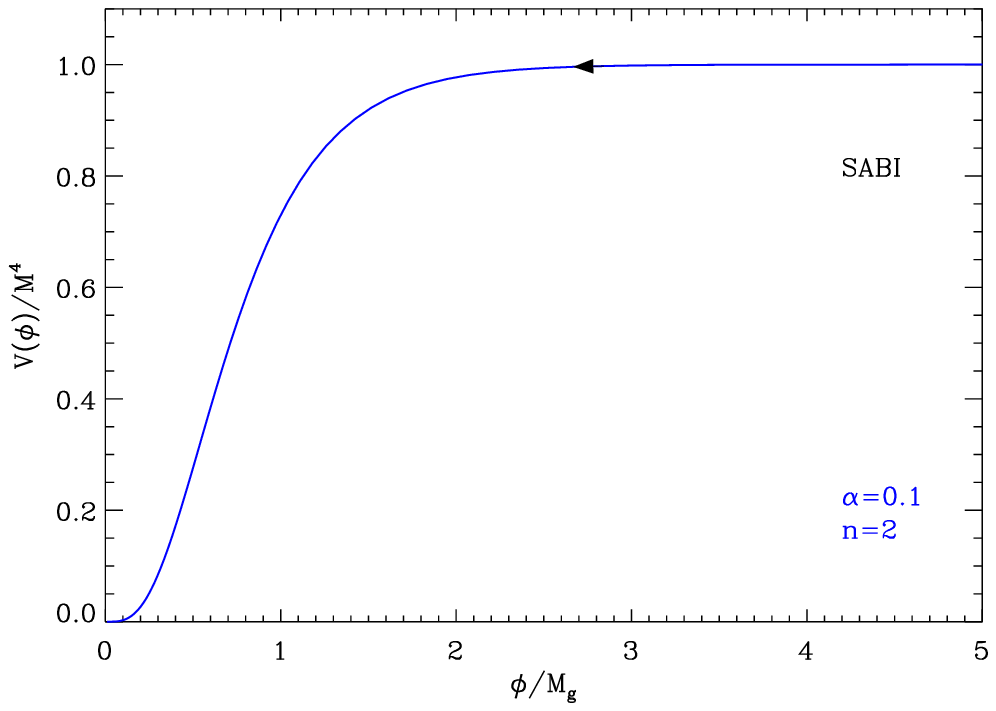}
\includegraphics[width=\wdblefig]{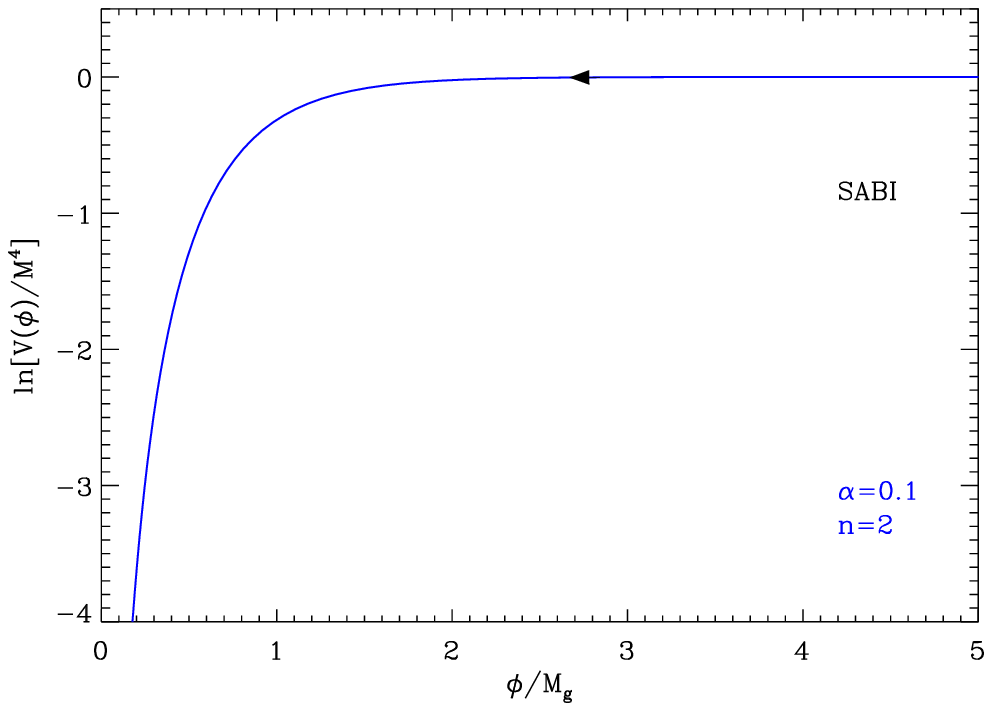}
\includegraphics[width=\wdblefig]{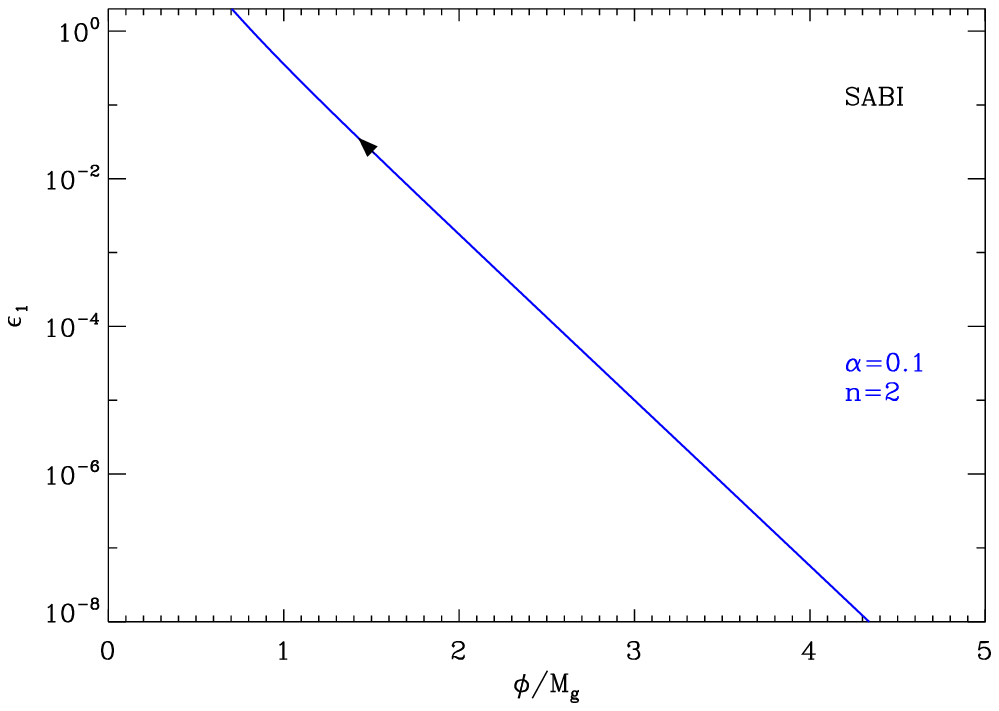}
\includegraphics[width=\wdblefig]{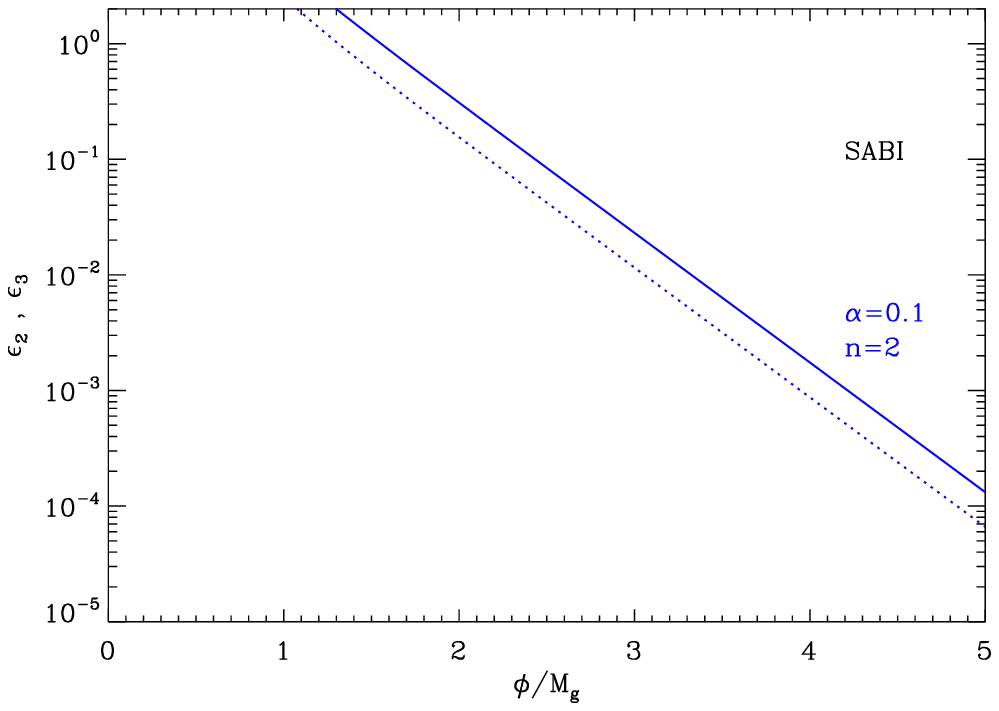}
\caption{Superconformal $\alpha$-attractor B Inflation (SABI). Top
  left panel: the potential as a function of $\phi/\Mg$.  Top right
  panel: logarithm of the potential. Bottom left panel: the first
  slow-roll parameter $\epsilon_1$. Bottom right panel: slow-roll
  parameters $\epsilon_2$ (solid line) and $\epsilon_3$ (dotted
  line).}
\label{fig:potsabi}
\end{center}
\end{figure}

As explained in the previous section, the potential of the SABI model
depend on two parameter, $\alpha$ and $n$, and can also be written as (we
have redefined the scale $M$)
\begin{equation}
V(\phi) = M^4 \left(1-e^{-\sqrt{\frac{2}{3\alpha}}x}\right)^{2n},
\label{eq:potsabi}
\end{equation}
where $x=\phi/\Mg$. For $n=1$, one recovers the $\alpha$-attractor
model, see \sectionc{sec:saai}, and for $n=1$ and $\alpha=1$, one has
the Starobinsky model, see \sectionc{sec:si}. These models may also be
referred to as $\alpha$-attractor E models in the literature. The
potential~\eqref{eq:potsabi} is represented in \Fig{fig:potsabi} for
different values of $\alpha$ and $n$. Since after inflation the field
settles at the minimum of the potential where it vanishes, the
numerical value of $\Mg \simeq \Mp$.

The three Hubble flow functions are given by
\begin{equation}
\begin{aligned}
\epsilon _1 &= \frac{4n^2}{3\alpha \left(e^{\sqrt{\frac{2}{3\alpha}}x} - 1\right)^{2}}\,,\quad
\epsilon_2= \frac{2n}{3\alpha
  \left[\sinh\left(\frac{x}{\sqrt{6\alpha}}\right)\right]^{2}}\,, \quad
\epsilon  _3 &=
\dfrac{4n}{3\alpha \tanh \left(\frac{x}{\sqrt{6\alpha}}\right)
  \left(e^{\sqrt{\frac{2}{3\alpha}}x} - 1\right)}\,.
\end{aligned}
\label{eq:epssabi}
\end{equation}
Evidently, when $\alpha=1$ and $n=1$, these expressions reduce to
\Eqs{eq:srparametershiggs} while if $n=1$ (and $\alpha$ unspecified),
one recovers the expressions of the Hubble flow functions given in
\sectionc{sec:saai}. The Hubble flow functions have been represented
in the lower panels of \Fig{fig:potsabi}.

In this scenario inflation gracefully ends when $\epsilon_1=1$, at a
field value $\xend$ given by
\begin{equation}
  \xend=\sqrt{\frac{3\alpha}{2}}\ln \left(1+
  \frac{2n}{\sqrt{3\alpha}}\right).
\label{eq:xendsabi}
\end{equation}
However, as it was the case for Higgs inflation in \sectionc{sec:hi},
and also for the $\alpha$-attractor scenario (see \sectionc{sec:saai}),
violation of the slow-roll conditions can occur before. The value of
the field for which $\epsilon_2=1$ can be expressed as
\begin{equation}
 \xepstwoOne = \sqrt{6\alpha}\,\arsinh
  \left(\sqrt{\frac{2n}{3\alpha}}\right),
\end{equation}
and the field value for which $\epsilon_3=1$ is
\begin{equation}
  \xepsthreeOne = \sqrt{6\alpha}\,
  \artanh \left(\frac{2}{1+\sqrt{1+6\alpha/n}}\right).
\end{equation}
In the case of the $\alpha$-attractor model, regardless of the value
of $\alpha$, the field reaches first the value $\xepstwoOne$, then
$\xepsthreeOne$ and, finally, $\xend$. It is interesting to notice
that, here, this hierarchy does no longer exist: when the parameter $n$ is
changed, the field value can reach $\xepstwoOne$ before inflation
stops, or not.

The slow-roll trajectory can be analytically derived and reads
\begin{equation}
\Nend-N=\frac{1}{2n}\sqrt{\frac{3\alpha}{2}}\left(\xend-x\right) +
\frac{3\alpha}{4n}\left(\ee^{\sqrt{\frac{2}{3\alpha}} x}
-\ee^{\sqrt{\frac{2}{3\alpha}}\xend}\right).
\label{eq:sabi:traj}
\end{equation}
As it was the case for the $\alpha $-attractor model, this trajectory
can be inverted and expressed in term of the ``$-1$-branch'' of the
Lambert function $\Lambert{-1}$. One finds
\begin{equation}
\begin{aligned}
  x =& \sqrt{\frac{3\alpha}{2}} \Biggl\lbrace -\frac{4n}{3\alpha}
  \Delta N +\sqrt{\frac{2}{3\alpha}}\xend
  -\ee^{\sqrt{\frac{2}{3\alpha}}\xend} 
\\ &-
  \Lambert{-1}
    \left[-\exp\left(-\frac{4n}{3\alpha} \Delta N +\sqrt\frac{2}{3\alpha}
        \xend-\ee^{\sqrt{\frac{2}{3\alpha}}\xend} \right)\right]\Biggr\rbrace,
\end{aligned}
\label{eq:sabi:trajinverted} 
\end{equation}
where, as usual, $\Delta N = \Nend - N$. The reason that inflation
proceeds along the $-1$ branch of the Lambert function has already
been explained in \sectionc{sec:saai}.

Finally, the value of $\xstar$, at which the pivot
mode crossed out the Hubble radius during inflation can be expressed as
\begin{equation}
\begin{aligned}
  \xstar & =\sqrt\frac{3\alpha}{2}\left[-\frac{4n}{3\alpha}\Delta
    \Nstar + \ln\left( 1+\frac{2n}{\sqrt{3\alpha}} \right) - \left(
    1+\frac{2n}{\sqrt{3\alpha}} \right) \right] \\ &
  -\sqrt{\dfrac{3\alpha}{2}}\, \Lambert{-1}\left\lbrace - \exp\left[
    -\frac{4n}{3\alpha}\Delta
    \Nstar+\ln\left(1+\frac{2n}{\sqrt{3\alpha}} \right) - \left(
    1+\frac{2n}{\sqrt{3\alpha}} \right) \right] \right \rbrace,
\end{aligned}
\label{eq:xstarsabi}
\end{equation}
where, in this expression, we have used the value of $\xend$ derived
above. From the knowledge of $\xstar$, the energy scale $M$ of the
potential can be inferred and one obtains
\begin{equation}
\frac{M^4}{\Mg^4}=\frac{1920 \pi^2 n^2}{\alpha}
\left(1-\ee^{-\sqrt{\frac{2}{3\alpha}}\xstar}\right)^{-2(n+1)}
\ee^{-2\sqrt{\frac{2}{3\alpha}}\xstar} \dfrac{\Qrms^2}{T^2}\,.
\label{eq:sabi:COBE}
\end{equation}

The reheating consistent slow-roll prediction for Superconformal
$\alpha$-attractor B Inflation have been represented in
\Figs{fig:CMBSABI_0} to \ref{fig:CMBSABI_2}.

\subsection{Superconformal \texorpdfstring{$\alpha$}{alpha}-Attactor
  T Inflation (SATI)}
\label{sec:sati}

\begin{figure}
\begin{center}
\includegraphics[width=\wdblefig]{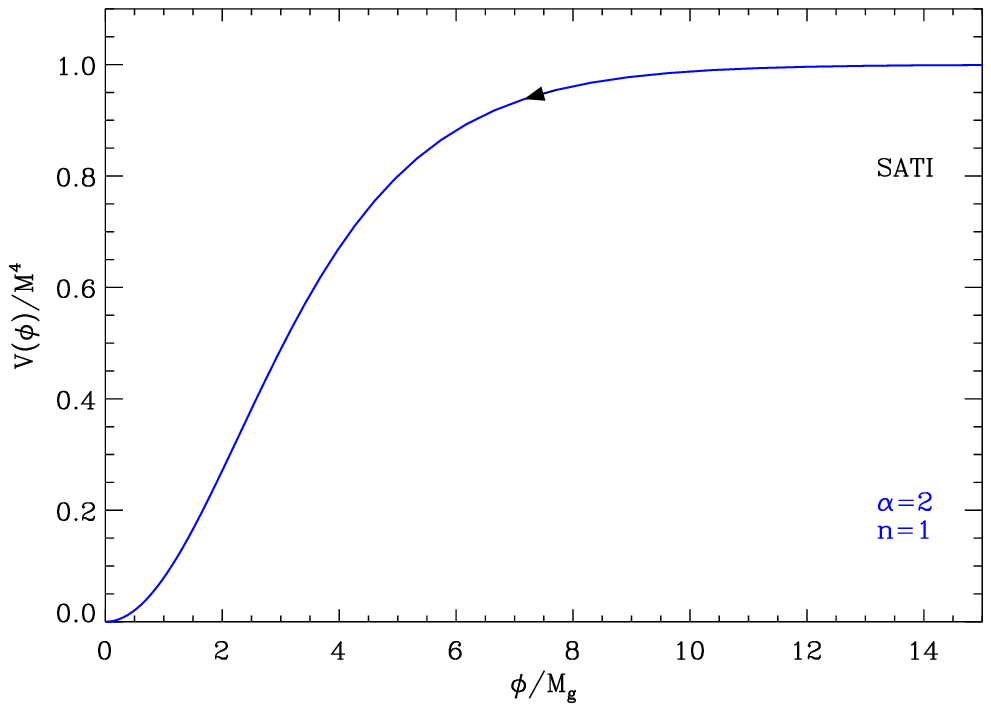}
\includegraphics[width=\wdblefig]{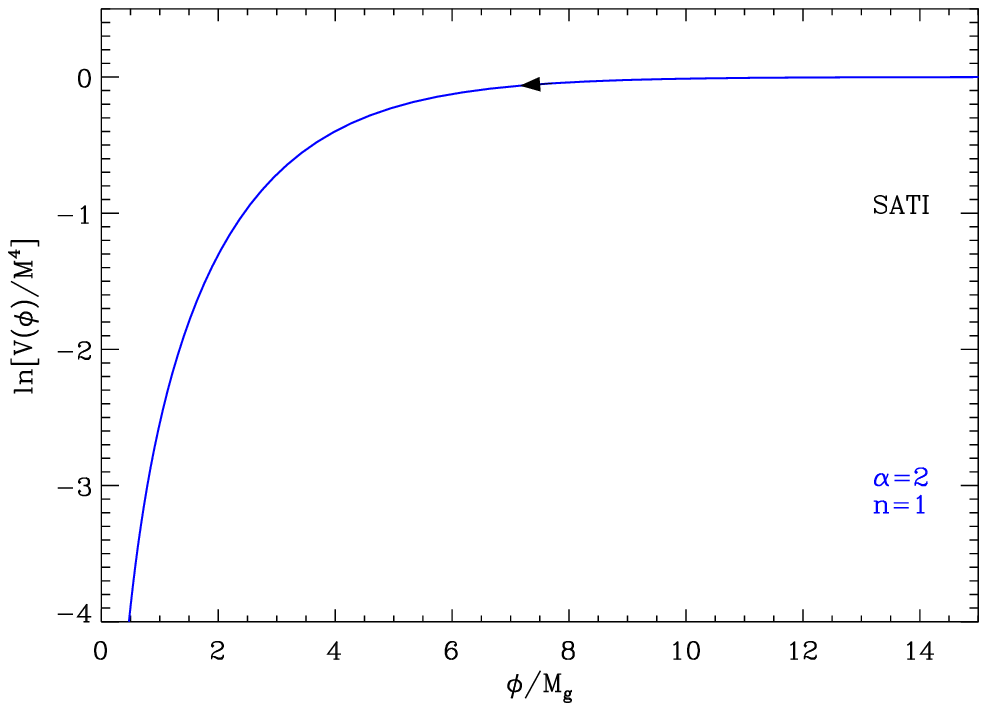}
\includegraphics[width=\wdblefig]{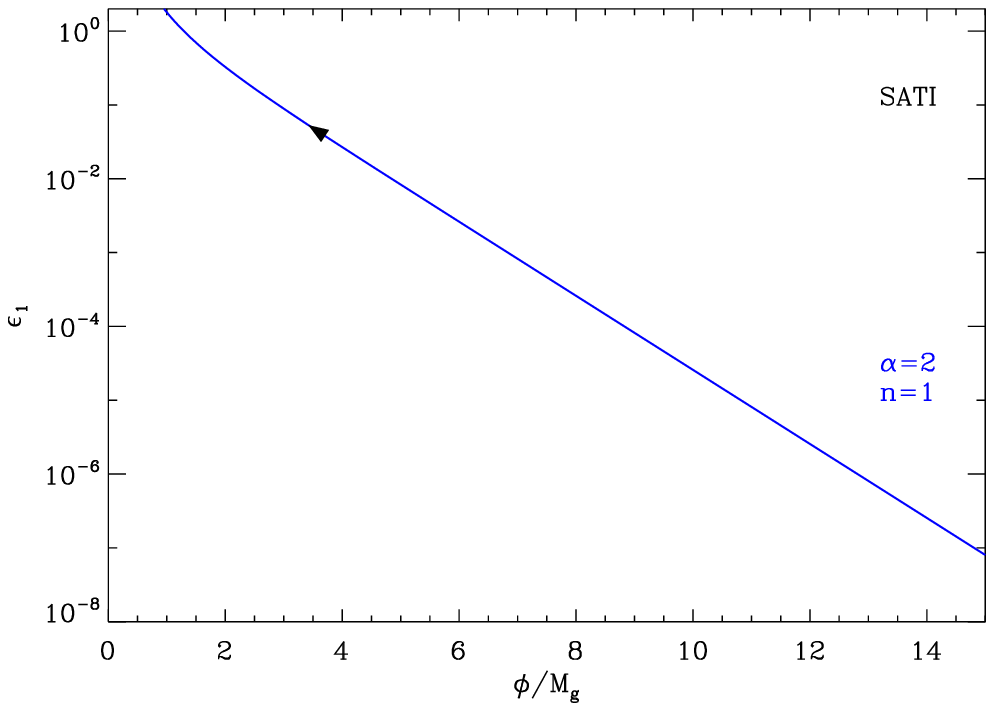}
\includegraphics[width=\wdblefig]{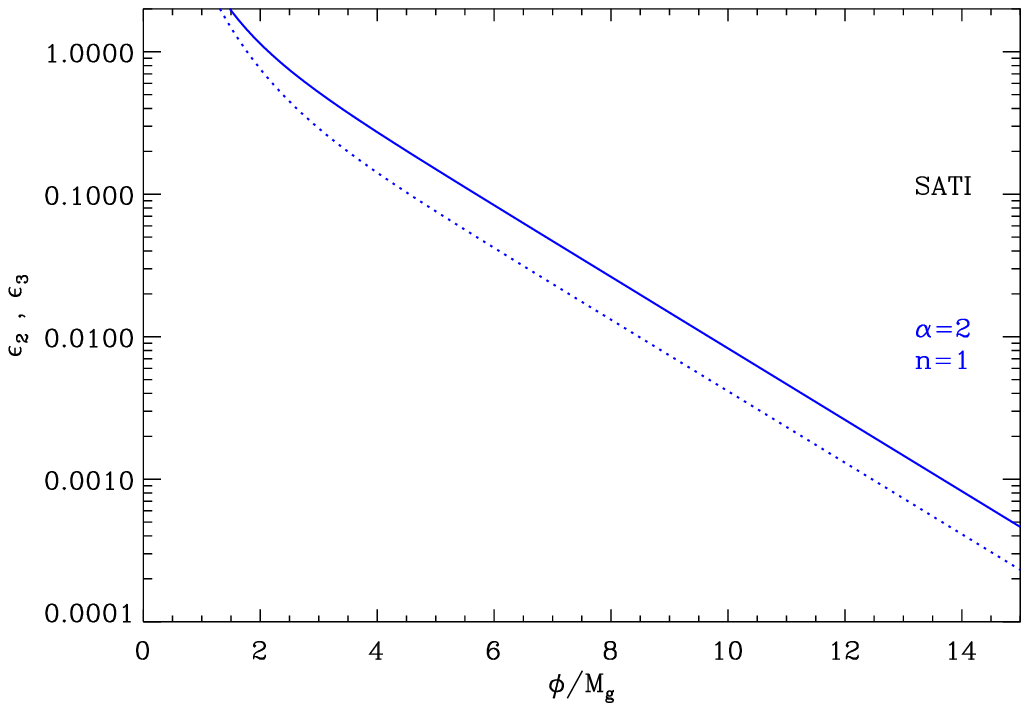}
\caption{Superconformal \texorpdfstring{$\alpha$}{alpha}-Attactor T
  Inflation (SATI). Top left panel: the potential as a function of
  $\phi/\Mg$.  Top right panel: logarithm of the potential. Bottom
  left panel: the first slow-roll parameter $\epsilon_1$. Bottom right
  panel: slow-roll parameters $\epsilon_2$ (solid line) and
  $\epsilon_3$ (dotted line).}
\label{fig:potsati}
\end{center}
\end{figure}

These models have been discussed in \Refc{Kallosh:2013yoa}. In
\sectionc{sec:sabi}, we have seen how to generate a class of
potentials that depend on $\tanh[\phi/(\Mg\sqrt{6\alpha})]$, see
\Eq{eq:pottanhal}. The precise shape of the potential then depends on
an arbitrary function $f\sim \FT(.)$. A specific choice was made in
\sectionc{sec:sabi} for this function and, here, we study another
choice. In fact, this choice of $\FT(.)$ was already considered in
\Eq{eq:ftpowerlaw} and is just a power-law. This directly leads to the
SATI potential
\begin{equation}
  V(\phi)=M^4\left[\tanh \left(\frac{\phi}
    {\sqrt{6 \alpha}\Mg}\right)\right]^{2n},
  \label{eq:potsati}
\end{equation}
which represents a generalization of the TMI model discussed in
\sectionc{sec:tmi}. It describes a two parameters model with $\alpha$
and $n$, matching TMI for $\alpha=1$.

Defining $x\equiv \phi/\Mg$, one obtains the Hubble flow
functions
\begin{equation}
\begin{aligned}
  \epsilon _1 &= \frac{4n^2}{3\alpha}\sinh^{-2}
  \left(\frac{2x}{\sqrt{6\alpha}}\right)
,\quad
\epsilon_2= \frac{8n}{3\alpha}
\frac{\cosh\left(\dfrac{2x}{\sqrt{6\alpha}}\right)}
     {\sinh^2\left(\dfrac{2x}{\sqrt{6\alpha}}\right)}
\,, \quad
\epsilon_3=\frac{2n}{3\alpha}\frac{3+\cosh\left(
  \dfrac{4 x}{\sqrt{6\alpha}}\right)}{\sinh^2
  \left(\dfrac{2 x}{\sqrt{6\alpha}}\right)
  \cosh\left(\dfrac{2 x}{\sqrt{6\alpha}}\right)}\,.
\end{aligned}
\label{eq:epssati}
\end{equation}
For $\alpha=1$, one can check that the Hubble flow function of the TMI
model of \sectionc{sec:tmi} are recovered. The potential, its
logarithm and the the Hubbel flow functions have been plotted in
\Fig{fig:potsati} as a function of $x$. The field value after
inflation is expected to be at the minimum of the potential and
vanishing. As a result, the numerical value of $\Mg \simeq \Mp$.

In this scenario, as shown in \Fig{fig:potsati} (bottom left panel),
the first slow-roll parameter increases as the vacuum expectation of
the field decreases and this implies that inflation stops by violation
of the slow-roll conditions, when $\epsilon_1=1$. The corresponding
vacuum expectation value of the field reads
\begin{equation}
  \xend=\frac{\sqrt{6\alpha }}{2}
  \arcsinh\left(\frac{2n}{\sqrt{3\alpha}}\right).
\end{equation}

The slow-roll trajectory can be integrated exactly and one obtains
\begin{equation}
  \Nend-N=\frac{3\alpha}{4n}\left[\cosh\left(\frac{2x}{\sqrt{6\alpha}}\right)
    -\cosh\left(\frac{2\xend}{\sqrt{6\alpha}}\right)\right].
\end{equation}
This formula can be inverted analytically and, as a consequence,
$\phi/\Mg$ during slow-roll inflation can be expressed as
\begin{equation}
  x=\frac{\sqrt{6\alpha }}{2}\arccosh\left(\sqrt{1+\frac{4n^2}{3\alpha}}
  +\frac{4n}{3\alpha}
    \Delta N\right),
\label{eq:satitraj}
\end{equation}
where $\Delta N = \Nend - N$. The value of $\xstar$ is just given by
the above expression with $\Delta N=\Delta N_*$. Again, one verifies
that the above formulas are equivalent to those presented in
\sectionc{sec:tmi} when $\alpha=1$.

Finally, the mass scale $M$ that normalizes the potential can be expressed as
\begin{equation}
\label{eq:sati:COBE}
\frac{M^4}{\Mg^4}=\frac{1920\pi^2n^2}{\alpha
\sinh^2\left(\displaystyle\frac{2\xstar}{\sqrt{6\alpha}}\right)
\left[\tanh\left(\displaystyle\frac{\xstar}{\sqrt{6\alpha}}\right)
  \right]^{2n}}
\frac{\Qrms^2}{T^2}\, .
\end{equation}

The reheating consistent observable predictions for SATI have been
represented in \Figs{fig:CMBSATI_0} to \ref{fig:CMBSATI_2} for various
values of $n$ and $\alpha$. As before, one notices that the dependence
of the spectral index and tensor-to-scalar ratio with respect to $n$
are very small. Indeed, if $n \Delta N$ dominates in \Eq{eq:satitraj},
one obtains
\begin{equation}
  \xstar \simeq \dfrac{\sqrt{6 \alpha}}{2}
  \arccosh\left(\dfrac{4n}{3\alpha }
  \Delta\Nstar\right).
\end{equation}
Plugging this approximation into \Eqs{eq:epssati} gives
\begin{equation}
\epsonestar \simeq \dfrac{3\alpha }{4 \Delta\Nstar^2}\,, \quad \epstwostar
\simeq \dfrac{2}{\Delta\Nstar}\,, \quad \epsthreestar \simeq
\dfrac{1}{\Delta\Nstar}\,,
\end{equation}
and the Hubble-flow functions are independent of $n$ in the large
$\Delta\Nstar$ limit. However, one notices that $\epsonestar$ retains
a dependence in $\alpha$ while the two other Hubble flow parameters
remains unaffected.

The reheating consistent slow-roll prediction for Superconformal
$\alpha$-attractor T Inflation have been represented in
\Figs{fig:CMBSATI_0} to \ref{fig:CMBSATI_2}.

\section{Three Parameters Models}
\label{sec:threep}

\subsection{Running-mass Inflation (RMI)}
\label{sec:rmi}

\subsubsection{Theoretical Justifications}
\label{subsubsec:theoryrmi}

This model has been derived and studied in \Refcs{Stewart:1996ey,
  Stewart:1997wg,Covi:1998jp,Covi:1998mb,
  Leach:2000ea,Lyth:2000ie,Covi:2000gx,Covi:2002th,Kadota:2003tn,
  Covi:2004tp}. Following \Refc{Covi:1998mb}, let us briefly discuss
its physical origin. At tree level, a potential can always be expanded
as $V(\phi)\simeq M^4+m^2\phi^2/2+\lambda\phi^4/4 +\cdots $. Since the
potential must be flat to support inflation, quantum corrections may
play an important role. Typically, they modify the potential with a
term of the form $\left(c_1+c_2\phi^2+c_4\phi^4\right)\ln
\left(\phi/\mu\right)$, where $\mu$ is the renormalization scale. In a
non-supersymmetric framework, the quartic term dominates and one is
led to models similar to RCMI, RCQI or CWI, see \sectionc{sec:rcmi},
\ref{sec:rcqi} and~\ref{sec:cwi}. On the other hand, in a
supersymmetric context, at least if supersymmetry is spontaneously
broken, the quadratic and the quartic terms cancel and one is left
with a model similar to LI, see \sectioncs{sec:li}. If, however,
supersymmetry is explicitly broken by the presence of soft terms, then
the most important term will be the quadratic one.

Concretely, the above reasoning leads to a specific shape for the
inflaton potential. We start from a flat direction in
supersymmetry. Then, we assume that supersymmetry is explicitly
broken and, as a consequence, that the potential receives corrections
$\propto m^2\phi^2$, where $m$ is a soft mass. Higher order terms are
supposed to be negligible since we assume $\phi/\Mp \ll 1$. We thus
have
\begin{equation}
\label{eq:treermi}
V=V_0+\frac{1}{2}m^2\phi^2 +\cdots,
\end{equation}
The one loop corrections to this tree potential will typically induces
a logarithmic dependence of the soft mass through the renormalization
group equation
\begin{equation}
\frac{\dd m^2}{\dd \ln \phi}=\betam,
\end{equation}
where $\betam$ is proportional to the inflaton couplings
with the other fields present in the theory. Therefore, by Taylor
expanding the solution of the previous equation aroused
$\phi=\bar{\phi}$, we can write
\begin{equation}
\label{eq:expansionm}
m^2=m^2(\bar{\phi})+\betam\ln \left(\frac{\phi}{\bar{\phi}}\right) +\cdots.
\end{equation}
As a consequence, the potential~(\ref{eq:treermi}) can be re-expressed
as
\begin{equation}
V(\phi)=V_0+\frac12m^2(\bar{\phi})\phi^2+\frac12\betam\phi^2
\ln\left(\frac{\phi}{\bar{\phi}}\right).
\end{equation}
As noticed
in \Refcs{Covi:1998mb,Covi:2002th,Covi:2004tp}, the beta function can
typically be expressed as
\begin{equation}
\betam=\frac{-2C}{\pi}\alpha \tilde{m}^2+\frac{D}{16\pi^2}
\vert \lambda\vert^2 \mloop^2,
\label{eq:rmi:beta}
\end{equation}
if we assume that the inflaton interacts with gauge bosons and
fermions. The quantity $\alpha$ is the coupling constant between
$\phi$ and the gauge boson, $\lambda $ is a Yukawa
coefficient, $\tilde{m}$ is the gaugino mass, $m$ the fermionic mass
and $C$ and $D$ are dimensionless numbers of order one.

In the next section, we explore the cosmological consequences of
this type of potential. In particular, we will see that it can lead to
four different kind of inflationary scenarios.

\subsubsection{Slow-Roll Analysis}
\label{subsubsec:srrmi}

\begin{figure}
\begin{center}
\includegraphics[width=\wdblefig]{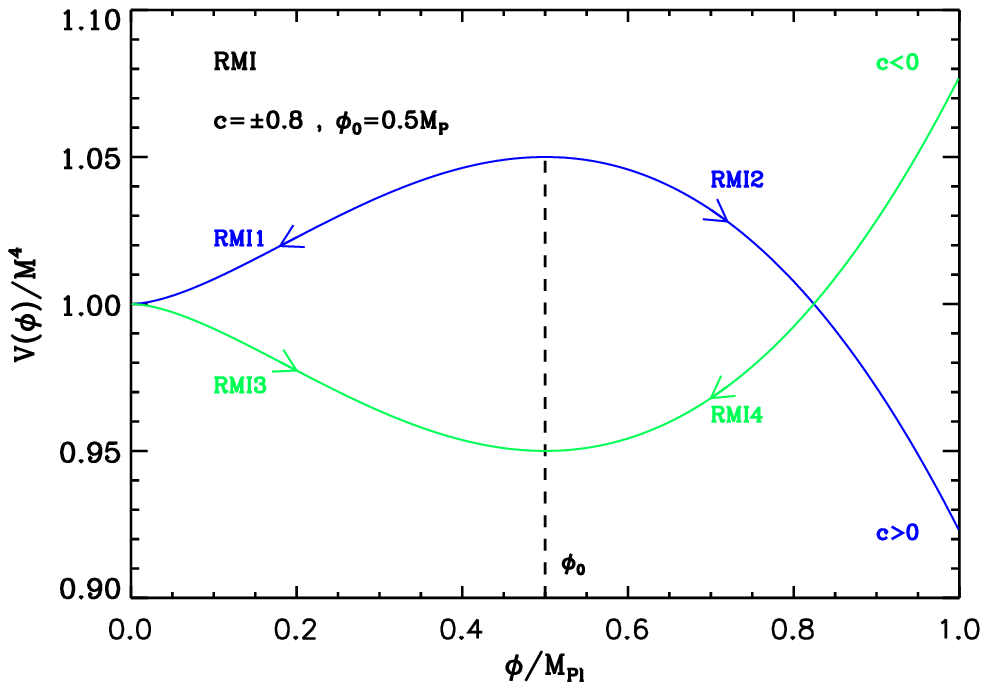}
\includegraphics[width=\wdblefig]{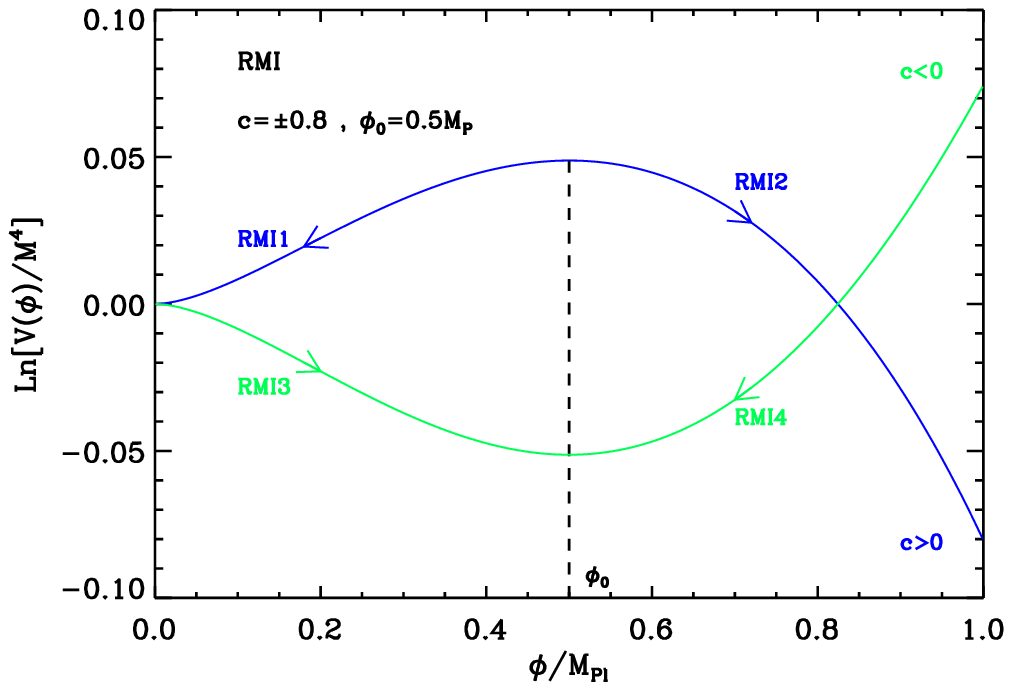}
\includegraphics[width=\wdblefig]{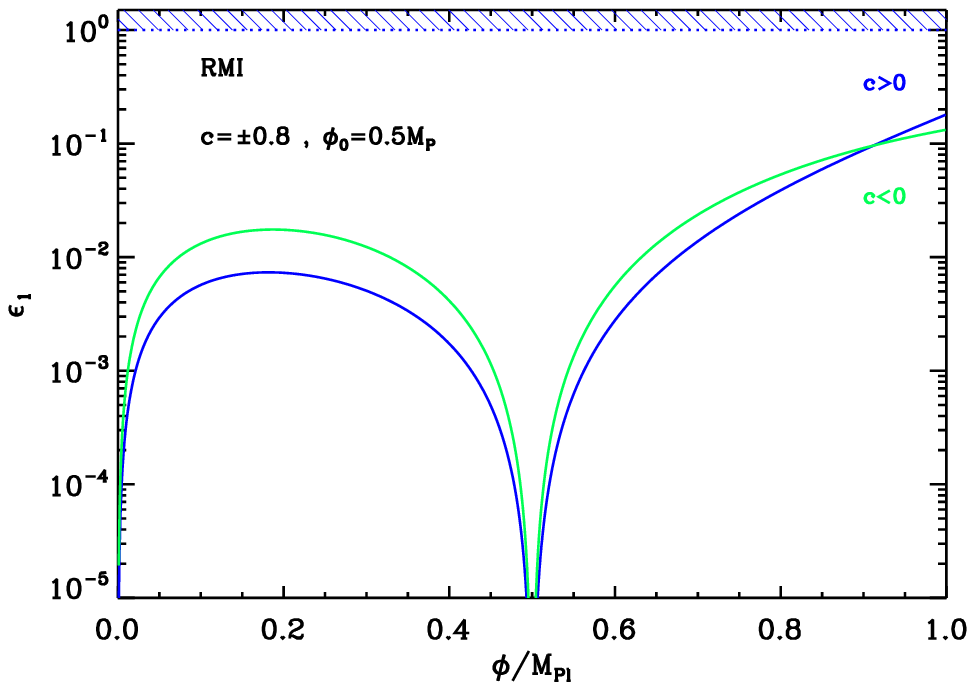}
\includegraphics[width=\wdblefig]{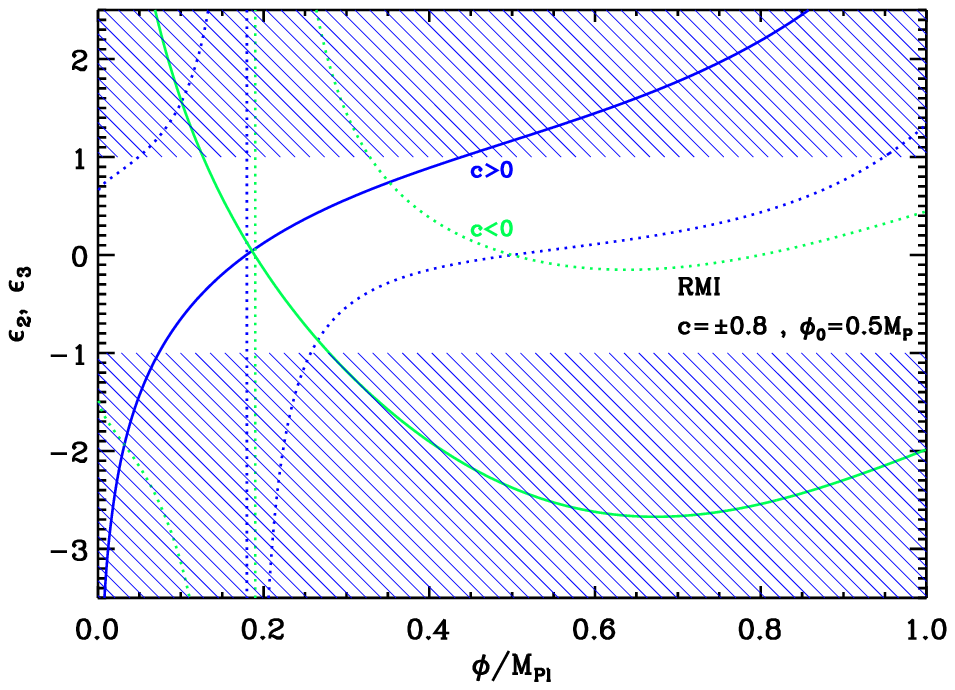}
\caption{Top left panel: running mass potential for $c=0.8$ (blue
  line) or $c=-0.8$ (green line) and $\phizero=0.5\Mp$. Top right
  panel: logarithm of the potentials for the same values of $c$ and
  $\phizero$. Bottom left panel: slow-roll parameter $\epsilon _1$ for
  a potential with $c=\pm 0.8$ and $\phizero=0.5\Mp$. Bottom right
  panel: slow-roll parameters $\epsilon _2$ (solid line) and $\epsilon
  _3$ (dotted line) for $c=\pm 0.8$ and $\phizero=0.5\Mp$. The value
  $c=\pm 0.8$ may not be physical and was chosen only in order to
  produce a clear plot.}
\label{potrmi}
\end{center}
\end{figure}

We now perform the slow-roll analysis of the potential previously
derived. In order to carry out this task, it is more convenient 
to re-write the potential as follows
\begin{equation}
\label{potentialrunning}
  V(\phi) = M^4\left[1-\frac{c}{2}\left(-\frac{1}{2} +\ln
\frac{\phi }{\phizero}\right)\frac{\phi ^2}{\Mp^2}\right],
\end{equation}
where we have defined the two parameters $c$ and $\phizero$ by
\begin{equation}
c=-\frac{\Mp^2 \betam}{2V_0}, \quad 
m^2(\bar{\phi})=-\betam\left[\frac12
+\ln \left(\frac{\phizero}{\bar{\phi}}\right)\right].
\end{equation}
In this expression, $M$, $c $ and $\phizero$ are free parameters. The
dimensionless parameter $c$ can be positive or negative. With the form
of the beta function given in \Eq{eq:rmi:beta}, the coefficient $c$ is
given by $\alpha m^2\Mp^2/V_0$. If one assumes that the soft masses
are of order $m\simeq H\simeq V_0^{1/2}/\Mp^2$, then $c\simeq
\alpha\simeq 10^{-2}$ to $10^{-1}$ or may be smaller depending on the
assumption on the couplings. This also mean that, in order for
\Eq{eq:expansionm} to be valid, one has
$\left\vert\ln\left(\phi/\phizero\right)\right\vert\ll 1$. Also, the
model is commonly worked out in the vacuum dominated regime (otherwise
it is equivalent to a large field model, LFI, see \sectionc{sec:lfi}),
which means that $c\phizero^ 2/\Mp^ 2\ll 1$.  The location $\phi=\phi
_0$ is an extremum of $V\left(\phi\right)$, a maximum if $c>0$ and a
minimum if $c<0$. The potential and its logarithm are represented in
\Fig{potrmi}.

Running mass inflation can be realized in four different
ways~\cite{Covi:1998mb}, denoted as RMI1, RMI2, RMI3 and RMI4 in what
follows. RMI1 corresponds to the case where $c>0$ and $\phi
<\phizero$, see \Fig{potrmi} (top panels). In this case, $\phi $
decreases during inflation which proceeds from the right to the
left. RMI2 also corresponds to $c>0$ but with $\phi >\phizero$ and
$\phi $ increases during inflation which now proceeds from the left to
the right. RMI3 refers to the situation where $c<0$ and $\phi
<\phizero$ all the time. In this case, $\phi $ increases during
inflation which proceeds from the left to the right. Finally, RMI4 has
$c<0$ and $\phi >\phizero$ decreases as inflation proceeds from the
right to the left.

Using the potential~(\ref{potentialrunning}), one can calculate the
three slow-roll parameters $\epsilon _1$, $\epsilon _2$ and $\epsilon
_3$. Defining $x\equiv \phi/\phizero$, one obtains the following
expressions
\begin{equation}
\epsilon _1 = \frac{c^2}{2} \left[\dfrac{\dfrac{\phizero}{\Mp} x \ln x
  }{1-\dfrac{c}{2}\dfrac{\phizero
      ^2}{\Mp^2} \left(-\dfrac12+\ln x \right) x^2 }\right]^{2},
\label{eq:eps1running}
\end{equation}
\begin{equation}
\epsilon _2 = 2c \, \dfrac{1+\dfrac{c}{4}\dfrac{\phizero ^2}{\Mp^2}
  x^2 +\left(1-\dfrac{c}{4} \dfrac{\phizero ^2}{\Mp^2} x^2 \right)\ln
  x + \dfrac{c}{2} \dfrac{\phizero ^2}{\Mp^2} x^2 \ln ^2
  x}{\left[1-\dfrac{c}{2} \dfrac{\phizero ^2}{\Mp^2} \left(-
    \dfrac12+\ln x \right) x^2 \right]^{2}}\,,
\label{eq:eps2running}
\end{equation}
and
\begin{equation}
\begin{aligned}
\epsilon _3 & = \dfrac{c \ln x}{\left[ 1-\dfrac{c}{2} \dfrac{\phizero
      ^2}{\Mp^2} \left( -\dfrac12 + \ln x \right) x^2 \right]^{2}}
\left[ 1+\dfrac{c}{4} \dfrac{\phizero^2} {\Mp^2} x^2 +\left(
  1-\dfrac{c}{4} \dfrac{\phizero^2}{\Mp^2} x^2 \right) \ln x
  +\dfrac{c}{2}\dfrac{\phizero^2}{\Mp^2} x^2 \ln^2 x \right]^{-1} \\ &
\times \left[1+\dfrac{c}{2} \dfrac{\phizero ^2}{\Mp^2} x^2 +
  \dfrac{c^2}{16} \dfrac{\phizero^4}{\Mp^4} x^4 + c
  \left(2\dfrac{\phizero^2}{\Mp^2} x^2 +
  \dfrac{c}{2}\dfrac{\phizero^4}{\Mp^4} x^4 \right) \ln x \right. \\ &
  + \left. c
  \left(3\dfrac{\phizero^2}{\Mp^2} x^2  -
  \dfrac{c}{2}\dfrac{\phizero^4}{\Mp^4} x^4  \right) \ln^2
  x  +  \dfrac{c^2}{2}\dfrac{\phizero^4}{\Mp^4} x^4 \ln^3 x
  \right].
\end{aligned}
\label{eq:eps3running}
\end{equation}
The slow-roll parameters are represented in the bottom panels in
\Fig{potrmi}.

Let us now examine how inflation ends in this model. The slow-roll
parameter $\epsilon_1$ has a maximum in the $x < 1$ region and a
maximum in the $x>1$ region, see \Fig{potrmi}. If these maxima were
larger than one, inflation could in principle stop by violation of the
slow-roll conditions. In the vacuum dominated approximation, however,
we see from \Eq{eq:eps1running}, that $\epsilon_1\simeq (c^2/2)
(\phizero^2/\Mp^2) x^2 \ln^ 2 x$. This means that the \vev $\xend$
satisfies $\xend \ln \xend = \pm (\sqrt{2}/c)(\Mp/ \phizero)$. But we
have established previously that the vacuum dominated condition
precisely implies that $c\Mp/ \phizero\gg 1$ and one would have $\ln
\xend \gg 1$. But for the model to be valid, we have already mentioned
that the condition $\left\vert\ln x \right\vert\ll 1$ should be
enforced. We conclude that the value of $\xend $ obtained above lies
outside the regime of validity of the potential. The end of inflation
either occurs by violation of slow-roll but in a regime where
additional unknown corrections arise and modify the shape of
$V(\phi)$, or by tachyonic instability. In this last case, inflation
stops in a regime where our calculations are valid. This also means
that we must consider an additional parameter in the model, namely
$\xend$. In this article, this is the assumption made which implies
that RMI is indeed a three parameters model.

We now turn to the calculation of the observable predictions. The
first step is to obtain the slow-roll trajectory. One obtains
\begin{equation}
\begin{aligned}
  N-\Nend & = \frac{1}{c} \left( \ln \left\vert\ln x \right\vert-\ln
  \left \vert\ln \xend \right\vert\right) -\frac{1}{4}
  \frac{\phizero^2}{\Mp^2} (x^2 - \xend^2) \\ & + \frac{1}{4}
  \left(\frac{\phizero}{\Mp^2}\right)^2\left[ \Ei\left(2\ln x \right)
    -\Ei\left(2\ln \xend \right) \right],
\end{aligned}
\label{eq:rmefold}
\end{equation}
where the exponential integral function $\Ei$ is defined by
$\Ei(x)\equiv -\int _{-x}^{+\infty} \dd t
\ee^{-t}/t$~\cite{Abramovitz:1970aa,Gradshteyn:1965aa}. This
expression cannot be inverted analytically. However, in the limit
$(c\phizero /\Mp) x \ll 1$ (the vacuum dominated regime), the above expression
can be approximated by
\begin{equation}
\label{Napproxrun}
N - \Nend \simeq \frac{1}{c}\left(\ln \left\vert \ln x \right \vert
-\ln \left \vert \ln \xend \right \vert \right),
\end{equation}
from which it follows that
\begin{equation}
\label{trajecrunning}
x(N) = \exp\left[\ee^{c(N-\Nend)}\ln \xend \right].
\end{equation}

The slow-roll predictions of the four models, RMI1, RMI2, RMI3 and
RMI4 are presented in \Figs{fig:CMBRMI1}, \ref{fig:CMBRMI2},
\ref{fig:CMBRMI3} and \ref{fig:CMBRMI4} for $\vert c\vert=10^{-2}$,
$\phizero/\Mp<1/\sqrt{\vert c\vert}$, and $1/e<\xend<e$,
respectively. In order to interpret them, it is interesting to use
some approximations. From the trajectory~(\ref{trajecrunning}), it is
straightforward to calculate $\xstar$. Recalling that inflation is
supposed to stop at $\xend$, one obtains $\xstar= \exp
\left(\ee^{-c\Delta \Nstar} \ln \xend \right)$. Then, using
\Eqs{eq:eps1running}, (\ref{eq:eps2running}) and
(\ref{eq:eps3running}) in the vacuum dominated limit, we find that
\begin{eqnarray}
\label{eq:approxe1rmi}
  \epsilon_{1*} &\simeq & \frac{c^2}{2} \left( \frac{\phizero}{\Mp} \right)^2
  \exp \left(2{\ue}^{-c\Delta \Nstar} \ln \xend \right)
\ee^{-2c \Delta \Nstar}\ln^2 \xend , \\ 
\label{eq:approxe2rmi}
  \epsilon_{2*} & \simeq & 2c \left(1+ \ee^{-c\Delta \Nstar} \ln \xend \right).
\end{eqnarray}
In fact, in order to compare with the existing literature, it turns
out to be convenient to define the following quantity
\begin{equation}
  s \equiv c \ln \xstar = -c\,
  \ee^{-c\Delta \Nstar} \ln \xend.
\label{eqs}
\end{equation}
For RMI1 and RMI4, $s>0$ while for RMI2 and RMI3 one has $s<0$. In
terms of $s$ \Eqs{eq:approxe1rmi} and~(\ref{eq:approxe2rmi}) can be
re-written as
\begin{equation}
\epsilon _{1*} \simeq \frac{s^2}{2}\left(\frac{\phi _0}{\Mp}\right)^2
\ee^{-2s/c}, \qquad \epsilon_{2*} \simeq 2c
\left(1-\frac{s}{c}\right).
\label{eps1eps2rm}
\end{equation}
These equations imply that the locus of the model predictions in the
plane $(\epsilon _1,\epsilon _2)$ are given by $\epsilon_2\simeq
2(c-s)+4\epsilon _1\Mp^2/\phizero^2$. If we neglect $\epsilon _{1*}$
(with respect to $\epsilon_{2*}$) one recovers the formula derived in
\Refcs{Covi:1998mb, Covi:2002th, Covi:2004tp}, namely $\nS-1\simeq
2(s-c)$. The same route for the third slow-roll parameter gives
$\epsilon _2\epsilon _3 \simeq -2cs$ and neglecting again $\epsilon_1$
gives the scalar running $\alphaS \simeq 2sc$. The above analytic
estimates agree well with the complete slow-roll predictions
represented in \Figs{fig:CMBRMI1}, \ref{fig:CMBRMI2},
\ref{fig:CMBRMI3} and \ref{fig:CMBRMI4}.

From the CMB normalization, we obtain the following expression for the
mass scale
\begin{equation}
\frac{M^4}{\Mp^4}=720 \pi^2 c^2\frac{\Qrms^2}{T^2} \dfrac{\phizero^2}{\Mp^2}
\frac{\xstar^2 \ln ^2\left(\xstar\right)}{
\left\{1-\dfrac{c}{2} \dfrac{\phizero^2}{\Mp^2} \left[-\dfrac{1}{2} +\ln \left(
\xstar \right)\right]\xstar ^2 \right\}^3}\,.
\end{equation}
In the vacuum dominated regime, this expression can be approximated by
\begin{equation}
\frac{M^4}{\Mp^4}\simeq 720 \pi^2 s^2\frac{\Qrms^2}{T^2}
\frac{\phizero^2}{\Mp^2}\ee^{s/c}.
\end{equation}
One can then easily deduce the mass scale $M$ for a given value of
$c$, $\phizero$ and $\xend$, the three parameters of the model.

\subsection{Valley Hybrid Inflation (VHI)}
\label{sec:vhi}

\subsubsection{Theoretical Justifications}
\label{subsubsec:theoryvhi}

Hybrid inflation is a two-fields model with the potential
given by the following expression~\cite{Linde:1991km, Linde:1993cn,
  Copeland:1994vg, Lyth:1998xn, Panagiotakopoulos:2000ky,
  Lazarides:2000ck, Covi:2000gx}
\begin{equation}
\label{eq:vhi:twofieldpot}
V\left(\phi ,\psi \right) = \frac12 m^2\phi ^2 + \frac{\lambda'}{4}
\left(\psi ^2-\Delta ^2\right)^2 + \frac{\lambda }{2}\phi^2 \psi ^2,
\end{equation}
where $\phi$ is the inflaton, $\psi$ the waterfall field, $\lambda$
and $\lambda'$ are two coupling constants and $\Delta$ a constant of
dimension one. A priori, given the above potential, inflation can
occur in different regimes. However, the standard lore is that
inflation can proceed along the valley given by $\psi=0$ and, in this
case, the potential reduces to an effective single field potential
that can be written as
\begin{equation}
\label{eq:vhi:pothep}
V(\phi)=M^4\left[1 +\left(\frac{\phi}{\mu} \right)^{p} \right],
\end{equation}
with $p=2$ and where one has used the following parameter redefinition
\begin{equation}
\label{paramhybrid}
M = \frac{\lambda'^{1/4}\Delta }{\sqrt{2}}\, ,\qquad \mu
    = \sqrt{\frac{\lambda'}{2}}\frac{\Delta^2}{m}\, .
\end{equation}
Inflation along the valley has been shown to be a dynamical attractor
of the two-field dynamics in \Refcs{Clesse:2008pf,
  Clesse:2009ur}. However, as recently shown in \Refc{Clesse:2010iz},
the hybrid potential can also support an inflationary phase along a
mixed valley-waterfall trajectory, which is genuinely a two-fields
dynamics. As we use a single field description here, those effects
cannot be described by the potential of \Eq{eq:vhi:pothep}. For this
reason, we will refer to the single field approximation as the
``valley hybrid regime''. Let us stress that, if the waterfall
inflationary regime occurs, then it will erase any observable effects
coming the valley hybrid regime. As a result, \Eq{eq:vhi:pothep} is a
good description of hybrid inflation only if the model parameters are
such that the waterfall regime remains sub-dominant. According to
\Refc{Clesse:2010iz, Kodama:2011vs}, this is the case provided
\begin{equation}
\sqrt{\lambda'} \dfrac{\Delta^3}{m} \ll \Mp^2,
\label{eq:vhi:nowater}
\end{equation}
a condition that will be assumed in the following. The effective
potential \eqref{eq:vhi:pothep} was also obtained in
\Refc{Bento:2002kp} in the context of supergravity brane inflation,
and in \Refc{Lin:2010zzk} in the context of hilltop supernatural
inflation. It depends on three parameters, namely $M$, $\mu $ and
$p$. In fact, as mentioned before, $p=2$ for the two-field model given
in \Eq{eq:vhi:twofieldpot} but we will consider the most general
situation with $p>0$ unspecified. Let us stress again that all multifield
effects such as the generation of isocurvature modes or cosmic strings
cannot be accounted within the single field
dynamics~\cite{Rocher:2004et, Ringeval:2005yn, BasteroGil:2006cm,
  Martin:2011ib}.

It is also worth mentioning that the potential~(\ref{eq:vhi:pothep})
with $p=2$ can also be obtained in the supergravity
context~\cite{Komargodski:2009rz, AlvarezGaume:2011xv,
  AlvarezGaume:2010rt, AlvarezGaume:2011db}. The main idea is to
consider a supergravity model which is not R-symmetry invariant and
described by the following K\"ahler and super-potentials:
\begin{eqnarray}
K &=& XX^{\dagger}+\frac{b}{6M^2}\left(XX^{\dagger}\right)^2
-\frac{c}{9M^2}XX^{\dagger}\left[X^2+\left(X^{\dagger}\right)^2\right], \\
W &=& fX,
\end{eqnarray}
Here $X$ is a superfield, $M<\Mp$ a mass scale and $b$, $c$ two
dimensionless constants, a priori of order one. The quantity $f$ is a
constant of dimension two that can be viewed as the supersymmetry
breaking scale. From these expressions, the scalar potential reads
\begin{equation}
V=f^2\left[1-\frac{2b}{3M^2}XX^{\dagger}
+\frac{c}{3M^2}\left(X^2+X^{\dagger}{}^2\right)
+\calO\left(\frac{1}{M^4}\right)\right],
\end{equation}
or, re-writing $X=\alpha+i\beta$, it reads
\begin{equation}
V\simeq f^2\left[1+\frac{2}{3M^2}(b-c)\alpha^2-\frac{2}{3M^2}
(b+c)\beta ^2\right].
\end{equation}
For a field evolution along the $\alpha$ direction, we recover a
potential of the VHI type with $p=2$ ($b-c$ must be positive). In this
setup, $\alpha/M\ll 1$ is required in order for the field $\alpha$ to
be approximately canonically normalized, the K\"ahler potential
being not minimal. It is also interesting to comment on the
$\eta$-problem in this model since this is a generic issue in
supergravity. If one calculates the slow-roll parameter $\eta\equiv
\Mp^2V_{\alpha \alpha}/V$, one finds that
\begin{equation}
\eta=\frac{4\Mp^2}{3M^2}(b-c).
\end{equation}
Therefore, one must take $M\lesssim \Mp$ and fine-tune the difference
$b-c$ to a small number.

\subsubsection{Slow-Roll analysis}
\label{subsubsec:srvhi}

\begin{figure}
\begin{center}
\includegraphics[width=\wdblefig]{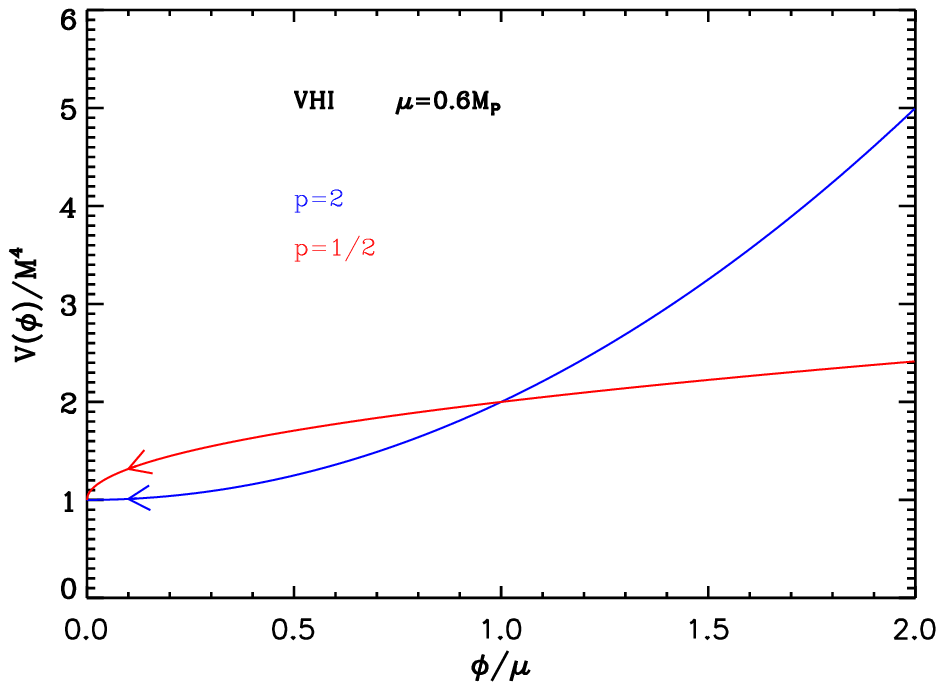}
\includegraphics[width=\wdblefig]{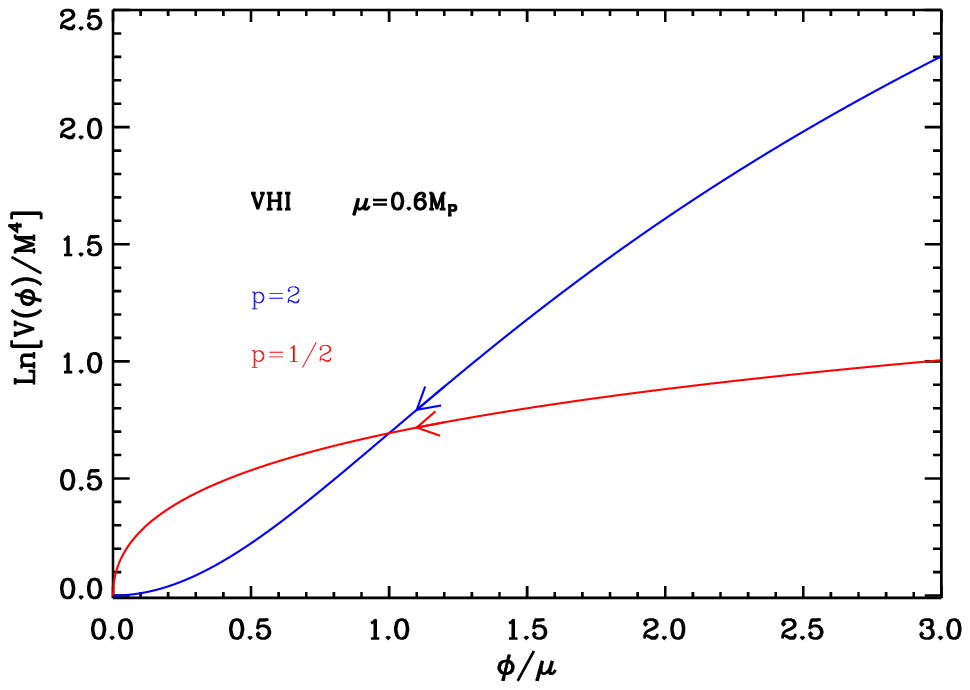}
\includegraphics[width=\wdblefig]{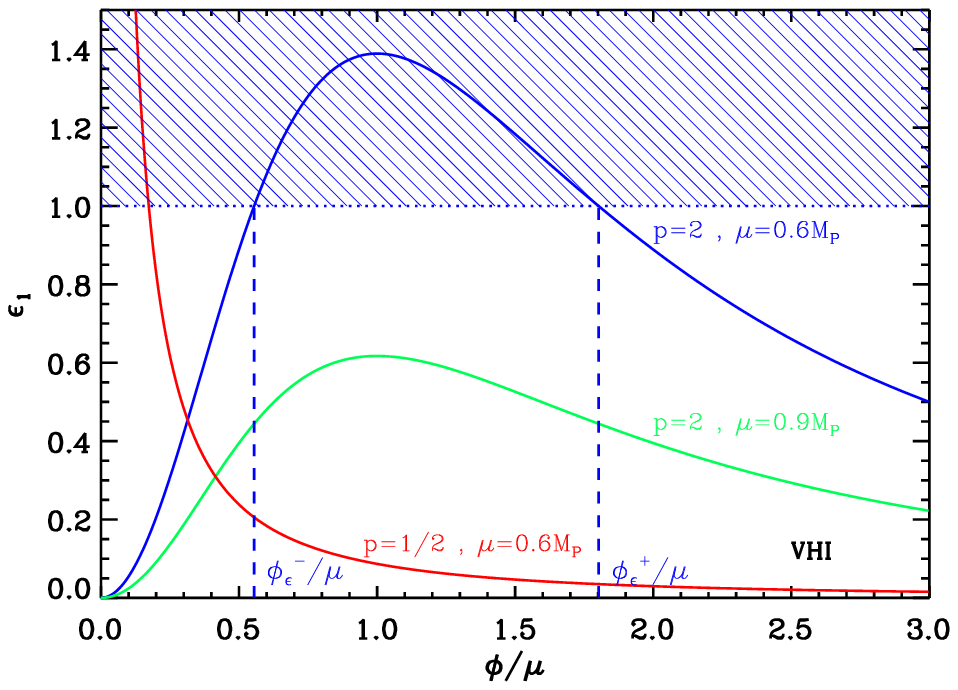}
\includegraphics[width=\wdblefig]{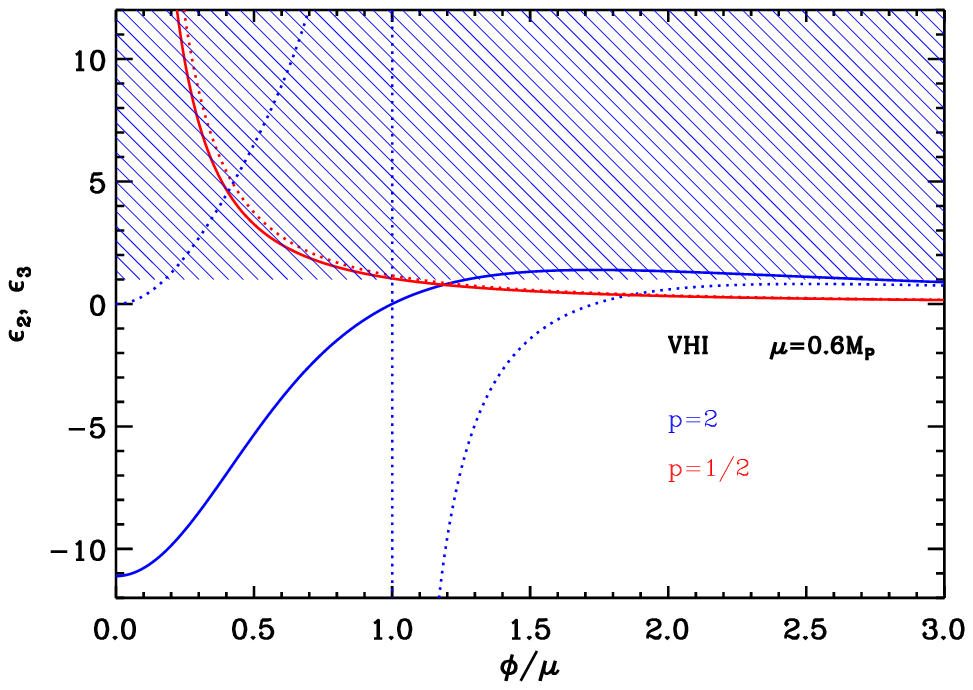}
\caption{Valley Hybrid inflation (VHI) for $p=1/2$ (red line) and
  $p=2$ (blue line). Upper panels: the potential and its logarithm for
  $\mu=0.6\Mp$. Bottom left panel: slow-roll parameter $\epsilon_1$
  for $p=1/2$, $\mu=0.6\Mp$ (red line), $p=2$, $\mu=0.6\Mp$ (blue
  line) and $p=2$, $\mu=0.9\Mp$ (green line). For small values of
  $\mu$ and $p>1$, the inflationary regions are separated into a large
  field one and the vacuum dominated one. The latter may not exist due
  to slow-roll violations if the field first rolls down the potential
  in the large field domain (see the text for a detailed
  discussion). The shaded area indicates the regions in which
  acceleration cannot occur. Bottom right panel: slow-roll parameters
  $\epsilon_2$ (solid line) and $\epsilon 3$ (dotted line) for
  $\mu=0.6\Mp$.}
\label{fig:vhi:pot}
\end{center}
\end{figure}

We now turn to the slow-roll analysis of the VHI scenario. Recall that
we consider the following potential
\begin{equation}
\label{eq:vhi:pot}
V(\phi)=M^4\left[1 +\left(\frac{\phi}{\mu} \right)^{p} \right],
\end{equation}
where the parameter $M$ and $\mu$ have been expressed in terms of the
parameters of the two-field model in \Eq{paramhybrid}. The first three
Hubble flow functions in the slow-roll approximation can be derived
from \Eq{eq:vhi:pot} in a straightforward fashion. Defining the
quantity $x$ by the following expression
\begin{equation}
x \equiv \dfrac{\phi}{\mu}\,,
\end{equation}
they read
\begin{equation}
  \epsilon_1 = \frac{p^2}{2} \left(\frac{\Mp}{\mu }\right)^2 \frac{
    x^{2p-2}}{\left( 1+ x^p \right)^2}\, ,\qquad \epsilon_2 = 2 p
  \left( \frac{\Mp}{\mu} \right)^2 x^{p-2} \frac{ x^{p} - p + 1}{
    \left( 1 + x^p \right)^2}\,,
\label{epsilonhyb}
\end{equation}
and
\begin{equation}
  \epsilon_3 = p\left( \dfrac{\Mp}{\mu } \right)^2 x^{p-2} \dfrac{
    2x^{2p} - (p-1)(p+4) x^p + (p-1)(p-2) }{ \left( 1+ x^p \right)^2
    \left( x^p -p+1 \right) }\, .
\label{epsilonhyb3}
\end{equation}
A specific feature of hybrid inflation in comparison to large and
small field models is that $\epsilon_2$ and $\epsilon_3$ can be
negative (see \Fig{fig:vhi:pot}). In particular
\begin{equation}
\epsilon _2 \underset{x \to 0}{\simeq} - 2p(p-1) \left(
\frac{\Mp}{\mu}\right)^2 x^{p-2},
\end{equation}
and $\epsilon_3$ blows up in the limit $x^p \to
p-1$. Together with the potential, the three Hubble flow functions
have been represented in \Fig{fig:vhi:pot}.

The slow-roll trajectory is obtained by integrating
\Eq{eq:srtrajectory} with the valley hybrid potential and reads
\begin{equation}
N-\Nend = \frac{1}{2p}\frac{\mu^2}{\Mp^2}\,\left[
- x^2 + \xend^2 + \frac{2}{2-p} \left( \xend^{2-p}
  - x^{2-p} \right) \right] ,
\label{eq:vhi:traj}
\end{equation}
which is, up to a sign, the same as for the SFI models [see
\Eq{eq:sfitraj}]. The case $p=2$ requires special attention, but as
for SFI, is recovered as the limit $p \to 2$ in the previous
equation. One obtains
\begin{equation}
\label{eq:vhi:trajp2}
N - \Nend = \frac14 \frac{\mu^2}{\Mp^2} \left[ -x^2 + \xend^2  -2 \ln \left(
  \frac{x}{\xend} \right) \right],
\end{equation}
which is again very similar to SFI, up to a sign. The trajectory
\eqref{eq:vhi:traj} cannot be inverted analytically in the general
case. It is however possible to perform this inversion for many
integer values of $p$, but those expressions will be omitted for the
sake of clarity. We simply give an approximate solution valid only in
the limit $x \ll 1$ and $p>2$
\begin{equation}
x \simeq \left[\xend ^{2-p} +p(p-2) \frac{\Mp^2}{\mu^2} (N - \Nend)
  \right]^{1/(2-p)}.
\end{equation}

If the waterfall inflation does not take place, \ie under the
condition \eqref{eq:vhi:nowater}, valley hybrid inflation ends by a
tachyonic instability in the small field regime $x < 1$, also referred
to as ``the vacuum dominated regime''. From the two-fields potential
\eqref{eq:vhi:twofieldpot}, one sees that the transverse direction
becomes tachyonic at the inflaton value
\begin{equation}
\phiend= \sqrt{\dfrac{\lambda'}{\lambda}} \Delta .
\label{eq:vhi:xcrit}
\end{equation}
In the single field approach, $\xend$ is therefore an extra-parameter
and VHI is a three parameters model according to our
classification. However, as can be seen in \Fig{fig:vhi:pot}, one
should pay attention to the various domains in which inflation can
take place. They are given by the behavior of $\epsilon_1(x)$.

If $p>1$, the slow-roll parameter $\epsilon_1$ vanishes when the field
goes to zero and at infinity while it reaches a maximum for
\begin{equation}
\xepsoneMax=\left(p-1\right)^{1/p}\,,
\end{equation}
equals to
\begin{equation}
\epsonemax = \frac{1}{2}\left(\frac{\Mp}{\mu} \right)^2
\left(p-1\right)^{\frac{2p-2}{p}} .
\end{equation}
Defining
\begin{equation}
\mueps \equiv \dfrac{\Mp}{\sqrt{2}} (p-1)^{1-1/p}\,,
\end{equation}
for all $\mu > \mueps$, one has $\epsilon_1(x) < 1$ and inflation can
proceed all over the domain $x>0$. On the contrary, if $\mu<\mueps$,
then inflation can, a priori, proceed in two disconnected
domains. Either $0 < x < \xepsoneOneMinus$ or $x > \xepsoneOnePlus$
where $\xepsoneOnePM$ are the two roots of $\epsilon_1 =1$, \ie the
solutions of
\begin{equation}
x^{2p} + 2 x^p - \frac{p^2}{2} \left( \frac{\Mp}{\mu} \right)^2
x^{2p-2} + 1 = 0 .
\label{eq:vhi:xeps1}
\end{equation}
This equation cannot be solved explicitly in the general case but, as
for the trajectory, there are explicit analytic expressions for many
integer values of $p$. For instance, for $p=2$, one gets
\begin{equation}
\xepsoneOnePMcase{p=2}=\frac{1}{\sqrt{2}}\frac{\Mp }{\mu } \left(1\pm
  \sqrt{1-2\frac{\mu^2}{\Mp^2}}\right) .
\end{equation}
The positive sign corresponds to the largest root while the minus one
to the smallest (see \Fig{fig:vhi:pot}). In the limit $\mu \ll \Mp$,
one has $\xepsoneOnePlus \simeq p \Mp /(\sqrt{2}\,\mu)$ which is also
the expression of $\xend$ for the large field model LFI (see
\sectionc{sec:lfi}). This does not come as a surprise since in that
situation \Eq{eq:vhi:pot} is indeed dominated by the monomial term. In
fact, the two above-mentioned domains precisely corresponds to a large
field one for $x > \xepsoneOnePlus$ and a vacuum dominated one for $x<
\xepsoneOneMinus$. It is a common mistake to assume that the large
field domain remains unobservable due to the existence of the vacuum
dominated one. In fact, as shown in \Refc{Clesse:2008pf}, the large
field regime becomes observable provided $\mu \ll \mueps$. In that
situation, after having crossed $\xepsoneOnePlus$, the field
fast-rolls in the region $\epsilon_1(x)>1$. Then, it enters the domain
$x<\xepsoneOneMinus$ with a strong initial velocity and, as a
consequence, crosses the whole vacuum dominated region, still in
fast-roll, to reach $\xend$. All observable predictions in such a
situation are therefore similar to that obtained in the LFI
models. Let us notice that, if there exists a mechanism that can
gently put the field without a strong initial velocity inside the
$x<\xepsoneOneMinus$ domain, then inflation can still occur in the
vacuum dominated region, even though $\mu < \mueps$. But if the field
is coming from the region $x>\xepsoneOnePlus$, then this regime does
not exist anymore.

For $p=1$, $\epsilon_1(x)$ is a decreasing function of the field and
takes a finite value $\Mp^2/(2 \mu^2)$ for $x \to 0$. The
behavior is similar to the case $p>1$ and if $\mu>\Mp/\sqrt{2}$
inflation can take place all over $x>\xend$. However, if $\mu <
\Mp/\sqrt{2}$ then the vacuum dominated region does not exist anymore
and $\xepsoneOne = \xepsoneOnePlus = \Mp/(\sqrt{2} \mu) -1$ One should
also notice that if $p=1$ the relation $\epsilon_2=4\epsilon_1$
applies.

Finally, for $p<1$, $\epsilon_1(x)$ is a decreasing function of the
field but it blows up when $x \to 0$. In that situation,
inflation stops at $x=\max(\xepsoneOneMinus,\xend)$ but the field will
still fast-roll till the tachyonic instability develops at $\xend$. As
a result, even if for some cases $\xepsoneOneMinus > \xend$, the
observable predictions remain mostly the same.

According to the previous discussion, for $p>1$, the VHI effective
potential is therefore adequate to describe the vacuum dominated
regime only, \ie for $\xend <x <\xepsoneOneMinus$ where $\xend$ is the
instability point given by \Eq{eq:vhi:xcrit}. In that situation,
solving \Eq{eq:phistarlnrrad} together with the trajectory
\eqref{eq:vhi:traj} gives the observable field value $\xstar$ at which
the pivot mode crossed the Hubble radius during inflation. The
potential parameter $M$ is fixed from the amplitude of the CMB
anisotropies
\begin{equation}
  \dfrac{M^4}{\Mp^4} = 720 \pi^2 p^2 \dfrac{\Mp^2}{\mu^2}
  \dfrac{ \xstar^{2p-2} }{ \left(1+\xstar^p\right)^3 } \dfrac{\Qrms^2}{T^2}\,.
\end{equation}
The reheating consistent slow-roll predictions are displayed in
\Figs{fig:CMBVHIpEQonehalf}, \ref{fig:CMBVHIpEQ1},
\ref{fig:CMBVHIpEQ1andahalf}, \ref{fig:CMBVHIpEQ2} and
\ref{fig:CMBVHIpEQ3} for $p=0.5$, $p=1$, $p=1.5$, $p=2$ and $p=3$,
respectively. For $p>1$ and $\xepsoneMax>1$, $\xend$ is varied between
$0$ and an upper bound such that $x_\uin<\xepsoneOneMinus$. One the
other hand, if $\xepsoneMax<1$, then one simply takes $\xend<10$. For
$p\leq 1$, $\xend$ is varied on a wider range, with no particular
constraints.  For $p=1$, the predictions lie on the line
$\epsilon_2=4\epsilon_1$ as expected whereas for $p>1$ one recovers a
blue spectral index when $\xepsoneMax>1$, while a red spectral index
can be obtained when $\xepsoneMax<1$ and $\xstar>\xepsoneMax$, with
$\xstar<1$ (that is to say, the large field regime).

\subsection{Dynamical Supersymmetric Inflation (DSI)}
\label{sec:dsi}

\subsubsection{Theoretical Justifications}
\label{subsubsec:theorydsi}

This model has been studied in \Refcs{Kinney:1997hm,
  Kinney:1998dv}. As for the IMI scenario, see \sectionc{sec:imi}, the
model is based on \Refc{Affleck:1984xz} which has shown that inverse
power law potentials naturally arise in supersymmetric theories. The
fact that we have an inverse power law behavior, rather than the usual
positive power law behavior, can be traced back to the presence of
non-perturbative effects, such as for instance gaugino condensation,
see \sectionc{sec:imi}. Based on the previous considerations, one can
write that
\begin{equation}
V=V_0+\frac{\Lambda_3^{p+4}}{\phi^p}+\frac{\phi^{q+4}}{\Mp^q},
\end{equation}
where the last term encodes a correction to $V(\phi)$ due to a
non-renormalizable operator. It is Planck suppressed since $\Mp$ is
the only explicit scale present in the theory. This term implies that
there is a minimum located at
\begin{equation}
\label{eq:dsi:phic}
\phiVmin = \left(\frac{p}{q+4}\Lambda_3^{p+4}\Mp^q\right)^{\frac{1}{p+q+4}}.
\end{equation}
This means that the extra term can be neglected in the region $\phi\ll
\phiVmin$ and, in the following, we assume that this is the case. The
difference with the IMI scenario is the presence of the constant term
$V_0$ which will be assumed to be dominant.

\subsubsection{Slow-Roll Analysis}
\label{subsubsec:srdsi}

In this sub-section, we now turn to the slow-roll analysis of the DSI
scenario. For this purpose, we rewrite the potential as
\begin{equation}
\label{dsi:eq:pot}
V(\phi) = M^4\left[ 1+\left(\frac{\phi}{\mu} \right)^{-p} \right],
\end{equation}
where $p$ is a free index parameter and where we defined
\begin{equation}
V_0=M^4, \qquad \mu^p=\frac{\Lambda_3^{p+4}}{M^4}.
\end{equation}
As already mentioned, in order for inflation to take place in
the vacuum dominated regime, we must assume that $\phi\gg\mu$. In
\Refcs{Kinney:1997hm, Kinney:1998dv}, it was argued that natural
values for $\Lambda_3$ and $M$ are $10^6\GeV$ and $10^{10}\GeV$,
respectively. This means that a scale of order $\mu\simeq
10^{6+14/p}\,\GeV$ is a reasonable prior for $\mu$.

\begin{figure}
\begin{center}
\includegraphics[width=\wdblefig]{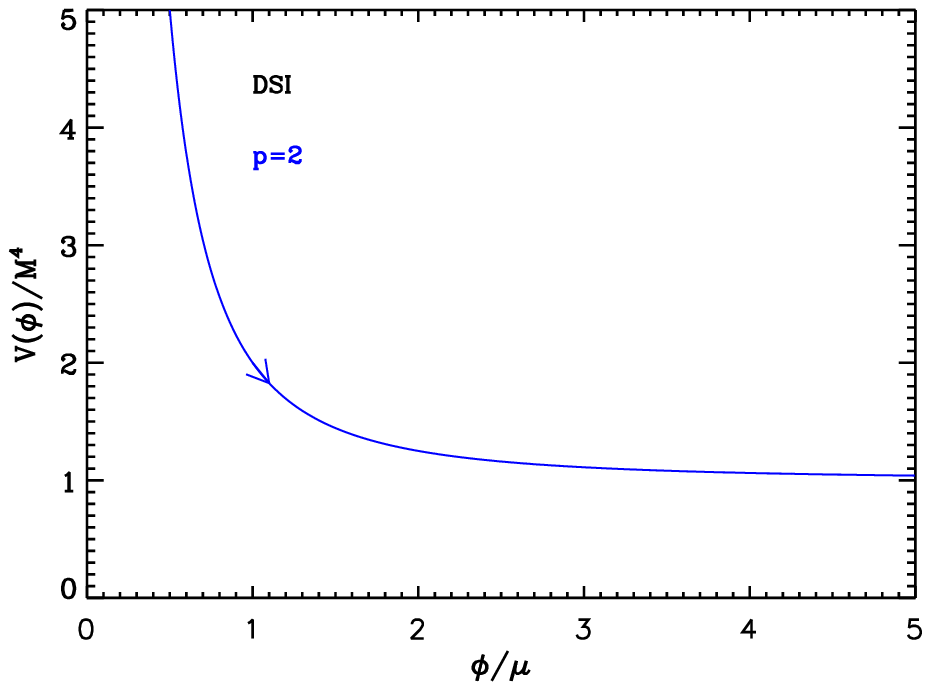}
\includegraphics[width=\wdblefig]{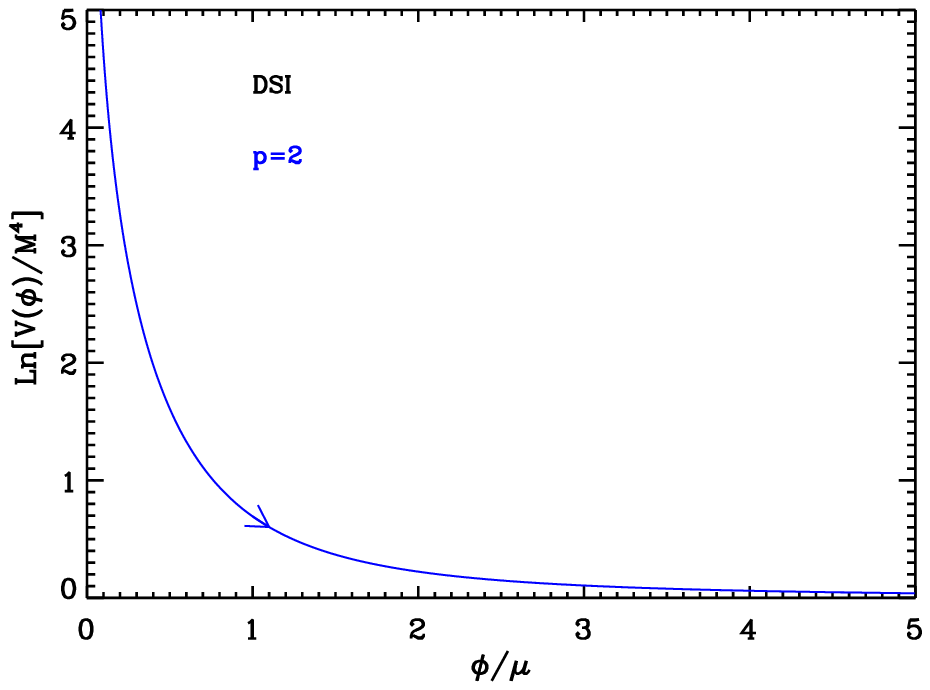}
\includegraphics[width=\wdblefig]{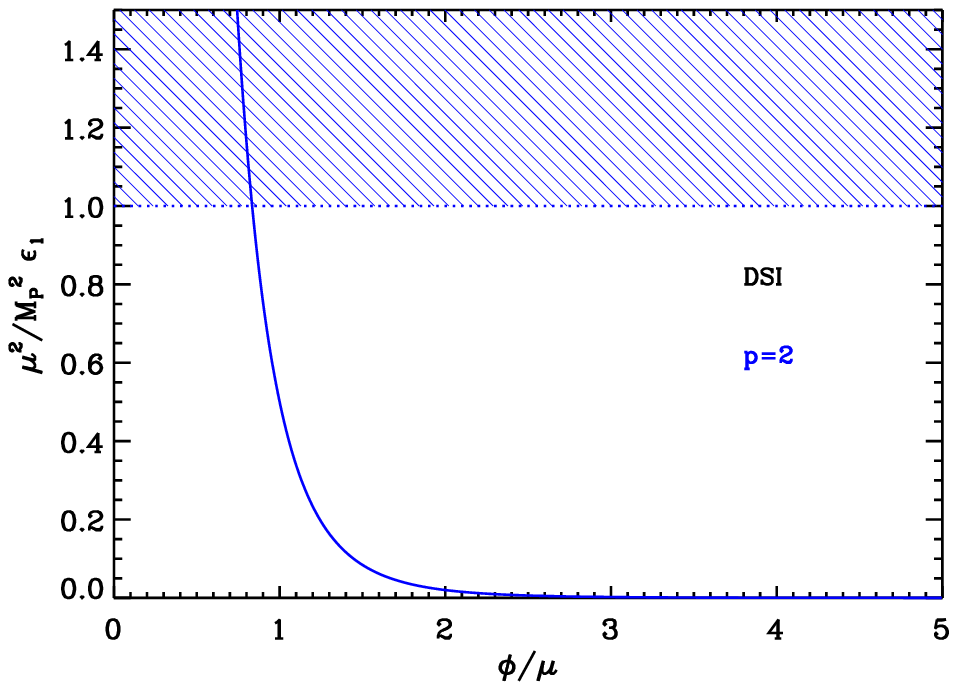}
\includegraphics[width=\wdblefig]{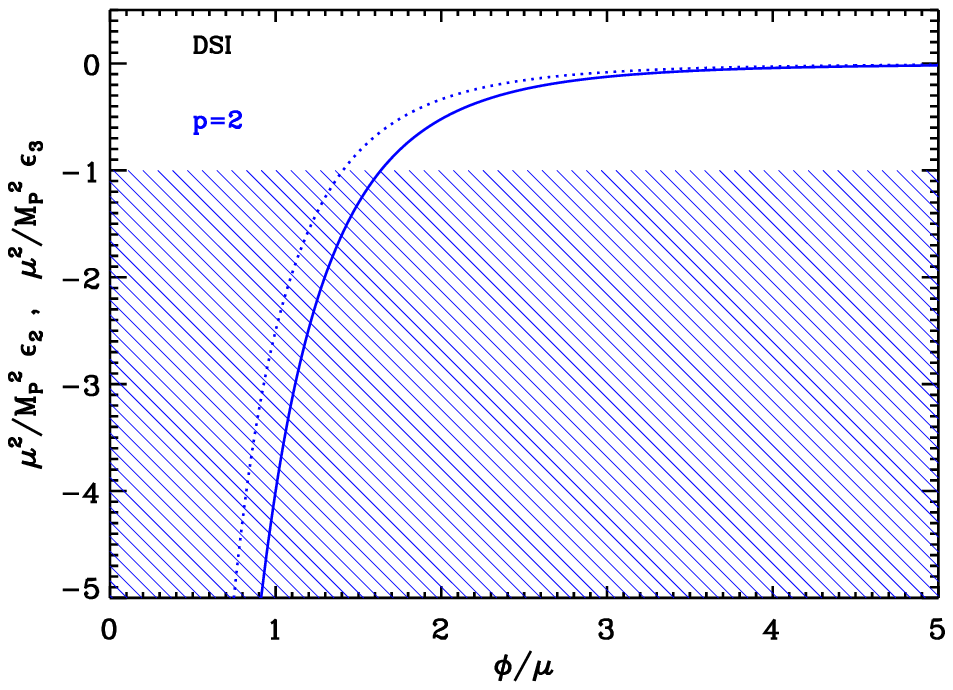}
\caption{Dynamical Supersymmetric Inflation (DSI) for $p=2$. Upper
  panels: the potential and its logarithm as a function of
  $\phi/\mu$. Bottom left panel: slow-roll parameter $\epsilon_1$
  rescaled by $\Mp^2/\mu^2$. The shaded area indicates the region in
  which inflation cannot occur for $\mu = \Mp$. Bottom right panel:
  slow-roll parameters $\epsilon_2$ (solid line) and $\epsilon_3$
  (dotted line), rescaled by $\Mp^2/\mu^2$.}
\label{potdsi}
\end{center}
\end{figure}

The potential~(\ref{dsi:eq:pot}), as well as its logarithm, is
displayed in \Fig{potdsi}. It is a decreasing function of the field,
hence inflation proceeds from the left to the right. Defining the
quantity
\begin{equation}
x \equiv \dfrac{\phi}{\mu}\,,
\end{equation}
the first three Hubble flow functions in the slow-roll approximation
read
\begin{equation}
\epsilon_1 = \dfrac{p^2}{2}\left(\dfrac{\Mp}{\mu }\right)^2 \dfrac{
  x^{-2p-2}}{\left(1+x^{-p} \right)^2}\,, \qquad
\epsilon_2 = -2p\left(\dfrac{\Mp}{\mu }\right)^2 x^{-p-2}
 \dfrac{x^{-p}+p+1}{\left(1 +x^{-p}\right)^2}\, ,
\end{equation}
and
\begin{equation}
\epsilon_3 = -p\left(\dfrac{\Mp}{\mu }\right)^2 x^{-p-2}
  \dfrac{\left[2x^{-2p} + \left(p+1\right)\left(p-4\right)x^{-p}
    +\left(p+1\right)\left(p+2\right)\right]}
    {\left(1+x^{-p}\right)^2 \left( x^{-p}+p+1\right)}\,.
\end{equation}
Let us already notice that, from these expressions, one has
\begin{equation}
-2\epsilon_1-\epsilon_2=\left(\frac{\Mp}{\mu}\right)^2
\frac{px^{-p-2}}{\left(1+x^{-p}\right)^2}
\left[px^{-p}+2p\left(p+1\right)x^{-p-2}\right]>0 ,
\end{equation}
which implies a blue spectral index for the scalar power spectrum
since, at first order, $\nS-1 = -2\epsilon_{1*} -\epsilon_{2*}$. The
three slow-roll parameters become very small at large fields $x\gg
1$. There is a value $\xepsoneOne$ such that $\epsilon_1=1$. For $x$
such that $x<\xepsoneOne$, $\epsilon_1>1$ and inflation cannot take
place. This value has to be determined numerically, but since the
natural values for $\mu$ are such that $\mu/\Mp\ll 1$, an approximate
expression can be derived
\begin{equation}
\xepsoneOne \simeq
\left(\frac{p}{\sqrt{2}}\frac{\Mp}{\mu}\right)^{1/(p+1)}.
\end{equation}
Because the potential is decreasing with $x$, inflation can only take
place in the domain $x> \xepsoneOne \gg 1$ if $\mu \ll \Mp$. It cannot
stop by slow-roll violation and another mechanism such as, \eg a
tachyonic instability, has to be introduced. We will denote by $\xend$
the field value at which this occurs. It represents an extra parameter
of the model. Obviously, it must be such that $\xepsoneOne< \xend \ll
\xVmin$.

\begin{figure}
\begin{center}
\includegraphics[width=\wsingfig]{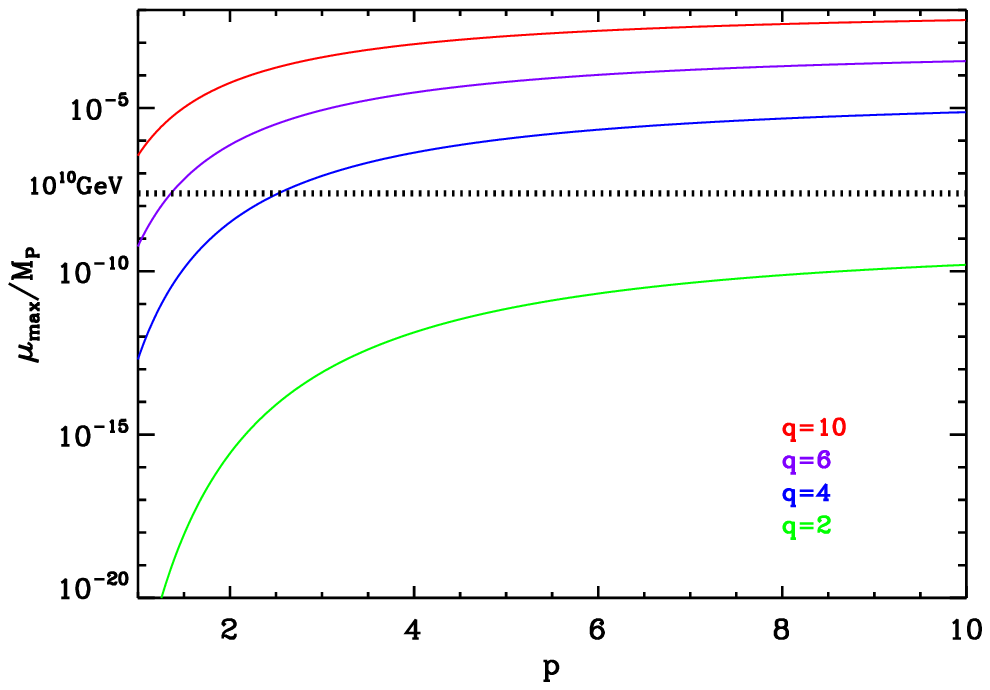}
\caption{Dynamical Supersymmetric Inflation. Maximal value of
  $\mu/\Mp$ with respect to $p$, and for different values of $q$, such
  that the condition $\xendmin < \xendmax$ is satisfied. We have fixed
  $\Delta\Nmax=50$. The black dotted line show a typical value for
  $\mu/\Mp\simeq 10^{10}\,\GeV$~\cite{Kinney:1997hm}.}
\label{fig:dsi:mumax}
\end{center}
\end{figure}

Let us now turn to the slow-roll trajectory. It can be integrated
explicitly from \Eq{eq:srtrajectory} and one obtains
\begin{equation}
\begin{aligned}
\Nend-N & = \frac{\mu^2}{2p\Mp^2}\left(\xend^2
  +\frac{2}{p+2} \xend^{p+2} - x^2 -\frac{2}{p+2} x^{p+2} \right).
\end{aligned}
\end{equation}
In the $\mu/\Mp\ll 1$ limit, one has $x > \xepsoneOne \gg 1$, and the previous 
trajectory can be approximated by
\begin{equation}
\begin{aligned}
\Nend-N \simeq \frac{\mu^2}{p(p+2)\Mp^2} \left(\xend^{p+2}-
x^{p+2}\right).
\end{aligned}
\end{equation}
This expression can be analytically inverted to get the observable
field value $\xstar$ in terms of $\Delta \Nstar = \Nend - \Nstar$ as
\begin{equation}
\label{dsi:eq:invertedtraj}
\xstar\simeq\left[\xend^{p+2}-\frac{\Mp^2}
{\mu^2}p\left(p+2\right)\Delta\Nstar\right]^{\frac{1}{p+2}}.
\end{equation}
One can notice that the total amount of \efolds is bounded because
$\xend \ll \xVmin$ and cannot take infinitely large values. In order
to get a number of \efolds, $\Delta N> \Delta\Nmin$, $\xend$ should be
sufficiently large with $\xend > \xendmin$. More precisely, setting
$\xini=\xepsoneOne$, one has
\begin{equation}
\begin{aligned}
\label{eq:dsi:xendmin}
\xendmin &\simeq\left[p\left(p+2\right)\frac{\Mp^2}{\mu^2}\Delta \Nmin
  + \left(\frac{p}{\sqrt{2}}\frac{\Mp}{\mu}\right)^{\frac{p+2}
    {p+1}}\right]^{\frac{1}{p+2}} \simeq
\left[p\left(p+2\right)\frac{\Mp^2}{\mu^2} \Delta \Nmin
  \right]^{\frac{1}{p+2}}.
\end{aligned}
\end{equation}
In practice one wants $\Delta \Nmin > 50$ to solve the problems of the
standard Big-Bang scenario. Whether this value is compatible, or not,
with the condition $\xend \ll \xVmin$ depends on the value of $M^4$
appearing in \Eq{eq:dsi:phic}, which is itself determined by the
amplitude of the CMB anisotropies. This one reads
\begin{equation}
  \left(\frac{M}{\Mp}\right)^4 = 720\pi^2p^2
  \left(\frac{\Mp}{\mu}\right)^2 \xstar^{-2p-2} \left(1+
    \xstar^{-p}\right)^{-3} \frac{\Qrms^2}{T^2}\, .
\end{equation} 
In the limit $\mu/\Mp\ll 1$, one has $\xstar\gg 1$ and this expression
can be approximated by
\begin{equation}
\dfrac{M^4}{\Mp^4} \simeq 720 \pi^2 p^2 \dfrac{\Mp^2}{\mu^2}
\xstar^{-2p-2} \dfrac{\Qrms^2}{T^2}\, .
\end{equation}
Therefore, from \Eq{eq:dsi:phic}, one has
\begin{equation}
\xVmin \simeq \left[720\pi^2\frac{p^3}{q+4}\left(\frac{\Mp}
{\mu}\right)^{6+q}\xstar^{-2p-2}\frac{\Qrms^2}{T^2}\right]^{\frac{1}
{p+q+4}} ,
\end{equation}
with $\xstar$ depending on $\xend$ through \Eq{dsi:eq:invertedtraj}.
One can see that the previous expression decreases with $\xstar$ and
the condition $\xend \ll \xVmin$ imposes an upper bound on $\xend <
\xendmax$ with
\begin{equation}
\label{eq:dsi:xendmax}
\xendmax \simeq \left[720\pi^2\frac{p^3}{q+4}\frac{\Qrms^2}
{T^2}\left(\frac{\Mp}{\mu}\right)^{q+6} \right]^{1/(3p+q+6)}.
\end{equation}
The prior condition on $\xend$ is therefore of the type $\xendmin <
\xend \ll \xendmax$, with $\xendmin$ defined by \Eq{eq:dsi:xendmin}
and $\xendmax$ defined by \Eq{eq:dsi:xendmax}.  For any $q>0$, these
two equations show that there exists an upper bound $\mu < \mumax$
under which the condition $\xendmin \ll \xendmax$ is satisfied. It
reads
\begin{equation}
\dfrac{\mumax}{\Mp} \simeq \frac{\left(720\pi^2\frac{p^3}
{q+4}\frac{\Qrms^2}{T^2}\right)^{(p+2)/(pq)}}
{\left[p(p+2)\Delta\Nmin\right]^{(3p+q+6)/(pq)}}\,,
\end{equation}
and has been represented in \Fig{fig:dsi:mumax}. One can see that a
typical value $\mu/\Mp \simeq 10^{10}\,\GeV$ (see
\Refc{Kinney:1997hm}) is not allowed for realistic values of $p$ and
$q$. As such, the prior space for $p$, $\mu$, and $\xend$ is
constrained and should be handled carefully.

The reheating consistent slow-roll predictions of the dynamical
supersymmetric models are displayed in \Figs{fig:CMBDSIpEQ2},
\ref{fig:CMBDSIpEQ3} and \ref{fig:CMBDSIpEQ4} for $p=2$, $p=3$ and
$p=4$, respectively, and with $10^{-10} \Mp < \mu<\mumax$ (where
$\mumax$ has been calculated taking $q=8$ and $\Delta\Nmin=60$ to
cover a large prior space). The reheating equation of state parameter
$\wrehbar$ has been taken to $0$ but since there is no potential
minimum around which the inflaton field can oscillate at the end of
inflation, this parameter is a priori unspecified and can take
different values. In any case the reheating temperature is strongly
degenerated with the parameter $\xendmin <\xend < \xendmax$ preventing
their inference. One can check that the spectral index is blue, as
announced earlier, making these models disfavored by the
observations. The typical amount of gravitational waves is very small,
in agreement with the results of \Refc{Kinney:1997hm}.

\subsection{Generalized Mixed Inflation (GMLFI)}
\label{sec:gmlfi}

\begin{figure}
\begin{center}
\includegraphics[width=\wdblefig]{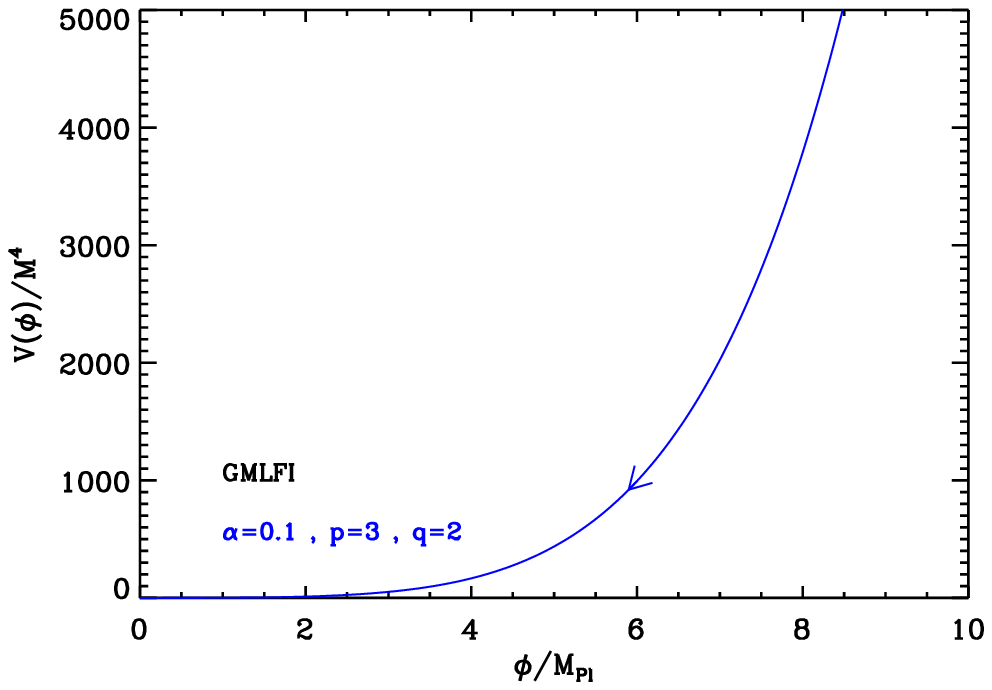}
\includegraphics[width=\wdblefig]{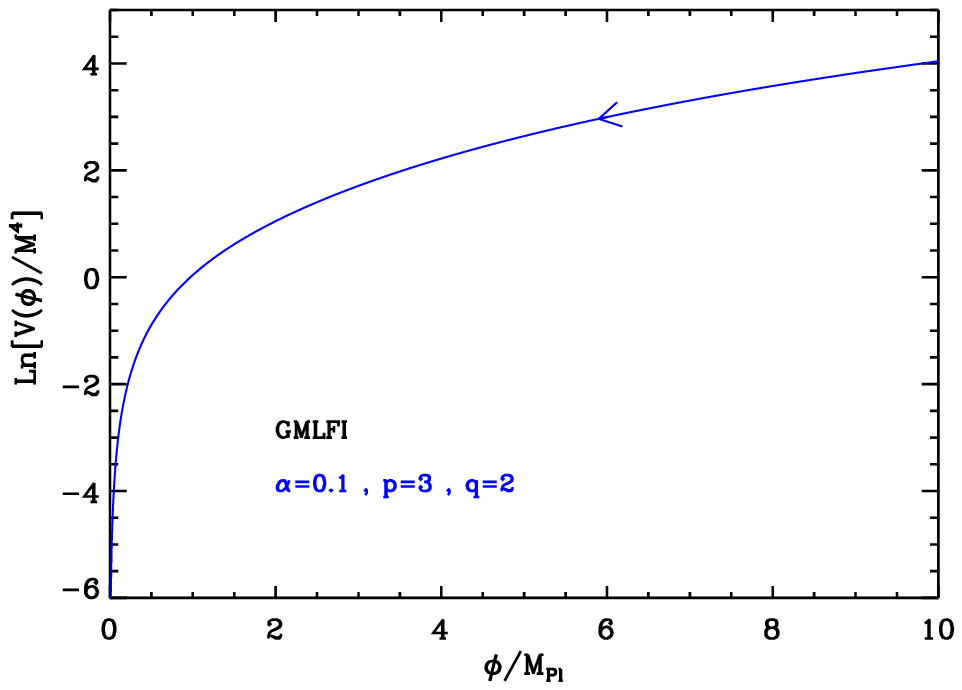}
\includegraphics[width=\wdblefig]{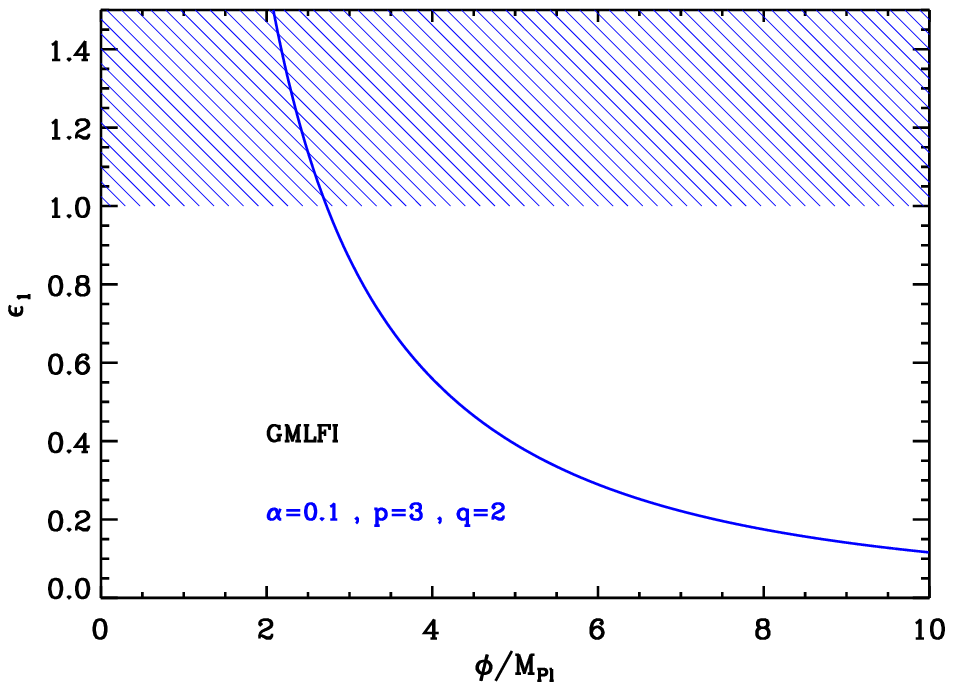}
\includegraphics[width=\wdblefig]{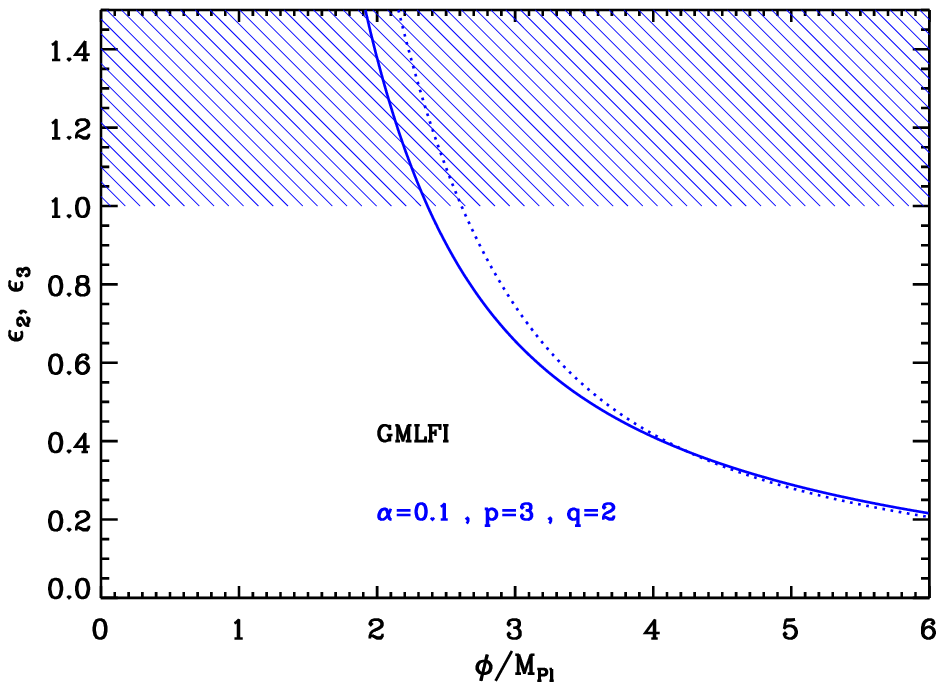}
\caption{Generalized Mixed Inflation (GMLFI) for $p=3, q=2$ and
  $\alpha=0.1$. Upper panels: the potential and its logarithm with
  respect the field value. Bottom left panel: slow-roll parameter
  $\epsilon_1$, the shaded region is where inflation stops. Bottom
  right panel: slow-roll parameters $\epsilon_2$ (solid line) and
  $\epsilon_3$ (dotted line).}
\label{potGMLFI}
\end{center}
\end{figure}

This model is a generalization of MLFI (see \sectionc{sec:mlfi}) and
is, by definition, the sum of two monomial functions with arbitrary
power indices. The corresponding potential can be written as
\begin{equation}
\label{eq:gmlfi:pot}
V = M^4\left(\frac{\phi}{\Mp}\right)^p \left[1 + \alpha\left(
  \frac{\phi}{\Mp} \right)^q \right] ,
\end{equation}
where $\alpha$, $p$ and $q$ are three dimensionless positive
parameters. It can be seen as a generalization of the large field
inflation potential (LFI, see \sectionc{sec:lfi}), which is recovered
when $\alpha\to 0$ or $\alpha\to\infty$. The parameter
$\alpha$ therefore controls the relative weight of the two
terms. Since the potential is an increasing function of the inflaton
\vev, inflation proceeds from the right to the left and occurs in the
large field regime $\phi/\Mp\gg 1$. Defining the quantity $x$ by
\begin{equation}
x \equiv \dfrac{\phi}{\Mp}\,,
\end{equation}
the first three Hubble flow functions in the slow-roll approximation
can be expressed as
\begin{equation}
  \epsilon_1 =\frac{1}{2x^2}
  \left[\dfrac{ p+\alpha \left( p+q \right) x^q }
  { 1+\alpha x^q}\right]^2\, ,
\label{eq:sr1gmlfi}
\end{equation}
\begin{equation}
\epsilon_2 = \frac{2}{x^2} \dfrac{p+\alpha^2\left(p+q\right)x^{2q}
  +\alpha \left(2p+q-q^2\right)x^{q}}{\left(1+\alpha x^q \right)^2}
,
\end{equation}
and
\begin{equation}
\begin{aligned}
& \epsilon_3 = \dfrac{1}{x^2\left(1+\alpha x^q\right)^2} 
    \left[pq^2+\alpha^2q^2\left(p+q\right) x^{2q} +\alpha q^2
    \left( 2p+q-q^2 \right) x^q\right]^{-1} 
    \\ & \times \Bigg\lbrace 2q^2 \left[p^2 + \alpha^4
    (p+q)^2\right] x^{4q} +\alpha^2q^2 \left[
    12p^2+6pq \left(2-q\right) + \left(q-2\right) \left(q-1\right) q^2
    \right] x^{2q}  \\ & +
   \alpha^3q^3 \left(p+q \right) \left[ 8\dfrac{p}{q} + \left(
    1-q \right) \left( 4+q \right) \right]
  x^{3q} + \alpha pq^2 \left[ 8p+q \left(
    4+q^2-3q \right) \right] x^q \Bigg\rbrace.
\end{aligned}
\end{equation}
They are decreasing functions of the field, vanishing when
$x\to\infty$ and diverging when $x\to 0$. Together
with the potential and its logarithm, the Hubble flow functions are
represented in \Fig{potGMLFI}.

In \Fig{potGMLFI}, one sees that inflation ends by slow-roll violation
at $x=\xend$, the solution of the equation $\epsilon_1(\xend)=1$. From
\Eq{eq:sr1gmlfi}, one obtains
\begin{equation}
\sqrt{2}\alpha\xend^{q+1}
+\sqrt{2}\xend=\pm \left[p+ \alpha
  \left(p+q\right)\xend^q\right].
\end{equation}
One can check that, for $\alpha=0$, one recovers the LFI-$p$ result
$\xend=p/\sqrt{2}$ (see \sectionc{sec:lfi}) and that, for
$\alpha\to\infty$, one gets $\xend =
\left(p+q\right)/\sqrt{2}$, which correspond again to the LFI-$p+q$
solution.  The above equation cannot be solved analytically for
arbitrary values of $p$, $q$. This is possible only in some particular
cases, namely $q=0$, $q=1$ or $q=2$. For $q=0$, this is LFI whereas
$q=2$ corresponds to MLFI, both solutions being given
in \sectionc{sec:lfi} and \sectionc{sec:mlfi}, respectively. For
$q=1$, one obtains
\begin{equation}
\xend = \frac{\sqrt{2}}{4} \left(p+1\right) -
\frac{1}{2\alpha} + \frac{ \sqrt{4+4\sqrt{2} \alpha \left(p-1\right) +
    2\alpha^2 \left(p+1 \right)^2}}{4\alpha}\,,
\end{equation}
but, in general, $\xend$ has to be determined numerically.

The slow-roll trajectory can be integrated explicitly using
\Eq{eq:srtrajectory} and this leads to
\begin{equation}
\begin{aligned}
\label{eq:gmlfi:traj}
  \Nend-N & =
  \frac{1}{2\left(p+q\right)}x^2 \left
  \lbrace 1+\frac{q}{p}
  \Fgauss\left[1,\frac{2}{q},1+\frac{2}{q},-\alpha q\left(\frac{1}{p}
    + \frac{1}{q}\right)x^q \right]
  \right \rbrace \\& - \frac{1}{2 \left(p+q \right)} \xend^2
   \left \lbrace 1+\frac{q}{p} \Fgauss
  \left[1,\frac{2}{q},1+\frac{2}{q},-\alpha q\left(\frac{1}
  {p}+\frac{1}{q}\right) \xend^q \right]\right\rbrace.
\end{aligned}
\end{equation}
Here, $\Fgauss$ stands for the Gauss hypergeometric
function~\cite{Abramovitz:1970aa,Gradshteyn:1965aa}. Since it is equal
to unity when its last argument vanishes, one can check that, in the
limit $\alpha \to 0$, one recovers the slow-roll trajectory
for the LFI-$p$ models while the limit $\alpha\to\infty$ leads
to the trajectory of the LFI-$(p+q)$ models. Finally, since $\Fgauss
\left(1,1,2,x\right) = -\ln\left(1-x\right)/x$, one can also check
that the MLFI case corresponds to $p=q=2$. The previous expression can
only be inverted for $q=0$ (LFI) and $q=2$ (MLFI), and they have been
already discussed in \sectionc{sec:lfi} and \sectionc{sec:mlfi},
respectively. The case $q=1$ can also be simplified using $\Fgauss
\left(1,2,3,x\right) = -2/x-2\ln(1-x)/x^2$.  In general, one has to
inverse this slow-roll trajectory numerically.

The parameter $M$ can be determined from the amplitude of
the CMB anisotropies and the Hubble crossing \vev $\xstar$. One
obtains
\begin{equation}
\frac{M^4}{\Mp^4} = 720\pi^2 \dfrac{\left[p+\alpha\left(p+q\right)
    \xstar^q\right]^2}{\xstar^{p+2}  \left( 1+\alpha \xstar^q \right)^{3}}
\frac{\Qrms^2}{T^2}\, .
\end{equation}

The reheating consistent slow-roll predictions for the generalized
mixed large field models are displayed in
Figs~\ref{fig:CMBGMLFIpEQ2qEQ1}, \ref{fig:CMBGMLFIpEQ2qEQ3}, and
\ref{fig:CMBGMLFIpEQ3qEQ2} for ($p=2$ and $q=1$), ($p=2$ and $q=3$)
and ($p=3$ and $q=2$), respectively. As for MLFI, the predictions lie
between the LFI-$p$ and LFI-$(p+q)$ models, but can actually exit this
region for large enough values of $\alpha$. This means that, if one
starts from a pure $V\propto\phi^{p+q}$ potential and adds a small
$\propto\phi^p$ term, then this extra term has the effect of
increasing the ``effective value'' of the power index of the
potential. Moreover, since for large field inflation models, the
$p$-model fits the data better than the $(p+q)$-one, it follows that
small values for the parameter $\alpha$ are favored, together with
high reheating temperatures.

\subsection{Logarithmic Potential Inflation (LPI)}
\label{sec:lpi}

\subsubsection{Theoretical Justifications}
\label{subsubsec:theorylpi}

This class of model assumes that inflation is driven by a composite
state in a strongly interacting theory, see
\Refcs{Bezrukov:2011mv,Channuie:2012bv,Channuie:2013lla}. Let us consider 
the following model, see \sectionc{sec:oi} for more details
\begin{equation}
\label{eq:lagrangegi}
\calL_{\mathrm{GI}}=-\varphi^{-3/2}\partial_{\mu}\varphi\partial^{\mu}\varphi
-\frac{\varphi}{2}\ln\left(\frac{\varphi}{\Lambda^4}\right),
\end{equation}
where $\Lambda$ is a mass scale. Moreover, let us consider 
the situation where the model has a general non-minimal coupling to 
gravity of the form 
\begin{equation}
S =\int \dd^4 \bmx\sqrt{-g}\left[-\frac12\left(M^2+\xi \varphi^{1/2}\right)R
+\calL_{\mathrm{GI}}\right].
\end{equation}
The coupling to gravity is characterized by the parameter $\xi$. Then,
the action in the Einstein frame
reads~\cite{Bezrukov:2011mv,Channuie:2012bv,Channuie:2013lla}
\begin{eqnarray}
S &=&\int \dd^4 \bmx\sqrt{-g}
\biggl[-\frac{1}{2}\Mp^2R-
\Omega^{-2}\left(1+\frac{3\xi^2\varphi^{1/2}}{4\Mp^2}\Omega^{-2}
\right)\varphi^{-3/2}
\partial_{\mu}\varphi\partial^{\mu}\varphi
-\Omega^{-4}V_{\mathrm{GI}}\biggr],
\end{eqnarray}
where $V_{\mathrm{GI}}$ refers to the potential in \Eq{eq:lagrangegi}
and $\Omega^2=\left(M^2+\xi\varphi^{1/2}\right)/\Mp^2$. If $\xi\neq 0$
and if we are in the large field limit, then $\Omega^2\simeq
\xi\varphi^{1/2}/\Mp^2$ and the canonically normalized field $\phi$ is
such that $\phi \propto \ln\varphi$. In that case the potential
reduces to $\Omega^{-4}V_{\mathrm{GI}}\propto \ln \varphi \propto
\phi$. Therefore, we have obtained a LFI model with $p=1$,
see \sectionc{sec:lfi}. On the other hand, if one assumes that $\xi=0$, then 
$\varphi=\phi^4/(4\sqrt{2})^4$ and 
\begin{equation}
V = 2\Lambda^4 \left( \frac{\phi}{\phizero}\right)^4 \ln
\left(\frac{\phi}{\phizero}\right),
\end{equation}
with $\phizero\equiv 4\sqrt{2}\Lambda$. This resembles the potential
found in \sectionc{sec:oi} which, for $\beta=0$ (see the precise
definition in that section), was such that $V\propto
\phi^4\ln^2\left(\phi/\phizero\right)$. These considerations motivate
the next section devoted to the slow-roll analysis of this class of
scenarios.

\subsubsection{Slow-Roll Analysis}
\label{subsubsec:srlpi}

\begin{figure}
\begin{center}
\includegraphics[width=\wdblefig]{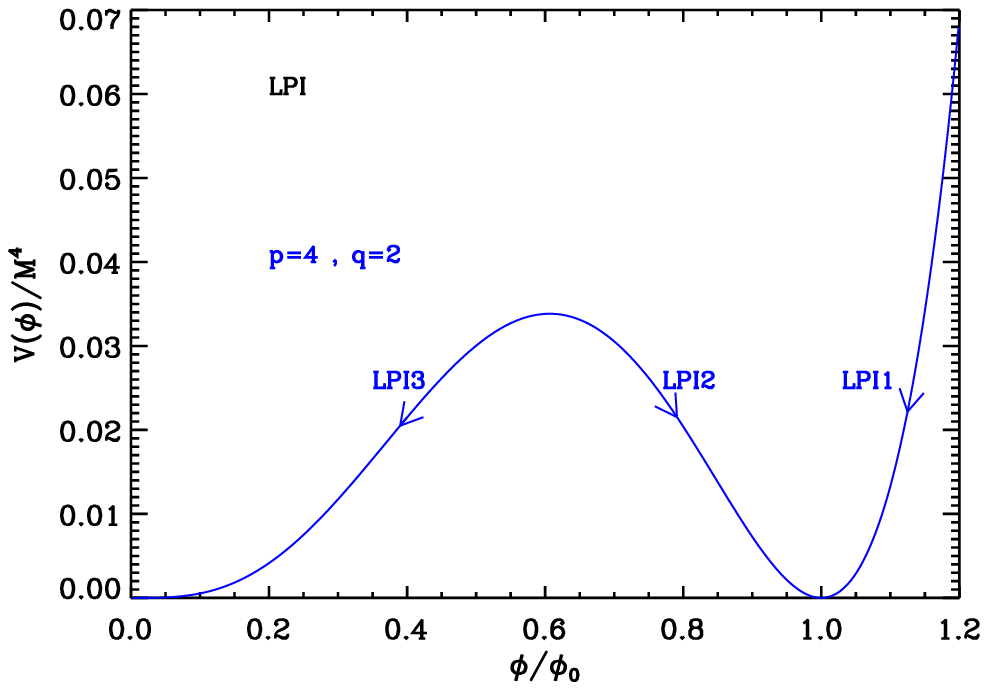}
\includegraphics[width=\wdblefig]{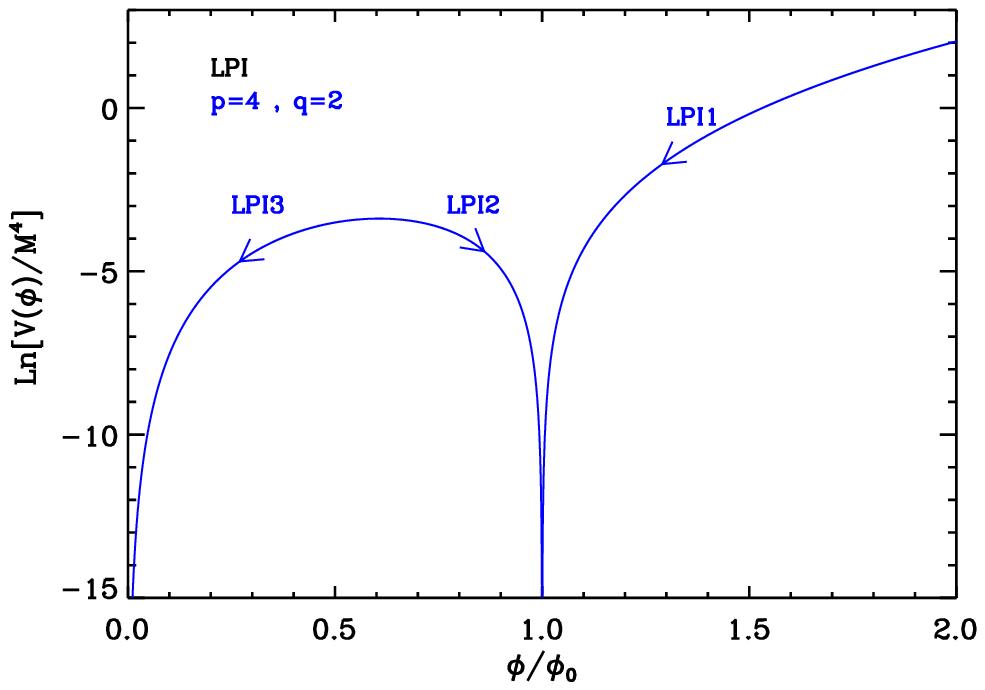}
\includegraphics[width=\wdblefig]{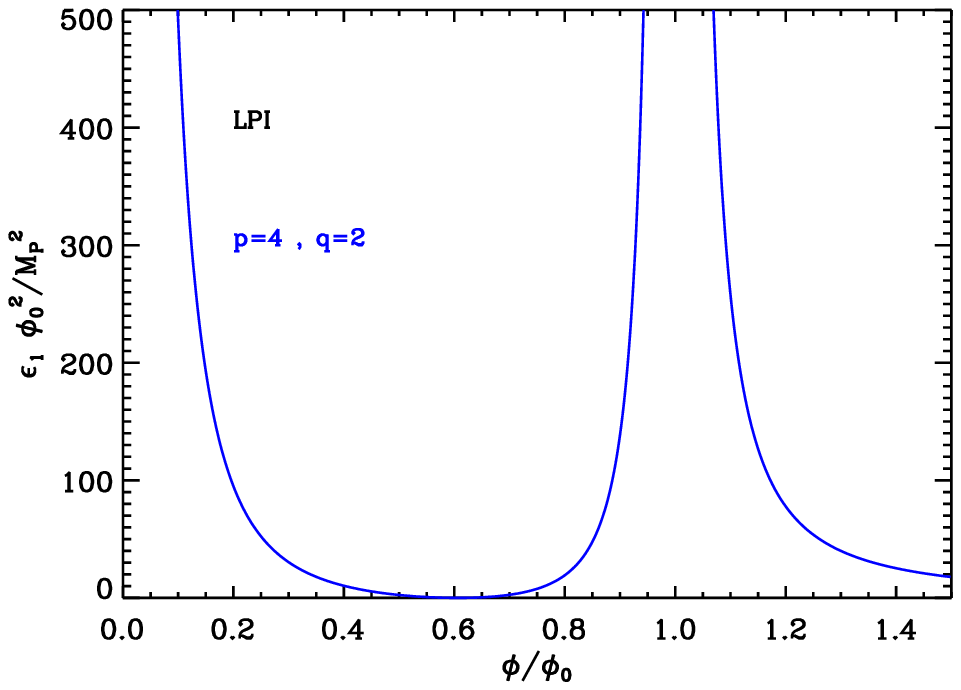}
\includegraphics[width=\wdblefig]{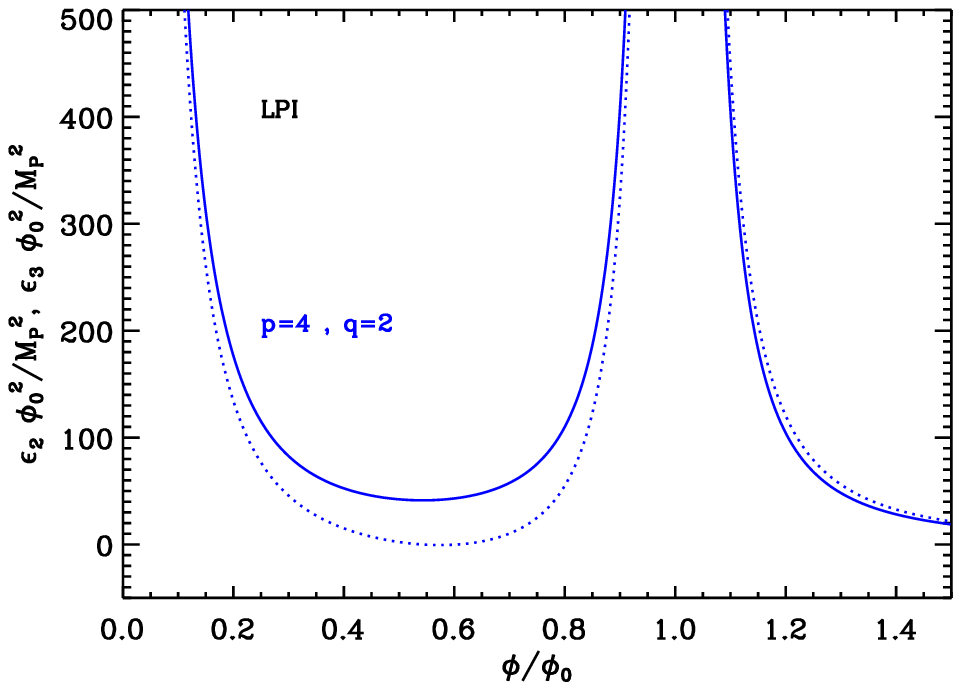}
\caption{Logarithmic Potential Inflation (LPI) for $p=4, q=2$. Upper
  panels: the potential and its logarithm. Bottom left panel:
  slow-roll parameter $\epsilon_1$. Bottom right panel: slow-roll
  parameters $\epsilon _2$ (solid line) and $\epsilon _3$ (dotted
  line).}
\label{fig:potlpi}
\end{center}
\end{figure}

Based on the previous discussion, we now turn to the slow-roll analysis 
of the models described by the following potential
\begin{equation}
  V\left(\phi\right) = M^4\left(\frac{\phi}{\phizero}\right)^{p}
  \left(\ln \frac{\phi}{\phizero}\right)^q.
\end{equation}
We have just seen that, for $p=4$ and $q=2$, the model discussed in
\Refc{Channuie:2012bv} is recovered, see \sectionc{sec:oi}, while for
$p=4$ and $q=1$, this model matches with the so-called Glueball
Inflation of \Refc{Bezrukov:2011mv}. This class of models has also
been studied on general grounds in \Refc{Barrow:1995xb}. In the
following, we keep $p$ and $q$ unspecified. Defining the quantity $x$
by the following relation
\begin{equation}
x \equiv \dfrac{\phi}{\phizero}\,,
\end{equation}
the potential has a local maximum at $x=\xVmax$ and a local minimum
(at which the potential vanishes) at $x=\xVzero$ with
\begin{equation}
\xVmax = \ee^{-q/p}, \qquad \xVzero = 1.
\end{equation}
For $x>\xVzero$, $V(x)$ increases and finally diverge when $x$ goes to
infinity. The potential is always definite positive in the $x>1$
branch, whereas it is definite positive in the $x<1$ branch only if
$q$ is an even integer. The first three Hubble flow functions in the
slow-roll approximation are given by
\begin{equation}
\epsilon_1 = \frac{\Mp^2}{\phizero^2} \frac{\left( q+p\ln x
    \right)^2}{ 2x^2\ln^2 x }\, ,\qquad
\epsilon_2 = 2\frac{\Mp^2}{\phizero^2} \frac{q+q \ln x  +
  p \ln^2 x }{ x^2 \ln^2 x }\, ,
\end{equation}
and
\begin{equation}
\epsilon_3 = \frac{\Mp^2}{\phizero^2} \left( q+p\ln x 
  \right) \frac{2q+3q \ln x  + 2q\ln^2 x  + 2p
  \ln^3 x }{x^2\ln^2 x  \left(
    q+q\ln x + p\ln^2 x \right)}\, .
\end{equation}
Together with the potential, they are displayed in \Fig{fig:potlpi}.

As can be checked on this figure, and assuming $q$ is even, the
behavior of $\epsilon_1(x)$ exhibits three domains in which inflation
can occur and can naturally end. Either $x>1$ and inflation proceeds
from the right to the left (LPI1), or $\xVmax <x <1$ and inflation
proceeds from the left to the right (LPI2), or $0< x< \xVmax$ and
inflation proceeds from the right to the left (LPI3), see the three
arrows in \Fig{fig:potlpi}. For these three cases, the slow-roll
trajectory can be integrated analytically and one has
\begin{equation}
N-\Nend = \left(\frac{\phizero}{\Mp}\right)^2 \left\lbrace -
\frac{x^2-\xend^2}{2p} + \frac{q}{p^2}\ee^{-2q/p}\left[
  \Ei\left(\frac{2q}{p}+2\ln\, x\right)- \Ei \left(\frac{2q}{p}+2\ln\,
  \xend\right)\right] \right \rbrace .
\label{eq:lpitraj}
\end{equation}
Let us remark that for $x\to +\infty$ (LPI1), one recovers the
large field inflation (LFI) trajectory of \sectionc{sec:lfi} with $p$
becoming the same parameter of LFI.

In the three above described regimes, inflation ends at the field
value $\xend$ solution of $\epsilon_1(\xend) = 1$, \ie verifying
\begin{equation}
p \ln(\xend) + q \mp \sqrt{2} \dfrac{\phizero}{\Mp} 
\xend \ln \xend = 0.
\end{equation}
This is a transcendental equation that cannot be solved analytically
for any values of $p$ and $q$. It can nevertheless be solved
numerically in each of the three above-mentioned situations. Together
with \Eq{eq:phistarlnrrad}, \Eq{eq:lpitraj} uniquely determines the
observable field value $\xstar$ at which the pivot scale crossed out
the Hubble radius during inflation. Therefore, according to our
classification, LPI is a three parameters model with $p$, $q$ and
$\phizero$.

Finally, the parameter $M$ is fixed by the amplitude of the CMB
anisotropies to
\begin{equation}
  \frac{M^4}{\Mp^4}=720\pi^2 \left(\frac{\Mp}{\phizero}\right)^2
  \dfrac{\left(q+p\ln \xstar  \right)^2}{\xstar^{2+p}
    \ln^{2+q} \xstar  } \frac{\Qrms^2}{T^2}\, .
\end{equation}
The reheating consistent slow-roll predictions for the LPI1 models
with $p=4$ are represented in Figs~\ref{fig:CMBLPI1pEQ4qEQ2},
\ref{fig:CMBLPI1pEQ4qEQ1}, and \ref{fig:CMBLPI1pEQ4qEQ3} for $q=2$,
$q=1$ and $q=3$, respectively. The predictions for LPI2 are displayed
in Figs~\ref{fig:CMBLPI2pEQ1qEQ2}, \ref{fig:CMBLPI2pEQ2qEQ2}, and
\ref{fig:CMBLPI2pEQ3qEQ4} for ($p=1$, $q=2$), ($p=2$, $q=2$) and
($p=3$, $q=4$), respectively.  For the LPI3 scenario, the predictions
have been plotted in Figs~\ref{fig:CMBLPI3pEQ1qEQ2},
\ref{fig:CMBLPI3pEQ2qEQ2}, and \ref{fig:CMBLPI3pEQ3qEQ4} for ($p=1$,
$q=2$), ($p=2$, $q=2$) and ($p=3$, $q=4$), respectively. One can see
that the current CMB data generically require LPI inflation to take
place with super-Planckian values for $\phizero$ while some
combinations of $p$ and $q$ are already disfavored at more than
two-sigma.

\subsection{Constant \texorpdfstring{$\nS$}{nS} D Inflation (CNDI)}
\label{sec:cndi}

\begin{figure}
\begin{center}
\includegraphics[width=\wdblefig]{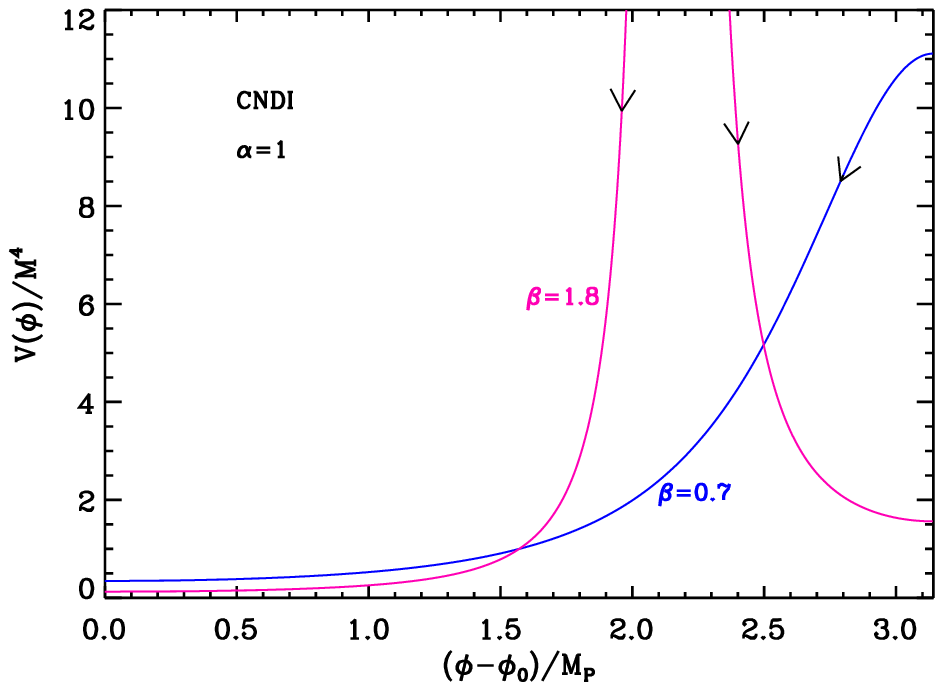}
\includegraphics[width=\wdblefig]{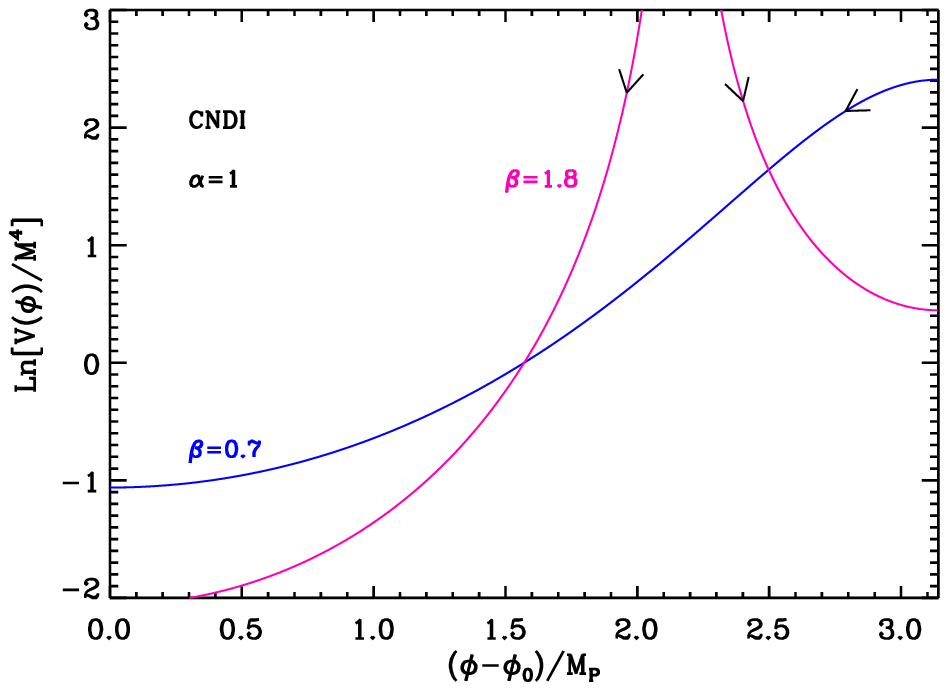}
\includegraphics[width=\wdblefig]{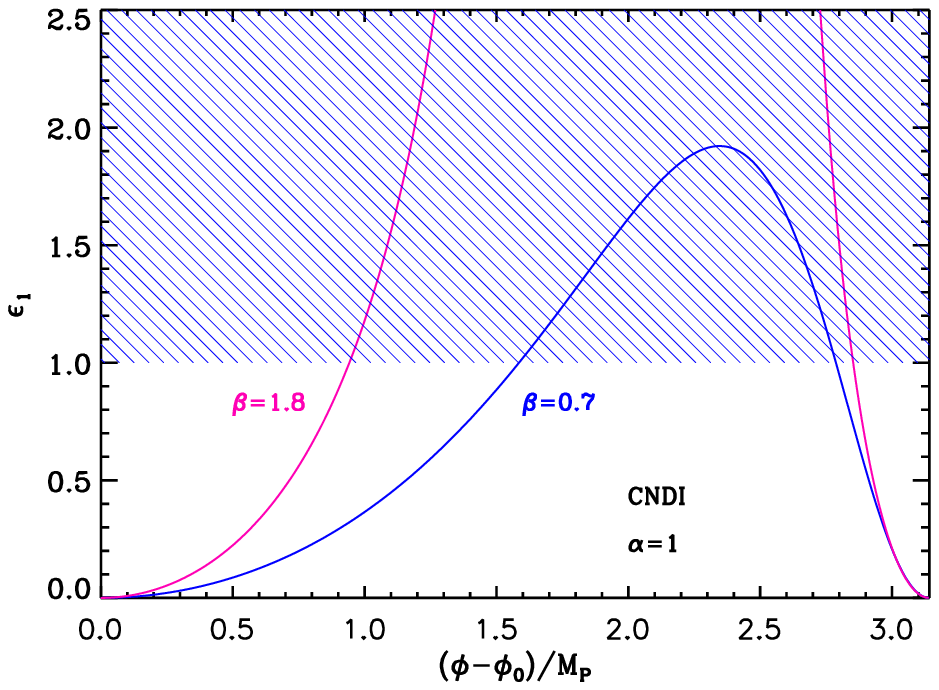}
\includegraphics[width=\wdblefig]{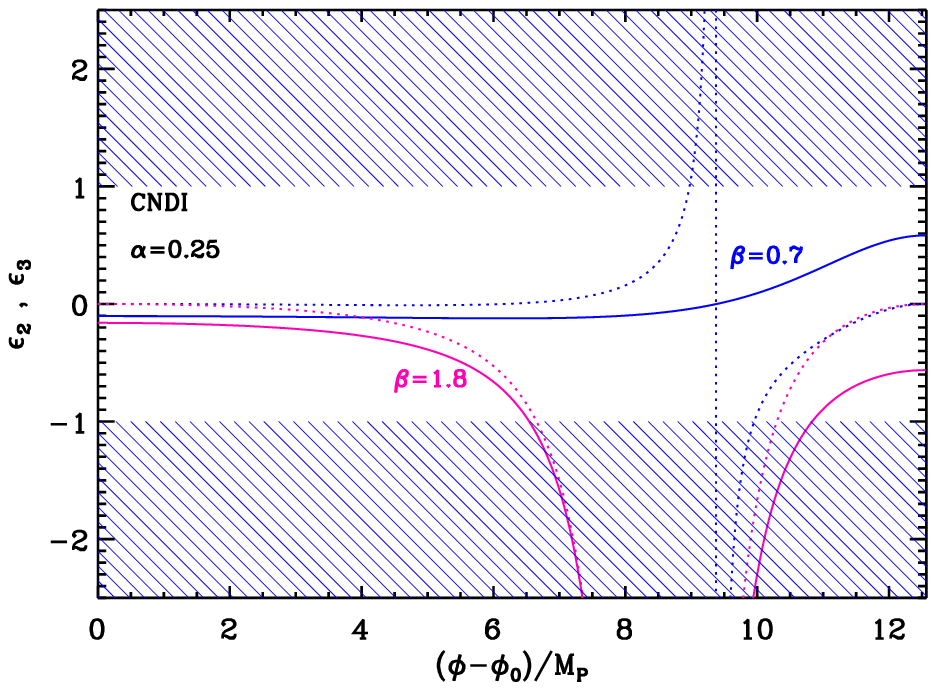}
\caption{Top left panel: constant $\nS$ D inflaton potential for
  $\alpha=1$ and two values of $\beta$, namely $\beta=0.7$ (solid blue
  line) and $\beta=1.3$ (solid pink line). Top right panel: logarithm
  of the potential for the same values of $\alpha$ and $\beta$ and
  with the same color code. Bottom left panel: first slow-roll
  parameter $\epsilon _1$ for a potential with $\alpha=1$ and
  $\beta=0.7$ (solid blue line), $\beta=1.8$ (solid pink line). The
  shaded area indicates the breakdown of slow-roll inflation (strictly
  speaking where acceleration cannot occur). Bottom right panel:
  second and third slow-roll parameters $\epsilon _2$ and $\epsilon
  _3$ for $\alpha=0.25$ and the same values of $\beta$ as in the other
  plots.}
\label{fig:potCNDI}
\end{center}
\end{figure}

This model has been studied in \Refc{Hodges:1990bf}. Its potential is
designed to produce a power law power spectrum $\propto k^n$
(where $n$ is a constant). In this sense, the approach followed here
is similar to the one investigated in \sectioncs{sec:cnai},
\ref{sec:cnbi} and~\ref{sec:cnci}. The potential studied in this
section is given by
\begin{equation}
V\left(\phi\right) = \frac{M^4}{ \left\lbrace 1 + \beta\cos\left[
    \alpha \left( \dfrac{\phi-\phizero}{\Mp} \right) \right] \right \rbrace^2}\, ,
\end{equation}
where $\alpha$ and $\beta$ are two dimensionless parameters. Since the
potential is an even function of
$x\equiv\left(\phi-\phizero\right)/\Mp$ and is $2\pi$-periodic, it can
be studied without loss of generality in the range
$x\in\left[0,\pi/\alpha\right]$ only (with $\alpha>0$,
$\beta>0$). The potential and its logarithm are displayed in
\Fig{fig:potCNDI} (top panels) for two different representative values
of $\beta$. If $\beta<1$ (blue curve), it is an increasing function of
the field, hence inflation proceeds from the right to the left. On the
contrary, if $\beta\geq 1$ (pink curve), it diverges at $\xVinfty =
\arccos \left(-1/\beta\right)/\alpha$. Then, for $x<\xVinfty$ it is an
increasing function of $x$ and inflation proceeds from the right to
the left, whereas for $x>\xVinfty$ it is an decreasing function of $x$
and inflation proceeds from the left to the right.

The three first slow-roll parameters are given by the following
expressions
\begin{equation}
  \epsilon_1 = \frac{2\alpha^2 \beta^2 \sin^2 \left(\alpha x\right)}{
    \left[ 1+\beta\cos\left( \alpha x \right) \right]^2 }\, ,\qquad
  \epsilon_2 = \frac{-4\alpha^2\beta\left[\beta+\cos\left(\alpha
      x\right)\right]}{\left[1+\beta\cos\left(\alpha
      x\right)\right]^2}\, ,
\end{equation}
and
\begin{equation}
  \epsilon_3 = \frac{-2\alpha^2\beta\left[2\beta^2-1+\beta\cos\left(\alpha
      x\right)\right]\sin^2\left(\alpha
    x\right)}{\left[\beta+\cos\left(\alpha
      x\right)\right]\left[1+\beta\cos\left(\alpha
      x\right)\right]^2}\, .
\end{equation}
They are displayed in \Fig{fig:potCNDI} (bottom panels). Let us now study
in more detail the behavior of $\epsilon_1$ and $\epsilon_2$. It
depends on whether $\beta $ is larger or smaller than $1$. If
$\beta<1$, the first slow-roll parameter $\epsilon_1$ vanishes at
$x=0$ and $x=\pi/\alpha$, and reaches a maximum in between at
$\xepstwoZero$. This maximum is larger than one provided
$\alpha>\alphamin \left(\beta\right)$, where
\begin{equation}
\alphamin \left(\beta\right)=\sqrt{\frac{1-\beta^2}{2\beta^2}}\, .
\end{equation}
In that case, inflation can stop by slow-roll violation, at the
position $\xend$ given by
\begin{equation}
\label{eq:endcndi}
  \xend=\xepsoneOnePlus = \frac{1}{\alpha} \arccos
  \left[\frac{\alpha\sqrt{2\beta^2 \left(1+2\alpha^2\right)-2}-1} {
      \beta +2\alpha^2\beta}\right],
\end{equation}
and proceeds in the range $\left[\xend,\pi/\alpha\right]$
(from the right to the left). On the other hand, the second slow-roll
parameter $\epsilon_2$ is a monotonic increasing function of $x$,
which vanishes at $\xepstwoZero=\arccos
\left(-\beta\right)/\alpha$. If $\beta\geq 1$, as can be seen in
\Fig{fig:potCNDI}, the first slow-roll parameter $\epsilon_1$ diverges
at $\xVinfty=\arccos(-1/\beta)/\alpha$, so that inflation cannot stop
by slow-roll violation in that case. This means that inflation must
end by another mechanism and, therefore, that the model depends on an
additional parameter. The second slow-roll parameter $\epsilon_2$ is
always negative and also diverges at $\xVinfty$. Let us notice that,
for $\beta<1$ and $\alpha>\alphamin \left(\beta\right)$, and for
$\beta>1$ (for any $\alpha$), we will need below the other solution of
$\epsilon_1=1$, namely
\begin{equation}
  \xepsoneOneMinus=\frac{1}{\alpha} \arccos
  \left[-\frac{\alpha \sqrt{2\beta^2 \left(1+2\alpha^2\right)-2}+1}{
      \beta + 2\alpha^2 \beta} \right].
\end{equation}

We are now in a position where the slow-roll trajectory can be
determined. It turns out that this one can be integrated analytically
and reads
\begin{equation}
\label{eq:trajeccndi}
  N-\Nend = \frac{1}{2\alpha^2} \left\{ -\ln \left[\sin\left(\alpha x\right)
    \right] - \frac{1}{\beta} \ln \left[\tan\left(\alpha
    \frac{x}{2}\right) \right] + \ln\left[\sin
    \left(\alpha \xend\right)\right] + \frac{1}{\beta} \ln
  \left[\tan \left(\alpha \frac{\xend}{2} \right) \right] \right\}.
\end{equation}
Because of the logarithmic functions, a sufficient number of \efolds
can be realized only if the initial conditions are fine-tuned and
$\xini$ is chosen to be extremely close to $\pi/\alpha$. 

Indeed, inserting \Eq{eq:endcndi} into \Eq{eq:trajeccndi}, the total
number of \efolds during inflation becomes a function of $\xini$ and
of the two parameters $\alpha $ and $\beta$. For given values of those
parameters, one can check that $(\Nend-\Nini)(\xini)$ remains always
small compared to unity, unless $\xini\to \pi/\alpha$ where it
blows up. Let us write $\xini $ as $\pi/\alpha+\delta \xini$ with
$\delta \xini\ll 1$ and defining $A\equiv \ln \left[\sin\left(\alpha
  \xend\right) \right]+\ln \left[\tan\left(\alpha \xend/2\right)
  \right]/\beta$, one arrives at
\begin{equation}
\delta \xini \simeq \left[\alpha \left(\frac{\alpha}{2}\right)^{-1/\beta}
\ee^{-A}\right]^{\beta/(1-\beta)} \ee^{-2\alpha^2\beta (\Nend-\Nini)/(1-\beta)}.
\end{equation}
The coefficient between the squared brackets only depends on $\alpha $
and $\beta$ which are, a priori, coefficients of order one. On the
other hand, the argument of the exponential is $2(\Nend-\Nini)>120$,
times a negative term of order one. This means that $\delta \xini$
must be exponentially small to obtain a significant number of \efolds
and one can question the physical relevance of such a fine-tuning.
The typical predictions of the model (taking $\xstar \simeq
\pi/\alpha$) actually are $\epsilon_1\simeq 0$, $\epsilon_2\simeq
4\alpha^2\beta/\left(1-\beta\right)$, and $\epsilon_3\simeq 0$. It
follows that the condition $\alpha>\alphamin\left(\beta\right)$
implies $\epsilon_2>2\left(1+\beta\right)/\beta>4$, which is
completely ruled out by the observations. Therefore, we conclude that
the case $\beta <1$ is not of cosmological interest.

The only remaining possibility is $\beta >1$. Inflation cannot end by
slow-roll violation and $\xend$ is an additional parameter, making the
model a three parameters one. In the range $\alpha \xend\ll 1$, one
has $\epsilon_1\ll 1$ and $\epsilon_2\simeq -4\alpha^2\beta/(1+\beta)$
such that the spectral index is given by $\nS\simeq 1 +
4\alpha^2\beta/ \left(\beta+1\right)$. Therefore, it is indeed a
constant.

The CMB normalization gives the mass scale $M$ as
\begin{equation}
\label{eq:cobecndi}
  \left(\frac{M}{\Mp}\right)^4=2880 \alpha^2 \beta^2 \pi^2 \sin^2
  \left(\alpha \xstar\right) \frac{\Qrms^2}{T^2}\,,
\end{equation}
which has to be numerically evaluated when if $\alpha \xstar$ is not
small. The predictions of CNDI inflation are displayed in
\Figs{fig:CMBCNDIbetaEQ0dot1} and \ref{fig:CMBCNDIbetaEQ5}. We see
that, in the regime $\alpha \xend\ll 1$, the spectral index is
constant, as expected. However, this occurs in a regime where the
predictions are not consistent with the observations (the spectrum is
too blue). On the other hand, when $\alpha \xend$ is no longer small,
we observe strong deviations from $\nS\simeq 1 + 4\alpha^2\beta/
\left(\beta+1\right)$ but, for intermediate values of $\alpha \simeq
0.3$, this renders the predictions compatible with the
data. Obviously, these considerations bear some resemblance with the
findings of \sectioncs{sec:cnai}, \ref{sec:cnbi} and~\ref{sec:cnci}.

\subsection{String Axion Inflation II (SAIII)}
\label{sec:saiii}

This model shares the same theoretical construction as String Axion
Inflation I (SAII) presented in \sectionc{sec:saii} and has been
proposed in \Refc{Kobayashi:2015aaa}. Compared to SAII, a mass
term coming from higher-order corrections associated with instanton
effects, proportional to $\phi^2$, appears in the potential. This mass
term could also be viewed as yet another correction to the potential
of Natural Inflation (NI), see \sectionc{sec:ni}, and this has been
further discussed in \Refc{Kobayashi:2010pz}. For reasons that
will be clearer later on, it is convenient to write the potential of
SAIII under the form
\begin{equation}
V(\phi) = M^4 \left\{ 1 - \cos\left(\dfrac{\phi}{\mu}\right) + \alpha
\left[ \dfrac{\phi}{\mu} \sin\left(\dfrac{\phi}{\mu}\right) + \dfrac{1}{2}
\beta \left(\dfrac{\phi}{\mu}\right)^2 \right] \right\},
\label{eq:pot:saiii}
\end{equation}
where $\mu$ is a {\vev}, and $\alpha$ and $\beta$ are two dimensionless
constants that are not required to be small. However, the new mass
term is required to be positive, which implies the
constraint $\alpha \beta \ge 0$, \ie $\alpha$ and $\beta$ must be of the same
sign.

\begin{figure}
\begin{center}
\includegraphics[width=\wdblefig]{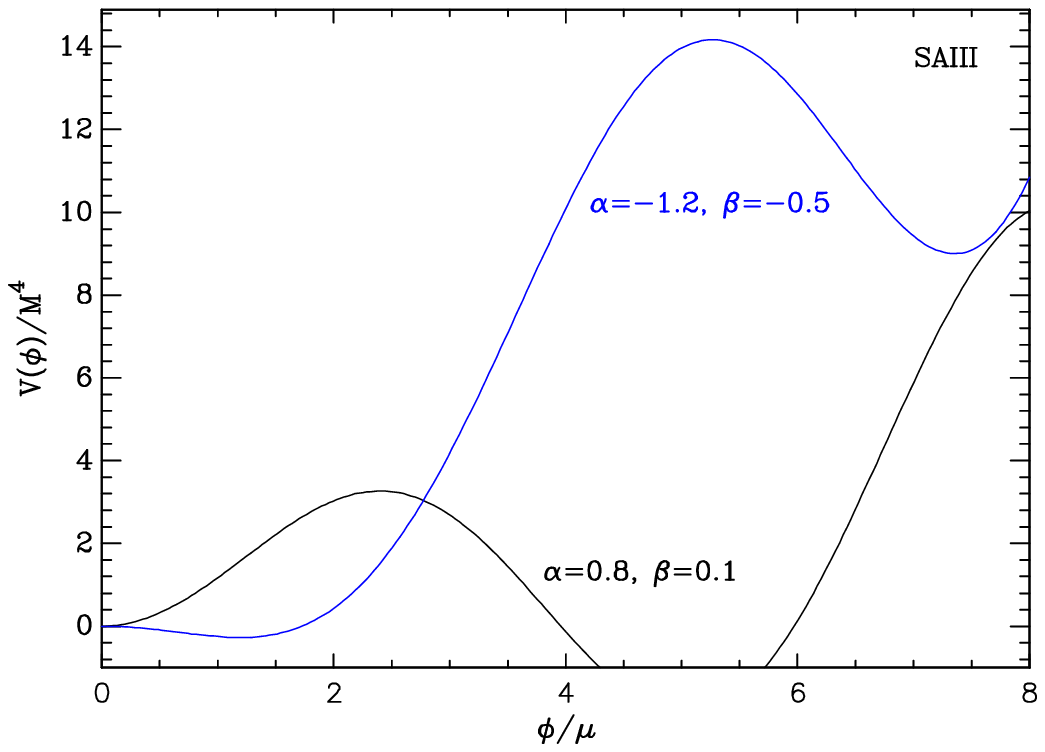}
\includegraphics[width=\wdblefig]{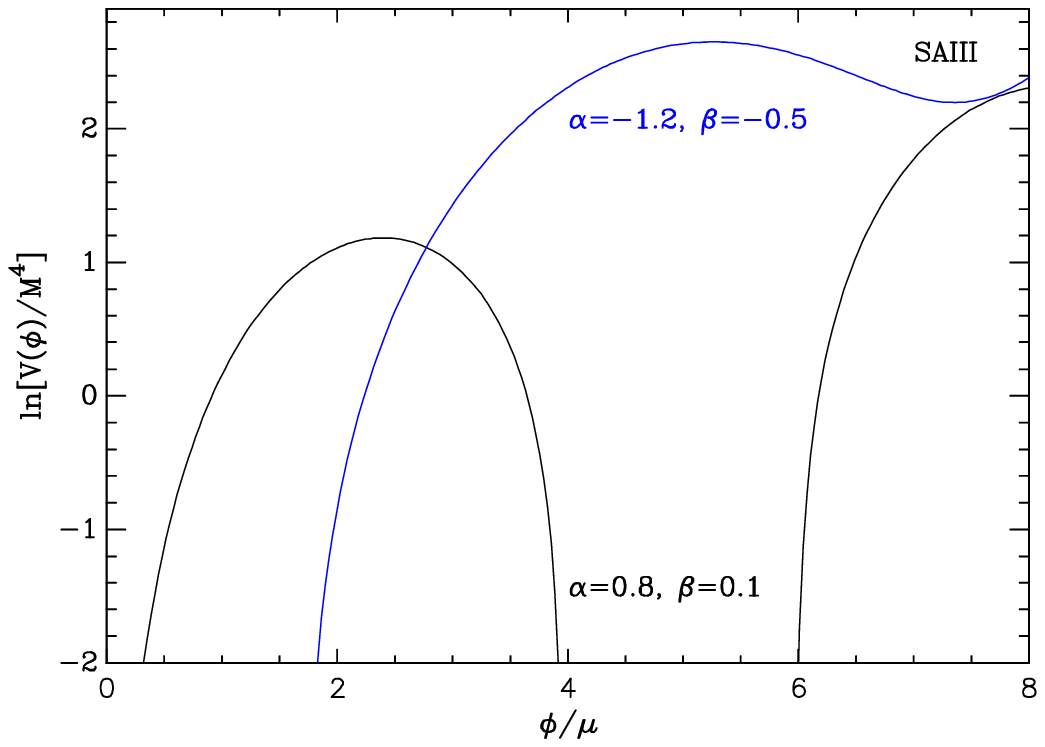}
\includegraphics[width=\wdblefig]{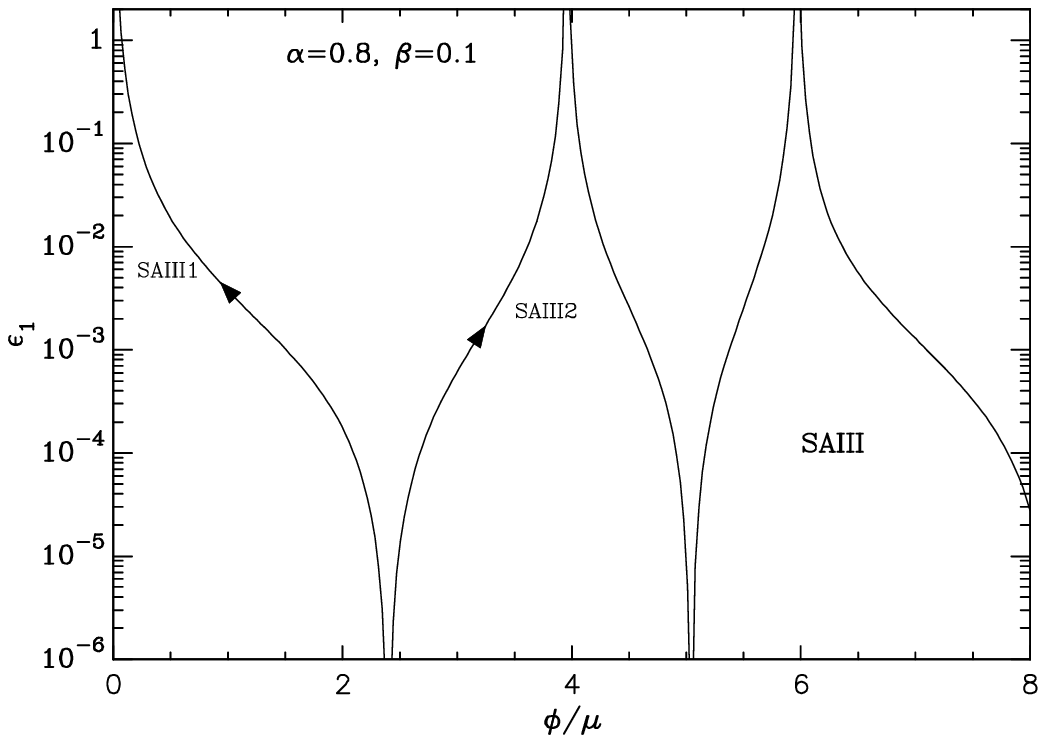}
\includegraphics[width=\wdblefig]{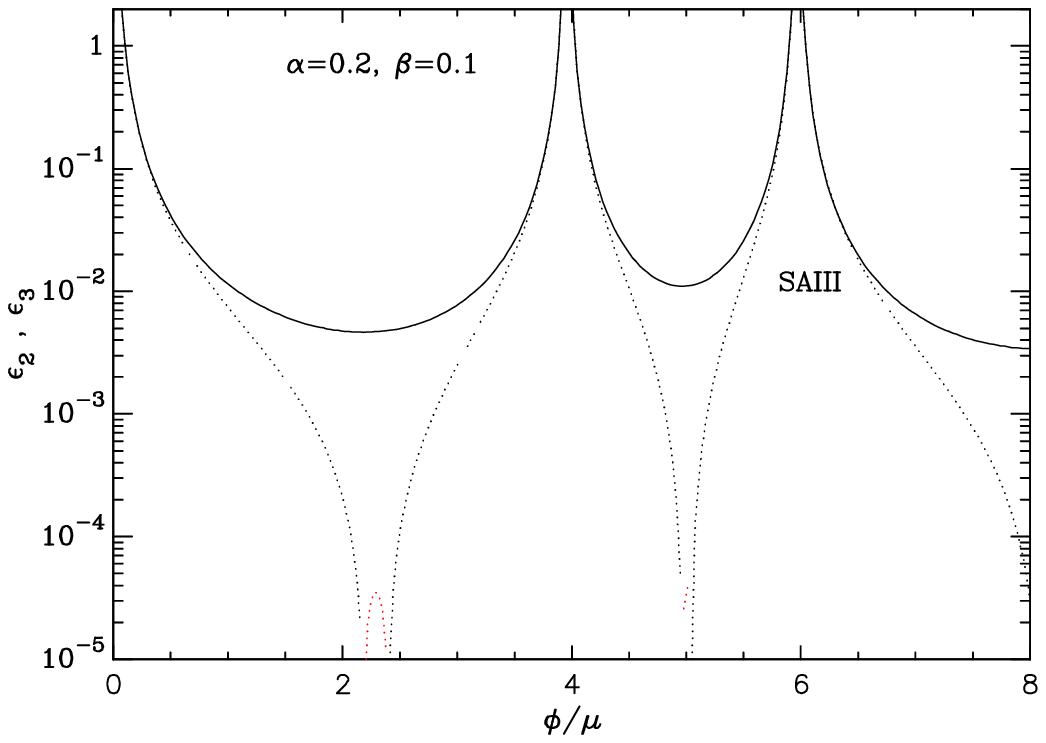}
\caption{String Axion Inflation II (SAIII) for $\alpha=0.8$,
  $\beta=0.1$ (black curve) and $\alpha=-1.2$, $\beta=-0.5$ (blue
  curve). Top panels: the potential and its logarithm. Only the SAIII1
  and SAIII2 regimes exist for $\alpha=0.8$, $\beta=0.1$ (black)
  whereas inflation gracefully ends only for SAIII1 in the case
  $\alpha=-1.2$, $\beta=-0.5$. Bottom left panel: slow-roll parameter
  $\epsilon_1$ for $\alpha=0.8$, $\beta=0.1$ and $\mu = 20\,\Mp$, with
  the two inflationary regimes annotated with an arrow indicating the
  direction to which the field evolves. Bottom right panel: slow-roll
  parameters $\epsilon_2$ (solid line) and $\epsilon_3$ (dotted line)
  for the same parameters value. When $\epsilon_3$ becomes negative,
  the plot shows $|\epsilon_3|$ as a red dotted line, the black dotted
  line corresponds to positive values. The regime SAIII3 is
  represented in \Fig{fig:potsaiii3}.}
\label{fig:potsaiii12}
\end{center}
\end{figure}

\begin{figure}
\begin{center}
\includegraphics[width=\wdblefig]{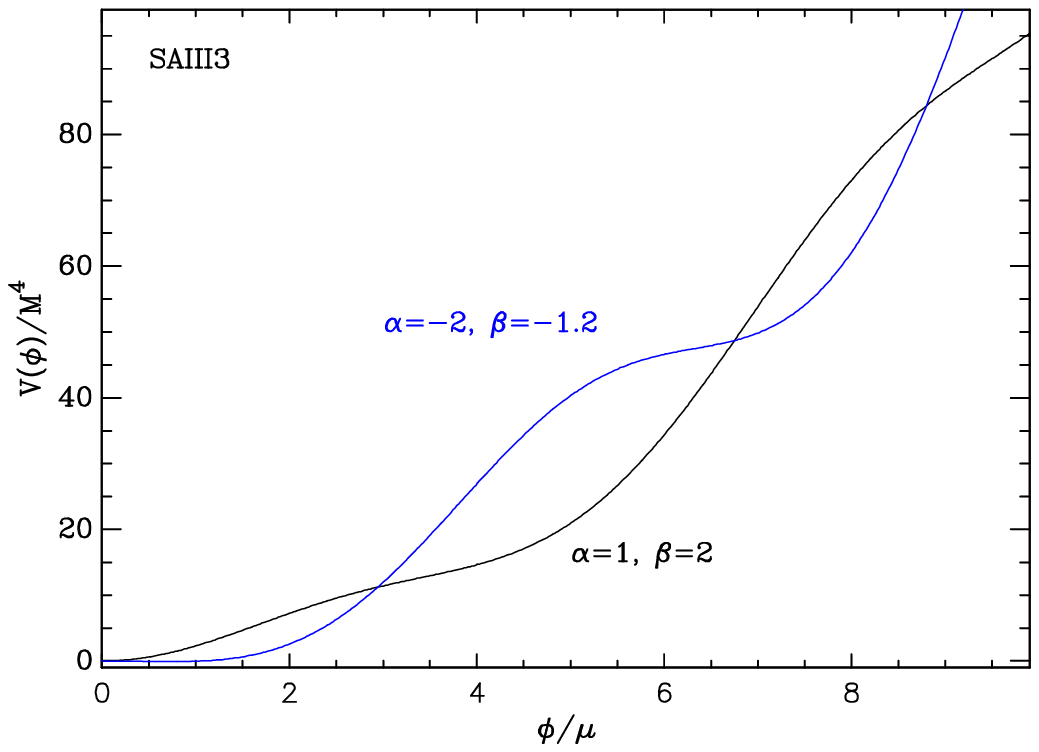}
\includegraphics[width=\wdblefig]{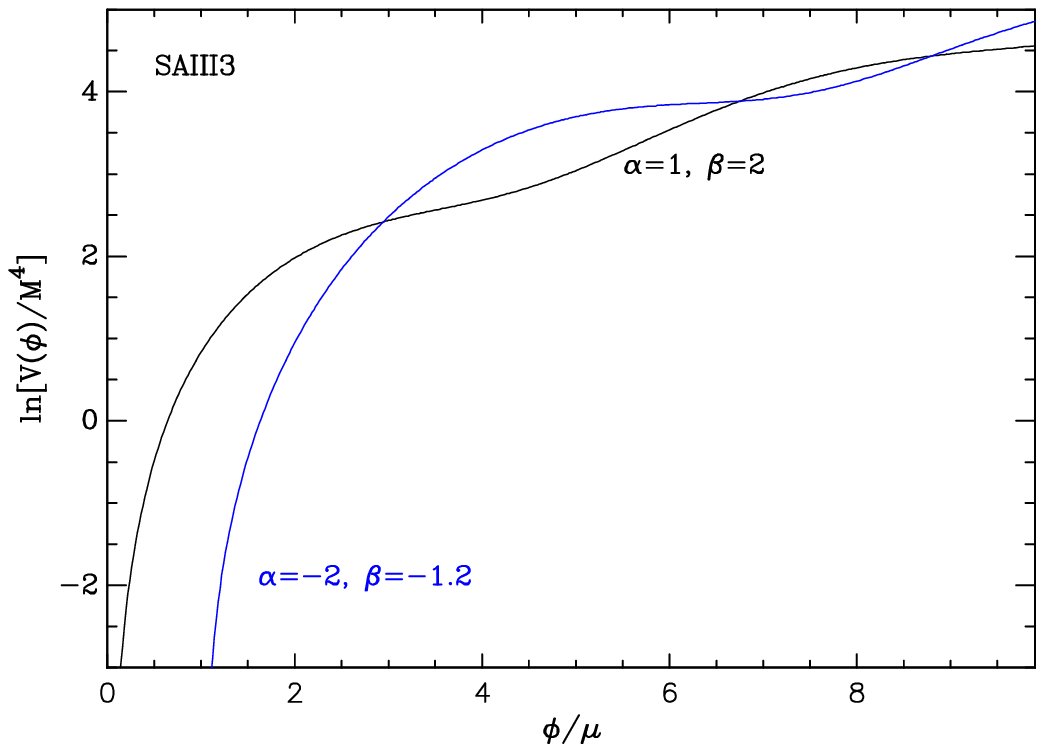}
\includegraphics[width=\wdblefig]{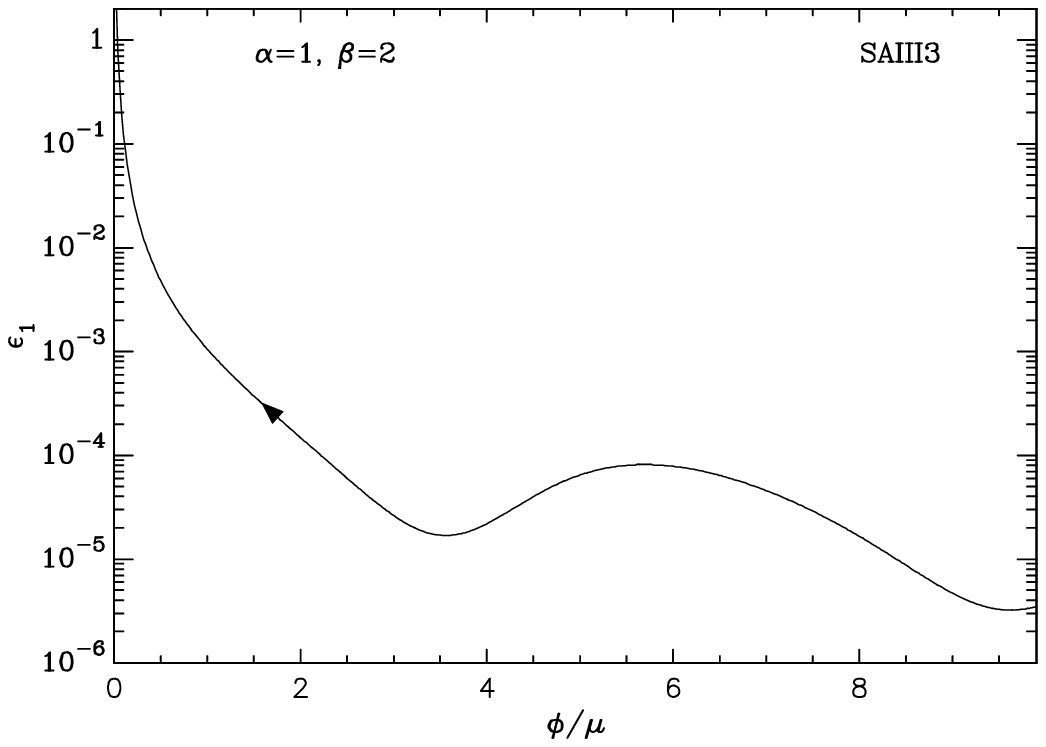}
\includegraphics[width=\wdblefig]{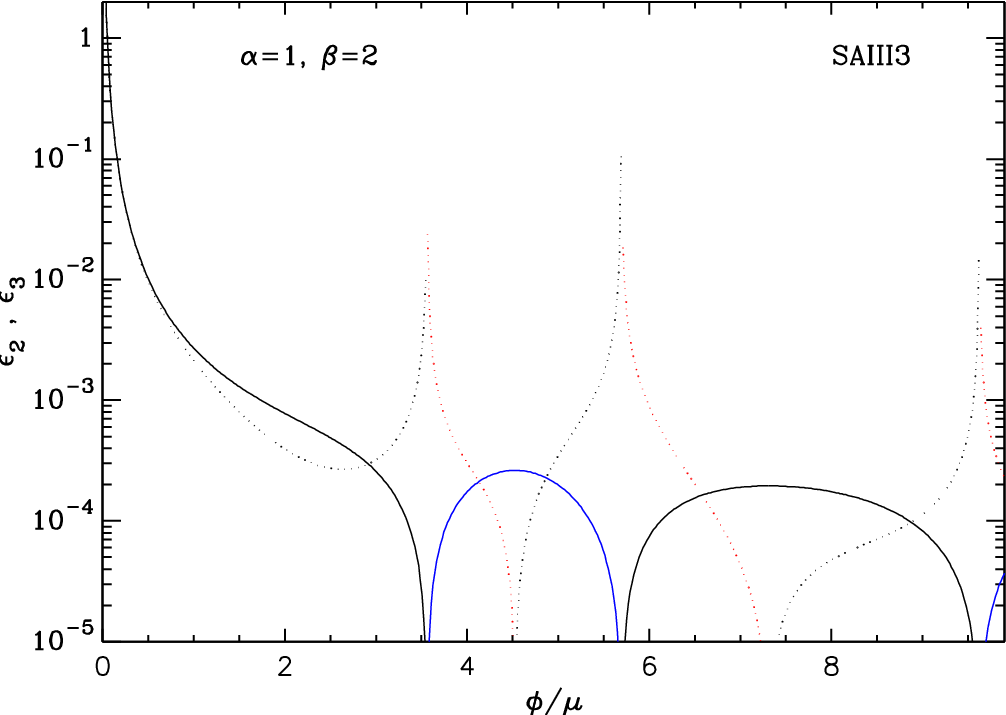}
\caption{String Axion Inflation II in the SAIII3 regime where the
  potential does not exhibit positive extrema, for $\alpha=1$,
  $\beta=2$ (black curve) and $\alpha=-2$, $\beta=-1.2$ (blue
  curve). Top panels: the potential and its logarithm. Bottom left
  panel: slow-roll parameter $\epsilon_1$ for $\alpha=0.8$,
  $\beta=0.1$ and $\mu = 40\,\Mp$, with the field evolution annotated
  with an arrow indicating the direction to which it evolves. Bottom
  right panel: slow-roll parameters $\epsilon_2$ (solid line) and
  $\epsilon_3$ (dotted line) for the same parameters value. When
  $\epsilon_3$ becomes negative, the plot shows $|\epsilon_3|$ as a
  red dotted line, the blue dotted line corresponds to positive
  values. Similarly, negative values of $\epsilon_2$ are represented
  as green solid curves. Notice that $\epsilon_3$ becomes singular at
  the points where $\epsilon_2=0$, but the product $\epsilon_2
  \epsilon_3$ remains finite. The other regimes SAIII1 and SAIII2 are
  represented in \Fig{fig:potsaiii12}.}
\label{fig:potsaiii3}
\end{center}
\end{figure}

The potential is symmetric with respect to $\phi=0$ and the analysis can thus be restricted to the domain $\phi\geq 0$. As for SAII, depending on the value
of $\alpha$ and $\beta$, the potential can become negative in some
domains that are separated by a local maximum. The inflationary domains
existing on both side of the first maximum of $V(\phi)$ will be
referred to as SAIII1 and SAIII2, by analogy with the treatment of
SAII carried out in \sectionc{sec:saii}. However, and as opposed to SAII,
the additional mass term implies that, for large
enough values of $\beta$, the potential can become a strictly
monotonic increasing function of $\phi$. In this regime, the model
becomes similar to Large-Field Inflation with $p=2$ (LFI${}_2$, see \sectionc{sec:lfi}), plus some small
modulations. This regime will be denoted SAIII3 and its existence is
mutually exclusive with SAIII1 and SAIII2. The potential and its
logarithm have been represented in \Figs{fig:potsaiii12} and
\ref{fig:potsaiii3} for those three inflationary regimes.

\subsubsection{Parameter Space Analysis}

\begin{figure}
\begin{center}
\includegraphics[width=\wsingfig]{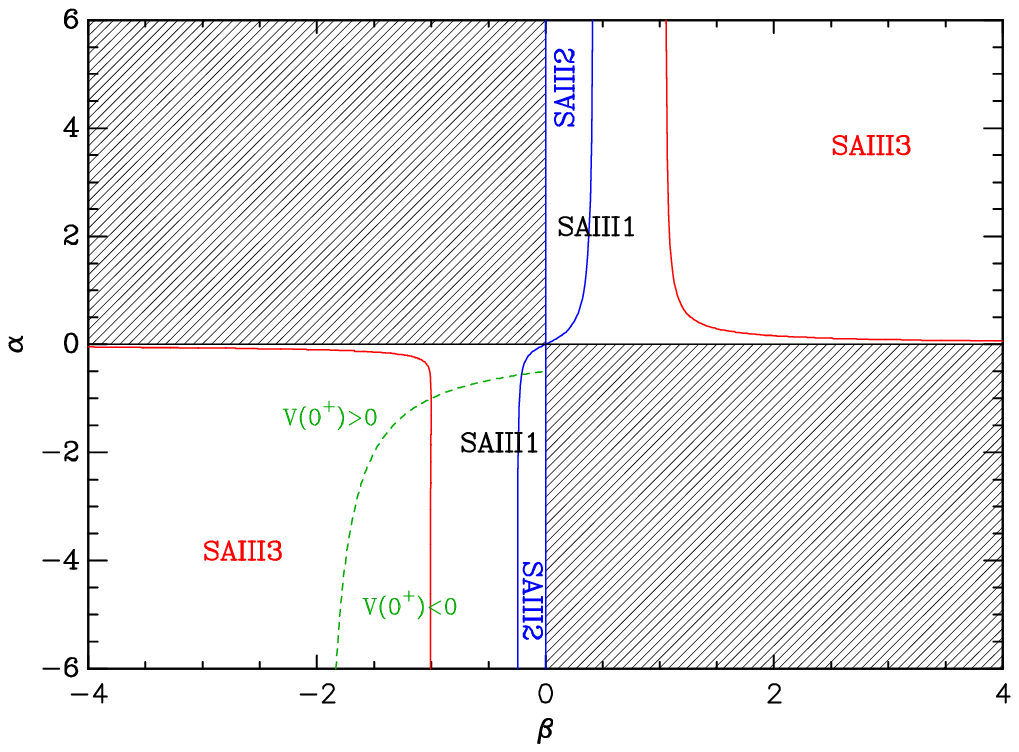}
\caption{The parameter space $(\beta,\alpha)$ of String Axion
  Inflation II and its various inflationary regimes. SAIII3 exists
  only in the upper right and lower left corner separated by the red
  curves, which are $\alpha_1(\beta)$, $\alpha_2(\beta)$ and
  $\alpha_3(\beta)$ (see text). SAIII3 is mutually exclusive with
  SAIII1, therefore SAIII1 always exists on the left of the positive
  red curve and on the right of the negative one. However, SAIII2 can
  take place only if the potential becomes negative at field values
  larger than the one corresponding to the first maximum of the
  potential. This occurs only within the central domain bounded by the
  blue curves. For convenience, we have also represented in dashed
  green the separatrix for which the potential is negative, or
  positive, around the origin. Parameters can be accomodated such that
  all the three inflationary regimes may encounter a positive or
  negative potential around $\phi=0$.}
\label{fig:absaiii}
\end{center}
\end{figure}

Let us first derive the conditions on $\alpha$ and $\beta$ under which
SAIII3 exists, and SAIII1/SAIII2 do not exist. From the above considerations, the
potential must not have any local extremum. Defining $x\equiv \phi/\mu$, and
deriving \Eq{eq:pot:saiii} with respect to $x$, one gets
\begin{equation}
\dfrac{V'(x)}{M^4} = \left(1+\alpha\right) \sin(x) + \alpha
x\left[\beta + \cos(x)\right].
\label{eq:saiii:Vprime}
\end{equation}
The potential possesses local extrema if solutions to $V'(x)=0$
exist. This is a transcendental equation that can be recast into the
form $f(x) = x$ by defining the function
\begin{equation}
f(x) \equiv - \dfrac{(1+\alpha) \sin(x)}{\alpha\left[\cos(x) + \beta \right]}\,.
\label{eq:saiii:f}
\end{equation}

The location of the separatrices in parameter space that delineate the
regions where solutions to $f(x)=x$ exist cannot be obtained
analytically, but they can be worked out numerically as follows. Along
such a separatrix, at the location $x=x_f$ where $f(x_f)=x_f$, the two
functions $x$ and $f(x)$ must tangent each other. In other words, one
should have $f'(x_f)=1$ along the separatrices. Given that
\begin{equation}
\label{eq:saiii:fprime}
f'(x) = -\dfrac{1+\alpha}{\alpha} \dfrac{1 + \beta
  \cos(x)}{\left[\beta + \cos(x)\right]^2}\,,
\end{equation}
this implies that 
\begin{equation}
\label{eq:saiii:xfexact}
\cos\left(x_f\right)=-\left(3+\frac{1}{\alpha}\right)
\frac{\beta}{2}\pm\sqrt{\left(3+\frac{1}{\alpha}\right)\frac{\beta}{2}-\beta^2-1-\frac{1}{\alpha}}\,
.
\end{equation}
This formula can be plugged into the relation $f(x_f)=x_f$, using the
fact that $x_f=\arccos(\cos x_f)$ or $2\pi-\arccos(\cos x_f)$ and that
$\sin x_f=\pm\sin[\arccos(\cos x_f)]$. One obtains an equation that
involves $\alpha$ and $\beta$ and that must be solved numerically in order
to obtain the functions $\alpha(\beta)$ that delineate the regions
where SAIII3 exists.

In order to gain further analytical insight, the following constraints
can be put on the location of the separatrices. First, from
\Eq{eq:saiii:f}, one notices that $f(x)$ has poles for
$\cos(x)=-\beta$ provided $|\beta|\le 1$. Since the sign of the
numerator in $f$ necessarily flips between two such poles (and given
that the sign of the denominator remains the same), $f$ interpolates
between $\pm\infty$ between two consecutive poles and necessarily
crosses $x$. This ensures that $V(x)$ is extremal somewhere, and that
SAIII3 does not exist. The separatrices need therefore to be looked
for at values of $\beta$ such that $\vert \beta \vert>1$ and we will
thus focus on this region hereafter.

Second, from \Eq{eq:saiii:fprime}, the function $f$ is extremal where
$f'(x)=0$, \ie where $\cos(x) = -1/\beta$, and the two extrema of $f$
in $[0,2\pi$] occur at
\begin{equation}
\xfminus = \arccos\left(-\dfrac{1}{\beta} \right), \qquad \xfplus = \pi +
\arccos\left(\dfrac{1}{\beta} \right).
\label{eq:saiii:xf}
\end{equation}
Because $f(x)$ is $2\pi$-periodic, and since $f(0)=0$, if there is no
solution to $f(x)=x$ within the domain $[0,2\pi]$, none can exist for
$x>2\pi$.

In order to identify which of $\xfminus$ and $\xfplus $ is a minimum
or a maximum, one can consider the sign of
\begin{equation}
f'(0) = - \dfrac{1+\alpha}{\alpha(1+\beta)} \, .
\end{equation}
If $\beta>1$ (hence $\alpha>0$, since $\alpha\beta>0$), $f'(0)<0$, the
function $f$ initially decreases away from the origin, and its first
extremum is a minimum: in this case, $\xfminus $ is a negative minimum
and $\xfplus $ is a positive maximum, given that $f(\pi)=0$.  If
$\beta<-1$ (hence $\alpha<0)$, $f'(0)<0$ if $\alpha>-1$, in which case
$\xfminus $ is a minimum and $\xfplus $ is a maximum, while $f'(0)>0$
if $\alpha<-1$, in which case $\xfminus $ is a maximum and $\xfplus $
is a minimum. In passing, let us note that $f'(0)>1$ if and only if
$-2<\beta<-1$ and $\alpha<-1/(2+\beta)$. When this happens, $f(x)>x$
when $x$ approaches $0$ while $f(x)<x$ at $x=\pi$ since $f(\pi)=0$,
hence solutions to the equation $f(x)=x$ can be found and SAIII3 does
not exist. Otherwise, let us derive the condition for the maximum of
the function $f(x)$ to lie above the $x$ function. This will provide a
sufficient condition for solutions to exist, which is however not a
necessary condition but that will still allow us to better bound the
location of the separatrices.  From the above considerations, three
subcases need to be distinguished.

If $\beta>1$, the condition $f(\xfplus) \ge \xfplus$ reads
\begin{equation}
\dfrac{1+\alpha}{\alpha \sqrt{\beta^2 - 1}} \ge \pi + \arccos\left(\dfrac{1}{\beta} \right).
\end{equation}
Solving for $\alpha$, the condition for $V$ to have an extremum
translates into $\alpha \le \alpha_2(\beta)$, where
\begin{equation}
\alpha_2(\beta) \equiv \dfrac{1}{\sqrt{\beta^2-1} \left[\pi +
    \arccos\left(\dfrac{1}{\beta} \right) \right] - 1}\,.
\label{eq:saiii:alphatwo}
\end{equation}
Conversely, the necessary (but in principle not sufficient) condition for SAIII3 to exist if $\beta > 1$
is then $\alpha > \alpha_2(\beta)$.

If $\beta<-1$ and $-1<\alpha <0$, the condition $f(\xfplus) \ge \xfplus$ now reads
\begin{equation}
-\dfrac{1+\alpha}{\alpha \sqrt{\beta^2 - 1}} \ge \pi + \arccos\left(\dfrac{1}{\beta} \right).
\end{equation}
Solving again for $\alpha$, this translates into the condition $\alpha
\ge \alpha_1(\beta)$ where
\begin{equation}
\alpha_1(\beta) \equiv \dfrac{-1}{\sqrt{\beta^2-1} \left[\pi +
    \arccos\left(\dfrac{1}{\beta} \right) \right] + 1}\,.
  \label{eq:saiii:alphaone}
\end{equation}
The corresponding necessary condition for SAIII3 to exist here is $-1< \alpha <
\alpha_1(\beta)$.

Finally, if $\beta<-1$ and $\alpha < -1$, the condition $f(\xfminus) \ge \xfminus$ reads
\begin{equation}
\dfrac{1+\alpha}{\alpha \sqrt{\beta^2 - 1}} \ge
\arccos\left(-\dfrac{1}{\beta} \right),
\end{equation}
or, $\alpha \le \alpha_3(\beta)$ where we have defined
\begin{equation}
\alpha_3(\beta) \equiv \dfrac{1}{\sqrt{\beta^2-1}
  \arccos\left(\dfrac{-1}{\beta}\right) - 1}\,.
\label{eq:saiii:alphathree}
\end{equation}
In this parameter space corner, for SAIII3 to exist one must ensure that
$\alpha_3(\beta) < \alpha <-1$.

In practice, one can check that the three curves $\alpha_1(\beta)$,
$\alpha_2(\beta)$ and $\alpha_3(\beta)$ provide very good
approximations to the exact solutions of $f(x_f)=x_f$ with $x_f$ given
by \Eq{eq:saiii:xf}, hence they can be used as proxies for the
separatrices in parameter space. They have been represented in
\Fig{fig:absaiii} as red curves, where the domains in which SAIII3 is
defined have been explicitly labeled.

Conversely, in all domains in which SAIII3 does not exist, the
potential has, at least, one positive maximum and this ensures that
the SAIII1 inflationary regime can occur for $x < \xVmax$, where
$\xVmax$ has to be numerically determined to solve $V'(x)=0$, see
\Eq{eq:saiii:Vprime}. The regime SAIII2, occurring for $x>
\xVmax$ at increasing field values, gracefully ends only if the
potential does not admit a de-Sitter minimum at a larger field
value. In other words, denoting by $\xVmin$ the value at which the
potential is minimum, with $\xVmin > \xVmax$, one should have
$V'(\xVmin)=0$ and $V(\xVmin) \le 0$, {\ie},
\begin{equation}
  1 - \cos\left(\xVmin\right) + \alpha\left[\xVmin \sin\left(\xVmin\right) + \dfrac{1}{2}
    \beta \xVmin^2 \right] \le 0.
\label{eq:saiii:Vneg}
\end{equation}
One cannot determine an analytical condition on $\alpha$ and $\beta$
such that this condition is fulfilled. However, in the limit $|\alpha|
\gg 1$, the condition $V(\xVzeroPlus)=0$ simplifies to
\begin{equation}
\dfrac{\sin(\xVzeroPlus)}{\xVzeroPlus} = - \dfrac{\beta}{2},
\end{equation}
and assessing if a solution exists boils down to comparing the amplitude
of the second and third extremum of the function $\sin(x)/x$ to
$|\beta/2|$. In the domain of interest, we find that a solution exists,
for $\alpha \to \infty$, only if $\beta\in]-0.257,0.434[$. In general,
\Eq{eq:saiii:Vneg} has to be solved numerically and the
 solutions have been represented as blue curves in
\Fig{fig:absaiii}.

Let us finally notice that, expanding the potential around $x=0$, one has
\begin{equation}
\dfrac{V(x)}{M^4} = \left(\alpha + \dfrac{1 + \alpha \beta}{2} \right)
x +\order{x^3}.
\end{equation}
The potential is therefore a negative decreasing function of $x$ around the
origin for $\alpha < -1/(\beta + 2)$. This is the same behavior as
discussed for SAII, which is recovered by taking $\beta=0$. When this
occurs, the inflationary domains are defined only for $x > \xVzeroMinus$,
where $\xVzeroMinus$ is the first positive solution of $V(x)=0$. The
separatrix $\alpha = -1/(\beta+2)$ has been represented as a green
dashed curve in \Fig{fig:absaiii}.

\subsubsection{Slow-Roll Analysis}

\begin{figure}
\begin{center}
\includegraphics[width=\wdblefig]{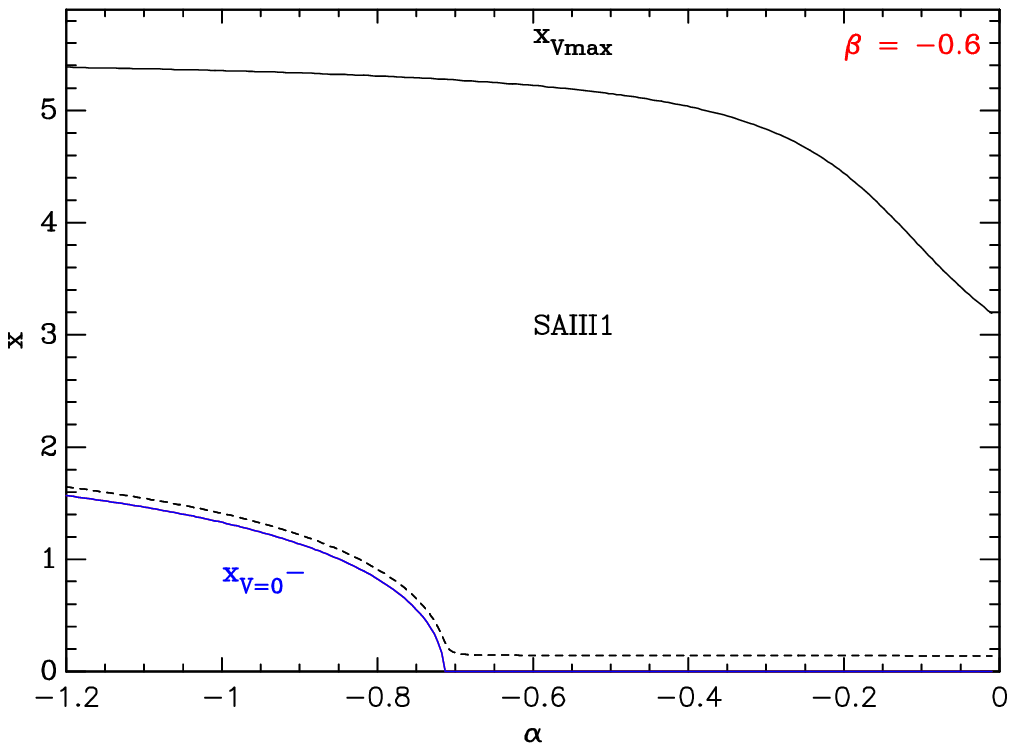}
\includegraphics[width=\wdblefig]{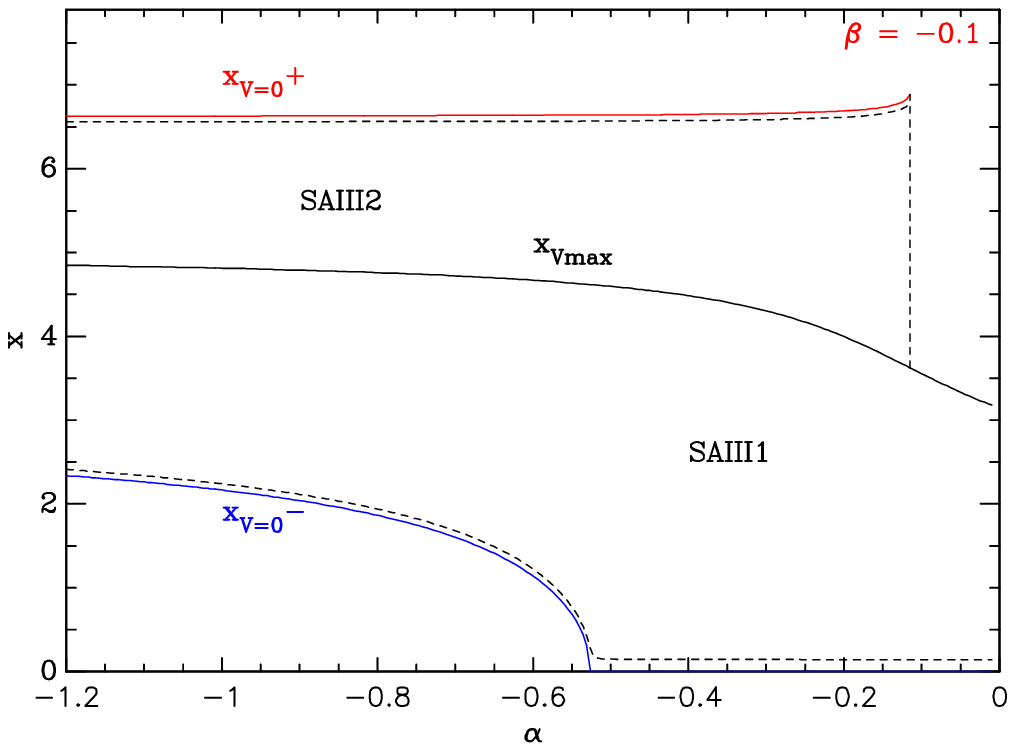}
\includegraphics[width=\wdblefig]{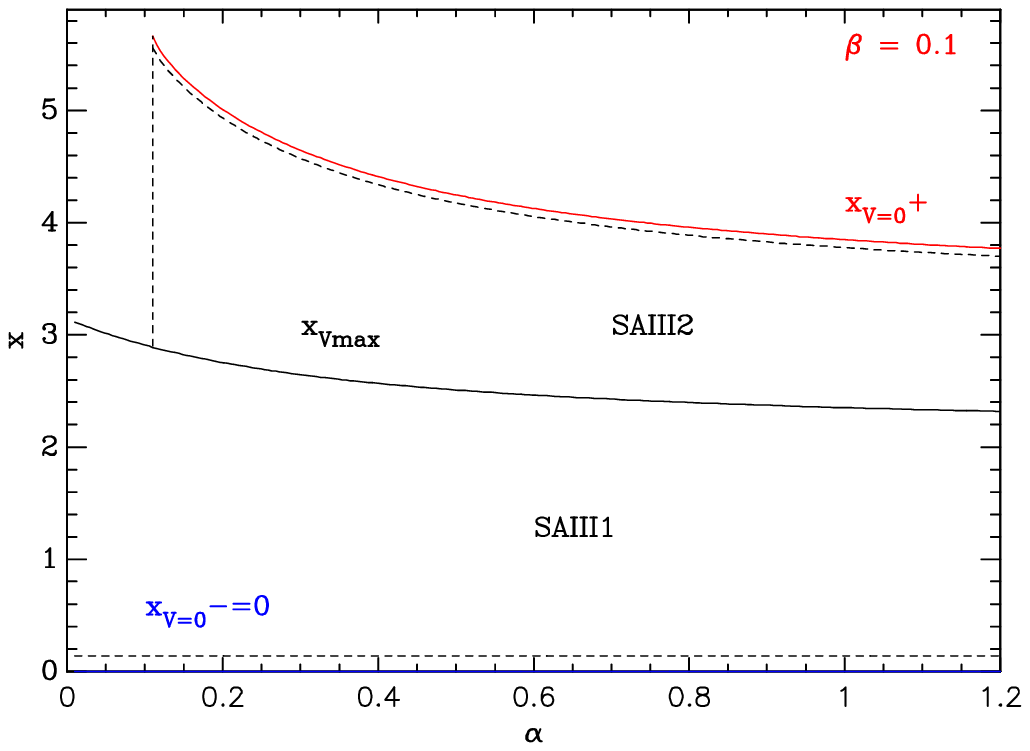}
\includegraphics[width=\wdblefig]{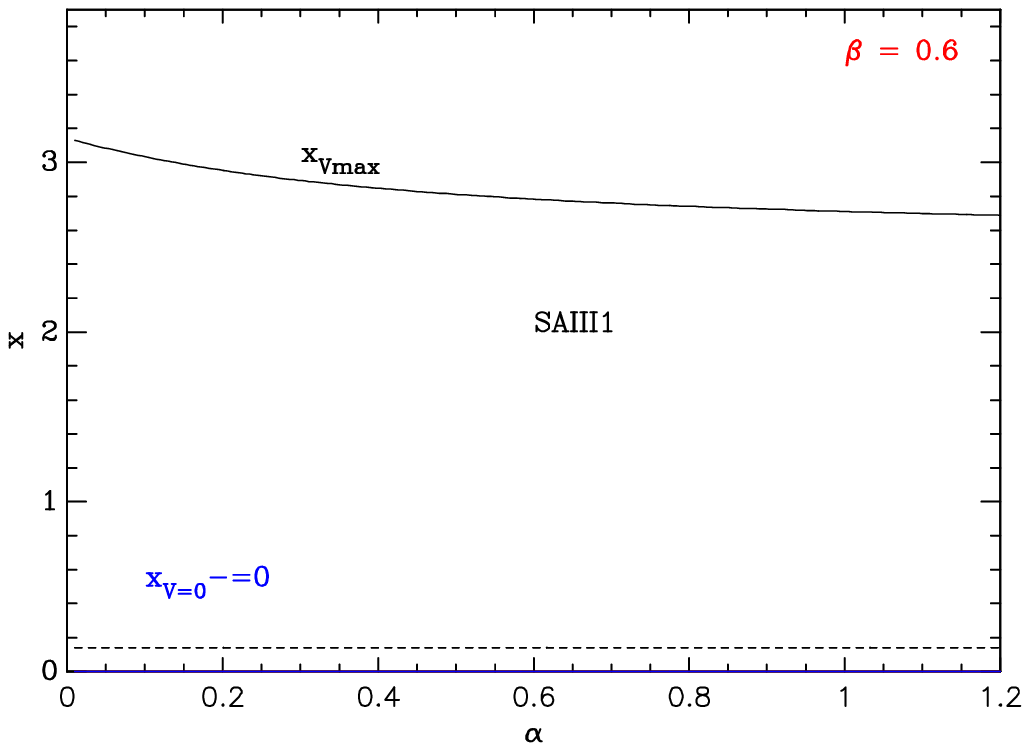}
\caption{Field domains over which SAIII1 and SAIII2 are defined, as a
  function of $\alpha$. Each panel corresponds to, clockwise,
  $\beta=-0.6$, $\beta=-0.1$, $\beta=0.6$ and $\beta=0.1$. The regime
  SAIII2 appears only in a limited domain of $\alpha$ for $\beta=\pm
  0.1$ and does not exist for $\beta=\pm 0.6$, see also
  \Fig{fig:absaiii}. When they appear, the blue and red curves
  labeled $\xVzeroPlus$ and $\xVzeroMinus$ determine the field range
  over which the potential is positive. The lower and upper dashed
  lines represent the field value (in unit of $\mu$, and for
  $\mu=10\,\Mp$) at which inflation gracefully ends, namely the
  relevant solution of $\epsilon_1(x)=1$.}
\label{fig:xvsaiii12}
\end{center}
\end{figure}

\begin{figure}
\begin{center}
\includegraphics[width=\wdblefig]{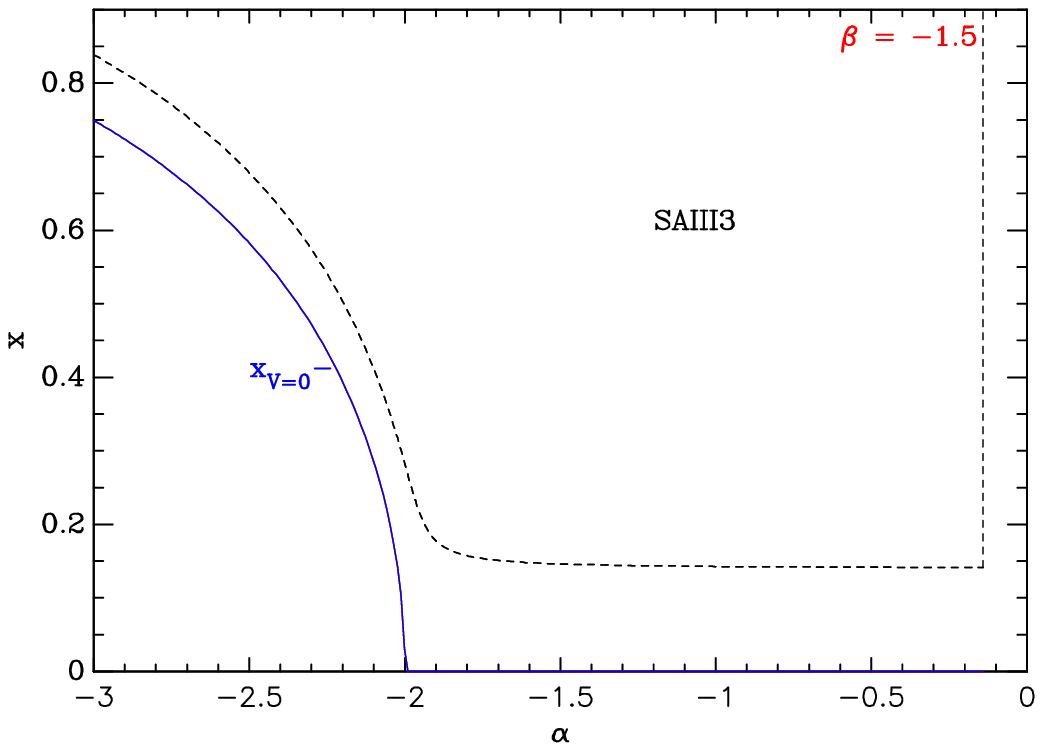}
\includegraphics[width=\wdblefig]{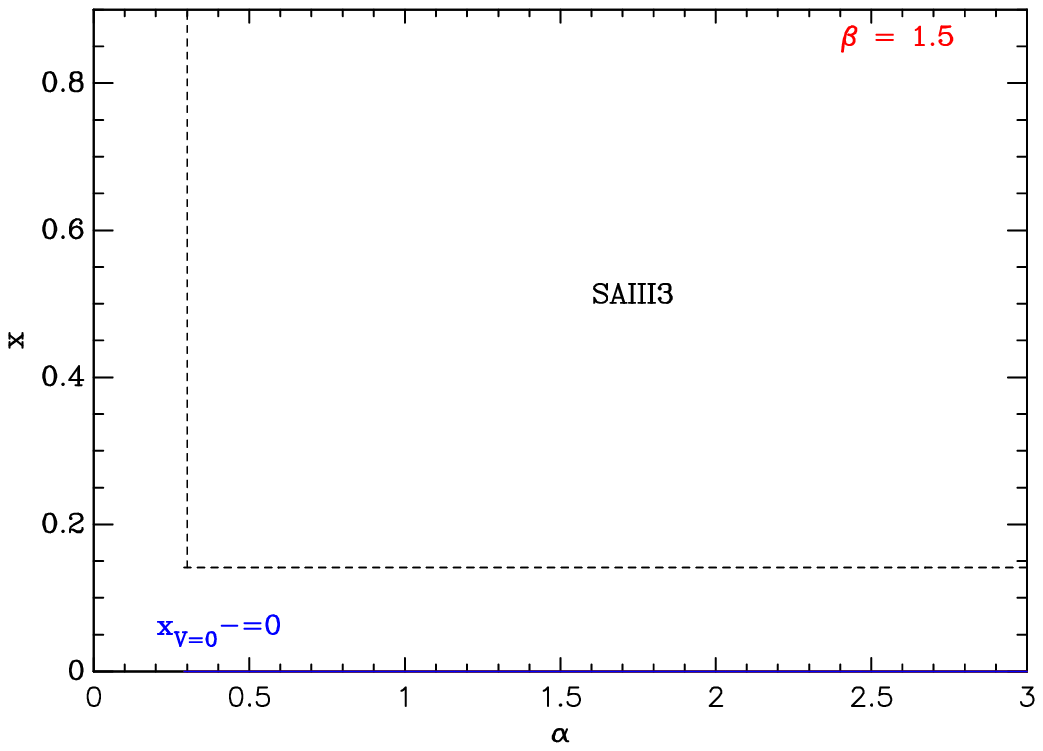}
\caption{Field domains over which SAIII3 is defined, as a function of
  $\alpha$. The left panel is for $\beta=-1.5$ while the right one is
  for $\beta=1.5$. When it is non-vanishing, the blue curve labeled
  $\xVzeroMinus$ gives the field value below which the potential is
  negative around the origin. The lower dashed line represents the
  field value (in unit of $\mu$, and for $\mu=10\,\Mp$) at which
  inflation gracefully ends, namely the smallest solution of
  $\epsilon_1(x)=1$. For $\beta=1.5$, $\xend(\alpha)$ is almost
  horizontal and weakly depends on $\alpha$ because the potential
  shape is quite close to the one of LFI2. In that case, to a very
  good approximation, one has $\xend \simeq \sqrt{2}/\mu$.}
\label{fig:xvsaiii3}
\end{center}
\end{figure}

The first slow-roll parameter reads
\begin{equation}
  \epsilon_1 = \dfrac{1}{2\mu^2} \left[ \dfrac{(1+\alpha) \sin(x) +
    \alpha x \cos(x) + \alpha \beta x}{1 - \cos(x) + \alpha x \sin(x)
      + \frac{1}{2} \alpha \beta x^2} \right]^2,
\label{eq:saiiieps1}
\end{equation}
while the second one is given by
\begin{equation}
  \begin{aligned}
    \epsilon_2  & = \dfrac{1}{\mu^2} \dfrac{1}{\left[1 - \cos(x) + \alpha x \sin (x) +
        \frac{1}{2} \alpha \beta x^2 \right]^2} \\ & \times
    \Big(2 + \alpha \left\{-2 \beta
    +\alpha \left[\left(\beta ^2+2\right) x^2+1\right]+4\right\} +
    \alpha x \sin(x) \left\{2 + \beta \left[\alpha
      \left(x^2+2\right)+4\right] \right\} \\ & + \cos(x) \left\{-2 -4 \alpha
    +\alpha \beta \left[ (2 \alpha -1) x^2+2\right]\right\}
    -\alpha ^2 -\cos(2x) \Big),
    \label{eq:saiiieps2}
  \end{aligned}
\end{equation}
and, finally, the third one is
\begingroup
\allowdisplaybreaks
\begin{align}
    \epsilon_3 & = -\dfrac{1}{2 \mu^2} \dfrac{(1+\alpha) \sin(x) +
      \alpha x \left[\beta +\cos(x)\right]}{\left[1 - \cos(x) + \alpha
        x \sin(x) + \frac{1}{2} \alpha \beta x^2 \right]^2} \nonumber \\ &
    \times \Big[ \left(\alpha x \cos(x) \left\{\beta \left[\alpha
        \left(x^2+6\right)+2\right] + 2\right\} + \sin(x) \left[6
        \alpha +(\alpha +1) \alpha \beta \left(x^2+2\right) + 2\right]
      \right. \\ & + \left. 2 \alpha ^2 \left[\left(\beta ^2+2\right)
        x + \sin(2 x)\right] \right)\left[\alpha \beta x^2+2 \alpha x
        \sin(x)-2 \cos(x)+2\right] \nonumber \\ & - 4 \left\{ \alpha x
      \left[\beta +\cos(x)\right] + (\alpha +1) \sin(x) \right\}
      \nonumber \\ &
      \times \left(\alpha \left\{-2 \beta +\alpha \left[\left(\beta
        ^2+2\right) x^2 +1 \right] +4 \right\} + \alpha x \sin(x)
      \left\{ \beta \left[\alpha \left(x^2+2\right) +4 \right] +
      2\right\} \right. \nonumber \\ &+ \left. \cos(x) \left\{-4 \alpha +\alpha
      \beta \left[ (2 \alpha -1) x^2 + 2 \right] -2 \right\} -
      \alpha^2 \cos(2 x) + 2\right) \Big] \nonumber \\ & \times \Big( \alpha
    \left\{-2 \beta +\alpha \left[\left(\beta ^2+2\right)x^2 +
      1\right] +4 \right\} + \alpha x \sin(x) \left\{\beta
    \left[\alpha \left(x^2+2\right) + 4\right] + 2\right\} \nonumber \\ & +
    \cos(x) \left\{-4 \alpha +\alpha \beta \left[ (2 \alpha -1) x^2 +
      2\right] - 2\right\} - \alpha ^2 \cos(2 x) +2 \Big)^{-1}.
    \label{eq:saiiieps3}
  \end{align}  
\endgroup
The three slow-roll parameters have been represented as a function of
$x$ for SAIII1 and SAIII2 in \Fig{fig:potsaiii12}, and for SAIII3
in \Fig{fig:potsaiii3}. In the latter regime, the modulation of
the potential implies that $\epsilon_2$ may change sign during
inflation (see lower right panel in \Fig{fig:potsaiii3}). As a
result, $\epsilon_3(x)$ becomes singular when $\epsilon_2(x)=0$ and we
are, a priori, in the presence of slow-roll violations at second order. In
fact, they are not really problematic as the product $\epsilon_2
\epsilon_3$, which is the quantity entering into the second-order
slow-roll power spectra, remains always finite, but they do signal
potentially large running of the spectral index.

For the three regimes inflation gracefully ends at the
field value $\xend$ solution of $\epsilon_1(\xend)=1$, in the domain of
interest. Because $\epsilon_1$ diverges for $V(x)\to 0$, this always
occurs in a domain in which the potential is positive definite, and we
have $\xend \in]\xVzeroMinus,\xVmax[$ for SAIII1,
$\xend\in]\xVmax,\xVzeroPlus[$ for SAIII2 and
$\xend\in]\xVzeroMinus,+\infty[$ for SAIII3 (if $\xVzeroMinus$
does not exist, it has to be replaced by $x=0$). The actual value for
$\xend$ has to be determined numerically, in the previous domains, by
solving
\begin{equation}
(1+\alpha) \sin(x) + \alpha x \cos(x) + \alpha \beta x = \pm \mu
  \sqrt{2} \left[1 - \cos(x) + \alpha x \sin(x) + \dfrac{1}{2} \alpha
    \beta x^2 \right].
\label{eq:saiiixend}
\end{equation}

In \Fig{fig:xvsaiii12}, we have represented, for various values
of $\beta$, the values of $\xVzeroMinus$, $\xVzeroPlus$ and
$\xend$ as functions of $\alpha$ for the regimes SAIII1 and
SAIII2. For $\beta=0$ one recovers the results of SAII and this is
displayed in \Fig{fig:xvsaii}. For SAIII3, the potential does
not have any maximum, but for $\beta<0$, there are some values of
$\alpha$ for which $\xVzeroMinus$ exists and this is represented in
\Fig{fig:xvsaiii3}. As soon as $\beta$ becomes large, the
potential of SAIII3 is essentially dominated by the mass term and a
very good approximation is $\xend(\alpha) \simeq \sqrt{2}/\mu$, the
field value at which Large Field Inflation with $p=2$ ends. This can
be seen on the right panel of \Fig{fig:xvsaiii3} where the dashed
curve representing the function $\xend(\alpha)$ is essentially a
horizontal line. A word of caution is however in order. There are some
values of $\alpha$ and $\beta$ within SAIII3 for which $\epsilon_1(x)$
may transiently exceed unity with $x> \xend$, precisely due to the
modulation around the LFI2 potential. In these situations, inflation
ends with ``hiccups'', it stops and restarts within a few e-folds
before definitely stopping at the value of $\xend$ we have
calculated. Therefore, slow-roll violations can occur, but being at
the end of inflation, they are unlikely to be directly observable. They could nonetheless
have other interesting effects, \eg for primordial black holes
formation.

The slow-roll trajectory cannot be analytically solved and requires a
numerical integration. It reads
\begin{equation}
\Nend - N = \mu^2 \int_{\xend}^{x} \dfrac{1-\cos(x) + \alpha x \sin(x)
+\frac{1}{2} \alpha \beta x^2}{(1+\alpha)\sin(x) + \alpha x
  \left[\cos(x) + \beta \right]} \, \ud x,
\end{equation}
where $\xend$ is obtained from the solution of \Eq{eq:saiiixend}
in the relevant field domain for the inflationary regime under
scrutiny (SAIII1, SAIII2 or SAIII3).

The normalization of the potential $M^4$ is obtained from the
amplitude of the CMB anisotropies once the field value $\xstar$ at
which the pivot mode crossed the Hubble radius is determined. One gets
\begin{equation}
\left(\dfrac{M}{\Mp}\right)^4 = \dfrac{720 \pi^2}{\mu^2}
\dfrac{\left[(1+\alpha)\sin(\xstar) + \alpha \xstar \cos(\xstar) +
    \alpha \beta \xstar\right]^2}{\left[1 - \cos(\xstar) + \alpha
    \xstar \sin(\xstar) + \frac{1}{2} \alpha \beta \xstar^2\right]^3}\frac{\Qrms^2}{T^2}\,.
\end{equation}
The reheating consistent slow-roll predictions for SAIII1, SAIII2 and
SAIII3 are represented in \Figs{fig:CMBSAIII1m} to
\ref{fig:CMBSAIII3p_5}. Because the parameter space in
$(\alpha,\beta)$ is disjoint between positive and negative values, the
models have been split into two sub-regimes according to the sign of
$\alpha \beta$. Notice the strong running of the predictions
associated with SAIII3, for some values of $\alpha$ and $\beta$, they
essentially explore the whole plane $(\nS,r)$ while varying $\mu$.

\subsection{Radiatively Corrected Large Field Infation (RCLFI)}
\label{sec:rclfi}

\begin{figure}
\begin{center}
\includegraphics[width=\wdblefig]{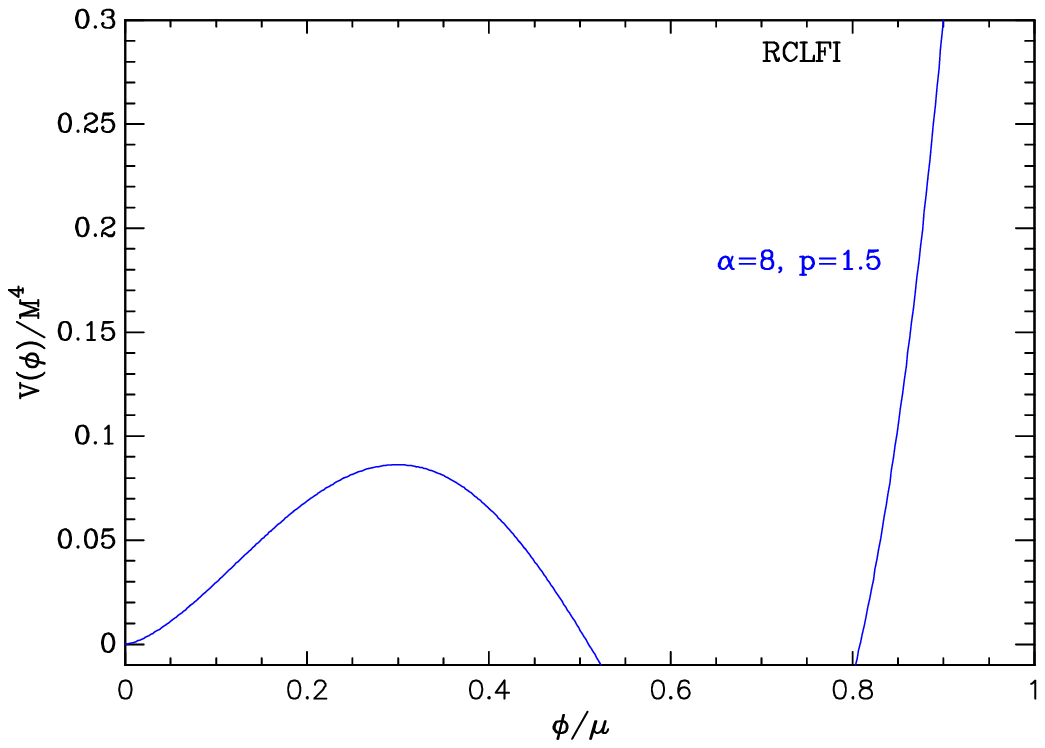}
\includegraphics[width=\wdblefig]{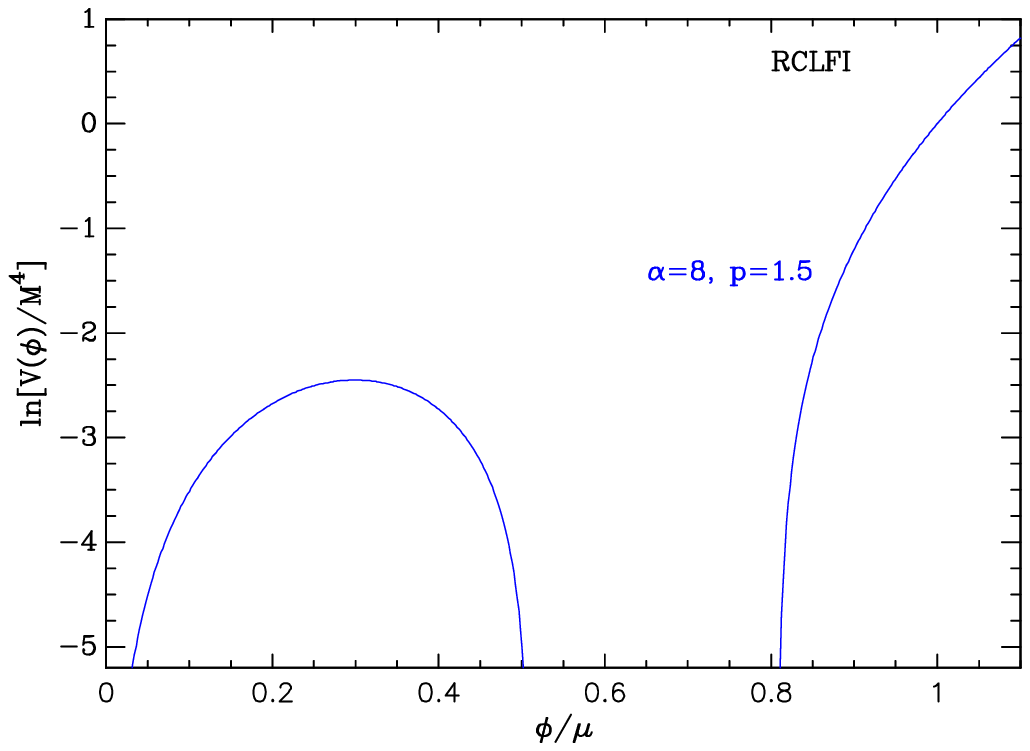}
\includegraphics[width=\wdblefig]{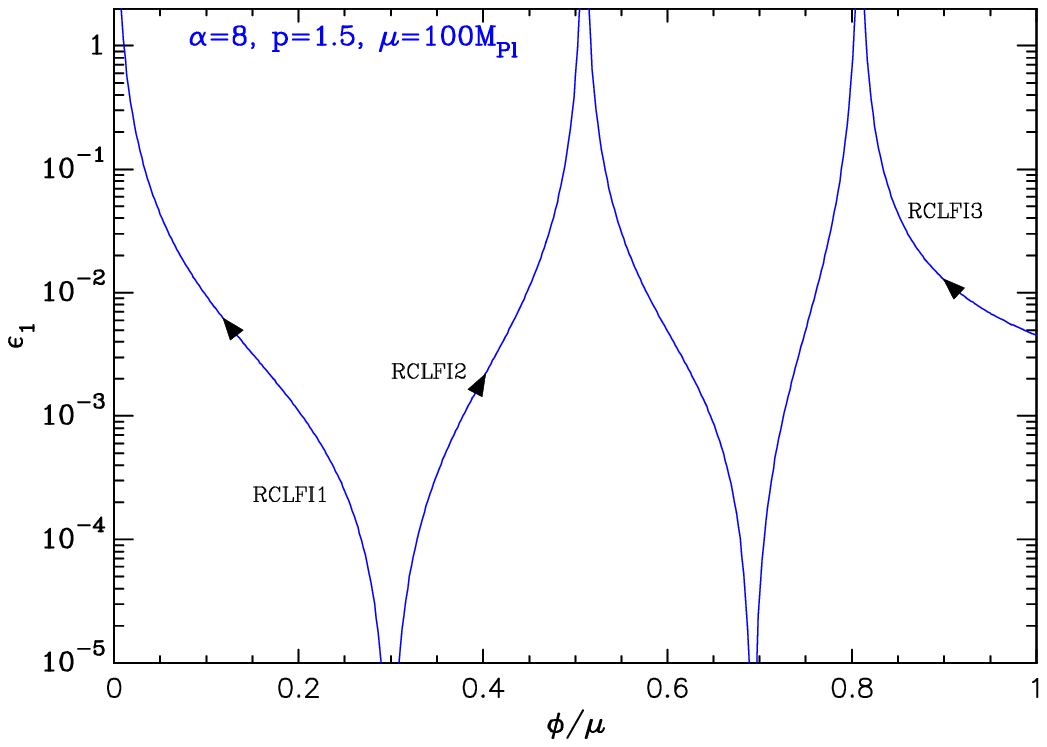}
\includegraphics[width=\wdblefig]{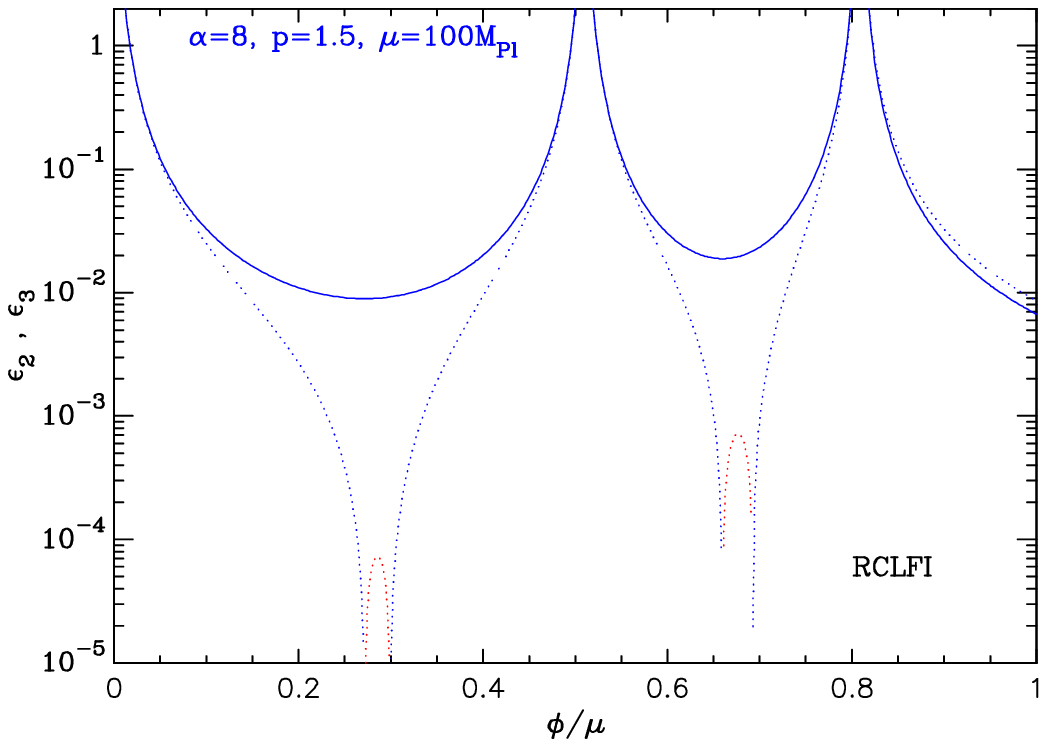}
\caption{Radiatively Corrected Large Field Inflation in the RCLFI1,
  RCLFI2 and RCLFI3 regimes for $\alpha=8$, $p=2.5$. Top panels: the
  potential and its logarithm. Bottom left panel: slow-roll parameter
  $\epsilon_1$ for $\mu = 100\,\Mp$, with the three inflationary
  regimes annotated with an arrow indicating the direction to which
  the field evolves. Bottom right panel: slow-roll parameters
  $\epsilon_2$ (solid line) and $\epsilon_3$ (dotted line) for the
  same parameters value. When $\epsilon_3$ becomes negative, the plot
  shows $|\epsilon_3|$ as a red dotted line, the black dotted line
  corresponds to positive values. The perturbative regime RCLFI4 is
  represented in \Fig{fig:potrclfi4}.}
\label{fig:potrclfi123}
\end{center}
\end{figure}

\begin{figure}
\begin{center}
\includegraphics[width=\wdblefig]{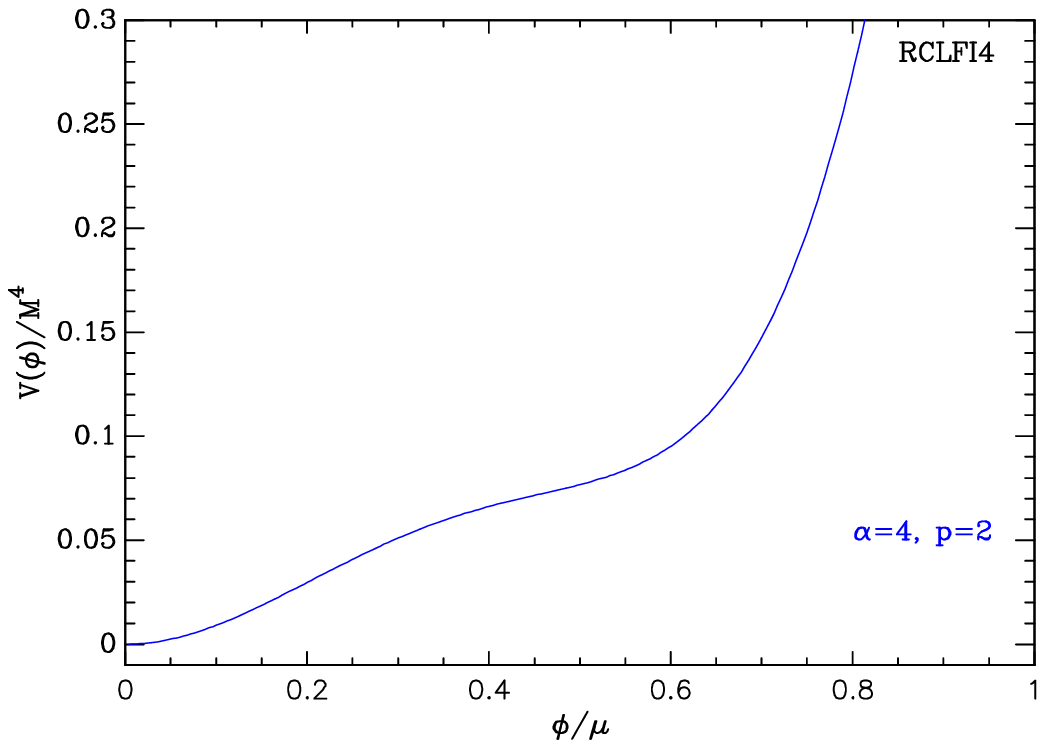}
\includegraphics[width=\wdblefig]{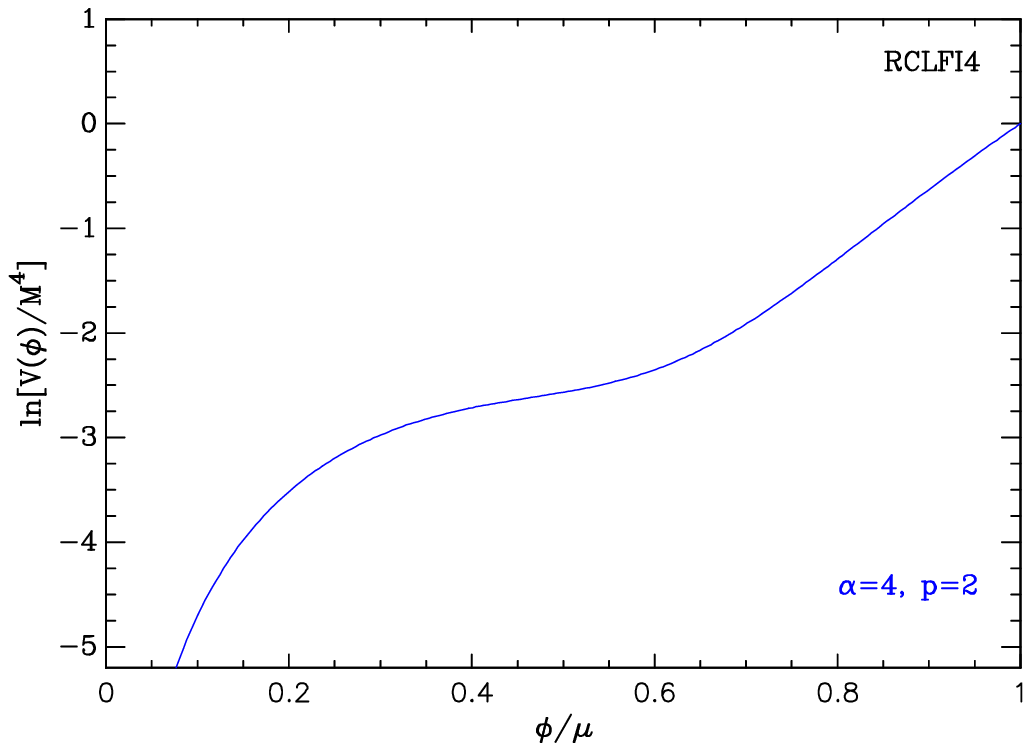}
\includegraphics[width=\wdblefig]{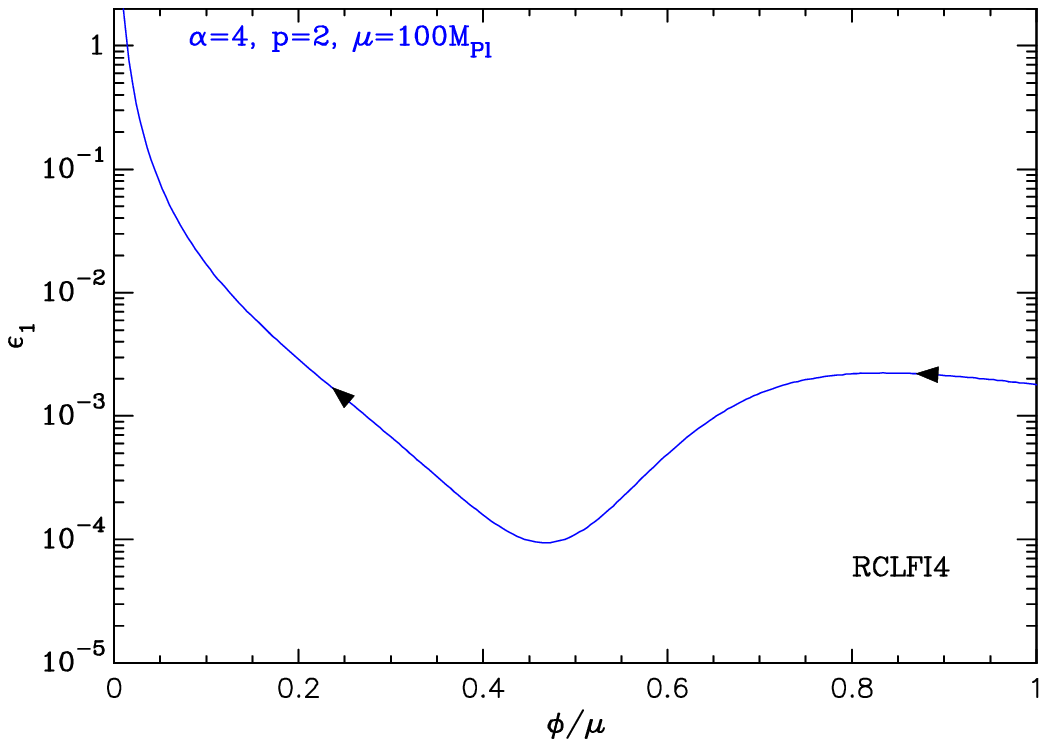}
\includegraphics[width=\wdblefig]{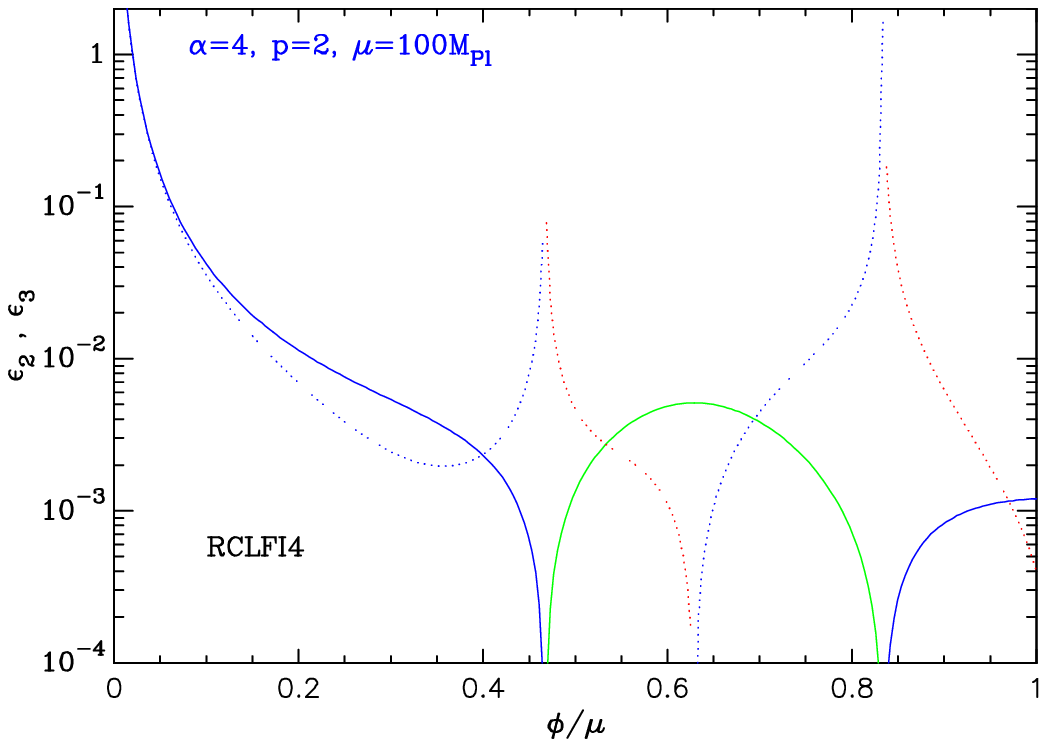}
\caption{Radiatively Corrected Large Field Inflation in the
  perturbative regime RCLFI4, where the potential does not exhibit
  extrema, for $\alpha=4$, $p=2$. Top panels: the potential and its
  logarithm. Bottom left panel: slow-roll parameter $\epsilon_1$ for
  $\mu = 100\,\Mp$, with the field evolution annotated with an arrow
  indicating the direction to which it evolves. Bottom right panel:
  slow-roll parameters $\epsilon_2$ (solid line) and $\epsilon_3$
  (dotted line) for the same parameters value. When $\epsilon_3$
  becomes negative, the plot shows $|\epsilon_3|$ as a red dotted
  line, the black dotted line corresponds to positive
  values. Similarly, negative values of $\epsilon_2$ are represented
  as blue solid curves. Notice that $\epsilon_3$ becomes singular at
  the points where $\epsilon_2=0$, but the product $\epsilon_2
  \epsilon_3$ remains finite. The other regimes RCLFI1 to RCLFI3 are
  represented in \Fig{fig:potrclfi123}.}
\label{fig:potrclfi4}
\end{center}
\end{figure}

This model is based on \Refc{Enqvist:2013eua} and considers the
radiative corrections of bosonic and fermionic fields onto a monomial
potential, {\ie}, a large-field model (see
\sectionc{sec:lfi}). Compared to the RCMI and RCQI models already
discussed in \sectioncs{sec:rcmi} and \ref{sec:rcqi} respectively, both
proposed in \Refc{NeferSenoguz:2008nn}, RCLFI is not restricted to the
quadratic and quartic monomial terms and the renormalization scale
$\mu$ is not fixed at the Planck mass but becomes a free
parameter. Bosonic and fermionic loop corrections yield a potential of
the Coleman-Weinberg type, see \sectionc{sec:cwi}, which
reads~\cite{Enqvist:2013eua}
\begin{equation}
V(\phi) = \dfrac{1}{2} \lambda \mpl^4 \left(\dfrac{\phi}{\mpl}\right)^p
+ \dfrac{g^4 - 4 h^4}{64 \pi^2} \phi^4 \ln \left(\dfrac{\phi^2}{\mu^2}\right).
\end{equation}
Here $g$ is the renormalized coupling constant coming from a bosonic
interaction term of the form $\phi^2 \chi^2$, and $h$ is the one coming
from a Yukawa coupling between $\phi$ and a fermionic field. Depending
on the relative strength of these couplings, the coefficient in front
of the logarithmic term can be positive or negative.

For our purpose, it is more convenient to recast the potential in a
simpler form
\begin{equation}
V(\phi) = M^4 \left[ \left(\dfrac{\phi}{\mu}\right)^p + \alpha
  \left(\dfrac{\phi}{\mu}\right)^4 \ln \left(\dfrac{\phi}{\mu}\right) \right],
\label{eq:potrclfi}
\end{equation}
where $\lambda = (M/\mpl)^4 (\mpl/\mu)^p$ and $g^4-4h^4 = 32 \pi^2
(M/\mu)^4 \alpha$.

Let us notice that the potential is renormalizable only for $p \le 4$,
and, if one wants to keep the loop corrections under control, the
coupling constants should be small, namely the combination $\alpha M^4
/\mu^4$ should not exceed unity. In the following, these
considerations are relaxed and we will also consider the predictions
coming from \Eq{eq:potrclfi} in full generality. Another point worth
mentioning concerns the peculiar value $p=4$. In that case, as one can
check in \Eq{eq:potrclfi}, the renormalization scale $\mu$ can be re-absorbed in the normalization of the potential $M^4$ and, up to a
redefinition of the parameter $\alpha$, the model ends up being
equivalent to RCQI, see \sectionc{sec:rcqi}.

The potential will be studied for $\phi>0$ only, and it is displayed in the top panels of \Figs{fig:potrclfi123} and~\ref{fig:potrclfi4} for $\alpha=8$ and $p=3/2$, and $\alpha=4$ and $p=2$, respectively. Depending on the value of $\alpha$, one can see that the potential can be negative in some
intermediate field domains separated by a maximum (see \Fig{fig:potrclfi123}). The appearance of a
local maximum implies that, in addition to the large-field
inflationary regime, inflation can also occurs close to the top of the
new local maximum, on both sides, and we will be referring to these
regimes as RCLFI1 and RCLFI2. The large-field regime will be referred
to as RCLFI3. Finally, there are parameter values for which the local
maximum disappears and all these three regimes become unified in what
will be referred to as RCLFI4 (see \Fig{fig:potrclfi4}). This one implicitly requires the
logarithmic term to be small everywhere and is prototypical of what
one should expect from perturbative loop corrections.

\subsubsection{Parameter Space Analysis}

\begin{figure}
\begin{center}
\includegraphics[width=\wsingfig]{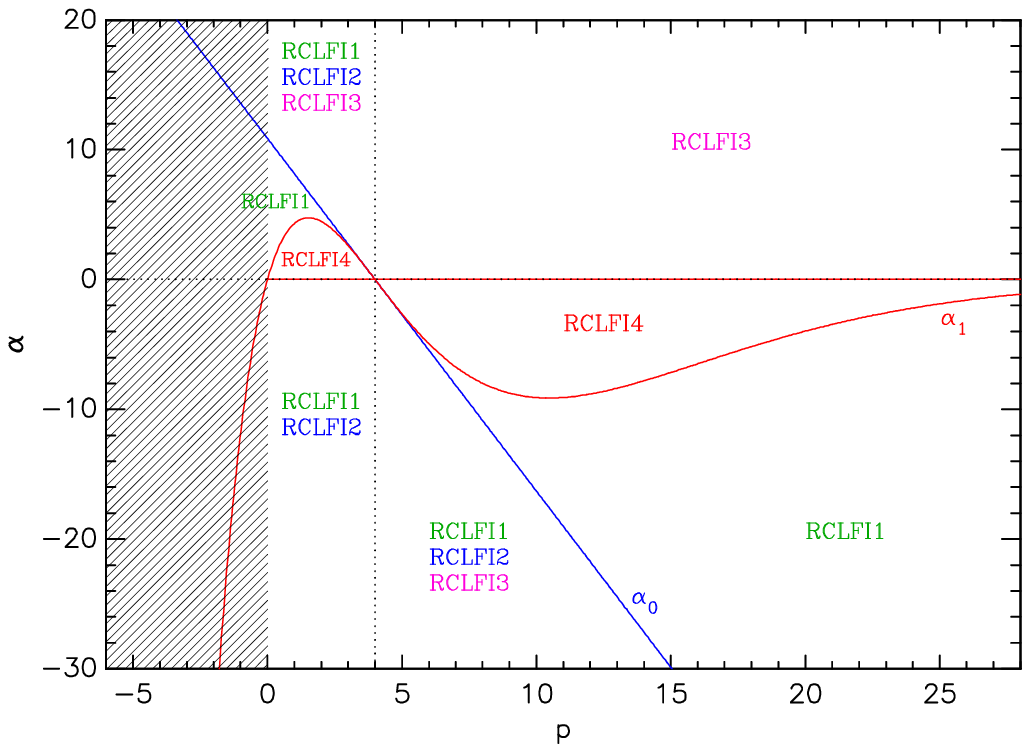}
\caption{The parameter space $(p,\alpha)$ of Radiatively Corrected
  Large Field Inflation and its various inflationary regimes. The
  perturbative regime, RCLFI4, exists only in the domain between the
  horizontal axis and $\alpha_1(p)$ represented by the red
  curve. RCLFI3 can occur if a large-field domain exists. Finally,
  RCLFI1 and RCLFI2 are hilltop-like models and can occur only if the
  potential has a local maximum. Moreover, for RCLFI2, inflation may
  never ends if the local minimum towards which the field evolves is
  positive and this case is excluded. The corresponding regions are
  represented with the green and blue labels, see also the main text.}
\label{fig:parclfi}
\end{center}
\end{figure}

Let us first discuss the existence of the four afore-mentioned
inflationary regimes with respect to the model parameters. As already
discussed, the case $p=4$ corresponds to RCQI and will thus not be considered.

Close to the origin, the potential behaves as $V(x) \propto x^p$ for
$p<4$ or $V(x) \propto \alpha x^4 \ln x$ for $p>4$, where we have introduced
\begin{equation}
x=\frac{\phi}{\mu}\, .
\end{equation} 
Therefore, it is
positive and increasing with $x$ for $p<4$, and also for $p>4$ and
$\alpha < 0$. For $p>4$ and $\alpha > 0$, it is negative and
decreasing with $x$ in a neighborhood of
$x=0$. In the regime $x\gg 1$, we have $V(x) \propto x^p$ for $p>4$ and $V(x)
\propto \alpha x^4 \ln x$ for $p<4$. The potential is thus positive
and growing in all situations but one: $p<4$ and $\alpha<0$ where it
becomes unbounded from below (RCLFI3 does not exist in this case).

The (non-vanishing) field values at which the potential vanishes are
solution of
\begin{equation}
1 + \alpha x^{4-p} \ln x = 0\, .
\end{equation}
The above equation has solutions for $(p-4)/\alpha > -1/e$, given in terms of the Lambert function $W$ by
\begin{equation}
\xVzeroPM = \left[\dfrac{\alpha}{p-4}
  \Lambert{}\left(\dfrac{p-4}{\alpha}\right) \right]^{1/(p-4)}.
\label{eq:rclfixvzero}
\end{equation}
At a fixed value of $p$, we can thus define the limiting values of
$\alpha$ for which these solutions may exist as
\begin{equation}
\alphazero \equiv - e(p-4).
\label{eq:rclfialpha0}
\end{equation}

There is only one solution given by the principal branch $\Lambert{0}$
for $(p-4)/\alpha >0$, and it will be referred to as
$\xVzeroPlus$. For $-1/e < (p-4)/\alpha < 0$, there are two
solutions, one still given by the principal branch $\Lambert{0}$, and
a new one given by the branch $\Lambert{-1}$ that will be referred to
as $\xVzeroMinus$. In view of the asymptotic behavior of the
potential, the cases where there is only one zero corresponds to the
potential transitioning from negative values close to the origin to
asymptotically positive growth, and conversely. The cases where two
zeros appear correspond to a potential with a local maximum and a
negative local minimum.

There is also the possibility that the local minimum is positive and in order
to discuss this situation one should determine the field values for
which $V'(x)=0$, {\ie},
\begin{equation}
p x^{p-4} + \alpha(1 + 4 \ln x) = 0.
\end{equation}
The solutions are again given in terms of the Lambert function, they
exist only for
\begin{equation}
\dfrac{p(p-4)}{4\alpha} e^{1-p/4} \ge -\dfrac{1}{e}\,.
\label{eq:rclfidVnull}
\end{equation}
This leads us to define another boundary function $\alpha_1(p)$ by
\begin{equation}
\alpha_1 \equiv -\dfrac{p(p-4)}{4} e^{2-p/4}.
\label{eq:rclfialpha1}
\end{equation}
When the condition \Eq{eq:rclfidVnull} is satisfied, the solutions are
given by
\begin{equation}
\xdVzeroPM = \left\{ \dfrac{4 \alpha}{p(p-4)}
\Lambert{}\left[\dfrac{p(p-4)}{4\alpha} e^{1-p/4}\right] \right\}^{1/(p-4)}.
\end{equation}
There is only one solution, $\xdVzeroPlus$, given by the principal
branch $\Lambert{0}$, for $[p(p-4)/(4\alpha)] e^{1-p/4} > 0$. Another
solution appears for $-1/e < [p(p-4)/(4\alpha)] e^{1-p/4} < 0$ that
will be referred to as $\xdVzeroMinus$ given by the branch
$\Lambert{-1}$. As for the zeros of the potential, when the solution
is unique, it represents a local maximum, or minimum, in a
transitioning regime between a negative potential close to the origin
to a positive one asymptotically, or the converse. When there are two
solutions, we have both a local maximum and a local minimum.

From these considerations we can determine the parameter space in
which RCLFI1, RCLFI2, RCLFI3 and RCLFI4 exist. RCLFI1 and RCLFI2 are
hilltop-like models and inflation takes place close to the local
maximum of the potential, at decreasing field values for RCLFI1 and at
increasing field values for RCLFI2. Moreover, in the case of RCLFI2,
the field evolves towards the local minimum of the potential and this
one can actually be positive: this would prevent inflation to
end. Therefore, we add the requirement that this local minimum is
negative, or null, to ensure a graceful exit of RCLFI2. Combining the above considerations, we
find that the RCLFI1 regime can take place for $p>4$ and $\alpha <
\alpha_1(p)$, or, $p<4$ and $\alpha < 0$, or, $p<4$ and $\alpha >
\alpha_1(p)$. For RCLFI2, the conditions are identical but one should
replace $\alpha_1$ by $\alphazero$ to ensure graceful exit. RCLFI3
describes inflation in the large-field domain and requires the
potential to grow positive at large-field values. It exists for $p>4$
and $\alpha>0$, or , $p>4$ and $\alpha < \alphazero(p)$, or, $p<4$ and
$\alpha > \alphazero(p)$. Finally, the perturbative regime demands that
the potential has no extrema and this corresponds to $p>4$ and
$\alpha_1(p)<\alpha < 0$, or, $p<4$ and $0<\alpha<\alpha_1(p)$. These
domains have been represented in \Fig{fig:parclfi}.

\subsubsection{Slow-Roll Analysis}

The slow-roll parameters associated with RCLFI read
\begin{equation}
\epsilon_1 = \dfrac{\Mp^2}{2 x^2 \mu^2} \left[\dfrac{p x^p + \alpha
    x^4\left(1 + 4\ln x\right)}{x^p + \alpha x^4 \ln x}\right]^2,
\label{eq:rclfieps1}
\end{equation}
together with
\begin{equation}
  \begin{aligned}
    \epsilon_2 & = \dfrac{2 \Mp^2}{x^2 \mu^2} \dfrac{- \alpha  x^{p+4}
      \left\{\left[p(p-9) + 12\right] \ln x  - 2
      p+7 \right\} +  p x^{2 p} + \alpha ^2 x^8 \left(4 \ln^2x + \ln x +
      1\right)}{\left(x^p+\alpha  x^4 \ln x \right)^2}\,,
  \end{aligned}
\end{equation}
and
\begin{equation}
  \begin{aligned}
    \epsilon_3 &=\dfrac{\Mp^2}{x^2 \mu^2 \left(x^p+\alpha  x^4 \ln
      x \right)^2} \\ & \times  \dfrac{p x^p+\alpha  x^4+4 \alpha  x^4 \ln
      x}{\alpha  x^4 \ln x
      \left[\alpha  x^4-\left(p^2-9 p+12\right)
        x^p\right]-7 \alpha  x^{p+4}+p x^p \left(x^p+2
      \alpha  x^4\right)+\alpha ^2 x^8+4 \alpha ^2 x^8
      \ln ^2 x} \\
    & \times \Big\{3 \alpha  p^2 x^{2 p+4}+\alpha 
    x^4 \ln x \left[-\alpha  \left(3 p^2-30
      p+68\right) x^{p+4}-\left(p^3-9 p^2+20 p-24\right)
      x^{2 p}+3 \alpha ^2 x^8\right] \\& + \alpha ^2 x^8 \ln^2 x \left[ \left(p^3-15 p^2+74 p-96\right) x^p+2
      \alpha  x^4\right]+2 p x^p \left(-9 \alpha 
    x^{p+4}+x^{2 p}+3 \alpha ^2 x^8\right) \\ & + \alpha  x^4
    \left(-21 \alpha  x^{p+4}+26 x^{2 p}+2 \alpha ^2
    x^8\right)+8 \alpha ^3 x^{12} \ln
    ^3(x)\Big\}.
  \end{aligned}
\end{equation}

The three slow-roll parameters have been plotted as a function of $x$
for RCLFI1, RCLFI2 and RCLFI3 in \Fig{fig:potrclfi123}, and for
RCLFI4 in \Fig{fig:potrclfi4}. For the perturbative regime, even
though the potential does not exhibit extrema, the logarithmic
correction modulates the shape of the potential and this implies that
$\epsilon_2$ may change sign during inflation (see lower right panel
in \Fig{fig:potrclfi4}). As a result, $\epsilon_3(x)$ becomes
singular when $\epsilon_2(x)=0$ and we are, a priori, in the presence of
slow-roll violations at second order. In fact, they are not really
problematic as the product $\epsilon_2 \epsilon_3$, which is the
quantity entering into the second-order slow-roll power spectra,
remains always finite, but they do signal potentially large running of
the spectral index.

For the four regimes, RCLFI inflation gracefully ends at the field
value $\xend$ solution of $\epsilon_1(\xend)=1$, in the domain of
interest. Because $\epsilon_1$ diverges for $V(x)\to 0$, this always
occur in a domain in which the potential is positive definite and
ensures that inflation remains confined in these domains. There is no
analytical solution of the equation $\epsilon_1(x)=1$ and $\xend$
has to be determined numerically by solving
\begin{equation}
p x^p + \alpha x^4 \left(1+ 4 \ln x \right) = \pm \sqrt{2} x \mu
\left(x^p + \alpha x^4 \ln x \right).
\label{eq:rclfixend}
\end{equation}

The slow-roll trajectory cannot be analytically solved and also
requires a numerical integration. It reads
\begin{equation}
\Nend - N = \dfrac{\mu^2}{\Mp^2} \int_{\xend}^{x} \dfrac{y^{p+1} + \alpha y^5 \ln
  y}{p y^p + \alpha y^4\left(1 + 4 \ln y\right)} \, \ud y,
\end{equation}
where $\xend$ is obtained from the solution of \Eq{eq:rclfixend}
in the relevant field domain for the inflationary regime under
scrutiny (RCLFI1, RCLFI2, RCLFI3 or RCLFI4).

The normalization of the potential $M^4$ is obtained from the
amplitude of the CMB anisotropies once the field value $\xstar$ at
which the pivot mode crossed the Hubble radius is determined. This one
obtained from the trajectory, the value of $\xend$, and the reheating
equation~\eqref{eq:phistarlnrrad}. One gets
\begin{equation}
\left(\dfrac{M}{\Mp}\right)^4 = \dfrac{720 \pi^2 \Mp^4}{\mu^4} \dfrac{\left[p \xstar^p
  + \alpha \xstar^4\left(1 + 4 \ln\xstar\right)\right]^2}{\xstar^2
  \left(\xstar^p + \alpha \xstar^4 \ln \xstar \right)^3} \frac{\Qrms^2}{T^2}
\,.
\end{equation}
The reheating consistent slow-roll predictions for the four RCLFI
regimes are represented in Figs.~\ref{fig:CMBRCLFI1pm_0} to
\ref{fig:CMBRCLFI4m_3}. Because the parameter space in $(p,\alpha)$ is
disjoint, the models have been splitted in two sub-regimes according
to the sign of $p-4$ and/or the sign of $\alpha$.

\subsection{Non-Renormalizable Corrected Loop Inflation (NCLI)}

\label{sec:ncli}

\begin{figure}
\begin{center}
\includegraphics[width=\wdblefig]{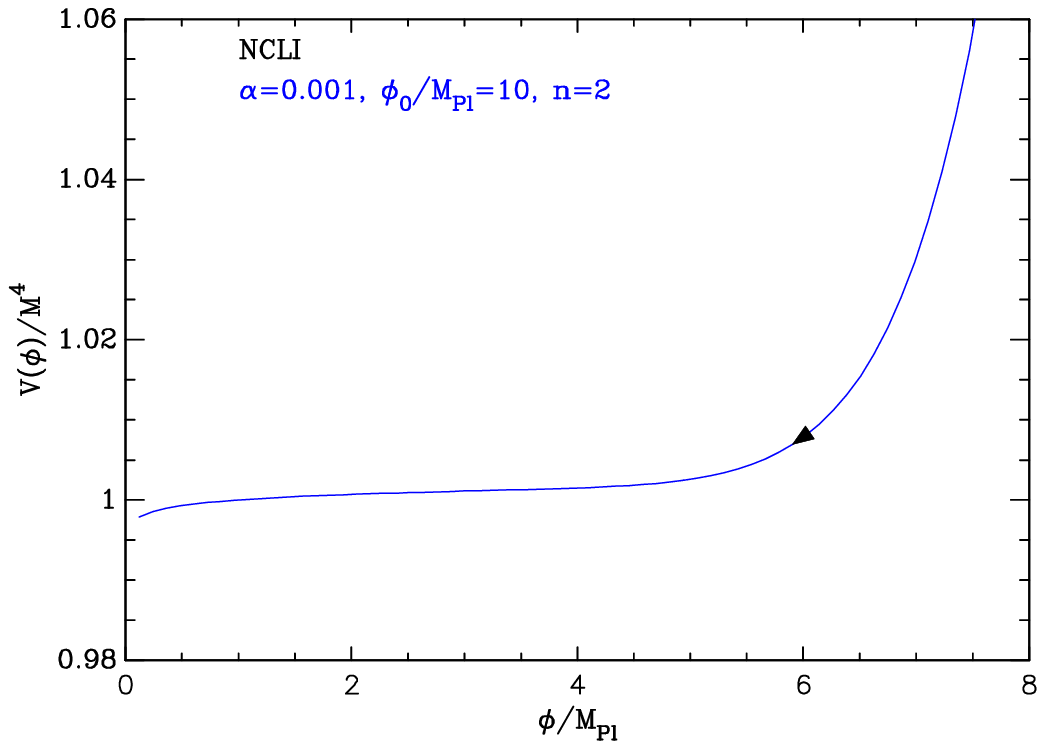}
\includegraphics[width=\wdblefig]{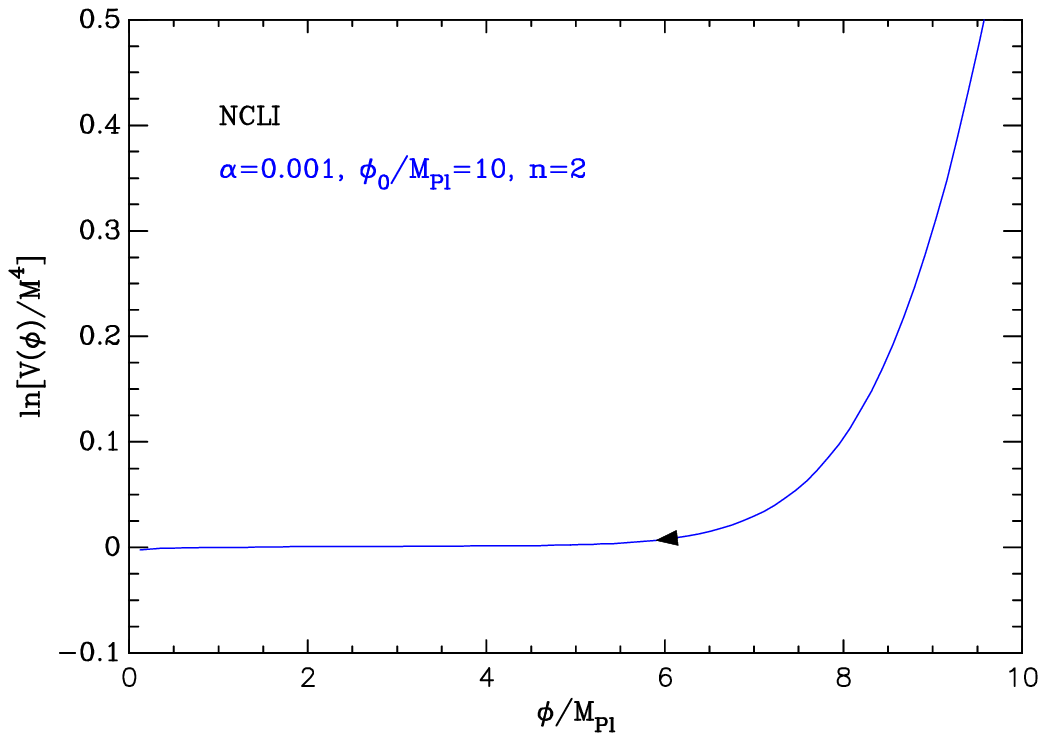}
\includegraphics[width=\wdblefig]{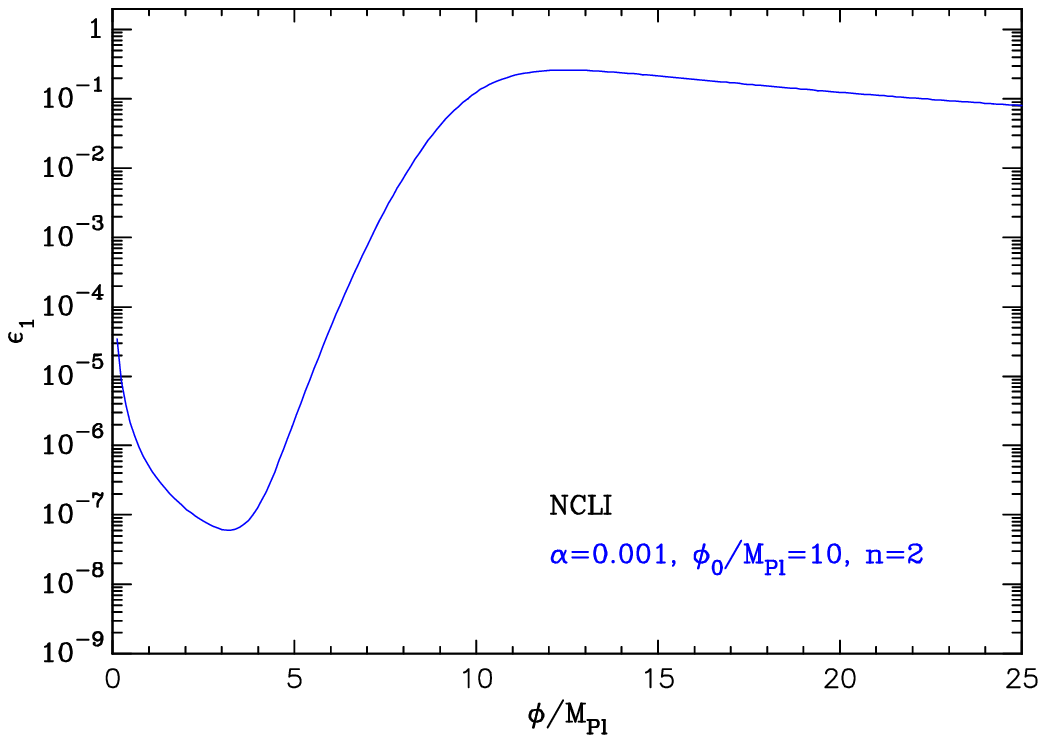}
\includegraphics[width=\wdblefig]{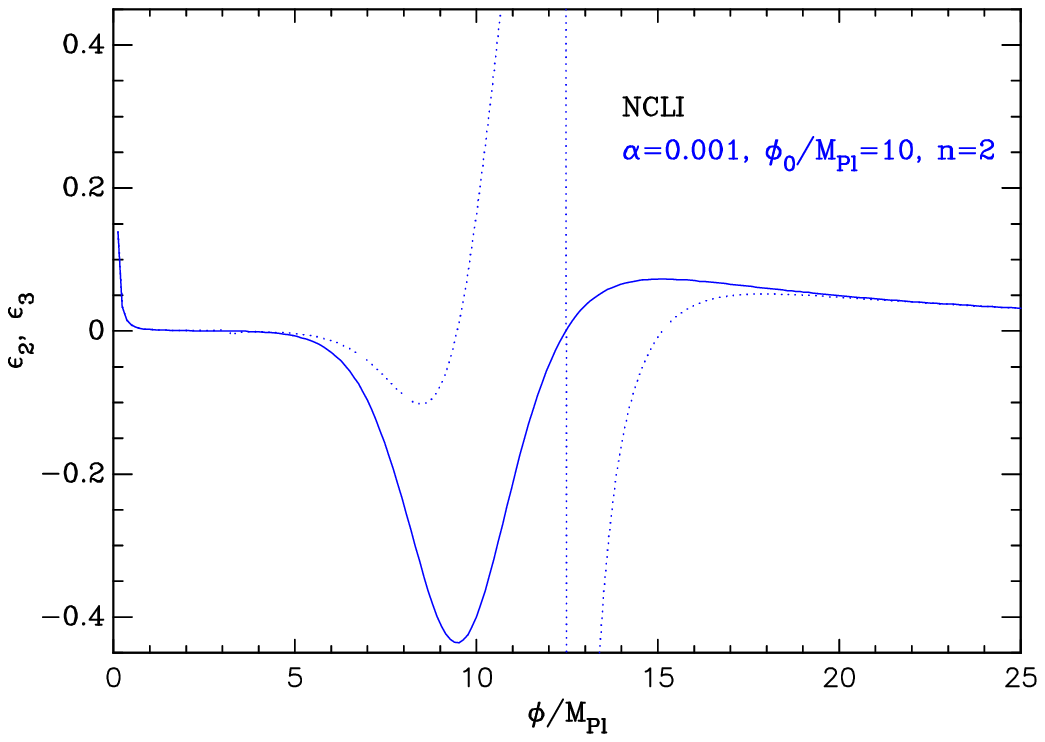}
\caption{Non-Renormalizable Corrected Loop Inflation (NCLI).  Top left
  panel: Non-Renormalizable Corrected Loop Inflation potential
  \Eq{eq:ncli:pot2} as a function of $\phi/\Mp$, for $\alpha=10^{-3}$,
  $\phizero=10\Mp$ and $n=2$. Top right panel: logarithm of the
  potential. Bottom left panel: slow-roll parameter $\epsilon _1$.
  Bottom right panel: slow-roll parameters $\epsilon_2$ (solid line)
  and $\epsilon_3$ (dotted line) for the same parameters value. Notice
  that $\epsilon_3$ becomes singular at the points where
  $\epsilon_2=0$, but the product $\epsilon_2 \epsilon_3$ remains
  finite.}
\label{potNCLI}
\end{center}
\end{figure}

\subsubsection{Theoretical Justifications}

This model is based on \Refc{Ballesteros:2005eg}, where the potential
is assumed to be exactly flat at tree level, with two types of
corrections considered: (i) corrections that do not spoil the flatness
of the potential and which correspond to radiative modulations of the
potential, and (ii) corrections that do spoil the flatness, and which
correspond to non-renormalizable operators in the tree-level
potential. These are suppressed by $(\phi/\Lambda)^{2n}$, where $n$ is
a positive integer and $\Lambda$ is the scale where new physics becomes
relevant and is assumed to be larger than the energy at which
inflation takes place. The potential reads~\cite{Ballesteros:2005eg}
\begin{equation}
V(\phi) = \rho+\beta
\ln\left[\frac{m(\phi)}{Q}\right]+\phi^4\frac{\phi^{2n}}{\Lambda^{2n}}\,.
\end{equation}
In this expression, $\rho$ corresponds to the tree-level flat
potential, $\beta$ is a positive coupling constant,
$m^2(\phi)=\lambda^2\phi^2/2$ is a mass term, and $Q$ corresponds to a
renormalization energy scale.

\subsubsection{Slow-Roll Analysis}

For our purpose, it is more convenient to recast the potential into
the form
\begin{equation}
\label{eq:ncli:pot2}
V(\phi) = M^4\left[1+\alpha\ln\left(\frac{\phi}{\Mp}\right)
+\left(\frac{\phi}{\phizero}\right)^{4+2n}\right],
\end{equation}
with $M^4=\rho+\beta\ln[\lambda^2\Mp^2/(2Q)]$, $\alpha=2\beta/M^4$ and
$\phizero^{4+2n}=\Lambda^{2n} M^4$. This model can be seen as a
correction to (or an extended version of) Loop Inflation, see
\sectionc{sec:li}, which is recovered in the limit $\phizero
\to\infty$. The potential is displayed in \Fig{potNCLI}. It is an
increasing function of the field value, and is positive for
$\phi>\phiVzero$, with
\begin{equation}
\phiVzero = \phizero \left\lbrace \frac{\alpha}{4+2n}
\Lambert{0}\left[\frac{4+2n}{\alpha}\ee^{-(4+2n)\left(\frac{1}{\alpha}+\ln\frac{\phizero}{\Mp}\right)}\right]
\right\rbrace^{\frac{1}{4+2n}} ,
\end{equation}
and where $\Lambert{0}$ is the $0$-branch of the Lambert function. The potential
has a concave shape below its inflection point
$\phi< \phiddVzero$, where
\begin{equation}
\phiddVzero=\phizero \left[\frac{\alpha}{\left(4+2n\right)\left(3+2n\right)}\right]^{\frac{1}{4+2n}} ,
\end{equation}
and is convex above.

In the slow-roll approximation, the first Hubble-flow function is
given by
\begin{equation}
\label{eq:ncli:eps1}
\epsilon_1 = \frac12 \frac{\Mp^2}{\phizero^2}
\frac{\left[\alpha \frac{\phizero}{\phi}+(4+2n)\left(\frac{\phi}{\phizero}\right)^{3+2n}
\right]^2}{\left[1+\alpha \ln \left(\frac{\phi}{\Mp}\right)+
\left(\frac{\phi}{\phizero}\right)^{4+2n}\right]^2}\,,
\end{equation}
while the second one reads
\begin{equation}
  \begin{aligned}
\label{eq:ncli:eps2}
\epsilon_2 &= \dfrac{2\dfrac{\Mp^2}{\phizero^2}}{\left[1+\alpha \ln \left(\frac{\phi}{\Mp}\right)+
  \left(\frac{\phi}{\phizero}\right)^{4+2n}\right]^{2}} 
\Biggl\{\left[\alpha\frac{\phizero}{\phi}+2(2+n)\left(\frac{\phi}{\phizero}
\right)^{3+2n}\right]^2
 \\ & 
+\left[\alpha \left(\frac{\phizero}{\phi}\right)^2-2(2+n)(3+2n)
\left(\frac{\phi}{\phizero}
\right)^{2+2n}\right]
\left[1+\alpha \ln \left(\frac{\phi}{\Mp}\right)+
\left(\frac{\phi}{\phizero}\right)^{4+2n}\right]\Biggr\},
  \end{aligned}
\end{equation}
and, finally,
\begingroup
\allowdisplaybreaks
  \begin{align}
    \label{eq:ncli:eps3}
\epsilon_3 &= \frac{\Mp^2}{\phizero^2}
\left[\alpha \frac{\phizero}{\phi}+2(2+n)
\left(\frac{\phi}{\phizero}
\right)^{3+2n}\right]
\Biggl\{2\left[\alpha \frac{\phizero}{\phi}+2(2+n)
\left(\frac{\phi}{\phizero}
\right)^{3+2n}\right]^3
\nonumber \\ &
+3\left[\alpha \frac{\phizero^2}{\phi^2}-2(2+n)(3+2n)
\left(\frac{\phi}{\phizero}
\right)^{2+2n}\right]
\left[\alpha \frac{\phizero}{\phi}+2(2+n)
\left(\frac{\phi}{\phizero}
\right)^{3+2n}\right]
\nonumber \\ & \times
\left[1+\alpha \ln \left(\frac{\phi}{\Mp}\right)+
\left(\frac{\phi}{\phizero}\right)^{4+2n}\right]
 \nonumber \\ & 
+
\left[2\alpha \frac{\phizero^3}{\phi^3}+4(1+n)(2+n)(3+2n)
\left(\frac{\phi}{\phizero}
\right)^{1+2n}\right]
\left[1+\alpha \ln \left(\frac{\phi}{\Mp}\right)+
\left(\frac{\phi}{\phizero}\right)^{4+2n}\right]^2\Biggr\}
\nonumber \\ & \times
\Biggl(\left[1+\alpha \ln \left(\frac{\phi}{\Mp}\right)+
\left(\frac{\phi}{\phizero}\right)^{4+2n}\right]^2
\Biggl\{
\left[\alpha \frac{\phizero}{\phi}+2(2+n)
\left(\frac{\phi}{\phizero}
\right)^{3+2n}\right]^2 \nonumber \\ & 
+
\left[\alpha \frac{\phizero^2}{\phi^2}-2(2+n)(3+2n)
\left(\frac{\phi}{\phizero}
\right)^{2+2n}\right]
\left[1+\alpha \ln \left(\frac{\phi}{\Mp}\right)+
\left(\frac{\phi}{\phizero}\right)^{4+2n}\right]\Biggr\}\Biggr)^{-1} .
  \end{align}
\endgroup
These parameters are displayed in the lower panels of \Fig{potNCLI},
and, clearly feature two different regimes. At large-field values, the
potential is dominated by the monomial correction, proportional to
$\phi^{4+2n}$, and the Hubble-flow parameters increase as the field
decreases, \ie as inflation proceeds. However, in that region of the
potential, higher-order corrections may become important, which is why
inflation is meant to take place at smaller-field values in that
model. At small-field values, the potential is dominated by the
logarithmic term and the constant term, and is therefore of the same
type as Loop Inflation, see \sectionc{sec:li}. In that region, the
Hubble-flow parameters again increase as inflation proceeds.

When transiting between these two regions, the behavior of the
Hubble-flow parameters is more involved. For instance, $\epsilon_1$
first reaches a maximum, then decreases for a transient period and
increases again. If $\phizero$ is large enough, this local maximum is
such that $\epsilon_1<1$ and inflation does not end before
$\epsilon_1$ increases again when $\phi$ approaches $0$. Otherwise,
inflation could terminate at the end of the first phase, and resume
afterwards, but we do not consider this possibility any further since,
as already stressed, the model is reliable only in the second phase.

Inflation ends when $\epsilon_1=1$, and the corresponding field value
$\phiend$ can be obtained by solving the equation
\begin{equation}
\alpha \dfrac{\phizero}{\phiend}+(4+2n)\left(\dfrac{\phiend}{\phizero}\right)^{3+2n}
= \sqrt{2}\dfrac{\phizero}{\Mp}\left[1+\alpha \ln \left(\dfrac{\phiend}{\Mp}\right)+
\left(\dfrac{\phiend}{\phizero}\right)^{4+2n}\right]
\end{equation}
Unfortunately, this equation does not have analytical solutions and it
must be solved numerically. Likewise, the slow-roll trajectory,
\begin{equation}
\Nend - N = \frac{\phizero}{\Mp} \int_{\phiend}^{\phi} \dfrac{1+\alpha
  \ln \left(\dfrac{\chi}{\Mp}\right)+
  \left(\dfrac{\chi}{\phizero}\right)^{4+2n}}{\alpha
  \dfrac{\phizero}{\chi}+(4+2n)\left(\dfrac{\chi}{\phizero}\right)^{3+2n}}
\dfrac{\ud\chi}{\Mp} \, .
\end{equation}
cannot be integrated analytically and must be computed numerically. 

The normalization of the potential $M^4$ is obtained from the
amplitude of the CMB anisotropies once the field value $\phistar$ at
which the pivot mode crossed the Hubble radius is determined. One gets
\begin{equation}
\left(\frac{M}{\Mp}\right)^4=720\pi^2\frac{\Mp^2}{\phizero^2}
\frac{\left[\alpha\dfrac{\phizero }{\phistar }+2 (n+2)
    \left(\dfrac{\phistar }{\phizero}\right)^{2 n+3}\right]^2}{
  \left[\alpha \log \left(\dfrac{\phistar}{\Mp}
    \right)+\left(\dfrac{\phistar }{\phizero}\right)^{2
      (n+2)}+1\right]^3} \dfrac{\Qrms^2}{T^2} \,.
\end{equation}
The reheating consistent slow-roll predictions for NCLI are
represented in \Figs{fig:CMBRNCLI_0} to \ref{fig:CMBRNCLI_3}, for
$n=2$ and $n=3$, plus a few values of $\alpha$ and $\phizero$. One can
check that when $\phizero$ is large, one recovers the same predictions
as in Loop Inflation. When $\phizero$ decreases, the spectral index
increases, and quickly leaves the region allowed by the data. As a
consequence, the amplitude of the corrective monomial term is bounded
from above in this model.

The way this upper bound varies with $\alpha$ can be understood as
follows. In Loop Inflation, in the regime $\alpha\ll 1$, the slow-roll
trajectory is approximately given by $\phistar/\Mp\simeq
\sqrt{2\alpha(\Nend-\Nstar)}$, see the relation given below
\Eq{eq:li:traj}. By performing an expansion of \Eqs{eq:ncli:eps1},
\eqref{eq:ncli:eps2} and~\eqref{eq:ncli:eps3} in $1/\phizero$, one
obtains for the first Hubble flow function(still at leading order in $\alpha$)
\begin{align}
\epsonestar & \simeq \frac{\alpha}{4\Delta
  \Nstar}+\alpha\left(4+2n\right)\left(2\alpha\Delta
\Nstar\right)^{n+1} \left(\frac{\Mp}{\phizero}\right)^{4+2n}.
\end{align}
The second reads
\begin{align}
\epstwostar&\simeq \frac{1}{\Delta
  \Nstar}-4\left(n+2\right)\left(3+2n\right) \left(2\alpha\Delta
\Nstar\right)^{n+1}\left(\frac{\Mp}{\phizero}\right)^{4+2n}
,
\end{align}
while the third one is
\begin{align}
\epsthreestar &\simeq \frac{1}{\Delta
  \Nstar}+4\left(n+2\right)\left(2n^2+7n+7\right) \left(2\alpha\Delta
\Nstar\right)^{n+1}\left(\frac{\Mp}{\phizero}\right)^{4+2n},
\end{align}
where $\Delta\Nstar = \Nend - \Nstar$. One can check that the leading
terms of these expressions match the approximate predictions of Loop
Inflation (LI), see \Eq{eq:li:predic}. The first-order corrections
have a relative amplitude controlled by $(\Delta
\Nstar)^{n+2}\alpha^{n+1}(\Mp/\phizero)^{4+2n} $ for all three
parameters, so one concludes that the predictions of NCLI are close to
the ones of LI provided
\begin{equation}
\label{eq:ncli:phi0:cond}
\frac{\phizero}{\Mp}\gg \alpha^{\frac{n+1}{2n+4}}\, .
\end{equation} 
From this expression, one can see that the smaller $\alpha$, the
smaller $\phizero$ can be. The above equation also provides a correct
estimate for the value of $\phizero$ below which the predictions of
NCLI deviate from the ones of LI (and are therefore disfavored by the
data), as can be seen in \Figs{fig:CMBRNCLI_0} to \ref{fig:CMBRNCLI_3}.

\subsection{Hybrid Natural Inflation (HNI)}
\label{sec:hni}

\begin{figure}
\begin{center}
\includegraphics[width=\wdblefig]{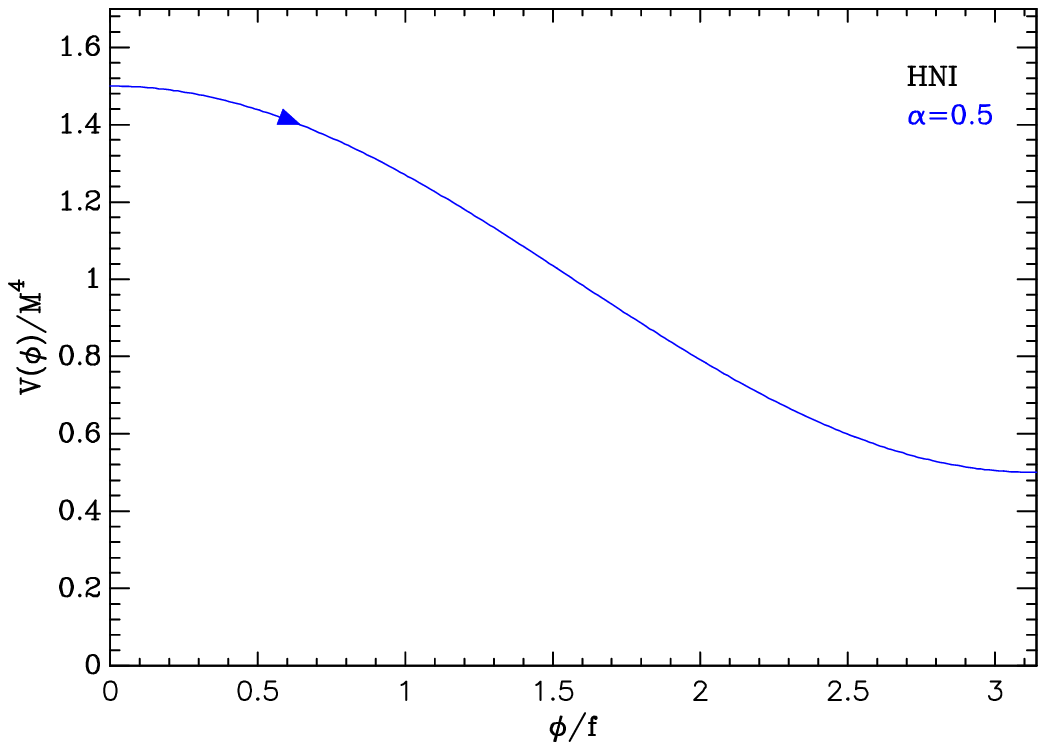}
\includegraphics[width=\wdblefig]{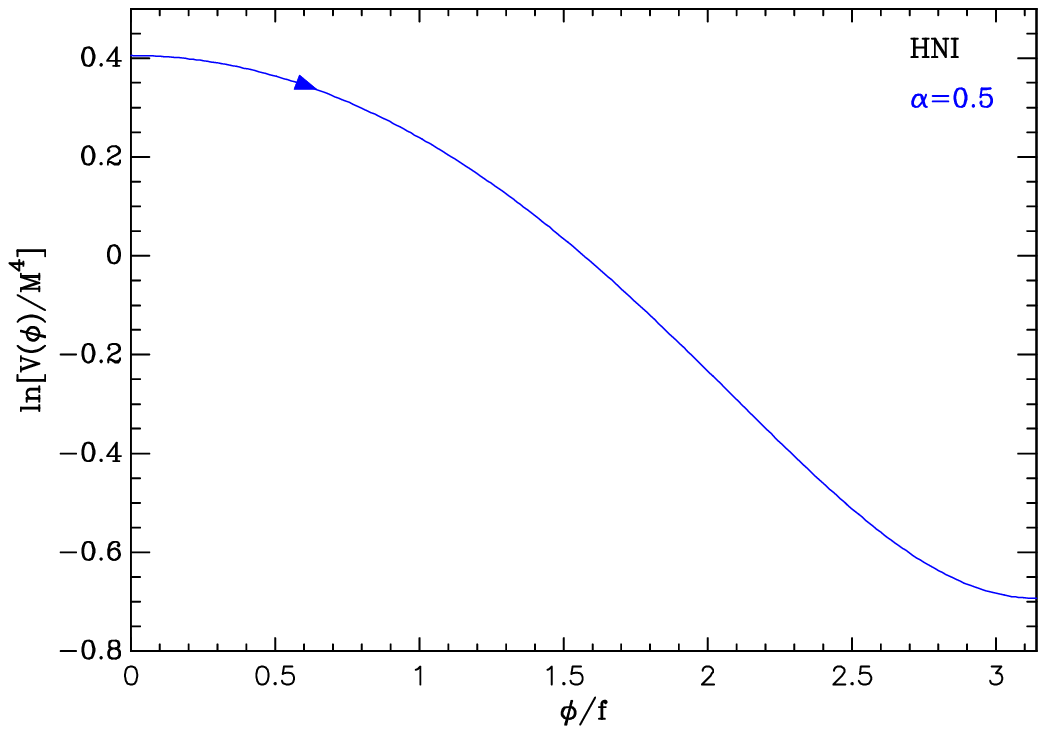}
\includegraphics[width=\wdblefig]{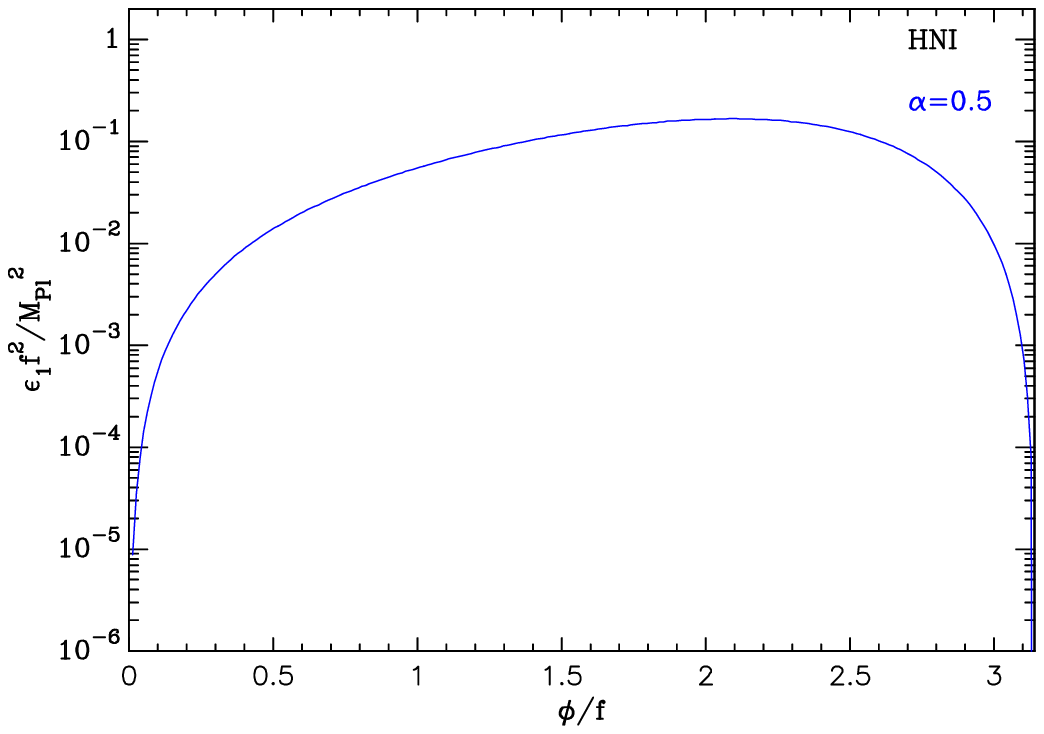}
\includegraphics[width=\wdblefig]{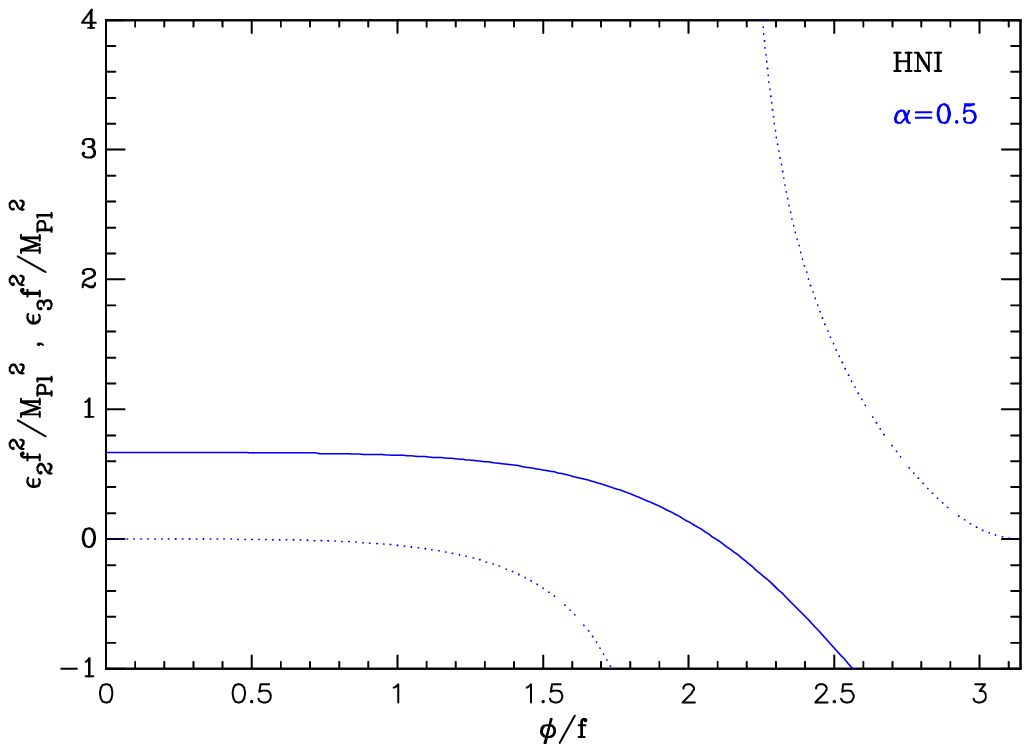}
\caption{Hybrid Natural Inflation for $\alpha=0.5$. Top panels: the
  potential and its logarithm. Bottom left panel: slow-roll parameter
  $\epsilon_1 f^2/\Mp^2$. Notice that, depending on $\alpha$ and $f$, the
  maximal value of $\epsilon_1$ can become larger than unity. Bottom
  right panel: slow-roll parameters $\epsilon_2 f^2/\Mp^2$ (solid line) and
  $\epsilon_3 f^2/\Mp^2$ (dotted line) for $\alpha=0.5$. As for SFI2 (see
  \sectionc{sec:sfi}), the potential at $\phi=0$ has a tachyonic mass
  and $\epsilon_2$ goes to a constant at small-field values, which 
  violates slow roll for sub-Planckian $f$.}
\label{fig:pothni}
\end{center}
\end{figure}

\subsubsection{Theoretical Justifications}

This scenario is an extension of Natural Inflation (NI, see
\sectionc{sec:ni}) in which the end of inflation can be triggered by a
waterfall mechanism as in Hybrid Inflation (see
\sectionc{sec:vhi}). This idea, as well as explicit supersymmetric and
non-supersymmetric two-fields constructions of the model, have been presented in
\Refcs{Stewart:2000pa, Cohn:2000hc, Antusch:2008gw, Ross:2009hg,
  Ross:2010fg, Vazquez:2014uca}. In addition to stopping inflation, the
waterfall field distorts the effective potential along the
inflationary direction such that the potential for HNI reads
\begin{equation}
V(\phi) = M^4 \left[1 + \alpha \cos\left(\dfrac{\phi}{f}\right) \right].
\label{eq:pothni}
\end{equation}
The typical vacuum expectation value for the inflaton, $f$,  can be
made super-Planckian as HNI describes a multiple-field model (see
\sectionc{sec:ni}). The parameter $0<\alpha<1$ encodes the
distortions induced by the waterfall field onto the inflationary
direction, and as can be seen in \Fig{fig:pothni}, this implies that
the minimum of the potential is non-vanishing. Note however that the true
minimum of the potential in the two-field space is elsewhere, and,
exactly as for Hybrid Inflation, the ground state of the waterfall
field is expected to cancel this apparent residual cosmological
constant. For our purpose, it means that the potential of
\Eq{eq:pothni} cannot be trusted for $\phi/f \simeq \pi$, and we will
only consider the inflationary domain connected to the maximum of the
potential at $\phi=0$.

Finally, let us remark that the potential of NI is recovered for
$\alpha \to 1$ but, even in this limit, both models may have different
observable predictions as HNI may end by waterfall instability instead
of slow-roll violations. As such, HNI has three parameters, $\alpha$,
$f$ and the field value at which inflation stops, $\xend$.

\subsubsection{Slow-roll Analysis}

Introducing $x=\phi/f$, the Hubble-flow functions in the slow-roll
approximation are
\begin{equation}
\epsilon_1 = \dfrac{\alpha^2 \Mp^2}{2 f^2}
\dfrac{\sin^2(x)}{\left[1+\alpha
    \cos(x)\right]^2}\,, \qquad
\epsilon_2=\dfrac{2 \alpha \Mp^2}{f^2} \dfrac{\alpha + \cos(x)}{\left[1+\alpha
    \cos(x)\right]^2}\,,
\label{eq:hnieps12}
\end{equation}
and
\begin{equation}
\epsilon_3 = -\dfrac{\alpha \Mp^2}{f^2} \, \dfrac{\sin^2(x)}{\left[1 +
    \alpha \cos(x)\right]^2} \dfrac{1 -
    2\alpha^2 - \alpha \cos(x)}{\alpha + \cos(x)}\,.
\label{eq:hnieps3}
\end{equation}
They have been represented in the lower panels of \Fig{fig:pothni} for
$\alpha=0.5$. The first Hubble-flow function exhibits a maximum within
the inflationary domain, at a field value that is solution of
$\epsilon_2(x)=0$, and given by
\begin{equation}
\xepsoneMax = \arccos({-\alpha}).
\label{eq:hnixepsonemax}
\end{equation}
Plugging this value into \Eq{eq:hnieps12} gives the maximal value
reached by $\epsilon_1(x)$ over the domain $x \in]0,\pi[$,
\begin{equation}
\epsonemax = \dfrac{\Mp^2}{2 f^2} \dfrac{\alpha^2}{1 - \alpha^2}\,.
\end{equation}
It is larger than unity if $\alpha > \alpha_1$, or,
equivalently, if $f < f_1$, where
\begin{equation}
\alpha_1 \equiv \dfrac{1}{\sqrt{1+\Mp^2/(2 f^2)}}\,, \qquad f_1 \equiv
\dfrac{\alpha}{\sqrt{2 \alpha^2 - 2}}\,.
\label{eq:alphafhni}
\end{equation}
When this happens, the inflationary domain becomes disconnected, and,
a priori, one could have inflation close to the top of the potential
(around $\phi=0$), or close to the bottom of the potential at $\phi/f
\lesssim \pi$. As we have mentioned in the previous section, the
latter situation will be discarded as one cannot trust anymore
\Eq{eq:pothni} to describe the two-field dynamics (see also the
discussion regarding this possibility for VHI in \sectionc{sec:vhi}). Nonetheless,
this creates the possibility that HNI gracefully ends if the field
value at which the waterfall instability develops is in the domain for
which $\epsilon_1 > 1$. This scenario will be referred to as HNI1 and
the smallest root of $\epsilon_1(x)=1$ gives the field value at which
inflation ends in that case,
\begin{equation}
\xepsoneOne = \arccos\left\{\dfrac{\Mp}{\alpha f}
\dfrac{\sqrt{\dfrac{\alpha^2 \Mp^2}{4f^2} + \dfrac{\alpha^2-1}{2}} -
  1}{1+\dfrac{\Mp^2}{2f^2}}\right\}.
\label{eq:xepsoneonehni}
\end{equation}
This solution exists provided $\alpha > \alpha_1$ (or $f < f_1$) and
one has $\xend = \xepsoneOne$, making HNI1 a two-parameter model.

In the other, and more generic, scenario, referred to as HNI2,
inflation will be assumed to always end by instability, for all values
of $\alpha$ and $f$. As a result, for $\alpha > \alpha_1$, there is an
upper bound for the field value at which the instability occurs,
namely $\xend < \xepsoneOne$. For $\alpha < \alpha_1$, $\xend$ can be
taken up to $\xendmax=\pi$, the minimum of the potential.

The slow-roll trajectory can be explicitly integrated using
\Eq{eq:srtrajectory} and reads
\begin{equation}
\Nend - N = \dfrac{f^2}{\alpha \Mp^2}\left\{ \left(1-\alpha\right)
\ln\left[\dfrac{\cos\left(\dfrac{x}{2}\right)}{\cos\left(\dfrac{\xend}{2}\right)}\right]
- \left(1 + \alpha
\right) \ln
\left[\dfrac{\sin\left(\dfrac{x}{2}\right)}{\sin\left(\dfrac{\xend}{2}
    \right)} \right] \right\}.
\label{eq:trajhni}
\end{equation}
The logarithmic divergence at $x \to 0$ shows that the number of \efolds~that can be realized around the top of the potential is unbounded, so $\xend$ can be arbitrarily small. From \Eq{eq:hnieps12}, one also sees that although
$\epsilon_1 \to 0$ at the top of the potential, the second slow-roll
parameter reaches a constant value
\begin{equation}
\epsilon_2(x=0) = \dfrac{2 \alpha}{1+\alpha} \dfrac{\Mp^2}{f^2}\,.
\end{equation}
As such, for $\alpha = \order{1}$, slow roll inflation at the top of
the HNI potential can only take place for super-Planckian values of
$f$, a situation in all points similar to the small-field model SFI2
(see \sectionc{sec:sfi}). It is however possible to accommodate
sub-Planckian values of $f$ with slow roll but only at the expanse of
having $\alpha \ll 1$. The scenario HNI2 in the limit $\xend \ll 1$ is
therefore a constant-spectral-index model with vanishing running and highly suppressed tensor-to-scalar ratio.

Using the reheating equation~\eqref{eq:phistarlnrrad}, together with
the field value at which inflation ends, $\xend$, \Eq{eq:trajhni}
uniquely determines $\xstar$, the field value at which the pivot mode
crossed the Hubble radius during inflation. The mass scale $M$ of the
potential is fixed by the CMB normalization and verifies
\begin{equation}
\left(\dfrac{M}{\Mp} \right)^4 = 720 \pi^2 \alpha^2 \dfrac{\Mp^2}{f^2}
\dfrac{\sin^2(\xstar)}{\left[1 + \alpha \cos(\xstar)
    \right]^3}\dfrac{\Qrms^2}{T^2}\, .
\end{equation}
The reheating consistent slow-roll predictions for the two HNI models
(HNI1 and HNI2) are represented in \Figs{fig:CMBHNI1_0} to
\ref{fig:CMBHNI2_7}.

\subsection{N-Formalism Inflation (NFI)}
\label{sec:nfi}

\subsubsection{Theoretical Justifications}

This model is phenomenological and motivated by the search of
``universality classes'' for slow-roll inflation as originally
proposed in \Refc{Roest:2013fha}. There, it was argued that most of
the observationally relevant inflationary models should lead to a first
Hubble flow function varying as
\begin{equation}
  \epsilon_1 \propto \dfrac{1}{\Delta N^\alpha}\,,
\label{eq:nfiidea}
\end{equation}
the higher-order terms of an expansion in $1/\Delta \Nstar$ being
neglected. This idea was later proven to not encompass not all relevant models, and to lead to
insufficiently accurate predictions in light of the precision of
the CMB data in \Refc{Martin:2016iqo}. However, it is still possible
to search for models verifying \Eq{eq:nfiidea} \emph{at leading order in
slow-roll}. Plugging \Eq{eq:nfiidea} into the definition of the first
Hubble-flow function given in \Eq{eq:defeps1}, one gets $\Delta \phi/\Mp
\propto \Delta N^{1-\frac{\alpha}{2}}$ with $\Delta \phi =
\phiend-\phi$, which implies
\begin{equation}
\dfrac{1}{\Mp} \dfrac{\ud \phi}{\ud N} \propto \dfrac{1}{\Delta N^{\frac{\alpha}{2}}} \propto
\left(\dfrac{\Delta \phi}{\Mp} \right)^{\frac{\alpha}{\alpha-2}}.
\end{equation}
From the slow-roll trajectory given in \Eq{eq:phidot2}, one gets
\begin{equation}
\dfrac{\ud \ln V}{\ud \phi} \propto \dfrac{1}{\Mp} \left(\dfrac{\Delta \phi}{\Mp} \right)^{\frac{\alpha}{\alpha-2}}.
\end{equation}
Ignoring the singular case $\alpha=2$, the potential verifies
\begin{equation}
  \ln\left(\dfrac{V}{\Mp^4}\right) \propto \left(\dfrac{\Delta \phi}{\Mp}\right)^{\frac{2(\alpha-1)}{\alpha-2}},
\end{equation}
and is an exponential function of some power of the field.

\subsubsection{Slow-roll Analysis}

\begin{figure}
\begin{center}
\includegraphics[width=\wdblefig]{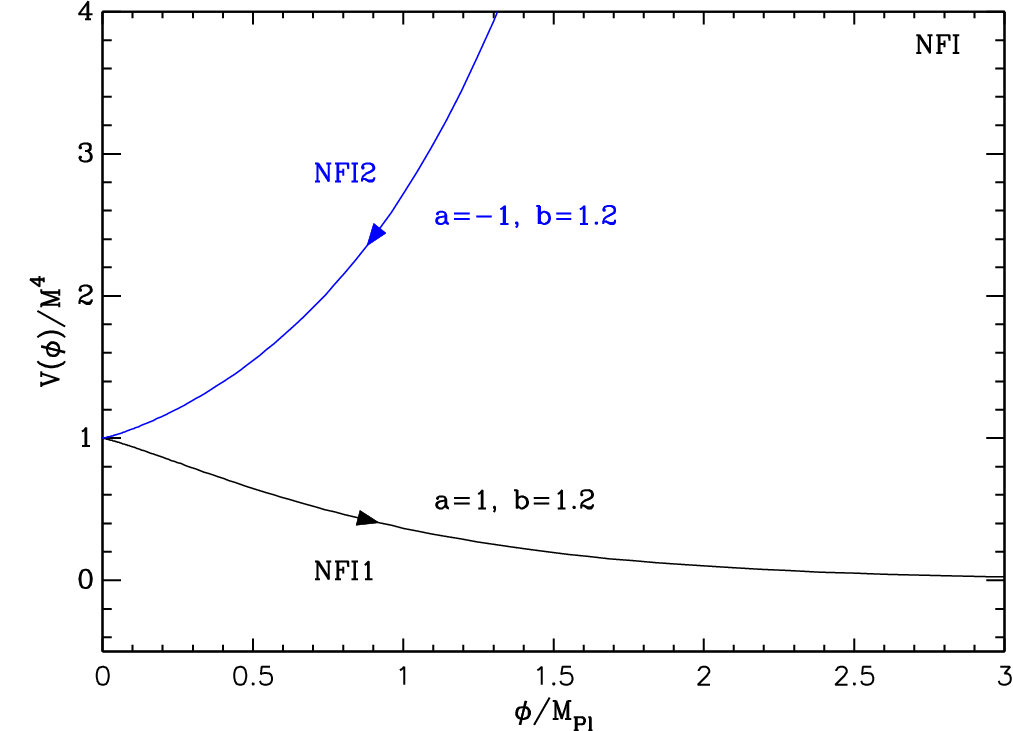}
\includegraphics[width=\wdblefig]{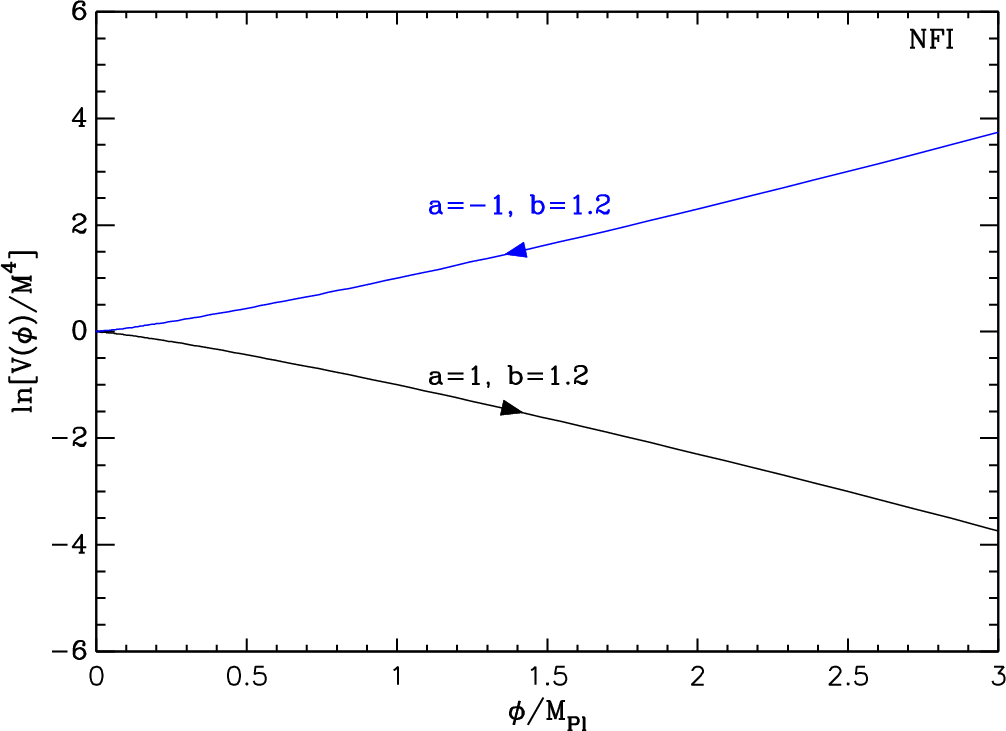}
\includegraphics[width=\wdblefig]{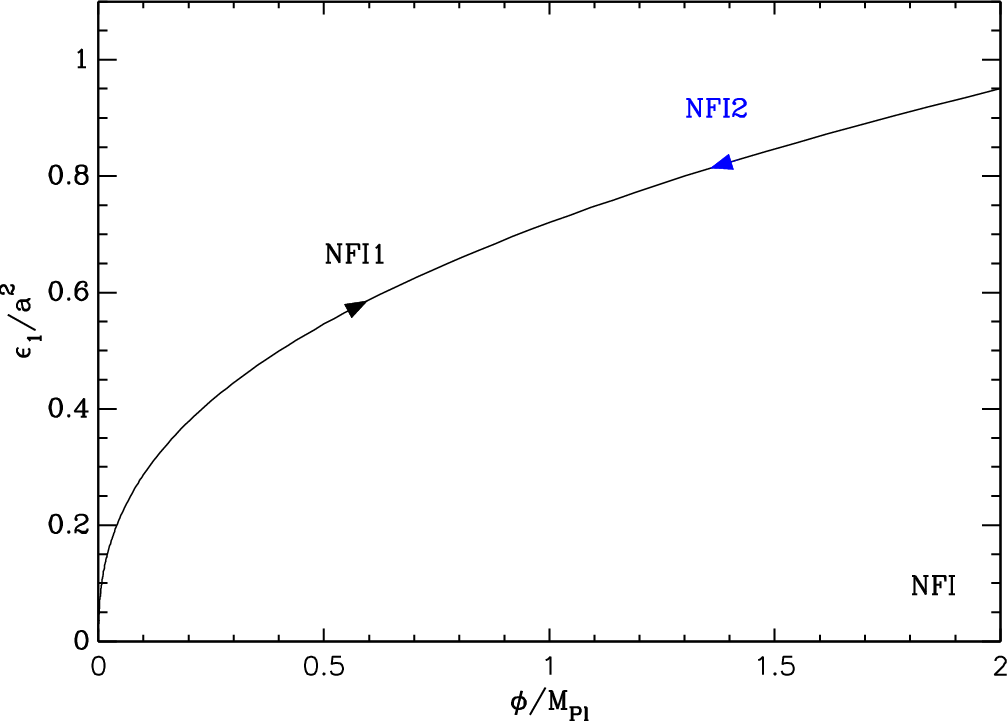}
\includegraphics[width=\wdblefig]{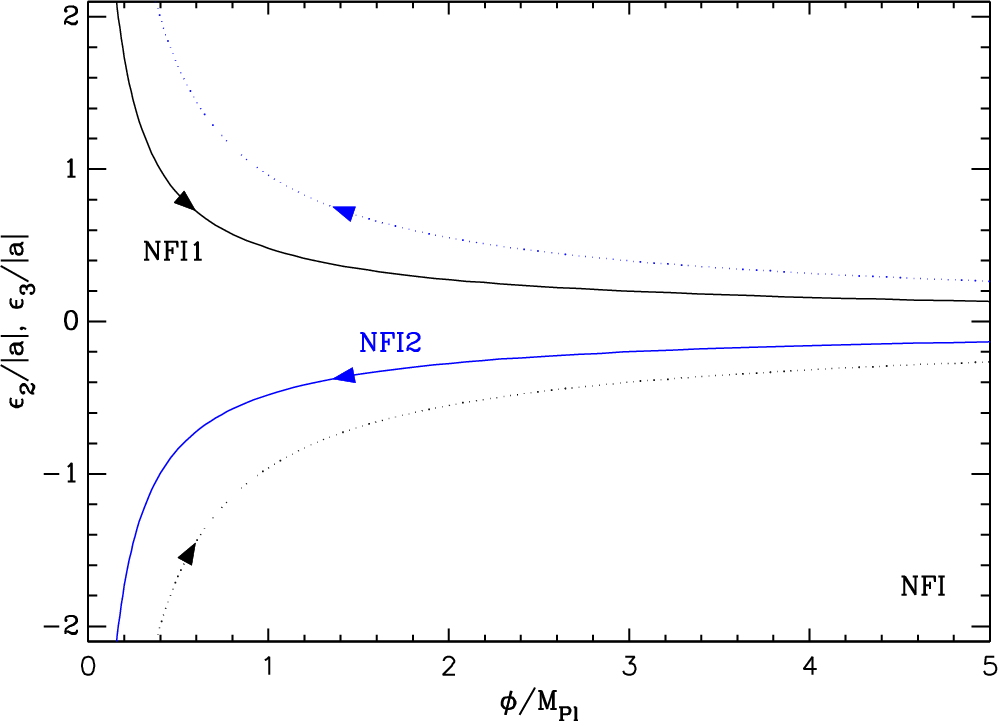}
\caption{N-Formalism Inflation in the regime NFI1 for $a=1$, $b=1.2$
  (black curves) and NFI2 for $a=-1$, $b=1.2$ (blue curves). Top
  panels: the potential and its logarithm. Bottom left panel:
  slow-roll parameter $\epsilon_1/a^2$, which is the same in both
  regime. Bottom right panel: slow-roll parameters $\epsilon_2 /|a|$
  (solid line) and $\epsilon_3 / |a|$ (dotted line). Their signs are
  different in the NFI1 and NFI2 regime. Within the range $1<b<2$,
  NFI2 exhibit slow-roll violations at small field values but only for
  $\epsilon_2$ and $\epsilon_3$. This does not stop inflation as
  $\epsilon_1 \to 0$ for $\phi \to 0$. For $b>2$, all slow-roll
  parameters are regular and vanish at the origin.}
\label{fig:potnfi12}
\end{center}
\end{figure}

\begin{figure}
\begin{center}
\includegraphics[width=\wdblefig]{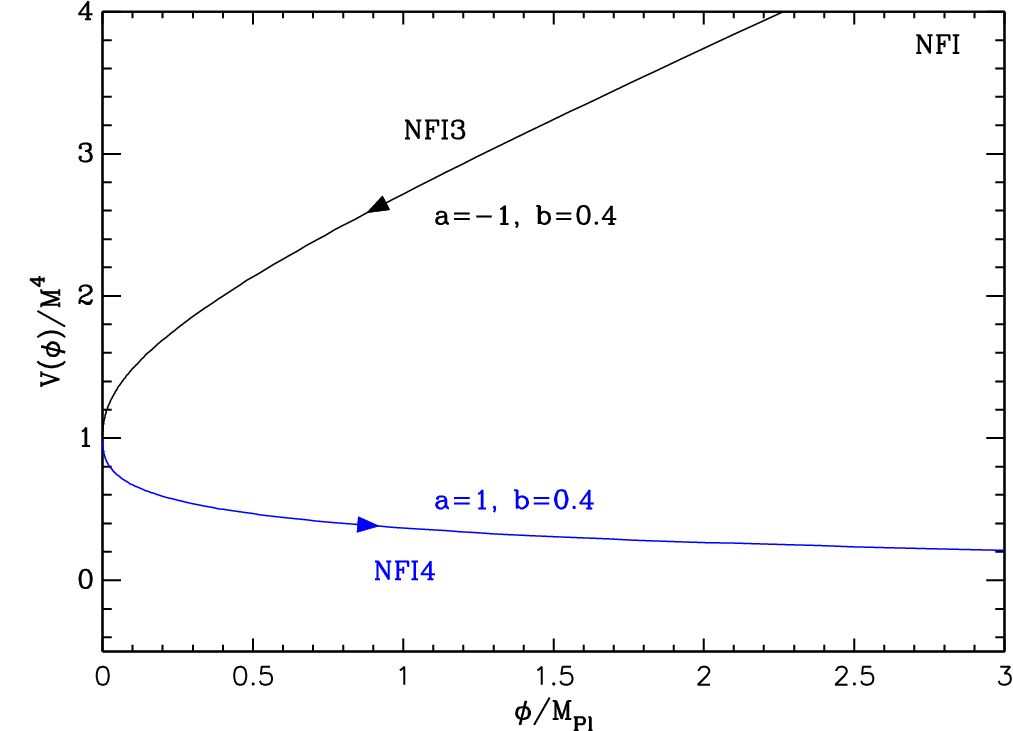}
\includegraphics[width=\wdblefig]{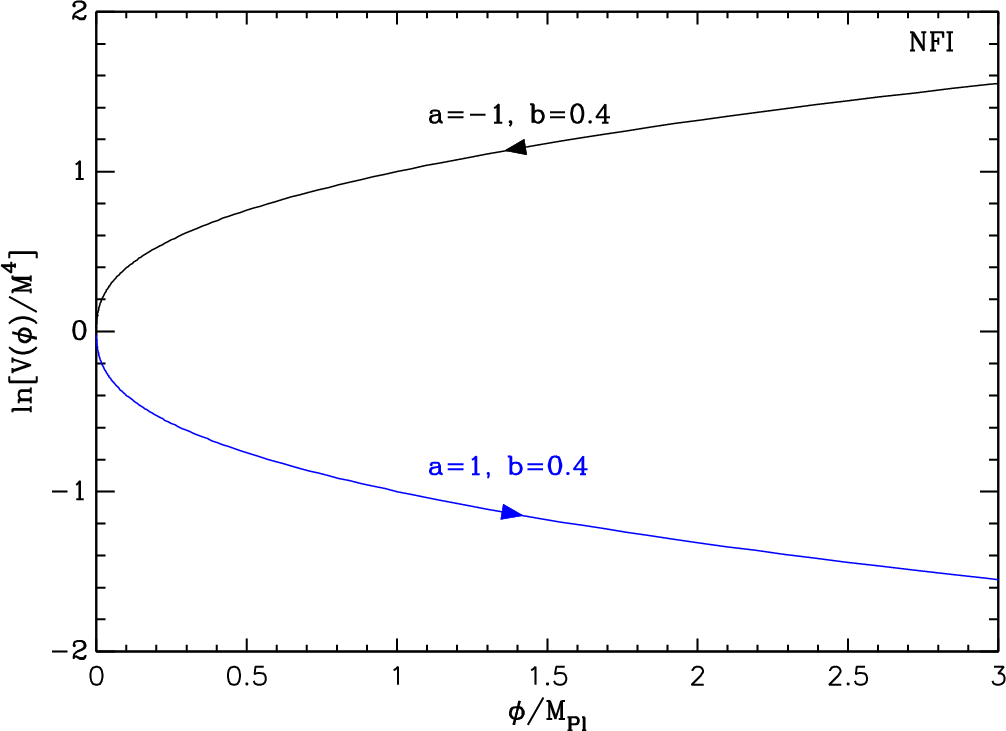}
\includegraphics[width=\wdblefig]{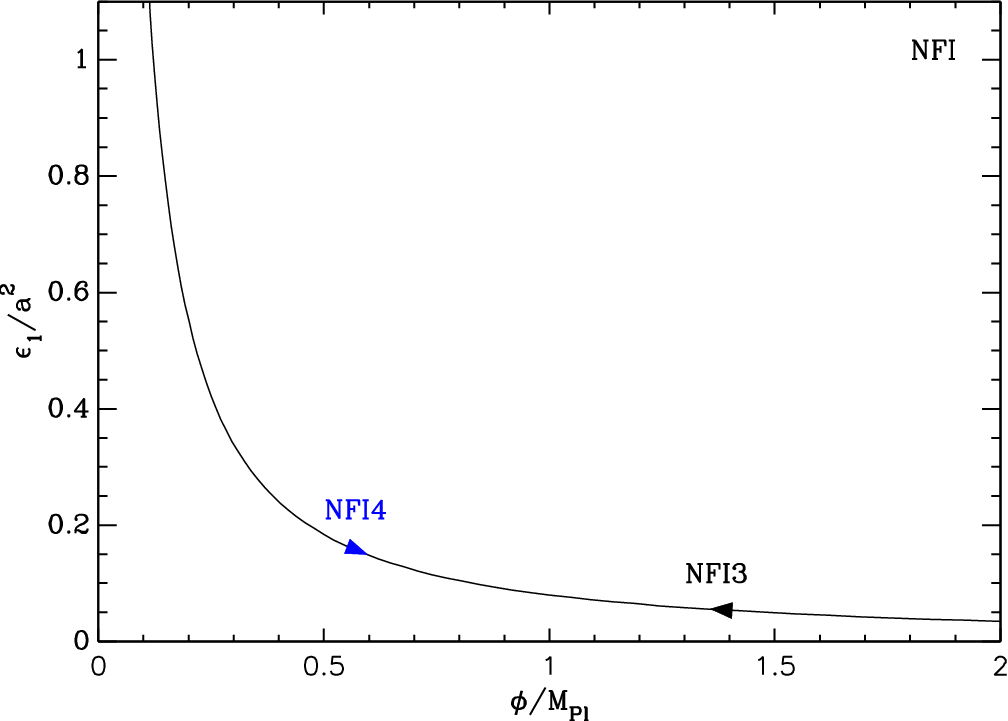}
\includegraphics[width=\wdblefig]{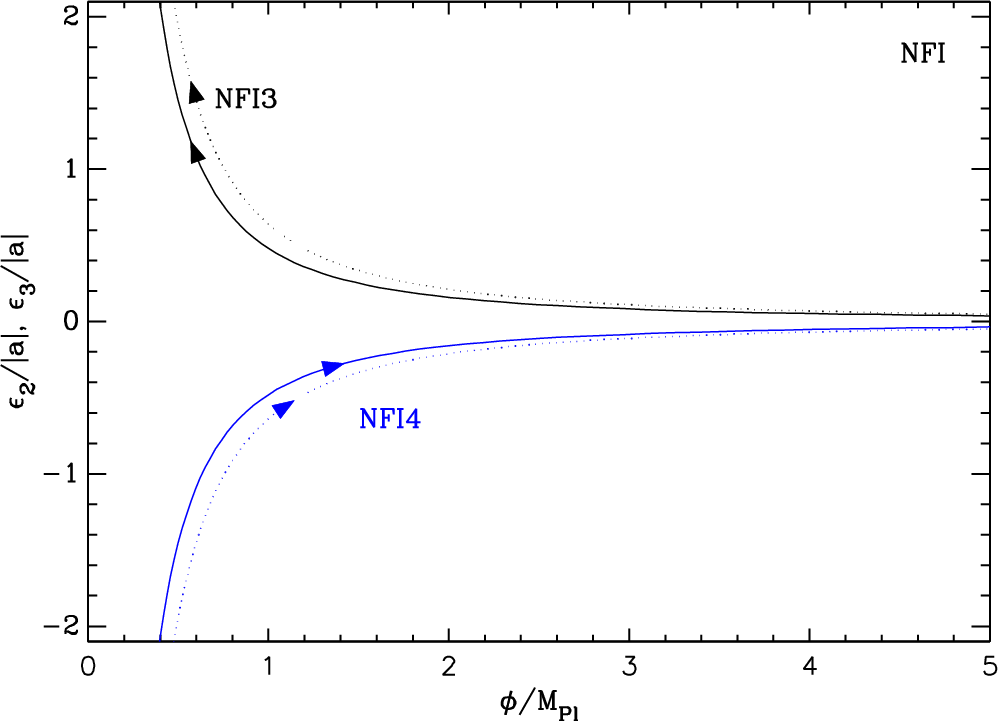}
\caption{N-Formalism Inflation in the regime NFI3 for $a=-1$, $b=0.4$
  (black curves) and NFI4 for $a=1$, $b=0.4$ (blue curves). Top
  panels: the potential and its logarithm. Bottom left panel:
  slow-roll parameter $\epsilon_1/a^2$, which is the same in both
  regime. Bottom right panel: slow-roll parameters $\epsilon_2 /|a|$
  (solid line) and $\epsilon_3 / |a|$ (dotted line). They differ in
  sign between the NFI3 and NFI4 regimes.}
\label{fig:potnfi34}
\end{center}
\end{figure}

Let us write the potential of NFI under the more convenient form
\begin{equation}
V(\phi) = M^4 e^{-a \left(\frac{\phi}{\Mp}\right)^b},
\label{eq:potnfi}
\end{equation}
where $a$ and $b$ are two dimensionless parameters. The peculiar case
$b=1$ is Power Law Inflation (PLI) discussed in \sectionc{sec:pli} and
will not be further discussed in the following. Using
\begin{equation}
x \equiv \dfrac{\phi}{\Mp},
\end{equation}
the Hubble flow functions in the slow-roll approximation read
\begin{equation}
\epsilon_1 = \dfrac{1}{2} a^2 b^2 x^{2(b-1)}, \qquad \epsilon_2 =
2ab(b-1)x^{b-2}, \qquad \epsilon_3 = ab(b-2)x^{b-2}.
\label{eq:nfieps123}
\end{equation}
When $b$ is not an integer, the potential is well-defined only for $x>0$
but could be extended in the negative domain otherwise (see
below). The solution to $\epsilon_1=1$ is given by
\begin{equation}
\xepsoneOne \equiv \left(\dfrac{2}{a^2 b^2}\right)^{\frac{1}{2(b-1)}},
\label{eq:nfi:xepsoneone}
\end{equation}
and one can see that inflation can only take
place in the domains $x < \xepsoneOne$ for $b>1$ and in the domains $x
> \xepsoneOne$ for $b<1$. Moreover, depending on the sign of both $a$
and $b$, the potential, within the positive domain $x>0$, can either
be an increasing or a decreasing function of the field value. If the
potential drives the field away from $\xepsoneOne$ then inflation
requires an additional mechanism to end. In this situation, $\phiend$,
the field value at which inflation ends is an additional model
parameter making NFI a three-parameter model. However, if the
potential drives the field towards $\xepsoneOne$ then inflation can
gracefully end and the model has only two parameters in that case.

\begin{figure}
\begin{center}
\includegraphics[width=\wsingfig]{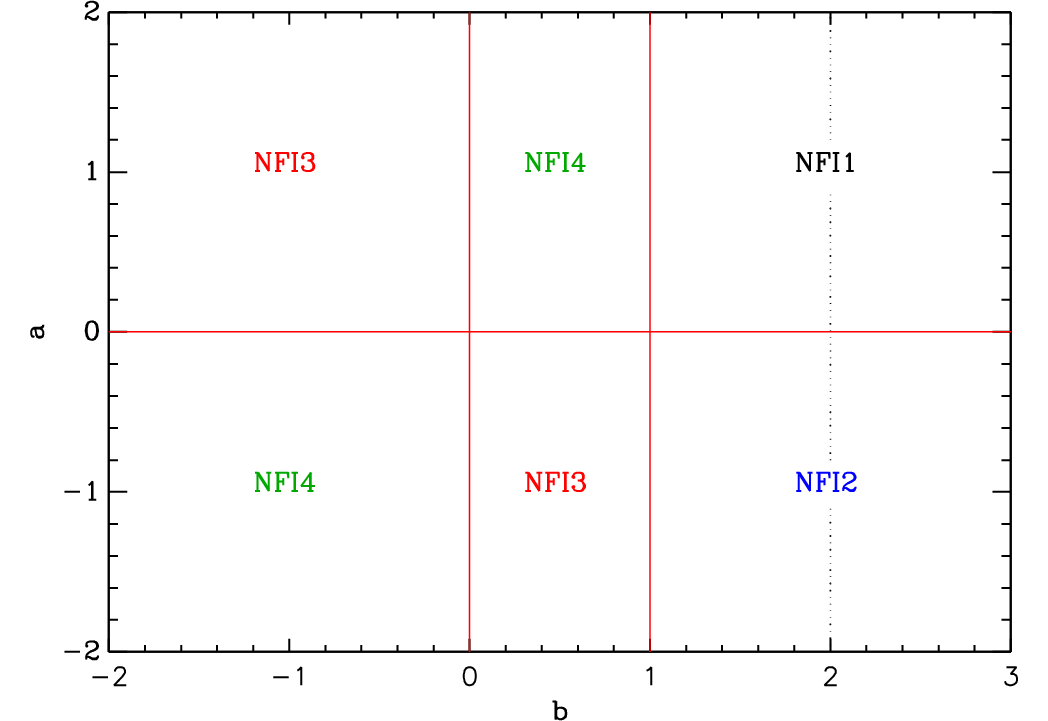}
\caption{The four inflationary regimes of N-Formalism Inflation in the
  parameter space $(a,b)$. Inflation naturally ends for NFI1 and NFI3
  whereas an additional mechanism has to be invoked for NFI2 and
  NFI4. The vertical dotted line at $b=2$ is a separatrix in the
  behavior of the Hubble flow functions at small-field values. For
  $b<2$, both $\epsilon_2$ and $\epsilon_3$ diverge for $x\to 0$ but
  vanish otherwise.}
\label{fig:abnfi}
\end{center}
\end{figure}

For these reasons we distinguish four regimes. The first, NFI1,
obtained for $a>0$ and $b>1$, is associated with a decreasing potential
with respect to $x$, inflation proceeds at increasing field values and
naturally ends at $\xepsoneOne$. The second regime, referred to as
NFI2, corresponds to $a<0$ and $b>1$. The potential is an increasing
function of the field, inflation proceeds at decreasing field
values moving away from $\xepsoneOne$. The third regime, NFI3, is
obtained for either $a<0$ and $0<b<1$, or, $a>0$ and $b<0$. It is
associated with a increasing potential with respect to $x$, inflation
occurring at decreasing field values and moving towards $\xepsoneOne$
with a graceful ending. Finally, NFI4 is obtained for either $a>0$ and
$0<b<1$, or, $a<0$ and $b<0$. Inflation proceeds at increasing field
values, moving away from $\xepsoneOne$ and requiring an additional
mechanism to stop. The potential and the Hubble-flow functions for the
regimes NFI1 and NFI2 have been represented in
figure~\ref{fig:potnfi12} while NFI3 and NFI4 are shown in
figure~\ref{fig:potnfi34}. Figure~\ref{fig:abnfi} shows the location
of these four inflationary regimes in the parameter space $(a,b)$.

The integer values of $b$ allow for the potential to be extended into
the domain $x<0$. For even values of $b$, the potential of
\Eq{eq:potnfi} is symmetric under the transformation $\phi \rightarrow
-\phi$ such that, at negative field values, inflation proceeds exactly
as in the positive domain. For odd values of $b$, the potential
remains unchanged under the transformation $\phi \rightarrow -\phi$
combined with $a \rightarrow -a$. As a result, inflation in the
negative domain of NFI1 is the same as inflation in the positive
domain of NFI2, and conversely. This symmetry also unifies the
negative domain of NFI3 with the positive domain of NFI4, and
conversely. All in all, we can restrict the analysis to $x>0$ in all
possible situations.

The slow-roll trajectory can be integrated analytically and reads
\begin{equation}
\Nend - N = \dfrac{\xend^{2-b} - x^{2-b}}{ab(2-b)}\,.
\label{eq:trajnfi}
\end{equation}
The case $b=2$ is included as the regular logarithmic limit of this
expression. The trajectory can be inverted into
\begin{equation}
x = \left[\xend^{2-b} - ab(2-b) \left(\Nend - N\right)\right]^{\frac{1}{2-b}},
\label{eq:xtrajnfi}
\end{equation}
which matches the one of Small Field Inflation (SFI) in the deep
sub-Planckian limit, see \Eq{eq:sfixtrajapprox}. The case $b=2$ can be
dealt by starting with the logarithmic limit of \Eq{eq:trajnfi} and one
gets a pure exponential term
\begin{equation}
x = \xend e^{-2a(\Nend-N)}.
\label{eq:xtraj2nfi}
\end{equation}
As already mentioned, for NFI1 and NFI3 we have $\xend = \xepsoneOne$
given in \Eq{eq:nfi:xepsoneone} whereas $\xend$ is an additional model
parameter for NFI2 and NFI4. Let us notice that for $b>2$,
\Eq{eq:trajnfi} diverges for $x \to 0$ and the maximal number of
{\efolds} achievable at the origin of the potential is
unbounded. In the opposite situation, for $b<2$, this implies that
there is a maximal acceptable value for $|a|$ to have enough
{\efolds} of NFI1 inflation between $x=0$ and
$\xend=\xepsoneOne$. Similarly, $|a|$ should be small enough to
have enough {\efolds} of NFI2 inflation between
$\xinimax=\xepsoneOne$ and $\xendmin=0$. In the regime NFI3 one
always has $b<1$ and the number of {\efolds} is unbounded. The
regime NFI4 proceeds at $x>\xepsoneOne$ and the number of {\efolds}
can be made as large as needed by choosing the free parameter
$\xend$ large enough.

From the trajectory of \Eq{eq:trajnfi}, combined with the reheating
equation~\eqref{eq:phistarlnrrad} and the field value $\xend$ at which
inflation ends, one obtains $\xstar$, the field value at which the
pivot mode crossed the Hubble radius during inflation. The mass scale
$M$ of the potential is then fixed by the CMB normalization and verifies
\begin{equation}
\left(\dfrac{M}{\Mp} \right)^4 = 720 \pi^2 a^2 b^2 e^{a\xstar^b} \xstar^{2(b-1)}\dfrac{\Qrms^2}{T^2}\, .
\end{equation}

The reheating consistent slow-roll predictions for the four NFI
regimes are represented in \Figs{fig:CMBNFI1_0} to
\ref{fig:CMBNFI4_5}. Let us remark that plugging \Eq{eq:xtrajnfi}
within the expression of the Hubble-flow functions given in
\Eq{eq:nfieps123}, one obtains
\begin{equation}
\epsonestar = \dfrac{a^2 b^2}{2 \left[ \xend^{2-b} + ab(b-2) \Delta
    \Nstar \right]^{\frac{2b-2}{b-2}}}\,.
\end{equation}
For $b>2$, in the regime NFI1, for $a$ small enough one can 
neglect $\xend=\xepsoneOne$ compared to the term containing
$\Delta\Nstar$. In this limit one recovers the power-law behavior of \Eq{eq:nfiidea}, namely
\begin{equation}
\epsonestar \propto \dfrac{1}{\Delta\Nstar^{\frac{2b-2}{b-2}}}\,.
\end{equation}
Notice that the case $\epsonestar \propto 1/\Delta\Nstar^2$ (which
appeared as singular in the introductory discussion of this section)
corresponds to the asymptotic limit $b \to + \infty$ while other
values of $b>2$ produce higher-than-two power-law
exponents. Less-than-unity positive exponents can be generated within
NFI2 with $0<b<1$ and $a<0$ while chosing $\xend$ very
small. Exponents between one and two can be generated within NFI3 for
$b<0$ and a small enough $a>0$. When the contribution from $\xend$ is
not negligible, the model predictions deviate from this simple power
law. It is also interesting to remark that the peculiar case $b=2$,
generating the exponential trajectory of \Eq{eq:xtraj2nfi}, yields
\begin{equation}
\epsonestar = \dfrac{2 a^2 \xend^2}{e^{4 a \Delta \Nstar}},
\end{equation}
which is quite different from a power-law dependence in
$1/\Delta\Nstar$. There is nothing surprising in this result as the
potential of NFI written in \Eq{eq:potnfi} has been derived from
\Eq{eq:nfiidea} as a leading-order approximation only. The proper
way to devise an inflationary model generating a given
functional shape for $\epsilon_1(N)$ exactly is what is presented for VFMI in \sectionc{sec:vfmi}.

\subsection{Radiatively Corrected Inflection Point Inflation (RCIPI)}
\label{sec:rcipi}

\subsubsection{Theoretical Justifications}

This class of models has been introduced in \Refc{Ballesteros:2015noa}
and constitutes yet another implementation of the radiative
corrections induced by bosonic and fermionic loops onto a monomial
potential of the large-field type (see \sectioncs{sec:lfi},
\ref{sec:rcmi}, \ref{sec:rcqi} and \ref{sec:rclfi}). More precisely,
\Refc{Ballesteros:2015noa} explores the possibility that the bosonic
and fermionic degrees of freedom are somehow compensated to generate an
inflection-point potential\footnote{In \Refc{Ballesteros:2015noa},
this inflection point is unfortunately referred to as a ``plateau'',
which it is not. See \Refc{Chowdhury:2019otk} for the difference between
plateau inflation and other models.} while never triggering an
instability at large-field values. In the case of a quartic tree-level
potential, $V_4(\phi) = \lambda \phi^4$, the Coleman-Weinberg
corrections (see \sectionc{sec:cwi}), at large values of the field
$\phi$, are expected to generate an effective quartic coupling
\begin{equation}
\lambda(\phi) = \lambda(\phizero) + \dfrac{c_1(\phizero)}{2} \ln
\left(\dfrac{\phi}{\phizero}\right)^2 +
\dfrac{c_2(\phizero)}{8}\left[\ln\left(\dfrac{\phi}{\phizero}\right)^2
  \right]^2 + \cdots,
\end{equation}
where $\phizero$ is the typical field value associated with the
renormalization scale. Requiring the corrected effective potential to have an
inflection point at $\phizero$, namely $V'(\phizero)=V''(\phizero)=0$,
demands that
\begin{equation}
c_2(\phizero) = -4 c_1(\phizero) = 16 \lambda(\phizero),
\label{eq:rcipic12}
\end{equation}
and the resulting inflationary potential reads
\begin{equation}
V_4(\phi) = \lambda(\phizero) \phi^4
\left\lbrace 1 - 2 \ln \left(\dfrac{\phi}{\phizero}\right)^2 + 2
\left[\ln\left(\dfrac{\phi}{\phizero}\right)^2 \right]^2 + \cdots \right\rbrace.
\end{equation}
Because the relation in \Eq{eq:rcipic12} requires some tuning between
the bosonic and fermionic loop corrections, it is not expected to be
exact, simply by the existence of higher order-corrections. One can
therefore introduce two distortion parameters, $b_1$ and $b_2$,
encoding by how much \Eq{eq:rcipic12} is violated. One finally obtains
an effective quartic potential
\begin{equation}
  V_4(\phi) = \lambda \phi^4
  \left\lbrace 1 - 2(1-b_1) \ln \left(\dfrac{\phi}{\phizero}\right)^2 + 2(1+b_2)
  \left[\ln\left(\dfrac{\phi}{\phizero}\right)^2 \right]^2\right\rbrace.
  \label{eq:pot:rpqti}
\end{equation}
As done in \Refc{Ballesteros:2015noa}, the same reasoning can be
applied to a monomial large-field potential being a pure mass term
$V_2(\phi) = \lambda \Mp^2 \phi^2$. In that case, the condition for the effective potential
to exhibit an inflection point at $\phizero$ is modified to
\begin{equation}
c_2(\phizero) = -2 c_1(\phizero) = 4 \lambda(\phizero).
\end{equation}
Including the distortion parameters, the effective quadratic potential becomes
\begin{equation}
  V_2(\phi) = \lambda \phi^2 \Mp^2
  \left\lbrace 1 - (1-b_1) \ln \left(\dfrac{\phi}{\phizero}\right)^2 + \dfrac{1}{2}(1+b_2)
  \left[\ln\left(\dfrac{\phi}{\phizero}\right)^2 \right]^2\right\rbrace.
  \label{eq:pot:rpqdi}
\end{equation}
In both \Eqs{eq:pot:rpqti} and \eqref{eq:pot:rpqdi}, the
renormalization scale $\phizero$ can be completely reabsorbed into a
redefinition of the coupling $\lambda$ and of the distortion parameters
$b_1$ and $b_2$. We therefore define the potential of RCIPI as
\begin{equation}
V(\phi) = M^4 \left(\dfrac{\phi}{\Mp}\right)^p \left[1 + \alpha
  \ln\left(\dfrac{\phi}{\Mp}\right) + \beta \ln^2
    \left(\dfrac{\phi}{\Mp}\right)  \right],
\label{eq:potrcipi}
\end{equation}
which has three parameters, $p$, $\alpha$ and $\beta$. It is defined
for positive field values $\phi \ge 0$ and we assume $p>0$.

Expanding the logarithmic terms of the effective potential $V_4(\phi)$
in \Eq{eq:pot:rpqti}, and matching them to \Eq{eq:potrcipi}, one gets
for $p=4$
\begin{equation}
M^4 = \lambda \Mp^4 \left[1 +
  4(1-b_1)\ln\left(\dfrac{\phizero}{\Mp}\right) + 8(1+b_2)
  \ln^2\left(\dfrac{\phizero}{\Mp}\right) \right],
\label{eq:M4:rpqti}
\end{equation}
with
\begin{equation}
  \begin{aligned}
\alpha & = 
  -\dfrac{4(1-b_1)+16(1+b_2)\ln\left(\dfrac{\phizero}{\Mp}\right)}{1+4(1-b_1)
    \ln\left(\dfrac{\phizero}{\Mp}\right) + 8(1+b_2)\ln^2\left(\dfrac{\phizero}{\Mp}\right)}\,,\\
\beta & = 
\dfrac{8(1+b_2)}{1+4(1-b_1)\ln\left(\dfrac{\phizero}{\Mp}\right)
  + 8(1+b_2)\ln^2\left(\dfrac{\phizero}{\Mp}\right)}\,.
\end{aligned}
\label{eq:alphabeta:rpqti}
\end{equation}
For the quadratic potential $V_2(\phi)$, expanding
\Eq{eq:pot:rpqdi} and identifying the resulting terms with \Eq{eq:potrcipi}
yields for $p=2$
\begin{equation}
  M^4 = \lambda \Mp^4 \left[1 + 2(1-b_1) \ln
    \left(\dfrac{\phizero}{\Mp}\right) + 2(1+b_2)
    \ln^2\left(\dfrac{\phizero}{\Mp} \right) \right],
\label{eq:M4:rpqdi}
\end{equation}
with
\begin{equation}
  \begin{aligned}
    \alpha &= - \dfrac{2(1-b_1) + 4(1+b_2)
      \ln\left(\dfrac{\phizero}{\Mp}\right)}{1 + 2(1-b_1) \ln
      \left(\dfrac{\phizero}{\Mp}\right) + 2(1+b_2)
      \ln^2\left(\dfrac{\phizero}{\Mp} \right)}\,,\\
   \beta &= \dfrac{2(1+b_2)}{1 + 2(1-b_1) \ln
      \left(\dfrac{\phizero}{\Mp}\right) + 2(1+b_2)
      \ln^2\left(\dfrac{\phizero}{\Mp} \right)}\,.
  \end{aligned}
\label{eq:alphabeta:rpqdi}
\end{equation}
Let us notice that a convenient renormalization scale for simplifying
these expressions is $\phizero=\Mp$, for which an inflection point
appears at $\alpha_4=-4$, $\beta_4=8$ and $\alpha_2=-2$,
$\beta_2=2$ (and, more generally, at $\alpha_p=-p$ and $\beta_p=p^2/2$). These expressions also show that, as soon as the
inflection point is detuned, a tachyonic mass term appears and
slow-roll violations are present in a very same fashion as
for SFI2 (see \sectionc{sec:sfi}).

In order for the potential of RCIPI to remain positive, as requested
from its desired stability, one has to impose some restrictions on the
parameters $\alpha$ and $\beta$. At large field values, the last term
of \Eq{eq:potrcipi} dominates and the potential is positive only for
$\beta \ge 0$. Moreover, to prevent the potential to become negative
in some intermediate domain, one has to impose
\begin{equation}
\alpha ^2 \le 4 \beta.
\end{equation}
In the following, we study the generic potential of RCIPI under these
conditions while its observable predictions can be narrowed down to
the quadratic and quartic cases by using \Eqs{eq:M4:rpqti} to
\eqref{eq:alphabeta:rpqdi}. Let us however notice that the regimes for
which the potential could become negative, when driven by the
first-order loop corrections, have already been studied within the
RCLFI scenario in \sectionc{sec:rclfi}.

\begin{figure}
\begin{center}
\includegraphics[width=\wsingfig]{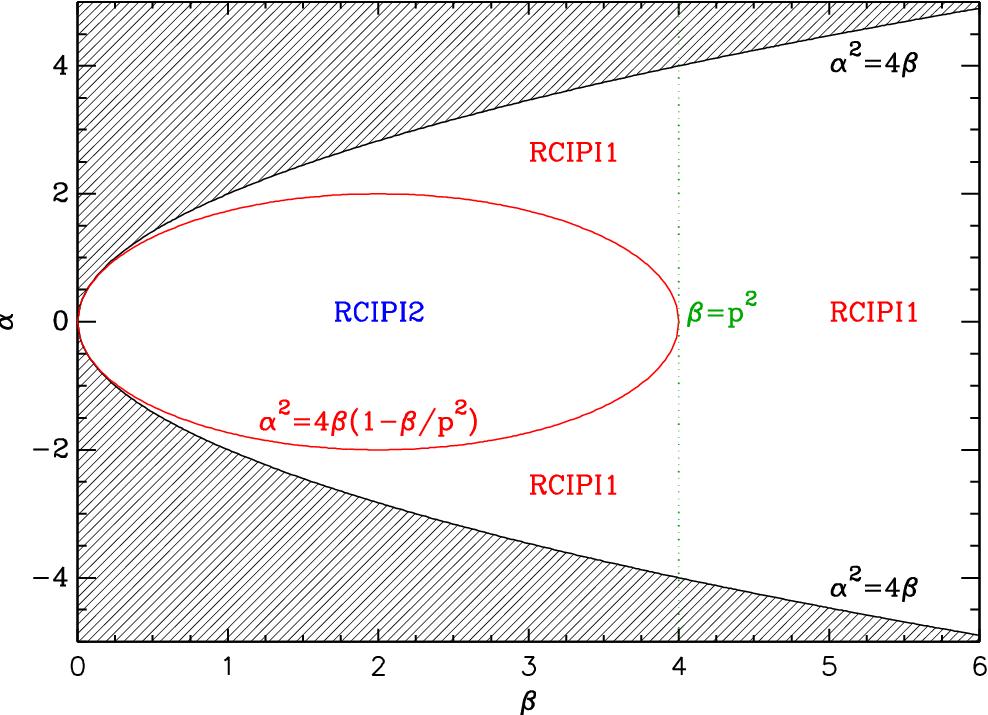}
\caption{Location of the inflationary regimes RCIPI1 and RCIPI2 in the
  parameter space $(\beta,\alpha)$ for $p=2$. The hatched region
  associated with $\alpha^2 > 4 \beta$ is not considered as the
  potential would become negative in some field domains. The red
  ellipse, defined by $\alpha = \pm \alphazero$, separates the domain in
  which the potential is a monotonic increasing function of the field
  (inside) from the ones in which it develops a false vacuum
  (outside). The RCIPI1 regimes are hilltop models starting from a
  local maximum of the potential and evolving towards the origin. The
  RCIPI2 regime is a large field-like inflationary model, running from
  large to small field values. On the ellipse, the potential has a
  flat inflection point and only RCIPI1 is considered. This is because, since the number of \efolds~diverges at the flat inflection point [see the discussion below \Eq{eq:rcipi:trajo}], inflation never ends in RCIPI2 on the ellipse.}
\label{fig:alphazerorcipi}
\end{center}
\end{figure}

\subsubsection{Parameter Space Analysis}

For $\alpha^2 \le 4 \beta$ and $\beta \ge 0$, the potential is
well-defined and positive for $\phi \ge 0$. It has been constructed to
possibly have an inflection point, and moving away from this
situation, one can have either a monotonic increasing potential with
respect to $\phi$, or the appearance of both a local maximum and
minimum. Defining
\begin{equation}
x \equiv \dfrac{\phi}{\Mp}\,,
\end{equation}
the potential is extremal when $V'(x)=0$, {\ie} for $x$ solution of
\begin{equation}
x^{p-1} \left[p + \alpha + (p \alpha +2 \beta) \ln(x) + p \beta
  \ln^2(x) \right] = 0.
\label{eq:rcipi:Vprime}
\end{equation}
For $p>1$, there is an obvious solution at $x=0$, which is also the
global vanishing minimum of the potential. We are rather interested in
the roots of the second term, which is a quadratic equation in
$\ln(x)$. The existence of these roots depends on the sign of the
determinant
\begin{equation}
\Delta = 4 \beta^2 - p^2\left(4 \beta - \alpha^2 \right).
\label{eq:rcipi:Delta}
\end{equation}
The inflection point is recovered when there is only one root to this
equation, namely for $\Delta=0$ or $\alpha = \pm \alphazero$ with
\begin{equation}
\alphazero \equiv 2 \sqrt{\beta} \, \sqrt{1- \dfrac{\beta}{p^2}}\,.
\end{equation}
As a result, the inflection point exists only in the domains for which
$\beta \le p^2$. If $|\alpha| < \alphazero$ then \Eq{eq:rcipi:Delta} becomes
negative and the potential is a monotonic increasing function of the
field values. For $|\alpha| > \alphazero$  we have $\Delta >0$ and \Eq{eq:rcipi:Vprime}
admits the two roots
\begin{equation}
\xdVzeroPM = \exp\left[\dfrac{-(p \alpha + 2 \beta) \pm \sqrt{4
      \beta^2 - p^2\left(4 \beta-\alpha^2\right)}}{2 p \beta} \right].
\label{eq:rcipi:xdVzero}
\end{equation}
For all the other cases, associated with $\beta > p^2$, $\Delta > 0$
for all values of $\alpha$ and the potential always develops two
extrema at the locations given by \Eq{eq:rcipi:xdVzero}. Let us notice
that for $\beta \le p^2$ and $\alpha=\pm \alphazero$ the position of the infection point is
also given by \Eq{eq:rcipi:xdVzero} in which the square root term
vanishes and reads
\begin{equation}
\xzero = \exp\left(- \dfrac{p \alpha + 2 \beta}{2 p \beta}\right).
\end{equation}
As soon as \Eq{eq:rcipi:xdVzero} has two distinct solutions, the
potential of RCIPI develops a false vacuum at $x=\xdVzeroPlus$ in which
inflation would become eternal. Within slow-roll, and without
resorting to an additional mechanism ending such a phase of inflation
(which would make RCIPI a four-parameter model), this implies that
slow-roll inflation may gracefully end only if it moves away from this
false vacuum by going towards the global minimum at $x=0$. We refer to
this regime as RCIPI1 and it exists only in the domain $x <
\xdVzeroMinus$. As such, it requires either $\beta > p^2$, or
$|\alpha| \ge \alphazero$ when $\beta \le p^2$. RCIPI1 also encompasses
the case of a flat inflection point. Let us notice that non-slow
roll solutions may exist, as for instance if the field acquires enough
kinetic energy in a previous inflationary phase to climb out of the
false vacuum, but these situations necessarily violate slow roll. The
other slow-roll inflationary regime corresponds to the situation in
which the potential is a monotonic increasing function of the
field. This happens only for $\beta \le p^2$ with $|\alpha| < \alphazero$
and inflation proceeds from large to small field values. This
regime will be referred to as RCIPI2. The location of these two regimes
in the parameter space $(\beta,\alpha)$ have been represented in
\Fig{fig:alphazerorcipi} for $p=2$.

\subsubsection{Slow-roll Analysis}

\begin{figure}
\begin{center}
\includegraphics[width=\wdblefig]{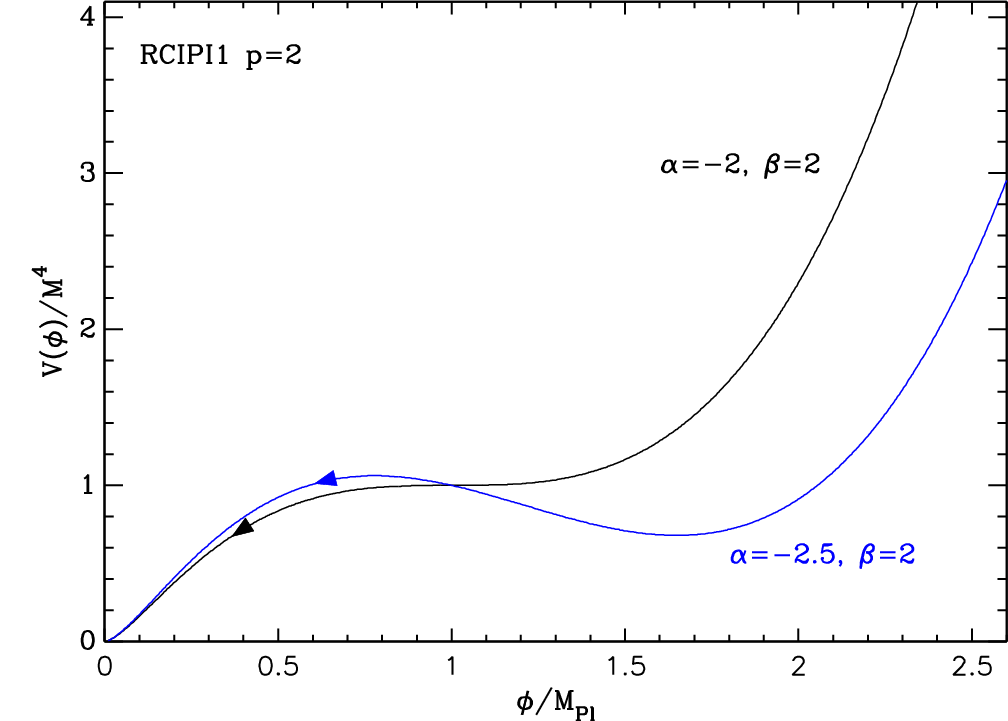}
\includegraphics[width=\wdblefig]{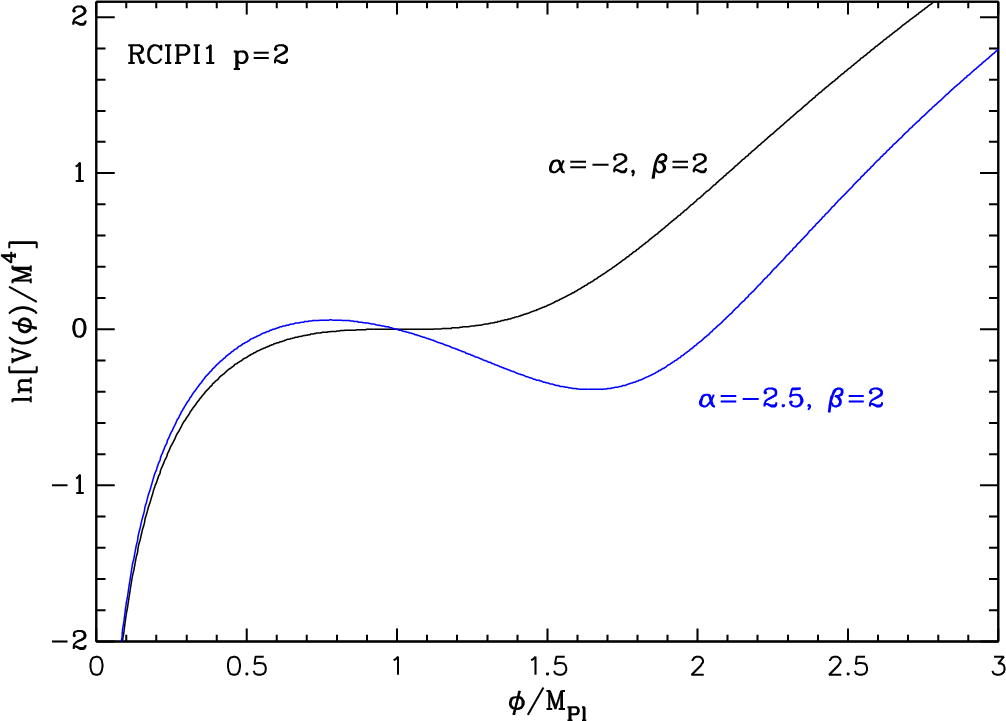}
\includegraphics[width=\wdblefig]{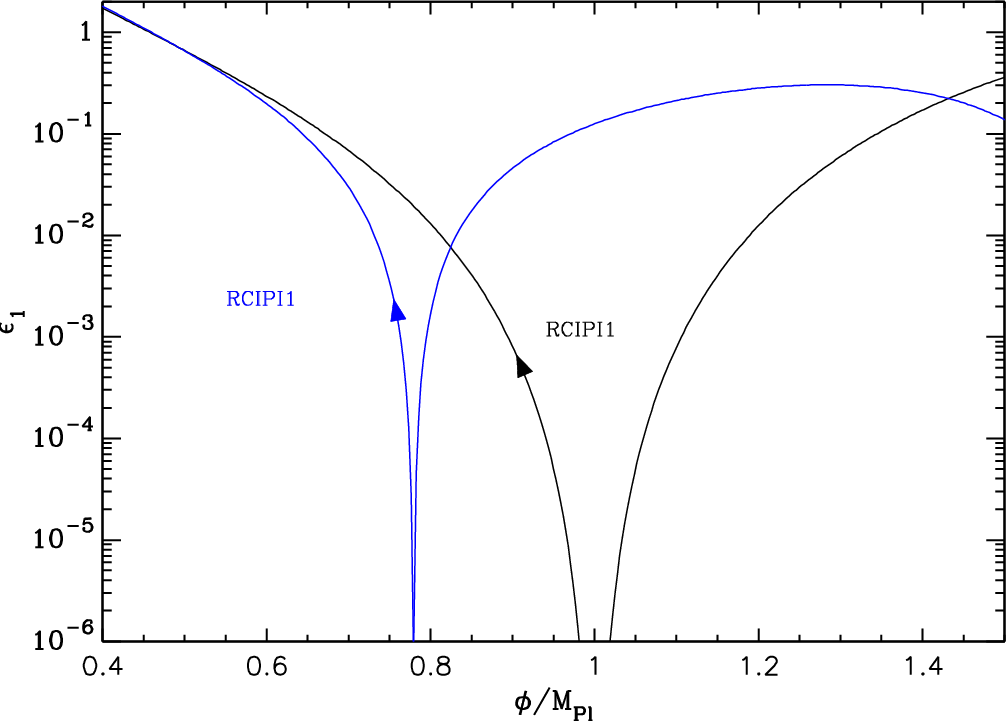}
\includegraphics[width=\wdblefig]{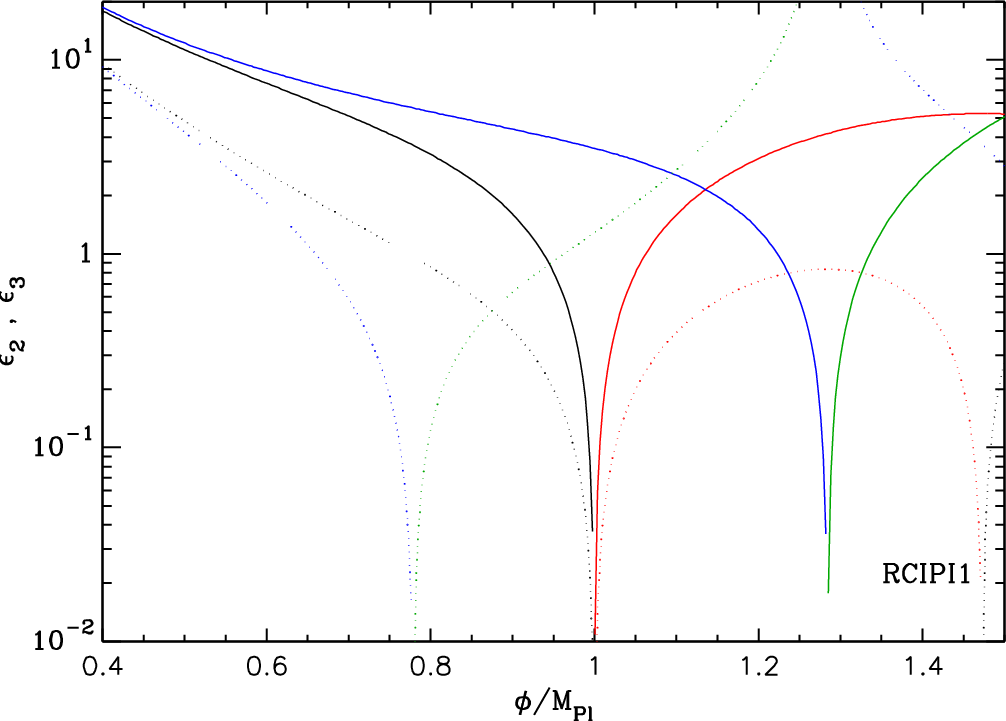}
\caption{Radiatively Corrected Inflection Point Inflation for $p=2$ in
  the regime RCIPI1. Inflation proceeds at decreasing field values
  from an inflection point (black curves), which can be detuned to a
  hilltop (blue curves). Top panels: the potential and its
  logarithm. Bottom left panel: slow-roll parameter
  $\epsilon_1$. Bottom right panel: slow-roll parameters $\epsilon_2$
  (solid curves) and $\epsilon_3$ (dotted curves). The opposite of the
  negative values of $\epsilon_2$ and $\epsilon_3$ have been
  represented in red for $\alpha=-2$, $\beta=2$ and in green for the
  detuned case with $\alpha=-2.5$, $\beta=2$. Notice that as soon as
  the inflection point becomes a hilltop, $\epsilon_2$ can become
  large due to the appearance of a tachyonic mass term (see also SFI2,
  \sectionc{sec:sfi}).}
\label{fig:potrcipi1}
\end{center}
\end{figure}

\begin{figure}
\begin{center}
\includegraphics[width=\wdblefig]{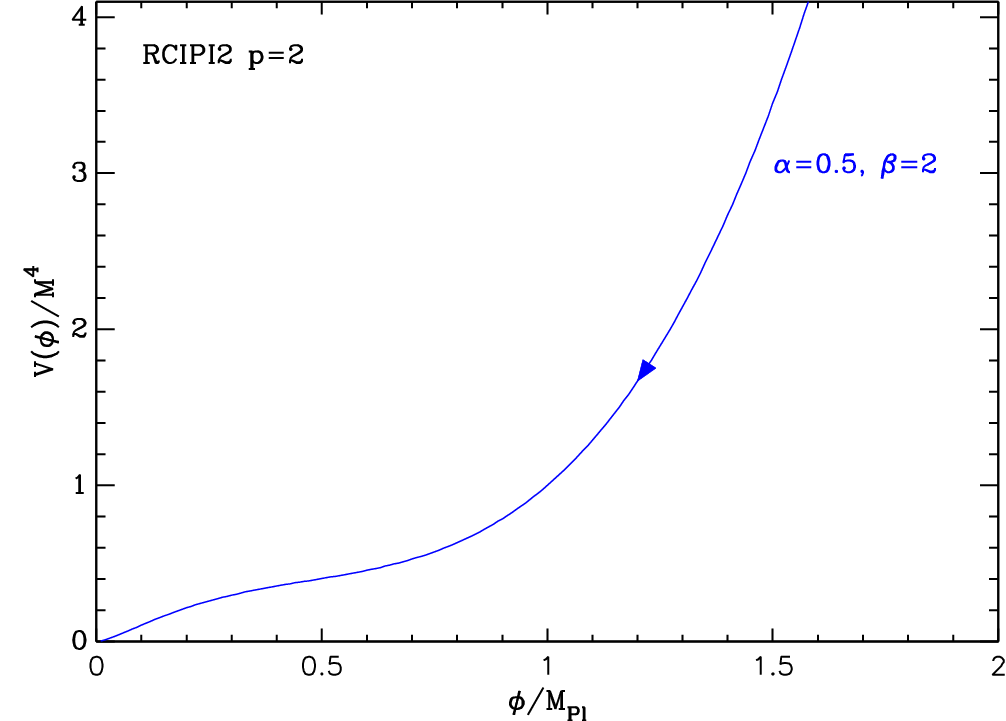}
\includegraphics[width=\wdblefig]{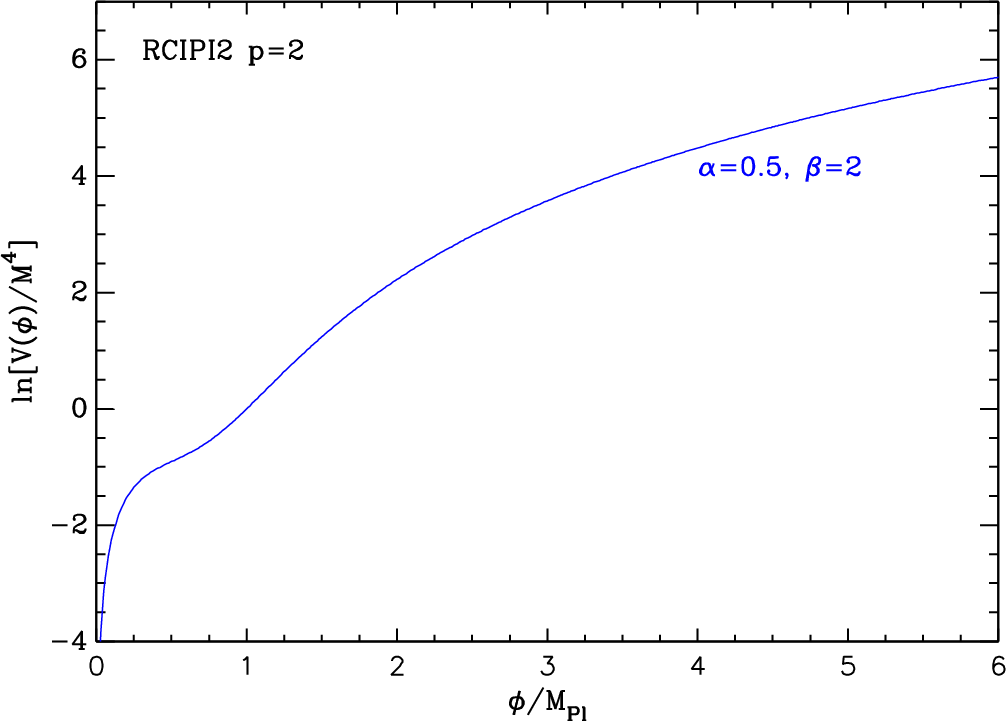}
\includegraphics[width=\wdblefig]{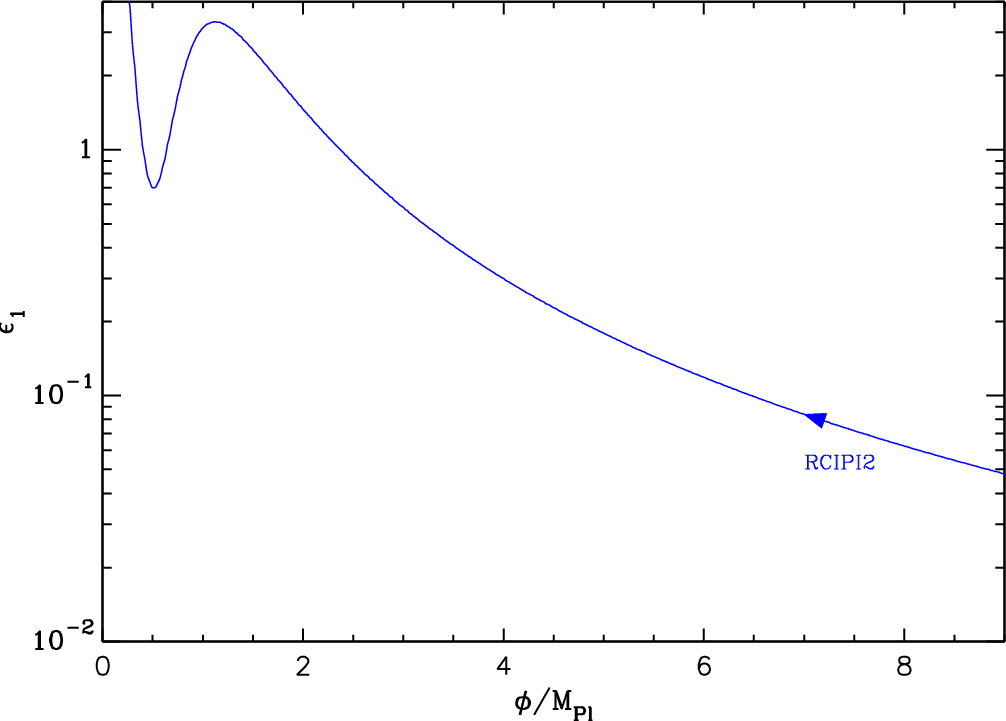}
\includegraphics[width=\wdblefig]{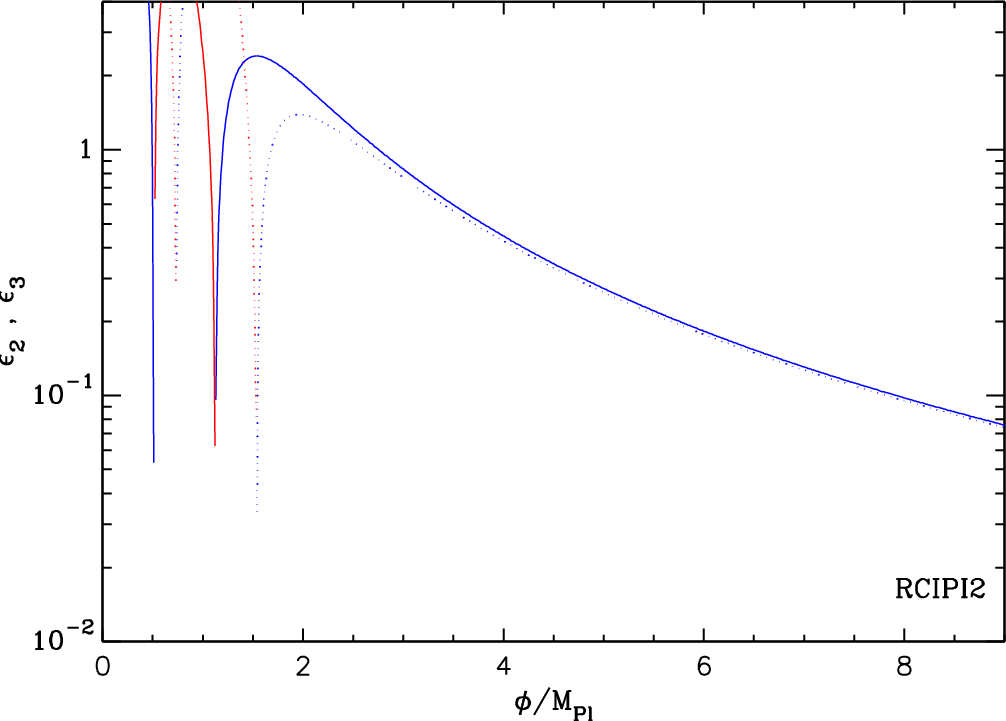}
\caption{Radiatively Corrected Inflection Point Inflation for $p=2$ in the
  regime RCIPI2. The potential is a monotonic increasing function of
  the field values and inflation may proceed at large field values. Top
  panels: the potential and its logarithm. Bottom left panel:
  slow-roll parameter $\epsilon_1$. Bottom right panel: slow-roll
  parameters $\epsilon_2$ (solid curves) and $\epsilon_3$ (dotted
  curves). The opposite of the negative values of $\epsilon_2$ and
  $\epsilon_3$ have been represented in red. Notice that inflation
  ends with ``hiccups'', slow-roll violations occur but remain confined
  at the end of inflation.}
\label{fig:potrcipi2}
\end{center}
\end{figure}

The first Hubble-flow function in the slow-roll approximation, stemming
from the potential of \Eq{eq:potrcipi}, reads
\begin{equation}
\epsilon_1 = \dfrac{\left[p+\alpha + (p \alpha + 2 \beta) \ln(x) +
    p\beta \ln^2(x)\right]^2}{2 x^2\left[1 + \alpha \ln(x) + \beta
    \ln^2(x) \right]^2}\,.
\label{eq:sr1rcipi}
\end{equation}
The second and third Hubble-flow functions are given by
\begin{equation}
\epsilon_2 = \dfrac{2}{x^2} \left\{p + \dfrac{\alpha^2 - 4
  \beta}{\left[1+ \alpha \ln(x)
    + \beta \ln^2(x) \right]^2} +
\dfrac{\alpha + 2\beta + 2\beta \ln(x)}{1 + \alpha \ln(x) + \beta
  \ln^2(x)} \right\},
\label{eq:sr2rcipi}
\end{equation}
and
\begin{equation}
\begin{aligned}
  \epsilon_3 &= \dfrac{p + \alpha + \ln(x)\left[p\alpha + 2\beta +
      p\beta \ln(x) \right]}{x^2 \left\{1 + \ln(x) \left[\alpha + \beta
        \ln(x) \right]\right\}^4 \left\{p + \dfrac{\alpha^2 - 4
      \beta}{\left[1
        + \alpha \ln(x) + \beta \ln^2(x) \right]^2} +
\dfrac{\alpha + 2\beta + 2\beta \ln(x)}{1 + \alpha \ln(x) + \beta
  \ln^2(x)} \right\}} \\
  & \times \Bigg\{ 2\left(\alpha^2 - 4\beta\right)\left[\alpha + 2\beta
  \ln(x) \right] + \left[\alpha + 2\beta \ln(x) \right] \left[\alpha +
    2\beta + 2\beta \ln(x) \right] \left[1 + \alpha \ln(x) + \beta
    \ln^2(x) \right] \\ & - 2 \beta \left[1 + \alpha \ln(x) + \beta
    \ln^2(x) \right]^2 + 2 \left(\alpha^2 - 4\beta\right) \left[1 + \alpha \ln(x) + \beta \ln^2(x)
    \right]  \\ & + 2 \left[\alpha + 2\beta + 2\beta
      \ln(x) \right] \left[1 + \alpha \ln(x) + \beta \ln^2(x) \right]^2 +
    2p \left[1 + \alpha \ln(x) + \beta \ln^2(x)\right]^3 \Bigg\}.
\end{aligned}
\label{eq:sr3rcipi}
\end{equation}
In the limit $x \to 0$ one has $\epsilon_1(x) \simeq p^2 \ln^2(x)/(2
x^2)$, which is divergent, and this ensures that inflation always
gracefully ends when approaching the global minimum of the potential
at $x=0$. The field value $\xend$ at which this occurs is solution of
the equation $\epsilon_1(x)=1$, {\ie}
\begin{equation}
p + \alpha + (p \alpha + 2\beta)\ln(x) + p\beta \ln^2(x) = \pm
\sqrt{2} x \left[1+\alpha \ln(x) + \beta \ln^2(x) \right].
\label{eq:rcipi:xepsoneone}
\end{equation}
It does not admit analytical solution and has to be solved
numerically. Moreover it has multiple roots in domains that can be
determined by studying the sign of $\epsilon_2(x)$. This one vanishes
for
\begin{equation}
p \left[1+ \alpha \ln(x) + \beta \ln^2(x) \right]^2 + \left[\alpha +
  2\beta + 2\beta \ln(x) \right] \left[1 + \alpha \ln(x) + \beta
  \ln^2(x) \right] + \alpha^2 - 4 \beta = 0.
\end{equation}
This is a quartic equation in $\ln(x)$ and it admits, at maximum, four
roots that can be determined analytically and can be found in the
{\ASPIC} library. When the potential develops a false vacuum one of
these roots lies between the locations of the two potential extrema because,
there, $\epsilon_1$ vanishes so it must have a local maximal (where $\epsilon_2$ vanishes) in between. The other roots correspond to other local extrema of $\epsilon_1(x)$, which can
potentially exceed unity. However, RCIPI1 proceeds at decreasing field
values, always with $x<\xdVzeroMinus$, and one can show that it never
encounters one of the local extrema of $\epsilon_1$. Therefore, only
the smallest of all roots of \Eq{eq:rcipi:xepsoneone} gives $\xend$,
the field value at which RCIPI1 ends. Concerning RCIPI2, the situation
is more complex. The potential is a monotonic increasing function of
$x$ and RCIPI2 inflation proceeds at decreasing field values, from
large to small values of $x$. Although there is no inflection point,
the potential exhibits a change of convexity and $\epsilon_2$ may
still vanish before the end of inflation. Because $\epsilon_1$ then
develops a local maximum, possibly exceeding unity, inflation can be
potentially halted before restarting for a few {\efolds} to definitely
end at $\xend$, this one being still defined as the lowest root of
\Eq{eq:rcipi:xepsoneone}. These ``hiccups'' are necessarily associated
with slow-roll violations, but, being located close to the end of
inflation, they are not directly observable in the CMB. As it is the
case for SAIII3 (see \sectionc{sec:saiii}), they might however be of some interest for the formation of primordial black holes.

The potential, and the Hubble-flow functions, in the RCIPI1 regimes
have been represented in \Fig{fig:potrcipi1} in the case of an
inflection point, for $p=2$, $\alpha=-2$, $\beta=2$, as well as when
it is strongly detuned to a hilltop ($p=2$, $\alpha=-2.5$ and
$\beta=2$). The potential and Hubble flow functions for RCIPI2, when
the potential is monotonic, are represented in \Fig{fig:potrcipi2}. As
can be seen on the lower left panel of this figure, $\epsilon_1$ may
transiently exceed unity before inflation definitely ends.

The slow-roll trajectory is given by the integral
\begin{equation}
\Nend - N = \int_{\xend}^x \dfrac{y + \alpha y \ln(y) + \beta y
  \ln^2(y)}{p+\alpha + (p \alpha + 2\beta)\ln(y) + p\beta \ln^2(y)}\,
\ud y.
\label{eq:rcipi:trajint}
\end{equation}
It can be determined analytically after expanding the denominator, which
is proportional to $V'(y)$, over its roots. Defining the, possibly complex, numbers
\begin{equation}
z_{\pm} \equiv \dfrac{-(p \alpha + 2 \beta) \pm \sqrt{\Delta}}{2p \beta}\,,
\end{equation}
one has for $\Delta \ne 0$
\begin{equation}
\begin{aligned}
  \Nend - N & = \dfrac{1}{\sqrt{\Delta}} \int_{\xend}^x \dfrac{y + \alpha y \ln(y) + \beta y
  \ln^2(y)}{\ln(y) - z_+}\,\ud y - \dfrac{1}{\sqrt{\Delta}}  \int_{\xend}^x \dfrac{y + \alpha y \ln(y) + \beta y
  \ln^2(y)}{\ln(y) - z_-}\,\ud y.
\end{aligned}
\label{eq:rcipi:trajexpand}
\end{equation}
Let us notice that, for $\Delta > 0$, one has $\xdVzeroPM =
\exp(z_\pm)$. All the terms obtained by expanding the numerators of
\Eq{eq:rcipi:trajexpand} can be expressed as exponential integrals,
and, after some algebra, one gets
\begin{equation}
\begin{aligned}
\Nend - N & = \dfrac{x^2-\xend^2}{2 p} + \dfrac{e^{2 z_+}\left(1 +
    \alpha z_+ + \beta z_+^2\right)}{\sqrt{\Delta}} \left\{\Ei\left[2
  \ln(x)- 2z_+\right] - \Ei\left[2 \ln(\xend)- 2z_+\right] \right\} \\
 & -  \dfrac{e^{2 z_-}\left(1 +
    \alpha z_- + \beta z_-^2\right)}{\sqrt{\Delta}} \left\{\Ei\left[2
  \ln(x)- 2z_-\right] - \Ei\left[2 \ln(\xend)- 2z_-\right] \right\}.
\end{aligned}
\label{eq:rcipi:trajpm}
\end{equation}
Let us notice that, for RCIPI1, when $x \to \xdVzeroMinus$ the
argument of the exponential integral becomes very small, and
negative. In this limit, one has $\Ei(x) \to \gamma + \ln(-x)$ and
the logarithmic divergence implies that an arbitrarily large number
of {\efolds} can be realized at the top of the local maximum. However,
the argument of the exponential integral being a logarithm of the
field value, it is very slowly divergent, not faster than $\Delta N
\propto -\ln[-2\ln(x/\xdVzeroMinus)]$. In other words, to obtain $60$
{\efolds} of inflation at the top of the local maximum, one
should fine tune the initial field values as $x/\xdVzeroMinus <
\exp[-e^{-60}] \simeq 1 - e^{-60}$.

The case of a flat inflection point for $\alpha = \pm \alphazero$
implies that $\Delta = 0$ and requires special treatment. Denoting
\begin{equation}
z_\zero \equiv \ln(\xzero) = - \dfrac{p \alpha + 2\beta}{2p \beta}\,,
\end{equation}
one can replace the denominator of \Eq{eq:rcipi:trajint} by
\begin{equation}
p+\alpha + (p \alpha + 2\beta)\ln(y) + p\beta \ln^2(y) = p \beta \left[\ln(y) - z_\zero \right]^2.
\end{equation}
After expressing all terms of \Eq{eq:rcipi:trajint} as exponential
integrals, one gets for $\Delta = 0$
\begin{equation}
  \begin{aligned}
    \Nend - N & = e^{2 z_\zero} \left[2 + \alpha +2(\alpha+\beta) z_\zero
      + 2 \beta z_\zero^2\right]\left\{
    \Ei\left[2\ln(x)-2z_\zero\right] -
    \Ei\left[2\ln(\xend)-2z_\zero\right] \right\} \\
    & + \dfrac{x^2\left[2 + (2\alpha+\beta)z_\zero +2 \beta
        z_\zero^2 -\beta \ln(x) \right]}{2 z_\zero -2 \ln(x)} \\ &-
    \dfrac{\xend^2\left[2 + (2\alpha+\beta)z_\zero +2 \beta z_\zero^2
        -\beta \ln(\xend) \right]}{2 z_\zero -2 \ln(\xend)}\,.
  \end{aligned}
\label{eq:rcipi:trajo}
\end{equation}
This time, for RCIPI1 and $x \to \xdVzeroMinus$, we see that $\Delta
N$ is divergent as $1/\ln(\xzero/x)$ and getting enough {\efolds} of
inflation close to the inflection point requires only $x/\xzero <
e^{-1/\Delta N} \simeq 1 - 1/\Delta N$.

From the numerical solution $\xend$, the analytical trajectories of
\Eqs{eq:rcipi:trajpm} and \eqref{eq:rcipi:trajo}, and the reheating
equation~\eqref{eq:phistarlnrrad}, one can determine $\xstar$, the
field value at which the pivot mode crosses the Hubble radius during
inflation. It fixes the normalization of the potential $M^4$ from the
amplitude of the CMB anisotropies and one has
\begin{equation}
\left(\dfrac{M}{\Mp}\right)^4 = 720 \pi^2 \dfrac{\left[p+\alpha +
    (p\alpha + 2\beta) \ln(\xstar) + p \beta
    \ln^2(\xstar)\right]^2}{\xstar^{p+2} \left[1 + \alpha \ln(\xstar)
    + \beta \ln^2(\xstar) \right]^3} \dfrac{\Qrms^2}{T^2}\,.
\end{equation}

The reheating-consistent slow-roll predictions for RCIPI1 are
represented in \Figs{fig:CMBRCIPI1_0} to \ref{fig:CMBRCIPI1_14} and
for RCIPI2 in \Figs{fig:CMBRCIPI2_0} to \ref{fig:CMBRCIPI2_10}. As
already noticed, successful inflation near the detuned inflection
point requires quite some fine-tuning and the presence of a tachyonic
mass induces a strong sensitivity of the observables with respect to
the model parameters. For both RCIPI1 and RCIPI2, small values of
$\beta$ allow for just enough inflation to make observable the
``distorted parts of the potential. Due to the potentially large
values of $\epsilon_2$ and $\epsilon_3$, the model predictions explore
a large part of the space $(\nS,r)$ even for small changes in the parameter
$\alpha$ and $\beta$.

\section{Conclusions}
\label{sec:conclusion}

\begin{figure}
\begin{center}
\includegraphics[width=\wappfig,clip=true]{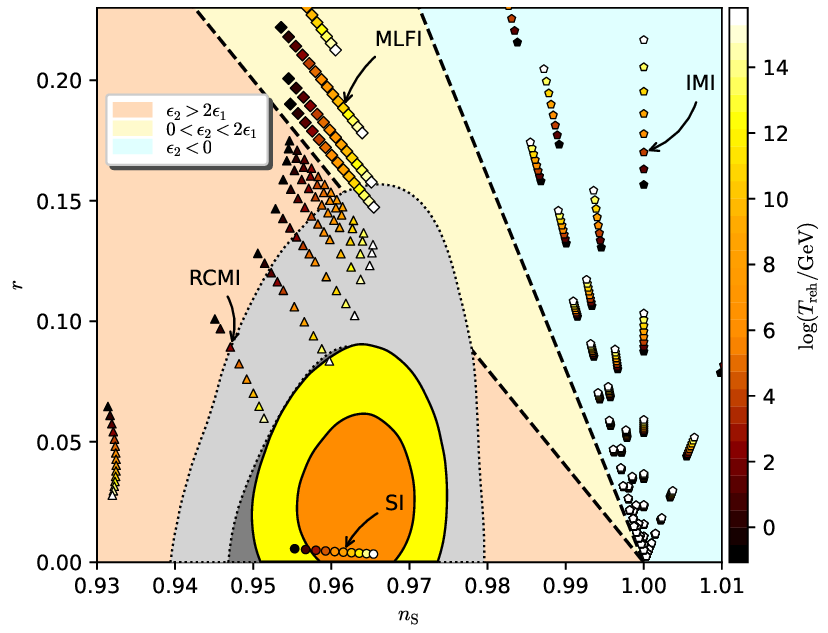}
\includegraphics[width=\wappfig,clip=true]{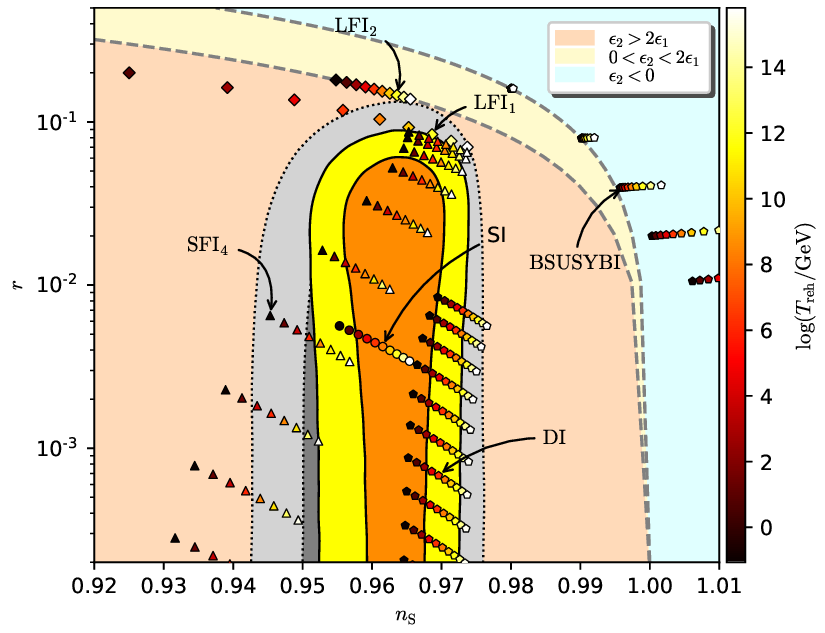}
\end{center}
\caption{Upper panel: various \ASPIC scenarios in the $(\nS,r)$ plane
  using the Schwarz-Terrero-Escalante
  classification~\cite{Schwarz:2004tz} and compared to the {\data}
  data (yellow contours) and the Planck 2013 data~\cite{Ade:2013ktc,
    Planck:2013kta, Ade:2013ydc, Ade:2013xla,Ade:2013uln,Ade:2013zuv}
  (light gray shading). Bottom panel: same plot in logarithmic scale
  for another sample of models.}
\label{fig:nsRconclusion}
\end{figure}

\begin{figure}
\begin{center}
\includegraphics[width=\wappfig]{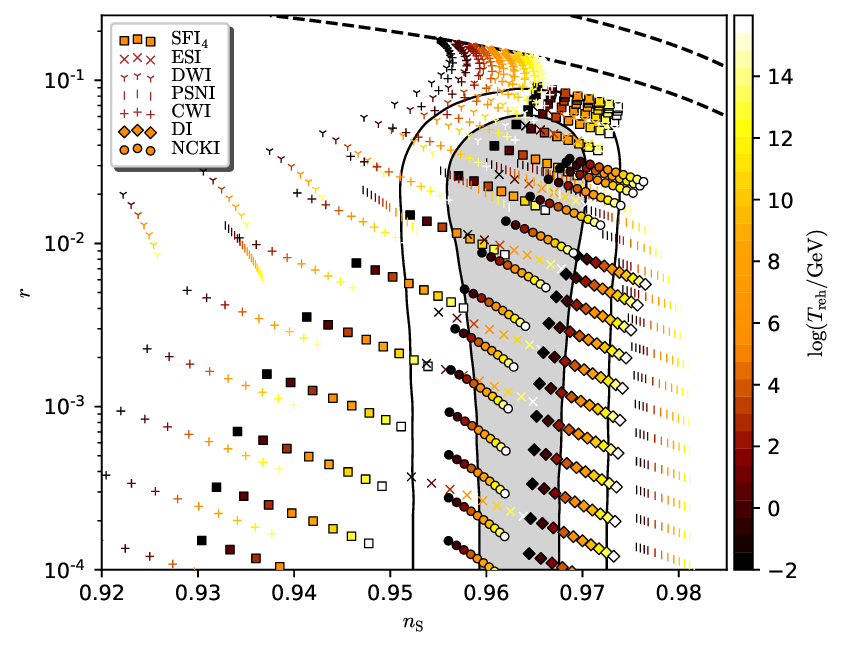}
\caption{Observable predictions in the $(\nS,r)$ plane for various
  models belonging to region $1$ of the Schwarz-Terrero-Escalante
  classification (see \Fig{fig:nsRconclusion}). Despite the fact that
  they are in the same broad class, the accuracy of the CMB data
  allows us to discriminate among them thereby justifying a detailed
  navigation within the inflationary landscape.}
\label{fig:nsRSFILike}
\end{center}
\end{figure}

Let us very briefly recap our main findings and present some
directions for future works.

In this article, we have discussed the question of how the
inflationary theory can be constrained given that we now have at our
disposal high-accuracy cosmological data. We have argued that this can
be done by means of the slow-roll power spectrum which has the
advantage of being relatively model independent. Concretely, it leads
to the Hubble flow posterior distributions $P\left(\epsilon_n |
\Clm\right)$. This is interesting since it gives a general constraint
on the derivatives of the inflaton potential. But, at the same time,
this does not answer some legitimate fundamental questions one might
have about the plethora of inflationary scenarios studied so far. For
instance, it does not tell us rigorously which constraints exist on
the parameters of a given model. Indeed, suppose that we are
interested in LFI, $V(\phi)\propto \phi^p$. It is obvious that we
would like to know which values of $p$ are favored by the data for
this class of models.

In order to complement the slow-roll approximated power spectra and to
address the above mentioned issues, we have argued that it is
interesting to scan the inflationary landscape model by model and have
provided the public code $\ASPIC$ to do so. Such a strategy has to be
done for all the inflationary scenarios since it would be arbitrary to
consider only a restricted class while ignoring the others. In fact,
this question deserves to be discussed in more detail. One could
indeed imagine that it is not necessary to consider all the models one
by one and that considering a representative for each class is
sufficient. Indeed, before this work, it was common to distinguish
three broad types of scenarios: large field models (LFI), small field
models (SFI) and Hybrid models (VHI). Such a classification is not
very precise and biased because it pushes to the front line these
three models. It could be reasonably argued that a better
classification is the one of Schwarz and Terrero-Escalante introduced
in \Refc{Schwarz:2004tz}. For a scalar field, the ratio of the kinetic
energy to the total energy density is given by
$\epsilon_1/3=\dot{\phi}^2/(2\rho)$. Because $\epsilon_2$ is, by
definition, the logarithmic derivative of $\epsilon_1$ with respect to
the \efold number, the kinetic contribution to the total energy
density increases if $\epsilon_2>0$ and decreases if
$\epsilon_2<0$. On the other hand, we also have
\begin{equation}
\dfrac{\dd (\dot{\phi}^2/2)}{\dd t}
=H \dfrac{\dot{\phi}^2}{2} \left(\epsilon_2-2\epsilon_1\right),
\end{equation}
and, therefore, the absolute value of the kinetic energy increases if
$\epsilon_2>2\epsilon_1$ whereas it decreases if
$\epsilon_2<2\epsilon_1$. This allows us to identify three different
regions: $\epsilon_2>0$ and $2\epsilon_1 < \epsilon_2$ (region $1$),
$\epsilon_2<2\epsilon_1$ (region $2$), $\epsilon_2<0<2\epsilon_1$
(region $3$).

These three regions are identified in \Fig{fig:nsRconclusion},
together with the Planck 2013 and the {\data} bounds\footnote{The slight shift
visible on the one- and two-sigma contours between the two plots come
from the different priors used, either flat on $\epsilon_1$ or flat on
$\log \epsilon_1$ (Jeffreys' prior).}. If we use the first-order
slow-roll expressions, the condition $\epsilon_2>0$ is equivalent to
$r<8(1-\nS)$ while $\epsilon_2>2\epsilon_1$ amounts to
$r<4(1-\nS)$. These two lines are also represented in
\Fig{fig:nsRconclusion} (solid black lines). These three regions
encompass the large field, small field and hybrid cases previously
mentioned. However, the correspondence is not perfect and we notice,
for instance, that LFI for $p=1$ belongs to region $1$ whereas for
$p>2$ it belongs to region $2$.

Having identified three broad classes of scenarios, the next question
is whether testing only a representative model for each class could be
sufficient. For instance, the confidence regions of
\Fig{fig:nsRconclusion} show that, over a decade of CMB measurements,
the cosmological data have essentially excluded region $2$ and region
$3$ and one may be tempted to summarize region $1$ to one of its
prototypical representative: Starobinsky Inflation (SI). In
\Fig{fig:nsRSFILike}, we have considered the predictions of seven
other different models that all belong to region $1$. This plot
clearly shows that these models make predictions spanning different
domains that are separated enough to be distinguishable within current
or future data. Given the quality of the current data, working only
with broad classes of models seems to be no longer
justified. Therefore, if one really wants to scan the inflationary
landscape, the approach advocated in this paper is well-suited.

With \ASPIC, we have provided a new tool to treat any model of
inflation and this has led us to derive observational predictions for
$\Nmodel$ models. \ASPIC is an evolutive project and therefore the
next steps will be to complete and upgrade it with new
models. Finally, the ultimate goal is to identify which $\ASPIC$
models are performing the best at explaining cosmological data. In
order to carry out this task, an appropriate method is to use Bayesian
evidence and model comparison. Then, we should be able to identify, in
a statistically well-defined manner, and for any data sets, what might
be called ``the best model of inflation''~\cite{Martin:2014rqa,
  Vennin:2015eaa, CORE:2016ymi, Hardwick:2018zry, Martin:2024nlo}.

\newpage

\appendix

\section{Reheating consistent slow-roll predictions}
\label{sec:predictions}

\subsection{Starobinksy and Higgs Inflation (\hyperref[sec:si]{SI}/\hyperref[sec:hi]{HI})}

\begin{figure}[H]
\begin{center}
\includegraphics[width=\wappfig,clip=true]{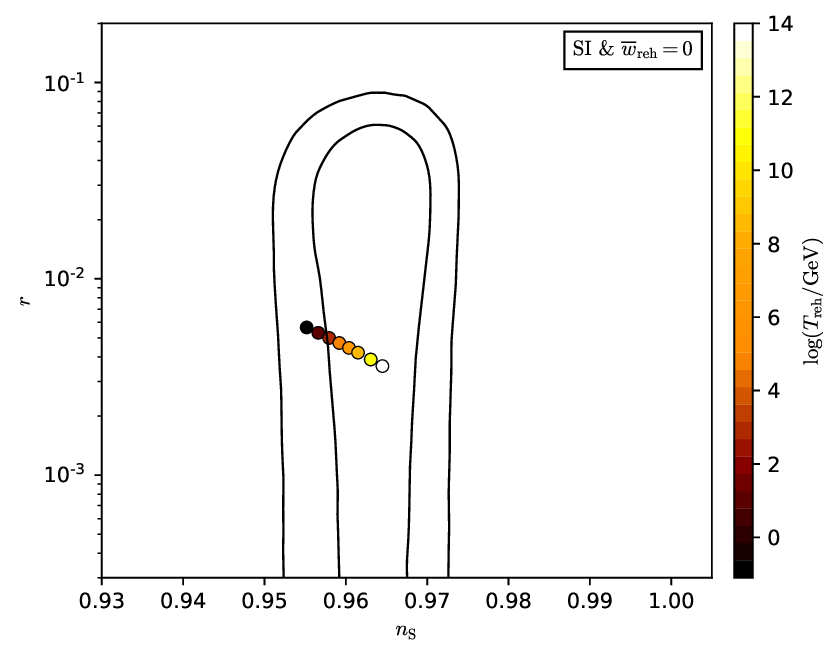}
\includegraphics[width=\wappfig,clip=true]{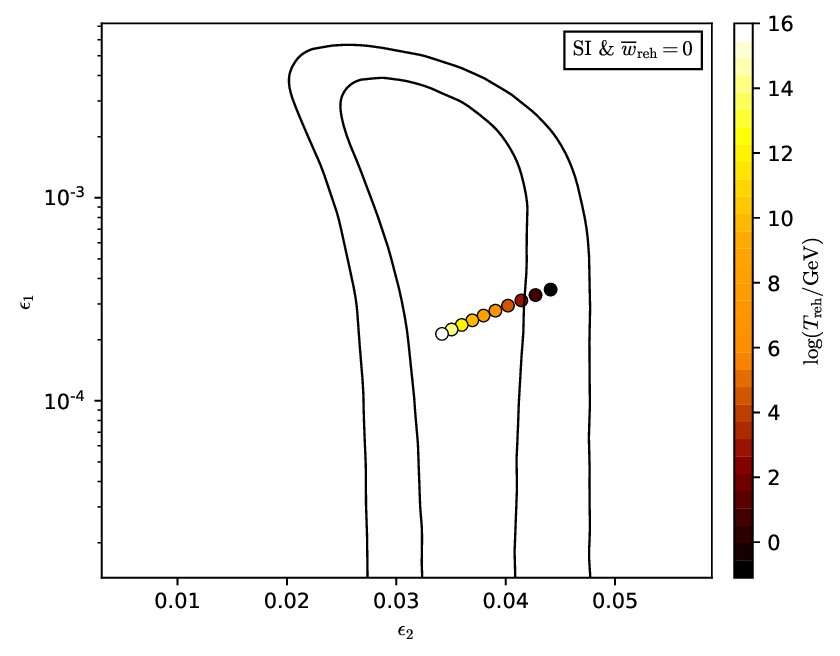}
\caption{Reheating consistent slow-roll predictions for the Starobinsky and the Higgs
  model in the plane $(\nS,r)$ (top panel) and the plane
  $(\epsilon_1,\epsilon_2)$ (bottom panel). The solid contours are the
  one and two-sigma {\data} confidence intervals (marginalized over
  second order slow-roll).}
\label{fig:CMBHI}
\end{center}
\end{figure}

\subsection{Radiatively Corrected Higgs Inflation (\hyperref[sec:rchi]{RCHI})}

\begin{figure}[H]
\begin{center}
\includegraphics[width=\wappfig,clip=true]{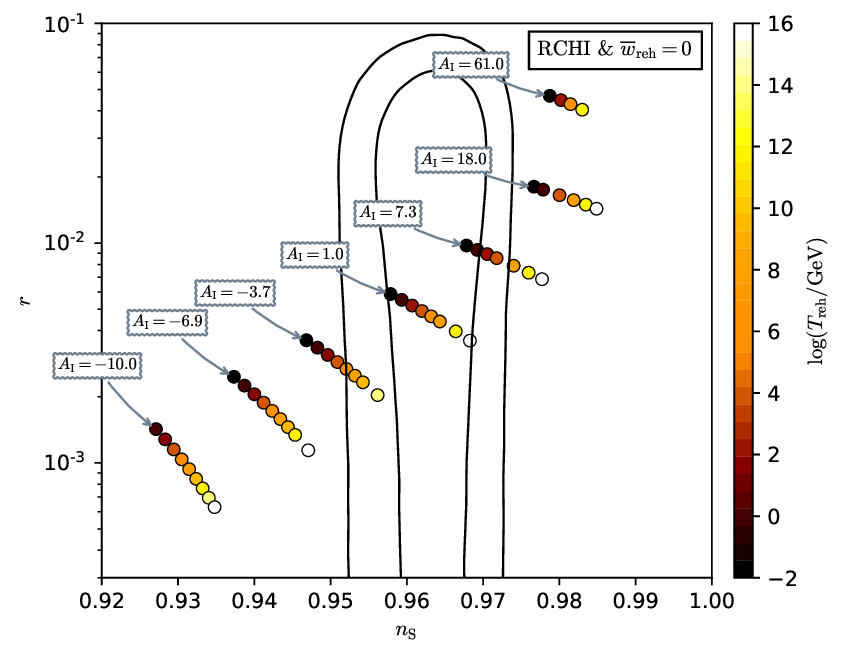}
\includegraphics[width=\wappfig,clip=true]{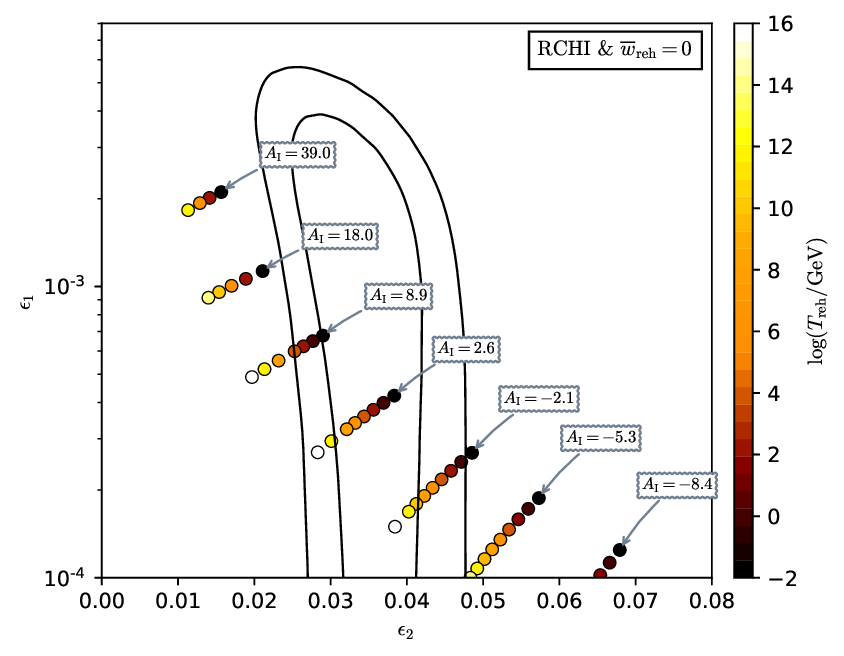}
\caption{Reheating consistent slow-roll predictions for the
  radiatively corrected Higgs model in the plane $(\nS,r)$ (top panel)
  and the plane $(\epsilon_1,\epsilon_2)$ (bottom panel). The solid
  contours are the one and two-sigma {\data} confidence intervals
  (marginalized over second order slow-roll).}
\label{fig:CMBRCHI}
\end{center}
\end{figure}

\subsection{Large Field Inflation (\hyperref[sec:lfi]{LFI})}

\begin{figure}[H]
\begin{center}
\includegraphics[width=\wappfig,clip=true]{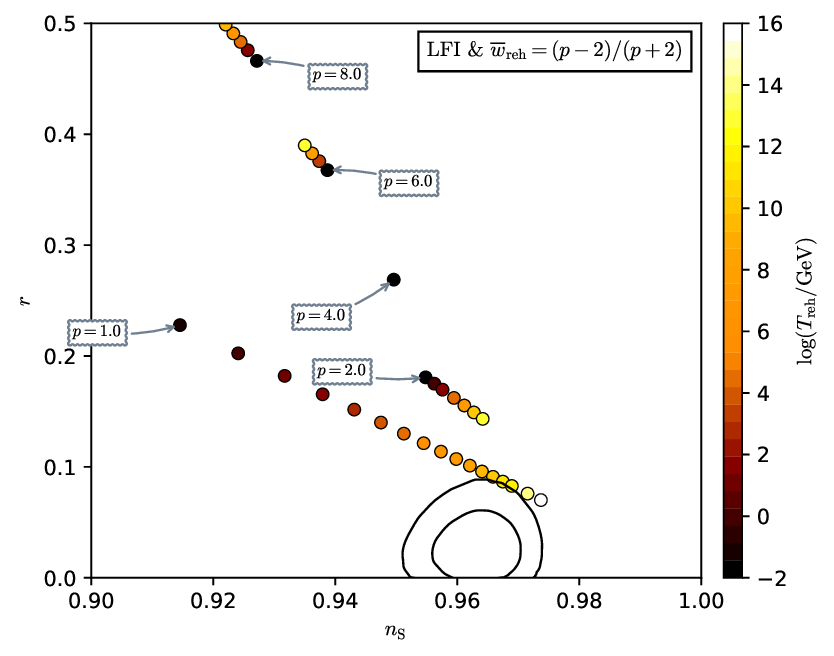}
\includegraphics[width=\wappfig,clip=true]{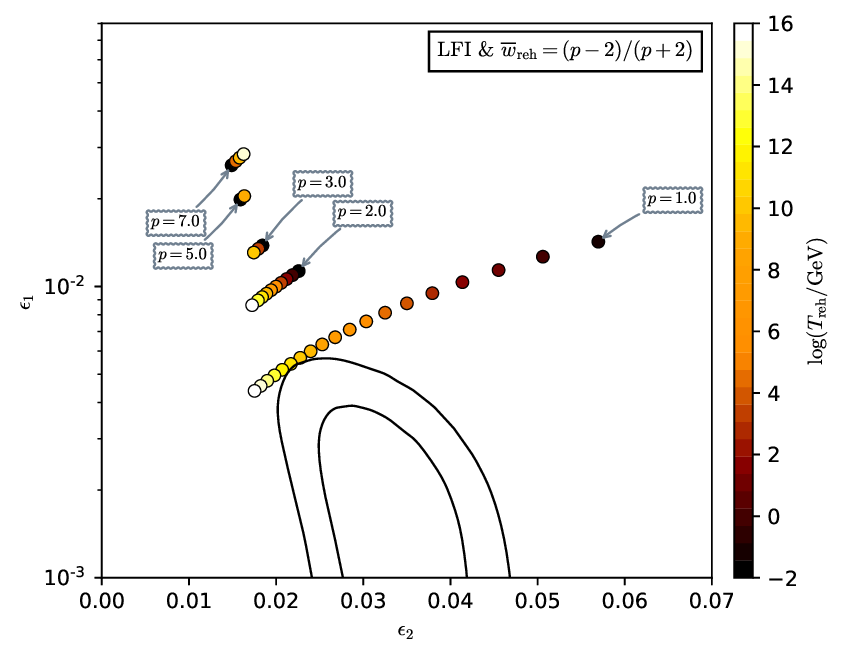}
\caption{Reheating consistent slow-roll predictions for the large
  field models in the plane $(\nS,r)$ (top panel) and the plane
  $(\epsilon_1,\epsilon_2)$ (bottom panel). The solid contours are the
  one and two-sigma {\data} confidence intervals (marginalized over
  second order slow-roll).}
\label{fig:CMBLFI}
\end{center}
\end{figure}

\subsection{Mixed Large Field Inflation (\hyperref[sec:mlfi]{MLFI})}

\begin{figure}[H]
\begin{center}
\includegraphics[width=\wappfig,clip=true]{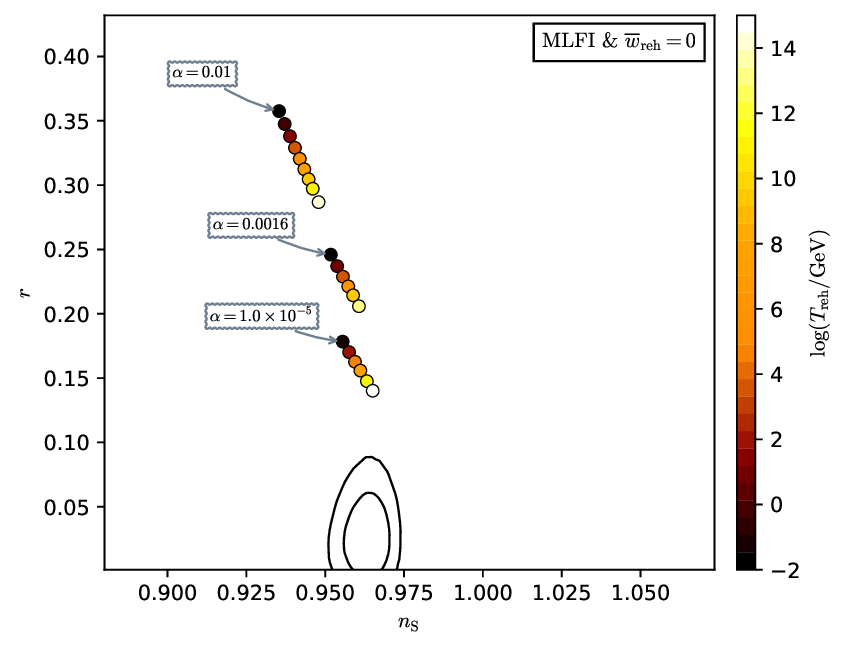}
\includegraphics[width=\wappfig,clip=true]{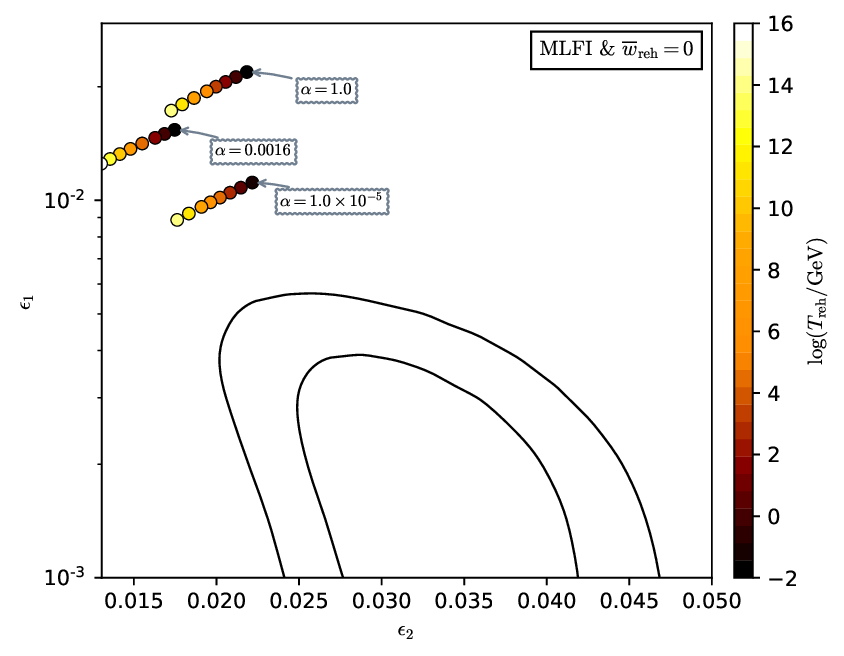}
\caption{Reheating consistent slow-roll predictions for the mixed
  large field models in the plane $(\nS,r)$ (top panel) and the plane
  $(\epsilon_1,\epsilon_2)$ (bottom panel). The solid contours are the
  one and two-sigma {\data} confidence intervals (marginalized over
  second order slow-roll). Predictions are within the one of the
  quadratic LFI ($p=2$) and quartic LFI ($p=4$) models.}
\label{fig:CMBMLFI}
\end{center}
\end{figure}

\subsection{Radiatively Corrected Massive Inflation (\hyperref[sec:rcmi]{RCMI})}

\begin{figure}[H]
\begin{center}
\includegraphics[width=\wappfig,clip=true]{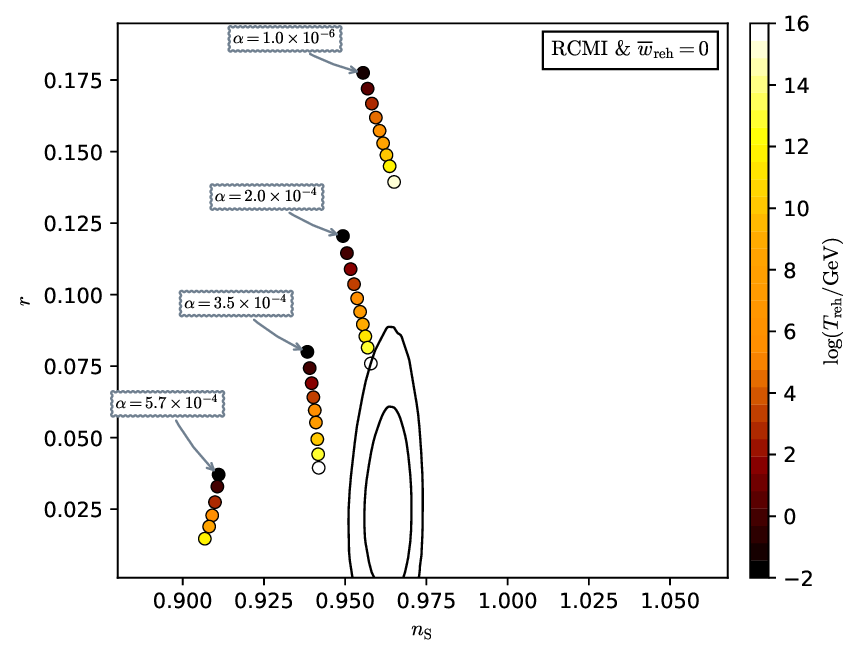}
\includegraphics[width=\wappfig,clip=true]{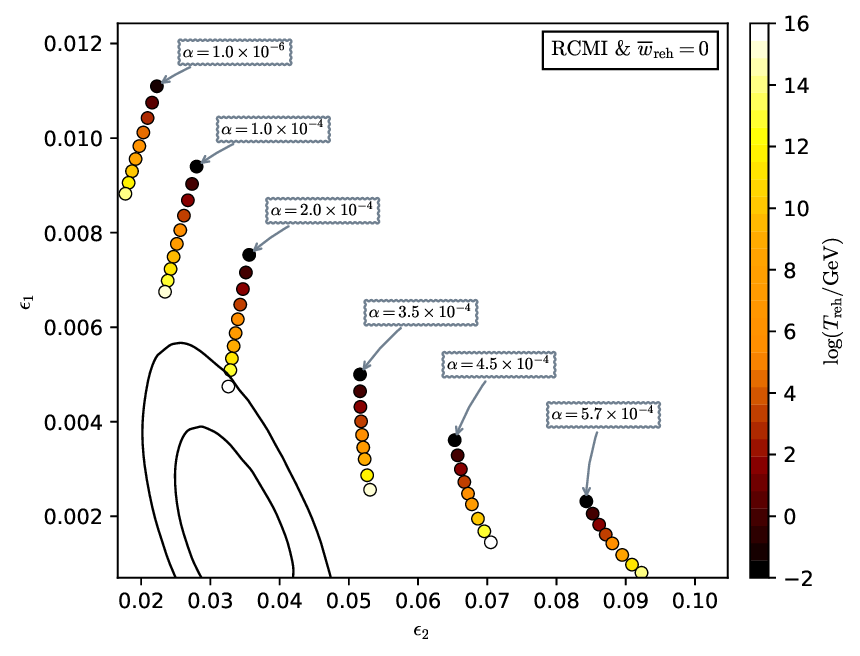}
\caption{Reheating consistent slow-roll predictions for the
  radiatively corrected massive models in the plane $(\nS,r)$. The
  solid contours are the one and two-sigma {\data} confidence
  intervals (marginalized over second order slow-roll). For $\alpha
  \to 0$, the predictions match the ones of the LFI quadratic model
  ($p=2$).}
\label{fig:CMBRCMI}
\end{center}
\end{figure}

\subsection{Radiatively Corrected Quartic Inflation (\hyperref[sec:rcqi]{RCQI})}

\begin{figure}[H]
\begin{center}
\includegraphics[width=\wappfig,clip=true]{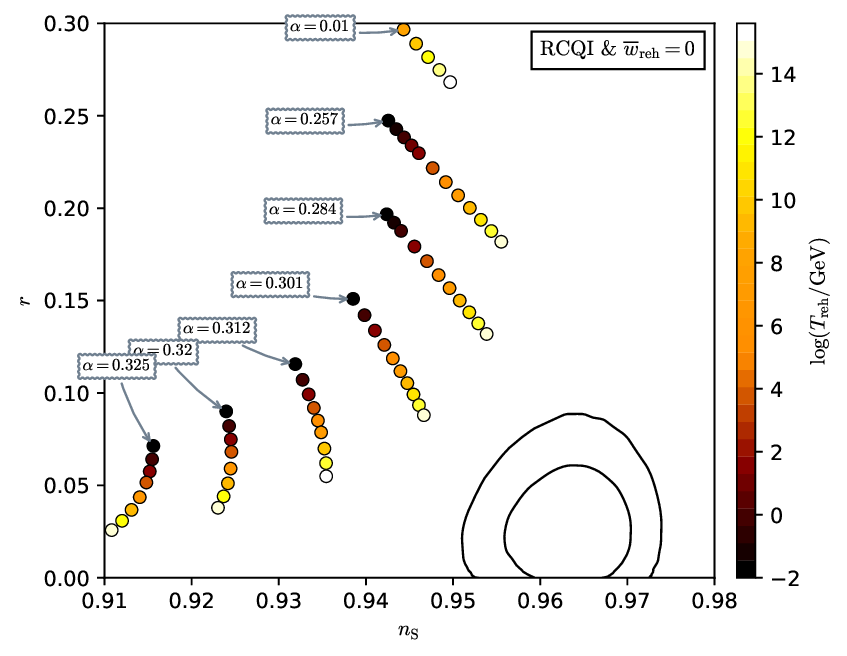}
\includegraphics[width=\wappfig,clip=true]{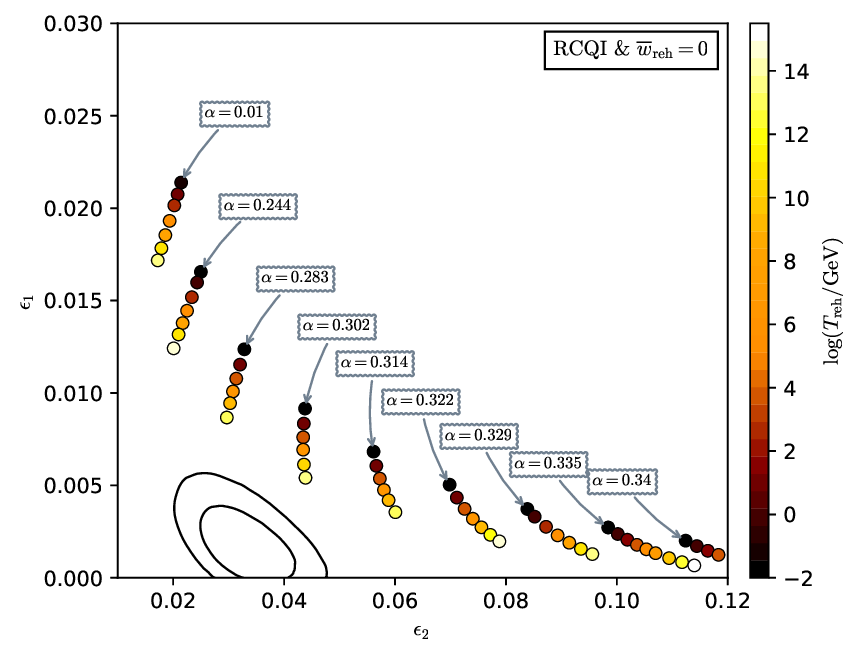}
\caption{Reheating consistent slow-roll predictions for the
  radiatively corrected quartic models in the plane $(\nS,r)$ (top
  panel) and the plane $(\epsilon_1,\epsilon_2)$ (bottom panel), with
  $\wrehbar=0$. The solid contours are the one and two-sigma
  {\data} confidence intervals (marginalized over second order
  slow-roll).}
\label{fig:CMBRCQI}
\end{center}
\end{figure}

\begin{figure}[H]
\begin{center}
\includegraphics[width=\wappfig,clip=true]{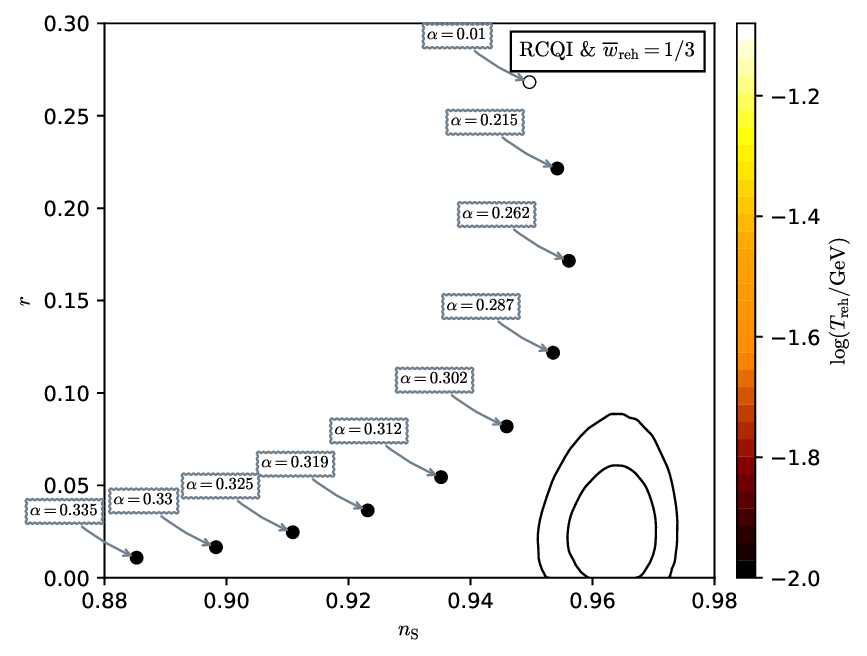}
\includegraphics[width=\wappfig,clip=true]{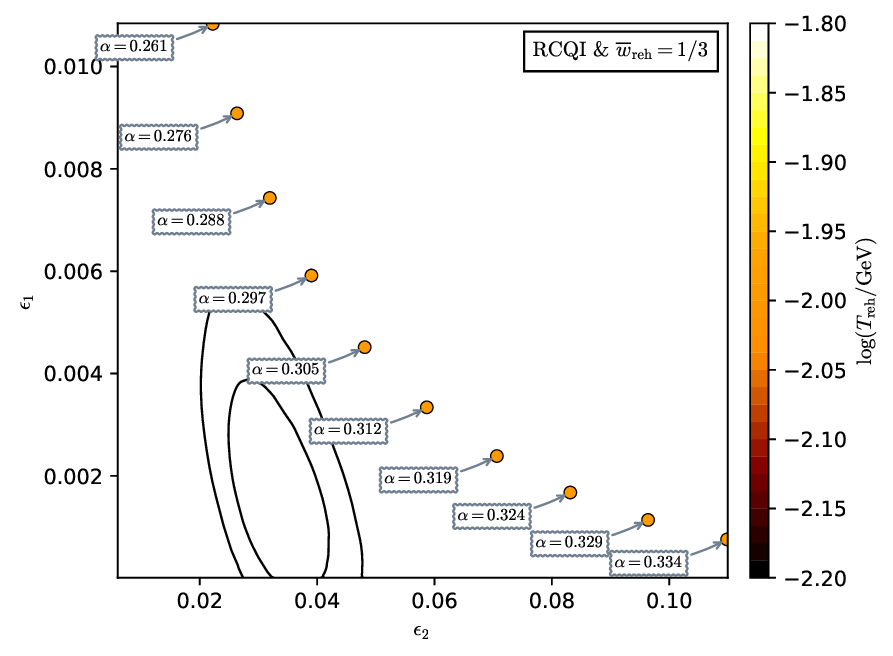}
\caption{Reheating consistent slow-roll predictions for the
  radiatively corrected quartic models in the plane $(\nS,r)$ (top
  panel) and the plane $(\epsilon_1,\epsilon_2)$ (bottom panel), with
  $\wrehbar=\frac{1}{3}$. This value of $\wrehbar$ may be more physically
  justified if the reheating phase takes place at the bottom of the
  potential, which is quartic in a good approximation. The solid contours are the one and
  two-sigma {\data} confidence intervals (marginalized over second order
  slow-roll).}
\label{fig:CMBRCQIb}
\end{center}
\end{figure}

\subsection{Natural Inflation (\hyperref[sec:ni]{NI})}

\begin{figure}[H]
\begin{center}
\includegraphics[width=\wappfig,clip=true]{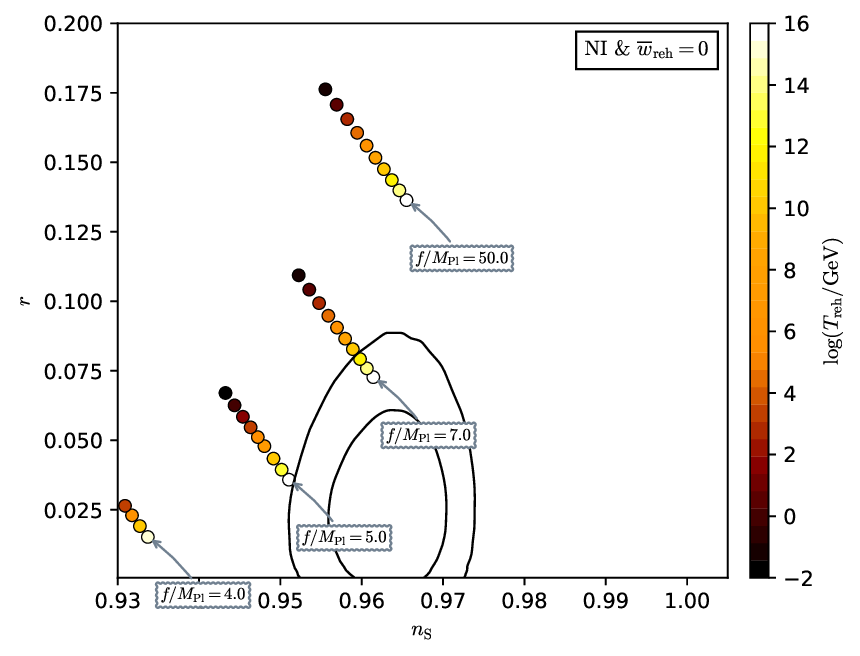}
\includegraphics[width=\wappfig,clip=true]{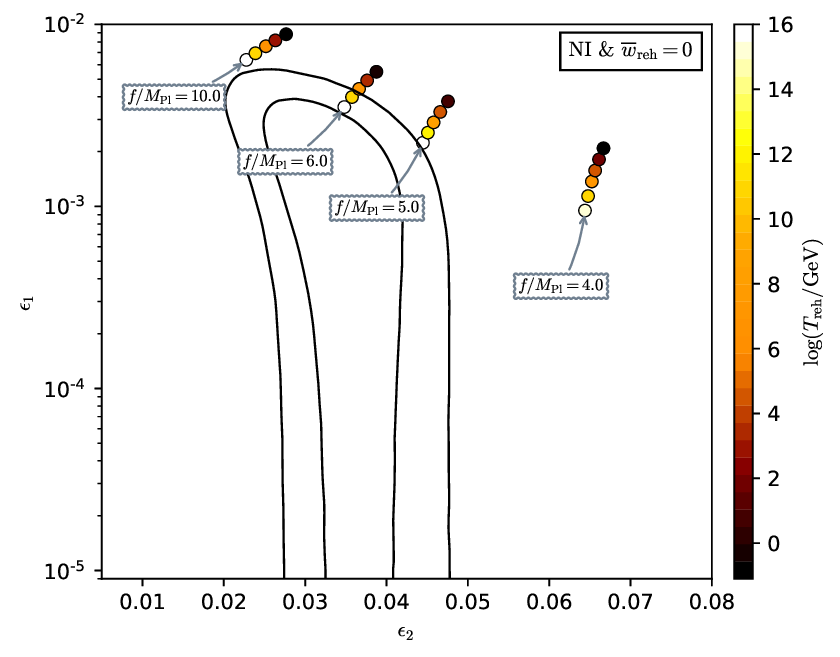}
\caption{Reheating consistent slow-roll predictions for the natural
  inflation models in the plane $(\nS,r)$ (top
  panel) and the plane $(\epsilon_1,\epsilon_2)$ (bottom panel). The
  solid contours are the one and two-sigma {\data} confidence
  intervals (marginalized over second order slow-roll).}
\label{fig:CMBNI}
\end{center}
\end{figure}

\subsection{Exponential SUSY Inflation (\hyperref[sec:esi]{ESI})}

\begin{figure}[H]
\begin{center}
\includegraphics[width=\wappfig,clip=true]{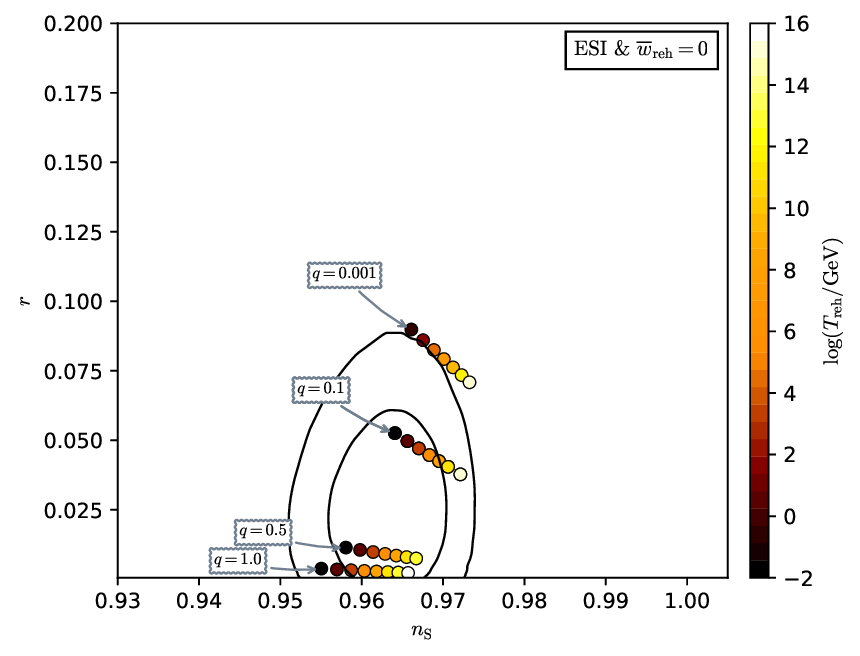}
\includegraphics[width=\wappfig,clip=true]{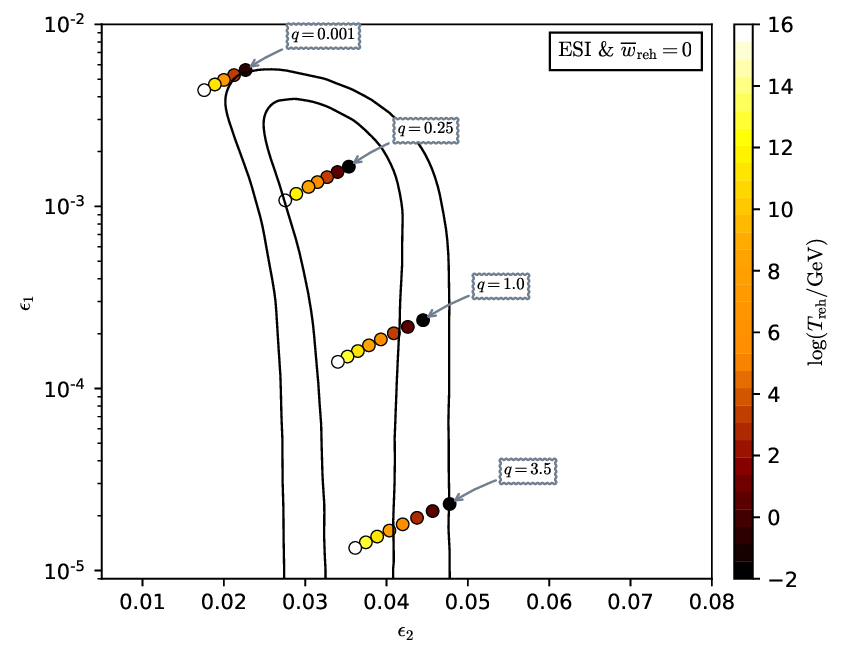}
\caption{Reheating consistent slow-roll predictions for the
  exponential Susy models in the plane $(\nS,r)$ (top panel) and the
  plane $(\epsilon_1,\epsilon_2)$ (bottom panel). The solid contours
  are the one and two-sigma {\data} confidence intervals (marginalized
  over second order slow-roll).}
\label{fig:CMBESI}
\end{center}
\end{figure}

\begin{figure}[H]
\begin{center}
\includegraphics[width=\wappfig,clip=true]{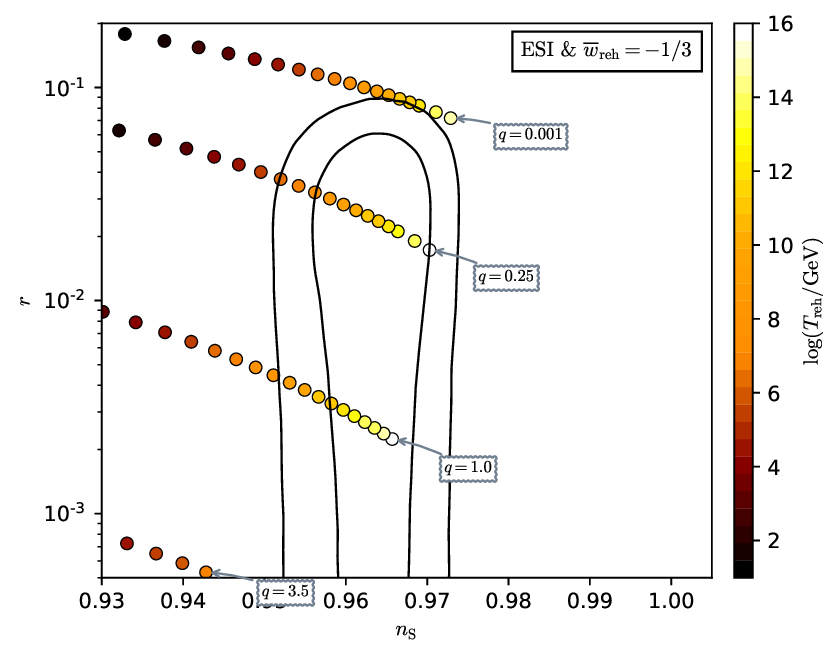}
\includegraphics[width=\wappfig,clip=true]{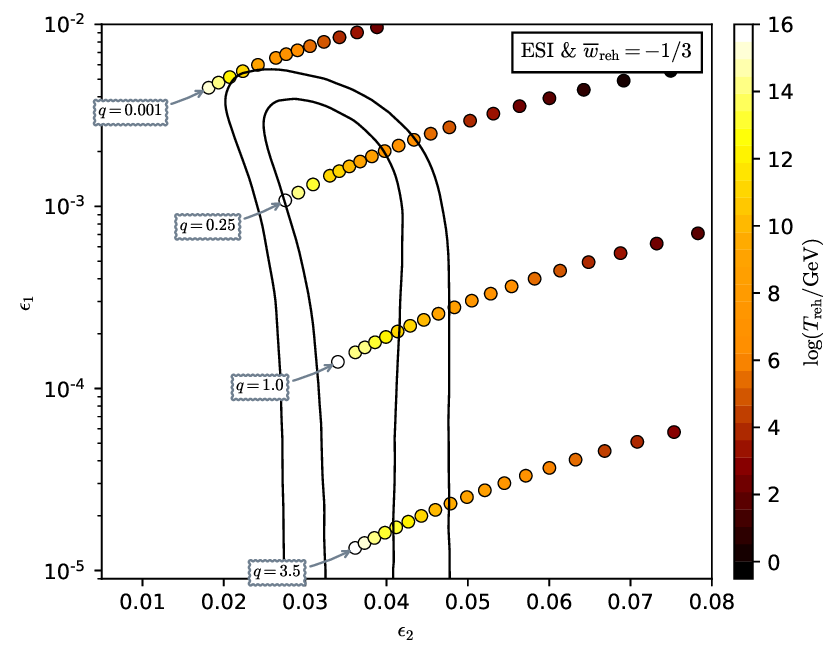}
\caption{Reheating consistent slow-roll predictions for the
  exponential Susy models in the plane $(\nS,r)$ (top panel) and the
  plane $(\epsilon_1,\epsilon_2)$ (bottom panel), with
  $\wrehbar=-1/3$. This value of $\wrehbar$ may be more physically
  justified (although rather extreme) if a parametric reheating feels
  the bottom of the potential, which is linear in a good
  approximation. The solid contours are the one and two-sigma {\data}
  confidence intervals (marginalized over second order slow-roll).}
\label{fig:CMBESIb}
\end{center}
\end{figure}

\subsection{Power Law Inflation (\hyperref[sec:pli]{PLI})}

\begin{figure}[H]
\begin{center}
\includegraphics[width=\wappfig,clip=true]{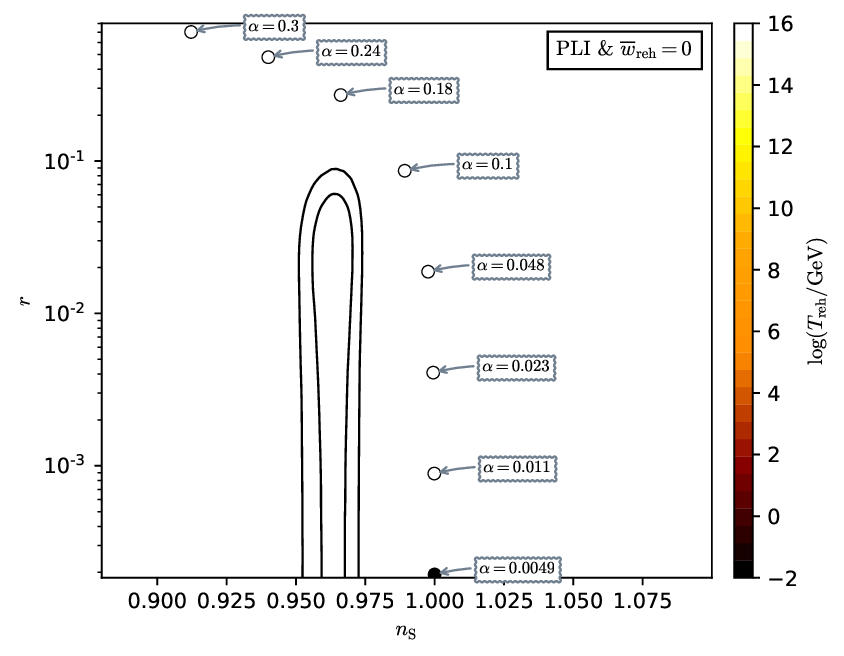}
\includegraphics[width=\wappfig,clip=true]{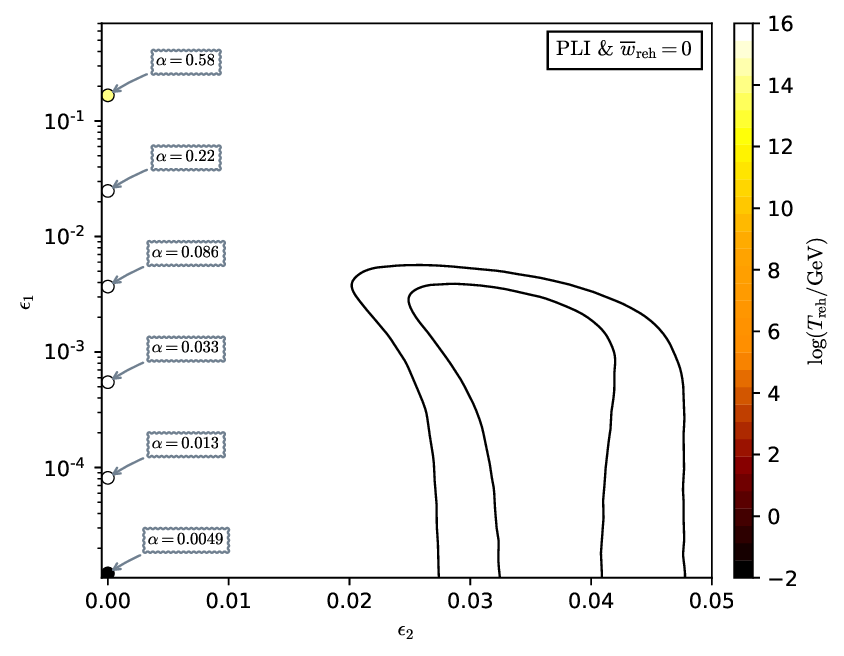}
\caption{Reheating consistent slow-roll predictions for the power law
  models in the plane $(\nS,r)$ (top panel) and the plane
  $(\epsilon_1,\epsilon_2)$ (bottom panel). The solid contours are the
  one and two-sigma {\data} confidence intervals (marginalized over
  second order slow-roll).}
\label{fig:CMBPLI}
\end{center}
\end{figure}

\subsection{K\"ahler Moduli Inflation I (\hyperref[sec:kmii]{KMII})}

\begin{figure}[H]
\begin{center}
\includegraphics[width=\wappfig,clip=true]{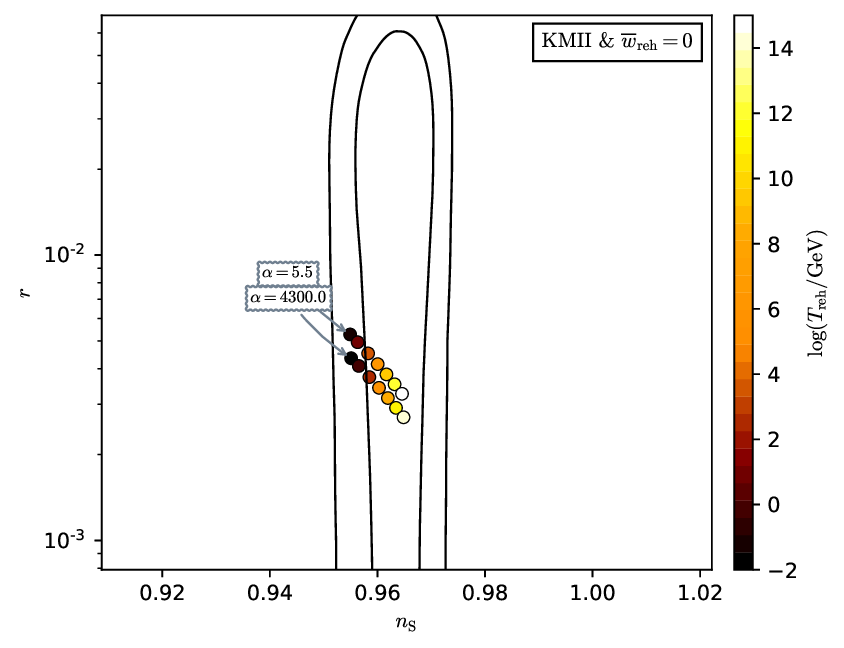}
\includegraphics[width=\wappfig,clip=true]{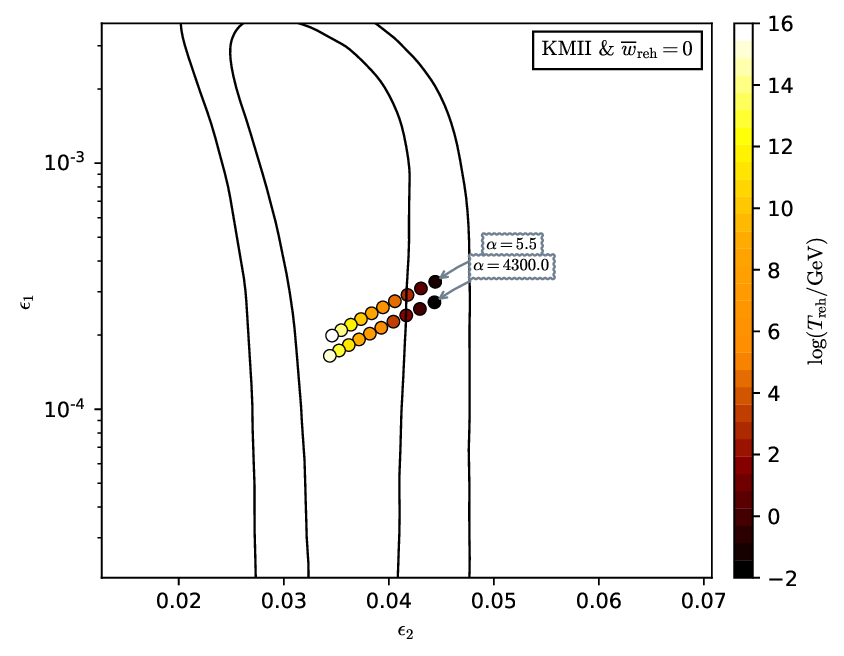}
\caption{Reheating consistent slow-roll predictions for the K\"ahler
  Moduli I models in the plane $(\nS,r)$ (top panel) and the plane
  $(\epsilon_1,\epsilon_2)$ (bottom panel). The solid contours are the
  one and two-sigma {\data} confidence intervals (marginalized over
  second order slow-roll). Predictions remain mostly insensitive to
  the value of $\alpha$..}
\label{fig:CMBKMII}
\end{center}
\end{figure}

\subsection{Horizon Flow Inflation at first order (\hyperref[sec:hf1i]{HF1I})}

\begin{figure}[H]
\begin{center}
\includegraphics[width=\wappfig,clip=true]{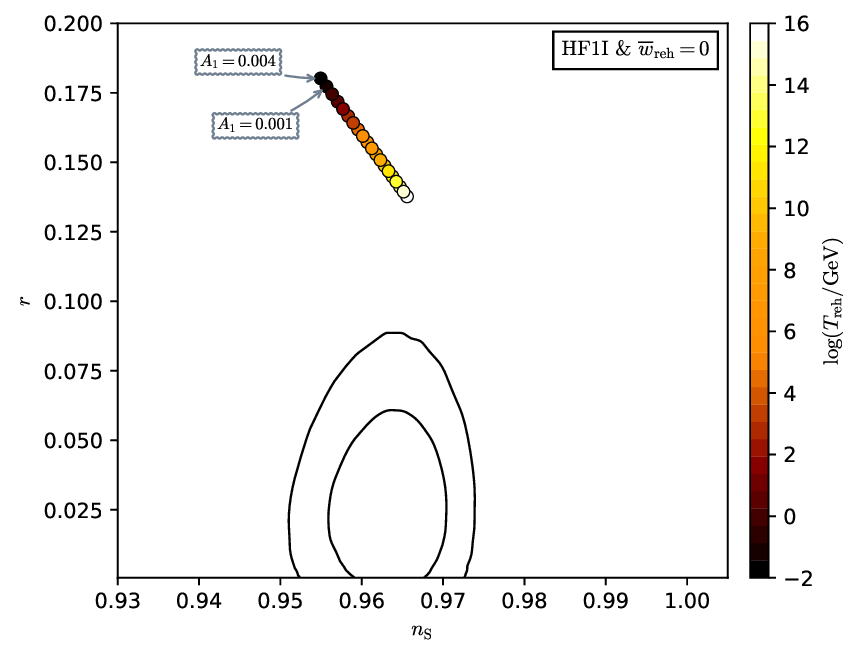}
\includegraphics[width=\wappfig,clip=true]{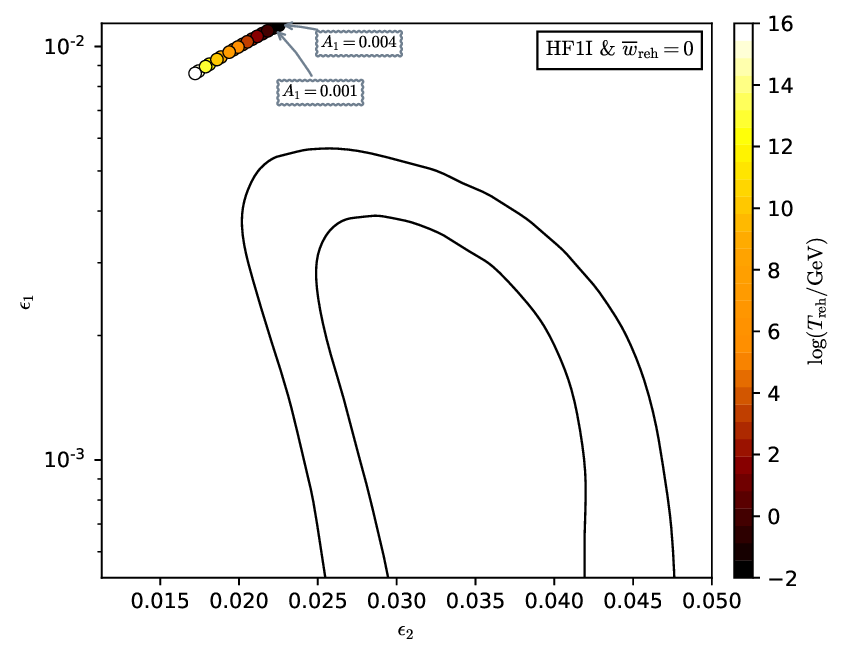}
\caption{Reheating consistent (exact) predictions for the horizon flow
  inflation at first order models in the plane $(\nS,r)$ (top panel)
  and the plane $(\epsilon_1,\epsilon_2)$ (bottom panel). The solid
  contours trace the two-sigma {\data} confidence intervals
  (marginalized over second order slow-roll). Notice that, up to the
  amplitude of the CMB anisotropies, the predictions do not depend
  much on $A_1$ as they are all superimposed.}
\label{fig:CMBHF1I}
\end{center}
\end{figure}

\subsection{Colemann-Weinberg Inflation (\hyperref[sec:cwi]{CWI})}

\begin{figure}[H]
\begin{center}
\includegraphics[width=\wappfig,clip=true]{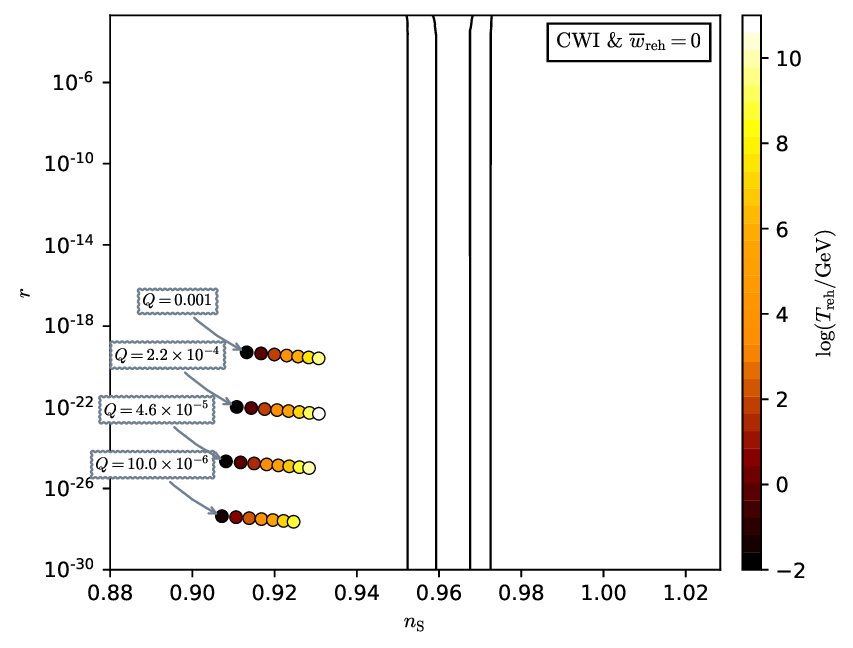}
\includegraphics[width=\wappfig,clip=true]{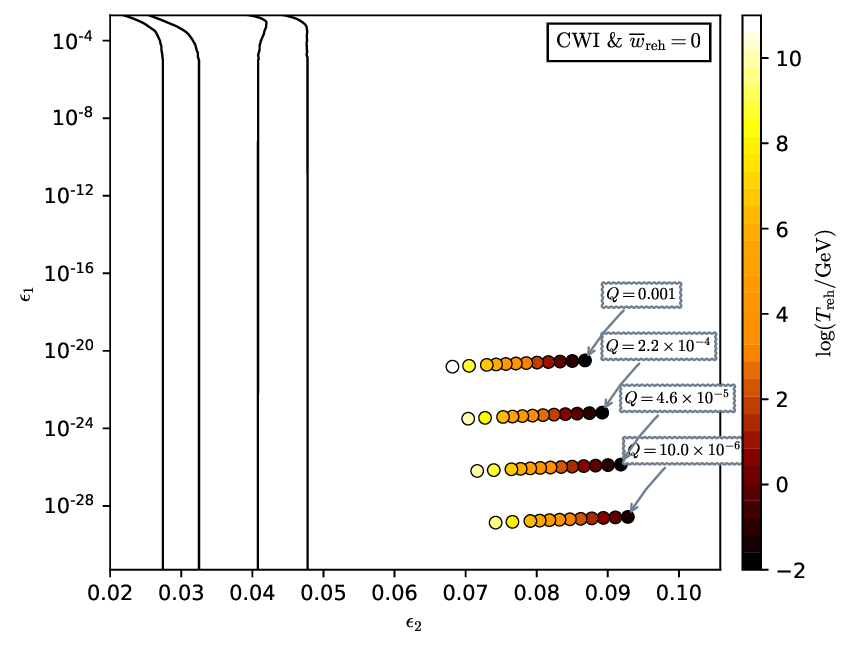}
\caption{Reheating consistent slow-roll predictions for the
  Colemann-Weinberg models in the plane $(\nS,r)$ (top panel) and the
  plane $(\epsilon_1,\epsilon_2)$ (bottom panel), in the physical
  domain $Q/\Mp\in[10^{-5},10^{-3}]$. The solid contours are the one
  and two-sigma {\data} confidence intervals (marginalized over second
  order slow-roll). The typical amount of gravitational waves is
  extremely small and dominated by second order effects (not accounted
  for in the figures).}
\label{fig:CMBCWI1}
\end{center}
\end{figure}

\begin{figure}[H]
\begin{center}
\includegraphics[width=\wappfig,clip=true]{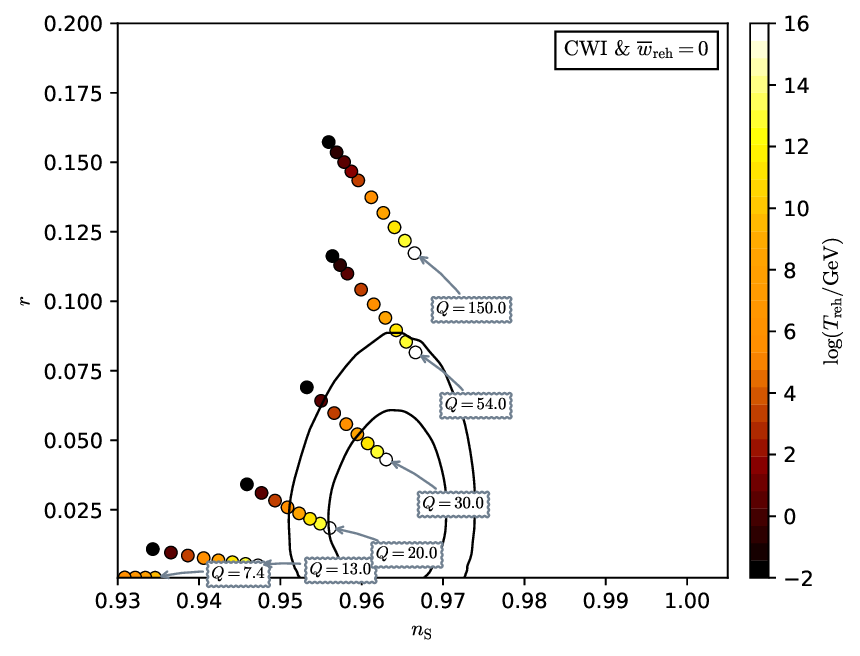}
\includegraphics[width=\wappfig,clip=true]{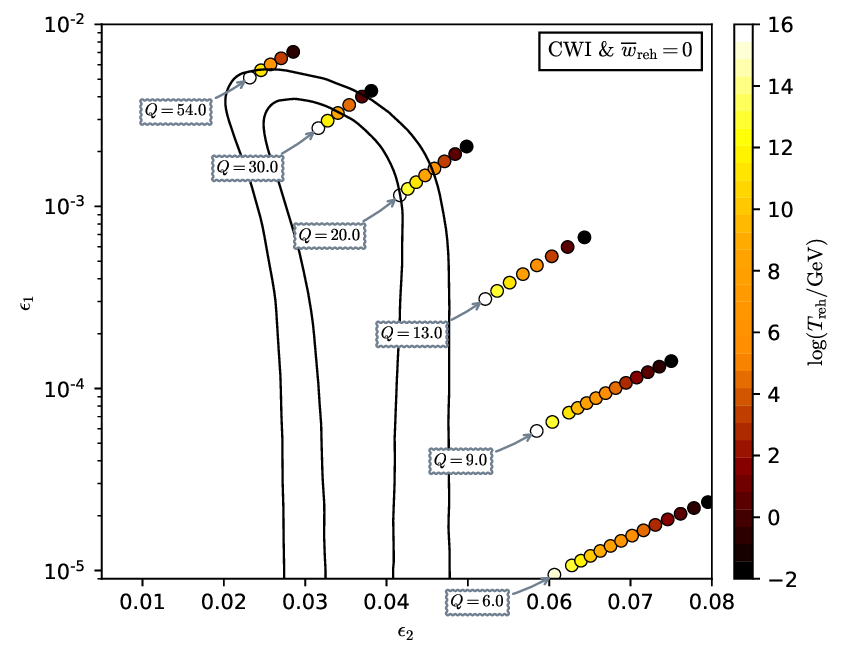}
\caption{Reheating consistent slow-roll predictions for the
  Colemann-Weinberg models in the plane $(\nS,r)$ (top panel) and the
  plane $(\epsilon_1,\epsilon_2)$ (bottom panel), in the domain
  $Q/\Mp\in[1,100]$ . The solid contours are the one and two-sigma
  {\data} confidence intervals (marginalized over second order
  slow-roll). When $Q/\Mp\gg 1$, the model is similar to a quadratic
  potential close to its minimum, and the predictions match the LFI
  $\epsilon_1=\epsilon_2/2$ relation (see \sectionc{sec:lfi}).}
\label{fig:CMBCWI2}
\end{center}
\end{figure}

\subsection{Loop Inflation (\hyperref[sec:li]{LI})}

\begin{figure}[H]
\begin{center}
\includegraphics[width=\wappfig,clip=true]{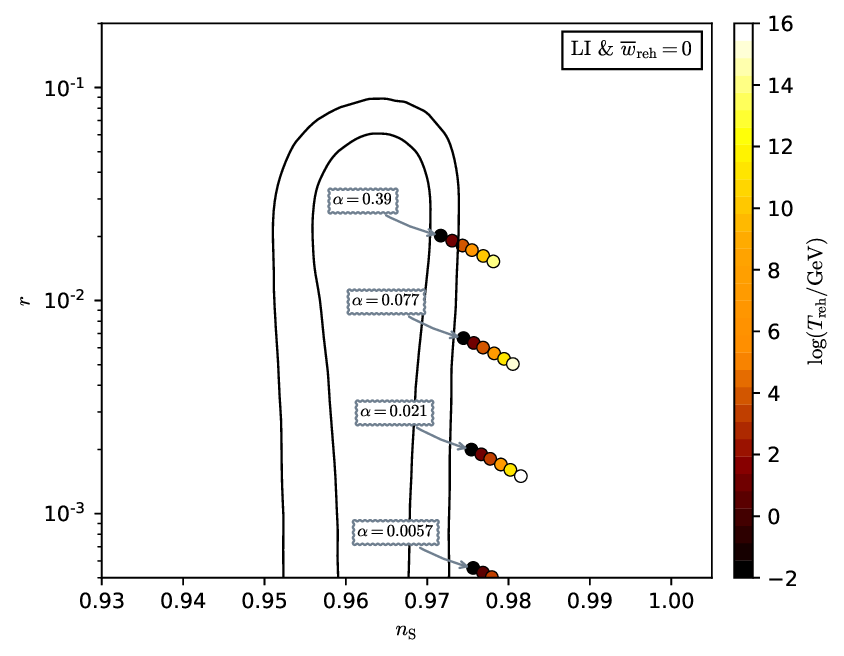}
\includegraphics[width=\wappfig,clip=true]{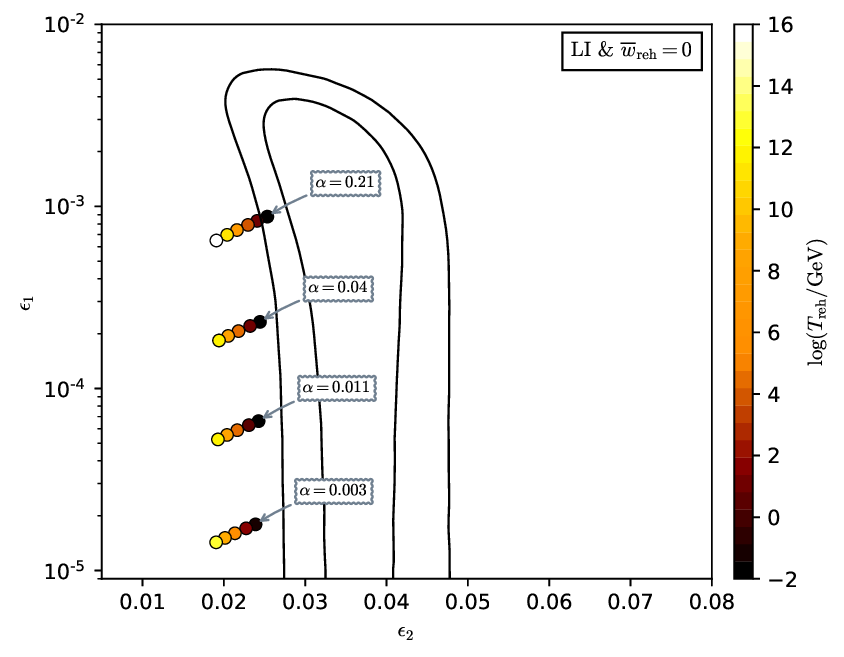}
\caption{Reheating consistent slow-roll predictions for the loop
  inflation models for $\alpha>0$, in the plane $(\nS,r)$ (top panel),
  and the plane $(\epsilon_1,\epsilon_2)$ (bottom panel). The solid
  contours are the one and two-sigma {\data} confidence intervals
  (marginalized over second order slow-roll).}
\label{fig:CMBLIalphaPositive}
\end{center}
\end{figure}

\begin{figure}[H]
\begin{center}
\includegraphics[width=\wappfig,clip=true]{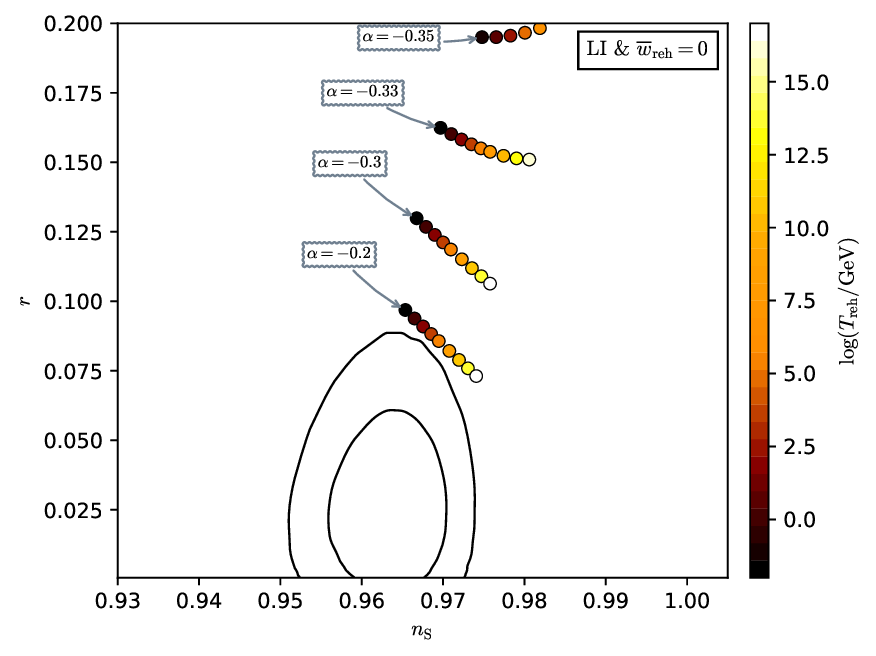}
\includegraphics[width=\wappfig,clip=true]{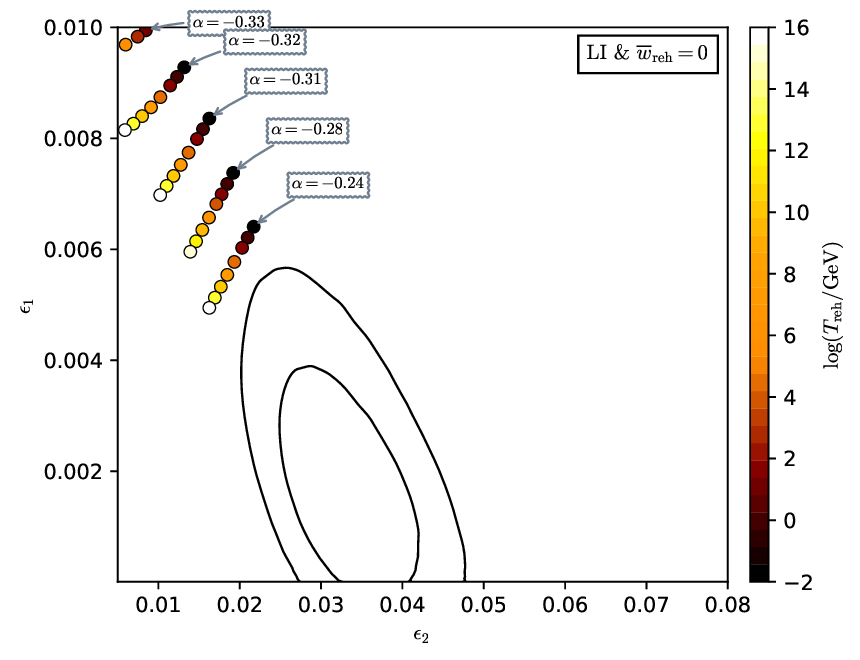}
\caption{Reheating consistent slow-roll predictions for the loop
  inflation models for $\alpha<0$, in the plane $(\nS,r)$ (top panel),
  and the plane $(\epsilon_1,\epsilon_2)$ (bottom panel). The solid
  contours are the one and two-sigma {\data} confidence intervals
  (marginalized over second order slow-roll).}
\label{fig:CMBLIalphaNegative}
\end{center}
\end{figure}

\subsection{\texorpdfstring{$R+R^{2p}$}{RpI} Inflation (\hyperref[sec:rpi]{RpI})}

\begin{figure}[H]
\begin{center}
\includegraphics[width=\wappfig,clip=true]{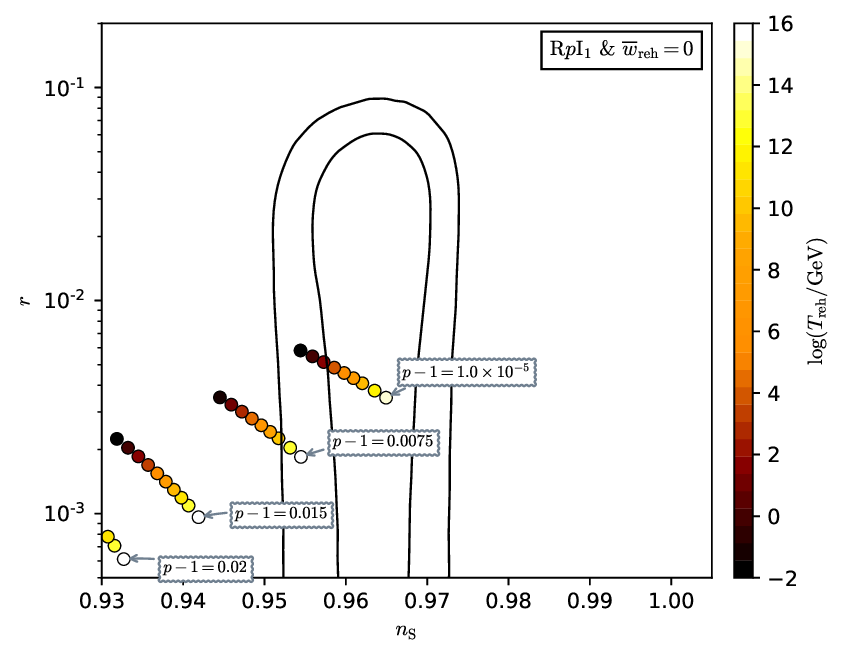}
\includegraphics[width=\wappfig,clip=true]{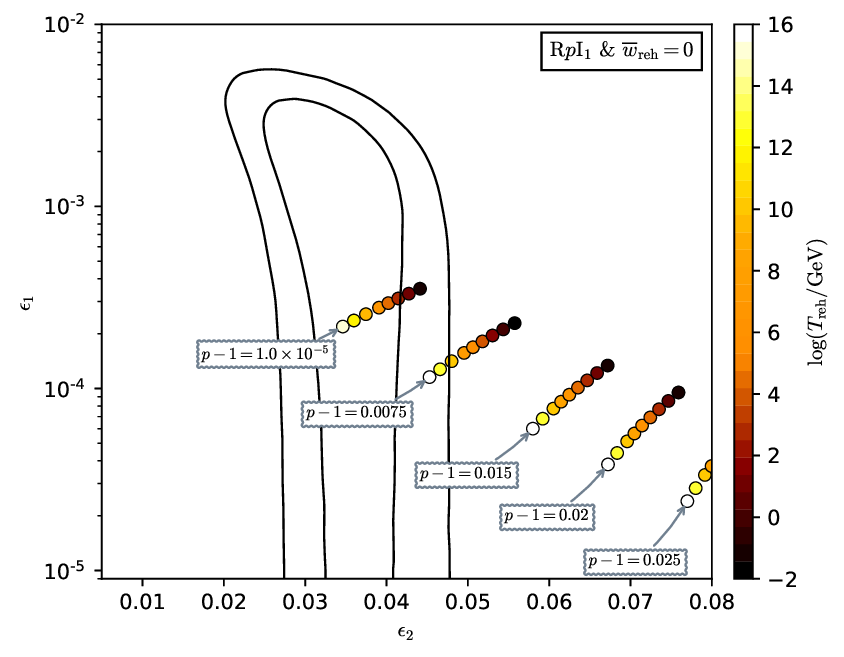}
\caption{Reheating consistent slow-roll predictions for the $R+R^{2p}$
  inflation models in the RpI1 regime, in the plane $(\nS,r)$ (top
  panel), and the plane $(\epsilon_1,\epsilon_2)$ (bottom panel). The
  solid contours are the one and two-sigma {\data} confidence
  intervals (marginalized over second order slow-roll). For $p\to0$,
  one recovers Starobinski Inflation.}
\label{fig:CMBRpI1}
\end{center}
\end{figure}

\begin{figure}[H]
\begin{center}
\includegraphics[width=\wappfig,clip=true]{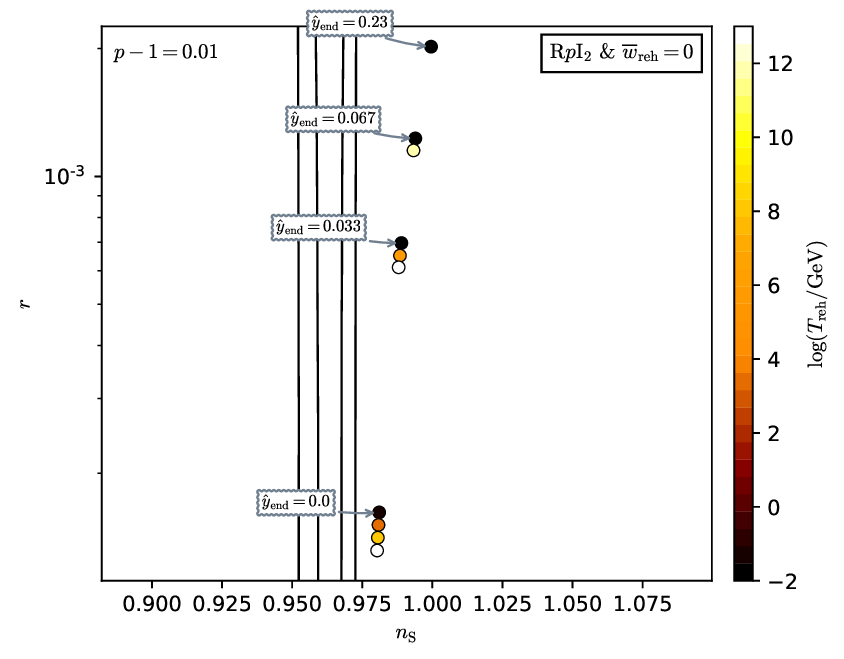}
\includegraphics[width=\wappfig,clip=true]{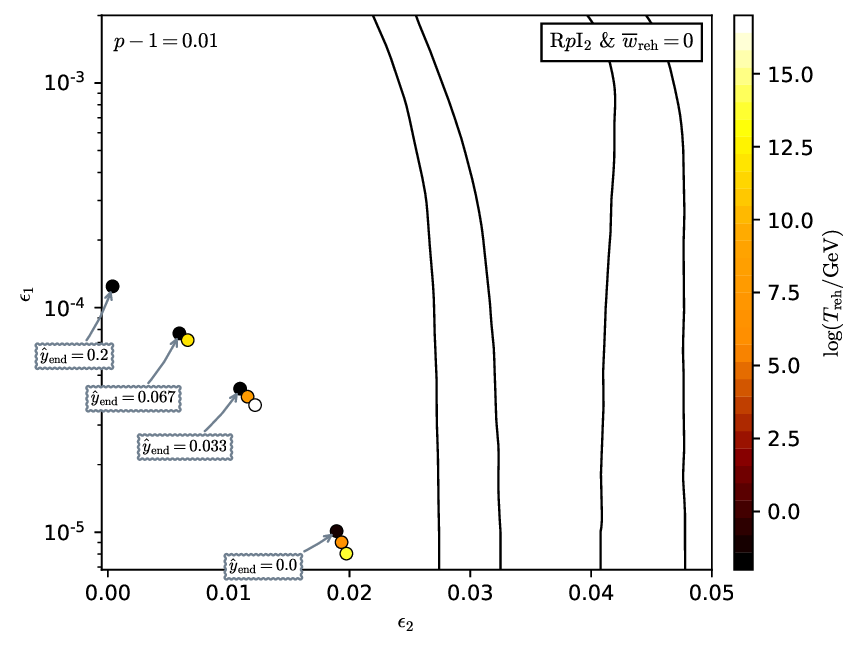}
\caption{Reheating consistent slow-roll predictions for the $R+R^{2p}$
  inflation models in the RpI2 regime and for $p-1=10^{-2}$, in the
  plane $(\nS,r)$ (top panel), and the plane $(\epsilon_1,\epsilon_2)$
  (bottom panel). The solid contours are the one and two-sigma {\data}
  confidence intervals (marginalized over second order
  slow-roll). When $\yend \gg 1$, one has $\epsilon_2\to 0$.}
\label{fig:CMBRpI2}
\end{center}
\end{figure}

\begin{figure}[H]
\begin{center}
\includegraphics[width=\wappfig,clip=true]{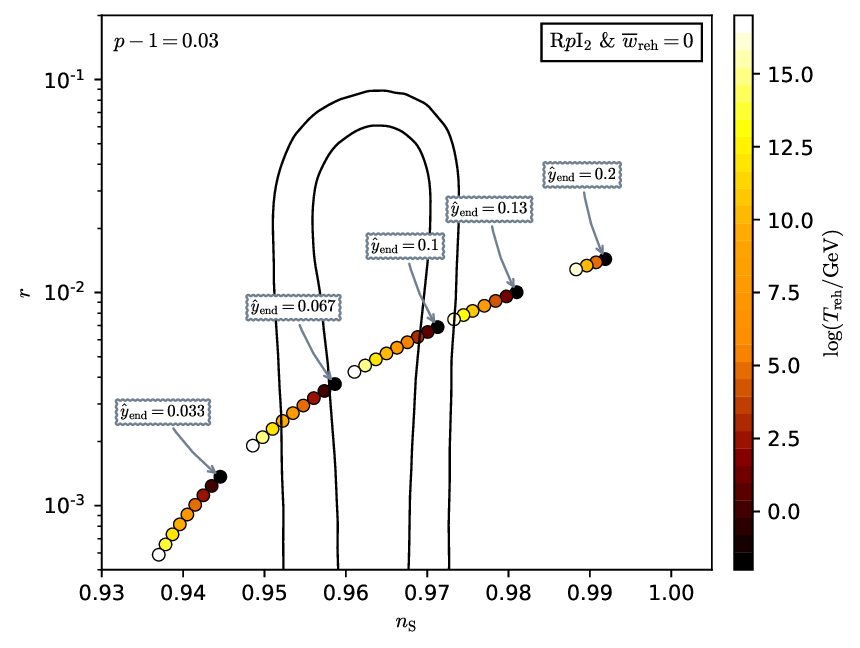}
\includegraphics[width=\wappfig,clip=true]{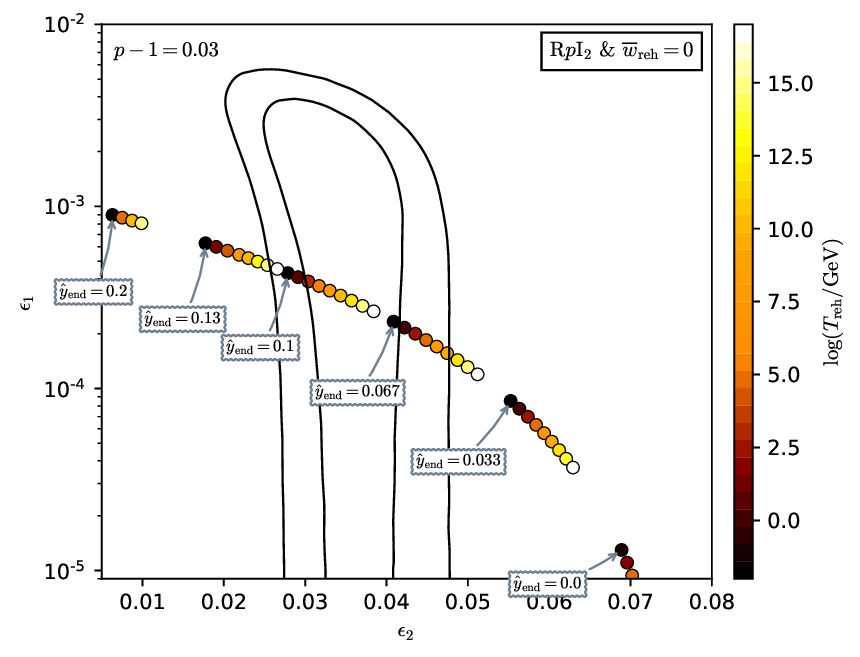}
\caption{Reheating consistent slow-roll predictions for the $R+R^{2p}$
  inflation models in the RpI2 regime and for $p-1=0.03$, in the plane
  $(\nS,r)$ (top panel), and the plane $(\epsilon_1,\epsilon_2)$
  (bottom panel). The solid contours are the one and two-sigma {\data}
  confidence intervals (marginalized over second order slow-roll). As
  for figure~\ref{fig:CMBRpI2}, the limit $\yend \gg 1$ corresponds to
  $\epsilon_2\to 0$.}
\label{fig:CMBRpI2_1}
\end{center}
\end{figure}

\begin{figure}[H]
\begin{center}
\includegraphics[width=\wappfig,clip=true]{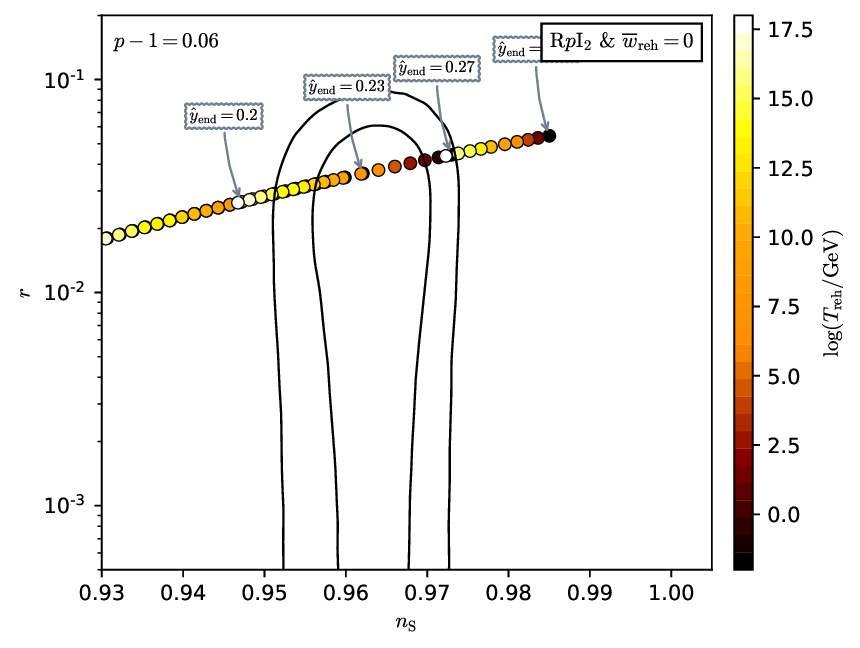}
\includegraphics[width=\wappfig,clip=true]{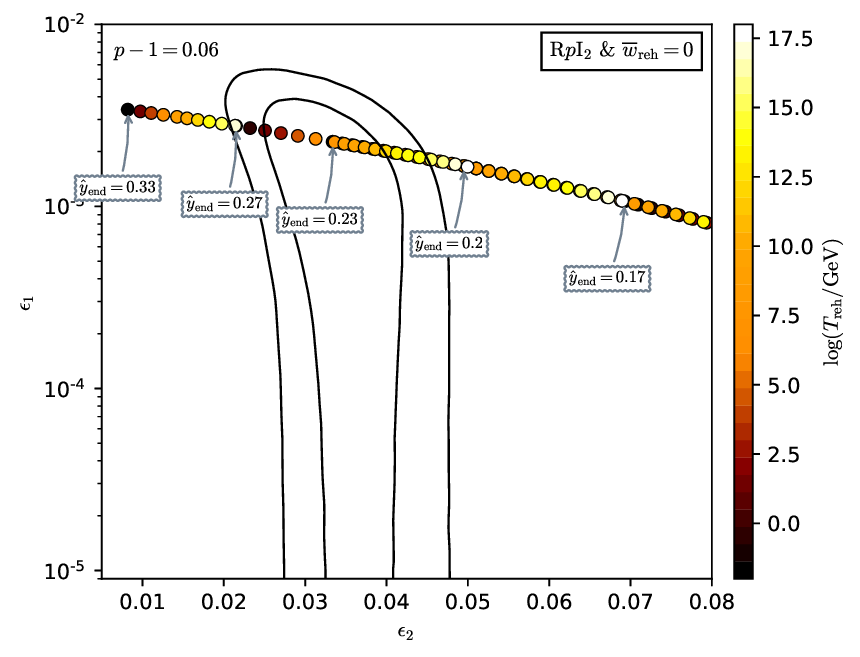}
\caption{Reheating consistent slow-roll predictions for the $R+R^{2p}$
  inflation models in the RpI2 regime and for $p-1=0.06$, in the plane
  $(\nS,r)$ (top panel), and the plane $(\epsilon_1,\epsilon_2)$
  (bottom panel). The solid contours are the one and two-sigma {\data}
  confidence intervals (marginalized over second order
  slow-roll). Increasing the values of $p-1$ within the RpI2 models
  boosts the productiuon of primordial gravitational waves.}
\label{fig:CMBRpI2_2}
\end{center}
\end{figure}

\begin{figure}[H]
\begin{center}
\includegraphics[width=\wappfig,clip=true]{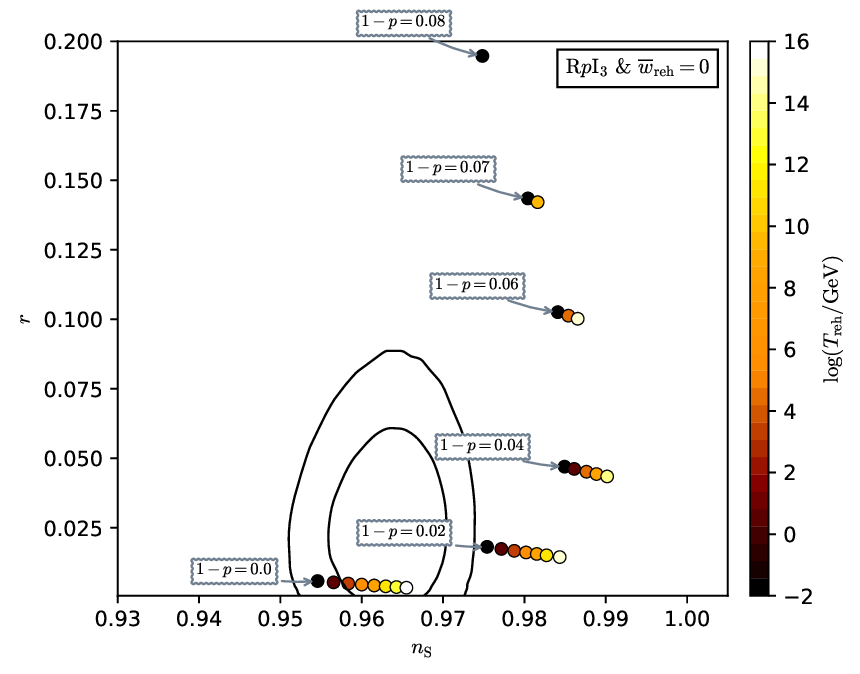}
\includegraphics[width=\wappfig,clip=true]{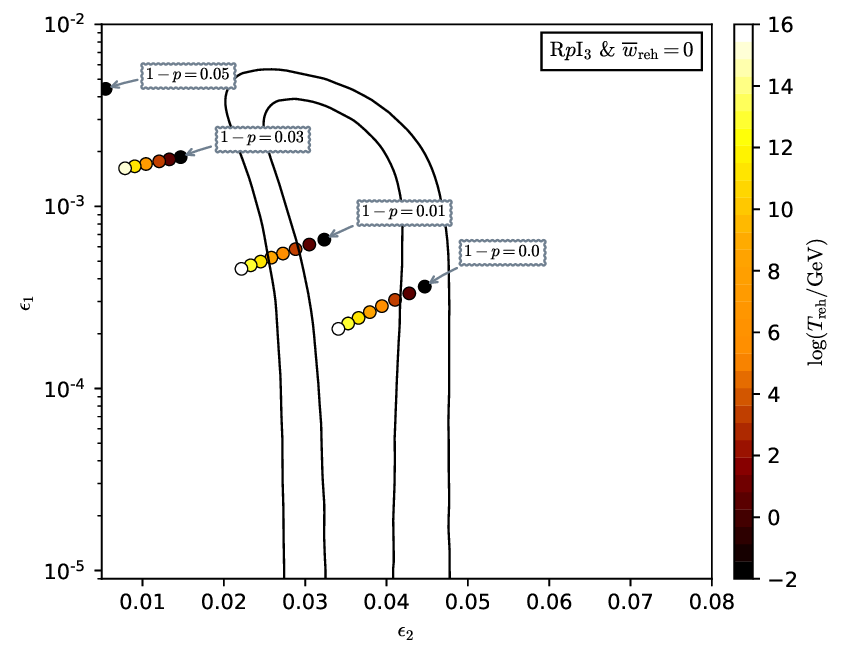}
\caption{Reheating consistent slow-roll predictions for the $R+R^{2p}$
  inflation models in the RpI3 regime, in the plane $(\nS,r)$ (top
  panel), and the plane $(\epsilon_1,\epsilon_2)$ (bottom panel). The
  solid contours are the one and two-sigma {\data} confidence
  intervals (marginalized over second order slow-roll).}
\label{fig:CMBRpI3}
\end{center}
\end{figure}

\subsection{Double Well Inflation (\hyperref[sec:dwi]{DWI})}

\begin{figure}[H]
\begin{center}
\includegraphics[width=\wappfig,clip=true]{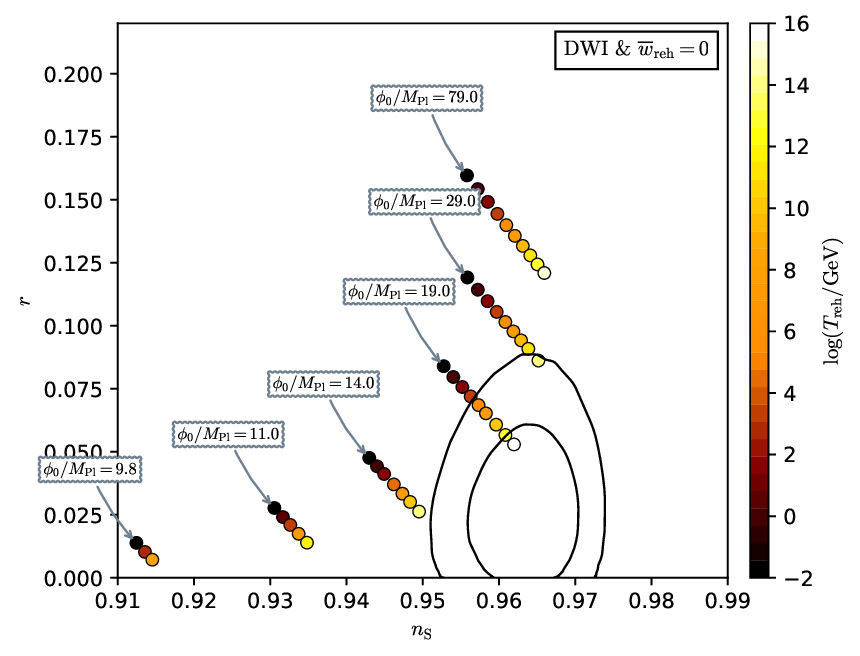}
\includegraphics[width=\wappfig,clip=true]{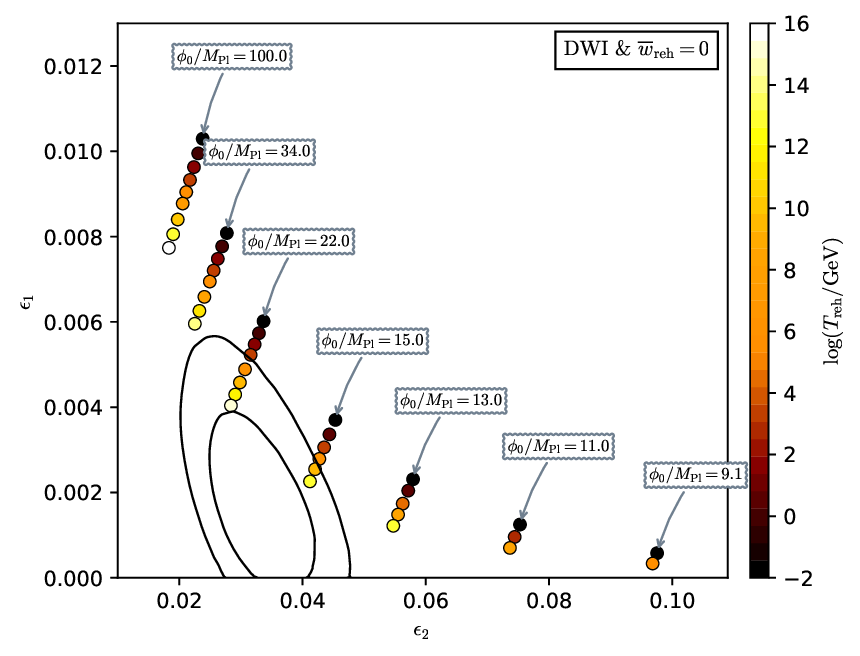}
\caption{Reheating consistent slow-roll predictions for the double
  well models in the plane $(\nS,r)$ (top panel) and the plane
  $(\epsilon_1,\epsilon_2)$ (bottom panel). The solid contours are the
  one and two-sigma {\data} confidence intervals (marginalized over
  second order slow-roll). The shape of the zone covered by the models
  predictions is similar to the one for Small Field Inflation (SFI,
  see \Fig{fig:CMBSFI2}), except in the domain $\phizero\gg\Mp$, which
  is the one favored by the observations. }
\label{fig:CMBDWI}
\end{center}
\end{figure}

\subsection{Mutated Hilltop Inflation (\hyperref[sec:mhi]{MHI})}

\begin{figure}[H]
\begin{center}
\includegraphics[width=\wappfig,clip=true]{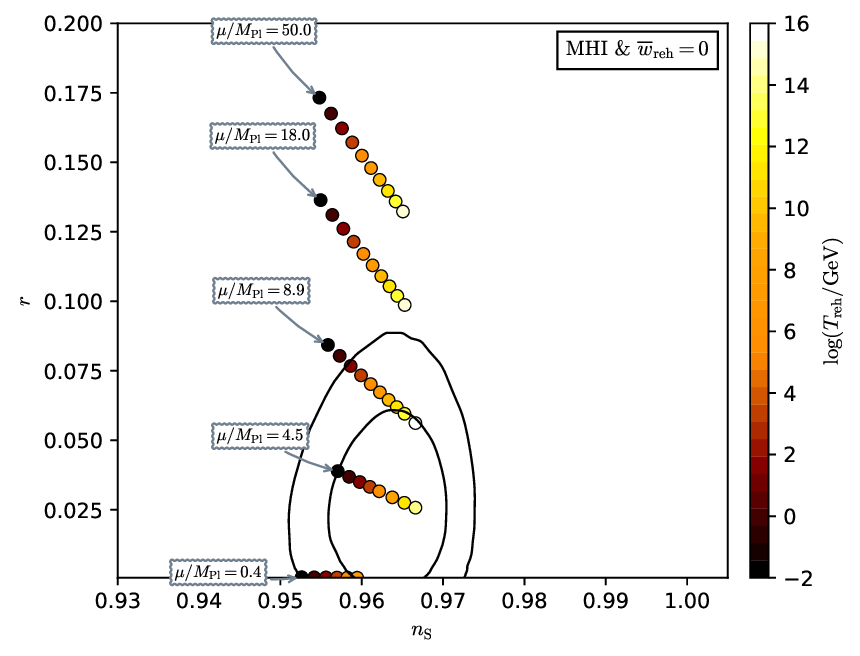}
\includegraphics[width=\wappfig,clip=true]{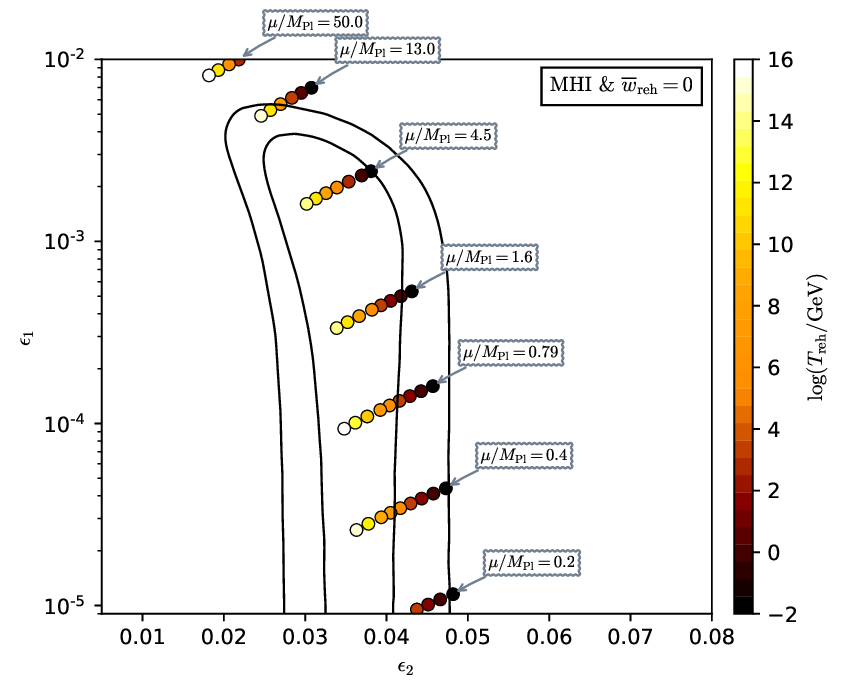}
\caption{Reheating consistent slow-roll predictions for the mutated
  hilltop models in the plane $(\nS,r)$ (top panel) and the plane
  $(\epsilon_1,\epsilon_2)$ (bottom panel). The solid contours are the
  one and two-sigma {\data} confidence intervals (marginalized over
  second order slow-roll).  For small values of $\mu/\Mp$, this model
  predicts a very small amount of primordial gravitational waves.}
\label{fig:CMBMHI}
\end{center}
\end{figure}

\subsection{Radion Gauge Inflation (\hyperref[sec:rgi]{RGI})}

\begin{figure}[H]
\begin{center}
\includegraphics[width=\wappfig,clip=true]{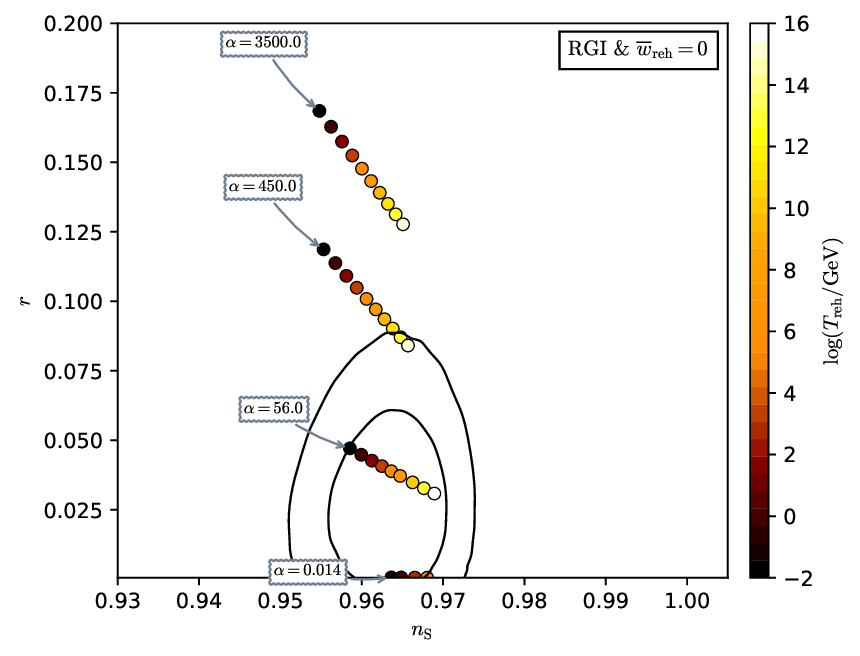}
\includegraphics[width=\wappfig,clip=true]{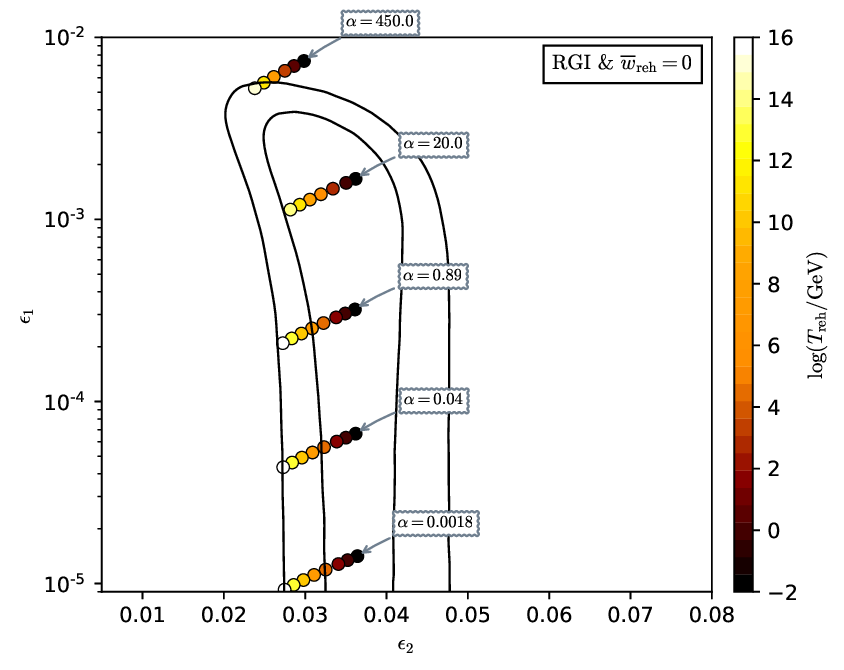}
\caption{Reheating consistent slow-roll predictions for the radion
  gauge models in the plane $(\nS,r)$ (top panel) and the plane
  $(\epsilon_1,\epsilon_2)$ (bottom panel). The solid contours are the
  one and two-sigma {\data} confidence intervals (marginalized over
  second order slow-roll). At large values of $\alpha$, the
  predictions are the same as the large field model with $p=2$ (see
  \Fig{fig:CMBLFI}) for which $\epsilon_2=2\epsilon_1$.}
\label{fig:CMBRGI}
\end{center}
\end{figure}

\subsection{MSSM Inflation (\hyperref[sec:mssmi]{MSSMI})}

\begin{figure}[H]
\begin{center}
\includegraphics[width=\wappfig,clip=true]{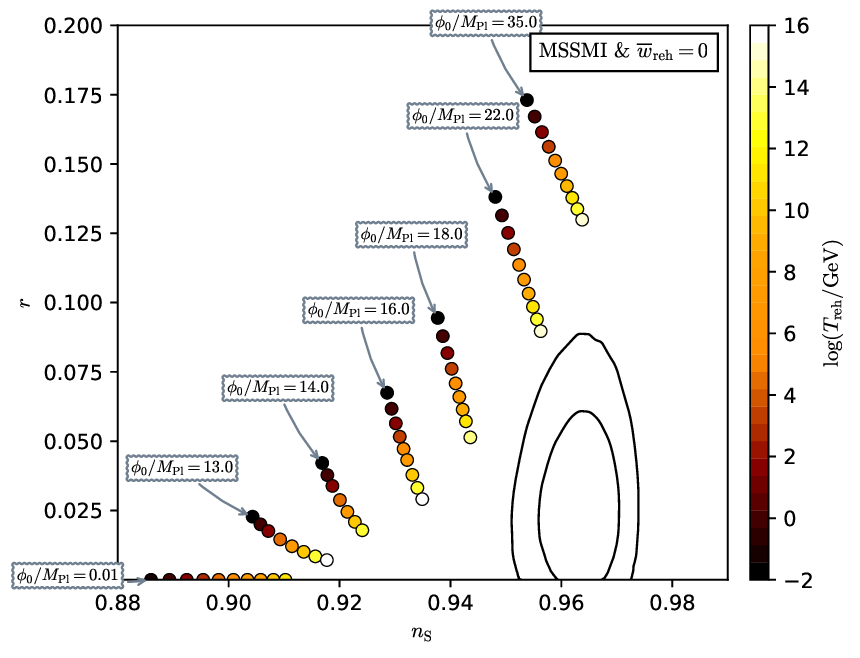}
\includegraphics[width=\wappfig,clip=true]{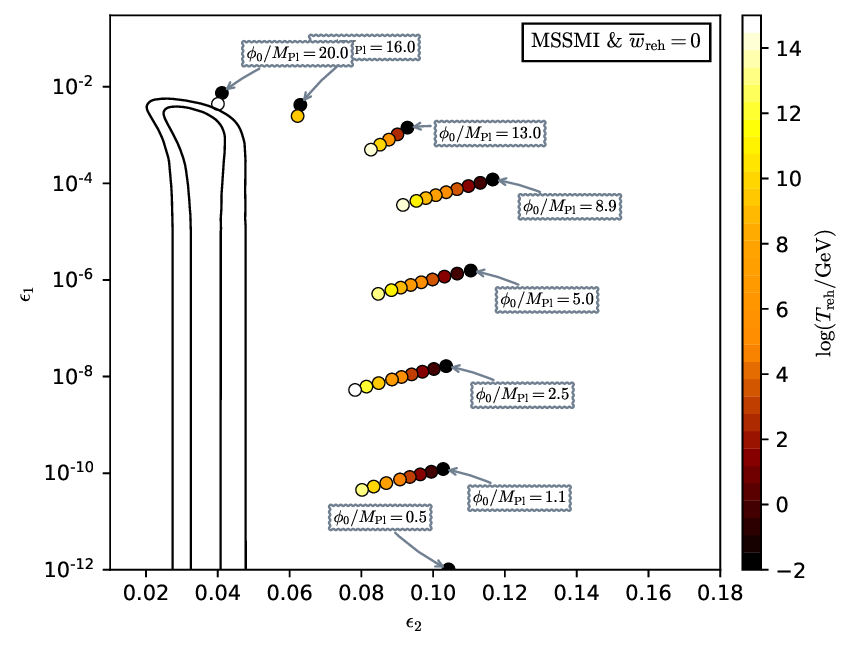}
\caption{Reheating consistent slow-roll predictions for the MSSMI
  models in the plane $(\nS,r)$ (top panel) and the plane
  $(\epsilon_1,\epsilon_2)$ (bottom panel). The solid contours are the
  one and two-sigma {\data} confidence intervals (marginalized over
  second order slow-roll). For large values of $\phizero\Mp$, the
  model predictions approach $r=4(1-\nS)$, i.e, $\epsilon_2 = 2 \epsilon_1$.}
\label{fig:CMBMSSMI}
\end{center}
\end{figure}

\subsection{Renormalizable Inflection Point Inflation (\hyperref[sec:ripi]{RIPI})}

\begin{figure}[H]
\begin{center}
\includegraphics[width=\wappfig,clip=true]{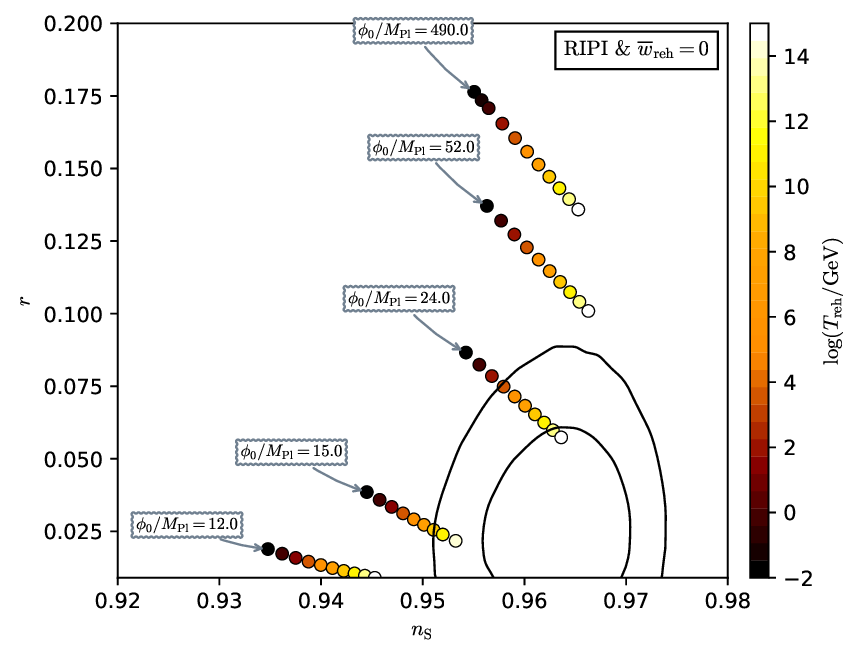}
\includegraphics[width=\wappfig,clip=true]{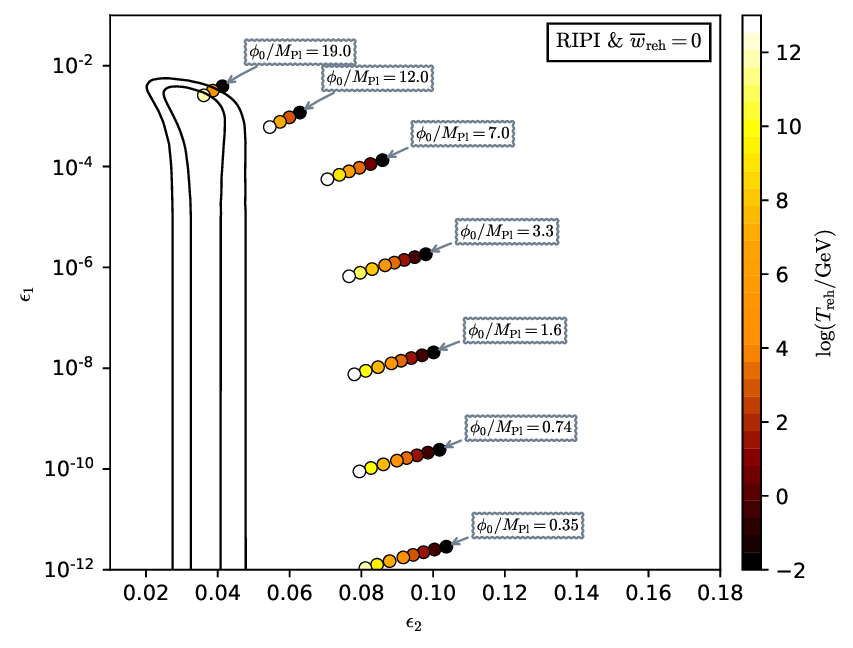}
\caption{Reheating consistent slow-roll predictions for the
  renormalizable inflection point models in the plane $(\nS,r)$ (top
  panel) and the plane $(\epsilon_1,\epsilon_2)$ (bottom panel). The
  solid contours are the one and two-sigma {\data} confidence
  intervals (marginalized over second order slow-roll).}
\label{fig:CMBRIPI}
\end{center}
\end{figure}

\subsection{Arctan Inflation (\hyperref[sec:ai]{AI})}

\begin{figure}[H]
\begin{center}
\includegraphics[width=\wappfig,clip=true]{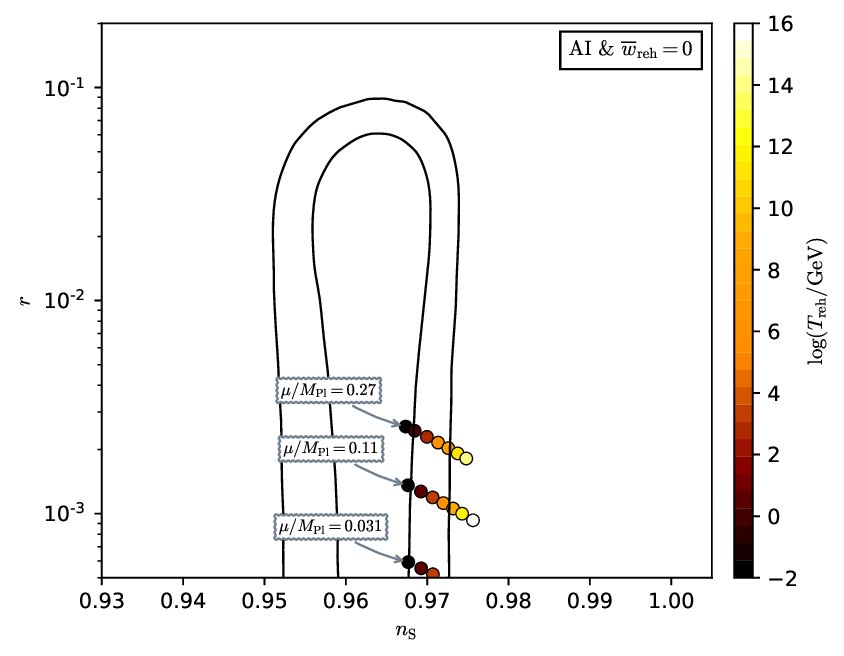}
\includegraphics[width=\wappfig,clip=true]{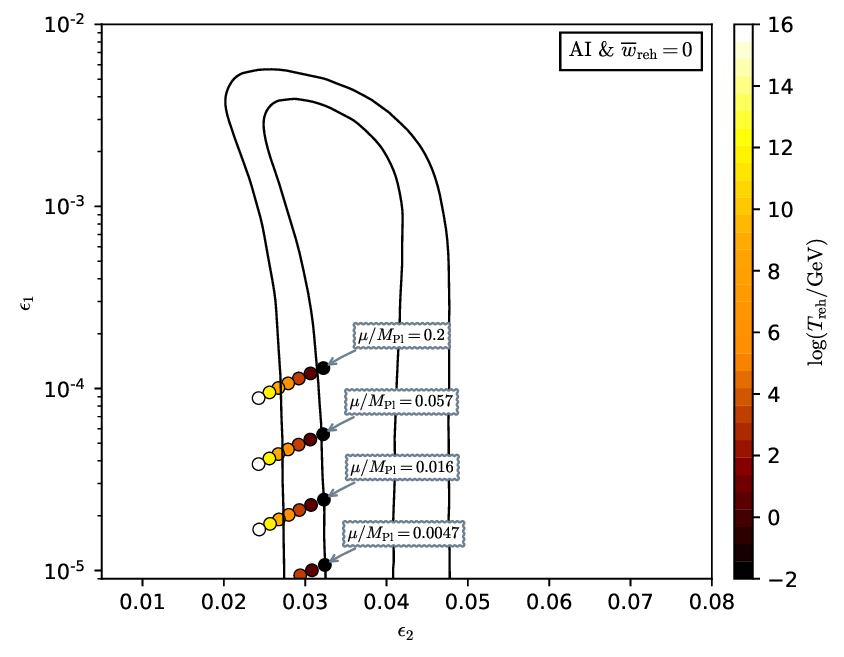}
\caption{Reheating consistent slow-roll predictions for the ArcTan
  models in the plane $(\nS,r)$ (top panel) and the plane
  $(\epsilon_1,\epsilon_2)$ (bottom panel). The solid contours are the
  one and two-sigma {\data} confidence intervals (marginalized over
  second order slow-roll).}
\label{fig:CMBAI}
\end{center}
\end{figure}

\subsection{Constant \texorpdfstring{$\nS$}{nS} A Inflation (\hyperref[sec:cnai]{CNAI})}

\begin{figure}[H]
\begin{center}
\includegraphics[width=\wappfig,clip=true]{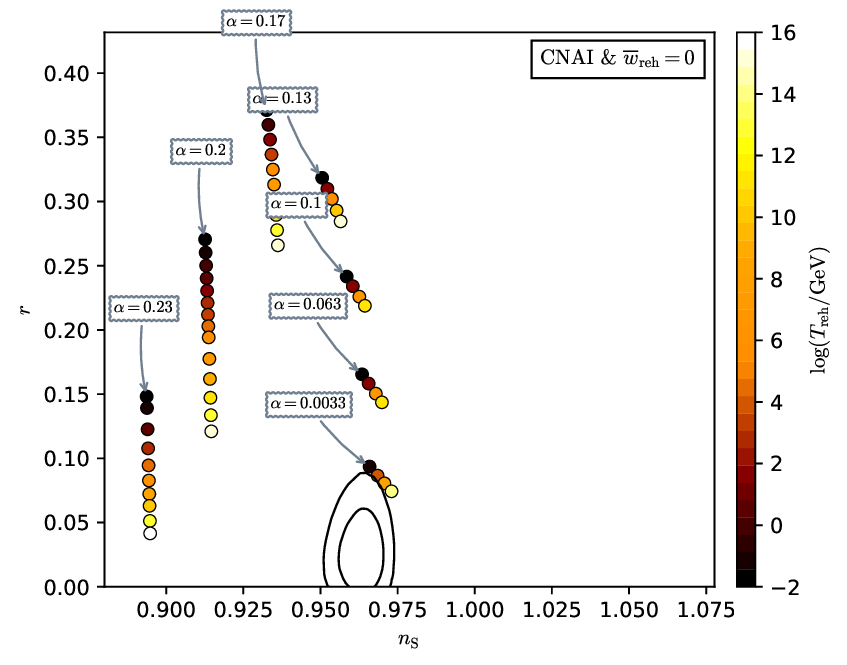}
\includegraphics[width=\wappfig,clip=true]{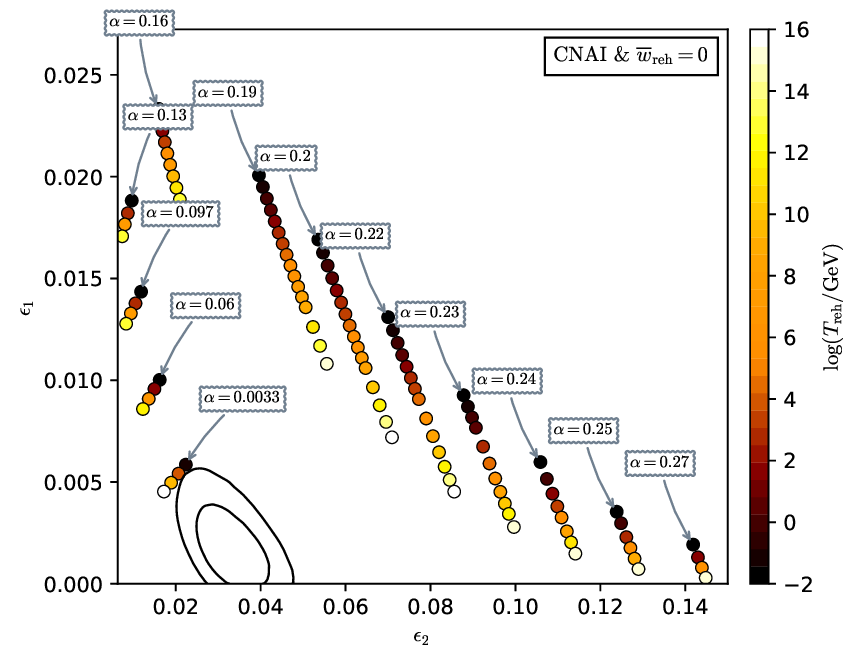}
\caption{Reheating consistent slow-roll predictions for the constant
  $\nS$ A models in the plane $(\nS,r)$ (top panel) and the plane
  $(\epsilon_1,\epsilon_2)$ (bottom panel). The solid contours are the
  one and two-sigma {\data} confidence intervals (marginalized over
  second order slow-roll).}
\label{fig:CMBCNAI}
\end{center}
\end{figure}

\subsection{Constant \texorpdfstring{$\nS$}{nS} B Inflation (\hyperref[sec:cnbi]{CNBI})}

\begin{figure}[H]
\begin{center}
\includegraphics[width=\wappfig,clip=true]{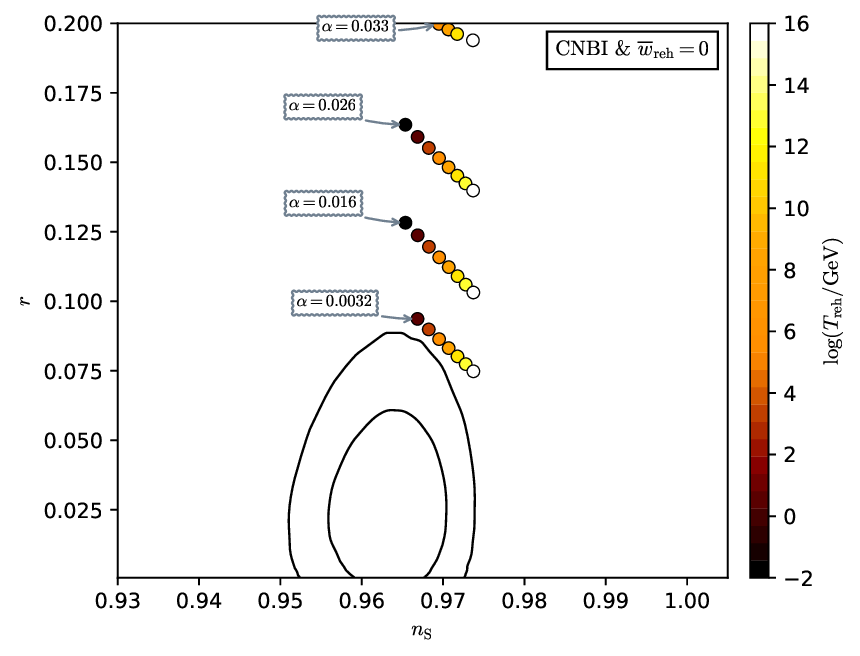}
\includegraphics[width=\wappfig,clip=true]{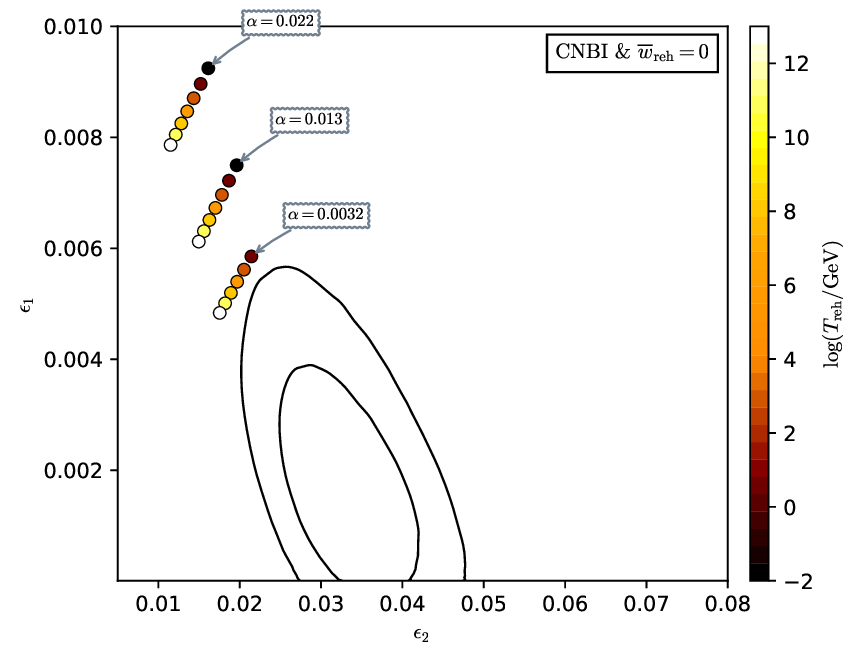}
\caption{Reheating consistent slow-roll predictions for the constant
  $\nS$ B models in the plane $(\nS,r)$ (top panel) and the plane
  $(\epsilon_1,\epsilon_2)$ (bottom panel). The solid contours are the
  one and two-sigma {\data} confidence intervals (marginalized over
  second order slow-roll).}
\label{fig:CMBCNBI}
\end{center}
\end{figure}

\subsection{Open String Tachyonic Inflation (\hyperref[sec:osti]{OSTI})}

\begin{figure}[H]
\begin{center}
\includegraphics[width=\wappfig,clip=true]{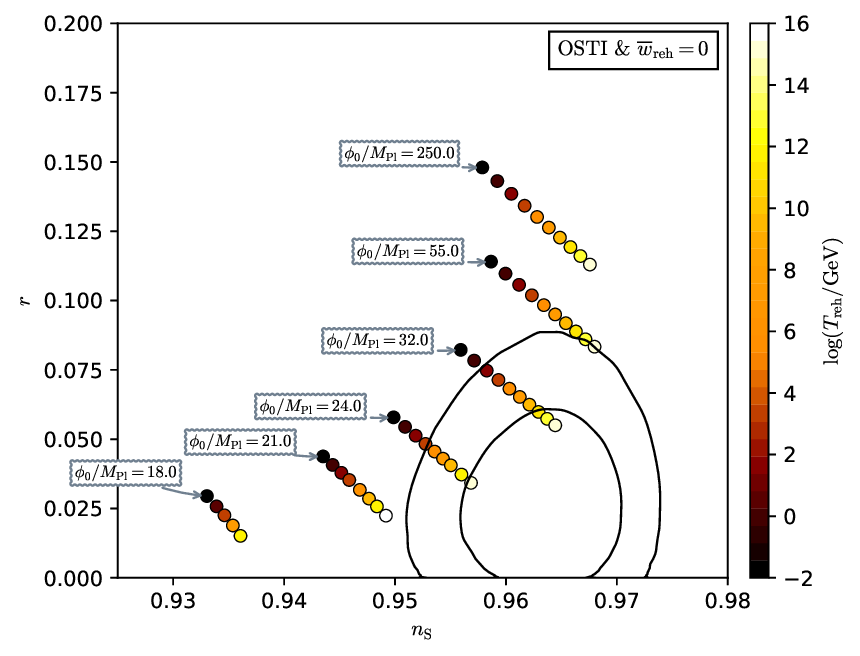}
\includegraphics[width=\wappfig,clip=true]{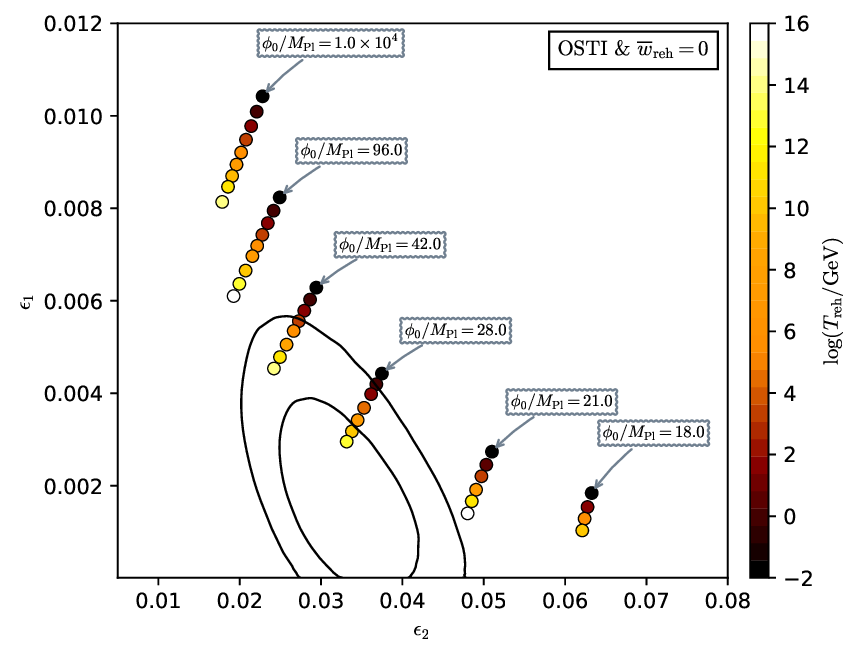}
\caption{Reheating consistent slow-roll predictions for the open
  string tachyonic models in the plane $(\nS,r)$ (top panel) and the
  plane $(\epsilon_1,\epsilon_2)$ (bottom panel). The solid contours
  are the one and two-sigma {\data} confidence intervals (marginalized
  over second order slow-roll). For $\phizero/\Mp\gg 1$, the model
  predictions approach $r=4\left(1-\nS\right)$, {\ie}
  $\epsilon_2=2\epsilon_1$.}
\label{fig:CMBOSTI}
\end{center}
\end{figure}

\subsection{Witten-O'Raifeartaigh Inflation (\hyperref[sec:wri]{WRI})}

\begin{figure}[H]
\begin{center}
\includegraphics[width=\wappfig,clip=true]{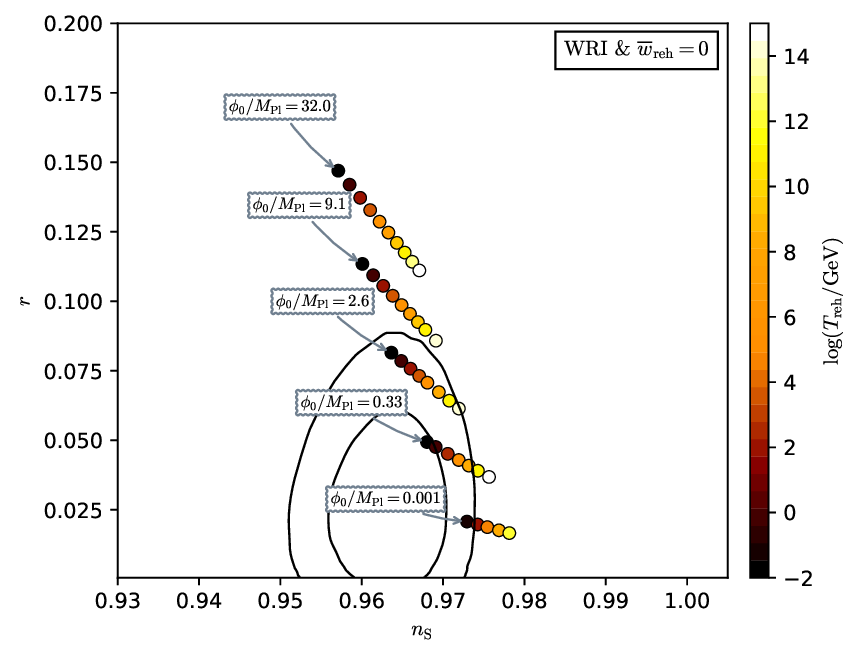}
\includegraphics[width=\wappfig,clip=true]{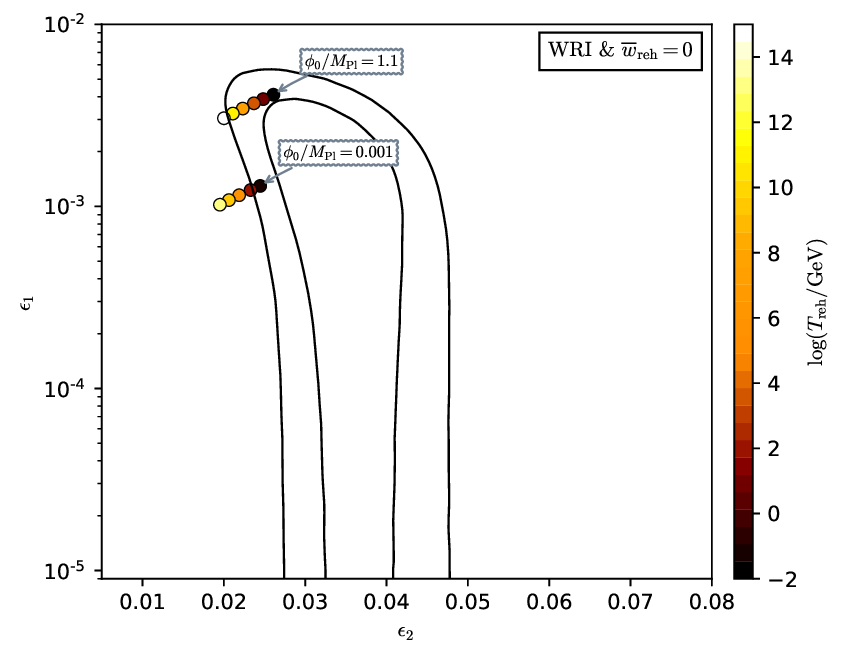}
\caption{Reheating consistent slow-roll predictions for the
  Witten-O'Raifeartaigh models in the plane $(\nS,r)$ (top panel) and
  the plane $(\epsilon_1,\epsilon_2)$ (bottom panel). The solid
  contours are the one and two-sigma {\data} confidence intervals
  (marginalized over second order slow-roll). At large field values
  $\phizero/\Mp\gg 1$, the model predictions approach
  $r=4\left(1-\nS\right)$, \ie $\epsilon_2=2\epsilon_1$. }
\label{fig:CMBWRI}
\end{center}
\end{figure}

\subsection{Dual Inflation (\hyperref[sec:di]{DI})}

\begin{figure}[H]
\begin{center}
\includegraphics[width=\wappfig,clip=true]{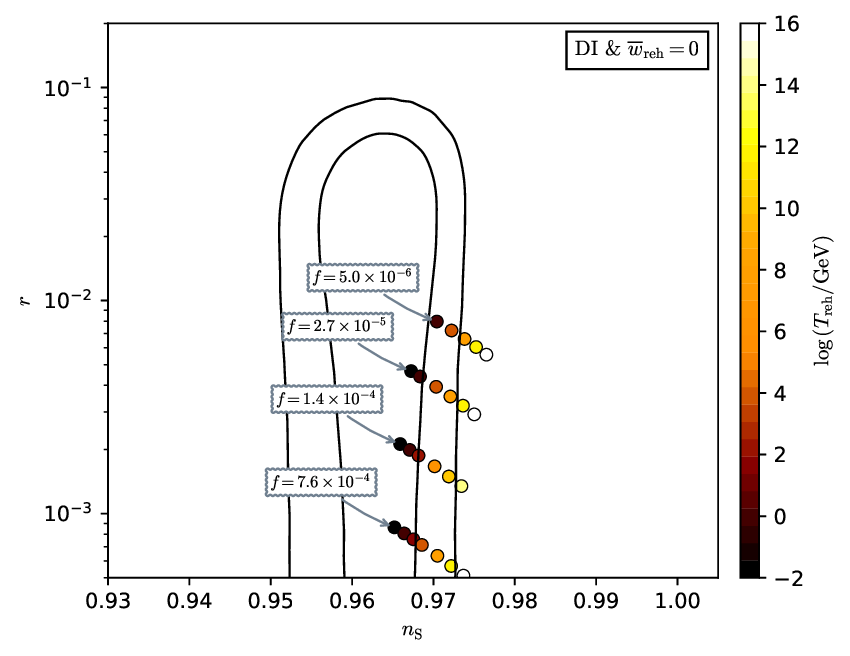}
\includegraphics[width=\wappfig,clip=true]{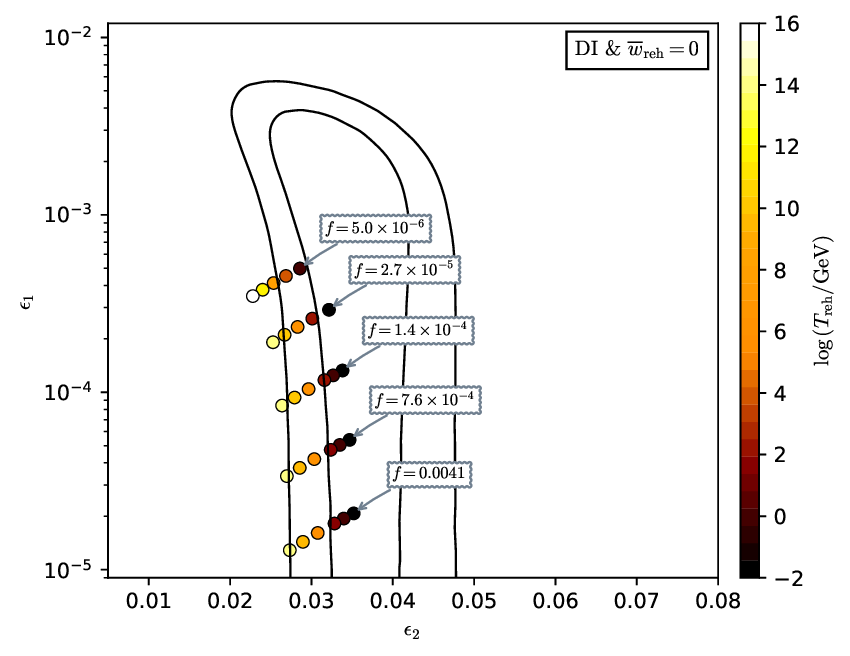}
\caption{Reheating consistent slow-roll predictions for the dual
  inflation in the plane $(\nS,r)$ (top panel) and the plane
  $(\epsilon_1,\epsilon_2)$ (bottom panel). The solid contours are the
  one and two-sigma {\data} confidence intervals (marginalized over
  second order slow-roll).}
\label{fig:CMBDI}
\end{center}
\end{figure}

\subsection{Cublicly Corrected Starobinsky Inflation (\hyperref[sec:ccsi]{CCSI})}

\begin{figure}[H]
\begin{center}
\includegraphics[width=\wappfig,clip=true]{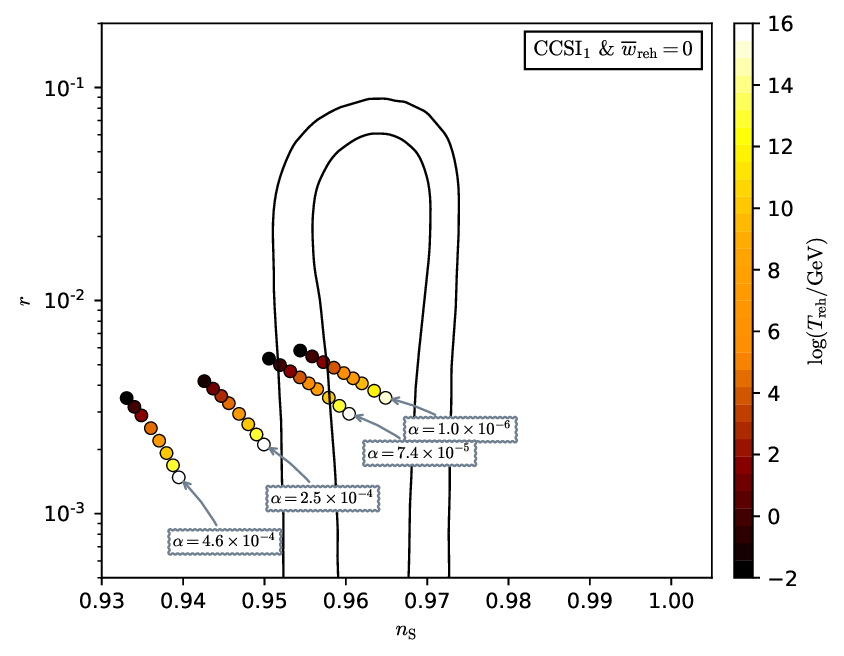}
\includegraphics[width=\wappfig,clip=true]{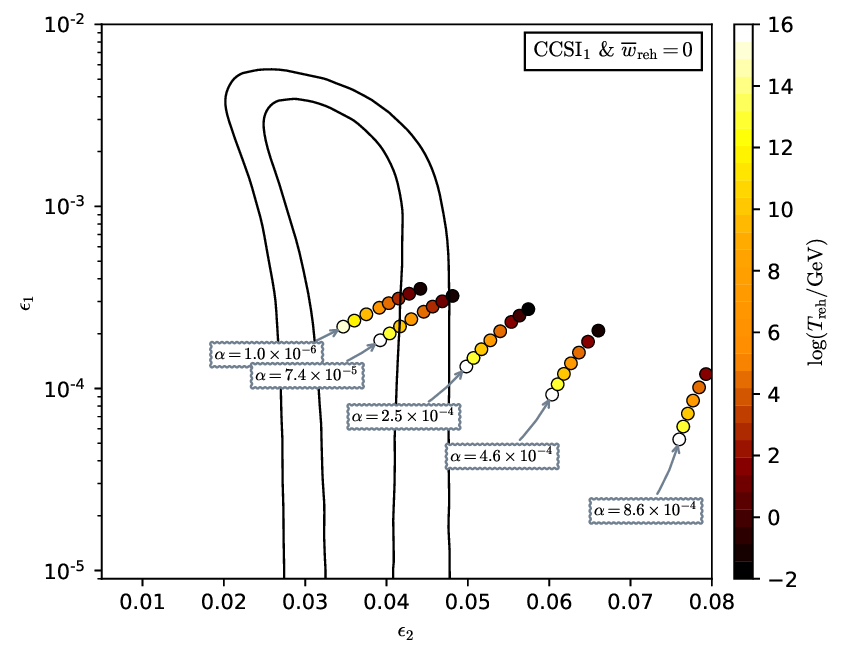}
\caption{Reheating consistent slow-roll predictions for cubicly
  corrected Starobinsky inflation with $\alpha >0$ and at small field
  values (CSSI1), in the plane $(\nS,r)$ (top panel) and the plane
  $(\epsilon_1,\epsilon_2)$ (bottom panel). The solid contours are the
  one and two-sigma {\data} confidence intervals (marginalized over
  second order slow-roll).}
\label{fig:CMBCCSI1}
\end{center}
\end{figure}

\begin{figure}[H]
\begin{center}
\includegraphics[width=\wappfig,clip=true]{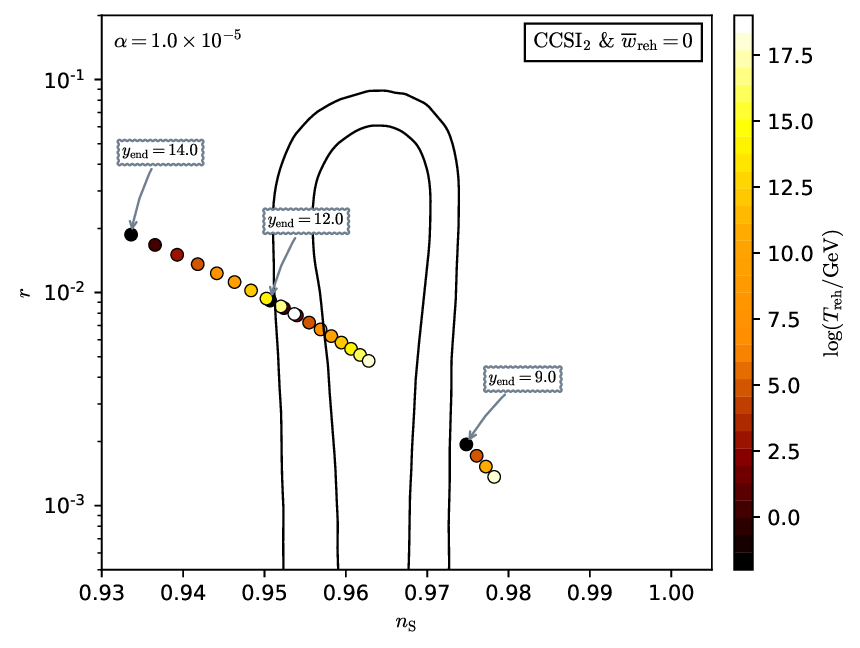}
\includegraphics[width=\wappfig,clip=true]{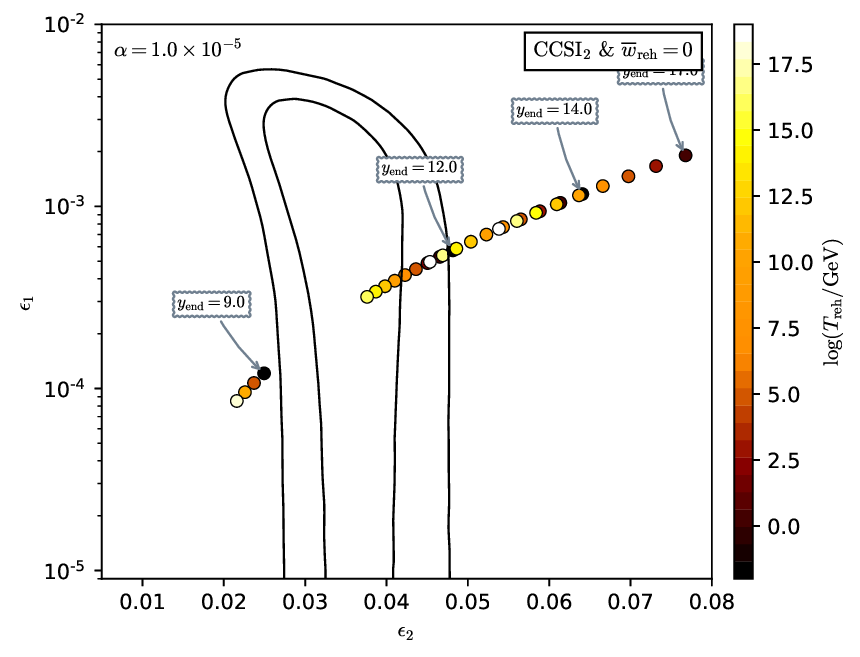}
\caption{Reheating consistent slow-roll predictions for cubicly
  corrected Starobinsky inflation with $\alpha=10^{-5}$ and at large
  field values (CSSI2), in the plane $(\nS,r)$ (top panel) and the
  plane $(\epsilon_1,\epsilon_2)$ (bottom panel). The solid contours
  are the one and two-sigma {\data} confidence intervals (marginalized
  over second order slow-roll). The dimensionless field values at
  which inflation ends, $\yend$, varies between the minimal possible
  value to obtain $120$ {\efolds} of inflation up to five times this
  number.}
\label{fig:CMBCCSI2}
\end{center}
\end{figure}

\begin{figure}[H]
\begin{center}
\includegraphics[width=\wappfig,clip=true]{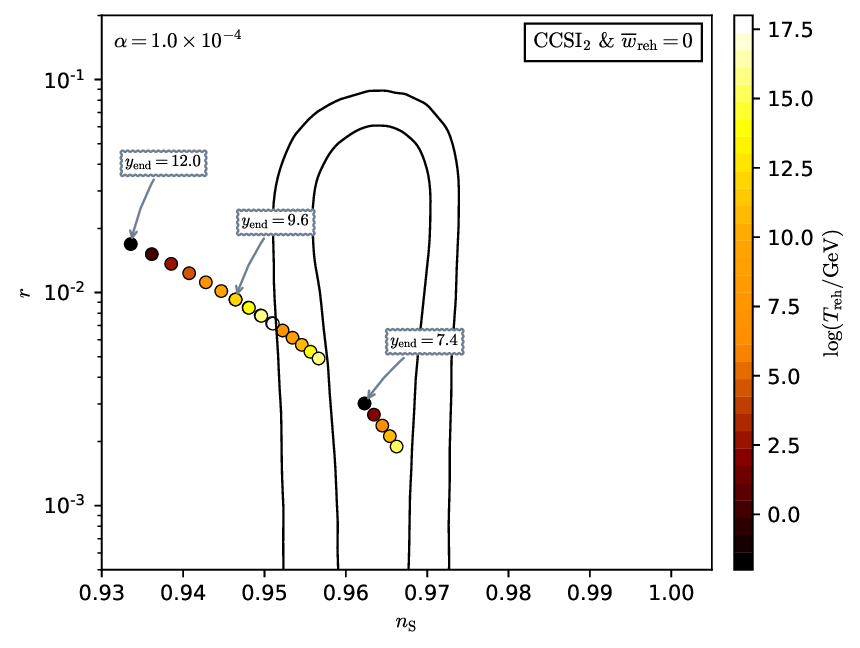}
\includegraphics[width=\wappfig,clip=true]{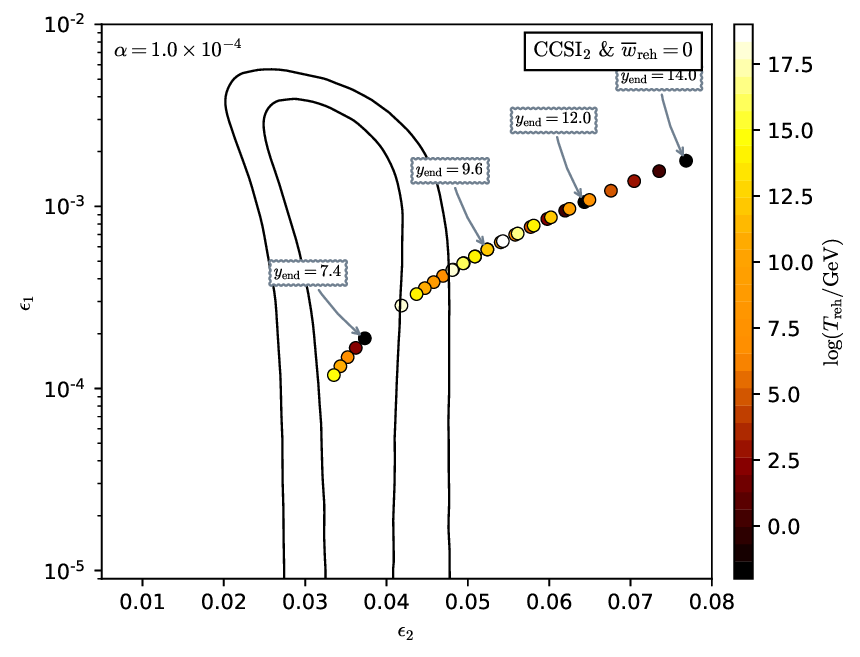}
\caption{Reheating consistent slow-roll predictions for cubicly
  corrected Starobinsky inflation with $\alpha=10^{-4}$ and at large field
  values (CSSI2), in the plane $(\nS,r)$ (top panel) and the plane
  $(\epsilon_1,\epsilon_2)$ (bottom panel). The solid contours are the
  one and two-sigma {\data} confidence intervals (marginalized over
  second order slow-roll). The dimensionless field values at
  which inflation ends, $\yend$, varies between the minimal possible
  value to obtain $120$ {\efolds} of inflation up to five times this
  number.}
\label{fig:CMBCCSI2_1}
\end{center}
\end{figure}

\begin{figure}[H]
\begin{center}
\includegraphics[width=\wappfig,clip=true]{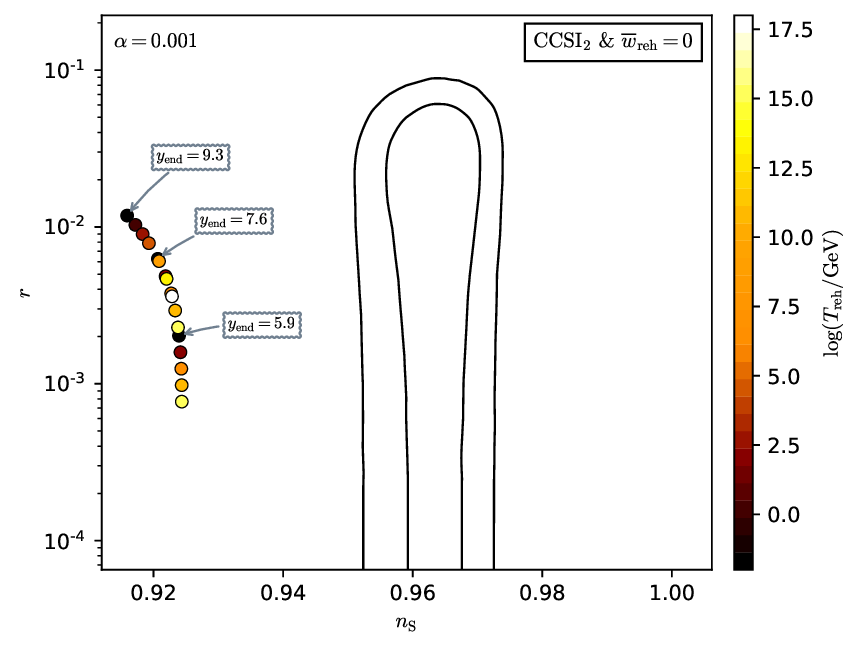}
\includegraphics[width=\wappfig,clip=true]{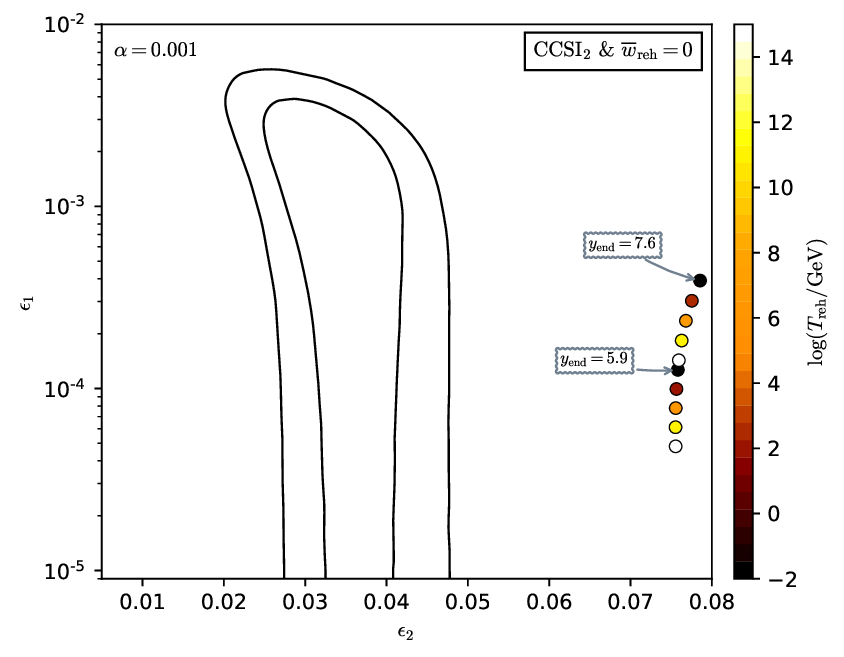}
\caption{Reheating consistent slow-roll predictions for cubicly
  corrected Starobinsky inflation with $\alpha=10^{-3}$ and at large
  field values (CSSI2), in the plane $(\nS,r)$ (top panel) and the
  plane $(\epsilon_1,\epsilon_2)$ (bottom panel). The solid contours
  are the one and two-sigma {\data} confidence intervals (marginalized
  over second order slow-roll). The dimensionless field values at
  which inflation ends, $\yend$, varies between the minimal possible
  value to obtain $120$ {\efolds} of inflation up to five times this
  number.}
\label{fig:CMBCCSI2_2}
\end{center}
\end{figure}

\begin{figure}[H]
\begin{center}
\includegraphics[width=\wappfig,clip=true]{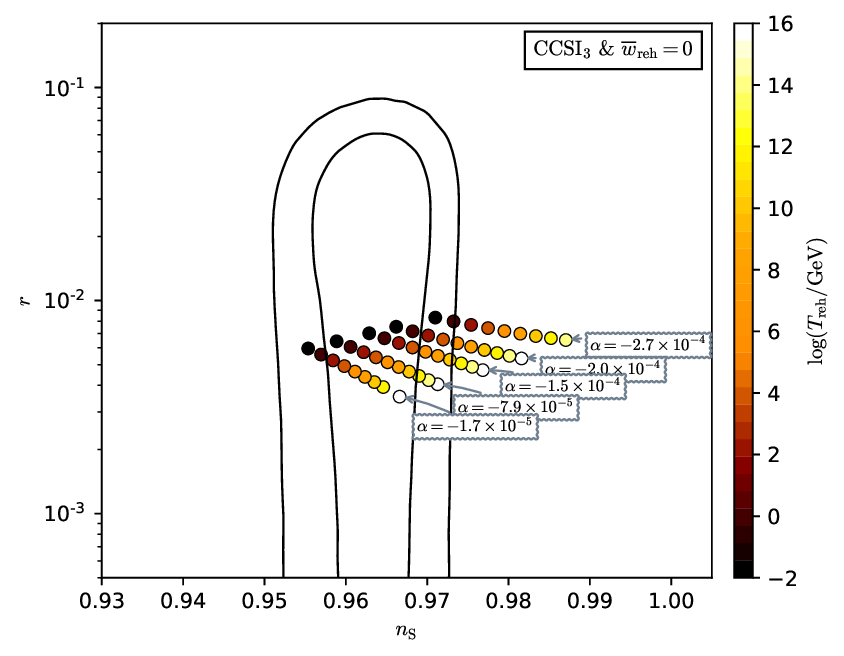}
\includegraphics[width=\wappfig,clip=true]{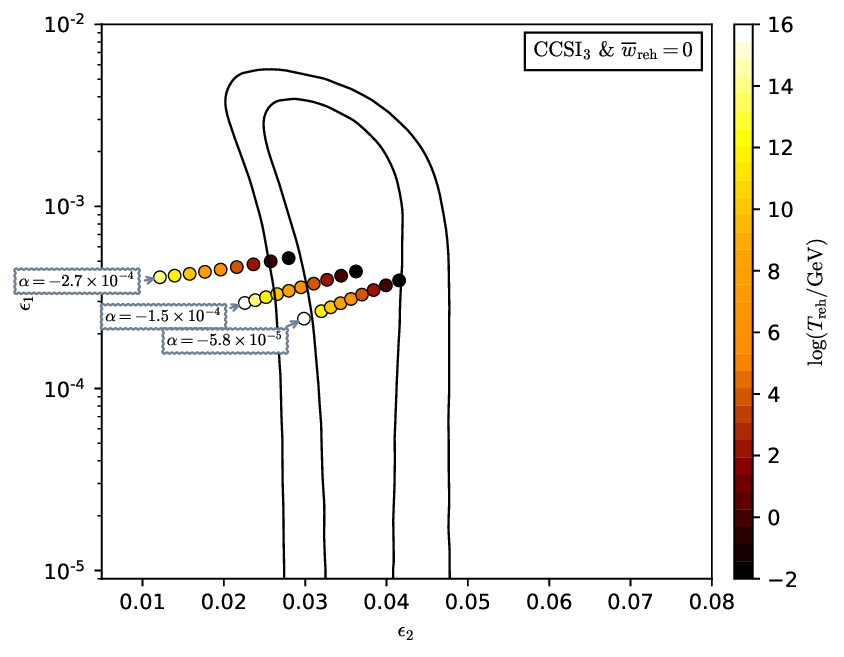}
\caption{Reheating consistent slow-roll predictions for cubicly
  corrected Starobinsky inflation with $\alpha<0$ and at small field
  values (CSSI3), in the plane $(\nS,r)$ (top panel) and the plane
  $(\epsilon_1,\epsilon_2)$ (bottom panel). The solid contours are the
  one and two-sigma {\data} confidence intervals (marginalized over
  second order slow-roll).}
\label{fig:CMBCCSI3}
\end{center}
\end{figure}

\subsection{Symmetry Breaking K\"ahler Inflation (\hyperref[sec:sbki]{SBKI})}

\begin{figure}[H]
\begin{center}
\includegraphics[width=\wappfig,clip=true]{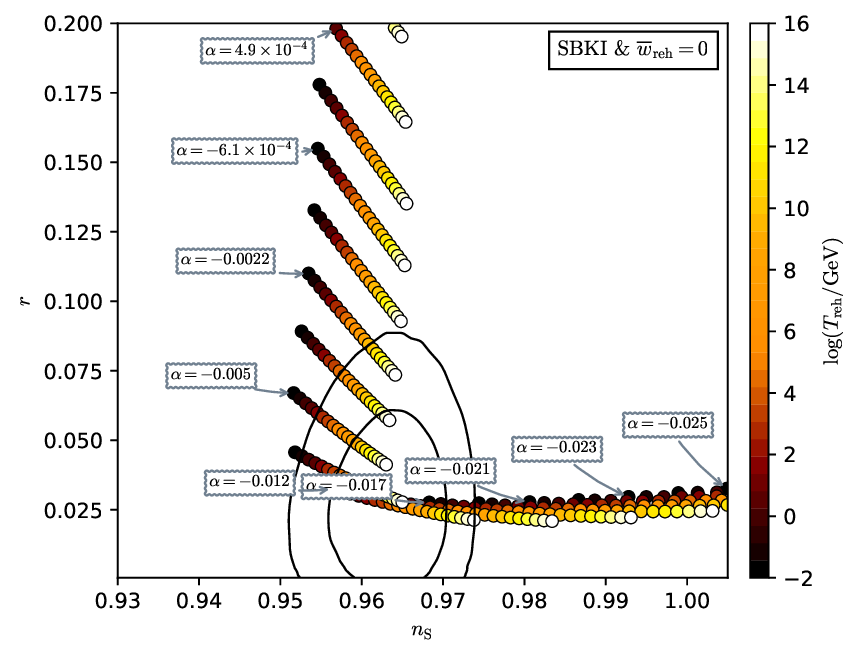}
\includegraphics[width=\wappfig,clip=true]{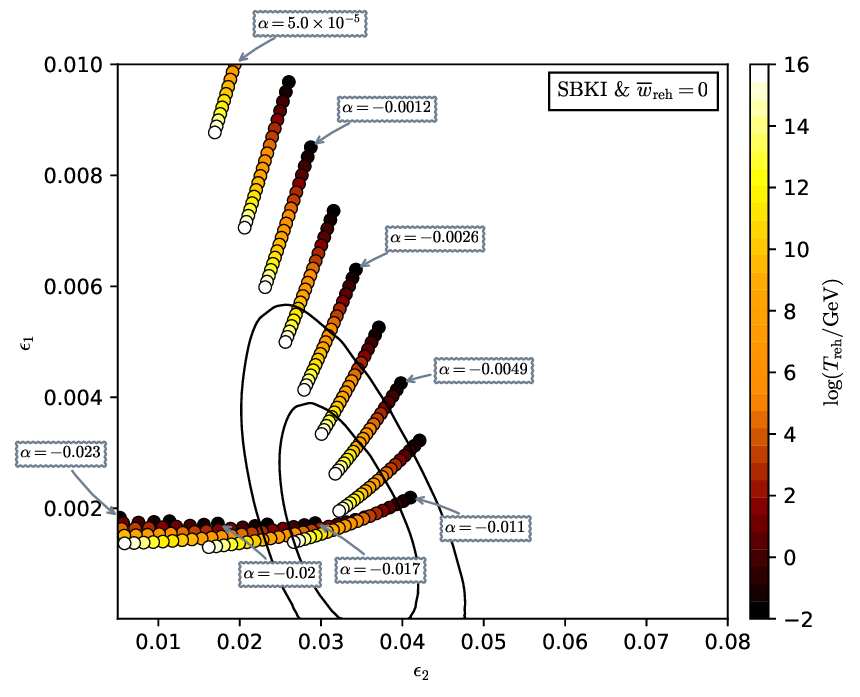}
\caption{Reheating consistent slow-roll predictions for the Symmetry
  Breaking K\"ahler Inflation models in the plane $(\nS,r)$ (top
  panel) and the plane $(\epsilon_1,\epsilon_2)$ (bottom panel). The
  solid contours are the one and two-sigma {\data} confidence
  intervals (marginalized over second order slow-roll).}
\label{fig:CMBSBKI}
\end{center}
\end{figure}

\subsection{Axion Hilltop Inflation (\hyperref[sec:ahi]{AHI})}

\begin{figure}[H]
\begin{center}
\includegraphics[width=\wappfig,clip=true]{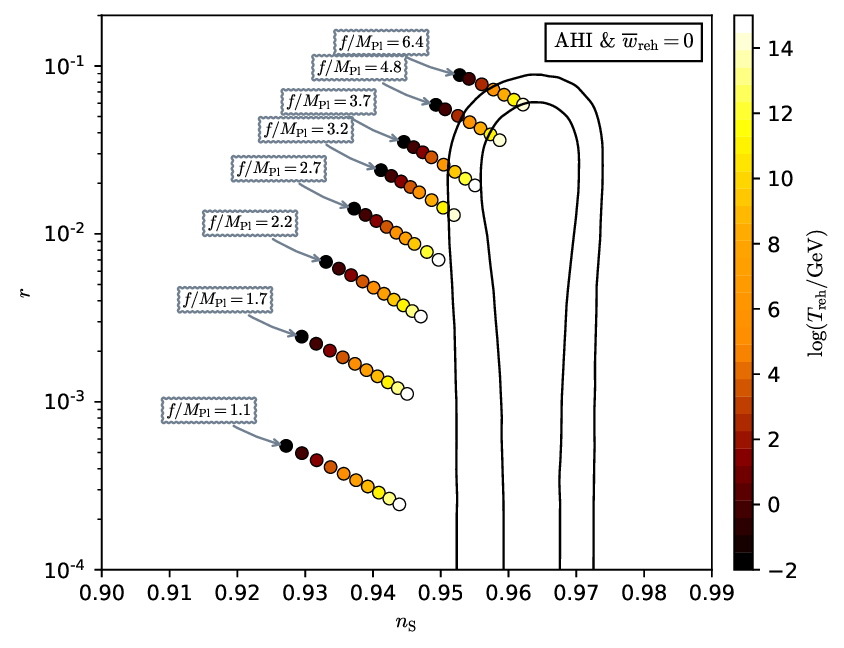}
\includegraphics[width=\wappfig,clip=true]{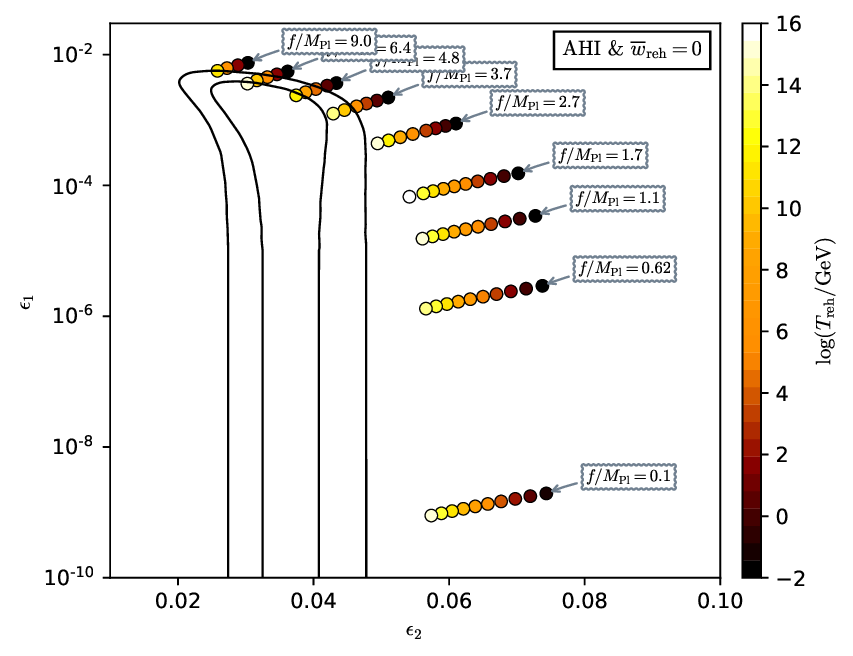}
\caption{Reheating consistent slow-roll predictions for the
  axion hilltop inflation model as a function of $f/\Mp$ in the plane $(\nS,r)$
  (top panel) and the plane $(\epsilon_1,\epsilon_2)$ (bottom
  panel). The solid contours are the one and two-sigma {\data}
  confidence intervals (marginalized over second order slow-roll).}
\label{fig:CMBAHI_0}
\end{center}
\end{figure}

\subsection{Pure Arctan Inflation (\hyperref[sec:pai]{PAI})}

\begin{figure}[H]
\begin{center}
\includegraphics[width=\wappfig,clip=true]{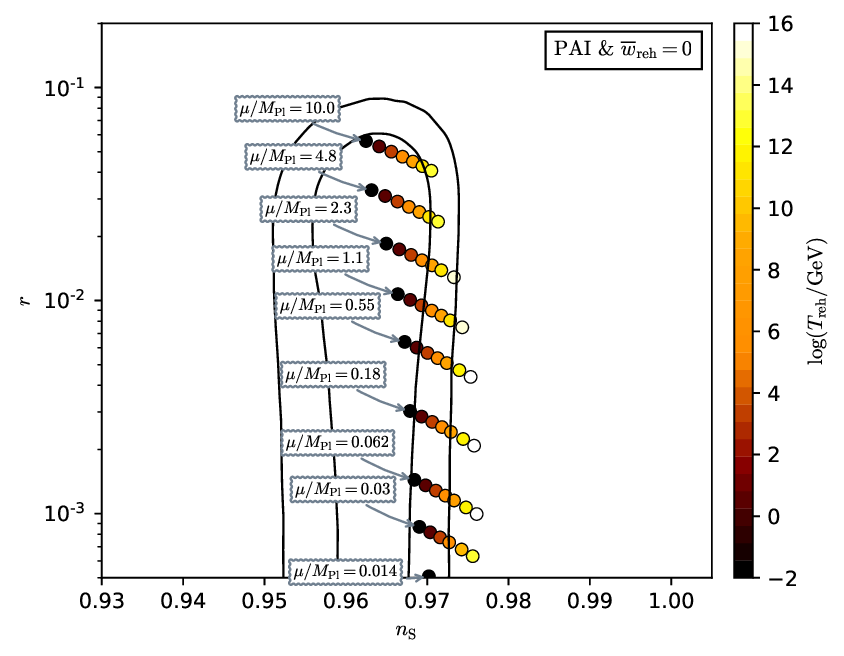}
\includegraphics[width=\wappfig,clip=true]{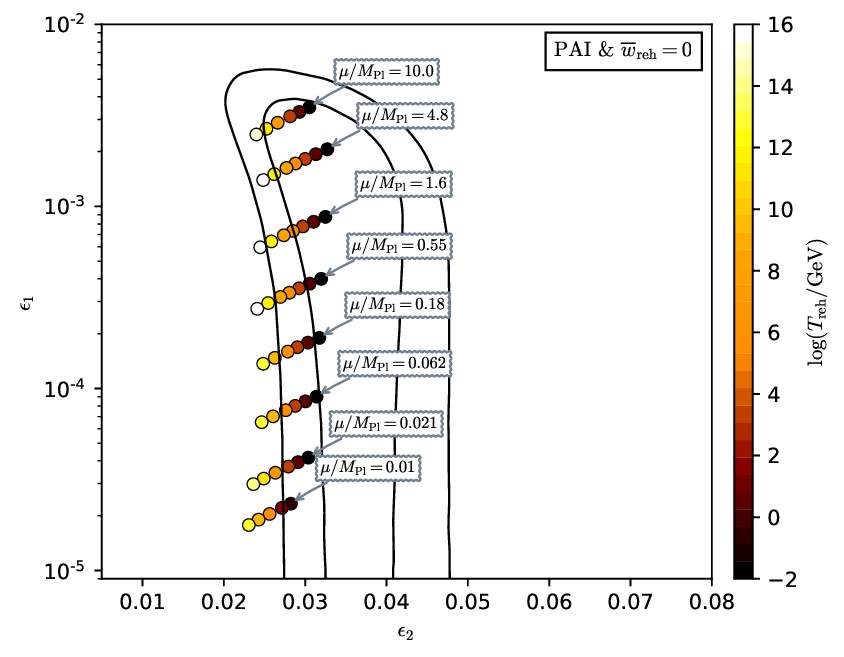}
\caption{Reheating consistent slow-roll predictions for the
  pure arctan inflation model as a function of $\mu/\Mp$ in the plane $(\nS,r)$
  (top panel) and the plane $(\epsilon_1,\epsilon_2)$ (bottom
  panel). The solid contours are the one and two-sigma {\data}
  confidence intervals (marginalized over second order slow-roll).}
\label{fig:CMBPAI_0}
\end{center}
\end{figure}

\subsection{Superconformal \texorpdfstring{$\alpha$}{alpha}-Attractor A Inflation (\hyperref[sec:saai]{SAAI})}

\begin{figure}[H]
\begin{center}
\includegraphics[width=\wappfig,clip=true]{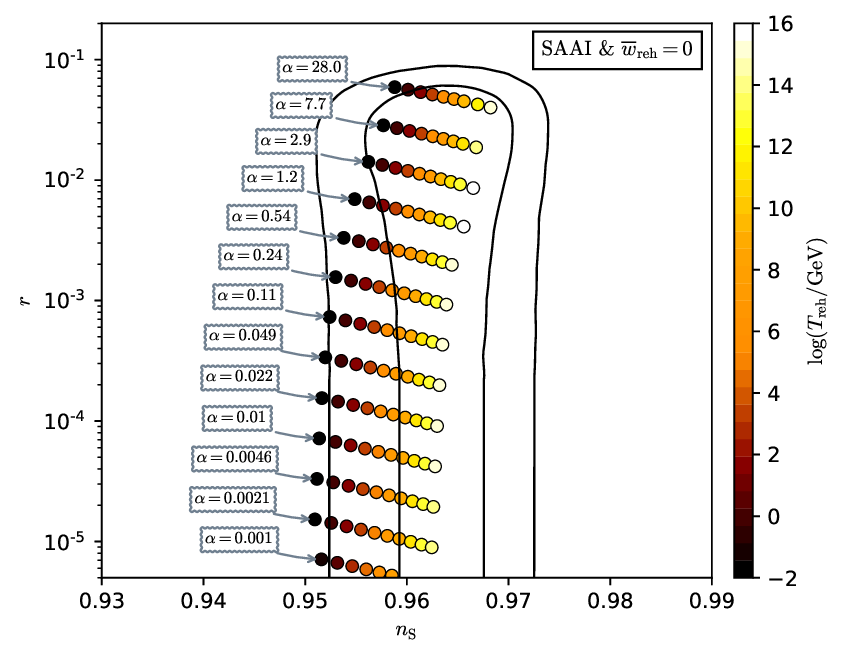}
\includegraphics[width=\wappfig,clip=true]{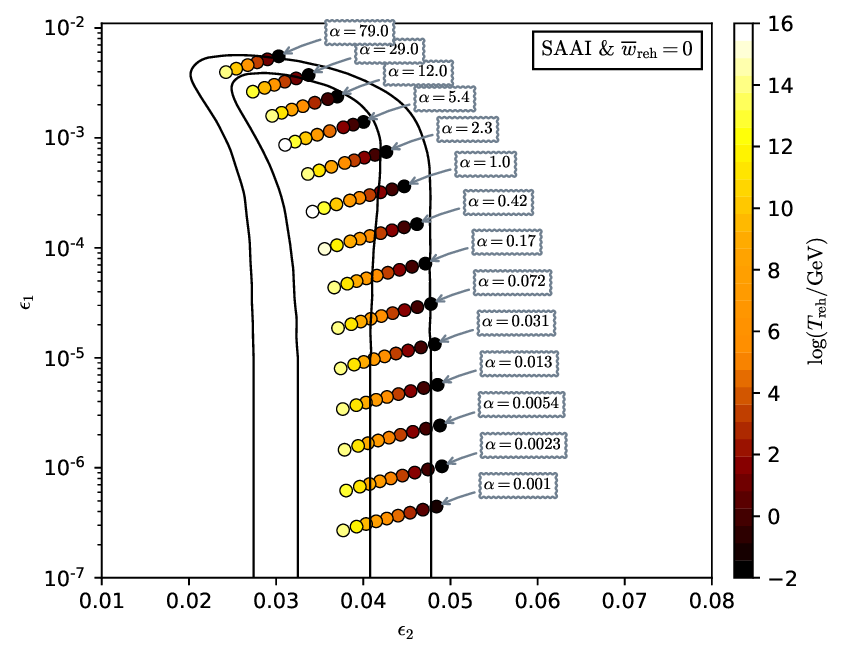}
\caption{Reheating consistent slow-roll predictions for the
  superconformal $\alpha$-attractor A inflation model as a function of
  $\alpha$ in the plane $(\nS,r)$ (top panel) and the plane
  $(\epsilon_1,\epsilon_2)$ (bottom panel). The solid contours are the
  one and two-sigma {\data} confidence intervals (marginalized over
  second order slow-roll).}
\label{fig:CMBSAAI_0}
\end{center}
\end{figure}

\subsection{T-Model Inflation (\hyperref[sec:tmi]{TMI})}

\begin{figure}[H]
\begin{center}
\includegraphics[width=\wappfig,clip=true]{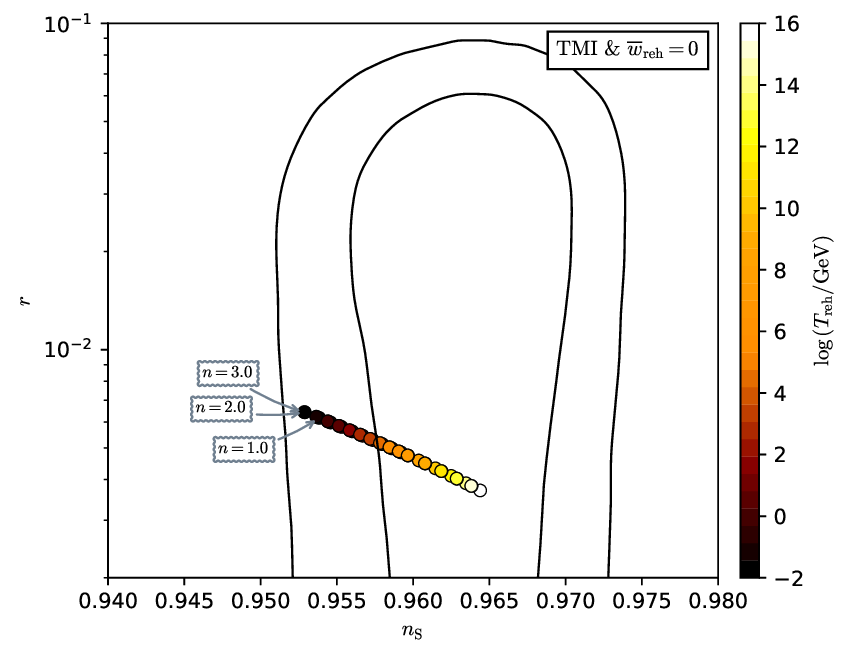}
\includegraphics[width=\wappfig,clip=true]{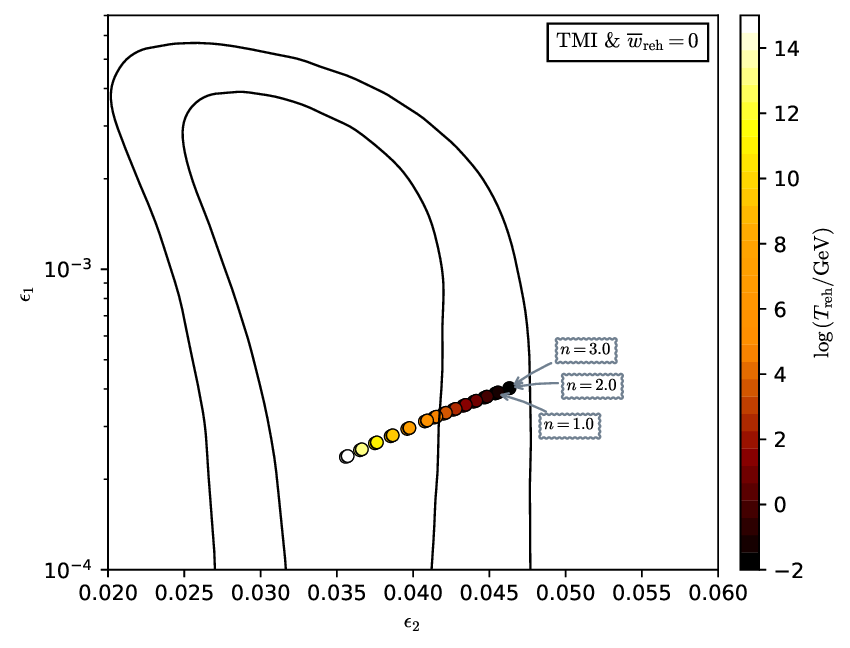}
\caption{Reheating consistent slow-roll predictions for T-Model
  Inflation. Predictions are represented as a function of $n$ in
  the plane $(\nS,r)$ (top panel) and in the plane
  $(\epsilon_1,\epsilon_2)$ (bottom panel). The solid contours are the
  one and two-sigma {\data} confidence intervals (marginalized over
  second order slow-roll).}
\label{fig:CMBTMI_0}
\end{center}
\end{figure}

\subsection{Small Field Inflation (\hyperref[sec:sfi]{SFI})}

\begin{figure}[H]
\begin{center}
\includegraphics[width=\wappfig,clip=true]{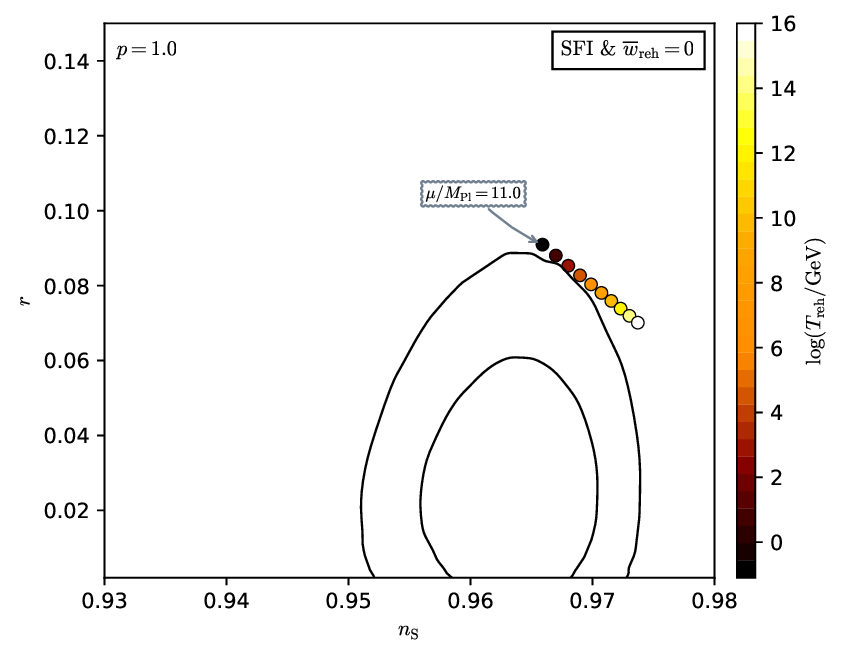}
\includegraphics[width=\wappfig,clip=true]{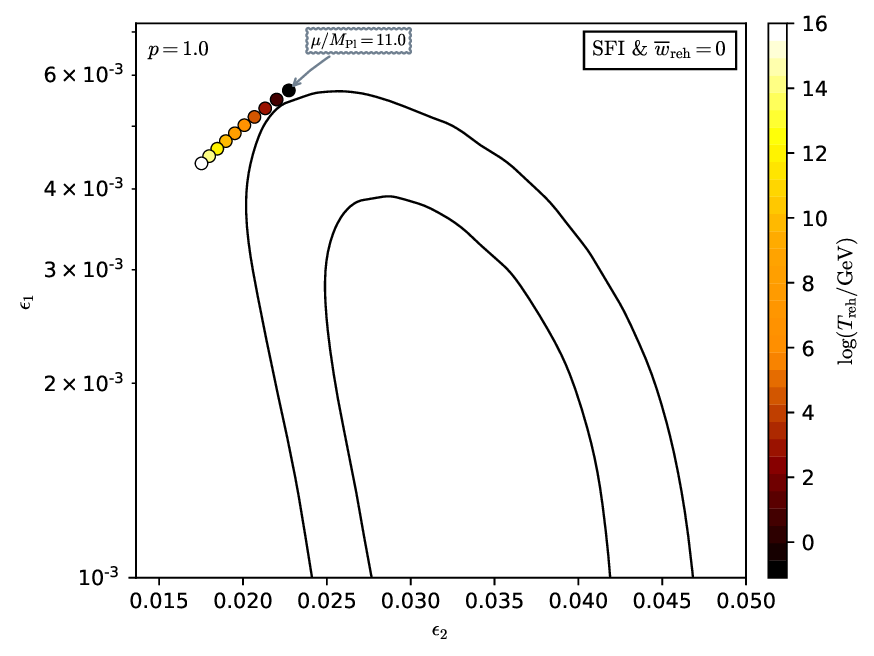}
\caption{Reheating consistent slow-roll predictions for the small
  field models with $p=1$ in the plane $(\nS,r)$ (top panel) and the
  plane $(\epsilon_1,\epsilon_2)$ (bottom panel). The solid contours
  are the one and two-sigma {\data} confidence intervals (marginalized
  over second order slow-roll). The model predictions are unsensitive
  to the value of $\mu$ and verify $r=(8/3)\left(1-\nS\right)$,
  \ie $\epsilon_2=4\epsilon_1$.}
\label{fig:CMBSFI1}
\end{center}
\end{figure}

\begin{figure}[H]
\begin{center}
\includegraphics[width=\wappfig,clip=true]{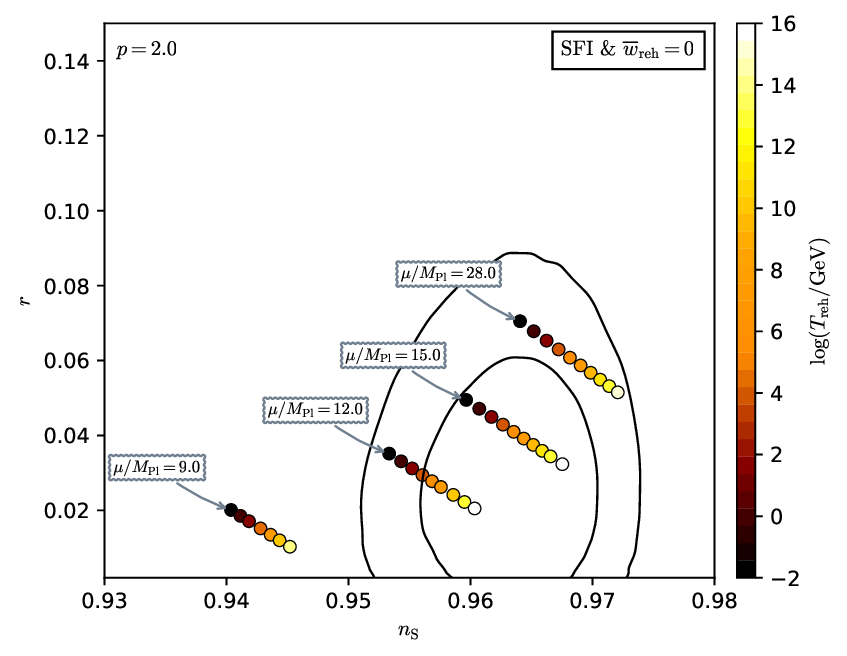}
\includegraphics[width=\wappfig,clip=true]{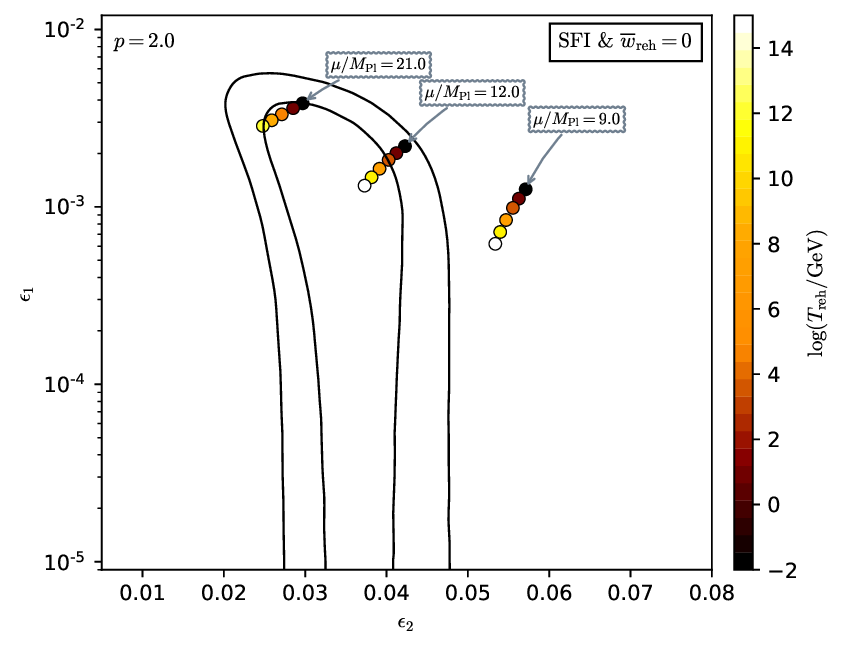}
\caption{Reheating consistent slow-roll predictions for the small
  field models with $p=2$ in the plane $(\nS,r)$ (top panel) and the
  plane $(\epsilon_1,\epsilon_2)$ (bottom panel). The solid contours
  are the one and two-sigma {\data} confidence intervals (marginalized
  over second order slow-roll). Clearly, if $\mu/\Mp$ is not too high
  these values are limited from below to stay inside the two-sigma
  contours, and $\mu/\Mp<10$ is disfavored by the data. For
  $\mu/\Mp\gg 1$, the model predictions approach
  $r=(8/3)\left(1-\nS\right)$, \ie $\epsilon_2=4\epsilon_1$.}
\label{fig:CMBSFI2}
\end{center}
\end{figure}

\begin{figure}[H]
\begin{center}
\includegraphics[width=\wappfig,clip=true]{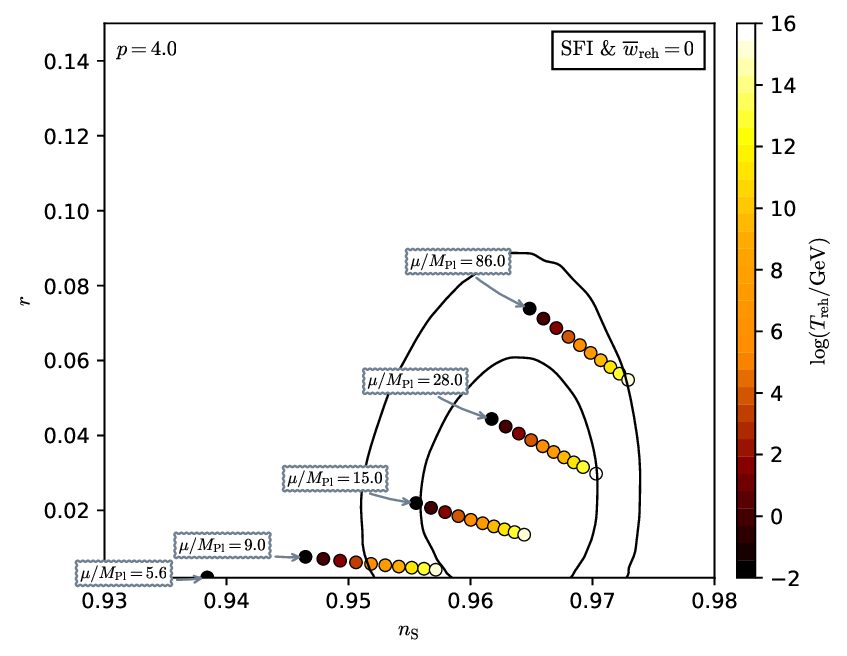}
\includegraphics[width=\wappfig,clip=true]{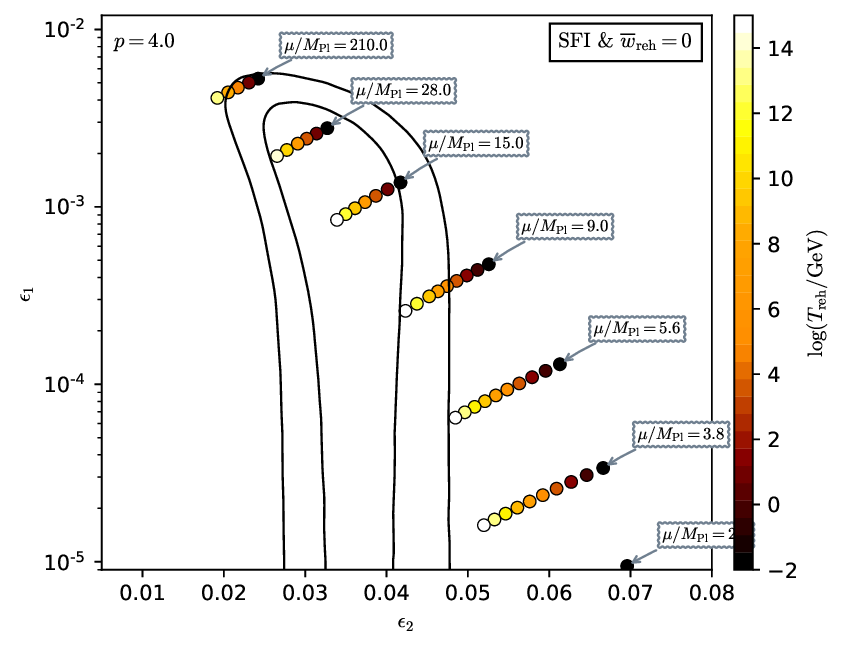}
\caption{Reheating consistent slow-roll predictions for the small
  field models with $p=4$ in the plane $(\nS,r)$ (top panel) and the
  plane $(\epsilon_1,\epsilon_2)$ (bottom panel). The solid contours
  are the one and two-sigma {\data} confidence intervals (marginalized
  over second order slow-roll). Clearly, if $\mu/\Mp$ is not too high
  these values are limited from below to stay inside the two-sigma
  contours. For $\mu/\Mp\gg 1$, the model predictions approach
  $r=(8/3)\left(1-\nS\right)$, \ie $\epsilon_2=4\epsilon_1$.}
\label{fig:CMBSFI4}
\end{center}
\end{figure}

\subsection{Intermediate Inflation (\hyperref[sec:ii]{II})}

\begin{figure}[H]
\begin{center}
\includegraphics[width=\wappfig,clip=true]{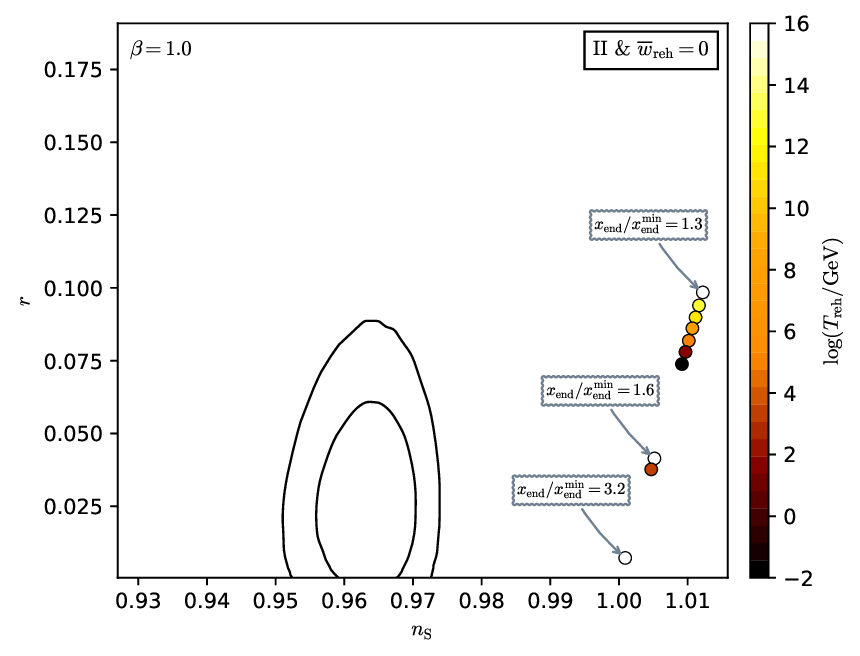}
\includegraphics[width=\wappfig,clip=true]{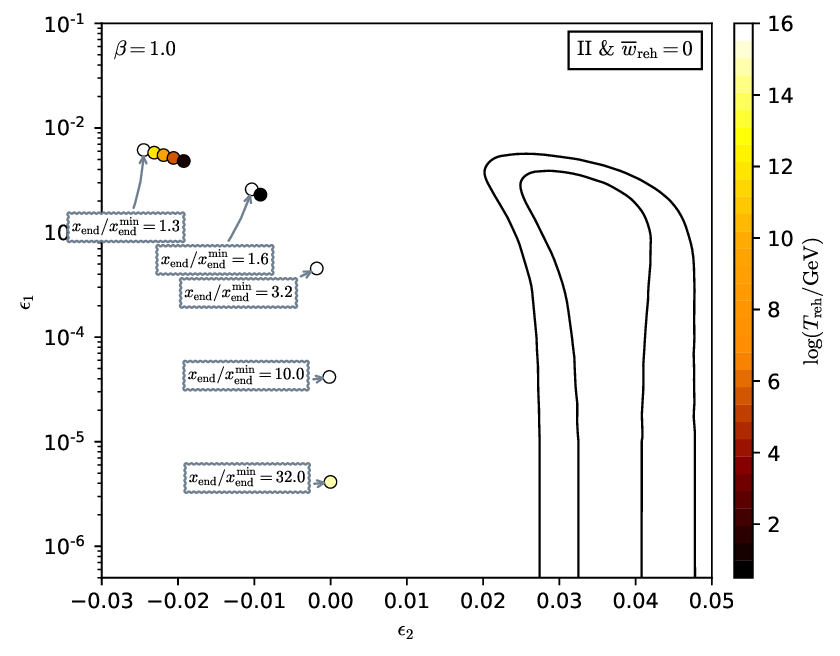}
\caption{Reheating consistent slow-roll predictions for the
  intermediate inflation models with $\beta=1$ in the plane $(\nS,r)$
  (top panel) and the plane $(\epsilon_1,\epsilon_2)$ (bottom
  panel). The solid contours are the one and two-sigma {\data}
  confidence intervals (marginalized over second order slow-roll). The
  model predictions for $\xend \gg 1$ correspond to the points such
  that $\epsilon_1=-(\beta/4) \epsilon_2$. Let us notice that the
  energy scale at which reheating ends is degenerated with $\xend$.}
\label{fig:CMBII}
\end{center}
\end{figure}

\begin{figure}[H]
\begin{center}
\includegraphics[width=\wappfig,clip=true]{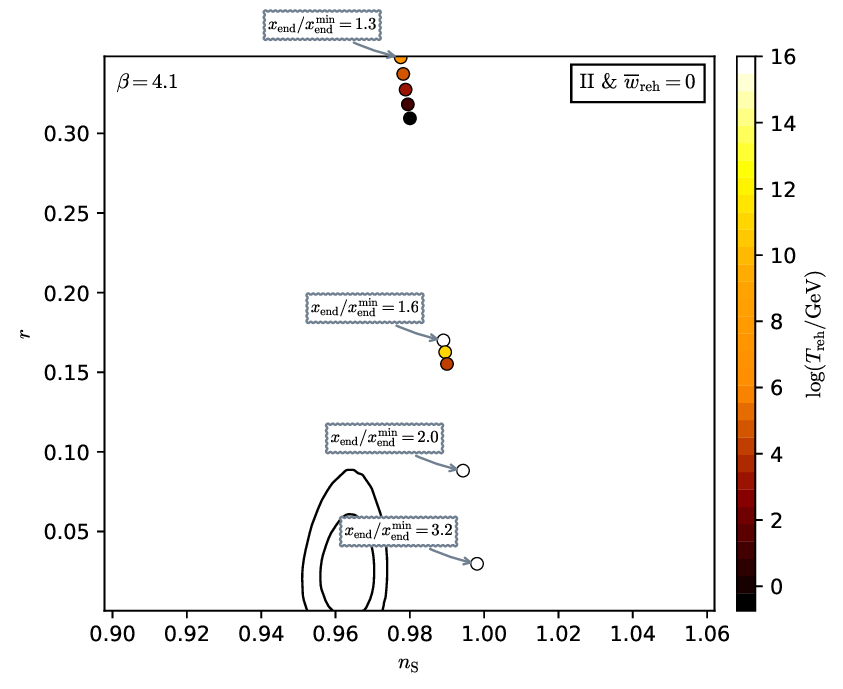}
\includegraphics[width=\wappfig,clip=true]{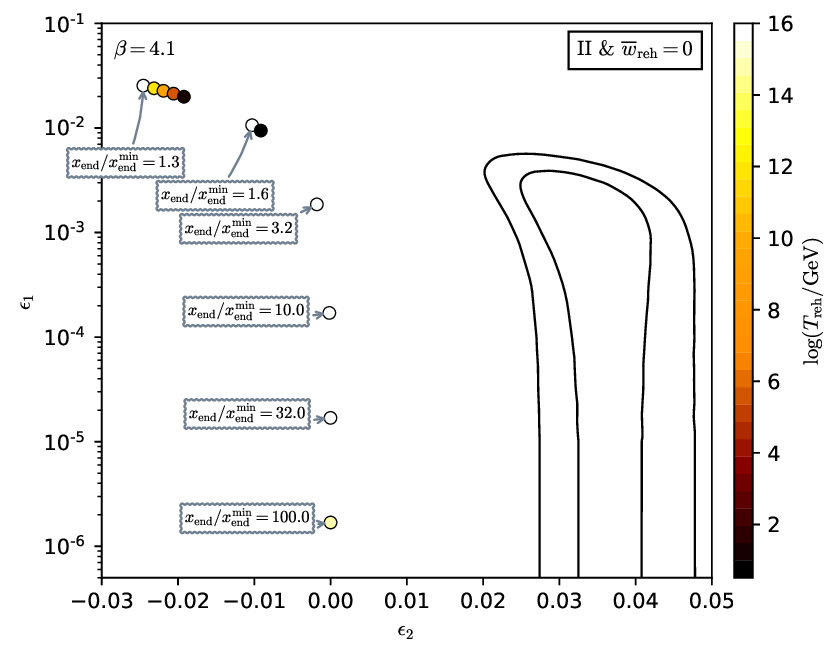}
\caption{Reheating consistent slow-roll predictions for the
  intermediate inflation models with $\beta=4.1$ in the plane $(\nS,r)$
  (top panel) and the plane $(\epsilon_1,\epsilon_2)$ (bottom
  panel). The solid contours are the one and two-sigma {\data}
  confidence intervals (marginalized over second order slow-roll).}
\label{fig:CMBII_1}
\end{center}
\end{figure}

\begin{figure}[H]
\begin{center}
\includegraphics[width=\wappfig,clip=true]{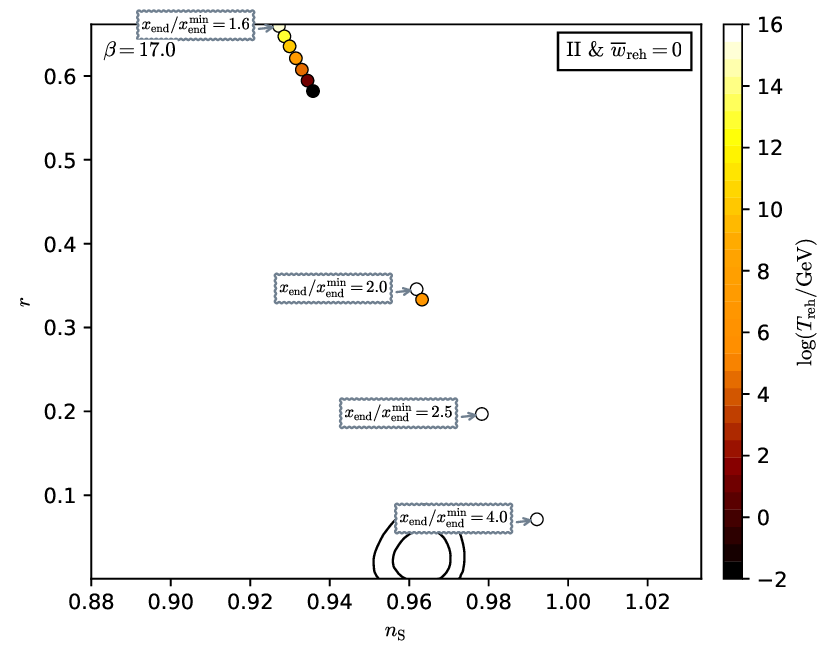}
\includegraphics[width=\wappfig,clip=true]{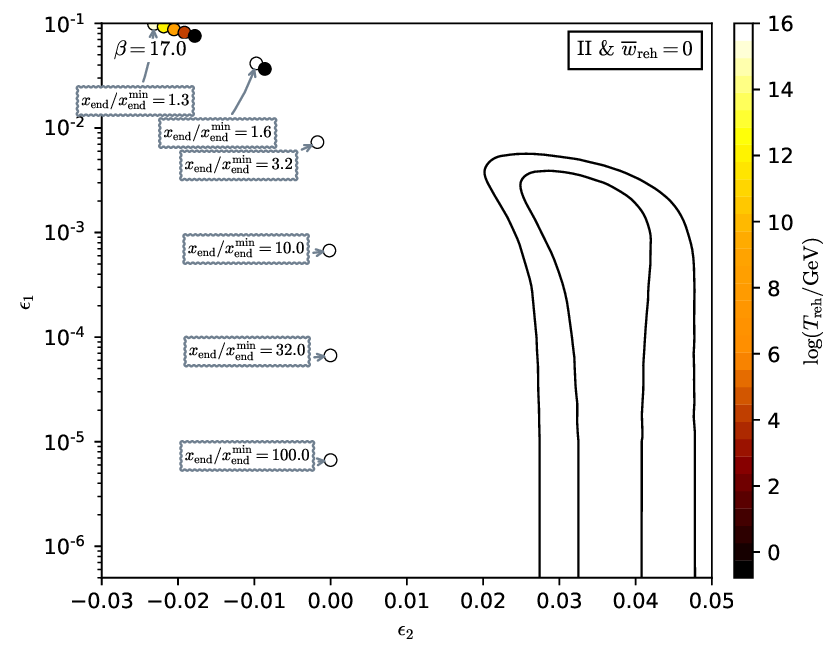}
\caption{Reheating consistent slow-roll predictions for the
  intermediate inflation models with $\beta=17$ in the plane $(\nS,r)$
  (top panel) and the plane $(\epsilon_1,\epsilon_2)$ (bottom
  panel). The solid contours are the one and two-sigma {\data}
  confidence intervals (marginalized over second order slow-roll).}
\label{fig:CMBII_2}
\end{center}
\end{figure}

\begin{figure}[H]
\begin{center}
\includegraphics[width=\wappfig,clip=true]{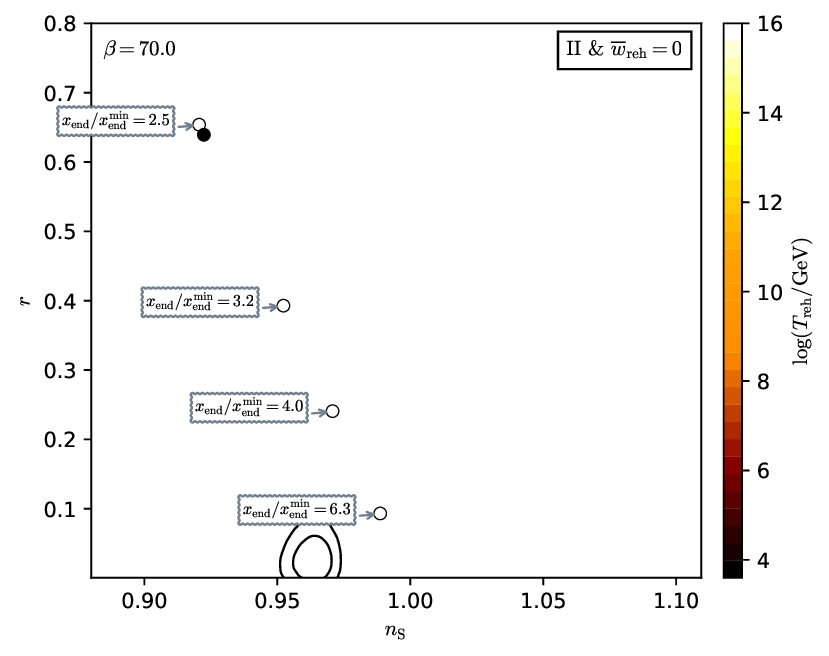}
\includegraphics[width=\wappfig,clip=true]{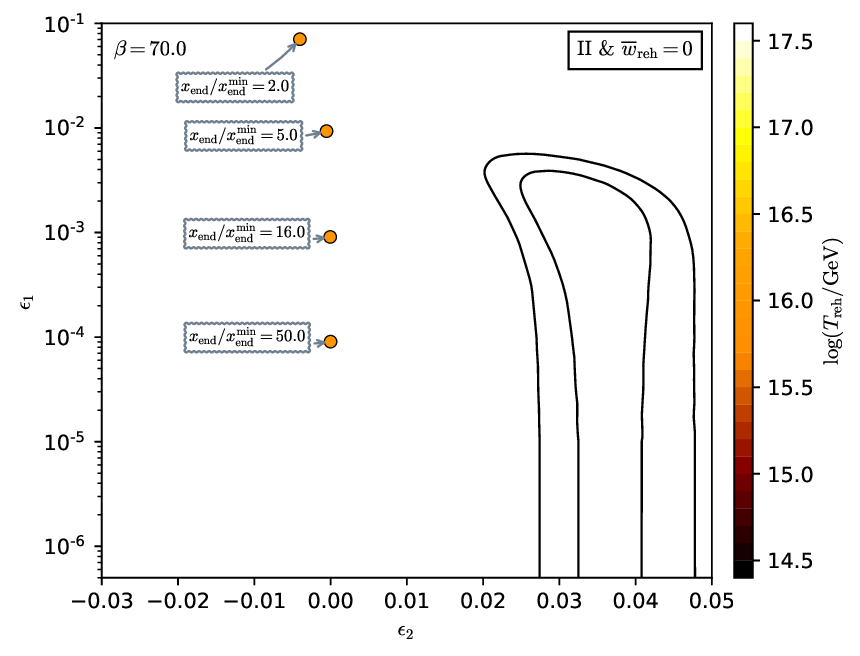}
\caption{Reheating consistent slow-roll predictions for the
  intermediate inflation models with $\beta=70$ in the plane $(\nS,r)$
  (top panel) and the plane $(\epsilon_1,\epsilon_2)$ (bottom
  panel). The solid contours are the one and two-sigma {\data}
  confidence intervals (marginalized over second order slow-roll). For
  large values of $\beta$, the spectral index is red ($\nS < 1$) but
  the at the expense of producing a significant amount of primordial
  gravitational waves.}
\label{fig:CMBII_3}
\end{center}
\end{figure}

\subsection{K\"ahler Moduli Inflation II (\hyperref[sec:kmiii]{KMIII})}

\begin{figure}[H]
\begin{center}
\includegraphics[width=\wappfig,clip=true]{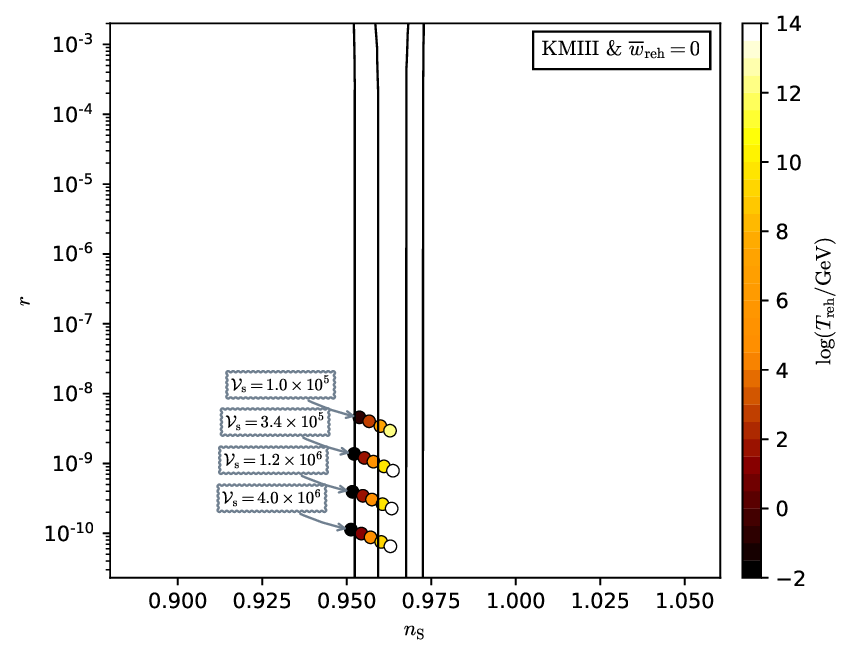}
\includegraphics[width=\wappfig,clip=true]{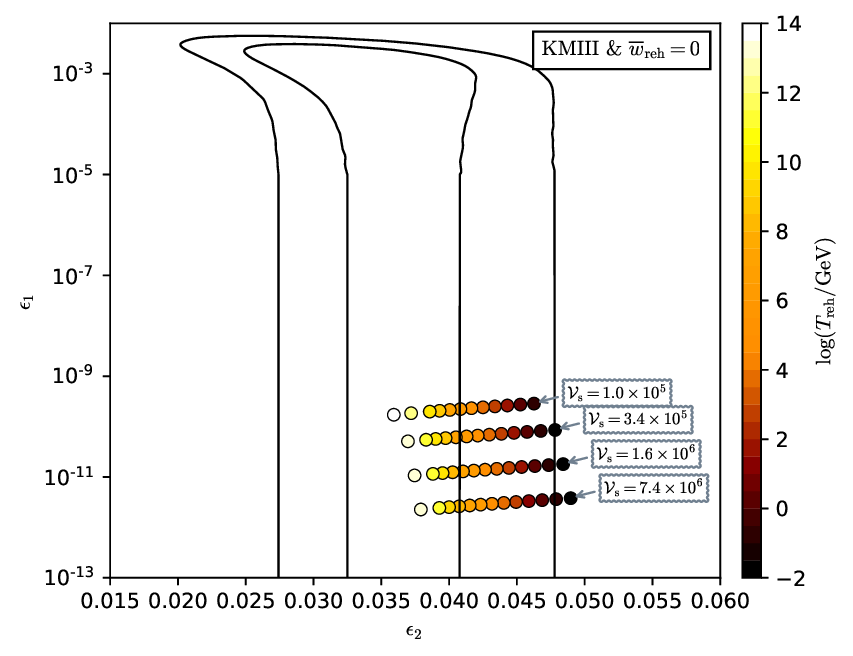}
\caption{Reheating consistent slow-roll predictions for the K\"ahler
  moduli III models in the plane $(\nS,r)$ (top panel) and the plane
  $(\epsilon_1,\epsilon_2)$ (bottom panel), for $10^{5}<\calV<10^{7}$,
  $\alpha=\calV^{5/3}$ and $\beta=\calV^{2/3}$. The solid contours are
  the one and two-sigma {\data} confidence intervals (marginalized
  over second order slow-roll).}
\label{fig:CMBKMIII}
\end{center}
\end{figure}

\subsection{Logamediate Inflation (\hyperref[sec:lmi]{LMI})}

\begin{figure}[H]
\begin{center}
\includegraphics[width=\wappfig,clip=true]{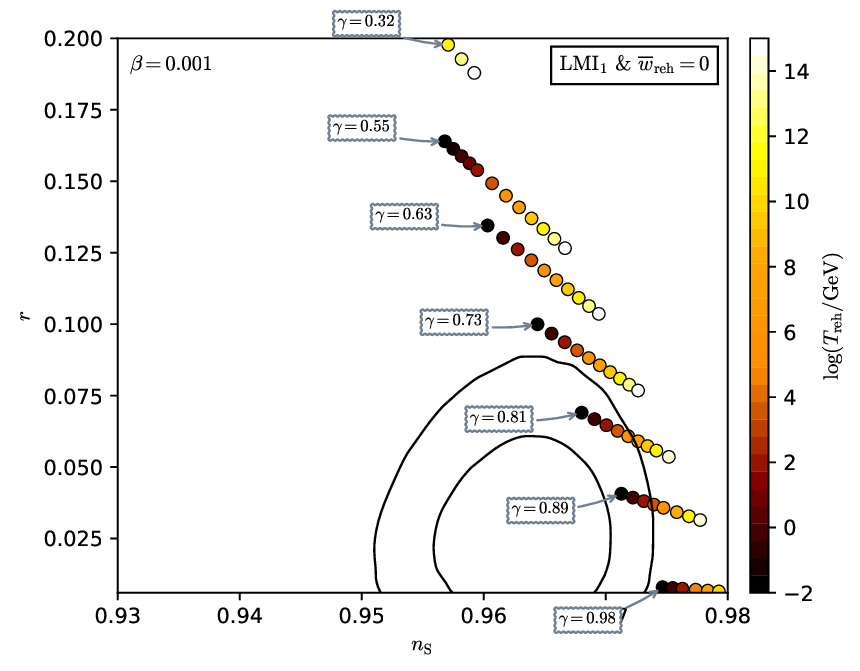}
\includegraphics[width=\wappfig,clip=true]{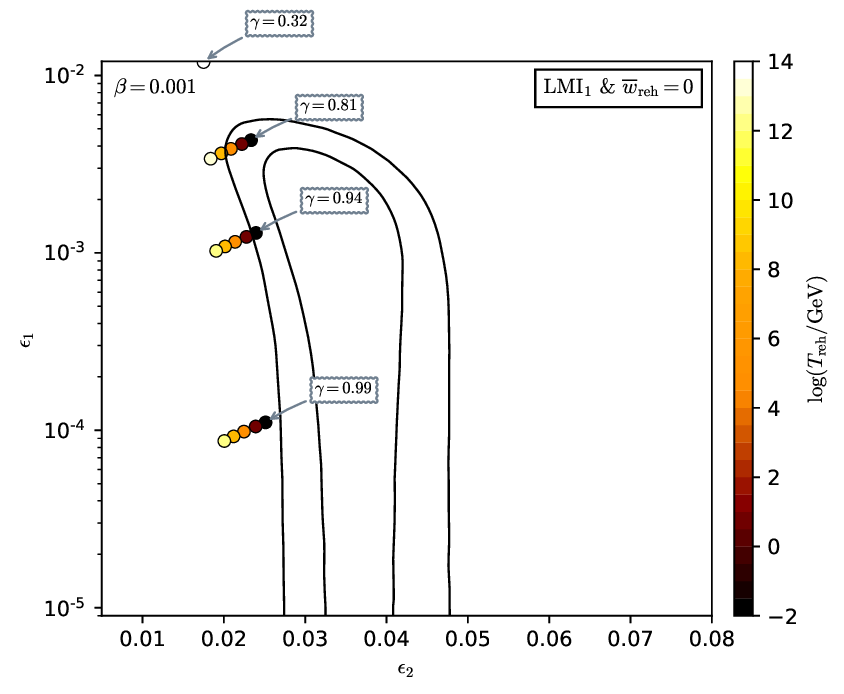}
\caption{Reheating consistent slow-roll predictions for the
  Logamediate Inflation 1 models with $\beta=10^{-3}$, in the plane
  $(\nS,r)$ (top panel) and the plane $(\epsilon_1,\epsilon_2)$
  (bottom panel). Inflation proceeds at decreasing field values
  $x<\xVmax$. The solid contours are the one and two-sigma {\data}
  confidence intervals (marginalized over second order slow-roll). For
  $\beta\ll 1$, the exponential term in the potential \Eq{eq:lmi:pot}
  is almost constant so that the model is close to large field
  inflation (LFI, see \sectionc{sec:lfi}). In that limit, one has
  $\epsilon_1=\left(1-\gamma\right)\epsilon_2$.}
\label{fig:CMBLMI1beta10PowerMinus3}
\end{center}
\end{figure}

\begin{figure}[H]
\begin{center}
\includegraphics[width=\wappfig,clip=true]{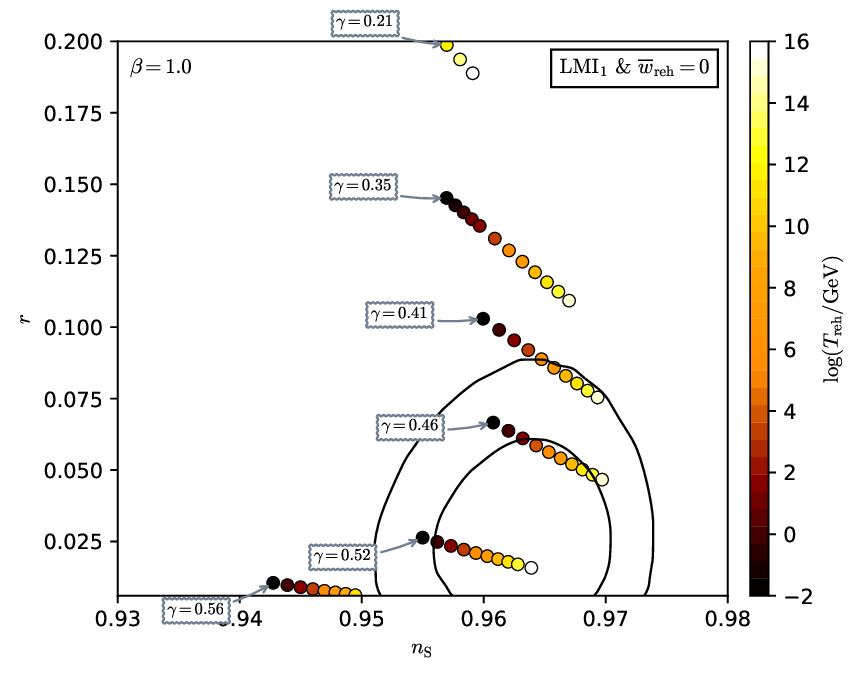}
\includegraphics[width=\wappfig,clip=true]{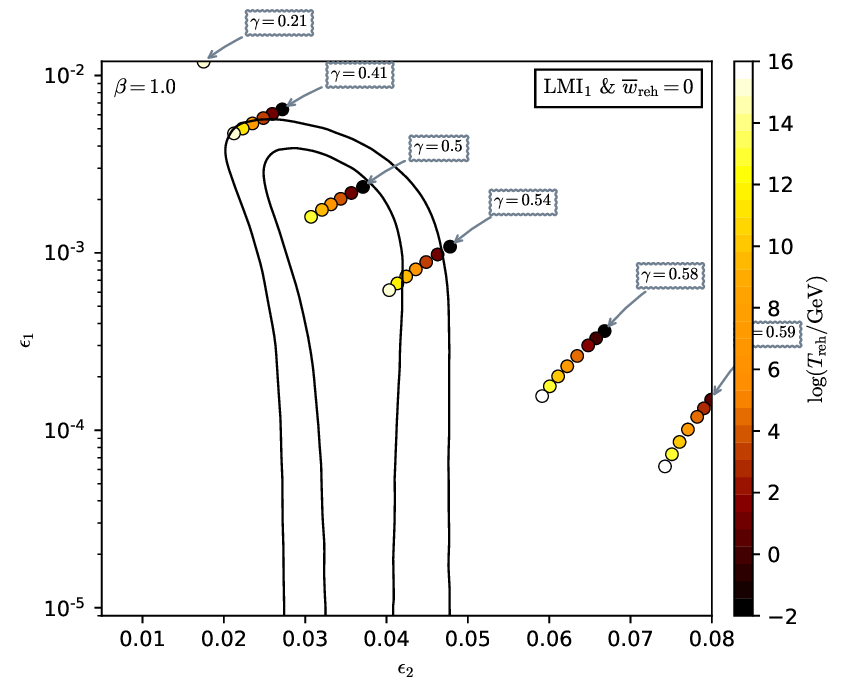}
\caption{Reheating consistent slow-roll predictions for the
  Logamediate Inflation 1 models with $\beta=1$ in the plane $(\nS,r)$
  (top panel) and the plane $(\epsilon_1,\epsilon_2)$ (bottom
  panel). Inflation proceeds as in \Fig{fig:CMBLMI1beta10PowerMinus3},
  at decreasing field values and with $x<\xVmax$. The solid contours
  are the one and two-sigma {\data} confidence intervals (marginalized
  over second order slow-roll).}
\label{fig:CMBLMI1beta1}
\end{center}
\end{figure}

\begin{figure}[H]
\begin{center}
\includegraphics[width=\wappfig,clip=true]{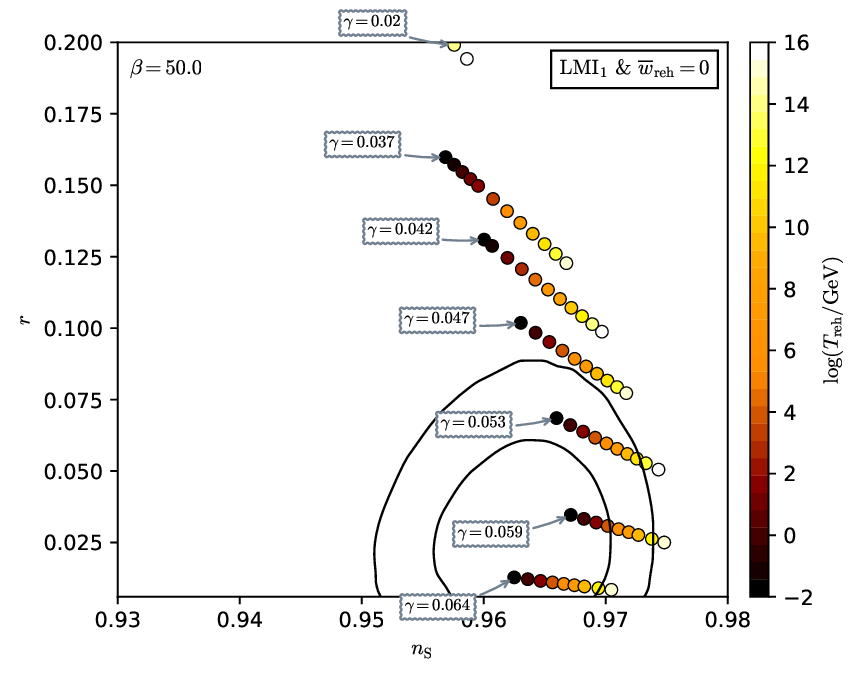}
\includegraphics[width=\wappfig,clip=true]{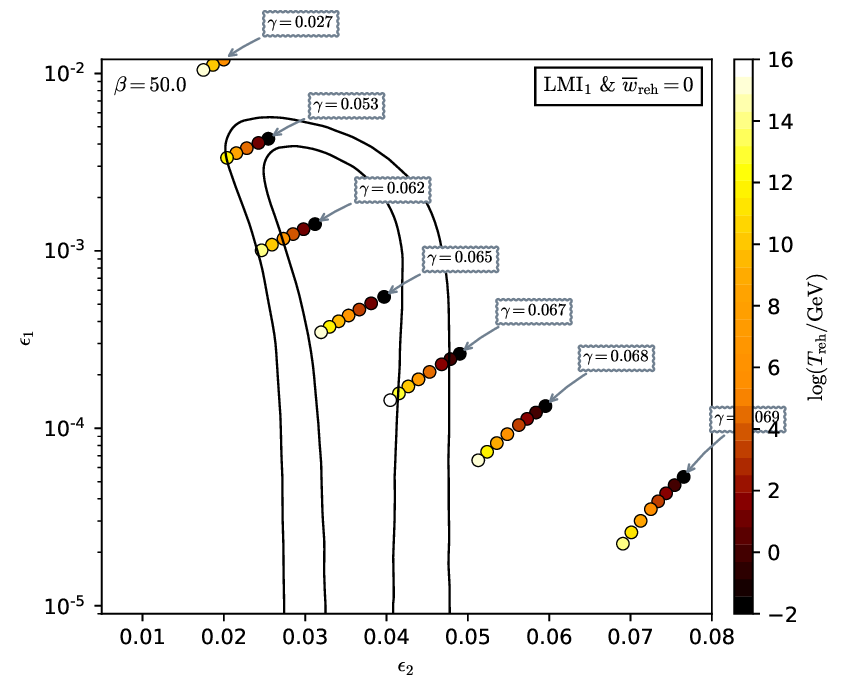}
\caption{Reheating consistent slow-roll predictions for the
  Logamediate Inflation 1 models ($x<\xVmax$) with $\beta=50$, in the
  plane $(\nS,r)$ (top panel) and the plane $(\epsilon_1,\epsilon_2)$
  (bottom panel). The solid contours are the one and two-sigma {\data}
  confidence intervals (marginalized over second order slow-roll). For
  such high values of $\beta$, only small values of $\gamma$ are in
  agreement with observations.}
\label{fig:CMBLMI1beta50}
\end{center}
\end{figure}

\begin{figure}[H]
\begin{center}
\includegraphics[width=\wappfig,clip=true]{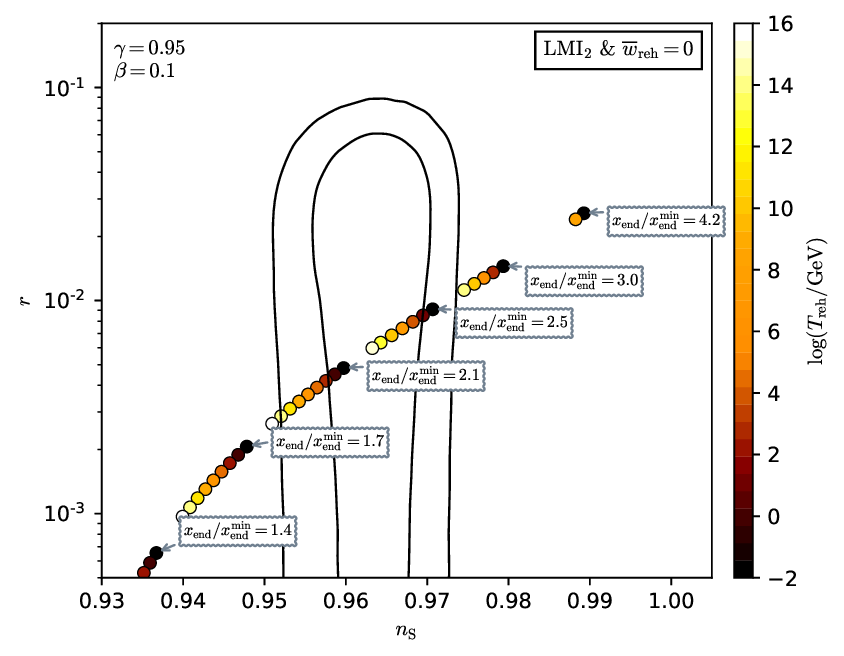}
\includegraphics[width=\wappfig,clip=true]{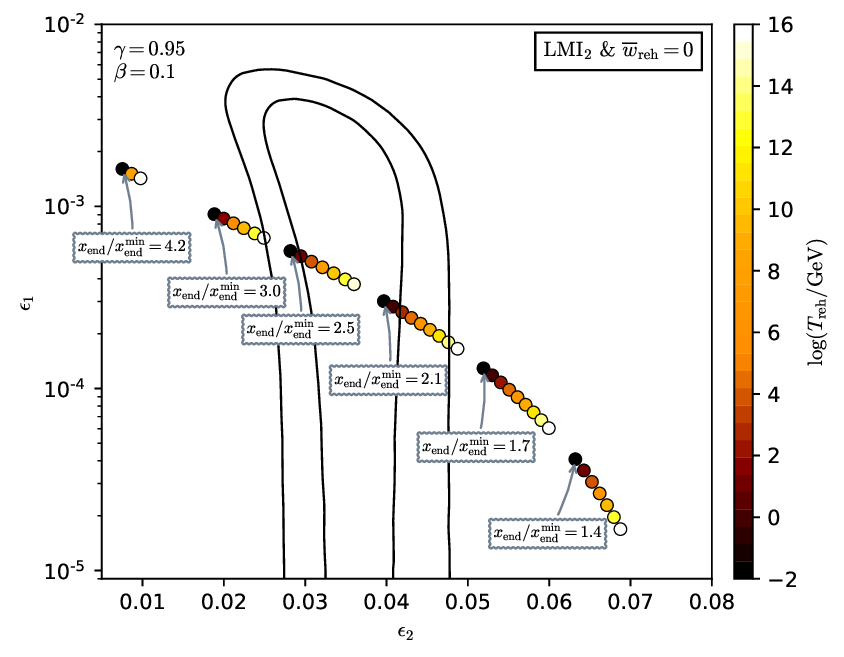}
\caption{Reheating consistent slow-roll predictions for the
  Logamediate Inflation 2 models with $\beta=0.1$ and $\gamma=0.95$,
  in the plane $(\nS,r)$ (top panel) and the plane
  $(\epsilon_1,\epsilon_2)$ (bottom panel). Inflation proceeds at
  increasing field values and with $x>\xVmax$. The solid contours are
  the one and two-sigma {\data} confidence intervals (marginalized
  over second order slow-roll).  }
\label{fig:CMBLMI2beta10PowerMinus1}
\end{center}
\end{figure}

\begin{figure}[H]
\begin{center}
\includegraphics[width=\wappfig,clip=true]{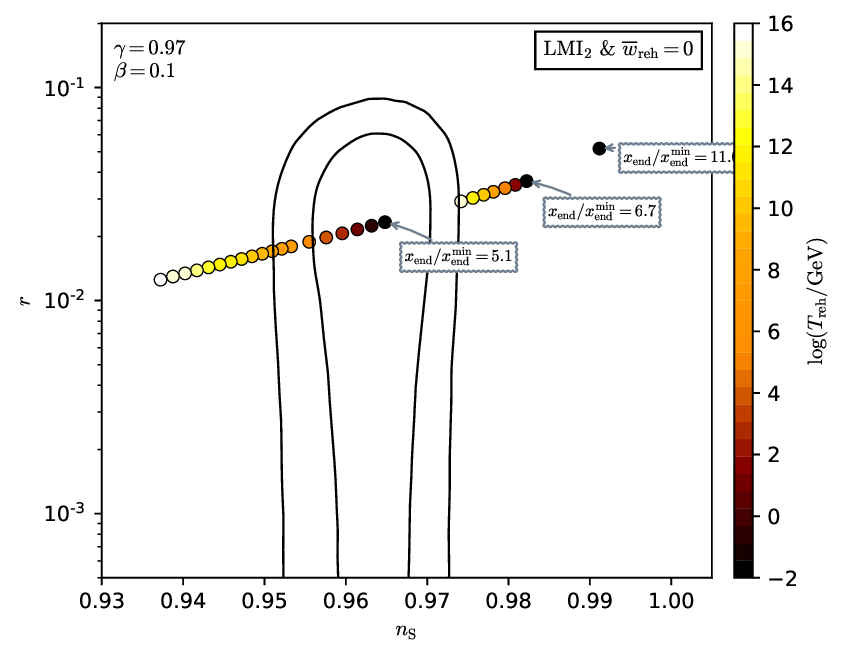}
\includegraphics[width=\wappfig,clip=true]{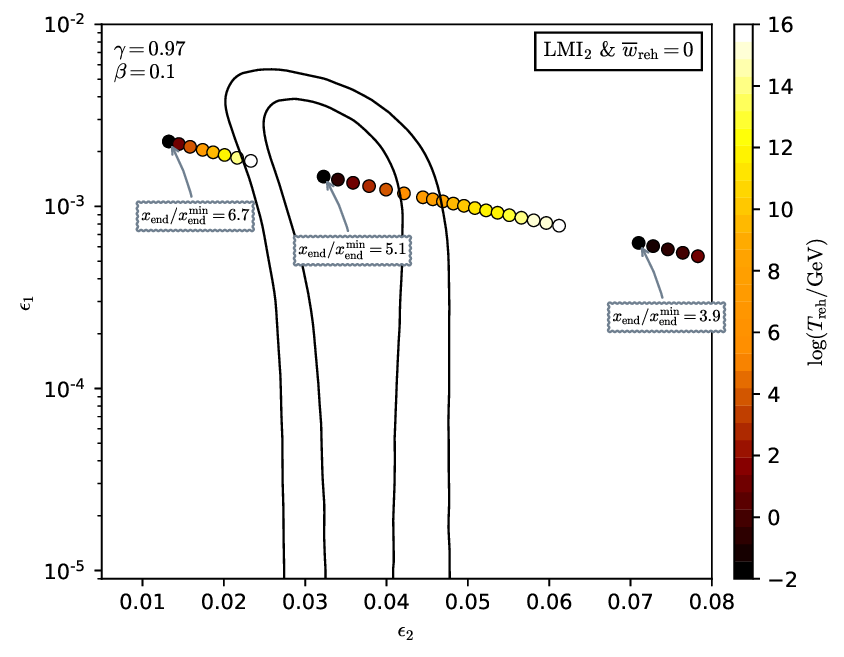}
\caption{Reheating consistent slow-roll predictions for the
  Logamediate Inflation 2 models with $\beta=0.1$ and $\gamma$
  slightly increased to $\gamma=0.97$ with respect to the previous
  figure~\ref{fig:CMBLMI2beta10PowerMinus1}. Inflation proceeds at
  increasing field values and with $x>\xVmax$. The solid contours are
  the one and two-sigma {\data} confidence intervals (marginalized
  over second order slow-roll).}
\label{fig:CMBLMI2beta10PowerMinus1_1}
\end{center}
\end{figure}

\begin{figure}[H]
\begin{center}
\includegraphics[width=\wappfig,clip=true]{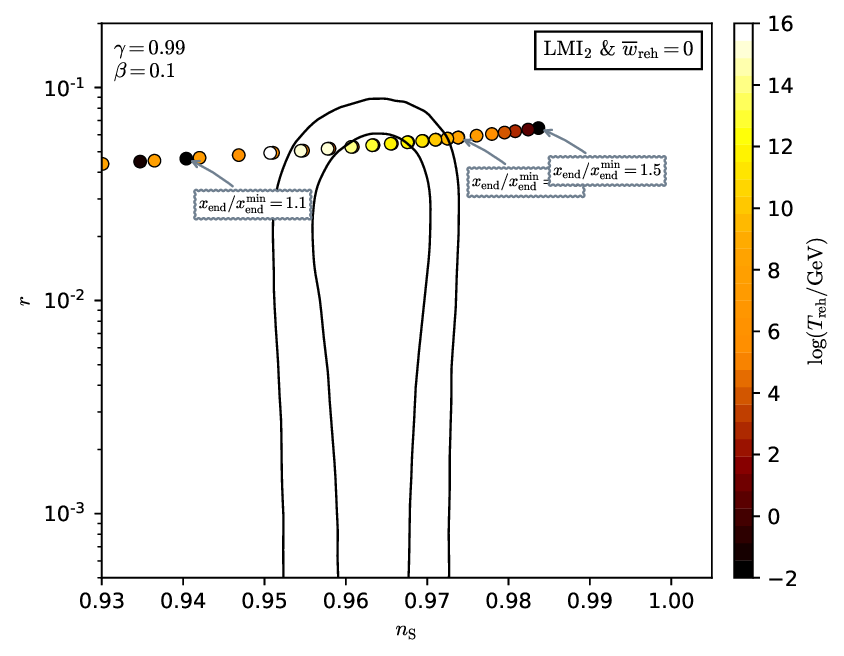}
\includegraphics[width=\wappfig,clip=true]{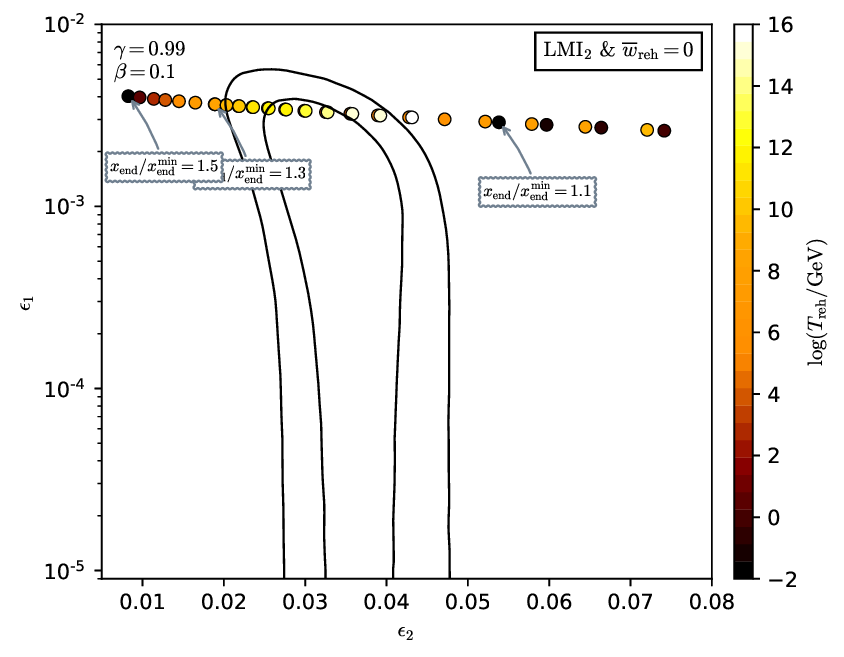}
\caption{Reheating consistent slow-roll predictions for the
  Logamediate Inflation 2 models with $\beta=0.1$ and another slightly
  larger value of $\gamma=0.99$ with respect to the two previous
  figures~\ref{fig:CMBLMI2beta10PowerMinus1} and
  \ref{fig:CMBLMI2beta10PowerMinus1_1}. Inflation proceeds at
  increasing field values and with $x>\xVmax$. The solid contours are
  the one and two-sigma {\data} confidence intervals (marginalized
  over second order slow-roll).}
\label{fig:CMBLMI2beta10PowerMinus1_2}
\end{center}
\end{figure}

\begin{figure}[H]
\begin{center}
\includegraphics[width=\wappfig,clip=true]{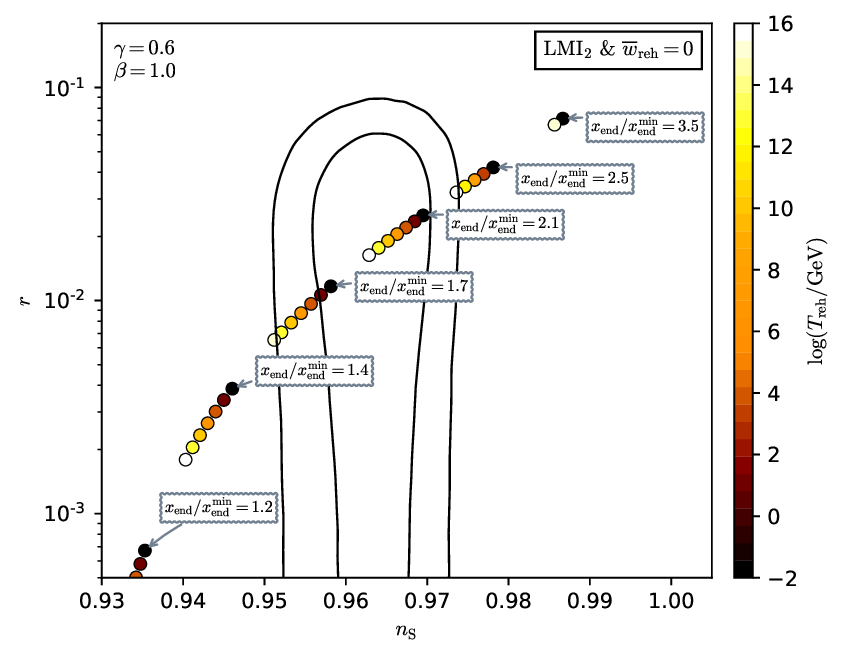}
\includegraphics[width=\wappfig,clip=true]{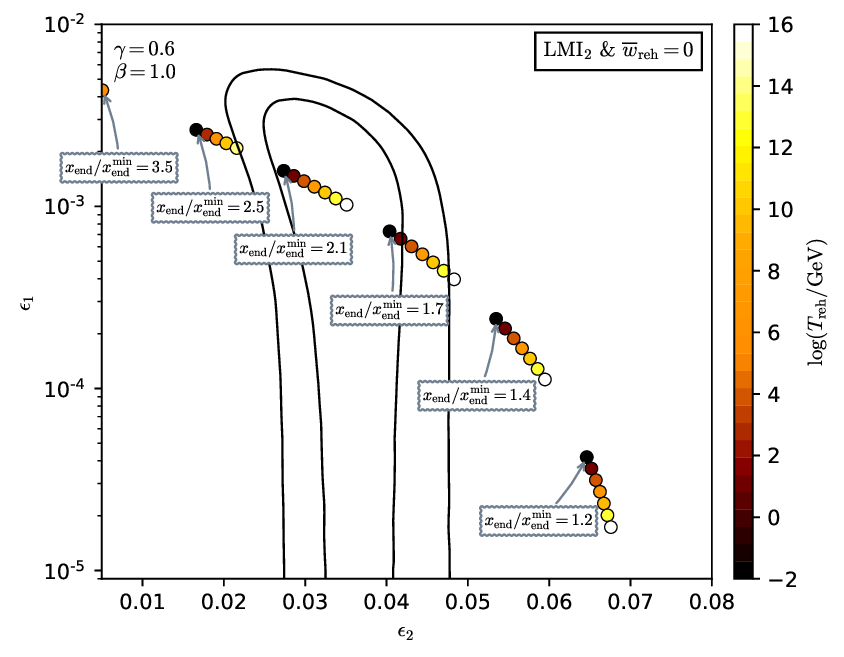}
\caption{Reheating consistent slow-roll predictions for the
  Logamediate Inflation 2 models ($x>\xVmax$) with $\beta=1$ and
  $\gamma=0.6$, in the plane $(\nS,r)$ (top panel) and the plane
  $(\epsilon_1,\epsilon_2)$ (bottom panel). The solid contours are the
  one and two-sigma {\data} confidence intervals (marginalized over
  second order slow-roll). When $\xend$ becomes large and lies in the
  fine-tuned region of LMI2, \ie $ \xVmax < x< \xepsoneMax$, the
  predictions approach the pure de Sitter case.}
\label{fig:CMBLMI2beta1}
\end{center}
\end{figure}

\begin{figure}[H]
\begin{center}
\includegraphics[width=\wappfig,clip=true]{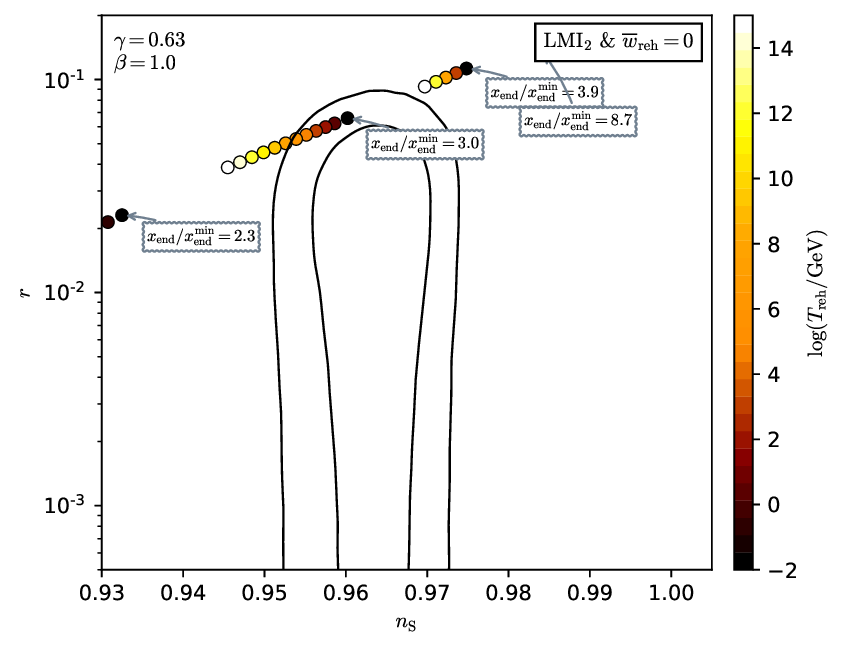}
\includegraphics[width=\wappfig,clip=true]{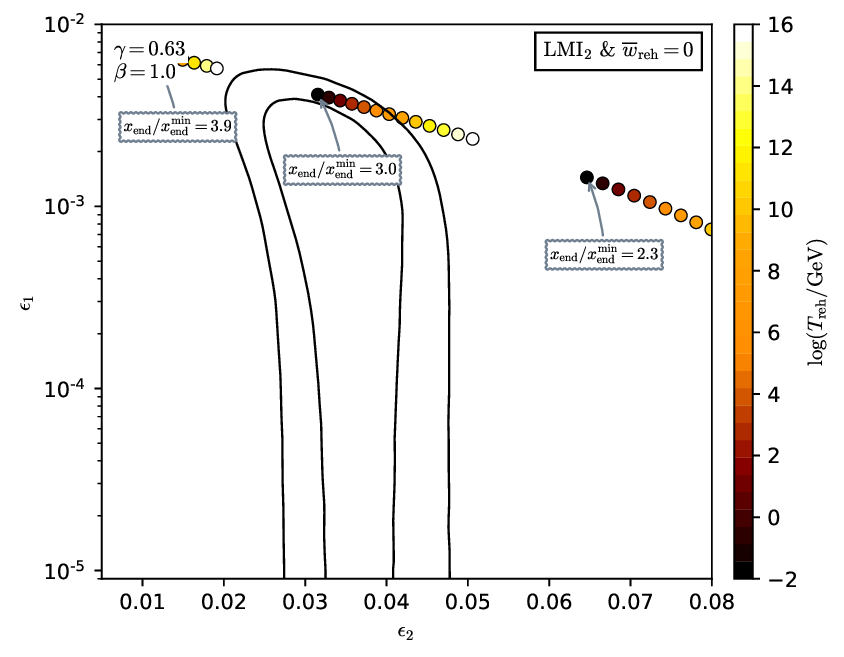}
\caption{Reheating consistent slow-roll predictions for the
  Logamediate Inflation 2 models ($x>\xVmax$) with $\beta=1$ and a
  small increase of $\gamma$ to $\gamma=0.63$ compared to
  figure~\ref{fig:CMBLMI2beta1}. The solid contours are the one and
  two-sigma {\data} confidence intervals (marginalized over second
  order slow-roll). When $\xend$ becomes large and lies in the
  fine-tuned region of LMI2, \ie $ \xVmax < \xend< \xepsoneMax$, the
  predictions approach the pure de Sitter case.}
\label{fig:CMBLMI2beta1_4}
\end{center}
\end{figure}

\begin{figure}[H]
\begin{center}
\includegraphics[width=\wappfig,clip=true]{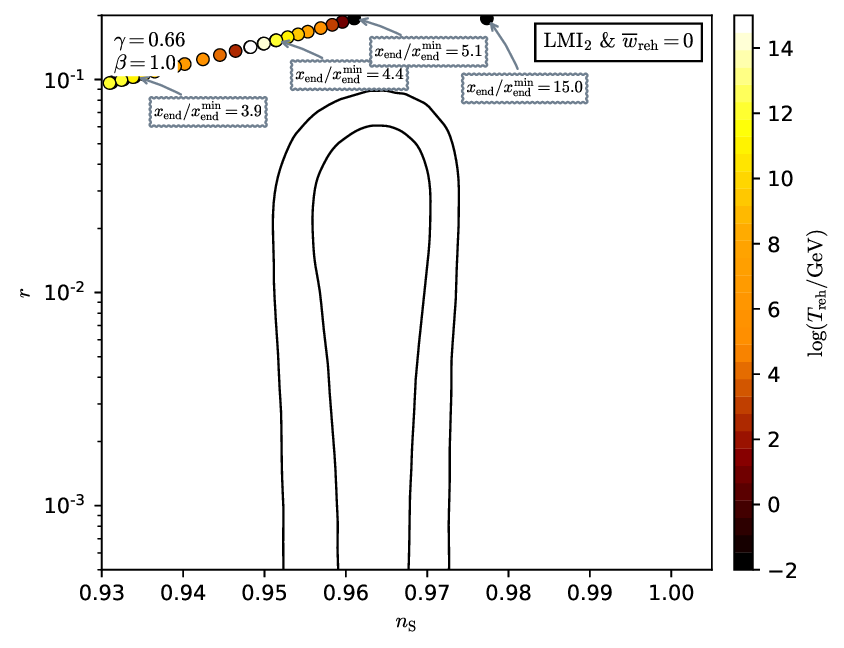}
\includegraphics[width=\wappfig,clip=true]{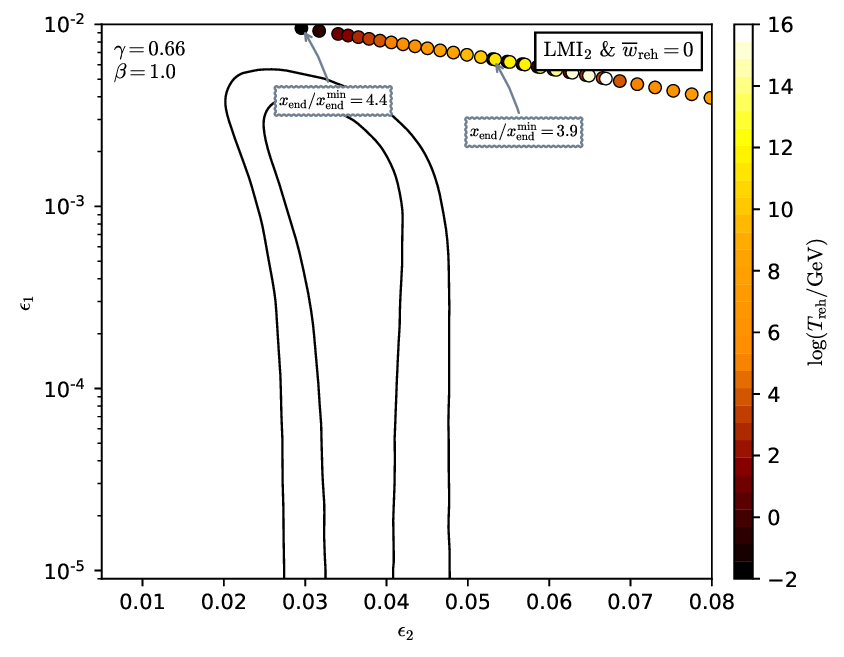}
\caption{Reheating consistent slow-roll predictions for the
  Logamediate Inflation 2 models ($x>\xVmax$) with $\beta=1$ and
  another increase of $\gamma$ to $\gamma=0.66$ compared to
  figures~\ref{fig:CMBLMI2beta1} and \ref{fig:CMBLMI2beta1_4}. The
  solid contours are the one and two-sigma {\data} confidence
  intervals (marginalized over second order slow-roll).}
\label{fig:CMBLMI2beta1_5}
\end{center}
\end{figure}

\begin{figure}[H]
\begin{center}
\includegraphics[width=\wappfig,clip=true]{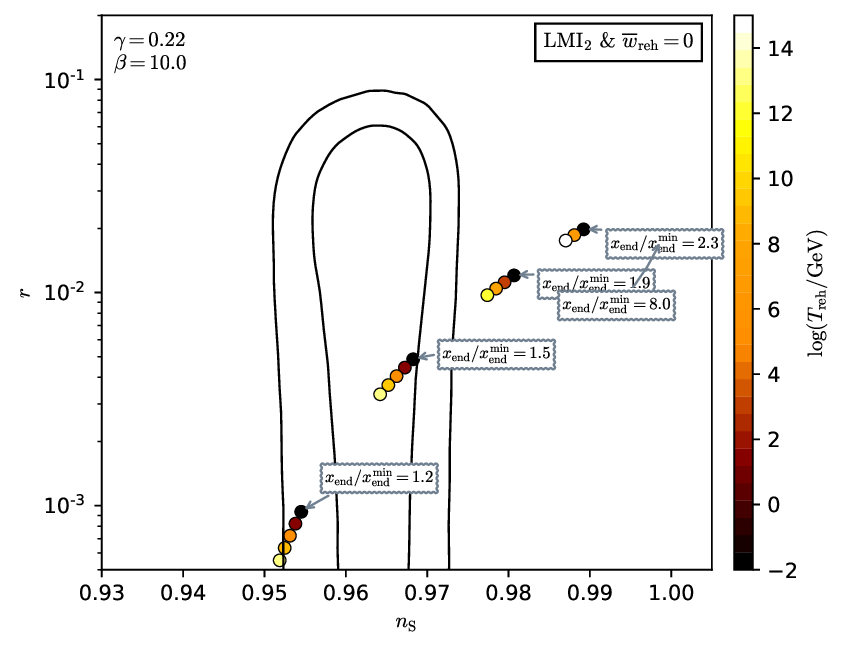}
\includegraphics[width=\wappfig,clip=true]{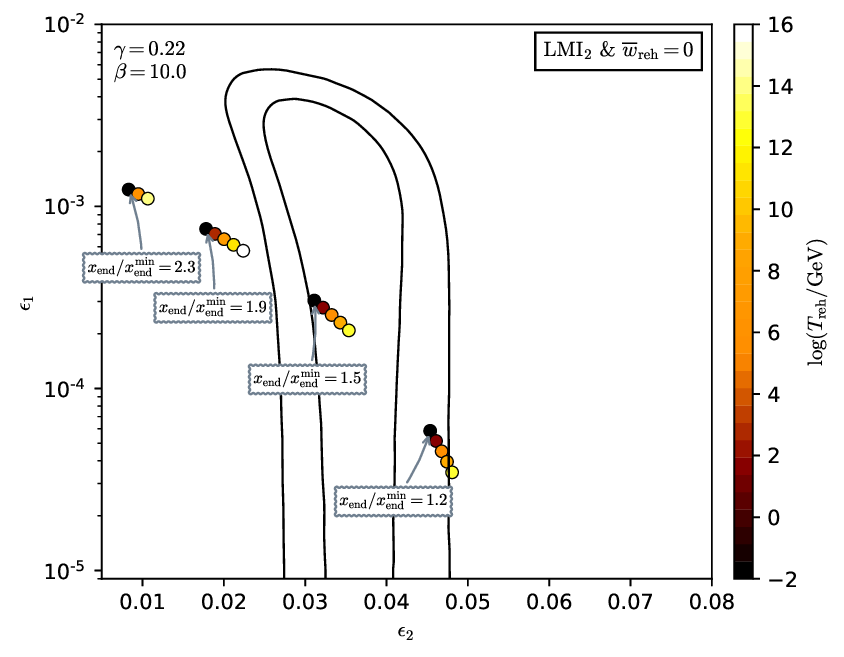}
\caption{Reheating consistent slow-roll predictions for the
  Logamediate Inflation 2 models ($x>\xVmax$) with $\beta=10$ and
  $\gamma=0.22$, in the plane $(\nS,r)$ (top panel) and the plane
  $(\epsilon_1,\epsilon_2)$ (bottom panel). The solid contours are the
  one and two-sigma {\data} confidence intervals (marginalized over
  second order slow-roll). When $\xend$ becomes large and lies in the
  fine-tuned region of LMI2, \ie $ \xVmax < \xend< \xepsoneMax$, the
  predictions approach the pure de Sitter case.}
\label{fig:CMBLMI2beta10}
\end{center}
\end{figure}

\begin{figure}[H]
\begin{center}
\includegraphics[width=\wappfig,clip=true]{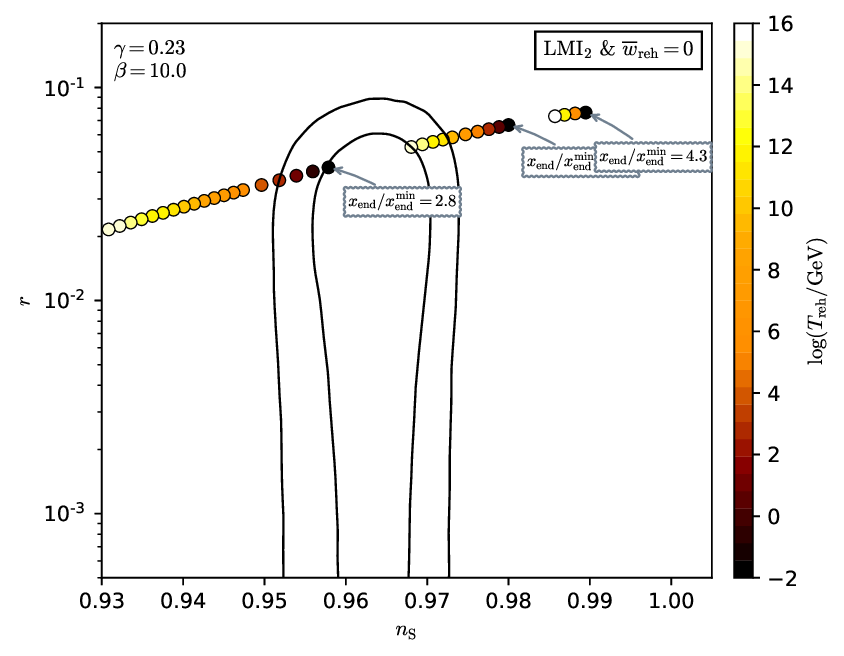}
\includegraphics[width=\wappfig,clip=true]{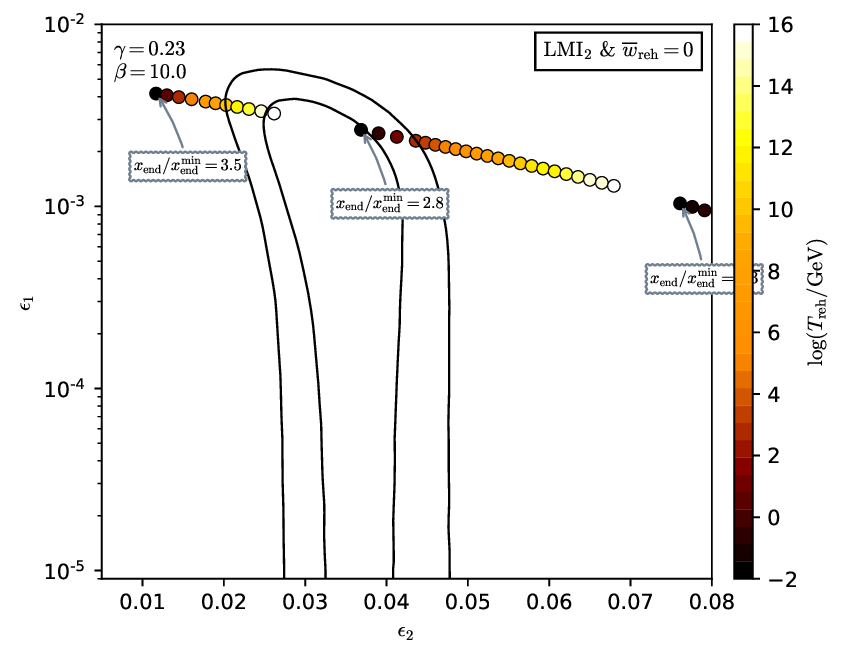}
\caption{Reheating consistent slow-roll predictions for the
  Logamediate Inflation 2 models ($x>\xVmax$) with $\beta=10$ and a
  slightly increased value of $\gamma=0.23$ (compared to
  figure~\ref{fig:CMBLMI2beta10}), in the plane $(\nS,r)$ (top panel)
  and the plane $(\epsilon_1,\epsilon_2)$ (bottom panel). The solid
  contours are the one and two-sigma {\data} confidence intervals
  (marginalized over second order slow-roll).}
\label{fig:CMBLMI2beta10_7}
\end{center}
\end{figure}

\begin{figure}[H]
\begin{center}
\includegraphics[width=\wappfig,clip=true]{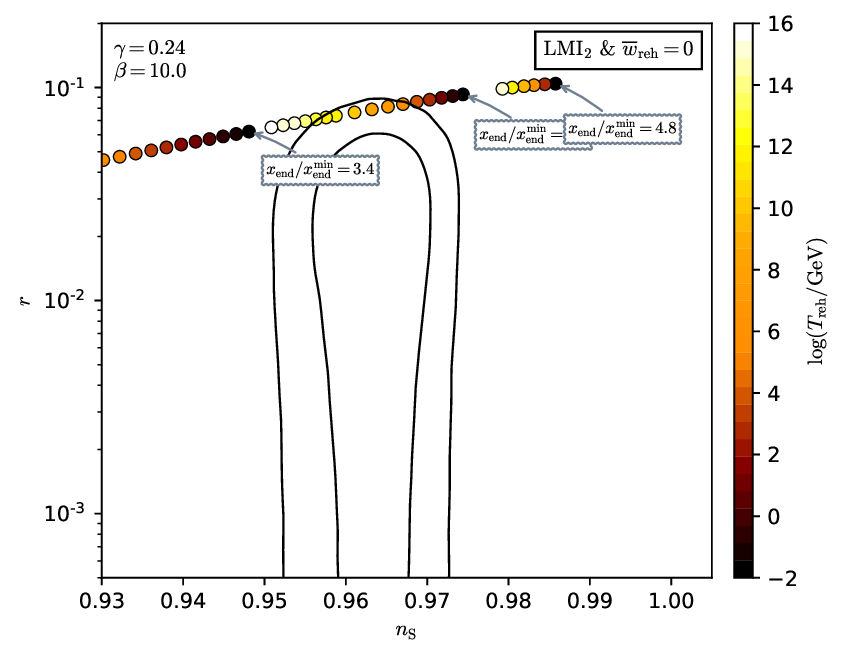}
\includegraphics[width=\wappfig,clip=true]{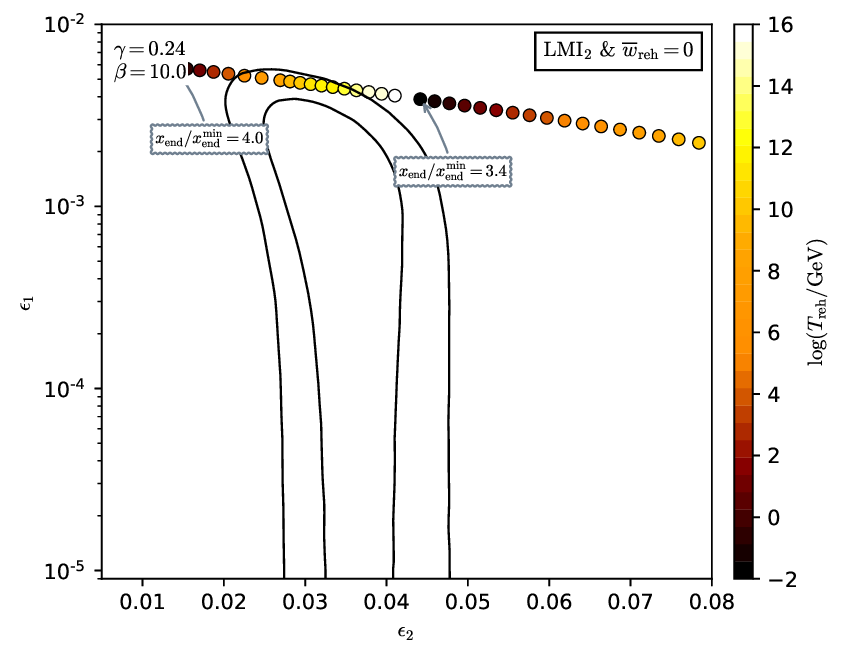}
\caption{Reheating consistent slow-roll predictions for the
  Logamediate Inflation 2 models ($x>\xVmax$) with $\beta=10$ and
  $\gamma$ increased $\gamma=0.24$ with respect to
  figures~\ref{fig:CMBLMI2beta10} and \ref{fig:CMBLMI2beta10_7}, in
  the plane $(\nS,r)$ (top panel) and the plane
  $(\epsilon_1,\epsilon_2)$ (bottom panel). The solid contours are the
  one and two-sigma {\data} confidence intervals (marginalized over
  second order slow-roll).}
\label{fig:CMBLMI2beta10_8}
\end{center}
\end{figure}

\subsection{Twisted Inflation (\hyperref[sec:twi]{TWI})}

\begin{figure}[H]
\begin{center}
\includegraphics[width=\wappfig,clip=true]{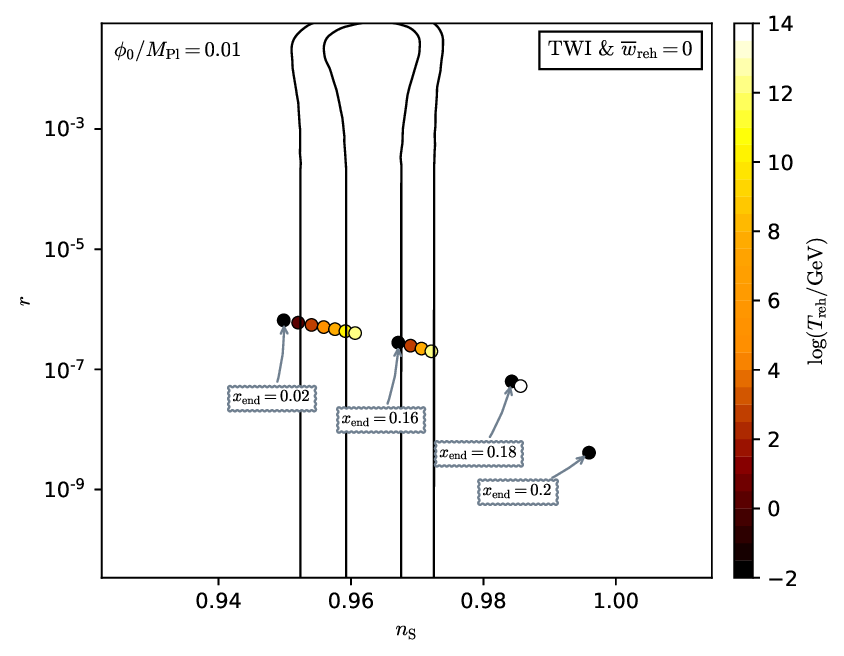}
\includegraphics[width=\wappfig,clip=true]{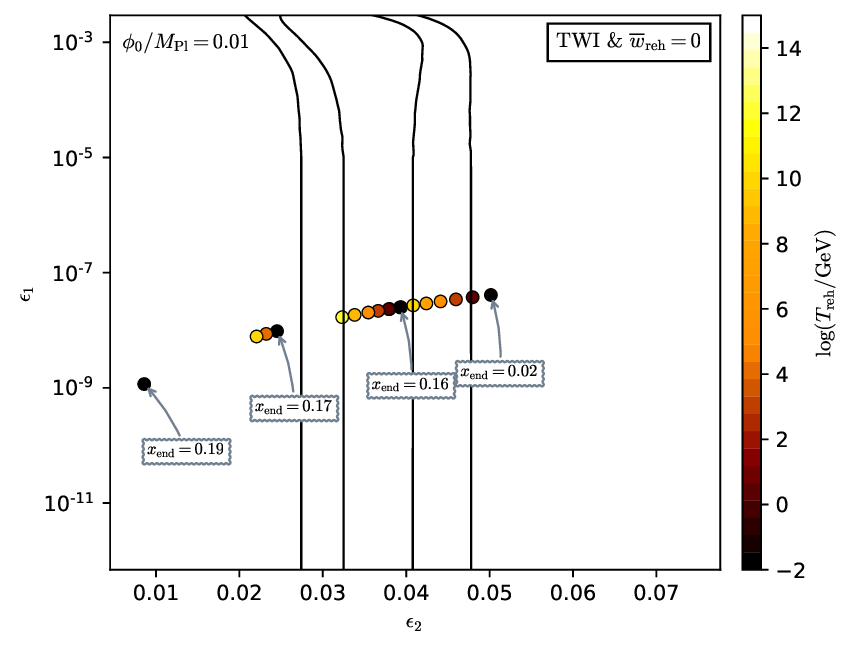}
\caption{Reheating consistent slow-roll predictions for the twisted
  models having $\phizero/\Mp=10^{-2}$ in the plane $(\nS,r)$ (top
  panel) and the plane $(\epsilon_1,\epsilon_2)$ (bottom panel). The
  solid contours are the one and two-sigma {\data} confidence
  intervals (marginalized over second order slow-roll).}
\label{fig:CMBTWI}
\end{center}
\end{figure}

\begin{figure}[H]
\begin{center}
\includegraphics[width=\wappfig,clip=true]{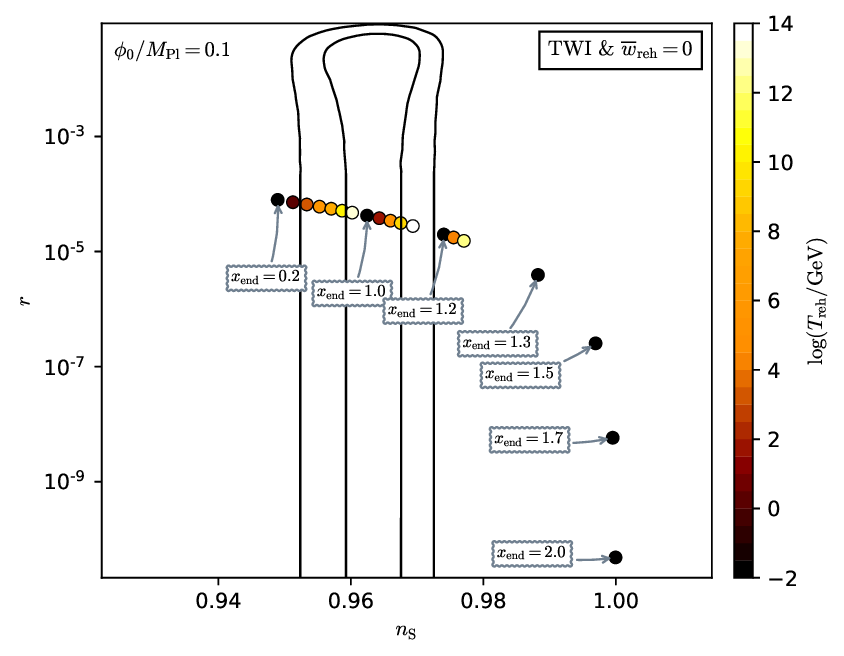}
\includegraphics[width=\wappfig,clip=true]{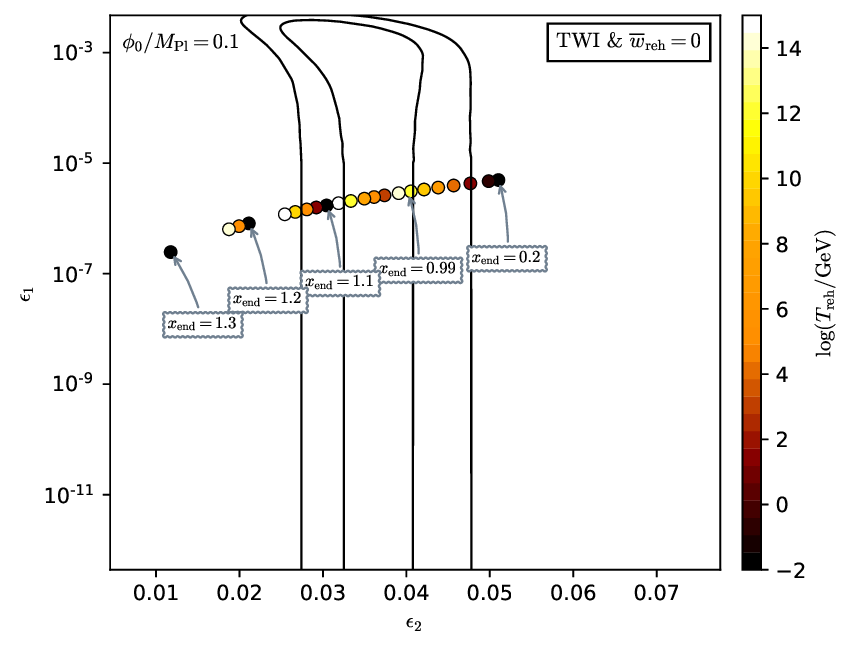}
\caption{Reheating consistent slow-roll predictions for the twisted
  models having $\phizero/\Mp=10^{-1}$ in the plane $(\nS,r)$ (top
  panel) and the plane $(\epsilon_1,\epsilon_2)$ (bottom panel). The
  solid contours are the one and two-sigma {\data} confidence
  intervals (marginalized over second order slow-roll).}
\label{fig:CMBTWI_1}
\end{center}
\end{figure}

\begin{figure}[H]
\begin{center}
\includegraphics[width=\wappfig,clip=true]{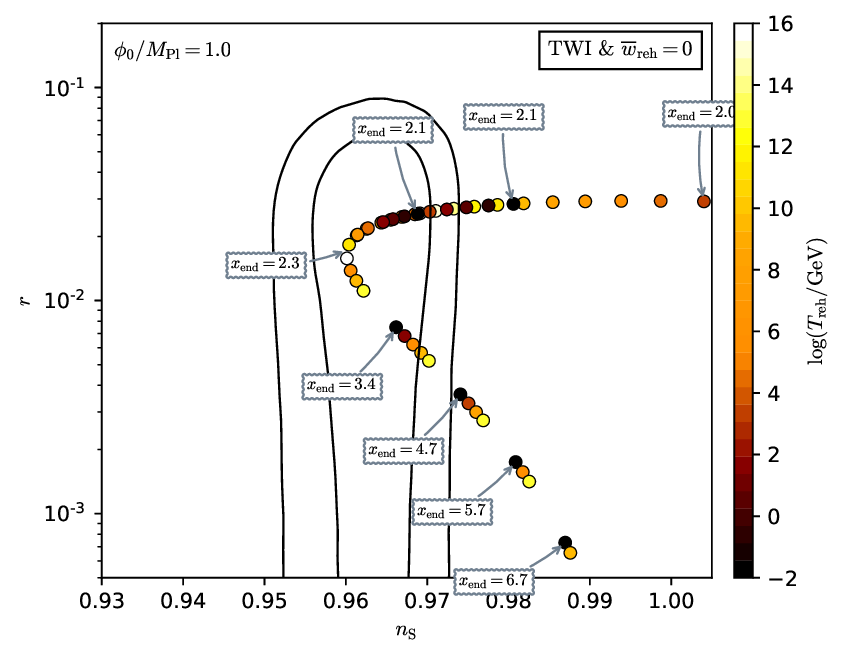}
\includegraphics[width=\wappfig,clip=true]{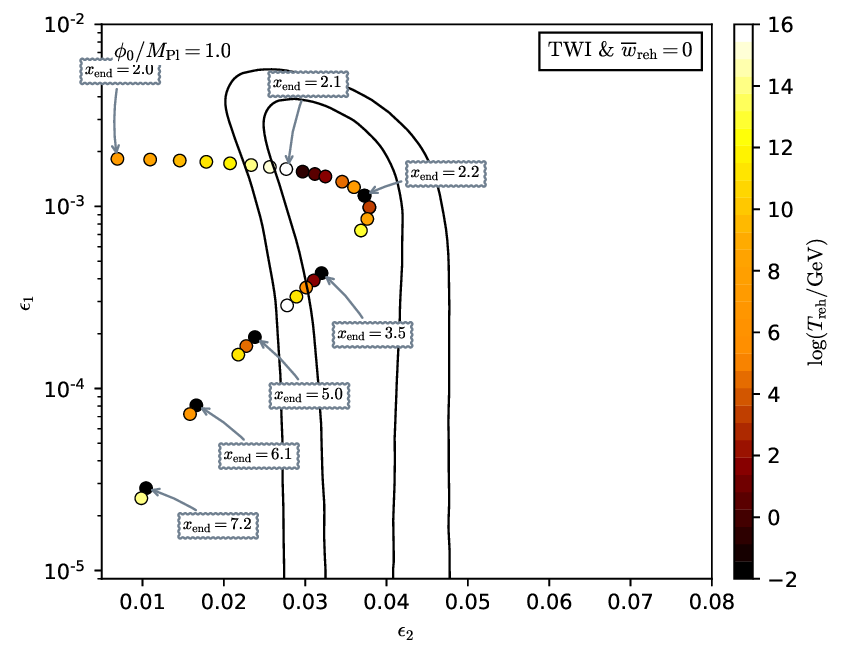}
\caption{Reheating consistent slow-roll predictions for the twisted
  models having $\phizero=\Mp$ in the plane $(\nS,r)$ (top
  panel) and the plane $(\epsilon_1,\epsilon_2)$ (bottom panel). The
  solid contours are the one and two-sigma {\data} confidence
  intervals (marginalized over second order slow-roll).}
\label{fig:CMBTWI_2}
\end{center}
\end{figure}

\subsection{GMSSM Inflation (\hyperref[sec:gmssmi]{GMSSMI})}

\begin{figure}[H]
\begin{center}
\includegraphics[width=\wappfig,clip=true]{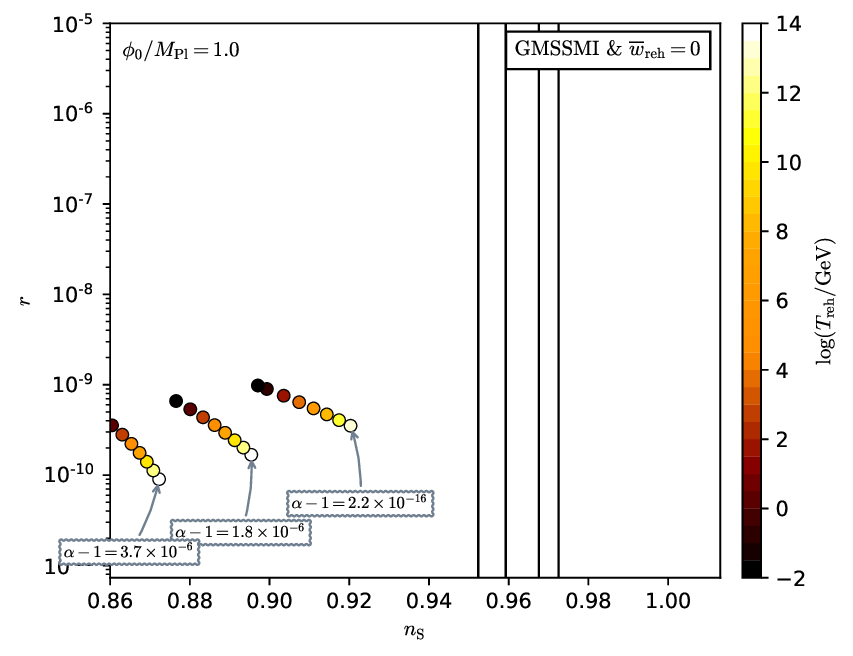}
\includegraphics[width=\wappfig,clip=true]{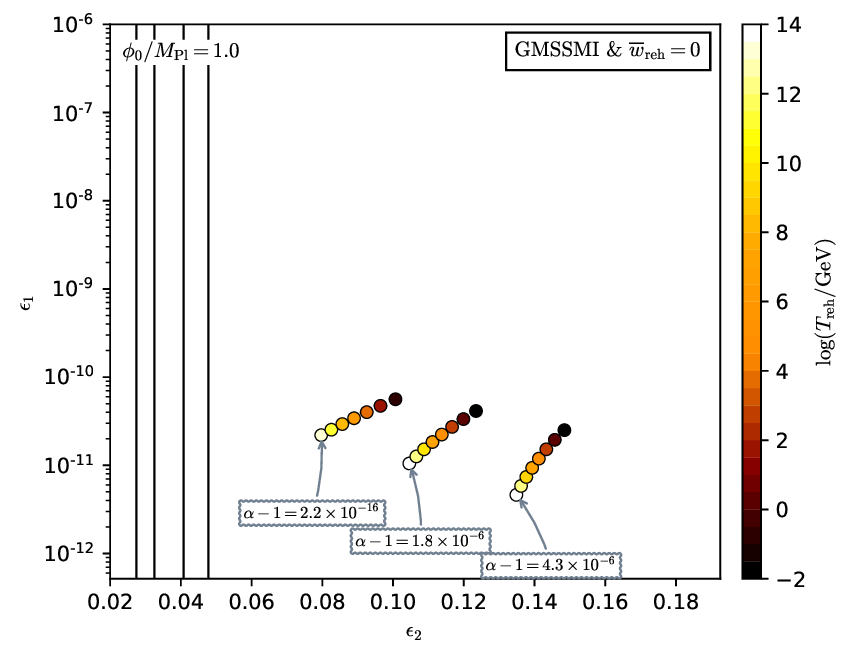}
\caption{Reheating consistent slow-roll predictions for the GMSSMI
  models in the plane $(\nS,r)$ (top panel) and the plane
  $(\epsilon_1,\epsilon_2)$ (bottom panel), with $\phizero = \Mp$ and
  for $1<\alpha<1+\phizero^4/\Mp^4\pi^2/900/(\Nend-\Nini)^2$.  The
  solid contours are the one and two-sigma {\data} confidence
  intervals (marginalized over second order slow-roll). When $\alpha
  \to 1$, one recovers the standard MSSM predictions, see
  \Fig{fig:CMBMSSMI}.}
\label{fig:CMBGMSSMIalpha>1}
\end{center}
\end{figure}

\begin{figure}[H]
\begin{center}
\includegraphics[width=\wappfig,clip=true]{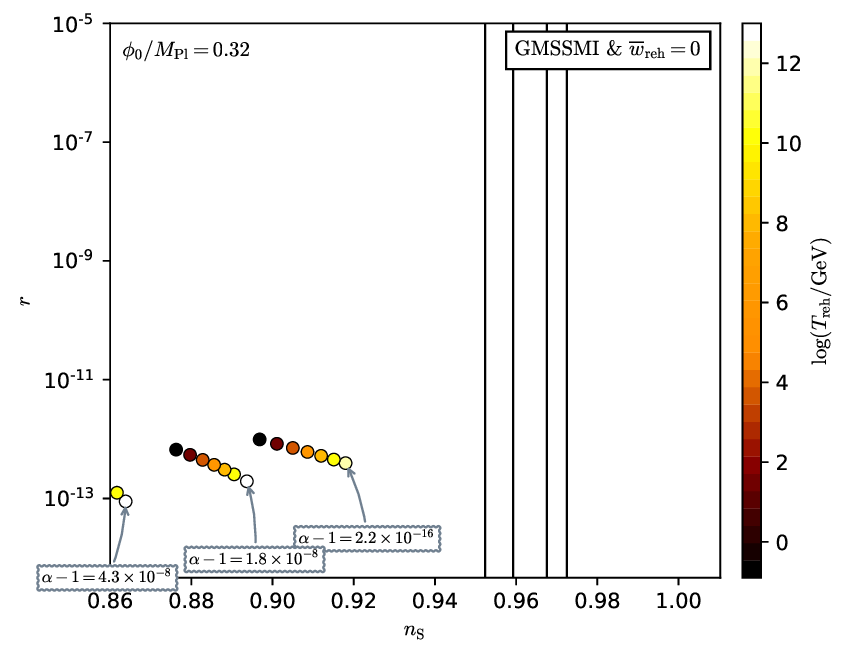}
\includegraphics[width=\wappfig,clip=true]{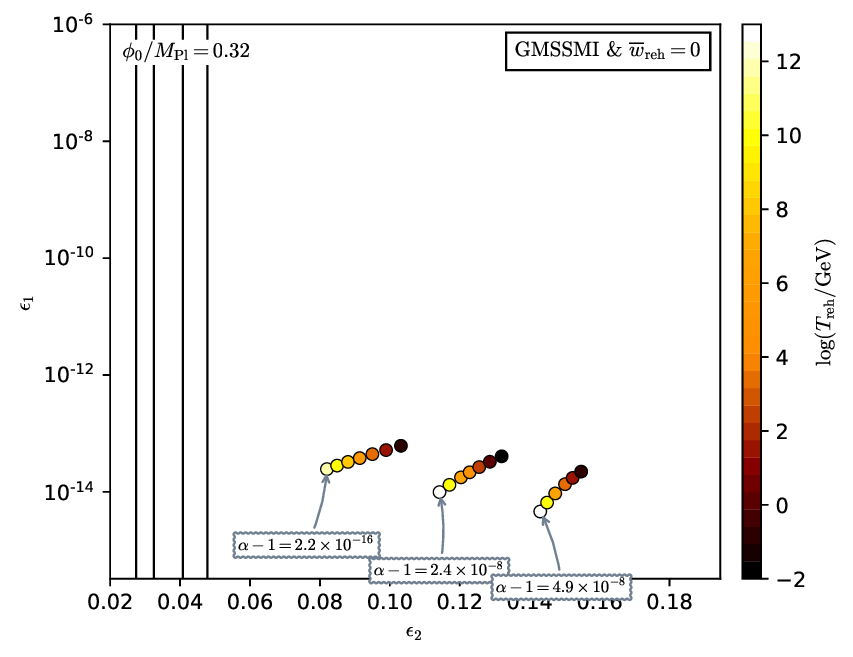}
\caption{Reheating consistent slow-roll predictions for the GMSSMI
  models in the plane $(\nS,r)$ (top panel) and the plane
  $(\epsilon_1,\epsilon_2)$ (bottom panel), with sub-Planckian
  $\phizero/\Mp=0.32$ and for
  $1<\alpha<1+\phizero^4/\Mp^4\pi^2/900/(\Nend-\Nini)^2$. Compared to
  figure~\ref{fig:CMBGMSSMIalpha>1}), the amount of primordial
  gravitational waves produced is much reduced. The solid contours are
  the one and two-sigma {\data} confidence intervals (marginalized
  over second order slow-roll).}
\label{fig:CMBGMSSMIalpha>1_1}
\end{center}
\end{figure}

\begin{figure}[H]
\begin{center}
\includegraphics[width=\wappfig,clip=true]{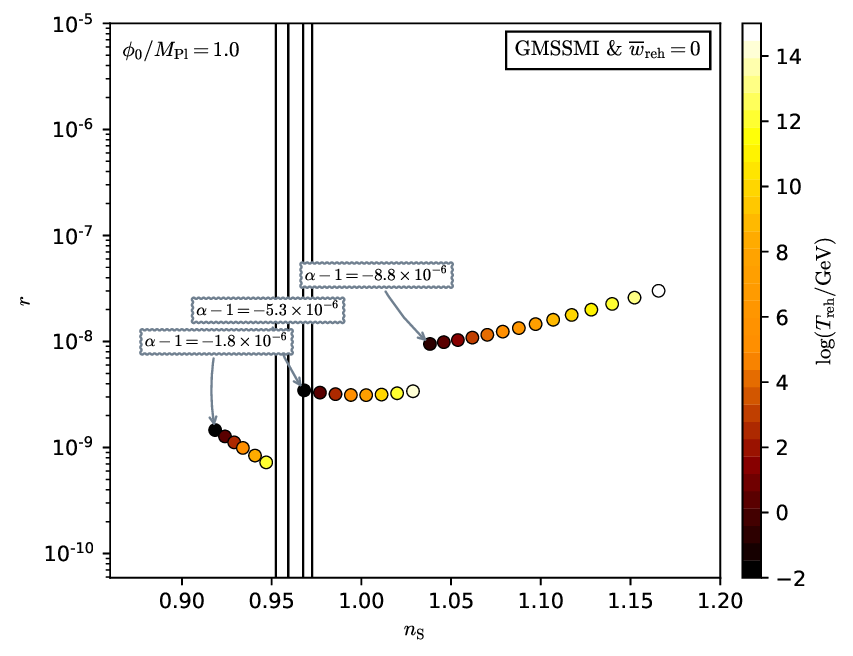}
\includegraphics[width=\wappfig,clip=true]{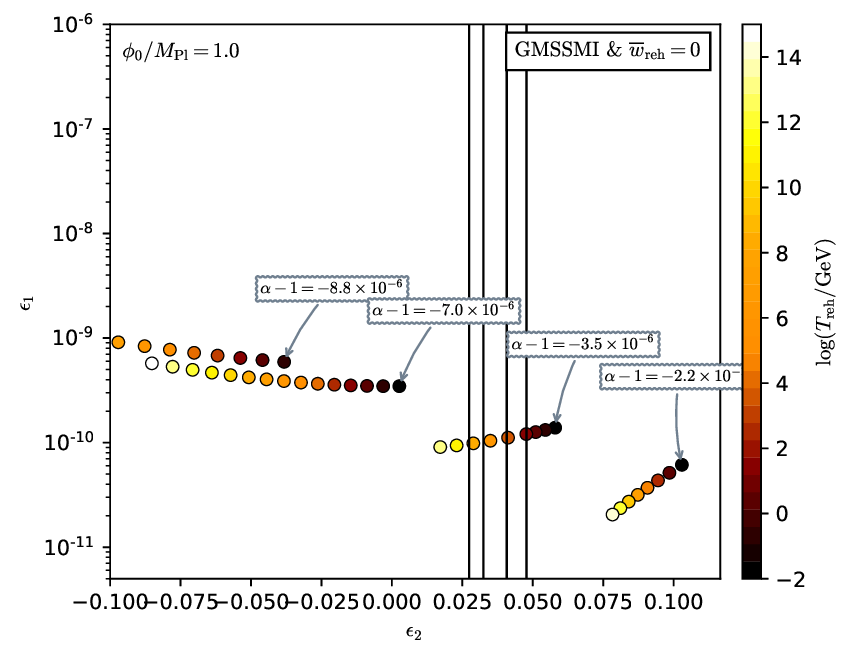}
\caption{Reheating consistent slow-roll predictions for the GMSSMI
  models in the plane $(\nS,r)$ (top panel) and the plane
  $(\epsilon_1,\epsilon_2)$ (bottom panel), with $\phizero=\Mp$ and
  for $1-\phizero^4/\Mp^ 4\pi^2/900/(\Nend-\Nini)^2<\alpha<1$.  The
  solid contours are the one and two-sigma {\data} confidence
  intervals (marginalized over second order slow-roll). When
  $\alpha\to 1$, one recovers the standard MSSM predictions,
  see \Fig{fig:CMBMSSMI}. Notice the rather strong fine-tuning on
  $\alpha$ for the model predictions to be compatible with
  observations.}
\label{fig:CMBGMSSMIalpha<1}
\end{center}
\end{figure}

\begin{figure}[H]
\begin{center}
\includegraphics[width=\wappfig,clip=true]{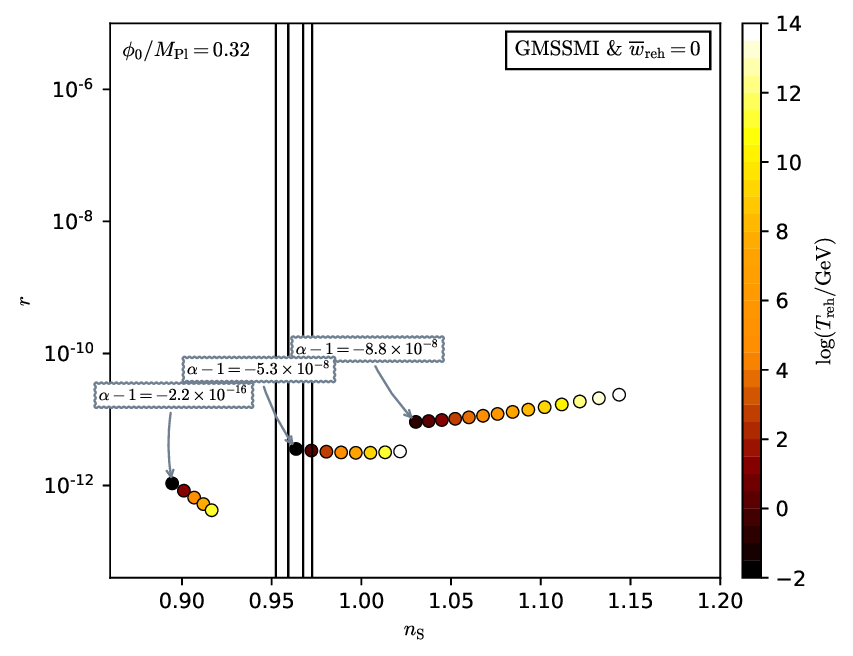}
\includegraphics[width=\wappfig,clip=true]{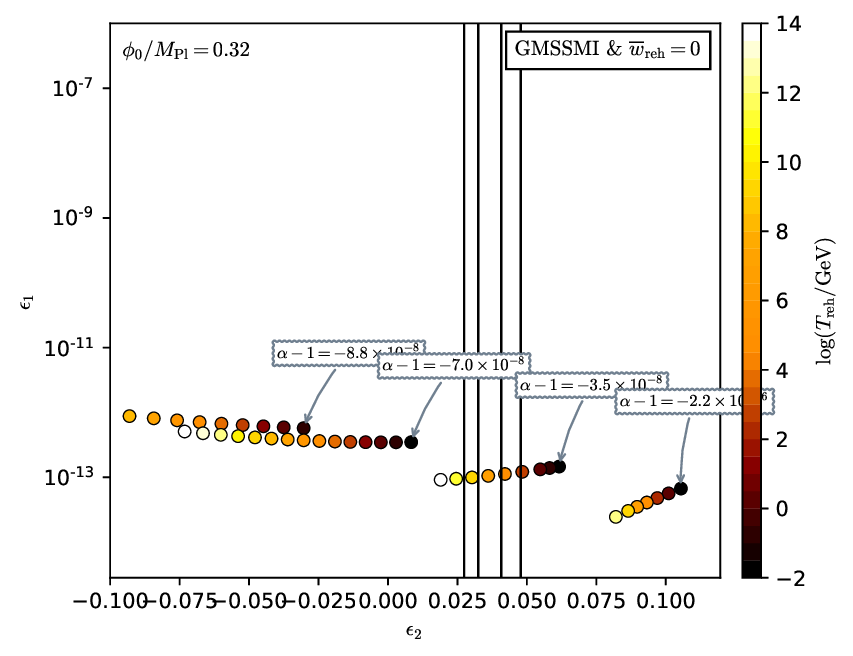}
\caption{Reheating consistent slow-roll predictions for the GMSSMI
  models in the plane $(\nS,r)$ (top panel) and the plane
  $(\epsilon_1,\epsilon_2)$ (bottom panel), with sub-Planckian
  $\phizero/\Mp-0.32$ and for $1-\phizero^4/\Mp^
  4\pi^2/900/(\Nend-\Nini)^2<\alpha<1$.  The solid contours are the
  one and two-sigma {\data} confidence intervals (marginalized over
  second order slow-roll). When $\alpha\to 1$. Compared to
  figure~\ref{fig:CMBGMSSMIalpha<1}, the amount of primordial
  gravitational waves is reduced.}
\label{fig:CMBGMSSMIalpha<1_3}
\end{center}
\end{figure}

\subsection{Generalized Renormalizable Inflection Point Inflation (\hyperref[sec:gripi]{GRIPI})}

\begin{figure}[H]
\begin{center}
\includegraphics[width=\wappfig,clip=true]{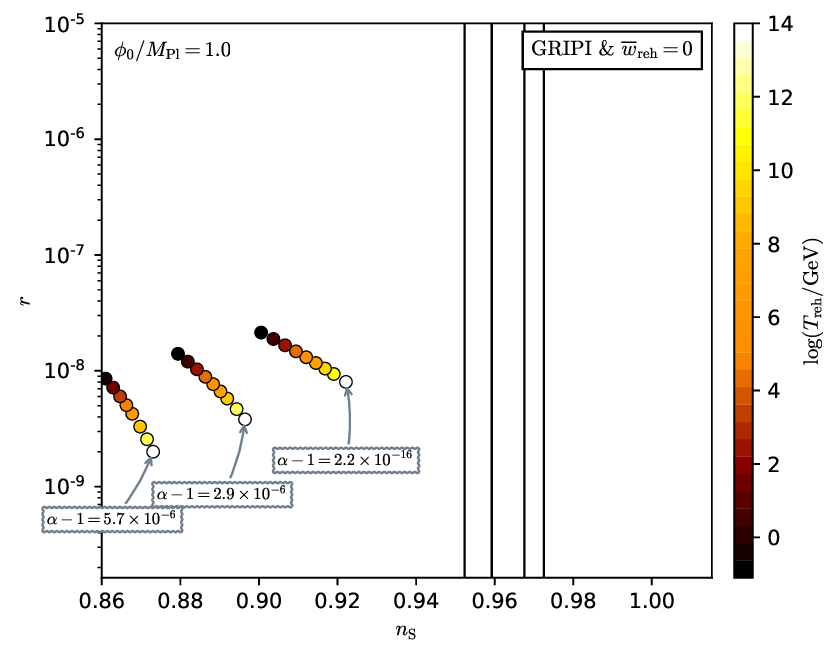}
\includegraphics[width=\wappfig,clip=true]{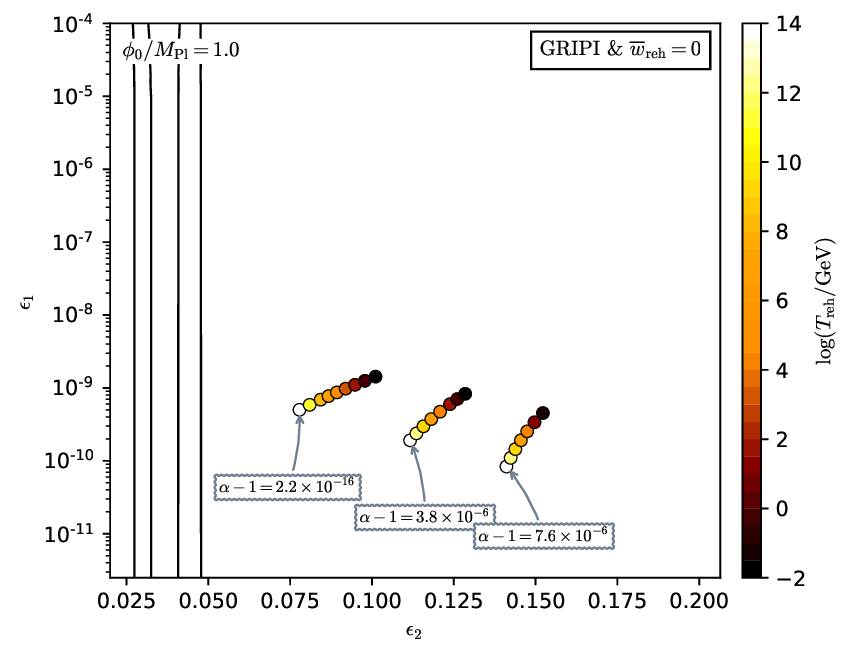}
\caption{Reheating consistent slow-roll predictions for the
  generalized renormalizable inflection point models in the plane
  $(\nS,r)$ (top panel) and the plane $(\epsilon_1,\epsilon_2)$
  (bottom panel), with $\phizero=\Mp$ and for
  $1<\alpha<1+\phizero^4/\Mp^ 4\pi^2/576/(\Nend-\Nini=60)^2$.  The
  solid contours are the one and two-sigma {\data} confidence
  intervals (marginalized over second order slow-roll). When $\alpha
  \to 1$, one recovers the standard RIPI predictions, see
  \Fig{fig:CMBRIPI}.}
\label{fig:CMBGRIPIalpha>1}
\end{center}
\end{figure}

\begin{figure}[H]
\begin{center}
\includegraphics[width=\wappfig,clip=true]{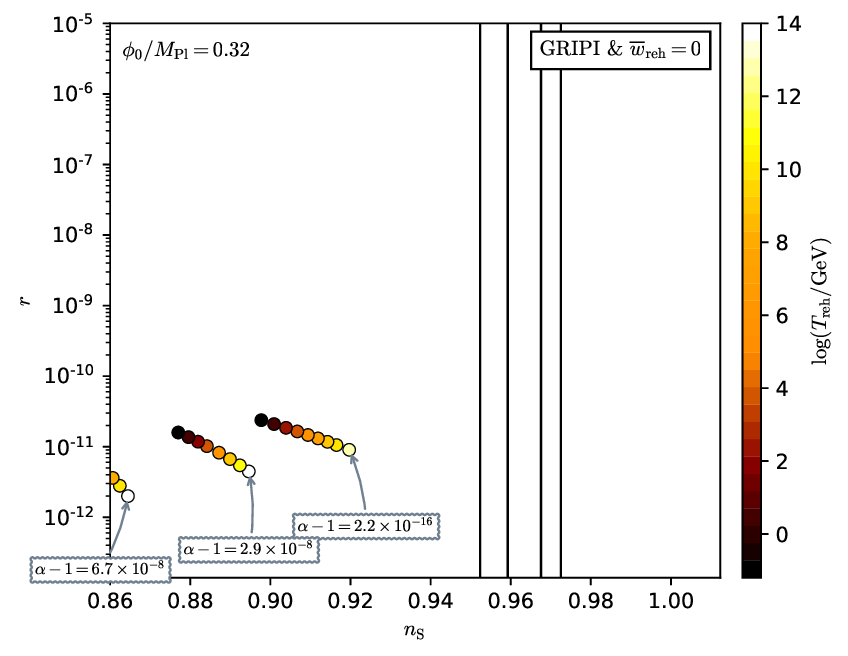}
\includegraphics[width=\wappfig,clip=true]{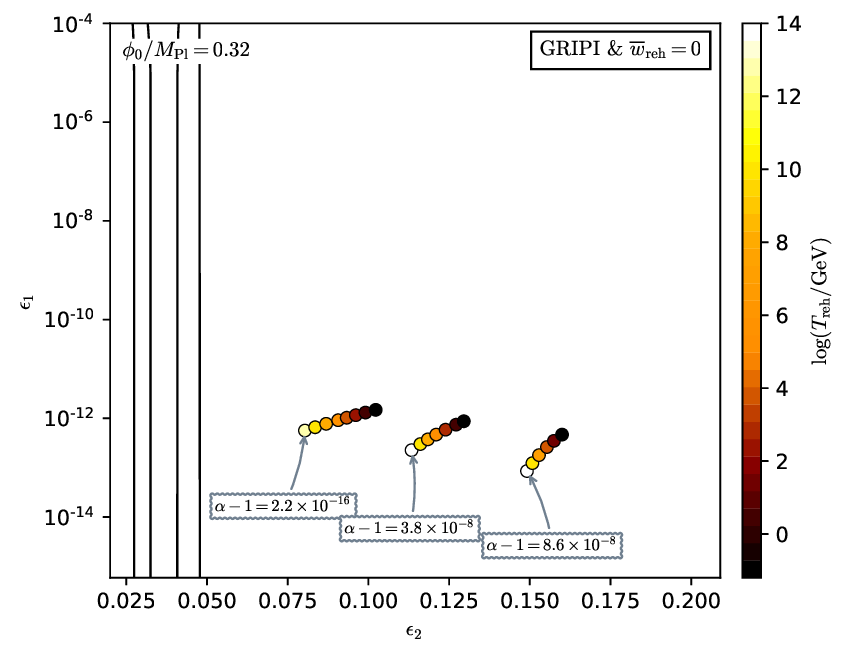}
\caption{Reheating consistent slow-roll predictions for the
  generalized renormalizable inflection point models in the plane
  $(\nS,r)$ (top panel) and the plane $(\epsilon_1,\epsilon_2)$
  (bottom panel), with sub-Planckian $\phizero=0.32\Mp$ and for
  $1<\alpha<1+\phizero^4/\Mp^ 4\pi^2/576/(\Nend-\Nini=60)^2$. The
  solid contours are the one and two-sigma {\data} confidence
  intervals (marginalized over second order slow-roll). Reducing
  $\phizero$ strongly lowers the amount of primordial gravitational
  wave produced, see figure~\ref{fig:CMBGRIPIalpha>1}.}
\label{fig:CMBGRIPIalpha>1_1}
\end{center}
\end{figure}

\begin{figure}[H]
\begin{center}
\includegraphics[width=\wappfig,clip=true]{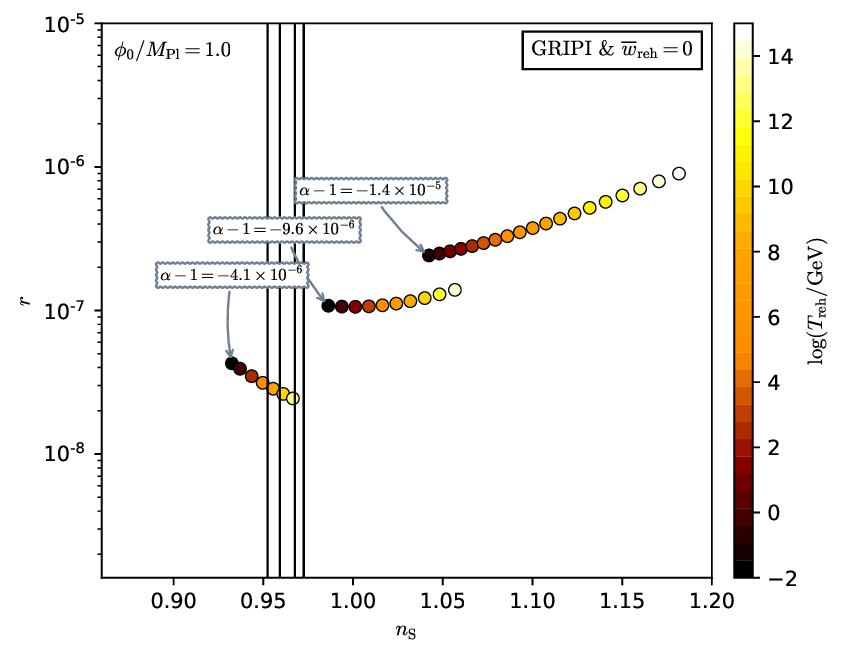}
\includegraphics[width=\wappfig,clip=true]{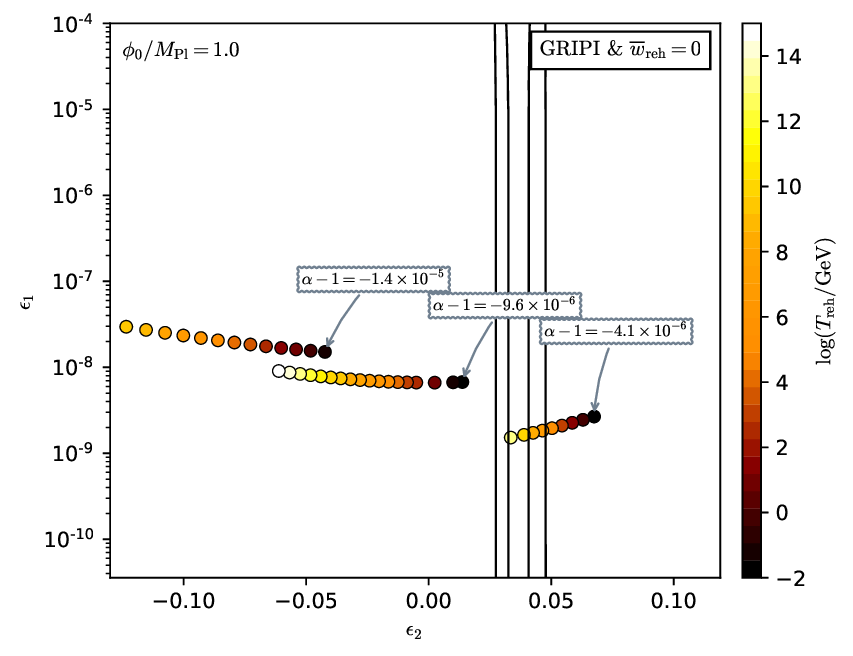}
\caption{Reheating consistent slow-roll predictions for the
  generalized renormalizable inflection point models in the plane
  $(\nS,r)$ (top panel) and the plane $(\epsilon_1,\epsilon_2)$
  (bottom panel), with $\phizero=\Mp$ and for $1-\phizero^4/\Mp^
  4\pi^2/576/(\Nend-\Nini=60)^2<\alpha<1$. The solid contours are the
  one and two-sigma {\data} confidence intervals (marginalized over
  second order slow-roll). Notice the strong fine-tuning required on
  $\alpha$ for the model to be compatible with the observations.}
\label{fig:CMBGRIPIalpha<1}
\end{center}
\end{figure}

\begin{figure}[H]
\begin{center}
\includegraphics[width=\wappfig,clip=true]{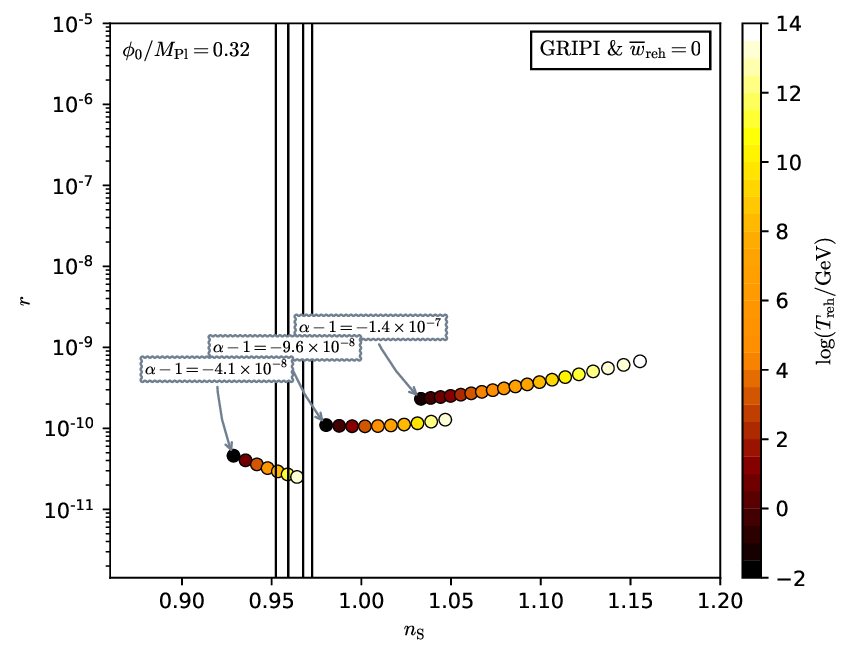}
\includegraphics[width=\wappfig,clip=true]{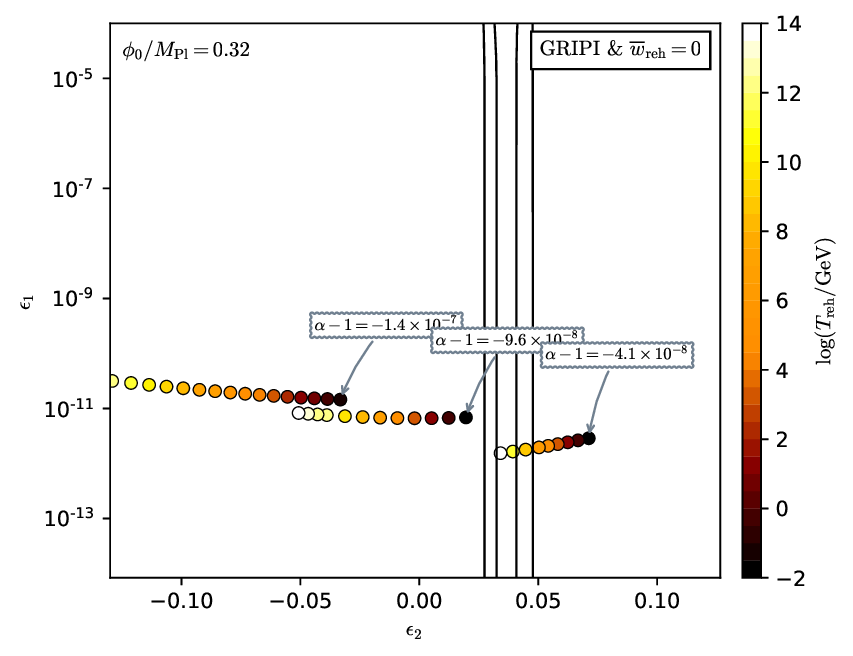}
\caption{Reheating consistent slow-roll predictions for the
  generalized renormalizable inflection point models in the plane
  $(\nS,r)$ (top panel) and the plane $(\epsilon_1,\epsilon_2)$
  (bottom panel), with sub-Planckian $\phizero=0.32\Mp$ and for
  $1-\phizero^4/\Mp^ 4\pi^2/576/(\Nend-\Nini=60)^2<\alpha<1$. The
  solid contours are the one and two-sigma {\data} confidence
  intervals (marginalized over second order slow-roll). Compared to
  figure~\ref{fig:CMBGRIPIalpha<1}, the amount of primordial
  gravitational waves is reduced.}
\label{fig:CMBGRIPIalpha<1_3}
\end{center}
\end{figure}

\subsection{Brane SUSY Breaking Inflation (\hyperref[sec:bsusybi]{BSUSYBI})}

\begin{figure}[H]
\begin{center}
\includegraphics[width=\wappfig,clip=true]{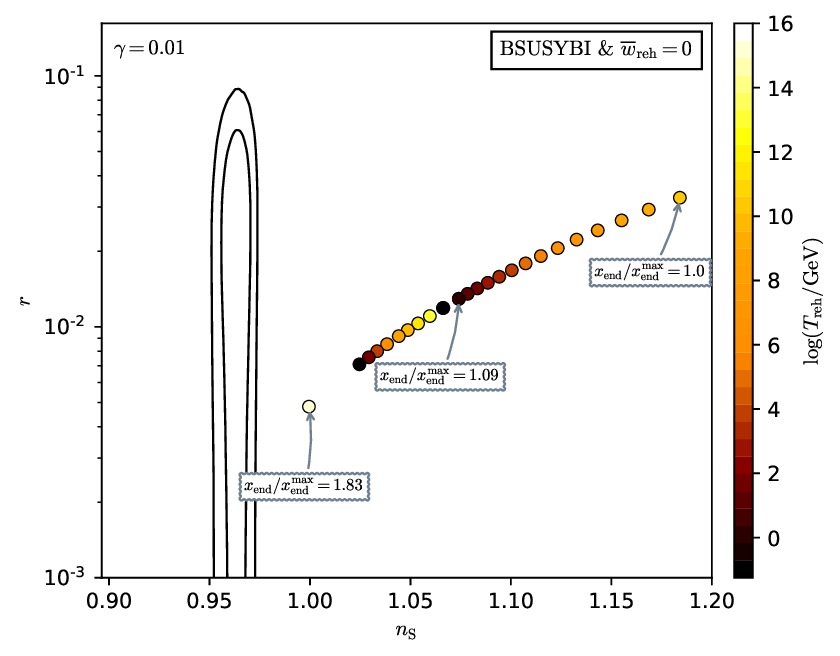}
\includegraphics[width=\wappfig,clip=true]{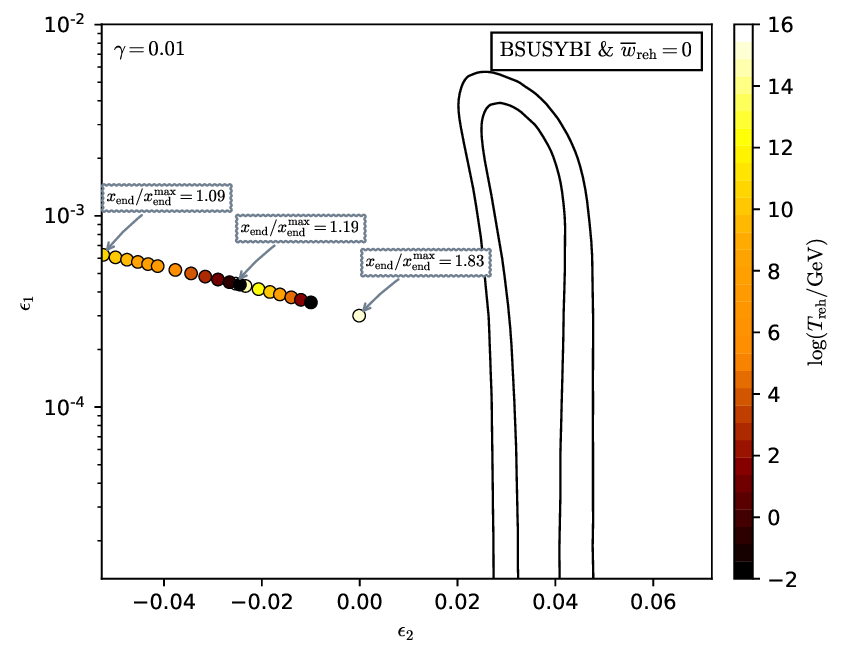}
\caption{Reheating consistent slow-roll predictions for the BSUSYBI
  models with $\gamma=10^{-2}$ in the plane $(\nS,r)$ (top panel) and
  the plane $(\epsilon_1,\epsilon_2)$ (bottom panel). The solid
  contours are the one and two-sigma {\data} confidence intervals
  (marginalized over second order slow-roll). The parameter $\xend$
  varies between $2 \xendmax < \xend< \xendmax$ ($\xendmax<0$), under
  which the predictions of the model coincide with the line
  $\epsilon_2=0$ (black solid), {\ie} PLI (see \sectionc{sec:pli}).}
\label{fig:CMBBSUSYBI}
\end{center}
\end{figure}

\begin{figure}[H]
\begin{center}
\includegraphics[width=\wappfig,clip=true]{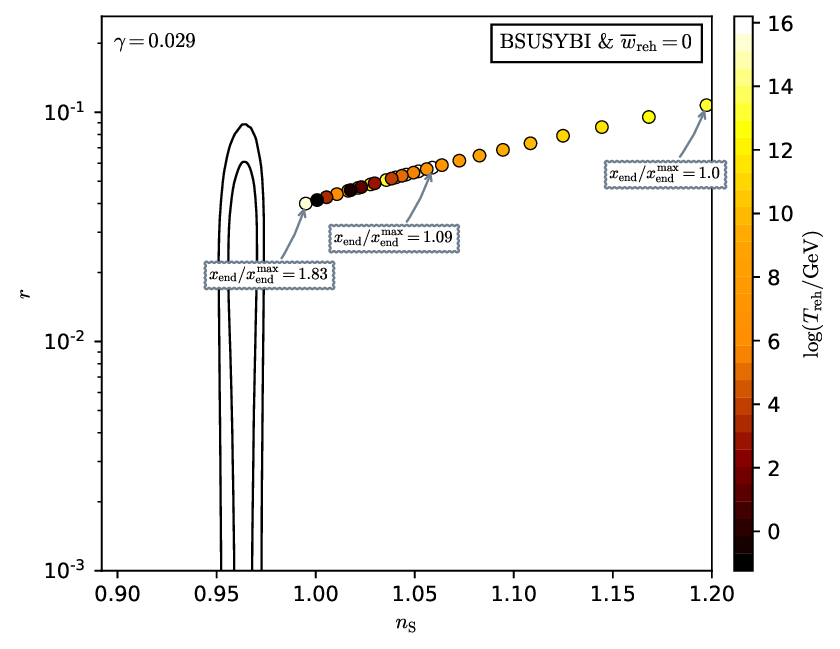}
\includegraphics[width=\wappfig,clip=true]{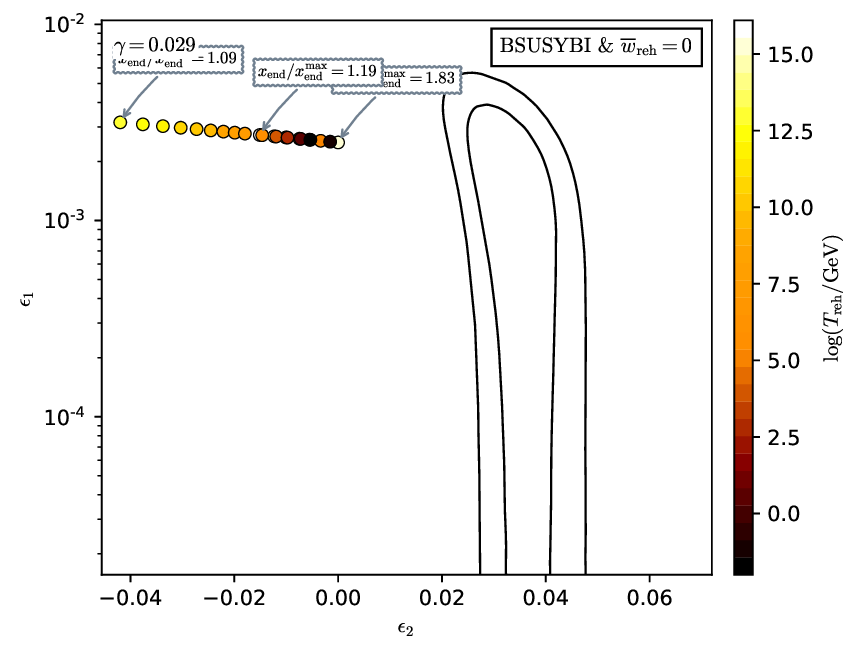}
\caption{Reheating consistent slow-roll predictions for the BSUSYBI
  models with $\gamma=0.029$, slightly increased compared to
  figure~\ref{fig:CMBBSUSYBI}, in the plane $(\nS,r)$ (top panel) and
  the plane $(\epsilon_1,\epsilon_2)$ (bottom panel). The solid
  contours are the one and two-sigma {\data} confidence intervals
  (marginalized over second order slow-roll). The parameter $\gamma$
  should be $\lesssim 5\times 10^{-2}$ to predict a reasonable amount
  of primordial gravitational waves.}
\label{fig:CMBBSUSYBI_1}
\end{center}
\end{figure}

\subsection{Tip Inflation (\hyperref[sec:ti]{TI})}

\begin{figure}[H]
\begin{center}
\includegraphics[width=\wappfig,clip=true]{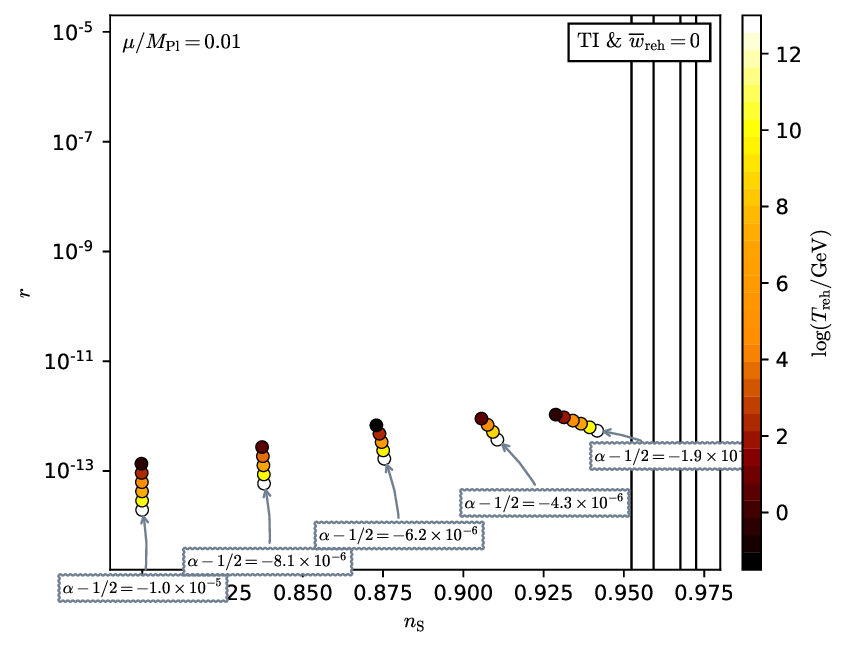}
\includegraphics[width=\wappfig,clip=true]{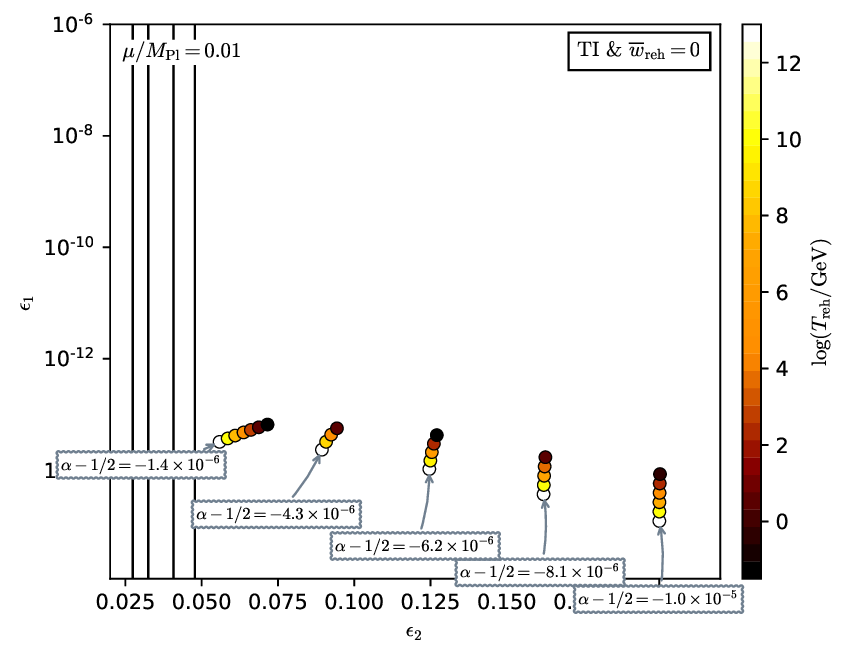}
\caption{Reheating consistent slow-roll predictions for the tip
  inflation models with $\alpha<1/2$, and for $\mu/\Mp=10^{-2}$ in the
  plane $(\nS,r)$ (top panel) and the plane $(\epsilon_1,\epsilon_2)$
  (bottom panel). The solid contours are the one and two-sigma {\data}
  confidence intervals (marginalized over second order slow-roll).}
\label{fig:CMBTIalphaLTonehalf}
\end{center}
\end{figure}

\begin{figure}[H]
\begin{center}
\includegraphics[width=\wappfig,clip=true]{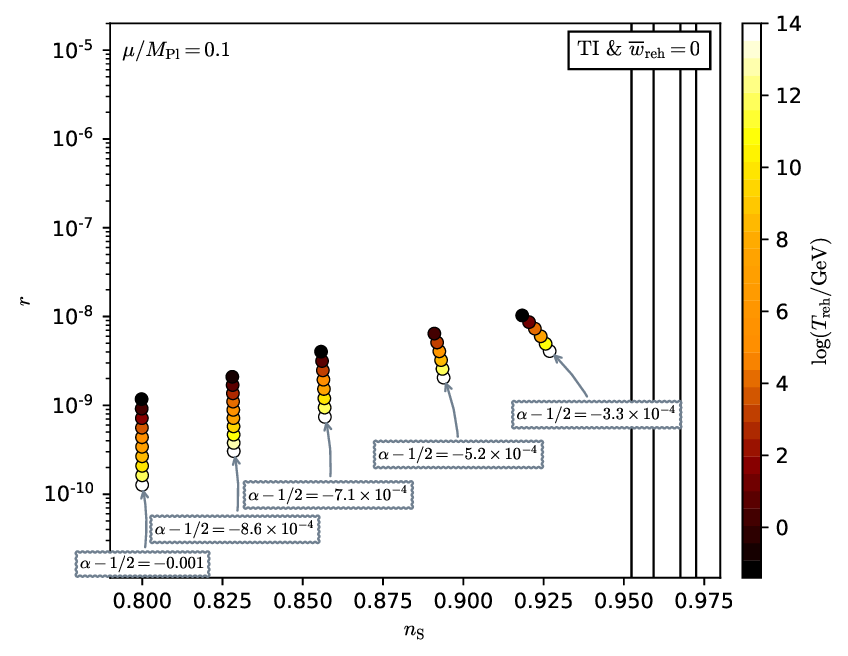}
\includegraphics[width=\wappfig,clip=true]{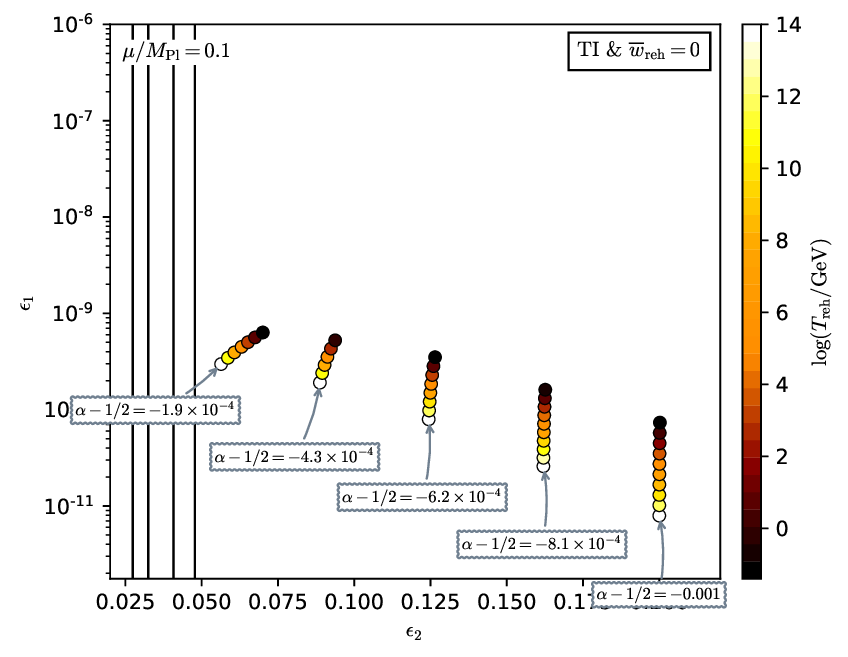}
\caption{Reheating consistent slow-roll predictions for the tip
  inflation models with $\alpha<1/2$, and for $\mu/\Mp=10^{-1}$ in the
  plane $(\nS,r)$ (top panel) and the plane $(\epsilon_1,\epsilon_2)$
  (bottom panel). The solid contours are the one and two-sigma {\data}
  confidence intervals (marginalized over second order slow-roll). See
  also figure~\ref{fig:CMBTIalphaLTonehalf}.}
\label{fig:CMBTIalphaLTonehalf_1}
\end{center}
\end{figure}

\begin{figure}[H]
\begin{center}
\includegraphics[width=\wappfig,clip=true]{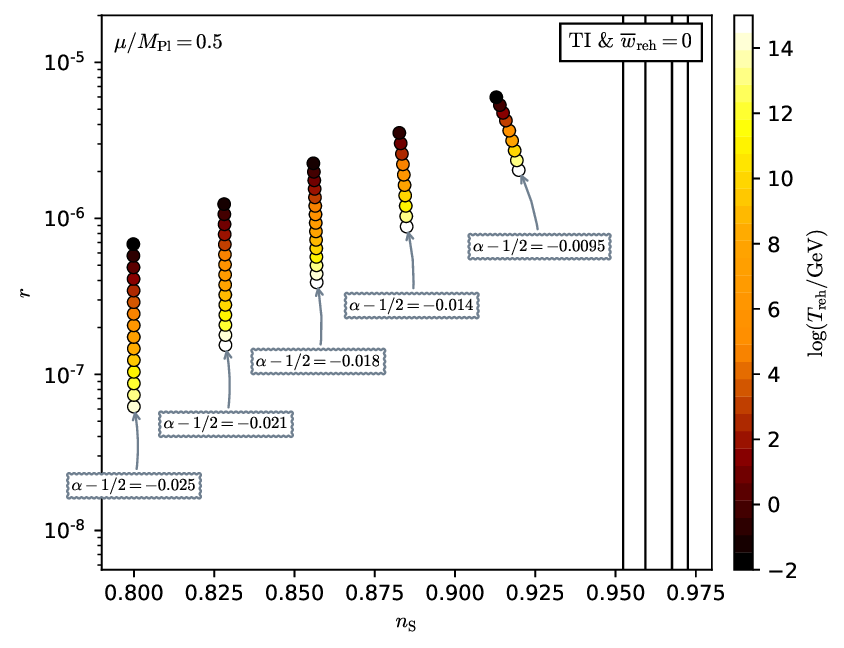}
\includegraphics[width=\wappfig,clip=true]{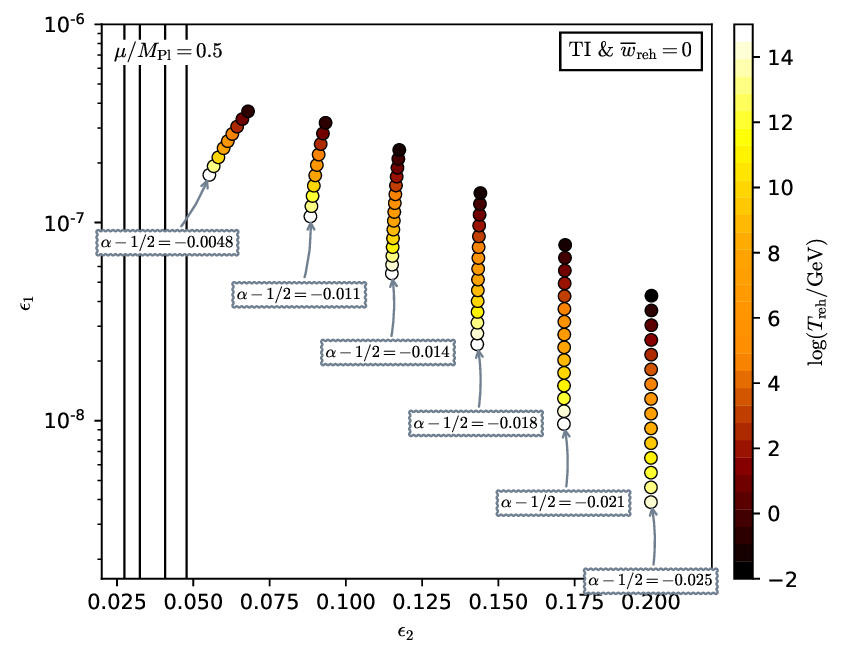}
\caption{Reheating consistent slow-roll predictions for the tip
  inflation models with $\alpha<1/2$, and for $\mu/\Mp=0.5$ in the
  plane $(\nS,r)$ (top panel) and the plane $(\epsilon_1,\epsilon_2)$
  (bottom panel). The solid contours are the one and two-sigma {\data}
  confidence intervals (marginalized over second order slow-roll). To
  be compared to smaller values of $\mu/\Mp$ in
  figures~\ref{fig:CMBTIalphaLTonehalf} and
  \ref{fig:CMBTIalphaLTonehalf_1}.}
\label{fig:CMBTIalphaLTonehalf_2}
\end{center}
\end{figure}

\begin{figure}[H]
\begin{center}
\includegraphics[width=\wappfig,clip=true]{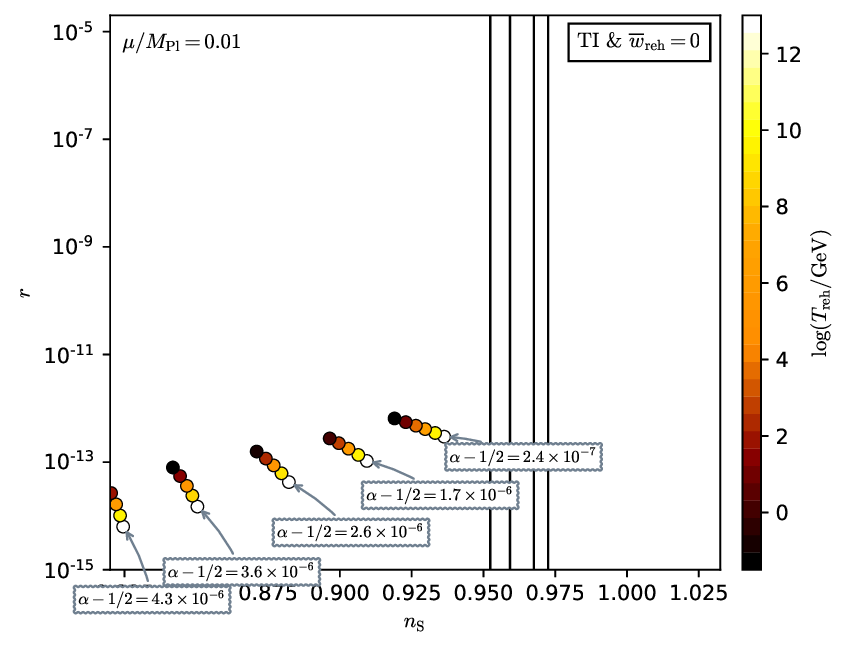}
\includegraphics[width=\wappfig,clip=true]{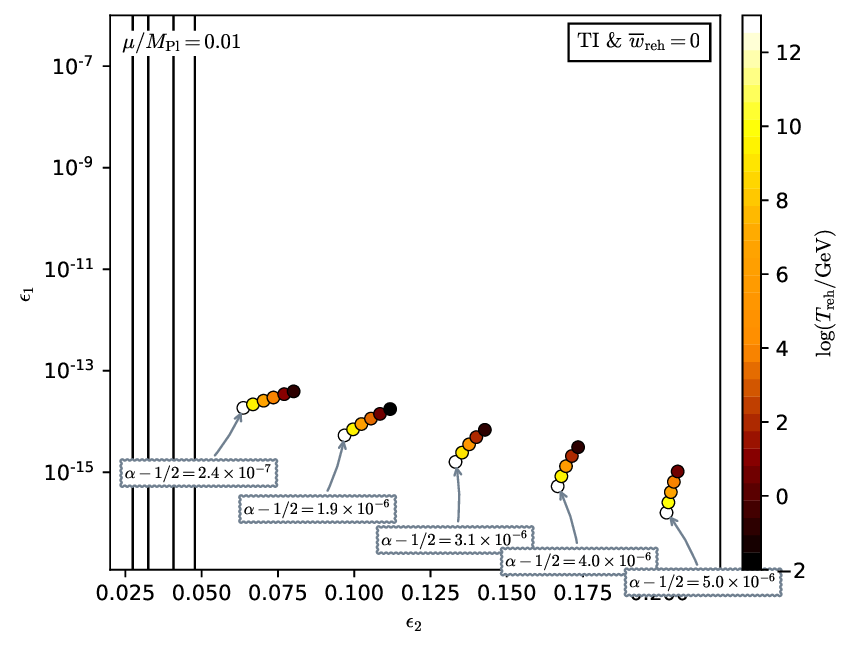}
\caption{Reheating consistent slow-roll predictions for the tip
  inflation models with $\alpha>1/2$, and for $\mu/\Mp=10^{-2}$ in the
  plane $(\nS,r)$ (top panel) and the plane $(\epsilon_1,\epsilon_2)$
  (bottom panel). The solid contours are the one and two-sigma {\data}
  confidence intervals (marginalized over second order slow-roll).}
\label{fig:CMBTIalphaGTonehalf}
\end{center}
\end{figure}

\begin{figure}[H]
\begin{center}
\includegraphics[width=\wappfig,clip=true]{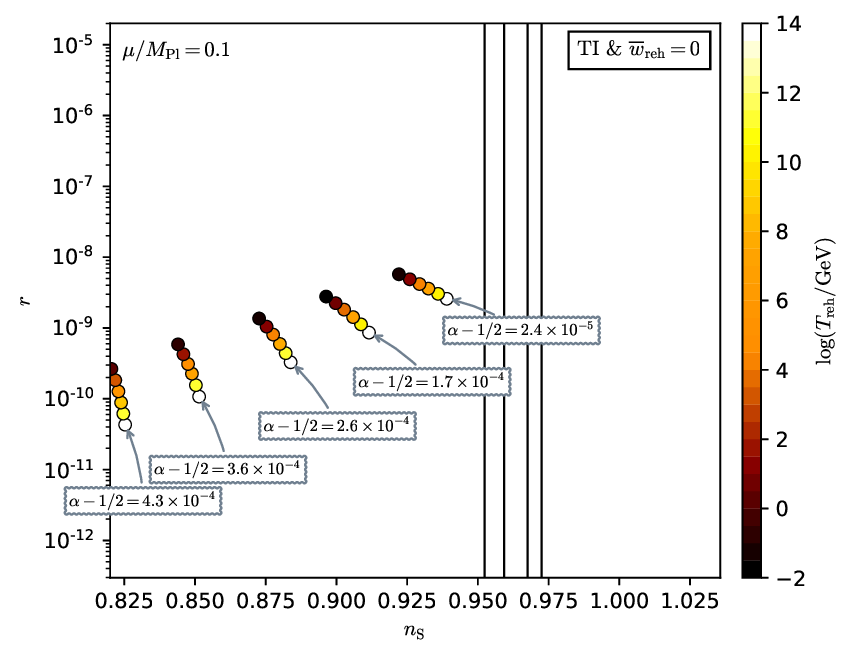}
\includegraphics[width=\wappfig,clip=true]{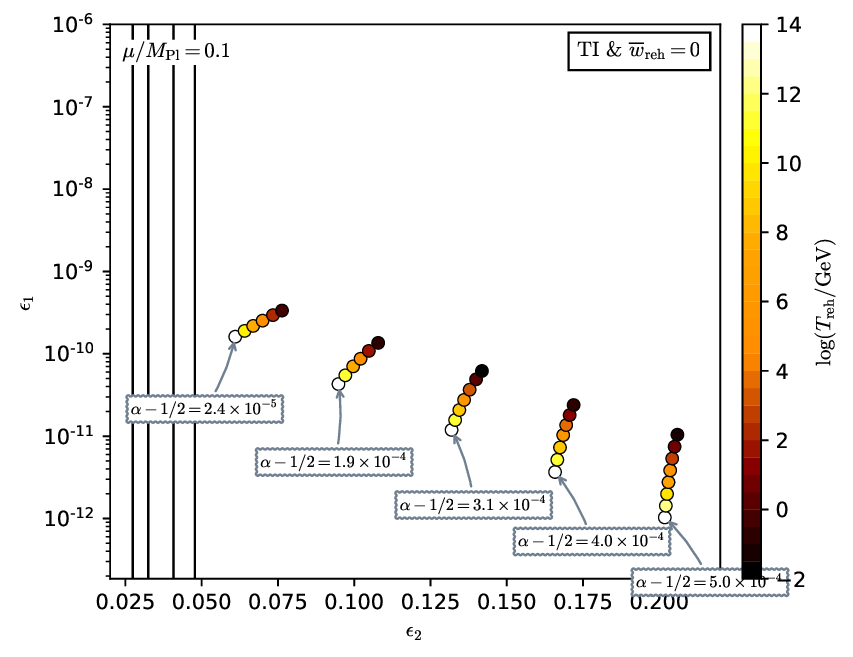}
\caption{Reheating consistent slow-roll predictions for the tip
  inflation models with $\alpha>1/2$, and for $\mu/\Mp=10^{-1}$ in the
  plane $(\nS,r)$ (top panel) and the plane $(\epsilon_1,\epsilon_2)$
  (bottom panel). The solid contours are the one and two-sigma {\data}
  confidence intervals (marginalized over second order slow-roll).}
\label{fig:CMBTIalphaGTonehalf_1}
\end{center}
\end{figure}

\begin{figure}[H]
\begin{center}
\includegraphics[width=\wappfig,clip=true]{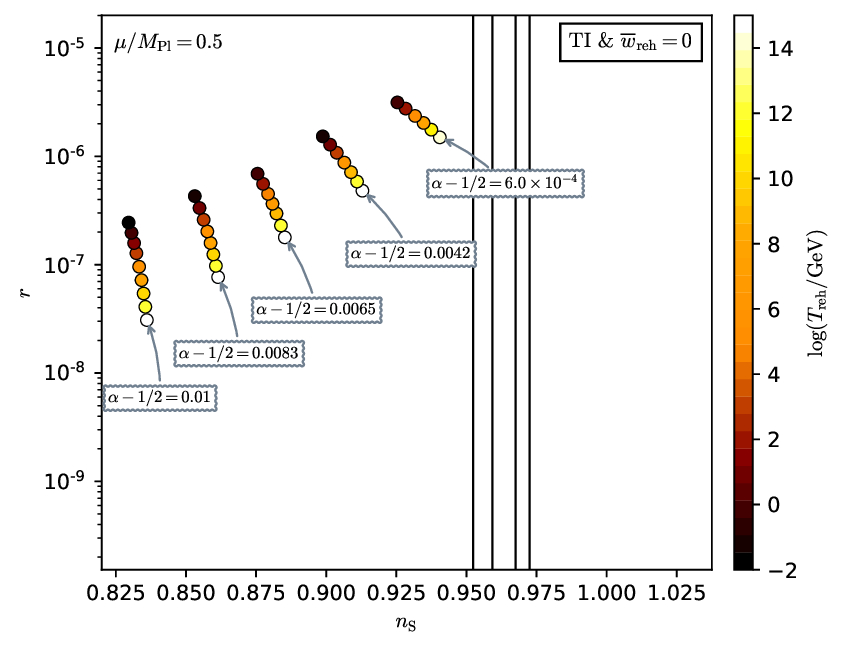}
\includegraphics[width=\wappfig,clip=true]{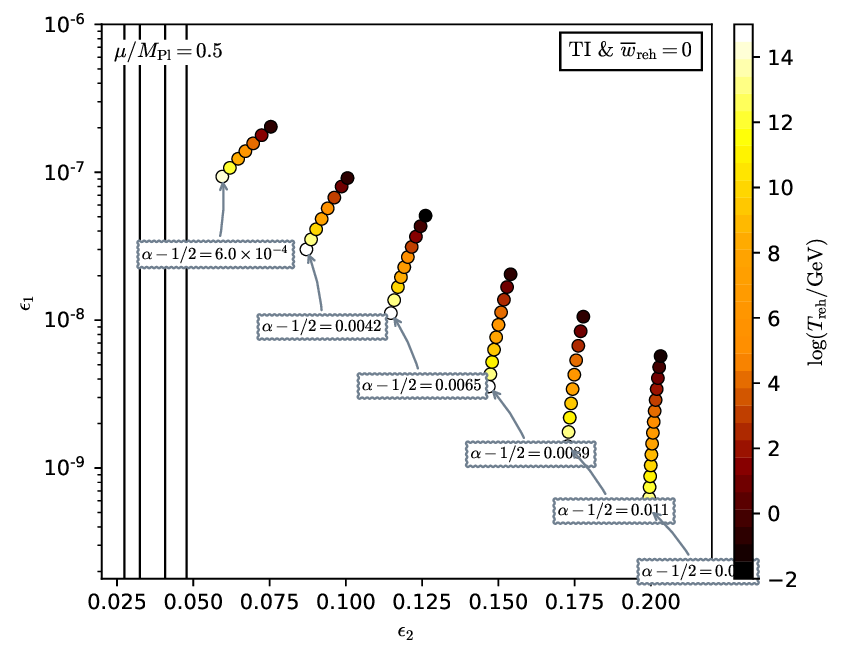}
\caption{Reheating consistent slow-roll predictions for the tip
  inflation models with $\alpha>1/2$, and for $\mu/\Mp=0.5$ in the
  plane $(\nS,r)$ (top panel) and the plane $(\epsilon_1,\epsilon_2)$
  (bottom panel). The solid contours are the one and two-sigma {\data}
  confidence intervals (marginalized over second order slow-roll). To
  be compared to smaller values of $\mu/\Mp$ in
  figures~\ref{fig:CMBTIalphaGTonehalf} and
  \ref{fig:CMBTIalphaGTonehalf_1}.}
\label{fig:CMBTIalphaGTonehalf_2}
\end{center}
\end{figure}

\begin{figure}[H]
\begin{center}
\includegraphics[width=\wappfig,clip=true]{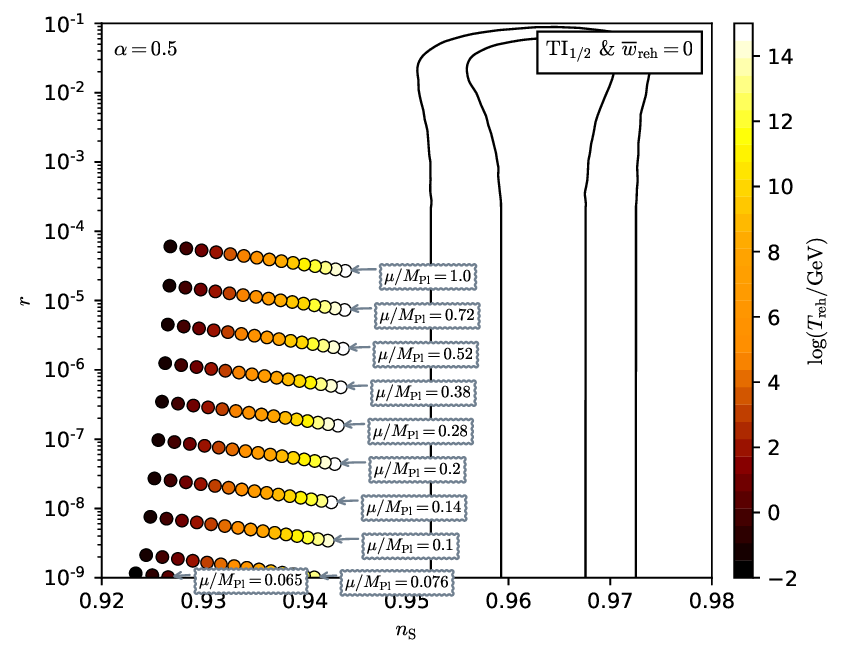}
\includegraphics[width=\wappfig,clip=true]{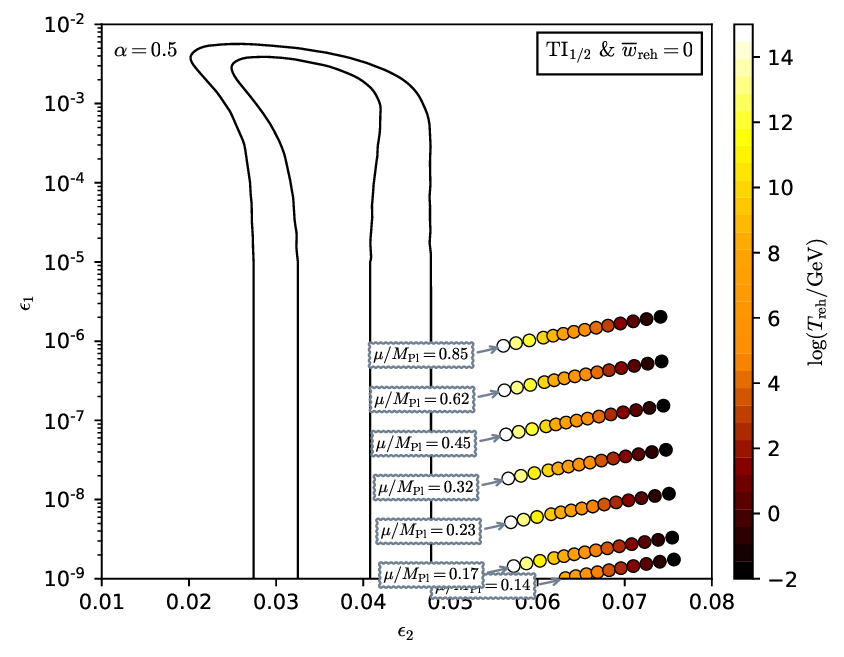}
\caption{Reheating consistent slow-roll predictions for the tip
  inflation models with $\alpha=1/2$ in the plane $(\nS,r)$ (top
  panel) and the plane $(\epsilon_1,\epsilon_2)$ (bottom panel). The
  solid contours are the one and two-sigma {\data} confidence
  intervals (marginalized over second order slow-roll).}
\label{fig:CMBTIalphaEQonehalf}
\end{center}
\end{figure}

\subsection{\texorpdfstring{$\beta$}{Beta} Exponential Inflation (\hyperref[sec:bei]{BEI})}

\begin{figure}[H]
\begin{center}
\includegraphics[width=\wappfig,clip=true]{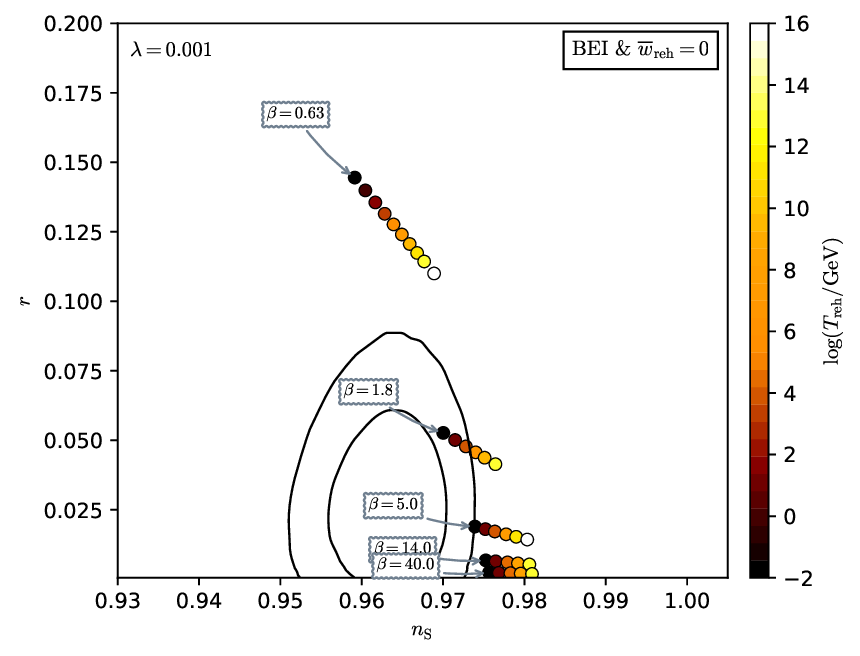}
\includegraphics[width=\wappfig,clip=true]{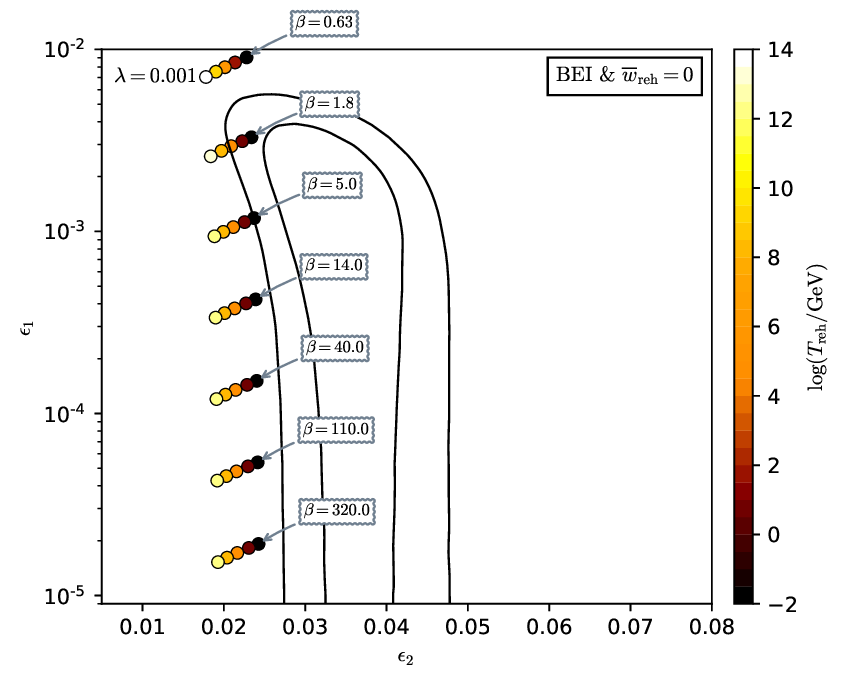}
\caption{Reheating consistent slow-roll predictions for the $\beta$
  Exponential Inflation models in the plane $(\nS,r)$ (top panel) and
  the plane $(\epsilon_1,\epsilon_2)$ (bottom panel). The parameter
  $\lambda$ has been fixed to $10^{-3}$ for this plot but the
  predictions almost do not depend on it (see
  figure~\ref{fig:CMBBEI_1}). The solid contours are the one and
  two-sigma {\data} confidence intervals (marginalized over second
  order slow-roll).}
\label{fig:CMBBEI}
\end{center}
\end{figure}

\begin{figure}[H]
\begin{center}
\includegraphics[width=\wappfig,clip=true]{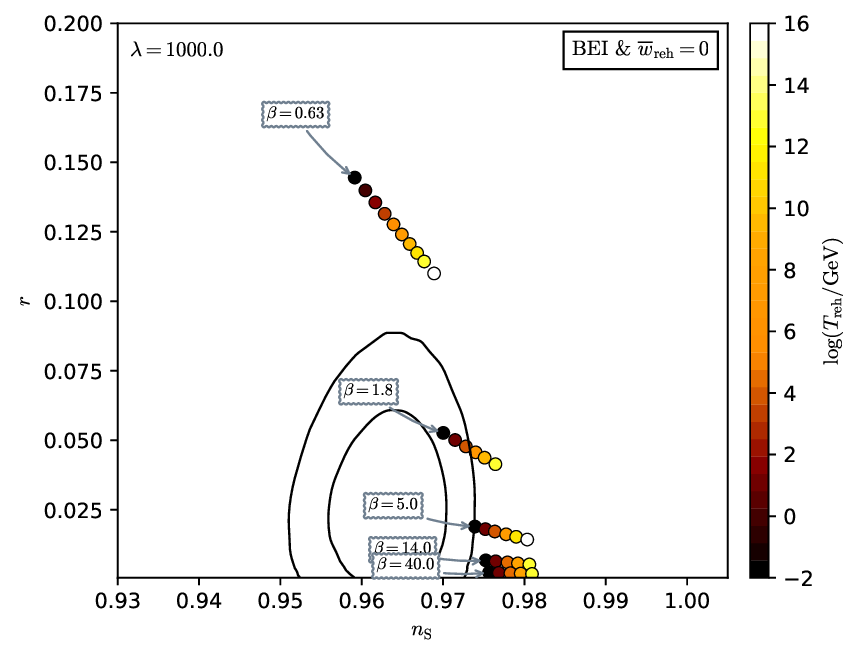}
\includegraphics[width=\wappfig,clip=true]{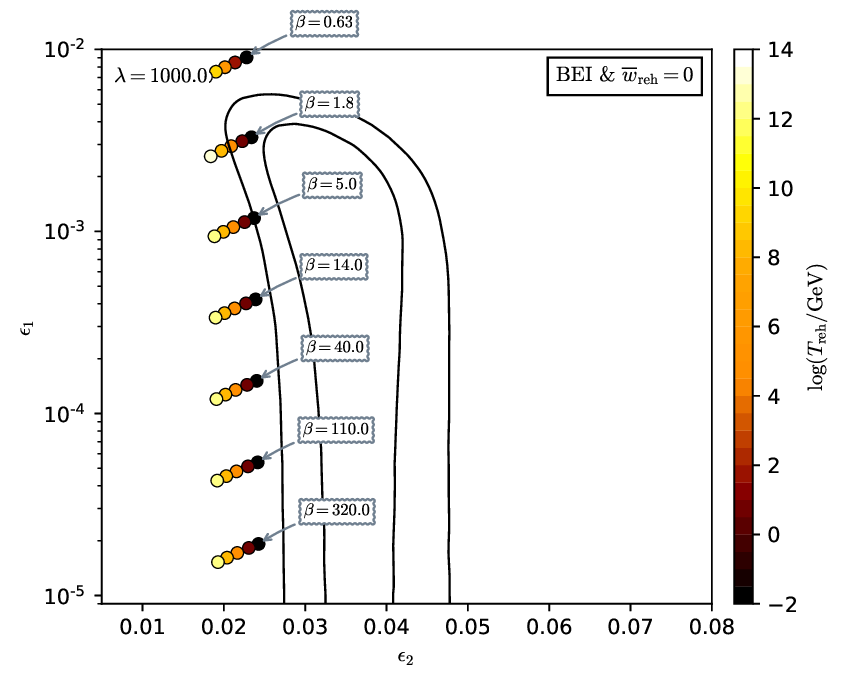}
\caption{Reheating consistent slow-roll predictions for the $\beta$
  Exponential Inflation models with a large value of $\lambda=10^{3}$
  in the plane $(\nS,r)$ (top panel) and the plane
  $(\epsilon_1,\epsilon_2)$ (bottom panel). The predictions are
  undistinguishable from the ones having $\lambda=10^{-3}$, see
  figure~\ref{fig:CMBBEI}. The solid contours are the one and
  two-sigma {\data} confidence intervals (marginalized over second
  order slow-roll).}
\label{fig:CMBBEI_1}
\end{center}
\end{figure}

\subsection{Pseudo Natural Inflation (\hyperref[sec:psni]{PSNI})}

\begin{figure}[H]
\begin{center}
\includegraphics[width=\wappfig,clip=true]{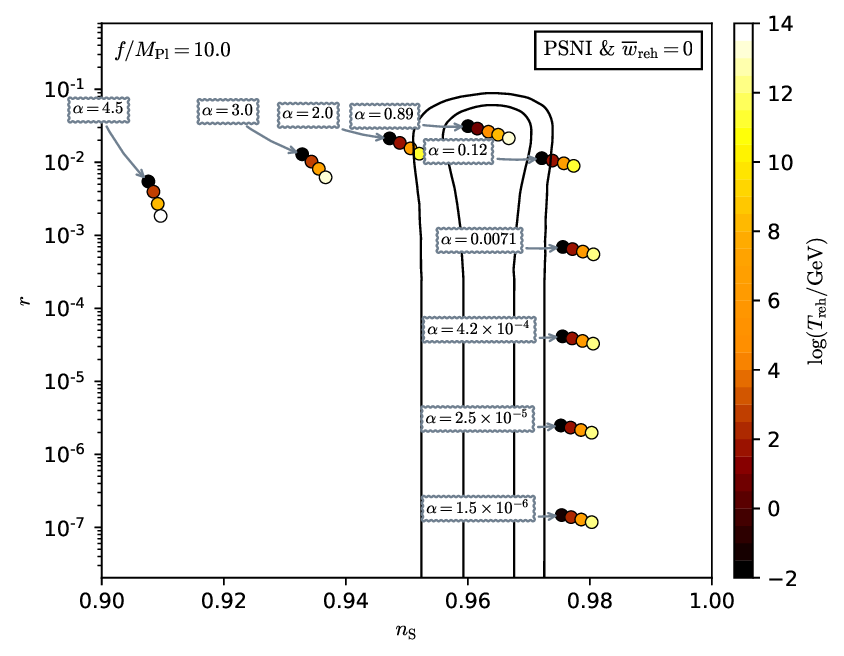}
\includegraphics[width=\wappfig,clip=true]{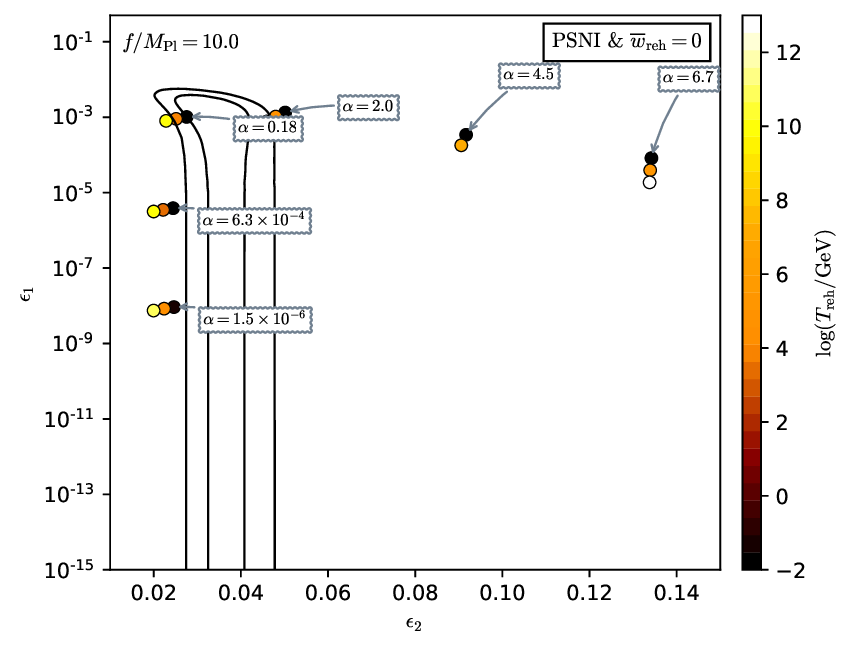}
\caption{Reheating consistent slow-roll predictions for the pseudo
  natural inflation models with $f/\Mp=10$, in the plane $(\nS,r)$
  (top panel) and the plane $(\epsilon_1,\epsilon_2)$ (bottom
  panel). The solid contours are the one and two-sigma {\data}
  confidence intervals (marginalized over second order slow-roll).}
\label{fig:CMBPSNI}
\end{center}
\end{figure}

\begin{figure}[H]
\begin{center}
\includegraphics[width=\wappfig,clip=true]{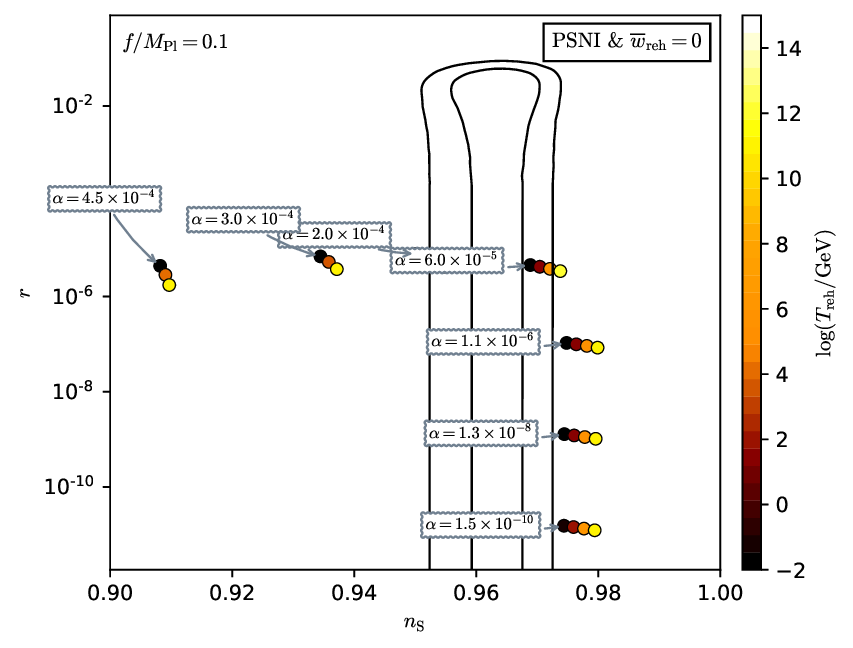}
\includegraphics[width=\wappfig,clip=true]{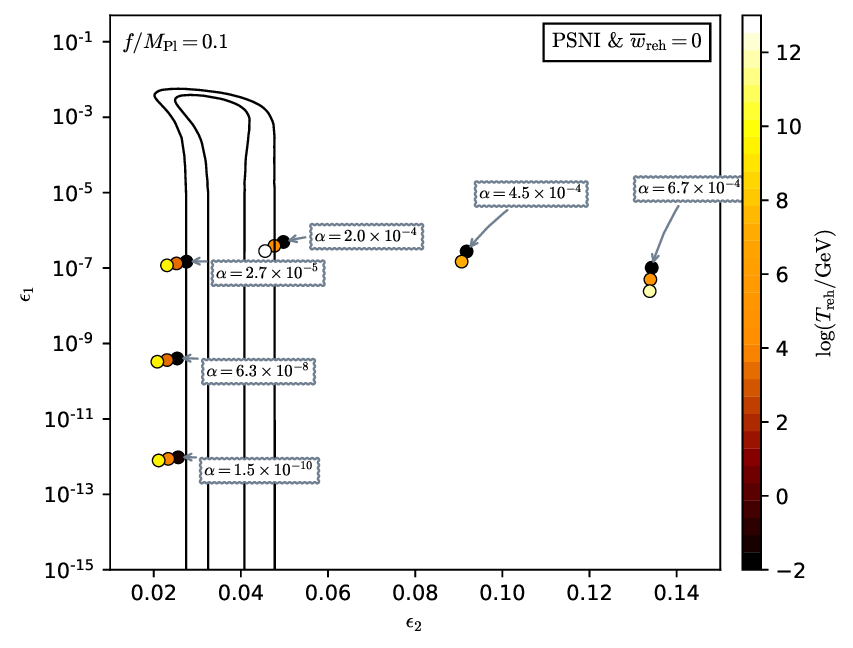}
\caption{Reheating consistent slow-roll predictions for the pseudo
  natural inflation models with sub-Planckian $f/\Mp=0.1$, in the
  plane $(\nS,r)$ (top panel) and the plane $(\epsilon_1,\epsilon_2)$
  (bottom panel). The solid contours are the one and two-sigma {\data}
  confidence intervals (marginalized over second order
  slow-roll). Compared to super-Planckian values of $f$, the amount of
  primordial gravitational waves is reduced, see
  figure~\ref{fig:CMBPSNI}.}
\label{fig:CMBPSNI_1}
\end{center}
\end{figure}

\begin{figure}[H]
\begin{center}
\includegraphics[width=\wappfig,clip=true]{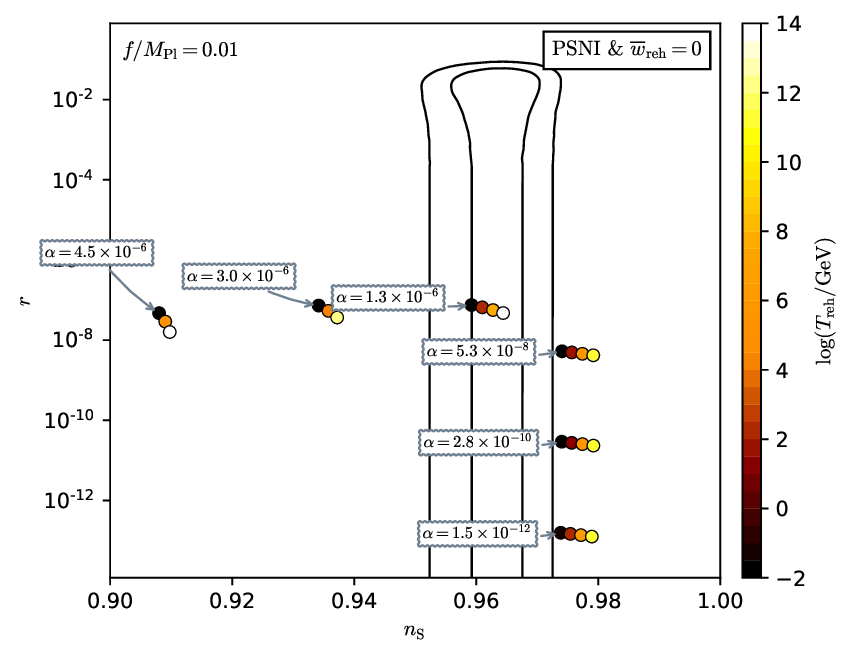}
\includegraphics[width=\wappfig,clip=true]{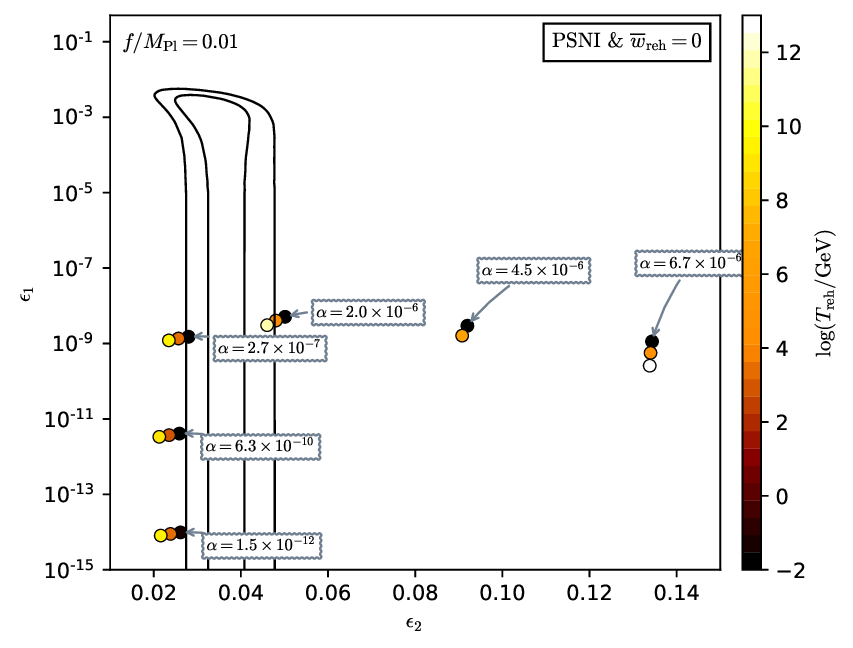}
\caption{Reheating consistent slow-roll predictions for the pseudo
  natural inflation models with a small $f/\Mp=10^{-2}$, in the plane
  $(\nS,r)$ (top panel) and the plane $(\epsilon_1,\epsilon_2)$
  (bottom panel). The solid contours are the one and two-sigma {\data}
  confidence intervals (marginalized over second order
  slow-roll). Predictions for larger values of $f$ are plotted in
  figures~\ref{fig:CMBPSNI} and \ref{fig:CMBPSNI_1}.}
\label{fig:CMBPSNI_2}
\end{center}
\end{figure}

\subsection{Non Canonical K\"ahler Inflation (\hyperref[sec:ncki]{NCKI})}

\begin{figure}[H]
\begin{center}
\includegraphics[width=\wappfig,clip=true]{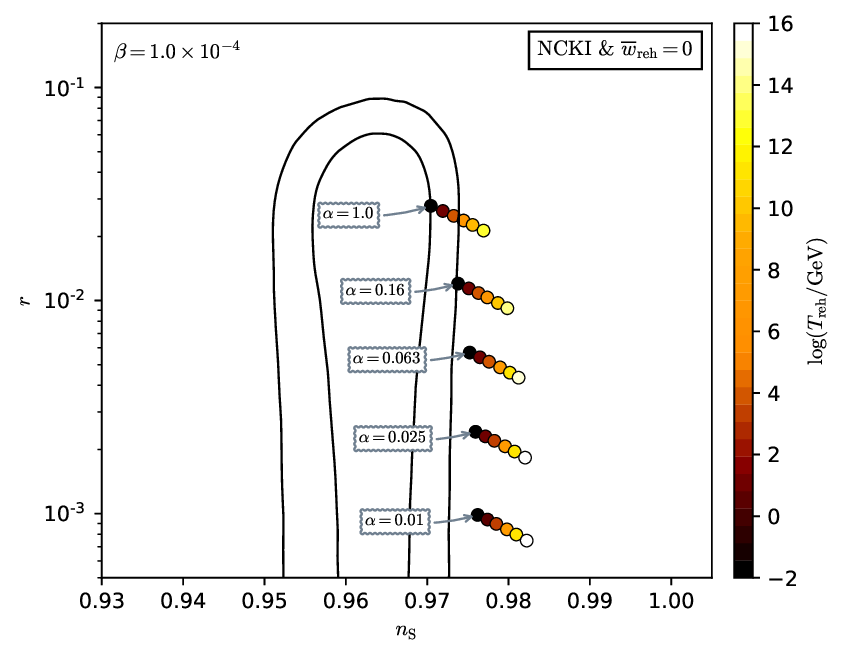}
\includegraphics[width=\wappfig,clip=true]{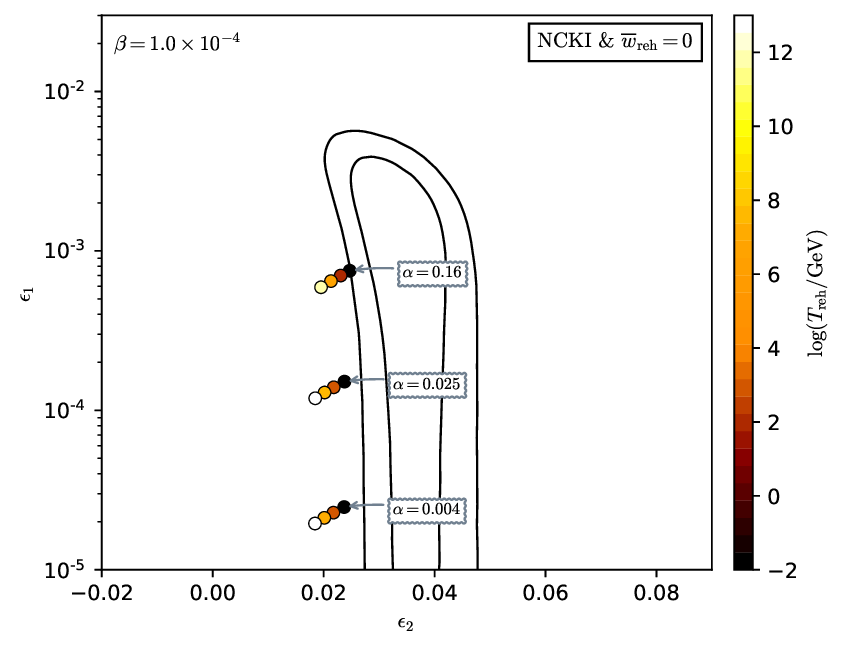}
\caption{Reheating consistent slow-roll predictions for the non
  canonical K\"ahler inflation models with positive $\beta=10^{-4}$ in
  the plane $(\nS,r)$ (top panel) and the plane
  $(\epsilon_1,\epsilon_2)$ (bottom panel). The solid contours are the
  one and two-sigma {\data} confidence intervals (marginalized over
  second order slow-roll).}
\label{fig:CMBNCKIbetaGT0}
\end{center}
\end{figure}

\begin{figure}[H]
\begin{center}
\includegraphics[width=\wappfig,clip=true]{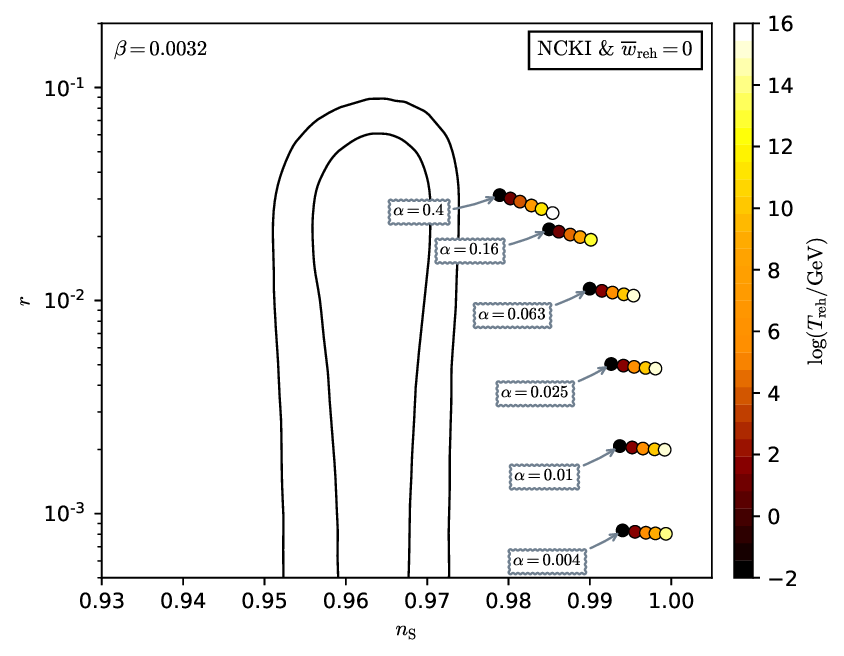}
\includegraphics[width=\wappfig,clip=true]{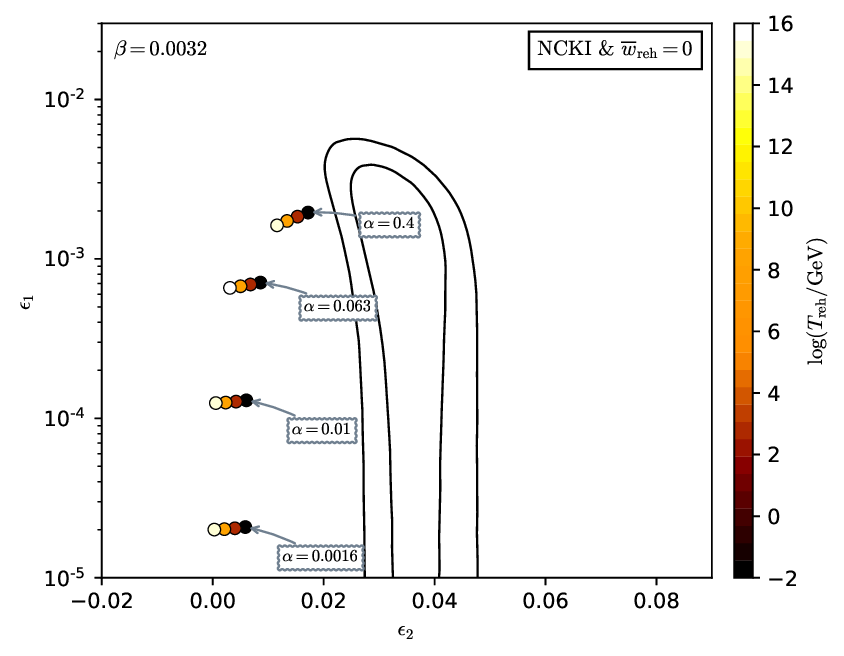}
\caption{Reheating consistent slow-roll predictions for the non
  canonical K\"ahler inflation models with positive $\beta=3.2 \times 10^{-3}$ in
  the plane $(\nS,r)$ (top panel) and the plane
  $(\epsilon_1,\epsilon_2)$ (bottom panel). The solid contours are the
  one and two-sigma {\data} confidence intervals (marginalized over
  second order slow-roll).}
\label{fig:CMBNCKIbetaGT0_1}
\end{center}
\end{figure}

\begin{figure}[H]
\begin{center}
\includegraphics[width=\wappfig,clip=true]{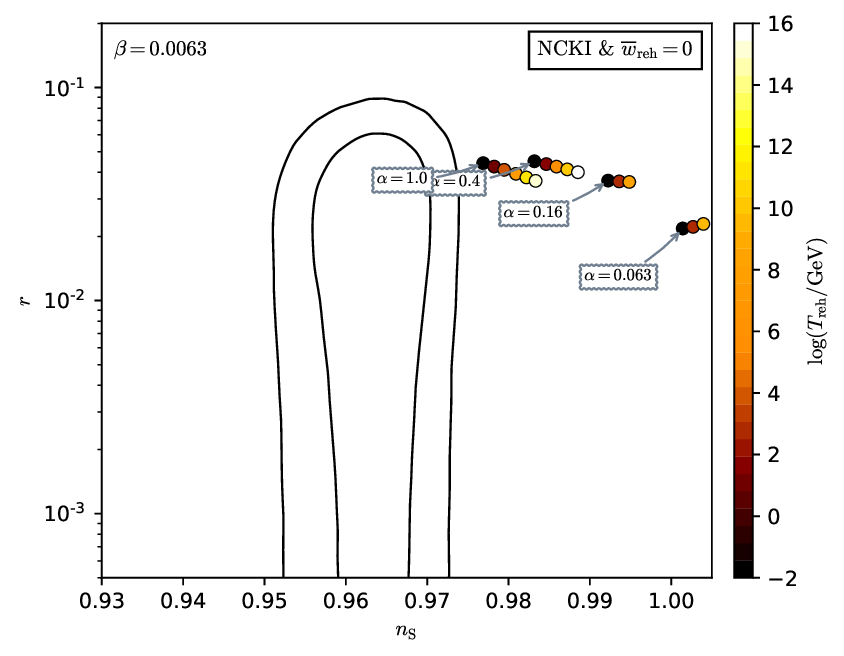}
\includegraphics[width=\wappfig,clip=true]{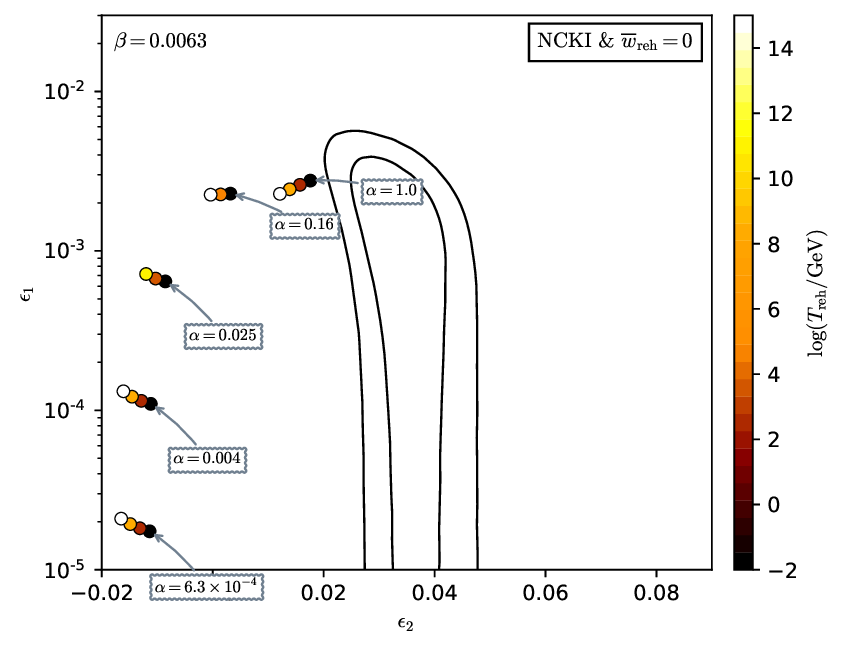}
\caption{Reheating consistent slow-roll predictions for the non
  canonical K\"ahler inflation models with positive $\beta=6.3 \times
  10^{-3}$ in the plane $(\nS,r)$ (top panel) and the plane
  $(\epsilon_1,\epsilon_2)$ (bottom panel). The solid contours are the
  one and two-sigma {\data} confidence intervals (marginalized over
  second order slow-roll).}
\label{fig:CMBNCKIbetaGT0_2}
\end{center}
\end{figure}

\begin{figure}[H]
\begin{center}
\includegraphics[width=\wappfig,clip=true]{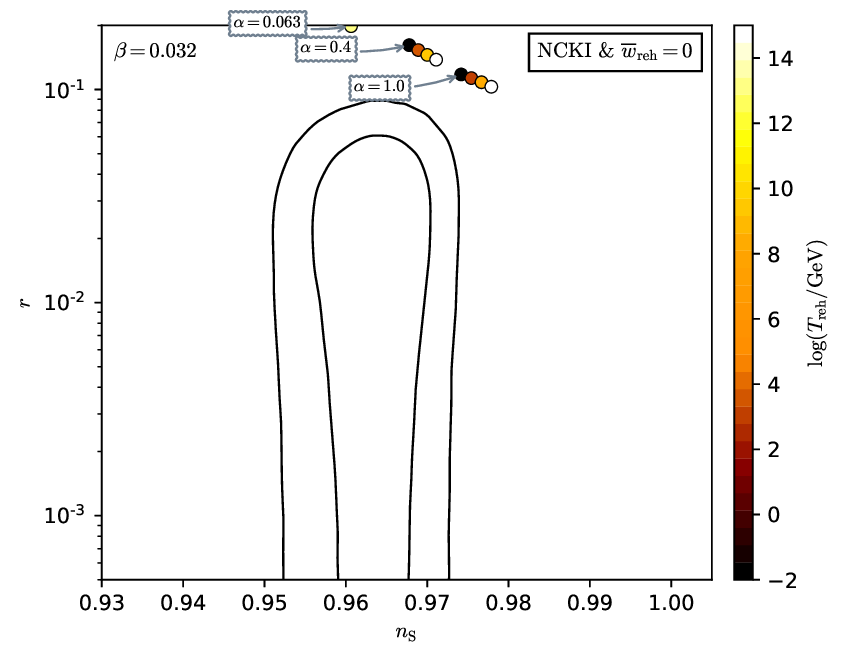}
\includegraphics[width=\wappfig,clip=true]{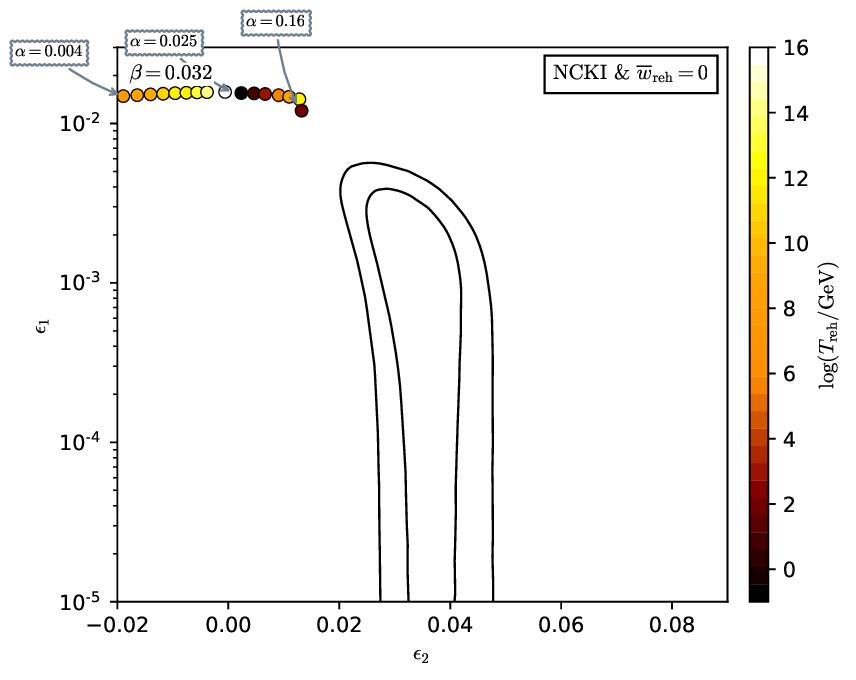}
\caption{Reheating consistent slow-roll predictions for the non
  canonical K\"ahler inflation models with positive $\beta=3.2\times 10^{-2}$
  in the plane $(\nS,r)$ (top panel) and the plane
  $(\epsilon_1,\epsilon_2)$ (bottom panel). The solid contours are the
  one and two-sigma {\data} confidence intervals (marginalized over
  second order slow-roll). For larger values of $\beta\gtrsim 1$, the
  predictions are almost identical to those displayed here. See
  figures~\ref{fig:CMBNCKIbetaGT0} to \ref{fig:CMBNCKIbetaGT0_2} for
  smaller values of $\beta$.}
\label{fig:CMBNCKIbetaGT0_3}
\end{center}
\end{figure}

\begin{figure}[H]
\begin{center}
\includegraphics[width=\wappfig,clip=true]{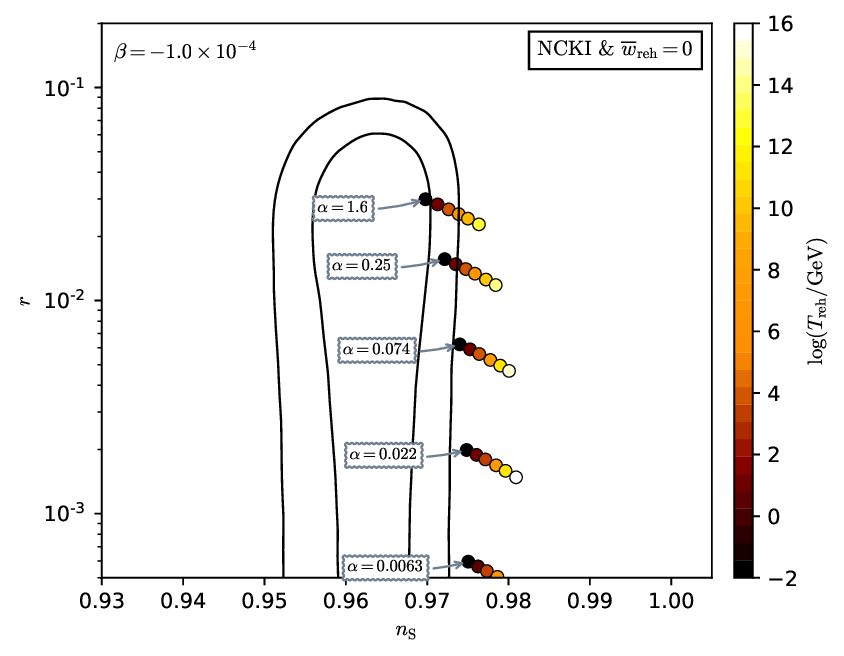}
\includegraphics[width=\wappfig,clip=true]{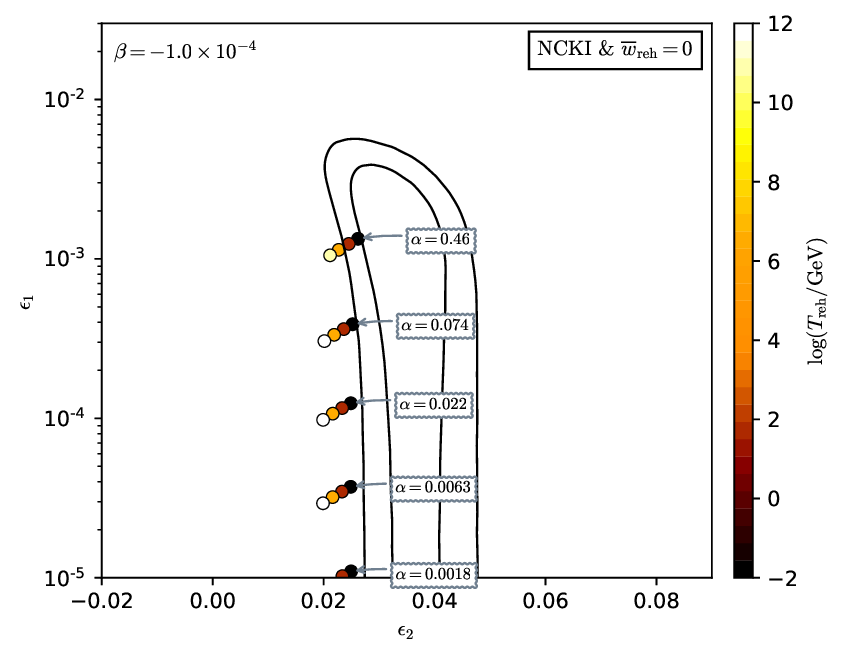}
\caption{Reheating consistent slow-roll predictions for the non
  canonical K\"ahler inflation models with negative $\beta=-10^{-4}$
  in the plane $(\nS,r)$ (top panel) and the plane
  $(\epsilon_1,\epsilon_2)$ (bottom panel). The solid contours are the
  one and two-sigma {\data} confidence intervals (marginalized over
  second order slow-roll). }
\label{fig:CMBNCKIbetaLT0}
\end{center}
\end{figure}

\begin{figure}[H]
\begin{center}
\includegraphics[width=\wappfig,clip=true]{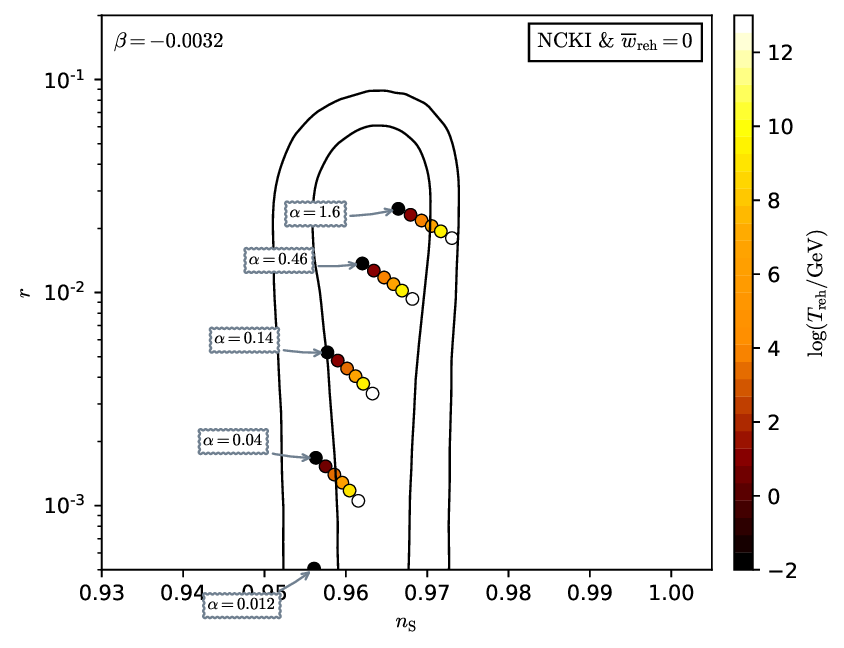}
\includegraphics[width=\wappfig,clip=true]{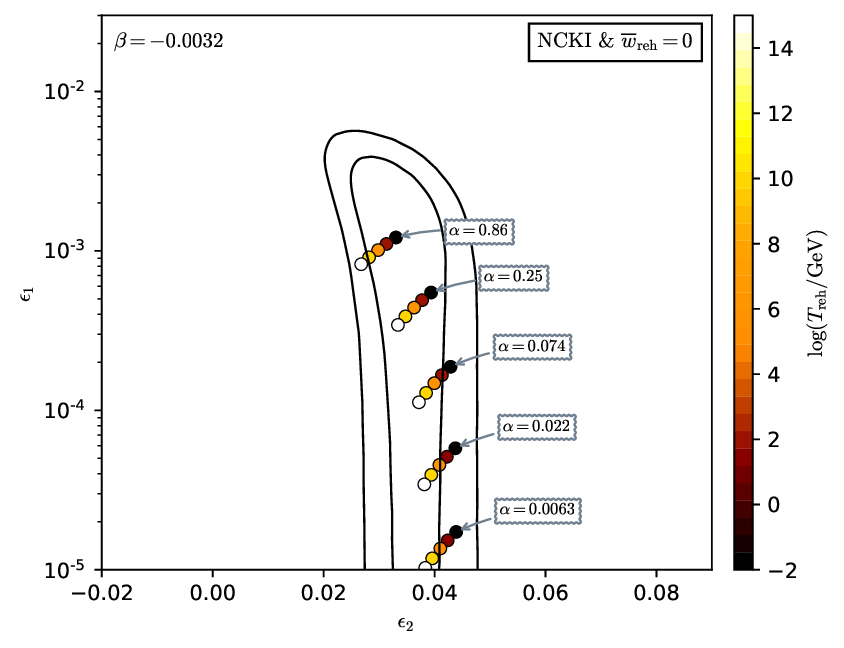}
\caption{Reheating consistent slow-roll predictions for the non
  canonical K\"ahler inflation models with negative $\beta=-3.2 \times
  10^{-3}$ in the plane $(\nS,r)$ (top panel) and the plane
  $(\epsilon_1,\epsilon_2)$ (bottom panel). The solid contours are the
  one and two-sigma {\data} confidence intervals (marginalized over
  second order slow-roll). }
\label{fig:CMBNCKIbetaLT0_5}
\end{center}
\end{figure}

\begin{figure}[H]
\begin{center}
\includegraphics[width=\wappfig,clip=true]{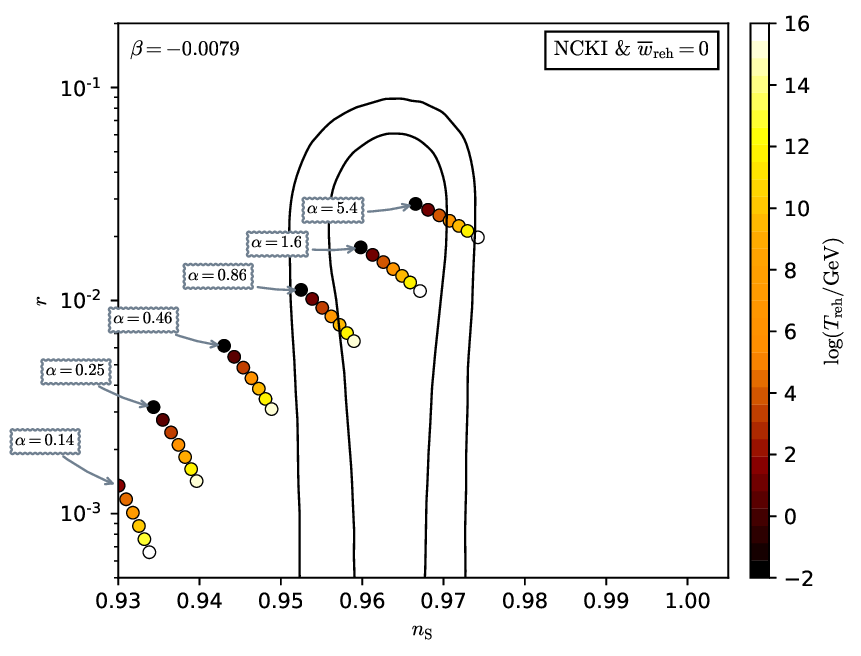}
\includegraphics[width=\wappfig,clip=true]{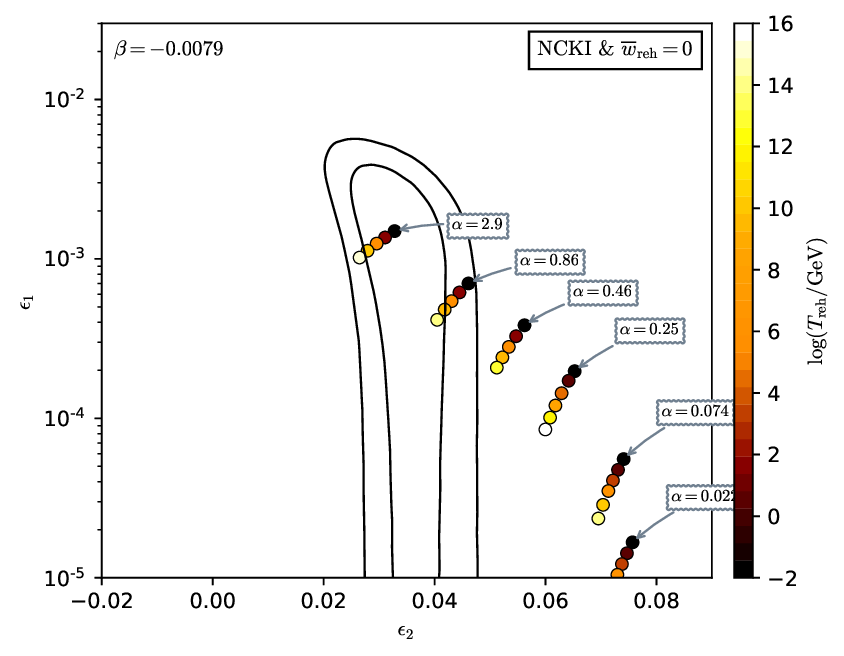}
\caption{Reheating consistent slow-roll predictions for the non
  canonical K\"ahler inflation models with negative $\beta=-7.9 \times
  10^{-3}$ in the plane $(\nS,r)$ (top panel) and the plane
  $(\epsilon_1,\epsilon_2)$ (bottom panel). The solid contours are the
  one and two-sigma {\data} confidence intervals (marginalized over
  second order slow-roll). }
\label{fig:CMBNCKIbetaLT0_6}
\end{center}
\end{figure}

\begin{figure}[H]
\begin{center}
\includegraphics[width=\wappfig,clip=true]{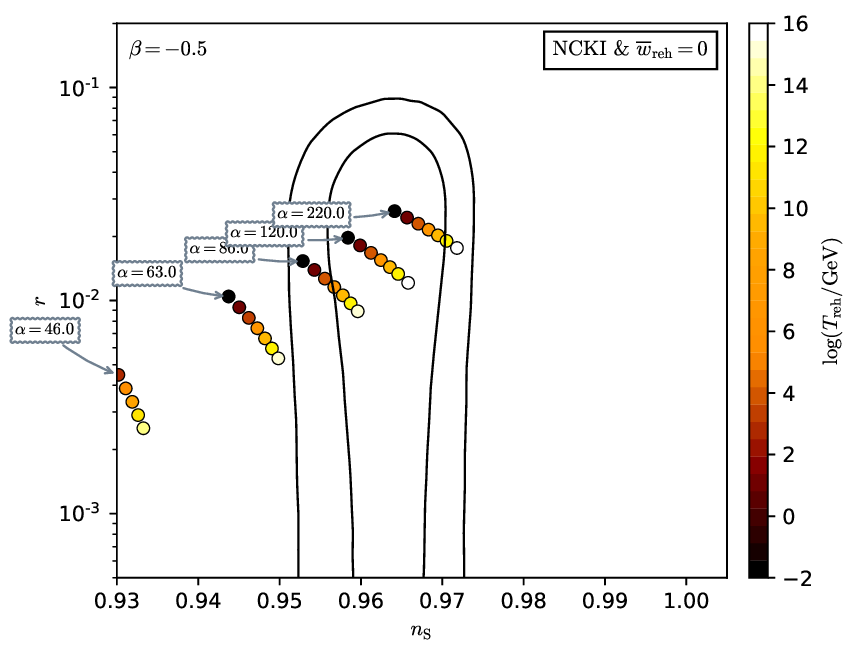}
\includegraphics[width=\wappfig,clip=true]{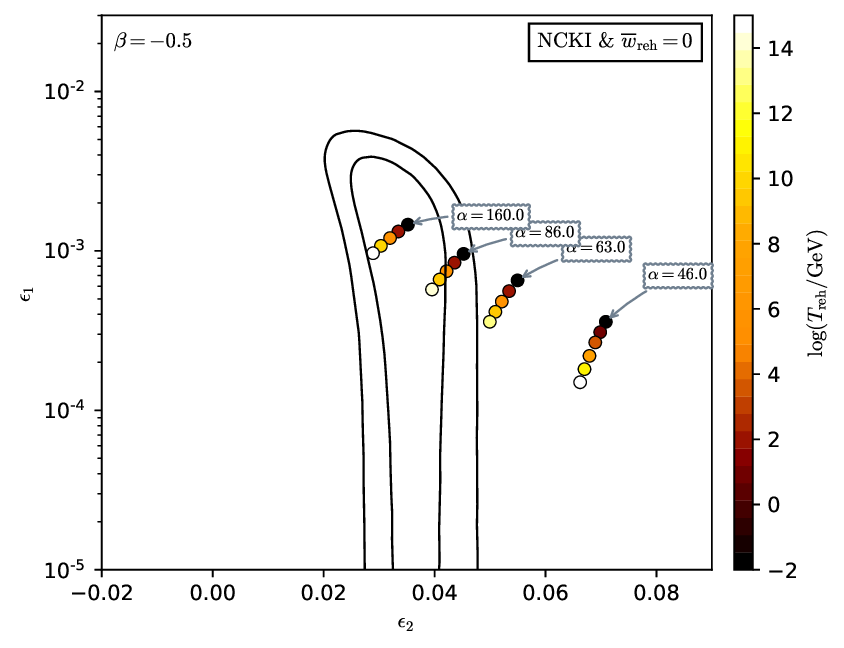}
\caption{Reheating consistent slow-roll predictions for the non
  canonical K\"ahler inflation models with negative $\beta=-0.5$
  in the plane $(\nS,r)$ (top panel) and the plane
  $(\epsilon_1,\epsilon_2)$ (bottom panel). The solid contours are the
  one and two-sigma {\data} confidence intervals (marginalized over
  second order slow-roll). For smaller values of $\beta \lesssim -1$, the
  predictions are almost identical to those displayed here. See
  figures~\ref{fig:CMBNCKIbetaLT0} to \ref{fig:CMBNCKIbetaLT0_6} for
  smaller values of $|\beta|$. }
\label{fig:CMBNCKIbetaLT0_7}
\end{center}
\end{figure}

\subsection{Constant Spectrum Inflation (\hyperref[sec:csi]{CSI})}

\begin{figure}[H]
\begin{center}
\includegraphics[width=\wappfig,clip=true]{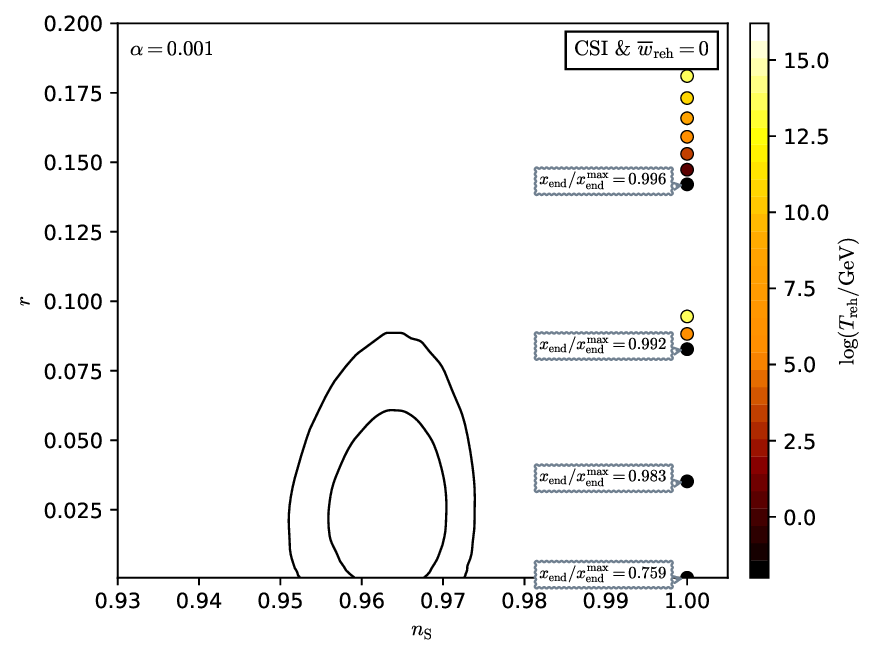}
\includegraphics[width=\wappfig,clip=true]{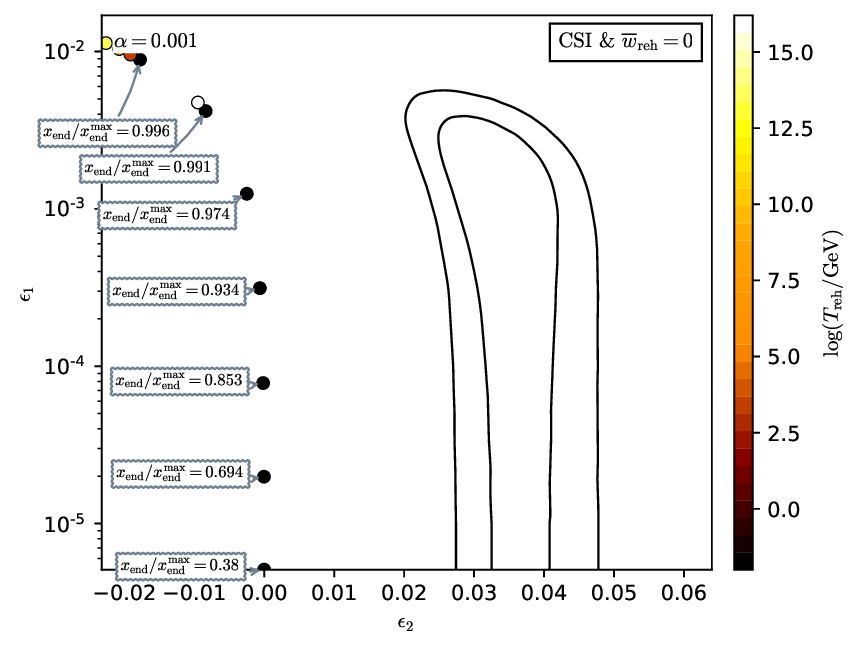}
\caption{Reheating consistent slow-roll predictions for the Constant
  Spectrum models in the plane $(\nS,r)$ (top panel) and the plane
  $(\epsilon_1,\epsilon_2)$ (bottom panel), for $\alpha=10^{-3}$. The
  solid contours are the one and two-sigma {\data} confidence
  intervals (marginalized over second order slow-roll). Model
  predictions verify $\nS=1$.}
\label{fig:CMBCSIalphaEQ10PowerMinus3}
\end{center}
\end{figure}

\begin{figure}[H]
\begin{center}
\includegraphics[width=\wappfig,clip=true]{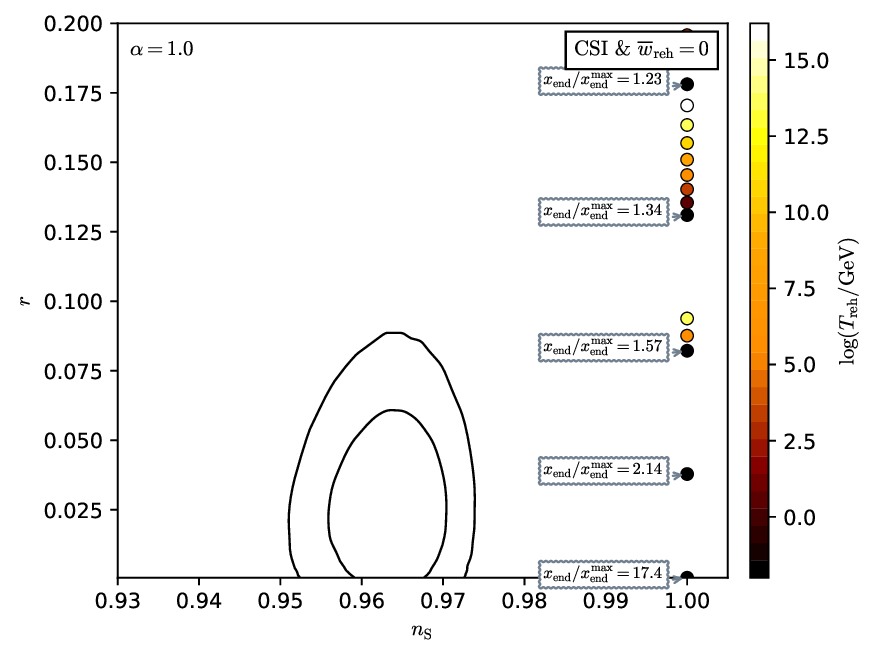}
\includegraphics[width=\wappfig,clip=true]{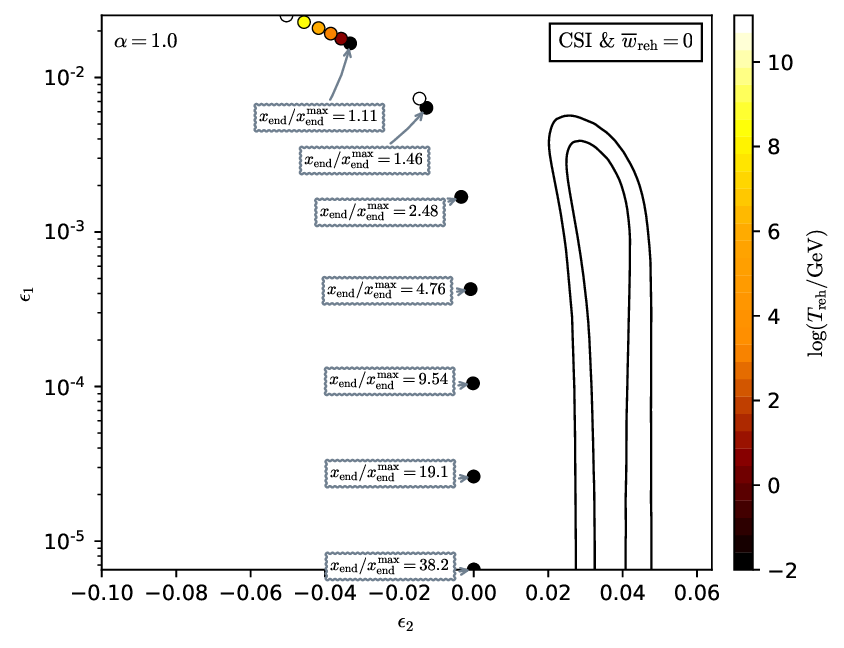}
\caption{Reheating consistent slow-roll predictions for the Constant
  Spectrum models in the plane $(\nS,r)$ (top panel) and the plane
  $(\epsilon_1,\epsilon_2)$ (bottom panel), for $\alpha=1$. The two
  solid contours are the one and two-sigma {\data} confidence
  intervals (marginalized over second order slow-roll). Independently
  of $\alpha$, one still has the predictions lying along $\nS=1$, see
  also figure~\ref{fig:CMBCSIalphaEQ10PowerMinus3}.}
\label{fig:CMBCSIalphaEQ1}
\end{center}
\end{figure}

\subsection{Orientifold Inflation (\hyperref[sec:oi]{OI})}

\begin{figure}[H]
\begin{center}
\includegraphics[width=\wappfig,clip=true]{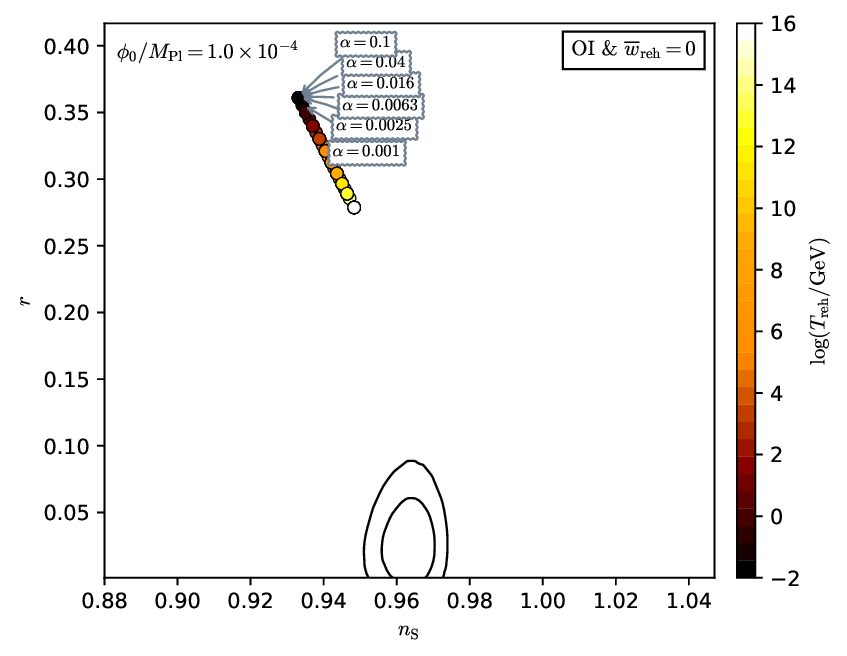}
\includegraphics[width=\wappfig,clip=true]{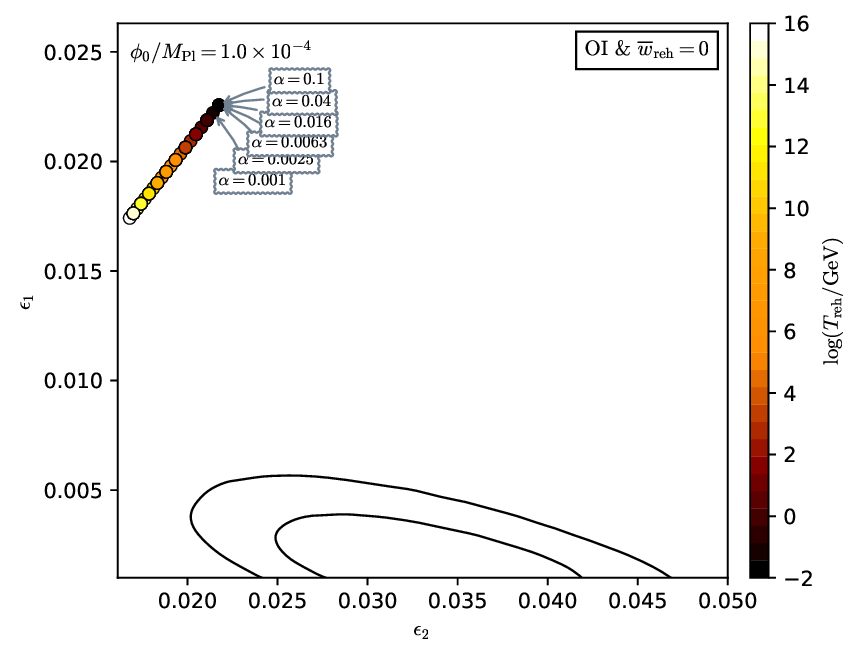}
\caption{Reheating consistent slow-roll predictions for the
  orientifold inflation models for $\phizero/\Mp=10^{-4}$ in the plane
  $(\nS,r)$ (top panel) and the plane $(\epsilon_1,\epsilon_2)$
  (bottom panel). The solid contours are the one and two-sigma {\data}
  confidence intervals (marginalized over second order slow-roll).
  The predictions of the model almost do not depend on its parameters,
  they are all superimposed and one cannot distinguish the different
  values of $\alpha$, and of $\phizero$ (see
  figure~\ref{fig:CMBOI_1}).}
\label{fig:CMBOI}
\end{center}
\end{figure}

\begin{figure}[H]
\begin{center}
\includegraphics[width=\wappfig,clip=true]{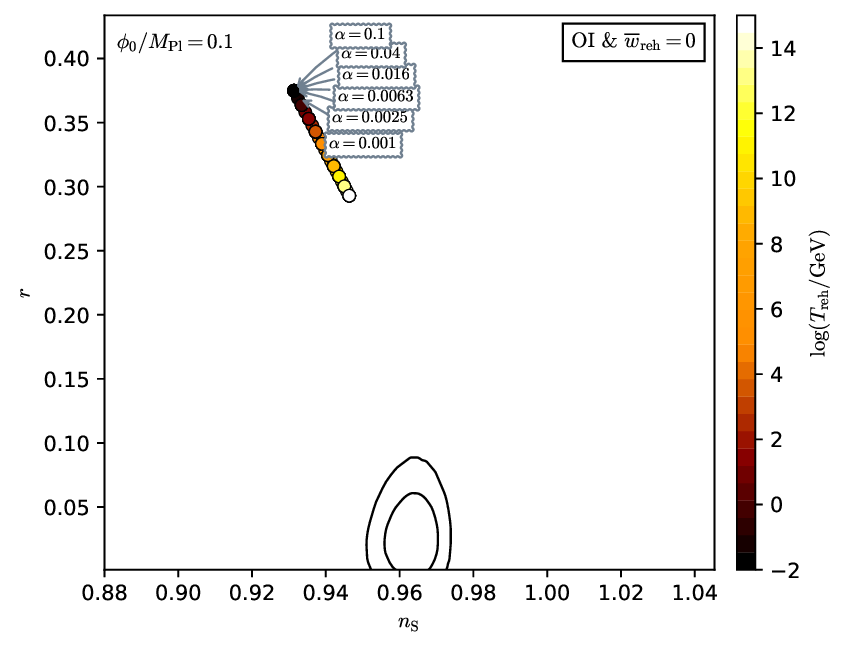}
\includegraphics[width=\wappfig,clip=true]{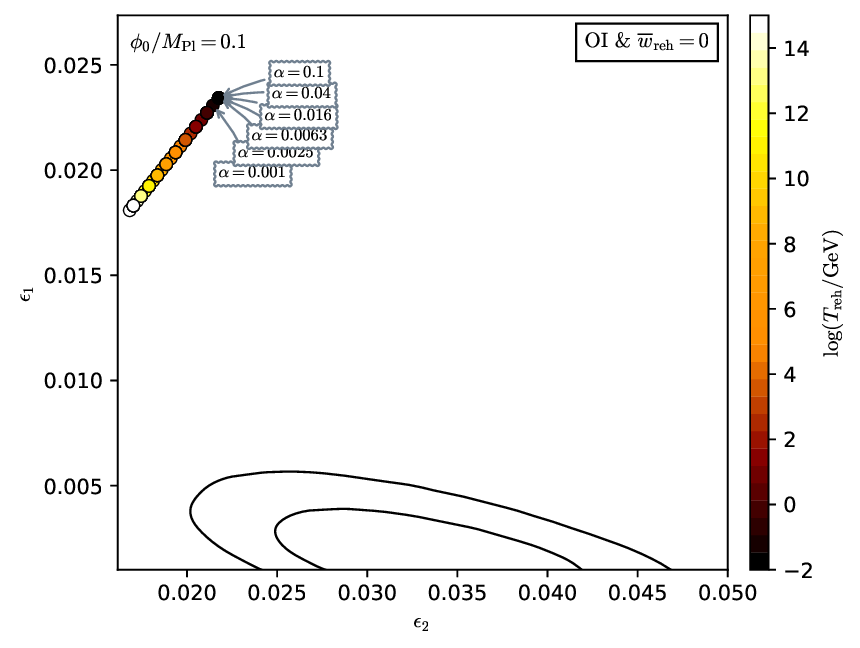}
\caption{Reheating consistent slow-roll predictions for the
  orientifold inflation models for $\phizero/\Mp=10^{-1}$ in the plane
  $(\nS,r)$ (top panel) and the plane $(\epsilon_1,\epsilon_2)$
  (bottom panel). The solid contours are the one and two-sigma {\data}
  confidence intervals (marginalized over second order slow-roll).
  The prediction of the model is the same as for $\phizero/\Mp=10^{-4}$
  (se figure~\ref{fig:CMBOI}) and almost do not depend on $\alpha$.}
\label{fig:CMBOI_1}
\end{center}
\end{figure}

\subsection{Constant \texorpdfstring{$\nS$}{nS} C Inflation (\hyperref[sec:cnci]{CNCI})}

\begin{figure}[H]
\begin{center}
\includegraphics[width=\wappfig,clip=true]{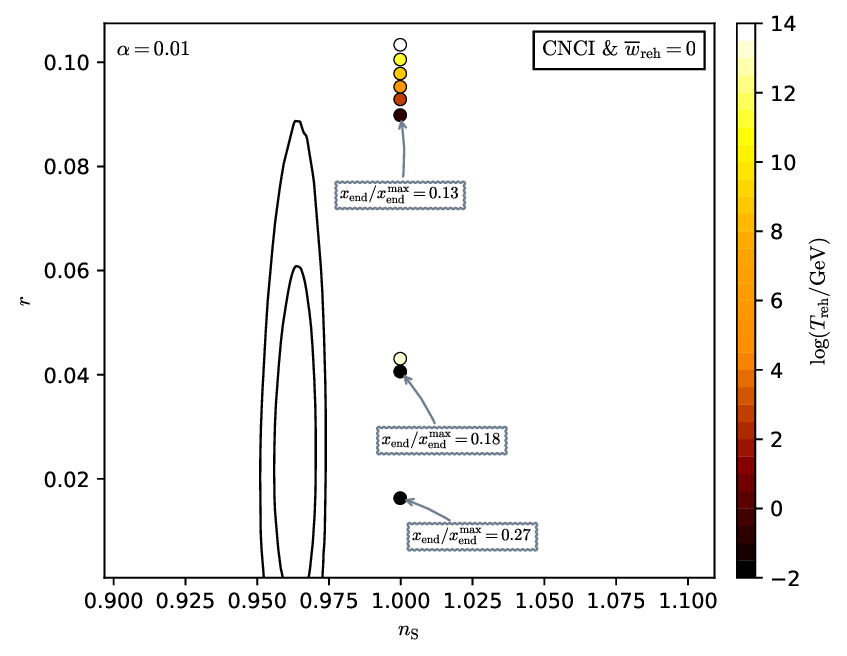}
\includegraphics[width=\wappfig,clip=true]{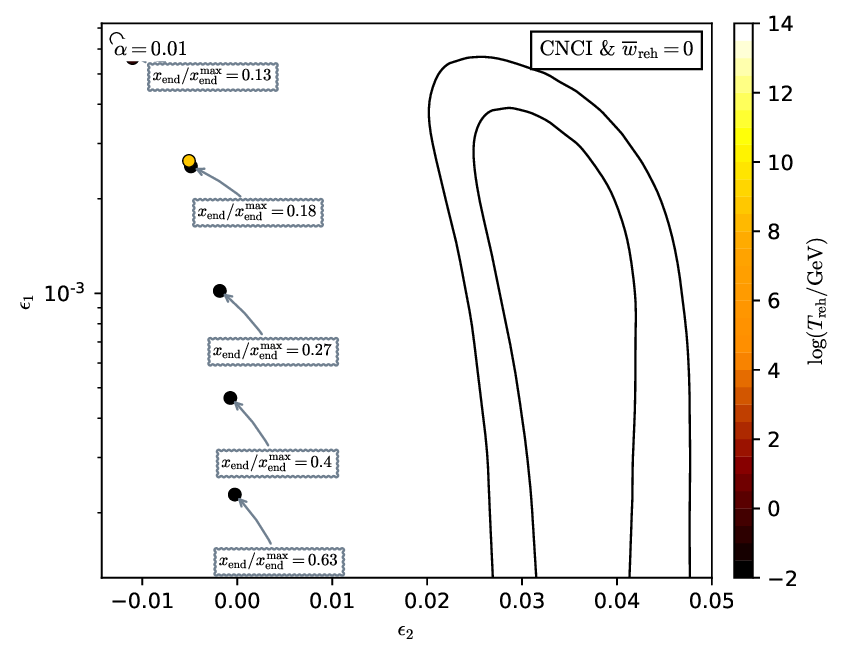}
\caption{Reheating consistent slow-roll predictions for the constant
  $\nS$ C inflation models for $\alpha=10^{-3}$ in the plane $(\nS,r)$
  (top panel) and the plane $(\epsilon_1,\epsilon_2)$ (bottom
  panel). The solid contours are the one and two-sigma {\data}
  confidence intervals (marginalized over second order slow-roll). For
  all values of $\xend$, the predictions are very cllose to the constant value
  $\nS=1$.}
\label{fig:CMBCNCI}
\end{center}
\end{figure}

\begin{figure}[H]
\begin{center}
\includegraphics[width=\wappfig,clip=true]{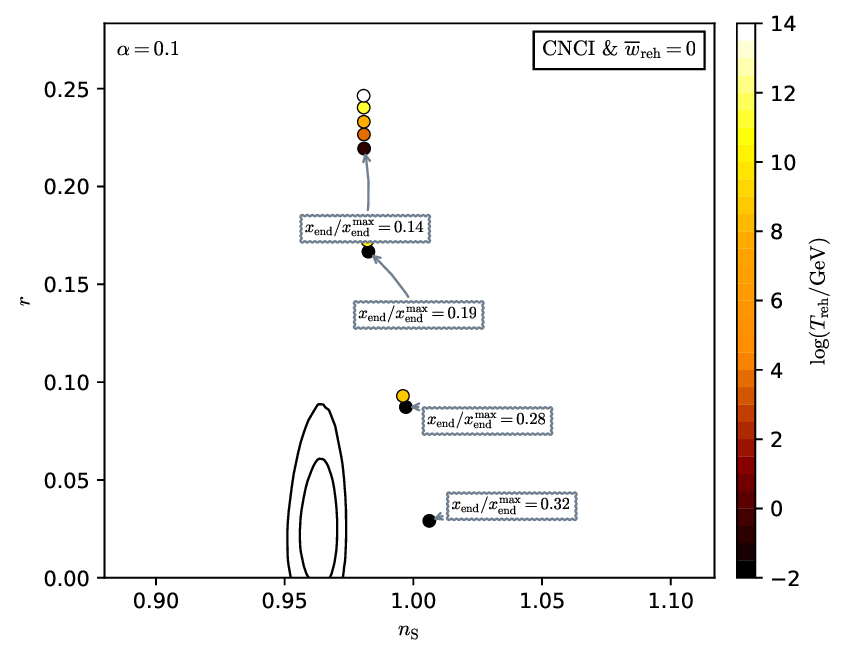}
\includegraphics[width=\wappfig,clip=true]{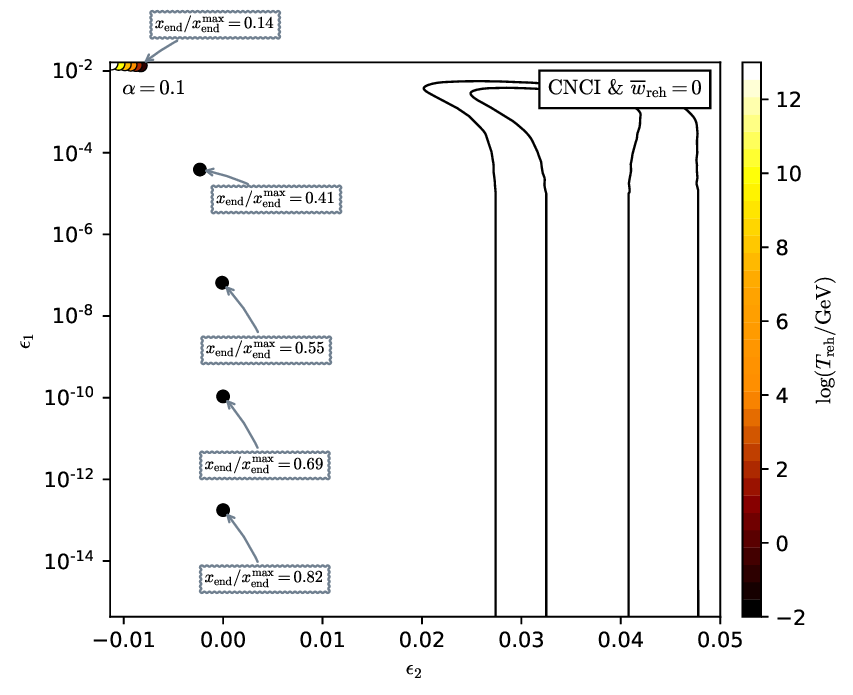}
\caption{Reheating consistent slow-roll predictions for the constant
  $\nS$ C inflation models for $\alpha=0.1$ in the plane $(\nS,r)$
  (top panel) and the plane $(\epsilon_1,\epsilon_2)$ (bottom
  panel). The solid contours are the one and two-sigma {\data}
  confidence intervals (marginalized over second order
  slow-roll). Compared the smaller values of $\alpha$ (see
  figure~\ref{fig:CMBCNCI}), at intermediate values of $\xend$, the
  model predictions deviate from $\nS=1$ while they approach $\nS=1 -
  2 \alpha^2$ for small values of $\xend$.}
\label{fig:CMBCNCI_1}
\end{center}
\end{figure}

\begin{figure}[H]
\begin{center}
\includegraphics[width=\wappfig,clip=true]{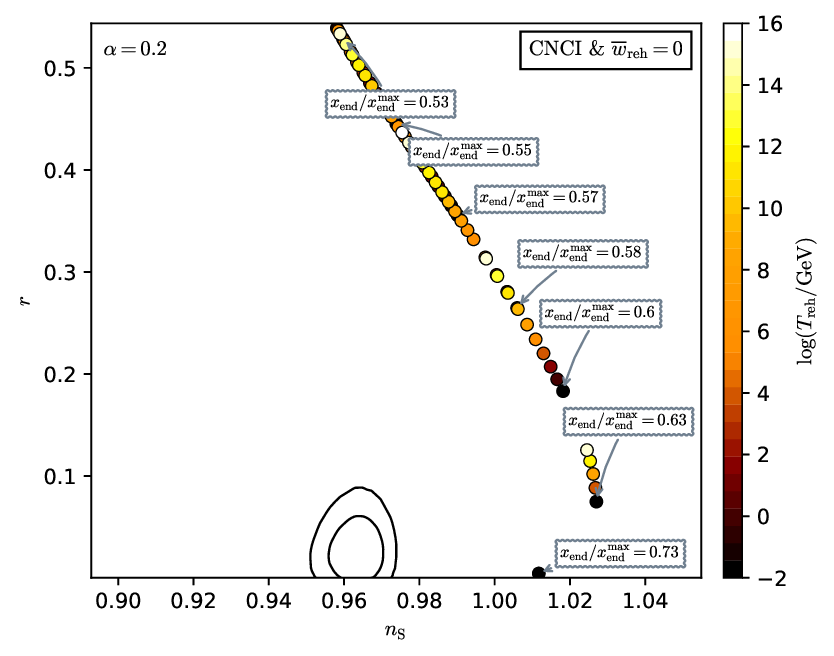}
\includegraphics[width=\wappfig,clip=true]{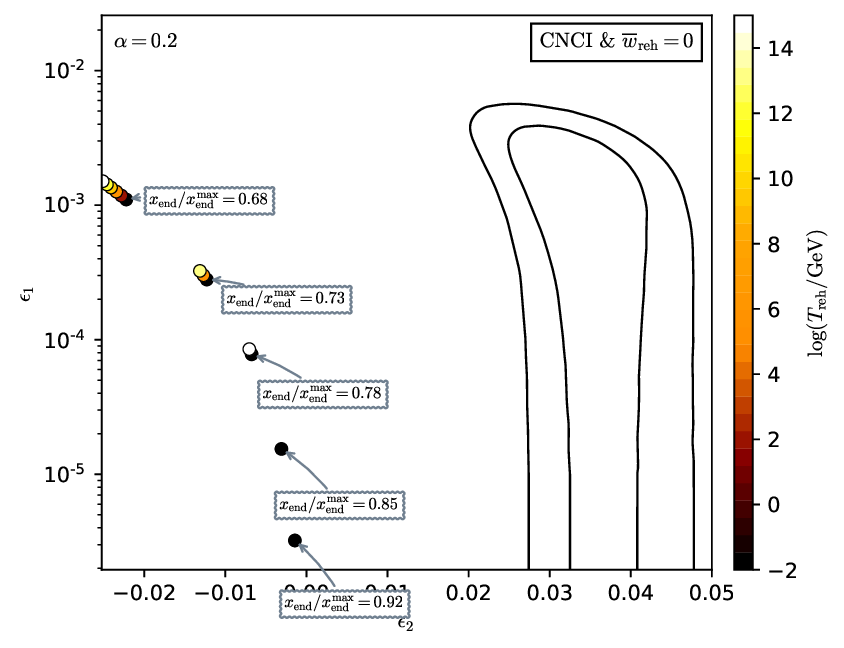}
\caption{Reheating consistent slow-roll predictions for the constant
  $\nS$ C inflation models for $\alpha=0.2$ in the plane $(\nS,r)$
  (top panel) and the plane $(\epsilon_1,\epsilon_2)$ (bottom
  panel). The solid contours are the one and two-sigma {\data}
  confidence intervals (marginalized over second order
  slow-roll). Compared to smaller values of $\alpha$ (see
  figures~\ref{fig:CMBCNCI} and \ref{fig:CMBCNCI_1}), for most of the
  $\xend$ values, the model predictions are not with a constant
  spectral index. Only for very large $\xend \to \xendmax$ they are
  along $\nS=1$ while the limit $\nS=1-2\alpha^2$ is reached for
  $\xend \ll \xendmax$, with, however, an unreasonable amount of
  primordial gravitational waves.}
\label{fig:CMBCNCI_2}
\end{center}
\end{figure}

\subsection{Supergravity Brane Inflation (\hyperref[sec:sbi]{SBI})}

\begin{figure}[H]
\begin{center}
\includegraphics[width=\wappfig,clip=true]{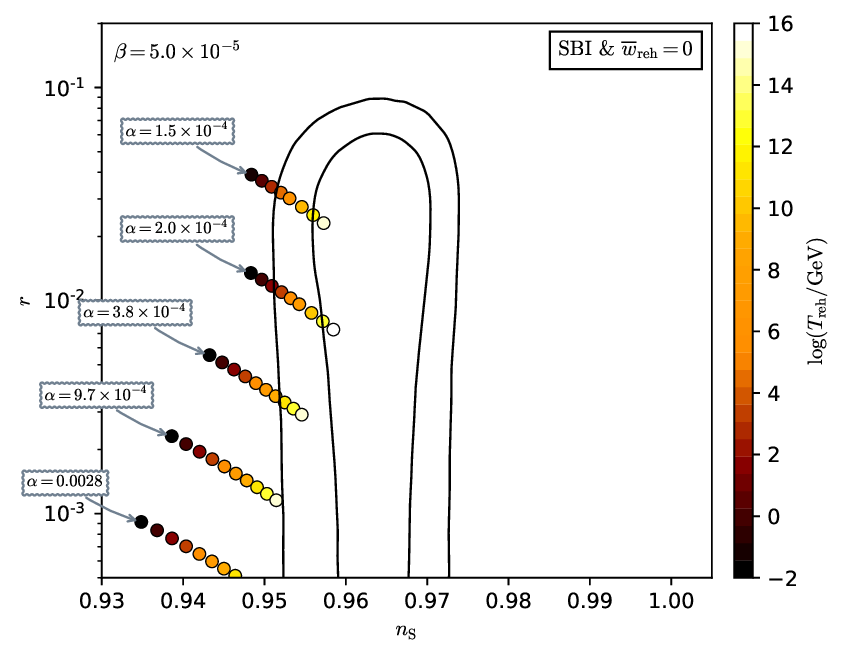}
\includegraphics[width=\wappfig,clip=true]{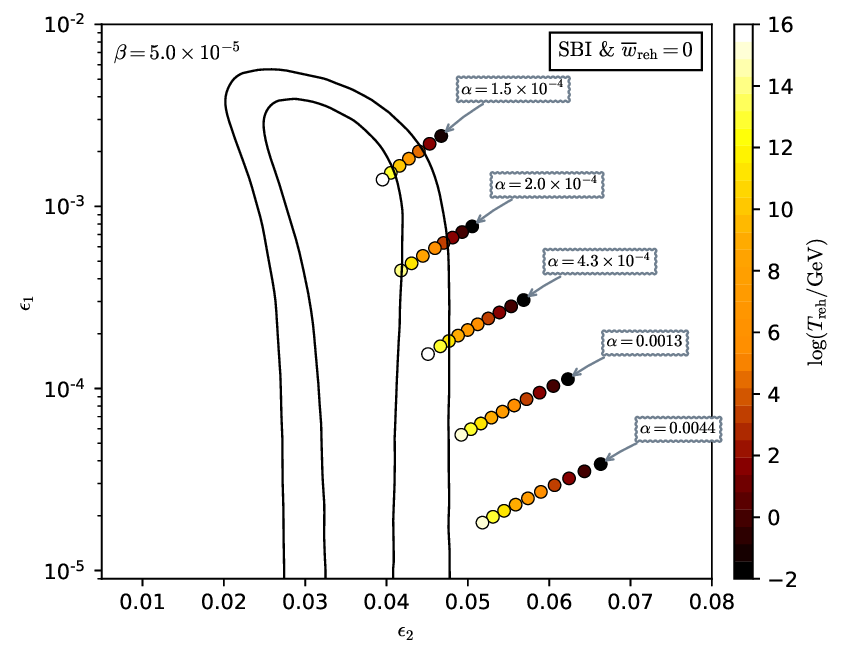}
\caption{Reheating consistent slow-roll predictions for the
  supergravity brane inflation models for $\beta=5\times 10^{-5}$ in
  the plane $(\nS,r)$ (top panel) and the plane
  $(\epsilon_1,\epsilon_2)$ (bottom panel). The solid contours are the
  one and two-sigma {\data} confidence intervals (marginalized over
  second order slow-roll).}
\label{fig:CMBSBIbetaEQ5x10PowerMinus5}
\end{center}
\end{figure}

\begin{figure}[H]
\begin{center}
\includegraphics[width=\wappfig,clip=true]{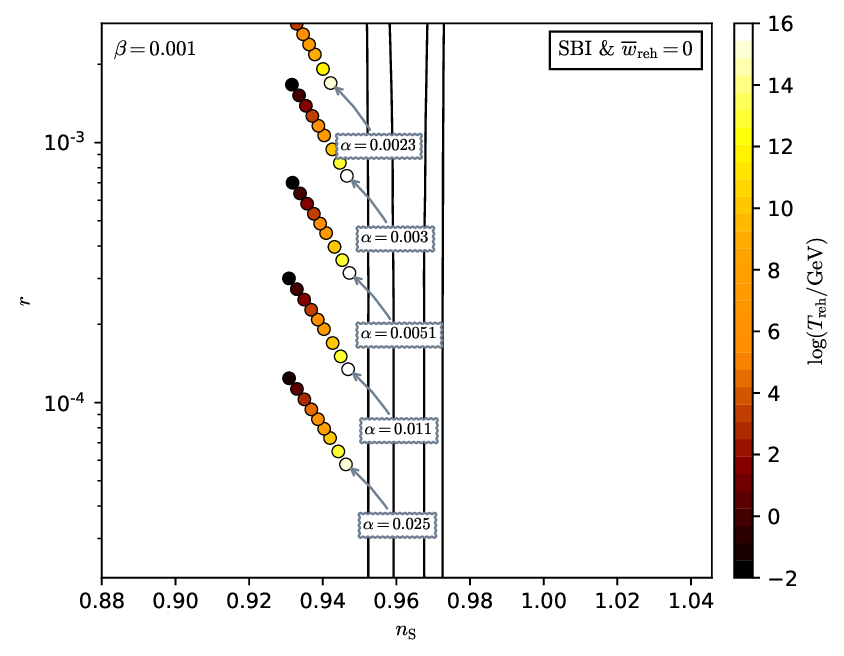}
\includegraphics[width=\wappfig,clip=true]{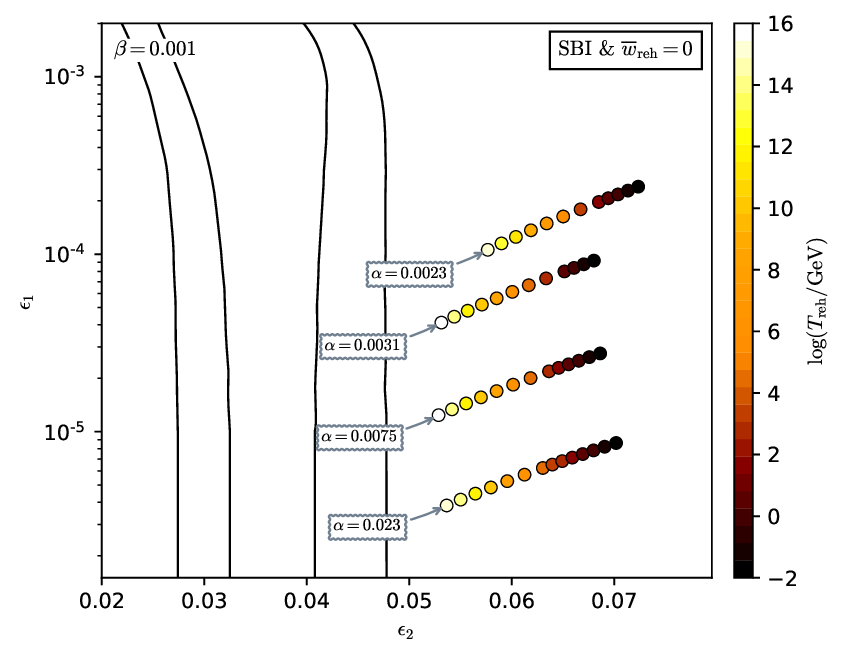}
\caption{Reheating consistent slow-roll predictions for the
  supergravity brane inflation models for $\beta=10^{-3}$ in the plane
  $(\nS,r)$ (top panel) and the plane $(\epsilon_1,\epsilon_2)$
  (bottom panel). The solid contours are the one and two-sigma {\data}
  confidence intervals (marginalized over second order slow-roll).}
\label{fig:CMBSBIbetaEQ10PowerMinus3}
\end{center}
\end{figure}

\begin{figure}[H]
\begin{center}
\includegraphics[width=\wappfig,clip=true]{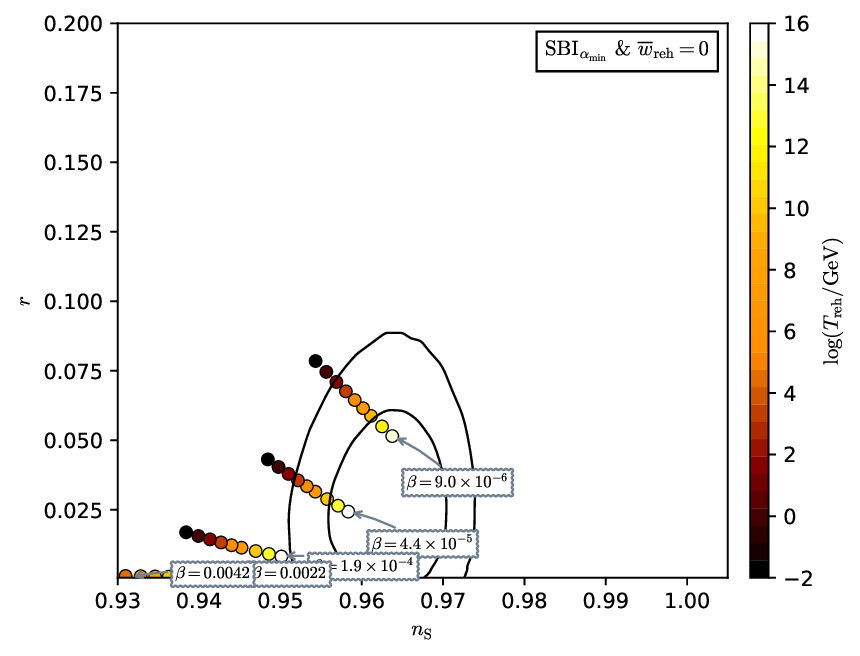}
\includegraphics[width=\wappfig,clip=true]{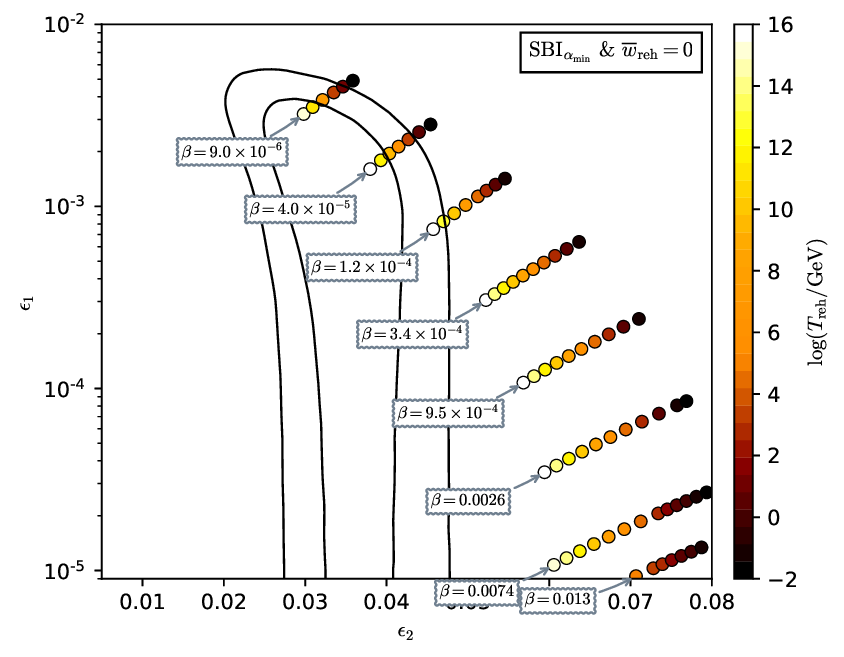}
\caption{Reheating consistent slow-roll predictions for the
  supergravity brane inflation models for $\alpha=\alphamin(\beta)$ in
  the plane $(\nS,r)$ (top panel) and the plane
  $(\epsilon_1,\epsilon_2)$ (bottom panel). The solid contours are the
  one and two-sigma {\data} confidence intervals (marginalized over
  second order slow-roll).}
\label{fig:CMBSBIalphamin}
\end{center}
\end{figure}

\subsection{Spontaneous Symmetry Breaking Inflation 1 (\hyperref[sec:ssbi]{SSBI1})}

\begin{figure}[H]
\begin{center}
\includegraphics[width=\wappfig,clip=true]{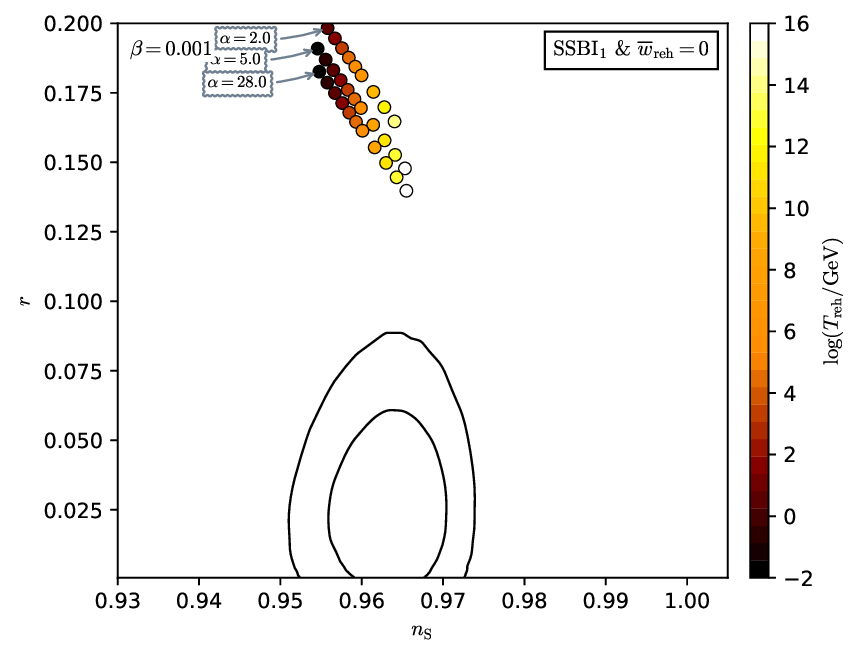}
\includegraphics[width=\wappfig,clip=true]{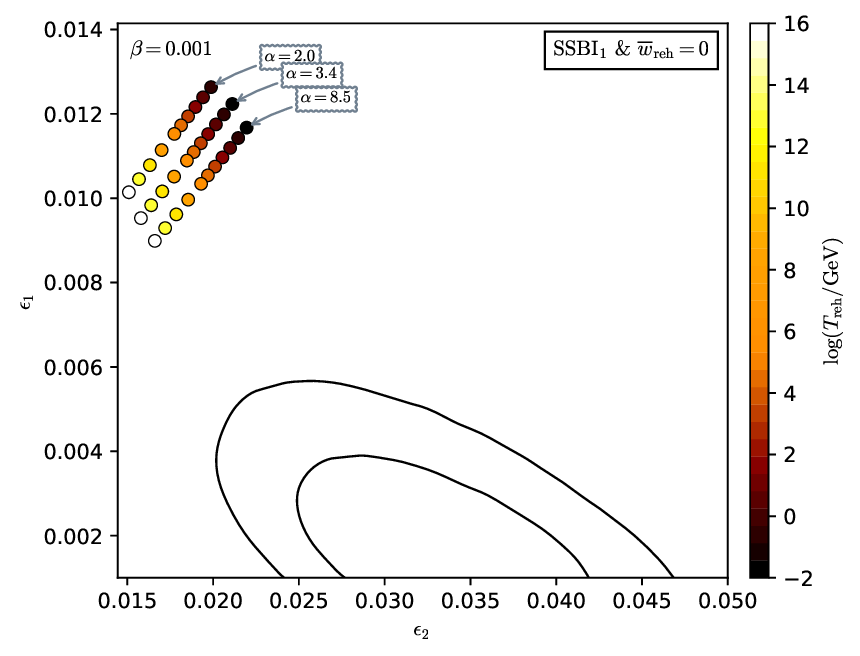}
\caption{Reheating consistent slow-roll predictions for the
  spontaneous symmetry breaking 1 inflation ($\alpha>0,\beta>0$)
  models with $\beta =10^{-3}$, in the plane $(\nS,r)$ (top panel) and
  the plane $(\epsilon_1,\epsilon_2)$ (bottom panel). The solid
  contours are the one and two-sigma {\data} confidence intervals
  (marginalized over second order slow-roll). The parameter $\alpha$
  is varied between $\alpha_{\min}\left(\beta\right)<\alpha<10^6
  \alpha_{\min}\left(\beta\right)$.}
\label{fig:CMBSSBI1betaEQ10PowerMinus3}
\end{center}
\end{figure}

\begin{figure}[H]
\begin{center}
\includegraphics[width=\wappfig,clip=true]{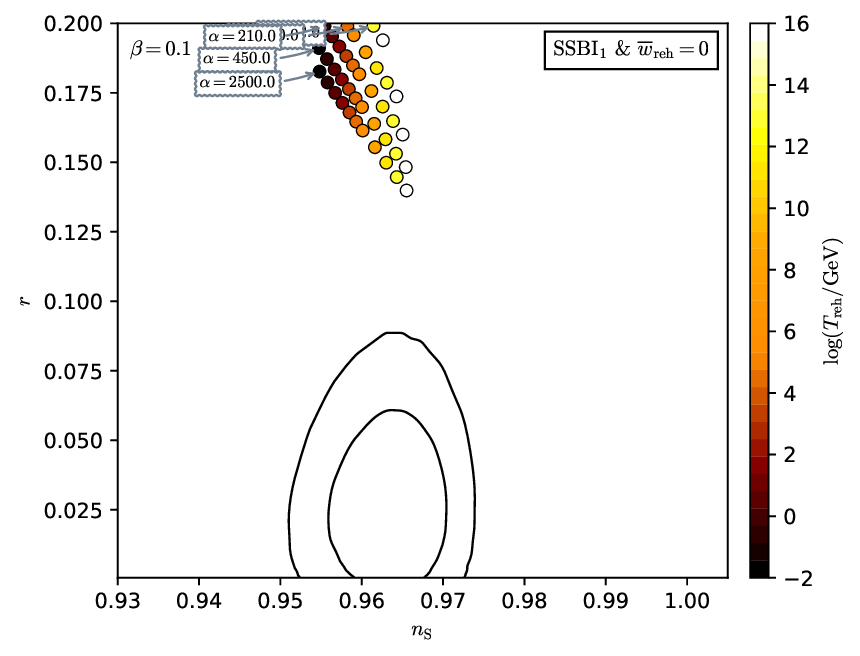}
\includegraphics[width=\wappfig,clip=true]{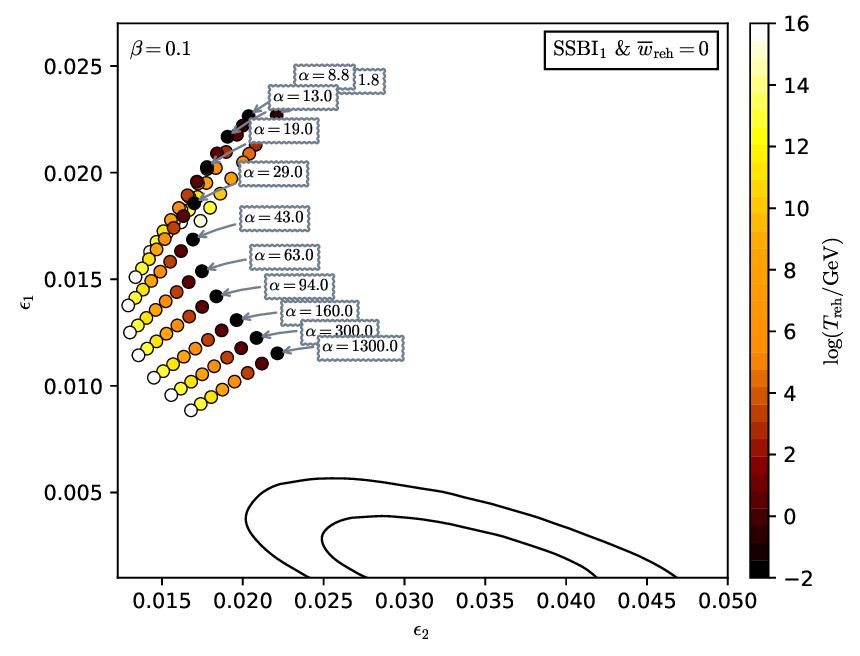}
\caption{Reheating consistent slow-roll predictions for the
  spontaneous symmetry breaking 1 inflation ($\alpha>0,\beta>0$)
  models with $\beta=10^{-1}$, in the plane $(\nS,r)$ (top panel) and
  the plane $(\epsilon_1,\epsilon_2)$ (bottom panel). The solid
  contours are the one and two-sigma {\data} confidence intervals
  (marginalized over second order slow-roll). The parameter $\alpha$
  is varied between $\alpha_{\min}\left(\beta\right)<\alpha<10^6
  \alpha_{\min}\left(\beta\right)$.}
\label{fig:CMBSSBI1betaEQ10PowerMinus1}
\end{center}
\end{figure}

\begin{figure}[H]
\begin{center}
\includegraphics[width=\wappfig,clip=true]{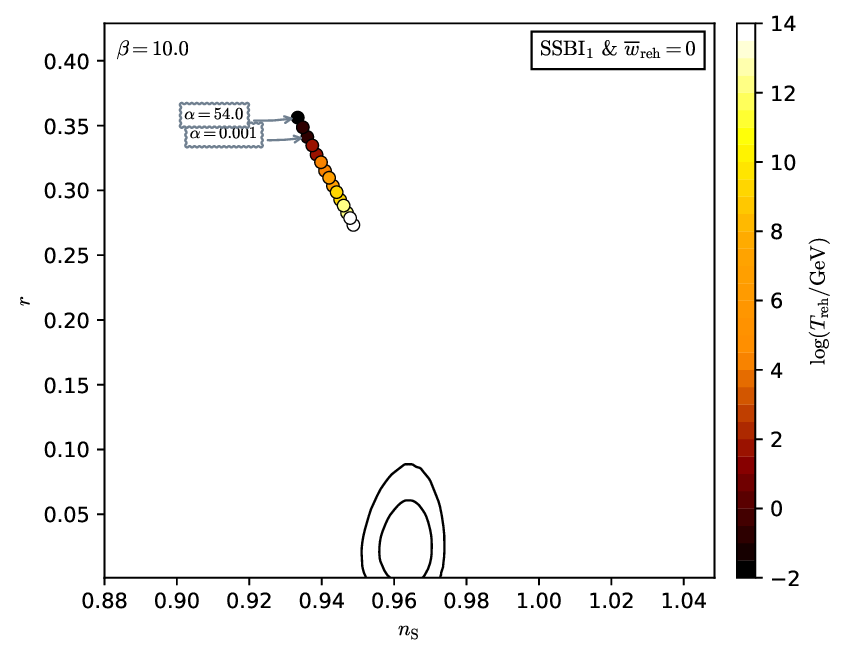}
\includegraphics[width=\wappfig,clip=true]{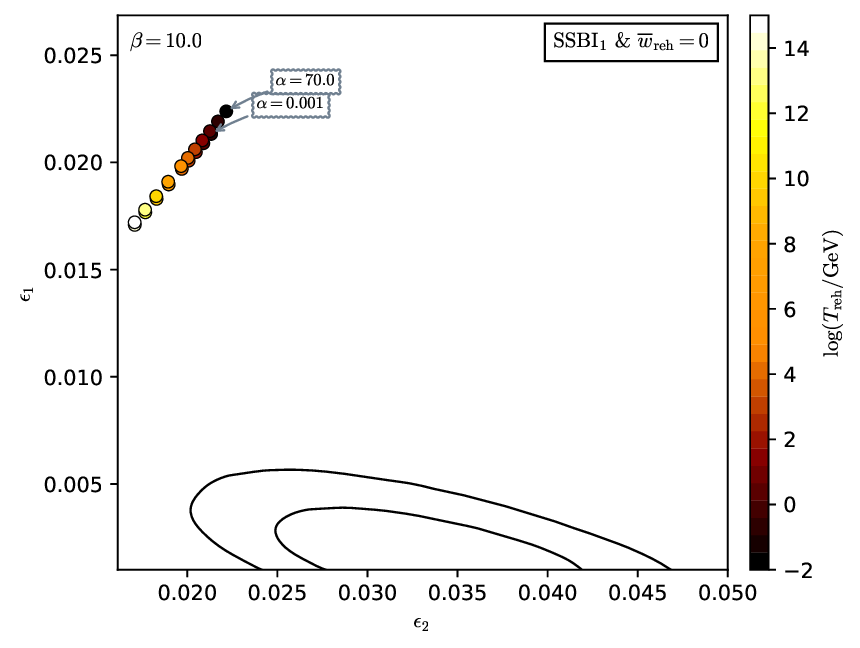}
\caption{Reheating consistent slow-roll predictions for the
  spontaneous symmetry breaking 1 inflation ($\alpha>0,\beta>0$)
  models with $\beta=10$, in the plane $(\nS,r)$ (top panel) and the
  plane $(\epsilon_1,\epsilon_2)$ (bottom panel). The solid contours
  are the one and two-sigma {\data} confidence intervals (marginalized
  over second order slow-roll). The parameter $\alpha$ is varied
  between $\alpha_{\min}\left(\beta\right)<\alpha<10^6
  \alpha_{\min}\left(\beta\right)$ but model predictions are weakly
  sensitive to its value.}
\label{fig:CMBSSBI1betaEQ10}
\end{center}
\end{figure}

\subsection{Spontaneous Symmetry Breaking Inflation 2 (\hyperref[sec:ssbi]{SSBI2})}

\begin{figure}[H]
\begin{center}
\includegraphics[width=\wappfig,clip=true]{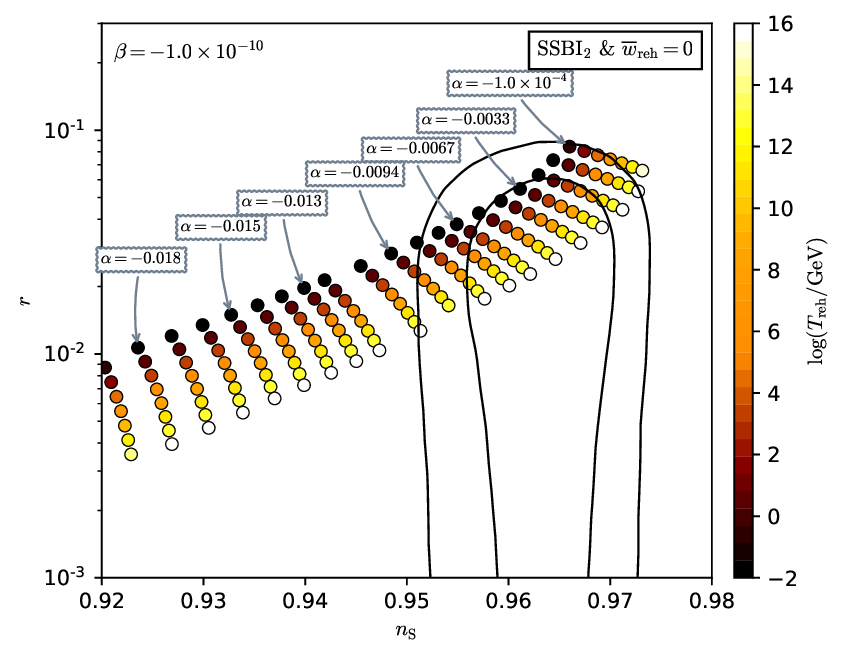}
\includegraphics[width=\wappfig,clip=true]{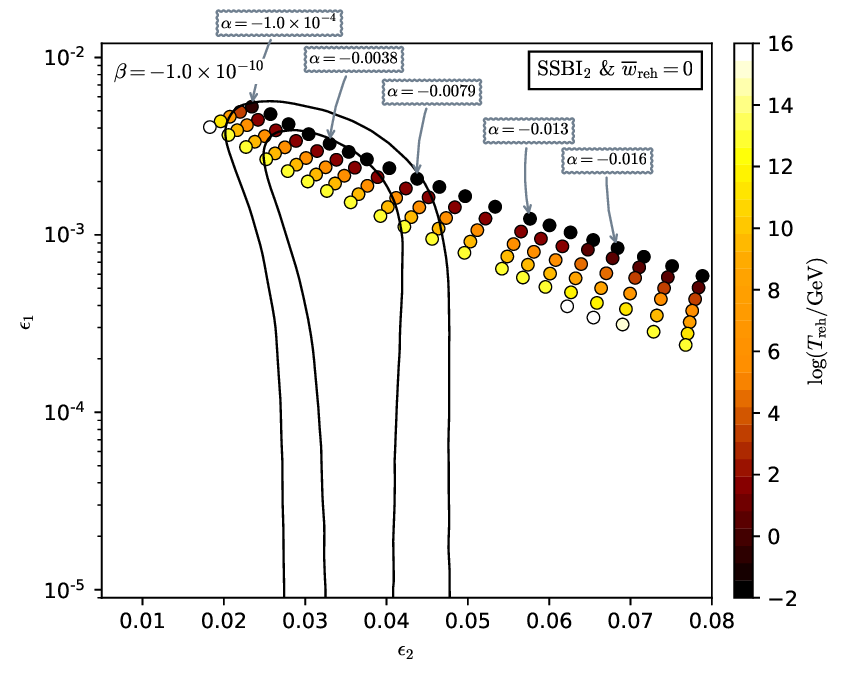}
\caption{Reheating consistent slow-roll predictions for the
  spontaneous symmetry breaking 2 inflation ($\alpha<0,\beta<0$)
  models with $\beta=-10^{-10}$, in the plane $(\nS,r)$ (top panel)
  and the plane $(\epsilon_1,\epsilon_2)$ (bottom panel). The solid
  contours are the one and two-sigma {\data} confidence intervals
  (marginalized over second order slow-roll). See
  figures~\ref{fig:CMBSSBI2_1} to \ref{fig:CMBSSBI2_3} for larger
  values of $|\beta|$.}
\label{fig:CMBSSBI2}
\end{center}
\end{figure}

\begin{figure}[H]
\begin{center}
\includegraphics[width=\wappfig,clip=true]{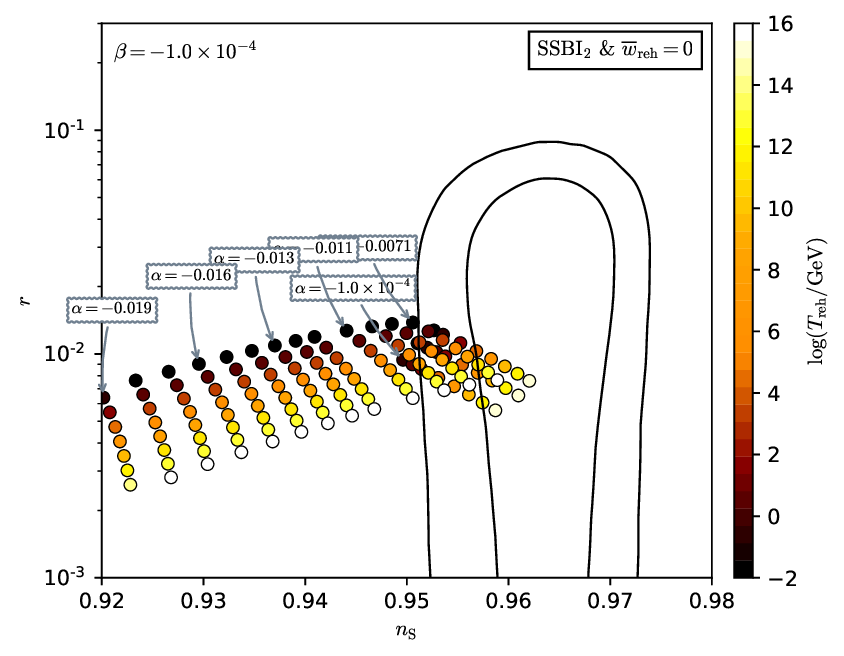}
\includegraphics[width=\wappfig,clip=true]{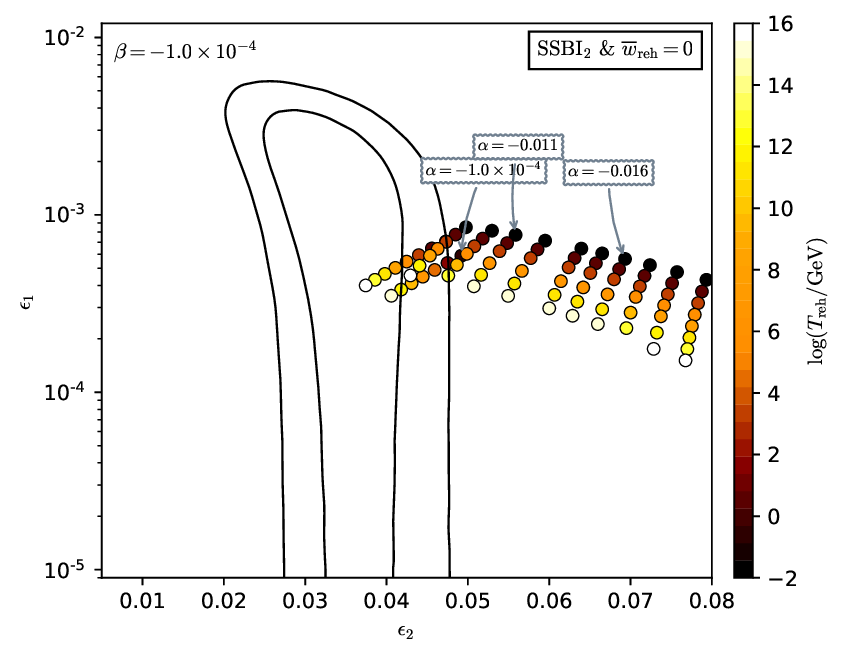}
\caption{Reheating consistent slow-roll predictions for the
  spontaneous symmetry breaking 2 inflation ($\alpha<0,\beta<0$)
  models with $\beta=-10^{-4}$, in the plane $(\nS,r)$ (top panel) and
  the plane $(\epsilon_1,\epsilon_2)$ (bottom panel). The solid
  contours are the one and two-sigma {\data} confidence intervals
  (marginalized over second order slow-roll).}
\label{fig:CMBSSBI2_1}
\end{center}
\end{figure}

\begin{figure}[H]
\begin{center}
\includegraphics[width=\wappfig,clip=true]{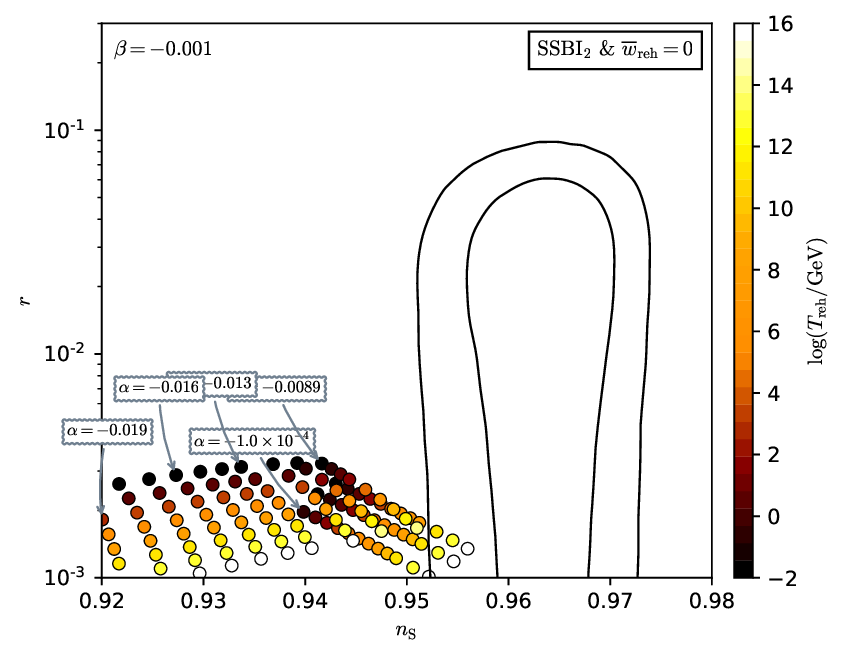}
\includegraphics[width=\wappfig,clip=true]{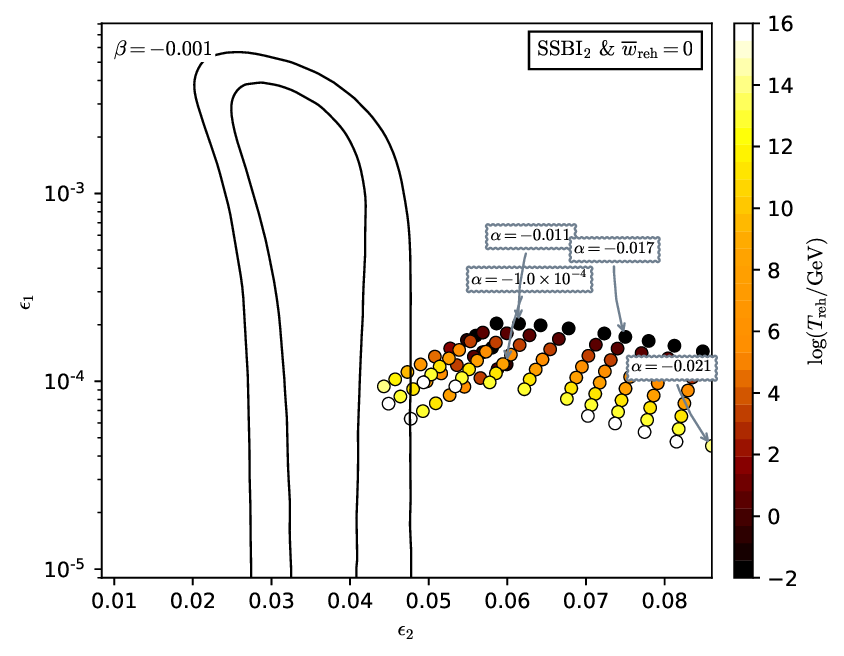}
\caption{Reheating consistent slow-roll predictions for the
  spontaneous symmetry breaking 2 inflation ($\alpha<0,\beta<0$)
  models with $\beta=-10^{-3}$, in the plane $(\nS,r)$ (top panel) and
  the plane $(\epsilon_1,\epsilon_2)$ (bottom panel). The solid
  contours are the one and two-sigma {\data} confidence intervals
  (marginalized over second order slow-roll).}
\label{fig:CMBSSBI2_2}
\end{center}
\end{figure}

\begin{figure}[H]
\begin{center}
\includegraphics[width=\wappfig,clip=true]{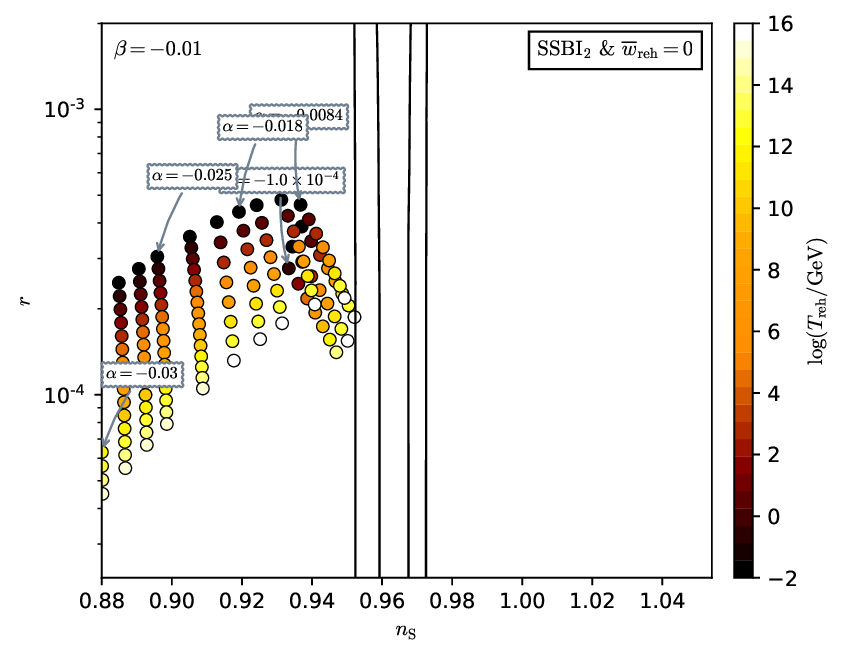}
\includegraphics[width=\wappfig,clip=true]{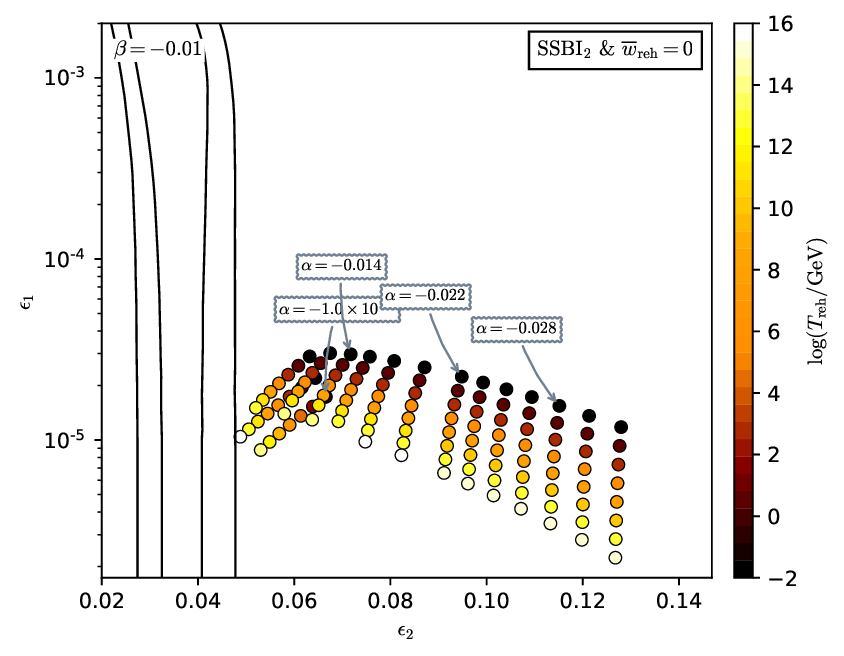}
\caption{Reheating consistent slow-roll predictions for the
  spontaneous symmetry breaking 2 inflation ($\alpha<0,\beta<0$)
  models with $\beta=-10^{-2}$, in the plane $(\nS,r)$ (top panel) and
  the plane $(\epsilon_1,\epsilon_2)$ (bottom panel). The solid
  contours are the one and two-sigma {\data} confidence intervals
  (marginalized over second order slow-roll).}
\label{fig:CMBSSBI2_3}
\end{center}
\end{figure}

\subsection{Spontaneous Symmetry Breaking Inflation 3 (\hyperref[sec:ssbi]{SSBI3})}

\begin{figure}[H]
\begin{center}
\includegraphics[width=\wappfig,clip=true]{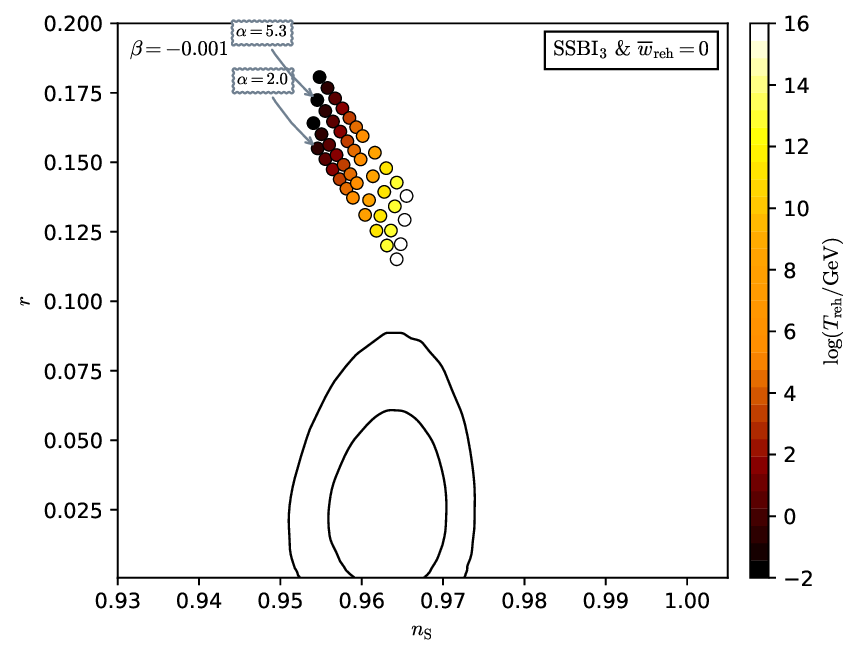}
\includegraphics[width=\wappfig,clip=true]{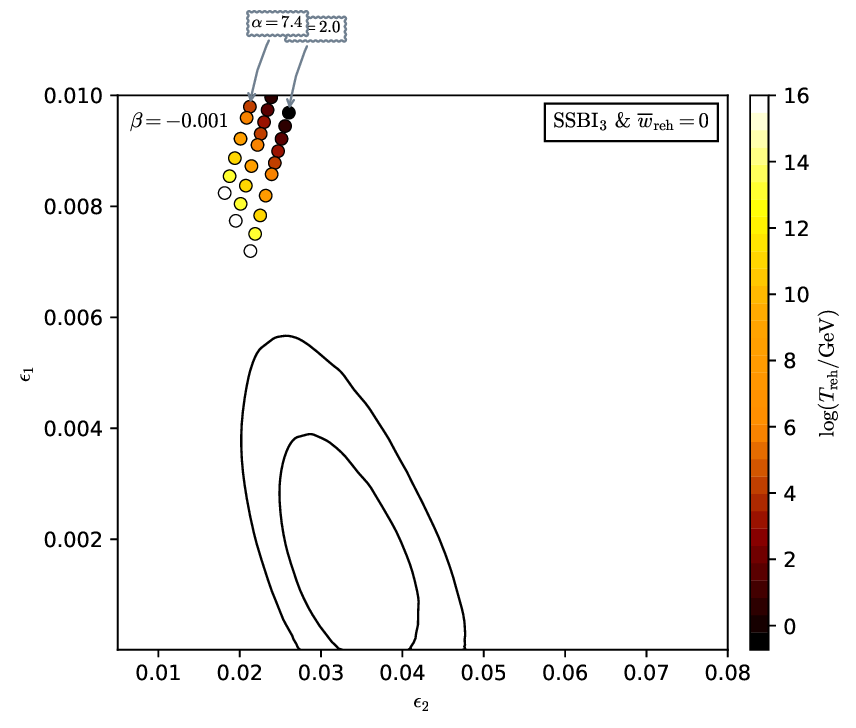}
\caption{Reheating consistent slow-roll predictions for the
  spontaneous symmetry breaking 3 inflation
  $[\alpha>0,\beta<0, x^2<-\alpha/\left(2\beta\right)]$ models for
  $\beta=-10^{-3}$, in the plane $(\nS,r)$ (top panel) and the plane
  $(\epsilon_1,\epsilon_2)$ (bottom panel). The solid contours are the
  one and two-sigma {\data} confidence intervals (marginalized over
  second order slow-roll). The parameter $\alpha$ is varied between
  $\alphamin\left(\beta\right)\simeq 2<\alpha<10^3
  \alphamin\left(\beta\right)$.}
\label{fig:CMBSSBI3betaEQMinus10PowerMinus3}
\end{center}
\end{figure}

\begin{figure}[H]
\begin{center}
\includegraphics[width=\wappfig,clip=true]{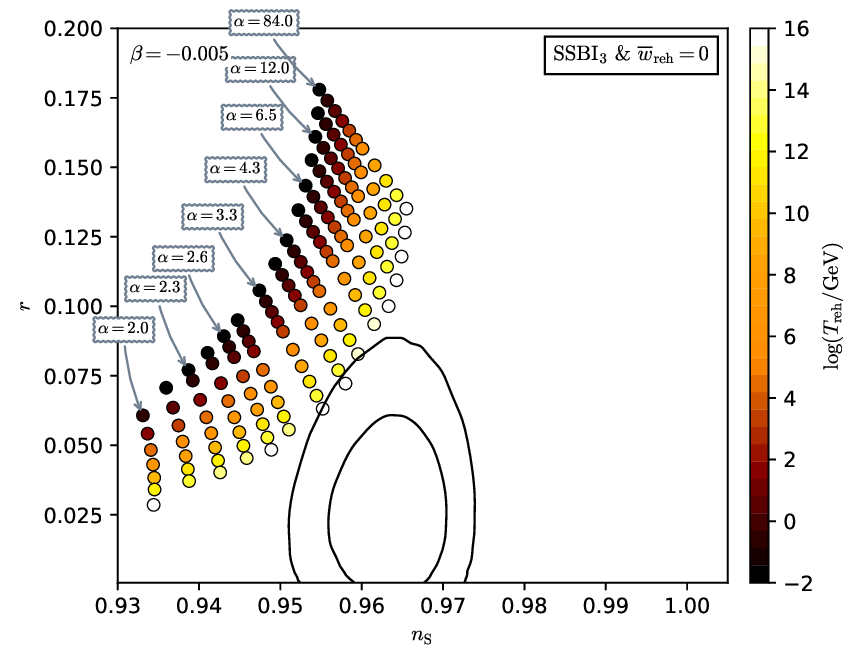}
\includegraphics[width=\wappfig,clip=true]{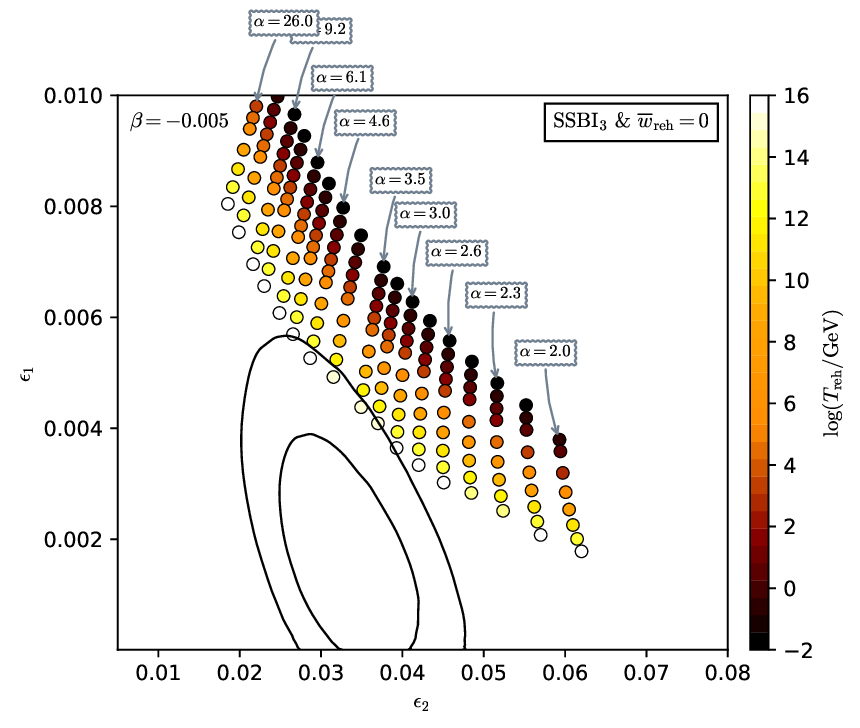}
\caption{Reheating consistent slow-roll predictions for the
  spontaneous symmetry breaking 3 inflation
  $[\alpha>0,\beta<0, x^2<-\alpha/\left(2\beta\right)]$ models for
  $\beta=-5\times 10^{-3}$, in the plane $(\nS,r)$ (top panel) and the
  plane $(\epsilon_1,\epsilon_2)$ (bottom panel). The solid contours
  are the one and two-sigma {\data} confidence intervals (marginalized
  over second order slow-roll). The parameter $\alpha$ is varied
  between $\alphamin\left(\beta\right)\simeq 2<\alpha<10^3
  \alphamin\left(\beta\right)$.}
\label{fig:CMBSSBI3betaEQMinus5x10PowerMinus3}
\end{center}
\end{figure}

\begin{figure}[H]
\begin{center}
\includegraphics[width=\wappfig,clip=true]{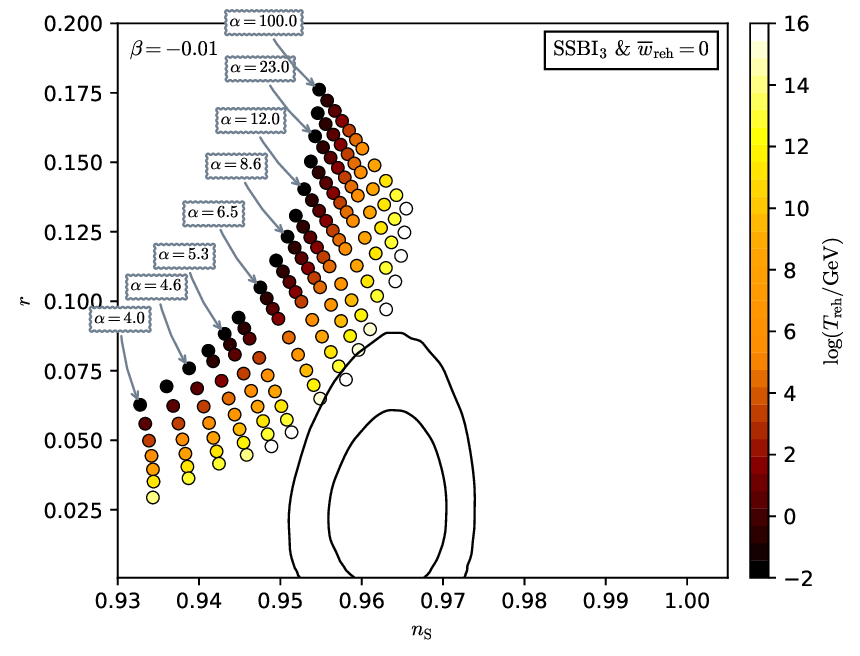}
\includegraphics[width=\wappfig,clip=true]{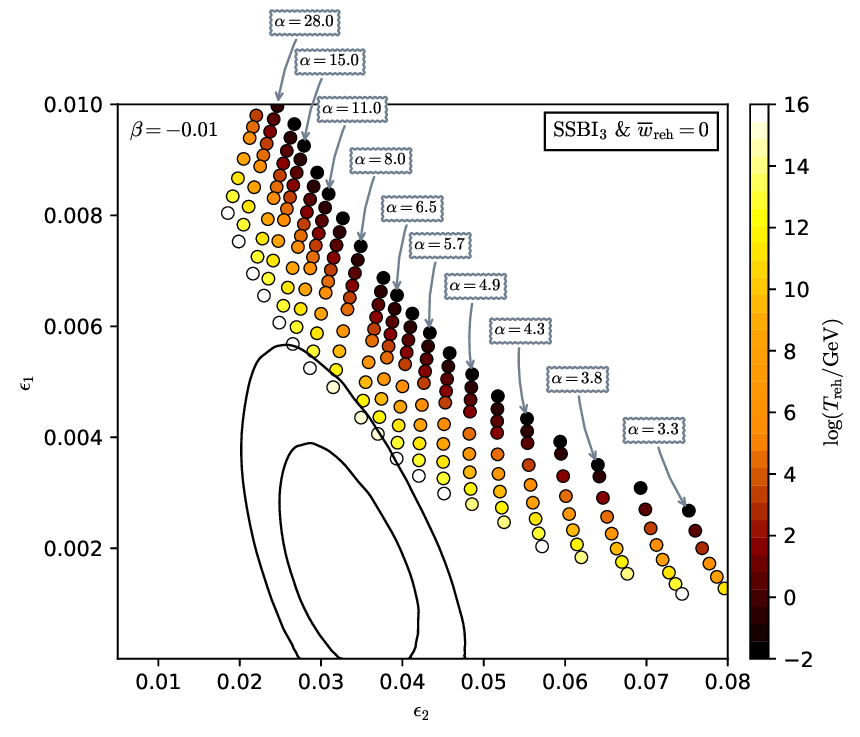}
\caption{Reheating consistent slow-roll predictions for the
  spontaneous symmetry breaking 3 inflation
  $[\alpha>0,\beta<0, x^2<-\alpha/\left(2\beta\right)]$ models for
  $\beta=-10^{-2}$, in the plane $(\nS,r)$ (top panel) and the plane
  $(\epsilon_1,\epsilon_2)$ (bottom panel). The solid contours are the
  one and two-sigma {\data} confidence intervals (marginalized over
  second order slow-roll). The parameter $\alpha$ is varied between
  $\alphamin\left(\beta\right)\simeq 2<\alpha<10^3
  \alphamin\left(\beta\right)$.}
\label{fig:CMBSSBI3betaEQMinus10PowerMinus2}
\end{center}
\end{figure}

\subsection{Spontaneous Symmetry Breaking Inflation 4 (\hyperref[sec:ssbi]{SSBI4})}

\begin{figure}[H]
\begin{center}
\includegraphics[width=\wappfig,clip=true]{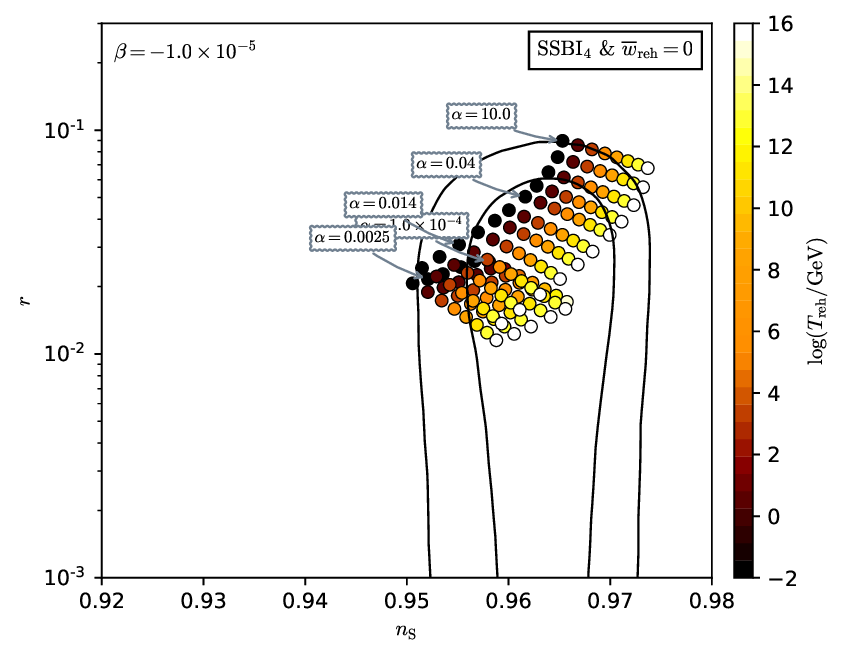}
\includegraphics[width=\wappfig,clip=true]{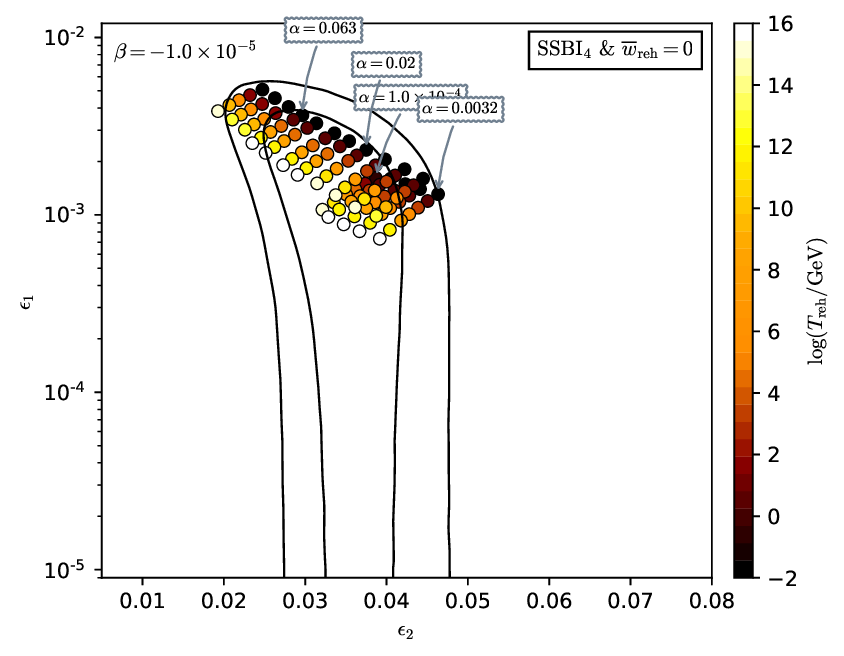}
\caption{Reheating consistent slow-roll predictions for the
  spontaneous symmetry breaking 4 inflation $[\alpha>0,\beta<0,
    x^2>-\alpha/\left(2\beta\right)]$ models for $\beta=-10^{-5}$, in
  the plane $(\nS,r)$ (top panel) and the plane
  $(\epsilon_1,\epsilon_2)$ (bottom panel). The solid contours are the
  one and two-sigma {\data} confidence intervals (marginalized over
  second order slow-roll).}
\label{fig:CMBSSBI4betaEQMinus10PowerMinus5}
\end{center}
\end{figure}

\begin{figure}[H]
\begin{center}
\includegraphics[width=\wappfig,clip=true]{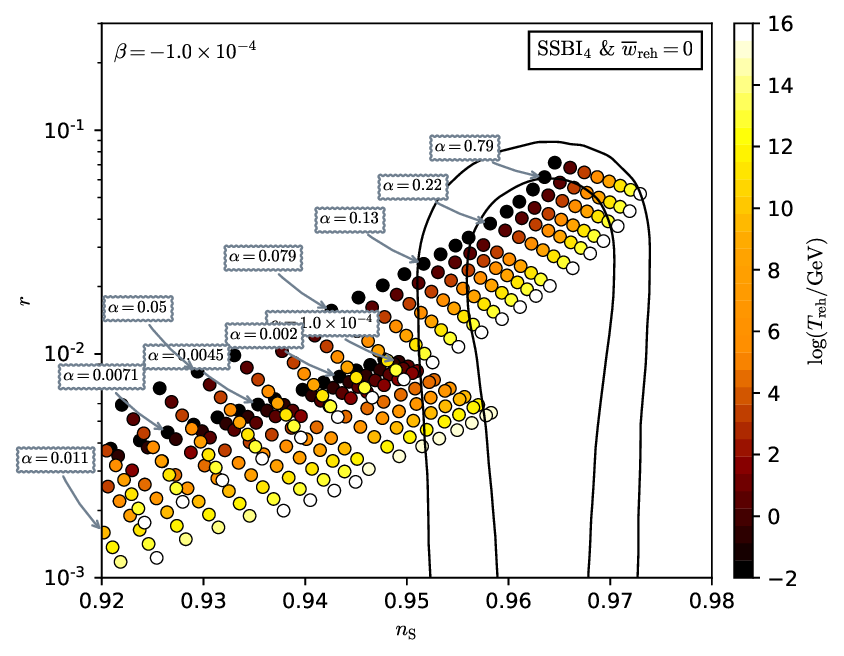}
\includegraphics[width=\wappfig,clip=true]{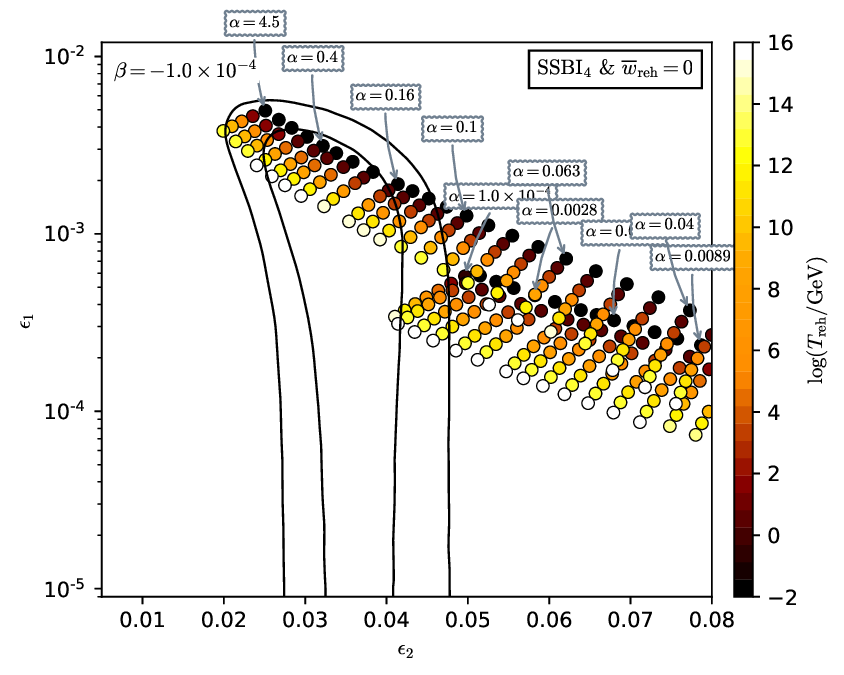}
\caption{Reheating consistent slow-roll predictions for the
  spontaneous symmetry breaking 4 inflation $[\alpha>0,\beta<0,
    x^2>-\alpha/\left(2\beta\right)]$ models for $\beta=-10^{-4}$, in
  the plane $(\nS,r)$ (top panel) and the plane
  $(\epsilon_1,\epsilon_2)$ (bottom panel). The solid contours are the
  one and two-sigma {\data} confidence intervals (marginalized over
  second order slow-roll).}
\label{fig:CMBSSBI4betaEQMinus10PowerMinus4}
\end{center}
\end{figure}

\begin{figure}[H]
\begin{center}
\includegraphics[width=\wappfig,clip=true]{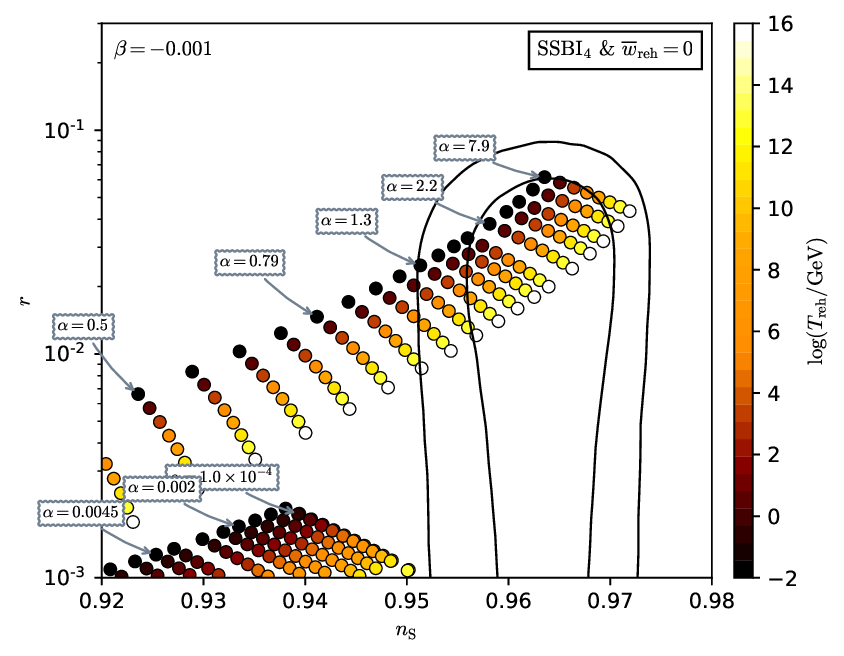}
\includegraphics[width=\wappfig,clip=true]{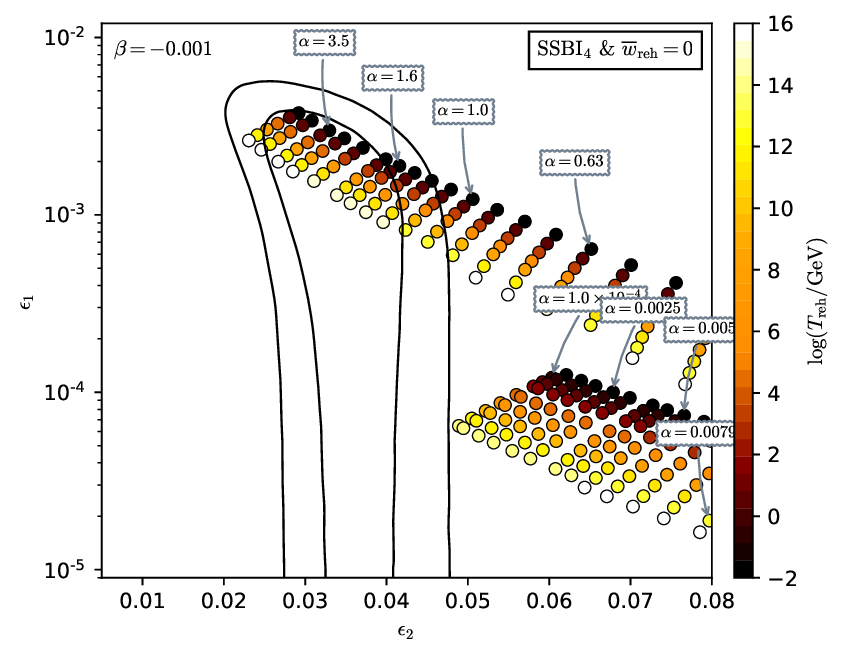}
\caption{Reheating consistent slow-roll predictions for the
  spontaneous symmetry breaking 4 inflation $[\alpha>0,\beta<0,
    x^2>-\alpha/\left(2\beta\right)]$ models for $\beta=-10^{-3}$, in
  the plane $(\nS,r)$ (top panel) and the plane
  $(\epsilon_1,\epsilon_2)$ (bottom panel). The solid contours are the
  one and two-sigma {\data} confidence intervals (marginalized over
  second order slow-roll).}
\label{fig:CMBSSBI4betaEQMinus10PowerMinus3}
\end{center}
\end{figure}

\subsection{Spontaneous Symmetry Breaking Inflation 5 (\hyperref[sec:ssbi]{SSBI5})}

\begin{figure}[H]
\begin{center}
\includegraphics[width=\wappfig,clip=true]{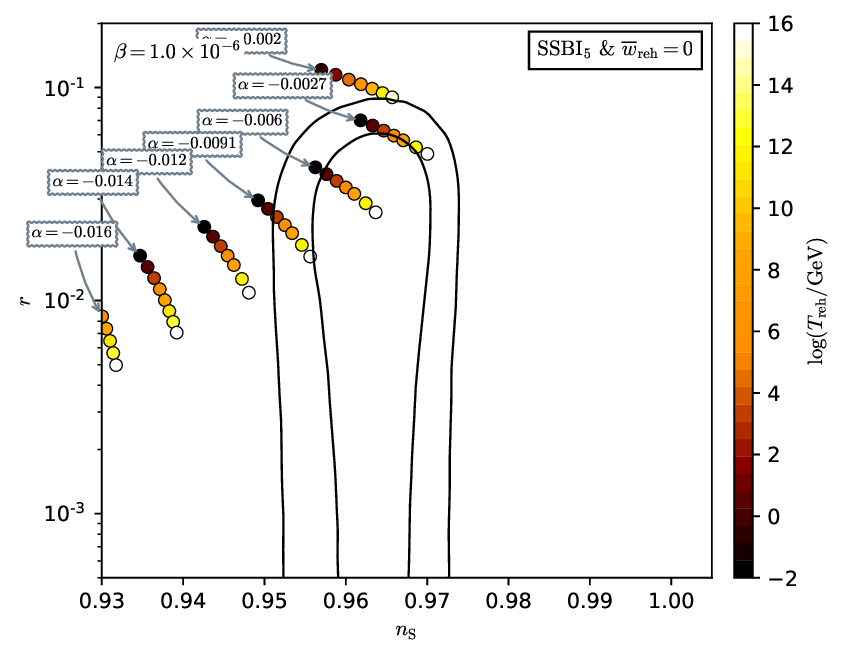}
\includegraphics[width=\wappfig,clip=true]{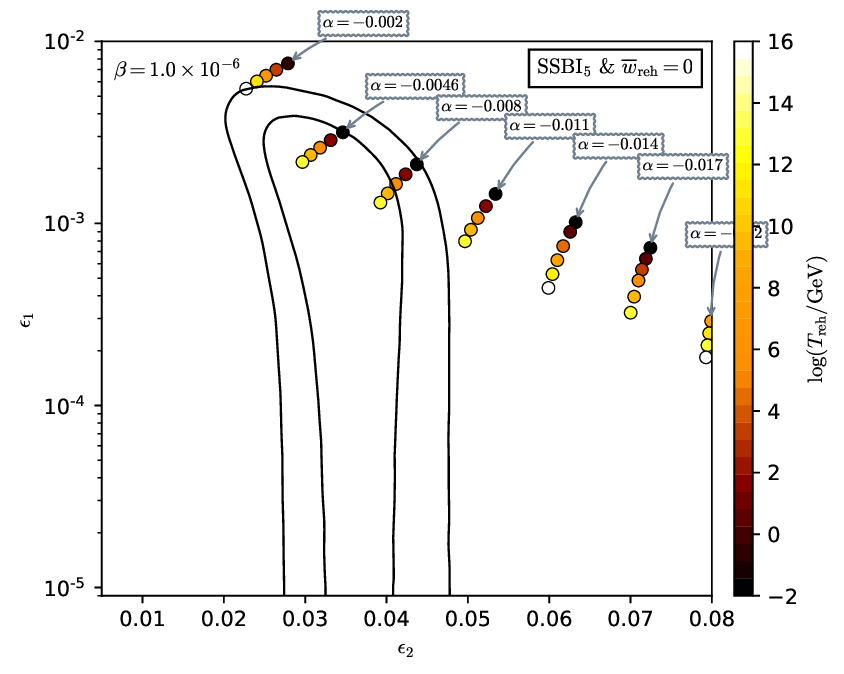}
\caption{Reheating consistent slow-roll predictions for the
  spontaneous symmetry breaking 5 inflation
  $[\alpha<0,\beta>0,x^2<-\alpha/\left(2\beta\right)]$ models for
  $\beta=10^{-6}$, in the plane $(\nS,r)$ (top panel) and the plane
  $(\epsilon_1,\epsilon_2)$ (bottom panel). The solid contours are the
  one and two-sigma {\data} confidence intervals (marginalized over
  second order slow-roll). The parameter $\alpha$ is varied between
  $|\alphamin(\beta)| <|\alpha| < 10 |\alphamin\left(\beta\right)|$.}
\label{fig:CMBSSBI5betaEQ10PowerMinus6}
\end{center}
\end{figure}

\begin{figure}[H]
\begin{center}
\includegraphics[width=\wappfig,clip=true]{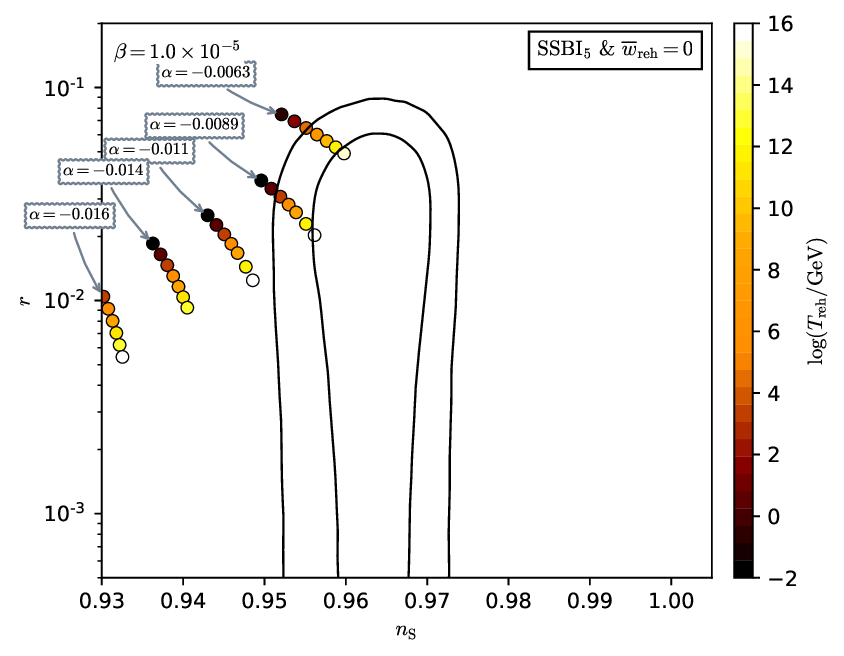}
\includegraphics[width=\wappfig,clip=true]{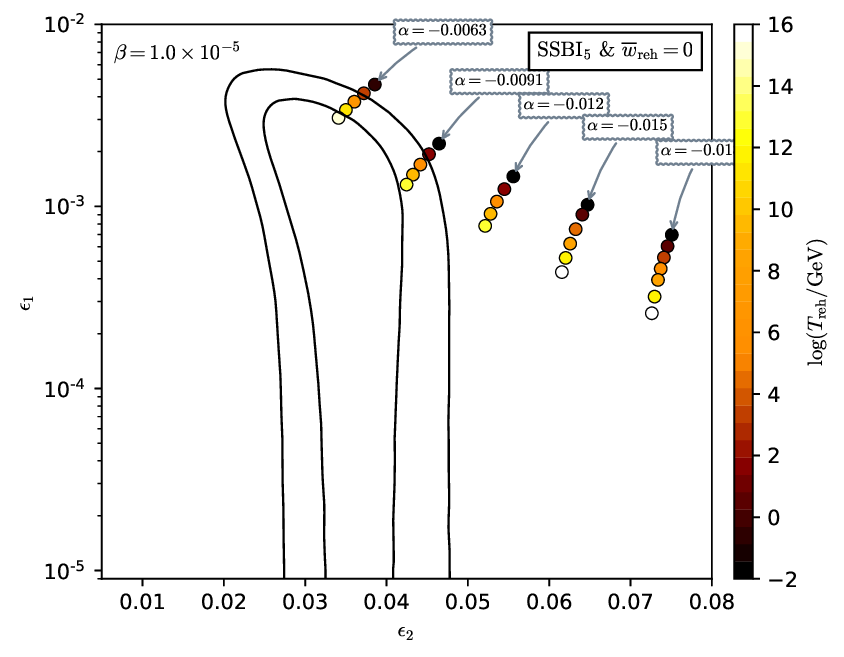}
\caption{Reheating consistent slow-roll predictions for the
  spontaneous symmetry breaking 5 inflation
  $[\alpha<0,\beta>0,x^2<-\alpha/\left(2\beta\right)]$ models for
  $\beta=10^{-5}$, in the plane $(\nS,r)$ (top panel) and the plane
  $(\epsilon_1,\epsilon_2)$ (bottom panel). The solid contours are the
  one and two-sigma {\data} confidence intervals (marginalized over
  second order slow-roll). The parameter $\alpha$ is varied between
  $|\alphamin(\beta)| <|\alpha| < 10 |\alphamin\left(\beta\right)|$.}
\label{fig:CMBSSBI5betaEQ10PowerMinus5}
\end{center}
\end{figure}

\begin{figure}[H]
\begin{center}
\includegraphics[width=\wappfig,clip=true]{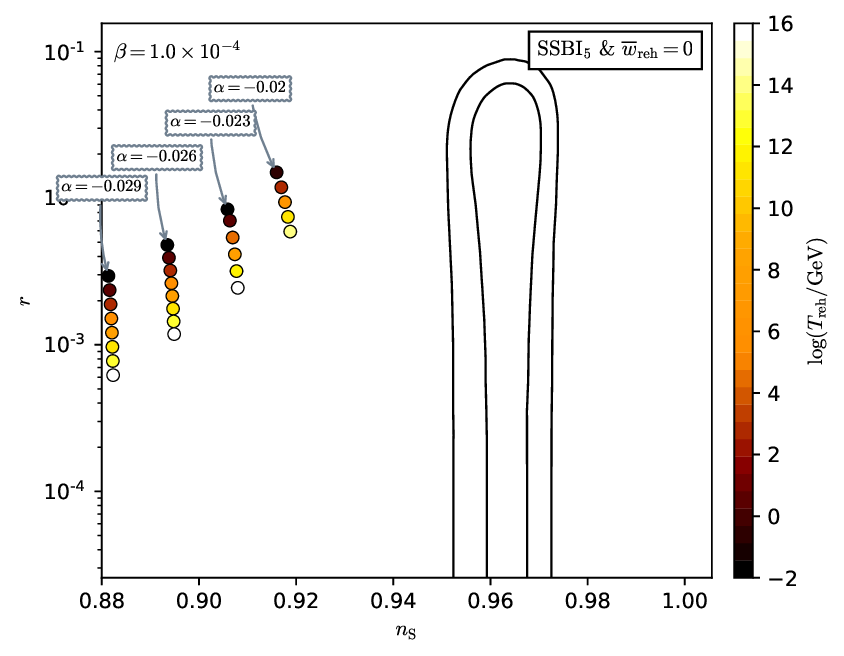}
\includegraphics[width=\wappfig,clip=true]{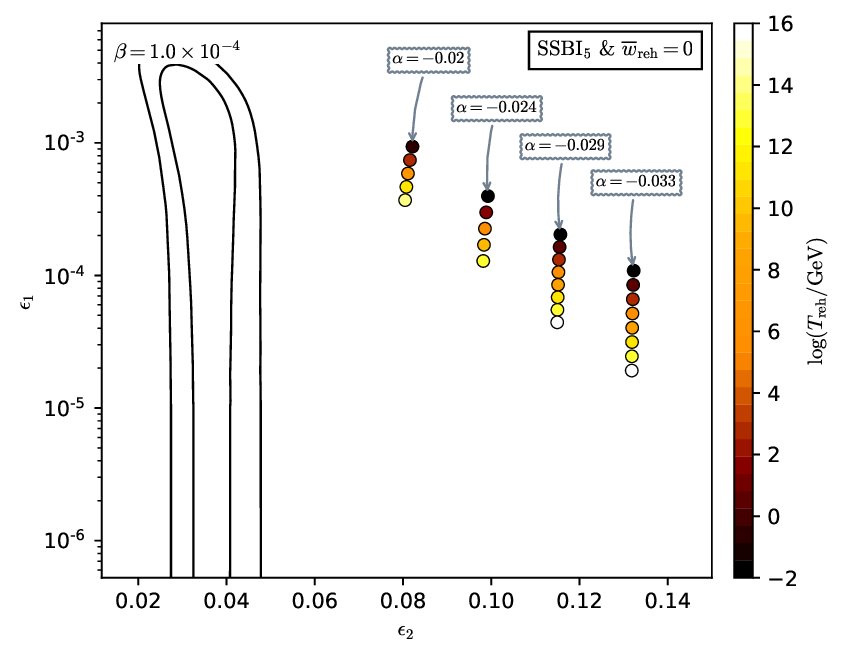}
\caption{Reheating consistent slow-roll predictions for the
  spontaneous symmetry breaking 5 inflation
  $[\alpha<0,\beta>0,x^2<-\alpha/\left(2\beta\right)]$ models for
  $\beta=10^{-4}$, in the plane $(\nS,r)$ (top panel) and the plane
  $(\epsilon_1,\epsilon_2)$ (bottom panel). The solid contours are the
  one and two-sigma {\data} confidence intervals (marginalized over
  second order slow-roll). The parameter $\alpha$ is varied between
  $|\alphamin(\beta)| <|\alpha| < 10 |\alphamin\left(\beta\right)|$.}
\label{fig:CMBSSBI5betaEQ10PowerMinus4}
\end{center}
\end{figure}

\subsection{Spontaneous Symmetry Breaking Inflation 6 (\hyperref[sec:ssbi]{SSBI6})}

\begin{figure}[H]
\begin{center}
\includegraphics[width=\wappfig,clip=true]{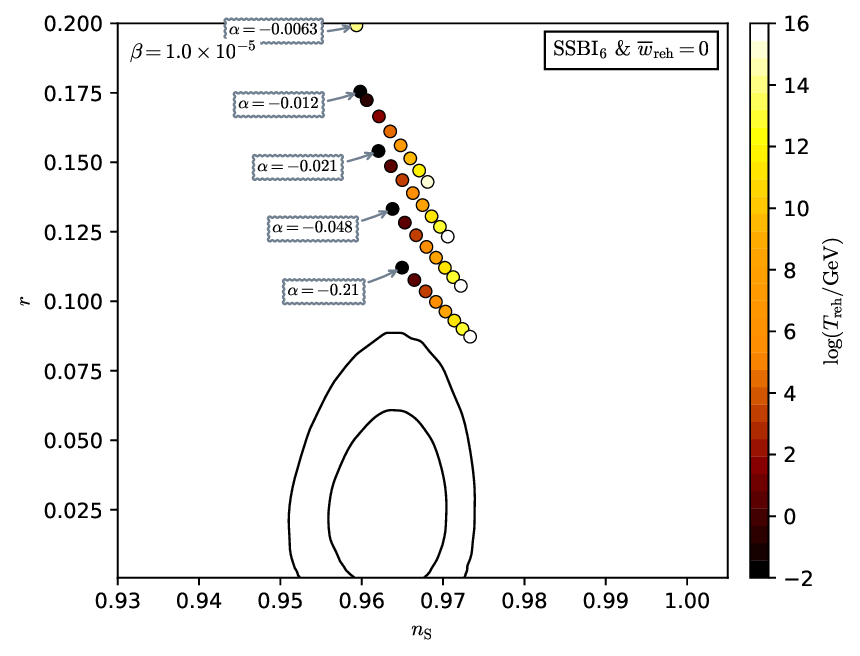}
\includegraphics[width=\wappfig,clip=true]{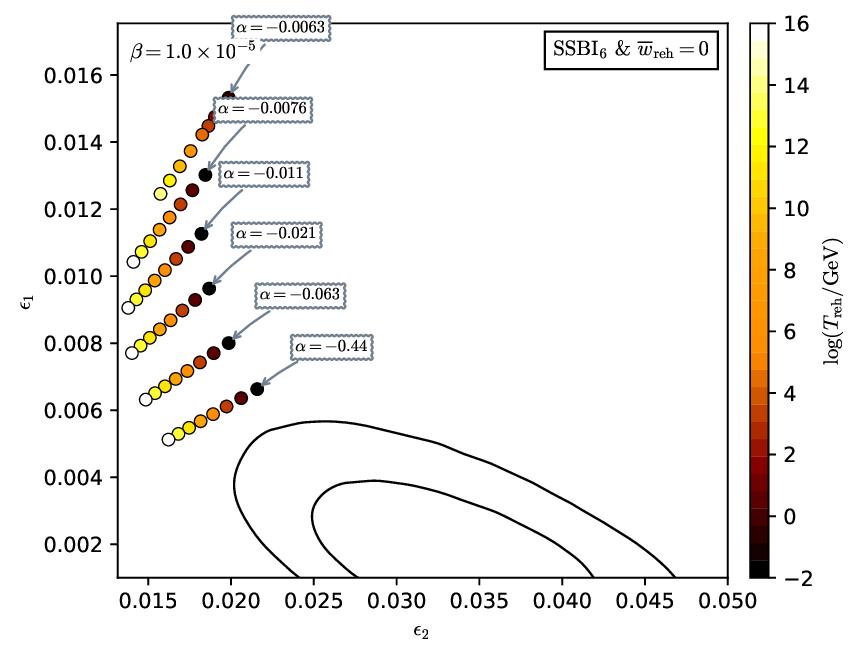}
\caption{Reheating consistent slow-roll predictions for the
  spontaneous symmetry breaking 6 inflation
  $[\alpha<0,\beta>0,x^2>-\alpha/\left(2\beta\right)]$ models for
  $\beta=10^{-5}$, in the plane $(\nS,r)$ (top panel) and the plane
  $(\epsilon_1,\epsilon_2)$ (bottom panel). The solid contours are the
  one and two-sigma {\data} confidence intervals (marginalized over
  second order slow-roll). The parameter $\alpha$ is varied between
  $|\alphamin(\beta)| <|\alpha| < 10^4
  |\alphamin\left(\beta\right)|$.}
\label{fig:CMBSSBI6betaEQ10PowerMinus5}
\end{center}
\end{figure}

\begin{figure}[H]
\begin{center}
\includegraphics[width=\wappfig,clip=true]{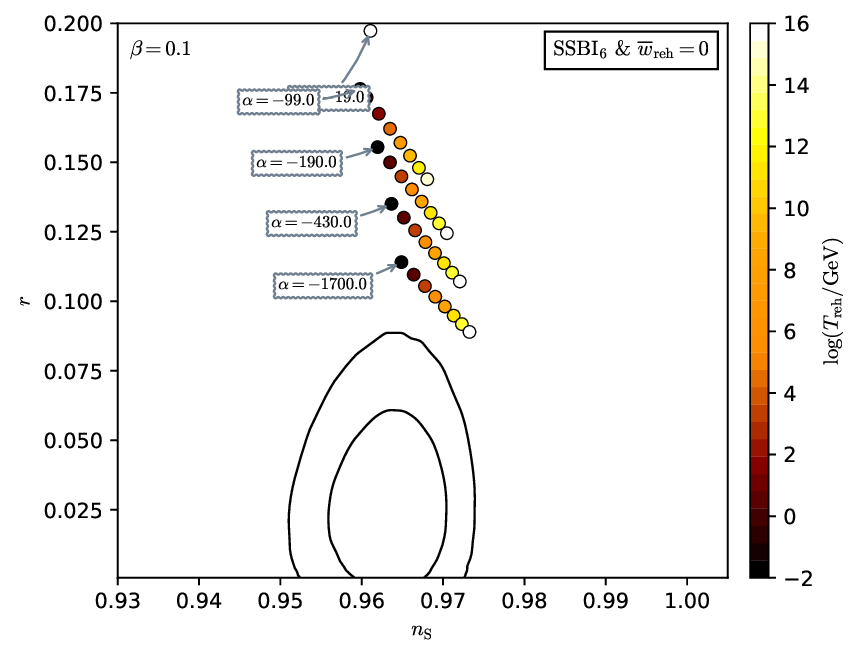}
\includegraphics[width=\wappfig,clip=true]{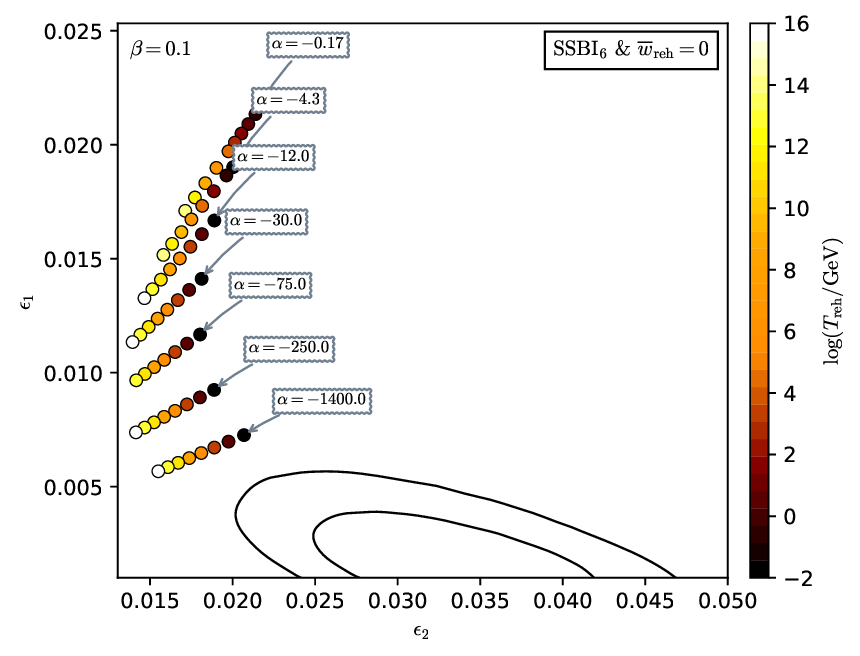}
\caption{Reheating consistent slow-roll predictions for the
  spontaneous symmetry breaking 6 inflation
  $[\alpha<0,\beta>0,x^2>-\alpha/\left(2\beta\right)]$ models for
  $\beta=10^{-1}$, in the plane $(\nS,r)$ (top panel) and the plane
  $(\epsilon_1,\epsilon_2)$ (bottom panel). The solid contours are the
  one and two-sigma {\data} confidence intervals (marginalized over
  second order slow-roll). The parameter $\alpha$ is varied between
  $|\alphamin(\beta)| <|\alpha| < 10^4
  |\alphamin\left(\beta\right)|$.}
\label{fig:CMBSSBI6betaEQ10PowerMinus1}
\end{center}
\end{figure}

\begin{figure}[H]
\begin{center}
\includegraphics[width=\wappfig,clip=true]{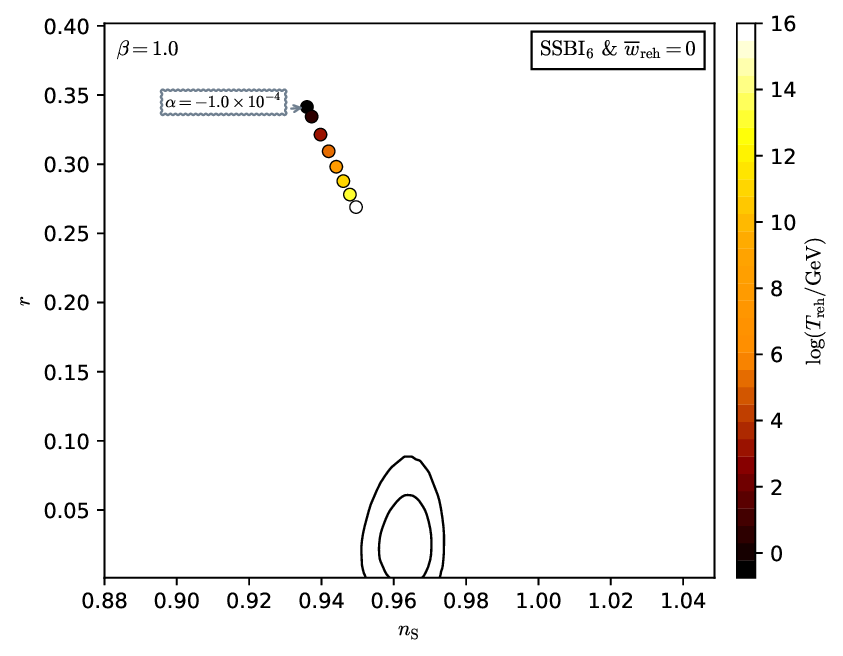}
\includegraphics[width=\wappfig,clip=true]{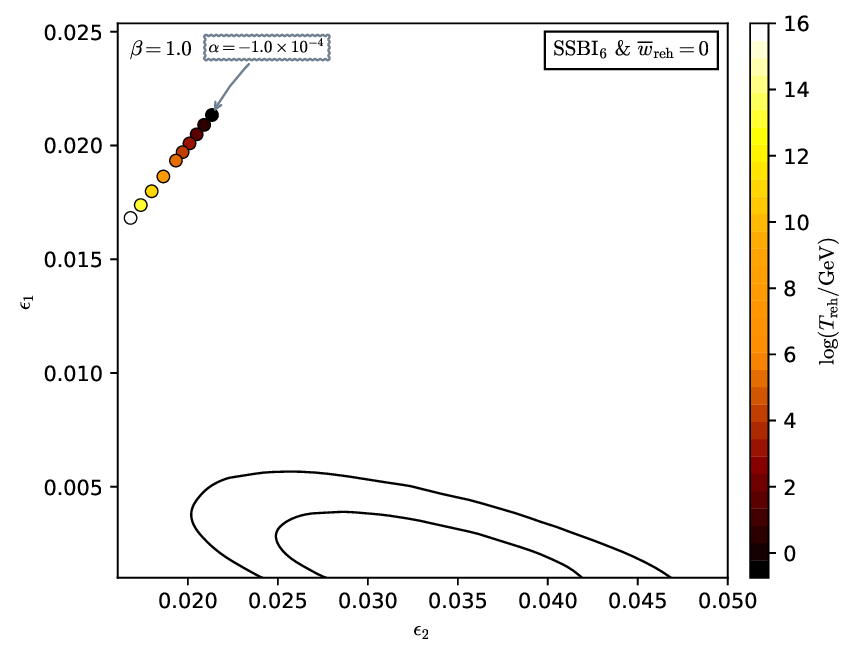}
\caption{Reheating consistent slow-roll predictions for the
  spontaneous symmetry breaking 6 inflation
  $[\alpha<0,\beta>0,x^2>-\alpha/\left(2\beta\right)]$ models for
  $\beta=1$, in the plane $(\nS,r)$ (top panel) and the plane
  $(\epsilon_1,\epsilon_2)$ (bottom panel). The solid contours are the
  one and two-sigma {\data} confidence intervals (marginalized over
  second order slow-roll). The parameter $\alpha$ is varied between
  $|\alphamin(\beta)| <|\alpha| < 10^4 |\alphamin\left(\beta\right)|$
  but the predictions are almost unsensitive to its value.}
\label{fig:CMBSSBI6betaEQ1}
\end{center}
\end{figure}

\subsection{Inverse Monomial Inflation (\hyperref[sec:imi]{IMI})}

\begin{figure}[H]
\begin{center}
\includegraphics[width=\wappfig,clip=true]{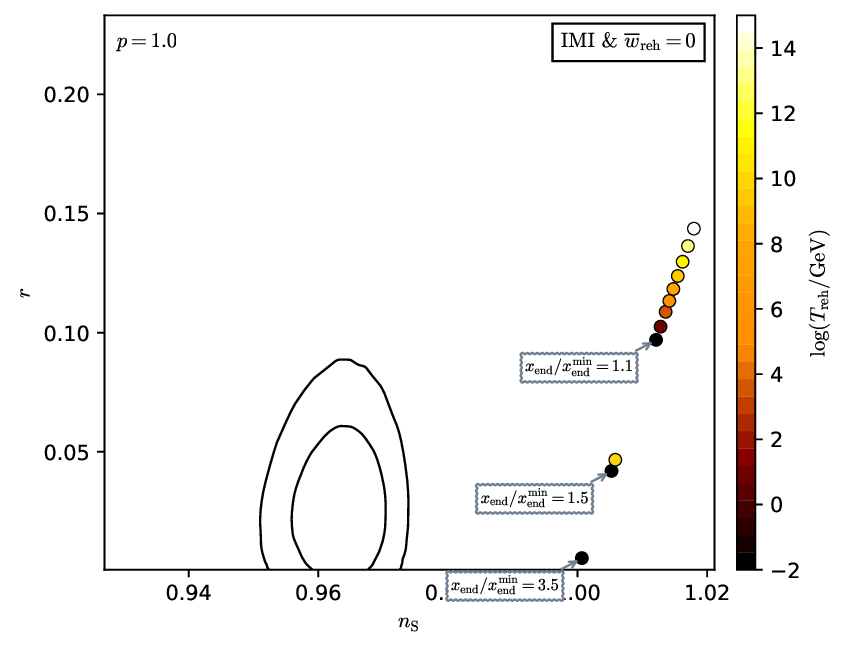}
\includegraphics[width=\wappfig,clip=true]{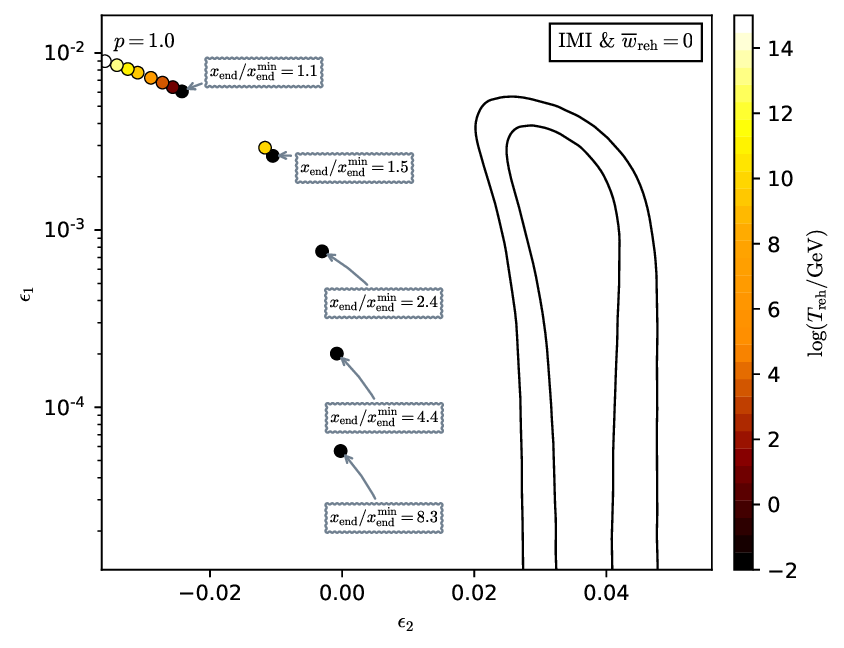}
\caption{Reheating consistent slow-roll predictions for the IMI models
  with $p=1$, in the plane $(\nS,r)$ (top panel) and the plane
  $(\epsilon_1,\epsilon_2)$ (bottom panel). The solid contours are the
  one and two-sigma {\data} confidence intervals (marginalized over
  second order slow-roll). The parameter $\xend$ varies above
  $\xendmin\left(\Delta N=65\right)$. The model predictions are along the
  curves $\left(1-2/p\right)r=8\left(1-\nS\right)$, \ie
  $\epsilon_1=-(p/4)\epsilon_2$. For other values of $p$, see
  figures~\ref{fig:CMBIMI_1} to \ref{fig:CMBIMI_3}.}
\label{fig:CMBIMI}
\end{center}
\end{figure}

\begin{figure}[H]
\begin{center}
\includegraphics[width=\wappfig,clip=true]{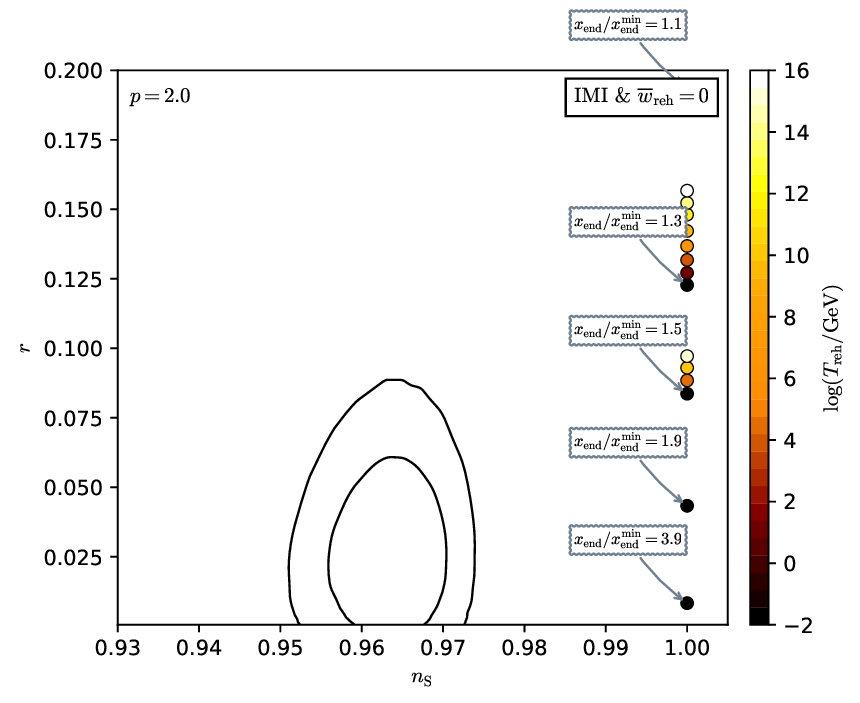}
\includegraphics[width=\wappfig,clip=true]{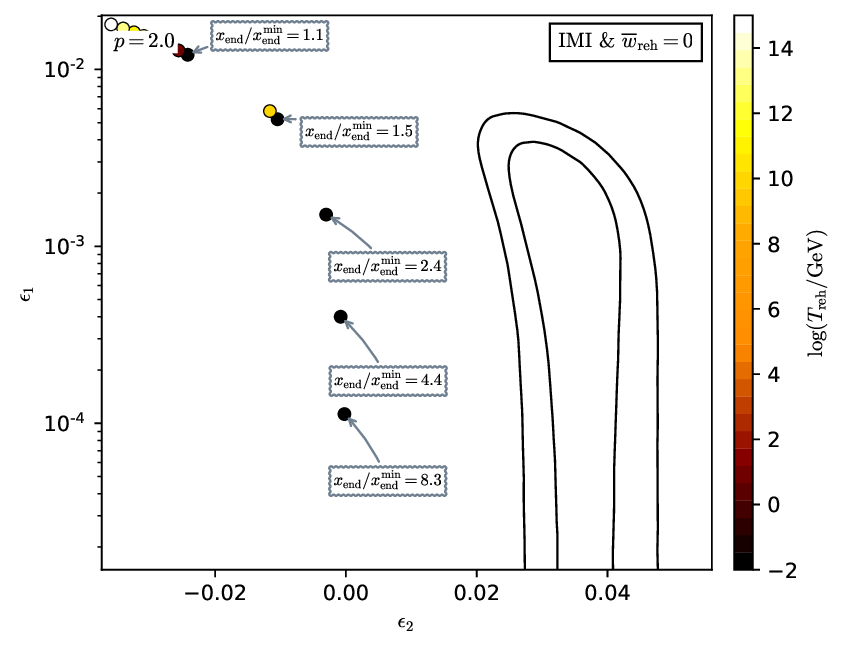}
\caption{Reheating consistent slow-roll predictions for the IMI models
  with $p=2$, in the plane $(\nS,r)$ (top panel) and the plane
  $(\epsilon_1,\epsilon_2)$ (bottom panel). The solid contours are the
  one and two-sigma {\data} confidence intervals (marginalized over
  second order slow-roll). The parameter $\xend$ varies above
  $\xendmin\left(\Delta N=65\right)$. The model predictions verify
  $\nS=1$.}
\label{fig:CMBIMI_1}
\end{center}
\end{figure}

\begin{figure}[H]
\begin{center}
\includegraphics[width=\wappfig,clip=true]{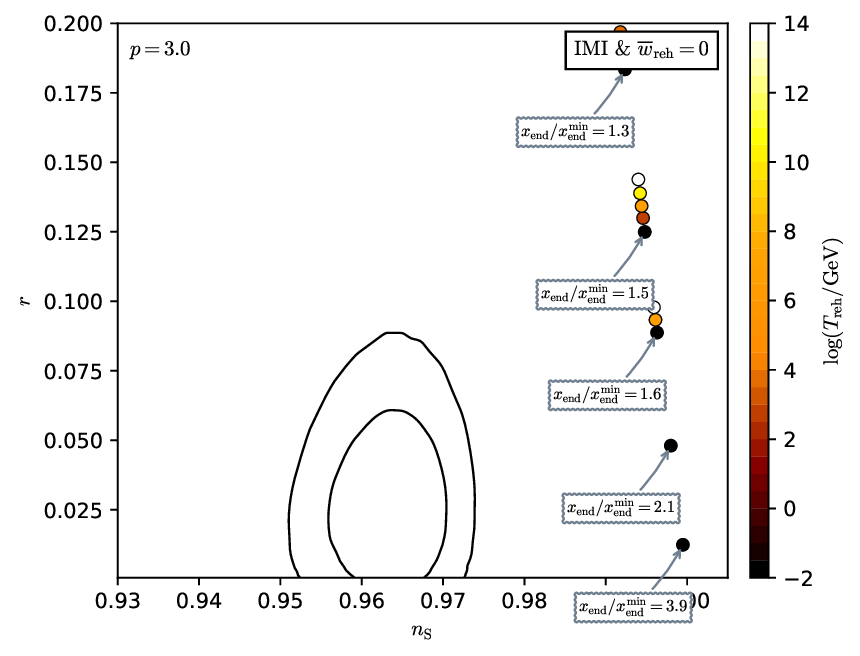}
\includegraphics[width=\wappfig,clip=true]{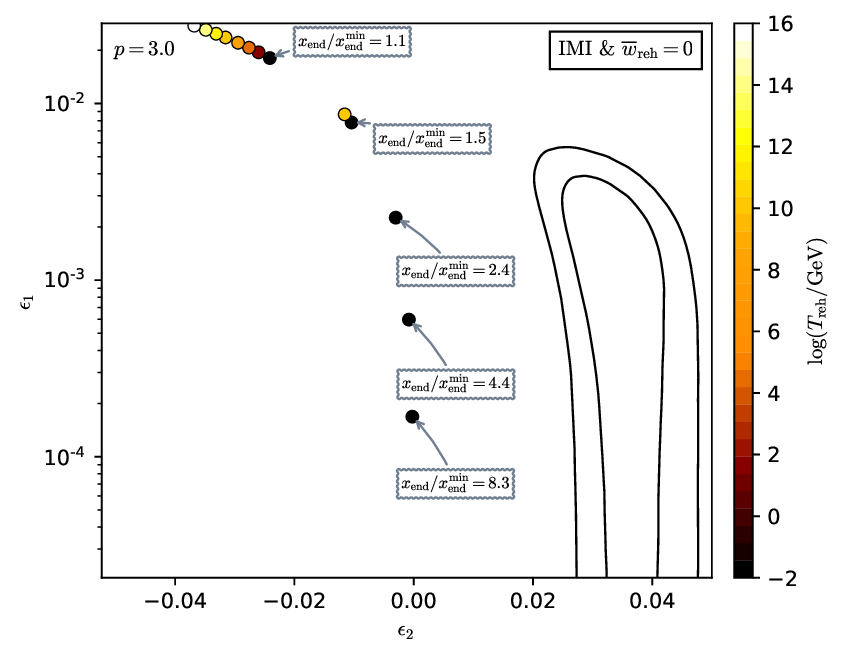}
\caption{Reheating consistent slow-roll predictions for the IMI models
  with $p=3$, in the plane $(\nS,r)$ (top panel) and the plane
  $(\epsilon_1,\epsilon_2)$ (bottom panel). The solid contours are the
  one and two-sigma {\data} confidence intervals (marginalized over
  second order slow-roll). The parameter $\xend$ varies above
  $\xendmin\left(\Delta N=65\right)$. The model predictions are along
  the curves $\left(1-2/p\right)r=8\left(1-\nS\right)$, \ie
  $\epsilon_1=-(p/4)\epsilon_2$.}
\label{fig:CMBIMI_2}
\end{center}
\end{figure}

\begin{figure}[H]
\begin{center}
\includegraphics[width=\wappfig,clip=true]{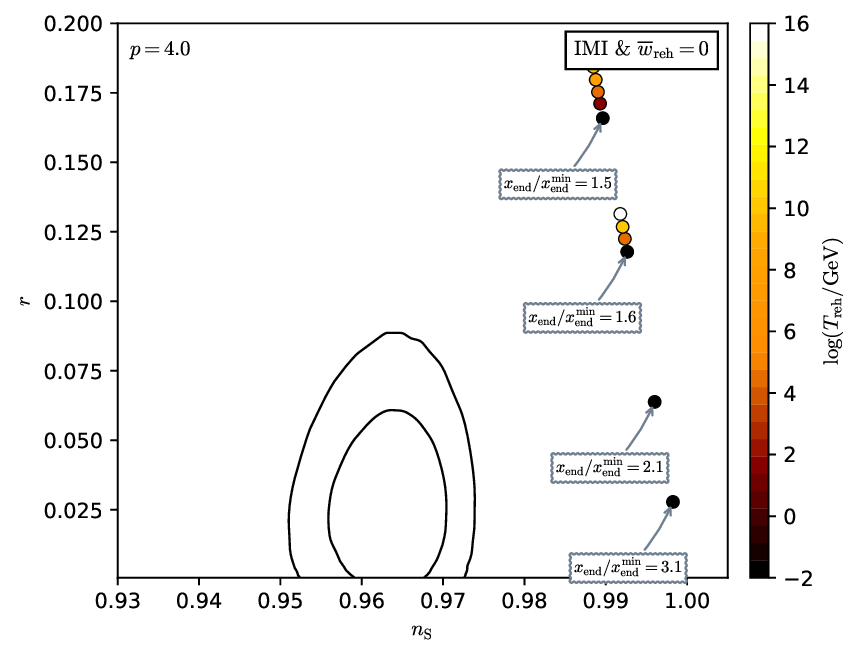}
\includegraphics[width=\wappfig,clip=true]{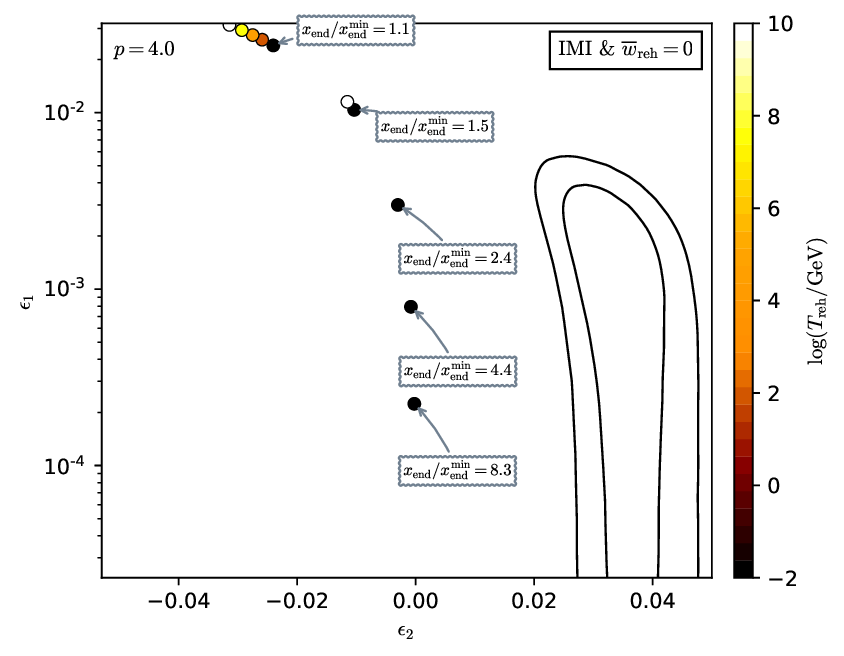}
\caption{Reheating consistent slow-roll predictions for the IMI models
  with $p=4$, in the plane $(\nS,r)$ (top panel) and the plane
  $(\epsilon_1,\epsilon_2)$ (bottom panel). The solid contours are the
  one and two-sigma {\data} confidence intervals (marginalized over
  second order slow-roll). The parameter $\xend$ varies above
  $\xendmin\left(\Delta N=65\right)$. The model predictions are along
  the curves $\left(1-2/p\right)r=8\left(1-\nS\right)$, \ie
  $\epsilon_1=-(p/4)\epsilon_2$.}
\label{fig:CMBIMI_3}
\end{center}
\end{figure}

\subsection{Brane Inflation (\hyperref[sec:bi]{BI})}

\begin{figure}[H]
\begin{center}
\includegraphics[width=\wappfig,clip=true]{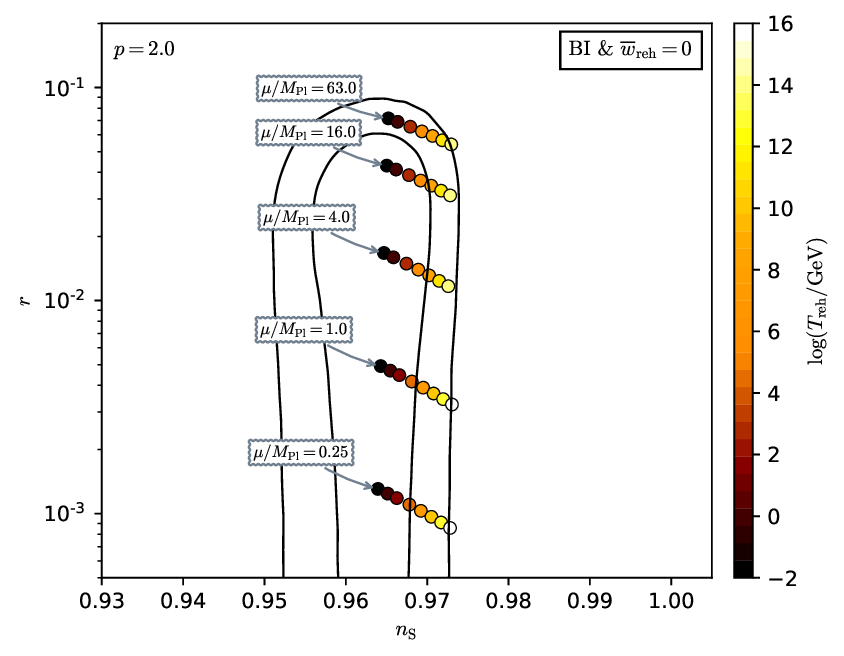}
\includegraphics[width=\wappfig,clip=true]{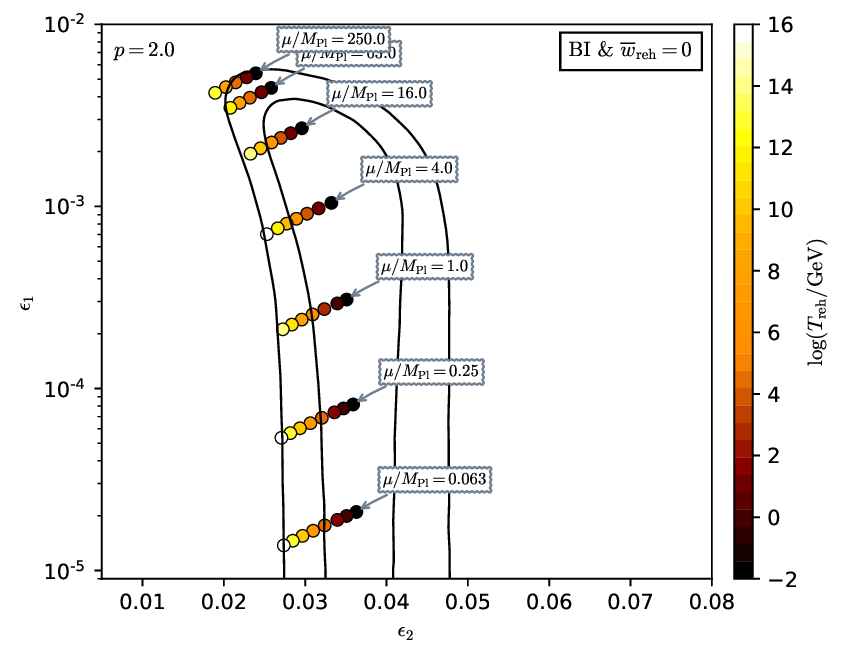}
\caption{Reheating consistent slow-roll predictions for the brane
  inflation models with $p=2$ in the plane $(\nS,r)$ (top panel) and
  the plane $(\epsilon_1,\epsilon_2)$ (bottom panel). The solid
  contours are the one and two-sigma {\data} confidence intervals
  (marginalized over second order slow-roll). For $\mu \gg \Mp$, the
  model predictions approach the curve $r=(8/3)\left(1-\nS\right)$,
  \ie $\epsilon_2=4\epsilon_1$. See figures~\ref{fig:CMBBI3} and
  \ref{fig:CMBBI4} for other values of $p$.}
\label{fig:CMBBI2}
\end{center}
\end{figure}

\begin{figure}[H]
\begin{center}
\includegraphics[width=\wappfig,clip=true]{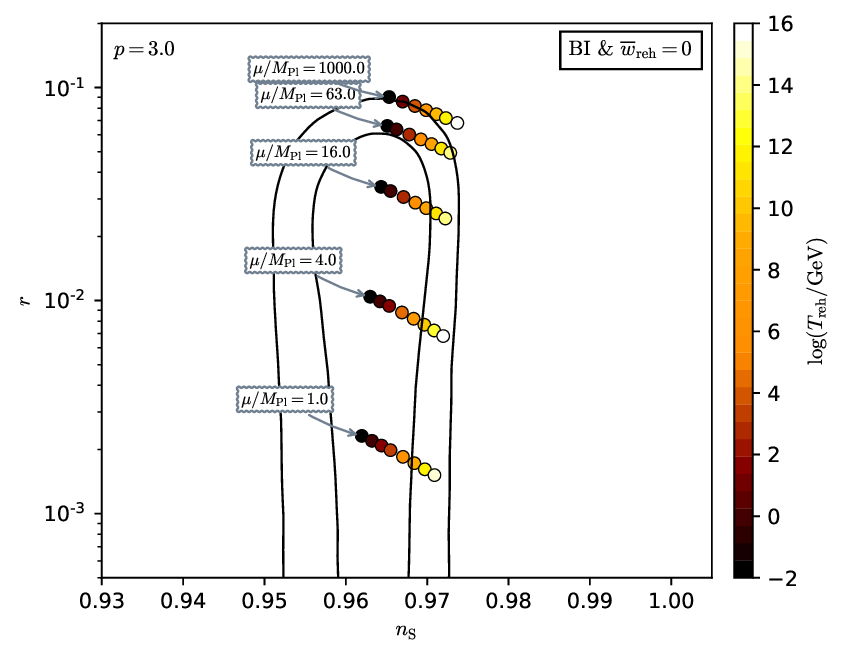}
\includegraphics[width=\wappfig,clip=true]{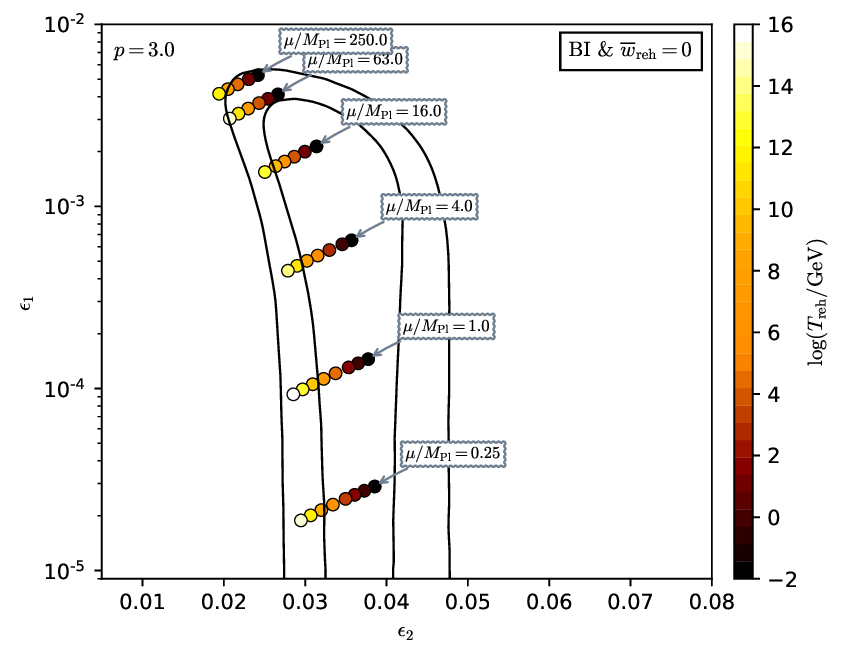}
\caption{Reheating consistent slow-roll predictions for the brane
  inflation models with $p=3$ in the plane $(\nS,r)$ (top panel) and
  the plane $(\epsilon_1,\epsilon_2)$ (bottom panel). The solid
  contours are the one and two-sigma {\data} confidence intervals
  (marginalized over second order slow-roll). For $\mu \gg \Mp$, the
  model predictions approach the curve $r=(8/3)\left(1-\nS\right)$,
  \ie $\epsilon_2=4\epsilon_1$.}
\label{fig:CMBBI3}
\end{center}
\end{figure}

\begin{figure}[H]
\begin{center}
\includegraphics[width=\wappfig,clip=true]{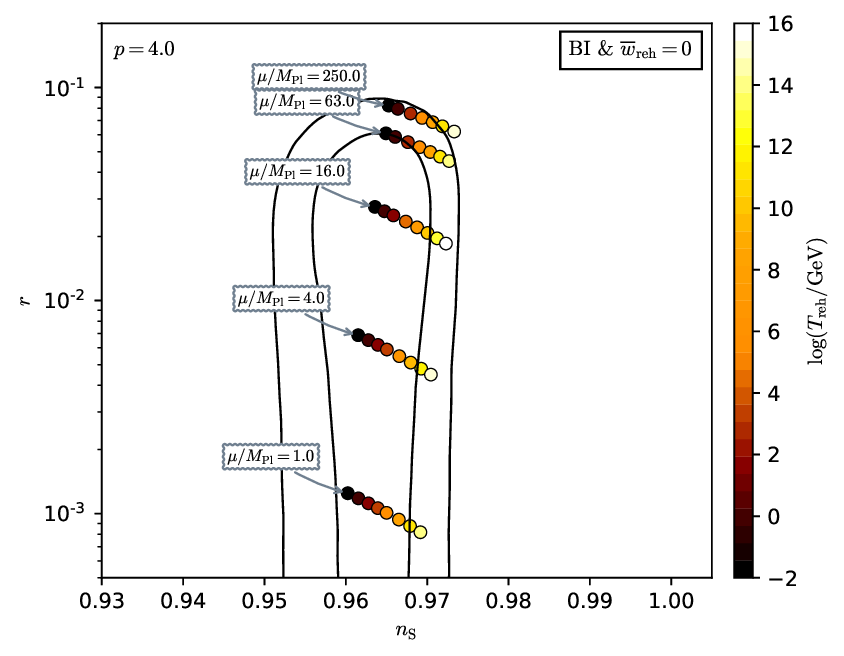}
\includegraphics[width=\wappfig,clip=true]{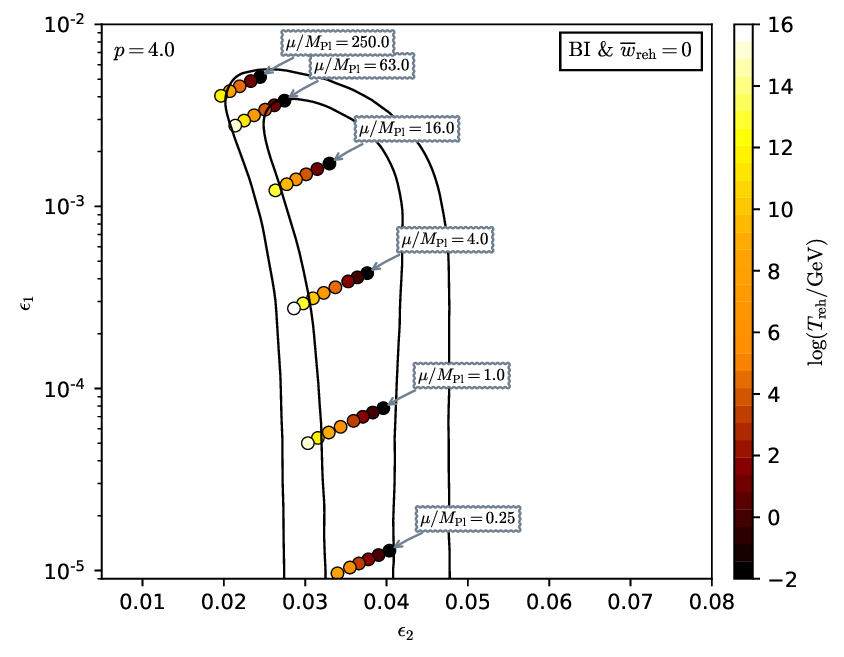}
\caption{Reheating consistent slow-roll predictions for the brane
  inflation models with $p=4$ in the plane $(\nS,r)$ (top panel) and
  the plane $(\epsilon_1,\epsilon_2)$ (bottom panel). The solid
  contours are the one and two-sigma {\data} confidence intervals
  (marginalized over second order slow-roll). For $\mu \gg \Mp$, the
  model predictions approach the curve $r=(8/3)\left(1-\nS\right)$,
  \ie $\epsilon_2=4\epsilon_1$.}
\label{fig:CMBBI4}
\end{center}
\end{figure}

\begin{figure}[H]
\begin{center}
\includegraphics[width=\wappfig,clip=true]{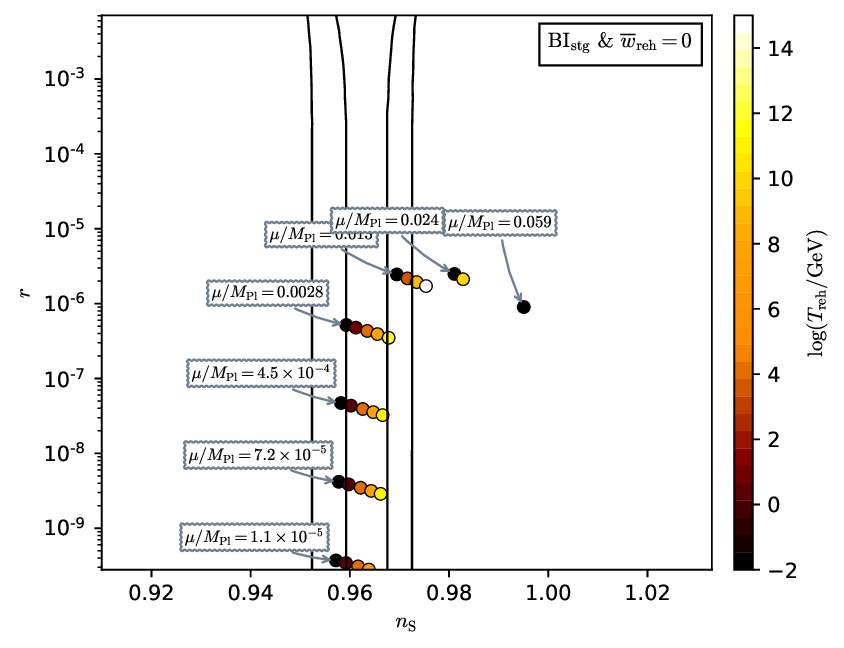}
\includegraphics[width=\wappfig,clip=true]{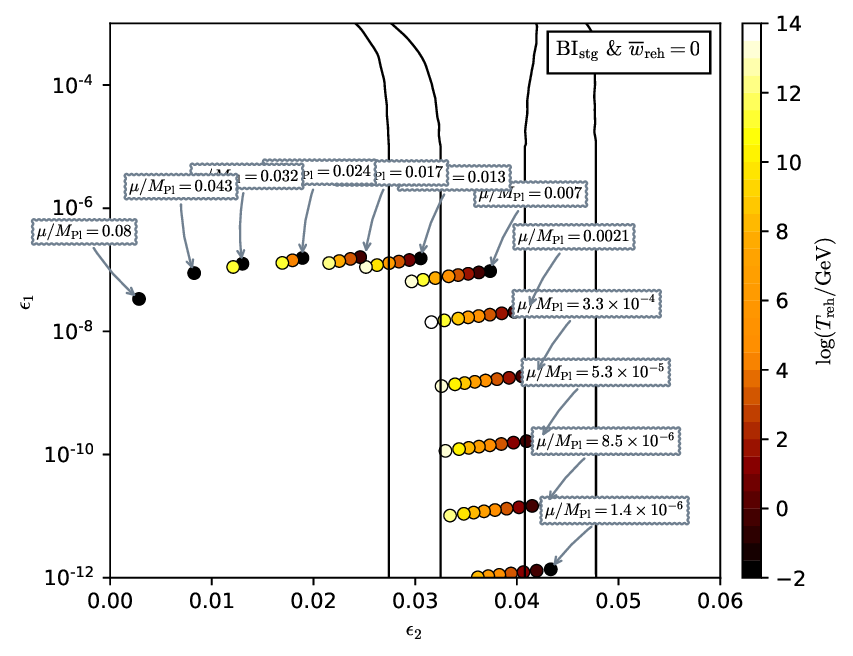}
\caption{Reheating consistent slow-roll predictions for the brane
  inflation models in the string framework ($p=4$, $\mu\ll\Mp$,
  $\calN=5$, $v=16/27$, $\gstrings= 5\times 10^{-3}$, $\alpha'=0.25$),
  in the plane $(\nS,r)$ (top panel) and the plane
  $(\epsilon_1,\epsilon_2)$ (bottom panel). The solid contours are the
  one and two-sigma {\data} confidence intervals (marginalized over
  second order slow-roll). For $\mu/\Mp>0.02$, inflation ends by slow
  roll violation as opposed to tachyonic instability for lower values
  of $\mu/\Mp$.}
\label{fig:CMBBIstg}
\end{center}
\end{figure}

\subsection{KKLT Inflation (\hyperref[sec:bi]{KKLTI})}

\begin{figure}[H]
\begin{center}
\includegraphics[width=\wappfig,clip=true]{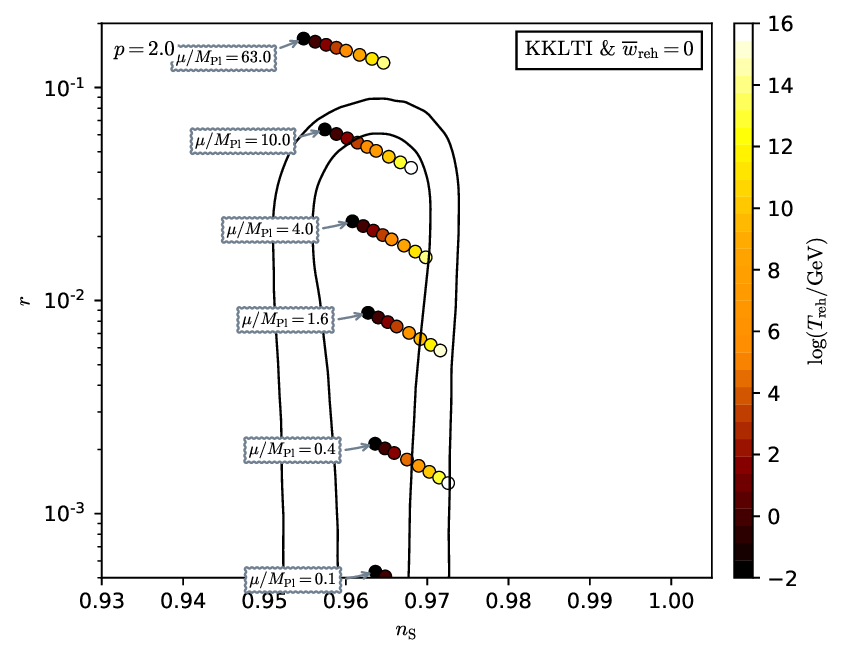}
\includegraphics[width=\wappfig,clip=true]{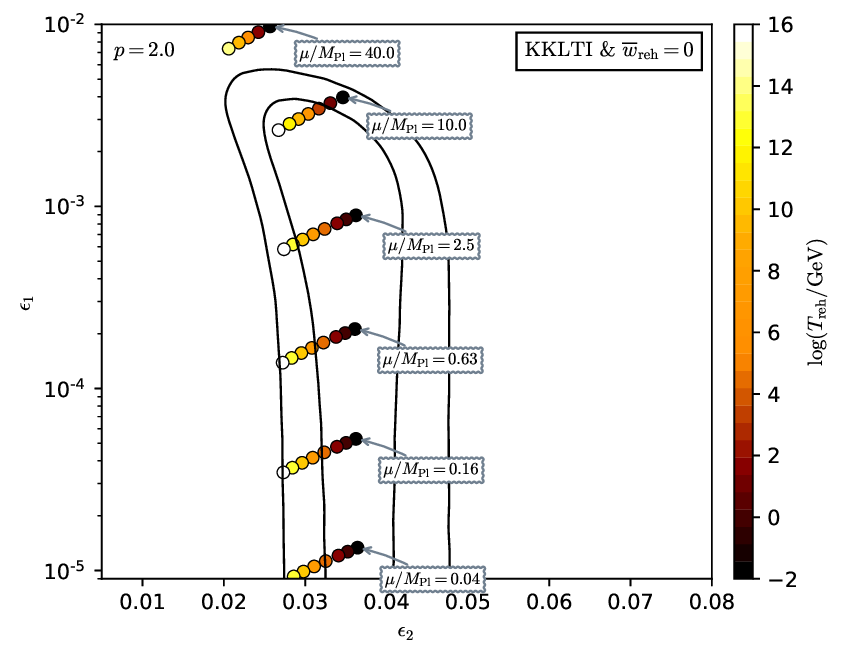}
\caption{Reheating consistent slow-roll predictions for the KKLT
  inflation models with $p=2$ in the plane $(\nS,r)$ (top panel) and
  the plane $(\epsilon_1,\epsilon_2)$ (bottom panel). The solid
  contours are the one and two-sigma {\data} confidence intervals
  (marginalized over second order slow-roll). For $\mu\gg\Mp$, the
  model predictions deviate from the BI's ones, the latter lying along
  the locus $r=(8/3)\left(1-\nS\right)$, \ie
  $\epsilon_2=4\epsilon_1$ (see figure~\ref{fig:CMBBI2}).}
\label{fig:CMBKKLTI2}
\end{center}
\end{figure}

\begin{figure}[H]
\begin{center}
\includegraphics[width=\wappfig,clip=true]{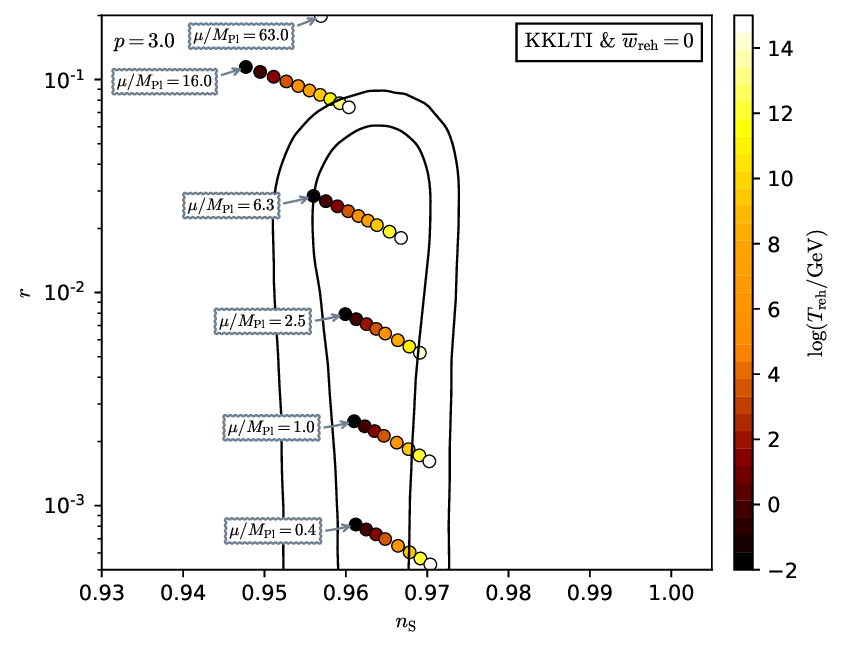}
\includegraphics[width=\wappfig,clip=true]{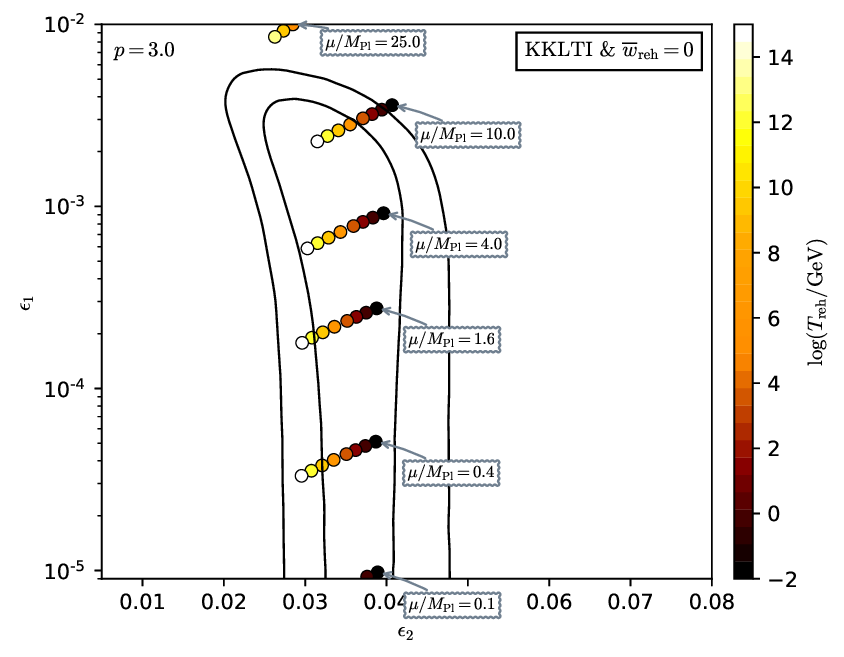}
\caption{Reheating consistent slow-roll predictions for the KKLT
  inflation models with $p=3$ in the plane $(\nS,r)$ (top panel) and
  the plane $(\epsilon_1,\epsilon_2)$ (bottom panel). The solid
  contours are the one and two-sigma {\data} confidence intervals
  (marginalized over second order slow-roll). For $\mu\gg\Mp$, the
  model predictions deviate from the BI's ones, the latter lying along
  the locus $r=(8/3)\left(1-\nS\right)$, \ie $\epsilon_2=4\epsilon_1$
  (see figure~\ref{fig:CMBBI3}).}
\label{fig:CMBKKLTI3}
\end{center}
\end{figure}

\begin{figure}[H]
\begin{center}
\includegraphics[width=\wappfig,clip=true]{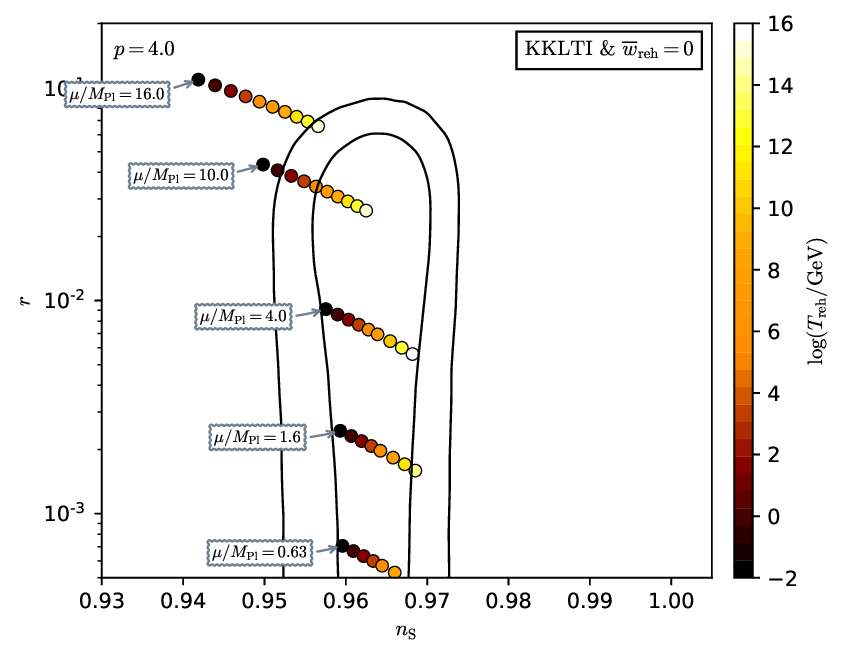}
\includegraphics[width=\wappfig,clip=true]{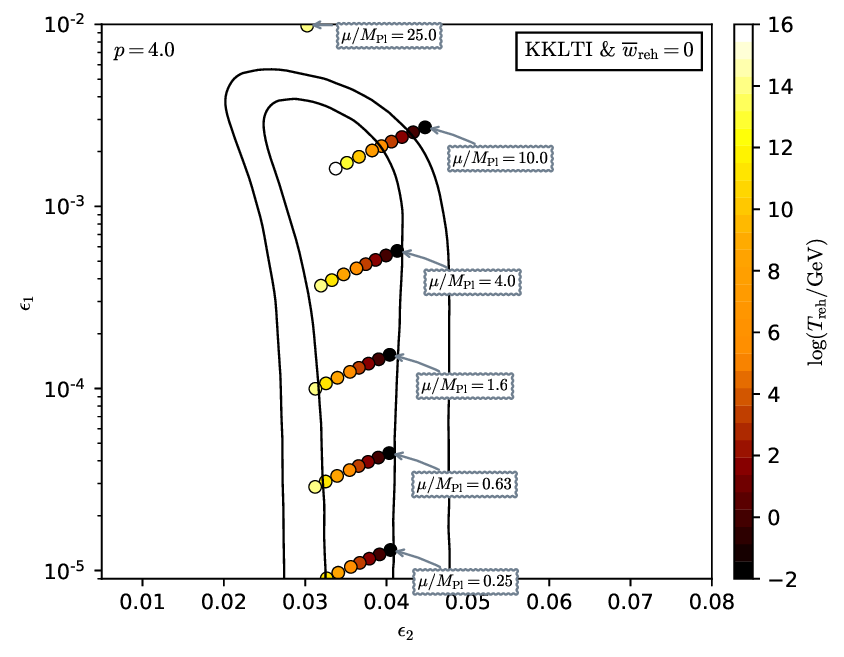}
\caption{Reheating consistent slow-roll predictions for the KKLT
  inflation models with $p=4$ in the plane $(\nS,r)$ (top panel) and
  the plane $(\epsilon_1,\epsilon_2)$ (bottom panel). The solid
  contours are the one and two-sigma {\data} confidence intervals
  (marginalized over second order slow-roll). For $\mu\gg\Mp$, the
  model predictions deviate from the BI's ones, the latter lying along
  the locus $r=(8/3)\left(1-\nS\right)$, \ie $\epsilon_2=4\epsilon_1$
  (see figure~\ref{fig:CMBBI4}).}
\label{fig:CMBKKLTI4}
\end{center}
\end{figure}

\begin{figure}[H]
\begin{center}
\includegraphics[width=\wappfig,clip=true]{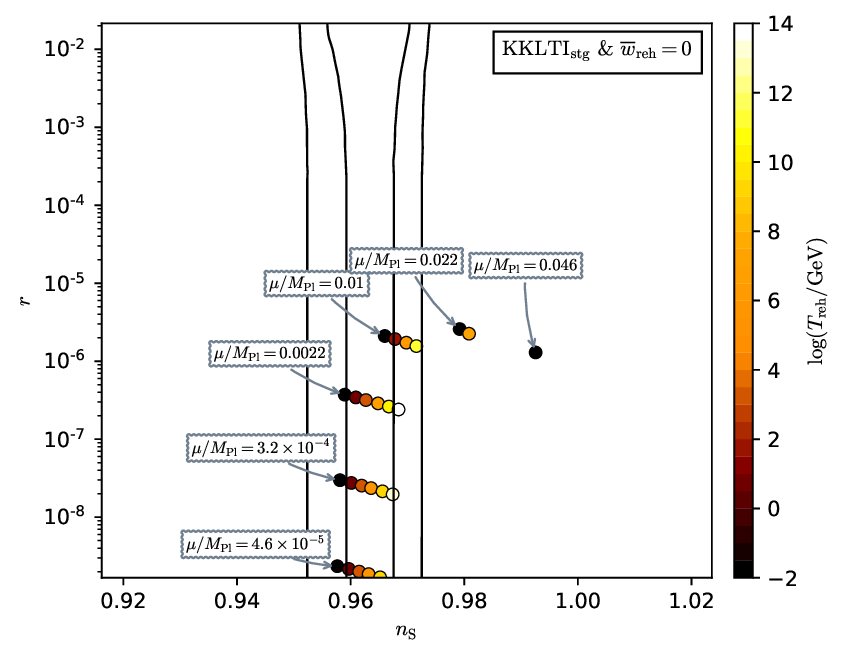}
\includegraphics[width=\wappfig,clip=true]{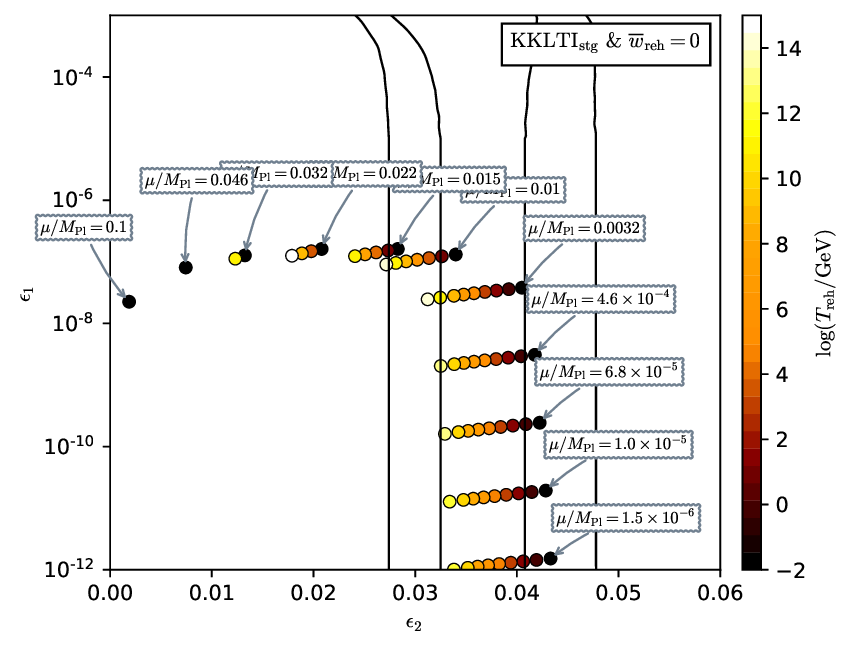}
\caption{Reheating consistent slow-roll predictions for the KKLT
  inflation models in the string framework ($p=4$, $\mu\ll\Mp$,
  $\calN=5$, $v=16/27$, $\gstrings=0.005$, $\alpha'=0.25$) for the, in
  the plane $(\nS,r)$ (top panel) and the plane
  $(\epsilon_1,\epsilon_2)$ (bottom panel). The solid contours are the
  one and two-sigma {\data} confidence intervals (marginalized over
  second order slow-roll). For $\mu/\Mp>0.02$, inflation ends by slow
  roll violation as opposed to tachyonic instability for lower values
  of $\mu$. Because $\mu \ll \Mp$, the model predictions are
  undistinguishable from the BI's ones (see
  figure~\ref{fig:CMBBIstg}).}
\label{fig:CMBKKLTIstg}
\end{center}
\end{figure}

\subsection{String Axion Inflation I 1 (\hyperref[sec:saii]{SAII1})}

\begin{figure}[H]
\begin{center}
\includegraphics[width=\wappfig,clip=true]{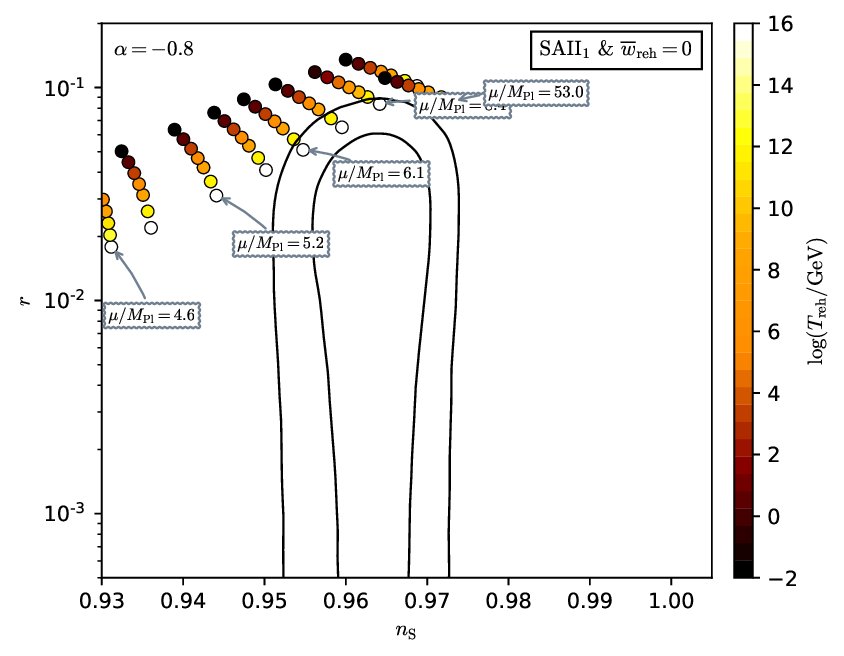}
\includegraphics[width=\wappfig,clip=true]{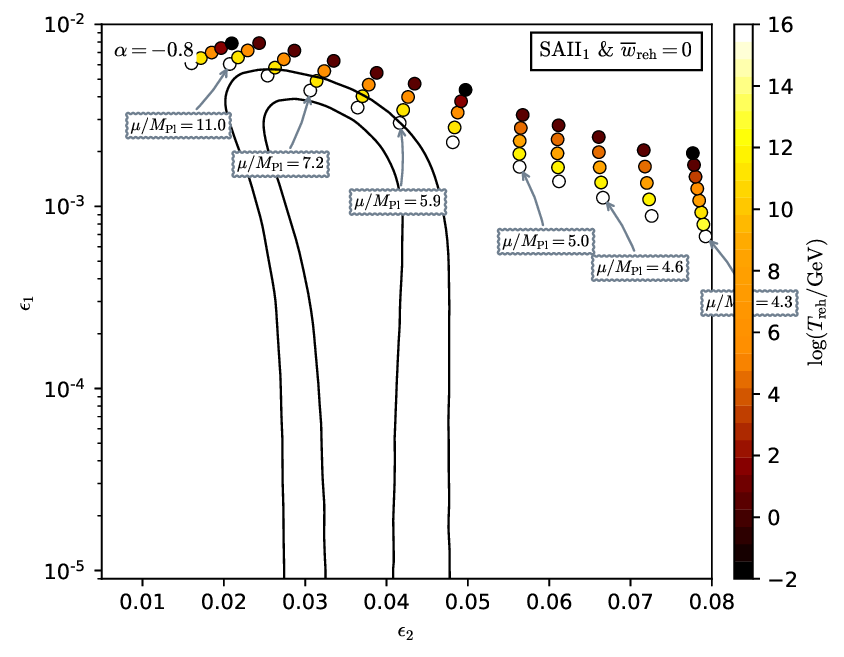}
\caption{Reheating consistent slow-roll predictions for the
  SAII1 inflation models with $\alpha=-0.8$
  in the plane $(\nS,r)$ (top panel) and the plane
  $(\epsilon_1,\epsilon_2)$ (bottom panel). The solid contours are the
  one and two-sigma {\data} confidence intervals (marginalized over
  second order slow-roll). The model predictions for larger values of
  $\alpha$ are represented in figures~\ref{fig:CMBSAII1_1} to
  \ref{fig:CMBSAII1_2}.}
\label{fig:CMBSAII1}
\end{center}
\end{figure}

\begin{figure}[H]
\begin{center}
\includegraphics[width=\wappfig,clip=true]{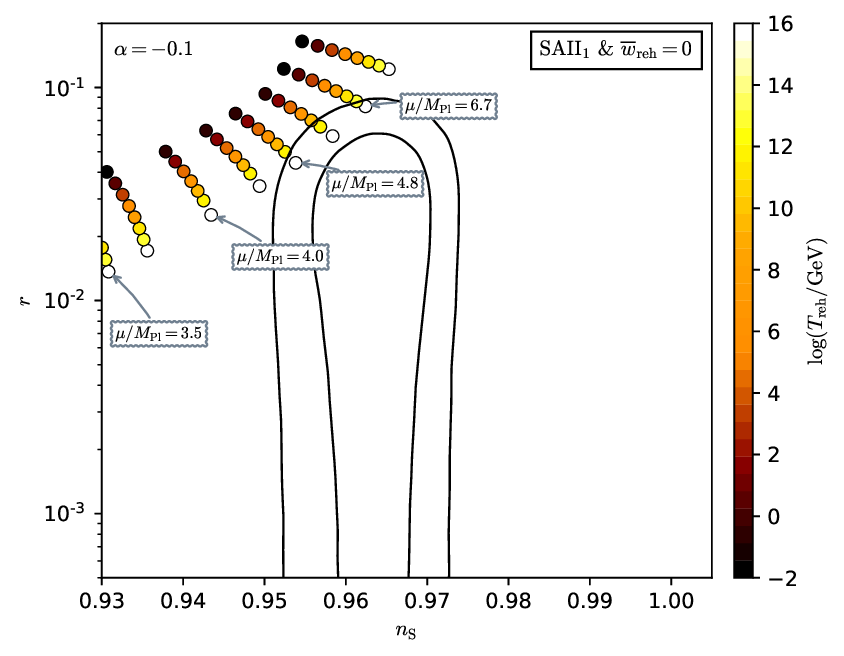}
\includegraphics[width=\wappfig,clip=true]{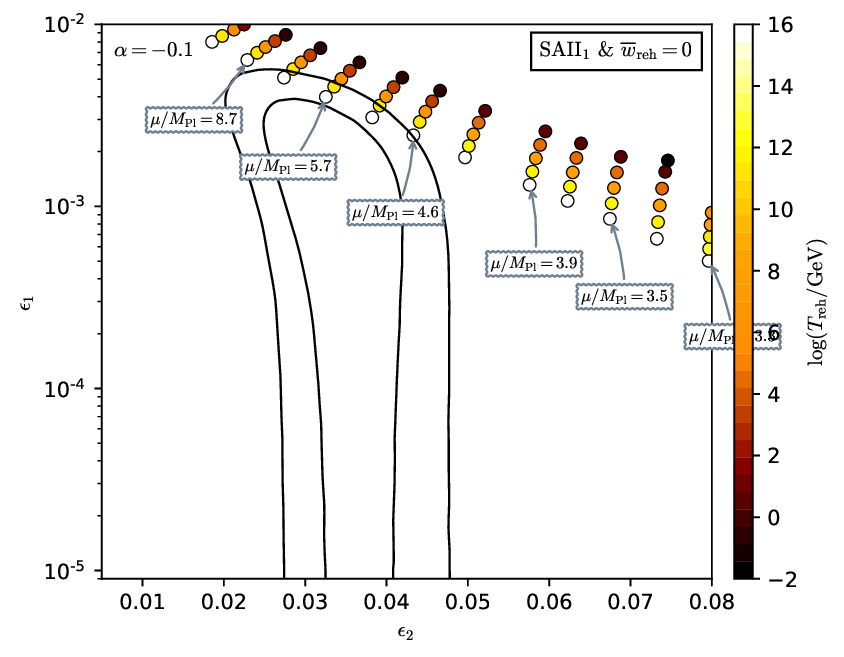}
\caption{Reheating consistent slow-roll predictions for the
  SAII1 inflation models with $\alpha=-0.1$
  in the plane $(\nS,r)$ (top panel) and the plane
  $(\epsilon_1,\epsilon_2)$ (bottom panel). The solid contours are the
  one and two-sigma {\data} confidence intervals (marginalized over
  second order slow-roll).}
\label{fig:CMBSAII1_1}
\end{center}
\end{figure}

\begin{figure}[H]
\begin{center}
\includegraphics[width=\wappfig,clip=true]{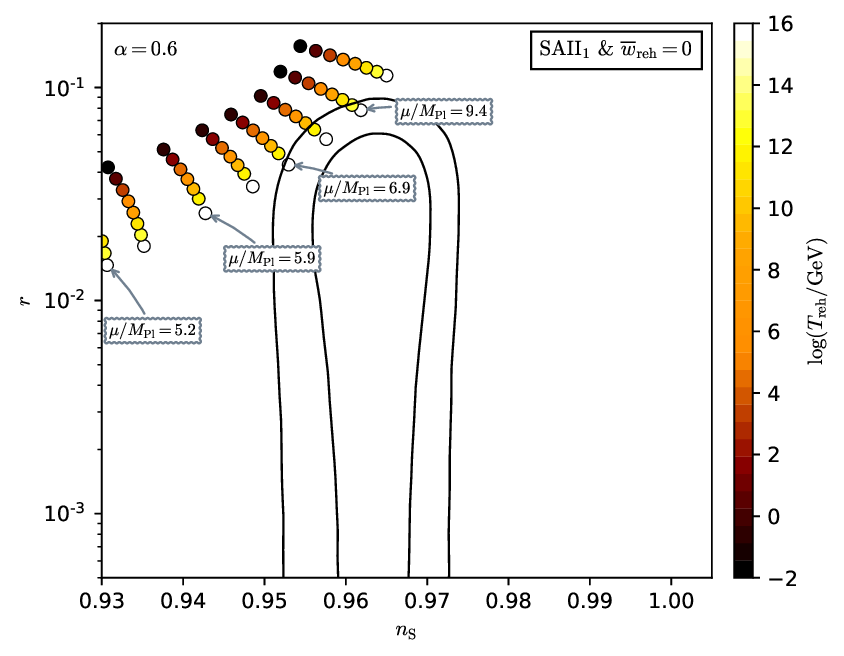}
\includegraphics[width=\wappfig,clip=true]{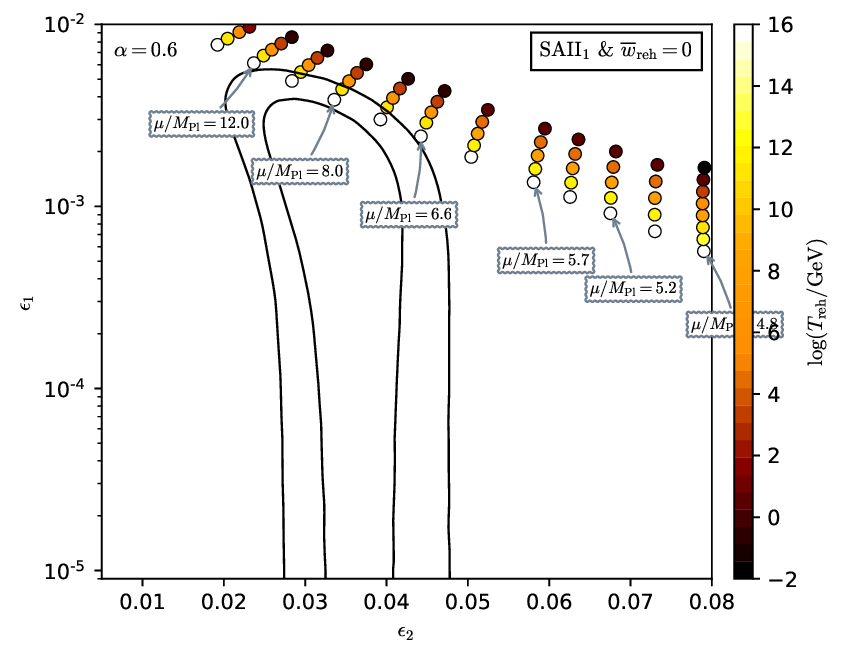}
\caption{Reheating consistent slow-roll predictions for the
  SAII1 inflation models with $\alpha=0.6$
  in the plane $(\nS,r)$ (top panel) and the plane
  $(\epsilon_1,\epsilon_2)$ (bottom panel). The solid contours are the
  one and two-sigma {\data} confidence intervals (marginalized over
  second order slow-roll).}
\label{fig:CMBSAII1_2}
\end{center}
\end{figure}

\subsection{String Axion Inflation I 2 (\hyperref[sec:saii]{SAII2})}

\begin{figure}[H]
\begin{center}
\includegraphics[width=\wappfig,clip=true]{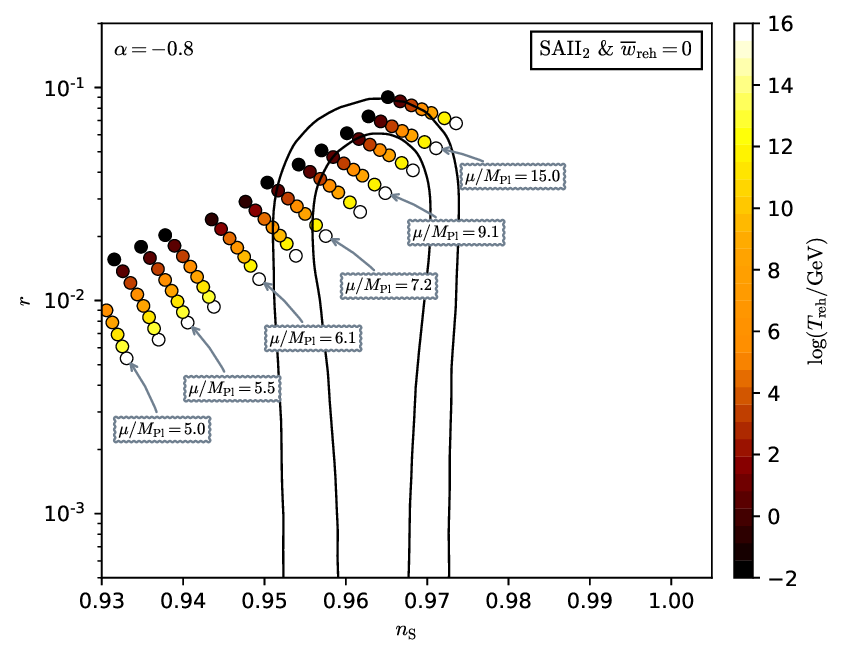}
\includegraphics[width=\wappfig,clip=true]{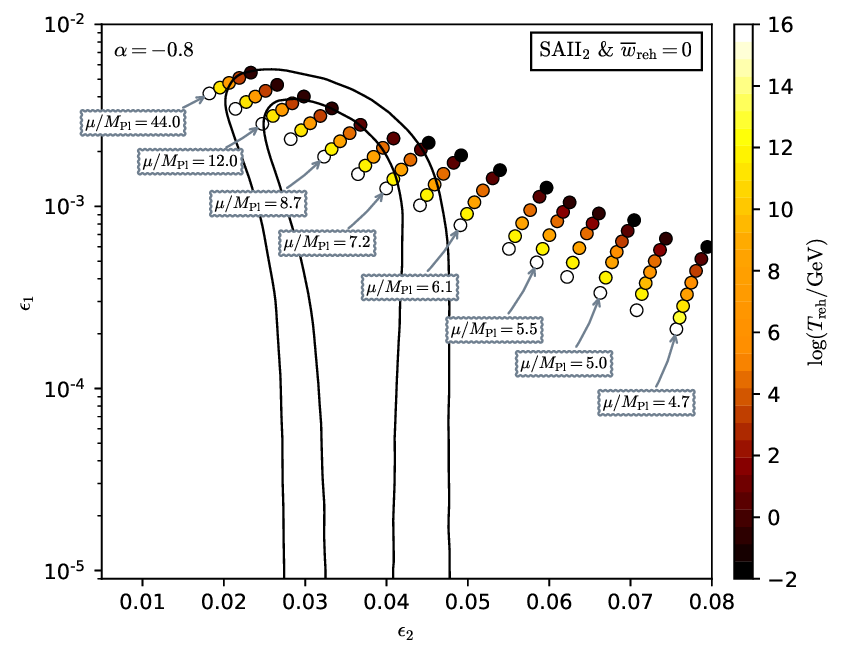}
\caption{Reheating consistent slow-roll predictions for the
  SAII2 inflation models with $\alpha=-0.8$
  in the plane $(\nS,r)$ (top panel) and the plane
  $(\epsilon_1,\epsilon_2)$ (bottom panel). The solid contours are the
  one and two-sigma {\data} confidence intervals (marginalized over
  second order slow-roll). The model predictions for larger values of
  $\alpha$ are represented in figures~\ref{fig:CMBSAII2_1} to
  \ref{fig:CMBSAII2_2}.}
\label{fig:CMBSAII2}
\end{center}
\end{figure}

\begin{figure}[H]
\begin{center}
\includegraphics[width=\wappfig,clip=true]{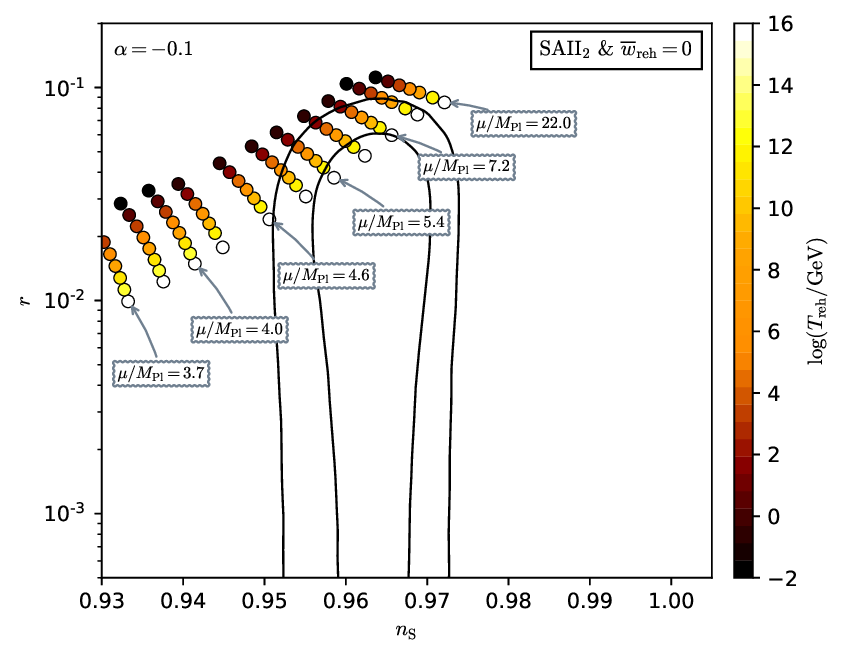}
\includegraphics[width=\wappfig,clip=true]{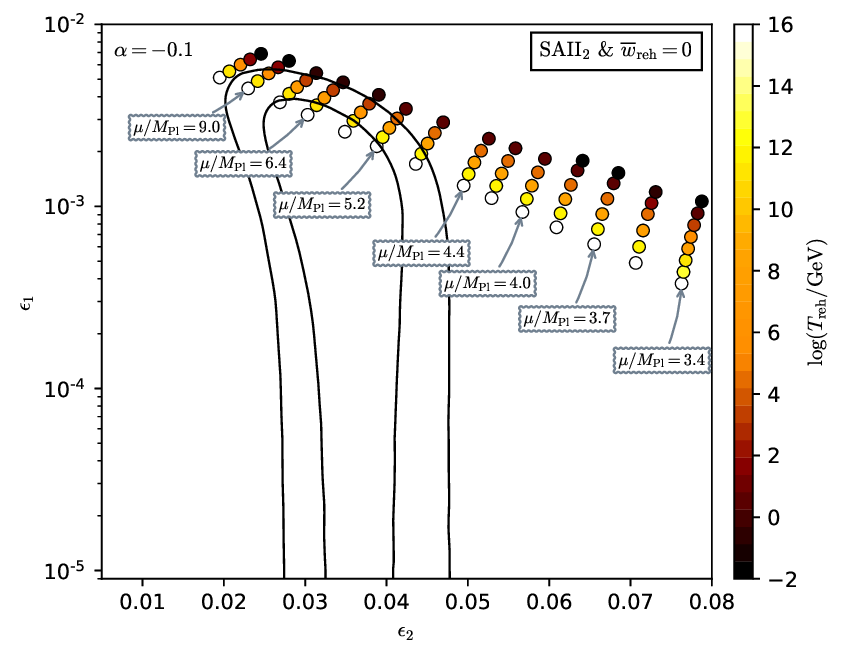}
\caption{Reheating consistent slow-roll predictions for the
  SAII2 inflation models with $\alpha=-0.1$
  in the plane $(\nS,r)$ (top panel) and the plane
  $(\epsilon_1,\epsilon_2)$ (bottom panel). The solid contours are the
  one and two-sigma {\data} confidence intervals (marginalized over
  second order slow-roll).}
\label{fig:CMBSAII2_1}
\end{center}
\end{figure}

\begin{figure}[H]
\begin{center}
\includegraphics[width=\wappfig,clip=true]{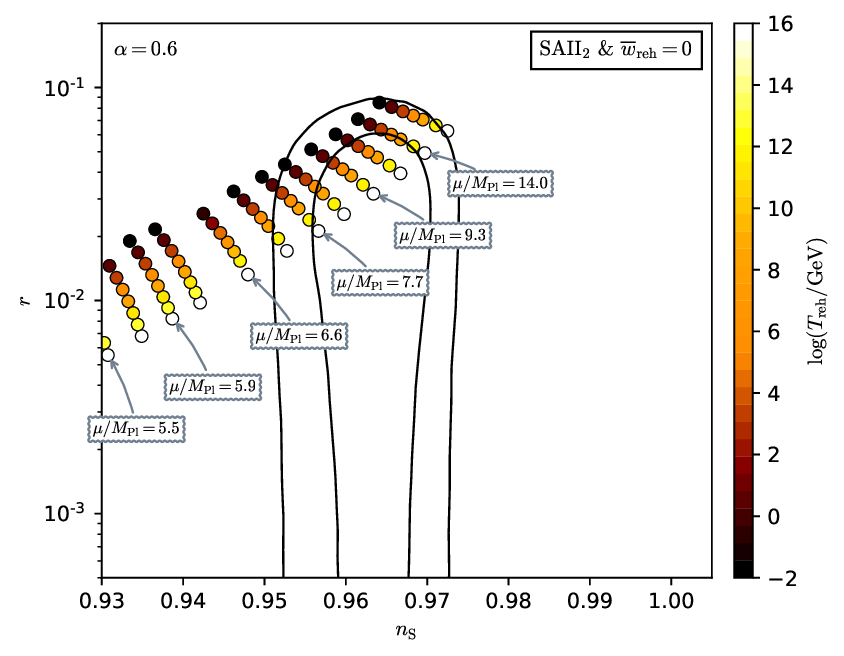}
\includegraphics[width=\wappfig,clip=true]{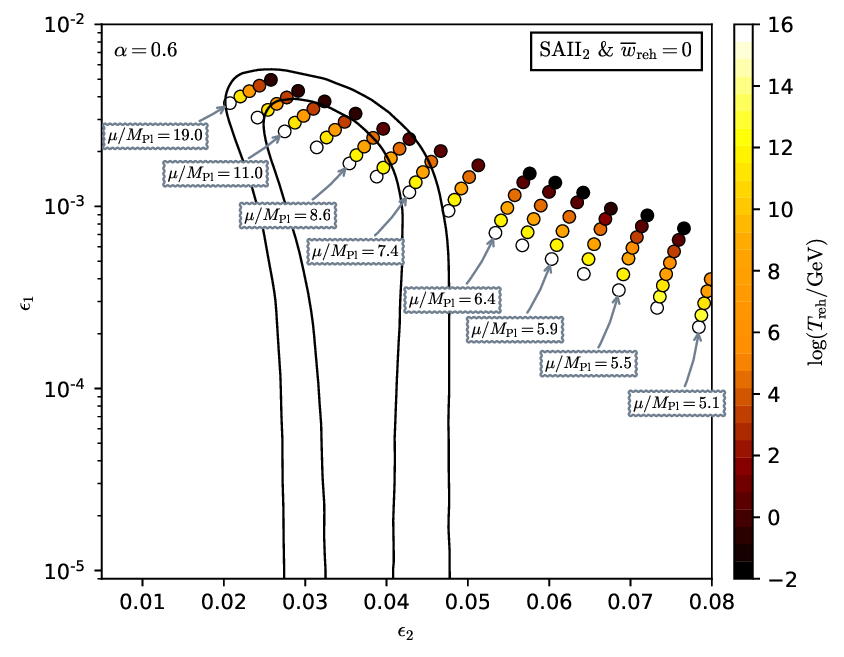}
\caption{Reheating consistent slow-roll predictions for the
  SAII2 inflation models with $\alpha=0.6$
  in the plane $(\nS,r)$ (top panel) and the plane
  $(\epsilon_1,\epsilon_2)$ (bottom panel). The solid contours are the
  one and two-sigma {\data} confidence intervals (marginalized over
  second order slow-roll).}
\label{fig:CMBSAII2_2}
\end{center}
\end{figure}

\subsection{Mukhanov Inflation (\hyperref[sec:vfmi]{VFMI})}

\begin{figure}[H]
\begin{center}
\includegraphics[width=\wappfig,clip=true]{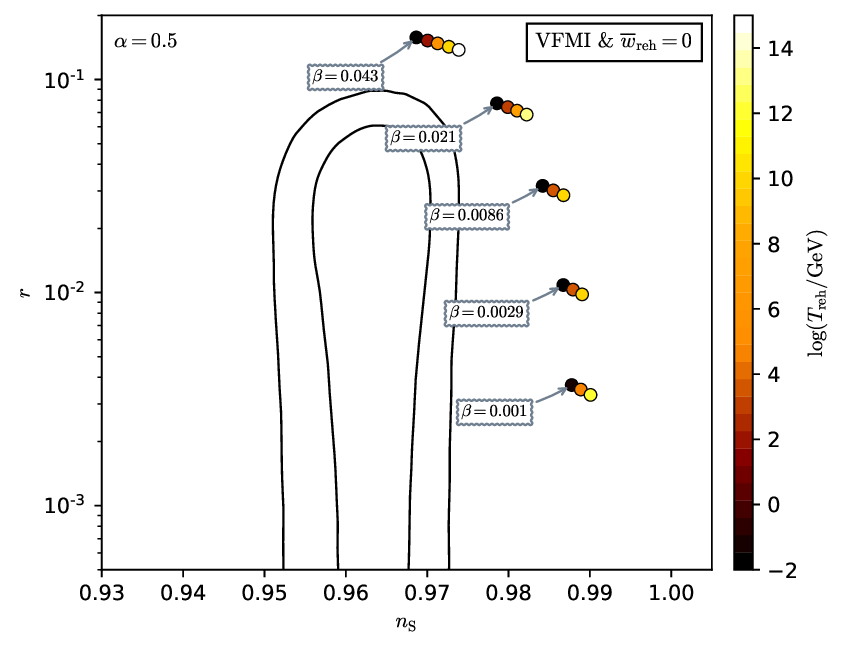}
\includegraphics[width=\wappfig,clip=true]{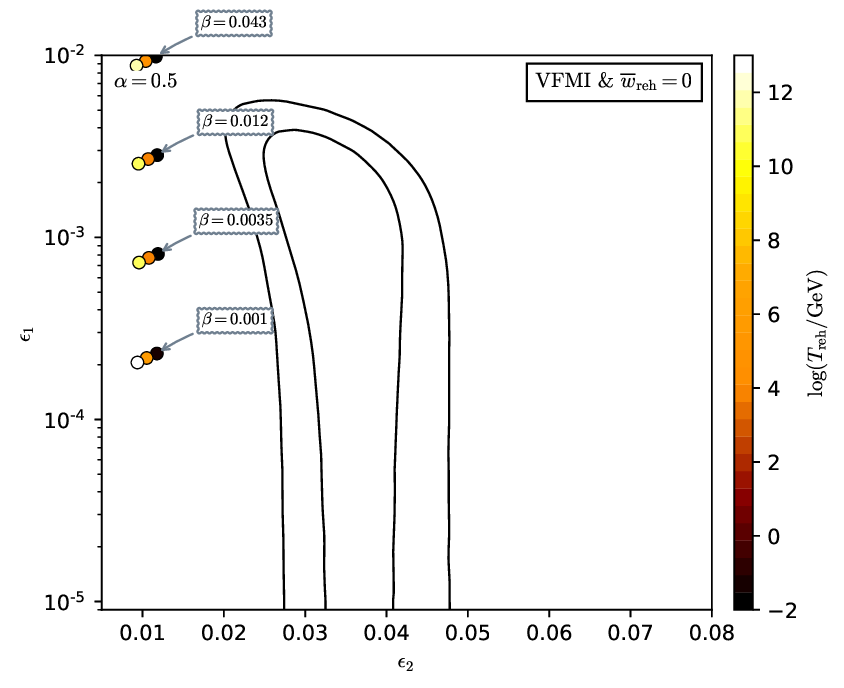}
\caption{Reheating consistent slow-roll predictions for the
  Mukhanov inflationary models with $\alpha=1/2$
  in the plane $(\nS,r)$ (top panel) and the plane
  $(\epsilon_1,\epsilon_2)$ (bottom panel). The solid contours are the
  one and two-sigma {\data} confidence intervals (marginalized over
  second order slow-roll). The model predictions for larger values of
  $\alpha$ are represented in figures~\ref{fig:CMBVFMI_1} to
  \ref{fig:CMBVFMI_2}.}
\label{fig:CMBVFMI}
\end{center}
\end{figure}

\begin{figure}[H]
\begin{center}
\includegraphics[width=\wappfig,clip=true]{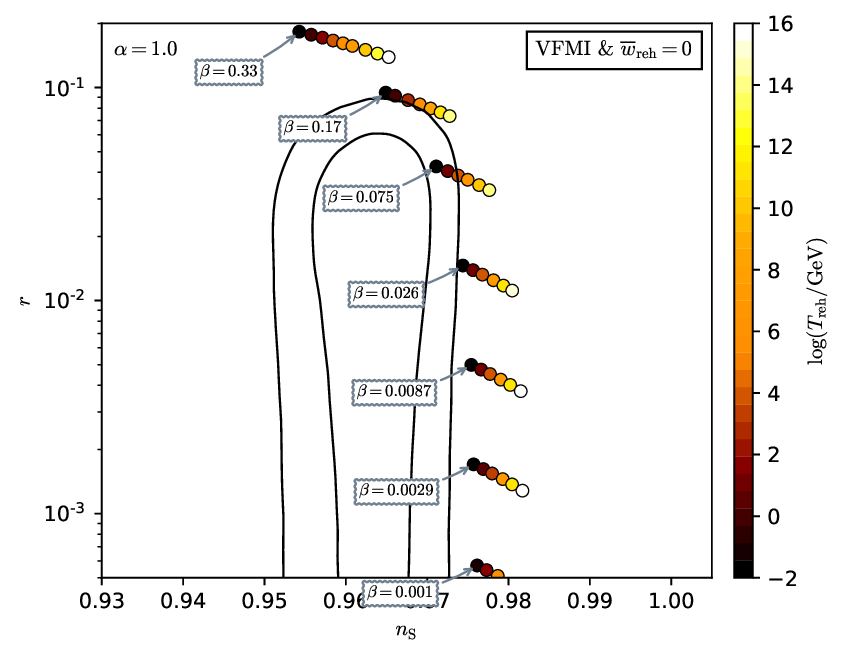}
\includegraphics[width=\wappfig,clip=true]{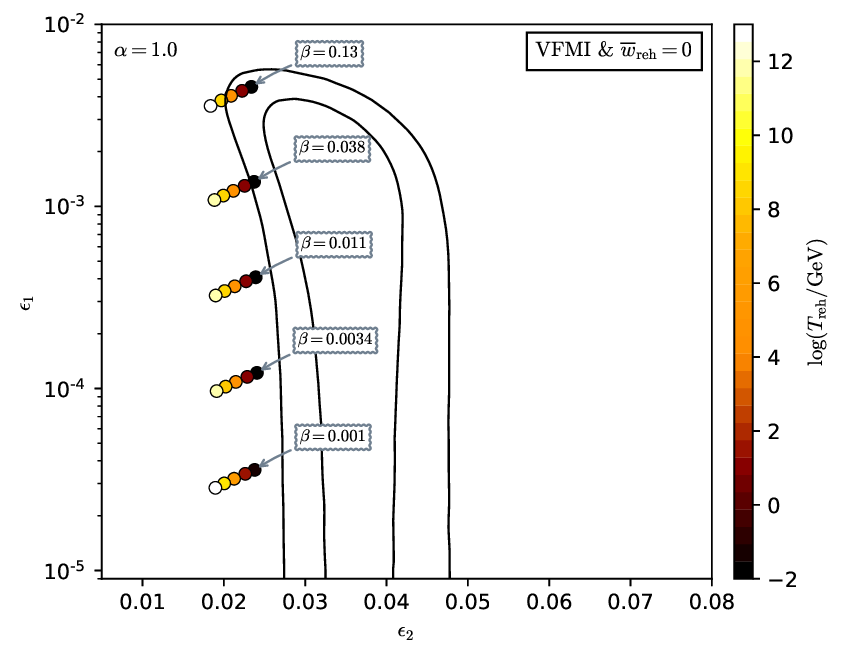}
\caption{Reheating consistent slow-roll predictions for the Mukhanov
  inflationary models with $\alpha=1$ in the plane $(\nS,r)$ (top panel)
  and the plane $(\epsilon_1,\epsilon_2)$ (bottom panel). The solid
  contours are the one and two-sigma {\data} confidence intervals
  (marginalized over second order slow-roll).}
\label{fig:CMBVFMI_1}
\end{center}
\end{figure}

\begin{figure}[H]
\begin{center}
\includegraphics[width=\wappfig,clip=true]{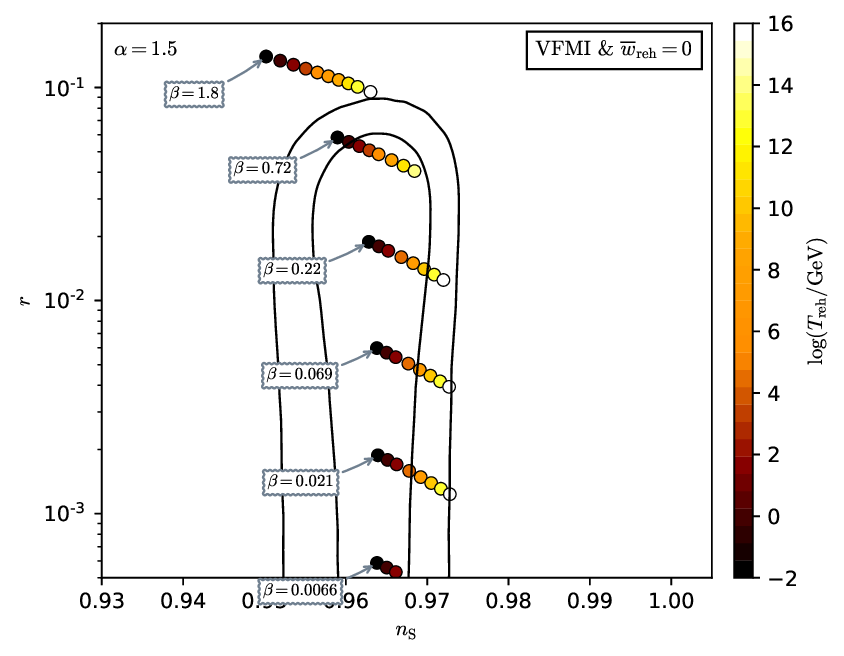}
\includegraphics[width=\wappfig,clip=true]{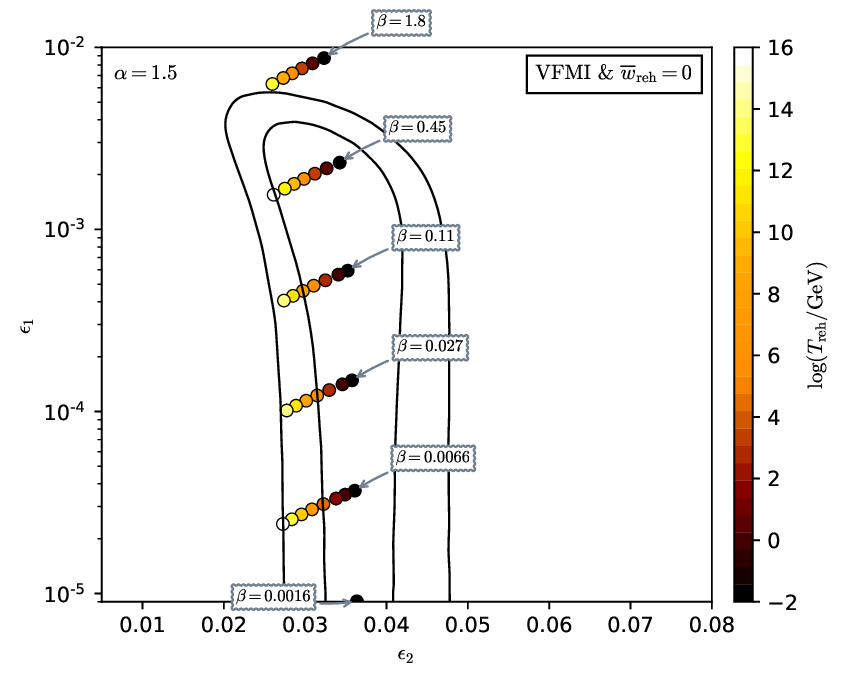}
\caption{Reheating consistent slow-roll predictions for the Mukhanov
  inflationary models with $\alpha=3/2$ in the plane $(\nS,r)$ (top
  panel) and the plane $(\epsilon_1,\epsilon_2)$ (bottom panel). The
  solid contours are the one and two-sigma {\data} confidence
  intervals (marginalized over second order slow-roll).}
\label{fig:CMBVFMI_2}
\end{center}
\end{figure}

\begin{figure}[H]
\begin{center}
\includegraphics[width=\wappfig,clip=true]{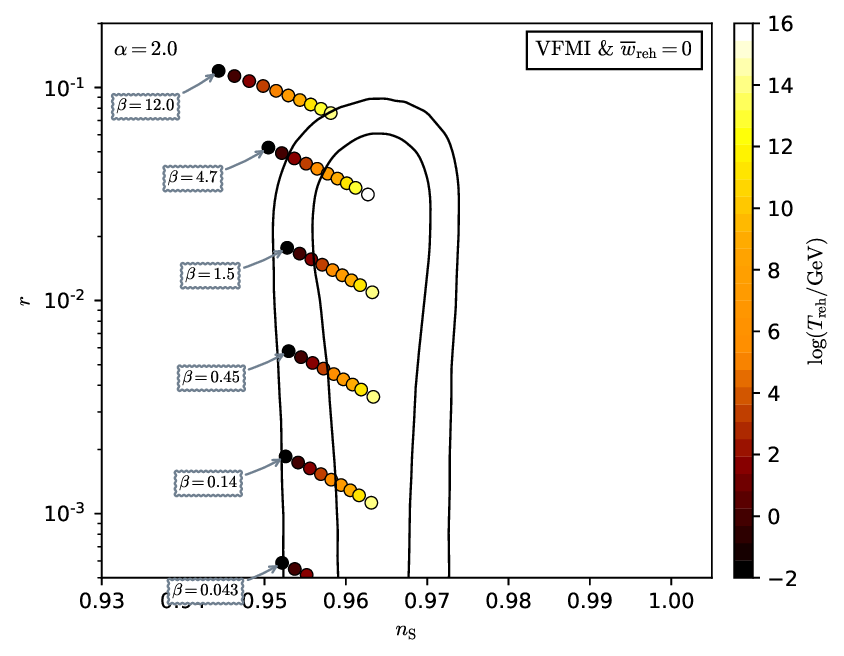}
\includegraphics[width=\wappfig,clip=true]{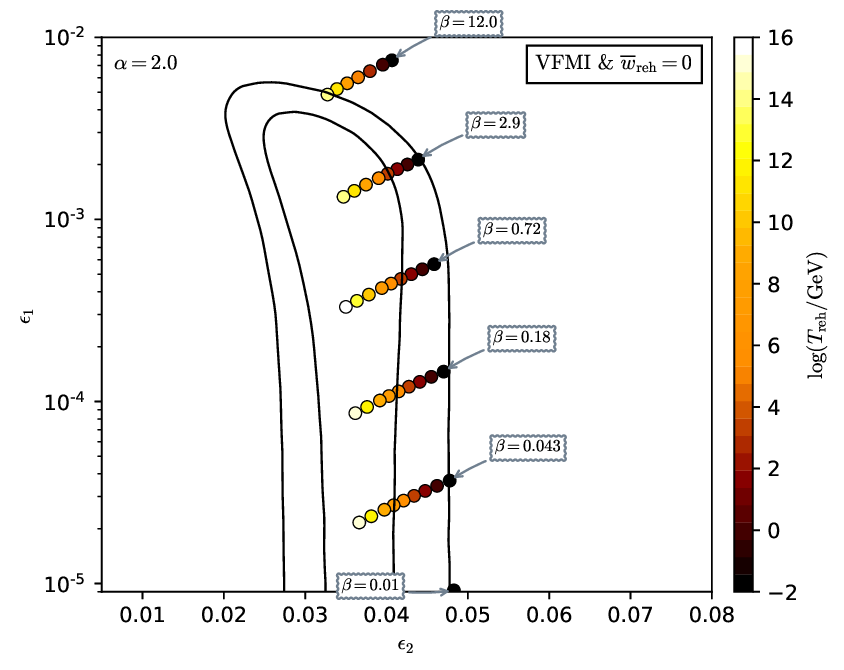}
\caption{Reheating consistent slow-roll predictions for the Mukhanov
  inflationary models with $\alpha=2$ in the plane $(\nS,r)$ (top panel)
  and the plane $(\epsilon_1,\epsilon_2)$ (bottom panel). The solid
  contours are the one and two-sigma {\data} confidence intervals
  (marginalized over second order slow-roll).}
\label{fig:CMBVFMI_3}
\end{center}
\end{figure}

\begin{figure}[H]
\begin{center}
\includegraphics[width=\wappfig,clip=true]{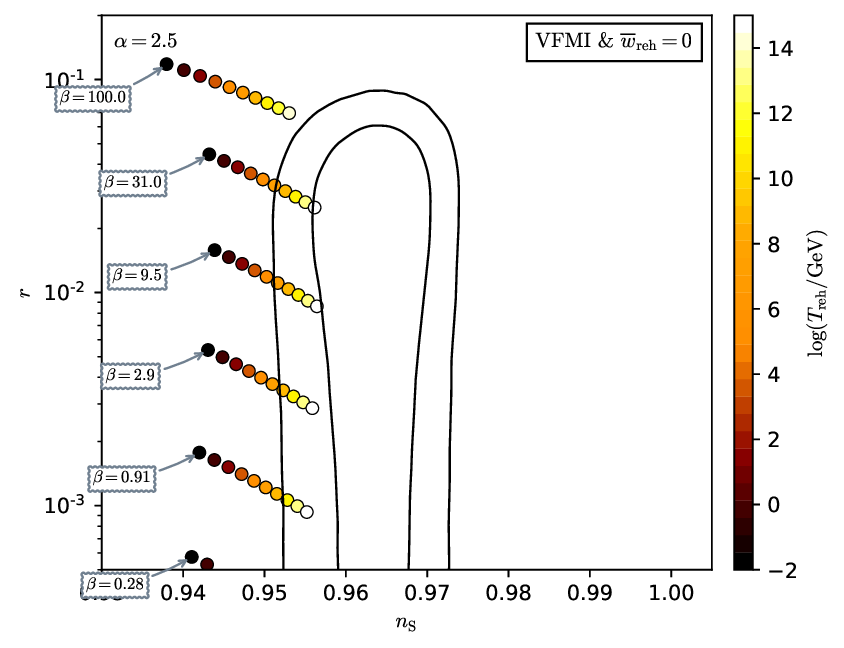}
\includegraphics[width=\wappfig,clip=true]{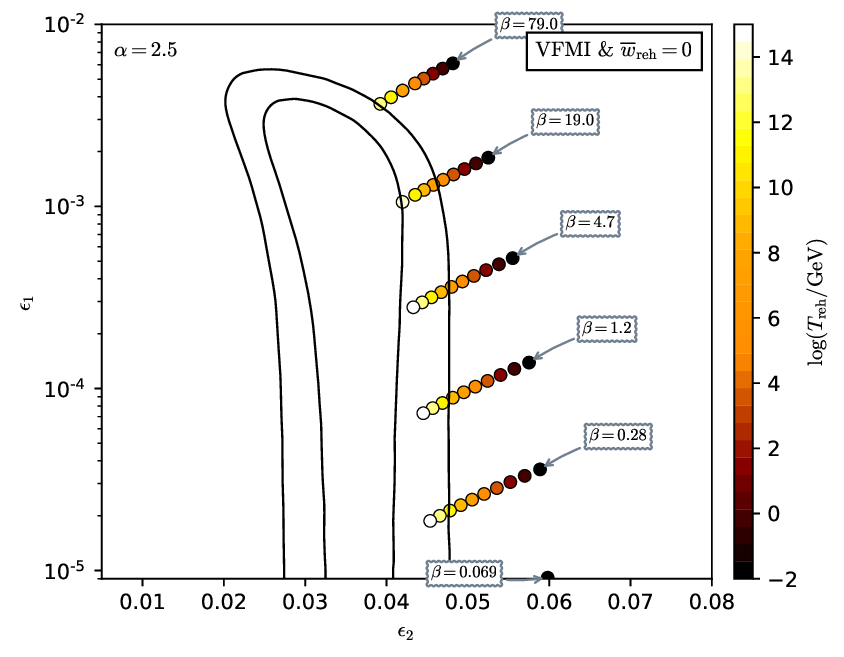}
\caption{Reheating consistent slow-roll predictions for the Mukhanov
  inflationary models with $\alpha=5/2$ in the plane $(\nS,r)$ (top
  panel) and the plane $(\epsilon_1,\epsilon_2)$ (bottom panel). The
  solid contours are the one and two-sigma {\data} confidence
  intervals (marginalized over second order slow-roll).}
\label{fig:CMBVFMI_4}
\end{center}
\end{figure}

\begin{figure}[H]
\begin{center}
\includegraphics[width=\wappfig,clip=true]{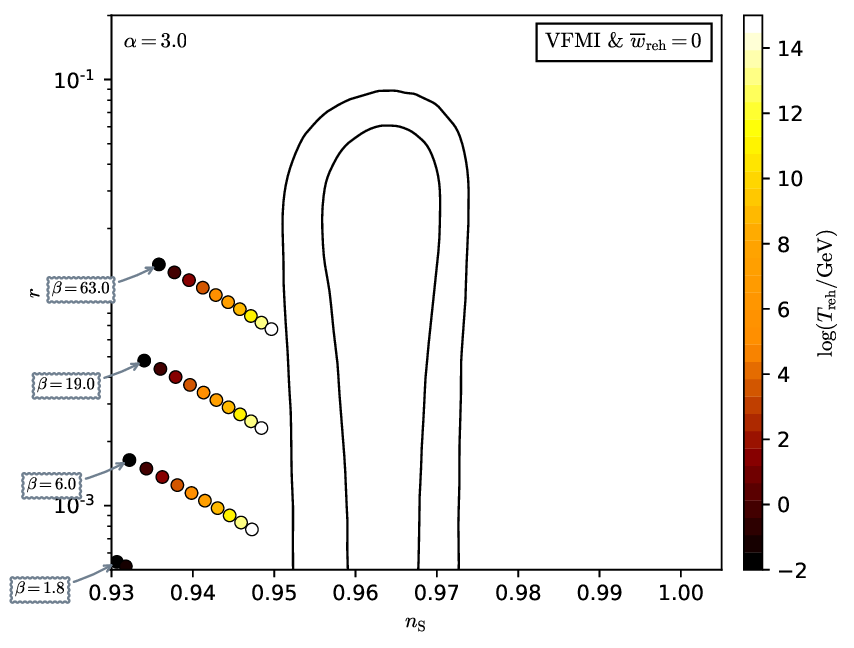}
\includegraphics[width=\wappfig,clip=true]{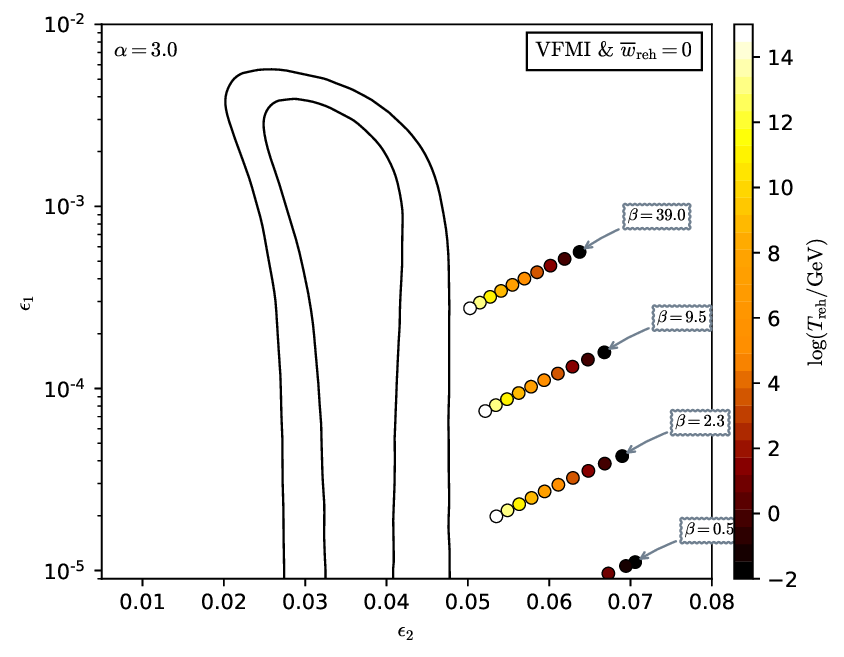}
\caption{Reheating consistent slow-roll predictions for the Mukhanov
  inflationary models with $\alpha=3$ in the plane $(\nS,r)$ (top panel)
  and the plane $(\epsilon_1,\epsilon_2)$ (bottom panel). The solid
  contours are the one and two-sigma {\data} confidence intervals
  (marginalized over second order slow-roll). See
  figures~\ref{fig:CMBVFMI} to \ref{fig:CMBVFMI_4} for smaller values
  of $\alpha$.}
\label{fig:CMBVFMI_5}
\end{center}
\end{figure}

\subsection{Fibre Inflation (\hyperref[sec:fi]{FI})}

\begin{figure}[H]
\begin{center}
\includegraphics[width=\wappfig,clip=true]{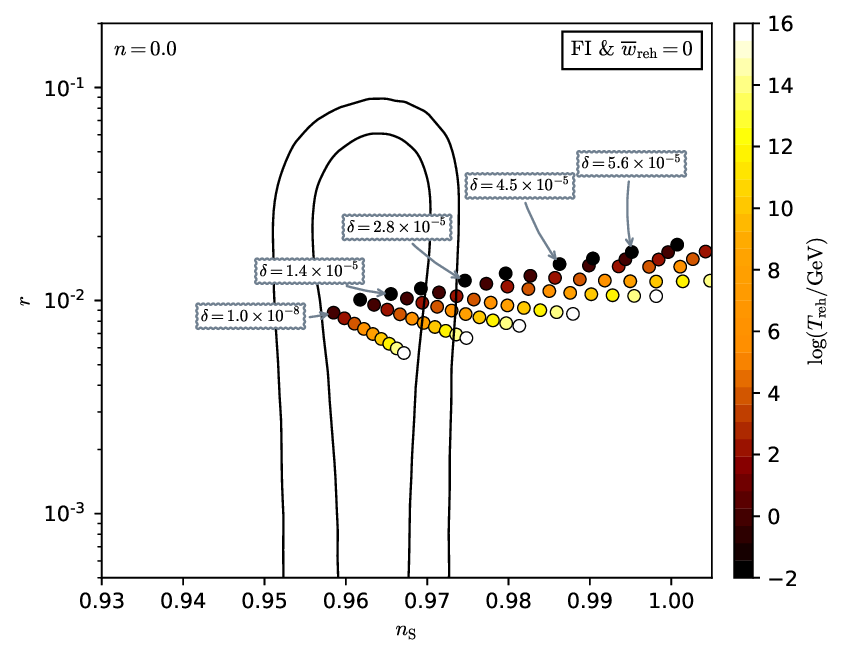}
\includegraphics[width=\wappfig,clip=true]{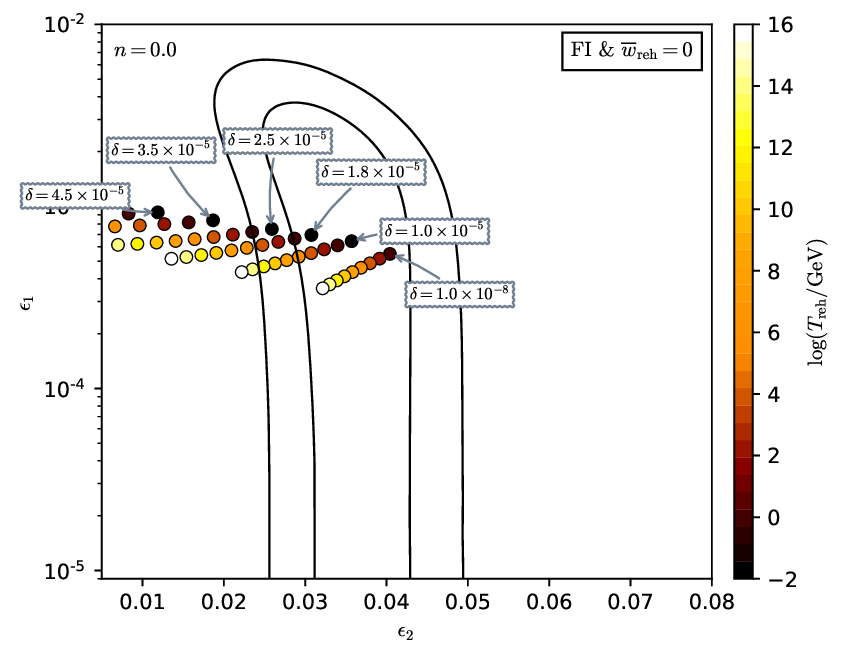}
\caption{Reheating consistent slow-roll predictions for the Fibre Inflation models with $n=0$, in the
  plane $(\nS,r)$ (top panel) and the plane $(\epsilon_1,\epsilon_2)$
  (bottom panel). The solid contours are the one and two-sigma {\data}
  confidence intervals (marginalized over second order slow-roll).}
\label{fig:CMBFI_0}
\end{center}
\end{figure}

\begin{figure}[H]
\begin{center}
\includegraphics[width=\wappfig,clip=true]{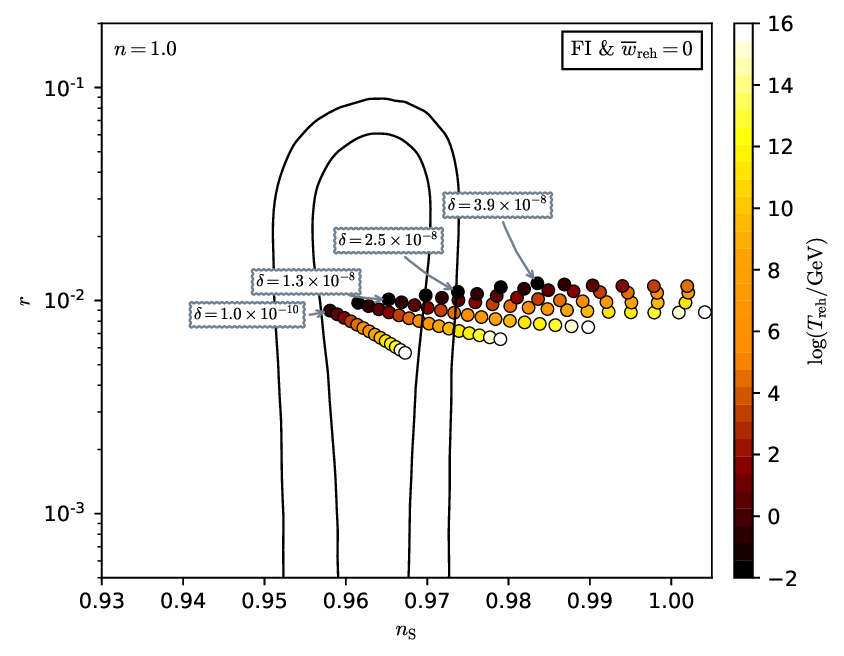}
\includegraphics[width=\wappfig,clip=true]{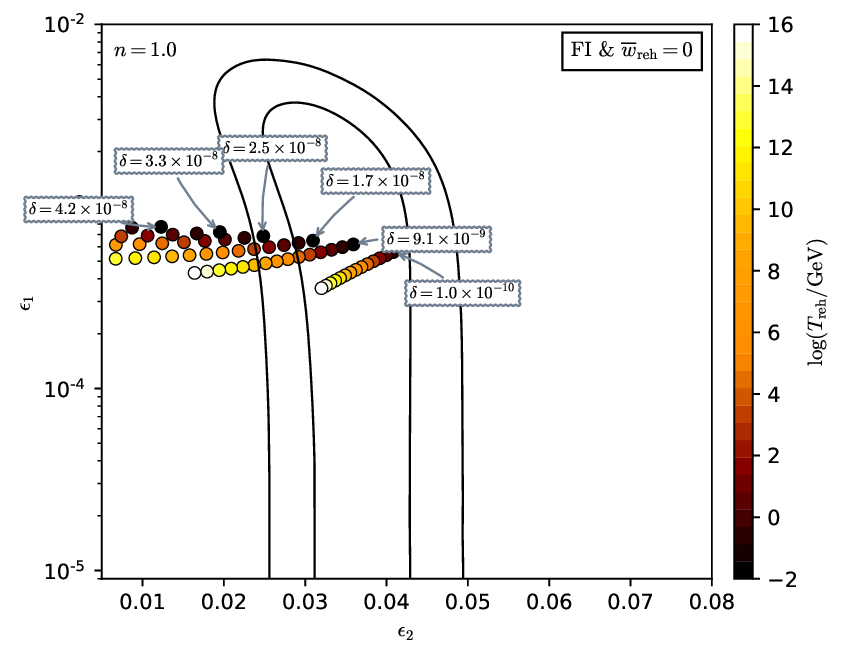}
\caption{Reheating consistent slow-roll predictions for the Fibre Inflation models with $n=1$, in the
  plane $(\nS,r)$ (top panel) and the plane $(\epsilon_1,\epsilon_2)$
  (bottom panel). The solid contours are the one and two-sigma {\data}
  confidence intervals (marginalized over second order slow-roll).}
\label{fig:CMBFI_1}
\end{center}
\end{figure}

\subsection{Hyperbolic Inflation (\hyperref[sec:hbi]{HBI})}

\begin{figure}[H]
\begin{center}
\includegraphics[width=\wappfig,clip=true]{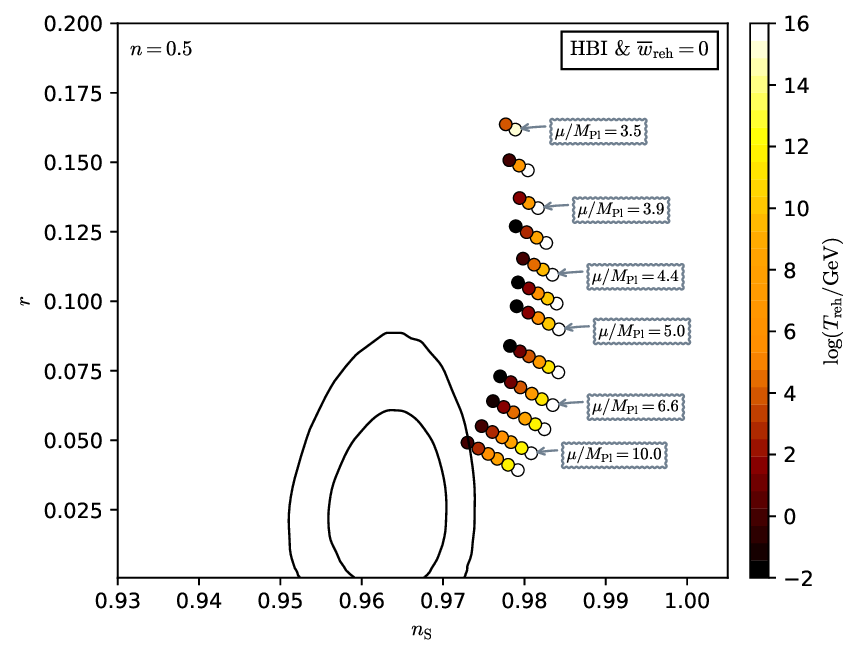}
\includegraphics[width=\wappfig,clip=true]{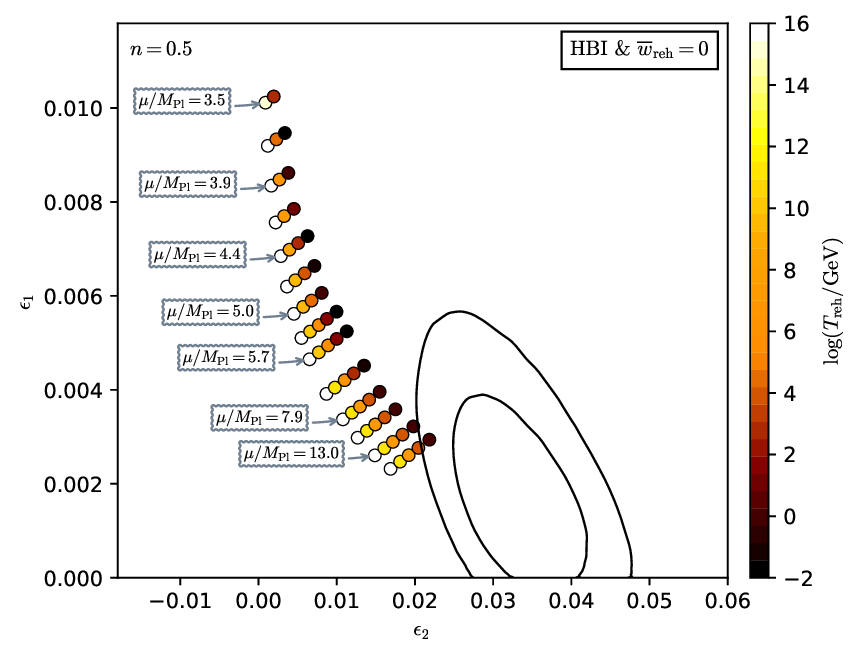}
\caption{Reheating consistent slow-roll predictions for the Hyperbolic
  Inflation models with $n=0.5$, in the plane $(\nS,r)$ (top panel)
  and the plane $(\epsilon_1,\epsilon_2)$ (bottom panel). The solid
  contours are the one and two-sigma {\data} confidence intervals
  (marginalized over second order slow-roll). See \Figs{fig:CMBHBI_1}
  and \ref{fig:CMBHBI_2} for other values of $n$.}
\label{fig:CMBHBI_0}
\end{center}
\end{figure}

\begin{figure}[H]
\begin{center}
\includegraphics[width=\wappfig,clip=true]{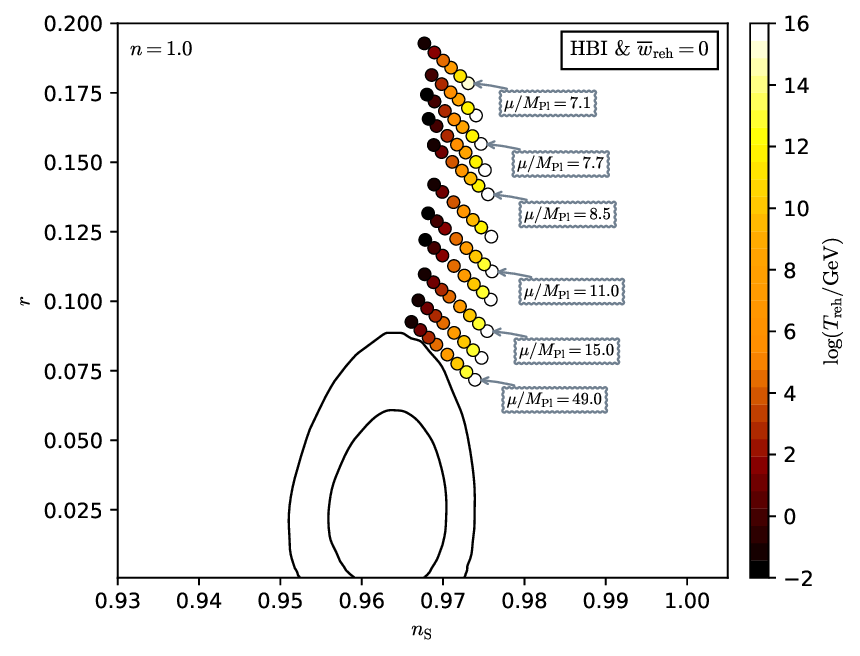}
\includegraphics[width=\wappfig,clip=true]{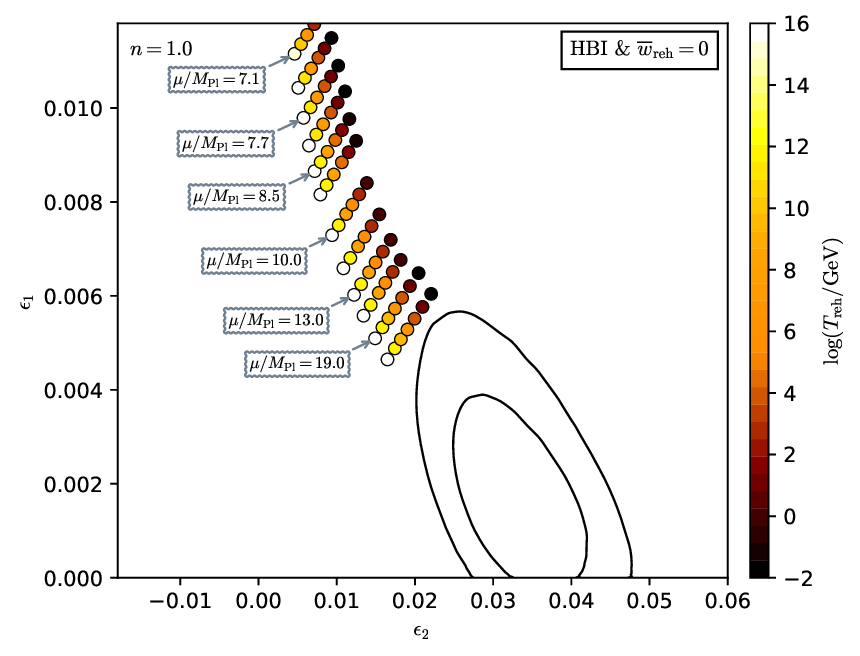}
\caption{Reheating consistent slow-roll predictions for the Hyperbolic Inflation models with $n=1$, in the
  plane $(\nS,r)$ (top panel) and the plane $(\epsilon_1,\epsilon_2)$
  (bottom panel). The solid contours are the one and two-sigma {\data}
  confidence intervals (marginalized over second order slow-roll).}
\label{fig:CMBHBI_1}
\end{center}
\end{figure}

\begin{figure}[H]
\begin{center}
\includegraphics[width=\wappfig,clip=true]{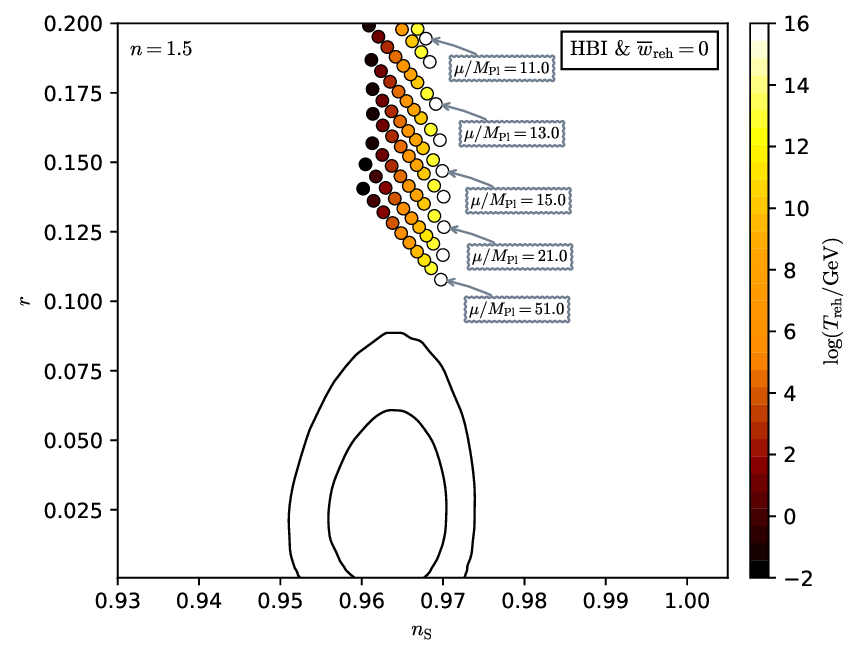}
\includegraphics[width=\wappfig,clip=true]{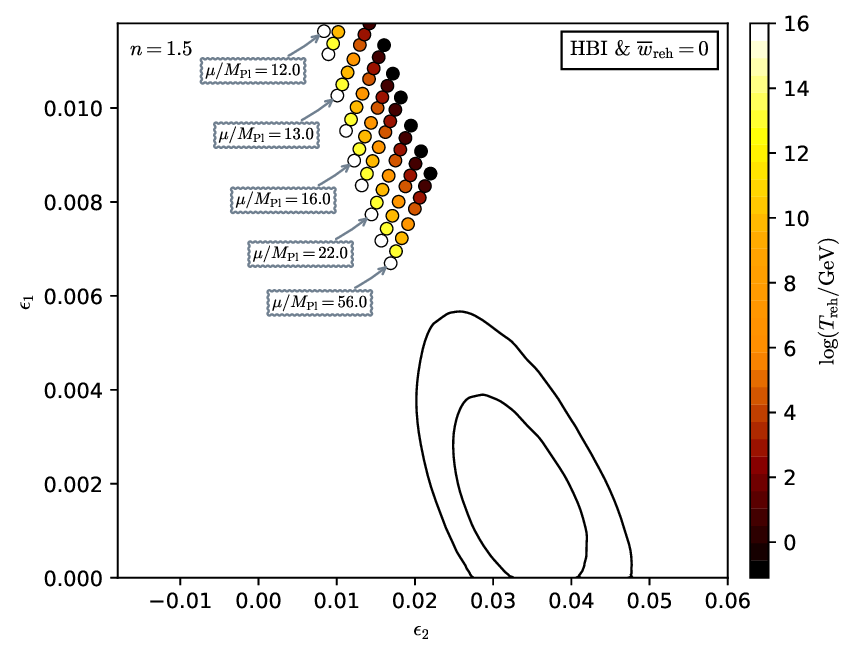}
\caption{Reheating consistent slow-roll predictions for the Hyperbolic Inflation models with $n=1.5$, in the
  plane $(\nS,r)$ (top panel) and the plane $(\epsilon_1,\epsilon_2)$
  (bottom panel). The solid contours are the one and two-sigma {\data}
  confidence intervals (marginalized over second order slow-roll).}
\label{fig:CMBHBI_2}
\end{center}
\end{figure}

\subsection{Smeared Higgs Inflation (\hyperref[sec:shi]{SHI})}

\begin{figure}[H]
\begin{center}
\includegraphics[width=\wappfig,clip=true]{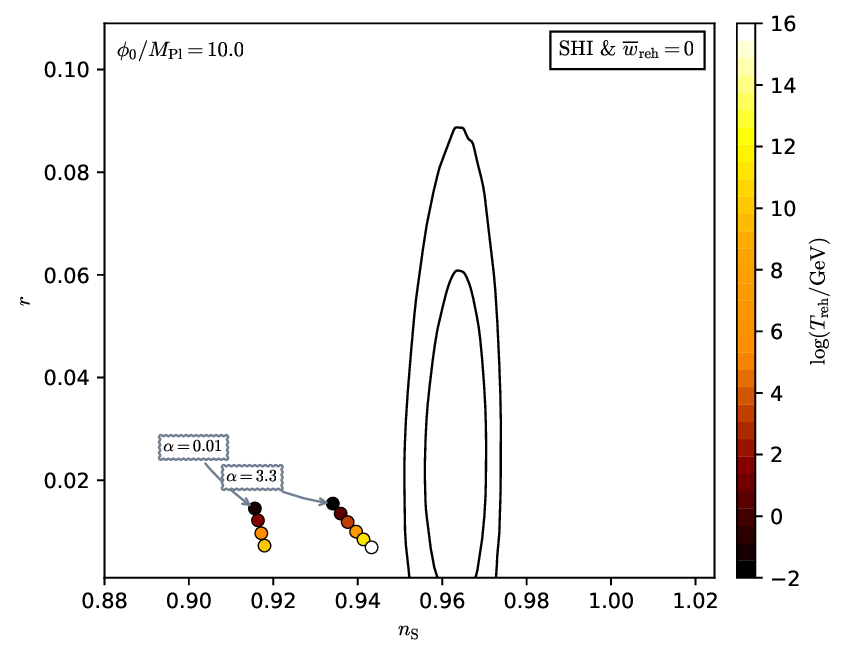}
\includegraphics[width=\wappfig,clip=true]{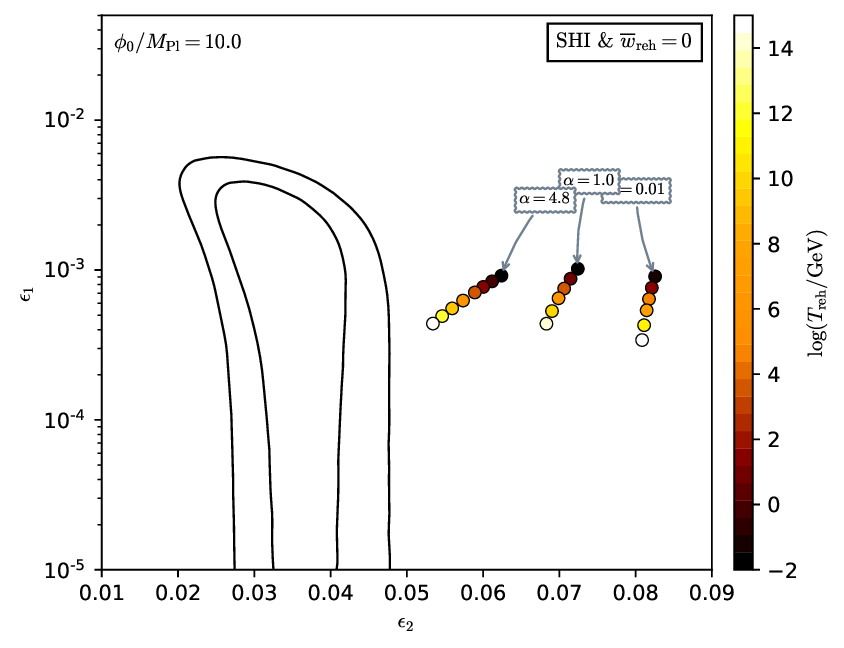}
\caption{Reheating consistent slow-roll predictions for the Smeared
  Higgs Inflation models with $\phizero=10\Mp$, in the plane $(\nS,r)$ (top panel)
  and the plane $(\epsilon_1,\epsilon_2)$ (bottom panel). The solid
  contours are the one and two-sigma {\data} confidence intervals
  (marginalized over second order slow-roll). See \Figs{fig:CMBSHI_1}, \ref{fig:CMBSHI_2} and \ref{fig:CMBSHI_3} for other values of $\phizero$.}
\label{fig:CMBSHI_0}
\end{center}
\end{figure}

\begin{figure}[H]
\begin{center}
\includegraphics[width=\wappfig,clip=true]{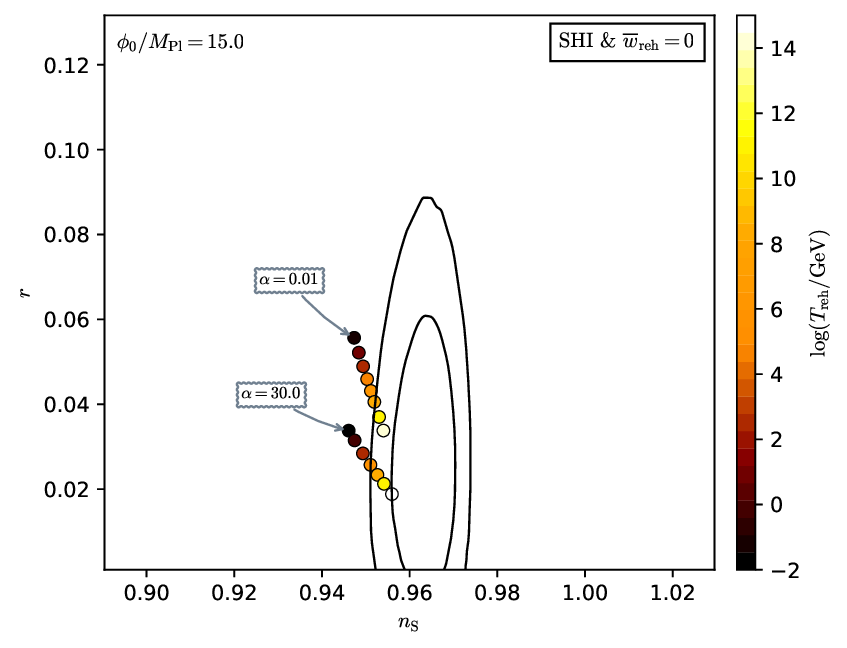}
\includegraphics[width=\wappfig,clip=true]{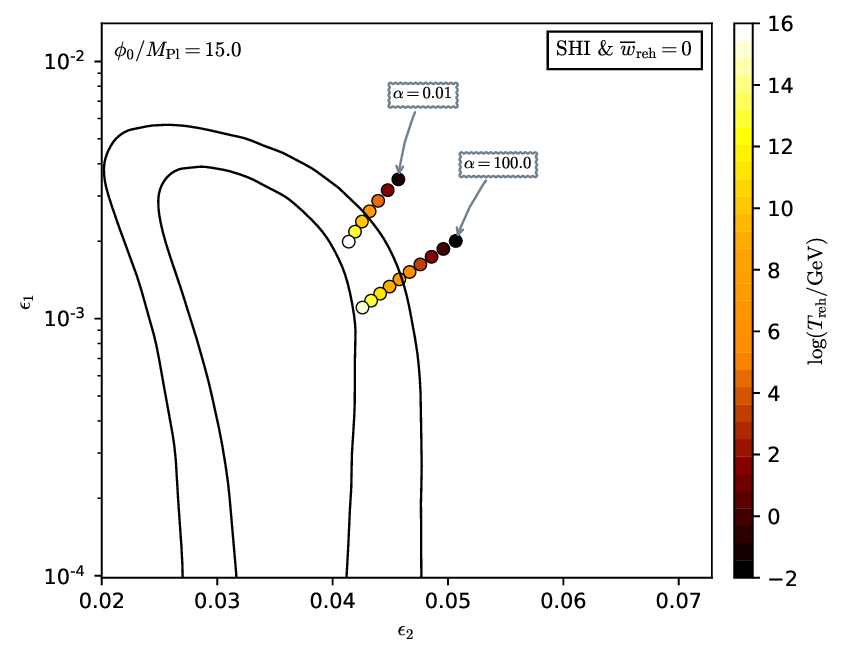}
\caption{Reheating consistent slow-roll predictions for the Smeared
  Higgs Inflation models with $\phizero=15\Mp$, in the plane $(\nS,r)$ (top panel)
  and the plane $(\epsilon_1,\epsilon_2)$ (bottom panel). The solid
  contours are the one and two-sigma {\data} confidence intervals
  (marginalized over second order slow-roll).}
\label{fig:CMBSHI_1}
\end{center}
\end{figure}

\begin{figure}[H]
\begin{center}
\includegraphics[width=\wappfig,clip=true]{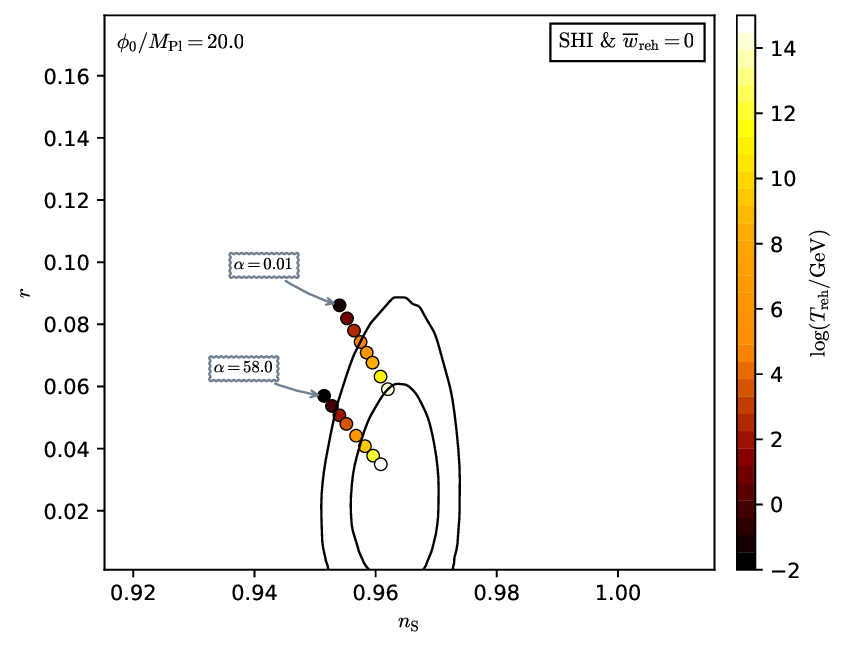}
\includegraphics[width=\wappfig,clip=true]{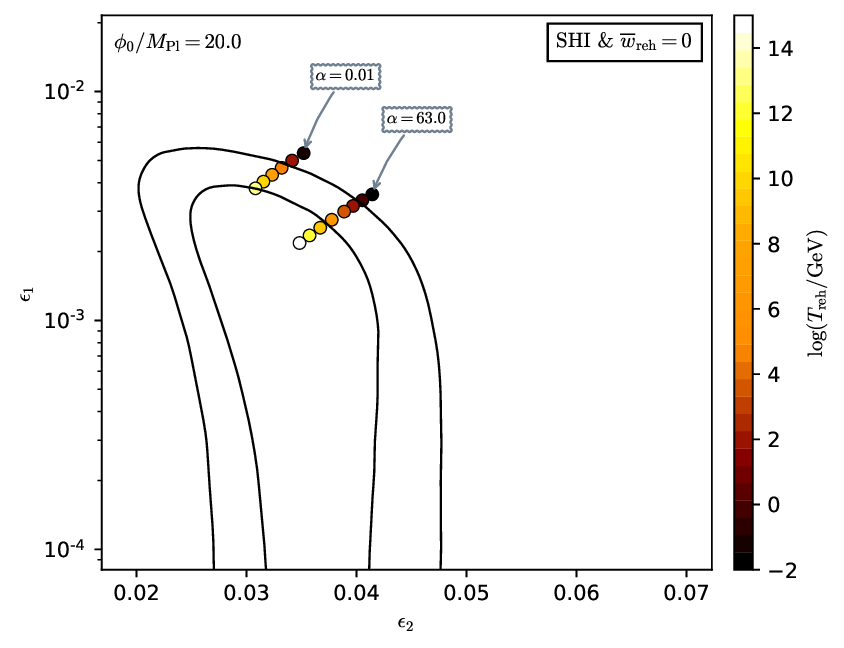}
\caption{Reheating consistent slow-roll predictions for the Smeared
  Higgs Inflation models with $\phizero=20\Mp$, in the plane $(\nS,r)$ (top panel)
  and the plane $(\epsilon_1,\epsilon_2)$ (bottom panel). The solid
  contours are the one and two-sigma {\data} confidence intervals
  (marginalized over second order slow-roll).}
\label{fig:CMBSHI_2}
\end{center}
\end{figure}

\begin{figure}[H]
\begin{center}
\includegraphics[width=\wappfig,clip=true]{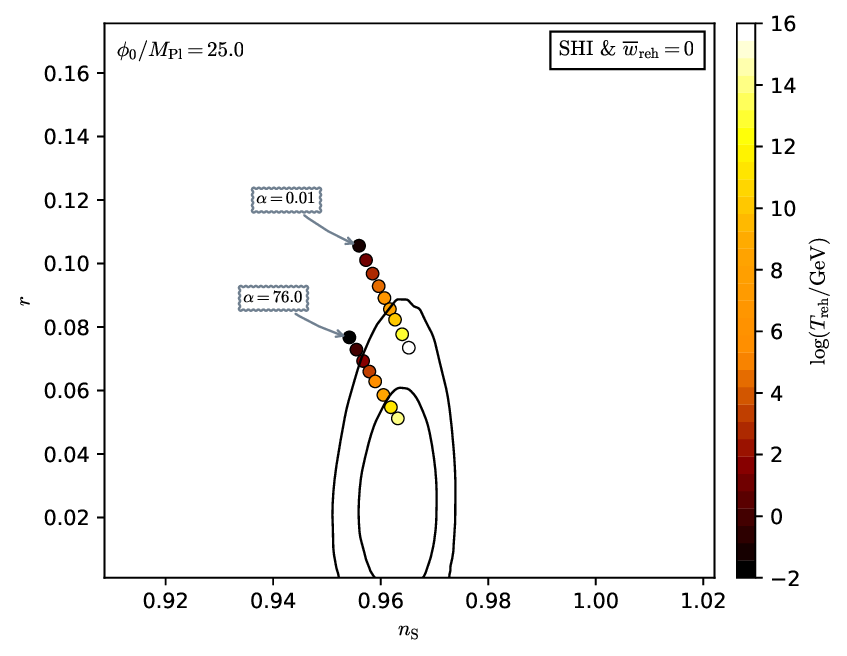}
\includegraphics[width=\wappfig,clip=true]{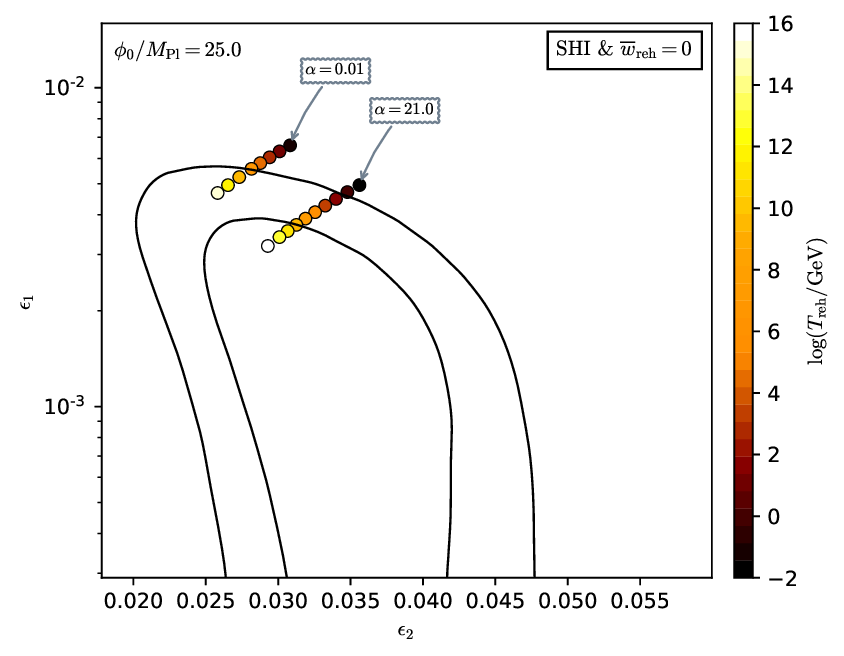}
\caption{Reheating consistent slow-roll predictions for the Smeared
  Higgs Inflation models with $\phizero=25\Mp$, in the plane $(\nS,r)$ (top panel)
  and the plane $(\epsilon_1,\epsilon_2)$ (bottom panel). The solid
  contours are the one and two-sigma {\data} confidence intervals
  (marginalized over second order slow-roll).}
\label{fig:CMBSHI_3}
\end{center}
\end{figure}

\subsection{Double Exponential Inflation (\hyperref[sec:dei]{DEI})}

\begin{figure}[H]
\begin{center}
\includegraphics[width=\wappfig,clip=true]{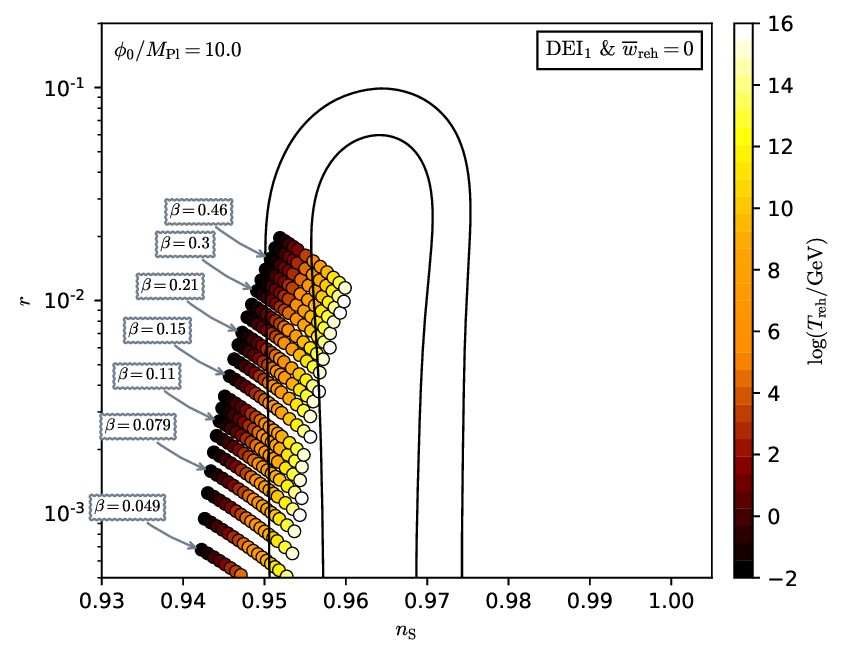}
\includegraphics[width=\wappfig,clip=true]{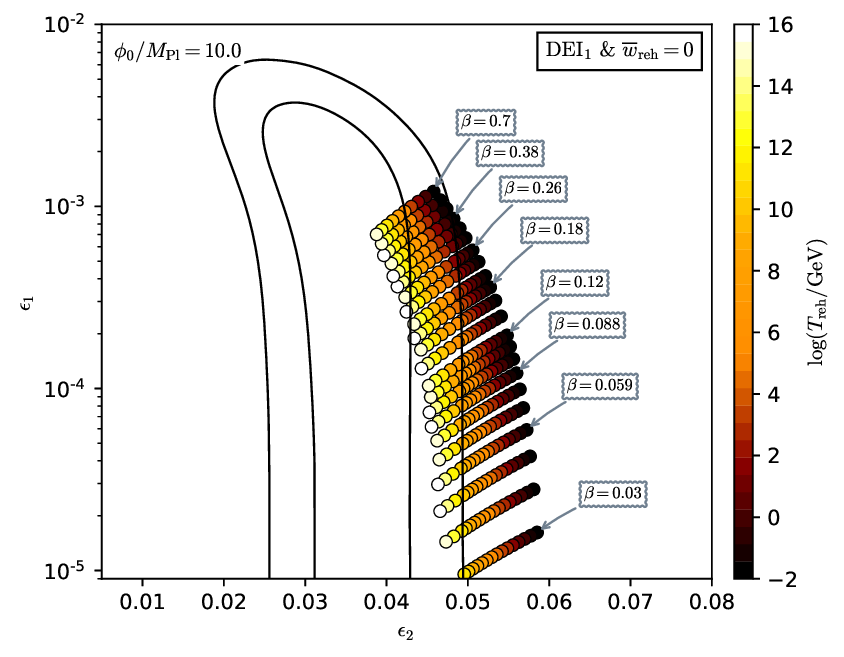}
\caption{Reheating consistent slow-roll predictions for the Double
  Exponential Inflation models with $\phizero=10\Mp$, in the plane
  $(\nS,r)$ (top panel) and the plane $(\epsilon_1,\epsilon_2)$
  (bottom panel). The solid contours are the one and two-sigma {\data}
  confidence intervals (marginalized over second order slow-roll). See
  \Figs{fig:CMBDEI_1}, \ref{fig:CMBDEI_2}, and \ref{fig:CMBDEI_3} for
  other values of $\phizero$.}
\label{fig:CMBDEI_0}
\end{center}
\end{figure}

\begin{figure}[H]
\begin{center}
\includegraphics[width=\wappfig,clip=true]{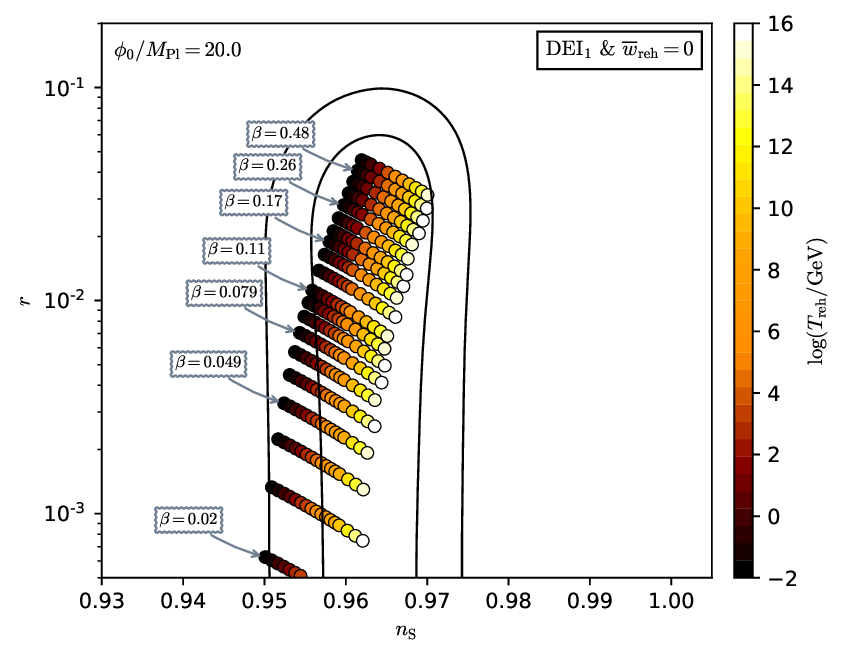}
\includegraphics[width=\wappfig,clip=true]{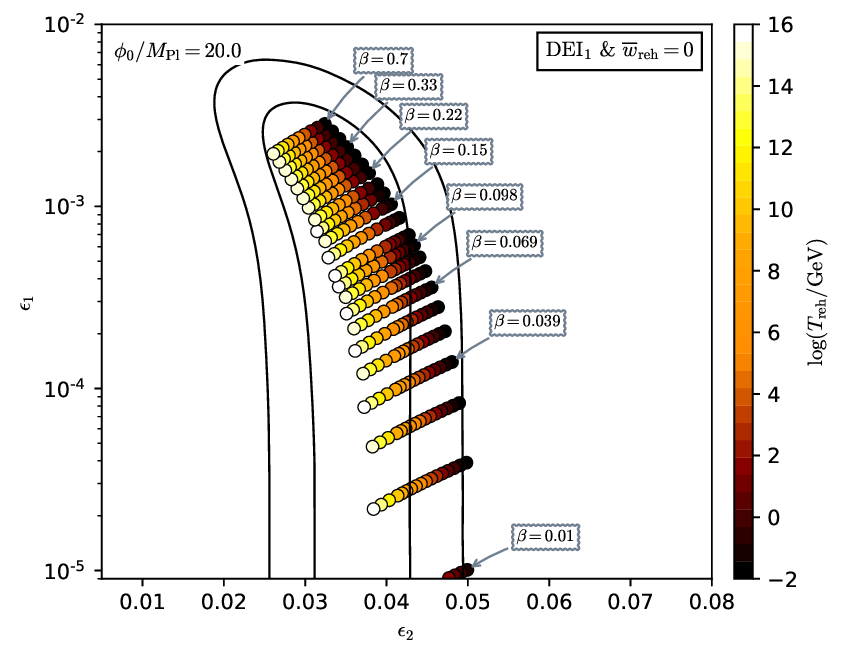}
\caption{Reheating consistent slow-roll predictions for the Double
  Exponential Inflation models with $\phizero=20\Mp$, in the plane
  $(\nS,r)$ (top panel) and the plane $(\epsilon_1,\epsilon_2)$
  (bottom panel). The solid contours are the one and two-sigma {\data}
  confidence intervals (marginalized over second order slow-roll).}
\label{fig:CMBDEI_1}
\end{center}
\end{figure}

\begin{figure}[H]
\begin{center}
\includegraphics[width=\wappfig,clip=true]{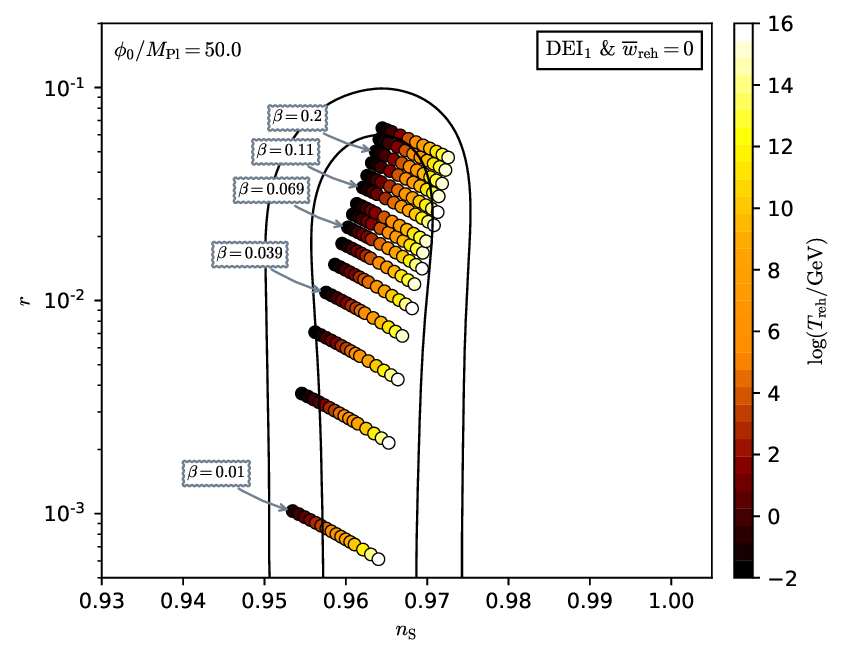}
\includegraphics[width=\wappfig,clip=true]{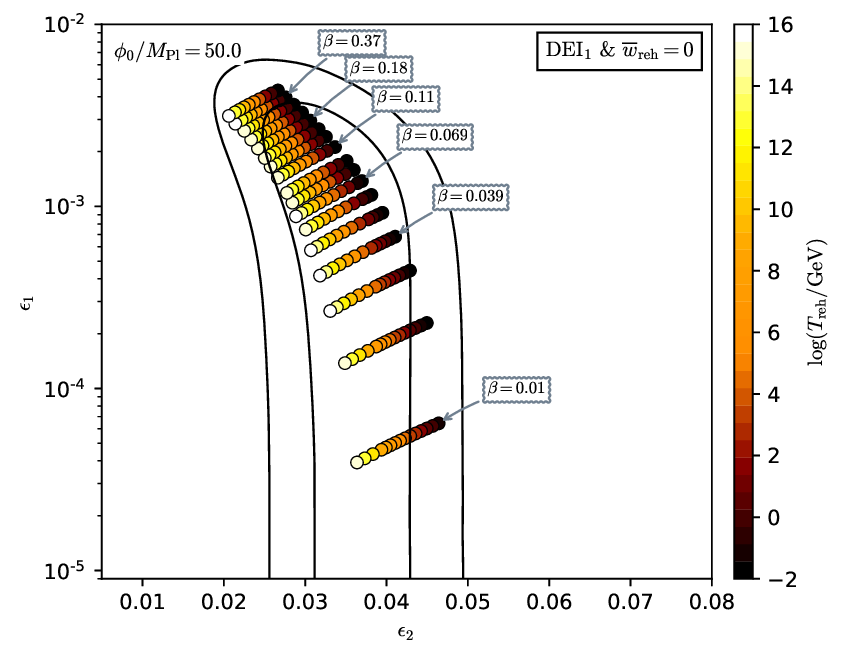}
\caption{Reheating consistent slow-roll predictions for the Double
  Exponential Inflation models with $\phizero=50\Mp$, in the plane
  $(\nS,r)$ (top panel) and the plane $(\epsilon_1,\epsilon_2)$
  (bottom panel). The solid contours are the one and two-sigma {\data}
  confidence intervals (marginalized over second order slow-roll).}
\label{fig:CMBDEI_2}
\end{center}
\end{figure}

\begin{figure}[H]
\begin{center}
\includegraphics[width=\wappfig,clip=true]{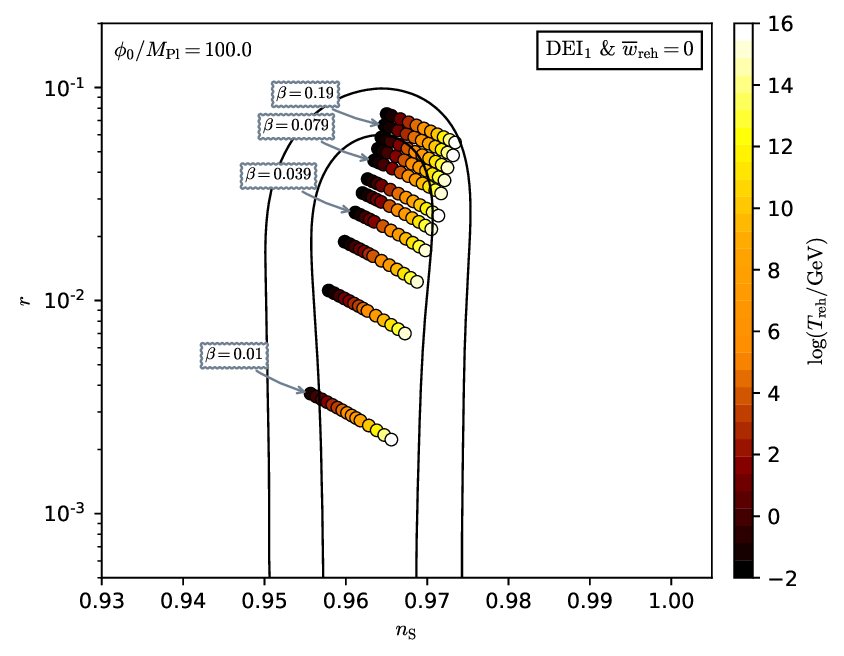}
\includegraphics[width=\wappfig,clip=true]{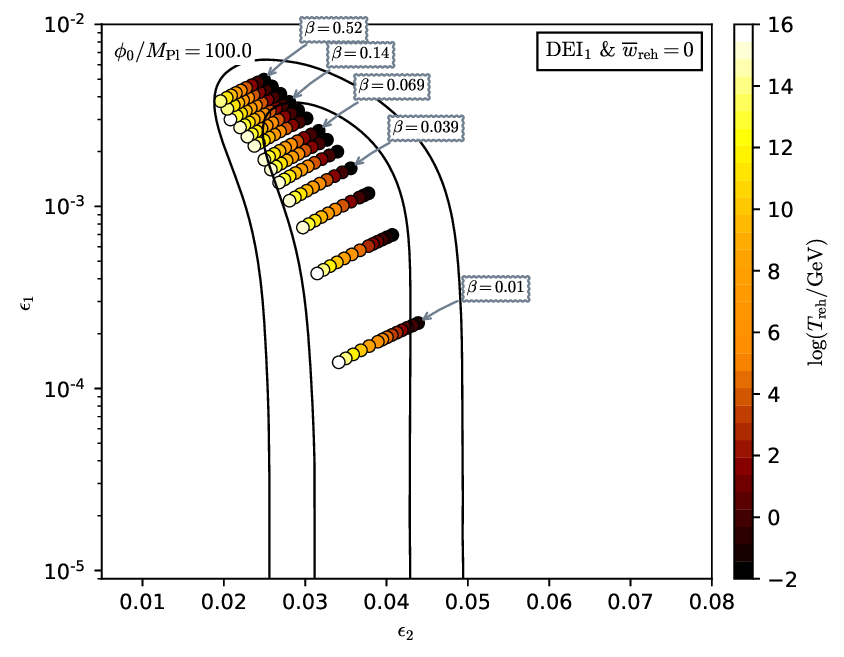}
\caption{Reheating consistent slow-roll predictions for the Double
  Exponential Inflation models with $\phizero=100\Mp$, in the plane
  $(\nS,r)$ (top panel) and the plane $(\epsilon_1,\epsilon_2)$
  (bottom panel). The solid contours are the one and two-sigma {\data}
  confidence intervals (marginalized over second order slow-roll).}
\label{fig:CMBDEI_3}
\end{center}
\end{figure}

\subsection{S-Dual Inflation (\hyperref[sec:sdi]{SDI})}

\begin{figure}[H]
\begin{center}
\includegraphics[width=\wappfig,clip=true]{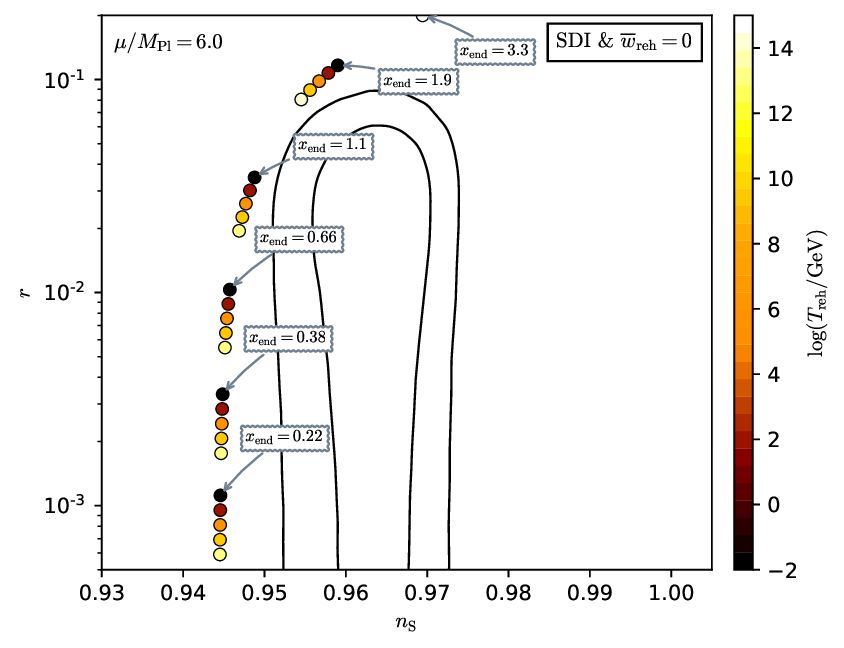}
\includegraphics[width=\wappfig,clip=true]{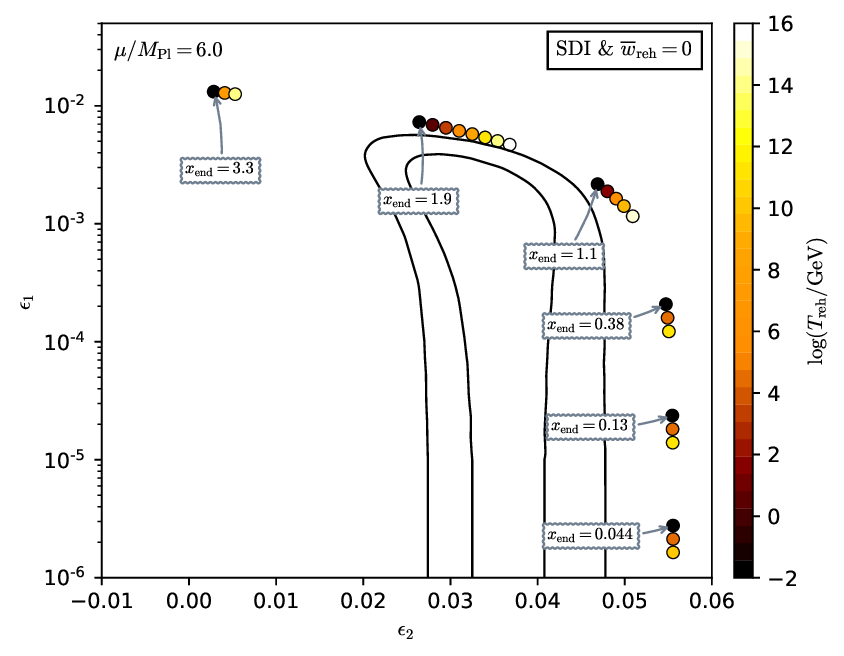}
\caption{Reheating consistent slow-roll predictions for S-Dual
  Inflation and for $\mu=6\Mp$. Predictions are represented in the
  plane $(\nS,r)$ (top panel) and in the plane
  $(\epsilon_1,\epsilon_2)$ (bottom panel) for various values of the
  field value at which inflation ends $\xend=\phiend/\mu$. The solid
  contours are the one and two-sigma {\data} confidence intervals
  (marginalized over second order slow-roll). See also
  Figs.~\ref{fig:CMBSDI_1} and \ref{fig:CMBSDI_2} for other values of
  $\mu$.}
\label{fig:CMBSDI_0}
\end{center}
\end{figure}

\begin{figure}[H]
\begin{center}
\includegraphics[width=\wappfig,clip=true]{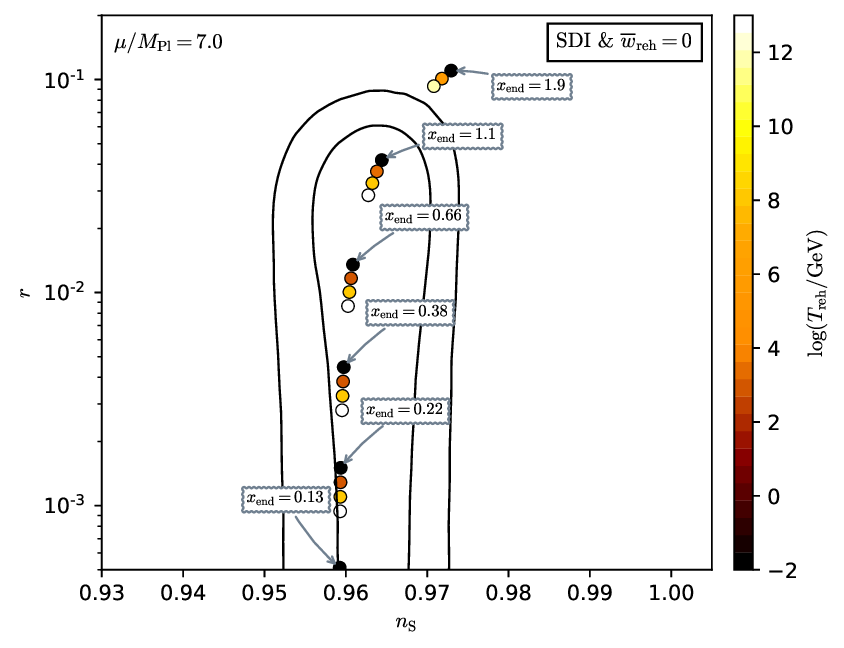}
\includegraphics[width=\wappfig,clip=true]{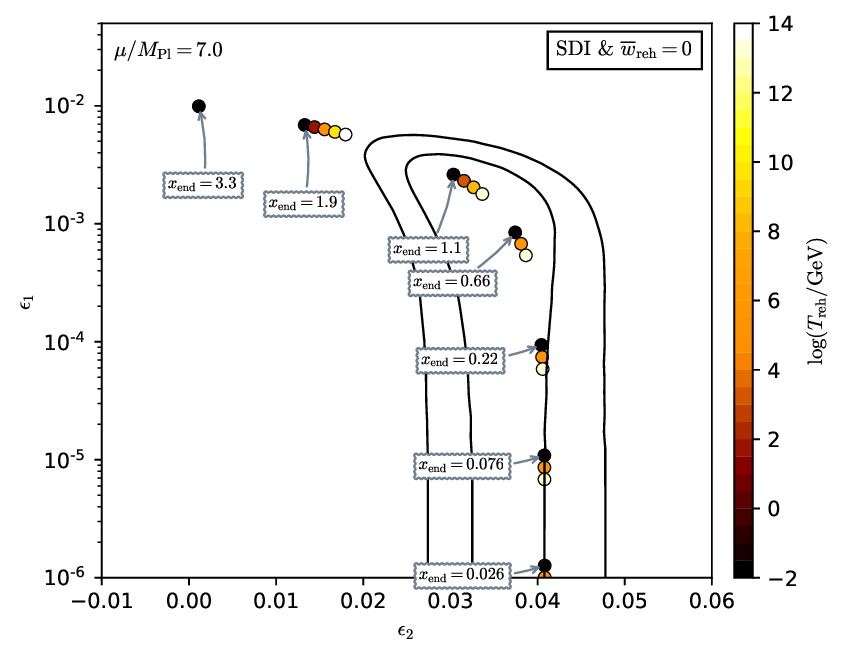}
\caption{Reheating consistent slow-roll predictions for S-Dual
  Inflation and for $\mu=7\Mp$. Predictions are represented in the
  plane $(\nS,r)$ (top panel) and in the plane
  $(\epsilon_1,\epsilon_2)$ (bottom panel) for various values of the
  field value at which inflation ends $\xend=\phiend/\mu$. The solid
  contours are the one and two-sigma {\data} confidence intervals
  (marginalized over second order slow-roll).}
\label{fig:CMBSDI_1}
\end{center}
\end{figure}

\begin{figure}[H]
\begin{center}
\includegraphics[width=\wappfig,clip=true]{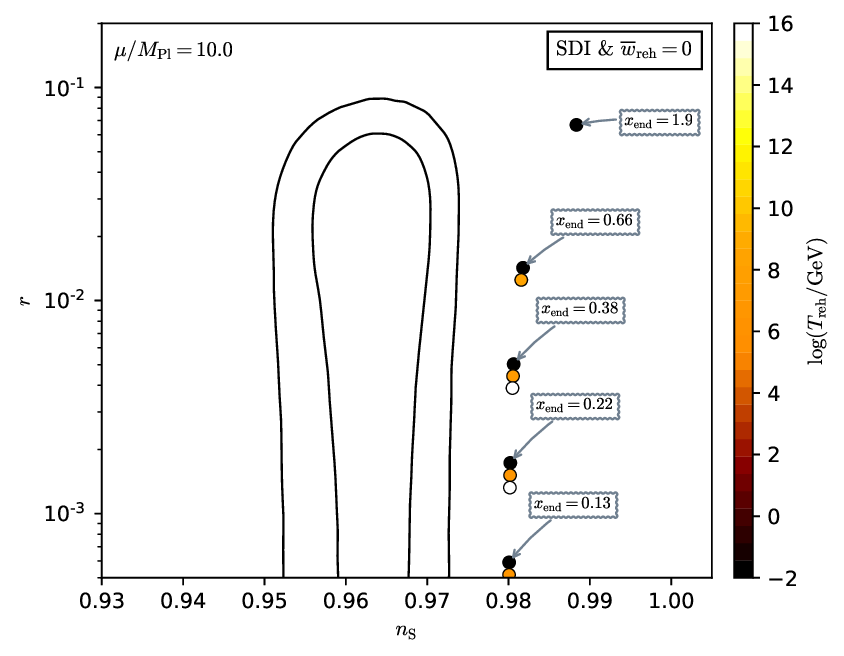}
\includegraphics[width=\wappfig,clip=true]{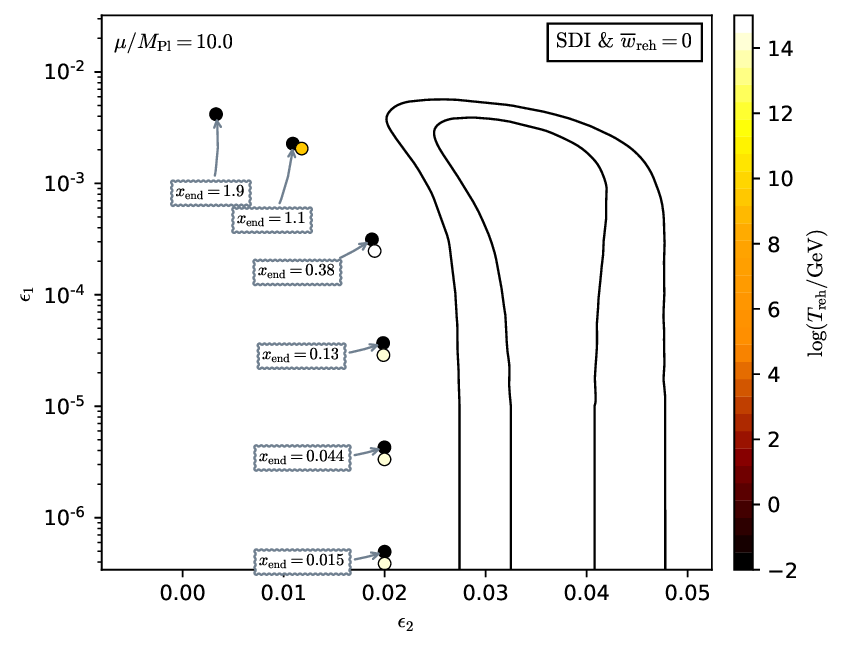}
\caption{Reheating consistent slow-roll predictions for S-Dual
  Inflation and for $\mu=10\Mp$. Predictions are represented in the
  plane $(\nS,r)$ (top panel) and in the plane
  $(\epsilon_1,\epsilon_2)$ (bottom panel) for various values of the
  field value at which inflation ends $\xend=\phiend/\mu$. The solid
  contours are the one and two-sigma {\data} confidence intervals
  (marginalized over second order slow-roll).}
\label{fig:CMBSDI_2}
\end{center}
\end{figure}

\subsection{Generalized Double Well Inflation (\hyperref[sec:gdwi]{GDWI})}

\begin{figure}[H]
\begin{center}
\includegraphics[width=\wappfig,clip=true]{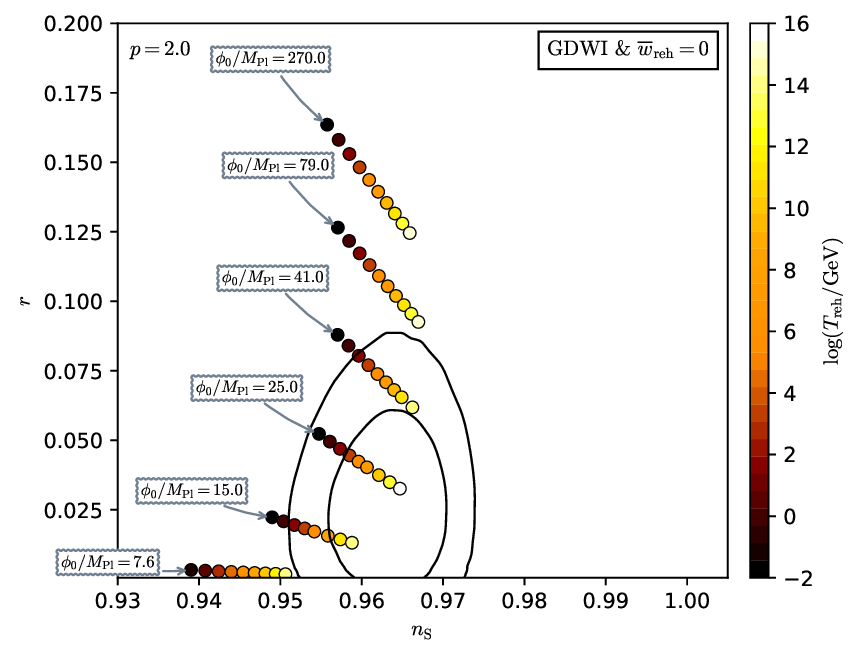}
\includegraphics[width=\wappfig,clip=true]{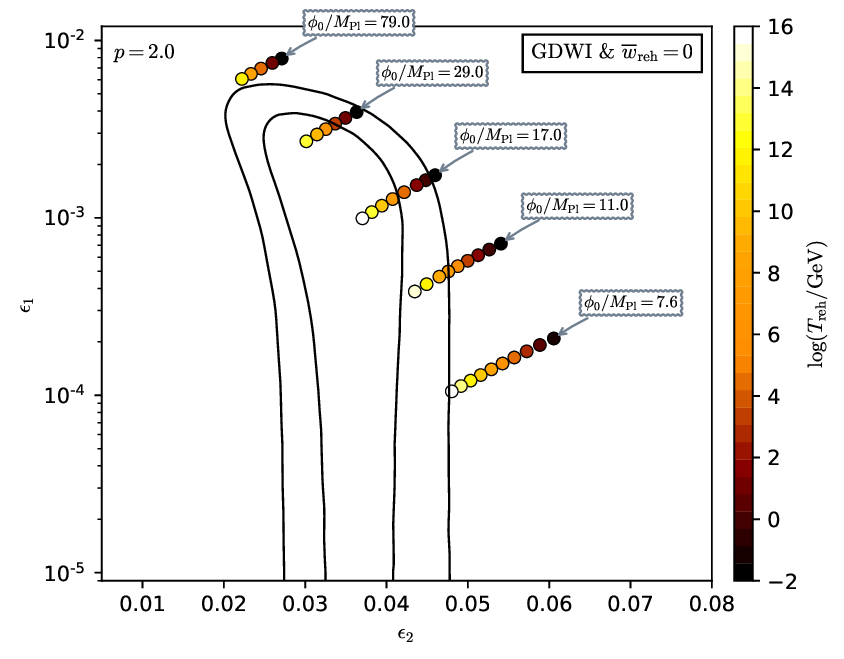}
\caption{Reheating consistent slow-roll predictions for the
  Generalized Double Well Inflation model with a power law index
  $p=2$. Predictions are represented as a function of the {\vev} $\phizero/\Mp$ in the
  plane $(\nS,r)$ (top panel) and in the plane
  $(\epsilon_1,\epsilon_2)$ (bottom panel). The solid contours are the
  one and two-sigma {\data} confidence intervals (marginalized over
  second order slow-roll). See also Figs.~\ref{fig:CMBGDWI_1} and
  \ref{fig:CMBGDWI_2} for other values of $p$.}
\label{fig:CMBGDWI_0}
\end{center}
\end{figure}

\begin{figure}[H]
\begin{center}
\includegraphics[width=\wappfig,clip=true]{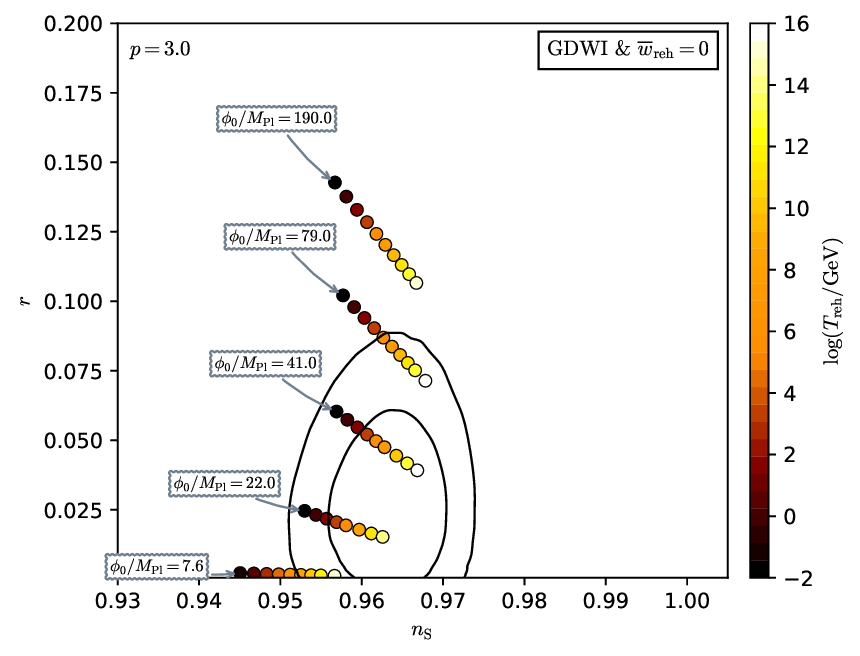}
\includegraphics[width=\wappfig,clip=true]{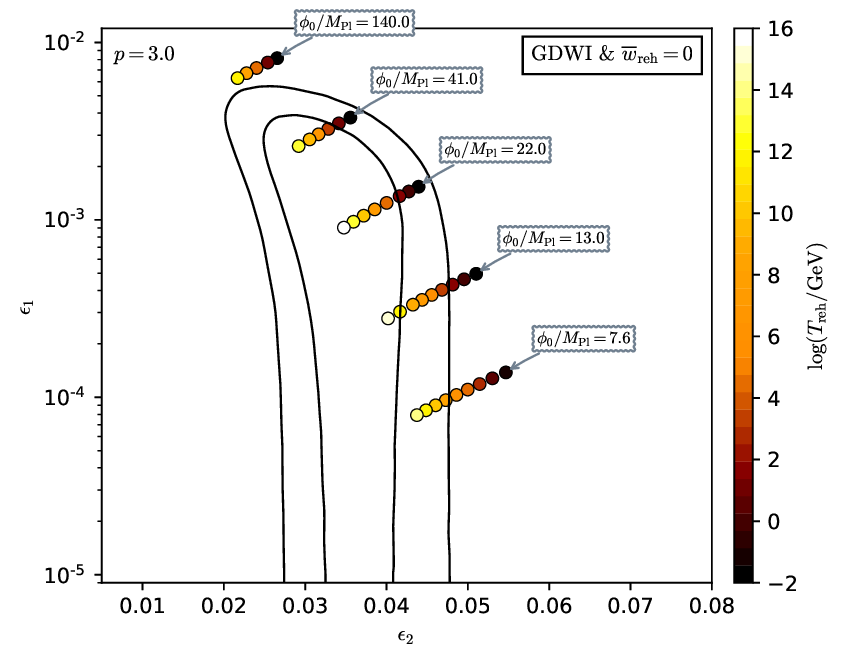}
\caption{Reheating consistent slow-roll predictions for the
  Generalized Double Well Inflation model with a power law index
  $p=3$. Predictions are represented as a function of the {\vev} $\phizero/\Mp$ in the
  plane $(\nS,r)$ (top panel) and in the plane
  $(\epsilon_1,\epsilon_2)$ (bottom panel). The solid contours are the
  one and two-sigma {\data} confidence intervals (marginalized over
  second order slow-roll).}
\label{fig:CMBGDWI_1}
\end{center}
\end{figure}

\begin{figure}[H]
\begin{center}
\includegraphics[width=\wappfig,clip=true]{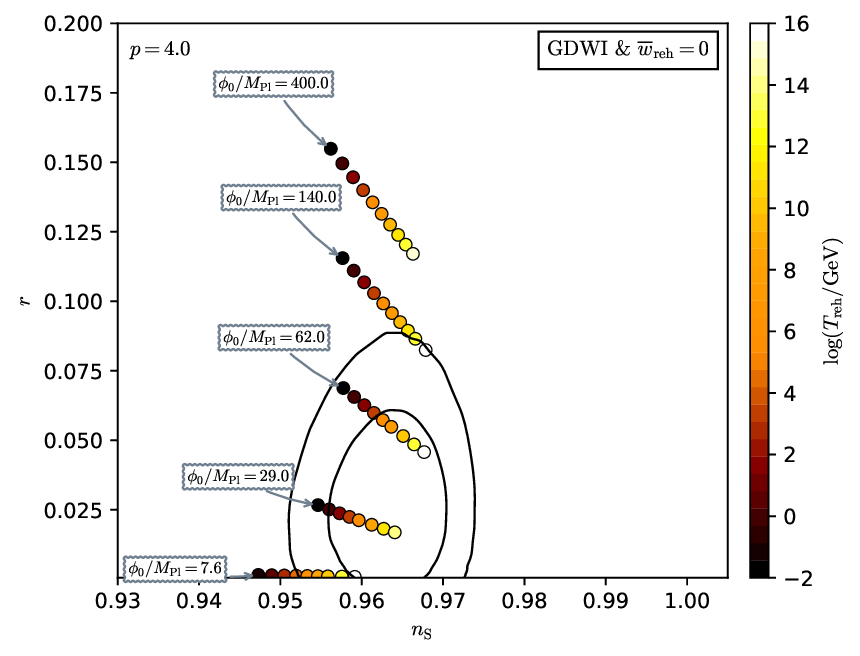}
\includegraphics[width=\wappfig,clip=true]{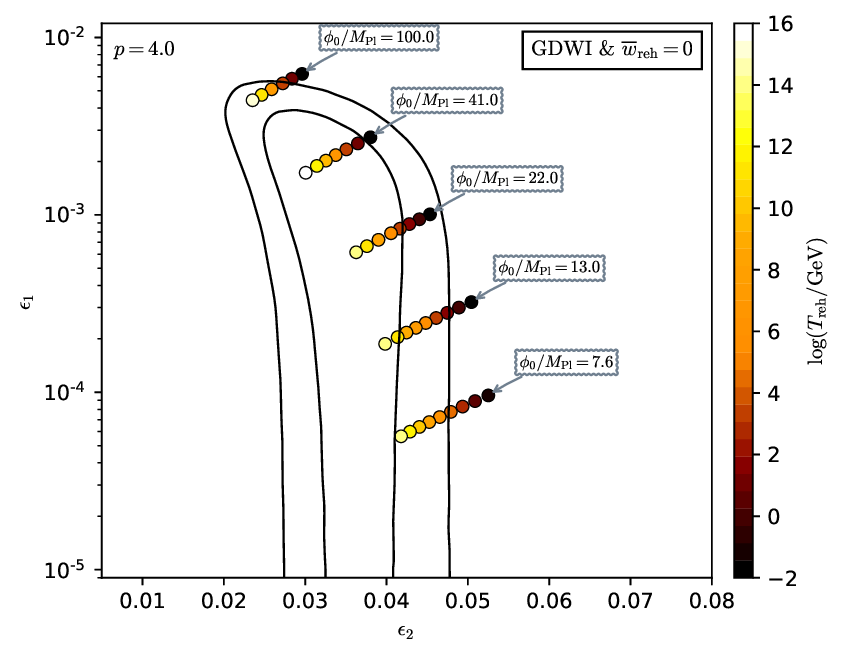}
\caption{Reheating consistent slow-roll predictions for the
  Generalized Double Well Inflation model with a power law index
  $p=4$. Predictions are represented as a function of the {\vev} $\phizero/\Mp$ in the
  plane $(\nS,r)$ (top panel) and in the plane
  $(\epsilon_1,\epsilon_2)$ (bottom panel). The solid contours are the
  one and two-sigma {\data} confidence intervals (marginalized over
  second order slow-roll).}
\label{fig:CMBGDWI_2}
\end{center}
\end{figure}

\subsection{Non-Minimal Large Field Inflation 1 (\hyperref[sec:nmlfi]{NMLFI1})}

\begin{figure}[H]
\begin{center}
\includegraphics[width=\wappfig,clip=true]{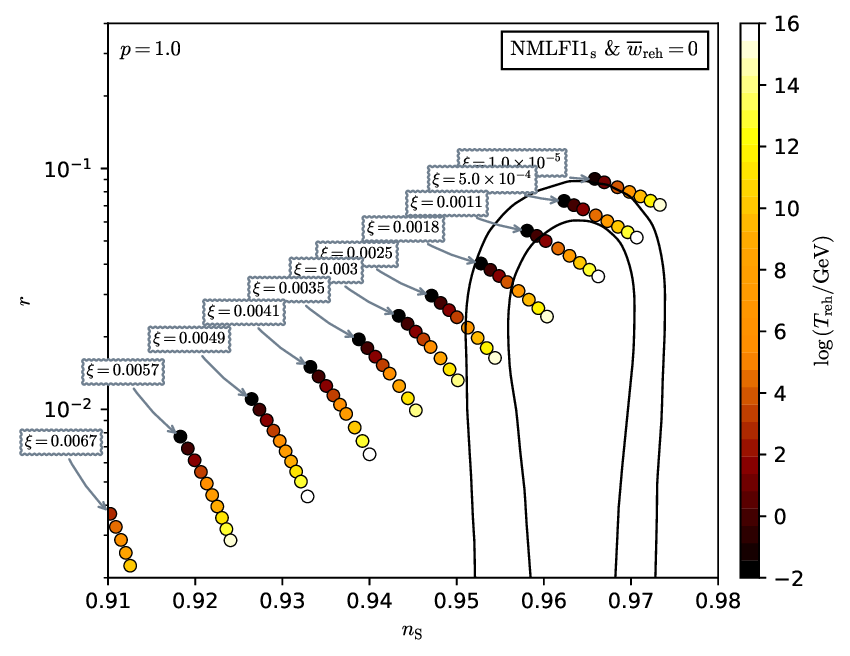}
\includegraphics[width=\wappfig,clip=true]{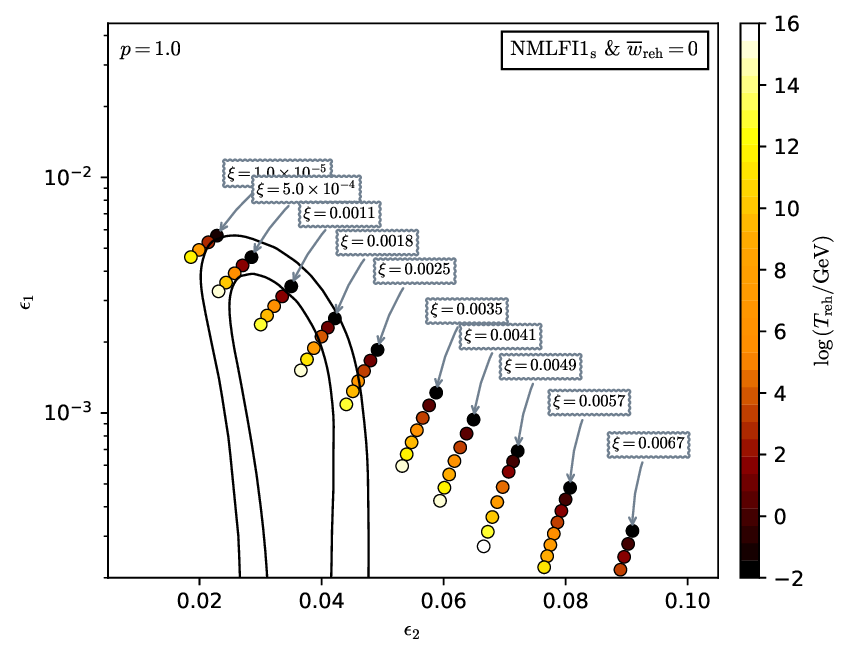}
\caption{Reheating consistent slow-roll predictions for the
  Non-Minimal Large Field Inflation 1 model, for $p=1$. Predictions
  are represented as a function of $\xi$ in the plane $(\nS,r)$ (top
  panel) and in the plane $(\epsilon_1,\epsilon_2)$ (bottom
  panel). The solid contours are the one and two-sigma {\data}
  confidence intervals (marginalized over second order slow-roll). See
  also \Figs{fig:CMBNMLFI1_1} to \ref{fig:CMBNMLFI1_2} for the
  other parameter values in the regime $p<4$ where NMLFI1 is an
  hilltop-like model at small field values.}
\label{fig:CMBNMLFI1_0}
\end{center}
\end{figure}

\begin{figure}[H]
\begin{center}
\includegraphics[width=\wappfig,clip=true]{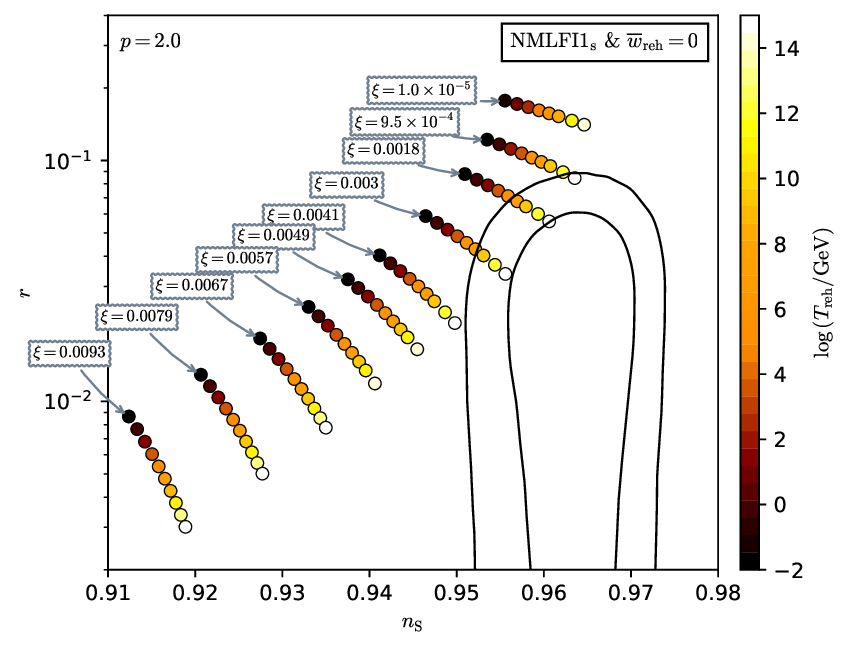}
\includegraphics[width=\wappfig,clip=true]{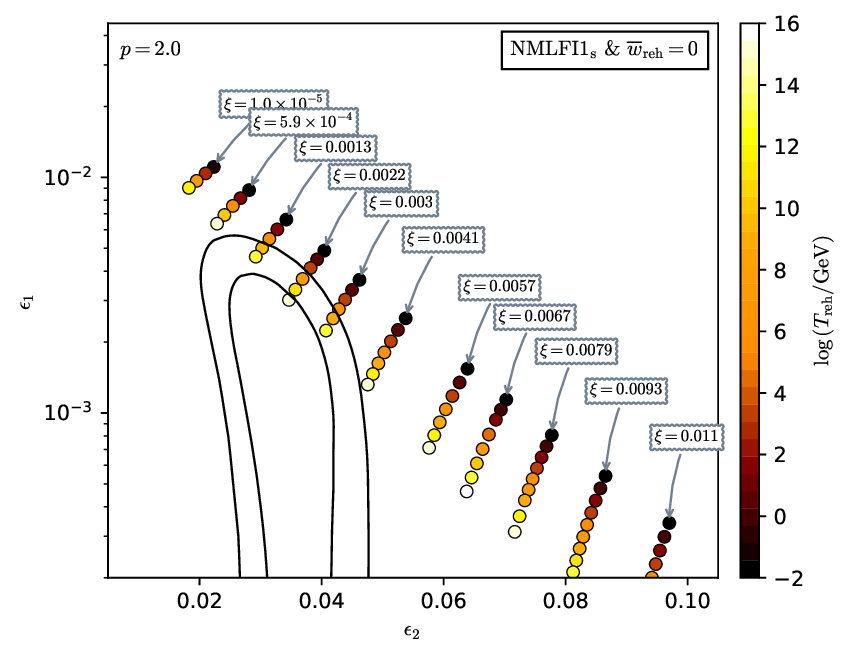}
\caption{Reheating consistent slow-roll predictions for the
  Non-Minimal Large Field Inflation 1 model, for
  $p=2$. Predictions are represented as a function of $\xi$ in the
  plane $(\nS,r)$ (top panel) and in the plane
  $(\epsilon_1,\epsilon_2)$ (bottom panel). The solid contours are the
  one and two-sigma {\data} confidence intervals (marginalized over
  second order slow-roll).}
\label{fig:CMBNMLFI1_1}
\end{center}
\end{figure}

\begin{figure}[H]
\begin{center}
\includegraphics[width=\wappfig,clip=true]{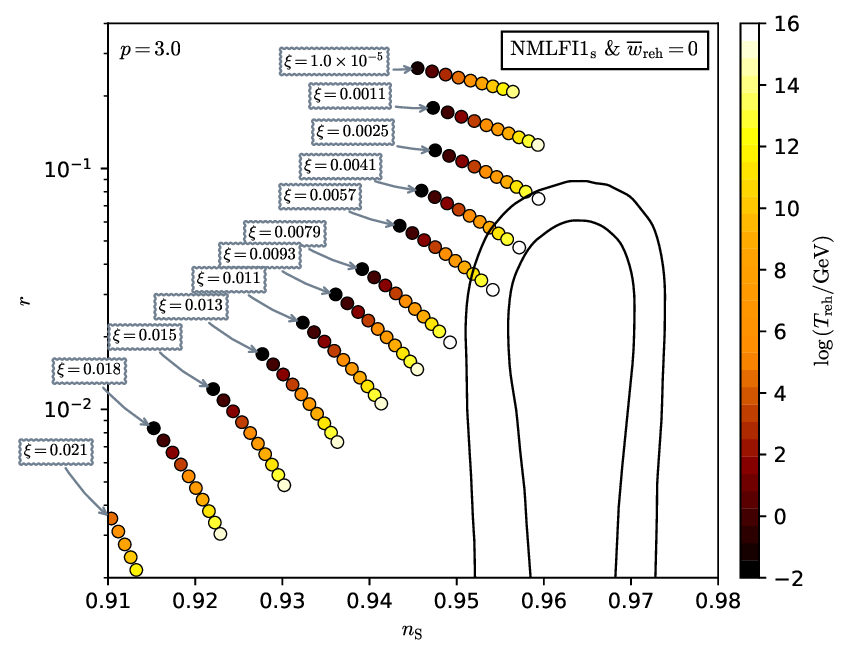}
\includegraphics[width=\wappfig,clip=true]{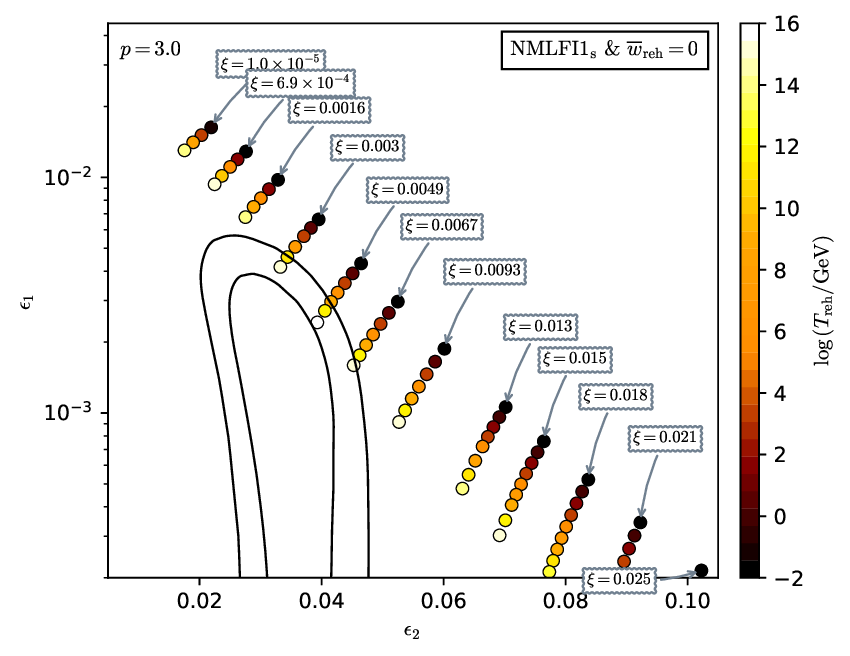}
\caption{Reheating consistent slow-roll predictions for the
  Non-Minimal Large Field Inflation 1 model, for $p=3$. Predictions
  are represented as a function of $\xi$ in the plane $(\nS,r)$ (top
  panel) and in the plane $(\epsilon_1,\epsilon_2)$ (bottom
  panel). The solid contours are the one and two-sigma {\data}
  confidence intervals (marginalized over second order slow-roll).}
\label{fig:CMBNMLFI1_2}
\end{center}
\end{figure}

\begin{figure}[H]
\begin{center}
\includegraphics[width=\wappfig,clip=true]{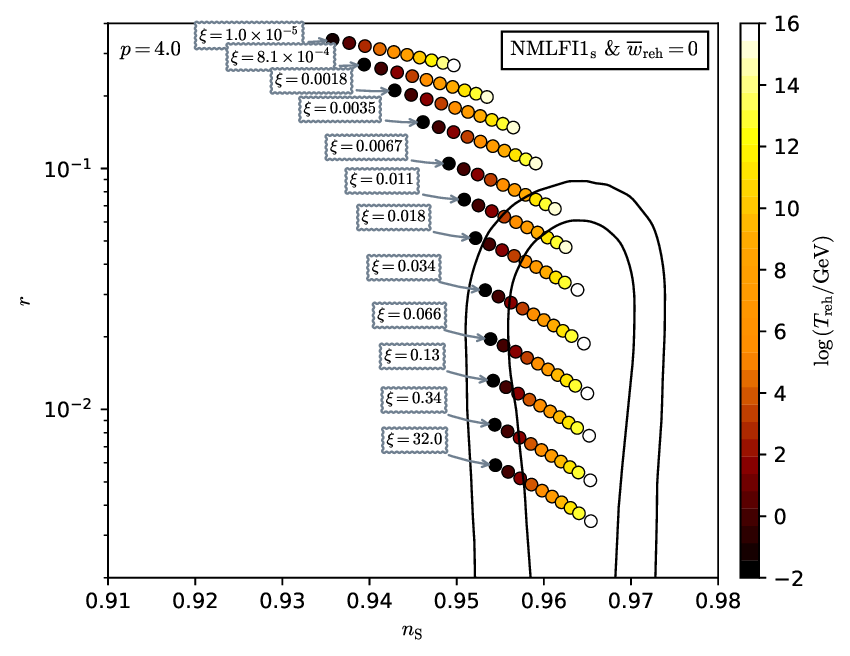}
\includegraphics[width=\wappfig,clip=true]{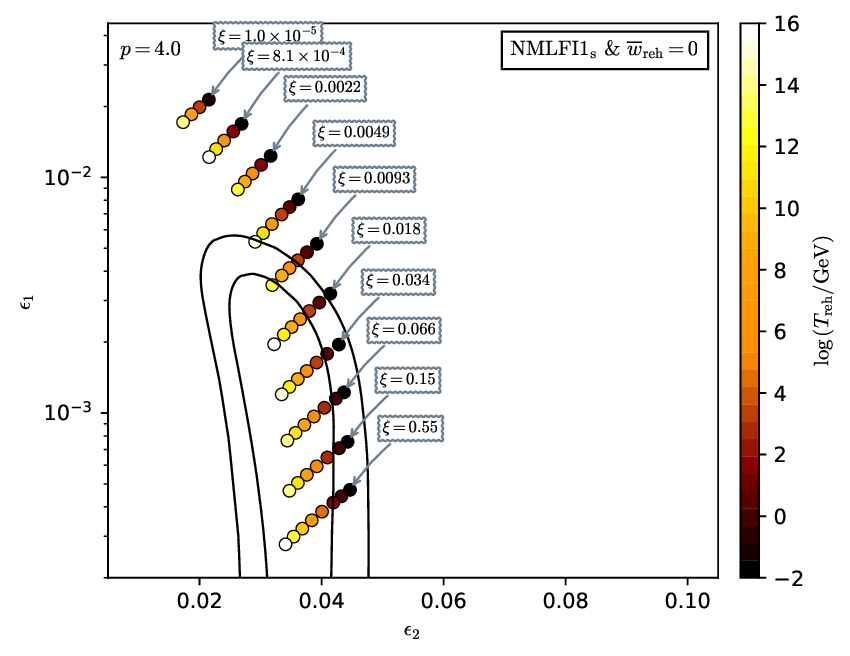}
\caption{Reheating consistent slow-roll predictions for the
  Non-Minimal Large Field Inflation 1 model, in the plateau regime, for
  $p=4$. Predictions are represented as a function of $\xi$ in the
  plane $(\nS,r)$ (top panel) and in the plane
  $(\epsilon_1,\epsilon_2)$ (bottom panel). For this value of $p=4$,
  NMLFI1 is the same model as Higgs Inflation (HI, with $v=0$) whose
  reheating predictions are plotted in \Fig{fig:CMBHI}. Notice that,
  for NMLFI, $\xi$ remains a free parameter whereas it is fixed by the
  amplitude of the CMB anisotropies for HI. The solid contours are the
  one and two-sigma {\data} confidence intervals (marginalized over
  second order slow-roll).}
\label{fig:CMBNMLFI1_3}
\end{center}
\end{figure}

\begin{figure}[H]
\begin{center}
\includegraphics[width=\wappfig,clip=true]{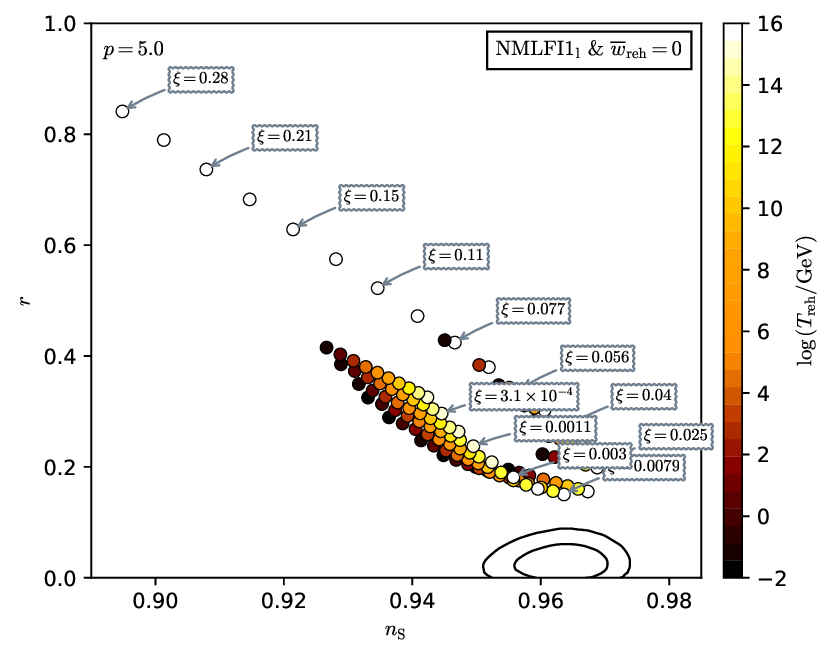}
\includegraphics[width=\wappfig,clip=true]{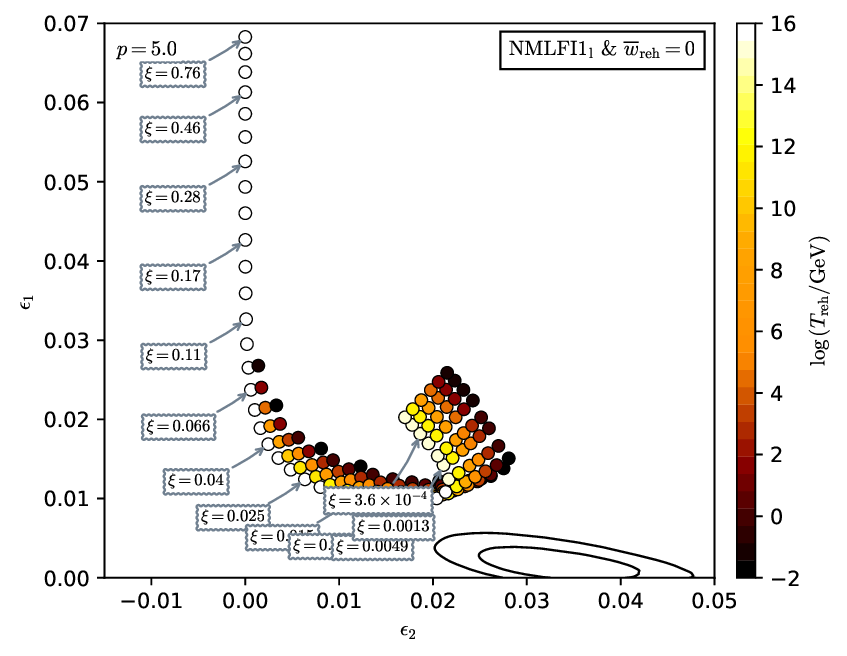}
\caption{Reheating consistent slow-roll predictions for the
  Non-Minimal Large Field Inflation 1 model, for $p=5$, in the plane
  $(\nS,r)$ (top panel) and in the plane $(\epsilon_1,\epsilon_2)$
  (bottom panel). The solid contours are the one and two-sigma {\data}
  confidence intervals (marginalized over second order slow-roll). For
  $p>4$, the potential of NMLFI is monotonously growing with $\chi$
  and NMLFI1 is a large field model. See also \Fig{fig:CMBNMLFI1_5}.}
\label{fig:CMBNMLFI1_4}
\end{center}
\end{figure}

\begin{figure}[H]
\begin{center}
\includegraphics[width=\wappfig,clip=true]{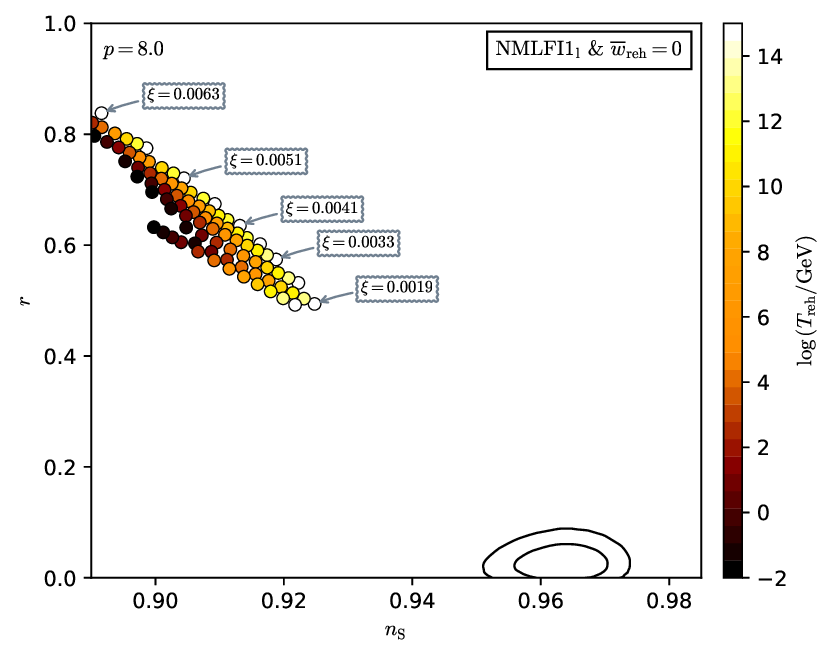}
\includegraphics[width=\wappfig,clip=true]{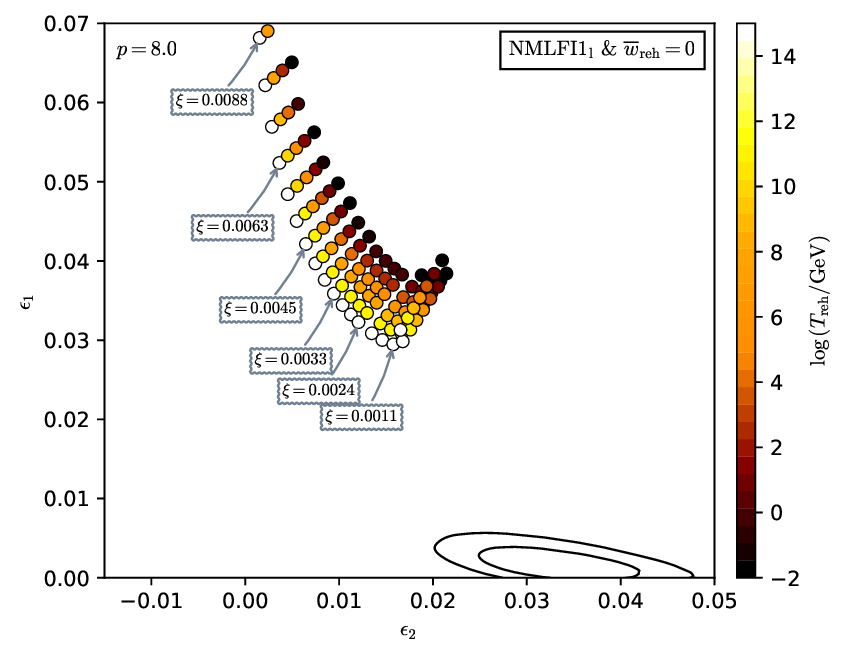}
\caption{Reheating consistent slow-roll predictions for the
  Non-Minimal Large Field Inflation 1 model, for $p=8$. Predictions
  are represented as a function of $\xi<\xizero(p)$ in the plane
  $(\nS,r)$ (top panel) and in the plane $(\epsilon_1,\epsilon_2)$
  (bottom panel). The solid contours are the one and two-sigma {\data}
  confidence intervals (marginalized over second order slow-roll). The
  condition $\xi<\xizero(p)$ is necessary for $p>p_+\simeq 7.46$ to
  ensure that the potential is not too steep to support inflation.}
\label{fig:CMBNMLFI1_5}
\end{center}
\end{figure}

\subsection{Non-Minimal Large Field Inflation 3 (\hyperref[sec:nmlfi]{NMLFI3})}

\begin{figure}[H]
\begin{center}
\includegraphics[width=\wappfig,clip=true]{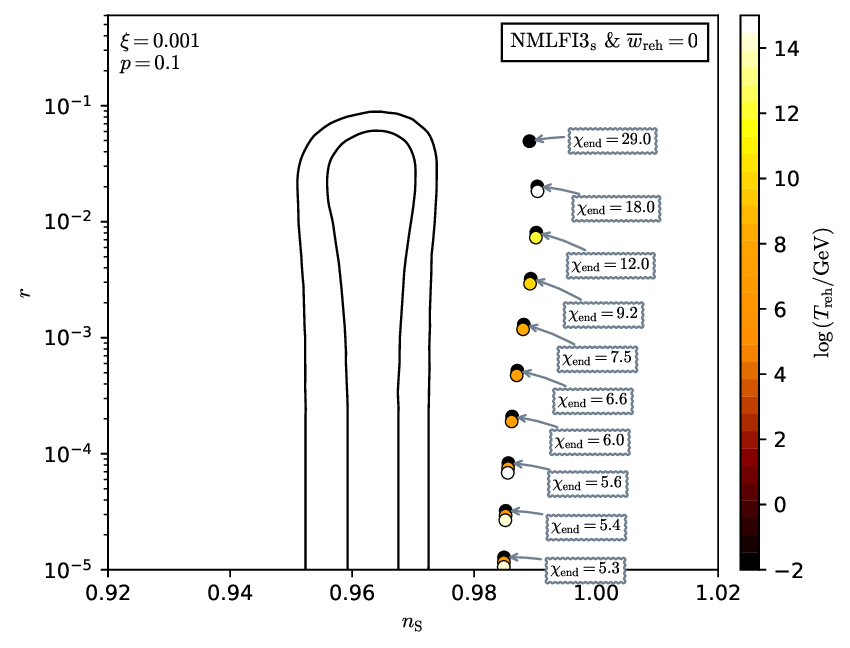}
\includegraphics[width=\wappfig,clip=true]{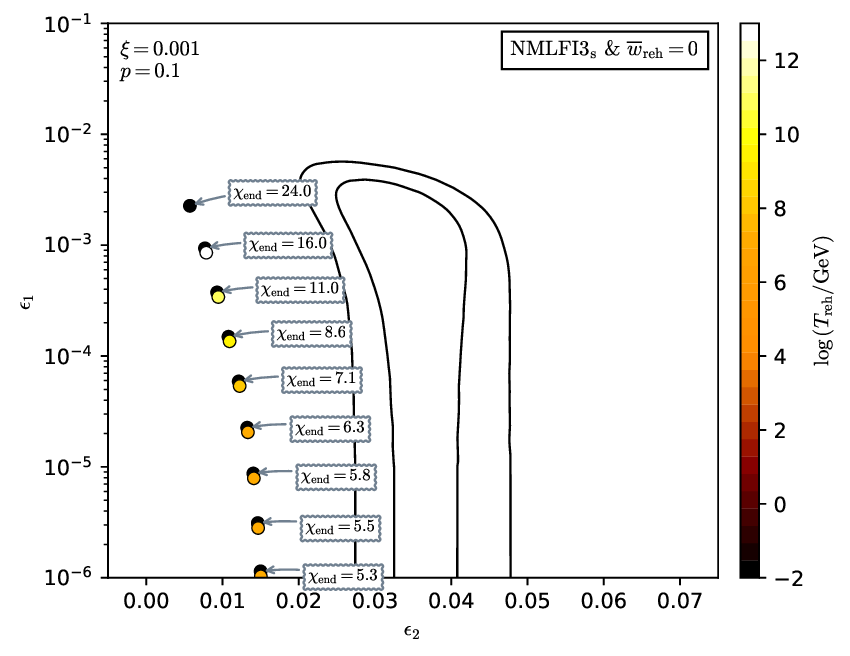}
\caption{Reheating consistent slow-roll predictions for the
  Non-Minimal Large Field Inflation 3 model, for $p=0.1$ and
  $\xi=10^{-3}$. Predictions are represented as a function of
  $\chiend$ in the plane $(\nS,r)$ (top panel) and in the plane
  $(\epsilon_1,\epsilon_2)$ (bottom panel). The solid contours are the
  one and two-sigma {\data} confidence intervals (marginalized over
  second order slow-roll). See also Figs.~\ref{fig:CMBNMLFI3_1} to
  \ref{fig:CMBNMLFI3_8} for the other ``small'' parameter values
  having $p<p_-$ and $\xi < \xizero(p)$.}
\label{fig:CMBNMLFI3_0}
\end{center}
\end{figure}

\begin{figure}[H]
\begin{center}
\includegraphics[width=\wappfig,clip=true]{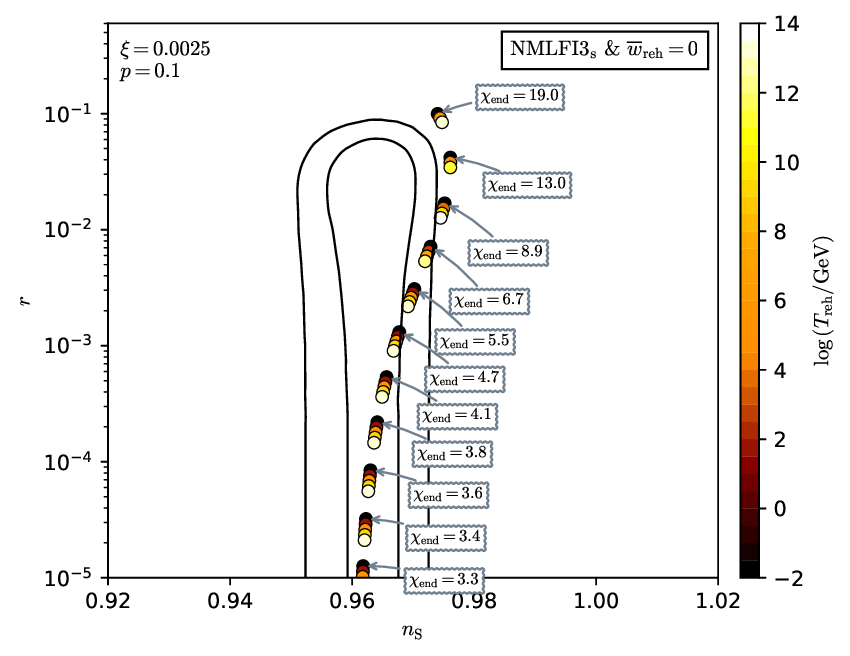}
\includegraphics[width=\wappfig,clip=true]{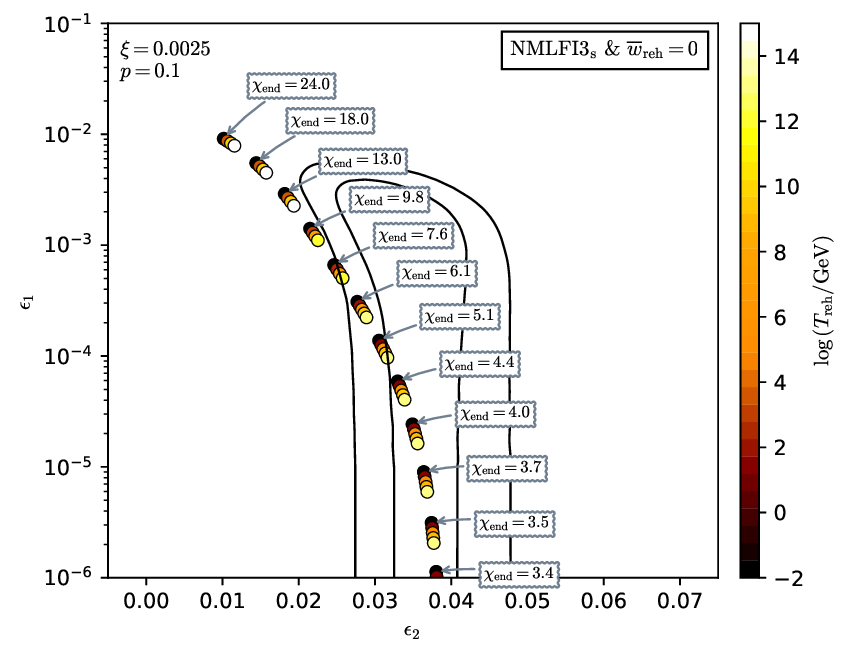}
\caption{Reheating consistent slow-roll predictions for the
  Non-Minimal Large Field Inflation 3 model, for $p=0.1$ and
  $\xi=2.5 \times 10^{-3}$. Predictions are represented as a function of
  $\chiend$ in the plane $(\nS,r)$ (top panel) and in the plane
  $(\epsilon_1,\epsilon_2)$ (bottom panel). The solid contours are the
  one and two-sigma {\data} confidence intervals (marginalized over
  second order slow-roll).}
\label{fig:CMBNMLFI3_1}
\end{center}
\end{figure}

\begin{figure}[H]
\begin{center}
\includegraphics[width=\wappfig,clip=true]{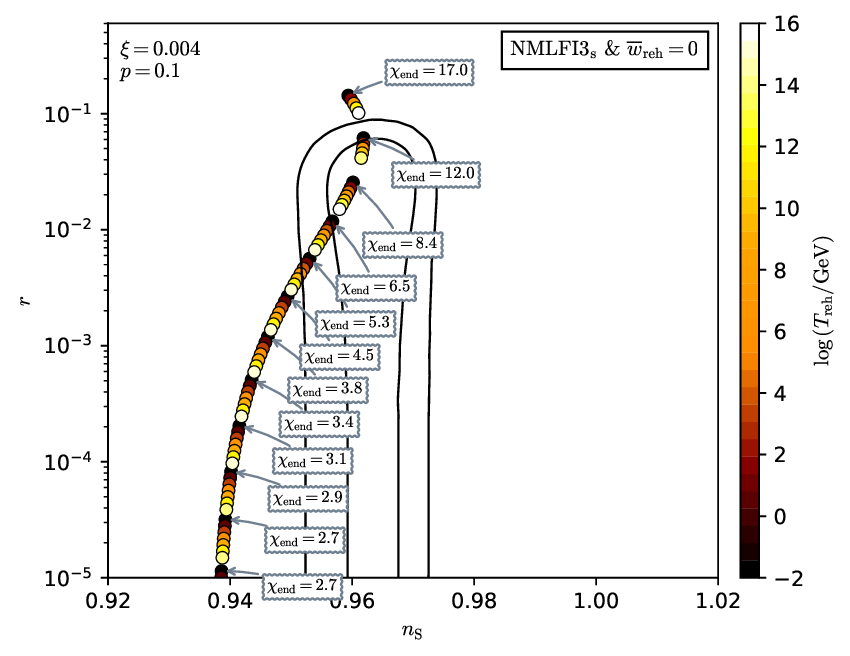}
\includegraphics[width=\wappfig,clip=true]{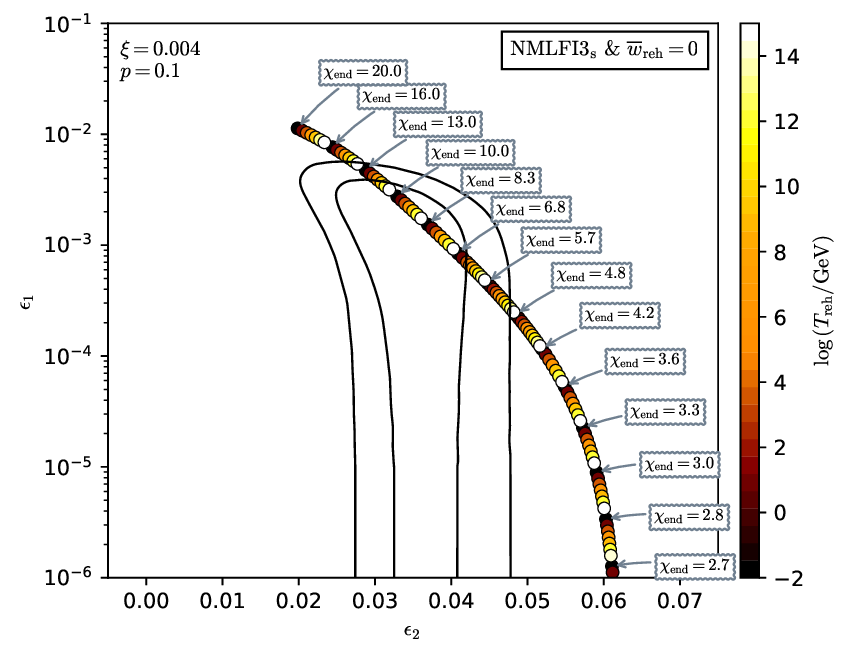}
\caption{Reheating consistent slow-roll predictions for the
  Non-Minimal Large Field Inflation 3 model, for $p=0.1$ and
  $\xi=4 \times 10^{-3}$. Predictions are represented as a function of
  $\chiend$ in the plane $(\nS,r)$ (top panel) and in the plane
  $(\epsilon_1,\epsilon_2)$ (bottom panel). The solid contours are the
  one and two-sigma {\data} confidence intervals (marginalized over
  second order slow-roll).}
\label{fig:CMBNMLFI3_2}
\end{center}
\end{figure}

\begin{figure}[H]
\begin{center}
\includegraphics[width=\wappfig,clip=true]{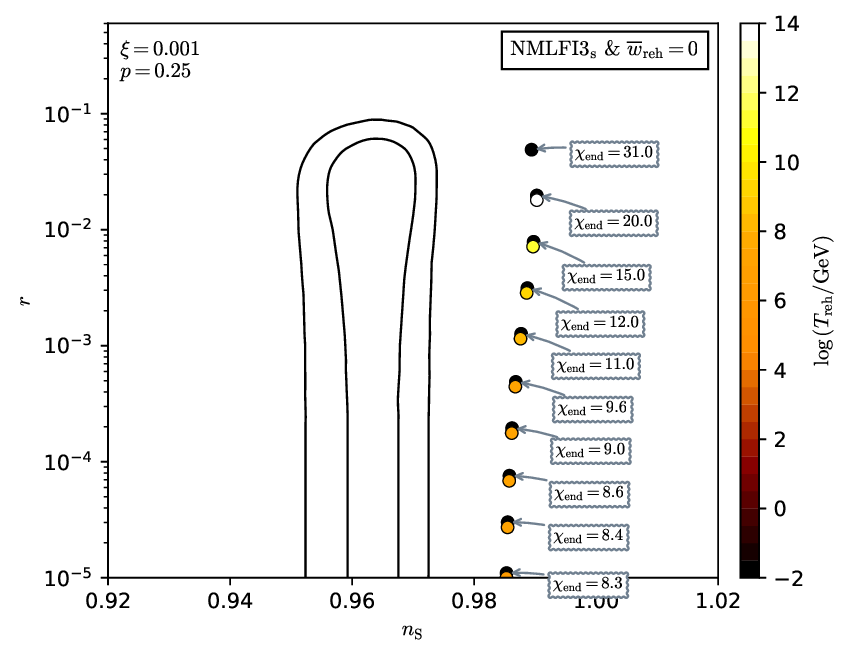}
\includegraphics[width=\wappfig,clip=true]{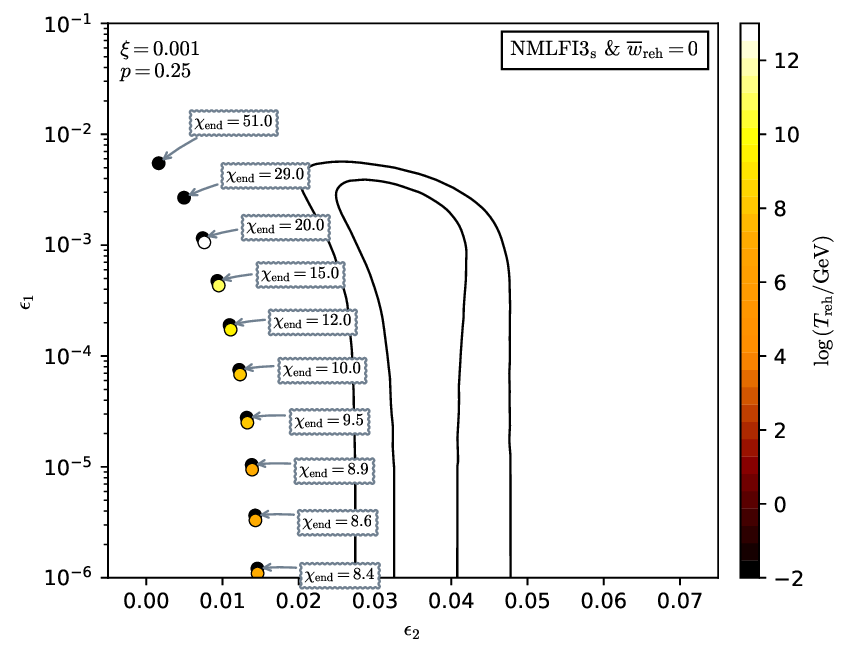}
\caption{Reheating consistent slow-roll predictions for the
  Non-Minimal Large Field Inflation 3 model, for $p=0.25$ and
  $\xi=10^{-3}$. Predictions are represented as a function of
  $\chiend$ in the plane $(\nS,r)$ (top panel) and in the plane
  $(\epsilon_1,\epsilon_2)$ (bottom panel). The solid contours are the
  one and two-sigma {\data} confidence intervals (marginalized over
  second order slow-roll).}
\label{fig:CMBNMLFI3_3}
\end{center}
\end{figure}

\begin{figure}[H]
\begin{center}
\includegraphics[width=\wappfig,clip=true]{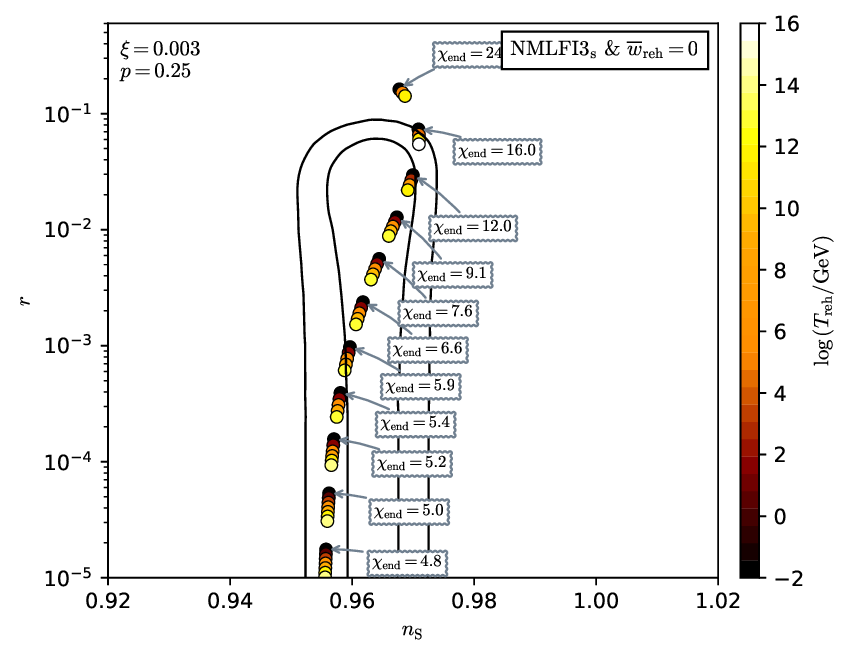}
\includegraphics[width=\wappfig,clip=true]{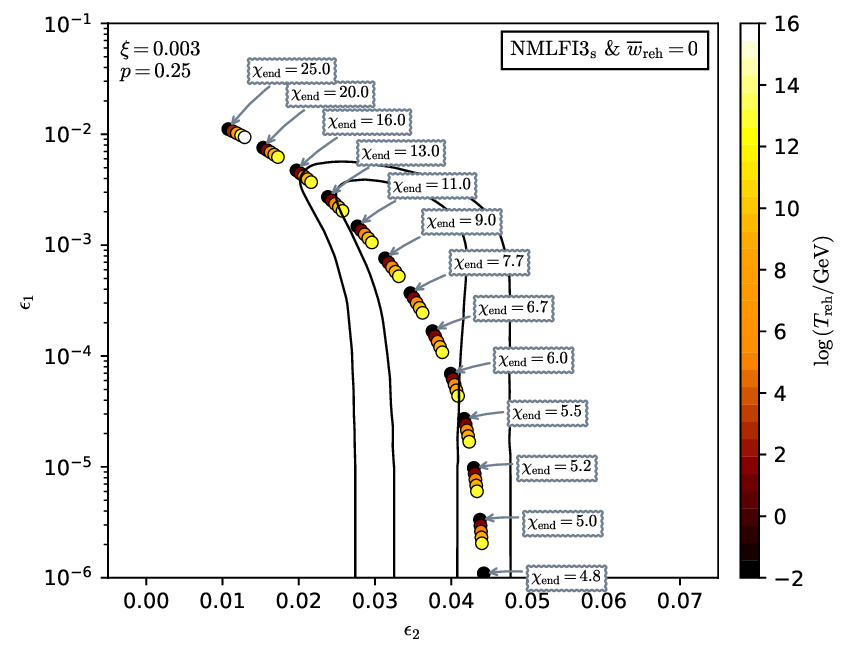}
\caption{Reheating consistent slow-roll predictions for the
  Non-Minimal Large Field Inflation 3 model, for $p=0.25$ and
  $\xi=3\times 10^{-3}$. Predictions are represented as a function of
  $\chiend$ in the plane $(\nS,r)$ (top panel) and in the plane
  $(\epsilon_1,\epsilon_2)$ (bottom panel). The solid contours are the
  one and two-sigma {\data} confidence intervals (marginalized over
  second order slow-roll).}
\label{fig:CMBNMLFI3_4}
\end{center}
\end{figure}

\begin{figure}[H]
\begin{center}
\includegraphics[width=\wappfig,clip=true]{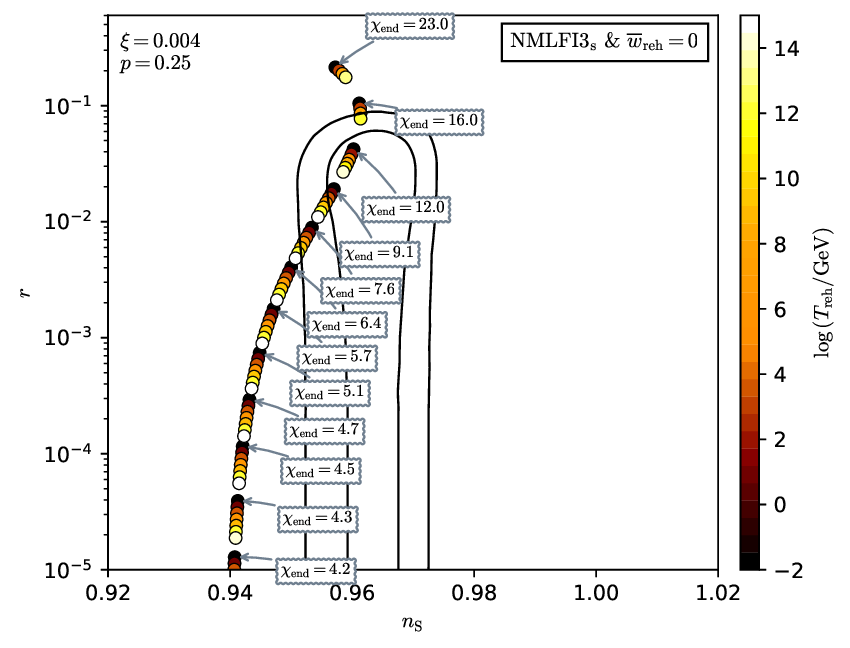}
\includegraphics[width=\wappfig,clip=true]{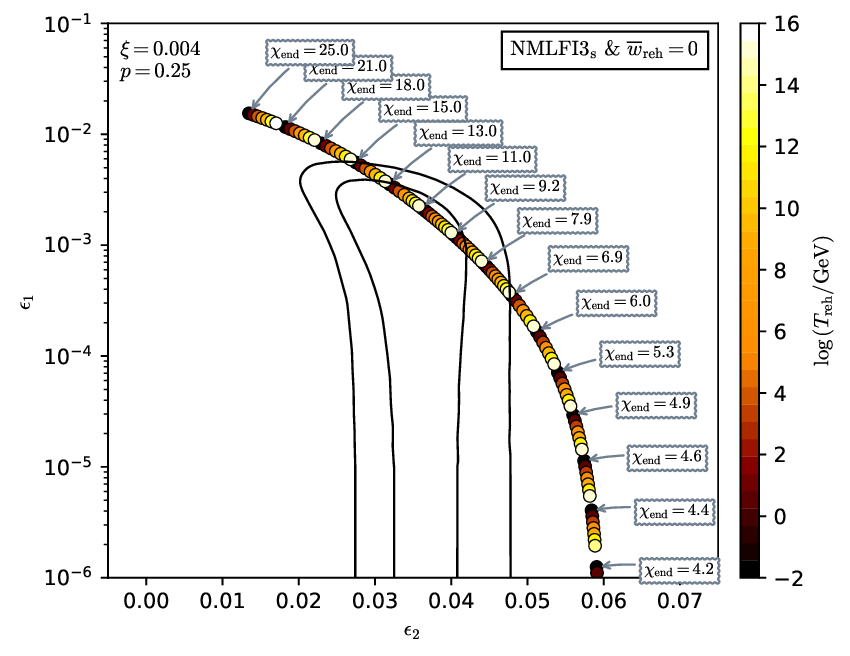}
\caption{Reheating consistent slow-roll predictions for the
  Non-Minimal Large Field Inflation 3 model, for $p=0.25$ and
  $\xi=4\times 10^{-3}$. Predictions are represented as a function of
  $\chiend$ in the plane $(\nS,r)$ (top panel) and in the plane
  $(\epsilon_1,\epsilon_2)$ (bottom panel). The solid contours are the
  one and two-sigma {\data} confidence intervals (marginalized over
  second order slow-roll).}
\label{fig:CMBNMLFI3_5}
\end{center}
\end{figure}

\begin{figure}[H]
\begin{center}
\includegraphics[width=\wappfig,clip=true]{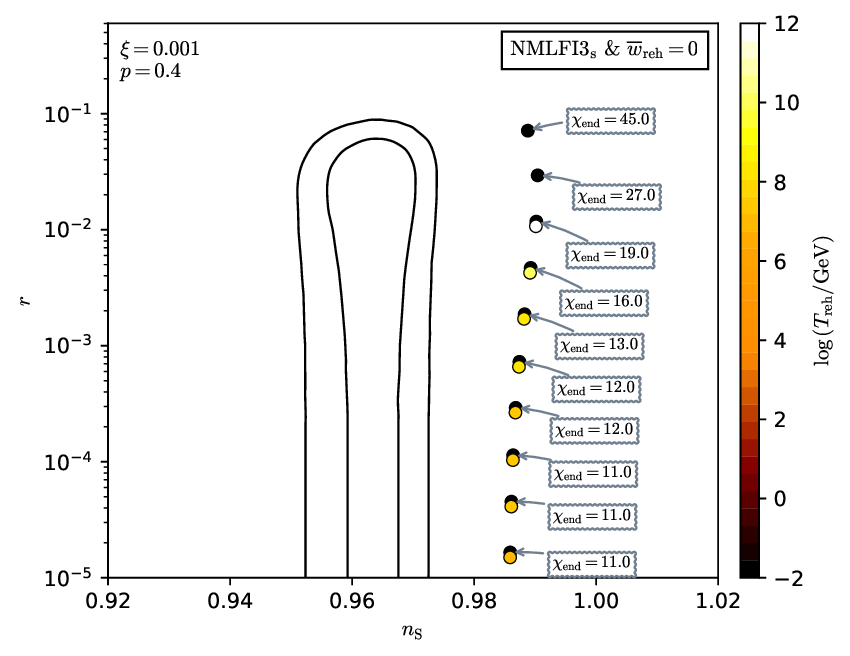}
\includegraphics[width=\wappfig,clip=true]{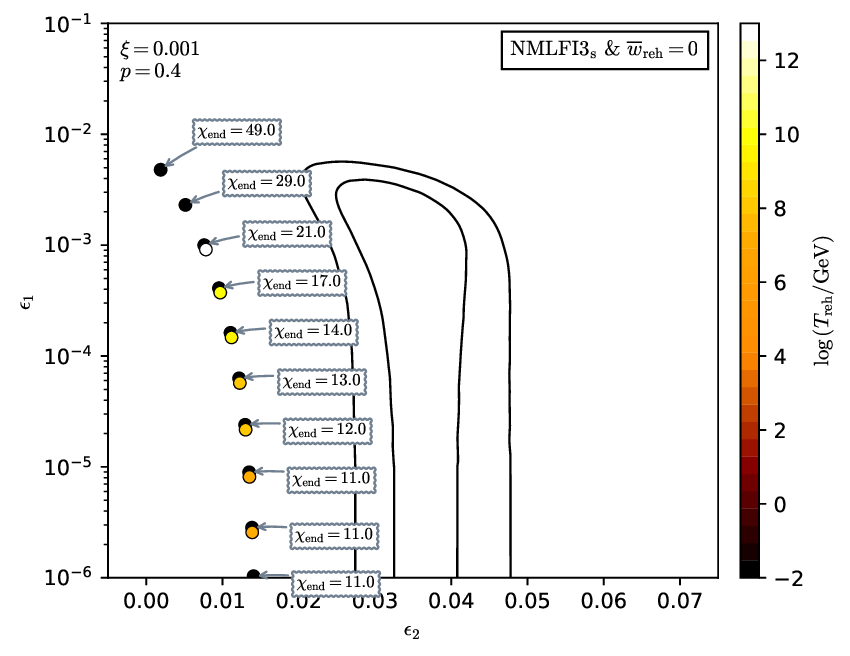}
\caption{Reheating consistent slow-roll predictions for the
  Non-Minimal Large Field Inflation 3 model, for $p=0.4$ and
  $\xi=10^{-3}$. Predictions are represented as a function of
  $\chiend$ in the plane $(\nS,r)$ (top panel) and in the plane
  $(\epsilon_1,\epsilon_2)$ (bottom panel). The solid contours are the
  one and two-sigma {\data} confidence intervals (marginalized over
  second order slow-roll).}
\label{fig:CMBNMLFI3_6}
\end{center}
\end{figure}

\begin{figure}[H]
\begin{center}
\includegraphics[width=\wappfig,clip=true]{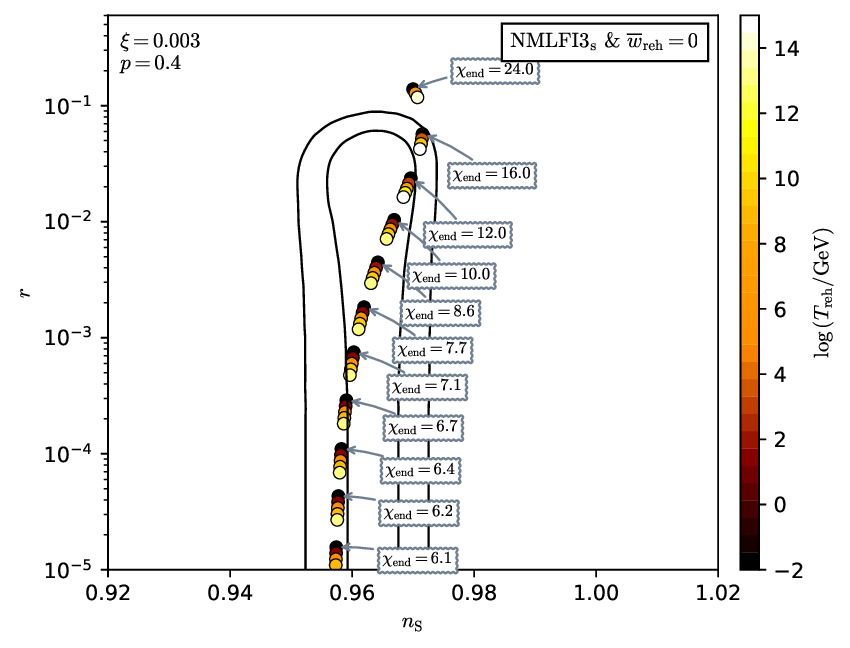}
\includegraphics[width=\wappfig,clip=true]{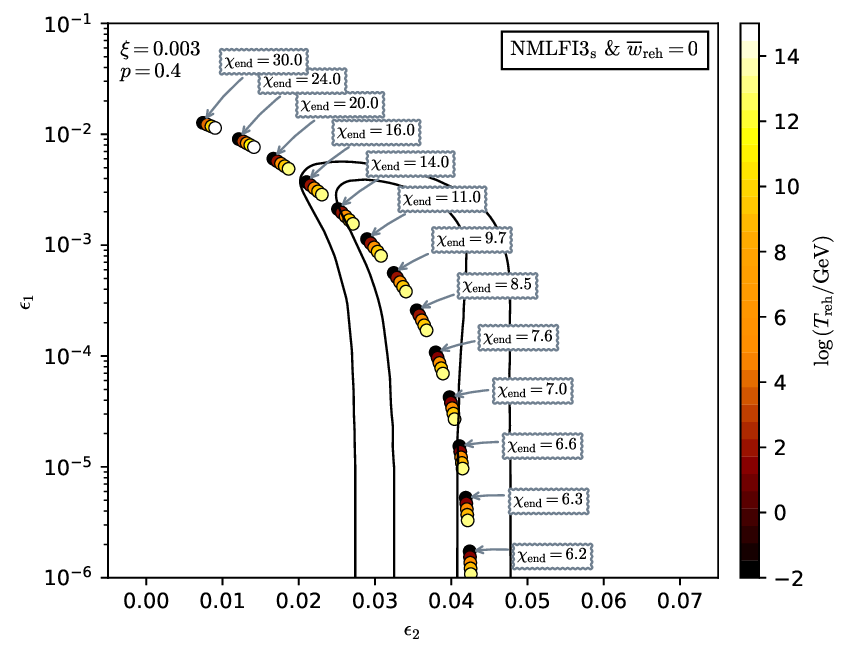}
\caption{Reheating consistent slow-roll predictions for the
  Non-Minimal Large Field Inflation 3 model, for $p=0.4$ and
  $\xi=3\times 10^{-3}$. Predictions are represented as a function of
  $\chiend$ in the plane $(\nS,r)$ (top panel) and in the plane
  $(\epsilon_1,\epsilon_2)$ (bottom panel). The solid contours are the
  one and two-sigma {\data} confidence intervals (marginalized over
  second order slow-roll).}
\label{fig:CMBNMLFI3_7}
\end{center}
\end{figure}

\begin{figure}[H]
\begin{center}
\includegraphics[width=\wappfig,clip=true]{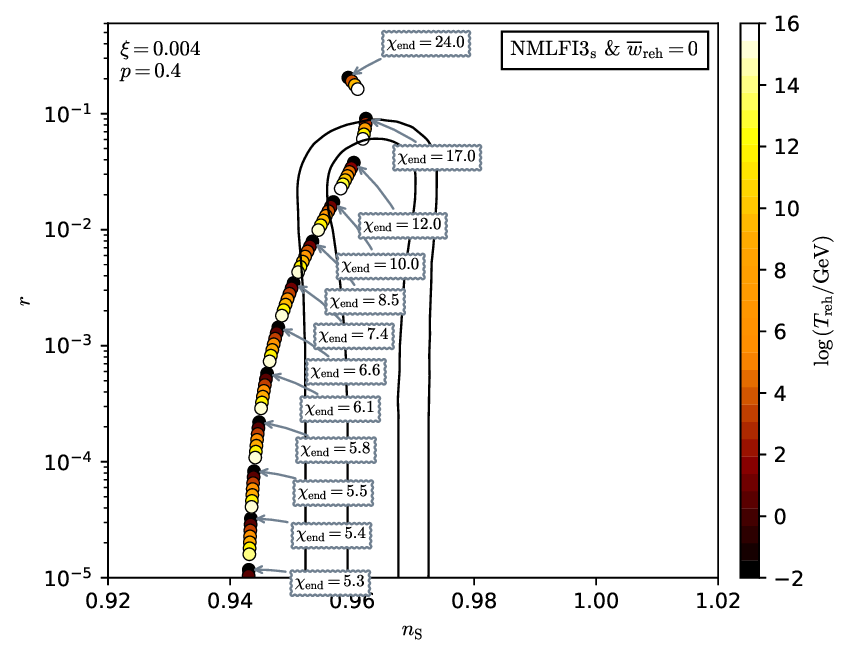}
\includegraphics[width=\wappfig,clip=true]{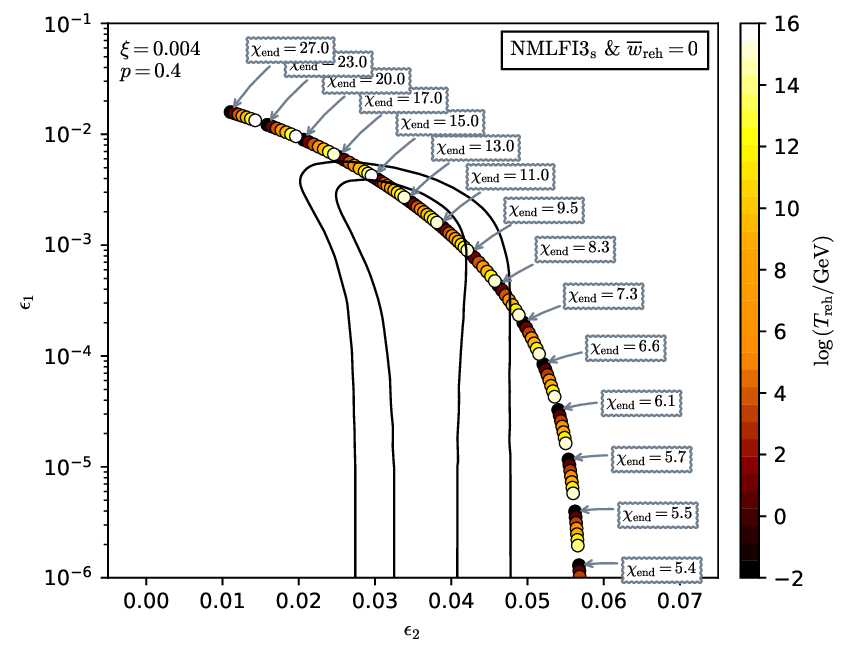}
\caption{Reheating consistent slow-roll predictions for the
  Non-Minimal Large Field Inflation 3 model, for $p=0.4$ and
  $\xi=4\times 10^{-3}$. Predictions are represented as a function of
  $\chiend$ in the plane $(\nS,r)$ (top panel) and in the plane
  $(\epsilon_1,\epsilon_2)$ (bottom panel). The solid contours are the
  one and two-sigma {\data} confidence intervals (marginalized over
  second order slow-roll).}
\label{fig:CMBNMLFI3_8}
\end{center}
\end{figure}

\begin{figure}[H]
\begin{center}
\includegraphics[width=\wappfig,clip=true]{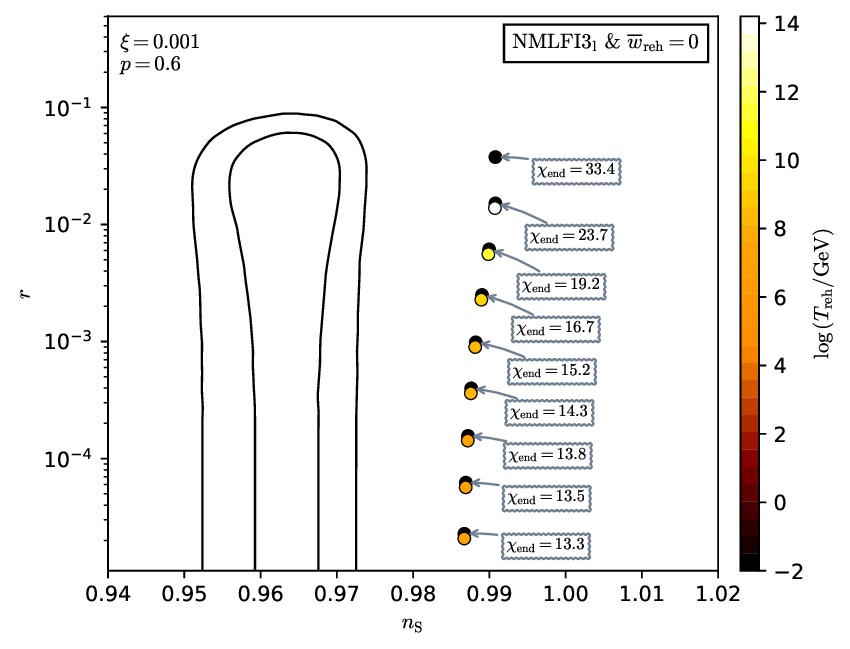}
\includegraphics[width=\wappfig,clip=true]{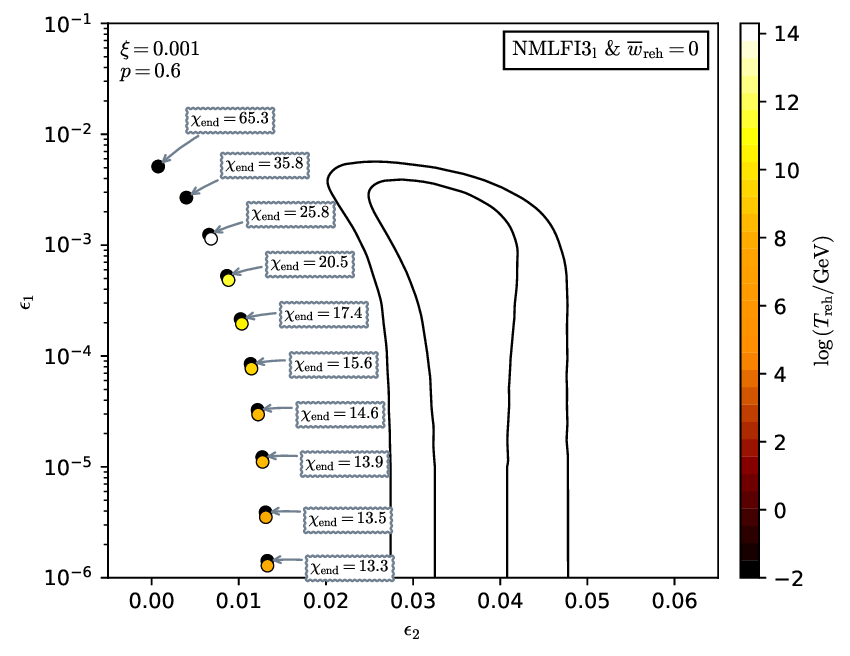}
\caption{Reheating consistent slow-roll predictions for the
  Non-Minimal Large Field Inflation 3 model, for $p=0.6$ and
  $\xi=10^{-3}$. Predictions are represented as a function of
  $\chiend$ in the plane $(\nS,r)$ (top panel) and in the plane
  $(\epsilon_1,\epsilon_2)$ (bottom panel). The solid contours are the
  one and two-sigma {\data} confidence intervals (marginalized over
  second order slow-roll). See also \Figs{fig:CMBNMLFI3_10} to
  \ref{fig:CMBNMLFI3_17} for the other ``large'' parameter values
  having $p>p_-$ (and $p<4$).}
\label{fig:CMBNMLFI3_9}
\end{center}
\end{figure}

\begin{figure}[H]
\begin{center}
\includegraphics[width=\wappfig,clip=true]{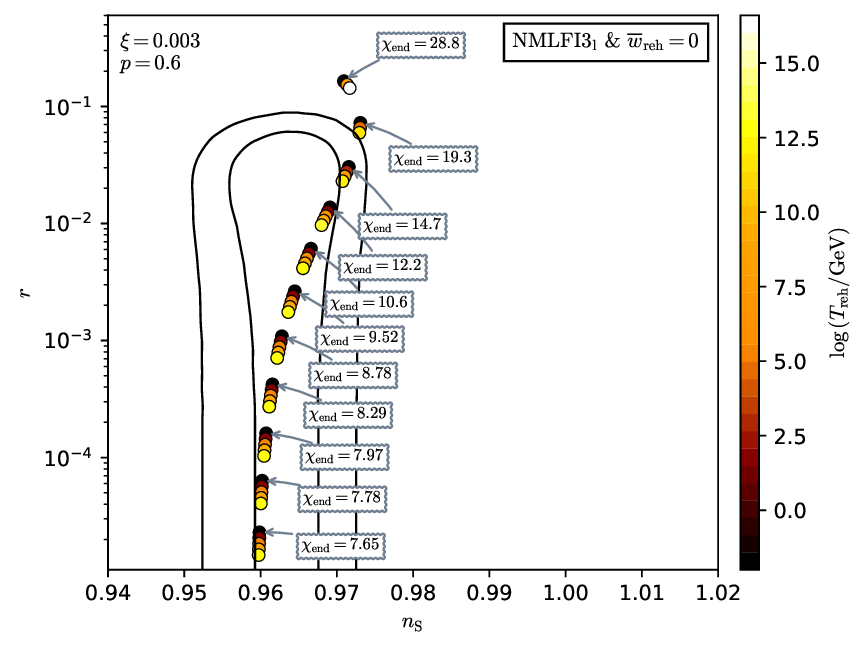}
\includegraphics[width=\wappfig,clip=true]{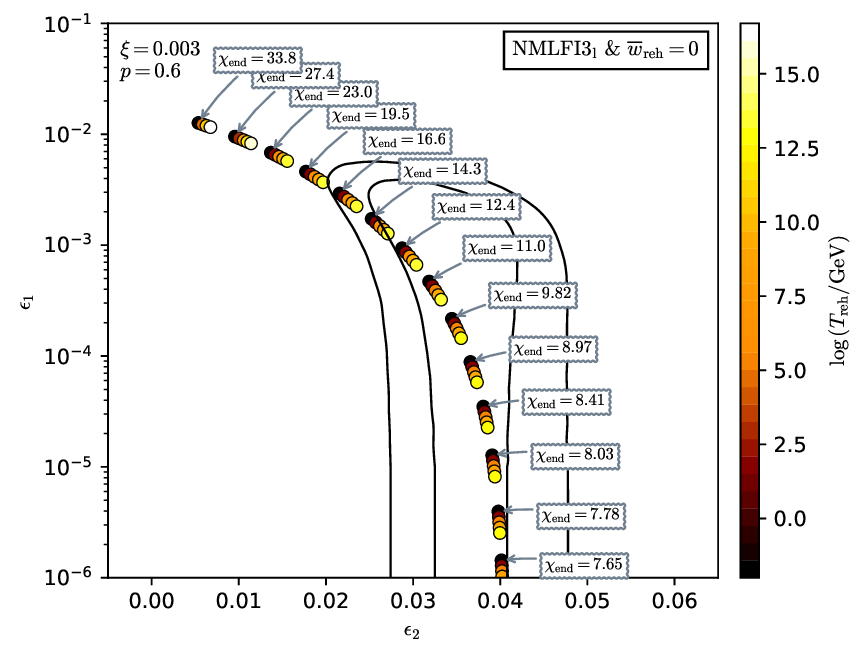}
\caption{Reheating consistent slow-roll predictions for the
  Non-Minimal Large Field Inflation 3 model, for $p=0.6$ and
  $\xi=3\times 10^{-3}$. Predictions are represented as a function of
  $\chiend$ in the plane $(\nS,r)$ (top panel) and in the plane
  $(\epsilon_1,\epsilon_2)$ (bottom panel). The solid contours are the
  one and two-sigma {\data} confidence intervals (marginalized over
  second order slow-roll).}
\label{fig:CMBNMLFI3_10}
\end{center}
\end{figure}

\begin{figure}[H]
\begin{center}
\includegraphics[width=\wappfig,clip=true]{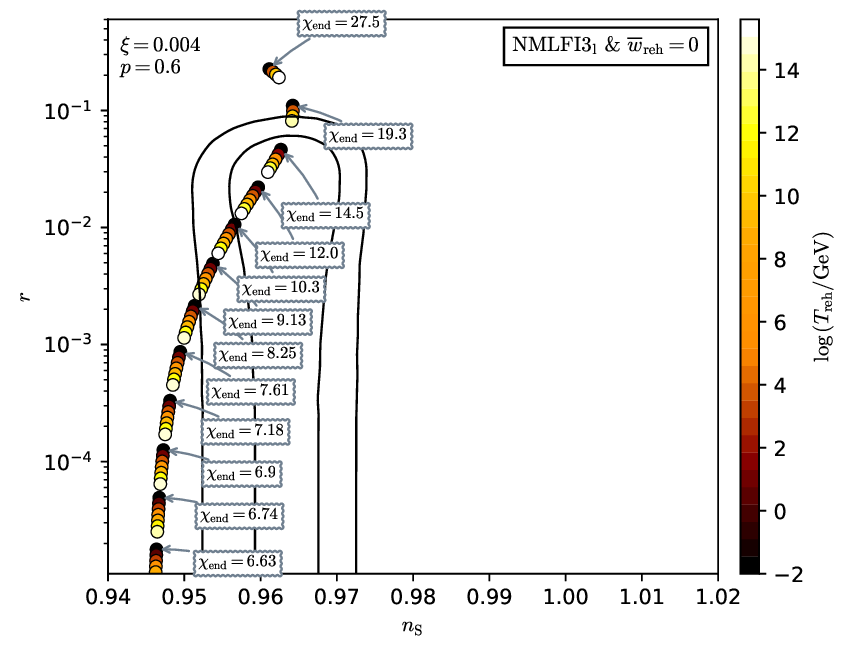}
\includegraphics[width=\wappfig,clip=true]{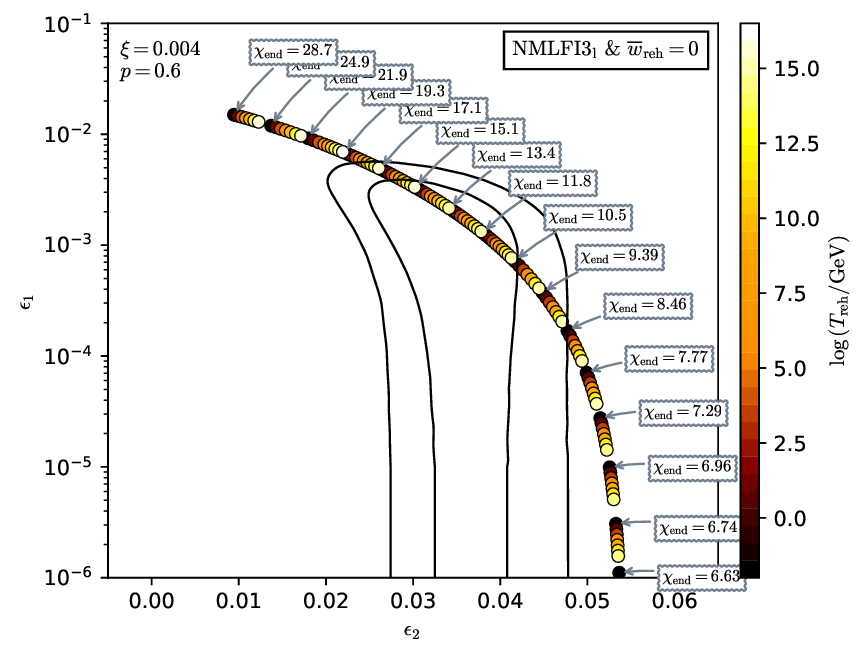}
\caption{Reheating consistent slow-roll predictions for the
  Non-Minimal Large Field Inflation 3 model, for $p=0.6$ and
  $\xi=4\times 10^{-3}$. Predictions are represented as a function of
  $\chiend$ in the plane $(\nS,r)$ (top panel) and in the plane
  $(\epsilon_1,\epsilon_2)$ (bottom panel). The solid contours are the
  one and two-sigma {\data} confidence intervals (marginalized over
  second order slow-roll).}
\label{fig:CMBNMLFI3_11}
\end{center}
\end{figure}

\begin{figure}[H]
\begin{center}
\includegraphics[width=\wappfig,clip=true]{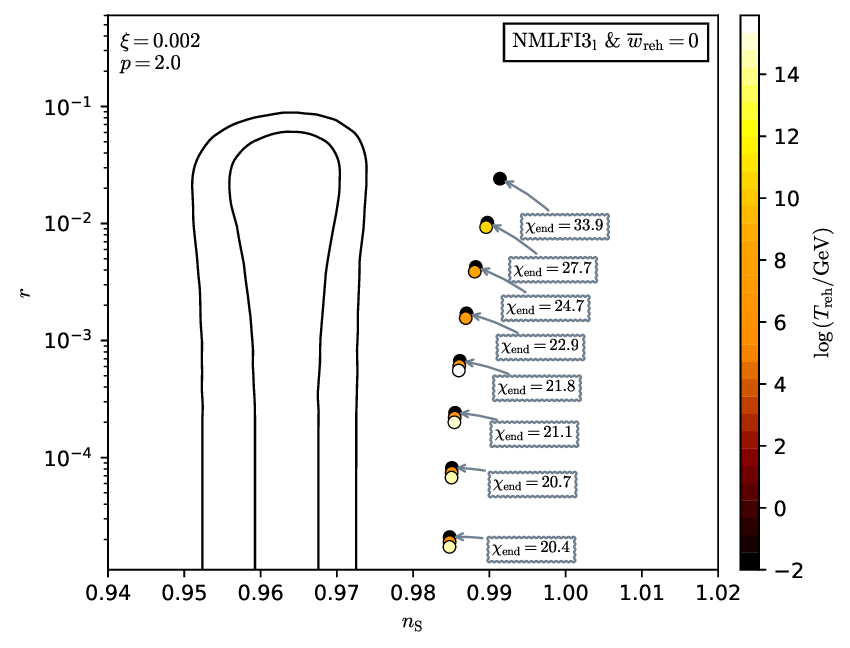}
\includegraphics[width=\wappfig,clip=true]{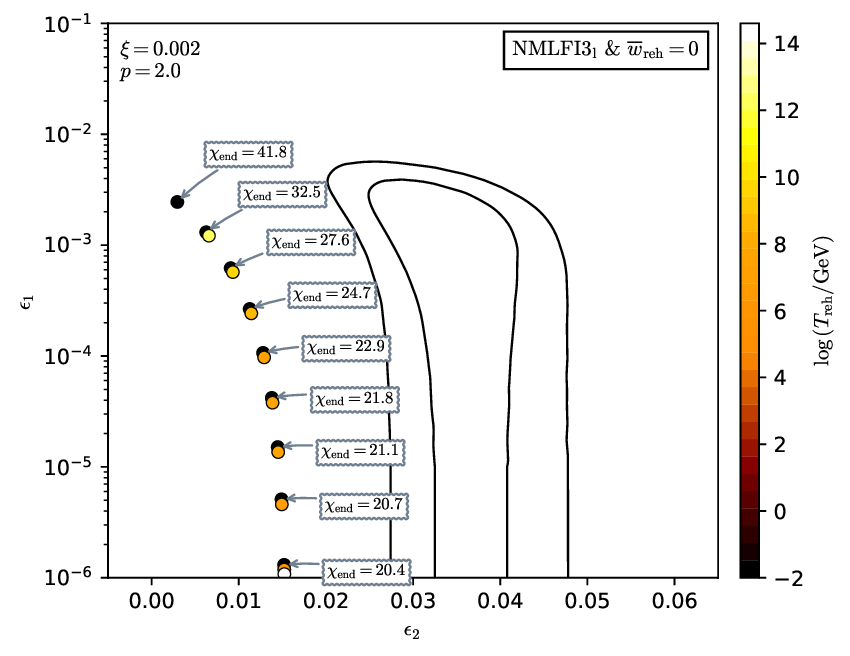}
\caption{Reheating consistent slow-roll predictions for the
  Non-Minimal Large Field Inflation 3 model, for $p=2$ and
  $\xi=2\times 10^{-3}$. Predictions are represented as a function of
  $\chiend$ in the plane $(\nS,r)$ (top panel) and in the plane
  $(\epsilon_1,\epsilon_2)$ (bottom panel). The solid contours are the
  one and two-sigma {\data} confidence intervals (marginalized over
  second order slow-roll).}
\label{fig:CMBNMLFI3_12}
\end{center}
\end{figure}

\begin{figure}[H]
\begin{center}
\includegraphics[width=\wappfig,clip=true]{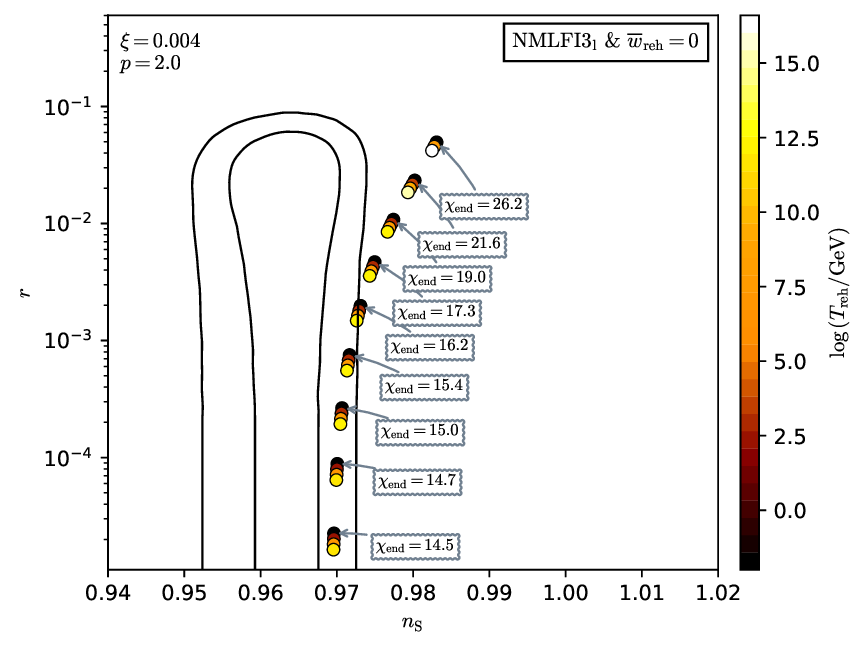}
\includegraphics[width=\wappfig,clip=true]{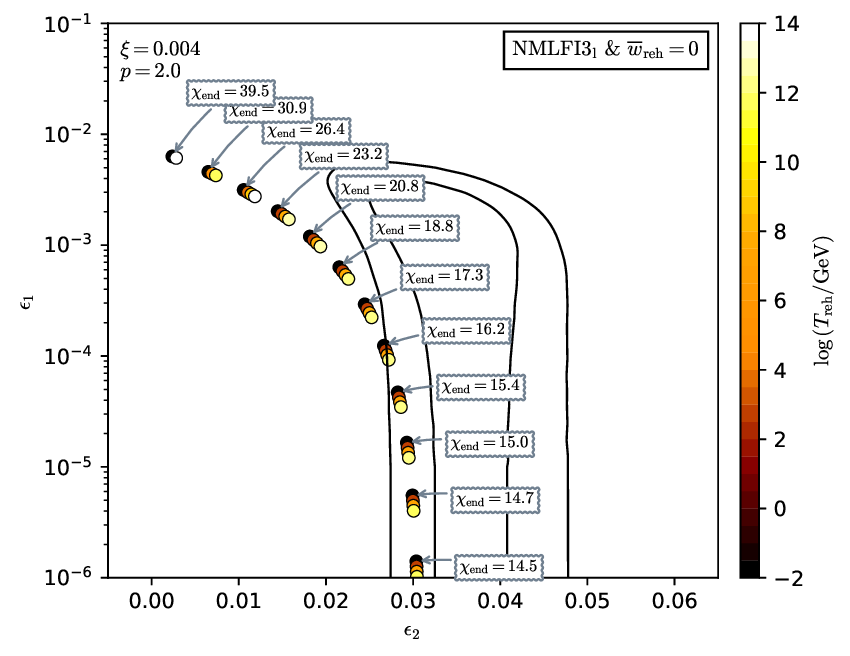}
\caption{Reheating consistent slow-roll predictions for the
  Non-Minimal Large Field Inflation 3 model, for $p=2$ and
  $\xi=4\times 10^{-3}$. Predictions are represented as a function of
  $\chiend$ in the plane $(\nS,r)$ (top panel) and in the plane
  $(\epsilon_1,\epsilon_2)$ (bottom panel). The solid contours are the
  one and two-sigma {\data} confidence intervals (marginalized over
  second order slow-roll).}
\label{fig:CMBNMLFI3_13}
\end{center}
\end{figure}

\begin{figure}[H]
\begin{center}
\includegraphics[width=\wappfig,clip=true]{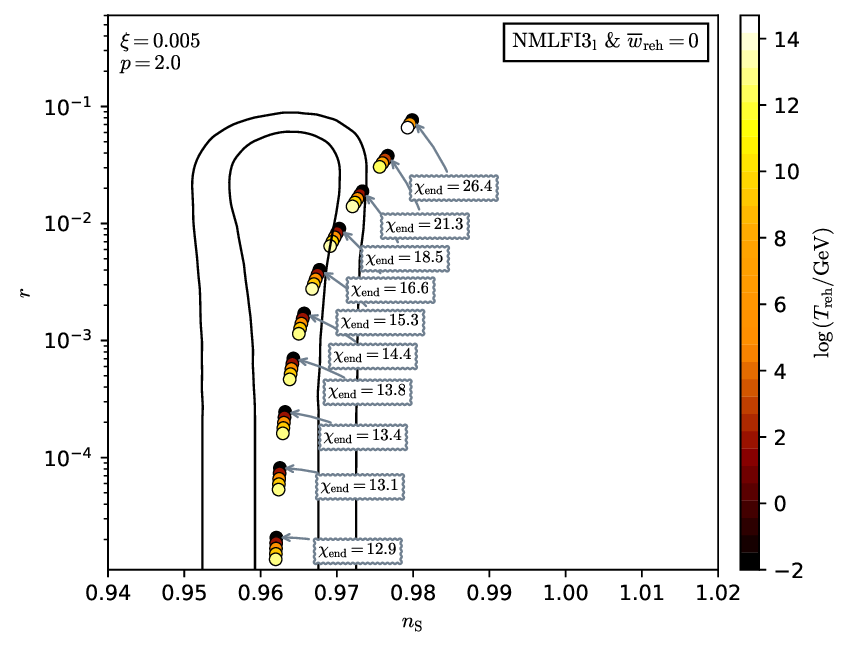}
\includegraphics[width=\wappfig,clip=true]{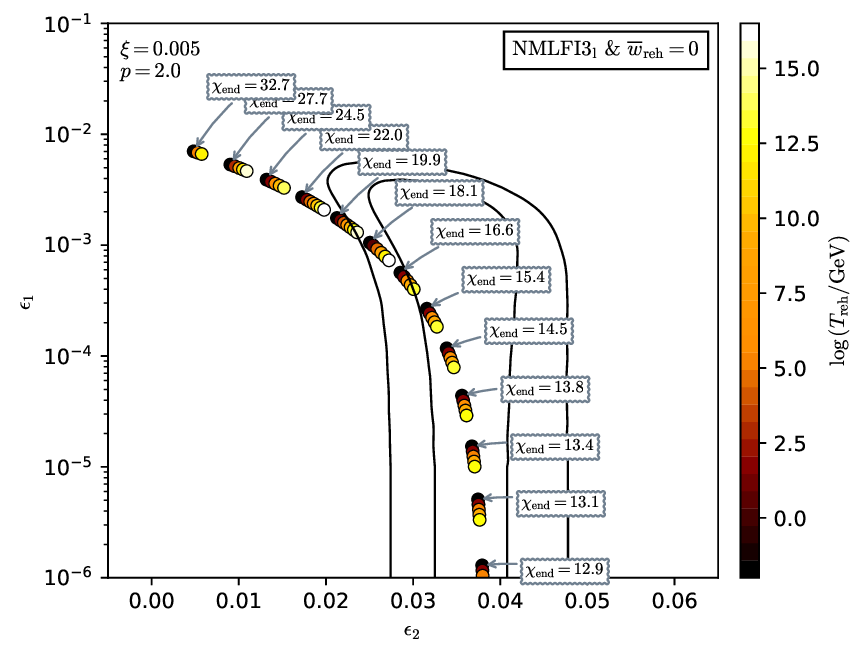}
\caption{Reheating consistent slow-roll predictions for the
  Non-Minimal Large Field Inflation 3 model, for $p=2$ and
  $\xi=5\times 10^{-3}$. Predictions are represented as a function of
  $\chiend$ in the plane $(\nS,r)$ (top panel) and in the plane
  $(\epsilon_1,\epsilon_2)$ (bottom panel). The solid contours are the
  one and two-sigma {\data} confidence intervals (marginalized over
  second order slow-roll).}
\label{fig:CMBNMLFI3_14}
\end{center}
\end{figure}

\begin{figure}[H]
\begin{center}
\includegraphics[width=\wappfig,clip=true]{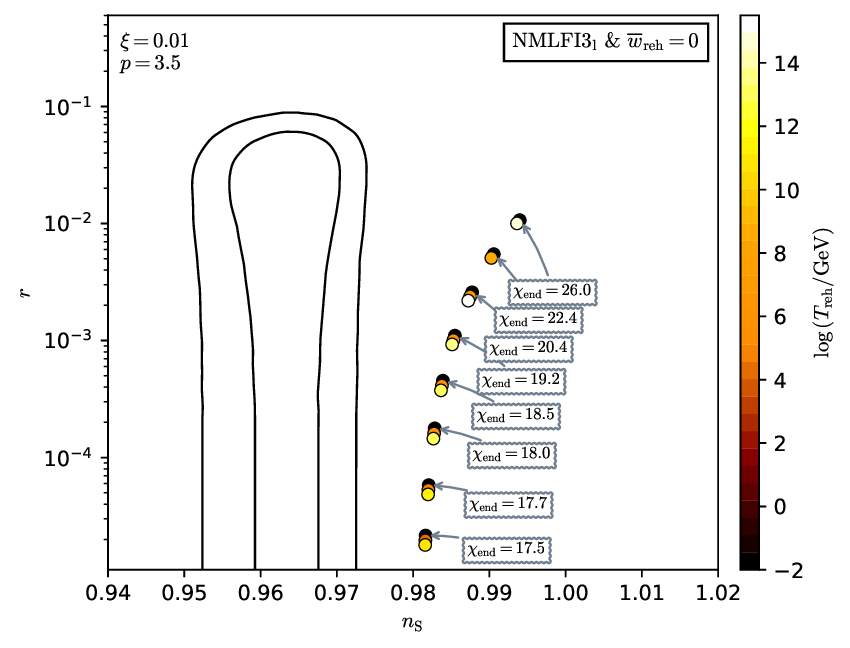}
\includegraphics[width=\wappfig,clip=true]{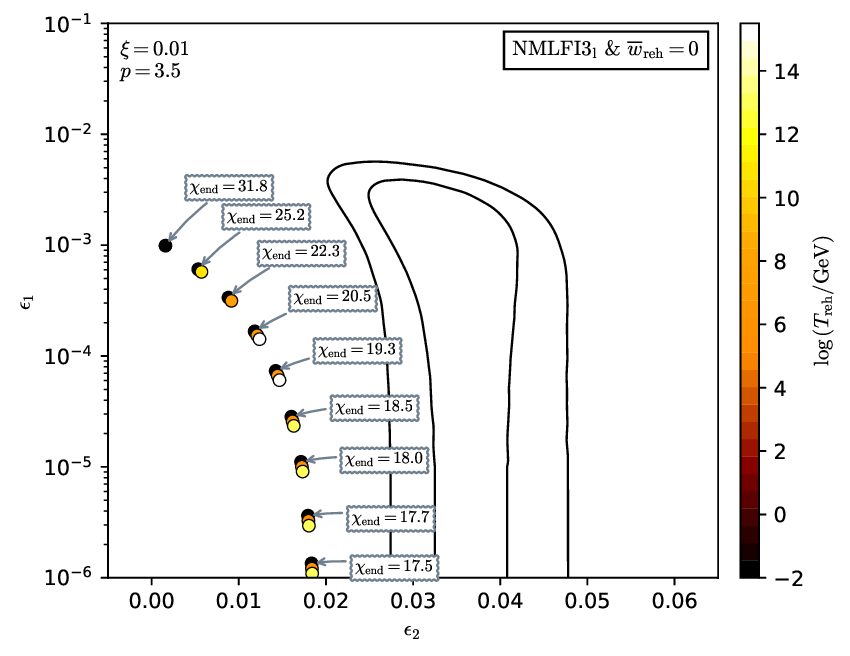}
\caption{Reheating consistent slow-roll predictions for the
  Non-Minimal Large Field Inflation 3 model, for $p=3.5$ and
  $\xi=10^{-2}$. Predictions are represented as a function of
  $\chiend$ in the plane $(\nS,r)$ (top panel) and in the plane
  $(\epsilon_1,\epsilon_2)$ (bottom panel). The solid contours are the
  one and two-sigma {\data} confidence intervals (marginalized over
  second order slow-roll).}
\label{fig:CMBNMLFI3_15}
\end{center}
\end{figure}

\begin{figure}[H]
\begin{center}
\includegraphics[width=\wappfig,clip=true]{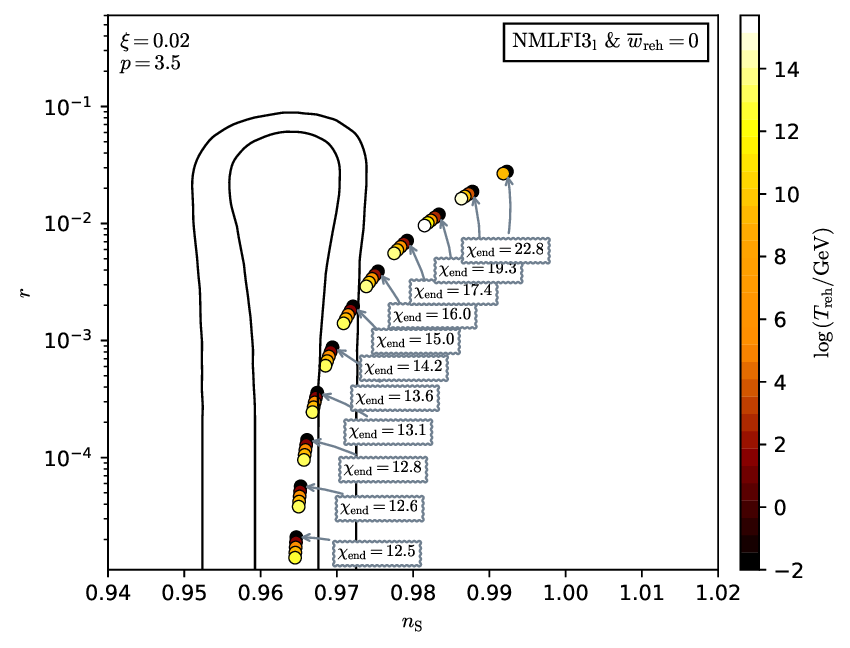}
\includegraphics[width=\wappfig,clip=true]{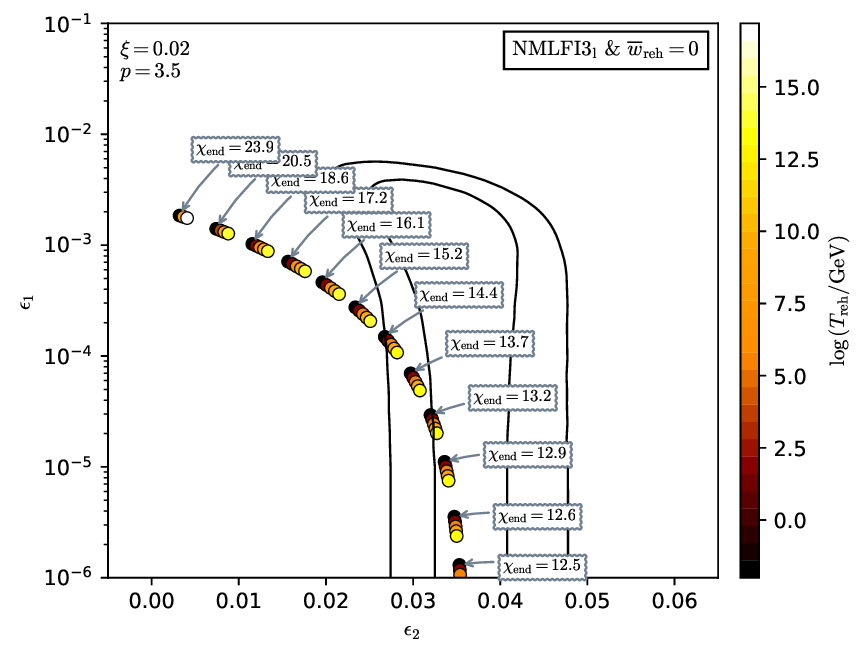}
\caption{Reheating consistent slow-roll predictions for the
  Non-Minimal Large Field Inflation 3 model, for $p=3.5$ and
  $\xi=2\times 10^{-2}$. Predictions are represented as a function of
  $\chiend$ in the plane $(\nS,r)$ (top panel) and in the plane
  $(\epsilon_1,\epsilon_2)$ (bottom panel). The solid contours are the
  one and two-sigma {\data} confidence intervals (marginalized over
  second order slow-roll).}
\label{fig:CMBNMLFI3_16}
\end{center}
\end{figure}

\begin{figure}[H]
\begin{center}
\includegraphics[width=\wappfig,clip=true]{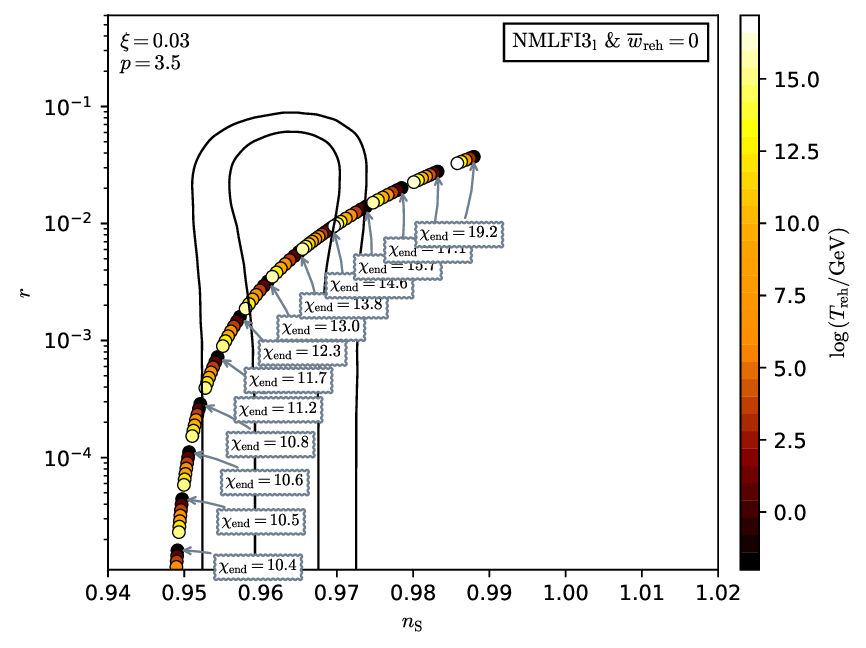}
\includegraphics[width=\wappfig,clip=true]{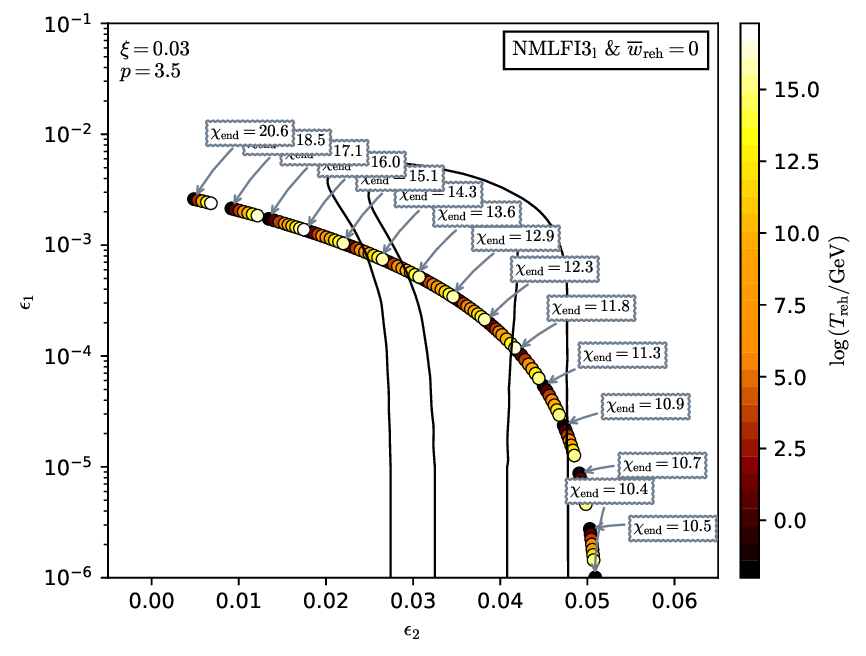}
\caption{Reheating consistent slow-roll predictions for the
  Non-Minimal Large Field Inflation 3 model, for $p=3.5$ and
  $\xi=3\times 10^{-2}$. Predictions are represented as a function of
  $\chiend$ in the plane $(\nS,r)$ (top panel) and in the plane
  $(\epsilon_1,\epsilon_2)$ (bottom panel). The solid contours are the
  one and two-sigma {\data} confidence intervals (marginalized over
  second order slow-roll).}
\label{fig:CMBNMLFI3_17}
\end{center}
\end{figure}

\subsection{Superconformal \texorpdfstring{$\alpha$}{alpha}-Attractor B Inflation (\hyperref[sec:sabi]{SABI})}

\begin{figure}[H]
\begin{center}
\includegraphics[width=\wappfig,clip=true]{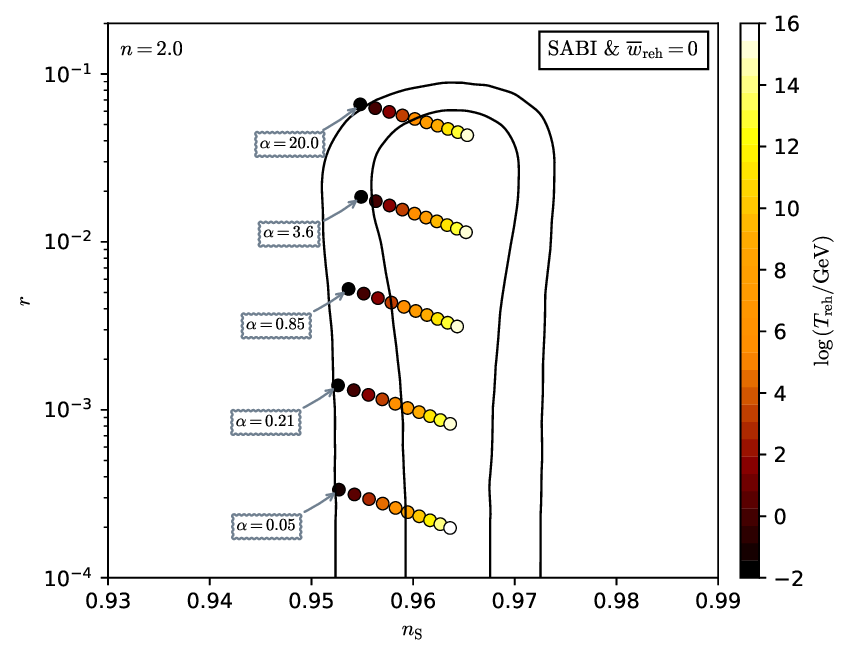}
\includegraphics[width=\wappfig,clip=true]{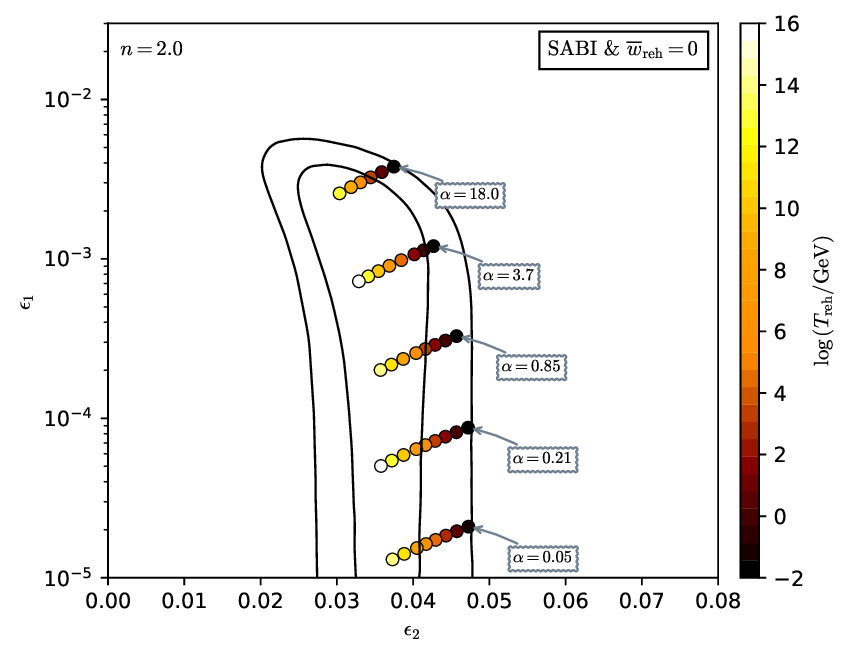}
\caption{Reheating consistent slow-roll predictions for the
  Superconformal $\alpha$-attractor B Inflation model for
  $n=2$. Predictions are represented as a function of $\alpha$ in the
  plane $(\nS,r)$ (top panel) and in the plane
  $(\epsilon_1,\epsilon_2)$ (bottom panel). The solid contours are the
  one and two-sigma {\data} confidence intervals (marginalized over
  second order slow-roll). See also Figs.~\ref{fig:CMBSABI_1} and
  \ref{fig:CMBSABI_2} for other values of $n$.}
\label{fig:CMBSABI_0}
\end{center}
\end{figure}

\begin{figure}[H]
\begin{center}
\includegraphics[width=\wappfig,clip=true]{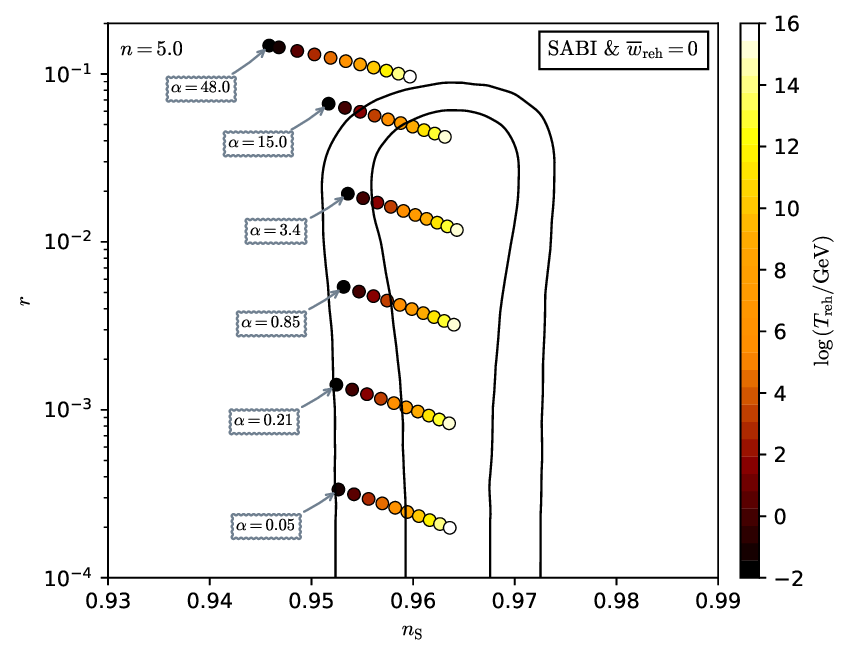}
\includegraphics[width=\wappfig,clip=true]{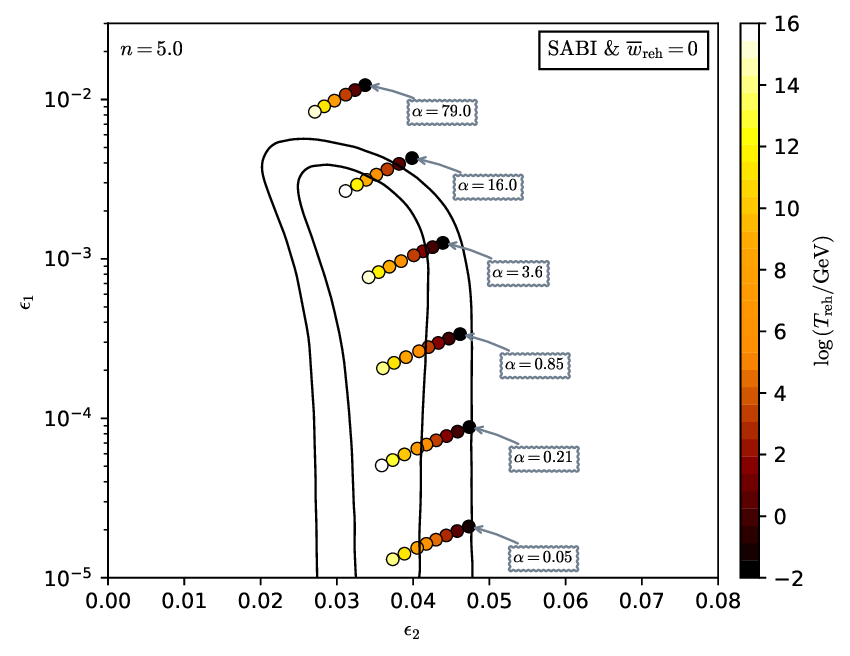}
\caption{Reheating consistent slow-roll predictions for the
  Superconformal $\alpha$-attractor B Inflation model for
  $n=5$. Predictions are represented as a function of $\alpha$ in the
  plane $(\nS,r)$ (top panel) and in the plane
  $(\epsilon_1,\epsilon_2)$ (bottom panel). The solid contours are the
  one and two-sigma {\data} confidence intervals (marginalized over
  second order slow-roll).}
\label{fig:CMBSABI_1}
\end{center}
\end{figure}

\begin{figure}[H]
\begin{center}
\includegraphics[width=\wappfig,clip=true]{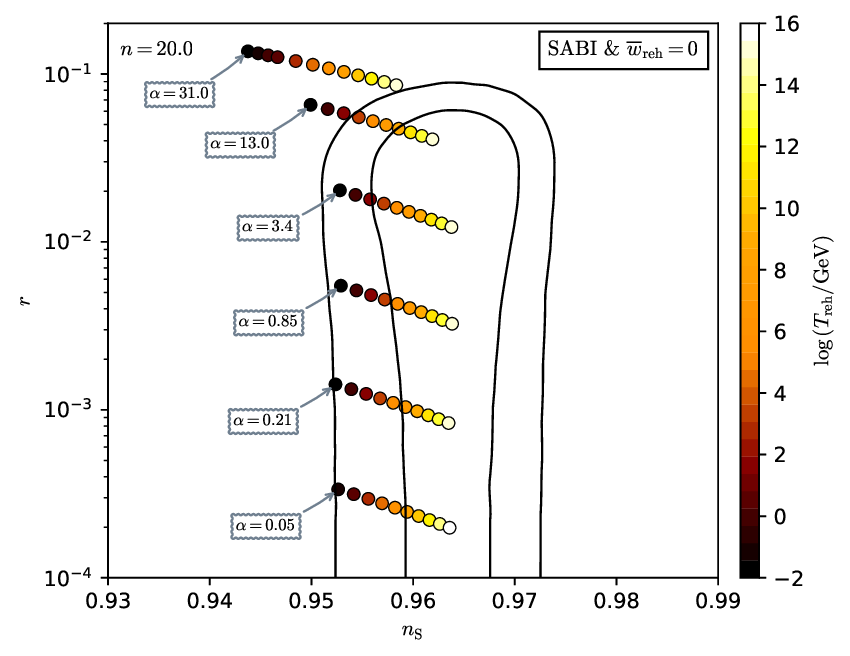}
\includegraphics[width=\wappfig,clip=true]{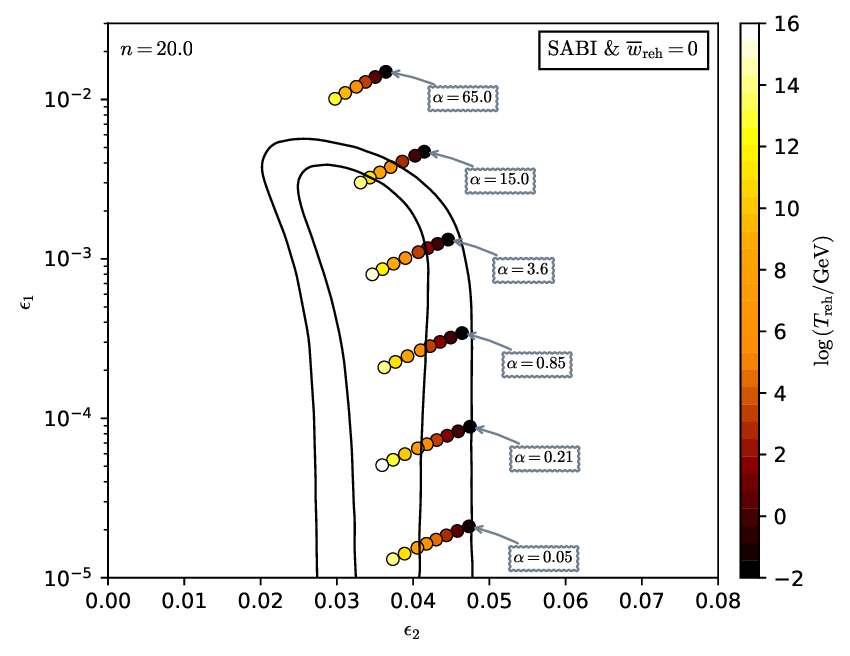}
\caption{Reheating consistent slow-roll predictions for the
  Superconformal $\alpha$-attractor B Inflation model for
  $n=20$. Predictions are represented as a function of $\alpha$ in the
  plane $(\nS,r)$ (top panel) and in the plane
  $(\epsilon_1,\epsilon_2)$ (bottom panel). The solid contours are the
  one and two-sigma {\data} confidence intervals (marginalized over
  second order slow-roll).}
\label{fig:CMBSABI_2}
\end{center}
\end{figure}

\subsection{Superconformal \texorpdfstring{$\alpha$}{alpha}-Attractor T Inflation (\hyperref[sec:sati]{SATI})}

\begin{figure}[H]
\begin{center}
\includegraphics[width=\wappfig,clip=true]{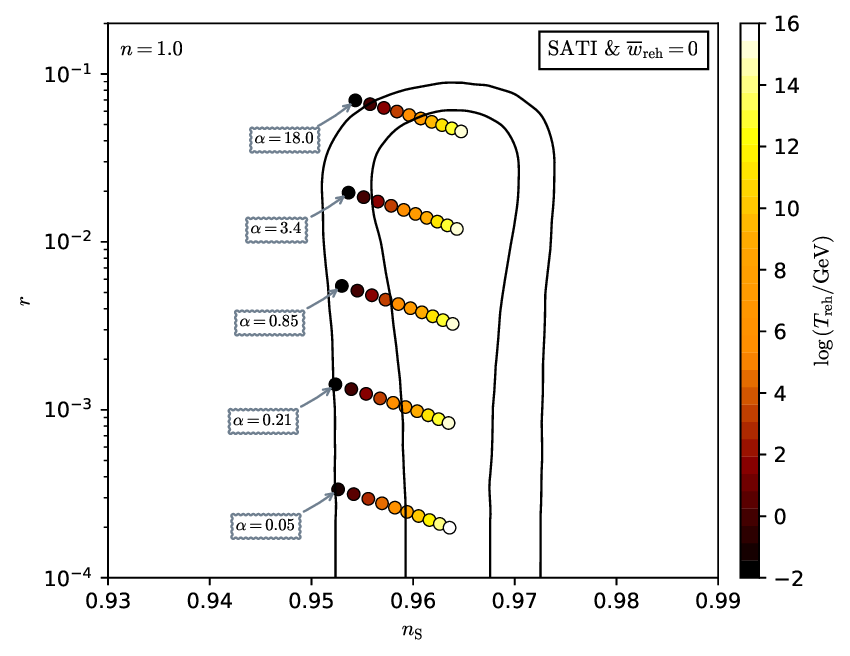}
\includegraphics[width=\wappfig,clip=true]{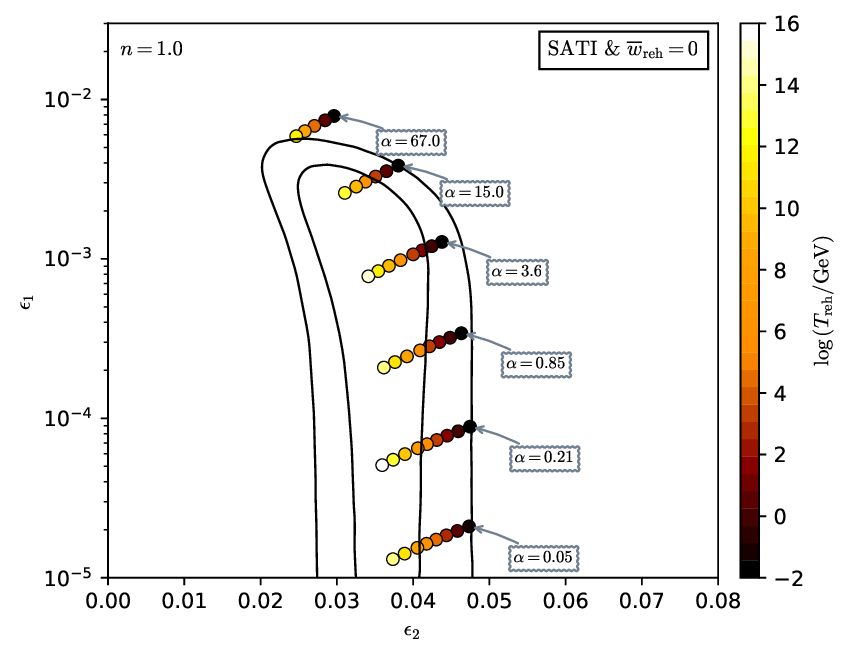}
\caption{Reheating consistent slow-roll predictions for the
  Superconformal $\alpha$-attractor T Inflation model for
  $n=1$. Predictions are represented as a function of $\alpha$ in the
  plane $(\nS,r)$ (top panel) and in the plane
  $(\epsilon_1,\epsilon_2)$ (bottom panel). The solid contours are the
  one and two-sigma {\data} confidence intervals (marginalized over
  second order slow-roll). See also Figs.~\ref{fig:CMBSATI_1} and
  \ref{fig:CMBSATI_2} for other values of $n$.}
\label{fig:CMBSATI_0}
\end{center}
\end{figure}

\begin{figure}[H]
\begin{center}
\includegraphics[width=\wappfig,clip=true]{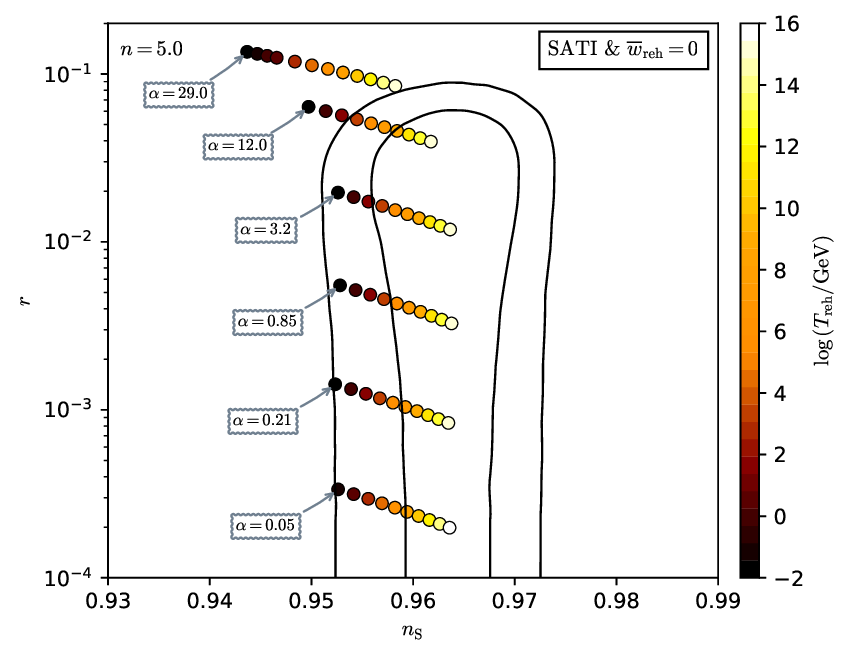}
\includegraphics[width=\wappfig,clip=true]{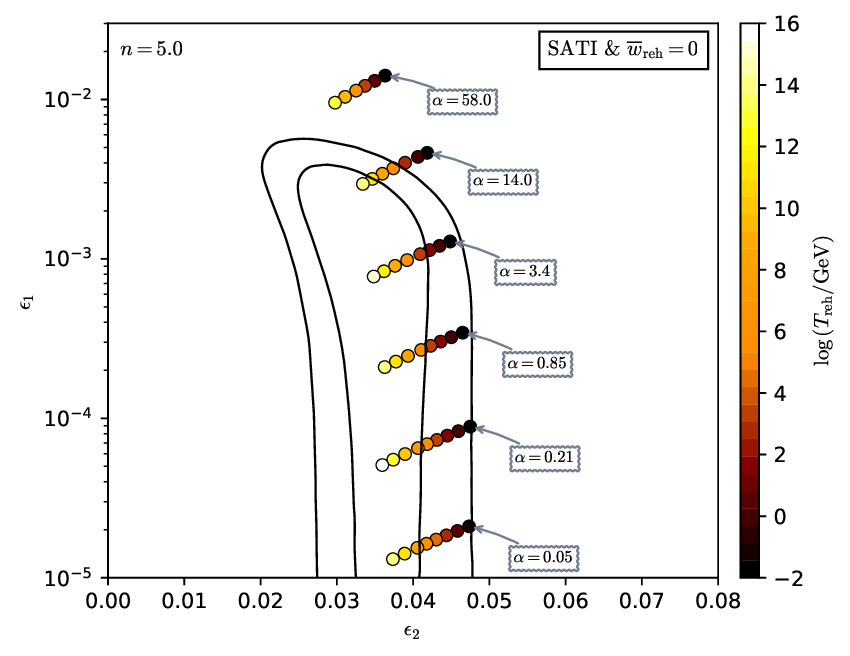}
\caption{Reheating consistent slow-roll predictions for the
  Superconformal $\alpha$-attractor T Inflation model for
  $n=5$. Predictions are represented as a function of $\alpha$ in the
  plane $(\nS,r)$ (top panel) and in the plane
  $(\epsilon_1,\epsilon_2)$ (bottom panel). The solid contours are the
  one and two-sigma {\data} confidence intervals (marginalized over
  second order slow-roll).}
\label{fig:CMBSATI_1}
\end{center}
\end{figure}

\begin{figure}[H]
\begin{center}
\includegraphics[width=\wappfig,clip=true]{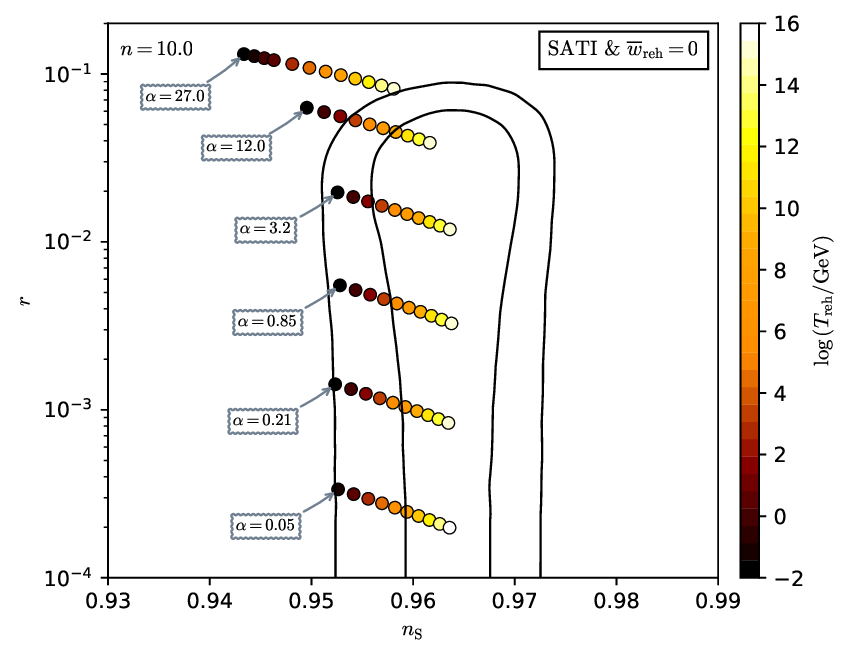}
\includegraphics[width=\wappfig,clip=true]{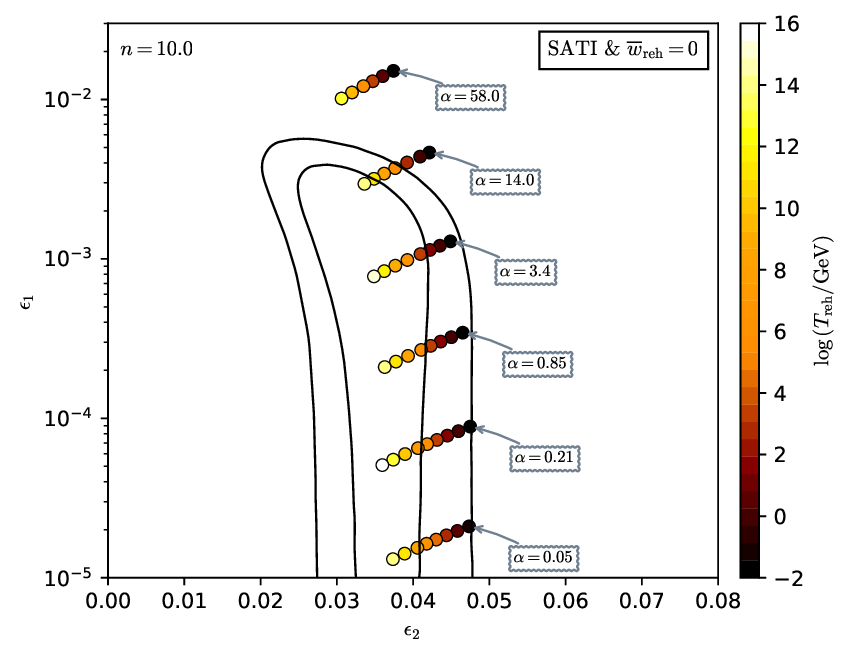}
\caption{Reheating consistent slow-roll predictions for the
  Superconformal $\alpha$-attractor T Inflation model for
  $n=10$. Predictions are represented as a function of $\alpha$ in the
  plane $(\nS,r)$ (top panel) and in the plane
  $(\epsilon_1,\epsilon_2)$ (bottom panel). The solid contours are the
  one and two-sigma {\data} confidence intervals (marginalized over
  second order slow-roll).}
\label{fig:CMBSATI_2}
\end{center}
\end{figure}

\subsection{Running Mass Inflation 1 (\hyperref[sec:rmi]{RMI1})}

\begin{figure}[H]
\begin{center}
\includegraphics[width=\wappfig,clip=true]{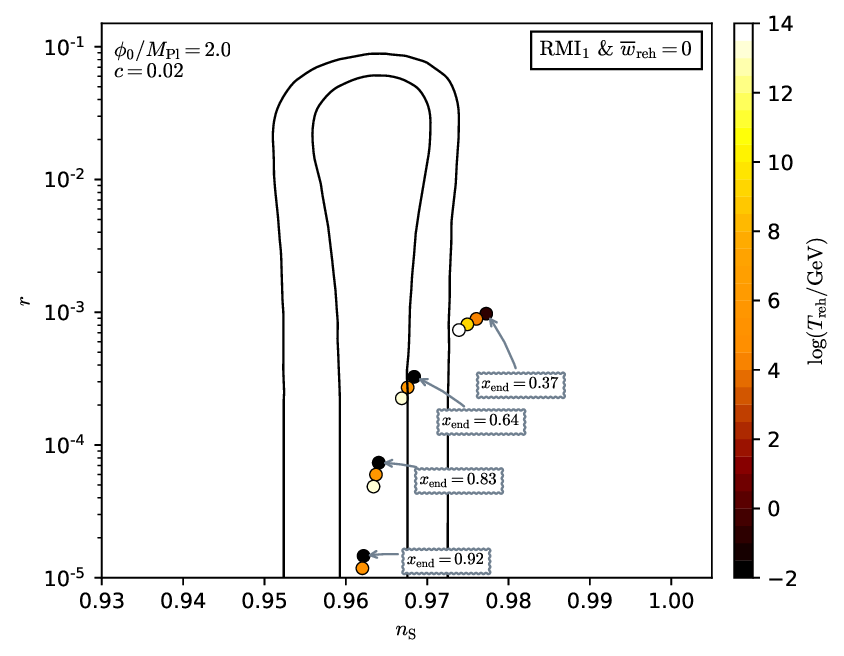}
\includegraphics[width=\wappfig,clip=true]{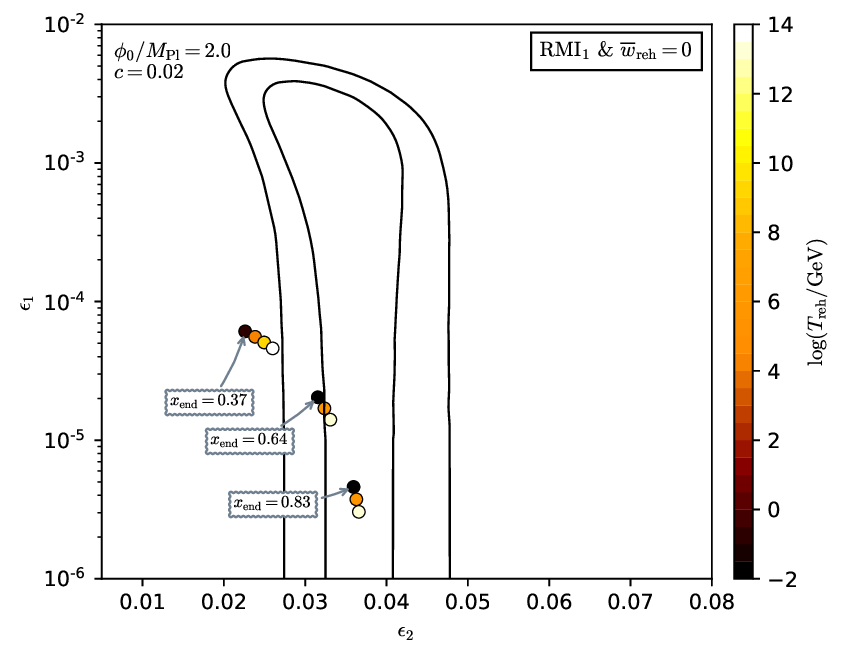}
\caption{Reheating consistent slow-roll predictions for the running
  mass inflation 1 models ($c>0$, $x<1$) with $c=0.02$ and
  $\phizero/\Mp=2$ (which satisfies $\phi/\Mp < 1/\sqrt{c}$), in the
  plane $(\nS,r)$ (top panel) and the plane $(\epsilon_1,\epsilon_2)$
  (bottom panel). The solid contours are the one and two-sigma {\data}
  confidence intervals (marginalized over second order slow-roll). The
  field value at which inflation ends is varied in the range
  $1/e<\xend<1$. See figures~\ref{fig:CMBRMI1_1} to
  \ref{fig:CMBRMI1_3} for other values of $c$ and $\phizero$.}
\label{fig:CMBRMI1}
\end{center}
\end{figure}

\begin{figure}[H]
\begin{center}
\includegraphics[width=\wappfig,clip=true]{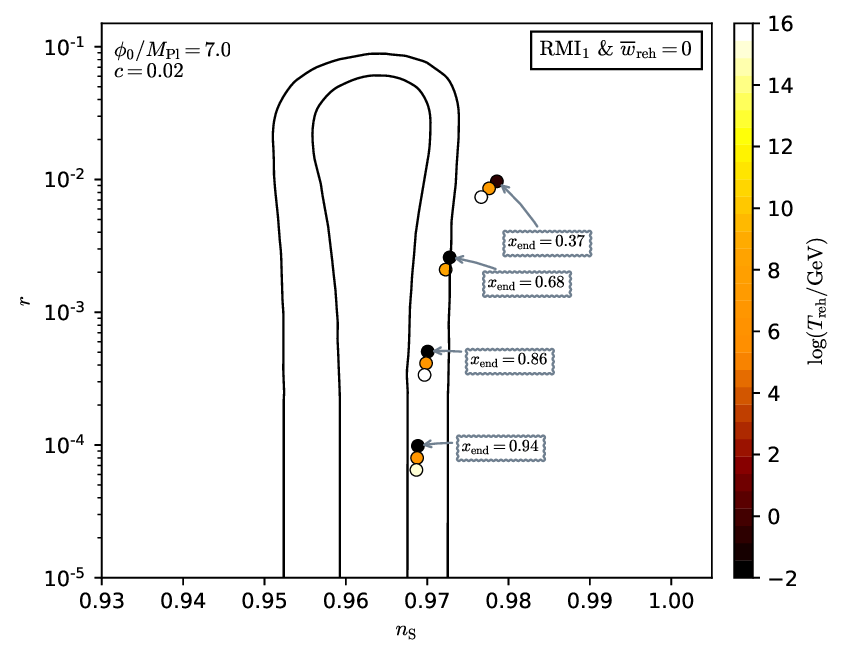}
\includegraphics[width=\wappfig,clip=true]{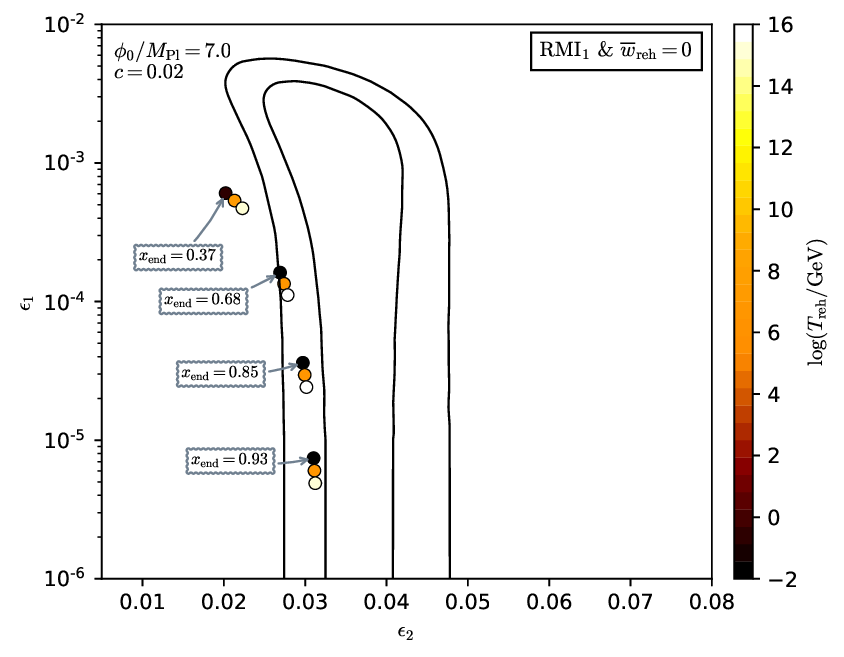}
\caption{Reheating consistent slow-roll predictions for the running
  mass inflation 1 models ($c>0$, $x<1$) with $c=0.02$ and
  $\phizero/\Mp=7$ (which satisfies $\phi/\Mp < 1/\sqrt{c}$), in the
  plane $(\nS,r)$ (top panel) and the plane $(\epsilon_1,\epsilon_2)$
  (bottom panel). The solid contours are the one and two-sigma {\data}
  confidence intervals (marginalized over second order slow-roll). The
  field value at which inflation ends is varied in the range
  $1/e<\xend<1$.}
\label{fig:CMBRMI1_1}
\end{center}
\end{figure}

\begin{figure}[H]
\begin{center}
\includegraphics[width=\wappfig,clip=true]{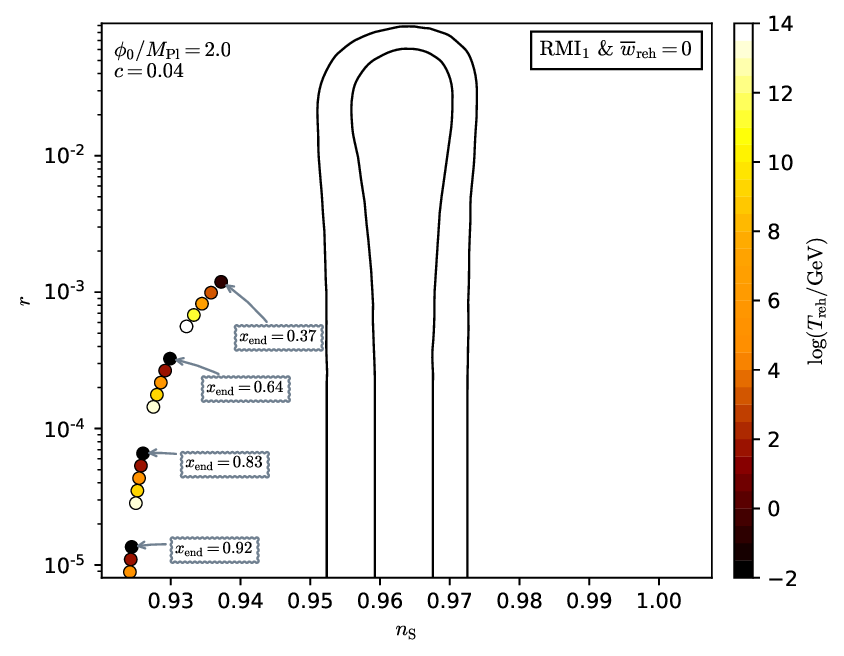}
\includegraphics[width=\wappfig,clip=true]{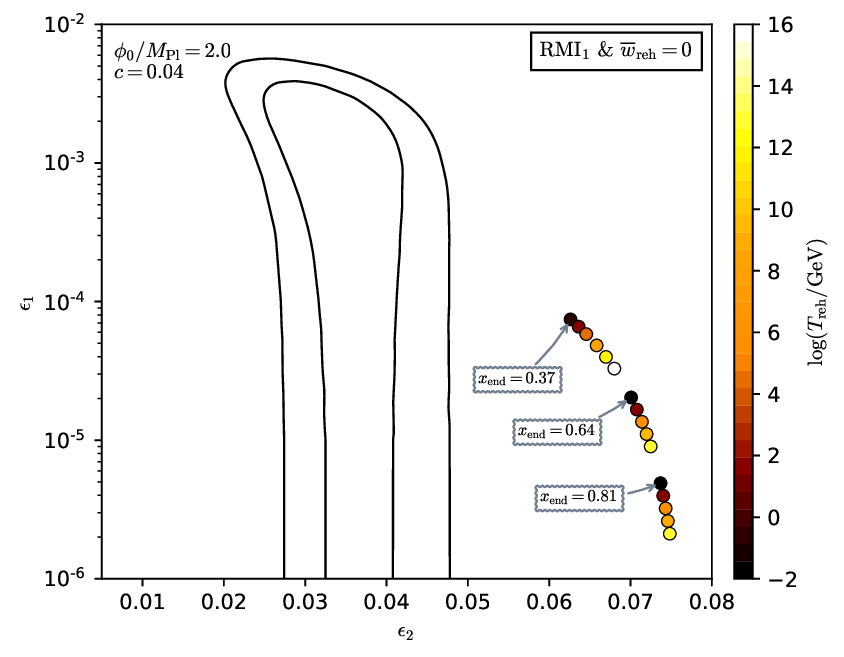}
\caption{Reheating consistent slow-roll predictions for the running
  mass inflation 1 models ($c>0$, $x<1$) with $c=0.04$ and
  $\phizero/\Mp=2$ (which satisfies $\phi/\Mp < 1/\sqrt{c}$), in the
  plane $(\nS,r)$ (top panel) and the plane $(\epsilon_1,\epsilon_2)$
  (bottom panel). The solid contours are the one and two-sigma {\data}
  confidence intervals (marginalized over second order slow-roll). The
  field value at which inflation ends is varied in the range
  $1/e<\xend<1$.}
\label{fig:CMBRMI1_2}
\end{center}
\end{figure}

\begin{figure}[H]
\begin{center}
\includegraphics[width=\wappfig,clip=true]{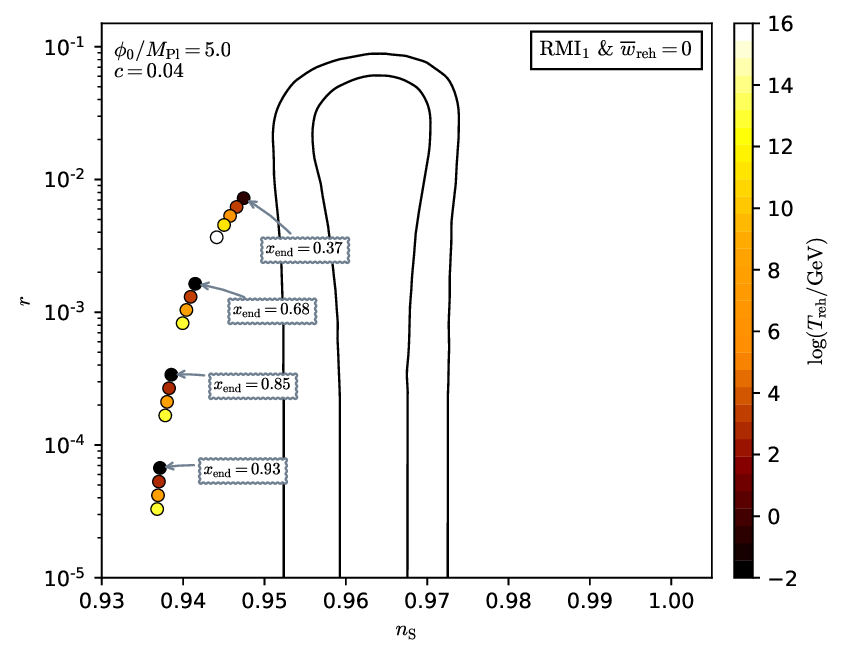}
\includegraphics[width=\wappfig,clip=true]{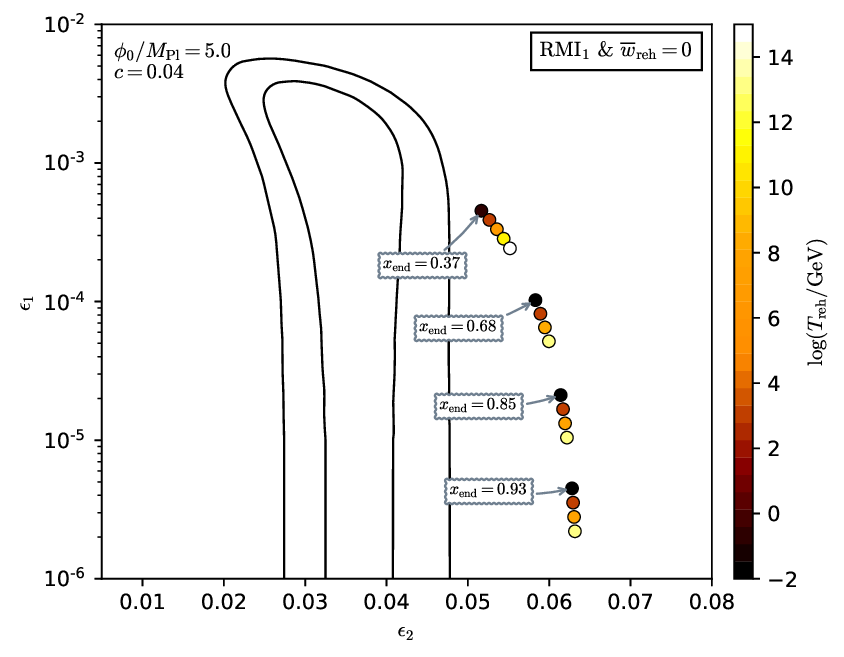}
\caption{Reheating consistent slow-roll predictions for the running
  mass inflation 1 models ($c>0$, $x<1$) with $c=0.04$ and
  $\phizero/\Mp \to 5$ (which saturates $\phi/\Mp < 1/\sqrt{c}$), in the
  plane $(\nS,r)$ (top panel) and the plane $(\epsilon_1,\epsilon_2)$
  (bottom panel). The solid contours are the one and two-sigma {\data}
  confidence intervals (marginalized over second order slow-roll). The
  field value at which inflation ends is varied in the range
  $1/e<\xend<1$.}
\label{fig:CMBRMI1_3}
\end{center}
\end{figure}

\subsection{Running Mass Inflation 2 (\hyperref[sec:rmi]{RMI2})}

\begin{figure}[H]
\begin{center}
\includegraphics[width=\wappfig,clip=true]{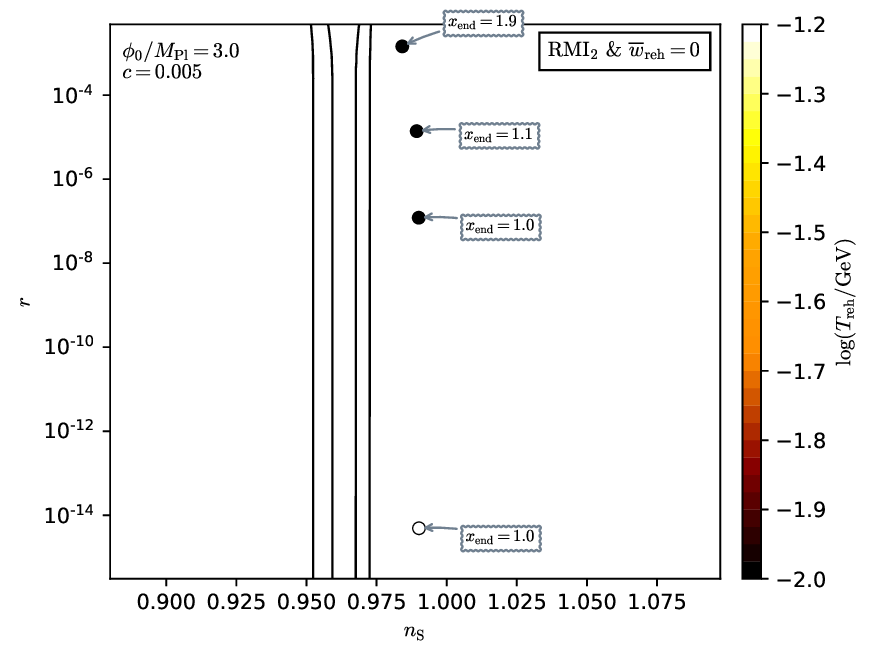}
\includegraphics[width=\wappfig,clip=true]{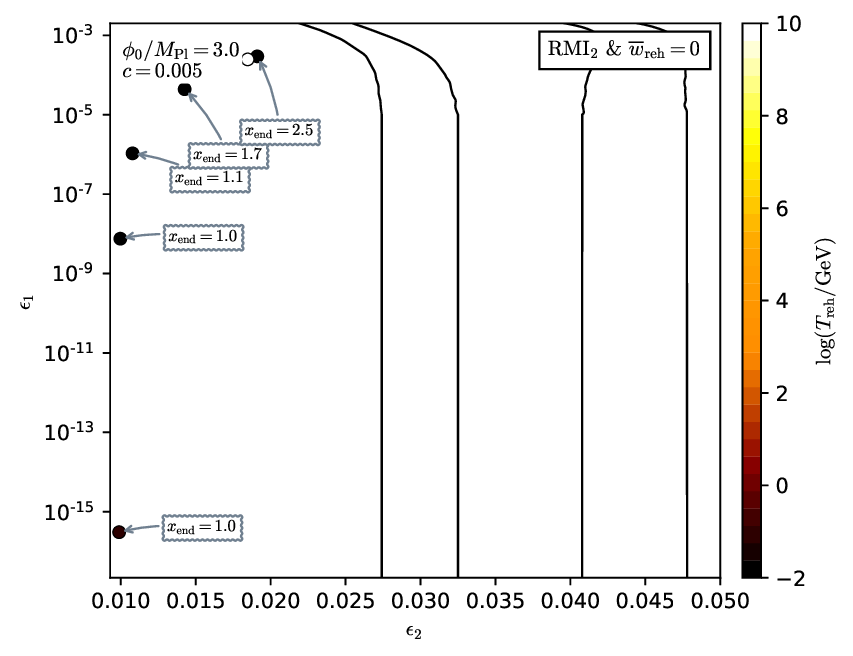}
\caption{Reheating consistent slow-roll predictions for the running
  mass inflation 2 models ($c>0$, $x>1$) with $c=0.005$ and
  $\phizero/\Mp=3$ (which satisfies $\phizero/\Mp<1/\sqrt{c}$), in the
  plane $(\nS,r)$ (top panel) and the plane $(\epsilon_1,\epsilon_2)$
  (bottom panel). The solid contours are the one and two-sigma {\data}
  confidence intervals (marginalized over second order slow-roll). The
  field value at which inflation ends is varied in the range
  $1<\xend<e$. See figures~\ref{fig:CMBRMI2_1} to \ref{fig:CMBRMI2_3}
  for other values of $c$ and $\phizero$.}
\label{fig:CMBRMI2}
\end{center}
\end{figure}

\begin{figure}[H]
\begin{center}
\includegraphics[width=\wappfig,clip=true]{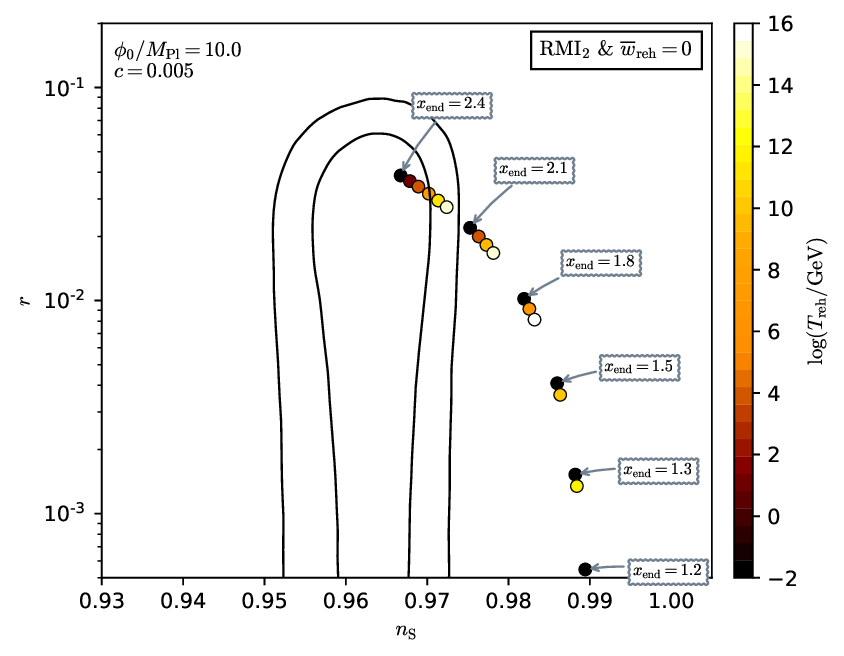}
\includegraphics[width=\wappfig,clip=true]{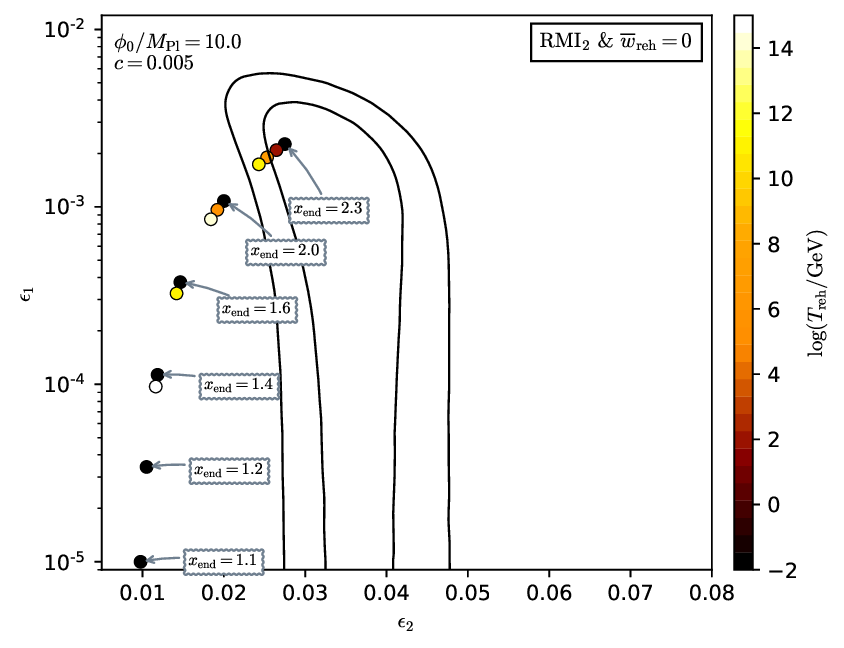}
\caption{Reheating consistent slow-roll predictions for the running
  mass inflation 2 models ($c>0$, $x>1$) with $c=0.005$ and
  $\phizero/\Mp=10$ (which satisfies $\phizero/\Mp<1/\sqrt{c}$), in the
  plane $(\nS,r)$ (top panel) and the plane $(\epsilon_1,\epsilon_2)$
  (bottom panel). The solid contours are the one and two-sigma {\data}
  confidence intervals (marginalized over second order slow-roll). The
  field value at which inflation ends is varied in the
  range $1<\xend<e$.}
\label{fig:CMBRMI2_1}
\end{center}
\end{figure}

\begin{figure}[H]
\begin{center}
\includegraphics[width=\wappfig,clip=true]{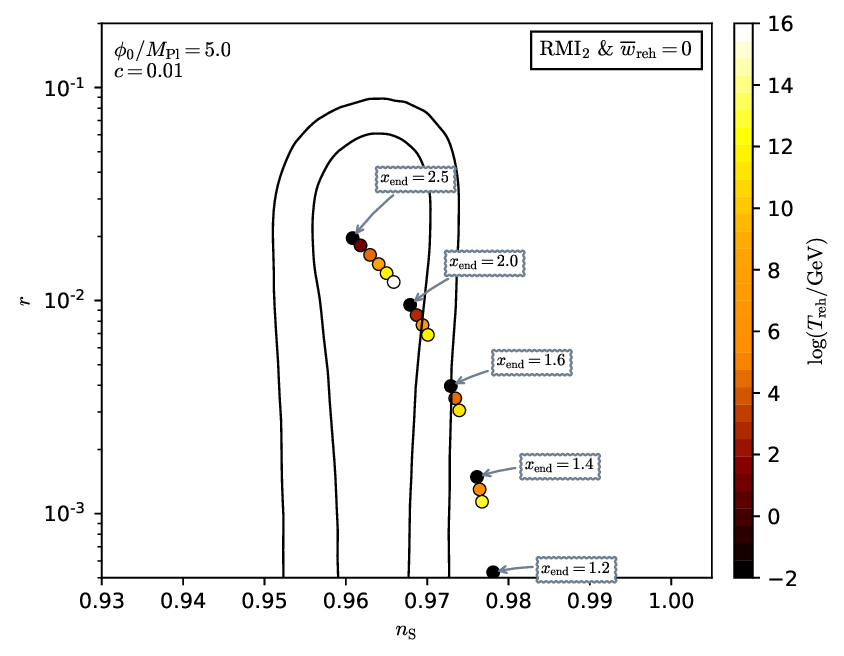}
\includegraphics[width=\wappfig,clip=true]{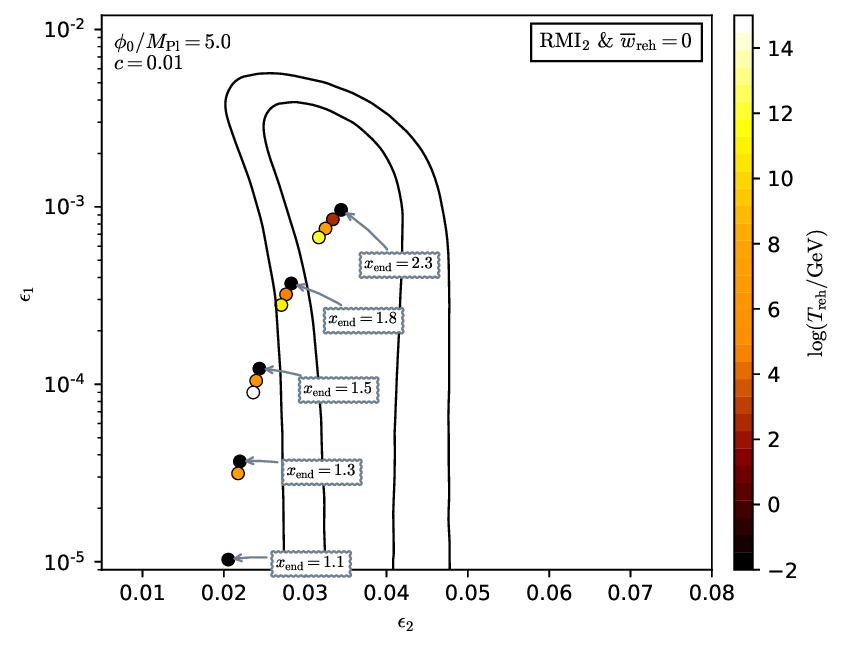}
\caption{Reheating consistent slow-roll predictions for the running
  mass inflation 2 models ($c>0$, $x>1$) with $c=0.01$ and
  $\phizero/\Mp=5$ (which satisfies $\phizero/\Mp<1/\sqrt{c}$), in the
  plane $(\nS,r)$ (top panel) and the plane $(\epsilon_1,\epsilon_2)$
  (bottom panel). The solid contours are the one and two-sigma {\data}
  confidence intervals (marginalized over second order slow-roll). The
  field value at which inflation ends is varied in the
  range $1<\xend<e$.}
\label{fig:CMBRMI2_2}
\end{center}
\end{figure}

\begin{figure}[H]
\begin{center}
\includegraphics[width=\wappfig,clip=true]{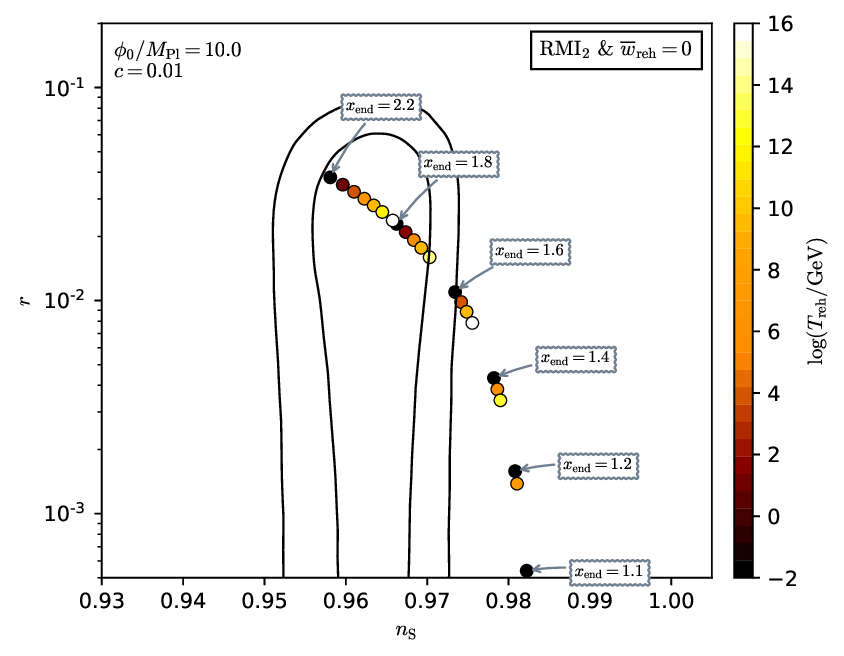}
\includegraphics[width=\wappfig,clip=true]{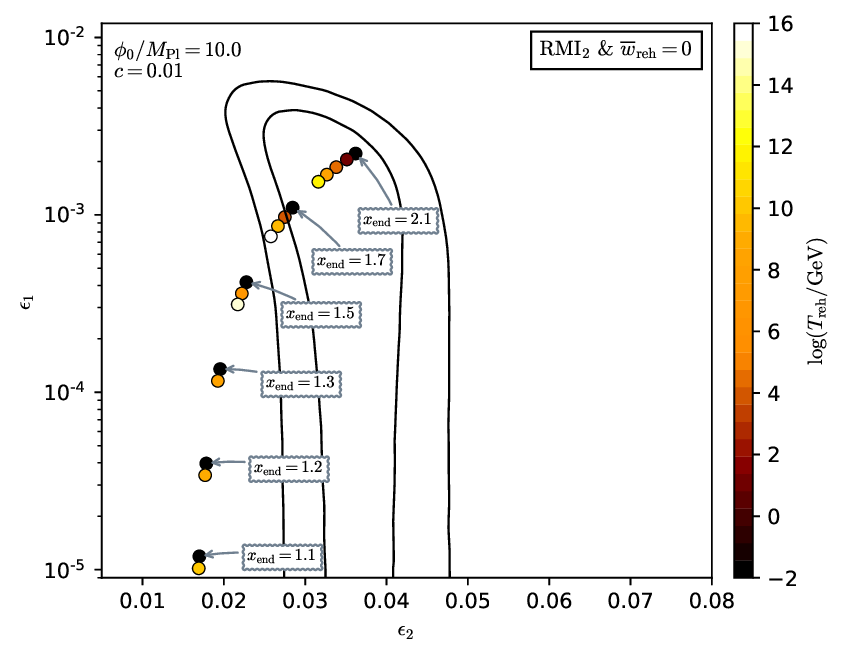}
\caption{Reheating consistent slow-roll predictions for the running
  mass inflation 2 models ($c>0$, $x>1$) with $c=0.01$ and
  $\phizero/\Mp  \to 10$ (which saturates $\phizero/\Mp<1/\sqrt{c}$), in the
  plane $(\nS,r)$ (top panel) and the plane $(\epsilon_1,\epsilon_2)$
  (bottom panel). The solid contours are the one and two-sigma {\data}
  confidence intervals (marginalized over second order slow-roll). The
  field value at which inflation ends is varied in the
  range $1<\xend<e$.}
\label{fig:CMBRMI2_3}
\end{center}
\end{figure}

\subsection{Running Mass Inflation 3 (\hyperref[sec:rmi]{RMI3})}

\begin{figure}[H]
\begin{center}
\includegraphics[width=\wappfig,clip=true]{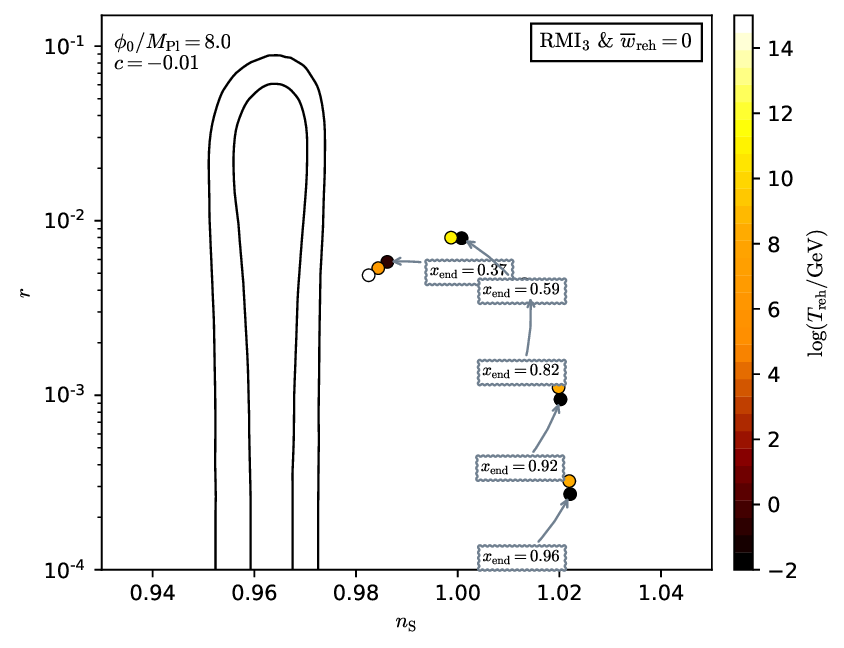}
\includegraphics[width=\wappfig,clip=true]{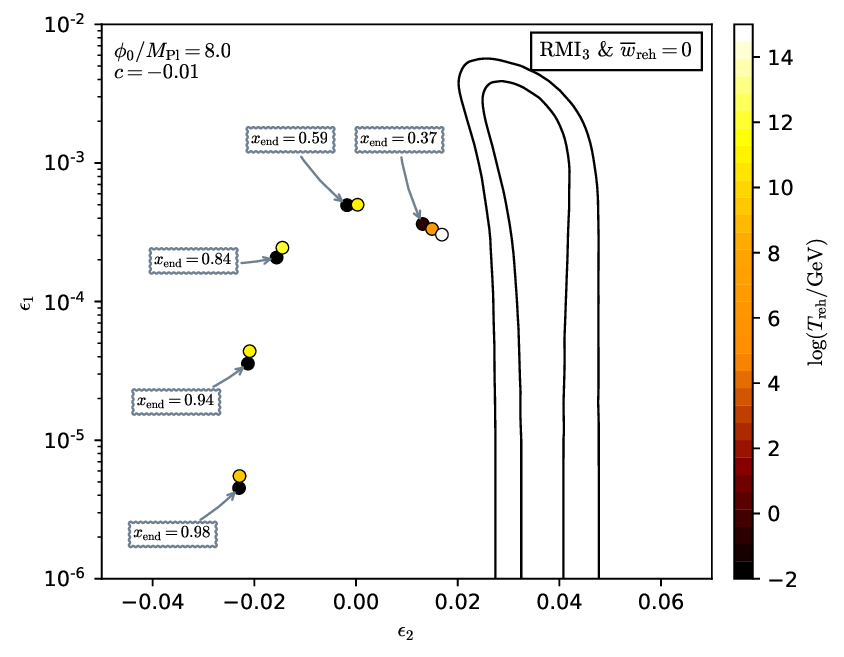}
\caption{Reheating consistent slow-roll predictions for the running
  mass inflation 3 models ($c<0$, $x<1$) with $c=-0.01$ and
  $\phizero/\Mp=8$ (which satisfies $\phizero/\Mp<1/\sqrt{-c}$) in the
  plane $(\nS,r)$ (top panel) and the plane $(\epsilon_1,\epsilon_2)$
  (bottom panel). The solid contours are the one and two-sigma {\data}
  confidence intervals (marginalized over second order slow-roll). The
  field value at which inflation ends is varied in the range
  $1/e<\xend<1$. See figures~\ref{fig:CMBRMI3_1} to
  \ref{fig:CMBRMI3_3} for other values of $c$ and $\phizero$.}
\label{fig:CMBRMI3}
\end{center}
\end{figure}

\begin{figure}[H]
\begin{center}
\includegraphics[width=\wappfig,clip=true]{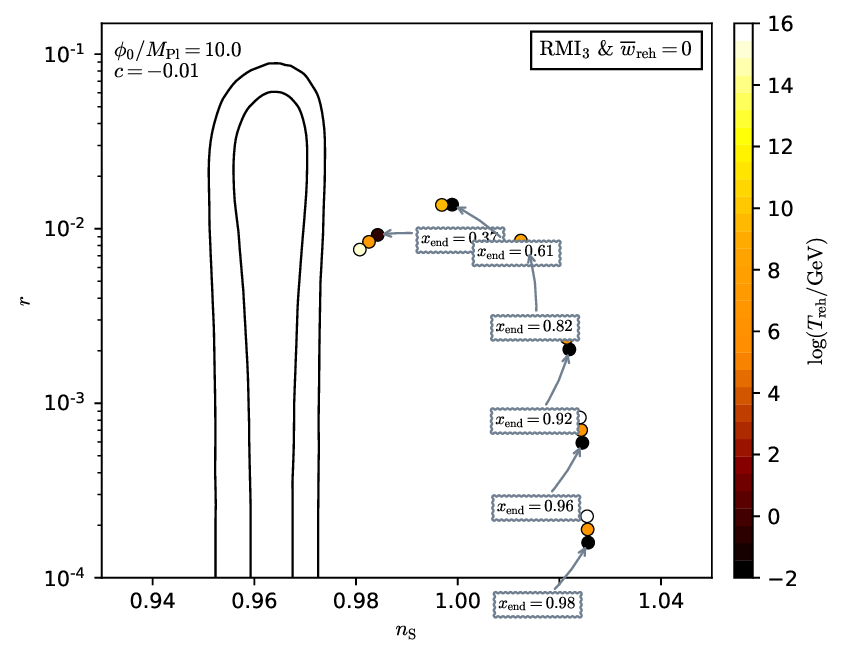}
\includegraphics[width=\wappfig,clip=true]{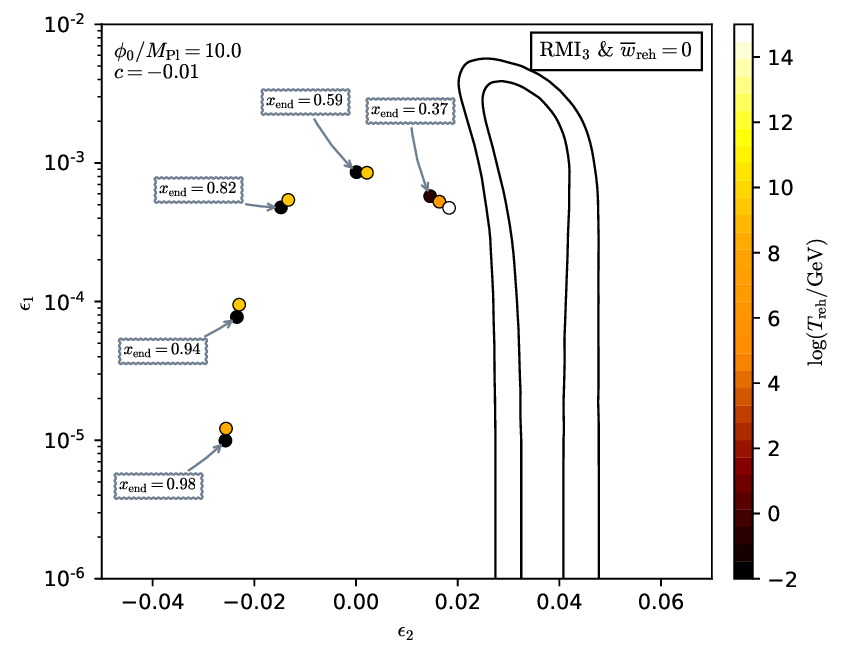}
\caption{Reheating consistent slow-roll predictions for the running
  mass inflation 3 models ($c<0$, $x<1$) with $c=-0.01$ and
  $\phizero/\Mp \to 10$ (which saturates $\phizero/\Mp<1/\sqrt{-c}$) in the
  plane $(\nS,r)$ (top panel) and the plane $(\epsilon_1,\epsilon_2)$
  (bottom panel). The solid contours are the one and two-sigma {\data}
  confidence intervals (marginalized over second order slow-roll). The
  field value at which inflation ends is varied in the range
  $1/e<\xend<1$.}
\label{fig:CMBRMI3_1}
\end{center}
\end{figure}

\begin{figure}[H]
\begin{center}
\includegraphics[width=\wappfig,clip=true]{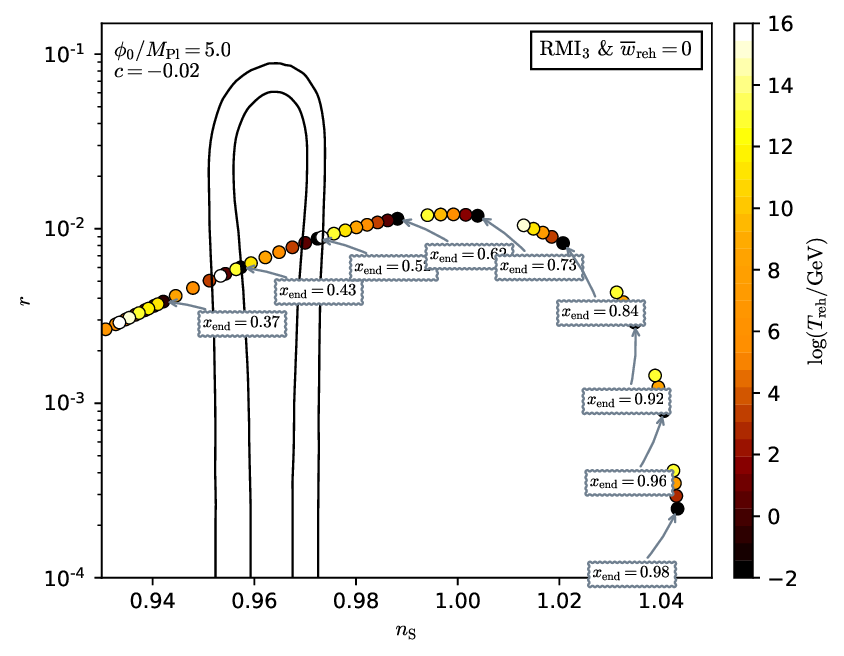}
\includegraphics[width=\wappfig,clip=true]{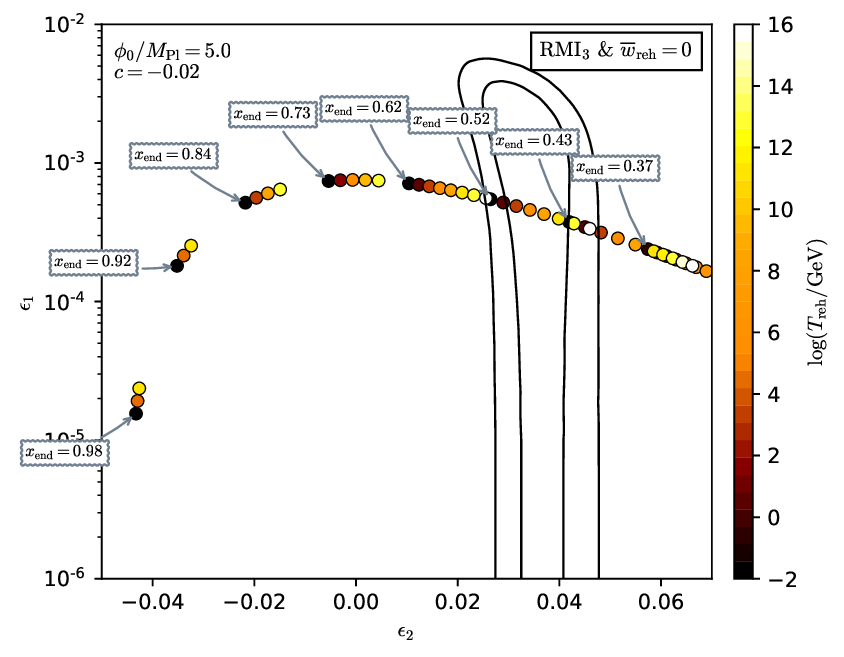}
\caption{Reheating consistent slow-roll predictions for the running
  mass inflation 3 models ($c<0$, $x<1$) with $c=-0.02$ and
  $\phizero/\Mp=5$ (which satisfies $\phizero/\Mp<1/\sqrt{-c}$) in the
  plane $(\nS,r)$ (top panel) and the plane $(\epsilon_1,\epsilon_2)$
  (bottom panel). The solid contours are the one and two-sigma {\data}
  confidence intervals (marginalized over second order slow-roll). The
  field value at which inflation ends is varied in the range
  $1/e<\xend<1$.}
\label{fig:CMBRMI3_2}
\end{center}
\end{figure}

\begin{figure}[H]
\begin{center}
\includegraphics[width=\wappfig,clip=true]{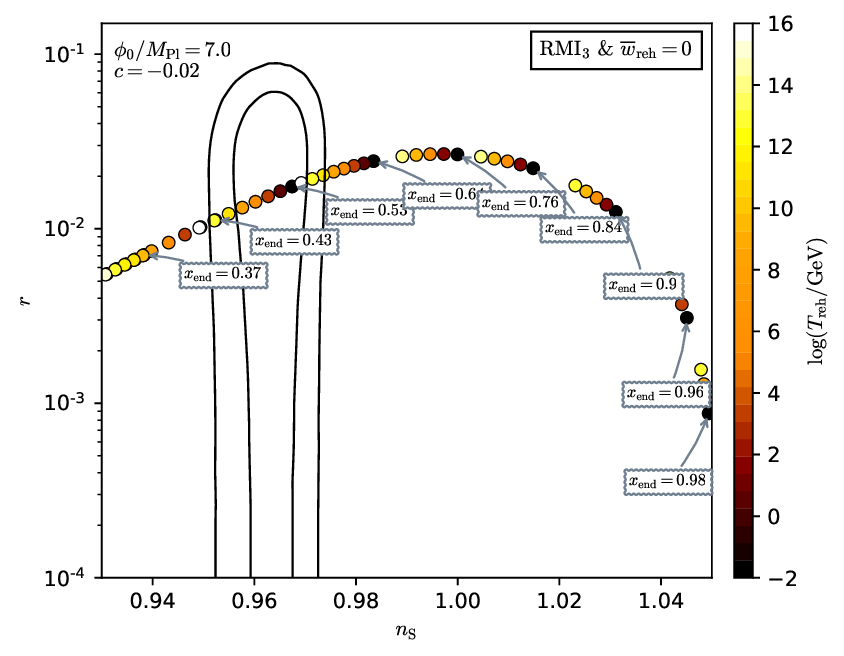}
\includegraphics[width=\wappfig,clip=true]{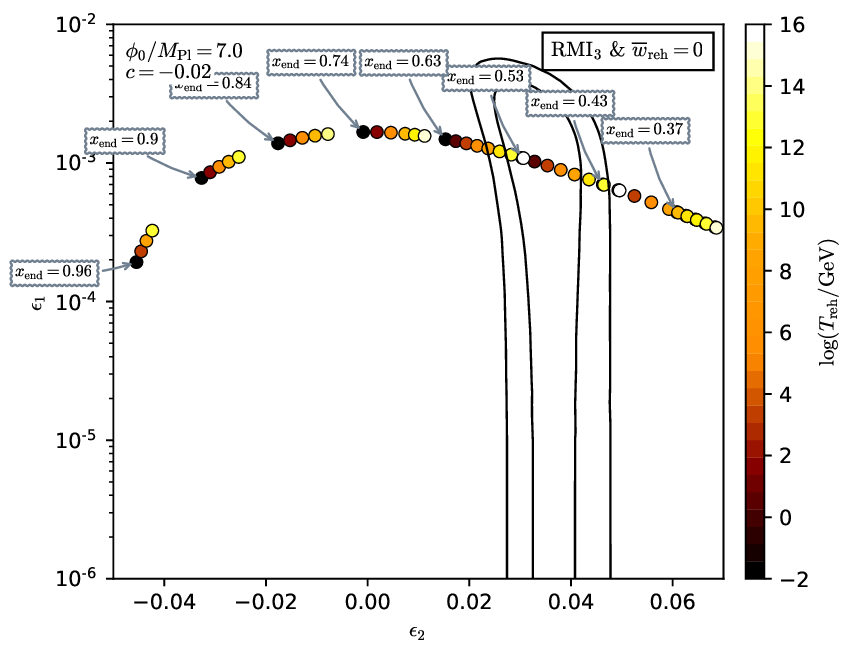}
\caption{Reheating consistent slow-roll predictions for the running
  mass inflation 3 models ($c<0$, $x<1$) with $c=-0.02$ and
  $\phizero/\Mp=7$ (which satisfies $\phizero/\Mp<1/\sqrt{-c}$) in the
  plane $(\nS,r)$ (top panel) and the plane $(\epsilon_1,\epsilon_2)$
  (bottom panel). The solid contours are the one and two-sigma {\data}
  confidence intervals (marginalized over second order slow-roll). The
  field value at which inflation ends is varied in the range
  $1/e<\xend<1$.}
\label{fig:CMBRMI3_3}
\end{center}
\end{figure}

\subsection{Running Mass Inflation 4 (\hyperref[sec:rmi]{RMI4})}

\begin{figure}[H]
\begin{center}
\includegraphics[width=\wappfig,clip=true]{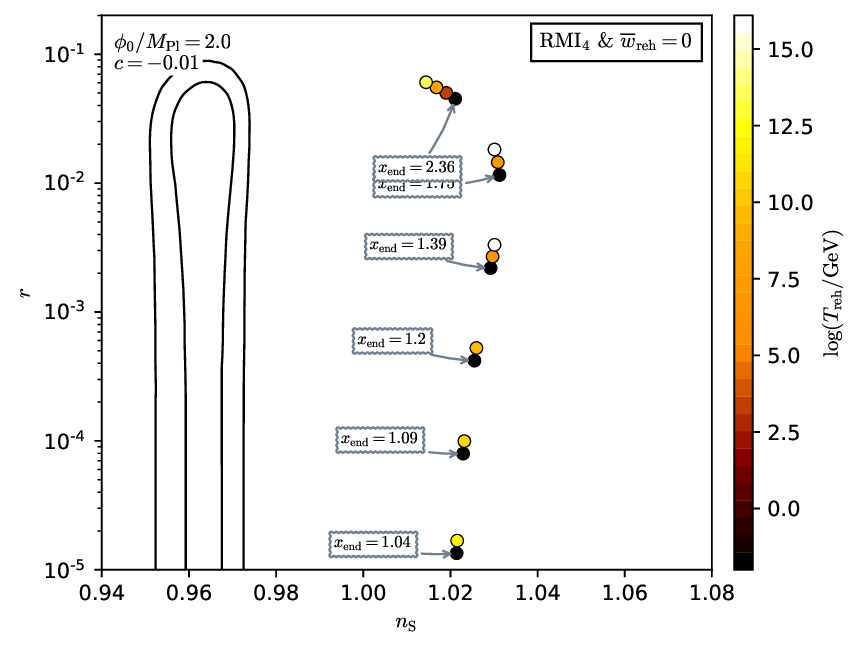}
\includegraphics[width=\wappfig,clip=true]{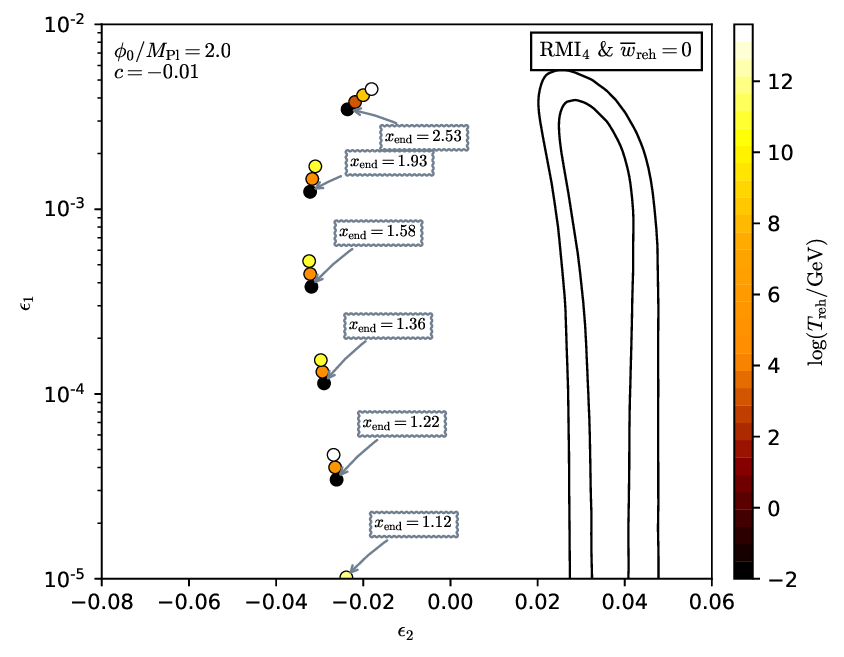}
\caption{Reheating consistent slow-roll predictions for the running
  mass inflation 4 models ($c<0$, $x>1$) with $c=-0.01$ and
  $\phizero/\Mp=2$ (which satisfies $\phizero/\Mp<1/\sqrt{-c}$) in the
  plane $(\nS,r)$ (top panel) and the plane $(\epsilon_1,\epsilon_2)$
  (bottom panel). The solid contours are the one and two-sigma {\data}
  confidence intervals (marginalized over second order slow-roll). The
  field value at which inflation ends is varied in the range
  $1<\xend<e$. See figures~\ref{fig:CMBRMI4_1} to \ref{fig:CMBRMI4_3}
  for other values of $c$ and $\phizero$.}
\label{fig:CMBRMI4}
\end{center}
\end{figure}

\begin{figure}[H]
\begin{center}
\includegraphics[width=\wappfig,clip=true]{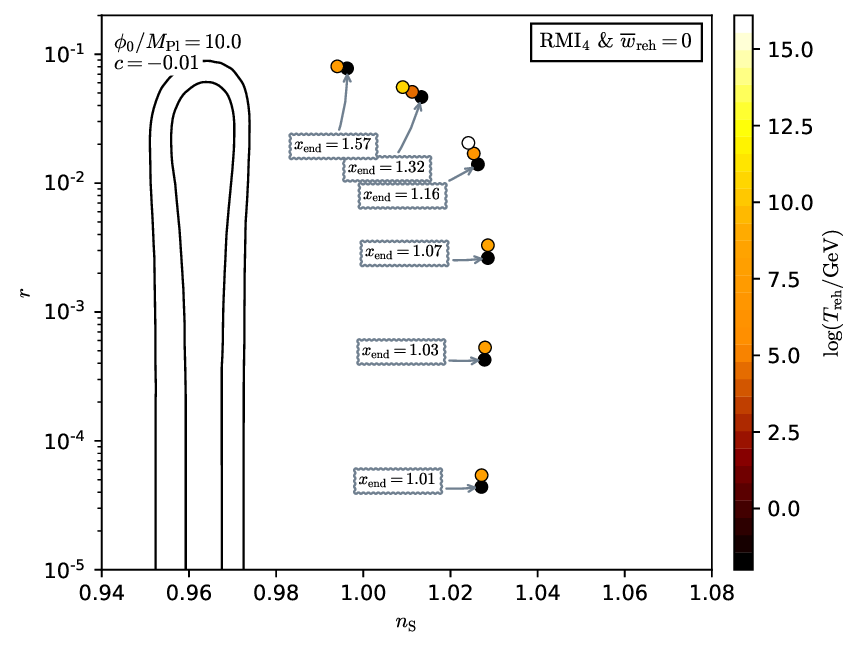}
\includegraphics[width=\wappfig,clip=true]{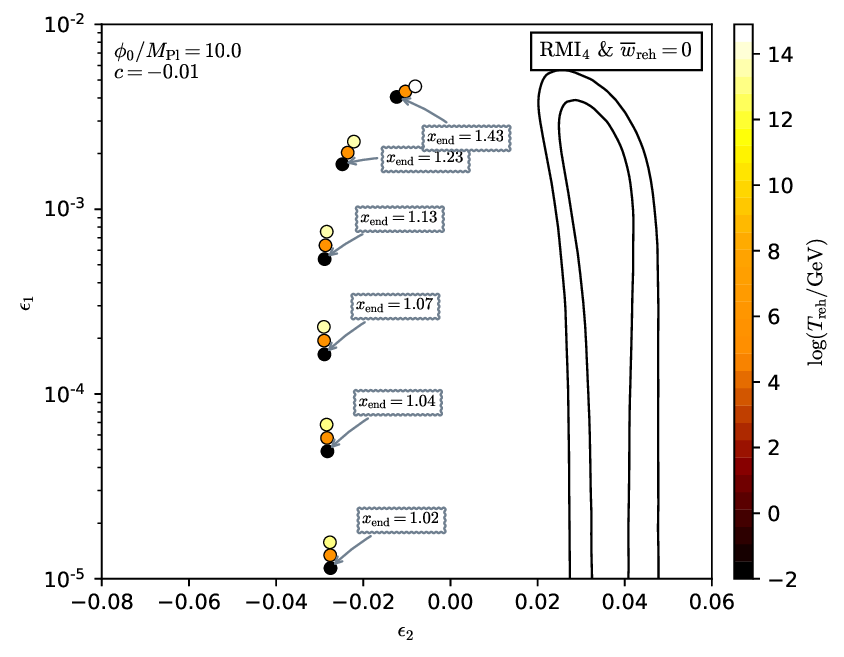}
\caption{Reheating consistent slow-roll predictions for the running
  mass inflation 4 models ($c<0$, $x>1$) with $c=-0.01$ and
  $\phizero/\Mp \to 10$ (which saturates $\phizero/\Mp<1/\sqrt{-c}$) in the
  plane $(\nS,r)$ (top panel) and the plane $(\epsilon_1,\epsilon_2)$
  (bottom panel). The solid contours are the one and two-sigma {\data}
  confidence intervals (marginalized over second order slow-roll). The
  field value at which inflation ends is varied in the range
  $1<\xend<e$. }
\label{fig:CMBRMI4_1}
\end{center}
\end{figure}

\begin{figure}[H]
\begin{center}
\includegraphics[width=\wappfig,clip=true]{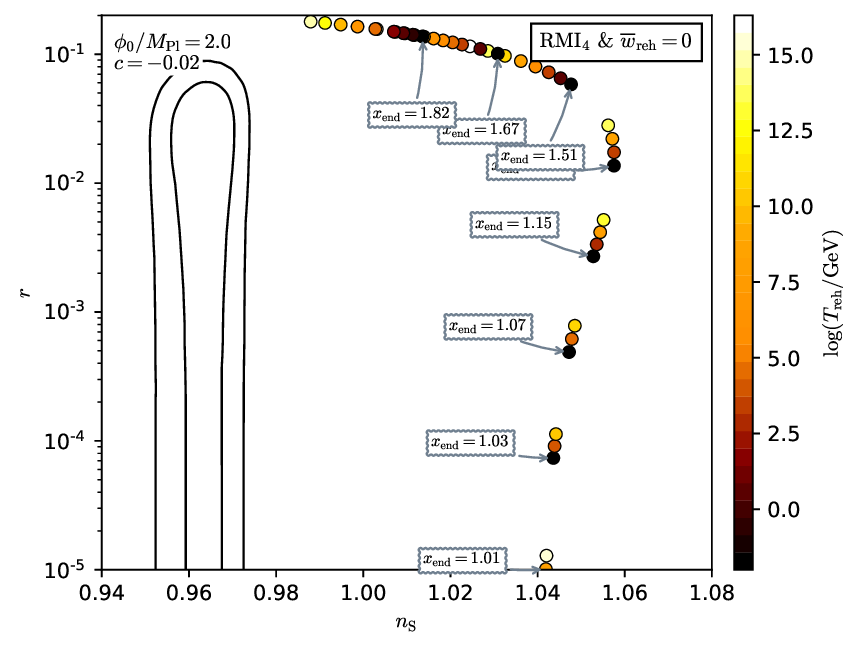}
\includegraphics[width=\wappfig,clip=true]{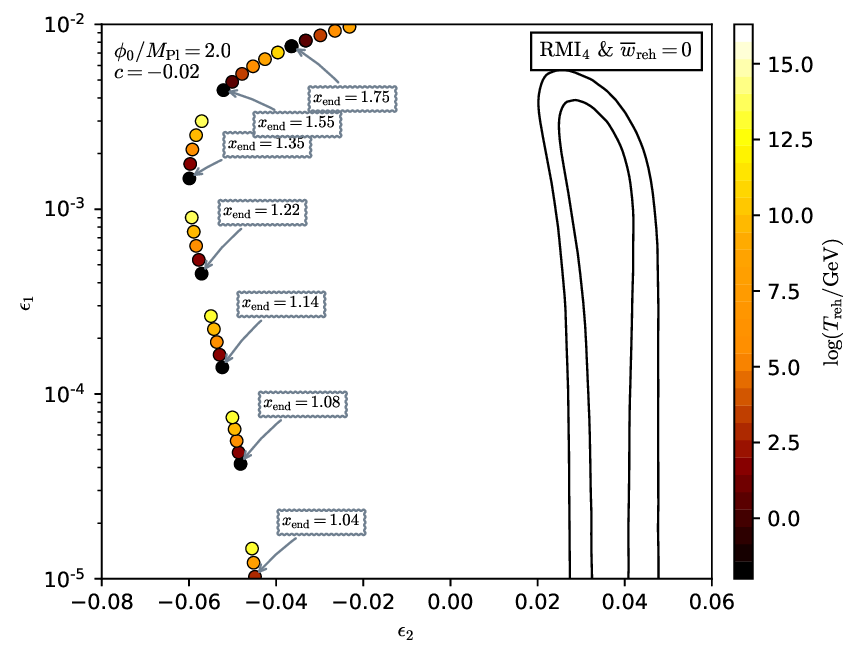}
\caption{Reheating consistent slow-roll predictions for the running
  mass inflation 4 models ($c<0$, $x>1$) with $c=-0.02$ and
  $\phizero/\Mp=2$ (which satisfies $\phizero/\Mp<1/\sqrt{-c}$) in
  the plane $(\nS,r)$ (top panel) and the plane
  $(\epsilon_1,\epsilon_2)$ (bottom panel). The solid contours are the
  one and two-sigma {\data} confidence intervals (marginalized over
  second order slow-roll). The field value at which inflation ends is
  varied in the range $1<\xend<e$. }
\label{fig:CMBRMI4_2}
\end{center}
\end{figure}

\begin{figure}[H]
\begin{center}
\includegraphics[width=\wappfig,clip=true]{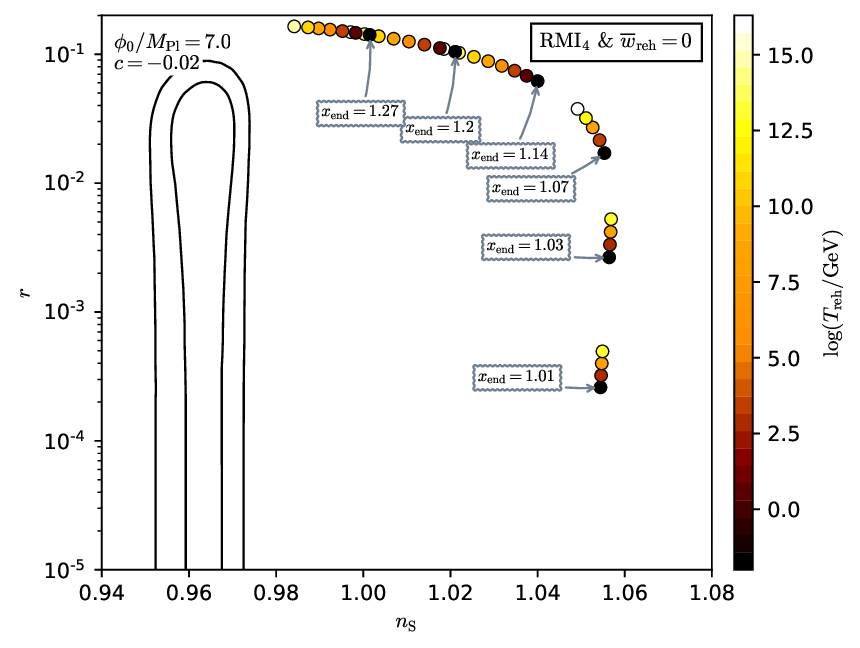}
\includegraphics[width=\wappfig,clip=true]{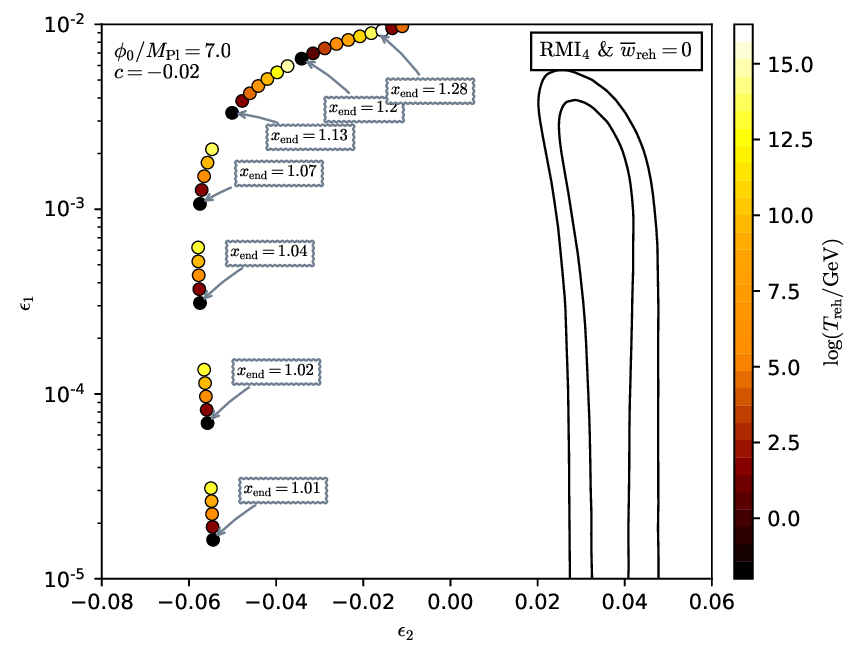}
\caption{Reheating consistent slow-roll predictions for the running
  mass inflation 4 models ($c<0$, $x>1$) with $c=-0.02$ and
  $\phizero/\Mp=7$ (which satisfies $\phizero/\Mp<1/\sqrt{-c}$) in
  the plane $(\nS,r)$ (top panel) and the plane
  $(\epsilon_1,\epsilon_2)$ (bottom panel). The solid contours are the
  one and two-sigma {\data} confidence intervals (marginalized over
  second order slow-roll). The field value at which inflation ends is
  varied in the range $1<\xend<e$. }
\label{fig:CMBRMI4_3}
\end{center}
\end{figure}

\subsection{Valley Hybrid Inflation (\hyperref[sec:vhi]{VHI})}

\begin{figure}[H]
\begin{center}
\includegraphics[width=\wappfig,clip=true]{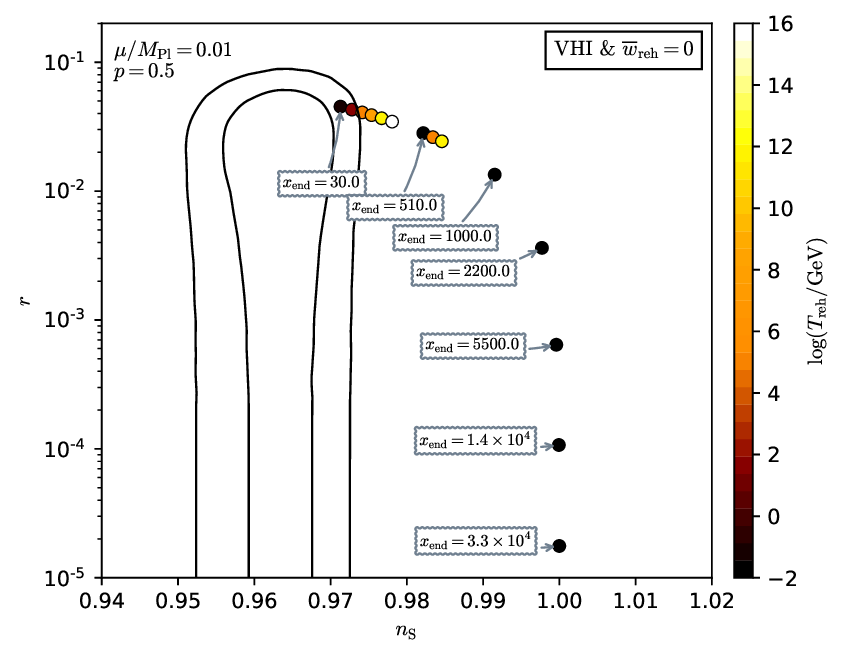}
\includegraphics[width=\wappfig,clip=true]{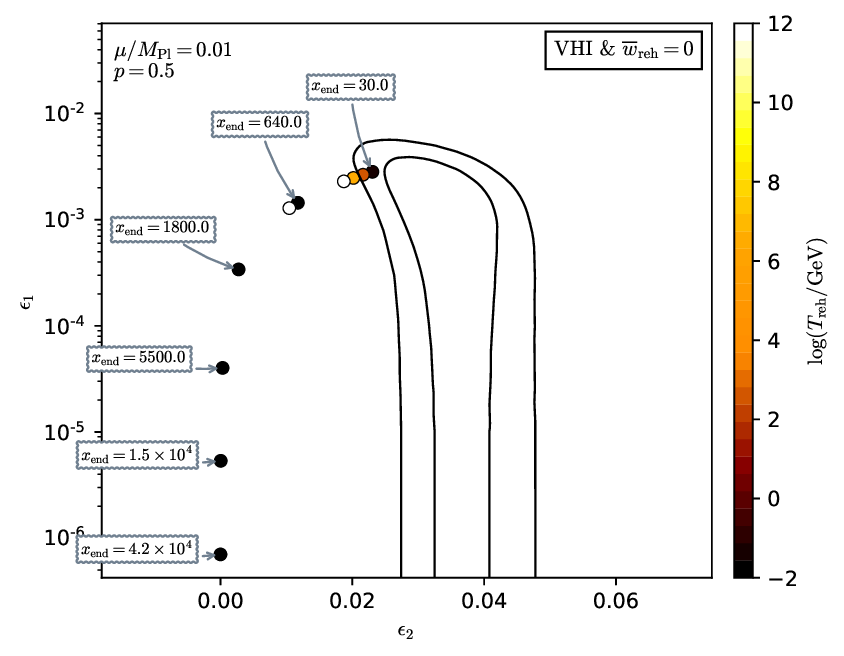}
\caption{Reheating consistent slow-roll predictions for the valley
  hybrid inflation models with $p=0.5$ and $\mu/\Mp=10^{-2}$, in the
  plane $(\nS,r)$ (top panel) and the plane $(\epsilon_1,\epsilon_2)$
  (bottom panel). The solid contours are the one and two-sigma {\data}
  confidence intervals (marginalized over second order slow-roll). See
  figures~\ref{fig:CMBVHIpEQonehalf_1} to \ref{fig:CMBVHIpEQ3_1} for
  other values of $p$ and $\mu$.}
\label{fig:CMBVHIpEQonehalf}
\end{center}
\end{figure}

\begin{figure}[H]
\begin{center}
\includegraphics[width=\wappfig,clip=true]{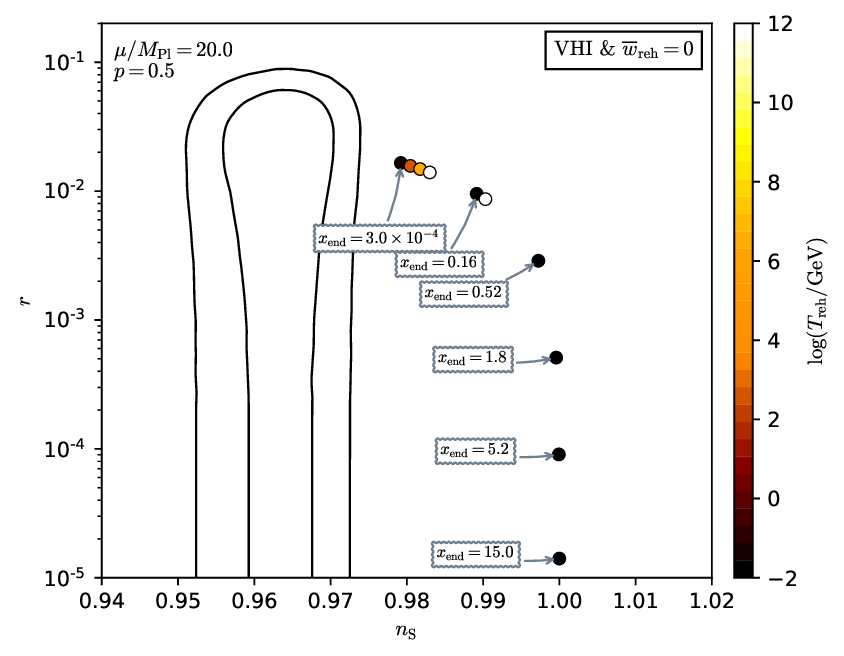}
\includegraphics[width=\wappfig,clip=true]{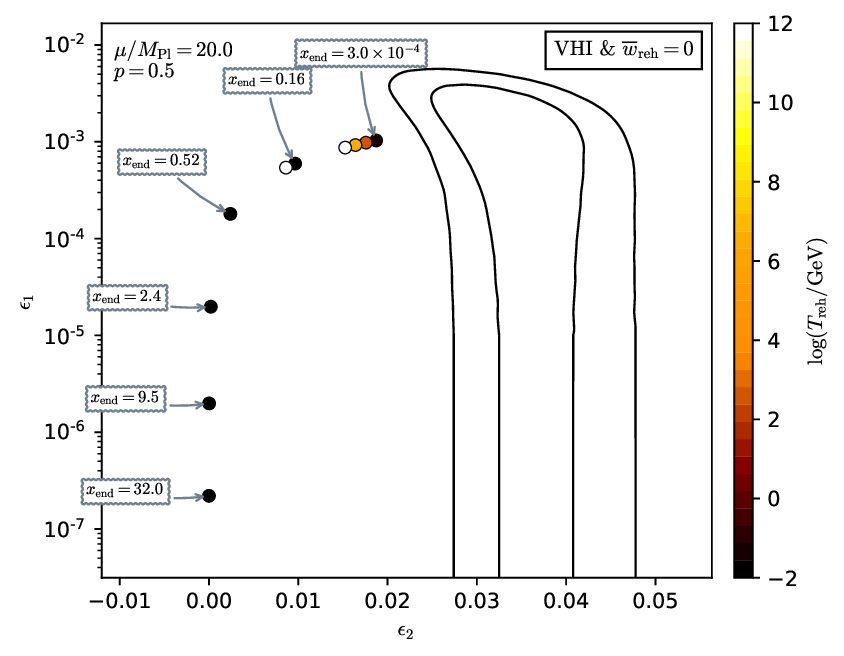}
\caption{Reheating consistent slow-roll predictions for the valley
  hybrid inflation models with $p=0.5$ and $\mu/\Mp=20$, in the
  plane $(\nS,r)$ (top panel) and the plane $(\epsilon_1,\epsilon_2)$
  (bottom panel). The solid contours are the one and two-sigma {\data}
  confidence intervals (marginalized over second order slow-roll).}
\label{fig:CMBVHIpEQonehalf_1}
\end{center}
\end{figure}

\begin{figure}[H]
\begin{center}
\includegraphics[width=\wappfig,clip=true]{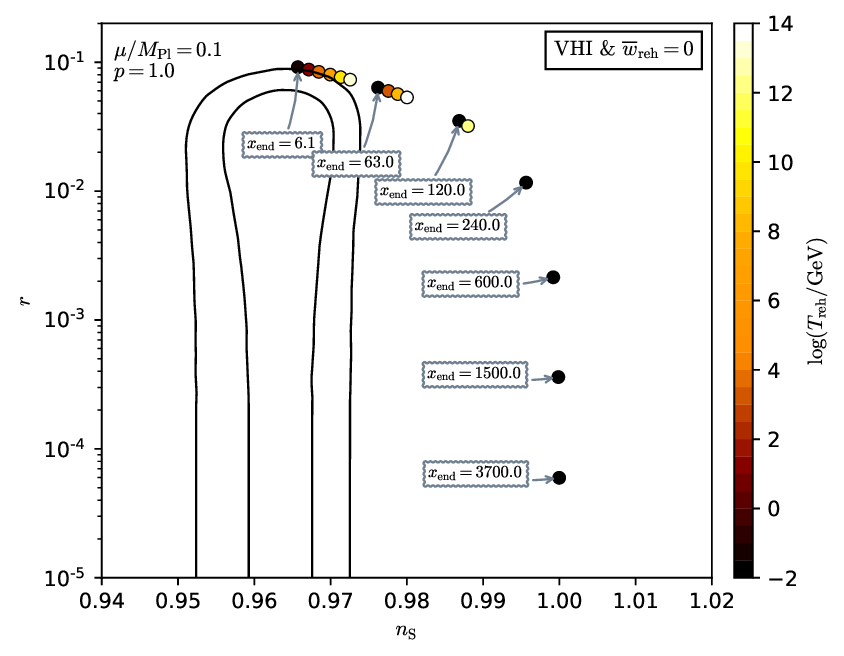}
\includegraphics[width=\wappfig,clip=true]{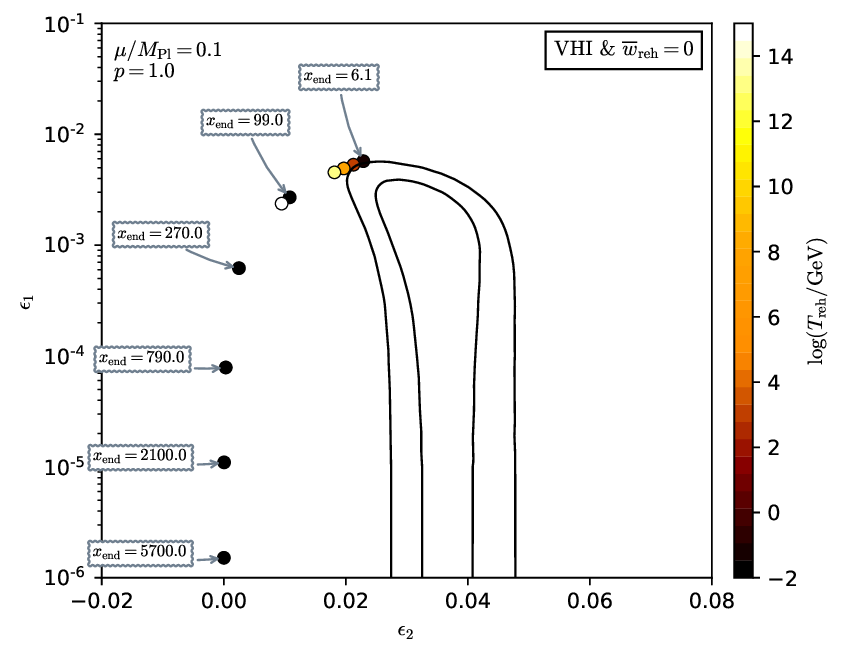}
\caption{Reheating consistent slow-roll predictions for the valley
  hybrid inflation models with $p=1$ and $\mu/\Mp=10^{-1}$, in the
  plane $(\nS,r)$ (top panel) and the plane $(\epsilon_1,\epsilon_2)$
  (bottom panel). The solid contours are the one and two-sigma {\data}
  confidence intervals (marginalized over second order slow-roll).
  The model predictions are degenerated along the curve
  $\epsilon_2=4\epsilon_1$, see also figure~\ref{fig:CMBVHIpEQ1_1}.}
\label{fig:CMBVHIpEQ1}
\end{center}
\end{figure}

\begin{figure}[H]
\begin{center}
\includegraphics[width=\wappfig,clip=true]{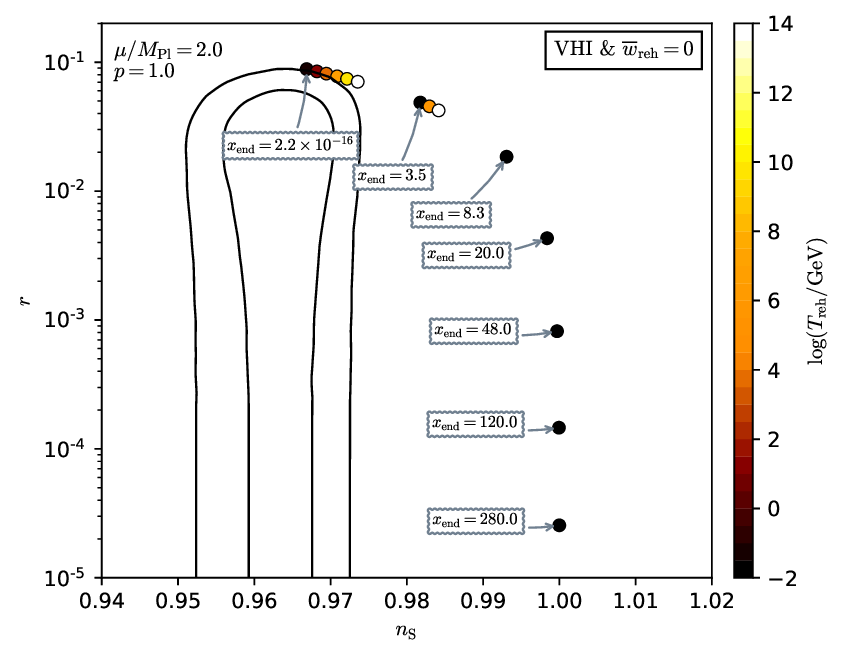}
\includegraphics[width=\wappfig,clip=true]{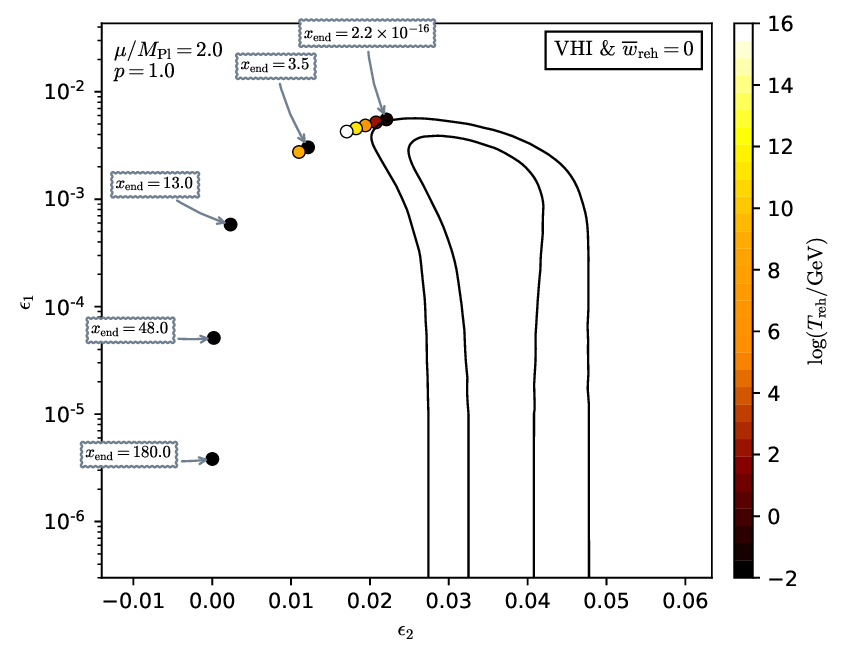}
\caption{Reheating consistent slow-roll predictions for the valley
  hybrid inflation models with $p=1$ and $\mu/\Mp=2$, in the plane
  $(\nS,r)$ (top panel) and the plane $(\epsilon_1,\epsilon_2)$
  (bottom panel). The solid contours are the one and two-sigma {\data}
  confidence intervals (marginalized over second order slow-roll).
  The model predictions are degenerated along the curve
  $\epsilon_2=4\epsilon_1$, see also figure~\ref{fig:CMBVHIpEQ1}.}
\label{fig:CMBVHIpEQ1_1}
\end{center}
\end{figure}

\begin{figure}[H]
\begin{center}
\includegraphics[width=\wappfig,clip=true]{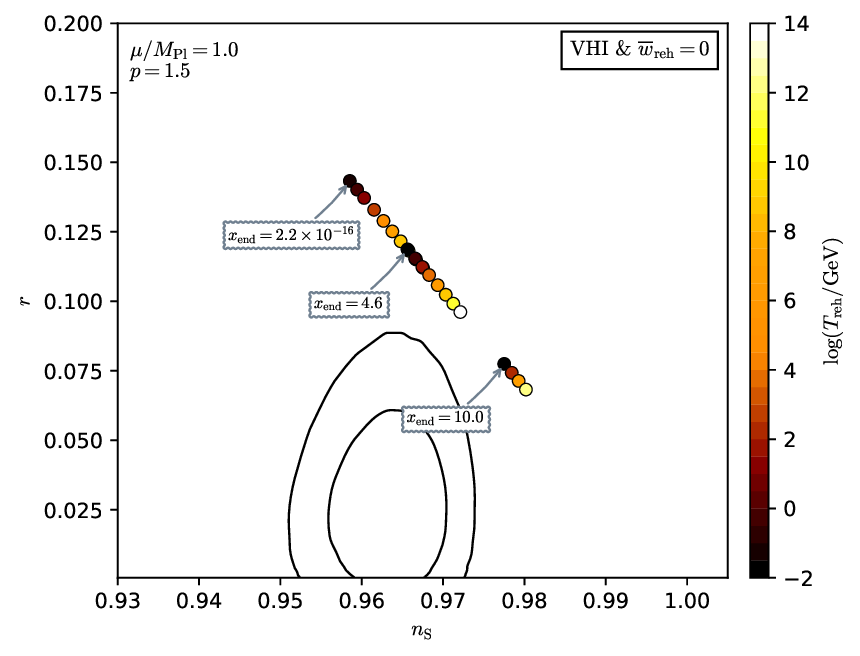}
\includegraphics[width=\wappfig,clip=true]{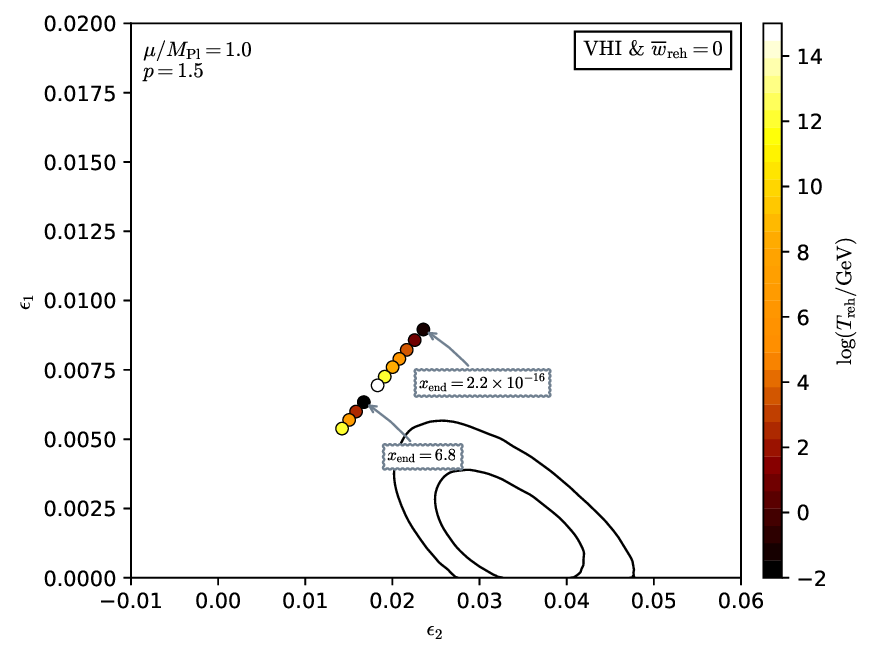}
\caption{Reheating consistent slow-roll predictions for the valley
  hybrid inflation models with $p=1.5$ and $\mu=\Mp$, in the plane
  $(\nS,r)$ (top panel) and the plane $(\epsilon_1,\epsilon_2)$
  (bottom panel). The solid contours are the one and two-sigma {\data}
  confidence intervals (marginalized over second order slow-roll).}
\label{fig:CMBVHIpEQ1andahalf}
\end{center}
\end{figure}

\begin{figure}[H]
\begin{center}
\includegraphics[width=\wappfig,clip=true]{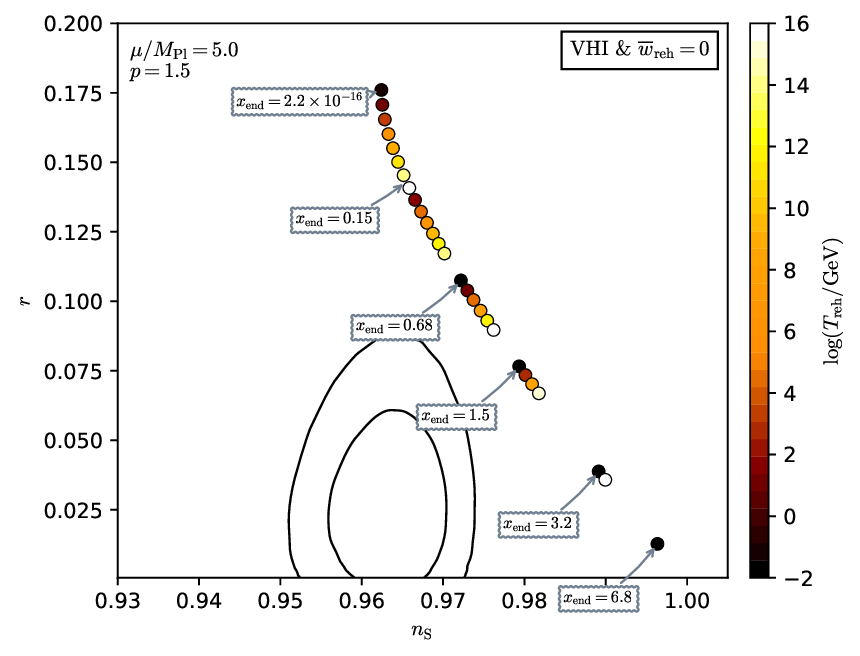}
\includegraphics[width=\wappfig,clip=true]{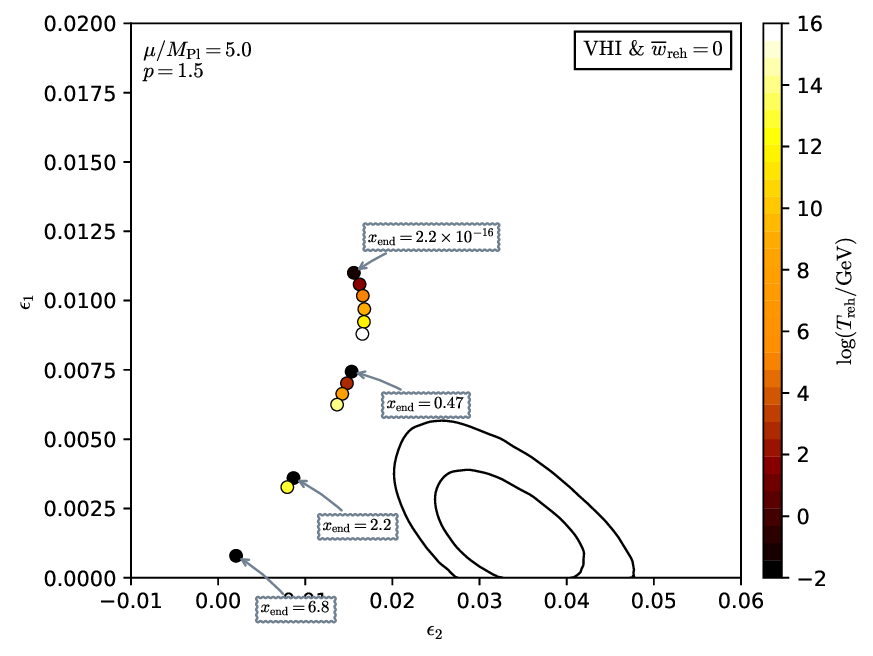}
\caption{Reheating consistent slow-roll predictions for the valley
  hybrid inflation models with $p=1.5$ and $\mu/\Mp=5$, in the plane
  $(\nS,r)$ (top panel) and the plane $(\epsilon_1,\epsilon_2)$
  (bottom panel). The solid contours are the one and two-sigma {\data}
  confidence intervals (marginalized over second order slow-roll).}
\label{fig:CMBVHIpEQ1andahalf_1}
\end{center}
\end{figure}

\begin{figure}[H]
\begin{center}
\includegraphics[width=\wappfig,clip=true]{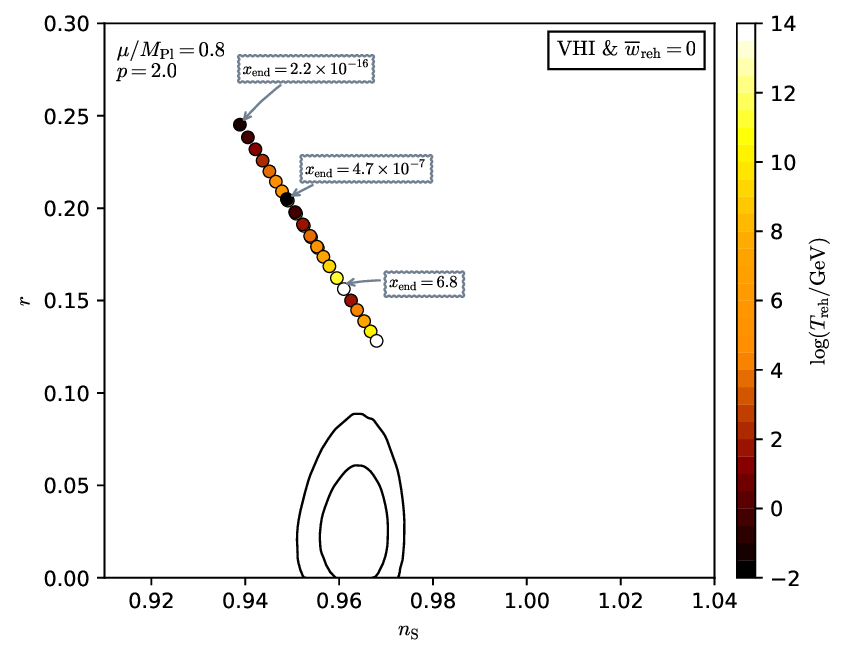}
\includegraphics[width=\wappfig,clip=true]{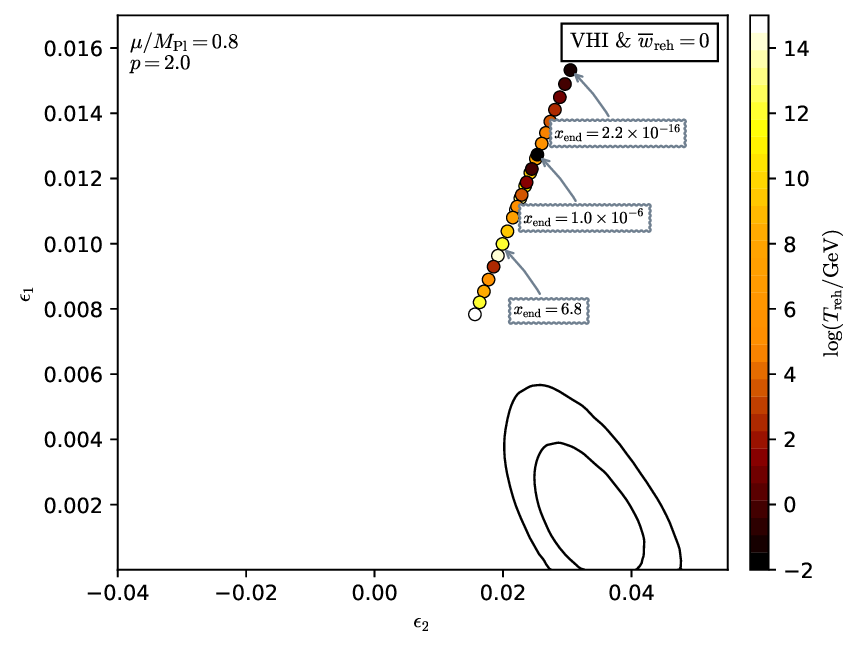}
\caption{Reheating consistent slow-roll predictions for the valley
  hybrid inflation models with $p=2$ and $\mu/\Mp=0.8$, in the plane
  $(\nS,r)$ (top panel) and the plane $(\epsilon_1,\epsilon_2)$
  (bottom panel). The solid contours are the one and
  two-sigma {\data} confidence intervals (marginalized over second
  order slow-roll).}
\label{fig:CMBVHIpEQ2}
\end{center}
\end{figure}

\begin{figure}[H]
\begin{center}
\includegraphics[width=\wappfig,clip=true]{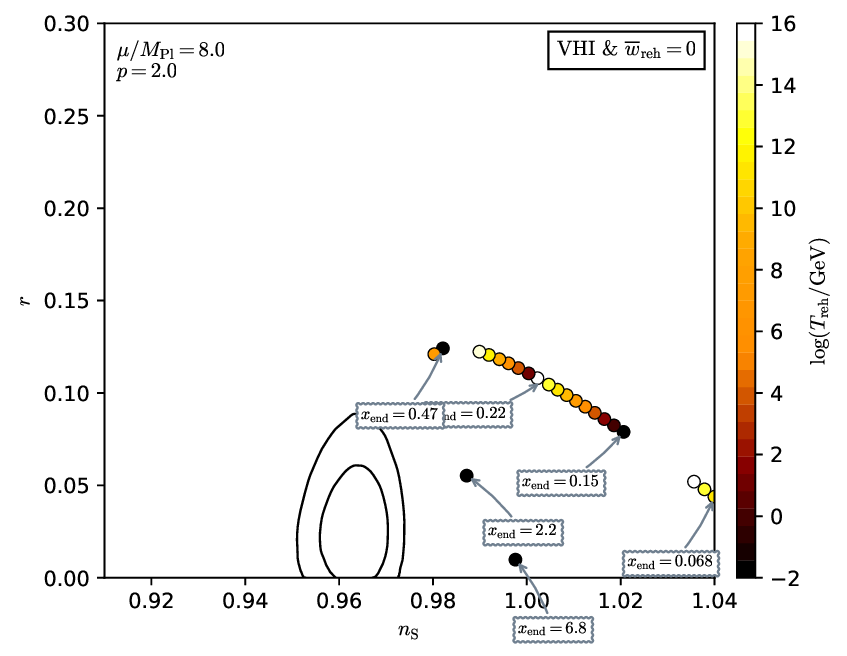}
\includegraphics[width=\wappfig,clip=true]{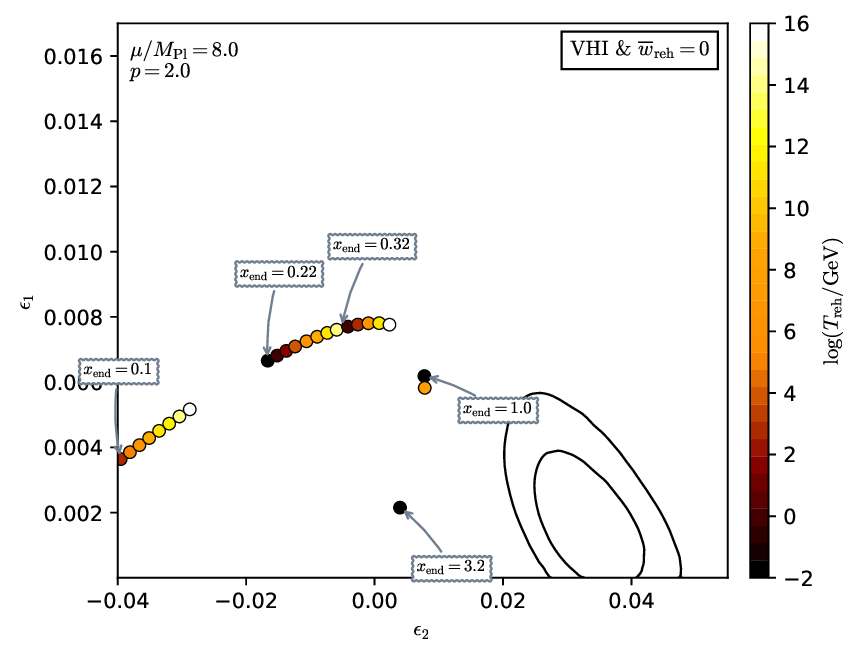}
\caption{Reheating consistent slow-roll predictions for the valley
  hybrid inflation models with $p=2$ and $\mu/\Mp=8$, in the plane
  $(\nS,r)$ (top panel) and the plane $(\epsilon_1,\epsilon_2)$
  (bottom panel). The solid contours are the one and
  two-sigma {\data} confidence intervals (marginalized over second
  order slow-roll).}
\label{fig:CMBVHIpEQ2_1}
\end{center}
\end{figure}

\begin{figure}[H]
\begin{center}
\includegraphics[width=\wappfig,clip=true]{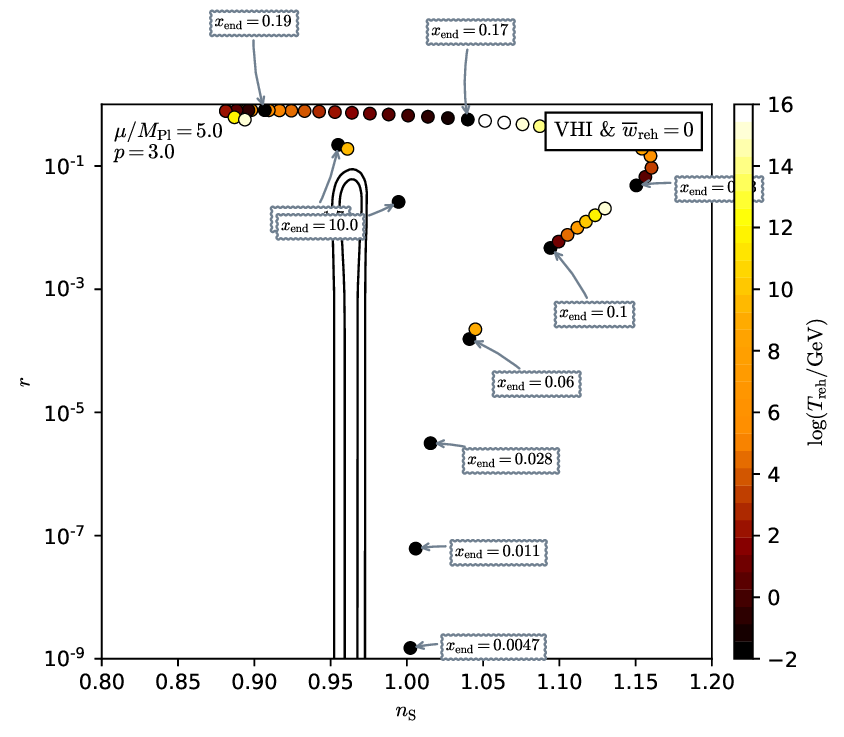}
\includegraphics[width=\wappfig,clip=true]{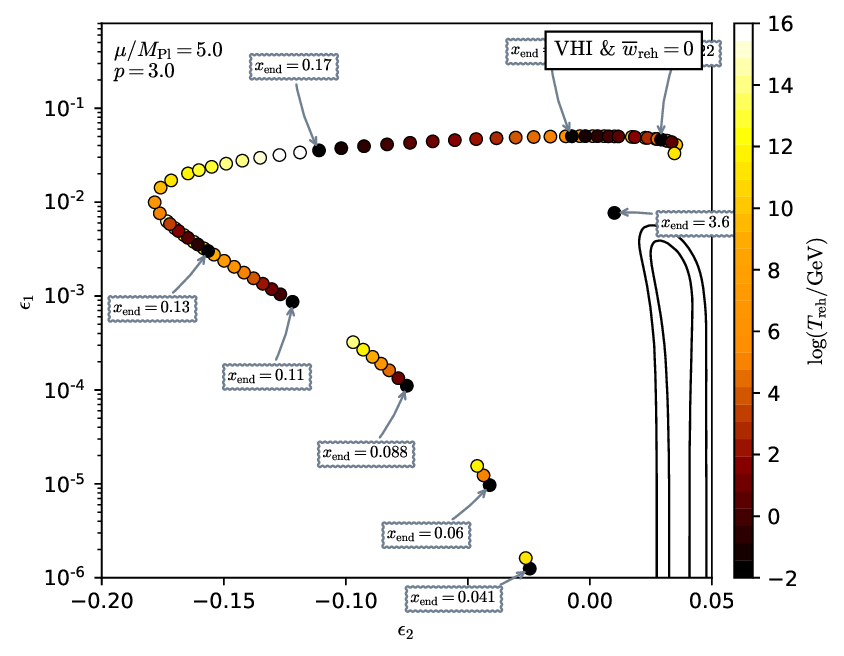}
\caption{Reheating consistent slow-roll predictions for the valley
  hybrid inflation models with $p=3$ and $\mu/\Mp=5$, in the plane
  $(\nS,r)$ (top panel) and the plane $(\epsilon_1,\epsilon_2)$
  (bottom panel). The solid contours are the one and two-sigma {\data}
  confidence intervals (marginalized over second order slow-roll).}
\label{fig:CMBVHIpEQ3}
\end{center}
\end{figure}

\begin{figure}[H]
\begin{center}
\includegraphics[width=\wappfig,clip=true]{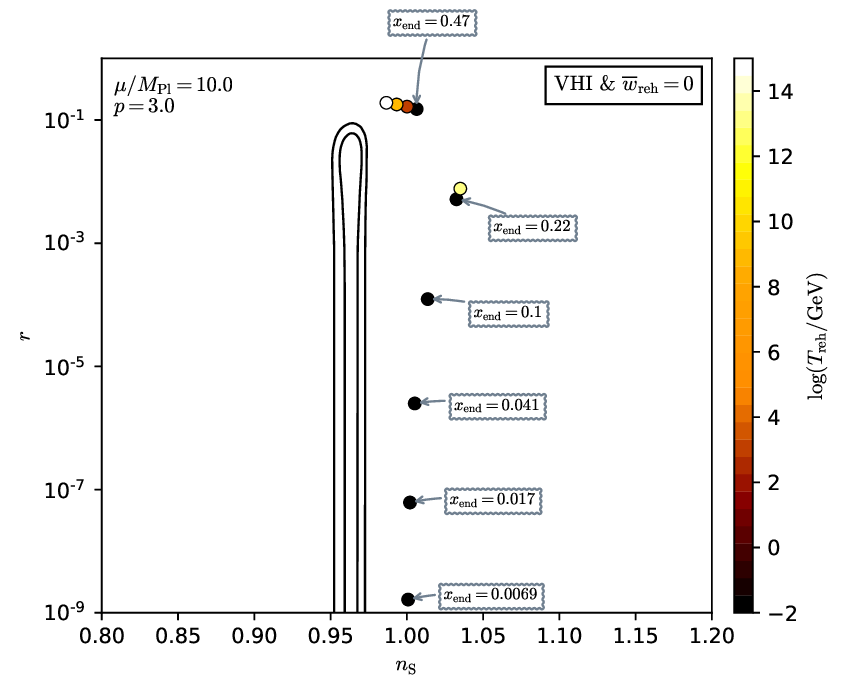}
\includegraphics[width=\wappfig,clip=true]{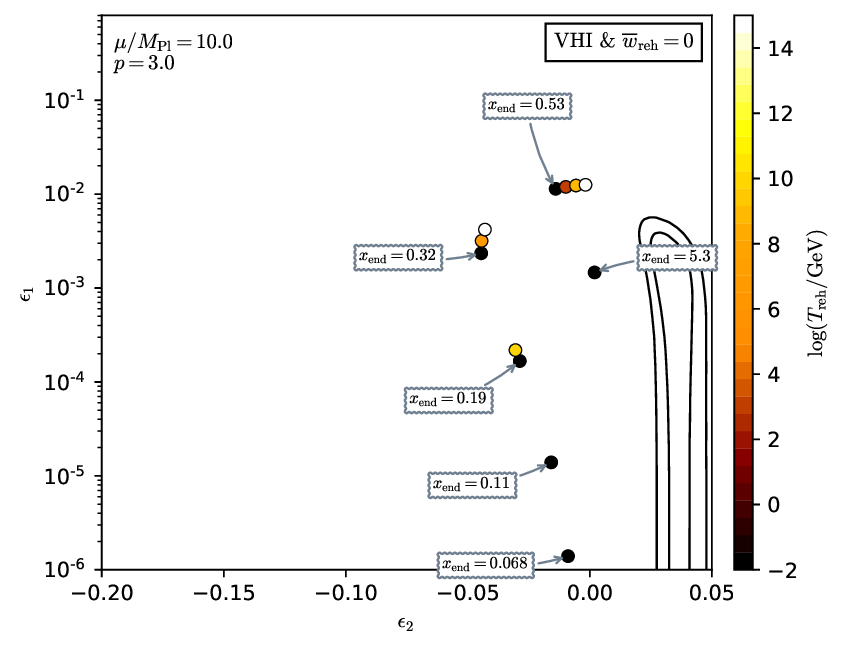}
\caption{Reheating consistent slow-roll predictions for the valley
  hybrid inflation models with $p=3$ and $\mu/\Mp=10$, in the plane
  $(\nS,r)$ (top panel) and the plane $(\epsilon_1,\epsilon_2)$
  (bottom panel). The solid contours are the one and two-sigma {\data}
  confidence intervals (marginalized over second order slow-roll).}
\label{fig:CMBVHIpEQ3_1}
\end{center}
\end{figure}

\subsection{Dynamical Supersymmetric Inflation (\hyperref[sec:dsi]{DSI})}

\begin{figure}[H]
\begin{center}
\includegraphics[width=\wappfig,clip=true]{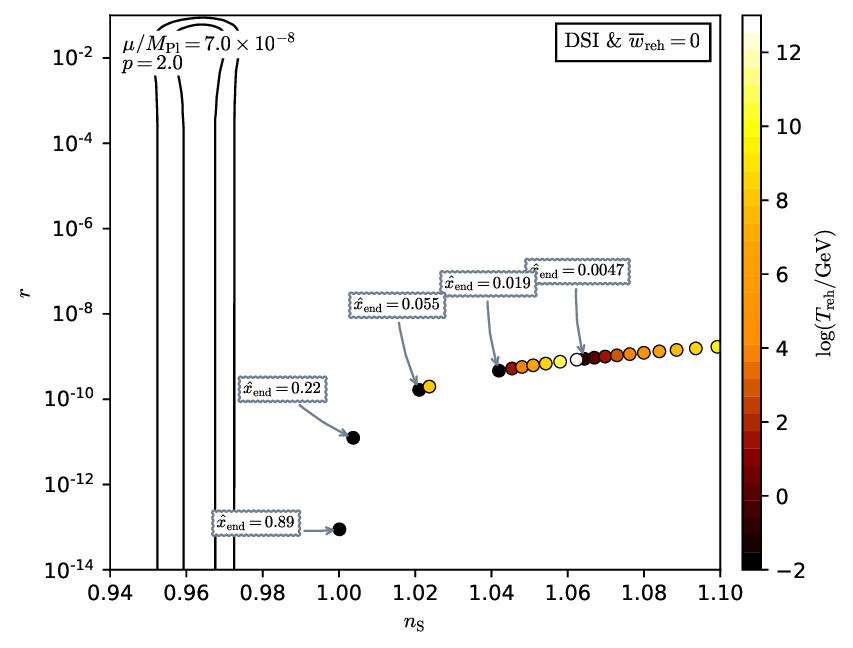}
\includegraphics[width=\wappfig,clip=true]{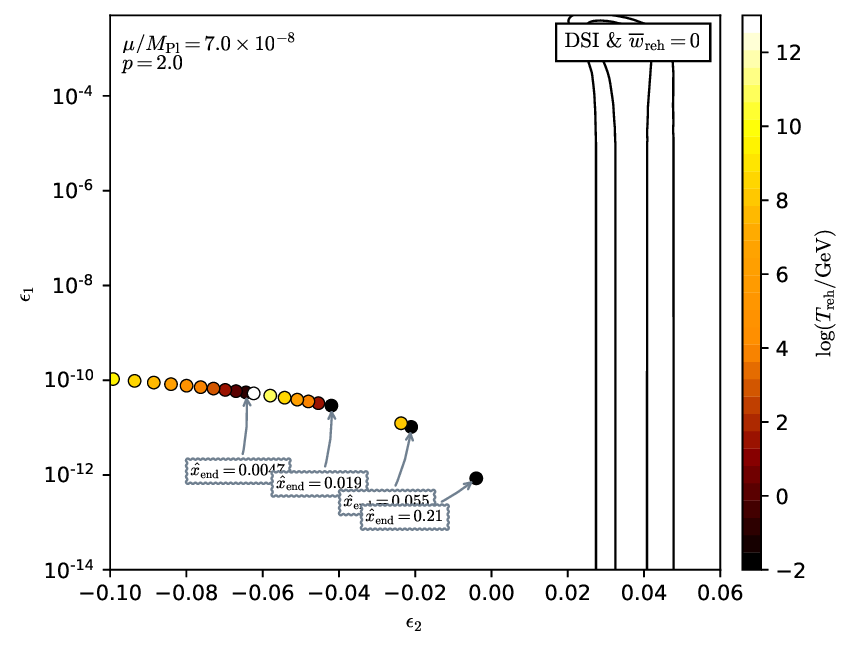}
\caption{Reheating consistent slow-roll predictions for the dynamical
  supersymmetric inflation models with $p=2$ and $\mu/\Mp=7 \times
  10^{-8}$ (which satisfies $\mu<\mumax$), in the plane $(\nS,r)$ (top
  panel) and the plane $(\epsilon_1,\epsilon_2)$ (bottom panel). The
  solid contours are the one and two-sigma {\data} confidence
  intervals (marginalized over second order slow-roll).  The field
  value at which inflation ends is varied within its maximal allowed
  range, {\ie} with $\xendhat\equiv (\xend -
  \xendmin)/(\xendmax-\xendmin)$ in the domain $[0,1]$. See
  figures~\ref{fig:CMBDSIpEQ2_1} to \ref{fig:CMBDSIpEQ4_5} for other
  values of $p$ and $\mu$.}
\label{fig:CMBDSIpEQ2}
\end{center}
\end{figure}

\begin{figure}[H]
\begin{center}
\includegraphics[width=\wappfig,clip=true]{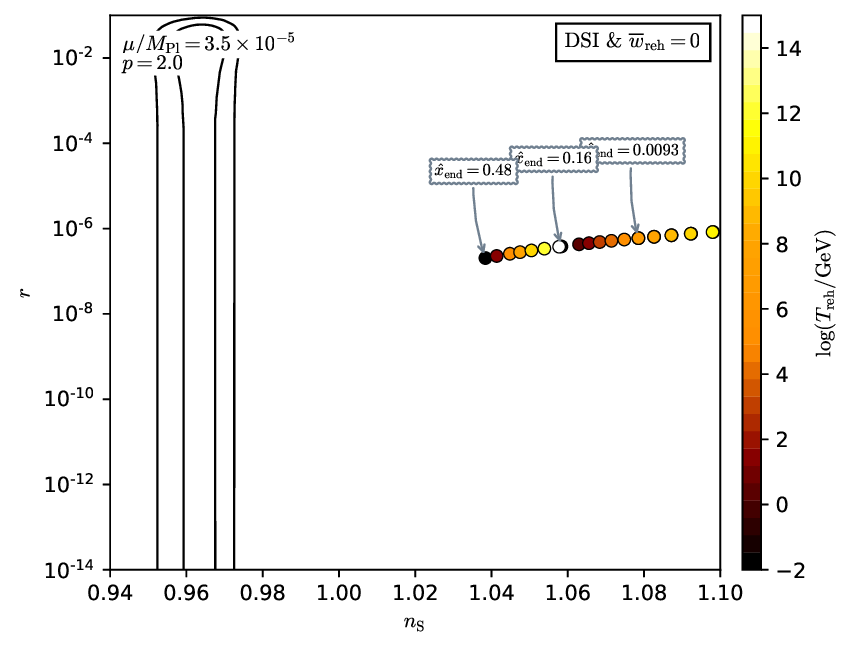}
\includegraphics[width=\wappfig,clip=true]{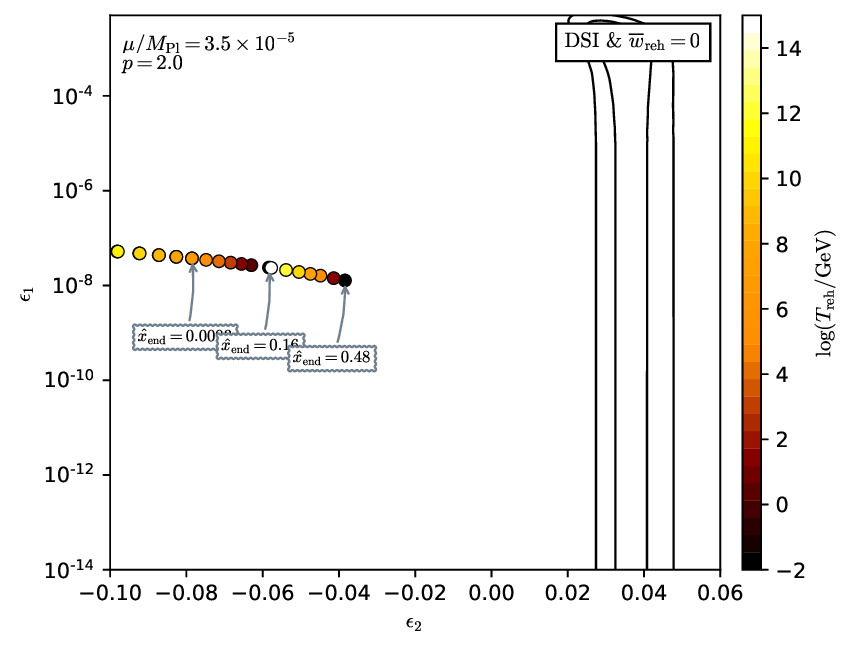}
\caption{Reheating consistent slow-roll predictions for the dynamical
  supersymmetric inflation models with $p=2$ and $\mu/\Mp=3.5 \times
  10^{-5}$ (which satisfies $\mu <\mumax$), in the plane
  $(\nS,r)$ (top panel) and the plane $(\epsilon_1,\epsilon_2)$
  (bottom panel). The solid contours are the one and two-sigma {\data}
  confidence intervals (marginalized over second order slow-roll).
  The field value at which inflation ends is varied within its maximal
  allowed range, {\ie} with $\xendhat\equiv (\xend -
  \xendmin)/(\xendmax-\xendmin)$ in the domain $[0,1]$.}
\label{fig:CMBDSIpEQ2_1}
\end{center}
\end{figure}

\begin{figure}[H]
\begin{center}
\includegraphics[width=\wappfig,clip=true]{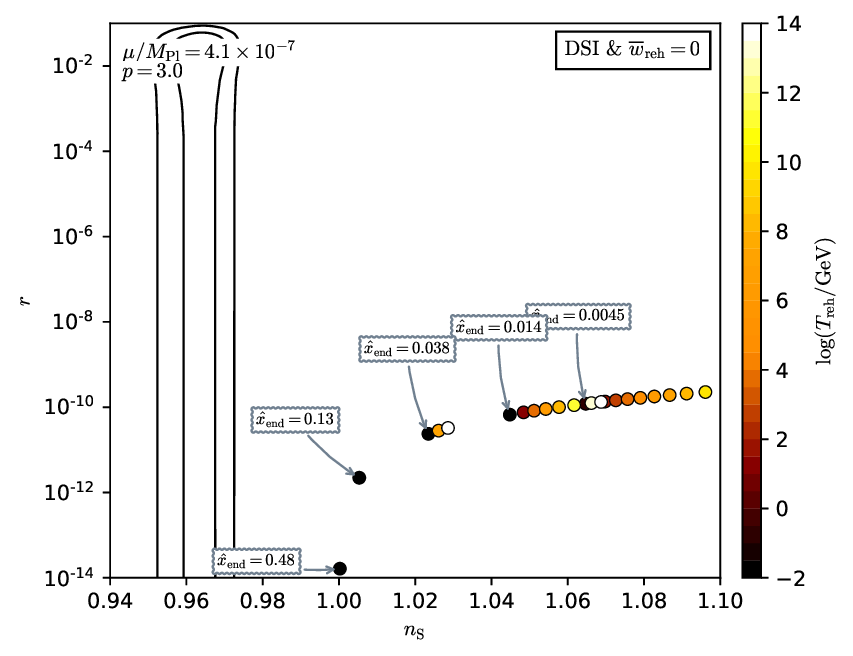}
\includegraphics[width=\wappfig,clip=true]{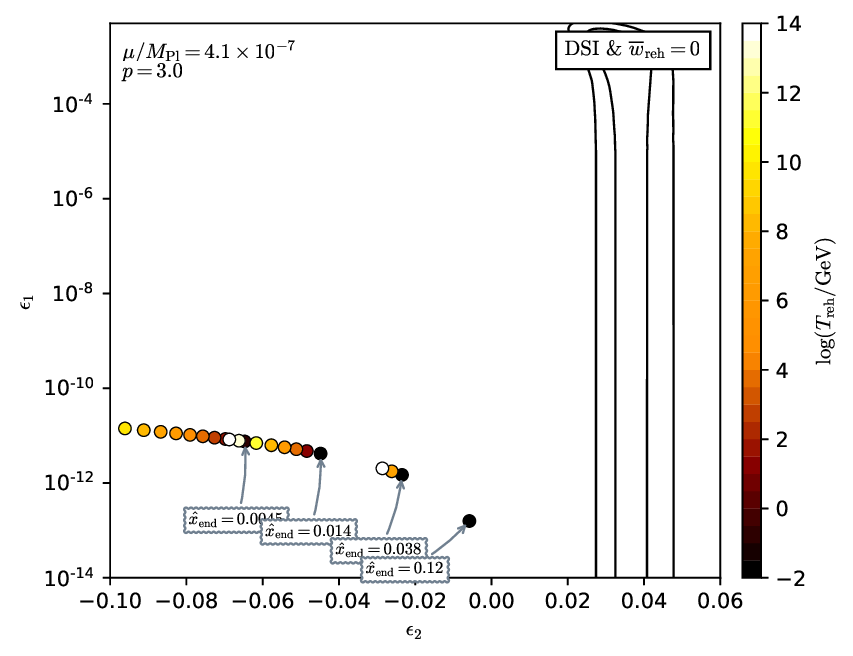}
\caption{Reheating consistent slow-roll predictions for the dynamical
  supersymmetric inflation models with $p=3$ and $\mu/\Mp=4.1 \times
  10^{-7}$ (which satisfies $\mu <\mumax$), in the plane
  $(\nS,r)$ (top panel) and the plane $(\epsilon_1,\epsilon_2)$
  (bottom panel). The solid contours are the one and two-sigma {\data}
  confidence intervals (marginalized over second order slow-roll). The
  field value at which inflation ends is varied within its maximal
  allowed range, {\ie} with $\xendhat\equiv (\xend -
  \xendmin)/(\xendmax-\xendmin)$ in the domain $[0,1]$.}
\label{fig:CMBDSIpEQ3}
\end{center}
\end{figure}

\begin{figure}[H]
\begin{center}
\includegraphics[width=\wappfig,clip=true]{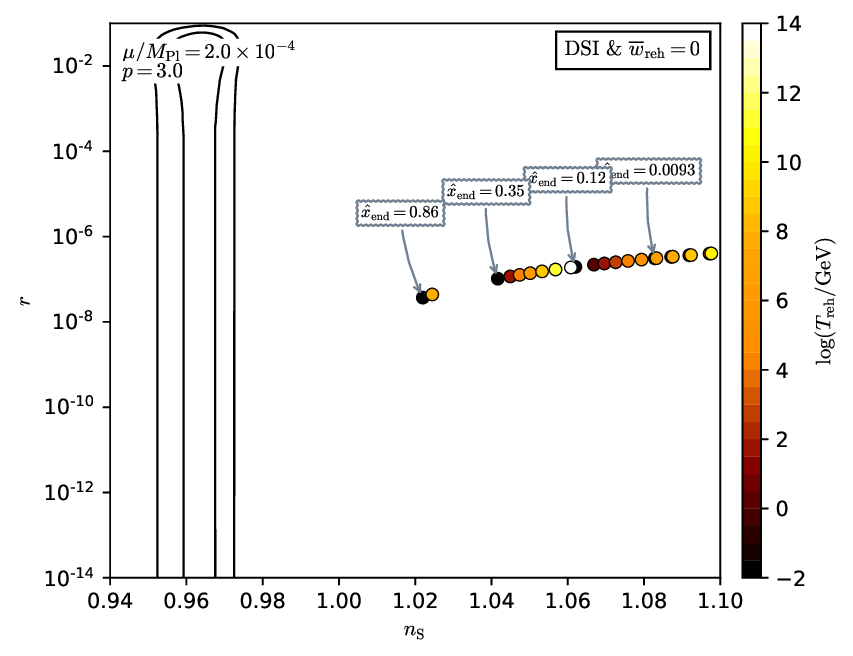}
\includegraphics[width=\wappfig,clip=true]{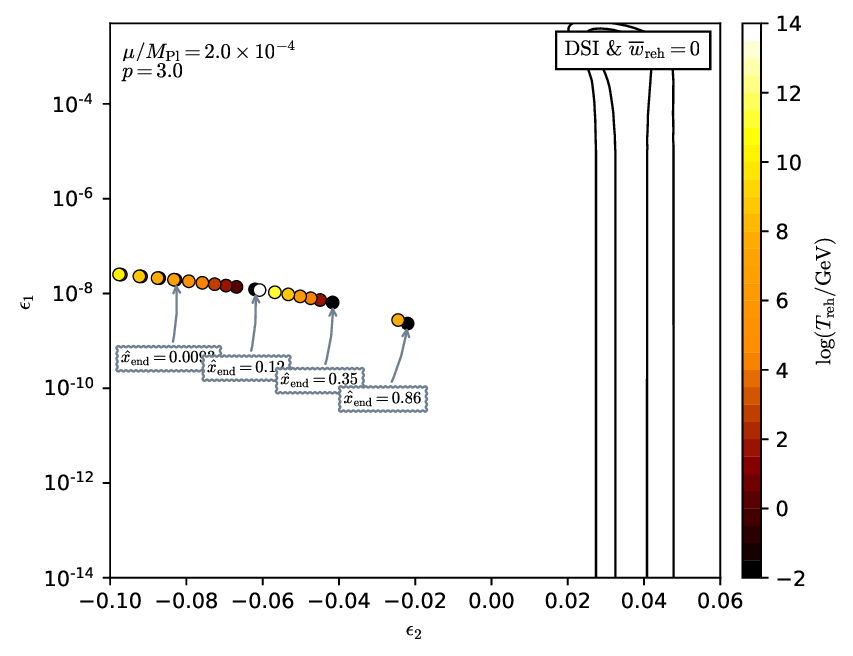}
\caption{Reheating consistent slow-roll predictions for the dynamical
  supersymmetric inflation models with $p=3$ and $\mu/\Mp=2 \times
  10^{-5}$ (which satisfies $\mu <\mumax$), in the plane
  $(\nS,r)$ (top panel) and the plane $(\epsilon_1,\epsilon_2)$
  (bottom panel). The solid contours are the one and two-sigma {\data}
  confidence intervals (marginalized over second order slow-roll). The
  field value at which inflation ends is varied within its maximal
  allowed range, {\ie} with $\xendhat\equiv (\xend -
  \xendmin)/(\xendmax-\xendmin)$ in the domain $[0,1]$.}
\label{fig:CMBDSIpEQ3_3}
\end{center}
\end{figure}

\begin{figure}[H]
\begin{center}
\includegraphics[width=\wappfig,clip=true]{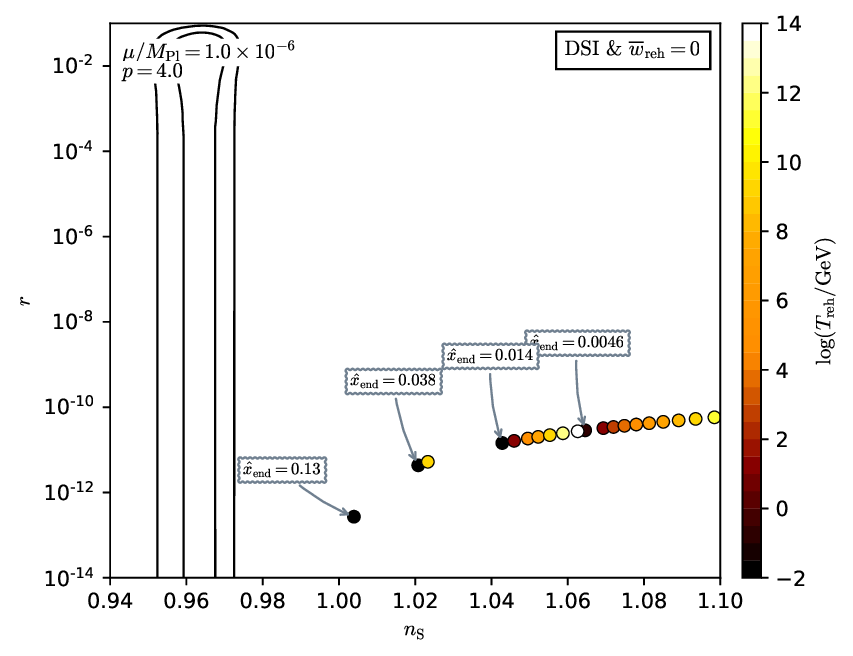}
\includegraphics[width=\wappfig,clip=true]{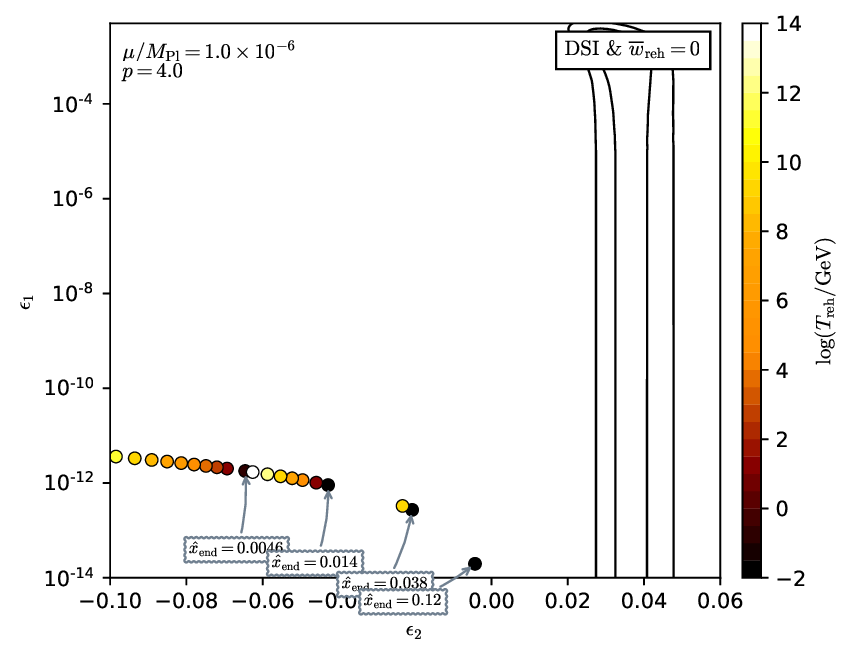}
\caption{Reheating consistent slow-roll predictions for the dynamical
  supersymmetric inflation models with $p=4$ and $\mu/\Mp=10^{-6}$
  (which satisfies $\mu <\mumax$), in the plane $(\nS,r)$ (top panel)
  and the plane $(\epsilon_1,\epsilon_2)$ (bottom panel). The solid
  contours are the one and two-sigma {\data} confidence intervals
  (marginalized over second order slow-roll). The field value at which
  inflation ends is varied within its maximal allowed range, {\ie}
  with $\xendhat\equiv (\xend - \xendmin)/(\xendmax-\xendmin)$ in the
  domain $[0,1]$.}
\label{fig:CMBDSIpEQ4}
\end{center}
\end{figure}

\begin{figure}[H]
\begin{center}
\includegraphics[width=\wappfig,clip=true]{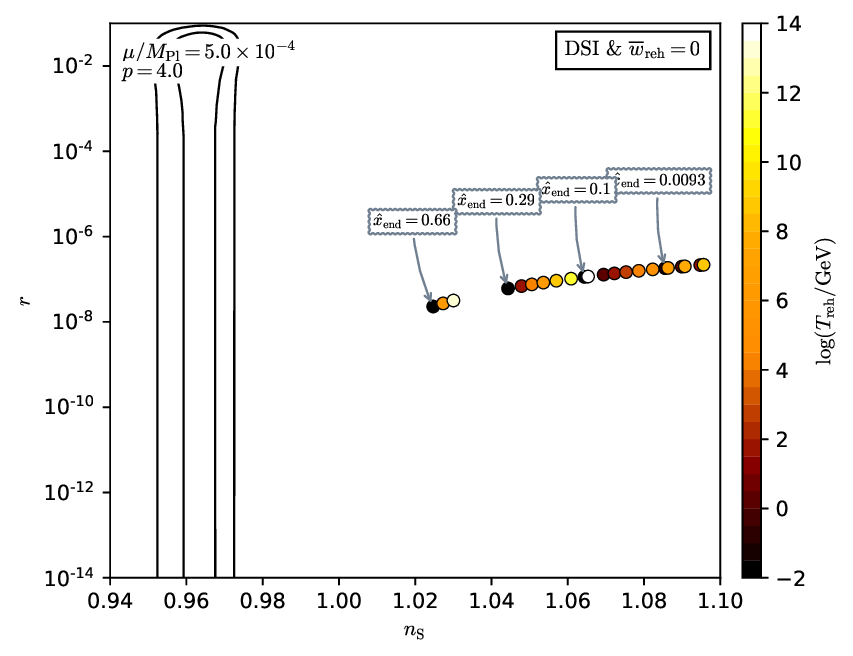}
\includegraphics[width=\wappfig,clip=true]{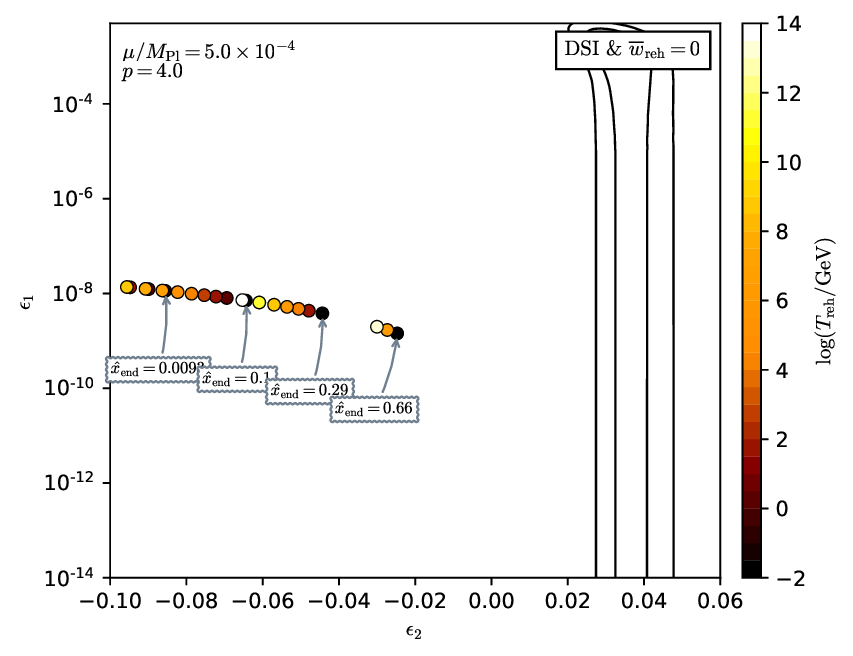}
\caption{Reheating consistent slow-roll predictions for the dynamical
  supersymmetric inflation models with $p=4$ and $\mu/\Mp=10^{-6}$
  (which satisfies $\mu <\mumax$), in the plane $(\nS,r)$ (top panel)
  and the plane $(\epsilon_1,\epsilon_2)$ (bottom panel). The solid
  contours are the one and two-sigma {\data} confidence intervals
  (marginalized over second order slow-roll). The field value at which
  inflation ends is varied within its maximal allowed range, {\ie}
  with $\xendhat\equiv (\xend - \xendmin)/(\xendmax-\xendmin)$ in the
  domain $[0,1]$.}
\label{fig:CMBDSIpEQ4_5}
\end{center}
\end{figure}

\subsection{Generalized Mixed Inflation (\hyperref[sec:gmlfi]{GMLFI})}

\begin{figure}[H]
\begin{center}
\includegraphics[width=\wappfig,clip=true]{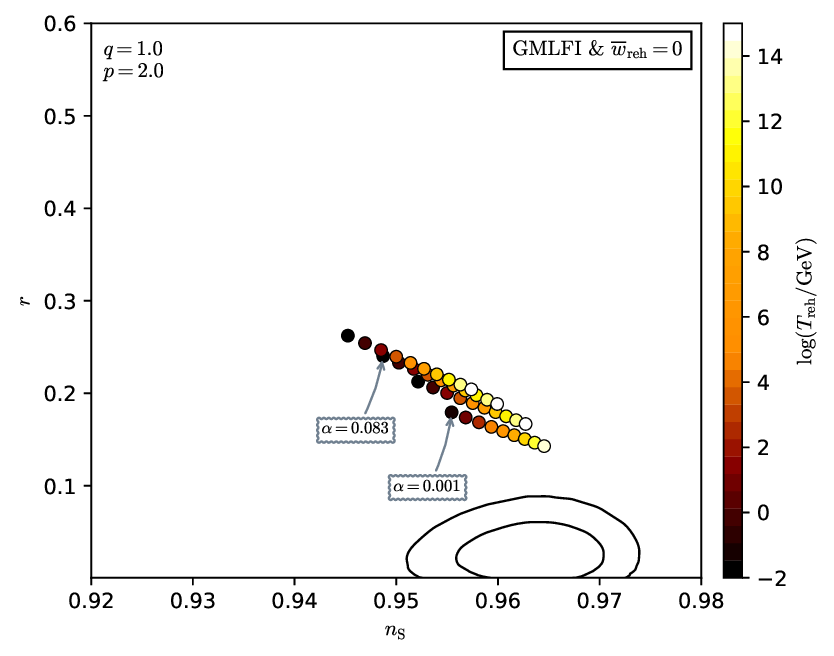}
\includegraphics[width=\wappfig,clip=true]{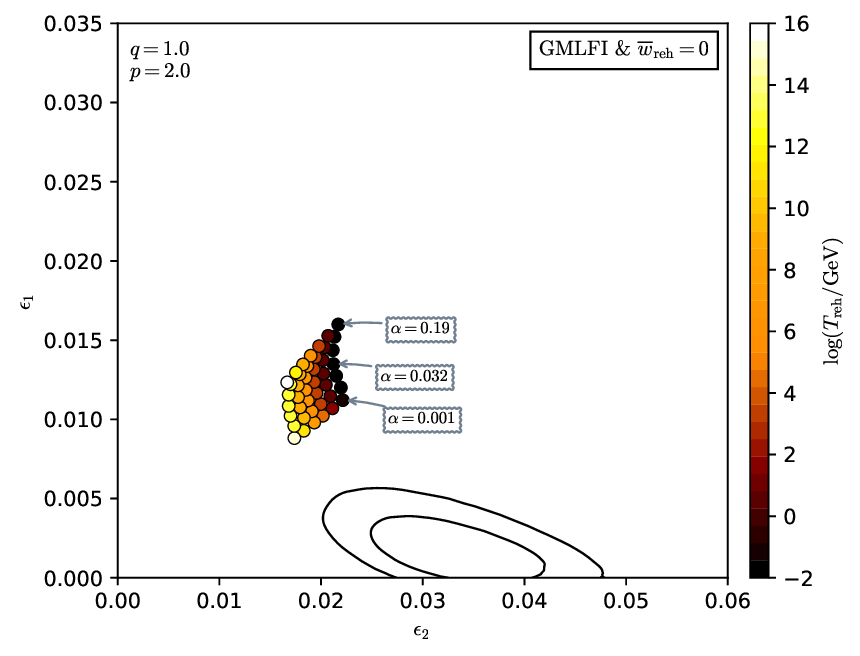}
\caption{Reheating consistent slow-roll predictions for the
  generalized mixed inflation models with $q=1$ and $p=2$, in the
  plane $(\nS,r)$ (top panel) and the plane $(\epsilon_1,\epsilon_2)$
  (bottom panel). The solid contours are the one and two-sigma {\data}
  confidence intervals (marginalized over second order slow-roll).}
\label{fig:CMBGMLFIpEQ2qEQ1}
\end{center}
\end{figure}

\begin{figure}[H]
\begin{center}
\includegraphics[width=\wappfig,clip=true]{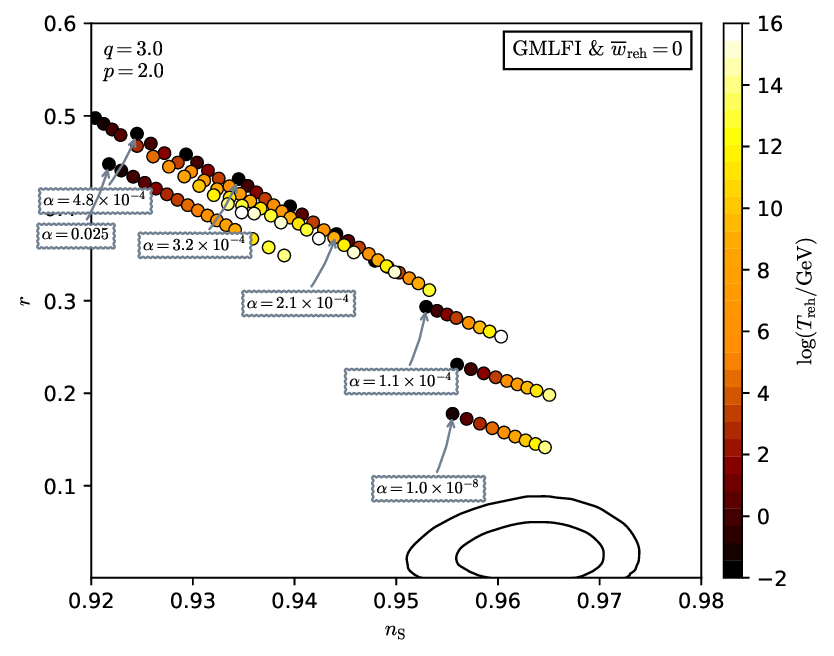}
\includegraphics[width=\wappfig,clip=true]{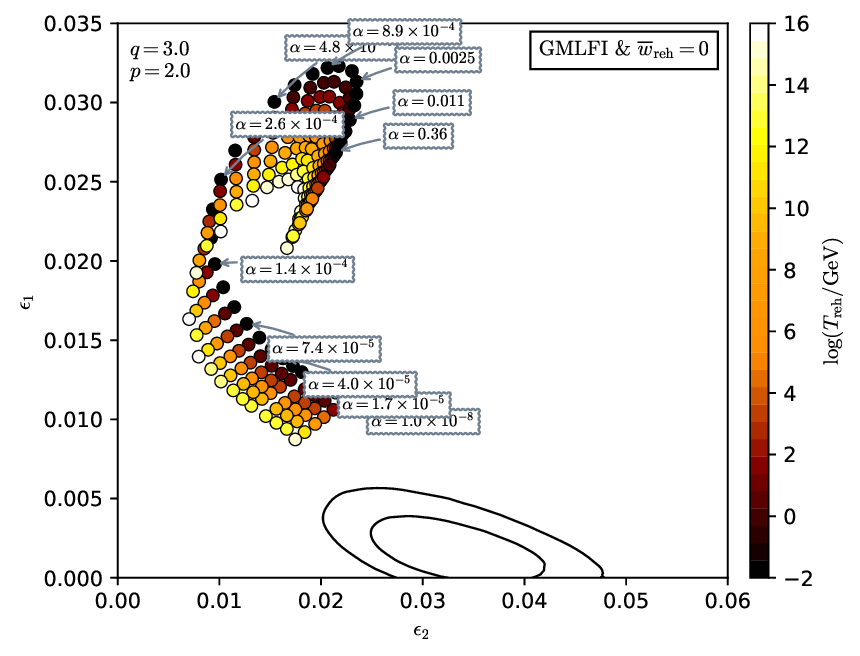}
\caption{Reheating consistent slow-roll predictions for the
  generalized mixed inflation models with $q=3$ and $p=2$, in the
  plane $(\nS,r)$ (top panel) and the plane $(\epsilon_1,\epsilon_2)$
  (bottom panel). The solid contours are the one and two-sigma {\data}
  confidence intervals (marginalized over second order slow-roll).} 
\label{fig:CMBGMLFIpEQ2qEQ3}
\end{center}
\end{figure}

\begin{figure}[H]
\begin{center}
\includegraphics[width=\wappfig,clip=true]{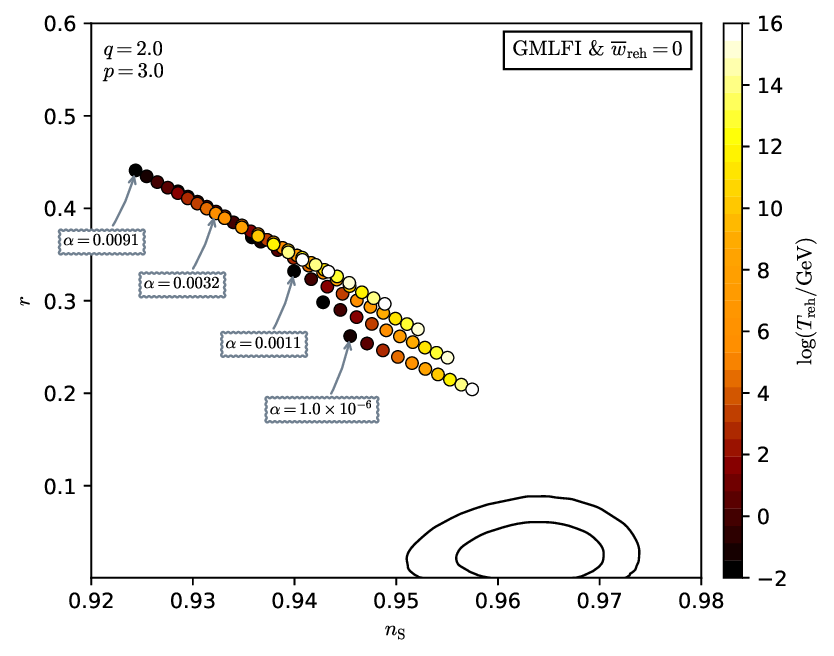}
\includegraphics[width=\wappfig,clip=true]{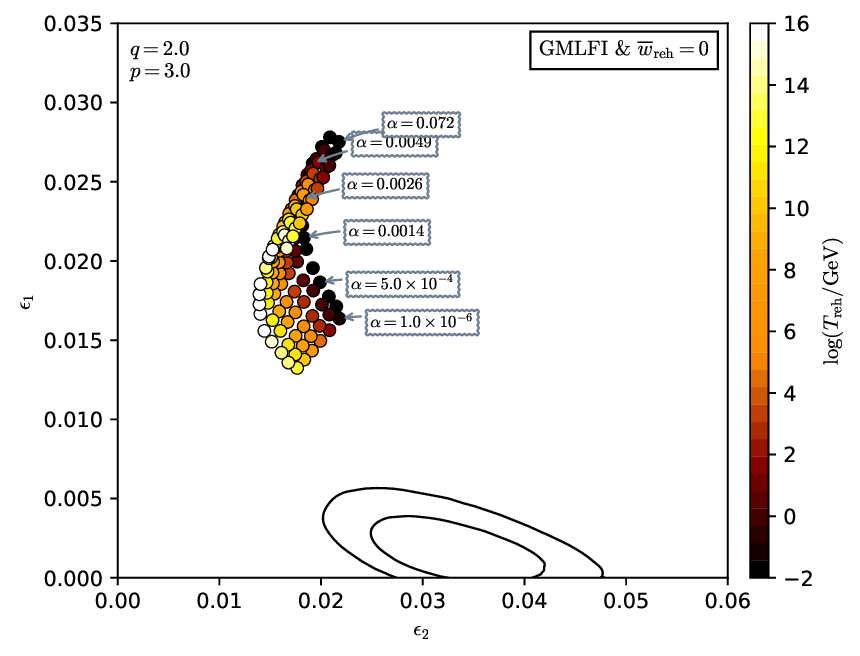}
\caption{Reheating consistent slow-roll predictions for the
  generalized mixed inflation models with $q=2$ and $p=3$, in the
  plane $(\nS,r)$ (top panel) and the plane $(\epsilon_1,\epsilon_2)$
  (bottom panel). The solid contours are the one and two-sigma {\data}
  confidence intervals (marginalized over second order slow-roll).}
\label{fig:CMBGMLFIpEQ3qEQ2}
\end{center}
\end{figure}

\subsection{Logarithmic Potential Inflation 1 (\hyperref[sec:lpi]{LPI1})}

\begin{figure}[H]
\begin{center}
\includegraphics[width=\wappfig,clip=true]{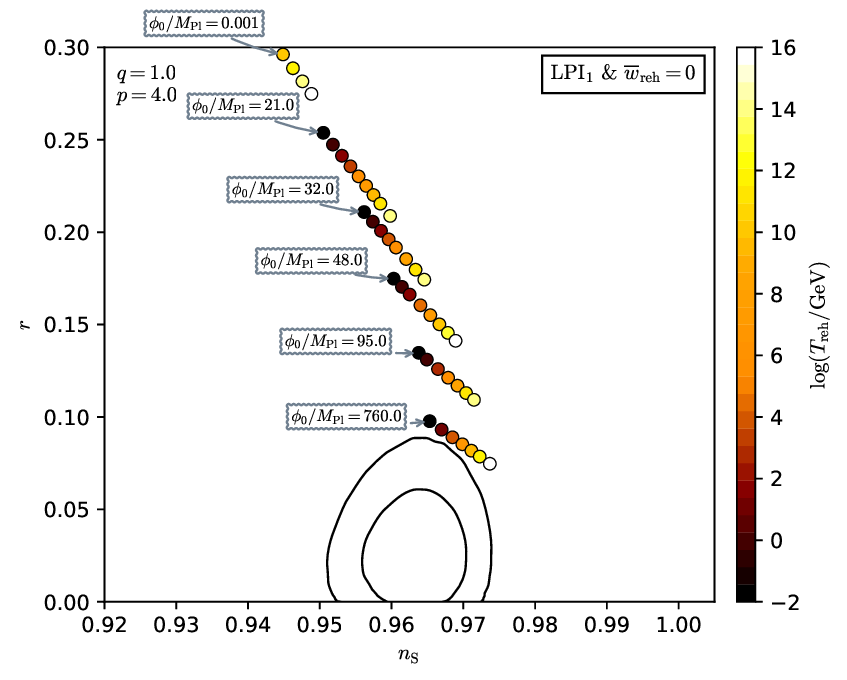}
\includegraphics[width=\wappfig,clip=true]{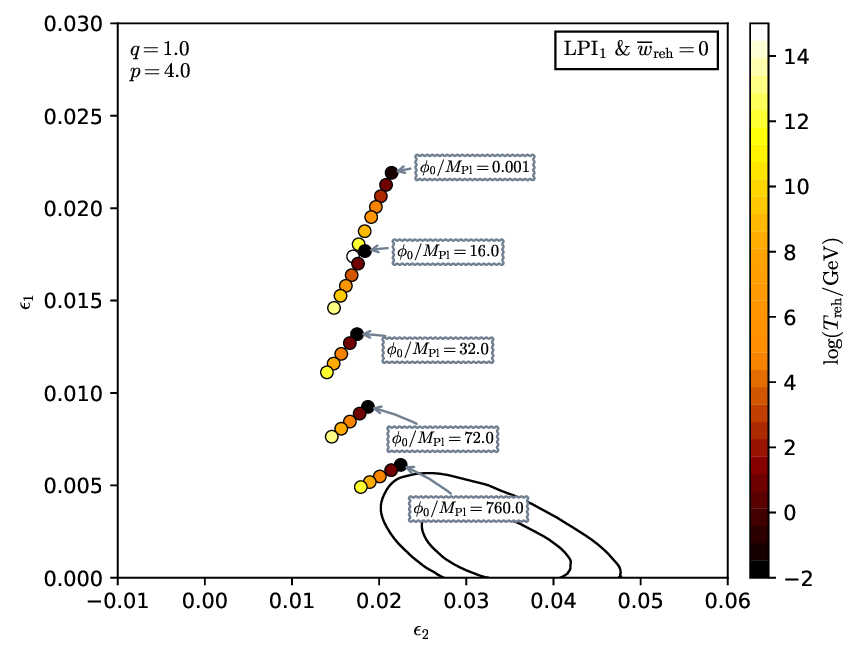}
\caption{Reheating consistent slow-roll predictions for the
  logarithmic potential inflation 1 models for $q=1$ and $p=4$ in the
  plane $(\nS,r)$ (top panel) and the plane $(\epsilon_1,\epsilon_2)$
  (bottom panel). The solid contours are the one and two-sigma {\data}
  confidence intervals (marginalized over second order slow-roll). See
  figures~\ref{fig:CMBLPI1pEQ4qEQ2} and \ref{fig:CMBLPI1pEQ4qEQ2} for
  other values of $q$.}
\label{fig:CMBLPI1pEQ4qEQ1}
\end{center}
\end{figure}

\begin{figure}[H]
\begin{center}
\includegraphics[width=\wappfig,clip=true]{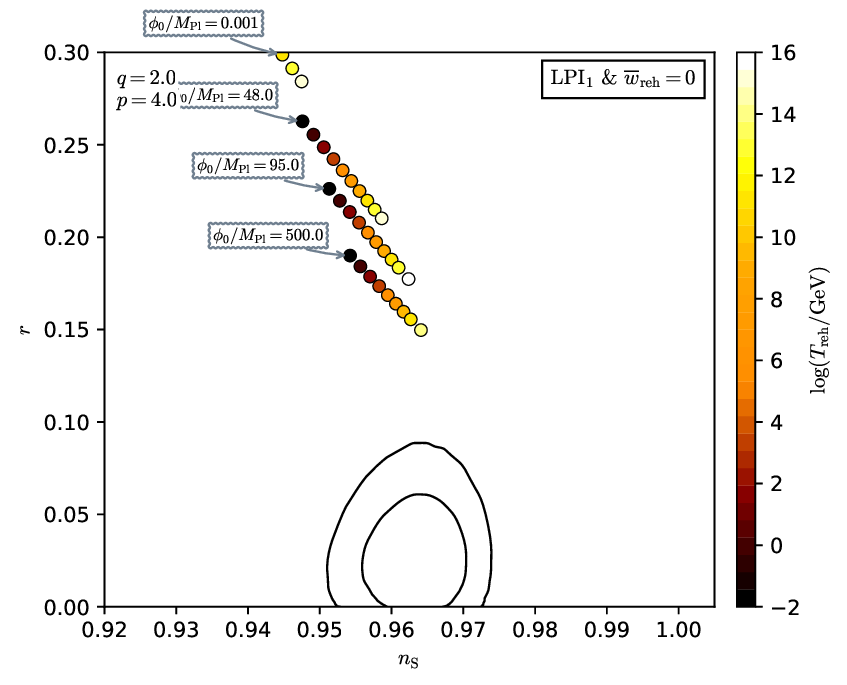}
\includegraphics[width=\wappfig,clip=true]{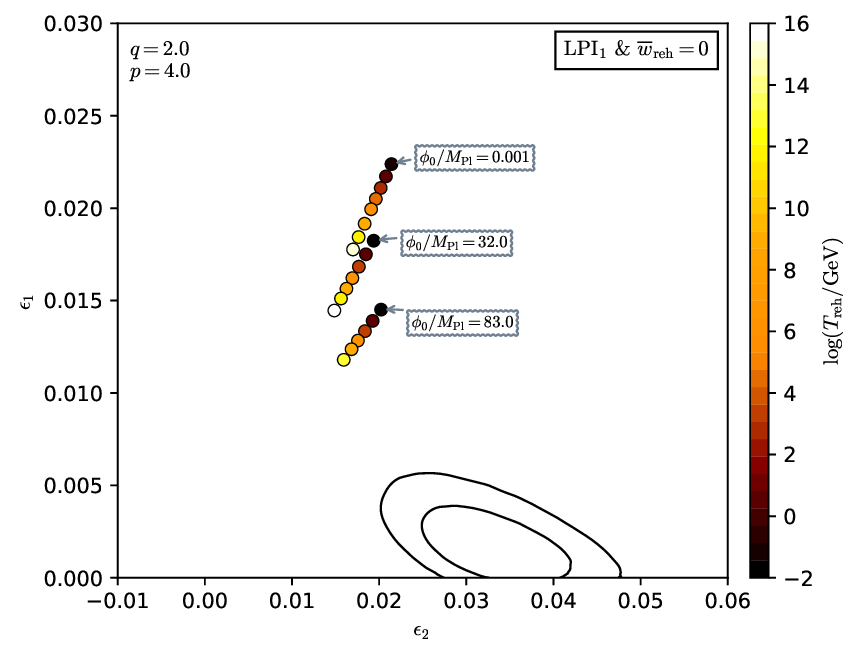}
\caption{Reheating consistent slow-roll predictions for the
  logarithmic potential inflation 1 models for $q=2$ and $p=4$ in the
  plane $(\nS,r)$ (top panel) and the plane $(\epsilon_1,\epsilon_2)$
  (bottom panel). The solid contours are the one and two-sigma {\data}
  confidence intervals (marginalized over second order slow-roll).}
\label{fig:CMBLPI1pEQ4qEQ2}
\end{center}
\end{figure}

\begin{figure}[H]
\begin{center}
\includegraphics[width=\wappfig,clip=true]{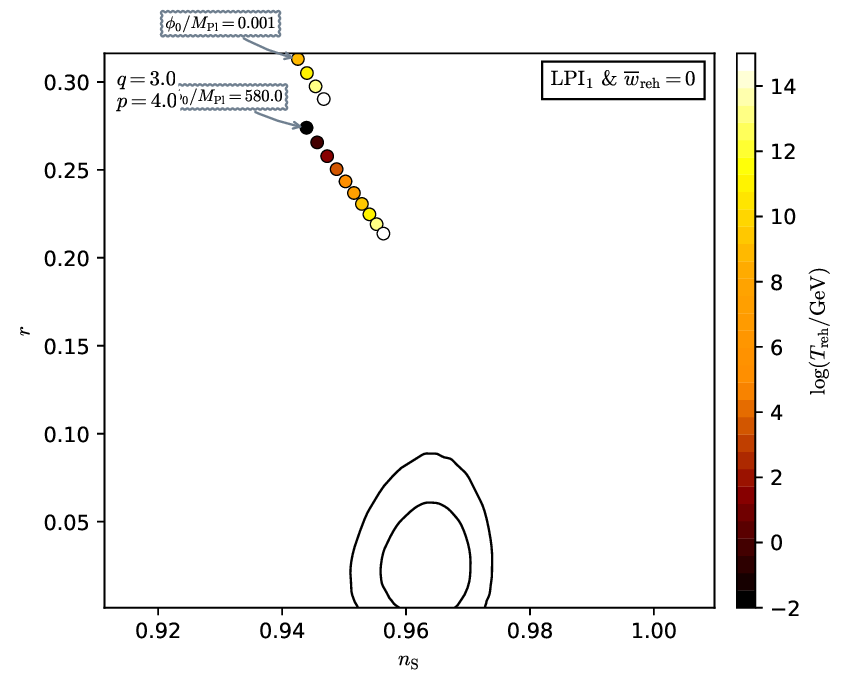}
\includegraphics[width=\wappfig,clip=true]{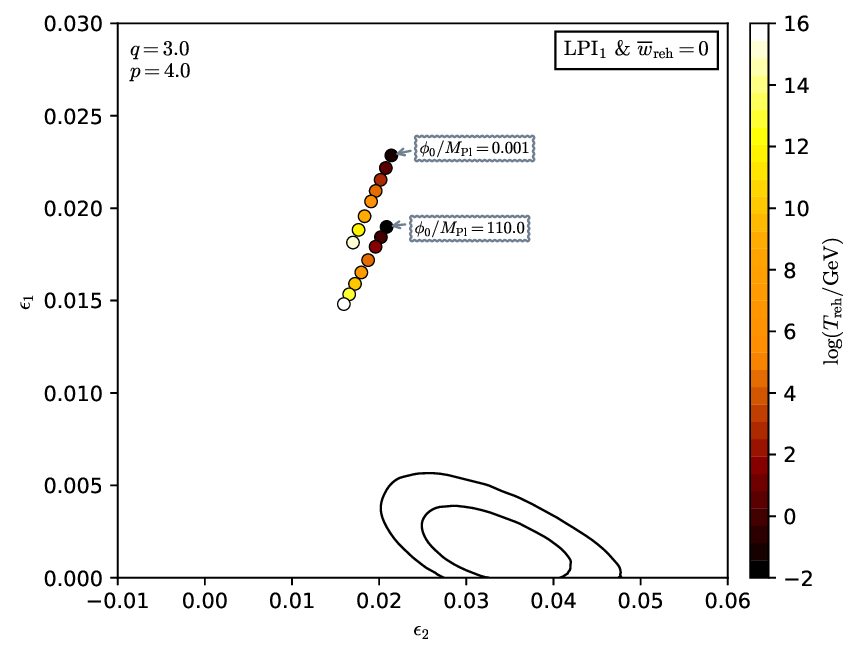}
\caption{Reheating consistent slow-roll predictions for the
  logarithmic potential inflation 1 models for $q=3$ and $p=4$ in the
  plane $(\nS,r)$ (top panel) and the plane $(\epsilon_1,\epsilon_2)$
  (bottom panel). The solid contours are the one and
  two-sigma {\data} confidence intervals (marginalized over second order
  slow-roll).}
\label{fig:CMBLPI1pEQ4qEQ3}
\end{center}
\end{figure}

\subsection{Logarithmic Potential Inflation 2 (\hyperref[sec:lpi]{LPI2})}

\begin{figure}[H]
\begin{center}
\includegraphics[width=\wappfig,clip=true]{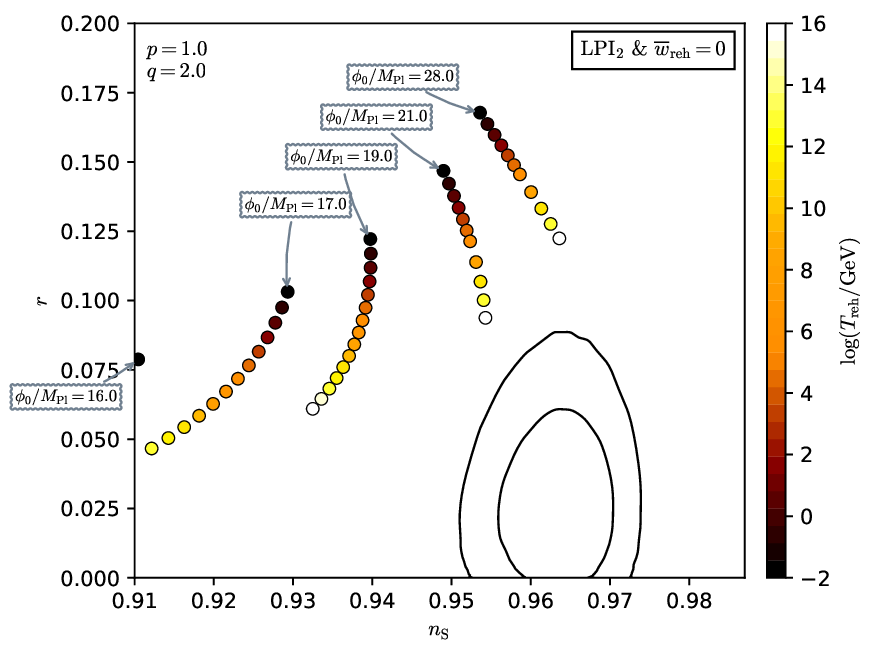}
\includegraphics[width=\wappfig,clip=true]{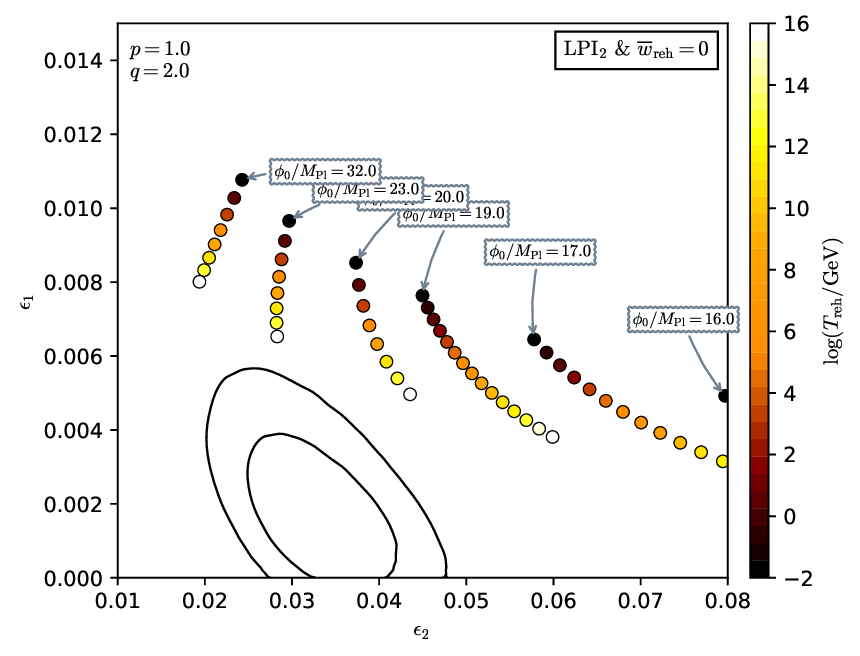}
\caption{Reheating consistent slow-roll predictions for the
  logarithmic potential inflation 2 models for $p=1$ and $q=2$ in the
  plane $(\nS,r)$ (top panel) and the plane $(\epsilon_1,\epsilon_2)$
  (bottom panel). The solid contours are the one and two-sigma {\data}
  confidence intervals (marginalized over second order slow-roll).}
\label{fig:CMBLPI2pEQ1qEQ2}
\end{center}
\end{figure}

\begin{figure}[H]
\begin{center}
\includegraphics[width=\wappfig,clip=true]{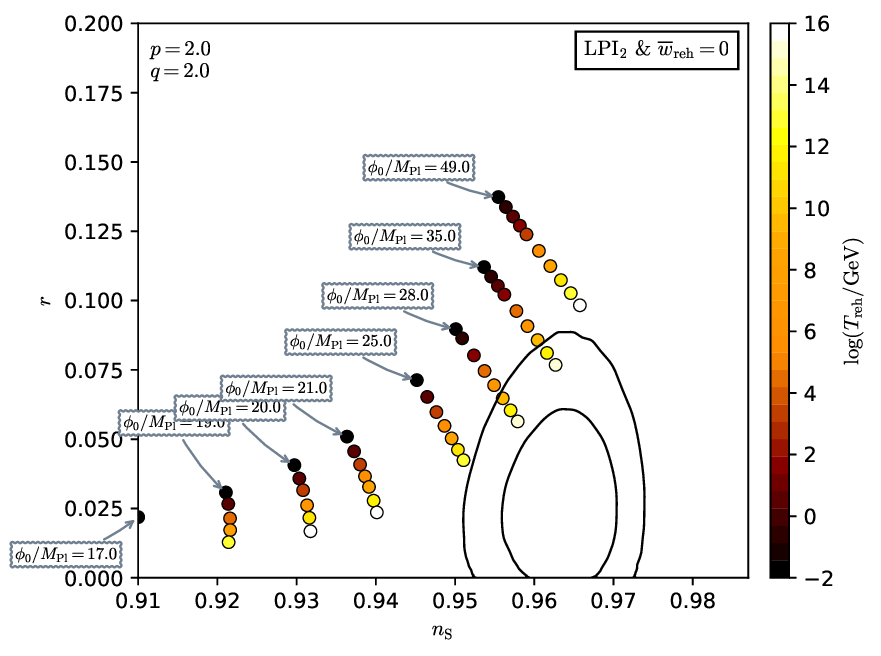}
\includegraphics[width=\wappfig,clip=true]{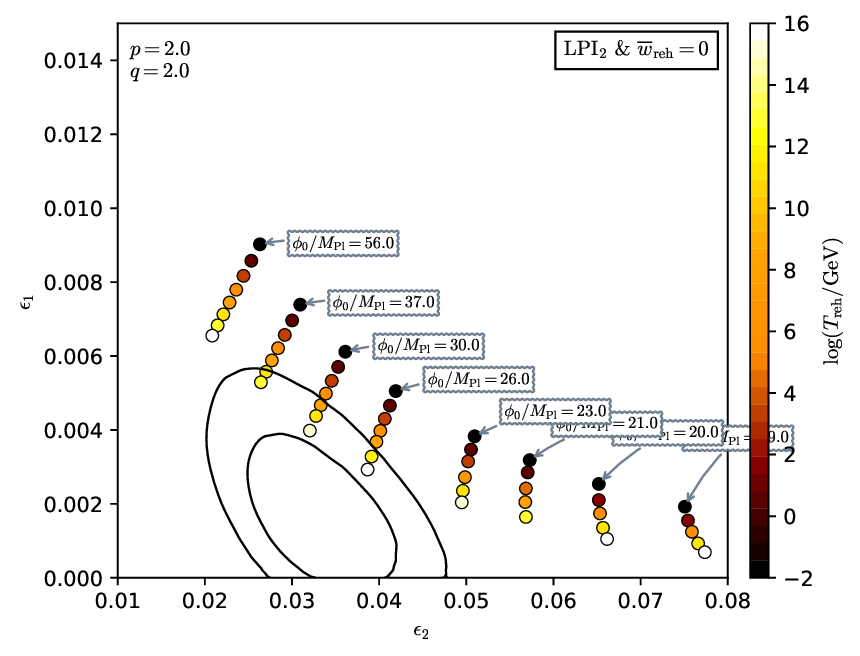}
\caption{Reheating consistent slow-roll predictions for the
  logarithmic potential inflation 2 models for $p=2$ and $q=2$ in the
  plane $(\nS,r)$ (top panel) and the plane $(\epsilon_1,\epsilon_2)$
  (bottom panel). The solid contours are the one and two-sigma {\data}
  confidence intervals (marginalized over second order slow-roll).}
\label{fig:CMBLPI2pEQ2qEQ2}
\end{center}
\end{figure}

\begin{figure}[H]
\begin{center}
\includegraphics[width=\wappfig,clip=true]{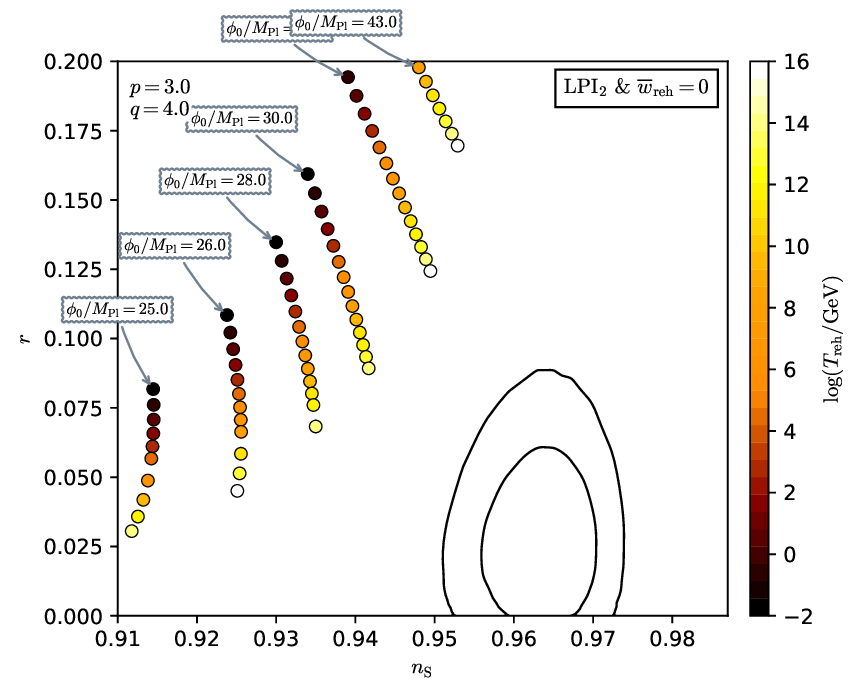}
\includegraphics[width=\wappfig,clip=true]{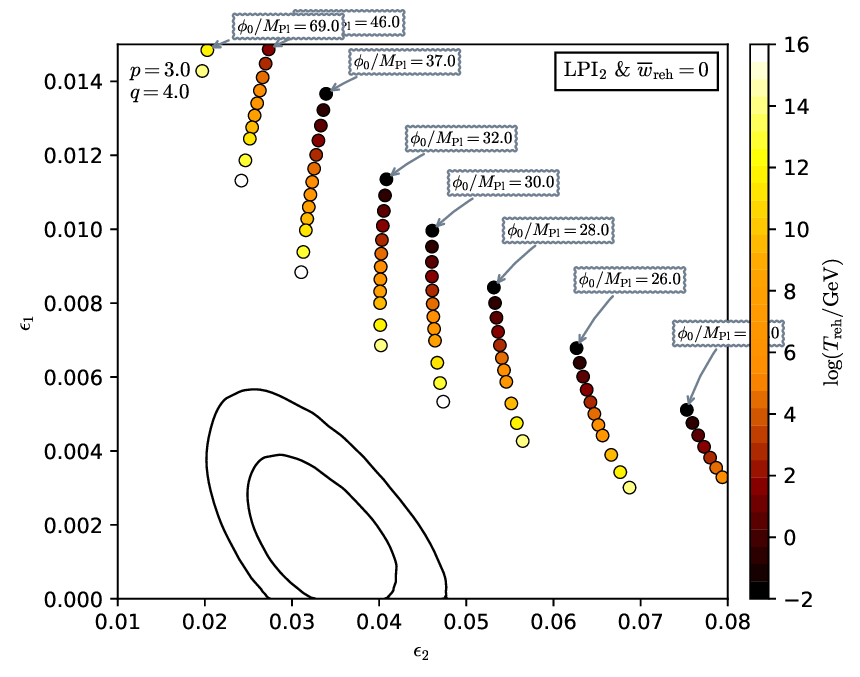}
\caption{Reheating consistent slow-roll predictions for the
  logarithmic potential inflation 2 models for $p=3$ and $q=4$ in the
  plane $(\nS,r)$ (top panel) and the plane $(\epsilon_1,\epsilon_2)$
  (bottom panel). The solid contours are the one and two-sigma {\data}
  confidence intervals (marginalized over second order slow-roll).}
\label{fig:CMBLPI2pEQ3qEQ4}
\end{center}
\end{figure}

\subsection{Logarithmic Potential Inflation 3 (\hyperref[sec:lpi]{LPI3})}

\begin{figure}[H]
\begin{center}
\includegraphics[width=\wappfig,clip=true]{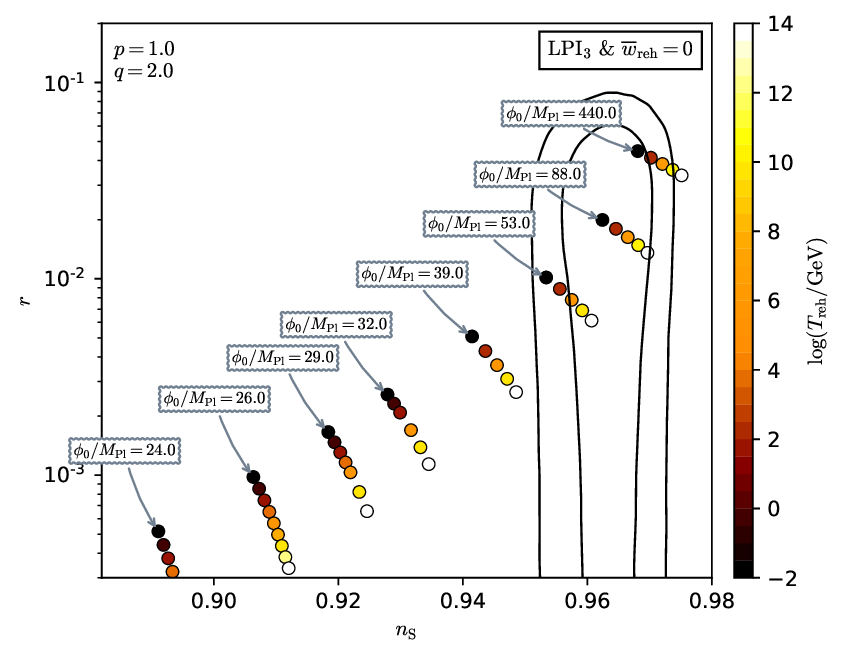}
\includegraphics[width=\wappfig,clip=true]{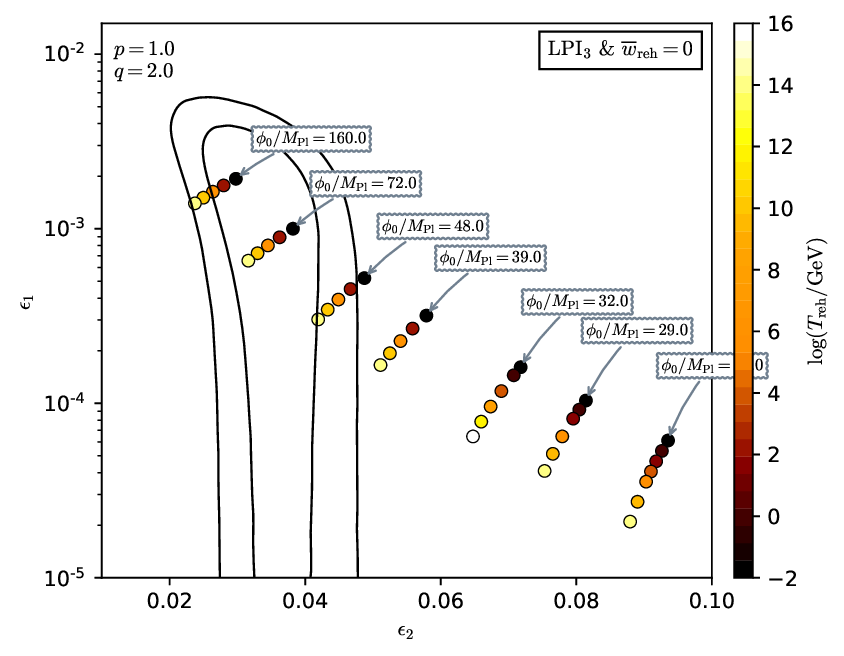}
\caption{Reheating consistent slow-roll predictions for the
  logarithmic potential inflation 3 models for $p=1$ and $q=2$ in the
  plane $(\nS,r)$ (top panel) and the plane $(\epsilon_1,\epsilon_2)$
  (bottom panel). The solid contours are the one and two-sigma {\data}
  confidence intervals (marginalized over second order slow-roll).}
\label{fig:CMBLPI3pEQ1qEQ2}
\end{center}
\end{figure}

\begin{figure}[H]
\begin{center}
\includegraphics[width=\wappfig,clip=true]{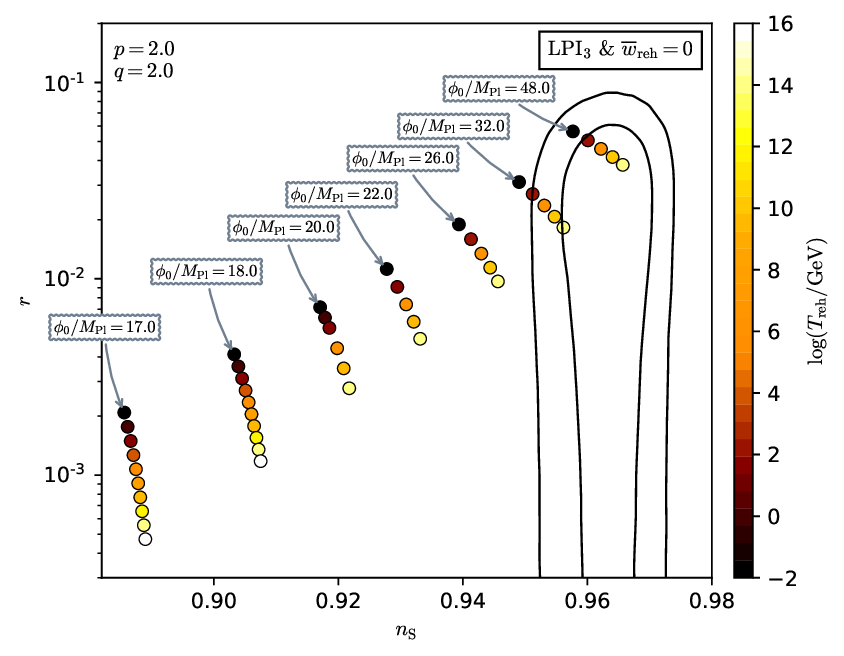}
\includegraphics[width=\wappfig,clip=true]{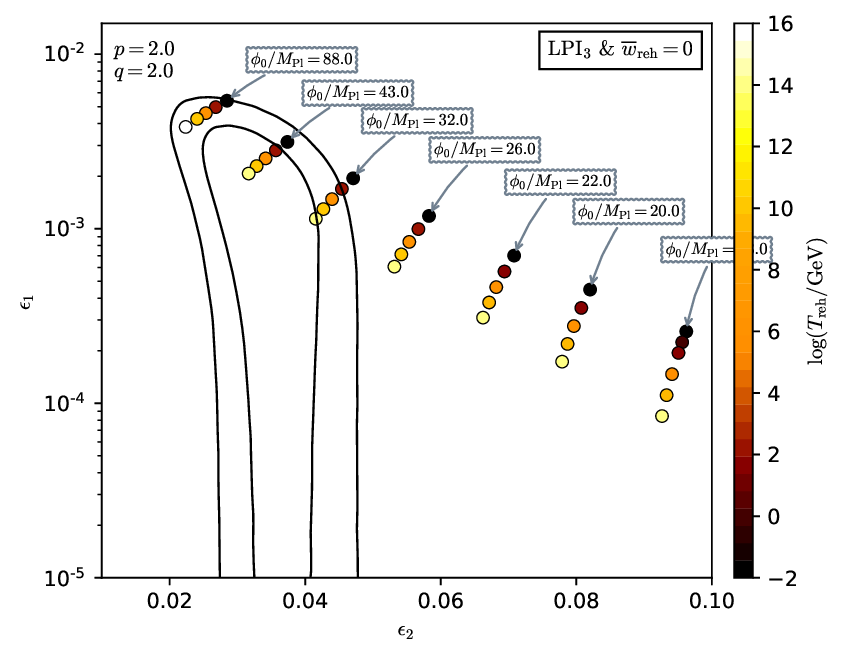}
\caption{Reheating consistent slow-roll predictions for the
  logarithmic potential inflation 3 models for $p=2$ and $q=2$ in the
  plane $(\nS,r)$ (top panel) and the plane $(\epsilon_1,\epsilon_2)$
  (bottom panel). The solid contours are the one and two-sigma {\data}
  confidence intervals (marginalized over second order slow-roll).}
\label{fig:CMBLPI3pEQ2qEQ2}
\end{center}
\end{figure}

\begin{figure}[H]
\begin{center}
\includegraphics[width=\wappfig,clip=true]{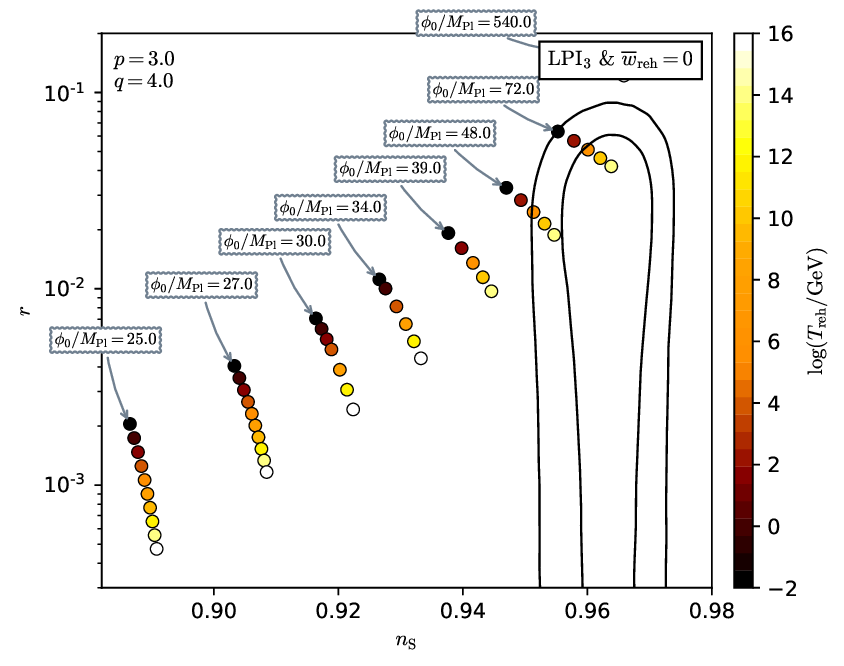}
\includegraphics[width=\wappfig,clip=true]{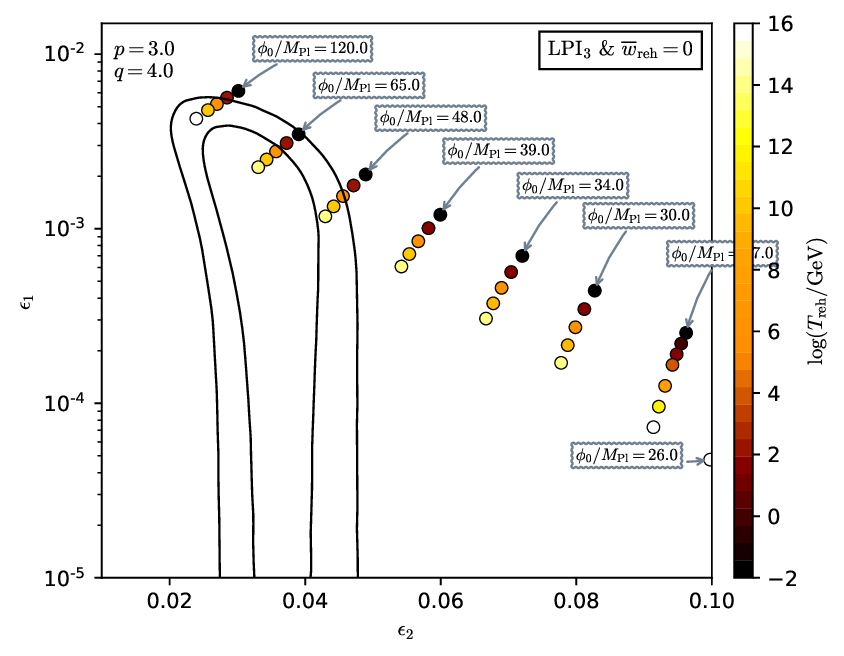}
\caption{Reheating consistent slow-roll predictions for the
  logarithmic potential inflation 3 models for $p=3$ and $q=4$ in the
  plane $(\nS,r)$ (top panel) and the plane $(\epsilon_1,\epsilon_2)$
  (bottom panel). The solid contours are the one and two-sigma {\data}
  confidence intervals (marginalized over second order slow-roll).}
\label{fig:CMBLPI3pEQ3qEQ4}
\end{center}
\end{figure}

\subsection{Constant \texorpdfstring{$\nS$}{nS} D Inflation (\hyperref[sec:cndi]{CNDI})}

\begin{figure}[H]
\begin{center}
\includegraphics[width=\wappfig,clip=true]{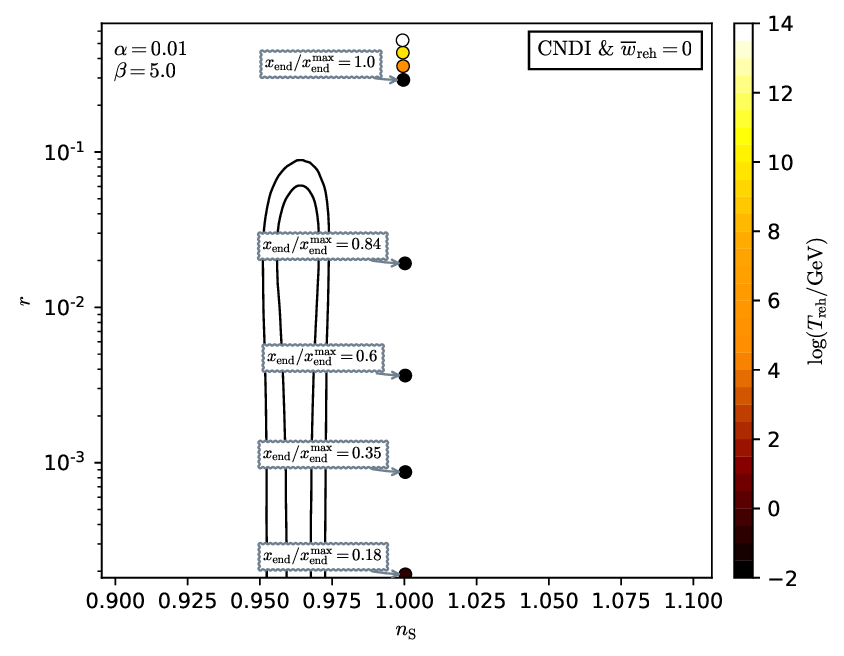}
\includegraphics[width=\wappfig,clip=true]{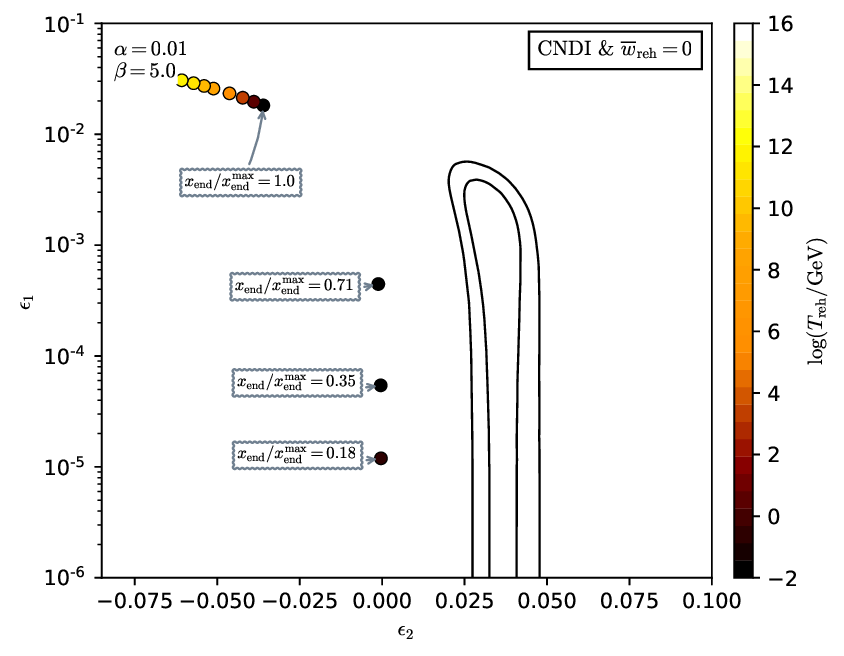}
\caption{Reheating consistent slow-roll predictions for the constant
  $\nS$ D inflation models for $\alpha=10^{-2}$ and $\beta=5$, in the
  plane $(\nS,r)$ (top panel) and the plane $(\epsilon_1,\epsilon_2)$
  (bottom panel). The solid contours are the one and two-sigma {\data}
  confidence intervals (marginalized over second order slow-roll). The
  model predictions match well the constant spectral index value
  $\nS=1+4\alpha^2\beta/\left(\beta+1\right)$, see also figure~\ref{fig:CMBCNDIbetaEQ5_1}.}
\label{fig:CMBCNDIbetaEQ5}
\end{center}
\end{figure}

\begin{figure}[H]
\begin{center}
\includegraphics[width=\wappfig,clip=true]{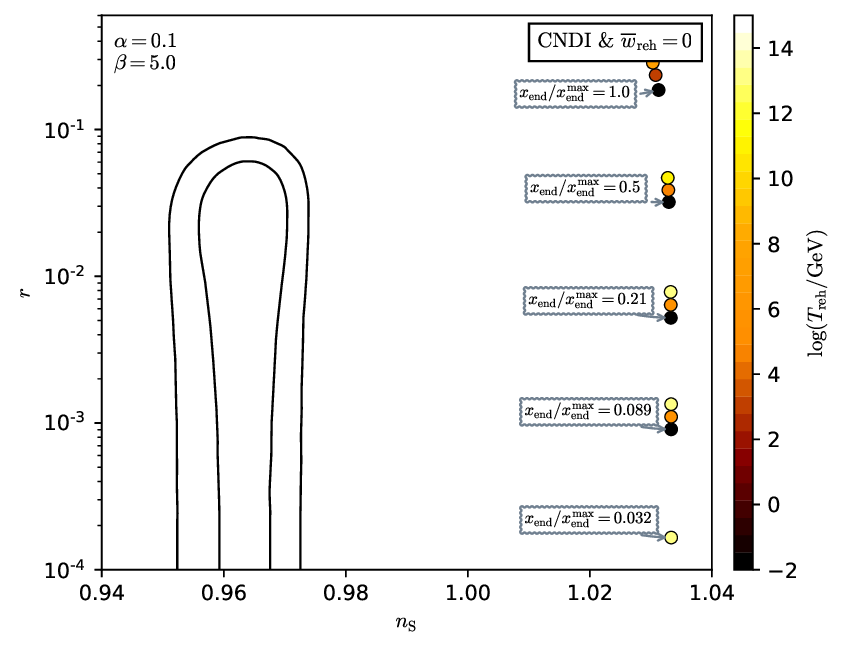}
\includegraphics[width=\wappfig,clip=true]{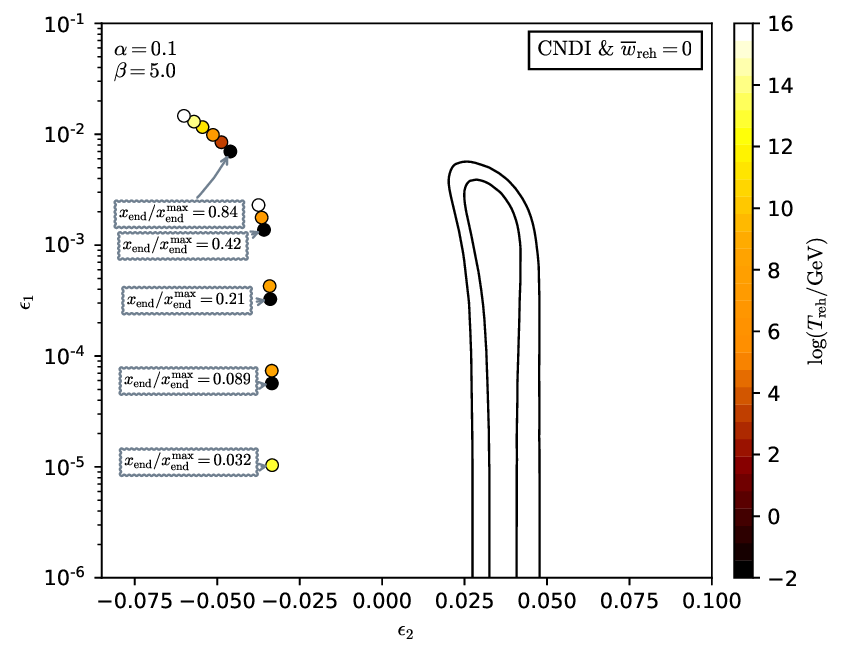}
\caption{Reheating consistent slow-roll predictions for the constant
  $\nS$ D inflation models for $\alpha=10^{-1}$ and $\beta=5$, in the
  plane $(\nS,r)$ (top panel) and the plane $(\epsilon_1,\epsilon_2)$
  (bottom panel). The solid contours are the one and two-sigma {\data}
  confidence intervals (marginalized over second order slow-roll). The
  model predictions match well the constant spectral index value
  $\nS=1+4\alpha^2\beta/\left(\beta+1\right)$, see also
  figure~\ref{fig:CMBCNDIbetaEQ5}.}
\label{fig:CMBCNDIbetaEQ5_1}
\end{center}
\end{figure}

\begin{figure}[H]
\begin{center}
\includegraphics[width=\wappfig,clip=true]{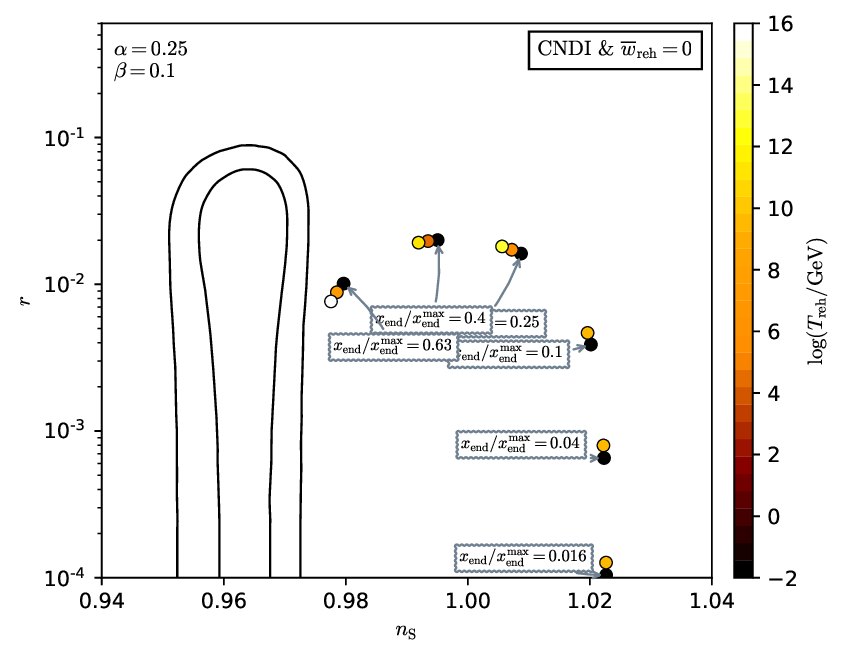}
\includegraphics[width=\wappfig,clip=true]{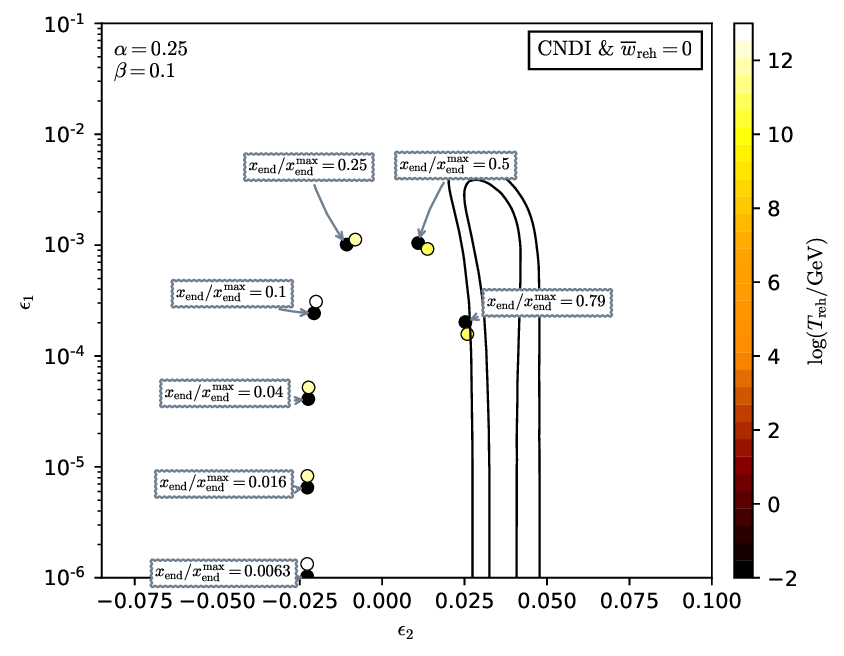}
\caption{Reheating consistent slow-roll predictions for the constant
  $\nS$ D inflation models for $\alpha=0.25$ and $\beta=0.1$ in the
  plane $(\nS,r)$ (top panel) and the plane $(\epsilon_1,\epsilon_2)$
  (bottom panel). The solid contours are the one and two-sigma {\data}
  confidence intervals (marginalized over second order slow-roll). At
  large values of $\xend$, the model predictions deviate significantly
  from $\nS=1+4\alpha^2\beta/\left(\beta+1\right)$. See
  figures~\ref{fig:CMBCNDIbetaEQ0dot1_3} to
  \ref{fig:CMBCNDIbetaEQ0dot1_5} for other values of $\alpha$.}
\label{fig:CMBCNDIbetaEQ0dot1}
\end{center}
\end{figure}

\begin{figure}[H]
\begin{center}
\includegraphics[width=\wappfig,clip=true]{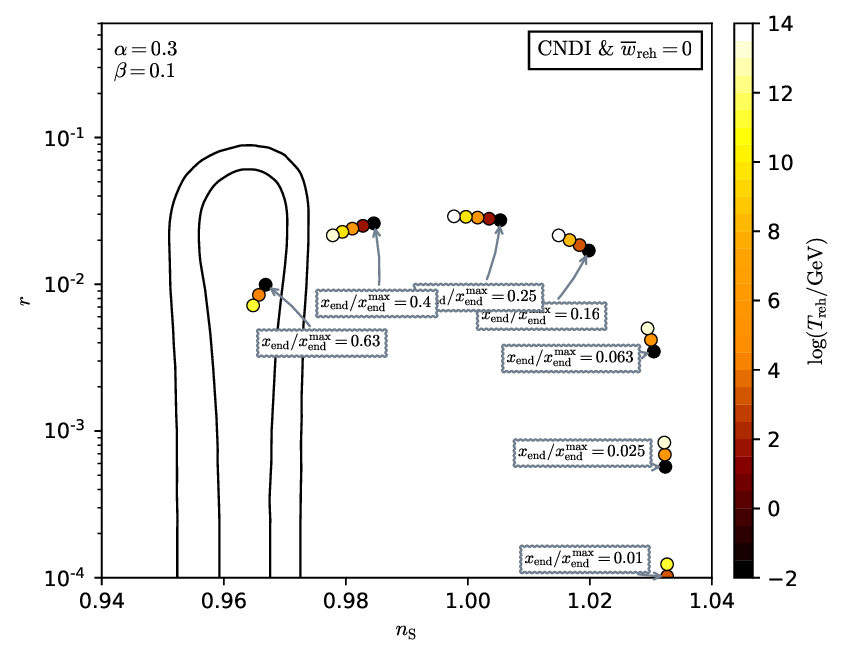}
\includegraphics[width=\wappfig,clip=true]{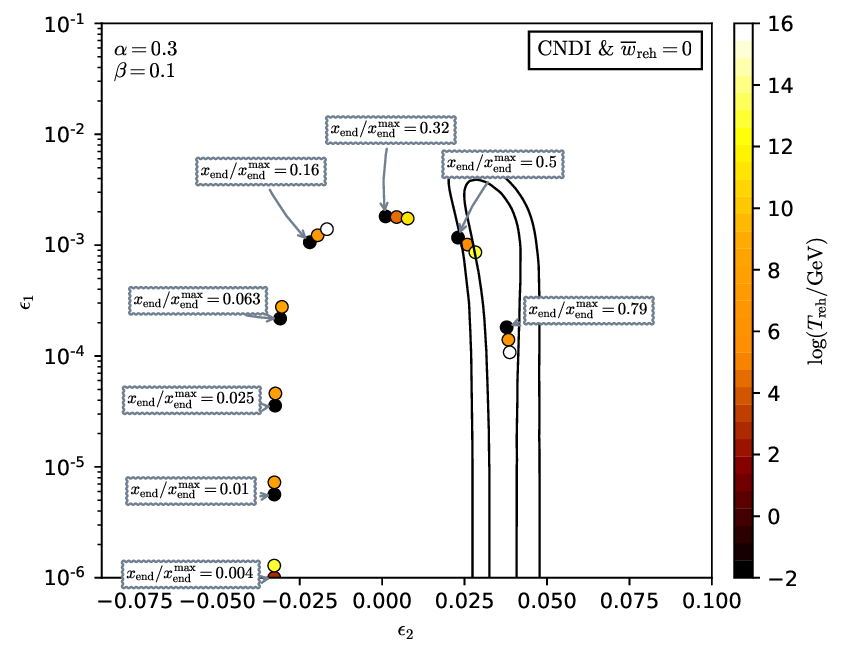}
\caption{Reheating consistent slow-roll predictions for the constant
  $\nS$ D inflation models for $\alpha=0.30$ and $\beta=0.1$ in the plane $(\nS,r)$ (top
  panel) and the plane $(\epsilon_1,\epsilon_2)$ (bottom panel). The
  solid contours are the one and two-sigma {\data} confidence
  intervals (marginalized over second order slow-roll). At large
  values of $\xend$, the model predictions deviate significantly from
  $\nS=1+4\alpha^2\beta/\left(\beta+1\right)$.}
\label{fig:CMBCNDIbetaEQ0dot1_3}
\end{center}
\end{figure}

\begin{figure}[H]
\begin{center}
\includegraphics[width=\wappfig,clip=true]{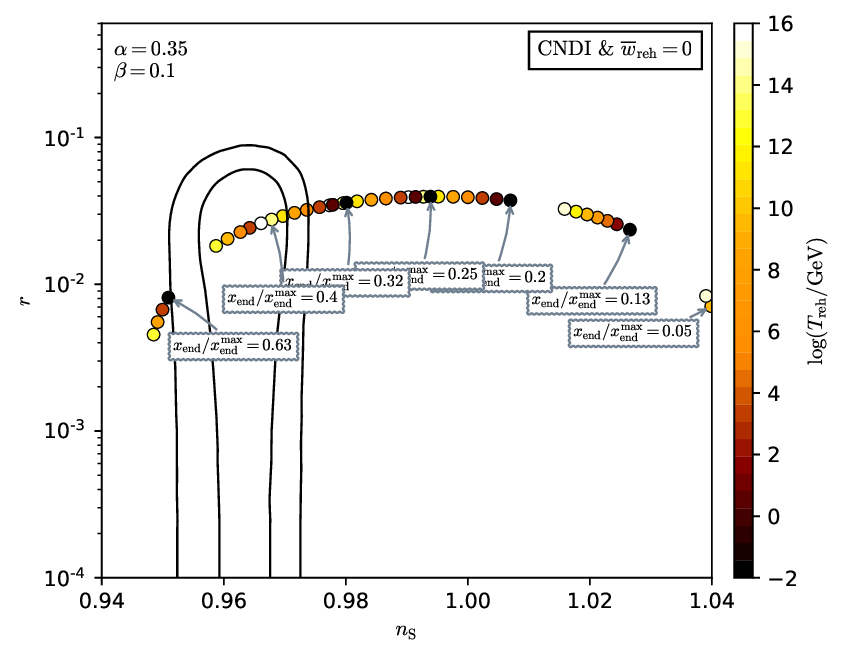}
\includegraphics[width=\wappfig,clip=true]{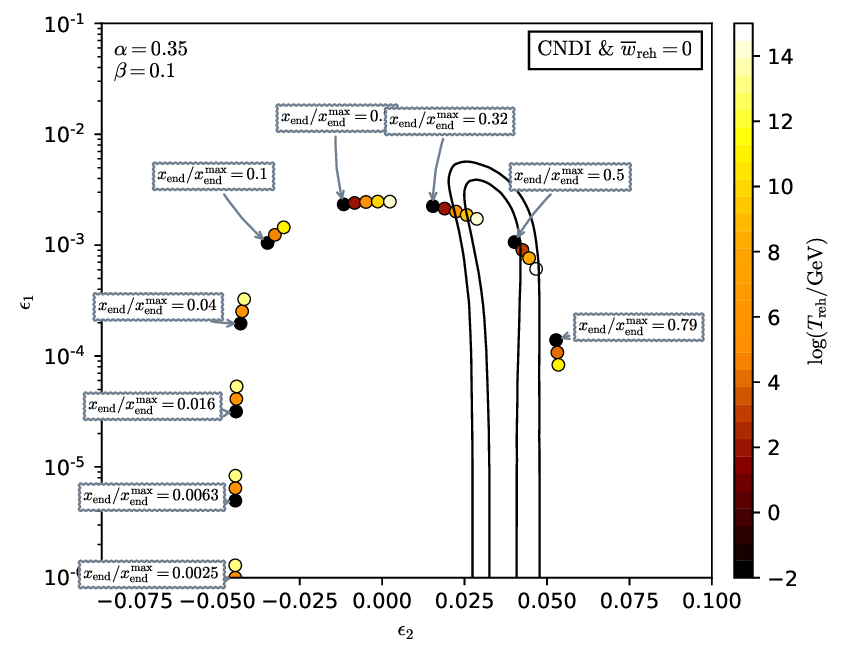}
\caption{Reheating consistent slow-roll predictions for the constant
  $\nS$ D inflation models for $\alpha=0.35$ and $\beta=0.1$ in the plane $(\nS,r)$ (top
  panel) and the plane $(\epsilon_1,\epsilon_2)$ (bottom panel). The
  solid contours are the one and two-sigma {\data} confidence
  intervals (marginalized over second order slow-roll). At large
  values of $\xend$, the model predictions deviate significantly from
  $\nS=1+4\alpha^2\beta/\left(\beta+1\right)$.}
\label{fig:CMBCNDIbetaEQ0dot1_4}
\end{center}
\end{figure}

\begin{figure}[H]
\begin{center}
\includegraphics[width=\wappfig,clip=true]{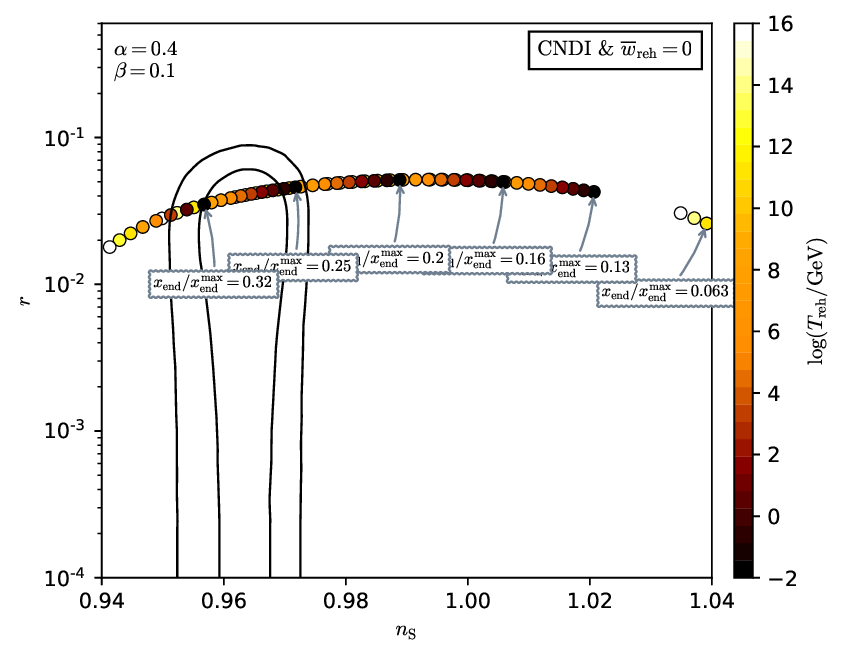}
\includegraphics[width=\wappfig,clip=true]{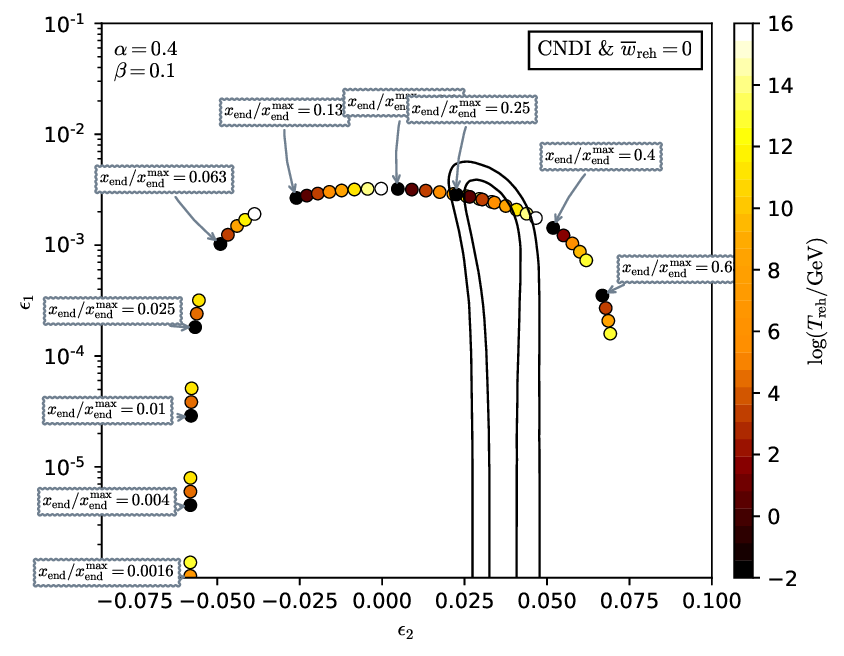}
\caption{Reheating consistent slow-roll predictions for the constant
  $\nS$ D inflation models for $\alpha=0.40$ and $\beta=0.1$ in the plane $(\nS,r)$ (top
  panel) and the plane $(\epsilon_1,\epsilon_2)$ (bottom panel). The
  solid contours are the one and two-sigma {\data} confidence
  intervals (marginalized over second order slow-roll). At large
  values of $\xend$, the model predictions deviate significantly from
  $\nS=1+4\alpha^2\beta/\left(\beta+1\right)$.}
\label{fig:CMBCNDIbetaEQ0dot1_5}
\end{center}
\end{figure}

\subsection{String Axion Inflation II 1 (\hyperref[sec:saiii]{SAIII1})}

\begin{figure}[H]
\begin{center}
\includegraphics[width=\wappfig,clip=true]{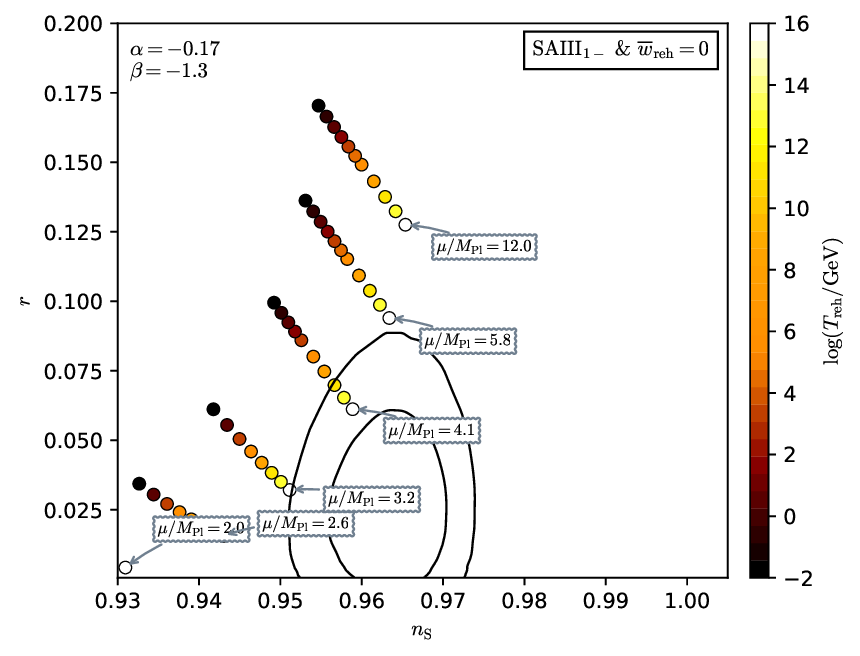}
\includegraphics[width=\wappfig,clip=true]{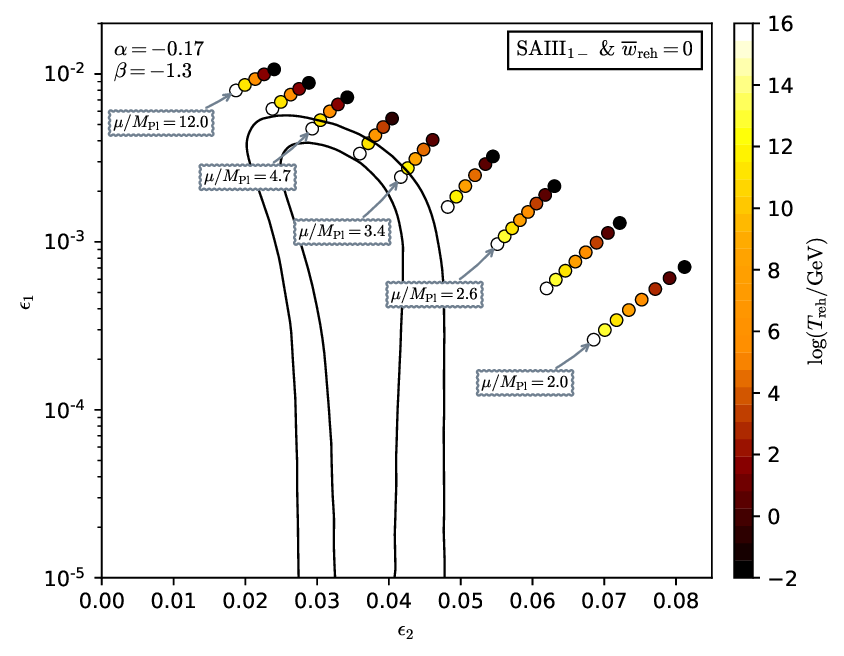}
\caption{Reheating consistent slow-roll predictions for the String
  Axion Inflation II models, in the SAIII1 regime and at a negative
  value of $\beta=-1.3$. Predictions are represented in the plane
  $(\nS,r)$ (top panel) and in the plane $(\epsilon_1,\epsilon_2)$
  (bottom panel). The solid contours are the one and two-sigma {\data}
  confidence intervals (marginalized over second order slow-roll), see
  also \Fig{fig:CMBSAIII1m_1}.}
\label{fig:CMBSAIII1m}
\end{center}
\end{figure}

\begin{figure}[H]
\begin{center}
\includegraphics[width=\wappfig,clip=true]{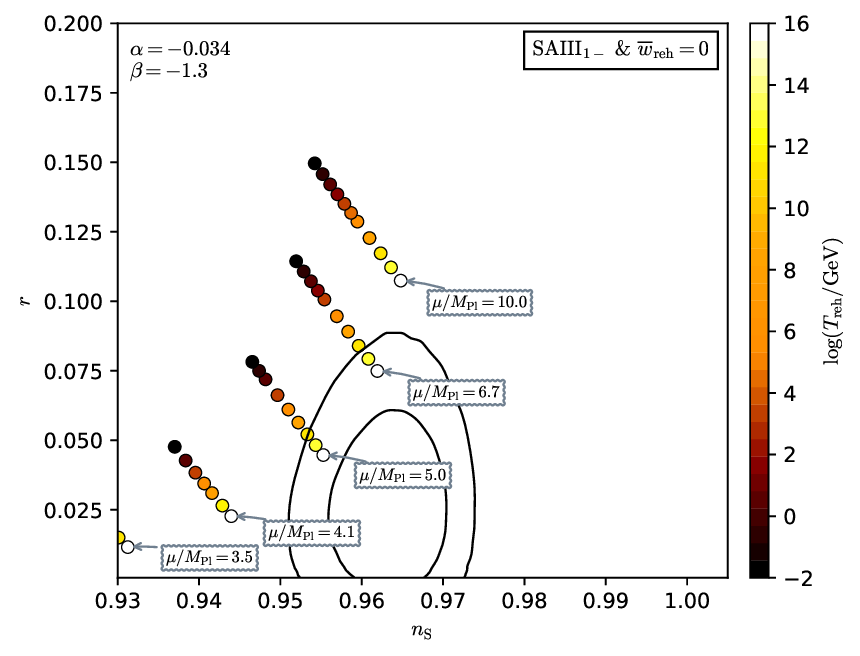}
\includegraphics[width=\wappfig,clip=true]{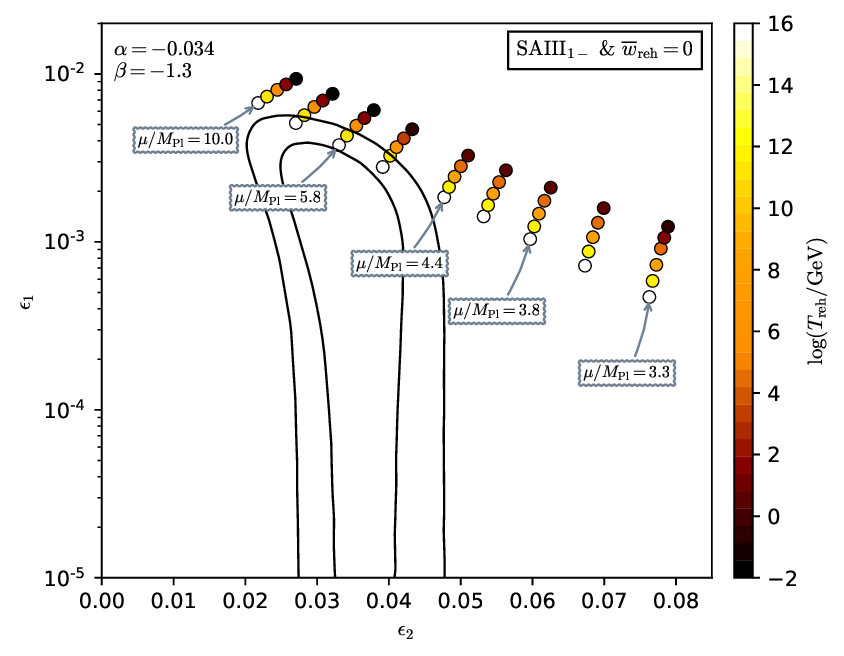}
\caption{Reheating consistent slow-roll predictions for the String
  Axion Inflation II models, in the SAIII1 regime and at a negative
  value of $\beta=-1.3$. Predictions are represented in the plane
  $(\nS,r)$ (top panel) and in the plane $(\epsilon_1,\epsilon_2)$
  (bottom panel). The solid contours are the one and two-sigma {\data}
  confidence intervals (marginalized over second order slow-roll).}
\label{fig:CMBSAIII1m_1}
\end{center}
\end{figure}

\begin{figure}[H]
\begin{center}
\includegraphics[width=\wappfig,clip=true]{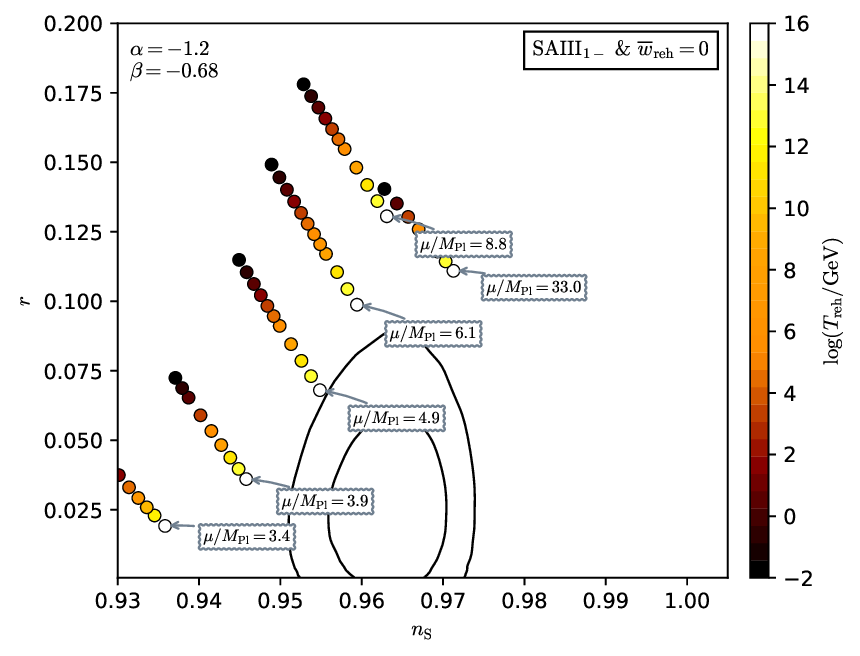}
\includegraphics[width=\wappfig,clip=true]{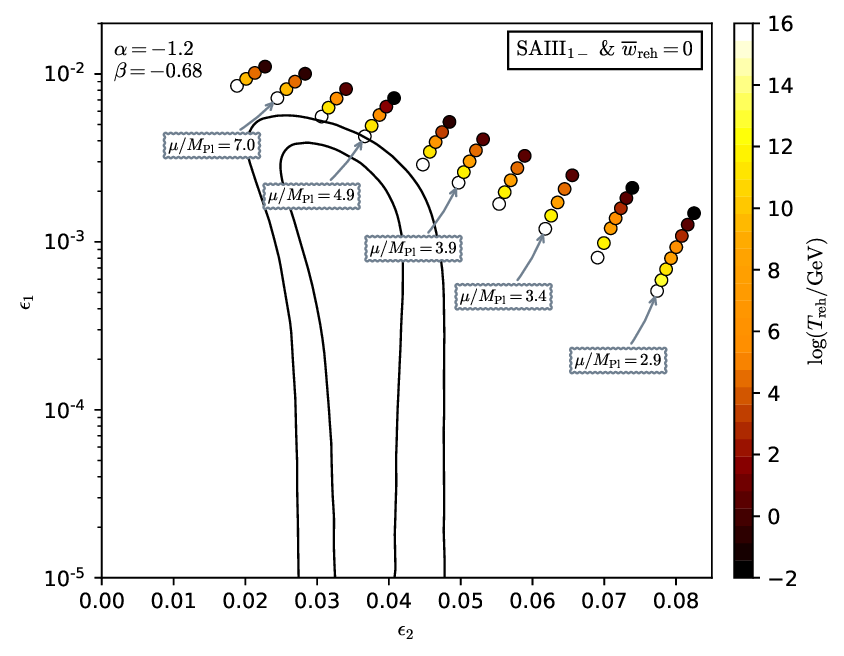}
\caption{Reheating consistent slow-roll predictions for the String
  Axion Inflation II models, in the SAIII1 regime and at a negative
  value of $\beta=-0.68$. Predictions are represented in the plane
  $(\nS,r)$ (top panel) and in the plane $(\epsilon_1,\epsilon_2)$
  (bottom panel). The solid contours are the one and two-sigma {\data}
  confidence intervals (marginalized over second order slow-roll), see
also \Fig{fig:CMBSAIII1m_3}.}
\label{fig:CMBSAIII1m_2}
\end{center}
\end{figure}

\begin{figure}[H]
\begin{center}
\includegraphics[width=\wappfig,clip=true]{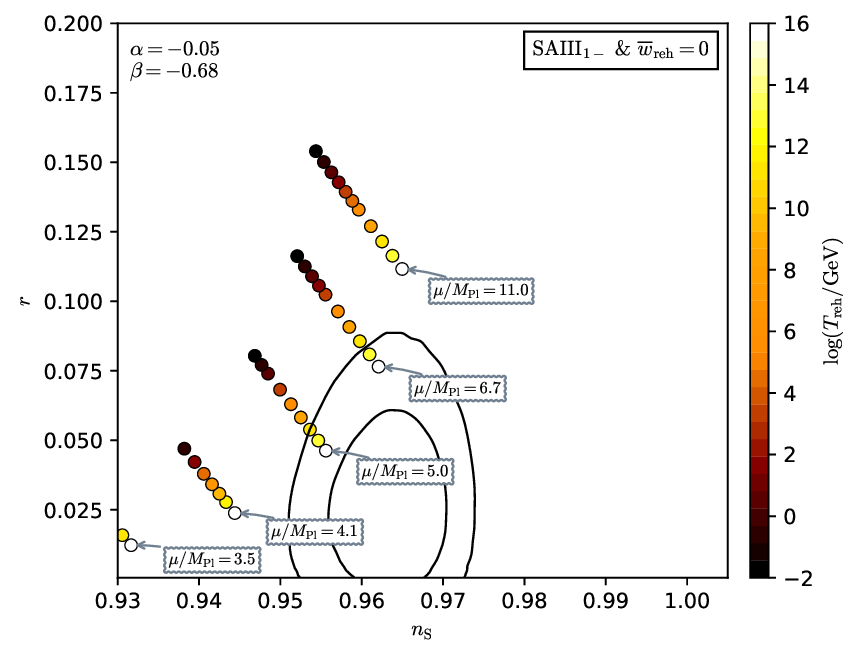}
\includegraphics[width=\wappfig,clip=true]{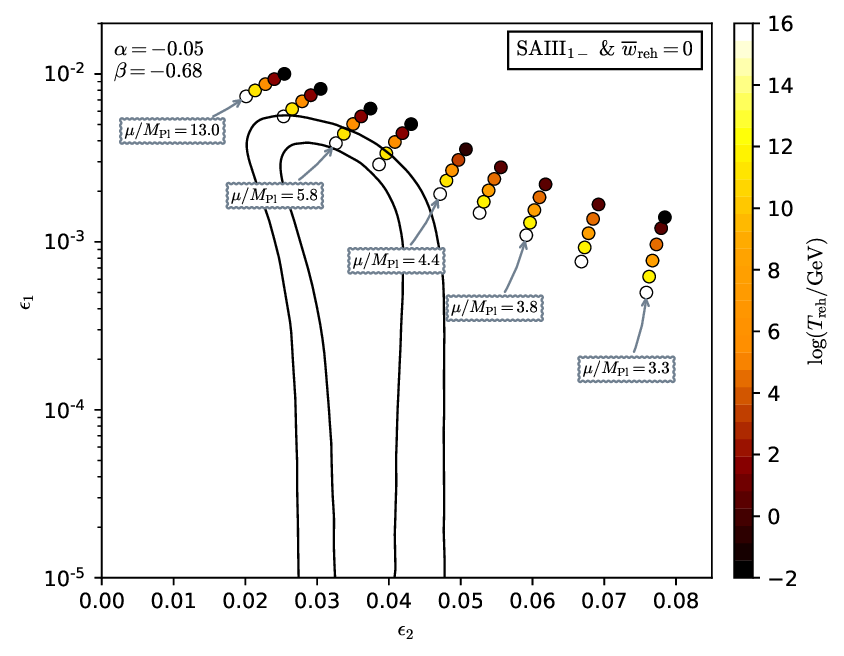}
\caption{Reheating consistent slow-roll predictions for the String
  Axion Inflation II models, in the SAIII1 regime and at a negative
  value of $\beta=-0.68$. Predictions are represented in the plane
  $(\nS,r)$ (top panel) and in the plane $(\epsilon_1,\epsilon_2)$
  (bottom panel). The solid contours are the one and two-sigma {\data}
  confidence intervals (marginalized over second order slow-roll).}
\label{fig:CMBSAIII1m_3}
\end{center}
\end{figure}

\begin{figure}[H]
\begin{center}
\includegraphics[width=\wappfig,clip=true]{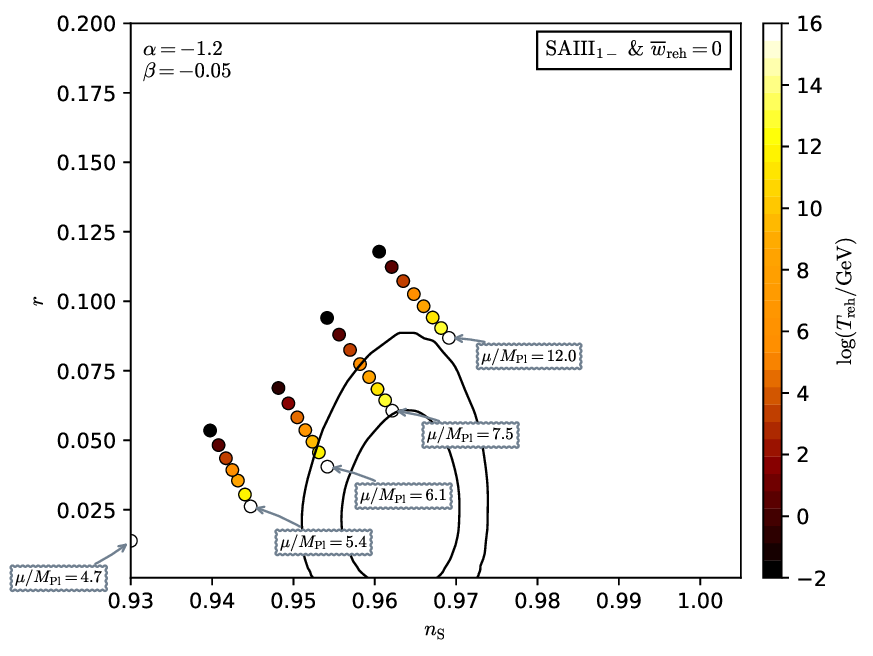}
\includegraphics[width=\wappfig,clip=true]{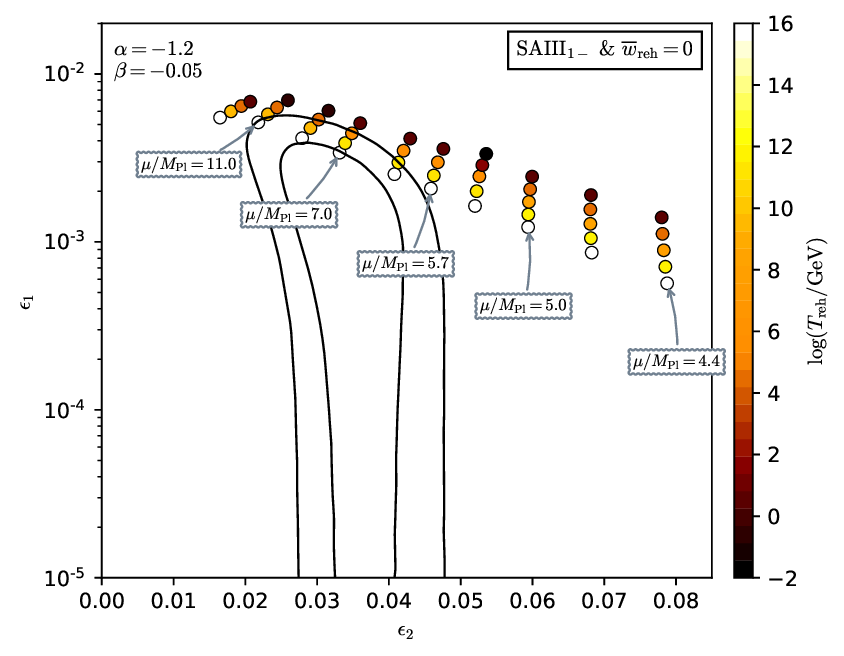}
\caption{Reheating consistent slow-roll predictions for the String
  Axion Inflation II models, in the SAIII1 regime and at a small negative
  value of $\beta=-0.05$. Predictions are represented in the plane
  $(\nS,r)$ (top panel) and in the plane $(\epsilon_1,\epsilon_2)$
  (bottom panel). The solid contours are the one and two-sigma {\data}
  confidence intervals (marginalized over second order slow-roll), see
  also \Fig{fig:CMBSAIII1m_5}.}
\label{fig:CMBSAIII1m_4}
\end{center}
\end{figure}

\begin{figure}[H]
\begin{center}
\includegraphics[width=\wappfig,clip=true]{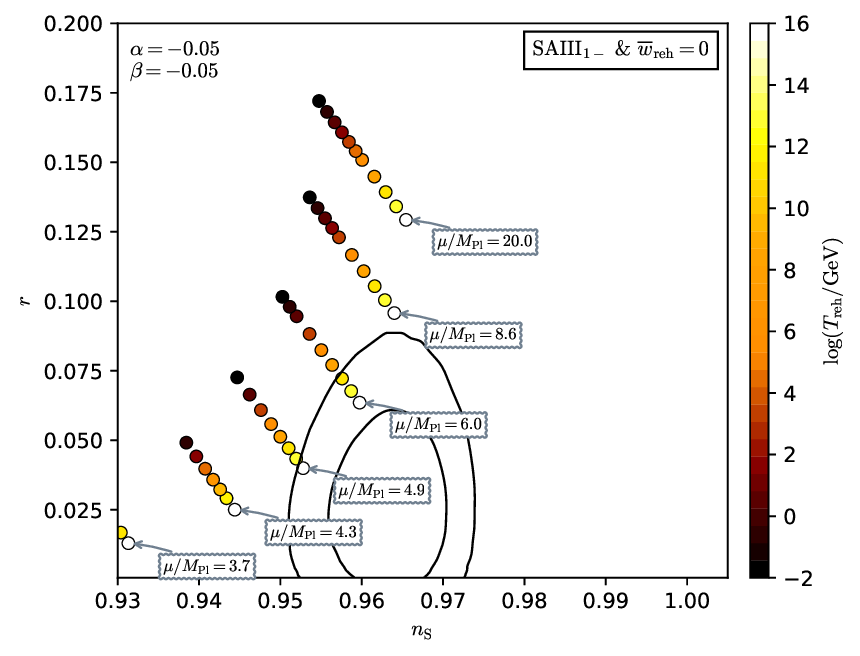}
\includegraphics[width=\wappfig,clip=true]{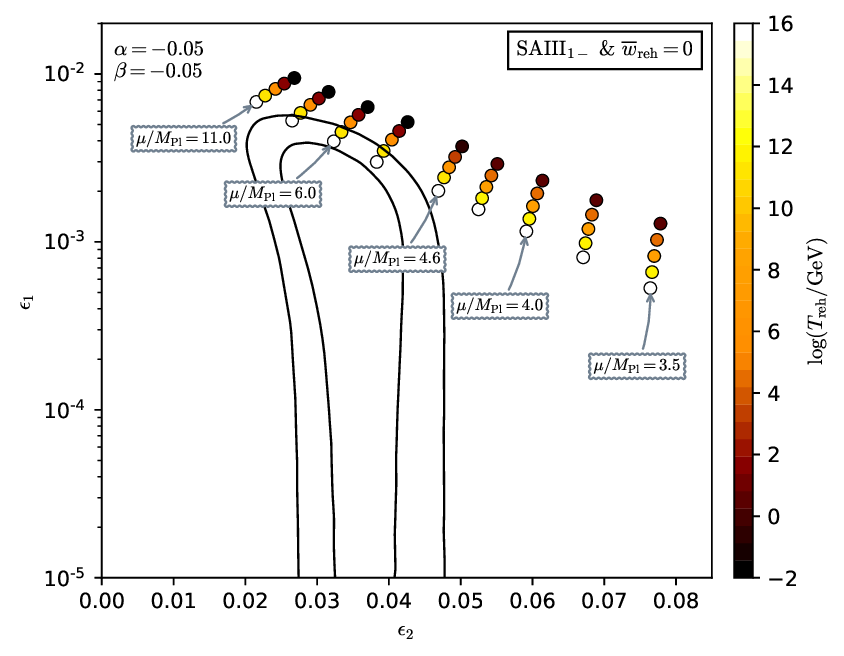}
\caption{Reheating consistent slow-roll predictions for the String
  Axion Inflation II models, in the SAIII1 regime and at a small
  negative value of $\beta=-0.05$. Predictions are represented in the
  plane $(\nS,r)$ (top panel) and in the plane
  $(\epsilon_1,\epsilon_2)$ (bottom panel). The solid contours are the
  one and two-sigma {\data} confidence intervals (marginalized over
  second order slow-roll).}
\label{fig:CMBSAIII1m_5}
\end{center}
\end{figure}

\begin{figure}[H]
\begin{center}
\includegraphics[width=\wappfig,clip=true]{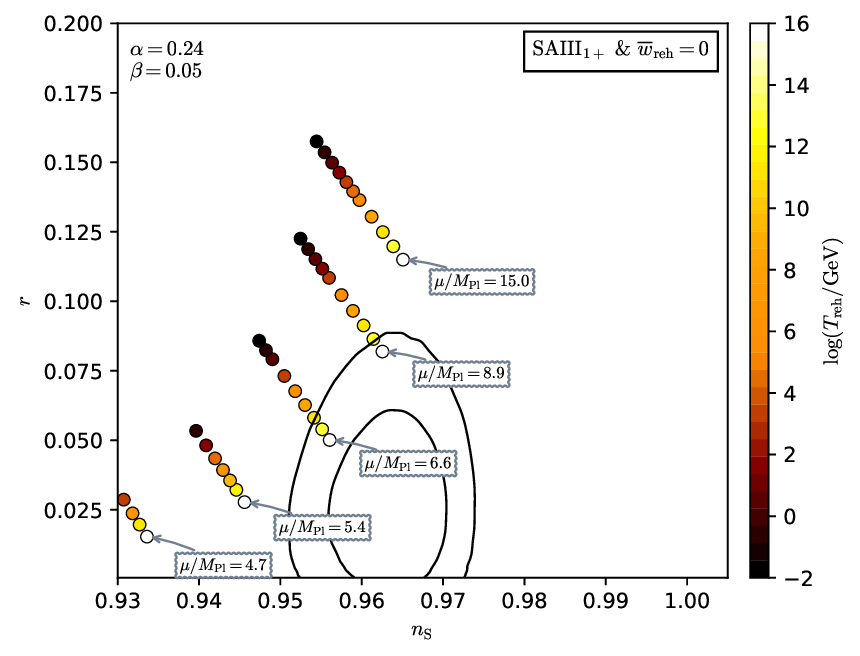}
\includegraphics[width=\wappfig,clip=true]{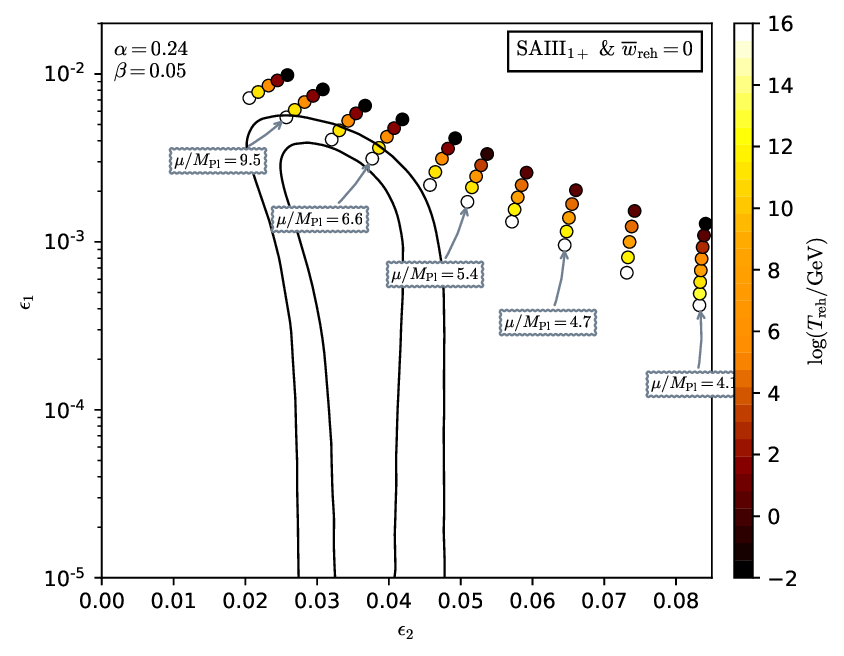}
\caption{Reheating consistent slow-roll predictions for the String
  Axion Inflation II models, in the SAIII1 regime and at a small positive
  value of $\beta=0.05$. Predictions are represented in the plane
  $(\nS,r)$ (top panel) and in the plane $(\epsilon_1,\epsilon_2)$
  (bottom panel). The solid contours are the one and two-sigma {\data}
  confidence intervals (marginalized over second order slow-roll), see
  also \Fig{fig:CMBSAIII1p_1}.}
\label{fig:CMBSAIII1p}
\end{center}
\end{figure}

\begin{figure}[H]
\begin{center}
\includegraphics[width=\wappfig,clip=true]{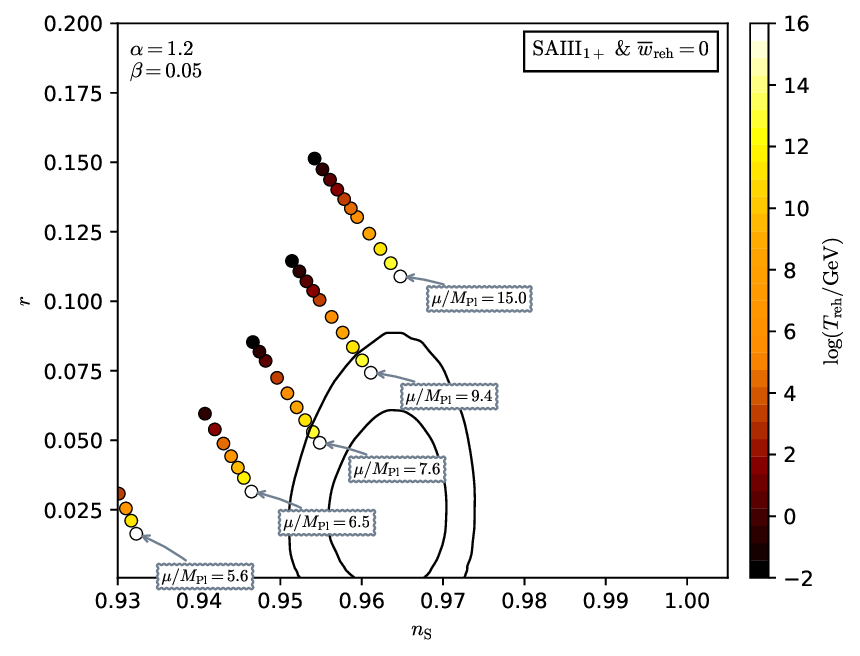}
\includegraphics[width=\wappfig,clip=true]{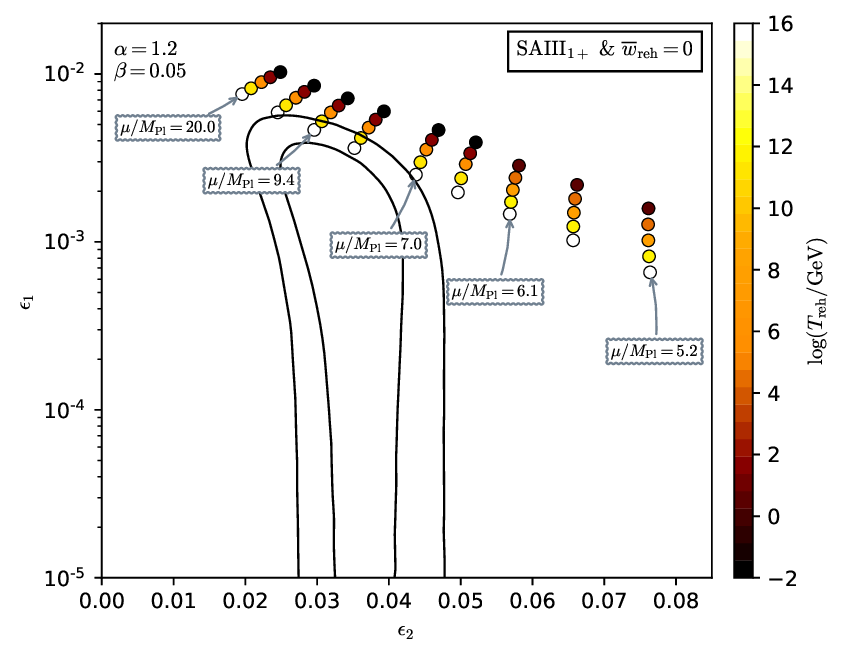}
\caption{Reheating consistent slow-roll predictions for the String
  Axion Inflation II models, in the SAIII1 regime and at a small positive
  value of $\beta=0.05$. Predictions are represented in the plane
  $(\nS,r)$ (top panel) and in the plane $(\epsilon_1,\epsilon_2)$
  (bottom panel). The solid contours are the one and two-sigma {\data}
  confidence intervals (marginalized over second order slow-roll).}
\label{fig:CMBSAIII1p_1}
\end{center}
\end{figure}

\begin{figure}[H]
\begin{center}
\includegraphics[width=\wappfig,clip=true]{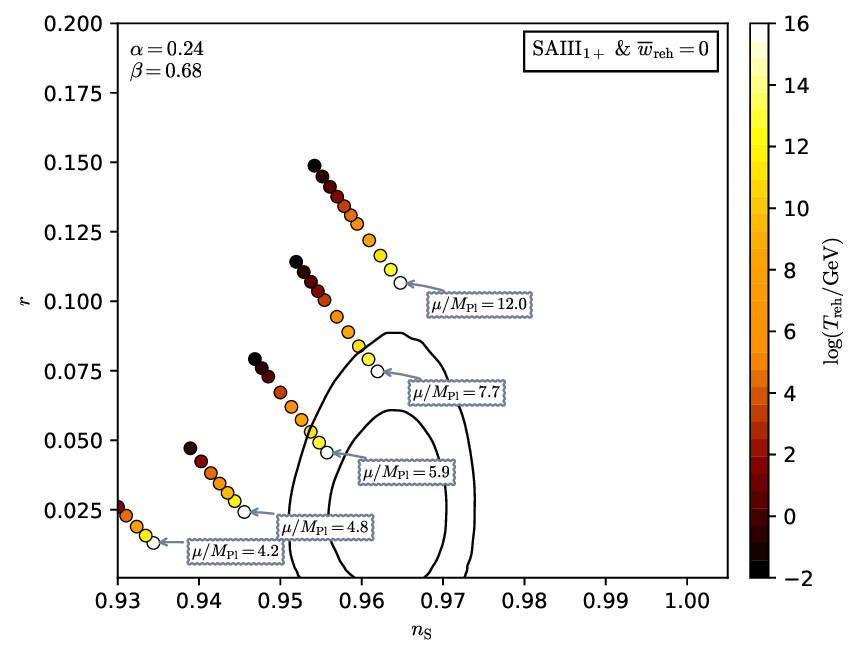}
\includegraphics[width=\wappfig,clip=true]{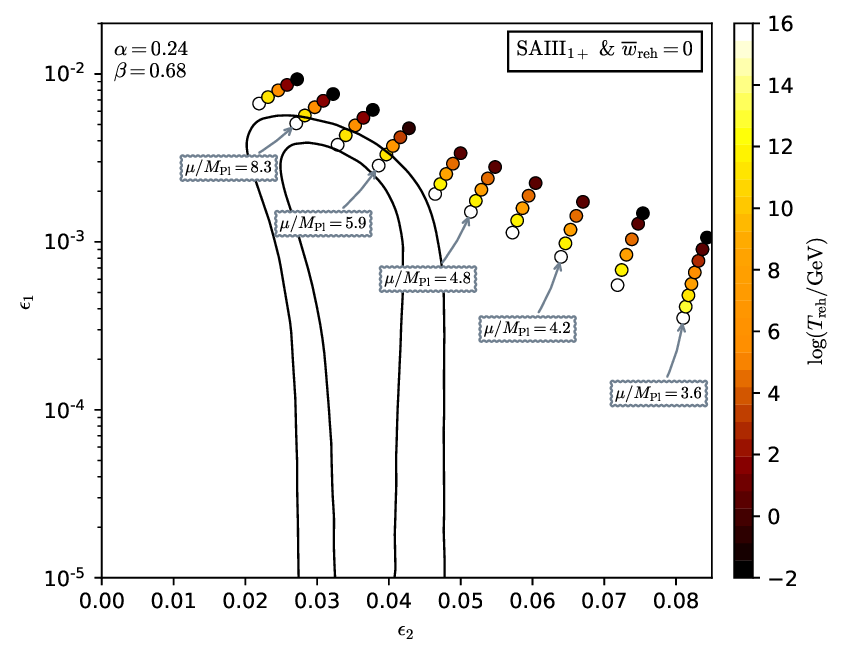}
\caption{Reheating consistent slow-roll predictions for the String
  Axion Inflation II models, in the SAIII1 regime and at a positive
  value of $\beta=0.68$. Predictions are represented in the plane
  $(\nS,r)$ (top panel) and in the plane $(\epsilon_1,\epsilon_2)$
  (bottom panel). The solid contours are the one and two-sigma {\data}
  confidence intervals (marginalized over second order slow-roll), see
  also \Fig{fig:CMBSAIII1p_3}.}
\label{fig:CMBSAIII1p_2}
\end{center}
\end{figure}

\begin{figure}[H]
\begin{center}
\includegraphics[width=\wappfig,clip=true]{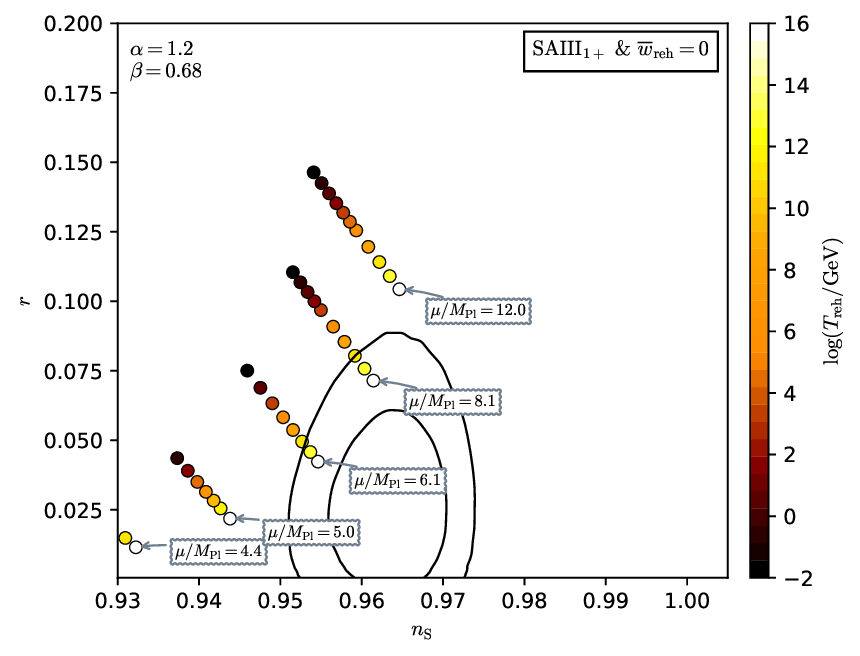}
\includegraphics[width=\wappfig,clip=true]{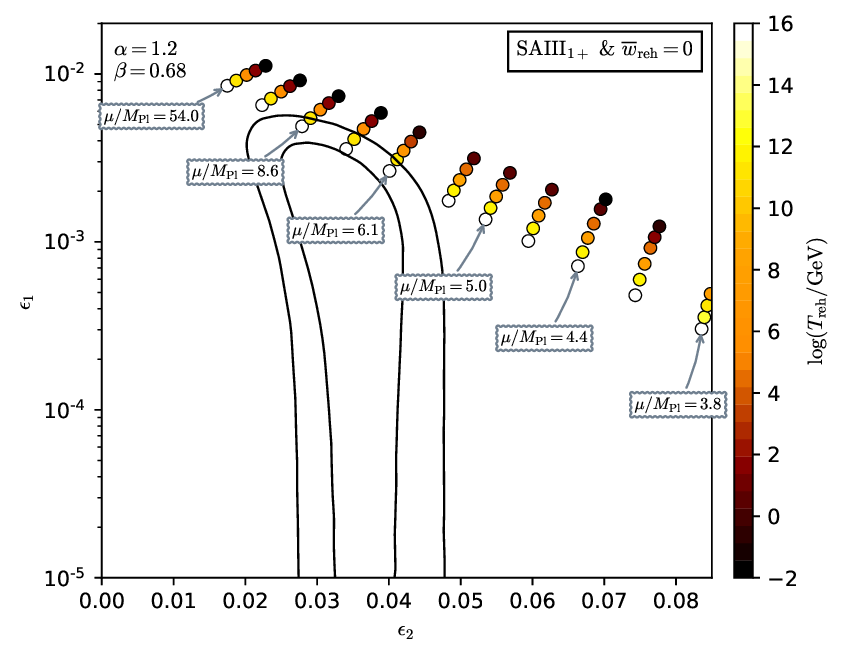}
\caption{Reheating consistent slow-roll predictions for the String
  Axion Inflation II models, in the SAIII1 regime and at a positive
  value of $\beta=0.68$. Predictions are represented in the plane
  $(\nS,r)$ (top panel) and in the plane $(\epsilon_1,\epsilon_2)$
  (bottom panel). The solid contours are the one and two-sigma {\data}
  confidence intervals (marginalized over second order slow-roll).}
\label{fig:CMBSAIII1p_3}
\end{center}
\end{figure}

\begin{figure}[H]
\begin{center}
\includegraphics[width=\wappfig,clip=true]{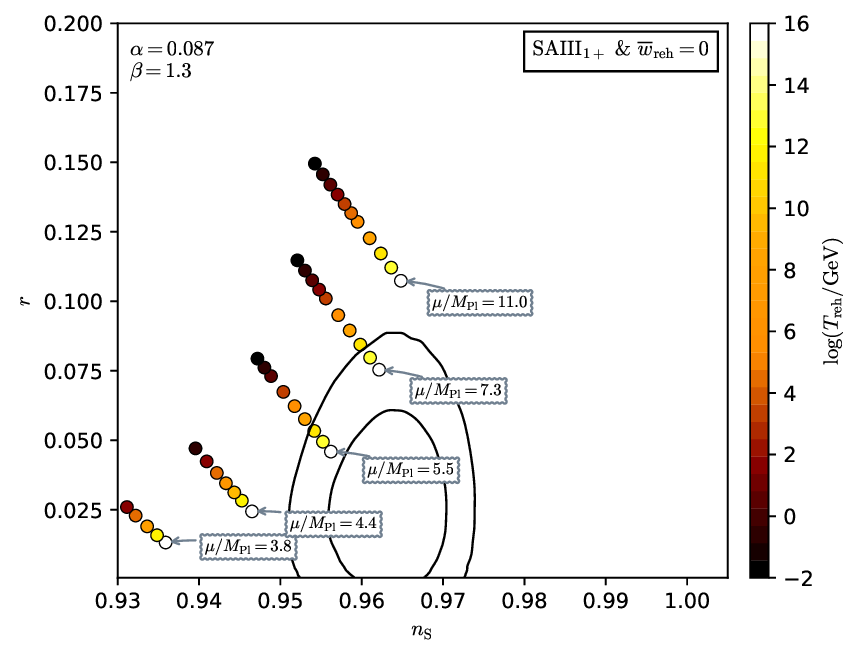}
\includegraphics[width=\wappfig,clip=true]{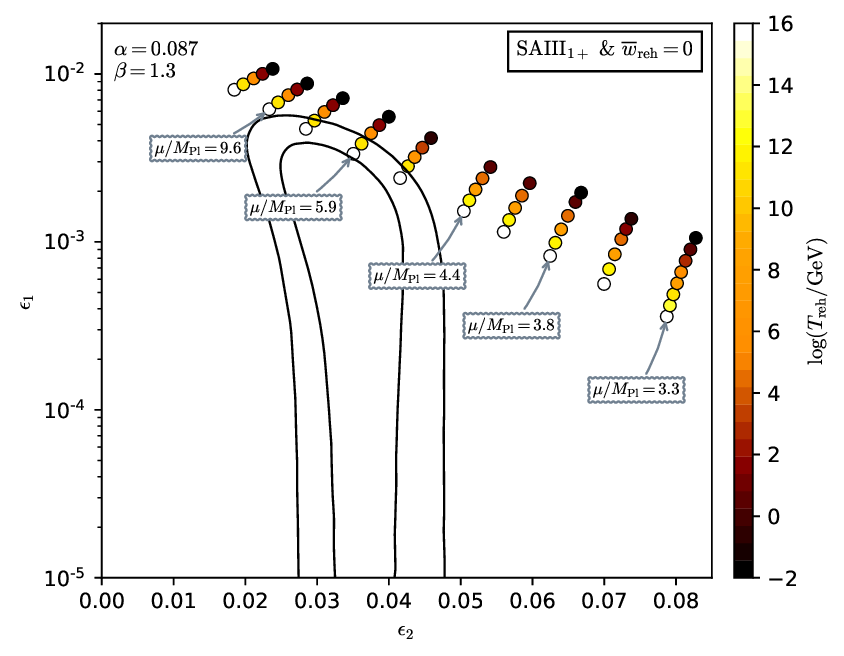}
\caption{Reheating consistent slow-roll predictions for the String
  Axion Inflation II models, in the SAIII1 regime and at a positive
  value of $\beta=1.3$. Predictions are represented in the plane
  $(\nS,r)$ (top panel) and in the plane $(\epsilon_1,\epsilon_2)$
  (bottom panel). The solid contours are the one and two-sigma {\data}
  confidence intervals (marginalized over second order slow-roll), see
  also \Fig{fig:CMBSAIII1p_5}.}
\label{fig:CMBSAIII1p_4}
\end{center}
\end{figure}

\begin{figure}[H]
\begin{center}
\includegraphics[width=\wappfig,clip=true]{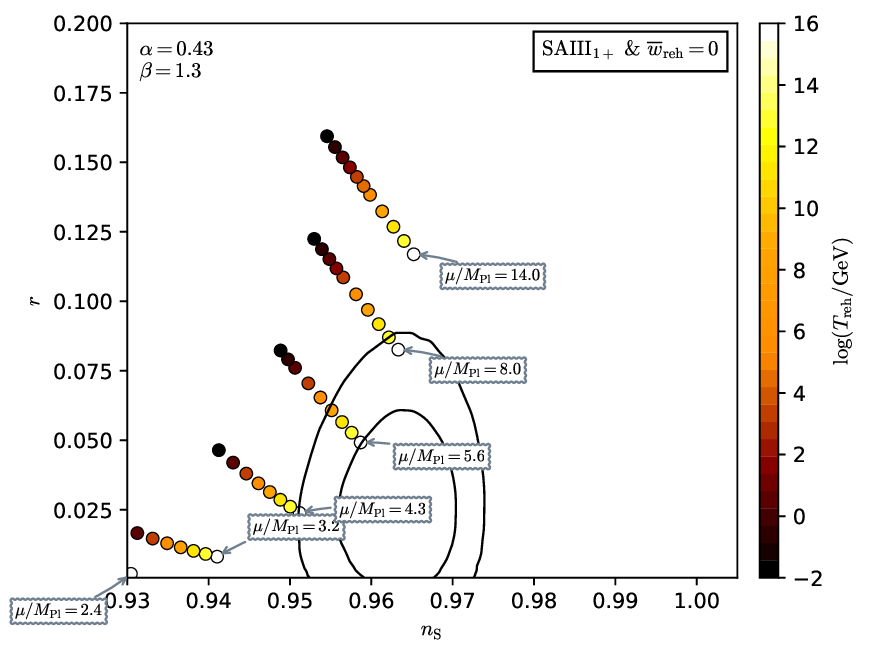}
\includegraphics[width=\wappfig,clip=true]{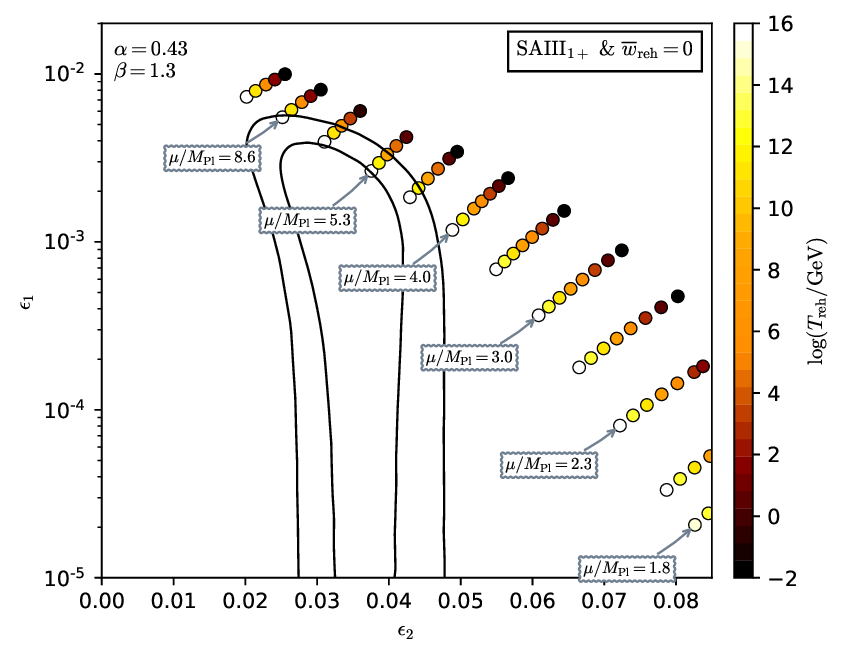}
\caption{Reheating consistent slow-roll predictions for the String
  Axion Inflation II models, in the SAIII1 regime and at a positive
  value of $\beta=1.3$. Predictions are represented in the plane
  $(\nS,r)$ (top panel) and in the plane $(\epsilon_1,\epsilon_2)$
  (bottom panel). The solid contours are the one and two-sigma {\data}
  confidence intervals (marginalized over second order slow-roll).}
\label{fig:CMBSAIII1p_5}
\end{center}
\end{figure}

\subsection{String Axion Inflation II 2 (\hyperref[sec:saiii]{SAIII2})}

\begin{figure}[H]
\begin{center}
\includegraphics[width=\wappfig,clip=true]{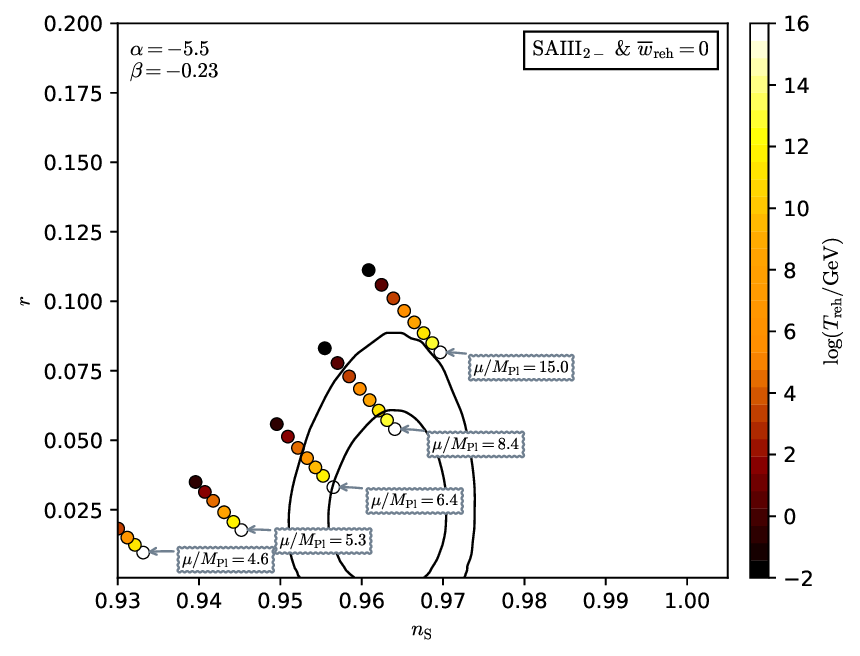}
\includegraphics[width=\wappfig,clip=true]{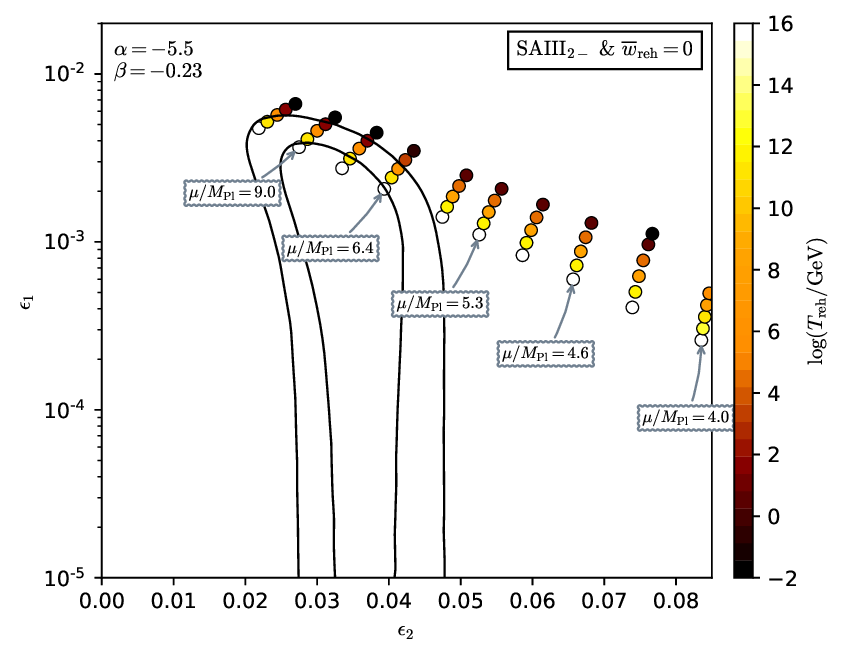}
\caption{Reheating consistent slow-roll predictions for the String
  Axion Inflation II models, in the SAIII2 regime and at a negative
  value of $\beta=-0.23$. Predictions are represented in the plane
  $(\nS,r)$ (top panel) and in the plane $(\epsilon_1,\epsilon_2)$
  (bottom panel). The solid contours are the one and two-sigma {\data}
  confidence intervals (marginalized over second order slow-roll), see
  also \Fig{fig:CMBSAIII2m_1}.}
\label{fig:CMBSAIII2m}
\end{center}
\end{figure}

\begin{figure}[H]
\begin{center}
\includegraphics[width=\wappfig,clip=true]{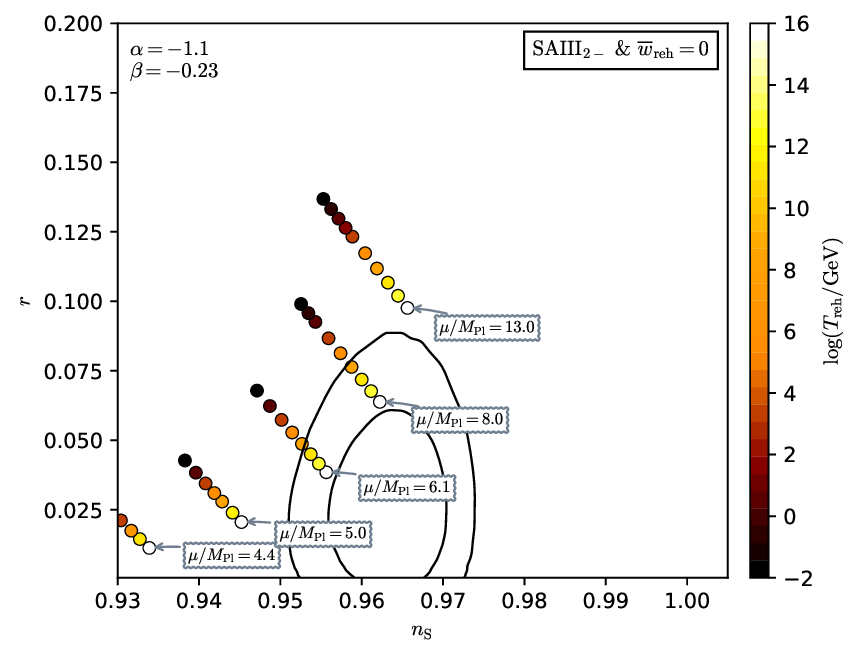}
\includegraphics[width=\wappfig,clip=true]{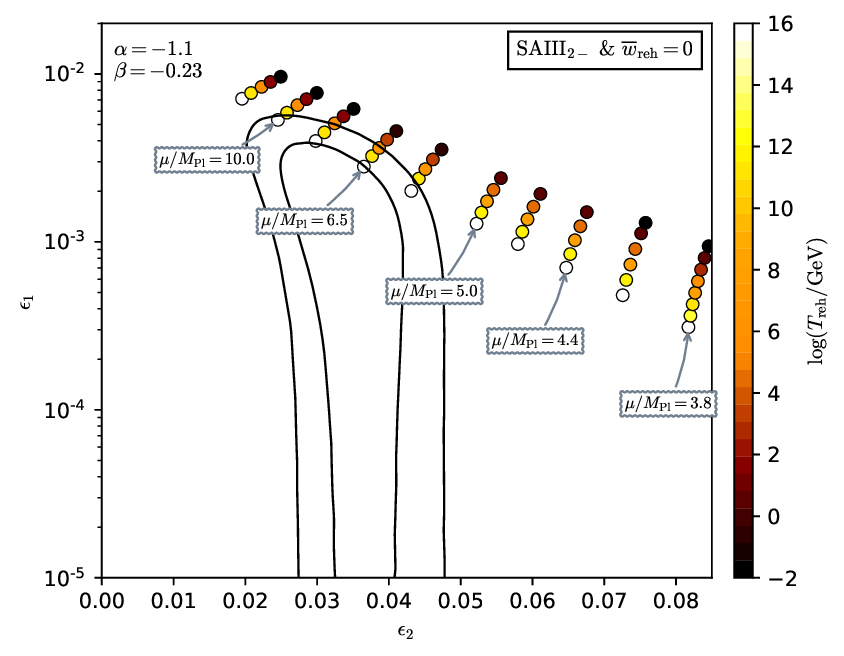}
\caption{Reheating consistent slow-roll predictions for the String
  Axion Inflation II models, in the SAIII2 regime and at a negative
  value of $\beta=-0.23$. Predictions are represented in the plane
  $(\nS,r)$ (top panel) and in the plane $(\epsilon_1,\epsilon_2)$
  (bottom panel). The solid contours are the one and two-sigma {\data}
  confidence intervals (marginalized over second order slow-roll).}
\label{fig:CMBSAIII2m_1}
\end{center}
\end{figure}

\begin{figure}[H]
\begin{center}
\includegraphics[width=\wappfig,clip=true]{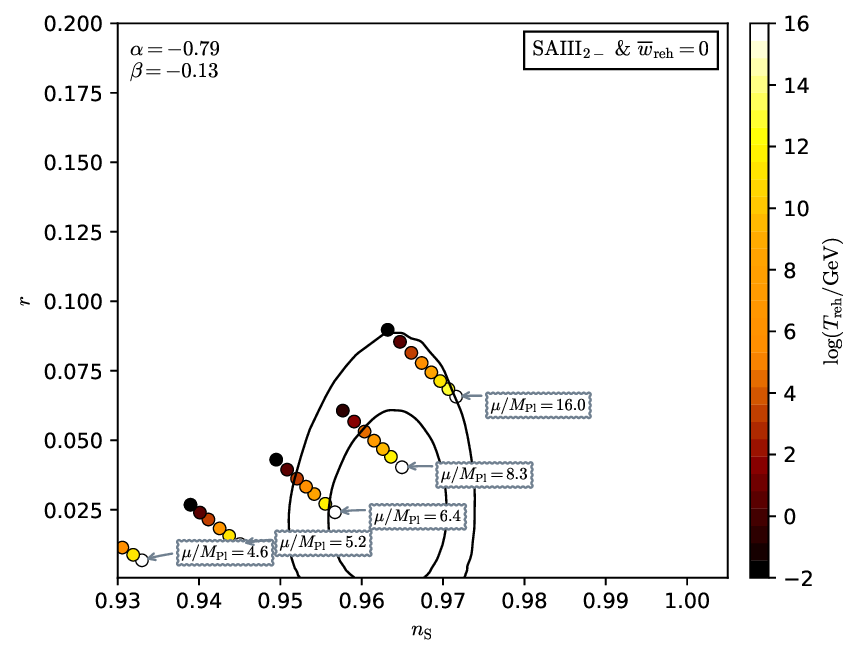}
\includegraphics[width=\wappfig,clip=true]{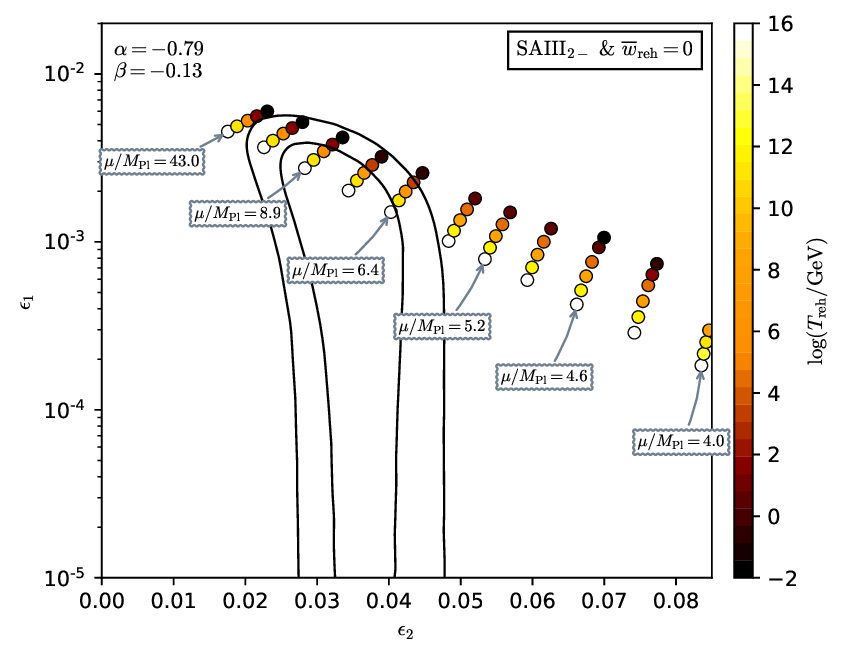}
\caption{Reheating consistent slow-roll predictions for the String
  Axion Inflation II models, in the SAIII2 regime and at a negative
  value of $\beta=-0.13$. Predictions are represented in the plane
  $(\nS,r)$ (top panel) and in the plane $(\epsilon_1,\epsilon_2)$
  (bottom panel). The solid contours are the one and two-sigma {\data}
  confidence intervals (marginalized over second order slow-roll), see
  also \Fig{fig:CMBSAIII2m_3}.}
\label{fig:CMBSAIII2m_2}
\end{center}
\end{figure}

\begin{figure}[H]
\begin{center}
\includegraphics[width=\wappfig,clip=true]{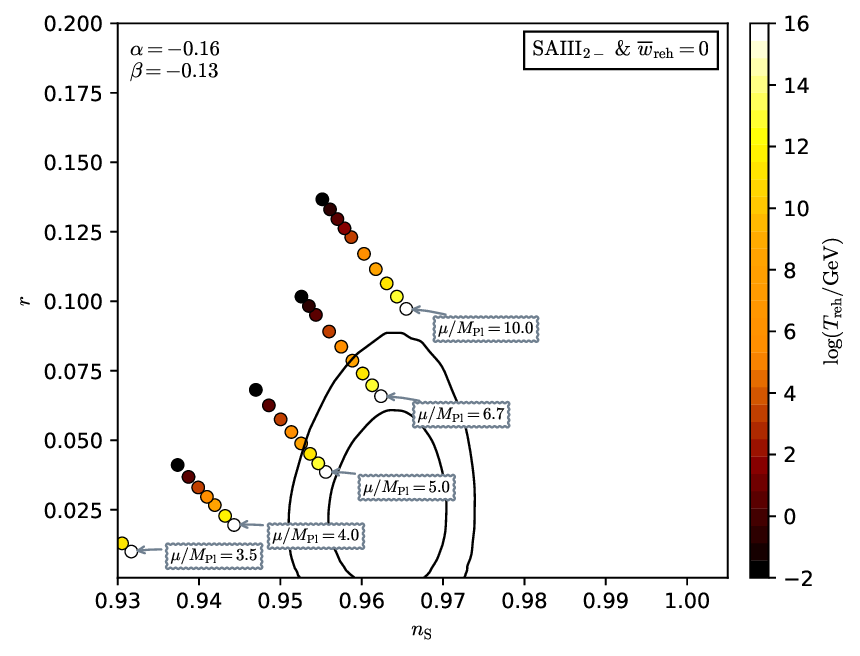}
\includegraphics[width=\wappfig,clip=true]{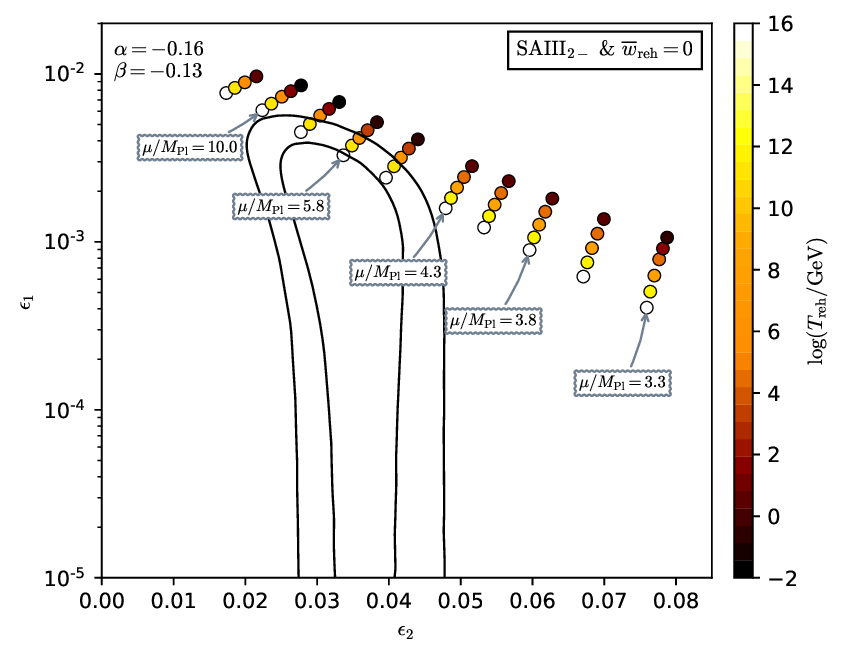}
\caption{Reheating consistent slow-roll predictions for the String
  Axion Inflation II models, in the SAIII2 regime and at a negative
  value of $\beta=-0.13$. Predictions are represented in the plane
  $(\nS,r)$ (top panel) and in the plane $(\epsilon_1,\epsilon_2)$
  (bottom panel). The solid contours are the one and two-sigma {\data}
  confidence intervals (marginalized over second order slow-roll).}
\label{fig:CMBSAIII2m_3}
\end{center}
\end{figure}

\begin{figure}[H]
\begin{center}
\includegraphics[width=\wappfig,clip=true]{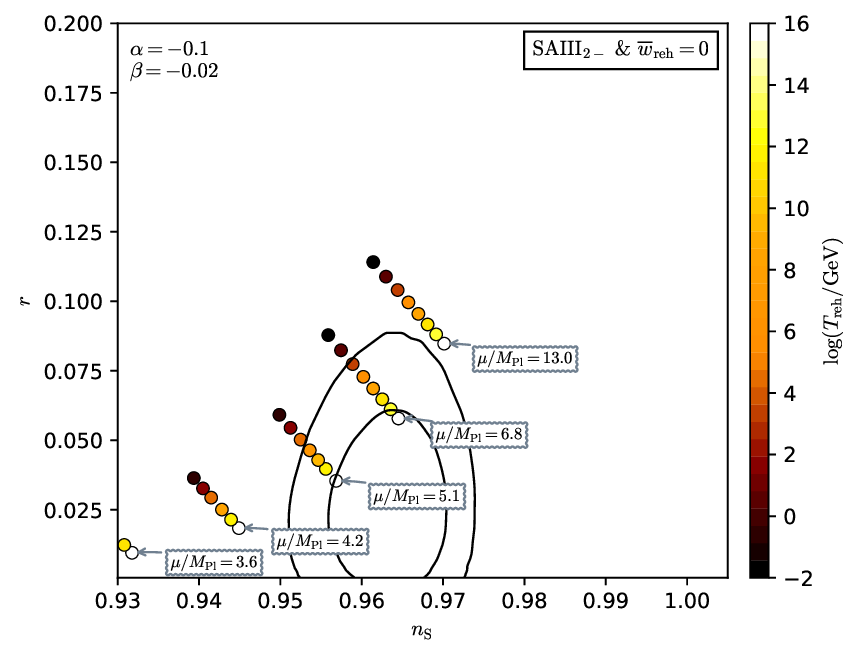}
\includegraphics[width=\wappfig,clip=true]{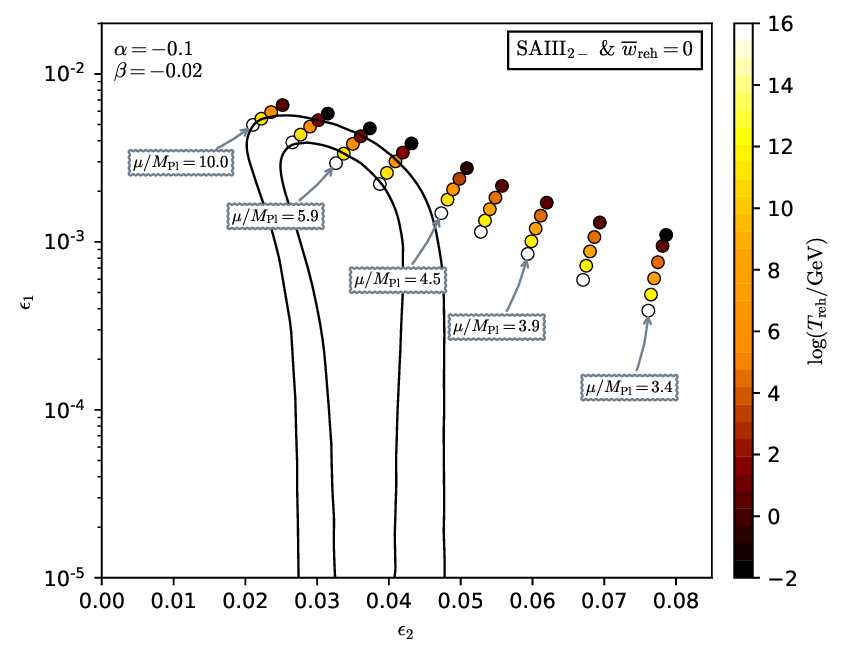}
\caption{Reheating consistent slow-roll predictions for the String
  Axion Inflation II models, in the SAIII2 regime and at a small negative
  value of $\beta=-0.02$. Predictions are represented in the plane
  $(\nS,r)$ (top panel) and in the plane $(\epsilon_1,\epsilon_2)$
  (bottom panel). The solid contours are the one and two-sigma {\data}
  confidence intervals (marginalized over second order slow-roll), see
  also \Fig{fig:CMBSAIII2m_5}.}
\label{fig:CMBSAIII2m_4}
\end{center}
\end{figure}

\begin{figure}[H]
\begin{center}
\includegraphics[width=\wappfig,clip=true]{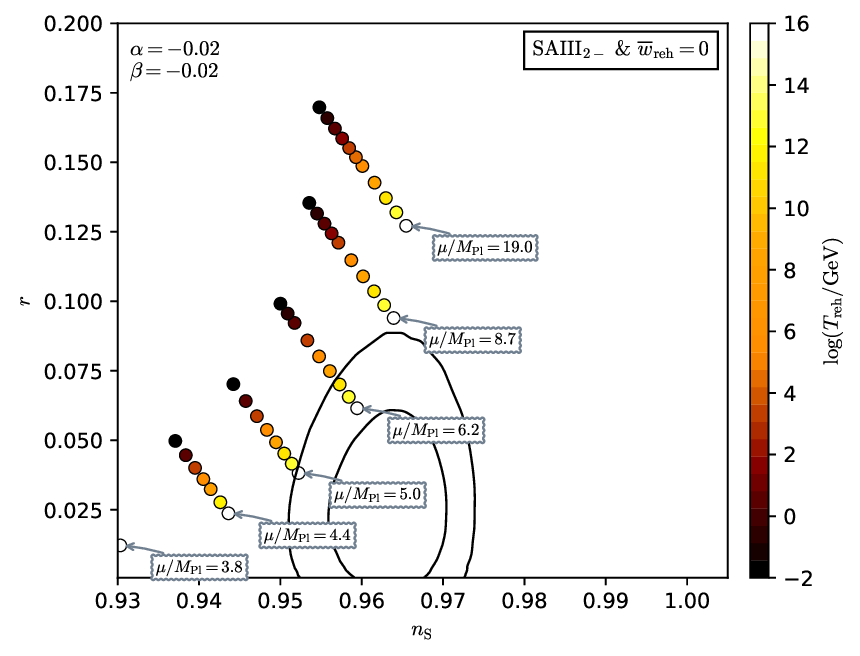}
\includegraphics[width=\wappfig,clip=true]{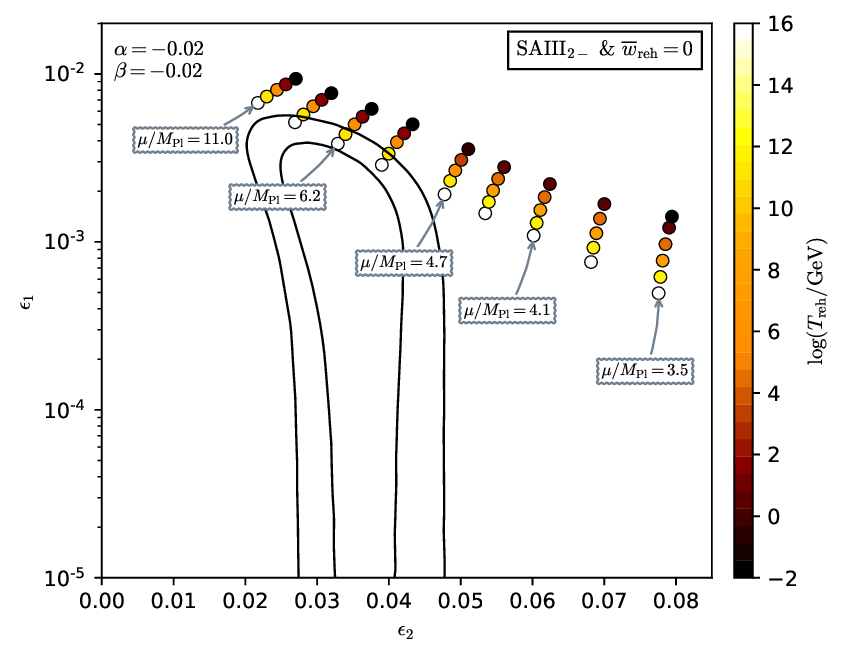}
\caption{Reheating consistent slow-roll predictions for the String
  Axion Inflation II models, in the SAIII2 regime and at a small negative
  value of $\beta=-0.02$. Predictions are represented in the plane
  $(\nS,r)$ (top panel) and in the plane $(\epsilon_1,\epsilon_2)$
  (bottom panel). The solid contours are the one and two-sigma {\data}
  confidence intervals (marginalized over second order slow-roll).}
\label{fig:CMBSAIII2m_5}
\end{center}
\end{figure}

\begin{figure}[H]
\begin{center}
\includegraphics[width=\wappfig,clip=true]{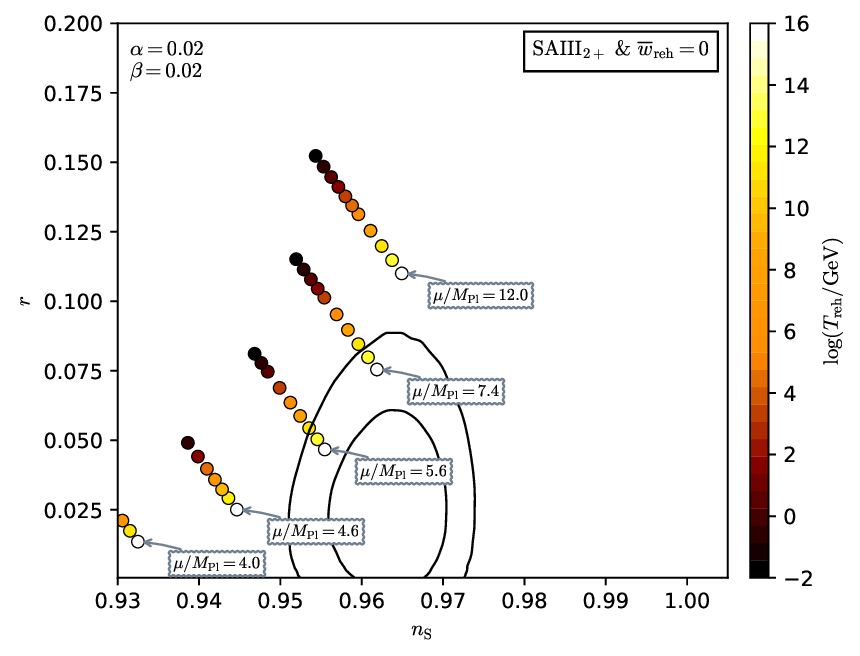}
\includegraphics[width=\wappfig,clip=true]{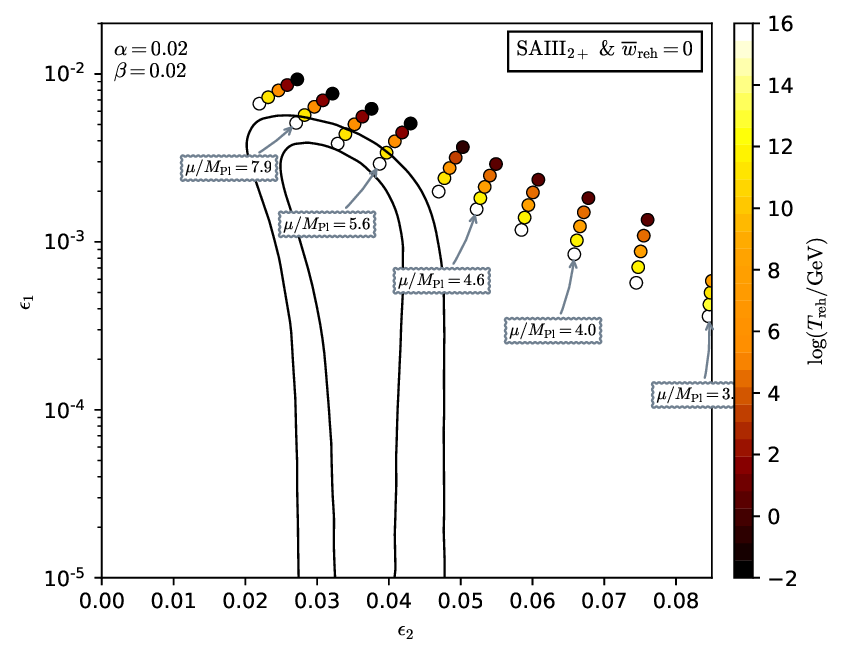}
\caption{Reheating consistent slow-roll predictions for the String
  Axion Inflation II models, in the SAIII2 regime and at a small positive
  value of $\beta=0.02$. Predictions are represented in the plane
  $(\nS,r)$ (top panel) and in the plane $(\epsilon_1,\epsilon_2)$
  (bottom panel). The solid contours are the one and two-sigma {\data}
  confidence intervals (marginalized over second order slow-roll), see
  also \Fig{fig:CMBSAIII2p_1}.}
\label{fig:CMBSAIII2p}
\end{center}
\end{figure}

\begin{figure}[H]
\begin{center}
\includegraphics[width=\wappfig,clip=true]{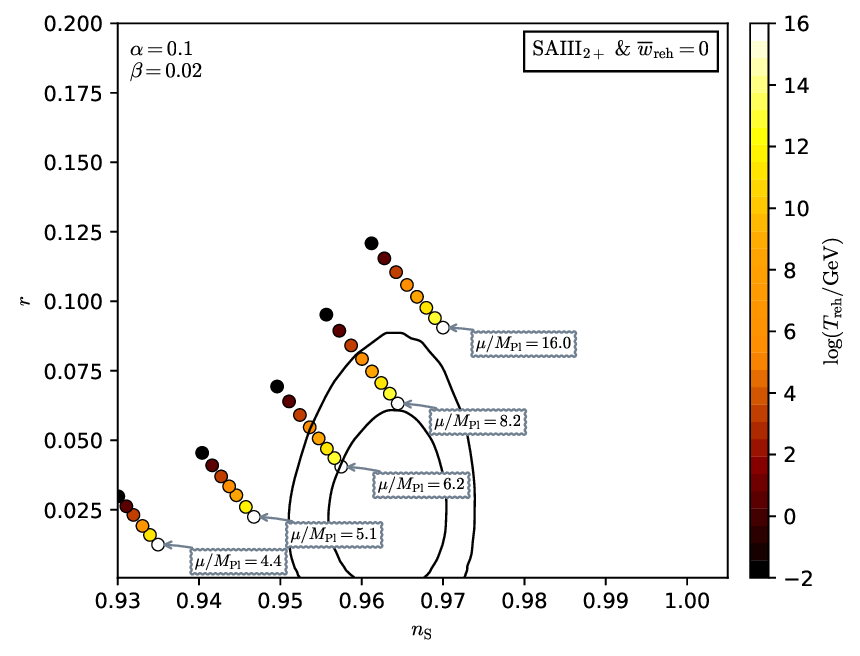}
\includegraphics[width=\wappfig,clip=true]{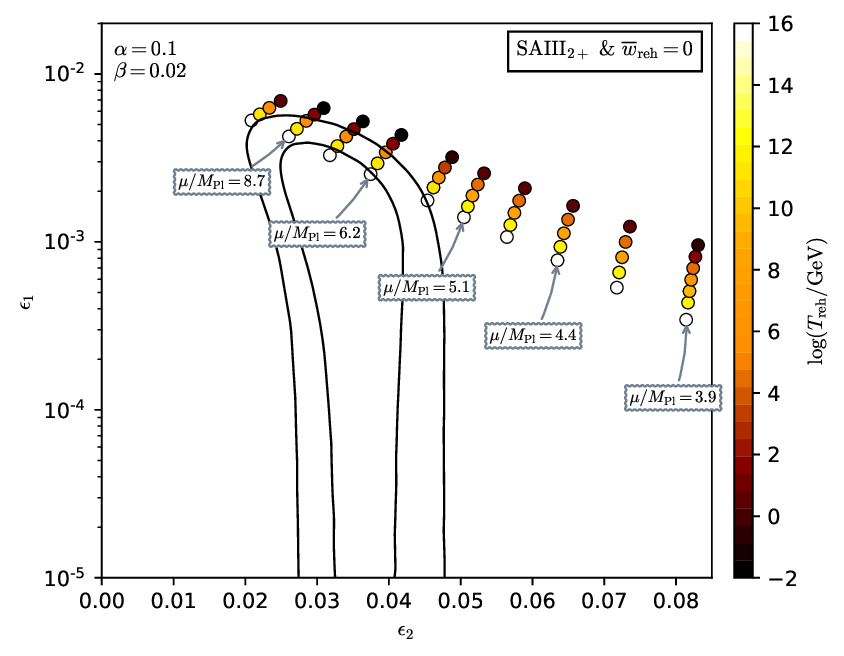}
\caption{Reheating consistent slow-roll predictions for the String
  Axion Inflation II models, in the SAIII2 regime and at a small positive
  value of $\beta=0.02$. Predictions are represented in the plane
  $(\nS,r)$ (top panel) and in the plane $(\epsilon_1,\epsilon_2)$
  (bottom panel). The solid contours are the one and two-sigma {\data}
  confidence intervals (marginalized over second order slow-roll).}
\label{fig:CMBSAIII2p_1}
\end{center}
\end{figure}

\begin{figure}[H]
\begin{center}
\includegraphics[width=\wappfig,clip=true]{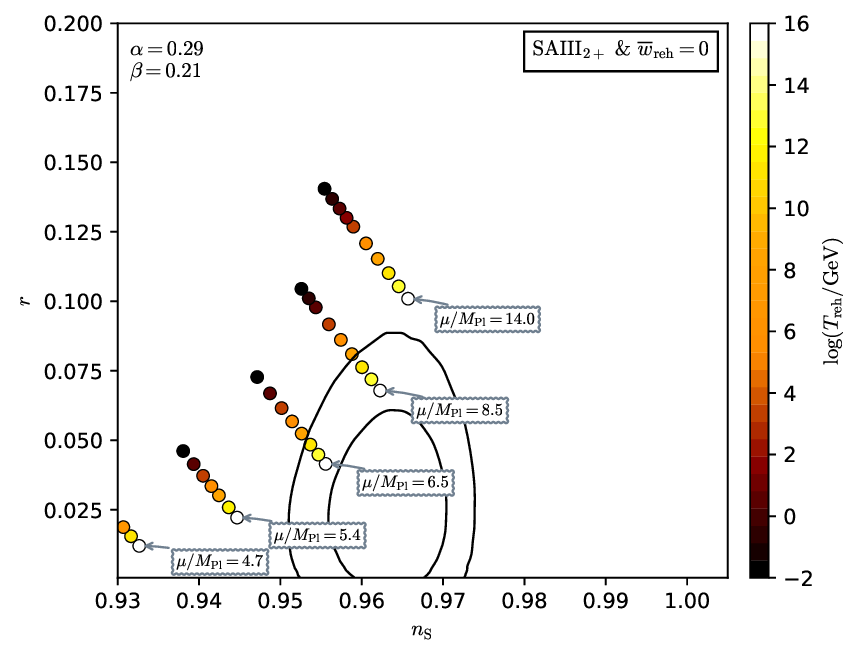}
\includegraphics[width=\wappfig,clip=true]{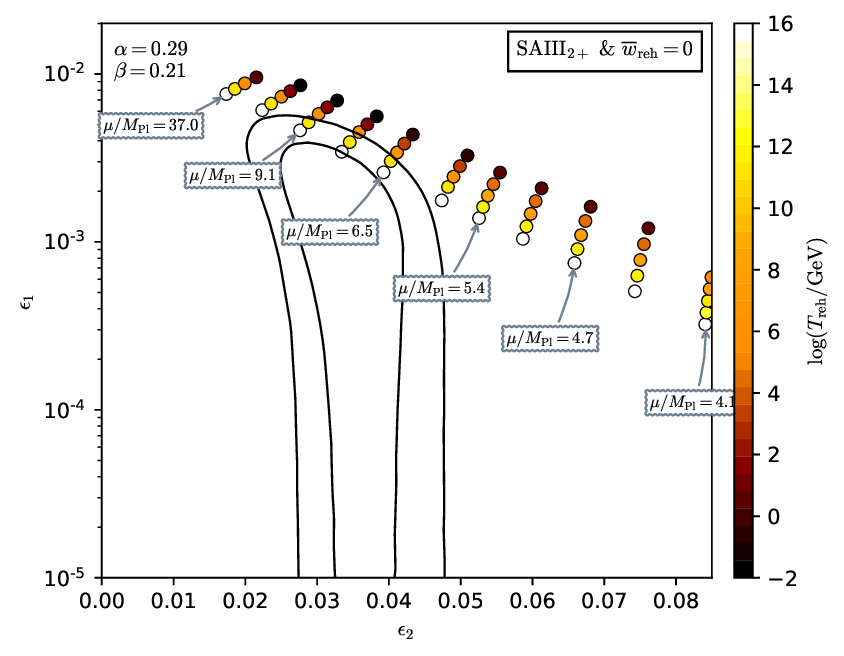}
\caption{Reheating consistent slow-roll predictions for the String
  Axion Inflation II models, in the SAIII2 regime and at a positive
  value of $\beta=0.21$. Predictions are represented in the plane
  $(\nS,r)$ (top panel) and in the plane $(\epsilon_1,\epsilon_2)$
  (bottom panel). The solid contours are the one and two-sigma {\data}
  confidence intervals (marginalized over second order slow-roll), see
  also \Fig{fig:CMBSAIII2p_3}.}
\label{fig:CMBSAIII2p_2}
\end{center}
\end{figure}

\begin{figure}[H]
\begin{center}
\includegraphics[width=\wappfig,clip=true]{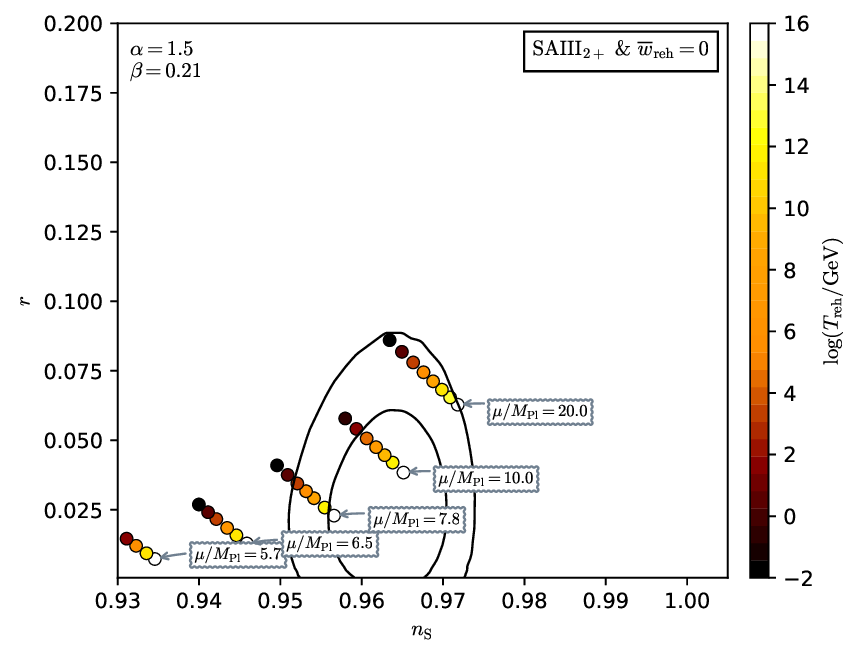}
\includegraphics[width=\wappfig,clip=true]{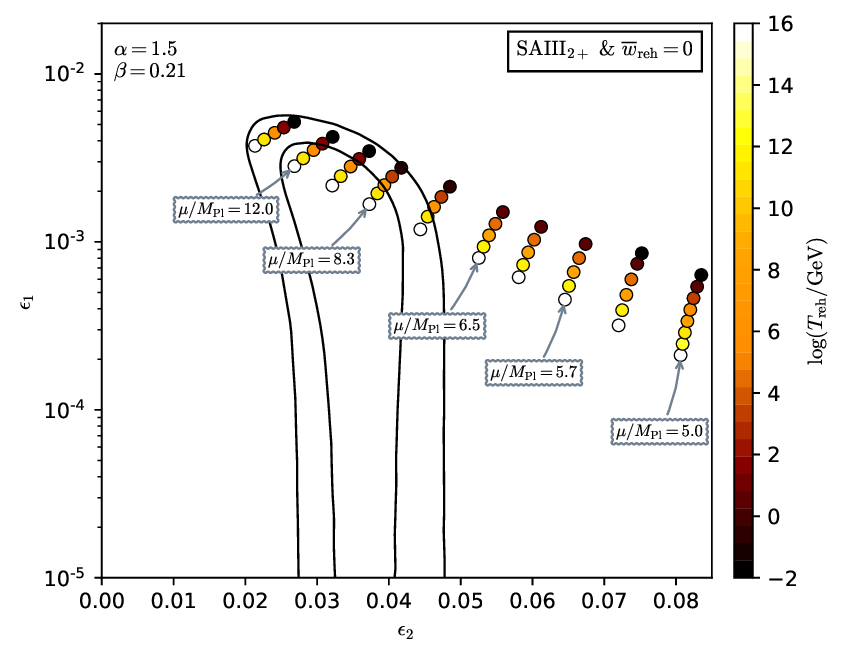}
\caption{Reheating consistent slow-roll predictions for the String
  Axion Inflation II models, in the SAIII2 regime and at a positive
  value of $\beta=0.21$. Predictions are represented in the plane
  $(\nS,r)$ (top panel) and in the plane $(\epsilon_1,\epsilon_2)$
  (bottom panel). The solid contours are the one and two-sigma {\data}
  confidence intervals (marginalized over second order slow-roll).}
\label{fig:CMBSAIII2p_3}
\end{center}
\end{figure}

\begin{figure}[H]
\begin{center}
\includegraphics[width=\wappfig,clip=true]{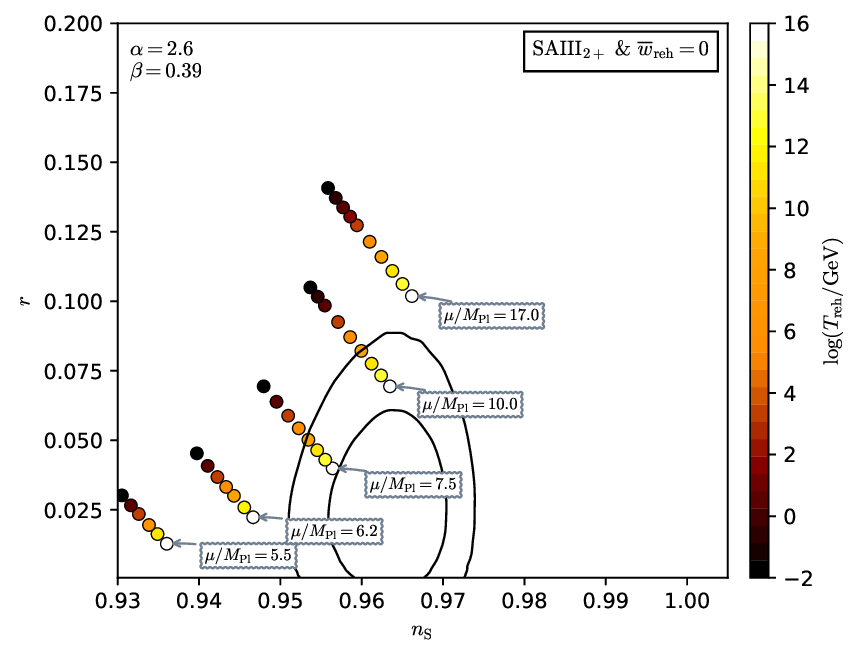}
\includegraphics[width=\wappfig,clip=true]{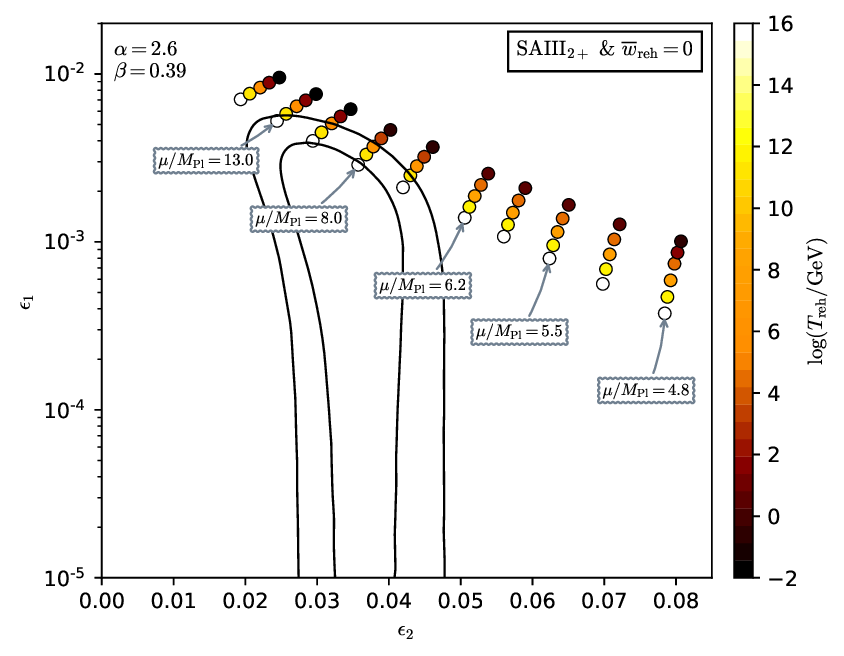}
\caption{Reheating consistent slow-roll predictions for the String
  Axion Inflation II models, in the SAIII2 regime and at a positive
  value of $\beta=0.39$. Predictions are represented in the plane
  $(\nS,r)$ (top panel) and in the plane $(\epsilon_1,\epsilon_2)$
  (bottom panel). The solid contours are the one and two-sigma {\data}
  confidence intervals (marginalized over second order slow-roll), see
  also \Fig{fig:CMBSAIII2p_5}.}
\label{fig:CMBSAIII2p_4}
\end{center}
\end{figure}

\begin{figure}[H]
\begin{center}
\includegraphics[width=\wappfig,clip=true]{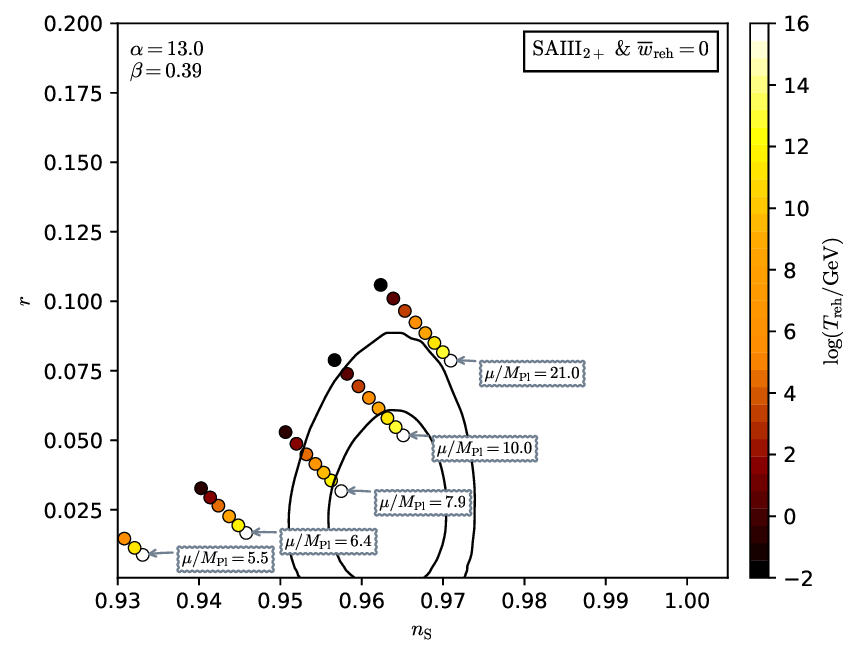}
\includegraphics[width=\wappfig,clip=true]{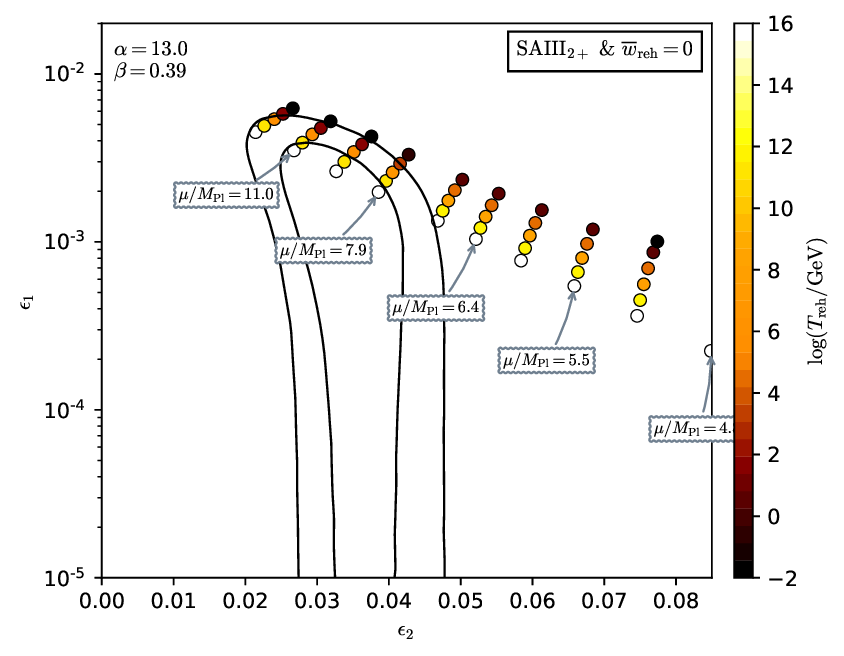}
\caption{Reheating consistent slow-roll predictions for the String
  Axion Inflation II models, in the SAIII2 regime and at a positive
  value of $\beta=0.39$. Predictions are represented in the plane
  $(\nS,r)$ (top panel) and in the plane $(\epsilon_1,\epsilon_2)$
  (bottom panel). The solid contours are the one and two-sigma {\data}
  confidence intervals (marginalized over second order slow-roll).}
\label{fig:CMBSAIII2p_5}
\end{center}
\end{figure}

\subsection{String Axion Inflation II 3 (\hyperref[sec:saiii]{SAIII3})}

\begin{figure}[H]
\begin{center}
\includegraphics[width=\wappfig,clip=true]{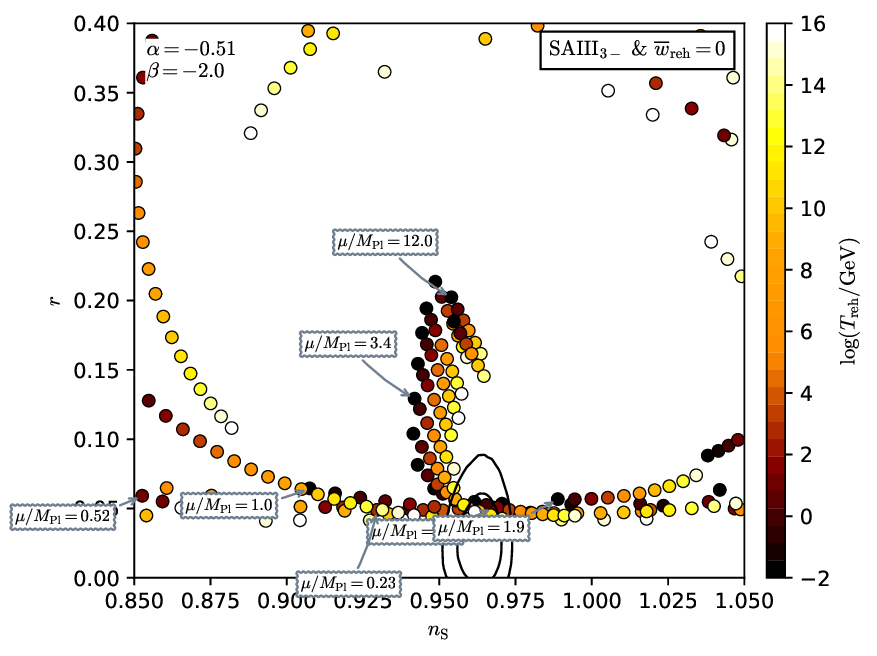}
\includegraphics[width=\wappfig,clip=true]{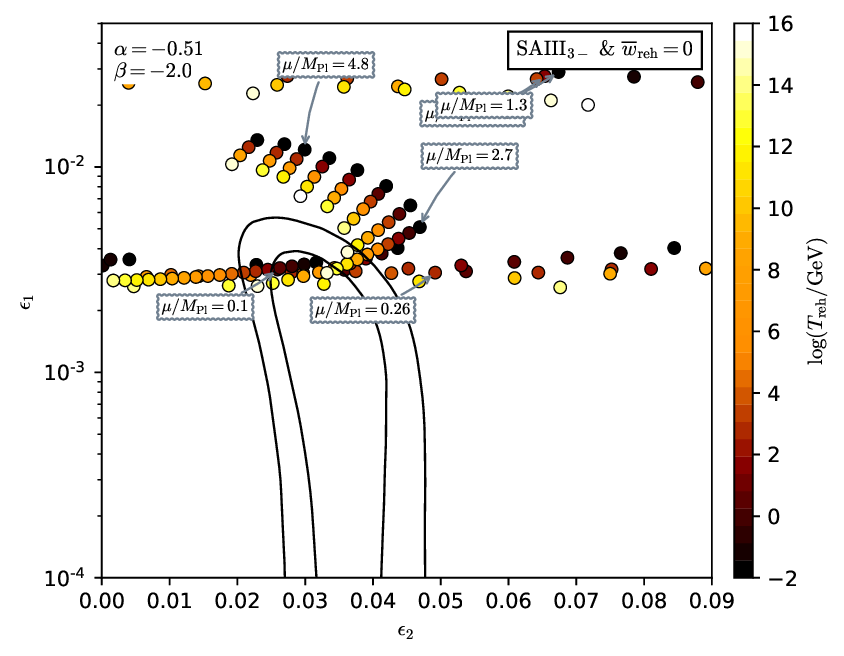}
\caption{Reheating consistent slow-roll predictions for the String
  Axion Inflation II models, in the SAIII3 regime and at a large
  negative value of $\beta=-2$. Predictions are represented in the
  plane $(\nS,r)$ (top panel) and in the plane
  $(\epsilon_1,\epsilon_2)$ (bottom panel). The solid contours are the
  one and two-sigma {\data} confidence intervals (marginalized over
  second order slow-roll). Notice the strong dependency of the
  predictions with respect to $\mu$. This is due to the modulations of
  the potential, a small change in the value of $\xend$ may produce
  drastic modifications of the observable model predictions. See also
  \Fig{fig:CMBSAIII3m_1}.}
\label{fig:CMBSAIII3m}
\end{center}
\end{figure}

\begin{figure}[H]
\begin{center}
\includegraphics[width=\wappfig,clip=true]{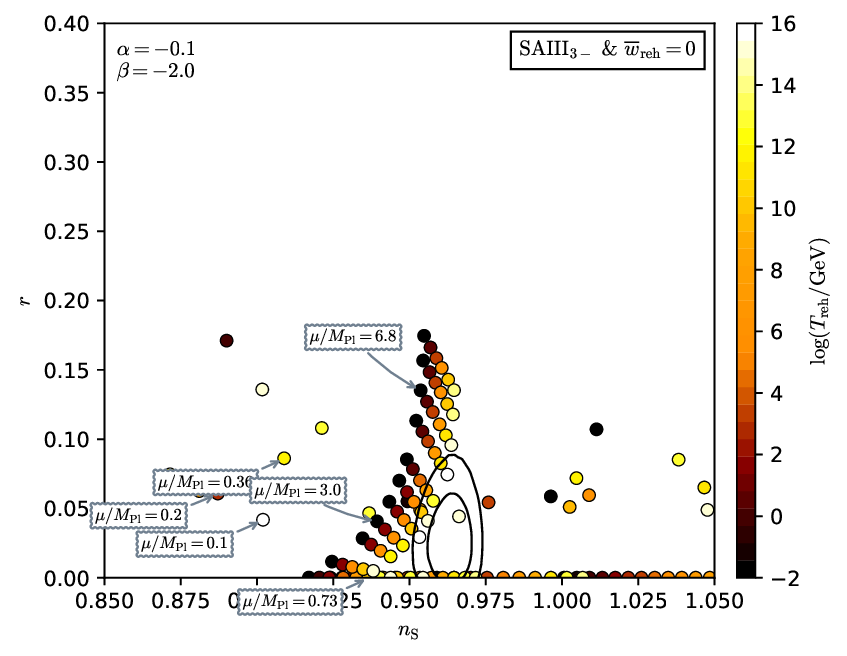}
\includegraphics[width=\wappfig,clip=true]{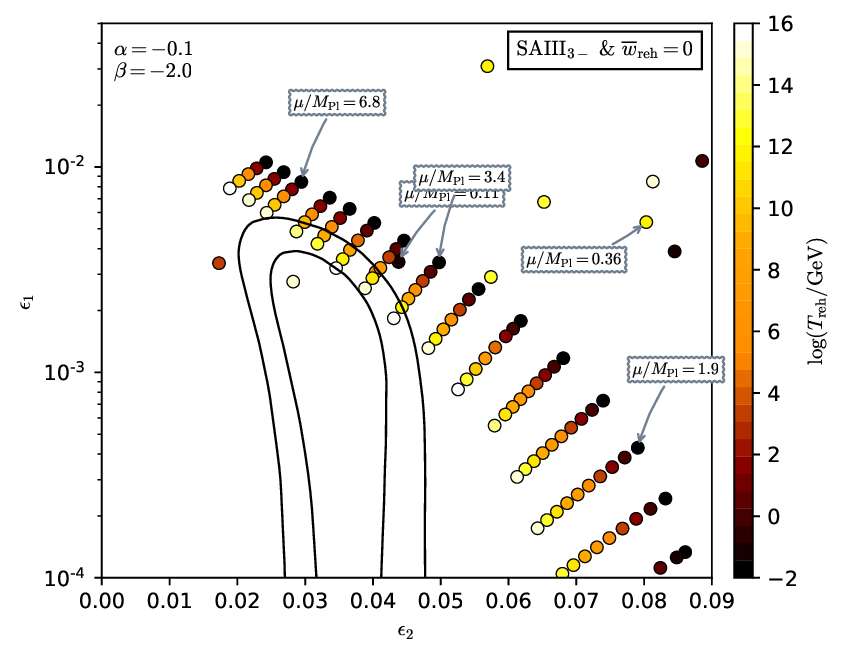}
\caption{Reheating consistent slow-roll predictions for the String
  Axion Inflation II models, in the SAIII2 regime and at a large negative
  value of $\beta=-2$. Predictions are represented in the plane
  $(\nS,r)$ (top panel) and in the plane $(\epsilon_1,\epsilon_2)$
  (bottom panel). The solid contours are the one and two-sigma {\data}
  confidence intervals (marginalized over second order slow-roll).}
\label{fig:CMBSAIII3m_1}
\end{center}
\end{figure}

\begin{figure}[H]
\begin{center}
\includegraphics[width=\wappfig,clip=true]{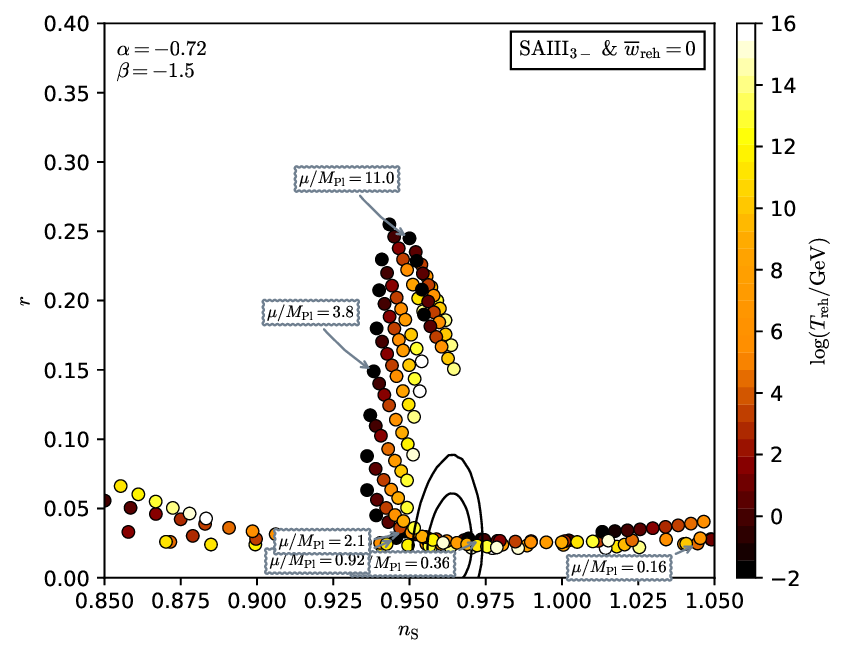}
\includegraphics[width=\wappfig,clip=true]{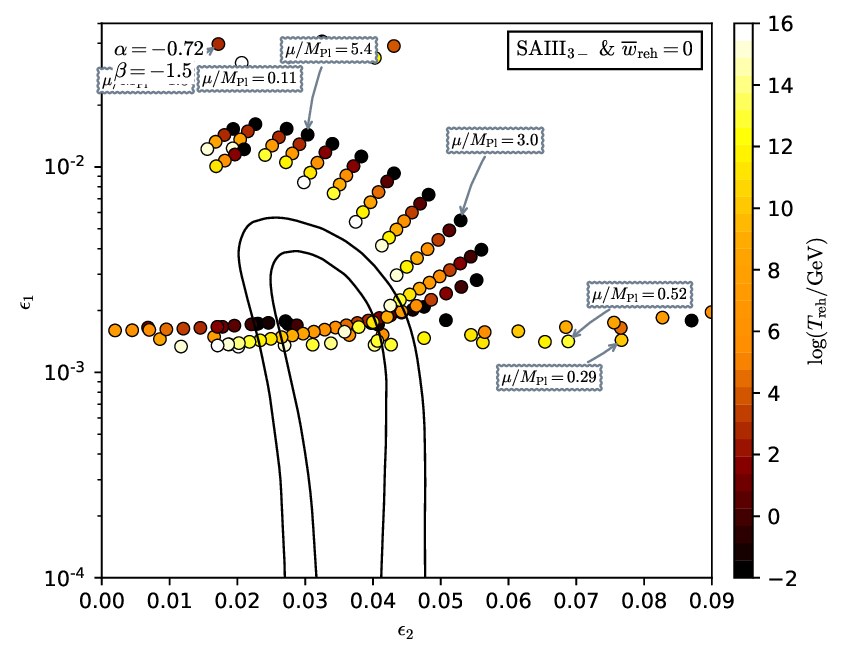}
\caption{Reheating consistent slow-roll predictions for the String
  Axion Inflation II models, in the SAIII2 regime and at a negative
  value of $\beta=-1.5$. Predictions are represented in the plane
  $(\nS,r)$ (top panel) and in the plane $(\epsilon_1,\epsilon_2)$
  (bottom panel). The solid contours are the one and two-sigma {\data}
  confidence intervals (marginalized over second order slow-roll), see
  also \Fig{fig:CMBSAIII3m_3}.}
\label{fig:CMBSAIII3m_2}
\end{center}
\end{figure}

\begin{figure}[H]
\begin{center}
\includegraphics[width=\wappfig,clip=true]{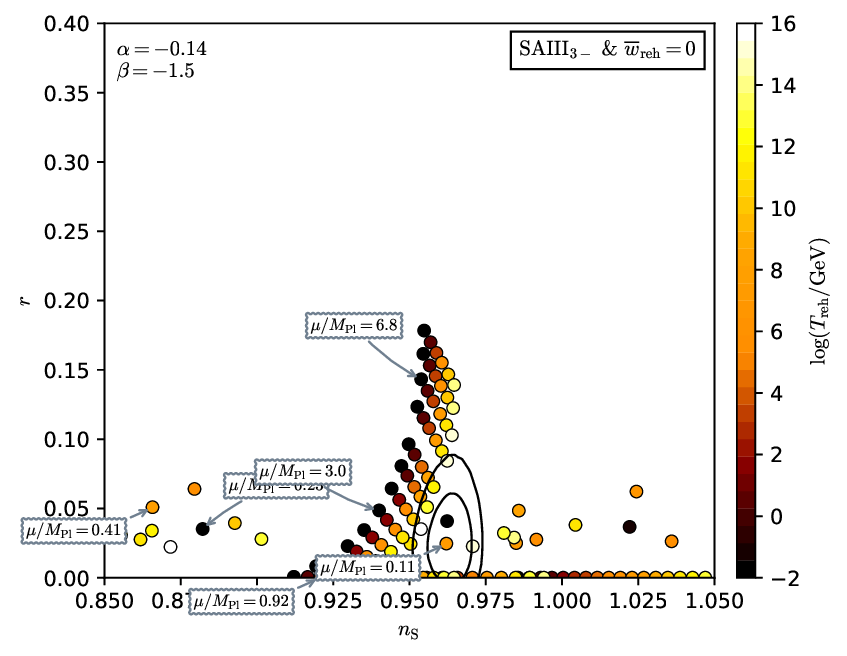}
\includegraphics[width=\wappfig,clip=true]{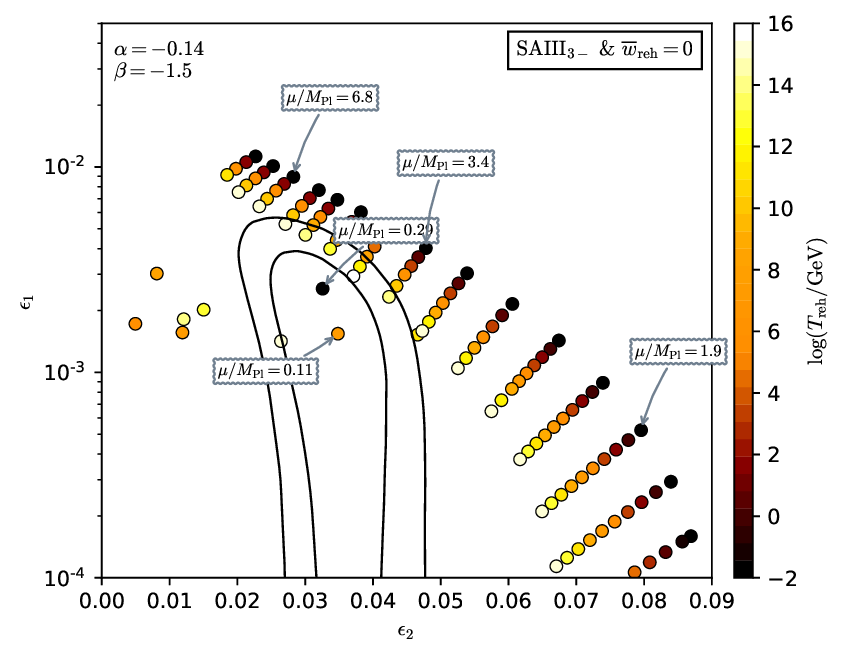}
\caption{Reheating consistent slow-roll predictions for the String
  Axion Inflation II models, in the SAIII2 regime and at a negative
  value of $\beta=-1.5$. Predictions are represented in the plane
  $(\nS,r)$ (top panel) and in the plane $(\epsilon_1,\epsilon_2)$
  (bottom panel). The solid contours are the one and two-sigma {\data}
  confidence intervals (marginalized over second order slow-roll).}
\label{fig:CMBSAIII3m_3}
\end{center}
\end{figure}

\begin{figure}[H]
\begin{center}
\includegraphics[width=\wappfig,clip=true]{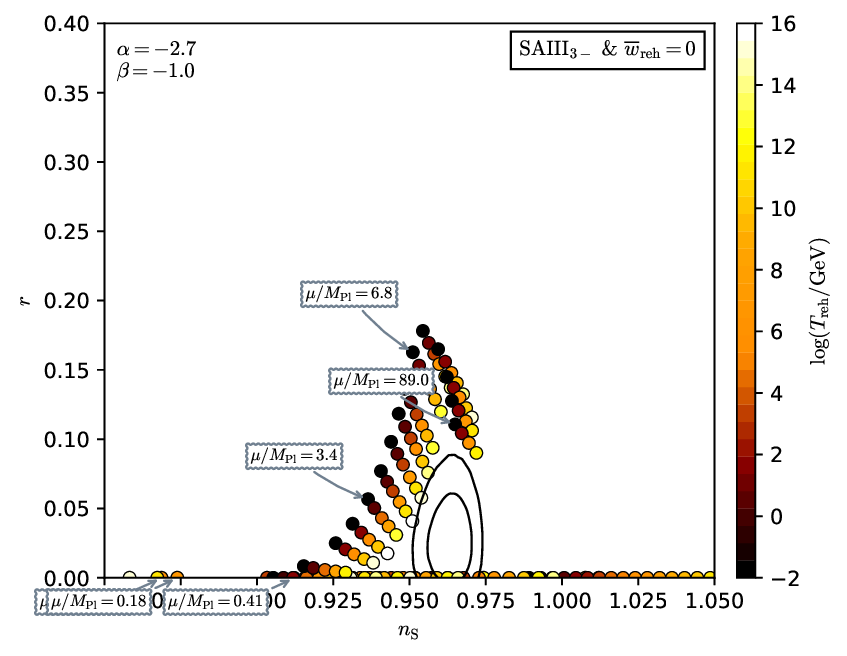}
\includegraphics[width=\wappfig,clip=true]{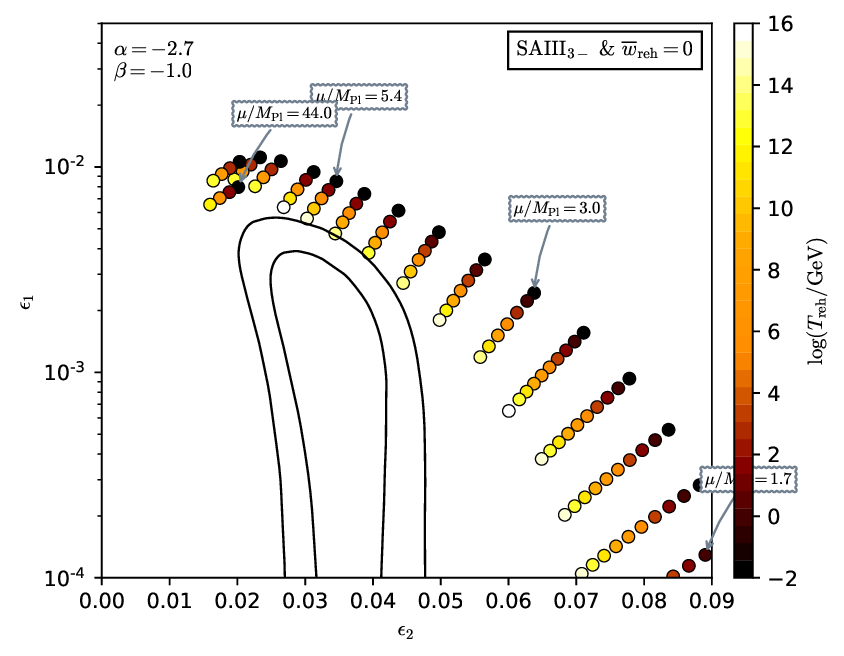}
\caption{Reheating consistent slow-roll predictions for the String
  Axion Inflation II models, in the SAIII2 regime and at a negative
  value of $\beta=-1$. Predictions are represented in the plane
  $(\nS,r)$ (top panel) and in the plane $(\epsilon_1,\epsilon_2)$
  (bottom panel). The solid contours are the one and two-sigma {\data}
  confidence intervals (marginalized over second order slow-roll), see
  also \Fig{fig:CMBSAIII3m_5}.}
\label{fig:CMBSAIII3m_4}
\end{center}
\end{figure}

\begin{figure}[H]
\begin{center}
\includegraphics[width=\wappfig,clip=true]{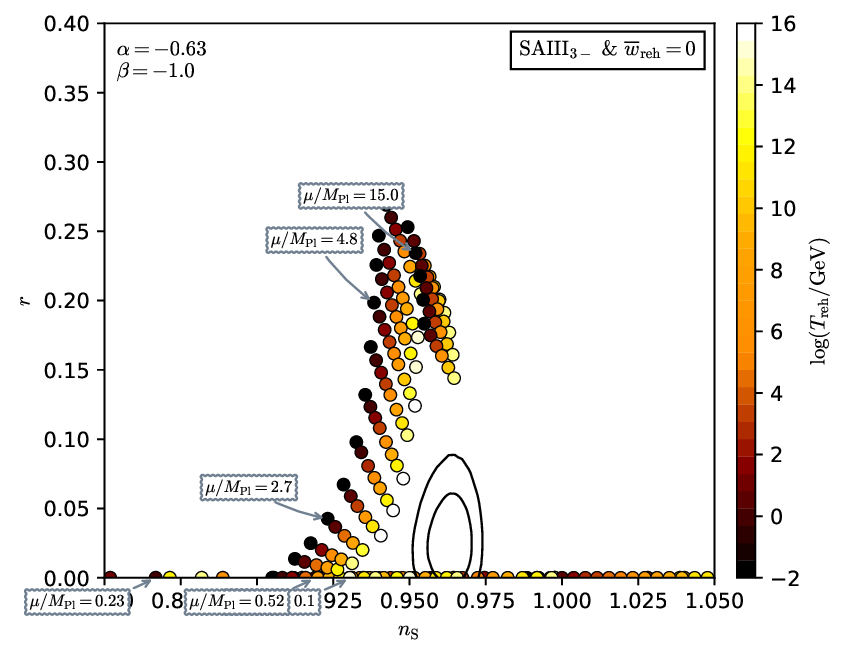}
\includegraphics[width=\wappfig,clip=true]{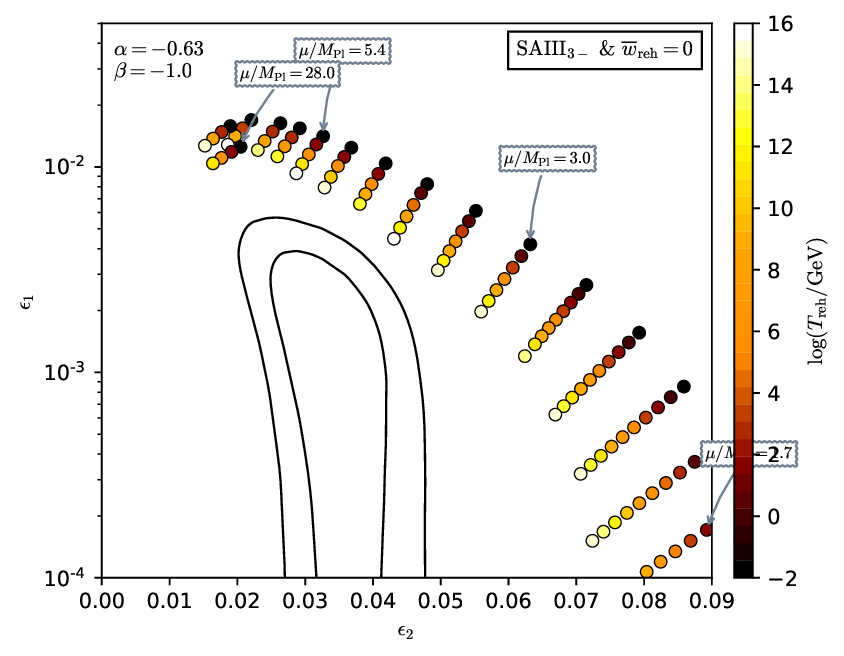}
\caption{Reheating consistent slow-roll predictions for the String
  Axion Inflation II models, in the SAIII2 regime and at a negative
  value of $\beta=-1$. Predictions are represented in the plane
  $(\nS,r)$ (top panel) and in the plane $(\epsilon_1,\epsilon_2)$
  (bottom panel). The solid contours are the one and two-sigma {\data}
  confidence intervals (marginalized over second order slow-roll).}
\label{fig:CMBSAIII3m_5}
\end{center}
\end{figure}

\begin{figure}[H]
\begin{center}
\includegraphics[width=\wappfig,clip=true]{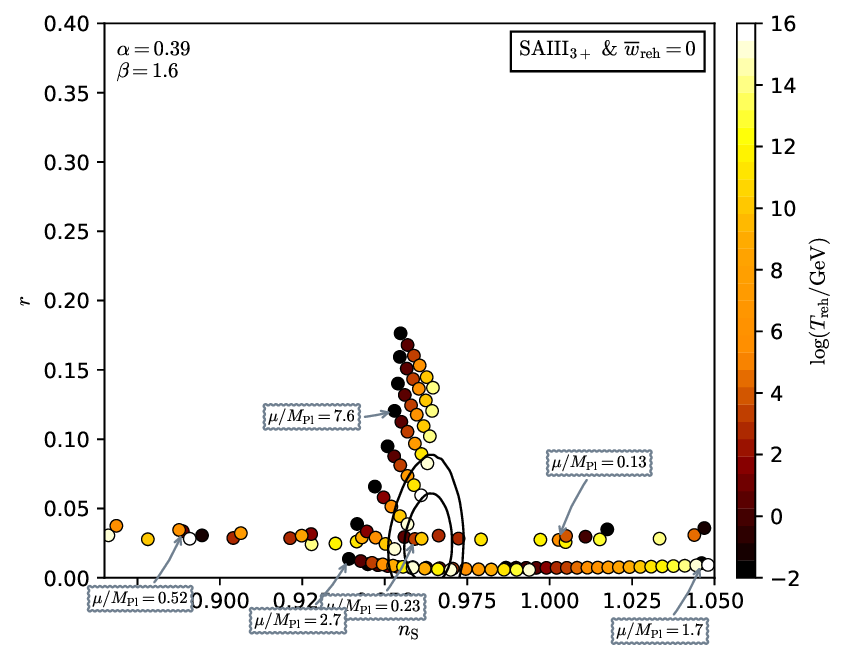}
\includegraphics[width=\wappfig,clip=true]{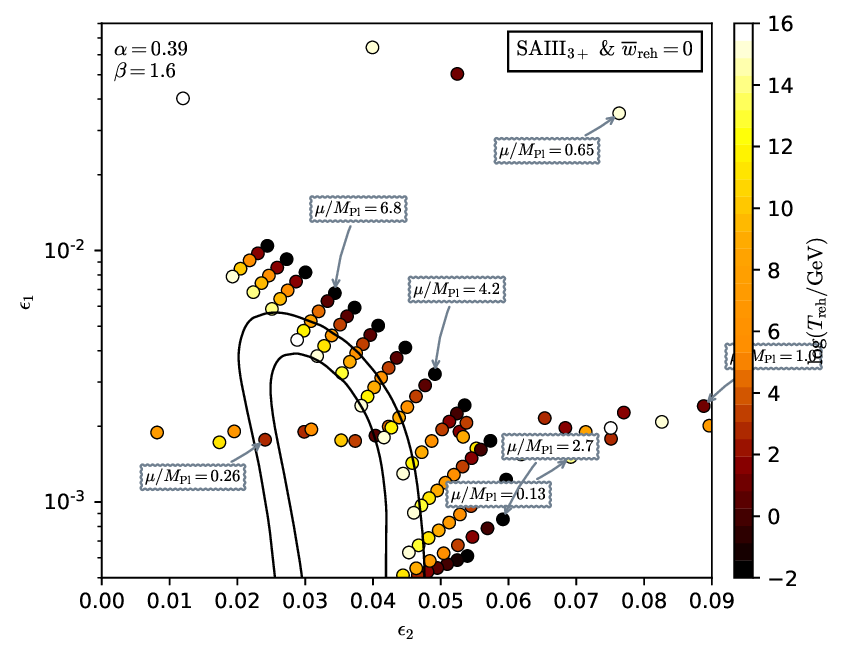}
\caption{Reheating consistent slow-roll predictions for the String
  Axion Inflation II models, in the SAIII3 regime and at a
  positive value of $\beta=1.6$. Predictions are represented in the
  plane $(\nS,r)$ (top panel) and in the plane
  $(\epsilon_1,\epsilon_2)$ (bottom panel). The solid contours are the
  one and two-sigma {\data} confidence intervals (marginalized over
  second order slow-roll). Notice the strong dependency of the
  predictions with respect to $\mu$. This is due to the modulations of
  the potential, a small change in the value of $\xend$ may produce
  drastic modifications of the observable model predictions. See also
  \Fig{fig:CMBSAIII3p_1}.}
\label{fig:CMBSAIII3p}
\end{center}
\end{figure}

\begin{figure}[H]
\begin{center}
\includegraphics[width=\wappfig,clip=true]{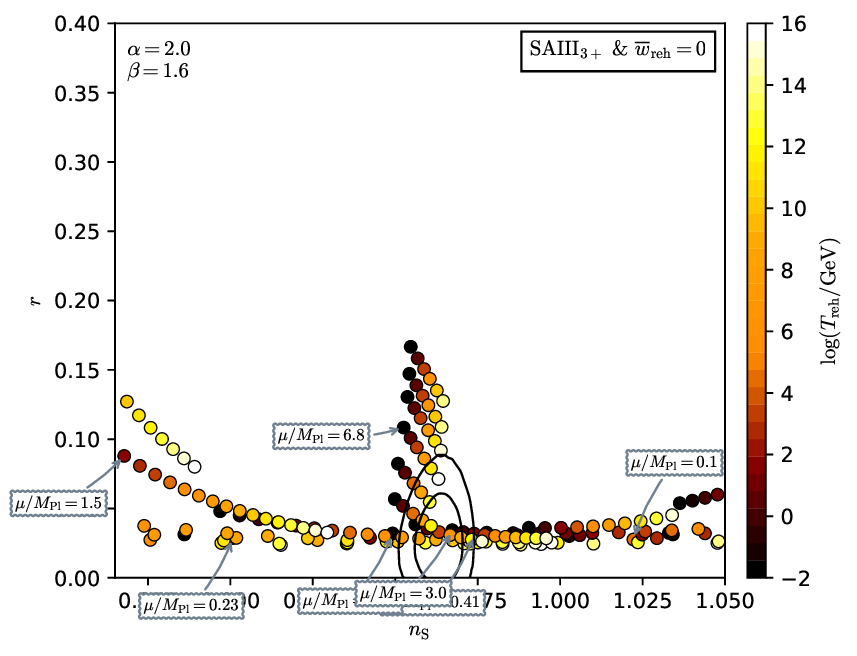}
\includegraphics[width=\wappfig,clip=true]{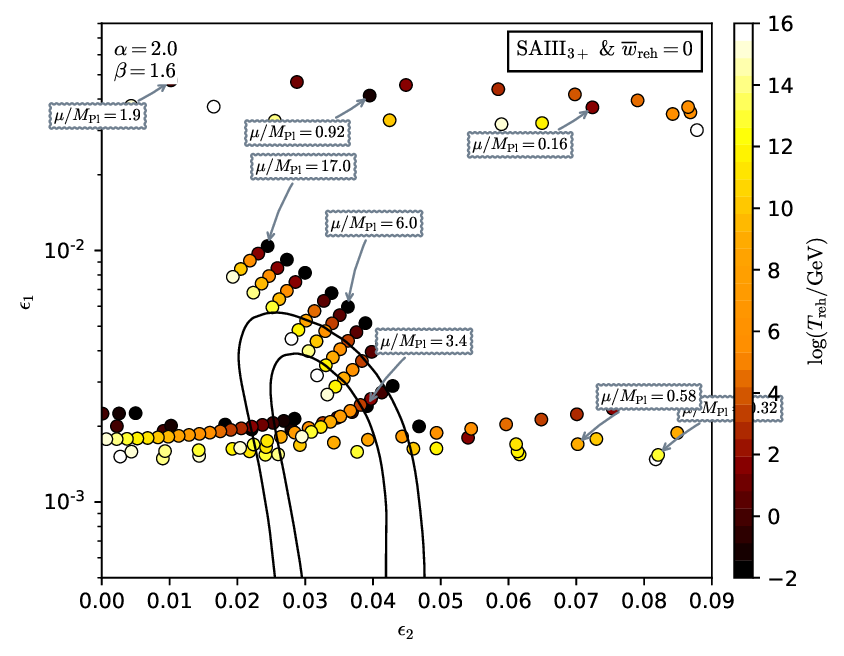}
\caption{Reheating consistent slow-roll predictions for the String
  Axion Inflation II models, in the SAIII3 regime and at a positive
  value of $\beta=1.6$. Predictions are represented in the plane
  $(\nS,r)$ (top panel) and in the plane $(\epsilon_1,\epsilon_2)$
  (bottom panel). The solid contours are the one and two-sigma {\data}
  confidence intervals (marginalized over second order slow-roll).}
\label{fig:CMBSAIII3p_1}
\end{center}
\end{figure}

\begin{figure}[H]
\begin{center}
\includegraphics[width=\wappfig,clip=true]{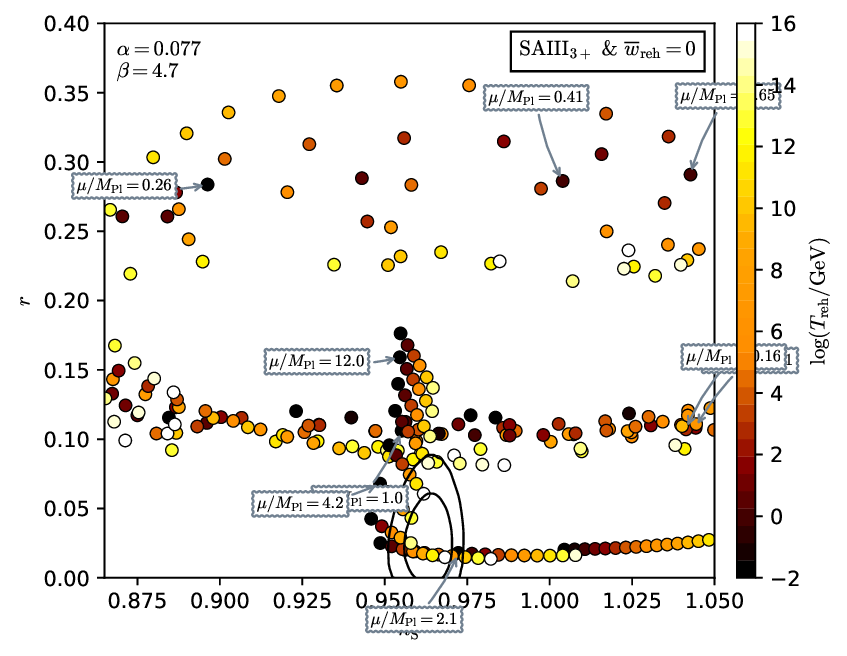}
\includegraphics[width=\wappfig,clip=true]{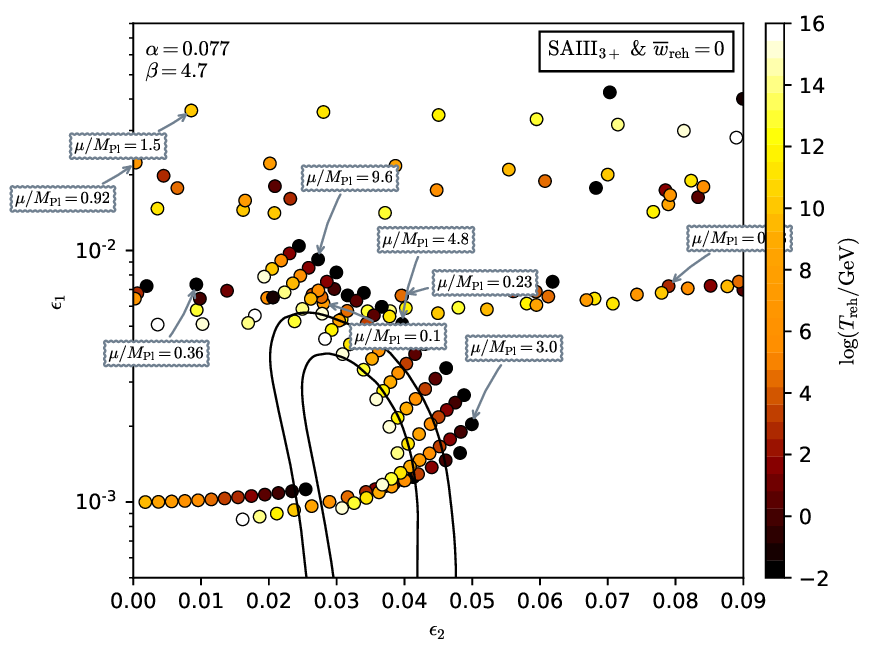}
\caption{Reheating consistent slow-roll predictions for the String
  Axion Inflation II models, in the SAIII3 regime and at a positive
  value of $\beta=4.7$. Predictions are represented in the plane
  $(\nS,r)$ (top panel) and in the plane $(\epsilon_1,\epsilon_2)$
  (bottom panel). The solid contours are the one and two-sigma {\data}
  confidence intervals (marginalized over second order slow-roll), see
  also \Fig{fig:CMBSAIII3p_3}.}
\label{fig:CMBSAIII3p_2}
\end{center}
\end{figure}

\begin{figure}[H]
\begin{center}
\includegraphics[width=\wappfig,clip=true]{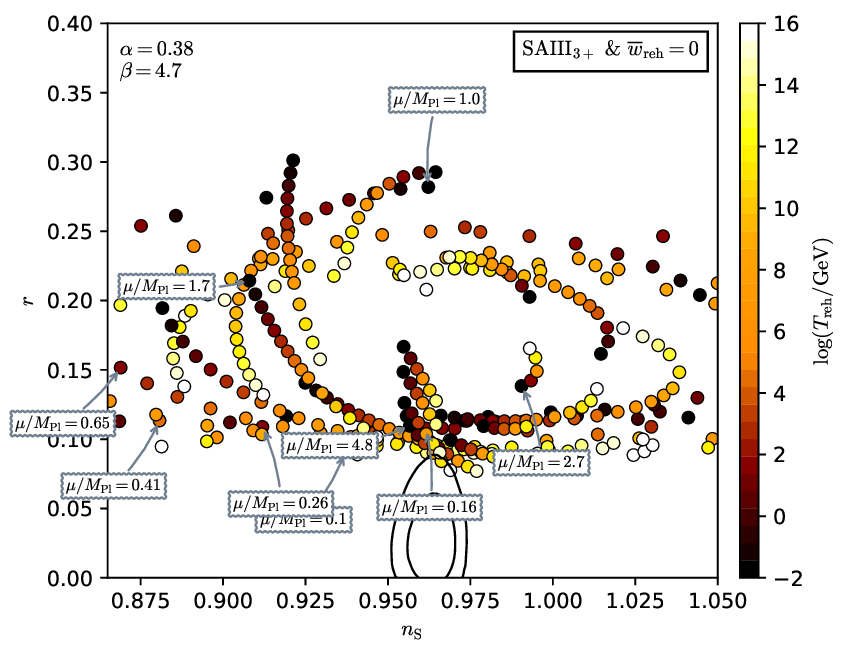}
\includegraphics[width=\wappfig,clip=true]{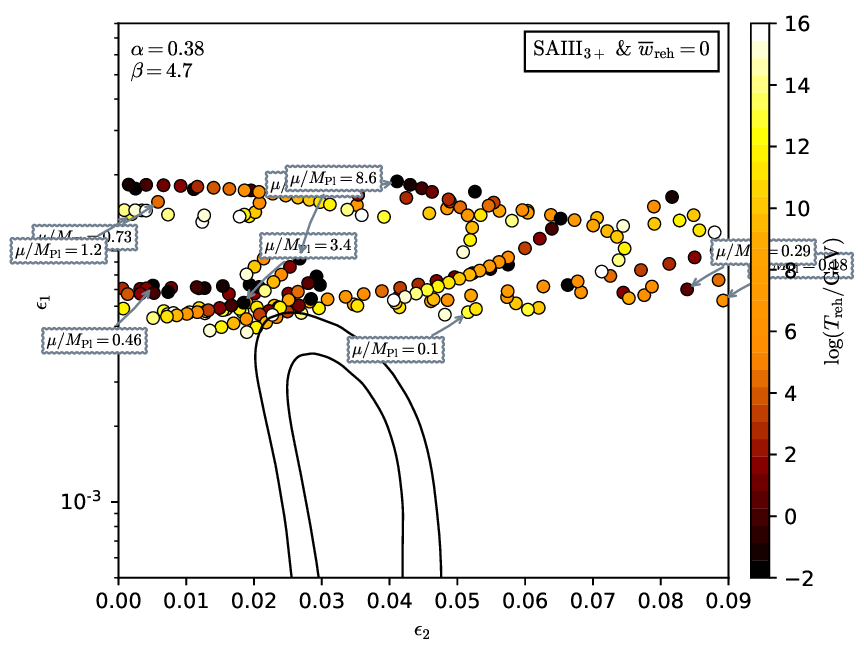}
\caption{Reheating consistent slow-roll predictions for the String
  Axion Inflation II models, in the SAIII3 regime and at a positive
  value of $\beta=4.7$. Predictions are represented in the plane
  $(\nS,r)$ (top panel) and in the plane $(\epsilon_1,\epsilon_2)$
  (bottom panel). The solid contours are the one and two-sigma {\data}
  confidence intervals (marginalized over second order slow-roll).}
\label{fig:CMBSAIII3p_3}
\end{center}
\end{figure}

\begin{figure}[H]
\begin{center}
\includegraphics[width=\wappfig,clip=true]{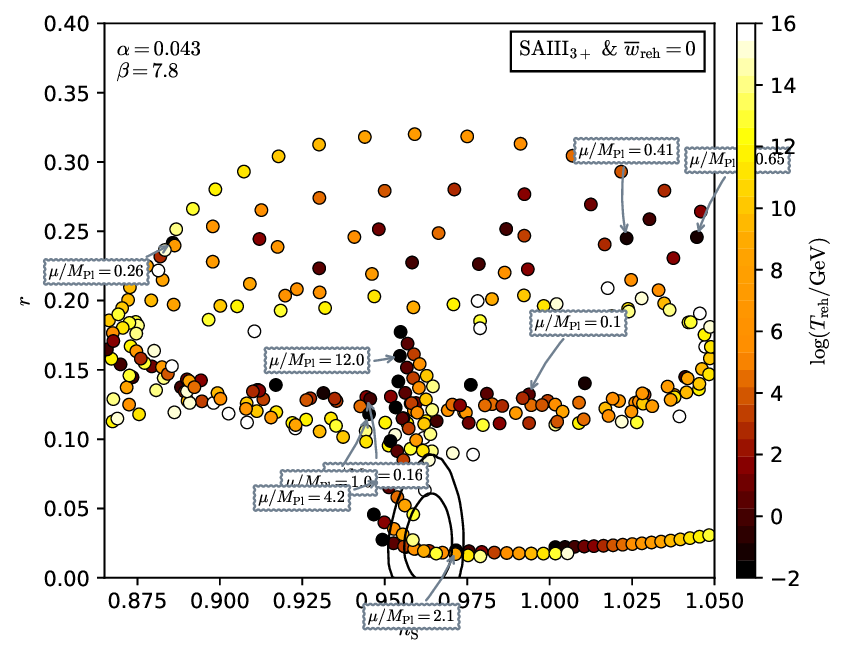}
\includegraphics[width=\wappfig,clip=true]{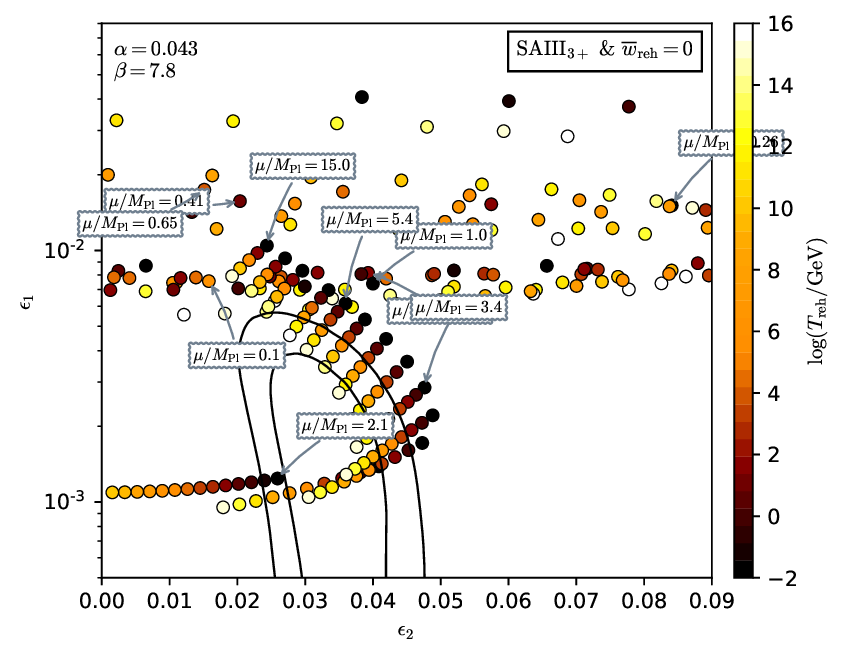}
\caption{Reheating consistent slow-roll predictions for the String
  Axion Inflation II models, in the SAIII3 regime and at a large positive
  value of $\beta=7.8$. Predictions are represented in the plane
  $(\nS,r)$ (top panel) and in the plane $(\epsilon_1,\epsilon_2)$
  (bottom panel). The solid contours are the one and two-sigma {\data}
  confidence intervals (marginalized over second order slow-roll), see
  also \Fig{fig:CMBSAIII3p_5}.}
\label{fig:CMBSAIII3p_4}
\end{center}
\end{figure}

\begin{figure}[H]
\begin{center}
\includegraphics[width=\wappfig,clip=true]{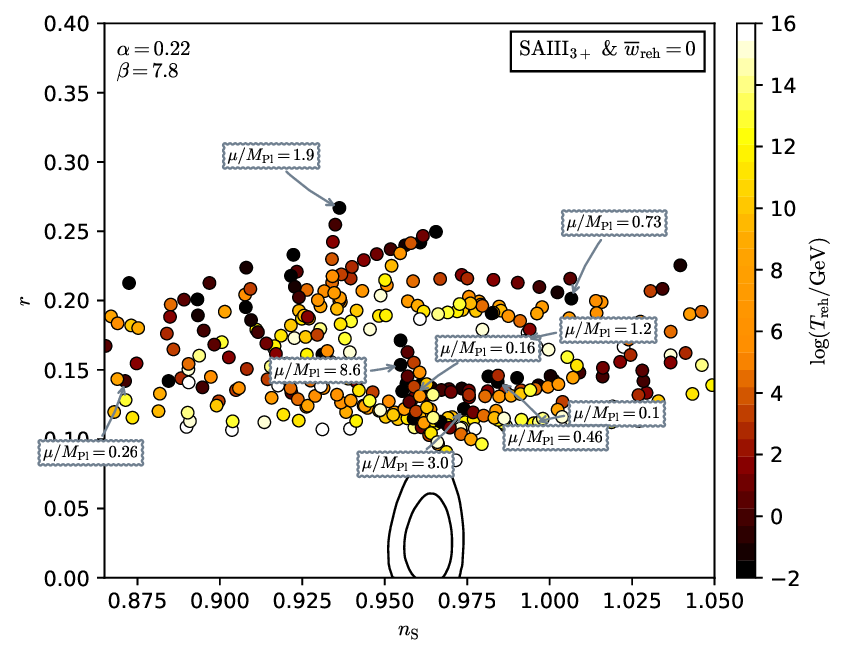}
\includegraphics[width=\wappfig,clip=true]{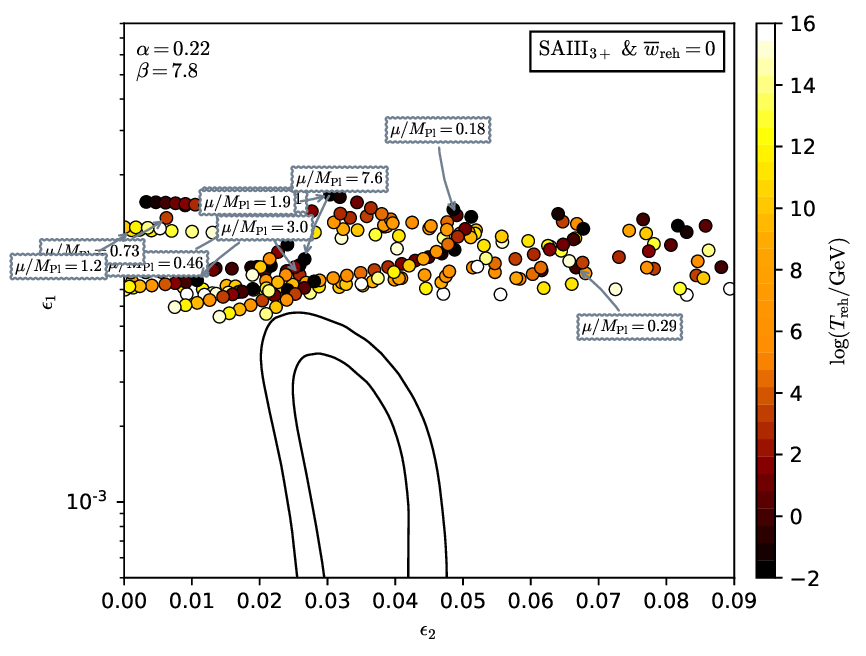}
\caption{Reheating consistent slow-roll predictions for the String
  Axion Inflation II models, in the SAIII3 regime and at a large
  positive value of $\beta=7.8$. Predictions are represented in the
  plane $(\nS,r)$ (top panel) and in the plane
  $(\epsilon_1,\epsilon_2)$ (bottom panel). The solid contours are the
  one and two-sigma {\data} confidence intervals (marginalized over
  second order slow-roll).}
\label{fig:CMBSAIII3p_5}
\end{center}
\end{figure}

\subsection{Radiatively Corrected Large Field Inflation 1 (\hyperref[sec:rclfi]{RCLFI1})}

\begin{figure}[H]
\begin{center}
\includegraphics[width=\wappfig,clip=true]{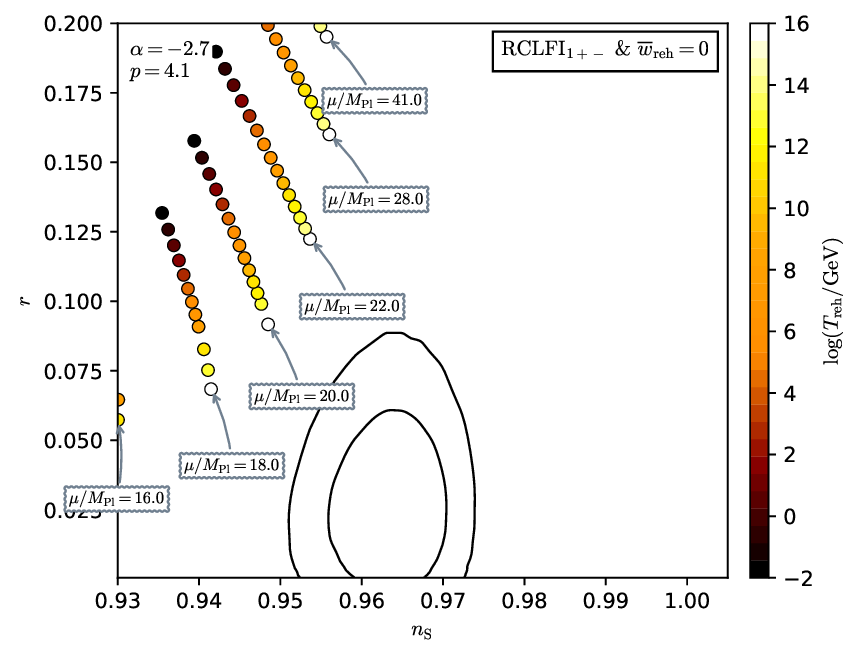}
\includegraphics[width=\wappfig,clip=true]{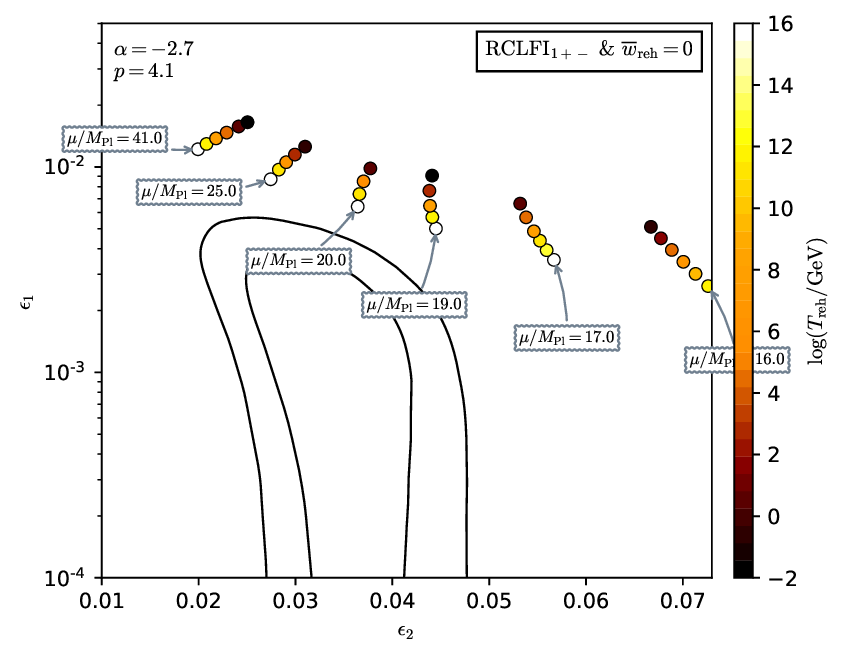}
\caption{Reheating consistent slow-roll predictions for the
  Radiatively Corrected Large Field Inflation models, in the RCLFI1
  regime, for $p>4$ and $\alpha < -[p(p-4)/4]
  \exp(2-p/4)<0$. Predictions are represented in the plane $(\nS,r)$
  (top panel) and in the plane $(\epsilon_1,\epsilon_2)$ (bottom
  panel) for various values of the field {\vev} $\mu$. The solid
  contours are the one and two-sigma {\data} confidence intervals
  (marginalized over second order slow-roll). See also
  Figs.~\ref{fig:CMBRCLFI1pm_1} to \ref{fig:CMBRCLFI1pm_3} for other
  values of $p$ and $\alpha$.}
\label{fig:CMBRCLFI1pm_0}
\end{center}
\end{figure}

\begin{figure}[H]
\begin{center}
\includegraphics[width=\wappfig,clip=true]{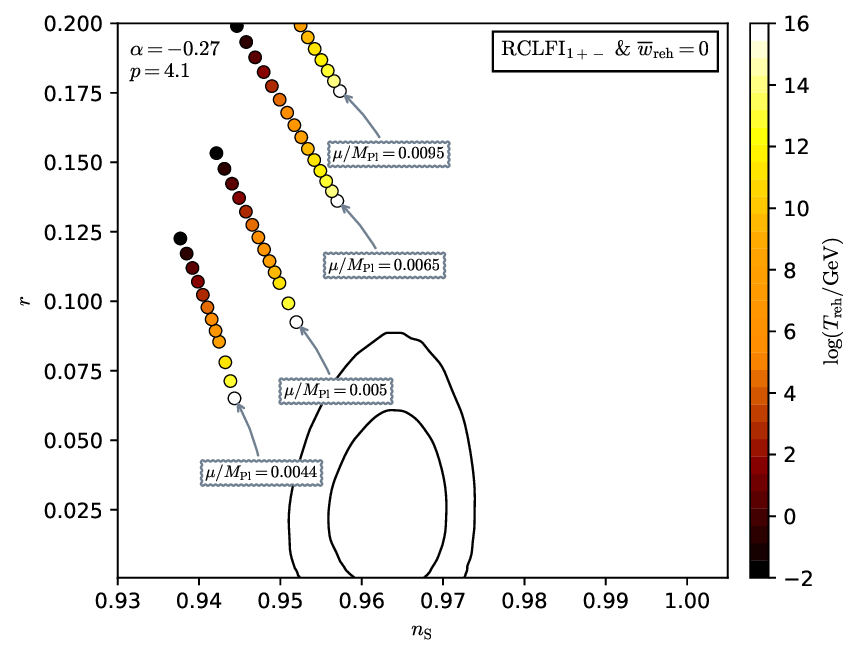}
\includegraphics[width=\wappfig,clip=true]{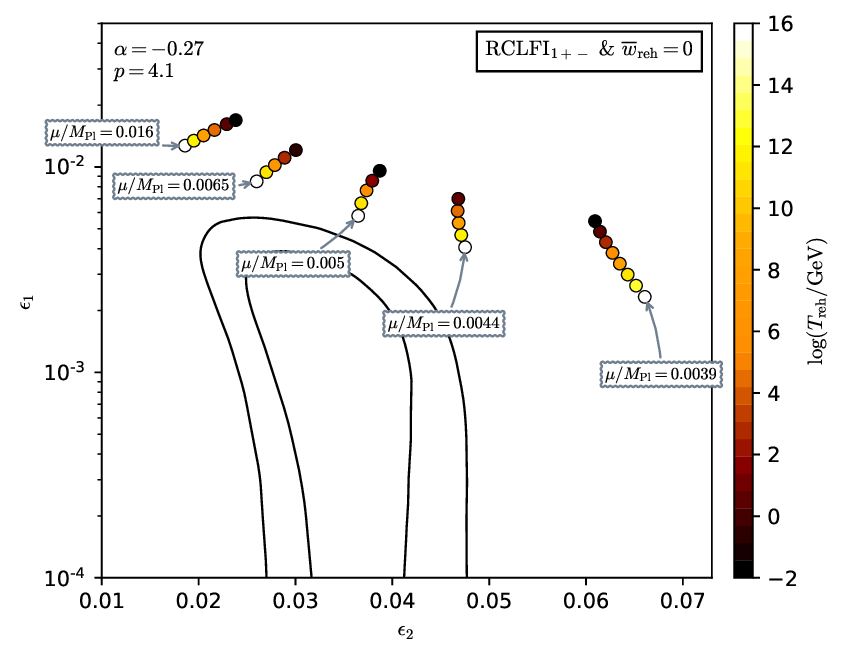}
\caption{Reheating consistent slow-roll predictions for the
  Radiatively Corrected Large Field Iinflation models, in the RCLFI1
  regime, for $p>4$ and $\alpha < -[p(p-4)/4]
  \exp(2-p/4)<0$. Predictions are represented in the plane $(\nS,r)$
  (top panel) and in the plane $(\epsilon_1,\epsilon_2)$. The solid
  contours are the one and two-sigma {\data} confidence intervals
  (marginalized over second order slow-roll).}
\label{fig:CMBRCLFI1pm_1}
\end{center}
\end{figure}

\begin{figure}[H]
\begin{center}
\includegraphics[width=\wappfig,clip=true]{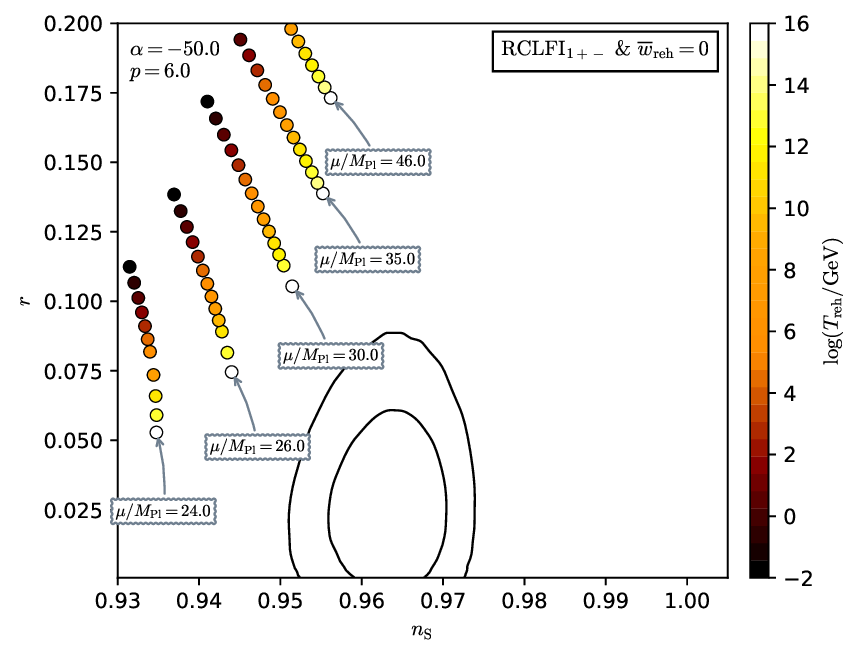}
\includegraphics[width=\wappfig,clip=true]{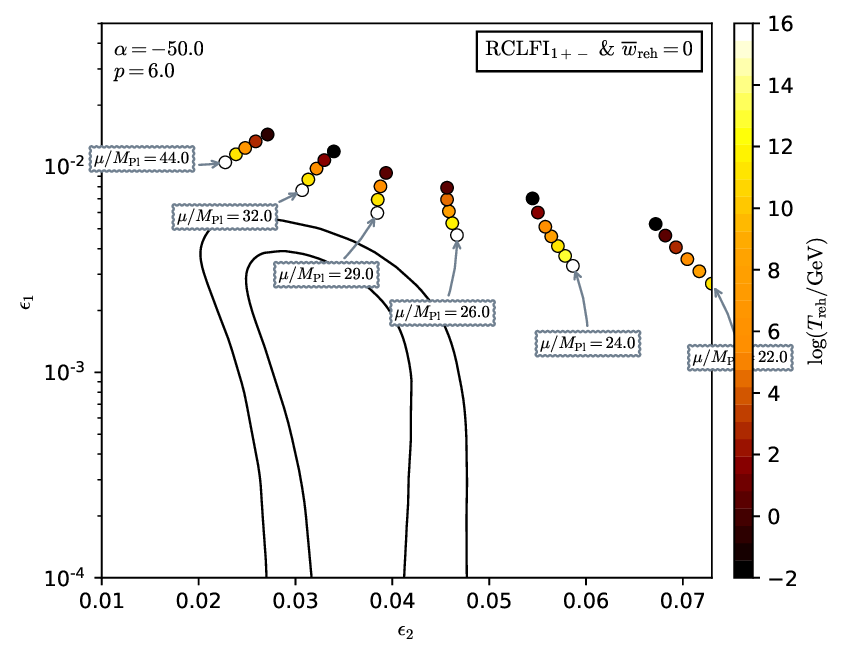}
\caption{Reheating consistent slow-roll predictions for the
  Radiatively Corrected Large Field Iinflation models, in the RCLFI1
  regime, for $p>4$ and $\alpha < -[p(p-4)/4]
  \exp(2-p/4)<0$. Predictions are represented in the plane $(\nS,r)$
  (top panel) and in the plane $(\epsilon_1,\epsilon_2)$. The solid
  contours are the one and two-sigma {\data} confidence intervals
  (marginalized over second order slow-roll).}
\label{fig:CMBRCLFI1pm_2}
\end{center}
\end{figure}

\begin{figure}[H]
\begin{center}
\includegraphics[width=\wappfig,clip=true]{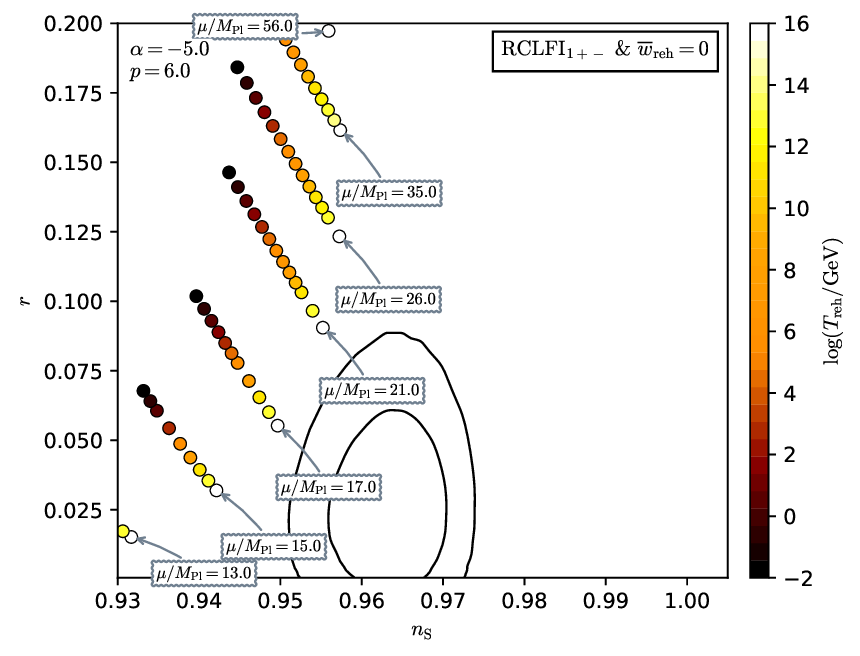}
\includegraphics[width=\wappfig,clip=true]{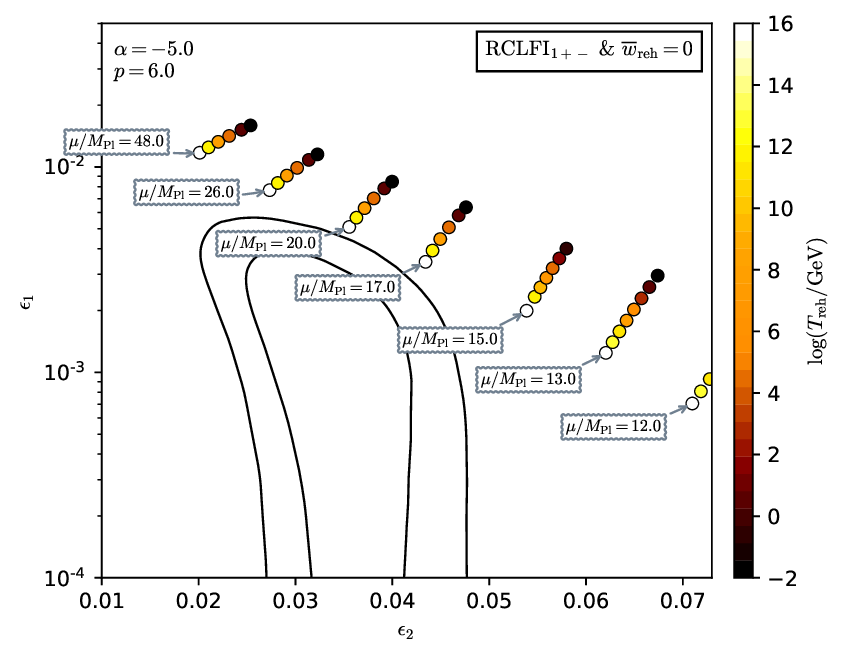}
\caption{Reheating consistent slow-roll predictions for the
  Radiatively Corrected Large Field Iinflation models, in the RCLFI1
  regime, for $p>4$ and $\alpha < -[p(p-4)/4]
  \exp(2-p/4)<0$. Predictions are represented in the plane $(\nS,r)$
  (top panel) and in the plane $(\epsilon_1,\epsilon_2)$. The solid
  contours are the one and two-sigma {\data} confidence intervals
  (marginalized over second order slow-roll).}
\label{fig:CMBRCLFI1pm_3}
\end{center}
\end{figure}

\begin{figure}[H]
\begin{center}
\includegraphics[width=\wappfig,clip=true]{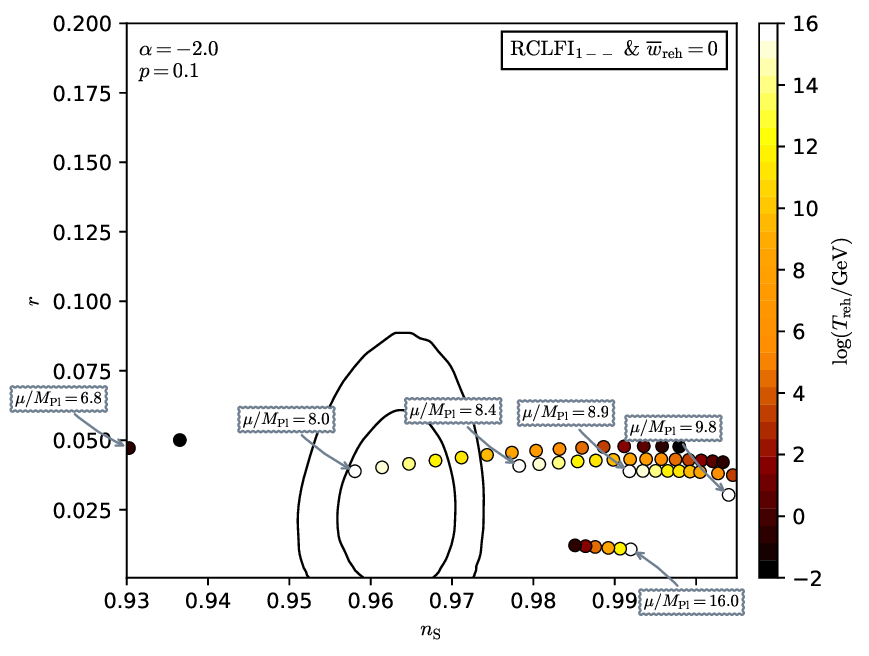}
\includegraphics[width=\wappfig,clip=true]{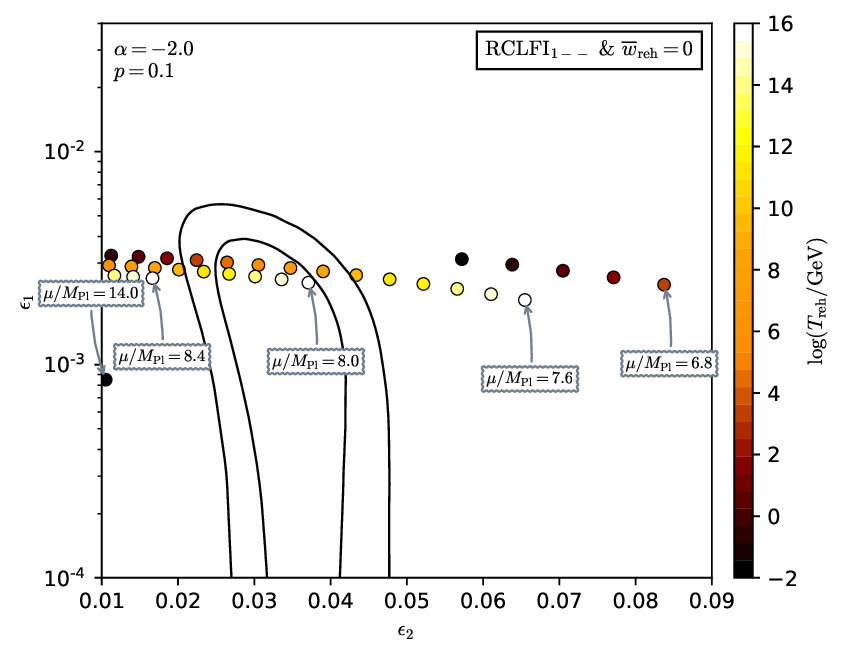}
\caption{Reheating consistent slow-roll predictions for the
  Radiatively Corrected Large Field Inflation models, in the RCLFI1
  regime, for $p<4$ and $\alpha<0$. Predictions are represented in the
  plane $(\nS,r)$ (top panel) and in the plane
  $(\epsilon_1,\epsilon_2)$ (bottom panel) for various values of the
  field {\vev} $\mu$. The solid contours are the one and two-sigma
  {\data} confidence intervals (marginalized over second order
  slow-roll). See also Figs.~\ref{fig:CMBRCLFI1mm_1} to
  \ref{fig:CMBRCLFI1mm_3} for other values of $p$ and $\alpha$.}
\label{fig:CMBRCLFI1mm_0}
\end{center}
\end{figure}

\begin{figure}[H]
\begin{center}
\includegraphics[width=\wappfig,clip=true]{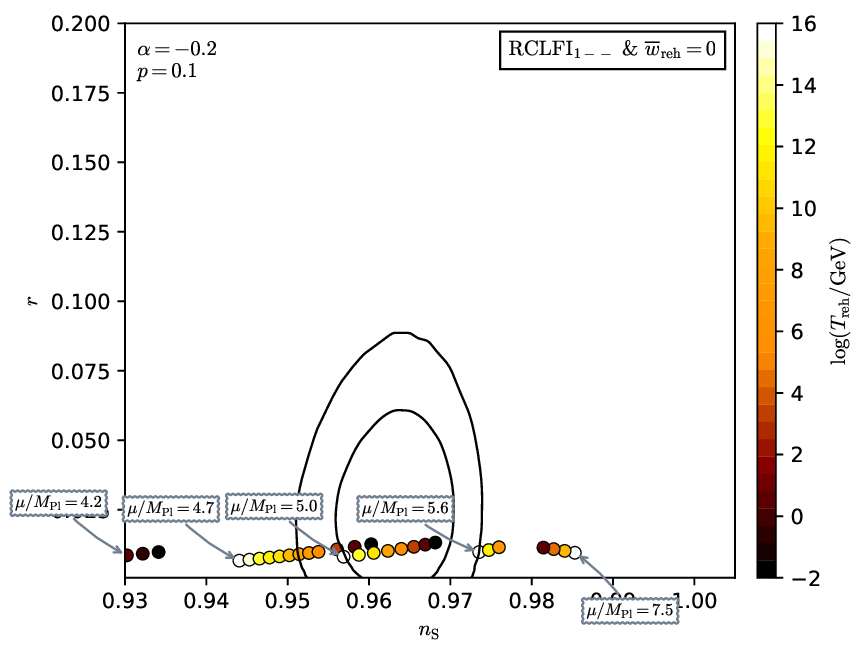}
\includegraphics[width=\wappfig,clip=true]{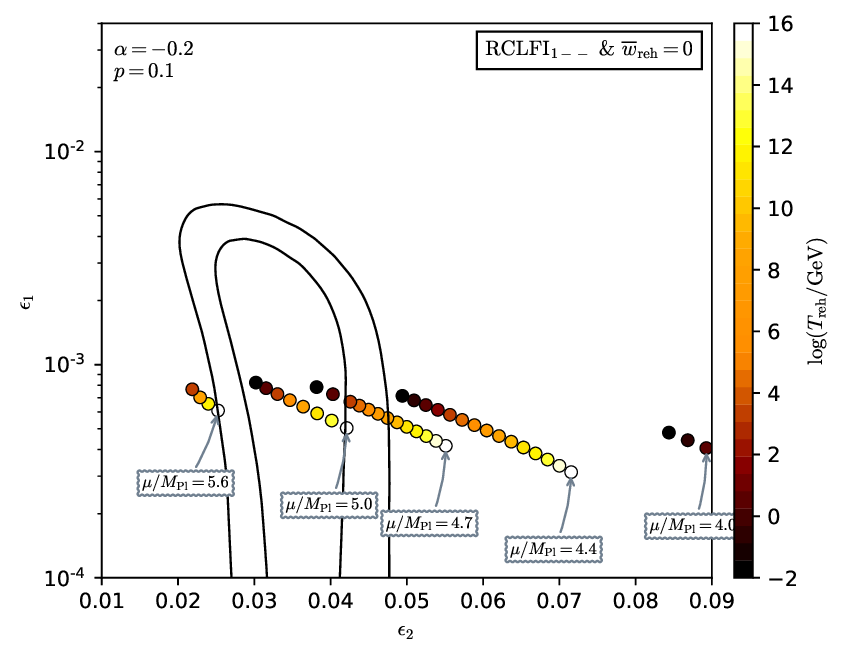}
\caption{Reheating consistent slow-roll predictions for the
  Radiatively Corrected Large Field Inflation models, in the RCLFI1
  regime, for $p<4$ and $\alpha<0$. Predictions are represented in the
  plane $(\nS,r)$ (top panel) and in the plane
  $(\epsilon_1,\epsilon_2)$ (bottom panel). The solid contours are the
  one and two-sigma {\data} confidence intervals (marginalized over
  second order slow-roll).}
\label{fig:CMBRCLFI1mm_1}
\end{center}
\end{figure}

\begin{figure}[H]
\begin{center}
\includegraphics[width=\wappfig,clip=true]{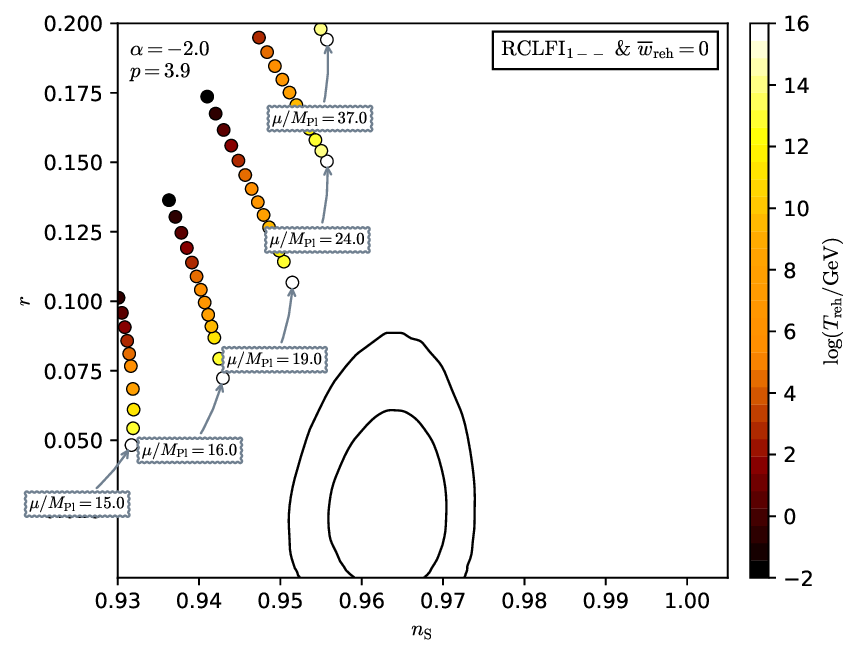}
\includegraphics[width=\wappfig,clip=true]{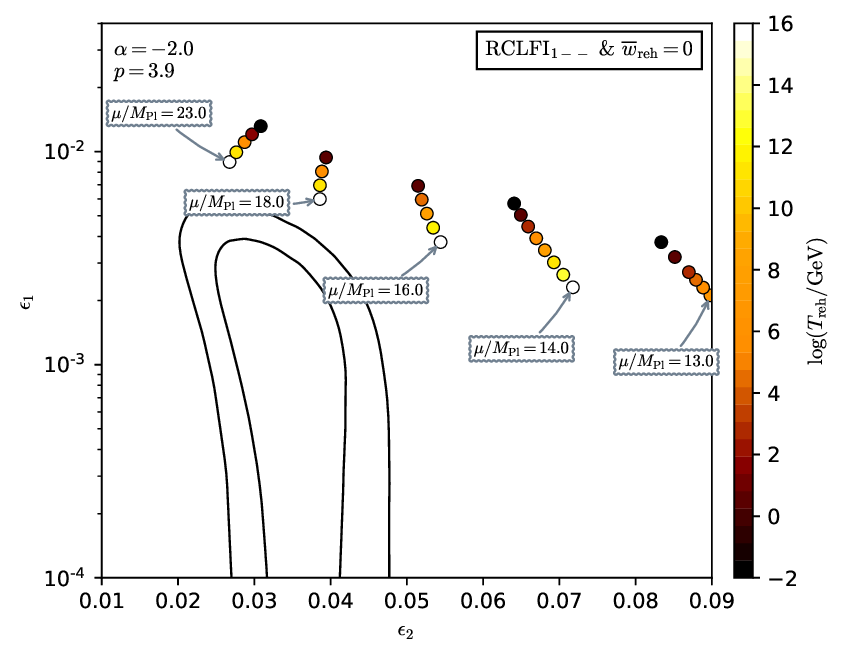}
\caption{Reheating consistent slow-roll predictions for the
  Radiatively Corrected Large Field Inflation models, in the RCLFI1
  regime, for $p<4$ and $\alpha<0$. Predictions are represented in the
  plane $(\nS,r)$ (top panel) and in the plane
  $(\epsilon_1,\epsilon_2)$ (bottom panel). The solid contours are the
  one and two-sigma {\data} confidence intervals (marginalized over
  second order slow-roll).}
\label{fig:CMBRCLFI1mm_2}
\end{center}
\end{figure}

\begin{figure}[H]
\begin{center}
\includegraphics[width=\wappfig,clip=true]{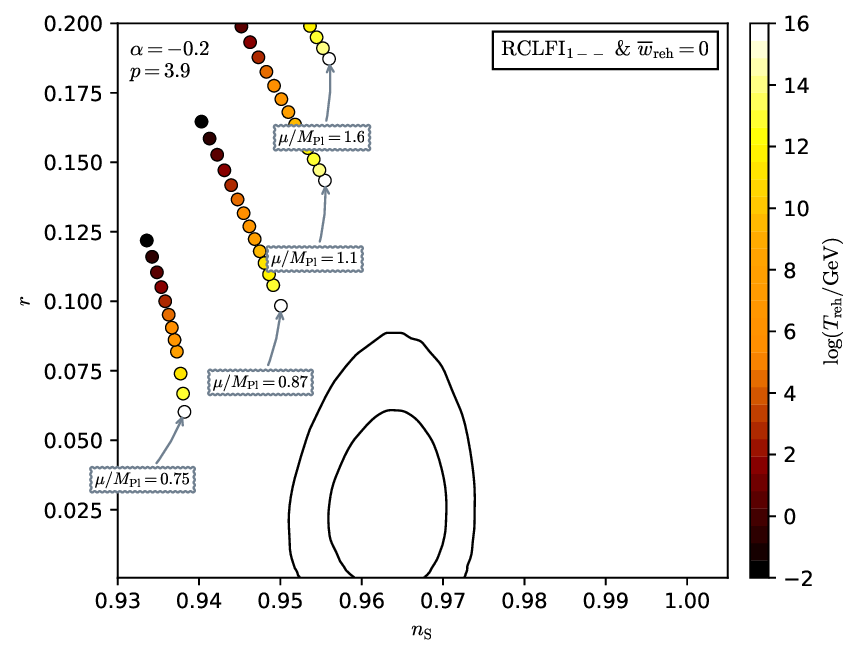}
\includegraphics[width=\wappfig,clip=true]{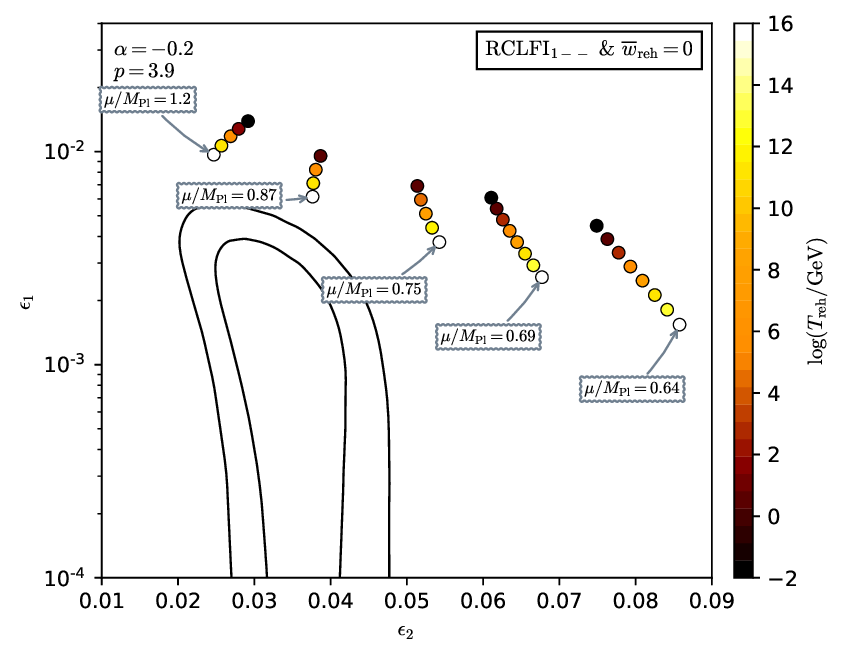}
\caption{Reheating consistent slow-roll predictions for the
  Radiatively Corrected Large Field Inflation models, in the RCLFI1
  regime, for $p<4$ and $\alpha<0$. Predictions are represented in the
  plane $(\nS,r)$ (top panel) and in the plane
  $(\epsilon_1,\epsilon_2)$ (bottom panel). The solid contours are the
  one and two-sigma {\data} confidence intervals (marginalized over
  second order slow-roll).}
\label{fig:CMBRCLFI1mm_3}
\end{center}
\end{figure}

\begin{figure}[H]
\begin{center}
\includegraphics[width=\wappfig,clip=true]{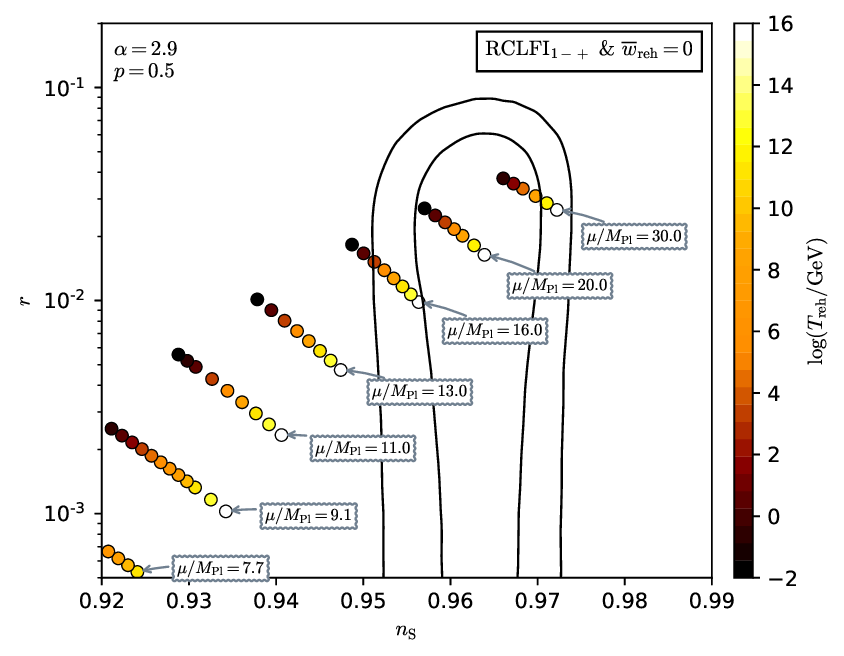}
\includegraphics[width=\wappfig,clip=true]{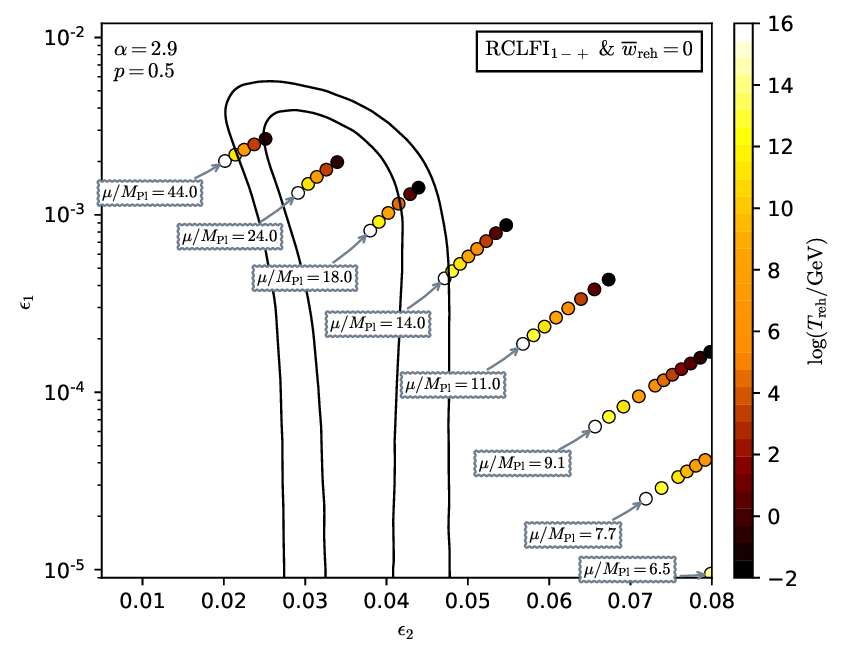}
\caption{Reheating consistent slow-roll predictions for the
  Radiatively Corrected Large Field Inflation models, in the RCLFI1
  regime, for $p<4$ and $\alpha>-[p(p-4)/4]\exp(2-4/p)>0$. Predictions
  are represented in the plane $(\nS,r)$ (top panel) and in the plane
  $(\epsilon_1,\epsilon_2)$ (bottom panel) for various values of the
  field {\vev} $\mu$. The solid contours are the one and two-sigma
  {\data} confidence intervals (marginalized over second order
  slow-roll). See also Figs.~\ref{fig:CMBRCLFI1mp_1} to
  \ref{fig:CMBRCLFI1mp_3} for other values of $p$ and $\alpha$.}
\label{fig:CMBRCLFI1mp_0}
\end{center}
\end{figure}

\begin{figure}[H]
\begin{center}
\includegraphics[width=\wappfig,clip=true]{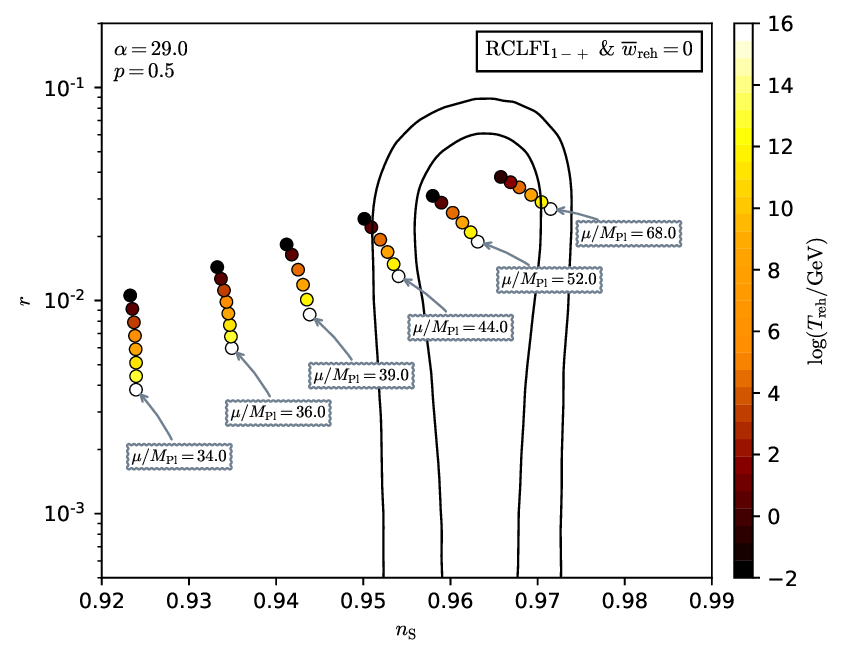}
\includegraphics[width=\wappfig,clip=true]{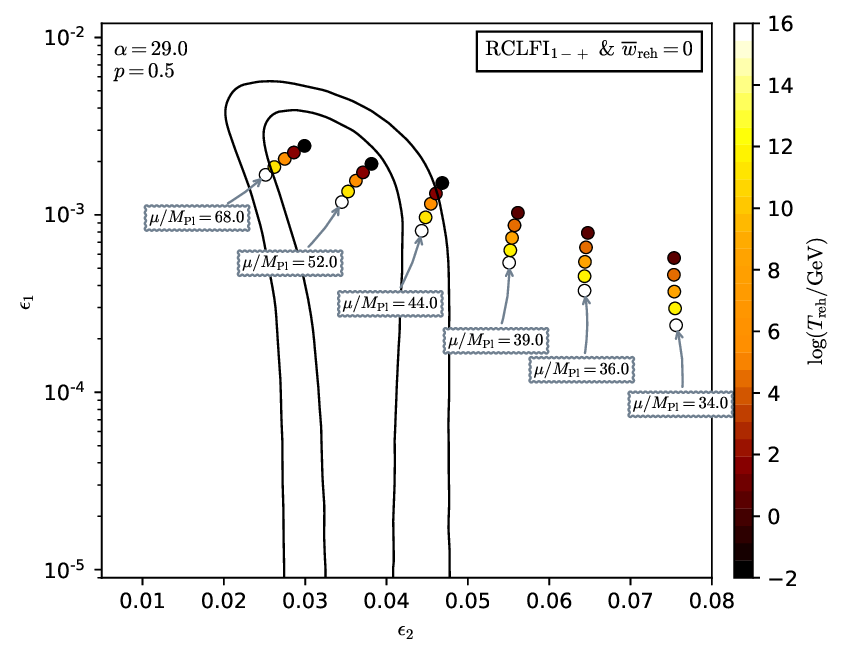}
\caption{Reheating consistent slow-roll predictions for the
  Radiatively Corrected Large Field Inflation models, in the RCLFI1
  regime, for $p<4$ and $\alpha>-[p(p-4)/4]\exp(2-4/p)>0$. Predictions
  are represented in the plane $(\nS,r)$ (top panel) and in the plane
  $(\epsilon_1,\epsilon_2)$ (bottom panel). The solid contours are the
  one and two-sigma {\data} confidence intervals (marginalized over
  second order slow-roll).}
\label{fig:CMBRCLFI1mp_1}
\end{center}
\end{figure}

\begin{figure}[H]
\begin{center}
\includegraphics[width=\wappfig,clip=true]{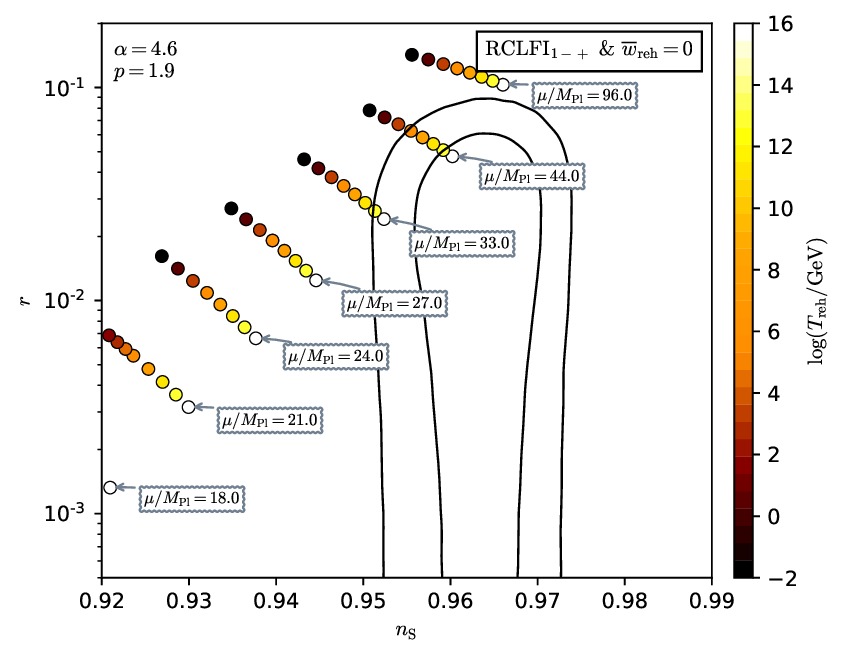}
\includegraphics[width=\wappfig,clip=true]{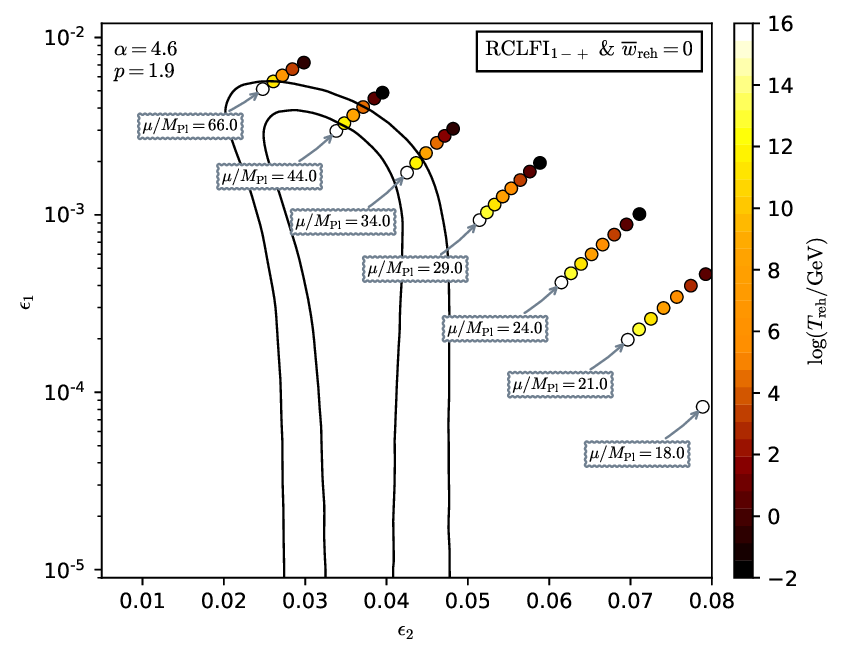}
\caption{Reheating consistent slow-roll predictions for the
  Radiatively Corrected Large Field Inflation models, in the RCLFI1
  regime, for $p<4$ and $\alpha>-[p(p-4)/4]\exp(2-4/p)>0$. Predictions
  are represented in the plane $(\nS,r)$ (top panel) and in the plane
  $(\epsilon_1,\epsilon_2)$ (bottom panel). The solid contours are the
  one and two-sigma {\data} confidence intervals (marginalized over
  second order slow-roll).}
\label{fig:CMBRCLFI1mp_2}
\end{center}
\end{figure}

\begin{figure}[H]
\begin{center}
\includegraphics[width=\wappfig,clip=true]{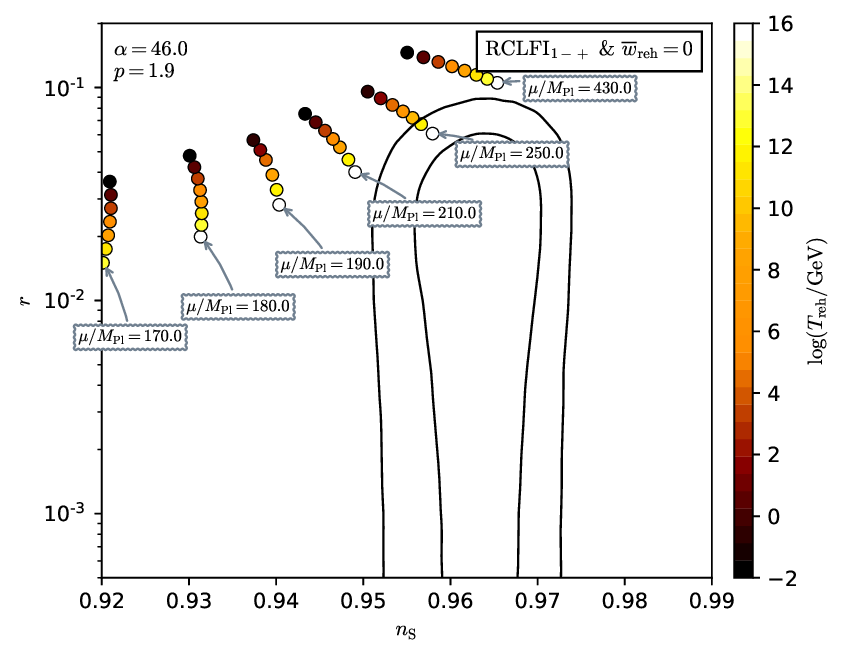}
\includegraphics[width=\wappfig,clip=true]{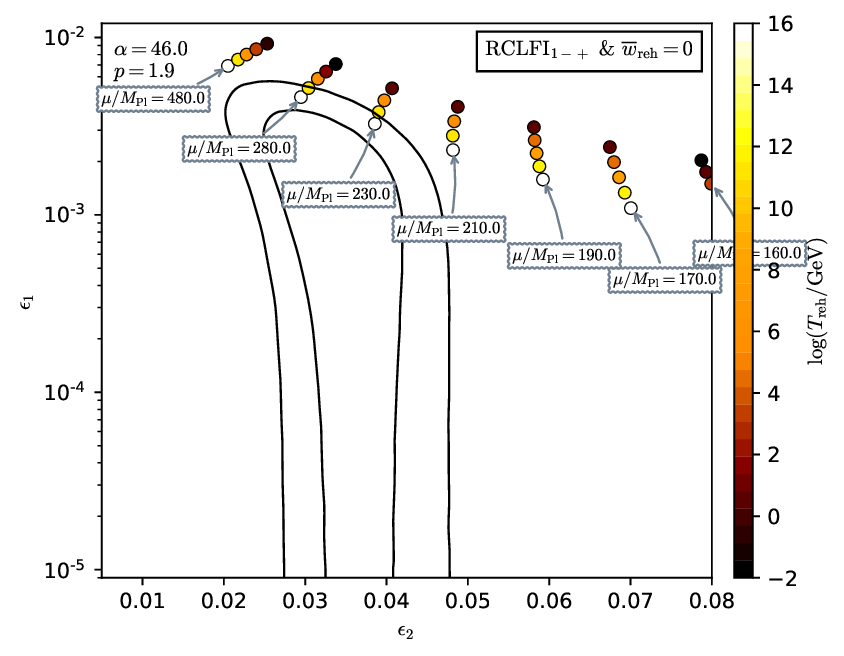}
\caption{Reheating consistent slow-roll predictions for the
  Radiatively Corrected Large Field Inflation models, in the RCLFI1
  regime, for $p<4$ and $\alpha>-[p(p-4)/4]\exp(2-4/p)>0$. Predictions
  are represented in the plane $(\nS,r)$ (top panel) and in the plane
  $(\epsilon_1,\epsilon_2)$ (bottom panel). The solid contours are the
  one and two-sigma {\data} confidence intervals (marginalized over
  second order slow-roll).}
\label{fig:CMBRCLFI1mp_3}
\end{center}
\end{figure}

\subsection{Radiatively Corrected Large Field Inflation 2 (\hyperref[sec:rclfi]{RCLFI2})}

\begin{figure}[H]
\begin{center}
\includegraphics[width=\wappfig,clip=true]{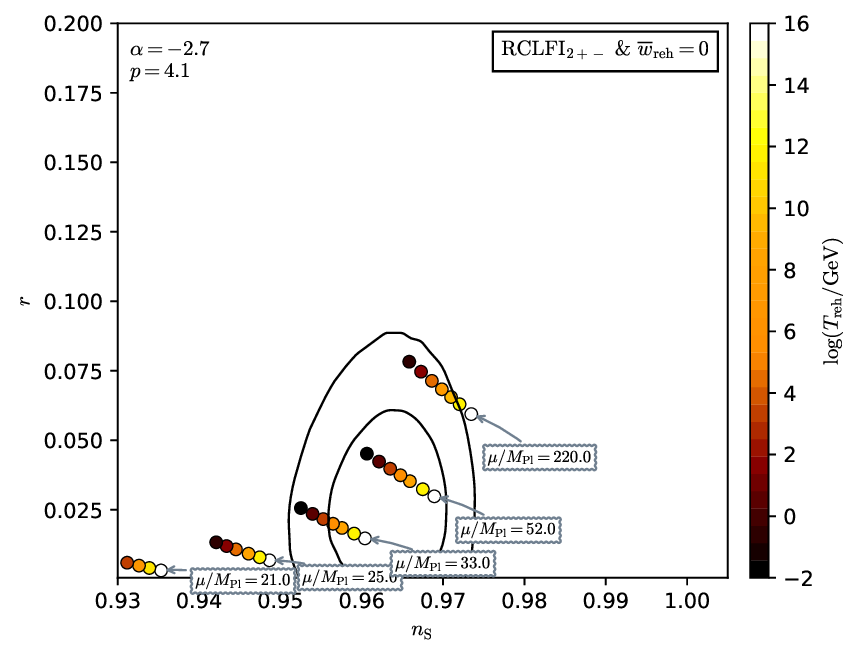}
\includegraphics[width=\wappfig,clip=true]{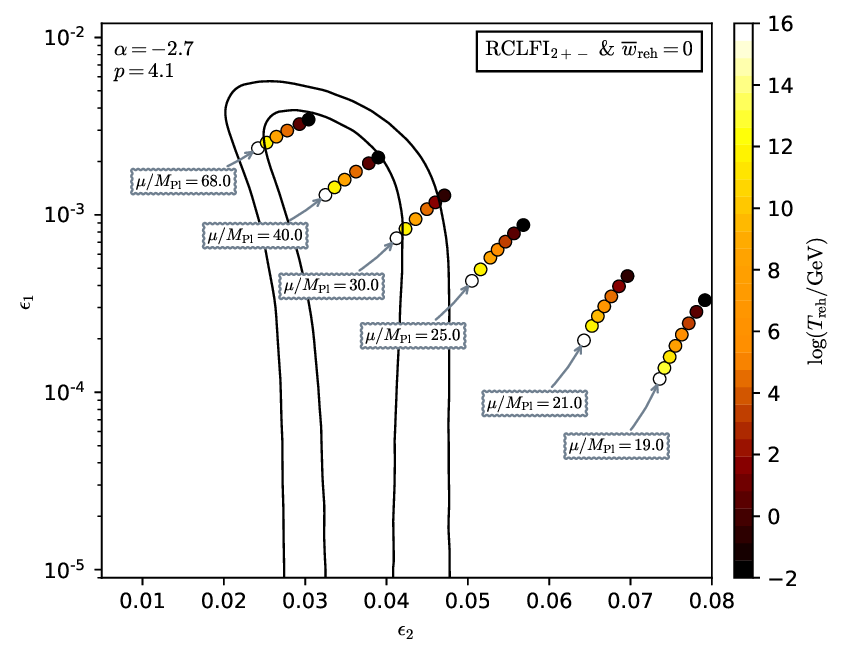}
\caption{Reheating consistent slow-roll predictions for the
  Radiatively Corrected Large Field Inflation models, in the RCLFI2
  regime, for $p>4$ and $\alpha < -e(p-4)<0$. Predictions are
  represented in the plane $(\nS,r)$ (top panel) and in the plane
  $(\epsilon_1,\epsilon_2)$ (bottom panel) for various values of the
  field {\vev} $\mu$. The solid contours are the one and two-sigma
  {\data} confidence intervals (marginalized over second order
  slow-roll). See also Figs.~\ref{fig:CMBRCLFI2pm_1} to
  \ref{fig:CMBRCLFI2pm_3} for other values of $p$ and $\alpha$.}
\label{fig:CMBRCLFI2pm_0}
\end{center}
\end{figure}

\begin{figure}[H]
\begin{center}
\includegraphics[width=\wappfig,clip=true]{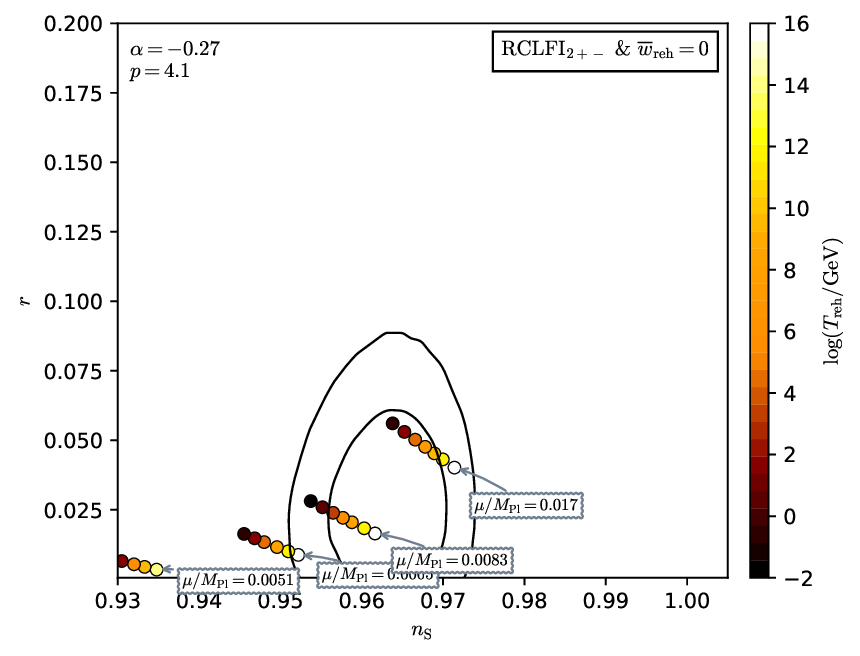}
\includegraphics[width=\wappfig,clip=true]{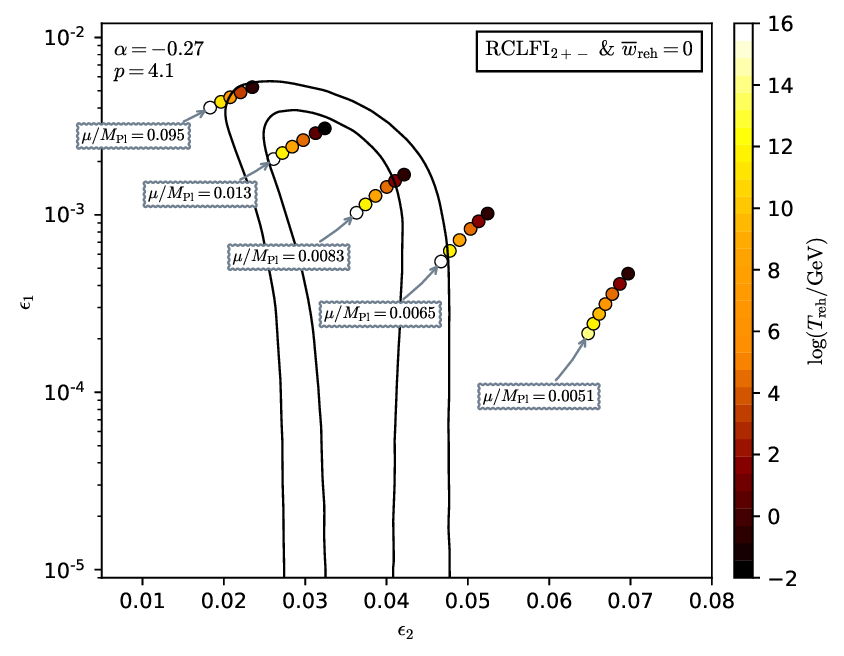}
\caption{Reheating consistent slow-roll predictions for the
  Radiatively Corrected Large Field Inflation models, in the RCLFI2
  regime, for $p>4$ and $\alpha < -e(p-4)<0$. Predictions are
  represented in the plane $(\nS,r)$ (top panel) and in the plane
  $(\epsilon_1,\epsilon_2)$ (bottom panel). The solid contours are the
  one and two-sigma {\data} confidence intervals (marginalized over
  second order slow-roll).}
\label{fig:CMBRCLFI2pm_1}
\end{center}
\end{figure}

\begin{figure}[H]
\begin{center}
\includegraphics[width=\wappfig,clip=true]{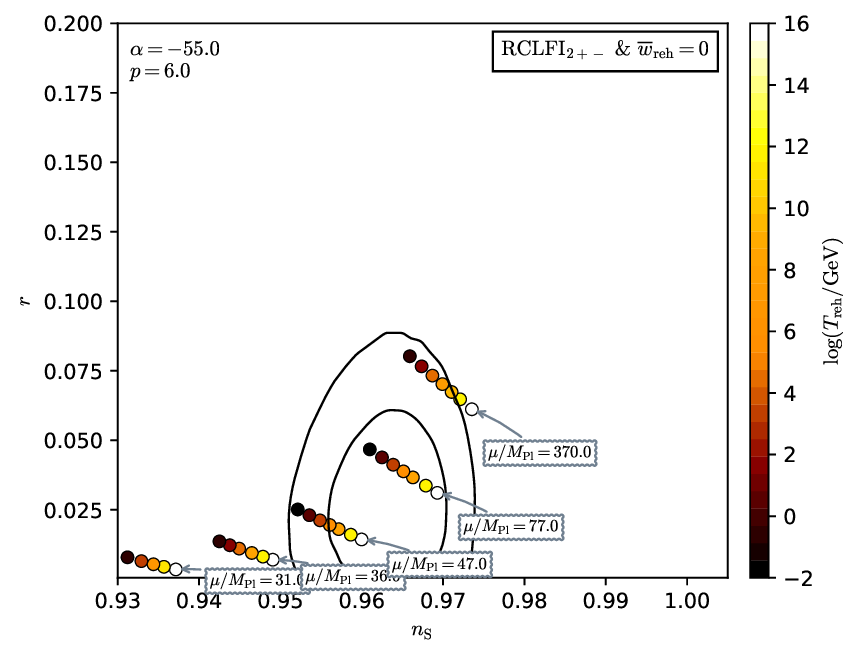}
\includegraphics[width=\wappfig,clip=true]{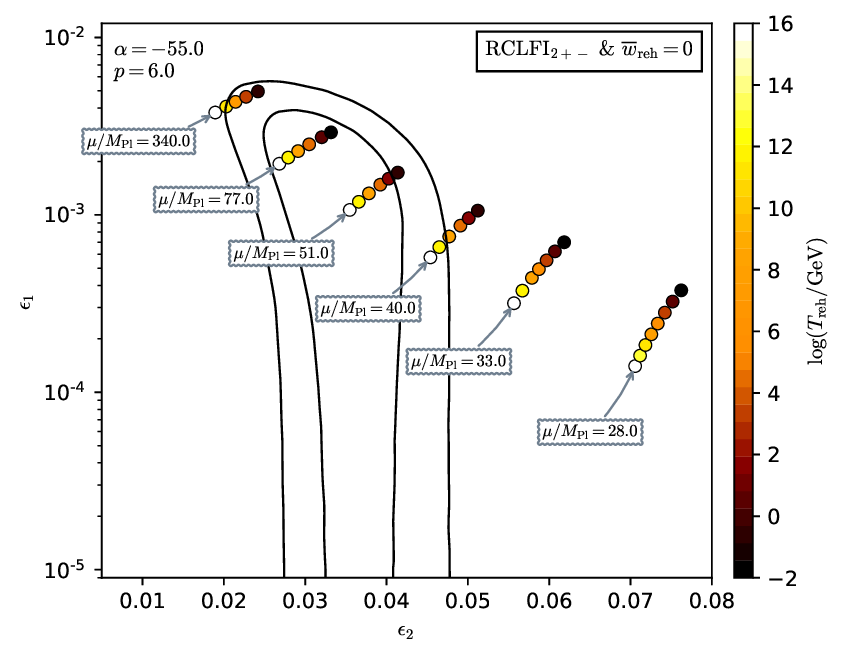}
\caption{Reheating consistent slow-roll predictions for the
  Radiatively Corrected Large Field Inflation models, in the RCLFI2
  regime, for $p>4$ and $\alpha < -e(p-4)<0$. Predictions are
  represented in the plane $(\nS,r)$ (top panel) and in the plane
  $(\epsilon_1,\epsilon_2)$ (bottom panel). The solid contours are the
  one and two-sigma {\data} confidence intervals (marginalized over
  second order slow-roll).}
\label{fig:CMBRCLFI2pm_2}
\end{center}
\end{figure}

\begin{figure}[H]
\begin{center}
\includegraphics[width=\wappfig,clip=true]{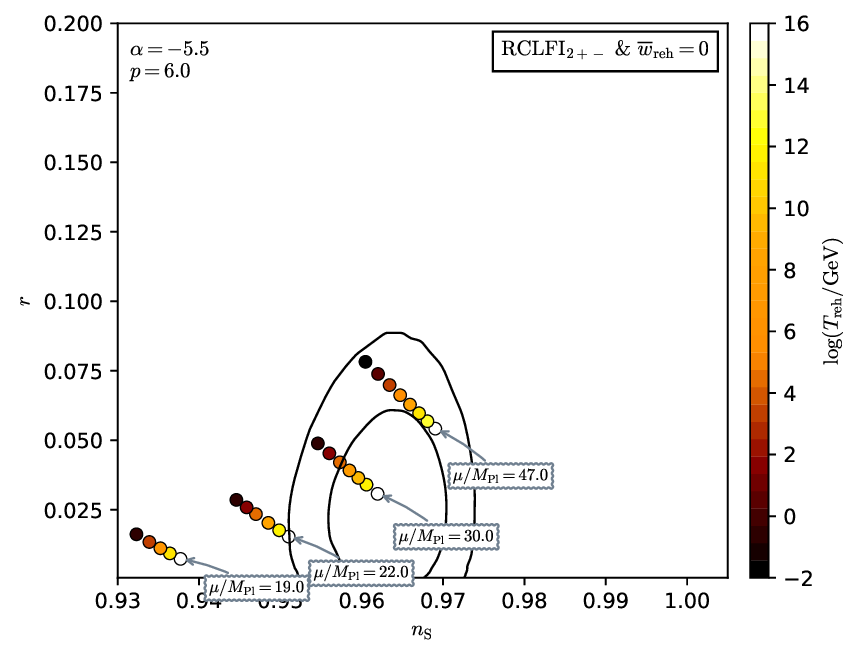}
\includegraphics[width=\wappfig,clip=true]{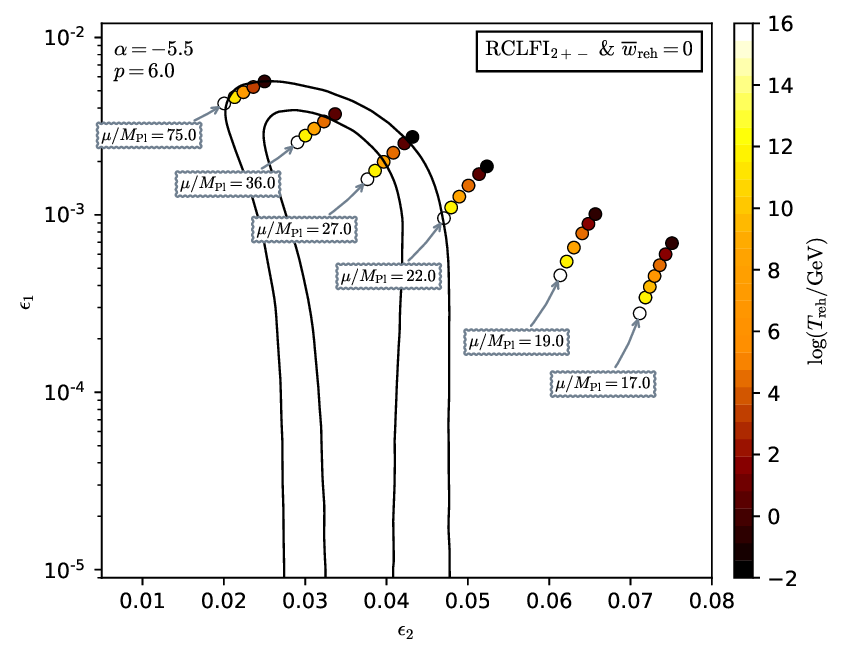}
\caption{Reheating consistent slow-roll predictions for the
  Radiatively Corrected Large Field Inflation models, in the RCLFI2
  regime, for $p>4$ and $\alpha < -e(p-4)<0$. Predictions are
  represented in the plane $(\nS,r)$ (top panel) and in the plane
  $(\epsilon_1,\epsilon_2)$ (bottom panel). The solid contours are the
  one and two-sigma {\data} confidence intervals (marginalized over
  second order slow-roll).}
\label{fig:CMBRCLFI2pm_3}
\end{center}
\end{figure}

\begin{figure}[H]
\begin{center}
\includegraphics[width=\wappfig,clip=true]{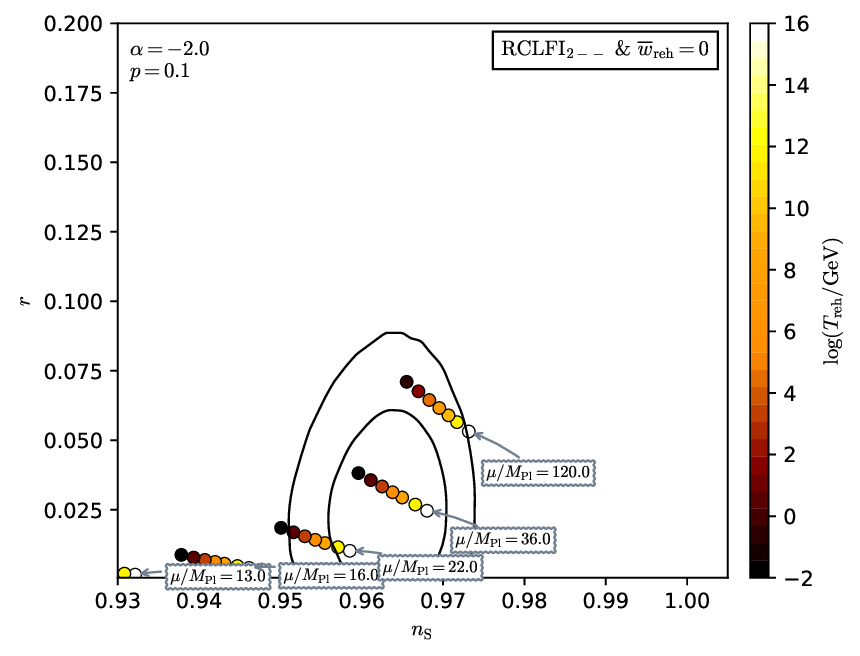}
\includegraphics[width=\wappfig,clip=true]{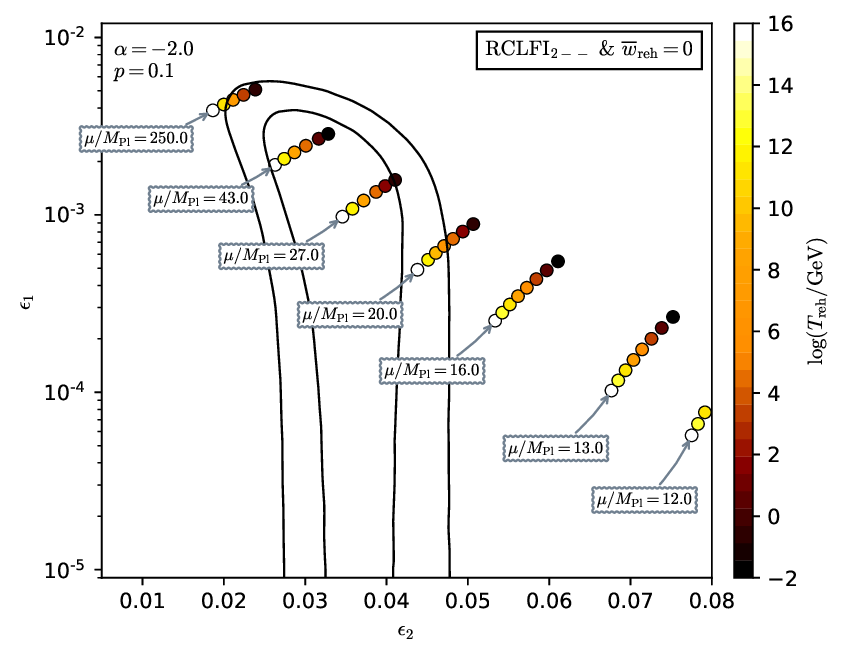}
\caption{Reheating consistent slow-roll predictions for the
  Radiatively Corrected Large Field Inflation models, in the RCLFI2
  regime, for $p<4$ and $\alpha<0$. Predictions are
  represented in the plane $(\nS,r)$ (top panel) and in the plane
  $(\epsilon_1,\epsilon_2)$ (bottom panel) for various values of the
  field {\vev} $\mu$. The solid contours are the one and two-sigma
  {\data} confidence intervals (marginalized over second order
  slow-roll). See also Figs.~\ref{fig:CMBRCLFI2mm_1} to
  \ref{fig:CMBRCLFI2mm_3} for other values of $p$ and $\alpha$.}
\label{fig:CMBRCLFI2mm_0}
\end{center}
\end{figure}

\begin{figure}[H]
\begin{center}
\includegraphics[width=\wappfig,clip=true]{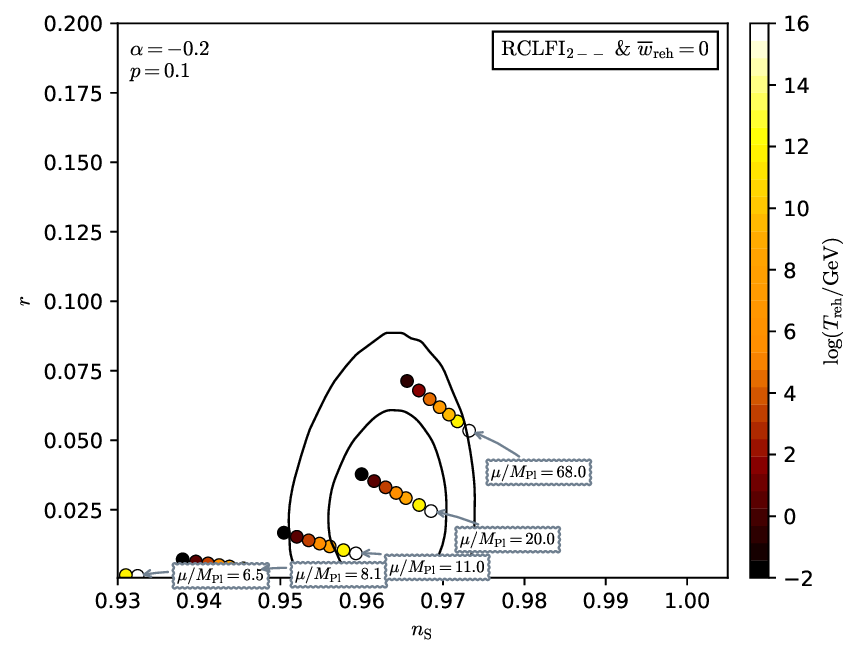}
\includegraphics[width=\wappfig,clip=true]{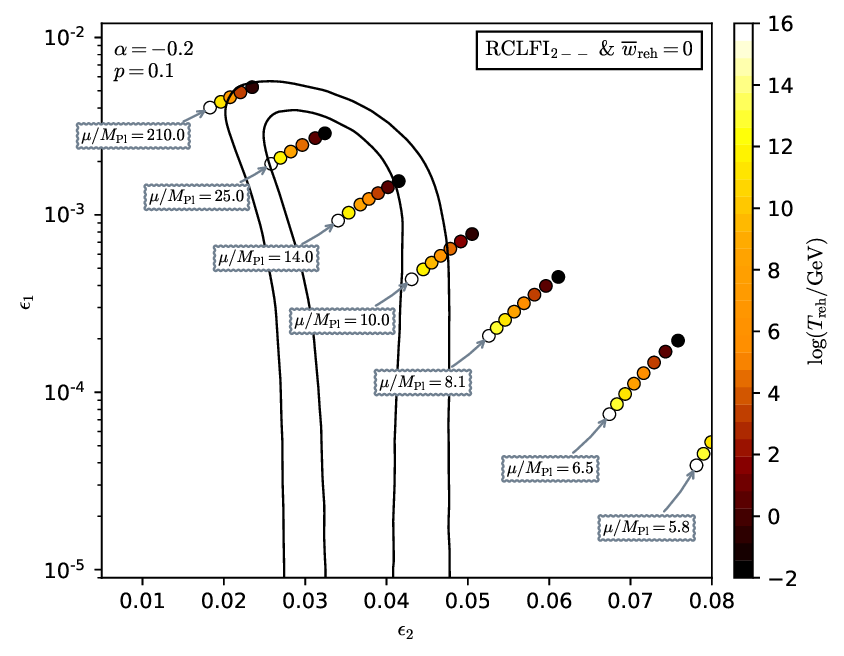}
\caption{Reheating consistent slow-roll predictions for the
  Radiatively Corrected Large Field Inflation models, in the RCLFI2
  regime, for $p<4$ and $\alpha<0$. Predictions are represented in the
  plane $(\nS,r)$ (top panel) and in the plane
  $(\epsilon_1,\epsilon_2)$ (bottom panel). The solid contours are the
  one and two-sigma {\data} confidence intervals (marginalized over
  second order slow-roll).}
\label{fig:CMBRCLFI2mm_1}
\end{center}
\end{figure}

\begin{figure}[H]
\begin{center}
\includegraphics[width=\wappfig,clip=true]{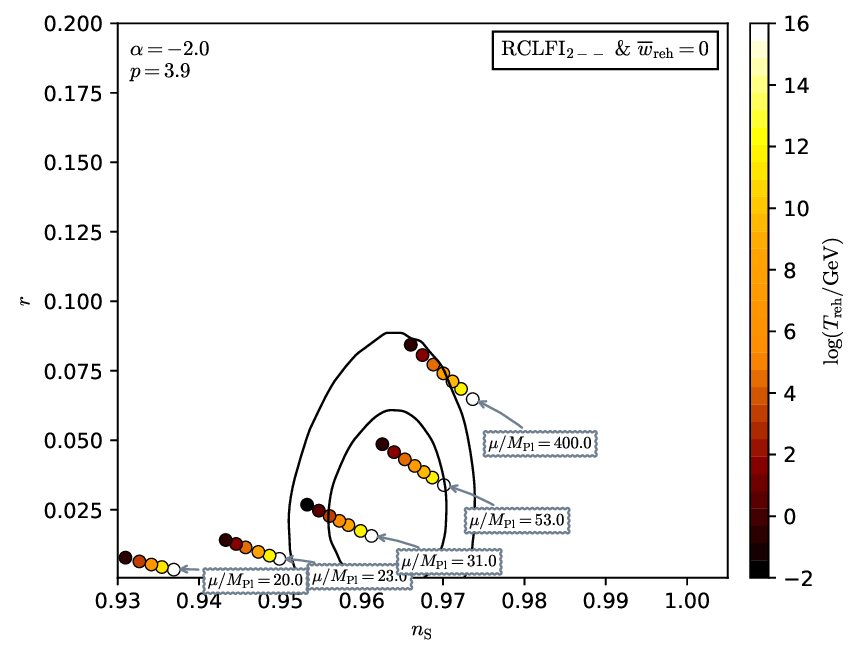}
\includegraphics[width=\wappfig,clip=true]{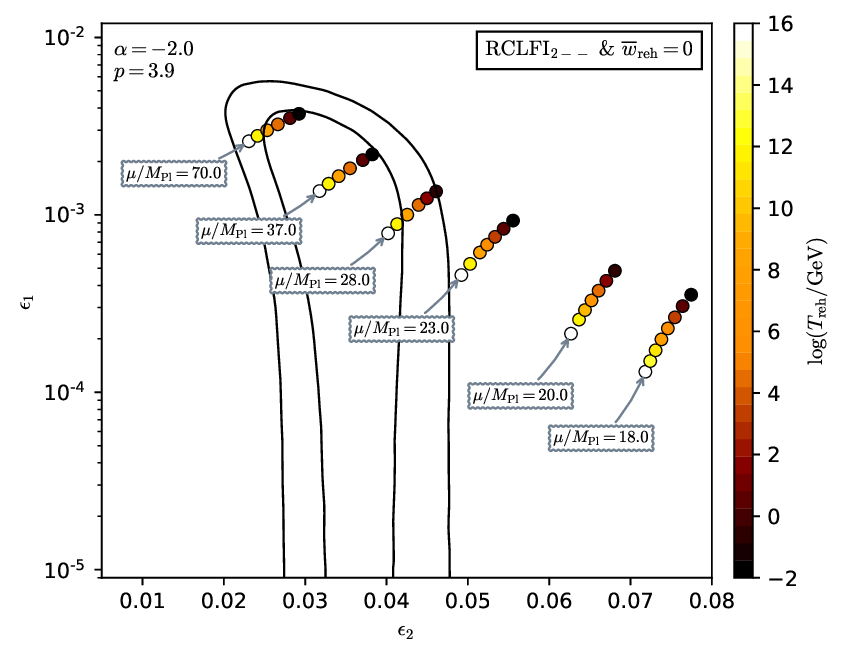}
\caption{Reheating consistent slow-roll predictions for the
  Radiatively Corrected Large Field Inflation models, in the RCLFI2
  regime, for $p<4$ and $\alpha<0$. Predictions are represented in the
  plane $(\nS,r)$ (top panel) and in the plane
  $(\epsilon_1,\epsilon_2)$ (bottom panel). The solid contours are the
  one and two-sigma {\data} confidence intervals (marginalized over
  second order slow-roll).}
\label{fig:CMBRCLFI2mm_2}
\end{center}
\end{figure}

\begin{figure}[H]
\begin{center}
\includegraphics[width=\wappfig,clip=true]{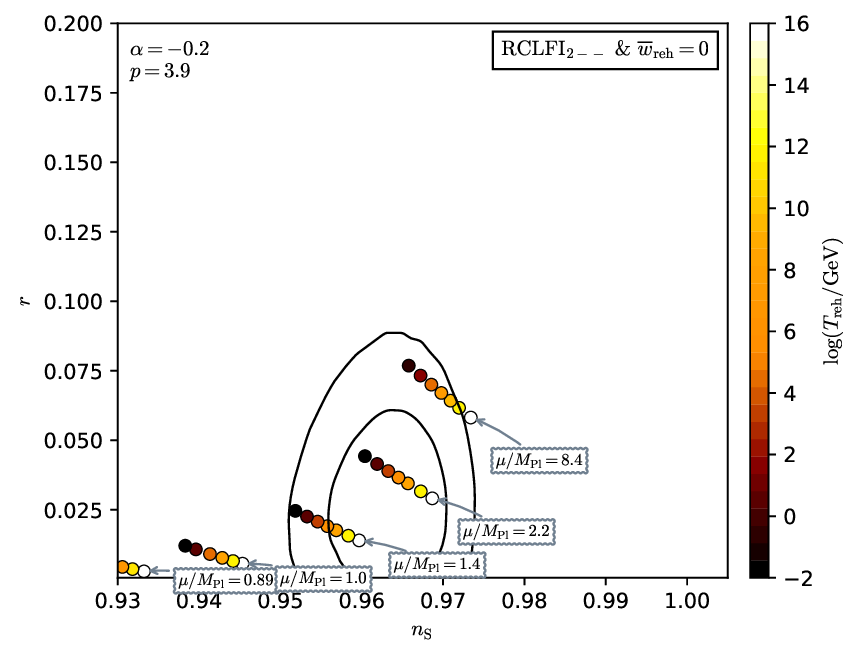}
\includegraphics[width=\wappfig,clip=true]{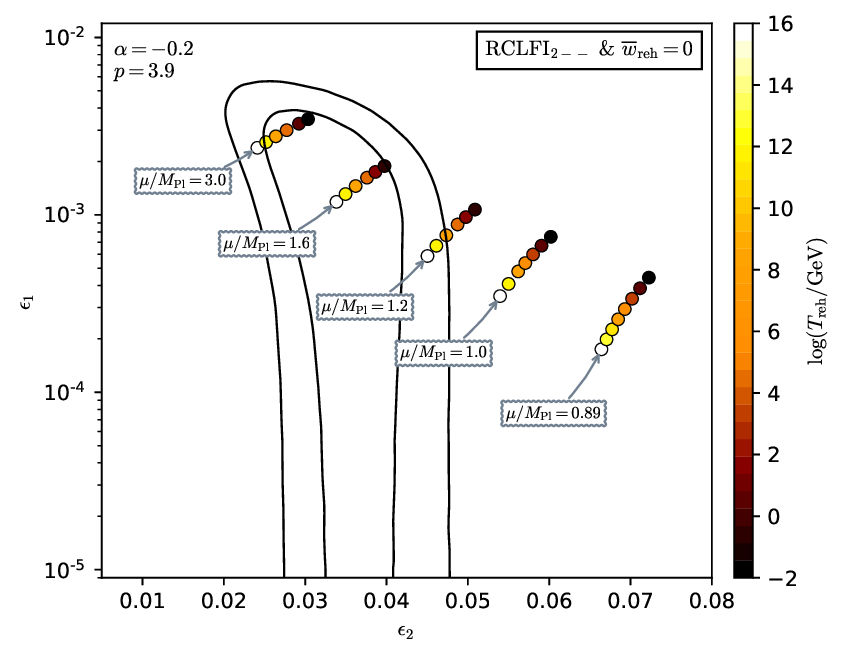}
\caption{Reheating consistent slow-roll predictions for the
  Radiatively Corrected Large Field Inflation models, in the RCLFI2
  regime, for $p<4$ and $\alpha<0$. Predictions are represented in the
  plane $(\nS,r)$ (top panel) and in the plane
  $(\epsilon_1,\epsilon_2)$ (bottom panel). The solid contours are the
  one and two-sigma {\data} confidence intervals (marginalized over
  second order slow-roll).}
\label{fig:CMBRCLFI2mm_3}
\end{center}
\end{figure}

\begin{figure}[H]
\begin{center}
\includegraphics[width=\wappfig,clip=true]{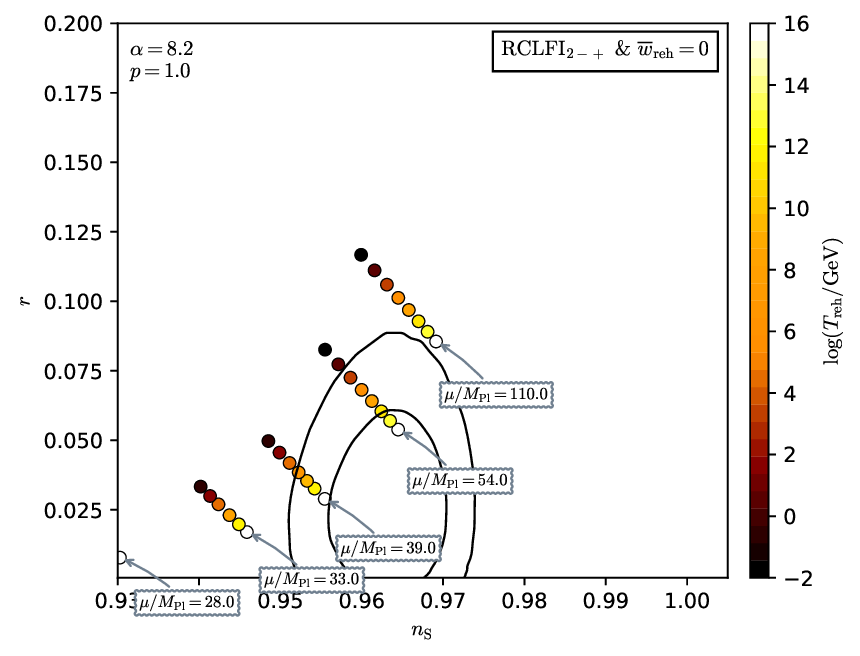}
\includegraphics[width=\wappfig,clip=true]{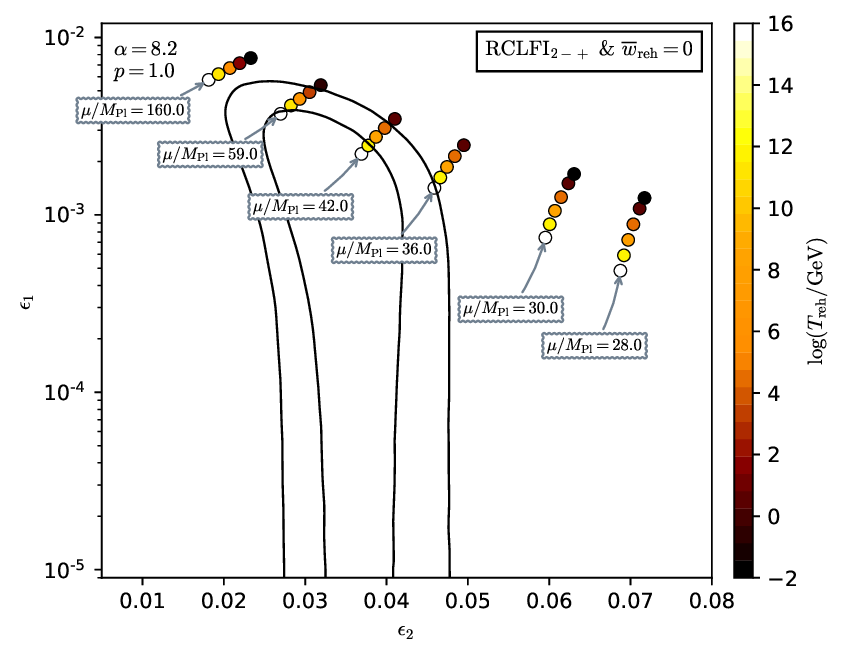}
\caption{Reheating consistent slow-roll predictions for the
  Radiatively Corrected Large Field Inflation models, in the RCLFI2
  regime, for $p<4$ and $\alpha > -e(p-4) > 0$. Predictions are
  represented in the plane $(\nS,r)$ (top panel) and in the plane
  $(\epsilon_1,\epsilon_2)$ (bottom panel) for various values of the
  field {\vev} $\mu$. The solid contours are the one and two-sigma
  {\data} confidence intervals (marginalized over second order
  slow-roll). See also Figs.~\ref{fig:CMBRCLFI2mp_1} to
  \ref{fig:CMBRCLFI2mp_3} for other values of $p$ and $\alpha$.}
\label{fig:CMBRCLFI2mp_0}
\end{center}
\end{figure}

\begin{figure}[H]
\begin{center}
\includegraphics[width=\wappfig,clip=true]{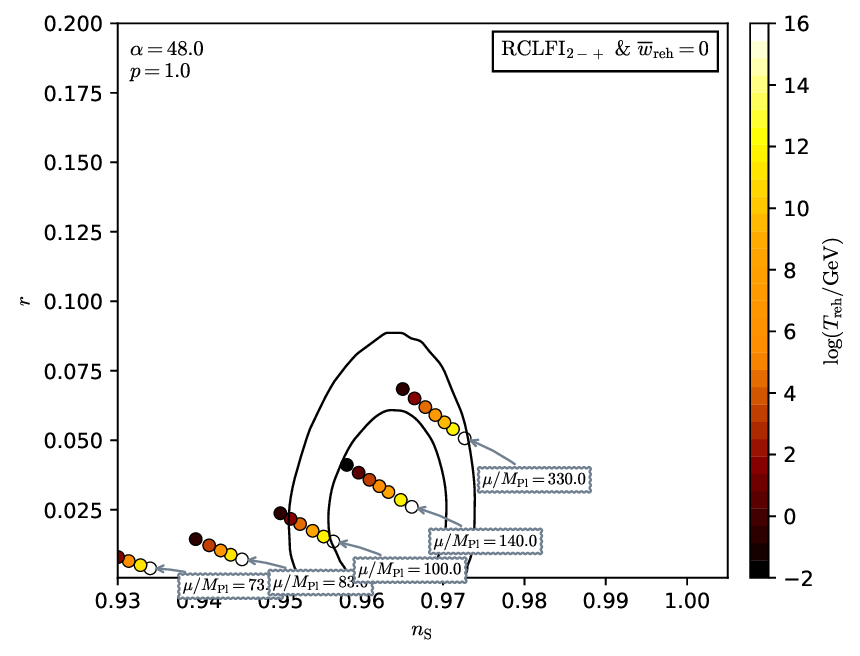}
\includegraphics[width=\wappfig,clip=true]{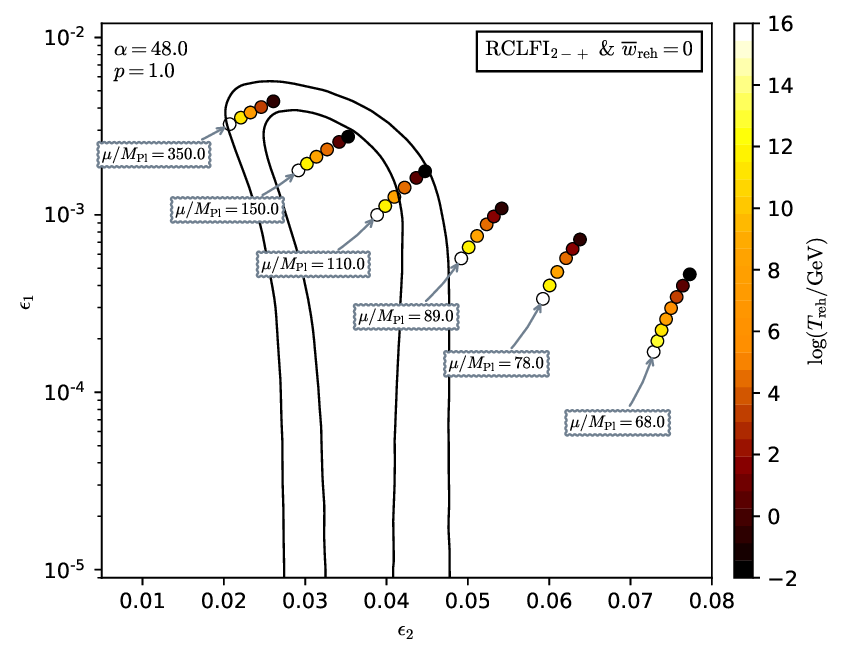}
\caption{Reheating consistent slow-roll predictions for the
  Radiatively Corrected Large Field Inflation models, in the RCLFI2
  regime, for $p<4$ and $\alpha > -e(p-4) > 0$. Predictions are
  represented in the plane $(\nS,r)$ (top panel) and in the plane
  $(\epsilon_1,\epsilon_2)$ (bottom panel). The solid contours are the
  one and two-sigma {\data} confidence intervals (marginalized over
  second order slow-roll).}
\label{fig:CMBRCLFI2mp_1}
\end{center}
\end{figure}

\begin{figure}[H]
\begin{center}
\includegraphics[width=\wappfig,clip=true]{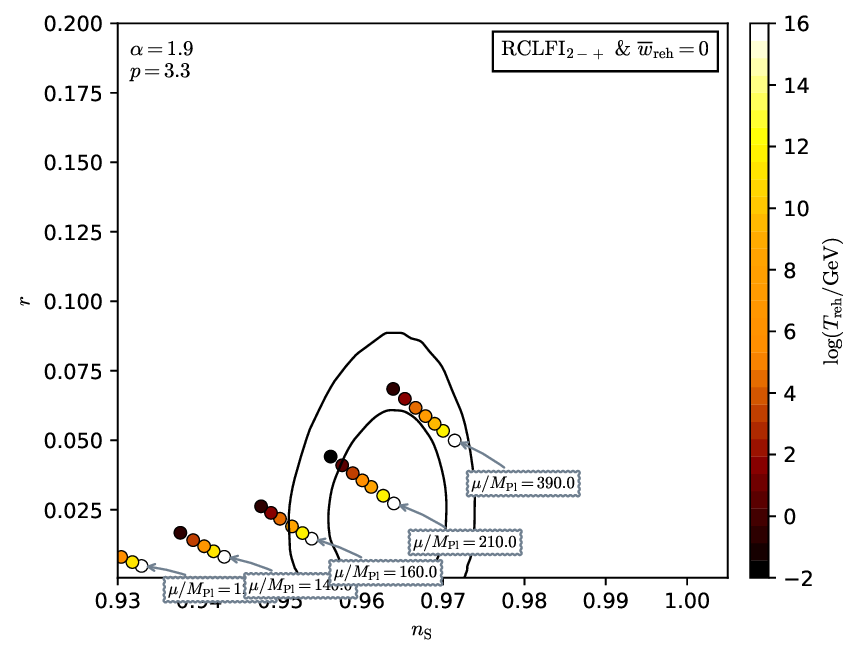}
\includegraphics[width=\wappfig,clip=true]{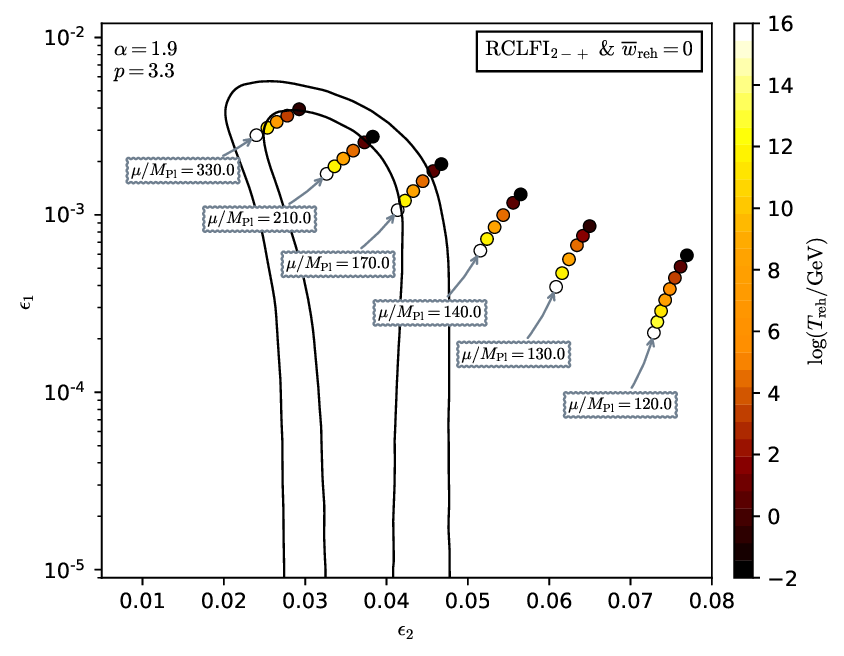}
\caption{Reheating consistent slow-roll predictions for the
  Radiatively Corrected Large Field Inflation models, in the RCLFI2
  regime, for $p<4$ and $\alpha > -e(p-4) > 0$. Predictions are
  represented in the plane $(\nS,r)$ (top panel) and in the plane
  $(\epsilon_1,\epsilon_2)$ (bottom panel). The solid contours are the
  one and two-sigma {\data} confidence intervals (marginalized over
  second order slow-roll).}
\label{fig:CMBRCLFI2mp_2}
\end{center}
\end{figure}

\begin{figure}[H]
\begin{center}
\includegraphics[width=\wappfig,clip=true]{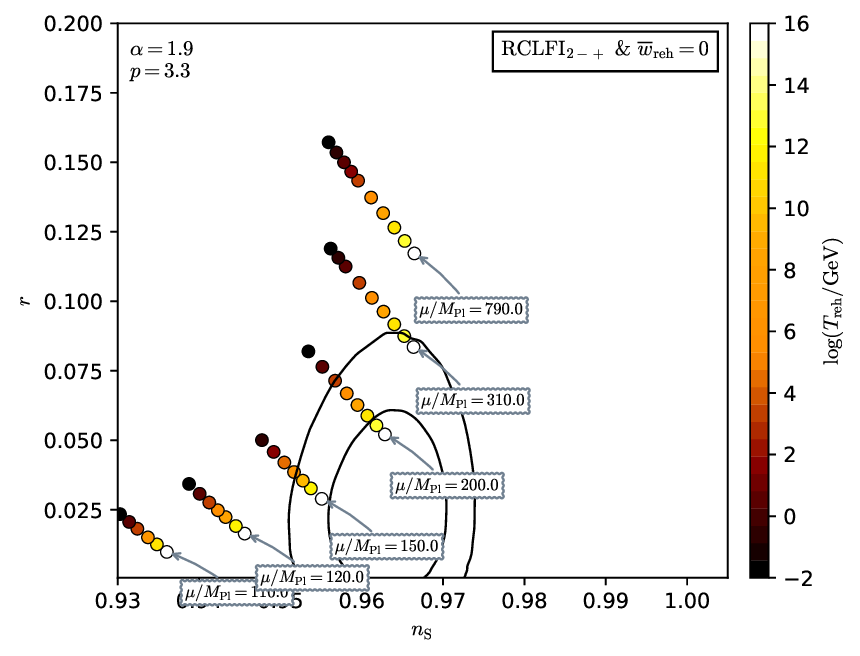}
\includegraphics[width=\wappfig,clip=true]{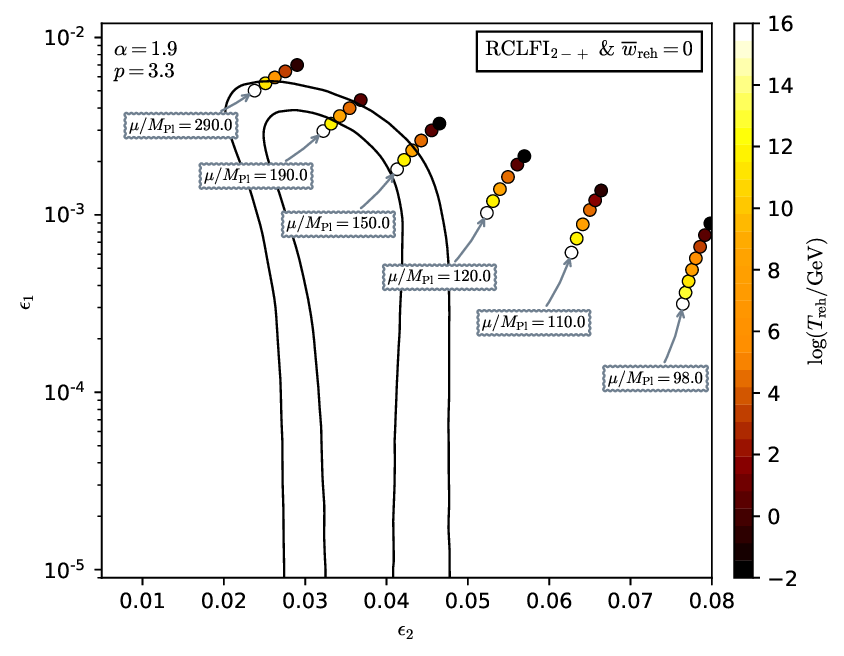}
\caption{Reheating consistent slow-roll predictions for the
  Radiatively Corrected Large Field Inflation models, in the RCLFI2
  regime, for $p<4$ and $\alpha > -e(p-4) > 0$. Predictions are
  represented in the plane $(\nS,r)$ (top panel) and in the plane
  $(\epsilon_1,\epsilon_2)$ (bottom panel). The solid contours are the
  one and two-sigma {\data} confidence intervals (marginalized over
  second order slow-roll).}
\label{fig:CMBRCLFI2mp_3}
\end{center}
\end{figure}

\subsection{Radiatively Corrected Large Field Inflation 3 (\hyperref[sec:rclfi]{RCLFI3})}

\begin{figure}[H]
\begin{center}
\includegraphics[width=\wappfig,clip=true]{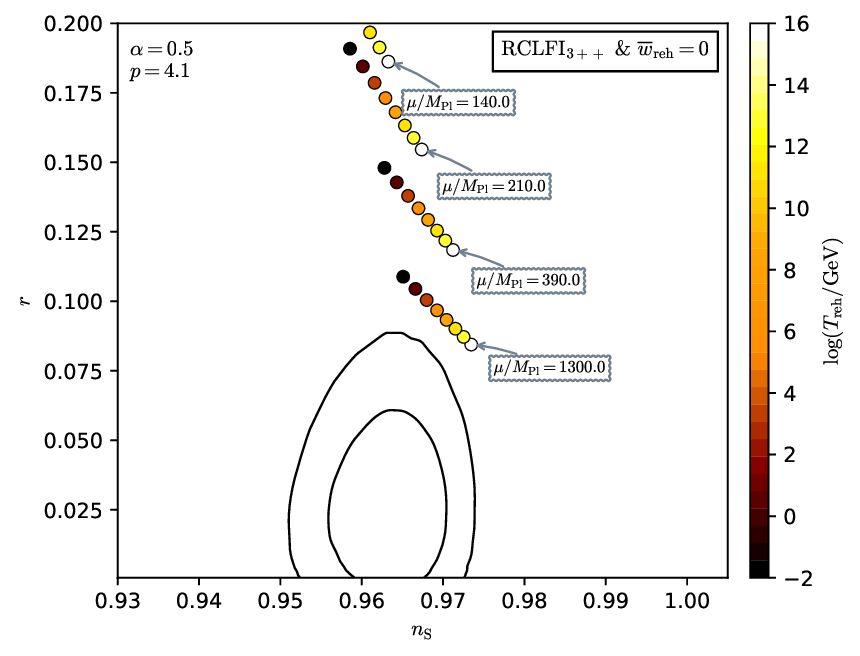}
\includegraphics[width=\wappfig,clip=true]{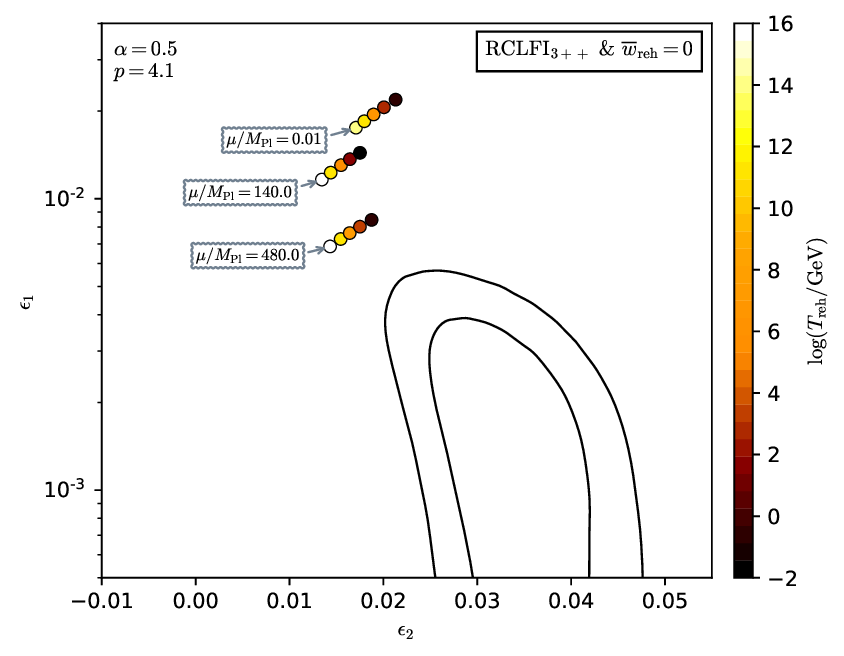}
\caption{Reheating consistent slow-roll predictions for the
  Radiatively Corrected Large Field Inflation models, in the RCLFI3
  regime, for $p>4$ and $\alpha > 0$. Predictions are
  represented in the plane $(\nS,r)$ (top panel) and in the plane
  $(\epsilon_1,\epsilon_2)$ (bottom panel) for various values of the
  field {\vev} $\mu$. The solid contours are the one and two-sigma
  {\data} confidence intervals (marginalized over second order
  slow-roll). See also Figs.~\ref{fig:CMBRCLFI3pp_1} to
  \ref{fig:CMBRCLFI3pp_3} for other values of $p$ and $\alpha$.}
\label{fig:CMBRCLFI3pp_0}
\end{center}
\end{figure}

\begin{figure}[H]
\begin{center}
\includegraphics[width=\wappfig,clip=true]{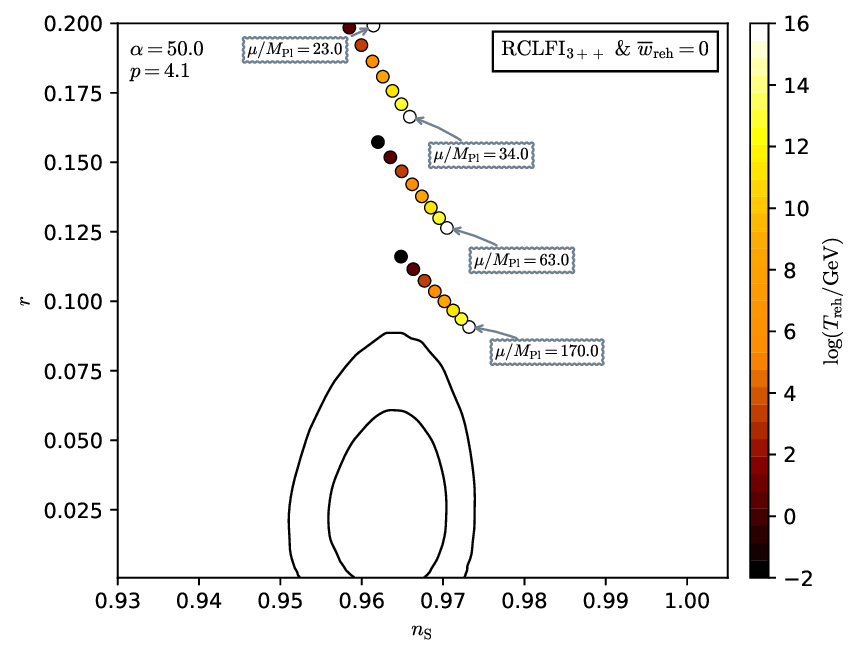}
\includegraphics[width=\wappfig,clip=true]{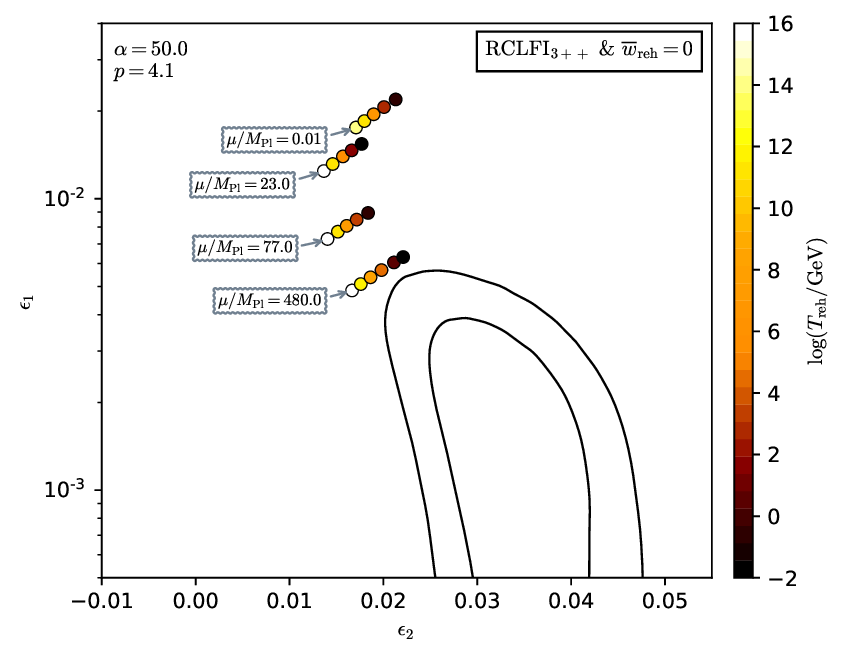}
\caption{Reheating consistent slow-roll predictions for the
  Radiatively Corrected Large Field Inflation models, in the RCLFI3
  regime, for $p>4$ and $\alpha > 0$. Predictions are represented in
  the plane $(\nS,r)$ (top panel) and in the plane
  $(\epsilon_1,\epsilon_2)$ (bottom panel). The solid contours are the
  one and two-sigma {\data} confidence intervals (marginalized over
  second order slow-roll).}
\label{fig:CMBRCLFI3pp_1}
\end{center}
\end{figure}

\begin{figure}[H]
\begin{center}
\includegraphics[width=\wappfig,clip=true]{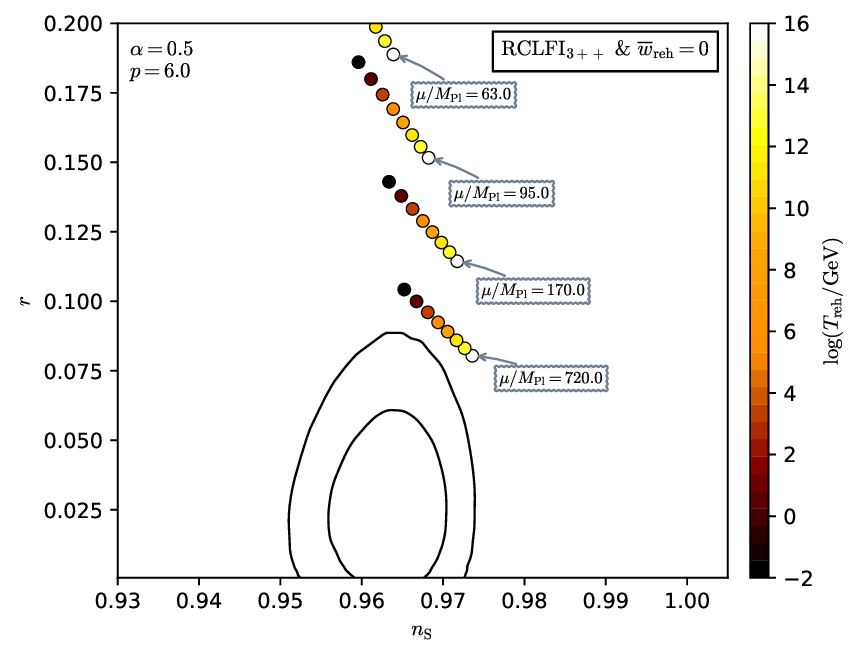}
\includegraphics[width=\wappfig,clip=true]{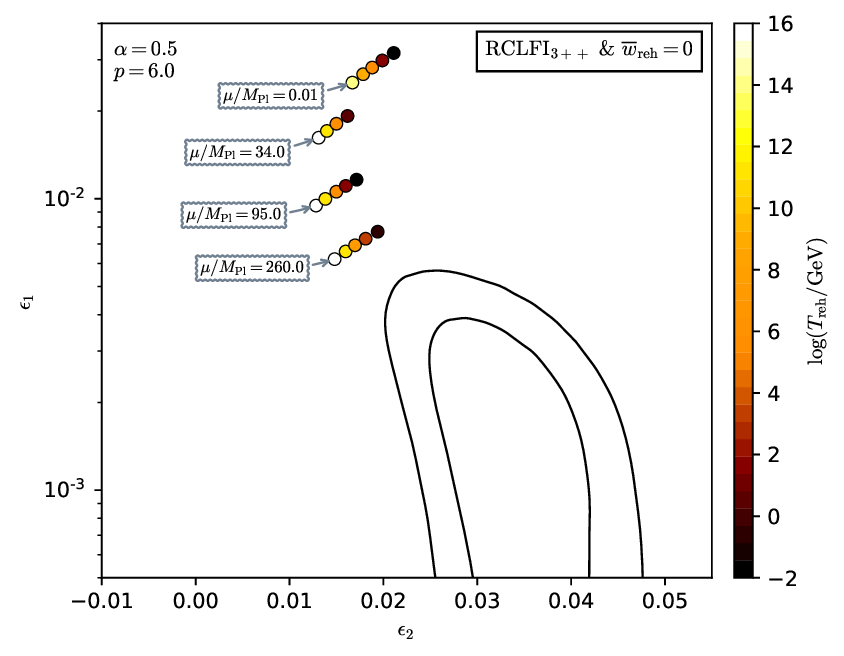}
\caption{Reheating consistent slow-roll predictions for the
  Radiatively Corrected Large Field Inflation models, in the RCLFI3
  regime, for $p>4$ and $\alpha > 0$. Predictions are represented in
  the plane $(\nS,r)$ (top panel) and in the plane
  $(\epsilon_1,\epsilon_2)$ (bottom panel). The solid contours are the
  one and two-sigma {\data} confidence intervals (marginalized over
  second order slow-roll).}
\label{fig:CMBRCLFI3pp_2}
\end{center}
\end{figure}

\begin{figure}[H]
\begin{center}
\includegraphics[width=\wappfig,clip=true]{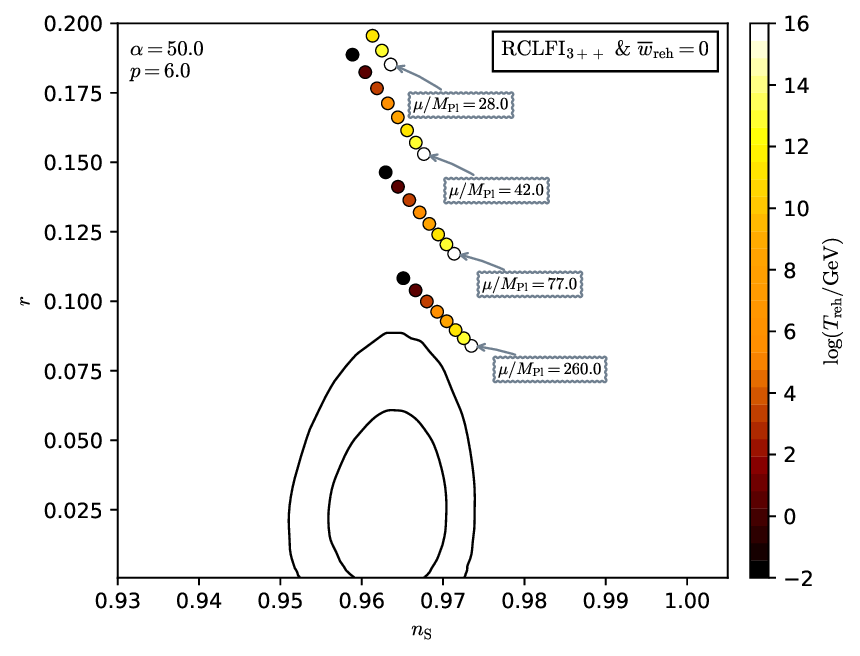}
\includegraphics[width=\wappfig,clip=true]{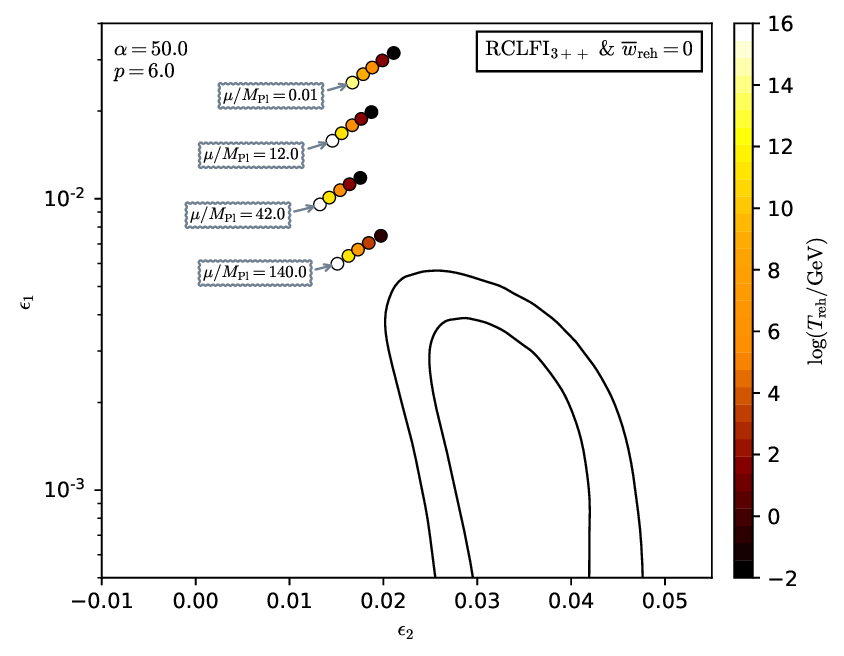}
\caption{Reheating consistent slow-roll predictions for the
  Radiatively Corrected Large Field Inflation models, in the RCLFI3
  regime, for $p>4$ and $\alpha > 0$. Predictions are represented in
  the plane $(\nS,r)$ (top panel) and in the plane
  $(\epsilon_1,\epsilon_2)$ (bottom panel). The solid contours are the
  one and two-sigma {\data} confidence intervals (marginalized over
  second order slow-roll).}
\label{fig:CMBRCLFI3pp_3}
\end{center}
\end{figure}

\begin{figure}[H]
\begin{center}
\includegraphics[width=\wappfig,clip=true]{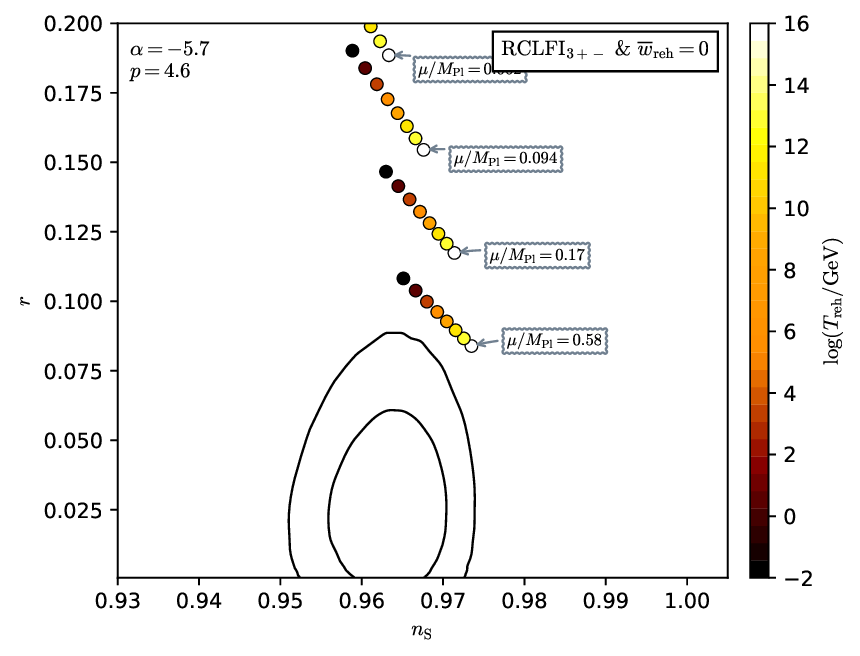}
\includegraphics[width=\wappfig,clip=true]{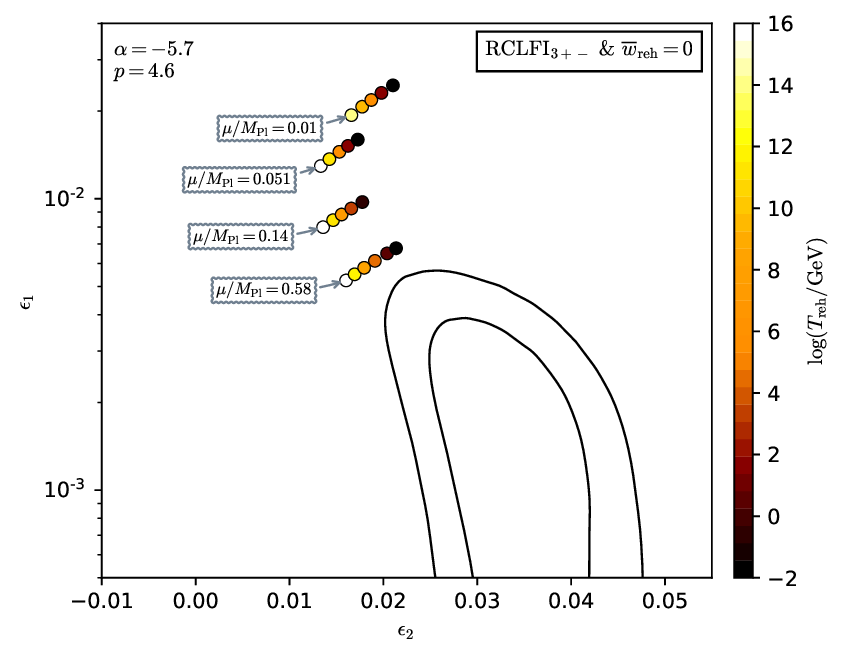}
\caption{Reheating consistent slow-roll predictions for the
  Radiatively Corrected Large Field Inflation models, in the RCLFI3
  regime, for $p>4$ and $\alpha < -e(p-4) < 0$. Predictions are
  represented in the plane $(\nS,r)$ (top panel) and in the plane
  $(\epsilon_1,\epsilon_2)$ (bottom panel) for various values of the
  field {\vev} $\mu$. The solid contours are the one and two-sigma
  {\data} confidence intervals (marginalized over second order
  slow-roll). See also Figs.~\ref{fig:CMBRCLFI3pm_1} to
  \ref{fig:CMBRCLFI3pm_3} for other values of $p$ and $\alpha$.}
\label{fig:CMBRCLFI3pm_0}
\end{center}
\end{figure}

\begin{figure}[H]
\begin{center}
\includegraphics[width=\wappfig,clip=true]{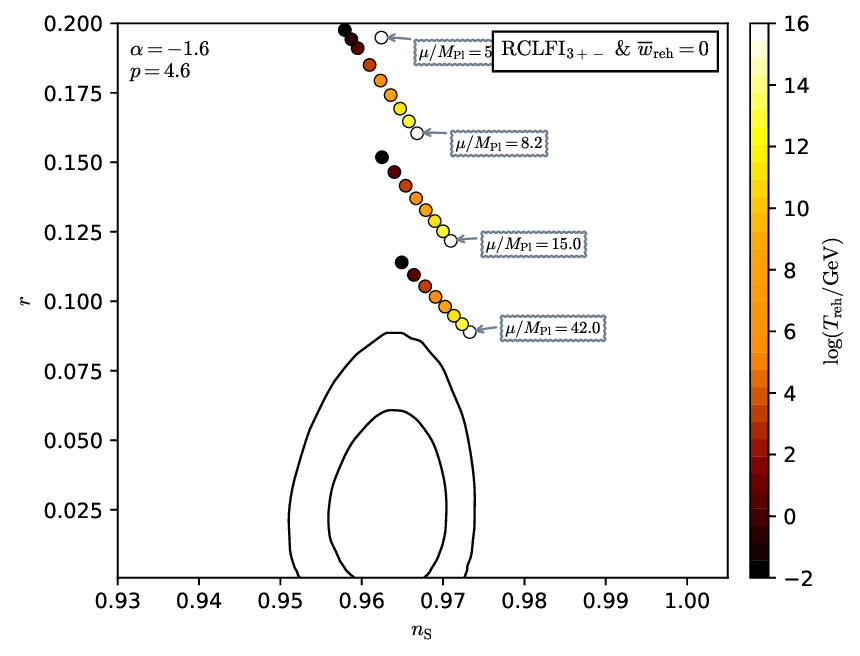}
\includegraphics[width=\wappfig,clip=true]{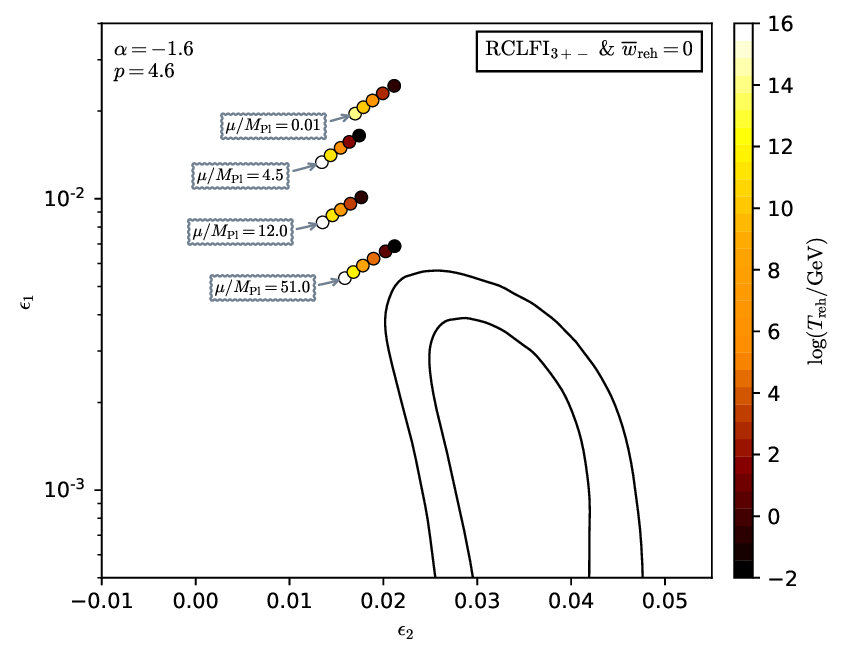}
\caption{Reheating consistent slow-roll predictions for the
  Radiatively Corrected Large Field Inflation models, in the RCLFI3
  regime, for $p>4$ and $\alpha < -e(p-4) < 0$. Predictions are
  represented in the plane $(\nS,r)$ (top panel) and in the plane
  $(\epsilon_1,\epsilon_2)$ (bottom panel). The solid contours are the
  one and two-sigma {\data} confidence intervals (marginalized over
  second order slow-roll).}
\label{fig:CMBRCLFI3pm_1}
\end{center}
\end{figure}

\begin{figure}[H]
\begin{center}
\includegraphics[width=\wappfig,clip=true]{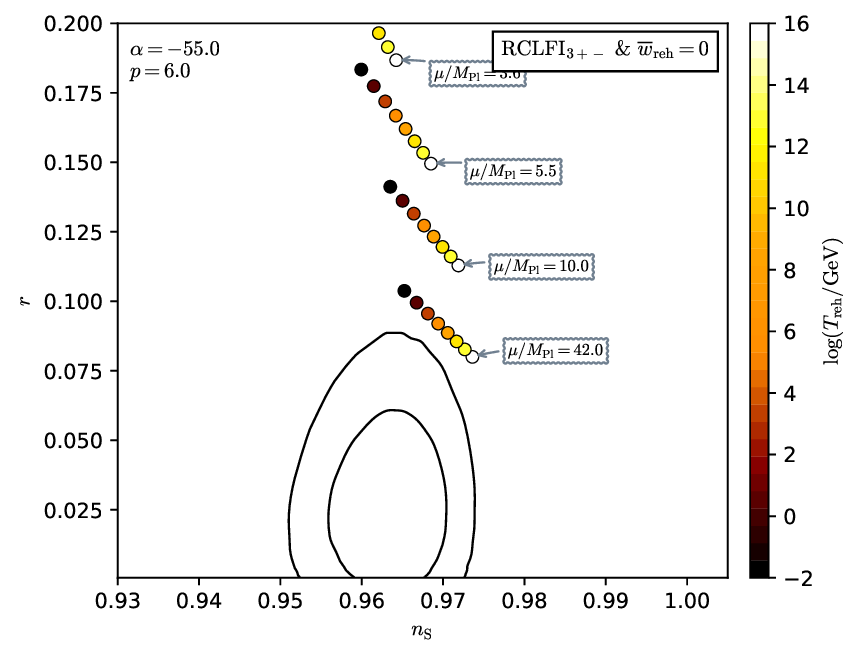}
\includegraphics[width=\wappfig,clip=true]{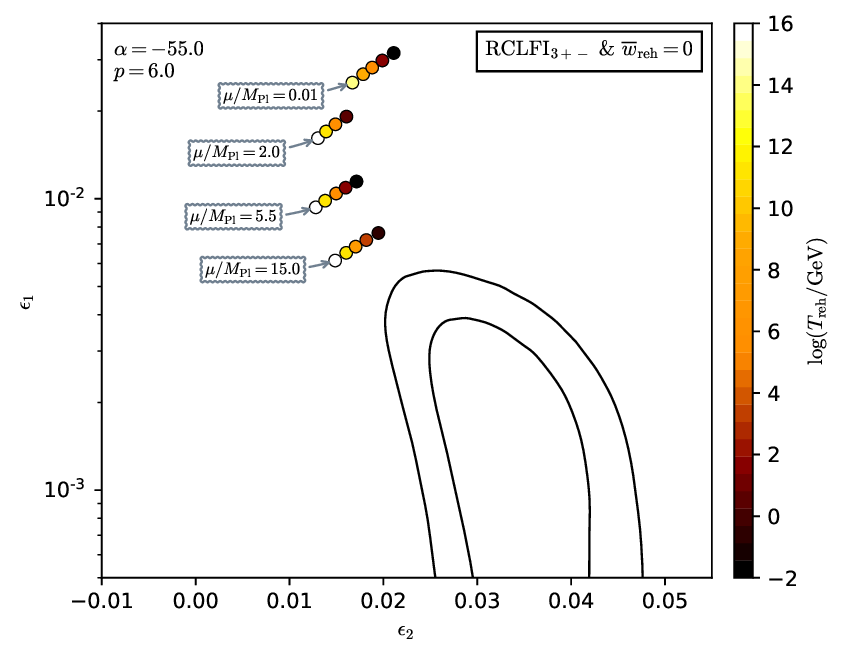}
\caption{Reheating consistent slow-roll predictions for the
  Radiatively Corrected Large Field Inflation models, in the RCLFI3
  regime, for $p>4$ and $\alpha < -e(p-4) < 0$. Predictions are
  represented in the plane $(\nS,r)$ (top panel) and in the plane
  $(\epsilon_1,\epsilon_2)$ (bottom panel). The solid contours are the
  one and two-sigma {\data} confidence intervals (marginalized over
  second order slow-roll).}
\label{fig:CMBRCLFI3pm_2}
\end{center}
\end{figure}

\begin{figure}[H]
\begin{center}
\includegraphics[width=\wappfig,clip=true]{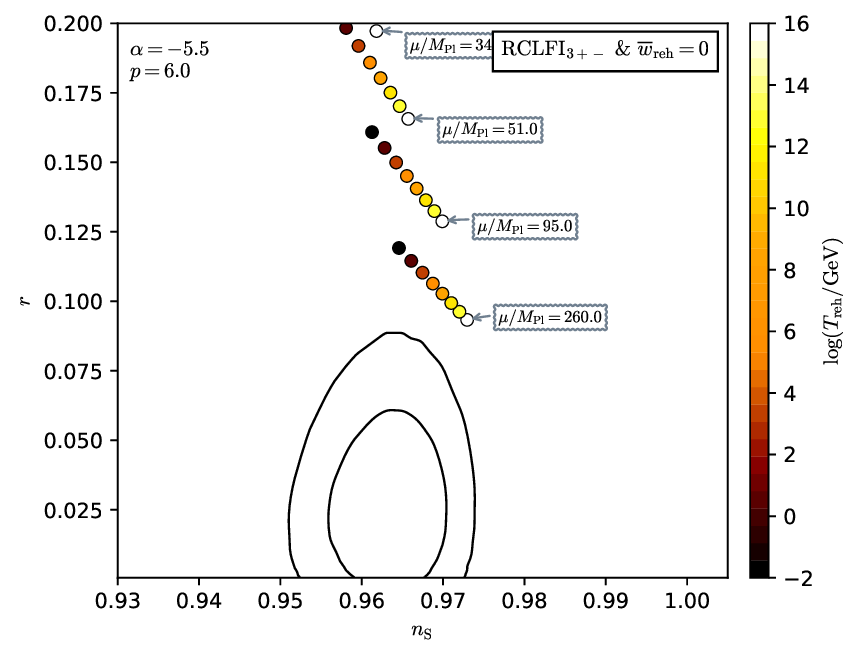}
\includegraphics[width=\wappfig,clip=true]{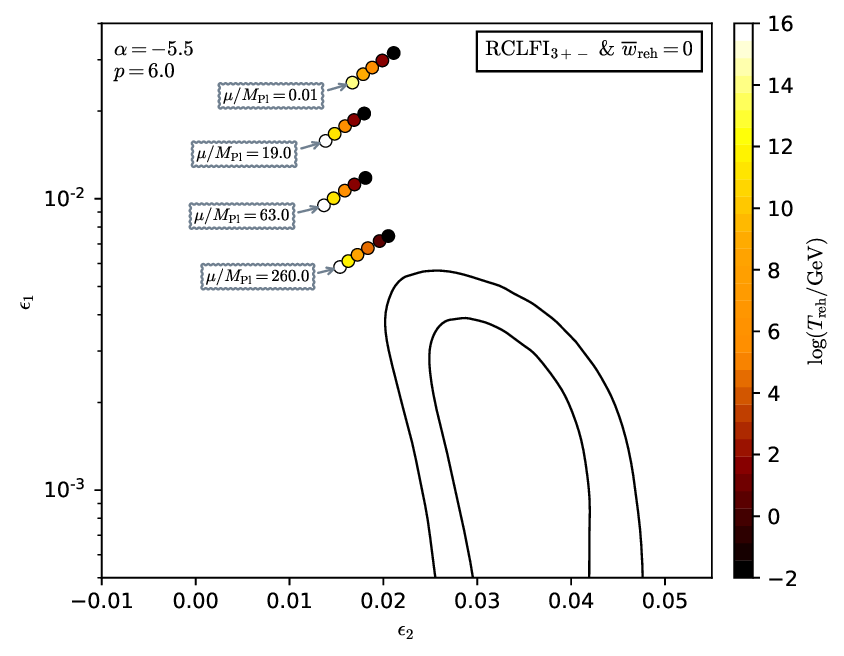}
\caption{Reheating consistent slow-roll predictions for the
  Radiatively Corrected Large Field Inflation models, in the RCLFI3
  regime, for $p>4$ and $\alpha < -e(p-4) < 0$. Predictions are
  represented in the plane $(\nS,r)$ (top panel) and in the plane
  $(\epsilon_1,\epsilon_2)$ (bottom panel). The solid contours are the
  one and two-sigma {\data} confidence intervals (marginalized over
  second order slow-roll).}
\label{fig:CMBRCLFI3pm_3}
\end{center}
\end{figure}

\begin{figure}[H]
\begin{center}
\includegraphics[width=\wappfig,clip=true]{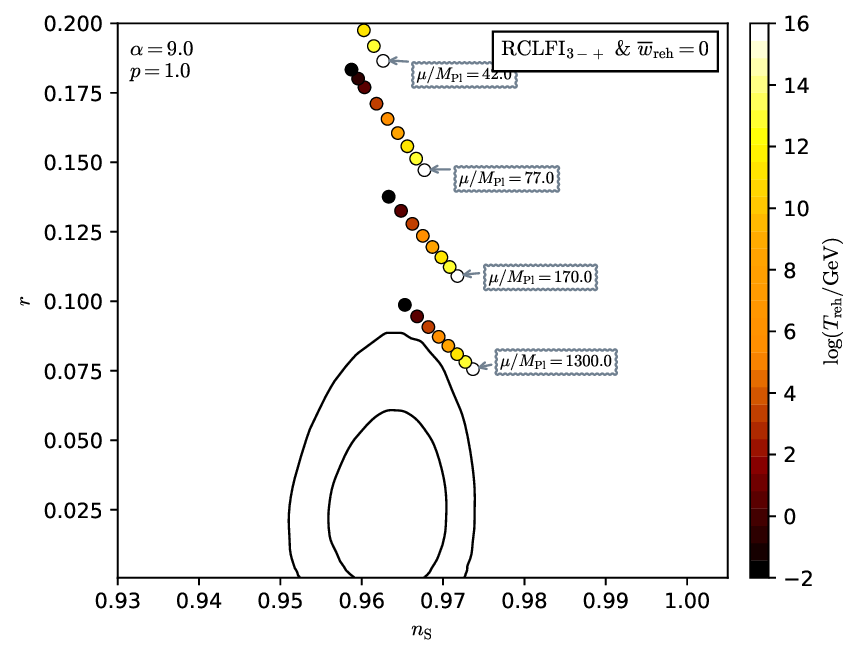}
\includegraphics[width=\wappfig,clip=true]{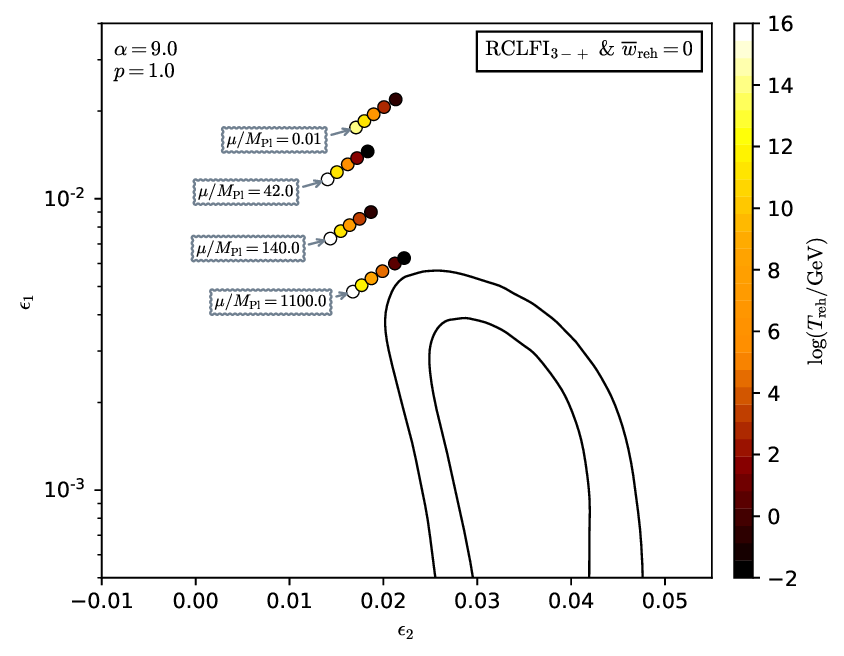}
\caption{Reheating consistent slow-roll predictions for the
  Radiatively Corrected Large Field Inflation models, in the RCLFI3
  regime, for $p<4$ and $\alpha > -e(p-4) > 0$. Predictions are
  represented in the plane $(\nS,r)$ (top panel) and in the plane
  $(\epsilon_1,\epsilon_2)$ (bottom panel) for various values of the
  field {\vev} $\mu$. The solid contours are the one and two-sigma
  {\data} confidence intervals (marginalized over second order
  slow-roll). See also Figs.~\ref{fig:CMBRCLFI3mp_1} to
  \ref{fig:CMBRCLFI3mp_3} for other values of $p$ and $\alpha$.}
\label{fig:CMBRCLFI3mp_0}
\end{center}
\end{figure}

\begin{figure}[H]
\begin{center}
\includegraphics[width=\wappfig,clip=true]{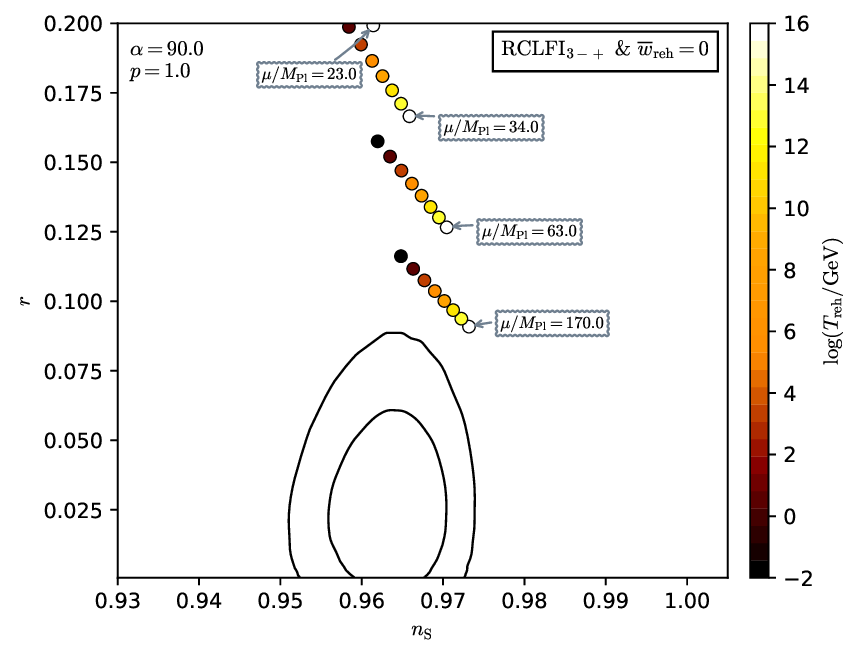}
\includegraphics[width=\wappfig,clip=true]{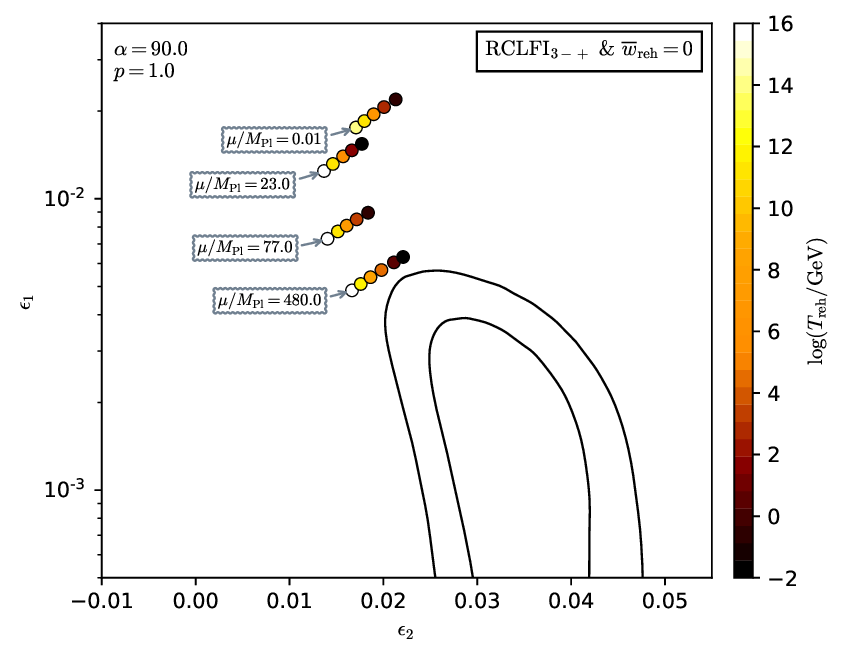}
\caption{Reheating consistent slow-roll predictions for the
  Radiatively Corrected Large Field Inflation models, in the RCLFI3
  regime, for $p<4$ and $\alpha > -e(p-4) > 0$. Predictions are
  represented in the plane $(\nS,r)$ (top panel) and in the plane
  $(\epsilon_1,\epsilon_2)$ (bottom panel). The solid contours are the
  one and two-sigma {\data} confidence intervals (marginalized over
  second order slow-roll).}
\label{fig:CMBRCLFI3mp_1}
\end{center}
\end{figure}

\begin{figure}[H]
\begin{center}
\includegraphics[width=\wappfig,clip=true]{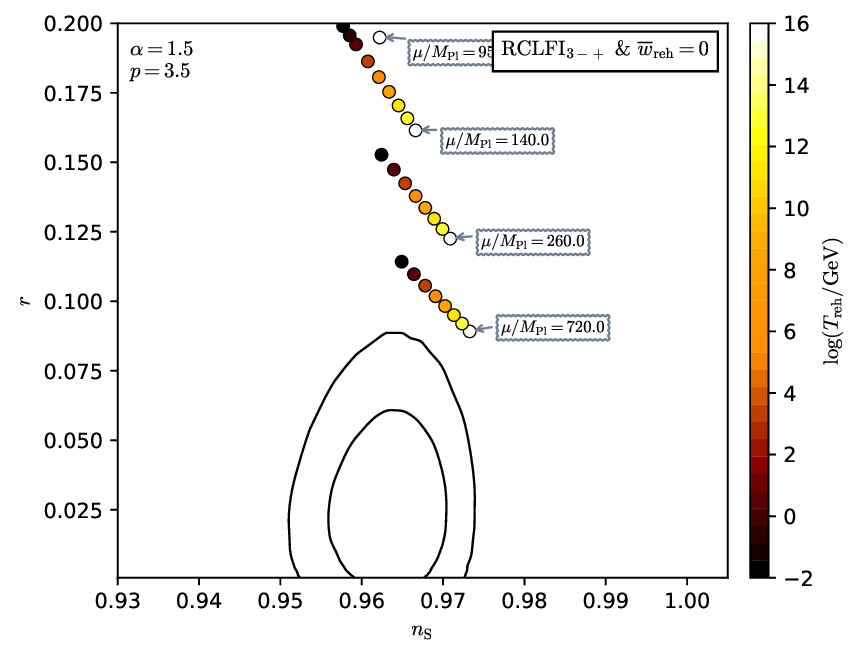}
\includegraphics[width=\wappfig,clip=true]{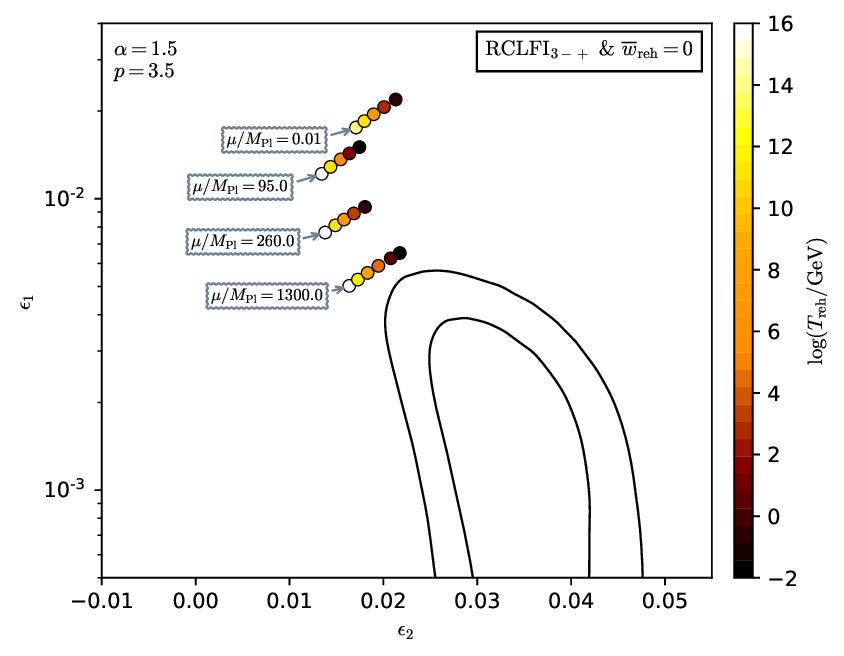}
\caption{Reheating consistent slow-roll predictions for the
  Radiatively Corrected Large Field Inflation models, in the RCLFI3
  regime, for $p<4$ and $\alpha > -e(p-4) > 0$. Predictions are
  represented in the plane $(\nS,r)$ (top panel) and in the plane
  $(\epsilon_1,\epsilon_2)$ (bottom panel). The solid contours are the
  one and two-sigma {\data} confidence intervals (marginalized over
  second order slow-roll).}
\label{fig:CMBRCLFI3mp_2}
\end{center}
\end{figure}

\begin{figure}[H]
\begin{center}
\includegraphics[width=\wappfig,clip=true]{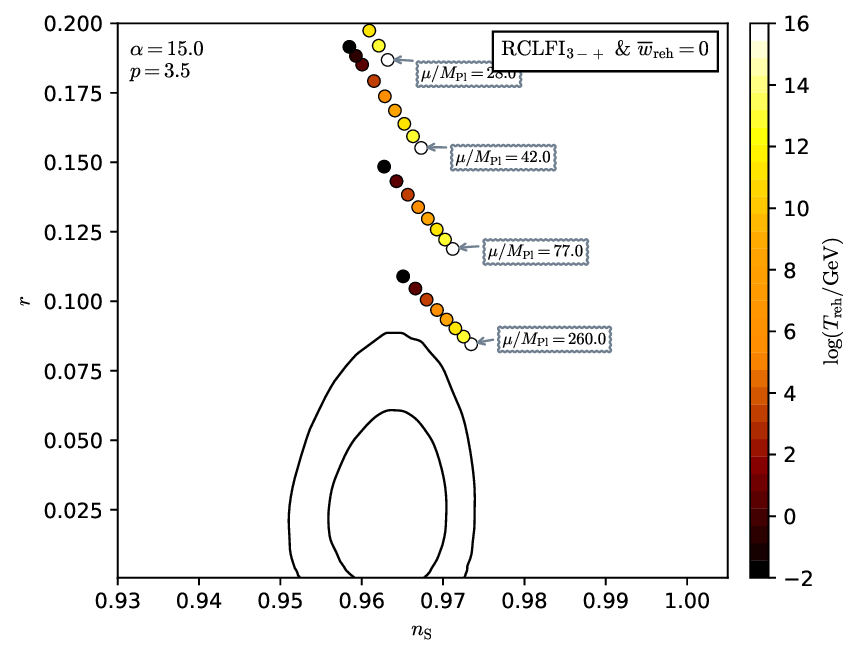}
\includegraphics[width=\wappfig,clip=true]{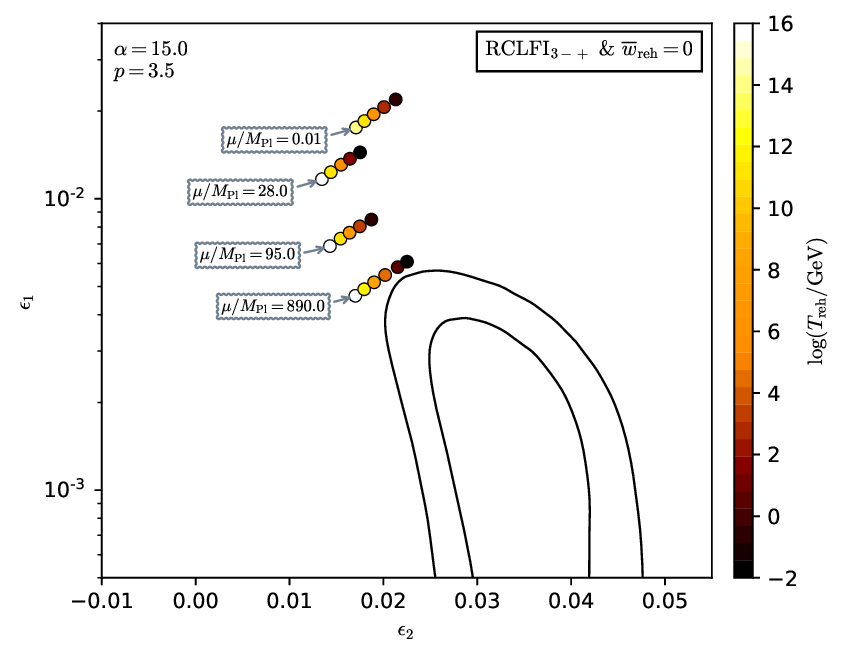}
\caption{Reheating consistent slow-roll predictions for the
  Radiatively Corrected Large Field Inflation models, in the RCLFI3
  regime, for $p<4$ and $\alpha > -e(p-4) > 0$. Predictions are
  represented in the plane $(\nS,r)$ (top panel) and in the plane
  $(\epsilon_1,\epsilon_2)$ (bottom panel). The solid contours are the
  one and two-sigma {\data} confidence intervals (marginalized over
  second order slow-roll).}
\label{fig:CMBRCLFI3mp_3}
\end{center}
\end{figure}

\subsection{Radiatively Corrected Large Field Inflation 4 (\hyperref[sec:rclfi]{RCLFI4})}

\begin{figure}[H]
\begin{center}
\includegraphics[width=\wappfig,clip=true]{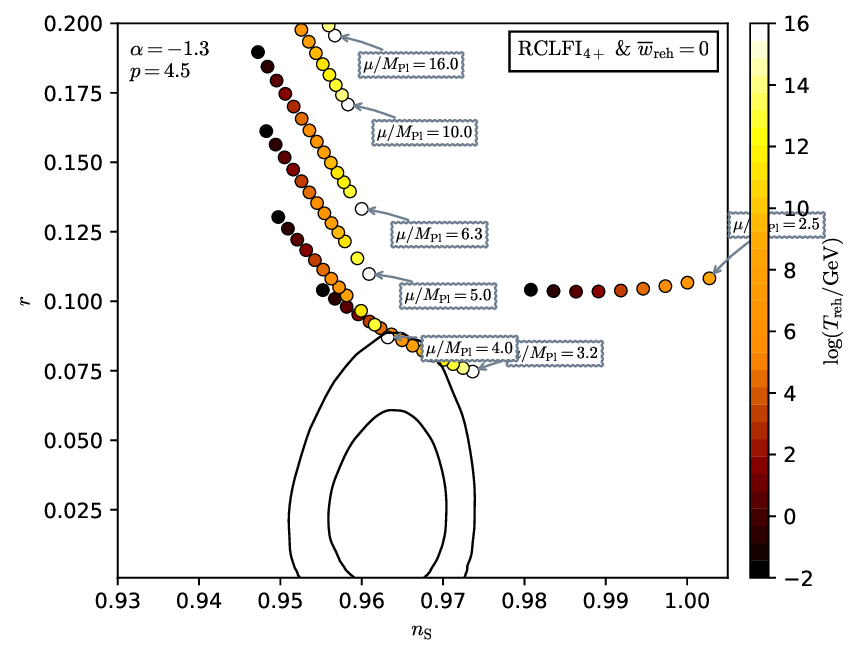}
\includegraphics[width=\wappfig,clip=true]{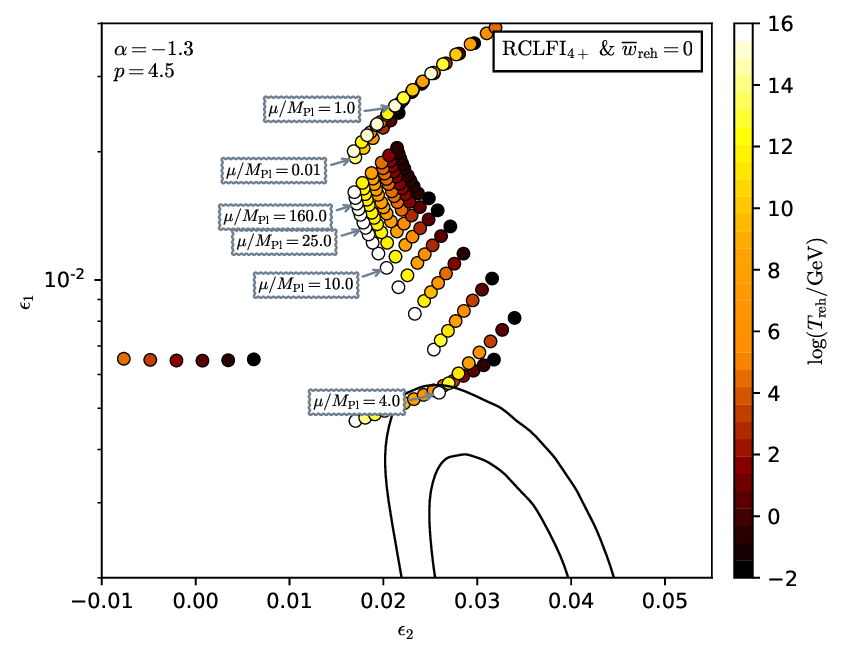}
\caption{Reheating consistent slow-roll predictions for the
  Radiatively Corrected Large Field Inflation models, in the RCLFI4
  regime, for $p>4$ and $-[p(p-4)/4]\exp(2-p/4)<\alpha < 0$. Predictions are
  represented in the plane $(\nS,r)$ (top panel) and in the plane
  $(\epsilon_1,\epsilon_2)$ (bottom panel) for various values of the
  field {\vev} $\mu$. The solid contours are the one and two-sigma
  {\data} confidence intervals (marginalized over second order
  slow-roll). See also Figs.~\ref{fig:CMBRCLFI4p_1} to
  \ref{fig:CMBRCLFI4p_3} for other values of $p$ and $\alpha$.}
\label{fig:CMBRCLFI4p_0}
\end{center}
\end{figure}

\begin{figure}[H]
\begin{center}
\includegraphics[width=\wappfig,clip=true]{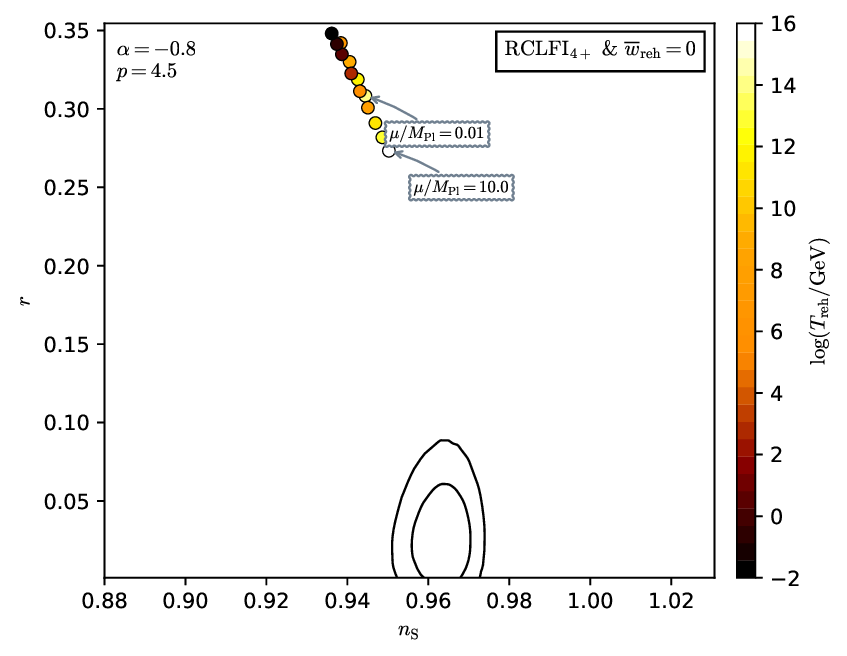}
\includegraphics[width=\wappfig,clip=true]{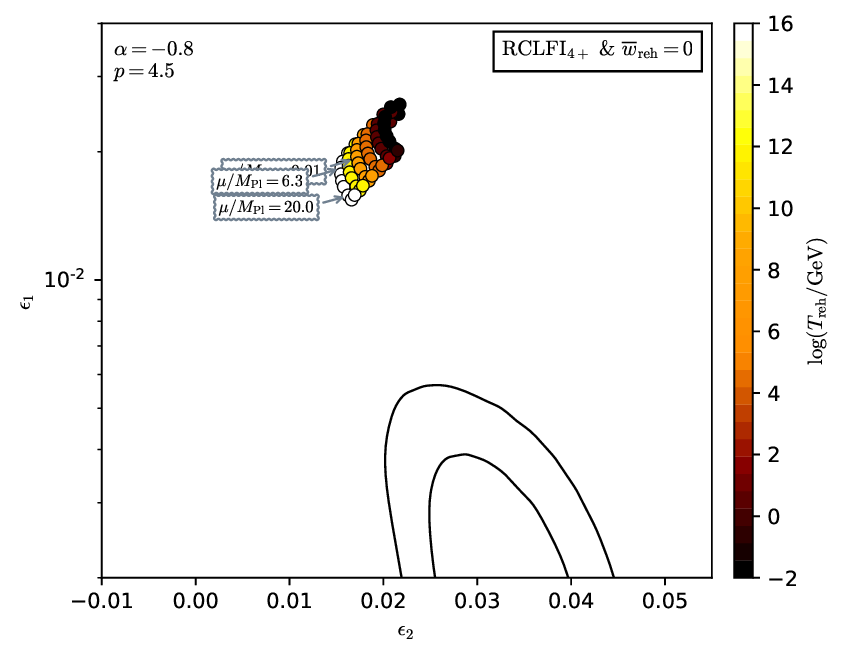}
\caption{Reheating consistent slow-roll predictions for the
  Radiatively Corrected Large Field Inflation models, in the RCLFI4
  regime, for $p>4$ and $-[p(p-4)/4]\exp(2-p/4)<\alpha <
  0$. Predictions are represented in the plane $(\nS,r)$ (top panel)
  and in the plane $(\epsilon_1,\epsilon_2)$ (bottom panel). The solid
  contours are the one and two-sigma {\data} confidence intervals
  (marginalized over second order slow-roll).}
\label{fig:CMBRCLFI4p_1}
\end{center}
\end{figure}

\begin{figure}[H]
\begin{center}
\includegraphics[width=\wappfig,clip=true]{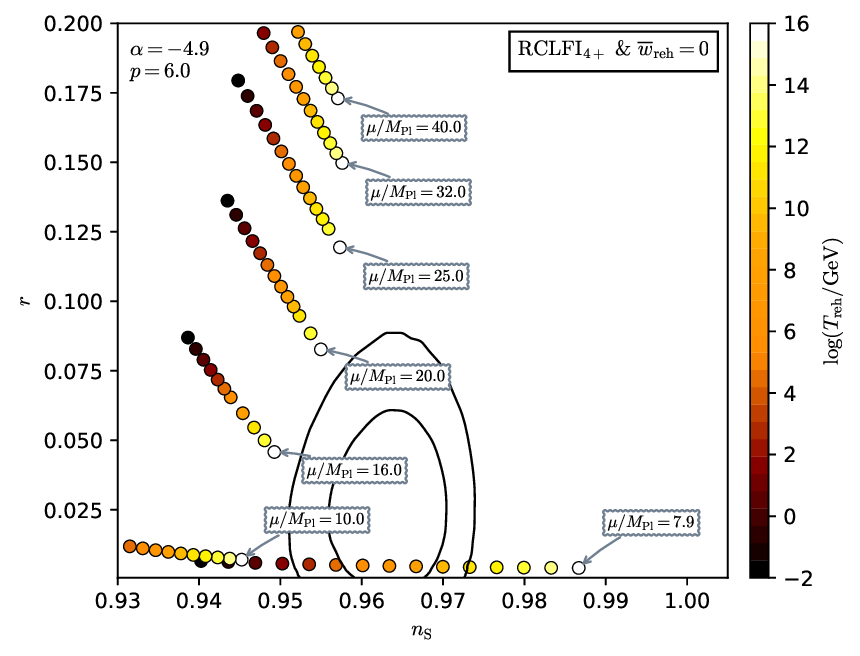}
\includegraphics[width=\wappfig,clip=true]{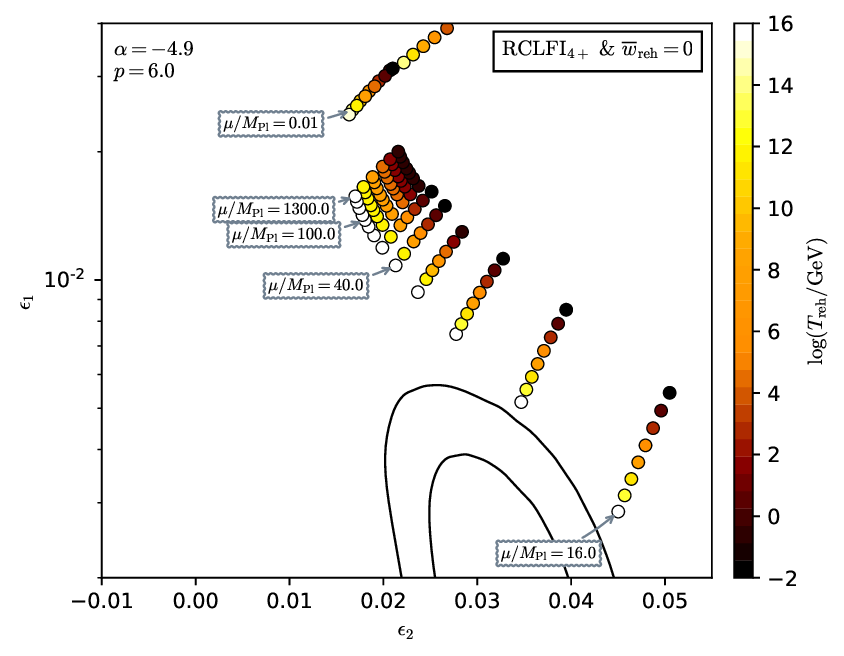}
\caption{Reheating consistent slow-roll predictions for the
  Radiatively Corrected Large Field Inflation models, in the RCLFI4
  regime, for $p>4$ and $-[p(p-4)/4]\exp(2-p/4)<\alpha <
  0$. Predictions are represented in the plane $(\nS,r)$ (top panel)
  and in the plane $(\epsilon_1,\epsilon_2)$ (bottom panel). The solid
  contours are the one and two-sigma {\data} confidence intervals
  (marginalized over second order slow-roll).}
\label{fig:CMBRCLFI4p_2}
\end{center}
\end{figure}

\begin{figure}[H]
\begin{center}
\includegraphics[width=\wappfig,clip=true]{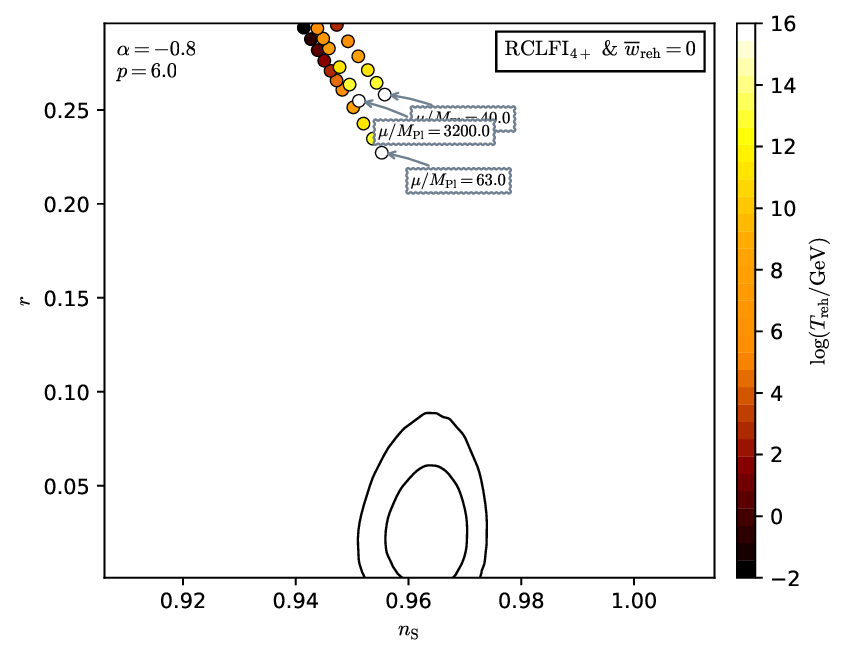}
\includegraphics[width=\wappfig,clip=true]{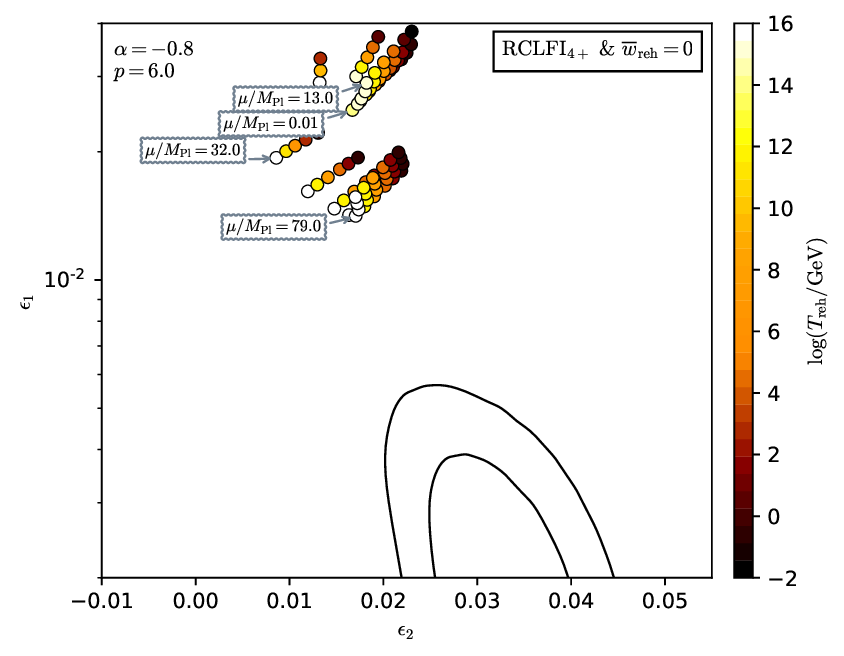}
\caption{Reheating consistent slow-roll predictions for the
  Radiatively Corrected Large Field Inflation models, in the RCLFI4
  regime, for $p>4$ and $-[p(p-4)/4]\exp(2-p/4)<\alpha <
  0$. Predictions are represented in the plane $(\nS,r)$ (top panel)
  and in the plane $(\epsilon_1,\epsilon_2)$ (bottom panel). The solid
  contours are the one and two-sigma {\data} confidence intervals
  (marginalized over second order slow-roll).}
\label{fig:CMBRCLFI4p_3}
\end{center}
\end{figure}

\begin{figure}[H]
\begin{center}
\includegraphics[width=\wappfig,clip=true]{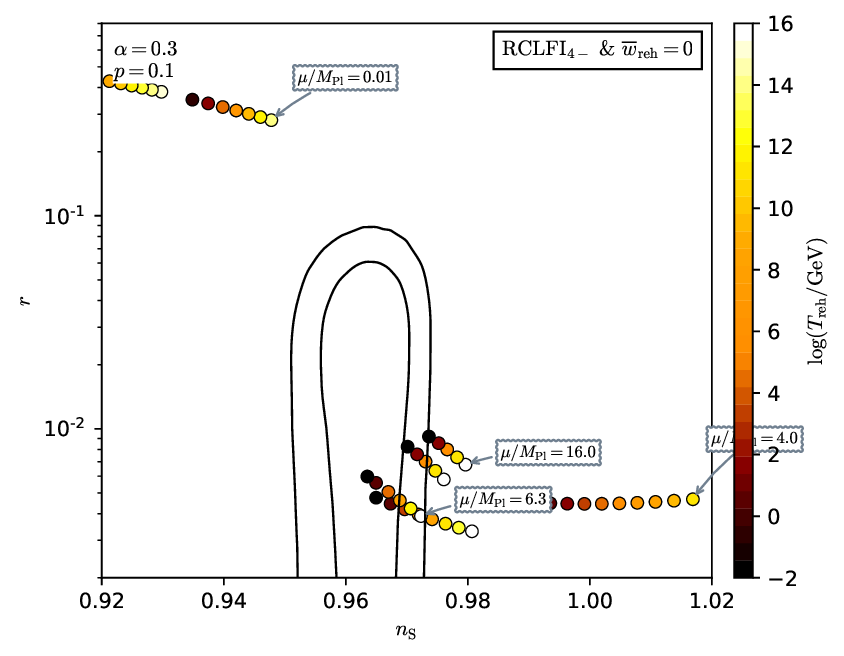}
\includegraphics[width=\wappfig,clip=true]{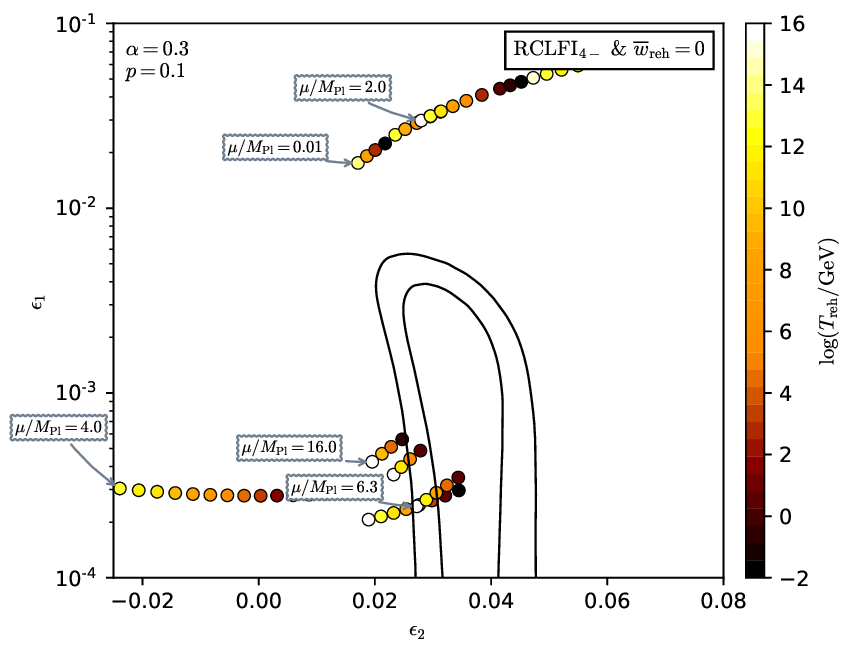}
\caption{Reheating consistent slow-roll predictions for the
  Radiatively Corrected Large Field Inflation models, in the RCLFI4
  regime, for $p<4$ and $0 < \alpha <
  -[p(p-4)/4]\exp(2-p/4)$. Predictions are represented in the plane
  $(\nS,r)$ (top panel) and in the plane $(\epsilon_1,\epsilon_2)$
  (bottom panel) for various values of the field {\vev} $\mu$. The
  solid contours are the one and two-sigma {\data} confidence
  intervals (marginalized over second order slow-roll). See also
  Figs.~\ref{fig:CMBRCLFI4m_1} to \ref{fig:CMBRCLFI4m_3} for other
  values of $p$ and $\alpha$.}
\label{fig:CMBRCLFI4m_0}
\end{center}
\end{figure}

\begin{figure}[H]
\begin{center}
\includegraphics[width=\wappfig,clip=true]{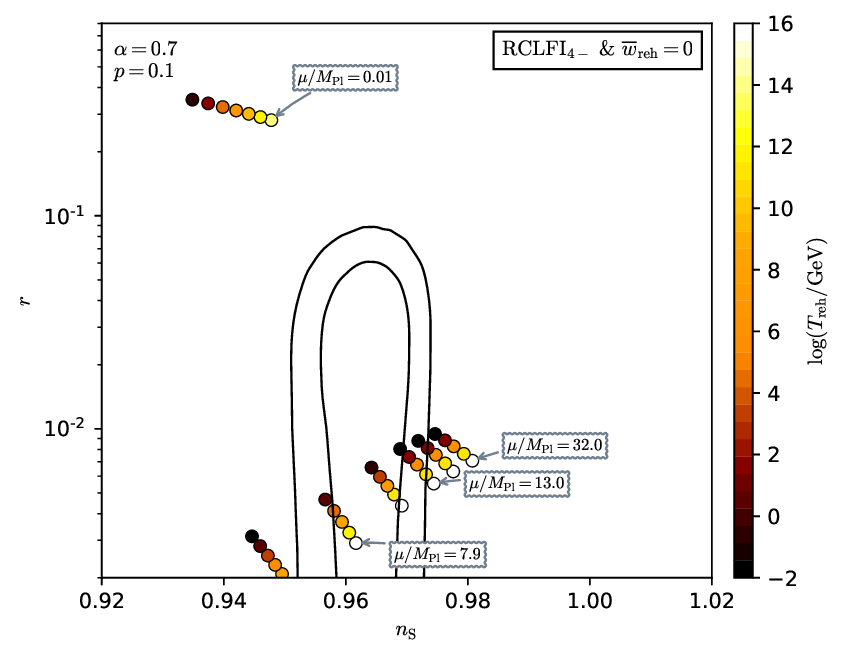}
\includegraphics[width=\wappfig,clip=true]{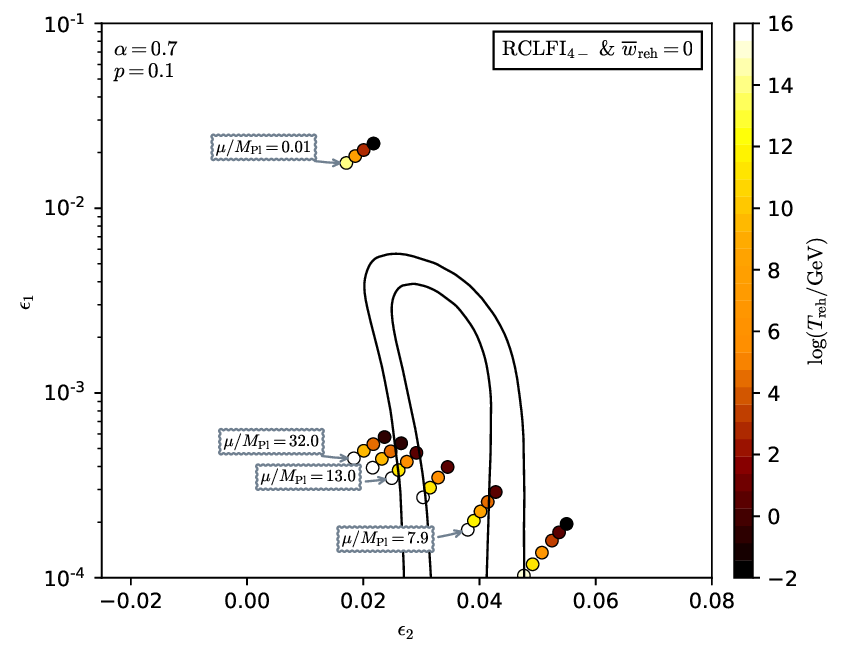}
\caption{Reheating consistent slow-roll predictions for the
  Radiatively Corrected Large Field Inflation models, in the RCLFI4
  regime, for $p<4$ and $0 < \alpha <
  -[p(p-4)/4]\exp(2-p/4)$. Predictions are represented in the plane
  $(\nS,r)$ (top panel) and in the plane $(\epsilon_1,\epsilon_2)$
  (bottom panel). The solid contours are the one and two-sigma {\data}
  confidence intervals (marginalized over second order slow-roll).}
\label{fig:CMBRCLFI4m_1}
\end{center}
\end{figure}

\begin{figure}[H]
\begin{center}
\includegraphics[width=\wappfig,clip=true]{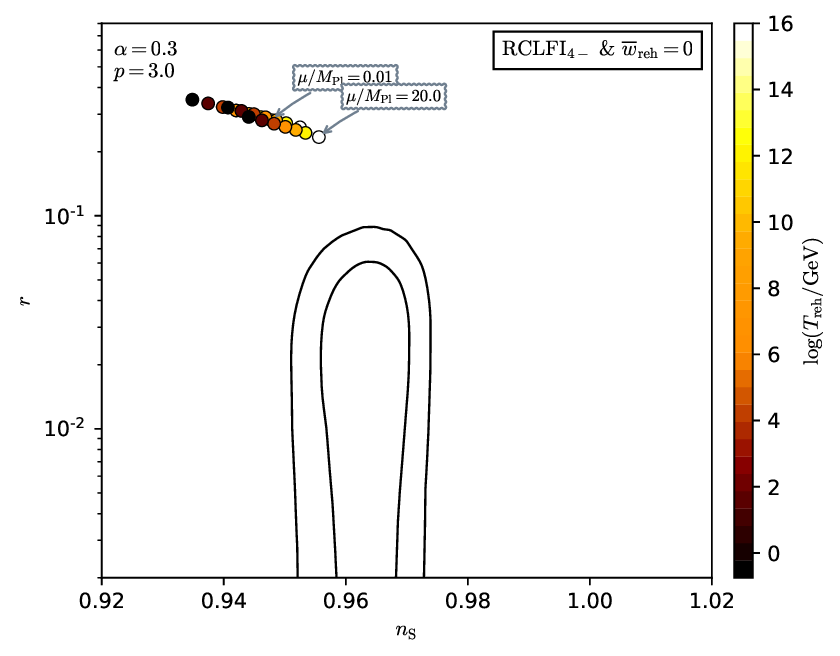}
\includegraphics[width=\wappfig,clip=true]{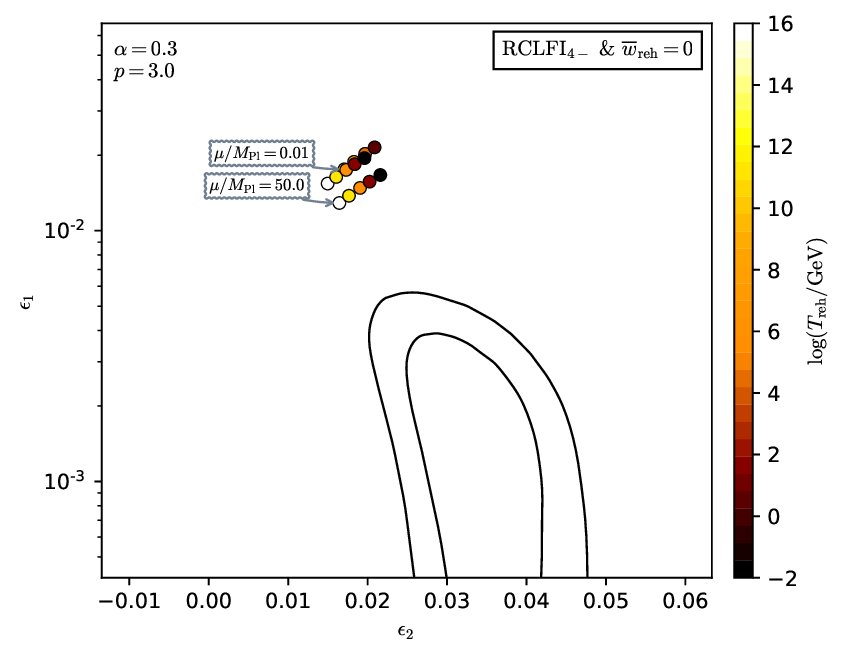}
\caption{Reheating consistent slow-roll predictions for the
  Radiatively Corrected Large Field Inflation models, in the RCLFI4
  regime, for $p<4$ and $0 < \alpha <
  -[p(p-4)/4]\exp(2-p/4)$. Predictions are represented in the plane
  $(\nS,r)$ (top panel) and in the plane $(\epsilon_1,\epsilon_2)$
  (bottom panel). The solid contours are the one and two-sigma {\data}
  confidence intervals (marginalized over second order slow-roll).}
\label{fig:CMBRCLFI4m_2}
\end{center}
\end{figure}

\begin{figure}[H]
\begin{center}
\includegraphics[width=\wappfig,clip=true]{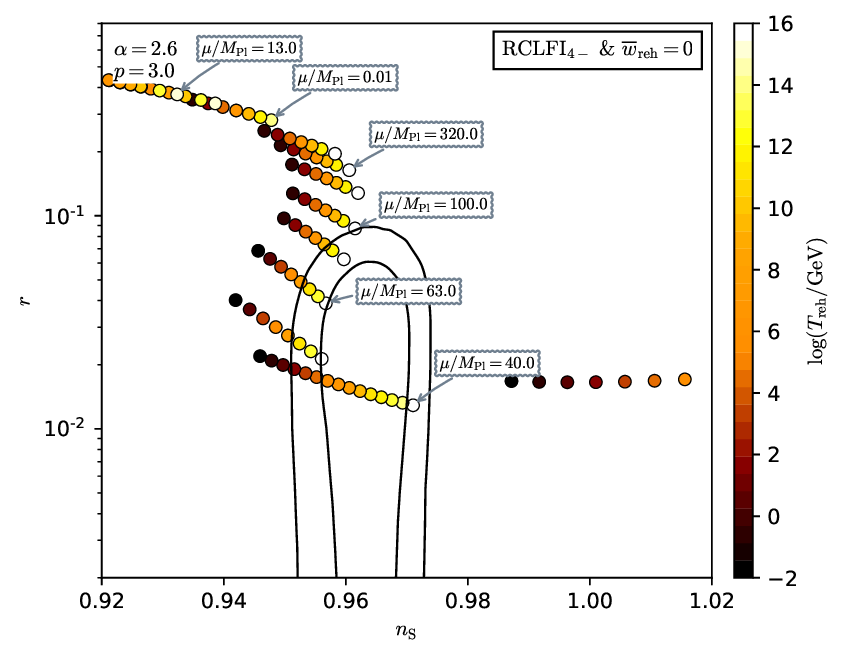}
\includegraphics[width=\wappfig,clip=true]{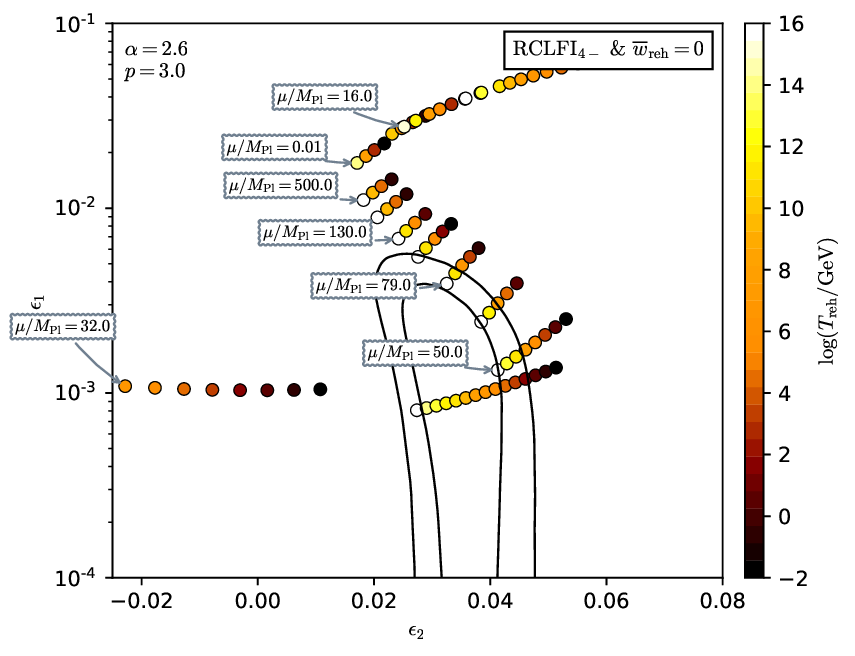}
\caption{Reheating consistent slow-roll predictions for the
  Radiatively Corrected Large Field Inflation models, in the RCLFI4
  regime, for $p<4$ and $0 < \alpha <
  -[p(p-4)/4]\exp(2-p/4)$. Predictions are represented in the plane
  $(\nS,r)$ (top panel) and in the plane $(\epsilon_1,\epsilon_2)$
  (bottom panel). The solid contours are the one and two-sigma {\data}
  confidence intervals (marginalized over second order slow-roll).}
\label{fig:CMBRCLFI4m_3}
\end{center}
\end{figure}

\subsection{Non-Renormalizable Corrected Loop Inflation  (\hyperref[sec:ncli]{NCLI})}

\begin{figure}[H]
\begin{center}
\includegraphics[width=\wappfig,clip=true]{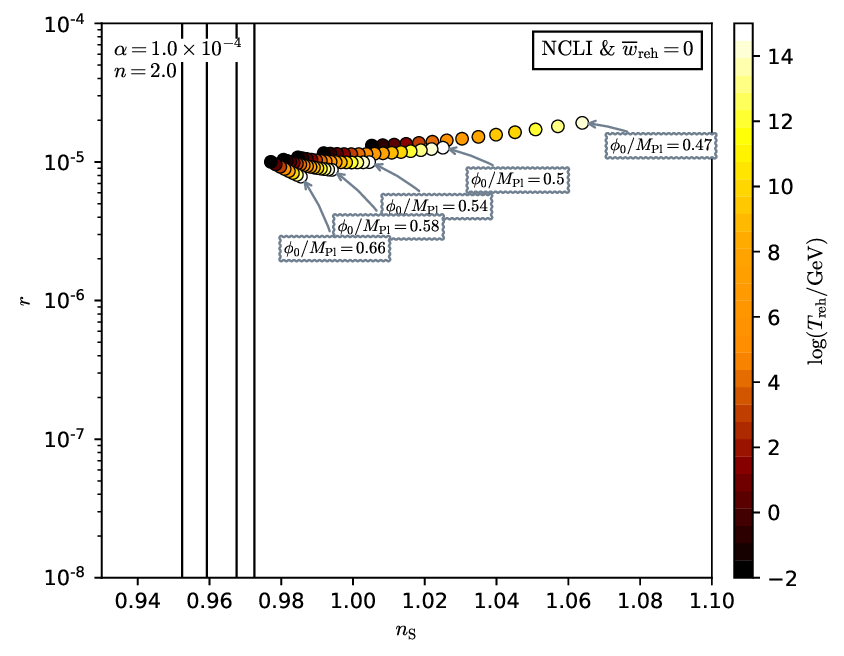}
\includegraphics[width=\wappfig,clip=true]{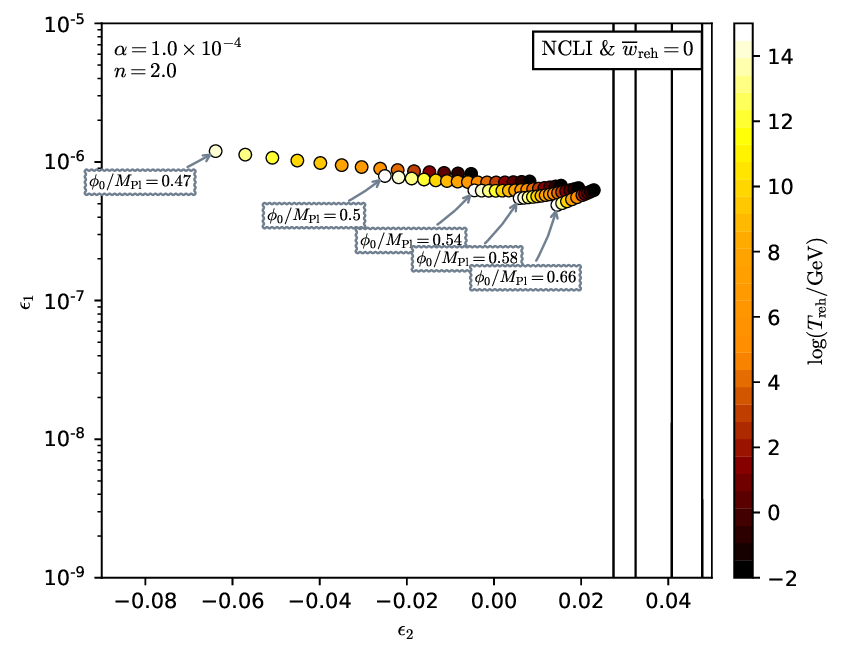}
\caption{Reheating consistent slow-roll predictions for the
  Non-Renormalizable Corrected Loop Inflation models, for $n=2$ and
  $\alpha=10^{-4}$. Predictions are represented in the plane $(\nS,r)$
  (top panel) and in the plane $(\epsilon_1,\epsilon_2)$ (bottom
  panel). The solid contours are the one and two-sigma {\data}
  confidence intervals (marginalized over second order slow-roll). See
  also Figs.~\ref{fig:CMBRNCLI_1} to \ref{fig:CMBRNCLI_3} for other
  values of $\alpha$ and $n$.}
\label{fig:CMBRNCLI_0}
\end{center}
\end{figure}

\begin{figure}[H]
\begin{center}
\includegraphics[width=\wappfig,clip=true]{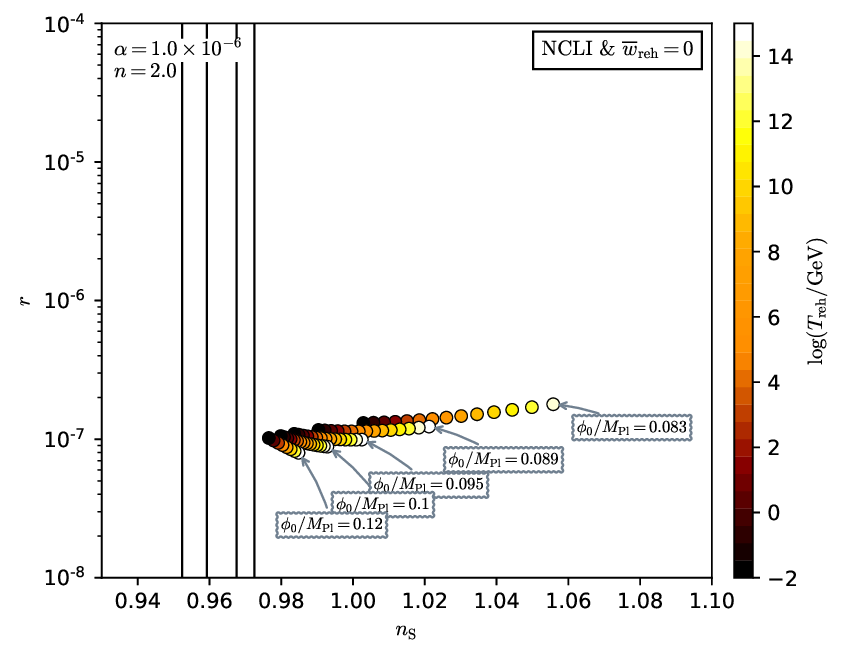}
\includegraphics[width=\wappfig,clip=true]{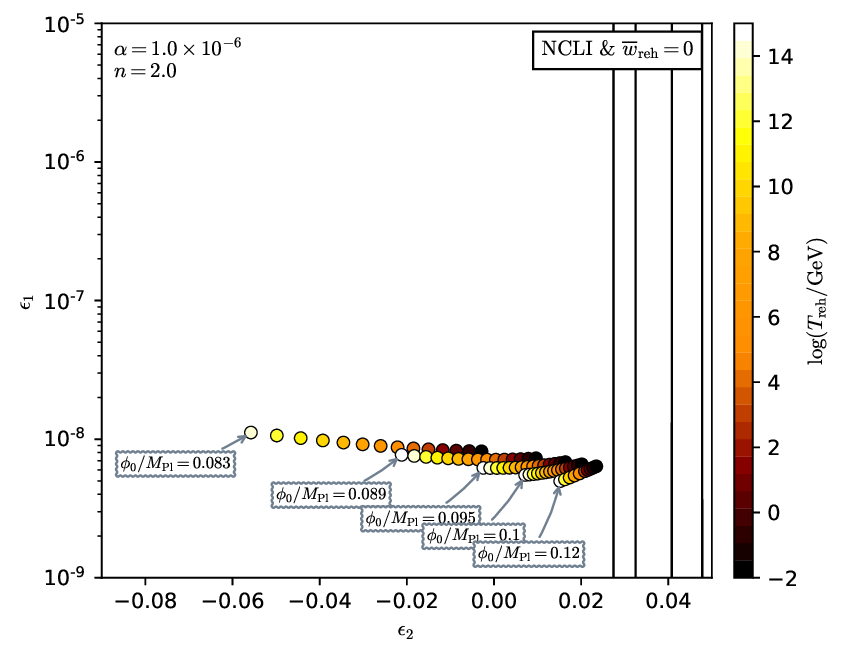}
\caption{Reheating consistent slow-roll predictions for the
  Non-Renormalizable Corrected Loop Inflation models, for $n=2$ and
  $\alpha=10^{-6}$. Predictions are represented in the plane $(\nS,r)$
  (top panel) and in the plane $(\epsilon_1,\epsilon_2)$ (bottom
  panel). The solid contours are the one and two-sigma {\data}
  confidence intervals (marginalized over second order slow-roll).}
\label{fig:CMBRNCLI_1}
\end{center}
\end{figure}

\begin{figure}[H]
\begin{center}
\includegraphics[width=\wappfig,clip=true]{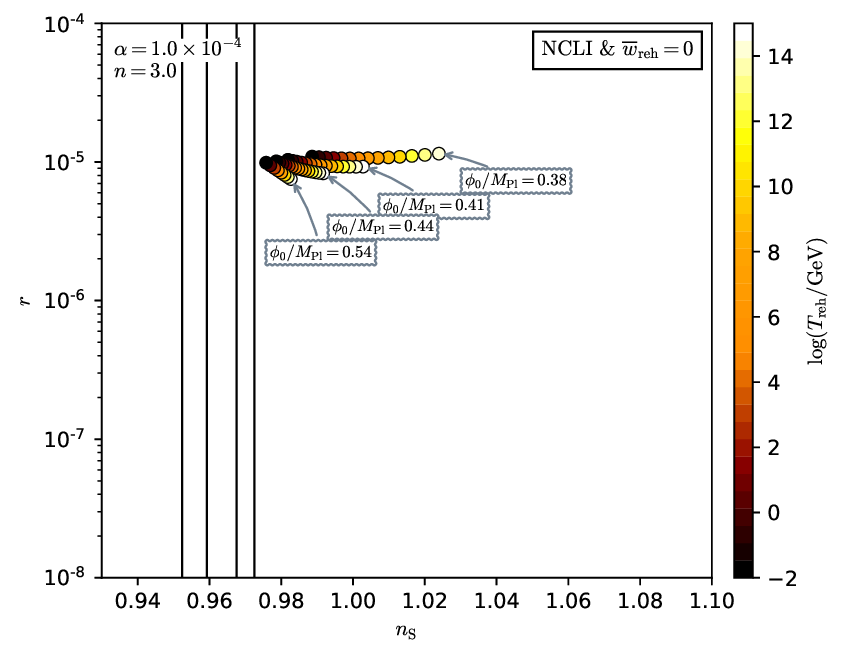}
\includegraphics[width=\wappfig,clip=true]{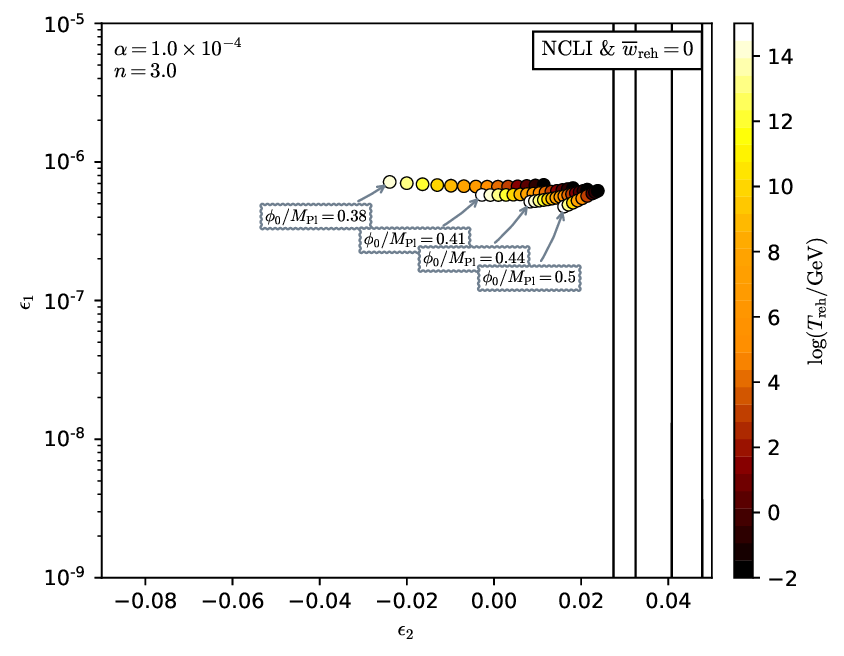}
\caption{Reheating consistent slow-roll predictions for the
  Non-Renormalizable Corrected Loop Inflation models, for $n=3$ and
  $\alpha=10^{-4}$. Predictions are represented in the plane $(\nS,r)$
  (top panel) and in the plane $(\epsilon_1,\epsilon_2)$ (bottom
  panel). The solid contours are the one and two-sigma {\data}
  confidence intervals (marginalized over second order slow-roll).}
\label{fig:CMBRNCLI_2}
\end{center}
\end{figure}

\begin{figure}[H]
\begin{center}
\includegraphics[width=\wappfig,clip=true]{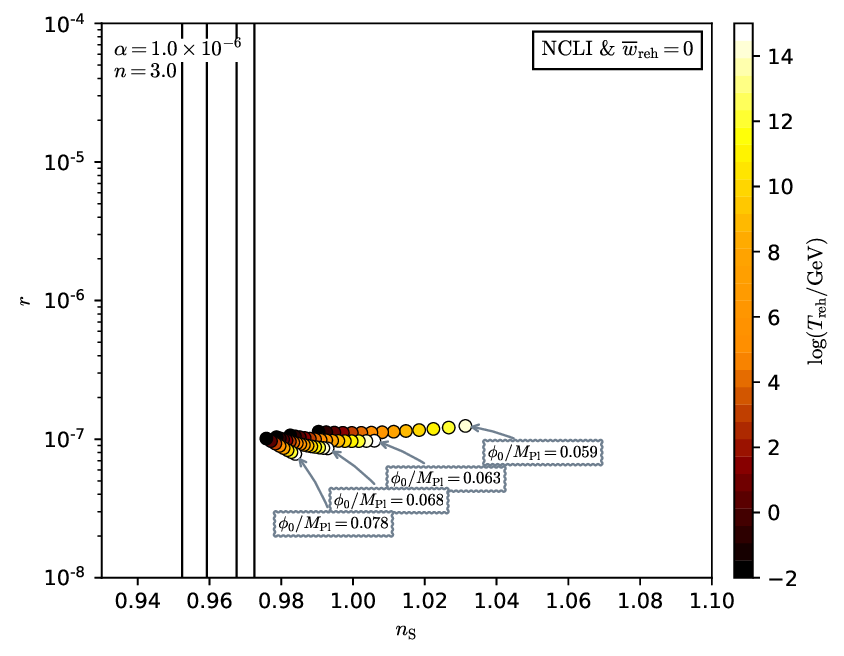}
\includegraphics[width=\wappfig,clip=true]{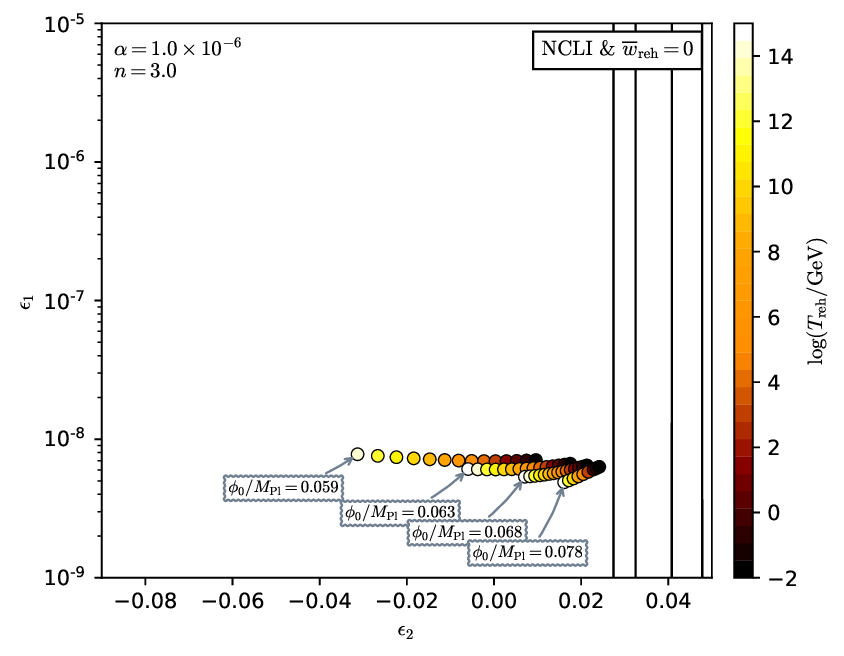}
\caption{Reheating consistent slow-roll predictions for the
  Non-Renormalizable Corrected Loop Inflation models, for $n=3$ and
  $\alpha=10^{-6}$. Predictions are represented in the plane $(\nS,r)$
  (top panel) and in the plane $(\epsilon_1,\epsilon_2)$ (bottom
  panel). The solid contours are the one and two-sigma {\data}
  confidence intervals (marginalized over second order slow-roll).}
\label{fig:CMBRNCLI_3}
\end{center}
\end{figure}

\subsection{Hybrid Natural Inflation 1 (\hyperref[sec:hni]{HNI1})}

\begin{figure}[H]
\begin{center}
\includegraphics[width=\wappfig,clip=true]{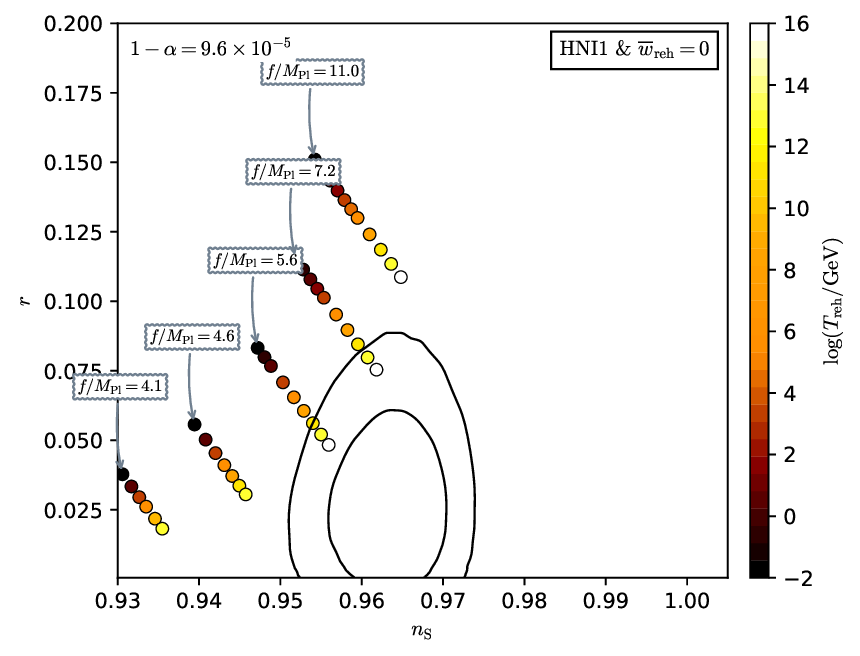}
\includegraphics[width=\wappfig,clip=true]{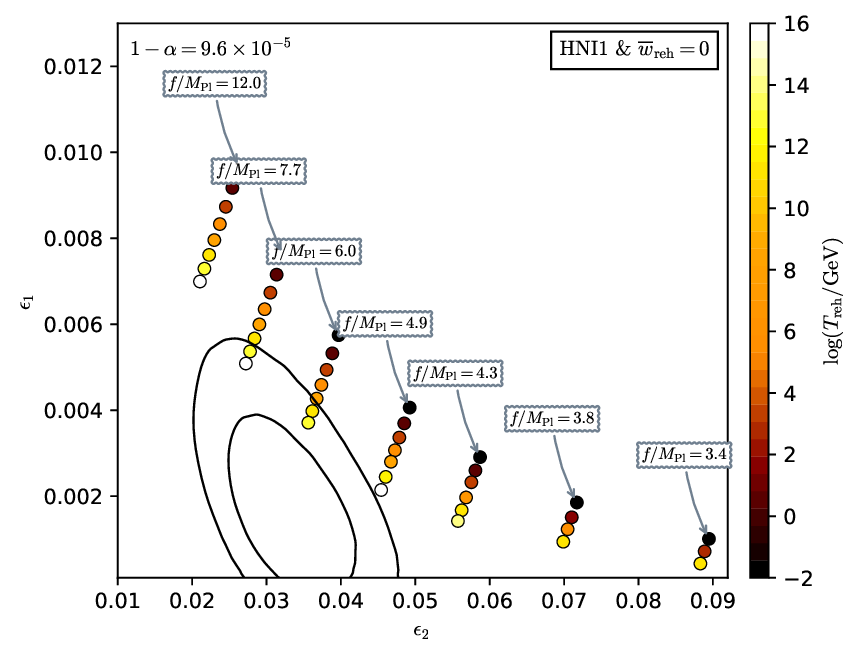}
\caption{Reheating consistent slow-roll predictions for Hybrid Natural
  Inflation 1 and for $1-\alpha=9.6\times10^{-5}$ (inflation
  gracefully ends $\alpha > \alpha_1$). Predictions are represented in
  the plane $(\nS,r)$ (top panel) and in the plane
  $(\epsilon_1,\epsilon_2)$ (bottom panel) for various values of the
  field {\vev} $f$. The solid contours are the one and two-sigma
  {\data} confidence intervals (marginalized over second order
  slow-roll). See also Figs.~\ref{fig:CMBHNI1_1} to
  \ref{fig:CMBHNI1_3} for other values of $\alpha$.}
\label{fig:CMBHNI1_0}
\end{center}
\end{figure}

\begin{figure}[H]
\begin{center}
\includegraphics[width=\wappfig,clip=true]{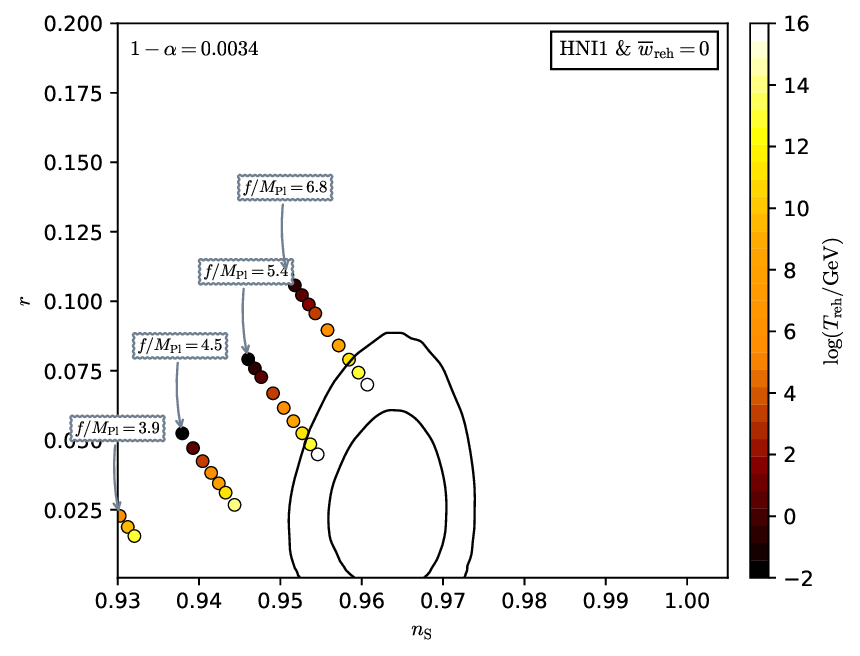}
\includegraphics[width=\wappfig,clip=true]{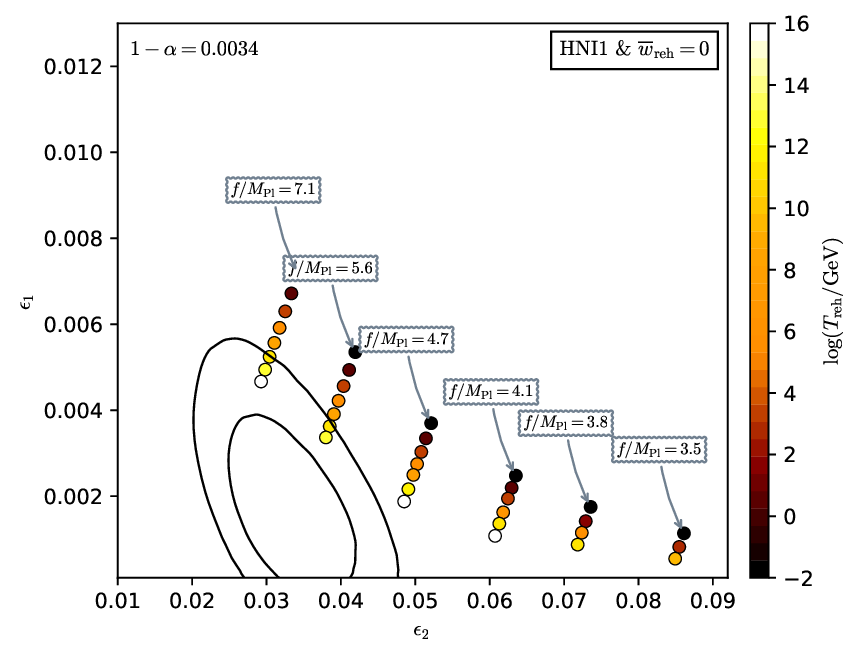}
\caption{Reheating consistent slow-roll predictions for Hybrid Natural
  Inflation 1 and for $1-\alpha=3.4\times 10^{-3}$ (inflation
  gracefully ends $\alpha > \alpha_1$). Predictions are represented in
  the plane $(\nS,r)$ (top panel) and in the plane
  $(\epsilon_1,\epsilon_2)$ (bottom panel) for various values of the
  field {\vev} $f$. The solid contours are the one and two-sigma
  {\data} confidence intervals (marginalized over second order
  slow-roll). See also Figs.~\ref{fig:CMBHNI1_1} to
  \ref{fig:CMBHNI1_3} for other values of $\alpha$.}
\label{fig:CMBHNI1_1}
\end{center}
\end{figure}

\begin{figure}[H]
\begin{center}
\includegraphics[width=\wappfig,clip=true]{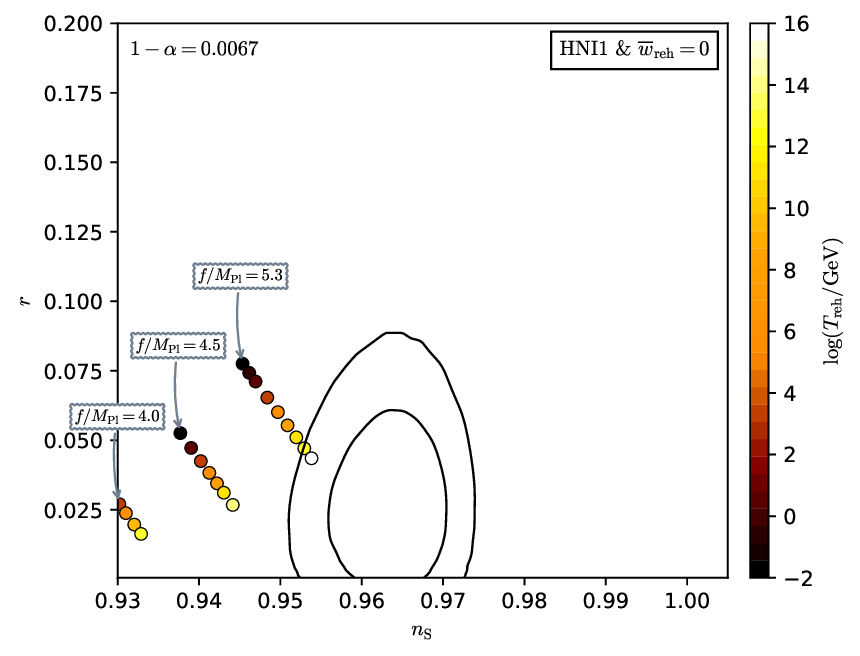}
\includegraphics[width=\wappfig,clip=true]{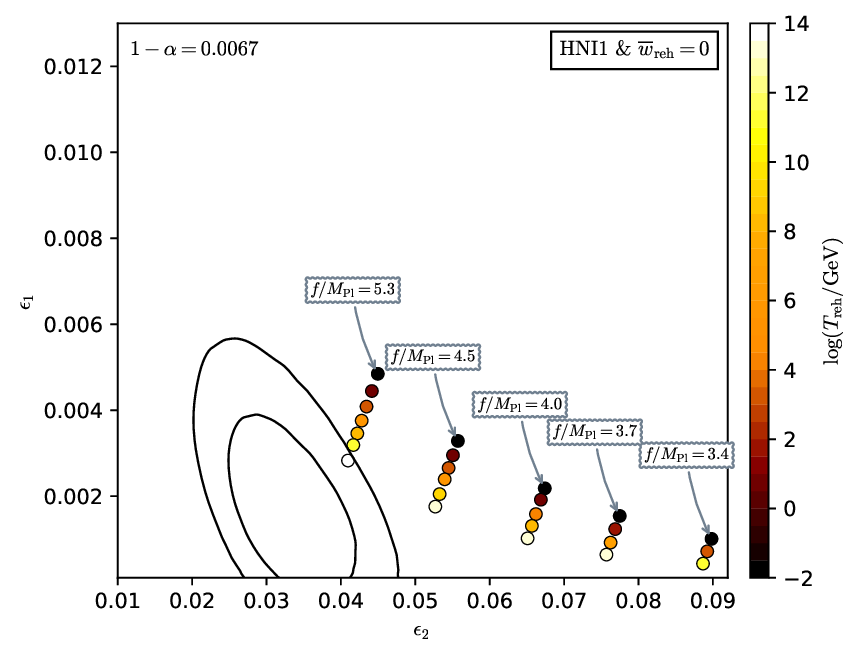}
\caption{Reheating consistent slow-roll predictions for Hybrid Natural
  Inflation 1 and for $1-\alpha=6.7\times10^{-5}$ (inflation
  gracefully ends $\alpha > \alpha_1$). Predictions are represented in
  the plane $(\nS,r)$ (top panel) and in the plane
  $(\epsilon_1,\epsilon_2)$ (bottom panel) for various values of the
  field {\vev} $f$. The solid contours are the one and two-sigma
  {\data} confidence intervals (marginalized over second order
  slow-roll).}
\label{fig:CMBHNI1_2}
\end{center}
\end{figure}

\begin{figure}[H]
\begin{center}
\includegraphics[width=\wappfig,clip=true]{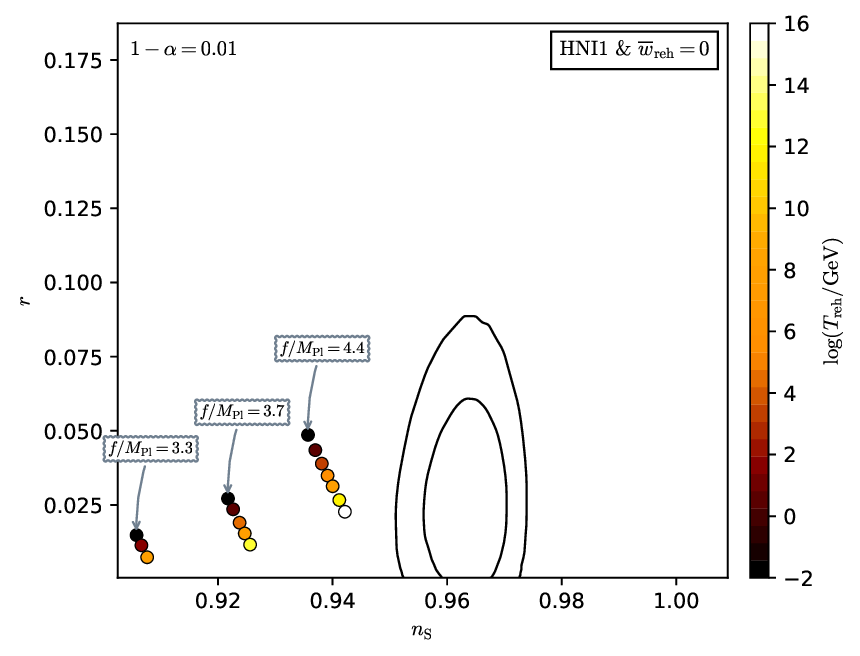}
\includegraphics[width=\wappfig,clip=true]{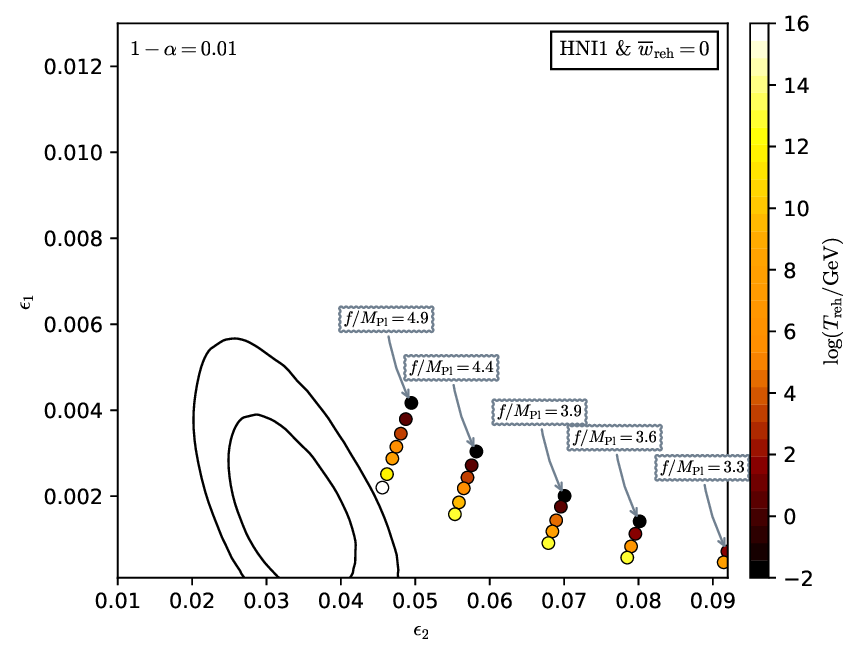}
\caption{Reheating consistent slow-roll predictions for Hybrid Natural
  Inflation 1 and for $1-\alpha=10^{-2}$ (inflation
  gracefully ends $\alpha > \alpha_1$). Predictions are represented in
  the plane $(\nS,r)$ (top panel) and in the plane
  $(\epsilon_1,\epsilon_2)$ (bottom panel) for various values of the
  field {\vev} $f$. The solid contours are the one and two-sigma
  {\data} confidence intervals (marginalized over second order
  slow-roll).}
\label{fig:CMBHNI1_3}
\end{center}
\end{figure}

\subsection{Hybrid Natural Inflation 2 (\hyperref[sec:hni]{HNI2})}

\begin{figure}[H]
\begin{center}
\includegraphics[width=\wappfig,clip=true]{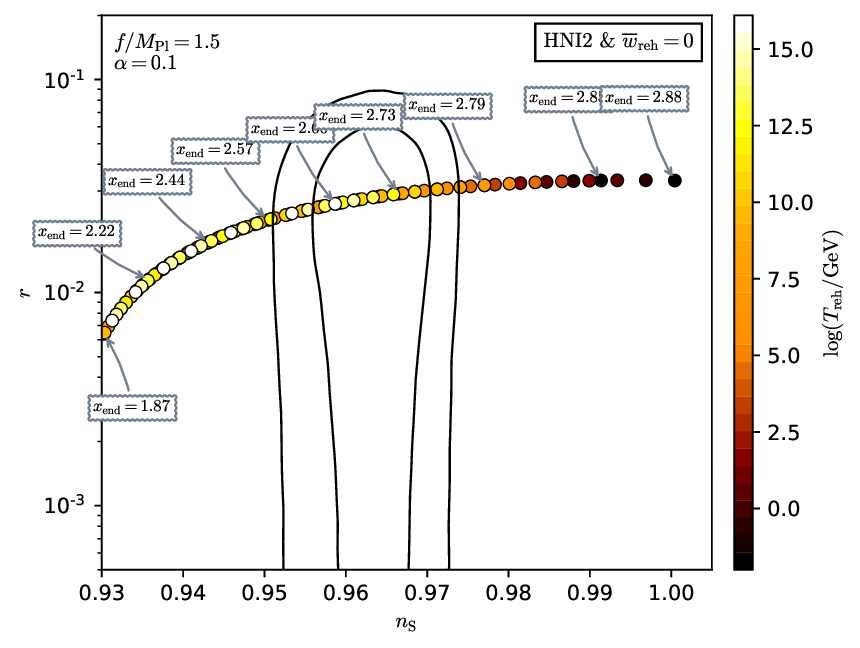}
\includegraphics[width=\wappfig,clip=true]{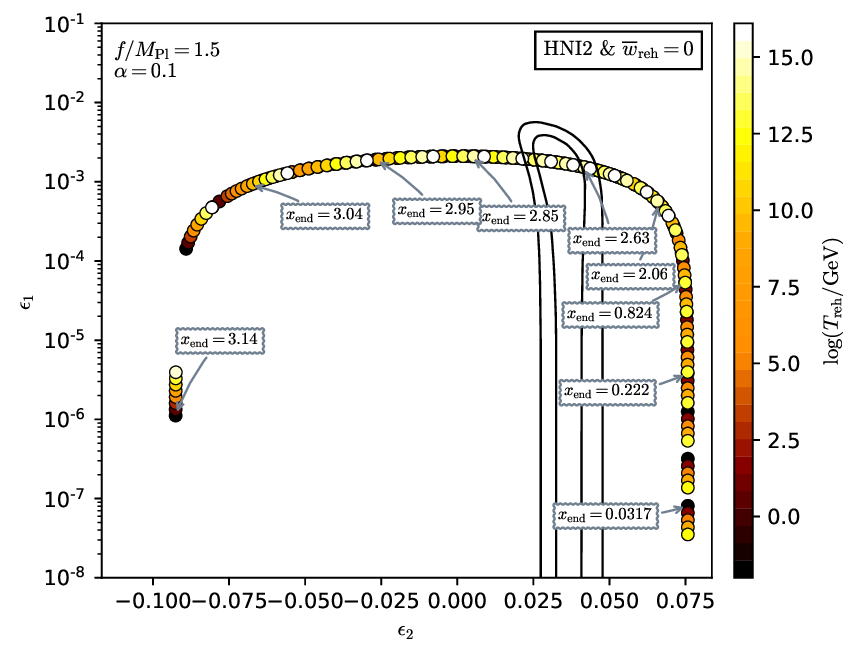}
\caption{Reheating consistent slow-roll predictions for Hybrid Natural
  Inflation 2 and for $\alpha=0.1$ and $f=1.5 \Mp$. Predictions are
  represented in the plane $(\nS,r)$ (top panel) and in the plane
  $(\epsilon_1,\epsilon_2)$ (bottom panel) for various field values
  $\xend$ at which the instability ends inflation. The solid contours
  are the one and two-sigma {\data} confidence intervals (marginalized
  over second order slow-roll). See also Figs.~\ref{fig:CMBHNI2_1} to
  \ref{fig:CMBHNI2_7} for other values of $\alpha$ and $f$.}
\label{fig:CMBHNI2_0}
\end{center}
\end{figure}

\begin{figure}[H]
\begin{center}
\includegraphics[width=\wappfig,clip=true]{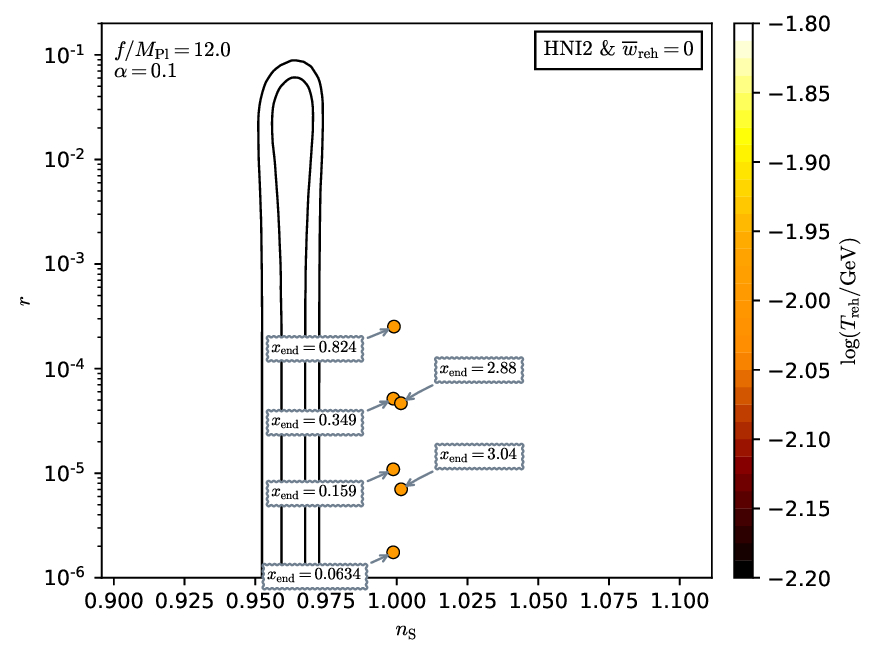}
\includegraphics[width=\wappfig,clip=true]{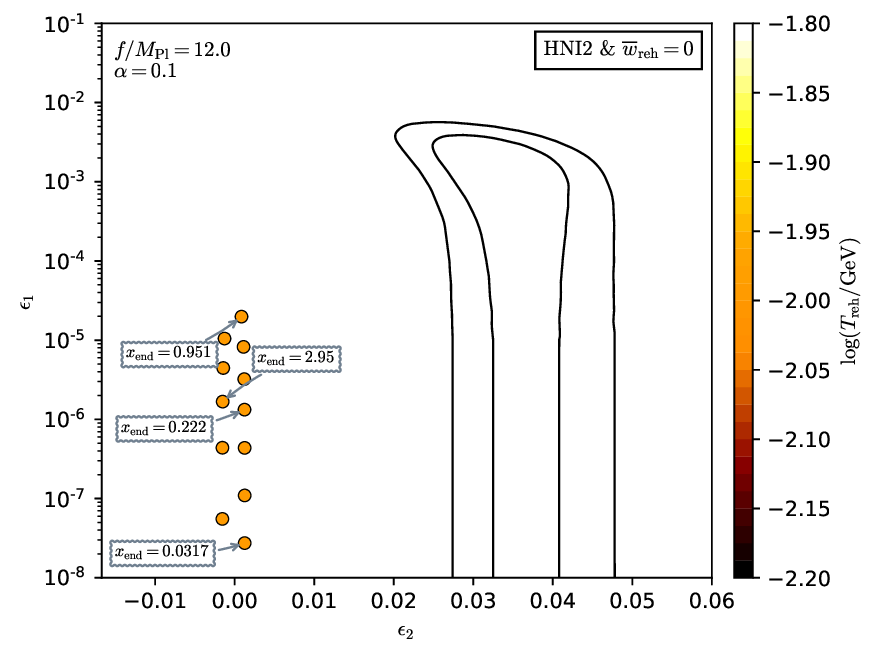}
\caption{Reheating consistent slow-roll predictions for Hybrid Natural
  Inflation 2 and for $\alpha=0.1$ and $f=4.3 \Mp$. Predictions are
  represented in the plane $(\nS,r)$ (top panel) and in the plane
  $(\epsilon_1,\epsilon_2)$ (bottom panel) for various field values
  $\xend$ at which the instability ends inflation. The solid contours
  are the one and two-sigma {\data} confidence intervals (marginalized
  over second order slow-roll).}
\label{fig:CMBHNI2_1}
\end{center}
\end{figure}

\begin{figure}[H]
\begin{center}
\includegraphics[width=\wappfig,clip=true]{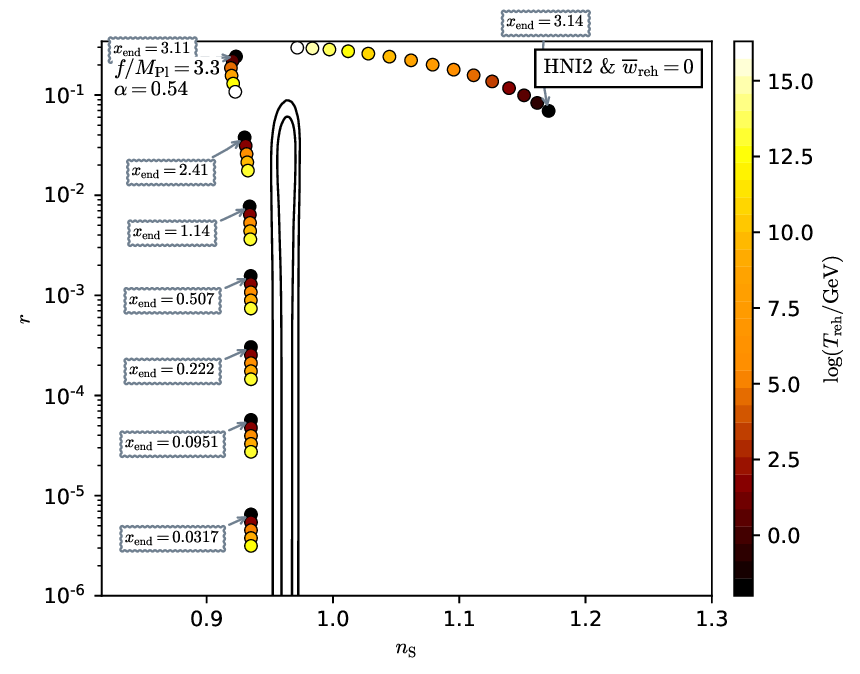}
\includegraphics[width=\wappfig,clip=true]{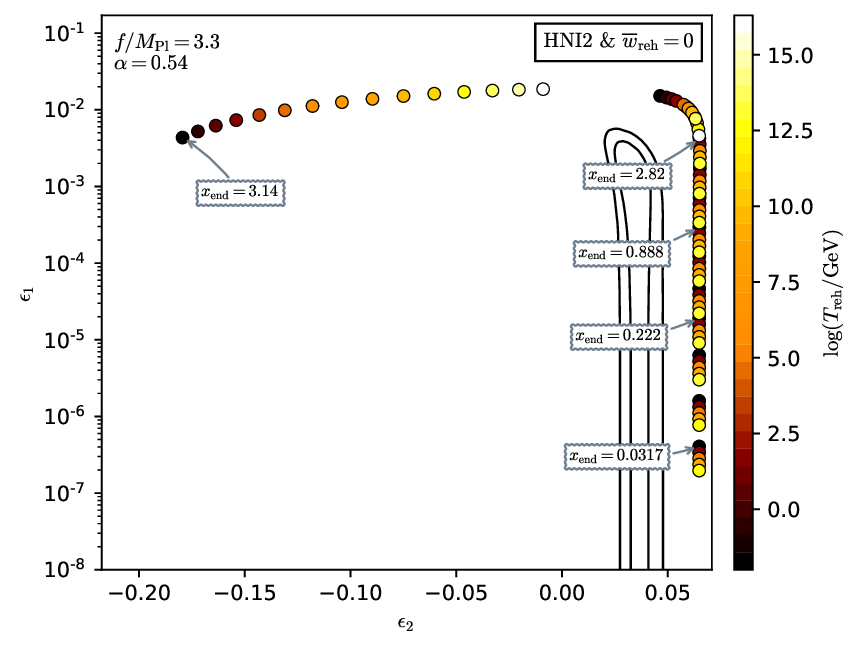}
\caption{Reheating consistent slow-roll predictions for Hybrid Natural
  Inflation 2 and for $\alpha=0.1$ and $f=12 \Mp$. Predictions are
  represented in the plane $(\nS,r)$ (top panel) and in the plane
  $(\epsilon_1,\epsilon_2)$ (bottom panel) for various field values
  $\xend$ at which the instability ends inflation. The solid contours
  are the one and two-sigma {\data} confidence intervals (marginalized
  over second order slow-roll).}
\label{fig:CMBHNI2_2}
\end{center}
\end{figure}

\begin{figure}[H]
\begin{center}
\includegraphics[width=\wappfig,clip=true]{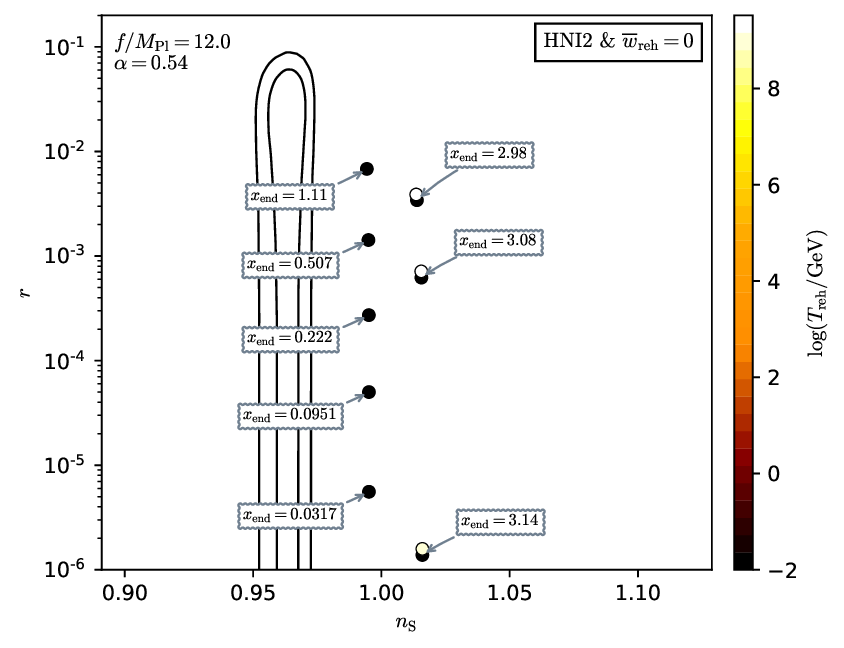}
\includegraphics[width=\wappfig,clip=true]{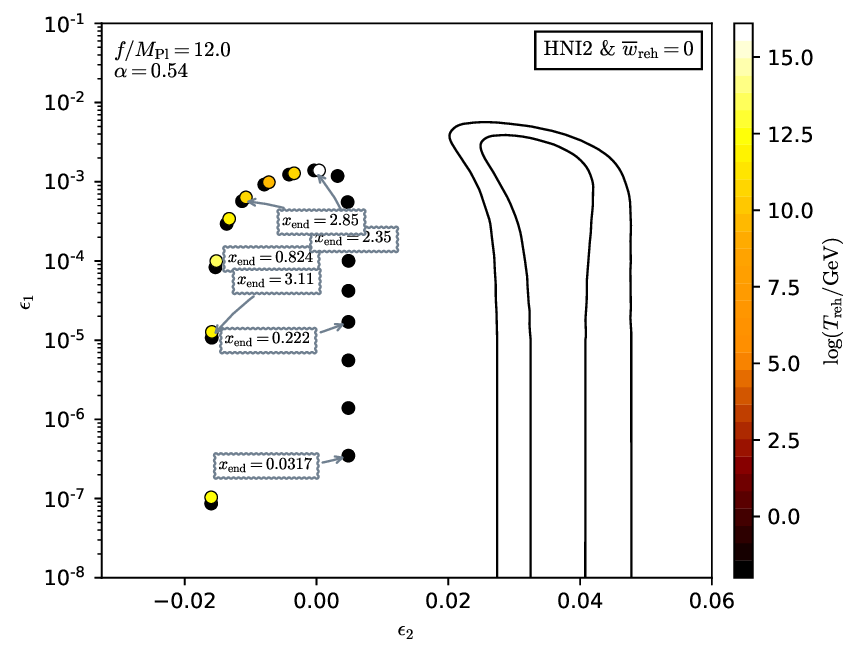}
\caption{Reheating consistent slow-roll predictions for Hybrid Natural
  Inflation 2 and for $\alpha=0.54$ and $f=1.5 \Mp$. Predictions are
  represented in the plane $(\nS,r)$ (top panel) and in the plane
  $(\epsilon_1,\epsilon_2)$ (bottom panel) for various field values
  $\xend$ at which the instability ends inflation. The solid contours
  are the one and two-sigma {\data} confidence intervals (marginalized
  over second order slow-roll).}
\label{fig:CMBHNI2_3}
\end{center}
\end{figure}

\begin{figure}[H]
\begin{center}
\includegraphics[width=\wappfig,clip=true]{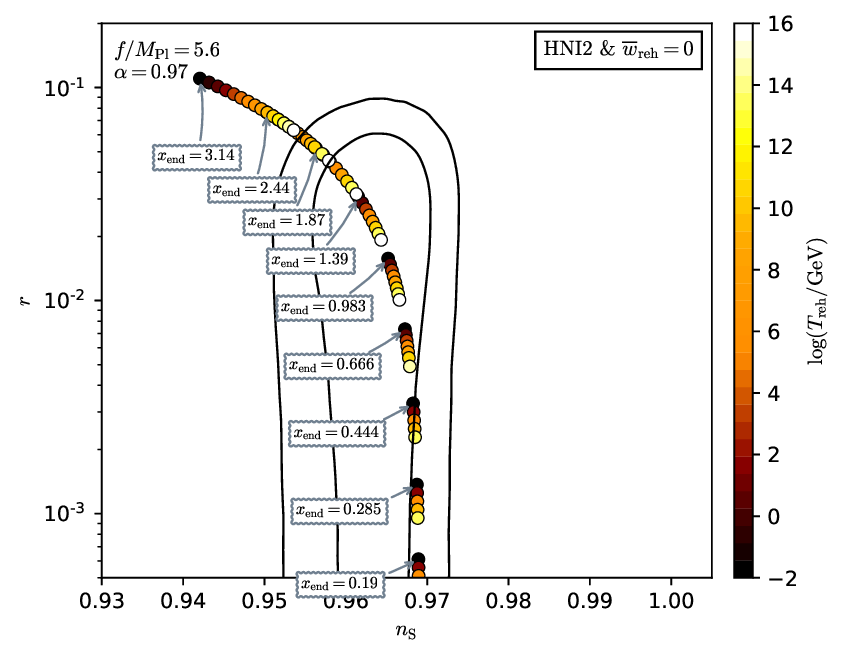}
\includegraphics[width=\wappfig,clip=true]{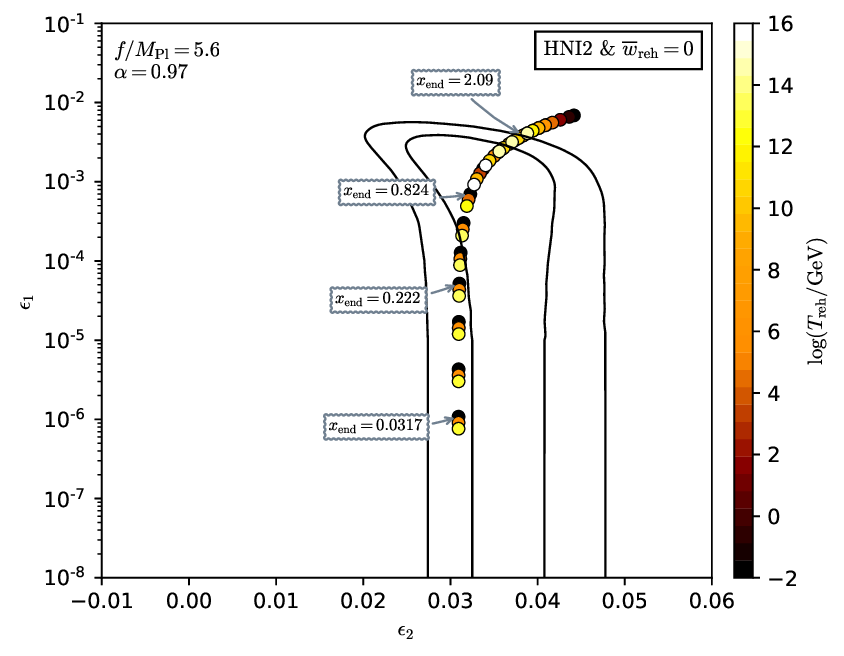}
\caption{Reheating consistent slow-roll predictions for Hybrid Natural
  Inflation 2 and for $\alpha=0.54$ and $f=4.3 \Mp$. Predictions are
  represented in the plane $(\nS,r)$ (top panel) and in the plane
  $(\epsilon_1,\epsilon_2)$ (bottom panel) for various field values
  $\xend$ at which the instability ends inflation. The solid contours
  are the one and two-sigma {\data} confidence intervals (marginalized
  over second order slow-roll).}
\label{fig:CMBHNI2_4}
\end{center}
\end{figure}

\begin{figure}[H]
\begin{center}
\includegraphics[width=\wappfig,clip=true]{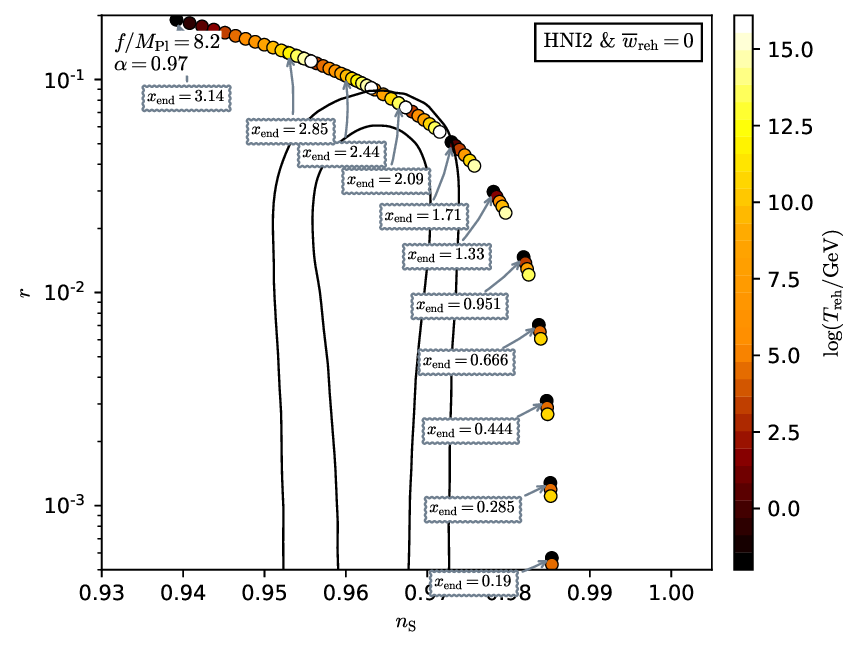}
\includegraphics[width=\wappfig,clip=true]{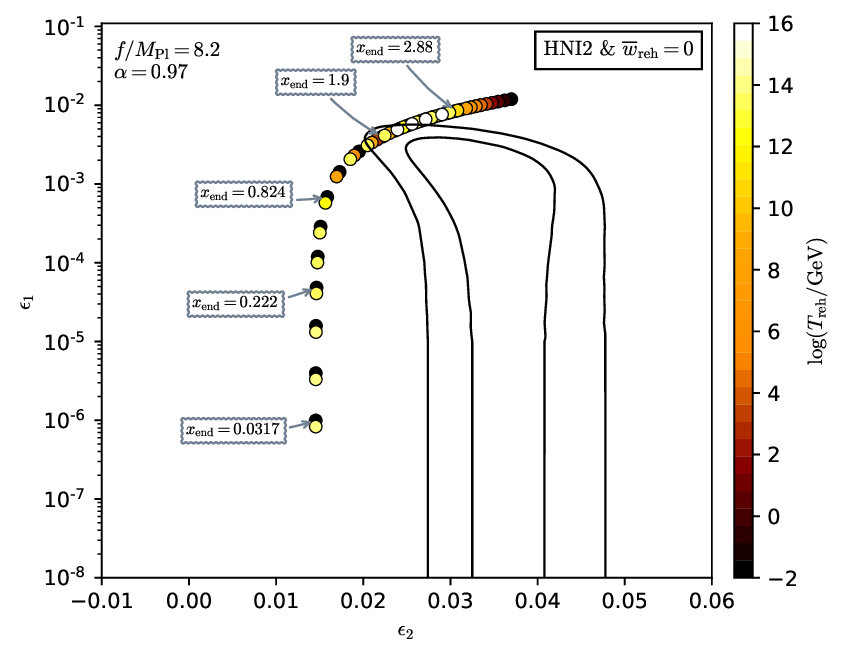}
\caption{Reheating consistent slow-roll predictions for Hybrid Natural
  Inflation 2 and for $\alpha=0.54$ and $f=12 \Mp$. Predictions are
  represented in the plane $(\nS,r)$ (top panel) and in the plane
  $(\epsilon_1,\epsilon_2)$ (bottom panel) for various field values
  $\xend$ at which the instability ends inflation. The solid contours
  are the one and two-sigma {\data} confidence intervals (marginalized
  over second order slow-roll).}
\label{fig:CMBHNI2_5}
\end{center}
\end{figure}

\begin{figure}[H]
\begin{center}
\includegraphics[width=\wappfig,clip=true]{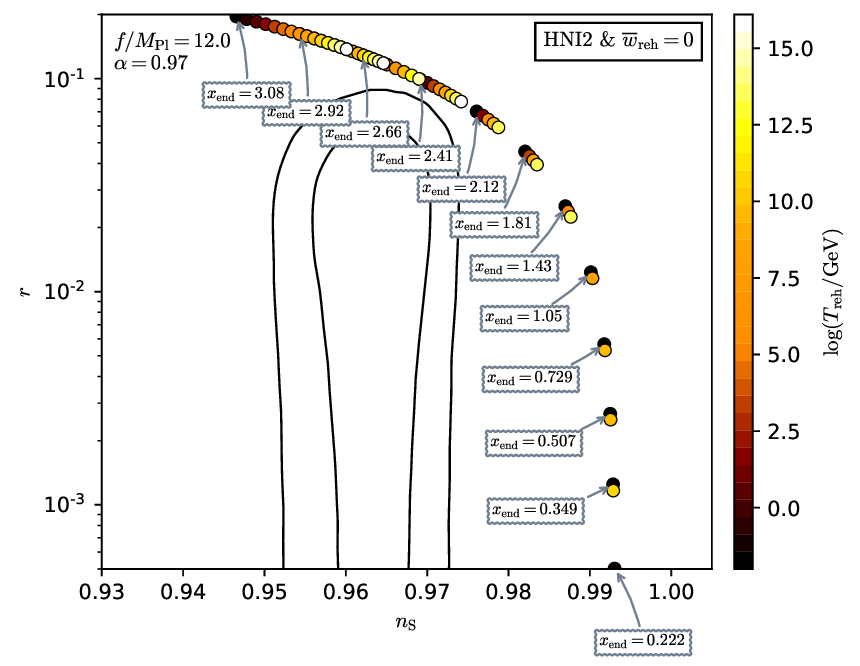}
\includegraphics[width=\wappfig,clip=true]{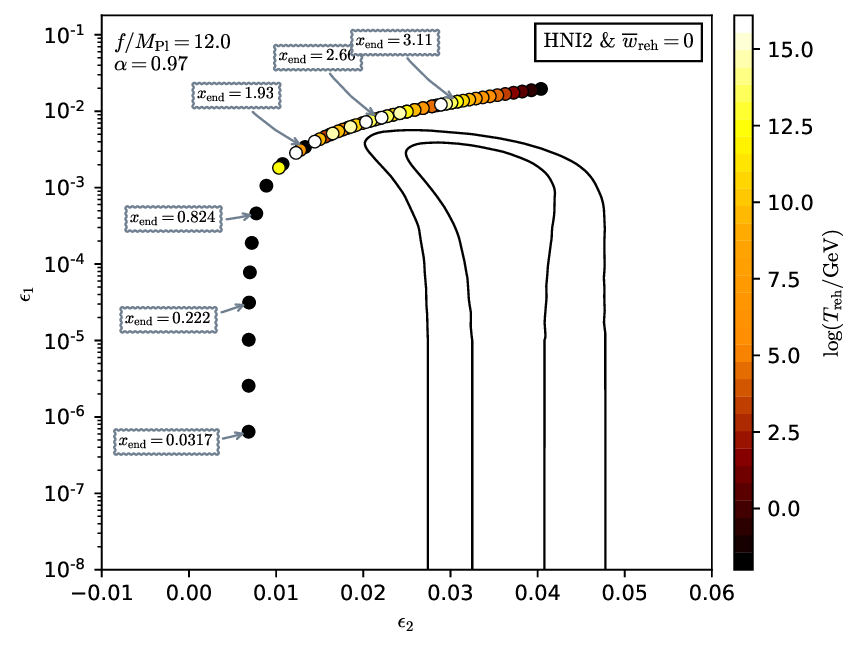}
\caption{Reheating consistent slow-roll predictions for Hybrid Natural
  Inflation 2 and for $\alpha=0.97$ and $f=4.3 \Mp$. Predictions are
  represented in the plane $(\nS,r)$ (top panel) and in the plane
  $(\epsilon_1,\epsilon_2)$ (bottom panel) for various field values
  $\xend$ at which the instability ends inflation. The solid contours
  are the one and two-sigma {\data} confidence intervals (marginalized
  over second order slow-roll).}
\label{fig:CMBHNI2_6}
\end{center}
\end{figure}

\begin{figure}[H]
\begin{center}
\includegraphics[width=\wappfig,clip=true]{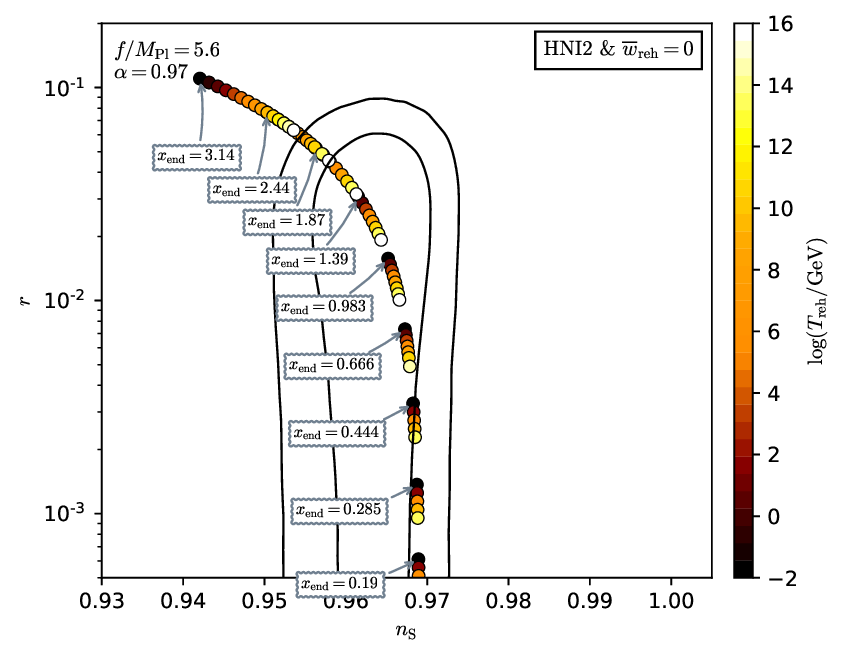}
\includegraphics[width=\wappfig,clip=true]{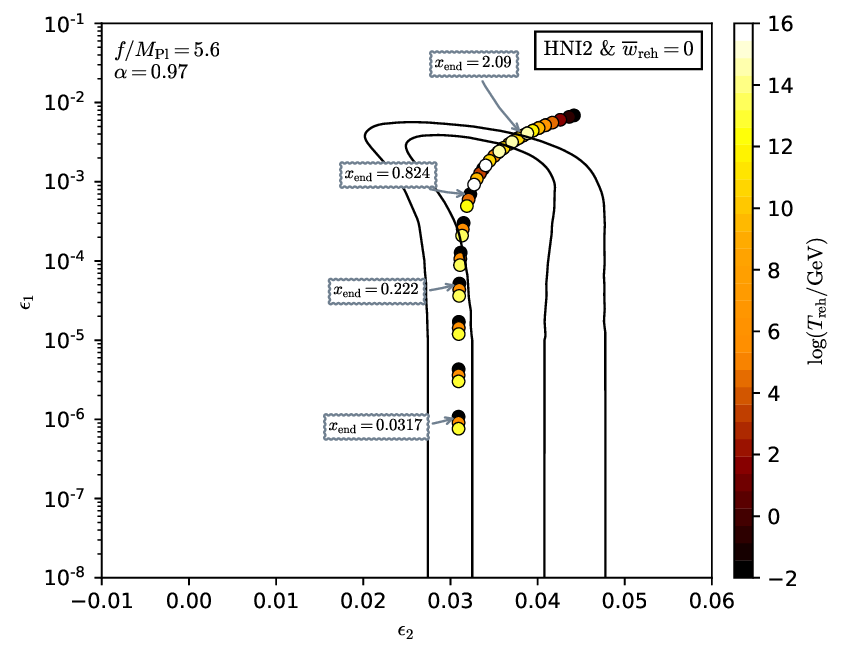}
\caption{Reheating consistent slow-roll predictions for Hybrid Natural
  Inflation 2 and for $\alpha=0.97$ and $f=12 \Mp$. Predictions are
  represented in the plane $(\nS,r)$ (top panel) and in the plane
  $(\epsilon_1,\epsilon_2)$ (bottom panel) for various field values
  $\xend$ at which the instability ends inflation. The solid contours
  are the one and two-sigma {\data} confidence intervals (marginalized
  over second order slow-roll).}
\label{fig:CMBHNI2_7}
\end{center}
\end{figure}

\subsection{N-Formalism Inflation 1 (\hyperref[sec:nfi]{NFI1})}

\begin{figure}[H]
\begin{center}
\includegraphics[width=\wappfig,clip=true]{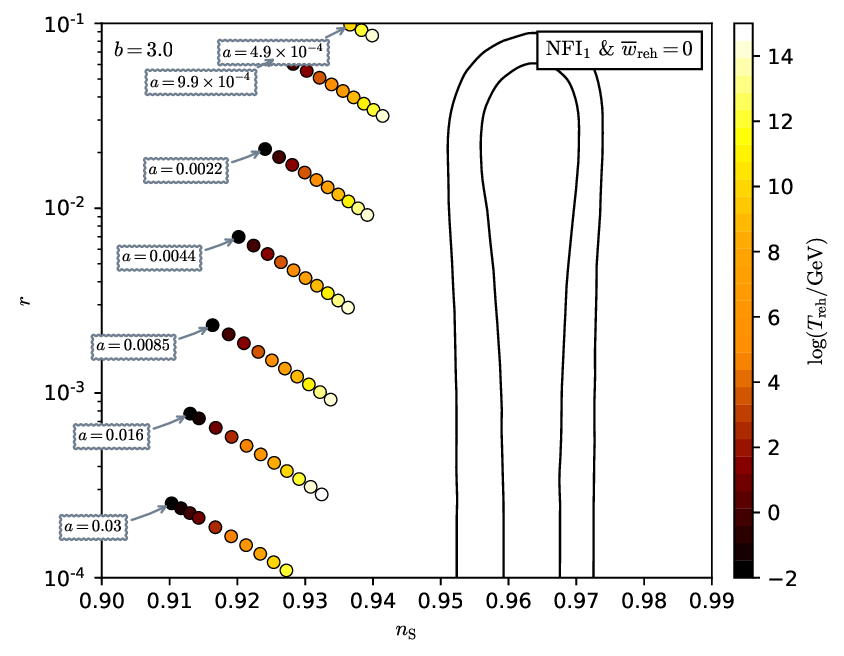}
\includegraphics[width=\wappfig,clip=true]{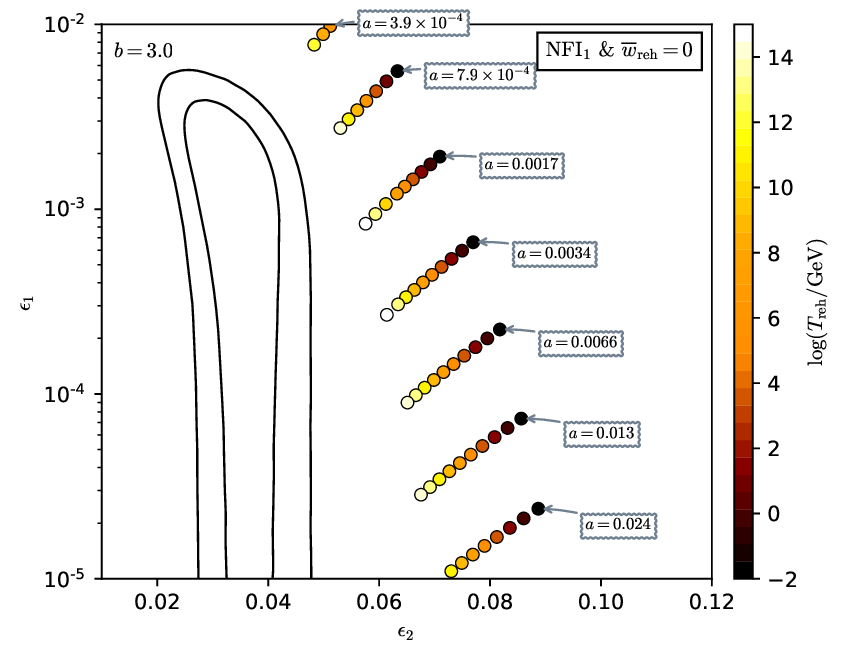}
\caption{Reheating consistent slow-roll predictions for N-Formalism
  Inflation 1 with $b=3$ (inflation gracefully ends). Predictions
  are represented in the plane $(\nS,r)$ (top panel) and in the plane
  $(\epsilon_1,\epsilon_2)$ (bottom panel) for various values of the
  parameter $a$. The solid contours are the one and two-sigma
  {\data} confidence intervals (marginalized over second order
  slow-roll). See also Figs.~\ref{fig:CMBNFI1_1} to
  \ref{fig:CMBNFI1_2} for other values of $b$.}
\label{fig:CMBNFI1_0}
\end{center}
\end{figure}

\begin{figure}[H]
\begin{center}
\includegraphics[width=\wappfig,clip=true]{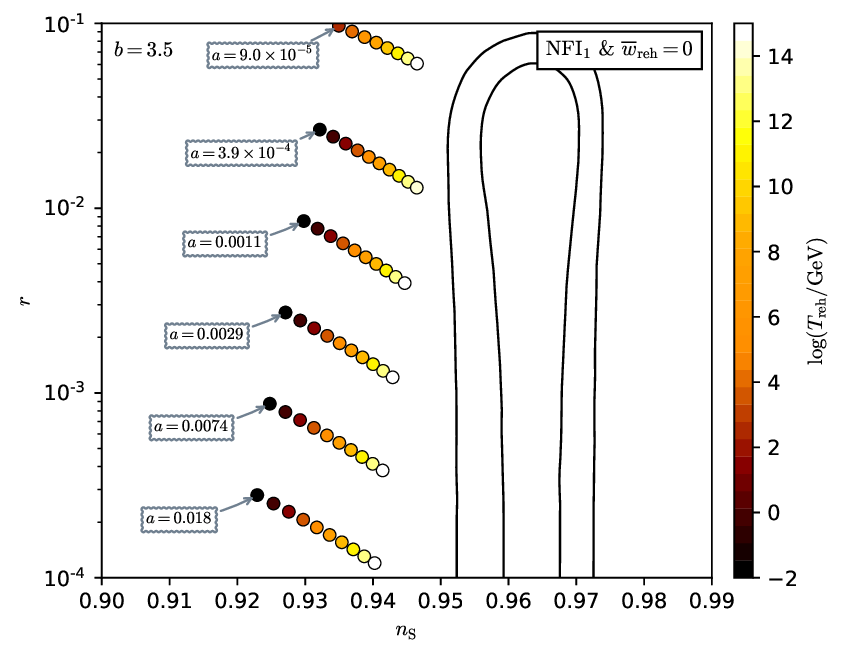}
\includegraphics[width=\wappfig,clip=true]{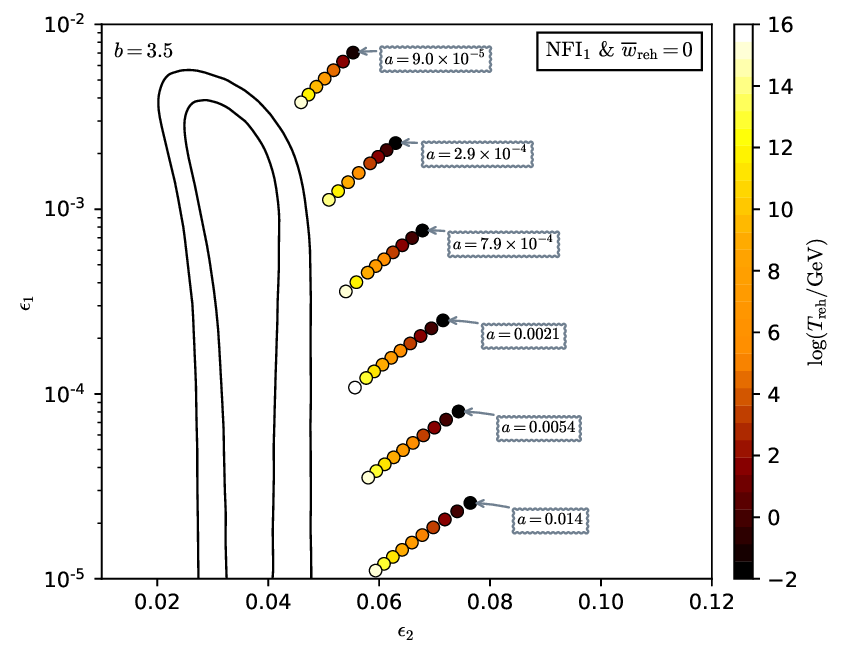}
\caption{Reheating consistent slow-roll predictions for N-Formalism
  Inflation 1 with $b=3.5$ (inflation gracefully ends). Predictions
  are represented in the plane $(\nS,r)$ (top panel) and in the plane
  $(\epsilon_1,\epsilon_2)$ (bottom panel) for various values of the
  parameter $a$. The solid contours are the one and two-sigma
  {\data} confidence intervals (marginalized over second order
  slow-roll).}
\label{fig:CMBNFI1_1}
\end{center}
\end{figure}

\begin{figure}[H]
\begin{center}
\includegraphics[width=\wappfig,clip=true]{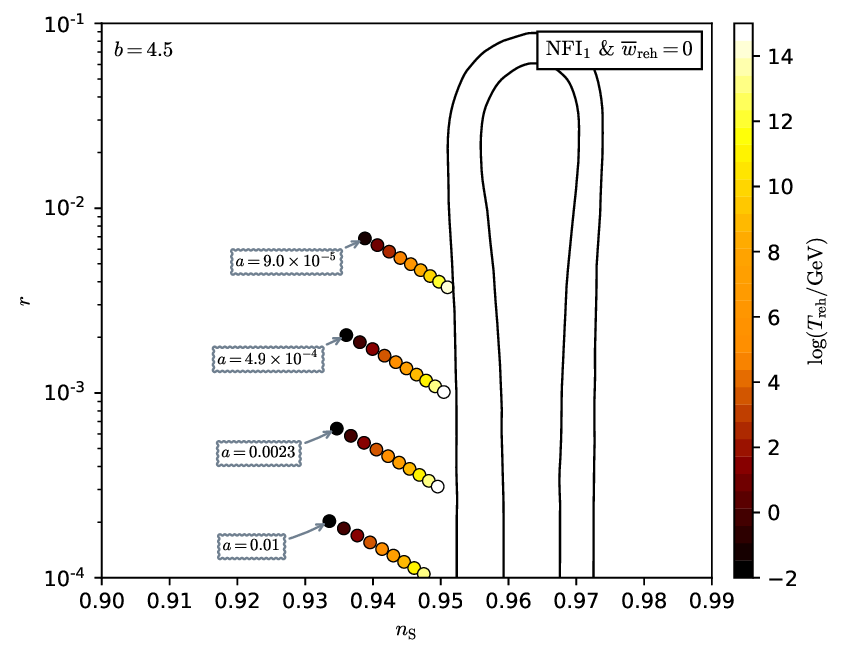}
\includegraphics[width=\wappfig,clip=true]{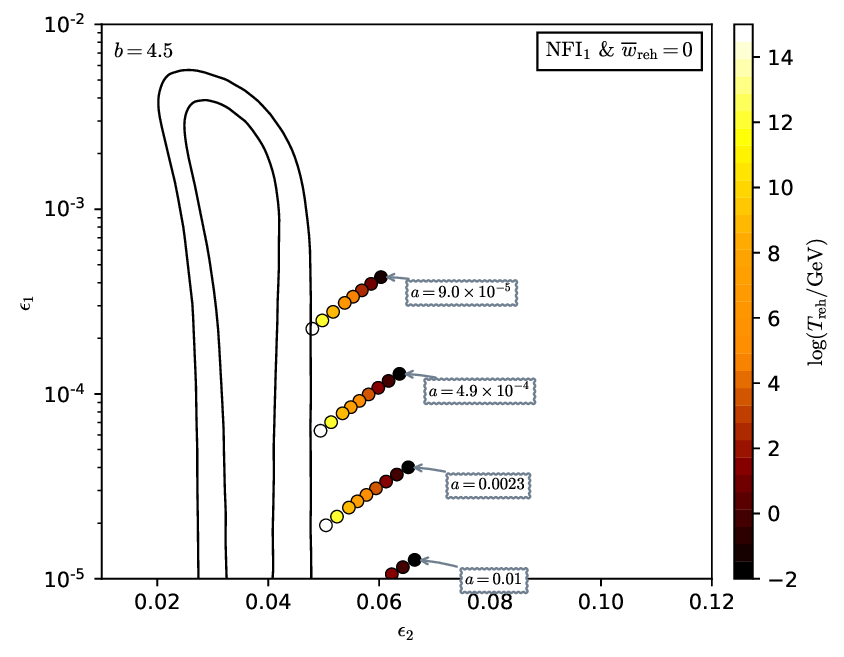}
\caption{Reheating consistent slow-roll predictions for N-Formalism
  Inflation 1 with $b=4.5$ (inflation gracefully ends). Predictions
  are represented in the plane $(\nS,r)$ (top panel) and in the plane
  $(\epsilon_1,\epsilon_2)$ (bottom panel) for various values of the
  parameter $a$. The solid contours are the one and two-sigma
  {\data} confidence intervals (marginalized over second order
  slow-roll).}
\label{fig:CMBNFI1_2}
\end{center}
\end{figure}

\subsection{N-Formalism Inflation 2 (\hyperref[sec:nfi]{NFI2})}

\begin{figure}[H]
\begin{center}
\includegraphics[width=\wappfig,clip=true]{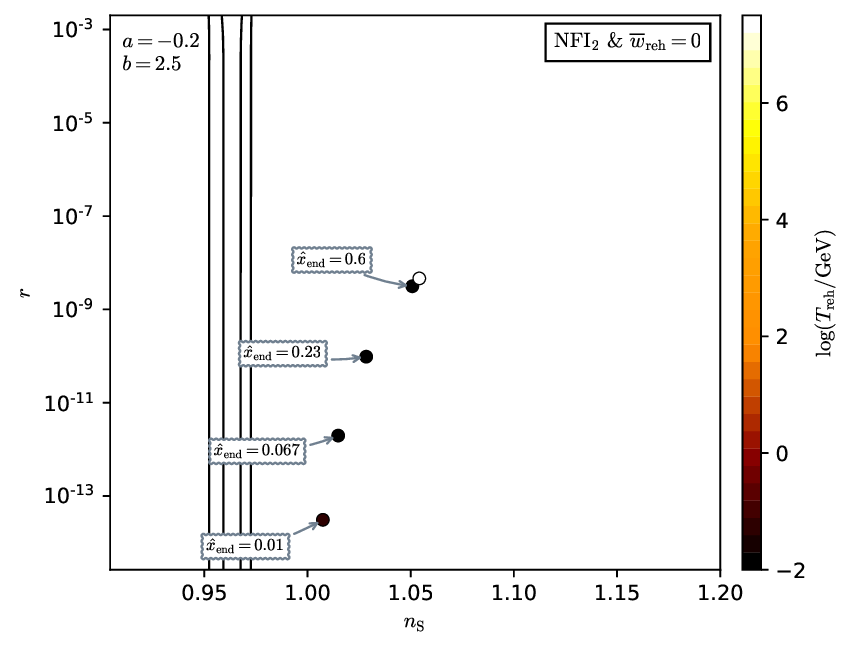}
\includegraphics[width=\wappfig,clip=true]{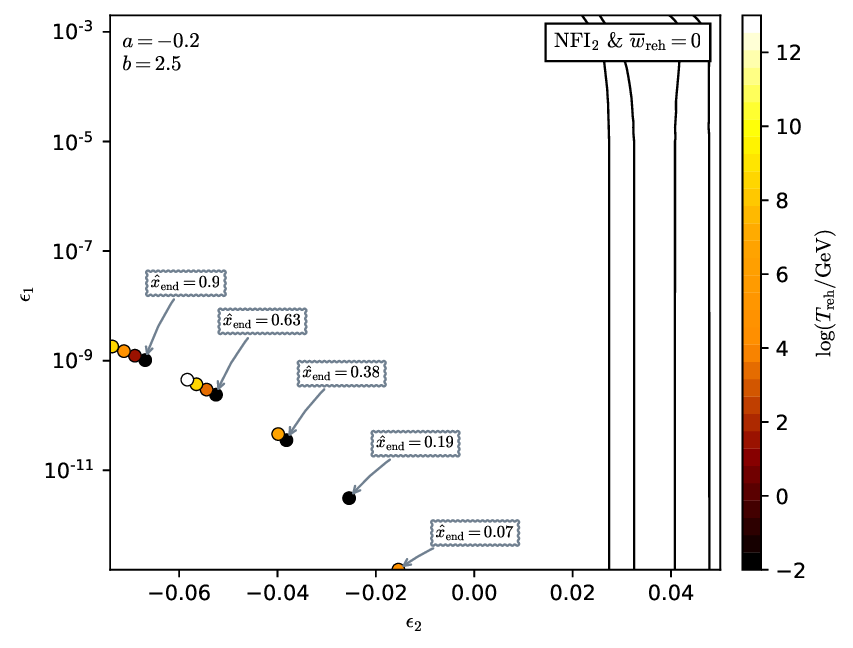}
\caption{Reheating consistent slow-roll predictions for N-Formalism
  Inflation 2 with $a=-0.2$ and $b=2.5$. Predictions are represented in
  the plane $(\nS,r)$ (top panel) and in the plane
  $(\epsilon_1,\epsilon_2)$ (bottom panel) for various values of the
  normalized field values $\xendhat$ at which inflation ends. This one
  is varied within its maximal allowed range, {\ie} with
  $\xendhat\equiv (\xend - \xendmin)/(\xendmax-\xendmin)$ in the
  domain $[0,1]$. The solid contours are the one and two-sigma {\data}
  confidence intervals (marginalized over second order slow-roll). See
  also Figs.~\ref{fig:CMBNFI2_1} to \ref{fig:CMBNFI2_3} for other
  values of $(a,b)$.}
\label{fig:CMBNFI2_0}
\end{center}
\end{figure}

\begin{figure}[H]
\begin{center}
\includegraphics[width=\wappfig,clip=true]{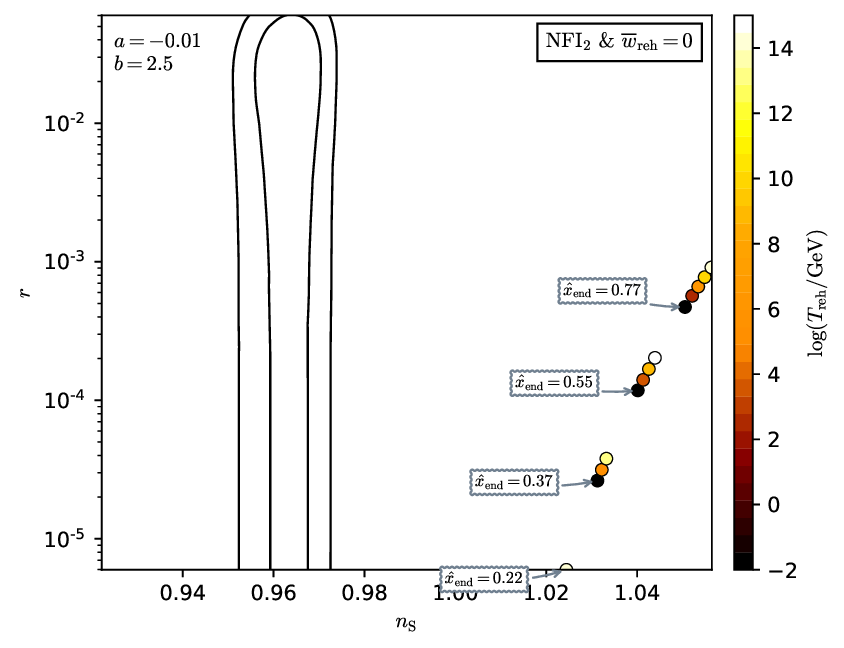}
\includegraphics[width=\wappfig,clip=true]{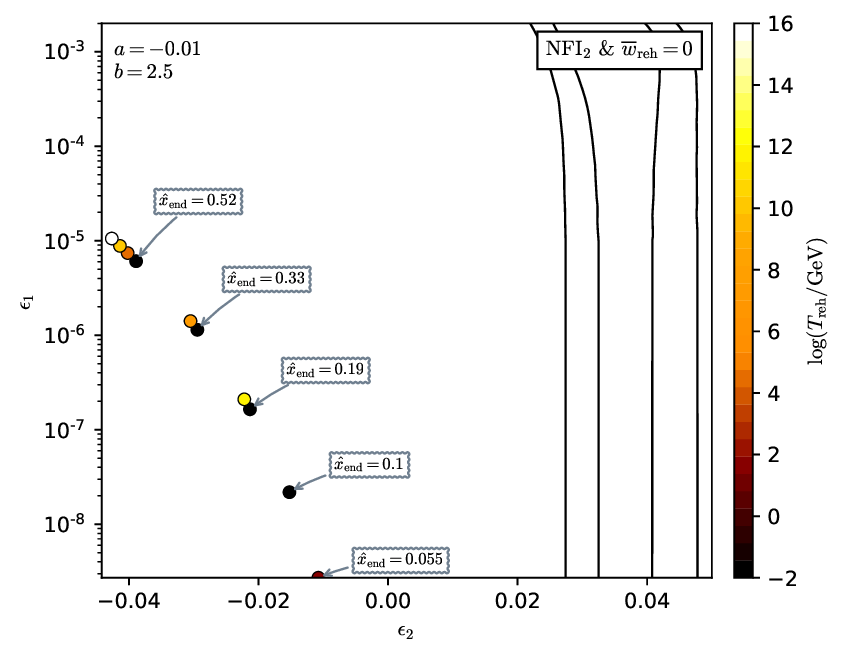}
\caption{Reheating consistent slow-roll predictions for N-Formalism
  Inflation 2 with $a=-0.01$ and $b=2.5$. Predictions are represented in
  the plane $(\nS,r)$ (top panel) and in the plane
  $(\epsilon_1,\epsilon_2)$ (bottom panel) for various values of the
  normalized field values $\xendhat$ at which inflation ends. This one
  is varied within its maximal allowed range, {\ie} with
  $\xendhat\equiv (\xend - \xendmin)/(\xendmax-\xendmin)$ in the
  domain $[0,1]$. The solid contours are the one and two-sigma {\data}
  confidence intervals (marginalized over second order slow-roll).}
\label{fig:CMBNFI2_1}
\end{center}
\end{figure}

\begin{figure}[H]
\begin{center}
\includegraphics[width=\wappfig,clip=true]{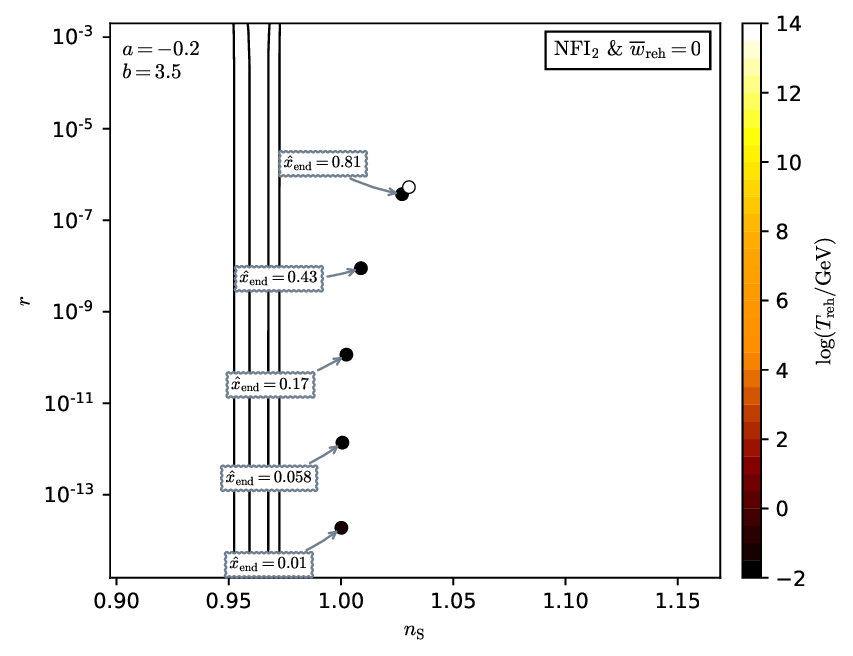}
\includegraphics[width=\wappfig,clip=true]{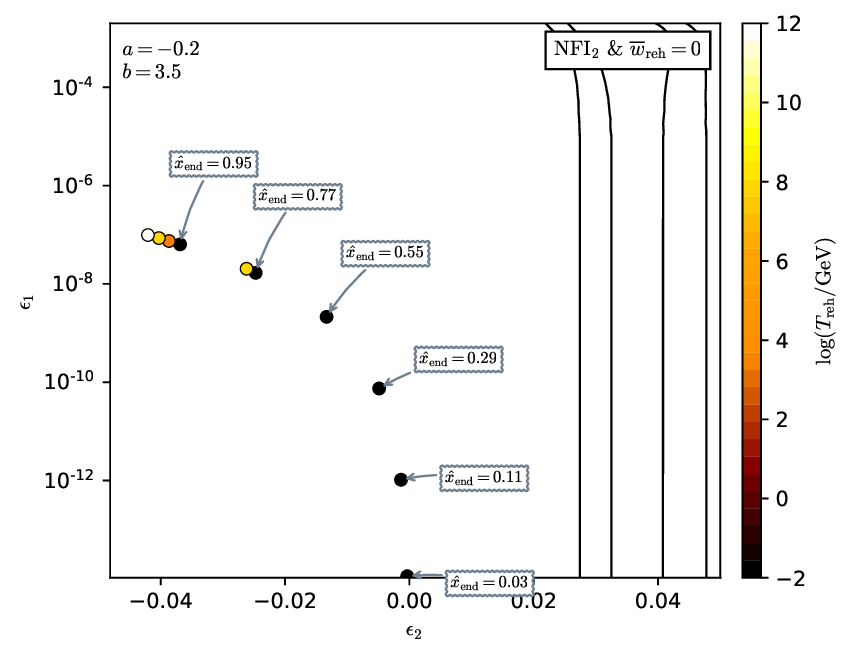}
\caption{Reheating consistent slow-roll predictions for N-Formalism
  Inflation 2 with $a=-0.2$ and $b=3.5$. Predictions are represented in
  the plane $(\nS,r)$ (top panel) and in the plane
  $(\epsilon_1,\epsilon_2)$ (bottom panel) for various values of the
  normalized field values $\xendhat$ at which inflation ends. This one
  is varied within its maximal allowed range, {\ie} with
  $\xendhat\equiv (\xend - \xendmin)/(\xendmax-\xendmin)$ in the
  domain $[0,1]$. The solid contours are the one and two-sigma {\data}
  confidence intervals (marginalized over second order slow-roll).}
\label{fig:CMBNFI2_2}
\end{center}
\end{figure}

\begin{figure}[H]
\begin{center}
\includegraphics[width=\wappfig,clip=true]{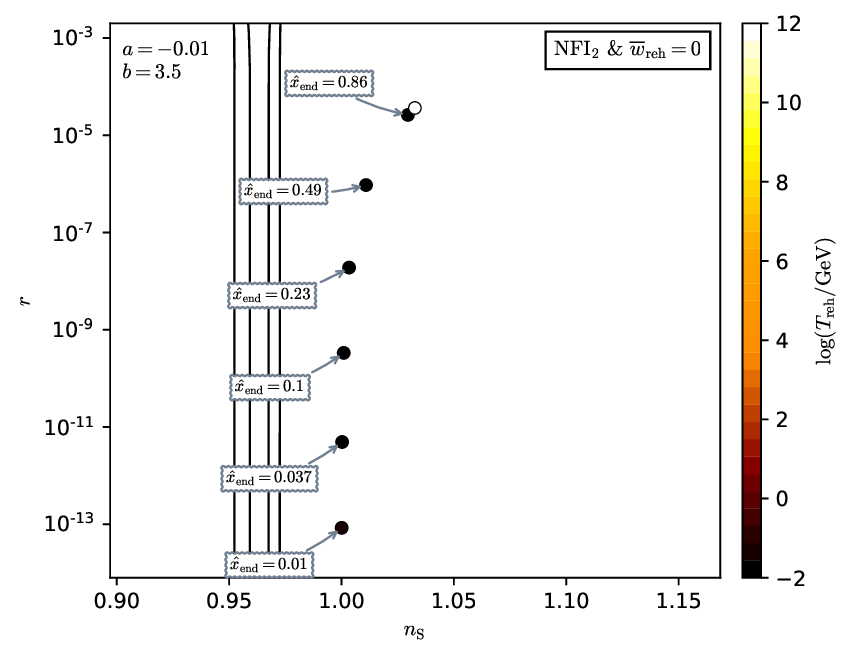}
\includegraphics[width=\wappfig,clip=true]{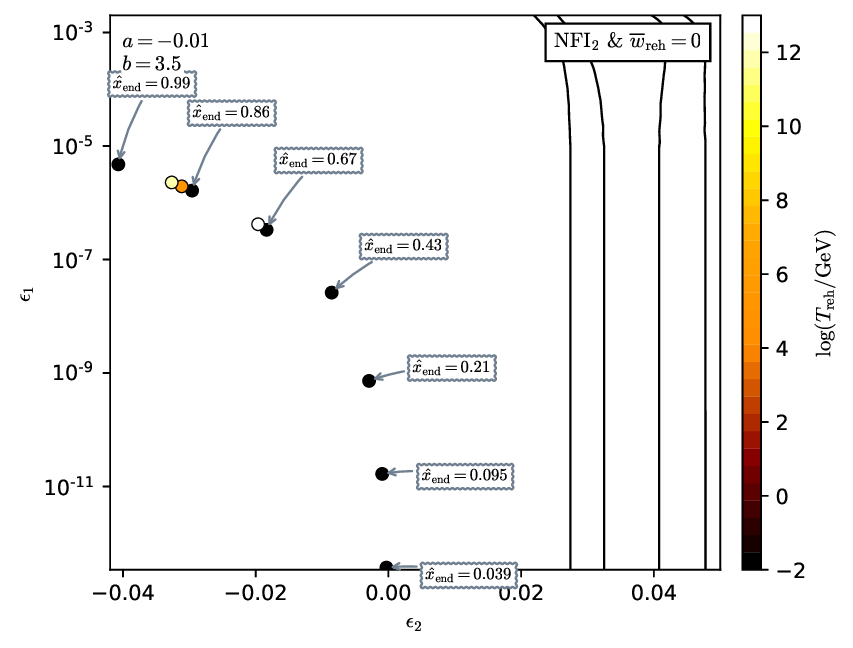}
\caption{Reheating consistent slow-roll predictions for N-Formalism
  Inflation 2 with $a=-0.01$ and $b=3.5$. Predictions are represented in
  the plane $(\nS,r)$ (top panel) and in the plane
  $(\epsilon_1,\epsilon_2)$ (bottom panel) for various values of the
  normalized field values $\xendhat$ at which inflation ends. This one
  is varied within its maximal allowed range, {\ie} with
  $\xendhat\equiv (\xend - \xendmin)/(\xendmax-\xendmin)$ in the
  domain $[0,1]$. The solid contours are the one and two-sigma {\data}
  confidence intervals (marginalized over second order slow-roll).}
\label{fig:CMBNFI2_3}
\end{center}
\end{figure}

\subsection{N-Formalism Inflation 3 (\hyperref[sec:nfi]{NFI3})}

\begin{figure}[H]
\begin{center}
\includegraphics[width=\wappfig,clip=true]{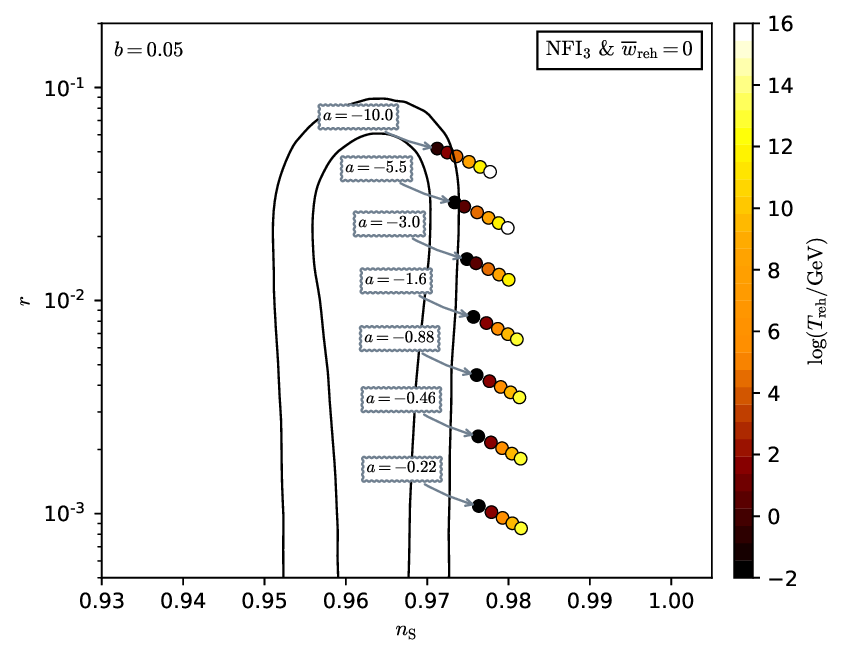}
\includegraphics[width=\wappfig,clip=true]{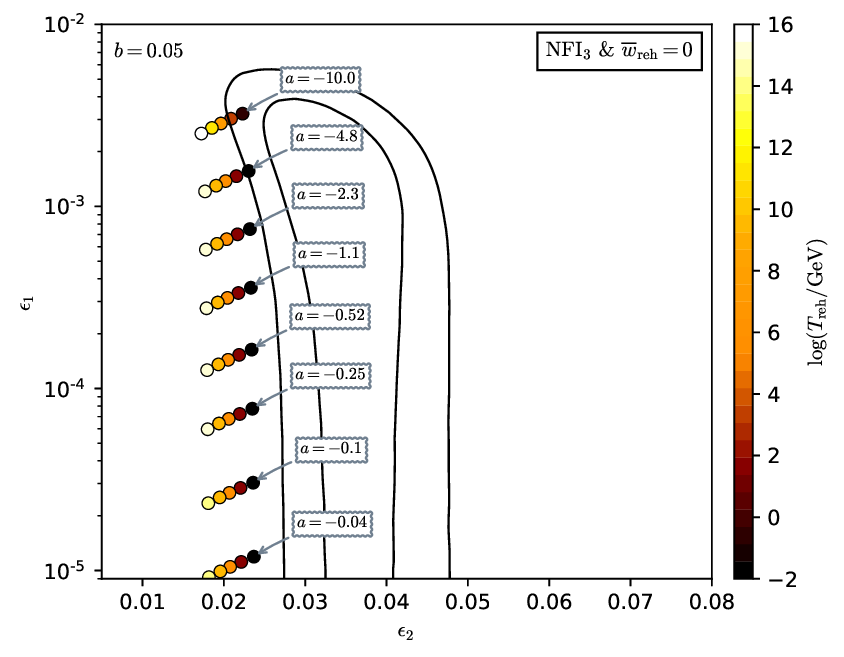}
\caption{Reheating consistent slow-roll predictions for N-Formalism
  Inflation 3 with $b=0.05$ (inflation gracefully ends). Predictions
  are represented in the plane $(\nS,r)$ (top panel) and in the plane
  $(\epsilon_1,\epsilon_2)$ (bottom panel) for various values of the
  parameter $a<0$. The solid contours are the one and two-sigma
  {\data} confidence intervals (marginalized over second order
  slow-roll). See also Figs.~\ref{fig:CMBNFI3_1} to
  \ref{fig:CMBNFI3_5} for other values of $b$.}
\label{fig:CMBNFI3_0}
\end{center}
\end{figure}

\begin{figure}[H]
\begin{center}
\includegraphics[width=\wappfig,clip=true]{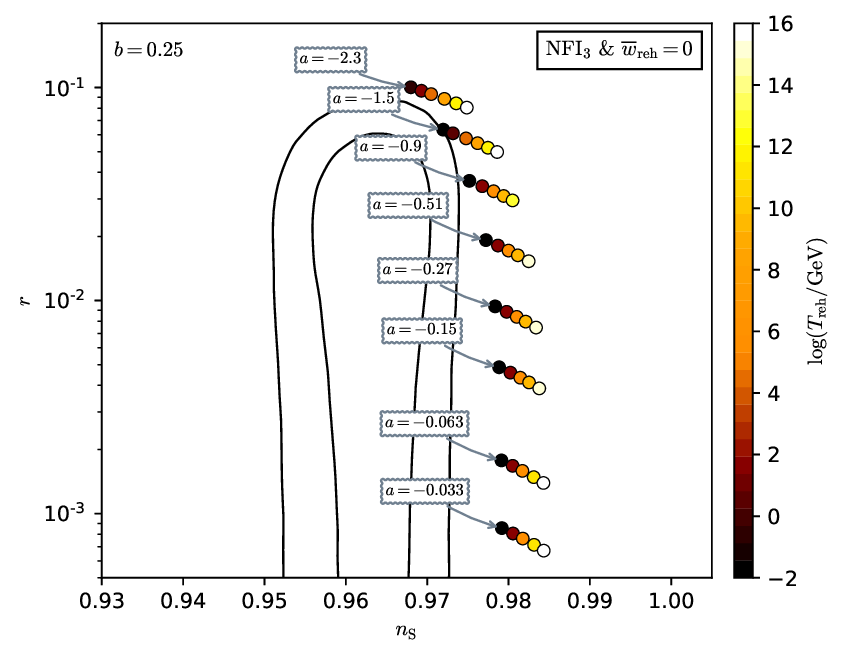}
\includegraphics[width=\wappfig,clip=true]{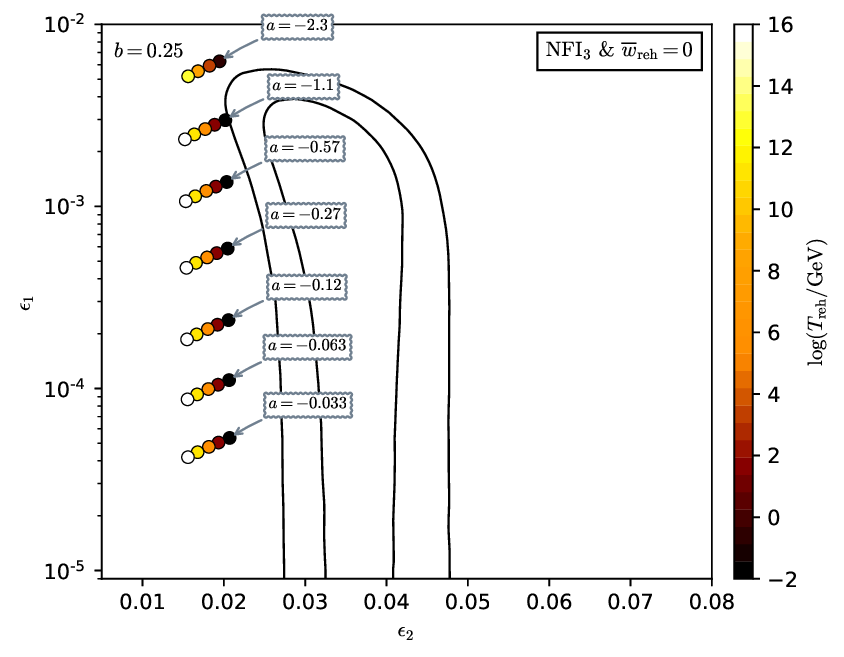}
\caption{Reheating consistent slow-roll predictions for N-Formalism
  Inflation 3 with $b=0.25$ (inflation gracefully ends). Predictions
  are represented in the plane $(\nS,r)$ (top panel) and in the plane
  $(\epsilon_1,\epsilon_2)$ (bottom panel) for various values of the
  parameter $a<0$. The solid contours are the one and two-sigma
  {\data} confidence intervals (marginalized over second order
  slow-roll).}
\label{fig:CMBNFI3_1}
\end{center}
\end{figure}

\begin{figure}[H]
\begin{center}
\includegraphics[width=\wappfig,clip=true]{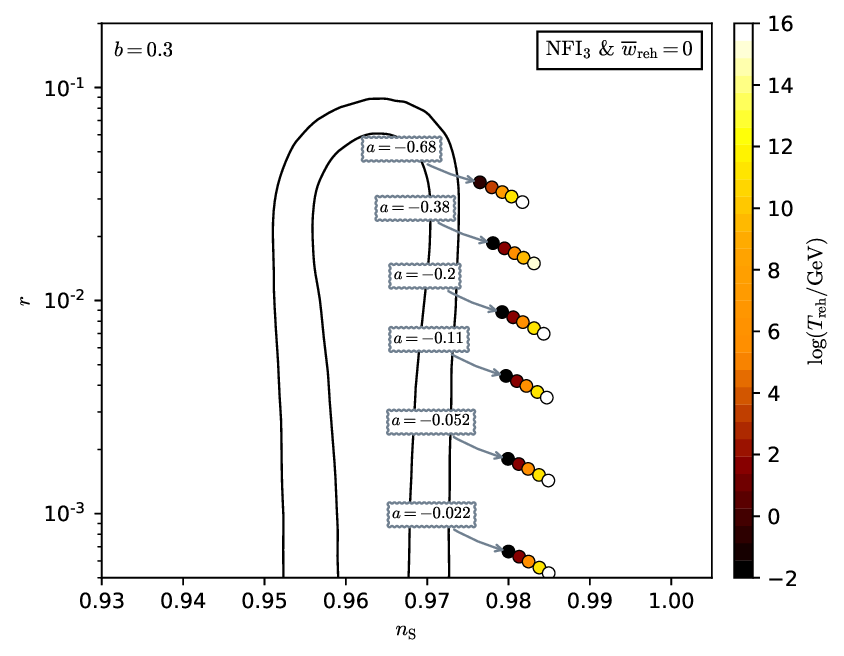}
\includegraphics[width=\wappfig,clip=true]{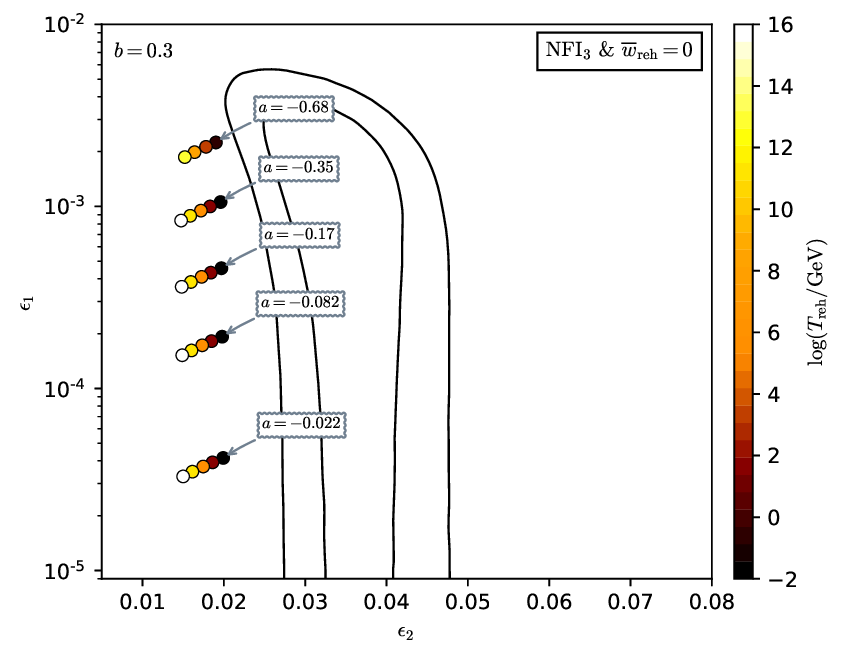}
\caption{Reheating consistent slow-roll predictions for N-Formalism
  Inflation 3 with $b=0.3$ (inflation gracefully ends). Predictions
  are represented in the plane $(\nS,r)$ (top panel) and in the plane
  $(\epsilon_1,\epsilon_2)$ (bottom panel) for various values of the
  parameter $a<0$. The solid contours are the one and two-sigma
  {\data} confidence intervals (marginalized over second order
  slow-roll).}
\label{fig:CMBNFI3_2}
\end{center}
\end{figure}

\begin{figure}[H]
\begin{center}
\includegraphics[width=\wappfig,clip=true]{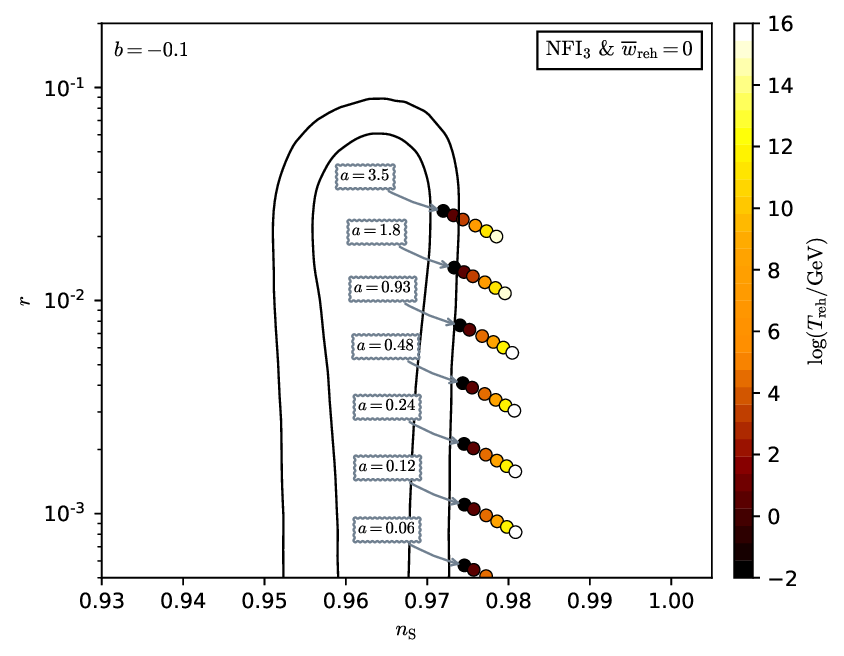}
\includegraphics[width=\wappfig,clip=true]{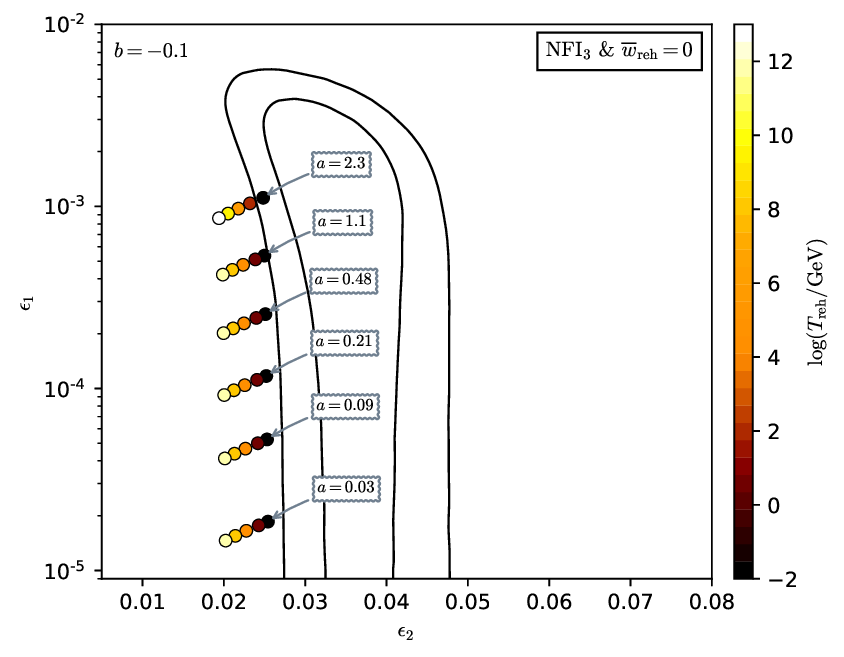}
\caption{Reheating consistent slow-roll predictions for N-Formalism
  Inflation 3 with $b=-0.1$ (inflation gracefully ends). Predictions
  are represented in the plane $(\nS,r)$ (top panel) and in the plane
  $(\epsilon_1,\epsilon_2)$ (bottom panel) for various values of the
  parameter $a>0$. The solid contours are the one and two-sigma
  {\data} confidence intervals (marginalized over second order
  slow-roll).}
\label{fig:CMBNFI3_3}
\end{center}
\end{figure}

\begin{figure}[H]
\begin{center}
\includegraphics[width=\wappfig,clip=true]{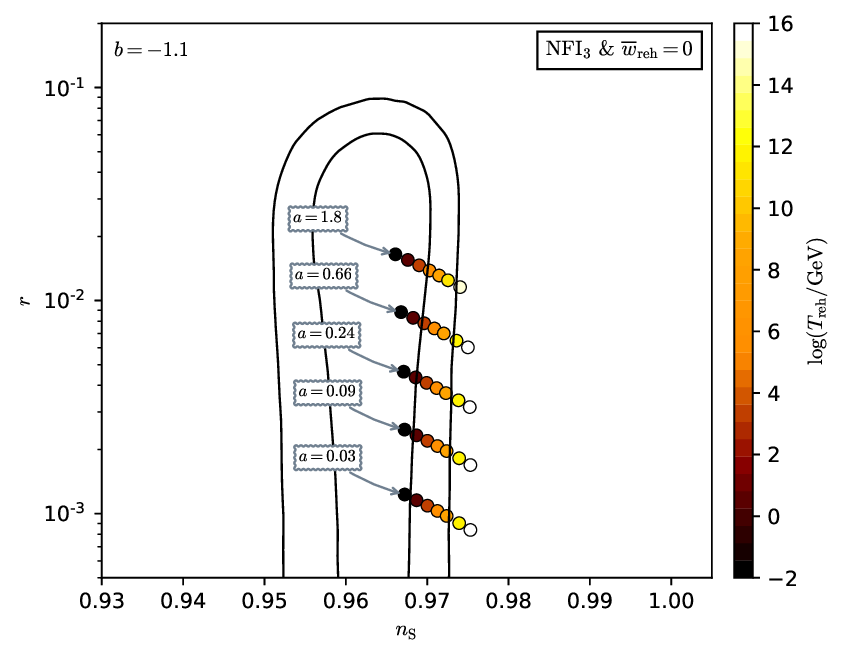}
\includegraphics[width=\wappfig,clip=true]{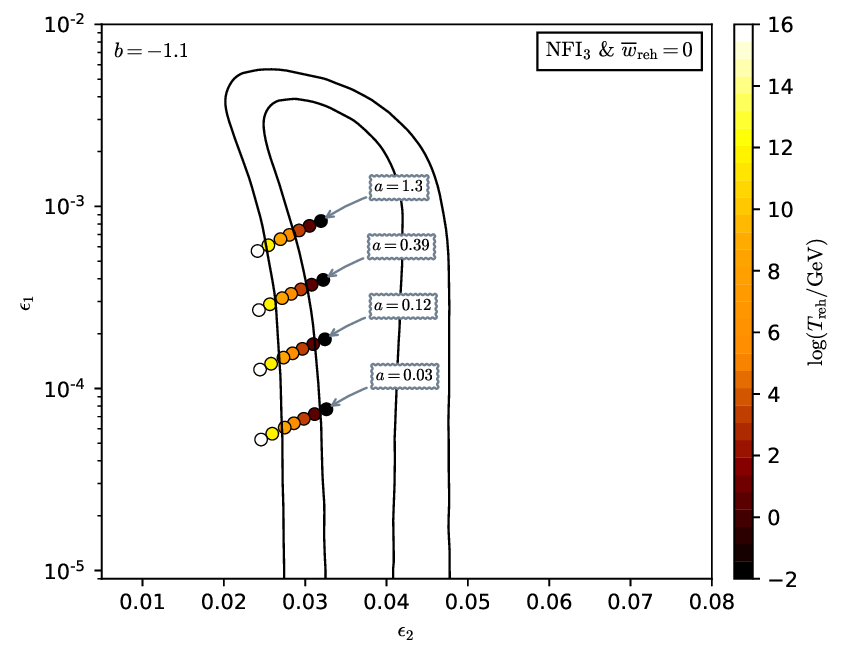}
\caption{Reheating consistent slow-roll predictions for N-Formalism
  Inflation 3 with $b=-1.1$ (inflation gracefully ends). Predictions
  are represented in the plane $(\nS,r)$ (top panel) and in the plane
  $(\epsilon_1,\epsilon_2)$ (bottom panel) for various values of the
  parameter $a>0$. The solid contours are the one and two-sigma
  {\data} confidence intervals (marginalized over second order
  slow-roll).}
\label{fig:CMBNFI3_4}
\end{center}
\end{figure}

\begin{figure}[H]
\begin{center}
\includegraphics[width=\wappfig,clip=true]{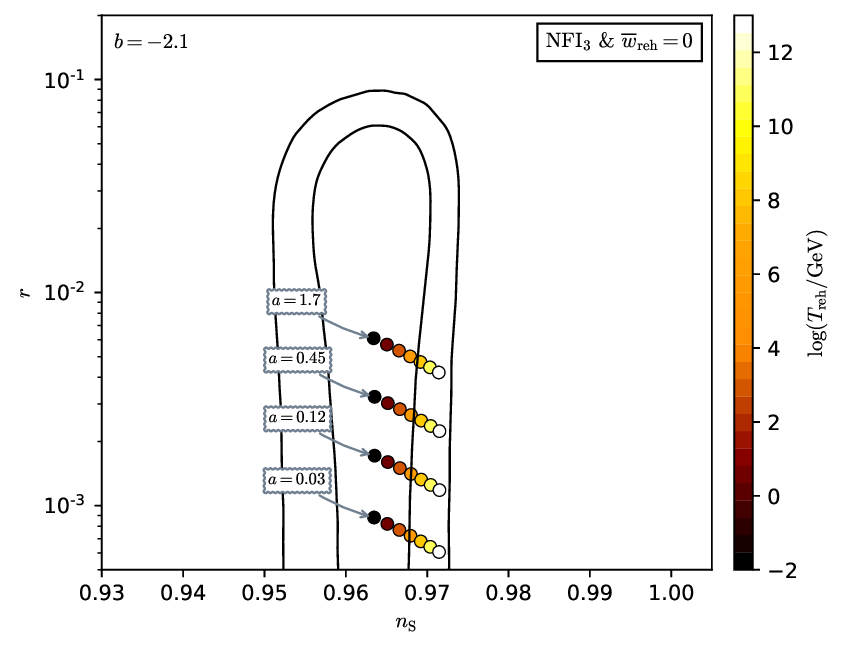}
\includegraphics[width=\wappfig,clip=true]{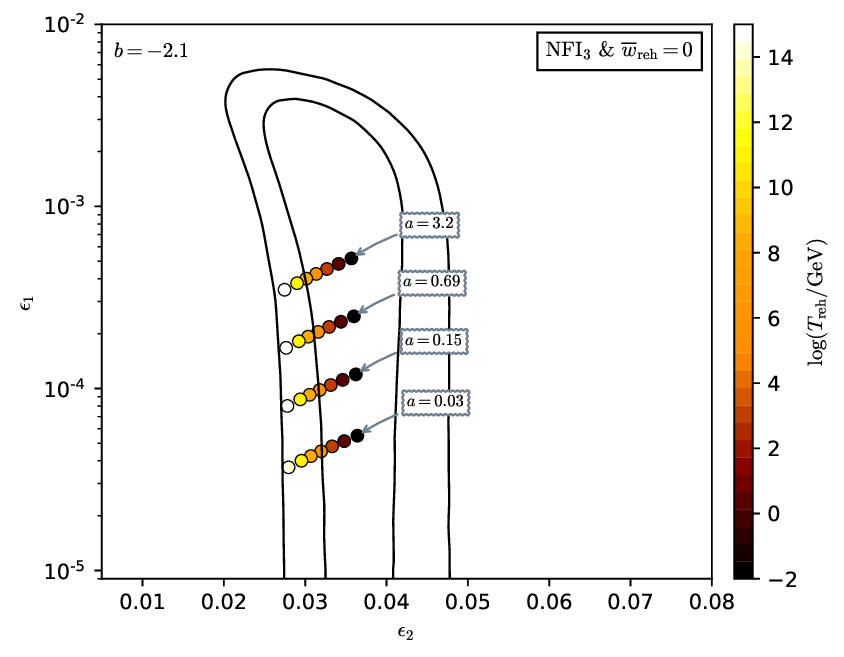}
\caption{Reheating consistent slow-roll predictions for N-Formalism
  Inflation 3 with $b=-2.1$ (inflation gracefully ends). Predictions
  are represented in the plane $(\nS,r)$ (top panel) and in the plane
  $(\epsilon_1,\epsilon_2)$ (bottom panel) for various values of the
  parameter $a>0$. The solid contours are the one and two-sigma
  {\data} confidence intervals (marginalized over second order
  slow-roll).}
\label{fig:CMBNFI3_5}
\end{center}
\end{figure}

\subsection{N-Formalism Inflation 4 (\hyperref[sec:nfi]{NFI4})}

\begin{figure}[H]
\begin{center}
\includegraphics[width=\wappfig,clip=true]{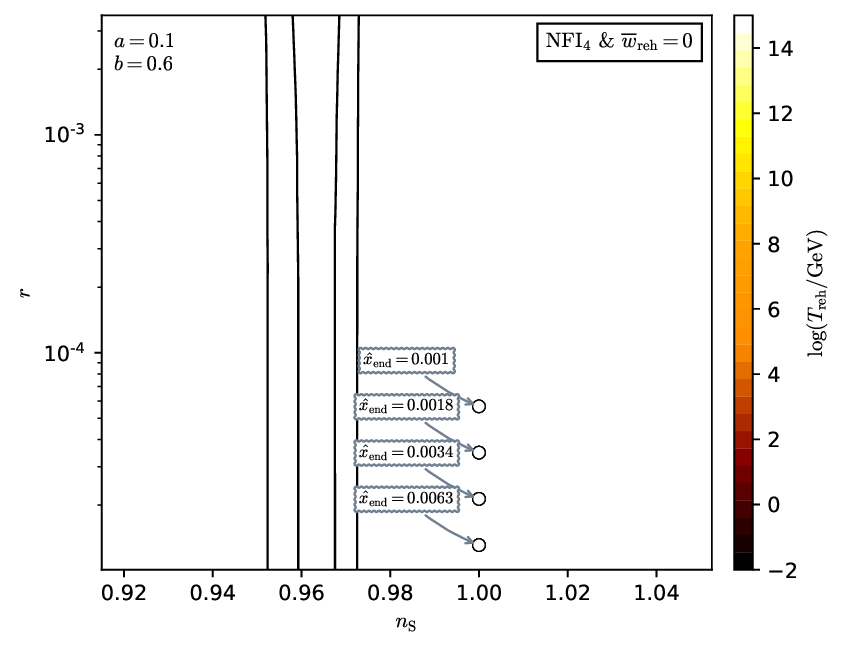}
\includegraphics[width=\wappfig,clip=true]{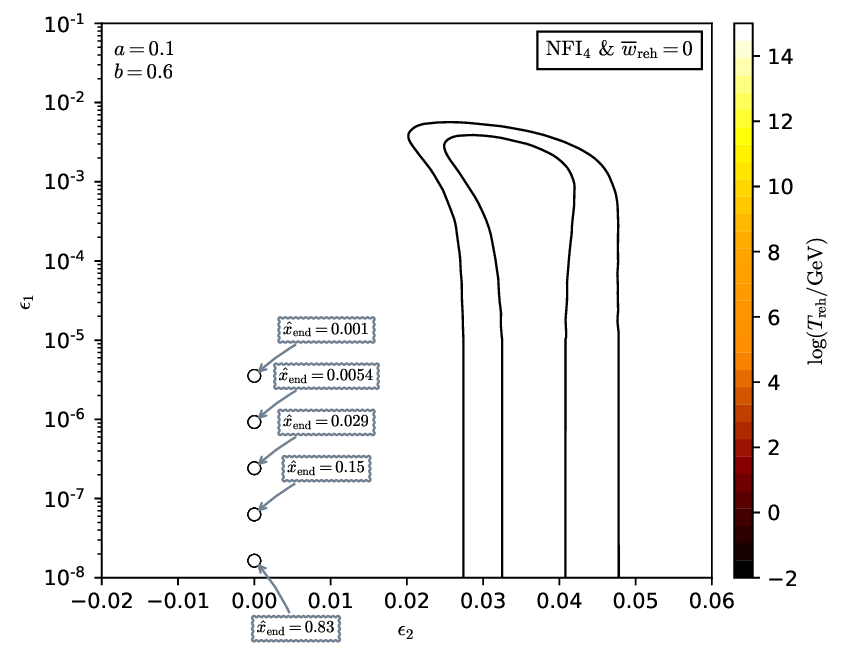}
\caption{Reheating consistent slow-roll predictions for N-Formalism
  Inflation 4 with $a=0.1$ and $b=0.6$. Predictions are represented in
  the plane $(\nS,r)$ (top panel) and in the plane
  $(\epsilon_1,\epsilon_2)$ (bottom panel) for various values of the
  normalized field values $\xendhat$ at which inflation ends. This one
  is varied within its maximal allowed range, {\ie} with
  $\xendhat\equiv (\xend - \xendmin)/(\xendmax-\xendmin)$ in the
  domain $[0,1]$. The solid contours are the one and two-sigma {\data}
  confidence intervals (marginalized over second order slow-roll). See
  also Figs.~\ref{fig:CMBNFI4_1} to \ref{fig:CMBNFI4_5} for other
  values of $a$ and $b$.}
\label{fig:CMBNFI4_0}
\end{center}
\end{figure}

\begin{figure}[H]
\begin{center}
\includegraphics[width=\wappfig,clip=true]{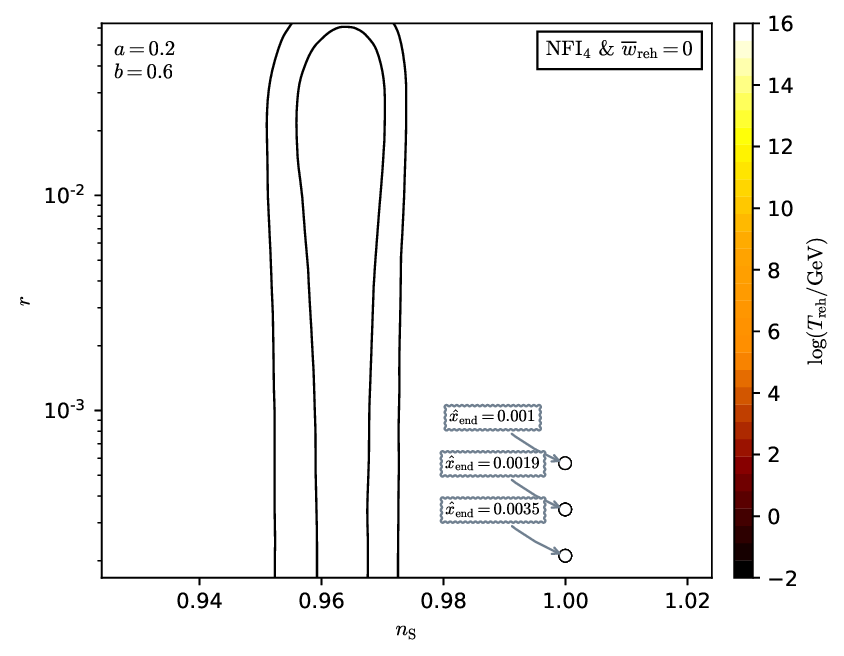}
\includegraphics[width=\wappfig,clip=true]{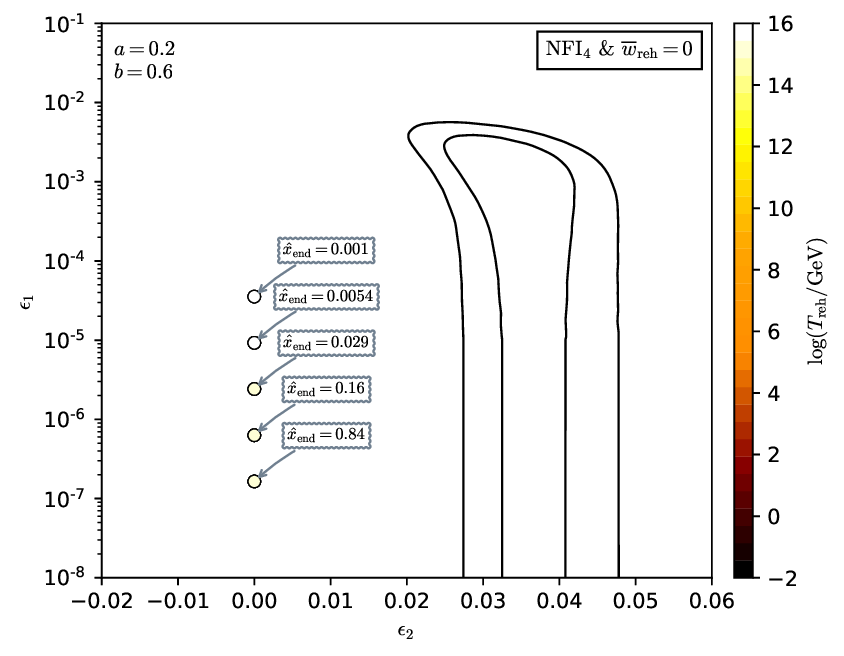}
\caption{Reheating consistent slow-roll predictions for N-Formalism
  Inflation 4 with $a=0.2$ and $b=0.6$. Predictions are represented in
  the plane $(\nS,r)$ (top panel) and in the plane
  $(\epsilon_1,\epsilon_2)$ (bottom panel) for various values of the
  normalized field values $\xendhat$ at which inflation ends. This one
  is varied within its maximal allowed range, {\ie} with
  $\xendhat\equiv (\xend - \xendmin)/(\xendmax-\xendmin)$ in the
  domain $[0,1]$. The solid contours are the one and two-sigma {\data}
  confidence intervals (marginalized over second order slow-roll).}
\label{fig:CMBNFI4_1}
\end{center}
\end{figure}

\begin{figure}[H]
\begin{center}
\includegraphics[width=\wappfig,clip=true]{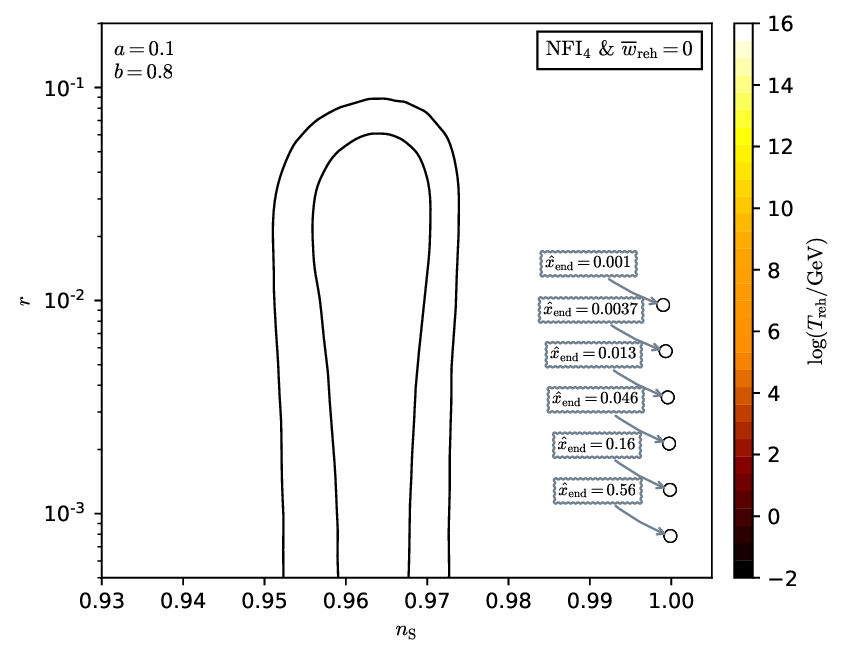}
\includegraphics[width=\wappfig,clip=true]{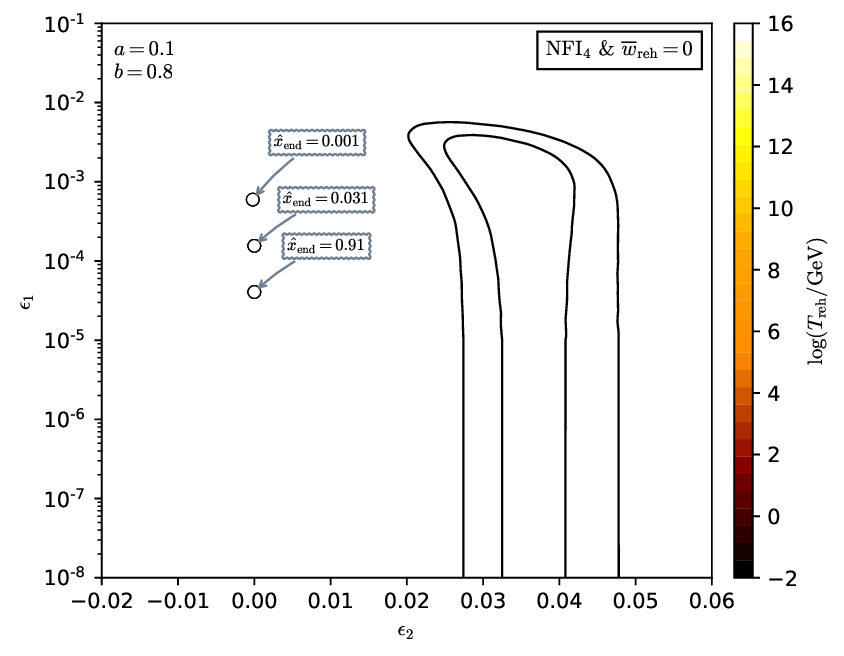}
\caption{Reheating consistent slow-roll predictions for N-Formalism
  Inflation 4 with $a=0.1$ and $b=0.8$. Predictions are represented in
  the plane $(\nS,r)$ (top panel) and in the plane
  $(\epsilon_1,\epsilon_2)$ (bottom panel) for various values of the
  normalized field values $\xendhat$ at which inflation ends. This one
  is varied within its maximal allowed range, {\ie} with
  $\xendhat\equiv (\xend - \xendmin)/(\xendmax-\xendmin)$ in the
  domain $[0,1]$. The solid contours are the one and two-sigma {\data}
  confidence intervals (marginalized over second order slow-roll).}
\label{fig:CMBNFI4_2}
\end{center}
\end{figure}

\begin{figure}[H]
\begin{center}
\includegraphics[width=\wappfig,clip=true]{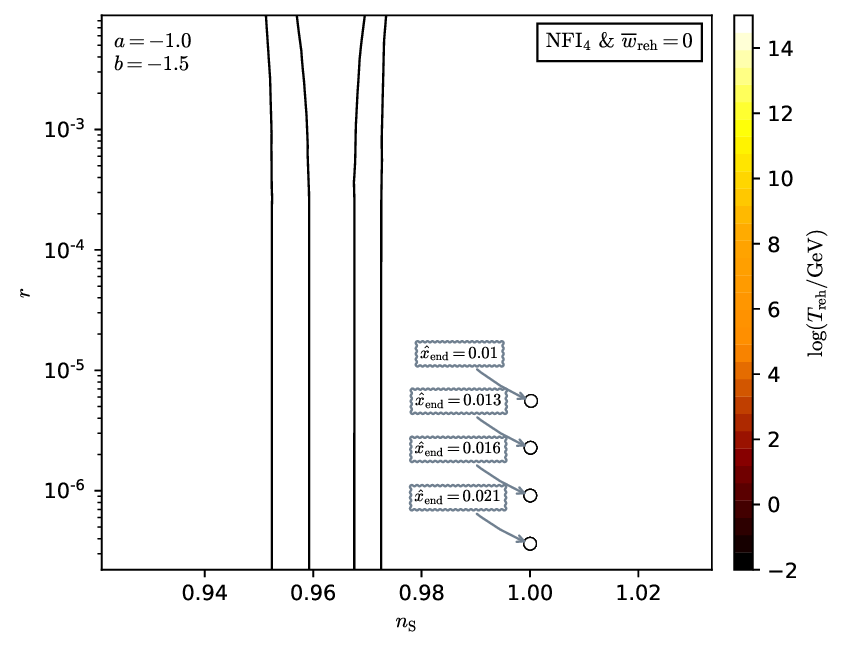}
\includegraphics[width=\wappfig,clip=true]{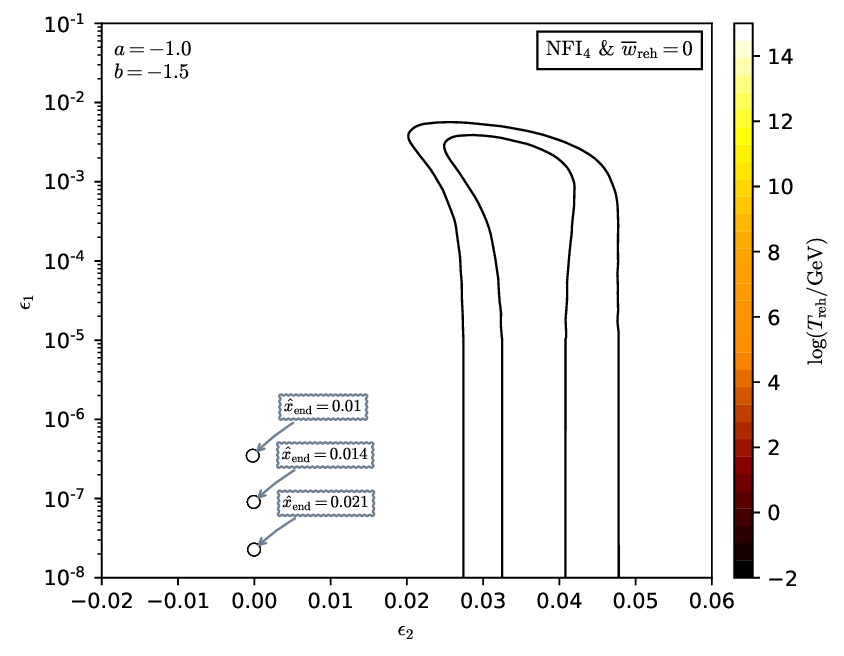}
\caption{Reheating consistent slow-roll predictions for N-Formalism
  Inflation 4 with $a=-1.0$ and $b=-1.5$. Predictions are represented in
  the plane $(\nS,r)$ (top panel) and in the plane
  $(\epsilon_1,\epsilon_2)$ (bottom panel) for various values of the
  normalized field values $\xendhat$ at which inflation ends. This one
  is varied within its maximal allowed range, {\ie} with
  $\xendhat\equiv (\xend - \xendmin)/(\xendmax-\xendmin)$ in the
  domain $[0,1]$. The solid contours are the one and two-sigma {\data}
  confidence intervals (marginalized over second order slow-roll).}
\label{fig:CMBNFI4_3}
\end{center}
\end{figure}

\begin{figure}[H]
\begin{center}
\includegraphics[width=\wappfig,clip=true]{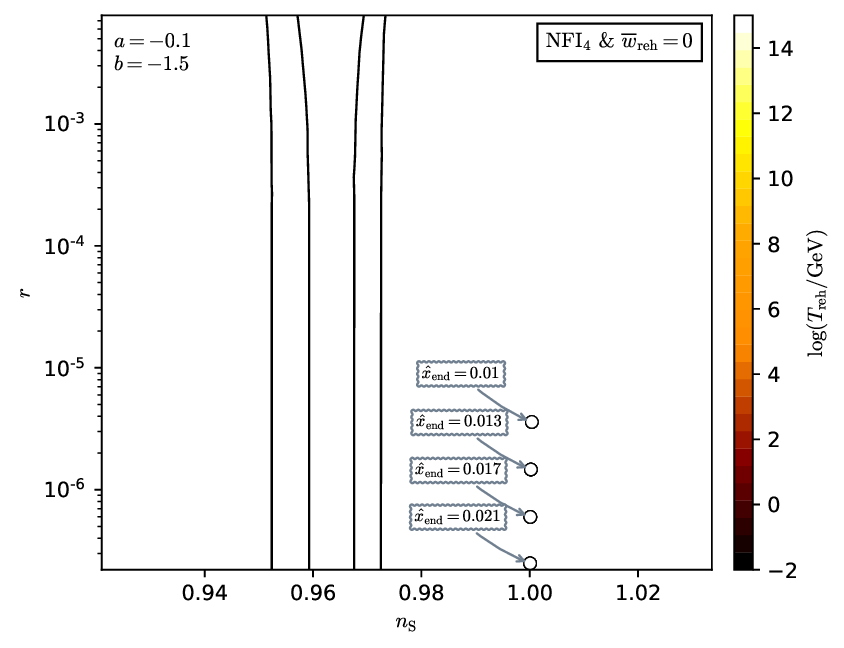}
\includegraphics[width=\wappfig,clip=true]{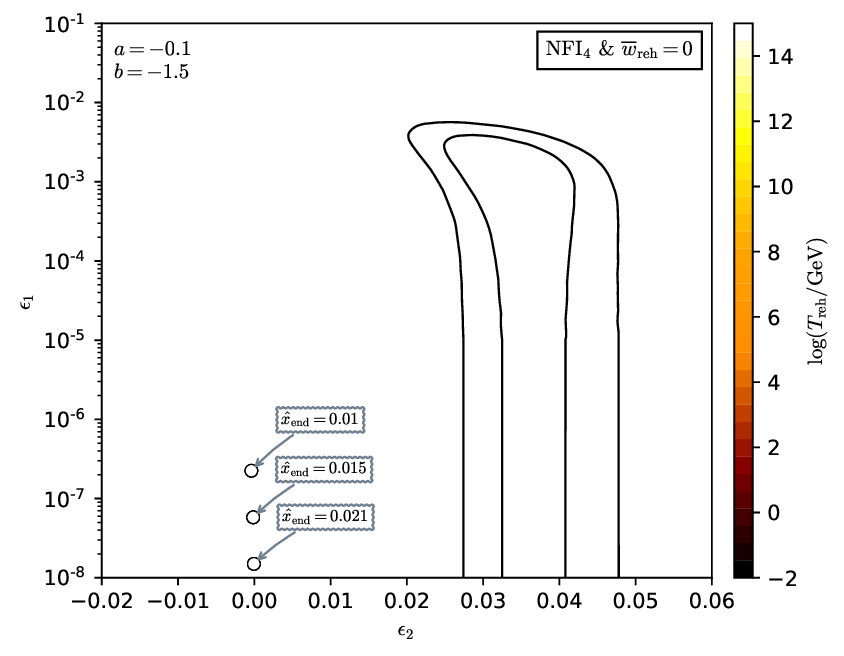}
\caption{Reheating consistent slow-roll predictions for N-Formalism
  Inflation 4 with $a=-0.1$ and $b=-1.5$. Predictions are represented in
  the plane $(\nS,r)$ (top panel) and in the plane
  $(\epsilon_1,\epsilon_2)$ (bottom panel) for various values of the
  normalized field values $\xendhat$ at which inflation ends. This one
  is varied within its maximal allowed range, {\ie} with
  $\xendhat\equiv (\xend - \xendmin)/(\xendmax-\xendmin)$ in the
  domain $[0,1]$. The solid contours are the one and two-sigma {\data}
  confidence intervals (marginalized over second order slow-roll).}
\label{fig:CMBNFI4_4}
\end{center}
\end{figure}

\begin{figure}[H]
\begin{center}
\includegraphics[width=\wappfig,clip=true]{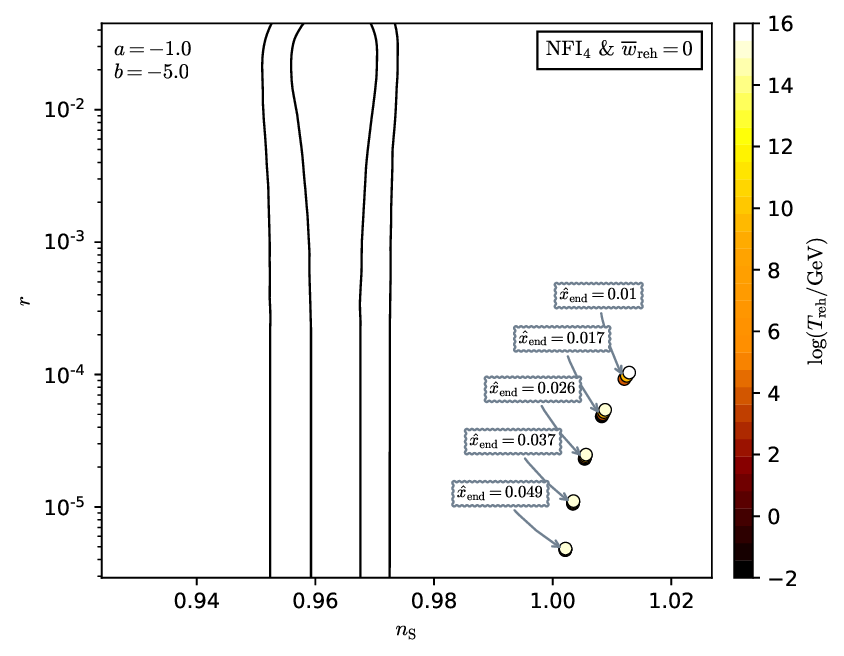}
\includegraphics[width=\wappfig,clip=true]{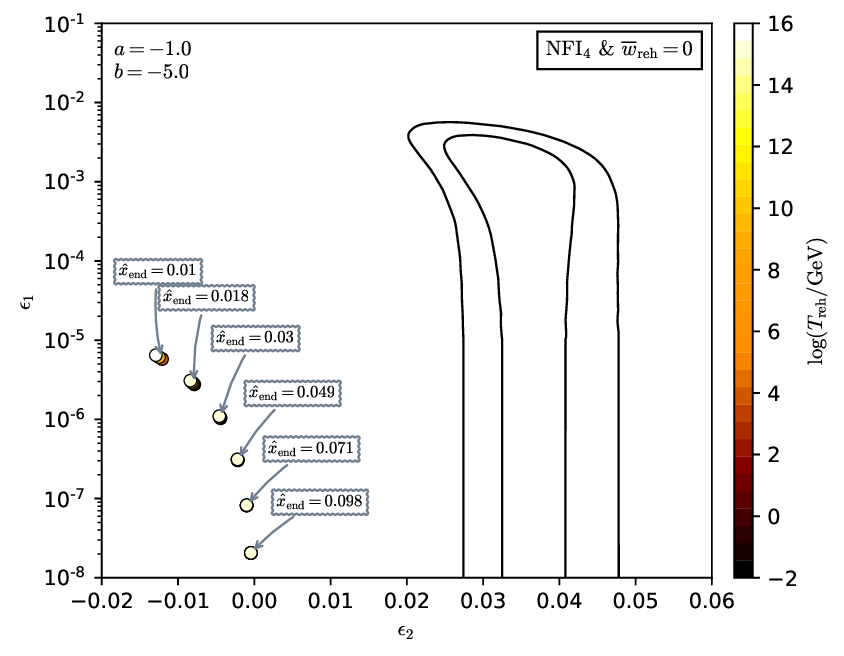}
\caption{Reheating consistent slow-roll predictions for N-Formalism
  Inflation 4 with $a=-1$ and $b=-5$. Predictions are represented in
  the plane $(\nS,r)$ (top panel) and in the plane
  $(\epsilon_1,\epsilon_2)$ (bottom panel) for various values of the
  normalized field values $\xendhat$ at which inflation ends. This one
  is varied within its maximal allowed range, {\ie} with
  $\xendhat\equiv (\xend - \xendmin)/(\xendmax-\xendmin)$ in the
  domain $[0,1]$. The solid contours are the one and two-sigma {\data}
  confidence intervals (marginalized over second order slow-roll).}
\label{fig:CMBNFI4_5}
\end{center}
\end{figure}

\subsection{Radiatively Corrected Inflection Point Inflation 1 (\hyperref[sec:rcipi]{RCIPI1})}

\begin{figure}[H]
\begin{center}
\includegraphics[width=\wappfig,clip=true]{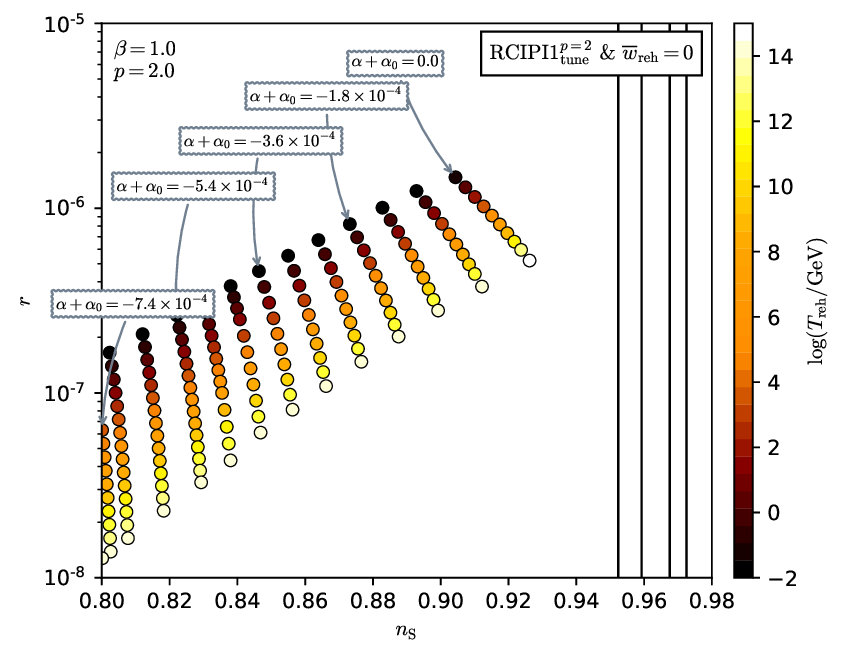}
\includegraphics[width=\wappfig,clip=true]{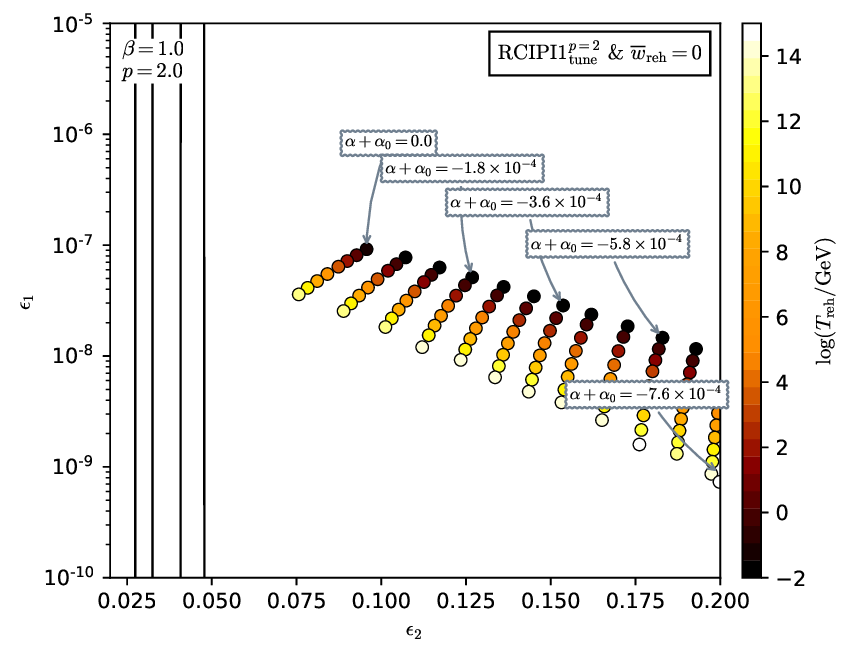}
\caption{Reheating consistent slow-roll predictions for Radiatively
  Corrected Inflection Point Inflation 1 for $p=2$, $\beta=1$ and when
  the potential has almost an inflection point. This one occurs for $\alpha =
  -\alphazero$. Predictions are represented in the plane $(\nS,r)$ (top
  panel) and in the plane $(\epsilon_1,\epsilon_2)$ (bottom panel) for
  various values of $\alpha+\alphazero$. The solid contours are the one
  and two-sigma {\data} confidence intervals (marginalized over second
  order slow-roll). See also Figs.~\ref{fig:CMBRCIPI1_1} to
  \ref{fig:CMBRCIPI1_5} for other values of $p$ and $\beta$.}
\label{fig:CMBRCIPI1_0}
\end{center}
\end{figure}

\begin{figure}[H]
\begin{center}
\includegraphics[width=\wappfig,clip=true]{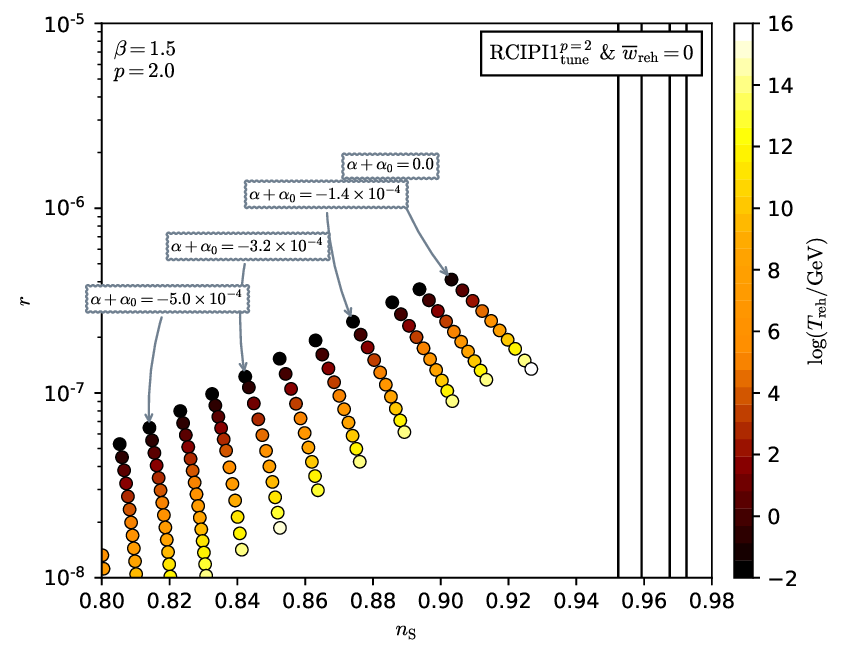}
\includegraphics[width=\wappfig,clip=true]{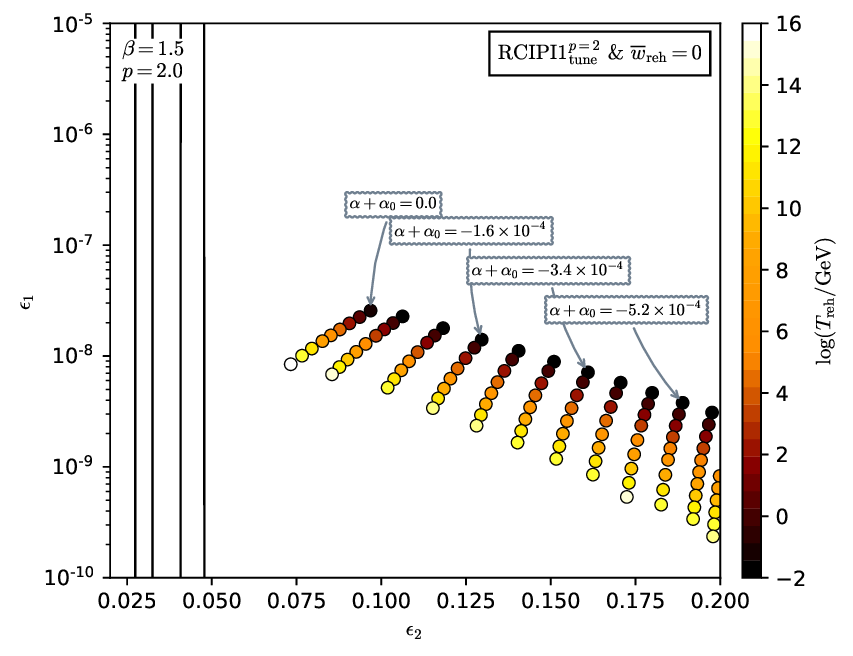}
\caption{Reheating consistent slow-roll predictions for Radiatively
  Corrected Inflection Point Inflation 1 for $p=2$, $\beta=1.5$ and when
  the potential has almost an inflection point. This one occurs for
  $\alpha = -\alphazero$. Predictions are represented in the plane
  $(\nS,r)$ (top panel) and in the plane $(\epsilon_1,\epsilon_2)$
  (bottom panel) for various values of $\alpha+\alphazero$. The solid
  contours are the one and two-sigma {\data} confidence intervals
  (marginalized over second order slow-roll).}
\label{fig:CMBRCIPI1_1}
\end{center}
\end{figure}

\begin{figure}[H]
\begin{center}
\includegraphics[width=\wappfig,clip=true]{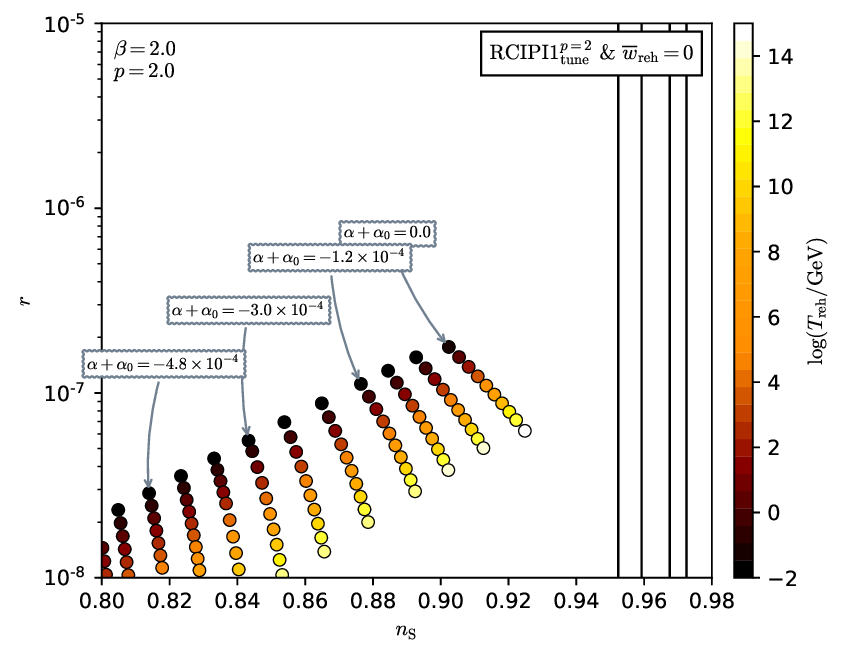}
\includegraphics[width=\wappfig,clip=true]{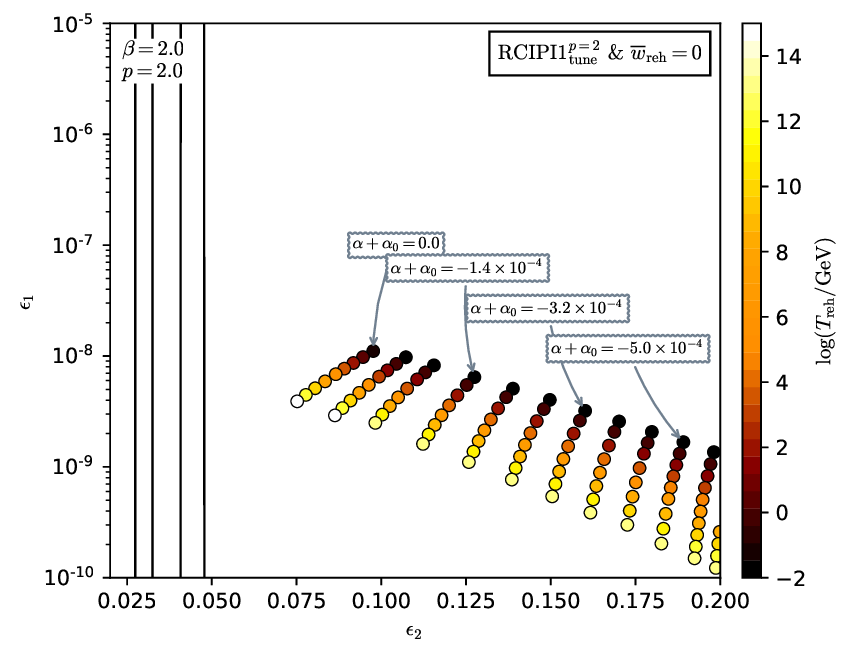}
\caption{Reheating consistent slow-roll predictions for Radiatively
  Corrected Inflection Point Inflation 1 for $p=2$, $\beta=2$ and when
  the potential has almost an inflection point. This one occurs for
  $\alpha = -\alphazero$. Predictions are represented in the plane
  $(\nS,r)$ (top panel) and in the plane $(\epsilon_1,\epsilon_2)$
  (bottom panel) for various values of $\alpha+\alphazero$. The solid
  contours are the one and two-sigma {\data} confidence intervals
  (marginalized over second order slow-roll).}
\label{fig:CMBRCIPI1_2}
\end{center}
\end{figure}

\begin{figure}[H]
\begin{center}
\includegraphics[width=\wappfig,clip=true]{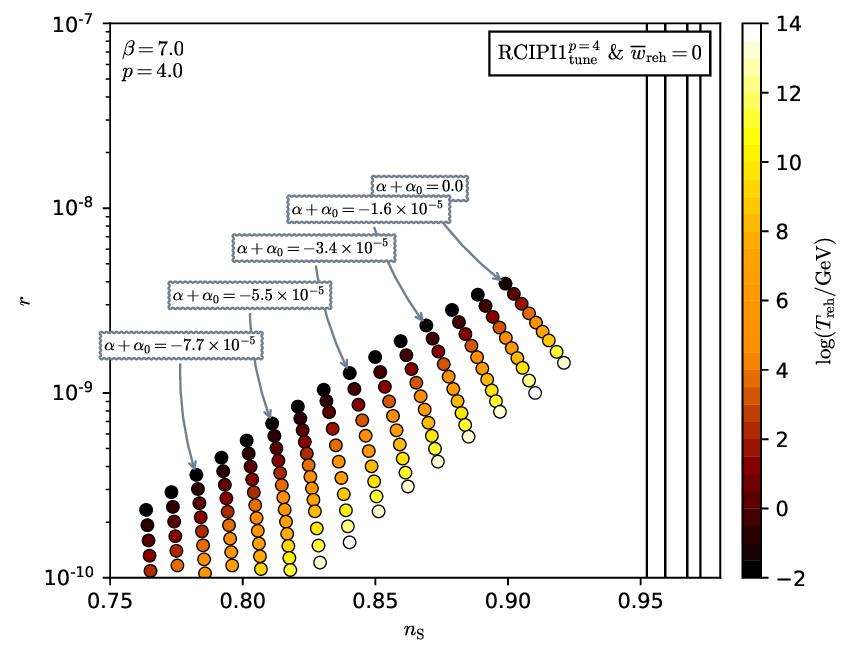}
\includegraphics[width=\wappfig,clip=true]{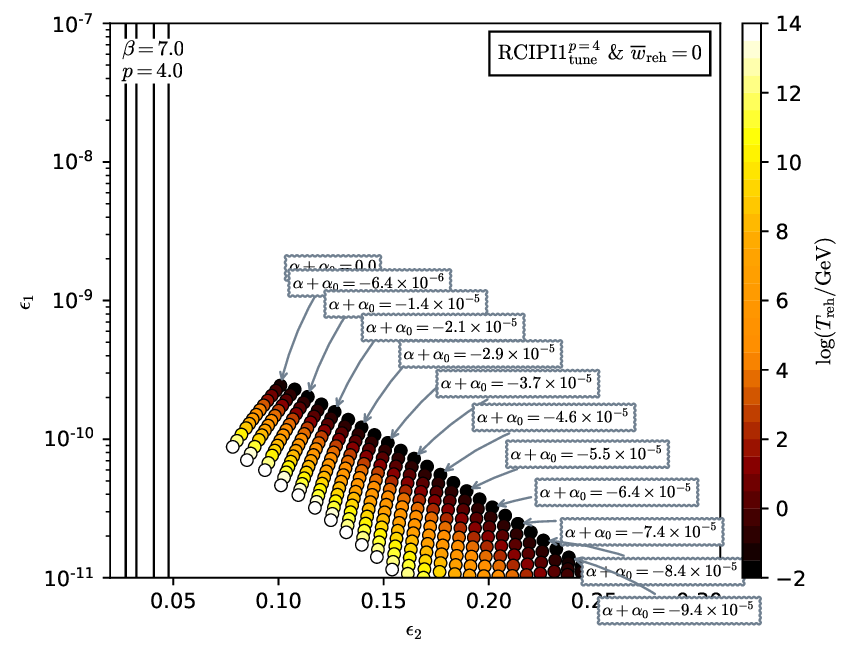}
\caption{Reheating consistent slow-roll predictions for Radiatively
  Corrected Inflection Point Inflation 1 for $p=4$, $\beta=7$ and when
  the potential has almost an inflection point. This one occurs for
  $\alpha = -\alphazero$. Predictions are represented in the plane
  $(\nS,r)$ (top panel) and in the plane $(\epsilon_1,\epsilon_2)$
  (bottom panel) for various values of $\alpha+\alphazero$. The solid
  contours are the one and two-sigma {\data} confidence intervals
  (marginalized over second order slow-roll).}
\label{fig:CMBRCIPI1_3}
\end{center}
\end{figure}

\begin{figure}[H]
\begin{center}
\includegraphics[width=\wappfig,clip=true]{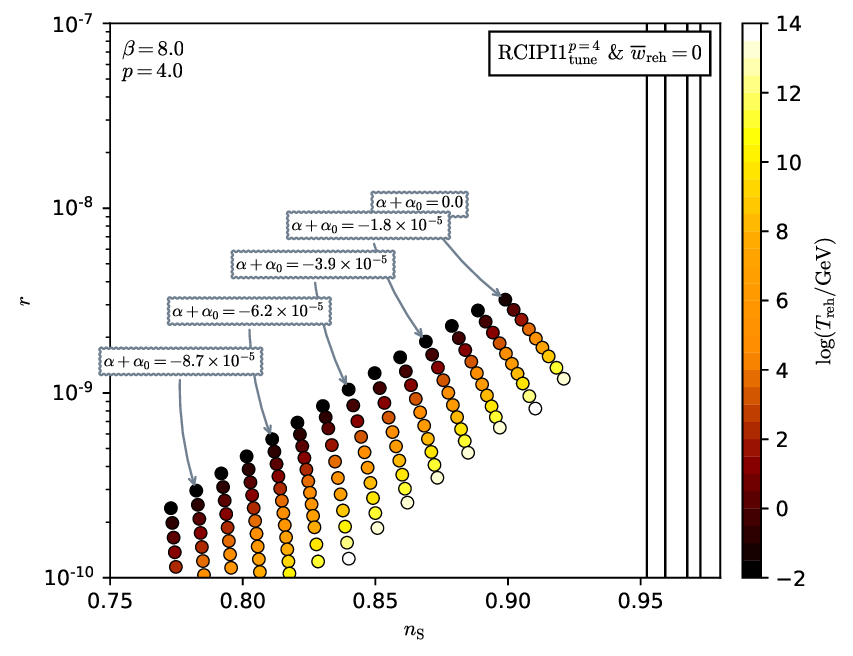}
\includegraphics[width=\wappfig,clip=true]{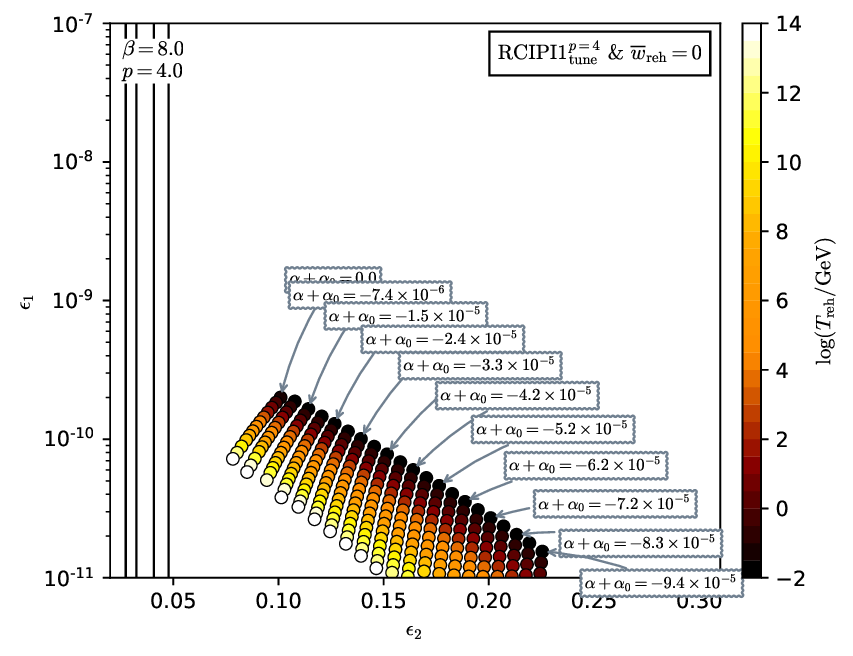}
\caption{Reheating consistent slow-roll predictions for Radiatively
  Corrected Inflection Point Inflation 1 for $p=4$, $\beta=8$ and when
  the potential has almost an inflection point. This one occurs for
  $\alpha = -\alphazero$. Predictions are represented in the plane
  $(\nS,r)$ (top panel) and in the plane $(\epsilon_1,\epsilon_2)$
  (bottom panel) for various values of $\alpha+\alphazero$. The solid
  contours are the one and two-sigma {\data} confidence intervals
  (marginalized over second order slow-roll).}
\label{fig:CMBRCIPI1_4}
\end{center}
\end{figure}

\begin{figure}[H]
\begin{center}
\includegraphics[width=\wappfig,clip=true]{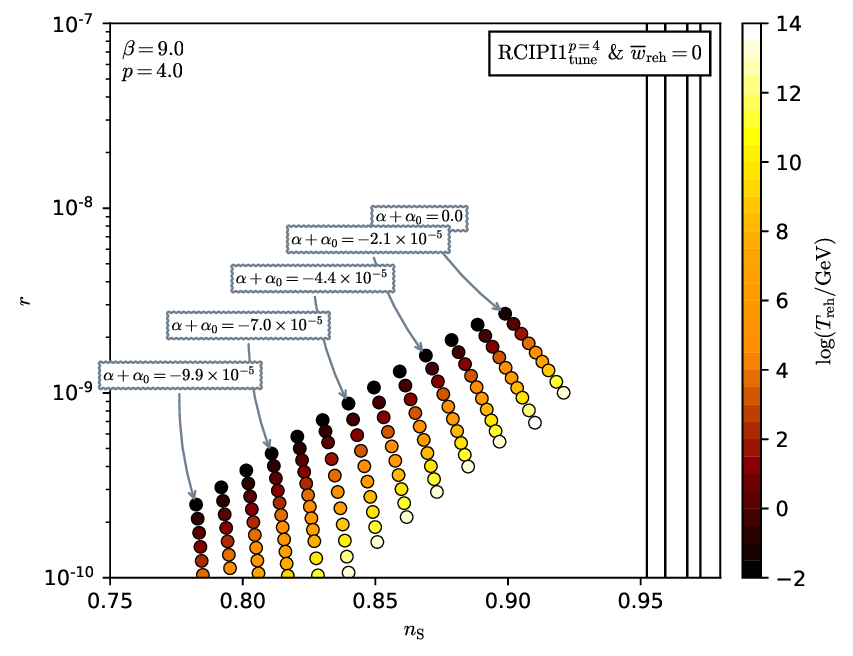}
\includegraphics[width=\wappfig,clip=true]{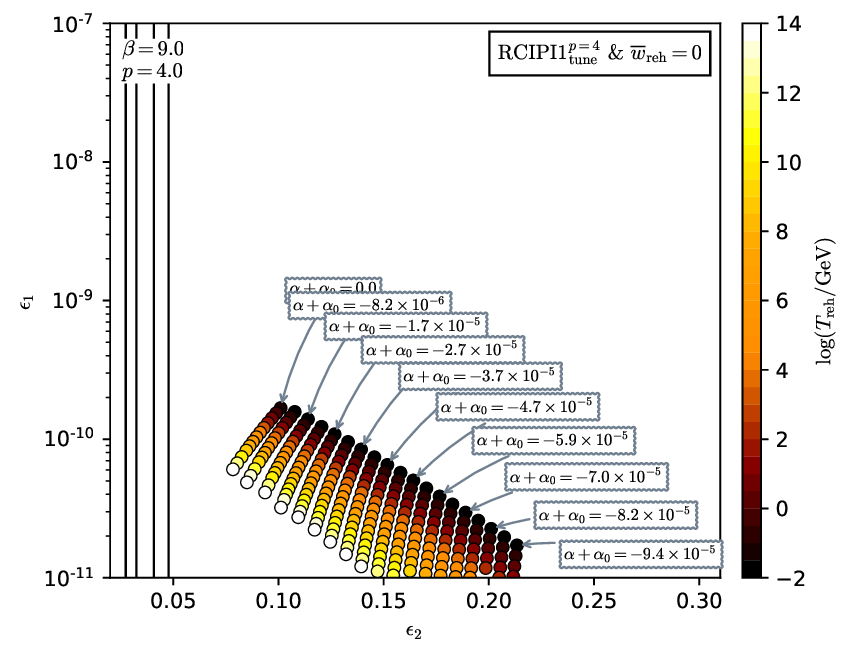}
\caption{Reheating consistent slow-roll predictions for Radiatively
  Corrected Inflection Point Inflation 1 for $p=4$, $\beta=9$ and when
  the potential has almost an inflection point. This one occurs for
  $\alpha = -\alphazero$. Predictions are represented in the plane
  $(\nS,r)$ (top panel) and in the plane $(\epsilon_1,\epsilon_2)$
  (bottom panel) for various values of $\alpha+\alphazero$. The solid
  contours are the one and two-sigma {\data} confidence intervals
  (marginalized over second order slow-roll).}
\label{fig:CMBRCIPI1_5}
\end{center}
\end{figure}

\begin{figure}[H]
\begin{center}
\includegraphics[width=\wappfig,clip=true]{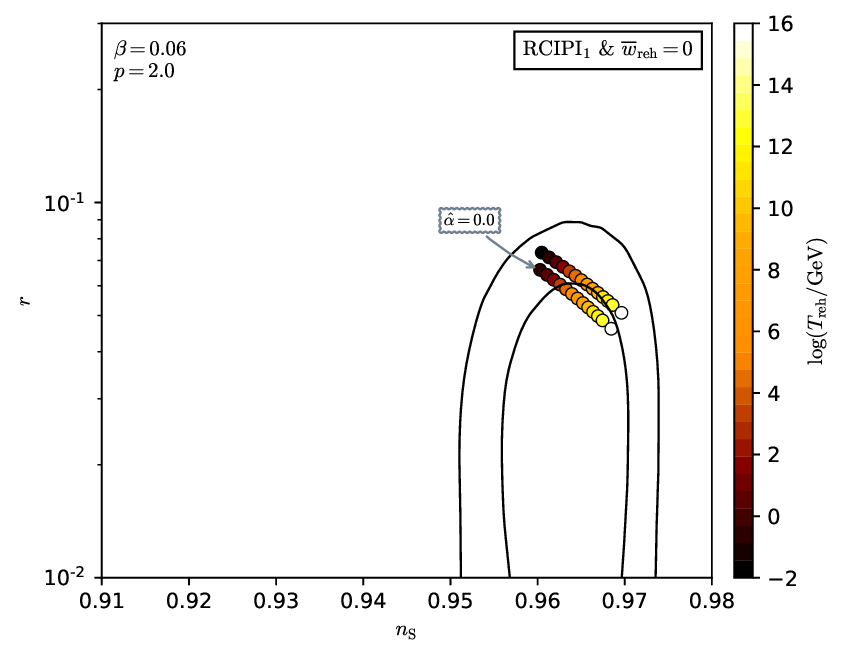}
\includegraphics[width=\wappfig,clip=true]{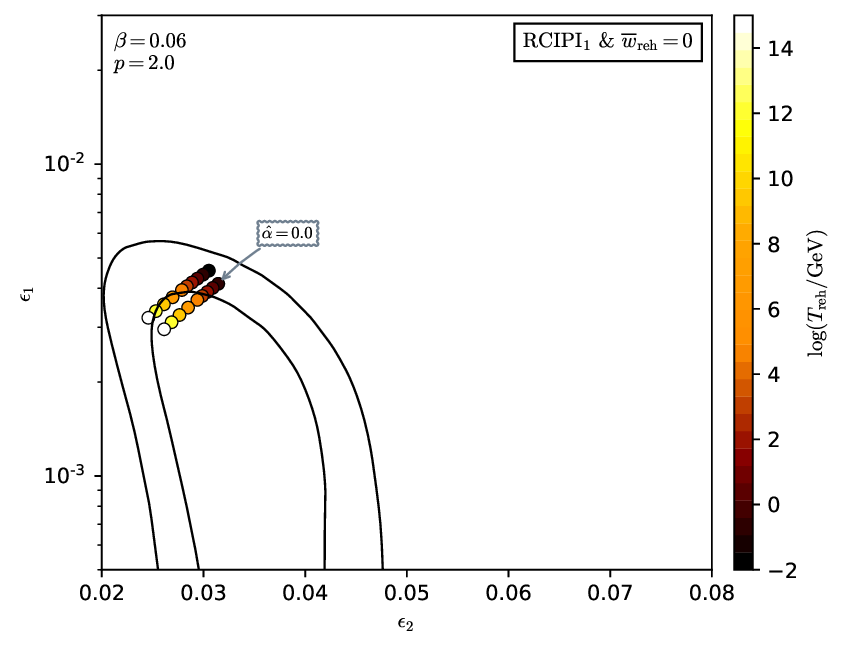}
\caption{Reheating consistent slow-roll predictions for Radiatively
  Corrected Inflection Point Inflation 1 for $p=2$ and
  $\beta=0.06$. Predictions are represented in the plane $(\nS,r)$
  (top panel) and in the plane $(\epsilon_1,\epsilon_2)$ (bottom
  panel) for various values of $\alphahat \equiv
  (2\sqrt{\beta} + \alpha)/(2\sqrt{\beta} -\alphazero)$. This one is
  varied within its maximal allowed range $]0,1]$ for triggering the
  RCIPI1 regime. The solid contours are the one and two-sigma {\data}
  confidence intervals (marginalized over second order slow-roll). See
  also Figs.~\ref{fig:CMBRCIPI1_7} to \ref{fig:CMBRCIPI1_14} for other
  values of $p$ and $\beta$.}
\label{fig:CMBRCIPI1_6}
\end{center}
\end{figure}

\begin{figure}[H]
\begin{center}
\includegraphics[width=\wappfig,clip=true]{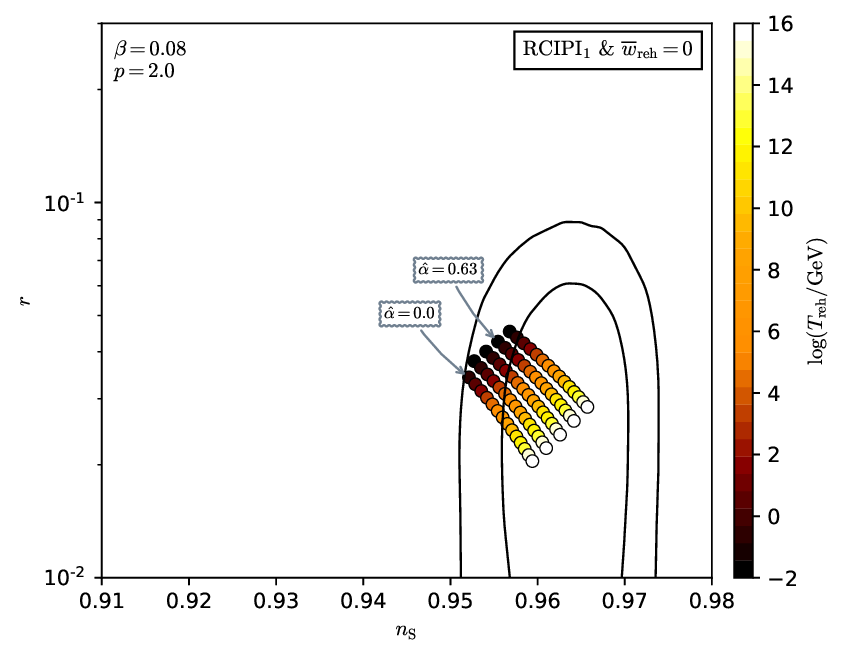}
\includegraphics[width=\wappfig,clip=true]{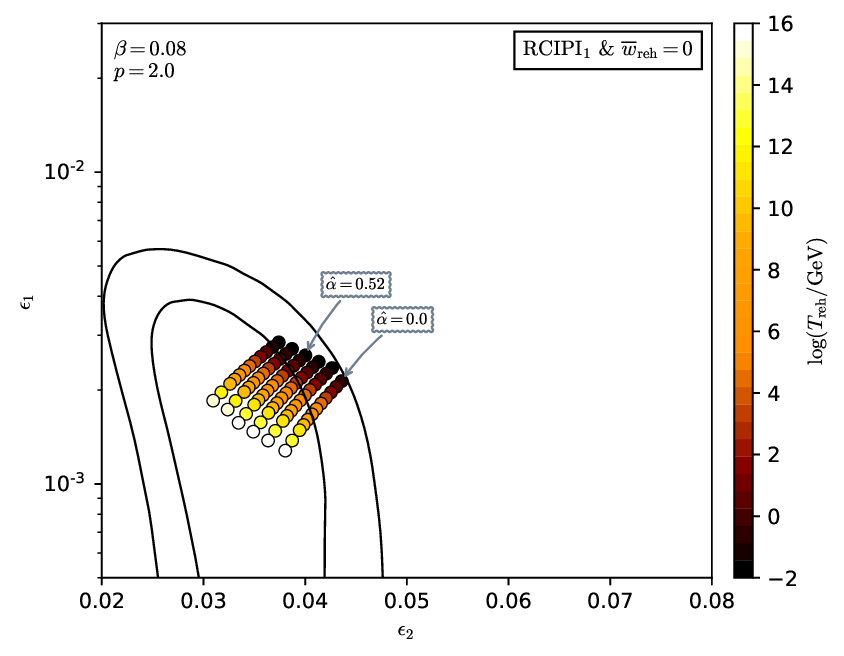}
\caption{Reheating consistent slow-roll predictions for Radiatively
  Corrected Inflection Point Inflation 1 for $p=2$ and
  $\beta=0.08$. Predictions are represented in the plane $(\nS,r)$
  (top panel) and in the plane $(\epsilon_1,\epsilon_2)$ (bottom
  panel) for various values of $\alphahat \equiv
  (2\sqrt{\beta} +\alpha)/(2\sqrt{\beta} -\alphazero)$. This one is
  varied within its maximal allowed range $]0,1]$ for triggering the
  RCIPI1 regime. The solid contours are the one and two-sigma {\data}
  confidence intervals (marginalized over second order slow-roll).}
\label{fig:CMBRCIPI1_7}
\end{center}
\end{figure}

\begin{figure}[H]
\begin{center}
\includegraphics[width=\wappfig,clip=true]{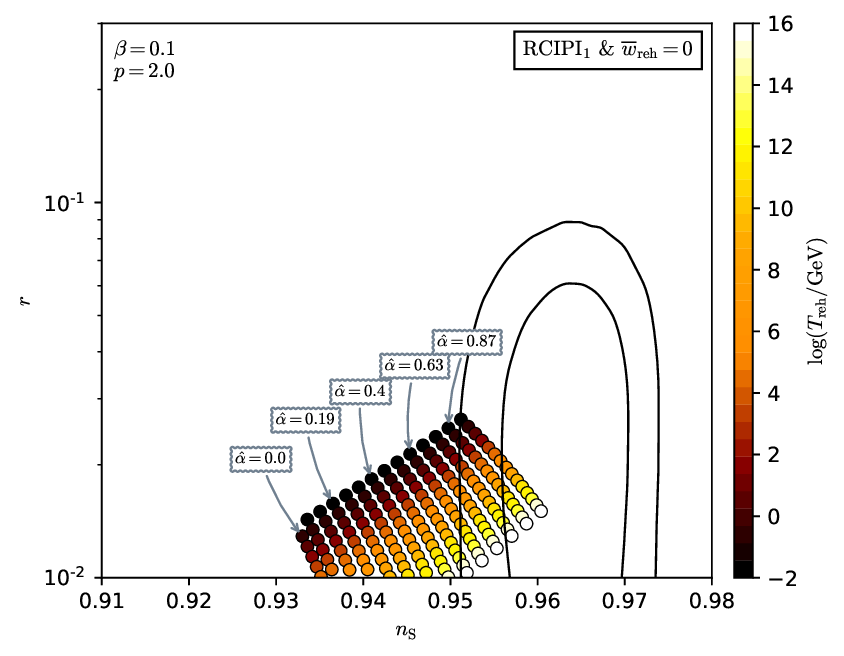}
\includegraphics[width=\wappfig,clip=true]{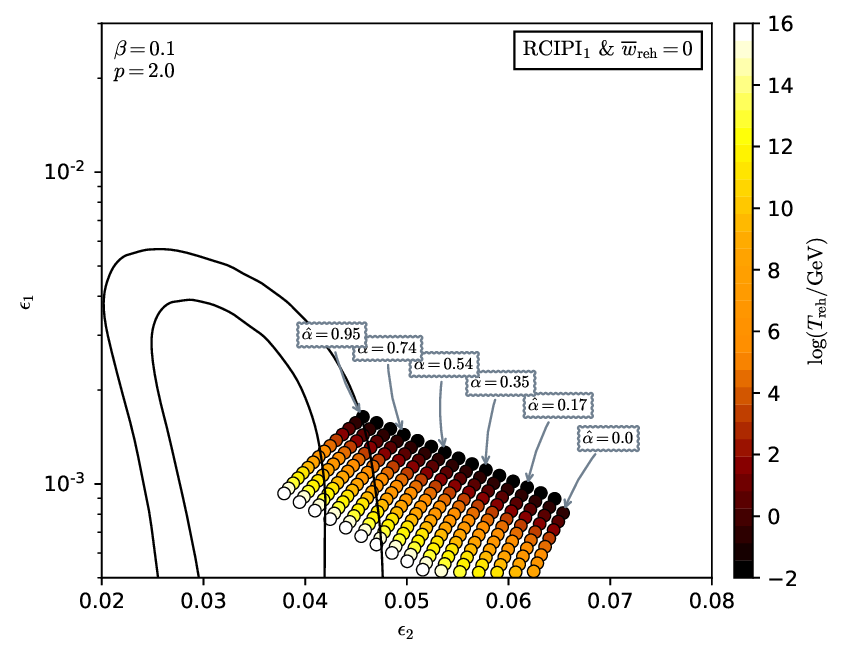}
\caption{Reheating consistent slow-roll predictions for Radiatively
  Corrected Inflection Point Inflation 1 for $p=2$ and
  $\beta=0.1$. Predictions are represented in the plane $(\nS,r)$
  (top panel) and in the plane $(\epsilon_1,\epsilon_2)$ (bottom
  panel) for various values of $\alphahat \equiv
  (2\sqrt{\beta} +\alpha)/(2\sqrt{\beta} -\alphazero)$. This one is
  varied within its maximal allowed range $]0,1]$ for triggering the
  RCIPI1 regime. The solid contours are the one and two-sigma {\data}
  confidence intervals (marginalized over second order slow-roll).}
\label{fig:CMBRCIPI1_8}
\end{center}
\end{figure}

\begin{figure}[H]
\begin{center}
\includegraphics[width=\wappfig,clip=true]{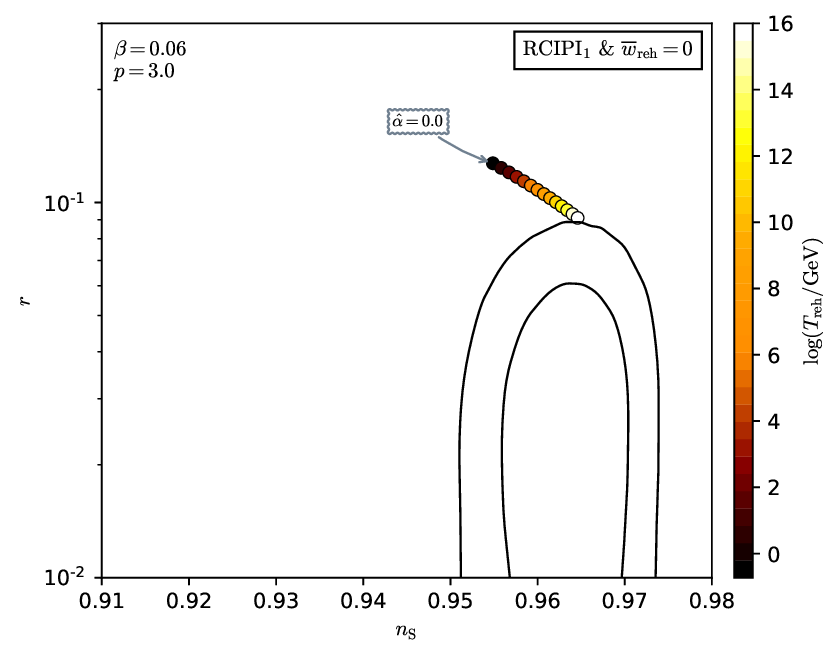}
\includegraphics[width=\wappfig,clip=true]{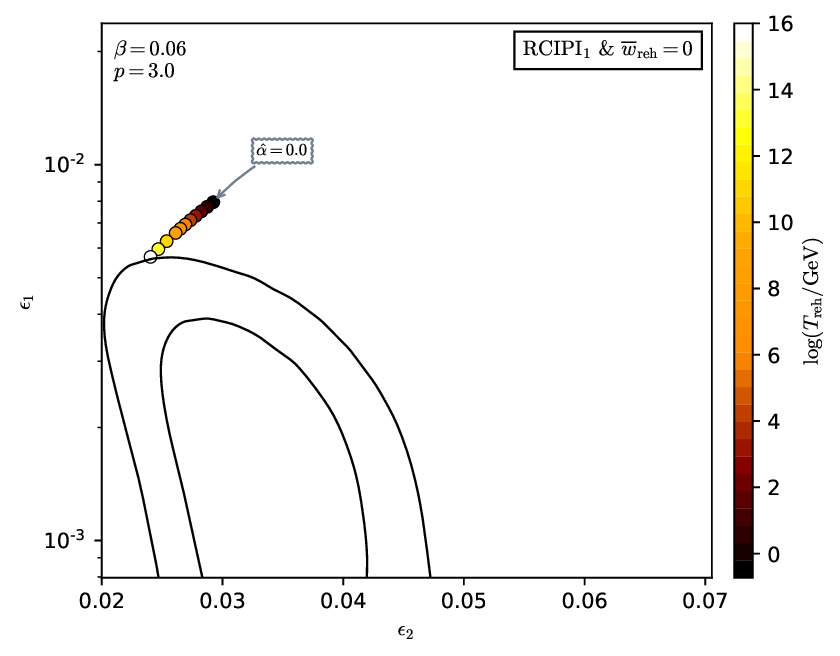}
\caption{Reheating consistent slow-roll predictions for Radiatively
  Corrected Inflection Point Inflation 1 for $p=3$ and
  $\beta=0.06$. Predictions are represented in the plane $(\nS,r)$
  (top panel) and in the plane $(\epsilon_1,\epsilon_2)$ (bottom
  panel) for various values of $\alphahat \equiv
  (2\sqrt{\beta}+\alpha)/(2\sqrt{\beta} -\alphazero)$. This one is
  varied within its maximal allowed range $]0,1]$ for triggering the
  RCIPI1 regime. The solid contours are the one and two-sigma {\data}
  confidence intervals (marginalized over second order slow-roll).}
\label{fig:CMBRCIPI1_9}
\end{center}
\end{figure}

\begin{figure}[H]
\begin{center}
\includegraphics[width=\wappfig,clip=true]{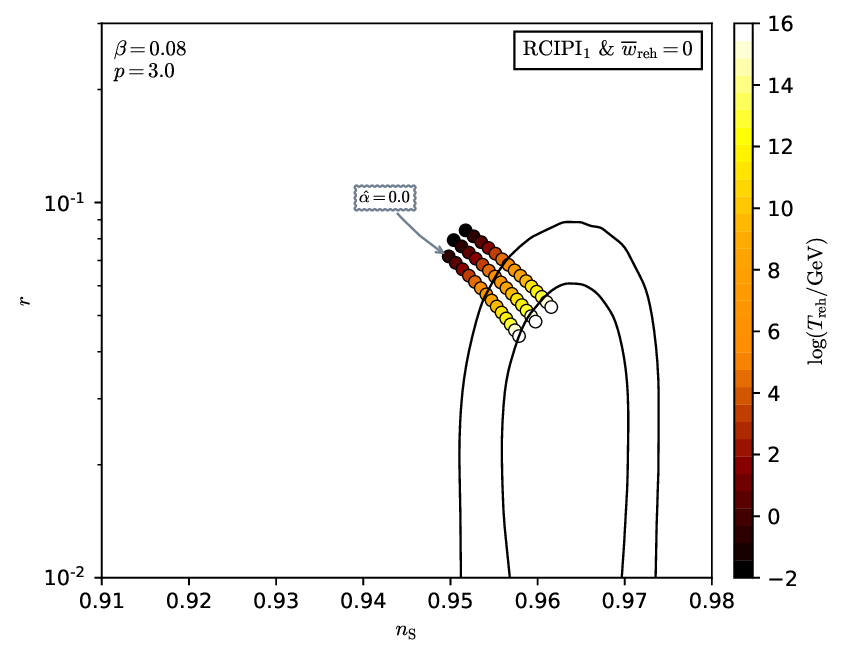}
\includegraphics[width=\wappfig,clip=true]{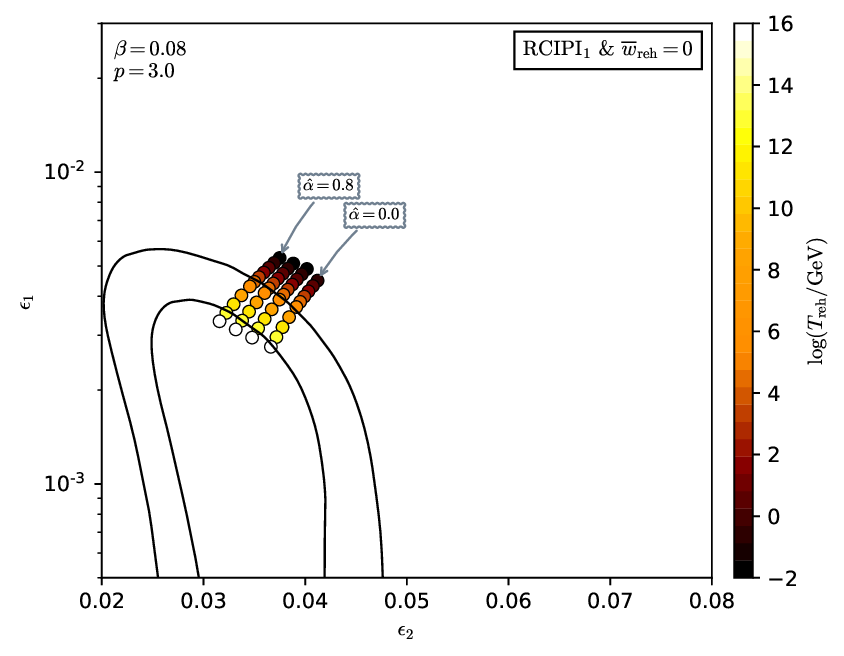}
\caption{Reheating consistent slow-roll predictions for Radiatively
  Corrected Inflection Point Inflation 1 for $p=3$ and
  $\beta=0.08$. Predictions are represented in the plane $(\nS,r)$
  (top panel) and in the plane $(\epsilon_1,\epsilon_2)$ (bottom
  panel) for various values of $\alphahat \equiv
  (2\sqrt{\beta}+\alpha)/(2\sqrt{\beta} -\alphazero)$. This one is
  varied within its maximal allowed range $]0,1]$ for triggering the
  RCIPI1 regime. The solid contours are the one and two-sigma {\data}
  confidence intervals (marginalized over second order slow-roll).}
\label{fig:CMBRCIPI1_10}
\end{center}
\end{figure}

\begin{figure}[H]
\begin{center}
\includegraphics[width=\wappfig,clip=true]{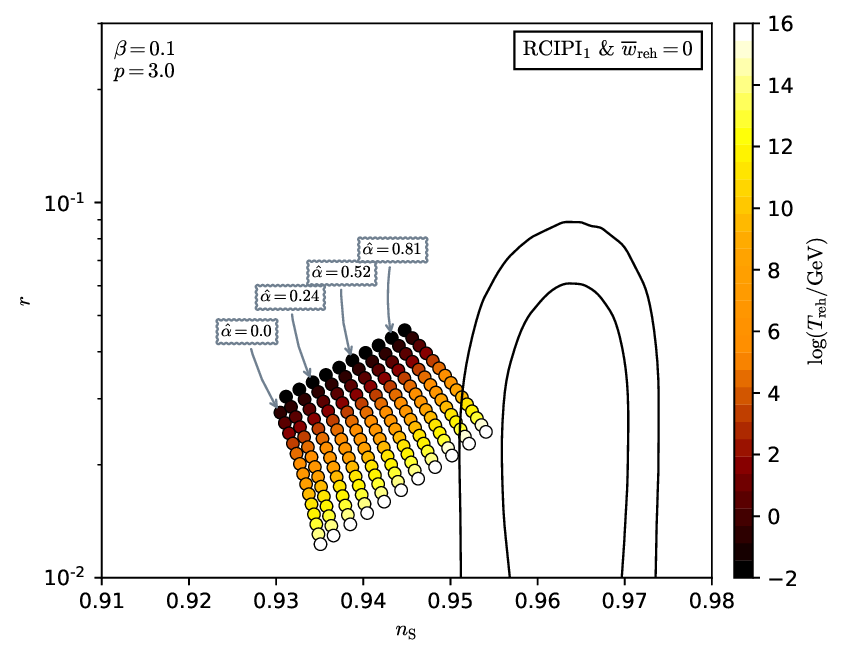}
\includegraphics[width=\wappfig,clip=true]{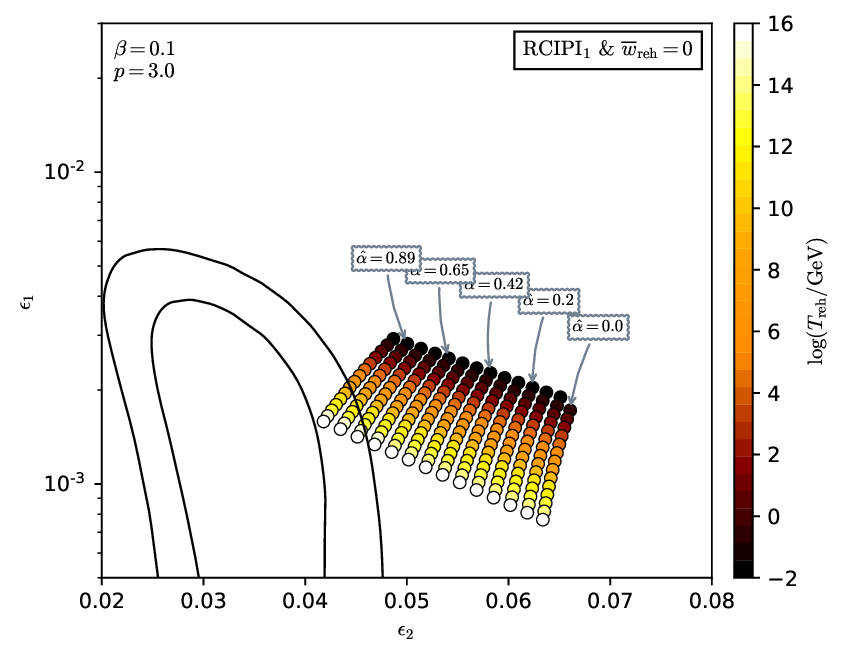}
\caption{Reheating consistent slow-roll predictions for Radiatively
  Corrected Inflection Point Inflation 1 for $p=3$ and
  $\beta=0.1$. Predictions are represented in the plane $(\nS,r)$
  (top panel) and in the plane $(\epsilon_1,\epsilon_2)$ (bottom
  panel) for various values of $\alphahat \equiv
  (2\sqrt{\beta}+\alpha)/(2\sqrt{\beta} -\alphazero)$. This one is
  varied within its maximal allowed range $]0,1]$ for triggering the
  RCIPI1 regime. The solid contours are the one and two-sigma {\data}
  confidence intervals (marginalized over second order slow-roll).}
\label{fig:CMBRCIPI1_11}
\end{center}
\end{figure}

\begin{figure}[H]
\begin{center}
\includegraphics[width=\wappfig,clip=true]{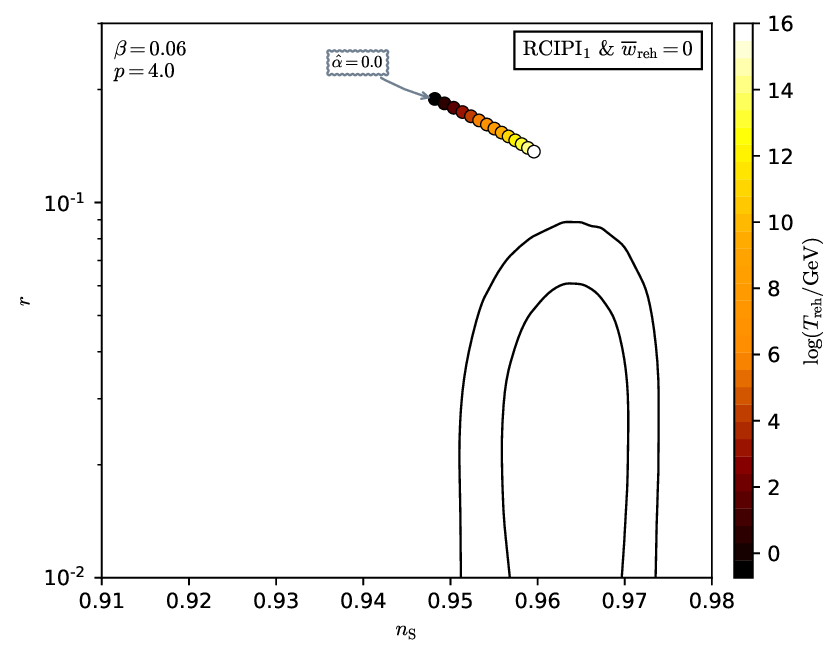}
\includegraphics[width=\wappfig,clip=true]{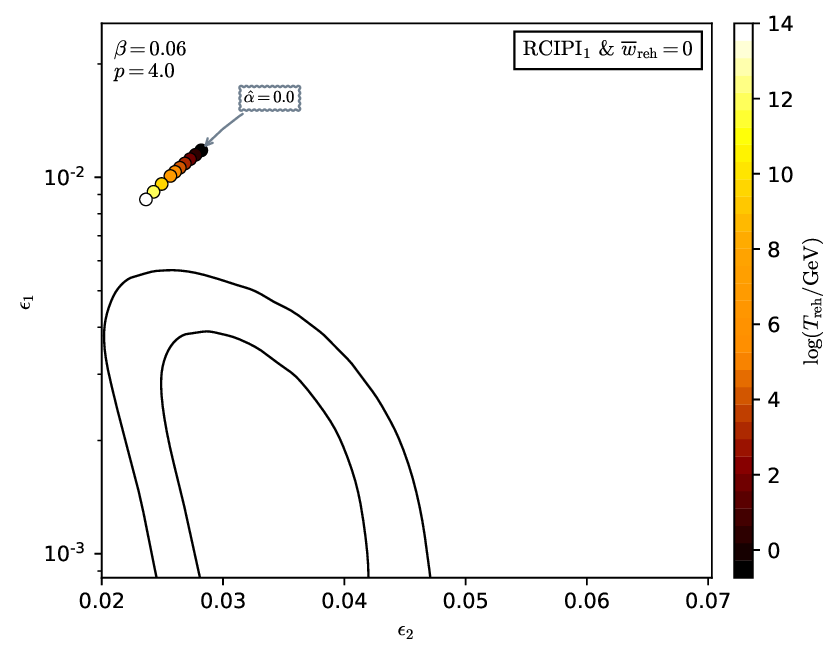}
\caption{Reheating consistent slow-roll predictions for Radiatively
  Corrected Inflection Point Inflation 1 for $p=4$ and
  $\beta=0.06$. Predictions are represented in the plane $(\nS,r)$
  (top panel) and in the plane $(\epsilon_1,\epsilon_2)$ (bottom
  panel) for various values of $\alphahat \equiv
  (2\sqrt{\beta}+\alpha)/(2\sqrt{\beta} -\alphazero)$. This one is
  varied within its maximal allowed range $]0,1]$ for triggering the
  RCIPI1 regime. The solid contours are the one and two-sigma {\data}
  confidence intervals (marginalized over second order slow-roll).}
\label{fig:CMBRCIPI1_12}
\end{center}
\end{figure}

\begin{figure}[H]
\begin{center}
\includegraphics[width=\wappfig,clip=true]{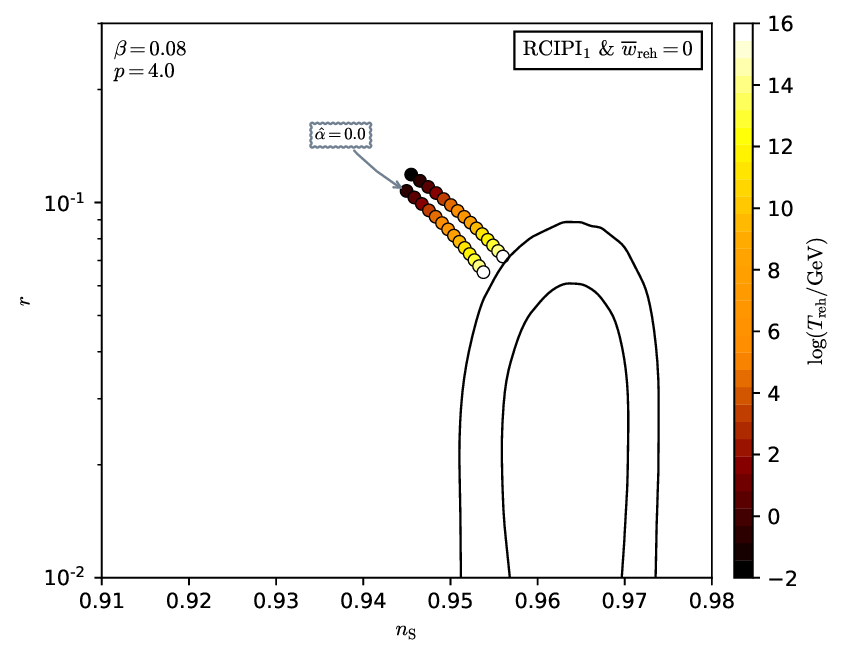}
\includegraphics[width=\wappfig,clip=true]{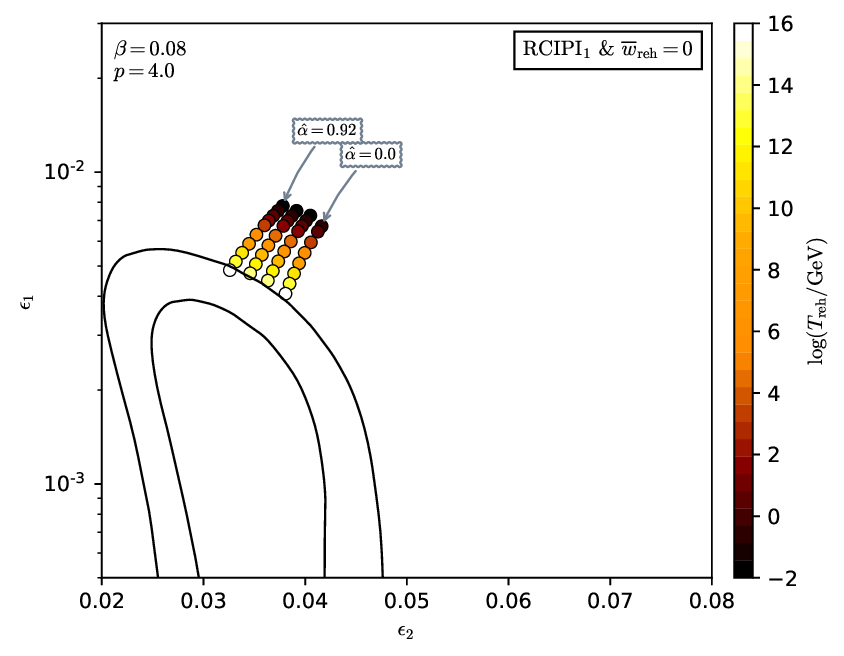}
\caption{Reheating consistent slow-roll predictions for Radiatively
  Corrected Inflection Point Inflation 1 for $p=4$ and
  $\beta=0.08$. Predictions are represented in the plane $(\nS,r)$
  (top panel) and in the plane $(\epsilon_1,\epsilon_2)$ (bottom
  panel) for various values of $\alphahat \equiv
  (2\sqrt{\beta}+\alpha)/(2\sqrt{\beta} -\alphazero)$. This one is
  varied within its maximal allowed range $]0,1]$ for triggering the
  RCIPI1 regime. The solid contours are the one and two-sigma {\data}
  confidence intervals (marginalized over second order slow-roll).}
\label{fig:CMBRCIPI1_13}
\end{center}
\end{figure}

\begin{figure}[H]
\begin{center}
\includegraphics[width=\wappfig,clip=true]{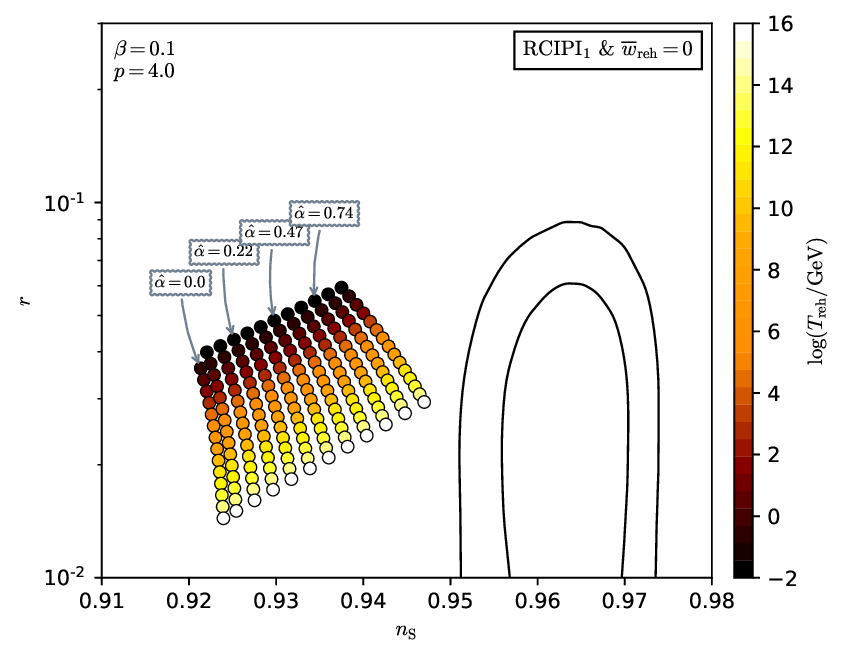}
\includegraphics[width=\wappfig,clip=true]{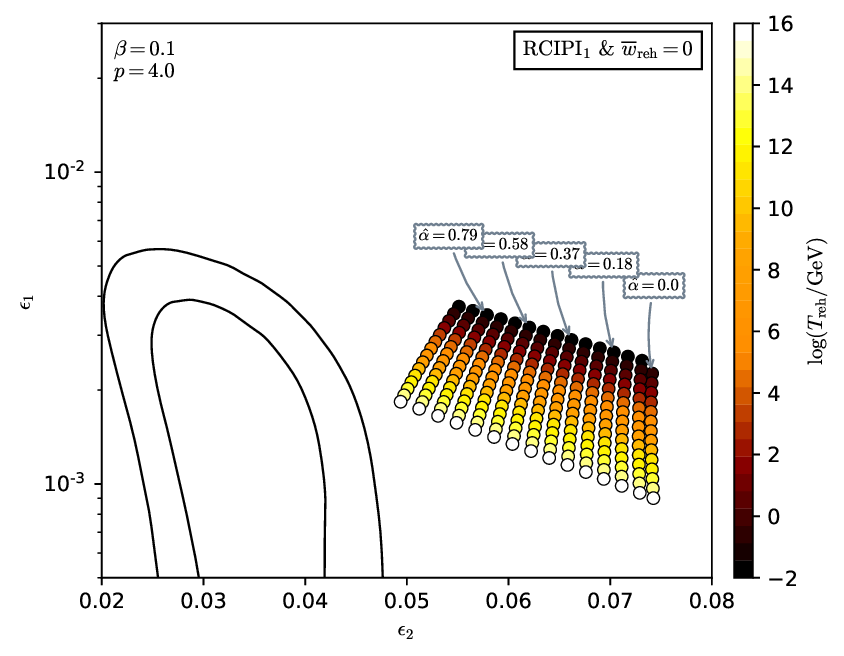}
\caption{Reheating consistent slow-roll predictions for Radiatively
  Corrected Inflection Point Inflation 1 for $p=4$ and
  $\beta=0.1$. Predictions are represented in the plane $(\nS,r)$
  (top panel) and in the plane $(\epsilon_1,\epsilon_2)$ (bottom
  panel) for various values of $\alphahat \equiv
  (2\sqrt{\beta}+\alpha)/(2\sqrt{\beta} -\alphazero)$. This one is
  varied within its maximal allowed range $]0,1]$ for triggering the
  RCIPI1 regime. The solid contours are the one and two-sigma {\data}
  confidence intervals (marginalized over second order slow-roll).}
\label{fig:CMBRCIPI1_14}
\end{center}
\end{figure}

\subsection{Radiatively Corrected Inflection Point Inflation 2 (\hyperref[sec:rcipi]{RCIPI2})}

\begin{figure}[H]
\begin{center}
\includegraphics[width=\wappfig,clip=true]{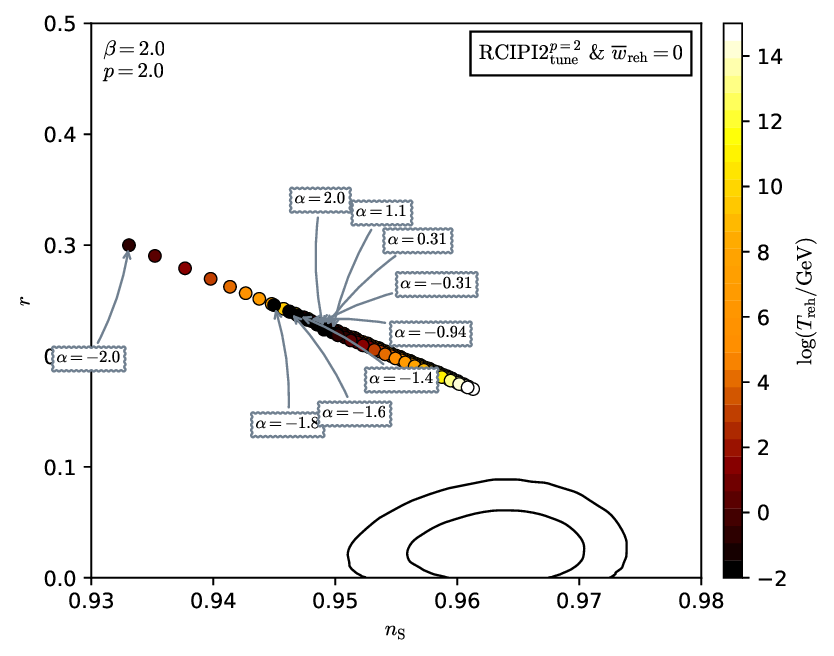}
\includegraphics[width=\wappfig,clip=true]{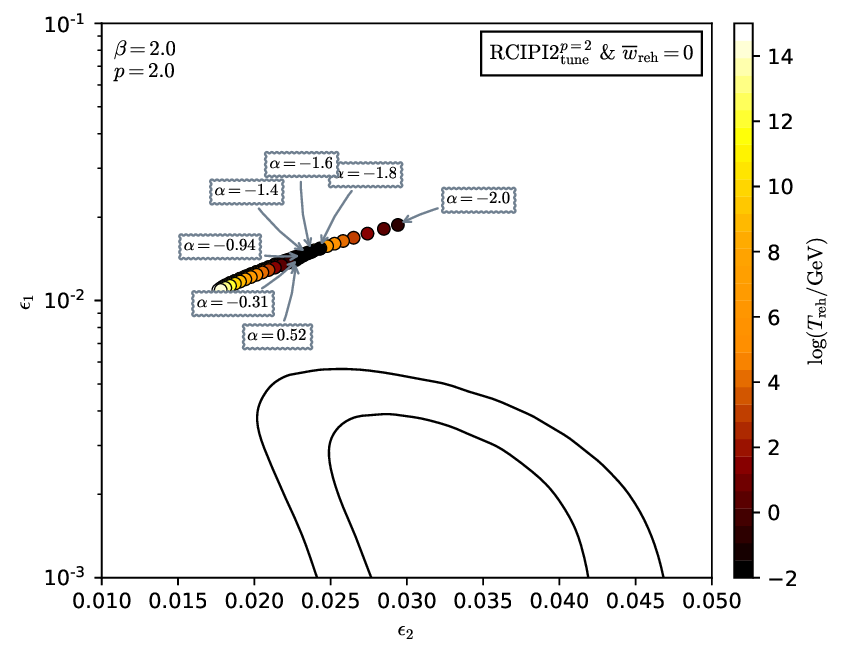}
\caption{Reheating consistent slow-roll predictions for Radiatively
  Corrected Inflection Point Inflation 2 for $p=2$, $\beta=2$ and when
  the potential is monotonic. The limiting case of an inflection point
  occurs for $\alpha = \pm 2$. Predictions are represented in the
  plane $(\nS,r)$ (top panel) and in the plane
  $(\epsilon_1,\epsilon_2)$ (bottom panel) for various values of
  $\alpha$ ranging within the maximal domain $]-2,2[$ supporting
  RCIPI2. The solid contours are the one and two-sigma {\data}
  confidence intervals (marginalized over second order slow-roll). See
  also \Fig{fig:CMBRCIPI2_1}.}
\label{fig:CMBRCIPI2_0}
\end{center}
\end{figure}

\begin{figure}[H]
\begin{center}
\includegraphics[width=\wappfig,clip=true]{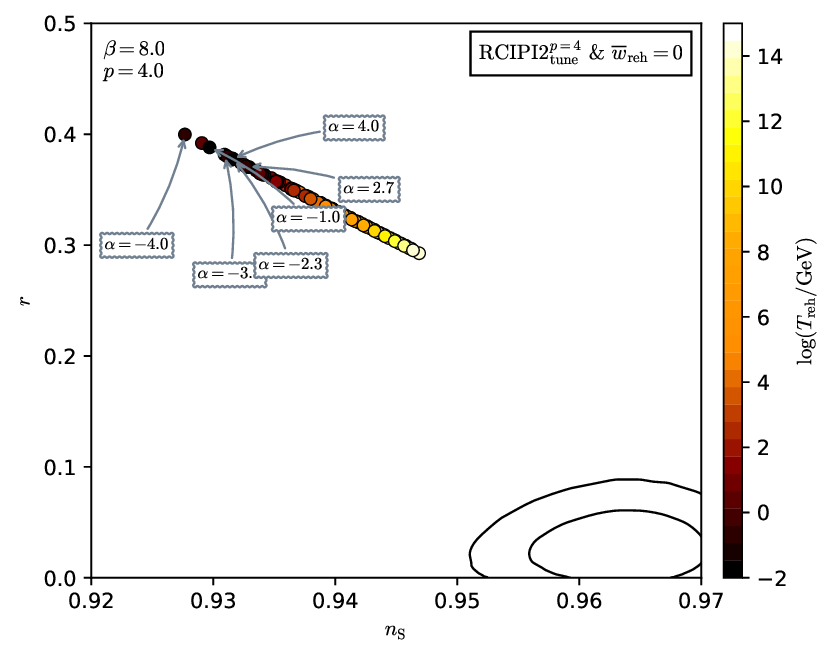}
\includegraphics[width=\wappfig,clip=true]{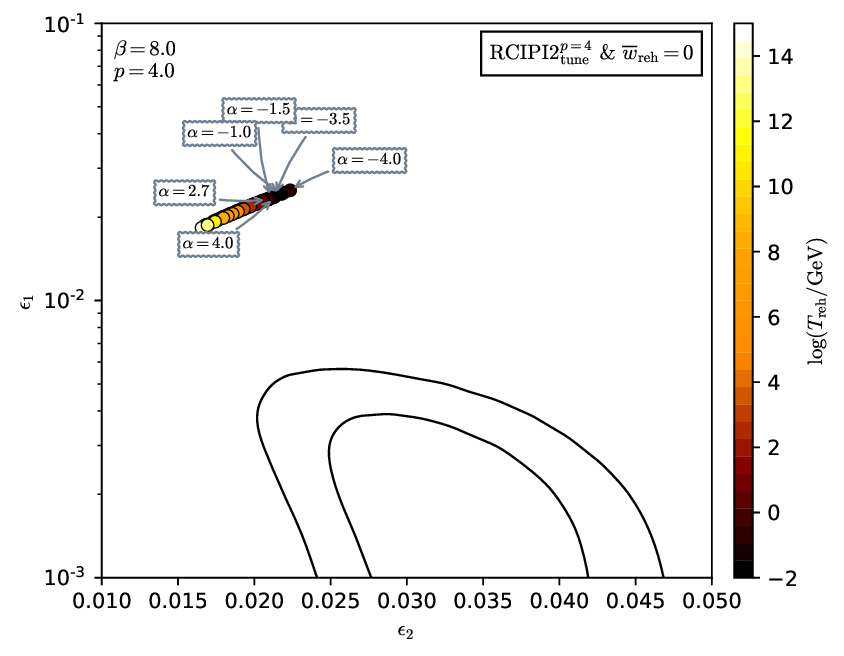}
\caption{Reheating consistent slow-roll predictions for Radiatively
  Corrected Inflection Point Inflation 2 for $p=4$, $\beta=8$ and when
  the potential is monotonic. The limiting case of an inflection point
  occurs for $\alpha = \pm 4$. Predictions are
  represented in the plane $(\nS,r)$ (top panel) and in the plane
  $(\epsilon_1,\epsilon_2)$ (bottom panel) for various values of
  $\alpha$ ranging within the whole domain $]-4,4[$ supporting
  RCIPI2. The solid contours are the one and two-sigma {\data}
  confidence intervals (marginalized over second order slow-roll).}
\label{fig:CMBRCIPI2_1}
\end{center}
\end{figure}

\begin{figure}[H]
\begin{center}
\includegraphics[width=\wappfig,clip=true]{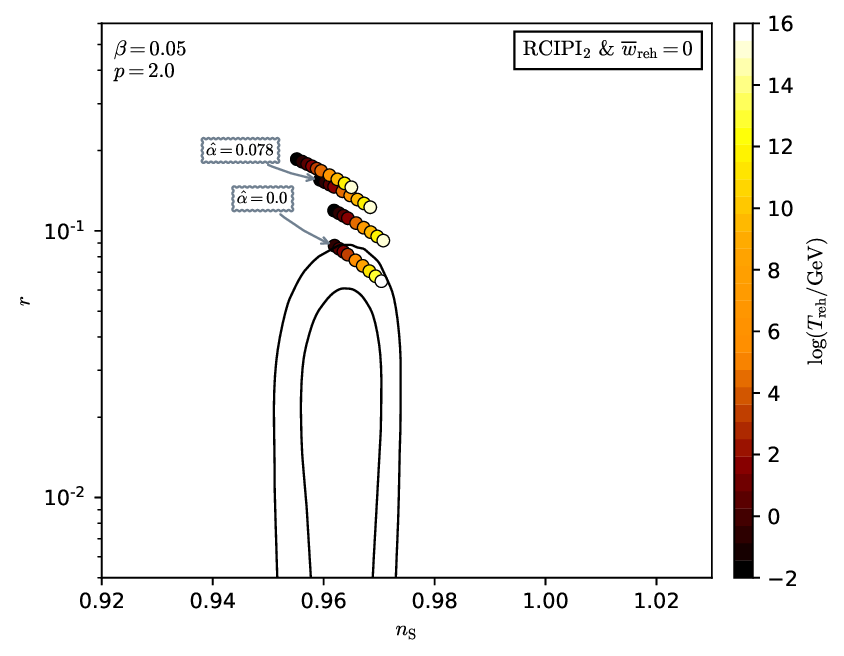}
\includegraphics[width=\wappfig,clip=true]{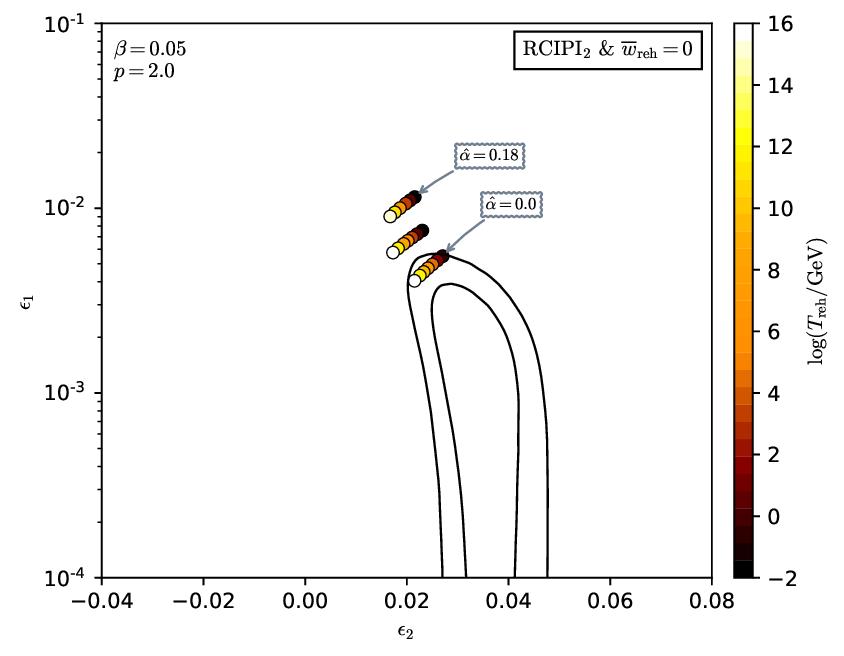}
\caption{Reheating consistent slow-roll predictions for Radiatively
  Corrected Inflection Point Inflation 2 for $p=2$ and
  $\beta=0.05$. Predictions are represented in the plane $(\nS,r)$
  (top panel) and in the plane $(\epsilon_1,\epsilon_2)$ (bottom
  panel) for various values of $\alphahat \equiv
  (\alpha + \alphazero)/(2\alphazero)$. This one is
  varied within its maximal allowed range $]0,1[$ for triggering the
  RCIPI2 regime. The solid contours are the one and two-sigma {\data}
  confidence intervals (marginalized over second order slow-roll). See
  also Figs.~\ref{fig:CMBRCIPI1_3} to \ref{fig:CMBRCIPI1_10} for other
  values of $p$ and $\beta$.}
\label{fig:CMBRCIPI2_2}
\end{center}
\end{figure}

\begin{figure}[H]
\begin{center}
\includegraphics[width=\wappfig,clip=true]{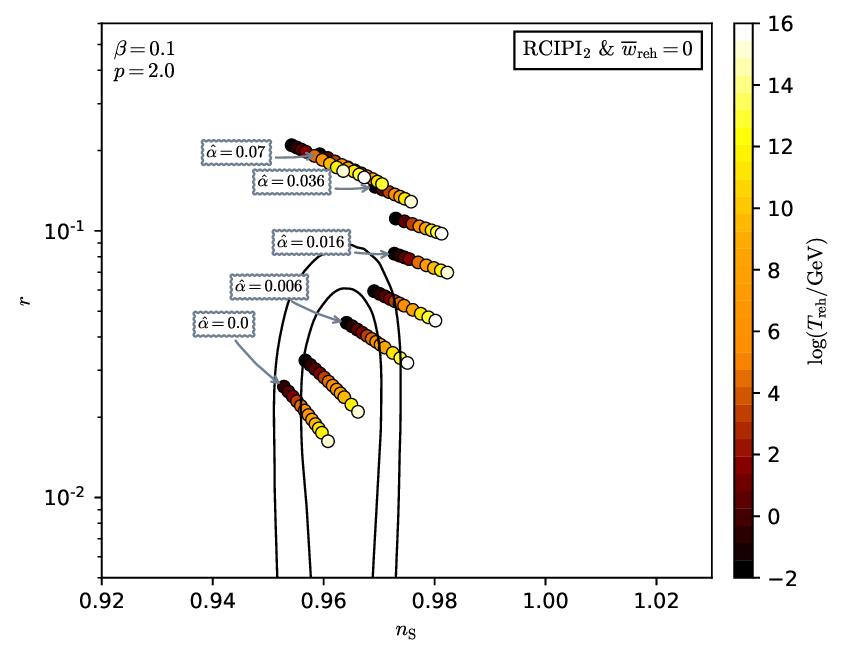}
\includegraphics[width=\wappfig,clip=true]{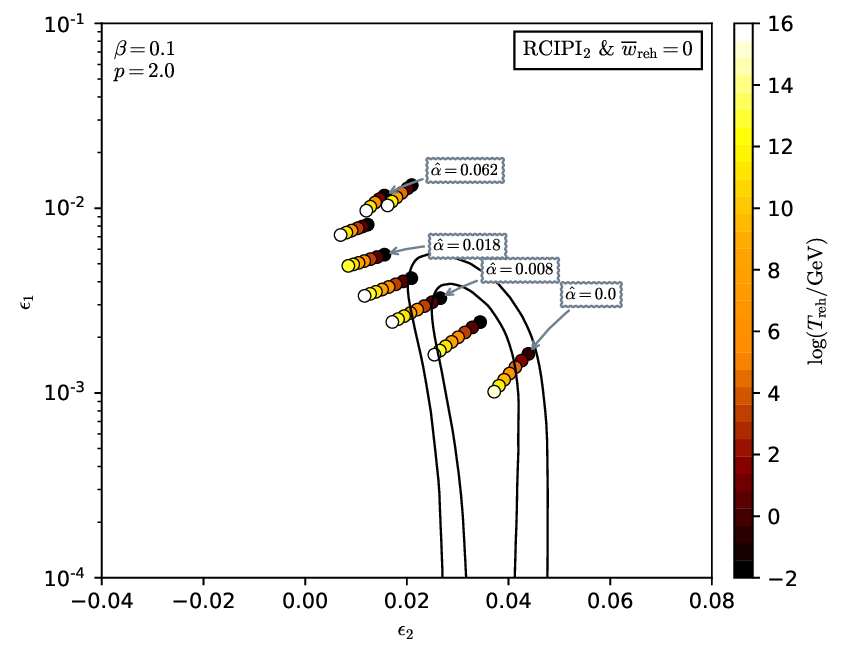}
\caption{Reheating consistent slow-roll predictions for Radiatively
  Corrected Inflection Point Inflation 2 for $p=2$ and
  $\beta=0.1$. Predictions are represented in the plane $(\nS,r)$
  (top panel) and in the plane $(\epsilon_1,\epsilon_2)$ (bottom
  panel) for various values of $\alphahat \equiv
  (\alpha + \alphazero)/(2\alphazero)$. This one is
  varied within its maximal allowed range $]0,1[$ for triggering the
  RCIPI2 regime. The solid contours are the one and two-sigma {\data}
  confidence intervals (marginalized over second order slow-roll).}
\label{fig:CMBRCIPI2_3}
\end{center}
\end{figure}

\begin{figure}[H]
\begin{center}
\includegraphics[width=\wappfig,clip=true]{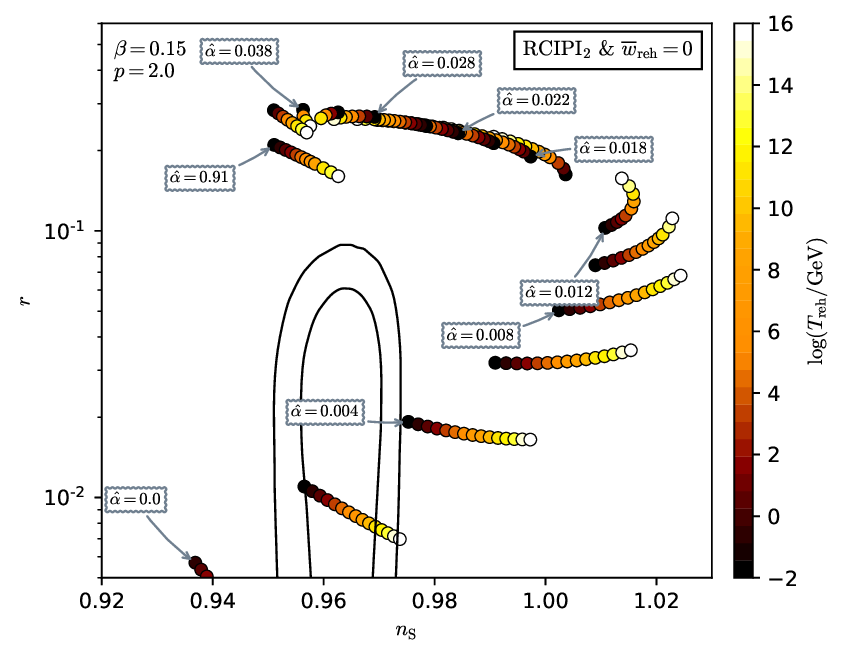}
\includegraphics[width=\wappfig,clip=true]{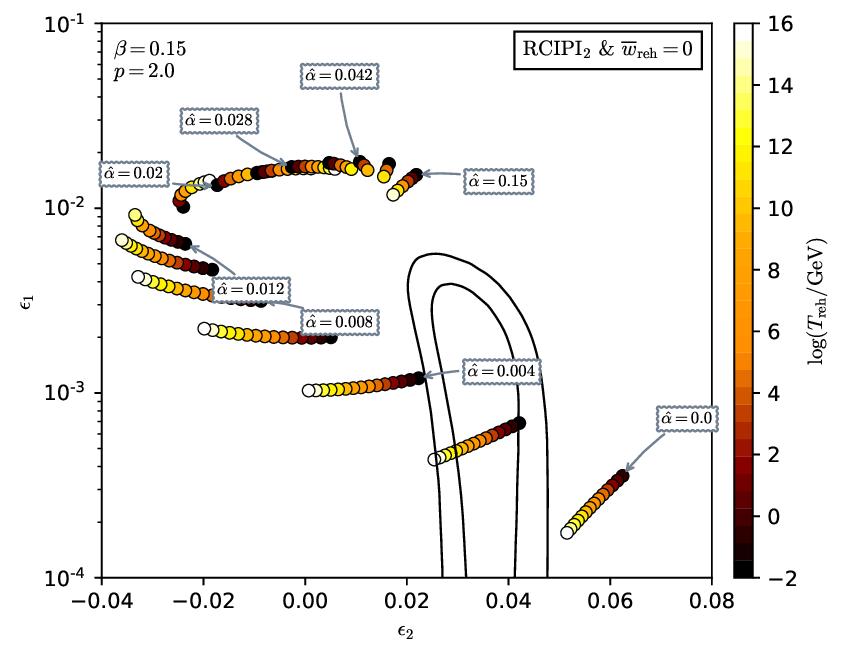}
\caption{Reheating consistent slow-roll predictions for Radiatively
  Corrected Inflection Point Inflation 2 for $p=2$ and
  $\beta=0.15$. Predictions are represented in the plane $(\nS,r)$
  (top panel) and in the plane $(\epsilon_1,\epsilon_2)$ (bottom
  panel) for various values of $\alphahat \equiv
  (\alpha + \alphazero)/(2\alphazero)$. This one is
  varied within its maximal allowed range $]0,1[$ for triggering the
  RCIPI2 regime. The solid contours are the one and two-sigma {\data}
  confidence intervals (marginalized over second order slow-roll).}
\label{fig:CMBRCIPI2_4}
\end{center}
\end{figure}

\begin{figure}[H]
\begin{center}
\includegraphics[width=\wappfig,clip=true]{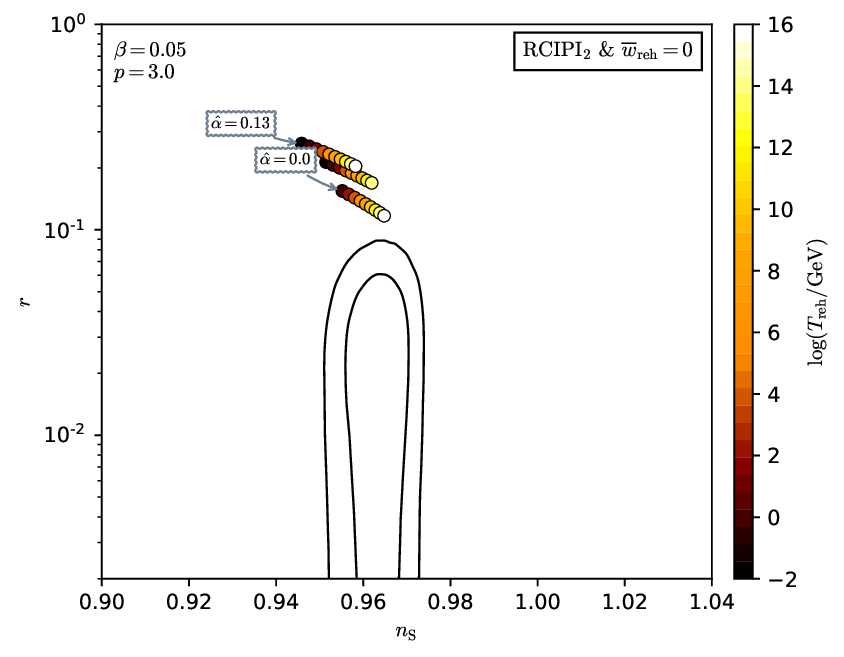}
\includegraphics[width=\wappfig,clip=true]{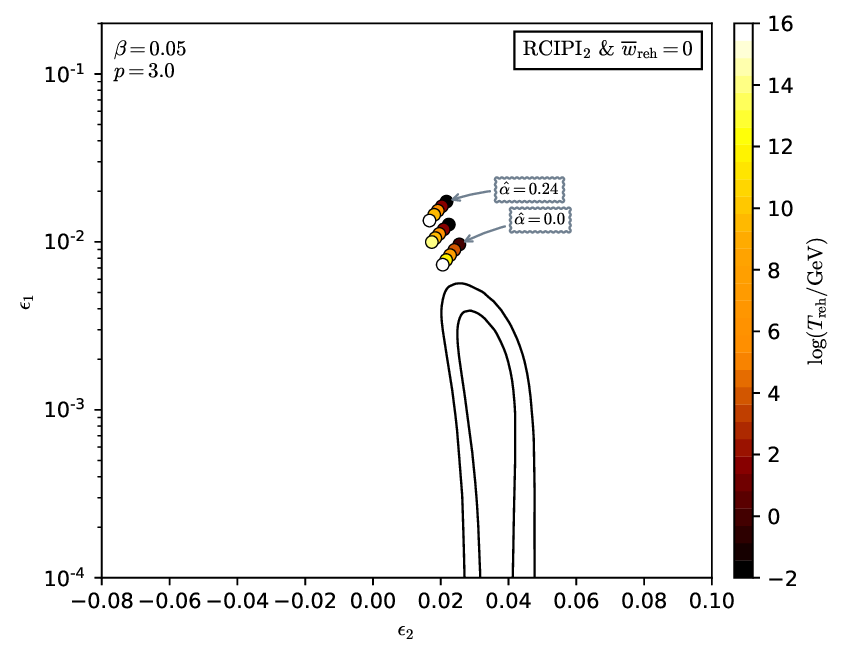}
\caption{Reheating consistent slow-roll predictions for Radiatively
  Corrected Inflection Point Inflation 2 for $p=3$ and
  $\beta=0.05$. Predictions are represented in the plane $(\nS,r)$
  (top panel) and in the plane $(\epsilon_1,\epsilon_2)$ (bottom
  panel) for various values of $\alphahat \equiv
  (\alpha + \alphazero)/(2\alphazero)$. This one is
  varied within its maximal allowed range $]0,1[$ for triggering the
  RCIPI2 regime. The solid contours are the one and two-sigma {\data}
  confidence intervals (marginalized over second order slow-roll).}
\label{fig:CMBRCIPI2_5}
\end{center}
\end{figure}

\begin{figure}[H]
\begin{center}
\includegraphics[width=\wappfig,clip=true]{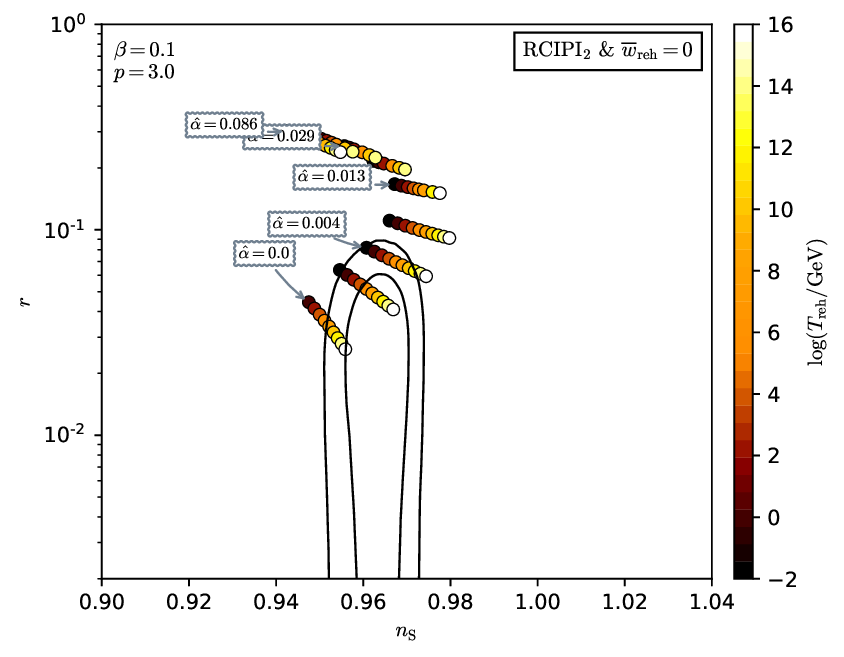}
\includegraphics[width=\wappfig,clip=true]{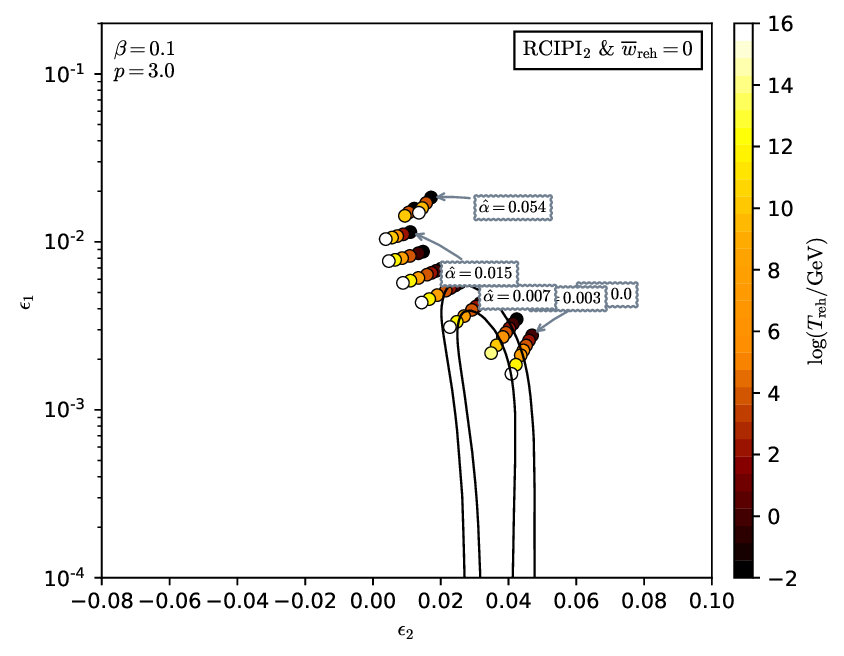}
\caption{Reheating consistent slow-roll predictions for Radiatively
  Corrected Inflection Point Inflation 2 for $p=3$ and
  $\beta=0.1$. Predictions are represented in the plane $(\nS,r)$
  (top panel) and in the plane $(\epsilon_1,\epsilon_2)$ (bottom
  panel) for various values of $\alphahat \equiv
  (\alpha + \alphazero)/(2\alphazero)$. This one is
  varied within its maximal allowed range $]0,1[$ for triggering the
  RCIPI2 regime. The solid contours are the one and two-sigma {\data}
  confidence intervals (marginalized over second order slow-roll).}
\label{fig:CMBRCIPI2_6}
\end{center}
\end{figure}

\begin{figure}[H]
\begin{center}
\includegraphics[width=\wappfig,clip=true]{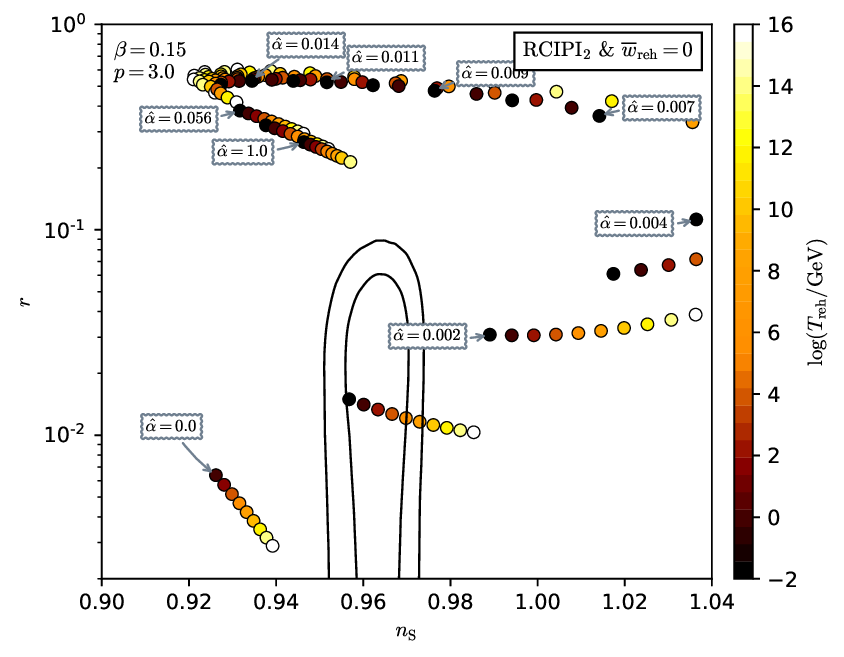}
\includegraphics[width=\wappfig,clip=true]{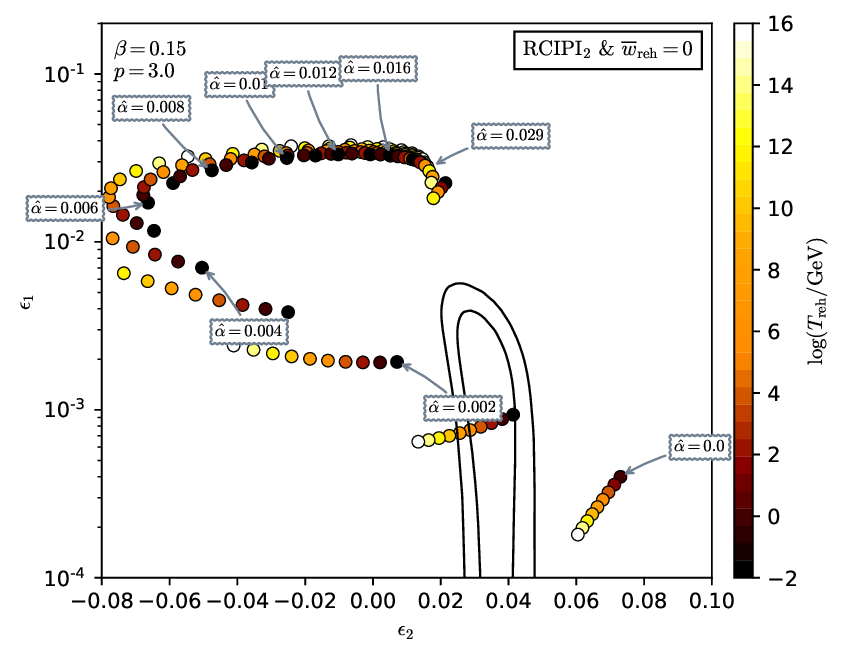}
\caption{Reheating consistent slow-roll predictions for Radiatively
  Corrected Inflection Point Inflation 2 for $p=3$ and
  $\beta=0.15$. Predictions are represented in the plane $(\nS,r)$
  (top panel) and in the plane $(\epsilon_1,\epsilon_2)$ (bottom
  panel) for various values of $\alphahat \equiv
  (\alpha + \alphazero)/(2\alphazero)$. This one is
  varied within its maximal allowed range $]0,1[$ for triggering the
  RCIPI2 regime. The solid contours are the one and two-sigma {\data}
  confidence intervals (marginalized over second order slow-roll).}
\label{fig:CMBRCIPI2_7}
\end{center}
\end{figure}

\begin{figure}[H]
\begin{center}
\includegraphics[width=\wappfig,clip=true]{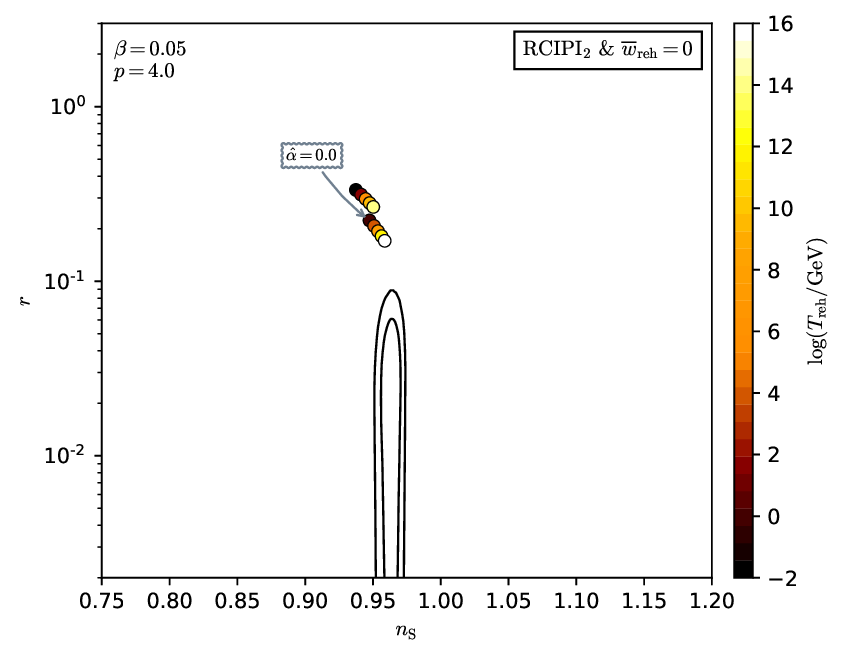}
\includegraphics[width=\wappfig,clip=true]{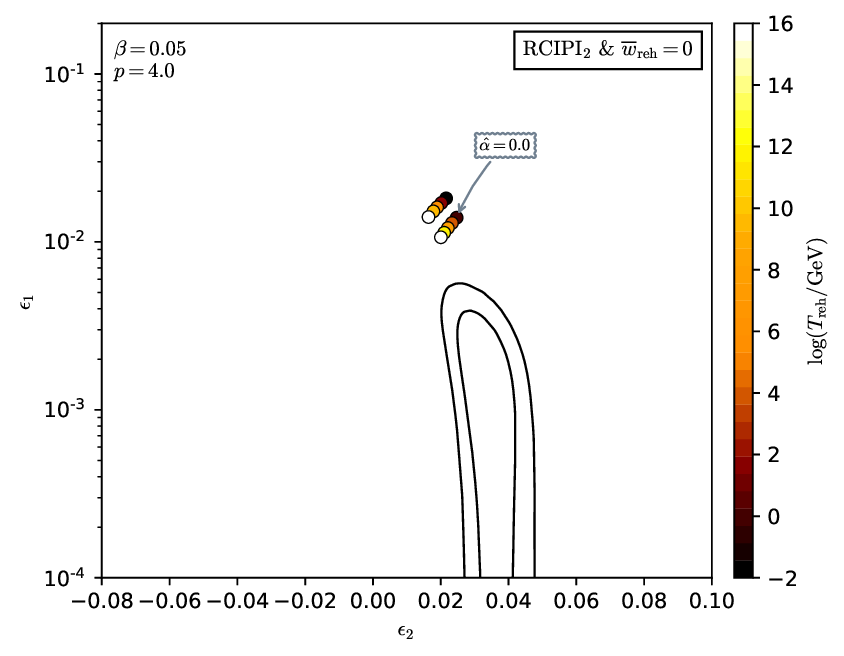}
\caption{Reheating consistent slow-roll predictions for Radiatively
  Corrected Inflection Point Inflation 2 for $p=4$ and
  $\beta=0.05$. Predictions are represented in the plane $(\nS,r)$
  (top panel) and in the plane $(\epsilon_1,\epsilon_2)$ (bottom
  panel) for various values of $\alphahat \equiv
  (\alpha + \alphazero)/(2\alphazero)$. This one is
  varied within its maximal allowed range $]0,1[$ for triggering the
  RCIPI2 regime. The solid contours are the one and two-sigma {\data}
  confidence intervals (marginalized over second order slow-roll).}
\label{fig:CMBRCIPI2_8}
\end{center}
\end{figure}

\begin{figure}[H]
\begin{center}
\includegraphics[width=\wappfig,clip=true]{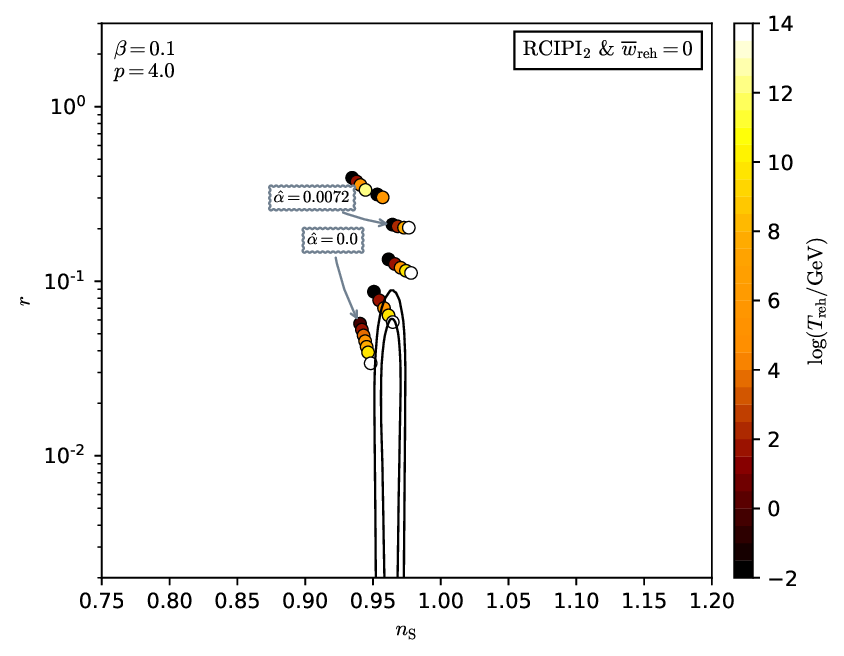}
\includegraphics[width=\wappfig,clip=true]{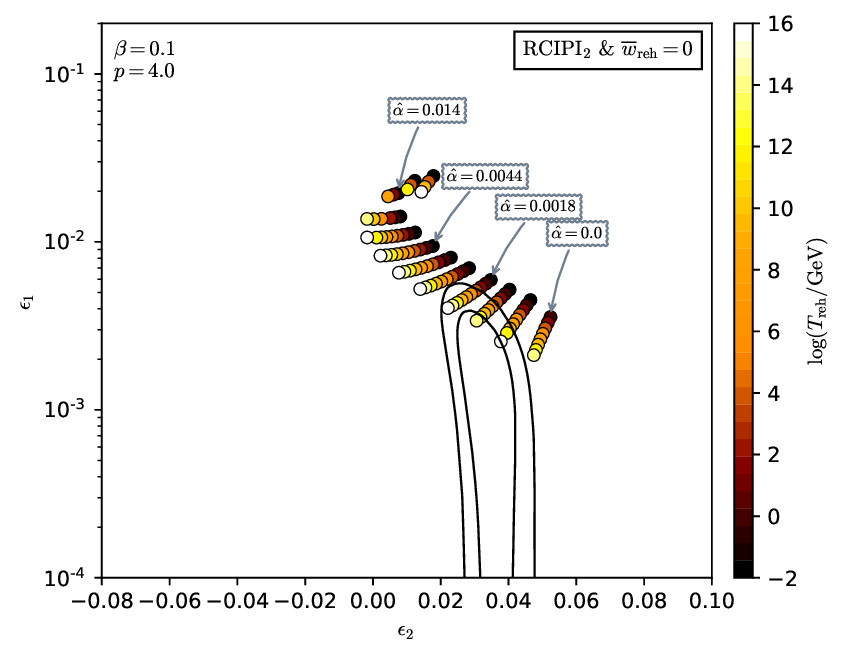}
\caption{Reheating consistent slow-roll predictions for Radiatively
  Corrected Inflection Point Inflation 2 for $p=4$ and
  $\beta=0.1$. Predictions are represented in the plane $(\nS,r)$
  (top panel) and in the plane $(\epsilon_1,\epsilon_2)$ (bottom
  panel) for various values of $\alphahat \equiv
  (\alpha + \alphazero)/(2\alphazero)$. This one is
  varied within its maximal allowed range $]0,1[$ for triggering the
  RCIPI2 regime. The solid contours are the one and two-sigma {\data}
  confidence intervals (marginalized over second order slow-roll).}
\label{fig:CMBRCIPI2_9}
\end{center}
\end{figure}

\begin{figure}[H]
\begin{center}
\includegraphics[width=\wappfig,clip=true]{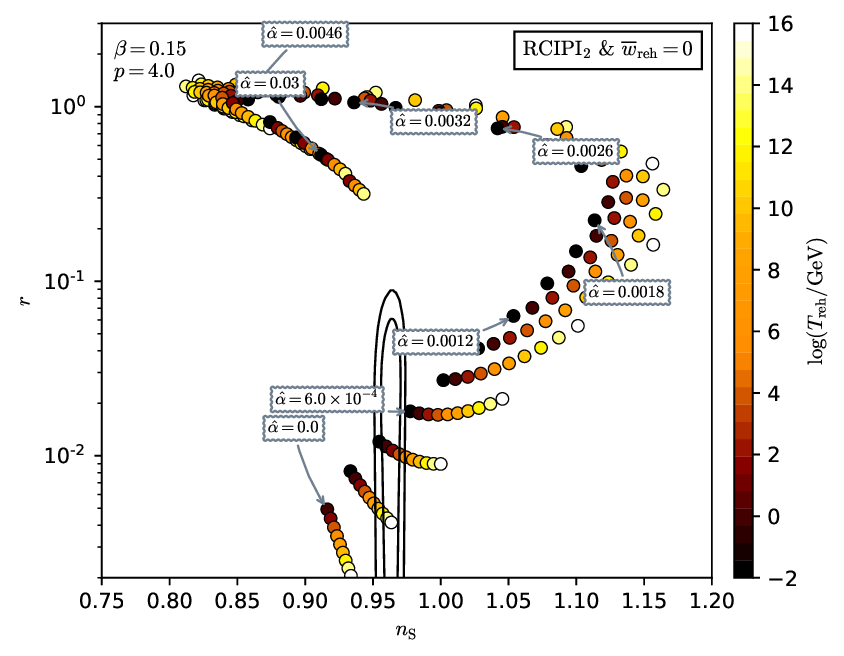}
\includegraphics[width=\wappfig,clip=true]{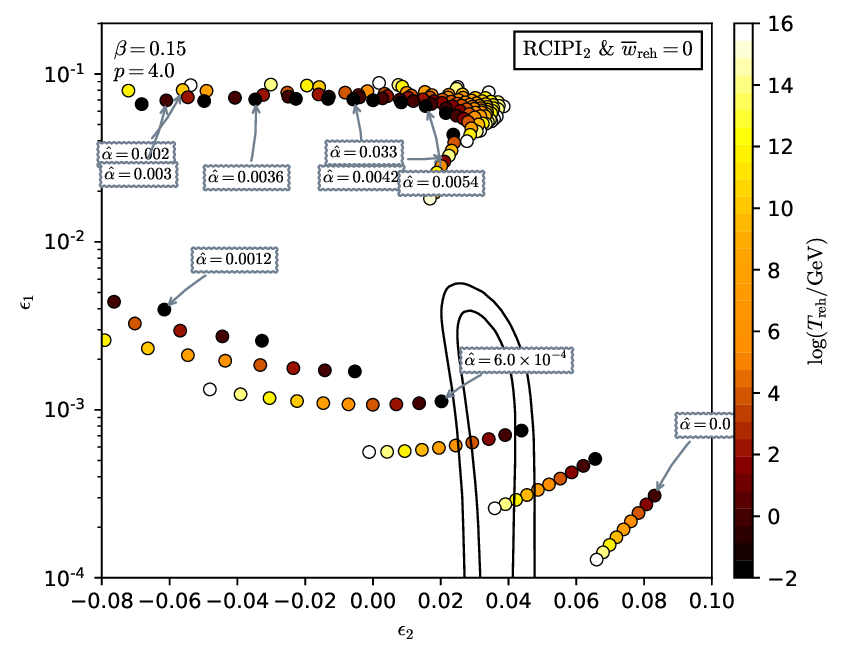}
\caption{Reheating consistent slow-roll predictions for Radiatively
  Corrected Inflection Point Inflation 2 for $p=4$ and
  $\beta=0.15$. Predictions are represented in the plane $(\nS,r)$
  (top panel) and in the plane $(\epsilon_1,\epsilon_2)$ (bottom
  panel) for various values of $\alphahat \equiv
  (\alpha + \alphazero)/(2\alphazero)$. This one is
  varied within its maximal allowed range $]0,1[$ for triggering the
  RCIPI2 regime. The solid contours are the one and two-sigma {\data}
  confidence intervals (marginalized over second order slow-roll).}
\label{fig:CMBRCIPI2_10}
\end{center}
\end{figure}

\acknowledgments We would like to thank the ten referees who have
accepted to review this work over the past decade. Their insightful
comments and careful checking of some calculations have been much
appreciated. This work is partially supported by the ESA Belgian
Federal PRODEX Grant No.~4000103071 and the Wallonia-Brussels
Federation grant ARC No.~11/15-040. The ``Opiparous Edition'' has
benefited from support by the ESA Belgian Federal PRODEX Grant
No.~4000143201, the ``Fonds de la Recherche Scientifique - FNRS''
under Grant $\mathrm{N^{\circ}T}.0198.19$ as well as by the
Wallonia-Brussels Federation Grant ARC
$\mathrm{N^{\circ}}19/24-103$. We would like to thank
\texttt{EvaluatorIAP} for his everyday and decade-long encouragements.

\bibliographystyle{JHEP}
\bibliography{aspic}

\end{document}